# 50 Years of Quantum Chromodynamics

**Franz Gross[a,1,2], Eberhard Klempt[b,3],**

**Stanley J. Brodsky[c,4], Andrzej J. Buras[c,5], Volker D. Burkert[c,1], Gudrun Heinrich[c,6], Karl Jakobs[c,7], Curtis A. Meyer[c,8], Kostas Orginos[c,1,2], Michael Strickland[c,9], Johanna Stachel[c,10], Giulia Zanderighi[c,11,12],**

**Nora Brambilla[5,12,13], Peter Braun-Munzinger[10,14], Daniel Britzger[11], Simon Capstick[15], Tom Cohen[16], Volker Crede[15], Martha Constantinou[17], Christine Davies[18], Luigi Del Debbio[19], Achim Denig[20], Carleton DeTar[21], Alexandre Deur[1], Yuri Dokshitzer[22,23], Hans Günter Dosch[10] Jozef Dudek[1,2], Monica Dunford[24], Evgeny Epelbaum[25], Miguel A. Escobedo[26], Harald Fritzsch[d,27], Kenji Fukushima[28], Paolo Gambino[11,29], Dag Gillberg[30,31], Steven Gottlieb[32], Per Grafstrom[33], Massimiliano Grazzini[34], Boris Grube[1], Alexey Guskov[35], Toru Iijima[36], Xiangdong Ji[16], Frithjof Karsch[37], Stefan Kluth[11], John B. Kogut[38,39], Frank Krauss[40], Shunzo Kumano[41,42], Derek Leinweber[43], Heinrich Leutwyler[44], Hai-Bo Li[45], Yang Li[46], Bogdan Malaescu[47], Chiara Mariotti[48], Pieter Maris[49], Simone Marzani[50], Wally Melnitchouk[1], Johan Messchendorp[51], Harvey Meyer[20], Ryan Edward Mitchell[52], Chandan Mondal[53], Frank Nerling[51,54,55], Sebastian Neubert[3], Marco Pappagallo[56], Saori Pastore[57], José R. Peláez[58], Andrew Puckett[59], Jianwei Qiu[1,2], Klaus Rabbertz[60], Alberto Ramos[61], Patrizia Rossi[1,62], Anar Rustamov[51,63], Andreas Schäfer[64], Stefan Scherer[65], Matthias Schindler[66], Steven Schramm[67], Mikhail Shifman[68], Edward Shuryak[69], Torbjörn Sjöstrand[70], George Sterman[71], Iain W. Stewart[72], Joachim Stroth[51,54,55], Eric Swanson[73], Guy F. de Téramond[74], Ulrike Thoma[3], Antonio Vairo[75], Danny van Dyk[40], James Vary[49], Javier Virto[76,77], Marcel Vos[78], Christian Weiss[1], Markus Wobisch[79] Sau Lan Wu[80], Christopher Young[81], Feng Yuan[82], Xingbo Zhao[53], Xiaorong Zhou[46]**

[a]e-mail: flgros@wm.edu
[b]e-mail: klempt@hiskp.uni-bonn.de
[c]Convenor
[d]Deceased

[1]Thomas Jefferson National Accelerator Facility, 12000 Jefferson Avenue, Newport News, VA 23606, USA
[2] Department of Physics, William and Mary, Williamsburg, VA 23187, USA
[3] Helmholtz-Institut für Strahlen- und Kernphysik, Universität Bonn, Nußallee 14-16, D 53115 Bonn, Germany
[4] Theoretical Physics, SLAC National Accelerator Laboratory, 2575 Sand Hill Road, Menlo Park, CA, 94025, USA
[5] Institute for Advanced Study, Technische Universität München, Lichtenbergstraße 2a, D-85748 Garching b. München, Germany
[6] Institut für Theoretische Physik, Karlsruher Institut für Technologie (KIT), D-76128 Karlsruhe, Germany
[7] Physikalisches Institut, Universität Freiburg, 79104 Freiburg, Germany
[8] Carnegie Mellon University, Pittsburgh, PA 15213, USA
[9] Department of Physics, Kent State University, 800 E Summit St, Kent, OH 44240, USA
[10] Physikalisches Institut, Universität Heidelberg, 69120 Heidelberg, Germany
[11] Max-Planck-Institut für Physik, Föhringer Ring 6, 80805 München, Germany
[12] Physik Department, Technische Universität München, James-Franck-Straße 1, D-85748 Garching b. München, Germany
[13] Munich Data Science Institute, Technische Universität München, Walther-von-Dyck-Straße-10, D-85748 Garching b. München, Germany
[14] Extreme Matter Institute EMMI, GSI, 64291 Darmstadt, Germany
[15] Department of Physics, Florida State University, Tallahassee, FL 32306, USA
[16] Department of Physics, University of Maryland, College Park, MD 20742, USA
[17] Physics Department, Temple University, 1925 N. 12th Street, Philadelphia, PA 19122, USA
[18] School of Physics and Astronomy, University of Glasgow, Glasgow, G12 8QQ, UK

[19] Higgs Centre for Theoretical Physics, School of Physics and Astronomy, The University of Edinburgh, Edinburgh EH9 3FD, UK
[20] PRISMA$^+$ Cluster of Excellence & Institut für Kernphysik and Helmholtz Institute Mainz, Johannes Gutenberg University Mainz, Mainz, D-55128, Germany
[21] Department of Physics and Astronomy, University of Utah, Salt Lake City, Utah, 84112, USA
[22] Riga Technical University Center of High Energy Physics and Accelerator Technologies, Riga, Latvia
[23] Petersburg Nuclear Physics Institute, Gatchina, Russia
[24] Kirchhoff-Institut für Physik, Ruprecht-Karls-Universität Heidelberg, Heidelberg, Germany
[25] Institut für Theoretische Physik II, Ruhr-Universität Bochum, 44780 Bochum, Germany
[26] Instituto Galego de Física de Altas Enería (IGFAE), Universidade de Santiago de Compostela, E-15782, Galicia, Spain
[27] Department für Physik der Universität München, Theresienstraße 37, D-80333 München, Germany
[28] School of Science, University of Tokyo, Bunkyo City, Tokyo 113-8654, Japan
[29] Dipartimento di Fisica, Università di Torino & INFN, Sezione di Torino, Via Pietro Giuria 1, I-10125 Turin, Italy
[30] Department of Physics, Carlton University, 1125 Colonel By Drive, Ottawa K1S 5B6 Ontario, Canada
[31] Department of Physics, Lund University, Lund, Sweden
[32] Department of Physics, Indiana University, Bloomington, IN 47405, USA
[33] CERN, Geneva, Switzerland and Università di Bologna, Dipartimento di Fisica, 40126 Bologna, Italy
[34] Department of Physics, University of Zurich, Winterthurerstrasse 190, 8057 Zurich, Switzerland
[35] Joint Institute for Nuclear Research, 141980 Dubna, Moscow region, Russia
[36] Kobayashi-Maskawa Institute (KMI)/Graduate School of Science Nagoya University, Furocho, Chikusa Ward, Nagoya, Aichi 464-8601, Japan
[37] Physics Department, Bielefeld University, D-33615 Bielefeld, Germany
[38] Department of Energy, Division of High Energy Physics, Washington, DC 20585, USA
[39] Department of Physics –TQHN, University of Maryland, 82 Regents Drive, College Park, MD 20742, USA
[40] Institute for Particle Physics Phenomenology, Physics Department, Durham University, Durham DH1 3LE, UK
[41] Department of Mathematics, Physics, and Computer Science, Faculty of Science, Japan Women's University, 2-8-1 Mejirodai, Bunkyo-ku, Tokyo 112-8681, Japan
[42] Theory Center, Institute of Particle and Nuclear Studies, High Energy Accelerator Research Organization (KEK), 1-1 Oho, Tsukuba, Ibaraki, 305-0801, Japan
[43] Centre for the Subatomic Structure of Matter (CSSM), Department of Physics, The University of Adelaide, SA 5005, Australia
[44] Albert Einstein Center for Fundamental Physics, Institute for Theoretical Physics, University of Bern, Sidlerstrasse 5, 3012 Bern, Switzerland
[45] Institute of High Energy Physics, Beijing 100049, PR of China, and University of Chinese Academy of Sciences, Beijing 100049, PR of China
[46] University of Science and Technology of China, No.96, JinZhai Road, Baohe District, Hefei, Anhui, 230026, PR of China
[47] LPNHE, Sorbonne Université, Université de Paris Cité, CNRS/IN2P3, Paris, France, 75252
[48] INFN, Sezione di Torino, Via Pietro Giuria 1, I-10125 Turin, Italy
[49] Deptarment of Physics and Astronomy, Iowa State University, Ames, IA 50011, USA
[50] Dipartimento di Física, Università di Genova and INFN, Sezione di Genova,Via Dodecaneso 33, 16146, Italy,
[51]GSI Helmholtzzentrum für Schwerionenforschung GmbH, Planckstraße 1, 64291 Darmstadt, Germany,
[52] Department of Physics, Indiana University Bloomington, 107 S. Indiana Avenue, Bloomington, IN 47405, USA
[53]Institute of Modern Physics, Chinese Academy of Sciences, Lanzhou, Gansu 730000, PR of China
[54]Helmholtz Forschungsakademie Hessen für FAIR (HFHF), GSI Helmholtzzentrum für Schwerionenforschung, Campus FrankfurtFrankfurt, Germany
[55]Goethe Universität, Institut für Kernphysik, Max-von-Laue-Str. 1, 60438 Frankfurt, Germany
[56]Dipartimento Interateneo di Fisica, Università di Bari and INFN, Sezione di Bari, Via Amendola 173, 70125 Bari, Italy
[57] Department of Physics and McDonnell Center for the Space Sciences, Washington University in Saint Louis, Saint Louis, MO 63130, USA
[58] Departamento de Física Teórica and IPARCOS. Universidad Complutense 28040 Madrid, Spain
[59] University of Connecticut, Storrs, CT, 06269, USA
[60] CERN, Geneva, Switzerland and ETP, KIT, Postfach 6980, D-76128 Karlsruhe, Germany
[61] IFIC (UVEG/CSIC) Valencia, C. del Catedrático José Beltrán 2, 46980 Paterna, Spain
[62] INFN, Laboratori Nazionali di Frascati, 00044 Frascati, Italy
[63] National Nuclear Research Center, AZ1000 Baku, Azerbaijan
[64] Institut für Theoretische Physik, Universität Regensburg, D-93040 Regensburg, Germany
[65]Institut für Kernphysik, Johannes Gutenberg-Universität Mainz, D-55099 Mainz, Germany
[66] Department of Physics and Astronomy, University of South Carolina, Columbia, SC 29208, USA
[67]Département de Physique Nucléaire et Corpusculaire, Université de Genève, 1205 Genève, Switzerland
[68] School of Physics and Astronomy, University of Minnesota, Minneapolis MN 55455, USA
[69] Department of Physics and Astronomy, Stony Brook University, Stony Brook NY 11794, USA
[70] Department of Astronomy and Theoretical Physics, Lund University, Box 43, SE-221 00 Lund, Sweden
[71] C. N. Yang Institute for Theoretical Physics and Department of Physics and Astronomy Stony Brook University, Stony Brook, New York 11794, USA

[72] Center for Theoretical Physics, Massachusetts Institute of Technology, Cambridge, MA 02139, USA
[73] Department of Physics and Astronomy, University of Pittsburgh, Pittsburgh PA 15260, USA
[74] Laboratorio de Física Teórica y Computacional, Universidad de Costa Rica, 11501 San José, Costa Rica
[75] Physik Department, Technische Universität München, James-Franck-Straße 1, D-85748 Garching b. München, Germany
[76] Departament de Física Quántica i Astrofísica, Universitat de Barcelona, Martí i Franqués 1, 08028 Barcelona, Catalunya
[77] Institut de Ciències del Cosmos (ICCUB), Universitat de Barcelona, Martí i Franqués 1, 08028 Barcelona, Catalunya
[78] IFIC (UVEG/CSIC) Valencia, 46980 Paterna, Spain
[79] Departmant of Physics, Louisiana Tech University, 201 Mayfield Ave, Ruston, LA 71272, USA
[80] Department of Physics, University of Wisconsin, Madison, WI 53706, USA
[81] Institute of Physics, Albert Ludwig University of Freiburg, Freiburg im Breisgau, Germany
[82] Nuclear Science Division, Lawrence Berkeley National Laboratory, 1 Cyclotron Rd, Berkeley, CA 94720, USA

**Abstract** Quantum Chromodynamics, the theory of quarks and gluons, whose interactions can be described by a local SU(3) gauge symmetry with charges called "color quantum numbers", is reviewed; the goal of this review is to provide advanced Ph.D. students a comprehensive handbook, helpful for their research. When QCD was "discovered" 50 years ago, the idea that quarks could exist, but not be observed, left most physicists unconvinced. Then, with the discovery of charmonium in 1974 and the explanation of its excited states using the Cornell potential, consisting of the sum of a Coulomb-like attraction and a long range linear confining force, the theory was suddenly widely accepted. This paradigm shift is now referred to as the *November revolution*. It had been anticipated by the observation of scaling in deep inelastic scattering, and was followed by the discovery of gluons in three-jet events.

The parameters of QCD include the running coupling constant, $\alpha_s(Q^2)$, that varies with the energy scale $Q^2$ characterising the interaction, and six quark masses. QCD cannot be solved analytically, at least not yet, and the large value of $\alpha_s$ at low momentum transfers limits perturbative calculations to the high-energy region where $Q^2 \gg \Lambda^2_{\rm QCD} \simeq (250\ {\rm MeV})^2$. Lattice QCD (LQCD), numerical calculations on a discretized space-time lattice, is discussed in detail, the dynamics of the QCD vacuum is visualized, and the expected spectra of mesons and baryons are displayed. Progress in lattice calculations of the structure of nucleons and of quantities related to the phase diagram of dense and hot (or cold) hadronic matter are reviewed. Methods and examples of how to calculate hadronic corrections to weak matrix elements on a lattice are outlined.

The wide variety of analytical approximations currently in use, and the accuracy of these approximations, are reviewed. These methods range from the Bethe-Salpeter, Dyson-Schwinger coupled relativistic equations, which are formulated in both Minkowski or Euclidean spaces, to expansions of multi-quark states in a set of basis functions using light-front coordinates, to the AdS/QCD method that imbeds 4-dimensional QCD in a 5-dimensional deSitter space, allowing confinement and spontaneous chiral symmetry breaking to be described in a novel way. Models that assume the number of colors is very large, i.e. make use of the large $N_c$-limit, give unique insights. Many other techniques that are tailored to specific problems, such as perturbative expansions for high energy scattering or approximate calculations using the operator product expansion are discussed. The very powerful effective field theory techniques that are successful for low energy nuclear systems (chiral effective theory), or for non-relativistic systems involving heavy quarks, or the treatment of gluon exchanges between energetic, collinear partons encountered in jets, are discussed.

The spectroscopy of mesons and baryons has played an important historical role in the development of QCD. The famous X,Y,Z states – and the discovery of pentaquarks – have revolutionized hadron spectroscopy; their status and interpretation are reviewed as well as recent progress in the identification of glueballs and hybrids in light-meson spectroscopy. These exotic states add to the spectrum of expected $q\bar{q}$ mesons and $qqq$ baryons. The progress in understanding excitations of light and heavy baryons is discussed. The nucleon as the lightest baryon is discussed extensively, its form factors, its partonic structure and the status of the attempt to determine a three-dimensional picture of the parton distribution.

An experimental program to study the phase diagram of QCD at high temperature and density started with fixed target experiments in various laboratories in the second half of the 1980's, and then, in this century, with colliders. QCD thermodynamics at high temperature became accessible to LQCD, and numerical results on chiral and deconfinement transitions and properties of the deconfined and chirally restored form of strongly interacting matter, called the Quark-Gluon Plasma (QGP), have become very precise by now. These results can now be confronted with experimental data that are sensitive to the nature of the phase transition. There is clear evidence that the QGP phase is created. This phase of QCD matter can already be characterized by some properties that indicate, within a temperature range of a few times the pseudocritical temperature, the medium behaves like a near ideal liquid. Experimental observables are presented that demonstrate deconfinement. High and ultrahigh density QCD matter at moderate and low temperaures shows interesting features and new phases that are of astrophysical relevance. They are reviewed here and some of the astrophysical implications are discussed.

Perturbative QCD and methods to describe the different aspects of scattering processes are discussed. The primary parton-parton scattering in a collision is calculated in perturbative QCD with increasing complexity. The radiation of soft gluons can spoil the perturbative convergence, this can be cured by resummation techniques, which are also described here. Realistic descriptions of QCD scattering events need to model the cascade of quark and gluon splittings until hadron formation sets in, which is done by parton showers. The full event simulation can be performed with Monte Carlo event generators, which simulate the full chain from the hard interaction to the hadronic final states, including the modelling of non-perturbative components. The contribution of the LEP experiments (and of earlier collider experiments) to the study of jets is reviewed. Correlations between jets and the shape of jets had allowed the collaborations to determine the "color factors" – invariants of the SU(3) color group governing the strength of quark-gluon and gluon-gluon interactions. The calculated jet production rates (using perturbative QCD) are shown to agree precisely with data, for jet energies spanning more than five orders of magni-tude. The production of jets recoiling against a vector boson, $W^{\pm}$ or $Z$, is shown to be well understood. The discovery of the Higgs boson was certainly an important milestone in the development of high-energy physics. The couplings of the Higgs boson to massive vector bosons and fermions that have been measured so far support its interpretation as mass-generating boson as predicted by the Standard Model. The study of the Higgs boson recoiling against hadronic jets (without or with heavy flavors) or against vector bosons is also highlighted. Apart from the description of hard interactions taking place at high energies, the understanding of "soft QCD" is also very important. In this respect, Pomeron – and Odderon – exchange, soft and hard diffraction are discussed.

Weak decays of quarks and leptons, the quark mixing matrix and the anomalous magnetic moment of the muon are processes which are governed by weak interactions. However, corrections by strong interactions are important, and these are reviewed. As the measured values are incompatible with (most of) the predictions, the question arises: are these discrepancies first hints for New Physics beyond the Standard Model?

This volume concludes with a description of future facilities or important upgrades of existing facilities which improve their luminosity by orders of magnitude. The best is yet to come!

# Contents

# Preface

Quantum Chromodynamics or QCD was developed and defined over a brief period from 1972-73. One of us (EK) wrote an article early in 2021 on the scalar glueball and searched the literature to find where glueballs were first mentioned. This was at the 16th International Conference on High-Energy Physics (ICHEP 72). In the winter of 2021/2022 he thought it was time to prepare a volume dedicated to 50 Years of QCD. He got approval from the EJPC, and asked FG to join the effort. Here is the result.

It's been quite an adventure to guide and prepare this volume. From the start it was to be published as a single article, organized and edited by the two coeditors, with integrated contributions from invited scientists familiar with all aspects of the subject. Our initial outline included only eight sections, but as we got advice from our conveners and early contributors, the number of sections grew to the 14 you see here, and in some cases the number of subsections in each section also grew. The subject is both beautiful and vast, and keeping this volume "limited" in length was a real challenge.

Our goal was to prepare a volume for young Ph.Ds and postdocs that could serve as a readable resource and introduction to specialties outside of their own field of research – a shortcut to acquiring the broad familiarity that usually takes time to acquire. We also invited our contributors to reflect on how they developed their ideas/insights, usually discouraged in scientific articles. We believe that what has resulted is truly unique.

The volume begins with the personal reflections of two scientists who were contributors to the foundations of QCD (Sec. 1), and follows with three early developments that quickly showed that QCD as on the right track (Sec. 2). Prominent among these was the "November revolution," where the discovery and explanation of the states of charmonium lead quickly to the Cornell potential and an early description of why quarks could not be seen, convincing may doubters that quarks were real.

After establishing that the QCD fine structure constant, $\alpha_s$, is too large at hadronic scales for perturbation theory to work (Sec. 3), we describe in some detail Lattice QCD (Sec. 4), believed now to be the only method that can give *exact* predictions for QCD (with numerical errors, of course, which are decreasing rapidly as the computations and computers improve). Unfortunately, Lattice QCD does not give much of an intuitive picture of how the physics works, so approximate analytic methods are needed (and will probably always be needed) and these are summarized in Secs. 5 and 6, including effective field theories, a powerful tool with many applications. Perhaps some day we will have exact analytic solutions, but not today.

From there our account turns to experimental manifestations of QCD (with theoretical support), starting with the exploration of the QCD phase diagram in heavy ion collisions and in dense matter (Sec. 7), followed by the study of mesons (Sec. 8) and baryons (Sec. 9) that reveal the existence of "exotic" states like glueballs, hybrids, hadronic molecules, and tetra- and pentaquarks. A special focus is given to the nucleon and its structure (Sec. 10). Then, collisions at high energies are discussed, from the hard scattering of two partons followed by their hadronization (Sec. 11); the production and identification of jets of particles culminated in the discovery of the Higgs boson and measurements of its properties (Sec. 12); weak decays, precision analyses of the quark mixing matrix, and the anomalous magnetic moment of muons that show the first hints of New Physics beyond the Standard Model (Sec. 13). The volume concludes with a brief account of experimental projects under construction or already funded (Sec. 14). We do not discuss the many exciting theoretical or experimental ideas that are currently in the drawing board, or as theorists sometimes say, on the "second sheet" (when they are joking about wonderful ideas still in an imaginative state). These we save for the next volume!

It has been a great experience for us to work on this volume; we hope you will find some pleasure in skimming through it.

# 1 Theoretical Foundations

*Conveners:*
**Franz Gross and Eberhard Klempt**

This section contains two personal accounts of the early development or "discovery" of QCD.

Leutwyler's contribution starts with a broad picture of the chaotic state of "theories" of the strong interactions in the 1960's and carries us through to the present day. He describes how many thought field theory could not work for the description of "nuclear forces." They thought the use of dispersion relations and unitarity would provide a better approach, but now we know that these are only useful tools. His discussion of how the exact and approximate symmetries of QCD lead to an understanding of the mass scales of the quarks shows how much the development of QCD and the standard model have brought order out of chaos, and have led to a deep understanding of the physics.

The second contribution by Fritzsch gives a more focused and personal account of how some issues that had to be surmounted before QCD became the accepted theory of the strong forces. He describes several arguments that them to the necessity for three colors of quarks (and the SU(3) color symmetry). He reminds us that QCD and the existence of quarks did not become widely accepted until the discovery of the $J/\psi$, among the topics discussed in the following Sec. 2.

Both of these accounts of the history and the physics are exciting to read, and a broad introduction to this volume. We hope you will enjoy them as much as we have.

## 1.1 The strong interactions[1]

**Heinrich Leutwyler**

### 1.1.1 Beginnings

The discovery of the neutron in 1932 [2] may be viewed as the birth of the strong interaction: it indicated that the nuclei consist of protons and neutrons and hence the presence of a force that holds them together, strong enough to counteract the electromagnetic repulsion. Immediately thereafter, Heisenberg introduced the notion of isospin as a symmetry of the strong interaction, in order to explain why proton and neutron nearly have the same mass [3]. In 1935, Yukawa pointed out that the nuclear force could be generated by the exchange of a hypothetical spinless particle, provided its mass is intermediate between the masses of proton and electron – a *meson* [4]. Today, we know that such a particle indeed exists: Yukawa predicted the pion. Stueckelberg pursued similar ideas, but was mainly thinking about particles of spin 1, in analogy with the particle that mediates the electromagnetic interaction [5].

In the thirties and fourties of the last century, the understanding of the force between two nucleons made considerable progress, in the framework of nonrelativistic potential models. These are much more flexible than quantum field theories. Suitable potentials that are attractive at large distances but repulsive at short distances do yield a decent understanding of nuclear structure: Paris potential, Bonn potential, shell model of the nucleus. In this framework, nuclear reactions, in particular the processes responsible for the luminosity of the sun, stellar structure, $\alpha$-decay and related matters were well understood more than sixty years ago.

These phenomena concern interactions among nucleons with small relative velocities. Experimentally, it had become possible to explore relativistic collisions, but a description in terms of nonrelativistic potentials cannot cover these. In the period between 1935 and 1965, many attempts at formulating a theory of the strong interaction based on elementary fields for baryons and mesons were made. In particular, uncountable PhD theses were written, based on local interactions of the Yukawa type, using perturbation theory to analyze them. The coupling constants invariably turned out to be numerically large, indicating that the neglect of the higher order contributions was not justified. Absolutely nothing worked even half way.

Although there was considerable progress in understanding the general principles of quantum field theory (Lorentz invariance, unitarity, crossing symmetry, causality, analyticity, dispersion relations, CPT theorem, spin and statistics) as well as in renormalization theory, faith in quantum field theory was in decline, even concerning QED (Landau pole). To many, the renormalization procedure – needed to arrive at physically meaningful results – looked suspicious, and it appeared doubtful that the strong interaction could at all be described by means of a local quantum field theory. Some suggested that this framework should be replaced by S-matrix theory – heated debates concerning this suggestion took place at the time [6]. Regge poles were considered as a promising alternative to the quantum fields (the Veneziano model is born in 1968 [7]). Sixty years ago, when I completed my studies, the quantum field theory of the strong interaction consisted of a collection of beliefs, prejudices and assumptions. Quite a few of these turned out to be wrong.

### 1.1.2 Flavor symmetries

Symmetries that extend isospin to a larger Lie group provided the first hints towards an understanding of the structure underneath the strong interaction phenomena. The introduction of the strangeness quantum number and the Gell-Mann-Nishijima formula [8, 9] was a significant step in this direction. Goldberger and Treiman [10] then showed that the *axial vector current* plays an important role, not only in the weak interaction (the pion-to-vacuum matrix element of this current – the pion decay constant $F_\pi$ – determines the rate of the weak decay $\pi \to \mu\nu$) but also in the context of the strong interaction: the nucleon matrix element of the axial vector current, $g_A$, determines the strength of the interaction between pions and nucleons:

$$g_{\pi N} = g_A M_N / F_\pi \,.$$

At low energies, the main characteristic of the strong interaction is that the energy gap is small: the lightest state occurring in the eigenvalue spectrum of the

[1] The present section is an extended version of my lecture notes *On the history of the strong interaction* [1]

Hamiltonian is the pion, with[2] $M_\pi \simeq 135$ MeV, small compared to the mass of the proton, $M_p \simeq 938$ MeV. In 1960, Nambu found out why that is so: it has to do with a hidden, approximate, continuous symmetry [11]. Since some of its generators carry negative parity, it is referred to as a *chiral symmetry*. For this symmetry to be consistent with observation, it is essential that an analog of spontaneous magnetization occurs in particle physics: for dynamical reasons, the state of lowest energy – the vacuum – is not symmetric under chiral transformations. Consequently, the symmetry cannot be seen in the spectrum of the theory: it is *hidden* or *spontaneously broken*. Nambu realized that the spontaneous breakdown of a continuous symmetry entails massless particles analogous to the spin waves of a magnet and concluded that the pions must play this role. If the strong interaction was strictly invariant under chiral symmetry, there would be no energy gap at all – the pions would be massless.[3] Conversely, since the pions are not massless, chiral symmetry cannot be exact – unlike isospin, which at that time was taken to be an exact symmetry of the strong interaction. The spectrum does have an energy gap because chiral symmetry is not exact: the pions are not massless, only light. In fact, they represent the lightest strongly interacting particles that can be exchanged between two nucleons. This is why, at large distances, the potential between two nucleons is correctly described by the Yukawa formula.

The discovery of the *Eightfold Way* by Gell-Mann and Ne'eman paved the way to an understanding of the mass pattern of the baryons and mesons [13, 14]. Like chiral symmetry, the group SU(3) that underlies the Eightfold Way represents an approximate symmetry: the spectrum of the mesons and baryons does not consist of degenerate multiplets of this group. The splitting between the energy levels, however, does exhibit a pattern that can be understood in terms of the assumption that the part of the Hamiltonian that breaks the symmetry transforms in a simple way. This led to the Gell-Mann-Okubo formula [14, 15] and to a prediction for the mass of the $\Omega^-$, a member of the baryon decuplet which was still missing, but was soon confirmed experimentally, at the predicted place [16].

### 1.1.3 Quark Model

In 1964, Gell-Mann [17] and Zweig [18] pointed out that the observed pattern of baryons can qualitatively be understood on the basis of the assumption that these particles are bound states built with three constituents, while the spectrum of the mesons indicates that they contain only two of these. Zweig called the constituents "aces". Gell-Mann coined the term "quarks", which is now commonly accepted. The Quark Model gradually evolved into a very simple and successful semi-quantitative framework, but gave rise to a fundamental puzzle: why do the constituents not show up in experiment? For this reason, the existence of the quarks was considered doubtful: "Such particles [quarks] presumably are not real but we may use them in our field theory anyway ..." [19]. Quarks were treated like the veal used to prepare a pheasant in the royal french cuisine: the pheasant was baked between two slices of veal, which were then discarded (or left for the less royal members of the court). Conceptually, this was a shaky cuisine.

If the flavor symmetries are important, why are they not exact? Gell-Mann found a beautiful explanation: *current algebra* [14, 19]. The charges form an exact algebra even if they do not commute with the Hamiltonian and the framework can be extended to the corresponding currents, irrespective of whether or not they are conserved. Adler and Weisberger showed that current algebra can be tested with the sum rule that follows from the nucleon matrix element of the commutator of two axial vector charges [20, 21]. Weinberg then demonstrated that even the strength of the interaction among the pions can be understood on the basis of current algebra: the $\pi\pi$ scattering lengths can be predicted in terms of the pion decay constant [22].

### 1.1.4 Behavior at short distances

Bjorken had pointed out that if the nucleons contain point-like constituents, then the *ep* cross section should obey scaling laws in the deep inelastic region [23]. Indeed, the scattering experiments carried out by the MIT-SLAC collaboration in 1968/69 did show experimental evidence for such constituents [24]. Feynman called these *partons*, leaving it open whether they were the quarks or something else.

The operator product expansion turned out to be a very useful tool for the short distance analysis of the theory – the title of the paper where it was introduced [25], "Nonlagrangian models of current algebra," reflects the general skepticism towards Lagrangian quantum field theory that I mentioned in Section 1.1.1.

### 1.1.5 Color

The Quark Model was difficult to reconcile with the spin-statistics theorem which implies that particles of spin $\frac{1}{2}$ must obey Fermi statistics. Greenberg proposed that the quarks obey neither Fermi-statistics nor Bose-statistics, but “para-statistics of order three” [26]. The proposal amounts to the introduction of a new internal quantum number. Indeed, Bogolyubov, Struminsky and Tavkhelidze [27], Han and Nambu [28] and Miyamoto [29] independently pointed out that some of the problems encountered in the quark model disappear if the $u$, $d$ and $s$ quarks occur in 3 states. Gell-Mann coined the term “color” for the new quantum number.

---

[2] I am using natural units where $\hbar = c = 1$.

[3] A precise formulation of this statement, known as the *Goldstone theorem*, was given later [12].

One of the possibilities considered for the interaction that binds the quarks together was an abelian gauge field analogous to the e.m. field, but this gave rise to problems, because the field would then interfere with the other degrees of freedom. Fritzsch and Gell-Mann pointed out that if the gluons carry color, then the empirical observation that quarks appear to be confined might also apply to them: the spectrum of the theory might exclusively contain color neutral states [30].

In his lectures at the Schladming Winter School in 1972 [31], Gell-Mann thoroughly discussed the role of the quarks and gluons: theorists had to navigate between Scylla and Charybdis, trying to abstract neither too much nor too little from models built with these objects. The basic tool at that time was *current algebra on the light cone.* He invited me to visit Caltech. I did that during three months in the spring break of 1973 and spent an extremely interesting period there. The personal recollections of Harald Fritzsch (see Section 1.2) describe the developments that finally led to *Quantum Chromodynamics.*

As it was known already that the electromagnetic and weak interactions are mediated by gauge fields, the idea that color might be a local symmetry as well does not appear as far fetched. The main problem at the time was that for a gauge field theory to describe the hadrons and their interaction, it had to be fundamentally different from the quantum field theories encountered in nature so far: all of these, including the electroweak theory, have the spectrum indicated by the degrees of freedom occurring in the Lagrangian: photons, leptons, intermediate bosons, ... The proposal can only make sense if this need not be so, that is if the spectrum of physical states in a quantum field theory can differ from the spectrum of the fields needed to formulate it: gluons and quarks in the Lagrangian, hadrons in the spectrum. This looked like wishful thinking. How come that color is confined while electric charge is free?

### 1.1.6 Electromagnetic interaction

The final form of the laws obeyed by the electromagnetic field was found by Maxwell, around 1860 – these laws survived relativity and quantum theory, unharmed. Fock pointed out that the Schrödinger equation for electrons in an electromagnetic field,

$$\frac{1}{i}\frac{\partial\psi}{\partial t}-\frac{1}{2m_e^2}\,(\vec{\nabla}+i\,e\vec{A})^2\psi-e\,\varphi\,\psi=0\,, \qquad (1.1.1)$$

is invariant under a group of local transformations:

$$\vec{A}'(x)=\vec{A}(x)+\vec{\nabla}\alpha(x),\qquad \varphi'(x)=\varphi(x)-\frac{\partial\alpha(x)}{\partial t}$$
$$\psi(x)'=e^{-ie\alpha(x)}\,\psi(x)\,, \qquad (1.1.2)$$

in the sense that the fields $\vec{A}', \varphi', \psi'$ describe the same physical situation as $\vec{A}, \varphi, \psi$ [32]. Weyl termed these *gauge transformations* (with gauge group U(1) in this case). In fact, the electromagnetic interaction is fully characterized by symmetry with respect to this group: gauge invariance is the crucial property of this interaction.

I illustrate the statement with the core of Quantum Electrodynamics: photons and electrons. Gauge invariance allows only two free parameters in the Lagrangian of this system: $e, m_e$. Moreover, only one of these is dimensionless: $e^2/4\pi = 1/137.035\,999\,084\,(21)$. U(1) symmetry and renormalizability fully determine the properties of the e.m. interaction, except for this number, which so far still remains unexplained.

### 1.1.7 Nonabelian gauge fields

Kaluza [33] and Klein [34] had shown that a 5-dimensional Riemann space with a metric that is independent of the fifth coordinate is equivalent to a 4-dimensional world with *gravity*, a *gauge field* and a *scalar field.* In this framework, gauge transformations amount to a shift in the fifth direction: $x^{5'} = x^5+\alpha(\vec{x},t)$. In geometric terms, a metric space of this type is characterized by a group of isometries: the geometry remains the same along certain directions, indicated by Killing vectors. In the case of the 5-dimensional spaces considered by Kaluza and Klein, the isometry group is the abelian group U(1). The fifth dimension can be compactified to a circle – U(1) then generates motions on this circle. A particularly attractive feature of this theory is that it can explain the quantization of the electric charge: fields living on such a manifold necessarily carry integer multiples of a basic charge unit.

Pauli noticed that the Kaluza-Klein scenario admits a natural generalization to higher dimensions, where larger isometry groups find place. Riemann spaces of dimension $> 5$ admit nonabelian isometry groups that reduce the system to a 4-dimensional one with *gravity*, *nonabelian gauge fields* and several *scalar fields.* Pauli was motivated by the isospin symmetry of the meson-nucleon interaction and focused attention on a Riemann space of dimension 6, with isometry group SU(2).

Pauli did not publish the idea that the strong interaction might arise in this way, because he was convinced that the quanta of a gauge field are massless: gauge invariance does not allow one to put a mass term into the Lagrangian. He concluded that the forces mediated by gauge fields are necessarily of long range and can therefore not mediate the strong interaction, which is known to be of short range. More details concerning Pauli's thoughts can be found in [35]. The paper of Yang and Mills appeared in 1954 [36]. Ronald Shaw, a student of Salam, independently formulated nonabelian gauge field theory in his PhD thesis [37]. Ten years later, Higgs [38], Brout and Englert [39] and Guralnik, Hagen and Kibble [40] showed that Pauli's objection is not valid in general: in the presence of scalar fields, gauge fields can pick up mass, so that forces mediated by gauge fields can be of short range. The work of Glashow [41], Weinberg [42] and Salam [43] then demonstrated that nonabelian gauge fields are relevant for physics: the framework discovered by Higgs et al. does accommodate a realistic description of the e.m. and weak interactions.

### 1.1.8 Asymptotic freedom

Already in 1965, Vanyashin and Terentyev [44] found that the renormalization of the electric charge of a vector field is of opposite sign to the one of the electron. In the language of SU(2) gauge field theory, their result implies that the $\beta$-function is negative at one loop.

The first correct calculation of the $\beta$-function of a nonabelian gauge field theory was carried out by Khriplovich, for the case of SU(2), relevant for the electroweak interaction [45]. He found that $\beta$ is negative and concluded that the interaction becomes weak at short distance. In his PhD thesis, 't Hooft performed the calculation of the $\beta$-function for an arbitrary gauge group, including the interaction with fermions and Higgs scalars [46, 47]. He demonstrated that the theory is renormalizable and confirmed that, unless there are too many fermions or scalars, the $\beta$-function is negative at small coupling.

In 1973, Gross and Wilczek [48] and Politzer [49] discussed the consequences of a negative $\beta$-function and suggested that this might explain Bjorken scaling, which had been observed at SLAC in 1969. They pointed out that QCD predicts specific modifications of the scaling laws. In the meantime, there is strong experimental evidence for these.

### 1.1.9 Arguments in favor of QCD

The reasons for proposing QCD as a theory of the strong interaction are discussed in [50]. The idea that the observed spectrum of particles can fully be understood on the basis of a theory built with quarks and gluons still looked rather questionable and was accordingly formulated in cautious terms. In the abstract, for instance, we pointed out that "...there are several advantages in abstracting properties of hadrons and their currents from a Yang-Mills gauge model based on colored quarks and color octet gluons." Before the paper was completed, the papers by Gross, Wilczek and Politzer quoted above circulated as preprints - they are quoted and asymptotic freedom is given as argument #4 in favor of QCD. Also, important open questions were pointed out, in particular, the U(1) problem.

Many considered QCD a wild speculation. On the other hand, several papers concerning gauge field theories that include the strong interaction appeared around the same time, for instance [51, 52].

### 1.1.10 November revolution

The discovery of the $J/\psi$ was announced simultaneously at Brookhaven and SLAC, on November 11, 1974. Three days later, the observation was confirmed at ADONE, Frascati and ten days later, the $\psi'$ was found at SLAC, where subsequently many further related states were discovered. We now know that these are bound states formed with the $c$-quark and its antiparticle which is comparatively heavy and that there are two further, even heavier quarks: $b$ and $t$.

At sufficiently high energies, quarks and gluons do manifest themselves as jets. Like the neutrini, they have left their theoretical place of birth and can now be seen flying around like ordinary, observable particles. Gradually, particle physicists abandoned their outposts in no man's and no woman's land, returned to the quantum fields and resumed discussion in the good old *Gasthaus zu Lagrange*, a term coined by Jost. The theoretical framework that describes the strong, electromagnetic and weak interactions in terms of gauge fields, leptons, quarks and scalar fields is now referred to as the Standard Model - this framework clarified the picture enormously.[4]

[4] Indeed, the success of this theory is amazing: Gauge fields are renormalizable in four dimensions, but it looks unlikely that the Standard Model is valid much beyond the explored energy range. Presumably it represents an effective theory. There is no reason, however, for an effective theory to be renormalizable. One of the most puzzling aspects of the Standard Model is that it is able to account for such a broad range of phenomena that are characterized by very different scales within one and the same renormalizable theory.

### 1.1.11 Quantum chromodynamics

If the electroweak gauge fields as well as the leptons and the scalars are dropped, the Lagrangian of the Standard Model reduces to QCD:

$$\mathcal{L}_{QCD} = -\tfrac{1}{4}F^A_{\mu\nu}F^{A\,\mu\nu} + i\bar{q}\gamma^\mu(\partial_\mu + ig_s\tfrac{1}{2}\lambda^A\mathcal{A}^A_\mu)q \\ -\bar{q}_{\rm R}\mathcal{M}q_{\rm L} - \bar{q}_{\rm L}\mathcal{M}^\dagger q_{\rm R} - \theta\,\omega\,. \quad (1.1.3)$$

The gluons are described by the gauge field $\mathcal{A}^A_\mu$, which belongs to the color group $\mathrm{SU_c}(3)$ and $g_s$ is the corresponding coupling constant. The field strength tensor $F^A_{\mu\nu}$ is defined by

$$F^A_{\mu\nu} = \partial_\mu\mathcal{A}^A_\nu - \partial_\nu\mathcal{A}^A_\mu - g_s f_{ABC}\mathcal{A}^B\mathcal{A}^C\,, \quad (1.1.4)$$

where the symbol $f_{ABC}$ denotes the structure constants of SU(3). The quarks transform according to the fundamental representation of $\mathrm{SU_c}(3)$. The compact notation used in (1.1.3) suppresses the labels for flavor, colour and spin: the various quark flavors are represented by Dirac fields, $q = \{u,\, d,\, s,\, c,\, b,\, t\}$ and $q_{\rm R} = \frac{1}{2}(1+\gamma_5)q$, $q_{\rm L} = \frac{1}{2}(1-\gamma_5)q$ are their right- and left-handed components. The field u(x), for instance, contains 3×4 components. While the 3×3 Gell-Mann matrices $\lambda^A$ act on the color label and satisfy the commutation relation

$$[\lambda^A, \lambda^B] = 2if_{ABC}\lambda^C\,, \quad (1.1.5)$$

the Dirac matrices $\gamma^\mu$ operate on the spin index. The mass matrix $\mathcal{M}$, on the other hand, acts in flavor space. Its form depends on the choice of the quark field basis. If the right- and left-handed fields are subject to independent rotations, $q_{\rm R} \to V_{\rm R}\,q_{\rm R}$, $q_{\rm L} \to V_{\rm L}\,q_{\rm L}$, where $V_{\rm R}, V_{\rm L} \in \mathrm{SU}(N_f)$ represent $N_f \times N_f$ matrices acting on the quark flavour, the quark mass matrix is replaced by $\mathcal{M} \to V_{\rm R}^\dagger\mathcal{M}V_{\rm L}$. This freedom can be used to not only diagonalize $\mathcal{M}$, but to ensure that the eigenvalues are real, nonnegative and ordered according to $0 \le m_u \le m_d \le \ldots \le m_t$.

The constant $\theta$ is referred to as the vacuum angle and $\omega$ stands for the winding number density

$$\omega = \frac{g_s^2}{32\pi^2}F^A_{\mu\nu}\tilde{F}^{A\,\mu\nu}\,, \quad (1.1.6)$$

where $\tilde{F}^{A\,\mu\nu} = \frac{1}{2}\epsilon^{\mu\nu\rho\sigma}F^A_{\rho\sigma}$ is the dual of the field strength.

As it is the case with electrodynamics, gauge invariance fully determines the form of the chromodynamic interaction. The main difference between QED and QCD arises from the fact that the corresponding gauge groups, U(1) and SU(3), are different. While the structure constants of U(1) vanish because this is an abelian group, those of $\mathrm{SU_c}(3)$ are different from zero. For this reason, gauge invariance implies that the Lagrangian contains terms involving three or four gluon fields: in contrast to the photons, which interact among themselves only via the exchange of charged particles, the gluons would interact even if quarks did not exist.

The term involving $\omega$ can be represented as a derivative, $\omega = \partial_\mu f^\mu$. Since only the integral over the Lagrangian counts, this term represents a contribution that only depends on the behavior of the gauge field at the boundary of space-time. In the case of QED, where renormalizability allows the presence of an analogous term, quantities of physical interest are indeed unaffected by such a contribution, but for QCD, this is not the case. Even at the classical level, nonabelian gauge fields can form instantons, which minimize the Euclidean action for a given nonzero winding number $\nu = \int d^4x\,\omega$.

### 1.1.12 Theoretical paradise

In order to briefly discuss some of the basic properties of QCD, let me turn off the electroweak interaction, treat the three light quarks as massless and the remaining ones as infinitely heavy:

$$m_u = m_d = m_s = 0\,, \qquad m_c = m_b = m_t = \infty\,. \quad (1.1.7)$$

The Lagrangian then contains a single parameter: the coupling constant $g_s$, which may be viewed as the net color of a quark. Unlike an electron, a quark cannot be isolated from the rest of the world – its color $g_s$ depends on the radius of the region considered. According to perturbation theory, the color contained in a sphere of radius $r$ grows logarithmically with the radius[5]:

$$\alpha_s \equiv \frac{g_s^2}{4\pi} = \frac{2\pi}{9\,|\ln(r\,\Lambda)|}\,. \quad (1.1.8)$$

Although the classical Lagrangian of massless QCD does not contain any dimensionful parameter, the corresponding quantum field theory does: the strength of the interaction cannot be characterized by a number, but by a dimensionful quantity, the intrinsic scale $\Lambda$.

The phenomenon is referred to as *dimensional transmutation*. In perturbation theory, it manifests itself through the occurrence of divergences – contrary to what many quantum field theorists thought for many years, the divergences do not represent a disease, but are intimately connected with the structure of the theory. They are a consequence of the fact that a quantum field theory does not inherit all of the properties of the corresponding classical field theory. In the case of massless Chromodynamics, the classical Lagrangian does not contain any dimensionful constants and hence remains invariant under a change of scale. This property, which is referred to as conformal invariance, does not survive quantization, however. Indeed, it is crucial for Quantum Chromodynamics to be consistent with what is known about the strong interaction that this theory does have an intrinsic scale.

[5] The formula only holds if the radius is small, $r\,\Lambda \ll 1$.

Massless QCD is how theories should be: the Lagrangian does not contain a single dimensionless parameter. In principle, the values of all quantities of physical interest are predicted without the need to tune parameters (the numerical value of the mass of the proton in kilogram units cannot be calculated, of course, because that number depends on what is meant by a kilogram, but the mass spectrum, the width of the resonances, the cross sections, the form factors, ... can be calculated in a parameter free manner from the mass of the proton, at least in principle).

### 1.1.13 Symmetries of massless QCD

The couplings of the $u$-, $d$- and $s$-quarks to the gauge field are identical. In the *chiral limit*, where the masses are set equal to zero, there is no difference at all – the Lagrangian is symmetric under SU(3) rotations in flavor space. Indeed, there is more symmetry: for massless fermions, the right- and left-handed components can be subject to independent flavor rotations. The Lagrangian of QCD with three massless flavors is invariant under $\mathrm{SU(3)}_R\times\mathrm{SU(3)}_L$. QCD thus explains the presence of the mysterious chiral symmetry discovered by Nambu: an exact symmetry of this type is present if some of the quarks are massless.

Nambu had conjectured that chiral symmetry breaks down spontaneously. Can it be demonstrated that the symmetry group $\mathrm{SU(3)}_R\times\mathrm{SU(3)}_L$ of the Lagrangian of massless QCD spontaneously break down to the subgroup $\mathrm{SU(3)}_{R+L}$? To my knowledge an analytic proof is not available, but the work done on the lattice demonstrates beyond any doubt that this does happen. In particular, for $m_u = m_d = m_s$, the states do form degenerate multiplets of $\mathrm{SU(3)}_{R+L}$ and, in the limit $m_u, m_d, m_s \to 0$, the pseudoscalar octet does become massless, as required by the Goldstone theorem.

### 1.1.14 Quark masses

The 8 lightest mesons, $\pi^+,\pi^0,\pi^-,K^+,K^0,\bar{K}^0,K^-,\eta$, do have the quantum numbers of the Nambu-Goldstone bosons, but massless they are not. The reason is that we are not living in the paradise described above: the light quark masses are different from zero. Accordingly, the Lagrangian of QCD is only approximately invariant under chiral rotations, to the extent that the symmetry breaking parameters $m_u$ , $m_d$ , $m_s$ are small. Since they differ, the multiplets split. In particular, the Nambu-Goldstone bosons pick up mass.

Even before the discovery of QCD, attempts at estimating the masses of the quarks were made. In particular, nonrelativistic bound state models for mesons and baryons where constructed. In these models, the proton mass is dominated by the sum of the masses of its constituents: $m_u + m_u + m_d \simeq m_p$, $m_u \simeq m_d \simeq 300\,\mathrm{MeV}$.

With the discovery of QCD, the mass of the quarks became an unambiguous concept: the quark masses occur in the Lagrangian of the theory. Treating the mass term as a perturbation, one finds that the expansion of $m^2_{\pi^+}$ in powers of $m_u$, $m_d$, $m_s$ starts with $m^2_{\pi^+} = (m_u + m_d)B_0 + \ldots$ The constant $B_0$ also determines the first term in the expansion of the square of the kaon masses: $m^2_{K^+} = (m_u + m_s)B_0 + \ldots$, $m^2_{K^0} = (m_d + m_s)B_0 + \ldots$ Since the kaons are significantly heavier than the pions, these relations imply that $m_s$ must be large compared to $m_u$, $m_d$.

The first crude estimate of the quark masses within QCD relied on a model for the wave functions of $\pi$, $K$, $\rho$, which was based on SU(6) (spin-flavor-symmetry) and led to $B_0 \simeq \frac{3}{2} m_\rho F_\rho/F_\pi$. Numerically, this yields $B_0 \simeq 1.8\,\mathrm{GeV}$. For the mean mass of the two lightest quarks, $m_{ud} \equiv \frac{1}{2}(m_u+m_d)$, this estimate implies $m_{ud} \simeq 5\,\mathrm{MeV}$, while the mass of the strange quark becomes $m_s \simeq 135\,\mathrm{MeV}$ [53]. Similar mass patterns were found earlier, within the Nambu-Jona-Lasinio model [54] or on the basis of sum rules [55].

### 1.1.15 Breaking of isospin symmetry

From the time Heisenberg had introduced isospin symmetry, it was taken for granted that the strong interaction strictly conserves isospin. QCD does have this symmetry if and only if $m_u = m_d$. If that condition were met, the mass difference between proton and neutron would be due exclusively to the e.m. interaction. This immediately gives rise to a qualitative problem: why is the charged particle, the proton, lighter than its neutral partner, the neutron?

The Cottingham formula [56] states that the leading contribution of the e.m. interaction to the mass of a particle is determined by the cross section for electron scattering on this particle. We evaluated the formula on the basis of Bjorken scaling and of the experimental data for electron scattering on protons and neutrons available at the time. Since we found that the electromagnetic self energy of the proton is larger than the one of the neutron, we concluded that the strong interaction does not conserve isospin: even if the e.m. interaction is turned off, $m_u$ must be different from $m_d$. In fact, the first crude estimate for the masses of the light quarks [57],

$$m_u \simeq 4\,\text{MeV}\,, \quad m_d \simeq 7\,\text{MeV}\,, \quad m_s \simeq 135\,\text{MeV}\,, \qquad (1.1.9)$$

indicated that $m_d$ must be almost twice as large as $m_u$.

It took quite a while before this bizarre pattern was generally accepted. The Dashen theorem [58] states that, in a world where the quarks are massless, the e.m. self energies of the kaons and pions obey the relation $m^{2\,em}_{K^+} - m^{2\,em}_{K^0} = m^{2\,em}_{\pi^+} - m^{2\,em}_{\pi^0}$. If the mass differences were dominated by the e.m. interaction, the charged kaon would be heavier than the neutral one. Hence the mass difference between the kaons cannot be due to the electromagnetic interaction, either. The estimates for the quark mass ratios obtained with the Dashen theorem confirm the above pattern [59].

### 1.1.16 Approximate symmetries are natural in QCD

At first sight, the fact that $m_u$ strongly differs from $m_d$ is puzzling: if this is so, why is isospin such a good quantum number? The key observation here is the one discussed in Section 1.1.12: QCD has an intrinsic scale, $\Lambda$. For isospin to represent an approximate symmetry, it is not necessary that $m_d - m_u$ is small compared to $m_u + m_d$. It suffices that the symmetry breaking parameter is small compared to the intrinsic scale, $m_d - m_u \ll \Lambda$.

In the case of the eightfold way, the symmetry breaking parameters are the differences between the masses of the three light quarks. If they are small compared to the intrinsic scale of QCD, then the Green functions, masses, form factors, cross sections ... are approximately invariant under the group $\text{SU}(3)_{R+L}$. Isospin is an even better symmetry, because the relevant symmetry breaking parameter is smaller, $m_d - m_u \ll m_s - m_u$. The fact that $m^2_{\pi^+}$ is small compared to $m^2_{K^+}$ implies $m_u + m_d \ll m_u + m_s$. Hence all three light quark masses must be small compared to the scale of QCD.

In the framework of QCD, the presence of an approximate chiral symmetry group of the form $\text{SU}(3)_\text{R} \times \text{SU}(3)_\text{L}$ thus has a very simple explanation: it so happens that the masses of $u$, $d$ and $s$ are small. We do not know why, but there is no doubt that this is so. The quark masses represent a perturbation, which in first approximation can be neglected – in first approximation, the world is the paradise described above.

### 1.1.17 Ratios of quark masses

The confinement of color implies that the masses of the quarks cannot be identified by means of the four-momentum of a one-particle state – the spectrum of the theory does not contain such states. As parameters occurring in the Lagrangian, they need to be renormalized and the renormalized mass depends on the regularization used to set up the theory. In the $\overline{\text{MS}}$ scheme [60–62], they depend on the running scale – only their ratios represent physical quantities. Among the three lightest quarks, there are two independent mass ratios, which it is convenient to identify with

$$S = \frac{m_s}{m_{ud}}\,, \qquad R = \frac{m_s - m_{ud}}{m_d - m_u}\,, \qquad (1.1.10)$$

where $m_{ud} \equiv \frac{1}{2}(m_u + m_d)$.

Since the isospin breaking effects due to the e.m. interaction are not negligible, the physical masses of the Goldstone boson octet must be distinguished from their masses in QCD, i.e. in the absence of the electroweak interactions. I denote the latter by $\hat{m}_P$ and use the symbol $\hat{m}_K$ for the mean square kaon mass in QCD, $\hat{m}^2_K \equiv \frac{1}{2}(\hat{m}^2_{K^+} + \hat{m}^2_{K^0})$. The fact that the expansion of the square of the Goldstone boson masses in powers of $m_u$, $m_d$, $m_s$ starts with a linear term implies that, in the chiral limit, their ratios are determined by $R$ and $S$. In particular, the expansion of the ratios of $\hat{m}^2_{\pi^+}$, $\hat{m}^2_{K^+}$ and $\hat{m}^2_{K^0}$ starts with

$$\frac{2\hat{m}^2_K}{\hat{m}^2_{\pi^+}} = (S+1)\{1 + \Delta_S\}\,, \qquad (1.1.11)$$

$$\frac{\hat{m}^2_K - \hat{m}^2_{\pi^+}}{\hat{m}^2_{K^0} - \hat{m}^2_{K^+}} = R\{1 + \Delta_R\}\,, \qquad (1.1.12)$$

where $\Delta_S$ as well as $\Delta_R$ vanish in the chiral limit – they represent corrections of $O(\mathcal{M})$. The left hand sides only involve the masses of $\pi^+$, $K^+$ and $K^0$. Invariance of QCD under charge conjugation implies that the masses of $\pi^-$, $K^-$ and $\bar{K}^0$ coincide with these. There are low energy theorems analogous to (1.1.11), (1.1.12), involving the remaining members of the octet, $\pi^0$ and $\eta$, but these are more complicated because the states $|\pi^0\rangle$ and $|\eta\rangle$ undergo mixing {at leading order, chiral symmetry implies that the mixing angle is given by $\tan(2\theta) = \sqrt{3}/2R$}. In the isospin limit, $\{m_u = m_d,\ e = 0\}$, the masses of $\pi^0$ and $\pi^+$ coincide and $\hat{m}_\eta$ obeys the Gell-Mann-Okubo formula, $(\hat{m}^2_\eta - \hat{m}^2_K)/(\hat{m}^2_K - \hat{m}^2_\pi) = \frac{1}{3}\{1 + O(\mathcal{M})\}$.

While the accuracy to which $S$ can be determined on the lattice is amazing, the uncertainty in $R$ is larger by almost an order of magnitude [63]:

$$S = 27.42(12)\,, \qquad R = 38.1(1.5)\,. \qquad (1.1.13)$$

The reason is that $R$ concerns isospin breaking effects. The contributions arising from QED are not negligible at this precision and since the e.m. interaction is of long range, it is more difficult to simulate on a lattice.

The difference shows up even more clearly in the corrections. The available lattice results [63] lead to $\Delta_S = 0.057(7)$, indicating that the low energy theorem (1.1.11) picks up remarkably small corrections from higher orders of the quark mass expansion. Those occurring in the Gell-Mann-Okubo formula are also known to be very small. The number $\Delta_R = -0.016(57)$ obtained from the available lattice results is also small, but the uncertainty is so large that even the sign of the correction remains open.

The quantities $\Delta_S$, $\Delta_R$ exclusively concern QCD and could be determined to high precision with available methods, in the framework of $N_f = 1+1+1$: three flavours of different mass. For isospin breaking quantities, the available results come with a large error because they do not concern QCD alone but are obtained from a calculation of the physical masses, so that the e.m. interaction cannot be ignored. A precise calculation of $\hat{m}_{\pi^+}$, $\hat{m}_{K^+}$, $\hat{m}_{K^0}$ within lattice QCD would be of considerable interest as it would allow to subject a venerable low energy theorem for the quark mass ratio $Q^2 \equiv (m_s^2 - m_{ud}^2)/(m_d^2 - m_u^2)$ [64] to a stringent test. The theorem implies that the leading contributions to $\Delta_R$ and $\Delta_S$ are equal in magnitude, but opposite in sign: $\Delta_R = -\Delta_S + O(\mathcal{M}^2)$ [65]. The available numbers are consistent with this relation but far from accurate enough to allow a significant test. There is no doubt that the leading terms dominate if the quark masses are taken small enough, but since the estimates for $\Delta_R$ and $\Delta_S$ obtained at the physical values of the quark masses turn out to be unusually small, it is conceivable that the corrections of $O(\mathcal{M}^2)$ are of comparable magnitude. For $m_u = m_d$, the masses of the Goldstone bosons have been worked out to NNLO of Chiral Perturbation Theory [66]. An extension of these results to $\hat{m}_{\pi^+}$, $\hat{m}_{K^+}$, $\hat{m}_{K^0}$ for $m_u \neq m_d$ should be within reach and would allow a much more precise lattice determination of $\Delta_R$.

### 1.1.18 U(1) anomaly, CP-problem

Even before the discovery of QCD, it was known that, in the presence of vector fields, the Ward identities for axial currents contain anomalies [67–69]. In particular, an external e.m. field generates an anomaly in the conservation law for the axial current $\bar{u}\gamma^\mu\gamma_5 u - \bar{d}\gamma^\mu\gamma_5 d$. The anomaly implies a low energy theorem for the decay $\pi^0 \to \gamma + \gamma$, which states that, to leading order in the expansion in powers of the momenta and for $m_u = m_d = 0$, the transition amplitude is determined by $F_\pi$, i.e. by the same quantity that determines the rate of the decay $\pi^+ \to \mu + \nu_\mu$.

In QCD, the conservation law for the singlet axial current contains an anomaly,

$$\partial_\mu(\bar{q}\gamma^\mu\gamma_5 q) = 2i\,\bar{q}\mathcal{M}\gamma_5 q + 2N_f\,\omega\,, \qquad (1.1.14)$$

where $N_f$ is the number of flavors and $\omega$ is specified in (1.1.6). The phenomenon plays a crucial role because it implies that even if the quark mass matrix $\mathcal{M}$ is set equal to zero, the singlet axial charge is not conserved. Hence the symmetry group of QCD with 3 massless flavors is $\mathrm{SU(3)_R}\times\mathrm{SU(3)_L}\times\mathrm{U(1)_{R+L}}$, not $\mathrm{U(3)_R}\times\mathrm{U(3)_L}$. QCD is not invariant under the chiral transformations generated by the remaining factor, $\mathrm{U(1)_{R\text{-}L}}$. This is why the paradise described above contains 8 rather than 9 massless Goldstone bosons.

The factor $\mathrm{U(1)_{R\text{-}L}}$ changes the phase of the right-handed components of all quark fields by the same angle, $q'_\mathrm{R} = e^{i\beta}q_\mathrm{R}$, while the left-handed components are subject to the opposite transformation: $q'_\mathrm{L} = e^{-i\beta}q_\mathrm{L}$. This change of basis can be compensated by modifying the quark mass matrix with $\mathcal{M}' = e^{2i\beta}\mathcal{M}$, but in view of the anomaly, the operation does not represent a symmetry of the system. The relation (1.1.14) shows, however, that current conservation is not lost entirely – it only gets modified. In fact, if the above change of the quark mass matrix is accompanied by a simultaneous change of the vacuum angle, $\theta' = \theta - 2\beta$, the physics does remain the same. Note that, starting from an arbitrary mass matrix, a change of basis involving the factor $\mathrm{U(1)_{R\text{-}L}}$ is needed to arrive at the convention where $\mathcal{M}$ is diagonal with real eigenvalues. In that convention, the vacuum angle does have physical significance – otherwise only the product $e^{i\theta}\mathcal{M}$ counts.

The Lagrangian of QCD is invariant under charge conjugation, but the term $-\theta\,\omega$ has negative parity. Accordingly, unless $\theta$ is very small, there is no explanation for the fact that CP-violating quantities such as the electric dipole moment of the neutron are too small to have shown up in experiment. This is referred to as the strong CP-problem.

There is a theoretical solution of this puzzle: if the lightest quark were massless, $m_u = 0$, QCD would conserve CP. The Dirac field of the $u$-quark can then be subject to the chiral transformation $u'_\mathrm{R} = e^{i\beta}u_\mathrm{R}$, $u'_\mathrm{L} = e^{-i\beta}u_\mathrm{L}$ without changing the quark mass matrix. As discussed above, the physics remains the same, provided the vacuum angle is modified accordingly. This shows that if one of the quarks were massless, the vacuum angle would become irrelevant. It would then be legitimate to set $\theta = 0$, so that the Lagrangian becomes manifestly CP-invariant.

This 'solution', however, is fake. If $m_u$ were equal to zero, the ratio $R$ would be related to $S$ by $R = \frac{1}{2}(S-1)$. The very accurate value for $S$ in equation (1.1.13) would imply $R = 13.21(6)$, more than 16 standard deviations away from the result quoted for $R$.

### 1.1.19 QCD as part of the Standard Model

In the Standard Model, the vacuum contains a condensate of Higgs bosons. At low energies, the manner in which the various other degrees of freedom interact with these plays the key role. Since they do not have color and are electrically neutral, their condensate is transparent for gluons and photons. The gauge bosons $W^{\pm}$, $Z$ that mediate the weak interaction, as well as the leptons and quarks do interact with the condensate: photons and gluons remain massless, all other particles occurring in the Standard Model are hindered in moving through the condensate and hence pick up mass. In cold matter only the lightest degrees of freedom survive: photons, gluons, electrons, $u$- and $d$-quarks – all other particles are unstable, decay and manifest their presence only indirectly, through quantum fluctuations.

At low energies, the Standard Model boils down to a remarkably simple theory: QCD + QED. The Lagrangian only contains the coupling constants $g_s$, $e$, $\theta$ and the masses of the quarks and leptons as free parameters, but describes the laws of nature relevant at low energies to breathtaking precision. The gluons and the photons represent the gauge fields that belong to color and electric charge, respectively. Color is confined, but electric charge is not: while electrons can move around freely, quarks and gluons form color neutral bound states – mesons, baryons, nuclei.

The structure of the atoms is governed by QED because the e.m. interaction is of long range. In particular, their size is of the order of the Bohr radius, $a_B = 4\pi/e^2 m_e$, which only involves the mass of the electron and the coupling constant $e$. The mass of the atoms, on the other hand, is dominated by the energy of the gluons and quarks that are bound in the nucleus. It is of the order of the scale $\Lambda_{QCD}$, which characterizes the value of $g_s$ in a renomalization group invariant manner. Evidently, the sum of the charges of the quarks contained in the nucleus also matters, as it determines the number of electrons that can be bound to it. The mass of the quarks, on the other hand, plays an important role only in so far as it makes the proton the lightest baryon – the world would look rather different if the neutron was lighter ...

The properties of the interaction among the quarks and gluons does not significantly affect the structure of the atoms, but from the theoretical point of view, the gauge field theory that describes it, QCD, is the most remarkable part of the Standard Model. In fact, it represents the first non-trivial quantum field theory that is internally consistent in four-space-time dimensions. In contrast to QED or to the Higgs sector, QCD is asymptotically free. The behavior of the quark and gluon fields at very short distances is under control. A cutoff is needed to set the theory up, but it can unambiguously be removed. In principle, all of the physical quantities of interest are determined by the renormalization group invariant quark mass matrix, by the vacuum angle $\theta$ and a scale. In the basis where the quark mass matrix is diagonal and real, the vacuum angle is tiny. We do not know why this is so, nor do we understand the bizarre pattern of eigenvalues.

## 1.2 The origins of QCD

**Harald Fritzsch**

Murray Gell-Mann and I started to collaborate in October 1970. We considered the results of the experiments on deep inelastic scattering at the Stanford Linear Accelerator Center. James Bjorken had predicted, using current algebra, that the cross sections showed at large values of the virtual photon mass and the energy transfer to the nucleon a scaling behavior, i.e. the cross section is a function of the ratio $x$, where $x$ is the ratio of the square of the virtual photon mass to the energy transfer to the nucleon, multiplied with the nucleon mass. This ratio $x$ varies from zero to one.

Since in the scaling region the cross sections were determined by the commutator of two electromagnetic currents at nearly lightlike distances, Gell-Mann and I assumed, that this commutator near the light cone is given by the free quark model. Thus the Bjorken scaling followed from this assumption.

The interaction between the quarks was assumed not to be present near the light cone. The cross section in the deep inelastic region determined the distribution functions of the three quarks and antiquarks, which are given by the proton matrix element of the commutator of the electromagnetic current.

In the free quark model the commutator near the light cone is given by a singular function, multiplied by a bilocal function of quark fields [70]. The matrix elements of these bilocal operators determined the quark distribution functions of the nucleon. The integral of the quark distribution functions gives the contribution of all the quark momenta to the nucleon momentum.

Gell-Mann and I expected that this integral would be +1, since inside the nucleon were only the three quarks and three antiquarks. However according to the experiments at SLAC this integral was only about 45%:

$$\int_0^1 x\left[u(x)+\overline{u}(x)+d(x)+\overline{d}(x)+s(x)+\overline{s}(x)\right]\mathrm{d}x \simeq 0.45. \quad (1.2.1)$$

Thus besides the quarks there must exist neutral quanta, which are relevant for the confinement of the quarks and which contribute about 55% to the momentum of a fast moving nucleon. This observation was the first indication that the strong interactions are described by a gauge theory. In such a theory there would be besides the quarks and antiquarks also neutral gluons.

Afterwards Gell-Mann and I considered several problems of the quark theory. The $\Omega^-$ particle was a bound state of three strange quarks. The three spin vectors of the quarks were symmetricalls arranged, and the space wave function was symmetric, since the $\Omega^-$ is the ground state of three strange quarks. Thus an interchange of two strange quarks was symmetric, but according to the Pauli principle it should be antisymmetric.

Another problem was related to the electromagnetic decay of the neutral pion. The decay rate, calculated in the quark model, is much smaller than the observed decay rate, only about 1/9 of the observed rate.

We also studied the cross section for the reaction electron-positron annihilation into hadrons. The ratio $R$ of the cross section for hadron production and the cross section for the production of a muon pair can be calculated in the quark model. It is given by the sum of the squares of the electric charges of the three quarks, i.e. 2/3. But according to the experiments at the CEA accelerator at Harvard university this ratio was about three times larger: $R \simeq 2$.

To solve theses problems, Murray Gell-Mann, William Bardeen and I introduced for the quarks a new quantum number, which we called "color". Each quark is described by a red, a green and a blue quark. The three colors can be transformed by the color group $\mathrm{SU}(3)_\mathrm{C}$, which is assumed to be an exact symmetry. Measurable quantities, e.g. cross sections or the wave functions of hadrons, are color singlets.

The quark wave function $\psi_\Omega$ of the $\Omega^-$ is also a color singlet:

$$\psi_\Omega \simeq (rgb - grb + brg - rbg + gbr - bgr). \quad (1.2.2)$$

This wave function is antisymmetric under the exchange of two quarks - there is no problem with the Pauli principle. The quark wave functions of mesons are also color singlets:

$$\psi_\mathrm{meson} \simeq (\bar{r}r + \bar{g}g + \bar{b}b). \quad (1.2.3)$$

The decay amplitude for the neutral pion decay is three times larger, if the quarks are colored. Thus the decay rate is nine times larger and agrees with the observed decay rate [71]. The ratio $R$ for electron-positron annihilaton, given by the sum of the squares of the quark charges, is now also three times larger: $R \simeq 2$. Thus the introduction of the color quantum number solved the three problems mentioned above.

The color quantum number also explains why mesons are quark-antiquark bound states and baryons are three quark bound states, since they must be color singlets. Thus the mesons and baryons could be considered to be "white" states, since a particular color cannot be seen from the outside - the color quantum number is only relevant inside the mesons and baryons.

In the spring of 1972 Gell-Mann and I tried to understand why a colored quark cannot be observed - it is confined inside a baryon or meson or inside an atomic nucleus. We considered to use the color symmetry group as a gauge group. The gauge bosons of such a gauge theory would be color octets. I proposed to call these gauge bosons "chromons", but Gell-Mann insisted to call them "gluons", mixing the English language and the Greek language.

We called this new gauge theory "Quantum Chromodynamiccs" (QCD). The Lagrangian of QCD is [30, 50]:

$$\mathcal{L} = \overline{q}\left[i\gamma^\mu\left(\partial_\mu + ig_s\frac{\lambda^A}{2}\mathcal{A}_\mu^A\right) - m\right]q - \frac{1}{4}F_{\mu\nu}^A F^{A\,\mu\nu}\,, \quad (1.2.4)$$

where the $\lambda^A$ are the Gell-Mann matrices, and

$$F_{\mu\nu}^A = \partial_\mu\mathcal{A}_\nu^A - \partial_\nu\mathcal{A}_\mu^A - g_s f_{ABC}\mathcal{A}_\mu^B\mathcal{A}_\nu^C\,. \quad (1.2.5)$$

$f_{ABC}$ are called SU(3) structure constants. This Lagrangian is very similar to the Lagrangian of Quantum Electrodynamics. The electromagnetic field is replaced by the eight gluon fields $\mathcal{A}^A$, the electron mass by the quark mass, and the charge $e$ is replaced by the strong coupling $g_s$. The strong interaction constant is defined by $\alpha_s = g_s^2/4\pi$.

However, the big difference between Quantum Electrodynamics and Quantum Chromodynamics is the presence of the $\mathcal{A}^2$ term in $F_{\mu\nu}^A$, not present in Quantum Electrodynamics. This term shows that a gluon interacts not only with a quark, but also with another gluon, and gives rise to 3- and 4-gluon couplings.

The quark masses, which appear in the Lagrangian of QCD, are not the masses of free quarks, but the masses, relevant inside the hadrons. The masses of the quarks depend on the energy scale. They are large at small energies and small at high energies. Here are the typical masses for the up-quark, the down-quark and the strange quark at the energy given by the mass of the $Z$-boson, $M_Z \simeq 91.2\,\text{GeV}$:

$$m_u \simeq 1.2\,\text{MeV},\ m_d \simeq 2.2\,\text{MeV},\ m_s \simeq 53\,\text{MeV}.$$

These masses describe the symmetry breaking of the $SU(3)_\text{F}$ flavor group. Interesting is the violation of the isospin symmetry. The down quark is heavier than the up quark. For this reason the neutron is heavier than the proton, and the proton is stable. If there would be no isospin violation, i.e. $m_u = m_d$, the proton would be heavier than the neutron due to the electromagnetic self-energy and it would decay into the neutron - life would not be possible.

Gell-Mann and I assumed that the interaction in QCD is zero at light-like distances. The light cone current algebra, which we had discussed in ref. [70], would not be changed. The confinement of colored states, i.e. the quarks and the gluons, would be due to the interaction at long distances.

Soon we realized that our assumption, that there is no interaction near the light-cone, was not correct. David Gross, Frank Wilczek and, independently, David Politzer calculated this interaction, which is the interaction, given by the Lagrangian, but near the light-cone the relevant coupling constant is not zero, but only very small.

The QCD Lagrangian describes a theory, which is asymptotically free. At small distances the interaction is very small, at large distances the interaction is strong. Thus the coupling constant is not constant, but a function of the energy. The sliding of the coupling constant $g_s$ as a function of the renormalization mass $\mu$ is given by the beta-function $\beta(g_s)$:

$$\mu \frac{\mathrm{d}}{\mathrm{d}\mu} g_s(\mu) = \beta(g_s). \tag{1.2.6}$$

This beta function is positive for many theories, for example quantum electrodynamics. The fine structure constant $\alpha$ is at the energy of 100 GeV about 10% larger than at low energies.

The beta function can be calculated in perturbation theory. One finds for QCD:

$$\mu \frac{\mathrm{d}}{\mathrm{d}\mu} g_s(\mu) \simeq -\frac{1}{16\pi^2}\left(11 - \frac{2}{3} n_f\right) g_s^3(\mu). \tag{1.2.7}$$

Here the coefficient "11" describes the contribution of the gluons to the beta function. The asymptotic freedom of QCD is due to this coefficient - it is related to the self-interactions of the gluons. The number $n_f$ is the number of the different quark flavors. For the three quarks up, down and strange one has $n_f = 3$.

In QCD one can describe the energy dependence of the coupling constant by introducing a scale parameter $\Lambda$:

$$\alpha_s(\mu^2) \simeq \frac{4\pi}{\left(11 - \frac{2}{3} n_f\right) \ln\left(\frac{\mu^2}{\Lambda^2}\right)}. \tag{1.2.8}$$

This scale parameter has been measured by many experiments (see Section 3.2):

$$\Lambda = (332 \pm 17)\,\text{MeV}. \tag{1.2.9}$$

In experiments one has measured the scale dependence of the coupling constant. It agrees very well with the theoretical prediction. We also mention the value of the coupling constant at the mass of the $Z$-boson, where it was possible to measure the coupling constant rather precisely (see Section 3.2):

$$\alpha_s = 0.1181 \pm 0.0011. \tag{1.2.10}$$

In QCD, Bjorken scaling in deep inelastic scattering is not an exact property of the strong interactions. The quark distribution functions change slowly at high energies. This change can be calculated in perturbation theory (see Section 2.3). The results agree rather well with the experimental results. Also the gluon distribution function $g(x)$ has been measured. Since the gluons and the quarks contribute to the momentum of a high energy proton, the following sum rule must be obeyed:

$$\int_0^1 x\left[g(x) + u(x) + \overline{u}(x) + d(x) + \overline{d}(x) + s(x) + \overline{s}(x)\right] \mathrm{d}x = 1. \tag{1.2.11}$$

Using the scale parameter $\Lambda$, one can in principle calculate many properties of the strong interactions, for example the masses of the hadrons like the proton mass: $m_p = \text{const} \times \Lambda$. The proton mass depends also on the quark masses, however the up and down quark masses are very small and can be neglected. The calculations of the hadron masses are complicated and are often carried out by discretizing space and time (see Section 4 on Lattice QCD).

In QCD one can also change the three quark masses. For example we can assume that the three quark masses are zero. In this case the flavor group $SU(3)_\text{F} \times SU(3)_\text{F}$ would be unbroken. The three pions, the four $K$-mesons and the $\eta$ - meson would be massless and the eight vector mesons would have the same mass. There is not a ninth massless pseudoscalar meson, since the singlet axial current has an anomaly:

$$\partial_\mu\left(\bar{u}\gamma^\mu\gamma_5 u + \bar{d}\gamma^\mu\gamma_5 d + \bar{s}\gamma^\mu\gamma_5 s\right) = const \times g_s^2\, \epsilon^{\mu\nu\rho\sigma} F^A_{\mu\nu} F^A_{\rho\sigma}. \tag{1.2.12}$$

where $\epsilon^{\mu\nu\rho\sigma}$ is the totally antisymmetric tensor. In ref. [30] Gell-Mann and I also studied what happens if the quarks are removed from the QCD Lagrangian. In this case only the eight gluons are present. At low energies there would be a discrete spectrum of particles, which consist of gluons - the glue mesons, gluonium particles or glueball (see Section 8.4). If the three quarks are introduced, the glue mesons would mix with the quark-antiquark mesons. The experimentalists have thus far not clearly identified a glue meson. Presumably in nature there are only mixtures of glue mesons and quark-antiquark mesons. But there might be mesons, which are essentially glue mesons, since the mixing is very small for these mesons.

It is useful to consider the theory of QCD with just one heavy quark Q. The ground-state meson in this hypothetical case would be a quark–antiquark bound state (see Sections 8.1, 8.6). The effective potential between the quark and its antiquark at small distances would be a Coulomb potential proportional to $1/r$, where $r$ is the distance between the quark and the antiquark. However, at large distances the self-interaction of the gluons becomes important. The gluonic field lines at large distances do not spread out as in electrodynamics. Instead, they attract each other. Thus the quark and the antiquark are connected by a string of gluonic field lines. The force between the quark and the antiquark is constant, i.e. it does not decrease as in electrodynamics. The heavy quarks are confined.

In the annihilation of electrons and positrons at very high energies it has been possible to test the theory of quantum chromodynamics rather precisely. If an electron and a positron collide, a quark and an antiquark are produced. The two quarks move away from each other almost with the speed of light. Since the two quarks do not exist as free particles, they fragment into two jets of hadrons, mostly pions. These particles form two narrow jets. These jets have been observed since 1979 at the collider at DESY, later at the LEP-collider at CERN. Sometimes a quark emits a high energy gluon, which also fragments into hadrons. Thus three jets are produced, two quark jets and one gluon jet. Such three jet events have been observed since 1979 at DESY, later at CERN (see Section 2.2).

Now we consider high energy collisions of atomic nuclei, for example collisions of lead nuclei. Such collisions are studied at the Relativistic Heavy Ion Collider (RHIC) in Brookhaven, at Fermilab and at the LHC in CERN. In such collisions a new state of matter is produced for a short time, a quark-gluon-plasma. Astrophysicists assume that such a plasma exists also for a long time near the center of a large neutron star (see Section 7.1).

Right after the Big Bang the matter was a quark-gluon-plasma. During the expansion of the universe the plasma changed later into a gas of protons and neutrons (see Section 7.2).

In the fall of 1973 I was convinced, that Gell-Mann and I had discovered the correct theory of the strong interactions: Quantum Chromodynamics. Almost every day I discussed this theory with Richard Feynman, and he also thought that it was correct. In 1974 Feynman gave lectures on QCD. But Gell-Mann still thought that the true theory of the strong interactions should be a theory based on strings.

In the years after 1973 it became clear that QCD is the correct theory of the strong interactions. I was proud that I had contributed to the birth of this theory, which is now a major part of the Standard Theory of particle physics.

# 2 Experimental Foundations

*Conveners:*
**Franz Gross and Eberhard Klempt**

Quantum Chromodynamics or QCD: What a gorgeous theory! You start with free colored quarks. You request invariance with respect to the exchange of colors at any time and any space point, and the quarks interact. That is all what QCD requires. It is a remarkable simple concept. But: is this the true theory of strong interactions? In this Section, the milestones are discussed which convinced even sceptical physicists of the quark model and of the new theory.

A breakthrough was achieved in the *November revolution:* Charmonium was discovered at SLAC, the $c$-quark was shown to exist, the GIM meachanism (proposed by Sheldon Glashow, John Iliopoulos and Luciano Maiani in 1964) explaining the absence of neutral currents in weak interactions found experimental confirmation.

John B. Kogut's contribution remembers the excitement in these days. A new spectroscopy came into life with a convincing interpretation based on the famous Cornell potential. The mediators of the strong interaction, called gluons, carry – unlike the electrically neutral photons – themselves the charge of the strong interaction and are confined. San Lau Wu recalls her personal contributions to the discovery of gluons at DESY where events were found in which $e^+e^-$ annihilate into three bunches of particles, three jets. The three jets were interpreted as processes in which the two quarks – observed as jets – radiate off a gluon which manifests itself as the third jet.

The evidence for the correctness of QCD grew rapidly. Yuri Dokshitzer reminds us of the most important steps.

Scaling, observed already in 1972, proved the existence of interaction centers – called partons by Feynman – inside of nucleons. From the ratio of the cross sections for $e^+e^-$ annihilation into hadrons over that for $\mu^+\mu^-$ the number of colors $N_c = 3$ was deduced. And the strong interaction constant $\alpha_s$ was shown to decrease with momentum transfer opening QCD to perturbative approaches. Dokshitzer introduces many basic concepts like jet finding algorithms, evolution, divergences and resummation, which will be discussed in more detail in later sections.

## 2.1 Discovery of heavy mesons as bound states of heavy quarks

**John B. Kogut**

### 2.1.1 SLAC, Light Quarks and Deep Inelastic Scattering

Many physicists and accelerators contributed to the establishment of the Standard model. But two accelerators were particularly important to US-based researchers. They were the 2-mile Linac and the 80 m diameter electron-positron ring, SPEAR (Stanford Positron-Electron Asymmetric Rings), of the Stanford Linear Accelerator Center (SLAC), Fig. 2.1.1. The Linac, which was built under the direction of SLAC's first director, W. Panofsky ("PIEP"), and started operations in 1965, discovered the light constituents of the protons, the u, d and s quarks, by measuring the inclusive deep inelastic cross section of $e^- + p \to e^{-'} + X$. The deep inelastic scattering program was critical to the founding of Quantum Chromodynamics (QCD) and is discussed extensively elsewhere in this journal review.

When I arrived at SLAC as an incoming graduate student in 1967, theoretical research revolved around Bjorken ("bj") scaling, and the parton model of bj and Feynman. One of the tools of the trade was the Infinite Momentum Frame (IMF). D. Soper, bj and I put the IMF on a firm foundation by quantizing Quantum Electrodynamics (QED) on the light cone [72]. This work initiated the program of light cone formulations of field theories (later called light front quantization by some advocates) that will be reviewed in other chapters of this journal review. Later, S. Berman, bj and I developed the parton picture of the final states of inclusive processes involving large momentum transfers [73] and introduced parton fragmentation functions. This work had to address the mysterious phenomenon of quark confinement, the fact that quarks were "observed" when their properties were measured in deep inelastic processes, but no quarks were found isolated in the debris of the collisions. Although considerable progress has been made and many field theoretic mechanisms have been studied and proposed, especially in the context of Lattice formulations of QCD, the quark confinement problem remains open. It certainly was on many physicists' minds in the early days.

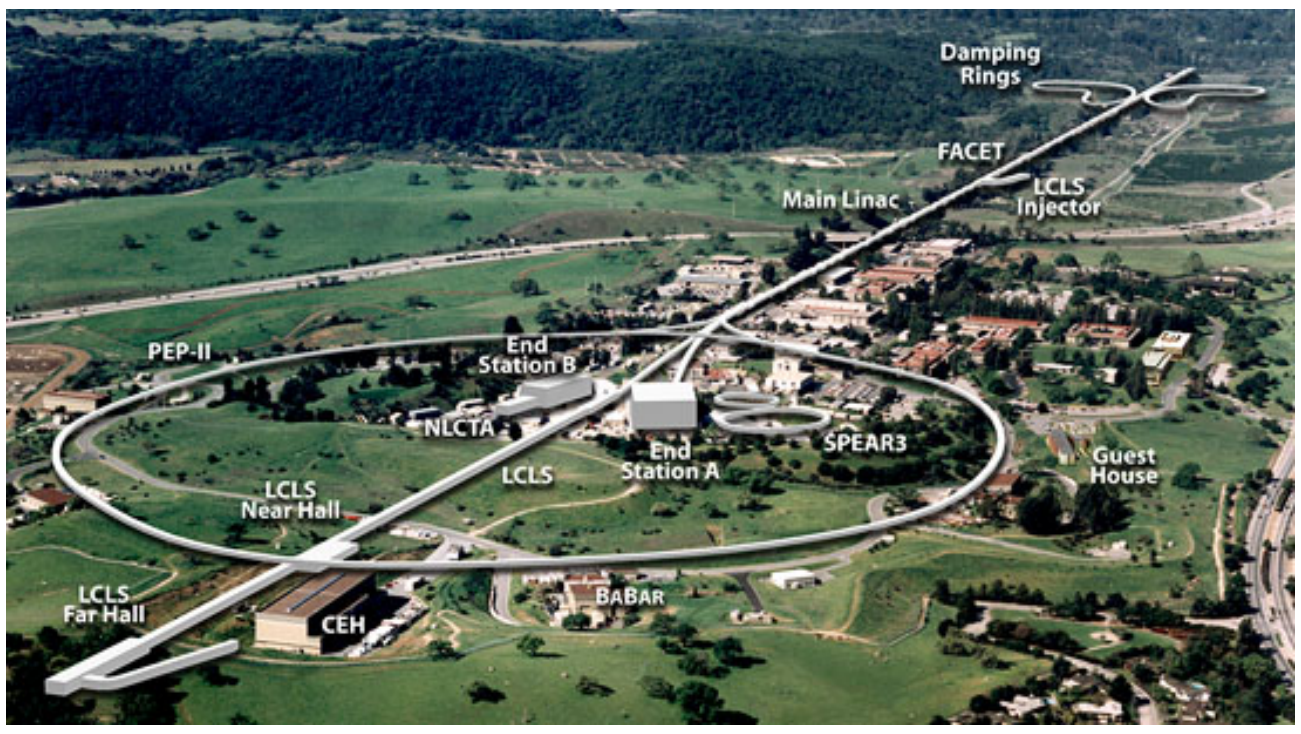

**Fig. 2.1.1** Aerial View of SLAC, 2020: The Linac, SPEAR and their descendents

### 2.1.2 Charmonium and The November Revolution

Several years later, during the summer of 1974, experimentalists from SLAC presented some intriguing data from the earliest runs of their very new electron-positron collider, SPEAR. A later section of this article will sketch the history of electron-positron colliders at SLAC since these machines were so central to the establishment of the Standard Model. The data of the summer of 1974 focused on the ratio $R$,

$$R = \frac{\sigma_{e^+e^- \to \text{hadrons}}}{\sigma_{e^+e^- \to \mu^+\mu^-}} \tag{2.1.1}$$

which, when plotted against the CM energy, showed a high, broad peak around 3.0-3.5 GeV. This suggested new interactions in the reaction's direct channel. One popular speculation was that a new quark threshold had been reached. A notable paper [74] stated that the reported broad peak in R should be accompanied by narrow resonant peaks at slightly lower energies. On November 11, 1974, SPEAR announced such a narrow peak at an energy 3.105 GeV [75] with an electronic width of $\Gamma_e \approx 5.5$ keV. Brookhaven also found this state in proton-proton collisions in fixed target experiments at the Alternating Gradient Synchrotron (AGS) [76] but that experiment didn't have the resolution of the relatively clean electron-positron collisions at SPEAR to measure its narrow width. With the news of a narrow state at 3.105 GeV, the high energy theory community exploded with speculations. The charmed quark hypothesis was just one of many competitors. Recall that the 1960-1970's was an era of discovery of many strong interaction states that were described by non-field theoretic approaches to high energy physics, such as Regge poles, bootstraps, etc. The field was stunned again two weeks later, on November 25, 1974, when SPEAR announced a second narrow peak at energy 3.695 GeV [77]! This challenged all the speculations circulating worldwide. The charm hypothesis was the most appealing to myself and collaborators since we were students of deep inelastic scattering and local field theory. The charm hypothesis was critical to the phenomenology of the electroweak sector of the Standard Model: the four quark model of $u, d, s$ and $c$ quarks solved the problem of neutral strangeness changing weak currents (the GIM mechanism [78]) of the three quark model. In addition, for the cancellations of the GIM mechanism to work effectively, the charm quark could not be too heavy. There were estimates that its mass $m_c \le 2.0 - 2.5\,\text{GeV}$ which put it inside the interesting range to explain the new resonances. In fact, the conventional quark model of mesons and baryons predicted that the charmed meson threshold of the SPEAR experiment, the minimum energy to produce two free charmed mesons, each consisting of a charmed quark and a light ($u, d$, or $s$) quark or anti-quark, should be $M_c = 2m_c + 0.7\,\text{GeV}$. Since the second state at 3.695 GeV was very narrow, $M_c$ had to be above 3.695 GeV. So, if $m_c$ lay in the range 1.5-2.0 GeV, the charm hypothesis appeared to be compatible with all the known data. The only "fly in the ointment" was that SPEAR had not announced the discovery of charmed mesons above 3.695 GeV. Nervous charm enthusiasts worried that maybe the charm idea was flawed! Following Ref. [74], the new states were tentatively called "charmonium", in analogy to positronium. Then the 3.105 state would be the $1^3S_1$ state of a $c$ and $\bar{c}$, and the 3.695 would be the $2^3S_1$. $S$-waves were required so that the $c$ and $\bar{c}$ would couple directly to the virtual photon created in the direct channel when the electron and positron annihilated. I recall that when these ideas were first discussed, many researchers sought to understand positronium better and ran off to their physics libraries and read Schwinger's classic works on the subject! Positronium spectroscopy had been calculated in great detail. This was possible because the static electron-positron interaction potential was just Coulomb's law. One needed the generalization of this interaction potential to strong interactions, QCD, to repeat those exercises for charmonium. At short distances it was plausible to assume a Coulomb-like formula with the fine structure constant replaced by $\alpha_s = g^2/4\pi$, where $g$ is the strong coupling constant of QCD. In fact, $g$ should be the running coupling, a scale dependent quantity, and $\alpha_s$ should be small, say $\alpha_s \sim 0.2$ for mass scales of $\sim 2\,\text{GeV}$ to accommodate the success of the parton model in deep inelastic scattering where experiments suggested that the parton distribution functions satisfy Bjorken scaling to good approximation for $Q^2 \approx 2 - 3\,\text{GeV}^2$. Next, one needed the potential at intermediate distances, where the $c\bar{c}$ pair feels the QCD forces of confinement but the system is below the charm threshold so that screening by light quarks is not yet active. Studies of model field theories of confinement [79] and the lattice version of QCD [80, 81] led to the idea that chromo-electric flux tubes form in this kinematic region and lead to a linear confining potential between heavy colored quarks. These ideas lead to the static $c\bar{c}$ potential [82],

$$V(r) = -\frac{\alpha_s}{r}\{1 - \frac{r^2}{a^2}\} \tag{2.1.2}$$

where $a$ sets the scale of the linear potential. The need for the linear term in Eqn. (2.1.3) was actually compelling in the original data. The ratio of the squares of the wave functions of the two charmonium states at the origin was called

$$\eta = |\frac{\psi(1^3S_1; r = 0)}{\psi(2^3S_1; r = 0)}|^2 = \frac{3.105\,\Gamma_e(3105)}{3.695\,\Gamma_e(3695)} \approx 1.4 - 1.7 \tag{2.1.3}$$

where we related the wave functions at the origin to the electronic width of each state and used early data to evaluate $\eta$. What do the values 1.4-1.7 imply about the potential? One can check that for a harmonic potential $\eta = 2/3$, for a linear potential $\eta = 1$ and for a Coulomb potential $\eta = 8$. So, to accommodate Eqn. (2.1.3), a combination of a linear confining potential and Coulomb potential was preferred. In Ref. [82] the parameters in the potential $(\alpha_s, a)$ were determined from the experimental data of the day by solving the radial Schrodinger equation and imposing the constraints: 1. The mass difference between the two charmonium states is 0.59 GeV, 2. $\Gamma_e(3105) = 5.5\,\text{keV}$, 3. $m_c$ should lie between 1.5 and 2.0 GeV, and 4. $\alpha_s$ should be between 0.2 and 0.3. At this point the authors of Ref. [82] needed a convenient computer program to solve the radial Schrodinger equation with a potential of the form Eqn. (2.1.2). Luckily, we had access to a skilled computational physicist with a trove of software programs! That computational physicist was K. G. Wilson who used numerical methods to teach undergraduate quantum mechanics. Remember that this was 1974 when universities had computer centers with IBM mainframes driven by punch cards! A good fit was found with his program for $m_c = 1.6\,\text{GeV}$, $\alpha_s = 0.2$, and $a = 2\,\text{fm}$. It was important to check that these parameters led to a non-relativistic description of the charmonium bound states. In fact, the average velocity-squared of the charmed quarks in the bound states was computed to be $(v/c)^2 \leq 1/25$. The bound states of the $c\bar{c}$ system that resulted are shown in Fig. 2.1.2.

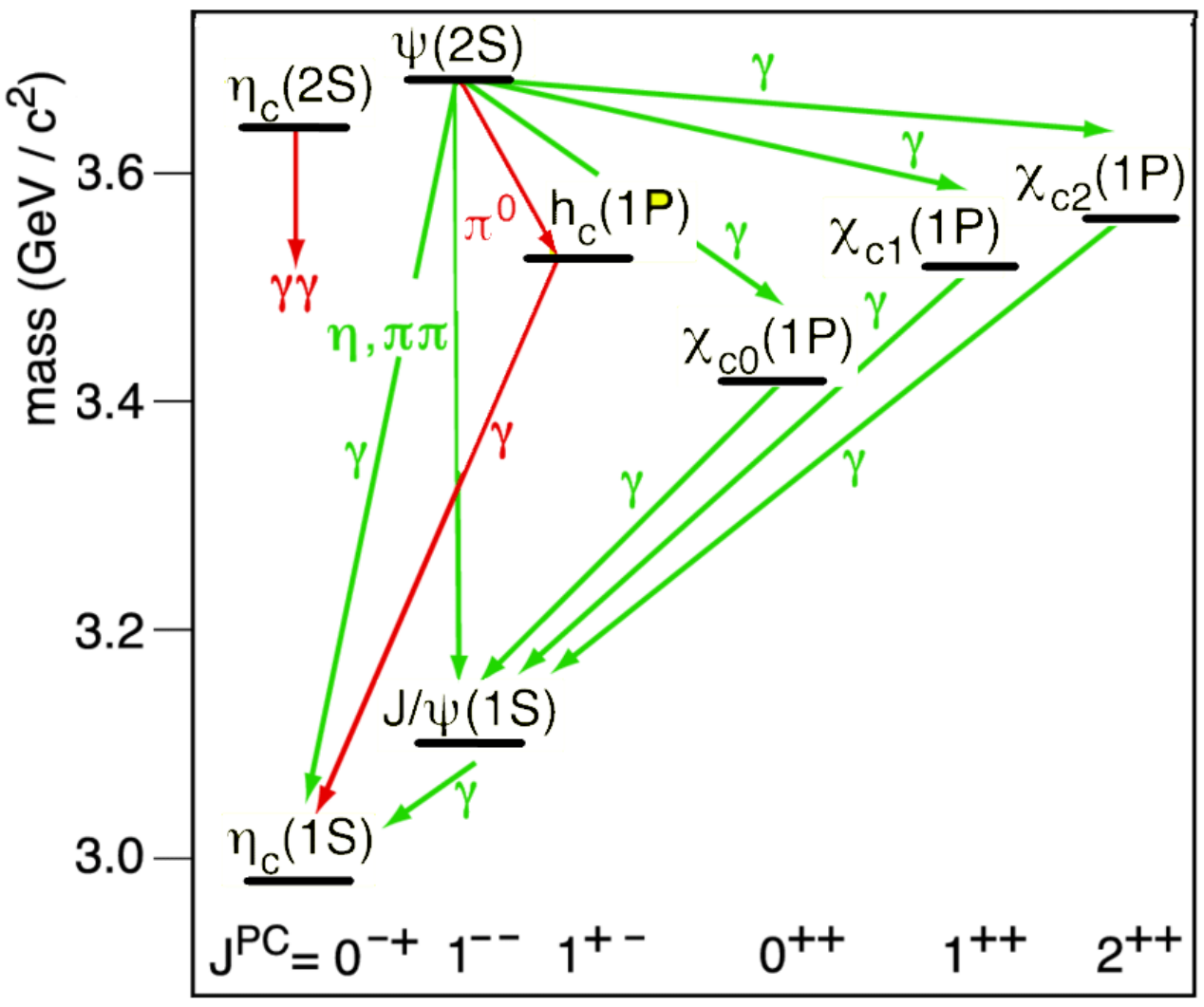

**Fig. 2.1.2** Charmonium Spectroscopy. Note the P-waves $^3P_J$ and the Radiative Transitions.

The most relevant result in Fig. 2.1.2 was the existence of the P wave states that lie between the 3.105 and 3.695 GeV states. For a pure Coulomb potential the P wave states would be degenerate with the 3.695 state. However, for a linear potential, the $2^3S_1$ state resides at higher energy than the P wave states, as shown in the figure, because the $2^3S_1$ has a radial node. The existence of these states led to the main point of Ref. [82]: there are additional states which could be found experimentally at SPEAR and they constitute strong, new evidence for the charm hypothesis! Strong E1, electric dipole, transitions would produce monochromatic photons when the 3.695 state decays to one of the P waves and then additional monochromatic photons should appear when each P wave decays to the 3.105 state! These monochromatic photons should be "easy" to find at SPEAR because it had a $4\pi$ general purpose detector, the Mark I. The energies of the P waves and the strengths of the E1 transitions followed from the wave functions found from the radial Schrodinger equation. These results were catalogued in Ref. [82] and were refined in later more ambitious publications. Of course, the wave functions and the radiative transition rates depend much more sensitively on the parameters in the potential than the energies of the P waves themselves. In any case, the predictions of Ref. [82] were reasonable guides for the experimental program which discovered the states and the radiative transitions in 1976, the same year that the charmed D mesons were also identified in the final states of the electron-positron collisions! Many more predictions and calculations were presented in Ref. [82]] and in similar works done by other groups [83]. Some of these points will be discussed in later chapters in this journal review. In addition, more sophisticated potentials than Eqn. (2.1.2) were eventually studied. Tensor interactions, fine and hyperfine interactions were added in, and their effects are shown in some of the splittings in Fig. 2.1.2 (Ref. [82, 83]). And the influence of the nearby threshold at $M_c$ on the bound states was also accounted for. All of these developments did not change the main thrust of Ref. [82]: the existence of the P wave states and their radiative transitions were special to the charm quark interpretation of the SPEAR experiment and gave additional motivation to the early acceptance of the Standard Model.

### 2.1.3 Electron-Positron Colliders at Stanford

Now let's change the viewpoint of this article and turn to the accelerator physicists and the experimentalists at SPEAR. There is a cliché that behind every invention there is a visionary. In the case of electron-positron colliders, one of the field's several visionaries was definitely Gerry O'Neil. Other visionaries were Burt Richter and Martin Perl. Professor O'Neil taught me physics in college, but he was more interested in building accelerators to collide electrons and positrons head-on in their center of mass frame to create pure electromagnetic energy and search for new states of matter. I recall that he traveled to Novosibirsk, where a collider was being constructed, several times during a one semester undergraduate course on modern physics. Upon each return he "debriefed" his class on the progress of his efforts. In 1965 Gerry O'Neil and others from Princeton and Stanford built two 300 MeV electron storage rings in the High Energy Physics Laboratory (HEPL) at Stanford. These rings resulted in electron-electron collisions which successfully increased the limits of validity of Quantum Electrodynamics. However, it was basically a "single experiment" machine, so during construction Gerry and his collaborators also sketched an outline of a 3ĠeV electron-positron colliding beam facility. These ideas evolved into the blueprints for the famous SPEAR collider at SLAC. To many persons' surprise, just as electron-positron collider ideas were gaining traction, Gerry's visionary ideas moved in a different direction: to outer space projects, such as a permanent space station in an earth orbit. He left the fledgling field of colliders just as it was about to yield great discoveries!

The construction of SPEAR began in 1970 under the direction of Burt Richter and John Rees, and it was completed quickly (in 20 months, four months ahead of schedule) in 1972 and at modest cost. The final SPEAR design was the result of several revisions, forced on the group by budget restrictions and engineering considerations. During one of the revisions, the two planned rings for the electrons and positrons became one and SPEAR was no longer asymmetric. Nonetheless, the inventors kept the appealing name "SPEAR"!

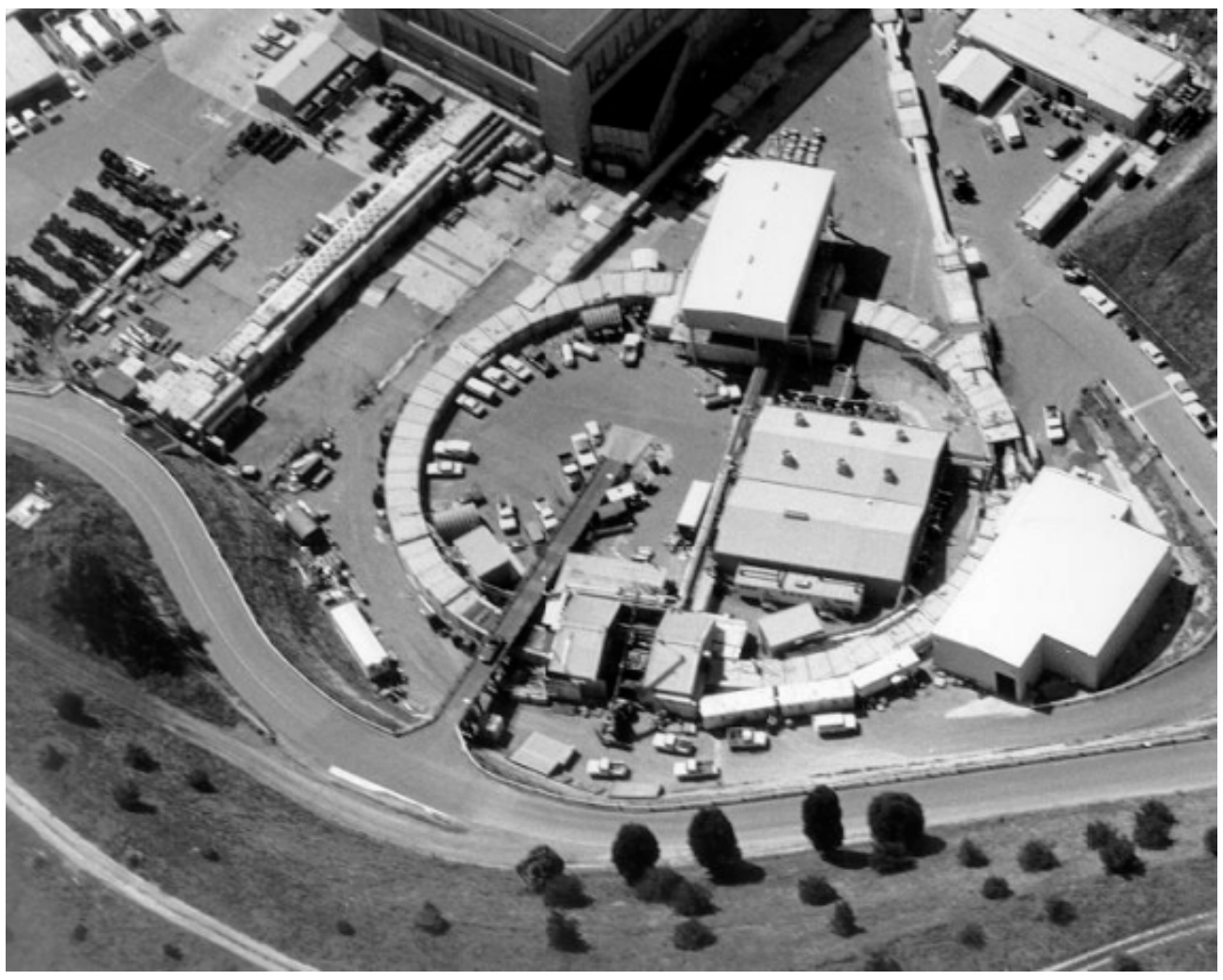

**Fig. 2.1.3** The 80 m SPEAR Ring in a parking lot at SLAC. The photo also shows the separate $e^+$ and $e^-$ beam lines and the detector hall.

Wolfgang Panofsky was still the Director of SLAC and had lobbied the US Congress and the funding agency, the Atomic Energy Commission (AEC), the predecessor of the Department of Energy, to fund the construction of SPEAR as a federal project. However, there were many projects competing with SPEAR at the time, and it did not achieve federal project status. However, Panofsky and Richter did not want to delay its construction, so the AEC allowed SPEAR to be built using ordinary laboratory operating funds! This meant that it had to be done cheaply. Some have estimated the cost between 2 to 5 million dollars. So, the usual idea of having the accelerator constructed underground within an enclosed building had to be abandoned. SPEAR was built outside on a parking lot (Fig. 2.1.3), with concrete blocks providing the shielding!

Of course, the accelerator needed a detector or two at its beam intersection regions. Richter and others formed a Berkeley/Stanford team to design and build a multipurpose detection system surrounding one of the SPEAR interaction regions (Fig. 2.1.4). The result was the Magnetic Detector or Mark I. This was the first $4\pi$ general purpose detector. It proved crucial in the coming discovery process. Other detector designs with limited angular apertures would have suffered from the relatively low statistics of the early machines and wouldn't have operated as well with diverse final states consisting of photons, leptons and various light mesons.

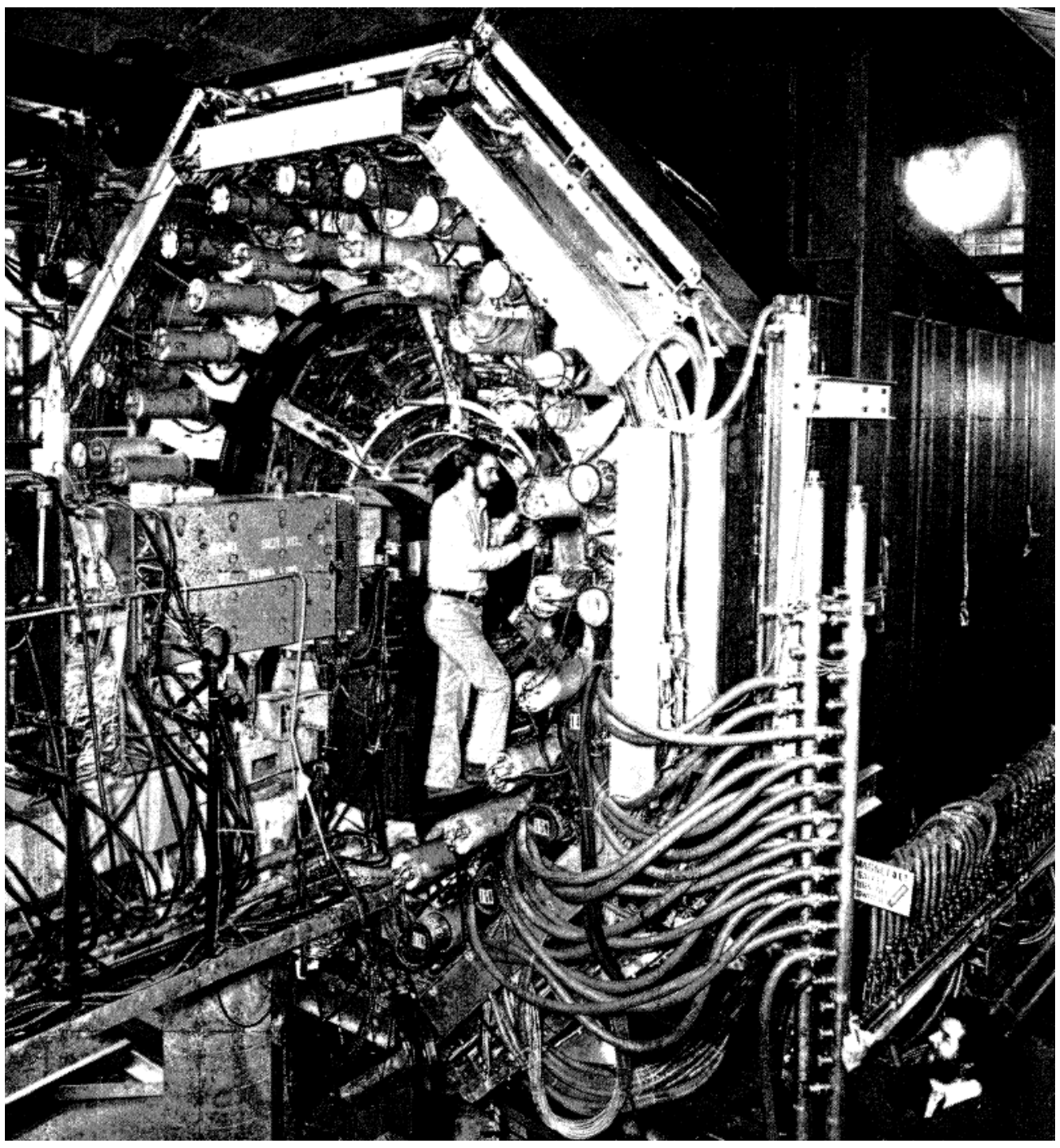

**Fig. 2.1.4** Photograph of the wide angle SLAC-LBL Mark I detector in 1974.

It was clear at the time that electron-positron colliders had many attractive properties: 1. All the energy of the beams goes into creating new particles, unlike fixed target machines, 2. The beams consist of point-like particles, so the interactions are simple and clean theoretically. However, they suffer from one limitation: radiation losses. However, it turned out that "one man's problem is another man's opportunity". From the beginning, several Stanford faculty members realized SPEAR's potential to produce useful synchrotron radiation, so they asked Panofsky and Richter to devise a way to form an X-ray beam out of SPEAR. The X-ray synchrotron radiation emitted by the circulating beams in the machine was much higher in intensity, by a factor of 10 to 100, than any other facility in the world. It could be used for imaging and structural analysis in many areas of research, from semiconductor materials to protein molecules. So, Richter's team attached an extra vacuum chamber to SPEAR and made provision for a hole in the shielding wall for the beamline. This was the start of The Stanford Synchrotron Radiation Project (SSRP). Even though it began as a parasitic operation, synchrotron radiation represented an unparalleled opportunity!

Richter also saw the SLAC Linac as a light source. These ideas led to the invention and development of a undulator so that the Linac's electron beam could be-come the source for the most intense Free electron Laser (FEL) on the planet. The LCLS (Linear Collider Light Source) was born in 2009. It has led to revolutions in our understanding of the temporal dynamics of atoms, molecules and condensed matter systems. This is another story which we can't cover here, but it is amusing to understand that a "problem" with circular colliders grew into a new generation of accelerator facilities!

### 2.1.4 The Revolution Begins

In the spring of 1973, SPEAR began to gather high-energy physics data. By the next year, the machine was measuring very erratic but generally much larger than expected values of R, Eqn. (2.1.1), the ratio of hadron production to lepton production. These early measurements were done with wide energy resolution, several hundred MeV, to produce and measure many interactions and final state particles. But there were "inconsistencies" in the data: small changes in the beam energies sometimes led to large changes in the observed value of $R$. These were the first signs of a new particle, which Richter's team called the "$\psi$". "Nobody dreamed that there was any state, particle, that was as narrow in width as the $W$ turned out to be," said Richter in 2003. "So the first question was what the hell was wrong with the apparatus, is there something wrong with the computers, is there something wrong with the data taking?" [84]. No-one could find any such errors, and some researchers on the Mark I collaboration pushed to rescan the region. In fact, by this time, SPEAR had been upgraded and Robert Hofstadter, who was running an experiment at SPEAR's other detector, wanted to move on to higher energies. Finally, Richter decided to go ahead with rechecking the anomalous results, but only for one weekend in November 1974.

### 2.1.5 Minute-by-minute Developments in the SPEAR Control Room

"During the night of 9-10 November, the hunt began, changing the beam energies in 0.5 MeV steps. By 11.00 a.m. Sunday morning the new particle had been unequivocally found. A set of cross section measurements around 3.1 GeV showed that the probability of interaction jumped by a factor of ten from 20 to 200 nanobarns. In a state of euphoria, the champagne was cracked open and the team began celebrating an important discovery. While Gerson Goldhaber retired to write up the findings 'on-line' for immediate publication, Fig. 2.1.5, it was decided to polish up the data by going slowly over the resonance again. The beams were nudged from 1.55 to 1.57 MeV and everything went crazy. The interaction probability soared higher; from around 20 nanobarns the cross section jumped to 2000 nanobarns and the detector was flooded with events producing hadrons. Pief Panofsky, the Director of SLAC, paced around the control room invoking the Deity in utter amazement at what was being seen. This heavy particle, displaying such extraordinary stability, they called '$\psi$' and they announced it in a paper beginning with the words 'We have observed a very sharp peak'. Within hours of the SPEAR measurements, the telephone wires across the Atlantic were humming as information, enquiries and rumors were exchanged" [85].

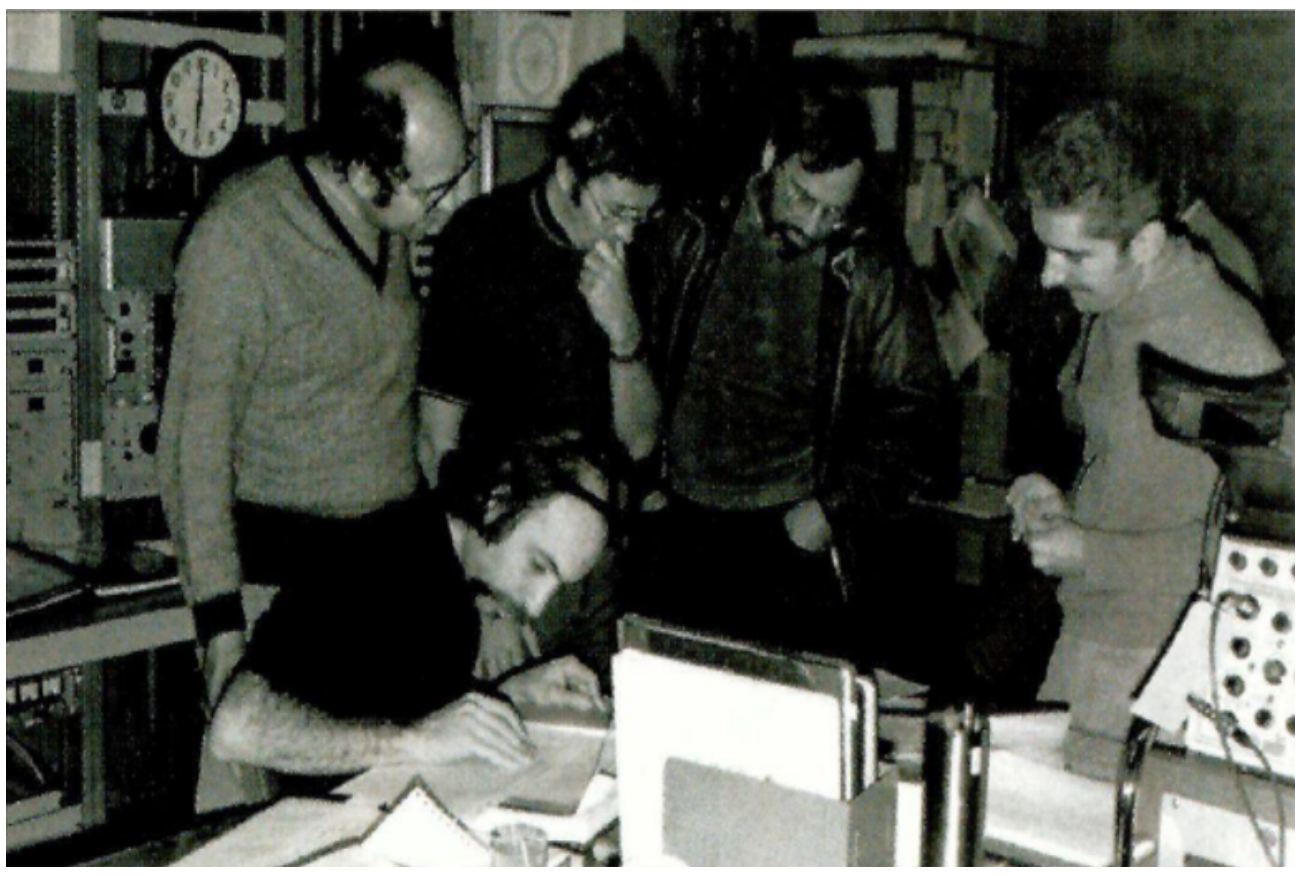

**Fig. 2.1.5** The SPEAR Control Room during the Big Night. SLAC and LBL physicists analyzing the raw data.

Just two weeks later, the scene repeated itself, except at a higher energy, 3.695 GeV. And the next S-wave charmonium state was found. Physicists around the country had "befriended" various members of the SLAC/ LBL group in the control room by now and news of the new resonance spread across the country within minutes. I heard about it in an early morning phone call with bj. I also learned that it was he who had suggested the high resolution scans in energy that led to both discoveries!

SPEAR meanwhile continued to yield breakthroughs. In 1976 the P-waves were discovered through their radiative transitions [86] and charmed mesons [87] were found above threshold as well.

Those were the days!

### 2.1.6 The Path Forward. Hamiltonian Lattice Gauge Theory and Statistical Field Theory.

After Ref. [82] was published, it was time to move on to more fundamental considerations. I believed that the most important implications of the potential model had been made and working through additional details was less important. Instead, there were major challenges in developing an approach to QCD that would lead to systematic, potentially exact, predictions of the theory. This was the thrust of Wilson's lattice formulation of QCD [80], which will be discussed at length elsewhere in this journal review. A Hamiltonian version of the theory [81] was also developed because it emphasized 1. The spectroscopy of the theory, and 2. The quantum character of the states. An added bonus of this development was a new formulation for strongly coupled systems for applications to condensed matter physics [88]. This development mirrors the past of SPEAR: SPEAR started out by establishing the Standard model of high energy physics, and now is pushing the frontiers of imaging, free electron lasers and quantum systems. In parallel, the lattice Hamiltonian form of strongly coupled gauge theories is playing a role in the development of Quantum Information Systems that may lead to new quantum computers and quantum detectors. These subjects are now the central themes in a new generation of studies and workshops on quantum physics [89, 90]. References [81] and [88], which were originally conceived for QCD, are proving useful here, and are, in fact, among the most cited publications in the 48 year history of lattice gauge theory. Perhaps, these contributions will inspire the next generation of theorists who will push the frontiers of strongly coupled gauge theories into the next era.

## 2.2 Experimental discovery of gluons

**Sau Lan Wu**

### 2.2.1 Yang-Mills non-Abelian gauge particles

It was in 1954 when Chen Ning Yang and Robert Mills, who was a graduate student, shared the same office at the Brookhaven National Laboratory and developed their non-Abelian gauge theory. Their office was shared with another famous physicist Burton Richter, who was also a graduate student at that time. Almost exactly twenty five years later, the first Yang-Mills non-Abelian gauge particle was observed at the German National Laboratory called Deutsches Elektronen-Synchrotron (DESY). Here are some of the interesting dates. The idea of Yang and Mills was first presented at the April 1954 meeting in Washington, DC of the American Physical Society and the full Yang-Mills paper was submitted for publication on June 28, 1954 [36]. The first public announcement for the experimental discovery of the first Yang-Mills gauge particle was made at the Neutrino 79 conference on June 18-22, 1979 [91], and the first full paper was received for publication on August 29, 1979 [92].

The word "gluon" was originally introduced by Murray Gell-Mann to designate a hypothetical neutral vector field [14] coupled strongly to the baryon current, without reference to color. Since then, the meaning of this word has changed: nowadays, this word "gluon" is used exclusively to mean the Yang-Mills non-Abelian gauge particle for strong interactions.

### 2.2.2 Harvard to M.I.T. to Wisconsin

After being awarded my Ph.D. degree at Harvard University, Samuel S. S. Ting of M.I.T. kindly offered me a postdoctoral position in his group. A few years later, I felt that, for the development of my career in physics, it was time for me to get a faculty position. Sam then helped me to look for a faculty position at the University of Michigan, where he received his own doctoral degree. I got into contact with Michael Longo, a professor of physics there, and he was very supportive. Therefore I applied to the University of Michigan. Since thanks to Longo I got on the so-called short list of candidates, I was invited to go to Ann Arbor for an interview.

In the meantime, I contacted David Cline, a professor of the University of Wisconsin I had met before. David told me that he would forward my name to Ugo Camerini, a colleague of his at Wisconsin. I contacted Ugo. Shortly before my scheduled interview at Ann Arbor, I got a telegram from the University of Michigan saying that the position had been given to somebody else. I hesitated about going to that interview, but my friends told me that I should nevertheless keep the appointment. In the meantime, I got an invitation from the University of Wisconsin for an interview. Thus I traveled from Europe for an interview at Michigan first, and then continued to Wisconsin for another one.

I remember very well that, when I had the interview at the University of Wisconsin in Madison, Don Reeder took me out to dinner at an Italian restaurant close to the University Square and we had a very nice discussion. Don was at that time not only a Professor of Physics but also the Principal Investigator for the funding of experimental high-energy physics. Afterwards, I met with a number of faculty members in high-energy physics, and they were all very supportive. Again through the effort of Cline, I also got an offer from Fermilab. I had to make a decision, and I finally chose the University of Wisconsin. It was one of the best decisions I have made.

### 2.2.3 DESY

After becoming an assistant professor at the University of Wisconsin-Madison in 1977, I had to make the decision of what important problem in physics to tackle.

Once again, I got wise advice from David Cline, who had helped me so much. He told me: "Sau Lan, you do not need to work with anybody, and you have no boss. You are your own boss, and you decide what to work on." At that time, the Department of Energy gave one lump sum of money to the University of Wisconsin for the faculty members in experimental high-energy physics to share. From this funding, Don Reeder gave me the positions of three post-docs and one graduate student.

I spent the first months of my assistant professorship thinking about what physics to work on.

At that time, we knew of four quarks: the up quark, the down quark, the strange quark, and the newly discovered charm quark from the $J/\psi$, which has led to the Nobel Prize for Sam Ting and Burt Richter. The immediate and important question is: how do these quarks interact with each? For this, we knew very little at that time besides that this interaction is likely to be mediated by a Yang-Mills non-Abelian gauge particle — the gluon. See Sec. 2.2.1. In other words, while the electromagnetic interaction is transmitted by the photon, which is an Abelian gauge particle, this additional interaction is transmitted by a Yang-Mills non-Abelian gauge particle.

Indirect indication of gluons had been first given by deep inelastic electron scattering and neutrino scattering. The results of the SLAC-MIT deep inelastic scattering experiment [93–96] on the Callan-Gross sum rule were inconsistent with parton models that involved only quarks. The neutrino data from Gargamelle [97] showed that 50% of the nucleon momentum is carried by isoscalar partons or gluons. Further indirect evidence for gluons was provided by the observation of scale breaking in deep inelastic scattering [98–100]. The very extensive neutrino scattering data from BEBC and CDHS Collaborations [101–103] at CERN made it feasible to determine the distribution functions of the quark and gluon by comparison with what was expected from QCD, and it was found that the gluon distribution function is sizeable. This information about the gluon is interesting but indirect. The discovery of the gluon requires direct observation.

During my first year as an assistant professor at the University of Wisconsin, I was fascinated by the Yang-Mills non-Abelian gauge theory. This was to be contrasted with the experimental situation at that time: while photons were everywhere in the detectors, no Yang-Mills gauge particle had been observed in any experiment.

From these considerations, I formulated the following problem for myself: how could I discover experimentally the first Yang-Mills gauge particle?

From my previous experience with electron accelerators and proton accelerators at DESY and BNL, it was soon clear to me that the experimental discovery of the first Yang-Mills gauge particle was more likely at an electron machine rather than a proton machine. At that time, two electron-positron colliding beam accelerators were being built: PEP at SLAC and PETRA at DESY; after visiting both SLAC and DESY, I decided that PETRA was a better choice for me.

At PETRA (Positron-Electron Tandem Ring Accelerator), there were five experiments: CELLO, JADE, MARK J, PLUTO, TASSO. I approached first the PLUTO Collaboration and then the JADE Collaboration, but nothing worked out. Then my luck changed completely: I ran into Björn Wiik, one of the two co-spokesman of the TASSO Collaboration, the other one being Günter Wolf. Björn asked me what I was doing; when I told him my situation, he was surprised and said to me: "Come to see me in my office this afternoon." When I went to his office, he asked me: "Why don't you join the TASSO Collaboration instead?" I said that I would love to do that. Björn said that he would talk to Günter and also to Paul Söding, a senior physicist in TASSO, and let me know. Thanks to Björn, this was how I became a member of the TASSO Collaboration at DESY. All three of them, Björn, Günter, and Paul, are excellent physicists.

After becoming a member of the TASSO Collaboration, the physics problem that I formulated for myself took on a concrete form: how could I discover experimentally the first Yang-Mills gauge particle with the TASSO detector?

A feature of the TASSO detector is the two-arm spectrometer, which leads to the name TASSO — Two - Arm Spectrometer SOlenoid. The end view of this detector, i.e., the view along the beam pipe of the completed detector, is shown in Fig. 2.2.1. When TASSO was first moved into the PETRA beams in 1978, not all of the detector components shown in Fig. 2.2.1 were in working order. For my purpose of the experimental discovery of the first Yang-Mills non-Abelian gauge particle, the most important component of the TASSO detector was the drift chamber, which was already functioning properly.

### 2.2.4 Three-jet events

One of the simplest ways to produce a photon — the Abelian gauge particle for electromagnetic interactions — is through electron bremsstrahlung process, i.e.,

$$e\, e \rightarrow e\, e\, \gamma.$$

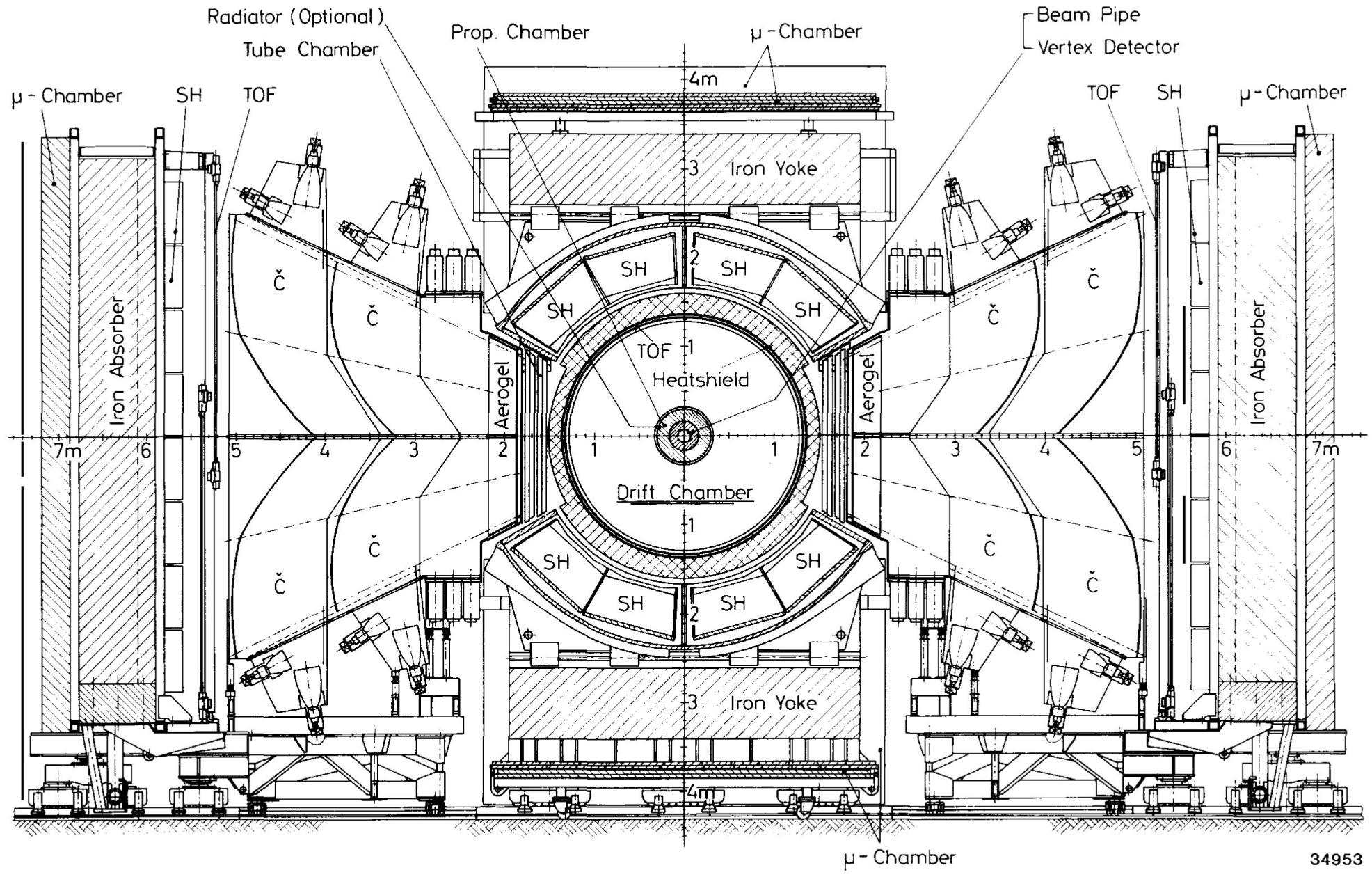

**Fig. 2.2.1** End view of the TASSO detector.

The analogous process for the production of the gluon is [104]

$$e^+e^- \to q\bar{q}g$$

where $q$ is the quark of Gell-Mann [17] and Zweig [18], and $g$ is the gluon — the Yang-Mills non-Abelian gauge particle for strong interactions.

Once I realized that this is the way to discover the gluon experimentally, I faced the following two major problems.

(1) How can these production processes $e^+e^- \to q\bar{q}g$ be found in the TASSO detector?

(2) How high does the center-of-mass $e^+e^-$ energy have to be for this process to be seen clearly?

A couple of years before I became a faculty member at the University of Wisconsin, the production process

$$e^+e^- \to q\bar{q}$$

was observed at the SPEAR $e^+e^-$ collider at SLAC [105]. In the MARK I detector at SPEAR, both the quark $q$ and the anti-quark $\bar{q}$ were observed as jets, i.e., groups of particles moving in nearly the same direction. With this experimental information from MARK I, I had to make my best guess as to how the gluon bremsstrahlung process $e^+e^- \to q\bar{q}g$ would look like in the TASSO detector. Since the gluon is the Yang-Mills non-Abelian gauge particle for strong interactions, it is itself a source for gluon fields. It therefore seemed reasonable to believe that the gluon in the gluon bremsstrahlung process would be seen in the detector also as a jet, just like the quark and the antiquark.

Therefore the gluon bremsstrahlung process $e^+e^- \to q\bar{q}g$ leads to three-jet events.

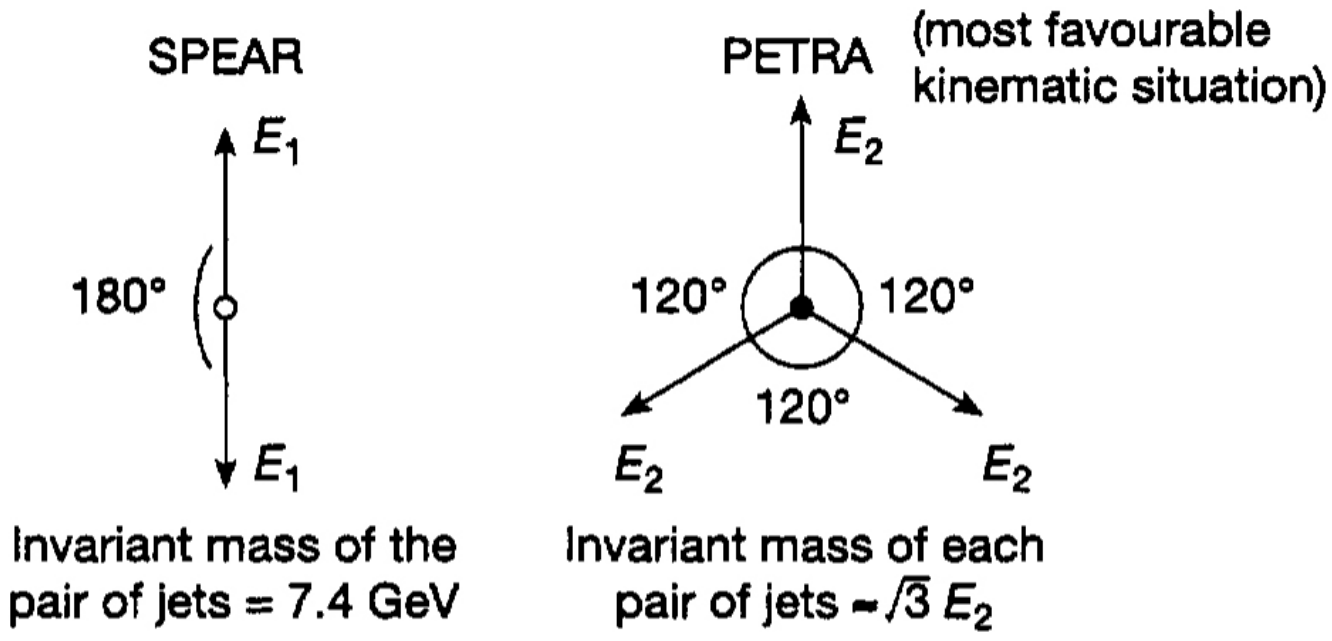

**Fig. 2.2.2** Two-jet and three-jet configurations at SPEAR and PETRA respectively.

Using the SPEAR information on the quark jets from the process, $e^+e^- \to q\bar{q}$, I convinced myself that three-jet events, if they were produced, could be detected once the PETRA energy went above three times the SPEAR energy i.e., $3 \times 7.4 \sim 22\,\text{GeV}$. The arguments were as follows:

Figure 2.2.2 shows a comparison of the two-jet configuration at SPEAR with the most favorable kinematic situation of the three-jet configuration at PETRA. If the two invariant masses are taken to be the same, i.e., $\sqrt{3}E_2 \approx 7.4\,\text{GeV}$, then the total energy of the three jets is $3E_2 \approx 13\,\text{GeV}$, which must be further increased because each jet has to be narrower than the SPEAR jets. This additional factor is estimated to be $180°/120° = 1.5$, leading to about 20 GeV. Phase space considerations further increase this energy to about 22 GeV.

This answers the question (2) above.

This estimate of 22 GeV was very encouraging because PETRA was expected to exceed it soon; indeed, it provided the main impetus for me to continue the project to discover the first Yang–Mills non-Abelian gauge particle.

At the same time, l had to address the question (1) above: how could I find three-jet events at PETRA? I made a number of false starts until I realized the power of the following simple observation. By energy-momentum conservation, the two jets in $e^+e^- \to q\bar{q}$ must be back-to-back. Similarly, the three jets in $e^+e^- \to q\bar{q}g$ must be coplanar. Therefore, the search for the three jets can be carried out in the two-dimensional event plane, the plane formed by the momenta of $q, \bar{q}$ and $g$. A few pages of my notes written in June 1978 and further historical details can be found in Ref. [106].

The procedure of mine did not identify which jet would be the gluon. Still, this procedure has a number of desirable features.

- First, all three jet axes are determined, and they are in the same plane. This is the feature that played a central role in the later determination of the spin of the gluon.
- Secondly, particle identification is not needed, since there is no Lorentz transformation.
- Thirdly, the computer time is moderate for the "slow" computers at that time even when all the measured momenta are used.
- Finally, it is not necessary to have the momenta of all the produced particles; it is only necessary to have at least one momentum from each of the three jets. Thus, for example, my procedure works well even when no neutral particles are included.

This last advantage is important, and it is the reason why this procedure is a good match to the TASSO detector at the time of the PETRA turn-on.

I had Georg Zobernig as my post-doc; he was and is excellent in working with computers. My procedure of identifying the three-jet events in order to discover the gluon, programmed by Zobernig on an IBM 370/168 computer, was ready before the turn-on of PETRA in September of 1978. For that time in 1978, the programming was highly non-trivial. In his later publications, he has used the name Haimo Zobernig.

### 2.2.5 Discovery of the Gluon

When we had obtained data for center-of-mass energies of 13 GeV and 17 GeV, Zobernig and I looked for three-jet events. It was not until just before the Neutrino 79 (International Conference on Neutrino, Weak Interactions and Cosmology at Bergen, Norway) in the late spring of 1979 that we started to obtain data at the higher center-of-mass energy of 27.4 GeV. We found one clear three-jet event from a total of 40 hadronic events at this center-of-mass energy. This first three-jet event of PETRA, as seen in the event plane, is shown in Fig. 2.2.3. When this event was found, Wiik had already left Hamburg to go to the Bergen Conference. Therefore, during the weekend before the conference, I took the display produced by my procedure for this event to Norway to meet Wiik at his house near Bergen. During this weekend, I also telephoned Günter Wolf at his home in Hamburg and told him of the finding. Wiik showed the event in his plenary talk "First Results from PETRA", acknowledging that it was my work with Zobernig by putting our names on his transparency of the three-jet event, and referred to me for questions. Donald Perkins of Oxford University took this offer and challenged me by wanting to see all forty TASSO events. I showed him all forty events, and, after we had spent some time together studying the events, he was convinced.

With these three-jet events, the question is: what are the three jets? Since quarks are fermions, and two fermions (electron and positron) cannot become three fermions, it immediately follows that these three jets cannot all be quarks and antiquarks. In other words, *a new particle has been discovered.*

The earliest papers related to the PETRA three-jet events are Refs. [91, 92, 107, 108] all by members of the TASSO Collaboration, and *TASSO Note 84*, June 26, 1979 (by Sau Lan Wu and Haimo Zobernig). Ref. [107] provides the method of analysis used in the four later papers, which all give experimental results.

Very shortly afterwards, the other experiments at PETRA — JADE, MARK J, and PLUTO Collaborations — published their own three-jet analyses. Their early papers related to the PETRA three-jet events are Refs. [109–111], and their results all confirm the earlier ones of TASSO. Since this discovery of the gluon was the highlight of the 1979 Lepton-Photon Conference at

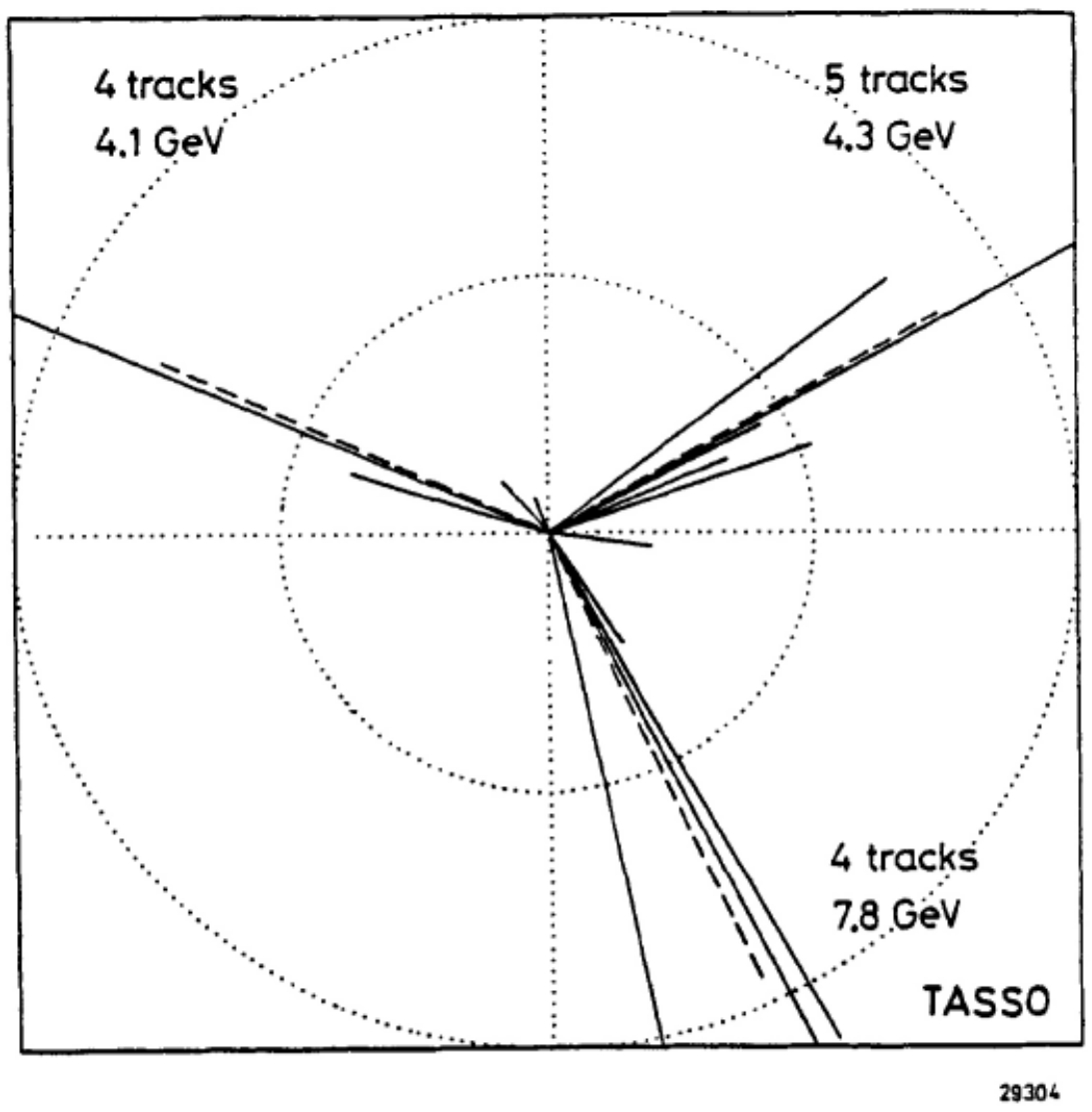

**Fig. 2.2.3** The first three-jet event from electron-positron annihilation, as viewed in the event plane. It has three well separated jets [91].

Fermi National Accelerator Laboratory (FNAL), Leon Lederman, Director of FNAL, called a press conference on the discovery of the gluon.

Because of the discovery of the gluon by the TASSO Collaboration, Söding, Wiik, Wolf, and I were awarded the 1995 European Physical Society High Energy and Particle Physics Prize. With my leading role in this discovery, I was chosen to give the acceptance speech at the EPS award ceremony.

This was how the first Yang-Mills non-Abelian gauge particle was discovered experimentally at DESY, Hamburg, Germany in the spring of 1979, a quarter of a century after the original paper of Chen Ning Yang and Robert Mills. Four years later, the second and third Yang-Mills non-Abelian gauge particles — the $W$ and $Z$ — were discovered at CERN by the UA1 and UA2 Collaborations [112–115].

The experimental discovery of these Yang-Mills non-Abelian gauge particles points to another prophetic feature of the original paper of Yang and Mills [36]: the mass of the first Yang-Mills gauge particle has been found to be nearly zero, while those of the second and third Yang-Mills gauge particle are quite high – about 80 GeV for the W and 91 GeV for the $Z$. The relevant sentence in the original paper [36] is the following: "We have therefore not been able to conclude anything about the mass of the **b** quantum." For further comments on this point, see pp.19-21 of [116].

### 2.2.6 Some later developments

The discovery of the gluon in 1979 was not only the discovery of a new elementary particle, but also the first elementary boson that has been seen experimentally as a jet. Indeed, it is so far the ONLY elementary boson seen this way. In principle, a scalar quark would share this property, but no scalar quark has ever been observed in any experiment.

The discovery of such a new type of elementary particle is guaranteed to lead to subsequent new understanding of fundamental physics, both experimental and theoretical. Here I will discuss one of the of the most important experimetal consequences of this 1979 discovery of the gluon; the role it plays in the 2012 discovery of the Higgs particle.

An important theoretical topic, the very recent understanding of the quark-gluon coupling constant $g_s$, is discussed in considerable detail in Sec. 3, and briefly in my Summary and Outlook, Sec. 2.2.8.

### 2.2.7 Role of gluon in the discovery of the Higgs particle [38–40]

Since the gluon is the Yang–Mills gauge particle for strong interactions, to a good approximation a proton consists of a number of gluons in addition to two $u$ quarks, one $d$ quark, and some sea-quarks. Since the coupling of the Higgs particle to any elementary particle is proportional to its mass, there is little coupling between the Higgs particle and these constituents of the proton. Instead, some heavy particle needs to be produced in a proton-proton collision, for example at LHC, and is then used to couple to the Higgs particle. Among all the known elementary particles, the top quark $t$, with a mass of 173 GeV$/c^2$, is the heaviest[117, 118].

The top quark, which may be virtual, is produced predominantly together with an anti-top quark or an anti-bottom quark [119]. Since the top quark has a charge of $+2/3$ and is a color triplet, such pairs can be produced by

(a) a photon: $\gamma \to t\bar{t}$;
(b) a $Z$: $Z \to t\bar{t}$;
(c) a $W$: $W^+ \to t\bar{b}$; or
(d) a $g$: $g \to t\bar{t}$.

As discussed in the preceding paragraph, there is no photon, or $Z$, or $W$ as a constituent of the proton. Since, on the other hand, there are gluons in the proton, (d) is by far the most important production process for the top quark.

Because of color conservation — the gluon has color but not the Higgs particle — the top and anti-top pair produced by a gluon cannot annihilate into a Higgs particle. In order for this annihilation into a Higgs particle to occur, it is necessary for the top or the anti-top quark to interact with a second gluon to change its color content. It is therefore necessary to involve two gluons, one each from the protons of the two opposing beams of LHC, and we are led to the diagram of Fig. 2.2.4 for Higgs production. This production process is called "gluon–gluon fusion" (also called "gluon fusion"). As expected from the large mass of the top quark, this gluon–gluon fusion is by far the most important Higgs production process, and shows the central role played by the gluon in the discovery of the Higgs particle in 2012.

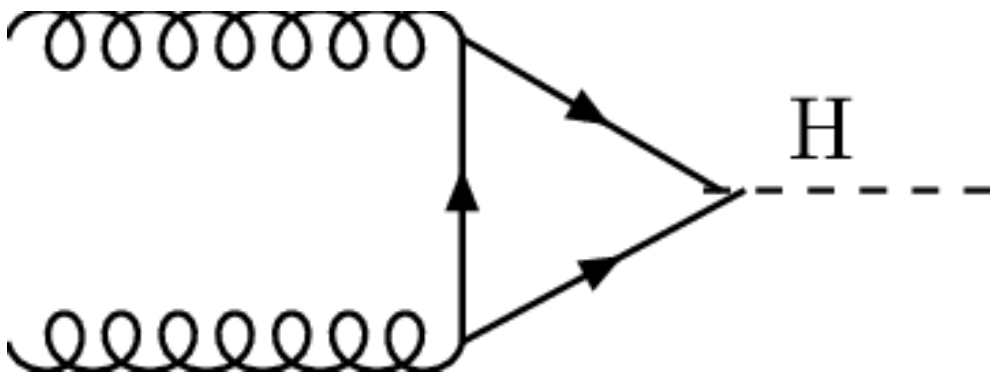

**Fig. 2.2.4** Feynman diagram for the Higgs (H) production by gluon–gluon fusion (also called gluon fusion).

The percentage of this gluon-gluon fusion contribution to the Higgs production cross section depends on the mass of the Higgs particle. For the actual mass of the Higgs particle, the gluon contributes, through this gluon-gluon fusion process, about 90% of Higgs production at the Large Hadron Collider. A more dramatic, but perhaps unfair, way of saying the same is that, if there were no gluon, the Higgs particle could not have been discovered for years!

### 2.2.8 Summary and outlook

One of the most influential papers in theoretical physics during the second half of the twentieth century — very likely the most important and influential one — is that of Yang and Mills published in 1954 [36]. The importance of this paper on the non-Abelian quantum gauge theory is due to that (1) it presents a completely new idea, and (2) it points out the direction for the later development of the understanding of particle physics.

Twenty five years later in 1979, the first such particle — the Yang-Mills non-Abelian gauge particle for strong interactions, later called the gluon, even though this word "gluon" refers originally to a different proposed particle — was experimentally discovered with the TASSO Collaboration at the German Laboratory DESY [91, 92].

Another thirty three years later in 2012, this gluon played a central role in the discovery of the Higgs particle [38–40] by the ATLAS Collaboration [120] and the CMS Collaboration [121] at CERN: this Higgs particle is produced predominantly through gluon fusion, i.e., the fusion of one gluon from one proton beam with another gluon from the opposing proton beam.

As soon as the gluon was discovered in 1979, the obvious question was immediately raised: What determines the strength of the gluon-quark coupling constant? I have kept this important question in my mind for forty years. The conventional answer is discussed in Sec. 3 below, but I have a novel idea about how the Standard Model might be modified to determine $g_s$. I refer to this idea as the "basic standard model." It is discussed in Refs. [122, 123].

## 2.3 Successes of perturbative QCD

**Yuri Dokshitzer**

Fifty years is a long time, though not for a theory as ambitious as QCD. To cover all the pQCD applications would be mission impossible. There are many review papers, both topical and anniversary, some good some excellent. My review is biased, focusing on issues that I personally find important and/or entertaining.

*QCD?*
Sure. It is undoubtedly the true microscopic theory of hadrons and their interactions. Whether it deserves a status of a well formulated Quantum Field Theory (QFT) is another matter. QCD is an ultimate proof of non-maliciousness of the God of physics. This theory is as amazing as it is embarrassing, in enabling us to predict so much while understanding so little.

*Perturbative?*
A perturbative (PT) approach means casting an answer as power series in a small expansion parameter. By calculating more terms of the series one aims at increasing accuracy of a theoretical prediction. The quark-gluon dynamics does offer such parameter: the QCD coupling. At small distances it becomes reasonably small thanks to asymptotic freedom, inviting us to draw and calculate Feynman diagrams for interacting quark and gluon fields.

*Successes?*
Countless experimental findings speak loudly and clearly in favor of pQCD. However, until the color confinement problem is solved, we have to invent hypotheses and build models linking quark-gluon dynamics and the hadron world. It is useful to keep this in mind when what is commonly referred to as a *QCD prediction* confronts reality.

By trial-and-error we learn.

### 2.3.1 pQCD: Domain of interest

The name of the pQCD kingdom is Hard Processes. We call “hard” any process involving hadrons where the energy-momentum that color objects exchange or acquire from (transfer to) colorless fields is much larger than the confinement scale $\mathcal{O}(\Lambda_{\rm QCD})$. Classical examples are $e^+e^-$ annihilation into hadrons, Deep Inelastic lepton-hadron Scattering (DIS), or the Drell–Yan process of production in hadron collisions of massive lepton pairs or any other heavy colorless objects like $W^\pm$, $Z^0$, $H$ bosons. To the same family belong production of heavy quarks and their bound states, as well as large-$p_T$ photons and hadron jets.

Heavy quarks are often thought to be more friendly towards pQCD than their light siblings. This is true, but not because a massive quark couples to the gluon field more weakly than a massless one. The QCD interaction strength is universal, as a matter of principle. An internal structure of a $D$ meson is as non-perturbative (NP) as that of $K$ or $\pi$. At the same time, heavy quarks are typically produced with relatively large transverse momenta $p_T \sim m_Q$ and are closer to one another inside the $Q\bar{Q}$ bound states. This is what actually explains that *friendliness* motto.

Sometimes pQCD applies even to light hadrons. This occurs when a hadron is put under a condition forcing its valence quarks to sit tight in order to hide their color. Small-size configurations dominate when an *initial state* hadron, in spite of having experienced a hit with large momentum transfer, is forbidden to break up and is asked to scatter elastically. Alternatively, a hadron can be squeezed by demanding its exclusive production in the *final state.*

This class of phenomena goes under the name of *color transparency.* Diffractive dissociation of an energetic pion on a nuclear target is a bright example. Normally a big nucleus would absorb the projectile. However, if an incident pion happens to be in a squeezed state, its valence quarks act as a small-size color dipole. Its interaction with the medium weakens and the pion gets a chance to penetrate the nucleus, defying the exponential attenuation wisdom. What one finds behind the target then is a pair of quark jets, because the probability for such a $q\bar{q}$ configuration to return back into a normal pion state is too small to be counted on.

Also pQCD unexpectedly finds its place in the $hA$ ($AA$) interaction environment where multiple scattering of a projectile effectively pushes up the characteristic hardness scale, $< k_T^2 > \propto A^{1/3}$, putting interesting physics like induced gluon radiation or jet quenching under pQCD control.

Whatever the hardness of the process, it is hadrons, not quarks or gluons, that hit the detectors. This makes the applicability of the pQCD approach, even to hard processes, far from obvious. One relies on plausible arguments (completeness, duality) and tries to learn from inclusive hadron observables that are *less vulnerable* to our ignorance about confinement.

### 2.3.2 pQCD: Domain of applicability

The main lesson we learned from confronting QCD expectations with reality is quite encouraging. The strong interaction that is supposed to hold color bearers inside hadrons turns out to be not so strong, if you think about it. The strong color force gets easily screened at large distances by light quarks that pop up from the vacuum. We have not yet mastered this mechanism quantitatively. Meanwhile, the very fact that the confinement happens to be “soft” dramatically enlarges the pQCD playing ground.

*Precocious pQCD*
. The parton model [124] pictured electron–nucleon interactions as *elastic* scattering of an incident electron that transfers, via virtual photon exchange, momentum $q$ to a point-like constituent of the target hadron — a parton. Inelasticity of the $ep$ collision is characterized by a dimensionless Lorentz-invariant parameter $x = -q^2/2(q\cdot P)$ which determines an invariant mass $W$ of the final hadronic system: $W^2 - M_P^2 = 2(q\cdot P)(1-x)$. The physical meaning of the Bjorken variable $x$ becomes transparent in a reference frame where the virtual photon has zero energy component, $q_0 = 0$, and collides with the proton head-on (Breit frame). Here $x$ becomes a fraction of the large proton momentum $P$ carried by the hit parton ($p_{\rm part} \simeq xP$).

This picture culminated in the Bjorken hypothesis: that the probability of finding a given parton inside the nucleon is independent from the momentum transfer $q^2$. The Bjorken scaling was expected to hold *asymptotically,* that is when $|q^2|$ is so large as to ensure insignificance of any re-interaction between constituents. In the Bjorken limit $|q^2| \to \infty$ the elastic $ep$ cross section dies out, while proton breakup into large-mass hadron systems dominates: hence *Deep* and *Inelastic*.

The first SLAC–MIT observation of DIS sent a striking message. Defying expectation, the scaling regime manifested itself surprisingly early, right above 1 GeV momentum transfer, as shown in Fig. 2.3.1. Charged constituents (read: quarks), probed with better than 0.2 fm resolution, behaved as free objects. And 50 years later they still do.

Another evidence in favor of *precocious freedom* comes from $e^+e^-$ annihilation into hadrons which provides the cleanest environment for exploring QCD. Here all the murky hadron dynamics is restricted to the final state, and we can watch what happens to a pair of bare quarks created in the annihilation point and moving apart with light speed.

Figure 2.3.2 shows the total hadroproduction cross section, normalized by the QED cross section $e^+e^- \to \mu^+\mu^-$ as a function of annihilation energy $\sqrt{s} = 2E_q$. We see that first the quark and the antiquark interact in the final state producing hadron resonances (vector mesons $\rho$, $\omega$, $\phi$). As soon as the quark energy exceeds 1 GeV, the stormy sea calms down abruptly and turns into still waters. Quarks with larger energies forget about one another and behave as free particles. They separate unimpeded and develop their private multi-hadron images — jets. (The story repeats above the charm threshold.)

This plot contains more than a mere counting of the number of families of colored quarks,

$$R(s) = \frac{\sigma_{e^+e^-\to\text{hadr.}}}{\sigma_{e^+e^-\to\mu^+\mu^-}}, \qquad R_{\text{q.m.}} = N_c \sum_f e_f^2.$$

Notice a slight non-linearity of the pQCD red line in Fig.2.3.2. Its origin — a QCD correction to the annihilation cross section due to gluon radiation:

$$\frac{R(s)}{R_{\text{q.m.}}(s)} = 1 + \frac{3C_F}{4}\frac{\alpha_s(s)}{\pi} + \ldots,$$

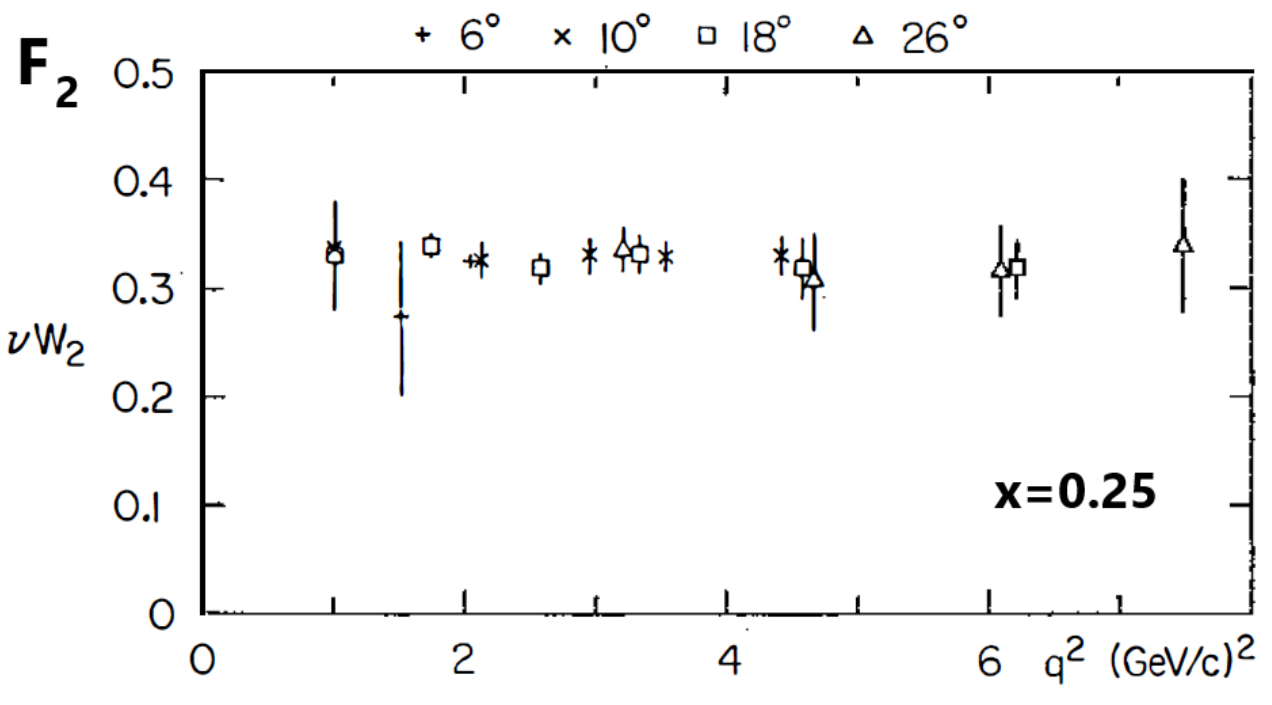

**Fig. 2.3.1** DIS structure function $F_2 = \nu W_2$ *precociously* scales with momentum transfer $q^2$ [95]

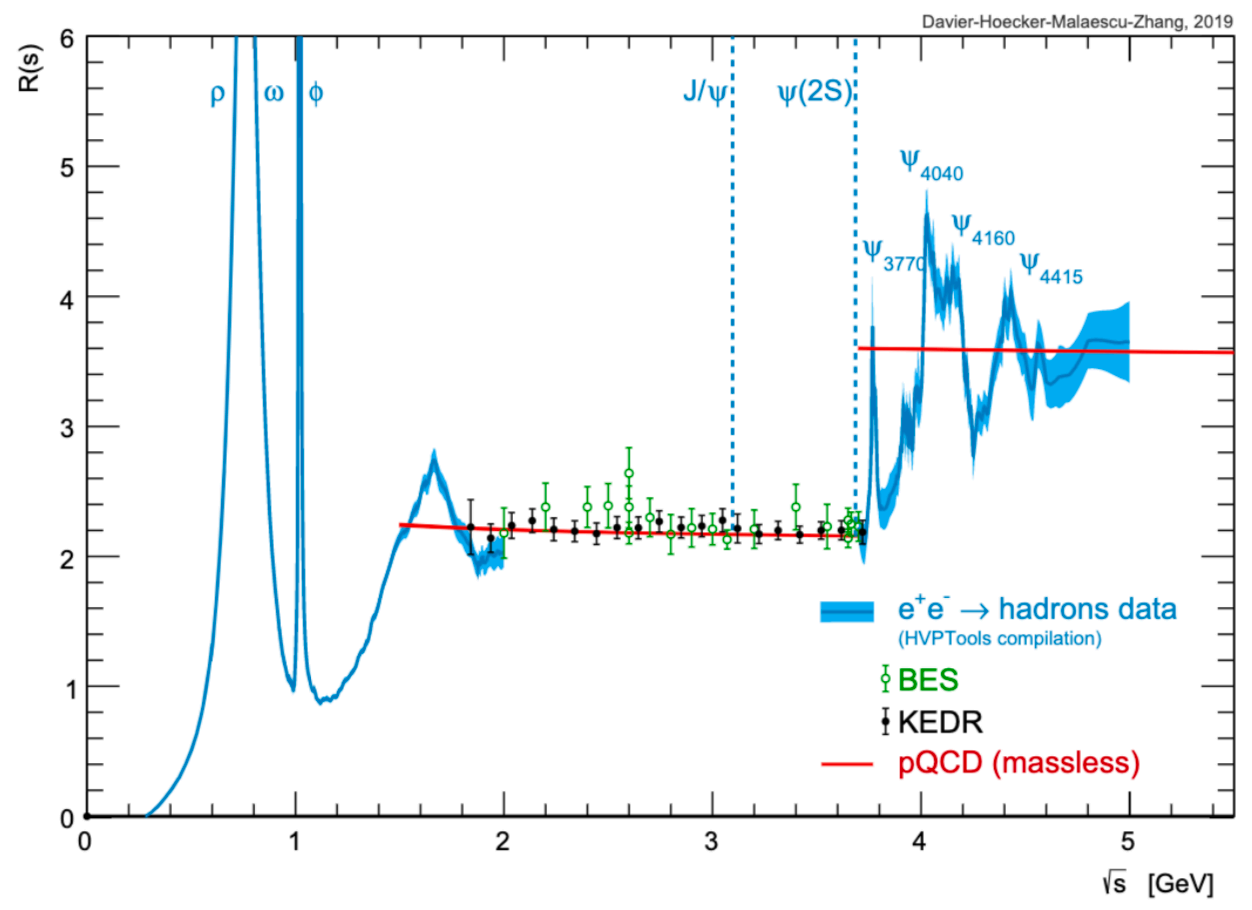

**Fig. 2.3.2** In $e^+e^-$ annihilation, a quark and an antiquark born with momenta above 1 GeV fly away as free partons

where $C_F = (N_c^2-1)/2N_c = 4/3$ is the quark "color charge" (quadratic Casimir operator of the fundamental representation of the $SU(N_c)$ group). The running coupling effect timidly winks at us.

*Tau-lepton as a pQCD blessing.*

Even at smaller momentum scales, pQCD can be successfully applied. It suffices to "properly place your eyes"[6], that is to choose the right question to ask.

An amusing and practically important example of precocious pQCD control is provided by hadronic decays of the $\tau$-lepton. Given lepton-quark universality of the weak interaction, by simply counting degrees of freedom one would expect

$$R_\tau = \frac{\tau \to \nu_\tau + \text{hadrons}}{\tau \to \nu_\tau + e^-\bar{\nu}_e} = N_c = 3.$$

Experimentally it is 20% higher: $R_\tau \simeq 3.64$. Quite a serious discrepancy. We could refuse to discuss it by presenting a legitimate excuse: the lepton mass $m_\tau \simeq 1.78$ GeV is too small for pQCD to apply.

Meantime, there is a more constructive way to address this discrepancy. In the spirit of the Bloom–Gilman duality idea that has emerged in the DIS context [125], it is tempting to explore whether hadron and quark languages would complement each other. The lepton $\tau$ decays via many hadronic channels with squared invariant mass $s = (P_\tau - p_\nu)^2 = m_\tau(m_\tau - 2E_\nu)$ ranging from $m_\pi^2 \simeq 0$ all the way up to $m_\tau^2$. Summing over all hadron states and integrating over $s$ one has a good chance to mimic the QCD prediction, should there be one.

On the QCD side, since the gluon interaction does not discriminate quark flavors ($W^- \to d\bar{u}, s\bar{u}$ in place of $\gamma^* \to u\bar{u}$ or $d\bar{d}$ ), formation of the final state via virtual

[6] M. B. Voloshin

$W^-$ is no different from that in the $e^+e^-$ annihilation case. This allows one to express the pQCD correction to the branching ratio $B_h$ via $\alpha_s(m_\tau^2)$ — the strong coupling at the tau-mass scale. Moreover, by employing the Shifman-Vainshtein-Zakharov (SVZ a.k.a. ITEP) sum rules (discussed in Sec. 5.8) designed to match theoretical quark-gluon calculations with hadron phenomenology via dispersion relations [126], it was possible to prove that the NP contributions are negligible [127] being suppressed as a high power of the $\tau$ mass, $(\Lambda/m_\tau)^6$ [128].

The creator has chosen the $\tau$ mass wisely. It lies conveniently inside a window where $\alpha_s(m_\tau^2)$ is *sufficiently large* as to make pQCD correction significant and well visible, and at the same time *not too large* to undermine the PT treatment. This resulted in [129, 130]

$$\alpha_s(m_\tau^2) = 0.345 \pm 0.010,$$

which value is three times larger that the reference QCD coupling at the $Z$-boson scale, $\alpha_s(M_Z^2)$, and is indispensable as a lever arm for visualizing asymptotic freedom, see Sec. 3 of this volume.

### 2.3.3 (p)QCD: Precursors and hints

QCD inherited quite a dossier of puzzles from the constituent quark model. It is worth recalling certain successes of the pre-QCD quark picture of hadrons, some of which are short of a miracle.

*Inheritance*
Among the first dynamical applications of the constituent quark model of hadrons were the 2–to–3 ratio of the total $\pi p$ and $pp$ cross sections [131] and an intriguingly simple additive pattern of magnetic moments of baryons (see [132] and references therein). In these and many other phenomena, well before QCD, quarks already demanded to be treated as independent quasi-free entities.

Probably the most amusing example of such inheritance is the so-called quark (or, more precisely, "constituent") counting rule [133, 134]. It links the exponent of the energy fall-off of large-angle $a + b \to c + d$ scattering cross sections with the number of "constituents" of the participating (initial and final) particles: $N = n_a + n_b + n_c + n_d$:

$$\frac{d\sigma}{dt} \propto s^{N-2}, \quad -t/s = \mathcal{O}(1). \tag{2.3.1}$$

A chilling example of this scaling law is provided by the process of photo-disintegration of a deuteron [135]. The scaling Eqn. (2.3.1) with $N = 13$ holds in the photon energy interval 1GeV $< E_\gamma <$ 4GeV while the cross section falls by whopping six orders of magnitude! (See [136] to enjoy the picture.)

*Hints*
. An approximate constancy of the total hadron-hadron scattering cross sections hinted at the presence of a vector field ($J = 1$) as a strong interaction mediator. Invention of gluons inspired a model of the Pomeron as a two-gluon $t$-channel exchange [137–139]). It was a little while before the Low-Nussinov Pomeron picture was confirmed and extended by rigorous analysis of high-energy scattering in a non-Abelian QFT [140], to become known as the BFKL Pomeron.

Another early benefit that QCD has offered was an (at least qualitative) explanation of the famous Okubo-Zweig-Iizuka (OZI) rule. It postulated that interacting hadrons do not mind exchanging constituent quarks but hate to allow a quark and its antiquark that are present in the initial state to annihilate. The OZI rule was forged to explain unwillingness of $\phi$ mesons ($\phi = s\bar{s}$) to decay into light $u, d$-built mesons. According to QCD, annihilation of a $q\bar{q}$ pair that constitutes a vector meson has to proceed via 3-gluons, so that the decay width becomes small: $\Gamma/M \propto \alpha_s^3$. It may look too brave to rely on the asymptotic freedom concept at scales as small as $M_\phi/2 = \mathcal{O}(0.5\,\text{GeV})$. However, bound states of heavier quarks ($J/\psi = c\bar{c}$ and $\Upsilon = b\bar{b}$ families) can be related with their QED counterpart — the $C$-odd $e^+e^-$ bound state — orthopositronium [74]. Constructing the ratio of the widths of hadronic and radiative decays

$$J/\psi, \Upsilon \to ggg \to X, \quad J/\psi, \Upsilon \to \gamma gg \to \gamma + X,$$

one arrives at a reasonable quantitative estimate of the QCD coupling at $m_c$ and $m_b$ scales, correspondingly. Here gluons manifested themselves as mediators of the strong interaction.

Gluons as hidden constituents of the proton also showed up indirectly in DIS as electrically neutral matter that carries about a half of the energy-momentum of the fast proton.

The last but not the least: the nature of multiparticle production in the processes involving hadrons also necessitates the presence of a vector field as interaction mediator.

Indeed, the bulk of inelastic high energy hadron-hadron collisions was long known to produce multiparticle final states with hadrons having finite transverse momenta and distributed uniformly in rapidity. In 1968 Gribov considered a fast proton with large energy $E \gg M_p$ fluctuating into a system of $\ln E$ quasi-real particles as an $s$-channel image of the $t$-channel vacuum pole (a.k.a. Pomeron) exchange [141]. Feynman has reverted the picture by prescribing $\ln E$ hadron multiplicity to a fragmenting quark with energy $E$ [142].

A uniform rapidity plateau is the key attribute of vector particles, hinting at gluon radiation underlying production of hadrons.

### 2.3.4 pQCD: Modus Operandi

Massless gluons and quarks are treated by pQCD as if they were photons and electrons. This is clearly not a nice thing to do. In the QED case electrons and photons are legitimate QFT objects. They know how to propagate freely, have a definite relation between energy and momentum and therefore can be prescribed a physical (measurable) mass. Causality and unitary unequivocally dictate the analytic structure of their respective Green functions and interaction amplitudes in general.

Quarks and gluons don't have this luxury. Being well aware of this complication, pQCD ignores it in a hope to be considered innocent until proven guilty.

*Renormalization: Scale*

To calculate probability of radiation, a gluon is put on mass-shell, $k^2 = 0$, as if it were a photon. Intensity of photon radiation is proportional to the fine structure constant $\alpha_{\text{e.m.}} \simeq 1/137.04$, whichever the process and its hardness. The on-mass-shell value of the QED coupling is a measurable quantity that determines multitude of macroscopic electromagnetic phenomena.

In QCD, on the contrary, the on-mass-shell coupling $\alpha_s(0)$ is undefinable simply because "on-mass-shell gluon" is an oxymoron, as is "on-mass-shell quark". One has to choose some sufficiently large momentum scale $\mu_R \gg \Lambda_{\text{QCD}}$ and employ $\alpha_s(\mu_R^2)$ as an expansion parameter to construct the PT series. This is called the *renormalization scale.*

The dependence of $\alpha_s$ on $\mu_R$ (hence, *running coupling*) is governed by the $\beta$-function, see Sec. 3.1. The first two coefficients $\beta_0$ and $\beta_1$ of the Taylor series of $\beta(\alpha_s)$ are driven by the ultraviolet (UV) behavior of the theory. Their values are universal, while $\beta_{n\geq 2}$ depend on the way $\alpha_s$ is defined.

Obviously, physical observables should not dependent on the choice of $\mu_R$. This enforces, through *running*, a definite $\mu_R$-dependence of the coefficients of higher order terms of the series, starting from the next-to-leading (NLO) one. In practice only a few terms of PT expansion are known for a given observable (say, Born + NLO + NNLO, with $\text{N}^3$LO becoming available in certain cases). One puts the residual $\mu_R$-dependence of truncated series to a good use. By varying $\mu_R$ in some interval (conventionally, between $\mu_R/2$ and $2\mu_R$) one gets an ad hoc estimate of theoretical uncertainty due to unknown higher orders.

*Renormalization: Scheme*

By slightly lowering the dimension of the world, $4 \to D = 4-2\epsilon$, one trades UV divergences for singularities in $\epsilon$ to tame the misbehaving integrals. Logarithmic UV-divergences of loop integrals that renormalize the coupling translate then into a pole at $\epsilon = 0$. By dropping it ("minimal subtraction") together with a boring constant (an artifact of the trick) one arrives at a finite answer — the $\overline{\text{MS}}$ coupling. Dimensional regularization (DREG) [143, 144] is a gentle procedure in that it respects and preserves internal symmetries of the problem (with gauge invariance the first to name).[7] Being well suited for multi-loop calculations, the $\overline{\text{MS}}$ scheme has become the standard of the trade.

Alternatively, one can introduce $\alpha_s$ directly from a physical observable without bothering about the UV problem [146]. Called *effective couplings*, many have been suggested since, emerging from $e^+e^-$ hadronic annihilation data [147], the Bjorken sum rule [148], static heavy quark interaction potential [149] or intensity of dipole gluon radiation off non-relativistic heavy quarks [150], etc. Efective couplings can be related to one another via the $\overline{\text{MS}}$ expansions (see, e.g. [151]).

There is one scheme that deserves special credit. Known as Monte Carlo (MC), Catani-Marchesini-Webber (CMW), bremsstrahlung, or simply "physical scheme", it first appeared implemented in the HERWIG MC parton cascades generator [152] and rediscovered in the context of an optimized pQCD description of inclusive heavy quark fragmentation functions [153]. The same coupling shows up in the anomalous dimension of a cusped Wilson line (Polyakov anomalous dimension) [154, 155].

This scheme adds to the $\overline{\text{MS}}$ coupling a definite $\mathcal{O}(\alpha_s^2)$ piece that keeps emerging in a multitude of observables. Among them the behavior of DIS parton distributions (pdf) and jet fragmentation functions (ff) in the quasi-elastic limit $1-x \ll 1$, threshold effects, quark and gluon Sudakov form factors and Regge trajectories, etc.

The reason is simple: it is the scheme that defines the coupling by the radiation intensity of gluons with relatively small energies. Radiation of *soft gluons* is classical by nature. In accord with the Low theorem it is fully determined by the classical trajectory of the charge

[7] When dealing with a sypersymmetric dynamics, one has to sharpen the DREG tool to preserve the fermion–boson symmetry. This is achieved by turning to the *dimensional reduction* (DRED) [145].

(be it electromagnetic or color one) and is insensitive to quantum properties of the particle that caries it [156].

*Infrared-finite coupling*
The QCD coupling grows with distance and becomes infinitely large at some point. This is true both at the one-loop level ($\beta_0$) where it develops a simple pole, c.f. (1.1.8), (1.2.8), and in the two-loop approximation when one takes into account the $\beta_1$ term in the running of the coupling. This is often referred to as the Landau pole/singularity in memory of the discovery 70 years ago by L. Landau and collaborators of the explosive behavior of running coupling in the context of QED.

Beyond the two loops, however, the situation changes. With the sign of $\beta_2$ depending on the scheme, some effective charges at this level stop suffering from the Landau singularity and instead freeze in the origin [147, 157, 158]. Actually, this freezing is as much an artifact as the Landau pole itself. To unambiguously define $\alpha_s$ and establish its behavior at small momenta is inconceivable without cracking the confinement problem.

At the same time, the very supposition that $\alpha_s(k^2)$ is finite for any $k^2 \geq 0$ (more accurately, is *integrable* over the infra-red domain) enhances the predictive power of pQCD. The Parisi–Petronzio analysis of the differential distribution of Drell–Yan pairs with very small transverse momenta $q_T$ and large invariant masses $q^2 \gg q_T^2$ provided a key example [159]. It was enough to assume that such a "good coupling" existed to get a PT prediction that actually *did not depend* on details of its behavior in the origin and agreed with the data.

Assuming the existence of a dispersion relation made it possible to quantify the leading power-suppressed NP contributions by expressing their magnitude via momentum integrals of the "good coupling" over the NP domain [160]. This approach proved to be especially productive in the realm of *jet shapes*, the majority of which suffer from significant $1/Q$ hadronization corrections [161] (see [136] for details).

### 2.3.5 Partons and Jets

The word *jets* appeared (though only once!) in a monumental parton-model study of inclusive production of a nucleon in $e^+e^-$ (the process related by crossing with lepton-nucleon DIS) [162].

The picture of quark jets has been elaborated [163], and the Feynman conjecture implemented as a working hypothesis to characterize the final state structure of hadroproduction processes with large transverse momenta [73]. In a footnote the authors remarked: "*The question of the ultimate fate of the fractional charge may be a difficulty of the quark-parton model*". Since "*the important longitudinal distances in configuration space for electroproduction*" increase linearly with energy and may become macroscopically large [164] "*This may imply that the active parton tends to travel a considerable distance without interaction before disintegrating into a jet of hadrons. Thus, there can be a separation of fractional charge over large distances in configuration space as well as momentum space*". The footnote ended with a prophetic remark: "*However, this does not mean that partons must "backflow" that distance to provide the necessary neutralization of fractional charge. This can be accomplished, for example, by a polarization current created by parton-antiparton pairs created from the vacuum by the field of the active parton*".

The worry was answered five years later when, with the advent of QCD, responsibility for confining fractional charges has been laid upon color.

In 1974 Kogut and Susskind came up with a picture of a flux tube (color string) that connects the quarks. With the color field strength increasing with quark separation, a chain of successive vacuum breakups, $q \to q + q'\bar{q}' \to (q\bar{q}')_{\text{meson}} + q' \to$ etc, contained fractional charges, together with the open color, inside colorless hadrons.

The authors have also remarked that hard gluon bremsstrahlung off the $q\bar{q}$ pair *may be expected* to give rise to three-jet events in the $e^+e^-$ annihilation into hadrons.

The time had come for pQCD to face the challenge.

*Gluon jets*
To unequivocally confirm QCD's claim to an honorable place of the theory of strong interactions, gluons had to be found manifesting as true particles.

Section 2.2 is devoted to the groundbreaking discovery of 3-jet $e^+e^- \to q\bar{q}g$ events. We'll stay on the theory side and peek into a seminal paper that set up the 3-jet quest [104]. What a shaky ground the authors were pushing off back in 1976! Quote:

- no direct experimental evidence yet exists for gluons (except possibly the fact that not all the nucleon's momentum is carried by known quark constituents),
- there is no direct evidence for asymptotic freedom (though there may be some deviations from scaling in DIS at high $Q^2$),
- fashion sets $\alpha_s(Q)$ to lie between 0.2 and 1 for $Q^2 \sim 10\,\text{GeV}^2$.

The authors professed *coplanar* structure of the final state, cross section scaling in $x_T = 2p_T/Q$, verified *asymptotic 2-jetness*, and rightly guessed a 10% fraction of 3-jet events.

Moreover, they drew a picture with two hadron chains stemming from the gluon fragmentation and remarked, without much ado:

> Looking at [this] one might naively expect more hadrons to be produced in gluon fragmentation than in quark fragmentation, and therefore that $f(x)$ for gluons should be more concentrated at low $x$.

That is, higher hadron multiplicity and softer energy spectrum in a gluon jet as compared to quark one. This little picture became a precursor of the Lund model interpretation of a gluon as a "kink" on the color string connecting the separating quark and antiquark [165].

*IRCS Ideology*

In 1977 Sterman and Weinberg drew an image of two-jet events as opposite cones of angular size $\delta$ containing all but a small fraction $\epsilon$ of the total annihilation energy.

In the Born approximation, $e^+e^- \to q\bar{q}$, the back to back quarks fit in with unit probability. In the next order in $\alpha_s$ there emerge a negative virtual correction to $\sigma_{q\bar{q}}$ and a new 3-particle production cross section $\sigma_{q\bar{q}+g}$, both infinite. However, the collinear divergence at $k_T \propto \Theta \to 0$ (present in all logarithmic QFTs with massless fields) and the soft divergence, $k_0 \to 0$ (specific for vector gluons and photons), cancel in the sum, leaving behind a finite correction $\propto \alpha_s \ln\delta \ln\epsilon$.

The SW construction became the first hadron observable that, after the total cross section $\sigma_{\text{tot}}(e^+e^- \to X)$, enjoyed the power of the Bloch-Nordsieck theorem. An ideology of Infrared-and-Collinear Stability (IRCS) was born:

> If radiative corrections to a given observable happen to be free from collinear and soft gluon divergences and thus the result is finite, feel free to confront the PT answer directly with experiment, without worrying about NP hadronization effects.

The flag got hoisted over the boot camp from where pQCD went on a rampage to conquer multiple production of hadrons in hard interactions: "*the detailed results of perturbation theory for production of arbitrary numbers of quarks and gluons can be reinterpreted in quantum chromodynamics as predictions for the production of jets*" [166].

*Defining and finding*

*A narrow bunch of hadrons* is not good enough: one needs an operational definition in order to work with jets, to predict, study and work with them. There emerged two major threads: 1) to look for a set of cones (of certain angular size) that would embed the final-state hadrons in an optimal way and 2) to look for a pair of particles closest in the momentum space and (if judged close enough) join them into one, thus recursively reducing an ensemble of $N$ hadrons to a few clusters — jets.

The original JADE clustering jet finder used an invariant mass of the pair as closeness measure. It did well experimentally, but did not satisfy theorists. By the time when the Workshop on Jet Studies at LEP and HERA was taking place in Durham in 1990, theorists became too greedy. To deal with respectful IRCS observables (which JADE finder's output are) was no longer enough for them.

Jet rates suffer from (or enjoy, up to you) large double-logarithmic corrections, and theorists were eager to make all-order resummed predictions. And the JADE finder did not allow that because of a weird way it was dealing with small-momenta particles (soft gluons). At a brainstorm session a proposal from the audience was made to replace the invariant mass *distance measure* $m_{ik}^2 \simeq 2E_iE_k(1-\cos\Theta_{ik})$ by the relative transverse momentum $k_T^2 \simeq 2\min\{E_i,E_k\}(1-\cos\Theta_{ik})$ to cure the problem. The next morning Siegfried Bethke who spent a sleepless night testing the new idea came up with encouraging news: the $k_T$ measure did well in yielding jets less affected by hadronization.

First reported in the summary of the Hard QCD working group [167], the "Durham" algorithm [168] has got a "Geneva" cousin [169], and then "Cambridge" [170] and "Aachen" [171] fraternal twins that have further reduced hadronization effects. The $k_T$-algorithm, generalized to DIS and hadron-hadron collisions [172], allowed theorists to produce all-order resummed expressions for the jet rates in $e^+e^-$ and elsewhere.

For 15 years or so the clustering algorithms lagged behind the cone-based ones. And for a good reason: $N^3$ operations needed to sort out a final state containing $N$ particles. Given that in the $pp$ environment (not mentioning $pA$ and $AA$) multiplicities are large, this made clustering procedures impractical.

The tables turned when an ingenious application of combinatorial geometry to the momenta clustering problem by Cacciari and Salam has reduced the calculation load down to $N\ln N$. Development of the "fast-$k_T$" clustering procedure permitted to analyze large multiplicity final states "in no time" [173]. This was especially welcome since all then-known cone-based finders were caught red-handed at violating the IRCS demand one way or another.

A long and turbulent history of competing jet-finders has terminated with invention of the "anti-$k_T$" jet find-ing algorithm [174]. It came in time — right before the start of the LHC operation. It satisfied both theorists (as pQCD-fiendly, IRCS respecting) and experimenters (fast and producing aesthetically pleasant roundish jets), and has established itself as the main (if not only) tool of the trade since. A full coverage of *Jetography* can be found in an excellent review [175].

*Heavy quark jet*
QCD expected the jets initiated by heavy quarks $Q$ to have a hole in the forward direction — *dead cone* of the size $\Theta_0 \simeq m_Q/E$. Indirect consequences of this specific feature have been experimentally confirmed a while ago: $Q$ loses little energy (leading particle effect), light hadron multiplicity in a $Q$-jet is reduced by a constant, $N_q(E) - N_Q(E) \simeq N_q(m_Q)$ [153].

A direct observation of the dead cone by the ALICE was recently reported in *Nature* [176].

### 2.3.6 Many jets, some loops

To construct a scattering amplitude at leading order (LO: Born approximation with the minimal power of the coupling constant) one sums up topologically different tree diagrams, each of which is a product of internal Feynman propagators and vertices. Momenta of all internal lines are fixed by kinematics so that no integration is involved. Because of heavier combinatorics and more complicated color structure, the complexity of the scattering amplitude increases with the number of external legs (read: jets).

*Loops and divergences*
QCD jets have become an indispensable tool for collider experiments in search for new physics. It is imperative to know the yield and structure of multi-jet final states with the best accuracy possible. One has to go beyond the Born approximation and calculate, step by step, higher order corrections. A virtual corrections (VC) generates a loop along with an integration over the 4-momentum flowing through the loop. With loops in the game, complexity of the task rises to all new level.

UV divergences being dealt with, VC is still divergent in the collinear and soft corners of the integration space. But so is the inclusive (integrated) cross section of the same order in $\alpha_s$. This time, due to real emission (RE) of an infinitely soft gluon or a collinear 2-parton configuration in the final state phase space. Combining VC with RE one gets rid of *almost all* divergences. The surviving collinear divergences hide into initial state pdf (and ff, should there be hadrons explicitly registered in the final state). Apart from that, the answer is finite. However it is difficult to get by subtracting infinities. One needs to regularize VC and RE separately and consistently or, better still, to perform subtraction at the level of the integrand to avoid divergences altogether.

*NLO*
Early NLO studies sent a rather disturbing message: large corrections were found both to Drell–Yan [177] and large-$p_T$ production [178] putting under question the very applicability of the PT approach. There is a good physical reason why those corrections turned out to be alarmingly large. I will hide it from you for lack of space-time. One way or another the initial shock was mitigated and a systematic attack on the NLO started.

The method that has been proposed for $e^+e^-$ annihilation, used DREG to deal with the VC+RE problem [179]. An idea to employ the notion of color dipoles to accurately treat collinear and soft singularities and cancel them at the integrand level gave more flexibility and allowed to construct a popular general purpose scheme for calculating the NLO jet cross sections in any hard process [180].

$L$-loop VCs are given by $4L$-dimensional Feynman integrals. They are analytic functions of external momentum invariants and can be reduced to a finite set of basic scalar integrals.

The problem has been fully solved for $L = 1$ [181]. This means that today all NLO amplitudes are known (with 6-gluon scattering marking the present-day complexity limit) [182]. Parton showers have been promoted to NLO as well [183].

*NNLO*
Since 2015, the number of important processes contolled in the following order of pQCD (NNLO) has been steadily increasing. In the bibliography titles of the Les Houches 2019 Summary [184] *next-to-next-to*, or *NNLO* appears 155 times. Drell-Yan/Higgs [185–188] and semi-inclusive DIS [189, 190] allowed to peek into $\mathrm{N^3LO}$.

Just enjoy the names that appear in the $\mathrm{N^{\geq 2}LO}$ context: CoLoRFulNNLO and Projection-to-Born methods, Nested soft-collinear and $N$-jettiness subtractions. A Shakespearean review [191] discusses pros and contras of DREG vs. subtraction regularization.

*Mathematical aspects*
To calculate Feynman integrals analytically is notoriously hard. General techniques for attacking loop amplitudes were listed and demonstrated in 1996 and are being used since: *spinor helicity formalism, color decompositions, supersymmetry, string theory, factorization and unitarity* [192].

The Loop-Tree duality approach (LTD) was initiated [193] and later generalized to become Four-dimensional Unsubtraction (FDU) [194].

Proceedings of the topical Florence workshop (cunningly named WorkStop/ThinkStart) [195] link to 200+ articles that cover the basics and the progress.

An all-in assault [196] resulted in an astonishing symbiosis of theoretical physics and pure mathematics. Particle theorists, maybe already familiar with integrability, now have to learn twisted cohomology groups, Hopf algebra, algebraic number theory and other scary things.

### 2.3.7 Resummation and Evolution

*Art of expansion*
Series in $\alpha_s$ can behave well, as for $R(e^+e^-)$, or look troubling as is the case of diphoton production, where moving from NLO to NNLO changes the cross section by 50% [197].

In fact, independent of the observable, PT series in QFT are *asymptotic*, so that beyond $\mathrm{N}^{1/\alpha}$LO things are bound to go haywire. This was not much trouble for QED, but it should be kept in mind for QCD, where the number of reliable terms in the expansion may be not so large.

Examining *how violently* a specific series diverges, hints at how much the NP physics affects a given observable (infrared renormalons [198]).

*Resummation*
Often $\alpha_s$ acquires one or even two enhancement factors: $\alpha_s \ln Q^2$ (SL), $\alpha_s \ln^2 Q^2$ (DL), and the PT expansion fails. When this happens, in order to get a reliable approximation one has to collect enhanced contributions and sum them in all orders. The Sterman-Weinberg 2-jet cross section acquires DLogs because of a veto imposed on accompanying gluon radiation. The $Q_T$-spectrum of a Drell–Yan pair or of a hadron registered in the current fragmentation of DIS in the kinematical region $Q_T \ll Q$, and an almost back-to-back energy–energy correlation in $e^+e^-$ were the first examples of inclusive observables which, in spite of not being subject to any explicit veto, are still affected by DLogs [199, 200].

In all these cases the origin of one of the logs is soft gluon radiation which is relatively easy to control. This makes resummation of DL-enhanced contributions straightforward and gives rise to Sudakov form factors. Quark and gluon form factors manifest themselves in a multitude of observables characterized by the presence of two different momentum scales. In particular, in distributions of various jet shapes, jet rates, etc.

It is important to emphasize that the very possibility of an all-order resummation depends on whether the operational definition of jets corresponds to the dynamics of the QCD parton multiplication picture ($k_T$-algorithms vs. JADE, as discussed above).

All-order resummation of single-logarithmic contributions (SLogs) becomes mandatory when we deal either with quasi-collinear configurations of partons with comparable energies (DGLAP physics) or with ensembles of soft gluons at large angles with respect to energetic emitters (radiative corrections to parton scattering amplitudes). In both cases particles involved are strongly ordered in *transverse momenta*.

*Factorization*
A particle with the smallest $k_T$ in the game *factors out*, in a sense that a singular contribution comes only from its attachment to an external leg. Generalization of the Low theorem from $\omega \ll m_e$ to arbitrary photon energies [201] was a precursor of the QCD $k_T$/collinear factorization.

Another arbitrary scale enters: factorization scale $\mu_F$. It sets a conventional border between PT and NP ingredients of the problem. In IRCS observables $\mu_F$ gets replaced by a variable related to resolution, rendering two well separated physical scales. For example, $yQ^2 \ll Q^2$ for jet rates, $(1-T)Q^2 \ll Q^2$ for the differential thrust distribution, etc.

Whenever there is *Factorization*, one can carry out *Resummation*, and interpret the results in terms of *Evolution* and corresponding *Evolution Equations*.

A few examples of the application of this idea, both well-known and lesser-known.

*KL*
The Kirschner–Lipatov equation resums DLogs in parton scattering amplitudes with quark exchange in the $t$-channel [202]. Such amplitudes fall as the energy $s$ increases, and higher-order DL contributions decelerate this fall. These DLog effects are inherently different from the DLog effects due to accompanying soft gluon radiation (Sudakov form factors).

By isolating the virtual particle with the lowest $k_T$ in the Feynman graph, and using gauge invariance and the unitarity relation, one can form the kernel of the evolution equation for the partial wave amplitudes, with $\ln k_T$ as the "evolution time".

*KOS*
Kidonakis, Oderda and Sterman have set the quest of resummation of SL radiative corrections to $2 \to 2$ QCD parton scattering amplitudes [203]. In QCD it becomes a multi-channel problem, since each gluon emission (either virtual or real) changes the color state of a parton pair. For gluon–gluon scattering, the anomalous dimension is a $6\otimes 6$ matrix (for the general $SU(N_c)$ case; which reduces to $5\otimes 5$ for $SU(3)$). It depends on the scattering angle and, obviously, on the rank of the color group, $N_c$. Three of the eigenvalues of the anomalous dimension matrix are proportional to $N_c$, and thus respect the so-called *Casimir scaling* (the perturbative expansion running in $N_c$), see e.g. [204]. The $N_c$-dependence of the other three eigenvalues is more involved [205]. They solve the cubic equation whose coefficients exhibit a weird symmetry between the number of colors and the scattering angle:

$$N_c \iff \pm\frac{\ln(s^2/tu)}{\ln(t/u)}.$$

This symmetry can hardly be accidental, but its origin remains a mystery.

*ERBL*

The ERBL equation applies to exclusive high-$Q^2$ reactions involving mesons and baryons, e.g. electromagnetic pion form factor [206, 207] or photo- (electro-) production of vector mesons like $J/\psi$ [208, 209]. Separate components of the valence quark wave function (distribution amplitude) acquire different $\log Q$ behavior — anomalous dimensions. The dominant component in the $Q^2\to\infty$ limit is called the asymptotic wave function: $\psi_\pi(z)\propto z(1–z)$ with $z$ the longitudinal momentum of the fast pion carried by a quark.

It is manifest in the distribution of energy between the two quark jets stemming from diffractive dissociation of a pion in $\pi A$ collisions [210].

*DGLAP*

The parton model implied limited transverse momenta. In logarithmic QFTs, instead, $k_T^2$ are broadly distributed up to the external momentum transfer scale $Q^2$, resulting in violation of the Bjorken scaling. The first systematic analysis of DIS structure functions and $e^+e^-$ fragmentation functions was carried out in the Leading Logarithmic Approximation (LLA) based on selection of enhanced contributions in each order of PT series, $\sum_n C_n(x)(g^2\log Q^2)^n$, in the framework of then-known QFT models [211, 212].

In 1974 the results were recast in the language of pdf evolving via Markov chain of independent $1\to 2$ parton splittings [213].

In 1977 arrived the QCD parton dynamics whose name was eventually settled as DGLAP [214, 215]. It was received with enthusiasm and gave rise to a host of new ideas: jet calculus, preconfinement, parton showers, to name a few.

With anomalous dimensions now known in 3 loops [216, 217], DGLAP does its job, predicting pdf evolution due to space-like cascades. Thanks to factorization, they describe the flux of initial-state partons as an input for any hard lepton–hadron or hadron–hadron interaction. The same universality applies to the final state (time-like cascades).

*Parton Cascades*

Partons have space-like momenta ($k^2<0$) in the initial state cascades; in the final state they are time-like ($k^2>0$). In the LLA, parton splitting functions in space-like ($S$) and time-like kinematics ($T$) are the same: $P_{ba}^{(S)}(z)=P_{ba}^{(T)}(z)$, and so are the anomalous dimensions — Mellin image of $P(z)$. Beyond LLA $P^{(T)}(z)$ departs from $P^{(S)}$ acquiring, in particular, $(\alpha_s\ln^2 z)^k$ terms in N$^k$LL.

Originally, the picture of QCD partons was treating the Bjorken/Feynman variable $x$ as being of the order one. Then $\alpha_s\ln^2 z=\mathcal{O}(\alpha_s)\ll 1$ and causes no trouble. However, when $x$ gets *parametrically* small so that $\alpha_s\ln^2 x\sim 1$, an entire tower of these enhanced terms has to be resummed.

This can be achieved by modifying the "time" in the evolution equation from $\ln k_T$ to $\ln\Theta$. In other words, by replacing the $k_T$-ordered cascades (S) by ordering of successive splitting angles (T). Angular Ordering (AO) takes care of destructive soft-gluon interference and affects particle production.

*BFKL*

The BFKL equation [140, 218] was derived in the LLA in $g^2\ln s=\mathcal{O}(1)$ to predict high-energy behavior of scattering amplitudes in Yang–Mills theory.

Gluons *reggeize* (spin of a $t$-channel gluon becomes effectively $t$-dependent, $J{=}J_g(t)$). In the vacuum channel ladder diagrams dominate with two Low–Nussinov gluons, now *reggeized*, connected by multiple gluon rungs strongly ordered in rapidity (*multiregge kinematics*). This yielded the growing total cross section $\sigma_{\rm tot}\propto s^{c\alpha_s}$. The NLL correction lowered the exponent. A power-like energy growth contradicts the asymptotic Froissart theorem, $\sigma_{\rm tot}\le A\ln^2 s$, but at available energies is legitimate. A need to rescue $s$-channel unitarity ignited new ideas and, correspondingly, equations: McLerran–Venugopalan Color Glass Condensate model of high-energy saturation (CGC), Balitsky–Kovchegov (BK) and Jalilian-Marian, Iancu, McLerran, Weigert, Leonidov and Kovner (JIMWLK) equations, for references [219].

The true problem is that the high energy scattering does not belong to the pQCD jurisdiction. This is not a hard process as long as no large-$k_T$ scale is involved. As a result, the "BFKL Pomeron" is sensitive to the behavior of the coupling in the NP domain [220, 221]. Strictly speaking it would be safer to apply to compact projectiles like bound states of heavy quarks, say $\sigma_{\mathrm{tot}}^{J/\psi\, J/\psi}(s)$.

Triggering a jet with $p_T \sim Q$ in DIS target fragmentation region should expose the BFKL dynamics (Mueller-Navelet jets [222]). DGLAP evolution gets suppressed over a large rapidity interval, leaving room for PT-controlled BFKL growth. Experimental data are not yet conclusive [223].

Applied to DIS, BFKL predicts a steep growth of pdf in the $x = Q^2/s \to 0$ limit, equivalent to $s \to \infty$. With DGLAP having its own way of making pdf rise, the two are difficult to disentangle.

*BFKL vs. DGLAP*
The meaning of *evolution* in the two cases is essentially different. Action $d/d\ln k_T^2$, dynamics in $x$ (DGLAP), vs. action $d/d\ln(1/x)$, dynamics in $\vec{k}_T$ (BFKL). The kernel of the DGLAP evolution equation is a function of the longitudinal momentum $P_{ba}(x)$, the BFKL kernel lives in the plane of transverse momenta $K(\vec{k}_T, \vec{q}_T)$. Eigenvalues of DGLAP are anomalous dimensions; the spectrum of BFKL — Regge trajectories. The origin of DGLAP evolution is the $k_T$-factorization [224]; BFKL rests upon $t$-channel unitarity. In spite of all the difference the two are intimately related [225].

### 2.3.8 Soft gluons and LPHD

It is soft gluon radiation that bears responsibility for faster-than-logarithmic growth of particle multiplicities in hard processes.

Hadron energy spectra in jets brought an exotic fruit. It was not poisonous, but still not easy to digest.

*Inside jet*
LEP [226], HERA [227] and Tevatron have found that the shape of single-inclusive energy spectra of all-charged hadrons (dominated by pions) is mathematically similar to that predicted by pQCD for soft gluons [228]. And this in spite of the fact that the characteristic *hump* that the spectrum develops because of soft-gluon coherence was situated as low as 1 GeV at LEP (and well below at TASSO energies).

CDF studies proved the origin of the hump due to parton cascading (as opposed to nonrelativistic finite mass effects) [229] and confirmed the pQCD expectation that the particle yield scales with maximal $k_T$ of partons, $E_{\mathrm{jet}} \sin\Theta_c$, with $\Theta_c$ the half-angle of the jet cone [230].

*Inter-jet particles*
Studies of hadron flows *in-between jets* added insult to injury. The message here is even more surprising. Information about the color structure of the ensemble of hard partons that form the jets is transmitted to pions with energies of 200–300 MeV, which make up the bulk of the hadrons produced away from the jets ("QCD Radiophysics") [231]. For example, a comparison of the hadron yield in the direction transverse to the 3-jet-event plane with the pQCD prediction of the soft gluon radiation pattern [232], yielded an independent measurement of the ratio of quark and gluon color charges [233], competing with results from hard gluon physics (scaling violation and 4-jet rates) [130].

From a theory standpoint, this similarity was not entirely unexpected. There was a premonition based on a semi-classical analysis of the structure of parton cascades in the configuration space which concluded that when the time comes for a given parton to hadronize, other partons are too far away, leaving no chance for cross-talk [234].

Local Parton–Hadron Duality (LPHD) as a Nature-approved supplement to pQCD sends a powerful message to the future quantitative theory of confinement: the Poynting vector of the color field should translate into Poynting vector of the hadron matter practically undamaged.

### 2.3.9 Conclusions

There are a number of pQCD-related stories I have left untold.

Why did it take almost 20 years for the inclusive energy–energy correlation in $e^+e^- \to h_1 h_2 X$, believed to be the most reliable IRCS pQCD prediction, to agree with the experimental data?

Why did the discovery of angular ordering - so important for understanding the coherent nature of particle production - remain unpublished for a long time?

What would make you submit to *Phys. Lett.* an article under the *wrong title* [235]?

How is it that a specific jet shape distribution turns out to be *narrower* than that of the underlying parton ensemble, in spite of usual smearing at the hadronization stage?

How tragic was a misprint in Ref. [236]?

I am confident that by the time *QCD-60* gets published, there will be many more pQCD success stories to tell, in addition to anecdotes.

# 3 Fundamental constants

*Conveners:*
**Eberhard Klempt and Giulia Zanderighi**

The fundamental constants of QCD are the strong coupling $g_s$ or $\alpha_s = \frac{g_s^2}{4\pi}$, and the six quark masses. Using lattice QCD (LQCD), Sec. 3.1 reviews how these parameters are defined and renormalized, and describes briefly what measurements are compared with lattice predictions in order to determine the values of these fundamental parameters. Section 3.2 reviews recent determinations of $\alpha_s$ and discusses systematic uncertainties and the procedure used by the current Particle Data Group (PDG) to obtain the world average value of $\alpha_s$. A precise knowledge of this coupling constant is needed to predict any background process in high-energy collisions. and to achieve precision in the calculation of signal processes.

Some may prefer to read Sec. 3.1 after the discussion of LQCD (in Sec. 4) and Sec. 3.2 after discussion of the measurements presented in Sec. 12. The editors decided to place these early in the volume in order to emphasize that the size of $\alpha_s(Q^2)$ at low $Q^2$ means that perturbation theory cannot work at the modest values of $Q^2$ characteristic of matter in its ground state. Non perturbative methods will be required.

## 3.1 Lattice determination of $\alpha_s$ and quark masses

**Luigi Del Debbio and Alberto Ramos**

Lattice QCD provides a first-principles, non-perturbative description of the strong interaction in the Standard Model (see section 4.1). Current state of the art simulations include sea quark effects, electromagnetic interactions, and isospin breaking, yielding accurate predictions for low-energy hadronic quantities that are not accessible in perturbation theory.

By discretizing space-time in a cubic lattice with spacing $a$, lattice QCD provides a non-perturbative regularization of QCD. Moreover this formulation is amenable to numerical simulations using Monte Carlo methods. A key ingredient in any lattice calculation consists in removing the regulator (i.e. taking the continuum limit $a \to 0$). This requires to tune the bare parameters of the lattice QCD action ($n_f$ bare quark masses in lattice units $am_i$, and the bare coupling $g_0$) in order to reproduce some hadronic input. Note that since the input of any simulation are dimensionless quantities, only dimensionless predictions can be made. Typically one uses meson masses ($\pi, K$ and $D$ in case that the simulation includes the charm quark) in units of a reference hadronic quantity to fix the values of the bare quark masses. The reference quantity, usually the mass of the omega baryon $M_\Omega$ or the $\pi/K$ meson decay constants $f_\pi, f_K$ is the quantity used to *set the scale*: all dimensionless predictions are computed in units of this reference scale. This tuning of the bare parameters in favor of physical observables constitutes the *renormalization* of the theory. Once this process is carried out one can make solid predictions for many other hadronic quantities, and also determine the values of the fundamental parameters of QCD. All in all, quark masses are computed in units of the reference scale. The running of the strong coupling is also computed at energy scales measured in units of the same reference scale. Using as input the experimental value of this reference scale ($M_\Omega$, $f_\pi$, $f_K$ or any other convenient choice), one can quote physical Dimension-full predictions[8]. In this way Lattice QCD is able to connect the experimentally observed hadron spectrum (meson and baryon masses) with the fundamental quark masses and strong coupling.

Here we address conceptually how the fundamental parameters of QCD are extracted from Lattice QCD computations, and what are the dominating sources of uncertainty. We will also comment on a few recent results. For a detailed overview on lattice determinations of the strong coupling, we point the reader to the recent review [237]. An exhaustive and critical list of lattice determinations both of quark masses and the strong coupling is available in the excellent FLAG review [63].

### 3.1.1 The scale of the strong interactions

It is convenient to frame the determination of the strong coupling constant as a determination of the intrinsic scale of QCD. We start from an observable $P$ that depends on a single scale $\mu$ (i.e. $P(\mu)$). Ideally this observable should be easy to determine from numerical lattice simulations and with a known perturbative expansion. As we will see later there are several possibilities. Once an observable is chosen, it can be used to define a renormalization scheme (renormalized coupling) via

$$\bar{g}_s^2(\mu) \propto P(\mu)\,, \tag{3.1.1}$$

where the proportionality factor (a simple normalization) is determined by the condition

$$\bar{g}_s^2(\mu) \overset{\mu\to\infty}{\sim} \bar{g}^2_{\overline{\mathrm{MS}}}(\mu) \tag{3.1.2}$$

[8] The interested reader can consult the section on scale setting in the review [237] and in the 2021 FLAG document [63] for a more detailed discussion.

with $\bar{g}^2_{\overline{\rm MS}}(\mu) \equiv (4\pi)\alpha_{\overline{\rm MS}}(\mu)$. It is convenient to work in mass independent renormalization schemes (i.e. the observable $P(\mu)$ is defined in the chiral limit $m_q = 0$). In these schemes the energy dependence of the coupling $\bar{g}_s(\mu)$ is described by the renormalization group (RG) function that has a known perturbative expansion

$$\beta_s(\bar{g}) = \mu\frac{d}{d\mu}\bar{g}_s(\mu) \stackrel{\bar{g}\to 0}{\sim} -\bar{g}_s^3\sum_{k=0}^{\infty} b_k\bar{g}_s^{2k}\,, \tag{3.1.3}$$

where the first two perturbative coefficients

$$b_0 = \tfrac{1}{(4\pi)^2}\left(11 - \tfrac{2n_f}{3}\right)\,, \tag{3.1.4a}$$

$$b_1 = \tfrac{1}{(4\pi)^4}\left(102 - \tfrac{38n_f}{3}\right)\,, \tag{3.1.4b}$$

$n_f$ is the number of fermions in the fundamental representation (*i.e.* quarks). Different renormalization schemes are related perturbatively by

$$\bar{g}^2_{s'}(\mu) \stackrel{\bar{g}_s\to 0}{\sim} \bar{g}^2_s(\mu) + c_{ss'}\bar{g}^4_s(\mu) + \ldots\,. \tag{3.1.5}$$

It is easy to check that the first two coefficients of the $\beta$-function eq. (3.1.4) are invariant under such changes of scheme (i.e. they are *scheme independent*).

Integrating the evolution equation (3.1.3) yields

$$\log\frac{\mu_1}{\mu_2} = \int_{\bar{g}_1}^{\bar{g}_2}\frac{{\rm d}x}{\beta_s(x)}\,, \tag{3.1.6}$$

where $\bar{g}_1 = \bar{g}_s(\mu_1)$ and $\bar{g}_1 = \bar{g}_s(\mu_1)$. The integral can be rewritten as

$$\int_{\bar{g}_1}^{\bar{g}_2}\frac{{\rm d}x}{\beta_s(x)} = \frac{1}{2b_0}\left(\frac{1}{\bar{g}_1^2} - \frac{1}{\bar{g}_2^2}\right) + \frac{b_1}{b_0^2}\log\frac{\bar{g}_1}{\bar{g}_2} + \int_{\bar{g}_1}^{\bar{g}_2}{\rm d}x\left[\frac{1}{\beta_s(x)} + \frac{1}{b_0x^3} - \frac{b_1}{b_0^2x}\right]\,. \tag{3.1.7}$$

Note that given the asymptotic form of the $\beta_s$ function (eq. (3.1.3)), the original integral in eq. (3.1.6) is divergent when either $\bar{g}_1\to 0$ or $\bar{g}_2\to 0$. On the other hand the integral in eq. (3.1.7) is finite in these limits (cf. the integrand is $\mathcal{O}(x)$). This observation allows us to split the integral in eq. (3.1.7) as $\int_{\bar{g}_1}^{\bar{g}_2} = \int_{\bar{g}_1}^{0} + \int_0^{\bar{g}_2}$ and write eq. (3.1.6) in the following way

$$\log\mu_1 - \frac{1}{2b_0\bar{g}_1^2} - \frac{b_1}{b_0^2}\log\bar{g}_1 \tag{3.1.8}$$

$$+\int_0^{\bar{g}_1}{\rm d}x\left[\frac{1}{\beta_s(x)} + \frac{1}{b_0x^3} - \frac{b_1}{b_0^2x}\right] = \tag{3.1.9}$$

$$\log\mu_2 - \frac{1}{2b_0\bar{g}_2^2} - \frac{b_1}{b_0^2}\log\bar{g}_2 \tag{3.1.10}$$

$$+\int_0^{\bar{g}_2}{\rm d}x\left[\frac{1}{\beta_s(x)} + \frac{1}{b_0x^3} - \frac{b_1}{b_0^2x}\right]\,. \tag{3.1.11}$$

Note that this last equation claims that a function of $\mu_1$ (the left hand side) is equal to a function of $\mu_2$ (the right hand side). The only solution is that both are constant. The constant is defined to be $\log\Lambda_s$ and we can write

$$\Lambda_s = \mu\left[b_0\bar{g}_s^2(\mu)\right]^{-\frac{b_1}{2b_0^2}} e^{-\frac{1}{2b_0\bar{g}_s^2(\mu)}}\times \exp\left\{-\int_0^{\bar{g}(\mu)}{\rm d}x\left[\frac{1}{\beta_s(x)} + \frac{1}{b_0x^3} - \frac{b_1}{b_0^2x}\right]\right\}\,. \tag{3.1.12}$$

Note that the integration of the renormalization group equation here is exact, valid beyond perturbation theory. The combination on the right-hand side of Eq. 3.1.12 has units of mass, and is independent of $\mu$. It is called the $\Lambda$-parameter and can be understood as the *intrinsic scale* of QCD. It is a free parameter, which provides a boundary condition for the evolution equation of the coupling.

Determining $\Lambda$ is equivalent to determining the coupling constant. It is customary to report the value of $\alpha_s(M_Z^2)$ in the $\overline{\rm MS}$ scheme, however the latter can be used together with the perturbative expansion of the beta function to compute the $\Lambda$-parameter. While the two pictures are clearly equivalent, there are some advantages in focussing on $\Lambda$ as the main character of our story:

- It makes clear that the determination of the strong coupling constant really amounts to the determination of one energy scale.
- Although the $\Lambda$-parameter depends on the renormalization scheme. The relation between $\Lambda$-parameters in two different schemes is exactly given by a one-loop computation. In order to see this we recall that by convention couplings in different schemes are normalized so that they agree to leading order (cf. Eq. (3.1.2)). This implies that renormalized couplings in two schemes $s$ and $s'$ are related perturbatively by

$$\bar{g}^2_{s'}(\mu) \stackrel{\bar{g}_s\to 0}{\sim} \bar{g}^2_s(\mu) + c_{ss'}\bar{g}^4_s(\mu) + \ldots\,, \tag{3.1.13}$$

with $c_{ss'}$ a finite number. This implies that the relation

$$\frac{\Lambda_{s'}}{\Lambda_s} = \exp\left(\frac{-c_{ss'}}{2b_0}\right) \tag{3.1.14}$$

is exact.
- The $\Lambda$-parameter is defined non-perturbatively. Even for schemes that are intrinsically defined in a perturbative context: $\overline{\rm MS}$ is a "perturbative scheme", but $\Lambda_{\overline{\rm MS}}$ is a meaningful quantity beyond perturbation theory thanks to Eq. (3.1.14).

– Even if the actual precision in the determination of the strong coupling looks impressive ($\approx 0.7\%$), this amounts to a determination of the $\Lambda$-parameter with approximately a 4% uncertainty. In particular some sub-percent effects (QED and isospin breaking corrections) are subdominant for lattice extractions of the strong coupling.

### 3.1.2 Challenges in extractions of the strong coupling

The extraction of the $\Lambda$-parameter in units of a well determined hadronic scale $\mu_{\rm had}$ (like the proton mass) via Eq. (3.1.12) requires the knowledge of the $\beta$-function in the scheme of choice, $\beta_s(x)$, for values $x \in [0, g_s(\mu_{\rm had})]$. Although in principle Lattice QCD can determine the running of $g_s(\mu)$ at any energy scale (it is just the scale dependence of the observable $O(\mu)$ in Eq. (3.1.1)), computational constraints impose that a typical lattice simulation can only resolve a certain range of scales. In particular if we want to describe hadronic physics, we can reach at most scales $\mu_{\rm PT} \sim 2-5$ GeV (see figure 3.1.1). For this reason, any lattice QCD extraction of the strong coupling uses the perturbative expansion

$$O(\mu) = \sum_{k=0}^{n_{\rm PT}} c_k \alpha^k_{\overline{\rm MS}}(\mu) + \mathcal{O}(\alpha^{n+1}_{\overline{\rm MS}}(\mu)) + \mathcal{O}\left(\frac{\Lambda^p}{\mu^p}\right) , \tag{3.1.15}$$

The known perturbative coefficients $c_i$ ($i = 1, \ldots, n_{\rm PT}$) together with the known 5-loop running of the beta function allow us to estimate the high-energy contribution

$$\int_0^{\bar{g}(\mu_{\rm PT})} {\rm d}x \left[\frac{1}{\beta_s(x)} + \frac{1}{b_0 x^3} - \frac{b_1}{b_0^2 x}\right] , \tag{3.1.16}$$

to $\Lambda_{\overline{\rm MS}}$. It is worthwhile to emphasize a few subtleties involved in this procedure. Since we only know a few terms in the perturbative expansion of the observable, the missing higher orders are a source of systematic error in the determination of $\Lambda$. In fact it easy to convince oneself that it introduces uncertainties of order

$$\mathcal{O}(\bar{g}^{2n_{\rm PT}}(\mu_{\rm PT})) . \tag{3.1.17}$$

A further source of systematic error comes from non-perturbative (power corrections) to the perturbative expansion. These corrections are suppressed as

$$\mathcal{O}\left(\frac{\Lambda^p}{\mu^p}\right) .$$

Both sources of systematic effect can be eliminated by just pushing $\mu_{\rm PT}$ to a high enough scale, but with data only available in a limited range of energies it is challenging to estimate the size of these corrections. Moreover, the perturbative corrections $O(\alpha^{n_{\rm PT}+1}_{\overline{\rm MS}}(\mu))$ decrease very slowly (i.e. logarithmically) with the scale $\mu_{\rm PT}$. This makes reducing perturbative uncertainties an *exponentially* difficult problem.

*The window problem*
The need to use low energies to determine the $\Lambda$ parameter in terms of a known, precise, hadronic input, is at odds with the need to reach large energy scales where perturbation theory is applicable with high enough accuracy. This is usually referred as the window problem. In practice scales of a few GeV are reached and the estimates of perturbative uncertainties remain the main source of error in most lattice calculatrions. Ref. [237] estimates that perturbative uncertainties alone amount to about 1-2% error in $\alpha_s(M_Z)$ for any method that suffers from the window problem.

*Dedicated approaches*
There exists however a known solution to overcome this intrinsic difficulty, and it comes under the name of finite size scaling [238]. The idea consists in decoupling the simulations where the hadronic input is determined and the simulations used to the determination of the running of the coupling. Each simulation can only resolve a limited range of scales, but a recursive procedure called finite size scaling allows us to relate the energy scales resolved in different simulations (see below for more details). Another recent proposal [239] does not provide a complete solution to the window problem, but reduces substantially the problem. In particular in this approach we are only concerned with power corrections, that decrease much faster with the energy scale than perturbative ones.

### 3.1.3 Lattice observables

There is a wide variety of lattice observables that are used for a determination of the strong coupling. This rich landscape allows for multiple independent determinations, providing a robust cross-check of the methodologies. Here we want to emphasise the broad range of observables. For a full review and combination of the results we refer the reader to Refs. [237] and [63].

We first review the dedicated strategies that aim at solving (or ameliorating) the window problem with a dedicated approach. Typically they require dedicated simulations and the uncertainties are statistically dominated.

*Finite size scaling.*
An ingenious solution to the window problem is obtained by separating the RG evolution, resolving only a limited range of scales in each single simulation, and adopting a recursive procedure to connect different simulations. The main idea is to use a finite-volume renormalization scheme, where the renormalization scale is identified with the inverse volume of the lattice. The renormalized coupling, denoted here as

$$\bar{g}^2_{\rm SF}(\mu), \qquad (\mu = 1/L)\,, \tag{3.1.18}$$

is extracted from observables computed in Monte Carlo simulations. The running of the coupling is encoded in the so-called *step scaling* function,

$$\sigma_s(u) = \bar{g}^2(\mu)\Big|_{\bar{g}^2(\mu/s)=u}, \qquad (\mu = 1/L)\,, \tag{3.1.19}$$

which yields the value of the renormalized coupling at the scale $\mu$ as a function of its value at the scale $\mu/s$, where $s$ a scaling factor. The step scaling function is evaluated numerically by computing $\bar{g}^2_{\rm SF}(\mu)$ on pairs of lattices of size $L$ and $sL$. Thereby multiple simulations on physical volumes much smaller than the typical hadronic scales are used to compute the non-perturbative evolution of the coupling from a hadronic scale $\mu_{\rm had}$ up to a high-energy scale, $\mu_{\rm PT}$, where the matching with perturbation theory is fully under control. While these volumes are too small to study hadronic physics, they are perfectly suitable to study the RG flow of the coupling. The only experimental input needed in this procedure is *one* dimensionful quantity that needs to be compared to one lattice measurement in a large volume in order to set the scale in physical units. It is interesting to remark that the strong coupling constant in this approach is determined from just *one* experimental dimensionful quantity. Perturbation theory is only used at scales larger than the perturbative scale, $\mu_{\rm PT} = s^n \mu_{\rm had}$. This scale can be made (almost) arbitrarily large with a modest (but dedicated) computational effort (typically $\mu_{\rm PT} \sim 100$ GeV). Finally it is worthwhile to emphasise that for the determinations based on finite size methods, the main source of uncertainty is statistical rather than systematic. Dedicated simulations will allow further improvements.

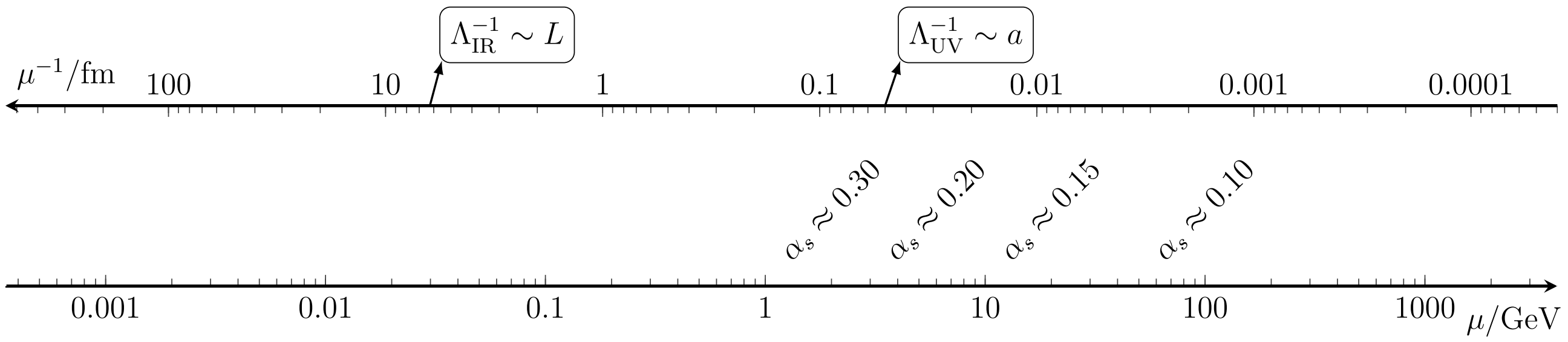

**Fig. 3.1.1** Any quantity determined in a lattice simulation must be determined at energy scales between the intrinsic UV (a few GeV) and IR (a few dozen MeV) cutoffs of the simulation, given by the total volume and lattice spacing respectively.

*Heavy quark decoupling*

Recently a new way to ameliorate the window problem has been proposed [239] (see also the review [240]). It is well known that the QCD coupling with $n_l$ massless quarks and $n_h$ heavy quarks (with mass $M \gg \Lambda$), can be matched using perturbation theory to the coupling of QCD with $n_l$ massless quarks. This matching is done in perturbation theory to high order and the perturbative and non-perturbative corrections are very small. These perturbative decoupling relations can also be understood as relations between the $\Lambda$ parameters with $n_l + n_h$ flavors and the $\Lambda$ parameter with $n_l$ flavors

$$\frac{\Lambda^{(n_l)}}{\Lambda^{(n_l+n_h)}} = P_{nl,n_l+n_h}(M/\Lambda)\,. \tag{3.1.20}$$

Both perturbative and non-perturbative uncertainties are very small in these relations even for quark masses of the order of the charm[241].

The main point of this new proposal consists in simulating $n_f$ fictitious heavy quarks. Since all quarks are heavy, a coupling computed in this setup $\bar{g}^{(n_f)}(\mu, M)$ is, up to heavy mass corrections just a pure gauge coupling

$$\frac{\bar{g}^{(n_f)}(\mu, M)}{g^{(0)}(\mu)} \overset{\Lambda/M\to 0}{\sim} 1 + \mathcal{O}(M^{-2})\,, \tag{3.1.21}$$

where $\mathcal{O}(M^{-2})$ represent corrections that can be $(M/\mu)^2$ or $(M/\Lambda)^2$. Conversely we can declare that both couplings are the same at slightly diferent values of the scale

$$\bar{g}^{(n_f)}(\mu^{(n_f)}, M) = g^{(0)}(\mu^{(0)})\,, \tag{3.1.22}$$

implying the relation between scales

$$\frac{\mu^{(n_f)}}{\mu^{(0)}} \overset{\Lambda/M\to 0}{\sim} 1 + \mathcal{O}(M^{-2})\,. \tag{3.1.23}$$

Together with the basic definition of the $\Lambda$ parameter Eq. (3.1.12), this last relation allows immediately to write a relation between $\Lambda$ parameters

$$\frac{\Lambda^{(0)}}{\mu^{(0)}} \overset{\Lambda/M\to 0}{\sim} P_{0,nf}(M/\Lambda)\frac{\Lambda^{(n_f)}}{\mu^{(n_f)}} + \mathcal{O}(M^{-2}) \tag{3.1.24}$$

This strategy allows us to determine the $n_f$-flavor $\Lambda$-parameter from the pure gauge one. One only needs the values of a massive coupling with either three or four flavors in order to apply the matching condition Eq. (3.1.22), and the pure gauge $\Lambda$-parameter

$$\Lambda^{(n_f)} = \lim_{M\to\infty} \frac{\mu^{(n_f)}}{P_{0,nf}(M/\Lambda)} \times \frac{\Lambda^{(0)}}{\mu^{(0)}} \tag{3.1.25}$$

The limit of infinite mass ensures that all corrections (both perturbative and power corrections) vanish, which makes this an exact relation.

Although this strategy does not completely solve the window problem, the slowly decreasing perturbative uncertainties are only present in the pure gauge determination of $\frac{\Lambda^{(0)}}{\mu^{(0)}}$. Note that the pure gauge theory is much more tractable: simulations are much cheaper, algorithms are better, and the step scaling strategy is much more straightforward. Perturbative uncertainties in the decoupling of heavy quarks are negligible, and only the power corrections $\mathcal{O}(M^{-2})$ have to be dealt with. A recent publication [242], using quarks in the range 2-12 GeV shows that precise results can be obtained with this strategy. The uncertainty is still dominated by statistical uncertainties, and in fact a substantial part of it comes from the pure gauge running, which can be further reduced.

Now we move to strategies that suffer from the window problem described above. In all these methods the uncertainties are dominated typically by the uncertainties associated with the truncation of the perturbative expansion Eq. (3.1.15), or the cutoff effects arising from the difficulty in performing a continuum extrapolation for quantities defined at scales of a few lattice spacing.

*Ghost-ghost-gluon vertex*
The QCD vertices are computed numerically and compared to their perturbative expansion. As the field correlators involved are not gauge-invariant, these calculations require a gauge-fixing procedure, which has potential extra uncertainties due to Gribov copies. Non-perturbative corrections and lattice cutoff effects are sizeable in the regime of current simulations.

*Static potential*
The interaction between static quarks is known to high orders in perturbation theory, and the data seems to follow to perturbative prediction down to scales of the order of 1.5 GeV. The main drawback comes from the fact that the observable is not IR-safe, which leads to the resummation of soft and ultra-soft divergences, and hence the introduction of an extra soft scale in the problem.

*Heavy-quark correlators*
The pseudoscalar density correlators are defined as

$$G(x_0) = a^6(am_0)^2 \sum_{\mathbf{x}} \langle \overline{\psi}\gamma_5\psi(\mathbf{x},x_0)\, \overline{\psi}\gamma_5\psi(\mathbf{0},0)\rangle \,. \tag{3.1.26}$$

Note that after summing over all spatial sites on the right-hand side, the correlator only depends on $x_0$. The normalization is fixed by multiplying the field correlator by the factor $a^6(am_0)^2$. Their moments have a well-defined perturbative expansion in powers of the strong coupling constant. These correlators are computed in lattice simulations, which yield a good statistical precision on the final result. The main drawback of this approach is the large cutoff effects that affect the quantities used. It is indeed very challenging to explore energy scales larger than the physical charm quark mass $m_c \sim 1.4\,\text{GeV}$, which is not clearly in the perturbative regime. The recent work in Ref. [243] explores different energy scales in the range $\bar{m}_c - 3\bar{m}_c$, but the continuum extrapolation is very challenging already at $\mu \gtrsim 2m_c$.

*Wilson loops*
The expectation values of Wilson loops of multiple sizes $m\times n$ are computed at the scale of the lattice cutoff $1/a$. While these quantities are not extrapolated to their continuum limit, they can be computed in bare lattice perturbation theory. The perturbative series can then be tranlsated into an expansion in the remormalized coupling $\alpha_{\overline{\text{MS}}}(\mu)$. The typical scale for these observables is $\mu \sim 1/a$. Unfortunately the known perturbative orders are not sufficient to describe the data and several coefficients of the expansion need to be fitted. While the statistical uncertainty of these determinations is excellent, they are plagued by the systematic errors due to the perturbative truncation.

*Hadron Vacuum Polarization (HVP)*
The strong coupling constant can be extracted from the correlators of vector and axial vector currents:

$$V_\mu^a(x) = \bar{\psi}_a\gamma_\mu\psi_a(x)\,,$$
$$A_\mu^a(x) = \bar{\psi}_a\gamma_5\gamma_\mu\psi_a(x)\,,$$

after a decomposition in Fourier space (with $J_\mu = V_\mu, A_\mu$)

$$\int \mathrm{d}^4x\, e^{\imath px}\, \langle J_\mu^a(x) J_\nu^a(0)\rangle = \\ = (\delta_{\mu\nu}p^2 - p_\mu p_\nu)\Pi_J^{(1)}(p^2) - p_\mu p_\nu \Pi_J^{(0)}(p^2)\,.$$

The quantity

$$\Pi(p^2) = \\ = \Pi_V^{(0)}(p^2) + \Pi_V^{(1)}(p^2) + \Pi_A^{(0)}(p^2) + \Pi_A^{(1)}(p^2)$$

is dimensionless and has a perturbative expansion

$$\Pi(p^2) \overset{p\to\infty}{\sim} c_0 + \sum_{k=1}^{4} c_k(s)\alpha^k_{\overline{\rm MS}}(\mu) + \mathcal{O}(\alpha^5_{\overline{\rm MS}}) ,$$
$$(s = p/\mu)$$

known up to 5-loops. The constant term $c_0(s)$ is divergent, so that the strong coupling is usually extracted from the difference $\Pi(p^2) - \Pi(p^2_{\rm ref})$, or the Adler function

$$D(p^2) = p^2 \frac{\mathrm{d}\Pi(p^2)}{\mathrm{d}p^2} . \tag{3.1.27}$$

The main issue with extractions based on the HVP is that power corrections are significant even for large momenta [244]. Ref. [245] pushes the determination to high energies, so that the data can be described without any power corrections, but then cutoff effects become larger and the window of scales to obtain the strong coupling decreases.

*Dirac Spectral Density*
The density of the eigenvalues of the Dirac operator,

$$\rho(\lambda) = \frac{1}{V}\left\langle \sum_k [\delta(\lambda - \imath\lambda_k) + \delta(\lambda + \imath\lambda_k)] \right\rangle , \tag{3.1.28}$$

has recently been used to determine the strong coupling via its perturbative expansion

$$\rho(\lambda) = \frac{3\lambda^3}{4\pi^2}\left(1 - \rho_1(s)\alpha_{\overline{\rm MS}}(\mu) - \rho_2(s)\alpha^2_{\overline{\rm MS}}(\mu) - \rho_3(s)\alpha^3_{\overline{\rm MS}}(\mu) + \mathcal{O}(\alpha^4_{\overline{\rm MS}})\right) ,$$
$$(s = \mu/\lambda) .$$

The extraction of the spectral density is usually performed at very low energy scales in order to keep the discretization effects under control. Recent work [246] imposes a cut $a\lambda < 0.5$ in order to avoid a substantial deviation from the continuum result. This restricts the energy scales that can be reached with their data-set (with lattice spacings $a^{-1} = 2.5, 3.6$ and $4.5\,\mathrm{GeV}$) to $\lambda < 1.2\,\mathrm{GeV}$.

### 3.1.4 Determinations of the quark masses

Because of confinement, only color-neutral states are observed as physical states and therefore the quark masses cannot be measured directly in experiments. On the other hand lattice QCD offers a unique opportunity to determine these quantities. In fact the $n_f$ bare quark masses appearing as parameters in the lattice QCD action have to be tuned using $n_f$ physical observables in order to make any meaningful prediction. Once this tuning is performed, we only need to renormalize its values to some convenient scheme. The scale dependence of renormalized values for quark masses in mass-independent renormalization schemes is described by the mass anomalous dimension, $\gamma(\bar{g})$, which only depends on the gauge coupling and obeys the RG equation

$$\mu\frac{d}{d\mu}\bar{m}_i(\mu) = \gamma(\bar{g})\bar{m}_i(\mu) \overset{\bar{g}\to 0}{\sim} -\bar{g}^2\sum_{k=0}^{\infty} d_k\bar{g}^{2k} , \tag{3.1.29}$$

where the leading perturbative coefficient

$$d_0 = \frac{1}{(4\pi)^2}\left(11 - \frac{2n_f}{3}\right) \tag{3.1.30}$$

is scheme-independent. As for the coupling, the quark masses are defined in a given renormalization scheme and at a given renormalization scale; the conventional practice is to quote a value for the masses in the $\overline{\rm MS}$ scheme, $\bar{m}_{\overline{\rm MS}}(\mu)$ (with $\mu = 2$ GeV for light quarks), but as in the case of the coupling we find more natural to work with renormalization group invariant (RGI) quantities. The evolution equation, Eq. (3.1.29), can be integrated exactly to yield

$$M_i = \bar{m}_i(\mu)\left(2b_0\bar{g}(\mu)^2\right)^{-d_0/(2b_0)} \times \exp\left\{-\int_0^{\bar{g}(\mu)} \mathrm{d}x \left[\frac{\gamma(x)}{\beta(x)} - \frac{d_0}{b_0 x}\right]\right\} . \tag{3.1.31}$$

Once again we can think of the RGI mass $M_i$ as a scale-independent energy that specifies the boundary condition for the mass evolution and hence fully determine the renormalized mass at all energies. An additional benefit of quoting RGI quark masses is that they are scheme independent (and therefore well defined beyond perturbation theory). On the other hand, the determination of RGI quark masses requires the knowledge of the evolution of the coupling. Given that the current precision of the $\Lambda$ parameter (about a 4%) is much lower than the precision of quark masses at low energies (about a 1% precision), the values of quark masses at a few GeV are much more precise than their RGI counterparts. Note however that this usually means that perturbation theory has been used at a few GeV, and all the caveats about the use of perturbation theory at medium energies raised in the previous section are also applicable here; the determination of quark masses is also plagued by a window problem. However from a practical point of view, the perturbative uncertainties in this case seem to be much better behaved than in the case of the extractions of the strong coupling.

Nowadays the most precise results available in the FLAG review[63] for light and heavy quark masses are obtained in the isosymmetric limit. There are few subtleties involved in these extractions; they originate from the fact that experimental inputs include QED and isospin-breaking corrections, while these effects are not included in the lattice simulations. These effects are small but they are relevant at the level of precision of state-of-the-art lattice computations. Ideally one would like to subtract the isospin breaking corrections from the experimental data. The problem is that electromagnetic interactions affect the RG functions (both $\beta(\bar{g})$ and $\tau(\bar{g})$) with $\mathcal{O}(\alpha_{\rm EM})$ contributions: quarks with different electric charges (like the $u$ and $d$ quarks) run differently. QED makes the isospin symmetric point ill defined. Even if we impose $\bar{m}_u(\mu) = \bar{m}_d(\mu)$ at $\mu = 2\,\mathrm{GeV}$, the $u$ and $d$ quarks will be non-degenerate at another generic renormalization scale. Since the subtraction of isospin breaking corrections depends on the definition of the isospin symmetric limit, it is clear that there are (small) ambiguities whatever convention one chooses. The FLAG review [63] contains a detailed discussion both in the quark mass section and in the scale setting section about this particular issue, and the reader is encouraged to consult it for more details.

We end this introduction by emphasizing that the inclusion of the leading QED and strong isospin breaking corrections (including quark loop effects) is an active area of research in lattice QCD. Results with a first principles description of the standard model at low energies, including QCD, QED and strong isospin breaking, are rapidly becoming the new standard for lattice computations where this level of precision is required.

### 3.1.5 Quark mass definitions

Here we consider the determination of quark masses in QCD alone (i.e. a sensible definition of the isospin symmetric point has been made). Quark currents play a central role in QCD In particular, the axial current and pseudo scalar density

$$A^a_\mu(x) = \bar{\psi}(x)\gamma_\mu\gamma_5\frac{\sigma^a}{2}\psi(x)\,, \tag{3.1.32}$$

$$P^a(x) = \bar{\psi}(x)\gamma_5\frac{\sigma^a}{2}\psi(x)\,, \tag{3.1.33}$$

are expected, in the continuum, to obey the PCAC relation

$$\partial_\mu A^a_\mu(x) = mP^a(x)\,. \tag{3.1.34}$$

This relation is often used to *define* renormalized quark masses. The reason is that we expect the same relation to hold in the lattice regularized theory after renormalization and up to cutoff effects[9]. The axial current and pseudo scalar density are renormalized multiplicatively

$$(A_R)^a_\mu(x) = Z_A A^a_\mu(x)\,, \tag{3.1.35}$$

$$(P_R)^a(x) = Z_P(\mu)P^a(x)\,. \tag{3.1.36}$$

Note that the axial current renormalization factor is scale independent. Quark masses are also expected to renormalize multiplicatively $\bar{m}(\mu) = Z_m(\mu)m_0$, leading to the lattice version of the PCAC relation

$$\partial_\mu A^a_\mu(x) = \frac{2Z_m(\mu)Z_P(\mu)}{Z_A}m_0P^a(x)\,. \tag{3.1.37}$$

This relation allows to determine the renormalized quark masses via the relation

$$\bar{m}(\mu) = Z_m(\mu)m_0 = \frac{Z_A\langle\partial_\mu A^a_\mu(x)O_{\rm ext}\rangle}{Z_P(\mu)\langle P^a(x)O_{\rm ext}\rangle}\,, \tag{3.1.38}$$

with much freedom to choose the probe $O_{\rm ext}$. Note that the *running* of the quark masses is given by the scale-dependent renormalization factor $Z_P(\mu)$. There are several methods to determine it on the lattice. Most recent works use nonperturbative renormalization schemes.

*RI-(S)MOM schemes*

These renormalization schemes are conceptually very similar to the one used in perturbation theory. There exists several possibilities, but all are based on imposing a suitable renormalization condition to some Green functions with external momenta playing the role of the renormalization scale. In principle the renormalization scheme is formulated in infinite volume and at zero mass. In this setup the connection with perturbation theory is known to high accuracy (up to 4-loops), but this setup cannot be simulated directly on the lattice, so the infinite volume and zero mass limit require a dedicated study. In particular these methods suffer from a window problem (the impossibility to keep the volume large and at the same time have access to high energy scales where perturbation theory can be trusted).

*Finite volume schemes*

In this schemes the renormalization condition is imposed in a finite volume $L$, which plays the role of the renormalization scale (i.e. $\mu \sim 1/L$). With a smart choice of boundary conditions one can directly simulate massless quarks. Contact with perturbation theory is typically only known up to 2-loops, but using the techniques of finite size scaling, this matching can be performed at very high energies (i.e. 100 GeV), where perturbative uncertainties are negligible.

[9] Depending on the type of fermion formulation used and other details, the cutoff effects can be $\mathcal{O}(a)$ or $\mathcal{O}(a^2)$. In practice most lattice determinations nowadays choose to eliminate the linear effects in $a$.

| | $M_u^{\mathrm{RGI}}$ | $M_d^{\mathrm{RGI}}$ | $M_s^{\mathrm{RGI}}$ | $M_c^{\mathrm{RGI}}$ | $M_b^{\mathrm{RGI}}$ | [MeV] |
|---|---|---|---|---|---|---|
| $N_{\mathrm{f}} = 2+1$ | 3.15(13) | 6.49(14) | 128(2) | 1526(17) | 6881(63) | |
| $N_{\mathrm{f}} = 2+1+1$ | 2.97(11) | 6.53(11) | 129.7(1.5) | 1520(22) | 6934(58) | |

**Table 3.1.1** FLAG averages of the RGI quark masses in MeV for the $u, d, s, c$ and $b$ quarks (see[63]). Several works contribute to these averages [243, 247–266] computed with either $N_{\mathrm{f}} = 2+1$ and $N_{\mathrm{f}} = 2+1+1$ lattice simulations have about a percent precision for all different quark masses.

### 3.1.6 Approaches for heavy quarks

Heavy quarks are difficult to simulate on the lattice. The reason is that in order to have discretization errors under control, the lattice cutoff $a^{-1}$ has to be much larger than all other scales considered in the problem. In particular we require $am \ll 1$. The lattice community has typically dealt with this problem using an effective description for the heavy quarks (see for example Refs. [267] and [268]). This topic is beyond the scope of this review. Here instead we will focus on some recent works that use a relativistic formulation for the heavy quarks. In particular the recent work [269] uses the expansion of a heavy-light meson mass $M_{hl}$ as a function of the heavy quark pole mass $m_h$

$$M_{hl} = m_h + \bar{\Lambda} + \frac{\mu_\pi - \mu_G(m_h)}{2m_h} + \mathcal{O}(1/m_h^2)\,. \qquad (3.1.39)$$

Here $\bar{\Lambda}$ is the binding energy, $\mu_\pi/2m_h$ is the kinetic energy and $\mu_G(m_h)$ is the hyperfine energy. This relation allows to fit meson masses to the heavy quark pole mass, and therefore to determine it by using the perturbative relation

$$m_h \sim \bar{m}_{\overline{\mathrm{MS}}} \left(1 + \sum_{k=0}^{\infty} r_n \alpha^{n+1}(\bar{m}_{\overline{\mathrm{MS}}})\right)\,. \qquad (3.1.40)$$

The problem of this approach is that the pole mass has a terribly behaved perturbative expansion. In fact

$$r_n = (2b_0)^n \Gamma(n+1+b_1/(2b_0^2))\,. \qquad (3.1.41)$$

Reference [269] uses instead the minimal renormalon subtraction scheme, that has better PT properties.

Making a long story short, heavy-light meson masses are related directly to quark masses, without the need of any non-perturbative renormalization. This approach is used to determine the $b$ meson mass. Masses of other quarks are extracted from appropriate quark mass ratios, that do not need the determination of any renormalization constant.

It has to be pointed out that the heavy quark masses used in this work are often of the order of the lattice UV cutoff, i.e. $aM \sim 1$, and that the direct connection between heavy-light meson masses and quark masses depends on the application of a particular resummed perturbative relation at relatively low energy scales. Despite these caveats, it is clear that this work has looked into the future by simulating relativistic heavy quarks close to the $b$ meson mass.

### 3.1.7 Conclusions

We conclude this section by summarising briefly the status of the determinations of the fundamental parameters of the SM from lattice QCD.

With the advent of dynamical quark simulations and new methods for non-perturbative renormalization, lattice QCD determinations of the strong coupling and quark masses have become both very accurate and very precise. Even if numerical simulations do not qualify as a proof, many of us believe that these computations have fulfilled the dream of connecting the fundamental quark masses and strong coupling to the well measured spectra of hadrons from first principles.

There are two challenges that lattice QCD computations face in this game. On one hand the strong coupling and quark masses are useful when quoted in the $\overline{\mathrm{MS}}$-scheme, requiring to make contact with perturbation theory while most lattice simulations are performed to explores hadronic low energy scales. On the other hand experimental input (hadron masses), have electromagnetic and strong isospin breaking corrections, while most lattice QCD simulations are performed in the isospin symmetric limit.

*The window problem*
Connecting the perturbative and hadronic regimes of QCD is hard. These two scales are separated by a large gap in energy scales, due to the logarithmic running of the strong coupling with the renormalization scale. It is very challenging to accommodate these disparate scales in a single lattice simulation, and if insists on doing so, compromises have to be made and perturbation theory has to be used at a few GeV.

*Isospin breaking corrections*
The simulation of electromagnetism on the lattice poses its own challenges (see [270] for a review), related with the description of charged states in presence of long range interactions. The simulation of non-degenerate light quarks is also numerically challenging. These reasons explain that most lattice computations are performed in the isospin symmetric limit.

The lattice community has made great progress on these fronts in recent years. The window problem has a known solution since the early 90's: finite size scaling [238]. It has been applied to $N_f = 0, 2, 3, 4$ QCD [271–273] and to the determination of quark masses [], but these determinations traditionally produced results for the strong coupling with large statistical uncertainties. Thanks to recent developments [274], finite size scaling studies can achieve a subpercent level of precision in the strong coupling [275]. These techniques have also been applied to the determination of quark masses [249, 276, 277]. Finite size scaling has been for a long time the only solution to the window problem, until a new method based on decoupling of heavy quarks has been proposed [239]. This new method largely reduces the window problem and recent results show that the strong coupling can also be determined using these techniques with a sub-percent precision [242]. This strategy has not yet been applied to the determination of quark masses, but the method should also lead to precise determinations of the running of quark masses.

With the advent of dynamical fermion simulations the precision of lattice determinations of quark masses has rapidly reached a very mature status. Renormalization is nowadays performed in a fully non-perturbative way, and using different strategies. Although contact with perturbation theory has to be made, and in principle there is also a window problem present in the extraction of quark masses, perturbative uncertainties in this case seem to be much better behaved than in the case of extractions of the strong coupling. All in all, at the current level of precision the presence of electromagnetism and strong isospin in nature is the main factor limiting the precision of many lattice computations. But the field evolves very quickly and there exist several lattice computations of the individual light quark masses $m_u, m_d$ that directly compute the QED effects in the quenched approximation. We are convinced that unquenched results will follow soon, and isospin breaking corrections will be applied to the determinations of all quark masses.

Only fifteen years after the first lattice QCD simulations with dynamical quarks, lattice QCD has been able to determine from first principles the strong coupling with a 0.7% error. Quark masses are determined with a percent error (see Table 3.1.1), and soon these computations will include full isospin breaking corrections. The implications of these calculations are far reaching in constraining the SM description of physical phenomena. Lattice determinations of $\alpha_s$ are the most precise (see the next section). The FLAG average based on $N_f = 2 + 1 + 1$ simulations of the $u$ quark mass is $M_u^{\mathrm{RGI}} = 2.97(11)$ MeV (based on the works [248, 269], see also [247]). This value disfavors a popular solution to the strong CP problem (a massless $u$ quark) by 30 standard deviations without any assumptions or model for the strong interactions.

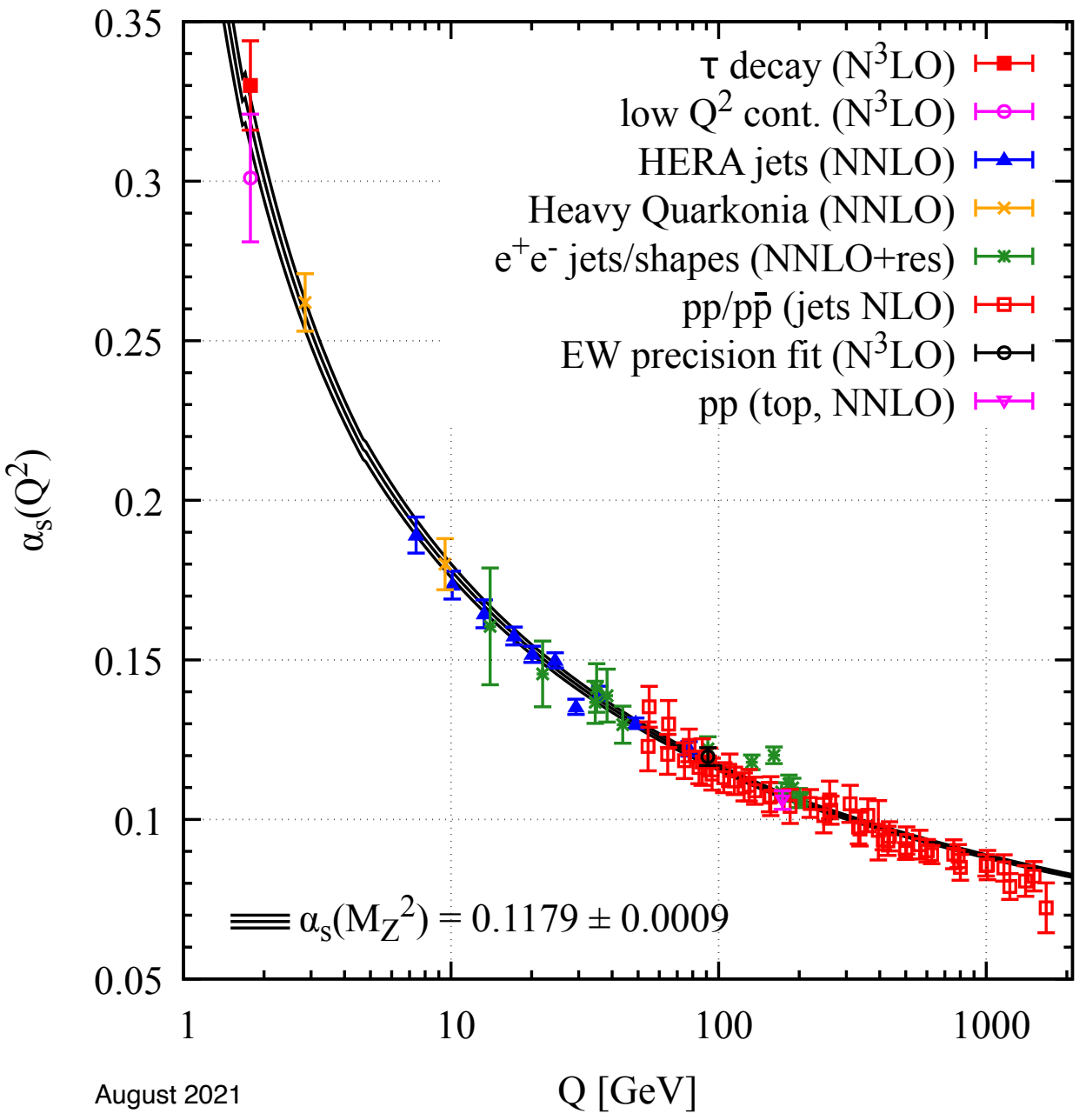

**Fig. 3.2.1** Measurements of the coupling constant $\alpha_s$, as a function of the energy scale $Q$. The level of precision of the perturbative prediction used in the measurement of $\alpha_s$ is indicated in brackets (NLO: next-to-leading order; NNLO: next-to-next-to-leading order; NNLO+res.: NNLO matched to a resummed calculation; N$^3$LO: next-to-next-to-leading order. Figure taken from Ref. [278].

## 3.2 The strong-interaction coupling constant

**Giulia Zanderighi**

### 3.2.1 The world average determination of $\alpha_s$

We summarize here the current procedure used in the PDG [278] to obtain the world average value of $\alpha_s(M_z^2)$ and its uncertainty, and we discuss future prospects for its improvement.

*Preliminary considerations*
All observables involving the strong interaction depend on the value of the strong coupling constant. This implies that a number of different observables can be used to determine the coupling constant, provided that a suitable theoretical prediction is available for that observable. The following considerations are used to assess if a particular observable is suitable for use in the determination of the strong coupling constant:

- The observable's sensitivity to $\alpha_s$ as compared to the experimental precision. For example, for the $e^+e^-$ cross section to hadrons (e.g. the $R$ ratio), QCD effects are only a small correction, since the perturbative series starts at order $\alpha_s^0$, but the experimental precision is high. Three-jet production, or event shapes, in $e^+e^-$ annihilation are directly sensitive to $\alpha_s$ since they start at order $\alpha_s$. Four- and five-jet cross-sections start at $\alpha_s^2$ and $\alpha_s^3$ respectively, and hence are very sensitive to $\alpha_s$. However, the precision of the measurements deteriorates as the number of jets involved increases.
- The accuracy of the perturbative prediction, or equivalently of the relation between $\alpha_s$ and the value of the observable. The minimal requirement is generally considered to be an NLO prediction. The PDG imposes now that at least NNLO accurate predictions be available. In certain cases where phase space restrictions require it, fixed-order predictions are supplemented with resummation. An improved perturbative accuracy is necessary to guarantee that the theoretical uncertainty is assessed in a robust way.
- The size of non-perturbative effects. Sufficiently inclusive quantities, like the $e^+e^-$ cross section to hadrons, have small non-perturbative contributions $\sim \Lambda^4/Q^4$. Other quantities, such as event-shape distributions, typically have contributions $\sim \Lambda/Q$. All other aspects being equivalent, observables with smaller non-perturbative corrections are preferable.
- The scale at which the measurement is performed. An uncertainty $\delta$ on a measurement of $\alpha_s(Q^2)$, at a scale $Q$, translates to an uncertainty $\delta' = \alpha_s^2(M_Z^2)/\alpha_s^2(Q^2) \times \delta$ on $\alpha_s(M_Z^2)$. For example, this enhances the already important impact of precise low-$Q$ measurements, such as from $\tau$ decays, in combinations performed at the $M_Z$ scale.

The PDG determination of $\alpha_s$ first separates measurements into a number of different categories, then calculates an average for each category. This average is then used as an input to the world average. The PDG procedure requires:

1. a specification of the conditions that a determination of $\alpha_s$ should fulfill in order be included in the average;
2. a specification of the separations of the different extractions of $\alpha_s(M_Z^2)$ into the separate categories;
3. a specification of the procedure within each category of the procedure to compute the average and its uncertainty;
4. a specification of the manner in which the different sub-averages and their uncertainties are combined to determine the final value of $\alpha_s(M_Z^2)$ and its uncertainty.

*Details of the PDG averaging procedure*

In the following, we summarize the procedure adopted in the last edition of the PDG [278]. There, the selection of results from which to determine the world average value of $\alpha_s(M_Z^2)$ is restricted to those that satisfy a well defined set of criteria. These are that the fit should be

1. accompanied by reliable estimates of all experimental and theoretical uncertainties;
2. based on the most complete perturbative QCD predictions of at least next-to-next-to leading order (NNLO) accuracy;
3. published in a peer-reviewed journal at the time of writing of the PDG report.

Note that the second condition to some extent follows from the first. In fact, determinations of the strong coupling from observables in $e^+e^-$ involving e.g. five or more jets are very sensitive to $\alpha_s$, and could provide additional constraints. However, these observables are currently described only at leading order (LO) or next-to-leading order (NLO), and the determination of the theoretical uncertainty is thus considered not sufficiently robust. It is also important to note that some determinations are included in the PDG, but the uncertainty quoted in the relevant publications is increased by the PDG authors to fulfill the first condition. Similarly, in some cases the central value used in the PDG differs from the one quoted in some publications, but can be extracted from the analysis performed in that work.

*Categories of observables*

All observables used in the determination of $\alpha_s(M_Z^2)$ in the PDG averaging procedure are classified in the following categories

- "Hadronic $\tau$ decays and low $Q^2$ continuum" ($\tau$ decays and low $Q^2$): the coupling constant is here determined at the $\tau$ mass, therefore once it is evolved up to the $Z$ mass the uncertainty shrink. Perturbative calculations for $\tau$ decays are available at N$^3$LO, however there are different approaches to treat the perturbative and non-perturbative contributions, which result in significant differences. These discrepancies are currently the limiting factor in reducing the uncertianty in this category.

– "Heavy quarkonia decays" ($Q\bar{Q}$ bound states): calculations are available at NNLO and N$^3$LO.
– "PDF fits" (PDF fits): this category include both global PDF fits and analyses of singlet and non-singlet structure functions. To quantify the theory uncertainty, half of the difference between results obtained with NNLO and NLO predictions is added in quadrature.
– "Hadronic final states of $e^+e^-$ annihilations" ($e^+e^-$ jets & shapes): these fits use measurements at PETRA and LEP. Non-perturbative corrections are important, going as $\Lambda/Q$ and can be estimated either via Monte Carlo simulations or analytic modeling.
– "Observables from hadron-induced collisions" (hadron colliders): NNLO calculations for $t\bar{t}$ or jet production at both the LHC and HERA, and $Z$+jet production at the LHC have allowed measurements for these processes to be used in $\alpha_s$ determinations. An important open question is whether a simultaneous PDF and $\alpha_s$ fit has to be carried out in order to avoid a potential bias.
– "Electroweak precision fit" (electroweak): $\alpha_s$ determinations are averaged from electroweak fits to data from the Tevatron, LHC, LEP and the SLC. These fits rely on the strict validity of the Standard Model.
– "Lattice": the average determined by the FLAG group in 2019 [279] from an input of 8 determinations was used in the last PDG determination; the subsequent 2021 $\alpha_s$ average is very consistent with that of 2019.

Detailed information about which observables are included in the different categories can be found in Ref. [278].

*Average and uncertainty in each category*
In order to calculate the world average value of $\alpha_s(M_Z^2)$, a preliminary step of pre-averaging results within the each category listed in Sec. 3.2.1 is carried out. For each sub-field, except for the "Lattice" category, the *unweighted average* of all selected results is taken as the pre-average value of $\alpha_s(M_Z^2)$, and the unweighted average of the quoted uncertainties is assigned to be the respective overall error of this pre-average. An unweighted average is used to avoid the situation in which individual measurements, which may be in tension with other measurements and may have underestimated uncertainties, can considerably affect the determination of the strong coupling in a given category. As an example, the determination of $\alpha_s(M_Z^2)$ from $e^+e^-$ jets & shapes currently averages ten determinations and arrives at $\alpha_s(M_Z^2) = 0.1171 \pm 0.0031$. Since two determinations [280, 281], both based on a similar theoretical framework, arrive at a small value of $\alpha_s(M_Z^2)$ and have a very small uncertainty, if one were to perform a weighted average one would arrive at $\alpha_s(M_Z^2)$ from $e^+e^-$ jets & shapes of $\alpha_s(M_Z^2) = 0.1155 \pm 0.0006$, which is not compatible with the current world average. This would, in fact, considerably change the world average because of the very small uncertainties. The current procedure is instead robust against $\alpha_s(M_Z^2)$ determinations that are outliers with small uncertainties as compared to the other determinations in the same category. For the "Lattice QCD" (lattice) sub-field, the PDG adopts the LAG2019 average value and uncertainty for this sub-field [279]. FLAG2019 also requires strict conditions on its own for a determination to be included in their average, which are in line with those used in the PDG. The results of the averages of the categories are given in table 3.2.1. From the table, it is clear that determinations from different categories are compatible with each other and accordingly can be combined to give rise to a final average.

*Final average*
Since the six sub-fields (excluding lattice) are largely independent of each other, the PDG determines a non-lattice world average value using a standard '$\chi^2$ *averaging*' method. This result in the final average of the six categories of

$$\alpha_s(M_Z^2) = 0.1175 \pm 0.0010\,, \qquad \text{(without lattice)}, \tag{3.2.1}$$

which is fully compatible with the lattice determination. In a last step the PDG performs an unweighted average of the values and uncertainties of $\alpha_s(M_Z^2)$ from the non-lattice result and the lattice result presented in the FLAG2019 report,which results in the final average of

$$\alpha_s(M_Z^2) = 0.1179 \pm 0.0009\,, \qquad \text{(final average)}. \tag{3.2.2}$$

Performing a weighted average of all seven categories would instead give rise to $\alpha_s(M_Z^2) = 0.1180 \pm 0.0006$. The PDG uncertainty is instead more conservative and about 50% larger. These final results are summarized in Fig. 3.2.2.

| category | $\alpha_s(M_Z^2)$ |
|---|---|
| $\tau$ decays and low $Q^2$ | $0.1178 \pm 0.0019$ |
| $Q\bar{Q}$ bound states) | $0.1181 \pm 0.0037$ |
| PDF fits | $0.1162 \pm 0.0020$ |
| $e^+e^-$ jets & shapes | $0.1171 \pm 0.0031$ |
| hadron colliders | $0.1165 \pm 0.0028$ |
| electroweak | $0.1208 \pm 0.0028$ |
| lattice | $0.1182 \pm 0.0008$ |

**Table 3.2.1** PDG average of the categories of observables. These are the final input to the world average of $\alpha_s$.

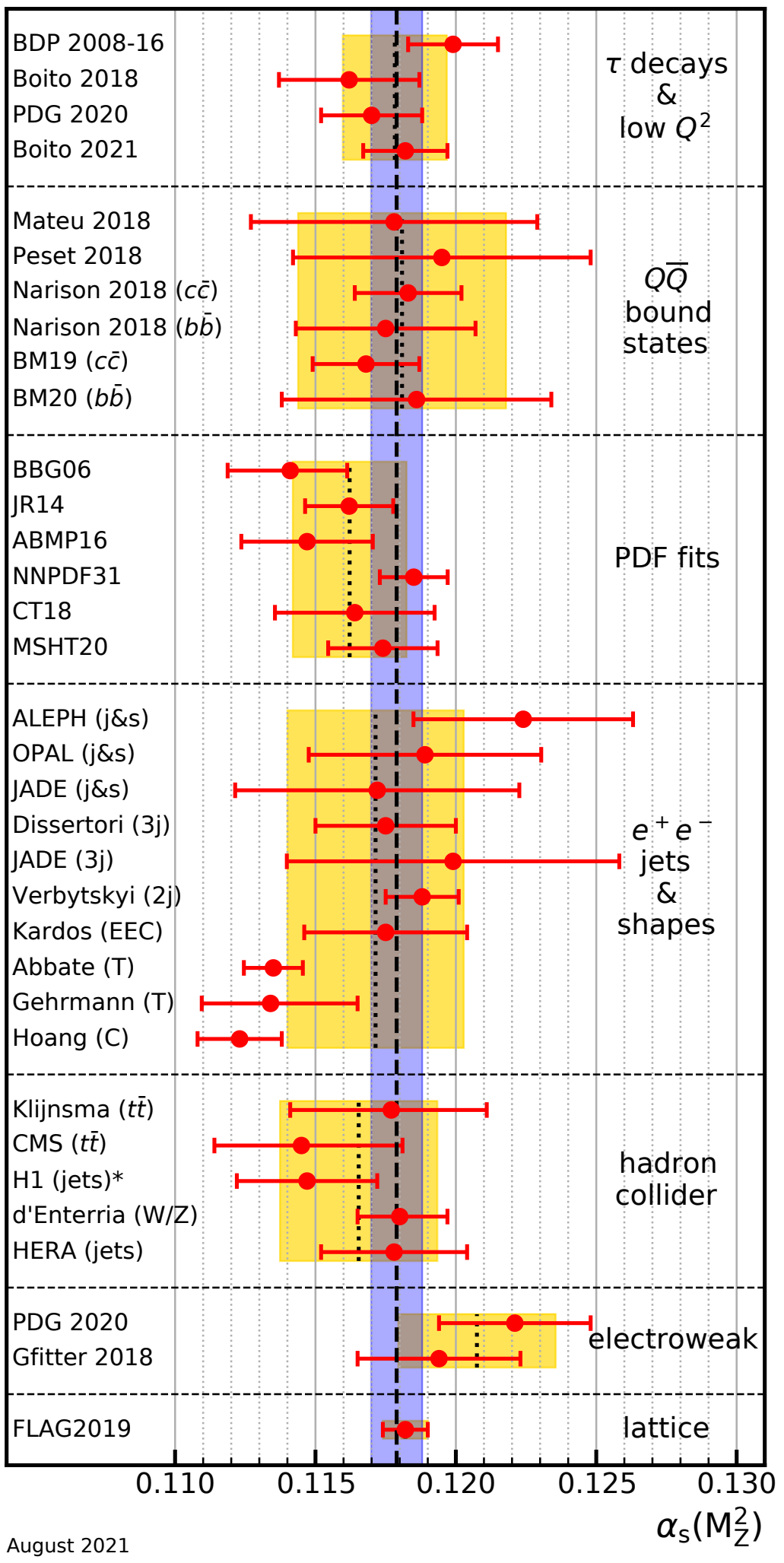

**Fig. 3.2.2** Summary of the determinations of $\alpha_s(M_Z^2)$ from the seven sub-fields used in the PDG [278], as discussed in the text. The yellow (light shaded) bands and dotted lines indicate the pre-average values of each sub-field. The dashed line and blue (dark shaded) band represent the final world average value of $\alpha_s(M_Z^2)$. The '*' symbol within the "hadron colliders" sub-field indicates a determination including a simultaneous fit of parton distribution functions. All other "hadron collider" determinations instead use a set of parton distribution function as input to the fit. Figure taken from Ref. [278].

### 3.2.2 Outlook

Despite the numerous determinations of the strong coupling constant, it remains to date the least well-known gauge coupling, with an uncertainty of about 1%. Still it is a remarkable success that all determinations from all categories agree well with each other, all within about one sigma. Future improvements are likely to be driven by those categories which today have the smallest uncertainties, i.e. lattice determinations, $\tau$ decays and low $Q^2$ measurements.

As far as the category "$\tau$ decays and low $Q^2$ measurements" are concerned, it is important to mention that the uncertainty quoted in the latter category includes the difference in the extractions that are obtained using contour improved perturbation theory (CIPT) and fixed order perturbation theory (FOPT). Recent arguments suggest that FOPT are to be preferred, see also dedicated discussions on this point in Ref. [282]. If this is confirmed, the value of $\alpha_s(M_Z^2)$ in this category would shift slightly to lower values, and would allow one to quote a reduced theoretical uncertainty since this additional source of uncertainty would be completely removed. Further improvements could also come from a better understanding of non-perturbative effects.

Important progress is also expected in the category "$e^+e^-$ jets & shapes", where the calculation of power corrections in the 3-jet region [161, 283] could have a sizeable impact, and improve fits of the coupling from event shapes. In fact, in current determinations that rely on an analytic computation of non-perturbative power corrections, these calculations are performed in the two-jet limit and applied to the kinematic region used in the fits where events typically have an additional hard emission, i.e. to three-jet configurations. A treatment of these corrections in the three-jet region is now possible, at least for some observables and the impact of this improved treatment of non-perturbative effects on $\alpha_s(M_Z^2)$ in this category is eagerly awaited.

As far as the hadron collider category is concerned, it is an open question if it is always preferred to fit $\alpha_s(M_Z^2)$ and the parton distribution functions simultaneously and how to best deal with correlations between the parton distribution function parameters and $\alpha_s(M_Z^2)$ in the cases where the fit is not performed simultaneously. In view of many more NNLO results to come and many more data from the LHC, we can expect theoretical work and advances in addressing this question. Ratios of cross sections are less sensitive to parton distribution functions and therefore could be considered more suitable to extract $\alpha_s$. For instance, NNLO predictions for 3-jet production will enable to perform fits of $\alpha_s(M_Z^2)$ from ratios with at least partial cancellation of some uncertainties. It is not clear whether this reduction in uncertainty also holds for the PDF dependence of such ratio predictions. Moreover, for predictions of ratios of cross sections, the natural central scale choice in numerator and denominator are in general not the same. Data from the "hadron collider category" at high $Q^2$ will also be crucial to test the running of the coupling to highest energy scales. Such tests are important since heavy states that couple strongly could modify the running of $\alpha_s$ at high $Q^2$.

Finally, it is important to mention that the recent years have seen remarkable advances in the determination of $\alpha_s(M_Z^2)$ from lattice calculations, also thanks to the the FLAG effort which imposes strict quality criteria for lattice determinations to be included in the FLAG average. This is now the single most precise result of all categories included in the PDG and agrees remarkably well (both in terms of central value and uncertainty) with the PDG world average of $\alpha_s$ without lattice data. Further improvements from lattice calculations are also expected in the coming decade. Given all the progress to be expected in the coming years in various aspects and categories, a determination of $\alpha_s$ with sub-percent precision seems finally within reach.

# 4 Lattice QCD

*Conveners:*
**Kostas Orginos and Franz Gross**

In 1974, shortly after Quantum Chromodynamics was established, Kenneth Wilson published a seminal paper in which he formulated the theory on a space-time lattice. This formulation had profound implications. It preserved the gauge invariance of the theory while regulating ultraviolet divergences and providing a definition of QCD as the continuum limit of the lattice theory. However, one may argue that the most crucial implication was the fact that it offered a pathway to non-perturbative computations. Quantities such as the spectrum of stable hadrons, decay constants, and Parton distribution functions to name a few, could now in principle be computed from the fundamental theory without the need for uncontrolled approximations. In the beginning, however, this formulation of QCD lent itself to a different type of analytic computations such as the strong coupling expansion where Wilson showed that color charges are indeed confined in the strong coupling limit.

Numerical investigations of Lattice QCD (LQCD) started a few years later with the pioneering work of Michael Creutz in 1980. There for the first time, the SU(2) pure Yang-Mills theory was investigated using Monte Carlo methods. Subsequently, many groups around the world started studying Lattice QCD, developed methods and algorithms, and investigated the efficacy of the available computer hardware for numerical calculations in LQCD. Although the fundamental principles of such calculations were clear, it was evident from the beginning that the computational cost for achieving phenomenologically relevant results was enormous. In addition, the limitations of Euclidean time formulation as well as the computational limitations imposed by finite volume and lattice spacing made it clear that computational power alone will not be enough. Therefore, intense theoretical research to develop methods and algorithms started in the '80s. Together with that effort, many groups devoted efforts to designing custom-made supercomputers that were best suited for the problem at hand. The idea of a massively parallel computer to solve scientific problems seemed at odds at the time with the vector machines that defined the commercially available high-performance computers. Yet in the '90s the rise of massively parallel computers, commercial or custom-made, led to major advances in LQCD. The new century brought a combination of powerful supercomputers, sophisticated numerical techniques, and advanced theoretical approaches that allowed for the first time to compute physical quantities at phenomenologically relevant accuracy.

Lattice QCD is now an established field that can provide results at unprecedented accuracy and can help move forward our fundamental understanding of particle physics. The impact of lattice QCD computations on strong interaction physics is evident throughout this volume. Nearly every section contains references to landmark lattice QCD computations. In this section, a brief introduction to the formulation of lattice QCD is given by Gottlieb, followed by De Tar's review of the basic LQCD algorithms. Leinweber discusses the structure of the QCD vacuum as it emerges from numerical experiments. Karsch reviews computations at non-zero temperatures and densities relevant to understanding quark-gluon plasma physics.

The discussion then continues with a focus on applications. Dudek reviews hadron spectroscopy with emphasis on finite volume methods that allow for the extraction of scattering amplitudes from Euclidean time correlation functions. Constantinou/Orginos discuss computations of the nucleon structure including modern approaches that allow for the extraction of momentum-fraction-dependent distributions from Euclidean time computations. Finally, Davies reviews computations for Weak matrix element computations which play a central role in the experimental program for probing physics beyond the standard model (BSM).

## 4.1 Lattice field theory

**Steve Gottlieb**

### 4.1.1 Introduction

In perturbative quantum field theories loop integrals lead to infinities. To deal with these infinities, a reg-ularization scheme must be introduced. Examples of regularization schemes are Pauli-Villars modification of particle propagators and dimensional regularization in which the number of space-time dimensions of the system becomes a variable. After regularization, calculations no longer suffer from infinities, but they do depend on a new parameter specific to the regularization scheme, e.g., $\Lambda$ a large mass in the Pauli-Villars scheme, or $\epsilon = 4 - d$ in the case of dimensional regularization. Since physical results should be independent of the regularization scheme, a renormalization procedure is introduced so that the so-called bare parameters of the theory depend on $\Lambda$ or $\epsilon$ in such a way that physical observables do not as there is a cancellation between the regularization dependence of the bare parameters and those of the loop integrals.

In lattice field theories (LFTs), the theory is modified so that (in finite volume) there are no longer an infinite number of degrees of freedom. For instance, in a scalar field theory instead of a real or complex value of the field at each of the infinite points of space time, there are only a finite number of real or complex degrees of freedom defined on a hypercubic grid of space time points. In this case, the parameter that characterizes the regulator is the distance between the space time points called the lattice spacing, usually denoted $a$. Usually, periodic boundary conditions in space and anti-periodic boundary conditions in time are used. As we will see in more detail below, the field can be Fourier transformed and in momentum space there is a maximum momentum as each component of the momentum is in the range $-\pi/a < p_i \leq \pi/a$. In a finite volume, there is also a minimum spacing between allowable momenta components that serves as an infrared regulator. To summarize, the lattice field theory regularizes the theory by introducing a maximum momentum, and the renormalization program is implemented by requiring that physical quantities be independent of the lattice spacing as $a \to 0$. Also, since the lattice theory only has hypercubic and not full rotational symmetry, we must demonstrate that the latter is restored for distances much larger than $a$.

*Actions for a free scalar theory*

To see how LFTs work, let's start with a free scalar field theory in the continuum, transform it to a Euclidean field theory and then put it on a lattice. Start with the Lagrangian density

$$\mathcal{L}(x) = \frac{1}{2}[\partial_\mu \phi(x)\partial^\mu \phi(x) - m^2\phi(x)^2] \tag{4.1.1}$$

and the action

$$\begin{aligned} S &= \int dtL = \int dt \int d^3x\mathcal{L}(x) \qquad (4.1.2)\\ &= \int d^4x \frac{1}{2}[\partial_\mu \phi(x)\partial^\mu \phi(x) - m^2\phi(x)^2] \\ &= \int d^4x \frac{1}{2}[\partial_t \phi(x)\partial_t \phi(x) \\ &\qquad - \nabla\phi(x)\cdot\nabla\phi(x) - m^2\phi(x)^2] \end{aligned}$$

where $\phi(x)$ is the scalar field and $m$ is its mass. The Feynman path integral is defined as

$$Z = \int [d\phi] \exp\{iS\}, \tag{4.1.3}$$

where $[d\phi]$ denotes the integration measure of all possible fields $\phi(x)$. To Euclideanize the theory let $t \to -i\tau$ which changes the sign of the time derivative term in the Lagrangian density. It also adds a factor $-i$ because of the change of integration variable in the action. So, the Euclidean action is defined to be

$$\begin{aligned} S_E = \int d^3x d\tau \frac{1}{2}[\partial_\tau \phi(x)\partial_\tau \phi(x) \\ + \nabla\phi(x)\cdot\nabla\phi(x) + m^2\phi(x)^2], \end{aligned} \tag{4.1.4}$$

and the path integral becomes

$$Z = \int [d\phi] \exp\{-S_E\}. \tag{4.1.5}$$

At this point, it is traditional to rename $\tau$ to $t$, the time variable with which we started, or let $\tau = x_4$. In any case, the field $\phi$ is defined on a 4-dimension Euclidean domain, $S_E$ is positive definite, and this looks like a partition function of a statistical mechanical system. The transformation to Euclidean time allows us to use the importance sampling techniques of statistical mechanics (Monte Carlo methods) introduced in the next section.

To convert to a lattice theory, introduce a spacing $a$ between the points of a hypercubic grid, so the lattice field $\phi_n$ is defined on a discrete set of points $n = (n_1, n_2, n_3, n_4)$ in $R^4$ and $x = an$. Typically, work is done in a finite volume so that $n_i$ is an integer between 0 and $N_i - 1$, where $N_i$ is the extent of the lattice in the $i$-th direction. The derivatives must be replaced by a finite difference approximation. There is more than one way to do this. Pretending for the moment that $\phi$ depends only on a single variable $x$, a forward difference is defined by

$$\Delta^+\phi(x) = \frac{\phi(x+a) - \phi(x)}{a}. \tag{4.1.6}$$

Taylor expanding $\phi(x+a)$ gives

$$\Delta^+\phi(x) = \phi'(x) + \frac{a}{2}\phi''(x) + \cdots . \tag{4.1.7}$$

Note that the symmetric finite difference operator

$$\Delta^S \phi(x) = \frac{\phi(x+a) - \phi(x-a)}{2a} \tag{4.1.8}$$

$$= \phi'(x) + \frac{a^2}{6}\phi'''(x) + \cdots \tag{4.1.9}$$

is a much better approximation of the continuum derivative since the correction is second order in the small lattice spacing $a$.

*Actions for a gauge invariant scalar theory with a $\phi^4$-type interaction*

To introduce gauge invariance, change the real scalar field to a complex field, and introduce a $\phi^4$-type interaction term

$$S = \int d^4x[\partial_\mu \phi^*(x)\partial^\mu \phi(x) - m^2\phi^*(x)\phi(x) \tag{4.1.10}$$

$$- \lambda(\phi^*(x)\phi(x))^2]. \tag{4.1.11}$$

A global gauge transformation is just a change $\phi \to \phi' = \Omega\phi$ where $\Omega$ is complex phase factor, $\Omega = \exp\{i\theta\}$, with $\theta$ a real number independent of $x$. The action is clearly invariant under this gauge transformation since $(\phi')^* = \Omega^*\phi^*$ and for every factor of $\Omega$ coming from transforming $\phi$, there is a corresponding factor of $\Omega^*$ from transforming $\phi^*$. A cubic term in the action would break this gauge invariance.

To generalize to local gauge invariance, allow $\theta$ to become a function of $x$. The mass and interaction terms are clearly still invariant because they only depend on $x$. However, the first term with derivatives transforms in a non-trivial way.

$$\partial_\mu \phi'(x) = \partial_\mu(\Omega(x)\phi(x))$$

$$= (\partial_\mu \Omega(x))\phi(x) + \Omega(x)(\partial_\mu \phi(x))\,. \tag{4.1.12}$$

To handle the extra term depending on $\partial_\mu \Omega(x)$, define a *covariant derivative* $D_\mu$ that has the property

$$D'_\mu \phi'(x) = \Omega(x) D_\mu \phi(x), \tag{4.1.13}$$

so that the covariant derivative $D_\mu \phi(x)$ transforms under a gauge transformation the same way that $\phi(x)$ does. To accomplish this, introduce a vector field $A_\mu(x)$, and define the covariant derivative to be

$$D_\mu = \partial_\mu + ieA_\mu\,. \tag{4.1.14}$$

Using this definition in (4.1.13) gives the constraint

$$(\partial_\mu + ieA'_\mu)(\Omega(x)\phi(x)) = \Omega(x)(\partial_\mu + ieA_\mu)\phi(x)\,. \tag{4.1.15}$$

Requiring that this hold for any field $\phi(x)$ gives the gauge transformation for the field $A_\mu$

$$A'_\mu \Omega = \Omega A_\mu + \frac{i}{e}\partial_\mu \Omega. \tag{4.1.16}$$

This derivation has preserved the order of the terms, so that this equation will hold even for non-Abelian theories in which $\Omega$ is a matrix. Solving for $A'_\mu$ in this most general case gives

$$A'_\mu = \Omega A_\mu \Omega^{-1} + \frac{i}{e}(\partial_\mu \Omega)\Omega^{-1}. \tag{4.1.17}$$

For the Abelian theory, this reduces to

$$A'_\mu = A_\mu - \frac{1}{e}\partial_\mu \theta. \tag{4.1.18}$$

This all works out very nicely in the continuum theory. Wilson's brilliant insight [80] was to define the lattice theory not with variables from the gauge algebra, but with variables that are elements of the gauge group, denoted $U(n,m)$. These are called *link variables.* or parallel transporters because they allow the comparison of a field at one point on the lattice with a neighboring point in a gauge covariant way. If $U(n,m)$ is associated with the link connecting nearest neighbor points $n$ and $m$, then

$$U(m,n) = U^\dagger(n,m) = U^{-1}(n,m) \tag{4.1.19}$$

where the second identity follows from the fact that $U$ is a unitary matrix. So, defining $U_{n\mu} = U(n, n+\hat{\mu})$, Eqn. (4.1.19) shows that $U(n+\hat{\mu}, n) = U^\dagger_{n\mu}$.

We want the product $U_{n\mu}\phi_{n+\hat{\mu}}$ to transform under a gauge transformation the same way that the field does at the point $n$. In other words, under a gauge transformation $U \to U'$ and $\phi_n \to \phi'_n = \Omega_n \phi_n$, so we must have

$$U'_{n\mu}\phi'_{n+\hat{\mu}} = \Omega_n U_{n\mu}\phi_{n+\hat{\mu}}. \tag{4.1.20}$$

Since $\phi'_{n+\hat{\mu}} = \Omega_{n+\hat{\mu}}\phi_{n+\hat{\mu}}$, this implies

$$U'_{n\mu} = \Omega_n U_{n\mu}\Omega^{-1}_{n+\hat{\mu}}. \tag{4.1.21}$$

Hence, the products of link variables along a path transform as $\Omega_n$ if the left-most point is $n$ and $\Omega^{-1}_m$, if the right-most point is $m$. With suitable products of link variables, we can transport a field as far as we wish and have it transform as a variable that 'lives' at the left-most point in the product.

The difference $U_{n\mu}\phi_{n+\hat{\mu}} - \phi_n$ transforms in a gauge covariant way, since under a gauge transformation it picks up a factor of $\Omega_n$. The relationship between the group element $U_{n\mu}$ and the gauge field $A_\mu(x)$ that takes a value in the Lie algebra is

$$U_{n\mu} = \mathcal{P}\exp\left\{ie\int_{an}^{an+a\hat{\mu}} dy_\nu\, A_\nu(y)\right\}$$

$$= \exp\left\{iea\Big[A_\mu(an + a\hat{\mu}/2) + \frac{a^2}{24}\partial^2_\mu A_\mu(an+a\hat{\mu}/2) + \ldots\Big]\right\} = 1 + iaeA_\mu(an+a\hat{\mu}/2) + \ldots \tag{4.1.22}$$

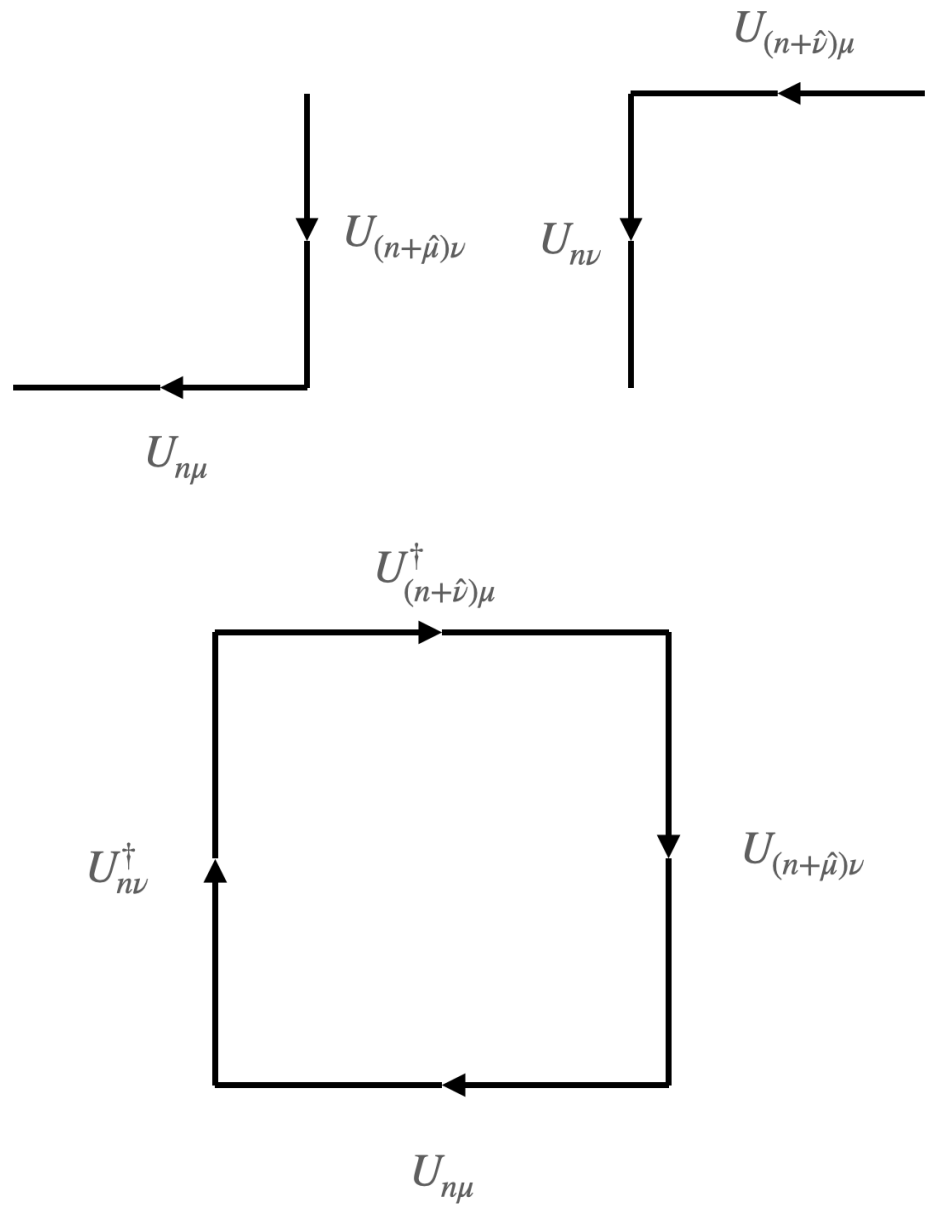

**Fig. 4.1.1** *Top:* The two paths that contribute to $[D_\mu, D_\nu]$. The $\mu$-direction is the horizontal axis, and $\nu$ is vertical. On the left, we have $D_\mu D_\nu$, on the right $D_\nu D_\mu$ *Bottom:* The links of a plaquette whose lower-left corner is site $n$, with directions same as in top figure.

where the lattice spacing $a$ is shown explicitly, and it is natural to relate the link variable $U$ to the (continuum) gauge field at the midpoint of the link. Note that in Wilson's original work, the lattice gauge field variables are $A_{n\mu}$, i.e., they are labeled by the left-hand site of the link.

Having defined the covariant derivative, the field strength tensor can be calculated. In the continuum:

$$F_{\mu\nu} = ie[D_\mu, D_\nu] \tag{4.1.23}$$

where the square brackets denote the commutator. This formula also holds in the case of non-Abelian gauge theory for which $A_\mu(x)$ is a matrix in the Lie algebra of the gauge group.

On the lattice, the covariant derivative involves parallel transport from a neighboring site. Since there are two covariant derivatives a field is transported from two sites away:

$$\begin{aligned} D_\mu D_\nu \phi_n &= U_{n\mu}(U_{(n+\hat{\mu})\nu}\phi_{n+\hat{\mu}+\hat{\nu}} \\ &\quad -\phi_{n+\hat{\mu}}) - U_{n\nu}\phi_{n+\hat{\nu}} - \phi_n \\ &= U_{n\mu}U_{(n+\hat{\mu})\nu}\phi_{n+\hat{\mu}+\hat{\nu}} \\ &\quad -U_{n\mu}\phi_{n+\hat{\mu}} - U_{n\nu}\phi_{n+\hat{\nu}} - \phi_n . \end{aligned} \tag{4.1.24}$$

The last three terms are symmetric under the interchange of $\mu$ and $\nu$, so only the first term contributes to the commutator. Thus, the field strength tensor is the difference between the product of the two two-link paths connecting sites $n$ and $n + \hat{\mu} + \hat{\nu}$. In one path we move in the $\mu$ direction first and in the other we move in the $\nu$ direction first. Also, the field strength tensor is gauge covariant as the common endpoints of the two paths determine how $F_{n\mu\nu}$ transforms:

$$F'_{n\mu\nu} = \Omega_n F_{n\mu\nu} \Omega^{-1}_{n+\hat{\mu}+\hat{\nu}}. \tag{4.1.25}$$

In the continuum the gauge action is proportional to $F^2_{\mu\nu}$, and is gauge invariant. Having just determined that $F_{\mu\nu}$ can be expressed in terms of a two-link path, we might expect that a four-link path would yield $F^2_{\mu\nu}$. It is easy to construct gauge invariant products of links. If we take the trace of the product of links along any closed path, it will be gauge invariant. The only closed four-link paths are those around the elementary squares of the lattice. The term plaquette is sometimes used to refer to the elementary squares of a hypercubic lattice. The plaquette is also used to refer to the product of the four link matrices around the square, or to the trace of this matrix. The context should make clear whether the author is referring to a shape, a matrix, or a number. Here the plaquette $U_{n\mu\nu}$ will be the trace of the product of the four links

$$U_{n\mu\nu} = \mathrm{Tr}(U_{n\mu}U_{(n+\hat{\mu})\nu}U^\dagger_{(n+\hat{\nu})\mu}U^\dagger_{n\nu}) \tag{4.1.26}$$

The Wilson plaquette gauge action is defined as

$$S_W = \frac{1}{g^2}\sum_n \sum_{\mu\neq\nu}(3 - \mathrm{Re}\, U_{n\mu\nu}) . \tag{4.1.27}$$

*Actions for fermions*

In the continuum, the free fermion action $S_F$ is given by:

$$S_F = \int d^4x \bar{\psi}(x)(i\gamma^\mu\partial_\mu - m)\psi(x), \tag{4.1.28}$$

where the gamma matrices obey $\{\gamma^\mu, \gamma^\nu\} = 2g^{\mu\nu}$. Going through the transformation to Euclidean space time, we introduce the Euclidean gamma matrices $\gamma^E_4 = \gamma^0$ and $\gamma^E_i = -i\gamma^i$. These gamma matrices obey $\{\gamma^E_\mu, \gamma^E_\nu\} = 2\delta_{\mu\nu}$. The Euclidean action is given by

$$S^E_F = \int d^4x \bar{\psi}(x)(\gamma^E_\mu \partial_\mu + m)\psi(x). \tag{4.1.29}$$

We simplify notation below by dropping the superscript E on the Euclidean gamma matrices. To include the interaction with the gauge field, the ordinary partial derivative in Eq. 4.1.28 is replaced by the covariant derivative. For $S^E_F$, a gauge covariant finite difference approximation is used

$$\partial_\mu\psi(x) \to \frac{1}{2a}(U_{n\mu}\psi_{n+\hat{\mu}} - U^\dagger_{(n-\hat{\mu})\mu}\psi_{n-\hat{\mu}}) \tag{4.1.30}$$

which is the analog of $\Delta^S$ introduced in Eq. 4.1.8. This action is called the naive fermion action, and we are about to see that it suffers from the so-called "fermion doubling problem."

To explore this, consider the case of a free fermion, so the link variables may be replaced by the unit matrix. Going to momentum space, let

$$\psi_n = \sum_p e^{(iap\cdot n)}\psi(p). \tag{4.1.31}$$

On the lattice there is maximum value for each momentum component because if $ap_\mu = 2\pi$ then the exponential will always be always be the same as for $p_\mu = 0$. Thus, the momentum components can be restricted to be less than $(2\pi)/a$ or more symmetrically,

$$-\frac{\pi}{a} < p_\mu \le \frac{\pi}{a}. \tag{4.1.32}$$

Because of the periodic boundary conditions on a lattice of finite extent, say $L$ in each direction, there is another restriction that $ap_\mu L = 2\pi j$ for some integer $j$. Thus the allowable momentum components are restricted to $(2\pi j)/(aL)$, so for finite $L$ the lattice provides an infrared as well as an ultraviolet cutoff. However, as $L$ goes to infinity, the momentum becomes a continuous variable, and in this case Eq. 4.1.31 becomes

$$\psi_n = \int_{-\pi/a}^{\pi/a} d^4p\, e^{(iap\cdot n)}\psi(p). \tag{4.1.33}$$

The fact that $\bar\psi$ and $\psi$ are displaced from each other on the lattice results in factors of $\exp\{\pm ip_\mu a\}$. The final result for the Euclidean action, written in momentum space, is

$$S_F^E = \int d^4p \left[\frac{i}{a}\sum_\mu \bar\psi(p)\gamma_\mu \sin(p_\mu a)\psi(p) + m\bar\psi(p)\psi(p)\right]$$
$$= \int d^4p\, \bar\psi(p)\, S^{-1}(p)\, \psi(p)\,, \tag{4.1.34}$$

*The fermion doubling problem*

At this point, most authors go on to solve for the free quark propagator and examine the pole structure. Let's just look at the current expression and compare with the continuum. When $p_\mu a$ is small, we may approximate $\sin(p_\mu a) \to p_\mu a$ so the factor of $a^{-1}$ before the sum is cancelled and this looks a lot like $i\not{p} + m$. As $p_\mu$ continues to grow toward $\pi/(2a)$, the sin function flattens out and then starts to return to zero at $p_\mu = \pi/a$. That means at the end of the Brillouin zone, there is again a region were there is linear dependence on the momentum. More concretely, let $p_\mu = \pi/a - k$ and note that $\sin(p_\mu a) = \sin(ka)$. We also need the region where $p_\mu = -\pi/a + k$ to have a region in momentum space just like the one at the origin. Since any component of $p$ can be near zero, or at the edge of the Brillouin zone there are $2^4$ regions in momentum space where the action takes the form of a free action. We wanted one fermion and we would up with 16! This is the crux of the doubling problem.

In his Erice lectures, Wilson provided a fix [284]. He added to the action a higher dimensional term, the lattice Laplacian, multiplied by the lattice spacing. This term vanishes as $a \to 0$. The covariant version of the second derivative $\nabla^2_\mu$ is defined

$$\nabla^2_\mu \psi_n = \frac{1}{a^2}\Big(U_{n\mu}\psi_{n+\hat\mu} + U^\dagger_{(n-\hat\mu)\mu}\psi_{n-\hat\mu}$$
$$-2\psi_n\Big)\,. \tag{4.1.35}$$

The Wilson fermion action is therefore

$$S_W^F = S_{naive} - \frac{ar}{2}\sum_x \bar\psi(x)\sum_\mu \nabla^2_\mu\psi(x)$$
$$= \bar\psi M_W(m)\psi\,, \tag{4.1.36}$$

where $r$ is a free parameter, usually set to $r = 1$, and $S_{naive}$ is given by Eq.4.1.29 after substituting Eq. 4.1.30. Fourier transforming, the free inverse propagator now is

$$aS^{-1}(p) = i\sum_\mu \gamma_\mu \sin(ap_\mu) + am - r\sum_\mu\left(\cos(ap_\mu) - 1\right)\,. \tag{4.1.37}$$

The last term, proportional to $r$, vanishes near $p = 0$, but near the edge of the Brillouin zone $\cos(ap_\mu) = -1$ and the doublers, with $n$ momentum components $p_\mu = \pm\pi/a$, now attain masses $m + 2nr/a$, and only one fermion, with $p \approx 0$, remains light. The Wilson term cures the doubling problem, but the action with $m = 0$ no longer has a chiral symmetry so there is an additive mass renormalization, and we must fine tune the parameters to determine where the fermion mass vanishes. The Wilson fermion action has errors $\mathcal{O}(a)$.

An important property of the Wilson Dirac operator is its $\gamma_5$ Hermiticity. That is

$$M_W^\dagger(m) = \gamma_5 M_W(m)\gamma_5. \tag{4.1.38}$$

We will see in the next section that $\det M_W(m)$, the fermion determinant, arises from integrating over the fermion fields. A consequence of $\gamma_5$ Hermiticity is that $\det M_W^\dagger(m) = \det M_W(m)$. If a theory has two equal mass fermions, the fermion determinant will be positive (semi-) definite as

$$\det(M_W(m)M_W(m)) = \det\Big(M_W^\dagger(m)M_W(m)\Big). \tag{4.1.39}$$

In addition to the dimension-5 operator Wilson introduced, there is a second operator introduced by Sheikholeslami and Wohlert [285] that can be adjusted to reduce the error to $\mathcal{O}(a^2)$. The operator is the lattice analog of $\bar{\psi}(x)\sigma_{\mu\nu}F_{\mu\nu}(x)\psi(x)$ where $\sigma_{\mu\nu} = \frac{i}{2}[\gamma_\mu, \gamma_\nu]$ is the commutator of the $\gamma$ matrices and $F_{\mu\nu}(x)$ is the field strength tensor defined in Eq. 4.1.23. Previously, we were considering electromagnetism, but the same formula applies to non-Abelian theories if we replace $e$ by $g$, the coupling constant for the non-Abelian group. A lattice expression for the field strength tensor can be constructed from four suitably oriented (uncontracted) plaquettes surrounding site $n$. This has come to be known as the *clover action* because the four plaquettes look like a four-leaf clover and clover is easy to spell. Thus, the Sheikholeslami-Wohlert or clover term in the action is

$$S_{SW} = \frac{iag}{4} c_{SW} \sum_{n,\mu,\nu} \bar{\psi}_n \sigma_{\mu\nu} \mathcal{F}_{n\mu\nu} \psi_n \ , \tag{4.1.40}$$

where $\mathcal{F}_{n\mu\nu}$ is the clover-like term discussed above. The coefficient $c_{SW}$ can be tuned either perturbatively [286, 287], or better yet, non-perturbatively [288, 289]. The addition of the clover term is an example of an improvement program introduced by Symanzik [290, 291].

### 4.1.2 Twisted mass quarks

One issue with the Wilson formulation is that for small mass, it is possible to encounter so-called ‘exceptional configurations’ for which it is very difficult, if not impossible, to construct the quark propagator [292]. This was particularly an issue in the quenched approximation in which the fermion determinant is neglected. It can also slow down generation of configurations with dynamical quarks. For a theory with two light flavors, such as $u$ and $d$, the twisted mass operator was invented to ensure that the fermion determinant is positive definite [293]. If the lattice Dirac operator is $D+m$, then

$$D_{\text{twist}} = D + m + i\mu\gamma_5\tau_3, \tag{4.1.41}$$

where $\tau_3$ operates on the two flavors of quarks. Then $\det D_{\text{twist}} = \det\big((D+m)^\dagger(D+m)+\mu^2\big)$. So, as long as $\mu$ is non-zero, $\det D_{\text{twist}}$ is positive and exceptional configurations are avoided. This action has been used by the European Twisted Mass Collaboration for over 15 years. The collaboration is now the Extended Twisted Mass Collaboration as there are non-European members.

### 4.1.3 Staggered quarks

Staggered quarks are an alternative to Wilson quarks that reduce the degree of doubling and retain some of the chiral properties of the continuum theory [81, 294–296]. One must be careful in reading the literature since some authors use $x_0$ for the time coordinate and others use $x_4$. This can have consequences for the field redefinition essential to the reduction in the number of fermions. Here we adopt the conventions in Refs. [297] and [298] rather than those in Ref. [299]. The key simplification is to rearrange the Dirac components at each site of the lattice in such a way that the action can be seen as comprised of four non-interacting fields. In this way, we may retain a single field component at each site and the doubling is reduced from 16 to 4. Initially, it was thought that this could be interpreted at four flavors or quarks, say, $u$, $d$, $s$, and $c$, but the modern interpretation is that each flavor has four ‘tastes.’ Tastes are not physical, so we must take a fourth root of the fermion determinant for each quark, and must be careful in constructing hadron operators to avoid mixing tastes as physical operators should really be constructed from a single taste. In the continuum limit, taste breaking vanishes so operators with mixed tastes should become degenerate with single taste operators.

Define a local redefinition of the Dirac components of the quark field by $\psi_n = \Omega_n \psi'_n$ and $\bar{\psi}_n = \bar{\psi}'_n \Omega^\dagger_n$. The $4\times 4$ matrix $\Omega_n$ is defined as

$$\Omega_n = \gamma_0^{n_0}\gamma_1^{n_1}\gamma_2^{n_2}\gamma_3^{n_3}. \tag{4.1.42}$$

This may appear more complicated than it really is. Note that as $\gamma_\mu^2 = 1$, each gamma matrix appears in $\Omega_n$ only when the corresponding coordinate is odd. There are only 16 distinct values for $\Omega_n$, and if we translate two sites in any direction, we have the same matrix. We will see that staggered quarks are naturally defined on $2^4$ sub-hypercubes of the lattice. The gamma matrices are unitary and Hermitian, so

$$\Omega^\dagger_n \gamma_\mu \Omega_{n+\hat{\mu}} = (-1)^{n_0+\cdots+n_{\mu-1}} \equiv \alpha_\mu(n). \tag{4.1.43}$$

The hopping term in the naive fermion action

$$\bar{\psi}_n \gamma_\mu U_{n\mu} \psi_{n+\hat{\mu}} \tag{4.1.44}$$

is transformed into

$$\bar{\psi}'_n \Omega^\dagger_n \gamma_\mu U_{n\mu} \Omega_{n+\hat{\mu}} \psi_{n+\hat{\mu}} = \bar{\psi}'_n \alpha_\mu(n) U_{n\mu} \psi_{n+\hat{\mu}}. \tag{4.1.45}$$

The same factor appears in the hopping term that involves $\psi_{n-\hat{\mu}}$ since $\Omega_{n+\hat{\mu}} = \Omega_{n-\hat{\mu}}$ as the two sites differ by two units in the $\mu$-direction. The gamma matrices have disappeared, and we are left with a unit matrix in Dirac index space, so there are four equivalent non-interacting components $\psi'_n$. We may discard three of the four components and write the staggered action in terms of a single component field $\chi$.

$$S_{\rm stag} = \frac{1}{2a}\sum_{n,\mu} \bar{\chi}_n \alpha_\mu(n)[U_{n\mu}\chi_{n+\hat{\mu}} - U^\dagger_{(n-\hat{\mu})\mu}\chi_{n-\hat{\mu}}] + m\sum_n \bar{\chi}_n\chi_n. \quad (4.1.46)$$

As mentioned above, because $\alpha_\mu(n)$ is periodic in each direction with period two, it is possible, perhaps natural, to interpret the 16 components on the sites of each $2^4$ as the components of four Dirac spinors, i.e., the four tastes.

For the free theory, the four tastes can be expressed in the following way. Let $y$ be a 4-component integer valued vector labeling the hypercubes. Let $\eta$ be a four component vector whose components may only take the value 0 or 1. That is, $\eta$ labels the 16 sites of a hypercube. For each hypercube $y$, the sites of the original lattice take the values $2y+\eta$ for one of the 16 values of $\eta$. Let $\alpha$ be a Dirac component index and $a$ be a taste label. Both $\alpha$ and $a$ range between 1 and 4. We have

$$\psi^{\alpha a}_y = \frac{1}{8}\sum_\eta \Omega^{\alpha a}_\eta \chi_{2y+\eta}. \quad (4.1.47)$$

This is not gauge covariant since we are adding together $\chi$ values from different lattice sites, so in the interacting case, $\chi$ at each site must be multiplied by suitable parallel transporter to move it to the origin of the hypercube. In practice, one really does not have to worry about this.

For Wilson quarks the action was improved by adding the clover term. For staggered quarks there have been similar advances by improving the action. For the most simple staggered action, the errors are $\mathcal{O}(a^2)$. Naik [300] introduced a 3-link hopping term. The gauge action was also improved by adding $2\times 1$ rectangles, and 6-link terms that circle a 3-dimensional cube, sometimes called the bent chair diagram, known as the Lüscher-Weisz gauge action [301, 302]. These terms are depicted in the top of Fig. 4.1.2. Essential benefits come from averaging or smearing the gauge fields in the 1-link hopping terms. These smearings are designed to reduce taste symmetry breaking. There have been two major rounds of these improvements, the first is known as the asqtad action [303–307] and the second is known as the highly improved staggered quark or HISQ action [308]. The paths for the fermion link smearings are shown in the bottom of Fig. 4.1.2. The HISQ action employs two levels of smearing. Reference [299] details the asqtad and HISQ actions and provides many physics results using the former action. The MILC Collaboration generates HISQ ensembles that are also used by the Fermilab Lattice and HPQCD Collaborations, and others. These improvements make the coding more complicated and require more floating point operations on a fixed grid size, but the payoffs can be enormous as the errors for the same lattice spacing are significantly reduced with the improved actions. If, say, an improvement would allow one to work at twice the lattice spacing as without the improvement there would be a significant reduction in computer time as halving the lattice spacing increases the work by a factor of 32 or more.

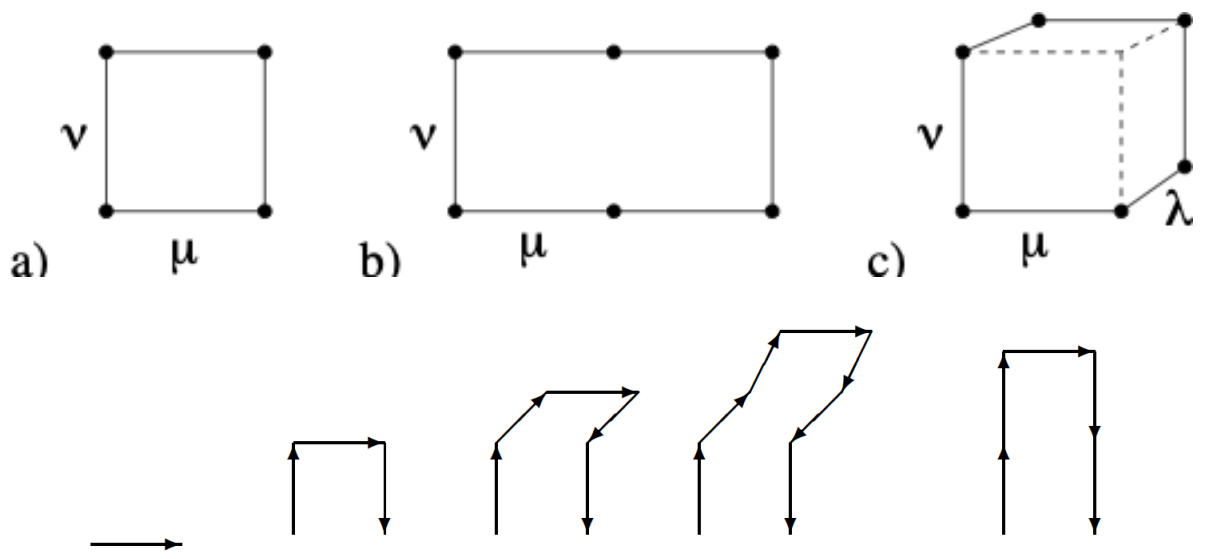

**Fig. 4.1.2** Top: Loops that are included in the gauge action for asqtad and HISQ quarks. Bottom: For the asqtad action, the one-link hopping term in the naive staggered quark action is replaced by a combination of 1-, 3-, 5-, and 7-link smearings. The right-most 5-link term is known as the Lepage term. This figure is adopted from Ref. [307]. There is also a straight 3-link term known as the Naik term. For the HISQ action, there are two levels of smearing, but no additional paths are involved.

### 4.1.4 Improving chiral symmetry

When the quark mass vanishes, the theory contains an important continuous symmetry known as chiral symmetry. The dynamical breaking of this symmetry is responsible for the pions being so light. The Wilson action explicitly breaks this symmetry, and the staggered actions discussed above only maintain some of the symmetry. However, there are other lattice actions that have much better chiral symmetry. These include the domain wall and overlap actions.

In the continuum, chiral symmetry follows from the fact that $\gamma_5$ anticommutes with the kinetic operator $D = \not{D}$. In 1982, Ginsparg and Wilson [312] considered the consequences of a generalized lattice chiral symmetry which is currently expressed as

$$D\gamma_5 + \gamma_5 D = aD\gamma_5 D. \quad (4.1.48)$$

Note the factor of the lattice spacing $a$ on the RHS. As $a \to 0$, we restore chiral symmetry; however, even at non-zero $a$ there is a more complicated chiral symmetry for operators that obey Eqn. 4.1.48.

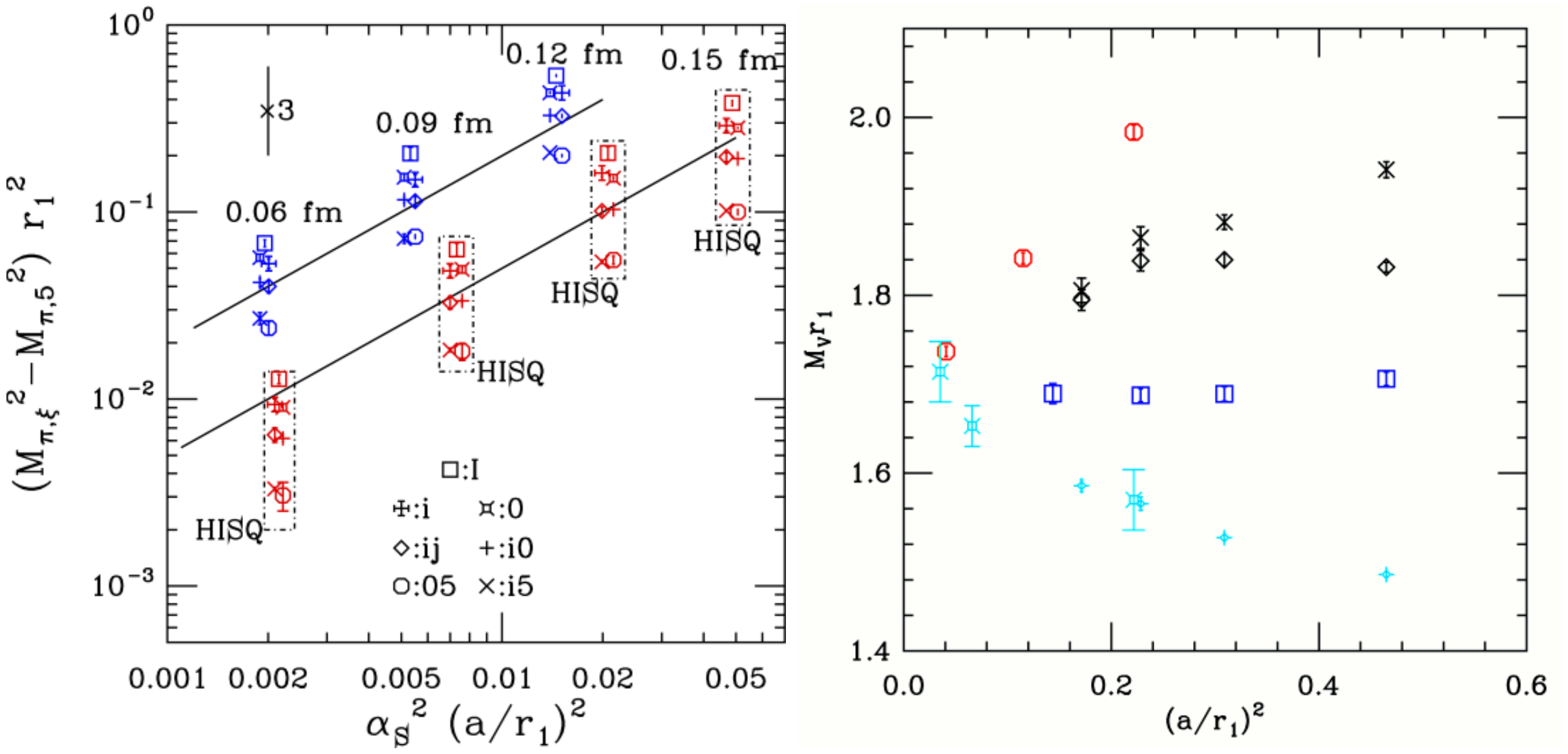

**Fig. 4.1.3** *Left*: We show how taste breaking of pseudoscalar mesons decreases as the lattice spacing is reduced for two types of improved staggered quarks asqtad (blue) and HISQ (red). Different symbols denote different taste mesons. The quantity plotted is the difference of squared mesons masses for the plotted meson mass ($\xi$) and the Goldstone taste combination ($\gamma_5$). The horizontal axis is the $\alpha_s^2 a^2$ in units determined by the heavy quark potential $r_1$. Taste symmetry is restored in the continuum limit and taste breaking is much smaller for HISQ than for asqtad. See Ref. [309] for details. *Right*; The $\rho$ meson mass as a function of lattice spacing for multiple actions shows a common continuum limit, but some actions have much more gentle lattice spacing dependence than others. Red octagons are unimproved staggered fermions with Wilson gauge action, diamonds are unimproved staggered fermions with Symanzik improved gauge action, crosses are Naik fermions and blue squares are asqtad fermions, both with Symanzik improved gauge action. For comparison we also show in light blue tadpole clover improved Wilson fermions with Wilson gauge action [310] (fancy squares) and with Symanzik improved gauge action [311] (fancy diamonds). (See Ref. [299] for details.)

$$\psi \to \psi' = \exp\Big(i\alpha\gamma_5(1 - \frac{a}{2}D)\Big)\psi \tag{4.1.49}$$

with a similar expression for $\bar{\psi}$. As $a \to 0$, Eqn. 4.1.49 approaches the usual expression for a chiral rotation, but at non-zero lattice spacing the transformation is more complicated as $D$ is a finite difference operator. Reference [312] was not heavily cited until 1998 when the overlap operator that was developed by Narayanan and Neuberger [313–316] was shown to obey Eq. 4.1.48 [317]. If $A$ obeys $\gamma_5$ Hermiticity, let $H = \gamma_5 A$, then

$$D_{ov} = \frac{1}{a}(1 + \gamma_5\,\mathrm{sign}[H]) \tag{4.1.50}$$

defines the overlap operator. An alternative expression is

$$D_{ov} = \frac{1}{a}(1 + \gamma_5 H(HH)^{-1/2}). \tag{4.1.51}$$

A suitable choice for $A$ is $D_W(0) - r$, with $0 < r < 2$. Numerically, it is difficult to compute the sign function or the inverse square root of a matrix. The $\chi QCD$ Collaboration uses overlap fermions.

In 1992, Kaplan introduced domain wall fermions in which chiral modes are bound to a defect in a 5-dimenensional (5D) theory [318]. The theory was further developed by Shamir [319, 320], and Furman and Shamir [321]. We adapt here the notation of Ref. [322]. Points in the five dimensional lattice are labeled by $m$ in the four dimensional space and $r$ in the 5th dimension, with $r = 0, \ldots, N_5 - 1$. The 5D fermion field is $\Psi(m, s)$. The 5D Dirac operator consists of two parts:

$$D^{\rm dw}(n, s; m, r) = \delta_{s,r} D(n; m) + \delta_{n,m} D_5^{\rm dw}(s, r). \tag{4.1.52}$$

The first term can be an ordinary Wilson operator with a modified mass:

$$D(n; m) = (4 - M_5)\delta_{n,m} - \frac{1}{2}\sum_{\mu=\pm1}^{\pm4}(1 - \gamma_\mu)U_{n;m}\delta_{n+\hat{\mu},m} \tag{4.1.53}$$

where we use notation $U_{n;m}$ to avoid having to specify hermitian conjugation for negative directions. Using $P_\pm = (1 \pm \gamma_5)/2$,

$$\begin{aligned} D_5^{\rm dw}(s; r) = \delta_{s,r} &- (1 - \delta_{s,N_5-1})P_-\delta_{s+1,r} \\ &-(1 - \delta_{s,0})P_+\delta_{s-1,r} + m(P_-\delta_{s,N_5-1}\delta_{0,r} \\ &+P_+\delta_{s,0}\delta_{N_5-1,r}). \end{aligned} \tag{4.1.54}$$

The physical 4D fields come from the boundaries of the 5D field:

$$\psi(n) = P_{-}\Psi(n,0) + P_{+}\Psi(n,N_5-1). \quad (4.1.55)$$

Domain wall fermions are used extensively for dynamical quarks, especially by the RBC/UKQCD and JLQCD Collaborations.

### 4.1.5 Continuum limit

To control systematic errors it is crucial to tune the quark masses to their physical value, to have a volume that is large enough to avoid finite volume errors, and to take the limit $a \to 0$. In the early days, it was too expensive to use physically light $u$ and $d$ quarks, so one also had to use chiral perturbation theory to extrapolate to those quark masses. Because QCD has the property of asymptotic freedom, the coupling constant goes to zero as the cutoff goes to infinity. On the lattice, the inverse lattice spacing plays the role of the cutoff. By dimensional transmutation, instead of expressing physical results in terms of the coupling, we do it in terms of the lattice spacing. In the left panel of Fig. 4.1.3, we show how taste breaking decreases as $a \to 0$ in accord with expected behavior for both asqtad and HISQ quarks. This also clearly shows that taste breaking is much smaller for HISQ (as it was designed with that in mind). In the right panel, we show how the $\rho$ meson mass depends on the lattice spacing. Some of these results are rather old, and some are in the quenched approximation; however, the point to be made is that different ways of putting quarks on the lattice have the same continuum limit, although the rate at which they approach that limit will vary.

### 4.1.6 Further Reading

I have made no attempt at a historically accurate account of lattice QCD, and due to space limitations much has been left out. Here I list some books on the topic. As far as I know, “Quarks, gluons and lattices” by Creutz is the first monograph[323]. Creutz also edited “Quantum Fields on the Computer,” which covers scalar and Yukawa theories in addition to QCD [324]. Proceedings from the 1989 TASI summer school edited by DeGrand and Toussaint [325] was an early essential reference. Books by Rothe [326] and by Montvay and Munster [327] appeared in the early 1990s. The former is now in its fourth edition and is available online via open access. Since 2000, at least three books have been published. Authors are Smit [298]; DeGrand and DeTar [297]; and Gattringer and Lang [322].

### 4.1.7 Personal remarks

In 1975, I had the opportunity to take my first European physics trip when I attended the Erice summer school in Sicily. Little did I know as I listened to Ken Wilson lecture on quark confinement and lattice gauge theory how profoundly his work would impact my own. (As an undergraduate, I only remember talking to Wilson once when he kindly gave me advice on which graduate schools I should apply to.) I recall being quite friendly with Michael Creutz during the school. Claudio Rebbi was one of the lecturers. I have had many great interactions with both of them. I was in awe of seeing Paul Dirac walking quietly around the school. Tom De Grand would later become a collaborator. Sidney Coleman, of course, gave a great series of lectures.

During my postdoc at Argonne, Creutz was kind enough to send me a printed copy of his code. I had a great title for a paper: “Looking for Glue in SU(2).” Unfortunately, I didn’t really know anything about glueballs, so I did not pursue that. While at Fermilab, I was looking for something new and worked on the Migdal-Kadanoff recursion relations with Khalil Bitar and Cosmas Zachos (who I had known when he was an undergrad and I was in graduate school). Don Weingarten visited and I started my career of Monte Carlo lattice calculations (using SU(3) not SU(2)). Hank Thaker, Paul Mackenzie, Weingarten and I used some of the VAX computers at Fermilab for our calculations to examine $\rho$ decay. We worked on a $6^2 \times 12 \times 18$ lattice and had so few configurations we joked that we knew each one by name. For no good reason, I still have some of the magnetic tapes on which we stored the configurations. This project continued when I moved to UC San Diego. A year later, my grad school housemate Doug Toussaint arrived as an assistant professor. I started working with him, and more senior people such as Bob Sugar and Julius Kuti. A few years later the MILC Collaboration started, and I would like to mention fellow founding members Claude Bernard and Carleton DeTar. Lattice gauge theory has been my life ever since then.

## 4.2 Monte-Carlo methods

**Carleton DeTar**

### 4.2.1 Introduction

In 1980, Michael Creutz pioneered the numerical simulation of lattice QCD [328, 329] with studies of Wilson’s lattice formulation of SU(2) Yang-Mills theory. This feasibility study started a vast enterprise devoted to "solving" QCD in the nonperturbative regime. Later on, as computing power grew, it became possible to include quarks, thus bringing simulations in contact with reality. This subsection introduces basic methods for carrying out the numerical simulation of lattice QCD using Monte Carlo methods. It concludes with a mention of ongoing improvements.

### 4.2.2 Lattice path integration

*Partition function*

The most widely used strategy for numerical simulation of QCD starts from a Feynman path integral formulation [330], which is based on the partition function

$$Z = \int [dU d\overline{\psi} d\psi] \exp\left[-S(U, \overline{\psi}, \psi)\right] \tag{4.2.1}$$

where

$$S(U, \overline{\psi}, \psi) = S_W + \overline{\psi} M \psi \tag{4.2.2}$$

is the Euclidean action for the lattice SU(3) gauge field $U$ and quark field $\psi$, as defined in Sec. 4.1 For simplicity here, we treat only one quark flavor, and we suppress the color ($c$), vector ($\mu$), and spatial ($n$) indices on $U_{cc',\mu}(n)$ and the color, spin ($\alpha$), and spatial indices on $\psi_{c,\alpha}(n)$. Note that for lattice volume $V$ (number of sites) there are $4V$ SU(3) matrices denoted by $U$ and $V$ spin/color vector fields denoted by $\psi$.

The integration over the gauge links $U$ is done over the classical SU(3) gauge field $U_\mu(n)$ on each lattice link. We use the invariant Haar measure $dU_\mu(n)$ on each link. (We won't need it, but there is an Euler-angle representation of the measure [331].) The integration is done without gauge fixing. Since the action $S$ is gauge invariant and the gauge group is compact, the integral over gauge choices is finite. In the Feynman formulation, fermion fields, in particular $\psi$, must be anticommuting Grassmann variables. This assures that they obey Fermi-Dirac statistics. It would be challenging to treat them directly in a computer simulation, but, fortunately, they can be integrated out using only the identities listed below and their analogs, leaving expressions involving only the classical gauge field. For a few more details, see Ref. [297].

In a Euclidean spacetime with finite time extent $T$, the quantity $Z$ in Eq. (4.2.1) is the thermal partition function for the theory defined by the action $S$ with hamiltonian $\mathcal{H}$ at inverse temperature $\beta = T$. Thus

$$Z(\beta) = \mathrm{Tr} \exp(-\beta \mathcal{H}) \,. \tag{4.2.3}$$

The zero temperature limit corresponds to an infinite time extent.

*Grassman calculus*

We need three important identities from the Grassmann calculus:

$$\int [d\overline{\psi} d\psi] \exp\left(-\overline{\psi} M \psi\right) = \det M \tag{4.2.4}$$

$$\begin{aligned} &\int [d\overline{\psi} d\psi] \overline{\psi}_{c,\alpha}(n) \psi_{c',\alpha'}(n') \exp\left(-\overline{\psi} M \psi\right) \\ &= M^{-1}_{c,\alpha;c',\alpha'}(n, n') \det M \end{aligned} \tag{4.2.5}$$

$$\begin{aligned} &\int [d\overline{\psi} d\psi] \psi_{c,\alpha}(n) \overline{\psi}_{c',\alpha'}(n') \overline{\psi}_{d,\beta}(m) \psi_{d',\beta'}(m') \\ &\exp\left(-\overline{\psi} M \psi\right) = [M^{-1}_{c,\alpha;c',\alpha'}(n, n') M^{-1}_{d,\beta;d',\beta'}(m, m') \\ &-M^{-1}_{c,\alpha;d',\beta'}(n, m') M^{-1}_{d,\beta;c',\alpha'}(m, n')] \det M \end{aligned} \tag{4.2.6}$$

The inverse of the fermion matrix $M$ is the fermion propagator. We see that each $\overline{\psi}$, $\psi$ pair in the integrand contributes a fermion propagator. All pairings can occur, as in the last example. The minus sign there arises from the anticommuting property of the fields.

*Observables*

Physical quantities are defined in terms of observables $\mathcal{O}(U, \overline{\psi}, \psi)$ constructed from the variables $U$, $\overline{\psi}$, and $\psi$. To obtain the expectation value of the observable, we calculate

$$\langle \mathcal{O} \rangle = Z^{-1} \int [dU d\overline{\psi} d\psi] \mathcal{O}(U, \overline{\psi}, \psi) \exp\left[-S(U, \overline{\psi}, \psi\right] \tag{4.2.7}$$

*Meson propagator*

For example, we might want to determine the mass of a pseudoscalar meson. To do so we work with an "operator" that "creates" or "destroys" the meson:

$$\mathcal{O}_{PS}(\mathbf{p}, t) = \sum_{\mathbf{r}} \exp(i\mathbf{p} \cdot \mathbf{r}) \overline{\psi}(\mathbf{r}, t) \gamma_5 \psi(\mathbf{r}, t) \,, \tag{4.2.8}$$

where $\mathbf{p}$ is the momentum. Note that if we replace the Grassmann field with a quantum field, the same operator in quantum field theory would create or destroy the meson. The sum over spatial sites $\mathbf{r} = (x, y, z)$ for fixed $t$ and $\mathbf{p}$ gives a meson of momentum $\mathbf{p}$ at Euclidean time $t$. To obtain the mass, we calculate at zero momentum and large $|t' - t|$

$$\begin{aligned} C_{PS}(t', t) &= \langle \mathcal{O}_{PS}(\mathbf{0}, t') \mathcal{O}_{PS}(\mathbf{0}, t) \rangle \\ &= z_{PS}^2 \exp[-M_{PS}|t' - t|] \end{aligned} \tag{4.2.9}$$

where $z_{PS}$ is the amplitude and $M_{PS}$ is the meson mass. In effect, we are creating the meson at time $t$ and destroying it at time $t'$. The meson propagates between these times. In Minkowski space the meson propagator would be proportional to the phase factor $\exp[-iM_{PS}|t' - t|]$. In Euclidean space here, it falls exponentially in the time separation at a rate controlled by the mass $M_{PS}$. This expression is strictly valid only for large time separations $|t'-t|$. At smaller separations, we would get additional, higher-mass contributions.

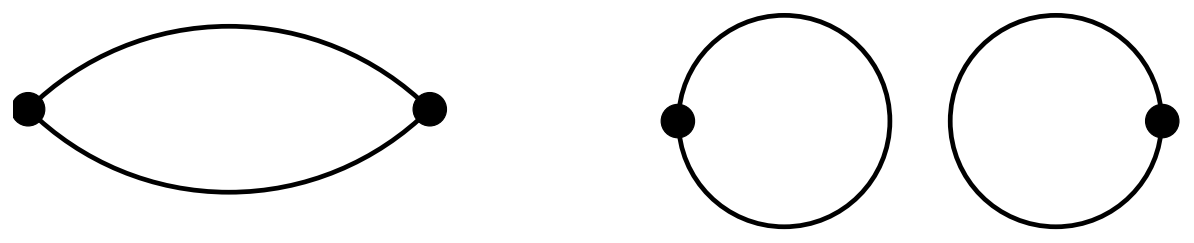

**Fig. 4.2.1** Quark line connected and disconnected diagrams.

The meson interpolating operators are sometimes called "source" and "sink". Which is which depends on the point of view, since they can serve a dual purpose.

*Integrating out the fermion fields*

Let's examine the expectation value in Eq. (4.2.6) in more detail. Note that we can integrate out the fermion fields exactly by making use of the identities in Eqs (4.2.4) and (4.2.6). When we insert the product of two interpolating operators from Eq. (4.2.8) into Eq. (4.2.6) we get a product of two Grassmann fields $\overline{\psi}$ and two Grassmann fields $\psi$. According to Eq. (4.2.6), we get

$$C_{PS}(t',t) = \Big\langle \sum_{\mathbf{r};\mathbf{r}'} \{\mathrm{Tr}_{cs}\, \gamma_5 M^{-1}(\mathbf{r},t;,\mathbf{r}',t')\gamma_5 M^{-1}(\mathbf{r}',t';\mathbf{r},t)) - \mathrm{Tr}_{cs}\, \gamma_5 M^{-1}(\mathbf{r},t;\mathbf{r},t)\, \mathrm{Tr}_{cs}\, \gamma_5 M^{-1}(\mathbf{r}',t');\mathbf{r}',t')\} \Big\rangle_G \tag{4.2.10}$$

where $\mathrm{Tr}_{cs}$ denotes a trace over color and spin indices and we have defined, for any function $E$ of the gauge field,

$$\langle E\rangle_G = Z^{-1} \int [dU] E \exp[-S_W] \det M\,. \tag{4.2.11}$$

With the fermion fields integrated out, the Feynman path integrals now involve only integration over the classical gauge field, which is amenable to numeric integration. The meson propagator in Eq. (4.2.10) has two terms that are represented diagrammatically in the two panels of Fig. 4.2.1. We call the two contributions quark-line "connected" and quark-line "disconnected". The loop in the disconnected diagram represents the annihilation of the quark and antiquark in the interpolating operator. It contributes only if the meson is a flavor singlet. With the addition of flavor we would find that the pion does not have this term.

*Form of the meson propagator*

On a finite lattice with Euclidean time extent $T$ and the usualperiodic/antiperiodic boundary conditions on the fields, the mesoncorrelation function $C_{PS}(t',t)$ gets another contribution as the meson propagates in the opposite direction from the source, exploiting the periodic/antiperiodic boundary condition, and arriving again at the sink. The distance traveled in Euclidean time is now $T - |t'-t|$. Thus we have

$$C_{PS}(t',t) = z_{PS}^2 \exp[-M_{PS}|t'-t|] + z_{PS}^2 \exp[-M_{PS}(T-|t'-t|)]\,. \tag{4.2.12}$$

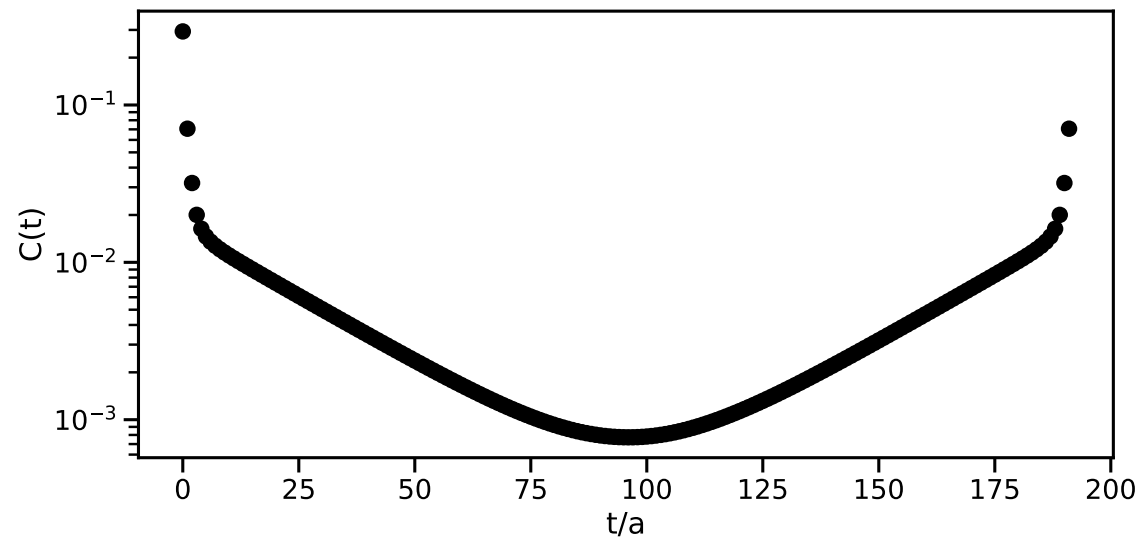

**Fig. 4.2.2** Zero momentum pion propagator at lattice spacing $a = 0.06$ fm as a function of Euclidean time expressed in units of the lattice spacing (courtesy William Jay). The source is at $t = 0$. In this case $T = 192a$.

Figure 4.2.2 illustrates the result of a calculation of the pion propagator showing both forward and backward propagation.

*Decay Constant*

The amplitude $z_{PS}$ is proportional to the meson decay constant $f_{PS}$:

$$z_{PS} = Z_{PS} f_{PS} \tag{4.2.13}$$

where $Z_{PS}$ is a renormalization constant ("matching factor") that relates the lattice interpolating operator to a physical continuum interpolating operator.

*Form factor*

Form factors give information about hadron structure and decay. Here we illustrate the construction of the electromagnetic form factor for the meson illustrated above. We calculate the three-point function

$$C_\mu(t,\mathbf{q},t',t'') = \langle \mathcal{O}_{PS}(t,\mathbf{q}) J_\mu(\mathbf{q},t') \mathcal{O}_{PS}(t'',\mathbf{0})\rangle\,, \tag{4.2.14}$$

where $J_\mu(t',\mathbf{q})$ is the current density projected onto spatial three momentum $\mathbf{q}$ and $\mathbf{0}$ denotes zero momentum. For simplicity we have chosen zero momentum for the meson interpolating operator at time $t''$, and we have enforced three-momentum conservation. (In Euclidean space-time, we don't have energy conservation, but the meson propagators are on shell.)

*Form of the three-point function*
For $t \ll t' \ll t''$ the three-point function has the form

$$C_\mu(t, \mathbf{q}, t', t'') = z_{PS}(\mathbf{q}) Z_V z_{PS}(\mathbf{0}) \exp(-M_{PS}|t' - t|) \\ F_\mu(\mathbf{q}) \exp(-M_{PS}|t'' - t'|) \tag{4.2.15}$$

where $F_\mu(\mathbf{q})$ is the desired form factor. The current renormalization constant $Z_V$ matches the lattice current $J_\mu$ to the continuum current.

*Integrating out the fermion fields*
There are a variety of choices for the current density. We could work with the conserved lattice Noether current. Or we could work with a "local" current

$$J_\mu(\mathbf{r}, t) = Q\overline{\psi}(\mathbf{r}, t)\gamma_\mu \psi(\mathbf{r}, t) \,. \tag{4.2.16}$$

where $Q$ is the charge. This current is not conserved at nonzero lattice spacing, but with suitable renormalization, it should give the same result as the conserved current in the continuum limit. We use the local current here for simplicity.

We integrate out the fermion fields following the same steps as for the meson propagator. We display, here, only the quark-line-connected contribution:

$$C_\mu(t, \mathbf{q}, t', t'') = \sum_{\mathbf{r}, \mathbf{r}', \mathbf{r}''} \exp(-i\mathbf{r} \cdot \mathbf{q}) \exp(i\mathbf{r}' \cdot \mathbf{q}) \\ \langle \mathrm{Tr}_{cs}\, \gamma_5 M^{-1}(\mathbf{r}, t; , \mathbf{r}', t')\gamma_\mu M^{-1}(\mathbf{r}', t'; \mathbf{r}'', t'')) \\ \gamma_5 M^{-1}(\mathbf{r}'', t'', \mathbf{r}, t)\rangle_G \tag{4.2.17}$$

The quark-line structure is the closed loop diagrammed in the left panel of Fig. 4.2.1.

### 4.2.3 Monte Carlo Methods

*Importance Sampling*
The path integral in Eq. (4.2.11) involves integration over so many variables that Monte-Carlo importance sampling becomes the only method of choice. A single point in the domain of integration is specified by the gauge field values $U$ on each link – called a gauge field configuration. The integrand is sampled over random gauge-field configurations with probability density $P$ of encountering a given configuration $U$. If the sampling is designed so that $P$ is proportional to the integrand weight factor

$$P \propto \exp[-S_W] \det M] \,, \tag{4.2.18}$$

then in an ensemble of such gauge configurations $U_i$ for $i = 1, \ldots, N$, the expectation value of an observable $E$ is simply the ensemble average in the limit $N \to \infty$.

$$\langle E \rangle = \lim_{N\to\infty} \frac{1}{N} \sum_{i=1}^{N} E(U_i) \,. \tag{4.2.19}$$

Of course, the weight factor must be positive definite in order to be treated as a probability density. This is usually the case, but there are important exceptions. One can use the same path-integral formalism to treat a grand-canonical ensemble of fermions at nonzero fermion number (or flavor) density; see Sec. 4.4. In this case the fermion determinant acquires a complex phase (the so-called "sign problem") that obviates a probabilistic treatment.

*Markov chain*
There are various methods for generating such an ensemble. They all involve creating a Markov chain of gauge configurations $U_i$, *i.e.*, a sequence generated by a stochastic rule that takes the previous configuration $U$ and produces a new configuration $U'$. The Markov chain proceeds from an arbitrary starting configuration. With a properly devised stochastic rule, after a sufficient number of steps the probability distribution approaches the desired distribution of Eq. (4.2.18). We say that the distribution has "reached equilibrium". Of course we must also take care that the distribution is "ergodic" in the sense that all important regions of the integrand are included in the ensemble – that the distribution isn't "frozen" around one local minimum of the effective action at the expense of other equally important minima.

*Heatbath algorithm*
The heatbath algorithm runs through the lattice updating each gauge link, one at a time. For the gauge link matrix $U_\mu(n)$, the integrand weight is regarded as defining a probability distribution $R[U_\mu(n)]$ for the gauge link being updated. One choses a new gauge link $U_\mu(n)'$ from that distribution and then moves on to the next gauge link. The name "heat bath" comes from early studies of $SU(2)$ pure gauge theory in which the effective action was proportional to a coupling constant that could be interpreted as an inverse Monte-Carlo temperature (not to be confused with the temperature of the partition function). So the update was analogous to exposing each link to a heat bath of that temperature. The heat bath method has fallen into disuse in lattice QCD now that more calculations include fermions, because the fermion determinant has a nontrivial dependence on the gauge links, which makes selecting a new link matrix $U_\mu(n)$ from a local probability distribution $R[U_\mu(n)]$ too expensive to implement.

*Metropolis-Hastings algorithm*
A classic method for generating the desired Markov chain uses the algorithm of Metropolis *et al.* and Hastings [332], usually abbreviated as the "Metropolis" algorithm. It works with a general class of stochastic rules for proposing a new gauge configuration $U'$ and then either accepts or rejects the new configuration based on a criterion designed to lead asymptotically to the desired ensemble:

– Propose a new configuration $U'$ with probability $Q(U' \leftarrow U)$. The transition must satisfy the reversibility condition:

$$Q(U \leftarrow U') = Q(U' \leftarrow U) \,. \tag{4.2.20}$$

Also, it must be possible after some number of steps to reach any configuration with nonzero probability.
– Choose a random number $\lambda$ distributed uniformly on $[0, 1]$.
– If the proposed change decreases the effective action $\Delta S_{\text{eff}} = S_{\text{eff}}(U') - S_{\text{eff}}(U) < 0$ then accept the change.
– Otherwise, accept the change if $\exp[-\Delta S_{\text{eff}}] > \lambda$. Otherwise, reject it.

The transition process defined by $Q(U' \leftarrow U)$ is quite general, which makes the algorithm particularly useful.

### 4.2.4 Molecular dynamics

By far the most common present-day method for generating the Markov chain uses a "molecular dynamics" method. We illustrate it for a scalar field $\phi$ with path-integral partition function

$$Z = \int [d\phi] \exp[-S(\phi)] \,. \tag{4.2.21}$$

We pair a dummy "momentum" $p(n)$ with the field $\phi(n)$ on each site of the lattice and rewrite the partition function as

$$Z' = \int [dp][d\phi] \exp\left[-p^2/2 - S(\phi)\right] \,. \tag{4.2.22}$$

The momentum integral is trivial and results in an immaterial constant factor. We then take a lesson from classical statistical mechanics and observe that this partition function describes a statistical ensemble of "particles" of unit mass, one per lattice site, and unit temperature $kT = 1$ moving in an interacting "potential" $S(\phi)$. The ensemble is microcanonical with total energy

$$E_{\text{tot}} = p^2/2 + S(\phi) \,. \tag{4.2.23}$$

The Hamilton equations of motion are, as usual,

$$d\phi(n)/d\tau = p(n) \tag{4.2.24}$$

$$dp(n)/d\tau = -\partial S/\partial\phi(n) \,, \tag{4.2.25}$$

where $\tau$ is a fictitious "Monte Carlo time". We then observe that if the system is large and the interactions are nontrivial, the classical motion of the system will lead to a Maxwell-Boltzmann distribution in the coordinates $\phi$ given by

$$P(\phi) \propto \exp[-S(\phi)] \,. \tag{4.2.26}$$

In standard practice, one chooses an arbitrary starting field configuration $\phi$ and sets the initial momenta according to the Gaussian distribution $\exp\left[-p^2/2\right]$. Using a numerical integrator, one integrates the equations of motion over some time interval $\Delta\tau$, at which time one saves an "updated" configuration $\phi_i$. Thus the Markov chain is defined by the values of $\phi$ at a series of time intervals or a series of what are called "molecular dynamics trajectories".

*Refreshed and hybrid Monte Carlo*
The total energy $E_{\text{tot}}$ is constant over a given trajectory. But it has no particular physical significance. To improve coverage of phase space it is common, at the beginning of each trajectory, to "refresh" the momenta $p$ by drawing new values from their Gaussian distribution. Thus each trajectory starts in a new direction with a new total energy, but the coordinates $\phi$ are kept continuous.

Another common variation of the method combines refreshed molecular dynamics with the Metropolis *et al.* method. This combination is called "hybrid Monte Carlo" [333]. That is, one starts a trajectory with coordinates $p, \phi$. At the end of the trajectory, one has coordinates $p', \phi'$. The transition $\phi' \leftarrow \phi$ is taken as a Metropolis move. The randomness in the refreshed initial momentum $p$ makes the move stochastic. Time-reversal invariance in $\tau$ assures detailed balance. If a trajectory is rejected, one reverts to the coordinate $\phi$ at the beginning of the trajectory, selects a new stochastic momentum, and tries again. The hybrid scheme helps compensate for possible inaccuracies in the numerical integration scheme, since it absolves many sins.

*Autocorrelations*
Markov chains have inherent correlations between successive members. These "autocorrelations" are undesirable, because they reduce the statistical independence of terms in the ensemble averages of Eq. (4.2.19) that give expectation values of physical observables. Autocorrelation is especially a concern with methods that make a series of small changes in the field configuration. With refreshed molecular dynamics one can adjust the trajectory length $\Delta\tau$ to help reduce correlations between successive terms $\phi_i$. One might expect that longer trajectories are better in this regard, but the "molecular motion" can contain cycles that bring parts of the system close to their original values. With hybrid schemes, longer trajectories can lead to lower Metropolis acceptance, which impedes progress. Shorter trajectories suffer from greater autocorrelation. Thus there is usually an optimum choice for the trajectory length that needs to be found empirically.

*Molecular dynamics for the gauge field*
The methods described above for a scalar field carry over to the SU(3) gauge field $U$. The gauge momentum, actually associated with the vector potential, $A_\mu(n)$, is given by a traceless antihermitian 3x3 matrix $H_\mu(n)$ for each $U_\mu(n)$. The molecular dynamics hamiltonian is, then,

$$\mathcal{H} = \frac{1}{2}\sum_{n,\mu} \mathrm{Tr}\, H_\mu(n)^2 + S_{\mathrm{eff}} \quad (4.2.27)$$

where we recall that

$$S_{\mathrm{eff}} = S_W + \ln \det[M] \quad (4.2.28)$$

To remain an SU(3) matrix, the equation of motion for $U_\mu(n)$ must be

$$dU_\mu(n)/d\tau = iH_\mu(n)U_\mu(n)\,. \quad (4.2.29)$$

The equation of motion for $H_\mu(n)$ can be found by requiring that the molecular dynamics Hamiltonian $\mathcal{H}$ remain constant in molecular-dynamics time [334]. For the sake of pedagogy, we first ignore the fermion determinant and consider the unimproved SU(3) gauge theory; see Sec. 4.1:

$$S_W = \frac{\beta}{6}\sum_{n,\mu\neq\nu} [3 - \mathrm{Re}\,\mathrm{Tr}\, U_{\mu\nu}(n)] \quad (4.2.30)$$

where $U_{\mu\nu}(n)$ is the plaquette product in the $\mu\nu$ plane with corner at site $n$. The plaquette can also be written as

$$\sum_\nu \mathrm{Re}\, U_{\mu\nu}(n) = U_\mu(n)V_\mu(n) + V_\mu(n)^\dagger U_\mu(n)^\dagger \quad (4.2.31)$$

where $V_\mu(n)$ is the sum of all "staples" attached to the link $U_\mu(n)$. Armed with this notation, we can write

$$0 = \dot{\mathcal{H}} = \sum_{n,\mu} \mathrm{Tr}\left[\dot{H}_\mu(n)H_\mu(n) + \frac{\beta}{6}\left(\dot{U}_\mu(n)V_\mu(n) + h.c.\right)\right], \quad (4.2.32)$$

and, using Eq. (4.2.29), we get

$$0 = \sum_{n,\mu} \mathrm{Tr}\left[\dot{H}_\mu(n)H_\mu(n) + \frac{\beta}{6}\left(iH_\mu(n)U_\mu(n)V_\mu(n) - h.c\right)\right], \quad (4.2.33)$$

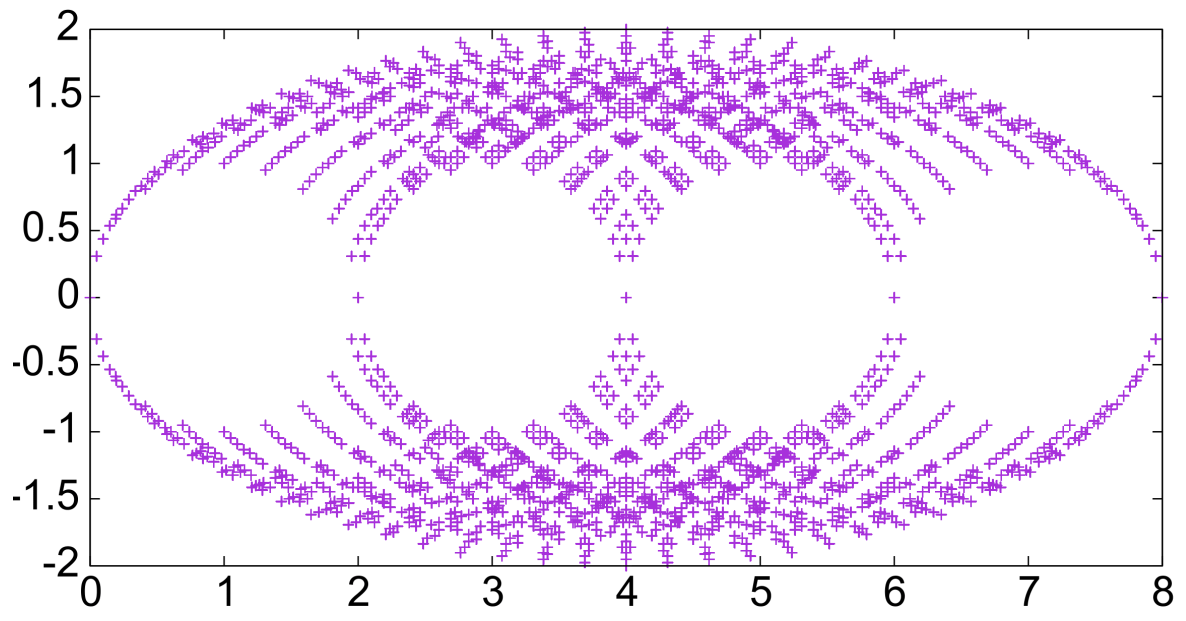

**Fig. 4.2.3** The spectrum of the free Wilson Dirac operator for a massless quark [335]

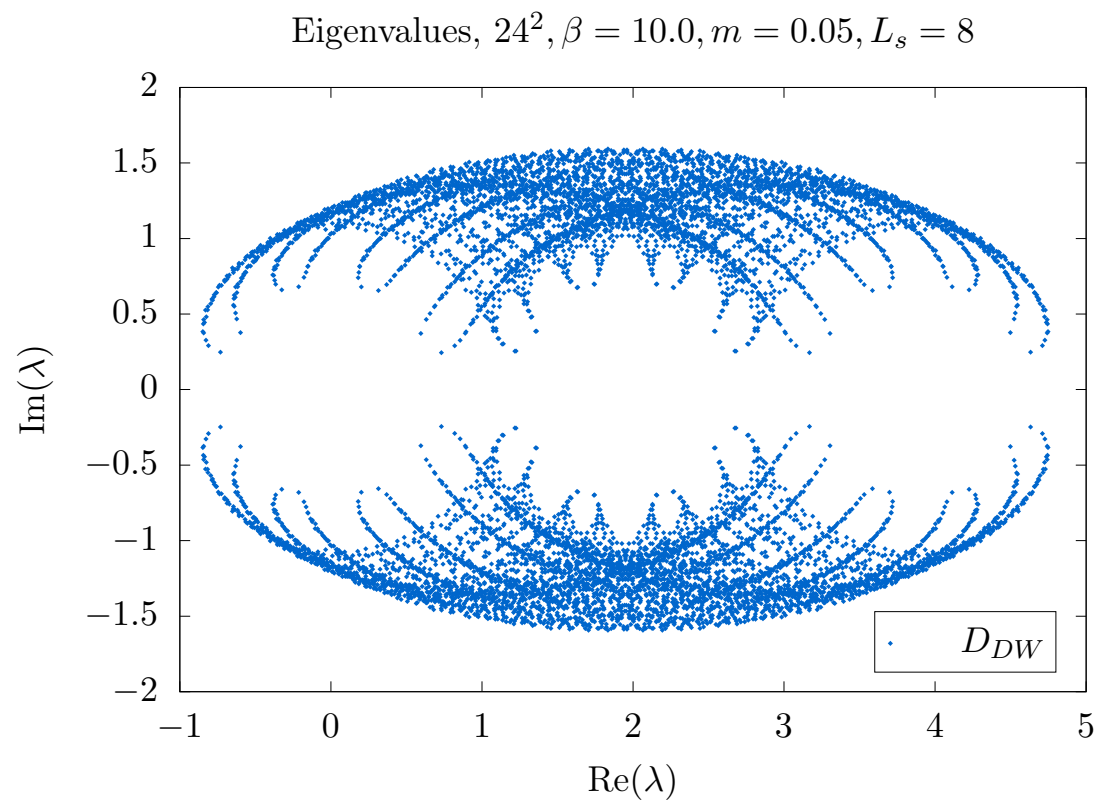

**Fig. 4.2.4** The spectrum of the domain wall operator at quark mass 0.05. [336].

or

$$0 = \sum_{n,\mu} \mathrm{Tr}\, H_\mu(n)[\dot{H}_\mu(n) + iF_\mu(n)] \quad (4.2.34)$$

where the gauge force is

$$F_\mu(n) = -\frac{\beta}{6}\left(U_\mu(n)V_\mu(n) - h.c.\right). \quad (4.2.35)$$

Since $H_\mu(n)$ in Eq. (4.2.34) is traceless the expression in brackets must be proportional to the identity matrix $cI$. But if it is to remain traceless, we must have $c = 0$. So, finally, we get

$$i\dot{H}_\mu(n) = F_\mu(n) = -\frac{\beta}{3}U_\mu(n)V_\mu(n)|_{\mathrm{TA}}\,, \quad (4.2.36)$$

where TA denotes the traceless, antihermitian part. The equations (4.2.29) and (4.2.36) form the basis for molecular dynamics evolution of the pure gauge theory.

*Spectrum of the Dirac matrix*
The Dirac matrix has the form (see Sec. 4.1)

$$M(U) = m + D(U)\,. \tag{4.2.37}$$

where $m$ is the quark mass. For all fermion formulations in common use today, the operator $D$ satisfies "$\gamma_5$ hermiticity", namely

$$D^\dagger = \gamma_5 D \gamma_5 \tag{4.2.38}$$

for some definition of $\gamma_5$. (For brevity, we drop the $(U)$ dependence of $M$ and $D$ in the following.) This implies that the complex eigenvalues of $D$ appear in complex conjugate pairs. Thus we can write the fermion determinant as

$$\det M = \prod_{\mathrm{Im}\,\lambda_i=0,i} (m+\lambda_i) \prod_{\mathrm{Im}\,\lambda_i>0} \left(m^2 + |\lambda_i|^2\right). \tag{4.2.39}$$

In order for $\det M$ to serve as a probability weight, it must be real and positive definite. Indeed, for all but the Wilson and clover actions, the real parts of the eigenvalues are nonnegative. For domain-wall and Wilson fermions, the eigenvalues $\lambda_i$ populate an ellipse in the right-half plane with voids, as illustrated in Figs. 4.2.3 and 4.2.4. For staggered fermions, they lie entirely on the imaginary axis (not shown). For overlap fermions, they lie on a circle in the right-half complex plane tangent to the imaginary axis (also not shown). For Wilson and clover fermions, they appear mostly in the right-half plane, but real, negative eigenvalues are possible, depending on the gauge configuration $U$. The eigenvalues of $M$ are $m+\lambda_i$, so negative $\lambda_i$ usually causes trouble for $M$ only for light quarks. As the lattice spacing is decreased, negative real parts become less frequent. Twisted-mass fermions (see Sec. 4.1) do not have this problem at maximal twist [337].

*The $\Phi$ algorithm*
The fermion determinant in Eq. (4.2.11) can be cast in a form compatible with the molecular-dynamics treatment of the gauge field. Perhaps the simplest approach is the "$\Phi$ algorithm". We introduce a complex lattice scalar field $\Phi$, often called a pseudofermion field, and first try

$$\det M = \int d\Phi d\Phi^* \exp\left[-\Phi^* M^{-1}\Phi\right]. \tag{4.2.40}$$

This works as long as the eigenvalues of $M$ have positive definite real parts. However, this form is awkward to implement. A more convenient form works with the normal operator $M^\dagger M$. From $\gamma_5$ hermiticity we have

$$\det M = \det\left[\gamma_5 M^\dagger \gamma_5\right] \tag{4.2.41}$$
$$\det M^2 = \det\left[M^\dagger M\right] \tag{4.2.42}$$

so

$$\det M^2 = \int d\Phi d\Phi^* \exp\{-\Phi^*[M^\dagger M]^{-1}\Phi\}\,. \tag{4.2.43}$$

Now the integral is always well defined for nonzero quark mass, but the square doubles the number of fermions. That could be acceptable if we are simulating up and down quarks in the isospin symmetric limit ($m_u = m_d$), but it would be a bad approximation for the other quarks. Remedies are discussed below. We continue with this form.

*Molecular dynamics with fermions*
To simulate Eq. (4.2.43), we note that the integrand has the form $\exp\left(-R^\dagger R\right)$ for $R = M^{-1}\Phi$, so if we draw $R$ from a Gaussian distribution, then $\Phi = MR$ is distributed according to the desired weight.

The $\Phi$ algorithm begins a short trajectory by constructing $\Phi = MR$ for a given starting gauge field $U$. The gauge field is then evolved with $\Phi$ fixed. The force exerted on the gauge field in Eq. (4.2.36) acquires a new contribution, namely, the fermion force:

$$\begin{aligned} iF_{F,\mu n} &= \frac{\partial}{i\partial A_{\mu,n}} \Phi^*[M^\dagger M]^{-1}\Phi \\ &= X^* \frac{\partial}{i\partial A_{\mu,n}} (M^\dagger M) X\,, \end{aligned} \tag{4.2.44}$$

where $M^\dagger M X = \Phi$. Typically, one refreshes the gauge momentum, evolves the gauge field at the initial fixed value of $\Phi$, and then repeats.

*Rational function approximation*
As we saw above, working with the normal operator $M^\dagger M$ doubles the number of fermion species. To eliminate the doubling, we should replace $M^\dagger M$ with $\sqrt{M^\dagger M}$. Similarly, for staggered fermions, we start with four tastes per flavor, which suggests $(M^\dagger M)^{1/8}$. With staggered fermions, the normal operator is checkerboard block-diagonal, so restricting the calculation to even lattice sites eliminates the normal-operator doubling. We then want $(M^\dagger M)^{1/4}|_{\mathrm{even}}$.

Such fractional powers are difficult to implement. A now commonly used remedy introduces a rational-function approximation for the fractional power [338]. Expanded in terms of its poles, the rational function approximation for a real function $f(x)$ of real $x$ has the form

$$f(x) \approx r(x) = \sum_{i=1}^{N} \frac{\alpha_i}{x - \beta_i}\,., \tag{4.2.45}$$

where $\alpha_i$ and $\beta_i$ are parameters of the rational function, and $N$ is a suitably high order. The approximation deteriorates for small $x$. It is designed to work over an interval $[x_{\min}, x_{\max}]$. The smaller $x_{\min}$ or the finer the desired accuracy, the larger the needed order $N$. The Zolotarev method [339] is widely used to obtain an efficient set of parameters $\alpha_n$ and $\beta_n$.

We note that $M^\dagger M = D^\dagger D + m^2$ for mass $m$. It is convenient to treat this expression as a function of $x = D^\dagger D$. So to apply the rational function approximation, we write

$$(M^\dagger M)^h \approx r_h(D^\dagger D) = \sum_{i=1}^{N} \frac{\alpha_{h,i}}{D^\dagger D - \beta_{h,i}} \,., \tag{4.2.46}$$

where we have labeled the coefficients of the expansion with the desired power $h$. So, finally, we have

$$\det\left(M^\dagger M\right)^h \approx \int d\Phi d\Phi^* \exp\left[-\Phi^* r_h(D^\dagger D)\right]\Phi] \tag{4.2.47}$$

To implement the $\Phi$ algorithm with fractional power $h$,

$$\int d\Phi d\Phi^* \exp\{-\Phi^* (M^\dagger M)^h \Phi\}\,, \tag{4.2.48}$$

we choose Gaussian random $R$ and calculate

$$\Phi = [M^\dagger M]^{-h/2} R \tag{4.2.49}$$

using a rational function approximation $r_{-h/2}(D^\dagger D)$. Then we calculate the fermion force with

$$\begin{aligned} iF_{F,\mu n} &= \frac{\partial}{i\partial A_{\mu,n}} \Phi^* (M^\dagger M)^h \Phi \\ &= \Phi^* \frac{\partial}{i\partial A_{\mu,n}} r_h(D^\dagger D) \Phi \qquad (4.2.50) \\ &= \sum_i X_i^* \alpha_{h,i} \frac{\partial}{i\partial A_{\mu,n}} [M^\dagger M] X_i \,, \qquad (4.2.51) \end{aligned}$$

where $X_i = [D^\dagger D - \beta_{h,i}]^{-1}\Phi$. Here the rational function parameters are appropriate for $r_h$. The $X_i$ are obtained using a multishift conjugate-gradient solver.

*Multiple flavors*

The rational function approximation can be extended to handle the products of determinants that arise with multiple flavors. For example, suppose we are simulating two degenerate light quarks (up and down) $m_l = m_u = m_d$ and one strange quark $m_s$. We use $f$ to distinguish the flavors in the fermion matrix $M_f$. After integrating out the Grassmann fields, the fermion integrand becomes

$$\det\left(M_l^\dagger M_l\right) \det\left(M_s^\dagger M_s\right)^{1/2}. \tag{4.2.52}$$

We could simulate this product by introducing a separate pseudofermion field for each flavor and proceeding as we did for a single flavor for each contribution. However, we can also simulate it using just one pseudofermion field:

$$\int d\Phi d\Phi^* \exp\{-\Phi^* (M_l^\dagger M_l)^{-1} (M_s^\dagger M_s)^{-1/2} \Phi\} \tag{4.2.53}$$

We construct a rational function that approximates the entire product.

$$(M_l^\dagger M_l)^{-1} (M_s^\dagger M_s)^{-1/2} = r_{-1,-1/2}(D^\dagger D)\,, \tag{4.2.54}$$

where we have added more labels to $r(x)$. The $\Phi$ algorithm is otherwise similar to that of the single-flavor case.

### 4.2.5 Improvements

*Hasenbusch term*

One popular and effective improvement [340] introduces a “preconditioning” determinant, a ”Hasenbusch term”, with moderately large mass $m_x$ together with its compensating inverse, for example, as

$$(M_l^\dagger M_l)^{-1} (M_s^\dagger M_s)^{-1/2} (M_x^\dagger M_x)^{3/2} (M_x^\dagger M_x)^{-3/2}\,. \tag{4.2.55}$$

The first three factors are then assigned a single pseudofermion field and approximated with a single rational function, and the fourth factor is assigned a separate pseudofermion field with a separate rational function. The Hasenbusch term tends to reduce the condition number of the product operator, thus reducing the needed rational function order and the associated computation time. The last (compensating) factor also has a lower condition number because of the larger mass.

*Multigrid solvers*

To evaluate the rational function in Eq. (4.2.51) requires solving a large linear system. As the lattice spacing decreases, the condition number of the linear system grows, making the conventional conjugate-gradient calculation more costly. This “critical slowing down” can be mitigated by using an adaptive geometric multigrid solver instead [341, 342]. So far the benefits of using multigrid solvers for gauge-field evolution have been demonstrated only for the Wilson-clover action [343]. Algorithms for multigrid solvers for staggered fermions [344] and domain-wall fermions [336, 345] are newer, so it remains to be seen whether they will lead to improvements in molecular dynamics evolution for those fermion formulations as well.

*Accelerating molecular dynamics*
As the lattice spacing decreases, the gauge-field evolution slows, and it gets trapped in a subset of gauge configurations with the same total topological charge. Thus it takes more computational time to obtain a new, statistically uncorrelated gauge configuration. Long-distance decorrelation is slower than short-distance. This observation suggests Fourier transforming Hamilton's equation for the gauge momentum,

$$idH_\mu(n) = F_\mu(n)d\tau \tag{4.2.56}$$

to (coordinate) momentum space, and, instead of using a common time step $d\tau$ for each momentum component, consider using a larger time step for the low-momentum modes [346] to move them farther. This method never proved effective enough to use in full-scale simulation. Modern versions of the Fourier acceleration scheme are under investigation. See, for example, [347].

*Trivializing map*
If we can find an invertible map of the gauge field $U$ to a new field $V$,

$$U = \mathcal{F}(V)\,, \tag{4.2.57}$$

such that the Jacobian of the transformation cancels the gauge action:

$$\det[\partial U_\mu(n)(V)/\partial V_\nu(m)]\exp[-S(U)] = 1\,, \tag{4.2.58}$$

then the path integral becomes trivial [348–350]. Lüscher describes this as a "map to the strong-coupling limit" and discusses possible maps for the pure gauge action. Of course, finding such a map is entirely nontrivial, but if one can at least find one that moves the action partially toward strong coupling, then one could construct a hybrid Monte Carlo scheme that updates the gauge field according to the recipe

$$U \to V \to V' \to U' \tag{4.2.59}$$

where the $V \to V'$ step uses standard gauge evolution for the transformed gauge field $V$. This stronger coupling evolution would suffer less from critical slowing down. Recently, there have been efforts to find such a map using machine-learning methods. See, for example, Ref. [351].

### 4.2.6 Personal remarks

I first learned about the lattice formulation of QCD and its virtues when Ken Wilson gave a seminar at the MIT Center for Theoretical Physics around the time he was developing his lattice formulation. I was quite impressed with how easily confinement, in the form of an area law for Wilson loops, emerged in the strong-coupling regime. But I wasn't as brave or savvy as Creutz in proceeding to develop numerical methods for working out the nonperturbative consequences of Wilson's formulation. I didn't turn to numerical lattice calculations until shortly after Creutz's seminal papers. For the rest of my career, I have enjoyed participating in and contributing to the remarkable progress in this field. As a graduate student schooled in the analytic S-matrix and bootstrap, I was pleased when I could make a strong-interaction prediction to an accuracy of 25%, based on phenomenological considerations. There was always the inevitable doubt about the validity of the methods. Today, in some cases, we are able to obtain per mille accuracy for aome hadronic properties. Furthermore, we have little doubt that our results are a correct prediction of the Standard Model, since our methods are grounded in first-principles. That has been enormously satisfying.

## 4.3 Vacuum structure and confinement

**Derek Leinweber**

### 4.3.1 Introduction

The self interactions of gluons make the empty vacuum unstable to the formation of quark and gluon field configurations which permeate spacetime. These ground-state QCD-vacuum field configurations form the foundation of matter. Lattice QCD simulations enable first principles explorations of this nontrivial vacuum field structure.

These gluon field configurations form the foundation of every lattice QCD calculation. Each field configuration on its own contains a rich diversity of emergent nonperturbative structure. It is the process of averaging over thousands of field configurations that restores the translational invariance of the vacuum. Each field configuration with its own rich structure is uncorrelated with other configurations considered in the averaging process.

Deep insight into the mechanisms giving rise to the observed quantum phenomena can be obtained through the visualization of these complex scientific data sets constructed in Lattice QCD simulations, insights that would otherwise remain hidden in the typical gigabyte data sets of modern quantum field theory.

The essential, fundamentally-important, nonperturbative features of the QCD vacuum fields are: the dynamical generation of mass through chiral symmetry breaking, and the confinement of quarks. But what are the fundamental mechanisms of QCD that underpin these phenomena? What aspect of the QCD vacuum causes quarks to be confined? Which aspect is responsible for dynamical mass generation? Do the underlying mechanisms share a common origin?

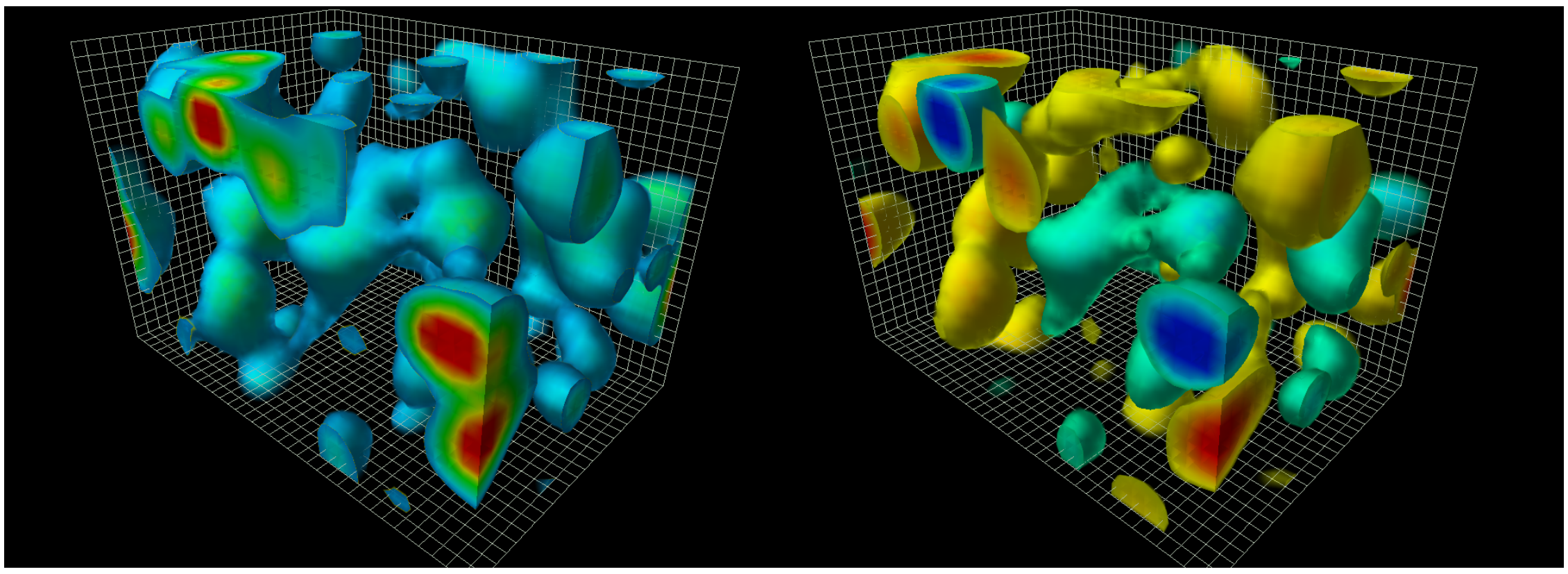

**Fig. 4.3.1** Frames from the animation of Ref. [352] illustrating the Euclidean action density or energy density of Eq. (4.3.2) (left) and the corresponding topological charge density of Eq. (4.3.4) (right) at an instant in time. The spatial volume is approximately 2.4 by 2.4 by 3.6 fm.

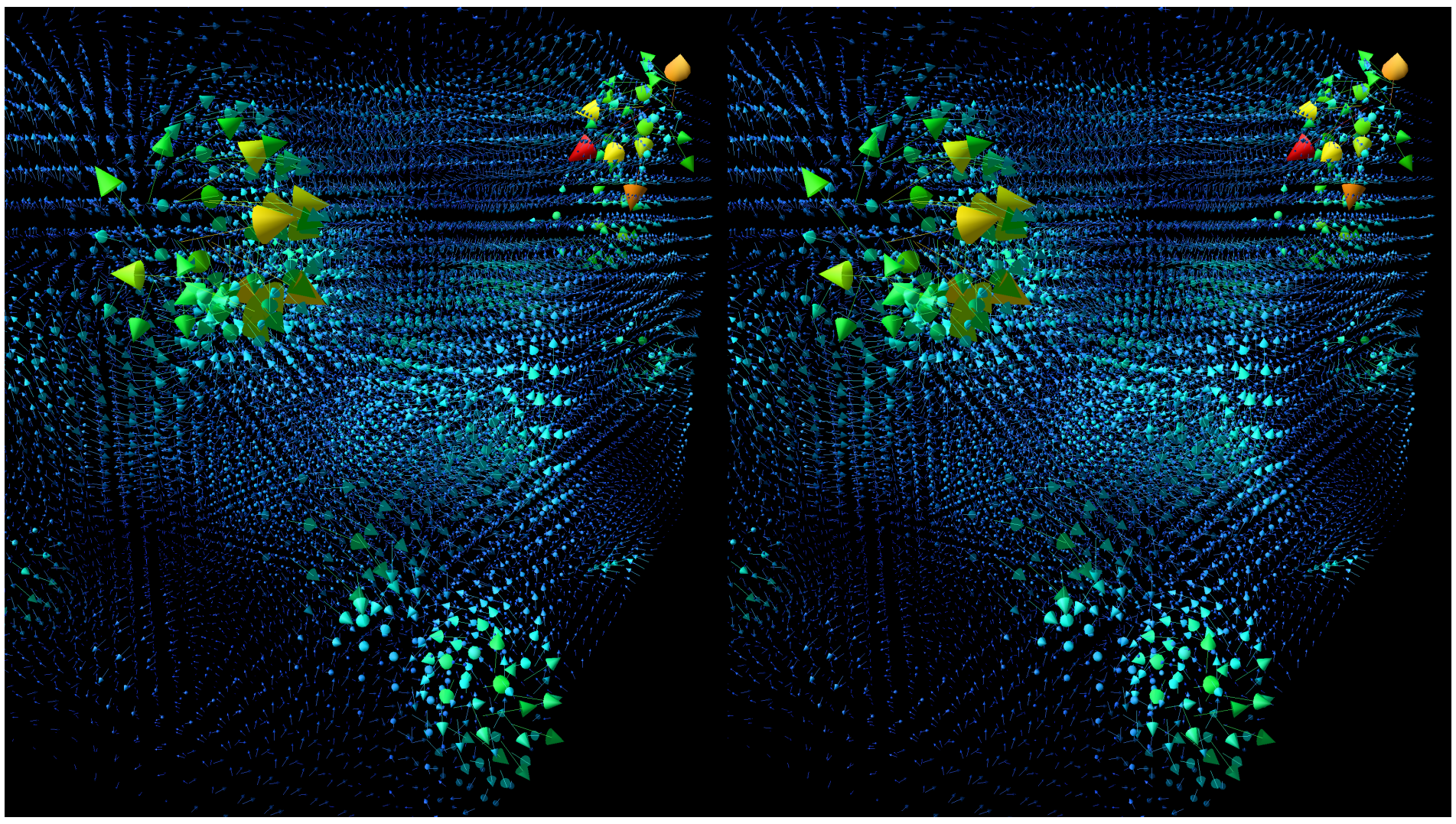

**Fig. 4.3.2** Stereoscopic image of one of the eight chromo-magnetic fields composing the nontrivial vacuum of QCD. Hints for stereoscopic viewing are provided in the text.

In this brief review, we will address these questions in a chronological manner to convey the progress in developing an understanding of the essential mechanisms underpinning the phenomena of QCD.

### 4.3.2 Nonperturbative vacuum structure

Among the earliest of vacuum-structure visualizations are images of the Euclidean action density, or energy density

$$S_E(\vec{x},t) = \frac{1}{2} F^{ab}_{\mu\nu}(\vec{x},t)\, F^{ba}_{\mu\nu}(\vec{x},t)\,, \tag{4.3.1}$$

$$= \mathrm{Tr}\left(\vec{E}^2(\vec{x},t) + \vec{B}^2(\vec{x},t)\right)\,, \tag{4.3.2}$$

where $F^{ab}_{\mu\nu}$ is the Euclidean field strength tensor

$$F^{ab}_{\mu\nu} = \partial_\mu A^{ab}_\nu - \partial_\nu A^{ab}_\mu + ig[A^{ab}_\mu, A^{ba}_\nu]\,, \tag{4.3.3}$$

with color indices $a$, $b$ = 1, 2, 3. The corresponding topological charge density proportional to $\vec{E}(\vec{x},t)\cdot\vec{B}^(\vec{x},t)$

$$q(\vec{x},t) = \frac{g^2}{32\,\pi^2}\,\epsilon_{\mu\nu\rho\sigma}\, F^{ab}_{\mu\nu}(\vec{x},t)\, F^{ba}_{\rho\sigma}(\vec{x},t)\,, \tag{4.3.4}$$

is also of interest as it characterizes the profile of instantons, nontrivial solutions of the classical Yang-Mills equations, discussed in further detail in Sec. 5.11.4.

Ref. [353] provides one of the earliest observations of instanton-like objects in lattice gauge-field configurations. Here cooling with the standard Wilson action was used to suppress short-distance field fluctuations enabling the observation of long-distance structures.

However a problem with the use of the standard Wilson action or even the $\mathcal{O}(a^2)$-improved plaquette plus rectangle action is that the lattice action of an instanton can be reduced by shrinking the size of an instanton [354] through lattice-spacing errors. Instantons shrink under cooling with these lattice actions and "fall through the lattice." This led to the development of highly-improved actions [355, 356] eliminating errors to $\mathcal{O}(a^4)$ or even over-improved actions where improvement terms are tuned to stabilize instantons, ensuring their stability under smoothing algorithms [354, 357].

The results presented in this section are based on pure SU(3) gluon fields created with the standard Wilson action at $\beta = 6.0$ on a $24^3 \times 36$ lattice with a lattice spacing, $a \simeq 0.1$ fm. The first coordinate of the Euclidean lattice was used for the time axis creating a $24^2 \times 36$ spatial volume. It is these calculations [358] that captured the attention of Prof Frank Wilczek as he prepared his 2004 Nobel Prize lecture. Ref. [359] provides a link to the *QCD Lava Lamp* animation that appeared in his Nobel Lecture [360]. In support of the Nobel Lecture a web page incorporating the best algorithms and visualization techniques of the time was created [361]. Parallel spatially-uniform $\mathcal{O}(a^4)$-improved smoothing algorithms [362] and an $\mathcal{O}(a^4)$-improved lattice field strength tensor [355] were formulated to accurately retain and present the long-distance nonperturbative properties of the ground-state vacuum fields. These images and animations [352, 361] have since appeared in popular-science publications, leading YouTube channels [363, 364], etc. [365].

Figure 4.3.1 displays two frames from the animation of Ref. [352]. Here 25 sweeps of three-loop, mean-field, $\mathcal{O}(a^4)$-improved cooling has been applied. Areas of high energy density are rendered in red and regions of moderate energy density are rendered in blue. The lowest energy densities are not rendered such that one can see into the volume. Similarly the topological charge density has regions of positive density rendered in red through yellow and regions of negative density rendered blue through cyan. While instanton-like objects are manifest, current research is examining the extent to which instanton-dyon degrees of freedom [366], *i.e.* fractionally charged regions, can be observed within these field configurations.

To directly view the the eight chromo-electric and eight chromo-magnetic gluon fields composing the vacuum, one must select a gauge. Figure 4.3.2 presents a stereoscopic illustration of one of the chromo-magnetic fields in Landau gauge [367]. Here the color and length of the arrows describe the magnitude of the vector fields. Animations of the fields are also available [352].

To see the 3D image of Figs. 4.3.2 and 4.3.9, try the following:

1. If you are viewing the image on a monitor, ensure the image width is 12 to 13 cm.
2. Bring your eyes very close to one of the image pairs.
3. Close your eyes and relax.
4. Open your eyes and allow the (blurry) images to line up. Tilting your head from side to side will move the images vertically.
5. Move back slowly until your eyes are able to focus. There's no need to cross your eyes!

With its lattice implementation of chiral symmetry, the overlap-Dirac operator provided a new approach to the exploration of the nonperturbative structure of the vacuum without resorting to smoothing algorithms [368]. Here low-lying Dirac eigenmode densities could be used to construct the topological charge density with the level of smoothness inversely related to the number of low-lying modes one considers [369]. Strong correlation with the instanton-like objects observed via smoothing algorithms was observed. The zero modes are chiral and are distributed across topological charge regions of a unique sign. Low-lying eigenmode densities are also highly correlated with the topological structures revealed under smoothing [369]. These correlations between gluonic and fermionic structures expose the dynamics underpinning dynamical chiral symmetry breaking and the origin of mass.

The manner in which the topological charge density is rendered can lead to rather different views on the nature of how topological charge is distributed in the vacuum. Figure 4.3.3 illustrates two different renderings of the same topological charge density. The sheet-like structure associated with the sign-changing nature of the topological charge density correlator $\langle\, q(0)\, q(x)\, \rangle$ [370, 371] is manifest when all magnitudes of the topological charge density are rendered down to zero. This is the celebrated sheet-like structure of the topological charge density [372]. However, when the rendering is restricted to larger values, one reveals a more lumpy structure with regions of significant coherent topological charge density.

More recently explorations of correlations between QCD phenomena and QED phenomena have commenced drawing on QCD+QED lattice simulations [373, 374]. First results and links to associated animations are reported in Ref. [365].

### 4.3.3 Center cluster structure of QCD vacuum fields

Further insight into the structure of QCD vacuum fields, their temperature dependence, and their evolution un-der Monte-Carlo evolution can be obtained through the consideration of the local Polyakov loop. The expectation value of the Polyakov loop is related to the finite temperature phase transition in QCD. It has an expectation value of zero in the confined phase and becomes nonzero in the deconfined phase.

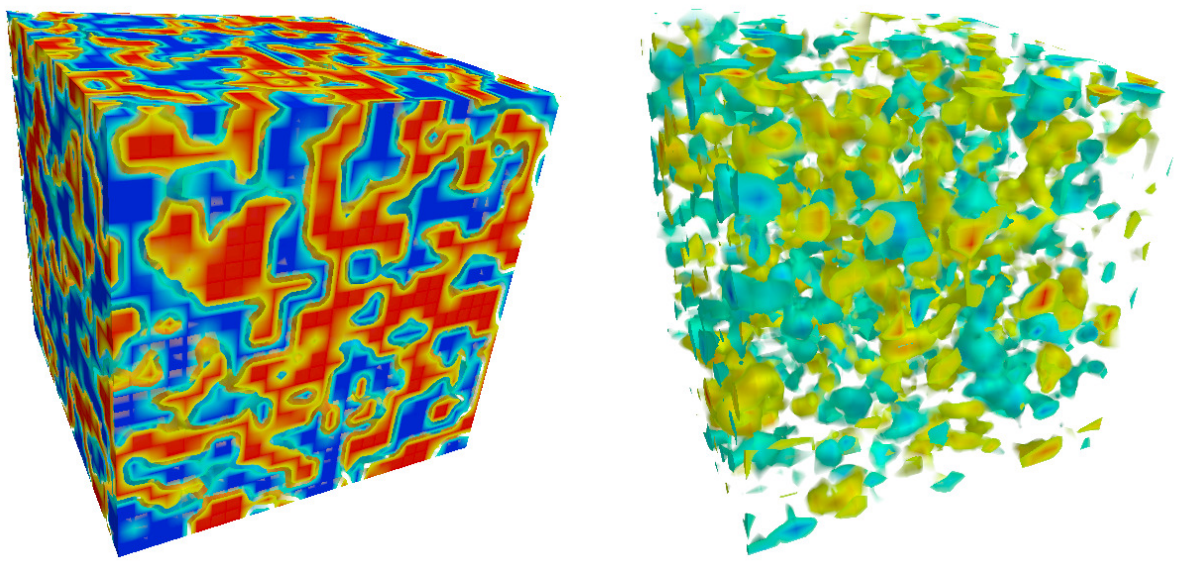

**Fig. 4.3.3** The short-distance sheet-like structure of the vacuum is made apparent in the left-hand illustration by rendering all magnitudes of the topological charge density down to zero. Negative charge density is rendered green to blue, and positive charge density is yellow to red. The same data is rendered in the right-hand plot, this time only rendering the regions having large topological charge density, revealing a structure of topological lumps.

The local Polyakov loop is the traced gauge-invariant product of time-oriented gauge links around the time extent of the lattice at each spatial point

$$L(\vec{x}) = \mathrm{Tr} \prod_{t=1}^{N_t} U_4(t, \vec{x}) = \rho(\vec{x})\, e^{i\phi(\vec{x})} \,, \tag{4.3.5}$$

Here, $U_4$ is the time-oriented link variable on a lattice with lattice spacing $a$, given by

$$U_\mu\,(x) = P \exp\left( ig \int_x^{x+\hat{\mu}a} dx^\mu A_\mu(x) \right) . \tag{4.3.6}$$

Center clusters [375, 376] are defined in terms of $L(\vec{x})$. They are regions of space where the local Polyakov loop prefers a single complex phase associated with the center of SU(3). The deconfinement transition occurs through the growth of a center cluster.

In the final expression of Eq. (4.3.5), the local Polyakov loop is decomposed into a phase, $\phi(\vec{x})$ and a magnitude, $\rho(\vec{x})$. Both the proximity of the phase to one of the cube-roots of one and the magnitude are considered in visualizing the structure of the center domains of the gluon field. In either case, the most proximal cube root of one to the phase is indicated by the use of color.

In Ref. [376] an anisotropic gauge action was used to explore the evolution of coherent center domains in the gluon field under both temperature and the Hybrid Monte Carlo (HMC) update algorithm. To investigate the larger-scale behavior of the clusters, small scale noise is removed from the visualization by performing four sweeps of stout-link smearing [377] prior to calculating the Polyakov loops.

In Fig. 4.3.4, clusters are rendered where the phase $\phi(\vec{x})$ is within a small window around each center phase, and the rest of the volume is rendered transparent. Within these coherent center domains, color-singlet quark-antiquark pairs or three-quark triplets have a finite energy and are spatially correlated. Thus, these fundamental domains govern the size of the quark cores of hadrons. As one domain dominates the vacuum above the critical temperature, the correlation length diverges and quarks become deconfined.

The evolution of these clusters with HMC simulation time is presented in Ref. [379], showing how center clusters are slowly moving with correlations in the center clusters persisting for approximately 5 seconds corresponding to 25 HMC trajectories. The temperature dependence of the center-cluster structure is also explored in these animations where a single phase eventually dominates above the critical temperature, as illustrated in Fig. 4.3.4.

### 4.3.4 Flux tubes in QCD ground-state vacuum fields

Flux tubes in the QCD vacuum are revealed by examining the correlation between ground-state field properties and the positions of static quarks within the fields. One begins with the standard approach of connecting static quark propagators by spatial-link paths in a gauge invariant manner. For mesonic systems, this is the standard Wilson loop. However, for baryonic systems one needs the structure illustrated in Fig. 4.3.5. The spatial link paths are typically broadened through a smearing algorithm to approximate the shape of the flux tube and thus obtain better overlap with the ground state potential of interest. While early calculations tuned the amount of smearing to provide optimal overlap with the ground state, more modern approaches create a basis of smeared sources and solve the generalized eigenvalue problem to obtain the optimal combination of sources. The static quark propagators are constructed from time directed link products at fixed spatial coordinate, $\prod_i U_t(\vec{x}, t_i)$, using the untouched "thin" links of the gauge configuration.

The correlation of the gluon field with the static quark positions is characterized by the gauge-invariant Euclidean action density $S_E(\vec{y}, t)$ observed at spatial coordinate $\vec{y}$ and Euclidean time $t$ measured relative to the origin of the three-quark Wilson loop. For the results presented herein, the action density is calculated using the highly-improved $\mathcal{O}(a^4)$ three-loop improved

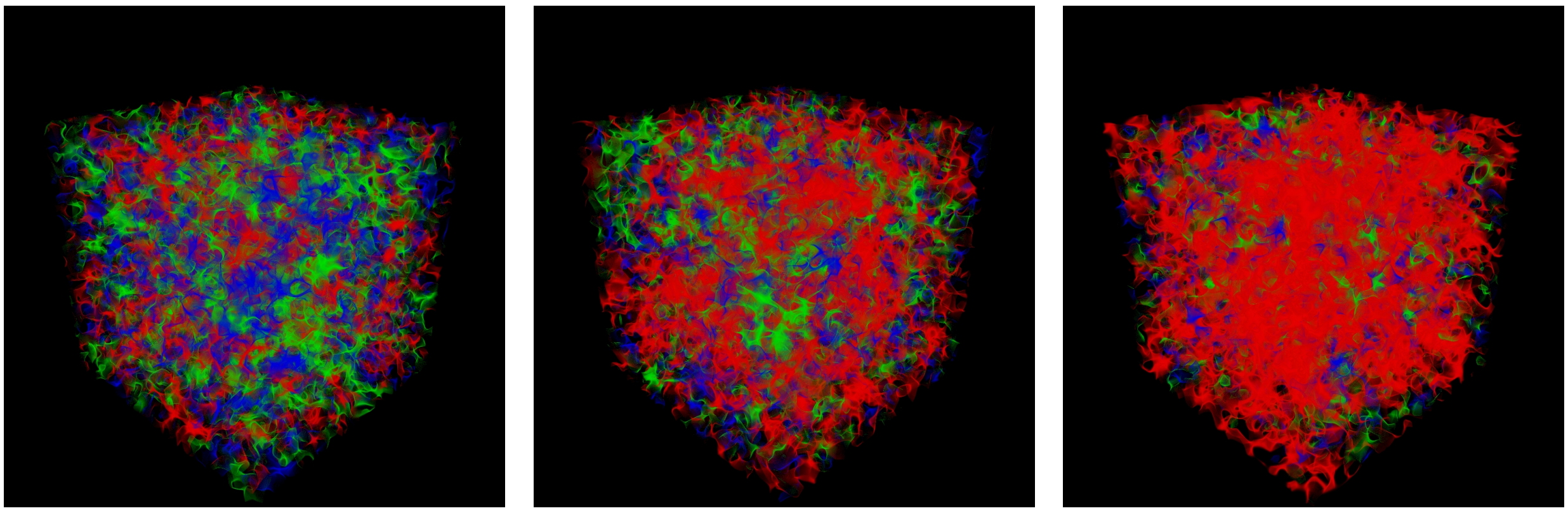

**Fig. 4.3.4** Center clusters on a gauge field configuration at $T = 0.89(1)\,T_C$ (left), $T = 1.14(2)\,T_C$ (middle), and $T = 1.36(2)\,T_C$ (right), This rendering from Ref. [376] is based on the proximity of the local Polyakov loop phase, $\phi(\vec{x})$, to one of the three center phases of SU(3). The length of each side of the cubic volume is 2.4 fm. The percolation of the red phase in the middle and right-hand plots illustrate the deconfinement of quarks above $T_C$.

lattice field-strength tensor [355] on four-sweep APE-smeared gauge links [378].

Defining the quark positions as $\vec{r}_1$, $\vec{r}_2$ and $\vec{r}_3$ relative to the origin of the three-quark Wilson loop, and denoting the Euclidean time extent of the loop by $\tau$, one evaluates the following correlation function

$$C(\vec{y};\vec{r}_1,\vec{r}_2,\vec{r}_3;\tau) = \frac{\langle W_{3Q}(\vec{r}_1,\vec{r}_2,\vec{r}_3;\tau)\, S_E(\vec{y},\tau/2)\rangle}{\langle W_{3Q}(\vec{r}_1,\vec{r}_2,\vec{r}_3;\tau)\rangle\,\langle S_E(\vec{y},\tau/2)\rangle}\,, \tag{4.3.7}$$

where $\langle\cdots\rangle$ denotes averaging over configurations and translational/reflection/rotational lattice symmetries [378]. Note that the correlation is examined at the midpoint in the time evolution of the static quark propagation to ensure the three quark state has relaxed to its ground state form. For fixed quark positions and Euclidean time, $C$ is a scalar field in three dimensions.

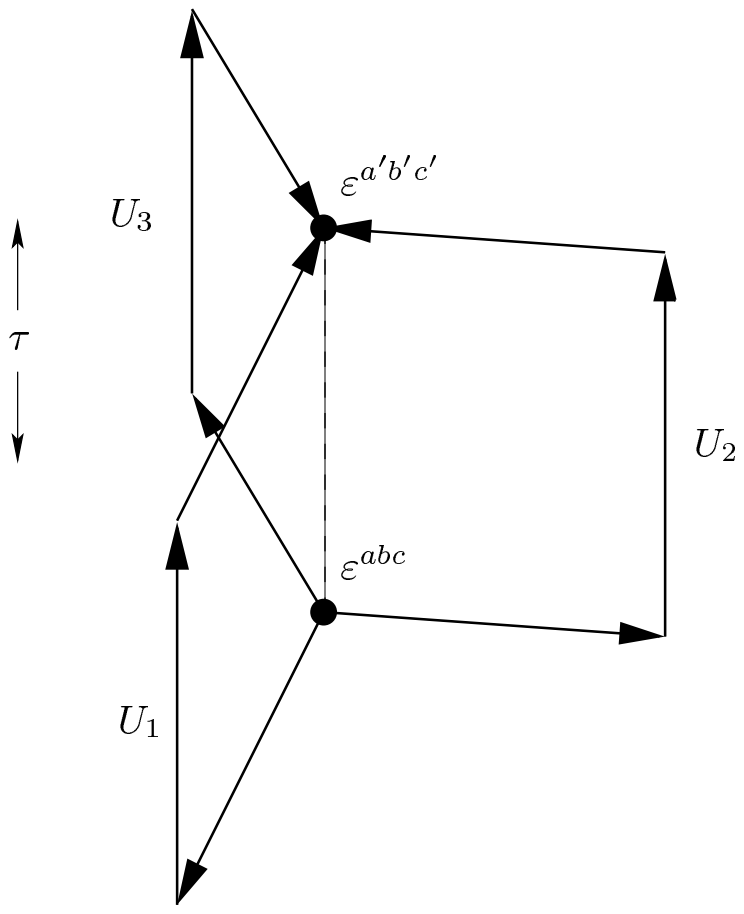

**Fig. 4.3.5** Gauge-link paths for three static quark propagators, $U_1$, $U_2$, and $U_3$, are connected in a gauge-invariant manner via spatially smeared link paths. $\varepsilon^{abc}$ and $\varepsilon^{a'b'c'}$ provide color anti-symmetrisation at the source and sink respectively, while $\tau$ indicates evolution of the three-quark system in Euclidean time [378].

This measure has the advantage of being positive definite, eliminating any sign ambiguity on whether vacuum field fluctuations are enhanced or suppressed in the presence of static quarks. The correlation, $C$, is generally less than 1, signaling the expulsion of vacuum fluctuations from the interior of heavy-quark hadrons. In other words, flux tubes represent the suppression of the vacuum field fluctuations that form the foundation of matter.

Figure 4.3.6 provides an illustration of the correlation $C(\vec{y})$. For values of $\vec{y}$ well away from the quark positions $\vec{r}_i$, there are no correlations and $C \to 1$. As the separation between the quark-antiquark pair changes, the flux tube of Fig. 4.3.6 (top) gets longer, but the diameter of the tube and the depth of the expulsion remain approximately constant. As it costs energy to expel the vacuum field fluctuations, the confinement potential grows linearly as the quark separation increases.

Of historical significance was the endeavor to determine whether baryon flux tubes are Y-shape or $\Delta$-shape (empty triangle) in nature. For the latter, the expectation was two-body tube-like structures around the edge of the three-quark system would dominate. Quantitative analyses of the static quark potential and the distribution of flux tubes led to a consensus [380] that the distribution is Y shape for large quark separations more than $\sim 0.5$ fm from the system center with the observation of filled $\Delta$ shapes at shorter-distance separations. The Y-shape ground state localizes at the Steiner point which minimizes the total string length.

The characteristic sizes of the flux-tube and node were quantified in Ref. [378]. The ground state flux-tube radius is $\sim 0.4$ fm with vacuum-field fluctuations suppressed by 7%. The node connecting the flux tubes

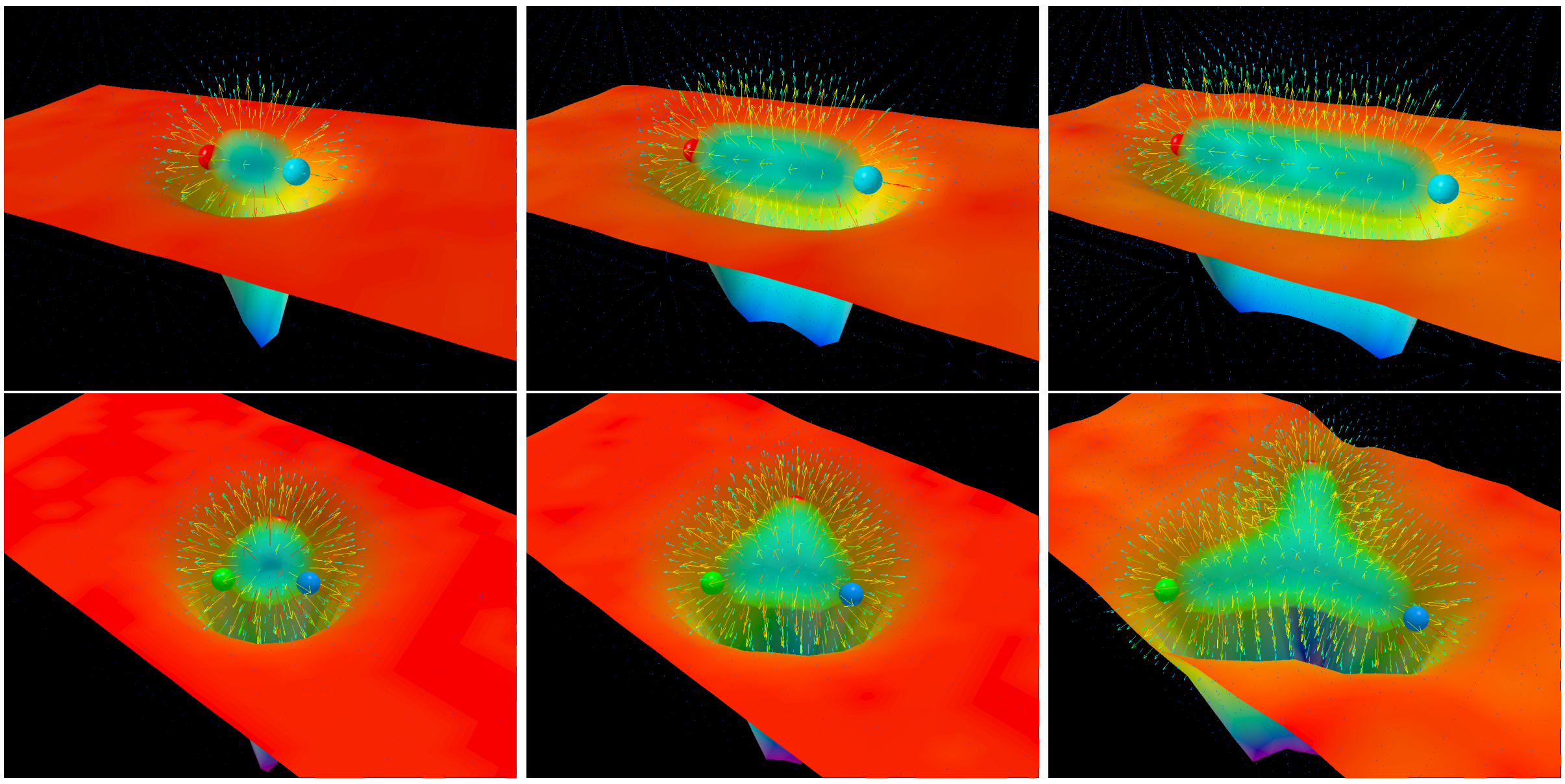

**Fig. 4.3.6** The suppression of QCD vacuum fields, as represented by the energy density, from the region between a quark-antiquark meson (top) or three-quark baryon (bottom). Quark positions are illustrated by the colored spheres. The separation of the quarks in the meson are 0.50 fm (left), 1.00 fm (middle), and 1.50 fm (right). The baryon frames illustrate the spherical cavity (or bag) observed at small quark separations of 0.27 fm from the center (left), the development of a filled-$\Delta$ shape at moderate separations of 0.42 fm (middle) and finally the emergence of a Y-shape flux tube (right) at large quark separations of 0.72 fm from the system center [378]. The surface plot illustrates the reduction of the vacuum energy density in a plane passing through the centers of the quarks. The vector field illustrates the gradient of this reduction. The tube joining the quarks reveals the positions in space where the vacuum energy density is maximally expelled and corresponds to the "flux tube" of QCD.

is larger at 0.5 fm with a suppression of the vacuum action at 8%.

### 4.3.5 Flux tube string breaking in QCD

With the advent of numerical simulations incorporating the dynamics of light fermion loops in the QCD vacuum, the observation of flux-tube breaking or string breaking was keenly anticipated. The idea is that for increasing quark separations, eventually there would be enough energy in the flux tube joining the two static $b$ quarks that it would become energetically favorable to break the string through the creation of a light quark-antiquark pair and the formation of two $B$ mesons. Even to this day, this *implicit* form of string breaking has yet to be observed. The difficulty lies in the extraordinarily poor overlap of the two-$B$ meson state with the spatial flux-tube operators used to create the string state.

This situation is in contrast to explorations of the structure of the $\Lambda(1405)$ baryon, where lattice-QCD calculations of the quark-sector contributions to the baryon magnetic moment indicate a molecular meson-baryon structure [381, 382]. Here a three-quark operator carrying the quantum numbers of the $\Lambda(1405)$ have *implicitly* excited quark-antiquark pairs to form the five-quark molecule.

In the absence of implicit string breaking, Bali *et al.* [383] led the breakthrough in observing string breaking in QCD via a variational method with explicit $B$-meson operators. These interpolating fields mix with the traditional flux-tube operators in a matrix of correlation functions. Upon solving for the energy eigenstates, mixed states with their associated avoided level crossings are observed.

Following the notation of Ref. [383], the calculation proceeds as follows. The $Q\overline{Q}$ static quark operator connected with an optimized spatially smeared flux-tube operator $V_t(\mathbf{r},\mathbf{0})$ from position $\mathbf{0}$ to $\mathbf{r}$ at Euclidean time $t$ is

$$\overline{Q}_{(\mathbf{r},t)}\,\frac{\boldsymbol{\gamma}\cdot\mathbf{r}}{r}\,V_t(\mathbf{r},\mathbf{0})\,Q_{(\mathbf{0},t)}\,, \tag{4.3.8}$$

where $\boldsymbol{\gamma}\cdot\mathbf{r}/r$ selects the spin-symmetric state to be combined with the symmetric gluonic string $V_t(\mathbf{r},\mathbf{0})$, enabling mixing with two pseudoscalar $B$ mesons. Note, the anti-symmetric spin-combination is obtained via $\boldsymbol{\gamma}\cdot\mathbf{r}/r \to \gamma_5$ and yields the same energy levels, as both spin cases are calculated from the same Wilson loop.

Similarly, the $B\overline{B}$ meson interpolating field for a pseudoscalar $\overline{B}$ meson at $\mathbf{r}$ and a $B$ meson at $\mathbf{0}$ at Euclidean time $t$ is

$$\overline{Q}_{(\mathbf{r},t)}\,\gamma_5\,q^i_{(\mathbf{r},t)}\,\overline{q}^i_{(\mathbf{0},t)}\,\gamma_5\,Q_{(\mathbf{0},t)}\,, \tag{4.3.9}$$

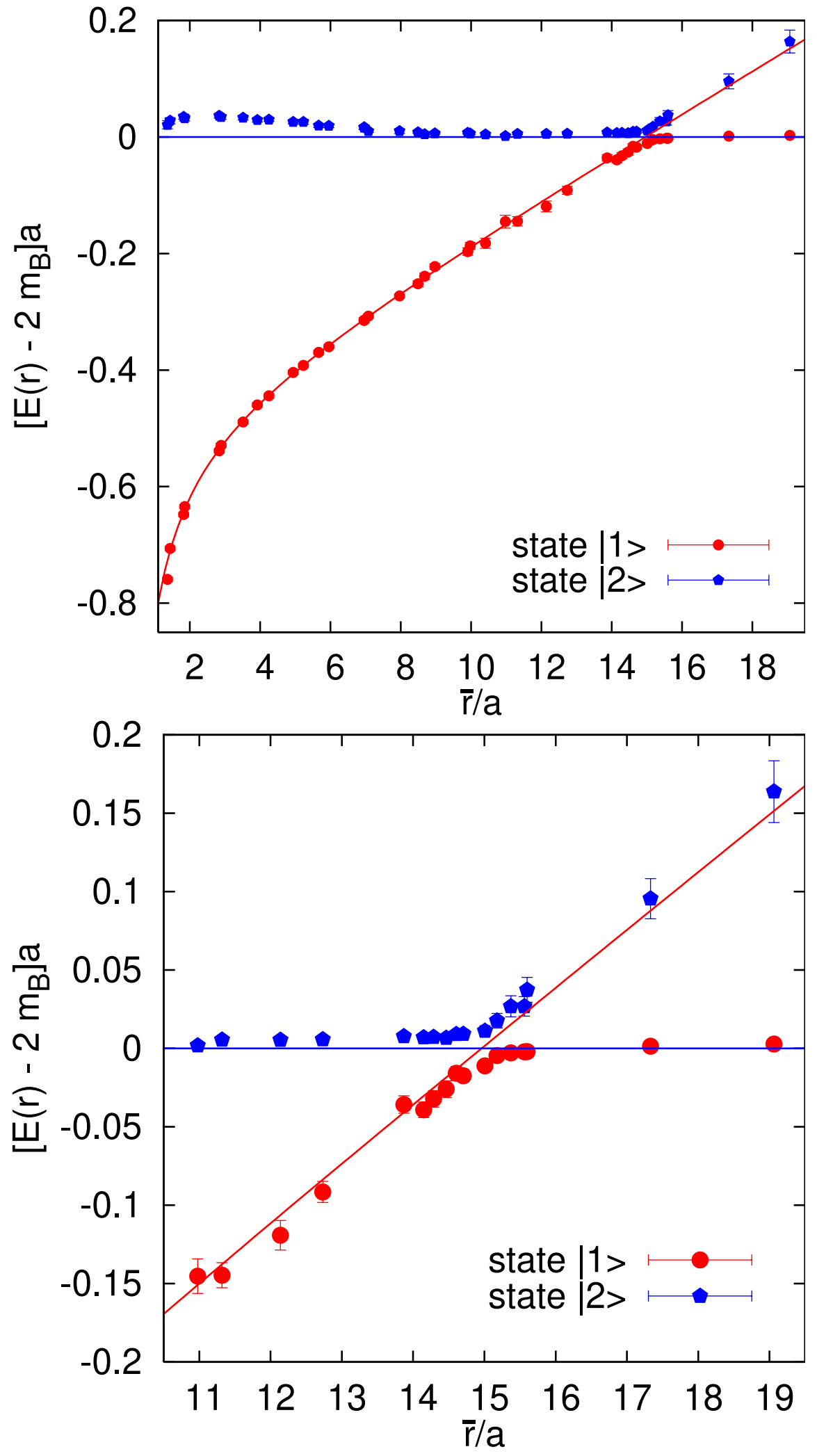

**Fig. 4.3.7** From Ref. [383], the two energy levels obtained in the variational analysis are plotted as a function of the static quark-antiquark separation $\bar{r}/a$ with lattice spacing $a \approx 0.083$ fm. Energy values are relative to twice the mass of the $B$-meson, $2\,m_B$ (horizontal line). The curve corresponds to the three parameter fit of $E_1(r) = V_0 + \sigma r - e/r$, for $0.2\,\mathrm{fm} \leq \bar{r} \leq 0.9\,\mathrm{fm} < r_c$ with $r_c \approx 15\,a \approx 1.25$ fm.. The bottom plot zooms into the avoided level crossing.

where $q^i_{(\mathbf{r},t)}$ annihilates the light-quark flavor, $i$. The four elements of the correlation matrix are obtained from the four combinations of these two operators.

Contracting the heavy-quark operators in the standard flux-tube operators provides

$$\left[\overline{Q}_{(\mathbf{r},t)} \frac{\boldsymbol{\gamma}\cdot\mathbf{r}}{r} V_t(\mathbf{r},\mathbf{0}) Q_{(\mathbf{0},t)}\right]^\dagger \overline{Q}_{(\mathbf{r},0)} \frac{\boldsymbol{\gamma}\cdot\mathbf{r}}{r} V_0(\mathbf{r},\mathbf{0}) Q_{(\mathbf{0},0)}$$
$$= 2\ \mathrm{tr}\left\{V_t^\dagger(\mathbf{r},\mathbf{0})\, U_\mathbf{r}(t,0)\, V_0(\mathbf{r},\mathbf{0})\, U_\mathbf{0}^\dagger(t,0)\right\} \equiv \Box \tag{4.3.10}$$

where the heavy-quark mass dependence has been suppressed for simplicity. Here $U_\mathbf{r}(t,0)$ denotes the product of time-oriented links at the position $\mathbf{r}$ from time 0 to $t$ and the trace is over color indices. This is the standard Wilson loop depicted by the $\mathbf{r}$ (horizontally) by $t$ (vertically) rectangle in Eq. (4.3.10).

Similarly, contracting out the quark field operators in the mixed correlator provides

$$\overline{Q}_{(\mathbf{0},t)}\, \gamma_5\, q^i_{(\mathbf{0},t)}\, \bar{q}^i_{(\mathbf{r},t)}\, \gamma_5\, Q_{(\mathbf{r},t)}\ \overline{Q}_{(\mathbf{r},0)}\, \frac{\boldsymbol{\gamma}\cdot\mathbf{r}}{r}\, V_0(\mathbf{r},0)\, Q_{(\mathbf{0},0)}$$
$$\equiv \Box = \Box \tag{4.3.11}$$

where the wavy line depicts a light quark operator. Finally, contraction of the quark operators in the $\overline{B}B$ correlator provides

$$\overline{Q}_{(\mathbf{0},t)}\, \gamma_5\, q^i_{(\mathbf{0},t)}\, \bar{q}^i_{(\mathbf{r},t)}\, \gamma_5\, Q_{(\mathbf{r},t)}\ \overline{Q}_{(\mathbf{r},0)}\, \gamma_5\, q^j_{(\mathbf{r},0)}\, \bar{q}^j_{(\mathbf{0},0)}\, \gamma_5\, Q_{(\mathbf{0},0)}$$
$$\equiv \left(\delta_{ij}\ \wr\ \wr - \Box\right). \tag{4.3.12}$$

Considering $n_f$ fermion flavors, one finally arrives at the correlation matrix

$$C(t) = \begin{pmatrix} \Box & \sqrt{n_f}\,\Box \\ \sqrt{n_f}\,\Box & -n_f\,\Box + \wr\ \wr \end{pmatrix}. \tag{4.3.13}$$

Calculation of the light-quark propagators demands the use of all-to-all techniques. Ref. [383] used a truncated eigenmode approach, complemented by a stochastic estimator technique, improved by hopping parameter acceleration. Through the use of a tuned flux-tube operator and tuned smeared-local quark propagators in the meson operators, the correlation matrix is parameterized in terms of two low lying energy eigenstates and solved.

Figure 4.3.7 illustrates the two energy levels obtained in the $n_f = 2$ analysis of Ref. [383]. Remarkably, the region of mixing is small and the energy associated with the mixing is subtle. The analysis has since been extended to $2+1$ light+strange fermion flavors in Ref. [384] where both $B$ and $B_s$ mesons participate in the mixing.

These results reflect the diverse nature of these two states. Indeed with so little overlap between the two states away from the avoided crossing region, a string-oriented system may evolve such that it maintains the string structure at very large separations [385]. In this "sudden approximation," the system evolves along the red lines of Fig. 4.3.7 providing a pathway to extraordinarily high energy excitations. The subsequent decay is considered "adiabatic" [385] where hadrons then follow the energy-eigenstate curves and split into fragments.

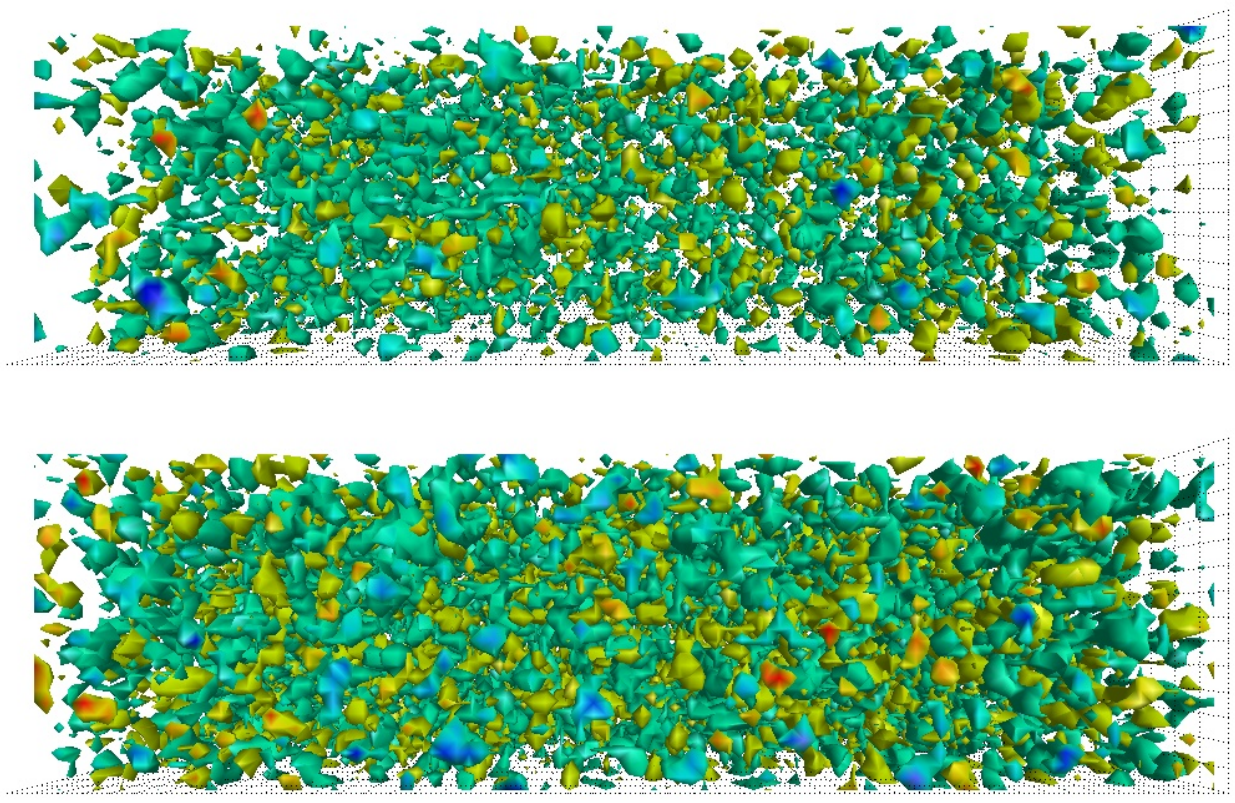

**Fig. 4.3.8** The topological charge density from Ref. [371] for the quenched (top) and the light-quark dynamical ensemble from the MILC Collaboration [386, 387], with dynamical masses of $am_{u,d} = 0.0062$, $am_{=}0.031$.

### 4.3.6 Impact of dynamical fermions on vacuum field structure

With the advent of full QCD simulations incorporating the effects of light dynamical-fermion flavors, attention turned to understanding how these light fermion loops in the vacuum changed the QCD ground-state structure. Drawing on gauge fields from the MILC collaboration [386, 387], advances in instanton-preserving smoothing algorithms [357] were deployed to reveal the impact of dynamical fermions on the topological charge density of the gauge fields [371].

The MILC simulations were performed using a one-loop Symanzik improved gauge action and an improved Kogut-Susskind quark action. Using the static quark potential, the lattice spacings were determined and tuned to be the same in all the runs to better expose differences due to dynamical fermions. At large distances, screening of the string tension was observed for light dynamical flavors [386, 387].

Figure 4.3.8 illustrates the topological-charge densities revealed following four sweeps of over-improved stout-link smearing [371]. The top illustration from quenched QCD, is qualitatively different from the lower illustration for a $2 + 1$ flavor dynamical-fermion configuration[10]. The zero modes associated with well-separated topological objects act to suppress the fermion determinant, such that the top configuration is improbable in full QCD. In the full-QCD simulations, the topological objects grow in size and number [371] to suppress the zero modes.

### 4.3.7 Center vortex structure of QCD vacuum fields

The essential, fundamentally-important, nonperturbative features of the QCD vacuum fields are the dynamical generation of mass through chiral symmetry breaking, and the confinement of quarks. But what is the fundamental mechanism of QCD that underpins these phenomena?

One of the most promising candidates is the center vortex perspective of QCD vacuum structure. While the ideas of a center-vortex dominated vacuum were laid down long ago [388–390], it wasn't until 1997 when Jeff Greensite, Manfried Faber, *et al.* demonstrated that lattice QCD techniques could be used to explore the importance of these ideas [391–396]. Indeed by the end of the millennium, the field had attracted broad interest with a comprehensive review in 2003 [397].

This perspective describes the nature of the nontrivial vacuum in terms of the most fundamental center of the gauge group. Herein our focus is on the $SU(3)$ gauge group where center vortices are characterized by the three center phases, $\sqrt[3]{1}$.

By identifying center vortices within the ground-state fields and then removing them, a deep understanding of their contributions can been developed. Removal of center vortices from the ground-state fields results in a loss of dynamical mass generation and restoration of chiral symmetry [398–400], a loss of the string tension [401–404], a suppression of the infrared enhancement in the Landau-gauge gluon propagator [402, 405–407], and the possibility that gluons are no longer confined [407].

One can also examine the role of the center vortices alone. Remarkably, center vortices produce both a linear static quark potential [401, 403, 404, 408, 409] and infrared enhancement in the Landau-gauge gluon propagator [406, 407]. The planar vortex density of center-vortex degrees of freedom scales with the lattice spacing providing a well defined continuum limit [401]. These results elucidate strong connections between center vortices and confinement.

[10] In the top illustration, one can see through the bulk of the topological charge distribution and observe the white background and the dotted lattice grid lines. This is not the case in the lower illustration where the topological charge fills out the space. Only a sprinkling of white space is observed. The quark-mass dependence of the dynamical-fermion illustration is subtle [371] indicating that the qualitative differences in the distributions comes about through the introduction of dynamical fermions in generating the configurations through Monte-Carlo methods. Not only are the objects in the quenched simulation further apart, a statistical analysis indicates there are fewer objects and the objects themselves are smaller in size when compared with the dynamical fermion distributions [371]. The physics underpinning these differences in the topological charge density distributions can be understood in terms of the modes of the Dirac operator generated by these distributions.

A connection between center vortices and instantons was identified through gauge-field smoothing [409]. An understanding of the phenomena linking these degrees of freedom was illustrated in Ref. [410]. In addition, center vortices have been shown to give rise to mass splitting in the low-lying hadron spectrum [398, 399, 411].

Still, the picture in pure $SU(3)$ gauge theory is not perfect. The vortex-only string tension obtained from pure Yang-Mills lattice studies has been consistently shown to be about $\sim 60\%$ of the full string tension. Moreover, upon removal of center vortices the gluon propagator showed a remnant of infrared enhancement [406]. In short, within the pure gauge sector, the removal of long-distance non-perturbative effects via center-vortex removal is not perfect.

Understanding the impact of dynamical fermions on the center-vortex structure of QCD ground-state fields is a contemporary focus of the center-vortex field [400, 403, 404, 407, 412, 413]. Herein, changes in the microscopic structure of the vortex fields associated with the inclusion of dynamical fermions are illustrated. The introduction of dynamical fermions brings the phenomenology of center vortices much closer to a perfect encapsulation of the salient features of QCD, confinement and dynamical mass generation through chiral symmetry breaking.

As such, it is interesting to ask, what do these center-vortex structures look like? To this end, we present visualizations of center vortices as identified on lattice gauge-field configurations. Some of these visualizations are presented as stereoscopic images. See the instructions provided in Sec. 4.3.2 for help in viewing these images.

*Center Vortex Identification*

Center vortices are identified through a gauge fixing procedure designed to bring the lattice link variables as close as possible to the identity matrix multiplied by a phase equal to one of the three cube-roots of 1. Here, the original Monte-Carlo generated configurations are considered. They are gauge transformed directly to Maximal Center Gauge [401, 414, 415]. This brings the lattice link variables $U_\mu(x)$ close to the center elements of $SU(3)$,

$$Z = \exp\left(\frac{2\pi i}{3}\, n\right) \mathbf{I}\,, \tag{4.3.14}$$

with $n = -1$, 0, or 1 enumerating the three cube roots of 1 such that the special property of $SU(3)$ matrices, $\det(Z) = 1$, is satisfied. One considers gauge transformations $\Omega$ such that,

$$\sum_{x,\mu} \left|\operatorname{tr} U_\mu^\Omega(x)\right|^2 \overset{\Omega}{\to} \max\,, \tag{4.3.15}$$

and then projects the link variables to the center

$$U_\mu(x) \to Z_\mu(x) \text{ where } Z_\mu(x) = \exp\left(\frac{2\pi i}{3}\, n_\mu(x)\right) \mathbf{I}\,. \tag{4.3.16}$$

Here, $n$ has been promoted to a field, $n_\mu(x)$, taking a value of $-1$, 0, or 1 for each link variable on the lattice. In this way, the gluon field, $U_\mu(x)$, is characterized by the most fundamental aspect of the $SU(3)$ link variable, the center, $Z_\mu(x)$. In the projection step, eight degrees of freedom are reduced to one of the three center phases. This "vortex-only" field, $Z_\mu(x)$, can be examined to learn the extent to which center vortices alone capture the essence of nonperturbative QCD.

The product of these center-projected links, $Z_\mu(x)$, around an elementary $1 \times 1$ square (plaquette) on the lattice also produces a centre element of $SU(3)$. The value describes the center charge associated with that plaquette

$$z = \prod_\Box Z_\mu(x) = \exp\left(2\pi i\, \frac{m}{3}\right), \; m = -1, 0, \text{ or } 1\,. \tag{4.3.17}$$

The most common value observed has $m = 0$ indicating that no centre charge pierces the plaquette. However, values of $m = \pm 1$ indicate that the center line of an extended three-dimensional vortex pierces that plaquette.

The complete center-line of an extended vortex is identified by tracing the presence of nontrivial center charge, $m = \pm 1$, through the spatial lattice. Figure 4.3.9 exhibits rich emergent structure in the nonperturbative QCD ground-state fields in a stereoscopic image. Here a 3D slice of the 4D space-time lattice is being considered at fixed time. Features include:

*Vortex Lines:*

The plaquettes with nontrivial center charge, characterized by $m = +1$ or $-1$, are plotted as jets piercing the center of the plaquette. Both the orientation and color of the jets reflect the value of the non-trivial center charge. Using a right-hand rule for the direction, plaquettes with $m = +1$ are illustrated by blue jets in the forward direction, and plaquettes with $m = -1$ are illustrated by red jets in the backward direction. Thus, the jets show the directed flow of $m = +1$ center charge, $z = e^{2\pi i/3}$, through spatial plaquettes. They are analogous to the line running down the center of a vortex in a fluid.

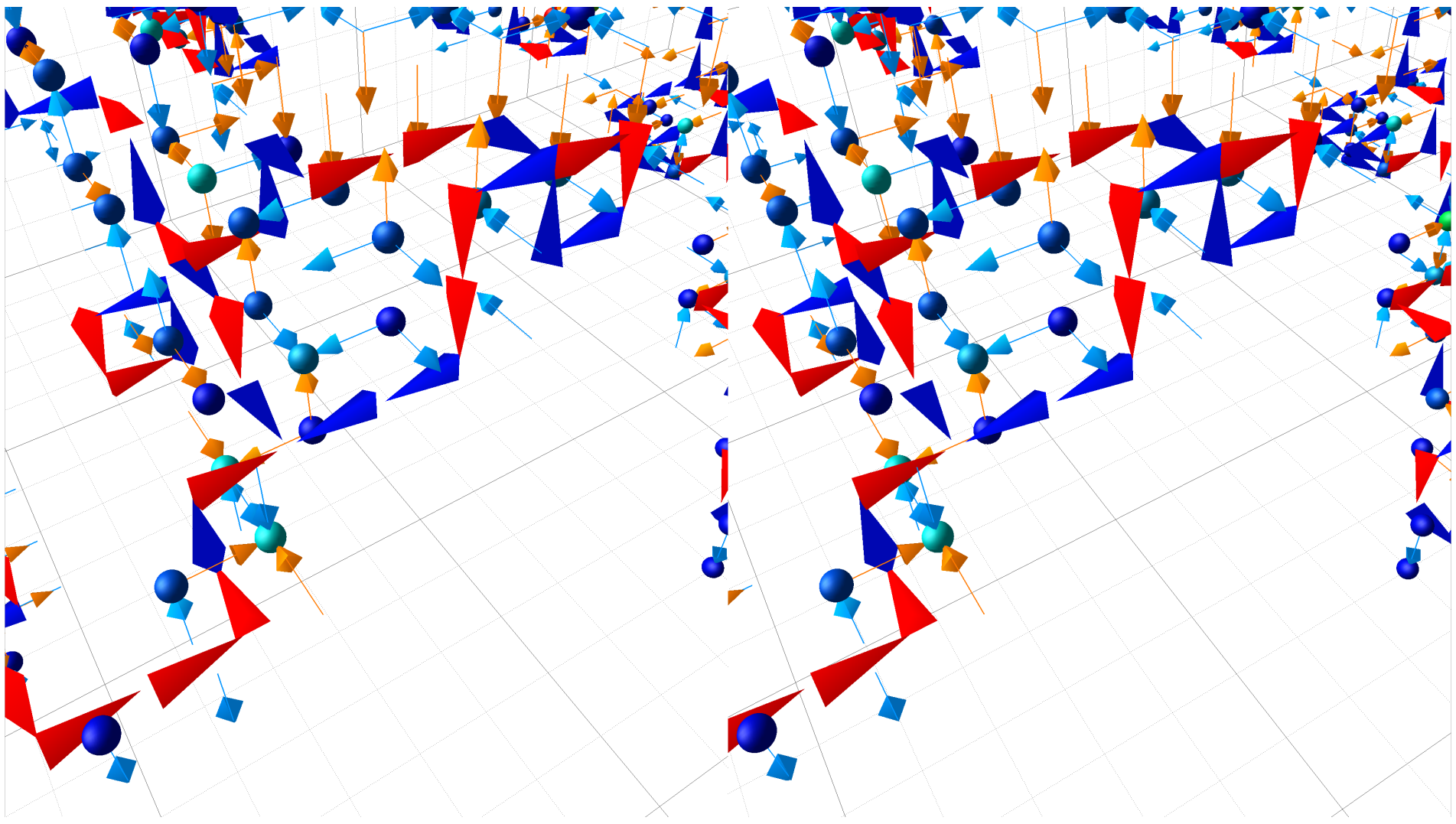

**Fig. 4.3.9** Stereoscopic image of center vortices as identified on the lattice from Ref. [412]. Vortex features including vortex lines (jets), branching points (3-jet combinations), crossing points (4 jets), indicator links (arrows) and singular points (spheres) are described in the text.

Vortices are somewhat correlated with the positions of significant topological charge density, but not in a strong manner [410]. However, the percolation of vortex structure is significant and the removal of these vortices destroys most instanton-like objects.

*Branching Points or Monopoles:*
In $SU(3)$ gauge theory, three vortex lines can merge into or emerge from a single point. Their prevalence is surprising, as is their correlation with topological charge density [410].

*Vortex Sheet Indicator Links:*
As the vortex line moves through time, it creates a vortex sheet in 4D spacetime. This movement is illustrated by arrows along the links of the lattice (shown as cyan and orange arrows in Fig. 4.3.9) indicating center charge flowing through space-time plaquettes in the suppressed time direction.

*Singular Points:*
When the vortex sheet spans all four space-time dimensions, it can generate topological charge. Lattice sites with this property are called singular points [396, 416–418] and are illustrated by spheres. The sphere color indicates the number of times the sheet adjacent to a point can generate a topological charge contribution [410].

Ref. [413] presents the first results demonstrating the impact of dynamical fermions on the center-vortex structure of QCD ground-state fields. There matched lattices were considered, one in pure-gauge and the other a 2+1-flavor dynamical-fermion lattice from the PACS-CS Collaboration [419]. These $32^3 \times 64$ lattice ensembles employ a renormalisation-group improved Iwasaki gauge action and non-perturbatively $\mathcal{O}(a)$-improved Wilson quarks, with $C_{\rm SW} = 1.715$.

The lightest $u$- and $d$-quark-mass ensemble identified by a pion mass of 156 MeV [419] is presented here. The scale is set using the Sommer parameter with $r_0 = 0.4921$ fm providing a lattice spacing of $a = 0.0933$ fm [419]. A matched $32^3 \times 64$ pure-gauge ensemble using the same improved Iwasaki gauge action with a Sommer-scale spacing of $a = 0.100$ fm was created [413] to enable comparisons with the PACS-CS ensembles.

The center-vortex structure of pure-gauge and dynamical fermion ground-state vacuum fields is illustrated in Fig. 4.3.10 from Ref. [413], where interactive 3D plots of this structure which can be activated in Adobe Reader. The impact of dynamical fermions on the center-vortex structure is much more significant than that discussed in Sec. 4.3.6.

In both illustrations, the vortex structure is dominated by a single large percolating structure. Whereas small loops will tend to pierce a Wilson loop twice with zero effect, it is this extended structure that gives rise to a net vortex piercing of a Wilson loop and the generation of an area law associated with confinement. These two illustrations are representative of the ensemble in that the vortex structure is typically dominated by a single large percolating cluster.

Closer inspection reveals a continuous flow of center charge, often emerging or converging to monopole or anti-monopole vertices where three jets emerge from or converge to a point. These are referred to as branching points, as a +1 center charge flowing out of a vertex is equivalent to +2 center charge flowing into the vertex and subsequently branching to two +1 jets flowing out of the vertex.

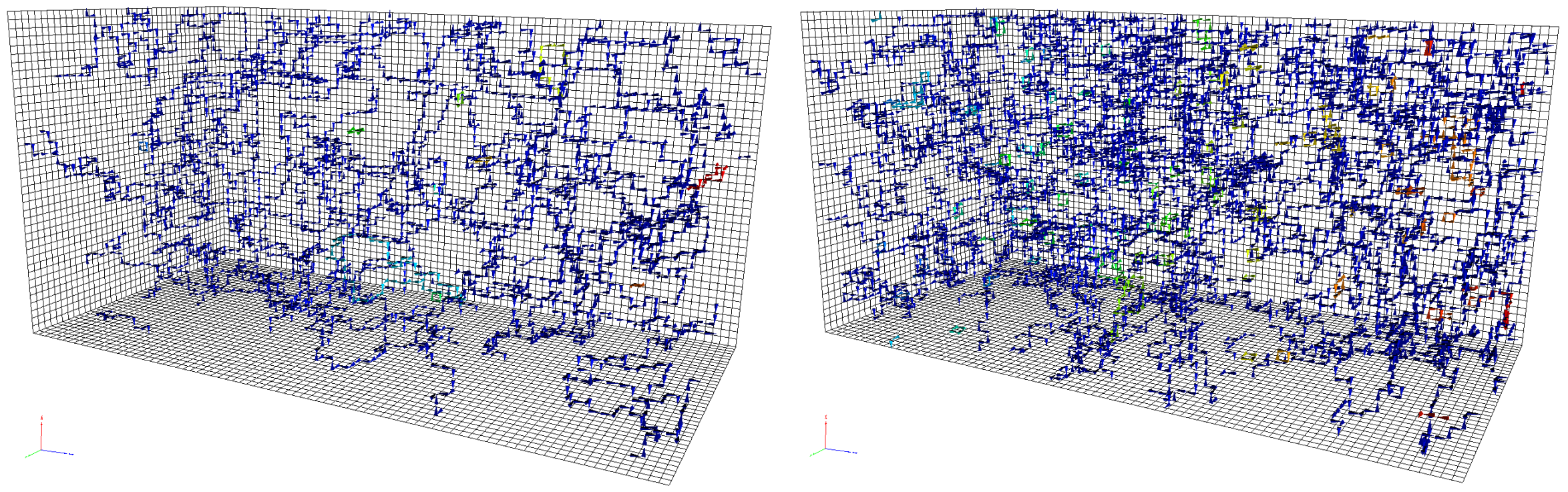

**Fig. 4.3.10** From Ref. [413], the center-vortex structure of a ground-state vacuum field configuration in pure SU(3) gauge theory (left) is compared with a field configuration in dynamical 2+1 flavor QCD corresponding to $m_\pi = 156$ MeV (right). The flow of +1 center charge through the gauge fields is illustrated by the jets. Blue jets are used to illustrate the single percolating vortex structure, while other colors illustrate smaller structures.

With the introduction of dynamical fermions, the structure becomes more complex, both in the abundance of vortices and branching points. The average number of vortices composing the primary cluster in these $32^2 \times 64$ spatial slices roughly doubles from $\sim 3,000$ vortices in the pure gauge theory to $\sim 6,000$ in full QCD. Still, there are $32^2 \times 64 \times 3 = 196,608$ spatial plaquettes on these lattices and thus the presence of a vortex is a relatively rare occurrence.

By counting the number of vortices between branching points one discovers the distribution is exponential, indicating a constant branching probability. This probability is higher in full QCD by a ratio of $\sim 3/2$.

With an understanding of the impact of dynamical-fermion degrees of freedom on the center-vortex structure of ground-state vacuum fields, attention has turned to understanding the impact on confinement. In a variational analysis of standard Wilson loops composed of several spatially-smeared sources to isolate the ground state potential, the static quark potential has been calculated on three ensembles including the original untouched links, $U_\mu(x)$, the vortex-only links, $Z_\mu(x)$, and vortex-removed links, $Z^\dagger_\mu(x)\, U_\mu(x)$ [407] where the multiplication of the conjugate of the centre-projected field ensure all plaquettes have $z = 0$.

For the original untouched configurations, the static quark potential is expected to follow a Cornell potential

$$V(r) = V_0 - \frac{\alpha}{r} + \sigma\, r\,. \tag{4.3.18}$$

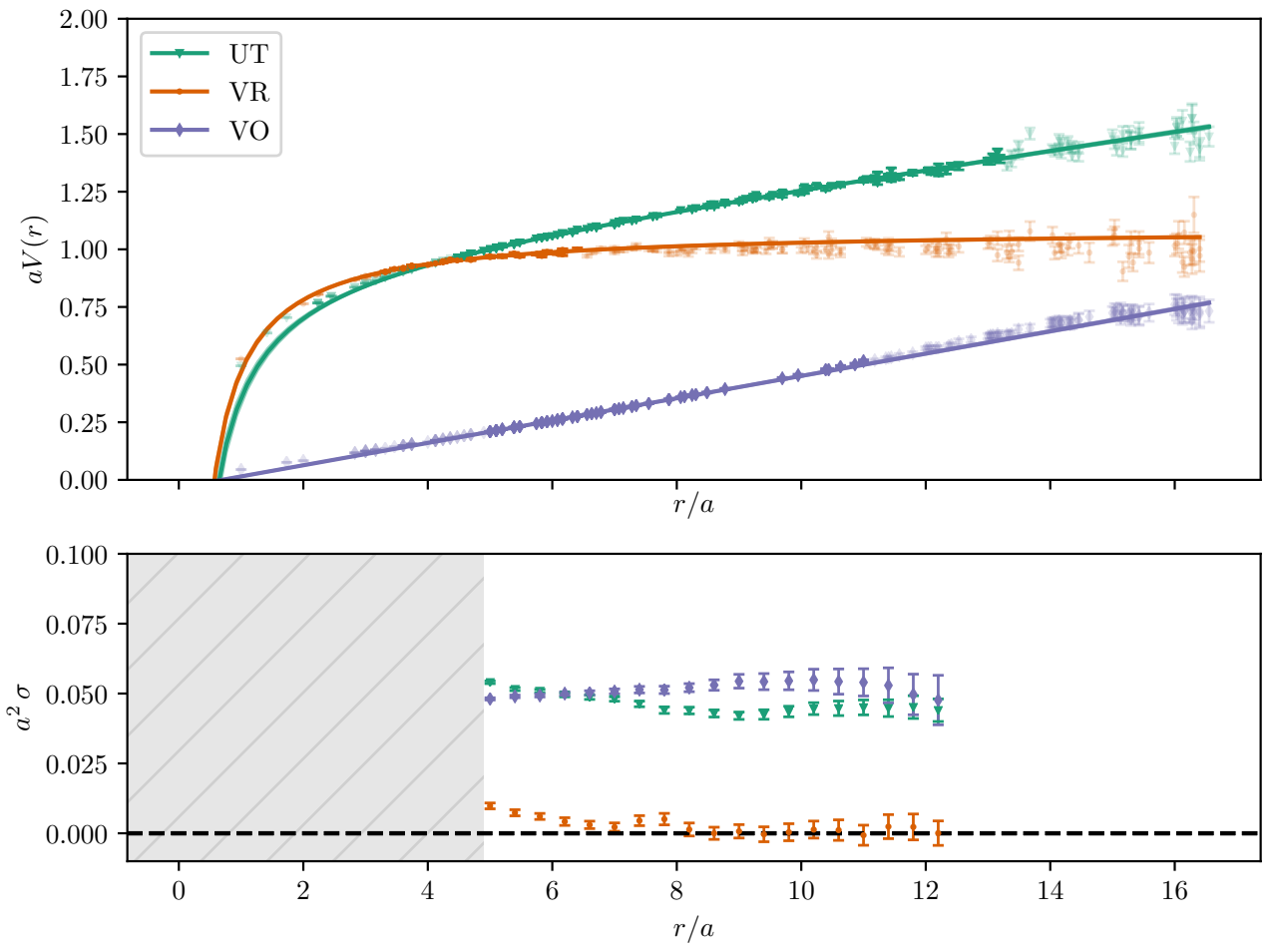

**Fig. 4.3.11** The static quark potential, as presented in Ref. [403], calculated on the vortex-modified dynamical-fermion ensemble, corresponding to a pion mass of 156 MeV. The lower plot shows the local slope from linear fits of the potentials in the upper plot over a forward-looking window from $r$ to $r + 4a$.

As center vortices are anticipated to encapsulate the non-perturbative long-range physics, the vortex-only results should give rise to a linearly rising potential. On the other hand, the vortex-removed results are expected to capture the short-range Coulomb behavior. Figure 4.3.11 from Ref. [407] illustrates the static quark potentials obtained from these three ensembles for the dynamical 2+1-flavor ensemble with a pion mass of 156 MeV [419].

Qualitatively, center vortices account for the long-distance physics. The removal of center vortices completely removes the confinement potential. And while the vortex-only string tension is typically 60 % of the original string tension in the pure gauge sector, the introduction of dynamical fermions has improved the vortex-only phenomenology significantly. Vortices alone capture both the screening of the pure-gauge string ten-sion and the full string tension of the original untouched ensemble. This result is associated with the significant modification of the center-vortex structure of ground-state vacuum fields induced by dynamical fermions.

The improved separation of perturbative and nonperturbative physics through the consideration of vortex-removed and vortex-only ensembles in full QCD is also manifest in the nonperturbative gluon propagator [407]. This time vortex removal removes the infrared enhancement of the gluon propagator, leaving a tree-level structure. Indeed the vortex-removed Euclidean correlator remains positive definite, admitting the possibility of a positive-definite spectral density associated with free gluons. The vortex-only ensembles capture the infrared enhancement of the gluon propagator and the screening of this enhancement in full QCD [407].

Similarly, dynamical mass generation in the nonperturbative quark propagator is suppressed under vortex removal in full QCD while the vortex-only ensemble provides dynamical mass generation [400]. While explicit chiral symmetry breaking through the quark mass, leaves a remnant of dynamical mass generation, it is anticipated that for sufficiently light current quark masses, chiral symmetry will be restored [399] and dynamical mass generation will be completely eliminated in the vortex-removed theory.

In summary, center-vortex structure is complex. Each ground-state configuration is dominated by a long-distance percolating center-vortex structure. In $SU(3)$ gauge field theory, a proliferation of branching points is observed, with further enhancement as light dynamical fermion degrees of freedom are introduced in simulating QCD. There is an approximate doubling in the number of non-trivial center charges in the percolating vortex structure as one goes from the pure-gauge theory to full QCD. Increased complexity in the vortex paths is also observed as the number of branching points is significantly increased with the introduction of dynamical fermions. In short, dynamical-fermion degrees of freedom radically alter the center-vortex structure of the ground-state vacuum fields. This change in structure acts to improve the phenomenology of center vortices better reproducing the string tension, dynamical mass generation and better removing nonperturbative physics under vortex removal. This represents a significant advance in the ability of center vortices to capture the salient nonperturbative features of QCD.

### 4.3.8 Summary

In the 50 years following the advent of QCD, the complexity of the nontrivial QCD vacuum has been exposed. Many theoretical ideas have been created and developed to explain the salient features of this nontrivial vacuum and their exploration continues. Numerical experiments within the realm of lattice QCD have been particularly useful in testing the veracity of the theoretical ideas proposed. Today, these numerical experiments are exploring the ideas of instanton-dyons and center-vortices as the essential features of QCD vacuum structure, confining color and dynamically generating mass through dynamical chiral symmetry breaking. The results are fascinating, and encourage further exploration of the essence of QCD vacuum structure.

## 4.4 QCD at non-zero temperature and density

**Frithjof Karsch**

### 4.4.1 QCD Thermodynamics on Euclidean lattices

The path integral formulation of QCD can easily be applied to cases of non-vanishing temperature ($T$) and other external control parameters, e.g. the chemical potentials ($\mu_f$) that couple to the conserved currents of quark-flavor number.

Using the lattice regularization scheme of QCD, introduced by K. Wilson [80], QCD thermodynamics is formulated on Euclidean space-time lattices of size $N_\sigma^3 N_\tau$ where, for a given lattice spacing ($a$), the lattice extent in Euclidean time controls the temperature $T = 1/N_\tau a$ and the spatial extent is related to the volume of the thermodynamic system, $V = (N_\sigma a)^3$. The chemical potentials enter directly in the fermion matrices, $M_f$, which arise from the QCD Lagrangian after integrating out the fermion fields.

Bulk thermodynamics can then be derived from the lattice regularized partition function,

$$Z = \int \prod_{x_0=1}^{N_\tau} \prod_{x_i=1}^{N_\sigma} \prod_{\hat{\nu}=0}^{3} \mathcal{D}U_{x,\hat{\nu}} \, e^{-S_G} \times \prod_{f=u,d,s..} \det M_f(m_f, \mu_f) \ , \quad (4.4.1)$$

where $x = (x_0, \vec{x})$ labels the sites of the 4-dimensional lattice, $S_G$ denotes the gluonic part of the Euclidean action, which is expressed in terms of $SU(3)$ matrices $U_{x,\hat{\nu}}$ and $M_f$ is the fermion matrix for quark flavor $f$. It is a function of quark mass, $m_f$ and flavor chemical potential $\hat{\mu}_f \equiv \mu_f/T$. Basic bulk thermodynamic observables (equation of state, susceptibilities, etc.) can then be obtained from the logarithm of the partition function, $Z$, which defines the pressure, $P$, as

$$P/T = \frac{1}{V} \ln Z(T, V, \vec{\mu}, \vec{m}) \ . \quad (4.4.2)$$

Applying standard thermodynamic relations one obtains other observables of interest; e.g. the energy density is related to the trace anomaly of the energy-momentum tensor, $\Theta^{\mu\mu}$,

$$\frac{\Theta^{\mu\mu}}{T^4} = \frac{\epsilon - 3P}{T^4} \equiv T\frac{\partial P/T^4}{\partial T} \ , \tag{4.4.3}$$

and the conserved charge densities are obtained as,

$$\frac{n_X}{T^3} = \frac{\partial P/T^4}{\partial \hat{\mu}_X} \ , \ \ X = B,\ Q,\ S \ . \tag{4.4.4}$$

While the framework of lattice QCD provides easy access to QCD thermodynamics at vanishing values of the chemical potentials, major difficulties arise at $\mu_f \neq 0$. The fermion determinants, $\det M_f(m_f, \mu_f)$, are no longer positive definite when the real part of the chemical potential is non-zero, $\mathrm{Re}\hat{\mu}_f \neq 0$. This includes the physically relevant case of strictly real chemical potentials. The presence of a complex valued integrand in the path integral makes the application of standard Monte Carlo techniques, which rely on a probabilistic interpretation of integration measures, impossible. The two most common approaches to circumvent this problem are to either (i) perform numerical calculations at imaginary values of the chemical potential, $\hat{\mu}_f^2 < 0$ [420, 421], or to (ii) perform Taylor series expansions around $\hat{\mu}_f = 0$ [422, 423]. In the former case numerical results need to be analytically continued to real values of $\mu_f$. In the latter case the QCD partition function is written as,

$$P/T^4 = \frac{1}{VT^3}\ln Z(T,V,\vec{\mu}) = \sum_{i,j,k=0}^{\infty} \frac{\chi_{ijk}^{BQS}}{i!j!k!}\hat{\mu}_B^i \hat{\mu}_Q^j \hat{\mu}_S^k \ , \tag{4.4.5}$$

with $\chi_{000}^{BQS} \equiv P(T,V,\vec{0})/T^4$ and expansion coefficients,

$$\chi_{ijk}^{BQS}(T) = \left. \frac{\partial P/T^4}{\partial \hat{\mu}_B^i \partial \hat{\mu}_Q^j \partial \hat{\mu}_S^k} \right|_{\hat{\mu}=0} \ , \tag{4.4.6}$$

can be determined in Monte Carlo simulations performed at $\hat{\mu}_X = 0$.

The phase structure of QCD can be explored using suitable observables that are sensitive to the spontaneous breaking and the eventual restoration of global symmetries. They can act as order parameters in certain limits of the parameter space spanned by the quark masses. In QCD exact symmetries exist either in the chiral limit, *i.e.* at vanishing values of $n_f$ quark masses, or for infinitely heavy quarks, *i.e.* in pure $SU(N_c)$ gauge theories, with $N_c$ denoting the number of colors.

In order to probe the restoration of the global chiral symmetries one analyzes the chiral condensate and its susceptibilities,

$$\langle\bar{\chi}\chi\rangle_f = \frac{T}{V}\frac{\partial}{\partial m_f}\ln Z = \frac{T}{V}\langle \mathrm{Tr} M_f^{-1}\rangle \ , \tag{4.4.7}$$

$$\chi_m^{fg} = \frac{\partial\langle\bar{\chi}\chi\rangle_f}{\partial m_g} \ , \quad \chi_t^f = T\frac{\partial\langle\bar{\chi}\chi\rangle_f}{\partial T} \ . \tag{4.4.8}$$

The former is an order parameter for the restoration of the $SU(n_f)_L \times SU(n_f)_R$ chiral flavor symmetry of QCD and distinguishes, in the limit of vanishing quark masses, a symmetry broken phase at low temperature from a chiral symmetry restored phase at high temperature,

$$\lim_{m_\ell \to 0}\langle\bar{\chi}\chi\rangle_\ell \begin{cases} > 0 & T < T_\chi \\ = 0 & T \geq T_\chi \end{cases} \ . \tag{4.4.9}$$

Similarly one considers the Polyakov loop $\langle L\rangle$ and its susceptibility $\chi_L$,

$$\langle L\rangle = \frac{1}{N_\sigma^3}\langle\sum_{\vec{x}} \mathrm{Tr} L_{\vec{x}}\rangle \ , \qquad L_{\vec{x}} = \prod_{x_0=1}^{N_\tau} U_{(x_0,\vec{x}),\hat{0}} \ ,$$
$$\chi_L = N_\sigma^3\left(\langle L^2\rangle - \langle L\rangle^2\right) \ , \tag{4.4.10}$$

to probe the breaking and restoration of the global $Z(N_c)$ center symmetry of pure $SU(N_c)$ gauge theories; *i.e.* $SU(N_c)$ gauge theories at finite temperature, formulated on Euclidean lattices, are invariant under global rotation of all temporal gauge field variables, $U_{\vec{x},\hat{0}} \ \to \ zU_{\vec{x},\hat{0}}$, with $z \in Z(N_c)$. The Polyakov loop expectation value vanishes as long as this center symmetry is not spontaneously broken.

The Polyakov loop expectation value also reflects the long distance behavior of Polyakov loop correlation functions,

$$|\langle L\rangle|^2 \equiv \lim_{|\vec{x}|\to\infty} G_L(\vec{x}) \begin{cases} = 0 \Leftrightarrow F_q = \infty \ , T \leq T_d \\ > 0 \Leftrightarrow F_q < \infty \ , T > T_d \end{cases} \tag{4.4.11}$$

where

$$G_L(\vec{x}) = \mathrm{e}^{-F_{\bar{q}q}(\vec{x},T)} = \langle \mathrm{Tr} L_{\vec{0}} \mathrm{Tr} L_{\vec{0}}^\dagger\rangle \tag{4.4.12}$$

is the correlation function of two Polyakov loops. It denotes the change in free energy (excess free energy, $F_{\bar{q}q}$), that is due to the presence of two static quark sources introduced in a thermal medium. At zero temperature this free energy reduces to the potential between static quark sources.

At least in the case of pure gauge theories this provides a connection between the confinement-deconfinement phase transition and the breaking of a global symmetry, the $Z(N_c)$ center symmetry of the $SU(N_c)$ gauge group. This symmetry, however, is explicitly broken in the presence of dynamical quarks with mass $m_f < \infty$. Unlike chiral symmetry restoration, deconfinement thus is not expected to be related to a phase transition in QCD with physical quark masses. Nonetheless, the consequences of deconfinement, related to the dissolution of hadronic bound states, becomes clearly visible in many thermodynamic observables.

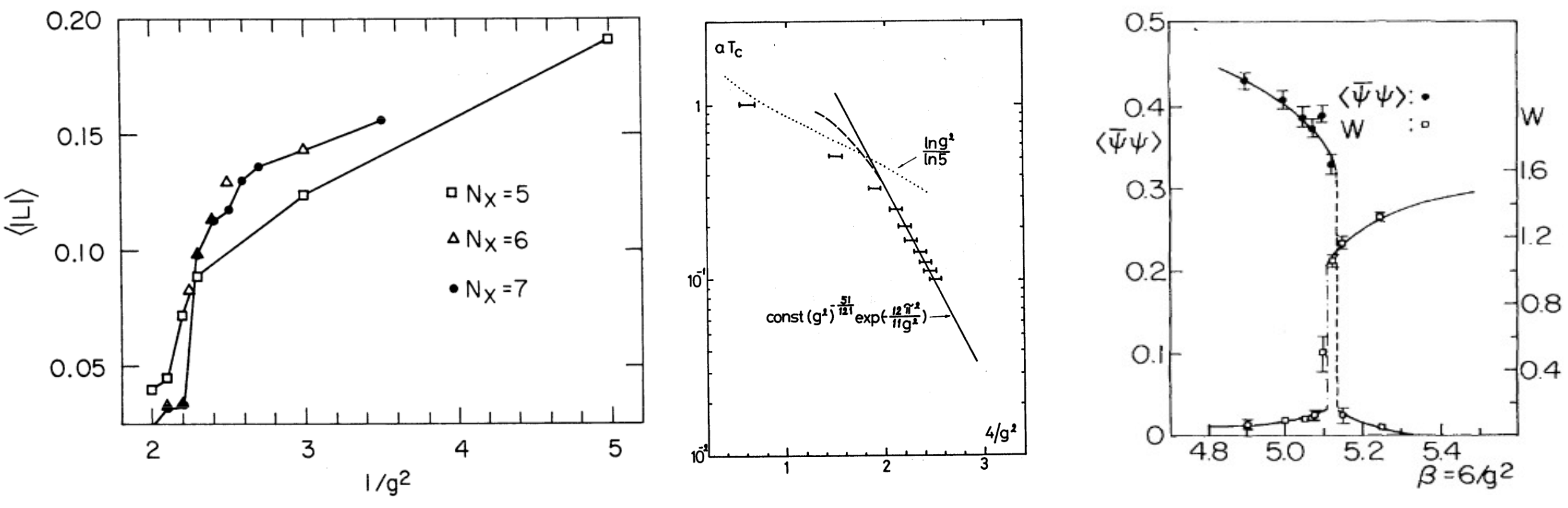

**Fig. 4.4.1** First evidence for the existence of a deconfinement phase transition in $SU(2)$ gauge theories using the Polyakov loop expectation value as an order parameter (left) [424] and a first extrapolation of the phase transition temperature to the continuum limit (middle) [425] . The right hand figure shows a first comparison of the temperature dependence of the Polyakov loop ($W \equiv \langle |L| \rangle$) and chiral condensate ($\langle \bar{\psi}\psi \rangle$) order parameters in a SU(3) gauge theory [426] .

### 4.4.2 Early lattice QCD calculations at non-zero temperature

Almost immediately after the formulation of QCD as the theory of strong interaction physics, its consequences for strong interaction matter at non-zero temperature were examined [427, 428]. It rapidly became obvious that fundamental properties of QCD, confinement and asymptotic freedom on the one hand [428, 429], and chiral symmetry breaking on the other hand [430], are likely to trigger a phase transition in strong interaction matter that separates a phase being dominated by hadrons as the relevant degrees of freedom from that of almost free quarks and gluons. The notion of a quark-gluon plasma was coined at that time [431].

Soon after these early, conceptually important developments it was realized that the formulation of QCD on discrete space-time lattices, which was introduced by K. Wilson as a regularization scheme in QCD [80], also provides a powerful framework for the analysis of non-perturbative properties of strong interaction matter through Monte-Carlo simulations [328]. This led to a first determination of a phase transition temperature in $SU(2)$ [424, 425] and $SU(3)$ [426, 432, 433] gauge theories, and a first determination of the equation of state of purely gluonic matter [434, 435]. The interplay between deconfinement on the one hand and chiral symmetry restoration on the other hand also was studied [426] early on and the question whether or not these two aspects of QCD may lead to two distinct phase transitions in QCD has been considered ever since. Some results from these first lattice QCD studies of the thermodynamics of strong interaction matter are shown in Fig. 4.4.1.

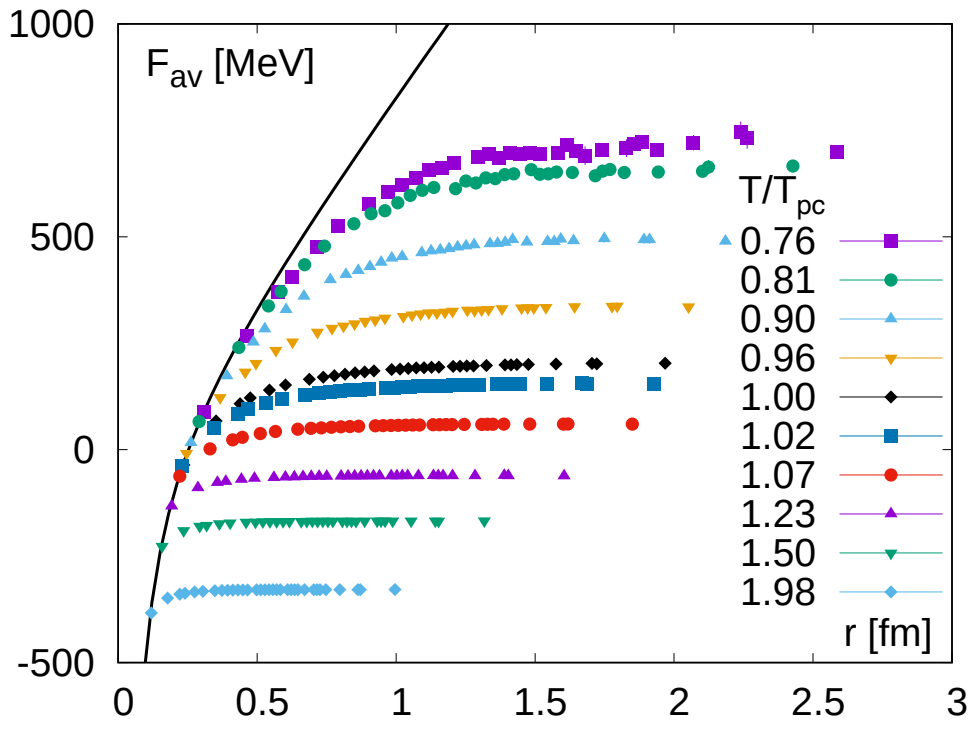

**Fig. 4.4.2** The so-called color averaged, heavy quark free energy ($F_{\rm av} \equiv F_{\bar{q}q}$) in the vicinity of the pseudo-critical transition temperature ($T_{pc}$) in 2-flavor QCD [436]. Results shown cover a temperature range from $T/T_{pc} \simeq 0.75$ to $T/T_{pc} \simeq 2$.

At physical values of the quark masses, neither deconfinement nor the effective restoration of chiral symmetry leads to a true phase transition. Still the transition from the low temperature hadronic to the high temperature partonic phase of QCD is clearly visible in the pseudo-critical behavior of the heavy quark free energy and the chiral condensate respectively. Some recent results on these observables, obtained in simulations of QCD with light, dynamical quark degrees of freedom, are shown in Figs. 4.4.2 and 4.4.3.

### 4.4.3 Global symmetries and the QCD phase diagram

The early studies of QCD thermodynamics made it clear that universality arguments and renormalization group techniques, successfully developed in condensed matter physics and applied in statistical physics to the analysis of phase transitions, also can be carried over to the analysis of the phase structure of quantum field theories [437, 438]. The renormalization group based arguments for the existence of a second order phase transition in the universality class of the 3-d Ising model in a $SU(2)$ gauge theory, and a first order transition for the $SU(3)$ color group of QCD [439] have been confirmed by detailed lattice QCD calculations [440, 441].

In the presence of $n_f$ light, dynamical quarks, distinguished by a flavor quantum number, it is the chiral symmetry of QCD that triggers the occurrence of phase transitions [430]. In addition to a global $U(1)$ symmetry that reflects the conservation of baryon number and is unbroken at all temperatures and densities, the massless QCD Lagrangian is invariant under the symmetry group

$$U(1)_A \times SU(n_f)_L \times SU(n_f)_R \; . \qquad (4.4.13)$$

The $SU(n_f)_L \times SU(n_f)_R$ symmetry corresponds to chiral rotations of $n_f$ massless quark fields in flavor space. This symmetry is spontaneously broken at low temperatures, giving rise to $n_f^2 - 1$ massless Goldstone modes, which for $n_f = 2$ are the three light pions of QCD. They have a non-vanishing mass only because of the explicit breaking of chiral symmetry by a mass term in the QCD Lagrangian that couples to the chiral order parameter field $\bar{\chi}_f \chi_f$. The axial $U(1)_A$ group corresponds to global rotations of quark fields for a given flavor $f$. Although it is an exact symmetry of the classical Lagrangian, it is explicitly broken in the quantized theory. This explicit breaking of a global symmetry, arising from fluctuations on the quantum level, is known as the $U(1)_A$ anomaly.

The renormalization group based analysis of the chiral phase transition, performed by Pisarski and Wilczek [430], made it clear that the chiral phase transition is sensitive to the number of light quark flavors that become massless. Furthermore, it has been argued in [430] that the order of the transition may be sensitive to the magnitude of the axial anomaly at non-zero temperature, which is closely related to the temperature dependence of topological non-trivial field configurations.

Although it was generally expected that the chiral phase transition in 3-flavor QCD becomes a first order phase transition in the chiral limit [430], there is currently no direct evidence for this from lattice QCD calculations. In fact, the current understanding is that the chiral phase transition is second order for all $n_f \leq 6$ [444].

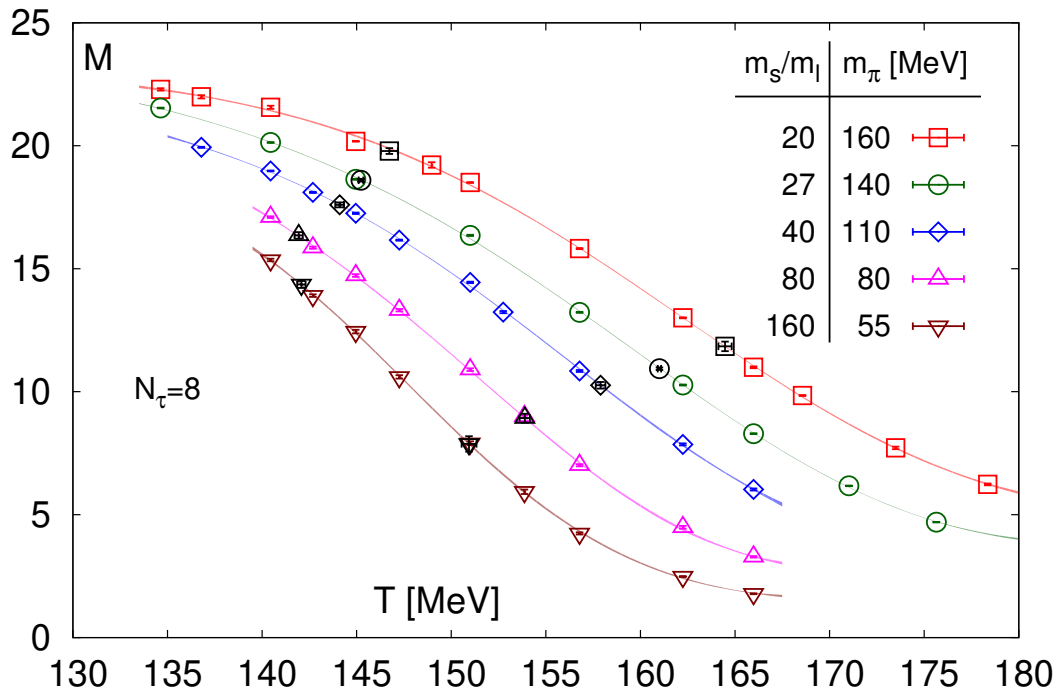

**Fig. 4.4.3** Quark mass dependence of chiral order parameter, $M$, defined in Eq. 4.4.14 for QCD with two degenerate light quark masses and a strange quark mass tuned to its physical value. Shown are results from calculations on lattices with temporal extent $N_\tau = 8$ performed for several values of the light quark masses [442, 443]. The light quark masses, $m_\ell$, are expressed in units of the strange quark mass, $H = m_\ell/m_s$. In the figure we give $1/H = m_s/m_\ell$ together with the corresponding values of the Goldstone pion mass.

In Fig. 4.4.4 (top) we show the original version of the QCD phase diagram in the plane of two degenerate light ($m_\ell$) and strange ($m_s$) quark masses, proposed in 1990 [445], together with an updated version from 2021 [444]. Here $m_\ell$ denotes the two degenerate up and down quark masses, $m_\ell \equiv m_u = m_d$. This sketch of our current understanding of the 3-flavor phase diagram also is supported by the increasing evidence for a non-singular crossover transition in QCD with physical light and strange quark masses and the absence of any evidence for a first order phase transition at lighter-than-physical values of the light and strange quark masses [444, 446]. In the chiral limit, *i.e.* for vanishing up and down quark masses[11], a second order phase transition will then occur.

### 4.4.4 The chiral phase transition at non-vanishing chemical potential

The occurrence of the chiral phase transition is signaled by the vanishing of the light quark chiral condensate. In order to remove multiplicative and additive divergences in $\langle\bar{\chi}\chi\rangle_\ell$ one considers instead the order parameter $M$ which is a combination of light and strange quark condensates,

$$M = 2\left(m_s\langle\bar{\psi}\psi\rangle_\ell - m_\ell\langle\bar{\psi}\psi\rangle_s\right)/f_K^4 \; , \qquad (4.4.14)$$

[11] Lattice QCD studies of the (2+1)-flavor phase diagram generally are performed with degenerate up and down quark masses.

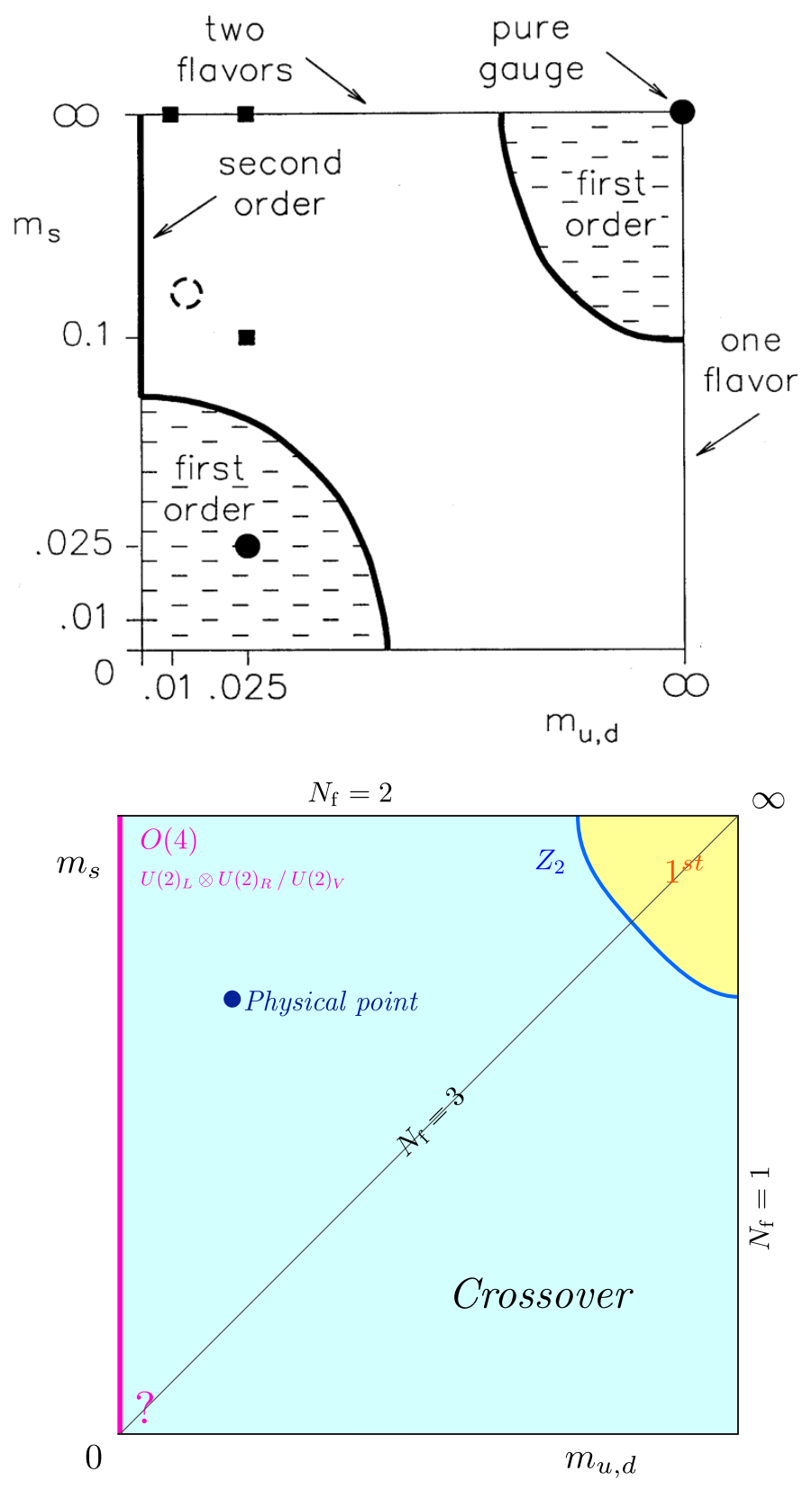

**Fig. 4.4.4** Sketch of the phase diagram of QCD in the plane of degenerate, light up and down quark masses and a strange quark mass (Columbia plot). The figure shows the original version from 1990 [445] (top) and an updated version from 2021 [444] (bottom).

and its derivative with respect to the light quark masses, i.e. the chiral susceptibility $\chi_M$

$$\chi_M = m_s \left( \frac{\partial M}{\partial m_u} + \frac{\partial M}{\partial m_d} \right)_{m_u = m_d \equiv m_\ell} . \tag{4.4.15}$$

Here the kaon decay constant $f_K = 156.1(9)/\sqrt{2}$ MeV, has been used to introduce a dimensionless order parameter. The scaling behavior of $M$ and $\chi_M$, have been used to characterize the chiral phase transition,

$$M \underset{m_\ell \to 0}{\sim} \begin{cases} A \left( \frac{T_c^0 - T}{T_c^0} \right)^\beta & , \ T < T_c^0 \\ 0 & , \ T \geq T_c^0 \end{cases} \tag{4.4.16}$$

$$\chi_M \underset{m_\ell \to 0}{\sim} \begin{cases} \infty & , \ T \leq T_c^0 \\ C \left( \frac{T - T_c^0}{T_c^0} \right)^{-\gamma} & , \ T > T_c^0 \end{cases} \tag{4.4.17}$$

where $\beta$, $\gamma$ are critical exponents.

We note that the low temperature behavior of the order parameter susceptibility, $\chi_M$, is quite different from that known, for instance, in the 3-$d$ Ising model. The susceptibility diverges in the massless limit at all

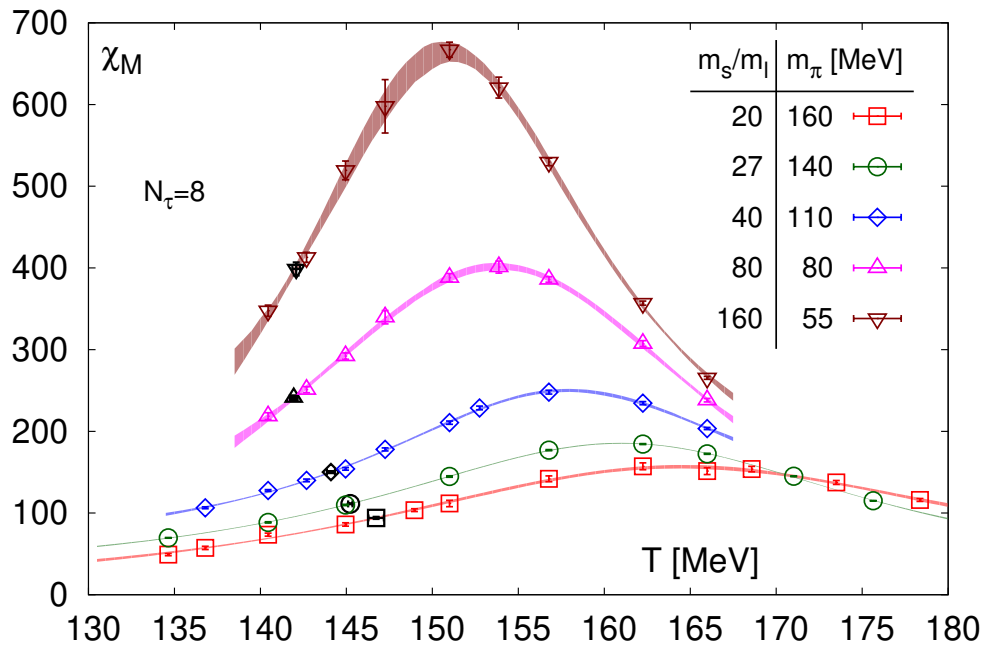

**Fig. 4.4.5** same as Fig. 4.4.3 but for the chiral susceptibility.

values of the temperature, $T \leq T_c^0$. This is a consequence of the breaking of a continuous rather than a discrete symmetry. The former gives rise to Goldstone modes, the pions in QCD, which contribute to the chiral condensate and as such to the order parameter $M$, *i.e.*,

$$M \sim a(T)\sqrt{m_\ell} \ , \ T < T_c^0 . \tag{4.4.18}$$

As a consequence the chiral susceptibility diverges below $T_c^0$, $\chi_M \sim 1/\sqrt{m_\ell}$, while at $T_c^0$ its divergence is controlled by the critical exponent $\delta = 1 + \gamma/\beta$,

$$\chi_M \sim \begin{cases} H^{-1/2} & T < T_\chi \\ H^{1/\delta - 1} & T = T_\chi \end{cases} , \tag{4.4.19}$$

with $H = m_\ell/m_s$. As $1 - 1/\delta > 1/2$ in all relevant universality classes $\chi_M$ develops a pronounced peak at small, but non-zero values of the quark masses,

$$\chi_M^{peak} \equiv \chi_M(T_{pc}(H)) \sim H^{1/\delta - 1} \ , \ H = m_\ell/m_s . \tag{4.4.20}$$

The location of such a peak in either $\chi_M$ or similarly in $T\partial M/\partial T$, defines pseudo-critical temperatures, $T_{pc}(H)$, which converge to the unique chiral phase transition, $T_c^0$, at $H = 0$. Some results on the quark mass dependence of $M$ and $\chi_M$ are shown in Figs. 4.4.3 and 4.4.5, respectively. A scaling analysis of these observables, performed in [442], led to the determination of the chiral phase transition temperature [442],

$$T_c^0 = 132^{+3}_{-6} \text{ MeV} . \tag{4.4.21}$$

Similar results have also been obtained in [447] where a quite different discretization scheme for the fermion sector of QCD has been used.

For physical light and strange quark masses, corresponding to $H \simeq 1/27$, one finds as a pseudo-critical temperature [448],

$$T_{pc} = 156.5(1.5) \text{ MeV} , \tag{4.4.22}$$

which is in good agreement with other determinations of pseudo-critical temperatures in $(2+1)$-flavor QCD [449–451].

The chiral symmetry group $SU(2)_L \times SU(2)_R$ is isomorphic to the rotation group $O(4)$. It thus is expected that the chiral phase transition for two vanishing light quark masses is in the same universality class as 3-$d$, $O(4)$ symmetric spin models. In fact, the rapid rise of $\chi_M$, shown in Fig. 4.4.5, is consistent with a critical exponent in this universality class, $\delta = 4.824$ [452]. However, a precise determination of this exponent in 2-flavor QCD is not yet possible. This leaves open the possibility for other symmetry breaking patterns and other universality classes playing a role in the chiral limit of 2-flavor QCD [453]. In fact, the discussion of such possibilities is closely related to the yet unsettled question concerning the influence of the axial $U(1)_A$ symmetry on the chiral phase transition. For a recent review on this question see, for instance [454].

*Thermal masses and screening masses*

The restoration of symmetries is reflected also in the modification of the hadron spectrum at non-zero temperature. Interactions in a thermal medium lead to modifications of resonance peaks that can modify the location of maxima and the width of spectral functions that control properties of hadron correlation functions. This gives rise to so-called thermal masses as well as thermal screening masses that control the long-distance behavior of hadron correlation functions in Euclidean time and spatial directions, respectively.

A consequence of $U(1)_A$ breaking in the vacuum or at low temperature is that masses of hadronic states that are related to each other through a $U(1)_A$ transformation differ, while they become identical, or close to each other, when the $U(1)_A$ symmetry is effectively restored. This is easily seen to happen at high temperature. The crucial question, of relevance for the QCD phase transition, however, is to which extent $U(1)_A$ symmetry breaking is reduced, or already disappeared at the chiral phase transition temperature. Settling this question requires the analysis of observables sensitive to $U(1)_A$ breaking close to $T_c^0$ and for smaller-than-physical light quark masses.

The calculation of in-medium modifactions of hadron masses is difficult, but has been attempted for quark masses close to their physical values [455]. Results for the temperature dependence of the mass-splitting of parity partners in the baryon octet [455] are shown in Fig. 4.4.6. These results suggest a strong temperature dependence of the negative parity states while the positive parity partners are not sensitive to temperature changes. At $T_{pc}$ the masses of parity partners are almost degenerate.

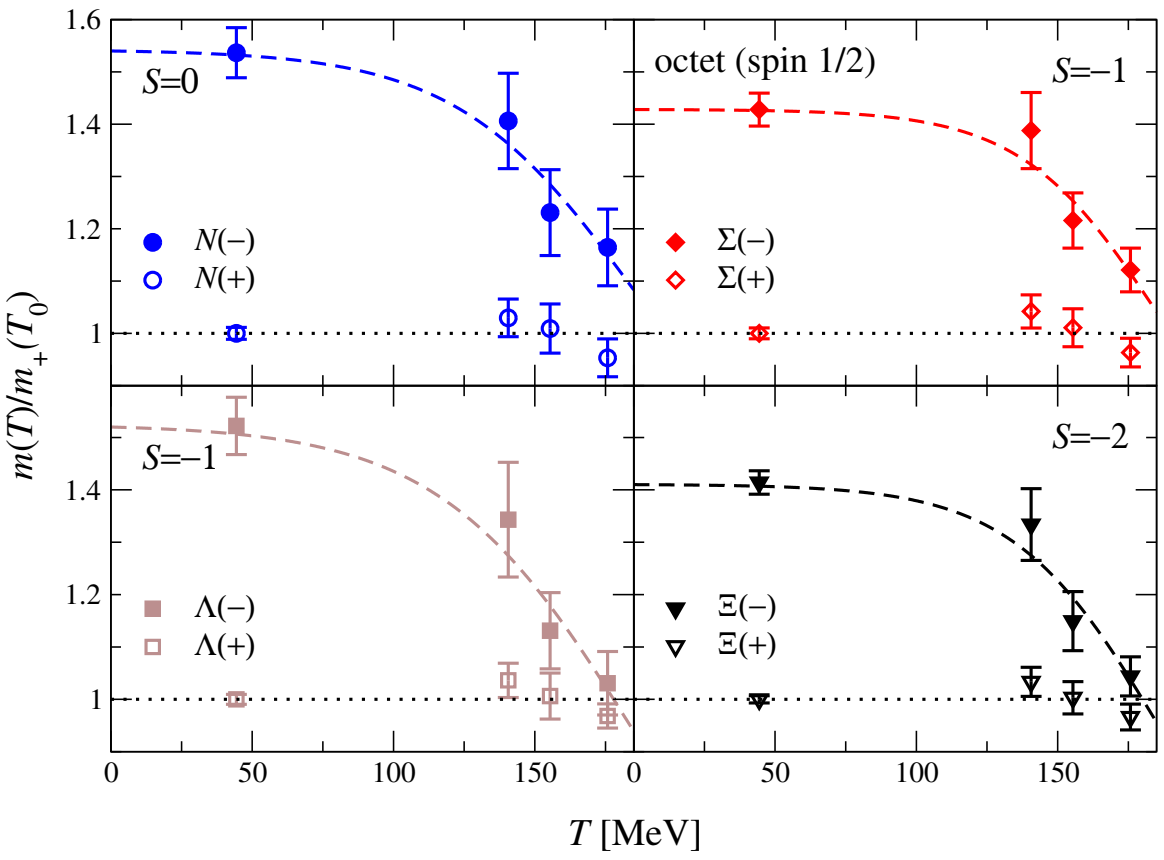

**Fig. 4.4.6** Temperature dependence of masses of parity partners in the baryon octet [455].

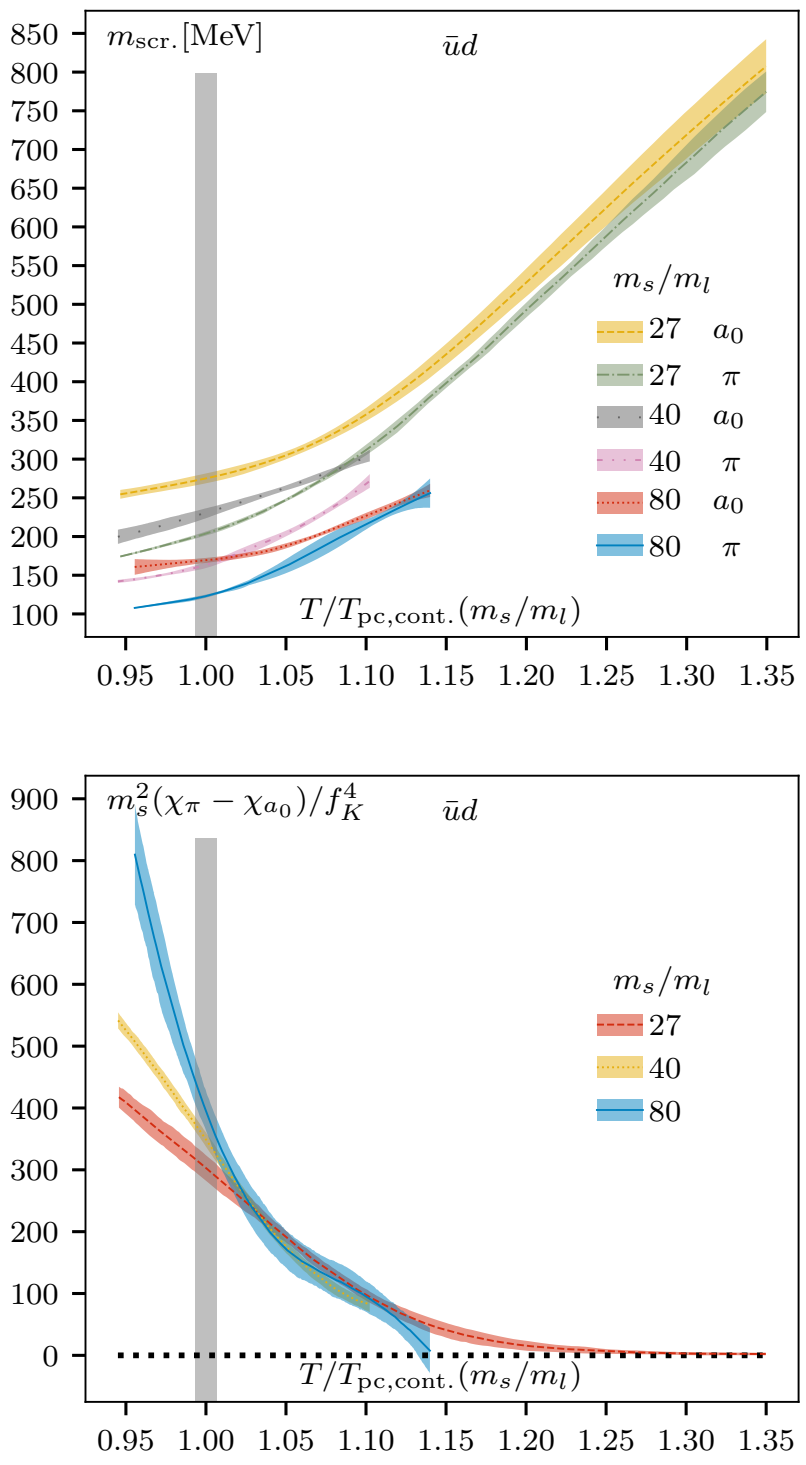

**Fig. 4.4.7** Screening masses (top) and the related susceptibilities (bottom) of scalar and pseudo-scalar mesons [456, 457].

More easily accessible are so-called screening masses, which also are obtained from ordinary hadron correlation functions and can be analyzed close to the chiral limit. Rather than analyzing the long-distance behavior of hadron correlation functions in Euclidean time, one extracts a so-called screening mass from the long-distance behavior in one of the spatial directions [458, 459]. Finite temperature meson screening correlators, projected onto lowest Matsubara frequency of a bosonic state, $p_0 \equiv \omega_0 = 0$, and zero transverse momentum, $\mathbf{p}_\perp \equiv (p_x, p_y) = 0$, are defined by

$$G_\Gamma(z,T) = \int_0^\beta d\tau \int dxdy \left\langle \mathcal{M}_\Gamma(\vec{r},\tau)\overline{\mathcal{M}_\Gamma}(\vec{0},0)\right\rangle \underset{z\to\infty}{\sim} \mathrm{e}^{-m_\Gamma(T)z} \ , \quad \vec{r}\equiv(x,y,z) \ , \tag{4.4.23}$$

where $\mathcal{M}_\Gamma \equiv \bar{\psi}\Gamma\psi$ is a meson operator that projects onto a quantum number channel that is selected through an appropriate choice of $\Gamma$-matrices [456, 458]. At large distances this permits the extraction of the screening mass, $m_\Gamma$, in the quantum number channel selected by $\Gamma$ from the exponential fall-off of these correlation functions. In Fig. 4.4.7 (left) we show results for the scalar and pseudo-scalar screening masses obtained in (2+1)-flavor QCD calculations for different values of the light to strange quark mass ratio. The integrated correlation functions define susceptibilities in these quantum number channels, which also should be degenerate, if $U(1)_A$ is effectively restored. Both observables seem to suggest that there remains a significant remnant of $U(1)_A$ breaking at the chiral phase transition temperature, $T_c^0$, which however reduces quickly above the chiral transition and gives rise to an effective restoration of $U(1)_A$ at $T \simeq 1.1T_c^0$.

In the region $T > T_c^0$ the difference between pseudo-scalar and scalar susceptibilities is related to the so-called disconnected part, $\chi_{dis}$, of the chiral susceptibility, $\chi_M = \chi_{dis} + \chi_{con}$, with

$$\chi_{dis} = \frac{1}{4N_\tau N_\sigma^3}\left(\langle(\mathrm{Tr}M_\ell^{-1})^2\rangle - \langle\mathrm{Tr}M_\ell^{-1}\rangle^2\right) \ , \tag{4.4.24}$$

$$\chi_{con} = \frac{1}{2N_\tau N_\sigma^3}\langle\mathrm{Tr}M_\ell^{-2}\rangle \ . \tag{4.4.25}$$

The disconnected chiral susceptibility can be expressed by an integral over the eigenvalue density, $\rho(\lambda)$, of the fermion matrix $M_f$,

$$\chi_{dis} = \int_0^\infty d\lambda\, \rho(\lambda)\frac{2m_\ell^2}{(\lambda^2+m_\ell^2)^2} \ . \tag{4.4.26}$$

In the chiral symmetric phase the density of vanishing eigenvalues, $\rho(0)$, vanishes. In order for $\chi_{dis}$ to be nonetheless non-zero in the chiral limit, the density of near-zero eigenvalues needs to converge to a non-vanishing value ($\delta$-function) at $\lambda = 0$ in the limit $m_\ell \to 0$ and $V \to \infty$. Controlling the various limits involved and also taking into account that the pseudo-critical transition temperature, $T_{pc}(H)$, has a sizeable quark mass dependence is difficult. Nonetheless, studies of the temperature dependence of the eigenvalue density of the Dirac matrix are crucial for a detailed understanding of the influence of the $U(1)_A$ anomaly on the QCD phase transition. Not surprisingly, it turns out that at non-zero values of the lattice spacing the spectrum of low lying eigenvalues is quite sensitive to the fermion discretization scheme. Using fermions with good chirality even at non-zero lattice spacing seems to be advantageous, although after having performed the extrapolation to the chiral limit, they should lead to results identical with those obtained, e.g. within the staggered discretization scheme. Current results are ambiguous. We show in Fig. 4.4.8 results from a calculation of eigenvalue distributions obtained from calculations with dynamical overlap fermions [460, 461]. These calculations provide evidence for a large density of near-zero eigenvalues and a non-zero eigenvalue density, possibly building up at $\lambda = 0$. This is in contrast to calculations performed with domain wall fermions [462] as well as so-called partially quenched calculations that use the overlap fermion operator to calculate eigenvalue distributions on gauge field configurations generated with dynamical staggered fermions [463]. Obviously this subtle aspect of the chiral phase transition is not yet resolved and the analysis of $U(1)_A$ restoration will remain to be a central topic in finite temperature QCD in the years to come.

### 4.4.5 The chiral phase transition at vanishing chemical potential

In the studies of QCD at non-vanishing baryon chemical potential the search for the existence of a second order phase transition at physical values of the quark masses, the critical end point (CEP), finds particular attention. It separates the crossover regime at small values of the chemical potential from a region of first order phase transitions, which is predicted in many phenomenologi-cal models to exist at high density. The CEP is searched for extensively in heavy ion experiments and, if confirmed, would provide a solid prediction for the existence of first order phase transitions in dense stellar matter, e.g. in neutron stars.

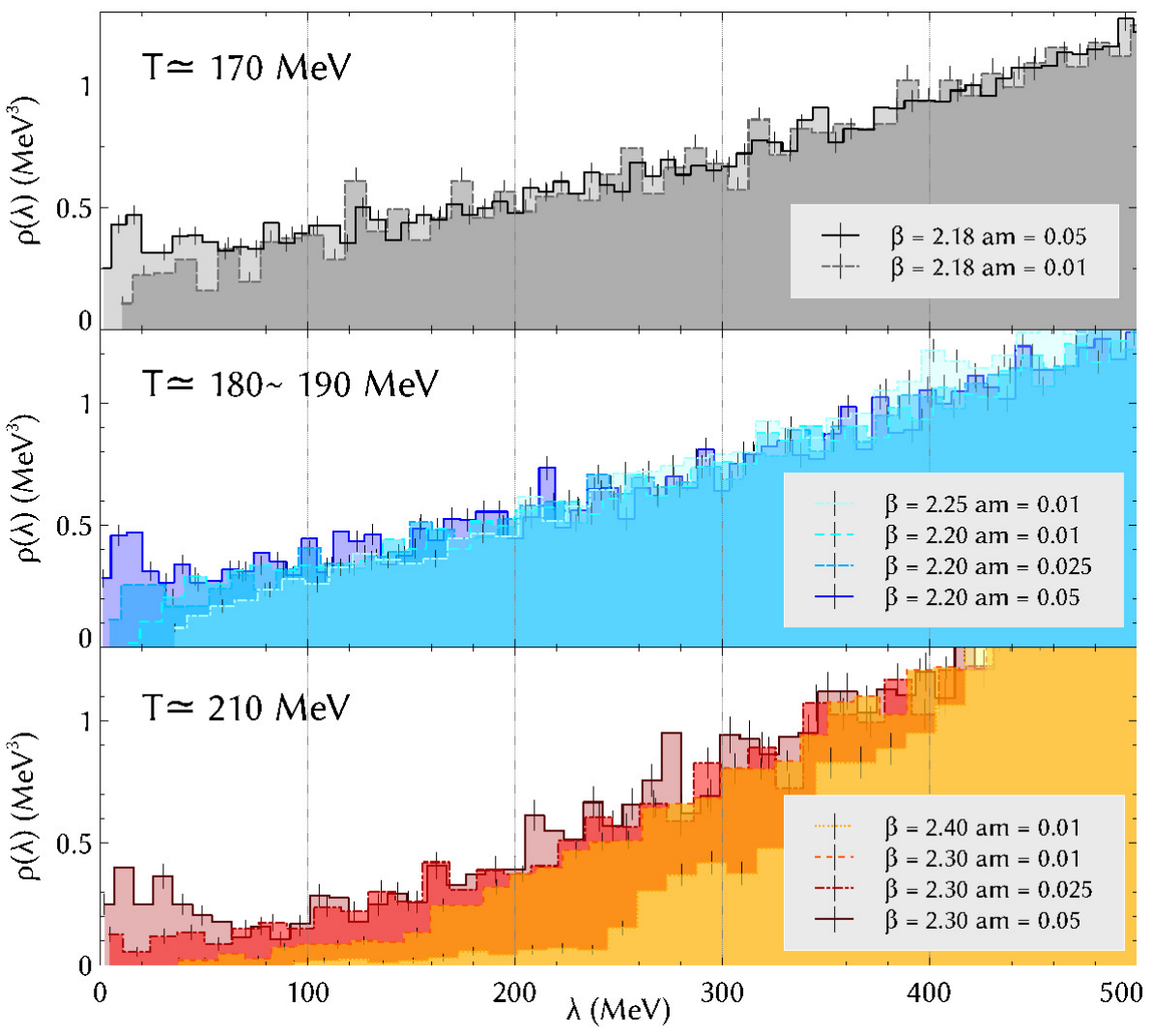

**Fig. 4.4.8** Eigenvalue density of the overlap fermion matrix obtained in calculations with dynamical overlap fermions [460].

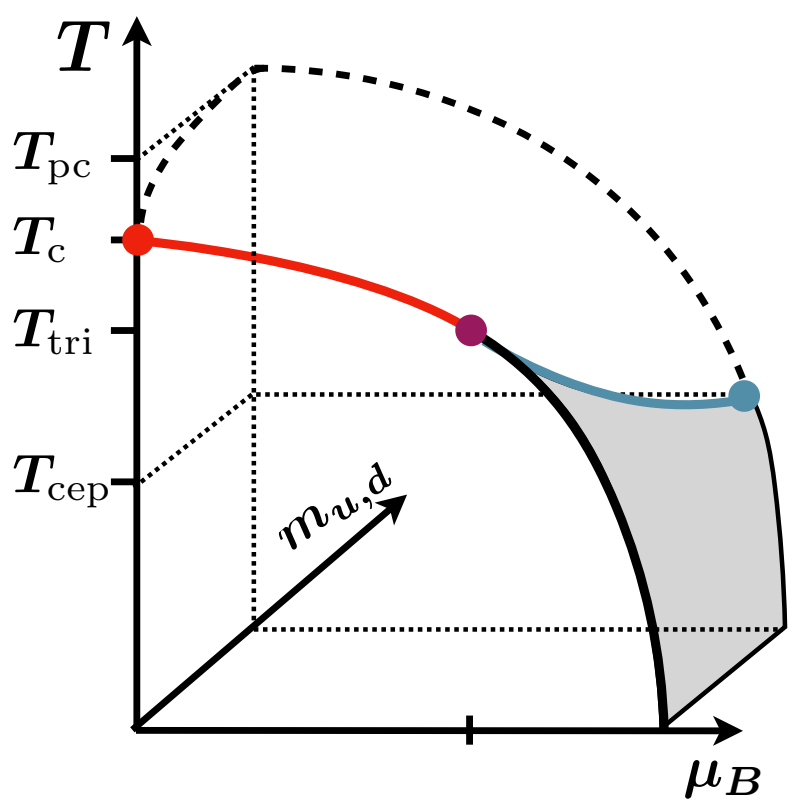

**Fig. 4.4.9** Sketch of a possible QCD phase diagram in the space of temperature ($T$), baryon chemical potential ($\mu_B$) and light quark masses ($m_{u,d}$).

The dependence of the transition temperature on the chemical potentials, e.g. $T_{pc}(\mu_B)$, can be deduced from the $\mu_B$-dependent shift of the peak in the chiral susceptibility. At non-vanishing values of the baryon chemical potential, $\mu_B$, the QCD phase transition temperature in the chiral limit as well as the region of pseudo-critical behavior in QCD with its physical quark mass values shifts to smaller values of the temperature. This shift has been determined in calculations with imaginary values of the chemical potentials as well as from Taylor series expansions of the order parameter $M$ and its susceptibility $\chi_M$. Using a Taylor series ansatz for $T_{pc}(\mu_B)$,

$$T_c(\mu_B) = T_c^0 \left(1 - \kappa_2^B \left(\frac{\mu_B}{T_c^0}\right)^2 - \kappa_4^B \left(\frac{\mu_B}{T_c^0}\right)^4\right) \quad (4.4.27)$$

one finds for the curvature coefficients $\kappa_2^B \simeq 0.012$ while the next correction is consistent with zero in all current studies, e.g. $\kappa_4^B = 0.00032(67)$ [451]. The pseudo-critical temperature $T_{pc}$ at physical values of the light and strange quark masses thus drops to about 150 MeV at $\mu_B \simeq 2T_{pc}$. This is still considerably larger than the chiral phase transition temperature, $T_c^0$, determined at $\mu_B = 0$. As various model calculations [464, 465] suggest that the CEP at non-zero $\mu_B$ is located at a temperature below $T_c^0$ one thus needs to get access to thermodynamics at large chemical potentials. Assuming that the curvature of the pseudo-critical line does not change drastically at large values of the chemical potentials, our current understanding of the QCD phase diagram in the $m_\ell$-$T$-$\mu_B$ space (see Fig. 4.4.9) suggests that a possible CEP in the phase diagram may exist only at a temperature,

$$T^{CEP}(\mu_B^{CEP}) < 130 \text{ MeV} \ , \ \mu_B^{CEP} > 400 \text{ MeV} \ . \quad (4.4.28)$$

Reaching the region $\mu_B/T > 3$ is a major challenge for any of the currently used approaches in lattice QCD calculations as well as for collider based heavy ion experiments that search for the CEP.

### 4.4.6 Equation of state of strongly interacting matter

The equation of state (EoS) of strongly interacting matter, *i.e.* the pressure and its derivatives with respect to temperature and chemical potentials provides the basic information on the phase structure of QCD. It is of central importance not only for the analysis of critical behavior in QCD but also for the analysis of experimental results on strong interaction thermodynamics that are obtained in relativistic heavy ion collision experiments.

At vanishing values of the chemical potentials the QCD EoS is well controlled and consistent results for pressure, energy and entropy densities, as well as derived observables such as the speed of sound or specific heat, have been obtained by several groups [466, 467]. We show results for some of these observables in Fig. 4.4.10. The figure on the right shows the square of the speed of sound, $c_s^2$, as function of the energy density. It can be seen that $c_s^2$ has a minimum in the transition region, sometimes called the softest point of the QCD EoS [469]. The energy density in the vicinity of the pseudo-critical temperature ($T_{pc} \simeq 155$ MeV) is found to be,

$$\epsilon_c \simeq (350 \pm 150) \text{ MeV/fm}^3 \ , \quad (4.4.29)$$

which is compatible with the energy density of the nucleon, $m_N/(4\pi r_N^3/3)$ for nucleon radii in the range $r_N = (0.8 - 1)$ fm. Also shown in the top figure is the trace of the energy-momentum tensor, $(\epsilon - 3P)/T^4$. Its deviation from zero gives some hint to the relevance of interactions in the medium (for an ideal gas as well as to leading order in high temperature perturbation theory one has $\epsilon = 3P$). Not unexpected this is largest close to the transition region and decreases only slowly in the high temperature regime. This large deviations from ideal gas or perturbative behavior is seen in many observables at temperature $T_{pc} < T < 2T_{pc}$.

Calculations of the equation of state as a function of $T$ and $\mu_B$ have been performed using direct simulations at imaginary chemical potentials, which then get analytically continued to real values of the chemical potentials [470], as well as calculations using up to eighth order Taylor expansions in $\mu_B$ [468]. Results of such calculations agree well for $\mu_B/T \leq (2-2.5)$. In Fig. 4.4.11 we show results for the $\mu_B$-dependent contribution to the pressure and net baryon number density. Comparing Fig. 4.4.11 (left) with Fig. 4.4.10 (left) shows that at $\mu_B/T \simeq 2$ and $T \simeq T_{pc}$ the pressure increases by about 30%, which is due to the increase in number of baryons in the medium.

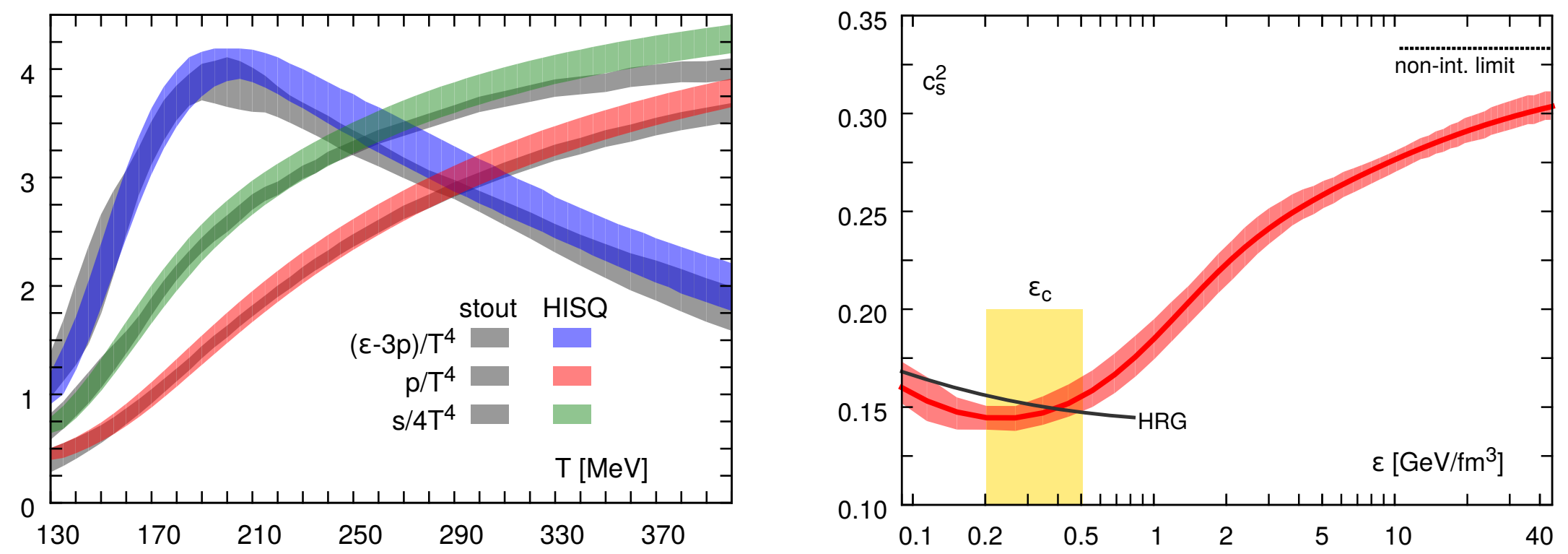

**Fig. 4.4.10** *Left:* Pressure, energy and entropy densities in (2+1)-flavor QCD at vanishing chemical potential. The figure is taken from [466]. Also shown in the figure are results obtained with the stout discretization scheme for staggered fermions [467]. *Right:* The speed of sound as function of energy density.

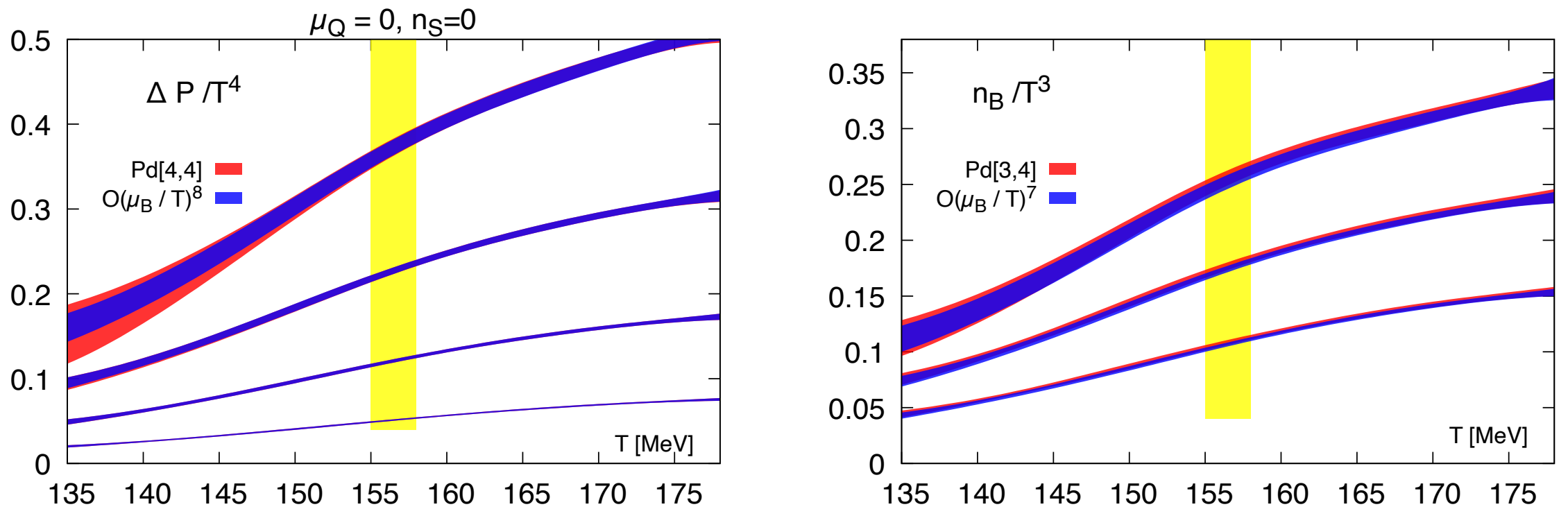

**Fig. 4.4.11** $\mu_B$-dependent contribution to the pressure (left) and net baryon number density (right) in (2+1)-flavor QCD at several values of the baryon chemical potential chemical potential, $\mu/T_B = 1.0$, 1.5, 2.0, 2.5, (bottom to top) and for $\hat{\mu}_B = 2.0$. Shown are results from Taylor expansion up to eighth order in $\hat{\mu}_B$ in the pressure series for isospin symmetric ($\mu_Q = 0$) strangeness neutral ($n_S = 0$) matter and corresponding Padé approximants obtained from these Taylor expansion coefficients. The figures are taken from [468].

At larger values of the baryon chemical potential the Taylor series will not convergence due to the presence of either poles in the complex $\mu_B$-plane or a real pole, that may correspond to the searched for CEP. The occurrence of poles in the complex plane also generates problems for the analytic continuation of results obtained in simulations at imaginary values of $\mu_B$ as a suitable ansatz for the continuation needs to be found. Many approaches to improve over straightforward Taylor series approaches or simulations at imaginary chemical potential are currently being discussed [471–474].

In the context of Taylor expansions a natural way to proceed is to use Padé approximants, which provide a resummation of the Taylor series and reproduce this series, when expanded for small $\mu_B$ [468, 475]. Results from [4,4] and [3,4] Padé approximants for the pressure and number density series, respectively, are also shown in Fig. 4.4.11. The good agreement with the Taylor series for $\mu_B/T \leq 2.5$ gives confidence in the validity of the Taylor series results and once more seems to rule out the occurrence of a CEP in this parameter range.

### 4.4.7 Outlook

Achieving better control over the influence of the axial anomaly on the QCD phase transition in the chiral limit at vanishing chemical potentials and getting better control over the dependence of the QCD phase diagram at large non-zero values of the chemical potentials certainly are the two largest challenges in studies of QCD thermodynamics for the next decade.

## 4.5 Spectrum computations

**Jozef Dudek**

### 4.5.1 Motivation for hadron spectroscopy

Many decades of experimental data collection has lead to a compendium of observed hadrons[476], most of which are short-lived resonances. The job of hadron spectroscopy is to understand the patterns in the spectrum, such as the distribution of states by spin, parity and flavor, and which decays are preferred by which states. These patterns are typically interpreted in terms of models or 'pictures' of hadron structure in which e.g. certain mesons are assigned status as $q\bar{q}$, as glueballs, as hybrids, as higher quark Fock states, or as molecular states of lighter hadrons[477].

For a long time, simplified dynamical models whose connection to QCD is often obscure have dominated the field, and through these significant intuition has been developed, but in recent years lattice QCD has matured to the level where it can address the physics of excited hadrons directly. Using this tool we aim to build an understanding of how QCD binds quarks and gluons into hadrons from first principles.

### 4.5.2 Precise mass determination for stable hadrons

As described in Sec. 4.2, hadron masses can be determined from the large time behavior of two-point correlation functions utilizing operators with the quantum numbers of hadrons constructed from quark and gluon fields. These correlation functions are calculated using quark propagators computed with a particular choice of discretization of the QCD action, and particular values of parameters which set the lattice spacing and the quark masses. When seeking precise determination of hadron masses, one can calculate with several quark masses and lattice spacings, and attempt to extrapolate to the physical limit where the quark mass takes its true value and where the lattice spacing becomes zero.

Figure 4.5.1 (taken from Ref. [478]) summarizes a number of efforts in this direction, showing the masses for low-lying mesons and baryons constructed from light, strange, charm and bottom quarks, comparing the computed values to measured values. Clear agreement is observed for many stable or nearly-stable hadrons. With increasing levels of precision on the mass estimates, the role of small effects like QED become important, and in recent years, these too have been estimated (e.g. Refs [479–482]) .

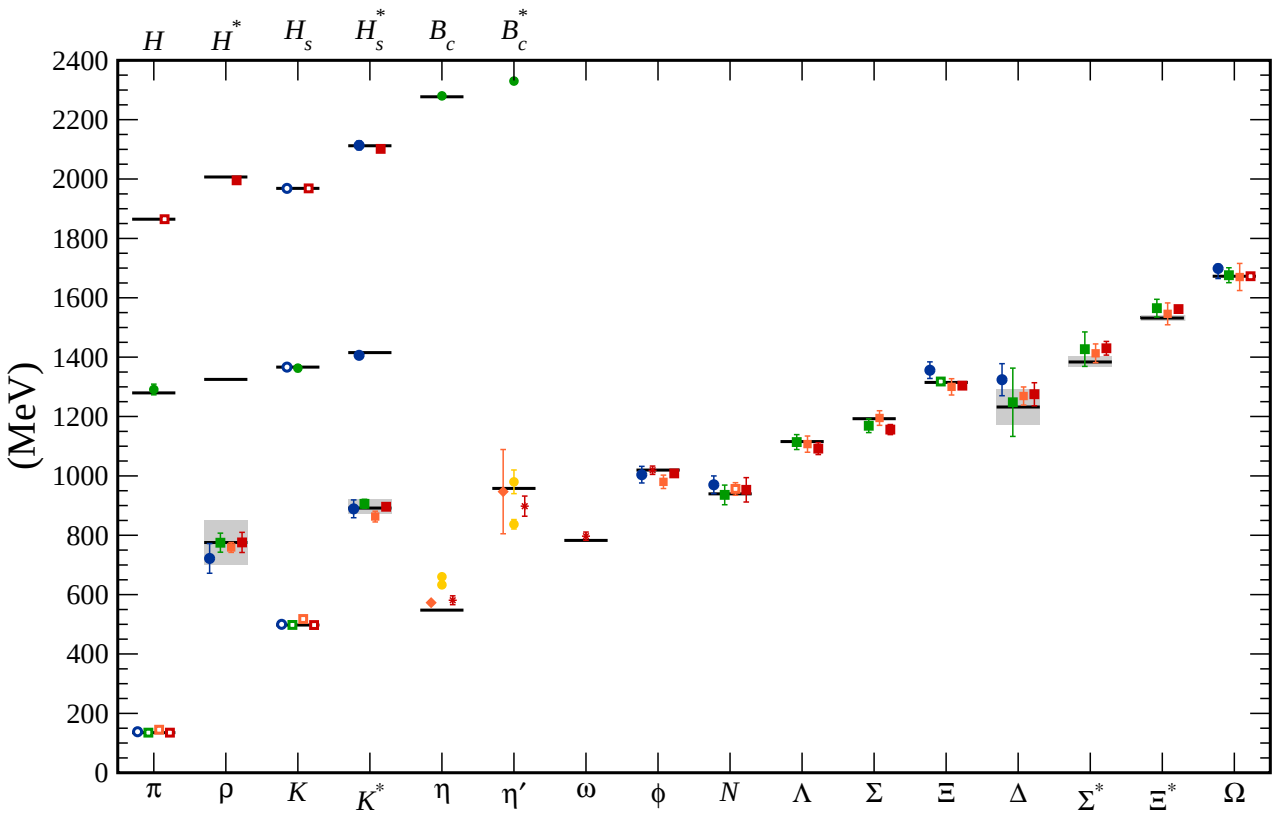

**Fig. 4.5.1** Summary of hadron spectrum calculations taken from Ref. [478]. Different symbol shapes indicate different quark discretizations, while the colors (red, orange, green, blue) indicate an increasing level of systematic control in the calculation. $b$-flavored meson masses are shifted down by 4000 MeV.

### 4.5.3 Expanding the scope of lattice spectroscopy

There are relatively few calculations in which hadron masses have been determined with a somewhat complete study of systematics, and they have been largely restricted to those situations where only a single completely-connected Wick contraction features in the relevant correlation function, and where the state of interest is the lightest with a given quantum number.

Examples which require something beyond this include *isoscalar mesons* in which quark-antiquark annhilation diagrams must be computed. Conventional propagator techniques cannot handle these diagrams, and while various stochastic techniques have been used, it was the introduction of the *distillation* approach [483] which not only opened up isoscalar meson spectroscopy but also the determination of multiple excited states.

Distillation is in essence a quark-field smearing implementation, where the smearing operator,

$$\Box(\mathbf{x}, \mathbf{y}) = \sum_{i=1}^{N} v_i(\mathbf{x})\, v_i^\dagger(\mathbf{y}),$$

is constructed from a limited number of low-lying eigenvectors of the gauge-covariant spatial Laplacian,

$$-\nabla^2 v_i(\mathbf{x}) = \lambda_i v_i(\mathbf{x}).$$

All quark fields in hadron interpolating operators are smeared by this operator, enhancing overlap with low-lying states. The unique advantage of this approach though is the way that the outer-product nature of the smearing operator allows a *factorization* of correlation functions into objects describing hadron operators, independent of objects called "permabulators" describing quark propagation,

$$\tau_{ij}(t,t') = \sum_{\mathbf{x},\mathbf{y}} v_i^\dagger(\mathbf{x}) M^{-1}(\mathbf{x},t;\mathbf{y},t') v_j(\mathbf{y}).$$

Annihilation contributions can be handled straightforwardly using timeslice-to-timeslice perambulators, $\tau_{ij}(t,t)$.

The factorization within distillation allows for massive re-use of the propagation objects, so that the inversion time cost of building a set of permabulators is amortized over a huge number of subsequent calculations[12]. In the context of determining excited states, it allows for the computation of many correlation functions using a large basis of interpolating operators.

While in principle any single correlation function

$$C(t,0) = \sum_n a_n \, e^{-M_n t}$$

contains information about the entire excited spectrum, $\{M_n\}$, in practice determining the spectrum by fitting subleading time-dependence is highly unstable. It is obvious for example, that degenerate or near-degenerate states cannot be distinguished by their time-dependence alone. A much more powerful approach makes use of orthogonality – if one considers a large basis of hadron interpolating operators all with the same overall quantum numbers, we expect there to be one linear combination that most effectively produces the ground state, another that produces the first-excited state and so on. It is straightforward to show that if one forms the *matrix* of two-point correlation functions

$$C_{ij}(t) = \langle 0|\mathcal{O}_i(t)\, \mathcal{O}_j^\dagger(0)|0\rangle,$$

with a basis of operators $\{\mathcal{O}_i\}_{i=1\ldots N}$, the optimal combinations correspond to the eigenvectors of the *generalized eigenvalue problem*,

$$C(t)\, v_n = \lambda_n(t,t_0)\, C(t_0)\, v_n \,,$$

where the eigenvalues give access to the corresponding mass or energy spectrum, $\lambda_n(t,t_0) \sim e^{-E_n(t-t_0)}$. This approach is typically referred to as *variational analysis* [486–488].

An example of a large basis of operators with the quantum numbers of mesons is the one presented in Ref [484], where smeared quark field bilinears featuring up to three gauge-covariant derivatives are used [489–492]. In order to respect the reduced rotational symmetry of the cubic lattice, operators of definite $J^P$ are *subduced* into irreducible representations (irreps) of the cubic symmetry. Using a basis like this, with the variational analysis approach presented above, can lead to results like those shown in Figure 4.5.2. The extracted spectrum shows many of the systematics of the experimental meson spectrum such as the $J^{PC}$ ordering of states and the presence of an "OZI-rule" in the hidden-light/hidden-strange composition of isoscalar mesons (dominantly ideal flavor mixing except for a few notable exceptions like $0^{-+}$). Also present in these extracted spectra are mesons with *exotic* $J^{PC} = 1^{-+}, 0^{+-}, 2^{+-}$, i.e. those not accessible to just a $q\bar{q}$ pair. Examining which interpolating operators are the largest components in the optimal operators for these states, we observe the presence of non-trivial gluonic structures, and it is natural to interpret these states as *hybrid mesons*. Non-exotic $J^{PC}$ states high in the spectrum are also observed to have these gluonic operator overlaps (states outlined in orange in Fig 4.5.2), and this leads to an identification of the lightest supermultiplet of hybrid mesons [493], ruling out certain previously reasonable models.

A closely related calculation using a large basis of operators with baryon quantum numbers appeared in Refs. [494, 495], with the spectra for $N^\star$ (isospin-1/2) and $\Delta^\star$ (isospin-3/2) excitations shown in Figure 4.5.3.

The calculation presented in Figure 4.5.2 was performed with a light quark mass heavier than physical, and at a single lattice spacing, and as such the results cannot be treated as precise, or suitable for direct comparison to experiment. But in the case of excited spectroscopy, precision is not the main aim, rather the intent is to build an understanding of the systematics of the hadron spectrum having a direct connection to QCD. In fact there is a more relevant problem with these results – they do not reflect the complete physics of excited states which lie above hadronic decay thresholds – these states should be unstable *resonances*, and resonances are not simply characterized by a mass.

### 4.5.4 Resonances and the finite-volume approach to scattering

The simplest context in which resonances appear is *elastic hadron-hadron scattering* in which the initial and final states are identical, and the amplitude can be expanded in partial-waves. Resonances of definite spin appear as enhancements in a single partial-wave in the continuous energy spectrum, and formally may be associated with pole singularities at complex values of the scattering energy.

In a finite spatial volume, such as that provided by the lattice, there can be no continuous energy spectrum, and instead only a discrete spectrum, but it is easy to see that this spectrum should be volume-dependent and sensitive to the infinite-volume scattering amplitudes. This can be illustrated in one-dimensional quantum mechanics [297] – a finite-length of $L$ can be implemented by applying periodic boundary conditions to a scattering wavefunction and its derivative. This leads to a quantization condition on possible allowed momenta, $p_n = \frac{2\pi}{L} n - \frac{2}{L}\delta(p_n)$, where $\delta(p)$ is the elastic phase-shift that describes scattering.

[12] There is also a stochastic implementation of distillation [485], which is argued to have a better cost-scaling with the volume of the lattice, but at the cost of somewhat less flexibility in re-use.

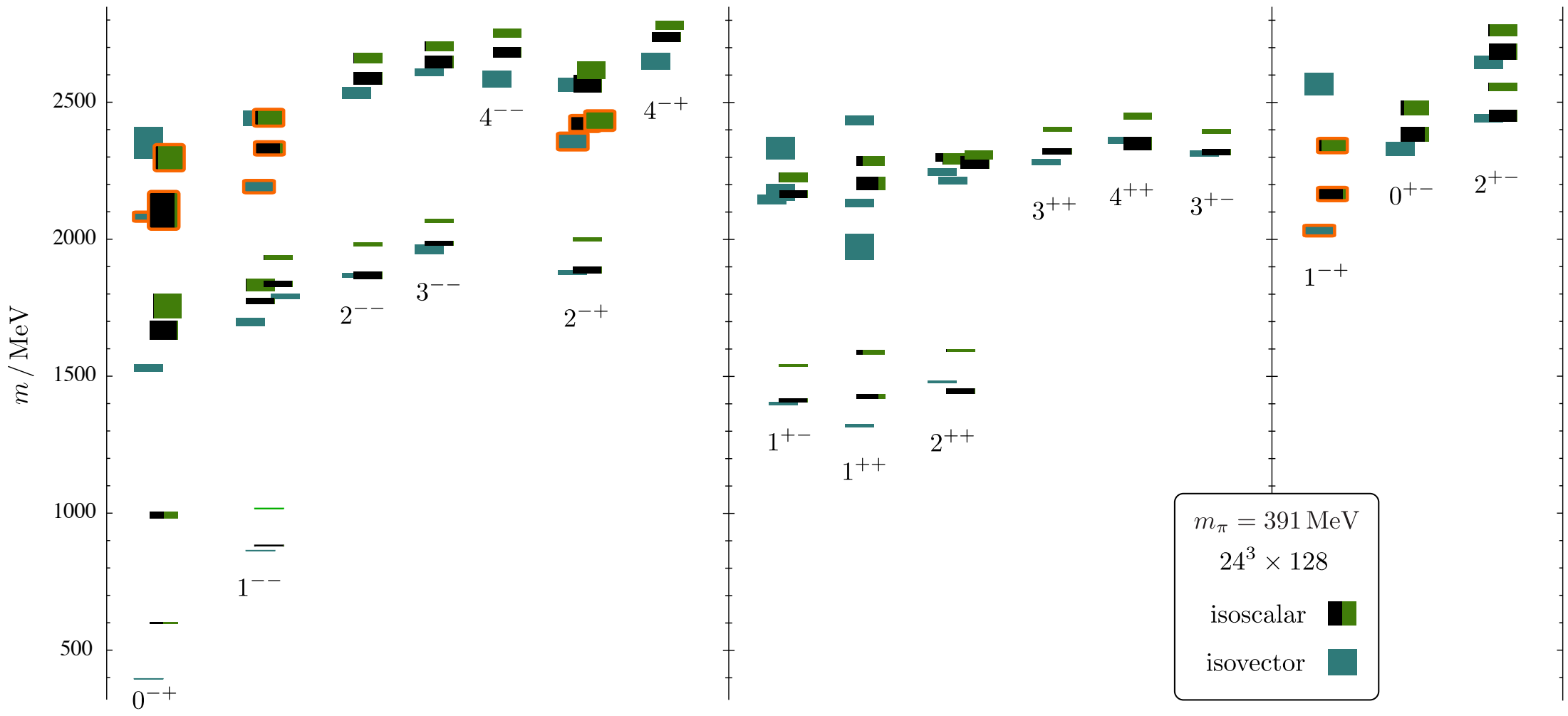

**Fig. 4.5.2** Spectrum of excited mesons extracted from lattice QCD calculation with heavier that physical light quarks. States labelled by their $J^{PC}$. Vertical height of each box represents the statistical uncertainty. Isoscalar meson boxes show the hidden-light (black) versus hidden-strange (green) composition. States with orange outlines have large overlap with operators featuring the chromomagnetic field, suggesting an identification as the lightest supermultiplet of *hybrid mesons*. Taken from Ref. [484].

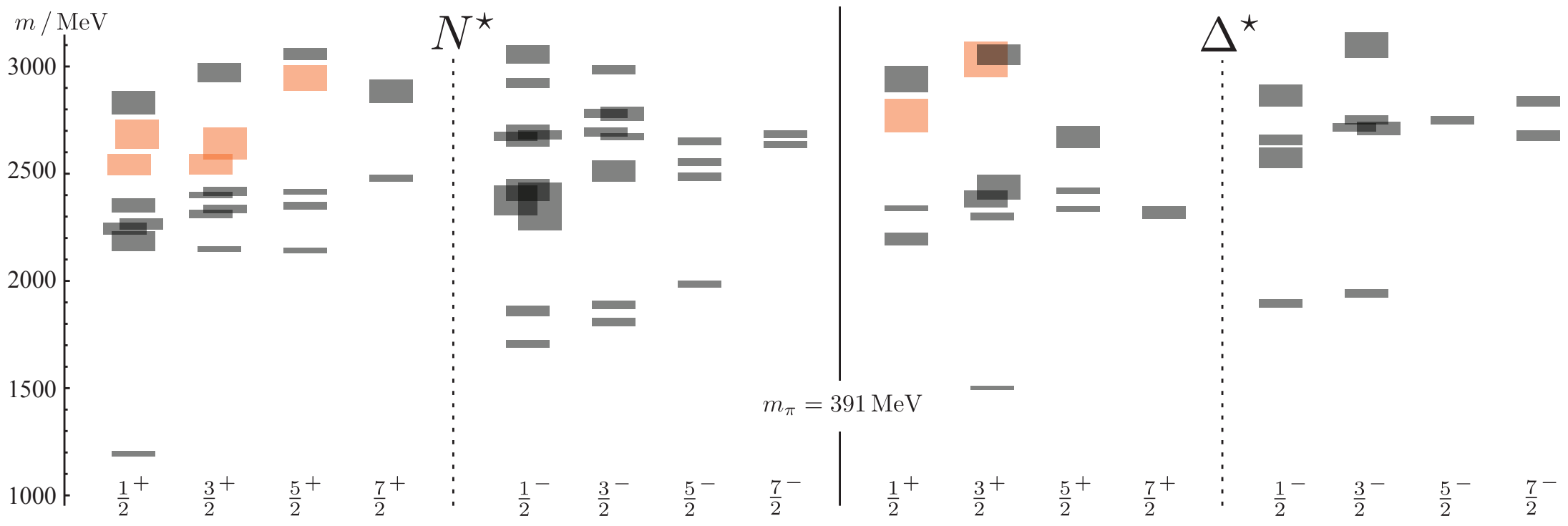

**Fig. 4.5.3** Spectrum of excited baryons extracted from lattice QCD calculation with heavier than physical light quarks. States labelled by their $J^P$. Vertical height of each box represents the statistical uncertainty. States colored orange have large overlap with operators featuring the chromomagnetic field, suggesting an identification as the lightest supermultiplet of *hybrid baryons*. Taken from Refs. [494, 495].

This observation is the core principle behind the lattice QCD approach to scattering. If the discrete spectrum of states in the finite spatial volume defined by the lattice can be obtained, it can be used to provide a set of constraints on the energy dependence of scattering amplitudes.

The analogous formalism for relativistic scattering in three spatial dimensions was derived in Refs [496, 497], and has been extended many times to now be in a form that is applicable to any number of coupled channels of two-body scattering (see the review, Ref [498]). One way of writing this *quantization condition* is

$$\det\left[\mathbf{1} + i\boldsymbol{\rho}(E)\,\mathbf{t}(E)\,\big(\mathbf{1} + i\boldsymbol{\mathcal{M}}(E,L)\big)\right] = 0\,, \tag{4.5.1}$$

where the scattering $t$-matrix is a dense matrix in the space of scattering channels, but block diagonal in angular momentum, $\ell$, while the matrix $\boldsymbol{\mathcal{M}}$, which features known functions (of essentially kinematic origin) of energy and box-size, is block-diagonal in channels, but dense in $\ell$.

The presence of multiple $\ell$ in the quantization condition is an important complicating factor that reflects the fact that the basis of *partial waves* of definite $\ell$, in which one naturally expands scattering, is not respected by the reduced rotational symmetry of the cu-bic boundary of the lattice. The angular momentum barrier at low energies ensures that in practice only a small finite number of $\ell$ values need to be considered.

Eqn. 4.5.1 can be interpreted as follows: if one knew the scattering amplitudes $\mathbf{t}(E)$, one would seek to find all the zero-crossings of the determinant function for a fixed value of $L$, and these would determine the finite-volume spectrum, $E_n(L)$, corresponding to this scattering amplitude. Of course in practice, lattice QCD will supply the discrete finite-volume spectra and one must work backwards to find the corresponding $\mathbf{t}(E)$.

One situation in which this is relatively straightforward is when we are in an energy region where only *elastic scattering* is kinematically allowed, and where one partial wave, $\ell$, is dominant. In this case Eqn. 4.5.1 reduces to the simple form $\cot\delta_\ell(E) = \mathcal{M}_{\ell,\ell}(E,L)$. In this case, given a lattice QCD determined finite-volume energy $E$, one simply plugs into the right-hand-side to obtain a value of the scattering phase-shift at that energy. If enough finite-volume energies are determined, in one or more lattice volumes, the energy dependence of $\delta_\ell(E)$ can be mapped out.

So the job of lattice QCD computation in studies of resonances is to provide accurate discrete finite-volume spectra. In order for calculations to resolve the *full* discrete spectrum of states (as opposed to the limited set described in the previous section) it proves necessary to include in the basis of operators a set which resemble pairs of mesons. These "meson-meson-like" operators are typically constructed from a product of two quark-bilinears, with each one being projected into a definite momentum. The important difference with respect to the single quark-bilinear operators described in the previous section, is that the "meson-meson-like" operators sample the entire spatial volume, causing them to have a much enhanced overlap with finite-volume eigenstates resembling a pair of mesons.

A basis of "meson-meson-like" operators can be constructed [499–501] and a natural guide to which are required in any given calculation comes from a *non-interacting energy* associated with each such operator. For example, operators resembling a pair of pions with $\ell = 0$ can be constructed as $\sum_{\hat{\mathbf{p}}} \mathcal{O}_\pi(\mathbf{p})\mathcal{O}_\pi(-\mathbf{p})$ where $\mathcal{O}_\pi(\mathbf{p})$ is a quark bilinear with the quantum numbers of a pion, and where the sum is over directions of momentum allowed on a cubic lattice. These operators naturally have a non-interacting energy $E_{\text{n.r.}} = 2\sqrt{m_\pi^2 + \mathbf{p}^2}$ associated with them that corresponds to the energy a state interpolated by this operator would have if there were no residual pion-pion interactions. Because there *are* interactions, the actual energy spectrum will differ from this, but it should be clear that operators with non-interacting energy far above the energy region under consideration will not need to be included.

Adding "meson-meson-like" operators to the basis increases the variety of Wick diagrams that need to be evaluated, and in general diagrams including quark-antiquark annihilation are present. Distillation is a very powerful tool to evaluate these diagrams using previously computed perambulators, without the need to make further approximations, or to introduce noise through stochastic approaches.

### 4.5.5 Elastic meson-meson scattering

An example of the approach described in the previous section is presented in Fig 4.5.4 which shows the $P$-wave of $\pi\pi$ scattering with isospin–1. The calculation, done with light-quark masses such that the pion mass is 391 MeV, computed the finite-volume spectrum in three lattice volumes. The panels on the left show the spectra in the rest frame ([000]) and several frames in which the $\pi\pi$ system has a net momentum $\mathbf{P} = \frac{2\pi}{L}[n_x n_y n_z]$. Each discrete energy is used to obtain a value of $\delta_1$ at the same energy, and these are plotted in the right panel, where the behavior is clearly that of a narrow resonance. The energy dependence can then be fitted using a Breit-Wigner or other suitable amplitude parameterization from which the mass and width of the $\rho$ resonance can be determined.

Calculations like this one, of the $\rho$ resonance, have become mainstream within the lattice community [499, 501–512][13], and the vector $K^*$ resonance in $K\pi$ scattering is similar (although in this case one has to deal with the effect of $S$-wave scattering in moving frames) [507, 513–517]. The elastic scattering amplitudes do not need to be resonant for this approach to be used, an example is $\pi\pi$ scattering with isospin–2 where the relatively weak effects can be resolved [500, 508, 518–522].

Pion-pion scattering with isospin–0 has received less attention [522–524]. In order to evaluate the relevant correlation functions, many diagrams featuring $q\bar{q}$ annihilation are required. One example calculation [523] that made use of distillation to evaluate all these diagrams is summarized in Fig 4.5.5, where a function of the phase-shift as a function of energy is shown for calculations at two different light quark masses. The behavior at the heavier quark mass is that of a system featuring a *stable bound state*, while at the lower quark mass, which much more closely resembles the experimental data, there appears to be a *broad resonance*.

[13] One calculation has considered the $\rho$ in $\pi\pi$ scattering using two lattice spacings [510], finding no statistically significant differences.

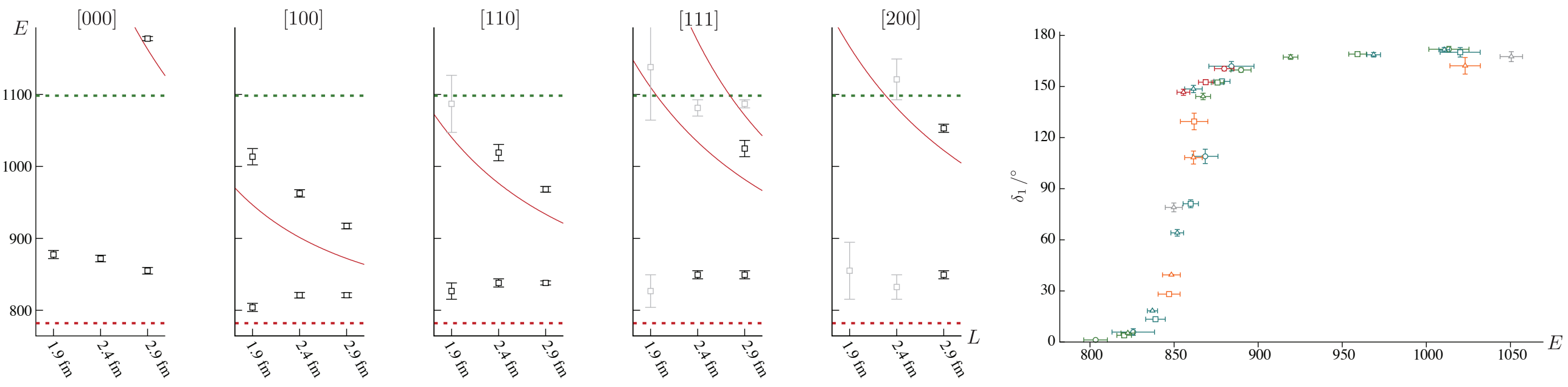

**Fig. 4.5.4** Isospin–1 $\pi\pi$ scattering with $J^P = 1^-$ from lattice QCD with $m_\pi = 391$ MeV taken from Ref. [499]. Left five panels show discrete spectrum of states in three lattice volumes, for five values of total $\pi\pi$ momentum. Red curves indicate the non-interacting $\pi\pi$ energies, and the green dashed line shows the $K\bar{K}$ threshold where scattering ceases to be elastic. Rightmost panel shows the $P$-wave elastic scattering phase-shift determined using the discrete spectrum points which is observed to correspond to a narrow $\rho$ resonance.

These results provide the first signs within QCD of the quark mass evolution of the $\sigma$ meson.

Scattering of mesons featuring charm or bottom quarks can be studied using the same technology [526–538]. Relatively few calculations have so far attempted to determine meson-baryon scattering and the baryonic resonances therein [539–541], largely because of the increased computational cost of such efforts above what is required for meson-meson scattering, and the fact that the lowest-lying resonance, the $\Delta(1232)$, only becomes unstable for decay to $N\pi$ at relatively low light-quark masses.

### 4.5.6 Coupled-channel scattering

The bulk of experimentally observed hadron resonances can decay into more than one hadronic final state, and as such can be considered to be resonances in *coupled-channel* scattering. Coupled-channel scattering (in a particular partial wave) can be described by a $t$-matrix, $t_{ij}(E)$, where the indices $i, j$ run over hadronic channels, e.g. $\pi\pi, K\bar{K}\ldots$.

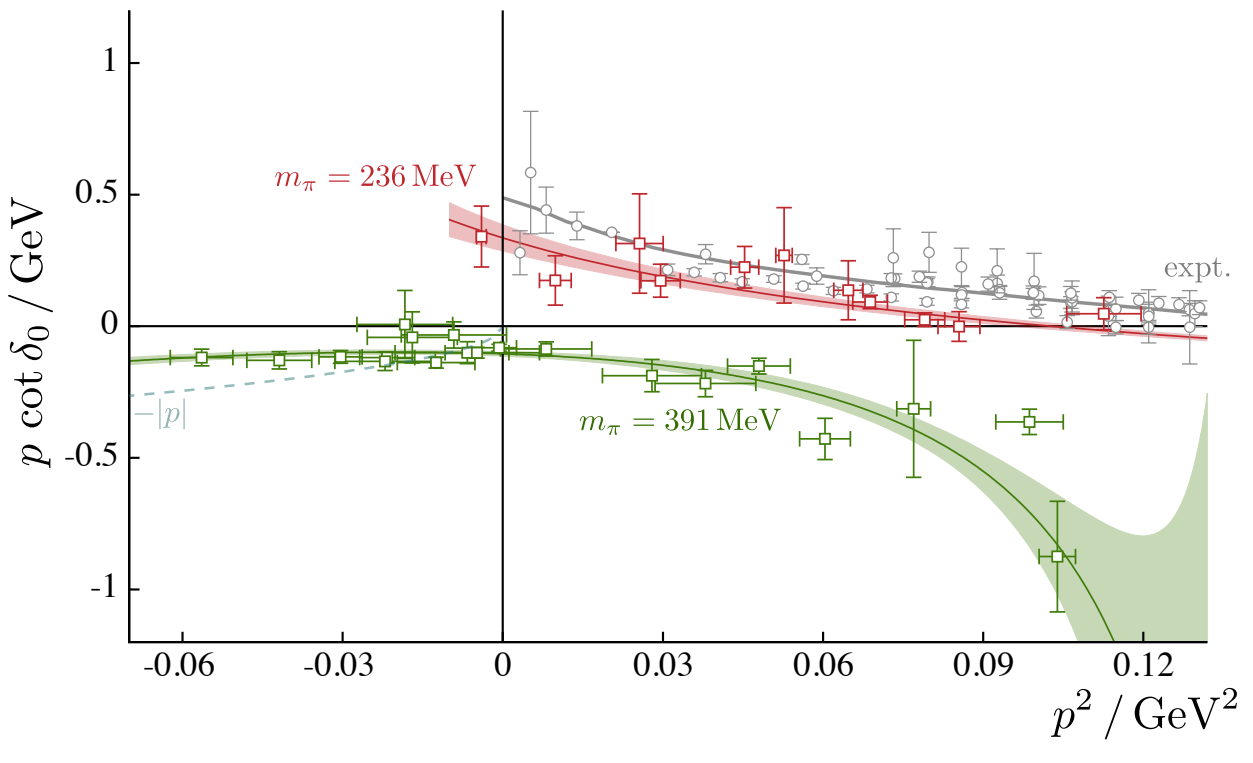

**Fig. 4.5.5** Isospin–0 $\pi\pi$ scattering with $J^P = 0^+$ from lattice QCD at two pion masses taken from Ref. [523]. Intersection of $p \cot \delta_0$ with $-|p|$ indicates the presence of a bound-state $\sigma$ at the heavier pion mass which is not present at the lower pion mass, or in experiment, where a broad resonance is believed to be present.

Eqn. 4.5.1 controls how the discrete spectrum in a finite volume is related to the $t$-matrix, but practical use of this equation when lattice QCD-obtained finite-volume spectra are in hand requires some thought. It is not possible to work energy-level by energy-level as we did for elastic scattering, as the $t$-matrix contains multiple unknowns at each energy. Rather, a successful approach has been to *parameterize* the energy-dependence of $\mathbf{t}(E)$, and to attempt to describe the entire finite-volume spectrum using this parameterization. A $\chi^2$ can be defined which quantifies the difference between the finite-volume spectrum obtained from solving Eqn. 4.5.1 for a particular parameterization and the lattice QCD obtained spectrum. This $\chi^2$ can be minimized by varying the free parameters to obtain a best fit.

In order to carry this out, it is necessary to construct appropriate parameterizations of $\mathbf{t}(E)$ which must include all kinematically open channels in the energy region being considered. They must also exactly respect *two-body unitarity* which is implicit in Eqn. 4.5.1. A rather general framework to achieve this is to use parameterizations of the $K$-matrix, which is flexible enough to handle both resonant and non-resonant cases in any number of channels.

The first lattice QCD calculation of coupled-channel scattering considered the $\pi K, \eta K$ system which was found to be almost decoupled, with resonances appearing coupled only to $\pi K$ [542, 543]. Since then there has been a steady stream of calculations of meson-meson scattering of gradually increasing complexity [501, 525, 531, 544–549].

An example of what can be extracted from lattice QCD for coupled-channel scattering is shown in Figure 4.5.6, taken from Ref. [525]. In this calculation of coupled $\pi\pi, K\bar{K}, \eta\eta$ scattering, performed with 391 MeV pions, the finite volume spectrum was found in three lattice volumes and several moving frames, leading to 57 energy levels to constrain the $S$-wave $t$-matrix and 36 levels to constrain the $D$-wave.

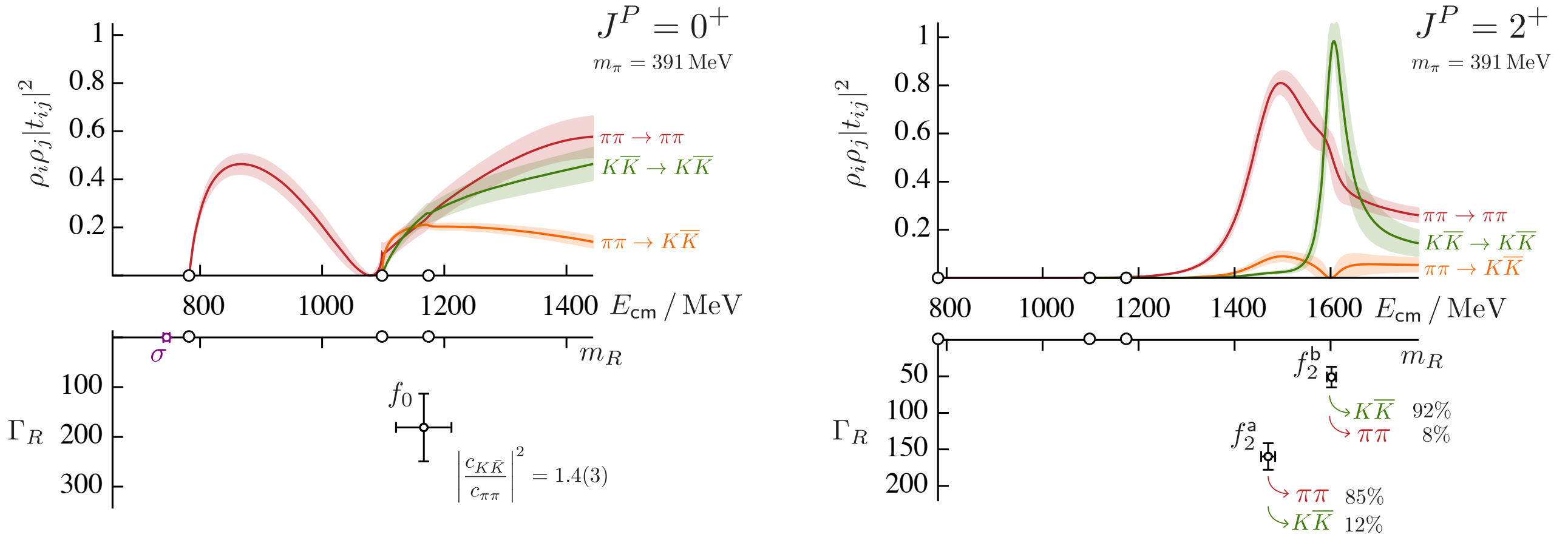

**Fig. 4.5.6** Coupled $\pi\pi, K\bar{K}$ scattering (also $\eta\eta$, not shown) computed on three lattice volumes with $m_\pi = 391$ MeV. Taken from Ref. [525]. Lower panels show resonance pole locations found by analytically continuing into the complex energy plane. In $J^P = 0^+$ case, ratio of couplings of $f_0$ resonance to $\pi\pi$, $K\bar{K}$ given. In $J^P = 2^+$ case, *branching fractions* of two resonances to $\pi\pi$, $K\bar{K}$ final states are given.

We observe a highly non-trivial energy-dependence in the $S$-wave where a broad enhancement at low energies is followed by a *dip* in the $\pi\pi \to \pi\pi$ amplitude at the $K\bar{K}$ threshold, while amplitudes leading to a $K\bar{K}$ final state turn on rapidly at threshold. While this energy dependence does not "by-eye" immediately suggest a simple resonance interpretation, the $t$-matrix can be analytically continued to complex energies, and two poles are found: one lies below $\pi\pi$ threshold and corresponds to the stable $\sigma$ discussed earlier, while the second lies close to the $K\bar{K}$ threshold, and might be associated with the experimental $f_0(980)$ resonance (which also appears as a sharp dip in $\pi\pi$ scattering). This resonance pole has large couplings to both $\pi\pi$ and $K\bar{K}$. These results prove to be robust to variations in the detailed form of the amplitude parameterization.

The $D$-wave result reflects more closely our intuitive picture of resonances, with two bumps appearing, associated to two pole singularities. The lighter state dominantly couples to $\pi\pi$, and a heavier narrower state is dominantly coupled to $K\bar{K}$, a situation that is very similar to the experimental $f_2(1270), f_2'(1525)$ states. The selective final state couplings reflect the 'OZI-rule' emerging dynamically from a non-perturbative calculation if we interpret the lighter state as dominantly $u\bar{u} + d\bar{d}$ and the heavier as dominantly $s\bar{s}$.

A different complication can occur when the scattering hadrons have non-zero spin. In this case, the same total $J^P$ can be constructed by more than one hadron-spin, orbital angular momentum combination. For example, if one scatters a vector $\omega$ meson from a pion, $J^P = 1^+$ can be constructed from $\ell = 0$ or from $\ell = 2$, or using the spectroscopic notation, ${}^3S_1, {}^3D_1$. In this case, even if $\pi\omega$ is the only channel accessible, one still has a system of coupled-partial-waves, and a two-dimensional $t$-matrix.

A version of Eqn. 4.5.1 still holds in such situations, and once again, provided enough energy levels can be computed in lattice QCD to provide sufficient constraint, the $t$-matrix can be determined. An example is shown in Figure 4.5.7 where coupled $\pi\omega, \pi\phi$ scattering was studied with pions of mass 391 MeV. With light quarks as heavy as this, the $\omega$ and $\phi$ mesons are absolutely stable. A clear resonant behavior is observed which can be associated with the experimental $b_1(1235)$ state, and the couplings at the pole yield a value for the $D/S$ amplitude ratio, a quantity that has been measured previously (references are listed in Ref. [476]).

The coupled channel technology has also been applied to scattering systems with charmed mesons [531, 549], and recently, for the first time to a scattering system housing an exotic $J^{PC}$ resonance believed to be a *hybrid meson* [547]. For meson resonances having decays only to one or more two-body final states, rigorous study within lattice QCD is today a reality, with observables being the mass and width of the resonance, as well as the couplings to decay channels, all of which follow from scattering amplitudes. Going beyond this, more information about resonances can be obtained if we generalize away from scattering to also consider processes in which an external current probes the system.

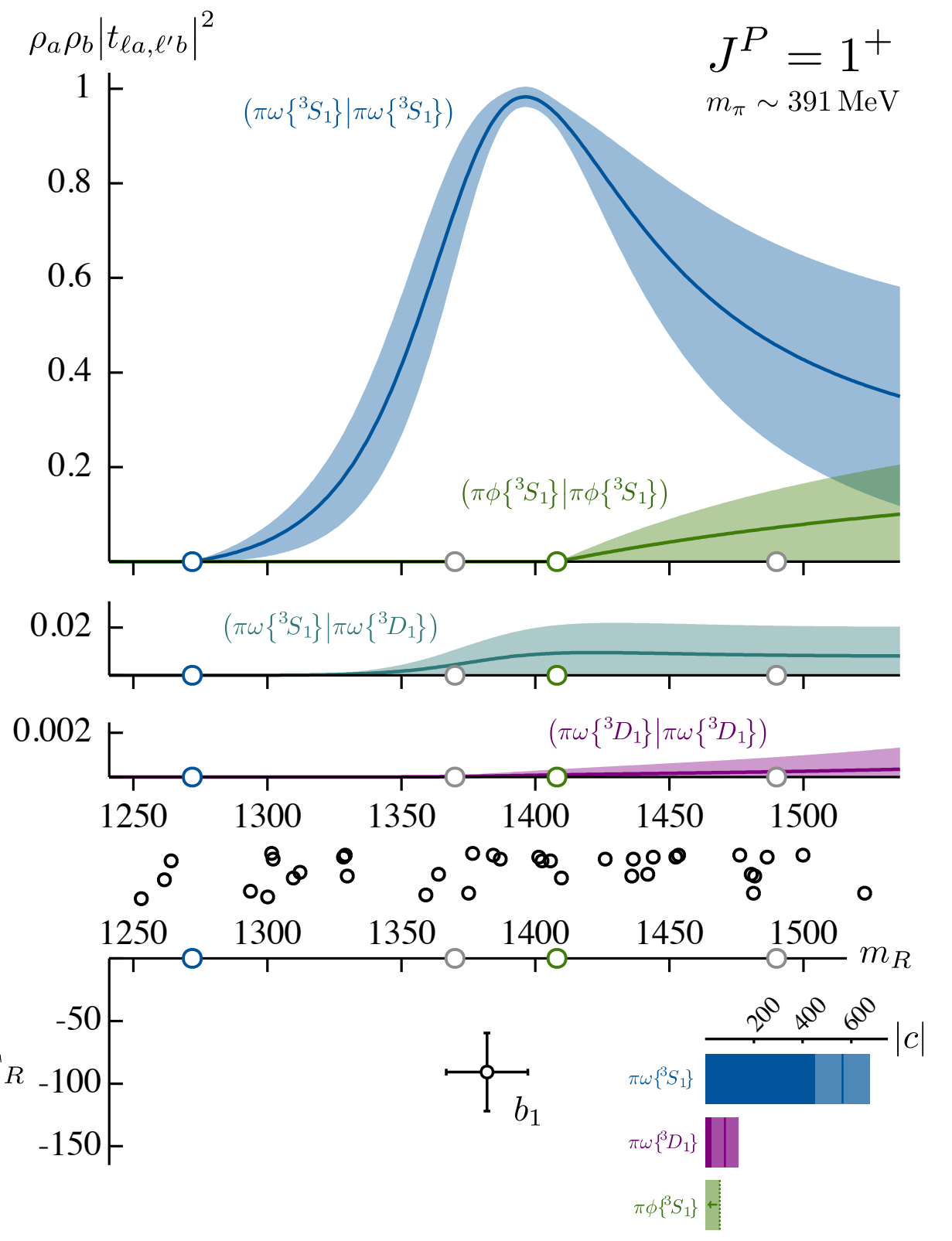

**Fig. 4.5.7** Coupled $\pi\omega, \pi\phi$ scattering, with $\pi\omega$ in coupled partial waves, $^3S_1$, $^3D_1$, computed on three lattice volumes with $m_\pi = 391$ MeV. Taken from Ref. [546]. $b_1$ resonance pole and coupling to channels shown in bottom panel.

### 4.5.7 Beyond scattering

An extension of the finite-volume formalism allows us to study systems in which a stable hadron emits or absorbs an electroweak current and transitions into a pair of strongly-interacting hadrons which may resonate. Applications include semileptonic heavy flavor decays with resonances in the final state, e.g. $B \to \ell^+\ell^- K^*$ where the $K^*$ decays to $K\pi$. To date the only application of this technology has been to a simpler reaction, $\gamma\pi \to \pi\pi$, where the final state features the $\rho$ resonance [550–552]. The approach requires first the determination of the $\pi\pi$ elastic scattering amplitude as described earlier, followed by computations of three-point correlation functions, from which transition matrix elements are extracted. The effect of the finite-volume is encoded in a correction to the normalization of the $\pi\pi$ state [553–555] that requires knowledge of the scattering amplitude. Figure 4.5.8 illustrates one result of such a calculation, showing the transition matrix element for $\pi\gamma \to \pi\pi$ (for two sample values of photon virtuality) along with the elastic $\pi\pi$ scattering amplitude – the clear $\rho$ resonance is present in both.

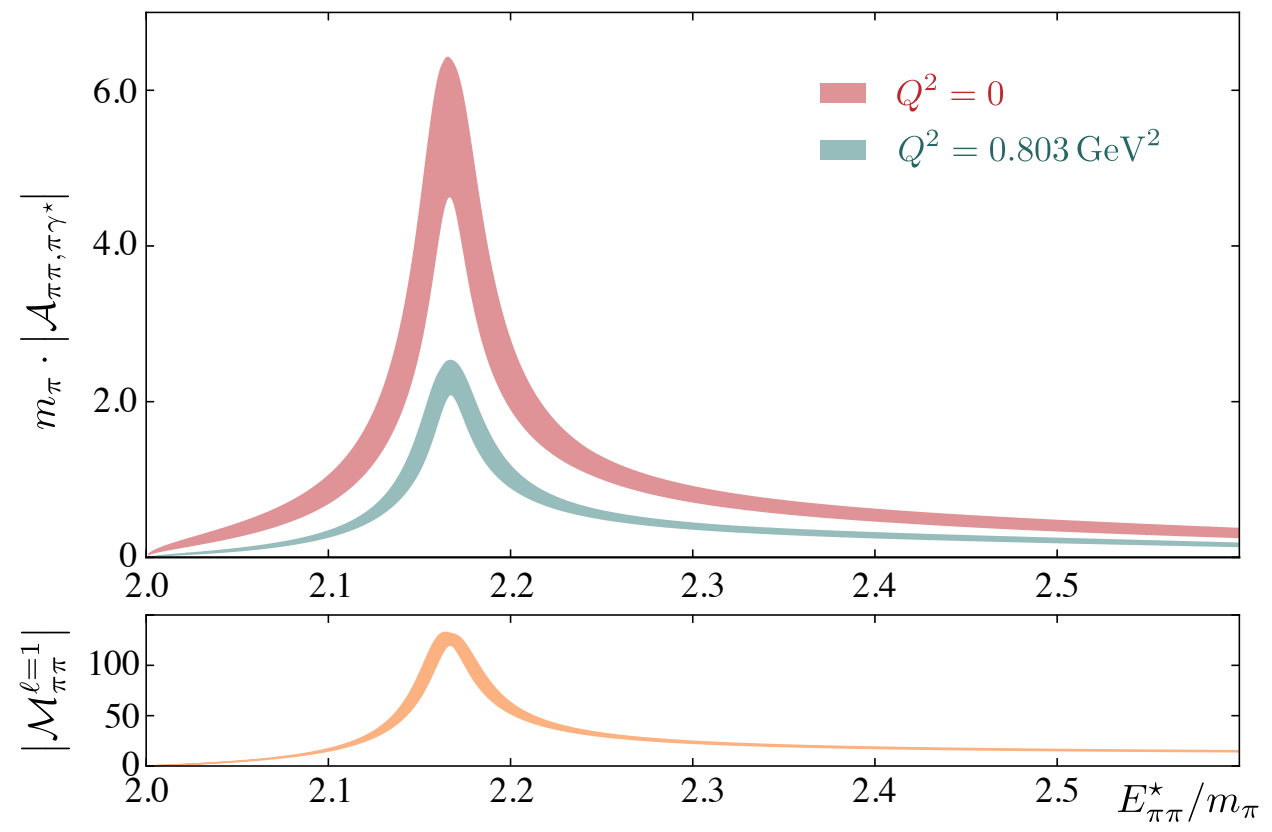

**Fig. 4.5.8** Upper panel shows the transition amplitude for $\pi\gamma \to \pi\pi$ with $J^P = 1^-$ computed from a lattice QCD calculation with $m_\pi = 391$ MeV for two sample values of the photon virtuality. The lower panel shows the corresponding $\pi\pi \to \pi\pi$ elastic scattering amplitude. Taken from Ref. [551].

As well as computation of experimentally measurable processes (such as the heavy flavor decays), this approach also allows us to compute in lattice QCD quantities that cannot be easily accessed in experiment. For example, analytically continuing the transition amplitude obtained above to the $\rho$ resonance pole, one obtains a *resonance* transition form-factor $\rho \to \pi\gamma^*$, whose virtuality dependence can be used to infer structural information about the $\rho$. A recent extension of the finite-volume formalism [557] to be able to handle processes like $\pi\pi\gamma \to \pi\pi$ will allow us to compute the true *resonance form-factors.*

### 4.5.8 The three-hadron frontier and other challenges

The progress reported above in the two-hadron sector has opened up the world of hadron resonance spectroscopy to first principles study using lattice QCD, but to go further an extension in formalism is required. The applicability of Eqn. 4.5.1 is limited to energies below the lowest *three-hadron threshold*, and this is particularly constraining as the light quark mass is decreased and the threshold for $\pi\pi\pi$ becomes very low, lower than the mass of most interesting resonances.

Development of finite-volume formalism to extend into the three-body sector has been underway for some time, making use of several approaches to three-body scattering, and they are now converging to a consensus, as reviewed in Ref. [558]. The resulting formalism is, as one might expect, significantly more complicated than in the two-body case, but the essential idea is still the same – the input from lattice is a set of discrete energy

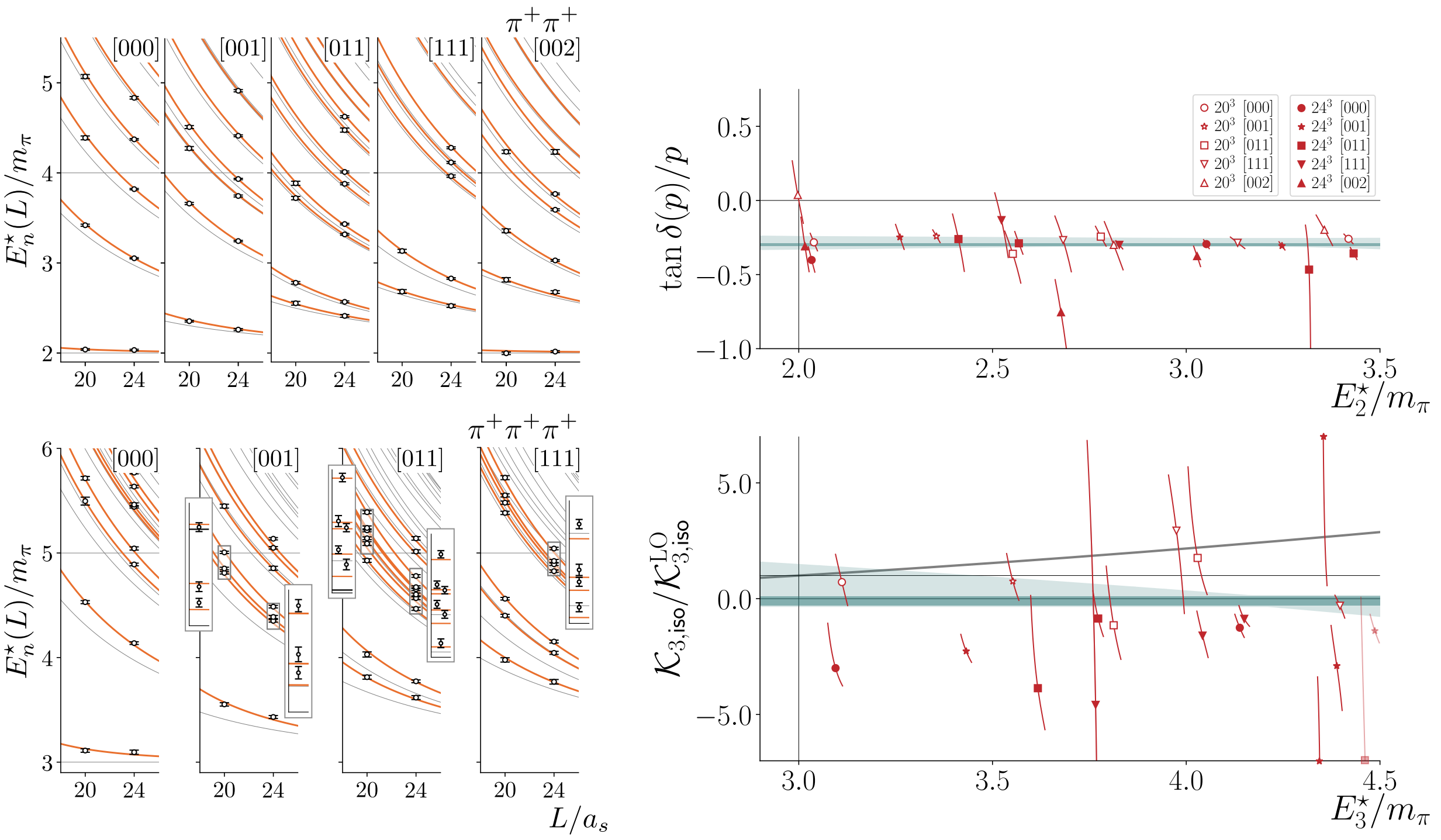

**Fig. 4.5.9** A lattice QCD determination of the spectra in two volumes of isospin–2 $\pi\pi$ and isospin–3 $\pi\pi\pi$ with $m_\pi = 391$ MeV. Orange curves show a description of these spectra using two-body and three-body finite-volume formalism with the amplitudes shown on the right. Taken from Ref. [556].

levels, now computed in channels with the quantum numbers of a three-hadron system.

The lattice QCD determinations of the finite-volume spectra follow a similar pattern to those described above, including now operators resembling systems of three-hadrons, but these are relatively straightforward to compute. The first investigations have focussed mostly on systems of maximal isospin [556, 559–562], e.g. $\pi\pi\pi$ with isospin–3, where there are no resonances either in the three-hadron system, nor in the two-hadron subsystems.

An example is presented in Figure 4.5.9 where we see discrete lattice QCD energy levels in two volumes for the $\pi\pi$ isospin–2 system and the $\pi\pi\pi$ isospin–3 system. These spectra can be described by two-body and three-body scattering amplitudes propagated through the finite-volume formalism, as shown by the orange curves. The amplitudes, as shown on the right of the figure (see the paper for the definition of the quantities plotted), are essentially structureless as expected in this non-resonant system. With proof-of-principle calculations like this one now done, the field is moving towards cases in which there are resonances, either in two-body subchannels, or in the three-body system, or both.

### 4.5.9 Summary

The progress in applying lattice to problems in hadron spectroscopy, as illustrated in this volume, suggests we have the beginnings of a rigorous foundation for the subfield, grounding it in first-principles QCD. The experimental hadron spectrum is already well studied, and there is a considerable corpus of model-based understanding, with which the lattice effort has to catch up. But already, with examples like the hybrid meson spectrum, lattice calculations are resolving long-standing conflicts. The ability to resolve excited hadrons as they truly are, as unstable resonances, makes a more direct connection to experiment possible, and the fact that calculations are possible of quantities which cannot be easily reached in experiments, like resonance form-factors, provides an opportunity to explore the internal structure of states that are otherwise poorly understood.

## 4.6 Hadron structure

**Martha Constantinou and K. Orginos**

The structure of the nucleon has been a central component to the development of QCD. Fundamental prop-erties of strong interactions, such as asymptotic freedom, were discovered while trying to unravel the nature of the nucleon. Hofstadter's elastic electron scattering experiments [563] discovered the first indications of a complex structure inside the proton. Later on, Deep Inelastic Scattering (DIS) discovered that partons, the constituents of the nucleon, are nearly free at short distances and led to the discovery of asymptotic freedom. Confinement, the fact that partons cannot break free from a hadron, is also a property of strong interactions that emerges from the study of hadronic structure. It was asymptotic freedom that eventually convinced theorists that QCD can describe the rich phenomenology of strong interactions.

Since its first exploration more than half a century ago, hadronic structure continues to be studied intensely both experimentally and theoretically. Theoretical studies include computations of various hadronic properties using lattice QCD, which offers a powerful non-perturbative, and systematically improvable way of computing fundamental properties of hadrons. This section summarizes the current status of lattice QCD calculations relevant to hadron structure. We start from simple observables such as nucleon charges which are important matrix elements for searches for physics beyond the standard model. We then proceed to a review of computations of nucleon form factors which are observables that give us information about the low energy structure of the hadron. Finally, we discuss modern methods for obtaining distribution functions from lattice QCD. Parton distribution functions are the simplest of such observables, which are relevant to understanding high-energy scattering experiments and give us a one-dimensional picture of the hadron. Generalized parton distribution functions (GPDs) and Transverse Momentum dependent distributions (TMDs) and their determination from lattice QCD will also be discussed.

### 4.6.1 Nucleon Charges

Nucleon matrix elements of local quark bi-linear operators of the form $\mathcal{O}_{\Gamma,\tau}(t) = \bar{q}(t)\Gamma\tau q(t)$ define the nucleon charges. Here $\Gamma$ is a general spin matrix and $\tau$ a flavor matrix. Isovector charges are obtained when $\tau = \tau_3$ the diagonal flavor Pauli matrix, while flavor diagonal charges are defined with an appropriate choice of $\tau$ that selects individual flavors. Nuclear matrix elements are obtained through computations of three-point functions of the form

$$C^{s,s'}_{\Gamma,\tau}(t',t) = \langle N^s(t)\mathcal{O}_{\Gamma,\tau}(t')\overline{N}^{s'}(0)\rangle\,, \tag{4.6.1}$$

where $N^s(t)$ is a nucleon interpolating field at time $t$, with helicity $s$ and projected to zero momentum. Typical nucleon interpolating fields can be written as $\sum_{abc,ijk}\epsilon_{abc}C^s_{ijk}q^a_i q^b_j q^c_k$ with $C^s_{ijk}$ appropriate weights. For a discussion of how these weights are obtained, see Ref. [564]. In the limit of $t \gg t' \gg 0$ the above correlator can be written as

$$C^{s,s'}_{\Gamma,\tau}(t',t) = z^s(z^{s'})^*\langle s|\mathcal{O}_{\Gamma,\tau}|s'\rangle e^{-M_N t} \tag{4.6.2}$$

where $\langle s|\mathcal{O}_{\Gamma,\tau}|s'\rangle$ is the desired nucleon matrix element and $M_N$ is the nucleon mass and $z^s$ is the overlap factor $\langle 0|N^s|s\rangle$. Using appropriate fitting procedure together with a nucleon two-point function

$$C(t) = \langle N^s(t)\overline{N}^s(0)\rangle = z^s(z^s)^* e^{-M_N t} \tag{4.6.3}$$

one obtains the desired matrix element. In general, these matrix elements require renormalization to obtain the matrix element at a given scale $\mu$ in a particular renormalization scheme. For a review of various methods used in lattice QCD to renormalize quark bi-linear operators, we refer the reader to Ref. [565]. Following this procedure, the nucleon charges have been obtained from lattice QCD. The isovector and flavor diagonal charges are essential quantities that, together with experimental observation, can constrain Beyond the Standard Model (BSM) theories. Therefore a significant effort in lattice QCD has been devoted to precise computations of the nucleon charges.

Establishing the lattice formulation of QCD requires that experimentally well-known quantities are correctly reproduced from numerical simulations. The axial charge of the nucleon, $g_A$, falls under this category and has been under investigation for several years. The field exhibits tremendous progress and among the highlights is the calculation of $g_A$ with controlled statistical uncertainties. The Flavour Lattice Averaging Group (FLAG) periodically reviews lattice results on several quantities, including $g_A$, and produces the FLAG averages. In Fig. 4.6.1, we provide a summary plot of lattice calculations [63] demonstrating that lattice results have improved in accuracy over the years and recent calculations at the physical point agree with the experimental average.

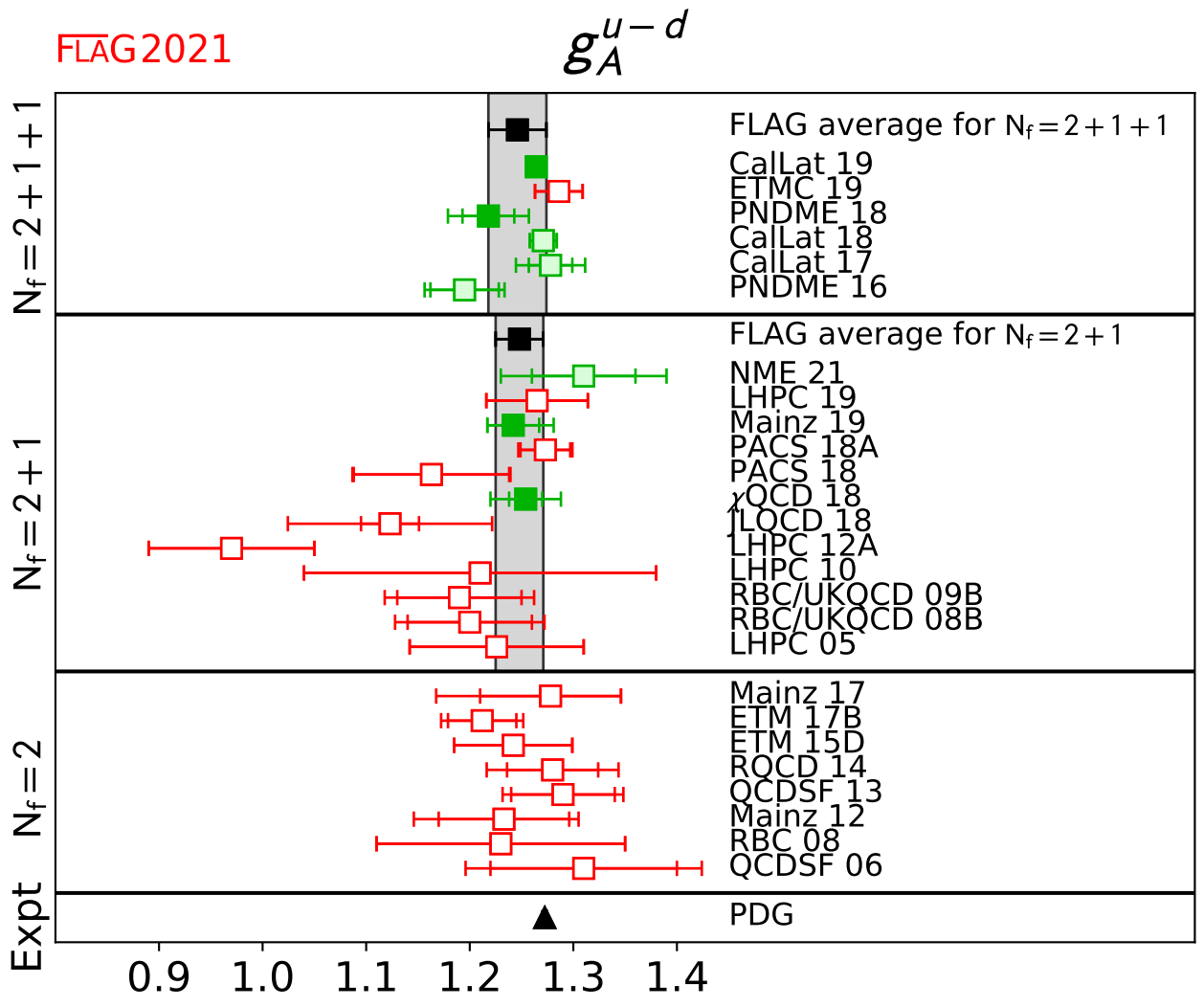

**Fig. 4.6.1** Lattice QCD determinations of the isovector axial charge compared to the experimental world average is taken from PDG. Figure from Ref. [63], and reprinted based on the arXiv distribution license.

The overall progress stimulated an intense activity in the field of hadron structure with the study of a large class of observables, some of which are known experimentally, but many that are still unexplored or difficult to measure [63, 566]. The investigations include nucleon charges such as the tensor and scalar and form factors for mesons and baryons. Selected results with simulations at physical quark masses can be found in Refs. [63, 566].

### 4.6.2 Nucleon Form factors

The Nucleon form factors are important properties of the nucleons that are essential for understanding their interactions in low-energy scattering experiments. They convey information about the internal structure of the hadron and their response to external probes, such as electromagnetic and weak currents. Properties such as the internal distribution of electric currents and charge and the size of the hadron can be deduced from electromagnetic form factors. Axial form factors describe the response of the hadron to external weak interaction probes. Future experiments, such as DUNE at Fermilab [567] and Hyper-Kamiokande [568], that aim to understand the properties of neutrinos, will require precise knowledge of the Nucleon axial form factors in order to achieve the precision they aim for. Therefore, lattice QCD computations of the Nucleon form factors are deemed essential and are vigorously pursued by several groups at this point. Advances in lattice QCD methods and computer hardware make such computations possible with sufficient precision to impact phenomenology [569].

Nucleon form factors are matrix element computations that require 3-point function computations

$$C^{s,s'}_{\Gamma,\tau}(t',t;\vec{p},\vec{p}') = \langle N^s(\vec{p},t)\mathcal{O}_{\Gamma,\tau}(t')\overline{N}^{s'}(\vec{p}',0)\rangle \,, \quad (4.6.4)$$

where $\vec{p}',\vec{p}$ are the initial and final momenta of the hadrons. In the limit of $t \gg t' \gg 0$, the above correlator can be written as the matrix element associated with the form factor, which emerges as:

$$C^{s,s'}_{\Gamma,\tau}(t',t;\vec{p},\vec{p}') = z(p)^s z(p')^{s'*} e^{-E(p)(t-t')} \times\langle s,\vec{p}|\mathcal{O}_{\Gamma,\tau}|s',\vec{p}'\rangle e^{-E(p')t'} \quad (4.6.5)$$

where $E(p)$ is the energy of the nucleon with momentum $p$ and $z^s$ is the overlap factor $\langle 0|N^s(\vec{p})|s,\vec{p}\rangle$. The matrix element $\langle s,\vec{p}|\mathcal{O}_{\Gamma,\tau}|s',\vec{p}'\rangle$ is related to the appropriate form factor for the operator $\mathcal{O}_{\Gamma,\tau}$ and is extracted with appropriate fitting methodology (see Refs. [569–571] for details of some of these methods).

In the case of the electromagnetic form factor where $\Gamma = \gamma_\mu$ and the flavor matrix combines the flavors of quarks with their appropriate charges, the matrix element is

$$\langle s,\vec{p}|\sum_f e_f\bar{q}_f\gamma_\mu q_f|s',\vec{p}'\rangle = \bar{U}(\vec{p})\left[F_1(Q^2) + \frac{i\sigma_{\mu\nu}}{2M}q^\nu F_2(Q^2)\right]U(\vec{p}') \quad (4.6.6)$$

where $U(\vec{p})$ is the spinor associated with the nucleon, $q_\mu = p_\mu - p'_\mu$, $Q^2 = -q^2$, and $F_1$, $F_2$ the two Lorentz invariant Dirac and Pauli form factors. The electric and magnetic form factors are defined as

$$G_E(Q^2) = F_1(Q^2) - \frac{Q^2}{4M^2}F_2(Q^2)$$
$$G_M(Q^2) = F_1(Q^2) + F_2(Q^2)\,. \quad (4.6.7)$$

With these form factors we can define the charge radius $\langle r_E^2\rangle$ and the magnetic radius $\langle r_M^2\rangle$ of the nucleon as

$$\langle r_E^2\rangle = -6\frac{dG_E(Q^2)}{dQ^2}\bigg|_{Q^2=0}$$
$$\langle r_M^2\rangle = -\frac{6}{G_M(0)}\frac{dG_M(Q^2)}{dQ^2}\bigg|_{Q^2=0}\,. \quad (4.6.8)$$

Because of the finite volume in lattice QCD computations, the form factors are only known on a set of discrete points. The full $Q^2$ dependence is recovered by fitting the data points to particular phenomenologically motivated forms. The simplest such form is the dipole:

$$F_{\text{dipole}}(Q^2) = \frac{r_F}{\left(1+\frac{Q^2}{M_F^2}\right)^2}\,, \quad (4.6.9)$$

where $r_F$ is the residue and $M_F^2$ is a mass parameter associated with the form factor at hand. This simple parametrization works well for the lattice calculations that are typically restricted to low $Q^2$. Recently the z-expansion [572] given by

$$F(Q^2) = \sum_{k=0}^{\infty} a_k\, z(Q^2)^k\,, \quad (4.6.10)$$

with

$$z(Q^2) = \frac{\sqrt{t_{\text{cut}}+Q^2}-\sqrt{t_{\text{cut}}-t_0}}{\sqrt{t_{\text{cut}}+Q^2}+\sqrt{t_{\text{cut}}-t_0}}\,, \quad (4.6.11)$$

has been employed for a more flexible parametrization. The position of the cut, $t_{\text{cut}}$, is the time-like kinematic threshold for particle production associated with the current whose form factor is discussed. The parameter $t_0$ is the point in $Q^2$ that is mapped to $z = 0$ and is chosen for convenience.

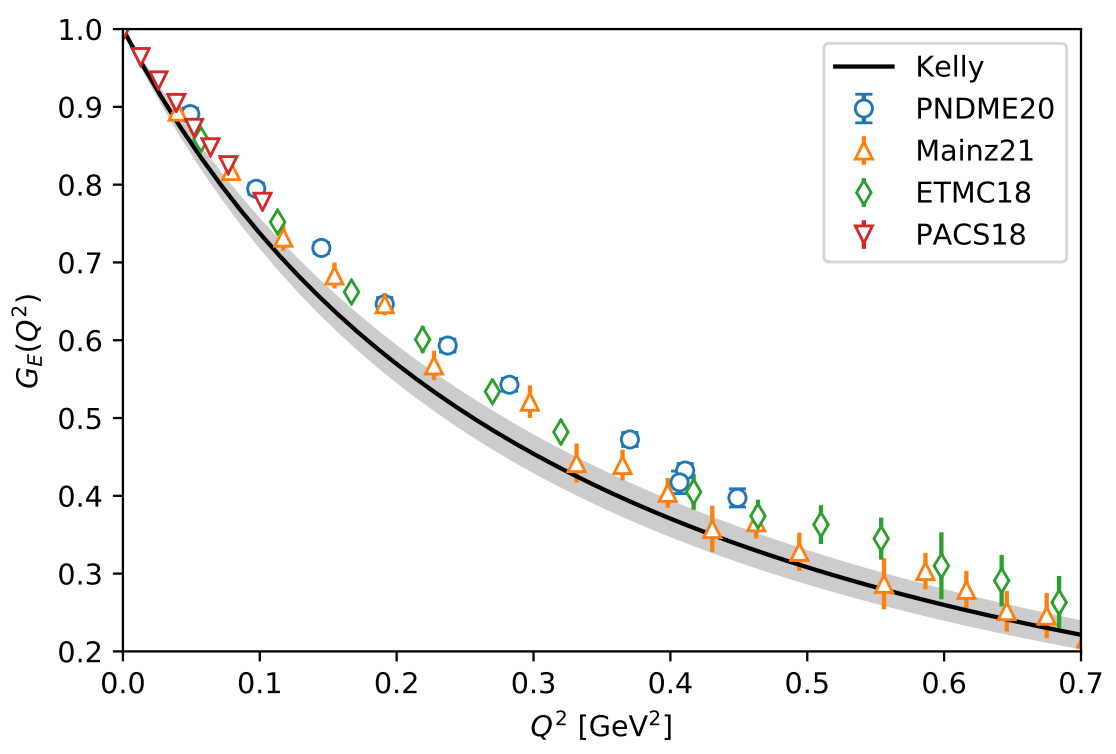

**Fig. 4.6.2** Status of recent lattice QCD results for the isovector nucleon electric form factor in comparison with the Kelly parametrization of experimental results (figure from Ref. [573]). Reprinted under the terms of the Creative Commons Attribution 4.0 International license.

Multiple lattice QCD collaborations have recently computed the nucleon vector form factors. Several lattice collaborations have recently computed the isovector electric form factor (i.e., the difference between the proton and the neutron form factors). After many years of study of various systematics involved, we now have computations with physical quark masses, careful analysis of excited state contamination of the ground state matrix element, and large enough volumes to avoid finite volume effects. In Fig. 4.6.2, the lattice data together with the Kelly parametrization [574] of experimental results are presented. The lattice data of PNDME20 [571] are plotted as blue circles, the Mainz21 [575] data are the orange triangles, the ETMC18 [576] data are the green diamonds, and the PACS18 [577] data are the red triangles. All these calculations are performed with different methodologies and approaches in treating excited state effects and varying fermion actions in both the sea and the valence sectors. PNDME20 uses the HISQ action in the sea sector and smeared Clover action in the valence sector. The ETMC18 calculations use the twisted mass action. Both the Mainz21 and the PACS18 collaborations use Clover fermion actions. Clearly, there are some tensions between various collaborations that will be resolved in future, more refined calculations. However, it should be noted that there is a fairly good agreement between the state-of-the-art calculations and experiment.

Lattice QCD computations of the form factors can lead to the determination of the radii of the nucleon. In addition, direct methods of determining the nucleon radii also exist. Lattice QCD calculation results for the magnetic and the charge isovector radius of the nucleon are presented in Fig 4.6.3. In this figure, the magenta right triangles are PNDME20 [571] using the mixed actions with Clover on HISQ, and the green triangles are from ETM18/20 [576, 578] using the twisted-mass action. Calculations using the Clover fermion action are represented by the maroon octagons [575] from Mainz21, the blue diamonds from PACS18/20 [577, 579], the red circle is from LHPC17, and the magenta left triangles from NME21 [570]. Note that results from [578, 579] are obtained with methods that directly estimate the slope of the form factor at $Q^2 = 0$. The vertical bands represent the phenomenological values for the radii obtained from the experiment by combining data from the proton and the neutron. In particular the isovector charge $r_E^{iv}$ and magnetic $r_M^{iv}$ radii are given by

$$r_E^{iv} = \sqrt{r_{Ep}^2 - r_{En}^2} \ , \quad r_M^{iv} = \sqrt{\frac{\mu_p r_{Mp}^2 - \mu_n r_{Mn}^2}{\mu_p - \mu_n}} \ , \tag{4.6.12}$$

where $r_{Ep}^2$, $r_{En}^2$ are the proton and neutron charge radii, $r_{Mp}^2$, $r_{Mn}^2$ are the proton and neutron magnetic radii, and $\mu_p$, $\mu_n$ are the proton and neutron magnetic moments. By combining results exclusively from the particle data group (PDG) [278] we obtain the red bands. For the charge radius, the cyan band is obtained by using the CODATA2018 value for the proton charge radius and the neutron charge radius from the recent work in [580]. The cyan band for the magnetic radius was obtained using the proton radius obtained by [581] and the rest of the needed quantities from PDG.

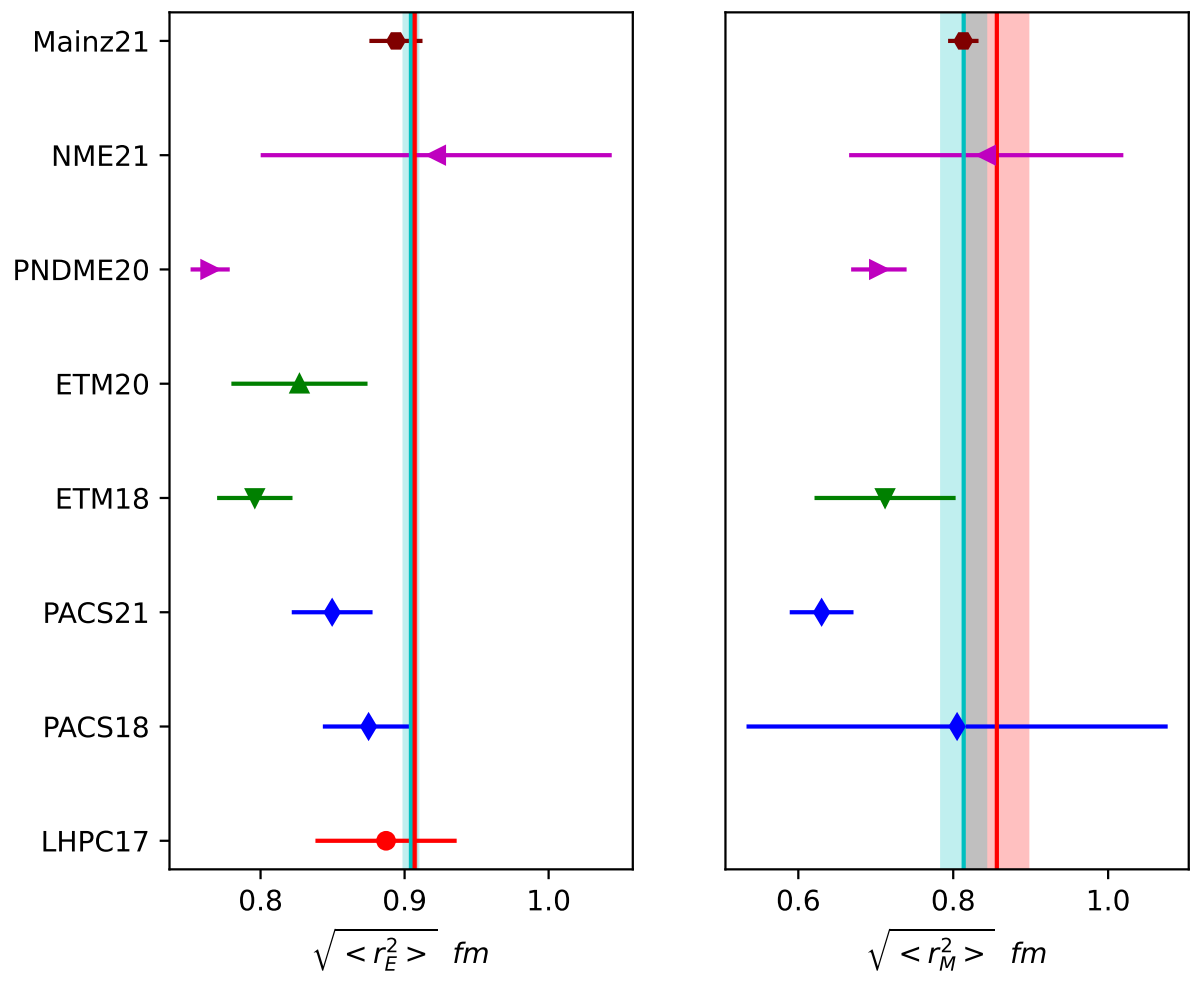

**Fig. 4.6.3** Lattice results for charge magnetic radii of the nucleon. The vertical bands are the estimates from experiment (see text for details).

In the case of the isovector axial form factors, one can take $\tau^+$ as the flavor matrix and $\Gamma = i\gamma_5\gamma_\mu$ and the resulting matrix element is

$$\langle s, \vec{p}|\bar{q}(i\gamma_5\gamma_\mu\tau^+)q|s', \vec{p}'\rangle = \bar{U}(\vec{p})\left[F_A(Q^2) + \frac{q_\mu}{M}\gamma_5 F_P(Q^2)\right]U(\vec{p}'), \quad (4.6.13)$$

with $F_A$ and $F_P$ being the corresponding invariant form factors. In Fig 4.6.4, recent lattice QCD computations of the axial form factor are presented. The red band denotes the parametrized experimental results from neutrino deuteron scattering [582]. The purple band are results from NME21 [570], and the green band are results from RQCD20 [583], where continuum, chiral and finite volume extrapolations have been performed. The rest contain results [577, 579, 584–587] from a few ensembles and are presented as points without the interpolating curves. It is clear that although there is tension between lattice QCD results and experiment, lattice QCD calculations are consistent with each other. As it is argued in Ref. [569] that lattice QCD calculations of axial nucleon form factors may play an essential role in future experiments and thus help us better understand neutrino physics.

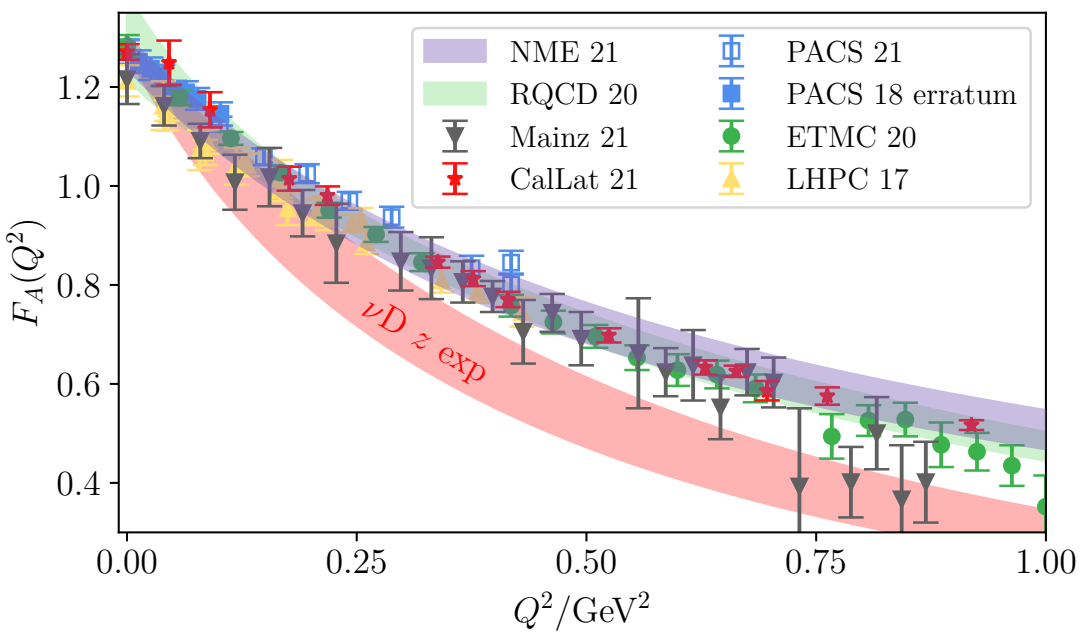

**Fig. 4.6.4** Lattice QCD results for the nucleon axial form factor compared to the experimental results from neutrino deuteron scattering. Figure from Ref. [569] and reprinted under the terms of the Creative Commons Attribution 4.0 International license.

### 4.6.3 Partonic Structure

Information on the internal structure of hadrons is obtained through their partonic content, particularly parton PDFs, GPDs, and TMDs (see Sec. 10). These quantities are light-cone correlation functions and cannot be calculated using the Euclidean formulation of lattice QCD due to the rotation $t \to i\tau$. The most common avenue to proceed is to calculate Mellin moments of distribution functions, which provide partial information on distribution functions.

Lattice QCD calculations have focused on proton charges, vector, and axial form factors, that are, the first Mellin moments of PDFs and GPDs, respectively. There are also limited studies of the scalar and tensor charges, as well as the second Mellin moments of PDFs and GPDs.

In theory, one can use a large number of Mellin moments to reconstruct the parton distributions using an operator product expansion (OPE). Practically, a proper and exact reconstruction is not possible due to the challenges of calculating reliably high moments; the signal-to-noise rapidly decreases, and an unavoidable power-law mixing occurs beyond the fourth moment [588–592]. Therefore, alternative methods are needed to obtain the $x$ dependence of distribution functions from a Euclidean formulation. The realization that matrix elements of momentum-boosted hadrons coupled with bilinear non-local operators can be related to light-cone distributions has transformed the field of PDF, GPDs, and TMDs calculations. The pioneering method of Large-Momentum Effective Theory (LaMET) that uses the aforementioned non-local operators has renewed the interest of the community to access the $x$ dependence of parton distributions. Over the years, there have been several methods proposed: a technique based on the hadronic tensor [593–595], auxiliary quark field approaches [596–598], a method to obtain high Mellin moments using smeared operators [599], LaMET [600, 601], pseudo-ITD [602], current-current correlators [603–605], and a method based on OPE [606].

In this review, we highlight selected results demonstrating the field's progress. More details can be found in the recent reviews [607–611].

*Isovector PDFs*

The isovector leading-twist PDFs have been the most well-studied and serve as a benchmark of the various methodologies to extract $x$ dependence from lattice data. Results with ensembles at physical quark masses have already been obtained for the unpolarized [612–615], helicity [612, 613, 616] and transversity [613, 617, 618] PDFs for the proton. Here we focus on the unpolarized case that has the most results allowing comparison between different methods and lattice formulations. The work of Ref. [613] uses a twisted-mass fermions ensemble with physical pion mass and employs the quasi-PDFs method. The lattice spacing is about 0.09 fm, and the nucleon momentum boost is up to 1.4 GeV. The unpolarized PDF of Ref. [614] has been obtained using the pseudo-ITD framework on three clover Wilson ensembles with pion mass 172, 278, and 358 MeV;

a chiral extrapolation has been performed to get the physical point. The pseudo-ITD methodology computes the Lorentz invariant amplitudes that contribute to the non-local matrix element and isolates the amplitude that contains the leading twist contribution. This amplitude is a function of the so-called Ioffe time $\nu$, which is the Fourier-dual of the momentum fraction $x$ [619–621]. The analysis of [614] includes lattice data up to Ioffe time $\nu = 8$ for the near-physical mass ensemble. Finally, the work of Ref. [615] extends and reanalyzes the data of Ref. [613] within the pseudo-ITD framework with up to $\nu = 8$. Having three independent calculations of the unpolarized PDF allows one to compare them and understand potentially systematic effects related to the method and computational setup. Such a comparison can be found in Ref. [614], which we include in Fig. 4.6.5. A good agreement is observed between the different calculations, which is very encouraging, as each methodology may suffer from different systematic effects.

*Gluon PDFs*

In general, gluon contributions are limitedly studied due to the enhanced gauge noise, the involvement of disconnected diagrams, and the challenges in the non-perturbative renormalization. In the case of $x$-dependent gluon PDFs, the renormalization cancels out using the pseudo-ITD method, which is a significant advantage. Recently, there have been calculations of the gluon PDF for the proton and the pion using the pseudo-ITD method [622, 623]. Ref. [622] presents a calculation using clover fermions at a pion mass $m_\pi = 358\,\text{MeV}$. One novelty of the calculation is the use of the momentum-smeared distillation technique [624] to suppress gauge noise. The work also employs Jacobi polynomials to reconstruct the $x$ dependence of the distribution [625]. The main results are shown in Fig. 4.6.6. The work of Ref. [623] presents a calculation of the gluon PDF for the pion using two HISQ coarse ensembles ($a = 0.12, 0.15$ fm) and pion masses $m_\pi = 220$, 310, 690 MeV. While the current status of gluon PDFs is exploratory, the available results are promising.

*Individual quark PDFs*

Calculations of individual-quark PDFs are challenging due to the involvement of disconnected diagrams that increases the statistical fluctuations of the correlators. The flavor decomposition of quark PDFs is interesting in its own right but is also needed to form the flavor-singlet combination to eliminate mixing with the gluon PDF. The mixing holds only for the unpolarized and helicity cases; there is no gluon transversity. Furthermore, the strange and charm quark PDFs are more susceptible to mixing as they enter the sea sector from gluon splitting. The effect of mixing is expected to be smaller for the light quarks that appear in the valence sector of the proton. The individual light quark unpolarized, helicity, and transversity PDFs were calculated in Refs. [629, 630] using an ensemble of twisted mass fermions at $m_\pi = 260$ MeV. The work shows that disconnected contributions to the unpolarized and transversity PDFs are tiny and can be neglected. However, calculations at the physical value of the quark masses are needed to confirm this. Refs. [629, 630] include the strange quark contributions, which may have increased systematic effects due to the mixing with the gluon PDFs. The same holds for Ref. [631] (clover on HISQ, $m_\pi = 220$, 310, 690 MeV), which calculates the strange and charm quark PDFs for the proton.

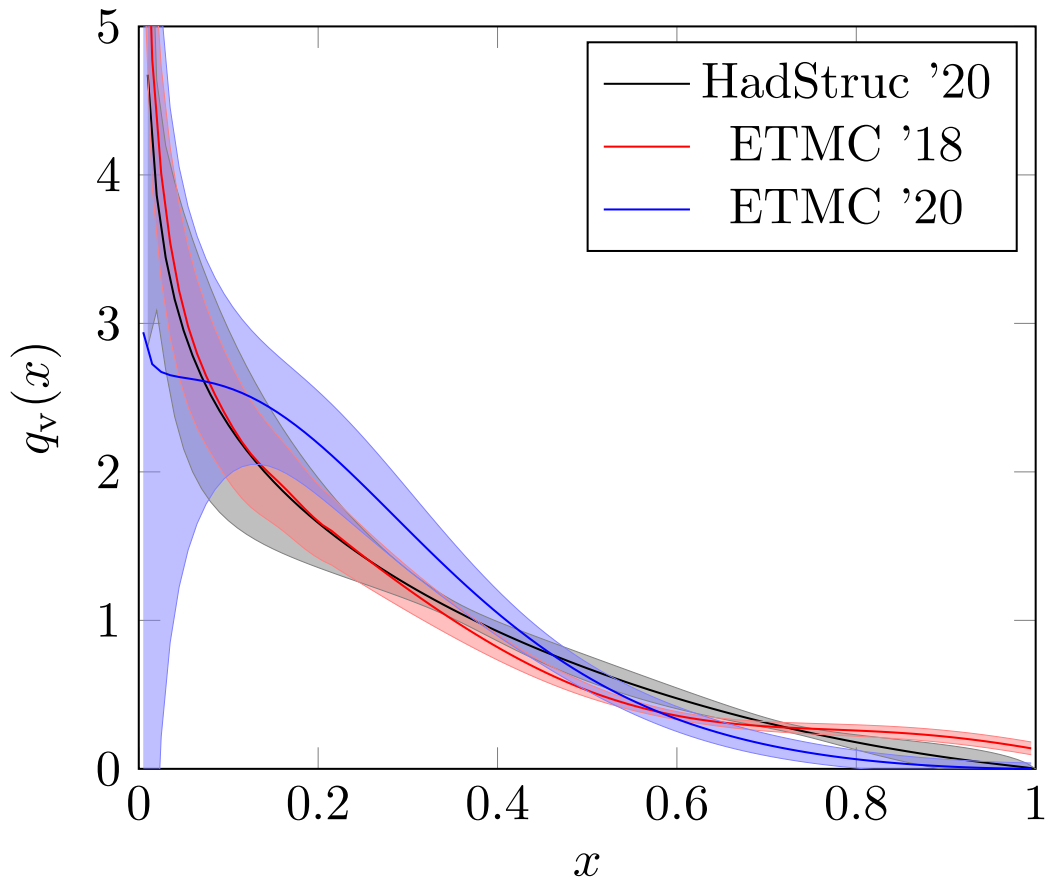

**Fig. 4.6.5** Lattice results for the unpolarized PDF using quasi-PDFs [613] (red band) and pseudo-ITDs from Ref. [614] (gray band) and Ref. [615] (blue band). Plot from Ref. [614]. Reprinted under the terms of the Creative Commons Attribution 4.0 International license.

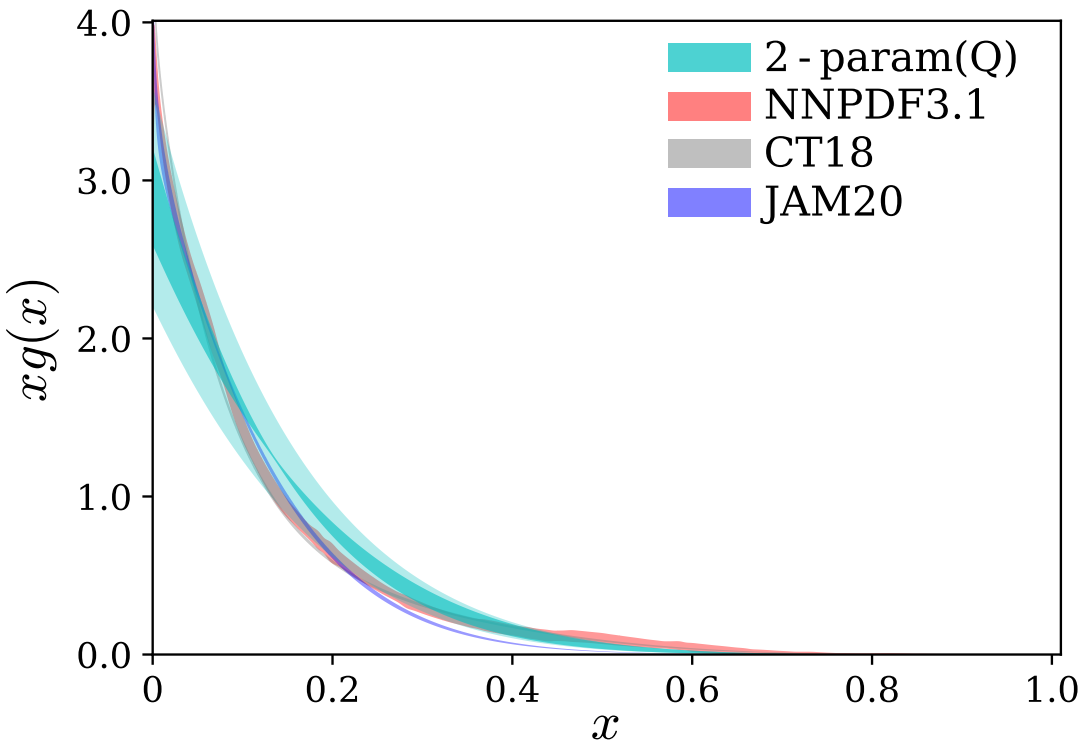

**Fig. 4.6.6** Lattice QCD results on the gluon PDF from Ref. [622] (cyan band) compared to estimates from global analyses [626–628]. Reprinted under the terms of the Creative Commons Attribution 4.0 International license.

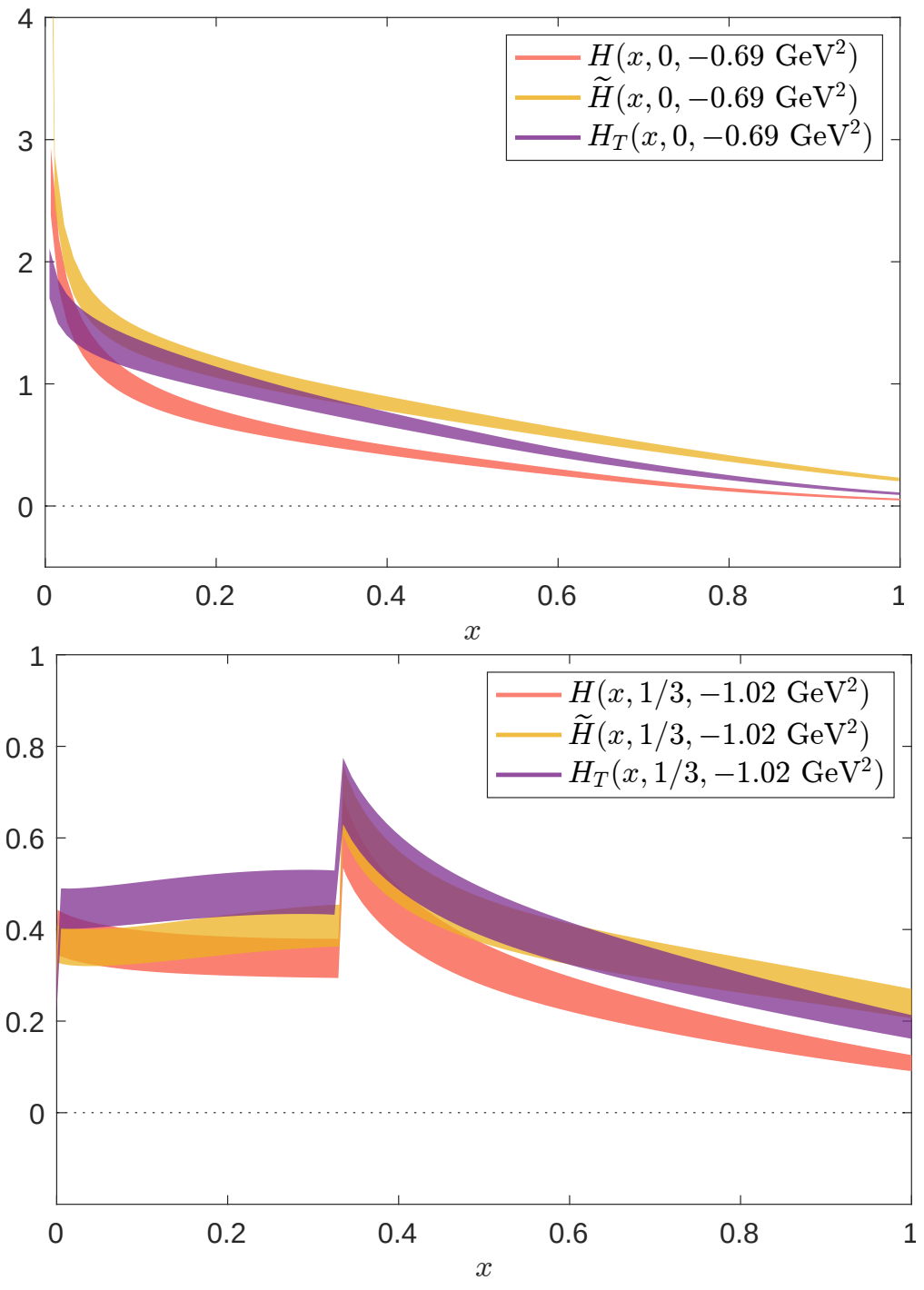

**Fig. 4.6.7** Top: $H$, $\widetilde{H}$, $H_T$ GPDs $t = -0.69\,\mathrm{GeV}^2$, $\xi = 0$. Bottom: $H$, $\widetilde{H}$, $H_T$ GPDs $t = -1.02\,\mathrm{GeV}^2$, $\xi = 1/3$. The unpolarized, helicity, and transversity data are shown with red, yellow, and purple bands, respectively. Figure from Ref. [633] and reprinted under the terms of the Creative Commons Attribution 4.0 International license.

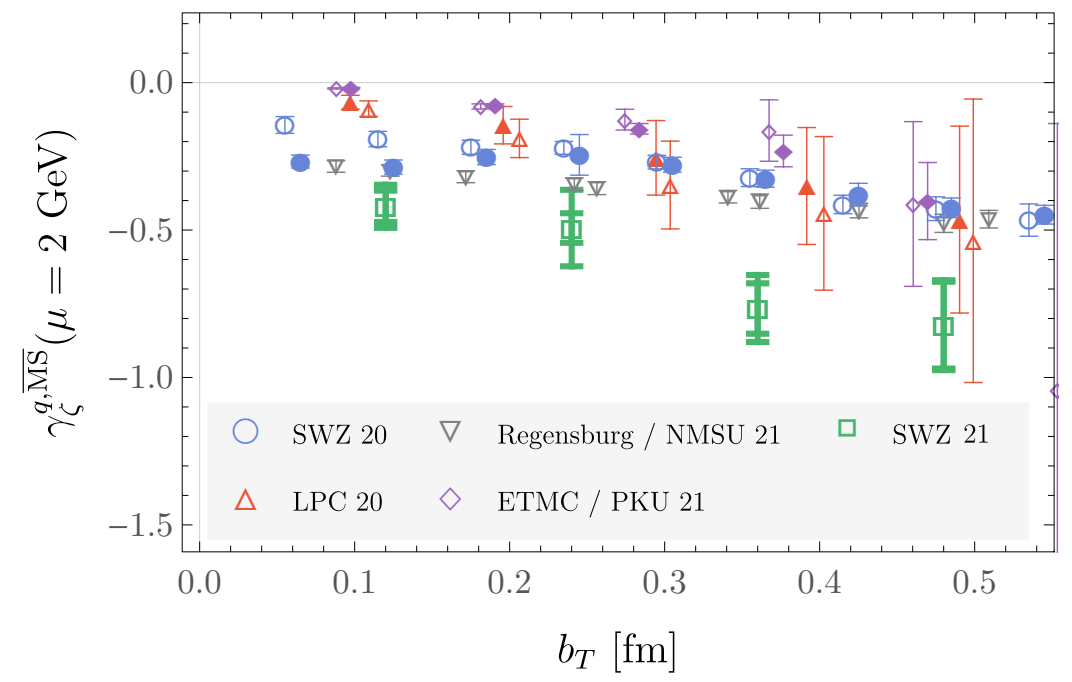

**Fig. 4.6.8** Lattice QCD determinations of the Collins-Soper evolution kernel obtained from Ref. [635] (SWZ 20), Ref. [636] (LPC 20), Ref. [637] (Regensburg/NMSU 21), and Ref. [638] (ETMC/PKU 21), and Ref. [639] (SWZ 21). Figure adapted from Ref. [639] and reprinted under the terms of the Creative Commons Attribution 4.0 International license.

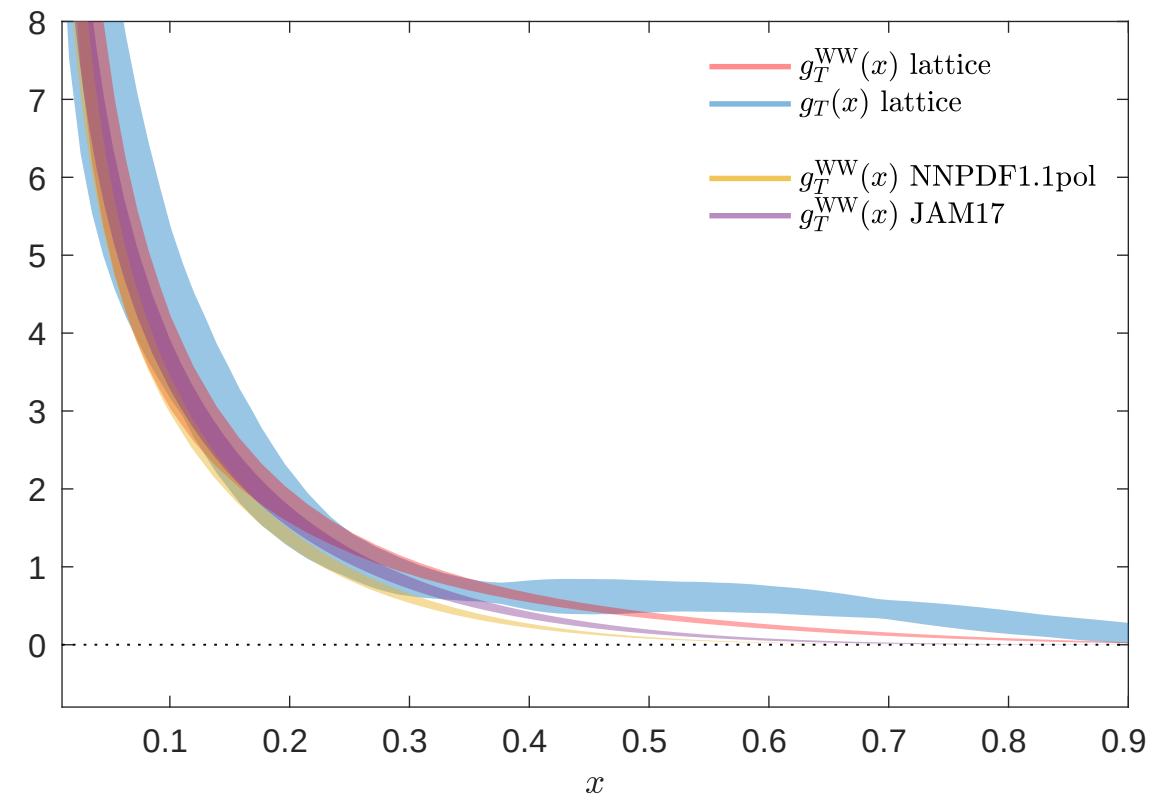

**Fig. 4.6.9** The Wandzura-Wilczek approximation for $g_T$. Figure from Ref. [640] and reprinted under the terms of the Creative Commons Attribution 4.0 International license.

*GPDs*

Another progress for lattice QCD is related to calculating $x$-dependent GPDs. These are computationally more expensive than PDFs due to the momentum transfer between the initial and final hadronic states. The momentum transfer must be equally split between the initial and final states, as the GPDs are defined in the symmetric frame; such a frame is computationally costly, preventing the extraction of GPDs for a dense set of values of $t$. A novel approach that related light-cone GPDs to Lorentz-invariant amplitudes has been recently proposed [632]. First results on the proton unpolarized and helicity GPDs have been obtained using the quasi-distribution approach [629]. The calculation is performed on a 260 MeV pion mass ensemble of twisted mass fermions. The work was extended for the chiral odd twist-2 GPDs in Ref. [633]. In Fig. 4.6.7, we compare the three types of GPDs for zero and nonzero skewness. As can be seen, the introduction of nonzero skewness leads to the appearance of a nontrivial ERBL region. Another calculation of the unpolarized GPDs can be found in Ref. [634], which was originally reported in a non-symmetric frame similar to the one used for frame-independent form factors.

*TMDs*

Unlike PDFs and GPDs, TMDs contain, in addition, rapidity divergences that require regularization. The regulator is encapsulated within the so-called soft function. The evolution in rapidity of the soft function can be studied separately through the Collins-Soper (CS) kernel. Aspects of the soft function are actively studied in lattice QCD [631, 635–639, 641], which is the ideal formulation as the soft-function is a non-perturbative quantity. A summary plot for the CS kernel is shown in Fig. 4.6.8.

*Higher-twist*

One of the latest developments in extracting $x$-dependent distribution functions is the exploration of twist-3 PDFs and GPDs that contain information on quark-gluon-quark correlations [642]. They are also related to the transverse force acting on transversely polarized quarks [643] and to the nuclear electric dipole moments [644]. First exploratory studies of twist-3 PDFs $e(x)$, $g_T(x)$, and $h_L(x)$ can be found in Refs. [640, 645–647], with numerical results for $g_T(x)$ and $h_L(x)$. An interesting investigation of twist-3 PDFs is the Wandzura-Wilczek (WW) approximation [648] according to which the twist-3 $g_T$ can be fully determined by its twist-2 counterpart, $g_1$. The WW approximation can also be studied for $h_L$. In Fig. 4.6.9 one can see $g_T^{\mathrm{WW}}$, demonstrating that the approximation holds in some regions of $x$, but an overall violation of up to 40% is permitted. Note that the 2-parton twist-3 PDFs mix with quark-gluon-quark correlations and the mixing should be addressed within the matching kernel [649, 650].

### 4.6.4 Outlook

Since the early days of lattice QCD in the 1980s, hadron structure calculations have been pursued vigorously. Over the years, the methods used to perform these calculations have improved steadily, and the Monte Carlo methods for sampling the QCD vacuum have reached the degree of efficiency required for such computations. Furthermore, computer hardware has now reached the Exaflop era. As a result, calculations for hadron structure are now achieving unprecedented precision in some cases (ex., nucleon charges). In other cases, new horizons open up, such as the ability to compute the momentum fraction x-dependence of distribution functions. In the future, lattice QCD computations of hadronic structure will continue to improve and provide us with the theoretical input needed to understand strong interaction physics better.

## 4.7 Weak matrix elements

**Christine Davies**

Quarks have the special property that they experience all of the fundamental forces in the Standard Model. As well as exchanging the gluons that keep them confined into hadrons, quarks can also occasionally emit weak interaction $W$ bosons or QED photons. Because W and $\gamma$ have no colour charge they escape cleanly from the hadron, carrying valuable information about the structure of the bound state. This structure is determined by strong interaction physics and so predictions from QCD can be tested against experimental information on these processes. The number of different quark flavours, and hadrons constructed from them, makes a rich mine for lattice QCD to work in.

In the bigger picture of the Standard Model we need to determine accurately the couplings between quarks and the $W$ boson given by the elements of the CKM matrix ([651], Sec. 13.2). This programme is a crucial ingredient in constraining the possibilities for new physics beyond the Standard Model. However, quarks are not free particles when they emit $W$ bosons. The experimental measurement of appropriate hadronic weak decay rates allows us to determine CKM elements but only if, as discussed above, we have understood the strong interaction physics that confines the quarks through calculation of the appropriate *hadronic* matrix elements of the weak current. As we will see below, some of the experimental information for weak (and electromagnetic) decay rates is very accurate and correspondingly accurate theoretical calculations in QCD are needed to make the most of it. These have always been a high priority for lattice QCD. Results have improved over time to the point where uncertainties are now below 1% in some cases. We will discuss the current status below, and briefly mention developments that will lead to improvements in future.

### 4.7.1 Decay constants

Decay constants are the hadronic parameters that encode the amplitude for finding the valence quark and anti-quark of a meson at the same point. This is then the parameter that is needed to determine the rate of annihilation of mesons with appropriate flavour quantum numbers to a $W$ or $\gamma$ (see Fig. 4.7.1). For a pseudoscalar meson the decay constant, $f$, is defined from the vacuum to meson matrix element of the axial current. For meson $P$ of quark content $a\bar{b}$

$$\langle 0|\bar{a}\gamma_\mu\gamma_5 b|P(p)\rangle \equiv f_P p_\mu \,. \tag{4.7.1}$$

For a meson at rest, applying the partially-conserved axial current (PCAC) relation $\partial_\mu A^\mu = (m_a + m_b)P_s$ to relate axial-vector and pseudoscalar currents gives

$$(m_a + m_b)\langle 0|\bar{a}\gamma_5 b|P(\vec{p} = 0)\rangle = f_P M_P^2 \,, \tag{4.7.2}$$

where $M_P$ is the meson mass.

In lattice QCD the matrix element on the l.h.s. of Eqs. 4.7.1 or 4.7.2 is obtained from the two-point correlation function between source and sink $a\bar{b}$ currents (Sec. 4.2) with Euclidean time separation, $t$, between them. The two-point function has contributions, exponential in $t$, from a tower of $a\bar{b}$ mesons. The exponential corresponding to the ground-state (lowest mass) meson dominates at large $t$ and this is the meson for which the parameters of the fit, the amplitude and mass, are most precisely determined. This mirrors experiment, where accurate meson weak or electromagnetic annihilation rates are possible when strong-interaction decay channels are heavily suppressed (not usually true for excited states). Note, however, that lattice QCD can determine $f$ for mesons which do not have the flavour quantum

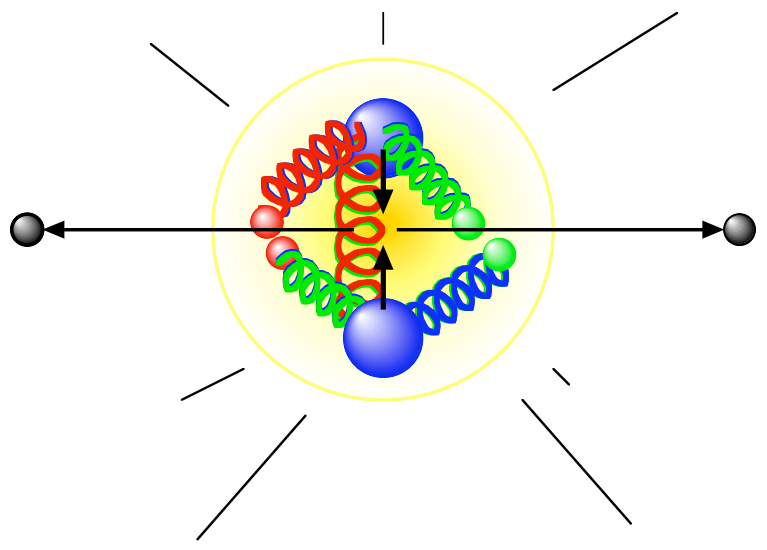

**Fig. 4.7.1** Schematic diagram of a meson annihilation to leptons via the coupling of the valence quark-antiquark pair to a $W$ or $\gamma$. The decay constant parameterises the amplitude to find the quark and antiquark at a point, the key hadronic information needed to determine the annihilation rate.

numbers to annihilate to $W$ or $\gamma$ – these results are still useful in other contexts.

The fit to the two-point function $C(t)$ gives both the ground-state meson mass and its decay constant. The contribution of the ground-state to $C(t)$ is

$$C(t) = a_0(e^{-M_0 t} + e^{-M_0(T-t)}) + \dots . \tag{4.7.3}$$

Here $T$ is the lattice time extent and … represents contributions from higher mass states. $M_0$ is the ground-state meson mass and the amplitude $a_0$ is given by

$$a_0 = (\langle 0|J|P_0\rangle)^2/(2M_0) \tag{4.7.4}$$

where $J$ is the current used at the source and sink of $C(t)$. The decay constant for $P_0$ can then be obtained from $a_0$ using Eqs. 4.7.1 or 4.7.2 as appropriate for $J$.

Decay constants for light pseudoscalar mesons ($f_\pi$ and $f_K$) have been calculable in lattice QCD with errors at the few percent level since 2004 [652]. This was one of the first calculations to be done once ensembles of gluon field configurations were available (from the MILC collaboration) that included $u$, $d$ and $s$ sea quarks with multiple values of the lattice spacing and light enough $u/d$ quarks for a reasonably well-controlled extrapolation to the physical continuum limit.

To achieve a small uncertainty in the result for the ground-state meson mass and decay constant it is important to have a large sample of correlators (to achieve small statistical errors) at multiple values of the lattice spacing using a discretisation of the QCD action with small discretisation errors (Secs. 4.2,4.1). An accurate determination of the lattice spacing (to convert $C(t)$'s fit parameters from lattice units to GeV) is needed. Attention must also be paid to the effect of the finite-volume of the lattice on the $\pi$ and $K$. Finite-volume (and discretisation) effects are incorporated into the chiral perturbation theory [653] used to fit the results as a function of $u/d$ quark mass (or $M_\pi$) to extrapolate to the continuum limit with physical quark masses. $M_\pi$

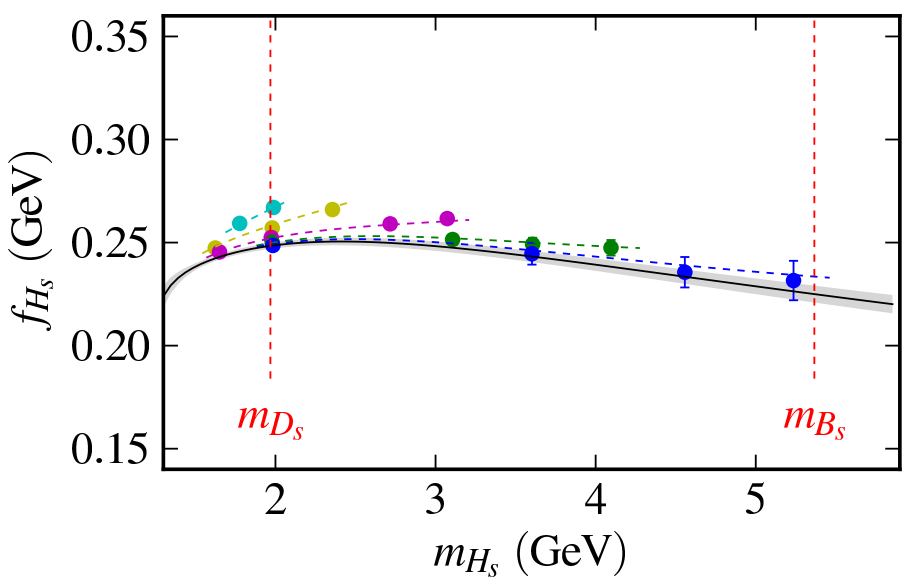

**Fig. 4.7.2** The decay constant of the heavy-strange pseudoscalar meson as a function of its mass from lattice QCD calculations [654] using the HISQ action [308]. Points with different colours are results for different lattice spacing values, with smaller lattice spacings having more reach to heavier masses. The grey curve is the continuum limit of an HQET-inspired fit to the results including discretisation effects. The result for $f_{B_s}$ can be read off at the mass of the $B_s$ meson.

is used to fix the average $u/d$ quark mass ($u$ and $d$ are taken to be degenerate in almost all calculations) and the physical value appropriate to a calculation in which the quark electric charges are ignored is the experimental value of the $\pi^0$ mass. $M_K$ fixes the $s$ quark mass and the physical value used is an average of the masses of $K^0$ and $K^+$ with an allowance for QED effects [652].

For decay constants a further important consideration is the normalisation of the axial vector current that appears in Eq. 4.7.1 so that it matches that of the continuum QCD current. For lattice QCD actions that have an exact PCAC relation (such as asqtad staggered quarks used in [652]) no renormalisation is needed. Rather than use the partially conserved axial current (which is a complicated point-split construction) it is easiest to use the pseudoscalar current, which is local, and calculate the decay constant directly from Eq. 4.7.2. The quark masses that appear in this expression are then the bare lattice quark masses being used in the calculation.

The key physics importance of the lattice QCD calculations of $f_\pi$ and $f_K$ is in determining the rate for $\pi^+/K^+$ annihilation to a $W$ boson, which can be measured accurately in experiment. The annihilation rate for meson $P$ with appropriate quark flavour quantum numbers is

$$\Gamma(P \to \ell\overline{\nu}) = \frac{G_F^2|V_{ab}|^2}{8\pi} f_P^2 m_\ell^2 M_P \left(1 - \frac{m_\ell^2}{M_P^2}\right)^2 \tag{4.7.5}$$

up to well-studied QED corrections. Only the $A$ of the $V-A$ weak interaction contributes in this case, so that $\Gamma \propto f_P^2$. $V_{ab}$ is the appropriate CKM element; this can be determined from the experimental measurement of $\Gamma$ given a value for $f_P$ from lattice QCD.

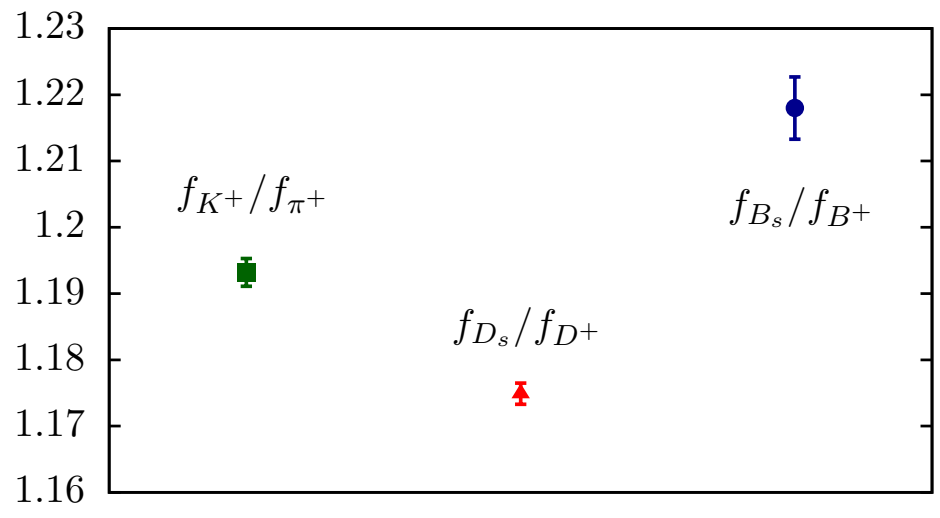

**Fig. 4.7.3** SU(3)-isospin breaking ratios of decay constants from lattice QCD. $f_K/f_\pi$ is from Eq. 4.7.6 [63], other results from Ref. [655].

Several systematic errors are reduced in an analysis of the ratio of widths for $K$ and $\pi$ [656]. This enables the ratio $|V_{us}|/|V_{ud}|$ to be determined and converted to a result for $|V_{us}|$ using accurate $|V_{ud}|$ values from superallowed nuclear $\beta$ decay [278]. Lattice QCD calculations have then largely concentrated on determining the ratio $f_K/f_\pi$, equivalent to fixing the lattice spacing from $f_\pi$. Following a great deal of work by the lattice community, current day results have improved to the point where the uncertainty on $f_{K^+}/f_{\pi^+}$ is reduced to 0.2%. The recent FLAG review [63] quotes an average of

$$f_{K^+}/f_{\pi^+} = 1.1932(21)\,, \quad n_f = 2+1+1 \qquad (4.7.6)$$

from lattice QCD results that include $u$, $d$, $s$ and $c$ quarks in the sea obtained in Ref. [655, 657–659]. The average is dominated by the result from the Fermilab Lattice/MILC collaborations [655]. The lattice calculations now include an analysis of the impact of the $u/d$ mass difference; work is ongoing to analyse QED effects on the lattice [660].

Heavier pseudoscalar mesons also annihilate to $W$s, giving access to other CKM elements. For example, the rate for $B \to \ell\overline{\nu}$ depends on $|V_{ub}|$ and $f_B$. The experimental determination of the decay rates is harder and they currently have larger uncertainties than for $K$ and $\pi$ [278]. On the lattice QCD side the heavier masses of the $c$ and $b$ quarks increase discretisation errors, since they take the form of powers of $ma$ for quark mass $m$. To counteract this lattice QCD theorists must improve the discretisation of the QCD (Dirac) action to increase the power of $ma$ (for $ma < 1$) with which these errors first appear. A very successful action in this regard is the Highly Improved Staggered Quark (HISQ) action [308] developed by the HPQCD collaboration, with tree-level discretisation errors starting at $(ma)^4$.

This discretisation allowed the first 1% accurate calculations for charmed meson decay constants [666]. The current state-of-the-art results are from the Fermilab Lattice/MILC collaborations using HISQ quarks and have 0.3% uncertainties [655]. The dominant uncertainty in the values of $V_{cs}$ and $V_{cd}$ from meson leptonic decay is then from the experimental decay rate [278].

For $b$ quarks discretisation errors are even more of a headache. During the 1990s methods were developed that exploited the nonrelativistic nature of the $b$ quark in its bound states, thus removing the $b$ quark mass as a dynamical scale (so that discretisation errors instead depend on the much smaller scales of the $b$ quark kinetic energy and momentum). These approaches are based on the discretisation onto a lattice of Heavy Quark Effective theory (HQET) [667] (for 'heavy-light' hadrons) and of non-relativistic QCD (NRQCD) [268] (applicable also to heavyonium). It was also shown that the large-mass limit of the clover-improved Wilson quark action [285] could be interpreted as a nonrelativistic effective theory [668]. A limitation of these formalisms is the need to normalise the weak current to match that of continuum QCD; this requires challenging calculations in lattice QCD perturbation theory and has only been done through $\mathcal{O}(\alpha_s)$ [669–671]. The ETM collaboration developed a ratio approach [672] to interpolate between results for quark masses around $c$ using the twisted mass quark formalism [293] and the infinite-mass (static) limit. These methods have been able to achieve a 2% uncertainty on $B$ decay constants [672, 673].

As increased computational power could be exploited to generate gluon field configurations with finer values of the lattice spacing, alternative methods became available. The MILC collaboration led the way including $2+1$ flavours of asqtad sea quarks with a range of lattice spacing values down to $a = 0.044\,\mathrm{fm}$. On these lattices the HPQCD collaboration showed that $b$ quarks could be treated with the relativistic HISQ formalism (with its absolute current normalisation) if calculations were done for a range of quark masses $> m_c$ and a range of lattice spacing values [654]. Fig. 4.7.2 shows the lattice results for the heavy-strange meson along with the joint fit of the dependence on the heavy meson mass and the lattice spacing. This enables a curve for the dependence of the decay constant on the heavy meson mass to be obtained in the continuum limit, from which the decay constant for the $B_s$ meson can be read. At the same time the dependence on heavy meson mass becomes clear; $f_{D_s} > f_{B_s}$ but only by about 10%, rather less than the leading order result from HQET, $f_P\sqrt{M_P} = \text{constant}$ [674] would suggest. The Fermilab Lattice/MILC collaborations have now extended this to $B$ mesons and including $2+1+1$ flavours of HISQ sea quarks for uncertainties on $f_B$ and $f_{B_s}$ below 1% [655].

The SU(3)-isospin-breaking ratio of decay constants, $f_{P_s}/f_P$, is calculated to better than 0.4% in Ref. [655] with results summarised in Fig. 4.7.3. The ratios are all close to 1.2 but there are small and significant differences as the mesons increase in mass from $K/\pi$ to $B_s/B$.

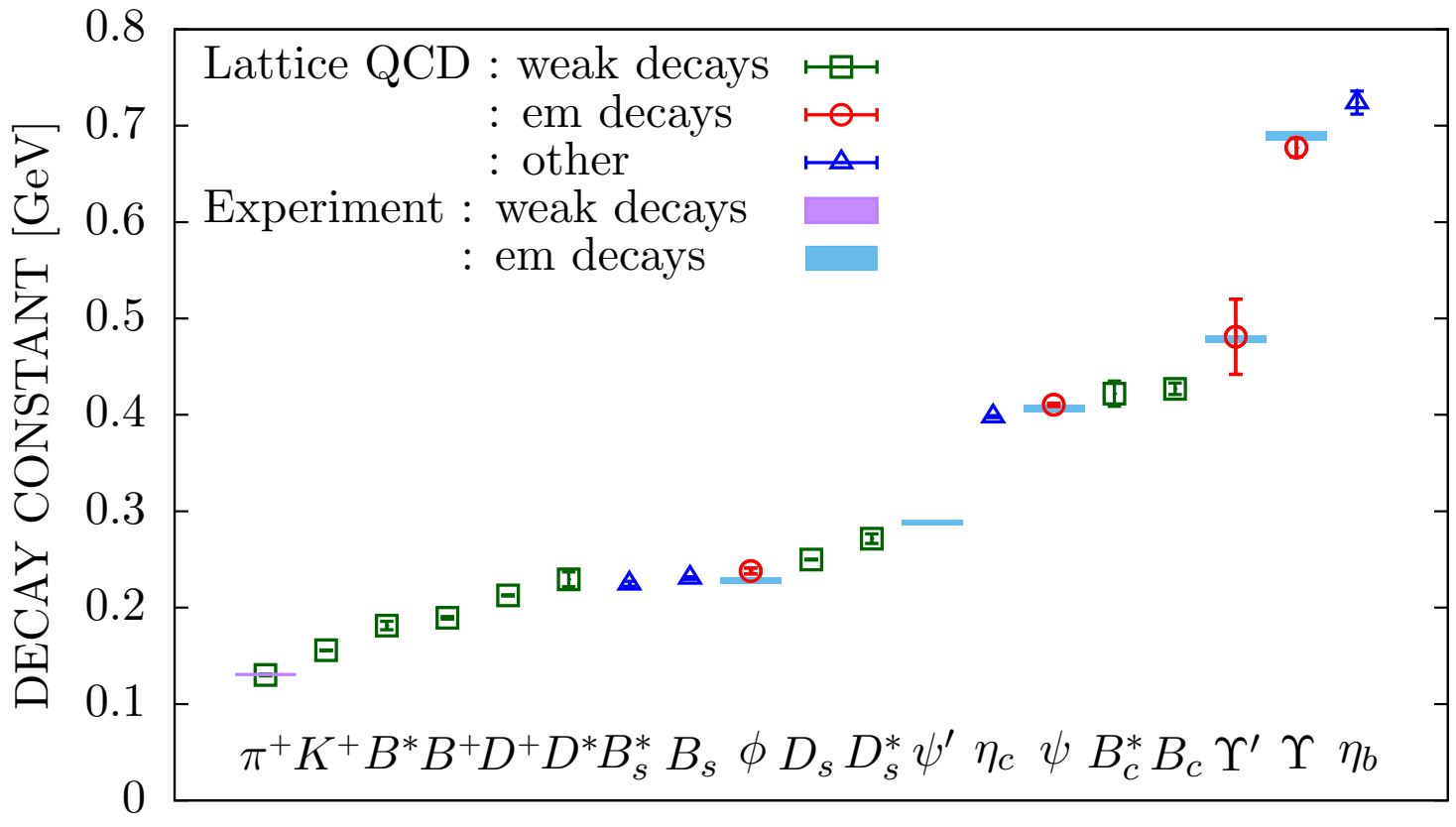

**Fig. 4.7.4** Summary of meson decay constant values calculated in lattice QCD and arranged in order of their size. Points with error bars use different symbols for values needed to determine weak or QED leptonic decay rates or those not linked to any simple decay mode. The decay constants inferred from experimental values for QED leptonic decay are given by blue bands. For weak decays, experimental results must be combined with lattice QCD to obtain CKM elements; $f_\pi$ can be inferred from the $\pi^+$ leptonic rate taking $|V_{ud}|$ from nuclear $\beta$ decay [278] and is shown by a purple band at 130.56(14) MeV. The lattice QCD result for $f_\pi$ comes from RBC/UKQCD [264], using the $\Omega$ baryon mass to fix the lattice spacing. Other results shown use $f_\pi$ to determine the lattice spacing, and so do not give a value for that quantity. $f_K$ is taken from Eq. 4.7.6; $f_B$, $f_D$, $f_{D_s}$ from Ref. [655]; $f_{B_c}$ [661], $f_{D^*_{(s)}}/f_{D_{(s)}}$ and $f_{B^*_{(s)}}/f_{B_{(s)}}$ [662]; $f_{B^*_c}/f_{B_c}$ [663]; $f_\phi$ [664] and charmonium and bottomonium results [258, 260, 665].

Vector mesons with appropriate quark flavour quantum numbers can also annihilate to leptons via a $W$ boson. Although the decay rate is not suppressed by lepton masses in that case (because of the meson spin) it is nevertheless hard to see experimentally because it is overwhelmed by the QED radiative decay $V \to P\gamma$; it may be possible in future for the $D^*_s$ [675]. The vector leptonic decay proceeds through the vector piece of the weak current and is determined by the corresponding vector decay constant. The lattice QCD vector current must again be normalised to match continuum QCD. Although in principle a conserved vector current can be used, it is easier to use a local vector current and renormalise it. There are a number of techniques to this (Sec. 4.2). The ratio of vector to pseudoscalar decay constants for heavy-light mesons has been calculated using NRQCD [663] (with perturbative renormalisation [676]) and using twisted-mass quarks [662] (using a MOM scheme [677]). Interestingly it is found that the ratio of $f_V/f_P$ is larger than 1 for $D$ mesons and less than 1 for $B$ mesons. Ref. [662] gives 1.078(36) for $f_{D^*}/f_D$ and 0.958(22) for $f_{B^*}/f_B$.

Vector $q\overline{q}$ mesons can annihilate to $\ell\overline{\ell}$ via a $\gamma$, and such decay rates have been determined experimentally to better than 2% for heavyonium mesons [278]. This provides an excellent opportunity for accurate comparison of lattice QCD and experiment for a decay rate free from CKM elements since

$$\Gamma(V \to \ell^+\ell^-) = \frac{4\pi\alpha^2 e_q^2}{3}\frac{f_V^2}{M_V}\,, \tag{4.7.7}$$

with $e_q$ the valence quark electric charge in units of $e$. Results for $f_{J/\psi}$ [258] and $f_\Upsilon$ [665] calculated with HISQ quarks, normalised via an SMOM scheme [678, 679] show good agreement with values inferred from the experimental decay rates, providing a solid underpinning for the other decay constants being discussed here.

Fig. 4.7.4 summarises the values of meson decay constants that are well-determined in lattice QCD, arranged by value order. It does not include values for mesons, such as the $\rho$ or $K^*$, that have a large decay width from a strong-interaction decay mode (Sec. 4.5). Notice that the range of decay constant values, from $f_{\pi^+} = 130.2(9)$ MeV [264] to $f_{\eta_b} = 724(12)$ MeV [665] is much smaller than the range of meson masses. As discussed above, decay constants reflect meson internal structure set by momenta inside the bound state rather than quark masses. For mesons containing $u/d$ quarks the range of variation is even smaller, less than a factor of two from $f_\pi$ to $f_{D^+} = 212.7(6)$ MeV [655], and the ordering is not intuitively obvious. Results are shown for decay constants relevant to weak leptonic decays (where comparison to experimental results yields a determination of the relevant CKM element) as well as those relevant to QED leptonic decays (where direct comparison to experimental rates is possible). It also includes decay constants that cannot be simply related to a decay process, but which nevertheless help to fill in the 'big picture' that we now have from lattice QCD for these simple matrix elements.

### 4.7.2 Mixing matrix elements and bag parameters

A fascinating phenomenon for neutral $K$ and $B$ mesons is that of 'oscillations', induced by the tiny weak interaction coupling between the mesons and their antiparticles. For exact CP invariance the eigenstates of the Hamiltonian are then $+/-$ combinations of the strong-interaction $P^0$ and $\overline{P}^0$ states, analogous to the eigenstates of two weakly coupled pendulums. An initial $P^0$ beam, created by a strong-interaction process, is equivalent to setting one pendulum swinging. At later times it becomes clear that the other pendulum is swinging/$\overline{P}^0$ is present (from interrogating the beam via suitable decay processes). The oscillation frequency is set by the eigenstate mass difference $\Delta M_P$ and can be measured very precisely in experiment. The coupling is a second-order weak process with the short-distance contribution given by the 'box diagram' of Fig. 4.7.5. As such it is sensitive to new physics that can be tested with accurate matrix elements for the box diagram between $P^0$ and $\overline{P}^0$, calculated in lattice QCD.

At the hadronic mass scales of the lattice the box diagram shrinks to an effective 4-quark operator (multiplied by a Wilson coefficient). For the SM case, the 'left-left' operator is

$$\mathcal{O}^{(1)} = \left[\overline{h}^{\alpha}\gamma_\mu(1-\gamma_5)\ell^{\alpha}\right]\left[\overline{h}^{\beta}\gamma_\mu(1-\gamma_5)\ell^{\beta}\right]. \quad (4.7.8)$$

$h$ is either $s$ or $b$ and $\alpha/\beta$ are colour indices. Matrix elements of further (BSM) 4-quark operators have also been calculated, see Ref. [63].

The matrix element of Eq. 4.7.8 between $P^0$ and $\overline{P}^0$, having a hadron on either end, is much harder to determine in lattice QCD than a decay constant, so results are not as mature and have larger uncertainties. The renormalisation of the 4-quark operator to match continuum QCD is also more challenging. Results are most usefully presented in terms of 'bag parameters' by removing factors of masses and decay constants from the matrix elements that would appear in the 'vacuum saturation approximation', i.e. inserting $|0\rangle\langle 0|$ between the two halves of the 4-quark operator. For $\mathcal{O}^{(1)}$ this gives [680]

$$\langle P^0|\mathcal{O}^{(1)}|\overline{P}^0\rangle = \frac{8}{3}f_P^2 M_P^2 B_P^{(1)}(\mu) \quad (4.7.9)$$

where the leftover 'fudge factor', $B_P$, is the bag parameter. It is normally quoted in the $\overline{\text{MS}}$ scheme; note its scale-dependence. Historically the assumption was then made that $B \approx 1$ but lattice QCD can achieve a much better result than this.

The bag parameter is often converted from $B^{(1)}(\mu)$ to its renormalisation-group-invariant (RGI) value,

$$\hat{B}^{(1)} = c_{\text{RGI}} B^{(1)}(\mu) \quad (4.7.10)$$

where $c_{\text{RGI}}$ is calculated to two-loops in perturbative QCD [63] and takes values 1.369 for $B_K$ (when $\mu = 2$ GeV) and 1.516 for $B_B$ (when $\mu = m_b$)).

Ref. [63] quotes an average for $\hat{B}_K^{(1)} = 0.7625(97)$ as an average of several lattice QCD results [264, 681–683] using different lattice QCD actions and renormalisation approaches with $n_f = 2+1$ sea quarks; Ref. [684] gives an $n_f = 2+1+1$ result. $B$ meson results are less accurate, because of a significantly worse signal/noise problem in the determination of the correlation functions [685]; direct determination of $B_B$ rather than $\mathcal{O}^{(1)}$ matrix elements cancels discretisation and light quark mass effects, however. Results including $n_f = 2+1+1$ sea quarks are available from HPQCD using NRQCD $b$ quarks (with $\mathcal{O}(\alpha_s)$ renormalisation [686]), giving $\hat{B}_{B_d}^{(1)} = 1.222(61)$ and $\hat{B}_{B_s}^{(1)} = 1.232(53)$) [685]; $n_f = 2+1$ results using other lattice QCD actions are given in [687–689]. Note that $\hat{B}_{B_s}/\hat{B}_{B_d}$ is consistent with 1 (1.008(25) from [685]), showing that the SU(3)-breaking in the 4-quark matrix elements is entirely that of the decay constants.

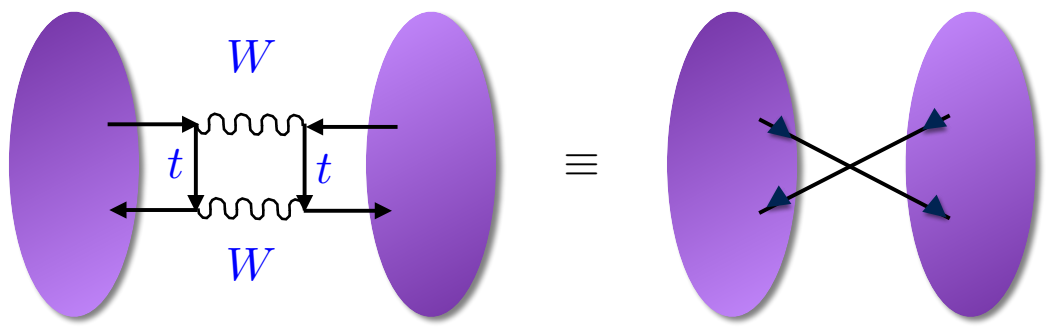

**Fig. 4.7.5** Schematic diagram of the short-distance contribution to neutral meson mixing via the 'box diagram' (left) involving $W$ bosons and top quarks. The matrix element that must be calculated in lattice QCD is that of the equivalent 4-quark operator (right).

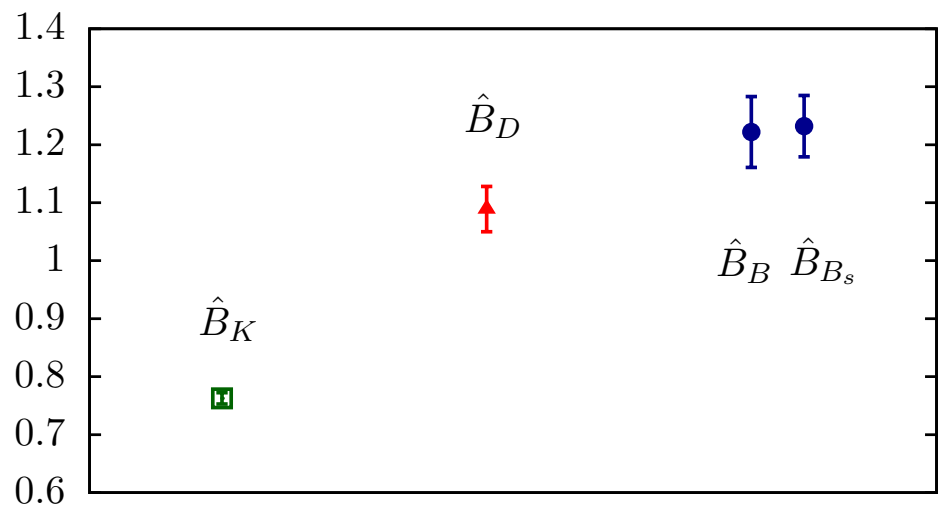

**Fig. 4.7.6** A comparison of RGI bag parameters from lattice QCD for $K^0$, $D^0$, $B^0$ and $B_s$, showing significant deviations from the naive vacuum saturation approximation estimates of 1 and a trend with meson mass.

Fig. 4.7.6 compares the results for $\hat{B}$, including a value for $\hat{B}_D$ [684] that lies between $\hat{B}_K$ and $\hat{B}_B$. The $D^0$ box diagram is mediated by down-type quarks and is expected to contribute only a small part of $\Delta M_D$, dominated by long-distance contributions. The short-distance results can be used to constrain new physics, however see Ref. [690].

For $B/B_s$ mesons the box diagram with top quarks of Fig. 4.7.5 dominates mixing (since $V_{tb} \approx 1$) so that

$$\Delta M_q = \frac{G_F^2 M_W^2 M_{B_q}}{6\pi^2} S_0(x_t)\eta_{2B} \left|V_{tq}^* V_{tb}\right|^2 f_{B_q}^2 \hat{B}_{B_q}^{(1)}, \tag{4.7.11}$$

and lattice QCD results for the bag parameters can be combined with (the very accurate) experimental results for the oscillation frequency to determine CKM elements $|V_{ts}|$ and $|V_{td}|$ that multiply the effective 4-quark operator. Agreement is seen within $2\sigma$ with CKM values from tree-level weak decays and unitarity [685].

For $K$ oscillations the situation is more complicated because of sizeable long-distance contributions to $\Delta M_K$ involving $u$- and $c$-mediated contributions. At the same time analysis of $K \to \pi\pi$ amplitudes [63] is also needed to determine the direct and indirect CP-violation parameters, $\epsilon'$ and $\epsilon$ that describe the CP-properties of the mass eigenstates and their decays. These are very hard calculations that have required the development of new techniques, and results are still at a fairly early stage, e.g. often only available at one value of the lattice spacing. The RBC/UKQCD collaboration has led the way here, exploiting the excellent chiral properties of the domain-wall quark action. They have calculated the amplitude $A_2$ to the isospin 2 two-pion state (the $\Delta I = 3/2$ amplitude) [691] and the amplitude $A_0$ to the isospin 0 state ($\Delta I = 1/2$) [692]. This enables a result of Ref. [692]

$$\epsilon'/\epsilon = 21.7(8.4) \times 10^{-4} \tag{4.7.12}$$

in good agreement with experiment ($16.6(2.3)\times10^{-4}$), suggesting no violation of the CKM paradigm at this level of accuracy. At the same time the lattice QCD results provide some insight into the observed $\Delta I = 1/2$ rule by which $A_0$ exceeds $A_2$ by a factor of 20. A factor of 2 is provided by perturbative QCD corrections to the coefficients of the appropriate 4-quark operators; lattice QCD shows that the other factor of 10 arises from the fact that, contrary to naive expectations, the contributions from different colour contractions of the dominant operator tend to cancel in $A_2$ and reinforce each other in $A_0$ [692, 693]. The development of methods to determine the long-distance contributions to $\Delta M_K$ [694] are also aimed at long-distance contributions to $K^+ \to \pi^+\nu\overline{\nu}$ and $K \to \pi\ell^+\ell^-$.

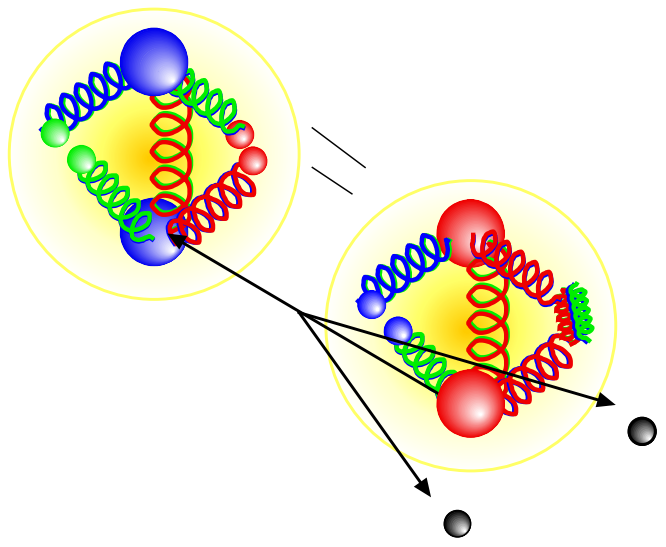

**Fig. 4.7.7** Schematic diagram of a meson to meson semileptonic decay. The hadronic information needed to determine the rate is parameterised by form factors.

Future improvements here require improved renormalisation techniques for lattice 4-quark operators. Gradient flow methods look promising here, see e.g. Ref. [695].

### 4.7.3 Form factors

Semileptonic weak decays of hadrons in which one of the constituent quarks changes flavour and the virtual $W$ boson emitted is seen as a $\ell\overline{\nu}_l$ pair (see Fig. 4.7.7) provide a huge range of possibilities for determining CKM elements and understanding hadron structure. The hadronic parameters that control the rate of these processes are known as form factors and they are functions of $q^2$, the squared 4-momentum transfer from parent hadron, $\varrho$, to child, $\chi$. The kinematic range of $q^2$ is from $q^2_{\max} = (M_\varrho - M_\chi)^2$ (where $\ell$ and $\overline{\nu}_\ell$ have maximum back-to-back momentum in the $\varrho$ rest-frame) to 0 (where $\chi$ and the $\ell\overline{\nu}_\ell$ pair are back-to-back). The form factors are largest at $q^2_{\max}$ and fall towards $q^2 = 0$, reflecting the internal momentum transfer via gluon exchange necessary to achieve the final state configuration.

Form factors are defined from matrix elements between $\varrho$ and $\chi$ of weak currents. The simplest situation is when both $\varrho$ and $\chi$ are pseudoscalar mesons. In that case only the vector current and vector form factor, $f_+(q^2)$, contribute to the the decay rate for $\varrho \to \chi\ell\overline{\nu}$ for zero lepton mass, with $m_\ell$-dependent corrections from the scalar form factor, $f_0(q^2)$. We have

$$\frac{d\Gamma}{dq^2} = \frac{G_F^2}{24\pi^3}|V_{ab}|^2(1-\epsilon)^2 \times \Big[|\vec{p}_\chi|^3(1+\frac{\epsilon}{2})|f_+(q^2)|^2 + |\vec{p}_\chi| M_\varrho^2 \left(1 - \frac{M_\chi^2}{M_\varrho^2}\right)^2 \frac{3\epsilon}{8}|f_0(q^2)|^2\Big] \tag{4.7.13}$$

for quark transition $a \to b$, $\epsilon = m_\ell^2/q^2$, and $\vec{p}_\chi$ is the 3-momentum of child $\chi$ in $\varrho$'s rest frame. The form factors

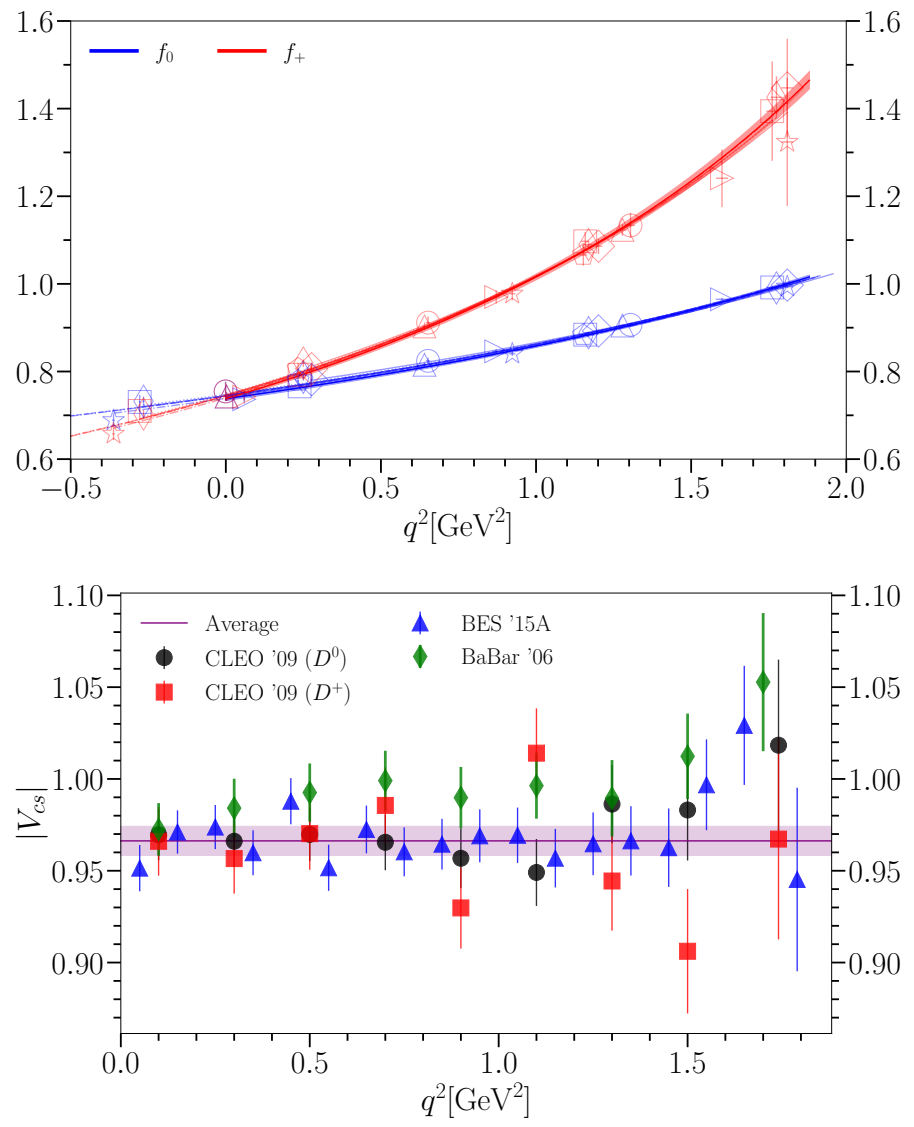

**Fig. 4.7.8** (Upper) Data points show lattice QCD results at multiple values of $q^2$ and multiple lattice spacings. Blue and red curves show the final determination of $f_0$ and $f_+$ in the continuum limit at physical quark masses. (Lower) Bin-by-bin values of $V_{cs}$ from combining these form factors with experimental data. The constancy of $V_{cs}$ shows that the $q^2$ dependence predicted by QCD matches that of experiment [697].

are defined from matrix elements

$$\langle\chi|V^\mu|\varrho\rangle = f_+^{\varrho\to\chi}(q^2)\left[p_\varrho^\mu + p_\chi^\mu - \frac{M_\varrho^2 - M_\chi^2}{q^2}q^\mu\right] + f_0^{\varrho\to\chi}(q^2)\frac{M_\varrho^2 - M_\chi^2}{q^2}q^\mu, \tag{4.7.14}$$

$$\langle\chi|S|\varrho\rangle = \frac{M_\varrho^2 - M_\chi^2}{m_a - m_b} f_0^{\varrho\to\chi}(q^2), \tag{4.7.15}$$

with kinematic constraint $f_+(0) = f_0(0)$. Eq. 4.7.15 makes use of the partially conserved vector current relation $\partial_\mu V^\mu = (m_a - m_b)S$ that means $f_0$ is correctly normalised in lattice QCD [696]. The renormalisation factor, $Z_V$, for the vector current can then be determined by, for example, matching $f_0(q^2_{\text{max}})$ from Eqs. 4.7.14 and 4.7.15 (see Ref. [697]).

To determine the form factors in lattice QCD requires the calculation of three-point correlation functions with appropriate source and sink operators for parent and child hadrons, and a current insertion at an intermediate time between them. Usually the parent hadron is taken to be at rest on the lattice and different spatial momenta are given to the child to map out the $q^2$ range. Fitting the three-point correlation function simultaneously with the two-point correlation functions for parent and child allows the parent-to-child matrix elements to be determined and Eqs. 4.7.14 and 4.7.15 applied. To obtain form factors in the continuum limit, interpolation in $q^2$ and extrapolation to $a = 0$ and physical quark masses is needed. Modern calculations (see, for example, Ref. [697]) transform $q^2$ into a region within the unit circle in $z$-space and then apply a polynomial fit in $z$ that allows for discretisation effects and mistuning of quark masses.

The channek $K \to \pi\ell\overline{\nu}$ is a key for the determination of $V_{us}$. The $q^2$ range for this decay is very small and so conventionally experiment accounts for the $q^2$ dependence of Eq. 4.7.13 and gives the final result as a value for $|V_{us}|f_+(0)$. Combining charged and neutral meson decay rates with QED radiative and strong-isospin-breaking corrections gives a result with 0.2% accuracy : $V_{us}f_+(0) = 0.21635(39)(3)$ [698], where the first, dominant, error is from the experiment and 0.2% accuracy is also now available from lattice QCD with 2+1+1 flavours. Ref. [63] gives $f_+(0) = 0.9698(17)$ from averaging [699, 700]. The two lattice QCD calculations take contrasting approaches. Ref. [699] determines $f_+(q^2)$ and $f_0(q^2)$, interpolating to $q^2 = 0$ and testing $q^2$ dependence against experiment; Ref. [700] tunes to $q^2 = 0$ using twisted boundary conditions [701] and calculates $f_0(0)$ since this needs no renormalisation. The result for $V_{us}$ from $K \to \pi\ell\overline{\nu}$ then shows an intriguing $3\sigma$ tension with CKM first row unitarity [698] and $2.5\sigma$ tension with $V_{us}$ from $K \to \ell\overline{\nu}$ [278].

$D$ meson decays (to $K$ or $\pi$) have a larger $q^2$ range and experimental data is available in bins of $q^2$. This provides the opportunity to test the $q^2$-dependence predicted by QCD against experiment as well as to determine $V_{cs}$ and $V_{cd}$. Fig. 4.7.8 shows how this is done [697]. The upper plot shows the determination of the $f_+$ and $f_0$ form factors and the lower plot shows the result of determining $V_{cs}$ bin-by-bin in $q^2$ using Eq. 4.7.13. A good fit is obtained to a constant with $V_{cs} = 0.9663(80)$, with errors from lattice QCD, experiment and QED corrections making similar contributions to the total uncertainty. See Ref. [702] for a determination of $V_{cd}$ using lattice QCD $D \to \pi$ form factors.

The semileptonic decays of $B$ mesons have a huge potential in searches for new physics as well as in giving access to key CKM elements $V_{ub}$ and $V_{cb}$. Form factors for these decays are challenging for lattice QCD, however, because the large $b$ quark mass means a large $q^2$ range. To reach $q^2 = 0$ the child spatial momentum must approximate $M_B/2$. Large values of $a|\vec{p}|$ induce poor signal/noise in correlation functions as well as discretisation effects, so early lattice QCD calculations worked close to $q^2_{\text{max}}$ with nonrelativistic formalisms for the $b$ quark.

To determine $V_{ub}$ from $B \to \pi\ell\overline{\nu}$, Ref. [63] performs a joint fit to lattice form factor results from Refs. [704, 705] (which use different variants of the improved Wilson action for the $b$ quark and different light quarks) and experimental data from BaBar and Belle, leaving $V_{ub}$ as a parameter. Such a fit allows experimental information on $q^2$-dependence to constrain the lattice results. The value for $V_{ub}$ obtained, $3.74(17) \times 10^{-3}$ is $1.7\sigma$ lower than that obtained from inclusive $b \to u$ determinations that do not specify the final state hadron.

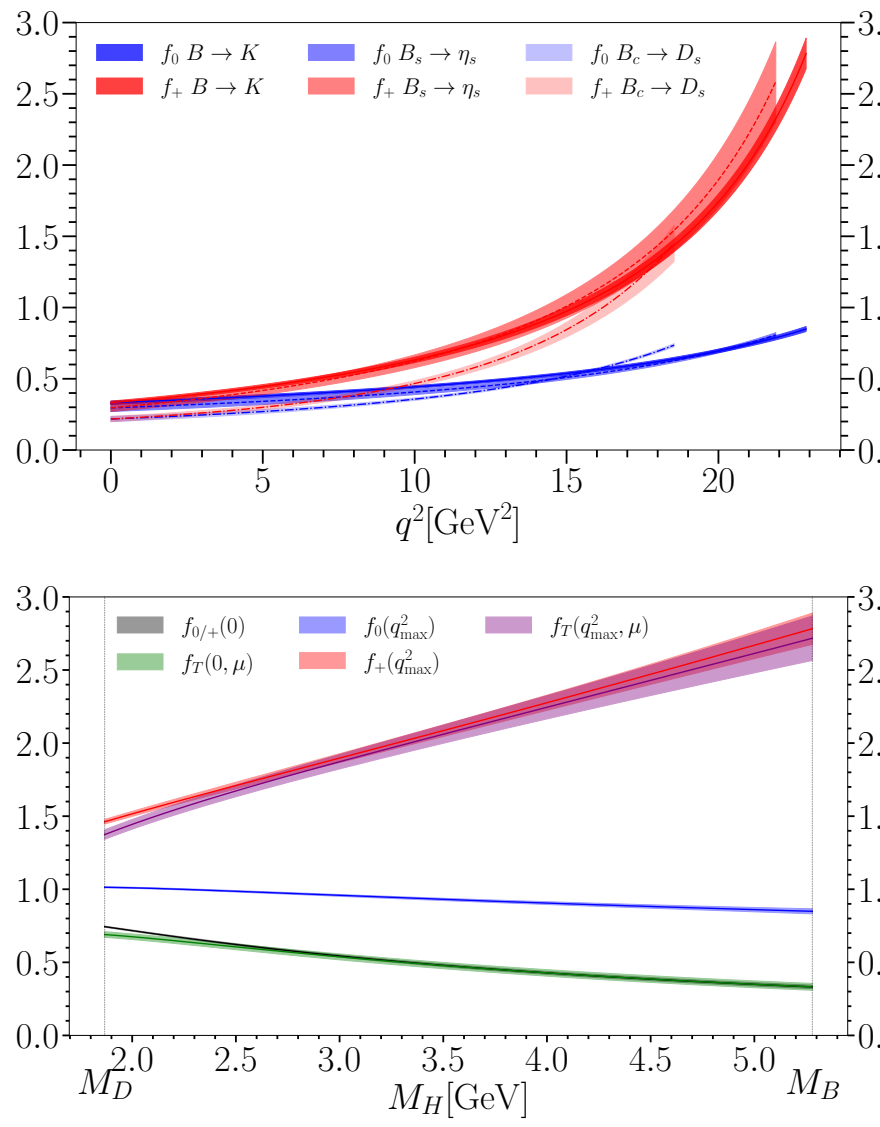

**Fig. 4.7.9** (Upper) Comparison of $b \to s$ form factors for meson transitions with different spectator quarks. Increasing the spectator mass to that of $c$ quarks reduces the form factors at low $q^2$ values [703]. (Lower) The dependence on heavy meson mass, $M_H$, of the form factors for $H \to K$ decay at $q^2_{\rm max}$ and $q^2 = 0$. Notice the slow downward drift at $q^2 = 0$ and for $f_0(q^2_{\rm max}$ as $H$ varies from $D$ to $B$, but much stronger variation for $f_+$ and $f_T$ (the tensor form factor) at $q^2_{\rm max}$ (remembering that $q^2_{\rm max}$ depends on $M_H$).

The transitions $b \to c$ have also shown a persistent tension between inclusive and exclusive results. Here the preferred exclusive method is to use $B \to D^*$ decay. Although a pseudoscalar to vector transition is more complicated, with 4 form factors, only the axial vector $A_1$ form factor contributes at $q^2_{\rm max}$. Lattice QCD therefore initially concentrated on this point [706, 707]. Now it has become clear that the $q^2$-dependence of the differential rate must be understood from the lattice QCD side and form factors have been calculated by the Fermilab Lattice/MILC collaboration [708] that cover more of the $q^2$ range using their improved-Wilson action for both $b$ and $c$. This does not resolve the inclusive/exclusive $V_{cb}$ tension but points the way to improved future analyses.

Recent $B$ form factors have been calculated using relativistic formalisms that can make use of nonperturbative current normalisation techniques discussed for Eqs 4.7.14 and 4.7.15. They obtain results for multiple heavy quark masses and lattice spacings and fit to obtain results for $B$ mesons in the continuum limit in a similar way to that for decay constants in Fig. 4.7.2. Calculations include HPQCD's form factors for $B_s \to D_s$ [709], $B_s \to D_s^*$ [710] and $B \to K$ [703] using HISQ quarks and JLQCD's form factors for $B \to \pi$ using domain-wall quarks [711]. This is likely to be the way forward for the future.

It is important to remember that QCD provides a smooth connection between different form factors as we change the mass for one or other of the participating quarks. In this way lattice QCD can provide 'a big picture' for form factors. Fig. 4.7.9 shows this connection for different spectator (not part of the weak current) quarks for the $b \to s$ transition. It also shows results for $H \to K$ decay where H is a meson containing a heavy quark with mass varying from $c$ to $b$ [703].

Future calculations will improve $B$ form factor uncertainties to the 1% level [712] for the increased datasets planned from LHC and Belle II. New developments include techniques for inclusive $B$ decays [713] and for handling final-state mesons that decay strongly (e.g. for $B \to K^* \ell \overline{\nu}$ analysis) [554]. An important focus will be improving lattice calculations needed to understand 'B anomalies' seen, for example in ratios of branching fractions to different flavours of leptons and differential rates for flavour-changing neutral current $b \to s$ transitions (e.g. $B \to K \ell^+ \ell^-$) that proceed through loops in the SM.

The lattice QCD calculation of form factors for weak decays of baryons is still in its infancy, because of the extra challenges provided by the poorer signal-noise. The nucleon axial coupling, $g_A$, has been a particular focus of attention and is discussed in Sec. 4.6. A notable success has been the use of lattice QCD form factors for $\Lambda_b \to \Lambda_c$ and $\Lambda_b \to \Lambda$ [714] to determine $V_{ub}/V_{cb}$ by LHCb [715]. This is clearly a developing area for the future.

# 5 Approximate QCD

*Conveners:*
**Stan Brodsky and Franz Gross**

The next two sections of this volume discus theoretical ways to model QCD. At the heart of all modern models are quarks, treated as elementary particles that interact both with single gluons and with a complex QCD vacuum containing condensates. Since numerical Lattice Gauge calculations discussed in the previous section are the only way known to treat these interactions exactly, all of these analytical methods are approximations.

Starting from a description of quarks (Sec. 5.1) and new virtual colored degrees of freedom multiple quark states could occupy at short distances (hidden color states discussed in Sec. 5.2), the section moves on to a discussion of the Bethe Salpeter (BS) and Dyson Schwinger (DS) equations (Sec. 5.3), where quark-gluon interactions are treated microscopically, much as pion-nucleon interactions were described in an earlier era. Here the multiple interactions make it impossible to treat them all systematically, and the equations must be truncated, introducing approximations with an accuracy that is sometimes hard to estimate. Light front coordinates are the preferred way to describe multi-quark systems, and Sec. 5.4 describes methods for expanding multi-quark quark wave functions in a light front basis that avoids some of the issues with the microscopic description, but also requires truncations of the expansion to a finite number of basis functions.

These methods handle the confinement of quarks in different ways with very different assumptions. In Sec. 5.5, recent developments based on superconformal quantum mechanics, light-front quantization, and its holographic embedding in a higher dimension classical gravity theory, known as AdS/QCD, have led to new analytic insights into the nonperturbative structure and dynamics of hadrons in physical spacetime, such as color confinement and chiral symmetry breaking. This contribution is followed by a short discussion (Sec. 5.6) of the model dependence of predictions of the behavior of the strong fine structure constant, $\alpha_s(Q^2)$ at small $Q$, where it becomes large. This discussion complements and completes the discussions of $\alpha_s(Q^2)$ in Sec. 3. Next, the interesting features that can be drawn from the study of QCD with a large number of colors, and the solvable 't Hooft model, are reviewed in Sec. 5.7.

The next three contributions in this section discuss approximations that treat specific issues: the use of sum rules based on the operator product expansion (OPE) to explain properties of mesons and other physical quantities (Sec. 5.8); approximations that work for high energy reactions which can be factorized into reaction specific high energy parts that can be computed perturbatively and low energy, reaction independent parts expressed in terms of unknown functions that are extracted from many experiments (Sec. 5.9); and the power counting rules that describe the behavior of exclusive processes at high energy (Se. 5.10).

The section concludes with Sec. 5.11 where a theoretical discussion of the complexity of the QCD vacuum needed to understand confinement and chiral symmetry breaking is presented. This discussion is complementary to the Lattice discussion of the same topic, Sec. 4.3.

Section 5 covers a very wide range of topics, but as you will see from what follows, is only part of the theoretical tool box developed to "solve" a theory based on a Lagrangian that can be written in one line!

## 5.1 Quark models

**Eric Swanson**

"It is more important to have the right degrees of freedom moving at the wrong speed, than the wrong degrees of freedom moving at the right speed."
— Gabriel Karl, as frequently quoted by Nathan Isgur.

### 5.1.1 Early Quark Models

The phrase "quark model" originally meant something like the "quark idea", referring to the introduction of quarks as the elements of the fundamental representation of $SU_F(3)$ by Gell-Mann and Zweig in 1964 [716–718]. Gell-Mann initially avoided attributing physical reality to the quark concept, and it was others, such as Dalitz [719], Becchi, Morpugo [720], Rubinstein, Scheck [721], and Lipkin [722] who developed the idea into a viable and predictive model in the sense we use now. That this was not a simple task is illustrated by a famous line from Kokkedee's review of the quark model, "The quark model should ... not be taken for more than it is, namely, the tentative and simplistic expression of an as yet obscure dynamics underlying the hadronic world."[723]

Kokkedee's pessimism was not misplaced. The inability to observe free quarks was originally explained by assuming that they had very high masses. The existence of relatively light hadrons then implied that the interquark binding force was "ultra-strong", which in turn requires relativistic and nonperturbative techniques. These technical problems were further exacerbated by the "statistics problem", wherein bound states of fermions must be antisymmetric. Thus, for example, the $\Delta^{++}$ requires an antisymmetric spatial wavefunction, in contrast with expectations for a low lying state. No satisfactory solution to the problem was found, in spite of the great contortions theorists invented.

Nevertheless, a few determined individuals persisted with the notion that quarks are "real". Early computations drew from long tradition in nuclear physics[720, 724, 725] and tended to focus on electroweak transitions since the couplings are weak and the effects of unknown spatial wavefunctions can be ignored (in magnetic dipole transitions) or simply modelled (in electric dipole transitions). These computations typically assumed nonrelativistic dynamics, factorized spatial wavefunctions, and electroweak currents coupling directly to quarks. The state of the art was formalized in a classic paper from 1967 by van Royen and Weisskopf, which placed the topic on firm footing (even though the quark model problems remained unresolved)[726]. By 1969, Copely, Karl, and Obryk had brought the quark model to a high level of predictiveness, introducing explicit simple harmonic oscillator wavefunctions and a "constituent" quark mass of roughly one third the proton mass, in line with its modern value[727].

### 5.1.2 QCD-improved Quark Models

It is no surprise that the advent of QCD revolutionized the conceptualization and application of the quark model, releasing a flood of research. QCD, of course, is the theory of hadrons; thus the quark model was no longer the first and final word for hadronic properties, and it quickly evolved into its current role as a computationally feasible model for QCD in the strong coupling regime.

Already by 1975 (November 1974), Appelquist and Politzer famously applied QCD to the $R$ ratio (proportional to the cross section for $e^+e^-$ to hadrons) and noted that ladder exchanges of gluons should give rise to "orthocharmonium" (the $J/\psi$) and "paracharmonium" (the $\eta_c$) states[74]. This was the time of the "November revolution" described in Sec. 2.1 above. These notions were greatly expanded by De Rujula, Georgi, and Glashow, who argued that one gluon exchange should dominate the short distance quark interaction and that it explained a wealth of experimental data, concluding that "The naive quark model, supplemented by color gauge theory, asymptotic freedom, and infrared slavery, is turning out to be not so naive, and more than just a model."[728]. In fact the results were successful enough that the authors initiated *and ended* the field in the same paper, declaring,

> Not until many of these predicted charmed states are discovered and measured can the subject of hadron spectroscopy join its distinguished colleagues, atomic and nuclear spectroscopy, as subjects certainly worthy of continued study, but understood (at some level) in principle.

Needless to say, such proclamations seem premature to modern eyes!

Amongst the first to join the fray were Isgur and Karl, who wrote a complete model Hamiltonian for baryons, assuming nonrelativistic dynamics, a quadratic confinement potential, and short distance spin-dependence as given by one gluon exchange[729]. (For a full discussion of baryon quark models, see Sec. 9.1) The resulting reasonably complete description of the low lying baryon spectrum and its properties caused a sensation, as it was realized that comprehensive and quantitative computations of hadronic properties were possible. However, there was a price to be paid: the good results were obtained only upon neglecting the spin-orbit interaction arising from one gluon exchange. It is, of course, difficult to argue in favor of one aspect of perturbative QCD while neglecting another! By way of defense, Isgur and Karl noted that the confinement interaction should contribute Thomas precession spin-orbit interactions, even though it is spin-independent, and that the long range spin-orbit interaction tends to cancel that due to one gluon exchange.

The issue of the spin-dependence of the long range (confinement) interaction reappeared in a nearly contemporary and seemingly disconnected area. At issue was the Dirac structure of a (presumed) relativistic long range two-body interaction for quarks,

$$1/2 \iint J(x)K(x-y)J(y),$$

where the current is written as $J = \bar{\psi}\Gamma\psi$, $\psi$ is a quark field, and $\Gamma$ is a four-by-four Dirac matrix. In 1978, Schnitzer realized that the masses of several newly discovered charmonia and bottomonia permitted settling the issue in favor of a scalar ($\Gamma = 1$) confinement interaction[730, 731].

Of course assuming that the interaction between quarks is specified by a current-current operator yields more than spin-dependence – it also gives the amplitude for quark pair creation, and therefore opens the field of strong hadronic transitions to investigation. (Such investigations actually date to the beginnings of the quark model, starting with Micu's hypothesis that quark pairs are produced in a spin triplet, angular momentum one, state[732, 733].)

In 1978, Eichten *et al.* produced the most famous version of such a model, the "Cornell model" (first introduced in 1975), in an ambitious attempt to understand the properties of charmonia, including their coupling to the open charm continuum[734]. Pragmatism forced compromise: the Cornell group had to assume a color density current to obtain agreement with the, by now well-established, one gluon exchange short range structure of the quark interaction, and in disagreement with the decay model of Micu (which is admittedly a guess) and Schnitzer's scalar confinement. Nevertheless, the model is well-constrained and does admirably well in predicting a wealth of charmonium properties.

By 1985 the field had progressed enough that comprehensive models capable of describing all mesons and baryons were being attempted. The most famous of these is that due to Godfrey and Isgur (mesons) and Capstick and Isgur (baryons)[735, 736]. The model has much in common with earlier ones such as Ref. [737]. The model assumes relativistic quark kinematics, the full one gluon exchange short range interaction, and a scalar confinement interaction (including its spin-orbit relativistic correction). All interactions were convoluted over a Gaussian to ameliorate the strength of the short range terms (which are not legal operators in quantum mechanics).

A model of the running strong coupling was used because there is strong evidence that weaker spin-dependent interactions are required for heavier quarks. The possibility of quark annihilation in isoscalar channels was allowed by including a phenomenological term. The model was "relativized" by including factors of $(m/E)^{\nu}$, where $\nu$ is a model parameter, in various matrix elements. Finally, additional factors of meson and quark mass were introduced to certain rates to bring their form into alignment with low energy theorems. The resulting masses, strong decays, and electroweak transitions have served as a benchmark in hadronic physics over the intervening 37 years.

### 5.1.3 Bag Models

The advent of QCD raised the possibility of inventing field-theoretic models of hadrons. The opportunity was seized first by Ken Johnson, who drew an analogy to bubble nucleation in first order phase transitions to imagine a hadron as perturbative fields confined to a vacuum bubble of size about 1 fm. The resulting model, developed with colleagues in 1974, became known as the "MIT bag model"[738]. The starting point is a postulated nontrivial QCD vacuum that exerts a pressure (described by the constant $B$) on a region of trivial space-time (called the "bag"). The model Hamiltonian is

$$L_{\text{bag}} = (L_{QCD} - B)\,\theta(bag) \tag{5.1.1}$$

where $\theta$ is zero outside the bag region. Because the action involves an integration over a finite region of space, the location of the bag surface is itself a dynamical field, related through the Euler-Lagrange equations to the quark and gluon fields by a complicated, nonlinear expression. As a result quantization is very difficult and semiclassical approximations are used to study the system. In particular, the "static bag approximation" is made, wherein quarks and gluons are presumed to be confined to a region of a given radius (it is possible to make more complicated models where small oscillations in the bag surface are permitted). The resulting equations of motion describe free fields subject to cavity boundary conditions, which can be obtained by summing cavity modes.

Almost simultaneously, similar ideas were being explored at Stanford, giving rise to the "SLAC bag model" [739]. In this case a scalar field played a role similar to the bag. Symmetry breaking in the scalar vacuum served to confine quarks to a small region where the scalar field exhibits soliton-like behavior. However, this implies that quarks are confined to a spherical shell, which contradicts experiment[740]. A subsequent model, called the "soliton bag model", is able to avoid this feature while interpolating the MIT and SLAC bag models[741]. Many variant bag models have been developed over the years that seek to address various shortcomings. For example, the MIT, SLAC, and soliton models all violate chiral symmetry. This can be overcome by explicitly introducing pion fields[742, 743] or topological features[744]. Other models will be discussed below.

A number of advantages of bag models are apparent: hadrons are bound systems of relativistic quarks and gluons, obey asymptotic freedom automatically, are confined to regions of order 1 fm in size, and respect color gauge invariance. These benefits spurred a large theoretical effort in hadronic modelling that lasted through the 1980s, and continues at a reduced level to the present. Unfortunately, the complexity of the model introduces a number of conceptual and technical difficulties. The cavity approximation, for example, is not translationally invariant and no projection onto momentum eigenstates exits. This has the practical demerit of introducing undesired center of mass degrees of freedom to the problem. Quark and gluon propagators can be formed by summing over appropriate cavity modes, but in practice this is difficult, and evaluating Feynman diagrams is technically cumbersome[740]. For example, self-energy diagrams are difficult to evaluate and are often ignored. Similarly, the expectation value of the bag Hamiltonian has a sum over zero point energies that diverges. Renormalizing this quantity is subtle, and the zero point energy is often replaced with a simple model. Lastly, the rigid cavity gives rise to spurious states that must be identified.

Early MIT bag model computations contained three parameters, the bag constant, the gauge coupling, and a zero-point energy parameter. Fits to the $\rho$, $N$, and $\Delta$ masses then fixed these constants. Unfortunately the resulting value for the strong coupling was $\alpha_S \approx 2.2$, which gives spin splittings that are too large in other hadrons. The resulting phenomenology is often of poor quality; for example, an early calculation of P-wave masses gives disappointing results[745]. Bag model phenomenology is clearly geared toward light hadrons. Heavy quark states, on the other hand, are surely described by nonrelativistic kinematics, a string-like confinement mechanism, and a value of the strong coupling that is set by $\alpha_S(m_Q)$. These features can be incorporated by allowing the bag to distort into a tube shape (in practice the distortion is small) and refitting the model parameters[746]. The resulting model does a reasonable job with the low lying charmonium and bottomonium vectors, predicts a $J/\psi - \eta_c$ splitting of 180 MeV (the measured value is 113 MeV), and $J^{PC} = 1^{-\pm}$ charmonium (bottomonium) hybrids at mass of approximately 4.0 (10.49) GeV.

| $J^P$ | color | flavor |
|---|---|---|
| $0^+$ | $\bar{\mathbf{3}}$ | $\bar{\mathbf{3}}$ |
| $1^+$ | $\bar{\mathbf{3}}$ | $\mathbf{6}$ |
| $0^-$ | $\bar{\mathbf{3}}$ | $\mathbf{6}$ |
| $1^-$ | $\bar{\mathbf{3}}$ | $\bar{\mathbf{3}}$ |

**Table 5.1.1** Diquark quantum numbers

One of the great advantages of bag models is that they made it clear that states incorporating gluonic degrees of freedom (glueballs and hybrids) should be considered seriously. Early contributions to the theory of glueballs include Jaffe and Johnson[747], who examined many novel states in the model, and Barnes, Close, and Monaghan, who computed spin-dependent mass shifts in the glueball spectrum[748]. These shifts are very large when common model parameters are used, giving glueball masses of $m(0^{++}) = 100$ MeV, $m(0^{-+}) = 400$ MeV, and $m(2^{++}) = 1300$ MeV, all of which are in strong disagreement with modern lattice values[745].

Studies of hybrid ($q\bar{q}g$) mesons originated in the MIT bag model[749] only a few years after the advent of both QCD and bag models, thereby raising interest in these novel states and highlighting the unusual ("exotic") quantum numbers that are available to these systems. Early computations in the MIT bag model worked to first order and focussed on light hybrid mesons[750, 751], obtaining, for example, a $J^{PC} = 1^{-+}$ light hybrid mass of 1300 MeV[752].

Problems with complexity and fidelity have caused bag models to largely fall out of favor as descriptions of hadrons. They do, however, continue to find applications in models of strongly interacting matter or other complex hadronic systems.

### 5.1.4 Diquark Models

Two quarks in a baryon experience a (perturbative) mutual attraction that is one half of the strength of that between a quark and an antiquark in a meson. If the third quark is isolated in some sense, it is fruitful to consider this quark-quark state as a compact object, called a *diquark*. More generally, a diquark is any system of two quarks considered collectively. The idea is already mentioned by Gell-Mann in 1964[716] and was introduced in Refs. [754] and [755] as a way to reduce three-body dynamics to the computationally simpler two-body dynamics.

In general a pair of quarks, denoted $[qq]$ in the following, can form $\bar{\mathbf{3}}$ and $\mathbf{6}$ color states, with the former being antisymmetric and the latter being symmetric under quark interchange. Because a pair of quarks in the $\mathbf{6}$ representation has a (perturbative) repulsive interaction ( $+\alpha_S/(6r)$), diquarks are only considered in the $\bar{\mathbf{3}}$ representation. In this case, possible quantum numbers for $[qq']$ are as listed in Table 5.1.3. The first two of these entries are often called "good" and "bad" diquarks respectively[753].

An early application of diquarks was to the description of light baryons[756]. The primary effect is a reduction in the number of degrees of freedom compared to a "symmetric" quark model, with commensurate decrease in the complexity of the excitation spectrum. For example, a symmetric quark model will feature orbital excitations in two relative coordinates (often taken to be the Jacobi coordinates $\vec{\rho}$ and $\vec{\lambda}$) while a quark-diquark bound state can only have orbital excitations in a single relative coordinate. It is telling that this simple diagnostic is difficult to apply since so little is known of the excited baryon spectrum (but see Sec. Baryons).

Light baryons do experience flavor-dependent correlations, which might be attributed to diquarks. For example, a neutron will have a negative charge radius because the $d$ quarks are in a spin-one state and are repelled via the hyperfine interaction, leaving the positive $u$ quark in the center (on average). Similarly, diquark overlaps (denoted by $I$) affect static observables like the ratio of magnetic moments and the ratio of axial and vector couplings:

$$\frac{\mu_p}{\mu_n} = -\frac{4+5I}{2+4I}, \qquad \left|\frac{G_A}{G_V}\right| = \frac{2+3I}{2+I}. \tag{5.1.2}$$

Unfortunately, the additional freedom (represented by $I$) does not permit a simultaneous fit to the experimental values of -1.46 and 1.25, respectively[756].

At a more formal level, the similarity of light quarks makes it difficult to separate one quark from the other two. In the extreme case of identical quarks, antisymmetrization of the state implies that such a separation is not feasible. This was noted long ago by Lichtenberg[757], who suggested including exchange forces to accommodate transitions of the form $q[qq] \to [qq]q$. Of course this implies that the diquark can no longer be

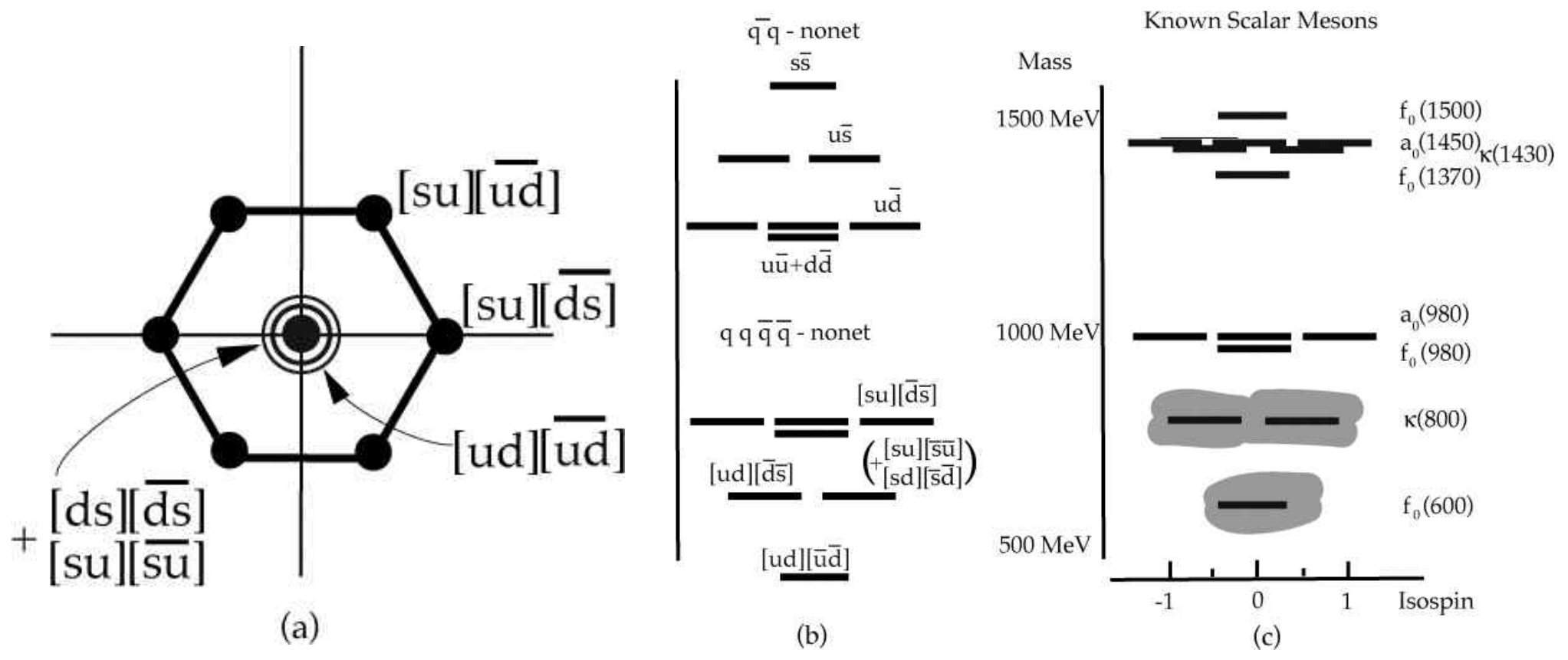

**Fig. 5.1.1** (a) Quark content of a diquark-antidiquark nonet. (b) Mass levels of ideally mixed $q\bar{q}$ nonet and diquark-antidiquark nonet. (c) Light scalar mesons. The shaded region indicates large widths. Figure from Ref. [753].

thought of a simple quasiparticle, but is rather something with internal structure that can be modified and excited.

Perhaps the most famous application of light diquarks is as a model of the scalar mesons. In the 1970's Jaffe noted that a good diquark and a good antidiquark naturally make a scalar nonet of states, as shown in Fig. 5.1.1(a). This nonet forms a spectrum as shown in panel (b) with counting that contrasts strongly with the "normal" $q\bar{q}$ scheme, shown at the top of panel (b). Remarkably this scheme agrees with the observed spectrum, as shown in panel (c)[753]. This ostensibly simple observation has a long and somewhat controversial history, as general acceptance of the existence of the light scalar mesons $f_0(600)$ and $\kappa$ has waxed and waned over the years.

More recently, the diquark simplification has been applied to Bethe-Salpeter approaches to the baryon spectrum with some success[758]. The concept has also found support in lattice computations that see evidence for the good light diquark[759].

The discovery of the $X(3872)$ prompted a surge in modelling of exotic hadronics, and led to renewed interest in diquarks. A prominent model, due to Maiani and collaborators[760], advocated that the $X(3872)$ is a $J^P = 1^+$ double diquark state with composition $[cq]_1[\bar{c}\bar{q}]_0 + [cq]_0[\bar{c}\bar{q}]_1$. This assignment sets the mass of the open charm diquark, $m_{[cq]} = 1933$ MeV, and implies a rich spectrum of exotic states. A novel prediction of the model is that *two* neutral vector exotic states should exist with a mass difference of approximately 8 MeV. Focussing on flavor quantum numbers, these are mixtures of $[cu][\bar{c}\bar{u}]$ and $[cd][\bar{c}\bar{d}]$. Amongst others, scalar states are predicted at 3723 MeV and 3832 MeV.

In spite of the explanatory power of the model, and reasonable agreement with properties of the $X(3872)$, none of these additional states have been observed. (For a complete discussion of this issue, see Sec. 8.5.2.)

Notwithstanding the checkered history of the diquark model, it must become relevant as quark masses become much greater than the QCD scale, $\Lambda$. In this case the quarks will sit deeply in a Coulombic well, are compact, and are described well by perturbative gluon exchange. It is widely believed that bottom quarks are sufficiently heavy for these phenomena to occur. If a pair of bottom quarks forms a hadron with light degrees of freedom (such as light quarks or gluons), then it is reasonable to model the bottom quarks as a $[bb]$ diquark, and this expectation becomes rigorous as the heavy quark mass becomes very large.

A consequence of this concerns spin splittings in heavy-light mesons and baryons, as first observed by Savage and Wise[761]. In the following $Q$ represents a quark with mass larger than the QCD scale, $\Lambda$ (thus $b$, $c$), while $q$ represents a quark with mass much less than $\Lambda$. The latter then refers to $u$ and $d$ quarks. The strange quark is ambiguous in this classification, and is sometimes grouped with the light quarks, and sometimes with heavy quarks. In practice heavy quark symmetries only become clear at the bottom mass and higher, while light quark (chiral) symmetry applies well to $u$ and $d$ quarks, and fairly well to $s$ quarks.

Heavy quark spin degrees of freedom interact via their color dipole moments, which permits relating spin splittings in $QQq$ baryons and $\bar{Q}'q$ states, with a relationship given by

$$m_{\Sigma^*(Q)} - m_{\Sigma(Q)} = \frac{3}{2}\frac{m_{Q'}}{m_Q}\left(\frac{\alpha_S(m_Q)}{\alpha_S(m_{Q'})}\right)^{9/33-2n_f} \times \left(m_{V(Q')} - m_{P(Q')}\right). \qquad (5.1.3)$$

Here $V$ and $P$ refer to vector and pseudoscalar mesons, while $\Sigma^*$ and $\Sigma$ refer to ground state and spin-excited $QQq$ baryons.

A slightly more model-dependent application establishes that the heavy $J^P = 1^+$ $ud\bar{b}\bar{b}$ tetraquark state must be strongly bound. The argument relies on the spin splittings, $\Sigma_b - \Lambda_b$ and $\Xi_b' - \Xi_b$, which indicate that the $(\bar{\mathbf{3}}_F, 0, \bar{\mathbf{3}}_c)$ light diquark lies approximately 100 MeV below the spin-averaged light diquark mass. This diquark interacts with a $b$ meson with quantum numbers $(\mathbf{1}_F, \frac{1}{2}, \mathbf{3}_c)$ to produce the relevant baryons. As argued above, and verified by small $B^* - B$ and $\Sigma_b^* - \Sigma_b$ mass splittings, the heavy (di)quark spin must decouple from the light degrees of freedom. Thus a light diquark has a similar mass when coupled to a heavy $[\bar{b}\bar{b}]$ diquark. Since the heavy diquark has quantum numbers $(\mathbf{3}_F, 0, \mathbf{3}_c)$, the $[ud][\bar{b}\bar{b}]$ tetraquark has quantum numbers $I = 0, 1/2$ and $J^P = 1^+$. Recent lattice field theory computations have proven these expectations correct[762].

Diquarks continue to find application in a variety of areas: reducing the daunting complexity that arises in Bethe-Salpeter equations for many-quark systems, Sec. 5.3, the operator-product expansion, Sec. 5.8, instanton vacuum modelling, Sec. 5.11, heavy quark effective field theory, Sec. 6.1, models of quark matter, Sec. 7.2, tetraquark models, Sec. 8.5, baryons, Secs. 9.1, 9.2, 9.4, and models of hadronization, Sec. 11.4.

### 5.1.5 Current Developments

The advent of new theoretical tools and the discovery of many novel hadrons have fueled the continued development of the constituent quark model. Amongst the latter are the $X(3872)$ that strongly hints at $qq\bar{q}\bar{q}$ structure and the importance of coupling mesons to the meson-meson continuum. Strong evidence for states consisting of $qqqq\bar{q}$, called "pentaquarks", also exists. At the same time, the maturation of lattice field theory has permitted the theoretical exploration of many nonperturbative hadronic properties and novel states involving glue, such as glueballs and hybrids. Such studies also inform the development of refined quark models that are capable of describing an ever greater range of phenomena. The development of effective field theory and its application to hadronic physics has also greatly expanded and strengthened the base upon which quark models are developed. Finally, field-theoretic nonperturbative methods, such as those based on the Schwinger-Dyson and Bethe-Salpeter methods, have served to expand the understanding and purview of quark models.

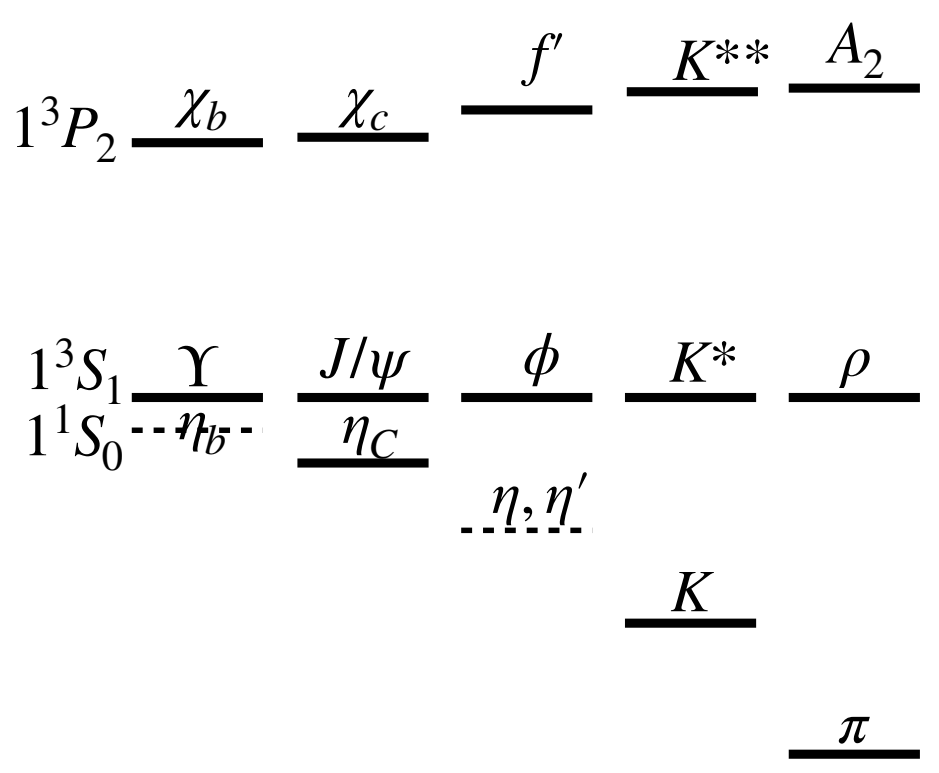

**Fig. 5.1.2** "A graphic illustration of the universality of meson dynamics". Figure taken from the original[763].

These new tools have helped to clarify several long-standing issues in the field. For example, it is well-known that the pion is anomalously light because it is the pseudo-Goldstone boson of QCD, reflecting the (broken) near chiral symmetry of the theory. Alternatively, the pion is light in quark models because the hyperfine interaction drives its mass well below that of the rho meson. The size of this mass splitting is *infinite* according to the one gluon exchange interaction (because it is proportional to $\delta(r)$)! In practice the hyperfine operator is smeared, which introduces a smearing parameter that can be fit to obtain the pion mass. This is hardly a satisfactory situation! In spite of this, Isgur has argued that the smooth evolution of hyperfine splitting from bottomonium to light quarks (Fig. 5.1.2) is a sign that the formalism is correct[763]. How these views can be made consistent is demonstrated in a specific model in Ref. [764], wherein it is shown how chiral symmetry breaking induced by a nontrivial vacuum and an effective hyperfine interaction mesh in a smooth fashion. Further insight is gained from the Schwinger-Dyson formalism, which convincingly demonstrates that chiral symmetry breaking gives rise to both a light pion and a dynamical quark mass that can be interpreted as the constituent quark[765].

Recent results from the lattice and other theoretical analyses indicate that long-held notions are likely incorrect. For example, scalar confinement cannot be correct–it has been known since the 1980s that a confining scalar $q\bar{q}$ interaction implies an *anti-confining* $qqq$ interaction because of the lack of an antiquark line.

(This disaster was avoided in, for example, the Godfrey-Isgur and Capstick-Isgur models by simply applying an extra sign.) The problem appears again in attempts at inducing chiral symmetry breaking in model field theories, where it is learned that scalar confinement interactions do not lead to a stable BCS-like vacuum[766]. In fact, it is not clear at all that the long range quark interaction need be described by a two-body interaction of the sort given above; QCD is much more complicated than this simple model[767].

Recent computations in lattice field theory have essentially settled the matter. This work relies on the model-independent expansion of the quark interaction in terms of nonperturbative matrix elements of gluonic operators[767, 768], which are evaluated numerically. The results disagree strongly with an assumed scalar long range interaction. They do agree in large part with a Dirac vector interaction, with the exceptions that the hyperfine interaction resembles a smeared delta function and the spin orbit interactions have effective string tensions that are reduced by a factor of approximately 77%[769]. The picture emerging is that perturbative gluon exchange dominates the interaction at very short distances (less than 0.1 fm) and an effective vector-like interaction dominates at intermediate ranges. At long range (greater than 1 fm), one must saturate gluon exchange with a sum over hybrid intermediate states. This brings in the nonperturbative matrix elements of chromoelectric and chromomagnetic fields (mentioned above) that give rise to the nontrivial structure seen in lattice field theory. It is somewhat ironic that early enthusiasm for perturbative gluon exchange has evolved in this fashion!

Other quark model lore from the 1980s has been swept away in a similar fashion. For example, the Godfrey-Isgur computation of meson decay to $\gamma\gamma$ employed a perturbative amplitude with a "mock meson" correction factor involving the meson mass. More sophisticated computations where the amplitude is computed with relativistic quark currents and a sum over intermediate states is made reveal good agreement with data and no need for artificial factors[770].

### 5.1.6 Open Problems

One of the major goals in modern quark modelling is incorporating the effects of nonperturbative gluonic degrees of freedom, which, of course, permits modelling glueballs and hybrid hadrons. Outright guesses from the past have been superseded by a body of lattice explorations of gluonic properties. Among these are the spectrum of adiabatic gluonic excitations[771, 772], the gluelump (bound states of gluons and a static adjoint color source) spectrum[486, 773–775], and properties of charmonium hybrids[776–778]. Of particular interest is the confirmation that the heavy quark multiplet structure anticipated in Ref. [772] is reflected in the charmonium spectrum[776]. It is interesting, and very suggestive, that this multiplet structure can be reproduced by degrees of freedom consisting of a quark, an antiquark, and an effective axial gluon with quantum numbers $J^{PC} = 1^{+-}$[493], pointing the way to possible future models.

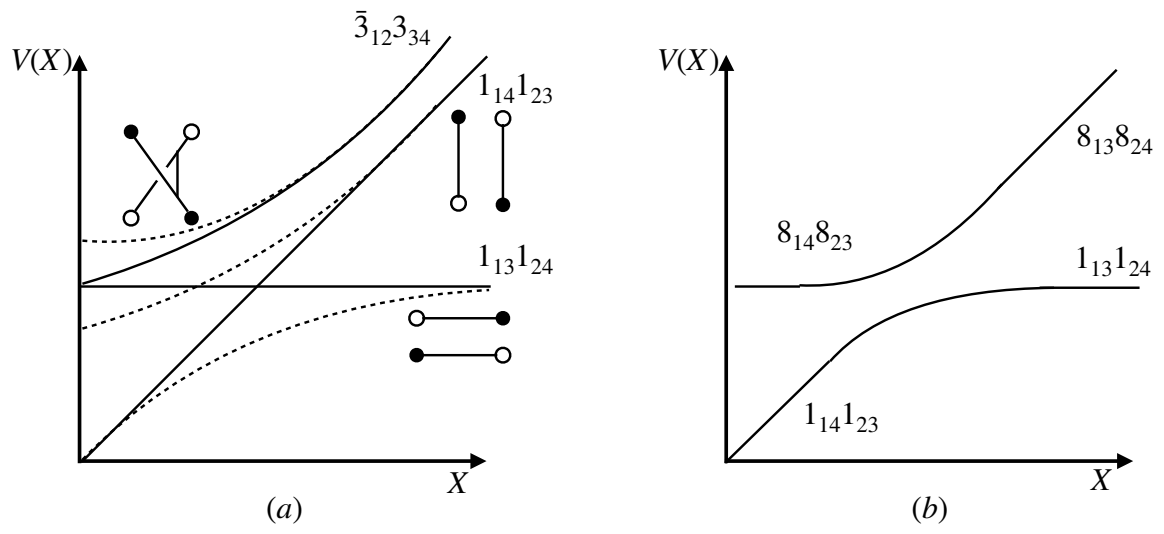

**Fig. 5.1.3** "The adiabatic potentials of the flux tube model (a) and of the $\vec{F}_i \cdot \vec{F}_j$ potential model (b) for two $qq\bar{q}\bar{q}$ geometries." Figure taken from the original[763].

The advent of compelling experimental evidence for tetra- and pentaquark states has heightened interest in modelling multiquark hadrons. This is an old field, which in the past suffered from sufficiently poor computations that Isgur dubbed it the "multiquark fiasco"[763]. Many technical problems were present in these calculations, but the chief physics problem is the nature of the quark interaction when more than three quarks are present. The issue, for example, is that a $qq\bar{q}\bar{q}$ can separate into two meson-meson channels and that the gluonic degrees of freedom should experience adiabatic surface crossing when transitioning between these configurations (see Fig. 5.1.3). Thus new gluonic interactions are necessarily introduced in multiquark states. Of course, one could always model these as a sum of two-body interactions with a perturbative color structure, but this seems unlikely to be viable. A widely accepted model of multiquark gluodynamics does not exist yet, and is urgently needed.

Multiquark states necessarily couple to systems of mesons and baryons, which makes it incumbent on modellers to understand the effects of coupled channels on hadronic properties. This requires knowing the effective quark pair operator. A common model, dating to 1969, has already been mentioned[732], but this can surely be improved. As a result, existing models of Fock sector mixing remain speculative. The problem is not amenable to effective field theory, so progress will likely rely on input from lattice field theory. Progress is ur-gent since channel coupling effects are expected to be important in many sectors of the spectrum, including the perpetually enigmatic light scalars mesons, and all states near thresholds, such as the $X(3872)$, the $P_c$ pentaquarks, and the $Z_c$ and $Z_b$ states.

It is perhaps a surprise that a model dating back nearly 60 years remains an active field of research. Such are the mysteries of QCD. On thing is certain: the quark model remains the de facto standard by which hadrons are interpreted.

## 5.2 Hidden Color

**Alexandre Deur**

Nuclear physics is one of the first rungs of the complexity ladder rising from our current fundamental understanding of Nature in terms of the Standard Model. The effective degrees of freedom (d.o.f) that emerge in nuclear physics are the hadrons, namely nucleons, mesons and their excited states. Yet, effective theories are intrinsically limited, their effective d.o.f being insufficient to account for peculiar phenomena, e.g., diffraction for geometrical optics. Then, more fundamental d.o.f are necessary. Likewise, certain nuclear phenomena are not reducible to hadronic d.o.f and either partonic d.o.f or new effective d.o.f are necessary. Hidden color (HC) is such a phenomenon. In conventional nuclear physics, a nucleus – such as the deuteron[14] – is effectively a bound state of individual nucleons. However, at the more fundamental level of QCD, the nuclear eigenstate can also have additional multi-quark Fock states which have zero color overall, but do not cluster as a collection of nucleons. These Fock states represent the HC d.o.f of nuclei.

The possibility of HC d.o.f [779–784] arises from observing that the representation of color singlet multihadron systems allows for colored cluster ($C_c$, colored "hadrons") components, e.g., a red-red-blue cluster bound to a green-green-blue cluster contributing to the deuteron wavefunction. Such a configuration can equivalently be reexpressed as a sum of singlet components, but without well-defined clustering properties since a given valence quark has a substantial probability to belong to any of the singlet states. Therefore, regardless of what (equivalent) representation is preferred, it cannot be expressed with singlet hadronic clusters, i.e., colorless hadronic d.o.f. This is HC. Clearly, HC goes beyond traditional nuclear physics but is a natural expectation of the underlying theory, QCD. HC predicts nuclear states not describable with usual hadronic d.o.f but with multiquark wavefunctions, e.g., 6-quark singlet states, or singlet systems made of $C_c$. The latter perspective renders intuitive that HC states are short-distance binding configurations.

For example, in a hadronic basis of nucleon $N$, $\Delta$ and $C_c$ d.o.f (for simplicity we ignore other $N^*$ isobars contributions), the deuteron is a sum of $NN$, $\Delta\Delta$ and $C_cC_c$ components, the latter dominating at short distance, *viz*, large $Q^2$ [785]:

$$|D\rangle = |NN\rangle + |\Delta\Delta\rangle + |C_cC_c\rangle$$

with

$$|NN\rangle = \tfrac{1}{3}\,|[6]\{33\}\rangle + \tfrac{2}{3}\,|[42]\{33\}\rangle - \tfrac{2}{3}\,|[42]\{51\}\rangle\,, \tag{5.2.1}$$

$$|\Delta\Delta\rangle = \sqrt{\tfrac{4}{45}}\,|[6]\{33\}\rangle + \sqrt{\tfrac{16}{45}}\,|[42]\{33\}\rangle + \sqrt{\tfrac{25}{45}}\,|[42]\{51\}\rangle\,, \tag{5.2.2}$$

$$|C_cC_c\rangle = \sqrt{\tfrac{4}{5}}\,|[6]\{33\}\rangle - \sqrt{\tfrac{1}{5}}\,|[42]\{33\}\rangle\,, \tag{5.2.3}$$

where [ ] and { } label respectively the orbital and spin-isospin symmetries which are characterized by the bracketed number in the usual Young tableau way, e.g.,

[6] ≡ 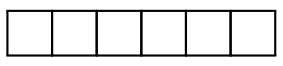

signifies 6 quarks in $s$-shell, or

[42] ≡

is for 4 quarks in $s$-shell and 2 in $p$-shell [786]. For $Q^2 \to \infty$, [6] dominates over [42]. Thus, the deuteron state is [6]{33} symmetric (and totally antisymmetric overall), from which 4/5 comes from the HC component, Eq. (5.2.3). The 80% dominance of HC at large $Q^2$ is therefore expected to control elastic scattering off the deuteron in this limit. In fact, the ratio of the reduced deuteron form factor (i.e., normalized to the nucleon form factor squared) to that of the pion is about 15% for $Q^2$ of a few GeV$^2$, indicating 15% of HC in $|D\rangle$ at this scale [785]. That $|NN\rangle$ and $|\Delta\Delta\rangle$ nearly vanish at large $Q^2$ means that two singlet hadrons tend to not be found close to each others, i.e., the traditional (*viz*, between singlet hadrons) nuclear force is repulsive at short distance. The rise with $Q^2$ of [6] over [42] tells us that the components of $|D\rangle$ behave differently with $Q^2$. Their evolutions come from gluon exchange and were calculated in Refs. [787–789]. It was shown that the singlet $pn$ state of the deuteron prevalent at small $Q^2$ evolves into 5 states: itself and 4 HC states.

The number of HC states quickly increases with the mass number $A$ of the system. For $A = 1$ there is 1 singlet state and no HC state:

$$3 \otimes 3 \otimes 3 = 10 \oplus 8 \oplus 8 \oplus 1,$$

[14] Throughout this section, deuteron is used as example of nuclear system, but the discussion is generic to multi-nucleon systems.

the last being the color singlet, the nucleon. For the deuteron, $A = 2$ and

$$3 \otimes 3 \otimes 3 \otimes 3 \otimes 3 \otimes 3$$
$$= 28 \oplus 5(35) \oplus 9(27) \oplus 15(10)$$
$$\oplus\, 16(8) \oplus 5(10^*) \oplus 5(1),$$

with the 5 last states 5(1) being the singlet states. Since there can be only one singlet state made of colorless 3-quark clusters –the traditional *pn* (or isobars) state– the four remaining singlet states are HC states. For $A = 3$, there are 41 HC states [790]. Calculating strictly within QCD the $Q^2$-evolution of nuclear amplitudes is presently not possible: Just $|D\rangle$ at leading order involves millions of Feynman graphs. Using a hadronic effective QFT is not helpful because adding the HC d.o.f negates the theory predicability [790]. A solution is to use the *reduced nuclear amplitude* technique [779, 791]. Based on LF QCD [792, 793], it models nuclear scattering amplitudes that obey QCD counting rules [134] (Section 5.10) and gauge invariance. The method neglects nuclear binding so that a nucleus is modeled as a cluster of collinear hadrons. Thus, the nuclear LFWF factorizes as a product of LFWF of nucleons in the nucleus times those of quarks in a nucleon: $\psi_A = \psi_{N/A} \prod_N \psi_{q/N}$, with the convenient LFWF probabilistic interpretation of the Fock states retained.

What are the possible signals for HC? An intuitive one is the yield ratio $(\gamma d \to \Delta^{++}\Delta^-)/(\gamma d \to pn)$; if $|D\rangle$ contained only a state of two weakly bound singlet hadrons,

$$\textcircled{\scriptstyle d\,u\,u}\;\textcircled{\scriptstyle d\,d\,u},$$

it would not break into a

$$\textcircled{\scriptstyle u\,u\,u}\,\textcircled{\scriptstyle d\,d\,d} = \Delta^{++}\Delta^-.$$

However, a 6-quark $|uuuddd\rangle$ state can well split into $\Delta^{++}$ and $\Delta^-$.

There are other possible HC signatures [785]: the dominance of HC at short distances makes large angle Compton scattering and pion photoproduction off the deuteron prime channels to search for HC. In electron scattering, the deuteron form factor at large $Q^2$ should not be explainable with hadronic d.o.f. Likewise, the deuteron inclusive tensor spin structure function $b_1$, a leading-twist quantity, is expected to be especially sensitive to HC [794]. Short range correlation (SRC) measurements can also provide a signal for HC as they probe the 2-nucleon potential at short distance. Thus, SRC data should be sensitive to the repulsion expected by HC and signaled by the vanishing of the $|NN\rangle$ and $|\Delta\Delta\rangle$ components. The quasi-elastic reaction (to access large $x$) at high $Q^2$ resolves the nucleons of a nucleus and provides the SRC of nucleon pairs. The ratio of *pn* over *pp* pairs was found to be 5 times larger than the standard hadronic expectation [795, 796]. This may stem from the repulsive core of the 2-nucleon potential. Furthermore, the measurement of the strength of 3-nucleon correlations in $A > 2$ nuclei indicates that their contribution is larger in heavy nuclei than initially expected, suggestive of the rapid increase of number of HC states with $A$. A challenge with SRC measurements is the fast $Q^2$ fall-off of form factors, so one may alternatively study, also at large $Q^2$ and high $x$, the behavior of inclusive structure functions which should obey in that regime the QCD dimensional counting rules based on the number $n_s$ of spectator partons [134] (see Section 5.10),

$$xF(x) \sim (1 - x/2)^{2n_s-1}.$$

In the maximum $x \to 2$ limit for the deuteron, $n_s = 5$ for HC (6-quark system) but $n_s = 2$ without a dominant HC state. HC evidence may come from indirect observations: without HC, the only process binding hadrons not sharing covalent quarks is glueball exchange. HC provides additional processes [787] which may be necessary to explain the structure of neutron stars [797, 798].

HC may have already been observed. We mentioned the SRC observations and that the deuteron form factor normalized to the nucleon form factor squared is 15% that of the pion. The $I(J^P) = 0(3^+)$ of the well-established $d^*(2380)$ (or $D_{03}$) *p-n* resonance [799–807] compellingly suggests that it is a 6-quark system with dominant HC [808–813]. Furthermore, while its dynamical decay properties can also be explained by a $\Delta\Delta$ state, the narrow 70 MeV width of the $d^*$ is 3 times smaller than expected for the $\Delta\Delta$ but agrees with a HC state. Refs. [814, 815] reviewed recently the $d^*(2380)$ properties. Similarly, the narrow de-excitation of $^4\mathrm{He}^*$ through $e^+e^-$ emission seen at ATOMKI [816] can be understood as the $^4$He nucleus being excited into a 12-quark HC state made of 5 colored *ud* pairs (hexadiquark) [817]: it was shown that the ATMOKI anomaly cannot be accounted for by standard electromagnetic decay without producing first a HC state [818]. The latter also explains the unusually strong binding of the $^4$He nucleus. Another possible observation of HC comes the $b_1$ data from HERMES [819]. They are positive for $x < 0.1$ but appear to become negative around $x \simeq 0.3$, which is expected of a 6-quark HC state [794].

These signals each hint at the existence of hidden-color degrees of freedom. By reaching higher $x$ and $Q^2$, the 12 GeV upgrade of JLab and the future EIC [820] will provide the opportunity to confirm this fundamental feature of QCD.

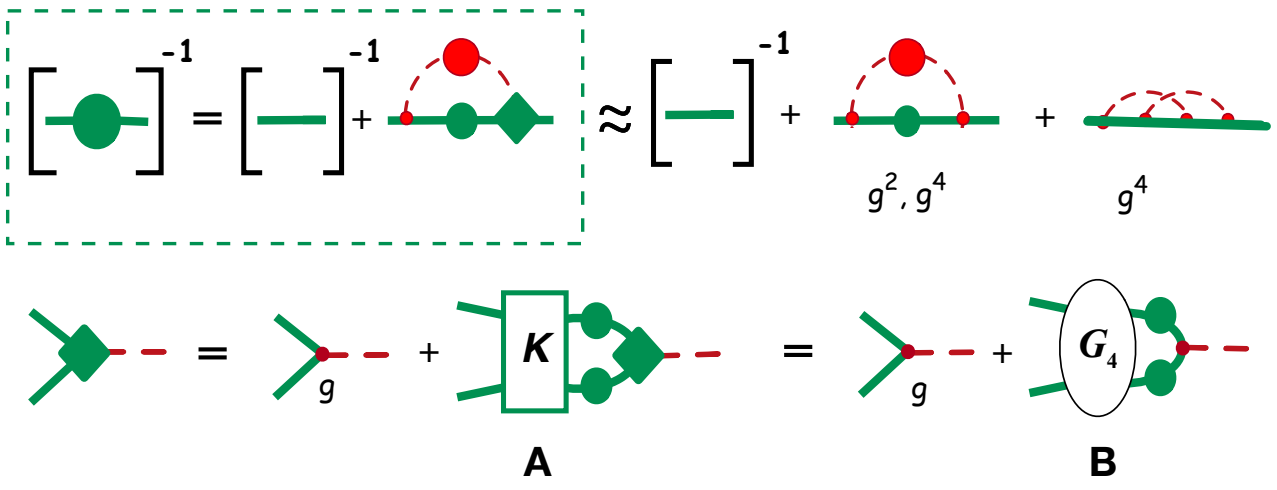

**Fig. 5.3.1** Top row: The exact DSE for the inverse dressed fermion propagator (in the dotted box), and its approximation to 4th order in QED. Bottom row: two versions of the exact DSE for the dressed QED vertex $\Gamma_\mu$ (green diamond): Diagram (A) in terms of the $q\bar{q}$ irreducible kernel $K$, and (B) in terms of the full scattering amplitude $G_4$. The thick green (dashed red) lines are the fermion (photon), solid green (red) circles are the fermion (photon) self energies so that a fully dressed propagator is a green (red) line with a green (red) circle; and small red dots label the point coupling $\gamma_\mu$ and have no structure (renormalization constants are ignored here).

## 5.3 DS/BS equations

**Franz Gross and Pieter Maris**

### 5.3.1 Introduction

In this section we look at two closely related approaches to treating the strong interactions that existed before 1972, and remained very useful, even after the onset of QCD. One of these originated with papers by Dyson (1949) [821] and Schwinger (1951) [822, 823], referred to as the Dyson-Schwinger equations (DSEs), and the second is the well known Bethe-Salpeter equation (BSE) [824], introduced in 1951[15].

In general, the DSEs form an infinite set of coupled integral equations for the Green's functions $G_n$ of a quantum field theory. [16] These equations are exact, but in practical calculations this set has to be truncated. The equations can be derived formally from the matrix elements of the Lagrangian density (as was done in the original papers), or in the path-integral formalism using functional derivatives [831], but Feynman diagrams can be used to provide a simple, pictorial way to understand them. Using QED as an example [17], Fig. 5.3.1 shows the exact DSEs needed to describe the self-energy of each fermion, and the dressed $\overline{\psi}_i\gamma^\mu\psi_i A_\mu$ vertex $\Gamma_i^\mu$.

[15] Although the BSE can be used to describe scattering, this seminal paper was entitled *A relativistic equation for bound state problems*, particularly serendipitous for applications to QCD, where all physical states are bound states of quarks, antiquarks and gluons.

[16] For recent reviews of the DSEs in the context of QCD and hadron physics, see Refs. [765, 825–830]

[17] When applied to QCD, with the photon replaced by a gluon, additional terms, such as the 3-gluon vertex, must be added.

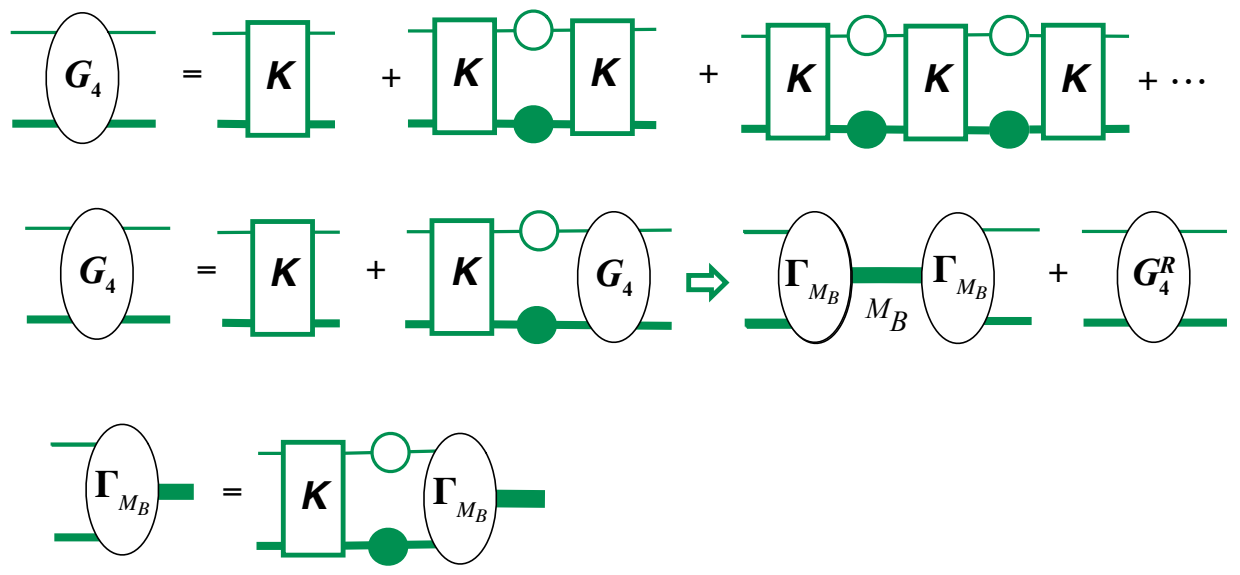

**Fig. 5.3.2** Diagrammatic representation of the BSE propagator for two unequal mass particles $m_1 > m_2$. The first line represents the iteration of an *irreducible* kernel $\mathcal{K}$, which is summed by the BSE (first part of the second line). If the propagator has a pole, then the BSE vertex function satisfies the homogeneous BSE shown in the last line.

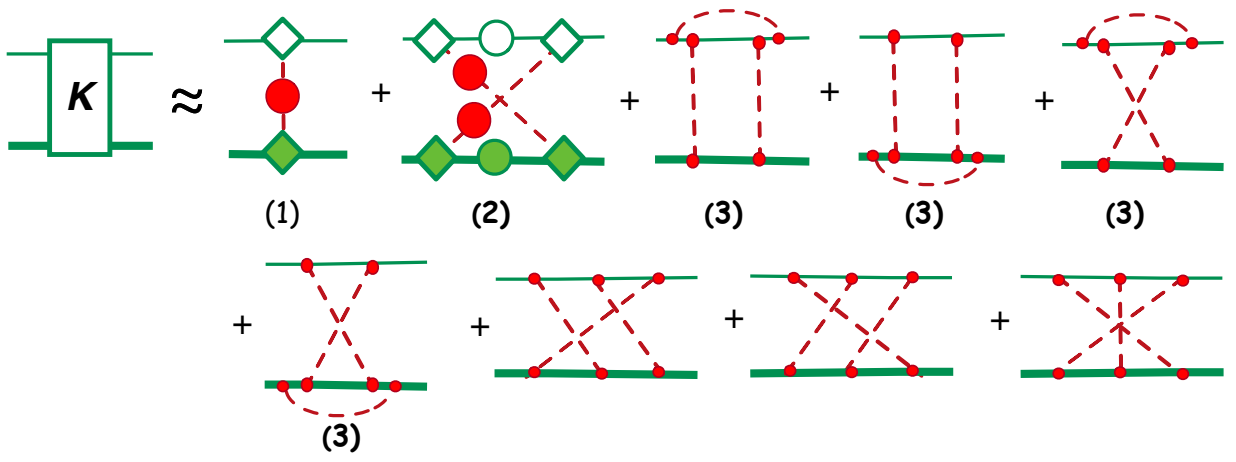

**Fig. 5.3.3** Diagrammatic representation of the BSE kernel up to 6th order in $g$. Diagram (2) is the dressed xbox diagram and diagrams (3) are irreducible photon dressings of the box and xbox.

The fermion-antifermion scattering amplitude $G_4$ of the two different fermions can be written as a series of interactions shown in the upper line of Fig. 5.3.2. Here the kernel $K$ is the sum of *irreducible* contributions to the off-shell scattering (*i.e.* diagrams that cannot be reduced by drawing an internal line through the diagram that intersects only the two fermions). The infinite series of *iterations* of the irreducible diagrams (each referred to as *reducible* because it can be cut into two pieces by an internal line which intersects only the two particles), connected by dressed propagators, is then summed by the equation shown in the LHS of the middle line. This is the DSE[18]. for the scattering amplitude $G_4$. If a bound state exists, it shows up as pole in this scattering amplitude, as illustrated in the RHS of the second line of Fig. 5.3.2. The BSE for the Bound State Amplitude (BSA) or vertex function, $\Gamma$ (shown in the bottom line), has the same kernel as $G_4$. Figure 5.3.3 shows contributions to the QED kernel up to order $g^6$. There is no known way to sum these contributions in closed form.

[18] For two particle scattering, DSE and BSE are used interchangeably.

*The bound state BSE*
As an example, the BSE for a $q\bar{Q}$ bound state in QED is

$$\Gamma_{M_B}(p;\widehat{P}) = \int \frac{d^4k}{(2\pi)^4} K_{ij}(p,k;\widehat{P})\mathcal{O}_i\chi_{M_B}(k;\widehat{P})\mathcal{O}_j \to 4\pi\alpha\int \frac{d^4k}{(2\pi)^4} D_{\nu\mu}(p,k;\widehat{P})\gamma^\nu\chi_{M_B}(k;\widehat{P})\gamma^\mu, \quad (5.3.1)$$

where $\chi_{M_B}(k;\widehat{P})$ is the BS wave function

$$\chi_{M_B}(k;\widehat{P}) = S_2(k_2)\Gamma_{M_B}(k;\widehat{P})S_1(k_1)\,, \quad (5.3.2)$$

with $S_i(k_i)$ the dressed propagator for particle $i$ and $\widehat{P}^2 = M_B^2$. The first line is exact, with the kernel written in the general form $K = K_{ij}\mathcal{O}_i \otimes \mathcal{O}_j$[19]; the second line is the *ladder truncation* with the kernel describing one photon exchange only, so $K \to 4\pi\alpha D_{\nu\mu}\gamma^\nu \otimes \gamma^\nu$. Dirac indices have been suppressed, and the four-momentum of the incoming $Q$ is $p_1 = p-(1-\eta)\widehat{P}$ and of the outgoing $q$ is $p_2 = p+\eta\widehat{P}$, reflecting the fact that the total momentum $\widehat{P}$ is conserved in relativistic equations. The physical observables do not depend on the choice of $\eta$, and the natural choice for mesons with equal-mass constituents (like a pion) is $\eta = \frac{1}{2}$. The canonical normalization condition for the BSE bound state vertex function can be derived directly from the inhomogeneous BSE (see e.g. Refs. [831, 832]).

Very soon after the BSE was introduced, Wick [833] showed that the equation could be transformed from Minkowski space to Euclidean space by rotating the time component to the imaginary axis $\{t,\mathbf{r}\} \to \{i\tau,\mathbf{r}\}$ (now referred to as a Wick rotation). Building on Wick's results, Cutkosky [834] found all the exact solutions to the bound state BSE in *ladder* truncation for a scalar theory of the $\chi^2\phi$ type where the exchange particle $\phi$ is massless. The solutions are symmetric under the $O(4)$ symmetry group, and hence have the same degeneracy as the nonrelativistic hydrogen atom. Some of the solutions correspond to excitations in the time direction that have no nonrelativistic analogues. Furthermore, these solutions have a negative norm (at least in QED and QCD), and are therefore unphysical. As far as we know, no other analytic solutions have been found, but in the last 25 years accurate solutions of the BSE in ladder truncation have been obtained numerically for both scalar and fermionic systems, discussed below.

Several facts about the BSE are sometimes overlooked:

- The equation shown in Fig. 5.3.2 is exact, but only if the *exact* kernel *and* self energies are known.
- All applications of the BSE are therefore approximations using an approximate kernel and self-energies.
- In addition to Eq. 5.3.1, which is a homogeneous equation, there is also a canonical normalization condition for the BSA; one should not normalize the BSA to a convenient observable.

*Methods to solve the BSE in Minkowski metric*
Due to the presence of poles in both the constituent propagators and in the kernel (coming from the exchanged bosons), it is highly nontrivial to solve the BSE numerically in Minkowski metric, even in ladder truncation. There are two methods to investigate the BSE directly in Minkowski metric, both dating back to the late 60s: the Covariant Spectator Theory (CST) which we discuss in Sec. 5.3.2, and use of the Nakanishi representation (also known as Perturbation Theory Integral Representation) [835].

The Nakanishi representation for the BSA is a spectral representation in which the singularities that arise from the poles in the propagators are isolated, allowing the BSE to be reduced to an integral equation for the (non singular) spectral function. This has been done initially for scalar field theory [836], and subsequently for fermion-antifermion bound states [837–839]. The obtained BSAs have been benchmarked against direct numerical solutions of the ladder BSE for Euclidean (spacelike) relative momenta.

Recently, the scalar BSE in ladder truncation has also been investigated in Minkowski metric by starting in the Euclidean formulation and rotating the $p_4$ axis to $ip_0$ (i.e. undoing the Wick rotation numerically) [840], and by using contour deformations in order to avoid singularities [841]. These methods give, within numerical precision, the same results for the BSA in the timelike region as the Nakanishi representation.

*Connection to the Light-Front wavefunction*
The use of the light-front (LF, first referred to as the infinite momentum frame) was introduced by Weinberg in 1966 [842], and the technique was developed very extensively in the 1980's by Lepage and Brodsky [207] and many others. It is now a standard method for describing the structure of hadrons and calculating a range of observables. Application of this technique will be extensively discussed in Sec. 5.4. Use of the LF is not manifestly rotationally invariant, but this can be handled by imposing the so-called angular conditions; see, for example, Ref. [843].

The LF wave function can be obtained from $\chi(p;P)$ by integrating over $p^- = p^0-p^3$, leaving $p^0+p^3 \equiv xP^+$ and $\mathbf{p}_\perp = \{p^x,p^y\}$ as independent variables. It turns out that the LF wave function, $\psi(x,p_\perp)$, is only nonzero

[19] Each of the operators $\mathcal{O}_i$ describes the structure of the dressed vertices, including possibilities like those illustrated in diagrams (2) and (3) of Fig. 5.3.3

for $0 < x < 1$, and vanishes outside this range, even though $p^+$ runs from minus infinity to plus infinity. This has been confirmed numerically for scalar theories in ladder truncation.

Instead of solving the BSE in Minkowski metric, and then projecting onto the light-front, one can also reconstruct the LF wave function (or e.g. parton distributions) from their moments, which can be evaluated directly from the BS wave function [844–846]. One caveat to keep in mind is that the BSE is typically solved in covariant gauges; the most commonly used gauge in the literature is the Landau gauge, though other gauges such as Feynman gauge are also used. On the other hand, the LF wave function is usually investigated in LF gauge. This makes it nontrivial to compare LF wave functions obtained from the explicitly covariant BSE to LF wavefuctions obtained within a LF approach.

### 5.3.2 The Covariant Spectator Theory (CST)

The CST, which is related, but not identical, to the BSE, can be obtained from the BSE if the internal loop energy is evaluated keeping only the pole contribution from the heaviest particle [847].[20] If $m_+ = m_1 > m_2 = m_-$, $\eta = \frac{1}{2}$, treating particle 1 as outgoing, and working in the rest frame where $P = \{W, \mathbf{0}\}$, then $p_1 = \frac{1}{2}P - p$, and the one-channel CST equation can be obtained from (5.3.1) using the prescription [21]

$$\Gamma(p;P) = -i\int \frac{d^4k}{(2\pi)^4}\frac{F(p,k;P)}{d_+(k)d_-(k)} \rightarrow \int \frac{d^3k}{(2\pi)^3 2E_k^+}\left[\frac{F(p,\widehat{k};P)}{\delta_m^2 + W(2E_k^+ - W)}\right], \quad (5.3.3)$$

where $d_\pm(k) = m_\pm^2 - (k \mp \frac{1}{2}P)^2 - i\epsilon$, $F$ is any covariant function, $\widehat{k}_1 = \frac{1}{2}W - \widehat{k} = \{E_k^+, \mathbf{k}\}$, $(E_k^+)^2 = m_+^2 + \mathbf{k}^2$, so that $(\widehat{k}_+)^2 = m_+^2$, and $\delta_m^2 = m_-^2 - m_+^2$. The CST equation is covariant in three dimensional space, and, unlike the LF, is rotationally invariant. The major motivation for the use of CST equations is that they have a smooth nonrelativistic limit, and in a few cases their ladder approximation is more accurate than the ladder approximation to the BSE. Their major disadvantage is that their kernels can be singular, and the treatment of these introduces an additional level of phenomenology (see below).

In scalar field theories when $m_1 \rightarrow \infty$, it has been shown that the sum of all ladders and crossed ladders (the *generalized* ladder sum) is given by the solution of the CST equation with *only* the one-boson-exchange (OBE) kernel (see Refs. [832] and [847]) [22]. This is referred to as the *cancellation theorem*.

While the complete cancellation holds only in an exceptional case, partial cancellations occur for other cases. Using the Feynman-Schwinger representation [848], it is possible to calculate the exact result for the generalized ladder sum without vertex or self-energy corrections. For scalar theories where $m_1 = m_2 \neq \infty$ and the exchanged mass $\mu = 0.15\,m$ [849], the BSE in ladder approximation gives only about one-quarter of the correct binding energy (at large coupling), while the one-channel CST equation, also in ladder approximation, gives a little more that half the correct result. The OBE approximation in the light-front approach gives the same result as the BSE in ladder approximation [850][23]. Another approach, the equal-time (ET) favored by Tjon[852] is slightly better than the CST, but only the CST (to our knowledge) uses the same two-body scattering amplitude in both the two-body and three-body systems. In a later paper [853], it was shown that the contributions of all self-energies and vertex corrections for scalar QED are very small, so that in this case the generalized ladders dominate (and are well approximated by the CST and ET). These remarkable results apply only to scalar theories, so the main justification for the use of the CST must rest on its simple nonrelativistic limit.

It turns out that the one-body CST prescription (5.3.3) must be generalized if it is to be used for all cases including $m_- = m_+$ and $W \rightarrow 0$. To treat these limits successfully, all four $k_0$ poles from the two fermion propagators must be included. There are two poles in the upper half $k_0$ plane ($r = -$) and two in the lower half ($r = +$), and if $s = \pm$ denotes the poles from particles $m_\pm$, then they can all be denoted by $k_{0r}^s = rE_k^s + \frac{1}{2}sW - ir\epsilon$. Since the contour can be closed in *either* half plane (but not both), we average over the two choices. This gives the new prescription

$$\Gamma(p;P) \rightarrow \frac{1}{2}\sum_{s,r}\int \frac{d^3k}{(2\pi)^3 2E_k^s}\left[\frac{F(p,\widehat{k}_r^s;P)}{s\delta_m^2 - W(2rsE_k^s + W)}\right], \quad (5.3.4)$$

where $\widehat{k}_r^s = \{k_{0r}^s, \mathbf{k}\}$ with $(E_k^\pm)^2 = m_\pm^2 + \mathbf{k}^2$.

The sum on the RHS of this equation has four terms, and substituting the four values $p \rightarrow \widehat{p}_r^s$ into the LHS

[20] This is sometimes referred to as "restricting the particle to its mass shell."

[21] With our choice of momenta, this is obtained by closing the $k_0$ contour in the upper half plane and keeping only the positive energy pole of particle 1, at $k_0 = \frac{1}{2}W - E_k^+$.

[22] In other words, when $m_1 \rightarrow \infty$, the CST in OBE approximation gives the same result as the BSE for a kernel containing *all* irreducible crossed ladders.

[23] This is not true for three-body systems, due to contributions with two (or more) exchange bosons in flight, which are included in the ladder BSE, but not in the OBE approximation on the light-front [851].

gives four coupled equations.[24] As discussed above, only one channel is needed when $m_+ \to \infty$. When the particles are identical, symmetry under interchange requires that the equation transform into itself when $p_1 \leftrightarrow p_2$, or $k_1 \leftrightarrow k_2$, and looking at $k^s_{0r}$ shows that this requires (if $P$ is not small) at least the channels where $\{r, s\} = \{+, +\}$ and $\{-.-\}$, so that $rs = +$ in both cases. Looking at (5.3.4, it is clear that it is symmetric under this transformation, remembering that for identical particles, $\delta^2_m = 0$ and $E^+_k - E^-_k$. Finally, when $W$ is small, there will be a singularity at $W = 0$ unless all four channels are kept.

Unfortunately, when a OBE kernel connects the channel with particle 1 on-shell to the channel with particle 2 is on-shell, the kernel will develop singularities. These are discussed in detail in Ref. [854], but the preferred way to remove them was only developed recently[25].

*Nuclear physics applications of the CST*

The two-channel CST has been used to give a high precision fit to the $np$ scattering data below 350 MeV ($\chi^2 = 1.12$ using only 15 parameters [854]), to explain the deuteron form factors (giving a quadrupole moment within 1% of its experimental value [856]), and to study the three nucleon system. All of these studies were done with two models. The simplest and most successful one uses a covariant OBE kernel consisting of the exchange of 6 mesons: $\pi$, $\eta$, $\sigma_0$ and $\sigma_1$ (scalar mesons with isospin 0 and 1), and $\rho$ and $\omega$. An interesting feature of these OBE models is that they include an off-shell coupling for the $\sigma$ mesons of the form

$$\Lambda_\sigma(p, k) = g_\sigma - \nu_\sigma \left[1 - \frac{\not p + \not k}{2m}\right], \qquad (5.3.5)$$

where the term proportional to $\nu_\sigma$ will give zero when the nucleons are on shell (with $\not p \to m \leftarrow \not k$). As it turns out (see below), this off-shell coupling is very important to the success of the model.

In the early days before the advent of QCD and powerful computers, the study of three nucleon systems posed special problems. The Alt-Grassberger-Sandhas equations [857], developed in 1967, introduced a systematic procedure for finding the solutions of $n$-body problems from the solution of the $n - 1$ body problem. Examples of early papers working directly with the the three nucleon equation are found in Ref. [858], which presents solutions with realistic potentials, and Ref. [859], which solves the 3-body BSE with separable kernels.

[24] This should be considered the correct form for the CST in all cases, but often some of the channels can be ignored.

[25] **FG**: These singularities troubled me for years. They are integrable, giving finite results, but only with the method described in Ref. [855] do I feel the problem is fully under theoretical control.

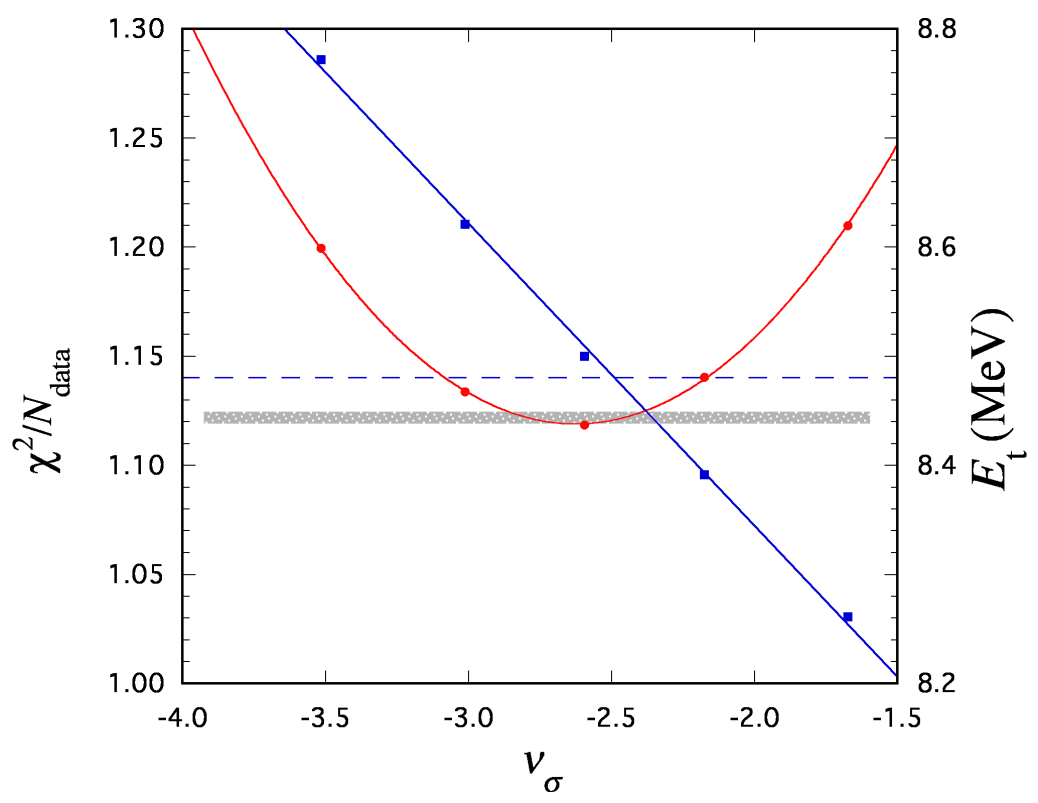

**Fig. 5.3.4** The red line (left-hand scale) shows how $\chi^2$ varies with $\nu_\sigma$, with the best fit at $\nu_\sigma \simeq -2.6$. The blue line (right-hand scale) shows the (linear) variation of the triton binding energy with $\nu_\sigma$, with the best fit also at $\nu_\sigma \simeq -2.6$. (From Ref. [860].)

The three-body CST equation given in Ref. [861] was used to compute the triton binding energy [862], and the three-nucleon form factor [863, 864]. During these studies a remarkable discovery was made: the best fits to the $np$ data require $\nu_\sigma \neq 0$ [26], and the same value of $\nu_\sigma$ also gives the best fit to the triton binding energy! This shows that three body-forces are not needed to explain this observable. This discovery, first found in 1996 [862], is shown with the latest (and best) fits in Fig. 5.3.4. It is a robust result that has continued to hold as the fits were improved, and is still not understood.

*Meson spectrum in the CST*

In the CST treatment, mesons are $q\bar{q}$ bound states with one quark confined to its mass-shell. States like the $\rho$, where $m_\rho > 2m_q$, could have both the quark and anti-quark on-shell at the same time unless the interaction forbids it. Fortunately, the structure of the CST equations permits an attractive relativistic generalization of linear confinement. This definition of confinement was first introduced in 1991 [865], and in 1999 it was shown explicitly that the confining interaction does indeed guarantee that meson vertex functions are zero when both quark and antiquark are on shell [866]. Subsequently, an improved definition [867] was found. For any smooth S-state function $\phi(p)$ the action of the linear confinement kernel is

$$\left\langle V_L \phi \right\rangle(p_1) = -\int_k \frac{m}{E_k} \frac{8\pi\sigma[\phi(\widehat{k}_1) - \phi(\widehat{p}_R)]}{(p_1 - \widehat{k}_1)^4} \qquad (5.3.6)$$

[26] **FG**: Originally we (Stadler and I) tried to fit the $np$ data without the off-shell coupling, and got the very high $\chi^2$ that an extrapolation of the curve shown in Fig. 5.3.4 suggests. Only after a frantic attempt to do better did we discover the importance of $\nu_\sigma$. Later, we were surprised to realize that the same mechanism also gave the correct triton binding energy.

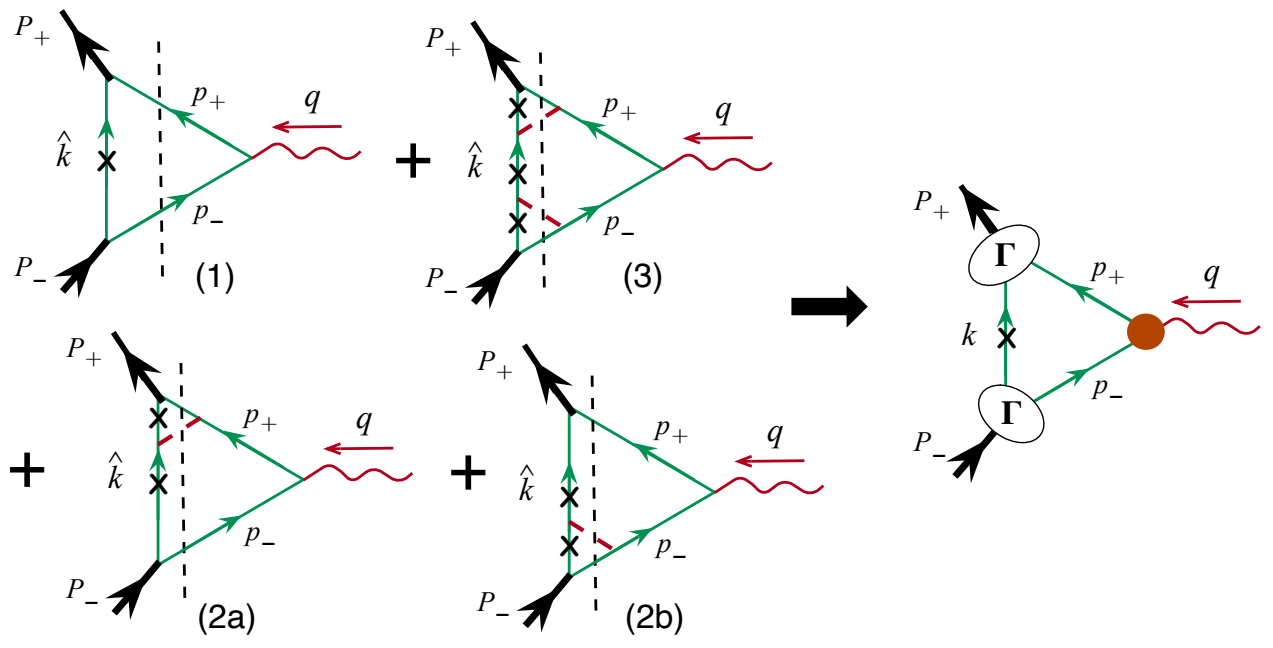

**Fig. 5.3.5** The form factor of a bound state (meson or deuteron). Left panel: the four diagrams that give the lowest anomalous thresholds for the dispersion integrals, with dispersion cuts shown by the dashed lines. Note the multiple spectators on shell. The dashed red line represents the exchanged particles that bind the state. Right panel: diagram showing the triangle (or impulse) contribution expressed in terms of the BS vertex function $\varGamma$ and dressed current (large red dot), needed to ensure gauge invariance (only in a CST calculation is the spectator on-shell).

where the spin dependence, and the form factors that provide convergence at large momenta have been omitted, $\sigma$ is the string tension, $p_1$ and $\widehat{k}_1$ are the momenta of particle 1, $\widehat{k}_1^2 = m_1^2$ is on-shell, and $\widehat{p}_R$ is chosen to reduce the singularity at $(p_1 - \widehat{k}_1)^4 = 0$ to an integrable principal value (for details see Ref. [867]). Extension of this definition to states with non-zero angular momentum is discussed in Ref. [868]. Using this confining kernel, together with a phenomenological constant plus a one-gluon-exchange (OGE) contribution in a 1-channel CST equation, gives a good account of the spectrum of heavy-heavy and heavy-light mesons [869, 870], as shown in Fig. 5.4.3 [27]. The 4-channel CST equation also provides a good description of the pion consistent with the axial-vector Ward–Takahashi identity (AV-WTI) [867].

*The origin of the CST –* **FG**
My involvement with this subject began in 1960 when nucleons and pions were thought to be fundamental particles and $S$-matrix theory was believed to be the best way to tackle the strong interactions. For my Ph.D. it was suggested that I look at the deuteron electromagnetic form factor. The upshot of my study lead to the realization that the form factor was dominated by a large number of meson-exchange processes, the first four of which are shown in the left panel of Fig. 5.3.5[28], and that these were best calculated by introducing a new equation that would sum these contributions – the one-channel CST equation.

If the internal propagators in the triangle diagram (right panel of Fig. 5.3.5) are dressed by form factors, then the off-shell nucleon current must also be dressed in order to ensure gauge invariance[29].

### 5.3.3 DSE for the quark propagator

We now turn to a discussion of the DSE for the quark propagator. The exact equation for the quark propagator is shown in the upper left-hand box in Fig. 5.3.1. In Euclidean metric ($\{\gamma_\mu, \gamma_\nu\} = 2\delta_{\mu\nu}$, $\gamma_\mu^\dagger = \gamma_\mu$ and $a \cdot b = \sum_{i=1}^4 a_i b_i$.) it is given by

$$S(p)^{-1} = i\, \not{p}\, Z_2 + m_q(\mu)\, Z_4 + Z_1\, g^2 \int \frac{d^4k}{(2\pi)^4}\, D_{\mu\nu}(q)\, \gamma_\mu \frac{\lambda^i}{2}\, S(k)\, \frac{\lambda^i}{2} \varGamma_\nu(k,p) \tag{5.3.7}$$

where $D_{\mu\nu}(q = k - p)$ is the renormalized dressed gluon propagator, and $\varGamma_\nu(k,p)$ is the renormalized dressed quark-gluon vertex. The solution of Eq. (5.3.7) can be written as

$$S(p) = \frac{1}{i\not{p}A(p^2) + B(p^2)} = \frac{Z(p^2)}{i\not{p} + M(p^2)}\,, \tag{5.3.8}$$

renormalized according to $S(p)^{-1} = i\not{p} + m_q(\mu)$ at a sufficiently large spacelike $\mu^2$, with $m_q(\mu)$ the current quark mass at the scale $\mu$. For divergent integrals a translationally-invariant regularization is necessary. Note that in the chiral limit, the current quark mass $m_q(\mu)$ is absent from Eq. (5.3.7) and there is no mass renormalization.

The most commonly used truncation is the rainbow truncation (analogous to the ladder truncation to the

[27] Since both the light front and CST are relativistic wave functions depending on only three variables, it has long been thought that, perhaps, they can be transformed into one other. The basis for such a comparison might be based on a connection between one of the components of the CST internal momentum (take $p_z$ for example) and the LF momentum fraction $x$, and a good candidate is $E_p + p_z = yD_0$, where $D_0$ is the energy of the bound state, and $y = x$. This transformation suggests an equivalence in some cases [871], but since $0 \le y \le \infty$, it is clear that $y \ne x$. Our conclusion is that CST and LF wave functions seem to describe the physics differently.

[28] A novel feature of the dispersion integrals describing these processes is the presence of anomalous thresholds starting at $s_i < 4m^2$. The imaginary part of the dispersion integral in the anomalous region (from $s_i$ to $4m^2$) is given entirely by the contributions from these diagrams when the four-momentum of *all* the spectators are on shell. For diagram (1) this threshold is at

$$s_0 = \frac{M_B^2}{m^2}\left(4m^2 - M_B^2\right) \approxeq 16m\epsilon$$

where $\epsilon = 2m - M_B$ is the binding energy[872]. For diagrams (2a) and (2b), one additional spectator is on shell, and for diagram (3), two additional spectators are on shell. The thresholds for these diagrams are larger than $s_0$ but still much less than the normal threshold of $4m^2$.

[29] D. O. Riska and I constructed such a current [873], which is used in all CST calculations. This current plays a role analogous to the BC or CP currents discussed below.

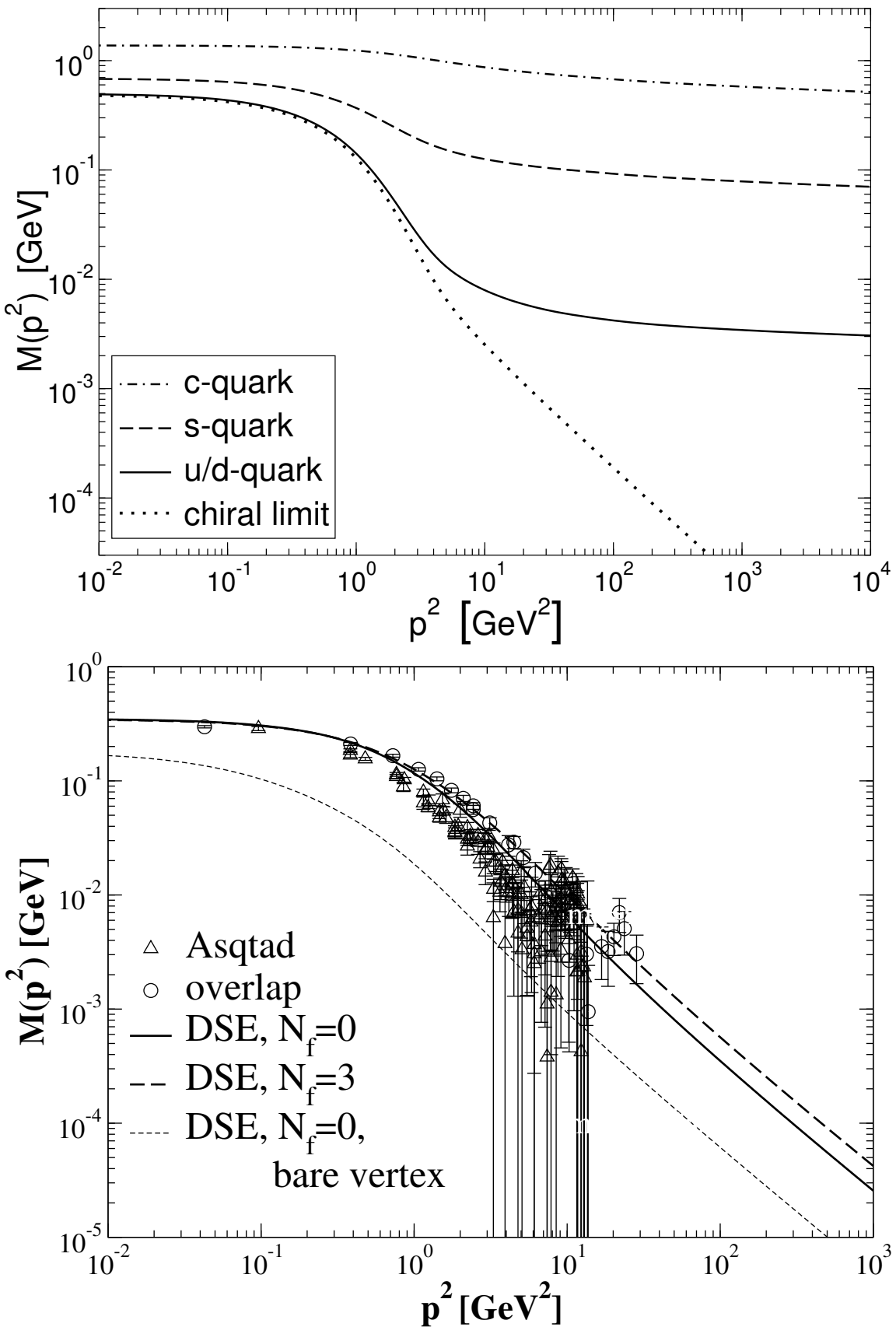

**Fig. 5.3.6** Dynamical quark mass function $M(p^2)$ for spacelike momenta: using the rainbow truncation with the Maris–Tandy model [876] (top, adapted from [877]), and from quenched ($N_f = 0$) and unquenched ($N_f = 3$ chiral quarks) DSEs using the CP vertex [878], as well as results obtained with a bare quark-gluon vertex, compared to quenched lattice data in the overlap [879] and Asqtad [880] formulations (bottom, adapted from [881])

BSE discussed above), in which the dressed gluon propagator and the quark-gluon vertex are replaced by their bare counter-parts, with a model effective running coupling

$$Z_1 g^2 D_{\mu\nu}(q)\gamma_\mu \otimes \Gamma_\nu(k,p) \to 4\pi\,\alpha(q^2)\,D^{\rm free}_{\mu\nu}(q)\gamma_\mu \otimes \gamma_\nu\,. \tag{5.3.9}$$

This truncation is the first term in a systematic expansion [874, 875]; furthermore, the preferred gauge for the fermion DSE is Landau gauge, which has the advantage that asymptotically, $Z(p^2) \to 1$.

By choosing a suitable model for the effective running coupling $\alpha$ that reduces asymptotically to leading-order perturbation theory, realistic quark mass functions, as shown in Fig. 5.3.6, are obtained. In particular, with a nonzero current quark mass, the dynamical mass function behaves at large $p^2$ like

$$M(p^2) \simeq \frac{\widehat{m}}{(\ln[p/\Lambda_{\rm QCD}])^{\gamma_m}}\,, \quad \gamma_m = \frac{12}{11N_c - 2N_f} \tag{5.3.10}$$

with the anomalous mass dimension, $\gamma_m$, in agreement with perturbation theory. In the chiral limit this model gives a nontrivial solution for the mass function that falls off like a power-law, modified by logarithmic corrections [882]

$$M_{\rm chiral}(p^2) \simeq \frac{2\pi^2\gamma_m}{3}\,\frac{-\langle\bar{q}q\rangle^0}{p^2\left(\ln[p/\Lambda_{\rm QCD}]\right)^{1-\gamma_m}}\,, \tag{5.3.11}$$

with $\langle\bar{q}q\rangle^0$ the quark condensate, in agreement with the Operator Product Expansion [883].

Of course, quantitative details of the quark propagator functions in the infrared region do depend on the truncation. The bottom panel of Fig. 5.3.6 shows the quark mass function $M(p^2)$ of the quark propagator in the chiral limit, obtained from the coupled quark, ghost, and gluon DSEs using the Curtis-Pennington (CP) vertex[30] suitably generalized for use in a non-Abelian QFT [878]. Qualitatively, these results agree with the quark mass functions shown in the top panel (both in the chiral limit, and with a nonzero current quark mass), though quantitatively they clearly do depend on both the details of the effective interaction and the vertex Ansatz.

*Do real quark mass poles exist?*

Knowledge of the behavior of the quark propagator in the complex momentum plane is necessary not only to solve the BSE at the bound state mass pole, but also because of possible connections to confinement, the CST, and the LF wave function. In QED, we know that real mass-poles must exist on the time-like axis, but early DSE studies of the fermion propagator in ladder truncation suggested the existence of complex "mass-like" singularities instead of real mass-poles at timelike momenta [885–887]. The absence of a mass-pole in the fermion propagator on the timelike axis would prevent the fermion from being on-shell, and could be an indication of confinement [888, 889].[31] More recently how-

30 The CP vertex [884] is a nonperturbative Ansatz for the electron-photon vertex that satifsies the Ward–Takahashi Identity.

31 **PM**: My interest in the fermion DSE started with my Masters research in the late 80s, with the question whether or not there was a dynamical mass generation in (2+1)-dimensional QED. In addition to dynamical chiral symmetry breaking, $QED_3$ also exhibits confinement; these two features make it an illustrative toy model for QCD. Consistent treatment of the photon propagator turns out to be crucial in $QED_3$: in the quenched approximation (no fermion loops, and hence no vacuum polatization), there is a logarithmically rising potential between a fermion and anti-fermion. This logarithmically confining potential persist in the presence of massive fermion loops in the vacuum polarization, but with massless fermions, this confining potential disappears. With the coupled DSEs for the fermion and photon propagator, it was found that there is a critical number of fermion flavors of about $N_f \sim 3$ to 4, below ever, is has been shown that, with proper regularization of potentially divergent integrals (e.g. using Pauli–Villars), that at least in weak-coupling quenched QED, the DSE for the electron propagator has the expected analytic structure, namely a mass-pole in the timelike region. This was obtained both in Feynman gauge and in Landau gauge; and using two independent numerical methods, explicitly rotating the spacelike region to the timelike region and using the Nakanishi formalism [890].

In QCD however, quarks (and gluons) are confined, and the quark propagator need not have a mass-pole at timelike momenta. A convenient way to study this is to use the Schwinger function, $\varDelta(t)$, defined by

$$\varDelta_{s,v}(t) = \int d^3x \int \frac{d^4p}{(2\pi)^4} e^{i(tp_4+\vec{x}\cdot\vec{p})} \sigma_{s,v}(p^2) = \frac{1}{\pi}\int_0^\infty dp_4 \cos(t\, p_4)\sigma_{s,v}(p_4^2) \; \geq 0 \qquad (5.3.12)$$

where $\sigma_{s,v}(p^2)$ is the scalar or vector part of the dressed quark propagator,

$$S(p) = i\not{p}\,\sigma_v(p^2) + \sigma_s(p^2)\,. \qquad (5.3.13)$$

For a propagator with a real mass-pole in the timelike region, this Schwinger function falls off like an exponential. In contrast, a propagator with a pair of complex-conjugate mass-like singularities, the Schwinger function is not positive-definite and exhibits an oscillatory behavior

$$\varDelta(t) \sim e^{-a\,t}\,\cos(bt+\delta)\,. \qquad (5.3.14)$$

In Ref. [881] a striking qualitative difference between the use of a bare quark-gluon vertex and the BC [891] or CP vertex was found: with a bare vertex, the Schwinger function behaves like a pair of complex-conjugate mass-like poles for the quark propagator, whereas the results with the BC and the CP vertex behave like a real mass-pole in the timelike region. Qualitatively similar results were found employing different models for the effective running coupling, including (3+1) dimensional QED. The existence of a pair of complex-conjugate mass-like singularities in the DSE solutions of the dressed quark propagator in rainbow truncation was also confirmed by direct analytic continuation of the quark DSE into the complex-momentum plane; the obtained real and imaginary parts of these singularities agree with those extracted from the Schwinger function. Whether or not confinement is realized through the absence of mass-like singularities on the real timelike axis remains to be seen. Note that these results are not inconsistent with the CST, which assumes the existence of real quark mass poles.

which there is both dynamical mass generation and a confining potential. Furthermore, it was found that in the presence of the logarithmically confining potential, the fermion propagator exhibits a pair of 'mass-like' singularities at complex conjugate momenta in the complex momentum plane, whereas in the absence of this logarithmically confining potential, the fermion propagator appears to have a real mass-pole at timelike momemta, as one would expect based on perturbation theory.

### 5.3.4 Pions: Goldstone bosons of QCD

Pions, and to some extent also kaons, are the pseudo-Goldstone bosons of QCD: in the chiral limit, $m_q = 0$, chiral symmetry is broken dynamically, which implies the existence of massless Goldstone bosons. In the flavor SU(2) chiral limit, there are three Goldstone bosons (three pions); and in the flavor SU(3) chiral limit, there would be eight Goldstone bosons. In the real world, the up, down, and strange quarks are not massless, but have a small current quark masses; in addition, one of the eight 'would-be' Goldstone bosons mixes with the isoscalar pseudoscalar meson (which is massive due to the axial anomaly) to form the $\eta$ and $\eta'$. This explains qualitatively why the three pions and four kaons are so much lighter than all other mesons, among other things. Therefore, in order to describe pions (and kaons), any truncation has to respect all constraints coming from chiral symmetry. Furthermore, it implies that the pion BSA is closely related to the (dynamically generated) scalar part of the quark self-energy, which can be made explicit by using the AV-WTI [892].

The axial-vector vertex $\varGamma_5^\mu$ satisfies a DSE as illustrated in the second row of Fig. 5.3.1, with an inhomogeneus term $\gamma^5\gamma^\mu$. But even without solving the DSE, one can relate this vertex directly to the dressed quark propagators via the AV-WTI

$$P_\mu \varGamma_5^\mu(p;P) = S^{-1}(p_2)\gamma_5 + \gamma_5 S^{-1}(p_1) - 2\,m_q(\mu)\,\varGamma_5(p;P)\,, \qquad (5.3.15)$$

where $\varGamma_5(p;P)$ is the pseudoscalar vertex, which also satisfies a DSE as shown in Fig. 5.3.1, with inhomogeneous term $\gamma^5$. This can be compared to the more familiar vector WTI for the quark-photon vertex (which satisfies the same DSE with inhomogeneous term $\gamma^\mu$),

$$P_\mu \varGamma^\mu(p;P) = S^{-1}(p_2) - S^{-1}(p_1) \qquad (5.3.16)$$

which ensures electromagnetic current conservation.

Meson poles in the quark-antiquark scattering amplitude, $G_4$, also appear in these vertices, depending on their quantum numbers. For the quark-photon vertex this automatically leads to Vector Meson Dominance (VMD), a model for the coupling of photons to hadrons that predates QCD [893] (see below). In the case of the axial-vector vertex, near a pseudoscalar meson pole at

$\widehat{P}^2 = -M_{\rm PS}^2$, we have[32]

$$\Gamma_5^\mu(p;P) \approx \frac{\Gamma_{\rm PS}(p;\widehat{P})}{P^2+M_{\rm PS}^2} Z_2 N_c \int \frac{d^4k}{(2\pi)^4} {\rm Tr}[\chi_{\rm PS}(k;\widehat{P})\,\gamma_5\,\gamma_\mu] = \frac{\Gamma_{\rm PS}(p;\widehat{P})}{P^2+M_{\rm PS}^2} f_{\rm PS}\widehat{P}^\mu \tag{5.3.17}$$

with $f_{\rm PS}$ the pseudoscalar decay constant, which governs the coupling of a pseudoscalar meson to the axial-vector current.

Similarly, pseudoscalar mesons appear as poles in the pseudoscalar vertex, and near $\widehat{P}^2 = -M_{\rm PS}^2$ this vertex behaves as

$$\Gamma_5(p;P) \approx \frac{\Gamma_{\rm PS}(p;\widehat{P})}{P^2+M_{\rm PS}^2} Z_4 N_c \int \frac{d^4k}{(2\pi)^4} {\rm Tr}[\chi_{\rm PS}(k;\widehat{P})\,\gamma_5] = \frac{\Gamma_{\rm PS}(p;\widehat{P})}{P^2+M_{\rm PS}^2} r_{\rm PS}(\mu) \tag{5.3.18}$$

with $r_{\rm PS}(\mu)$ the (renormalization-scale dependent) residue in the pseudoscalar channel. The AV-WTI relates the residues at these poles

$$f_{\rm PS}\, M_{\rm PS}^2 = -2\, m_q(\mu)\, r_{\rm PS}(\mu)\,, \tag{5.3.19}$$

which holds for any pseudoscalar meson. Therefore, in the chiral limit, $m_q(\mu) = 0$, either $f_{\rm PS}$ or $M_{\rm PS}$ must be zero. (If they are both zero, chiral symmetry will not be dynamically broken; see below.)

Furthermore, expanding the AV-WTI in powers of $M_{\rm PS}^2$ in the chiral limit, $m_q(\mu) = 0$, and using the most general Dirac decomposition of $\Gamma_{\rm PS}$ [33]

$$\Gamma_{\rm PS}(k;\widehat{P}) = \gamma_5 \left[ iE + \widehat{\not P}\, F + \not k\, G + \sigma_{\mu\nu}\, k_\mu \widehat{P}_\nu\, H \right] \tag{5.3.20}$$

one finds, to leading order in $M_{\rm PS}$,

$$f_{\rm PS} E(p;0) = B(p^2) \tag{5.3.21}$$

where $B(p)$ is the scalar part of the quark self-energy.

Thus, if chiral symmetry is dynamically broken, that is, if $m_q(\mu) = 0$ but $B(p^2) \neq 0$, $f_{\rm PS}$ is nonzero, see Eq. (5.3.21), and pions necessarily emerge as massless Goldstone bosons, see Eq. (5.3.19). Furthermore, the pseudoscalar component of the pion BSA is proportionally to the (dynamically generated) scalar self-energy of the quarks. In addition, the AV-WTI implies that the decay constant of excited pions (which necessarily have nonzero mass) has to vanish in the chiral limit.

These relations are exact, and the asymptotic behavior of the canonical pion BSA component can be obtained from the asymptotic behavior of the mass functions shown in Fig. 5.3.6. The same asymptotic behavior of the canonical BSA component also holds with nonzero current quark masses; as well as for excited pseudoscalar mesons.

Finally, with the definition of $r_{\rm PS}$ implicitly given in Eq. (5.3.18, and the relation (5.3.21), we arrive at the well-known Gell-Mann–Oakes–Renner relation

$$f_\pi^2\, m_\pi^2 = 2\, m_q(\mu)\, \langle \bar{q}q \rangle^\mu_{\rm chiral}\,, \tag{5.3.22}$$

with the chiral condensate

$$\langle \bar{q}q \rangle^\mu_{\rm chiral} = Z_4 N_c \int \frac{d^4k}{(2\pi)^4} \frac{4\, B_{\rm chiral}(k^2)}{k^2 A^2(k^2) + B^2_{\rm chiral}(k^2)} \tag{5.3.23}$$

Note that the renormalization scale dependence of the current quark mass, $m_q(\mu)$, exactly cancels that of the chiral condensate.

### 5.3.5 Mesons in Rainbow-Ladder (RL) truncation

Different types of mesons, such as pseudoscalar (pions, kaons) or vector mesons ($\rho$, $\phi$), are obtained by considering the most general Dirac and flavor (isospin) structure for the meson of interest, and solving the BSE, Eq. 5.3.1, at the bound state pole [34].

To obtain practical solutions from the exact BSE, Eq. 5.3.1, the kernel $K$ must be truncated; furthermore, one needs to approximate the dressed quark propagators. The most commonly used truncation is the ladder truncation, in which the BSE kernel $K$ in Eq. (5.3.1) is replaced by an one-gluon exchange (or, in the case of QED, a one-photon exchange)

$$K_{ij}(p,k;\widehat{P})\; \mathcal{O}_i \otimes \mathcal{O}_j \;\to\; 4\pi\, \alpha(q^2)\, D^{\rm free}_{\mu\nu}(q) \tfrac{\lambda^i}{2}\gamma_\mu \otimes \tfrac{\lambda^i}{2}\gamma_\nu\,, \tag{5.3.24}$$

with a model for the effective running coupling $\alpha(q^2)$. Here we use the ladder truncation, in combination with quark propagators that are the solution of the DSE in rainbow truncation – hence we refer to it as the Rainbow-Ladder (RL) truncation.

The resulting approximate BSE is solved numerically, starting from the Euclidean metric, and analytically continuing $\widehat{P}^2$ to negative values while keeping the integration variable Euclidean. This leads to complex momenta for the quark propagators, which is trivial with bare constituent propagators; it is also well-defined and straightforward to implement for (nonperturbatively) dressed propagators as long as there are no singularities in either the (dressed) propagators or the model for the effective interaction over a well-defined domain in the complex momentum plane, depending on the meson mass, choice of $\eta$, and choice of frame[35] – though one may have to solve the quark DSE numerically over this domain.

[32] Remember we are using Euclidean metric here.

[33] Here $E$, $F$, $G$, and $H$ scalar functions of $k^2$ and $k\cdot\widehat{P}$; for equal-mass mesons with $\eta = \frac{1}{2}$, the functions $E$, $F$, and $H$ are even in $k\cdot\widehat{P}$, whereas $G$ is odd in $k\cdot\widehat{P}$.

[34] The bound state mass is not known a priori; therefore one has to vary $\widehat{P}^2$ until one finds a solution. This is most conveniently done by introducing a fictitious eigenvalue $\lambda$ in front of the LHS of Eq. 5.3.1 to turn it into an eigenvalue problem, and search for a solution with $\lambda = 1$ by varying $\widehat{P}^2$.

In the previous section we showed in detail that the pion is the Goldstone boson associated with chiral symmetry breaking; it becomes massless in the chiral limit; and its canonical BSA component is given by the scalar self-energy of the quark. The ladder truncation by itself, in combination with bare propagators, does not preserve these features of the pion. However, the RL truncation with consistent dressed quark propagators does preserve the Goldstone nature of the pion, which one can prove analytically using Eq. 5.3.15 and performing a shift in integration variables.[36]

The RL truncation has been used extensively over the past 25 years, not only for pions, but also for other quantum numbers, and both for light systems, heavy systems, and heavy-light systems. A commonly used model for the interaction is the Maris–Tandy model [876]. This model is finite in the infrared region, with sufficient strength for dynamical chiral symmetry breaking, and agrees perfectly with pQCD for $q^2 > 25$ GeV$^2$. The dynamical mass function of the up/down quarks, strange quarks, and charm quarks were shown in Fig. 5.3.6.

For the light pseudoscalar and vector mesons, consisting of $u$, $d$, and $s$ quarks, we find excellent agreement with the experimental data, not only for the spectrum, but also for the decay constants. For the charmed mesons (both charmonium, and heavy-light systems) we also find agreement with experiment, within our numerical precision which is dominated by the need to solve the quark propagator over a large domain in the complex momentum plane. Results for axial-vector and scalar mesons are much less in agreement with experiment, but it is known that leading-order corrections to the RL truncation are significantly larger in the axial-vector and scalar channels than in the pseudoscalar and vector channels. Furthermore, the scalar mesons are notoriously difficult to describe, and are likely to have a significant 4-quark content (in particular the broad $\sigma$ meson, if it can be called a meson).

*Meson form factors and scattering*
With the BSA we can evaluate a range of other physical observables. We have already mentioned the electroweak decay constant, but more interesting are processes with three external probes such as mesons and/or photons. Consider the elastic form factor of a meson: the right panel of Fig. 5.3.5 shows the coupling of a photon to a meson in impulse approximation. One can show analytically that if one considers the dressed quark-photon vertex as the solution of its inhomogeneous BSE using the same RL kernel as for the quark propagators and the meson BSE, current conservation is automatically guaranteed. Another advantage of using such a dressed quark-photon vertex, instead of a bare vertex, is that vector meson poles will automatically appear as poles at $Q^2 = -M_V^2$ in the dressed vertex; thus, VMD is unambiguously included in this approach [895].

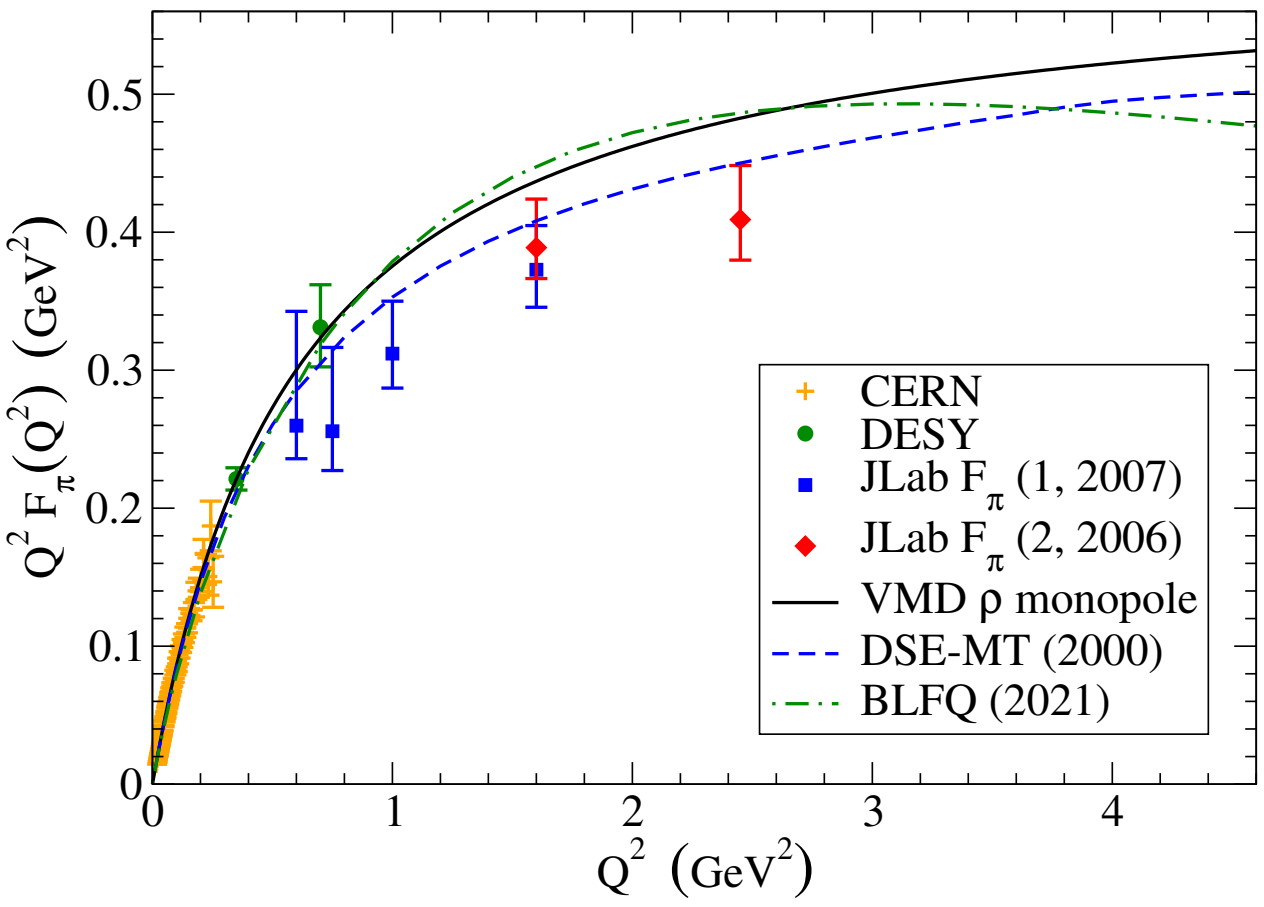

**Fig. 5.3.7** Spacelike pion form factor: Comparison between experiment and a VMD model, DSE in RL truncation [877, 895], and a recent LF calculation [896]. For the experimental data, see Refs. [897, 898] and references therein.

A practical challenge is that at least one of the mesons in Fig. 5.3.5 has to be in a moving frame. For small values of $Q^2$ one can use a Taylor expansion of the BSA in the rest frame, but explicitly solving the BSE in a moving frame greatly improves the accessible domain in $Q^2$ and reduces numerical uncertainties associated with e.g. a Taylor expansion. Figure 5.3.7 shows the predictions from the Maris–Tandy model in RL truncation for the pion elastic form factor, which are in perfect agreement with the data. For comparison, we also include a simple VMD model, as well as a recent LF calculation [896] discussed in more detail in Sec. 5.4.

Similar diagrams can and have also been used for electroweak transition form factors and the anomalous $\pi^0 \to 2\gamma$ process [899]. One finds generally good agreement with experimental data, thanks to the fact that this approach satisfies all constraints coming from electromagnetic current conservation, chiral symmetry, and dynamical chiral symmetry breaking; furthermore, it includes unambigously VMD effects, and it also agrees with perturbative QCD at large momenta. This is not to say that there are no short-comings in this approach: obviously there is physics beyond the RL truncation that is important, some of which are discussed below.

[35] The BSA is frame independent, but in Euclidean metric, $k \cdot P$ is purely imaginary in the restframe (remember $\widehat{P}^2$ is negative), and becomes generally complex in a moving frame. It has been shown that physical observables are indeed frame independent by solving the BSE in RL truncation explicitly in a moving frame [894].

[36] Hence the need for translationally-invariant regularization of potentially divergent integrals – this is also necessary for ensuring current conservation in electromagnetic interactions.

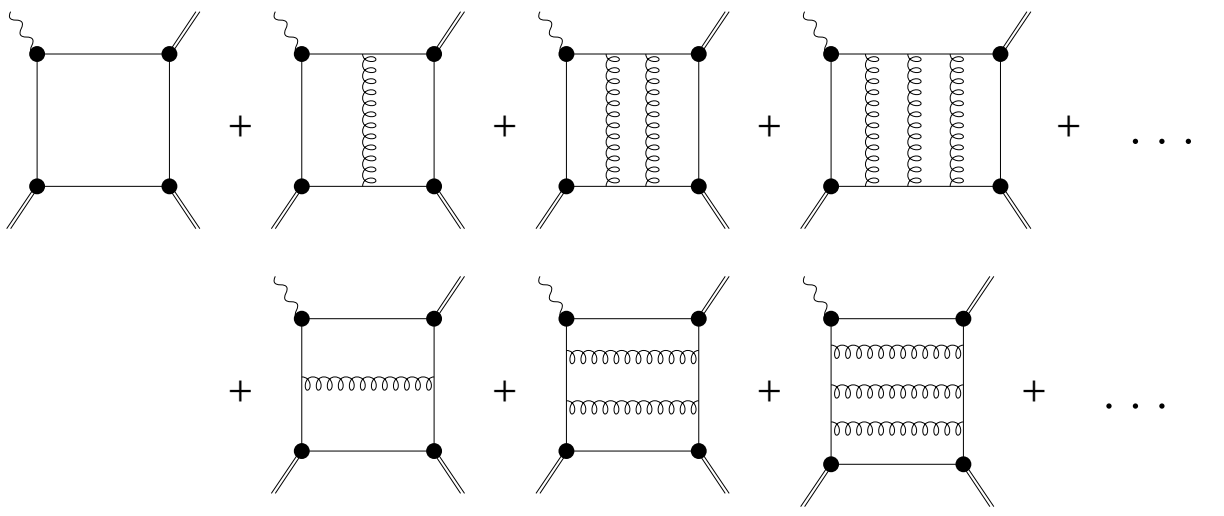

**Fig. 5.3.8** RL truncation for $\gamma\, 3\pi$ consistent with chiral symmetry and electromagnetic current conservations: quark propagators, vertices and box-diagram all dressed with the same RL kernel (adapted from [900]).

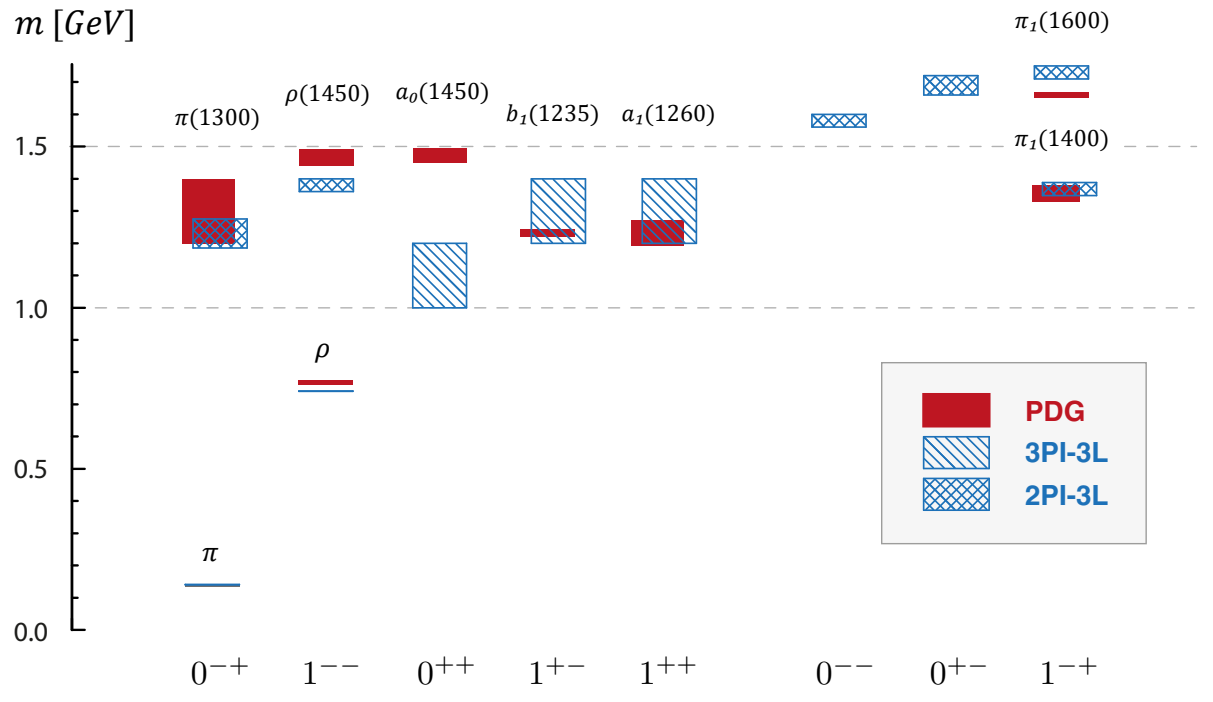

**Fig. 5.3.9** Light meson spectrum beyond RL truncation (Figure adapted from [904]).

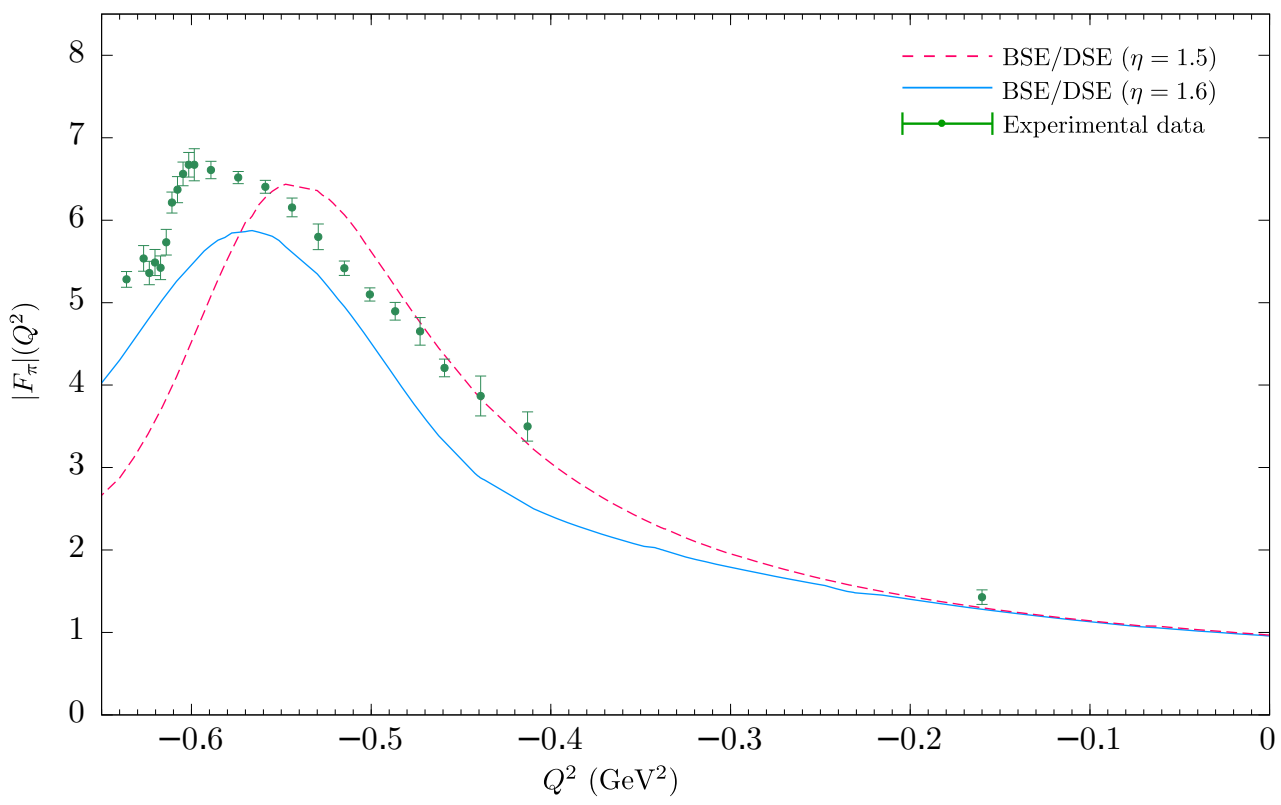

**Fig. 5.3.10** Pion form factor with pion loops in the timelike region (Figure adapted from [905]).

More challenging are scattering observables involving four external mesons and/or electroweak probes. Based on the success of describing form factors in impulse approximation, one might consider just the box diagram with dressed vertices and propagators for such processes. However, it turns out that this is insufficient, and does not reproduce the expected results for e.g. $\pi\pi$ scattering or $\gamma\, 3\pi$ coupling – which are both constrained by chiral symmetry. For a consistent description of scattering observables involving four external probes, one needs to include the same RL kernel inside the box diagram as well, resummed to all orders, as indicated in Fig. 5.3.8. With these ladder diagrams added to the box diagram, it has been shown explicitly that both the anomalous $\gamma\, 3\pi$ process [900] and $\pi\pi$ scattering [901] are in perfect agreement with chiral symmetry and electromagnetic current conservation. The same approach can in principle also be used for other processes, involving other mesons, and it would be very interesting to extend this approach in the future to e.g. Compton scattering on hadrons, as well as pion-nucleon scattering.

### 5.3.6 Beyond the RL approximation

Over the past two decades significant progress has been made in improving the RL truncation while preserving the relevant vector and axial-vector WTIs [875, 902, 903]. Although the details of these investigations differ, the general conclusion is that corrections beyond RL are relatively small in the pseudoscalar and vector channels, but can be significantly larger in the axial-vectors and scalar channels. This makes it understandable why the pseudoscalar and vector meson masses and decay constants are in such good agreement with data, but at the same time an accurate description of mesons with other quantum numbers requires going beyond RL.

One of the more promising methods to go beyond the RL truncation is based on the $n$-Particle Irreducible ($n$-PI) effective action, in particular the 2-PI and 3-PI effective action up to 3 loops [830, 904]. This generally leads to coupled integral equations for the quark, gluon, and ghost propagators, the quark-gluon vertex (and possibly other vertices), and possibly higher $n$-point functions. Computationally, solving these coupled sets of integral equations in multiple variables is significantly more complicated and time consuming than the RL truncation, but with current (and future) computational resources, the resulting integral equations can be solved for selected cases. The spectrum obtained for the light mesons (including the axial-vector mesons) is in good agreement with available data, see Fig. 5.3.9; the only obvious disagreement is in the scalar channel, where pion loops play an important role.

*Higher Fock components*

Although the RL truncation appears to be quite successful for a range of meson observables, it has its limitations. Consider the pion form factor: Fig. 5.3.7 shows this form factor in the spacelike region, but we can also extend these calculations to the timelike region. In the timelike region, we find a pole at $Q^2 = -M_\rho^2$; exactly as one would expect, because we already know that the homogeneous BSE for the vector channel has a solution at $\widehat{P}^2 = -M_\rho^2$. However, this pole is above the $2\pi$ threshold – in the real world, this pole is shifted to the second Reimann sheet, and there is a resonance peak with non-zero width at $Q^2 = -M_\rho^2$. Indeed, incorporating pion loops in the dressed quark-photon vertex in the timelike region changes the vector-meson pole to a resonance peak, and the resulting form factor is in good agreement with the data [905], see Fig. 5.3.10. Although the center of the peak is slightly shifted compared to the data, the peak height and width are in good agreement with the data in the timelike region

Similarly, pion loops are likely to be important for the scalar mesons, which can be included by incorporating configurations with two quarks and two anti-quarks in the BSE. This leads to a set of coupled equations between the usual quark-antiquark components, as well as ’meson-meson’ contributions and ’diquark-diquark’ contributions. This has recently been implemented for the scalar channel [906], which reveals that the $\sigma$ meson is indeed dominated by two-pion contributions, as one might expect. This approach will also be very useful to investigate exotic mesons, tetraquarks, and in the future also pentaquarks, all within the same framework.

### 5.3.7 Baryons

The notion of diquarks has been around for almost as long as QCD; see e.g. Ref. [756]. Initial DSE studies of baryons were therefore formulated in terms of bound states of a quark and a diquark; specifically a scalar and an axialvector diquark.

However, as described in Sec. 5.3.2, three fermion states can also be described by a 3-body BSE, and in recent years there has been significant progress in describing and understanding baryons as three-quark bound states using the DSE with essentially the same RL approximations as used for the mesons. An effective interaction is modeled using the Dirac structure of a one-gluon exchange between two quarks, see Eq. 5.3.24, in combination with consistent nonperturbatively dressed quark propagators. Figure 5.3.11 shows the calculated spectrum for nucleons and delta resonances together with the experimental spectrum. The results for the ground state nucleons, as well as their radial excitations, are in good agreement with experiment, both in the quark-diquark and the three-quark bound state pictures. For the other quantum numbers we see noticable differences between the quark-diquark and three-quark bound state results (and note that not all quantum numbers have been done as a three-quark bound state). The obtained bound state amplitudes can be used for the evaluation of nucleon form factors, see e.g. [907] and references therein, analogous to the calculation of the pion form factor discussed earlier.

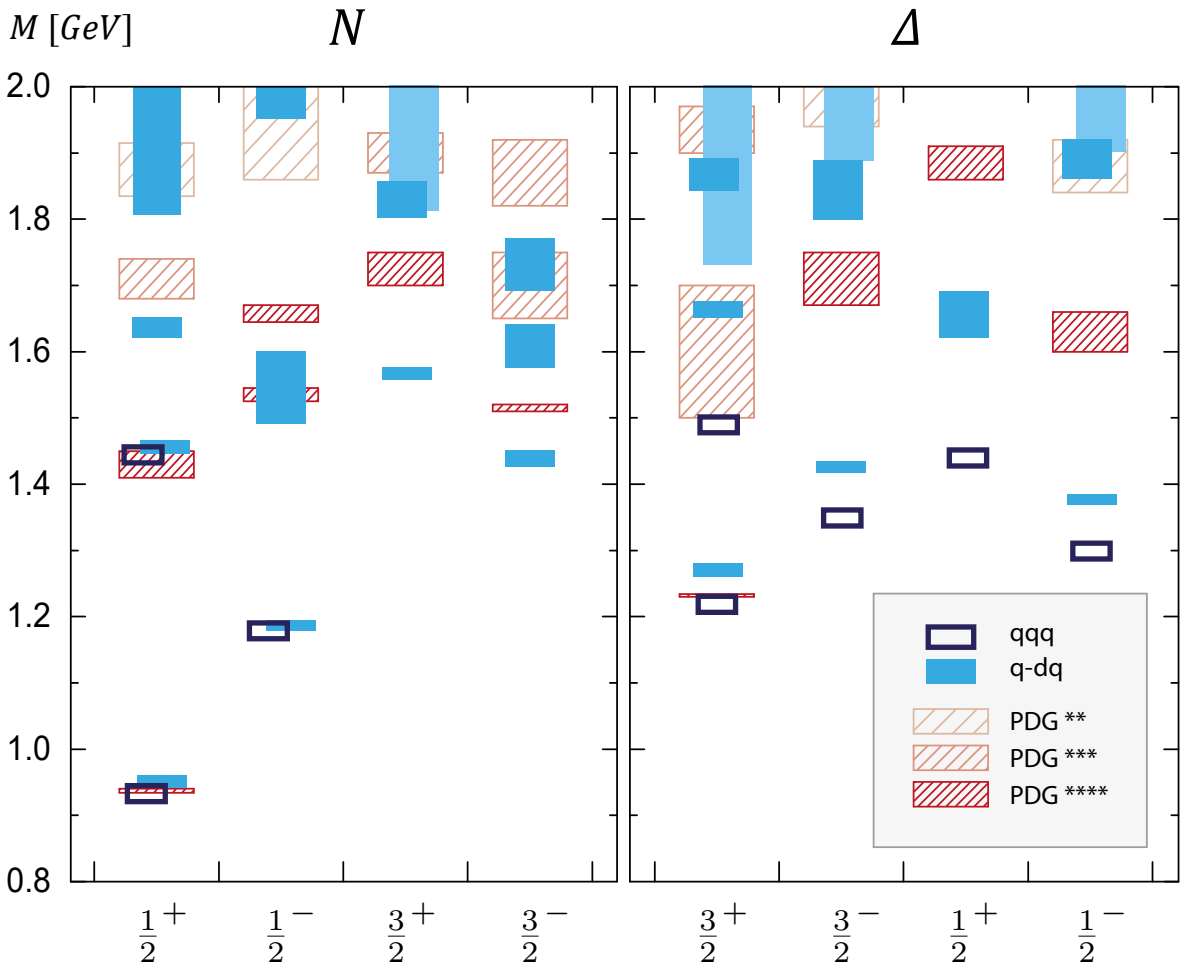

**Fig. 5.3.11** Baryon spectrum in RL truncation in the quark-diquark picture (blue bars) and as three-quark bound state (open boxes), compared to experimental spectrum (Figure adapted from [907]).

### 5.3.8 Conclusions

At the energy scales of mesons and barons, nonperturbative methods are needed, and the DSEs and BSE (or the CST) work very well. The main shortcoming of these methods is that the kernels needed to solve for the self energies of wave functions are unknown, and must be modeled. The combination of the ladder (L) truncation of the BSE with the closely related rainbow (R) truncation of the DSE for self energies are reasonably successful, in particular in describing chiral symmetry breaking and the role of the pion as the Goldstone boson of QCD. The few calculations beyond the RL truncation that exist show that the additional effects are not large, except in particular spin-isospin channels.

We expect this technique to develop in the years ahead and to remain an attractive method for theoretical study of QCD.

## 5.4 Light-front quantization

**James Vary, Yang Li, Chandan Mondal and Xingbo Zhao**

In this section, we discuss non-perturbative light-front Hamiltonian quantization methods. We primarily focus on introducing the Hamiltonians for QED and QCD derived in the light-cone gauge (for extensive reviews, see Refs [792, 908]). We introduce methods of solution and results for mesons and baryons. We focus on the Discretized Light Cone Quantization (DLCQ) and Basis Light Front Quantization (BLFQ) methods due to their ability to include gluons and sea quarks dynamically.

Light-front quantization is the natural language for describing the partonic degrees of freedom of QCD at high energies. This connection has been extensively exploited in phenomenological approaches to hard inclusive and exclusive processes (see Secs. 5.9,5.10). In these approaches, instead of solving the QCD dynamics, the symmetries and properties of QCD are employed to construct phenomenological partonic amplitudes or densities on the light front.

Before introducing specific light-front Hamiltonian methods of solution, let us recap the key concepts of the light-front Hamiltonian approach that spring from Dirac's formulation of Poincare' invariant quantum frameworks [909]. Our choice of light-front variables can be summarized in relation to equal-time variables by introducing

$$P = (P^0+P^3, P^0-P^3, P^\perp) = (P^+, \frac{M^2+(P^\perp)^2}{P^+}, P^\perp),$$

where $P$ and $M$ represent the 4-momentum and mass of the hadron, respectively. For the hadron's constituents (quarks, antiquarks, gluons), which we refer to as partons, we adopt $p_i^\perp$ as the transverse momentum of the $i$th parton, $x_i = \frac{p_i^+}{P^+}$ is its longitudinal momentum fraction, $\lambda_i$ is its light-front helicity [910], and roman alphabet subscripts run through the partons of the hadron.

The Hamiltonian eigenvalue problem for the mass-squared eigenstates and their associated light-front wave functions (LFWFs) begins with defining the light-front Schrödinger equation for the system's eigenstates. Taking $P^\perp = 0$ and $H = P^-$ (using dimensionless units for the conserved $P^+$)

$$H\,|P,\Lambda\rangle = M^2\,|P,\Lambda\rangle \tag{5.4.1}$$

where $\Lambda$ is the hadron's light-front helicity and $H$ contains kinetic, interaction and Lagrange multiplier terms

$$H = \sum_i \frac{p_i^{\perp 2}+m_i^2}{x_i} + H_{int} + \lambda_{CM} H_{CM}\,. \tag{5.4.2}$$

Here, the sum is over all partons and $m_i$ is the mass of the $i^{th}$ parton. The role of the Lagrange multiplier term ensures factorization of the state vector's transverse component into an internal, boost invariant, component times a center of mass (CM) component [911].

We note that this eigenvalue problem applies to systems with arbitrary baryon number so that, for example, it applies to atomic nuclei as well. An eigenstate of a system can be written in terms of a Fock-space expansion over sectors with $N$-partons as

$$|P,\Lambda\rangle = \sum_N \sum_{\lambda_1,\ldots,\lambda_N} \int \frac{\prod_{i=1}^N \mathrm{d}x_i \mathrm{d}p_i^\perp}{[2(2\pi)^N]^2\,\sqrt{x_1 x_N}} \delta(1-\sum_{i=1}^N x_i) \times \delta^2(\sum_{i=1}^N p_i^\perp) \psi^\Lambda_{\{\lambda_i\}_N}(\{p_i\}_N)\,|\{\lambda_i, p_i\}_N\rangle\,, \tag{5.4.3}$$

where $\psi^\Lambda_{\lambda_1,\ldots,\lambda_N}(p_1,\ldots,p_N)$ is the light-front helicity amplitude for each component. Each of the multi-parton basis states $|\{\lambda_i, p_i\}_N\rangle$ is defined as a properly normalized string of $N$ fermion, anti-fermion and gluon creation operators acting on the vacuum. Eq. 5.4.3 is schematic since, for fixed $N$, there can be many subcases with the same net fermion number. We note that the kinetic term in Eq. 5.4.2 is diagonal in this multi-parton basis. In the following sections, we introduce the discretized and basis function alternatives to Eq. 5.4.3.

For gauge theories, a traditional approach is to adopt the light-front gauge, $A^+ = 0$, and to reduce the Hamiltonian to the minimum number of dynamical degrees of freedom using constraint equations. For QED and QCD this produces the $H_{int}$ term of Eq. 5.4.2 expressed in terms of Pauli spinors with the boson-fermion vertices (QED and QCD) as well as boson-boson vertices (QCD only). In addition to these vertices, the gauge-fixing and reduction procedures lead to higher-order instantaneous interactions which manifest divergences. The resulting 3(7) vertices for QED [72](QCD [912, 913]) are deceptively simple and are shown in Fig. 5.4.1

Like its Lagrangian counterpart, Hamiltonian field theory needs to be regularized and renormalized. Dimensional regularization is only available for perturbative calculations. In non-perturbative solutions, the invariant mass cutoff and the Pauli-Villars regularization are often adopted. Since non-perturbative eigenvalue problems have to be solved numerically, finite discretization schemes are also needed. One can choose to use the discretization to define the regularization. DLCQ and BLFQ are such schemes. Alternatively, the discretization can be used purely as the numerical method. The problem remains to take the continuum limit. Thanks to the kinematical nature of the light-front boosts, cluster decomposition remains available in the continuum scheme. Hence perturbative type renormalization can be extended to this scheme, as realized in Fock sector dependent renormalization [914].

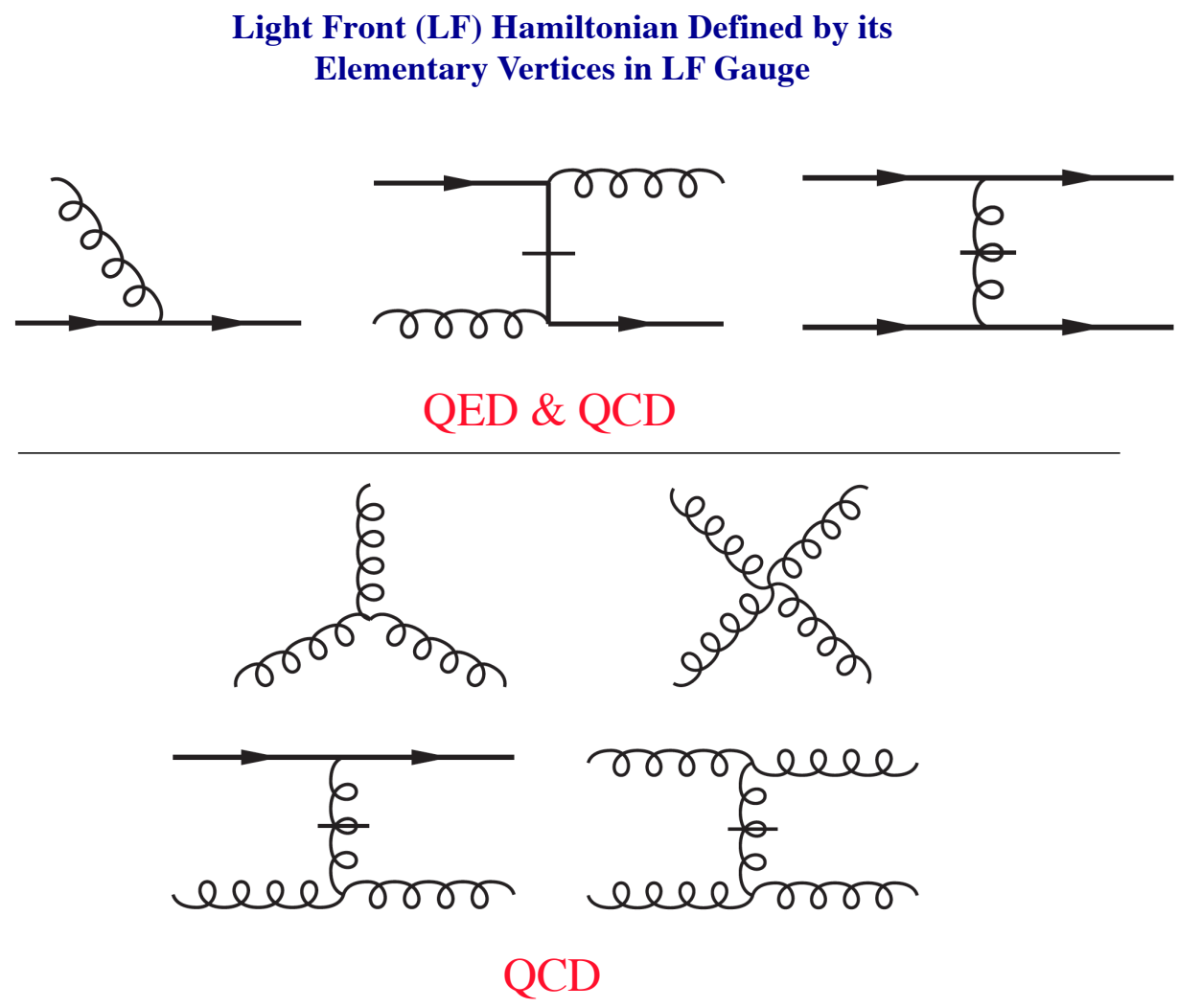

**Fig. 5.4.1** Vertices appearing in the LF Hamiltonian term $H_{int}$ of Eq. 5.4.2 upon choosing the LF gauge $A^+ = 0$ for QED [72] and additional vertices for QCD [912, 913]. See [792] for a recent review. Solid lines represent fermions (vertices with antifermions are obtained by reversing a fermion line) and wavy lines represent gauge bosons. A graph that includes a fermion or boson with a horizontal line through it represents an instantaneous interaction term. Though one LF time ordering is pictured (increasing LF time flows to the right), all allowed LF time orderings are included in $H_{int}$. Thus, for example, an incoming line can be switched to an outgoing line at any vertex and vice-versa.

The similarity renormalization group (SRG) approach is another non-perturbative approach based on Wilson's renormalization group evolution [915, 916]. Thanks to asymptotic freedom, the SRG transformation can be evaluated perturbatively up to some scale, say, a few GeV. Different schemes were designed for implementing SRG, notably the Bloch-Wilson formulation [917] and the renormalization group procedure for effective particles (RGPEP [918]). An RGPEP effective Hamiltonian for heavy flavor hadrons is derived using a gluon mass ansatz [919]. In the gluon sector, it successfully reproduces asymptotic freedom in the 3-gluon effective vertex [920].

The Fock space expansion (Eq. 5.4.3) provides the most straightforward representation of the eigenvalue problem Eq. 5.4.1. Within this basis, the eigenvalue equation becomes an infinite tower of coupled integral equations. The integrations can be evaluated using standard numerical techniques; however, truncation is needed to obtain numerical solutions. The situation is similar to the Dyson-Schwinger/Bethe-Salpeter equations approach in the covariant formulation (see Sec. 5.3). The light-front Tamm-Dancoff approximation (LFTDA) truncates the Fock sections in terms of the particle number [921]. LFTDA can be systematically renormalized using the Fock sector dependent renormalization [914]. This was used to investigate various theories within few-body truncation (see Ref. [908] for a review). Typically, the convergence of the Fock sector expansion can be checked numerically [922], although the numerical complexity increases dramatically as the number of Fock sectors increases. The light-front coupled cluster method was proposed to improve the convergence and pathology associated the hard Fock sector truncation by adopting a coherent basis [923].

Another major development in light-front quantization is the discovery of the remarkable connection between light-front dynamics, its holographic mapping to gravity in a higher-dimensional anti-de Sitter (AdS) space, and conformal quantum mechanics, known as light-front holography (LFH). This approach introduces a remarkably simple yet universal confining potential, which underlines the various phenomenological applications in light-front QCD. See Sec. 5.5 for details.

### 5.4.1 Discretized Light-Cone Quantization

While lattice calculations (see Sec. 4) solve QCD in Euclidean spacetime, DLCQ formulates the problem directly in Minkowski spacetime using a discretized momentum basis (see Ref [792] and references therein).

In DLCQ, one defines a mesh in momentum space that corresponds to standing waves in a box of length $L$ in each transverse direction and a similar set of modes in the longitudinal direction. Either periodic or anti-periodic boundary conditions are applied. Early applications of DLCQ to gauge theories included solving QED for positronium at strong coupling [924].Similarly, early successes include solving QCD in 1+1 dimensions [925]. Moving to QCD in 3+1 dimensions with DLCQ revealed formal and numerical challenges but produced many valuable results as reviewed in Ref. [908].

A hybrid light-front DLCQ/lattice formulation was introduced and employed to evaluate parton distribution functions for a sample set of meson states over a range of coupling strengths [926, 927]. These applications of DLCQ motivated the quest for an approach that both preserves the LF kinematic symmetries and provides a computational path with improved numerical efficiency.

### 5.4.2 Basis Light Front Quantization

The quest to develop LF Hamiltonian approaches in Minkowski-space that retain all available kinematic symmetries began with adoption of basis function methods for solving light front wave equations [928]. Later, the BLFQ approach [929] was introduced to treat gauge theory Hamiltonians using basis-functions that satisfied very general mathematical conditions and respected the LF kinematic symmetries. In addition, the BLFQ framework is well-suited for a longer-term goal of developing basis functions that approximated anticipated dynamical features of QCD such as confinement and chiral symmetry breaking for applications to hadron spectra. Such basis functions have the promise of facilitating convergence in non-perturbative LF QCD calculations.

In BLFQ, one introduces an alternative to the momentum space representation of the LF eigenstate presented in Eq. 5.4.3. Instead of working with LF plane waves, BLFQ introduces a superposition of orthonormal N-parton Fock space states expressed as independent partons in some convenient orthonormal single-parton basis. That is, we replace the conventional quantization in terms of LF plane waves with LF quantization in modes of a solvable single-parton LF Schrödinger equation akin to Eq. 5.4.1. Thus, the LF many-parton basis states can be written as strings of fermion, antifermion and boson creation operators that populate independent modes of the single-parton LF Schrödinger equation. All applications described below elect the 2D Harmonic Oscillator for the transverse modes owing to the ability to preserve transverse boost invariance. This choice is further motivated by holographic light-front QCD (see Sec. 5.5 for details) and has been our default choice for practical calculations. For the longitudinal modes there have been a number of choices including DLCQ. In principle, the basis is arbitrary within general mathematical restrictions so convenience and numerical efficiency are the key drivers for the choices represented in applications to date.

Let us label the set of quantum numbers for each single-parton mode with a lower-case Greek letter. This Greek label symbolizes the collection of all space-spin-color-flavor degrees of freedom of a single parton in QCD. Fermion and boson single-parton states are orthonormal and complete. Their creation operators satisfy the conventional anti-commutation (commutation) relations for fermions (bosons).

In BLFQ, an eigenstate of a system can then be written in terms of a Fock-space expansion over sectors with $N$-partons as

$$|P, \Lambda\rangle = \sum_N \sum_{\{\alpha_i\}_N} \psi^{\Lambda}_{\{\alpha_i\}_N} |\{\alpha_i\}_N, \Lambda\rangle \ . \tag{5.4.4}$$

where the inner sum includes all allowed configurations of $N$-partons satisfying global symmetry constraints such as baryon number, charge, total helicity projection on the $x^-$ direction, total LF momentum, flavor, etc. For states with two or more bosons, an additional factor is applied to maintain normalization when bosons occupy the same mode.

Up to this point, the Hamiltonian eigenvalue problem of Eq. 5.4.1 is infinite dimensional in both the number of single-parton modes and the number of Fock sectors. With a well-chosen BLFQ single-parton basis (see Sec. 5.4.7 for recent advances) and the vertices of QCD from Fig. 5.4.1, one hopes to achieve reasonable bound state properties with practical cutoffs in these sums suitable for low-resolution applications of QCD for spectra, electoweak transitions, form factors at low-$Q^2$, etc.

### 5.4.3 BLFQ with QED applications

Early applications of BLFQ aimed at solving strong coupling QED problems in order to establish computational techniques and validate BLFQ for achieving converged results in agreement with other methods. These test cases were demanding since they employed the transverse 2D harmonic oscillator and DLCQ for the longitudinal direction to form a basis space that, while suitable for bound state problems in QCD, is far from ideal for these QED applications.

The first application successfully solved for the electron anomalous magnetic moment in an external 2D harmonic trap and took the limit of removing the trap to verify agreement with the well-known Schwinger result [930]. For this application, the first and second vertices in Fig. 5.4.1 are included and sector-dependent renormalization [914] was successfully employed.

The next major advance successfully calculated the electron anomalous magnetic moment directly in free space and at the physical coupling [931, 932] using the same LF Hamilonian and renormalization procedures as Ref. [930] except that the instantaneous vertex was omitted. The demands on the numerical procedures increased dramatically due, in large part, to the slow convergence rate with increasing basis cutoff. The extrapolated result agrees with the Schwinger result to within 0.06% which approximately corresponds to the level of agreement expected between a non-perturbative and a perturbative calculation.

Moving ahead from these early applications, the goals of BLFQ were extended to evaluate additional observables familiar to hadronic physics using the resulting LFWFs. In particular, the BLFQ approach was applied to evaluate the GPDs [933] and the TMDs of the dressed electron [934]. In all cases, the non-perturbative BLFQ results compared favorably with results from perturbation theory at weak coupling.

The next major application was to solve for the low-lying spectrum of positronium at strong coupling

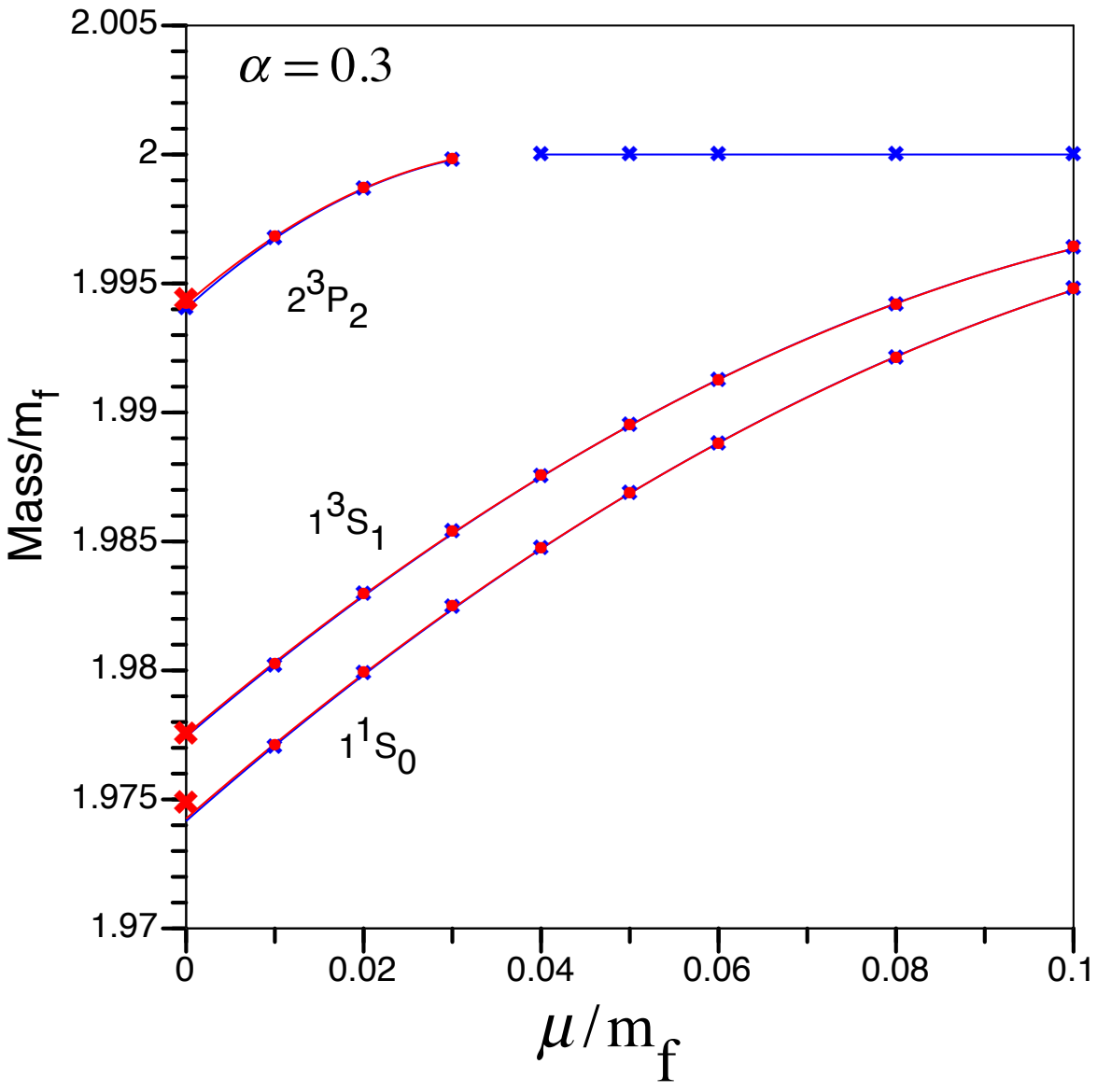

**Fig. 5.4.2** Positronium spectrum extracted from a BLFQ calculation of QED with an unphysically large coupling $\alpha = 0.3$ [935]. The positronium masses are expressed in terms of the electron mass $m_f$. The photon mass, $\mu$, serves as an infrared regulator. The positronium states are labeled by the spectroscopic notation $N\,^{2S+1}L_J$. The O($\alpha^4$) perturbative results are marked by red crosses on the vertical axis [936]. The blue crosses are obtained from extrapolating $N_{\max} \to \infty$ at fixed and sufficiently large $K$. For comparison, the results with extrapolated $K$ are shown in solid red disks. The blue and red curves are second order polynomials used to fit and extrapolate the regulator $\mu$ to zero.

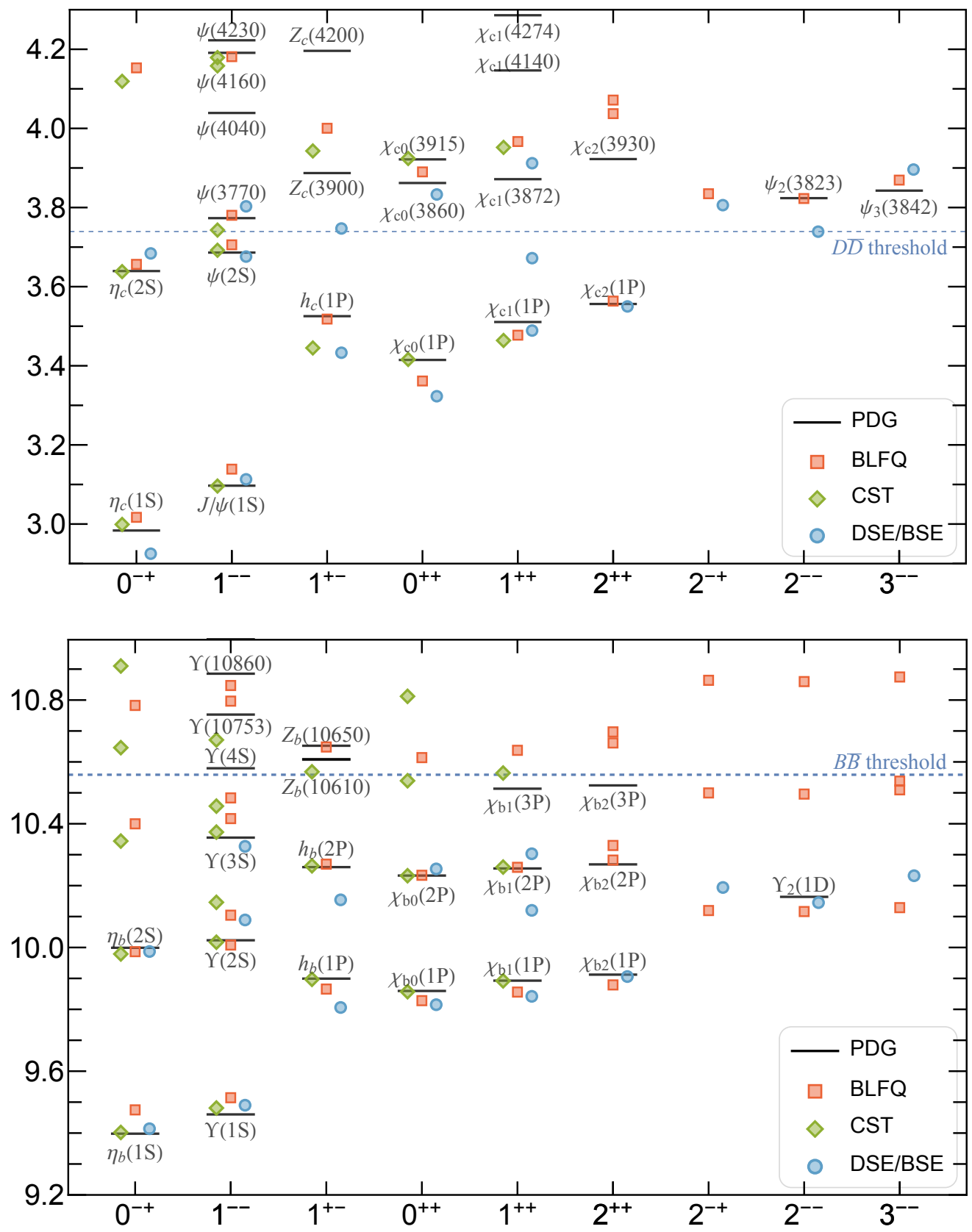

**Fig. 5.4.3** Charmonium (upper panel) and bottomonium (lower panel) spectra obtained from BLFQ [937], CST [869] and DSE/BSE [938] and compared with the PDG data [939]. See also Sec. 5.3. The vertical axis is the hadron mass in GeV. The horizontal axis is the quantum numbers $J^{PC}$, where $J$ is the total spin, $P$, $C$ are the parity and charge conjugation, respectively.

($\alpha = 0.3$) in the valence space of the electron and the positron using a derived effective interaction [940]. The application of BLFQ to positronium adopted the LF effective one-photon exchange interaction of Ref. [941], where they achieved a delicate cancellation of the instantaneous photon interaction term through a suitable choice of energy denominators in second order perturbation theory. These calculations were performed in the fermion single-particle basis with the 2D transverse harmonic oscillator and DLCQ for the longitudinal basis. Convergence was achieved directly in $K$ and by extrapolation in $N_{\max}$, the regulators introduced above. The results for the lowest bound states of positronium as a function of the photon regulator mass are shown in Fig. 5.4.2. At zero regulator mass, one obtains good agreement with results from perturbation theory. The resulting LFWFs were employed to demonstrate methods of calculating GPDs [942] and reveal relativistic effects in strongly-coupled positronium.

More recently, the BLFQ approach has been successfully applied to solve for the structure of the photon [943]. The basis space consists of the photon sector and the electron-positron sector so that only the first interaction term from Fig. 5.4.1 is retained in solving the Hamiltonian eigenvalue problem of Eq. 5.4.1. The basis space is defined as for the positronium application above with the addition of the Fock sector for the photon as a single-particle state. Factorization of the CM motion from the LFWFs is addressed using the Lagrange multiplier term in Eq. 5.4.2 as was accomplished in Ref. [935]. Using sector-dependent renormalization, one achieves the real photon eigenstate to be massless as desired.

The LFWFs obtained for the massless photon are therefore a superposition of a bare photon and an electron-positron pair. These LFWFs provide non-trivial Transverse Momentum Distributions (TMDs) and Parton Distribution Functions (PDFs) which are, in principle, experimentally measureable. Ref [943] provides BLFQ results for TMDs and PDFs in addition to comparisons with results from perturbation theory showing reasonable agreement is obtained as expected.

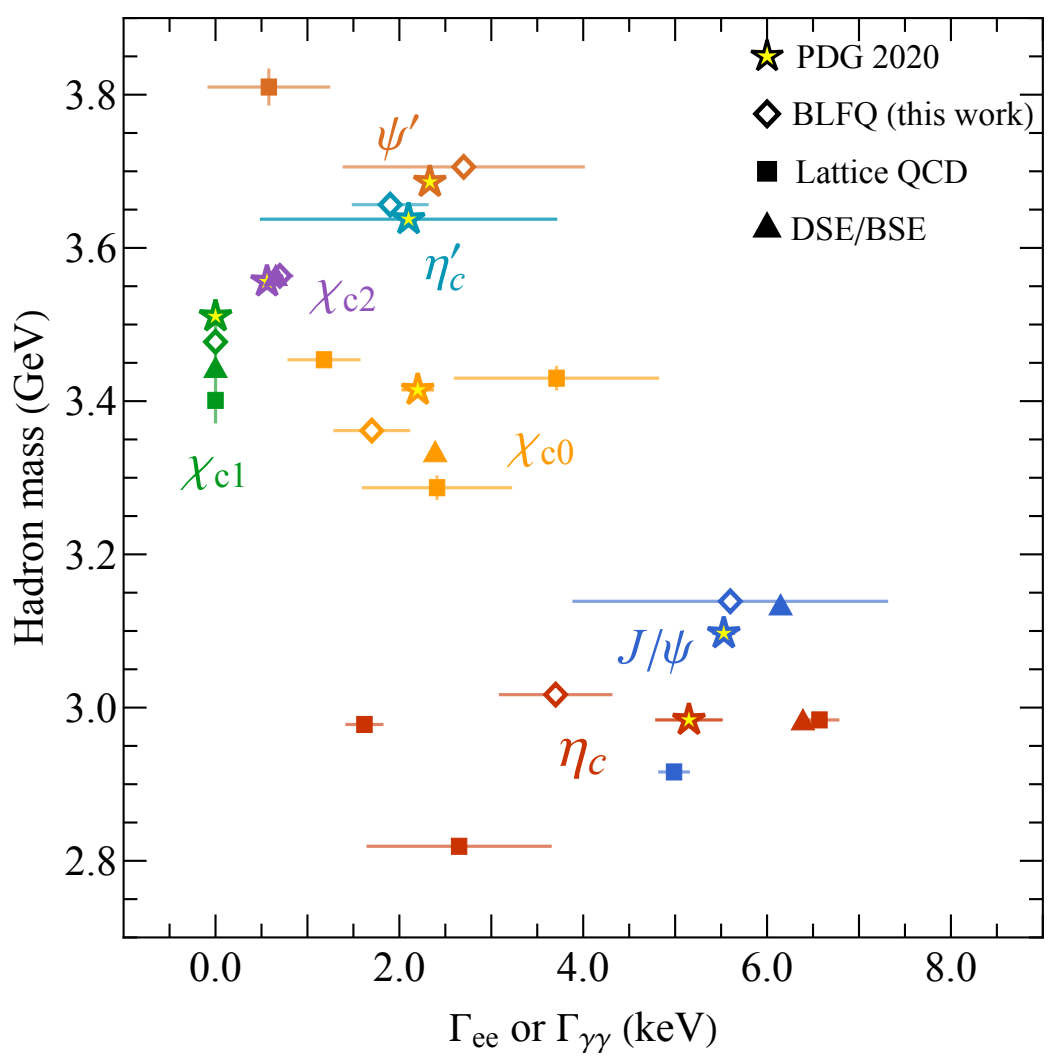

**Fig. 5.4.4** The BLFQ predictions of the charmonium dilepton (for the vectors) or diphoton (for the rest) widths in combination with the mass spectrum. The experimental data as compiled by the PDG are shown in stars. Lattice and DSE/BSE predictions are shown for comparison (see Ref. [944] and ther references therein)

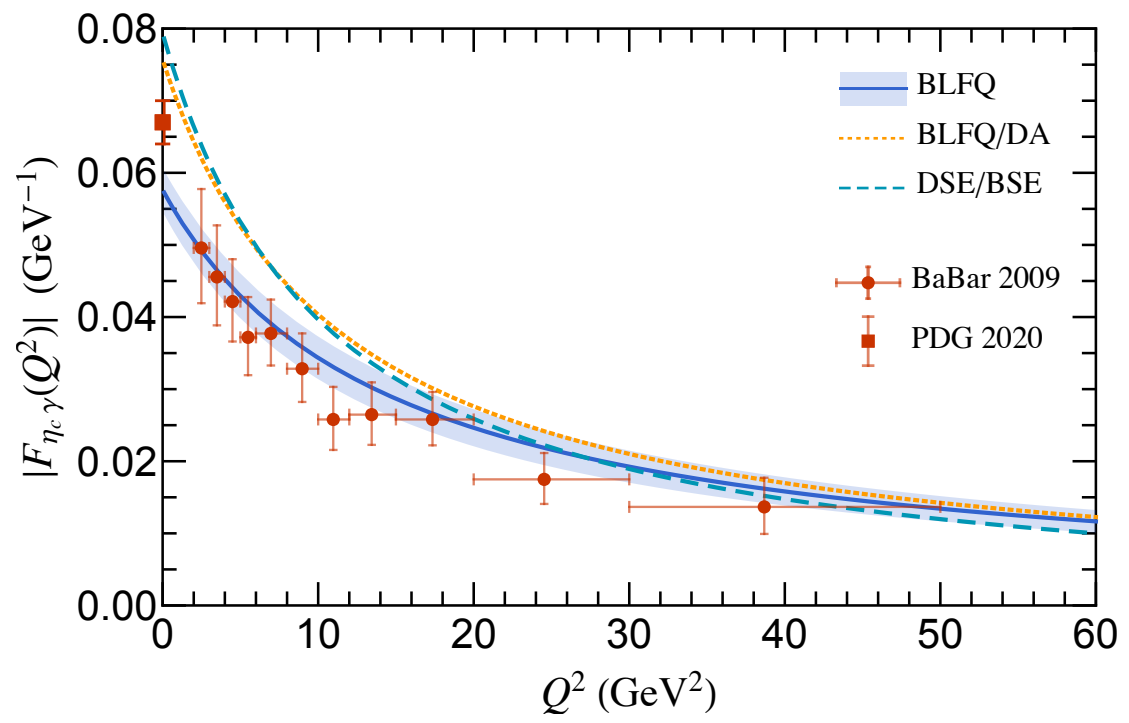

**Fig. 5.4.5** The singly-virtual two-photon transition form factors of $\eta_c$ from BLFQ as compared with the BABAR measurement [945] and the predictions from DSE/BSE. The BLFQ/DA result is obtained from pQCD predictions with LCDA obtained from the BLFQ light-front wave functions. The TFF at $Q^2 = 0$ is extracted from the diphoton width. See Ref. [944] and the references therein.

### 5.4.4 BLFQ for QCD with effective interactions

The high-precision results from the BLFQ treatment of QED problems (Sec. 5.4.3) provide an avenue to treat the one-gluon-exchange interaction between fermions in QCD ($H_{OGE}$), which is the dominant short-distance physics for hadrons. The confining interaction from Light-Front Holography (Sec. 5.5), supplemented by a convenient form for confinement in the longitudinal direction, form the long-distance part of the physics ($H_{con}$). The short distance and long distance terms then lead to the total LF effective interaction, $H_{int} = H_{con} + H_{OGE}$. Similar to the nuclear Shell Model, the solvable part of the Hamiltonian can be chosen to be the kinetic energy plus the confining interaction, $H_0 = H_{kin} + H_{con}$, to implement LFH, augmented with longitudinal confinement, in the zero-th order.

The first application was to compute the spectra and wave functions of heavy quarkonia [937, 950]. Figure 5.4.3 shows the charmonium and bottomonium spectra obtained from BLFQ. Two parameters, the quark mass and the confining strength, were tuned to fit the available experimental measurements, resulting an r.m.s. deviation of the masses about 40 MeV in each system.

The obtained LFWFs were used to evaluate a wide range of observables, including the decay constants [937, 950], light-cone distribution amplitudes [937], form factors [951], radiative transitions [944, 946, 952], semi-leptonic transitions [953], parton distributions [947] and GPDs [951]. Fig. 5.4.4 shows the BLFQ results of the charmonium dilepton (for vector mesons, e.g. $J/\psi$) or diphoton (for the rest) widths in combination with the masses [944], and compared with the available experiments as well as other theoretical approaches whenever available. Fig. 5.4.5 shows the diphoton transition form factor of $\eta_c$ from BLFQ, and compared with the *BABAR* measurement. The M1 widths of the radiative transitions across the heavy quarkonium systems are shown in Fig. 5.4.6, and compared with the PDG values. The PDF of the hadron at the initial scale $\mu_0$ can be obtained by integrating out the transverse momentum. The PDFs of $\eta_c$ obtained from BLFQ are shown in Fig. 5.4.7.

Applications to heavy-light quarkonia have also been achieved [953–956]. Here, the bottomonia and charmonia results were used to determine the quark masses and the confining strength was calculated using the relationship of heavy-quark effective theory as the r.m.s. of the strengths from the corresponding pure flavor systems. This led to successful applications to the spectra, decay constants and other properties of mixed flavor heavy quarkonia without adjustable parameters.

A major step forward was to apply BLFQ with effective interactions to light mesons [957–960]. In addition to the confining interactions as well as the one-gluon-exchange interaction, a Nambu-Jona-Lasinio (NJL) interaction was incorporated to generate the well-known $\rho$-$\pi$ splitting [957]. The obtained LFWFs were used to investigate the partonic structures of the pion. The pion PDF from BLFQ with the effective interactions including the NJL interaction is shown in the top panel of Fig. 5.4.8 where the PDF is compared with the PDF from BLFQ calculations that include one dynamical gluon (see Sec. 5.4.5).

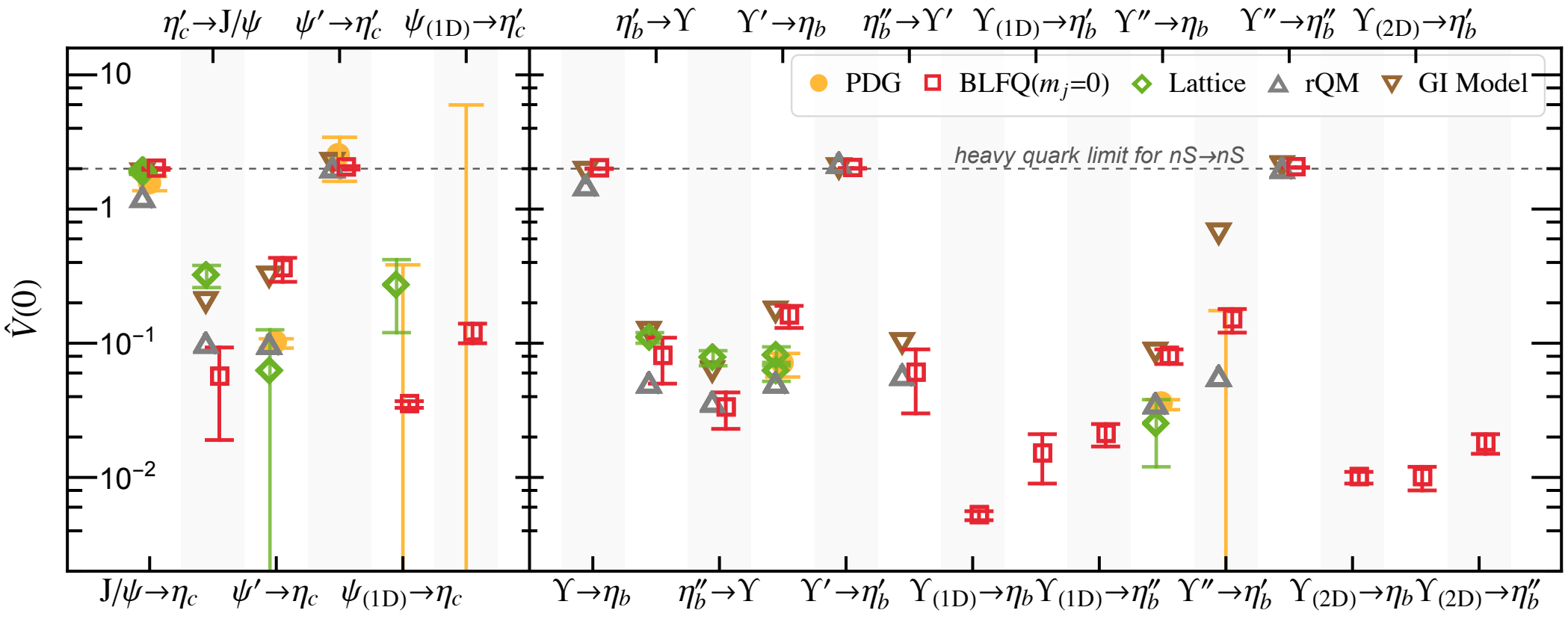

**Fig. 5.4.6** M1 transition form factor at $Q^2 = 0$ for charmonia and bottomonia obtained from BLFQ and compared with several theoretical predictions as well as the experimental data (see Ref. [946] and the references therein)

More recently, the BLFQ formalism has been successfully applied to solving for the structure of the nucleon [949, 962, 963] as well as $\Lambda$, $\Lambda_c$, and their isospin triplet baryons, i.e, $\Sigma^0$, $\Sigma^+$, $\Sigma^-$ and $\Sigma_c^0$, $\Sigma_c^+$, $\Sigma_c^{++}$ [964]. The investigated observables include the electromagnetic and axial form factors, transverse densities, PDFs, GPDs, radii, axial and tensor charges of the baryons. The electromagnetic form factors of the nucleons are compared with the experimental data as well as other approaches in Fig. 10.1.9 in Sec. 10.1. Overall, the theoretical predictions are in good agreement with the experimental measurement for the proton, while the neutron results somewhat deviate from experimental data. The neutron's charge form factor falls well below the data at low $Q^2$, where both experimental and theoretical uncertainties are large. The magnetic moment of the nucleon is related to the nucleon magnetic form factor at $Q^2 = 0$. We obtained the magnetic moment of the nucleon close to the recent lattice QCD results as shown in Table 5.4.1. From the electromagnetic form factors, one can also compute the electromagnetic radii of the nucleon. We summarize our predictions in Table 5.4.1. These results are in reasonable agreement with experiment (see Sec. 10.1). Figure 5.4.9 shows the nucleon axial form factor (see Sec. 10 for details), $G_A = G_A^u - G_A^d$ as a function of $Q^2$, while the contributions from up and down quarks to $G_A(Q^2)$ are also displayed. Our results are compared with the available data from (anti)neutrino scattering off protons or nuclei and charged pion electroproduction experiments and the lattice QCD simulations. Considering the experimental uncertainties and our treatment of the BLFQ uncertainties, we found good agreement with experiment.

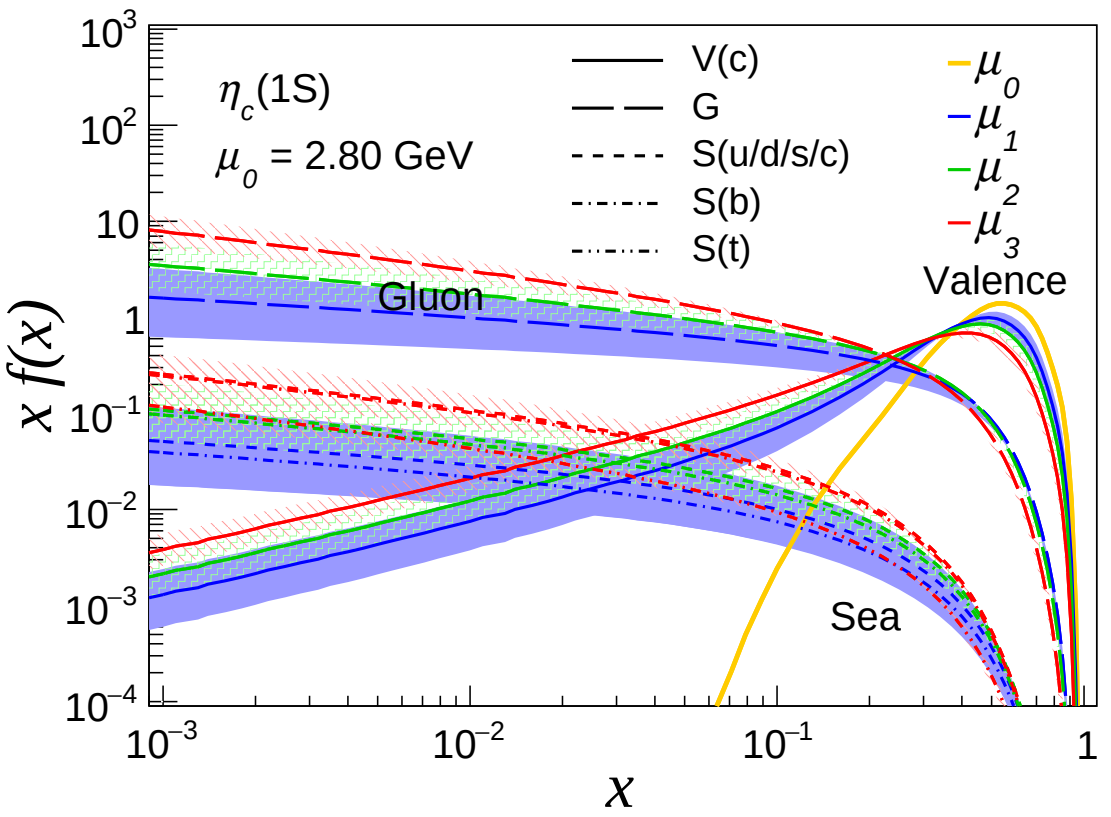

**Fig. 5.4.7** The PDFs of $\eta_c(1S)$ obtained from BLFQ [947]. The bands represent the range of the distributions for the initial scales $\mu_0 = m_q$ to $2\mu_h$. The lines with different color correspond to the different final scales: $\mu_1 = 20$ GeV (blue), $\mu_2 = 80$ GeV (green), and $\mu_3 = 1500$ GeV (red). The solid, thick long-dashed, dashed, dashed-dot, and dashed double-dot lines represent the $x$-PDFs of the valence quark, gluon, sea quark $(u/d/s/c)$, sea quark $(b)$, and sea quark $(t)$, respectively.

**Table 5.4.1** The electromagnetic properties (magnetic moments in units of nuclear magnetons and radii in units of fm), axial charge, axial radius, tensor charge, and the first moment of transversity PDFs. The BLFQ results are compared with the data extracted from experiments and the lattice QCD simulations (see Ref. [949] and the references therein).

| Quantity | BLFQ | Experiments | Lattice |
|---|---|---|---|
| $\mu_p$ | 2.44(3) | 2.79 | 2.43(9) |
| $\mu_n$ | −1.40(3) | −1.91 | −1.54(6) |
| $r_E^p$ | 0.802(40) | 0.833(10) | 0.742(13) |
| $r_M^p$ | 0.834(29) | 0.851(26) | 0.710(26) |
| $(r_E^n)^2$ | −0.033(198) | −0.116(2) | −0.074(16) |
| $r_E^n$ | 0.861(20) | 0.864(9) | 0.716(29) |
| $g_A^u$ | 1.16(4) | 0.82(7) | 0.830(26) |
| $g_A^d$ | −0.248(27) | −0.45(7) | −0.386(16) |
| $g_A^{u-d}$ | 1.41(6) | 1.2723(23) | 1.237(74) |
| $r_A$ | 0.680(70) | 0.667(12) | 0.512(34) |
| $g_T^u$ | 0.94(15) | 0.39(15) | 0.784(28) |
| $g_T^d$ | −0.20(4) | −0.25(20) | −0.204(11) |
| $\langle x\rangle_T^{u-d}$ | 0.229(48) | – | 0.203(24) |

At $Q^2 = 0$, the axial form factor is identified as the axial charge, $g_{\mathrm{A}} = G_{\mathrm{A}}(0)$. Our prediction, presented in Table 5.4.1, is somewhat higher than the extracted data. This discrepancy suggests the need to incorporate higher Fock sectors, which have a significant effect on the quark contribution to the nucleon spin. The corresponding axial radius $r_{\mathrm{A}}$ is in excellent agreement with the extracted data from the analysis of neutrino-nucleon scattering experiments [582, 965].

At leading twist, the complete spin structure of the nucleon is explained in terms of three independent PDFs, namely, the unpolarized, the helicity, and the transversity. The obtained LFWFs were also used to evaluate these leading twist quark PDFs. Figure 5.4.10 (pink bands) shows the unpolarized PDFs of the valence quarks at $\mu^2 = 10$ GeV$^2$ for valence-only space results [949] compared with the global fits. The error bands in our PDFs are due to the 10% uncertainties in the initial scale $\mu_0^2 = 0.195 \pm 0.020$ and the coupling constant $\alpha_s$. Our unpolarized valence PDFs for both the up and the down quarks agree well with the global fits. According to the Drell-Yan-West relation [966, 967], at large scale the valence quark distributions fall off at large $x$ as $(1-x)^p$, where $p$ denotes the number of valence quarks and for the nucleon $p = 3$. In our BLFQ approach, we observed that the up quark unpolarized PDF falls off at large $x$ as $(1-x)^{2.99}$, whereas for the down quark the PDF goes as $(1-x)^{3.24}$. These are in accord with the Drell-Yan-West relation and favour the perturbative QCD prediction [968].

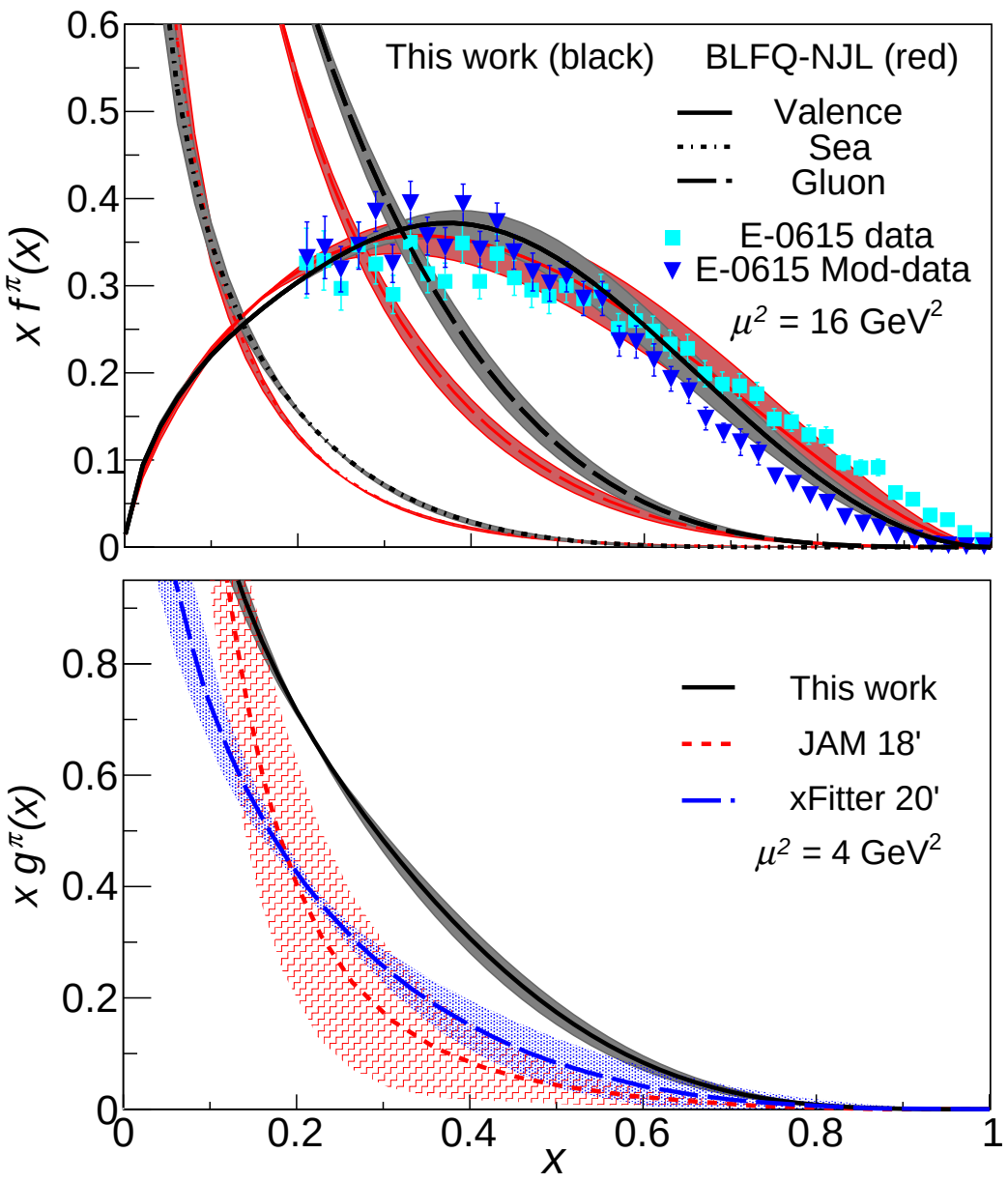

**Fig. 5.4.8** The PDFs of the pion from BLFQ including one dynamical gluon labeled as "This work" [896]. Upper panel: the black lines are the BLFQ results evolved from the initial scale ($0.34 \pm 0.03$ GeV$^2$) using the NNLO DGLAP equations to the experimental scale of 16 GeV$^2$. The red lines correspond to BLFQ-NJL predictions [948]. Results are compared with the original analysis of the FNAL-E615 experiment data and with its reanalysis (E615 Mod-data). Lower panel: the BLFQ result for the pion gluon PDF at $\mu^2 = 4$ GeV$^2$ is compared with the global fits, JAM and xFitter. See Ref. [896] and the references therein for details.

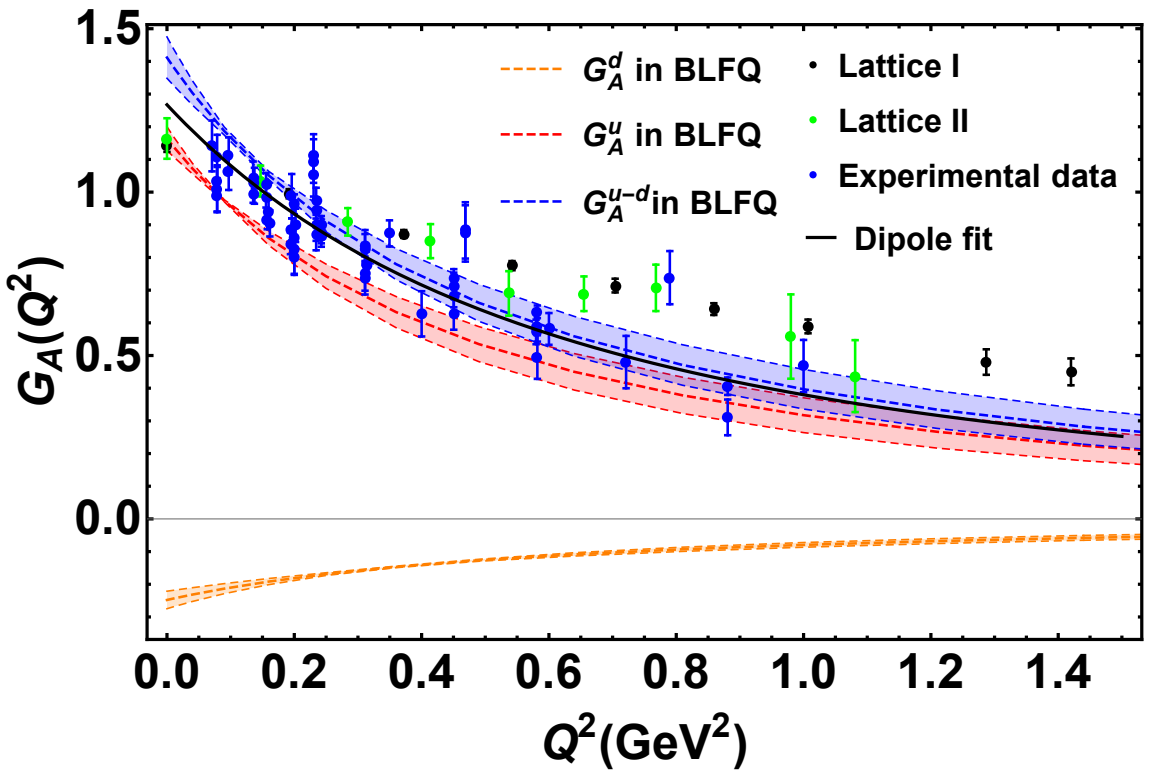

**Fig. 5.4.9** The axial form factors $G_{\mathrm{A}} = G_{\mathrm{A}}^u - G_{\mathrm{A}}^d$ and $G_{\mathrm{A}}^u$, $G_{\mathrm{A}}^d$ as the function of $Q^2$ from BLFQ. The blue band ($G_{\mathrm{A}}$), pink band ($G_{\mathrm{A}}^u$), and orange band ($G_{\mathrm{A}}^d$) are the BLFQ results, which are compared with the experimental measurements as well as the lattice results. The black line represents the dipole fit of the experimental data. See Ref. [949] and the references therein.

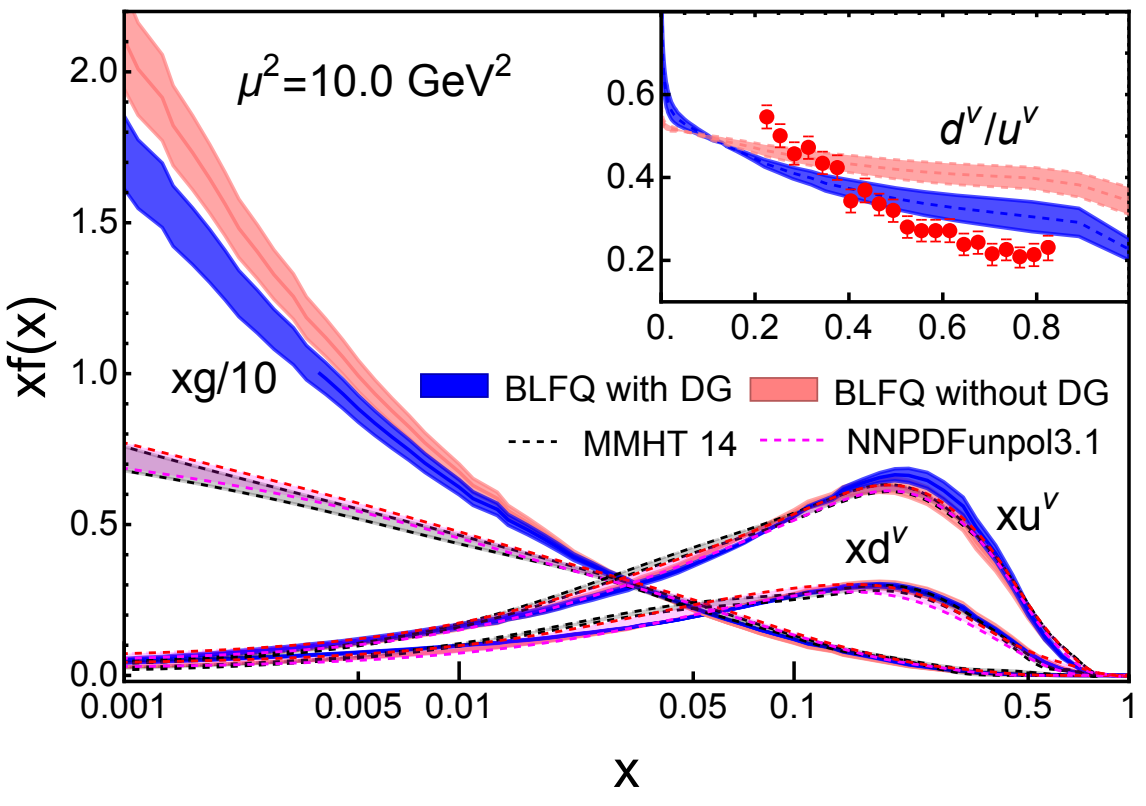

**Fig. 5.4.10** The unpolarized valence quark and gluon PDFs of the proton. The BLFQ results (blue bands: obtained with one dynamical gluon; pink bands: obtained from a light-front effective Hamiltonian based on only a valence Fock representation [949]) are compared with the NNPDF3.1 and MMHT global fits. (The inset) the ratio of the valence quark PDFs is compared with the extracted data from JLab MARATHON experiment. See Ref. [961] and the references therein.

The helicity PDFs are displayed in Fig. 5.4.11 (upper panel: pink bands), at the scale $\mu^2 = 3$ GeV$^2$, for the up and down quarks in the proton. Our BLFQ predictions are compared with the measured data from COMPASS [969]. We found that our down quark helicity PDF agrees reasonably well with the experimental data from COMPASS [969]. For the up quark, the $g_1(x)$ solved in the valence-only space is however overestimated at low $x$, whereas it tends to agree with the data above $x \sim 0.25$ regime.

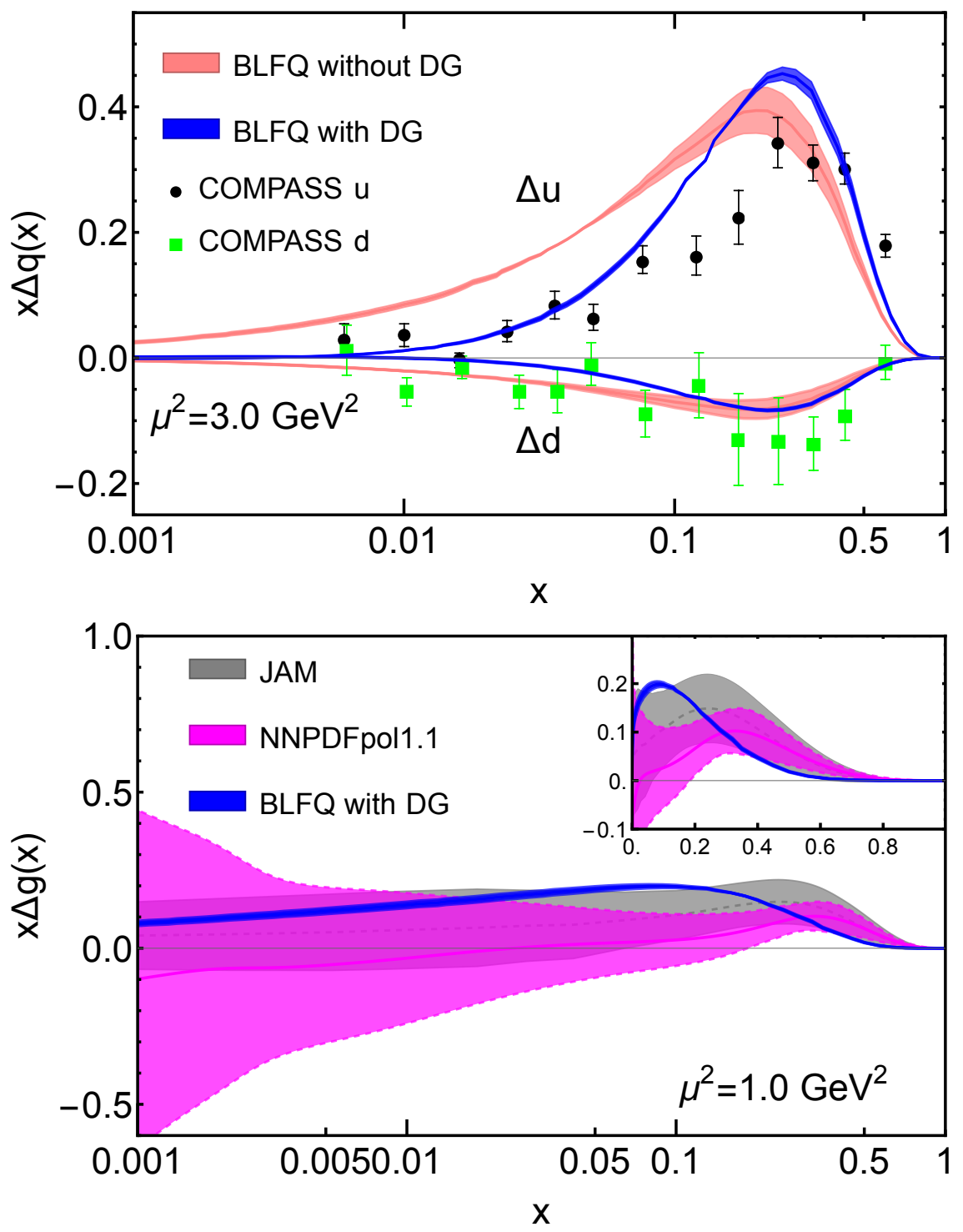

**Fig. 5.4.11** Upper panel: the helicity PDFs for the valence quarks and the gluon in the proton. We compare BLFQ predictions (blue bands: obtained with one dynamical gluon [961]; pink bands: obtained from a light-front effective Hamiltonian based on only a valence Fock representation [949]) with data from COMPASS Collaboration [969]. Lower panel: the gluon helicity PDF in the proton. We compare the BLFQ prediction (blue bands) with global analyses by JAM (gray band) and NNPDFpol1.1 (magenta band). The inset shows the gluon helicity PDF on a linear scale. See Ref. [961] and the reference therein.

The obtained LFWFs were also employed to compute the valence quark GPDs for zero skewness [949] and to study quark angular momentum densities inside the proton [971]. The helicity non-flip unpolarized GPD in impact parameter space, $\mathcal{H}^q(x, b_\perp)$, can be interpreted as the number density of quarks with longitudinal momentum fraction $x$ at a given transverse distance $b_\perp$ in the nucleon [972]. One can then define the $x$ dependent squared radius of the quark density in the transverse plane as [970]:

$$\langle b_\perp^2 \rangle^q(x) = \frac{\int d^2\vec{b}_\perp b_\perp^2 \mathcal{H}^q(x, b_\perp)}{\int d^2\vec{b}_\perp \mathcal{H}^q(x, b_\perp)}. \tag{5.4.5}$$

Figure 5.4.12 shows the $x$-dependent squared radius of the proton, $\langle b_\perp^2 \rangle(x) = 2e_u \langle b_\perp^2 \rangle^u(x) + e_d \langle b_\perp^2 \rangle^d(x)$ and compares the BLFQ prediction with the available extracted data within the range $0.05 \lesssim x \lesssim 0.2$ from the DVCS process [970]. As can be seen from Fig. 5.4.12, the BLFQ prediction for $\langle b_\perp^2 \rangle(x)$ is consistent with the extracted data. We also evaluated the proton's transverse squared radius [970]

$$\langle b_\perp^2 \rangle = \sum_q e_q \int_0^1 dx\, f^q(x)\, \langle b_\perp^2 \rangle^q(x). \tag{5.4.6}$$

In our BLFQ approach, we obtained the squared radius of the proton, $\langle b_\perp^2 \rangle = 0.40 \pm 0.04$ fm$^2$, close to the experimental data [970]: $\langle b_\perp^2 \rangle_{\text{exp}} = 0.43 \pm 0.01$ fm$^2$.

BLFQ has been recently applied to investigate the all-charm tetraquark system [973]. The results suggest that the lowest two-charm-two-anticharm state is not a tightly bound tetraquark. In particular, the lowest tetraquark mass extrapolated to the continuum limit in longitudinal resolution $K$ lies above the extrapolated threshold for two separated mesons.

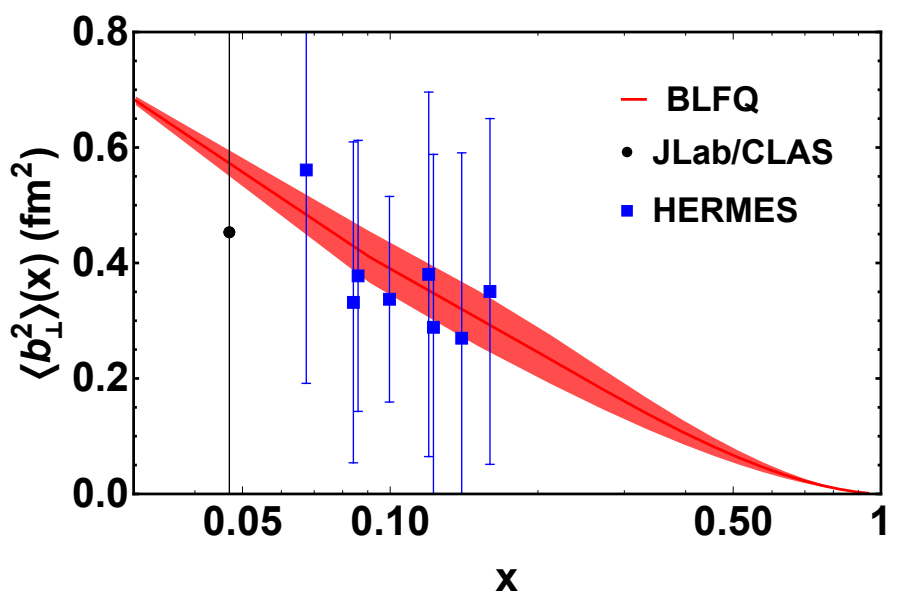

**Fig. 5.4.12** $x$-dependence of $\langle b_\perp^2 \rangle$ for quarks in the proton from BLFQ [949]. The line corresponds to the BLFQ predictions and the band indicates its uncertainty. The data points are taken from Ref. [970]

### 5.4.5 BLFQ beyond the valence Fock sector

In this section, we review more recent applications of BLFQ with the inclusion of dynamical gauge degrees of freedom: to positronium at strong coupling ($\alpha = 0.3$) with one dynamical photon earlier in DLCQ [974] and now in BLFQ [975, 976]; to mesons with one dynamical gluon [896] and to the proton with one dynamical gluon [961].

For the BLFQ application to QED, the positronium system with one dynamical photon presents valuable challenges with respect to non-perturbative renormalization [975–977]. The dynamics of the single fermion system must first be obtained and then embedded in the positronium system with consistent counting of the basis space quanta. That is, within a given Fock sector of positronium and within a given configuration, the distribution of quanta for that configuration dictates the renormalized mass of the fermion to be applied and the basis space in which that mass was determined.

With this dynamical approach, the leading self-energy divergence is taken into account which opens a path to proceed to larger basis spaces.

Going beyond the leading Fock component for QCD, BLFQ has been successfully employed to solve the unflavored light mesons and nucleon with one dynamical gluon [896, 961]. In particular, we adopted an effective light-front Hamiltonian and solved for their mass eigenvalues and eigenstates at the scales suitable for low-resolution probes. Our Hamiltonian incorporates light-front QCD interactions [792] relevant to constituent $|q\bar{q}\rangle$ and $|q\bar{q}g\rangle$ Fock sectors of the mesons and $|qqq\rangle$ and $|qqqg\rangle$ Fock sectors of the nucleon with a complementary 3D confinement [950]. By solving this Hamiltonian in the leading two Fock components and fitting the constituent parton masses and coupling constants as the model parameters [896], we obtained a good quality description of light meson mass spectroscopy[896].

We computed the pion electromagnetic form factor and the PDFs from our Hamiltonian's LFWFs. The BLFQ prediction of the electromagnetic form factor of the charged pion is compared with the experimental data in Fig. 5.3.7 in Sec. 5.3. Figure 5.4.8 shows our results for the pion PDFs and compares the valence quark distribution after QCD evolution with the data from the E615 experiment as well as the reanalysis of the E615 experiment. The pion PDFs previously obtained in BLFQ-NJL model [948, 957, 959] based on a valence Fock representation have also been included for comparison. The error bands in our evolved PDFs are manifested from an adopted 10% uncertainty in our initial scale, $\mu_0^2 = 0.34 \pm 0.03$ GeV$^2$, which we determined by requiring the result after evolution to generate the total first moments of the valence quark and the valence antiquark distributions from the global QCD analysis, $\langle x\rangle_{\text{valence}} = 0.48 \pm 0.01$ at $\mu^2 = 5$ GeV$^2$ [978]. We found a good agreement between our prediction for the pion valence quark PDF and the reanalyzed E615 data, while the BLFQ-NJL model favours the original E615 data.

The lower panel of Fig. 5.4.8 shows the gluon PDF in the pion. Including one dynamical gluon, the gluon density in the pion significantly increases compared to that in the BLFQ-NJL model as well as to the global fits [979]. The BLFQ-NJL model is based on the pion valence Fock component and gluons are produced solely from the scale evolution. However, the model, which includes a dynamical gluon at the initial scale, results in a larger gluon PDF at large-$x$ ($> 0.2$) after scale evolution.

We produced the unpolarized and polarized valence quark and gluon distributions in the proton using the resulting LFWFs for the proton with one dynamical gluon. We evolved our initial PDFs from the model scale, $\mu_0^2 = 0.23 \sim 0.25$ GeV$^2$, to the relevant experimental scales. The blue bands in Figures 5.4.10 and 5.4.11 show our results for the proton unpolarized and polarized PDFs, respectively. We obtained a good consistency between our prediction for the valence quark PDFs and the global fits. The ratio $d^v(x)/u^v(x)$ reasonably agrees with the extracted data from the MARATHON experiment at JLab [980]. At the endpoint, we predicted that $\lim_{x\to 1} d^v/u^v = 0.225 \pm 0.025$.

We found that the down quark unpolarized PDF falls off at large $x$ as $(1-x)^{3.5\pm 0.1}$, whereas for the up quark PDF exhibits $(1-x)^{3.2\pm 0.1}$. These findings support the perturbative QCD prediction [968]. We observed that the gluon PDF is suppressed at small-$x$ and shifts towards the global fits [627, 981] with the addition of a dynamical gluon, whereas the PDF for $x > 0.05$ agrees with the global fits.

Our helicity PDFs for both the up and down quarks (Fig. 5.4.11: upper panel) are reasonably consistent with the experimental data from COMPASS [969]. We noticed that the up quark polarized PDF improves significantly at small-$x$ region with the treatment for the nucleon with dynamical gluon. We observed a fair agreement between our prediction for the gluon helicity PDF (Fig. 5.4.11: lower panel) and the global analyses by the JAM [982] and the NNPDF Collaborations [983]. Note that there still remain huge uncertainties both in the large-x region and especially in the small-x region, where even the sign is uncertain. [984]. The partonic spin contributions to the proton spin are given by the first moment of the polarized PDFs. We found that the gluon carries 26% of the proton spin [961], which is likely to increase when more dynamical gluons are included.

### 5.4.6 Full BLFQ

The applications of BLFQ to hadron structures demonstrated so far have adopted explicit Fock sector truncations. The incorporation of a dynamical gauge boson (Sec. 5.4.5) has shown promising improvements in comparison with valence Fock sector only. A major next step is the full BLFQ [929], in which the Hamiltonians are solved non-perturbatively with basis regulators only and without additional Fock space truncation. The elimination of the additional Fock space truncation positions BLFQ on the path to a genuine *ab initio* approach to QCD. Initial applications which qualify as full BLFQ include solving scalar 1+1 D field theories without Fock space truncation [985].

The full BLFQ is posed as a quantum many-body problem while the number of partons is not fixed. The single-particle harmonic oscillator basis with the lon-gitudinal discretized momentum basis is the preferred choice of basis, together with the $N_{\max}$-$K$ regularization,

$$\sum_i [2n_i + |m_i| + 1] \leq N_{\max}, \qquad \sum_i p_i^+ = \frac{2\pi K}{L}. \tag{5.4.7}$$

As such, all kinematical symmetries of the LFQCD Hamiltonian, including the the factorization of the center-of-mass motion, are preserved in the many-body Hilbert space. This basis corresponds to a pair of soft IR & UV resolutions and a collinear resolution,

$$b^2/(N_{\max}-1) \lesssim \sum_i \frac{k_{i\perp}^2}{x_i} \lesssim b^2(N_{\max}-1), \tag{5.4.8}$$

$$\Delta x \gtrsim K^{-1}. \tag{5.4.9}$$

Here, $b = \sqrt{P^+ \Omega}$. $P^+$ is the longitudinal momentum of the bound state. $\Omega$ is the scale parameter of the transverse harmonic oscillator functions. Note that, if zero modes are omitted as is conventional, the $N_{\max}$-$K$ regularization renders the number of partons finite, and no further Fock sector truncation is needed.

A fundamental challenge of the full BLFQ is the exponential increase of the dimensionality of the Hilbert space, $\dim \mathcal{H} = N^{dN}$, $(N = \max\{N_{\max}, K\})$, a property shared by all strong coupling non-perturbative quantum many-body problems. Nevertheless, meaningful results may be achievable with continuing advances in high-performance computers at and beyond exascale ($10^{18}$ floating point operations per second). On the other hand, future quantum computers offer the promise to provide supremacy over even the best high-performance computers, in particular for non-perturbative quantum many-body problems such as posed by full BLFQ [986].

### 5.4.7 BLFQ with chiral symmetry breaking

Due to the light quark mass, $m_{\{u,d\}} \ll \Lambda_{\text{QCD}}$, chiral symmetry plays an important role on the light meson spectrum and structures. In particular, the pion is the Goldstone boson of the spontaneously broken chiral symmetry. Formally, chiral symmetry implies a partially conserved axial-vector current (PCAC). In BSE, this relation leads to a set of relations between the pion Bethe-Salpeter amplitudes and the quark self-energy (see Sec. 5.3). Recently, it was revealed that PCAC also leads to a chiral sum rule for the pion LFWFs [987].

It was long pointed out that chiral symmetry breaking in LFQCD is manifested in a different way from the instant form (see Ref. [988] for a recent review). In the instant form, chiral symmetry breaking is due to the condensate of quark-antiquark, viz. $\langle \bar{q}q \rangle \neq 0$. The light-front vacuum is trivial due to the positivity of the longitudinal momenta. Therefore, the vacuum condensate on the light front can only happen through the zero modes. Indeed, the wee parton condensate is long conjectured to be the mechanism for symmetry breaking on the light front, which is supported by 1+1D theories and has shown to be a useful starting point for BLFQ applications [989].

On the other hand, the axial charge on the light front annihilates the light-front vacuum, $Q_5|0\rangle = 0$, which suggests that the chiral condensate should be encoded within the hadron LFWFs [990]. One of the traces of the chiral symmetry breaking in the pion LFWFs is the chiral sum rule [987]. Taking advantage of light-front holography, this sum rule has been shown to be also consistent with the chiral symmetry breaking in AdS/QCD.

### 5.4.8 Nonperturbative reactions in BLFQ

One major advantage of the Hamiltonian formalism of quantum field theory is that it allows for tracking time evolution of quantum field configurations in real time.

As an extension of BLFQ, the time-dependent Basis Light-front Quantization (tBLFQ) has been developed as a time-dependent nonperturbative approach to quantum field theory [991]. In tBLFQ the light-front Schrödinger equation is solved to simulate the time evolution of quantum field configurations:

$$i\frac{\partial}{\partial x^+} \left|\psi; x^+\right\rangle = \frac{1}{2} P^-(x^+) \left|\psi; x^+\right\rangle, \tag{5.4.10}$$

where $|\psi; x^+\rangle$ represents the quantum field system under consideration and $P^-(x^+)$ is the light-front Hamiltonian, which includes the interactions among the fields under consideration. The tBLFQ approach is suitable for studying particle evolution in a strong and possibly time-dependent background field. The tBLFQ approach motivated a nonperturbative approach simulating nuclear reactions in low energy nuclear physics, named time-dependent Basis Function (tBF) [992].

One of the major goals of tBLFQ is to understand the nonperturbative dynamics in QCD, as in hadron scattering. The investigations of quark scattering with a nucleus constitute a first step toward this goal. In Ref. [993], tBLFQ is employed to simulate the scattering of an ultrarelativistic quark off a heavy nucleus at high energies. The color glass condensate, a classical effective theory of QCD, is adopted as a model for the color field of the heavy nucleus. The results can significantly reduce the theoretical uncertainties in the small $p_\perp$ region of the differential cross section which has important implications for the phenomenology of the hadron-nucleus and deep inelastic scattering at high energies. One important feature of tBLFQ is that it allows one to take "snapshots" of the system at intermediate times of the evolution, which provide physical insights into the nonperturbative mechanism in time-dependent processes. For example, Fig. 5.4.13 shows the evolution of the probability distribution of a quark in the transverse direction at different light-front times $x^+$. In Ref. [995] a calculation is performed in an extended Fock space where one dynamical gluon is included, which paves the way for studying partons' radiational energy loss in nuclear matter.

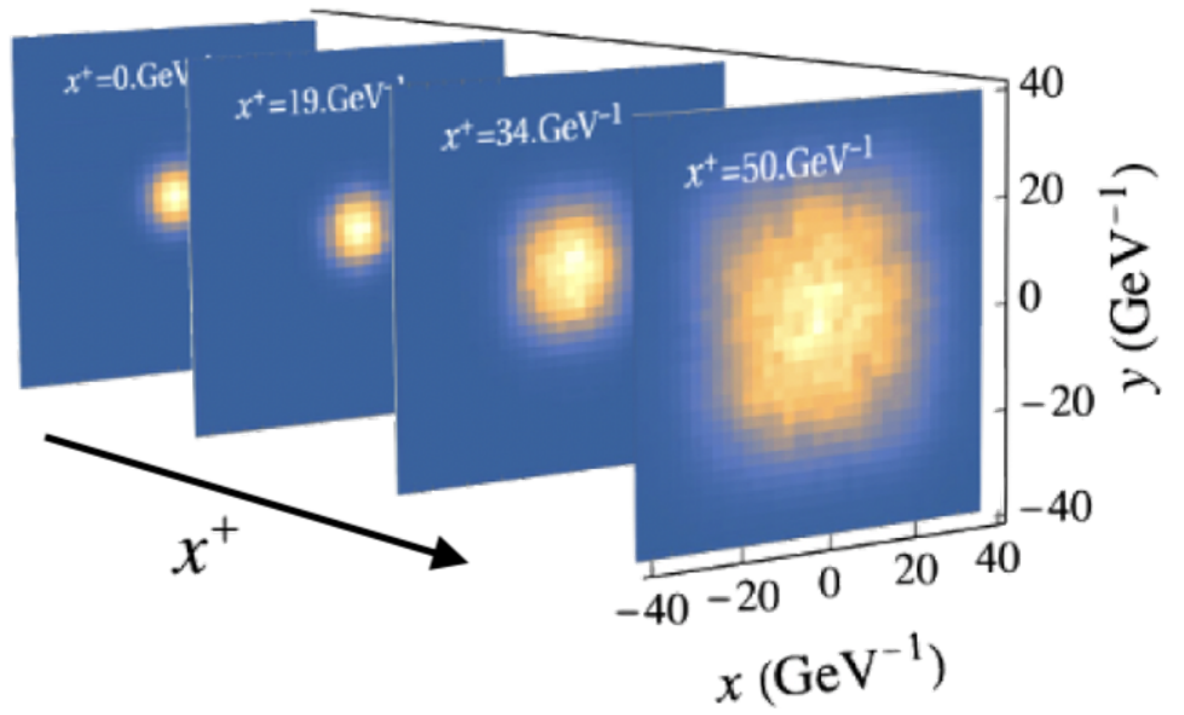

**Fig. 5.4.13** The evolution of the transverse density of a quark within a classical color field of a heavy nucleus at different light-front time $x^+$. The four "snapshots" are from Ref. [993]

In addition to the applications in QCD, tBLFQ has also been employed to study various nonperturbative processes in strong field QED [991, 996–999].

The tBLFQ approach can be further improved in three directions: i) increase in the level of complexity and realism of the background field: ii) expansion to reaction processes of a wider class; iii) the expansion of the Fock space in the description of quantum field configurations. While this can lead to more accurate simulation of dynamical processes, it will dramatically increase the required computational resources. Therefore, it is desirable to explore numerical algorithms for tBLFQ on next-generation advanced computational platforms.

### 5.4.9 Comparisons between BLFQ and BSE

The similarities and differences of the light-front and the BSE (see Sec. 5.3) approaches motivate a direct comparison of the amplitudes obtained from these two approaches [870]. Fig. 5.4.3 shows the comparison of quarkonia spectra obtained from BLFQ and CST. In both approaches, the model parameters were fixed by fitting to the experimentally measured quarkonia masses. Then, the obtained wave functions were used to compute physical observables. Fig. 5.4.14 compares the axial-vector LFWFs obtained from BLFQ and CST. The Brodsky-Huang-Lepage prescription [994] was used to convert the CST amplitude to the LFWFs [870]. Qualitatively, the wavefunctions are similar. However, some spin components show different characteristics due to the different implementation of discrete symmetries, which can be discerned in high-energy exclusive processes.

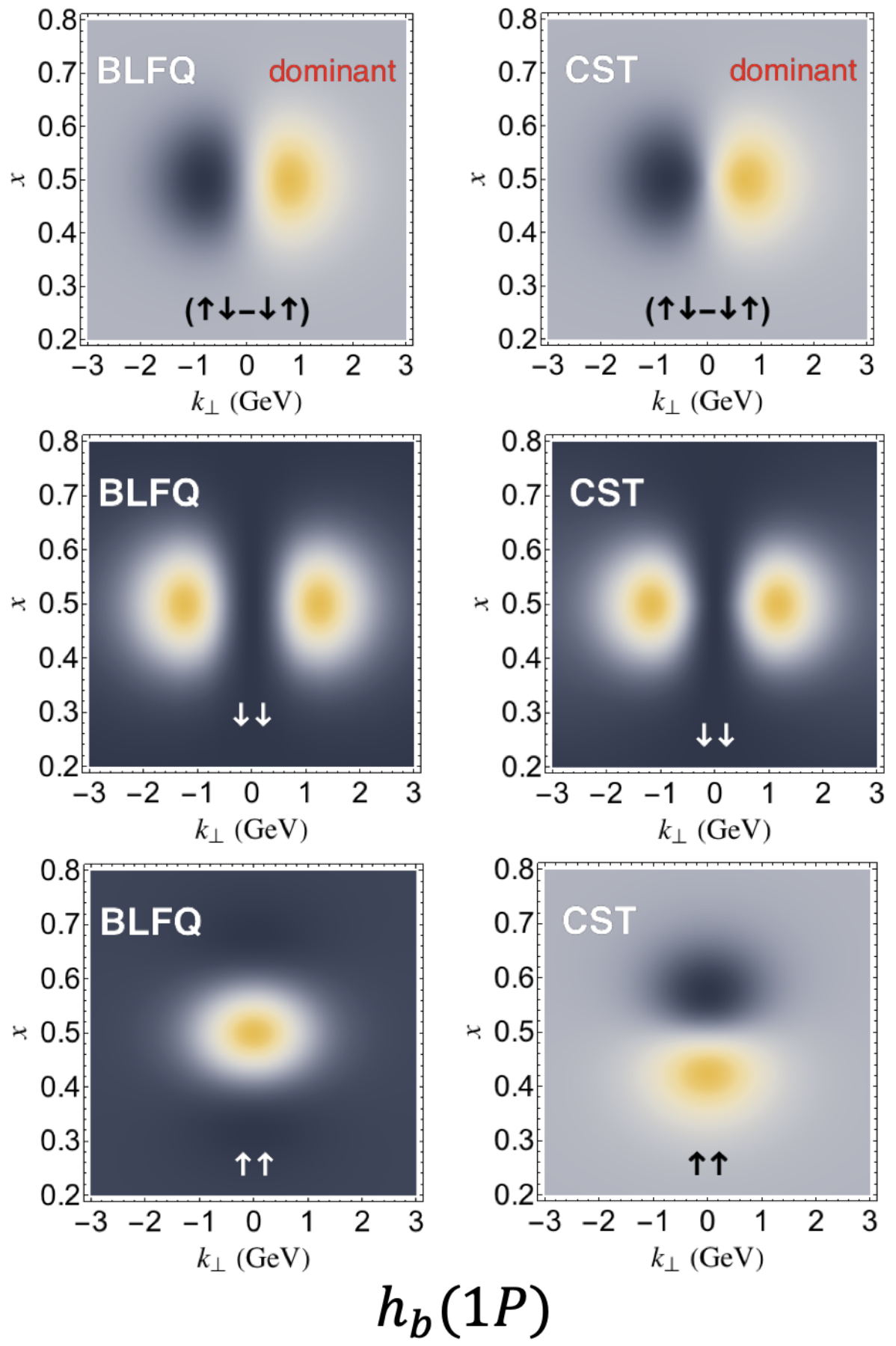

**Fig. 5.4.14** Comparison of selected LFWFs of $h_b$ obtained from BLFQ and CST [870]. The latter were converted from the BSA with the Brodsky-Huang-Lepage prescription [994]. The non-relativistic dominant spin components from both approaches (top panels) are qualitatively the same. However, subdominant LFWFs may appear dramatically different, some of which are in leading twist (bottom panels). This can be tested in high-energy exclusive processes. The discrepancy stems from the different implementation of the discrete symmetries on the light cone in BLFQ and CST.

## 5.5 AdS/QCD and light-front holography

**Stanley J. Brodsky, Guy F. de Téramond, and Hans Güenter Dosch**

### 5.5.1 Introduction

In spite of the important progress of Euclidean lattice QCD [80] and other nonperturbative approaches, a basic understanding of fundamental features of hadron physics from first principles, such as the mechanism of color confinement and the origin of the hadron mass scale, as well as general features of hadron structure, spectroscopy and dynamics, have remained among the most important unsolved challenges of the last 50 years in particle physics. Furthermore, other essential properties of the strong interactions, which were manifest in dual models and developed before QCD, are also not explicit properties of the QCD Lagrangian.

Recent theoretical developments for understanding features of hadronic physics are based on AdS/CFT – the correspondence between classical gravity in a higher-dimensional anti-de Sitter (AdS) space and conformal field theories (CFT) in physical space-time [1000–1002]. AdS/CFT has provided a semiclassical approximation for strongly-coupled quantum field theories, giving physical insights into nonperturbative dynamics. In practice, the AdS/CFT duality provides an effective weakly coupled description in a $(d + 1)$-dimensional $\mathrm{AdS}_{d+1}$ space in terms of a flat $d$-dimensional superconformal, strongly coupled, quantum field theory defined on the AdS asymptotic boundary, the physical four-dimensional Minkowski spacetime, where boundary conditions are imposed [1003]. This is illustrated in Fig. 5.5.1 for $d = 4$, where the asymptotic surface of the 5-dimensional $\mathrm{AdS}_5$ space is the physical four-dimensional Minkowski spacetime.

Anti-de Sitter $\mathrm{AdS}_{d+1}$ is the maximally symmetric $d+1$ space with negative constant curvature and a $d$-dimensional flat spacetime boundary. In Poincaré coordinates $x^M = \left(x^0, x^1, \cdots, x^d, z\right)$, where the asymptotic border of AdS space is given by $z = 0$, the line element is

$$\begin{aligned} ds^2 &= g_{MN} dx^M dx^N \\ &= \frac{R^2}{z^2}\left(\eta_{\mu\nu} dx^\mu dx^\nu - dz^2\right), \end{aligned} \tag{5.5.1}$$

where $\eta_{\mu\nu}$ is the usual Minkowski metric in $d$ dimensions, and $R$ is the AdS radius. The group of transformations leaving the $\mathrm{AdS}_{d+1}$ metric invariant, the isometry group $SO(2, d)$, has dimension $(d + 1)(d + 2)/2$. Five-dimensional anti-de Sitter space $\mathrm{AdS}_5$ has thus 15 isometries, which induce in the Minkowski-space boundary the symmetry under the conformal group with 15 generators in four dimensions: 6 Lorentz transformations plus 4 spacetime translations plus 4 special conformal transformations plus 1 dilatation [1004]. This conformal symmetry implies that there can be no scale

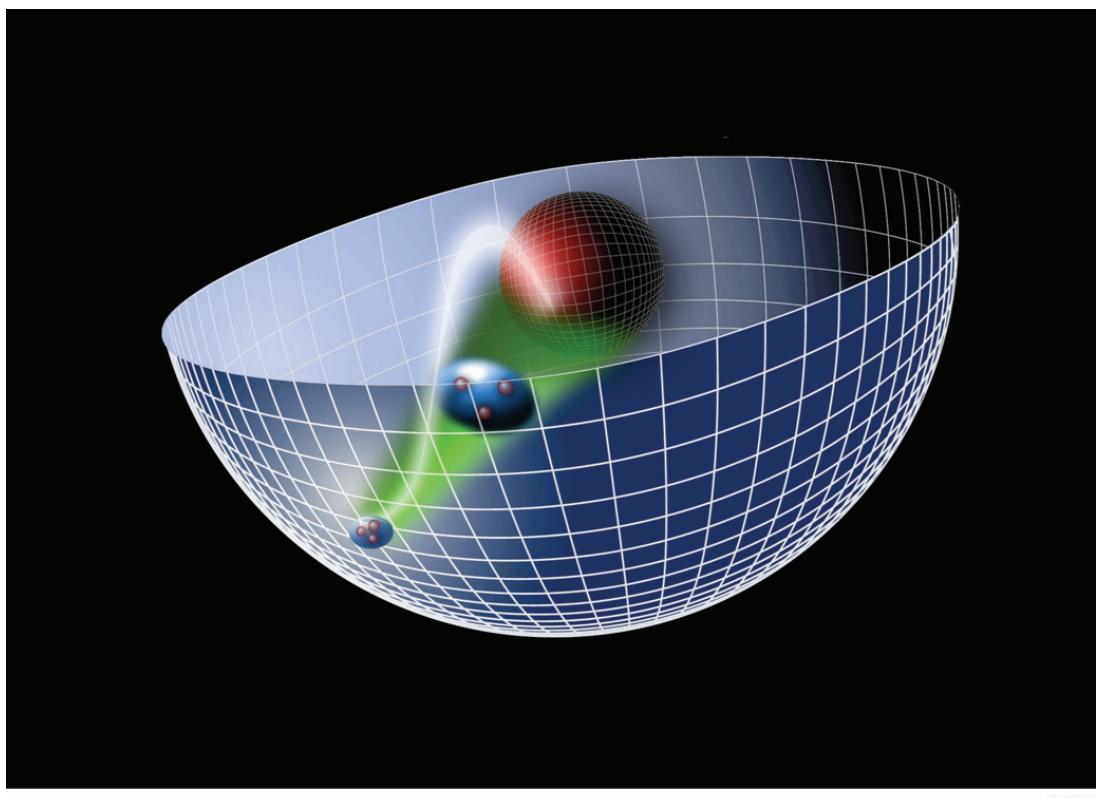

**Fig. 5.5.1** Different values of the AdS radius $z$ correspond to different scales at which the proton is examined. The inner sphere (where AdS space terminates) is represented by the red sphere. The green "cone" represents the warping of AdS space (Its negative curvature characteristic of a hyperbolic space). Therefore a proton "the small blue ball" appears as a smaller object near the boundary of AdS as perceived by an observer in physical Minkowski space.

in the boundary theory and therefore no discrete spectrum.

The relation between the dilatation symmetry and the symmetries in $\mathrm{AdS}_5$ can be seen directly from the AdS metric since (5.5.1) is invariant under a dilatation of all coordinates: A dilatation of the Minkowski coordinates $x^\mu \to e^\sigma x^\mu$ is compensated by a dilatation of the holographic variable $z \to e^\sigma z$. Therefore, the variable $z$ acts like a scaling variable in Minkowski space: different values of $z$ correspond to different energy scales at which a measurement is made. As a result, short spacetime intervals map to the boundary in AdS space-time near $z = 0$. This corresponds to the ultraviolet (UV) region of AdS space. On the other hand, a large four-dimensional object of confinement dimensions $1/\Lambda^2_{\mathrm{QCD}}$ maps to the large IR region of AdS space $z \sim 1/\Lambda_{\mathrm{QCD}}$. Thus, in order to incorporate confinement in the gravity dual the conformal invariance must be broken by modifying AdS space in the large $z$ IR region. For example, a simple way to obtain confinement and discrete normalizable modes (Fig. 5.5.1) is to introduce a sharp cut-off at the IR border $z_0 \sim 1/\Lambda_{QCD}$, as in the "hard-wall" model of Ref. [1005].

In general, one can deform the original AdS background geometry, giving rise to a less symmetric gravity dual. This approach provides useful tools for constructing dual gravity models in higher dimensions which incorporates confinement and basic QCD properties in physical spacetime. The resulting gauge/gravity duality is broadly known as the AdS/QCD correspondence, or simply holographic QCD, which has become an extensive field of research. The extent to which the full theory of QCD can be described in such a framework remains unclear. However, it is clear that holographic models motivated by the AdS/CFT correspondence can capture essential features of QCD and may give important insights into how QCD works. Different models can be derived via a top-down approach from brane configurations in string theory, as well as from more phenomenological bottom-up models, which are not constrained by string theory, and are therefore more flexible for incorporating key aspects of QCD. The best known example of the first category is the Witten-Sakai-Sugimoto model [1006], which contains vector mesons and pions in its spectrum arising from the breaking of chiral symmetry. Conversely, in the bottom-up hard-wall model of Refs. [1007, 1008], the global $SU(2)\times SU(2)$ chiral symmetry of QCD becomes a gauge invariant symmetry on the gravity side. The AdS/QCD model of Refs. [1007, 1008] has also been extended by using the "soft-wall" model introduced in Ref. [1009] in order to reproduce the observed linearity of Regge trajectories.

A third approach to AdS/QCD, *holographic light-front QCD* (HLFQCD) [1010], is based on the holographic embedding of Dirac's relativistic *front form* of dynamics [909] into AdS space. In the front form, the initial surface is the tangent plane to the light cone $x^0 + x^3 = 0$, the null plane, thus without reference to a specific Lorentz frame, in contrast with the usual *instant form* where quantization is defined at $x^0 = 0$. This precise mapping between semiclassical LF Hamiltonian equations in QCD and wave equations in AdS space, [1011] leads to relativistic wave equations in physical space-time (similar to the Schrödinger or Dirac wave equations in atomic physics) and provides an effective computational framework of hadron structure and dynamics [1010] [37]

A remarkable property of HLFQCD is the embodiment of a superconformal algebraic structure which not only introduces a mass scale within the algebra, but also determines the interaction completely [1017–1022].[38] Further extensions of HLFQCD provides nontrivial interconnections between the dynamics of form factors and quark and gluon distributions [1026–1028] with pre-QCD nonperturbative approaches such as Regge theory and the Veneziano model.

In this section we give an overview of relevant aspects of the holographic embedding of QCD quantized in the light front, with an emphasis on the underlying superconformal structure for hadron spectroscopy and hadron duality for amplitude dynamics. Introductory reviews are given in Refs. [1010, 1029–1031]. Other reviews describing distinct approaches and aspects of holographic QCD in the context of the gauge/gravity correspondence in addition to Refs. [1003, 1006], are given in Refs. [1032–1034] and in the book [1035], with applications to other topics such as holographic renormalization group flows, QCD at finite temperature and density, hydrodynamics and strongly coupled condensed matter systems.

### 5.5.2 Semiclassical approximation to light-front QCD

A semiclassical approximation to QCD has been obtained using light-front (LF) physics, where the quantization surface is the null plane, $x^+ = x^0 + x^3 = 0$ [909]. Evolution in LF time $x^+$ is given by the Hamiltonian equation [792]

$$i\frac{\partial}{\partial x^+}|\psi\rangle = P^-|\psi\rangle, \quad P^-|\psi\rangle = \frac{\mathbf{P}_\perp^2 + M^2}{P^+}|\psi\rangle, \qquad (5.5.2)$$

for a hadron with 4-momentum $P = (P^+, P^-, \mathbf{P}_\perp)$, $P^\pm = P^0 \pm P^3$, where $P^-$ is a dynamical generator and $P^+$ and $\mathbf{P}_\perp$ are kinematical. The simple structure of the LF vacuum allows a quantum-mechanical probabilistic interpretation of hadron states in terms of the eigenfunctions of the LF Hamiltonian equation (5.5.2) in a constituent particle basis, $|\psi\rangle = \sum_n \psi_n |n\rangle$, similar to usual Schrödinger equation. The LF wave functions (LFWFs), $\psi_n$ underly the physical properties of hadrons in terms of their quark and gluon degrees of freedom. For a $q\bar{q}$ bound state we factor out the longitudinal $X(x)$ and orbital $e^{iL\theta}$ dependence from $\psi$,

$$\psi(x, \zeta, \theta) = e^{iL\theta} X(x) \frac{\phi(\zeta)}{\sqrt{2\pi\zeta}},$$

where $\zeta^2 = x(1-x)\mathbf{b}_\perp^2$ is the invariant transverse separation between two quarks, with $\mathbf{b}_\perp$, the relative impact variable, conjugate to the relative transverse momentum $\mathbf{k}_\perp$ with longitudinal momentum fraction $x$.

[37] The origins of the light-front holographic approach can be traced back to the original article of Polchinski and Strassler [1005], where the exclusive hard-scattering counting rules [134, 1012], a property of hadrons in physical spacetime, can be derived from the warped geometry of five-dimensional $AdS_5$ space. Indeed, one can show that a precise mapping between the hadron form factors in AdS space [1013] and physical spacetime [966, 967] can be carried out for an arbitrary number of quark constituents [1014–1016]. The key holographic feature is the identification of the invariant transverse impact variable $\zeta$ for the $n$-parton bound state in physical 3+1 spacetime with the holographic variable $z$, the fifth dimension of AdS.

[38] The idea to apply an effective supersymmetry to hadron physics is certainly not new [1023–1025], but failed to account for the special role of the pion. In contrast, in the HLFQCD approach, the zero-energy eigenmode of the superconformal quantum mechanical equations is identified with the lightest meson which has no baryonic supersymmetric partner.

In the ultra-relativistic zero-quark mass limit the invariant LF Hamiltonian $P_\mu P^\mu|\psi\rangle = M^2|\psi\rangle$, with $P^2 = P^+P^- - \mathbf{P}_\perp^2$ can be systematically reduced to the wave equation [1011]:

$$\left(-\frac{d^2}{d\zeta^2} - \frac{1-4L^2}{4\zeta^2} + U(\zeta)\right)\phi(\zeta) = M^2\phi(\zeta), \qquad (5.5.3)$$

where the effective potential $U$ comprises all interactions, including those from higher Fock states. The critical value of the LF orbital angular momentum $L = 0$ corresponds to the lowest possible solution. The LF equation (5.5.3) is relativistic and frame-independent; It has a similar structure to wave equations in AdS provided that one identifies $\zeta = z$, the holographic variable [1011].

### 5.5.3 Higher integer-spin wave equations in AdS

We start with the AdS action for a tensor-$J$ field $\Phi_J = \Phi_{N_1\dots N_J}$ in the presence of a dilaton profile $\varphi(z)$ responsible for the confinement dynamics

$$S = \int d^4x\, dz \sqrt{g}\, e^{\varphi(z)} \left(D_M\Phi_J D^M\Phi_J - \mu^2\Phi_J^2\right), \quad (5.5.4)$$

where $g$ is the determinant of the metric tensor $g_{MN}$ and $D_M$ is the covariant derivative which includes the affine connection. [39][40] The variation of the AdS action leads to the wave equation

$$\left[-\frac{z^{d-1-2J}}{e^{\varphi(z)}}\partial_z\left(\frac{e^{\varphi(z)}}{z^{d-1-2J}}\partial_z\right) + \frac{(\mu R)^2}{z^2}\right]\Phi_J(z) = M^2\Phi_J(z), \qquad (5.5.5)$$

after a redefinition of the AdS mass $\mu$, plus kinematical constraints to eliminate lower spin from the symmetric tensor $\Phi_{N_1\dots N_J}$ [1036]. By substituting

$$\Phi_J(z) = z^{(d-1)/2-J}e^{-\varphi(z)/2}\,\phi_J(z)$$

in (5.5.5), we find the semiclassical light-front wave equation (5.5.3) with

$$U_J(\zeta) = \frac{1}{2}\varphi''(\zeta) + \frac{1}{4}\varphi'(\zeta)^2 + \frac{2J-d+1}{2\zeta}\varphi(\zeta), \quad (5.5.6)$$

as long as $\zeta = z$. The precise mapping allows us to write the LF confinement potential $U$ in terms of the dilaton profile which modifies the IR region of AdS space to incorporate confinement [1010], while keeping the theory conformal invariant in the ultraviolet boundary of AdS, namely $\varphi(z) \to 0$ for $z \to 0$. The separation of kinematic and dynamic components, allows us to determine the mass function in the AdS action in terms of physical kinematic quantities with the AdS mass-radius $(\mu R)^2 = L^2 - (d/2 - J)^2$ and $d$, the number of transverse coordinates [1011, 1036], consistent with the AdS stability bound [1037].

### 5.5.4 Higher half-integer-spin wave equations in AdS

A similar derivation follows from the Rarita-Schwinger action for a spinor field $\Psi_J \equiv \Psi_{N_1\dots N_{J-1/2}}$ in AdS [1036] for half-integral spin $J$. In this case, however, the dilaton term does not lead to an interaction [1038], and an effective Yukawa-type coupling to a potential $V$ in the action has to be introduced instead [1039–1041]:

$$S = \int d^4x\, dz\sqrt{g}\,\bar\Psi_J\left(i\Gamma^A e_A^M D_M - \mu + \frac{z}{R}V(z)\right)\Psi_J, \qquad (5.5.7)$$

where $e_A^M$ is the vielbein and the covariant derivative $D_M$ on a spinor field includes the affine connection and the spin connection. The tangent space Dirac matrices obey the usual anticommutation relations $\{\Gamma^A, \Gamma^B\} = 2\eta^{AB}$. The variation of the AdS action leads to a system of linear differential equations which is equivalent to the second order equations [1036]

$$\left(-\frac{d^2}{d\zeta^2} - \frac{1-4L^2}{4\zeta^2} + U^+(\zeta)\right)\psi_+ = M^2\psi_+, \qquad (5.5.8)$$

$$\left(-\frac{d^2}{d\zeta^2} - \frac{1-4(L+1)^2}{4\zeta^2} + U^-(\zeta)\right)\psi_- = M^2\psi_-, \qquad (5.5.9)$$

with $\zeta = z$, $|\mu R| = L + 1/2$ and equal probability $\int d\zeta\, \psi_+^2(\zeta)^2 = \int d\zeta\, \psi_-^2(\zeta)$. The semiclassical LF wave equations for $\psi_+$ and $\psi_-$ correspond to LF orbital angular momentum $L$ and $L+1$ with

$$U^\pm(\zeta) = V^2(\zeta) \pm V'(\zeta) + \frac{1+2L}{\zeta}V(\zeta), \qquad (5.5.10)$$

a $J$-independent potential, in agreement with the observed degeneracy in the baryon spectrum.

### 5.5.5 Superconformal algebraic structure and emergence of a mass scale

Embedding light-front physics in a higher dimension gravity theory leads to important insights into the nonperturbative structure of bound state equations in QCD for arbitrary spin, but it does not answer the question of how the effective confinement dynamics is actually determined, and how it can be related to the symmetries of QCD itself. An important clue, however, comes from the realization that the potential $V(\zeta)$ in Eq. (5.5.10) plays the role of the superpotential in supersymmetric (SUSY) quantum mechanics (QM) [1042].

[39] The affine connection, the vielbein and the spin connection are important elements in curved spaces, particularly if higher spins are involved. A brief introduction, useful for actual computations in AdS space, is given in Appendix A of Ref. [1010].

[40] In the present holographic approach the gluon field emerges as a constituent of a spin-2 field in AdS dual to the Pomeron in the 4-dimensional physical space (see Sec. 5.5.15).

Supersymmetric QM is based on a graded Lie algebra consisting of two anticommuting supercharges $Q$ and $Q^\dagger$, $\{Q, Q\} = \{Q^\dagger, Q^\dagger\} = 0$, which commute with the Hamiltonian $H = \frac{1}{2}\{Q, Q^\dagger\}$, $[Q, H] = [Q^\dagger, H] = 0$. If the state $|E\rangle$ is an eigenstate with energy $E$, $H|E\rangle = E|E\rangle$, then, it follows from the commutation relations that the state $Q^\dagger|E\rangle$ is degenerate with the state $|E\rangle$ for $E \neq 0$, but for $E = 0$ we have $Q^\dagger|E = 0\rangle = 0$, namely the zero mode has no supersymmetric partner [1042]; a key result for deriving the supermultiplet structure and the pattern of the hadron spectrum.

Following Ref. [1018] we consider the scale-deformed supercharge operator $R_\lambda = Q + \lambda S$, with $K = \frac{1}{2}\{S, S^\dagger\}$ the generator of special conformal transformations. The generator $R_\lambda$ is also nilpotent, $\{R_\lambda, R_\lambda\} = \{R_\lambda^\dagger, R_\lambda^\dagger\} = 0$, and gives rise to a new scale-dependent Hamiltonian $G$, $G = \frac{1}{2}\{R_\lambda, R_\lambda^\dagger\}$, which also closes under the graded algebra, $[R_\lambda, G] = [R_\lambda^\dagger, G] = 0$. The new supercharge $R_\lambda$ has the matrix representation

$$R_\lambda = \begin{pmatrix} 0 & r_\lambda \\ 0 & 0 \end{pmatrix}, \quad R_\lambda^\dagger = \begin{pmatrix} 0 & 0 \\ r_\lambda^\dagger & 0 \end{pmatrix}, \tag{5.5.11}$$

with $r_\lambda = -\partial_x + \frac{f}{x} + \lambda x$, $r_\lambda^\dagger = \partial_x + \frac{f}{x} + \lambda x$. The parameter $f$ is dimensionless and $\lambda$ has the dimension of $[M^2]$, and thus, a mass scale is introduced in the Hamiltonian without leaving the conformal group. The Hamiltonian equation $G|E\rangle = E|E\rangle$ leads to the wave equations

$$\left[-\frac{d^2}{dx^2} - \frac{1 - 4(f + \frac{1}{2})^2}{4x^2} + \lambda^2 x^2 + 2\lambda\left(f - \tfrac{1}{2}\right)\right]\phi_+ = E\phi_+, \tag{5.5.12}$$

$$\left[-\frac{d^2}{dx^2} - \frac{1 - 4(f - \frac{1}{2})^2}{4x^2} + \lambda^2 x^2 + 2\lambda\left(f + \tfrac{1}{2}\right)\right]\phi_- = E\phi_-, \tag{5.5.13}$$

which have the same structure as the Euler-Lagrange equations obtained from the AdS/CFT correspondence, but here, the form of the LF confinement potential, $\lambda^2 x^2$, as well as the constant terms in the potential are completely fixed by the superconformal symmetry [1021, 1022].

### 5.5.6 Light-front mapping and baryons

Upon mapping (5.5.12) and (5.5.13) to the semiclassical LF wave equations (5.5.8) and (5.5.9) using the substitutions $x \mapsto \zeta$, $E \mapsto M^2$, $f \mapsto L + \frac{1}{2}$, $\phi_+ \mapsto \psi_-$ and $\phi_- \mapsto \psi_+$, we find the expression $U^+ = \lambda^2\zeta^2 + 2\lambda(L+1)$ and $U^- = \lambda^2\zeta^2 + 2\lambda L$ for the confinement potential for baryons [1021]. The solution of the LF wave equations for this potential gives the eigenfunctions

$$\psi_+(\zeta) \sim \zeta^{\frac{1}{2}+L} e^{-\lambda\zeta^2/2} L_n^L(\lambda\zeta^2) \tag{5.5.14}$$

$$\psi_-(\zeta) \sim \zeta^{\frac{3}{2}+L} e^{-\lambda\zeta^2/2} L_n^{L+1}(\lambda\zeta^2) \tag{5.5.15}$$

with eigenvalues $M^2 = 4\lambda(n + L + 1)$. The polynomials $L_n^L(x)$ are associated Laguerre polynomials, where the radial quantum number $n$ counts the number of nodes in the wave function. We compare in Fig. 5.5.2 the model predictions with the measured values for the positive parity nucleons [476] for $\sqrt{\lambda} = 0.485$ GeV.

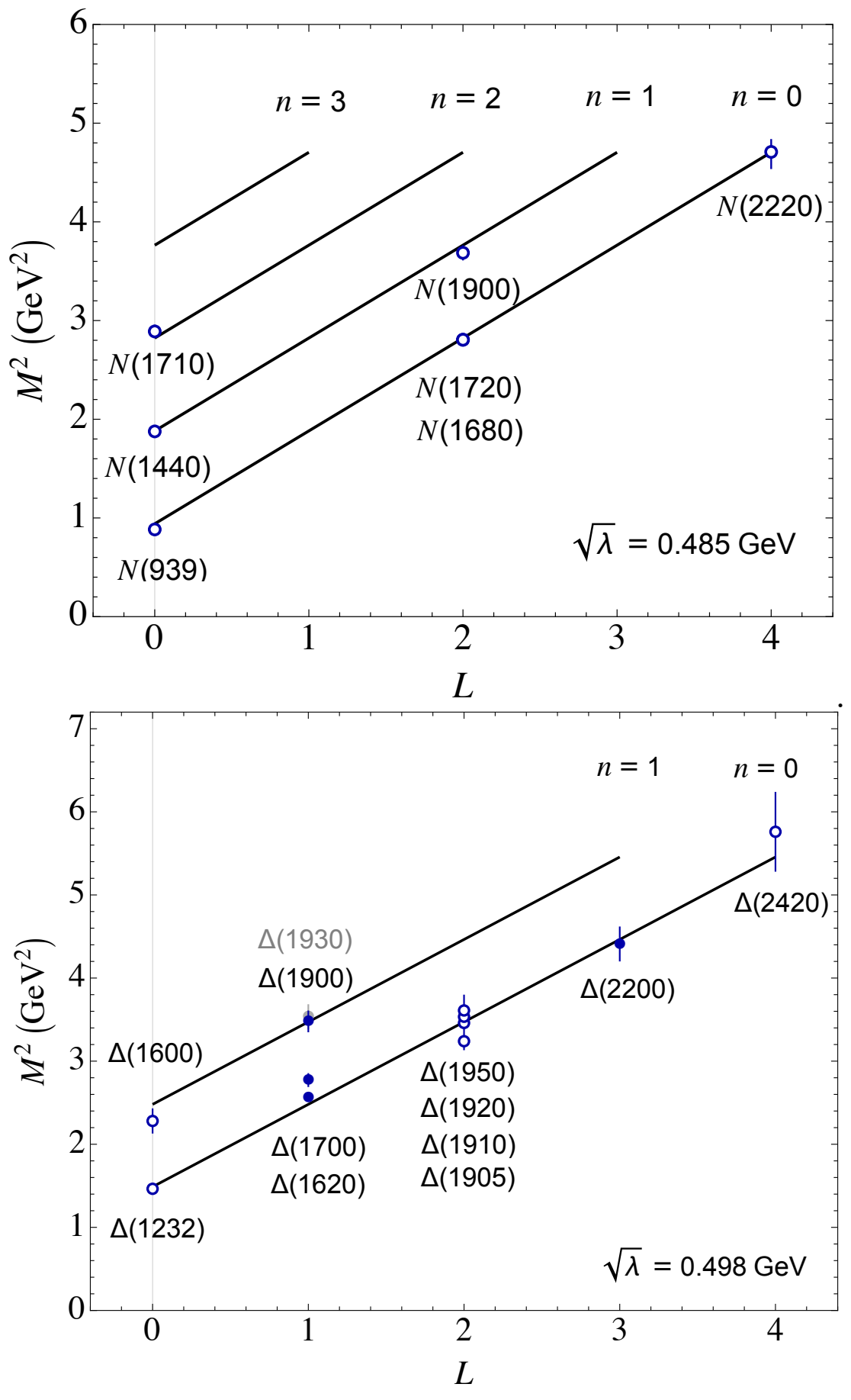

**Fig. 5.5.2** Model predictions for the orbital and radial positive-parity nucleons (up) and positive and negative parity $\Delta$ families (down) compared with the data from Ref. [476]. The values of $\sqrt{\lambda}$ are $\sqrt{\lambda} = 0.485$ GeV for nucleons and $\sqrt{\lambda} = 0.498$ GeV for the deltas.

### 5.5.7 Superconformal meson-baryon symmetry

Superconformal quantum mechanics also leads to a connection between mesons and baryons [1022] underlying the $SU(3)_C$ representation properties, since a diquark cluster can be in the same color representation as an antiquark, namely $\bar{3} \in 3 \times 3$. The specific connection follows from the substitution $x \mapsto \zeta$, $E \mapsto M^2$, $\lambda \mapsto \lambda_B = \lambda_M$, $f \mapsto L_M - \frac{1}{2} = L_B + \frac{1}{2}$, $\phi_+ \mapsto \phi_M$ and $\phi_2 \mapsto \phi_B$ in the superconformal equations (5.5.12) and (5.5.13). We find the LF meson (M) – baryon (B) bound-state equations

$$\left(-\frac{d^2}{d\zeta^2} - \frac{1-4L_M^2}{4\zeta^2} + U_M\right)\phi_M = M^2\,\phi_M, \qquad (5.5.16)$$

$$\left(-\frac{d^2}{d\zeta^2} - \frac{1-4L_B^2}{4\zeta^2} + U_B\right)\phi_B = M^2\,\phi_B, \qquad (5.5.17)$$

with the confinement potentials $U_M = \lambda_M^2\,\zeta^2 + 2\lambda_M(L_M - 1)$ and $U_B = \lambda_B^2\,\zeta^2 + 2\lambda_B(L_B + 1)$.

The superconformal structure imposes the condition $\lambda = \lambda_M = \lambda_B$ and the remarkable relation $L_M = L_B + 1$, where $L_M$ is the LF angular momentum between the quark and antiquark in the meson, and $L_B$ between the active quark and spectator cluster in the baryon. Likewise, the equality of the Regge slopes embodies the equivalence of the $3_C - \bar{3}_C$ color interaction in the $q\bar{q}$ meson with the $3_C - \bar{3}_C$ interaction between the quark and diquark cluster in the baryon. The mass spectrum from (5.5.16) and (5.5.17) is

$$M_M^2 = 4\lambda(n + L_M) \text{ and } M_B^2 = 4\lambda(n + L_B + 1). \qquad (5.5.18)$$

The pion has a special role as the unique state of zero mass and, since $L_M = 0$, it does not have a baryon partner.

AdS space is effectively modified in the IR by the dilaton profile in Eq. (5.5.5), while retaining conformal invariance in the UV (near the boundary of AdS space): It leads to the effective confinement potential $U(z)$ in Eq. (5.5.6). The dilaton profile can be determined from the superconformal algebra by integrating Eq. (5.5.6) for the effective potential $U$ in (5.5.16), $U(z) = \lambda^2 z^2 + 2\lambda(L-1)$. We obtain $\varphi(z) = \lambda z^2$. The dilaton is uniquely determined, provided that it depends only on the modification of AdS space [1043].

### 5.5.8 Spin interaction and diquark clusters

Embedding the LF bound-state equations in AdS space allows us to extend the superconformal Hamiltonian to include the spin-spin interaction, a problem not defined in the chiral limit by standard procedures. Since the dilaton profile $\varphi(z) = \lambda z^2$ is valid for arbitrary $J$, it leads to the additional term $2\lambda\mathcal{S}$ in the LF Hamiltonian for mesons and baryons, $G = \frac{1}{2}\{R_\lambda, R_\lambda^\dagger\} + 2\lambda\mathcal{S}$, which maintains the meson-baryon supersymmetry [1044]. The spin $\mathcal{S} = 0, 1$, is the total internal spin of the meson, or the spin of the diquark cluster of the baryon partner. The effect of the spin term is an overall shift of the quadratic mass,

$$M_M^2 = 4\lambda(n + L_M) + 2\lambda\mathcal{S}, \qquad (5.5.19)$$

$$M_B^2 = 4\lambda(n + L_B + 1) + 2\lambda\mathcal{S}, \qquad (5.5.20)$$

as depicted in Fig. 5.5.3 for the spectra of the $\rho$ mesons and $\Delta$ baryons by shifting one unit the value of $L_B$ [1022]. This shift leads to a degeneracy of meson and baryons states, a property known as the MacDowell symmetry [1045, 1046].

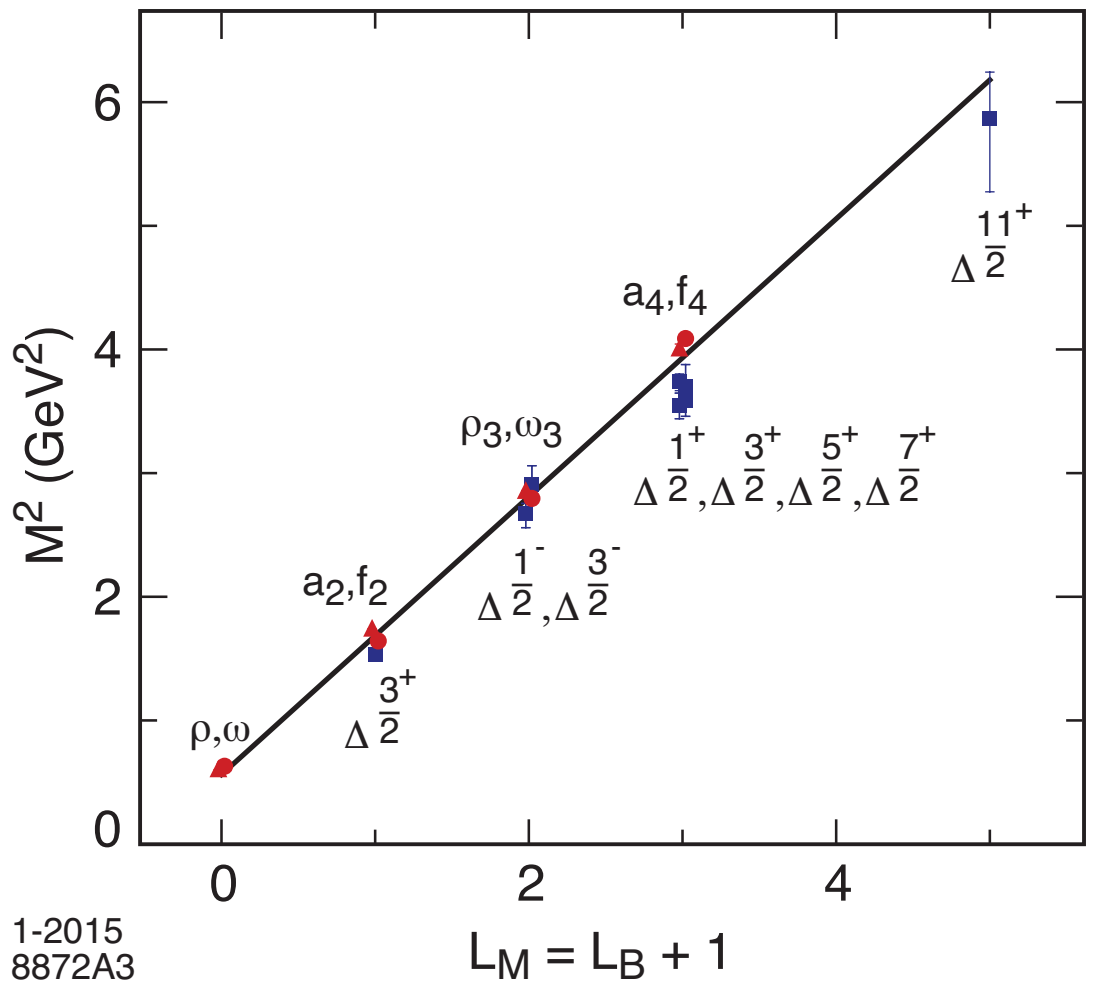

**Fig. 5.5.3** Supersymmetric vector meson and $\Delta$ partners from Ref. [1022]. The experimental values of $M^2$ from Ref. [476] are plotted vs $L_M = L_B + 1$ for $\sqrt{\lambda} \simeq 0.5$ GeV. The $\rho$ and $\omega$ mesons have no baryonic partner, since it would imply a negative value of $L_B$.

For the $\Delta$ baryons the total internal spin $S$ is related to the diquark cluster spin $\mathcal{S}$ by $S = \mathcal{S} + \frac{1}{2}(-1)^L$, and therefore, positive and negative $\Delta$ baryons have the same diquark spin, $\mathcal{S} = 1$. As a result, all the $\Delta$ baryons lie, for a given $n$, on the same Regge trajectory, as shown in Fig. 5.5.2. Plus parity nucleons are assigned $\mathcal{S} = 0$ and are well described by the holographic model as shown in Fig. 5.5.2. For negative parity nucleons both $\mathcal{S} = 0$ and $\mathcal{S} = 1$ are possible, but their precise comparison with data is not as successful as for the $\Delta$ baryons and positive parity nucleons.

### 5.5.9 Inclusion of quark masses and longitudinal dynamics

Finite quark masses break conformal invariance and pose a special challenge for all AdS/CFT approaches since the dual quantum field theory is inherently conformal. In the usual formulation of bottom-up holographic models one identifies quark mass and chiral condensates as coefficients of a scalar background field $X_0(z)$ in AdS space [1007, 1008]. A heuristic way to take into account the occurrence of quark mass terms, is to include the quark mass dependence in the invariant mass (IM) squared which controls the off-shell dependence of the LF wave function [1010, 1047]: $\chi_{\rm IM} = \mathcal{N} \exp\left(-\frac{1}{2\lambda}\sum_i \frac{m_i^2}{x_i}\right)$, where $\mathcal{N}$ is a normalization factor. This simple procedure leads, to a factorization of the transverse, $\phi(\zeta)$, and the longitudinal, $\chi(x)$, wave functions with the quadratic mass correction [1010, 1044, 1047] $\Delta M^2 = \langle\chi| \sum_i \frac{m_i^2}{x_i} |\chi\rangle$.

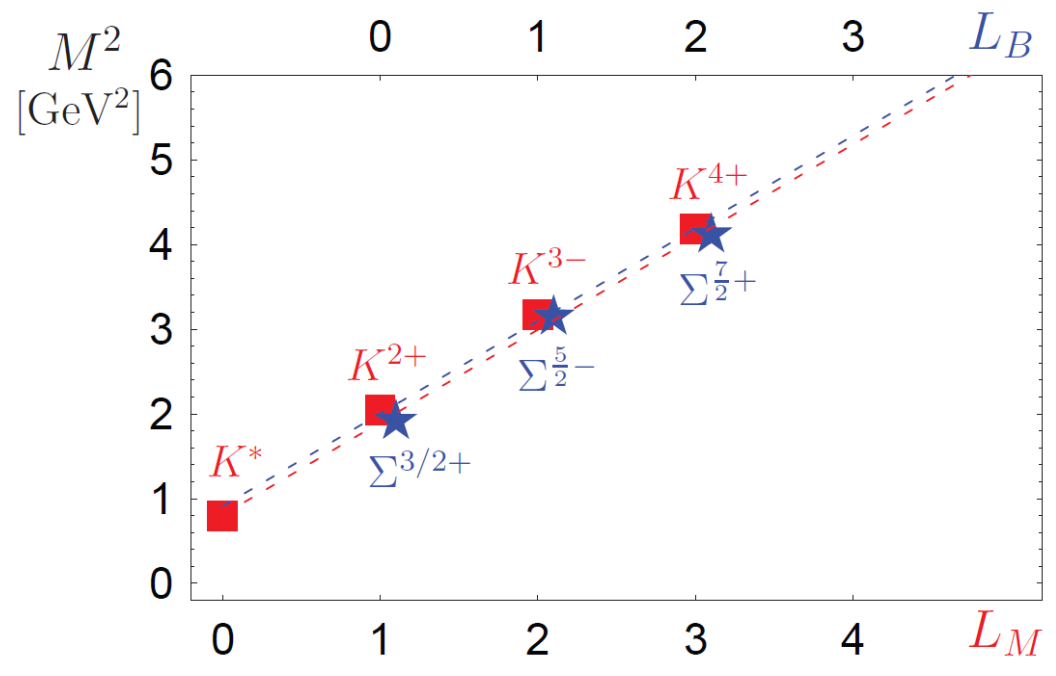

**Fig. 5.5.4** The $K^*$ and $\Sigma^*$ trajectories from supersymmetric HLFQCD in Ref. [1048] with $\sqrt{\lambda} = 0.51$ GeV. The error bars are smaller that the symbols in the figure and were not included.

The effective quark masses can be obtained by comparing the holographic results with the observed pseudoscalar masses. One obtains $m_q = 0.046$ GeV for the light quark mass and $m_s = 0.350$ GeV for the strange mass, with values between the Lagrangian and the constituent masses [1010, 1044, 1047]. The analysis has been consistently applied to the radial and orbital excitation spectra of the light meson and baryon families, giving the value $\sqrt{\lambda} = 0.523 \pm 0.024$ GeV [1044]. The comparison of the predicted $K^*$ and $\Sigma^*$ trajectories with experiment shown in Fig. 5.5.4 is a clear example of the validity of the supersymmetric meson-baryon connection including light quark masses. Starting with Ref. [1049], the application of the light-front holographic wave functions to diffraction physics has also been successful.

For heavy quarks the mass breaking effects are large. The underlying hadronic supersymmetry, however, is still compatible with the holographic approach and gives remarkable connections across the entire spectrum of light and heavy-light hadrons [1043, 1048]. In particular, the lowest mass meson of every family has no baryon partner, conforming to the SUSY mechanism. Compatibility with heavy quark symmetry [1043, 1048, 1050–1054] predicts a dependence of the holographic mass scale $\lambda$ on the quark mass.[41]

The extension of the LF holographic framework to incorporate longitudinal dynamics and chiral symmetry breaking, inspired in the original work of 't Hooft [1057], has recently attracted much interest [950, 989, 1055, 1058–1068]; however, in contrast with the transverse dynamics, the longitudinal confinement potential is not uniquely determined by the symmetries of the model.

**Table 5.5.1** Predicted masses for double heavy bosons from Ref. [1054]. Exotics which are predicted to be stable under strong interactions are marked by $^{(!)}$.

| quark content | $J^P$ | predicted Mass [MeV] | strong decay | threshold [MeV] |
|---|---|---|---|---|
| $cq\bar{c}\bar{q}$ | $0^+$ | 3660 | $\eta_c\pi\pi$ | 3270 |
| $cc\bar{q}\bar{q}^{(!)}$ | $1^+$ | 3870 | $D^*D$ | 3880 |
| $bq\bar{b}\bar{q}$ | $0^+$ | 10020 | $\eta_b\pi\pi$ | 9680 |
| $bb\bar{q}\bar{q}^{(!)}$ | $1^+$ | 10230 | $B^*B$ | 10800 |
| $bc\bar{q}\bar{q}^{(!)}$ | $0^+$ | 6810 | $BD$ | 7150 |

## 5.5.10 Completing the supersymmetric hadron multiplet

Besides the mesons and the baryons, the supersymmetric multiplet $\Phi = \{\phi_M, \phi_B^+, \phi_B^-, \phi_T\}$ contains a further bosonic partner, a tetraquark, which, follows from the action of the SUSY operator $R_\lambda^\dagger$ on the negative-chirality component of a baryon [1044]. A clear example is the SUSY positive parity $J^P$ multiplet $2^+, \frac{3}{2}^+, 1^+$ of states $f_2(1270)$, $\Delta(1232)$, $a_1(1260)$ where the $a_1$ is interpreted as a tetraquark.

Unfortunately, it is difficult to disentangle conventional hadronic quark states from exotic ones and, therefore, no clear-cut identification of tetraquarks for light hadrons, or hadrons with hidden charm or beauty, is possible [1044, 1053, 1069]. The situation is, however, more favorable for tetraquarks with open charm and beauty which may be stable under strong interactions and therefore easily identified [1070]. In Table 5.5.1, the computed masses from Ref. [1054] are presented. Our prediction [1054] for a doubly charmed stable boson $T_{cc}$ with a mass of 3870 MeV (second row) has been observed at LHCb a year later at 3875 MeV [1071], and it is a member of the positive parity $J^P$ multiplet $2^+, \frac{3}{2}^+, 1^+$ of states $\chi_{c2}(3565)$, $\Xi_{cc}(3770)$, $T_{cc}(3875)$. The occurrence of stable doubly beautiful tetraquarks and those with charm and beauty is well established, see also Ref. [1070].

41 For a relation with linear confinement see Refs. [1055, 1056].

### 5.5.11 Holographic QCD and Veneziano Amplitudes

The hadronic mass spectrum (5.5.18), which follows from the scale deformed superconformal equations (5.5.12) and (5.5.13), shows remarkable features which were essential ingredients to the pre-QCD physics of strong interactions. Starting from the $S$-matrix, Chew and Frautschi [1072] proposed to extend the concept of Regge trajectories [1073], $\alpha(t) = \alpha_0 + \alpha' t$, also to positive $t$-values. It led to a quadratic mass spectra, linear in the angular momentum, just as the spectra of Eq. (5.5.18). The analogy goes further: Veneziano [7] constructed a hadronic scattering amplitude

$$A(s,t) \sim B(1-\alpha(s), 1-\alpha(t)), \tag{5.5.21}$$

based on Euler's Beta function $B(u,v) = \frac{\Gamma(u)\Gamma(v)}{\Gamma(u+v)}$, which incorporates the duality in strong interactions [1074] and linear Regge trajectories. It is easy to see that this amplitude leads to particle poles at masses exactly matching Eq. (5.5.18), if one identifies the slope of the trajectory with the scale $\lambda$: $\alpha' = \frac{1}{4\lambda}$. In fact, from the analytic structure of the Beta function, particle poles appear at each value where $\alpha(t)$ is a negative integer. This leads to "Regge-daughter trajectories", which are identified with the radial excitations numbered by the integer $n$ in (5.5.18). But there is an important difference in the theoretical foundation: in the Veneziano approach, linear trajectories were assumed to exist, whereas here they are a consequence of the model, especially of the superconformal model, where the Regge intercept $\alpha_0$ is also determined, and expressed in terms of quark masses.

### 5.5.12 Electromagnetic form factors in holographic QCD

Holographic QCD incorporates important elements for the study of hadronic form factors, such as the connection between the twist of the hadron to the fall-off of its current matrix elements for large $Q^2$, and important aspects of vector meson dominance which are relevant at lower energies. The expression for the electromagnetic (EM) form factor (FF) in AdS space has been given by Polchinski and Strassler [1013]

$$F(Q^2) = \int \frac{dz}{z^3} V(Q^2, z)\, \Phi^2(z), \tag{5.5.22}$$

in their influential article describing deep inelastic scattering (DIS) using the gauge / gravity correspondence [42].

[42] For recent DIS studies examining various holographic QCD models see Ref [1075] and references therein.

It is written as the overlap of a normalizable mode $\Phi(z)$, representing a bound-state wave function in AdS for the initial and final states, with a non-normalizable solution $V(Q^2, z)$ of the wave equation (5.5.5) for a spin one conserved current in AdS, with $\mu = 0$ and $M^2 \to -Q^2$. The bulk-to boundary propagator, $V(Q^2, z)$ carries momentum $Q^2 = -t > 0$ from the external EM current. A precise mapping can be carried out to physical spacetime provided that the invariant transverse impact variable $\zeta$ for an arbitrary number of quarks is identified with the holographic variable $z$ [1014].

For the soft-wall model (SWM) of Ref. [1009] $\Phi_\tau(z) \sim z^\tau e^{-\lambda z^2/2}$, and $V(Q^2, z)$ is given in terms of the Tricomi function, $V(Q^2, z) \sim U(Q^2/4\lambda, 0, \lambda z^2)$. It corresponds to a conserved vector current with vanishing mass $\mu = 0$ in AdS. The result for the FF [1010] can be brought into the form of an Euler Beta function

$$F_\tau^{SWM}(t) \sim B(\tau - 1, 1 - t/4\lambda). \tag{5.5.23}$$

It generates a series of poles located at $M_n^2 = 4\lambda(n+1)$, and thus to the Regge intercept $\alpha_0 = 0$ [1076]. Therefore, one has to perform a pole shift [1010, 1026, 1027, 1077] in the expression (5.5.23) in order to bring the analytical structure of the FF in accordance with the spectra predicted by HLFQCD, which is in perfect agreement with observations. This shift leads to [1026]

$$F_\tau^{HLF}(t) \sim B\big(\tau - 1,\, 1/2 - t/4\lambda\big), \tag{5.5.24}$$

for the EM form factors in HLFQCD.

### 5.5.13 Form factors in dual models and holographic QCD

In a model extending the duality concept incorporated in Eq. (5.5.21) to reactions involving external currents, Ademollo and Del Giudice [1078], and Landshoff and Polkinghorne [1079], proposed a a Veneziano-like amplitude

$$F_\gamma(t) \sim B\big(\gamma, 1 - \alpha_\rho(t)\big), \tag{5.5.25}$$

in order to describe the electromagnetic FF; here $\alpha_\rho(t)$ is the Regge trajectory of the $\rho$ meson which couples to the quark current in the hadron, and the parameter $\gamma$ controls the rate of decrease of the FF. In fact, from Stirling's formula we find the asymptotic behavior $F_\gamma(Q^2) \sim \left(\frac{1}{Q^2}\right)^\gamma$ for large $Q^2 = -t$.

In LF QCD the parameter $\gamma$ has a well defined interpretation. To see this, we compare the asymptotic expression for $F_\gamma(Q^2)$ with the result from hard scattering counting rules at large $Q^2$ [134], $F_\tau(Q^2) \sim \left(\frac{1}{Q^2}\right)^{\tau-1}$, where the twist $\tau$ is the number of constituents $N$ in a given Fock component of the hadron. Thus, one has to choose in Eq. (5.5.25) $\gamma = \tau - 1$, in order to incorporate the scaling counting rule. This brings us to our final result for the analytical expression of the electromagnetic FF in the extended duality model [1026]

$$F_\tau(t) = \frac{1}{N_\tau} B\left(\tau - 1, 1 - \alpha(t)\right), \tag{5.5.26}$$

with $N_\tau = B(\tau - 1, 1 - \alpha(0))$, a remarkable expression which incorporates, at tree level, both the nonperturbative pole structure of the form factor and the hard scattering behavior.

For $\tau = N$, the number of constituents, the FF (5.5.26) is an $N - 1$ product of poles located at [1010]

$$-Q^2 = t = M_n^2 = \frac{1}{\alpha'}\big(n + 1 - \alpha(0)\big) > 0\,.$$

It generates the radial excitation spectrum of the exchanged vector mesons in the $t$-channel. For example, the $\rho$ trajectory has Regge intercept $\alpha_0 = 1/2$ and slope $\alpha \equiv 1/4\lambda$, with $\lambda \simeq (0.5 \text{ GeV})^2$. Thus $M_n^2 = 4\lambda(n + \frac{1}{2})$, corresponding to the $\rho$ vector meson and its radial excitations for $n = 0, 1, 2, \ldots, \tau - 2$ in agreement with Eq. (5.5.19). In general, the hadron wave function is a superposition of an infinite number of Fock components, and thus the full form factor should be written as a superposition $F(Q^2) = \sum_\tau C_\tau F_\tau(Q^2)$, with $\sum_\tau C_\tau = 1$, if all possible states are included. In practice, one expects a rapid convergence in the number of poles, with a dominant contribution from the $\rho$ vector meson plus contributions from the higher resonances $\rho'$, $\rho''$, ..., etc.

As a simple example, consider the valence contribution to the nucleon EM (spin non-flip) Dirac form factors by writing the flavor FFs as

$$F^u(t) = \left(2 - \frac{r}{3}\right) F_3(t) + \frac{r}{3} F_4(t), \tag{5.5.27}$$

$$F^d(t) = \left(1 - \frac{2r}{3}\right) F_3(t) + \frac{2r}{3} F_4(t), \tag{5.5.28}$$

where $F_\tau(Q^2)$ is given by (5.5.26). The holographic constraint of equal probability for nucleon states with LF orbital angular momentum $L$ and $L + 1$ (Sec. 5.5.4) determines the value $r = 3/2$, since the probability of the total quark spin along the plus $z$-direction for $L = 0$ (twist 3) should be identical to the probability of having total quark spin along the minus $z$-direction for $L = 1$. Actually, the values found in the recent analysis in Ref. [1027] deviate by $10 \sim 15$ % for the $u$-flavor FF and remain almost identical for the $d$ quark in the valence approximation. This leads to the results show in Fig. 5.5.5 for the nucleon isospin FF combination, $F^{I=0,1} = F_p(t) \pm F_n(t)$, where we compare the model predictions with the analysis of Ye et al. [1080]. Detailed studies show the importance of higher (large distance meson cloud) Fock components for the spin-flip Pauli FF [1077].

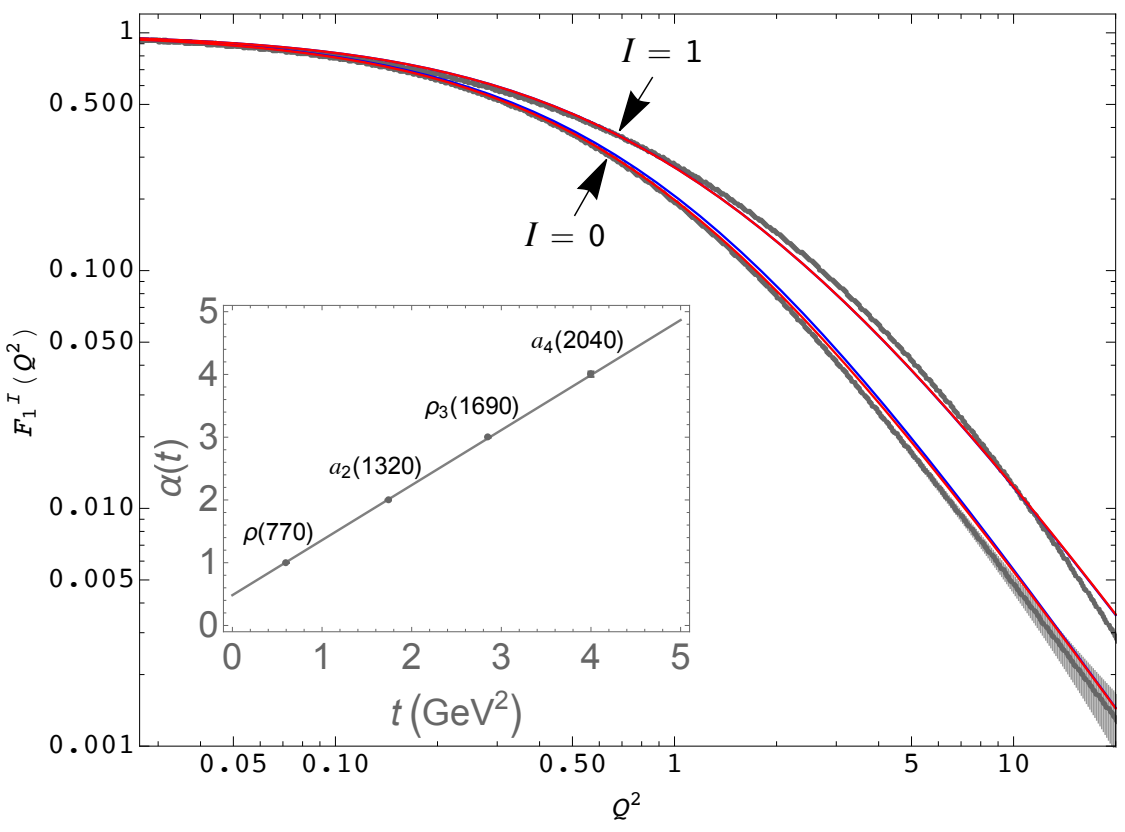

**Fig. 5.5.5** The LFHQCD prediction for the $I = 0, 1$ isospin combinations of nucleon factors compared with the $z$-expansion data analysis of Ye et al. [1080] (grey band): (blue line) valence contribution only, (red line) including $u\bar{u}$ and $d\bar{d}$ pairs. The inset from Ref. [1081] represents the $\rho$ Regge trajectory in Eq. (5.5.26) for $\sqrt{\lambda} = 0.534$ GeV and $\alpha_\rho(0) = 0.483$.

### 5.5.14 Quark distribution functions and the exclusive-inclusive connection

The mathematical structure of the Veneziano-type FFs (5.5.26), not only incorporates the hard scattering amplitude's dependence on twist, but it also gives important insights into the structure of the parton distributions since it becomes possible to include the Regge behavior at small values of $x$, as well as the exclusive-inclusive connection [966, 1082] at large values of the longitudinal momentum $x$ [1026]. In fact, the relation between the behavior of the structure function near $x = 1$ with the falloff of the FF at large $t$, described in the article of Landshoff and Polkinghorne [1079], is very close to the Drell-Yan "exclusive-inclusive" connection, also formulated in 1970 [966].

Using the integral representation of the Beta function, the FF (5.5.26) can be expressed in a reparametrization invariant form

$$F(t)_\tau = \frac{1}{N_\tau} \int_0^1 dx\, w(x) w(x)^{-\alpha(t)} \left[1 - w(x)\right]^{\tau - 2}. \tag{5.5.29}$$

The trajectory $\alpha(t)$ of the vector current can be computed within the superconformal LF holographic framework, and the intercept, $\alpha(0)$, incorporates the quark masses [1021, 1022]. The function $w(x)$ is a flavor-independent function with $w(0) = 0$, $w(1) = 1$ and $w'(x) \geq 0$.

The flavor FF can be written in terms of its generalized parton distribution (GPD) [1083–1085], $H^q(x,t) \equiv H^q(x,\xi=0,t)$, at zero skewness, $\xi$,

$$
\begin{aligned}
F^q(t) &= \int_0^1 dx\, H^q(x,t) \\
&= \int_0^1 dx\, q(x) \exp\left[t f(x)\right], \qquad (5.5.30)
\end{aligned}
$$

with the profile function, $f(x)$, and the particle distribution function (PDF), $q_\tau(x)$, both determined by $w(x)$:

$$f(x) = \frac{1}{4\lambda} \log\left(\frac{1}{w(x)}\right), \qquad (5.5.31)$$

$$q_\tau(x) = \frac{1}{N_\tau} w'(x) w(x)^{-\alpha(0)} [1 - w(x)]^{\tau-2}, \qquad (5.5.32)$$

with $\alpha = 1/4\lambda$. Boundary conditions follow from the Regge behavior at $x \to 0$, $w(x) \sim x$, and at $x \to 1$ from the exclusive-inclusive counting rule [966, 1082], $q_\tau(x) \sim (1-x)^{2\tau-3}$, which fixes $w'(1) = 0$. A simple ansatz for $w(x)$, $w(x) = x^{1-x} \exp\big(-a(1-x)^2\big)$, fulfills all conditions mentioned above. The flavor independent parameter $a$ has the value $a \simeq 0.5$ [1026].

Using the expression (5.5.30) at $t = 0$ and Eqs. (5.5.27-5.5.28), we obtain for the unpolarized quark distributions in the valence approximation

$$u_v(x) = \left(2 - \frac{r}{3}\right) q_3(x) + \frac{r}{3} q_4(x), \qquad (5.5.33)$$

$$d_v(x) = \left(1 - \frac{2r}{3}\right) q_3(x) + \frac{2r}{3} q_4(x), \qquad (5.5.34)$$

with normalization $\int dx\, u_v(x) = 2$ and $\int dx\, d_v(x) = 1$. The PDF $q_\tau(x)$ is given by (5.5.32) and $r = 3/2$. Our PDF results for the nucleon, Eqs. (5.5.33-5.5.34), and for the pion [1026], are compared with the global data analysis in Fig. 5.5.6. If the reparametrization function $w(x)$ is fixed by the nucleon PDFs, then the pion PDF is a prediction. pQCD evolution is performed from an initial scale determined from $\mu_0 \simeq 1$ GeV from the soft-hard matching procedure described in Ref. [1091]. Our result for the pion PDF in Fig. 5.5.6 is in good agreement with the data analysis in Ref. [1088] and consistent with the nucleon global fit through the GPD universality introduced in [1026]. It leads to a $1-x$ falloff, in contrast with the $(1-x)^2$ pQCD result at large-$x$ [1089, 1092], an issue much debated recently [845, 1093, 1094].

An analysis of the polarized quark distribution in the proton has been performed in Ref. [1027], assuming the Veneziano-type FF (5.5.26), with the separation of chiralities from the axial current. The model predictions for the ratio of polarized to unpolarized quark

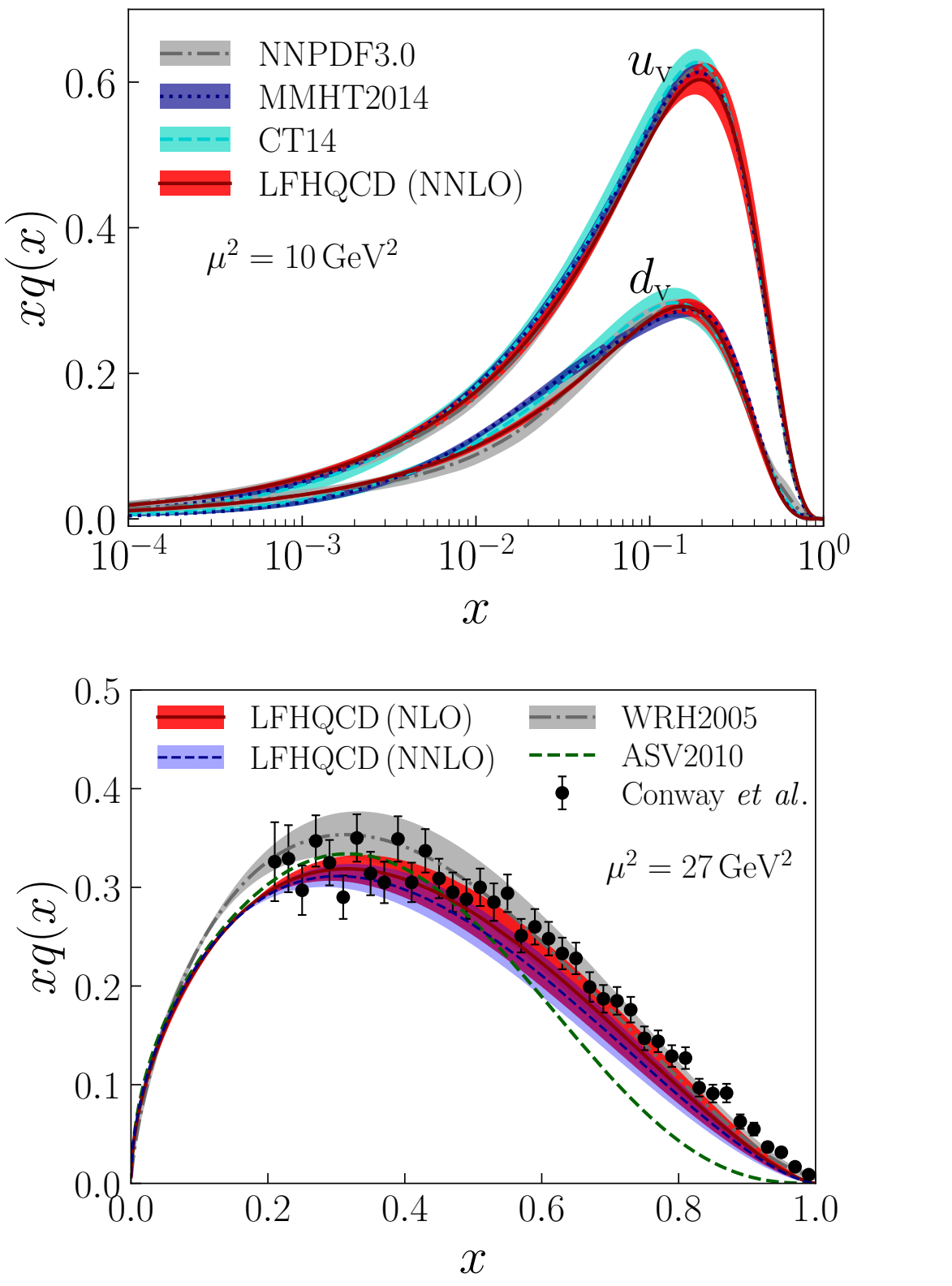

**Fig. 5.5.6** Comparison for $xq(x)$ in HLFQCD with global fits from [1026]. Up: proton valence approximation (red band). Data analysis from MMHT2014 (blue bands) [981], CT14 [1086] (cyan bands), and NNPDF3.0 (grey bands) [1087]. Down: pion results (red and light blue bands). NLO global fits from [1088, 1089] (gray band and green curve) and the LO data extraction [1090]. HLFQCD results are evolved from the initial scale $\mu_0 \simeq 1$ GeV at NLO and NNLO.

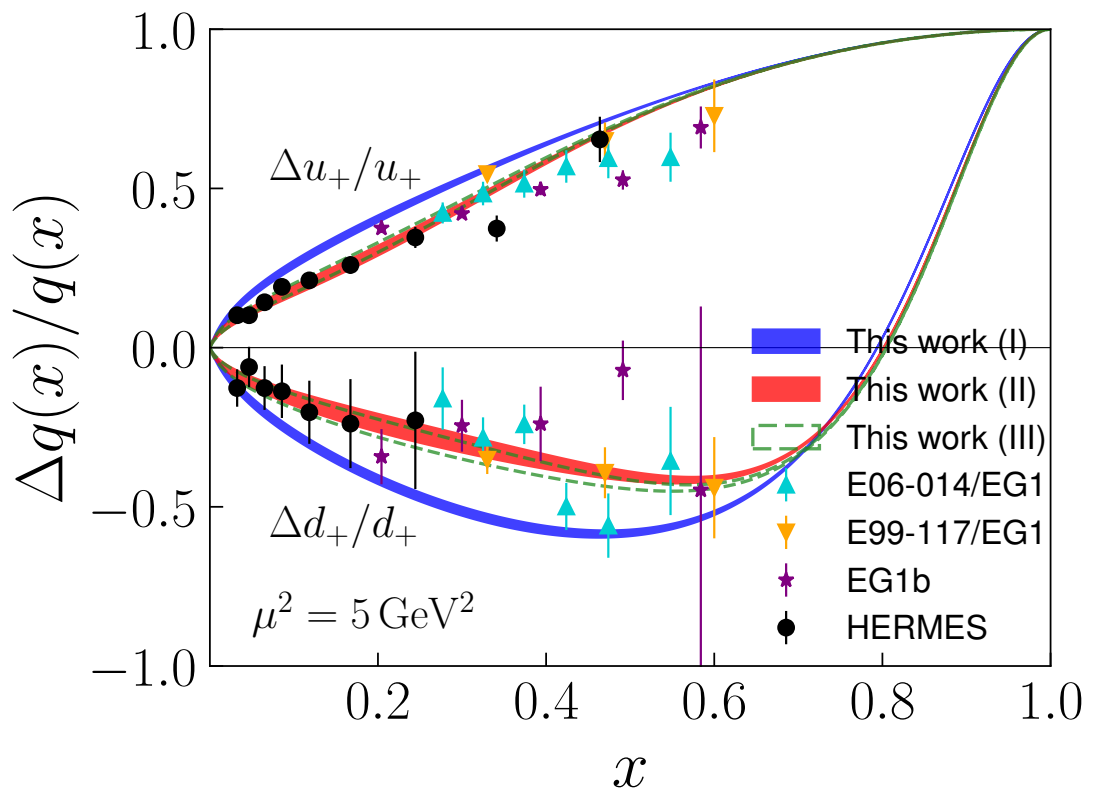

**Fig. 5.5.7** HLFQCD predictions from Ref. [1027] for the quark helicity asymmetry ratio $\Delta q_+/q_+$, $q_+ = q + \bar{q}$, are compared with existing data. The blue band is the valence contribution, the red band includes $q\bar{q}$ components and the dashed green band also includes the intrinsic sea contribution.

distribution functions is compared with available data in Fig. 5.5.7.

Another application of the LF holographic ideas is the computation of the intrinsic charm-anticharm asymmetry in the proton [1095], $c(x)-\bar{c}(x) = \sum_\tau c_\tau\big(q_\tau(x) - q_{\tau+1}(x)\big)$, with $\int_0^1 dx[c(x) - \bar{c}(x)] = 0$. The normalization of the charm form factor was computed using lattice QCD [1095], and the $J/\Psi$ trajectory in the GPDs from HLFQCD and heavy quark effective theory [1053]. A similar procedure was used to determine the intrinsic strange-antistrange asymmetry in the proton with the Regge trajectory in the holographic expressions corresponding to the $\phi$ meson current [1081], and most recently to study color transparency in nuclei [1096] (see Sec. 5.10), and to model the EMC effect in various nuclei [1097].

### 5.5.15 Gravitational form factors, gluon distributions and the Pomeron trajectory

Gravitational form factors (GFFs) are the hadronic matrix elements of the energy momentum tensor (EMT) and describe the coupling of a hadron to the graviton, thus providing key information on the dynamics of quarks and gluons within hadrons. In holographic QCD Pomeron exchange is identified as the graviton of the dual AdS theory [1098–1103]. The Pomeron couples as a rank-two tensor to hadrons and interacts strongly with gluons. Since we are interested in obtaining the intrinsic gluon distribution in the nucleon, we use the soft Pomeron of Donnachie and Landshoff [1104] with the Regge trajectory $\alpha_P(t) = \alpha_P(0) + \alpha'_P t$, with intercept $\alpha_P(0) \simeq 1.08$ and slope $\alpha_P \simeq 0.25\,\mathrm{GeV}^{-2}$ [476].

To actually compute the GFF one considers the perturbation of the gravity action by an arbitrary external source at the AdS asymptotic boundary which propagates inside AdS space and couples to the EMT [1016, 1105]. In analogy to the EM FF (5.5.22), the spin non-flip GFF $A(t)$ is written as the overlap of a normalizable mode $\Phi(z)$, representing a bound-state wave, with a non-normalizable mode $H(Q^2, z)$, the bulk-to-boundary propagator, corresponding to the gravitational current in AdS. We obtain [1016, 1105]

$$A(t) = \int \frac{dz}{z^3}\, H(Q^2, z)\, \Phi^2(z). \tag{5.5.35}$$

For the soft-wall profile introduced in Ref. [1009], the propagator in AdS, $H(Q^2, z)$, is also given by a Tricomi function [1010, 1105], $H(Q^2, z) \sim U\left(Q^2/4\lambda_g, -1, \lambda_g z^2\right)$. The effective physical scale $\lambda_g$ is the scale of the Pomeron, $\lambda_g = 1/4\alpha'_P \simeq 1\,\mathrm{GeV}^2$, which couples to the constituent gluon over a distance $\sim 1/\sqrt{\alpha_P} \sim \sqrt{4\lambda_g}$, described by

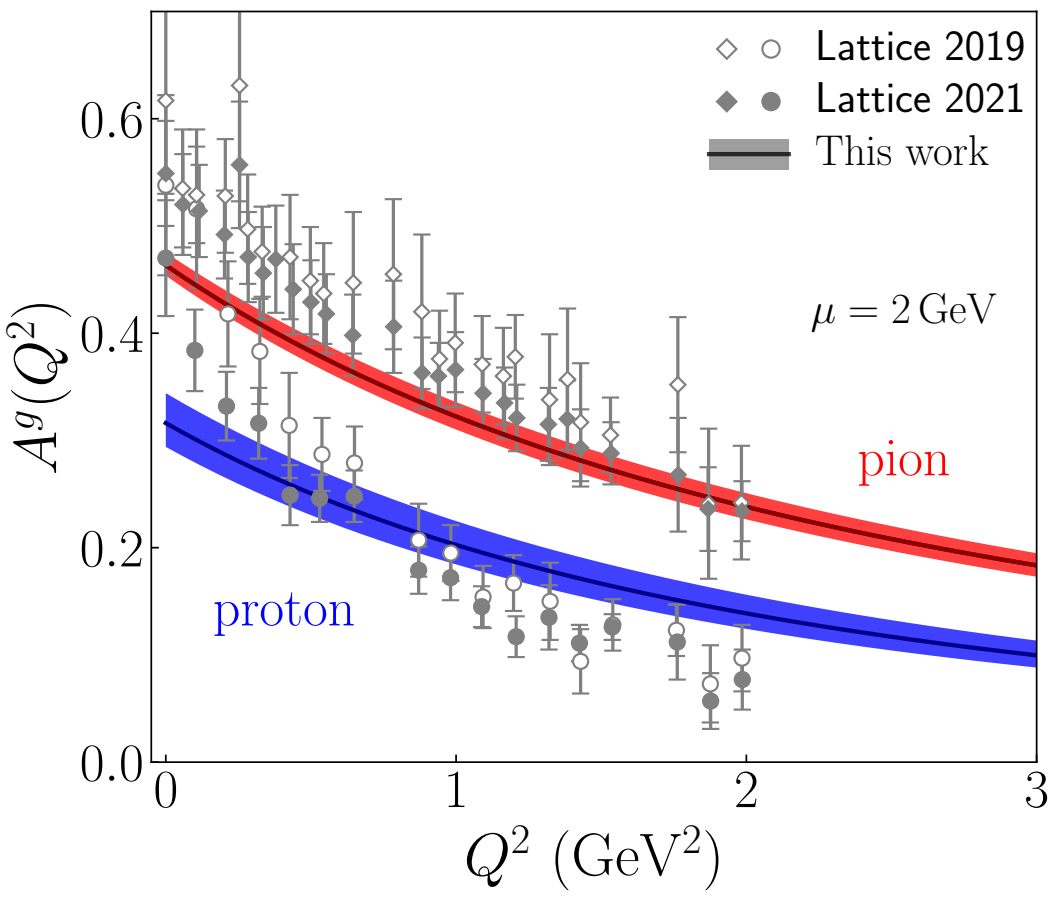

**Fig. 5.5.8** Gluon gravitational form factor $A^g(Q^2)$ of the proton (blue) and the pion (red) in comparison with lattice QCD computations [1108, 1109]. The value $A^g(0)$ corresponds to the momentum fraction carried by gluons at the scale $\mu = 2\,\mathrm{GeV}$. The bands indicate the uncertainty of $\lambda_g$ by $\pm 5\%$ and the normalization from the momentum sum rule.

the wave function $\Phi^g_\tau(z) \sim z^\tau e^{-\lambda_g z^2/2}$. Our final result is [1028]

$$A^g_\tau(Q^2) = \frac{1}{N_\tau} B\left(\tau - 1, 2 - \alpha_P(Q^2)\right), \tag{5.5.36}$$

with $N_\tau = B\left(\tau - 1, 2 - \alpha_P(0)\right)$. As for the EM FF, in writing (5.5.36) we have also shifted the Pomeron intercept to its physical value $\alpha_P(0) \approx 1$, since the holographic result (5.5.35) leads to a zero intercept. For integer twist, the GFF (5.5.36) is expressed as a product of $\tau - 1$ timelike poles located at

$$-Q^2 = M_n^2 = \frac{1}{\alpha'_P}\left(n + 2 - \alpha_P(0)\right), \tag{5.5.37}$$

the radial excitation spectrum of the spin-two Pomeron. The lowest state in this trajectory, the $2^{++}$, has the mass $M \simeq 1.92$ GeV, compared with the lattice results $M \simeq (2.15 - 2.4)$ GeV [476] [43]. We notice that Eq. (5.5.36) is the Veneziano amplitude of the FF for a spin-two current [1078, 1079].

The lowest twist contributions to the GFF corresponds to the $\tau = 4$ Fock state $|uudg\rangle$ in the proton and the $\tau = 3$ component $|u\bar{d}g\rangle$ in the pion, both containing an intrinsic gluon. The results for $A^g(t)$ are compared in Fig. 5.5.8 with recent lattice computations. We find for the gluon mass squared radius $\langle r_g^2 \rangle_p = 2.93/\lambda_g = (0.34\,\mathrm{fm})^2$ for the proton and $\langle r_g^2 \rangle_\pi = 2.41/\lambda_g = (0.31\,\mathrm{fm})^2$ for the pion. The model predictions in Fig. 5.5.8 have no free parameters [1028].

[43] There exist many computations of glueballs in top-down holographic models, see for example, [1106]; and also in bottom-up models. For a recent computation, see for example [1107], and references therein.

The intrinsic gluon distributions in the proton and the pion can be determined from the gravitational form factor (5.5.36) following the same procedure used in Sec. 5.5.14. The results are given in [1028] and agree very well with the data analysis from [626, 627, 979, 1110, 1111]. The model uncertainties for large $x$-values are smaller than those from the phenomenological analysis.

By using the gauge/gravity a simultaneous description of the BFKL hard Pomeron [140, 218, 1112] and the soft Regge domain has been proposed in Ref. [1098]. This model, however, did not solve the problem of the large difference of intercept values between both Pomerons. Using the scale dependence of the gluon distribution functions, our results give strong support to a single Pomeron with a scale dependent intercept [1113], which was proposed in Ref. [1114] in order to explain the diffractive scattering data at LHC energies [1115, 1116].

### 5.5.16 Summary

Holographic light front QCD is a nonperturbative analytic approach to hadron physics. It originates from the precise mapping of light front expressions of form factors in AdS space for an arbitrary number of partons [1014]. The holographic embedding in AdS also leads to semiclassical relativistic wave equations, similar to the Schrödinger equation in atomic physics, for arbitrary integer or half-integer spin [1011, 1036]. The model embodies an underlying superconformal algebraic structure responsible for the introduction of a mass scale within the superconformal group, and determines the effective confinement potential: It is not SUSY QCD. There is a zero eigenmode which is identified with the pion: It is massless in the chiral limit. The new framework leads to relations between the Regge trajectories of mesons, baryons, and tetraquarks. It also incorporates features of pre QCD, such as Veneziano model and Regge theory. Further extensions incorporate the exclusive-inclusive connection in QCD and provide nontrivial relations between hadron form factors and quark and gluon distributions. Measurements of the strong coupling in the nonperturbative domain [1117] are remarkably consistent with the predicted form in holographic QCD [158], a relevant issue in QCD which is discussed in the next section 5.6. Holographic light front QCD has led to significant advances in understanding hadron phenomena by incorporating emerging QCD properties in an effective computational framework of hadron structure.

## 5.6 The nonperturbative strong coupling

**Alexandre Deur**

The perturbative framework of QCD (pQCD) has been remarkably successful in describing the interactions between the fundamental constituents of hadrons in high energy experiments, thus establishing QCD as the theory of the strong force at small distances [278]. Most of nature's strong force phenomena, however, are governed by large-distance, nonperturbative physics [748, 1118–1122] where the methods of pQCD are not applicable. The Landau pole at low-energies in the running of the QCD coupling is an example of the expected failure of perturbation theory as the coupling increases. A nonperturbative treatment is necessary and allows us to define renormalization scheme dependent coupling constants.

Studying $\alpha_s(\mu)$ at low energy has been challenging: not only do nonperturbative calculations represent a difficult problem to solve, but more generally, we only know in the pQCD framework how to relate the $\alpha_s$ calculated in different schemes. Worst, there is no obvious prescription of how to define the coupling. One reason why a variety of definitions is possible is that $\alpha_s(\mu)$ need not be an observable. In fact, in most approaches –including the standard pQCD treatment– it is not an observable. For example $\alpha_s^{\rm pQCD}$ depends on the choice of renormalization scheme, generally taken to be $\overline{\rm MS}$. Such arbitrary dependence on a human convention shows that $\alpha_s(\mu)$ is not an observable. In addition, the quark-gluon, 3-gluon, 4-gluon or ghost-gluon vertices may have different couplings [44] i.e., several couplings, with distinct magnitudes and $\mu$-dependence, may be necessary to characterize QCD. This happens because the Slavnov-Taylor Identities (STI) [1123, 1124], the QCD equivalent of QED's Ward-Takahashi relation [1125, 1126], may not hold under certain choices of gauges and renormalization schemes, such as the MOM scheme. With the oft-used $\overline{\rm MS}$ scheme, the STI hold, *viz,*

$$\alpha_s^{\rm qg,\overline{MS}} = \alpha_s^{\rm 3g,\overline{MS}} = \alpha_s^{\rm 4g,\overline{MS}} = \alpha_s^{\rm gh,\overline{MS}}$$

but $\overline{\rm MS}$ is not practical for nonperturbative methods such as Lattice QCD and in the nonperturbative domain, the difference between

$$\alpha_s^{\rm qg,MOM}, \alpha_s^{\rm 3g,MOM}, \alpha_s^{\rm 4g,MOM}, \alpha_s^{\rm gh,MOM}$$

[44] When needed, we will use superscripts to qualify the coupling. For examples, $\alpha_s^{\rm pQCD,\overline{MS}}$ is the perturbative coupling in the $\overline{\rm MS}$ scheme, or $\alpha_s^{\rm qg,MOM}$, $\alpha_s^{\rm 3g,MOM}$, $\alpha_s^{\rm 4g,MOM}$ and $\alpha_s^{\rm gh,MOM}$ are the couplings for the quark-gluon, 3-gluon, 4-gluon or ghost-gluon vertices, respectively, computed in the MOM scheme.

is conspicuous. A wholly different approach is to define $\alpha_s(\mu)$ to be an observable [146], in analogy with the observable QED coupling $\alpha$ [1127], but while this circumvents the issues of breaking the STI and of scheme and gauge dependence, the prescription is rarely used in pQCD.

Many definitions of $\alpha_s$ have been considered, resulting in a range of values of $\alpha_s(\mu \ll \Lambda_s)$ from 0 to $\infty$, generating much confusion. Adding to this is the fact that, unlike the high-energy domain where pQCD rules, there is no obviously superior method to study the nonperturbative behavior of $\alpha_s(\mu)$. This is, of course, due to the challenge of solving QCD nonperturbatively. All major non-perturbative approaches, See Secs. 4, 5.3, 5.5 have been used (with the conspicuous exception of chiral effective field theory, Sec. 6.2, since its hadronic degrees of freedom do not couple with $\alpha_s$) as well as many models. These methods using different type of approximations, and the models being not directly based on QCD's Lagrangian or its symmetries, results have often differed. Yet a number of studies have converged toward a fruitful definition of $\alpha_s(\mu)$ which allows us to account for low energy phenomenology [1128]. Before describing it, we will first recall in broad brushstrokes the history of this endeavor, referring only to pioneering attempts and not the important body of subsequent works that clarified and refined these attempts.

Soon after the advent of QCD, it was realized that $\alpha_s(\mu)$ may display a plateau when $\mu \to 0$ (it is said to *freeze* at low $\mu$) [1129–1131], *viz*, the $\beta$ function of QCD, Eq. (1.2.6) may obey $\beta(\mu \to 0) \to 0$. The actual freezing value $\alpha_s(0)$ was debated and ranged from typically 0.5 to 5 [1128]. A pioneering and influential work in this context is due to Cornwall [1132] who used the Dyson-Schwinger equations (DSE), the gluon self-energy and initiated a method (the *Pinch* technique, PT) that allows to obtain gauge-independent results. The ensuing coupling $\alpha_s^{\mathrm{gse,PT}}$ displays a freezing behavior in qualitative agreement with quark models (e.g., Ref. [735] and Sec. 5.1) and quarkonium spectrum models (e.g., Ref. [82]).

A freezing of $\alpha_s(\mu)$ was by no mean the only proposal: others reasoned that it should diverge as $1/\mu^2$ [1133], that it should monotonically increase with $1/\mu$, but without diverging [1134], or that it should vanish as $\mu \to 0$ [153, 1135]. In all these case, $\beta(\mu \to 0) \neq 0$. As we alluded to, multiple reasons caused these widely varying expectations [1128]: differences in the basic definition of $\alpha_s(\mu)$; choice of vertex used to compute it; calculation artifacts from approximations (e.g., discretization in lattice QCD or truncation prescription for the DSE and other functional methods); choice of gauge and renormalization scheme; [45] or the existence of multiple solutions to the QCD equations providing $\alpha_s(\mu)$ without a decisive argument on which one is realized in nature. A prominent example is the *decoupling* [1135] and the *scaling* [1138] solutions that yield a vanishing or a freezing $\alpha_s^{\mathrm{gh,MOM}}$, respectively. Functional methods and lattice QCD have produced both solutions, albeit well-controlled lattice calculations appear to yield only the decoupling solution. In these calculations $\alpha_s^{\mathrm{gh}}$, called the *Taylor coupling* [1123], is most often used because it is the simplest coupling that can be computed from QCD correlation functions.

It is generally believed, after much discussion, that the decoupling solution is the one realized in nature, which simply means that in the particular gauges where ghost fields appear, the gluon and ghost fields decouple at low $\mu$. This is an important finding regarding the behavior of gluons and ghosts but it does not directly illuminate the strength of the strong force at low energy. Besides using correlation functions, other prevalent approaches to define $\alpha_s(\mu)$ are effective charges [146] and analytic approaches [160, 1134] –both methods promote $\alpha_s(\mu)$ to an observable quantity– or direct use of phenomenology, for example, using constituent quark models, the Q–$\overline{\mathrm{Q}}$ potential or the hadronic spectrum [82, 735, 1133, 1139–1142]. Like the DSE that must choose a truncation prescription, and lattice QCD with finite size lattices, discretization or pion mass approximations, the other methods also use approximations or/and include model-dependencies whose effects are often not well controlled. The synergy arising from methods with very distinct approximations allowed for a better understanding of the latter.

After many studies and developments, of which the aforesaid narrative is too a laconic cartoon, a coupling was identified and computed using a formalism guarantying that the STI hold in the nonperturbative domain [1143]. Therefore, QCD is here characterized by a single coupling, independent of the choice of vertex or process used to define it (process-independent, PI). In addition, the Pinch technique [1132] is used to guaranty gauge-independence. The calculation, using either the DSE or lattice QCD results on correlation functions, yields a coupling $\alpha_s^{\mathrm{PI,Pinch}}$ in agreement with the phenomenological coupling [1117, 1145, 1146] derived from the Bjorken sum rule [23] using the effective charge (EC) method [146], $\alpha_s^{\mathrm{EC,g_1}}$, and with $\alpha_s^{\mathrm{AdS/QCD}}$ obtained using AdS/QCD [158, 1091], See Sec. 5.5.

The latter is derived starting from the observation that for strongly coupled systems with a gravity dual,

[45] This issue has become negligible with the developments of methods removing renormalization scale ambiguities [1136, 1137].

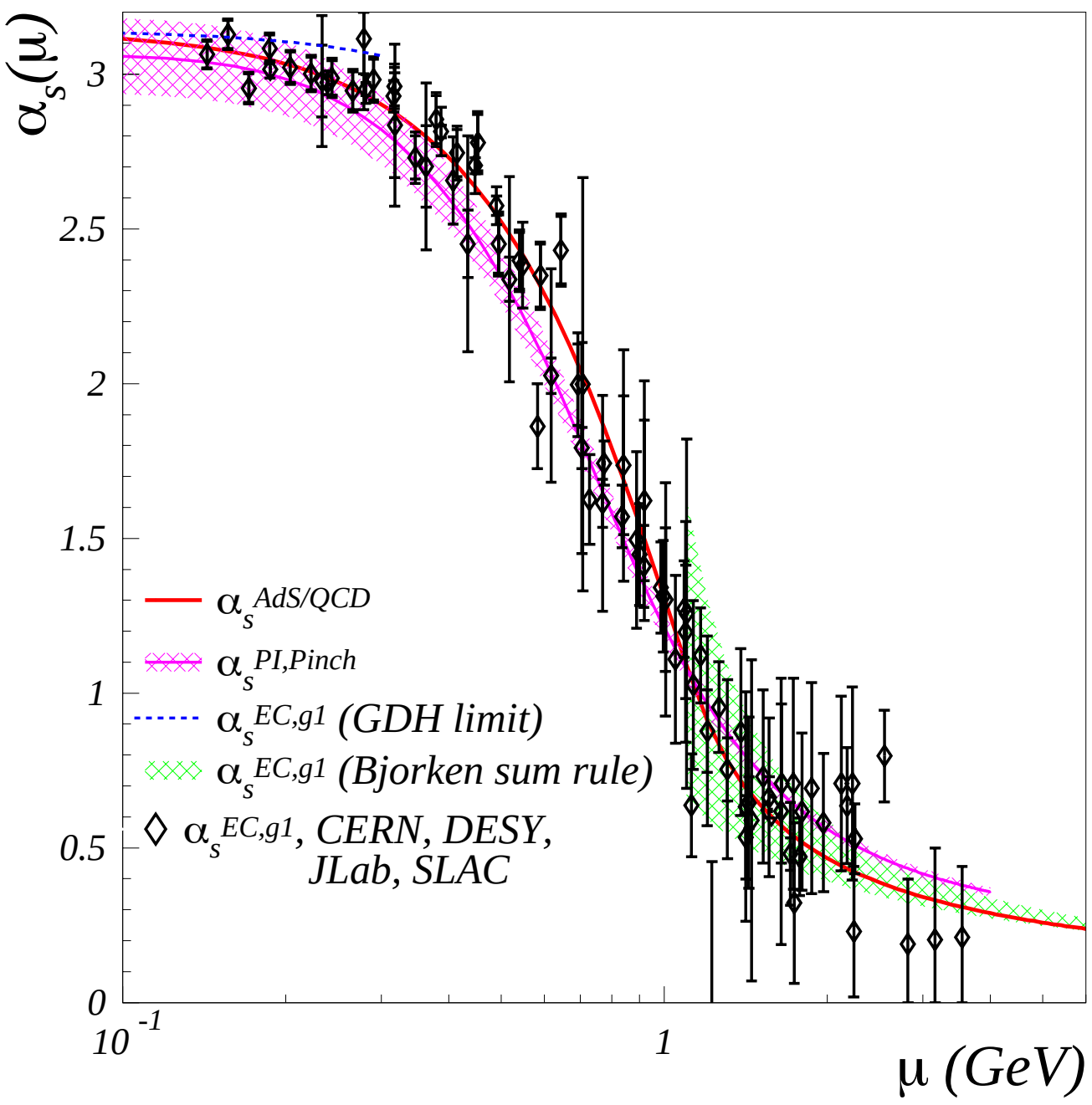

**Fig. 5.6.1** Nonperturbative strong couplings calculated with the holographic QCD framework ($\alpha_s^{\mathrm{AdS/QCD}}$, red line) [158], and Dyson-Schwinger formalism using the lattice determinations of correlations functions ($\alpha_s^{\mathrm{PI,Pinch}}$, magenta band) [1143, 1144]. The experimental extractions of $\alpha_s^{\mathrm{EC,g_1}}$ [1117, 1145, 1146] following the effective charge definition [146] are shown by the symbols. The green band and the dashed line is $\alpha_s^{\mathrm{EC,g_1}}$ deduced from the Bjorken [23] and Gerasimov-Drell-Hearn [1147, 1148] sum rules, respectively.

the radial direction in the bulk can be associated with the energy scale of the boundary theory [1149]: Large values of $\mu$ correspond to small values of $z$ near the high-momentum conformal boundary of AdS, $\mu \sim 1/z$. Conversely, large-$z$ distances in the low-momentum region of AdS correspond to low energy scales in the physical theory. The dilaton factor $\exp(\varphi(z))$ is a measure of the departure from conformality at the asymptotic AdS boundary, $z \to 0$, and should grow for large values of $z$, signaling confinement: It acts as an effective coupling in AdS space. We can use the procedure introduced in [158] to obtain the $\mu$ dependence $\alpha_s^{\mathrm{AdS/QCD}}$ from the Hankel transform of the dilaton factor [158]

$$\alpha_s^{\mathrm{AdS/QCD}}(\mu) \sim \int_0^\infty z dz J_0(z\mu) e^{-\lambda z^2} \sim e^{-\mu^2/4\lambda}, \tag{5.6.1}$$

where the overall normalization is not provided within AdS/QCD. The freezing value of the effective coupling $\alpha_s^{\mathrm{EC,g_1}}(0) = \pi$ is used. The dilaton profile $\lambda z^2$ is determined by the superconformal structure (Sec. 5.5.7). The transition between the predicted Gaussian form (5.6.1) and the log behavior expected from pQCD is determined from the matching of the perturbative and nonperturbative couplings and their derivatives for $\sqrt{\lambda} = 0.534$ GeV. The specific matching allows us to determine the perturbative QCD scale $\Lambda$ in terms of the hadronic mass scale $\sqrt{\lambda}$ [1150] for any choice of renormalization scheme, including the $\overline{MS}$ scheme [1091].

The couplings $\alpha_s^{\mathrm{AdS/QCD}}$, $\alpha_s^{\mathrm{PI,Pinch}}$ and $\alpha_s^{\mathrm{EC,g_1}}$ are shown in Fig. 5.6.1. When compared in the same renormalization scheme, they agree reasonably well with earlier determinations, such as $\alpha_s^{\mathrm{gse,PT}}$ or that of the Godfrey-Isgur quark model [735], see Sec. 5.1 and Refs. [1091, 1128]. The couplings in Fig. 5.6.1 are in close agreement and have been used in the derivation of many crucial nonperturbative quantities, including the QCD scale $\Lambda_s^{\overline{\mathrm{MS}}}$ [1151], as well as elastic and transition form factors [1152–1154], parton distributions (including generalized ones) [844, 1026, 1155–1158], the hadron mass spectrum [1150, 1159], or the pion decay constant [1159].

In summary, several definitions of the strong coupling in the nonperturbative domain are possible. Most are scheme and gauge dependent. They tend to vanish as $\mu \to 0$ in a non-freezing behavior, *viz* the QCD $\beta$-function itself does not vanish. This informs us on how quark, gluon and ghost fields interact at low energy in the chosen scheme, but does not directly provide a universal parameter reflecting QCD's strength. In contrast, a set of calculations [158, 1143] and phenomenological extractions [1117, 1145, 1146] based on the effective charge prescription [146], following that of QED [1127], provide observable couplings that agree which each other. The consistency of these various approaches in determining a single coupling

$$\alpha_s^{\mathrm{SE,g_1}} \simeq \alpha_s^{\mathrm{AdS/QCD}} \simeq \alpha_s^{\mathrm{PI,Pinch}}$$

and its success in computing a wide range of nonperturbative quantities suggest that a compelling candidate for a canonical nonperturbative QCD coupling has been identified. It freezes at low energy, a satisfactory behavior since in the nonperturbative domain, the coupling should be finite and non-vanishing, determined by the physics of color confinement and its scale parameter should be set by a typical hadronic mass, e.g., that of the nucleon. An infrared fixed point is in fact a natural consequence of color confinement: since the propagators of the colored fields have a maximum wavelength, all loop integrals in the computation of the gluon self-energy should decouple at $Q^2 \to 0$ [1160].

## 5.7 The 't Hooft model and large $N$ QCD

**Tom Cohen**

In 1973 the QCD Lagrangian was first written down[50]. In the same year, the one-loop function was calculated [48, 49, 1161, 1162] indicating that that the theory was asymptotically free, but also implying that the coupling constant grew at low momenta. This meant that perturbation theory in the coupling à la QED is inapplicable for low momentum observables such hadron masses, charge radii and the like. The following year 't Hooft[1163] proposed an entirely new expansion for the theory — an expansion in $1/N_c$ where $N_c$ is the number of colors—which, it was hoped, would allow for a systematic computation of these observables.

While the dream of using the $1/N_c$ expansion to compute these quantities directly for QCD in 3+1 dimensions has been elusive, the $1/N_c$ expansion and the associated large $N_c$ limit has played a significant role during the past half century in at least three ways: it has provided a tool for the theoretical exploration of models beyond QCD, including most famously, the AdS/CFT connection[1000, 1003] for $\mathcal{N} = 4$ super Yang-Mills; it has has provided a qualitative and occasionally semi-quantitative tool to understand a significant amount of phenomenology (for example in Ref. [1164]); and, it has provided an organizing principle for deciding which terms should be large in phenomenological models or effective field theory treatments (for example in Ref. [1165]).

The underlying idea of the $1/N_c$ expansion is that three is sufficiently large so that a multicolored world with arbitrarily many colors is sufficiently close to the physical world — at least for some observables of interest — that the $N_c \to \infty$ world is a good starting point for an expansion and that systematic $1/N_c$ corrections are controllable. This section will provide an elementary introduction to the large $N_c$ limit and $1/N_c$ expansion with an emphasis on the underlying foundational ideas of the subject. An excellent review of these foundational ideas can be found in Sidney Coleman's Erice lectures[1166]; a more modern review of the large $N_c$ limit and $1/N_c$ expansion for field theories with an emphasis on lattice results can be found in Ref. [1167], while a review of large $N_c$ baryon spectroscopy can be found in Ref. [1168].

### 5.7.1 Large $N_c$ scaling

The key to 't Hooft's analysis [1163] were two related insights. The first is that a smooth large $N_c$ limit depends on the QCD coupling, $g$, scaling with $N_c$ as

$$g^2 = \lambda/N_c \tag{5.7.1}$$

where $\lambda$ is independent of $N_c$. Superficially, this might seem like a weak coupling limit that justifies standard perturbation theory. However, it does not: color factors in loops grow with $N_c$ and can compensate for the small coupling. The second key insight was related to the color factors in loops. 't Hooft developed a clever double line notation for gluons that allows one to easily analyze the scaling behavior of Feynman diagrams. The notation exploits the fact that that gluona are in the adjoint representation: they are associated with color matrices with two indices–one carrying one fundamental color and the other anti-fundamental color. Thus if one ignores the fact that the matrices are traceless (a $1/N_c^2$ effect), the color carried by a gluon propagator is identical to that of a quark line side-by side with an anti-quark line. For the purposes of counting color factors at leading order in $1/N_c$ — and for that purpose only — it is legitimate to replace gluon propagators in Feynman diagrams with parallel quark-antiquark lines. A closed loop of fundamental or antifundamental color in a diagram corresponds to one factor of $N_c$ since there are $N_c$ fundamental colors.

Armed with this, it is straightforward to deduce the following asymptotic scaling behavior for connected diagrams with no external lines:

- Planar connected diagrams of gluons (diagrams in which, except at vertices gluon lines do not cross when written in a plane ) and with no external lines grow asymptotically with $N_c$ as $N_c^2$.
- A diagram containing a non-planar gluon line reduces the asymptotic $N_c$ scaling of a planar diagram by a factor of $N_c^{-2}$. Multiple non-planar gluons reduce the $N_c$ counting by a factor of at least $N_c^{-2}$ per non-planar gluon.
- A planar diagram that contains quark loops that form the boundary of the diagram, reduces the asymptotic $N_c$ scaling by a factor of $N_c^{-1}$ per quark loop relative to a purely gluonic diagram. Quark loops that cannot be written in this form reduce the $N_c$ scaling by larger amounts.

Note that planar diagrams containing gluons can still be very complicated and can contain arbitrarily many gluon propagators. The fact that planar diagrams of gluons generically scale as $N_c^2$ can be understood in the following way: a closed loop consisting of a single gluon line scales as $N_c^2$: in double line form, it has two loops. Any planar diagram of gluons can be constructed starting from this single gluon loop: simply add planar gluons to it one-by-one until one has the diagram of interest. It is easy to see that any planar gluon added to a previous planar diagram in this construction adds one additional color loop (for a factor of $N_c$) but also two factors of the coupling constant $g$ at the vertices where the new gluons couple to the old diagram; since $g^2 \sim 1/N_c$ this cancels the additional color loop factor preserving the asymptotic scaling as $N_c^2$. By induction all diagrams of this class scale asymptotically as $N_c^2$.

The fact that adding a non-planar diagram reduces the scaling by a factor of $N_c^{-2}$ can be understood in a similar way. If one starts with a planar diagram of gluons and adds a non-planar gluon to it, the number of color loop factors decreases by one for a suppression factor of $1/N_c$ while two additional factors of $g$ must be added for another factor of $1/N_c$. Thus for example a diagram with a single non-planar gluon will scale asymptotically as $N_c^0$.

Similarly the scaling of diagrams containing quark loops that form the boundary of the diagram can be understood by noting that such diagrams can always be obtained starting from a planar diagram of gluons and then inserting a quark loop into a gluon propagator. Doing so does not change the number of color loop factors but adds two coupling coupling constants for each quark loop which together scale as $N_c^{-1}$ per quark loop.

The scaling rules for diagrams allow one to deduce the asymptotic scaling for the properties of glueballs and mesons[1163, 1169]. This can be done via the study of correlators of local gauge-invariant operators, $J$, that carry the quantum numbers of the glueballs or mesons of interest. For concreteness, consider $J$ to be a quark bilinear such as $J = \bar{q}q$ for the case of mesons (where for simplicity spin and flavor will be neglected in the discussion as they do not affect the $N_c$ scaling) and an operator such as $J = F^a_{\mu\nu} F^{a\,\mu\nu}$ for the case of glueballs. The correlator can be obtained by inserting these operators into closed loop diagrams. Doing so does not alter the leading $N_c$ scaling of the diagram. Thus if one is studying correlators carrying glueball quantum numbers, then the leading diagrams scale as $N_c^2$; similarly if one is studying a correlator carrying meson quantum numbers, then one needs to have a quark loop in the diagram and the leading diagram scales as $N_c$.

Consider the two-point correlation function:

$$\begin{aligned} \mathit{\Pi}_J(q^2) &= -i \int d^4x e^{-iq_\mu x^\mu} \left\langle T\left(J(x)J(0)\right)\right\rangle \\ &= \int ds \frac{\rho(s)}{q^2 - s + i\epsilon} \end{aligned} \tag{5.7.2}$$

where up to an overall factor $\rho(s)$, the spectral density, is given by the imaginary part of the correlator. It scales with $N_c$ in the same way as the correlator. The contributions to the spectral density from a given diagram can be extracted from its imaginary part. Moreover, cutting a diagram at various points between the sources reveals the gluon and quark contributions to the imaginary part, which by construction will form color singlet combinations. Using the double line notation, it is easy to see that no matter where the diagrams are cut between the sources, at leading order in $1/N_c$ all of the quark and gluons indices contract into a single color singlet combination — *i.e.* one that cannot be broken into multiple color singlet combinations.

If additionally one assumes confinement in the most basic sense that all asymptotic states are color singlets, this means that at leading order in the $1/N_c$ expansion, the operator $J$ creates single hadron states. By matching the $N_c$ counting of the leading diagram to the propagation of a single hadron one sees that

$$\begin{aligned} \langle \text{meson}|J_\text{meson}|\text{vac}\rangle &\sim N_c \\ \langle \text{glueball}|J_\text{glueball}|\text{vac}\rangle &\sim N_c^2 \end{aligned} \tag{5.7.3}$$

With this one can deduce numerous properties[1163, 1169] of QCD as a theory of hadrons by matching correlators at the quark-gluon level to descriptions at the hadronic level. One finds:

- The masses of mesons and glueballs become independent of $N_c$ as $N_c \to \infty$.
- Mesons and glueballs become stable as $N_c \to \infty$.
- The physics of mesons and glueballs can be described by an effective tree-level theory with vertices that scale at leading order as $N_c^{2-n_\text{g}-\frac{1}{2}n_\text{m}}$, where $n_\text{m}$ and $n_\text{g}$ are the number of meson lines and gluon lines respectively at the vertex. This implies that
  1. Interactions between these hadrons are weak.
  2. Meson decay amplitudes scale asymptotically as $N_c^{-1/2}$ and their widths as $N_c^{-1}$. Glueball decay amplitudes scale as $N_c^{-1}$ and their widths as $N_c^{-2}$
  3. Meson-meson scattering amplitudes scale as $1/N_c$. Glueball-glueball scattering amplitudes scale as $1/N_c^2$, while meson-glueball scattering amplitudes scales as $1/N_c$.
  4. In decays of hadrons into mesons, when all else is equal, processes with fewer mesons as decay products are favored by powers of $N_c$. Thus for example the partial decay width of a meson into a $\rho$-meson and a pion scales as $1/N_c$ while the rate into three pions directly scales as $1/N_c^2$.
- There are an infinite number of distinct mesons for each quantum number. This can be seen by matching the correlator to one at large space-like $q^2$ which can be computed perturbatively [1169].
- Quantum number exotic hybrid mesons (states whose quantum numbers cannot be obtained as a quark-antiquark state in a simple quark model but require at least one additional gluon) behave like ordinary mesons in $N_c$ scaling[1170]. At large $N_c$ they are narrow, there are an infinite number of them for any quantum number and their interactions with each other and with other mesons and glueballs scale according to the same rules as ordinary mesons.

– The OZI rule[18, 1171, 1172] becomes exact in the large $N_c$ limit. This implies glueball-meson mixing is suppressed.
– Tetraquark states do not exist at large $N_c$ [1169, 1173].

These properties can be viewed as predictions of QCD: they specify which quantities are dominant assuming that the large $N_c$ world is a reasonable proxy for our world. But, at best they make qualitative predictions since the coefficients multiplying the leading terms in the expansion are not specified by this analysis. Moreover in the physical world $N_c$ is only three so one might expect that assuming dominance of the leading-order predictions of the $1/N_c$ expansion would at best be a crude description of the phenomenology. In addition, the extent to which the phenomenology is qualitatively described by the leading-order behavior depends on the observable in question.

In the meson sector the large $N_c$ world might well be considered as a crude but recognizable caricature of much of the observed $N_c = 3$ phenomenology, at least for meson constructed from light quarks. There are numerous mesons that are comparative narrow — with widths much smaller than masses. There are often several identified mesons in a single spin-isospin-parity channels; presumably the number of identifiable meson would increase if $N_c$ were made larger. The OZI rule is typically well satisfied phenomenologically; indeed it was proposed based on phenomenological grounds before the formulation of QCD[18, 1171, 1172].

While many qualitative aspects aspects of meson physics can be deduced from the behavior of the theory at large $N_c$, there are observables in the meson sector for which subleading effects are sufficiently large that the leading behavior in a $1/N_c$ expansion does not describe the physical world even qualitatively. For example, the would-be nonet of pseduo-Goldstone bosons; a nonet would exist if the OZI-rule held — as it does at large $N_c$. However experimentally there is an octet split from a much heavier $\eta'$ meson. Of course this splitting is related to topology and the axial anomaly[1174], but in a large $N_c$ world these effects would be suppressed by an overall factor of $N_c^{-2}$[1175]. The fact that in the physical world the splitting is large shows that the large $N_c$ world is quite different from ours for this observable .

In fact, there are large classes of observables for which the the large $N_c$ world appears to be quite different from the $N_c = 3$ world. At large $N_c$ there should be a very large number of species of narrow glueballs that are weakly mixed with mesons. However, in the physical of world of $N_c = 3$ there are comparatively few glueball candidates[278] and the evidence for such states is typically somewhat murky, either because the evidence of the resonance is weak or because of mixing with ordinary mesons makes their “glueball” status unclear. Indeed, the identification of a resonance as a glueball may depend on there being an “extra” isoscalar state compared to what one expects from a naive quark model. Nevertheless, large $N_c$ analysis of glueballs is of value at a theoretical level and to a limited extent also acts to inform phenomenology: by providing a regime in which narrow, weakly mixed glueballs must exist, minimally it demonstrates that there is nothing in the basic structure of gauge theories containing both light quark and gluon degrees of freedom that forbids the existence of glueball states.

In a a similar way, the spectrum of quantum number exotic hybrid mesons in nature look quite different than in a large $N_c$ world: there are few candidates for such hadrons carrying light quarks quantum numbers[278]. Moreover, the evidence for those candidates is also typically murky due to inconclusive evidence for a resonance. Again the large $N_c$ analysis demonstrates that there is nothing in the basic structure of gauge theories forbidding hybrid mesons. Large $N_c$ also predicts that there should not be resonances in tetraquark channels. However, a clear signal for a quantum number exotic tetraquark has recently been found[1071]. As it happens this state is associated with heavy quarks — it is a doubly charmed state — and it is easy to see that heavy quark limit and the large $N_c$ limits are not expected to commute for these channels. If one increased $N_c$ while keeping the quark masses fixed, it is expected that this state would disappear.

### 5.7.2 The ’t Hooft model

The $N_c$ scaling rules presented above can be thought of as predictions about the physical world, but only in a qualitative sense — and they fail, even qualitatively, for many observables. Initially it was hoped that the $1/N_c$ expansion could be used as the basis of a quantitative treatment that was largely analytic at low order, in much the same as a expansion in $\alpha$ provided a quantitative treatment of QED. However, for QCD in 3+1 dimensions this has not worked out: even at lowest order in the expansion, the theory has proved to be intractable. Interestingly, however, QCD in 1+1 dimension, the so-called ’t Hooft model[1057] was solved (initially for one flavor) at leading order in the expansion in the early 1970s.

Note that the large $N_c$ scaling arguments given above did not depend on QCD being in 3+1 dimensions; they should hold in 1+1 dimension as well. Thus, one can use the explicit solutions of the 't Hooft model as a way to check the self-consistency of these rules.

The 't Hooft model has a critical property in common with QCD — confinement. It is useful to recall, however, that the mechanism of confinement in 1+1 dimensions is very different than in 3+1 dimensions. It occurs for a rather trivial reason — electric flux lines cannot spread out and thus even electrodynamics is confining in 1+1 dimension. The physics of gauge fields in 1+1 dimensions is also very simple: the field strength tensor, $F_{\mu\nu}$, has an electric component $E = F_{01} = -F_{10}$ but no magnetic component. Thus in 1+1 dimensional QED, the gauge field is not associated with a propagating photon; the Euler-Lagrange equation for the gauge field is not dynamical, but simply an equation of constraint fixing the electric field from the charge density $j_0 = \overline{\psi}\gamma_0\psi$. This is because the Gauss law (plus some conditions at infinity) fully determines the electric field. Something completely analogous occurs in the 't Hooft model.

While the color electric field in QCD in 1+1 dimension can be fixed given the color charge density of the quarks, the gauge field, $A_\mu$, itself depends on making a gauge choice. Certain gauges, such as the axial gauge of $A_1 = 0$ or the light-cone gauge have a particular useful property: they automatically suppress gluon-gluon couplings. In the axial gauge this is clear since all of the nonlinear terms involve products of $A_1$ and $A_0$. Gluon-gluon couplings vanish in the light-cone gauge for similar reasons. Since it is these non-linear couplings that make QCD complicated, QCD in 1+1 dimensions greatly simplifies.

The 't Hooft model simplifies further at leading order in the $1/N_c$ expansion. The leading diagrams for the $j_\mu$ correlator (which carries meson quantum numbers) are planar with a single quark loop bounding the diagram. This means that no gluon lines can either cross (due to the large $N_c$ constraint ) nor interact (due to the lack of gluon-gluon interactions. Accordingly the correlator is given by the so-called rainbow-ladder approximation: each quark propagator has a self-energy given by the sum of rainbow diagrams, while the interactions between quark lines is the the sum of ladder diagrams. The sum of these diagrams can be reduced analytically to integral equation between spinor-valued objects.

These simplify further into simple integral equations if one uses the light-cone gauge, which is based on light-cone coordinates:

$$x^{\pm} = \frac{x^0 \pm x^1}{\sqrt{2}} \tag{5.7.4}$$

and has a metric given by $g^{+-} = g^{-+} = 1$ and $g^{++} = g^{--} = 0$. The light cone gauge condition is

$$A_- = A^+ = 0 \; ; \tag{5.7.5}$$

among other things it has the virtue of being Lorentz invariant.

At leading order in large $N_c$ the spectral function for this correlator is expected to be saturated by arbitrarily narrow meson states. Since the explicit form of the correlator is calculable, one can develop a light-cone Bethe-Salpeter type eigenvalue equation for $\mu^2$ the meson mass, and $\psi(K)$, the light-cone Bethe-Salpeter amplitude for the meson. It is given in terms of a light-cone momentum, $K$ appropriately scaled so that $\psi(K)$ vanishes at $K = 0$ and $K = 1$ and $\psi(K)$ is only defined for $0 \leq K \leq 1$. It is given by the following integral equation

$$\begin{aligned}\mu_n^2 \psi_n(K) = & \frac{m^2 - \frac{g^2}{\pi}}{K(1-K)} \psi_n(K) \\ & - \frac{g^2}{\pi} \int_0^1 dK' \frac{\mathcal{P}}{(K-K')^2} \psi_n(K')\end{aligned} \tag{5.7.6}$$

where $\mathcal{P}$ indicates principal value, $\mu_n$ is the meson mass for the $n_{\text{th}}$ meson, $\psi_n(K)$ is the Bethe-Salpeter amplitude for that state, $m$ is the quark mass and $g$ is the coupling constant (which has dimensions of mass in 1+1 dimensions).

While there is no analytic solution to this integral eigenvalue equation, it can easily be solved numerically to give the meson spectrum for the model. Note that $N_c$ is not present in this expression showing self-consistently that meson masses are independent of $N_c$ at large $N_c$ as deduced from general scaling rules.

The fact that $\psi(K)$ vanishes at the $K = 0$ and $K = 1$ implies that the spectrum will be discrete — there are no solutions corresponding to two free quarks; the model correctly incorporates confinement. It is easy to show that for all values of $m$ and $g$, $\mu^2$ is real. This shows self-consistently that mesons are stable at large $N_c$ and verifies the general analysis discussed above. Moreover it can be shown that $\mu^2$ is always positive, showing that no matter how large the coupling, $g$, there are no tachyonic states that would signal an instability.

For asymptotically large values of n, it is easy to find the eigenvectors and Bethe-Salpeter amplitudes:

$$\mu_n = g^2 \pi n \;\; , \;\; \psi(K) = \sin(n\pi K) \; . \tag{5.7.7}$$

This asymptomatic form shows that solutions exist for arbitrarily high $n$, indicating the self-consistency of the large $N_c$ analysis, which predicted that there are an infinite number of mesons at large $N_c$.

The limit of zero quark mass in the 't Hooft model at large $N_c$ is interesting as it provides an opportunity to study chiral symmetry and its spontaneous breaking[1176]. The regime in which chiral symmetry breaking takes place requires that care be taken in the ordering of limits. One must take the limit of $N_c \to \infty$ (with the 't Hooft coupling, $g^2 N_c$, held fixed), prior to the $m \to 0$ limit. This limiting procedure insures that the ratio $\frac{g}{m}$ goes to zero in the combined limit. In this limit, it can be shown[1176], that chiral condensate is given by

$$\langle \overline{q} q \rangle = -N_c \sqrt{\frac{g^2 N_c}{12\pi}} \ . \tag{5.7.8}$$

Thus the 't Hooft model provides a simple illustration of how chiral symmetry breaking can work in a gauge theory.

However, the nature of spontaneous chiral symmetry breaking in the 't Hooft model is rather subtle. Note that the spontaneous breaking of chiral symmetry is a violation of Coleman's theorem[1177] which rules out spontaneous symmetry breaking of a continuous symmetry for theories in 1+1 dimensions. Thus, spontaneous chiral symmetry breaking seems paradoxical.

The resolution of the paradox was provided by Witten[1178] in his analysis of an analogous problem: spontaneous chiral symmetry breaking in the Thirring model at large $N_c$. It turns out that that the spontaneous chiral symmetry breaking is an artifact of working at infinitely large $N_c$ from the outset; it is absent for any finite $N_c$, no matter how large. Thus, as the large $N_c$ limit is approached the condensate is always strictly zero and there are no Goldstone bosons. However, the theory is in a Berezinski-Kosterlitz-Thouless phase[1179, 1180] in which the symmetry is "almost broken" and correlation functions of $\overline{q}q$ behave in a nontrival way. For space-like separations

$$\langle T[\overline{q}q(x,t)\overline{q}q(0,0)]\rangle \sim (x^2 - t^2)^{\frac{\text{const}}{N_c}} \ . \tag{5.7.9}$$

One sees that for any finite $N_c$ correlation functions $\overline{q}q$ fall off at large distance and thus do not saturate as they would if a condensate had formed. However, they also do not fall off exponentially as they would if $\overline{q}q$ created massive particles. Instead, there are long-range correlations: the correlation functions fall as a power law with distance. Moreover, the power depends on $N_c$ in such a way that it goes to zero at infinite $N_c$. Thus if one takes $N_c$ to be infinite at the outset, the systems acts as though spontaneous symmetry breaking had occurred.

The large $N_c$ properties of glueballs deduced earlier cannot be checked in the 't Hooft model for a very simple reason: in 1+1 dimension there are no glueballs.

### 5.7.3 Baryons

Of course mesons, glueballs and hybrids are not the only hadrons, there are also baryons. Unfortunately, the direct study of correlation functions via diagrammatic methods as was done for meson and glueballs does not work for baryons. This is for an obvious reason: a baryon contains (at least) $N_c$ quarks so that the number of quark lines in diagrams must grow with $N_c$. Among other things, this destroys the dominance of planar diagrams.

Witten argued that one can deduce the correct scaling behavior of large $N_c$ baryons by first considering the case in which all of the quarks are heavy (with masses much larger than the QCD scale) [1169]. In that situation, quark-antiquark pairs are suppressed and the propagation of quarks is non-relativistic. At the most trivial level, it ought to be apparent that in this regime $M_{\text{baryon}} \approx N_c M_Q$ where $M_Q$ is the mass of a heavy quark: the dominant term in the mass of a nonrelativistic system is the mass of the constituents and the baryon contains $N_c$ quarks. Thus the mass of the baryon scaling of the baryon mass with $N_c$ is

$$M_{\text{baryon}} \sim N_c \ . \tag{5.7.10}$$

Of course this result is from the leading term in a combined expansion built around the heavy quark and large $N_c$ limits with the heavy quark limit taken first; one might worry that the limits do not commute for the baryon mass. However, it is straightforward to see that subleading terms in a $1/M_Q$ expansion of the baryon mass also have a leading-order term in the $N_c$ expansion that scales like $N_c$. This suggests that this scaling could be general and hold independently of the quark mass. To see how this comes about, recall that in a heavy quark expansion for the baryon mass, the leading term — the direct quark mass contribution — is essentially not dynamical; the dominant subleading terms overall are the leading dynamical ones. The effective heavy quark lagrangian includes a nonrelativistic kinetic energy for the quarks and a color-Coulomb interaction between them. Witten[1169] demonstrated that at large $N_c$, the Hartree mean-field approximation to the non-relativistic color Coulomb problem becomes exact. In the Hartree approximation, correlations are neglected and each quark sits in an effective 1-body potential derived from interactions with the other $N_c - 1$ quarks (which sit in the ground state of the same potential).

Since the color-Coulomb interaction between two quarks has two factors of the coupling constant $g$, it scales as $1/N_c$. The mean-field Hamiltonian between one quark and the remainder has that quark interacting with $N_c - 1$ quarks and interactions add coherently. Thus, the mean-field Hamiltonian scales as $(N_c-1)/N_c$ and at asymptotically large $N_c$ becomes independent of $N_c$. The one-body equation for a single quark is then independent of $N_c$ at large $N_c$ and the quark's ground state wave function is also independent of $N_c$. This means that the spatial extent of the Hartree potential is itself independent of $N_c$. The contribution of the kinetic energy to the mass scales as is $N_c$ since there are $N_c$ quarks. The potential energy contributes $\frac{1}{2}N_c\langle V_{\rm Hartree}\rangle$, where $\langle V_{\rm Hartree}\rangle$ is the expectation value of the mean-field potential for a single quark; the factor of $\frac{1}{2}$ is because the interaction energy in a pair of quarks is split between them. Thus the direct quark mass term, the kinetic energy term and interaction term all scale linearly with $N_c$, strongly suggesting that $M_{\rm baryon} \sim N_c$ independent of the quark mass.

Moreover there is a very powerful argument from Witten[1169] that the results deduced from this mean-field behavior should persist when the quarks are light. Formally one would need to start with a relativistic many-body equation for bound states — a type of Bethe-Salpeter equation generalized to many particles — and show that the analog of the Hartree approximation becomes exact in the large $N_c$ limit. While that would be technically quite complicated, it seems apparent that all of the scaling from the Hartree approximation for heavy quarks should go through provided that irreducible n-body interactions between quarks scales as $N_c^{n-1}$. If this is true it is easy to see that the analog of the Hartree potential will be independent of $N_c$: at asympotically large $N_c$: there are $N_c$ 2-body interactions that each scale as $1/N_c$, $N_c^2$ three-body interactions that each scale as $1/N_c^2$, $N_c^3$ four body-interactions that each scale as $1/N_c^3$ and so forth. Each of these has a net contribution that is independent of $N_c$ indicating that this generalized mean-field interaction for a single quark is independent of $N_c$. Moreover demonstrating that n-body interactions between quarks scales as $N_c^{n-1}$ is straightforward using diagrammatic arguments similar to those used for the glueball and meson sectors.

Using this Hartree picture it is possible to deduce [1169] the asymptotic scaling of numerous baryon properties:

- Ground state baryon masses scale asymptotically as $N_c$.
- The size of ground state baryons generically is independent of $N_c$. Explicitly this means that form factors of external currents for baryons (such as electric factors) generically scale as $N_c^0 f(q^2/N_c^0)$; for $q^2$ of order $N_c^0$ the form factor is independent of $N_c$. This in turn means the moments of distributions (which are related to derivatives of form factors) such as $\langle r^2\rangle$, $\langle r^4\rangle$ are independent of $N_c$ at large $N_c$.
- Generic couplings between a ground state baryon and $n$ mesons scale as $N_c^{1-n/2}$. Among other things this means that
  1. Meson-baryon couplings scale generically as $N_c^{1/2}$.
  2. Meson-baryon scattering amplitudes are generically independent of $N_c$ for large $N_c$
- Couplings between a meson, a ground state baryon and an excited baryon are generically independent of $N_c$ and excited baryons have widths that are independent of $N_c$. Unlike in the glueball and meson sectors, these states are not narrow at large $N_c$, nor can you can conclude that there an infinite number of them.

Witten observed[1169] an interesting pattern to the scaling properties for baryons given above. They scale asymptotically with $1/N_c$ in the same way as analogous properties of solitons scale a with coupling constants squared. This insight lead to a renaissance of interest[1181–1183] in the Skyrme model[1184] as a model as baryons

The scaling laws given above are generic. Spin and flavor considerations may act to suppress certain couplings below these generic results. Moreover, for the case of two or more degenerate flavors, the notion of "ground state baryon" becomes a bit involved. Both of these issues are related to an emergent spin-flavor symmetry — a symmetry that is not manifest in the QCD langrangian but emerges at large $N_c$. In general, this symmetry is a contracted $SU(2N_f)$ where $N_f$ is the number of degenerate light flavors — it reduces to $SU(4)$ if one considers the up and down quarks to be effectively degenerate and the strange quark much heavier.

An initial hint that a new symmetry beyond mere isospin symmetry was emergent at large $N_c$ could be seen in the 2-flavor Skyrme model[1181], treated classically (with requanitzed collective coordinates to restore broken symmetries). This treatment corresponds to leading order in the $1/N_c$ expansion. It was found that rather than having the nucleon as the sole ground state, one had a tower of states with $I = J$ (the first two being the nucleon ($I = \frac{1}{2}, J = \frac{1}{2}$) and the $\Delta$ ($I = \frac{3}{2}, J = \frac{3}{2}$) with the levels in the tower degenerate at leading order in $1/N_c$ [1181]; the splittings can be shown to be $\mathcal{O}(N_c^{-1})$. Moreover, it was found that the ratios of the values of certain observables held independently of the parameters of the model or even the precise form of the Skyrme Lagrangian[1185]. It was realized that this behavior was not a property of Skyrme models *per se* but rather reflected an underlying symmetry of baryons[1186–1188].

The symmetry can be seen to be required for the consistency[1187] of large $N_c$ scaling provided that the pion-nucleon coupling scales with $N_c$ generically—*i.e.* as $N_c^{1/2}$. With this scaling, the Born approximation for pion-nucleon nucleon would scale linearly with $N_c$. However, unitarity constrains the scattering amplitudes to scale no faster than $N_c^0$. Clearly, something must cancel the Born amplitude in any channel where the meson-baryon coupling scales generically. In the case of scalar-isoscalar mesons, it is easy to show that the heavy mass of the baryon at large $N_c$ implies that at leading order, the contribution of the cross-Born diagram cancels the contribution of the Born diagram. However, pions are derivatively coupled and hence couple to the spin of the nucleon and are isovector so they also couple to the isospin. The various components of spin do not commute with each other and similarly with the various components of isospin and as a result, the cancellation between the Born and cross-Born contributions to $\pi - N$ scattering appears to be spoiled. However, the cancellation between the Born and cross-Born contributions at the level of pion-nucleon scattering will be restored provided that the $\Delta$ is treated as being degenerate (at this order) with the nucleon and the ratio of $g_{\pi N \Delta}$ (the transition coupling between the pion the nucleon and the $\Delta$ ) is taken to be a prescribed number times $g_{\pi NN}$[1187]. Applying the same logic to the process $\pi + N \to \pi + \Delta$, requires $g_{\pi \Delta\Delta}$ to be a fixed multiple of $g_{\pi N \Delta}$. At this order in $1/N_c$, the $\Delta$ and the nucleon are degenerate and the $\Delta$ should be treated as stable. Thus one can legitimately consider $\pi - \Delta$ scattering. Applying the same logic, one deduces the existence of a degenerate $I = \frac{5}{2}, J = \frac{5}{2}$ baryon and so forth generating a tower of states that become degenerate at large $N_c$. Presumably the the nucleon and $\Delta$ correspond approximately to the observed states in the $N = 3$ world, while the $I = \frac{5}{2}, J = \frac{5}{2}$ is a large $N_c$ artifact.

It is possible to show that the structure described above is encoded in a contracted $SU(4)$ Lie algebra for two-flavor QCD. The fixed ratio of the coupling constants are given by the Clebsch-Gordan coefficients of the group. The same logic that gives rise to the contracted $SU(4)$ symmetry, gives a contracted $SU(6)$ for 3-flavor QCD to the extent that one can approximate the strange quark as being nearly degenerate with the up and down quarks[1188]. Moreover, it is possible to show that for certain observables the leading corrections to the the contracted $SU(2N_f)$ symmetry is of order $1/N_c^2$ rather than $1/N_c$[1189]. This fact allows one to make some semi-quantitative predictions based on the emergent symmetry encoded in the large $N_c$ limit for baryons. A good example of this are the mass relations of Ref. [1164].

### 5.7.4 Nucleon-nucleon interactions and nuclear physics

The study of nucleon-nucleon interactions is complicated for kinematical reasons associated with the large nucleon mass. There are two kinematic regimes of interest: one in which the momentum transfers are independent of $N_c$ and the other in which the momentum transfers are of order $N_c$ — *i.e.* in which the velocities are independent of $N_c$. Physical observables associated with nucleon-nucleon scattering do not have a smooth large $N_c$ in the regime in which momentum transfers are of order $N_c^0$, but an analysis based on a time-dependent Hartree picture suggests that some scattering observables will have smooth large $N_c$ limits[1169] in the regime of momentum transfers of order $N_c$. These observables do not include many standard scattering observables such as phase shifts; the ones that have smooth limits appear to be those in which one follows the bulk flow of quantities of interest[1190]. Presumably the total cross-section also has a smooth limit[1191]. There is some predictive power for the spin and flavor dependence of such observables owing to the contracted $SU(4)$ symmetry[1190, 1191].

In the regime in which momentum transfers are of order unity — the regime of relevance to nuclear structure — the logic of Ref. [1169] implies that the nucleon-nucleon interaction strength is of order $N_c$, which is formally of the same order as the nucleon mass, while its range is independent of $N_c$. This implies that nuclear matter would be crystalline at large $N_c$, with nucleons constrained to be near the minimum of the potential from other nucleons. This is radically different from what is seen nature, suggesting that a $1/N_c$ expansion around the large $N_c$ limit is not a useful approach to nuclear structure. Interestingly, however, if one focuses solely on the spin-flavor structure of the nucleon-nucleon potential — a quantity that is not directly physical — there is a hierarchy in the strength of various spin-flavor contributions. This hierarchy is qualitatively similar to what one would obtain from the contracted $SU(4)$ spin-flavor symmetry of large $N_c$ QCD[1192, 1193]. This behavior is consistent with what one would expect if the nucleon-nucleon force was described via meson exchanges, as has been typically done in nuclear physics. Since the overall potential strength at the one-meson exchange level is large in some channels, consistency requires subtle cancellations when multiple-meson exchange are included. Such cancellations naturally occur due to the contracted $SU(4)$ symmetry[1194].

### 5.7.5 Other large $N_c$ limits

The large $N_c$ limit of QCD is an extrapolation from our world at $N_c = 3$ to a large $N_c$ world. However, that extrapolation is not unique. The standard approach discussed above involves keeping the number of flavors fixed while letting $N_c$ go to infinity. However, there is an alternative, the Venziano limit[1195] in which the ratio of the number of colors to the number of flavors is held fixed as $N_c \to \infty$. The large $N_c$ world for these two limits are quite different.

There is yet another large $N_c$ limit that exploits the fact that at $N_c = 3$, the representation for fundamental color and for the antisymmetric combination of two antifundamental colors are identical (*i.e.* $r$ is indistinguishable from $\left(\bar{g}\bar{b} - \bar{b}\bar{g}\right)/\sqrt{2}$). However quarks with fundamental color and with two-index antisymmetric color extrapolate to large $N_c$ quite differently — there are $N_c$ distinct quark colors for the former and $N_c(N_c - 1)/2 \sim N_c^2$ for the latter.

Large $N_c$ QCD(AS), the limit based on quarks in the two-index antisymmetric representation has remarkable formal connections supersymmetric QCD[1196–1198]. Phenomenologically, QCD(AS) has scaling of meson properties with $N_c$ similar to those of glueballs; one important difference between QCD(AS) at large $N_c$ and the conventional large $N_c$ limit is that in QCD(AS) quantum number exotic tetraquarks are not forbidden; indeed, they are required [1199]. The description of baryons for QCD(AS) is in analogy to Witten's but a somewhat new type of analysis is required [1200]. Formally, the predictions for baryon spectroscopy are distinct in QCD(AS) and QCD with quarks in the fundamental[1201], but phenomenological predictions for both expansions work to the order expected in describing real world data.

## 5.8 OPE-based sum rules:

### SVZ sum rules, $\frac{1}{M_Q}$ expansion and all that

**Mikhail Shifman**

### 5.8.1 Preamble

Rewind to autumn of 1971. I am a student at ITEP in Moscow, working on my Masters degree. The famous paper of Gerhard 't Hooft [47] was published in Nuclear Physics in October, but neither myself nor anybody else in ITEP immediately noticed this groundbreaking publication. At that time I did not even know what Yang-Mills theories meant. Now, when I think of the inception of QCD, the memories of this paper and its sequel [46] (issued in December of 1971) always come to my mind. For me, psychologically this was the beginning of the QCD era.

To give an idea of the scientific atmosphere at that time (1972) I looked through the Proceedings of the 1972 International Conference On High-Energy Physics [1202]. Theoretical talks were devoted to dual models (a precursor to string theory), deep inelastic scattering and Bjorken scaling, current algebra, $e^+e^- \to$ hadrons, etc. In three talks – by Zumino, Bjorken and Ben Lee – the Weinberg-Salam model (a precursor to the present-day Standard Model) was reviewed.[46] Ben Lee was the only person to refer to 't Hooft's publications [46, 47]. The last talk of the conference summarizing its major topics was delivered by Murray Gell-Mann. In this talk Gell-Mann discusses, in particular, whether quarks are physical objects or abstract mathematical constructs. Most interesting for us is his analysis of the $\pi^0 \to 2\gamma$ decay. Gell-Mann notes that if quarks are fermions then the theoretically predicted amplitude is a factor of 3 lower than the corresponding experimental result, but makes no statement of the inevitability of the quark color.[47]

In October 1972 I was accepted to the ITEP graduate school. My first paper on deep inelastic scattering in the Weinberg-Salam model was completed in early 1973; simultaneously, I started studying Yang-Mills theories (in particular, the Faddeev-Popov quantization [1203] [48]) in earnest. At the same time, somewhere far away, behind the Iron Curtain, Callan and Gross searched for a theory with an ultraviolet fixed point at zero. In July of 1973 Coleman and Gross submitted to PRL a paper asserting that "no renormalizable field theory that consisted of theories with arbitrary Yukawa, scalar or Abelian gauge interactions could be asymptotically free" [1204]. Damn Iron Curtain! If Gross asked anyone from the ITEP Theory Department he would have obtained the answer right away. The above theorem was known to the ITEP theorists from the Landau time. For brevity I will refer to it as the Landau theorem, although it was established by his students rather than Landau himself. The general reason why this theorem holds was also known – the Källen-Lehman (KL) representation of the polarization operator plus unitarity.

[46] There is a curious anecdote I heard later: In December 1979, after the Glashow-Weinberg-Salam Nobel Prize ceremony, a program was aired on Swedish radio. At some point, Weinberg quoted a phrase from the Bible. Salam remarked that it exists in the Quran too, to which Weinberg reacted: "Yes, but we published it earlier!"

[47] For me personally the following remark in his talk was a good lesson for the rest of my career: "Last year the rate of $K_L^0 \to \mu^+\mu^-$ decay was reported to be lower than allowed by unitarity unless fantastic hypotheses are concocted. Now the matter has become experimentally controversial." Alas... concocting fantastic hypotheses was the core of my Masters thesis.

[48] A longer and more comprehensible version appeared in Russian as Kiev preprint ITP 67-36. In the beginning of the 1970s, it was translated in English by B. Lee (NAL-THY-57, 1972). Apparently, in [47], [46] 't Hooft used the short version while I could use the longer one.

An explanatory remark concerning the Landau theorem might be helpful here. For asymptotic freedom to take place the first coefficient of the $\beta$ function must be *negative*. The sign of the one-loop graphs which determine the coupling constant renormalization is in one-to-one correspondence with the sign of their imaginary parts (this is due to the dispersion KL representation for these graphs). Unitarity implies the positivity of the imaginary parts which inevitably leads to the *positive* first coefficients in the $\beta$ functions in renormalizable four-dimensional field theories based on arbitrary Yukawa, scalar or Abelian gauge interactions. This situation is that of the Landau zero charge in the infrared rather than asymptotic freedom. In Yang-Mills theories in physical ghost-free gauges some graphs have no imaginary parts which paves the way to asymptotic freedom (see e.g. [1205]).

In fact, it is quite incomprehensible why asymptotic freedom had not been discovered at ITEP after 't Hooft's 1971 publication. In Ref. [1205] the reader can find a narrative about this historical curiosity.

May 1973 should be viewed as the discovery of asymptoric freedom [48, 49]. That's when the breakthrough papers of Gross, Wilczek and Politzer were submitted – simultaneously – to PRL. David Gross recollects [1204]:

> *We completed the calculation in a spurt of activity. At one point a sign error in one term convinced us that [Yang-Mills] theory was, as expected, non-asymptotically free. As I sat down to put it together and to write up our results, I caught the error. At almost the same time Politzer finished his calculation and we compared, through Sidney, our results. The agreement was satisfying.*

It took a few extra months for QCD to take off as *the* theory of strong interactions. The events of the summer of 1973 that led to the birth of QCD are described by H. Leutwyler in Sec. 1.1 of this Volume. To my mind, the final acceptance came with the November Revolution of 1974 – the discovery of $J/\psi$ and its theoretical interpretation as ortho-charmonium.[49] In the fall of 1973 we submitted a paper [1207] explaining why the Landau theorem in four dimensions fails only in Yang-Mills theory.

QCD and its relatives are special because QCD is the theory of *nature*. QCD is strongly coupled in the infrared domain where it is impossible to treat it quasiclassically – perturbation theory fails even qualitatively. It does not capture the drastic rearrangement of the vacuum structure related to confinement. The Lagrangian is defined at short distances in terms of gluons and quarks, while at large distances of the order of $\gtrsim \Lambda_{\rm QCD}^{-1}$ (where $\Lambda_{\rm QCD}$ is the dynamical scale of QCD, which I will refer to as $\Lambda$ below) we deal with hadrons, e.g. pions, $\rho$ mesons, protons, etc. Certainly, the latter are connected with quarks and gluons in a divine way, but this connection is highly nonlinear and non-local; even now, 50 years later, the full analytic solution of QCD is absent.

Non-perturbative methods were desperately needed.

### 5.8.2 Inception of non-perturbative methods

Four years before QCD Ken Wilson published a breakthrough paper [25] on the operator product expansion (OPE) whose pivotal role in the subsequent development of HEP theory was not fully appreciated until much later. What is now usually referred to as Wilsonian renormalization group (RG), or Wilsonian RG flow, grew from this paper. The Wilsonian paradigm of separation of scales in quantum theory was especially suitable for asymptotically free theories. Wilson's formulation makes no reference to perturbation theory, it has a general nature and is applicable in the non-perturbative regime too. The focus of Wilson's work was on statistical physics, where the program is also known as the block-spin approach. Starting from microscopic degrees of freedom at the shortest distances $a$, one "roughens" them, step by step, by constructing a sequence of effective (composite) degrees of freedom at distances $2a$, $4a$, $8a$, and so on. At each given step $i$ one constructs an effective Hamiltonian, which fully accounts for dynamics at distances shorter than $a_i$ in the coefficient functions.

QCD required a number of specifications and adjustments. Indeed, the UV fixed point in QCD is at $\alpha_s = 0$; hence, the approach to this fixed point at short distances is very slow, logarithmic rather than power-like, characteristic for the $\alpha_s \neq 0$ fixed point. In fact, it is not the critical regime at the UV fixed point *per se* we are interested in but rather the regime of approach to this critical point. Moreover, it was not realized that (in addition to the dynamical scale $\Lambda$) the heavy quarks provide an extra scale – the heavy quark mass $m_Q$ – which must be included in OPE where necessary.

[49] I should also mention a highly motivating argument due to S. Weinberg who proved [1206] that (in the absence of the U(1) current gluon anomaly) $m_{\eta'} \leq \sqrt{3} m_\pi$. This argument seemingly was discussed during ICHEP 74 in July 1974.

**Table 5.8.1** The lowest-dimension operators in OPE. $\Gamma$ is a generic notation for combinations of the Dirac $\gamma$ matrices.

| Normal dim | 3 | 4 | 5 | 6 | 6 |
|---|---|---|---|---|---|
| Operator | $O_q = \bar{q}q$ | $O_G = G^2_{\mu\nu}$ | $O_{qG} = \bar{q}\sigma^{\mu\nu}G_{\mu\nu}q$ | $O_{4q} = (\bar{q}\Gamma q)^2$ | $O_{3G} = GGG$ |

Surprisingly, in high-energy physics of the 1970s the framework of OPE was narrowed down to a very limited setting. On the theoretical side, it was discussed almost exclusively in perturbation theory. On the practical side, its applications were mostly narrowed down to deep inelastic scattering, where it was customary to work in the *leading*-twist approximation.

The fact that the UV fixed point is at zero makes OPE both more simple and more complicated than in the general case. On one hand, the anomalous dimensions of all composite local operators which might be relevant in the given problem scale only logarithmically. On the other hand, slow (logarithmic) fall off of "tails" instead of desired power-like – makes analytic separation of scales technically difficult.

I believe that we – Arkady Vainshtein, Valentin Zakharov and myself – were the first to start constructing a QCD version of OPE. The first step in this direction was undertaken in 1974 in the problem of strangeness-changing weak decays [1208, 1209] (currently known as the penguin mechanism in flavor-changing decays). A mystery of $\Delta I = \frac{1}{2}$ enhancement in $K$ decays had been known for years (for a review see [1210]). A suggestion of how one could apply OPE to solve this puzzle was already present in Wilson's paper [25]. Wilson naturally lacked particular details of QCD. The first attempt to implement Wilson's idea in QCD was made in [1211, 1212]. Although these papers were inspirational, they missed the issue of a "new" OPE needed for QCD realities. Seemingly, we were the first to address this challenge, more exactly two of its features: mixed quark-gluon operators (in [1208, 1209] we introduced

$$O_{\rm peng} = \bar{s}_L\gamma^\mu\left(\mathcal{D}_\nu G^{\mu\nu}\right)d_L$$

which is purely $\Delta I = \frac{1}{2}$) and coefficients logarithmically depending on the charmed (*i.e.* heavy at that time) quark mass. Currently, $c, b, t$ quark masses appear in the penguin operators (illustrated in Fig. 5.8.1), the latter two being genuinely heavy. Through equations of motion the operator $O_{\rm peng}$ reduces to a four-quark operator but its chiral structure is different from conventional, namely, it contains both left-handed and right-handed quark fields since $\mathcal{D}_\nu G^{\mu\nu} \sim \sum_q \bar{q}\gamma^\mu q$. Combined with another revolutionary finding of QCD, the extraordinary smallness of the $u$ and $d$ quark masses, $m_{u,d} \sim 5\,$MeV (see Sec. 1.1.15), the mixed chiral structure of the emerging four-fermion operator provides the desired enhancement of the $\Delta I = \frac{1}{2}$ amplitude. It took us over two years to fight a succession of referees for publication of Ref. [1209]. One after another, they would repeat that mixed-chirality four-fermion operator in the considered theory was complete nonsense. Currently, the penguin mechanism in flavor changing weak transitions is a basic theoretic element for a large variety of such decays. As Vainshtein put it [1210], "Penguins spread out but have not yet landed."

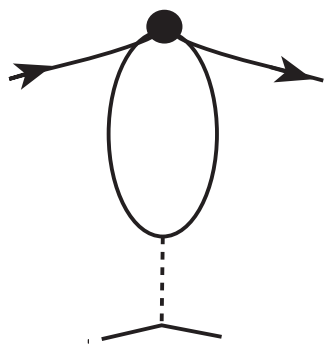

**Fig. 5.8.1** The penguin mechanism in flavor-changing decays. Any of three heavy quarks $c, b$ or $t$ can appear in the loop.

Systematic studies of Wilsonian OPE in QCD can be traced back to the summer of 1977 – that is when the gluon condensate $O_G$ (see Table 5.8.1) was first introduced [1213]. Vacuum expectation values of other gluon and quark operators were introduced in Ref. [126], which allowed one to analyze a large number of vacuum two- and three-point functions, with quite nontrivial results for masses, coupling constants, magnetic moments and other static characteristics of practically all low-lying hadronic states of mesons and baryons. A consistent Wilsonian approach requires an auxiliary normalization point $\mu$ which plays the role of a regulating parameter separating hard contributions included in the coefficient functions and soft contributions residing in local operators occurring in the expansion. The degree of locality is regulated by the same parameter. "Hard" versus "soft" means coming from the distances shorter than $\mu^{-1}$ in the former case and larger than $\mu^{-1}$ in the latter.

After setting the foundation of OPE in QCD [126] we were repeatedly returning to elaboration of various issues, in particluar, in the following works: [1214],[1215], and [1216].

### 5.8.3 SVZ sum rules. Concepts

The 1998 review [1215] summarizes for the reader foundations of the Shifman-Vainshtein-Zakharov (SVZ) sum rules in a pedagogical manner. At short distances QCD is the theory of quarks and gluons. Yang-Mills theory of gluons confines. This means that if you have a heavy probe quark and an antiquark at a large separation, a flux tube with a constant tension develops between them, preventing their "individual" existence. In the absence of the probe quarks, the flux tube can form closed contours interpreted as glueballs. This phenomenon is also referred to in the literature as the area law or the dual Meißner effect. Until 1994 the above picture was the statement of faith. In 1994 Seiberg and Witten found an analytic proof [1217, 1218] of the dual Meißner effect in $\mathcal{N} = 2$ super-Yang-Mills.[50] The Seiberg-Witten solution does *not* apply to QCD, rather to its distant relative. The real world QCD, with quarks, in fact has no area law (the genuine confinement is absent) since the flux tubes break through the quark-antiquark pair creation. Moreover, light quarks are condensed, leading to a spontaneous breaking of chiral symmetry, a phenomenon shaping the properties of the low-lying hadronic states, both mesonic and baryonic. The need to analytically understand these properties from first principles led us to the development of the SVZ method.

The quarks comprising the low-lying hadronic states, e.g. classical mesons or baryons, are not that far from each other, on average. The distance between them is of order of $\Lambda^{-1}$. Under the circumstances, the string-like chromoelectric flux tubes, connecting well-separated color charges, do not develop and details of their structure are not relevant. Furthermore, the valence quark pair injected in the vacuum – or three quarks in the case of baryons – perturb it only slightly. Then we do not need the full machinery of the QCD strings[51] to approximately describe the properties of the low-lying states. Their basic parameters depend on how the valence quarks of which they are built interact with typical vacuum field fluctuations.

We endowed the QCD vacuum with various condensates – approximately a half-dozen of them – in the hope that this set would be sufficient to describe a huge variety of the low-lying hadrons, mesons and baryons. The original set included[52] the gluon condensate $G^2_{\mu\nu}$, the quark condensate $\bar{q}q$, the mixed condensate $\bar{q}\sigma^{\mu\nu}G_{\mu\nu}q$, various four-quark condensates $\bar{q}\Gamma q\bar{q}\Gamma q$, and a few others (see Table 5.8.1). Later this set had to be expanded to address such problems as, say, the magnetic moments of baryons.

Our task was to determine the regularities and parameters of the classical mesons and baryons from a limited set of the vacuum condensates. Figure 5.8.2 graphically demonstrates the SVZ concept. On the theoretical side, an appropriate $n$-point function is calculated as an OPE expansion truncated at a certain order. In most problems only condensates up to dimension 6 (Table 5.8.1) are retained. In the "experimental" part the lowest-lying meson (or baryon) is singled out, while all higher states are represented in the quasiclassical approximation. Above an effective "threshold" $s_0$, where the spectral density becomes smooth, we apply quark-hadron duality to replace it by perturbative. Then the parameter $s_0$ is fitted along with the parameters of the lowest lying state – its mass and residue.

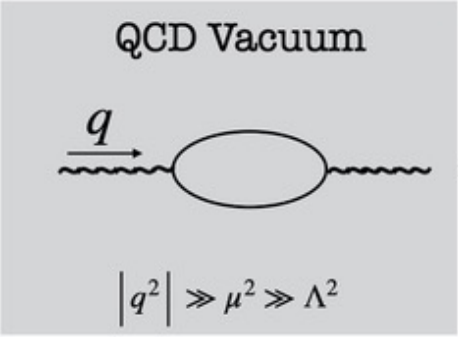

$$= \sum_{\dim_n} C_n(Q,\mu)\langle O_n(\mu)\rangle = \frac{f_0^2}{m_0^2 - q^2} + \sum_k \cdots$$

**Fig. 5.8.2** A two-point correlation function in the QCD vacuum. The left side is the OPE sum with a finite number of the lowest-dimension operators ordered according to their normal dimensions. The right side is the sum over mesons with the appropriate quantum numbers. The ground state in the given channel is singled out. The excited states are accounted for in the quasiclassical approximation. We define a positive variable $Q^2 = -q^2$ and a sliding $\mu^2$ parameter used as a separation parameter in OPE. For better convergence a Borel transformation is applied as explained below.

Acting in this way, one can determine the parameters $f_0$ and $m_0$ defined in Fig. 5.8.2 and their analogs in other problems. Of course, without invoking the entire infinite set of condensates one can only expect to obtain the hadronic parameters in an admittedly approximate manner.

### 5.8.4 Borelization

Analyzing the sum rules displayed in Fig. 5.8.2 we realized that their predictive power was limited – summation on both sides of the equation does not converge fast enough. On the right-hand side the contribution of high excitations is too large – the lowest lying states are "screened" – because the weight factors fall off rather slowly. Likewise, to achieve reasonable accuracy on the left-hand side one would need to add operators other than those collected in Table 5.8.1. At that time we knew next to nothing about higher-dimension operators, of dimension $\gtrsim 7$. The Borel transform came to the rescue.

The Borel transformation is a device well-known in mathematics. If one has a function $f(x)$ expandable in the Taylor series, $f(x) = x\sum_n a_n x^n$ with the coefficients $a_n$ which do not fall off sufficiently fast, one can

[50] More exactly, confinement through the flux tube formation was proven in the low-energy limit of this theory upon adding a small deformation term breaking $\mathcal{N} = 2$ down to $\mathcal{N} = 1$.

[51] Still unknown.

[52] A meticulous writer would have used the notation $\langle G^2_{\mu\nu}\rangle$, etc. but I will omit bra and ket symbols where there is no menace of confusion.

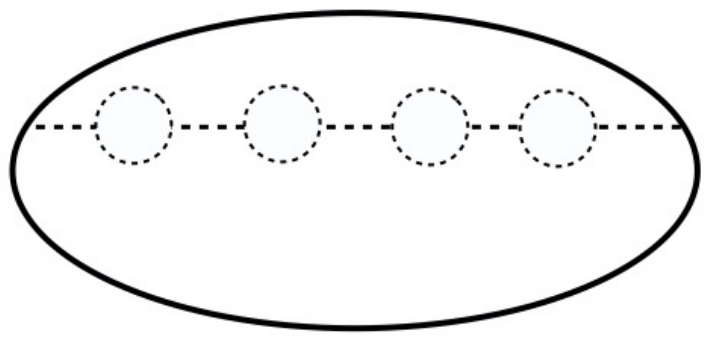

**Fig. 5.8.3** Graph showing four loops renormalizing a gluon line (represented by the dotted line). A renormalon is the sum over $n$ of such diagrams with $n$ loops.

instead introduce its Borel transform

$$\mathcal{B}f = x\sum_n \frac{a_n}{n!}x^n \tag{5.8.1}$$

and then, if needed, reconstruct $f$.[53]

If we apply this procedure to the sum rule in Fig. 2 we obtain for a given hadronic state $i$

$$\begin{aligned}\mathcal{B}\frac{f_i^2}{m_i^2+Q^2} &= \mathcal{B}\frac{f_i^2}{Q^2}\sum_n(-1)^n\left[\frac{m_i^2}{Q^2}\right]^n\\ &\to \frac{f_i^2}{Q^2}\sum_n\frac{(-1)^n}{n!}\left[\frac{m_i^2}{Q^2}\right]^n\\ &= \frac{f_i^2}{Q^2}\exp\left(-\frac{m_i^2}{Q^2}\right)\\ &\to \frac{f_i^2}{M^2}\exp\left(-\frac{m_i^2}{M^2}\right)\end{aligned} \tag{5.8.2}$$

where, in the final step (for historical reasons), I replaced $Q^2$ by a Borel parameter $M^2$. If $M^2$ can be chosen sufficiently small, higher excitations are *exponentially* suppressed.

Simultaneously, we improve the convergence of OPE on the left-hand side by applying the same operator $\mathcal{B}$. If the operator $\langle O_n\rangle$ has dimension $2d_n$, then the Borell transformation of the left hand side yields

$$\mathcal{B}\sum_n\frac{1}{(Q^2)^{d_n}}\langle O_n\rangle \to \sum_n\frac{1}{(d_n-1)!}\frac{1}{(M^2)^{d_n}}\langle O_n\rangle\,, \tag{5.8.3}$$

where I have again replaced $Q^2$ by the Borel parameter $M^2$. Since the expansion (5.8.3) goes in inverse powers of $M^2$, it is necessary to keep $M^2$ large enough. The two requirements on $M^2$ seem contradictory. However, for all "typical" resonances, such as say $\rho$ mesons, they can be met simultaneously [126, 1219, 1220] in a certain "window." The only exception is the $J^P = 0^{\pm}$ channel. There are special reasons why $0^{\pm}$ mesons are exceptional, see [1221].

### 5.8.5 Practical version of OPE

At the early stages of the SVZ program the QCD practitioners often did not fully understand the concept of scale separation in the Wilsonian OPE. It was generally believed that the coefficients are fully determined by perturbation theory while non-perturbative effects appear only in the OPE operators.[54] This belief led to inconsistencies which revealed themselves e.g. in the issue of renormalons (see below). A set of graphs represented by renormalons is constructed from a single gluon exchange by inserting any number of loops in the gluon line like beads in a necklace ([Ref. [1222]). Being treated *formally* this contribution, shown in Fig. 5.8.3, diverges factorially at high orders. I vividly remember that after the first seminar on SVZ in 1978 Eugene Bogomol'nyi asked me each time we met: "Look, how can you speak of power corrections in the $n$-point functions at large $Q^2$ if even the perturbative expansion (*i.e.* the expansion in $1/\log(Q^2/\Lambda^2)$) is not well defined? Isn't it inconsistent?" I must admit that at that time my answer to Eugene was somewhat evasive.

The basic principle of Wilson's OPE – the scale separation principle – is "soft versus hard" rather than "perturbative versus non-perturbative." Being defined in this way the condensates are explicitly $\mu$ dependent. All physical quantities are certainly $\mu$ independent; the normalization point dependence of the condensates is compensated by that of the coefficient functions – see Fig. 5.8.2.

The problem of renormalons disappears once we introduce the normalization point $\mu$. With $\mu \gg \Lambda$, there is no factorial divergence in high orders of perturbations theory. Renormalons conspire with the gluon condensates to produce, taken together, a well-defined OPE. The modern construction goes under the name of the "renormalon conspiracy"; it is explained in detail in my review [1216]. I hasten to add, though, that the renormalons acquire a life of their own in those cases in which OPE does not exist. Qualitatively, they can shed light on scaling dimensions of non-perturbative effects. The most clear-cut example of this type is the so-called "pole mass of the heavy quarks" [1223, 1224] and its relation to a theoretically well-defined mass parameter [1225].

In some two-dimensional solvable models exact OPE can be constructed which explicitly demonstrates the $\mu$ dependence of both the coefficient functions and the condensates in the Wilsonian paradigm and its cancellation in the physical quantities (for a recent study see e.g. [1226]). Needless to say, if QCD was exactly solved we would have no need in the SVZ sum rules.

We had to settle for a reasonable compromise, known as *the practical version of OPE.* In the practical version we calculate the coefficient functions perturbatively keeping a limited number of loop corrections. The conden-sate series is truncated too. The condensates are not calculated from first principles; rather a limited set is determined from independent data.

[53] The Borel transform is closely related to the Laplace transform.

[54] Unfortunately, this misconception lasted through the 1980s and was visible in the literature even in the 1990s and later.

The practical version is useful in applications only provided $\mu^2$ can be made small enough to ensure that the "perturbative" contributions to the condensates are much smaller than their genuine (mostly non-perturbative) values. At the same time, $\alpha_s(\mu^2)/\pi$ must be small enough for the expansion in the coefficients to make sense. The existence of such "$\mu^2$ window" is not granted *a priori* and is a very fortunate feature of QCD. We did observe this feature empirically in almost all low-lying hadrons [1227, 1228] [55]. At the same time, we identified certain exceptional channels revealing unforeseen nuances in hadronic physics [1221].

### 5.8.6 Implementation of the idea and results

After the strategic idea of quark and gluon interaction with the vacuum medium became clear we delved into the uncharted waters of microscopic hadronic physics. Remember, in 1977 nobody could imagine that basic hadronic parameters for at least some hadrons could be analytically calculated, at least approximately. As a show-case example we chose the most typical mesons, $\rho$ and $\phi$, to calculate their couplings to the electromagnetic current and masses. The agreement of our results with experiment was better than we could *a priori* expect. At first we were discouraged by a wrong sign of the gluon condensate term in the theoretical part of the appropriate SVZ sum rule. We suddenly understood that this sign could be compensated by the four-quark condensate – a real breakthrough. In November of 1977 we published a short letter [1213] which still missed a number of elements (e.g. Borelization) developed and incorporated later, one by one. We worked at a feverish pace for the entire academic year, accumulating a large number of results for the hadronic parameters. All low-lying meson resonances built from the $u, d, s$ quarks and gluons were studied and their static properties determined from SVZ: masses, coupling constants, charge radii, $\rho$-$\omega$ mixing, and so on, with unprecedented success. In summer of 1978, inspired by our progress, we prepared a number of preprints (I think, eight of them simultaneously [56]) and submitted to ICHEP-78 in Tokyo. Clearly none of us were allowed to travel to Tokyo to present our results.

I cannot help mentioning an incident that occurred in the spring of 1978 when we were mostly done with this work. The episode may have been funny were it not so nerve-wracking. When we decided that the calculational stage of the work was over, I collected all my drafts (hundreds of sheets of paper with derivations and math expressions), I organized them in proper order, selected all expressions we might have needed for the final draft of the paper and the future work, meticulously rewrote them in a voluminous notebook (remember, we had no access to photocopying machines), destroyed the original drafts, put the notebook in my briefcase and went home. It was about midnight, and I was so exhausted that I fell asleep while on the metro train. A loud voice announcing my stop awoke me, and I jumped out of the train, leaving the briefcase were it was, on the seat. By the time I realized what had happened the train was gone, and gone with it forever my calculations ... I have never recovered my briefcase with the precious notebook... After a few agonizing days it became clear that the necessary formulas and expressions had to be recovered anew. Fortunately, Vainshtein and Zakharov had kept many of their own derivations. Vainshtein never throws away anything as a matter of principle. Therefore, the problem was to dig out "informative" sheets of paper from the "noise" (this was hindered by the fact that Vainshtein was in Novosibirsk while we were in Moscow). Part of my drafts survived in the drawers of a huge desk that I had inherited from V. Sudakov. Better still, many crucial calculations were discussed so many times by us, over and over again, that I remembered them by heart. Nevertheless, I think it took a couple of uneasy weeks to reconstruct in full the contents of the lost notebook.

The SVZ method was further developed by many followers (e.g. the so-called light-cone sum rules for form-factors), see [1229] and [1230]. A broad picture of the hadronic world was obtained by the 1980s and later [1231]. Today the pioneering SVZ paper is cited 6000+ times. Until 1990s, when lattice QCD based on numeric calculations, started approaching its maturity, the SVZ method was the main tool for analyzing static hadronic properties.

### 5.8.7 Reliability and predictive power

The SVZ method is admittedly approximate. Yet, it is not a model in the sense that it cannot be arbitrarily bent to accommodate "wrong" data. It is instructive to narrate here the story of an alleged discovery of an alleged "paracharmonium" referred to as $X(2.83)$ in January of 1977 [1232]. It was widely believed then that $X(2.83)$ was the $0^-$ ground sate of $\bar{c}c$ quarks, $\eta_c$. If this was the case the mass difference between $J/\psi$ and $\eta_c$ would be close to 270 MeV. Shortly after, the interpretation of $X(2.83)$ as $\eta_c$ was categorically ruled out by the SVZ analysis [1233] which *predicted* that the above mass difference must be 100± 30 MeV. Two years later, a new experiment [1234] negated the existence of the $X(2.83)$ state. In the very same experiment the genuine paracharmonium was observed at $2.98 \pm 0.01$ GeV, in perfect agreement with [1233]. For us this was a triumph and a lesson – if one believes in a theory one should stand for it!

[55] Theoretical understanding of the roots of this phenomenon remains unclear. Seemingly, it has no known analogues in two-dimensional models.

[56] In the journal publication they were combined in three articles occupying the whole issue of Nucl. Phys. B147, №5, 1979.

### 5.8.8 OPE-based construction of heavy quark mass expansion

In the 1980s and early 1990s OPE was generalized to cover theoretical studies of mixed heavy-light hadrons, *i.e.* those built from light, $q$, and heavy, $Q$, flavors. In the 1990s those who used $1/m_Q$ expansion in theoretical analysis of $Q\bar{q}$ and $Qqq$ systems numbered in the hundreds. A large range of practical physics problem related to $Q\bar{q}$ and $Qqq$ systems were solved. Lattice analyses of such systems even now remain hindered, and in many instances the $1/m_Q$ expansion remains the only reliable theoretical method.

As I have mentioned in the second paragraph of Sec. 5.8.2, heavy quarks in QCD introduce an extra scale, $m_Q$. To qualify as a heavy quark $Q$ the corresponding mass term $m_Q$ must be much larger than $\Lambda_{\rm QCD}$. The charmed quark $c$ can be considered as heavy only with some reservations while $b$ and $t$ are *bona fide* heavy quarks. The hadrons composed from one heavy quark $Q$, a light antiquark $\bar{q}$, or a "diquark" $qq$, plus a gluon cloud (which also contains light quark-antiquark pairs) – let us call them $H_Q$ – can be treated in the framework of OPE. The role of the cloud is, of course, to keep all the above objects together, in a colorless bound state. The light component of $H_Q$, its light cloud, has a complicated structure; the soft modes of the light fields are strongly coupled and strongly fluctuate. Basically, the only fact which we know for sure is that the light cloud is indeed light; typical excitation frequencies are of order of $\Lambda$. One can try to visualize the light cloud as a soft medium.[57] The heavy quark $Q$ is then submerged in this medium. The latter circumstance allows one to develop a formalism similar to SVZ in which the soft QCD vacuum medium is replaced by that of the light cloud. As a result, an OPE-based expansion in powers of $1/m_Q$ emerges (see Fig. 5.8.4). When heavy quarks are in soft medium the heavy quark-antiquark pair creation does not occur and the field-theoretic description of the heavy quark becomes redundant. A large "mechanical" part in the $x$ dependence of $Q(x)$ can be *a priori* isolated, $Q(x) = \exp(-im_Q t)\tilde{Q}(x)$. The reduced bispinor field $\tilde{Q}(x)$ describes a residual heavy quark motion inside the soft cloud; the heavy quark mass appears only in the form of powers of $1/m_Q$ (first noted in 1982).

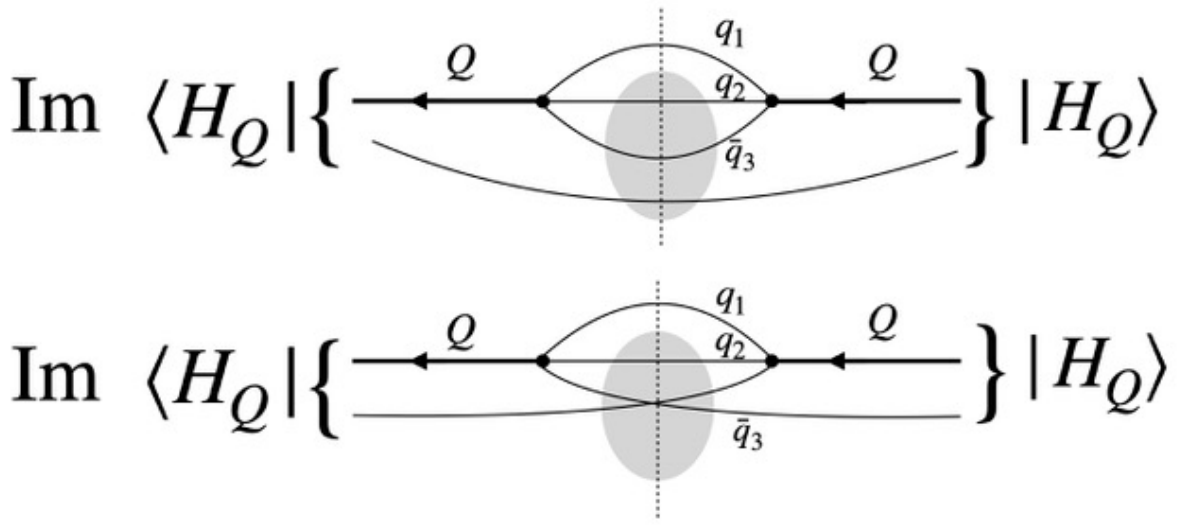

**Fig. 5.8.4** $1/m_Q$ expansion for a $H_Q$ weak inclusive decay rate (see Eq. (5.8.5)). Depicted are two operators, the leading $\bar{Q}Q$ and a subleading $(\bar{Q}q_3)(\bar{q}_3 Q)$. Both are sandwiched between the heavy hadron states $\langle H_Q|$ and $|H_Q\rangle$ and the decay rate is determined by the imaginary part. The grey area depicts the soft quark-gluon cloud. Adapted from Ref. [1235, 1236].

Comprehensive reviews on the OPE-based heavy quark theory exist [1225], [1237],[1238], [674]. There the reader will find exhaustive lists of references to original publications. Therefore, in my presentation below I will be brief, with a focus on a historical aspect, as I remember it, and limit myself to a few selected references.

In the early 1980s abundant data on the meson and baryon $H_Q$ states started to appear. Theoretical understanding of the total decay rates beyond the free-quark calculations became a major goal. This challenge paved the way to the beginning of the $1/m_Q$ expansion in $H_Q$ hadron physics in the mid 1980s. The decay rate into an inclusive final state $f$ can be written in terms of the imaginary part of a forward scattering operator (the so-called transition operator) evaluated to second order in the weak interactions [1235, 1236],

$$\mathrm{Im}\hat{T}(Q\to f\to Q) = \mathrm{Im}\int d^4x\; iT\left(\mathcal{L}_W(x)\mathcal{L}_W^{\dagger}(0)\right) \tag{5.8.4}$$

where $T$ denotes the time ordered product and $\mathcal{L}_W$ is the relevant weak Lagrangian at the normalization point $\mu \sim m_Q$. The factor $\exp(-im_Q t)$ mentioned above is implicit in Eq. (5.8.4). Descending to $\mu \ll m_Q$ one

[57] Hard gluons do play a role too. They have to be taken into account in the coefficient functions as will be mentioned In Sec. 5.8.10.

arrives at the OPE expansion

$$\begin{aligned}\Gamma(H_Q\to f) &= G_F^2|V_{\rm CKM}|^2 m_Q^5 \\ &\quad\times\sum_i \tilde{c}_i^{(f)}(\mu)\frac{\langle H_Q|O_i|H_Q\rangle_\mu}{2M_{H_Q}} \\ &\propto \Big[c_3^{(f)}(\mu)\frac{\langle H_Q|\bar{Q}Q|H_Q\rangle_{(\mu)}}{2M_{H_Q}} \\ &\quad + c_5^{(f)}(\mu)m_Q^{-2}\frac{\langle H_Q|\bar{Q}\frac{i}{2}\sigma G Q|H_Q\rangle_{(\mu)}}{2M_{H_Q}} \\ &\quad + \sum_i c_{6,i}^{(f)}(\mu)m_Q^{-3}\frac{\langle H_Q|(\bar{Q}\Gamma_i q)(\bar{q}\Gamma_i Q)|H_Q\rangle_{(\mu)}}{2M_{H_Q}} \\ &\quad + \mathcal{O}(1/m_Q^4)+...\Big]\,, \end{aligned} \tag{5.8.5}$$

where $\Gamma_i$ represent various combinations of the Dirac $\gamma$ matrices, see also Table 5.8.1. In SVZ we dealt with the vacuum expectation values of relevant operators while in the heavy quark physics the relevant operators are sandwiched between $H_Q$ states.

### 5.8.9 Applications

The expansion (5.8.5) allowed us to obtain [1235, 1236] the first *quantitative* predictions for the hierarchies of the lifetimes of $Q\bar{q}$ mesons and $Qqq$ baryons ($Q$ was either $c$ or $b$ quark) in the mid-1980s – another spectacular success of the OPE-based methods. The dramatic story of $\eta_c$ narrated in Sec. 5.8.7 repeated itself. With the advancement of experiment in the late 1990s, a drastic disagreement was allegedly detected in the ratio $\tau(\Lambda_b)/\tau(B_d)_{\rm exp} = 0.77\pm 0.05$ compared to the theoretical prediction

$$\tau(\Lambda_b)/\tau(B_d)_{\rm theor} = 0.9\pm 0.03$$

(e.g. [1225]). In the 2010s the $\Lambda_b$ lifetime was remeasured shifting the above experimental ratio up to $0.93\pm 0.05$. Hurrah!

In the mid-1980s, at the time of the initial theoretical studies of the $H_c$ and $H_b$ lifetime hierarchies [1235, 1236], next to nothing was known about heavy baryons. Since then enormous efforts were invested in improving theoretical accuracy both in *mesons and baryons* in particular by including higher-dimension operators in the inverse heavy quark mass expansion and higher-order $\alpha_s$ terms in the OPE coefficients. The status of the Inverse Heavy Quark Mass Expansion (IHQME) for $H_Q$ lifetimes as of 2014 was presented in the review [1239]. The advances reported there and in more recent years cover more precise determination of the matrix elements of four-quark operators via HQET sum rules [1240], calculations of the higher $\alpha_s$ corrections, in particular, $\alpha_s^3$ corrections to the semileptonic $b$ quark decay [1241], the first determination of the Darwin coefficient for non-leptonic decays [1242, 1243], etc. Comparison with the current set of data on $\tau(H_c)$ can be found in [1244]. In this context I should also mention an impressive publication [1245] (see also references therein) which, in addition to a comprehensive review of the OPE-based analysis of the $H_c$ lifetimes, acquaints the reader with a dramatic story of the singly charmed *baryon* hierarchy. Indeed, according to PDG-2018 the lifetime of $\Omega_c^0$ is 69±12 fs while PDG-2020 yields $\tau(\Omega_c^0) = 268\pm 24\pm 10$ fs! The jump in the $\Omega_c^0$ lifetime by a factor of 3 to 4 compared to the previous measurements was reported by LHCb [1246–1248].[58] With these new data the observed hierarchy of lifetimes changes: $\Omega_c^0$ moves from the first place (the shortest living $H_c$ baryon) to the third. The question arises whether the OPE-based theory can explain the current experimental situation $\tau(\Xi^0) < \tau(\Lambda_c^+) < \tau(\Omega_c^0) < \tau(\Xi^+)$. In [1245] it is argued that the answer is "yes, it is possible" (see Fig. 5 in [1245]) provided one takes into account $1/m_c^4$ contributions due to four-quark operators and $\alpha_s$ corrections in the appropriate coefficient functions.[59]

I should emphasize that the theoretical accuracy in the $H_c$ family is limited by the fact that the expansion parameter $\Lambda_{\rm QCD}/m_c$ is not small enough. Even including sub-leading contributions will hardly provide us with high-precision theoretical predictions. For $H_c$ states IHQME at best provides us with a semi-quantitative guide. On the other hand, in the theory of $H_b$ decays one expects much better precision.

### 5.8.10 Around 1990s and beyond

(1) Heavy quark symmetry when $\boldsymbol{m_Q\to\infty}$

The light-cloud interpretation as in Fig. 5.8.4 immediately implies that at zero recoil the (appropriately normalized) $B\to D$ formfactors reduce to unity. This is called the "small velocity (SV) limit theorem" [1250], [1251]. The above"unification" is similar to the vector charge non-renormalization theorem at zero momentum transfer, say, for the $\bar{u}\gamma^\mu d$ current. The $D$ and $B$ masses are very far from each other. One has to subtract the mechanical part of the heavy quark mass in order to see that all dynamical parameters are insensitive to the substitution $Q_1\leftrightarrow Q_2$ in the limit $m_{Q_{1,2}}\to\infty$, with the SV limit ensuing at zero recoil. Perhaps, this is the reason why it was discovered so late. The next step was made by Isgur and Wise who generalized this symmetry of the zero-recoil point by virtue of the Isgur-Wise function [1252, 1253].

[58] Of course, this could happen only because (presumably) statistical and/or systematic errors in the previous measurements were grossly underestimated. It is also curious to note that 30 years ago Blok and I argued [1249] (Secs. 4.2 and 6) that $\Omega_c^0$ could be the longest living singly charmed baryon due to its $ss$ spin-1 diquark structure.

[59] The four-quark operators introduced in [1235, 1236] responsible for the Pauli interference yield corrections $O(1/m_c^3)$, see Eq. (5.8.5). The authors of [1245] go beyond this set.

(2) HQET

Heavy quark effective theory which emerged in the 1990s [667, 1254] formalizes and automates a number of aspects of the generic $1/m_Q$ expansion. In fact, it immediately follows from the construction similar to (5.8.5). Simplified rules of behavior proved to be very helpful for QCD practitioners in the subsequent development of various applications. In HQET the reduced field $\tilde{Q}$ is treated quantum-mechanically, its non-relativistic nature is built in, and the normalization point $\mu$ is $\ll m_Q$ from the very beginning.[60] Applying the Dirac equation to eliminate small (lower) components in favor of the large components it is easy to derive the expansion of $\mathcal{L}^0_{\rm heavy}$, up to terms $1/m_Q^2$,

$$
\begin{aligned}
\mathcal{L}^0_{\rm heavy} &= \bar{Q}(i \not{D} - m_Q)Q \\
&= \bar{\tilde{Q}}\frac{1+\gamma_0}{2}\left(1+\frac{(\boldsymbol{\sigma}\boldsymbol{\pi})^2}{8m_Q^2}\right)\left[\pi_0 - \frac{1}{2m_Q}(\boldsymbol{\pi}\boldsymbol{\sigma})^2 - \right. \\
&\left. - \frac{1}{8m_Q^2}\left(-(\vec{D}\vec{E}) + 2\boldsymbol{\sigma}\cdot\vec{E}\times\boldsymbol{\pi}\right)\right] \\
&\times\left(1+\frac{(\boldsymbol{\sigma}\boldsymbol{\pi})^2}{8m_Q^2}\right)\frac{1+\gamma_0}{2}\tilde{Q} + \mathcal{O}\left(\frac{1}{m_Q^3}\right), \qquad (5.8.6)
\end{aligned}
$$

where $\boldsymbol{\sigma}$ denote the Pauli matrices and

$$(\boldsymbol{\pi}\boldsymbol{\sigma})^2 = \boldsymbol{\pi}^2 + \boldsymbol{\sigma}\vec{B}\,,$$

$\vec{E}$ and $\vec{B}$ denote the background chromoelectric and chromomagnetic fields, respectively. Moreover, the operator $\pi_\mu$ is defined through

$$
\begin{aligned}
iD_\mu Q(x) &= e^{-im_Q v_\mu x_\mu}\left(m_Q v_\mu + iD_\mu\right)\tilde{Q}(x) \\
&\equiv e^{-im_Q v_\mu x_\mu}\left(m_Q v_\mu + \pi_\mu\right)\tilde{Q}(x) \qquad (5.8.7)
\end{aligned}
$$

where $v_\mu$ is the heavy quark four-velocity. The set of operators presented in (5.8.6) plays the same basic role in $1/m_Q$ expansion as the set in Table 5.8.1 in SVZ sum rules.

In the remainder of this section I will briefly mention some classic problems with heavy quarks which were successfully solved in the given paradigm.

(3) CGG/BUV theorem

Up to order $1/m_Q^2$ all inclusive decay widths of the $H_Q$ mesons coincide with the parton model results for the $Q$ decay [1255], [1256],

$$\Gamma = \Gamma_0\left(1 - \frac{\mu_\pi^2}{2m_Q^2}\right), \quad \mu_\pi^2 = \frac{1}{2M_{H_Q}}\langle H_Q|\bar{\tilde{Q}}\vec{\pi}^2\tilde{Q}|H_Q\rangle \qquad (5.8.8)$$

where $\Gamma_0$ is the parton model result. There are no corrections $O(1/m_Q)$. This is known as the CGG/BUV theorem.

(4) Spectra and line shapes

Lepton spectra in semileptonic $H_Q$ decays were derived in [1257]. The leading corrections arising at the $1/m_Q$ level were completely expressed in terms of the difference in the mass of $H_Q$ and $Q$. Nontrivial effects appearing at the order $1/m_Q^2$ were shown to affect mainly the endpoint region; they are different for meson and baryon decays as well as for beauty and charm decays.

The theory of the line shape in $H_Q$ decays, such as $B \to X_s\gamma$ where $X_s$ denotes the inclusive hadronic state with the $s$ quark, resembles that of the Mössbauer effect. It is absolutely remarkable that for 10 years there were no attempts to treat the spectra and line shapes along essentially the same lines as it had been done in deep inelastic scattering (DIS) in the 1970s. Realization of this fact came only in 1994; technical implementation of the idea was carried out in [1258], [1259], and [1260].

(5) Hard gluons

Hard-gluon contributions special for the heavy quark theory result in powers of the logarithms $\alpha_s \log(m_Q/\mu)$. They determine the coefficients $c_i$ in Eq. (5.8.5) through the anomalous dimensions of the corresponding operators. They were discovered in [1261, 1262] and were called the *hybrid* logarithms. In HQET they are referred to as matching logarithms.

(6) In conclusion

Concluding the heavy quark portion I should add that Kolya Uraltsev (1957-2013), one of the major contributors in heavy quark theory died in 2013 at the peak of his creative abilities (see [1222]).

Concerning the OPE-based methods in QCD in general, I would like to make an apology to the many authors whose works have not been directly cited. The size limitations are severe. The appropriate references are given in the review papers listed in the text above.

Just for the record, a couple of reviews which are tangentially connected to the topic of the present article are given in Refs. [1263] and [1264].

[60] I personally prefer to consider the heavy quark expansions directly in *full QCD* in the framework of the Wilson OPE bypassing the intermediate stage of HQET.

### 5.8.11 Recent developments unrelated to the OPE-based methods

Quantum field theories from the same class as QCD are now experiencing dramatic changes and rapid advances in a deeper understanding of anomalies. I want to mention two crucial papers: [1265] and [1266]. The latter demonstrates that at $\theta = \pi$ there is a discrete 't Hooft anomaly involving time reversal and the center symmetry. It follows that at $\theta = \pi$ the vacuum cannot be a trivial non-degenerate gapped state.

## 5.9 Factorization and spin asymmetries

**Jianwei Qiu**

### 5.9.1 QCD Factorization

Hadrons, such as the proton, neutron and pion, are relativistic bound states of strongly interacting quarks and gluons of QCD. Without being able to see any quark or gluon directly in isolation, owing to the color confinement of QCD, it has been an unprecedented intellectual challenge to explore and quantify internal structure of hadrons in terms of their constituents, quarks and gluons, and the emergence of hadrons from quarks or gluons. Actually, the QCD color interaction is so strong at a typical hadronic scale $\mathcal{O}(1/R)$ with a hadron radius $R \sim 1$ fm that any scattering cross section with identified hadron(s) cannot be calculated fully in QCD perturbation theory.

QCD factorization [224] has been developed to describe high energy hadronic scattering with a large momentum transfer $Q \gg 1/R \sim \Lambda_{\mathrm{QCD}}$ by taking the advantage of the asymptotic freedom of QCD by which the color interaction becomes weaker and calculable perturbatively at short distances. QCD factorization provides a controllable and consistent way to *approximate* QCD contributions to *good* or factorizable hadronic cross sections by demonstrating

- all process-dependent nonperturbative contributions to these *good* cross sections are suppressed by powers of $\Lambda_{\mathrm{QCD}}/Q$, which could be neglected if the hard scale $Q$ is sufficiently large,
- all factorizable nonperturbative contributions are process independent, representing the characteristics of identified hadron(s), and
- the process dependence of factorizable contributions is perturbatively calculable from partonic scattering at the short-distance.

With our ability to calculate the process-dependent short distance partonic scatterings perturbatively at the hard scale $Q$, the prediction of QCD factorization follows when cross sections with different hard scatterings but the same nonperturbative long-distance effect of identified hadron are compared. QCD Factorization also supplies physical content to these perturbatively uncalculable, but universal long-distance effects of identified hadrons by matching them to hadronic matrix elements of active quark and/or gluon operators, which could be interpreted as parton distribution or correlation functions of the identified hadrons, and allows them to be measured experimentally or by numerical simulation.

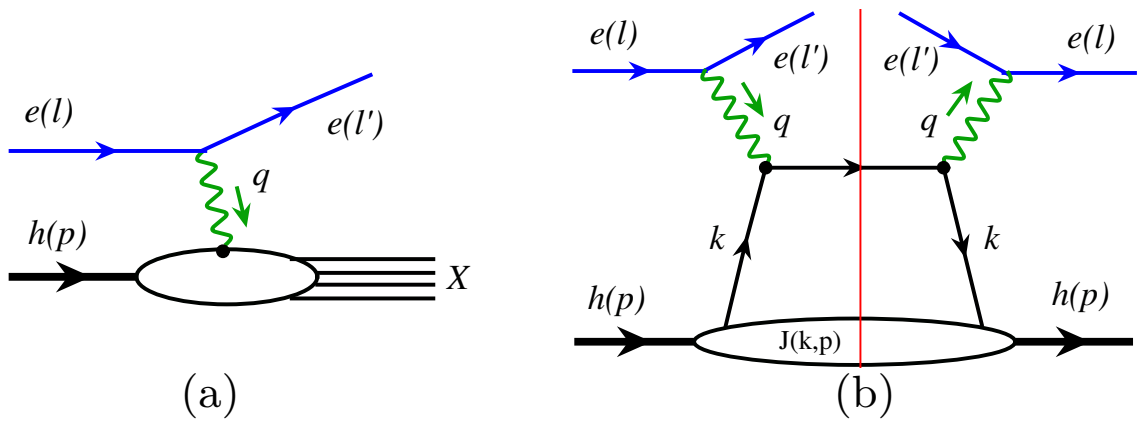

**Fig. 5.9.1** (a) Sketch for scattering amplitude of inclusive DIS. (b) Leading order contribution to inclusive DIS cross section in its cut diagram notation.

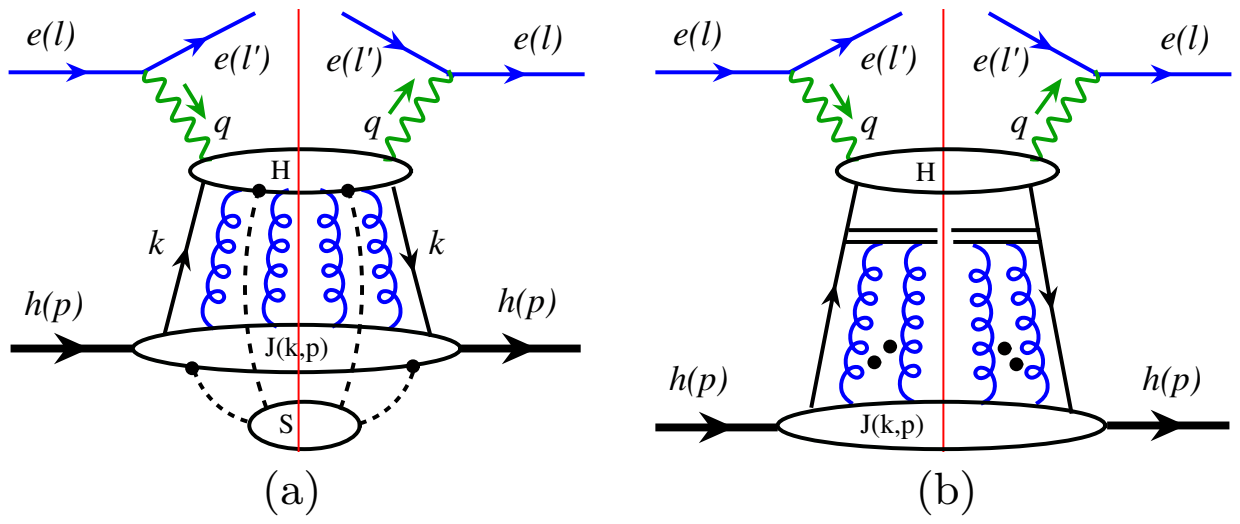

**Fig. 5.9.2** (a) Pinch surface for inclusive DIS with collinear and longitudinally polarized gluons (curly lines) and soft gluons (dashed lines). (b) Leading power factorized contribution to inclusive DIS with all collinear and longitudinally polarized gluons detached from the hard part $H$ and reconnected to the gauge links.

*Inclusive scattering with one identified hadron*

The deeply inelastic scattering (DIS) between a lepton $e$ of momentum $l$ and a hadron $h$ of momentum $p$, $e(l) + h(p) \to e(l') + X$, as shown in Fig. 5.9.1(a) where $l'$ is scattered lepton momentum and $X$ represents all possible final states, is an inclusive scattering with one identified hadron. With a large momentum transfer, $q = l - l'$ and $Q \equiv \sqrt{-q^2} \gg \Lambda_{\rm QCD}$, the DIS experiment at SLAC in 1969 discovered the point-like spin-1/2 partons/quarks inside a proton [93], which helped the discovery and formulation of QCD.

For inclusive DIS with two characteristic scales: $Q \gg \Lambda_{\rm QCD}$, QCD factorization is to consistently separate QCD dynamics taking place at these two distinctive scales by examining scattering amplitudes in terms of general properties of Feynman diagrams in QCD perturbation theory, leading to a factorization formalism, which is an approximation up to corrections suppressed in powers of $\Lambda_{\rm QCD}/Q$. For example, considering the leading order (LO) contribution to the inclusive DIS, as presented in Fig. 5.9.1(b) in its cut diagram notation, in which graphical contributions to the cross sections are represented by the scattering amplitude to the left of the final state cut (the red thin line) and the complex conjugate amplitude to the right, the scattering between the lepton of momentum $l$ and a quark (or a parton) of momentum $k$, $\hat{\sigma}^{\rm LO}(Q,k)$, is taken place at the hard scale $Q$, while the dynamics describing the quark inside the hadron, $J(k,p)$, is at the hadronic scale $1/R \sim \Lambda_{\rm QCD}$. The validity of such perturbative QCD factorization requires the suppression of quantum interference between the dynamics taking place at these two different momentum scales, which requires that the dominant contributions to the factorized formalism should necessarily come from the phase space where the active parton(s) linking the dynamics at two different scales are forced onto their mass shells, and are consequently long-lived compared to the time scale of the hard collision at the scale $Q$. This requirement is naturally satisfied for the LO contribution in Fig. 5.9.1(b),

$$
\begin{aligned}
\sigma^{\rm LO}_{\rm DIS} &\propto \int d^4k \left[\hat{\sigma}^{\rm LO}(Q,k)\frac{1}{k^2+i\varepsilon}J(k,p)\frac{1}{k^2-i\varepsilon}\right] \\
&\approx \int \frac{dk^+}{2k^+}d^2k_T\;\hat{\sigma}^{\rm LO}(Q,\hat{k}) \qquad\qquad (5.9.1)\\
&\quad\times \int dk^2\frac{1}{k^2+i\varepsilon}J(k,p)\frac{1}{k^2-i\varepsilon} + \mathcal{O}\left[\frac{\Lambda^2_{\rm QCD}}{Q^2}\right]
\end{aligned}
$$

where light-quark mass was neglected, and active quark of momentum $k$ is perturbatively pinched to be on-shell, $k^2 \approx \hat{k}^2 = 0$ with

$$
\hat{k} = (k^+, \frac{k_T^2}{2k^+}, \vec{k}_T)
$$

in the notation of light-cone coordinates, leading to a factorization formalism in Eq. (5.9.1) with all perturbatively pinched poles absorbed into the nonperturbative function of the identified hadron.

However, beyond the LO inclusive DIS, all internal loop momentum integrals to any scattering amplitude are defined by contours in complex momentum space, and it is only at momentum configurations where some subset of loop momenta are pinched that the contours are forced to or near mass-shell poles that correspond to long-distance behavior. The importance of such *pinched surfaces* in multidimensional momentum space was identified in the Libby-Sterman analysis [1267, 1268] that categorized all loop momenta into three groups: hard, collinear, and soft, along with the *reduced diagrams* by contracting off-shell lines to points, from which factorization formalisms can be derived. As shown in Fig. 5.9.2(a) for inclusive DIS, the identified hadron is associated with a group of collinear parton lines, and at the leading power, one physically polarized collinear parton plus infinite longitudinally polarized collinear gluons (curly lines) link the identified hadron to the hard part, $H$, in which all parton lines are off-shell by the hard scale $Q$. At the same time, the soft gluon lines (dashed lines in Fig. 5.9.2(a)) can attach to both the hard and collinear lines of the identified hadron. Since all parton propagators in $H$ are off-shell by $Q$, a soft gluon attachment to any of these lines in $H$ is necessarily to increase the number of off-shell propagators in $H$, and effectively suppresses the hard part by an inverse power of $Q$, making the contribution power suppressed. Therefore, we do not need to consider soft contributions to the inclusive DIS cross section at the leading power in $1/Q$ expansion.

The collinear and longitudinally polarized gluons have their polarization vectors proportional to their momenta in a covariant gauge. By applying the Ward Identity, all attachments of collinear and longitudinally polarized gluons to the hard part $H$ can be detached and reconnected to the gauge link pointing to the "−" light-cone direction if the identified hadron is moving in the "+" light-cone direction [224, 1269], as sketched in Fig. 5.9.2(b). After taking the leading power contribution from the spinor trace of the active quark line in Fig. 5.9.2(b) [1269, 1270], the inclusive DIS cross section at the leading power can be factorized as [1271–1273]

$$E'\frac{d\sigma^{\rm DIS}_{eh\to eX}}{d^3l'}(l,p;l') = \sum_{f=q,\bar{q},g}\int dx\;\phi_{f/h}(x,\mu^2) \times E'\frac{d\hat{\sigma}_{ef\to eX}}{d^3l'}(l,\hat{k};l',\mu^2) + \mathcal{O}\left[\frac{\Lambda^2_{\rm QCD}}{Q^2}\right] \tag{5.9.2}$$

where $\hat{k}\equiv xp^+$, $l'_T\sim Q\gg\Lambda_{\rm QCD}$, and $E'd\hat{\sigma}_{ef\to eX}/d^3l'$ is the short-distance part of DIS cross section on a parton state of flavor $f$ and collinear momentum fraction $x$ of the colliding hadron, with its long-distance contributions to the cross section systematically absorbed into the non-perturbative functions $\phi_{f/h}(x,\mu^2)$, which are defined in terms of hadronic matrix elements of active parton operators [1274]. For example, for an unpolarized active quark,

$$\phi_{q/h}(x,\mu^2) = \int\frac{d\xi^-}{2\pi}e^{ixp^+\xi^-} \times \langle h(p)|\overline{\psi}_q(0)\frac{\gamma^+}{2}\mathcal{W}_{[0,\xi^-]}\psi_q(\xi^-)|h(p)\rangle\,, \tag{5.9.3}$$

where $\mathcal{W}_{[0,\xi^-]}=\mathcal{P}\exp\left[ig\int_0^{\xi^-}d\eta^-A^+(\eta^-)\right]$ is the gauge link. The $\phi_{f/h}(x,\mu^2)$ carries nonperturbative information of the identified hadron, and is referred as an universal parton distribution function (PDF) for finding a parton of flavor $f$ inside a colliding hadron $h$, carrying its momentum fraction $x$, probed at a hard factorization scale $\mu\sim Q$. PDFs are discussed in more detail in Sec. 10.2.

With the precise definition of $\phi_{f/h}(x,\mu^2)$, the QCD factorization formalism, such as the one in Eq. (5.9.2), provides a systematic way to calculate the short-distance partonic scattering, $E'd\hat{\sigma}_{ef\to eX}/d^3l'$, in QCD perturbation theory. By applying the factorization formalism in Eq. (5.9.2) to a parton state of flavor $f$, $|h(p)\rangle\to|f(p)\rangle$, we can use perturbation theory to calculate the short distance partonic scattering order-by-order in powers of the strong coupling constant $\alpha_s$ by perturbatively calculating the DIS cross section on a parton of flavor $f$ on the left of Eq. (5.9.2), and PDFs of the same parton on the right, with the collinear divergence regularized. This leads to

$$E'\frac{d\hat{\sigma}^{(\rm n)}_{ef\to eX}}{d^3l'} = E'\frac{d\sigma^{\rm DIS(n)}_{ef\to eX}}{d^3l'}(l,p;l') - \sum_{\rm m=0}^{\rm n-1}\left[\sum_{f'=q,\bar{q},g}E'\frac{d\hat{\sigma}^{(\rm m)}_{ef'\to eX}}{d^3l'}\otimes\phi^{(\rm n-m)}_{f'/f}(x,\mu^2)\right] \tag{5.9.4}$$

where superscripts, n and m, indicate the order in power of $\alpha_s$. QCD factorization ensures that the collinear sensitivities from scattering off a parton on the left of Eq. (5.9.2) to be exactly cancelled by the same sensitivities from the PDFs of the same parton on the right [1269]. That is, the subtraction term in Eq. (5.9.4) cancels all long-distance physics of the partonic scattering cross section on a parton state of flavor $|f(p)\rangle$.

The inclusive DIS cross section can be physically measured in experiments and should not depend on how we describe it in terms of QCD factorization, or the choice of factorization scale $\mu$. That is, we require

$$d\sigma_{eh\to eX}/d\log\mu^2 = 0,$$

which implies evolution equations of PDFs, known as the DGLAP equations [211, 215]

$$\frac{d\phi_{f/h}(x,\mu^2)}{d\log\mu^2} = \sum_{f'}\int_x^1\frac{dx'}{x'}P_{f/f'}\left(\frac{x}{x'},\alpha_s(\mu^2)\right) \times \phi_{f'/h}(x',\mu^2) \tag{5.9.5}$$

where the evolution kernels $P_{f/f'}(x/x',\alpha_s(\mu^2))$ are calculable in perturbative QCD when the strong coupling constant $\alpha_s(\mu)$ is sufficiently small [216, 217]. Although PDFs are nonperturbative, their factorization scale dependence is a QCD prediction, which has been confirmed to great accuracy [626, 627].

Another example of factorizable inclusive cross section with one identified hadron is single inclusive hadron production in high energy electron-positron collision, $e^-(l)+e^+(l')\to h(p)+X$ with the observed hadron energy $E_p\gg\Lambda_{\rm QCD}$, as sketched in Fig. 5.9.3(a). Like the inclusive DIS in Eq. (5.9.1), the active parton momentum $k$, in Fig. 5.9.3(b), linking the hard $e^+e^-$ annihilation that produces this active parton, and describes how it hadronizes into the observed hadron, is perturbatively pinched to its mass-shell, which is necessary

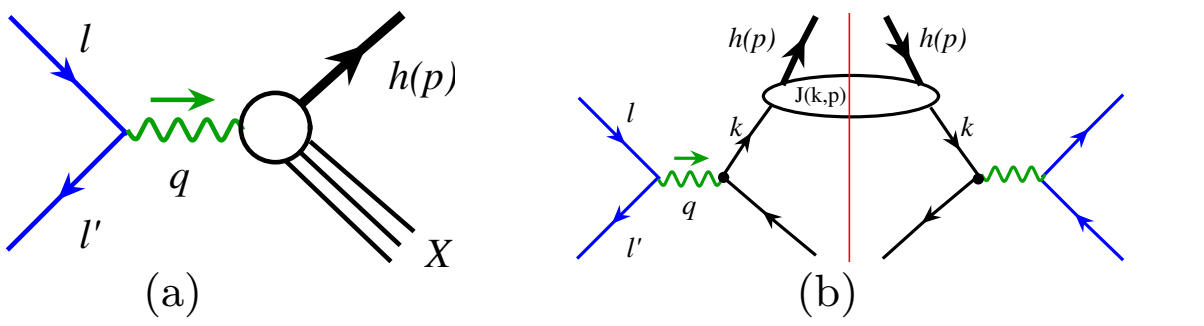

**Fig. 5.9.3** (a) Sketch for scattering amplitude of inclusive single hadron production in high energy $e^+e^-$ collisions. (b) Leading order contribution to inclusive single hadron production in its cut diagram notation.

for the factorization. For the leading power contribution beyond the LO in Fig. 5.9.3(b), similar to inclusive DIS, we do not need to worry about soft interaction between the hard part and the collinear partons along the direction of the produced hadron. By applying the Ward Identity, in the same way as in the factorization of inclusive DIS, the attachment of collinear and longitudinally polarized gluons from the observed hadron to the hard part, $H$, can be detached and reconnected to the gauge link to become a part of the nonperturbative, but universal, fragmentation functions (FFs) of the identified hadron, leading to the factorization formalism,

$$E_p \frac{d\sigma_{e^+e^-\to hX}}{d^3p}(s,p) = \sum_f \int \frac{dz}{z^2} D_{h/f}(z,\mu^2) \times E_k \frac{d\hat{\sigma}_{e^+e^-\to \hat{k}X}}{d^3\hat{k}}(s,\hat{k},\mu^2) + \mathcal{O}\left[\frac{\Lambda^2_{\rm QCD}}{Q^2}\right] \qquad (5.9.6)$$

where active parton momentum is $\hat{k} = p/z$,

$$\sqrt{s} = \sqrt{(l+l')^2}$$

is the collision energy, and $D_{h/f}(z,\mu^2)$ is the FF to find a hadron $h$ emerged from a produced parton of flavor $f$ while carrying the parton's momentum fraction $z$ [1274]. The fact that such a physical cross section should not depend on how we factorized it implies evolution equations for the FFs, like DGLAP for PDFs.

From the QCD factorization formalisms involving one identified hadron in Eqs. (5.9.2) and (5.9.6), extracting the universal PDFs and FFs from experimental data is a challenging inverse problem. Although the scale dependence of PDFs and FFs is a prediction of QCD dynamics, measurements of such cross sections with one identified hadron are not sufficient to disentangle the flavor and momentum fraction dependence of all PDFs and FFs, which are necessary for the predictive power of the QCD factorization approach to describe high energy hadronic cross sections.

*Inclusive scattering with two identified hadrons*

The Drell-Yan (DY) production of lepton pairs via a vector boson in hadron-hadron collisions, $A(p)+B(p') \to V(q) + X$ with $V(q)[= \gamma^*, W/Z, H^0, ...] \to l + l'$, as sketched in Fig. 5.9.4(a), is an ideal example of the study of QCD factorization for inclusive observables with two identified hadrons [224].

From the LO contribution in Fig. 5.9.4(b), both active partons (quark or antiquark) of momentum $k$ and $k'$ coming from colliding hadrons $A(p)$ and $B(p')$, respectively, are perturbatively pinched to their mass-shell, which is necessary for being able to factorize the nonperturbative hadronic information of colliding hadrons from the hard collision to produce the massive lepton pairs. Beyond the LO, each colliding hadron is associated with a group of collinear partons, and for the leading power contribution, only one physically polarized active parton plus infinite collinear and longitudinally polarized gluons from each hadron should attach to the hard part, $H$, with the remaining collinear partons forming a (spectator) jet function, which is the same as the inclusive scattering with one identified hadron. The key difference for QCD factorization of inclusive scattering with two identified hadrons from that with one hadron, according to the Libby-Sterman analysis [1267, 1268], is the soft interaction between the collinear partons of two different hadrons, as shown by the dashed lines in Fig. 5.9.5(a). Still the soft interaction between the collinear partons and the hard part can be neglected when calculating the leading power contributions. However, these long-distance soft interactions between hadrons have the potential to break the universality of the factorizable nonperturbative contribution from each identified hadron, and invalidate the predictive power of the QCD factorization approach for studying hadronic cross sections with identified hadrons.

When the colliding hadrons $A(p)$ and $B(p')$ are moving in the $+z$ and $-z$ direction, respectively, the factorization of collinear and longitudinally polarized gluons from the hard part $H$ is effectively the same as what was done for the case of single identified hadron. Since collinear and longitudinally polarized gluons have their polarization vectors proportional to their momenta in a covariant gauge, by applying the Ward Identity all collinear and longitudinally polarized gluons from hadron $A(p)$ can be detached from the hard part and reconnected to the gauge link in the "$-$" light-cone direc-

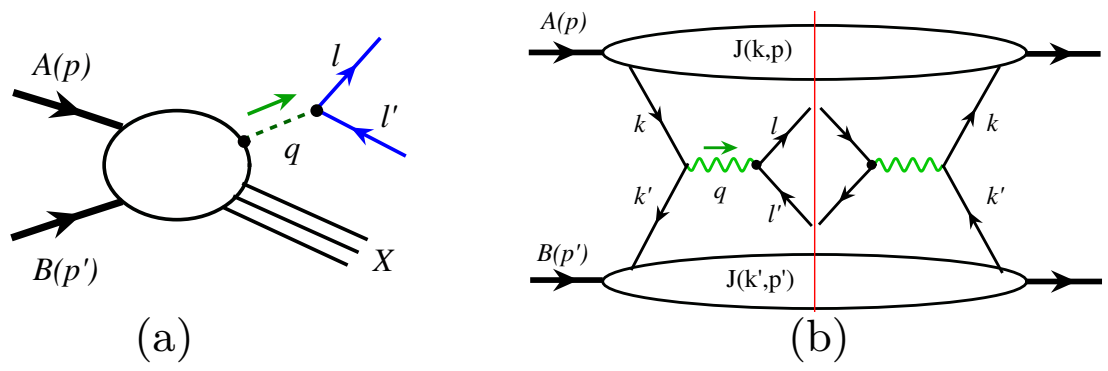

**Fig. 5.9.4** (a) Sketch for scattering amplitude of Drell-Yan production of a massive lepton pair. (b) Leading order contribution to the Drell-Yan cross section in its cut diagram notation.

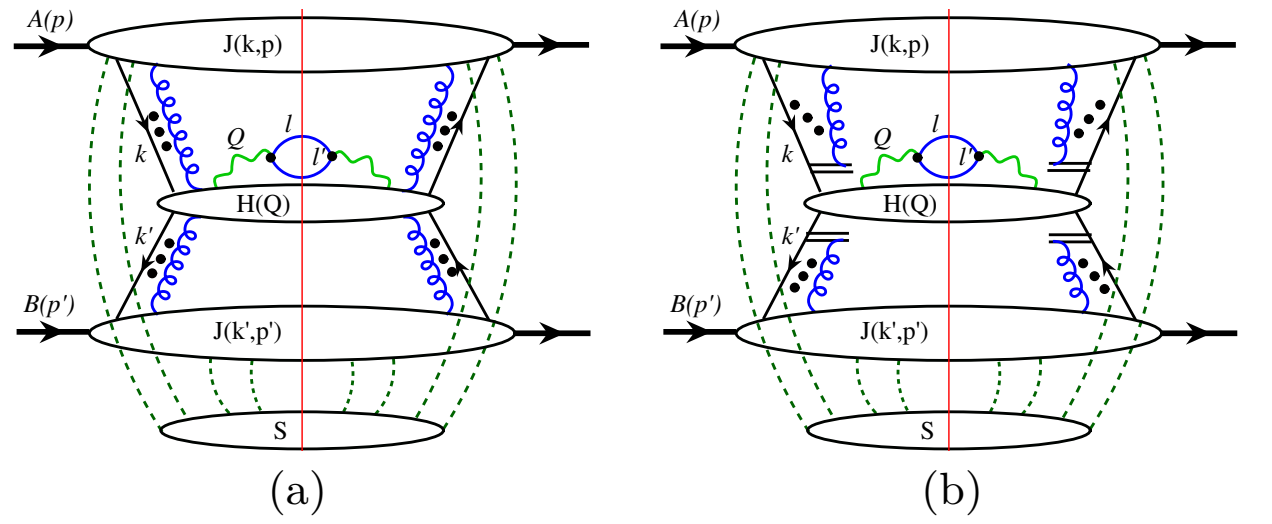

**Fig. 5.9.5** (a) Sketch for the leading QCD pinch surface for Drell-Yan production of lepton pair with collinear and longitudinally polarized gluons in curly lines and soft gluons in dashed lines. (b) QCD contribution to Drell-Yan process with all collinear and longitudinally polarized gluons detached from the hard part and reconnected to the gauge lines.

tion, while those from hadron $B(p')$ can be reconnected to the gauge link in the "+" light-cone direction, as sketched in Fig. 5.9.5(b).

In order to achieve the factorization, we need to get rid of the soft gluon interactions, the dashed lines in Fig. 5.9.5(b). If we scale collinear parton momenta from colliding hadron $A$, $k_i = (k_i^+, k_i^-, k_i^T) \sim (1, \lambda^2, \lambda)Q$ with $\lambda \sim \mathcal{O}(\Lambda_{\rm QCD}/Q)$, we maintain $k_i^2 \sim \mathcal{O}(\lambda^2 Q^2) \to 0$ as the loop momenta approach to the pinch surface. If we can choose soft gluon loop momenta to have the scaling behavior, $l_s \sim (\lambda_s, \lambda_s, \lambda_s)Q$, where $\lambda_s \sim \lambda^2$ (or $\lambda$) to have all components vanishing at the same rate, we have $(k_i + l_s)^2 \sim 2k_i^+ l_s^- \sim \mathcal{O}(\lambda^2/Q^2)$. That is, we only need to keep the "−" component of soft gluon momenta to flow into the jet of collinear partons from the colliding hadron $A$, whose leading components of Lorentz indices that interact with the soft gluons are in the "+" direction in a covariant gauge. Therefore, we can use the Ward Identity to detach the soft gluons from the jet of collinear partons from colliding hadron $A$ and reconnect them into a gauge link or an eikonal line. Applying the same reasoning with the role of the "±" components switched, we can detach all soft gluon interactions to the jet of collinear partons from colliding hadron $B$, and to factorize all soft gluon interactions with two colliding hadrons into an overall soft factor, as shown in Fig. 5.9.6.

However, this factorization can fail if the soft gluon momenta are trapped in the Glauber region. In this region the "±" components of the soft gluons are small compared to their transverse components, *i.e.* $l_s^{\pm}/l_s^T \sim \mathcal{O}(\lambda)$, so that we cannot neglect the transverse components, keeping only one "+" or "−" components [224]. It is the soft gluon interaction between the spectators of two colliding hadrons that can trap the ± components of the soft gluon momenta in the Glauber region. For example, in Fig. 5.9.7, the pair of propagators of momenta, $p-k-l$ and $k+l$, pinches the "−" component of $l$ to be, $l^- \propto l_T^2$, while the pair of propagators of momenta, $p'-k'+l$ and $k'-l$, pinches the "+" component of $l$ to be, $l^+ \propto l_T^2$, such that the soft gluon interaction between two jets of collinear partons from the colliding hadrons is pinched in the Glauber region; in this case the leading soft gluon interactions could break the universality of PDFs and the predictive power of the QCD factorization approach.

Removal of the trapped Glauber gluons might be the most difficult part of the QCD factorization proof [224]. It was achieved by three key steps: (1) all poles in one-half plane cancel after summing over all final-states (no more pinched poles), (2) all $l_s^{\pm}$-type integrations can be deformed out of the trapped soft region, and (3) all leading power spectator interactions can be factorized and summed into an overall unitary soft factor of gauge links (or eikonal lines) as argued above and shown in Fig. 5.9.6. The soft factor is process independent and made of four gauge links, along the light-cone directions conjugated to the directions of two incoming hadrons in the scattering amplitude, and the two in the complex conjugate scattering amplitude, respectively. For the collinear factorization, the soft factor = 1 due to the unitarity, and we have the corresponding factorization formalism for inclusive Drell-Yan production at the leading power,

$$\frac{d\sigma^{(\rm DY)}_{A+B\to ll'+X}}{dQ^2 dy} = \sum_{ff'} \int dx\, dx'\, \phi_{f/A}(x,\mu)\, \phi_{f'/B}(x',\mu) \times \frac{d\hat{\sigma}_{f+f'\to ll'+X}(x,x',\mu,\alpha_s)}{dQ^2 dy} + \mathcal{O}\left[\frac{\Lambda_{\rm QCD}^2}{Q^2}\right], \quad (5.9.7)$$

where $\sum_{ff'}$ runs over all parton flavors including quark and antiquark, as well as gluon.

To help separate the flavor dependence of PDFs, the lepton-hadron semi-inclusive DIS (SIDIS), $e(l)+h(p) \to$

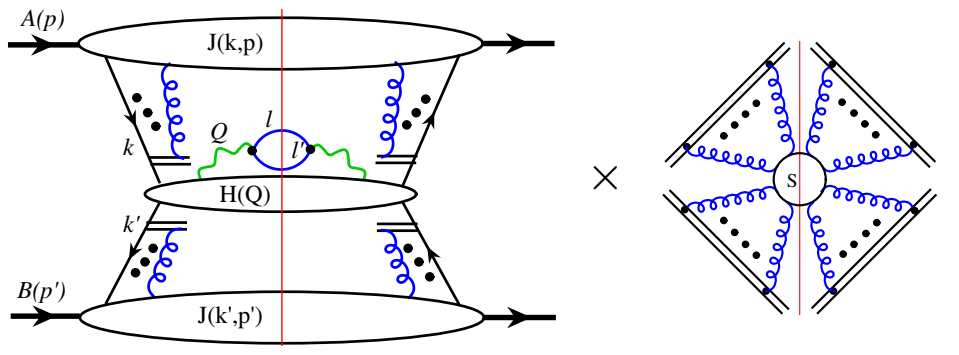

**Fig. 5.9.6** Sketch for factorized Drell-Yan production of a massive lepton pair at the leading power with all soft gluon interactions factorized into a multiplicative soft factor.

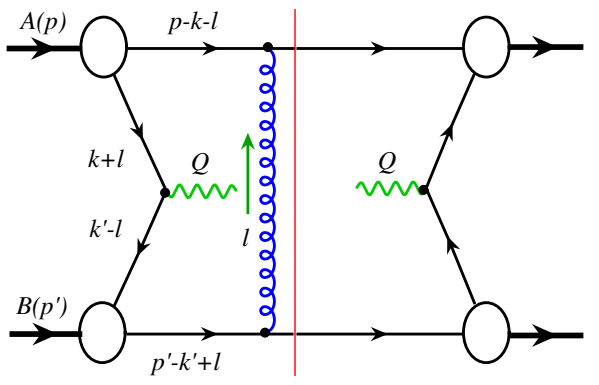

**Fig. 5.9.7** Sample diagram responsible for soft gluon interaction to have its momentum pinched in the Glauber region.

$e(l') + h'(p') + X$, as shown in Fig. 5.9.8(a), is another example of QCD factorization with two identified hadrons. From the LO contribution in Fig. 5.9.8(b), both active partons of momentum $k$ and $k'$ are perturbatively pinched to their mass-shell, leading to a potential factorization of PDF from colliding hadron and FF of the fragmenting parton to the observed hadron. Beyond the LO, like the Drell-Yan process, there could be soft interactions between the jet of collinear partons of the hadron $h$ and the jet of collinear partons along the direction of observed hadron $h'$.

Integrating over the transverse momentum of the observed final-state hadron to keep the SIDIS as a process with a single hard scale $Q$, and following the same factorization arguments for inclusive Drell-Yan processes, the SIDIS cross section can be factorized as

$$E' \frac{d\sigma^{\rm SIDIS}_{eh\to eh'X}}{d^3l'dz}(l,p;l',z) = \sum_{f,f'=q,\bar{q},g} \int dz'\, dx\, D_{h'/f'}(z',\mu^2)\,\phi_{f/h}(x,\mu^2) \times E'\frac{d\hat{\sigma}_{ef\to ef'X}}{d^3l'dz'}(l,\hat{k};l',z',\mu^2) + \mathcal{O}\left[\frac{\Lambda^2_{\rm QCD}}{Q^2}\right] \quad (5.9.8)$$

where $\hat{k} = xp$, $z' = p'/k'$ and $z = p\cdot p'/p\cdot q$.

Inclusive jet production in hadronic collisions: $A(p)+B(p') \to \sum_j J_j(p_j) + X$ is another observable with two identified hadrons although many hadrons were measured in the final-state when jets were constructed. When final-state jets are well-separated, the cross section for jets with large transverse energy has the same factorized formula as that in Eq. (5.9.7) except the perturbatively calculated hard part, $\hat{\sigma}_{ff'\to ll'X}$ is replaced

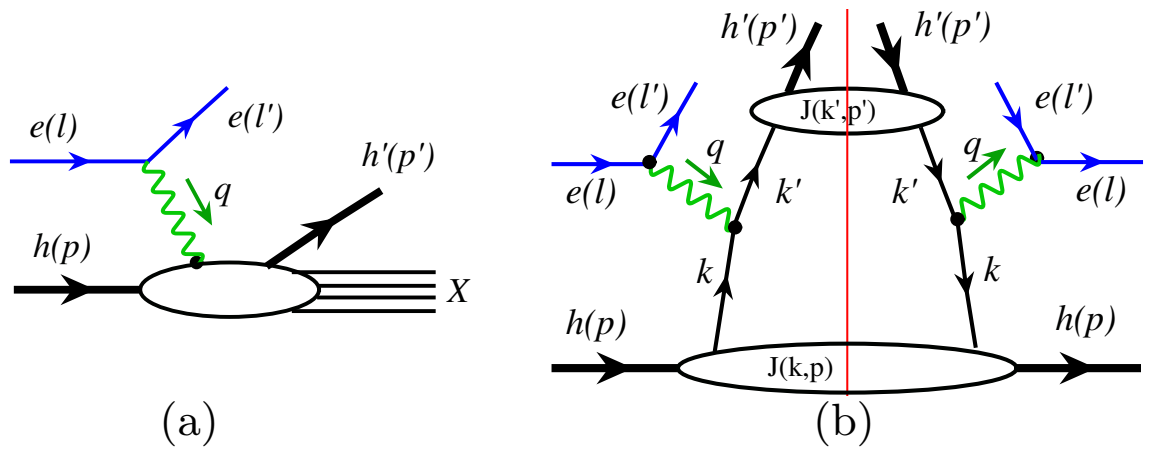

**Fig. 5.9.8** (a) Sketch for scattering amplitude of lepton-hadron SIDIS. (b) Leading order contribution to SIDIS cross section in its cut diagram notation.

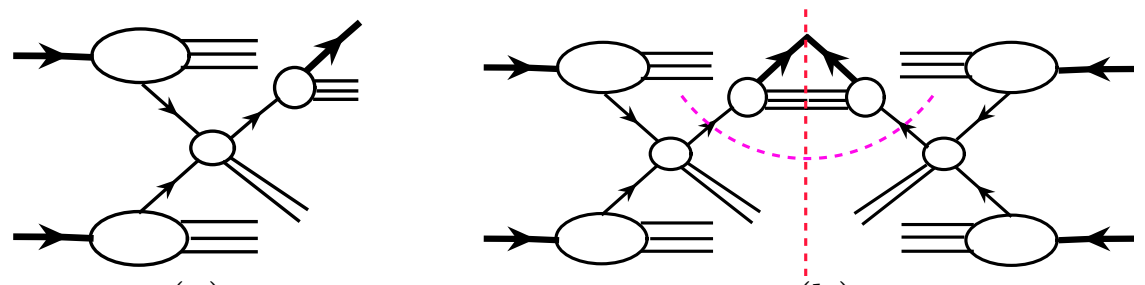

**Fig. 5.9.9** (a) Sketch for scattering amplitude of hadronic production of single hadron at large transverse momentum. (b) Its contribution to the cross section in the cut diagram notation.

by corresponding short-distance hard part, $\hat{\sigma}_{ff'\to {\rm Jet}}$ [1275].

$$\frac{d\sigma^{\rm (Jet)}_{A+B\to {\rm Jet}+X}}{dp_T dy} = \sum_{ff'}\int dx\, dx'\,\phi_{f/A}(x,\mu)\,\phi_{f'/B}(x',\mu) \times \frac{d\hat{\sigma}_{f+f'\to{\rm Jet}+X}(x,x',\mu,\alpha_s)}{dp_T dy}$$
$$= \sum_{ff'}\int dx\, dx'\,\phi_{f/A}(x,\mu)\,\phi_{f'/B}(x',\mu) \times \left[\sum_c \int \frac{dz}{z} J_c(z,p_T R,\mu)\frac{d\hat{\sigma}_{f+f'\to c+X}}{dp_{cT}dy_c} + \tilde{\sigma}(p_T,y)\right] \quad (5.9.9)$$

where $p_T$ and $y$ are the transverse momentum and rapidity of the observed jet, respectively. Like all perturbatively calculable hard parts of QCD factorization, the hard part for the jet production, $\hat{\sigma}_{ff'\to{\rm Jet}}$ is process-dependent, depending on whether the jet is produced in hadron-hadron or lepton-hadron collisions, as well as the choice of the jet algorithms. In Eq. (5.9.9), the process-dependent short-distance hard part for the jet production was reorganized into a process-independent jet function, $J_c$ from a single parton of flavor $c$, leaving all process-dependence into the production of this parton, $\hat{\sigma}_{f+f'\to c+X}$ and $\tilde{\sigma}(p_T,y)$ which might be neglected if logarithms of the jet production dominates [1276].

*Inclusive scattering with three identified hadrons*

Inclusive single hadron production at large transverse momentum $p_T$ in hadronic collisions: $A(p)+B(p') \to h(p_h)+X$ is a well-measured observable involving three identified hadrons, as shown in Fig. 5.9.9. Due to the additional identified hadron in the final-state, the unitarity sum of final-state hadrons used to prove the factorization of DY-type two-hadron observables needs to be modified.

Luckily, because of the large $p_T$ of the observed final-state hadron, the potentially dangerous gluon interactions between the observed hadron and the spectators of colliding hadrons are suppressed by the power of $1/p_T$, and the leading power pQCD factorization does

hold [1277],

$$\frac{d\sigma_{AB\to hX}(p,p,p_h)}{dy\,dp_T^2} = \sum_{f,f',c}\int \frac{dz}{z^2}\,dx\,dx'\,D_{h/c}(z,\mu^2) \times \phi_{f/A}(x,\mu^2)\,\phi_{f'/B}(x',\mu^2) \times \frac{d\hat{\sigma}_{ff'\to cX}(x,x',p_c=p_h/z)}{dy_c\,dp_{cT}^2}. \quad (5.9.10)$$

With proper PDFs and FFs, the NLO pQCD calculations for single hadron production gave an excellent description of RHIC data [1278]. However, the same formalism consistently underestimates the production rate at the fixed target energies [1279]. It was shown that high order corrections at the fixed target energies are very important, and the threshold resummation significantly improves the comparison between the theory and experimental data [1280].

*QCD global analysis and predictive power*

Much of the predictive power of QCD factorization for cross sections involving identified hadron(s) relies on the universality of the PDFs and/or FFs and our ability to solve the inverse problem to demonstrate the existence of one set of PDFs and FFs that are capable of describing all data of *good* (e.g. factorizable) cross sections with properly calculated short-distance partonic scattering cross sections in QCD perturbation theory.

The QCD global analysis is a program to test the consistency of QCD factorization by fitting all existing data from high energy scatterings with universal PDFs and/or FFs and corresponding factorization formalisms, from which the best set of PDFs and/or FFs will be extracted. The QCD factorization formalism has been extremely successful in interpreting high energy experimental data from all facilities around the world, covering many orders in kinematic reach in both parton momentum fraction $x$ and momentum transfer of the hard collision $Q$, and as large as 15 orders of magnitude in difference in the size of observed scattering cross sections, which is a great success story of QCD and the Standard Model at high energy. It has given us the confidence and the tools to discover the Higgs particle in proton-proton collisions [120, 121], and to search for new physics [1281].

*QCD factorization for two-scale observables*

The hard probe with a single large momentum transfer $Q\,(\gg 1/R)$ is so localized in space that it is not very sensitive to the details of confined three-dimensional (3D) internal structure of the colliding hadron, in which a confined parton should have a characteristic transverse momentum scale $\langle k_T\rangle \sim 1/R \ll Q$ and an uncertainty in transverse position $\langle b_T\rangle \sim R \gg 1/Q$. Recently, new and more precise data are becoming available for *two-scale* observables with a hard scale $Q$ to localize the collision to probe the partonic nature of quarks and gluons along with a soft scale to be sensitive to the dynamics taking place at $\mathcal{O}(1/R)$. At the same time, theory has made major progresses in the development of QCD factorization formalism for two types of two-scale observables, distinguished by their inclusive or exclusive nature, which enables quantitative matching between the measurements of such two-scale observables and the 3D internal partonic structure of a colliding hadron.

For inclusive two-scale observables, one well-studied example is the Drell-Yan production of a massive boson that decays into a pair of measured leptons in hadron-hadron collisions as a function of the pair's invariant mass $Q$ and transverse momentum $q_T$ in the Lab frame [1282]. When $Q \gg q_T \gtrsim 1/R$, the measured transverse momentum of the pair is sensitive to the transverse momenta of the two colliding partons before they annihilate into the massive boson, providing the opportunity to extract the information on the active parton's transverse motion at the hard collision, which is encoded in transverse momentum dependent (TMD) PDFs (or simply, TMDs), $\phi_{f/h}(x,k_T,\mu^2)$ [1282].

Like PDFs, TMDs are universal distribution functions describing how a quark (or gluon) with a momentum fraction $x$ and transverse momentum $k_T$ interacts with a colliding hadron of momentum $p$ with $xp \sim \mu \sim Q \gg k_T$. Another well-studied example is the SIDIS when the produced hadron is almost back-to-back to the scattered lepton in the Lab frame, or in the Breit frame, the transverse momentum of the produced hadron $p_{hT}$ is much smaller than the hard scale $Q$ [1283, 1284].

A necessary condition for QCD factorization of observables with identified hadron(s) is that the active parton linking the process-dependent short-distance dynamics and the process-independent nonperturbative physics of identified hadron(s) is perturbatively pinched to its mass-shell so that it is long-lived compared to the time scale of the hard collision. In this case the quantum interference between the perturbatively calculable hard collisions at the hard scale $Q$ and the process-independent part of leading nonperturbative information of the identified hadron(s) is strongly suppressed by the power of $\Lambda_{\rm QCD}/Q$. The pinch does not require the active parton's momentum to be collinear to the hadron momentum. The necessary condition is satisfied if the active parton momentum has a transverse component with $\langle k_T\rangle \ll Q$; the same condition that should be satisfied by the TMD factorization of Drell-Yan and SIDIS process for the leading power contribution in $q_T/Q$ or $p_{hT}/Q$, respectively. Although this condition is not necessarily sufficient, the TMD factorization for Drell-Yan process at the leading power of $q_T/Q \to 0$ was justified [1269, 1282], and the same for the SIDIS at leading power of $p_{hT}/Q$ [1285–1287]. More discussion on the impact of TMD factorization for the spin asymmetries will be given in Sec. 5.9.2.

Without breaking the colliding hadron, the exclusive observables could provide different aspects of the hadron's internal structure. Exclusive lepton-nucleon scattering with a virtual photon of invariant mass $Q \gg 1/R$ could provide various two-scale observables, such as the deeply virtual Compton scattering (DVCS) [1288], where the hard scale is $Q$ and the soft scale is $t \equiv (p-p')^2$. When $Q \gg \sqrt{|t|}$, such two-scale exclusive processes are dominated by the exchange of an active $q\bar{q}$ or $gg$ pair and can be systematically treated using the QCD factorization approach; factorized in terms of generalized PDFs or GPDs [1289–1292]. Recently, a new class of *single diffractive hard exclusive processes (SDHEP)* was introduced [1293, 1294]. This approach is not only sufficiently generic to cover all known processes for extracting GPDs, but also well-motivated for the search of new processes for the study of GPDs. It was demonstrated that many of those new processes can be factorized in terms of GPDs and could provide better sensitivity to the parton momentum fraction $x$ dependence of GPDs.

### 5.9.2 Spin Asymmetries

A measured cross section is always a positive and classical probability even though its underlying dynamics could be sensitive to quantum effects. On the other hand, a spin asymmetry, defined to be proportional to a difference of two cross sections with one (or more) spin vector(s) flipped, can probe QCD dynamics that a spin-averaged cross section is not sensitive to, and provide a better chance to explore the dynamics of quantum effects. It also provides opportunities to explore the origin of proton spin by carrying out scattering experiments with polarized protons.

*Quark and gluon contributions to proton spin*
The leading power collinear factorization formalisms can also apply to asymmetries of cross sections between two longitudinally polarized particles [1269]. Instead of measuring nonperturbative PDFs of a hadron, the double longitudinal spin asymmetry

$$A_{LL} \equiv \frac{\Delta\sigma}{\sigma} = \frac{\sigma(++) - \sigma(+,-)}{\sigma(+,+) + \sigma(+,-)}, \tag{5.9.11}$$

where $\pm$ indicates the helicity of the active parton compared to the longitudinal spin direction of the colliding particle, is sensitive to the active parton's helicity distribution inside a polarized colliding hadron. The double longitudinal spin-dependent cross sections, $\Delta\sigma$ is given by the same factorization formalisms introduced in the Sec. 5.9.1 with the spin-averaged collinear PDFs replaced by corresponding helicity distributions,

$$\begin{aligned}\phi_{f/h}(x,\mu^2) &\to \Delta\phi_{f/h}(x,\mu^2)\\ &= \frac{1}{2}\Big[\phi_{+/+}(x,\mu^2) - \phi_{-/+}(x,\mu^2)\Big].\end{aligned}$$

The same leading power collinear factorization formalisms introduced in the Sec. 5.9.1 can also apply to parity violating single longitudinal spin asymmetries of cross sections between one unpolarized and one longitudinally polarized particles,

$$A_L \equiv \frac{\sigma(+) - \sigma(-)}{\sigma(+) + \sigma(-)}. \tag{5.9.12}$$

The single longitudinal spin-dependent cross section, $\Delta\sigma = \sigma(+) - \sigma(-)$ with spin direction of the polarized parton flipped is also given by the same factorization formalisms by replacing one of the spin-averaged collinear PDFs, corresponding to the hadron that is replaced by a polarized colliding particle, by corresponding helicity distribution. With the flavor sensitivities of the weak interaction, the single longitudinal spin asymmetries measured by the RHIC spin program have provided important information on the flavor separation of quark helicity distributions [820, 1278].

The double and single longitudinal spin asymmetries, defined in Eqs. (5.9.11) and (5.9.12), respectively, have been studied in both hadron-hadron collisions at RHIC [1278] and lepton-hadron collisions [1295, 1296], and will be a major program at the future EIC [820].

After over 30 years since the discovery of EMC collaboration, and many experiments carried out worldwide, the RHIC spin program in particular, we learned from the momentum fractions $x$ that these experiments could access, that the proton spin gets about 30% from quark helicity and 40% from gluon helicity, and the rest could come from the region of $x$ that we have not been able to explore and/or orbital or transverse motion of quarks and gluons inside the bound proton [820]. (See the discussion in Sec. 10.2.)

*Double transverse-spin asymmetries*
The double transverse spin asymmetries are,

$$A_{NN} = \frac{\sigma(\uparrow,\uparrow) - \sigma(\uparrow,\downarrow)}{\sigma(\uparrow,\uparrow) + \sigma(\uparrow,\downarrow)},$$

where $\uparrow$ and $\downarrow$ indicate the direction of spin vectors transverse to the momentum direction of the colliding particles. Since QCD factorization requires that the factorized short-distance dynamics is not sensitive to the details of hadronic physics, the spin asymmetries are proportional to the difference of hadronic matrix elements of parton fields with the hadron spin flipped,

$$A \propto \sigma(Q,\vec{s}) - \sigma(Q,-\vec{s}) \\ \propto \langle p,\vec{s}|\mathcal{O}(\psi_q, A_g^\mu)|p,\vec{s}\rangle - \langle p,-\vec{s}|\mathcal{O}(\psi_q, A_g^\mu)|p,-\vec{s}\rangle. \tag{5.9.13}$$

The parity and time-reversal invariance of QCD requires

$$\langle p,\vec{s}|\mathcal{O}(\psi_q, A_g^\mu)|p,\vec{s}\rangle \\ = \langle p,-\vec{s}|\mathcal{P}\mathcal{T}\mathcal{O}^\dagger(\psi_q, A_g^\mu)\mathcal{T}^{-1}\mathcal{P}^{-1}|p,-\vec{s}\rangle. \tag{5.9.14}$$

Therefore, only partonic operators $\mathcal{O}(\psi_q, A_g^\mu)$ satisfying

$$\langle p,-\vec{s}|\mathcal{P}\mathcal{T}\mathcal{O}^\dagger(\psi_q, A_g^\mu)\mathcal{T}^{-1}\mathcal{P}^{-1}|p,-\vec{s}\rangle \\ = \pm\langle p,-\vec{s}|\mathcal{O}(\psi_q, A_g^\mu)|p,-\vec{s}\rangle \tag{5.9.15}$$

or

$$\langle p,\vec{s}|\mathcal{O}(\psi_q, A_g^\mu)|p,\vec{s}\rangle = \pm\langle p,-\vec{s}|\mathcal{O}(\psi_q, A_g^\mu)|p,-\vec{s}\rangle \tag{5.9.16}$$

contribute to the factorizable spin asymmetries. Those operators that lead to a "+" sign should contribute to spin-averaged cross sections, while those lead to a "−" sign should contribute to spin asymmetries. Only the leading twist quark operator that defines the quark transversity distribution $\delta q(x,\mu^2)$

$$\delta q(x,\mu^2) = \overline{\psi}_q(0)\gamma^+\gamma^\perp\gamma_5\psi_q(\xi^-),$$

(or $h_1(x,\mu^2)$), is relevant to the double transverse spin asymmetries of observables with a single large momentum transfer $Q$ in proton-proton collisions of transversely polarized protons.

The QCD factorization for the leading power contribution to the Drell-Yan production of a massive lepton pair in a collision with two transversely polarized protons should follow the same arguments that led to that in Fig. 5.9.6. Here all collinear and longitudinally polarized gluons factorized into gauge links, and soft gluon interactions are factorized into an overall soft-factor. The factorization of spinor traces of the Fermion lines needs to be modified to reflect the transverse-spin projector $\gamma^\pm\gamma^\perp\gamma_5$ (where $\pm$ indicates the two possibilities due to two colliding hadrons) instead of the $\gamma^\pm$ and $\gamma^\pm\gamma_5$ for unpolarized and longitudinally polarized active quarks. Therefore, the QCD factorization formalism for the numerator of the double transverse-spin asymmetries is the same as that in Eq. (5.9.7), except the unpolarized PDFs are replaced by the quark transversity distributions of various flavors (no gluon transversity distribution in a spin-1/2 transversely polarized proton), and the hard part is calculated with $\gamma^\pm\gamma^\perp\gamma_5$ spin projection for transversely polarized quarks. The collinear transversity distribution has the same definition as the quark distribution in Eq. (5.9.3) with the quark operator replaced by $\overline{\psi}_q(0)\gamma^+\gamma^\perp\gamma_5\,\mathcal{W}_{[0,\xi^-]}\psi_q(\xi^-)/2$ and the unpolarized hadron state $|h(p)\rangle$ is replaced by a transversely polarized hadron state $|h(p),\vec{s}_\perp\rangle$.

*Single transverse-spin asymmetries*

The transverse single-spin asymmetry (SSA),

$$A_N \equiv \frac{\sigma(s_T) - \sigma(-s_T)}{\sigma(s_T) + \sigma(-s_T)},$$

is defined as the ratio of the difference and the sum of the cross sections when the spin of one of the identified hadron $s_T$ is flipped. Two complementary QCD-based approaches have been proposed to analyze the physics behind the measured SSAs: (1) the TMD factorization approach [1283, 1284, 1297–1300], and (2) the collinear factorization approach [1301–1309].

In the TMD factorization approach, the asymmetry was attributed to the spin and transverse momentum correlation between the identified hadron and the active parton, and represented by the TMD parton distribution or fragmentation function. For example, the Sivers effect [1283] describes how hadron spin influences the parton's transverse motion inside a transversely polarized hadron, while the Collins effect [1284] describes how the parton's transverse spin affects its hadronization.

The TMD factorization approach is more suitable for evaluating the SSAs of scattering processes with two observed and very different momentum scales: $Q_1 \gg Q_2 \gtrsim \Lambda_{\rm QCD}$ where $Q_1$ is the hard scale while $Q_2$ is a soft scale sensitive to the active parton's transverse motion or momentum. For example, the Drell-Yan lepton pair production when $Q \gg q_T$ is a process that can be studied in terms of the TMD factorization [1269]. In addition, the SIDIS when the transverse momentum of observed final-state hadron $p_h \ll Q$ in the photon-hadron Breit frame is an ideal observable for studying $A_N$, since the leading power contribution to the TMD factorization of SIDIS is known to be valid [1269, 1285]. Although the $A_N$ in SIDIS can receive contribution from various sources, including the Sivers effect (Sivers function $f_{1T}^\perp$) and Collins effect (Collins function $H_1^\perp$), as well as contribution from the pretzelosity distribution $h_{1T}^\perp$ [1286], it is the choice of angular modulation that allows us to separate these three sources of contributions in SIDIS,

$$A_N^{\rm Sivers} \propto \langle\sin(\phi_h - \phi_s)\rangle_{UT} \propto f_{1T}^\perp \otimes D \tag{5.9.17}$$

$$A_N^{\rm Collins} \propto \langle\sin(\phi_h + \phi_s)\rangle_{UT} \propto h_1 \otimes H_1^\perp \tag{5.9.18}$$

$$A_N^{\rm Pretzelosity} \propto \langle\sin(3\phi_h - \phi_s)\rangle_{UT} \propto h_{1T}^\perp \otimes H_1^\perp \tag{5.9.19}$$

where $D$ is the normal unpolarized FF, the subscript "UT" stands for unpolarized lepton and transversely polarized hadron, $\phi_h$ is an angle between the leptonic plane and the hadronic plane in SIDIS and $\phi_s$ is the angle between the hadron transverse spin vector and the leptonic plane.

The predictive power of TMD factorization leads one to expect that the TMDs will be process-independent. However, it was found that the Sivers function measured in SIDIS and that in Drell-Yan process could differ by a sign. Such simple and generalized universality should preserve the predictive power of TMD factorization approach. Theoretically, such sign change can be better verified from the operator definition of the Sivers function. The quark Sivers function is defined as the spin-dependent part of the TMD parton distributions [1298, 1310],

$$f_{q/h\uparrow}(x, k_\perp, s_\perp) = \int \frac{dy^- d^2y_\perp}{(2\pi)^3} e^{ixp^+y^-} e^{-i\vec{k}_\perp\cdot\vec{y}_\perp} \times\langle p, s_\perp|\bar{\psi}(0)W_{[0,y]}\psi(y)|p, s_\perp\rangle|_{y^+=0}, \tag{5.9.20}$$

where $W_{[0,y]}$ is the gauge link for the leading power initial- and final-state interactions between the struck parton and the spectators or the remnant of the polarized hadron. The form of the gauge links including the phase of the interactions depends on the color flow of the scattering process and is process dependent. Luckily, the parity and time-reversal invariance of QCD removes almost all process dependence of the TMDs. By applying Eq. (5.9.14) to the matrix element in Eq. (5.9.20), we have

$$f^{\rm SIDIS}_{q/h\uparrow}(x, k_\perp, S_\perp) = f^{\rm DY}_{q/h\uparrow}(x, k_\perp, -S_\perp)\,. \tag{5.9.21}$$

Therefore, the Sivers function has an opposite sign in SIDIS and DY [1308, 1311]. Experimentally, it is important to verify such relationship.

In the collinear factorization approach, all active partons' transverse momenta are integrated into the collinear distributions, and the explicit spin-transverse momentum correlation in the TMD approach is now included in the high twist collinear parton distributions or fragmentation functions. Since the massless quark in short-distance hard collisions cannot flip the spin in QCD, the SSAs in the collinear factorization approach are generated by quantum interference between a scattering amplitude with one active parton and an amplitude with two active partons. The necessary spin-flip for SSAs is achieved by angular momentum flip between single active parton state and the state of two active partons. Such nonperturbative effect is represented by twist-3 collinear parton distributions or fragmentation functions, which has no probability interpretation, and the spin flip was made possible by QCD color Lorentz force [1302, 1303]. The collinear factorization approach is more relevant to the SSAs of scattering cross sections with a single hard scale $Q \gg \Lambda_{\rm QCD}$. The validity of QCD factorization for SSA in the collinear factorization approach requires study of the collinear factorization beyond the leading power (or twist-2) contribution.

It was demonstrated that QCD factorization works for the first sub-leading power contribution to the hadronic cross section, but, not beyond [1312]. That is, QCD factorization should work for the $1/Q^2$ power correction to inclusive and unpolarized Drell-Yan cross section [1313], $1/p_T^2$ corrections to unpolarized single high-$p_T$ particle production in hadron-hadron collisions [1314], and $1/p_T$ power correction to single high-$p_T$ particle production in hadron-hadron collisions with one of them transversely polarized [1302–1304, 1315]. It is the QCD factorization for the $1/p_T$ power correction to single high transverse momentum $p_T$ particle production in hadron-hadron collisions with one of them transversely polarized that enables the systematic collinear factorization approach to study $A_N$. For example, the SSA of single high-$p_T$ hadron production in hadronic collisions, $A(p, s_T) + B(p') \to h(P_h) + X$, can be factorized [1302, 1304]

$$A_N(s_T) \propto T^{(3)}(x, x, s_T) \otimes \hat{\sigma} \otimes D_f(z) +\delta q(x, s_T) \otimes \hat{\sigma}_D \otimes D^{(3)}(z, z) + \dots\,, \tag{5.9.22}$$

where $T^{(3)}$ and $D^{(3)}$ are twist-3 three-parton correlation functions and fragmentation functions, respectively, and $\delta q$ (or $h_1$) is the leading power transversity distribution, with "…" representing a small contributions [1316]. Various extractions of $T^{(3)}$ and $D^{(3)}$ from experimental data have been carried out [1305, 1317].

The SSA is a physically measured quantity and should not depend on how we describe it from QCD factorization or the choice of factorization scheme or scale, which leads to evolution equations of factorized nonperturbative distributions or twist-3 quark-gluon correlation functions relevant to the SSA [1318]. A complete set of the correlation functions was generated by inserting (1) the field operator $\int dy_1^- \left[iS_{T\rho}\, i\epsilon_T^{\rho\sigma} F_\sigma{}^+(y_1^-)\right]$ into the matrix element of twist-2 PDFs, and (2) the operator $\int dy_1^- \left[iS_T^\sigma F_\sigma{}^+(y_1^-)\right]$ into the matrix element of twist-2 helicity distributions [1318]. A close set of evolution equations of these twist-3 correlation functions as well as the leading order evolution kernels were derived [1318–1320].

Although the two approaches each have their own kinematic domain of validity, they are consistent with each other in the perturbative regime to which they both apply[1321, 1322].

## 5.10 Exclusive processes in QCD

George Sterman

### 5.10.1 Exclusive amplitudes for hadrons: geometry and counting rules

The analysis of exclusive reactions played a role in the development of quantum chromodynamics, and became a subject of ongoing research within QCD. This section reviews some of the early history, landmark developments and ongoing research in this lively topic, concentrating on wide-angle scattering. The reader is referred especially to the preceding contribution on factorization in cross sections, to Sec. 10 on the structure of the nucleon and Sec. 11 on QCD at high energy for closely related subject matter.

*Prehistory*
For many years, exclusive reactions were the language of experimental strong interaction physics at accelerators. In such reactions, up to low GeV energies (BeV at the time), new resonances were found, whose quantum numbers were revealed in the analysis of their decays. As energies increased, the analysis of exclusive reactions gave rise to theoretical advances like Regge theory, and the Veneziano amplitude [7], resulting eventually in string theory. Around the same time, the quark model for hadron spectroscopy was developed.

With the advent of multi-GeV hadronic and leptonic accelerators, any nonforward exclusive final state became a small part of the cross section. Nevertheless, if we assume that elastic scattering results directly from pairwise scattering amplitudes for constituent quarks, simple counting combined with the optical theorem leads to successful predictions on the ratios of total cross sections [1323]. Other pioneering concepts introduce a geometrical picture of colliding hadrons, whose interactions extend over their entire overlap during the scattering [1324]. This picture is agnostic on the dynamical nature of the strong interactions that mediate momentum transfer. The dual amplitudes of Ref. [7] are exponentially suppressed for fixed-angle scattering, and indeed, exponential fall-off in $|t|$ is characteristic of near-forward cross sections at high energy [1325]. For $|t|$ in the range of a few GeV, however, this decrease moderates to a power. This, along with the observation of power-law fall-off for form factors [563] suggested that fixed-angle amplitudes might, indeed must, reflect a point-like substructure for nucleons and mesons. This section will review some of the guiding developments in this area, which grew along with QCD, and which continue to shape contemporary theoretical and experimental programs.

*Hadrons in the language of partons.*
Hadrons are bound states, whose fine-grained properties are nonperturbative, yet based in the interactions of the quarks and gluons that appear in the Lagrangian density of QCD. To describe how partons can mediate the scattering of hadrons, we introduce a Fock space picture of the hadronic state with on-shell momentum, in terms of $P^+ = (1/\sqrt{2})(P^0 + P^3)$, mass $m_H$ and spin $s_H$, as [792]

$$|H, P^+, s_H\rangle \;=\; \sum_{F_H} c_F \, | \{f_i, x_i, \mathbf{k}_{i,\perp}, \lambda_i\} \, F_H \rangle \,, \quad (5.10.1)$$

where the infinite sum is over partonic Fock states, $F_H$, each consisting of a set of constituents, $\{f_i \dots\}$, labelled by flavors, $f_i$, by the fraction $x_i$ of $\vec{P}_H$, transverse momenta $\mathbf{k}_{i,\perp}$ and helicity $\lambda_i$. In QCD, the Fock states are labelled as well by the manner in which the colors of constituents combine to form color singlets. From these states, in principle, we can construct any of the universal quantities of perturbative QCD that can be written as expectation values of the hadronic state, including collinear and transverse momentum parton distributions. Here, however, we will for the most part make use of only the valence state, $F_{\rm val}$, with three constituents for a nucleon, two for a meson. Of course, we assume that $c_{F_{\rm val}}$ is nonzero in Eq. (5.10.1). The Fock state formalism puts this approximation in context, pointing the way to systematic expansions.

*Constituent counting.*
Influenced by the success of the parton model applied to quarks, and assuming a constituent expansion like the one just described, Brodsky and Farrar [1326], and Matveev, Muradian and Tavkhelidze [1012] realized that under broad assumptions on the strong interactions, the behavior in momentum transfer of a wide range of exclusive processes can be summarized by a simple rule, which goes under the name of quark, or more generally constituent, counting. We can see how this works by considering the very high-energy elastic scattering of two hadrons, in the first instance assumed to consist of a fixed set of “valence” partons, specified by the quark model ([uud] for the proton, for example), moving within a limited region of space, which we can think of a sphere of radius $R_H$ for hadron $H$.

Following the intuitive analysis of partons in deep-inelastic scattering, we imagine that hadrons can be thought of as Lorentz contracted and time dilated. Large momentum transfer requires all $n_i$ valence (anti-)quarks of the initial-state hadrons $i$ to arrive within a region of area $1/Q^2$, where $Q$ is the momentumm transfer. Now the hadrons don't know they are going to collide, so we assume their partons are more or less randomly scattered about within the areas of their Lorentz-contracted wave functions. Then the likelihood for them all to be within this small area is of order

$$\left(\frac{1}{Q^2} \times \frac{1}{\pi R_H^2}\right)^{n_i-1}$$

for each hadron of radius $R_H$. But this must also be true of both incoming and outgoing states, so that their wave functions may overlap.

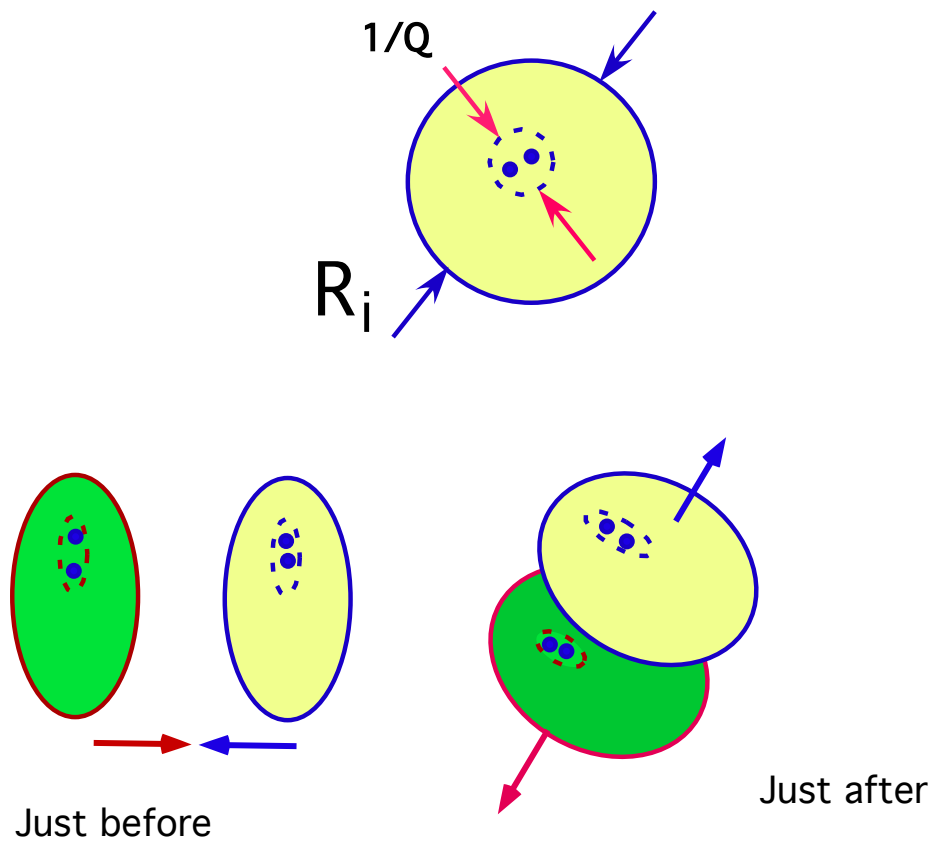

**Fig. 5.10.1** The geometry of constituent counting for $\pi - \pi$ scattering ($n_i = 2$). The top represents the pion in a valence state that can contribute to an exclusive reaction, as seen along the collision axis by an oncoming hadron. From Ref. [1327].

At the moment of collision, we don't have to make an assumption on the details of the hard scattering that redirects the partons, but we assume that otherwise the amplitude is a function only of the scattering angle. Then, at fixed $t/s$ (that is fixed center of mass scattering angle), we find the quark counting rules of Refs. [1012] and [1326],

$$\frac{d\sigma}{dt} = \frac{f(t/s)}{s^2} \left(\frac{1}{s\,\pi R_H^2}\right)^{\sum_{i=1}^{4}(n_i-1)} . \tag{5.10.2}$$

Figure 5.10.1 illustrates the scales involved, and the system just before and after the hard scattering. This relation provides a set of predictions for power-behavior, for example $d\sigma_{pp\to pp}/dt \propto s^{-10}$, which are generally successful [1328]. The determination of normalizations would require, of course, control over the short-distance interactions of the constituents, to which we will return below. For applications of these ideas to nuclei, see Sec. 5.2.

*Quark exchange, spin and transparency.*
Before going further into the technical status of exclusive amplitudes, it is natural to observe several fundamental consequences of this picture. First, assuming that the integrals over fractional momenta are insensitive to the endpoints, the rules of quark counting follow immediately by dimensional counting in the (in principle) calculable partonic scattering amplitudes. The picture is quite general, and applies as well to lepton-hadron elastic scattering. The constituent rules then determine the power behavior of hadronic form factors in momentum transfer, $Q$: $Q^{-2}$ for mesons and $Q^{-4}$ for baryons. In all processes, any scattering mediated by larger numbers of constituents is power-suppressed.

In the scattering of hadrons, there are generally many ways in which quarks can flow from the initial to the final state. Almost all of these describe quark exchange, whether in elastic scattering like $\pi^+ p \to \pi^+ p$, but especially for charge-exchange exclusive processes, like $\pi^- p \to \pi^0 n$. The valence Fock states described above, considered as functions of the transverse momenta of the constituents, can be used to construct a picture of $2 \to 2$ exclusive amplitudes based on the overlaps of incoming and outgoing states. These considerations lead to a variety of quite successful predictions for dependence on momentum transfer [1329]. A particularly striking example is the difference between proton-proton and antiproton-proton scattering, where the latter provides no opportunity for quark exchange. The ratio of these cross sections is about forty to one [1329].

For hadrons with light-quark valence structure (pions, nucleons) we anticipate that the scatterings will be computed with zero quark masses. Then, in any theory based on the exchange of vector gluons, the helicities of the quarks are conserved, and since the scattering is in valence states at small transverse sizes, the helicities of the valence states directly determine the spins of the external hadrons. This feature leads to many predictions for amplitudes in which spins are prepared and measured [1328]. Unlike constituent counting rules, however, predictions for spin more often fail; for the example of proton-proton scattering, see Ref. [785].

Finally, specializing to color-singlet hadrons in a theory with colored quarks, another fundamental prediction of this picture is *transparency* [1330], which refers to predictions for exclusive hard-scattering in nuclei. On the one hand, exclusive scattering emerges only from valence parton configurations, with all partons in a small regions of coordinate space. On the other hand, at high energies, the lifetime of such a virtual state is dilated by a large factor. Thus we anticipate that *both* the incoming and outgoing hadrons in an exclusive reaction propagate as effectively point-like particles through the surrounding medium, in particular, through a nucleus. For proton-nucleon elastic scattering with momentum transfer $Q$, the incoming proton must be in a state of effective area $1/Q^2$ on its way into the nucleus, and will be invisible to the color fields of nucleons it encounters, whose partons are typically spread out over scales of the order of the proton's radius. Only when it encounters a constituent nucleon that happens to be in a corresponding tiny valence state can it undergo elastic scattering, producing again a pair of "stealth" nucleons that are just as invisible on the way out. While the amplitude for this to happen remains just as small as for free proton-proton or proton-neutron scattering, it is not suppressed by initial- or final-state interactions, in contrast to most cross sections on nuclei. These considerations are summarized in the elegant prediction for scattering on a nucleus of atomic number $Z$,

$$\frac{d\sigma}{dt}\,[p+Z \to p+p+(Z-1)] \overset{s\to\infty,\,t/s\text{ fixed}}{\to} Z\frac{d\sigma}{dt}\,[p+p\to p+p]\ . \tag{5.10.3}$$

This is the case, at least asymptotically, and the manner in which asymptotic behavior is reached for varied elastic reactions is a subject of ongoing experimental (see for example, Refs. [210, 1331]) and theoretical investigation [1332, 1333].

*Splitting the hard scattering: Landshoff mechanism.* Without further assumptions, the same geometric - partonic considerations sketched above can lead to an alternative picture and prediction for asymptotic behavior, first formulated by Landshoff [1334]. To be specific, let's consider meson-meson elastic scattering ($n_i = n_f = 2$). Then, instead of a single short-distance scattering involving all four incoming and outgoing partons, we imagine two independent hard scatterings of parton pairs, each resulting in two pairs of partons travelling in the same direction, and forming the outgoing mesons. The geometric picture is shown in Fig. 5.10.2. We assume that the separation $b$ between the short-distance collisions of individual pairs of partons is generically of order $R_H$, the hadronic radius[61] Relative to the strict short-distance picture of Fig. 5.10.1, this reaction is enhanced by the ratio $R_H/(1/Q) = R_H Q$ in the amplitude for mesons, which is the ratio of the scale of the hard scattering to the size of the overlap between the hadrons, as shown in the figure. Similarly, there is an enhancement of $(R_H Q)^2$ for baryons, for which

$$\frac{d\sigma}{dt} = \frac{f(t/s)}{s^2}\left(\frac{1}{s\,\pi R_H^2}\right)^6 . \tag{5.10.4}$$

In the forward region with a still-large momentum transfer, $s \gg -t \gg \Lambda_{\rm QCD}$, we anticipate a factor $1/Q^2 \sim 1/t$

[61] We will come back to this assumption below.

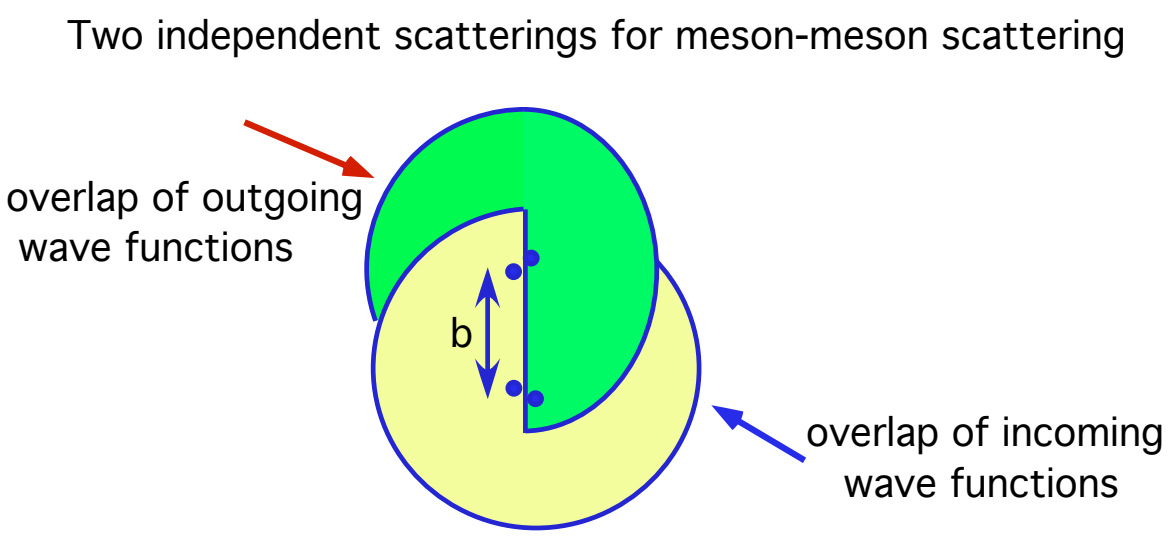

**Fig. 5.10.2** Geometric enhancement in the Landshoff mechanism. The pairs of colliding partons (within each pair, one from each colliding hadron) are separated by distance $b$. Within each pair, partons are separated by a much smaller distance of order $1/Q$. From Ref. [1327].

for each hard scattering, and we find

$$\frac{d\sigma}{dt} = \frac{f(0)}{t^2}\left(\frac{1}{t\,\pi R_H^2}\right)^6 . \tag{5.10.5}$$

Experimentally, at wide angles, data appear to prefer the direct counting behavior of Eq. (5.10.2), but at large $t$ and even higher $s$, a behavior like Eq. (5.10.5) is observed [1335, 1336].

### 5.10.2 Computing hard exclusive amplitudes in QCD

The considerations described above are based in the parton model, although they are a significant step beyond the classical parton model results, because the hard scattering is itself a strong interaction. With these concepts in hand, the next great step was to apply field theoretic analysis to elastic scattering, relying on asymptotic freedom to calculate short distance interactions where large momenta are exchanged, and on ideas of factorization to separate the dynamics binding each hadron from the short distance scattering and from each other. Before we review this landmark analysis for exclusive processes with hadrons, it is useful to touch on elastic scattering amplitudes for partons. These, of course, are not directly physical, but they play an important role in the factorized hadronic analysis that follows, and also in other areas, particularly jet cross sections.

*Partons: exclusive amplitudes in QCD.* We consider partonic scattering amplitudes at "wide angles", labelling the combination of incoming and outgoing (massless) partons and their momenta as $f$,

$$f\ :\ f_1(p_1)+f_2(p_2) \to f_3(p_3)+f_4(p_4)+\cdots+f_{n+2}(p_{n+2})\,. \tag{5.10.6}$$

To define such an amplitude in perturbation theory requires the regulation of infrared singularities associated with the virtual states that include zero-momentum lines and/or lines collinear to the external particles. This is conventionally done by dimensional regularization, that is, by treating the number of dimensions as a parameter, $d = 4 - 2\varepsilon$, and continuing $\varepsilon$ away from zero. Starting at one loop, infrared singularities manifest themselves as poles in $\varepsilon$, generally two per loop. Despite the growing order of the poles, the amplitude can be written in a factorized form, [1337–1339]

$$\mathcal{M}_L^{[f]}\left(v_i, \frac{Q^2}{\mu^2}, \alpha_s(\mu^2), \varepsilon\right) = \prod_{i\in f} J^{[i]}\left(\frac{Q'^2}{\mu^2}, \alpha_s(\mu^2), \varepsilon\right)$$
$$\times S_{LI}^{[f]}\left(v_i, \frac{Q^2}{\mu^2}, \alpha_s(\mu^2), \varepsilon\right) H_I^{[f]}\left(\beta_i, \frac{Q^2}{\mu^2}, \alpha_s(\mu^2), s\right) \quad (5.10.7)$$

In this expression, the functions $J^{[i]}$ contain all poles in $\varepsilon$ due to virtual lines collinear to the velocities, denoted $v_i$ ($v_i^2 = 0$) of the massless external partons $i$. These infrared poles are universal among the amplitudes of different partonic scattering processes. That is, they only depend on the whether the external parton is an (anti)quark or gluon. The infrared factors diverge very rapidly as $\varepsilon \to 0$, that is, in four dimensions. Many details can be found in Ref. [1340], but to get an idea of the strength of the infrared singularities, it is sufficient to see leading poles of the two-loop exponent of a jet function, given in terms of its expansion in terms anomalous dimensions $\gamma_K^{[i]}$,

$$J^{[i]}\left(\frac{Q^2}{\mu^2}, \alpha_s(\mu^2), \varepsilon\right) \sim \exp\left\{ - \left(\frac{\alpha_s}{8\pi}\right)\left(\frac{1}{\varepsilon^2}\gamma_K^{[i]\,(1)}\right) + \left(\frac{\alpha_s}{\pi}\right)^2 \left[\frac{\beta_0}{8}\frac{1}{\varepsilon^2}\frac{3}{4\varepsilon}\gamma_K^{[i](1)} - \frac{1}{2}\left(\frac{\gamma_K^{[i](2)}}{4\varepsilon^2}\right)\right] + \dots \right\}. \quad (5.10.8)$$

Here $\gamma_K^{[i]} = \sum_n \gamma_K^{[i](n)}(\alpha_s/\pi)^n$ is the coefficient of the $1/[1-x]_+$ term of the DGLAP evolution kernel for parton $i$, often denoted $A_i(\alpha_s)$, with $\gamma_K^{[q](1)} = C_F$, and $\beta_0$ is the lowest-order coefficient of the QCD beta function. The analysis that leads to the exponentiation of double infrared poles for partonic amplitudes relies on enhancements of radiation by accelerated massless charged particles at low angle and energy in gauge theories. The systematic treatment of these effects often goes by the name "Sudakov resummation", a term we will encounter below when we return to the Landshoff mechanism.

In Eq. (5.10.7), $S_L^{[f]}$ is a matrix in the space of color exchanges, labelled by color tensor $L$ (for example, octet or singlet exchange), which contains the remaining poles, all due to virtual lines with vanishing momenta. The soft matrix, $S_L^{[f]}$ also has an expression in terms of calculable "soft" anomalous dimensions, which have wide uses in inclusive as well as exclusive cross sections. The remaining set of functions, $H_l^{[f]}$ are free of infrared poles and contain all dependence on momentum transfers.

*Hadrons: Factorization and evolution for form factors and exclusive amplitudes.*
Historically, the analysis of hadronic exclusive amplitudes in QCD predated that for partonic amplitudes just discussed. This was possible because in these amplitudes external particles are, by construction, color singlets. We assume that the picture given above for quark counting still applies, that the elastic amplitudes result from redirecting valence quarks and antiquarks into collinear configurations in the final state, and that those configurations are color singlets. Then purely soft, as opposed to collinear, singularities disappear. Comparing to the partonic amplitude, Eq. (5.10.7), we derive an expression for the hadronic amplitude without a soft matrix, and with dimensionally-regularized jet fuctions replaced by hadronic wave functions [206, 207, 1272]. A short-distance, hard-scattering function denoted $H$ describes the short-distance scattering of $n_i$ valence quarks/antiquarks from each external hadron, $i$. The general form, in this case for $2 \to 2$ scattering, is

$$\mathcal{M}(s,t;\lambda_i) = \int \prod_{i=1}^{4} [dx_i]\ \phi_i(x_{i,m}, \lambda_i, \mu) \times H\left(\frac{x_{i,n}x_{j,m}p_i \cdot p_j}{\mu^2}; \lambda_i\right). \quad (5.10.9)$$

In contrast to partonic scattering, which describes the short-distance scattering of a single physical parton for each direction, hadronic wave functions, $\phi_i(x_{i,m}, \lambda_i, \mu)$, depend on how their valence partons share the momentum of their external hadron, labelled by fractions $x_{i,m}$, $\sum_m x_{i,m} = 1$. Hadronic helicities, labelled by $\lambda_i$, determine spin projections for the quark constituents of the valence state. The integrals over fractional momenta are denoted (here, for baryons) by the notation,

$$[dx_i] \ = \ dx_{i,1}dx_{i,2}dx_{i,3}\,\delta\left(1 - \sum_{n=1}^{3} x_{i,n}\right). \quad (5.10.10)$$

The factorization requires the choice of a factorization scale, $\mu$, which is naturally of the order of the renormalization scale for the matrix element that defines the wave functions $\phi(x_i, \lambda_i, \mu)$. A representative example is for $\pi^+$, whose wave function is the matrix element of the valence quark operators that absorb an up quark and an anti-down quark, between the single-pion state and the QCD vacuum. In this case, defining $x_1 = 1 - x_2 \equiv x$ as the fraction of the up quark, the expression (in a

physical gauge) is

$$\phi_\pi(x,\mu) = p\cdot n \int_{-\infty}^{\infty} \frac{d\lambda}{4\pi}\, e^{i(xp)\cdot(\lambda n)} \times \langle 0|\bar{d}(0)\, \frac{n\cdot\gamma\,\gamma_5}{2\sqrt{2n_c}}\, u(\lambda n)|\pi^+(p)\rangle\,, \quad (5.10.11)$$

where the vector $n^\mu$ is light-like and oppositely directed to the pion's momentum $p^\mu$, and $n_c$ is the number of colors. The matrix element requires renormalization because its fields are separated by a light-like distance, proportional to $n^\mu$.

We note the many similarities between the exclusive amplitudes Eq. (5.10.9) and factorized forms of inclusive cross sections in deep-inelastic and hadron-hadron scattering. The role of wave functions here is played by parton distributions there, and in both cases there is a convolution in partonic momentum fraction(s). In both cases also, the presence of a factorization scale, $\mu$, implies evolution equations, there for parton distributions and here for wave functions,

$$\mu\frac{\partial}{\partial\mu}\,\phi(x,\mu) = \int_0^1 dy\, V(x,y,\alpha_s(\mu))\;\phi(y,\mu)\,. \quad (5.10.12)$$

The evolution kernel $V(x,y,\alpha_s)$ incorporates cancellations between constituent self-energies and diagrams with gluons exchanged between constituents. In general, the factorization scale is proportional to the momentum transfer, and these evolution equations make it possible to extrapolate wave functions (and parton distributions) from one scale to another. While space does not allow a review of the kernel and the solutions of these equations here, an especially beautiful consequence of the particular evolution equations for pion wave functions is that at asymptotically large $\mu$ the wave functions approach known, fixed, finite expressions,

$$\lim_{\mu\to\infty} \phi_\pi(x,\mu) = \frac{3f_\pi}{\sqrt{n_c}}\; x(1-x)\,, \quad (5.10.13)$$

where $f_\pi$ is the pion decay constant and again $n_c$ the number of colors (3 for QCD of course). Again, this is a consequence of the detailed nature of the kernel in the evolution equation, (5.10.12), which follows in turn from the underlying factorization for hard exclusive processes, Eq. (5.10.9).

*Exceptional momentum configurations.*
In their original form, the factorized amplitudes of Eq. (5.10.9) apply to a very wide set of processes, including elastic form factors for pions and mesons, for which the external leptons can be counted as if they were hadrons with a single parton. Like any such factorized expression, however, its predictive power depends on its stability under higher-order corrections. Of particular interest are the limits where one fractional momenta $x_i$ approaches unity and the others vanish, a configuration for elastic scattering often referred to as the Feynman mechanism (see Lecture 29 of Ref. [1341]). Noting the example of Eq. (5.10.13), we generally expect, and in case of pions in the valence state can prove, that wave functions vanish sufficiently rapidly in these limits to preserve the stability of the factorized amplitude in Eq. (5.10.9). The onset of this limit is not easy to determine, however, and has been the subject of discussion in the literature. For form factors particularly, alternative treatments based on dispersion relations and QCD sum rules, provide an alternative picture for currently accessible momentum transfers [1342]. The situation for baryonic wave functions is even more complex, because the Feynman mechanism is not suppressed at fixed orders [1343]. At high momentum transfers, this may be resolved by higher-order corrections [1344] (see below), but phenomenological analyses based on the Feynman mechanism are also of interest [1345].

Another point of concern is the Landshoff mechanism identified above, in which subsets of the partons scatter elastically at different points in the space transverse to the beam directions, as in Fig. 5.10.2. This process is actually lower order in $\alpha_s$, but more importantly it is sensitive to the transverse structure of the external hadrons, that is, on information that is not included in the wave functions discussed above. However, the resummation of higher-order QCD corrections shows that large transverse separations are suppressed, returning us to expectations very similar to those of Eq. (5.10.9).

*Sudakov resummation and asymptotic behavior.*
As we have seen in Fig. 5.10.2 and Eq. (5.10.4), the Landshoff enhancement to inclusive amplitudes is due to the assumed possibility of separating hard scatterings between subsets of valence partons. As noted above, to estimate the enhancement we assume that the separation is generically of the order of the hadronic radius. The analysis through Sudakov resummation follows from the observation that the separation of partonic hard scatterings in an overall hadronic exclusive amplitude requires the scattering of isolated non-singlet color charges without radiation. In isolation, these accelerated charges would result in infrared singularities, as in Eq. (5.10.8) above, which would make the amplitude vanish in four dimensions. In our case, however, the outgoing configurations of the scattered partons are almost collinear, and the divergences (infrared poles) cancel. The larger the separation $b$ between the hard scatterings, however, the larger the finite remainder. The result is that any process with separated hard scatterings is suppressed relative to the acceleration of locally singlet charge configurations, which shows that the assumption of separated hard scatterings among pairs of partons made in our analysis of the Landshoff mechanism was not in fact warranted.

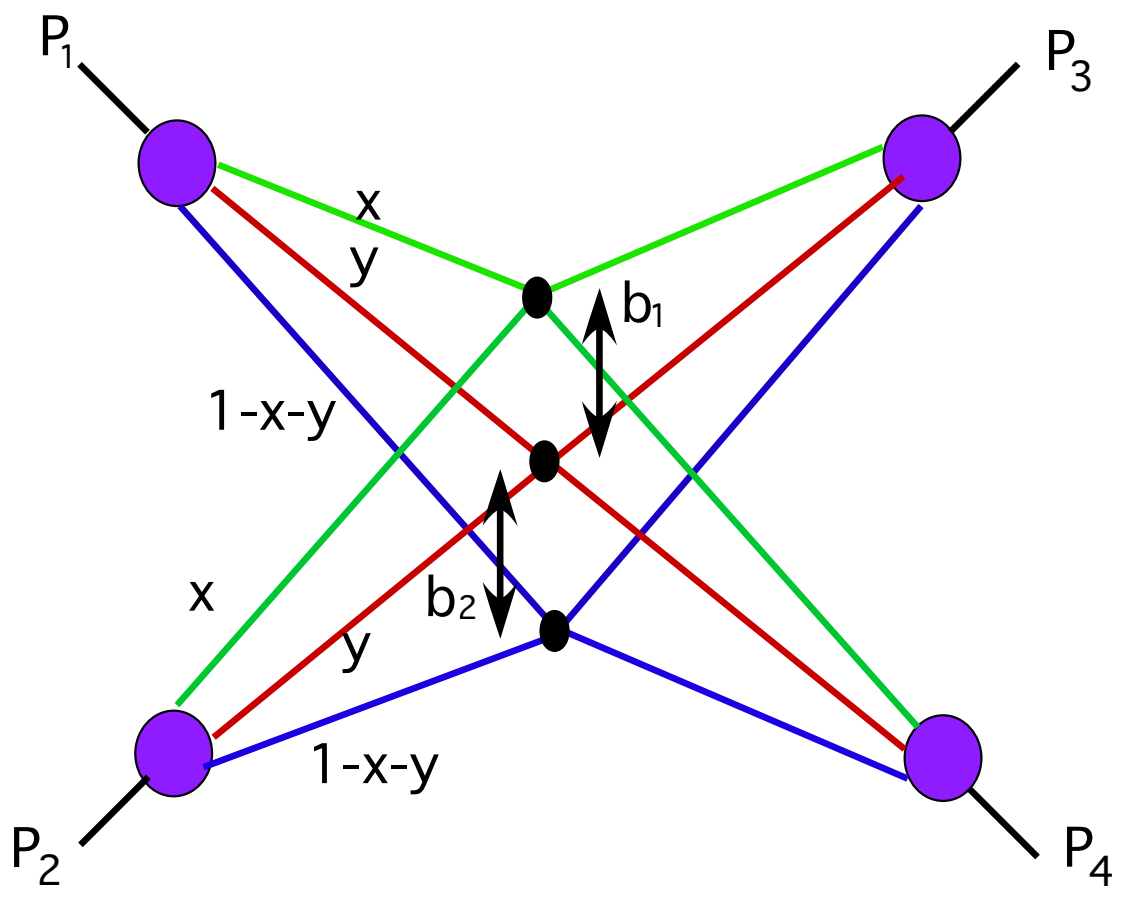

**Fig. 5.10.3** Transverse separations in a multiple hard scattering. Note that the eight potentially independent integrals over momentum fractions are replaced by only two integrals, the same for each external hadron. From Ref. [1327].

The observations above, which are the basis of transparency, can be quantified, by treating the distance between the hard scatterings in Fig. 5.10.2 as an impact factor, $b$, conjugate to transverse momentum. An analysis treating both transverse and longitudinal momenta of quarks leads to a factorized expression for hadronic scattering amplitude in terms of a wave function that depends on both the quark transverse momentum and longitudinal momentum fraction. As with the classic form, Eq. (5.10.9), there is a close analogy to parton distributions encountered in inclusive cross sections, in this case transverse momentum distributions (TMDs). The necessary wave functions generalize the light-cone matrix elements like Eq. (5.10.11) by displacing the fields in transverse (impact parameter) directions relative to the opposite-moving light cone.

This factorization in impact parameter space requires a soft matrix, which ties together soft radiation from the two (or three) separated hard scatterings in Fig. 5.10.2. Referring to the diagram in Fig. 5.10.3 for a baryonic exclusive process, we anticipate a perturbative suppression whenever the distances between hard scatterings, $b_1$ and $b_2$ in the figure, increase beyond the scale of the momentum transfer. For this process, we note that all four partons external to each hard scattering must carry the same momentum fraction. So the eight integrals over momentum fractions are reduced to two, which we label $x$ and $y$ here.

The form of factorization corresponding to Fig. 5.10.3 is then given at scattering angle $\theta$ and momentum transfer $Q$ by [1346]

$$\begin{aligned}\mathcal{M}(s,t) =& \frac{1}{2\pi^2\sin^2\theta}\sum_f\int_0^1 dx dy\,\theta(1-x-y)\\ &\times\int db_1 db_2\,\mathrm{Tr}_{\mathrm{color}}\left[U(b_iQ)H^1H^2H^3\right]\\ &\times\prod_{i=1,2,3,4}\mathcal{R}_i(x,y,b_1,b_2)\,,\end{aligned} \tag{5.10.14}$$

where the color Trace $\left[U(b_iQ)H^1H^2H^3\right]$ ties color together and includes $\epsilon_{abc}$ for colors of three quarks, with possible color exchange in each hard scattering,

$$H^i(x_ip_1,x_ip_2,x_ip_3,x_ip_4)\sim 1/(x_iQ)^2.$$

In Eq. (5.10.14) we may define $x_1 = x$, $x_2 = y$ and $x_3 = 1-x-y$.

The wave functions, $\mathcal{R}(x,y,b_1,b_2)$ drive the suppression of large $b_i$, and behave as

$$\begin{aligned}&\mathcal{R}_i(x,y,b_i)\sim\phi_i(x,y,b_1,b_2,\mu\sim 1/\langle b\rangle)\\ &\quad\times\exp\left[-\frac{\alpha_s}{\pi}\gamma_K^{[q][1]}\sum_{a=1}^{3}\ln^2\left(\frac{1}{x_aQb_a}\right)\right]\,,\end{aligned} \tag{5.10.15}$$

where $\gamma_K^{[q]}$ is the same anomalous dimension as for the quark jets in the partonic amplitude, Eq. (5.10.8). The $\phi_i(x,y,1/\langle b\rangle)$ are normal partonic wave functions of the form encountered above, now evaluated at a renormalization scale set by the inverse of the average impact parameter spacing between the hard scatterings. The exponential suppression by double logarithms of $b$ in Eq. (5.10.15) is the result of the systematic treatment of states with soft and collinear virtual radiation, and is thus an example of Sudakov resummation [1347]. It forces the impact parameters to vanish, on a scale that is for all intents and purposes of order $1/\sqrt{-t}$. Combined with the $1/t$ behaviors of the three partonic hard scatterings, the full amplitude behaves as nearly $1/t^4$, consistent with the original constituent counting rules of Eq. (5.10.2). The momentum transfer at which this behavior sets in, however, may be quite large, especially given the factors of $x$ and $y$, which are always less than unity, in the arguments of logarithms.

### 5.10.3 Toward the future

The true asymptotic behavior of many exclusive reactions in QCD is by now well characterized, but much remains to be understood. In particular, it is not fully clear to what extent the success of constituent counting rules provides us with a quantitative understanding of the normalizations of amplitudes at accessible momentum transfers, and when to expect predictions based on helicity conservation and transparency to apply. Progress in these directions will be part of the future of QCD, a future in which the gap between partonic and hadronic degrees of freedom is bridged.

## 5.11 Color confinement, chiral symmetry breaking, and gauge topology

**Edward Shuryak**

### 5.11.1 Overview

Nontrivial topological structures of non-Abelian gauge fields were discovered in the 1970's, starting with the 't Hooft-Polyakov monopole [1348, 1349] and Belavin-Polyakov-Schwartz-Tyupkin (BPST) instanton [1350]. These two sets of objects were soon related to two main nonperturbative phenomena – *confinement* and *chiral symmetry breaking*. Confinement was connected to the so called "dual superconductor" model [1348, 1351] magnetically charged monopole condensate expelling color-electric fields into flux tubes. Instantons, describing vacuum tunneling between topological barriers, have fermionic bound states – technically called *zero modes* – and in the QCD vacuum are "collectivised" into *quark condensates* which break the $SU(N_f)_A$ and $U(1)_A$ chiral symmetries of massless QCD. For decades, theory and phenomenology of monopoles and instantons were developed separately, but in the last two decades, following a breakthrough paper by Kraan and van Baal [1352] studies of $deconfinement$ and *chiral symmetry restoration* phase transitions, based on new semiclassical objects, called *instanton-monopoles* or *instanton - dyons* lead to a united quantitative description of both phase transitions, in QCD and even in its "deformed" versions.

### 5.11.2 Color confinement and deconfinement

Discovery of QCD 50 years ago put into motion many important developments in the 1970's. Asymptotic freedom led to a weak coupling regime at small distances and a flourishing "perturbative QCD" describing hard processes. Going in the opposite direction (small momenta or large distance, also called "infrared" or IR), one finds growing QCD coupling. In pure gauge theories the potential energy of a static quark and anti-quark pair grows linearly with increasing separation, $V(R_{q\bar{q}}) \sim \sigma R_{q\bar{q}}$. Therefore, with a finite amount of energy one cannot separate color charges: they are "confined". Furthermore, all electric fields are expelled from the vacuum and get confined as well, into so called "electric flux tubes" (also known as "QCD strings"). Their "tension" (energy per length) is $\sigma \approx 1\,GeV/$fm. In QCD with dynamical quarks, a new $q\bar{q}$ pair can be created, breaking the flux tube into two. Yet it is still true that any objects with nonzero color charge – such as quarks and gluons – do not exist as independent physical objects in the QCD vacuum. This is one of the definitions of "color confinement."

This attractive picture of course needed to be tested. K. Wilson [80] promoted the statement about a linear potential to a more abstract mathematical form: the vacuum expectation value of the Wilson line

$$W = \left\langle \frac{1}{N_c} Tr P \exp\left( i \int_C dx_\mu A^a_\mu T^a \right) \right\rangle, \tag{5.11.1}$$

over some contour $C$ of sufficiently large size with color gauge fields. Here $T^a$ are color algebra generators, and $P\exp$ means products of exponents along a given contour $C$. Wilson's criterium states that in confining theories

$$W = \exp[-\sigma * Area(C)] \tag{5.11.2}$$

falls exponentially with the area of a surface inclosed by the contour $C$. If it is a rectangular contour $T * L$ in 0 - 1 plane, the $area = T * L$ and $\sigma$ is then identified with the string tension. The very first numerical studies of non-Abelian gauge theory on the lattice, by M.Creutz [329] indeed found that the area law holds for large enough loops, and that $\sigma$ is indeed physical, that is it has correct dependence on the coupling as dictated by asymptotic freedom. (Needless to say, numerical evidence is not taken for a proof by mathematically inclined folks, and an analytic proof is still missing. A million dollar prize for such a proof still waits to be awarded.)

In Quantum Electrodynamics (QED) charge renormalization makes the coupling *larger* at small distances (large momenta transfers or UV limit), but *small* at large distances, which is explained by very intuitive "vacuum polarization" picture, in which virtual $e^+e^-$ pairs screen the charges. Screening of the charges by a QED medium – e.g. plasma of the Sun – is well known and tested.

One may now ask what happens in a "QCD medium". Asymptotic freedom tells us that, contrary to QED, at small distances the coupling *decreases*. But what would happen at large distances? Calculation of the polarization tensor [1353] had shown that, like in QED, the medium *screens* the charges. Therefore, at high enough temperature the interaction becomes weak at all distances. Therefore hot/dense QCD matter must be in a phase called a Quark-Gluon Plasma (QGP). It is the "normal phase" of QCD in which fields in the QCD

Lagrangian – quarks and gluons – correspond to quasi-particles which move relatively freely. It must be distinct from the QCD vacuum and low-$T$ hadronic phase, as there is no place for confinement, chiral condensate and other nonperturbative phenomena there. The "confining" QCD vacuum and the QGP must therefore be separated by a phase transition: and it is indeed seen in experiment and lattice studies, which now put the critical temperature at $T_{\rm deconfinement} \approx 155\,MeV$.

As discussed in detail in section on symmetries of QCD, at vanishing quark masses it has additional *chiral symmetries* . Without mass terms, in the Lagrangian the left and right-polarized components do not directly interact with each other and independent flavor rotations become possible. Such doubled flavor symmetry can be decomposed into a *vector* (the sum) and the axial (the L-R difference) symmetries. One of them, called axial $SU(N_f)_A$ symmetry ($N_f = 3$ is the number of light quark flavors, $u, d, s$), is *spontaneously broken* in the QCD vacuum, which possesses a nonzero *quark condensate* $\langle \bar{q}q \rangle \neq 0$. The melting (disappearance) of this condensate should happen at another transition $T > T_{chiral}$. Although in various settings $T_{\rm deconfinement} \neq T_{\rm chiral}$, in QCD they seem to coincide, again based on numerical lattice evidence.

Another chiral symmetry called $U(1)_A$ is broken by the *quantum anomaly* and is not actually a symmetry at all. ("Anomaly" means that while it is a symmetry of the Lagrangian, it is not a symmetry of the quantum partition function.)

### 5.11.3 Electric-magnetic duality and monopoles

Already our brief discussion above should have convinced the reader that the QCD vacuum is quite complicated, with one outstanding feature being the expulsion of color-electric fields into the flux tube. Already, in the 1970's [1351, 1354, 1355], an analogy between this phenomenon and an expulsion of magnetic fields from superconductors lead to the so called *"dual superconductor"* model of confinement.

In superconductors of the second kind there exist the so called magnetic flux tubes or *fluxons*. Magnetic fields are confined inside the tubes because of solenoidal (super)current of Cooper pairs on their surface. QCD flux tubes transfer flux of *electric* field instead. The word "dual" is used indicating that one has to interchange electric and magnetic fields. If so, the current in the solenoid needs to be magnetic. What can it be made of?

The apparent asymmetry of Maxwellian electrodynamics bothered theorists since late 1800's: can one allow magnetic charges, by adding a nonzero r.h.s. to the $\nabla \cdot \vec{B}$ equation? An interesting motion for a set of electric and magnetic charges was predicted by J.J.Thomson and H.Poincare. With discovery of quantum mechanics, Dirac [1356] famously observed that if they exist, then consistency of the theory requires that the product f electric and magnetic coupling to be quantized. As he emphasized, the existence of one monopole in the Universe would be enough to demand quantization of all electric charges, an empirical fact to which no other explanation existed. QED magnetic monopoles have been looked for in exceedingly more sensitive experiments, but so far none have been found.

Yet certain Non-Abelian gauge theories with adjoint scalars do possess solitonic magnetic monopole solutions of the equations of motion, as discovered independently by 't Hooft and Polyakov [1348, 1349] . Their prominent feature is that their *magnetic charges* comply with earlier ideas by Dirac about special conditions, making "invisible" Dirac strings and allowing coexistence of magnetic and electric charges in quantum settings. Here we cannot give justice to the explicit solution and its properties: the interested reader can find a detailed pedagogical description in books such as [1357]. Now, monopoles made of glue and scalars are bosons, so at low enough temperature their ensemble should undergo Bose-Einstein Condensation (BEC). If that happens, a "magnetically charged" monopole condensate would expel the (color)electric field into *electric confining flux tubes*, and explain confinement!

Seiberg and Witten [1217] have given an analytic proof in theories with more than one supersymmetry (which possess the needed adjoint scalars). They were able to get the exact dependence of the effective electric coupling on the vacuum expectation value (VEV) of the scalar $g^2(\langle\phi\rangle)$. When the VEV is large, the theory is similar to electroweak theory, with gluons and gluinos being light and weakly interacting, and monopoles very heavy. When the VEV decreases, the coupling increases to $O(1)$, and magnetic monopoles and dyons (particles with both electric and magnetic charges) have masses comparable to that of gluons and gluinos. Finally, near certain singular points the electric coupling goes infinitely strong, with gluons and gluinos much heavier than monopoles. An effective desciption in this regime is dual QED describing magnetic interactions of light monopoles. The remarkable fact is that opposite motion of electric and magnetic couplings follows exactly the "consistency condition" of QED $g_{electric} \cdot g_{magnetic} = const$ pointed out by Dirac [1356] nearly a century ago!

All this is very beautiful, creating significant theoretical activity at the turn of the century, but we need to return to QCD. It does *not* have adjoint scalar fields, so one cannot directly build 't Hooft- Polyakov monopoles.

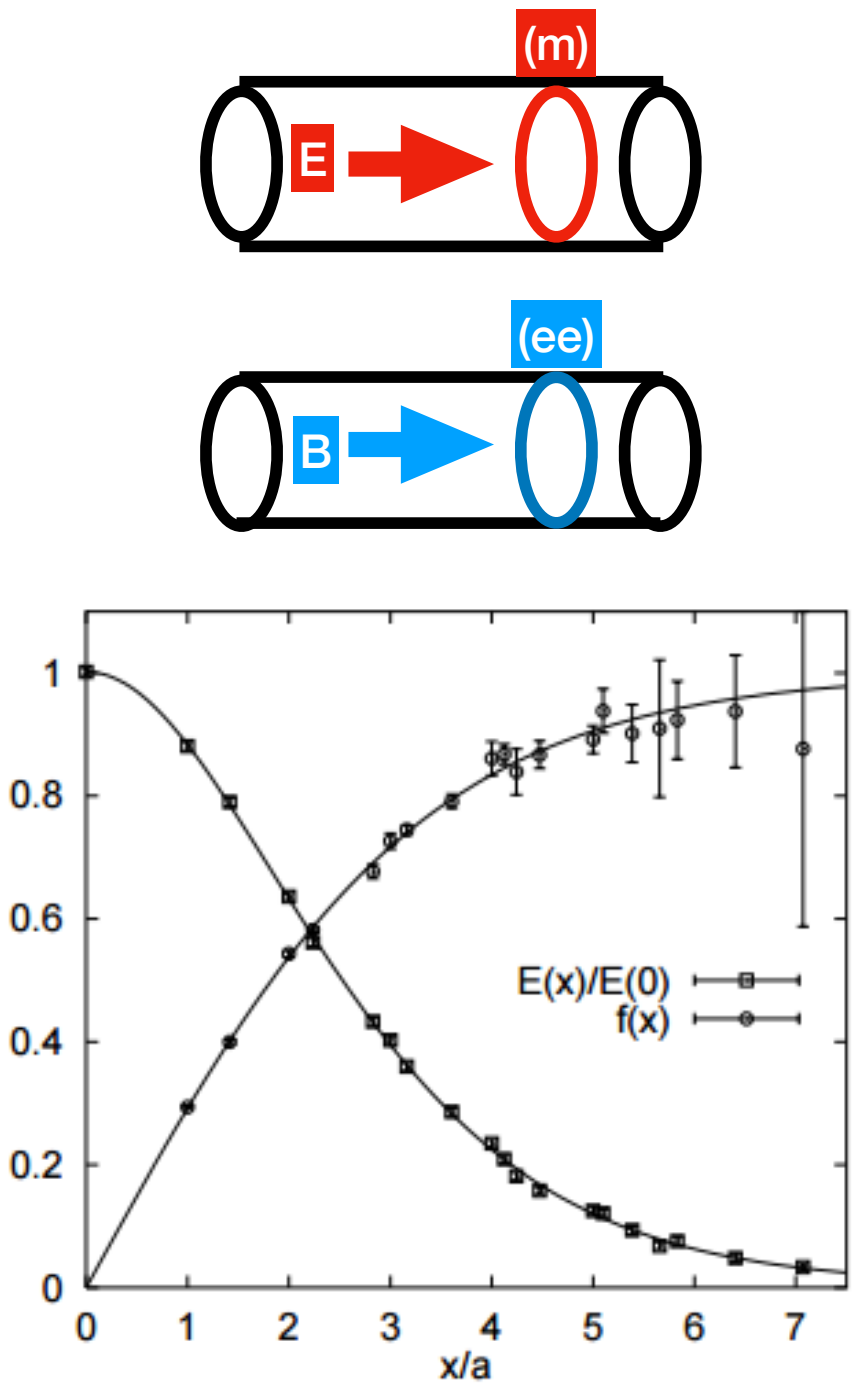

**Fig. 5.11.1** Upper panel: QCD electric flux tube in QCD vacuum (upper) and magnetic flux tube in superconductor (lower). The current rotated around is made of monopoles (upper) and Cooper pairs (lower), respectively. Lower panel: plot shows the lattice data on the distribution of the electric field strength (squares) and the monopole Bose condensate (discs) in cylindrical coordinates versus the distance in the transverse plane. As one can see, the field is maximal at the center where the monopole condensate vanishes. The flux tube is generated by two static quark-antiquark external sources (not shown). The lines correspond to a solution to (dual) Ginzburg-Landau equations.

However, by special procedures, it was possible to identify monopoles on the lattice, and locate their paths and correlations. It was observed, in particular, that these monopoles do indeed rotate around the confining flux tubes, producing solenoidal magnetic currents needed to stabilize them. The picture turns out to be a *dual copy* (meaning interchange electric ↔ magnetic) to well known magnetic flux tubes in superconductors. Fig.5.11.1 (displaying the result of lattice simulations summarized in the review [1358]) shows the distribution of the electric field and magnetic monopole condensate in a plane transverse to the electric flux tube. Furthermore, it has been shown[1359] that BEC phase transition of monopoles does coincide with the *deconfinement transition* at finite temperature $T_c$ of (pure gauge) theories.

Ensembles of monopoles in QCD were studied, with important applications. Monopole correlations reveal Coulomb-like forces between monopoles [1360], with their charges "running" in the direction opposite to that of electric charges [1361], exactly as predicted by Dirac! It has been shown [1362] that monopoles also play important role in deconfined QGP phase at $T > T_c$: in particular they dominate jet quenching in quark-gluon plasmas created in heavy ion collisions, and explain unusually small viscosities observed.

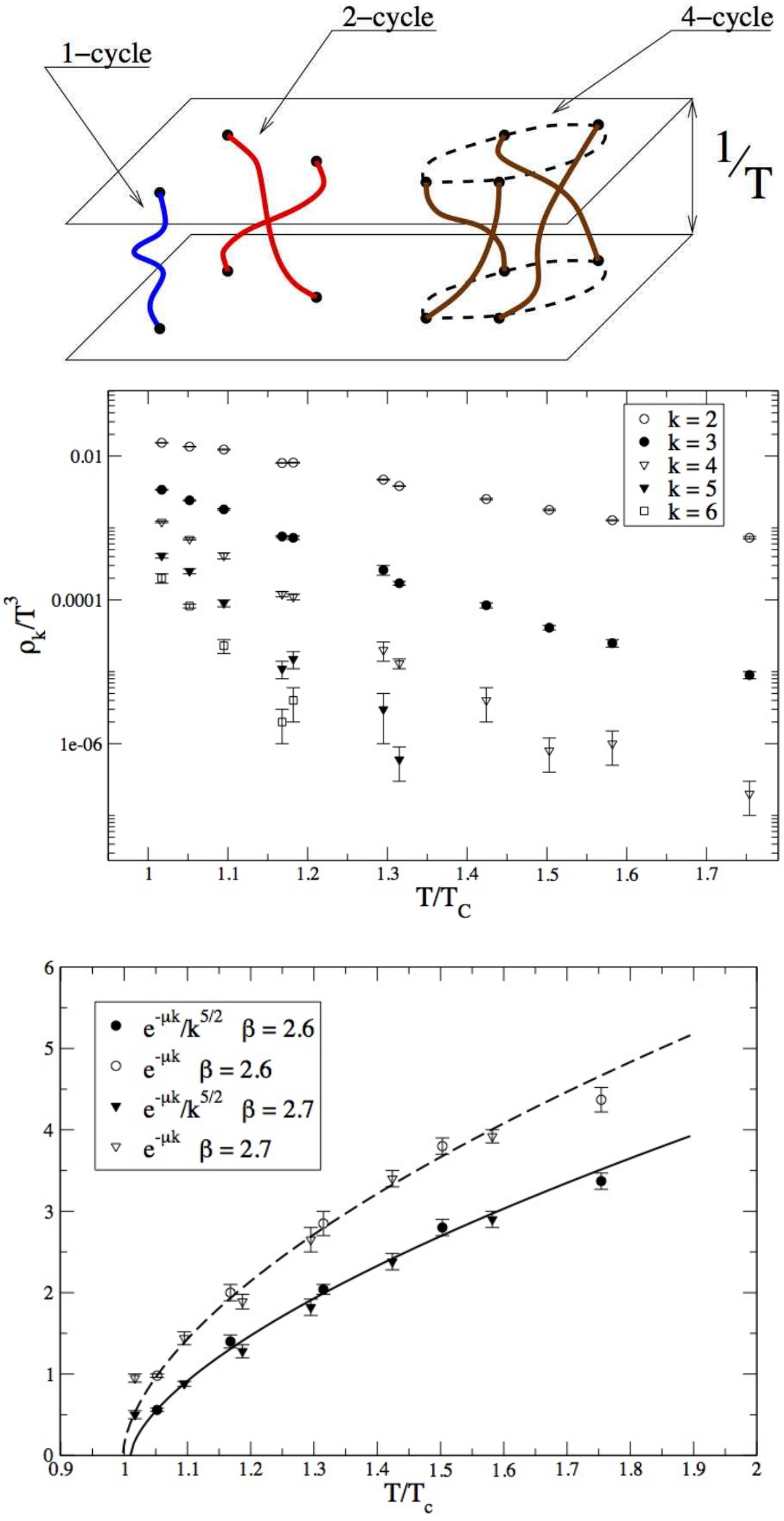

**Fig. 5.11.2** (top) Example of paths of 7 identical particles which undergo a permutation made up of a 1-cycle, a 2-cycle and a 4-cycle. (middle) Normalized densities $\rho_k/T^3$ as a function of $T/T_c$ . (bottom) Effective chemical potential $\mu_{\rm eff}(T)$ is versus temperature, vanishes at the critical temperature.

The idea of Bose-clusters is explained in the top pane of Fig.5.11.2: identical bosons may have "periodic paths" in which some number $k$ of them exchange places. Such clusters are widely known to the community doing many body path integral simulations for bosons, e.g. liquid $He^4$. Feynman argued that in order for the statistical sum to be singular at $T_c$, the sum over $k$ must diverge. In other words, one may see how the probability to observe $k$-clusters $P_k$ grows as $T \to T_c$ from above. In Fig.5.11.2(middle) from [1359] one sees the corresponding data for the cluster density. Their dependence on $k$ was fitted by two expressions, $\rho_k \sim \exp\left(-k\mu_{\rm eff}(T)\right)/k^{5/2}$ or the same without the $k^{-5/2}$ factor, to show that the critical $T$ is not sensitive to these details of the fit. The effective chemical potential $\mu_{\rm eff}(T)$ plotted versus temperature in the bottom panel of Fig. 5.11.2. vanishes exactly at the deconfinement temperature $T = T_c$ (defined by different methods). This means that monopoles indeed undergo Bose-Einstein condensation at exactly $T = T_c$.

### 5.11.4 Topological landscape

Magnetic monopoles were only the first of the solitons (solutions to nonlinear classical equations of motion, stable in the sector with fixed topology). In fact there exist a whole zoo of them, even in pure gauge theory without any scalar fields.

Gauge symmetry of QCD allow transformations of fields with arbitrary $SU(3)$ matrices $\Omega(x)$, with arbitrary dependence on space-time point $x$. Those matrices can be divided into topologically distinct classes. Introducing the Chern-Simons number $N_{CS}$ [1363] for the gauge potentials

$$N_{CS} \equiv \frac{\epsilon^{\alpha\beta\gamma}}{16\pi^2} \int d^3x \left( A^a_\alpha \partial_\beta A^a_\gamma + \frac{1}{3}\epsilon^{abc} A^a_\alpha A^b_\beta A^c_\gamma \right), \tag{5.11.3}$$

one may prove that if it is an integer, then the gauge configuration with minimal energy is "pure gauge", the field strength $G^a_{\mu\nu}$=0 and the minimal energy is zero. Thus the values of $N_{CS}$ numerate "classical vacua" with different topologies.

Yet when $N_{CS}$ is in between these integers, the field strength and the minimal energy is *nonzero*. This creates a "topological landscape", an infinite sequence of classical vacua separated by *barriers*, see Fig. 5.11.3. By minimizing the energy at fixed $N_{CS}$ (and r.m.s. size $\rho$) of the configurations, one can find [1364] the shape of

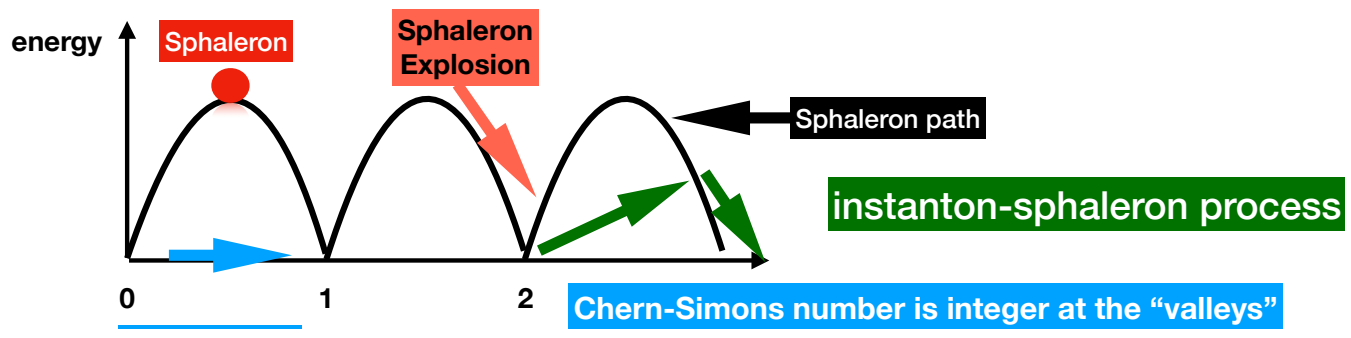

**Fig. 5.11.3** The "topological landscape": minimal potential energy $U_{\rm min}$ (in units of $1/g^2\rho$) versus the Chern-Simons number $N_{CS}$. Valleys at integer values are separated by barriers. The terminology and arrows are described in the text.

this barrier in parametric form

$$U_{\rm min}(\kappa,\rho) = (1-\kappa^2)^2 \frac{3\pi^2}{g^2\rho},$$
$$N_{CS}(\kappa) = \frac{1}{4}{\rm sign}(\kappa)(1-|\kappa|)^2(2+|\kappa|)\,. \tag{5.11.4}$$

Here $\kappa = 0$ corresponds to the top of the barrier: this configuration is called the "sphaleron" (which in Greek means "ready to fall"). It is a solution of the classical equations of motion, a magnetic ball in which field lines of $\vec{B}^a$ ($a = 1, 2, 3$ since it is restricted to the SU(2) subgroup of SU(3)) rotate around the $x, y, z$ axes. Since it corresponds to an energy maximum (rather than minimum), a small perturbation would force it to fall down the barrier profile: this process (also studied analytically and numerically) is called "the sphaleron explosion". (We indicated it on the right side of Fig. 5.11.3 by red downward arrow.)

Sphalerons were originally discovered in electroweak theory [1365, 1366]: in this case the sphaleron energy is very large, about 8 TeV. There were long debates whether those can be produced at LHC or future colliders: so far not a single event of this kind has been observed. Production of sphaleron-like hadronic clusters with various sizes and masses, in pp collisions at RHIC and LHC, are under consideration, see more in review [1367]. Green arrows on the r.h.s. of Fig. 5.11.3 indicate the instanton-sphaleron process in which vacuum is excited to a "turning point" magnetic configuration at the side of the barrier, from which it explodes (rolls downward).

Quantum mechanics allows potential barriers to be *penetrable* due to "tunneling". So, at any energy, even zero, *tunneling events* occur, changing $N_{CS}$ spontaneously. Under the barrier the potential energy is larger than the total, and the kinetic energy is negative $E-U = K < 0$. Since it is proportional to momentum squared $K \sim \pi^2$, the motion should occur with imaginary momentum. That lead to the idea to describe this motion in imaginary time $\tau = it$, or Euclidean space-time. Explicit solutions describing tunneling have been found [1350], and are known as the BPST *instantons* (indicated by the horizontal blue line on the left of Fig. 5.11.3). To find them one assumes the solution is spherically symmetric in 4-d, and can be described by scalar trial radial function $f$, with

$$gA^a_\mu = \eta_{a\mu\nu}\partial_\nu F(y), \quad F(y) = 2\int_0^{\xi(y)} d\xi' f(\xi') \tag{5.11.5}$$

with $\xi = \ln\big(x^2/\rho^2\big)$ and $\eta$ the 't Hooft symbol defined by

$$\eta_{a\mu\nu} = \begin{cases} \epsilon_{a\mu\nu} & \mu,\, \nu = 1,\, 2,\, 3, \\ \delta_{a\mu} & \nu = 4, \\ -\delta_{a\nu} & \mu = 4. \end{cases} \tag{5.11.6}$$

We also define $\bar{\eta}_{a\mu\nu}$ by changing the sign of the last two equations. Putting this expression into the gauge Lagrangian one finds that it takes the form

$$L_{\rm eff} = \int d\xi \left[\frac{\dot{f}^2}{2} + 2f^2(1-f)^2\right] \tag{5.11.7}$$

where the dot is the derivative with respect to $\xi$. This corresponds to the motion of a particle in a double-well potential. Note that, since $L = K - U$, the sign in front of the potential is *inverted*, giving two maxima rather than minima. The instanton solution is the one "sliding" from one maximum, at $\xi = 0$, to the other at $\xi = 1$.

As an individual instanton is basically a 4d ball of $G^a_{\mu\nu}$ fields, the gauge field vacuum (in Euclidean time) can be described by an ensemble of instantons and antiinstantons (those with $\bar{\eta}^a_{\mu\nu}$). The so called instanton liquid model (ILM) [1368] concluded that the instanton size and density

$$\rho = \frac{1}{3}\,{\rm fm}, \qquad n_{I+\bar{I}} = \frac{1}{R^4} = 1\,{\rm fm}^{-4} \tag{5.11.8}$$

led to chiral symmetry breaking, reproducing parameters of chiral perturbation theory and pion properties. Note that the 4d ball volume is $\pi^2\rho^4/2$, and the diluteness $n_{I+\bar{I}}\ \pi^2\rho^4/2 \sim 1/20 \ll 1$ of the ensemble is quite small. Yet, they are interacting with each other strongly, thus the use of the word "liquid" in the name. Many years later, numerical simulations on the lattice have shown what it looks like, see Fig.5.11.4 from [1369] . Technically, this is a lattice gauge field "deeply cooled" (with the action minimized) which removes gluons but keeps the gauge topology intact. One can find more on lattice topology in section 4.3.2.

### 5.11.5 Instantons bind quarks, and by this generates chiral dynamics

G. 't Hooft [1370] has found that instantons bind massless fermions at zero energy. Technically, these are solutions of the Dirac equation in the instanton field, called

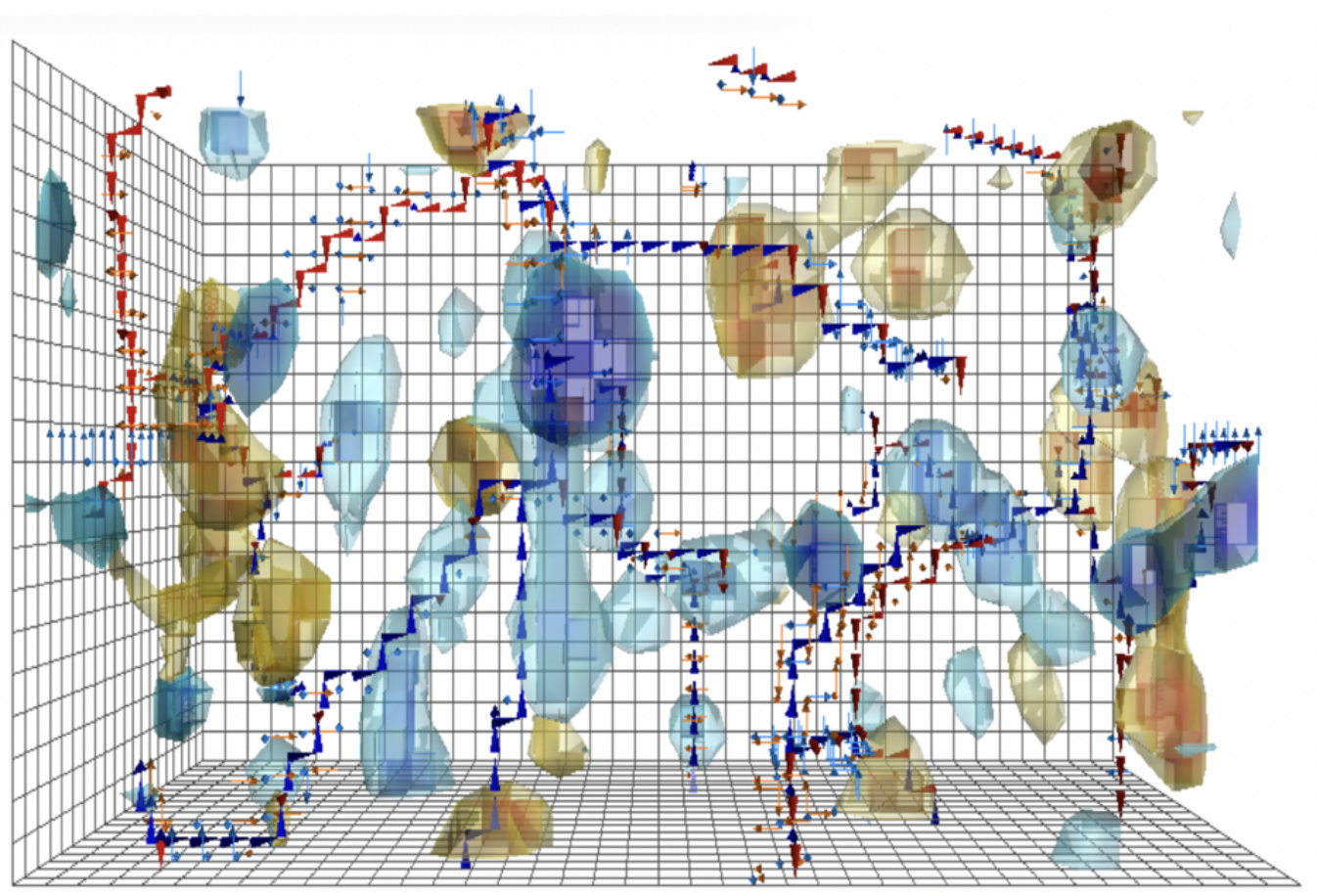

**Fig. 5.11.4** QCD lattice configuration under "deep cooling": blue and yellow regions are locations of instantons and anti-instantons. One can also see a few magnetic flux tubes.

*femionic zero modes.* The Pauli principle applies, and each instanton (a gauge field ball) binds one of each light quark, $u, d, s$. Therefore an "instanton liquid" contains "collectivized" light quark states. It is analogous to a ensemble of atoms: while each has its own electrons, at a finite density of atoms, these electrons can be in different phases, e.g. "insulating" or "conducting," depending on whether collective electron states do or do not have nonzero density of states on the Fermi surface. Similarly, an ensemble of instantons can have a spectrum of Dirac eigenvalues $\lambda$, either *with* or *without* a gap at $\lambda = 0$: in the latter case (analogous to a conductor) the chiral symmetry is spontaneously broken. With the ILM parameters mentioned above, one can prove that this is indeed the case in the QCD vacuum, and in fact it correctly reproduces the density of Dirac eigenvalues at zero (proportional to "vacuum quark condensate") $\langle\bar{q}q\rangle \approx -(240\,MeV)^3$ known from phenomenology.

This physics can be described in different, simpler terms. Massless quark fields in QCD have left and right-polarized components which, according to the QCD Lagrangian, have independent flavor symmetries. Yet, as quarks get dressed by nontrivial vacuum fields they *may* get mixed together so that the quarks develop nonzero "constituent quark masses" $M_{\rm eff} \sim$ 350-400 MeV. The nucleon mass is about $3M_{\rm eff}$: so the phenomenon of chiral symmetry breaking explains the "mystery of our mass".

Furthermore, gauge theory in Euclidean time can naturally describe the properties at finite temperature $T$: just define $\tau$ to be on a circle with a circumference $\hbar/T$ (known as Matsubara time). Then the instanton solution can easily be made periodic. Although zero fermionic modes are still there at any $T$, the spectrum changes at $T > T_\chi$ and the Dirac eigenvalue spectrum contains a nonzero gap, and chiral symmetry gets *unbroken* at high $T$. For a review on the chiral dynamics induced by an interacting ensemble of instantons see [1371].

### 5.11.6 QCD correlation functions: from quarks to mesons and baryons

Physics of QCD correlation functions using the so called QCD sum rule method and lattice numerical simulations is described in other sections. For a general pedagogical review see e.g. Ref. [1372]. At small distances between the operators the natural description is provided by perturbative diagrams, defined in terms of quarks and gluons. At large distances they are described in terms of the lowest hadrons with appropriate quantum numbers.

Of great interest however is their behavior at intermediate distances, at which a transition from one language to another takes place. As summarized in Ref. [1372], using diagrams with a *single* instanton one can explain the scale of this transition in "problematic" channels. In particular, it is attraction in the pion channel and repulsion in $\eta'$, attraction for scalar glueball and repulsion for pseudoscalar one, etc.

Furthermore, experimentally known correlation functions were quantitatively reproduced by the interacting instanton liquid model even at large distances, first for many mesonic channels [1373, 1374] and subsequently for baryonic correlators [1375]. As a result, the predictive power of the model has been explored in substantial depth. Many of the coupling constants and hadronic masses were calculated, with good agreement with experiment and lattice. (This was shown to be the case, in spite of the fact that instanton models did *not* explain confinement.)

Subsequent calculations of baryonic correlators [1375] have revealed further surprising facts. In the instanton vacuum the nucleon was shown to be made of a "constituent quark" plus a deeply bound *diquark*, with a mass nearly the same as that of constituent quarks. On the other hand, decuplet baryons (like $\Delta^{++}$) had shown no such diquarks, remaining weakly bound set of three constituent quarks. To my knowledge, this was the first dynamical explanation of deeply bound scalar diquarks. Deeply bound scalar diquarks are a direct consequence of the 't Hooft Lagrangian, a mechanism that is also shared by the Nambu-Jona-Lasinio interaction [1376], but ignored for a long time. This subsequently lead to the realization that diquarks can become Cooper pairs in dense quark matter; see [1377] for a review on "color superconductivity".

### 5.11.7 instanton - dyons lead to semiclassical theory of the deconfinement and chiral transitions

We have described *monopoles* and *instantons*, and have shown how they can help us understand such important nonpertubative properties as *confinement* and *chiral symmetry breaking*, respectively. Yet neither of them were able to describe both of them in a natural way.

This was achieved only during the last decade, using what are called *instanton − dyons* (kind of a hybrid of these two topological animals, also known as instanton-monopoles). Technically, if they are far from each other, they can be described as monopoles, which use the $A_0$ component of the gauge field instead of the adjoint scalar of the Georgi-Glashow model, involved in 't Hooft-Polyakov monopole construction. When they overlap, they can still be followed analytically. When their centers happen to be at the same spatial point, their superposition turns out to be nothing else but the well known instanton [1352, 1378]!

A hybrid often inherits good properties of both parents – but maybe some bad properties as well. In order to sort these out, we need to start explaining from special role of $A_0$ in the finite-temperature theory. We have mentioned that finite temperature theory is defined on a circle $\tau \in C^1$ with the Matsubara period. In such cases there exist a phenomenon known in mathematics as "holonomy": there are non-contractable contours. The so called *Polyakov line*

$$P = P\exp\left[i\int_C d\tau A_0^a T^a\right] \tag{5.11.9}$$

($T^a$ is a color generator) is a gauge invariant operator. (Because $A_0$ must be periodic on (Euclidean time circle) $C^1$, its gauge factors cancel out.) Therefore, if it has certain values, it cannot be undone and thus, at finite $T$, one cannot use the $A_0 = 0$ gauge. And indeed, the average of $P$ has some well defined expectation value $\langle P(T)\rangle$, extensively studied on the lattice (see Fig. 5.11.5). Since it is a unitary $SU(3)$ matrix, it can be defined by three eigenvalues $\exp(i\mu_i), i = 1, 2, 3$. The phases $\mu_i$ are called *holonomies* $\mu_i(T)$: they prescribe the magnitude of the fields $A_0^a(T)$. Physically $\langle P(T)\rangle \sim exp[-F_Q/T]$ is related to the free energy of a static quark: in the confining phase the latter is infinite and $\langle P(T)\rangle = 0$ while in quark-gluon plasma phase it is finite and $\langle P(T)\rangle \neq 0$: so it is the *order parameter* of deconfinement.

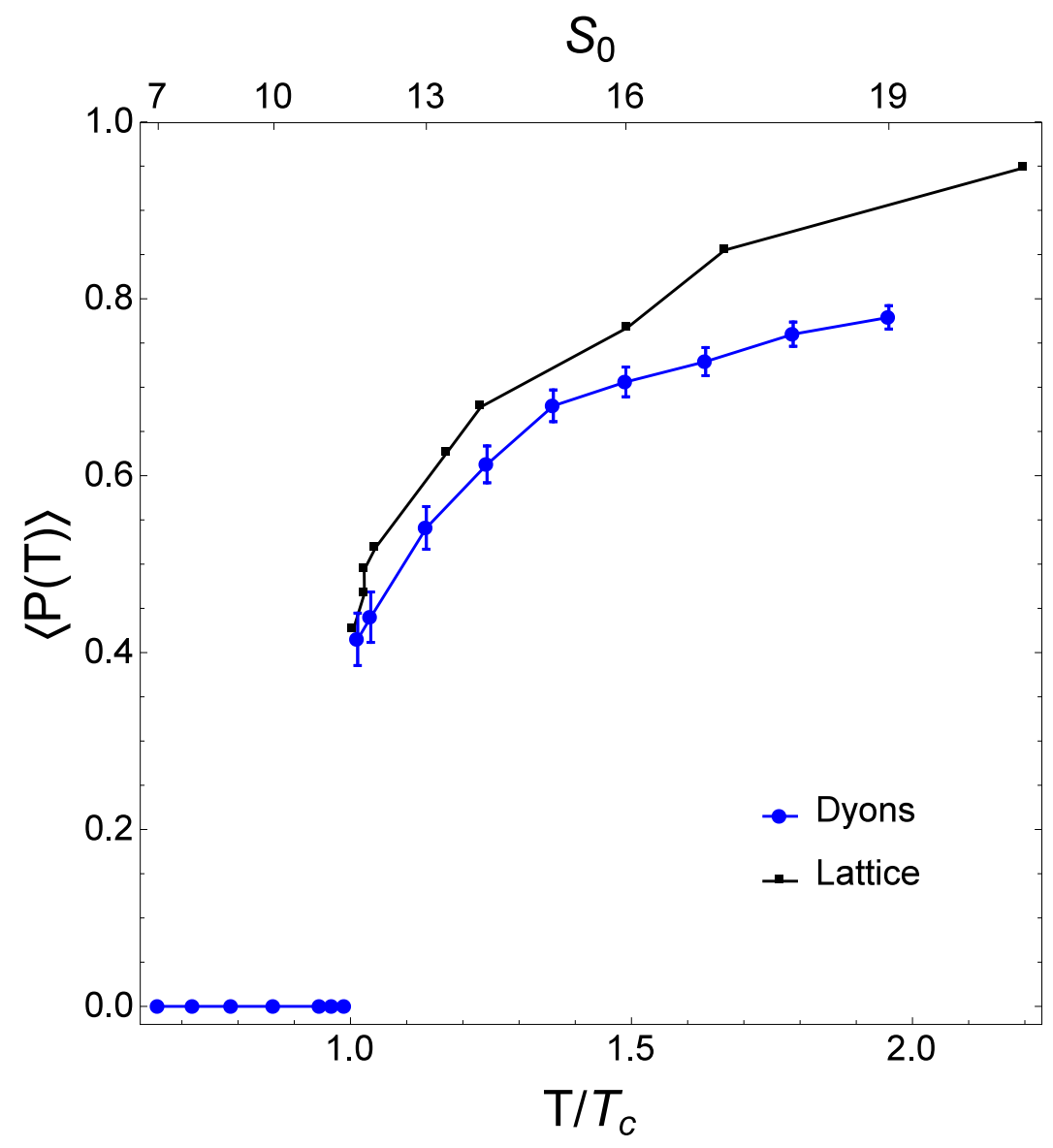

**Fig. 5.11.5** Temperature dependence of the mean Polyakov line in pure $SU(3)$ gauge theory, from lattice and instanton-dyon statistical simulations, displays a clear first order phase transition in which $\langle P\rangle$ jumps from zero below $T_c$ to a finite value in the quark-gluon plasma phase at high $T$.

Recognizing that $A_0$ may have a nonzero constant value all over the system, which cannot be gauged away and is thus physical, one has to look for solutions of the YM equations at finite temperatures which at distance $\vec{r} \to \infty$ go to such values of $A_0$. (Rather complicated) solutios of this type [1352] for instantons were found, and it was recognized (only after its actions were plotted) that it describes a continuous deformation, from one spherical instanton into $N_c$ independent bumps. If $N_c = 3$, one can follow how the triplet of *instanton − dyons* is born!

Now let us summarize their properties. Like instantons, they are (anti) selfdual $\vec{E} = \pm\vec{B}$ and live in Euclidean space time. So, they are not really particles, since they do not exist in the Minkowski world. Like instantons, they have nonzero topological charges $Q \sim \int d^4x(\vec{E}\cdot\vec{B})$. Unlike instantons, however, those charges are *not quantized to integers*: $Q_i, i = 1, 2, 3$ can take any values, except that their sum is still $|\sum_1^3 Q_i| = 1$. These $Q_i$ (equal to their actions $S_i$) are proportional to differences $\nu_i = \mu_{i+1} - \mu_i$ of the eigenvalues of the Polyakov line (the holonomies). So, $S_i = \nu_i S$ where the coefficient is the "instanton action"

$$S = \frac{8\pi^2}{g(T)^2} = \left(\frac{11}{3}N_c - \frac{2}{3}N_f\right) log\left(\frac{T}{\Lambda_{QCD}}\right)$$

Non-integer $Q_i$ is only possible because they inherited properties of another parent, the magnetic monopoles. These objects are connected by Dirac strings (this connection undoes the topological classification theorems which require that the fields be smooth at infinities.) They are called "dyons" because a magnetic charge plus a selfduality implies also presence of an *electric* charge (although real only in the Eucldean world and thus not quite physical).

Before we can proceed, we need to clarify one more puzzle related to fermionic zero modes of instanton - dyons . An instanton has *one* fermionic zero mode, and if it gets split into three instanton - dyons , one may ask how this zero mode be shared between them. The answer, also due to van Baal and collaborators, is that the zero mode is centered near *one of the three*: which one depends on the interrelation between holonomy phases $\mu_i$ and quark periodicity phases called $z_f$ where index $f$ means flavor, $u, d, s....$ Further details on instanton-dyons, their interaction and fermionic zero modes can be found in references mentioned.

This information should be sufficient to understand how one can "hunt" for these objects on the lattice. One method is "cooling" of vacuum fields, like that used in Fig.5.11.4. Better still is "constrained minimization" [1379] preserving the value of $\langle P\rangle$: it revealed selfdual clusters of topological charges which integrate to non-integer values. But the best is the "fermionic filter", developed by Gattringer *et. al.* and Ilgenfritz *et. al.*, based on the zero modes of the quark Dirac operator. In Fig.5.11.6 we show an example from [1380, 1381] in which it was used. QCD simulations with realistic masses, performed by large collaborations on supercomputers, provide a set of configurations to these authors. These calculations are especially expensive since they use the so called *domain wall fermions* providing very accurate chiral symmetry of lattice fermions. Yet Larsen *et. al.* used even better ones, called *overlap fermions*, for which chiral symmetry is exact even for finite lattices (without the continuum limit $a \to 0$ taken). Those possess *exact* zero modes $\lambda = 0$ and configurations have exactly integer topological charges.

Fig.5.11.6 shows a typical landscape of the zero mode densities $|\psi_0(x)|^2$ in two spatial dimensions. Red, blue and green colors show those for three different fermionic periodicity phases, identifying three instanton - dyon types (for $N_c = 3$) that they want to locate. The peaks correspond to locations and sizes of the individual zero modes in these field configurations. One can be convinced that the peaks are instanton - dyons because their shapes are well described by analytic formulae as derived by van Baal and collaborators, within a few percent accuracy. Furthermore, this is true not only for well separated ones, but also for overlapping ones! The gauge field configurations are for $T$ a little bit above deconfinement $T_c$, in a quark-gluon plasma possessing zillions of thermal quarks and gluons: and yet, the instanton - dyons are apparently undisturbed by them! (For clarity: we do not mean here the *values of the topological charge Q or number of zero modes*, protected by certain mathematical theorems. The observed space-time shapes of the Dirac eigenmodes are not protected by any theorems known to us.)

Previous works however have not analyzed the "topological clusters", the situations in which two or three dyons overlap strongly. The Kraan-van Baal solution allows to study these cases, and good agreement was also found in the numerical analysis of instanton-dyon ensembles in [1380, 1381]. The *semiclassical descriptio* of zero and near-zero Dirac modes on the lattice is quite accurate, at least in terms of the zero mode shapes. While the very existence of zero modes was required by topological theorems, good correspondence of their shapes (in physical thermal vacuum versus pure semiclassical dyons) was a good unexpected news.

This (and many similar plots) extracted from simulations of the QCD vacuum should convince the reader that instanton - dyons are well identified objects, in terms of which one can try to describe the underlying gauge field configurations. If so, perhaps a dream being alive for half a century since 1970's to develop consitent semiclassical theory of deconfinement and chiral transitions can still be realized.

Following this idea, ensembles of instanton - dyons were studied by a number of methods, including the mean field (solving certain "gap equations") or straightforward statistical Monte-Carlo simulations. Those were performed first for the SU(2) gauge theory [1382], then for the SU(3) [366], first without dynamical fermions, then with them [1383, 1384]. In Fig.5.11.5 we have shown one example of a comparison between a semiclassical instanton-dyon ensemble and lattice simulations. We cannot present here other results of these works but just state that they compare well with the location and properties of QCD phase transitions which we now know from lattice simulations. Note that those works were done on laptops or ordinary PCs, not supercomputers, and yet the number of topological objects in them are counted by hundreds, while very expensive lattice simulations have only few of them (as one can see from the example above).

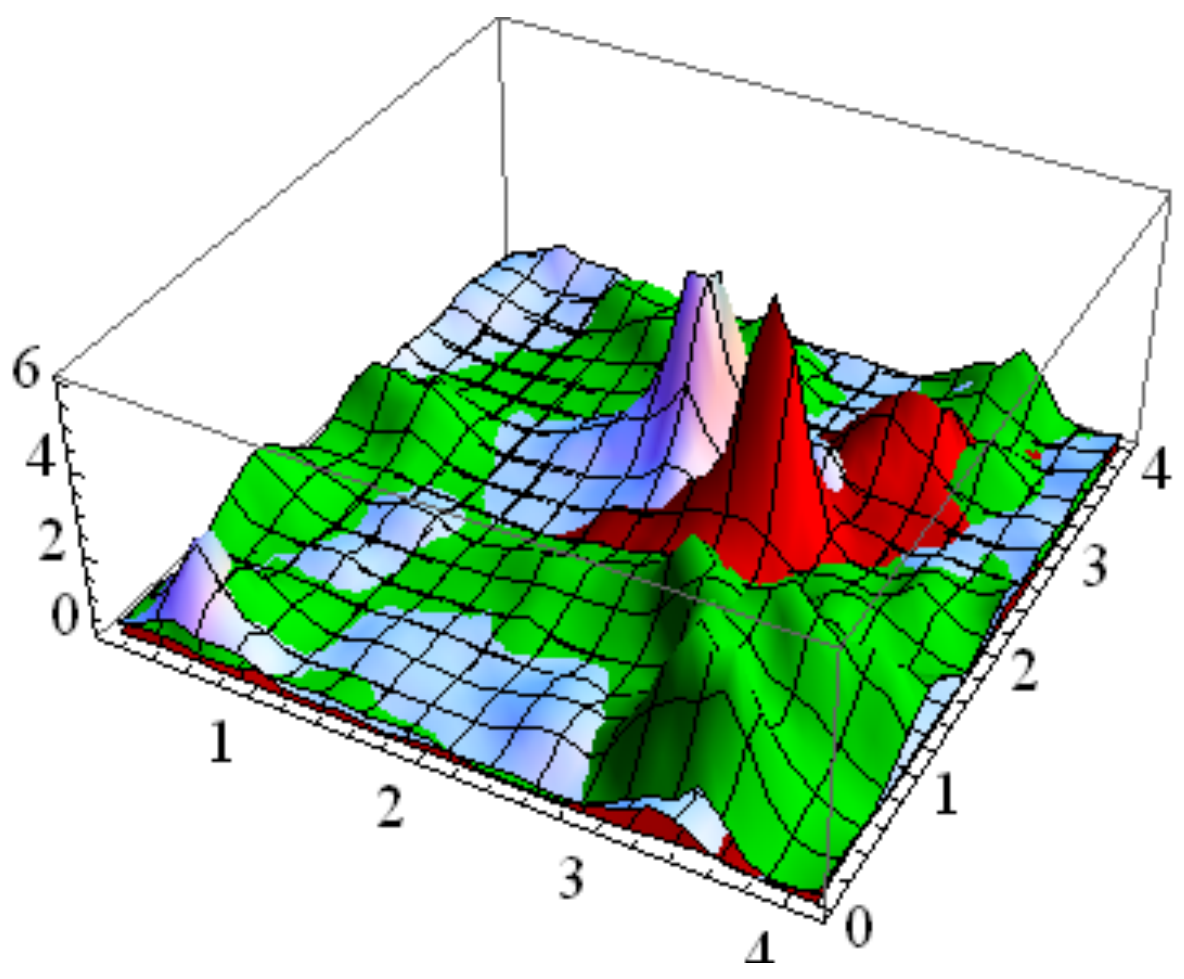

**Fig. 5.11.6** Space slice of density of an exact zero mode from QCD simulaiton at $T = T_c$. The three colors refer to dyons of three different types.

Furthermore, in such studies people used not just QCD, but also two types of "QCD deformations". One of them adds extra operators with powers of the Polyakov line to the gauge action. By changing their strength one can affect the location and strength of the deconfinement phase transition. Another type of QCD deformation makes quarks obeying modified periodicity condition on the Matsubara circle, making quark statistics to be intermediate between fermions and bosons. This deformation affects the location and strength of the chiral phase transition. What these deformations tell us is that these two phase transitions should not generically be coincident, as they are in QCD. Again, one can apply such deformations on the lattice or in the instanton - dyon semiclassical theory, and compare the results. So far, the agreement between them is quite good, which is encouraging.

### 5.11.8 Conclusions and discussion

The main thrust of this section is to convince the reader that *topological solitons* play an important role in the understanding of such nonperturbative phenomena in QCD as *confinement* and *chiral symmetry breaking* in vacuum, as well as *deconfinement* and *chiral symmetry restoration* at high temperatures. A wider view on that should include the *deformed versions* of QCD, or even other gauge theories, electroweak or supersymmetric theories.

It would be nice to have just *one* type of those: but in fact the history of the field we followed in this section included (at least) three: *the particle-monopoles, instantons and instanton - dyons.* All of those were found on the lattice, by different "filters", and were shown to be strongly correlated with certain physical phenomena we would like to understand.

The *particle−monopole* behavior convincingly shows that confinement is a Bose-Einstein condensation, explaining both the confining flux tubes and their disappearance at high $T$.

The *instantons* have fermionic clouds bound to them, and their "collectivization" into a "conductor" without a gap explains how a "quark condensate" is formed, the physics of massless pions, and (unlike earlier theories) why $\eta'$ is so heavy. They explain the value of the "constituent quark mass" as well as that of the nucleon (and thus ourselves). While instanton ensembles do not explain confinement, they do have most of the lowest mesons and baryons (nucleon included) as bound states.

The *instanton − dyons* (being a hybrid of the first two) connect topology with *holonomy* (the Polyakov line, or nonzero $A_0$ VEV, in Euclidean formulation) in a way, which produces a nice semiclassical theory of both deconfinement and chiral transition. It was shown to work quantitatively, not only for QCD, but for its deformations as well.

Taken together, those facts and observations are impressive. The reader is reminded that they constitute the result of five decades of work by multiple theorists. But still, the reader is perhaps a bit confused by the very richness of the story told. One would probably prefer a simpler and more uniform picture.

Such feelings are shared by some active participants in this process, and some light at the end of the tunnel is, in fact, now showing. At the end of the section, let us briefly describe these later developments.

It started with Ref. [1385], using the well controlled setting of the most-supersymmetric gauge theory, with $\mathcal{N}$=4 supersymmetries. This theory has adjoint scalars and 't Hooft-Polyakov monopoles as classical solutions, and the partition function in terms of these monopoles can be calculated. The same theory in a $R^3C^1$ setting (preserving supersymmetries) also has holonomies and instanton - dyons, and the partition function written in terms of these was calculated as well. The two expressions are completely different, one better converges at small and another at large radius of the circle $C^1$. Nevertheless, as Dorey *et. al.* observed, using Poisson summation formula, both produce *the same answer* for the statistical sum! This unexpected result was called "the Poisson duality". The importance of this paper was not noticed promptly. Indeed, such duality is very nontrivial: it is enough to remember that monopoles are particles moving in Minkowskian space-times, while instanton - dyons can only be defined in Euclidean periodic formulation. And yet, they apparently describe the same dynamics!

In fact this phenomenon has nothing to do with supersymmetry or gauge theories, and is present in a much broader domain. In Ref. [1386], the existence of the *Poisson duality* was demonstrated for a simple quantum mechanical rotator. The duality means that a partition sum can either be written using Hamiltonian and stationary states, or using Lagrangian and periodic Euclidean paths. Further elucidation of this duality regarding QCD monopoles [1386] shed light on their density and the long-known absence of classical solutions for them. All of these hint that the different faces of "gauge topology" we discussed will "asymptotically" converge into a single semiclassical theory.

# 6 Effective field theories

*Conveners:*
**Franz Gross and Mike Strickland**

In this second section on approximate methods, we discuss effective field theories (EFTs), a powerful technique that can be used when there are widely separated energy scales appearing in a problem. A classic example of this is non-relativistic QCD (NRQCD), which emerges in the limit of a large quark mass $M$ (Sec. 6.1). For $v = p/M \ll 1$, there is a large separation between 'hard' modes, with energy on the order of $M$; soft modes, with energy on the order of $Mv$; and ulrasoft (potential) modes, with energy/momentum on the order $Mv^2$. Using EFT methods, one can write an effective NRQCD Lagrangian that includes all terms allowed by QCD symmetries. The coefficients in this effective Lagrangian can be computed systematically by a matching procedure, which ensures that quantities calculated in the EFT are the same as those computed in QCD itself up to a given order in $v$. The NRQCD EFT can be extended by further integrating out the soft scale to obtain an effective theory called potential NRQCD (pNRQCD), which is written entirely in terms of singlet and octet quark-antiquark pairs. As a result, at leading order in pNRQCD, the physics of heavy quarkonium reduces to solving a Schrödinger equation for bound state wave functions.

This is but one example. The use of EFTs applied to QCD has allowed systematic progress on many fronts in the last decades. These include a systematically extendable model of low-energy hadronic physics called chiral perturbation theory (Sec. 6.2), which can be used as a foundation for nuclear physcs (Sec. 6.3) giving both a successful description of the NN interaction up to 200 MeV, and the properties of light nuclei up to $A \leq 12$. In the realm of jets, soft-collinear effective theory implements power counting in the transverse momentum of gluon radiation (Sec. 6.4).

EFT methods have also been used to understand high-temperature QCD thermodynamics, in which case the hard, soft, and ultrasoft scales are $T$, $gT$, and $g^2T$, respectively (Sec. 6.5). The resulting EFTs, called electrostatic QCD (EQCD) and magnetostatic QCD (MQCD) allow one to systematically calculate the equation of state of a high-temperature quark-gluon plasma. Together with other finite-temperature resummation schemes such as hard-thermal-loop perturbation theory these methods have provided a way to calculate the QCD equation state that agrees well with lattice calculations down to temperatures just above the quark-gluon plasma phase transition temperature. Finally, Sec. 6.6 describes how EFTs have recently been applied to non-equilibrium QCD physics such as the quantum transport of bottomonium through the quark-gluon plasma.

## 6.1 Nonrelativistic effective theory

**Antonio Vairo**

In QCD, quarks may be divided into two fundamental sets: heavy quarks (charm, bottom, top) whose masses $m_h$ are much larger than the typical hadronic scale $\Lambda_{\rm QCD}$ and light quarks (up, down, strange) whose masses $m_\ell$ are much smaller than $\Lambda_{\rm QCD}$. Both the hierarchies, $m_h \gg \Lambda_{\rm QCD}$ and $m_\ell \ll \Lambda_{\rm QCD}$, allow for an effective field theory (EFT) treatment of hadrons that exploits the symmetries that the hadrons manifest in the large and small mass limits. Because these symmetries are not manifest in QCD, the EFT is typically simpler and more predictive than the full QCD treatment, at least at the lowest orders in the effective expansion. At higher orders in the effective expansion the original symmetries of QCD are restored. We discuss EFTs for heavy quarks in this section, while EFTs for light quarks, i.e., chiral EFTs, are reviewed in the following sections.

In general, an effective field theory of QCD is constructed as an expansion in the ratio $\Lambda_\ell/\Lambda_h$ of a low energy scale $\Lambda_\ell$, e.g. $\Lambda_{\rm QCD}$, and a high energy scale $\Lambda_h$, e.g. $m_h$. Each term in the expansion is made of the fields describing the system at the low-energy scale; these terms may have any form consistent with the symmetries of QCD. The low-energy fields are the effective degrees of freedom. The resulting scattering matrix is the most general one consistent with analyticity, perturbative unitarity, cluster decomposition and the symmetry principles [1387].

It is said that the large energy scale "has been integrated out" from QCD. Analytic terms in the expansion parameter $\Lambda_\ell/\Lambda_h$ are accounted for by the operators of the EFT. Non-analytic terms, carrying the contributions of the high-energy modes in QCD, which are no longer dynamical in the EFT, are encoded in the parameters multiplying the EFT operators. These parameters are the Wilson coefficients of the EFT, also called matching coefficients, or low-energy constants in the chiral EFT. Hence, EFTs automatically factorize, for any observable, high-energy from low-energy contributions. The Wilson coefficients of the EFT Lagrangian are fixed by matching to the fundamental theory, i.e., by requiring that the EFT and the fundamental theory describe the same physics (observables, Green functions, scattering matrices, ...) at any given order in the expansion parameter $\Lambda_\ell/\Lambda_h$.

The advantage of dealing with heavy quarks is that the matching coefficients associated with the heavy quark mass scale are guaranteed to be computable in perturbative QCD, i.e., order by order in $\alpha_s(m_h)$, as a consequence of asymptotic freedom. This is not the case for matching coefficients associated with lower energy scales or for the low-energy constants that need to be computed either numerically in lattice QCD or fixed on experimental data.

To allow for controlled calculations based on the effective Lagrangian, operators, as well as the quantum corrections, are organized according to their expected importance. Operators in the Lagrangian are counted in powers of the small expansion parameter $\Lambda_\ell/\Lambda_h$, whereas quantum corrections are either computed exactly or counted in powers of the coupling constant. For example, a strict expansion in terms of the coupling is possible, as remarked above, when integrating out the heavy quark mass.

EFTs are renormalizable at each order in the expansion parameter. Hence, the EFT produces finite and controlled expansions for any observable of the effective degrees of freedom that may be computed respecting the energy scale hierarchy upon which the EFT is based. The power counting of the EFT, i.e., the way to assess the size of the different terms in the effective expansion, may or may not be obvious. The power counting turns out to be obvious if the system is characterized by just one dynamical energy scale. Reducing the description of a system to that one of an effective one scale system is the ultimate goal of any effective field theory.

In this section, we restrict ourself to EFTs for heavy quarks, where the heavy quark mass is the largest scale. These EFTs are called nonrelativistic EFTs, because requiring the heavy quark mass to be the largest scale implies that it is also larger than the momentum $p$ of the heavy quark in the hadron reference frame: the condition $m_h \gg p$ qualifies the quark as nonrelativistic. The presentation of this section follows the one of Ref. [1388].

For hadrons made of one heavy quark, like heavy-light mesons and baryons, the proper nonrelativistic EFT is called Heavy Quark Effective Theory (HQET). Heavy-light hadrons are systems characterized by just two relevant energy scales, $m_h$ and $\Lambda_{\rm QCD}$. HQET fol-lows from QCD by integrating out modes associated with the heavy quark mass and exploiting the hierarchy $m_h \gg \Lambda_{\rm QCD}$. In the context of HQET one deals with heavy-light hadrons made of either a charm or a bottom quark (the top quark has no time to form a bound state before decaying weakly into a $b$ quark). HQET is discussed in section 6.1.1.

Systems made of more than one heavy quark, like quarkonia (e.g. charmonium and bottomonium) or quarkonium-like states or doubly-heavy baryons are characterized by more energy scales. The typical distance between the heavy quarks is of the order of $1/(m_h v)$, $v \ll 1$ being the relative velocity of the heavy quark, which implies that the typical momentum transfer between the heavy quarks is of order $m_h v$, and the typical binding and kinetic energy is of order $m_h v^2$. The inverse of $m_h v^2$ sets the time scale of the bound state. These systems are to some extent the QCD equivalent of positronium in QED. In a positronium, an electron and a positron move with a relative velocity $v \sim \alpha$, where $\alpha$ is the fine structure constant, at a typical distance given by the Bohr radius, which is proportional to $1/(m\alpha)$, and are bound with the energy given by the Bohr levels, which are proportional to $m\alpha^2$.

At each of the energy scales one can construct an EFT, specifically, nonrelativistic QCD (NRQCD) at the scale $m_h v$, which is discussed in Sec. 6.1.2, and potential NRQCD (pNRQCD) at the scale $m_h v^2$, which is discussed in Sec. 6.1.3.

### 6.1.1 Heavy Quark Effective Theory

Heavy Quark Effective Theory was the first nonrelativistic EFT of QCD with a fully developed nonrelativistic expansion, computation of higher-order radiative corrections, renormalization group equations, and a wide range of physical applications [667, 1051, 1252, 1389] (for an early review see, for instance, Ref. [1390], for a textbook see Ref. [674]). This despite the fact that nonrelativistic QCD and QED, the EFTs for two nonrelativistic particles that we discuss in section 6.1.2, were suggested before [1391].

In a sense, HQET describes QCD in the opposite limit of the chiral EFT, however, it is important to realize that HQET is not the large mass limit of QCD, but the EFT suited to describe heavy-light hadrons, i.e., hadrons made of one heavy particle and light degrees of freedom. The heavy particle may be taken to be a heavy quark, but also a composite particle made by more than one heavy quark when the internal modes of the composite heavy particle can be ignored. The light degrees of freedom are made by light quarks and gluons. Among the light quarks we may distinguish between valence quarks and sea quarks, where the first ones are those that establish, together with the heavy degrees of freedom, the quantum numbers of the heavy-light hadron.

The HQET Lagrangian is made of low-energy degrees of freedom living at the low-energy scale $\Lambda_{\rm QCD}$. These are the low-energy modes of the heavy quark (antiquark), described by a Pauli spinor $\psi$ ($\chi$) that annihilates (creates) the heavy quark (antiquark), and low-energy gluons and light quarks. The HQET is constructed as an expansion in $1/m_h$: the heavy quark expansion. Matrix elements of operators of dimension $d$ are of order $\Lambda^d_{\rm QCD}$, hence the higher the dimension of the operator the higher the suppression in $\Lambda_{\rm QCD}/m_h$. In the rest frame of the heavy-light hadron, the HQET Lagrangian density for a heavy quark reads up to order $1/m_h^2$ and including the $1/m_h^3$ kinetic operator (HQET up to order $1/m_h^4$ can be found in Refs. [1392, 1393])

$$\mathcal{L}_{\rm HQET} = \mathcal{L}^{\psi} + \mathcal{L}^{\ell}, \tag{6.1.1}$$

with

$$\begin{aligned}\mathcal{L}^{\psi} =& \psi^\dagger \Bigg\{ iD_0 + \frac{\boldsymbol{D}^2}{2m_h} + \frac{\boldsymbol{D}^4}{8m_h^3} - c_F\,\frac{\boldsymbol{\sigma}\cdot g\boldsymbol{B}}{2m_h} \\ & - c_D\,\frac{[\boldsymbol{D}\cdot, g\boldsymbol{E}]}{8m_h^2} - ic_S\,\frac{\boldsymbol{\sigma}\cdot[\boldsymbol{D}\times, g\boldsymbol{E}]}{8m_h^2}\Bigg\}\psi,\end{aligned} \tag{6.1.2}$$

$$\begin{aligned}\mathcal{L}^{\ell} =& -\frac{1}{4}F^A_{\mu\nu}F^{A\,\mu\nu} + \frac{d_2}{m_h^2}F^A_{\mu\nu}D^2F^{A\,\mu\nu} \\ & - \frac{d_3}{m_h^2}gf_{ABC}F^A_{\mu\nu}F^B_{\mu\alpha}F^C_{\nu\alpha} \\ & + \sum_{\ell=1}^{n_\ell}\bar{q}_\ell\left(i\not{D} - m_\ell\right)q_\ell,\end{aligned} \tag{6.1.3}$$

where $[\boldsymbol{D}\cdot, g\boldsymbol{E}] = \boldsymbol{D}\cdot g\boldsymbol{E} - g\boldsymbol{E}\cdot\boldsymbol{D}$ and $[\boldsymbol{D}\times, g\boldsymbol{E}] = \boldsymbol{D}\times g\boldsymbol{E} - g\boldsymbol{E}\times\boldsymbol{D}$, $\boldsymbol{E}^i = F^{i0}$ is the chromoelectric field, $\boldsymbol{B}^i = -\epsilon_{ijk}F^{jk}/2$ the chromomagnetic field, and $\boldsymbol{\sigma}$ are the Pauli matrices. The fields $q_\ell$ are $n_\ell$ light-quark fields. The heavy-quark mass, $m_h$, has to be understood as the heavy quark pole mass, hence not the mass that is in the QCD Lagrangian. The coefficients $c_F$, $c_D$, $c_S$, $d_2$, and $d_3$ are Wilson coefficients of the EFT. They encode the contributions of the high-energy modes that have been integrated out from QCD. Since the high-energy scale, $m_h$, is larger than $\Lambda_{\rm QCD}$, the Wilson coefficients may be computed in perturbation theory and organized as an expansion in $\alpha_s$ at a typical scale of order $m_h$. The coefficients $c_F$, $c_D$, and $c_S$ are 1 at leading order, while the perturbative series of the coefficients $d_2$ and $d_3$ starts at order $\alpha_s$. The one-loop expression of the coefficients may be found in Ref. [1394]. Some of the coefficients are known far beyond one loop. For instance, the Fermi coefficient $c_F$, which plays a crucial role in the spin splittings, is known up to three loops [1395]. In Eq. (6.1.3) we have not considered $1/m_h^2$ suppressed operators involving light quarks [1396, 1397] since their impact is negligible in most hadronic observables. The HQET Lagrangian for a heavy antiquark may be obtained from Eqs. (6.1.1) and (6.1.2) by charge conjugation.

The HQET Lagrangian provides a description of heavy-light hadrons that is the same as QCD order by order in $\Lambda_{\rm QCD}/m_h$. Because QCD is a Lorentz invariant theory, this symmetry must be somehow maintained in HQET, although HQET breaks manifest Lorentz invariance by the nonrelativistic expansion. Indeed, Lorentz invariance is realized in HQET by constraining the Wilson coefficients [1394, 1398–1400]. For instance, Lorentz invariance relates $c_F$ and the spin-orbit coefficient $c_S$: $c_S = 2c_F - 1$. This relation is exact, which means that it holds to any order in $\alpha_s$.

The impact of HQET on the physics involving heavy-light hadrons and, in particular, their weak decays has been enormous. The reason is that the leading-order HQET Lagrangian,

$$\mathcal{L}^{(0)}_{\rm HQET} = \psi^\dagger iD_0\psi - \frac{1}{4}F^A_{\mu\nu}F^{A\,\mu\nu} + \sum_{\ell=1}^{n_\ell}\bar{q}_\ell\left(i\not{D} - m_\ell\right)q_\ell, \tag{6.1.4}$$

makes manifest a hidden symmetry of heavy-light hadrons. This symmetry is the heavy-quark symmetry and stands for invariance with respect to the heavy-quark flavor and spin. A consequence of the heavy-quark symmetry is that electroweak transitions in the heavy-light meson sector depend on only one form factor, the Isgur–Wise function $\xi(w)$, whose absolute normalization is $\xi(0) = 1$ [1252, 1389]. Moreover, the leading-order HQET Lagrangian is exactly renormalizable.

Higher-order operators in Eq. (6.1.1) break the heavy-quark symmetry (and exact renormalizability), however, they do it in a perturbative way controlled by powers of $\Lambda_{\rm QCD}/m_h$. Hence, observables computed up to some order in the HQET expansion depend on fewer and more universal nonperturbative matrix elements than they would in a full QCD calculation. This makes the heavy quark expansion more predictive than a full QCD calculation.

As an application, let us consider the heavy-light meson masses. Expressed in the HQET as an expansion up to order $1/m_h$, they read [1401]

$$M_{H^{(*)}} = m_h + \bar{\Lambda} + \frac{\mu_\pi^2}{2m_h} - d_{H^{(*)}}\frac{\mu_G^2(m_h)}{2m_h} + \mathcal{O}\left(\frac{1}{m_h^2}\right), \tag{6.1.5}$$

where $M_{H^{(*)}}$ is the spin singlet (triplet) meson mass, $m_h$ the heavy quark pole mass, $\bar{\Lambda}$ the binding energy in the static limit, of order $\Lambda_{\rm QCD}$, $\mu_\pi^2/2m_h$ the kinetic energy of the heavy quark ($\mu_\pi^2$ is the matrix element of $\psi^\dagger\boldsymbol{D}^2\psi$), of order $\Lambda_{\rm QCD}^2/m_h$, $d_{H^{(*)}}$ is 1 for $H$ and $-1/3$ for $H^*$, and $d_{H^{(*)}}\mu_G^2(m_h)/2m_h$ is the matrix element of $c_F\,\psi^\dagger\boldsymbol{\sigma}\cdot g\boldsymbol{B}/(2m_h)\psi$, of order $\Lambda_{\rm QCD}^2/m_h$. The heavy quark symmetry manifests itself through the universality of the leading term $M_{H^{(*)}} - m_h \approx \bar{\Lambda}$, and of the matrix elements $\mu_\pi^2$ and $\mu_G^2(m_h)/c_F(m_h)$, which depend neither on the heavy quark flavor nor on the heavy quark spin. The flavor dependence of $\mu_G^2(m_h)$ comes entirely from the Wilson coefficient $c_F$, which depends on $m_h$ through the running of the strong coupling. Equation (6.1.5) can be used to extract the heavy quark masses from the measured meson masses. One can also use lattice QCD data to determine meson masses for fictitious heavy quarks of any mass $m_h$, so to reconstruct $M_{H^{(*)}}$ as a function of $m_h$. One general difficulty in this kind of study is that the relation between the $\overline{\rm MS}$ mass, which is the short distance quantity that appears in the renormalized QCD Lagrangian, and the pole mass, which is the quantity that appears in the HQET Lagrangian, is plagued by a poorly convergent perturbative series (at present, the relation between the $\overline{\rm MS}$ mass and the pole mass is known up to four loops [1402, 1403]). The large terms in the perturbative series trace back to a renormalon singularity in the Borel plane of order $\Lambda_{\rm QCD}$. This singularity may be subtracted from the pole mass and reabsorbed into a redefinition of the other nonperturbative parameters appearing in Eq. (6.1.5). There are many possible subtraction schemes [1404–1409]. For illustration, we present the heavy quark masses and matrix elements appearing in (6.1.5) obtained from lattice QCD data set to reproduce the physical $D_s$ and $B_s$ masses in Ref. [269]:

$$\begin{aligned}
\overline{m}_c &= 1273(10)\ \text{MeV},\\
\overline{m}_b &= 4195(14)\ \text{MeV},\\
\bar{\Lambda} &= 555(31)\ \text{MeV},\\
\mu_\pi^2 &= 0.05(22)\ \text{GeV}^2,\\
\mu_G^2(m_b) &= 0.38(2)\ \text{GeV}^2,
\end{aligned}$$

where $\overline{m}_h$ is the $\overline{\rm MS}$ mass of the quark $h$ at the scale of its $\overline{\rm MS}$ mass, $\bar{\Lambda}$ is in the renormalon subtraction scheme adopted in Ref. [269, 1409] and the quantity $\mu_G^2$ has been evaluated for the $b$ quark. Note the approximate scaling of the nonperturbative parameters according to the power counting of HQET (with a somewhat smaller $\mu_\pi^2$).

Equation (6.1.5) can be immediately extended to heavy-light baryons. What changes is the explicit value of the nonperturbative matrix elements, as the light degrees of freedom are different from the mesonic case. Also doubly-heavy baryons may be described by the same mass formula if the typical distance between the two heavy quarks is much smaller than the typical size of a heavy-light meson, which is of order $1/\Lambda_{\rm QCD}$. In this case, at a distance of order $1/\Lambda_{\rm QCD}$ one cannot resolve the inner structure of the heavy diquark system, which effectively behaves as a pointlike particle in an antitriplet color configuration, i.e., as a heavy antiquark of mass $2m_h$; under some conditions, effects due to the heavy quark-quark interaction my be added perturbatively in the framework of the nonrelativistic EFTs developed in the following sections [761, 1410–1419]. Finally, the heavy quark symmetry may be also applied to link doubly-heavy tetraquarks (tetraquarks made of two heavy and two light quarks) with heavy-light baryons sharing the same light-quark content [1415, 1420, 1421]. Many of the new charmonium- and bottomonium-like states observed at colliders in the last decades have a doubly-heavy tetraquark content [1388].

### 6.1.2 Nonrelativistic QCD

Hadrons made of two or more nonrelativistic particles, like two heavy quarks or a heavy quark and a heavy antiquark, or more generally just heavy quark-antiquark pairs near threshold, are multiscale systems characterized by a hierarchy of dynamically generated scales:

$$m_h \gg m_h v \gg m_h v^2. \tag{6.1.6}$$

We discussed the origin of these energy scales at the beginning of the section. The nonrelativistic energy scales are correlated. To reach a situation like in HQET, i.e. an EFT with just one dynamical low-energy scale, we need to construct at least two nonrelativistic EFTs: one following from integrating out from QCD modes associated with the energy scale $m_h$ and one following from integrating out modes associated with the energy scale $m_h v$, ending up with an ultimate EFT at the energy scale $m_h v^2$ [1422]. An illustration of the tower of energy scales and corresponding EFTs is in Fig. 6.1.1. In the last twenty years, the development of such nonrelativistic EFTs of QCD has been the major theoretical breakthrough in the description of quarkonium and quarkonium-like systems [1423–1425]. For a more historical perspective, see Ref. [1426].

NRQCD is the EFT suited to describe systems made of a heavy quark and (anti)quark pair near threshold that follows from QCD by integrating out the energy scale associated with the heavy quark mass, $m_h$ [1391]. In a heavy quark-antiquark bound state, the virial theorem constrains the kinetic energy of the heavy particles to be of the same order as the binding energy.

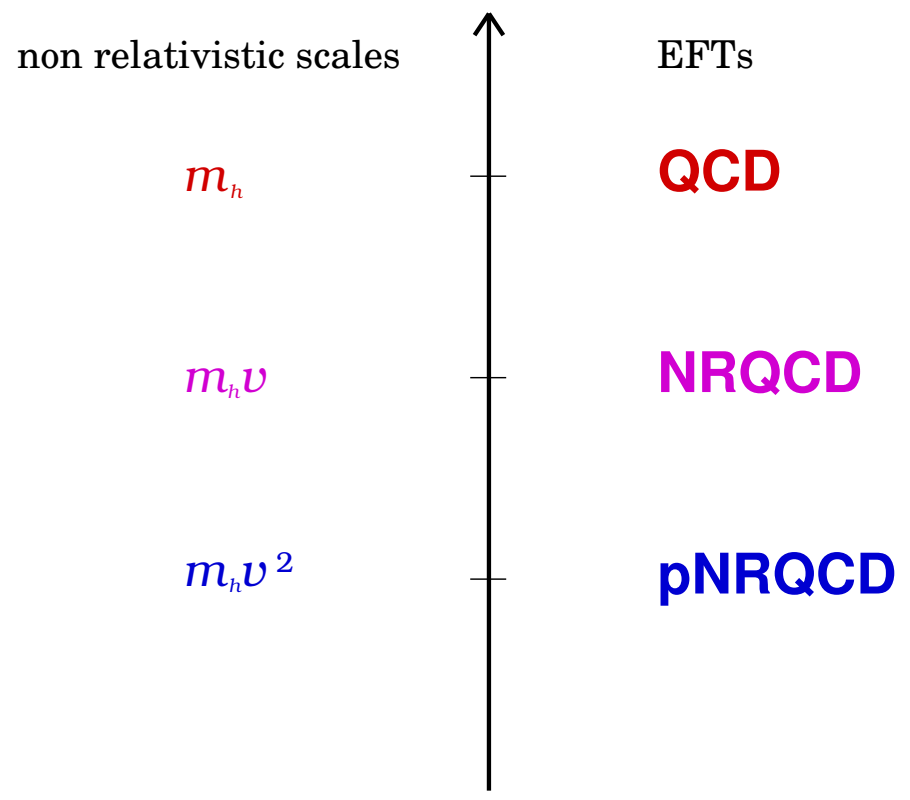

**Fig. 6.1.1** Hierarchy of energy scales and EFTs for systems made of a heavy quark and (anti)quark pair near threshold.

As a consequence, the power counting of NRQCD must be such that the leading-order NRQCD Lagrangian includes the kinetic energy operators, $\psi^\dagger\, \boldsymbol{\nabla}^2/(2m_h)\, \psi - \chi^\dagger \boldsymbol{\nabla}^2/(2m_h)\chi$, making the NRQCD Lagrangian, even at leading order, non renormalizable. This is different from HQET.

NRQCD posed initially also some difficulties in finding a computational scheme to integrate consistently over the different momentum and energy regions in dimensional regularization. NRQCD or its QED equivalent were therefore used for a long time either for analytical calculations in QED with a hard cut off [1427, 1428] or for lattice QCD calculations involving heavy quarks [268, 1429]. The advantage for lattice NRQCD calculations is that, once the heavy quark mass has been integrated out, the lattice spacing, $a$, is not constrained, as in full lattice QCD calculations, to be smaller than $1/m_h$, which would amount to requiring a very fine lattice if the quark is heavy. In lattice NRQCD the constraint is relaxed to $a < 1/(m_h v)$. Since at the same time the lattice size has to be large enough to include distances of the order of $1/\Lambda_{\rm QCD}$ for quenched calculations and $1/M_\pi$ for full calculations, simulations with heavy quarks in full QCD are computationally quite demanding. Lattice NRQCD has been, for a long time, the sole way to compute nonperturbatively observables involving bottom quarks in full QCD (see, for instance, Refs. [260, 1430–1433]). Only recently the first full lattice QCD calculations of bottomonium-like systems have become available [1434].

After the development of HQET, NRQCD became a systematic tool for analytical calculations of quarkonium observables. NRQCD is well suited to the description of heavy quark-antiquark annihilation, because this happens at the scale $m_h$, which is the energy scale that has been integrated out from QCD to construct NRQCD. In NRQCD, the information about annihi-lation goes into the (imaginary part) of the Wilson coefficients associated with the four-fermion operators. These four fermion operators, a novel feature of NRQCD with respect to HQET, are not only essential to the description of the annihilation processes, but also to the correct description of all short-distance interactions between the heavy particles. In NRQCD, annihilation processes factorize therefore into a short-distance part, which may be computed in perturbative QCD and is encoded in the Wilson coefficients, and into matrix elements of the NRQCD operators that encode the low-energy dynamics of the heavy quark-antiquark bound state. Processes involving heavy quark-antiquark annihilations are quarkonium inclusive and electromagnetic decay [1435, 1436] and quarkonium production [1436]. The large amount of data on quarkonium production in hadron and lepton colliders, together with the predictive power of NRQCD and its success in most of the predictions, has established NRQCD as a standard tool for studying quarkonium annihilation [1423–1425, 1437–1439].

Because four-fermion operators projecting onto color octet quark-antiquark states are possible, NRQCD naturally allows for production and decay of quark-antiquark states in a color octet configuration. These states constitute a suppressed, in $v$, component of the Fock state describing a physical quarkonium, but are necessary in the quarkonium phenomenology [1423–1425]. They are also necessary for the consistency of the theory, as they cancel infrared divergences in quarkonium decay and production observables and eventually provide finite, physical results [1435, 1436]. It should be noted, however, that the NRQCD factorization has been rigorously proved only for quarkonium decay but not for quarkonium production [1277, 1314, 1440–1442].

A last crucial progress in establishing NRQCD as a valuable tool for analytical calculations came when it was shown that the computation of the Wilson coefficients of NRQCD in dimensional regularization requires expanding in the heavy quark mass to avoid integrating over the high momentum region. This means that, even if the power countings of NRQCD and HQET are different, the matching to QCD proceeds in the same way, leading to the same operators and Wilson coefficients in the two-fermion and gauge sectors [1394].

The NRQCD Lagrangian density for systems made of a heavy quark and a heavy antiquark of equal masses $m_h$ up to order $1/m_h^2$, and including the $1/m_h^3$ kinetic operator, is given by

$$\mathcal{L}_{\text{NRQCD}} = \mathcal{L}^{\psi} + \mathcal{L}^{\chi} + \mathcal{L}^{\psi\chi} + \mathcal{L}^{\ell}, \tag{6.1.7}$$

where $\mathcal{L}^{\psi}$ and $\mathcal{L}^{\chi}$ are the HQET Lagrangian densities for the quark (see Eq. (6.1.2)) and antiquark, respectively, $\mathcal{L}^{\ell}$ is the Lagrangian density (6.1.3) for the light degrees of freedom, and $\mathcal{L}^{\psi\chi}$ is the four-fermion sector, which up to order $1/m_h^2$ reads

$$\begin{aligned}\mathcal{L}^{\psi\chi} = & \frac{f_1(^1S_0)}{m_h^2}\psi^\dagger\chi\chi^\dagger\psi + \frac{f_1(^3S_1)}{m_h^2}\psi^\dagger\boldsymbol{\sigma}\chi\cdot\chi^\dagger\boldsymbol{\sigma}\psi \\ & + \frac{f_8(^1S_0)}{m_h^2}\psi^\dagger \mathrm{T}^A\chi\chi^\dagger \mathrm{T}^A\psi + \frac{f_8(^3S_1)}{m_h^2}\psi^\dagger \mathrm{T}^A\boldsymbol{\sigma}\chi\cdot\chi^\dagger \mathrm{T}^A\boldsymbol{\sigma}\psi\,.\end{aligned} \tag{6.1.8}$$

As in the HQET case, $m_h$ is the pole mass. The four-fermion Lagrangian in Eq. (6.1.8) is made of all possible four-fermion operators of dimension 6. The corresponding Wilson coefficients are $f_1(^1S_0)$, $f_1(^3S_1)$, $f_8(^1S_0)$, and $f_8(^3S_1)$. The the first (second) four-fermion operator projects on a heavy quark-antiquark pair in a color singlet configuration with quantum numbers $^1S_0$ ($^3S_1$), whereas the third (fourth) four-fermion operator projects on a heavy quark-antiquark pair in a color octet configuration with quantum numbers $^1S_0$ ($^3S_1$). The matrices $\mathrm{T}^A$ are the SU(3) generators $\lambda^A/2$. The four-fermion Wilson coefficients have been computed in Refs. [1436, 1443]. They have a real part that starts at order $\alpha_s$ for $f_8(^3S_1)$ and at order $\alpha_s^2$ for the other coefficients, and they have an imaginary part, coming from one loop or higher annihilation diagrams, which is of order $\alpha_s^2$ for $\mathrm{Im}\, f_1(^1S_0)$, $\mathrm{Im}\, f_8(^1S_0)$, and $\mathrm{Im}\, f_8(^3S_1)$, and of order $\alpha_s^3$ for $\mathrm{Im}\, f_1(^3S_1)$. A list of imaginary parts of four-fermion Wilson coefficients in NRQCD and related bibliography can be found in Ref. [1444].

The four-fermion sector of the NRQCD Lagrangian has been derived up to order $1/m_h^4$ (complete) and orders $1/m_h^5$ and $1/m_h^6$ (partial) in Refs. [1445–1447]. Like for the Wilson coefficients in the two-fermion sector, also the coefficients in the four-fermion sector are not all independent: some are related by Poincaré invariance [1446, 1447].

Sometimes it is useful to isolate the electromagnetic component of the four-fermion operator and its corresponding Wilson coefficient. This is the case when computing electromagnetic decay widths and photoproduction cross sections in NRQCD. The electromagnetic operators are obtained by projecting on an intermediate QCD vacuum state, $|0\rangle$, e.g., $\psi^\dagger\chi\chi^\dagger\psi \to \psi^\dagger\chi\,|0\rangle\langle 0|\,\chi^\dagger\psi$.

Unlike in HQET, the power counting of NRQCD is not unique. The reason is that, while HQET is a one-scale EFT, its only dynamical scale being $\Lambda_{\text{QCD}}$, NRQCD is still a multiscale EFT. The dynamical scales of NRQCD are, at least, $m_h v$, $m_h v^2$, and $\Lambda_{\text{QCD}}$. In more complicated settings even more scales may be relevant. Hence, one can imagine different power countings: some more conservative, like assuming that the matrix elements all scale according to the largest dynamical scale, i.e., $m_h v$, [1448], and some less conservative or closer to a perturbative counting [1436]. All the power countings have in common that the kinetic energy scales like the binding energy and that therefore $\psi^\dagger i\partial_0\psi$ is of the same order as $\psi^\dagger\boldsymbol{\nabla}^2/(2m_h)\psi$, and analogously for the antiquark. As we have mentioned above, this reflects the virial theorem, an unavoidable consequence of the dynamics of a nonrelativistic bound state.

The leading-order NRQCD Lagrangian reads in Coulomb gauge [1436]

$$\mathcal{L}^{(0)}_{\rm NRQCD} = \psi^\dagger\left(iD_0+\frac{\boldsymbol{\nabla}^2}{2m_h}\right)\psi+\chi^\dagger\left(iD_0-\frac{\boldsymbol{\nabla}^2}{2m_h}\right)\chi - \frac{1}{4}F^A_{\mu\nu}F^{A\,\mu\nu}+\sum_{\ell=1}^{n_\ell}\bar{q}_\ell\left(i\not{D}-m_\ell\right)q_\ell. \quad (6.1.9)$$

Note that this Lagrangian contains the heavy quark mass, and therefore violates the heavy-quark flavor symmetry; hence the bottomonium binding energy is different, even at leading order, from the charmonium one. In the power counting of Ref. [1436] one further assumes: $D_0\sim m_hv^2$ (when acting on $\psi$ or $\chi$), $\boldsymbol{D}\sim m_hv$ (when acting on $\psi$ or $\chi$), $g\boldsymbol{E}\sim m_h^2v^3$, and $g\boldsymbol{B}\sim m_h^2v^4$. A consequence is that the heavy-quark spin symmetry is a symmetry of the leading-order NRQCD Lagrangian. Another consequence is that the order $1/m_h^3$ kinetic energy operator $\psi^\dagger\boldsymbol{D}^4/(8m_h^3)\psi$ and its charge conjugated are of the same order as the $1/m_h$ and $1/m_h^2$ operators in $\mathcal{L}^\psi$ and $\mathcal{L}^\chi$. Matrix elements of octet operators on quarkonium states are further suppressed by the fact that they project on subleading components of the quarkonium Fock state, the ones made of a heavy quark-antiquark pair in a color octet configuration and gluons. The amount of suppression depends on the adopted power counting.

### 6.1.3 potential Nonrelativistic QCD

Nonrelativistic bound states involve energy scales, $m_hv$, $m_hv^2$, and $\Lambda_{\rm QCD}$, that are still dynamical and entangled in NRQCD. A consequence of this is that, although the equations of motion that follow from the NRQCD Lagrangian (6.1.9) resemble a Schrödinger equation for nonrelativistic bound states, they are not quite that. They involve gauge fields and do not supply a field theoretical definition and derivation of the potential that would appear in a Schrödinger equation. Nevertheless, we expect that, in some nonrelativistic limit, a Schrödinger equation describing the quantum mechanics of the nonrelativistic bound state should emerge from field theory, since field theory may be understood as an extension of quantum mechanics that includes relativistic and radiative corrections. Another consequence already remarked in the previous section is that the power counting of NRQCD is not unique.

Since the scales $m_hv$ and $m_hv^2$ are hierarchically ordered, they may be disentangled by systematically integrating out modes associated with scales larger than the smallest scale, $m_hv^2$, and matching to a lower energy EFT, where only degrees of freedom resolved at distances of order $1/(m_hv^2)$ are left dynamical [1422]. This EFT is pNRQCD [1449, 1450]. Because the scale $m_hv$ has been integrated out, the power counting of pNRQCD is less ambiguous than the one of NRQCD. In situations where we can neglect the hadronic scale $\Lambda_{\rm QCD}$, the power counting of pNRQCD is indeed unique, as its only dynamical scale is $m_hv^2$.

Having integrated out the scale $m_hv$ associated with the inverse of the distance $r$ between the heavy quark and antiquark, implies that pNRQCD is constructed as an expansion in $r$, with Wilson coefficients encoding non-analytic contributions in $r$. This is analogous to how HQET and NRQCD are constructed; there the heavy quark mass, $m_h$, is integrated out and the EFTs are organized as expansions in $1/m_h$, with Wilson coefficients encoding the non-analytic contributions in the form of logarithms of $m_h$. Some of the Wilson coefficients of pNRQCD may be identified with the potentials in the Schrödinger equation of quarkonium.

The specific form of pNRQCD depends on the scale $\Lambda_{\rm QCD}$. If $\Lambda_{\rm QCD}\lesssim m_hv^2$, then one deals with weakly-coupled bound states and the EFT is called weakly-coupled pNRQCD. At distances of the order of or smaller than $1/(m_hv^2)$, one may still resolve colored degrees of freedom (gluons, quarks, and antiquarks), as color confinement has not yet set in. Hence gluons, quarks, and antiquarks are the degrees of freedom of weakly-coupled pNRQCD. Weakly-coupled pNRQCD is well suited to describe tightly bound quarkonia, like bottomonium and (to a less extent) charmonium ground states, the $B_c$ ground state, and threshold effects in $t\bar{t}$ production. If $\Lambda_{\rm QCD}\gtrsim m_hv^2$, then one deals with strongly-coupled bound states and the EFT is called strongly-coupled pNRQCD. At distances of the order of $1/(m_hv^2)$, confinement has set in and the only available degrees of freedom are color singlets. These are, in principle, all, ordinary and exotic, heavy, heavy-light and light hadrons that we might have in the spectrum. Strongly-coupled pNRQCD is suited to describe higher states in the bottomonium and charmonium spectra, as well as quarkonium exotica. If $m_hv\gg\Lambda_{\rm QCD}\gg m_hv^2$, the matching to pNRQCD may be done in two steps, first integrating out (perturbatively) $m_hv$ then (nonperturbatively) $\Lambda_{\rm QCD}$. Contributions coming from these two scales will be automatically factorized in pNRQCD observables.

*Weakly-coupled pNRQCD*

The degrees of freedom of weakly-coupled pNRQCD are heavy quarks and antiquarks of momentum $m_h v$ and energy $m_h v^2$, gluons of momentum and energy $m_h v^2$ (sometimes called ultrasoft gluons), and light quarks of momentum and energy $m_h v^2$; these remain after gluons (sometimes called soft gluons) and light quarks of energy or momentum $m_h v$ have been integrated out from NRQCD. Because a single heavy quark and antiquark cannot be resolved at the scale $m_h v^2$, it is useful to cast heavy quark and antiquark fields into bilocal fields that depend on time, $t$, the center of mass coordinate, $\boldsymbol{R}$, and the relative coordinate, $\boldsymbol{r}$. We call the color singlet component of the quark and antiquark field S, and its color octet component O, normalized to $\mathrm{S} = \mathbb{1}_{3\times 3} S/\sqrt{3}$ and $\mathrm{O} = \sqrt{2} O^A T^A$. The distance $r$ scales like $1/(m_h v)$, while the center of mass coordinate, $R$, and the time, $t$, scale like $1/(m_h v^2)$, because the quark-antiquark pair may only recoil against ultrasoft gluons. To ensure that gluons are ultrasoft in the pNRQCD Lagrangian, gauge fields are multipole expanded in $\boldsymbol{r}$. Hence gauge fields in the pNRQCD Lagrangian only depend on time and the center of mass coordinate. The pNRQCD Lagrangian is organized as a double expansion in $1/m_h$ and $r$. At order $r$ in the multipole expansion, the weakly-coupled pNRQCD Lagrangian density has the form [1449, 1450]

$$\mathcal{L}^{\text{weak}}_{\text{pNRQCD}} = \mathcal{L}^{S,O} + \mathcal{L}^{\ell}, \tag{6.1.10}$$

with

$$\begin{aligned}\mathcal{L}^{S,O} = \int d^3r\, \mathrm{Tr}\left\{\mathrm{S}^\dagger (i\partial_0 - h_s)\mathrm{S} + \mathrm{O}^\dagger (iD_0 O - h_o)\mathrm{O}\right\} \\ - V_A \mathrm{Tr}\left\{\mathrm{O}^\dagger \boldsymbol{r}\cdot g\boldsymbol{E}\,\mathrm{S} + \mathrm{S}^\dagger \boldsymbol{r}\cdot g\boldsymbol{E}\,\mathrm{O}\right\} \\ - \frac{V_B}{2}\mathrm{Tr}\left\{\mathrm{O}^\dagger \boldsymbol{r}\cdot g\boldsymbol{E}\,\mathrm{O} + \mathrm{O}^\dagger \mathrm{O}\boldsymbol{r}\cdot g\boldsymbol{E}\right\},\end{aligned} \tag{6.1.11}$$

where, up to order $1/m_h^2$, and including the $1/m_h^3$ terms in the kinetic energies,

$$h_s = \frac{\boldsymbol{p}^2}{m_h} + \frac{\boldsymbol{P}^2}{4m_h} - \frac{\boldsymbol{p}^4}{4m_h^3} + \cdots + V_s, \tag{6.1.12}$$

$$h_o = \frac{\boldsymbol{p}^2}{m_h} + \frac{\boldsymbol{P}^2}{4m_h} - \frac{\boldsymbol{p}^4}{4m_h^3} + \cdots + V_o. \tag{6.1.13}$$

The covariant derivative acting on the octet field is defined as $iD_0\mathrm{O} = i\partial_0\mathrm{O} + g[A_0(\boldsymbol{R},t),\mathrm{O}]$, $\boldsymbol{P} = -i\boldsymbol{D_R}$ is the (gauge covariant) center of mass momentum, $\boldsymbol{p} = -i\boldsymbol{\nabla_r}$ is the relative momentum, and $h_s$ and $h_o$ may be interpreted as the Hamiltonians of the color singlet and color octet heavy quark-antiquark fields. The dots in Eqs. (6.1.12) and (6.1.13) stand for higher-order terms in the nonrelativistic expansion of the kinetic energy. The trace in Eq. (6.1.11) is in spin and in color space. The Lagrangian $\mathcal{L}^{\ell}$ is the Lagrangian of the light degrees of freedom (6.1.3) inherited from NRQCD.

The quantities $V_s$, $V_o$, $V_A$, and $V_B$ are Wilson coefficients of pNRQCD. They encode contributions from the soft degrees of freedom that have been integrated out from NRQCD. Because (under the hierarchy of weakly-coupled pNRQCD) the soft scale, $m_h v$, is larger than $\Lambda_{\text{QCD}}$, the Wilson coefficients may be computed in perturbation theory, order by order in $\alpha_s$. They are, in general, functions of $\boldsymbol{r}$, as well as of the spin and momentum. At leading order, $V_A$ and $V_B$ are 1; they get possible corrections at order $\alpha_s^2$ [1451]. The coefficients $V_s$ and $V_o$ may be identified with the color singlet and octet potentials, respectively. To leading order $V_s^{(0)} = -4\alpha_s/(3r)$ and $V_o^{(0)} = \alpha_s/(6r)$, which are the Coulomb potentials in the color SU(3) fundamental and adjoint representation, respectively. The potentials $V_s$ and $V_o$ contain, however, also momentum- and spin-dependent corrections. For the singlet case (the octet case is analogous) we can write, up to order $1/m_h^2$:

$$V_s = V_s^{(0)}(r) + \frac{V_s^{(1)}(r)}{m_h} + \frac{V_{\text{SI}}^{(2)}}{m_h^2} + \frac{V_{\text{SD}}^{(2)}}{m_h^2}, \tag{6.1.14}$$

where, at order $1/m_h^2$ we have distinguished between spin-independent (SI) and spin-dependent (SD) terms. In turn, they can be organized as

$$\begin{aligned}V_{\text{SI}}^{(2)} =& V_r^{(2)}(r) + \frac{1}{4}V_{p^2,\text{CM}}^{(2)}(r)\boldsymbol{P}^2 + \frac{1}{4}\frac{V_{L^2,\text{CM}}^{(2)}(r)}{r^2}(\boldsymbol{r}\times\boldsymbol{P})^2 \\ &+ \frac{1}{2}\left\{V_{p^2}^{(2)}(r), \boldsymbol{p}^2\right\} + \frac{V_{L^2}^{(2)}(r)}{r^2}\boldsymbol{L}^2,\end{aligned} \tag{6.1.15}$$

$$\begin{aligned}V_{\text{SD}}^{(2)} =& \frac{1}{2}V_{LS,\text{CM}}^{(2)}(r)\,(\boldsymbol{r}\times\boldsymbol{P})\cdot(\boldsymbol{S}_1 - \boldsymbol{S}_2) + V_{LS}^{(2)}(r)\,\boldsymbol{L}\cdot\boldsymbol{S} \\ &+ V_{S^2}^{(2)}(r)\,\boldsymbol{S}^2 + V_{S_{12}}^{(2)}(r)\,S_{12},\end{aligned} \tag{6.1.16}$$

where { , } stands for the anticommutator,

$$\boldsymbol{S} = \boldsymbol{S}_1 + \boldsymbol{S}_2 = (\boldsymbol{\sigma}_1 + \boldsymbol{\sigma}_2)/2$$

is the total spin ($\boldsymbol{S}_i = \boldsymbol{\sigma}_i/2$ is the spin of the particle $i$), $\boldsymbol{L} = \boldsymbol{r}\times\boldsymbol{p}$ is the relative orbital angular momentum, and

$$S_{12} = 3(\hat{\boldsymbol{r}}\cdot\boldsymbol{\sigma}_1)(\hat{\boldsymbol{r}}\cdot\boldsymbol{\sigma}_2) - \boldsymbol{\sigma}_1\cdot\boldsymbol{\sigma}_2\,.$$

The potential $V_s^{(0)}$ is the static potential, the potential proportional to $V_{LS}^{(2)}$ may be identified with the spin-orbit potential, the potential proportional to $V_{S^2}^{(2)}$ with the spin-spin potential and the potential proportional to $V_{S_{12}}^{(2)}$ with the spin tensor potential. The potentials that contribute in the center of mass reference frame are, at leading (non-vanishing) order in perturbation theory (see, e.g., Ref. [1422]):

$$V^{(1)}(r) = -\frac{2\alpha_s^2}{r^2}\,, \tag{6.1.17}$$

$$V_r^{(2)}(r) = \frac{4}{3}\pi\alpha_s\delta^{(3)}(\boldsymbol{r})\,, \quad V_{p^2}^{(2)}(r) = -\frac{4\alpha_s}{3r}\,, \tag{6.1.18}$$

$$V_{L^2}^{(2)}(r) = \frac{2\alpha_s}{3r}\,, \qquad V_{LS}^{(2)}(r) = \frac{2\alpha_s}{r^3}\,, \tag{6.1.19}$$

$$V_{S^2}^{(2)}(r) = \frac{16\pi\alpha_s}{9}\delta^{(3)}(\boldsymbol{r})\,, \quad V_{S_{12}}^{(2)}(r) = \frac{\alpha_s}{3r^3}\,. \tag{6.1.20}$$

Potentials that depend on the center of mass momentum are relevant only if the quarkonium is recoiling, like in hadronic and electromagnetic transitions.

Beyond leading order, the static potential is known up to three-loop accuracy [1452–1454], and also subleading logarithms showing up at four loops have been computed [1455]; the $1/m_h$ potential is known up to order $\alpha_s^3$ [1456], and $1/m_h^2$ potentials up to order $\alpha_s^2$ (these potentials have a long history, see Ref. [1457] and references therein). We have assumed that the heavy quark and antiquark have equal masses; for the case of a quark and an antiquark of different masses, we refer, for instance, to Refs. [1399, 1422, 1458–1460].

The Wilson coefficients of pNRQCD inherit the Wilson coefficients of NRQCD. Hence, some of the couplings appearing in the expansion of the Wilson coefficients are naturally computed at the scale of NRQCD, $m_h$, while others, encoding the soft contributions, are naturally computed at the soft scale, $m_h v$. In weakly-coupled pNRQCD, because the leading potential is the Coulomb potential, the Bohr radius is proportional to $1/(m_h\alpha_s)$ and $v \sim \alpha_s$. Finally, like in any non relativistic EFT, also the Wilson coefficients of pNRQCD are related through constraints imposed by the Poincaré invariance of QCD, as we have seen at work in HQET and NRQCD. These constraints set the coefficients of the kinetic terms appearing in Eqs. (6.1.12) and (6.1.13) to be the ones coming from expanding the relativistic kinetic energies of a free quark and antiquark. Furthermore they fix the potentials depending on the center of mass momentum by expressing them in terms of the static potential,

$$V_{LS,\mathrm{CM}} = -\frac{1}{2r}\frac{dV_s^{(0)}}{dr}, \qquad V_{L^2,\mathrm{CM}} = -\frac{r}{2}\frac{dV_s^{(0)}}{dr},$$
$$V_{p^2,\mathrm{CM}} = \frac{r}{2}\frac{dV_s^{(0)}}{dr} - \frac{1}{2}V_s^{(0)}.$$

These and other constraints have been derived in Refs. [1399, 1447, 1459, 1461, 1462]. These relations are exact, i.e., valid at any order in perturbation theory and, when applicable, also nonperturbatively.

In the pNRQCD Lagrangian the relative coordinate $\boldsymbol{r}$ plays the role of a continuous parameter labeling different fields. The dynamical spacetime coordinates of the Lagrangian density are the time $t$ and the coordinate $\boldsymbol{R}$, which, in the case of the fields S and O, is the center of mass coordinate. Having written the Lagrangian in terms of singlet and octet fields has made each term in Eq. (6.1.11) explicitly gauge invariant.

The power counting of weakly-coupled pNRQCD is straightforward. We have already found that $r \sim 1/(m_h v)$ and $t$, $R \sim 1/(m_h v^2)$. Momenta scale like $p \sim m_h v$ and $P \sim m_h v^2$. Gluon fields and light quark fields are ultrasoft and scale like $m_h v^2$ or $\Lambda_{\mathrm{QCD}}$ to their dimension. The leading-order singlet Hamiltonian, $\boldsymbol{p}^2/m_h + V_s^{(0)}$, scales like $m_h v^2$ (and analogously in the octet case), which is the order of the Bohr levels. The potentials listed in Eqs. (6.1.17)-(6.1.20) contribute to $V_s$ at order $m_h v^4$, as $\alpha_s \sim v$.

The first correction to a pure potential picture of the quarkonium interaction comes from the chromoelectric dipole interaction terms in the second line of Eq. (6.1.11). These operators are of order $g(m_h v^2)^2/(m_h v) \sim g m_h v^3$. In order to project on color singlet states, the chromoelectric dipole interaction may enter only in loop diagrams, which at leading order is a self-energy diagram with two chromoelectric dipole vertices. Such a self-energy diagram is of order $g^2(m_h v^2)^3/(m_h v)^2 \sim g^2 m_h v^4$. The coupling $g^2$ is computed at the ultrasoft scale. Hence, if $\Lambda_{\mathrm{QCD}} \ll m_h v^2$, the coupling is perturbative and the self-energy diagram with two chromoelectric dipole vertices is suppressed with respect to the contributions coming from the potentials in Eqs. (6.1.17)-(6.1.20). Elsewhere, if $\Lambda_{\mathrm{QCD}} \sim m_h v^2$, it is of the same order.

At leading order in the multipole expansion, the equation of motion for the singlet field is

$$i\partial_0 \mathrm{S} = h_s \mathrm{S}, \tag{6.1.21}$$

which is the Schrödinger equation that in quantum mechanics describes the evolution of a nonrelativistic bound state. Potential NRQCD provides therefore a field theoretical foundation of the Schrödinger equation, a rigorous QCD definition and derivation of its potential, and the range of validity of the quantum mechanical picture. Ultrasoft gluons start contributing, and therefore correcting the potential picture, at order $m_h v^4$ (for $\Lambda_{\mathrm{QCD}} \sim m_h v^2$) or $m_h v^5$ (for $\Lambda_{\mathrm{QCD}} \ll m_h v^2$) in the spectrum.

Not only the static potential is derived from first principles in pNRQCD, but all higher-order corrections in the nonrelativistic expansion, including the spin-orbit, spin-spin and Darwin term as well. In general, the potentials factorize soft contributions from radiative corrections at the scale $m_h$. These last ones are encoded in the matching coefficients inherited from NRQCD, e.g., all $1/m_h^2$ spin-dependent potentials contain the Fermi coefficient $c_F$. Since the potentials are Wilson coefficients of an EFT, they are regularized, undergo renormalization and satisfy renormalization group equations that allow to resum potentially large logarithms in their expressions [1451, 1452, 1463–1469]. The scale dependence of the Wilson coefficients cancels in physical observables. For instance, the QCD static potential is infrared sensitive at three loops [1470], a sensitivity that stems from the fact that a static quark-antiquark pair may change its color status by emitting an ultrasoft gluon. The infrared sensitivity of the static potential cancels in the computation of the static energy against the self-energy diagram with two chromoelectric dipole vertices considered above, in a sort of non-Abelian Lamb shift mechanism [1452].

Weakly-coupled pNRQCD requires the fulfillment of the condition $\Lambda_{\rm QCD} \lesssim m_h v^2$. The condition $\Lambda_{\rm QCD} \ll m_t v^2$ is certainly fulfilled by top-antitop quark pairs near threshold. The production of $t\bar{t}$ pairs near threshold is expected to be measured with precision at future linear colliders, providing, among others, a determination of the top mass with an uncertainty well below 100 MeV, which is a crucial input to test the Standard Model. This requires a comparable theoretical accuracy, which has led in the last decades to several high-order calculations of the near threshold cross section in the framework of nonrelativistic EFTs of QCD [1471–1477]. The condition $\Lambda_{\rm QCD} \lesssim m_h v^2$ is also fulfilled by compact and Coulombic quarkonia. Examples are the bottomonium ground state, the ground state of the $B_c$ system, and, to a somewhat lesser extent, the charmonium ground state, and the first bottomonium excited states. We recall that in a Coulombic system the size is proportional to the inverse of the mass and to the principal quantum number. A review on applications of weakly-coupled pNRQCD to several tightly bound quarkonia can be found in Ref. [1478].

Weak-coupling determinations of the bottomonium ground state masses are typically used to extract the charm and bottom masses [1407, 1479–1487]. Hence, they provide alternative observables for the extraction of the heavy quark masses to the heavy-light meson masses discussed in Sec. 6.1.1. The results are consistent with the ones presented in Sec. 6.1.1. The present precision is N$^3$LO; the determination of the bottom mass includes the effects due to the charm mass at two loops. Once the heavy quark masses have been established for one set of spectroscopy observables, they can be used for others like the $B_c$ mass or the $B_c$ spectrum (see Ref. [1458] for an early work and Ref. [1460] for a state of the art calculation at N$^3$LO). Fine and hyperfine splittings of charmonium and bottomonium have been computed perturbatively in Refs. [1488, 1489] and to NLL accuracy in Ref. [1490], similarly for the $B_c^*$-$B_c$ hyperfine splitting in Ref. [1491]. After more than one decade of work the whole perturbative heavy quarkonium spectrum has been computed at N$^3$LO [1464, 1492–1497]. Recently, this result has been further improved reaching N$^3$LL accuracy up to a missing contribution of the two-loop soft running [1468, 1469]. The N$^3$LL order represents the presently achievable precision of these calculations. Going beyond this precision will require a major computational effort, like the four-loop determination of the static potential. Electromagnetic decays of the bottomonium lowest levels have been computed including N$^2$LL corrections in Refs. [1473, 1498]. A different power counting that includes at leading order the exact static potential has been used for these quantities in Ref. [1499]. Corrections to the wave function and leptonic decay width of the $\Upsilon(1S)$ at N$^3$LO have been computed in Refs. [1500, 1501]. Nonperturbative corrections in the form of condensates have been included in Refs. [1502, 1503]. Radiative quarkonium decays have been analyzed in Refs. [1504–1509]. Radiative and inclusive decays of the $\Upsilon(1S)$ may also serve as a determination of $\alpha_s$ at the bottom mass scale [1510]. Radiative transitions, M1 and E1, at relative order $v^2$ in the velocity expansion have been computed in Refs. [1511–1514]; noteworthily, pNRQCD may explain the observed tiny $\Upsilon(2S) \to \gamma\, \eta_b(1S)$ branching fraction. Finally, the photon line shape in the radiative transition $J/\psi \to \gamma\, \eta_c(1S)$ has been studied in Ref. [1515].

*Strongly-coupled pNRQCD*

When the hierarchy of scales is $\Lambda_{\rm QCD} \gg m_h v^2$, pNRQCD is a strongly-coupled theory. This condition may be appropriate to describe higher quarkonium states, and quarkonium exotica. Strongly-coupled pNRQCD is obtained by integrating out the hadronic scale $\Lambda_{\rm QCD}$, which means that all colored degrees of freedom are absent [1422, 1448, 1516–1519].

Let us consider the case of strongly-coupled pNRQCD for ordinary quarkonia. Lattice QCD shows evidence that the quarkonium static energy is separated by a gap of order $\Lambda_{\rm QCD}$ from the energies of the gluonic excitations between the static quark-antiquark pair [1520]. If, in addition, the binding energies of the states that can be constructed out of the quarkonium static energy are also separated by a gap of order $\Lambda_{\rm QCD}$ from the binding energies of the states that can be constructed out of the static energies of the gluonic excitations, and from open-flavor states, then one can integrate out all these latter states. The resulting EFT is made of a quark-antiquark color singlet field, whose modes are the quarkonium states, and light hadrons. The coupling of quarkonia with light hadrons has been considered in the framework of pNRQCD in Ref. [1521]. It impacts very mildly spectral properties (masses, widths) of quarkonia that lie well below the open-flavor threshold. For such quarkonia we may neglect their couplings with light hadrons and the pNRQCD Lagrangian density assumes the particularly simple form:

$$\mathcal{L}_{\text{pNRQCD}}^{\text{strong}} = \int d^3r \, \text{Tr}\left\{ \text{S}^\dagger \left( i\partial_0 - h_s \right) \text{S} \right\} . \qquad (6.1.22)$$

The Hamiltonian, $h_s$, has the same form as in Eqs. (6.1.12) and (6.1.14)-(6.1.16). The equation of motion is the Schrödinger equation (6.1.21).

The nonperturbative dynamics is encoded in the potentials, which at order $1/m_h$ is $V_s^{(1)}$ and at order $1/m_h^2$ are the spin-independent and spin-dependent terms identified in Eqs. (6.1.15) and (6.1.16). What distinguishes the EFT from phenomenological potential models can be summarized as follows:

(i) The potentials are products of Wilson coefficients, factorizing contributions from the high-energy scale, $m_h$, and low-energy matrix elements, encoding contributions coming from the scales $m_h v$ and $\Lambda_{\text{QCD}}$. The exact expressions follow from matching pNRQCD with its high-energy completion, which is NRQCD.

(ii) The high-energy Wilson coefficients of pNRQCD are inherited from NRQCD. These are the Wilson coefficients in the NRQCD Lagrangian (6.1.7). Because the NRQCD Wilson coefficients have a real and an imaginary part, also the pNRQCD potentials develop a real part, responsible for the quarkonium binding, and an imaginary part, responsible for the quarkonium annihilation. At higher orders, also contributions coming from the scale $\sqrt{m_h \Lambda_{\text{QCD}}}$ may become relevant [1519].

(iii) The low-energy matrix elements are nonperturbative. Their field-theoretical expressions, relevant for potentials up to order $1/m_h^2$, are known.

The static potential is equal to $\lim_{T\to\infty} i \ln W/T$, where $W$ is the expectation value of a rectangular Wilson loop of spatial extension $r$ and temporal extension $T$ [80, 1522–1524]. Similarly, the low-energy real parts of the other potentials can be expressed in terms of Wilson loops and field insertions on them [768, 1516, 1517]. These Wilson loops may be computed in weakly-coupled QCD giving back the weak-coupling potentials listed at leading order in Eqs. (6.1.17)-(6.1.20) [1525] or nonperturbatively via lattice QCD. Indeed, the computation of these potentials has a long history that begins with the inception of lattice QCD. Their most recent determinations can be found in Refs. [769, 1526, 1527], see also Ref. [1528]. Noteworthily the long-distance behaviour of the potentials agrees with the expectations of the effective string theory [1525, 1529–1531].

The low-energy contributions to the imaginary parts of the potential are matrix elements of the NRQCD four-fermion operators. Hence they are local terms proportional to $\delta^3(\boldsymbol{r})$ or derivatives of it. Nonperturbative contributions are encoded into constants that may be expressed in terms of momenta of correlators of chromoelectric and/or chromomagnetic fields [1448, 1518], and eventually fixed on data or computed with lattice QCD.

(iv) Finally, pNRQCD is renormalizable order by order in the expansion parameters in both its weak-coupling and strong-coupling versions. In particular, quantum-mechanical perturbation theory can be implemented at any order without incurring uncanceled divergences like in purely phenomenological potential models.

Strongly-coupled pNRQCD has been used to compute quarkonium inclusive and electromagnetic decay widths [1423, 1448, 1518, 1519], and hadronic and electromagnetic production cross sections [1532–1536]. The advantage with respect to the NRQCD approach is that, while the NRQCD four-fermion matrix elements depend on the quarkonium state, their pNRQCD expression factorizes all the quarkonium dependence into the wave function at the origin (or its derivatives) squared. The wave function at the origin squared gets multiplied by momenta of correlators of field-strength tensors, $F$, that are universal, quarkonium independent, constants. Schematically, one obtains for the expression of a generic NRQCD four-fermion matrix element in pNRQCD, entering either quarkonium production or decay, that

$$\langle \text{4-fermion} \rangle \sim |\text{wave-function}(0)|^2 \times \int dt \cdots \langle F(t) \cdots F(0) \rangle . \qquad (6.1.23)$$

This leads to a significant reduction in the number of nonperturbative parameters and allows to use information gained in the charmonium sector to make predictions in the bottomonium sector. Finally, pNRQCD conbined with the multipole expansion has been used to compute quarkonium hadronic transitions in Ref. [1537].

*pNRQCD for systems other than quarkonia*

Potential NRQCD can be used to describe systems with three valence quarks, two of them heavy [1411, 1412, 1418, 1419, 1538, 1539]. This is because the nonrelativistic hierarchy of scales given in Eq. (6.1.6) is preserved, which allows to systematically integrate out these scales to describe eventually the baryon with a suitable EFT. If the heavy quark-quark distance is of the order of $1/\Lambda_{\text{QCD}}$, then the valence light-quark affects the quark-quark potential. Elsewhere, if the heavy quark-quark distance is smaller than $1/\Lambda_{\rm QCD}$, then we may disentangle the quark-quark dynamics, described by a perturbative quark-quark potential, from the coupling of the heavy-quark pair with the light quark. Since in this last case, the light quark sees the heavy-quark pair as a pointlike particle, its interaction with the heavy-quark pair is described by HQET.

One can devise EFTs for describing low-energy modes of baryons made of three heavy quarks. These states have not been discovered yet in experiments, but they offer a unique tool to study confinement and the transition region from the Coulombic regime to the confined one in non-trivial geometrical settings [1538], using, for instance the three-quark static potential computed in lattice QCD [1540–1542].

Possible bound states made of two quarkonia or of a quarkonium and a nucleon (hadroquarkonium) may be characterized by even lower energy scales than those characterizing the binding in quarkonia or baryons made of at least two heavy quarks. These lower energy scales are those associated with pion exchanges responsible for the long-range interaction. One can treat these systems in an EFT framework by starting from the pNRQCD description of the quarkonium and the heavy-baryon chiral effective theory description of the nucleon. The long-range pion exchange interaction sets the scale of the typical size of the system to be of the order of $1/M_\pi$, i.e., much larger than the size of the quarkonium and even larger than its typical time scale, which is of the order of the inverse of the binding energy.

Once modes associated with the quarkonium binding energy and $M_\pi$ have been integrated out, the quarkonium-quarkonium or the quarkonium-nucleon interaction is described by a potential that, in this way, has been systematically computed from QCD. The coupling of quarkonium with the pions is encoded in a Wilson coefficient that may be identified with the quarkonium chromoelectric polarizability [1543]. In the quarkonium-quarkonium system, the lowest energy EFT describing modes of energy and momentum of order $M_\pi^2/(2m_h)$ is called van der Waals EFT (WEFT) [1521, 1544]. The resulting potential is, in fact, the van der Waals potential. In the quarkonium-nucleon system, the lowest energy EFT describing modes of energy and momentum of order $M_\pi^2/(2\Lambda_\chi)$ has been dubbed potential quarkonium-nucleon EFT (pQNEFT) [1545]. Such frameworks may be relevant to describe heavy pentaquarks.

Quarkonium-like multiparticle systems, where the light degrees of freedom remain adiabatically in a stationary state with respect to the heavy quark motion, can be studied within the Born–Oppenheimer approximation that may be implemented in a suitable version of pNRQCD called Born–Oppenheimer effective field theory (BOEFT) [772, 1546, 1547]. This framework has been applied to quarkonium hybrids, quarkonium tetraquarks and to heavy-meson threshold effects [1421, 1548–1551]. Finally, nonrelativistic EFTs like pNRQCD are also advantageous in describing the propagation of quarkonium in a thermal medium either in equilibrium [1552–1557] or out of equilibrium [1558–1562]; see also Sec. 6.6.

## 6.2 Chiral perturbation theory

**Stefan Scherer and Matthias Schindler**

Chiral perturbation theory (ChPT) is an effective field theory that describes the properties of strongly-interacting systems at energies far below typical hadron masses. The degrees of freedom are hadrons instead of the underlying quarks and gluons. ChPT is a systematic and model-independent approximation method based on an expansion of amplitudes in terms of light-quark masses and momenta. The following is a brief overview of ChPT that is largely based on Ref. [1563], which can be referred to for a more detailed introduction.

### 6.2.1 QCD and chiral symmetry

The QCD Lagrangian—obtained by applying the gauge principle with respect to the SU(3) color group to the free Lagrangians of six quark flavors with masses $m_f$—reads

$$\mathcal{L}_{\rm QCD} = \sum_{f=u,\ldots,t} \bar{q}_f \left(i\not{D} - m_f\right) q_f - \frac{1}{2}{\rm Tr}_c\left(\mathbf{F}_{\mu\nu}\mathbf{F}^{\mu\nu}\right). \tag{6.2.1}$$

For each quark flavor $f$, the quark field $q_f$ is a color triplet, transforming in the triplet representation,

$$q_f(x) \mapsto U(x) q_f(x), \tag{6.2.2}$$

where $U(x)$ denotes a smooth space-time-dependent SU(3) matrix. Using the Gell-Mann matrices [1564], the eight gluon fields $\mathcal{A}_\mu^A$ are collected in a traceless, Hermitian, $3\times 3$ matrix $\mathbf{A}_\mu = \lambda^A \mathcal{A}_\mu^A/2$ (summation over repeated indices implied), transforming inhomogeneously under a gauge transformation,

$$\mathbf{A}_\mu(x) \mapsto U(x)\mathbf{A}_\mu(x)U^\dagger(x) + \frac{i}{g_s}\partial_\mu U(x) U^\dagger(x), \tag{6.2.3}$$

where $g_s$ denotes the SU(3) gauge coupling constant. In terms of $\mathbf{A}_\mu$, the covariant derivative of the quark fields is defined as

$$\mathbf{D}_\mu q_f = \left(\partial_\mu + i g_s \mathbf{A}_\mu\right) q_f. \tag{6.2.4}$$

Finally, the field strength tensor is given by

$$\mathbf{F}_{\mu\nu} = \partial_\mu \mathbf{A}_\nu - \partial_\nu \mathbf{A}_\mu + i g_s [\mathbf{A}_\mu, \mathbf{A}_\nu]. \tag{6.2.5}$$

By construction, the Lagrangian of Eq. (6.2.1) is invariant under the combined transformations of Eqs. (6.2.2) and (6.2.3). From the point of view of gauge invariance, the strong-interaction Lagrangian could also involve a term of the type (c.f. Eq. (1.1.6) from Sec. 1.1)

$$\mathcal{L}_\theta = \frac{g_s^2 \bar{\theta}}{32\pi^2} \epsilon_{\mu\nu\rho\sigma} \mathrm{Tr}_c \left( \mathbf{F}^{\mu\nu} \mathbf{F}^{\rho\sigma} \right), \quad \epsilon_{0123} = 1, \tag{6.2.6}$$

where $\epsilon_{\mu\nu\rho\sigma}$ denotes the totally antisymmetric Levi-Civita tensor. The so-called $\theta$ term of Eq. (6.2.6) implies an explicit $P$ and $CP$ violation of the strong interactions. The present empirical information on the neutron electric dipole moment [1565] indicates that the $\theta$ term is small and, in the following, we will omit Eq. (6.2.6) from our discussion.

Since the covariant derivative of the quark fields is flavor independent, the Lagrangian of Eq. (6.2.1) has additional, accidental, and in this case global, symmetries aside from the gauge symmetry. Both the dynamics of the theory (via spontaneous symmetry breaking) and the values of the quark masses impact how these symmetries are (approximately) realized in nature. Dynamical chiral symmetry breaking introduces the scale $\Lambda_\chi = 4\pi F_0$ (see below) of the order of 1 GeV [1566]. In this context it is common to divide the six quark flavors into the three light quarks $u$, $d$, and $s$ with $m_l < \Lambda_\chi$ and the three heavy flavors $c$, $b$, and $t$ with $m_h > \Lambda_\chi$ (as discussed in the previous subsection). As a theoretical starting point, one may consider two limits, namely, sending the light-quark masses to zero (chiral limit) and the heavy-quark masses to infinity. In Ref. [1567], this situation is referred to as a "theoretician's paradise." In the following, we exclusively concentrate on the chiral limit for either two $(u, d)$ or three $(u, d, s)$ light quarks and omit the heavy quarks from our discussion. Introducing left-handed and right-handed quark fields (color and flavor indices omitted) as

$$q_L = \frac{1}{2}(\mathbb{1} - \gamma_5) q, \;\; q_R = \frac{1}{2}(\mathbb{1} + \gamma_5) q, \;\; \gamma_5 = i\gamma^0\gamma^1\gamma^2\gamma^3, \tag{6.2.7}$$

the QCD Lagrangian in the chiral limit decomposes into

$$\mathcal{L}^0_{\mathrm{QCD}} = \sum_{l=u,d,s} \left( \bar{q}_{L,l} i \not{D} q_{L,l} + \bar{q}_{R,l} i \not{D} q_{L,R} \right) - \frac{1}{2} \mathrm{Tr}_c \left( \mathbf{F}_{\mu\nu} \mathbf{F}^{\mu\nu} \right). \tag{6.2.8}$$

In the massless limit, the helicity of a quark is a good quantum number which is conserved in the interaction with gluons. Moreover, the classical Lagrangian in the chiral limit has a global $\mathrm{U}(3)_L \times \mathrm{U}(3)_R$ symmetry, *i.e.* , it is invariant under independent unitary flavor transformations of the left-handed and the right-handed quark fields,

$$q_L \mapsto U_L q_L, \quad q_R \mapsto U_R q_R.$$

At the classical level, this chiral symmetry results in $2 \times (8 + 1) = 18$ conserved currents:

$$L^\mu_a = \bar{q}_L \gamma^\mu \frac{\lambda_a}{2} q_L, \quad R^\mu_a = \bar{q}_R \gamma^\mu \frac{\lambda_a}{2} q_R, \quad a = 1, \ldots, 8,$$
$$V^\mu = \bar{q}_R \gamma^\mu q_R + \bar{q}_L \gamma^\mu q_L, \quad A^\mu = \bar{q}_R \gamma^\mu q_R - \bar{q}_L \gamma^\mu q_L.$$

Here, the Gell-Mann matrices act in flavor space, since $q_R$ and $q_L$ are flavor triplets.[62] Because of quantum effects the singlet axial-vector current $A^\mu = \bar{q}\gamma^\mu\gamma_5 q$ develops a so-called anomaly, resulting in the divergence equation

$$\partial_\mu A^\mu = \frac{3 g_s^2}{16\pi^2} \epsilon_{\mu\nu\rho\sigma} \mathrm{Tr}_c \left( \mathbf{F}^{\mu\nu} \mathbf{F}^{\rho\sigma} \right). \tag{6.2.9}$$

The factor of three originates from the number of flavors. In the large $N_c$ (number of colors) limit of Ref. [1163] the singlet axial-vector current is conserved, because the strong coupling constant behaves as $g_s^2 \sim N_c^{-1}$.

In the quantized theory, the spatial integrals over the charge densities of the symmetry currents give rise to the charge operators $Q_{La}$, $Q_{Ra}$ $(a = 1, \ldots, 8)$, and $Q_V$. They are generators of the group $\mathrm{SU}(3)_L \times \mathrm{SU}(3)_R \times \mathrm{U}(1)_V$, acting on the Hilbert space of QCD, and satisfy the commutation relations

$$[Q_{La}, Q_{Lb}] = i f_{abc} Q_{Lc}, \tag{6.2.10a}$$
$$[Q_{Ra}, Q_{Rb}] = i f_{abc} Q_{Rc}, \tag{6.2.10b}$$
$$[Q_{La}, Q_{Rb}] = 0, \tag{6.2.10c}$$
$$[Q_{La}, Q_V] = [Q_{Ra}, Q_V] = 0, \tag{6.2.10d}$$

where the $f_{abc}$ are the totally antisymmetric structure constants of the Lie algebra of SU(3) [1564]. In the chiral limit, these operators are time independent, *i.e.* , they commute with the Hamiltonian in the chiral limit,

$$[Q_{La}, H^0_{\mathrm{QCD}}] = [Q_{Ra}, H^0_{\mathrm{QCD}}] = [Q_V, H^0_{\mathrm{QCD}}] = 0. \tag{6.2.11}$$

It is convenient to consider the linear combinations $Q_{Aa} \equiv Q_{Ra} - Q_{La}$ and $Q_{Va} \equiv Q_{Ra} + Q_{La}$, which transform as $Q_{Aa} \mapsto -Q_{Aa}$ and $Q_{Va} \mapsto Q_{Va}$ under parity. The hadron spectrum can be organized in multiplets belonging to irreducible representations of $\mathrm{SU}(3)_V$ with a given baryon number. If not only the vector subgroup, but the full group were realized linearly by the spectrum of the hadrons, one would expect a so-called parity doubling of mass-degenerate states. The absence of such a doubling in the low-energy spectrum is an indication that the $\mathrm{SU}(3)_L \times \mathrm{SU}(3)_R$ chiral symmetry is dynamically broken in the ground state. One then assumes that the axial generators $Q_{Aa}$ do not annihilate the ground state of QCD,

$$Q_{Aa}|0\rangle \neq 0. \tag{6.2.12}$$

[62] Lower case Roman letters denote SU(3) flavor indices.

As a consequence of the Goldstone theorem [12], each axial generator $Q_{Aa}$ not annihilating the ground state corresponds to a *massless* Goldstone-boson field $\phi_a$ with spin 0, whose symmetry properties are tightly connected to the generator in question. The Goldstone bosons have the same transformation behavior under parity as the axial generators,

$$\phi_a(t,\vec{x}) \stackrel{P}{\mapsto} -\phi_a(t,-\vec{x}), \tag{6.2.13}$$

*i.e.* , they are pseudoscalars. From Eqs. (6.2.10a) and (6.2.10b) one obtains $[Q_{Va}, Q_{Ab}] = i f_{abc} Q_{Ac}$ and thus the Goldstone bosons transform under the subgroup $\mathrm{SU}(3)_V$, which leaves the vacuum invariant, as an octet:

$$[Q_{Va}, \phi_b(x)] = i f_{abc}\phi_c(x). \tag{6.2.14}$$

The members of the pseudoscalar octet $(\pi, K, \eta)$ of the real world are identified as the Goldstone bosons of QCD and would be massless for massless quarks.

After turning on the quark masses in terms of the mass term

$$\mathcal{L}_{\mathcal{M}} = -\bar{q}\mathcal{M}q = -\left(\bar{q}_R\mathcal{M}q_L + \bar{q}_L\mathcal{M}^\dagger q_R\right), \quad \mathcal{M} = \mathrm{diag}(m_u, m_d, m_s),$$

the Goldstone bosons will no longer be massless (see below). Moreover, the symmetry currents are no longer conserved. In terms of the vector currents $V_a^\mu = R_a^\mu - L_a^\mu$ and the axial-vector currents $A_a^\mu = R_a^\mu - L_a^\mu$, the corresponding divergences read

$$\partial_\mu V_a^\mu = i\bar{q}\left[\mathcal{M}, \frac{\lambda_a}{2}\right]q, \quad \partial_\mu A_a^\mu = i\bar{q}\gamma_5\left\{\frac{\lambda_a}{2}, \mathcal{M}\right\}q. \tag{6.2.15}$$

The properties of the currents corresponding to the approximate chiral symmetry of QCD can be summarized as follows:

1. In the limit of massless quarks, the sixteen currents $L_a^\mu$ and $R_a^\mu$ or, alternatively, $V_a^\mu = R_a^\mu + L_a^\mu$ and $A_a^\mu = R_a^\mu - L_a^\mu$ are conserved. The same is true for the singlet vector current $V^\mu$, whereas the singlet axial-vector current $A^\mu$ has an anomaly (see Eq. (6.2.9)).
2. For any values of quark masses, the individual flavor currents $\bar{u}\gamma^\mu u$, $\bar{d}\gamma^\mu d$, and $\bar{s}\gamma^\mu s$ are always conserved in the strong interactions reflecting the flavor independence of the strong coupling and the diagonal form of the quark-mass matrix. Of course, the singlet vector current $V^\mu$, being the sum of the three flavor currents, is always conserved.
3. In addition to the anomaly, the singlet axial-vector current has an explicit divergence due to the quark masses:
$$\partial_\mu A^\mu = 2i\bar{q}\gamma_5\mathcal{M}q + \frac{3g_s^2}{16\pi^2}\epsilon_{\mu\nu\rho\sigma}\mathrm{Tr}_c\left(\mathbf{F}^{\mu\nu}\mathbf{F}^{\rho\sigma}\right).$$
4. For equal quark masses, $m_u = m_d = m_s$, the eight vector currents $V_a^\mu$ are conserved, because $[\lambda_a, \mathbb{1}] = 0$. Such a scenario is the origin of the SU(3) symmetry originally proposed by Gell-Mann and Ne'eman [1568]. The eight axial-vector currents $A_a^\mu$ are not conserved. The divergences of the octet axial-vector currents of Eq. (6.2.15) are proportional to pseudoscalar quadratic forms. This can be interpreted as the microscopic origin of the PCAC relation (partially conserved axial-vector current) [19, 1569] which states that the divergences of the axial-vector currents are proportional to renormalized field operators representing the lowest-lying pseudoscalar octet.
5. Taking $m_u = m_d \neq m_s$ reduces SU(3) flavor symmetry to SU(2) isospin symmetry.
6. Taking $m_u \neq m_d$ leads to isospin-symmetry breaking.

Besides the conservation properties of the currents, one may also calculate their commutators (current algebra), which may then be used to derive certain relations among QCD Green functions analogous to the Ward identities of Quantum Electrodynamics. The set of all QCD Green functions involving color-neutral quark bilinears is very efficiently collected in a generating functional,

$$\exp\left(iZ_{\mathrm{QCD}}[v,a,s,p]\right) = \langle 0|T\exp\left[i\int d^4x\,\mathcal{L}_{\mathrm{ext}}(x)\right]|0\rangle_0, \tag{6.2.16}$$

where [64, 1570]:

$$\begin{aligned}\mathcal{L}_{\mathrm{ext}} &= \sum_{a=1}^{8} v_a^\mu\,\bar{q}\gamma_\mu\frac{\lambda_a}{2}q + v_{(s)}^\mu\,\frac{1}{3}\bar{q}\gamma_\mu q + \sum_{a=1}^{8} a_a^\mu\,\bar{q}\gamma_\mu\gamma_5\frac{\lambda_a}{2}q \\ &\quad - \sum_{a=0}^{8} s_a\,\bar{q}\lambda_a q + \sum_{a=0}^{8} p_a\,i\bar{q}\gamma_5\lambda_a q \\ &= \bar{q}\gamma_\mu\left(v^\mu + \frac{1}{3}v_{(s)}^\mu + \gamma_5 a^\mu\right)q - \bar{q}(s - i\gamma_5 p)q,\end{aligned} \tag{6.2.17}$$

where $\lambda_0 = \sqrt{\frac{2}{3}}\mathbb{1}$. A particular Green function is then obtained through a partial functional derivative with respect to the corresponding external fields. Note that both the quark field operators $q$ in $\mathcal{L}_{\text{ext}}$ and the ground state $|0\rangle$ refer to the chiral limit, indicated by the subscript 0 in Eq. (6.2.16). The quark fields are operators in the Heisenberg picture and have to satisfy the equation of motion and the canonical anticommutation relations. From the generating functional, we can even obtain Green functions of the "real world," where the quark fields and the ground state are those with finite quark masses. To that end one needs to evaluate the functional derivative of Eq. (6.2.16) at $s = \text{diag}(m_u, m_d, m_s)$. The chiral Ward identities result from an invariance of the generating functional of Eq. (6.2.16) under a *local* transformation of the quark fields and a simultaneous transformation of the external fields [64, 1570],

$$q_L \mapsto \exp\left(-i\frac{\Theta(x)}{3}\right) V_L(x) q_L, \tag{6.2.18a}$$

$$q_R \mapsto \exp\left(-i\frac{\Theta(x)}{3}\right) V_R(x) q_R, \tag{6.2.18b}$$

where $V_L(x)$ and $V_R(x)$ are independent space-time-dependent SU(3) matrices, provided the external fields are subject to the transformations

$$l_\mu \mapsto V_L l_\mu V_L^\dagger + i V_L \partial_\mu V_L^\dagger, \tag{6.2.19a}$$

$$r_\mu \mapsto V_R r_\mu V_R^\dagger + i V_R \partial_\mu V_R^\dagger, \tag{6.2.19b}$$

$$v_\mu^{(s)} \mapsto v_\mu^{(s)} - \partial_\mu \Theta, \tag{6.2.19c}$$

$$s + ip \mapsto V_R (s + ip) V_L^\dagger, \tag{6.2.19d}$$

$$s - ip \mapsto V_L (s - ip) V_R^\dagger. \tag{6.2.19e}$$

The derivative terms in Eqs. (6.2.19a)-(6.2.19c) serve the same purpose as in the construction of gauge theories, *i.e.* , they cancel analogous terms originating from the kinetic part of the quark Lagrangian.

### 6.2.2 Chiral perturbation theory for mesons

Effective field theory (EFT) is a powerful tool for describing the strong interactions at low energies. The essential idea behind EFT was formulated by Weinberg in Ref. [1387] as follows:

> "... if one writes down the most general possible Lagrangian, including all terms consistent with assumed symmetry principles, and then calculates matrix elements with this Lagrangian to any given order of perturbation theory, the result will simply be the most general possible S–matrix consistent with analyticity, perturbative unitarity, cluster decomposition and the assumed symmetry principles."

In the present context, we want to describe the low-energy dynamics of QCD in terms of its Goldstone bosons as effective degrees of freedom rather than in terms of quarks and gluons. The resulting low-energy approximation is called (mesonic) chiral perturbation theory (ChPT). Its foundations are discussed in Ref. [1571]. Since the interaction strength of the Goldstone bosons vanishes in the zero-energy limit and the quark masses are regarded as small perturbations around the chiral limit, the mesonic Lagrangian is organized in a simultaneous derivative and a quark-mass expansion. This Lagrangian is expected to have exactly eight pseudoscalar degrees of freedom transforming as an octet under flavor $\text{SU}(3)_V$. Moreover, taking account of spontaneous symmetry breaking, the ground state should only be invariant under $\text{SU}(3)_V \times \text{U}(1)_V$. Finally, in the chiral limit, we want the effective Lagrangian to be invariant under $\text{SU}(3)_L \times \text{SU}(3)_R \times \text{U}(1)_V$.

Our goal is to approximate the "true" generating functional $Z_{\text{QCD}}[v, a, s, p]$ of Eq. (6.2.16) by a sequence

$$Z_{\text{eff}}^{(2)}[v, a, s, p] + Z_{\text{eff}}^{(4)}[v, a, s, p] + \dots ,$$

where the effective generating functionals are obtained using the effective field theory. The rationale underlying this approach is the assumption that including all of the infinite number of effective functionals $Z_{\text{eff}}^{(2n)}[v, a, s, p]$ will, at least in the low-energy region, generate a result which is equivalent to that obtained from $Z_{\text{QCD}}[v, a, s, p]$. Because of spontaneous symmetry breaking, the chiral group $\text{SU}(3)_L \times \text{SU}(3)_R$ is realized nonlinearly on the Goldstone-boson fields [1387, 1572]. We define the SU(3) matrix

$$U(x) = \exp\left(i\frac{\phi(x)}{F_0}\right), \tag{6.2.20}$$

where the field matrix $\phi$ is a Hermitian, traceless $3 \times 3$ matrix,

$$\phi(x) = \sum_{a=1}^{8} \phi_a \lambda_a \equiv \begin{pmatrix} \pi^0 + \frac{1}{\sqrt{3}}\eta & \sqrt{2}\pi^+ & \sqrt{2}K^+ \\ \sqrt{2}\pi^- & -\pi^0 + \frac{1}{\sqrt{3}}\eta & \sqrt{2}K^0 \\ \sqrt{2}K^- & \sqrt{2}\bar{K}^0 & -\frac{2}{\sqrt{3}}\eta \end{pmatrix}, \tag{6.2.21}$$

and the parameter $F_0$ is the chiral limit of the pion-decay constant. Under *local* chiral transformations, $U(x)$ transforms as [64]

$$U(x) \mapsto V_R(x) U(x) V_L^\dagger(x). \tag{6.2.22}$$

In particular, Eq. (6.2.22) implies for the field matrix $\phi$ the transformation behavior $\phi(x) \mapsto V\phi(x)V^\dagger$ under global flavor $\text{SU}(3)_V$, *i.e.* , the Goldstone bosons indeed form an SU(3) octet [see Eq. (6.2.14)]. The most general Lagrangian with the smallest (nonzero) number of external fields is given by [64]

$$\mathcal{L}_2 = \frac{F_0^2}{4}\mathrm{Tr}[D_\mu U (D^\mu U)^\dagger] + \frac{F_0^2}{4}\mathrm{Tr}(\chi U^\dagger + U\chi^\dagger), \quad (6.2.23)$$

where

$$D_\mu U \equiv \partial_\mu U - i r_\mu U + i U l_\mu \mapsto V_R D_\mu U V_L^\dagger, \quad (6.2.24a)$$

$$\chi \equiv 2B_0(s + ip) \mapsto V_R \chi V_L^\dagger. \quad (6.2.24b)$$

If we denote a small four momentum as of $\mathcal{O}(q)$, the covariant derivative counts as $\mathcal{O}(q)$ and $\chi$ as $\mathcal{O}(q^2)$ (see below), such that the lowest-order Lagrangian is of $\mathcal{O}(q^2)$, indicated by the subscript 2. Using the cyclic property of the trace, $\mathcal{L}_2$ is easily seen to be invariant under the transformations of Eqs. (6.2.19a)-(6.2.19e) and (6.2.22). Moreover, $\mathcal{L}_2$ is invariant under the simultaneous replacements $U \leftrightarrow U^\dagger$, $l_\mu \leftrightarrow r_\mu$, and $\chi \leftrightarrow \chi^\dagger$. It is said to be of even intrinsic parity.

At lowest order, the effective field theory contains two parameters $F_0$ and $B_0$. In order to pin down the meaning of $F_0$, we consider the axial-vector current $J^\mu_{Aa}$ associated with $\mathcal{L}_2$:

$$J^\mu_{Aa} = -i\frac{F_0^2}{4}\mathrm{Tr}\left(\lambda_a\{U, \partial^\mu U^\dagger\}\right). \quad (6.2.25)$$

Expanding $U$ in terms of the field matrix $\phi$, and using $\mathrm{Tr}(\lambda_a\lambda_b) = 2\delta_{ab}$ results in

$$J^\mu_{Aa} = -F_0\partial^\mu\phi_a + \mathcal{O}(\phi^3), \quad (6.2.26)$$

from which we conclude that the axial-vector current has a nonvanishing matrix element when evaluated between the vacuum and a one-Goldstone-boson state:

$$\langle 0|J^\mu_{Aa}(x)|\phi_b(p)\rangle = ip^\mu F_0 \exp(-ip\cdot x)\delta_{ab}. \quad (6.2.27)$$

Equation (6.2.27) holds at leading order (LO) in ChPT. It is the current-density analog of Eq. (6.2.12), *i.e.* , a nonvanishing value of $F_0$ is a necessary and sufficient criterion for spontaneous symmetry breaking in QCD.

The expansion of the first term of Eq. (6.2.23) in the field matrix $\phi$ yields

$$\frac{1}{4}\mathrm{Tr}\left(\partial_\mu\phi\partial^\mu\phi\right) + \frac{1}{48F^2}\mathrm{Tr}\left([\phi, \partial_\mu\phi][\phi, \partial^\mu\phi]\right) + \ldots . \quad (6.2.28)$$

The first term of Eq. (6.2.28) describes the kinetic term of the eight Goldstone bosons and the second term contributes to the scattering of Goldstone bosons. The second term of Eq. (6.2.23) is an example how the explicit symmetry breaking by the quark masses is transferred from the QCD level to the EFT level. Both, $\mathcal{L}^0_{\mathrm{QCD}} + \mathcal{L}_{\mathrm{ext}}$ and $\mathcal{L}_2$ are invariant under *local* chiral transformations. Inserting $\mathcal{L}_{\mathrm{ext}} = \mathcal{L}_\mathcal{M}$ corresponds to $s = \mathrm{diag}(m_u, m_d, m_s)$ and it is the same $s$ that is to be used in the effective Lagrangian. The expansion of the $\chi$ term gives rise to

$$F_0^2 B_0(m_u+m_d+m_s) - \frac{B_0}{2}\mathrm{Tr}\left(\phi^2\mathcal{M}\right) + 2B_0\mathrm{Tr}\left(\mathcal{M}\phi^4\right) + \ldots \quad (6.2.29)$$

Even though the first term of Eq. (6.2.29) is of no dynamical significance for the interaction among the Goldstone bosons, it represents an interesting effect. Its negative is the energy density of the vacuum, $\langle\mathcal{H}_{\mathrm{eff}}\rangle_{\mathrm{min}}$, which is shifted relative to the chiral limit because of the nonzero quark masses. We compare the partial derivative of $\langle\mathcal{H}_{\mathrm{eff}}\rangle_{\mathrm{min}}$ with respect to (any of) the light-quark masses $m_l$ with the corresponding quantity in QCD,

$$\left.\frac{\partial\langle 0|\mathcal{H}_{\mathrm{QCD}}|0\rangle}{\partial m_l}\right|_{m_u=m_d=m_s=0} = \frac{1}{3}\langle 0|\bar{q}q|0\rangle_0 = \frac{1}{3}\langle\bar{q}q\rangle_0, \quad (6.2.30)$$

where $\langle\bar{q}q\rangle_0$ is the scalar singlet quark condensate. Within the framework of the lowest-order effective Lagrangian, the constant $B_0$ is thus related to the scalar singlet quark condensate by

$$3F_0^2 B_0 = -\langle\bar{q}q\rangle_0. \quad (6.2.31)$$

For an overview of recent lattice QCD determinations of $\langle\bar{q}q\rangle_0$ see Ref. [1573]. Because of the second term of Eq. (6.2.29), the Goldstone bosons are no longer massless. If, for the sake of simplicity, we consider the isospin-symmetric limit $m_u = m_d = \hat{m}$ (so that there is no $\pi^0$-$\eta$ mixing), we obtain for the masses of the Goldstone bosons, to lowest order in the quark masses ($\mathcal{O}(q^2)$, denoted by the subscript 2),

$$M^2_{\pi,2} = 2B_0\hat{m}, \quad (6.2.32a)$$

$$M^2_{K,2} = B_0(\hat{m} + m_s), \quad (6.2.32b)$$

$$M^2_{\eta,2} = \frac{2}{3}B_0\left(\hat{m} + 2m_s\right). \quad (6.2.32c)$$

These results, in combination with Eq. (6.2.31), correspond to relations obtained in Ref. [1574] and are referred to as the Gell-Mann, Oakes, and Renner relations. Because of the on-shell condition $p^2 = M^2$, Eqs. (6.2.32a)-(6.2.32c) justify the assignment $\chi = \mathcal{O}(q^2)$. Inserting the empirical values $M_\pi = 135$ MeV, $M_K = 496$ MeV, and $M_\eta = 548$ MeV for the lowest-order predictions provides a first estimate for the ratio of the quark masses,

$$\frac{M_K^2}{M_\pi^2} = \frac{\hat{m} + m_s}{2\hat{m}} \Rightarrow \frac{m_s}{\hat{m}} = 25.9, \quad (6.2.33a)$$

$$\frac{M_\eta^2}{M_\pi^2} = \frac{2m_s + \hat{m}}{3\hat{m}} \Rightarrow \frac{m_s}{\hat{m}} = 24.3. \quad (6.2.33b)$$

A remarkable feature of Eq. (6.2.23) is the fact that, once $F_0$ is known (from pion decay), chiral symmetry allows us to make absolute predictions about other processes. For example, the lowest-order results for the scattering of Goldstone bosons can be derived straightforwardly from the $\mathcal{O}(\phi^4)$ contributions of Eqs. (6.2.28) and (6.2.29). In particular, the $s$-wave $\pi\pi$-scattering lengths for the isospin channels $I = 0$ and $I = 2$ are obtained as [1570]

$$a_0^0 = \frac{7M_\pi^2}{32\pi F_\pi^2} = 0.160, \quad a_0^2 = -\frac{M_\pi^2}{16\pi F_\pi^2} = -0.0456, \tag{6.2.34}$$

where we replaced $F_0$ by the physical pion-decay constant and made use of the numerical values $F_\pi = 92.2$ MeV and $M_\pi = M_{\pi^+} = 139.57$ MeV. These results are identical with the current-algebra predictions of Ref. [22]. Actually, they serve as an illustration of the fact that the results of current algebra can (more easily) be reproduced from lowest-order perturbation theory in terms of an effective Lagrangian [1575]—in the present case the lowest-order mesonic ChPT Lagrangian.

However, ChPT is much more powerful than the effective Lagrangians of the 1960s, which, by definition, were meant to be applied only in lowest-order perturbation theory (see, e.g., the second footnote in Ref. [1576]). In ChPT, a systematic improvement beyond the tree-level of the lowest-order Lagrangian of Eq. (6.2.23) is accomplished by calculating loop corrections in combination with tree-level contributions from Lagrangians of higher order. For a long time it was believed that performing loop calculations using the Lagrangian of Eq. (6.2.23) would make no sense, because it is not renormalizable (in the traditional sense [821]). However, as emphasized by Weinberg [1387, 1577], the cancellation of ultraviolet divergences does not really depend on renormalizability; as long as one includes every one of the infinite number of interactions allowed by symmetries, the so-called non-renormalizable theories are actually just as renormalizable as renormalizable theories [1577]. This still leaves open the question of how to organize a perturbative description of observables. For that purpose, one needs a power-counting scheme to assess the importance of various diagrams calculated from the most general effective Lagrangian. Using Weinberg's power counting scheme [1387], one may analyze the behavior of a given diagram of mesonic ChPT under a linear re-scaling of all *external* momenta, $p_i \mapsto tp_i$, and a quadratic re-scaling of the light-quark masses, $m_l \mapsto t^2 m_l$, which, in terms of the Goldstone-boson masses, corresponds to $M^2 \mapsto t^2 M^2$. The chiral dimension $D$ of a given diagram with amplitude

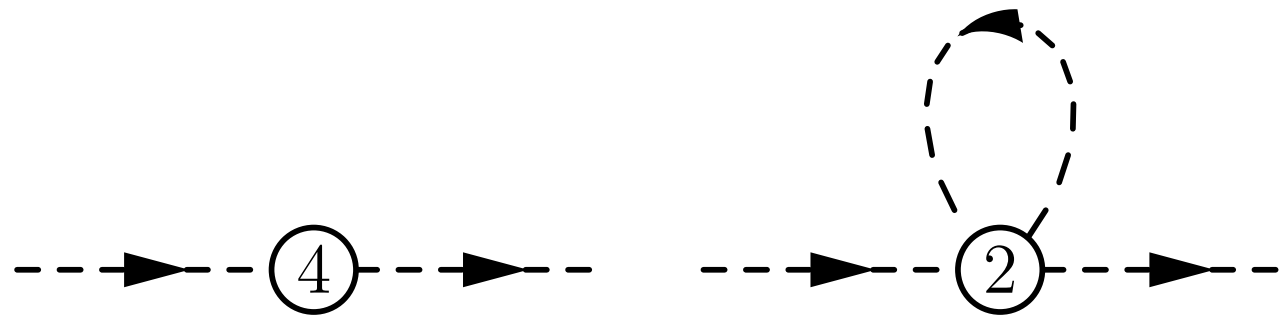

**Fig. 6.2.1** Self-energy diagrams at $\mathcal{O}(q^4)$. Vertices derived from $\mathcal{L}_{2n}$ are denoted by $2n$ in the interaction blobs.

$\mathcal{M}(p_i, m_l)$ is defined by

$$\mathcal{M}(tp_i, t^2 m_l) = t^D \mathcal{M}(p_i, m_l), \tag{6.2.35}$$

where, in $n$ dimensions,

$$D = nN_L - 2N_I + \sum_{k=1}^{\infty} 2kN_{2k} \tag{6.2.36}$$

$$= 2 + (n-2)N_L + \sum_{k=1}^{\infty} 2(k-1)N_{2k} \tag{6.2.37}$$

$$\geq 2 \text{ in 4 dimensions.}$$

Here, $N_L$ is the number of independent loops, $N_I$ the number of internal Goldsone-boson lines, and $N_{2k}$ the number of vertices originating from $\mathcal{L}_{2k}$. A diagram with chiral dimension $D$ is said to be of order $\mathcal{O}(q^D)$. Clearly, for small enough momenta and masses, diagrams with small $D$, such as $D = 2$ or $D = 4$, should dominate. Of course, the re-scaling of Eq. (6.2.35) must be viewed as a mathematical tool. While external three-momenta can, to a certain extent, be made arbitrarily small, the re-scaling of the quark masses is a theoretical instrument only. Note that, for $n = 4$, loop diagrams are always suppressed due to the term $2N_L$ in Eq. (6.2.37). In other words, we have a perturbative scheme in terms of external momenta and masses which are small compared to some scale (here $4\pi F_0 \approx 1$ GeV).

The most general Lagrangian at $\mathcal{O}(q^4)$ was constructed by Gasser and Leutwyler [64] and contains twelve low-energy constants (LECs) $(L_1, \ldots, L_{10}, H_1 H_2)$,

$$\mathcal{L}_4 = L_1 \left\{ \mathrm{Tr}[D_\mu U (D^\mu U)^\dagger] \right\}^2 + \ldots + H_2 \mathrm{Tr}\left(\chi\chi^\dagger\right). \tag{6.2.38}$$

The numerical values of the low-energy constants $L_i$ are not determined by chiral symmetry. In analogy to $F_0$ and $B_0$ of $\mathcal{L}_2$ they are parameters containing information on the underlying dynamics. For an extensive review of the status of these coupling constants, see Refs. [1578] as well as [1573].

As an example of a one-loop calculation let us consider the $\mathcal{O}(q^4)$ corrections to the masses of the Goldstone bosons. For that purpose one needs to evaluate the self-energy diagrams shown in Fig. 6.2.1. The corresponding expressions for the masses were first given in Ref. [64], of which we show the squared pion mass as a representative example:

$$M_{\pi,4}^2 = M_{\pi,2}^2\left\{1+\frac{M_{\pi,2}^2}{32\pi^2F_0^2}\ln\left(\frac{M_{\pi,2}^2}{\mu^2}\right)-\frac{M_{\eta,2}^2}{96\pi^2F_0^2}\ln\left(\frac{M_{\eta,2}^2}{\mu^2}\right)\right. \left.+\frac{16}{F_0^2}\left[(2\hat{m}+m_s)B_0(2L_6^r-L_4^r)+\hat{m}B_0(2L_8^r-L_5^r)\right]\right\}. \quad (6.2.39)$$

Because of the overall factor $M_{\pi,2}^2$, the pion stays massless as $m_l \to 0$. This is, of course, what we expected from QCD in the chiral limit, but it is comforting to see that the self interaction in $\mathcal{L}_2$ (in the absence of quark masses) does not generate Goldstone-boson masses at higher order. The ultraviolet divergences generated by the loop diagram of Fig. 6.2.1 are cancelled by a suitable adjustment of the parameters of $\mathcal{L}_4$. This is Weinberg's argument on renormalizability at work; as long as one works with the most general Lagrangian all ultraviolet divergences can be absorbed in the parameters of the theory. At $\mathcal{O}(q^4)$, the squared Goldstone-boson masses contain terms which are analytic in the quark masses, namely, of the form $m_l^2$ multiplied by the renormalized low-energy constants $L_i^r$. However, there are also nonanalytic terms of the type $m_l^2\ln(m_l)$—so-called chiral logarithms—which do not involve new parameters. Such a behavior is an illustration of the mechanism found by Li and Pagels [1579], who noticed that a perturbation theory around a symmetry, which is realized in the Nambu-Goldstone mode, results in both analytic as well as nonanalytic expressions in the perturbation. Finally, by construction, the scale dependence of the renormalized coefficients $L_i^r$ entering Eq. (6.2.39) is such that it cancels the scale dependence of the chiral logarithms [64]. Thus, physical observables do not depend on the scale $\mu$.

In terms of Fig. 6.2.1 and the result of Eq. (6.2.39), we can also comment on the so-called chiral-symmetry-breaking scale $\Lambda_\chi$ to be $\Lambda_\chi = 4\pi F_0$ [1566]. In a loop correction, every endpoint of an internal Goldstone-boson line is multiplied by a factor of $1/F_0$, since the SU(3) matrix of Eq. (6.2.20) contains the Goldstone-boson fields in the combination $\phi/F_0$. On the other hand, external momenta $q$ or Goldstone-boson masses produce factors of $q^2$ or $M^2$ (see Eqs. (6.2.28) and (6.2.29)). Together with a factor $1/(16\pi^2)$ remaining after integration in four dimensions they combine to corrections of the order of $[q/(4\pi F_0)]^2$ for each independent loop. Strictly speaking, this particular integral generates an additional factor of 2, and the factor of $1/(16\pi^2)$ should be considered an estimate.

The Lagrangians discussed so far are of even intrinsic parity. At $\mathcal{O}(q^4)$, they are incomplete, because they do not describe processes such as $K^+K^- \to \pi^+\pi^-\pi^0$ or $\pi^0 \to \gamma\gamma$. The missing piece is the effective Wess-Zumino-Witten (WZW) action [1580, 1581], which accounts for the chiral anomaly. The chiral anomaly results in the so-called anomalous Ward identities that give a particular form to the *variation* of the generating functional [1570, 1580]. At leading order, $\mathcal{O}(q^4)$, and in the absence of external fields, the WZW action reads [1580, 1581],

$$S_{\text{ano}}^0 = N_c\, S_{\text{WZW}}^0,$$
$$S_{\text{WZW}}^0 = -\frac{i}{240\pi^2}\int_0^1 d\alpha\int d^4x\epsilon^{ijklm}\text{Tr}(\mathcal{U}_i^L\mathcal{U}_j^L\mathcal{U}_k^L\mathcal{U}_l^L\mathcal{U}_m^L). \quad (6.2.40)$$

For the construction of the WZW action, the domain of definition of $U$ needs to be extended to a (hypothetical) fifth dimension,

$$U(y) = \exp\left(i\alpha\frac{\phi(x)}{F_0}\right), \quad (6.2.41)$$

where $y^i = (x^\mu,\alpha)$, $i = 0,\ldots,4$, and $0 \le \alpha \le 1$. Minkowski space is defined as the surface of the five-dimensional space for $\alpha = 1$. The indices $i,\ldots,m$ in Eq. (6.2.40) run from 0 to 4, $y_4 = y^4 = \alpha$, $\epsilon_{ijklm}$ is the completely antisymmetric (five-dimensional) tensor with $\epsilon_{01234} = -\epsilon^{01234} = 1$, and $\mathcal{U}_i^L = U^\dagger\partial U/\partial y^i$.

In contrast to $\mathcal{L}_2$ and $\mathcal{L}_4$, $S_{\text{ano}}^0$ is of odd intrinsic parity, *i.e.* , it changes sign under $\phi \to -\phi$. Expanding the SU(3) matrix $U(y)$ in terms of the Goldstone-boson fields, $U(y) = \mathbb{1} + i\alpha\phi(x)/F_0 + O(\phi^2)$, one obtains an infinite series of terms, each involving an odd number of Goldstone bosons. For example, after some rearrangements, the term with the smallest number of Goldstone bosons reads

$$S_{\text{WZW}}^{5\phi} = \frac{1}{240\pi^2F_0^5}\int d^4x\,\epsilon^{\mu\nu\rho\sigma}\text{Tr}(\phi\partial_\mu\phi\partial_\nu\phi\partial_\rho\phi\partial_\sigma\phi). \quad (6.2.42)$$

In particular, the WZW action without external fields involves at least five Goldstone bosons [1580]. Again, once $F_0$ is known, after inserting $N_c = 3$ one obtains a parameter-free prediction for, e.g., the process $K^+K^- \to \pi^+\pi^-\pi^0$.

In the presence of external fields, the anomalous action receives an additional term [1581–1583]

$$S_{\text{ano}} = N_c(S_{\text{WZW}}^0 + S_{\text{WZW}}^{\text{ext}}) \quad (6.2.43)$$

given by

$$S_{\text{WZW}}^{\text{ext}} = -\frac{i}{48\pi^2}\int d^4x\epsilon^{\mu\nu\rho\sigma}\text{Tr}\left[Z_{\mu\nu\rho\sigma}(U,l,r) -Z_{\mu\nu\rho\sigma}(\mathbb{1},l,r)\right]. \quad (6.2.44)$$

where the explicit form of $Z_{\mu\nu\rho\sigma}(U, l, r)$ can be found in [1582, 1583]. At leading order, the action of Eq. (6.2.44) is responsible for the two-photon decays of the $\pi^0$ or the $\eta$. Quantum corrections to the WZW classical action do not renormalize the coefficient of the WZW term. The counter terms needed to renormalize the one-loop singularities at $\mathcal{O}(q^6)$ are of a conventional chirally invariant structure. In the three-flavor sector, the most general odd-intrinsic-parity Lagrangian at $\mathcal{O}(q^6)$ contains 23 independent terms [1584, 1585]. For an overview of applications in the odd-intrinsic-parity sector, we refer to Ref. [1583].

### 6.2.3 ChPT for baryons

ChPT was first extended to the baryon sector in Ref. [1586], which considered a variety of matrix elements with single-nucleon incoming and outgoing states. While the general approach is analogous to that in the mesonic sector, *i.e.* , one considers the most general Lagrangian consistent with the symmetries of QCD and expands observables in a quark-mass and low-momentum expansion, the baryon sector exhibits some new features. In particular, unlike the Goldstone-boson masses, the baryon masses do not vanish in the chiral limit. This has important consequences for obtaining a proper power counting of diagrams containing baryon lines and for the regularization and renormalization of loop diagrams. In the following we restrict the discussion to $\mathrm{SU}(2)_L \times \mathrm{SU}(2)_R$ chiral symmetry; for the extension to $\mathrm{SU}(3)_L \times \mathrm{SU}(3)_R$ see, e.g., the reviews of Refs. [1587, 1588] and references therein. To construct the pion-nucleon Lagrangian, the proton ($p$) and neutron ($n$) fields are combined into an SU(2) doublet $\Psi$,

$$\Psi = \begin{pmatrix} p \\ n \end{pmatrix}. \tag{6.2.45}$$

The nucleon fields are chosen to transform under local $\mathrm{SU}(2)_L \times \mathrm{SU}(2)_R$ transformations as

$$\Psi \to K(V_L, V_R, U)\Psi, \tag{6.2.46}$$

where the SU(2) matrix $K$ depends on the left- and right-handed transformations as well as on the pion fields collected in $U$,

$$K(V_L, V_R, U) = \sqrt{V_R U V_L^\dagger}^{-1} V_R \sqrt{U}. \tag{6.2.47}$$

The baryon Lagrangian also contains the covariant derivative of the nucleon field given by

$$D_\mu \Psi = (\partial_\mu + \Gamma_\mu - i v_\mu^{(s)})\Psi, \tag{6.2.48}$$

with the connection [1586, 1589]

$$\Gamma_\mu = \frac{1}{2}\left[u^\dagger(\partial_\mu - i r_\mu)u + u(\partial_\mu - i l_\mu)u^\dagger\right], \tag{6.2.49}$$

where $u^2 = U$, and the isoscalar vector field $v_\mu^{(s)}$. Further, it is convenient to define

$$u_\mu = i\left[u^\dagger(\partial_\mu - i r_\mu)u - u(\partial_\mu - i l_\mu)u^\dagger\right]. \tag{6.2.50}$$

The LO Lagrangian can be written as [1586]

$$\mathcal{L}_{\pi N}^{(1)} = \bar{\Psi}\left(i\not{D} - \mathtt{m} + \frac{\mathtt{g}_A}{2}\gamma^\mu\gamma_5 u_\mu\right)\Psi. \tag{6.2.51}$$

It contains two LECs: $\mathtt{m}$ and $\mathtt{g}_A$. These correspond to the nucleon mass ($\mathtt{m}$) and the nucleon axial-vector coupling constant ($\mathtt{g}_A$), both taken in the chiral limit. The corresponding physical values will be denoted as $m_N$ and $g_A$ in the following. The superscript (1) in Eq. (6.2.51) denotes that the Lagrangian is of first order in the power counting. While neither the nucleon energy nor the chiral-limit nucleon mass are small parameters, the combination $i\not{D} - \mathtt{m}$ can be assumed to be a small quantity as long as the nucleon three-momentum is $\mathcal{O}(q)$.

This Lagrangian can be used to calculate the first loop contribution to the nucleon mass. The power counting predicts this contribution to be $\mathcal{O}(q^3)$. However, the application of dimensional regularization and the minimal subtraction scheme of ChPT ($\widetilde{\mathrm{MS}}$) as used in the meson sector results in terms that are of lower order than predicted by the power counting. Analogous issues also arise for other observables and higher-order contributions. The authors of Ref. [1586] pointed out that the failure of the power counting is related to the regularization and renormalization schemes and that the "same phenomenon would occur in the meson sector, if one did not make use of dimensional regularization." Several methods to address the power counting issue have been proposed [1590–1595].

One commonly used method is Heavy Baryon ChPT (HBChPT) [1590], which was inspired by Heavy Quark Effective Theory [667, 1254] (see the discussion in Sec. 6.1). Because the nucleon mass is large compared to the pion mass, an additional expansion of the pion-nucleon Lagrangian is performed in inverse powers of the nucleon mass. In this formalism, application of dimensional regularization in combination with $\widetilde{\mathrm{MS}}$ to loop diagrams, as in the meson sector, leads to a consistent power counting, connecting the chiral to the loop expansion. The heavy-baryon Lagrangian up to and including order $q^4$ is given in Ref. [1596]. For an introduction to, and applications of, this method see, e.g., Refs. [1587, 1597].

While the heavy-baryon formalism makes it possible to use techniques from the meson sector, the additional expansion in powers of the inverse nucleon mass results in a large number of terms in the higher-order Lagrangians. Some of the higher-order terms are related to those at lower orders by Lorentz invariance [1398].

Calculated amplitudes can be expressed in Lorentz-invariant forms, but Lorentz invariance is not manifest throughout intermediate steps of the calculations. Further, issues with analyticity arise in some specific cases because the heavy-baryon expansion results in a shift of the poles in the nucleon propagator [1592].

A manifestly Lorentz-invariant approach to baryon ChPT that addresses these issues was formulated in Ref. [1592], referred to as infrared regularization. While infrared regularization also uses dimensional regularization, the renormalization procedure is different from minimal subtraction. Loop integrals are separated into infrared-singular and infrared-regular parts. The infrared-singular parts contain the same infrared singularities as the original integral and they satisfy the power counting. The infrared-regular parts are analytic in small parameters for arbitrary spacetime dimensions and contain the power-counting-violating terms. Since the infrared-regular parts are analytic, they can be absorbed in the LECs of the baryon Lagrangian. Infrared regularization in its original formulation was applicable to one-loop diagrams. It has been widely used in the calculation of baryon properties, see, e.g., Ref. [1598] for a review.

The expansion of the infrared-regular parts in small parameters contains not only the terms violating the power counting, but also an infinite set of terms that satisfy the power counting. The extended on-mass-shell (EOMS) scheme [1595] provides a method to isolate the terms that violate the power counting and to absorb only these terms in the LECs of the Lagrangian. The EOMS scheme was also shown to be applicable to multi-loop diagrams [1599] and diagrams containing particles other than pions and nucleons [1600]. By reformulating infrared regularization analogously to the EOMS scheme [1601], it can be applied beyond one-loop pion-nucleon diagrams [1599]; see also Ref. [1602] for a different extension of infrared regularization.

The nucleon mass presents an example of the application of baryon ChPT. It has been determined to one-loop order in several approaches, including HBChPT [1603], infrared regularization [1592], and the EOMS scheme [1595]. Up to and including order $q^3$, the chiral expansion of the nucleon mass is given by

$$m_N = \mathtt{m} - 4c_1 M^2 - \frac{3\mathtt{g}_A^2}{32\pi F^2} M^3 + \ldots, \tag{6.2.52}$$

where $F$ denotes the pion-decay constant in the two-flavor chiral limit, $F_\pi = F[1 + \mathcal{O}(\hat{m})] = 92.2$ MeV and $M^2 = 2B\hat{m}$ is the lowest-order expression for the squared pion mass.

The result of Eq. (6.2.52) exhibits some general features of baryon ChPT: The expansion contains not just even powers in the small parameter $q$ like the meson sector, but also odd powers. As a result, the convergence of chiral expansions is expected to be slower in the baryon sector. The second-order contribution is proportional to the LEC $c_1$ from the second-order Lagrangian. On the other hand, the coefficient of the nonanalytic term proportional to $M^3$ is given entirely in terms of the LO LEC $\mathtt{g}_A$ and $F$. Similar features also appear at higher orders. The general form of the chiral expansion of the nucleon mass to higher orders is given by

$$\begin{aligned} m_N &= \mathtt{m} + k_1 M^2 + k_2 M^3 + k_3 M^4 \ln\left(\frac{M}{\mu}\right) + k_4 M^4 \\ &\quad + k_5 M^5 \ln\left(\frac{M}{\mu}\right) + k_6 M^5 \\ &\quad + k_7 M^6 \ln^2\left(\frac{M}{\mu}\right) + k_8 M^6 \ln\left(\frac{M}{\mu}\right) + k_9 M^6 + \ldots, \end{aligned} \tag{6.2.53}$$

where $\mu$ is the renormalization scale and the ellipsis denotes higher-order terms. The coefficients $k_i$ are linear combinations of various LECs. $k_1$ through $k_4$ can be determined by considering at most one-loop diagrams, while $k_5$ through $k_9$ receive contributions from two-loop diagrams. Using estimates of the LECs entering the $k_i$, Ref. [1604] estimated the nucleon mass in the chiral limit from an EOMS calculation to order $q^4$ to be

$$\begin{aligned} \mathtt{m} &= [938.3 - 74.8 + 15.3 + 4.7 - 0.7]\,\text{MeV} \\ &= 882.8\,\text{MeV}. \end{aligned} \tag{6.2.54}$$

Two-loop contributions to order $q^5$ were considered in Ref. [1605], while Refs. [1606, 1607] determined $m_N$ to order $q^6$. Because several currently undetermined LECs enter the expressions for several of the higher-order $k_i$, no reliable estimate of the complete two-loop contributions is possible. However, the coefficient $k_5$ of the leading nonanalytic contribution at order $q^5$ only depends on $\mathtt{g}_A$ and the pion-decay constant $F$ and can therefore be compared to lower-order terms. At the physical pion mass and with $\mu = m_N$, $k_5 M^5 \ln(M/m_N) = -4.8$ MeV.

Chiral expansions like that of Eq. (6.2.53) are also important at nonphysical pion masses in the extrapolation of lattice QCD results (for an introduction see, e.g., Ref. [1608]). The fifth-order term $k_5 M^5 \ln(M/m_N)$ becomes as large as the third-order term $k_2 M^3$, where $k_2$ also only depends on $\mathtt{g}_A$ and $F$, for a pion mass of about 360 MeV. While this comparison includes only one part of the two-loop contributions, it indicates a limit to the applicability of the power counting. This estimate agrees with others found using different methods in Refs. [1609, 1610].

Even though the nucleon mass is a static quantity, it is not entirely surprising that a combined chiral and momentum expansion in the baryon sector does not converge well for energies beyond about 300 MeV. This roughly corresponds to the mass gap between the nucleon and the $\Delta(1232)$ resonance. At the physical point, treating the $\Delta$ as an explicit degree of freedom has limited impact on the nucleon mass [1611, 1612]. However, the $\Delta(1232)$ also couples strongly to the $\pi N$ channel and has relatively large photon decay amplitudes, resulting in important contributions to processes such as pion-nucleon scattering, Compton scattering, and electromagnetic pion production. These issues were already pointed out in Ref. [1590], which advocated for treating $\Delta$ degrees of freedom as dynamic. In baryon ChPT with only pions and nucleons as degrees of freedom, effects of the $\Delta(1232)$ enter implicitly through the values of the LECs. However, these contributions can be proportional to powers of $M/\delta$, where $\delta = (m_\Delta - m)$. This ratio is small as the quark masses approach the chiral limit, but it is a rather large expansion parameter at the physical values, especially when combined with the strong coupling of the $\Delta$. By formulating a theory that also includes the $\Delta$ as an active degree of freedom, one hopes to improve the convergence of the perturbative expansion and potentially to increase the kinematic range of applicability.

The inclusion of $\Delta$ degrees of freedom poses additional challenges to the construction of the most general Lagrangian and to the power counting. The covariant description of spin-$\frac{3}{2}$, isospin-$\frac{3}{2}$ fields introduces unphysical degrees of freedom [1613, 1614]. For the free Lagrangian, these can be eliminated by subsidiary equations and projection operators. The correct number of degrees of freedom also has to be preserved when including interactions with pions, nucleons, and external fields. Various approaches addressing this issue have been considered, see, e.g., Refs. [1615–1620].

The main issue for the power counting is how to count the $\Delta$-nucleon mass difference $\delta$. In one version of the power counting [1617], it is a small quantity of the same order as the pion mass, $\delta \sim \mathcal{O}(q)$. In a different approach [1621], it is argued that (for physical quark masses) $M_\pi < \delta$ and that $M_\pi/\delta \sim \delta/\Lambda$, where $\Lambda \sim 1\,\text{GeV}$ is the breakdown scale of the EFT. Denoting $\bar{\delta} \equiv \delta/\Lambda$ implies that $M_\pi/\Lambda \sim \bar{\delta}^2$, *i.e.* , the pion mass is of higher order than the $\Delta$-nucleon mass difference in this power counting.

### 6.2.4 Conclusions

Over the last few decades, ChPT has developed into a mature and comprehensive approach to the low-energy interactions between Goldstone bosons, nucleons, and external fields, with numerous successful applications. ChPT has played an important role in interpreting lattice QCD calculations performed at unphysical pion masses. It has also served as a prototype for semi-phenomenological approaches in other systems. The application of ChPT methods to the interactions between two and more nucleons is discussed in the contribution by Epelbaum and Pastore.

## 6.3 Chiral EFT and nuclear physics

**Evgeny Epelbaum and Saori Pastore**

As explained in the previous section, ChPT allows one to describe the low-energy interactions between hadrons in the Goldstone-boson and single-baryon sectors by means of a perturbative expansion in light-quark masses and particle momenta in line with the symmetries of QCD. In this section we briefly review the extension and application of this systematic and model-independent method to systems with several baryons, focusing on the non-strange sector. This extension goes beyond strict perturbation theory and is commonly referred to as chiral effective field theory, or ChEFT, in order to make the distinction with ChPT clear.

### 6.3.1 The foundations of ChEFT

EFT methods enjoy increasing popularity in nuclear physics[63]. A unified description of few-nucleon systems, medium mass and heavy nuclei as well as nuclear matter up to the saturation density calls for an EFT applicable at nucleon momenta $p \sim M_\pi$, which must include pions as dynamical DoF. The corresponding framework, commonly referred to as chiral EFT (ChEFT), was pioneered by Weinberg [1626, 1627] and represents the most widely used EFT approach in nuclear physics applications. The method relies on the spontaneously broken approximate chiral symmetry of QCD and makes use of the effective Lagrangian for pions and nucleons already introduced in the previous section. Specifically, the $\mathcal{O}(q^2)$ and $\mathcal{O}(q^4)$ mesonic Lagrangians are given in Eqs. (6.2.23) and (6.2.38), respectively, while the LO pion-nucleon ($\pi$N) Lagrangian is written in Eq. (6.2.51). Most of the applications to few-nucleon systems are carried out using the heavy-baryon (HB) Lagrangian for the velocity-dependent nucleon field $N(x) = e^{imv\cdot x} P_v^+ \Psi(x)$, with $P_v^+ = (1 + v\cdot\gamma)/2$ being the velocity projection operator [1590]. The LO HB $\pi$N Lagrangian obtained from the covariant expression in Eq. (6.2.51) takes the form

$$\mathcal{L}_{\pi N}^{(1)} = N^\dagger (iv\cdot D + g_A S\cdot u)N\,, \tag{6.3.1}$$

where $S_\mu = -\gamma_5[\gamma_\mu,\, \gamma_\nu]v^\nu/4$ is the covariant spin-operator that is given by the usual Pauli matrices $S^\mu = (0,\, \vec\sigma/2)$ in the rest-frame system of the nucleon with $v_\mu = (1, \vec{0}\,)$. Higher-order terms in the HB $\pi$N Lagrangian can be found in Refs. [1587, 1596]. Finally, one also needs to include in the effective Lagrangian terms with more than two nucleon fields. The corresponding LO Lagrangian has the form [1626, 1627]

$$\mathcal{L}_{NN}^{(0)} = -\frac{1}{2} C_S (N^\dagger N)^2 + 2C_T N^\dagger S_\mu N\, N^\dagger S^\mu N\,, \tag{6.3.2}$$

with $C_S$, $C_T$ being low-energy constants (LECs).

[63] In the past decades, a variety of EFTs utilizing different degrees of freedom (DoF) have been developed to describe phenomena characterized by specific energy scales. For example, an EFT description of rotational bands of deformed heavy nuclei with excitation energies $E \ll 1$ MeV can be efficiently achieved in terms of collective coordinates with no need to resolve the internal structure of a nucleus under consideration [1622]. Low-energy properties of nuclei consisting of a dense core, surrounded by weakly bound nucleons, have been studied in halo-EFT [1623]. This framework treats the core nucleus as a point-like particle and utilizes the expansion in powers of $p/p_{\text{core}}$, with $p$ and $p_{\text{core}}$ denoting the binding momenta of the nucleons and of the core nucleus, respectively. Another EFT approach, the so-called pion-less EFT, is formulated in terms of nucleons as the only dynamical DoF and is well suited to describe the dynamics of few-nucleon systems at momenta $p \ll M_\pi$. This framework has proven to be particularly efficient for uncovering universal features of few-body systems around the unitary limit [1624, 1625].

While both ChPT and ChEFT rely on the same effective Lagrangian, the two frameworks are applied to describe rather different phenomenological situations. Contrary to the meson and single-baryon sectors, the scattering amplitudes for few-nucleon systems exhibit low-lying poles corresponding to bound (and virtual) states, which signal the breakdown of perturbation theory at very low momenta. For example, in the ${}^3S_1$ and ${}^1S_0$ channels of neutron-proton scattering, the poles are located at $p_{\rm cms} \sim 45i$ MeV and $p_{\rm cms} \sim -8i$ MeV, respectively, which is well within the validity domain of chiral (and even pion-less) EFT. This is in strong contrast to pion-pion scattering, where the lowest-lying resonances reside at momenta of the order of the breakdown scale of ChPT, and the scattering amplitude admits a perturbative expansion in powers of momenta for $p \sim M_\pi$. It is worth emphasizing that while the spontaneously broken chiral symmetry of QCD leads to a strong suppression of the interactions between Goldstone bosons (pions) at low energy, which is at the heart of ChPT, it does not constrain the strength of the interaction between the nucleons for $|\vec p\,| \to 0$, see Eq. (6.3.2).

So how can ChEFT be reconciled with the nonperturbative nature of the two-nucleon interaction? To answer this question one needs a power-counting scheme that determines the importance of *renormalized* contributions to the scattering amplitude. The power counting of mesonic ChPT was already given in Eqs. (6.2.35)-(6.2.37). Using the HB framework to avoid the appearance of positive powers of the nucleon mass in renormalized expressions as explained in the previous section, the power counting can be straightforwardly extended to single- and few-nucleon scattering amplitudes. A connected contribution to the scattering amplitude for $N$ nucleons with generic momenta $|\vec p\,| \sim M_\pi$ involving $N_L$ independent loop integrals is found to scale as $\mathcal{M} \sim q^D$, where $q \in \{|\vec p\,|/\Lambda_b,\, M_\pi/\Lambda_b\}$ with $\Lambda_b$ being the breakdown scale of ChEFT. In four space-time dimensions, the power $D$ is given by [1626, 1627]

$$D = 2 - N + 2N_L + \sum_i V_i \Delta_i\,, \tag{6.3.3}$$

where $V_i$ denotes the number of vertices of type $i$, whose dimension $\Delta_i$ is given by

$$\Delta_i = -2 + \frac{1}{2} n_i + d_i\,. \tag{6.3.4}$$

Here, $n_i$ is the number of nucleon fields while $d_i$ refers to the number of derivatives and/or insertions of $M_\pi$. Using Eq. (6.3.3), one can draw the relevant Feynman diagrams contributing to the multi-nucleon scattering amplitude at increasing orders in chiral EFT, see Fig. 6.3.1. The terms LO, NLO, N$^2$LO, N$^3$LO and N$^4$LO refer to the ChEFT orders $q^0$, $q^2$, $q^3$, $q^4$ and $q^5$, respectively. Notice that contributions at order $q^1$ are forbidden by parity conservation. However, the above classification of Feynman diagrams implies a perturbative nature of multi-nucleon scattering amplitudes, which is in contradiction with the empirical evidence. The key insight of Weinberg was the observation that certain contributions to the amplitude are enhanced beyond what is expected based on Eq. (6.3.3) [1626, 1627]. Consider, for example, the two-pion exchange planar box diagram (the last diagram in the second line of Fig. 6.3.1):

$$\mathcal{M} = i\int \frac{d^4 l_1}{(2\pi)^4} l_1^i l_1^j l_2^k l_2^l\, \hat O_{ijkl}\, \frac{i}{l_1^2 - M_\pi^2 + i\epsilon}\, \frac{i}{l_2^2 - M_\pi^2 + i\epsilon} \times \frac{2im}{(p_1 - l_1)^2 - m^2 + i\epsilon}\, \frac{2im}{(p_2 + l_1)^2 - m^2 + i\epsilon}\,, \tag{6.3.5}$$

where $p_1^\mu = (\sqrt{\vec p^{\,2} + m^2},\, \vec p\,)$ and $p_2^\mu = (\sqrt{\vec p^{\,2} + m^2},\, -\vec p\,)$ are the initial four-momenta of the nucleons, $l_1$ and $l_2 = p_1' - p_1 + l_1$ are pion momenta and we have used the relativistic rather than the strict HB expressions for the nucleon propagators for reasons to be given below. The spin-isospin operator $\hat O_{ijkl}$ with $i, j, k, l = 1, \ldots, 3$ emerges from four $\pi$N vertices $\propto g_A$ with $\Delta = 0$. Assuming $|\vec p\,|, l_1, l_2 \sim M_\pi \ll m$ and applying naive dimensional analysis (NDA) to the integrand in Eq. (6.3.5), the renormalized amplitude for the box diagram is expected to be of the order of $\mathcal{M} \sim M_\pi^2$ in agreement with the power counting formula in Eq. (6.3.3). On the other hand, performing the integration over $l_1^0$ using the residue theorem, one obtains

$$\mathcal{M} = \int \frac{d^3 l_1}{(2\pi)^3} \hat{O}_{ijkl} \left[ \frac{l_1^i l_1^j}{\omega_1^2} \frac{m}{\vec{p}^{\,2} - (\vec{p} - \vec{l}_1)^2 + i\epsilon} \frac{l_2^k l_2^l}{\omega_2^2} + \frac{\omega_1^2 + \omega_1\omega_2 + \omega_2^2}{2\omega_1^3 \omega_2^3 (\omega_1 + \omega_2)} l_1^i l_1^j l_2^k l_2^l + \mathcal{O}\left(\frac{1}{m}\right) \right], \tag{6.3.6}$$

where $\omega_i = \sqrt{\vec{l}_i^{\,2} + M_\pi^2}$ are the energies of the exchanged pions. Remarkably, the first term in the square brackets is enhanced by the factor $m/M_\pi$ compared to the power counting estimation. The origin of this enhancement can be traced back to the pinch singularity in the $m \to \infty$ limit [1626, 1627], which is why we used the relativistic expressions for the nucleon propagators[64]. Notice that infrared divergences of this kind do not appear in the single-baryon sector of ChPT.

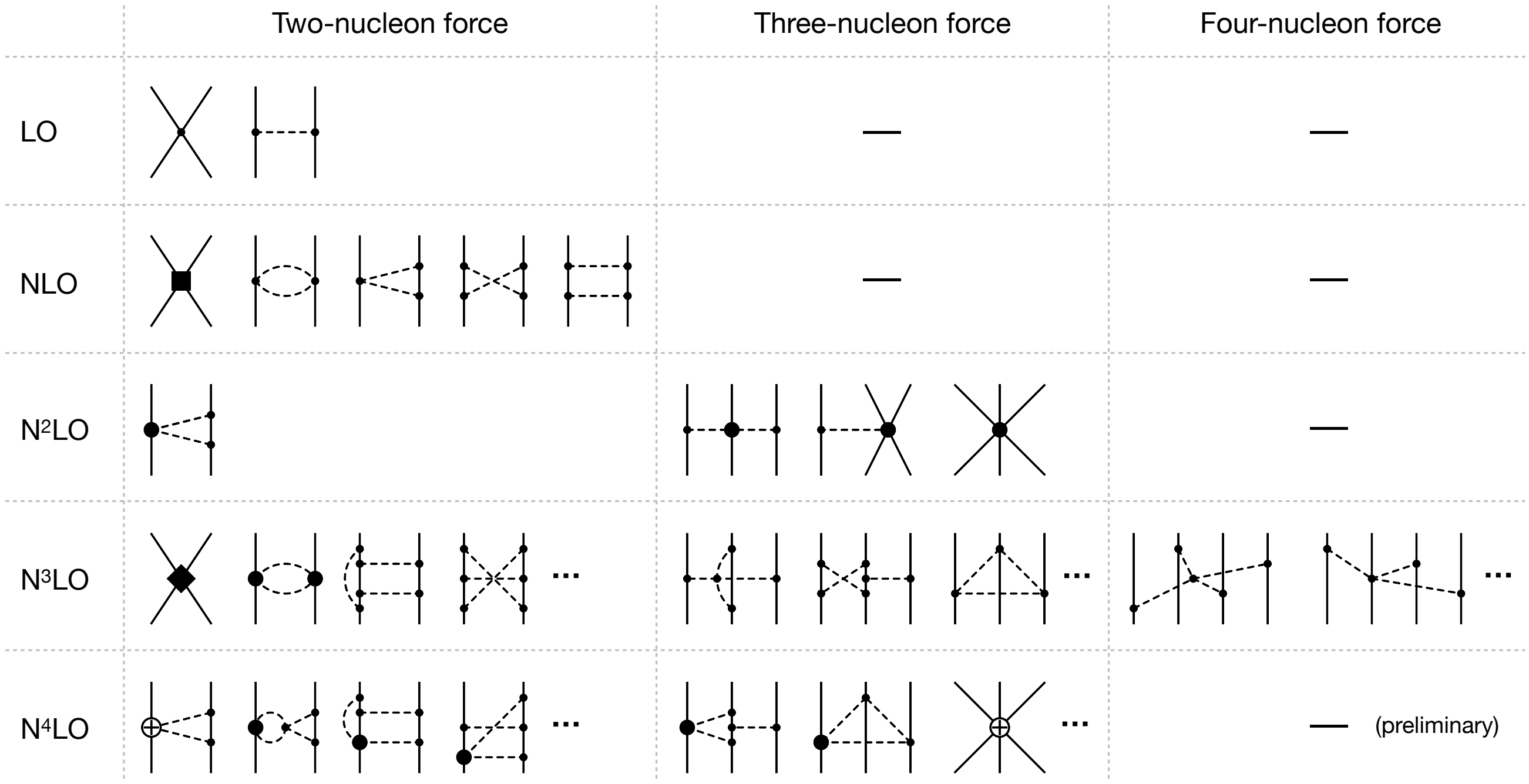

**Fig. 6.3.1** Diagrams contributing to the two-, three- and four-nucleon forces up to fifth order $\mathcal{O}(q^5)$ in ChEFT. Solid and dashed lines denote nucleons and pions, respectively. Solid dots, filled circles, filled squares, crossed circles and filled diamonds denote vertices with $\Delta = 0$, 1, 2, 3 and 4, respectively.

To identify all enhanced types of contributions to the amplitude it is useful to recall that performing the integration over $l_0$ leads to a decomposition of Feynman diagrams into a sum of diagrams emerging in old-fashioned time-ordered perturbation theory (TOPT). Indeed, the first (enhanced) term in the square brackets in Eq. (6.3.6) stems from two-nucleon-reducible TOPT diagrams which have an intermediate state involving two nucleons and no pions. Energy denominators associated with such purely nucleonic intermediate states of TOPT diagrams involve only nucleon kinetic energies $\sim M_\pi^2/m \ll M_\pi$ and are smaller than what is expected from NDA. This leads to the enhancement of reducible-type diagrams beyond the power counting estimation in Eq. (6.3.3)[65]. In contrast, the second term in the square brackets of Eq. (6.3.6) emerges from irreducible two-pion exchange diagrams with intermediate states involving at least one pion and results in the contribution $\mathcal{M} \sim M_\pi^2$ in agreement with Eq. (6.3.3).

In his seminal work [1626, 1627], Weinberg has argued that the breakdown of perturbation theory for the scattering amplitude in the few-nucleon sector of ChEFT can be traced back to the enhancement of reducible diagrams, which need to be resummed to all orders. He also noticed that ladder-type reducible TOPT diagrams automatically get resummed by solving the Lippmann-Schwinger-type integral equations for the amplitude

$$\mathcal{M} = V + V G_0 \mathcal{M} = V + V G_0 V + V G_0 V G_0 V + \ldots .$$

Indeed, the terms on the right-hand side of Eq. (6.3.6) can be easily identified with the iterated one-pion ex-change potential (OPEP) and the leading two-pion exchange potential (TPEP), $\mathcal{M} = V_{1\pi} G_0 V_{1\pi} + V_{2\pi} + \ldots$. Thus, low-energy processes involving several nucleons can be calculated in a systematically improvable way by applying ChPT to the kernel of the dynamical equation, defined as a sum of all possible few-nucleon-irreducible time-ordered diagrams, rather than to the scattering amplitude. The contributions to the nuclear forces depicted in Fig. 6.3.1 are to be understood as (sums of) the corresponding few-nucleon-irreducible time-ordered-like graphs rather than Feynman diagrams. Switching on external classical sources in the effective Lagrangian as explained in the previous section, the same framework can be used to derive nuclear current operators and to analyze low-energy electroweak processes (see the discussion below).

[64] This singularity is the basis of the covariant spectator theory discussed in Sec. 5.3.

[65] Reducible and irreducible diagrams also play a central role in the derivation of the Bethe-Salpeter equation; see Sec. 5.3.

It is worth emphasizing that the enhancement of reducible diagrams mentioned above is insufficient to justify the need for a non-perturbative resummation of the amplitude if one counts $m \sim \Lambda_b$ as done in ChPT. For example, the iterated OPEP contributes at order $V_{1\pi} G_0 V_{1\pi} \sim m M_\pi / \Lambda_b^2$ (assuming that all intermediate momenta are $\sim M_\pi$ after renormalization) and is thus suppressed relative to the tree-level term $V_{1\pi} = \mathcal{O}(1)$. To have a self-consistent non-perturbative approach, Weinberg proposed an alternative counting scheme for the nucleon mass by assigning $m \sim \Lambda_b^2 / M_\pi \gg \Lambda_b$, which is supported by the large-$N_c$ arguments given that $\Lambda_b \sim M_\rho = \mathcal{O}(1)$ while $m = \mathcal{O}(N_c)$. On the other hand, it is shown in Ref. [1628] that Weinberg's power counting can be realized via a suitable choice of renormalization conditions with no need to depart from the standard ChPT counting for the nucleon mass, see also Ref. [1629] for a related discussion.

Weinberg's power counting suggests that the LO potential stemming from the derivativeless contact interactions $\propto C_{S,T}$, see Eq. (6.3.2), and the OPEP as shown in Fig. 6.3.1 has to be iterated to all orders. For the contact interactions alone, the scattering amplitude resulting from solving the Lippmann-Schwinger (LS) equation can be calculated analytically and is renormalizable in the usual sense[66]. In contrast, iterations of the OPEP in spin-triplet channels lead to ultraviolet divergences whose cancellation requires counterterms with an increasing power of momenta. This feature, along with the numerical nature of the calculations in the presence of the OPEP, make renormalization of chiral EFT a complicated matter; see Ref. [1630] for a collection of perspectives.

[66] That is, all ultraviolet divergences emerging from the iterations of the LS equation can be absorbed into a redefinition of $C_{S,T}$.

Notice that the existence of shallow bound states alone does not necessarily imply a nonperturbative nature of the OPEP, but merely indicates a fine tuning of the LECs $C_{S,T}$ [1627, 1631]. An alternative approach based on a perturbative treatment of the OPEP was proposed by Kaplan, Savage and Wise (KSW) in the late nineties of the last century [1632, 1633]. This framework allows one to compute the NN scattering amplitude analytically and to implement the renormalization program in a straightforward way with no need to introduce a finite cutoff. However, extensive calculations performed in the KSW approach have revealed poor convergence (at least) in certain spin-triplet channels [1634, 1635], see also [1636–1638] for a related discussion, indicating that the OPEP should indeed be treated nonperturbatively in low partial waves.

The most advanced applications of chiral EFT to nuclear systems are carried out utilizing the finite-cutoff formulation of Ref. [1639]. In essence, it amounts to solving the quantum-mechanical $A$-body problem using the nuclear potentials calculated in ChPT and regularized with some finite cutoff $\Lambda$ taken of the order of $\Lambda \sim \Lambda_b$. The calculated scattering amplitudes are implicitly renormalized by tuning the bare LECs $C_S(\Lambda)$, $C_T(\Lambda)$, etc., of multi-nucleon vertices to low-energy observables. The resulting (renormalized) scattering amplitudes depend on the physical parameters and the cutoff $\Lambda$. The residual $\Lambda$-dependence of the calculated observables is expected to introduce an uncertainty beyond the order one is working at and offers a nontrivial *a posteriori* consistency check. For more details on the foundations and applications of the finite-cutoff formulation of chiral EFT see Refs. [1639, 1640]. Finally, a first step towards a formal renormalizability proof of the finite-cutoff scheme to all orders in the iterated OPEP using the Bogoliubov-Parasiuk-Hepp-Zimmermann (BPHZ) subtraction technique can be found in Ref. [1641].

### 6.3.2 Nuclear interactions from ChEFT

In ChPT, the S-matrix is usually obtained by applying the Feynman graph technique to the effective chiral Lagrangian. To derive nuclear forces, it is more natural and convenient to employ non-covariant old-fashioned perturbation theory as already mentioned above. This approach is based on the Hamiltonian rather than Lagrangian, so the first step amounts to using the canonical formalism for constructing the Hamiltonian $H = H_0 + H_I$ for interacting pions and nucleons from the effective chiral Lagrangian [1626, 1627]. The NN scattering amplitude between the initial and final states $|i\rangle$ and $|f\rangle$, respectively, can be written as

$$\langle f|\mathcal{M}|i\rangle = \langle f|H_I \sum_{n=0}^{\infty} \left( \frac{1}{E_i - H_0 + i\epsilon} H_I \right)^n |i\rangle \,, \qquad (6.3.7)$$

where $E_i$ is the energy of the nucleons in the state $|i\rangle$. Notice that the intermediate states in the above equation include both pions and nucleons. Let $\eta$ and $\lambda$ denote the projection operators on the purely nucleonic subspace and the rest of the Fock space, respectively. Eq. (6.3.7) can be cast into the form of the LS equation

$$\langle f|\mathcal{M}|i\rangle = \langle f|V \sum_{n=0}^{\infty} \left( \frac{\eta}{E_i - H_0 + i\epsilon} V \right)^n |i\rangle \,, \qquad (6.3.8)$$

where the potential $V$ can e.g. be chosen in the energy-dependent form as done in Refs. [1626, 1627, 1642, 1643]:

$$V(E_i) = \eta H_I \sum_{n=0}^{\infty} \left( \frac{\lambda}{E_i - H_0 + i\epsilon} H_I \right)^n \eta \,. \qquad (6.3.9)$$

The explicit energy dependence of $V$ is a higher-order effect, see e.g. Eq. (6.3.6), and can be eliminated yielding an energy independent hermitian NN potential. The method can be applied to many-body forces and has also been used to derive nuclear currents starting from the effective Lagrangian with external sources.

It is important to keep in mind that nuclear potentials, in contrast to the on-shell amplitude $\langle f|\mathcal{M}|i\rangle$, are not directly observable and represent scheme-dependent quantities. This intrinsic ambiguity reflects the arbitrariness in making off-shell extensions of the scattering amplitude. Clearly, such off-shell ambiguities cannot lead to measurable effects. Being a quantum-field-theory-based method, chiral EFT by construction maintains consistency between many-body interactions and current operators and ensures that calculated observables are independent of the off-shell ambiguities (up to higher-order corrections).

The method of deriving nuclear forces and currents by matching to the scattering amplitude as outlined above was used e.g. in Refs. [1644–1648] and is usually referred to as TOPT. Another closely related approach amounts to block-diagonalizing the pion-nucleon Hamiltonian via a suitable unitary transformation [1649]

$$H \to H' = U^\dagger H U = \begin{pmatrix} \eta H' \eta & 0 \\ 0 & \lambda H' \lambda \end{pmatrix} . \qquad (6.3.10)$$

Both the unitary operator $U$ and the nuclear potential $V = \eta(H' - H_0)\eta$ are calculated perturbatively using the standard power counting of ChPT as explained in Ref. [1650]. The method of unitary transformation (MUT) to derive nuclear forces and currents was applied e.g. in Refs [1651–1659]. A pedagogical discussion of methods outlined above can be found in Ref. [1660].

So far, we have left out renormalization of nuclear potentials. In contrast to the scattering amplitude, renormalizability of nuclear forces and currents derived in ChPT is not guaranteed by construction and was shown to impose severe constraints on their off-shell behavior starting from N$^3$LO [1650, 1655, 1657–1659, 1661].

Having introduced various methods to derive nuclear potentials from the effective chiral Lagrangian, we are now in the position to discuss the ChEFT expansion of the long-range NN force. The one- and two-pion exchange contributions up to N$^2$LO depend solely on the momentum transfer $\vec{q}$ and are, therefore, local. The resulting potentials have a clear and intuitive interpretation in coordinate space. Using the decomposition

$$\begin{aligned} V(\vec{r}) &= V_C(r) + V_S(r)\vec{\sigma}_1 \cdot \vec{\sigma}_2 + V_T(r) S_{12} \qquad (6.3.11) \\ &\quad + \big[W_C(r) + W_S(r)\vec{\sigma}_1 \cdot \vec{\sigma}_2 + W_T(r) S_{12}\big] \vec{\tau}_1 \cdot \vec{\tau}_2 \,, \end{aligned}$$

where $S_{12} = 3\vec{\sigma}_1 \cdot \hat{r}\vec{\sigma}_2 \cdot \hat{r} - \vec{\sigma}_1 \cdot \vec{\sigma}_2$ is the tensor operator while $\vec{\tau}_i$ refer to the isospin Pauli matrices of the nucleon $i$, the LO contribution due to the OPEP is given by

$$\begin{aligned} W^{(0)}_{T,\,1\pi}(r) &= \frac{g_A^2}{48\pi F_\pi^2} \frac{e^{-x}}{r^3} (3 + 3x + x^2) \,, \\ W^{(0)}_{S,\,1\pi}(r) &= \frac{g_A^2 M_\pi^2}{48\pi F_\pi^2} \frac{e^{-x}}{r} \,, \qquad (6.3.12) \end{aligned}$$

where the superscript of the potentials gives the ChEFT order. Further, $x \equiv M_\pi r$ while $g_A$ and $F_\pi$ denote the physical values of the nucleon axial-vector coupling and pion decay constant, respectively. Notice that only the $W^{(0)}_{T,\,1\pi}(r) \propto r^{-3}$ part of the tensor potential survives in the chiral limit of $M_\pi \to 0$. It is precisely this singular interaction that leads to the already mentioned non-renormalizability of the OPEP in all spin-triplet channels of NN scattering. The NLO contributions to the long-range NN interaction stem from the TPEP and are given by [1642, 1644, 1649, 1662]:

$$\begin{aligned} W^{(2)}_{C,\,2\pi}(r) &= \frac{M_\pi}{128\pi^3 F_\pi^4} \frac{1}{r^4} \big\{ K_1(2x) \\ &\quad \times \big[1 + 2g_A^2(5 + 2x^2) - g_A^4(23 + 12x^2)\big] \\ &\quad + xK_0(2x)\big[1 + 10g_A^2 - g_A^4(23 + 4x^2)\big]\big\} , \\ V^{(2)}_{T,\,2\pi}(r) &= -\frac{g_A^4 M_\pi}{128\pi^3 F_\pi^4} \frac{1}{r^4} \\ &\quad \times \big\{12xK_0(2x) + (15 + 4x^2)K_1(2x)\big\} , \\ V^{(2)}_{S,\,2\pi}(r) &= \frac{g_A^4 M_\pi}{32\pi^3 F_\pi^4} \frac{1}{r^4} \\ &\quad \times \big\{3xK_0(2x) + (3 + 2x^2)K_1(2x)\big\} , \qquad (6.3.13) \end{aligned}$$

where $K_{0,1}(x)$ denote the modified Bessel functions. To arrive at these expressions, one first needs to evaluate the three-dimensional loop integrals for the corresponding TOPT diagrams[67] using e.g. dimensional regularization. The resulting $p$-space potentials cannot be Fourier transformed to $r$-space directly since the Fourier integrals diverge at high momenta. Eq. (6.3.13) is obtained by Fourier transforming the regularized momentum-space potentials and subsequently removing the regulator.

Similarly, at N$^2$LO, the TPEP receives contributions given by [1644, 1649]

$$V_{C,2\pi}^{(3)}(r) = \frac{3g_A^2}{32\pi^2 F_\pi^4}\frac{e^{-2x}}{r^6}\big\{2c_1x^2(1+x)^2 + c_3(6+12x+10x^2+4x^3+x^4)\big\},$$
$$W_{T,2\pi}^{(3)}(r) = -\frac{g_A^2 c_4}{48\pi^2 F_\pi^4}\frac{e^{-2x}}{r^6}(1+x)(3+3x+x^2),$$
$$W_{S,2\pi}^{(3)}(r) = \frac{g_A^2 c_4}{48\pi^2 F_\pi^4}\frac{e^{-2x}}{r^6}(1+x)(3+3x+2x^2) \quad (6.3.14)$$

where $c_i$ are LECs accompanying the subleading $\pi\pi$NN vertices with $\Delta = 1$.

The expressions for the OPEP and TPEP, Eqs. (6.3.12) to (6.3.14), illustrate the general features of the chiral expansion of the long-range nuclear interactions:

- The chiral expansion of the $N$-pion exchange potential generally corresponds to the expansion in powers of $M_\pi/\Lambda_\chi$, where the chiral symmetry breaking scale $\Lambda_\chi$ is given by $4\pi F_\pi$ and/or the scale that governs the $\pi$N LECs starting from the subleading ones. The expansion pattern is the same as for ChPT in the meson and single-baryon sectors. The chiral expansion for $V(\vec{r})$ is expected to converge at distances $r \gtrsim 1/M_\pi$ and larger. In contrast, at short distances $r \ll 1/M_\pi$, the expansion diverges yielding highly singular van der Waals-like behaviour $V_{N\pi}^{(s)}(\vec{r}) \sim 1/r^{3+s}$; see also Ref. [1663] for further insights and examples. In the finite-cutoff formulation of chiral EFT, this unphysical short-distance behavior is removed by the regulator.
- Since all relevant $\pi$N LECs can nowadays be reliably determined from the pion-nucleon scattering amplitude in the subthreshold region, obtained from the dispersive Roy-Steiner-equation analysis [1664–1666], ChEFT yields parameter-free predictions for the long-range behavior of the nuclear forces and currents. These predictions are model-independent and represent non-trivial manifestations of the spontaneously broken chiral symmetry of QCD.
- Eqs. (6.3.12)-(6.3.14) also point towards some limitations of ChPT, which relies on NDA and cannot capture possible enhancements due to large dimensionless prefactors. In the NN sector, this especially affects the N$^2$LO contributions to the TPEP. The corresponding triangle diagram, see Fig. 6.3.1, leads to the contribution enhanced by a factor of $4\pi$ relative to what is expected based on the power counting, so that $\Lambda_\chi$ is in this case better estimated as $\Lambda_\chi \sim \sqrt{4\pi}F_\pi$ than $\Lambda_\chi \sim 4\pi F_\pi$. Enhancements of this kind are also not uncommon in the single-nucleon sector of ChPT. For the subleading central potential $V_{C,2\pi}^{(3)}(r)$, this enhancement combines with the large numerical coefficients and a large value of the LEC $c_3$ driven by the intermediate $\Delta(1232)$ excitation [1667]. Altogether, this results in $V_{C,2\pi}^{(3)}(r)$ being by far the dominant TPE component, whose strength is comparable to that of the OPEP even at $r \sim 2$ fm. The strongly attractive nature of the isoscalar central potential at intermediate distances is supported by phenomenology and often attributed to the $\sigma$-meson exchange in traditional nuclear physics jargon. The chiral expansion of the TPEP has been extended to N$^4$LO [1668–1670] and even beyond and was shown to yield converged results [1671, 1672].

The current status of the derivation of nuclear potentials in ChEFT is visualized in Fig. 6.3.1[68], see Refs. [1673, 1674] for comprehensive review articles. On the qualitative level, ChEFT provides a justification of the observed hierarchy of nuclear forces with $V_{2\mathrm{N}} \gg V_{3\mathrm{N}} \gg V_{4\mathrm{N}} \gg \ldots$ [1626, 1627].

The leading contributions to the three-nucleon force (3NF) at N$^2$LO have been known for a long time [1675, 1676]. The expressions for the N$^3$LO and (most of the) N$^4$LO corrections have been worked out in Refs. [1651–1654, 1677–1679]. The four-nucleon force is further suppressed relative to the 3NF and appears first at N$^3$LO [1650, 1661]. Isospin-breaking as well as parity- and time-reversal-violating nuclear potentials have also been worked out, see Refs. [1673, 1680] and references therein.

The first application of ChEFT to study nuclear current operators goes back to the pioneering papers by Park et al. [1681, 1682]. In the past decade, the vector [1645–1647, 1655, 1656, 1658], axial-vector [1648, 1657], pseudoscalar [1657] and scalar [1659, 1683] current operators have been worked out to the leading one-loop-order accuracy for the two-body contributions (i.e., to N$^3$LO using the counting scheme with $m \sim \Lambda_b^2/M_\pi$). As an example, the ChEFT expansion of the electro-magnetic nuclear currents is shown in Fig. 6.3.2. Similarly to the case of the nuclear forces, the chiral power counting leads, in general, to a suppression of many-body operators. On the other hand, the leading contributions to the single- and two-nucleon current density both appear at NLO. In contrast, the exchange charge density contributions are strongly suppressed relative to the LO term (the charge operator of the nucleon), with both two- and three-nucleon contributions appearing at N$^3$LO. A comprehensive review of nuclear currents in ChEFT, including a detailed comparison of results obtained by different groups and a thorough discussion of the differences between them, can be found in Ref. [1684].

[67] E.g., the second term in the square brackets in Eq. (6.3.6) gives the TPEP $\propto g_A^4$ stemming from the last diagram in the second row of Fig. 6.3.1 (planar box diagram).

[68] In some approaches, NN contact interactions are promoted to orders different than those derived by NDA.

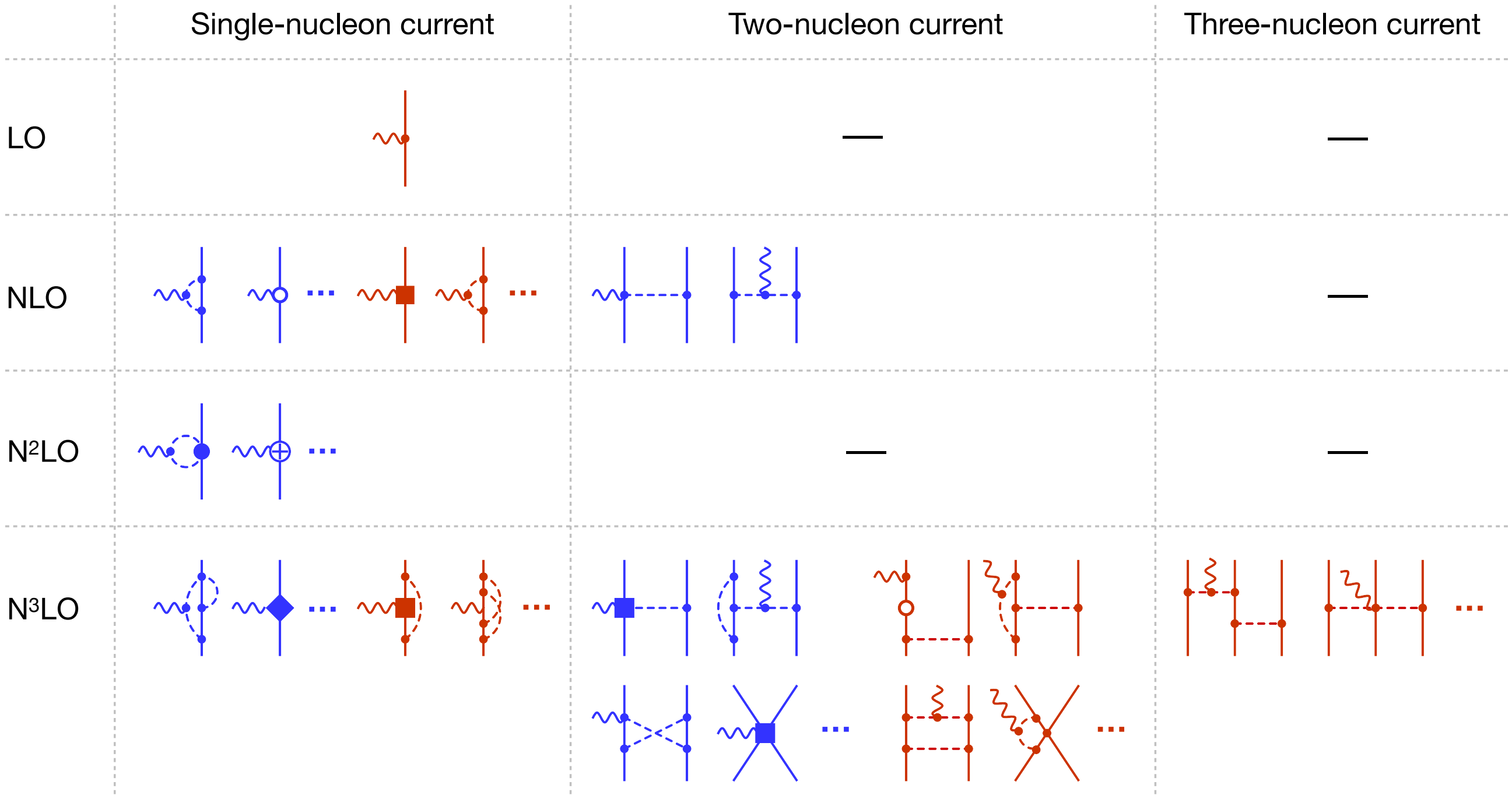

**Fig. 6.3.2** Diagrams contributing to the single-, two- and three-nucleon electromagnetic current operators at lowest orders of chiral EFT using the counting scheme with $m \sim \Lambda_b^2/M_\pi$. Wiggly lines denote photons. Blue and red diagrams depict the contributions to the current and charge densities, respectively. An open circle shows an insertion of the kinetic energy term with $\Delta = 2$. For remaining notations see Fig. 6.3.1.

All results described above are based on the effective chiral Lagrangian involving pions and nucleons as the only explicit DoF. As already emphasized in the previous section, given the low excitation energy of the $\Delta$-resonance and its strong coupling to the $\pi$N system, it might be advantageous to also treat the $\Delta$ DoF as dynamic. This formulation of ChEFT was already applied to derive the NN force and most of the 3NF contributions up through N$^3$LO [1643, 1685–1688]. The explicit treatment of the $\Delta$ leads to a reshuffling of certain contributions to lower orders in the EFT expansion. In particular, a part of the unnaturally strong N$^2$LO TPEP is shifted to NLO, and the LECs $c_{3,4}$ take more natural numerical values [1666]. These results indeed support the expected better convergence pattern of ChEFT with explicit $\Delta$ DoF.

Last but not least, ChEFT has also been extended to the SU(3) sector and applied to study the interactions between nucleons and hyperons, see e.g. Refs. [1689–1691] and Ref. [1692] for a recent review article.

### 6.3.3 Applications

As already pointed out, nuclear interactions derived in ChEFT are singular at short distances and need to be regularized prior to solving the dynamical equation. A broad range of regulators featuring different functional dependence on momenta and relative distances have been proposed in the literature, see Refs. [1672, 1693–1697] for some examples and Ref. [1698] for a related discussion. For the long-range OPEP and TPEP, it is advantageous to use a local regularization in order to preserve the analytic structure of the amplitude [1672, 1696]. For short-range terms, angle-independent non-local regulators maintain a one-to-one correspondence between the plane-wave and partial-wave bases, which simplifies the determination of the corresponding LECs. This choice is utilized in both available N$^4$LO implementations of the NN potentials [1672, 1699] which, however, differ in their way of regularizing the long-range terms. In both cases, the LECs accompanying the NN short-range interactions were determined solely from the neutron-proton and proton-proton data. Alternative fitting strategies, which include information about light and medium-mass nuclei and even nuclear matter, are also being explored [1700].

The very accurate and precise NN potentials of [1672, 1701], derived in chiral EFT with pions and nucleons as the only active DoF, provide an outstanding description of NN data up to the pion production threshold[69]. In fact, the results of Ref. [1701] comprise a full-fledged partial wave analysis of NN scattering data based solely on chiral EFT. For more details and comparison between different NN potentials see Ref. [1702].

To give an impression about the convergence pattern of ChEFT consider the total cross section for neutron-proton scattering at $E_{\rm lab} = 100$ MeV as a representative example. Using the potentials from Ref. [1701] one obtains for the cutoff $\Lambda = 450$ MeV (in mb)

$$\sigma_{\rm tot} = 84.0_{[q^0]} - 10.2_{[q^2]} + 0.4_{[q^3]} - 0.4_{[q^4]} + 0.6_{[q^5]} - 0.0_{[q^6]},$$

where the last term gives the contribution of the order-$q^6$ F-wave contact interactions. Given that the expansion parameter is $q = p_{\rm cms}/\Lambda_b \sim 1/3$, where we have used $\Lambda_b = 650$ MeV [1696, 1703, 1704], one observes that the order-$q^3$ and $q^4$ contributions appear to be smaller, while the order-$q^5$ correction is somewhat larger than naively expected. The truncation error of the calculated value can be estimated using a Bayesian approach by inferring the information about the convergence pattern of the ChEFT from the results at all available orders [1703]; see also Ref. [1696] for a related earlier work. Using the Bayesian model from Ref. [1705], the N$^4$LO truncation error for the case at hand is estimated to be $\delta\sigma_{\rm tot} = 0.14$ mb at 68% confidence level. The final result then reads $\sigma_{\rm tot} = 74.35(14)(17)(1)$ mb, where the last two errors refer to the statistical error and uncertainty in the $\pi$N LECs.

The sub-percent accuracy level of ChEFT has also been reached for other low-energy observables in the NN sector [1702]. In particular, the charge and quadrupole form factors of the deuteron were analyzed to N$^4$LO in Refs. [1706, 1707]. The predicted value for the deuteron structure radius, $r_{str} = 1.9729^{+0.0015}_{-0.0012}$ fm, was used, in combination with the very precise measurement of the charge radius difference between $^2$H and the proton [1708], to determine the neutron radius. The obtained value of the quadrupole moment $Q_d = 0.2854^{+0.0038}_{-0.0017}$ fm$^2$ [1707] is in a very good agreement with the spectroscopy determination $Q_d = 0.285699(15)(18)$ fm$^2$ [1709].

The spontaneously broken approximate chiral symmetry of QCD, together with the experimental information about the $\pi$N system, allow one to predict the long-range behavior of the nuclear forces. In the NN sector, these predictions have been verified from experimental data. For example, the only order-$q^3$ contribution to the NN force comes from the TPEP in Eqs. (6.3.14) (since the contact interactions contribute at orders $q^{2i}$, $i = 0, 1, 2, \ldots$). Adding these parameter-free contributions to the potential was demonstrated to very significantly improve the description of the data [1696, 1710, 1711]. A similar improvement is observed by adding the order-$q^5$ TPEP [1670, 1672, 1712]. It is also worth mentioning that the potentials of [1672] achieve a comparable precision to that of the available high-precision phenomenological potentials while having a much smaller number of adjustable parameters[70] This is yet another evidence of the important role played by chiral symmetry. Finally, the convergence of the chiral EFT expansion can be further improved by the inclusion of $\Delta$'s as explicit DoF of the theory. This is supported by the recently developed Norfolk chiral many-body interactions [1713]; see also Ref. [1714] for a related discussion.

Beyond the two-nucleon system, the results are presently limited to the N$^2$LO accuracy level due to the lack of *consistently regularized* many-body interactions and exchange currents starting from N$^3$LO. As discussed in Refs. [1640, 1684, 1702], using dimensional regularization in the derivation of nuclear interactions in combination with a cutoff regularization of the Schrödinger equation leads, in general, to violations of chiral symmetry. This issue affects all loop contributions to the 3NF and exchange current operators, which therefore need to be re-derived using symmetry-preserving cutoff regularization.

At the N$^2$LO level, the results for three-nucleon scattering observables [1705, 1715–1717] and the spectra of light- and medium-mass nuclei [1715, 1717–1724] are mostly consistent with experimental data within errors; see also Refs. [1725, 1726] for review articles. As a representative example, we show in Fig. 6.3.3 the calculated ground state energies of p-shell nuclei from Ref. [1715].

ChEFT interactions and associated currents have been vigorously utilized in the past ten years to study both static and dynamical electroweak properties of nuclei, including electromagnetic form factors[860, 1707, 1728], electromagnetic moments[1728–1730], electroweak decays[1731, 1732], and low-energy reactions such as electroweak captures[1733, 1734]. ChEFT currents were first used in calculations of nuclei with $A > 3$ in Ref. [1735] where they are used to study magnetic moments and electromagnetic transitions in $A \leq 10$ systems. Two-body currents were found to improve the agreement between experimental data and theoretical calculations. For example, a long standing under-prediction [1736] of the measured $^9$C magnetic moment by less sophisticated theoretical calculations is explained by the $\sim 40\%$ correction generated by two-body electromagnetic currents in Ref. [1735]. This enhancement can be appreciated in Fig. 6.3.4 by comparing blue dots (representing calculations based on the single nucleon paradigm) and red diamonds (representing calculations with two-body electromagnetic currents).

[69] This requires the inclusion of four order-$q^6$ contact interactions that contribute to F-waves [1672, 1699].

[70] The N$^4$LO potentials of [1672] depend on 27 LECs fitted to NN data, while the realistic potentials typically involve 40-50 adjustable parameters.

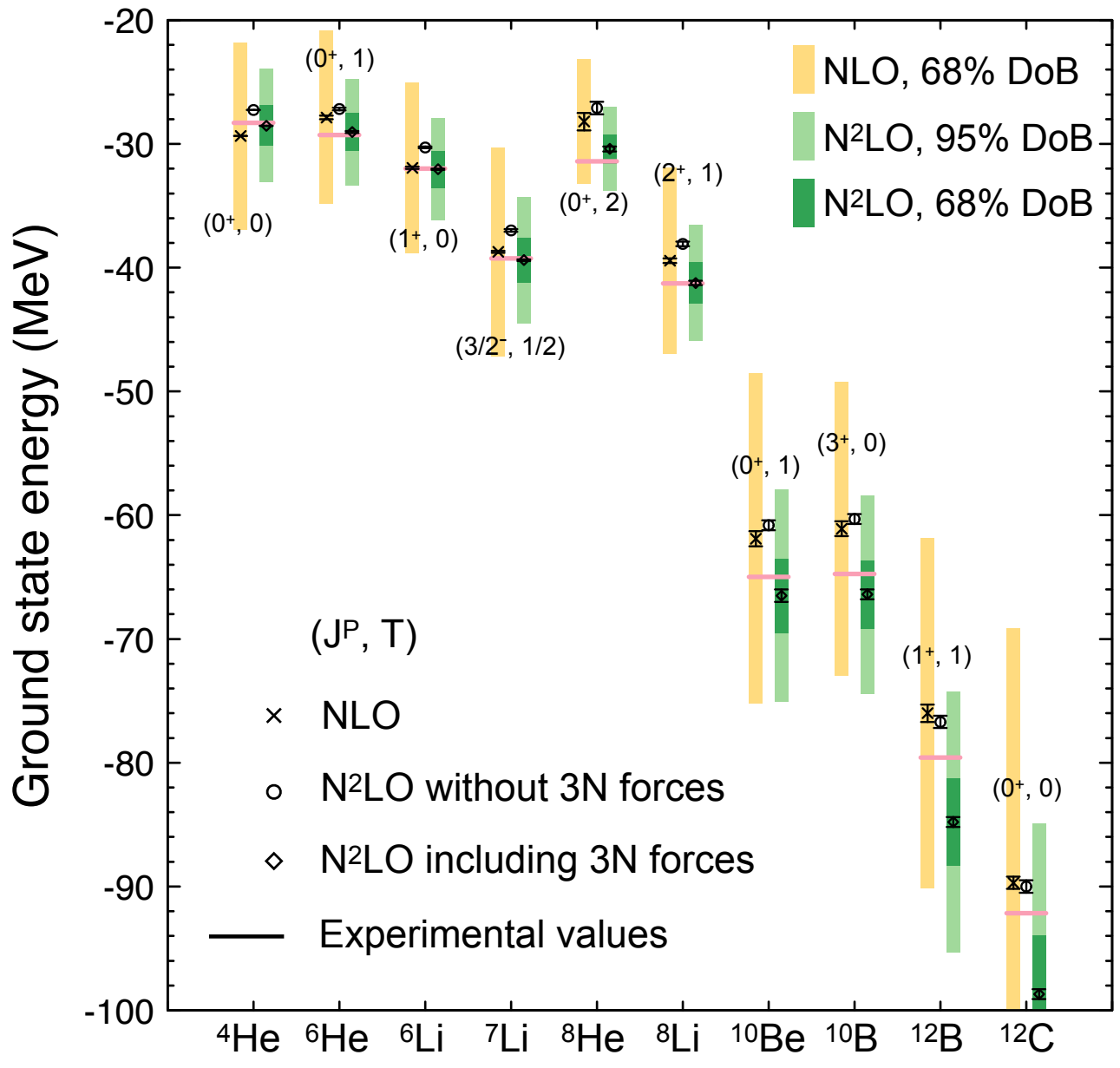

**Fig. 6.3.3** Predictions for ground state energies of selected p-shell nuclei at NLO and N$^2$LO using the chiral EFT NN potentials from Ref. [1672] together with the consistently regularized 3NF for $\Lambda = 450$ MeV. Black error bars indicate the uncertainties from the employed many-body method, while shaded bars refer to the EFT truncation errors (not shown for incomplete N$^2$LO calculations based on the NN force only). Figure adapted from Ref. [1715].

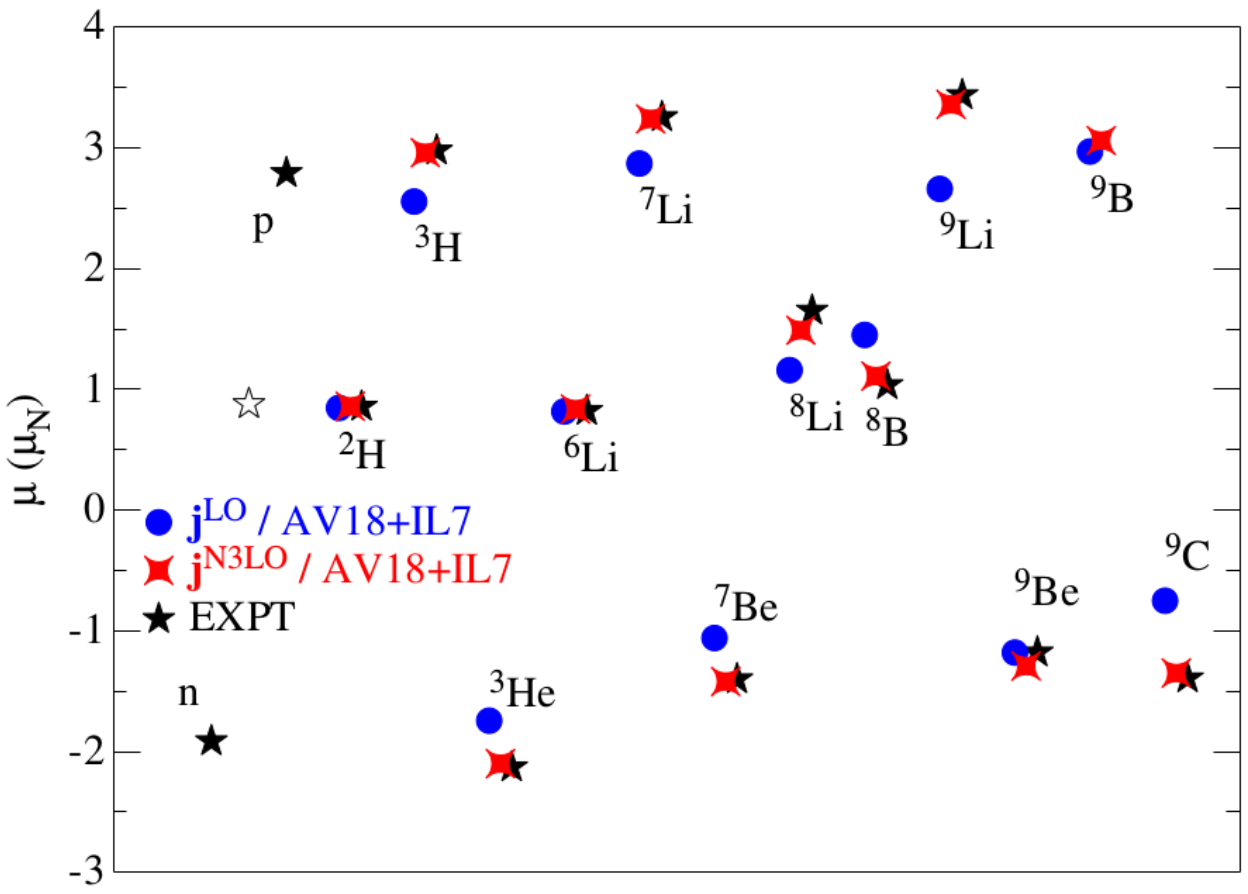

**Fig. 6.3.4** Magnetic moments in nuclear magnetons for $A \leq 9$ nuclei from Ref. [1727]. Black stars indicate the experimental values while blue dots (red diamonds) represent Green's Function Monte Carlo calculations which include the LO one-body currents (one-body plus two-body currents at N3LO) from ChEFT. For more details and references to the experimental data see [1727].

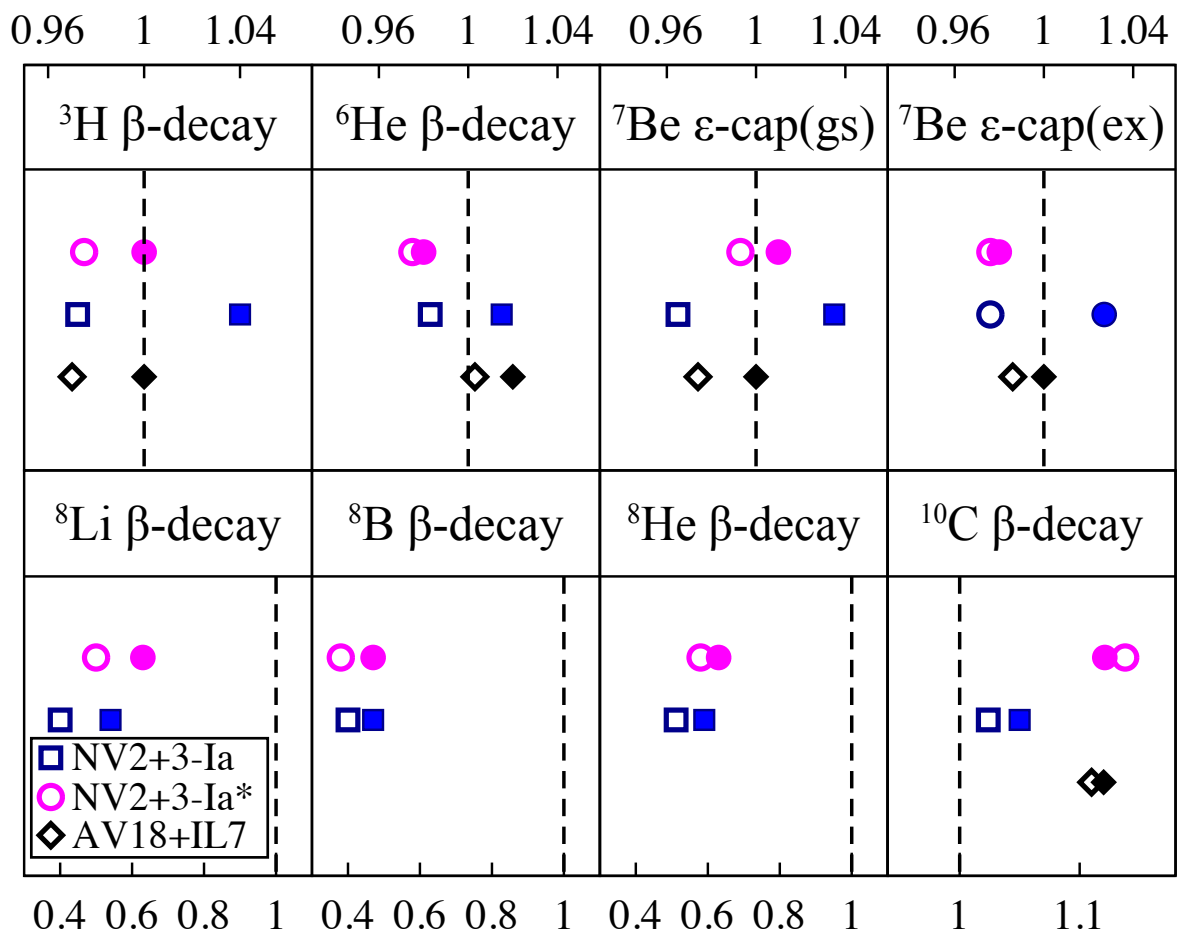

**Fig. 6.3.5** Ratios of Green's function Monte Carlo calculations to experimental values of the Gamow-Teller reduced matrix elements in the $^3$H, $^6$He, $^7$Be, $^8$B, $^8$Be, $^8$He and $^{10}$C weak transitions from Refs. [1732, 1737]. Theory predictions correspond to the ChEFT axial current at LO (empty symbols) and up to N3LO (filled symbols).

Axial currents are tested primarily in beta decays and electron capture processes for which data are readily available and known for the most part with great accuracy. The long-standing problem of the systematic over-prediction of Gamow-Teller beta decay matrix elements [1738] in simplified nuclear calculations, also known as the '$g_A$ problem', has been recently addressed by several groups [1732, 1737, 1739]. The authors of Refs. [1732, 1737] calculated the Gamow-Teller matrix elements in $A = 6-10$ nuclei accounting systematically for many-body effects in nuclear interactions and coupling to the axial current, both derived in ChEFT. The agreement of the calculations with the data is excellent for $A = 3, 6$ and 7 systems, with two-body currents providing a small ($\sim 2\%$) contribution to the matrix elements. Decays in the $A = 8$ and 10 systems, instead, require further developments of the nuclear wave functions [1737, 1739]. The '$g_A$-problem' can be resolved in light nuclei largely by correlation effects in the nuclear wave functions. A summary of these calculations is reported in Fig. 6.3.5. Similar results for these light nuclei obtained using the No-core shell model are reported in Ref. [1739].

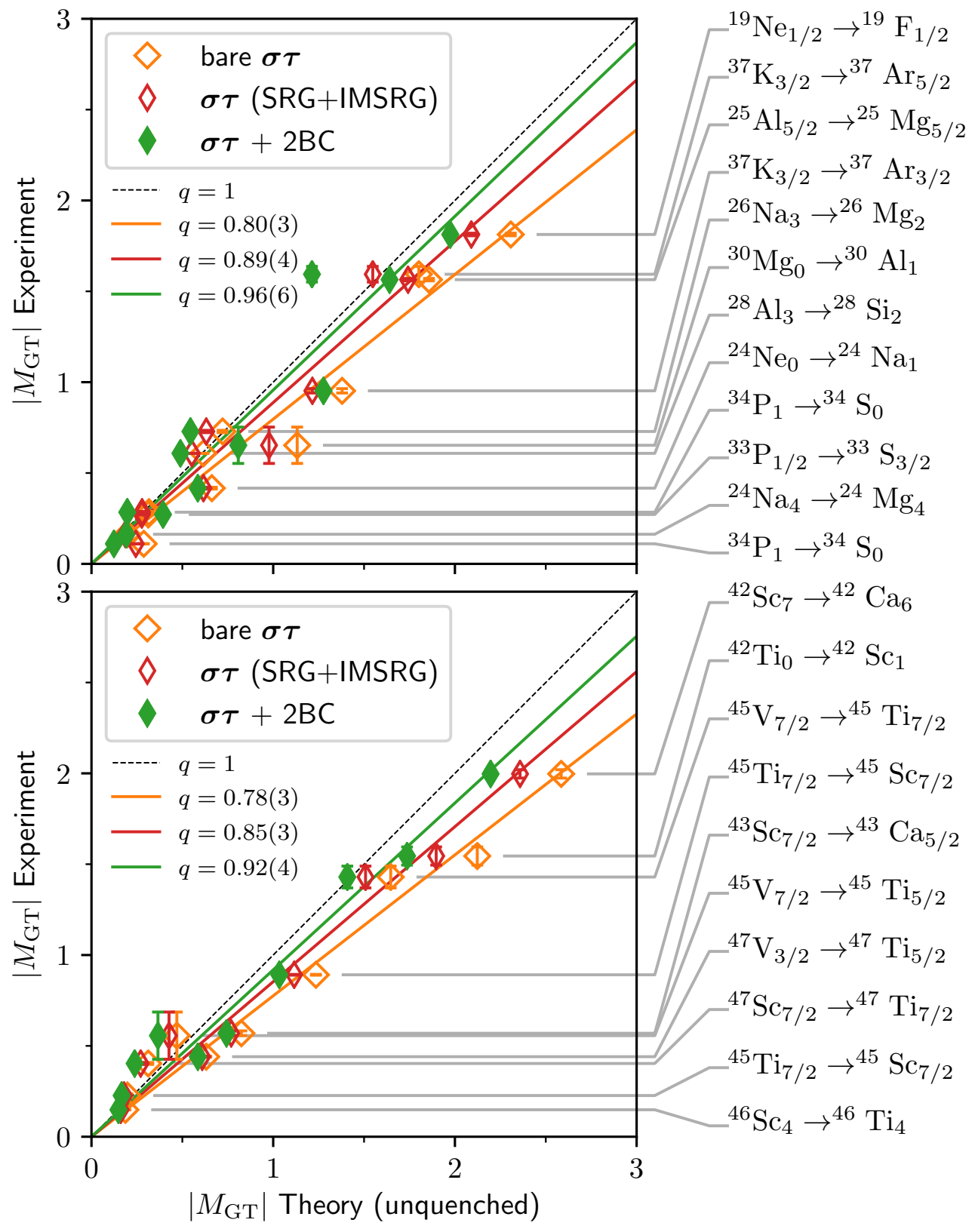

**Fig. 6.3.6** Comparison of experimental ($y$-axis) and theoretical ($x$-axis) Gamow-Teller matrix elements for medium-mass nuclei. The theoretical results were obtained using (i) a bare Gamow-Teller one-body operator, (ii) Gamow-Teller one-body operator consistently evolved with the Hamiltonian [1739], and (iii) a consistently-evolved Gamow-Teller operator that includes both one- and two-body currents. See Ref. [1739] for details.

The ChEFT approach is also being implemented in studies of medium-mass nuclei [1739]. As a representative of this class of electroweak calculations we show the results of Ref. [1739] on beta decay matrix elements visualized in Fig. 6.3.6. Here, the authors demonstrate that the quenching in the nuclear matrix elements arises primarily from ChEFT axial two-body currents and strong correlations in the nucleus. Nuclei from $A = 3$ to $^{100}$Sn are calculated based on ChEFT in agreement with experimental data.

To summarize, there has been exceptional progress in studying nuclear physics using ChEFT. In the last two decades this framework, rooted in the symmetries of QCD and their breaking pattern, has allowed for the calculation of many low-energy nuclear processes, such as electromagnetic reactions and $\beta$ decays in both light and medium-mass nuclei, has reached a remarkable agreement with experiment, and has contributed to solving long-standing anomalies in nuclear theory. As chiral interactions and currents are being refined and pushed to higher orders, we have entered the precision era of this powerful framework.

### 6.3.4 Connections to lattice QCD

Lattice QCD (LQCD) offers a first-principles approach to study hadronic and nuclear systems. Several LQCD groups have studied baryon-baryon systems as well as light (hyper-) nuclei at unphysically heavy pion masses using different methods. For non-strange nuclear systems, the current status of LQCD remains controversial, see [1740] for a review. On the EFT side, efforts concentrated on extrapolating lattice QCD results as follows:

– *Chiral extrapolations* of few-nucleon observables have been studied using a variety of ChEFT formulations, see e.g. Refs. [1741–1746]. Currently, the main limiting factor for constraining the quark mass dependence of the nuclear interactions is the lack of reliable LQCD results for not-too-heavy quark masses within the applicability domain of ChEFT.
– *Extrapolations of the NN scattering amplitude in energy* at fixed values of the quark masses were performed [1747, 1748] by exploiting the knowledge of the longest-range interaction due to the OPEP.
– *Infinite-volume extrapolations* of LQCD results for heavy pion masses were carried out in both pion-less [1749–1751] and chiral [1752] EFT.
– Finally, *extrapolations of LQCD results to heavier systems* were considered in Ref. [1753] using the framework of pion-less EFT and in Ref. [1754] utilizing a discretized formulation of ChEFT.

These studies demonstrate remarkable synergy between LQCD and EFT. In the future, LQCD is expected to provide valuable input for EFT calculations of systems and processes where scarce experimental data exist such as e.g. strange multi-baryon systems and nuclear matrix elements for BSM searches [1740].

### 6.3.5 Challenges and outlook

To summarize, ChEFT has revolutionized the field of nuclear physics over the past three decades by providing a systematically improvable and theoretically well founded approach to low-energy nuclear interactions, which relies on the symmetries of QCD (and their breaking pattern). The method has proven to be phenomenologically successful and has led to new research directions such as e.g. nuclear lattice simulations [1755–1757]. In the two-nucleon sector, ChEFT has already reached maturity to become a precision tool.

One of the most pressing remaining challenges is the development of accurate and precise three-nucleon interactions needed to shed light onto the long-standing discrepancies in the three-nucleon continuum [1725]. Pushing the ChEFT expansion for many-body forces and exchange currents to N$^3$LO and beyond calls for a symmetry preserving regularization [1702], and it will also require new ideas to overcome computational challenges related to the determination of LECs; see Refs. [1758–1760] for recent steps along these lines. Other frontiers include the derivation of consistently regularized electroweak currents, better understanding of renormalization in ChEFT, precision studies of nuclear structure, reactions and the equation of state of nuclear matter as well as applications to searches for BSM physics in processes involving nuclear systems.

## 6.4 Soft collinear effective theory

**Iain Stewart**

### 6.4.1 Introduction

Effective field theory is a powerful tool which enables the organization of QCD dynamics at different momentum scales. The most well known examples of EFTs involve the dynamics of massive particles, like integrating out the heavy electroweak $W$ and $Z$ bosons to obtain the Electroweak Hamiltonian, or systematically treating the mass scale of heavy quarks like the $t$, $b$, and $c$ in HQET or NRQCD. On the other hand, much of our knowledge about strong interactions comes from hard scattering interactions of light quarks and gluons, which are the most important processes in $pp$, $e^-p$, or $e^+e^-$ colliders. Such processes are the way we search for new particles or fundamental interactions at short distances, and indeed were key to the discovery of the $c$, $b$, and $t$ quarks, the $W$ and $Z$ bosons, and the Higgs $H$. In these processes we must simultaneously deal with perturbative QCD dynamics at the hard interaction scale $Q$ governing the dynamics of the high energy collision, as well as nonperturbative physics at the scale $\Lambda \ll Q$, which is responsible for the confinement and hadronization of partons. Many processes studied at colliders also have additional important intermediate scales $\Delta$, with $\Lambda \ll \Delta \ll Q$. Examples of $\Delta$ include the transverse momentum of particles inside an energetic jet produced from the collimated shower of a high energy quark or gluon, or the measurement of differential distributions of a kinematic variable $\Delta$, where the largest cross section contributions typically arise from the $\Lambda \ll \Delta \ll Q$ kinematic situation. The appropriate effective field theory for these processes is the Soft Collinear Effective Theory (SCET) [1761–1764]. Traditional QCD methods, outside the framework of EFT, have a long tradition for describing the physics of hard processes, including the Brodsky-Lepage/Efremov-Radyushkin formalism [206, 207, 1765] for exclusive hadronic processes, and the Collins-Soper-Sterman formalism [224, 1282, 1347, 1766] for inclusive cross sections. SCET builds naturally on this foundation.

SCET is an effective theory which systematically describes the infrared QCD dynamics in hard collisions, including the associated dynamics of soft and collinear degrees of freedom. Its popularity stems in part from the fact that it enables the description of a huge variety of collider processes [1767]. This includes processes that involve energetic hadrons such as large $Q^2$ form factors $\gamma^*\gamma \to \pi^0$, $\gamma^*\pi^+ \to \pi^+$, or fragmentation to one or more hadrons $h_i$ in processes like $e^+e- \to h_1h_2X$ and $pp \to h_1X$. Other examples include energetic hadronic collisions like at the Large Hadron Collider, including Higgs production $pp \to HX$ and Drell-Yan $pp \to X\ell^+\ell^-$, Deep Inelastic Scattering (DIS) $e^-p \to e^-X$ or $e^-$-ion $\to e^-X$, and Semi-Inclusive DIS $e^-p \to e^-hX$ (for the latter see Ref. [1768]). SCET also describes processes that produce energetic jets instead of (or in addition to) energetic hadrons, such as $e^+e^- \to$ 2-jets [280, 1769–1772], $pp \to H +$ 1-jet [1773, 1774], or $pp \to$ 2-jets [1775, 1776]. In addition it can be used to describe jet-substructure, the dynamics of particles and sub-jets inside an identified jet [1777–1790]. Finally, it can also be used to describe the dynamics of heavy particle production and decay. Indeed some of the original applications of SCET were to processes like $B \to \pi\ell\nu$ [1762, 1791–1793], $B \to D\pi$ [1794, 1795], $B \to \pi\pi$ [1792, 1796], and $B \to X_s\gamma$ [1761, 1763, 1797–1801] (where SCET is combined with HQET), as well as $e^+e^- \to J/\Psi X$ [1802–1805] and $\Upsilon \to X\gamma$ [1505, 1506, 1806–1809] (where SCET is combined with NRQCD). Recent applications of SCET include its extension to forward scattering and Regge phenomena [1810–1813], heavy-ion collisions [1814–1819], gravitational effects [1820–1825], the resummation of large electroweak logarithms [1826–1832], large logs in dark matter annihilation cross sections [1833–1837], and radiative corrections in neutrino-nucleon scattering [1838, 1839].

Features of SCET that people find useful include: the universal steps in deriving *factorization*, whereby observables split themselves into independent functions governing the hard, collinear and soft dynamics of a process, the transparency in carrying out higher order resummation of large logarithms, the ability to generalize factorization to more complicated processes and multiscale observables, and the capability to systematically study power corrections.

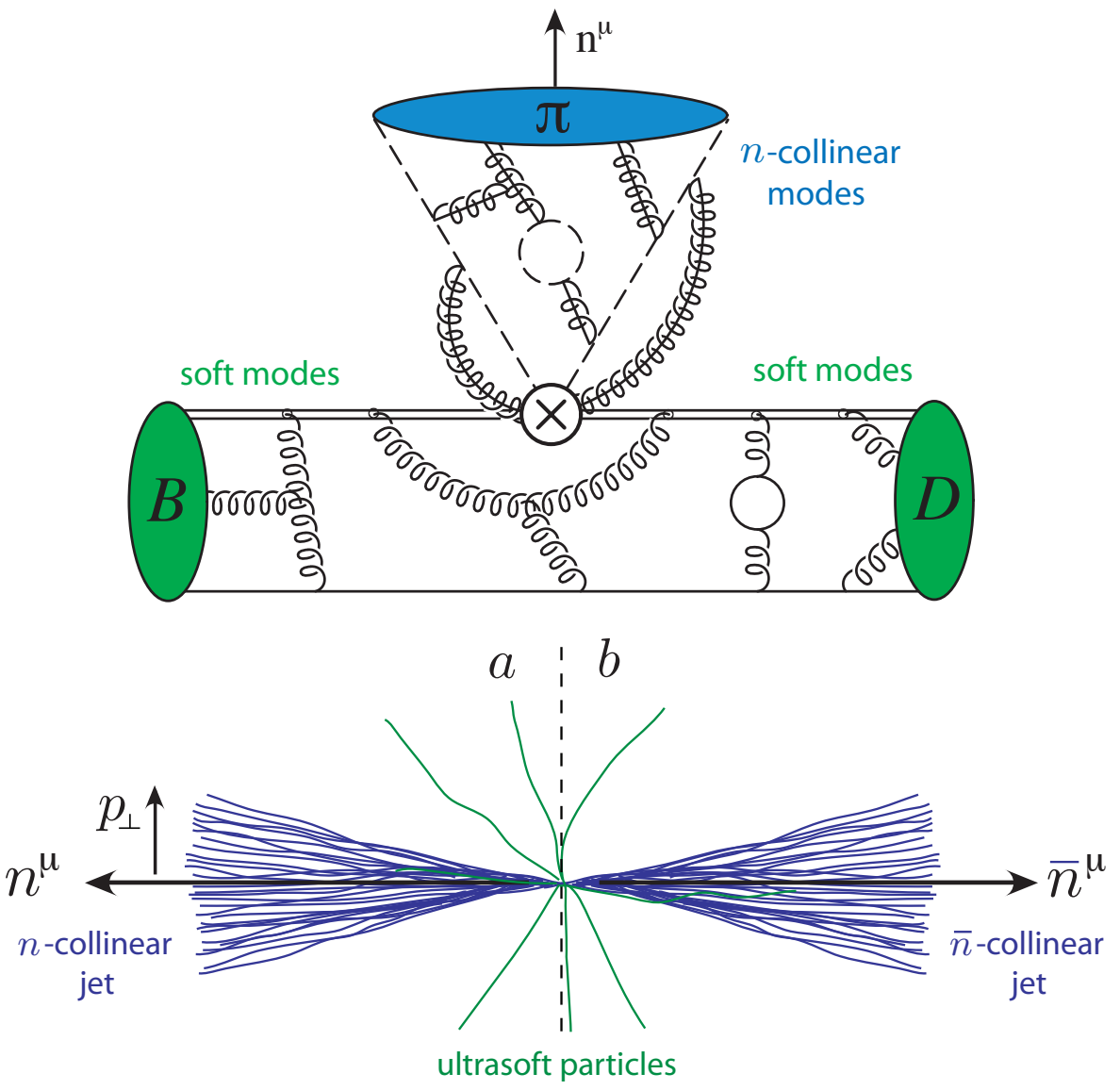

**Fig. 6.4.1** Example processes $B \to D\pi$ and $e^+e^- \to$ 2-jets.

### 6.4.2 Degrees of Freedom

SCET describes collinear particles that are constituents of energetic hadrons or jets and have a large momentum along a particular light-like direction $n_i^\mu$. For each collinear direction we have two reference vectors $n_i^\mu$ and $\bar{n}_i^\mu$ such that $n_i^2 = \bar{n}_i^2 = 0$ and $n_i \cdot \bar{n}_i = 2$. A common choice is $n_i^\mu = (1, \hat{n}_i)$ and $\bar{n}_i^\mu = (1, -\hat{n}_i)$, with $\hat{n}_i$ a unit three-vector in the collinear direction. Any four-momentum $p$ can be decomposed in terms of these as

$$p^\mu = \bar{n}_i \cdot p \, \frac{n_i^\mu}{2} + n_i \cdot p \, \frac{\bar{n}_i^\mu}{2} + p_{n_i\perp}^\mu \quad . \qquad (6.4.1)$$

Particles with $p^\mu$ close to $n_i^\mu$ are referred to as $n_i$-collinear and have $(n_i \cdot p, \bar{n}_i \cdot p, p_{n_i\perp}) = (p^+, p^-, p_\perp) \sim Q\,(\lambda^2, 1, \lambda)$, where $\lambda \ll 1$ is the small SCET power counting parameter, determined by scales and kinematics or by measurements restricting QCD radiation. SCET also describes particles with soft momenta

$$p^\mu \sim Q(\lambda, \lambda, \lambda)$$

and with ultrasoft (usoft) $p^\mu \sim Q(\lambda^2, \lambda^2, \lambda^2)$.

Examples are shown in Fig. 6.4.1. In the $B \to D\pi$ process, $Q = \{m_b, m_c, m_b - m_c\}$ and $\lambda = \Lambda/Q$, with the $B$ and $D$ composed of a heavy quark, light soft quarks, and soft gluons. The pion has $E_\pi = 2.3\,\text{GeV} = Q \gg \Lambda$, and has collinear quark and gluon constituents. In the $e^+e^- \to$ 2-jets process, we have back-to-back jets with energy $Q$, where $Q^2$ is the invariant mass of the $e^+e^-$ pair, and $\lambda = \Delta/Q$ with $\Lambda \ll \Delta \ll Q$. Here $\Delta$ is a scale that characterizes the transverse size of the jet, and associated to measurements made on the jets. For example, if a hemisphere jet mass $m_J$ is measured, then $\Delta = m_J$, while if thrust $1 - \tau$ is measured, $\Delta^2 = Q^2\tau$.

To ensure that collinear directions $n_i$ and $n_j$ are distinct, we must have $n_i \cdot n_j \gg \lambda^2$ for $i \neq j$. Since distinct reference vectors, $n_i$ and $n_i'$, with $n_i \cdot n_i' \sim \lambda^2$ both describe the same collinear physics, one can label a collinear sector by any member of an equivalence class of vectors, $\{n_i\}$. This freedom manifests as a symmetry of the effective theory known as reparametrization invariance (RPI) [1840, 1841]. Three classes of RPI transformations are

| RPI-I | RPI-II | RPI-III |
|---|---|---|
| $n_{i\mu} \to n_{i\mu} + \Delta_\mu^\perp$ | $n_{i\mu} \to n_{i\mu}$ | $n_{i\mu} \to e^{\alpha} n_{i\mu}$ |
| $\bar{n}_{i\mu} \to \bar{n}_{i\mu}$ | $\bar{n}_{i\mu} \to \bar{n}_{i\mu} + \epsilon_\mu^\perp$ | $\bar{n}_{i\mu} \to e^{-\alpha} \bar{n}_{i\mu}$, |

(6.4.2)

where $\alpha \sim \lambda^0$ and infinitesimal parameters $\Delta^\perp \sim \lambda$ and $\epsilon^\perp \sim \lambda^0$. These parameters satisfy $n_i \cdot \Delta^\perp = \bar{n}_i \cdot \Delta^\perp = n_i \cdot \epsilon^\perp = \bar{n}_i \cdot \epsilon^\perp = 0$.

The effective theory is constructed by separating collinear momenta into large (label) $\tilde{p}$ and small (residual) $p_r$ components

$$p^\mu = \tilde{p}^\mu + p_r^\mu = \bar{n}_i \cdot \tilde{p} \, \frac{n_i^\mu}{2} + \tilde{p}_{n_i\perp}^\mu + p_r^\mu \, , \qquad (6.4.3)$$

with $\bar{n}_i \cdot \tilde{p} \sim Q$, $\tilde{p}_{n_i\perp} \sim \lambda Q$. The small $p_r^\mu \sim \lambda^2 Q$ describes fluctuations about the label momentum. To simultaneously describe different regions of momentum space with operators that have manifest power counting, it is necessary to have multiple fields for the same fundamental particle. Namely, for each collinear direction we have collinear quark fields $\xi_{n_i} \sim \lambda$ and collinear gluon fields $A_{n_i}^\mu \sim (\lambda^2, \lambda^0, \lambda)$, as well as soft quark $q_s \sim \lambda^{3/2}$ and soft gluon $A_s^\mu \sim \lambda$ fields, and/or usoft quark $q_{us} \sim \lambda^3$ and usoft gluon $A_{us}^\mu \sim \lambda^2$ fields. These power counting assignments ensure that the corresponding kinetic terms in the action are $\mathcal{O}(\lambda^0)$.

The precise degrees of freedom depend on the process. Often only usoft or soft fields are present, in which case the theories are referred to as $\text{SCET}_\text{I}$ and $\text{SCET}_\text{II}$ respectively [1792]. $\text{SCET}_\text{I}$ is relevant for measurements sensitive to the small $n_i \cdot p \sim Q\lambda^2$ momentum, such as jet mass in $e^+e^- \to$ 2-jets, see Fig. 6.4.2; while $\text{SCET}_\text{II}$ is relevant for measurements that involve transverse momenta or collinear and soft modes with the same invariant mass. Examples also exist that require mixed soft-collinear modes with $p^\mu \sim Q'(\lambda^2, 1, \lambda)$ where $Q' \ll Q$, in which case the theory is referred to as $\text{SCET}_+$; see Ref. [1787]. Independent collinear, soft, and usoft gauge symmetries are also enforced for each set of fields [1764]. A general SCET $\lambda$ power counting formula can be used to determine the order of any diagram entirely from operators inserted at its vertices plus topological factors [1810, 1842].

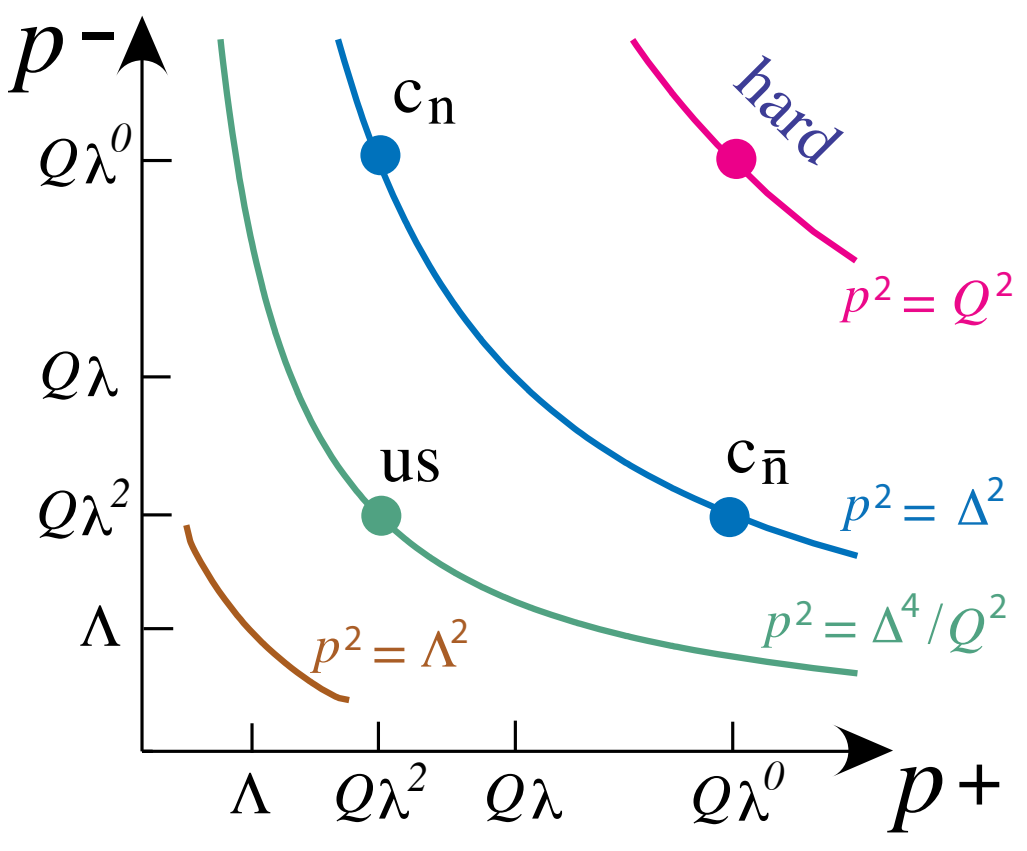

**Fig. 6.4.2** Degrees of freedom for jet mass in $e^+e^- \to$ 2-jets.

To fully expand in $\lambda$ one must carry out a multipole expansion for the fields in SCET. There are two equivalent ways that this expansion has been constructed in the literature, either in a combination with momentum space for large label momenta and position space for the residuals, with fields written as $\xi_{n_i,\tilde{p}}(x)$ [1763], or with the multipole expansion carried out entirely in position space [1791]. We will use the former, and facilitate the expansion by defining two derivative operators, a label momentum operator $\mathcal{P}^\mu_{n_i}$ giving large momentum components, such as $\mathcal{P}^\mu_{n_i}\,\xi_{n_i,\tilde{p}} = \tilde{p}^\mu\,\xi_{n_i,\tilde{p}}$, and a residual momentum operator giving residual small components, such as $i\partial^\mu \xi_{n_i,\tilde{p}}(x) \sim Q\lambda^2 \xi_{n_i,\tilde{p}}(x)$. The shorthand $\bar{\mathcal{P}} = \bar{n}_i \cdot \mathcal{P}_{n_i}$ is used for the largest $\mathcal{O}(\lambda^0)$ label momentum. Useful covariant derivatives include

$$
\begin{aligned}
i\bar{n}\cdot\mathbf{D}_n &= \bar{\mathcal{P}} + g\bar{n}\cdot\mathbf{A}_n\,, \quad i\mathbf{D}^\mu_{n\perp} = \mathcal{P}^\mu_\perp + g\mathbf{A}^\mu_{n\perp}\\
in\cdot\mathbf{D}_n &= in\cdot\partial + gn\cdot\mathbf{A}_n\,, \quad i\mathbf{D}^\mu_{us} = i\partial^\mu + g\mathbf{A}^\mu_{us}\\
in\cdot\mathbf{D} &= in\cdot\partial + gn\cdot\mathbf{A}_{us} + gn\cdot\mathbf{A}_n \quad , 
\end{aligned}
\tag{6.4.4}
$$

where $\mathbf{A}^\mu_n \equiv A^{A\mu}_n T^A$ and $igF^{A\mu\nu}_n T^A = [i\mathbf{D}^\mu_n, i\mathbf{D}^\nu_n] = ig\mathbf{F}^{\mu\nu}_n$. This is the standard sign convention for $g$ used in the SCET literature. It differs from the QCD summary above ($g \to -g$).

### 6.4.3 SCET Lagrangian and Factorization

The SCET Lagrangian is

$$
\mathcal{L}_{\rm SCET} = \mathcal{L}_{\rm hard} + \mathcal{L}_{\rm dyn} = \sum_{i\geq 0}\Big(\mathcal{L}^{(i)}_{\rm hard} + \mathcal{L}^{(i)}_{\rm dyn}\Big) + \mathcal{L}^{(0)}_G\,,
\tag{6.4.5}
$$

where the superscript $(i)$ indicates terms suppressed by $\mathcal{O}(\lambda^i)$ relative to the leading power Lagrangian. Here the hard short distance interactions are encoded in $\mathcal{L}^{(i)}_{\rm hard}$ with only one of these appearing in each amplitude (unless we study multiple hard scatterings). They contain multiple types of collinear (and soft) fields. The dynamic Lagrangians $\mathcal{L}^{(i)}_{\rm dyn}$ describe the evolution and interactions of collinear and (u)soft particles. We have singled out the so-called Glauber Lagrangian $\mathcal{L}^{(0)}_G$ for special treatment since it is the only term that violates factorization of collinear and (u)soft modes [1810].

At leading power the dynamic SCET$_{\rm I}$ and SCET$_{\rm II}$ Lagrangians are [1764]

$$
\begin{aligned}
\mathcal{L}^{\rm I(0)}_{\rm dyn} &= \sum_n \mathcal{L}^{(0)}_n + \mathcal{L}^{(0)}_{us}\,,\\
\mathcal{L}^{\rm II(0)}_{\rm dyn} &= \sum_n \mathcal{L}^{(0)}_n + \mathcal{L}^{(0)}_{s}\,,
\end{aligned}
\tag{6.4.6}
$$

where the first terms sum over all needed independent collinear sectors. In SCET$_{\rm II}$ each of $\mathcal{L}^{(0)}_n$ and $\mathcal{L}^{(0)}_s$ only involves collinear or soft fields, so the sectors are immediately factorized by the power expansion. In SCET$_{\rm I}$ the $n\cdot\mathbf{A}_{us}$ fields still interact with collinear fields since they are $\mathcal{O}(\lambda^2)$ just like $n\cdot\partial$ and $n\cdot\mathbf{A}_n$, and do not knock the collinear particles offshell (meaning that initial and final particles have momenta satisfying the collinear power counting). These $n\cdot\mathbf{A}_{us}$ interactions can be decoupled by the BPS field redefinition [1764]

$$
\xi_n(x) \to Y_n(x)\xi_n(x), \qquad \mathbf{A}^\mu_n(x) \to Y_n(x)\mathbf{A}^\mu_n(x)Y^\dagger_n(x),
\tag{6.4.7}
$$

where $Y_n$ is an ultrasoft Wilson line

$$
Y_n(x;-\infty,0) = P\exp\Big(ig\int_{-\infty}^0 {\rm d}s\, n\cdot\mathbf{A}_{us}(x+ns)\Big),
\tag{6.4.8}
$$

and $P$ is path ordering of color matrices with $s$. This transformation moves usoft interactions into the hard scattering operators, and leaves factorized Lagrangians $\mathcal{L}^{(0)}_n$ and $\mathcal{L}^{(0)}_{us}$, which only depend on collinear or usoft fields respectively. For example, for collinear quarks in SCET$_{\rm I}$ we have

$$
\begin{aligned}
\mathcal{L}^{\rm I(0)}_{n\xi} &= e^{-ix\cdot\mathcal{P}}\;\bar{\xi}_n\Big(in\cdot\mathbf{D} + i\not{D}_{n\perp}\frac{1}{i\bar{n}\cdot\mathbf{D}_n}i\not{D}_{n\perp}\Big)\frac{\not{\bar{n}}}{2}\,\xi_n\\
&\to e^{-ix\cdot\mathcal{P}}\;\bar{\xi}_n\Big(in\cdot\mathbf{D}_n + i\not{D}_{n\perp}\frac{1}{i\bar{n}\cdot\mathbf{D}_n}i\not{D}_{n\perp}\Big)\frac{\not{\bar{n}}}{2}\,\xi_n.
\end{aligned}
\tag{6.4.9}
$$

The first few Feynman rules prior to the field redefinition are shown in Fig. 6.4.3, and the one in the second line is removed from $\mathcal{L}^{\rm I(0)}_{n\xi}$ after implementing Eq.(6.4.7). After the transformation the $\mathcal{L}^{(0)}_n$ Lagrangian has the same form in SCET$_{\rm I}$ and SCET$_{\rm II}$.

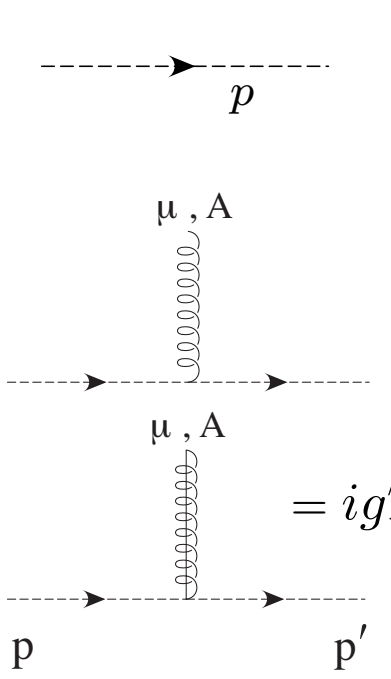

$$= i\,\frac{\bar{n}\!\!\!/}{2}\,\frac{\bar{n}\cdot p}{n\cdot p\,\bar{n}\cdot p\,+\,p_\perp^2\,+i0}$$

$$= ig\,T^A\,n_\mu\,\frac{\bar{n}\!\!\!/}{2}$$

$$= igT^A\Big[n_\mu + \frac{\gamma_\mu^\perp p\!\!\!/_\perp}{\bar{n}\cdot p} + \frac{p\!\!\!/'_\perp\gamma_\mu^\perp}{\bar{n}\cdot p'} - \frac{p\!\!\!/'_\perp p\!\!\!/_\perp\bar{n}_\mu}{\bar{n}\cdot p\,\bar{n}\cdot p'}\Big]\frac{\bar{n}\!\!\!/}{2}$$

**Fig. 6.4.3** $\mathcal{O}(\lambda^0)$ Feynman rules for collinear quarks (dashed) interacting with a soft gluon (spring) or collinear gluon (spring with a line through it). Rules with more collinear gluons are not shown.

The construction of SCET hard scattering Lagrangians $\mathcal{L}^{(i)}_{\rm hard}$ requires integrating out offshell fields, which have larger $p^2$ than the collinear and (u)soft fields (see Fig. 6.4.2 for example). When collinear particles in two different sectors interact, the resulting particles are hard and offshell with $p^2 \sim Q^2$. Likewise when collinear and soft particles interact this results in offshell hard-collinear particles with $p^2 \sim Q^2\lambda$. Systematically integrating out the corresponding offshell fields results in collinear and soft Wilson lines appearing in operators [1762–1764]. This involves an infinite number of gluon attachments and can be carried out analytically with background field techniques [1764, 1767]. In label momentum space the resulting collinear Wilson lines are defined as

$$W_{n_i}(x) = \Big[\sum_{\rm perms} \exp\Big(-\frac{g}{\bar{\mathcal{P}}}\,\bar{n}\!\cdot\!\mathbf{A}_{n_i}(x)\Big)\Big]. \qquad (6.4.10)$$

Note that it is the $\bar{n}_i\cdot\mathbf{A}_n \sim \lambda^0$ component of the gluon field that appears in these Wilson lines. In general all $\mathcal{O}(\lambda^0)$ gluon components can be traded for Wilson lines using $i\bar{n}_i\!\cdot\!\mathbf{D}_{n_i} = W_{n_i}\bar{\mathcal{P}}W^\dagger_{n_i}$. Unlike $Y_n$, the subscript on $W_{n_i}$ refers to the collinear fields it is built out of, not the Wilson line direction (which is $\bar{n}_i$). For zero residual momentum $x = (0, x^-, x_\perp)$, the $W_n(x)$ is simply the Fourier transform ($b^+ \leftrightarrow p^-$) of a standard position-space Wilson line ending at $b = (b^+, x^-, x_\perp)$:

$$W_n(b; -\infty, 0) = P\,\exp\Big(ig\int_{-\infty}^0 {\rm d}s\,\bar{n}\!\cdot\!\mathbf{A}_n(b+\bar{n}s)\Big). \qquad (6.4.11)$$

Since the construction of hard-collinear interactions in $\mathrm{SCET_{II}}$ can be facilitated by matching QCD$\to\mathrm{SCET_I}$ $\to\mathrm{SCET_{II}}$ [1792], it suffices to primarily focus on matching for $\mathrm{SCET_I}$. The definition for the soft Wilson line $S_n(x; -\infty, 0)$ appearing in $\mathrm{SCET_{II}}$ is identical to Eq. (6.4.8) with $\mathbf{A}_{us} \to \mathbf{A}_s$.

| Operator | $\mathcal{B}^\mu_{n_i\perp}$ | $\chi_{n_i}$ | $\mathcal{P}^\mu_\perp$ | $q_{us}$ | $\mathbf{D}^\mu_{us}$ |
|---|---|---|---|---|---|
| Power Counting | $\lambda$ | $\lambda$ | $\lambda$ | $\lambda^3$ | $\lambda^2$ |

**Table 6.4.1** Power counting for building block operators in $\mathrm{SCET_I}$.

Hard interactions involving collinear fermions provide a frame of reference that allows us to simplify the Dirac structures that appear, since so-called good fermion components dominate over bad components in the $\lambda$ expansion. In SCET this is encoded by the projection relations $(n\!\!\!/_i\bar{n}\!\!\!/_i/4)\xi_{n_i} = \xi_{n_i}$, which also implies $n\!\!\!/_i\xi_{n_i} = 0$. The same formulae also hold for $\chi_{n_i}$. Only the good components are needed to construct operators in SCET at any order in the power expansion, and indeed we have already written $\mathcal{L}^{(0)}_{n\xi}$ in Eq.(6.4.9) using them. Note that on its own, Eq.(6.4.9) is equivalent to a QCD Lagrangian for collinear quarks (indeed it has the same form as the light-cone QCD Lagrangian [1843]), with a distinction made only by which fermion components are sourced in the path integral.

Integrating out offshell fluctations also results in Wilson coefficients that depend on the large $\mathcal{O}(\lambda^0)$ momenta of collinear fields. It is straightforward to see why this is the case, since if we annihilate or produce two collinear particles with $p^\mu_n = \omega_1 n^\mu/2$ and $p^\mu_{\bar{n}} = \omega_2\bar{n}^\mu/2$, then $q = p_n + p_{\bar{n}}$ has $q^2 = \omega_1\omega_2 \sim Q^2$. Thus offshell fluctuations that depend on $Q^2$ also depend on the large momenta $\omega_i \sim \lambda^0$ of collinear fields. Two other constraints on the form of hard operators are SCET gauge invariance and the ability to use the equations of motion to reduce the operators basis to a minimal set. This is summarized by the fact that all operators can be constructed out of a minimal set of building blocks, formed from combinations of fields and Wilson lines [1762, 1763, 1844]. The collinearly gauge-invariant quark and gluon building block fields are defined as

$$\chi_{n_i,\omega}(x) = \Big[\delta(\omega - \bar{\mathcal{P}}_{n_i})\,W^\dagger_{n_i}(x)\,\xi_{n_i}(x)\Big], \qquad (6.4.12)$$
$$\mathcal{B}^\mu_{n_i\perp,\omega}(x) = \frac{1}{g}\Big[\delta(\omega + \bar{\mathcal{P}}_{n_i})\,W^\dagger_{n_i}(x)\,i\mathbf{D}^\mu_{n_i\perp}W_{n_i}(x)\Big].$$

The Wilson lines $W_{n_i}(x)$ are localized with respect to the position $x$, and we can therefore treat $\chi_{n_i,\omega}(x)$ and $\mathcal{B}^\mu_{n_i,\omega}(x)$ as local quark and gluon fields from the perspective of ultrasoft derivatives $\partial^\mu$ that act on $x$. Our conventions for $\chi_{n_i,\omega}$ have $\omega > 0$ for an incoming quark and $\omega < 0$ for an outgoing antiquark at lowest order. For $\mathcal{B}_{n_i\perp,\omega}$, $\omega > 0$ ($\omega < 0$) corresponds to outgoing (incoming) gluons at lowest order.

For $\mathrm{SCET_I}$ the complete set of building blocks and their power counting is summarized in Table 6.4.1. Both the $\chi_n$ and $\mathcal{B}_{n\perp}$ building block fields scale as $\mathcal{O}(\lambda)$. For the majority of jet processes there is a single collinear field operator for each collinear sector at leading power. For exclusive processes that directly produce energetic hadrons at the hard interaction (rather than by fragmentation) there are multiple building blocks from the same sector in the leading power operators, since we must form a color singlet in each sector in order to directly produce a color singlet hadron. The $\mathcal{P}_\perp \sim \lambda$ is not typically present at leading power. At subleading power, operators for all processes can involve multiple collinear fields in the same collinear sector, as well as $\mathcal{P}_\perp$ operator insertions. The power counting for an operator is obtained by simply adding up the powers for the building blocks it contains. To ensure consistency under renormalization group evolution the operator basis in SCET must be complete, namely all operators consistent with the symmetries of the problem must be included. The counting of subleading power operators is greatly facilitated by spinor-helicity SCET techniques [1845–1849].

A few examples of hard scattering operators can help clarify the above points. For $\text{SCET}_\text{I}$ processes like thrust, jet mass, or other dijet event shapes in $e^+e^-$ collisions, or for threshold resummation in Drell-Yan or DIS, the leading power Lagrangian from the electromagnetic current is

$$\mathcal{L}^{\text{I}(0)}_{\text{hard}}(0) = \frac{ie^2}{Q^2} J^\mu_{\bar{\ell}\ell} \int \mathrm{d}\omega_1 \mathrm{d}\omega_2\, C_f^{(0)}(\omega_1\omega_2,\mu) \times \big[(\bar{\chi}^f_{\bar{n},\omega_2})\,(Y^\dagger_{\bar{n}}Y_n)\gamma^\perp_\mu(\chi^f_{n,\omega_1})\big]_\mu\,, \tag{6.4.13}$$

where $C_f^{(0)}$ is the Wilson coefficient encoding virtual hard interactions at any order in $\alpha_s$, and renormalization is carried out in the $\overline{\text{MS}}$ scheme, inducing dependence on the renormalization scale $\mu$. In Eq.(6.4.13) the usoft Wilson lines $Y^\dagger_{\bar{n}}Y_n$ appear from the BPS field redefinition in Eq.(6.4.7). Also, the leptonic vector current is $J^\mu_{\bar{\ell}\ell} = (-1)\bar{\ell}\gamma^\mu\ell$, and we sum over quark flavors $f$. At any order in $\alpha_s$ the Wilson coefficient $C_f^{(0)}(\omega_1\omega_2)$ encodes virtual corrections from the hard scale $\omega_1\omega_2 \sim Q^2$. For hard Lagrangians with only a single field in a given collinear direction, the large collinear momentum factors $\omega_i$ are fixed by the overall kinematics of the hard process, and thus remain unchanged by perturbative corrections. For example, $\omega_1 = \omega_2 = Q$ for $e^+e^- \to$ 2-jets. At tree level $C_f^{(0)} = Q_f + \mathcal{O}(\alpha_s)$, where the quarks have charge $Q_f|e|$. To calculate $C_f^{(0)}$ at higher orders we carry out loop level matching calculations, comparing hard scattering Feynman diagrams separately computed and renormalized in full QCD and in $\text{SCET}_\text{I}$, while using the same states and infrared (IR) regulators. Since SCET captures all the IR physics, the difference between these calculations determines $C_f^{(0)}$ order by order, and implies it encodes hard effects. For the particular example in Eq.(6.4.13), $C_f^{(0)}$ is related to the IR finite part of the $\overline{\text{MS}}$ massless quark form factor with $Q^2 \gg \Lambda^2$. (In general when carrying out loop calculations in SCET with both (u)soft and collinear loops, one must include 0-bin subtractions which ensure there is not double counting of IR regions [1850]. For some choices of IR regulators these subtractions are scaleless in dimensional regularization, and hence can be dropped, up to interpreting the divergence structure.)

For $\text{SCET}_\text{II}$ processes like the broadening event shape for $e^+e^- \to$ 2-jets, or transverse momentum dependent (TMD) distributions for Drell-Yan, SIDIS, or $e^+e^- \to h_1h_2X$, the leading hard scattering Lagrangian is

$$\mathcal{L}^{\text{II}(0)}_{\text{hard}}(0) = \frac{ie^2}{Q^2} J^\mu_{\bar{\ell}\ell} \int \mathrm{d}\omega_1 \mathrm{d}\omega_2\, C_f^{(0)}(\omega_1\omega_2,\mu) \times \big[(\bar{\chi}^f_{\bar{n},\omega_2})\,(S^\dagger_{\bar{n}}S_n)\gamma^\perp_\mu(\chi^f_{n,\omega_1})\big]_\mu\,, \tag{6.4.14}$$

with the same Wilson coefficient $C_f^{(0)}$ as Eq.(6.4.13). The only difference is the appearance of soft Wilson lines $S$ instead of usoft $Y$. This operator can be obtained immediately from Eq.(6.4.13) by matching $\text{SCET}_\text{I}$ $\to\text{SCET}_\text{II}$.

As a final example we consider $\bar{B}^0 \to D^+\pi^-$ mediated by the weak $W$-boson flavor changing transition $b \to c\bar{u}d$. Here the matching is from the electroweak Hamiltonian $\mathcal{H}_W = 2\sqrt{2}G_F V^*_{ud}V_{cb}\sum_{i=0,8} C_i^{\text{F}} O_i$, with 4-quark operators $O_0 = [\bar{c}\gamma^\mu P_L b][\bar{d}\gamma_\mu P_L u]$ and $O_8 = [\bar{c}\gamma^\mu T^A P_L b][\bar{d}\gamma_\mu P_L T^A u]$, onto coefficients and operators in SCET. The heavy quark fields are matched onto HQET fields $h_v^{(Q)}$ for $Q = b, c$, while the light quarks become collinear. The leading power hard scattering Lagrangian in SCET is [1794]

$$\mathcal{L}^{\text{II}(0)}_{\text{hard}} = \int \mathrm{d}\omega_1 \mathrm{d}\omega_2\, C^{j(0)}_{BD\pi}(\omega_1,\omega_2,m_b,m_c,\mu) \times \big\{\big[\bar{h}^{(c)}_{v'}\Gamma^j_h h^{(b)}_v\big]\big[(\bar{\chi}^{\bar{d}}_{n,\omega_2})\,\Gamma_\xi(\chi^u_{n,\omega_1})\big]\big\}_\mu\,, \tag{6.4.15}$$

where we sum over $j = 1, 5$ with Dirac structures $\Gamma_h^{1,5} = \not{n}\{1,\gamma_5\}/2$ and $\Gamma_\xi = \not{\bar{n}}(1-\gamma_5)/4$. Here the hard coefficients $C^{j(0)}_{BD\pi}$ depend on multiple hard scales as in Eq.(6.4.15). There are no soft Wilson lines because the $n$-collinear quark pair is a color singlet and $S_n^\dagger S_n = 1$. An analogous SCET operator with color structure $T^A \otimes T^A$ exists and does involve soft Wilson lines. Since it can be factorized into a product of soft and collinear octet operators, it does not contribute to the physical process: a factorized octet collinear bilinear operator can not produce a color singlet pion.

Let us return to the leading power Glauber Lagrangian. It involves interactions between soft and collinear modes in the form of potentials, and has the form [1810]

$$\mathcal{L}_G^{(0)} = \sum_{n,\bar{n}} \mathcal{O}_n^{iB} \frac{1}{\mathcal{P}_\perp^2} \mathcal{O}_s^{BC} \frac{1}{\mathcal{P}_\perp^2} \mathcal{O}_{\bar{n}}^{jC} + \sum_n \mathcal{O}_n^{iB} \frac{1}{\mathcal{P}_\perp^2} \mathcal{O}_s^{j_n B} , \tag{6.4.16}$$

in both $\mathrm{SCET_I}$ and $\mathrm{SCET_{II}}$. Further details and the definitions for the operators $\mathcal{O}$ can be found in Ref. [1810]. Many of the steps involved in deriving factorization at leading power are manifest in the construction of SCET; in particular we arrive at hard scattering Lagrangians $\mathcal{L}_{\rm hard}^{(0)}$ that can be written as products of gauge invariant collinear and soft operators, and we have a direct sum of independent Lagrangians for soft and collinear fields in $\mathcal{L}_{\rm dyn}^{(0)}$. With just these terms the SCET Hilbert space of states factorizes as direct products, and matrix elements of collinear and soft operators with their Wilson coefficients define independent collinear, soft and hard functions (examples given below). Since $\mathcal{L}_G^{(0)}$ can be inserted any number of times without power suppression, and couples different sectors, it breaks factorization.

Thus proving factorization reduces to demonstrating that contributions from $\mathcal{L}_G^{(0)}$ either cancel out, or can be absorbed into other interactions. Both of these occur. For example, in $e^+e^- \to$ 2-jets the non-trivial interactions from $\mathcal{L}_G^{(0)}$ can be absorbed into the *direction* of the (u)soft and collinear Wilson lines, which in that case then run from $[0,\infty)$ rather than $(-\infty,0]$, see Ref. [1810]. The same absorption is true for the exclusive $B \to D\pi$ process, with the common feature being that these processes involve only active partons and do not involve forward scattering configurations (see also Refs. [1851, 1852]). In a process like Drell-Yan the cancellation of $\mathcal{L}_G^{(0)}$ is much more complicated due to interactions involving spectator partons in the initial protons, but these still cancel out. Low order demonstrations can be found in Refs. [1810, 1853, 1854], while the all order statement was made in the classic CSS proof of Glauber region cancellations in Ref. [1766]. For cases where factorization is known to be violated [1852, 1855–1859], it is not possible to absorb or cancel the effects of $\mathcal{L}_G^{(0)}$ in this manner. The factorization of Glauber effects in SCET can also be used to sum so-called superleading logarithms [1860].

It is worth noting that in SCET the proof of factorization for cross sections and decay rates at subleading power follows the same steps as at leading power. Higher power $\mathcal{L}_{\rm hard}^{(i\geq1)}$ simply involve more complicated products of factorized soft and collinear operators. While terms in $\mathcal{L}_{\rm dyn}^{(i\geq1)}$ also involve products of soft and collinear fields, they are always inserted only a finite number of times at any given order in the power counting, and hence still lead to factorized matrix elements, albeit with time ordered products of operators. For a gauge invariant description of power suppressed $\mathrm{SCET_I}$ operators see Refs. [1861, 1862]. Many of these observations go back to the beginnings of SCET, since the processes that people focused on at the time involved exclusive $B$ decays that only start at subleading power [1791–1793, 1795–1797, 1863–1867], primarily because the soft spectator quark in the $B$ had to be converted into a collinear quark, a subleading power process.

Finally, we remark that $\mathcal{L}_G^{(0)}$ is interesting in its own right, because for processes involving forward scattering rather than hard scattering, it does not cancel but instead provides the dominant contributions, yielding Reggeization, BFKL evolution, and the shockwave picture. For more work in this direction see Refs. [1810–1813]. It is also worth noting that this implies that SCET can potentially provide a framework to parameterize and describe spectator factorization violating contributions to certain hard scattering processes from first principles, though so far very little work has been done in this direction.

### 6.4.4 Examples of Factorization

To connect theory and experiment, consider a few examples of factorization formulae that have been derived or studied with SCET. A key attribute of these formulas is that they are determined using only the SCET power expansion, and do not rely on any $\alpha_s$ expansion. First consider $e^+e^- \to$ 2-jets, with a measurement of $\tau = 1 - T$ where $T$ is thrust, working in the dijet limit $\tau \ll 1$. We can relate $\tau$ to the sum of the two hemisphere jet masses, $\tau = (m_{Ja}^2 + m_{Jb}^2)/Q^2$, where $m_{Ja}^2$ and $m_{Jb}^2$ are each determined by the particles on one side of the plane perpendicular to the thrust axis. Thus $\tau \ll 1$ restricts the invariant mass of the radiation in both hemispheres and forces us into a dijet configuration. Squaring the $\mathrm{SCET_I}$ leading power amplitudes obtained from $\mathcal{L}_{\rm hard}^{(0)}$ in Eq.(6.4.13), Fierzing the fields of distinct types into independent matrix elements, integrating over phase space with the measurement function, and renormalizing the resulting factorized functions, gives [280, 1769–1772]

$$\frac{\mathrm{d}\sigma}{\mathrm{d}\tau} = \sigma_0 H(Q,\mu) Q \int \mathrm{d}\ell \mathrm{d}\ell' J_T(Q^2\tau - Q\ell, \mu) \times S_T(\ell-\ell',\mu) F(\ell',\Lambda) \, . \tag{6.4.17}$$

Here $H(Q,\mu) = |C^{(0)}(Q,\mu)|^2$ is a hard function encoding virtual corrections (magenta line in Fig. 6.4.2), the thrust jet function $J_T = J \otimes J$ combines two jet functions $J$ obtained from the $n$-collinear or $\bar{n}$-collinear matrix elements (dots on the blue line in Fig. 6.4.2), and the full soft function is defined from a vacuum matrix element of usoft Wilson lines. This soft function can be further factorized into two parts, $S_T \otimes F$, where $S_T$ is perturbative (green line in Fig. 6.4.2) and $F$ is nonperturbative (brown line in Fig. 6.4.2) [1800, 1868]. Renormalization group evolution of $H_T$, $J_T$, and $S_T$ enables a summation of large Sudakov double logarithms, $\alpha_s \ln^2 \tau$. The state-of-the-art for this resummation is next-to-next-to-next-to-leading logarithmic order (N$^3$LL), and was first achieved with SCET [1772]. For $\Lambda \ll Q\tau \ll Q$ the nonperturbative effects from $F$ are power corrections, so the spectrum is dominated by perturbation theory, and is used to obtain high precision fits for $\alpha_s(M_Z)$ [280, 281, 1869].

DIS, $e^- p \to e^- X$, provides another useful SCET factorization example [1767]. It is simplest to consider in the Breit frame where the virtual photon has $q^\mu = (0,0,0,Q)$. Here the factorization theorem is between hard and collinear modes with $\lambda = \Lambda/Q$, and soft contributions cancel out. A feature of this process is that the hard contributions come from both virtual effects as well as real radiation in $X$. Therefore matching onto Wilson coefficients $C^{i(0)}_{j\mathrm{DIS}}(\omega_i, Q, \mu)$ takes place at the level of the amplitude squared, and so does the construction of the appropriate SCET operators. These operators involve collinear quarks in $\bar\chi^q_{n,\omega_1} \bar{n}\!\!\!/ \chi^q_{n,\omega_2}$ with flavor $q$, or collinear gluons in $\mathcal{B}^\nu_{n\perp,\omega_1}\mathcal{B}_{n\perp,\omega_2\nu}$. The proton matrix elements $\langle p| \cdots |p\rangle$ of these operators define the well known quark parton distribution functions (PDFs) $f_{q/p}(\xi,\mu)$ and gluon PDFs $f_{g/p}(\xi,\mu)$, respectively. Carrying out the same steps listed above to arrive at Eq.(6.4.17) now gives factorization theorems for DIS structure functions. For example

$$W_1(x,Q^2) = \frac{-1}{x}\int_x^1 \mathrm{d}\xi\; H_1^{(i)}(\xi/x, Q, \mu)\, f_{i/p}(\xi,\mu^2)\,, \tag{6.4.18}$$

where we sum over parton types $i$, and the hard function $H_1^{(i)} = (1/\pi)\,\mathrm{Im}\, C^{i(0)}_{1\mathrm{DIS}}$. There is a similar formula for $W_2(x,Q^2)$. Eq.(6.4.18) factorizes perturbative short distance contributions in $H_1^{(i)}$ at the scale $Q$ from the nonperturbative PDFs $f_{i/p}$ at the scale $\Lambda$. Here the renormalization group evolution (RGE) sums up single logs $\alpha_s \ln(Q/\mu_0)$, for a hadronic scale $\mu_0 \simeq 1\,\mathrm{GeV} > \Lambda$. Thus SCET reproduces classic DIS results in a very simple manner. For example, the fact that $\bar\chi^q_{n,\omega_1} \bar{n}\!\!\!/ \chi^q_{n,\omega_2} \sim \lambda^2$ is related to the PDFs being built from twist-2 operators. The operator with Wilson lines in SCET captures the full tower of twist-2 operators simultaneously.

To provide a SCET$_{\mathrm{II}}$ example, we consider the Higgs transverse momentum $q_T$ in $pp \to H + X$ in the region where $Q = m_H \gg q_T \gg \Lambda$. Due to the measurement of $q_T \sim Q\lambda$ there is a restriction on the final state $X$. It can involve collinear and soft particles which individually have $p_T \sim Q\lambda$, but can no longer involve hard particles. Due to this restriction, the hard matching takes place at the amplitude level in this case, giving $\mathcal{L}^{(0)}_{\mathrm{hard}} \propto C^{(0)}_H(\omega_i,\mu)\mathrm{tr}[\mathcal{B}^\nu_{n\perp,\omega_1}\mathcal{S}^T_n \mathcal{S}_{\bar n}\mathcal{B}_{\bar n\perp,\omega_2\nu}]_\mu$, where $\mathcal{S}$ are soft Wilson lines in the adjoint representation. Since this only involves one field of each collinear type, the $\omega_i$ momenta are fixed by $Q$ and the Higgs rapidity $Y$ to be $\omega_1 = Qe^Y$ and $\omega_2 = Qe^{-Y}$. Here the factorization is simplest in Fourier space

$$\begin{aligned}
\frac{\mathrm{d}\sigma}{\mathrm{d}Q\mathrm{d}Y\mathrm{d}^2\vec p^{\,H}_T} &= 2H_{ggH}(Q,\mu)\int \mathrm{d}^2\vec b_T\, e^{i\vec b_T\cdot \vec p^{\,H}_T}\,\mathcal{S}_H(b_T,\mu,\nu) \\
&\times B^{\alpha\beta}_{g/p}(x_a,\vec b_T,\mu,\zeta_a/\nu^2) B_{g/p\alpha\beta}(x_b,\vec b_T,\mu,\zeta_b/\nu^2) \\
&= H_{ggH}(Q,\mu)\int \mathrm{d}^2\vec b_T\, e^{i\vec b_T\cdot \vec p^{\,H}_T} \\
&\quad\times \big[f_{1g/p}(x_a,b_T,\mu,\zeta_a) f_{1g/p}(x_b,b_T,\mu,\zeta_b) \\
&\qquad + h^\perp_{1g/p}(x_a,b_T,\mu,\zeta_a) h^\perp_{1g/p}(x_b,b_T,\mu,\zeta_b)\big]\,,
\end{aligned} \tag{6.4.19}$$

where $x_a = Qe^Y/\sqrt{s}$, $x_b = Qe^{-Y}/\sqrt{s}$, $s$ is the invariant mass of the colliding protons, and $\zeta_{a,b}$ are Collins-Soper parameters satisfying $\zeta_a\zeta_b = Q^4$. Here the hard function is $H_{ggH} \propto |C^{(0)}_H|^2$ (leaving out simple kinematic prefactors), the squared $\langle p|\cdots|p\rangle$ matrix element of $n$-collinear fields yields the beam function $B^{\alpha\beta}_{g/p}$ (and likewise for $\bar n$), and the squared vacuum matrix element of soft Wilson lines yields the soft function $\mathcal{S}_H$. In the final line of Eq.(6.4.19) we did two things in one step: i) grouped a $\sqrt{\mathcal{S}_H}$ together with each beam function to absorb the soft function symmetrically, and ii) decomposed the Lorentz indices $\alpha\beta$ into two possible structures, $g^{\alpha\beta}_T f_{1g/p}$ and $(b^\alpha_T b^\beta_T + \vec b^{\,2}_T g^{\alpha\beta}_T/2)h^\perp_{1g/p}$. This yields definitions for the TMD PDFs $f_{1g/p}$ (unpolarized gluon TMD PDF) and $h^\perp_{1g/p}$ (linearly polarized gluon TMD PDF).

A novel feature of this factorization theorem is the appearance of the rapidity scale $\nu$ in the collinear and soft functions, which is associated to the need to regulate rapidity divergences in many SCET$_{\mathrm{II}}$ processes [1850, 1870–1873], and the presence of the associated rapidity renormalization group equations [1872, 1874]. The result in the first line of Eq.(6.4.19) is presented with the rapidity regulator defined in [1872] and may look somewhat different with other choices of the rapidity regulator, such as in the original Collins construction [1269]. However the result in the final line will be the same. Evolution in both $\mu$ and $\nu$ is needed to sum the large logs, $\alpha_s \ln^2(Q/q_T)$, in this process, and the state of the art is resummation at N$^3$LL. This resummation may also be done at the level of the TMD PDFs, where the rapidity RGE is replaced by the Collins-Soper evolution in $\zeta_{a,b}$ [1347].

As our final example, we consider the measurement of jet mass in inclusive jet production, $pp \to$jet+$X$, where the jet has radius $R$ and is defined with the anti-$k_T$ algorithm. To make this example more interesting (and more phenomenologically relevant) we also carry out jet grooming to remove soft contaminating radiation in the jet, using the soft drop algorithm [1875, 1876]. Examples of contaminating radiation in the jet include initial state radiation from the protons, underlying event effects due to radiation from spectator partons, and pileup effects due to radiation from the interaction among other protons in the colliding beams. The soft-drop grooming is defined by iteratively applying a test on transverse momentum $p_T$ and angular separations $\varDelta R$ of branches $i$ and $j$ in an angular ordered tree formed from particles in the jet: $\min(p_{Ti}, p_{Tj})/(p_{Ti} + p_{Tj}) > z_{\rm cut}(\varDelta R_{ij}/R_0)^\beta$ where $z_{\rm cut}$, $\beta$, and $R_0$ are soft drop parameters. Branches that fail this test are removed from the tree, thus grooming soft radiation. This causes the soft function for this process to split itself into two parts [1786]: a global soft function sensitive to the scale $Qz_{\rm cut}$ associated with the groomed soft radiation, and a collinear-soft function, $S_c^\kappa$, that describes soft radiation that is collimated enough with the jet axis to have been retained by the grooming. The groomed jet mass cross section can be factorized as [1786, 1877, 1878]

$$\frac{{\rm d}\sigma}{{\rm d}m_J^2{\rm d}\varPhi_J} = N_\kappa(\varPhi_J, R, z_{\rm cut}, \beta, \mu) Q_{\rm cut}^{\frac{1}{1+\beta}} \int {\rm d}\ell \times J_\kappa(m_J^2 - Q\ell, \mu) S_c^\kappa\left[\ell Q_{\rm cut}^{\frac{1}{1+\beta}}, \beta, \mu\right], \tag{6.4.20}$$

with a sum on $\kappa = q, g$ for quark and gluon jets and $Q_{\rm cut} = p_T R z_{\rm cut}(R/R_0)^\beta$. Here $J_\kappa$ is the usual jet function since the collinear radiation is not affected by the grooming. The normalization factor $N_\kappa$ is a short hand for a combination of terms that include PDFs, a hard-collinear function describing the production of the parton $\kappa$, and the global soft function. This is an example of a SCET$_+$ factorization formula due to the presence of soft-collinear modes that make up the Wilson lines that appear in $S_c^\kappa$. Groomed observables have become widely used in predictions at hadron colliders due to the fact that they are much more robust to contamination, and have reduced hadronization corrections. Other examples of soft drop groomed calculations with SCET are found in Refs. [1788, 1790, 1878–1893].

### 6.4.5 State-of-the-Art and Attractive Directions

The nature of a short review is that key ideas can be highlighted, but it is hard to do credit to the depth of work in the field. Let me close by giving a brief overview of some of the interesting centers of activity currently going on with SCET, with an eye to the future.

SCET continues to have a significant impact on the field of high precision calculations for collider cross sections, in particular for the resummation of large logarithms. This activity is motivated by the clear universality of anomalous dimensions and factorized functions in SCET, giving results of broad utility. Below I summarize the highest order results achieved to date for various processes which exploit these perturbative achievements, referring to references therein for further background and details. This list includes: $e^+e^-$ thrust to N$^3$LL$'$ [280, 1772] and massive thrust to N$^3$LL [1894], $e^+e^-$ heavy jet mass to N$^3$LL [1895], $e^+e^-$ C-parameter to N$^3$LL$'$ [1896], $e^+e^-$ Energy-Energy-Correlator (EEC) to N$^3$LL$'$ [1897], $e^+e^-$ oriented event shapes to N$^3$LL [1898], $e^+e^-$ groomed jet mass to N$^3$LL [1899], $e^+e^- \to t\bar{t}$ thrust to N$^3$LL [1900], $e^-p$ DIS thrust to N$^3$LL [1901–1904], the Drell-Yan $p_T^{\ell\ell}$ spectrum to N$^3$LL$'$ [1905, 1906], the $pp$ Higgs $p_T^H$ spectrum [1907–1909] and rapidity spectrum [1910] to N$^3$LL$'$, and LHC processes with a jet-veto [1773, 1911–1916]. Recently the first N$^4$LL resummed calculation has been carried out for the EEC [1917] (with an approximation for the 5-loop cusp anomalous dimension). Key ingredients are the four-loop hard (collinear) anomalous dimensions [1918, 1919], the four-loop rapidity anomalous dimension for TMDs [1917, 1920], the four-loop cusp anomalous dimension [1921] and five-loop approximation [1922], and calculations of three loop boundary conditions [1923–1926]. Many more processes have been resummed to NNLL or NNLL$'$ order with SCET; for example in Refs. [1474, 1774, 1775, 1778, 1878, 1891, 1923, 1927–1959]. Factorized functions remain important targets for future perturbative calculations, with the anticipated reward of simultaneously impacting multiple processes.

Power corrections are another lively topic in SCET, from the continued activity around $B$-decays, to recent significant results for collider physics. A key strength of SCET is its systematic nature, ensuring one can target the desired terms without missing contributions. Recent collider physics literature on subleading power results in SCET includes: formalism such as enumerating operator bases [1846–1848, 1960, 1961], hard renormalization and evolution [1960, 1962–1964], collinear and soft renormalization and evolution [1965–1968], subleading power factorization [1768, 1969–1973], and resummation for collider observables, including for event shapes [1965, 1974, 1975], for threshold resummation [1976–1978], and for the EEC [1979]. These results provide bright prospects for the future, with the ultimate goal of building a complete story for the structure of gauge theories like QCD beyond leading power, and thus generalizing the leading power picture of collinear splittings and soft eikonal radiation.

One popular method for carrying out fixed order calculations at higher orders, is that of slicing, whereby a resolution variable is used to act as a physical regulator for infrared divergences, enabling analytic and numerical calculations to be combined in a systematic way. SCET has contributed to this program with the invention of N-jettiness subtractions [1980, 1981] based on the N-jettiness event shape variable [1776]. It has also been used to calculate power suppressed large logarithms, enabling order-of-magnitude improvements to slicing techniques [1849, 1966, 1982–1989]. Further improvements to such techniques will be important as theorists continue to move towards calculating experimentally accessible fiducial cross sections.

Other interesting applications of SCET include: the generalization of threshold factorization formulae to include collinear limits [1990, 1991], the computation of non-global logarithms and associated effects [1784, 1992–2005], the parametrization of hadronization corrections with field theory matrix elements [2006–2008], studying fragmentation inside a jet [1881, 1885, 1957, 2009–2021], and to studying double parton distributions and fragmentation [2022–2025]. A particularly interesting direction with many connections to other fields is the study of EECs. Results from SCET include deriving factorization for the back-to-back limit [2026], and collinear limit [2027, 2028], jet analyses with charged tracks [2029], generalizing factorization to the back-to-back limit at hadron colliders [2030], and deriving factorization formula for jet substructure applications of the EEC [2031]. The prospects for new applications of SCET technology remain bright.

A final hallmark of SCET is the use of the physical picture it provides to construct novel observables. Past examples of this type include: beam thrust and functions [1804, 2032, 2033], N-jettiness [1776], N-subjettiness [2034], jet substructure for disentangling color and spin in $J/\Psi$ production [2014, 2035], $D_2$ and related jet-substructure observables [1781, 2036–2038], the winner-takes-all-axis for jets [1889, 2039–2041], track functions [2042–2045], the XCone jet algorithm [2046], collinear drop [1887, 2047], an EEC probe of top mass [2048], and measuring initial state tomography with a Nuclear EEC [2049]. I look forward to many more examples of such new observables in the future.

## 6.5 Hard thermal loop effective theory

**Michael Strickland**

In this section we review progress in understanding QCD at finite temperature and density. Unlike QCD in vacuum new classes of physical infrared divergences appear which cause naive perturbation theory to break down. Luckily, at least at leading order in the coupling constant, it is possible to identify a class of diagrams that must be resummed in order to cure these divergences.

### 6.5.1 The breakdown of naive perturbation theory at finite temperature

There are two fundamental formalisms for computing the properties of QCD at high temperature: (1) the real-time formalism and (2) the imaginary-time formalism [2050–2052]. The former is necessary when considering systems that are out of equilibrium, while the second is more convenient for computing bulk thermodynamic quantities. Here we will focus on the imaginary-time formalism and progress that has been made in understanding how to reorganize the perturbative expansion of finite temperature QCD in order to deal with infrared singularities which emerge in this case using self-consistent inclusion of Debye screening and Landau damping. This is accomplished through an all-orders resummation of a class of diagrams referred to as the hard-thermal-loop (HTL) diagrams. For an introduction to the real-time formalism and applications to real-world calculations we refer the reader to Sec. 6.6 and Ref. [2051].

In thermal and chemical equilibrium with temperature $T$ and quark chemical potentials $\mu_i$ with $\pi T \gg \mu_i$, one finds that the naive loop expansion of physical quantities is ill-defined and diverges beyond a given loop order, which depends on the quantity under consideration. In the calculation of QCD thermodynamics, this stems from uncanceled infrared (IR) divergences that enter the expansion of the partition function at three-loop order. These IR divergences are due to long-distance interactions mediated by static gluon fields and result in contributions that are non-analytic in $\alpha_s = g^2/4\pi$, e.g., $\alpha_s^{3/2}$ and $\log(\alpha_s)$, unlike vacuum perturbation expansions which involve only powers of $\alpha_s$.

A simple way to understand at which perturbative orders terms non-analytic in $\alpha_s$ appear is to start from the contribution of non-interacting static gluons to a given quantity. For the pressure of a gas of gluons one has $P_{\rm gluons} \sim \int d^3p\, p\, n_B(E_p)$, where $n_B$ denotes a Bose-Einstein distribution function and $E_p$ is the energy of the in-medium gluons. The contributions from the momentum scales $\pi T$, $gT$ and $g^2T$ can be expressed as

$$P^{p\sim\pi T}_{\text{gluons}} \sim T^4 n_B(\pi T) \;\sim\; T^4 + \mathcal{O}(g^2)\,, \tag{6.5.1}$$
$$P^{p\sim gT}_{\text{gluons}} \sim (gT)^4 n_B(gT) \;\sim\; g^3T^4 + \mathcal{O}(g^4)\,, \tag{6.5.2}$$
$$P^{p\sim g^2T}_{\text{gluons}} \sim (g^2T)^4 n_B(g^2T) \;\sim\; g^6T^4\,, \tag{6.5.3}$$

where we have using the fact that $n_B(E) \sim T/E$ if $E \ll T$. This fact is of fundamental importance since it implies that when the energy/momentum are *soft*, corresponding to electrostatic contributions, $p_{\text{soft}} \sim gT$, one receives an enhancement of $1/g$ compared to contributions from *hard* momenta, $p_{\text{hard}} \sim T$, due to the bosonic nature of the gluon. For *ultrasoft* (magnetostatic) momenta, $p_{\text{ultrasoft}} \sim g^2T$, the contributions are enhanced by $1/g^2$ compared to the naive perturbative order. As the Eqs. (6.5.1)-(6.5.3) demonstrate, it is possible to generate contributions of the order $g^3 \sim \alpha_s^{3/2}$ from soft momenta and, in the case of the pressure, although perturbatively enhanced, ultrasoft momenta only start to play a role at order $g^6 \sim \alpha_s^3$.

Note that the expansion parameters in Eqs. (6.5.1) to (6.5.3) are of order $g^2 n_B(\pi T) \sim g^2$, $g^2 n_B(gT) \sim g$, and $g^2 n_B(g^2T) \sim 1$, implying in particular that the contribution of magnetostatic gluons to the pressure is fundamentally non-perturbative in nature at $\mathcal{O}(\alpha_s^3)$, which for the pressure corresponds to four-loop order. This complete breakdown of the loop expansion at the ultrasoft scale is called the *Linde problem* [2053, 2054]. The specific order at which the expansion breaks down depends on the quantity under consideration and is not universal. For example, in Ref. [2055], the authors demonstrated that a certain second-order transport coefficient, $\lambda_1$, receives a leading-order contribution from the ultrasoft scale. We also note that, in the case of the $\mathcal{O}(\alpha_s^3)$ contribution to the pressure it is possible to isolate the purely non-perturbative contribution and compute this numerically using a three-dimensional lattice calculation [2056]. Paradoxically, the difficult part then becomes computing the perturbative contributions at this order [2057]. Beyond four-loop order, all contributions are once again perturbatively computable.

As a result of the infrared enhancement of electrostatic contributions it was shown that a class of diagrams called hard-thermal-loop (HTL) graphs which have soft external and hard internal momenta need to be resummed to all orders in the strong coupling coupling [2058–2060]. In the high temperature limit, there exist several schemes for carrying out such resummations, see e.g. [2061–2079]. Here we will briefly review the method of dimensionally reduced effective theories (EFTs), which take advantage of the scale hierarchies and the manifestly gauge-invariant hard thermal loop perturbation theory (HTLpt) resummation. This makes use of the HTL effective action to reorganize the perturbative expansion of finite temperature and density QCD [2080].

### 6.5.2 Dimensional reduction and QCD EFT

The method of dimensional reduction is based on the fact that, at weak-coupling, there is a hierarchy of scales between the three energy scales (hard, soft, and ultrasoft or, equivalently, hard, electric, and magnetic) which contribute to bulk thermodynamic observables. Specifically, if $g \ll 1$, one has

$$m_{\text{magnetic}} \sim g^2T \ll m_{\text{electric}} \sim gT \ll m_{\text{hard}} \sim \pi T\,. \tag{6.5.4}$$

Above we have denoted the magnetostatic and electrostatic screening scales by $m_{\text{magnetic}}$ and $m_{\text{electric}}$, respectively, and the hard or thermal one, corresponding to the lowest non-zero Matsubara frequency, by $m_{\text{hard}}$. To leading order, the electrostatic screening mass can be computed from the IR limit of the $A_0$ one-loop self energy, however, the magnetic screening mass cannot be computed perturbatively [2064, 2065]. In the high temperature limit with $\pi T \gg \mu_i, m_i, \Lambda_{\text{QCD}}$, the above three scales are the only ones appearing and the two non-trivial scales $m_{\text{magnetic}}$ and $m_{\text{electric}}$ are connected to the static sector corresponding to the zero Matsubara mode ($n = 0$). As a result, in the effective field theory language it is natural to integrate out the hard scale, yielding a three-dimensional effective field theory which is valid for long-distance static field modes. Another way to see this is recognize that, in four-dimensional Euclidean space, a system in thermal equilibrium has its time direction compactified to a circle of radius $1/T$ [2050]. In the high-temperature limit, the Euclidean time direction has zero extent and the parent field theory becomes effectively three dimensional. Since fermionic modes have odd Matsubara frequencies, they become super massive and decouple from the theory in this limit, as do all non-zero gluonic Matsubara modes.

The construction of dimensionally reduced effective theories for high-temperature field theory began with the work of Ginsparg [2081] and was quickly followed by Appelquist and Pisarski [2082]. In the mid-1990s, Kajantie et al. were the first to apply this formalism to the study of the electroweak phase transition [2083]. Around the same time Braaten and Nieto demonstrated how to apply these ideas to thermal QCD [2064, 2065]. Recently, these methods have been extended to the computation of the thermodynamics of $\mathcal{N} = 4$ supersymmetric Yang-Mills theory to order $\lambda^2$, where $\lambda = g^2N_c$ is the t'Hooft coupling [2084].

In the EFT technique, the Lagrangian densities of the three- and four-dimensional theories can be obtained by writing down the most general local Lagrangians respecting all necessary symmetries. One then orders all operators in terms of their dimensionality and truncates the Lagrangians at the desired order. For electrostatic QCD (EQCD), this procedure results in [2064, 2065]

$$\begin{aligned}\mathcal{L}_\mathrm{E} = &\frac{1}{2}\mathrm{Tr}\mathbf{F}_{ij}^2 + \mathrm{Tr}[\mathbf{D}_i, \mathbf{A}_0]^2 + m_\mathrm{E}^2\mathrm{Tr}\mathbf{A}_0^2 \\ &+\lambda_\mathrm{E}^{(1)}(\mathrm{Tr}\mathbf{A}_0^2)^2 + \lambda_\mathrm{E}^{(2)}\mathrm{Tr}\mathbf{A}_0^4 \\ &+i\lambda_\mathrm{E}^{(3)}\mathrm{Tr}\mathbf{A}_0^3 + \cdots\,, \end{aligned} \tag{6.5.5}$$

where the adjoint fields $\mathbf{A}_i \equiv A_i^A T^A$, $\mathbf{A}_0 \equiv A_0^A T^A$ are three dimensional, $F_{ij}^A = \partial_i A_j^A - \partial_j A_i^A + g_\mathrm{E} f^{ABC} A_i^B A_j^C$, $\mathbf{F}_{ij} \equiv F_{ij}^A T^A$, and $\mathbf{D}_i = \partial_i - i g_\mathrm{E}\mathbf{A}_i$. Integrating out the temporal gauge field, one can obtain the magnetostatic effective theory (MQCD) with $\mathcal{L}_\mathrm{M} = \frac{1}{2}\mathrm{Tr}\mathbf{F}_{ij}^2 + \cdots$ and $F_{ij}^A = \partial_i A_j^A - \partial_j A_i^A + g_\mathrm{M} f^{ABC} A_i^B A_j^C$ [2064, 2065].

At leading order in $g$, the degrees of freedom in the above effective theories are the $n = 0$ Matsubara modes of the four-dimensional $A_i$ and $A_0$ fields. The former transforms as a three-dimensional adjoint gauge field and the latter as a scalar in the adjoint representation of SU($N_c$). By computing the contributions from the hard scale in the four-dimensional theory (non-resummed), the massless two-loop self-energy in the four-dimensional theory, and the massive three-loop vacuum graphs and matching the two theories, one obtains the following result for the QCD free energy through $\mathcal{O}(\alpha_s^{5/2})$

$$\begin{aligned}\frac{\mathcal{F}_\mathrm{QCD}}{\mathcal{F}_\mathrm{ideal}} = &1 - \frac{15}{4}\frac{\alpha_s}{\pi} + 30\left(\frac{\alpha_s}{\pi}\right)^{3/2} \\ &+\frac{135}{2}\left(\log\frac{\alpha_s}{\pi} - \frac{11}{36}\log\frac{\Lambda}{2\pi T} + 3.51\right)\left(\frac{\alpha_s}{\pi}\right)^2 \\ &+\frac{495}{2}\left(\log\frac{\Lambda}{2\pi T} - 3.23\right)\left(\frac{\alpha_s}{\pi}\right)^{5/2} + \mathcal{O}(\alpha_s^3\log\alpha_s)\,,\end{aligned} \tag{6.5.6}$$

where $\mathcal{F}_\mathrm{ideal} = -(8\pi^2/45)T^4$ is the free energy of an ideal gas of massless gluons and $\alpha_s = \alpha_s(\Lambda)$ is the running coupling constant in the $\overline{\mathrm{MS}}$ scheme. Note, importantly, the appearance of non-analytic terms in $\alpha_s$. Logarithms of $\alpha_s$ appear as ratios of the electric screening scale over the temperature. In order to avoid notational overlap with the chemical potential $\mu$, here $\Lambda$ is used to indicate the renormalization scale. There is also a residual dependence on the renormalization scale $\Lambda$ at orders $\alpha_s^2$ and $\alpha_s^{5/2}$. The result obtained when this expression is truncated at various orders in the coupling constant is shown in Fig. 6.5.1. As can be seen from this figure, at phenomenologically relevant temperatures the resulting weak coupling expansion shows poor convergence and an increasing sensitivity to the renormalization scale as the perturbative truncation order is increased. The reason for this poor convergence is that one is expanding around the $T = 0$ QCD vacuum, which does not include the effects of Debye screening and Landau damping. In order to improve the convergence of this series, HTLpt was introduced to reorganize the calculation instead around the $T \to \infty$ limit. We will discuss this reorganization in the next subsection.

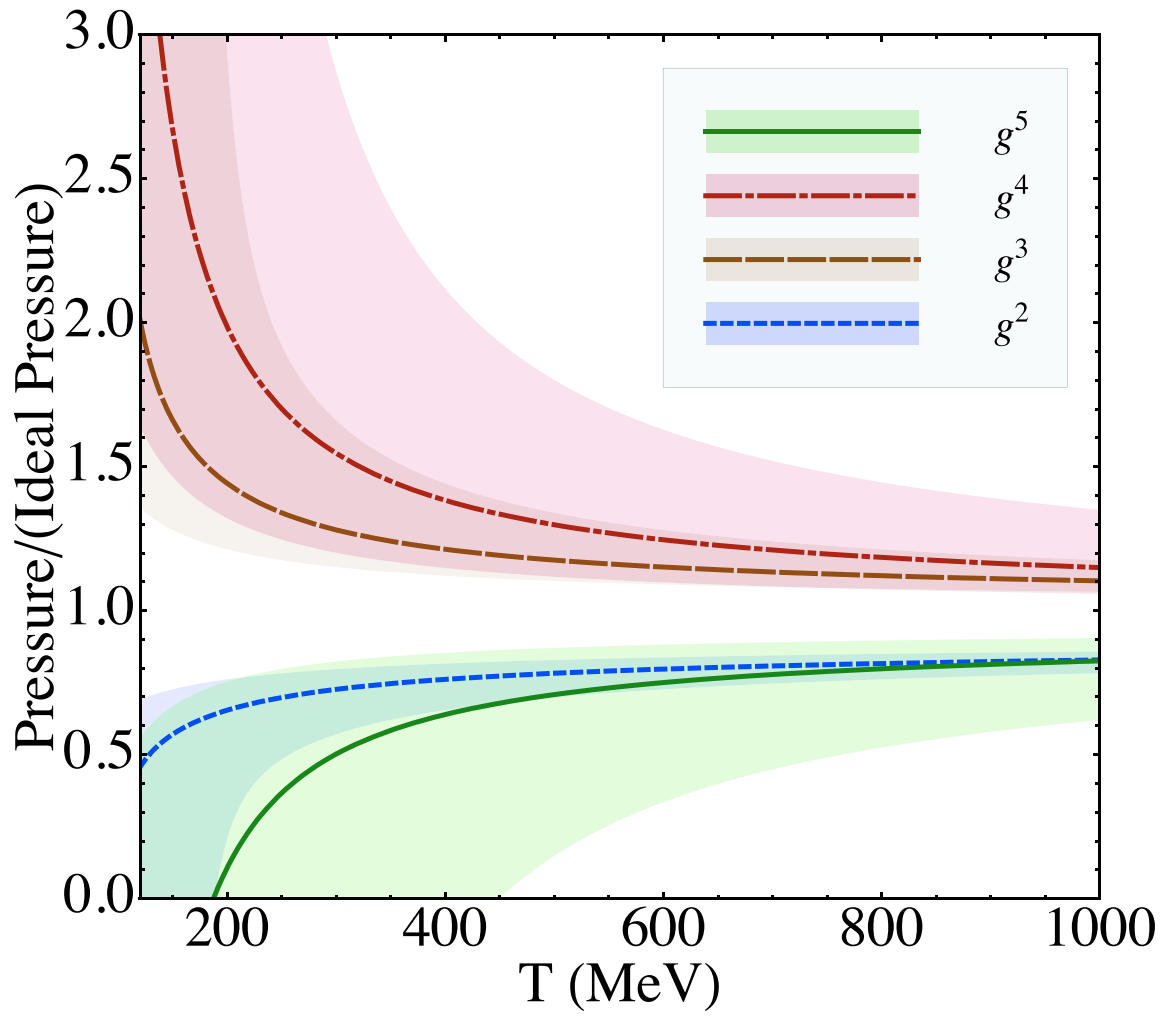

**Fig. 6.5.1** Naive weak-coupling expansion of the scaled QCD pressure for $N_f = 3$. Shaded bands show the result of varying the renormalization scale $\Lambda$ by a factor of 2 around the central renormalization scale $\Lambda = 2\pi T$.

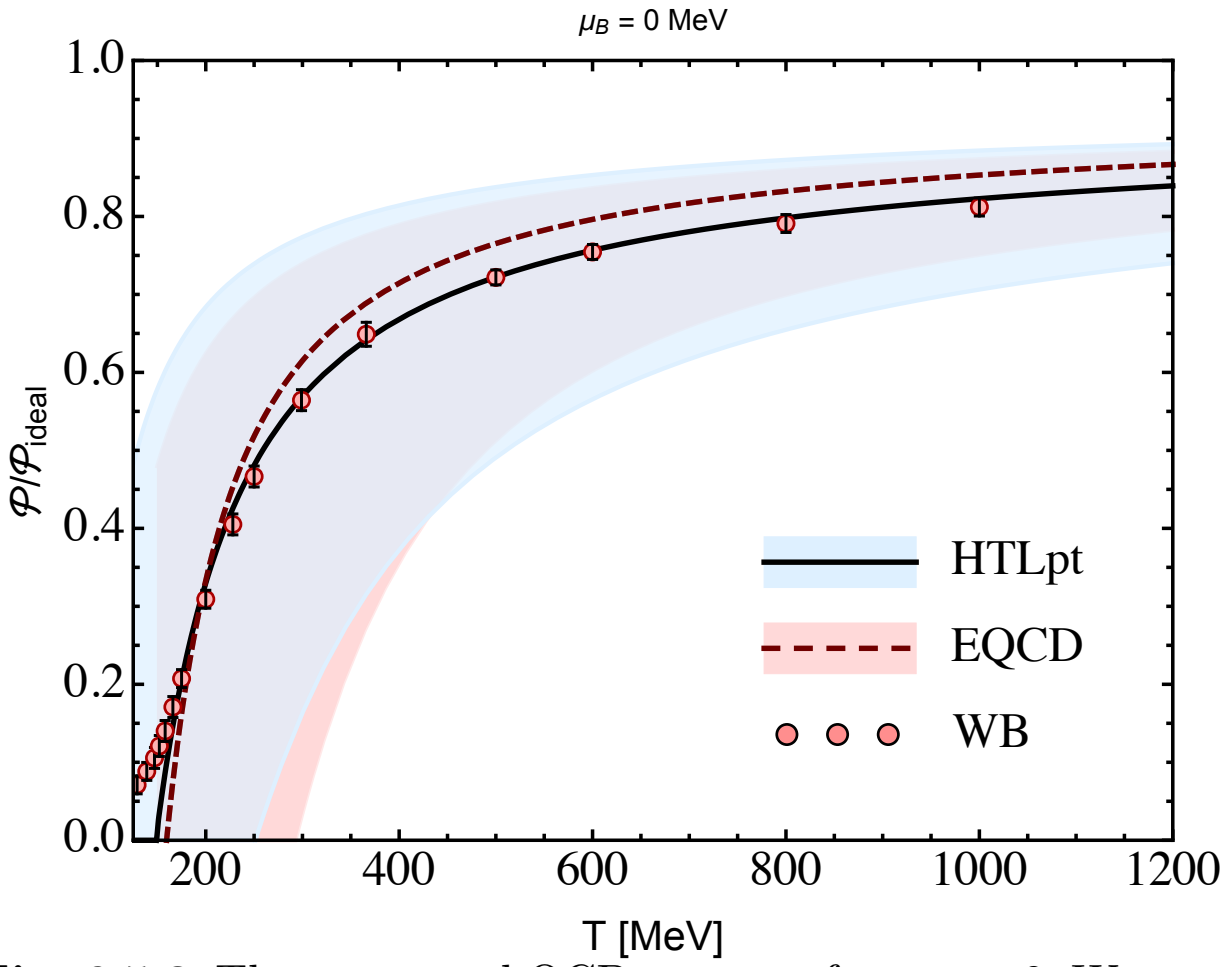

**Fig. 6.5.2** The resummed QCD pressure for $\mu_B = 0$. We compare the three-loop EQCD and HTLpt results with lattice data from the Wuppertal-Budapest (WB) collaboration [2085].

### 6.5.3 Hard-thermal loop perturbation theory

Hard-thermal-loop perturbation theory is a reorganization of perturbative QCD. The HTLpt Lagrangian density is written as [2074, 2075]

$$\mathcal{L} = (\mathcal{L}_\mathrm{QCD} + \mathcal{L}_\mathrm{HTL})\big|_{g\to\sqrt{\delta}g} + \Delta\mathcal{L}_\mathrm{HTL}\,, \tag{6.5.7}$$

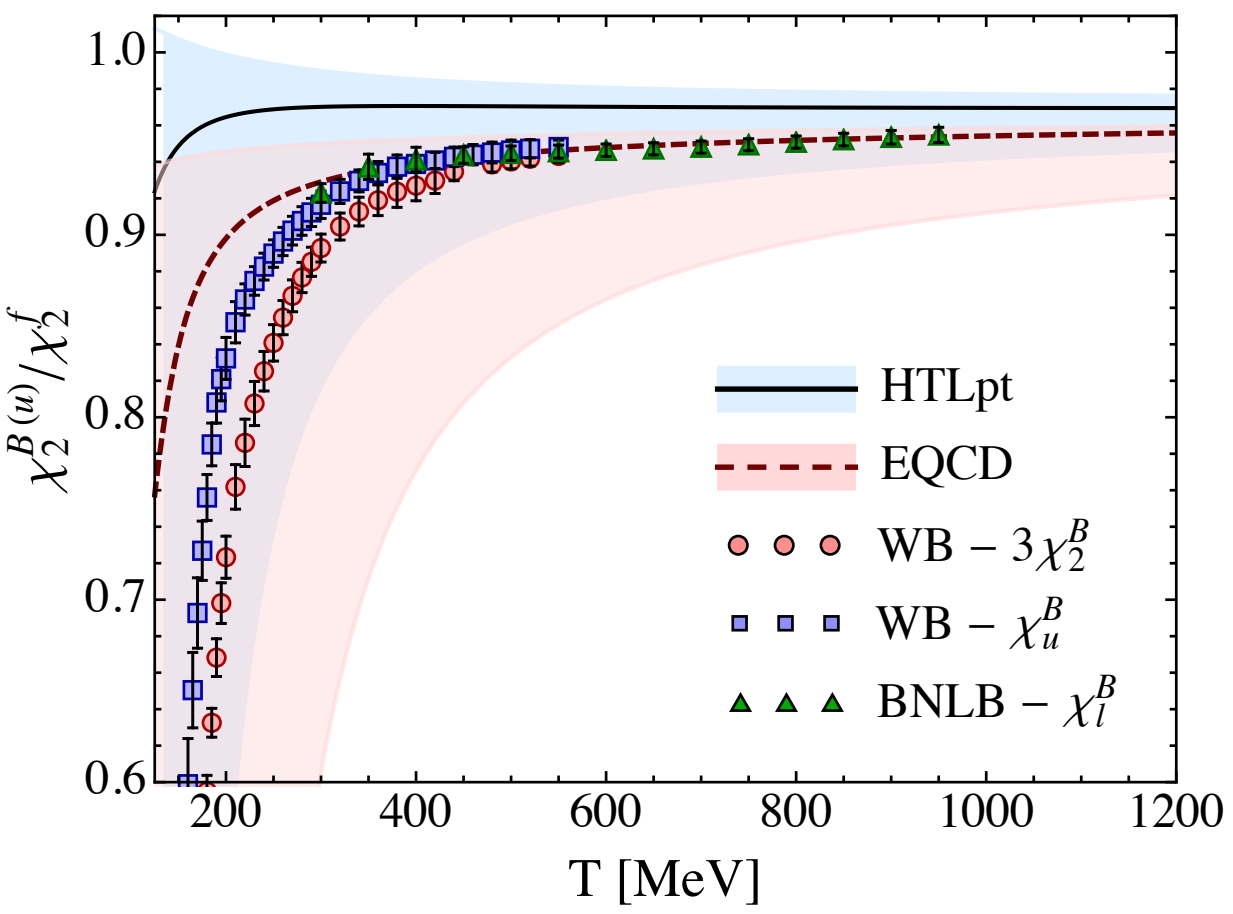

**Fig. 6.5.3** The second-order light quark (and baryon) number susceptibilities. Lattice data are from the Wuppertal-Budapest (WB) [2086, 2087] and BNLB collaborations [2088].

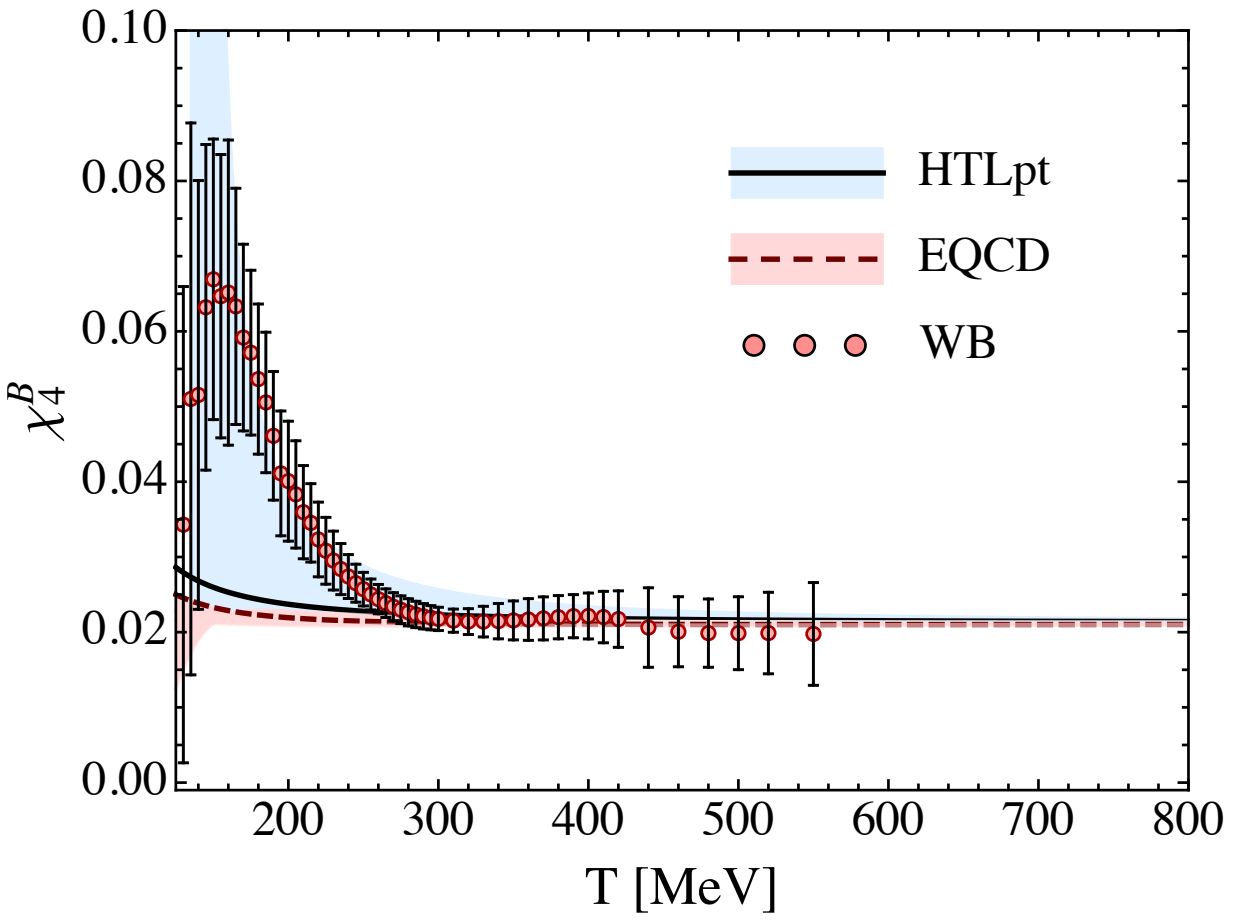

**Fig. 6.5.4** The 4th baryon number susceptibility. Lattice data sources are the same as Fig. 6.5.3.

where $\Delta\mathcal{L}_{\rm HTL}$ collects all necessary renormalization counterterms and $\delta$ is a formal expansion parameter, which will be taken to be unity in the end of the calculation. The HTL improvement term appearing above is

$$\mathcal{L}_{\rm HTL} = (1-\delta) i m_q^2 \bar{\psi}\gamma^\mu \left\langle \frac{y_\mu}{y\cdot\mathbf{D}} \right\rangle_{\hat{\mathbf{y}}} \psi \\ -\frac{1}{2}(1-\delta) m_D^2 \mathrm{Tr}\left[\mathbf{F}_{\mu\alpha} \left\langle \frac{y^\alpha y_\beta}{(y\cdot\mathbf{D})^2} \right\rangle_{\hat{\mathbf{y}}} \mathbf{F}^{\mu\beta}\right]. \quad (6.5.8)$$

Above $y^\mu = (1, \hat{\mathbf{y}})$ is a light-like four-vector with $\hat{\mathbf{y}}$ being a three-dimensional unit vector and the angular bracket indicates an average over the direction of $\hat{\mathbf{y}}$. The parameters $m_D$ and $m_q$ can be identified with the gluonic screening mass and the thermal quark mass. In HTLpt one treats $\delta$ as a formal expansion parameter. By including the HTL improvement term (6.5.8) HTLpt shifts the perturbative expansion from being around an ideal gas of massless particles to being around a gas of massive quasiparticles. This shift dramatically improves the convergence of the successive loop approximations to QCD thermodynamics [2067–2069, 2074–2079].

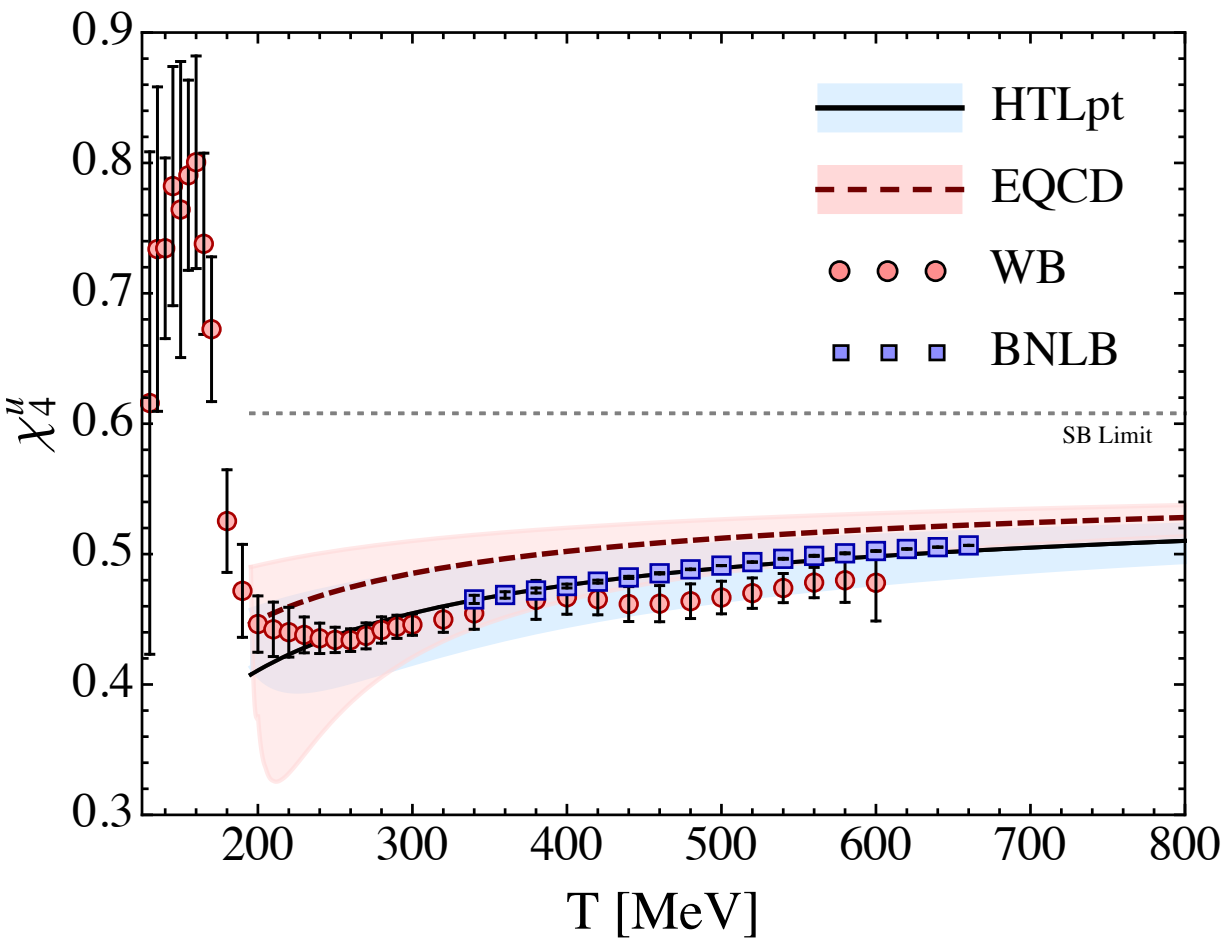

**Fig. 6.5.5** The 4th light quark number susceptibility. Lattice data sources are the same as Fig. 6.5.3.

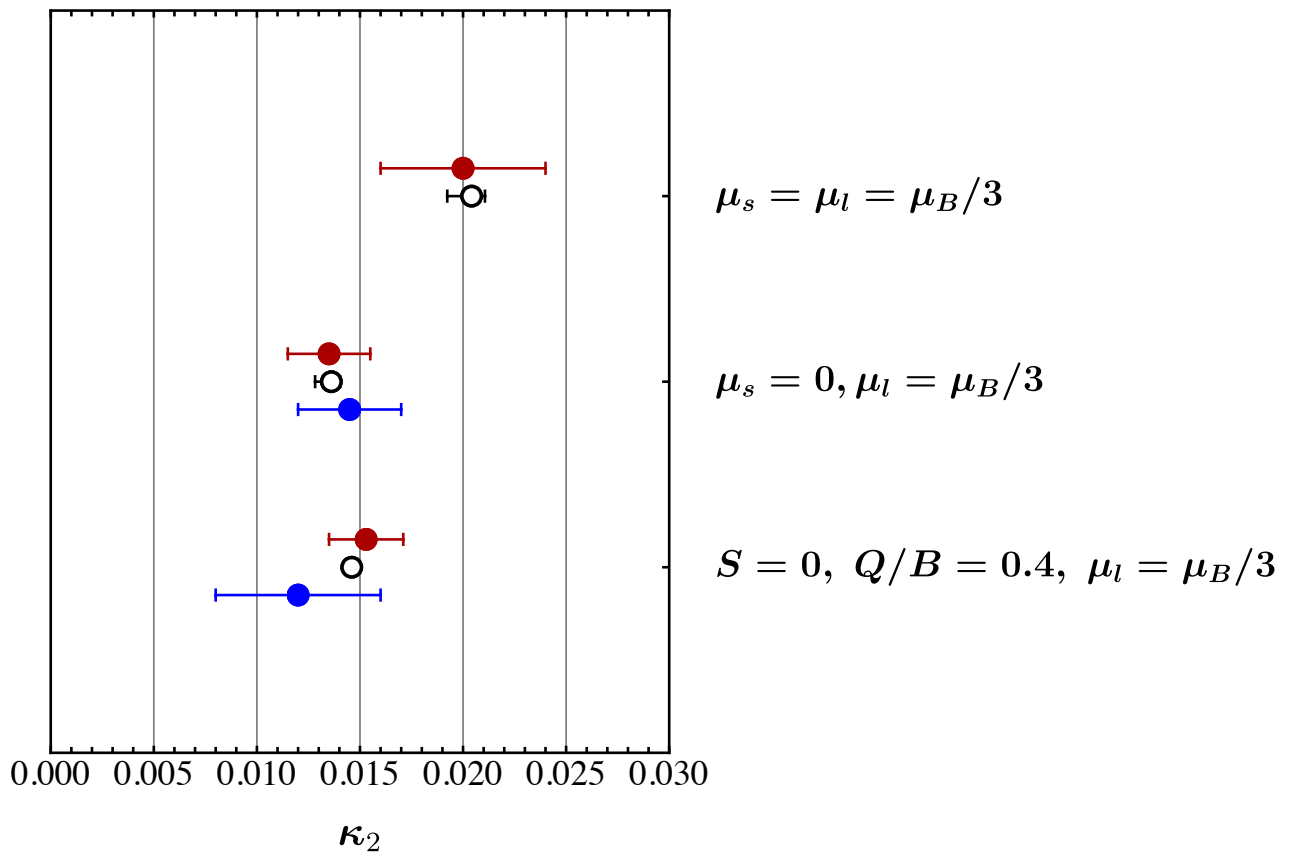

**Fig. 6.5.6** Filled circles are lattice calculations of $\kappa_2$ [449–451, 2089, 2090], from top to bottom, respectively. Red filled circles are results obtained using the imaginary chemical potential method and blue filled circles are results obtained using Taylor expansions around $\mu_B = 0$. Black open circles are the NNLO HTLpt predictions. The error bars associated with the HTLpt predictions result from variation of the assumed renormalization scale.

The HTLpt Lagrangian (6.5.7) reduces to the QCD Lagrangian when $\delta = 1$. Physical observables are calculated in HTLpt by expanding in powers of $\delta$, truncating at some specified order in $\delta$, and then setting $\delta = 1$. This defines a reorganization of the perturbative series in which the effects of $m_D^2$ and $m_q^2$ terms in (6.5.8) are included to leading order but then systematically subtracted out at higher orders in perturbation theory by the $\delta m_D^2$ and $\delta m_q^2$ terms in (6.5.8). To obtain leading order (LO), next-to-leading order (NLO), and next-to-next-leading order (NNLO) results for the QCD pressure, one expands to orders $\delta^0$, $\delta^1$, $\delta^2$, respec-tively. Note, importantly, that HTLpt is gauge invariant order-by-order in the $\delta$ expansion.

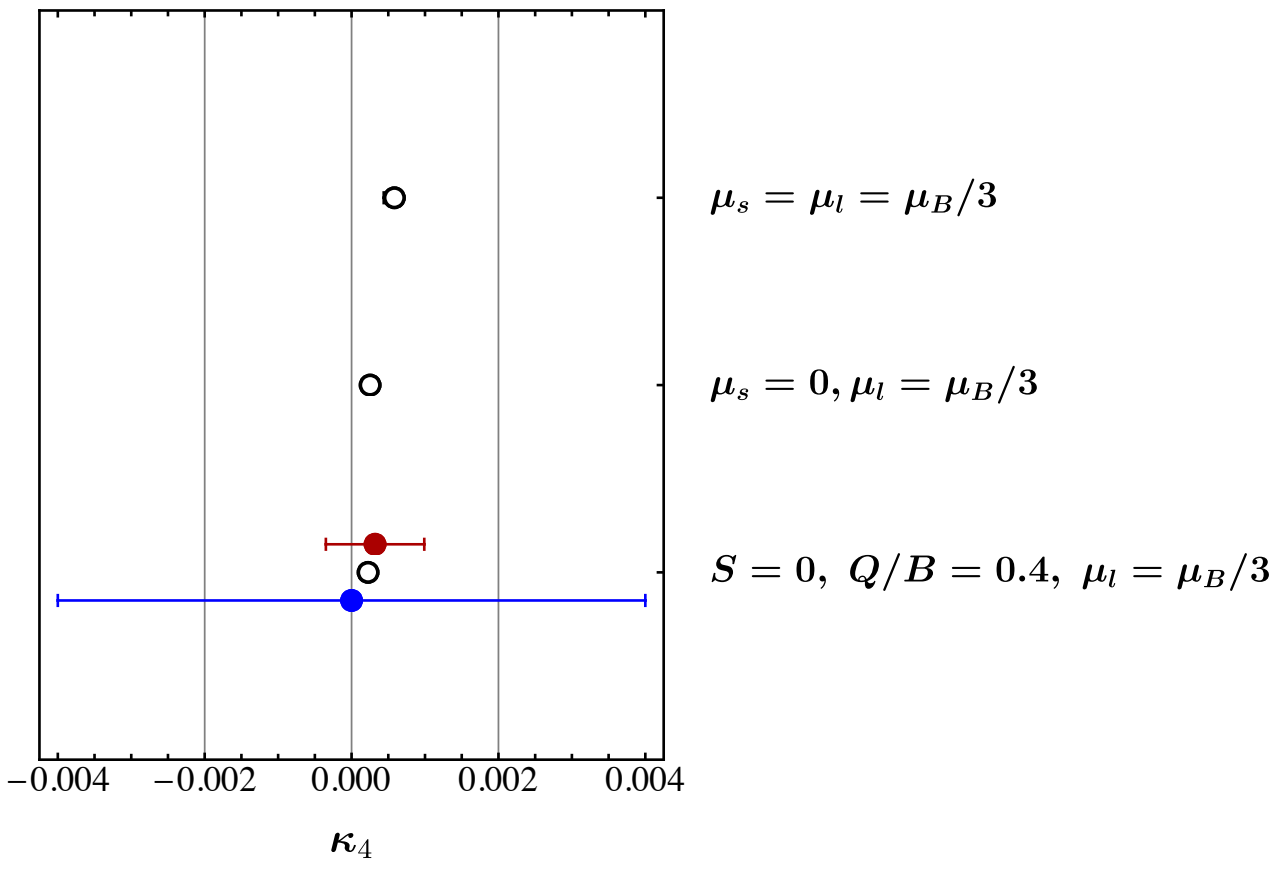

**Fig. 6.5.7** Filled circles are lattice calculations of $\kappa_4$ from Refs. [451, 2090], from top to bottom, respectively. The color coding etc. for the symbols is the same as in Fig. 6.5.6.

In order to obtain analytically tractable sum-integrals, in addition to the $\delta$ expansion, one must also make a Taylor expansion in the mass parameters scaled by the temperature, $m_D/T$ and $m_q/T$. The final result obtained at NNLO is completely analytic, however, it is too lengthy to list here, instead we refer the reader to the most recent works using HTLpt, which apply this technique at finite temperature and quark chemical potentials [2078, 2079, 2091]. In Fig. 6.5.2 we compare the NNLO EQCD and HTLpt results for the scaled pressure (negative of the free energy). As can be seen from this figure, for the central choice of the renormalization scale, namely $\Lambda_g = 2\pi T$ and $\Lambda_q = \pi T$, there is excellent agreement between HTLpt and the lattice data. The same is true to a lesser extent for EQCD. Both, however, have a large uncertainty related to the variation with respect to the renormalization scale. This sensitivity is particularly large for the free energy; however other quantities show much less renormalization scale dependence. From the NNLO results, one can obtain predictions for various quark and baryon number susceptibilities.

In Figs. 6.5.3-6.5.5 we present the NNLO resummed perturbative predictions for the second-order baryon number susceptibility, the fourth-order baryon number susceptibility, and the fourth-order light quark susceptibility, respectively. As these figures demonstrate HTLpt and EQCD to a only slightly lesser extent, have reasonable agreement with lattice extractions of these susceptibilities down to temperatures on the order of $T \sim$ 250 MeV which is only slightly higher than the QCD phase transition temperature of $T_c \sim$ 155 MeV. The lone exception is $\chi_2^{B(u)}$ where EQCD seems to perform better than HTLpt, although the results are consistent within the scale uncertainties. Finally, in Figs. 6.5.6 and 6.5.7 we present results recently presented in Ref. [2091] for the second- and fourth-order curvatures of the QCD phase transition line obtained from the analytical NNLO HTLpt result and the world's compiled lattice QCD data. We display three different physical cases which correspond to (1) equal quark chemical potentials, (2) zero strange quark chemical potential, and (3) the case $\langle S \rangle = 0$ and $Q/B = 0.4$, which corresponds to the case appropriate to heavy-ion collisions. As can be seen from these figures, NNLO HTLpt agrees quite well with the existing lattice data in each case. The horizontal error bars (which are sometimes not even visible) indicate the renormalization scale dependence of these curvatures.

To close this section, we have demonstrated that although naive perturbative expansions applied to QCD thermodynamics fail dramatically, it is possible reorganize the calculation of the QCD free energy in such a way as to achieve improved convergence at phenomenologically relevant temperatures. Interestingly, we find excellent agreement between the resummed approaches and lattice data down to rather low temperatures and are even able to predict the curvature of the QCD phase transition line using perturbation theory.

## 6.6 EFT methods for nonequilibrium systems

**Miguel Escobedo**

### 6.6.1 Introduction

There are many situations in which we are interested in describing non-equilibrium phenomena that involve the strong interaction. An example is the study of the medium created when colliding heavy ions at ultrarelativistic speeds. This kind of experiment is nowadays performed at facilities like the Large Hadron Collider (LHC) in Geneva and the Relativistic Heavy Ion Collider (RHIC) in Brookhaven. The motivation is to study a new state of matter that appears at high temperatures and densities, the Quark Gluon Plasma (QGP). More details are given in section 4.4. The medium created in heavy ion collisions can be regarded as an out-of-equilibrium system. Soft particles in the medium are able to approximately thermalize [2092, 2093]; however, this thermalization is only local. Looking at length and time scales much larger than the inverse of the temperature the bulk properties of the medium are well described by relativistic hydrodynamics [2094–2096].

One important way to obtain information about the QGP created in heavy ion collisions is by studying its effects on hard probes, for example, heavy quarkonium suppression and jet quenching [1561, 2097]. We can regard these particles as out-of-equilibrium probes interacting with a thermal equilibrium environment of soft particles. Precisely because they do not have time to thermalize inside of the medium, they allow us to characterize the QGP in a way that would not be possible otherwise. For example, they are sensitive to transport properties of the medium such as the heavy quark diffusion coefficient [2098] and the jet broadening parameter $\hat{q}$ [2097]. Note also that the problem of a hard probe interacting with a soft medium is one in which a hierarchy of well-separated energy scales appear. This is precisely the situation in which EFTs are useful. In summary, the study of hard probes in heavy ion collisions provides a clear motivation to study EFTs far from equilibrium.

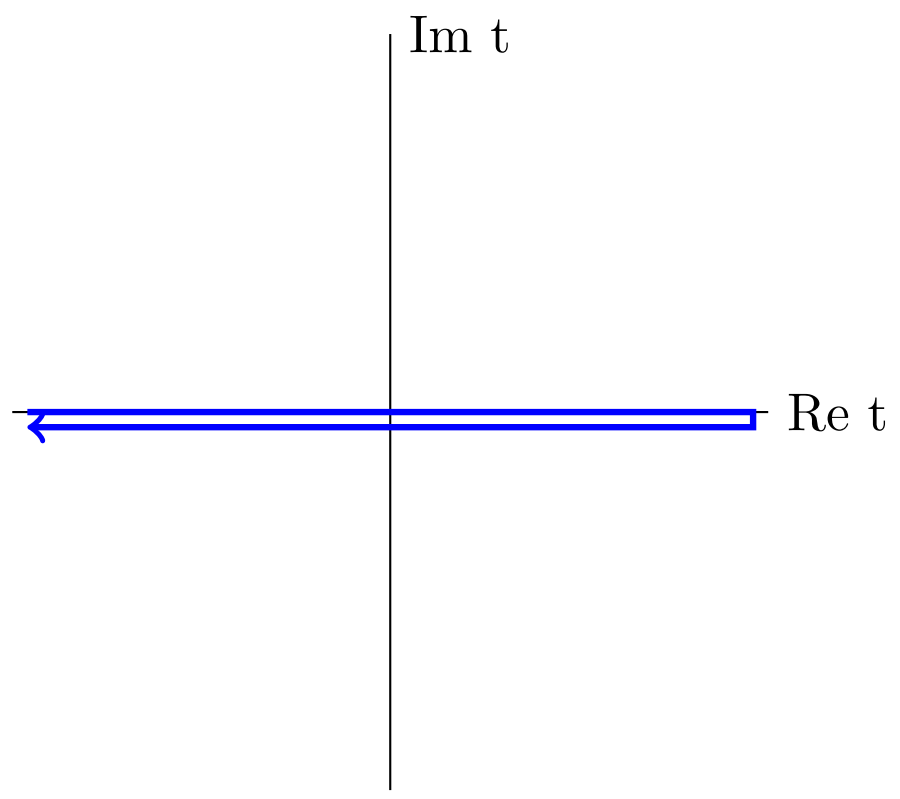

**Fig. 6.6.1** Schwinger-Keldysh contour of the real time formalism

The theoretical description of a QFT out of equilibrium requires the use of the real-time formalism [2099]. When dealing with $T = 0$ scattering process, we are used to assuming that the system is in the ground state both at the remote past and in the distant future. This is what is done to obtain the LSZ reduction formula [2100]. The consequence of this is that, when computing amplitudes, all field insertions are chronologically ordered. The situation is completely different when the initial state of the system is described by a given density matrix. In this case, the state of the system in the distant future is unknown, so we have to average over all the possible outcomes imposing that the system is described by the initial density matrix in the remote past. For this reason, the real-time formalism is sometimes called an *in-in* formalism while the formalism leading to the LSZ reduction formula is called an *in-out* formalism. As a consequence, the path integral needs to go from the remote past to the distant future and back again around a path called the Schwinger-Keldysh contour (see Fig. 6.6.1) [2099]. Fields in the upper (lower) branch of the contour are chronologically (anti-chronologically) ordered and are customary labelled fields of type 1 (2).[71]

The doubling of degrees of freedom discussed in the previous paragraph can affect the construction and use of an EFT in two different ways depending on whether or not medium degrees of freedom are integrated out when going from the full theory to the EFT. If the matching is not affected by the medium, then we can apply the real time formalism in exactly the same way as it is done for a normal QFT. However, if the matching is affected by the medium, we can not assume that the EFT does not contain terms mixing the two branches of the Schwinger-Keldysh contour. Recently, this issue has been discussed in detail in the context of the construction of an EFT for hydrodynamics [2101–2104]. However, regarding the study of hard probes, the complications arising from the doubling of degrees of freedom are substantially diminished when we take into account that only few of them are created in each heavy ion collision. We will discuss in detail how the dilute nature of heavy quarks and high-energy partons simplify their study in the real time formalism.

Lastly, there is another aspect of the application of EFTs to the study of hard probes out of equilibrium that we would like to highlight. This is the connection that naturally appears with the formalism of open quantum systems (OQSs) [2105]. The OQS formalism studies the evolution of quantum systems interacting with an environment, that at the same time is also a quantum system. The central object of study is the reduced density matrix, obtained from the density matrix of the combination of the system plus the environment after performing a trace over the degrees of freedom of the environment. The evolution of the reduced density matrix is not necessarily of the quantum Liouville type as there might appear terms that increase its von Neumann entropy. It happens that, when studying hard probes interacting with a medium using EFTs, one typically finds equations that are well known in the context of OQSs. This is not surprising since when we compute how thermal propagators influence the evolution of a hard probe we are actually making a trace over environment degrees of freedom.

In summary, in this section we are going to discuss the application of EFTs to study nonequilibrium phenomena. In particular, we will focus on interesting problems that appear in the study of heavy ion collisions in which a large separation of energy scales appear. First, we will review the open quantum system formalism. This will allow us to discuss the Lindblad equation, which will play a key role in the later discussion. Then, we will discuss the application of EFTs to the study of quarkonium suppression. More specifically, we will study the evolution of the reduced density matrix of heavy quarks using pNRQCD. In another subsection, we will review the description of jet broadening based on the study of the reduced density matrix using SCET. Finally, we will review applications of the EFT to study hydrodynamics and the interesting structure regarding the doubling of degrees of freedom that have been discovered in this context.

[71] From a complementary point of view, fields of type 1 (2) act on the left (right) of the density matrix.

### 6.6.2 Open quantum systems

Let us consider a *universe* formed by a system plus an environment. Let us assume that at some initial time the density matrix of the *universe* $\rho_U(t_0)$ fulfils

$$\rho_U(t_0) = \rho_S(t_0) \otimes \rho_E(t_0)\,, \tag{6.6.1}$$

where $S$ corresponds to the system and $E$ to the environment. The motivation for this assumption is twofold. On one hand, when studying a dilute hard probe interacting with a medium, it is natural to assume that the medium acts as a thermal reservoir that is not affected by the probe. On the other hand, even if the assumption is not true, any density matrix for the *universe* can be decomposed as a sum of density matrices that do fulfil this structure. The reduced density matrix at time $t_0$ is also $\rho_S(t_0)$.

$$\rho_S(t_0) = \mathrm{Tr}_E(\rho_U(t_0))\,. \tag{6.6.2}$$

Let us now look at what happens at $t > t_0$. If $U(t, t_0)$ is the time evolution operator of the *universe*, then

$$\rho_U(t) = U(t, t_0)\rho_U(t_0)U^\dagger(t, t_0)\,, \tag{6.6.3}$$

and it follows that

$$\rho_S(t) = \mathrm{Tr}_E(U(t, t_0)\rho_U(t_0)U^\dagger(t, t_0))\,. \tag{6.6.4}$$

However, in general it is not true that

$$\rho_S(t) \neq U_S(t, t_0)\rho_S(t_0)U_S^\dagger(t, t_0)\,, \tag{6.6.5}$$

so in this sense we can say that the evolution is non-unitary.

The equation that describes the time evolution of $\rho_S$ is called a master equation. In general, it is not trivial to determine the form of this equation. However, an important result of the OQS formalism is that the master equation of a Markovian evolution that preserves the fundamental properties of a density matrix (Hermitian positive-definite operator with trace equal to 1) takes the form of a GKSL or Lindblad equation [2106, 2107]

$$\frac{d\rho_S}{dt} = -i[H, \rho_S] + \sum_i \left( C_i\rho_S C_i^\dagger - \frac{1}{2}\{C_i^\dagger C_i, \rho_S\} \right)\,, \tag{6.6.6}$$

where $C_i$ are the collapse or Lindblad operators. They are operators that encode the dissipative part of the Lindblad equation and will depend on the problem we are studying. Let us note that it is very computationally expensive to solve the GKSL equation, as it is generally the case for any master equation. The reason is that the cost scales with $N^2$, where $N$ is the dimension of the Hilbert space. This means that, if we discretize the QCD system in a lattice, doubling the lattice size multiplies the numerical cost by four. This problem can be solved by using techniques called unravelling of the master equation. Examples of unravellings used to study quarkonium suppression are the Quantum State Diffusion [2108] and the Quantum Trajectories method [1562, 2109–2111].

### 6.6.3 EFTs for quarkonium suppression

Quarkonium suppression was proposed as a probe of the formation of a QGP in the pioneering work of Matsui and Satz [2112]. The original proposal was based on the phenomenon of colour screening. Chromoelectric fields are screened at large distances in the presence of a QGP. This modifies the heavy quarkonium potential and, if the screening length is smaller than the size of the bound state, inhibits bound state formation. Later on, it was realized that the potential develops an imaginary part in the presence of a QGP [2113]. This is related to the appearance of a thermal induced decay width which can dissociate quarkonium in many cases more efficiently than screening. However, before asking which phenomenon more substantially modifies the heavy quarkonium potential, we should understand whether quarkonium's evolution follows a Schrödinger equation at all in the presence of a medium and what is the definition of the potential. In Sec. 6.1, we have seen that similar issues can be addressed using non-relativistic EFTs such as NRQCD and pNRQCD at $T = 0$. Therefore, it is reasonable to expect that the finite temperature versions of these EFTs will allow us to answer the previous questions.

In order to construct an EFT, we should first discuss the energy scales and the symmetries of the problem. In addition to the hard, soft and ultrasoft scales that already appear when studying quarkonium at $T = 0$, we should also consider the energy scales induced by the presence of the medium. One of the energy scales that obviously appears is the temperature itself. However, in a weakly-coupled plasma ($g \ll 1$), other dynamically generated energy scales appear. For example, the Debye mass (of order $gT$) and the non-perturbative magnetic mass (of order $g^2T$). More details about these scales can be found in Sec. 6.5. Depending on the relation between the medium induce energy scales and those that already appear at $T = 0$, we will find different physical situations. For example, if the Debye mass is much larger than the inverse of the Bohr radius, there would be no bound state formation due to screening. On the other hand, if the temperature is smaller than the inverse of the Bohr radius, thermal effects are a perturbation compared with the binding energy because the medium sees quarkonium as a small colour dipole.

Regarding the symmetries of the problem, we will focus on the scenario in which quarkonium is co-moving with the medium. Note, however, that there are EFT studies considering the finite velocity case [2114, 2115]. In the co-moving case, the medium only breaks Lorentz symmetry. Note that in $T = 0$ NRQCD and pNRQCD, Lorentz symmetry is not explicit. It manifests through relations between the Wilson coefficients of different operators [1398, 1399]. These relations are broken in the presence of a medium [2116].

Now, let us discuss how the doubling of degrees of freedom influences the use of non-relativistic EFTs. First, consider the thermal equilibrium case. Since the mass of the heavy quark $M$ is much larger than the temperature $T$, it follows that the thermal modifications of the heavy quark propagator in NRQCD or the singlet propagator in pNRQCD are suppressed by the Boltzmann factor $e^{-M/T}$. This reflects the fact that physically heavy particles are dilute in a thermal equilibrium medium that has a temperature much lower than $M$. We are interested in the more general case in which the heavy particles are not in thermal equilibrium. However, we will still consider that heavy particles are dilute. This is clearly the case for bottom quarks at LHC since only a few of them are produced in each heavy ion collision.[72]. A direct consequence of the dilute limit is that the 12 propagator of a heavy particle is suppressed. This corresponds to a propagator involving a field of type 1 (upper branch of the Schwinger-Keldysh contour) and a field of type 2 (lower branch). Therefore, if we are interested in Green's functions involving only heavy quark fields of type 1, we can ignore the doubling of degrees of freedom and proceed in the same way as we would do at $T = 0$ (the doubling of degrees of freedom still affects the propagators of light particles). The reason is that, in any Green's function in which they appear at the same time heavy fields of type 1 and 2, there will appear at least one 12 propagator. In conclusion, if we are interested in spectroscopy at finite temperature, we can ignore the doubling of degrees of freedom. This is what we will do for the moment. Later on, we will discuss the evolution of reduced density matrix of quarkonium that involves discussing the 12 propagator.

The first applications of NRQCD and pNRQCD at finite temperature can be found in Reds. [1552, 1553]. Ref. [1552] considers the infinite mass limit while Ref. [1553] discusses the Abelian analogue of quarkonium, the hydrogen atom. In both works the issue of the doubling of degrees of freedom is discussed in detail. Later on, the results were generalized to the case of real quarkonium [2119]. Let us summarize the main results found by studying quarkonium spectroscopy at finite temperature using EFTs

- The leading thermal effect can only be encoded as a modification of the potential when the Debye mass $m_D$ is much larger than $E$. In the EFT framework, we only talk about a potential when we are dealing with an interaction that is non-local in space but local in time. When the condition $m_D \gg E$ is not fulfilled, thermal corrections are sensitive to $E$ in a non-polynomial way and this signals that the interaction is non-local in time. In summary, potential models are suitable when $m_D \gg E$.
- We can consider thermal effects a perturbation if $1/r \gg T$ (where $r$ is the radius). In this case, the medium does not modify the matchings from QCD to NRQCD and from NRQCD to pNRQCD. The medium sees quarkonium as a small color dipole. This manifests in the pNRQCD Lagrangian in the following way. The coupling between the singlet fields and the ultrasoft gluons of the medium is proportional to r. This implies that thermal effects are always multiplied by a factor of $rT$.
- In a qualitative way, we can say that quarkonium dissociates at the temperature at which thermal effects are of the same order of magnitude as the binding energy. The logic behind this statement is the following. If thermal effects are smaller than the binding energy, then they are a perturbation. If thermal effects are much larger than the binding energy it is impossible for a bound state to exist. Therefore, the transition between these two regimes must be found when the thermal effects and $E$ are of the same size. In the weakly-coupled scenario, the imaginary part of the potential is larger than the screening corrections to the real part. Therefore, dissociation occurs when $T \sim Mg^{4/3}$. At this temperature, screening is a perturbation as it only becomes important when $T \sim mg$. This is at odds with the original proposal of Matsui and Satz [2112] in which the mechanism responsible for quarkonium suppression was believed to be color screening.

[72] The situation could be different for charm quarks. For pNRQCD studies of the non-dilute limit for charmonium see [2117, 2118]

– There are two processes that contribute to the thermal decay width of quarkonium: gluo-dissociation and inelastic scattering with medium partons. Gluo-dissociation is the process in which a singlet state absorbs a medium gluon and becomes an octet state. It was first computed in Ref. [2120] using the Operator Product Expansions and the large-$N_c$ limit. Within pNRQCD, this process was studied in detail in Ref. [1555], where the expression of Ref. [2120] was generalized to a finite number of colors. Inelastic scattering with medium partons is a process in which a singlet scatters with a medium quark or gluon through the exchange of an off-shell gluon [1556]. Gluo-dissociation is a leading-order process in the coupling constant expansion but it has a smaller phase space since the gluon is required to be on-shell. The pNRQCD power counting correctly predicts that gluo-dissociation is the dominant process if $E \gg m_D$. On the contrary, if $m_D \gg E$, it is inelastic parton scattering that dominates.

### 6.6.4 The master equation in pNRQCD

Previously, we have discussed the information that can be obtained from the time-ordered propagator of quarkonium. This includes the values of the binding energies and decay widths. However, since we were using the dilute limit, we did not obtain any information about how the density of heavy quarkonium evolves inside of a medium. This is needed in order to compute the probability that a bound state is detected in a heavy-ion collision.

The information about the density of heavy quarkonium is contained in the 12 singlet propagator of pNRQCD. This is zero at leading order in the dilute limit; therefore, we need to go to next-to-leading order in this expansion; i.e. we need to consider all diagrams in which the 12 propagator appears only once.

Until now, all of the studies concerning the evolution of the density of heavy quarkonium inside a medium using non-relativistic EFTs have focused on the $1/r \gg T$ regime. In this case, we can use the $T = 0$ pNRQCD Lagrangian as a starting point. It has been demonstrated that computing the evolution of the 12 singlet and octet propagator gives a system of coupled equations that resembles very closely the master equations that appear in the OQS framework [1559, 1560]. This is not surprising, because we can regard $\langle S_1(t, \mathbf{r}_1) S_2^\dagger(t, \mathbf{r}_2)\rangle$ as the reduced density matrix of heavy quarks projected into the sub-space in which there is a singlet state. In general, the master equation is a complex non-Markovian equation. However, there are two limits in which simpler Markovian equations can be obtained. These limits are the ones that have been studied up to now in phenomenological applications.

In the limit $1/r \gg T, m_D \gg E$, we obtain a Lindblad equation in which all of the information about the medium is encoded in two non-perturbative parameters, $\kappa$ and $\gamma$. This equation has been used to predict the nuclear modification factor in heavy ion collisions using as additional input the initial distribution of heavy quarkonium previous to the formation of the QGP and how the temperature evolves with time. However, early studies were limited due to the high computational cost. This problem was solved by the application of the Monte Carlo wave function method [2109]. Thanks to this, it was possible to combine the solution of the master equation with state-of-the-art modelling of the time evolution of the medium to obtain results compatible with the observations at LHC [1562, 2109].

Another interesting limit is the one in which thermal effects are much smaller than the binding energy. In this case, we can use the rotating wave approximation,which assumes only the diagonal elements of the density matrix in the basis that diagonalizes the leading order Hamiltonian need be considered. Using this, the master equation simplifies into a Boltzmann equation [2117, 2118]. Moreover, using the molecular chaos assumption, it is possible to use the derived formulas outside of the dilute limit. Thanks to this, the authors of Refs. [2117, 2118] were able to successfully reproduce experimental data for charmonium suppression at LHC.

The application of pNRQCD to the computation of the nuclear modification factor has been a very active and successful approach in recent years. However, at the moment, all of the studies have focused on the case $1/r \gg T$ for the reasons discussed in the introduction. This limits the applicability of the approach to excited states that are expected to be of larger size. In subsection 6.6.6 we are going to discuss some recent developments that might be used to improve the situation.

### 6.6.5 EFT description of jet broadening

A jet is a collimated ensemble of particles with a large momentum and a small opening angle. They are useful in the context of QCD because the definition of a jet is constructed in such a way that the sensitivity to non-perturbative low-energy physics is minimized. More details can be found in Secs. 6.4 and 11. The interest in jets in heavy-ion collisions is due to a phenomenon called jet quenching [2097]. Jets lose energy when traversing a QGP. Therefore, by observing how opaque the medium is to high-energy particles allows us to infer some of its properties.

Jets might lose energy due to two different mechanisms: collisional and radiative energy loss. In the first case, the jet loses energy because it collides with the particles forming the medium. In the case of radiative energy loss, the collisions in the medium provide the high energy parton with additional transverse momentum and virtuality (a process called jet broadening). Due to this increase, the high energy parton is more likely to radiate energy outside of the jet cone. The amount of virtuality that a parton gains while traversing a given length in the medium is controlled by the transport coefficient $\hat{q}$. At the moment, it is generally believed that radiative energy loss is the dominant mechanism at high momentum while at low momentum both processes have to be taken into account.

The problem of a high-energy parton traversing a medium is one in which widely separated energy scales appear. First, we have the energy $Q$ of the high energy particle. This is the highest energy scale that appears in the problem. Additionally, we have the transverse momentum of the particle $p_\perp$. If we use light-cone coordinates, with $p_\pm = (p^0 \pm p^3)/\sqrt{2}$, and we choose the 3 direction such that $p^+ \sim Q$, then an on-shell particle must have $p_- \sim p_\perp^2/Q \ll p_\perp$. On top of this, we have to consider the energy scales induced by the presence of the medium, which by construction are always much smaller than $Q$. The EFT that is suitable to study this problem is SCET (see Sec. 6.4). Note that Glauber gluons (those with momentum $p = (p_+, p_-, p_\perp)$ of order $(T, T^2/Q, T)$) play a prominent role in the physics of a jet traversing a medium. Inclusion of Glauber gluons in the SCET formalism was discussed in Refs. [1815, 2121]. A more recent and extended discussion can be found in Ref. [1810].

There have been many studies of jet quenching using SCET [2122] and jet broadening [1814, 1817]. In contrast to the case of quarkonium suppression, at the moment all applications use SCET as a starting point, without constructing an EFT in which medium degrees of freedom have been integrated out. This may be due to the fact that there is no information relevant to jet quenching in the time ordered propagator of a high energy particle. Instead, we need to focus on the distribution of high-energy particles that requires an approach similar to the study of the 12 propagator of heavy quarkonium. Some of the results that have been obtained from the application of SCET to the study of jet quenching are the following:

- The non-perturbative expression of $\hat{q}$ in terms of an expectation value of gauge fields was re-obtained in Ref. [1814] for the case of a Feynman or Coulomb gauge and generalized to a gauge invariant expectation value in Ref. [1817]. This result is important because it allows one to compute $\hat{q}$ using non-perturbative approaches such as lattice QCD.
- The use of SCET including Glauber gluons made it possible to derive a medium-modified parton shower in a model in which the medium is approximated as an ensemble of static scattering centers [1815, 2121].

In recent years, SCET has been combined with the OQS approach to study jet quenching [1818, 2123] similarly to how pNRQCD was combined with OQS to study quarkonium suppression. In this case, one considers a high-energy particle (system) that is interacting with the soft particles that comprise the medium (environment). The interaction between the two is mediated by the Glauber part of the SCET Hamiltonian [1810]. The evolution of the reduced density matrix of the system (high-energy particle) has been studied first ignoring all radiation (only considering jet broadening) [1818] and, later on, incorporating the leading-order radiative corrections [2123]. In both cases, a master equation of the Lindblad type is found. The advantages of proceeding in this way is that the information about the medium is encoded in expectation values of gauge invariant operators of soft fields. This allows separating the physics of jet-medium interaction from the way in which the medium is modeled. In addition, it opens the way for future determinations of the influence of the medium using lattice QCD.

### 6.6.6 EFTs for hydrodynamics

We have previously discussed the difficulties encountered when constructing an EFT in which medium degrees of freedom are integrated out. In a few words, terms that mix the two branches of the Schwinger-Keldysh contour appear and this changes the properties of the EFT in a profound way in comparison with the EFT at $T = 0$. Let us summarize how this challenge has been avoided until now in the study of hard probes of the QGP:

- In the case of quarkonium we could use the dilute limit and focus on the time-ordered propagator. In this case, we know that the terms that mix the two branches of the SK contour give a small contribution and proceed as it is done at $T = 0$. The problem with this is that there is valuable information that can not be obtained from the time-ordered propagator in the dilute limit, as for example, the nuclear modification factor.
- We can choose to integrate out only the energy scales higher than the temperature. This is what has been done to study quarkonium suppression in the limit $1/r \gg T$ and jet quenching using SCET. However, this limits the applicability of the approach. Moreover, many of the simplifications introduced by the EFT framework come from being able to threat each energy scale separately from the others. This can not always be done if we are unable to integrate out medium-induced energy scales.

Recently, this issue has been addressed in the context of the construction of an EFT for hydrodynamics [2101–2104]. Going from a $T = 0$ EFT to an EFT living in the SK contour implies a doubling of degrees of freedom, but this is compensated by the fact that additional symmetries must be fulfilled. There are two symmetries that have been largely discussed:

– The SK symmetry. This symmetry must be fulfilled by any system, in or far from thermal equilibrium. It implies that the *largest time equation* [2124] must be fulfilled. This means that the difference of two Green's functions that only differ in the SK subindex of the field evaluated at the latest time must be zero. It is obvious that this must be the case because the trace of a commutator is zero. For example, in the case of a two-point Green's function

$$\langle \phi_1(t)\phi_1(0)\rangle - \langle \phi_2(t)\phi_1(0)\rangle = \mathrm{Tr}([\phi(t), \phi(0)\rho]) = 0\,. \tag{6.6.7}$$

One consequence of this symmetry is that in the limit of exactly classical fields ($\phi_1 = \phi_2$) the action of the EFT must be zero [2101].

– The KMS symmetry. This is a symmetry that must be fulfilled by system in thermal equilibrium. A well-known consequence of this symmetry is the fluctuation-dissipation theorem. It is akin to an *earliest time equation* in which, if $t$ is the earliest time, a Green's function in which the operator that appears just at the right of the density matrix is evaluated at time $t - i/T$ is equal to another Green's function that is equal except that the operator appears now just at the right of the density matrix and evaluated at time $t$. For the case of a two-point Green's function

$$\begin{aligned}&\langle \phi_1(t_2)\phi_1(t)\rangle - \langle \phi_1(t_2)\phi_2\left(t - i/T\right)\rangle = \\ &\mathrm{Tr}\left(\phi(t_2)\left(\phi(t)\rho - \rho\phi\left(t - i/T\right)\right)\right) = 0\,.\end{aligned} \tag{6.6.8}$$

Note that the previous equation is only valid if $\rho = \mathrm{e}^{-H/T}$.

At tree level it is relatively easy to write an EFT that fulfills these conditions. However, it is more difficult to ensure them when higher-order quantum loops are involved. A solution to this is to expand the theory by introducing ghost fields and using the BRST formalism.

We note that, apart from the theoretical importance as an example of an EFT in which medium degrees of freedom are integrated out, hydrodynamics is also very important in the field of heavy ion collisions. Among other important predictions, it describes the evolution of the soft medium in which the hard probes discussed in this subsection evolve [2094–2096].

# 7 QCD under extreme conditions

*Conveners:*
**Eberhard Klempt and Johanna Stachel**

In nucleus-nucleus collisions at ultra-relativistic energies a new kind of matter is created, the Quark-Gluon Plasma. Peter Braun-Munzinger, Anar Rustamov and Johanna Stachel report on the phase diagram of hadronic matter at high temperature and low net baryon density. A connection is made between the experimentally determined chemical freeze-out points and the pseudo-critical temperature for the chiral cross over transition computed in lattice QCD. The role of fluctuations giving experimental access to the nature of the chiral phase transition will be summarized. Azimuthal anisotropies of hadron distributions show that the Quark-Gluon Plasma formed in high energy collisions is strongly coupled, allowing to deduce bulk and shear viscosities. In the hot and dense plasma partons lose a large fraction of their energy and this observation leads to the determination of another medium parameter, a jet transport coefficient. Quarkonia and their role as a probe of deconfinement form the final topic of their contribution.

The phase structure of strongly interacting matter a low temperature and high density is discussed by Kenji Fukushima. In this region of the phase diagram that is probed e.g. in neutron stars, different phases and phase transitions are expected on theoretical grounds. Astrophysical observations and the observation of gravitational waves lead to important constraints for calculations modeling the transitions into a quarkyonic regime, into quark matter or color-superconducting states. The theoretical challenges to locate a conjectured critical end point in the QCD phase diagram are discussed.

## 7.1 QGP

**Peter Braun-Munzinger, Anar Rustamov and Johanna Stachel**

The infrared slavery and asymptotic freedom properties of QCD, discussed in previous sections, form the theoretical basis that strongly interacting matter at finite temperature and/or density exists in different thermodynamic phases. This was realized [427, 428] already shortly after these properties of QCD were introduced. The term quark-gluon plasma (QGP) was coined soon after by Shuryak [1353] for the high temperature/density phase where confinement is lifted and a global symmetry of QCD, the chiral symmetry, is restored. The first lattice QCD (lQCD) calculations of the equation of state were performed soon thereafter [434]. Already in early lQCD calculations a close link between deconfinement and restoration of chiral symmetry was found [426].

For deconfinement there is an order parameter for the phase transition, the so-called Polyakov loop, in the limit without dynamical quarks. For chiral symmetry restoration the chiral condensate $\langle\bar{\psi}\psi\rangle$ forms an order parameter for vanishing but also for finite quark masses. Indeed, recent numerical lQCD calculations [442] provide, in the limit of massless u and d quarks, strong indications for a genuine second-order chiral transition between a hadron gas and a QGP at a critical temperature of $T_c \approx 132^{+3}_{-6}$ MeV. For realistic u,d,s-quark masses, chiral symmetry is restored in a crossover transition at vanishing net-baryon density and a precisely determined pseudo-critical temperature of $T_{pc} = 156.5 \pm 1.5$ MeV [448]. Consistent with this result, a transition temperature of 158.0 ± 0.6 MeV was recently reported in [451]. This pseudo-critical temperature is found as a maximum in the susceptibility (derivative with respect to mass) of the chiral condensate as displayed in Fig 7.1.1. Contrary to early ideas, the system remains strongly coupled over a rather large temperature range above $T_{pc}$. This is reflected in the interaction measure computed in lQCD as the difference between the energy density and three times the pressure, $I = \epsilon - 3P$, which by definition vanishes for an ideal gas of massless quarks and gluons. Fig. 7.1.2 shows that this interaction measure, normalized to the fourth power of the temperature, peaks at about 20% above $T_{pc}$ and falls off only slowly towards higher temperature values.

The lQCD calculations have been extended into the region of finite net baryon density quantified by a baryon chemical potential $\mu_B$ [448, 451]. Current lQCD expansion techniques are valid in the regime of $\mu_B/T \leq 3$. The so obtained line of pseudo-critical temperatures is shown in the QCD phase diagram displayed in Fig 7.1.9 below. Because of the sign problem, the lQCD technique cannot be applied for still larger values of $\mu_B$, see e.g. [2125], and one has to resort to models of QCD for theoretical guidance in the high net baryon density region.

Experimentally, this regime of the QCD phase transitions is accessible by investigating collisions of heavy nuclei at high energy. It was conjectured already in [431] that, in such hadronic collisions, after some time local thermal equilibrium is established and all properties of the system (fireball) are determined by a single parameter, the temperature $T$, depending on time and spatial coordinates. This is exactly the regime probed by collisions of nuclei at the Large Hadron Collider (LHC), as will be outlined in the following. The region of finite to large $\mu_B$ is accessed by nuclear collisions at lower energies.

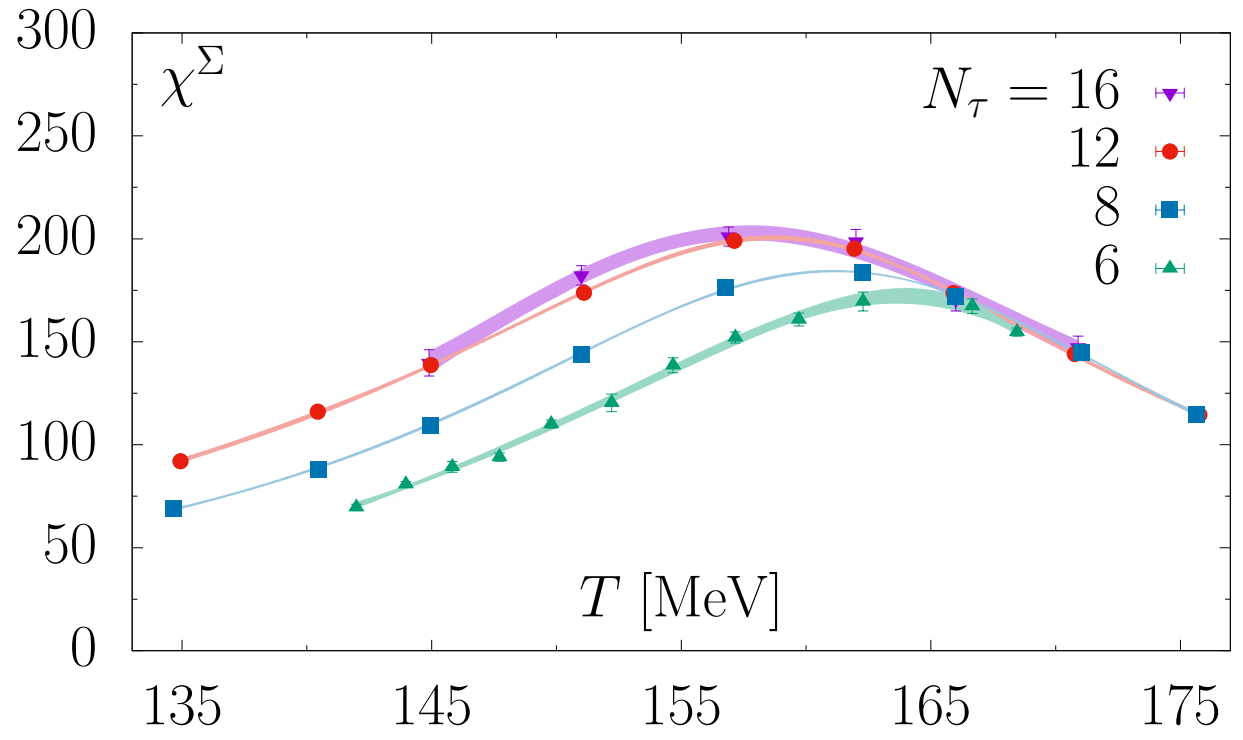

**Fig. 7.1.1** Susceptibility of the chiral u,d- and s-quark condensate as a function of temperature computed in 2+1 flavor lQCD (Fig. from [448]).

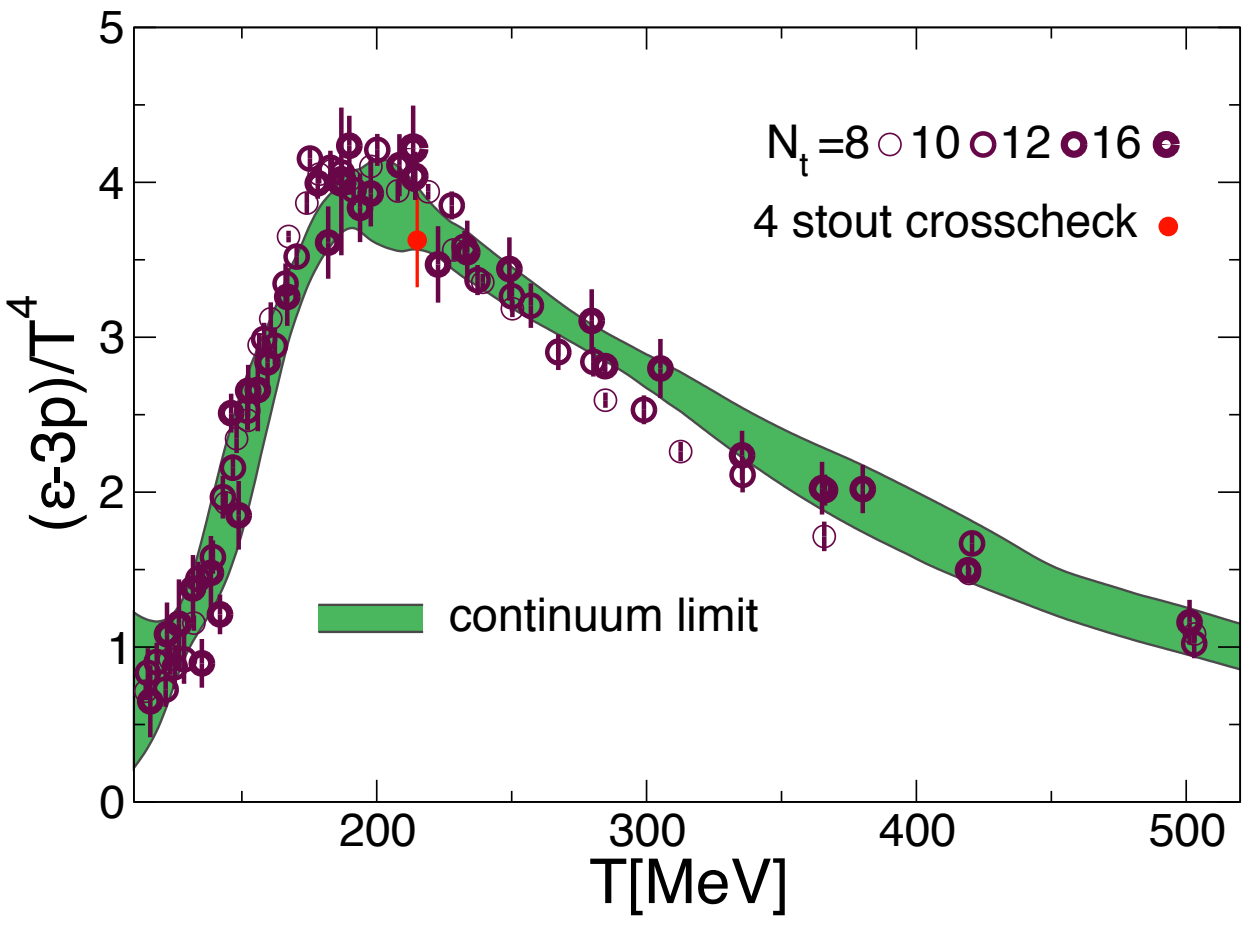

**Fig. 7.1.2** The interaction measure or trace anomaly normalized to the fourth power of the temperature as a function of temperature, computed in 2+1 flavor lQCD (Fig. from [467]).

In the following, we describe the experimental efforts, principally at the LHC and at RHIC (Relativistic Heavy Ion Collider), to provide from analysis of relativistic nuclear collision data quantitative information on the QCD phase diagram by studying hadron production as a function of the nucleon-nucleon center of mass energy $\sqrt{s_{\rm NN}}$. We can only touch a small fraction of the physics of the quark-gluon plasma (QGP) in this brief review. Excellent summaries of the many other in-teresting topics can be found in recent review articles [2126–2129].

In the early phase of the collision, the incoming nuclei lose a large fraction of their energy leading to the creation of a hot fireball characterized by an energy density $\epsilon$ and a temperature $T$. This stopping is characterized by the average rapidity shift of the incident nucleons, with $\Delta y$ = -ln$(E/E_0)$. Quantitative information is contained in the experimentally measured net-proton rapidity distributions (i.e. the difference between proton and anti-proton rapidity distributions). These distributions are presented for different collision energies from the SPS to RHIC energy range in [2130]. There it can be seen that the rapidity shift saturates at approximately two units from $\sqrt{s_{NN}} \approx 17.3$ GeV upwards, implying a fractional energy loss of $1 - \exp(-\Delta y) \approx 86\%$. In fact, the same rapidity shift was already determined for p–nucleus collisions at Fermilab for 200 GeV/c proton momentum [2131]. With increasing collision energy, the target and projectile rapidity ranges are well separated, leaving at central rapidity a net-baryon depleted or even free high energy density region. Fig. 7.1.3 shows the distribution of slowed down beam nucleons, after subtracting the tail of the target distribution and plotted against rapidity minus beam rapidity. It is apparent that up to $\sqrt{s_{NN}}$ = 62.4 GeV the concept of limiting fragmentation [2132] is well realized. At higher energies, this rapidity region is very hard to reach experimentally for identified particles.

The rapidity shift of the incident nucleons leads to high energy densities at central rapidity, i.e., in the center of the fireball. These initial energy densities can be estimated, after fixing the kinetic equilibration time scale $\tau_0$, using the Bjorken model [2135]:

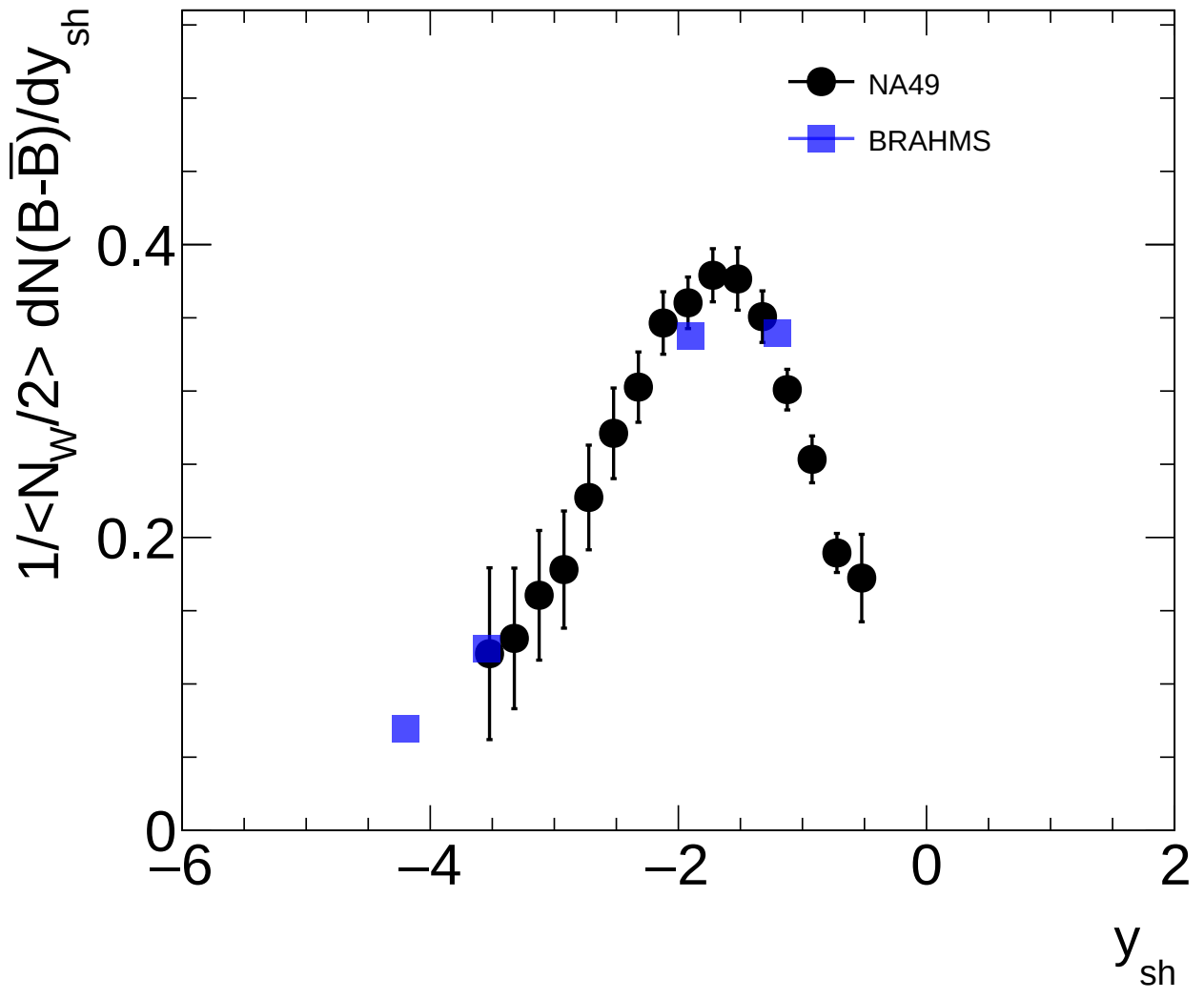

**Fig. 7.1.3** Normalized net-baryon rapidity densities for $\sqrt{s_{NN}}$ = 17.3 GeV [2133] and 62.4 GeV [2134] after subtracting the corresponding target contributions using the limiting fragmentation concept. Here $y_{sh} = y - y_b$ with $y_b$ the beam rapidity.

| | $\sqrt{s_{NN}}$ [GeV] | $dE_t/d\eta$ [GeV] | $\epsilon_{BJ}$ [GeV$/fm^3$] | T [GeV] |
|---|---|---|---|---|
| AGS | 4.8 | 200 | 1.9 | 0.180 |
| SPS | 17.2 | 400 | 3.5 | 0.212 |
| RHIC | 200 | 600 | 5.5 | 0.239 |
| LHC | 2760 | 2000 | 14.5 | 0.307 |

**Table 7.1.1** Collision energy, measured transverse energy pseudo-rapidity density at mid-rapidity [2136–2139], energy density, and initial temperature estimated as described in the text for central Pb–Pb and Au–Au collisions at different accelerators.

$$\epsilon_{BJ} = \frac{1}{A\tau_0}\frac{d\eta}{dy}\frac{dE_T}{d\eta}, \tag{7.1.1}$$

where $A = \pi r^2$ is the overlap area of two nuclei. Eq. 7.1.1 is evaluated at a time $\tau_0$ = 1 fm and the resulting energy densities are displayed in Table 7.1.1 for central Au–Au and Pb–Pb collisions. For central Pb–Pb collisions ($A$ = 150 fm$^2$) at $\sqrt{s_{NN}}$ = 2.76 TeV this yields an energy density of about 14 GeV/fm$^3$ [2136], more than a factor of 30 above the critical energy density for the chiral phase transition as determined in lQCD calculations. In fact, for all collision energies shown the initial energy density significantly exceeds the energy density computed in lQCD at the pseudo-critical temperature, indicating that the matter in the fireball is to be described with quark and gluon degrees of freedom rather than as hadronic matter. The corresponding initial temperatures can be computed using the energy density of a gas of quarks and gluons with two quark flavors, $\epsilon = 37\frac{\pi^2}{30}T^4$, yielding $T \approx 307$ MeV. Temperature values for lower collision energies are also quoted in the Table[73]. It can be seen that already at AGS energies the estimated values of $\epsilon$ and $T$ are significantly above the values for the chiral cross over transition.

Depending on energy, collisions of heavy ions populate different regimes falling into two categories: (i) the stopping or high baryon density region reached at $\sqrt{s_{NN}} \approx$ 3-20 GeV and (ii) the transparency or baryon-free region reached at $\sqrt{s_{NN}}$ >100 GeV. The net-baryon-free QGP presumably existed in the early Universe after the electro-weak phase transition and up to a few microseconds after the Big Bang[74]. On the other hand, a baryon-rich QGP could be populated in neutron star mergers or could exist, at very low temperatures, in the center of neutron stars[2140, 2141].

[73] The values reported in the table are all for vanishing chemical potentials. We have evaluated the differences if one assumes values for chemical potentials as determined at chemical freeze-out, see below. The resulting temperature values differ by less than 5% from those reported in Table 7.1.1. Owing to the proportionality of energy density to the fourth power of temperature, inclusion of a bag pressure only mildly changes the calculated temperature values.

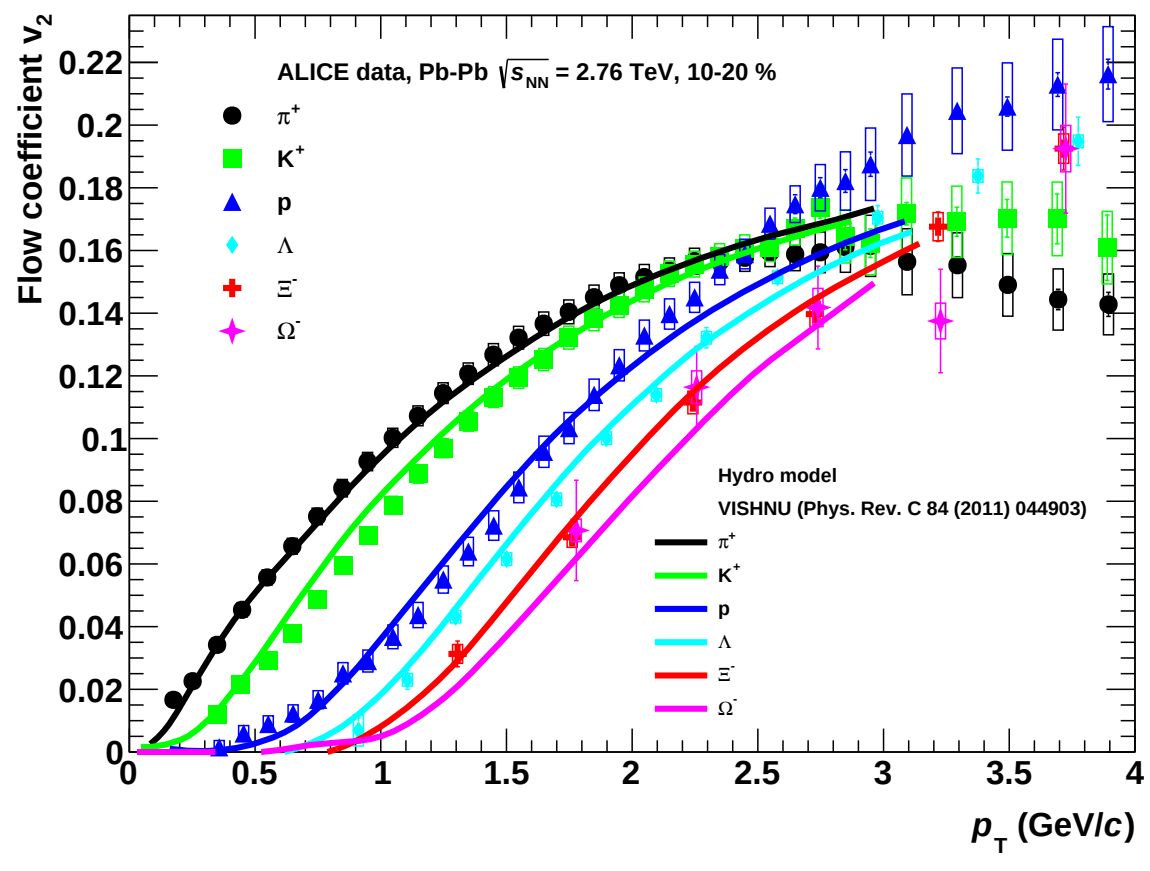

**Fig. 7.1.4** Elliptic flow coefficient $v_2$ for identified hadrons as a function of transverse momentum measured by ALICE and compared to results from viscous hydrodynamics calculations [2142].

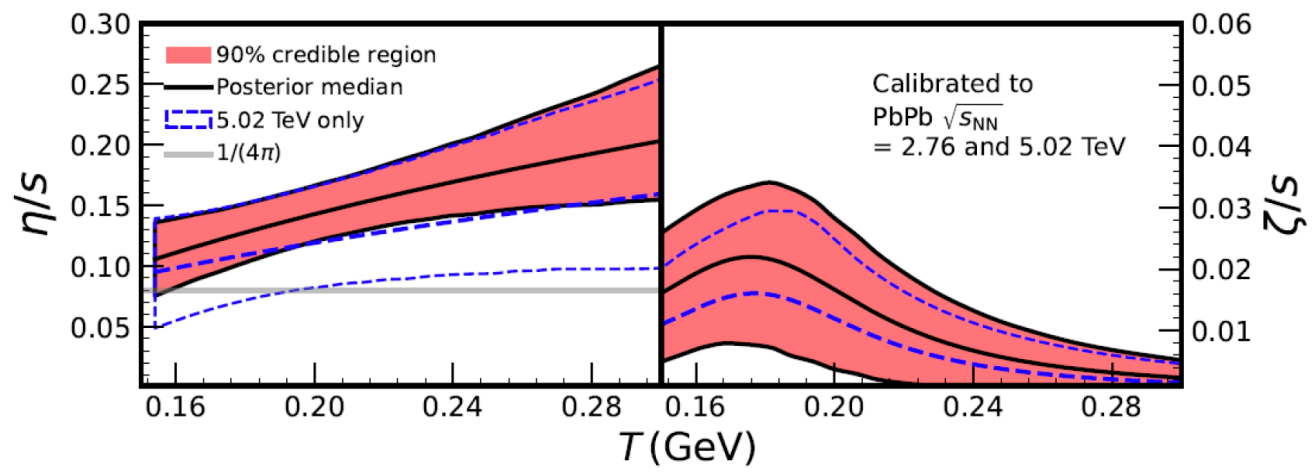

**Fig. 7.1.5** Temperature dependence of the shear (left panel) and bulk (right panel) viscosity to entropy density ratios. Figure taken from [2143].

For the system considered to come into local thermal equilibrium and, more importantly, for the development of a phase transition, the presence of interactions is necessary. In fact, close to the phase transition, the system has to be strongly coupled. As mentioned above, quarks and gluons under the extreme conditions reached in nuclear collisions are indeed strongly coupled. The large values of the interaction measure from lQCD calculations $(\epsilon - 3P)/T^4$, introduced above in Fig. 7.1.2, lend support to the strong coupling scenario. Further, the energy and entropy densities $\epsilon/T^4$ and $s/T^3$, as calculated in lQCD, fall significantly short (by about 20 %) of the Stefan-Boltzmann limit for an ideal gas of quarks and gluons up to a few times the pseudo-critical temperature. The conclusion about a strongly coupled QGP close to $T_{pc}$ also follows from experimental results at the colliders, and even at the SPS, on the coefficients of azimuthal anisotropies of hadron distributions in combination with a viscous hydrodynamic description.

For non-central nuclear collisions the distributions in transverse momentum $p_T$ of hadrons exhibit modulations with respect to the azimuthal angle $\phi$ in the reaction plane. These anisotropies can be characterized by $p_T$ dependent Fourier coefficients. The dominant term is the 2nd order Fourier coefficient $v_2$, also called the elliptic flow coefficient. This modulation has been predicted to arise from the anisotropy of the gradient of the pressure P in the early phase of the collision due to the geometry of the nuclear overlap region, leading to correspondingly larger expansion velocities in the reaction plane and hence large $v_2$ coefficients.

The strength of the coupling can be quantified by introducing transport parameters for the QGP such as the shear viscosity $\eta$, which is related to the mean free path of quarks and gluons inside the QGP, and the bulk viscosity $\zeta$, with its connection to QGP expansion dynamics and speed of sound. The smaller the transport coefficients the stronger the coupling. Larger values of the shear viscosity, e.g., suppress the magnitude of the elliptic flow.

For a strongly coupled system with small enough values of mean free path (comparable to or lower than the corresponding de Broglie wavelength of particles), treatment as a fluid is more appropriate. One then describes its properties by solving hydrodynamic equations. The shear viscosity enters the hydrodynamic equations as $\eta/(\epsilon + P) = \eta/(Ts)$, hence the quantity characterizing the medium is $\eta/s$. By comparing flow observables measured in experiments at RHIC [2144, 2145] and LHC [2146] to the corresponding calculations in viscous hydrodynamics, accompanied with converting the fluid into thermal distributions of hadrons at the freeze-out hyper-surface, remarkably low values for $\eta/s$ are obtained. Fig. 7.1.4 shows as an example the elliptic flow coefficients $v_2$ for different identified hadrons at the LHC. A mass ordering characteristic for a hydrodynamically expanding medium is observed very clearly. And indeed, the mass ordering and its $p_T$ dependence are described quantitatively by a relativistic viscous hydrodynamic calculation [2142] as indicated by the lines in Fig. 7.1.4 employing a small ratio of $\eta/s$.

In fact, a lower bound of $\eta/s = 1/(4\pi)$ (in units of $\hbar = k_B = 1$) can be obtained for a large class of strongly coupled field theories from the quantum mechanical uncertainty principle [2147] and using the AdS/CFT correspondence [2055, 2148, 2149]. Recently, the values and the temperature dependence of the shear and bulk viscosities employed in hydrodynamic codes were extracted by fitting spectra and azimuthal anisotropies of hadrons measured at the LHC and RHIC using Bayesian estimation methods [2143, 2150]. An example is shown in Fig. 7.1.5. Inspection of this figure indicates that, at $T_{pc}$, the estimated value of $\eta/s$ is close to the lower bound of $1/(4\pi)$ , indicating that the observed matter is a nearly perfect fluid. Above the transition temperature, the extracted band for $\eta/s$ is rising, reflecting a weakening of the coupling, although even at twice $T_{pc}$ the medium is still strongly coupled. On the other hand, as presented in Fig. 7.1.2, near the phase transition the lQCD results exhibit a maximum in the interaction measure, which is an indication for interactions in the system. In the hydrodynamic calculations the breaking of scale invariance is accounted for by introducing a bulk viscosity $\zeta$ along with the shear viscosity. While increasing sheer viscosity reduces the momentum anisotropy, hence lowering the elliptic flow coefficients, the bulk viscosity reduces the overall rate of the radial expansion. The right panel of Fig. 7.1.5 shows the temperature dependence of $\zeta/s$, which exhibits a peak just above the transition temperature [2143]. This location of the maximum is consistent with the temperature dependence of the interaction measure from lQCD.

[74] In the QGP of the early universe, particles interacting via the strong and electro-weak force are part of the system, while an accelerator-made QGP only contains strongly interacting particles.

Important information on the structure of the QGP is also obtained by studying the interaction of high-momentum partons with the thermalized quarks and gluons in the QGP. A strongly coupled QGP is opaque to high momentum partons, leading to the phenomenon of 'jet quenching' [2129]. In fact, the theoretical foundation for strong jet quenching by QCD bremsstrahlung was laid by [2151]. There it was shown that, for sufficiently energetic quarks and gluons, such that the radiation does not decohere, the radiative energy loss scales quadratically with the length traversed, leading to very large values. An important experimental observable linked to jet quenching is the observed suppression ('quenching') of high-momentum hadrons in central nuclear collisions at high collision energy. This suppression is quantified by the $p_T$ dependence of the ratio $R_{AA}$ of inclusive hadron production in collisions of nuclei with mass number A to that in proton-proton collisions, taking into account the collision geometry by scaling to the number of binary collisions [2129].

In Fig. 7.1.6 we present the evolution with cm energy of the transverse momentum dependence of $R_{AA}$ for leading particles as obtained from measurements at the SPS, RHIC, and LHC accelerators. Note that, by construction, $R_{AA} = 1$ for hard binary collisions in the absence of nuclear effect such as jet quenching. At very low $p_T$ one observes $R_{AA}$ values less than unity and increasing with $p_T$ since soft particle production scales with the number of participating nucleons and not the number of binary collisions. For RHIC and LHC energies the jet quenching is born out by a decreasing trend observed for $p_T > 2.5$ GeV/c reaching a broad minimum near $p_T = 7$ GeV/c of $R_{AA} = 0.1 - 0.2$: high momentum hadrons are quenched by about a factor of 5 or more. At LHC energies $R_{AA}$ increases again for higher $p_T$ values until a plateau is reached above $p_T \approx 100$ GeV/c. Measurements for fully reconstructed jets have been performed by the ATLAS collaboration. The results demonstrate [2153] that the quenching by about a factor of 2 persists to the highest available jet $p_T$ values of 1 TeV/c.

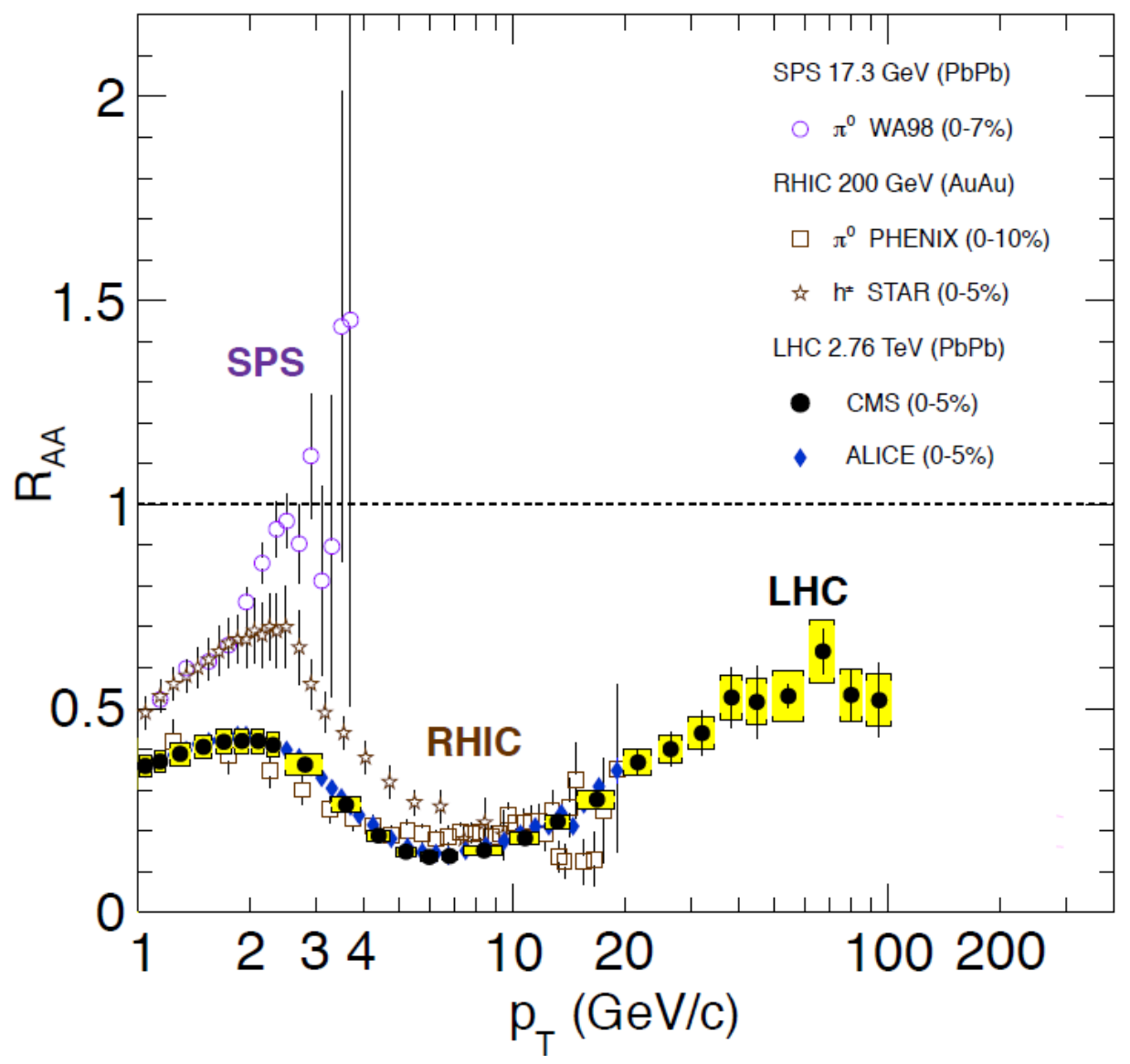

**Fig. 7.1.6** Evolution of the transverse momentum dependence of $R_{AA}$ for leading particles for central nuclear collisions with collision energy[2152].

The data on jet quenching have been modeled in terms of elastic and inelastic collisions of partons in the dense QGP, taking into account important coherence effects [2154, 2155]. For a recent summary see [2156] and ref. cited there.

To model experimental data with QCD-based jet quenching calculations one has to take into account that the jet is created as a product of an initial hard parton-parton collision with large momentum transfer Q. That implies that the parton initiating the jet is highly virtual. The magnitude of its 4-momentum Q as reflected in the total jet energy E can be hundreds of GeV (or even a few TeV at the LHC) while, for a real parton, $Q^2 \approx 0$. The highly virtual parton will evolve into a parton shower which eventually hadronizes to form a collimated jet of hadrons. During the entire evolution the highly virtual initial parton and the parton shower components lose energy by interactions with the QGP constituents, resulting in a medium-modification of the entire parton fragmentation pattern, i.e. the jet [2154]. The most modern jet quenching analyses take into account the different regimes of parton virtuality as described in [2156]. The calculations have as leading input parameter a jet transport coefficient q̂ that is determined by the differential mean squared momentum transfer $\langle k_t^2 \rangle$ between jet parton and the QGP constituents with respect to the length traversed, i.e. $\hat{q} = d\langle k_t^2 \rangle / dL$.

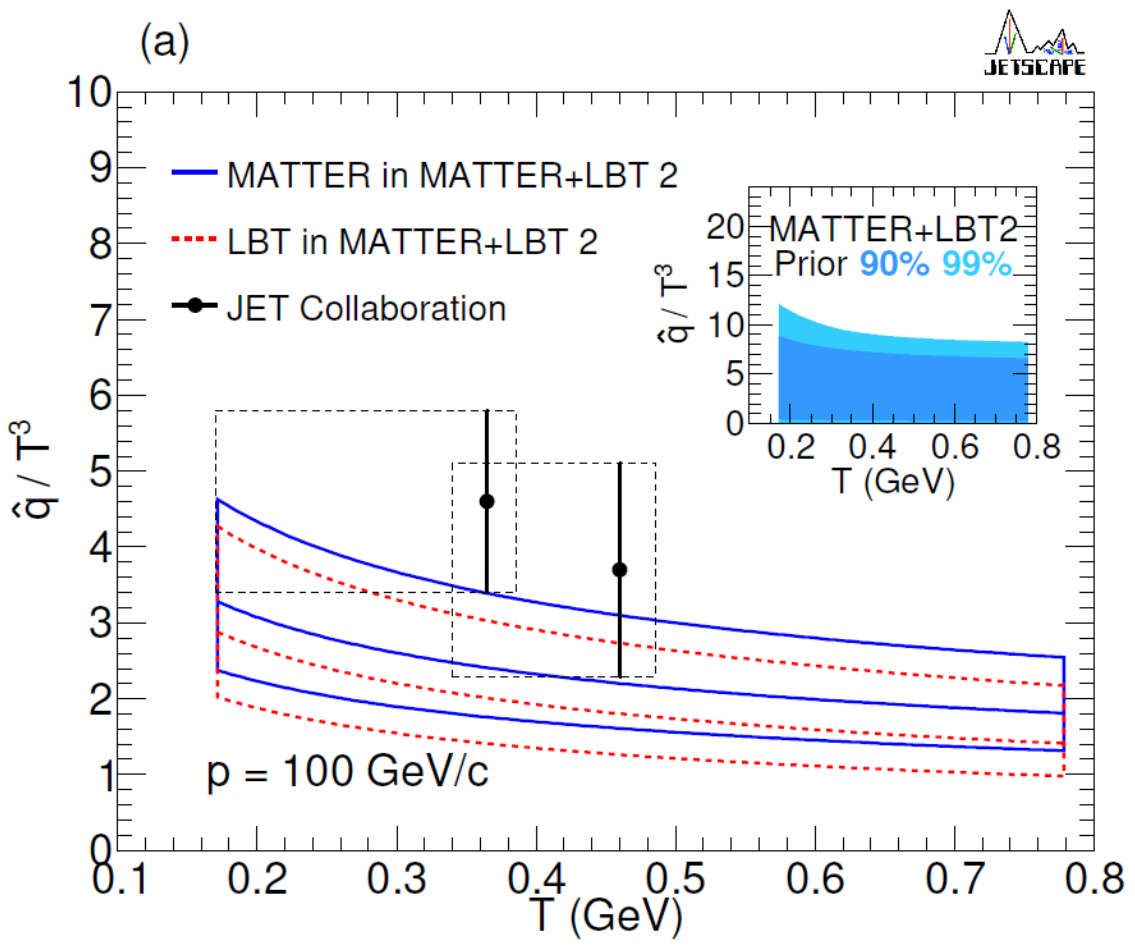

**Fig. 7.1.7** QGP jet transport parameter $\hat{q}/T^3$ obtained by the JETSCAPE collaboration[2156].

The recent analysis by the JETSCAPE collaboration [2156] uses data on inclusive hadron suppression from central Au-Au collisions at RHIC and Pb-Pb collisions at LHC, applying a Bayesian parameter estimation to determine the temperature dependence of the dimensionless, renormalized jet transport parameter $\hat{q}/T^3$. The calculations are based on two different models for parton energy loss, called MATTER and LBT, to effectively cover the large range of parton virtualities. A switch-over between the virtuality-ordered splitting dominated regime and the time-ordered transport dominated regime happens at low virtualities of $Q_0 = 2 - 2.7$ GeV. The results are shown in Fig. 7.1.7. Note that the plot is for a parton momentum of 100 GeV/c, but as demonstrated in [2156] the momentum dependence is rather mild. To put the results into context, a value of $\hat{q}/T^3 = 4$ implies that, at temperature T = 0.4 GeV, $\hat{q} \approx 1.3\,\mathrm{GeV}^2/\mathrm{fm}$. This value should be compared to what was determined for parton energy loss in cold nuclear matter. Analysis of data for deep inelastic scattering off large nuclei [2157] yielded a value of $\hat{q} = 0.024 \pm 0.008\,\mathrm{GeV}^2/\mathrm{fm}$. A global analysis of the jet transport coefficient for cold nuclear matter was performed recently in [2158]. These authors obtain values of $\hat{q} < 0.03\,\mathrm{GeV}^2/\mathrm{fm}$ over a wide range of $(x_B, Q^2)$ values (here, $x_B$ is the Bjorken x parameter). We conclude that, for high energy partons, the stopping power of a QGP formed at RHIC or LHC energy is increased by more than a factor of 40 compared to that for cold nuclear matter. The dramatic jet quenching observed experimentally as displayed in Fig. 7.1.6 finds its natural explanation in the large values of the transport coefficient q̂ of the QGP.

Direct experimental access to the QCD phase diagram is obtained from the measurement of the yields of hadrons produced in (central) high energy nuclear collisions. Analysis of these data in terms of the Statistical Hadronization Model (SHM), see [2159] and refs. given there, established that, at hadronization, the fireball formed in the collision is very close to a state in full (hadro-)chemical equilibrium.

The essential idea in the SHM is to approximate the partition function of the system by that of an ideal gas composed of all stable hadrons and resonances, hence also referred to as the Hadron Resonance Gas (HRG) model, see [2159]. From this partition function one can calculate the first moments (mean values) of densities of hadrons as a function of a pair of thermodynamic parameters, the temperature $T_{chem}$ and the baryon chemical potential $\mu_B$ at chemical freeze-out. To go beyond the ideal gas approximation, attractive and repulsive interactions between hadrons can be taken into account in the S-matrix formulation of statistical mechanics [2160] by including the first term in the virial expansion. Ideally, the relevant coefficients are obtained from measured phase shifts. For the pion-nucleon interaction this was implemented in [2161] and the proton yield for LHC energy was corrected accordingly [2162]. The predictions of the SHM for hadron yields are compared to experimental data at LHC energy for $T_{chem} = 156.5$ MeV in Fig. 7.1.8. The agreement is excellent for the yields of all measured hadrons, nuclei and hyper-nuclei and their anti-particles, with yields varying over 9 orders of magnitude. Remarkably, the description works equally well for loosely bound states. This has led to the conjecture of hadronization into compact multi-quark bags with the right quantum numbers evolving into the final nuclear wave functions in accordance with quantum mechanics [2159].

The values of the hadro-chemical freeze-out parameters at lower collisions energies are similarly obtained by fitting the SHM results to the measured hadron yields. The extracted freeze-out parameters $T_{chem}$ and $\mu_B$ [2159, 2163] are presented as red symbols in the QCD phase diagram shown in Fig. 7.1.9. Also included is a freeze-out point from the HADES collaboration in Au-Au collisions at $\sqrt{s_{\mathrm{NN}}} \approx 2.4$ GeV [2164]. They can be compared to the crossover chiral phase transition line as computed in lQCD (blue band). From LHC energies down to about $\sqrt{s_{\mathrm{NN}}} = 12$ GeV, i.e., over the entire range covered by lQCD, there is a remarkable agreement between $T_{chem}$ and the pseudo-critical temperature for the chiral cross over transition $T_{pc}$. We note that, along this phase boundary, the energy density computed (for 2 quark flavors) from the values of $T_{chem}$ and $\mu_B$ exhibits a nearly constant value of $\epsilon_{crit} \approx 0.46$ GeV/fm$^3$.

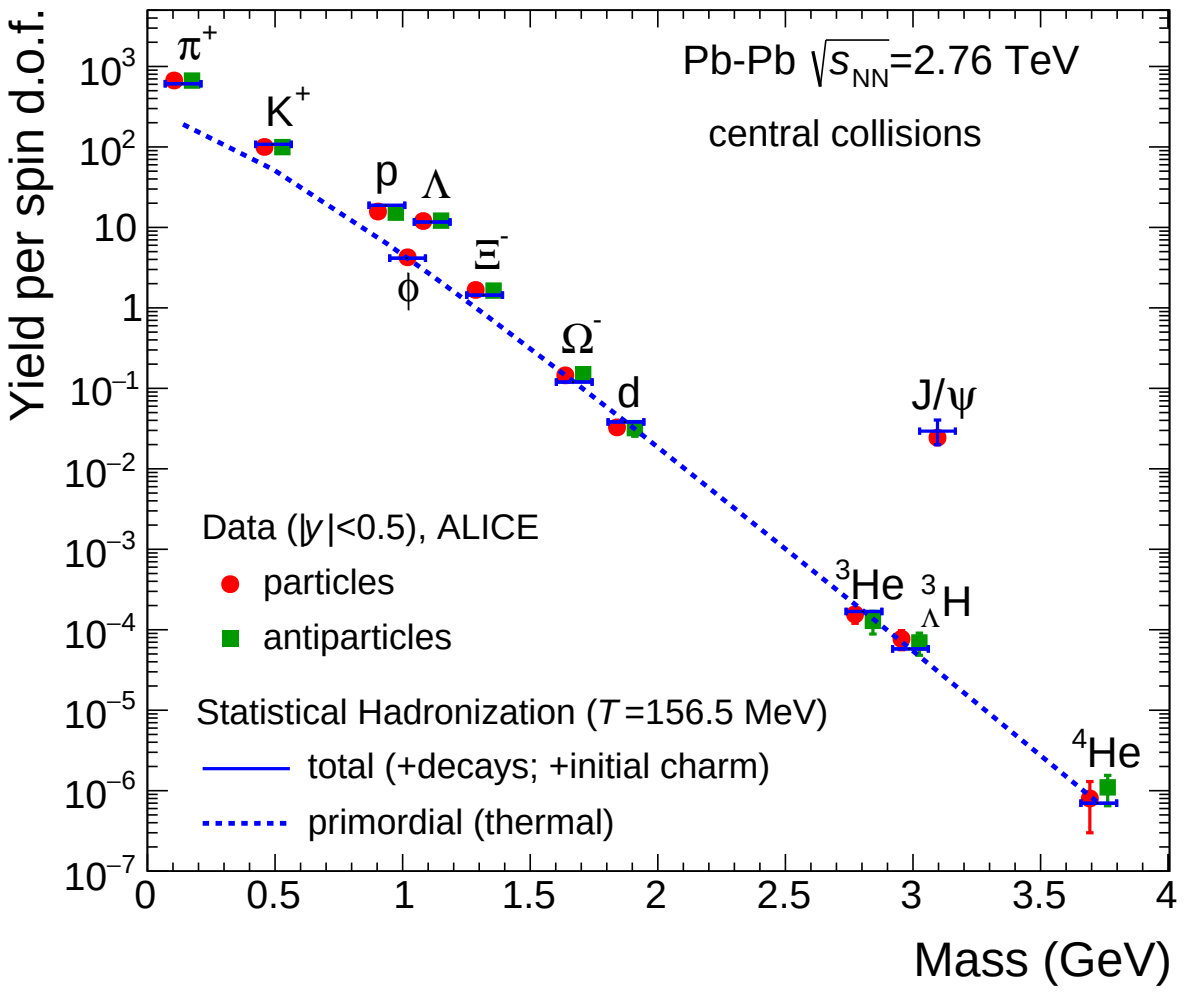

**Fig. 7.1.8** Primordial and total (anti-)particle yields, normalized to the spin degeneracy, as calculated within the SHMc [2159].

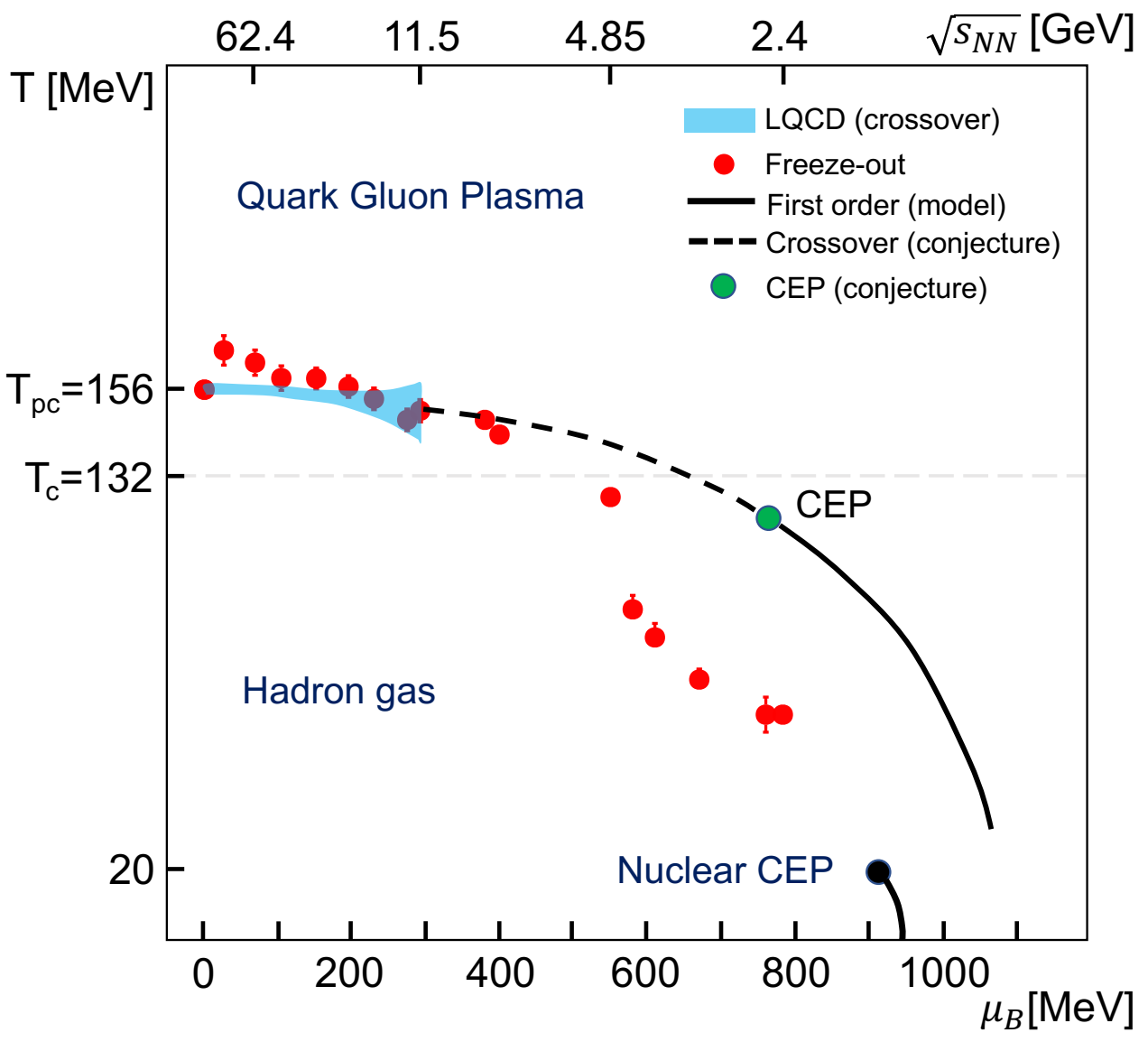

**Fig. 7.1.9** Phase diagram of strongly interacting matter. The red symbols correspond to chemical-freezeout parameters, temperature $T_{chem}$ and baryon chemical potential $\mu_B$ determined from experimental hadron yields [2159, 2163, 2164]. The blue band represents the results of lQCD computations of the chiral phase boundary [448, 451]. Also shown are a conjectured line of first order phase transition with a critical end point as well as the nuclear liquid-gas phase boundary.

The finding that the hadro-chemical freeze-out temperature is very close to $T_{pc}$ has a fundamental consequence: because of the very rapid temperature and density change across the phase transition and the resulting low hadron densities in the fireball combined with its size, the produced hadrons cease to interact inelastically within a narrow temperature interval [2165] after hadron formation.

This is very different from particle freeze-out in the early universe where for temperatures $T > 10$ MeV even the mean free path for neutrinos is much smaller than its size, see section 22.3 of [476].

For large values of baryon chemical potential, experimental data for hadron-chemical freeze-out exist but the phase structure of strongly interacting matter remains uncertain; various model calculations suggest the appearance of a line of first order phase transition, which in combination with the crossover transition at smaller values of $\mu_B$, would imply the existence of a critical end point (CEP) in the QCD phase diagram as indicated in Fig. 7.1.9. The experimental discovery of the CEP would mark a major break-through in our understanding of the QCD phase structure. The location of the CEP is most likely in the region $\mu_B > 470$ MeV, based mostly on results from lQCD. Searching for the CEP is the subject of a very active research program, at RHIC and the future FAIR facility at GSI. The importance of this research is underlined by the realization that we have currently no experimental evidence for the order of the chiral phase transition at any value of baryon chemical potential.

Important further information on the phase structure of QCD matter is expected by measuring, in addition to the first moments of hadron production data, also higher moments as such data can be directly connected to the QCD partition function via conserved charge number susceptibilities in the Grand Canonical Ensemble (GCE) [2166, 2167]. For a thermal system of volume $V$ and temperature $T$ the susceptibilities in the GCE are defined as the coefficients in the Maclaurin series of the reduced pressure $\hat{P} = P(T, V, \vec{\mu})/T^4$

$$\chi_n^q \equiv \frac{\partial^n \hat{P}}{\partial \hat{\mu}_q^n} = \frac{1}{VT^3} \frac{\partial^n lnZ(V,T,\vec{\mu})}{\partial \hat{\mu}_q^n} = \frac{\kappa_n(N_q)}{VT^3}, \qquad (7.1.2)$$

where $\vec{\mu} = \{\mu_B, \mu_Q, \mu_S\}$ is the chemical potential vector that is introduced to conserve, on average, baryon number, electric charge and strangeness. Here, $\hat{\mu}_q = \mu_q/T$ is the reduced chemical potential for the conserved charges $q \in \{B, Q, S\}$. The partition function $Z(V, T, \vec{\mu})$ encodes the Equation of State (EoS) of the system under consideration. Eq. 7.1.2 establishes a direct link between susceptibilities and fluctuations of conserved charge numbers. By measuring cumulants $\kappa_n(N_q)$ of net-charge number $(N_q)$ distributions one can, using Eq. 7.1.2, further probe and quantify the nature of the QCD phase transition.

Important at this point is to define a non-critical baseline, which is done by using the ideal gas EoS, extended such as to account for event-by-event charge conservation and correlations in rapidity space [2130, 2168, 2169], see also [2170]. In addition, non-critical contributions arising, e.g., from fluctuations of wounded nucleons [2171, 2172] need to be corrected for. Deviations from this non-critical baseline, for example leading to negative values of $\kappa_6$ for net-baryons would arise due to the closeness of the cross over transition to the O(4) 2nd order critical phase transition for vanishing light quark masses [2173].

In Fig. 7.1.10 the ALICE results on the normalized second order cumulants of net-proton distributions are presented as function of the experimental acceptance. The acceptance is quantified via the pseudo-rapidity coverage around mid-rapidity $\Delta\eta$ [2174–2176]. The measured cumulant values approach unity at small values of $\Delta\eta$, essentially driven by small number Poisson statistics. With increasing acceptance, the data progressively decrease from unity. For small but finite acceptance the decrease can be fully accounted for by overall baryon number conservation in full phase space. Hence, after correcting for baryon number conservation, the experimental data would be consistent with unity over the range of the experimental acceptance.

This observation has three important consequences. (i) It shows that, up to second order, cumulants of the baryon number distribution functions follow a poissonian distribution, a posteriori justifying the assumptions underlying the construction of the partition function used in the SHM. (ii) This is the first experimental verification of lQCD results which also predict unity for the second order scaled cumulants of baryon distributions. (iii) Compared to the different calculations, the data imply long range correlations in rapidity space, calling into question the baryon production mechanism implemented in string fragmentation models. Indeed, the results from the HIJING event generator based on the Lund String Fragmentation model shown in Fig. 7.1.10, due to the typical correlation over about one unit of rapidity, grossly overpredict the suppression due to baryon number conservation [2177].

Contrary to the detailed predictions for signals in the cross-over region of the transition covered by lQCD, no quantitative signals are available for the existence of a possible critical end point in the phase diagram.

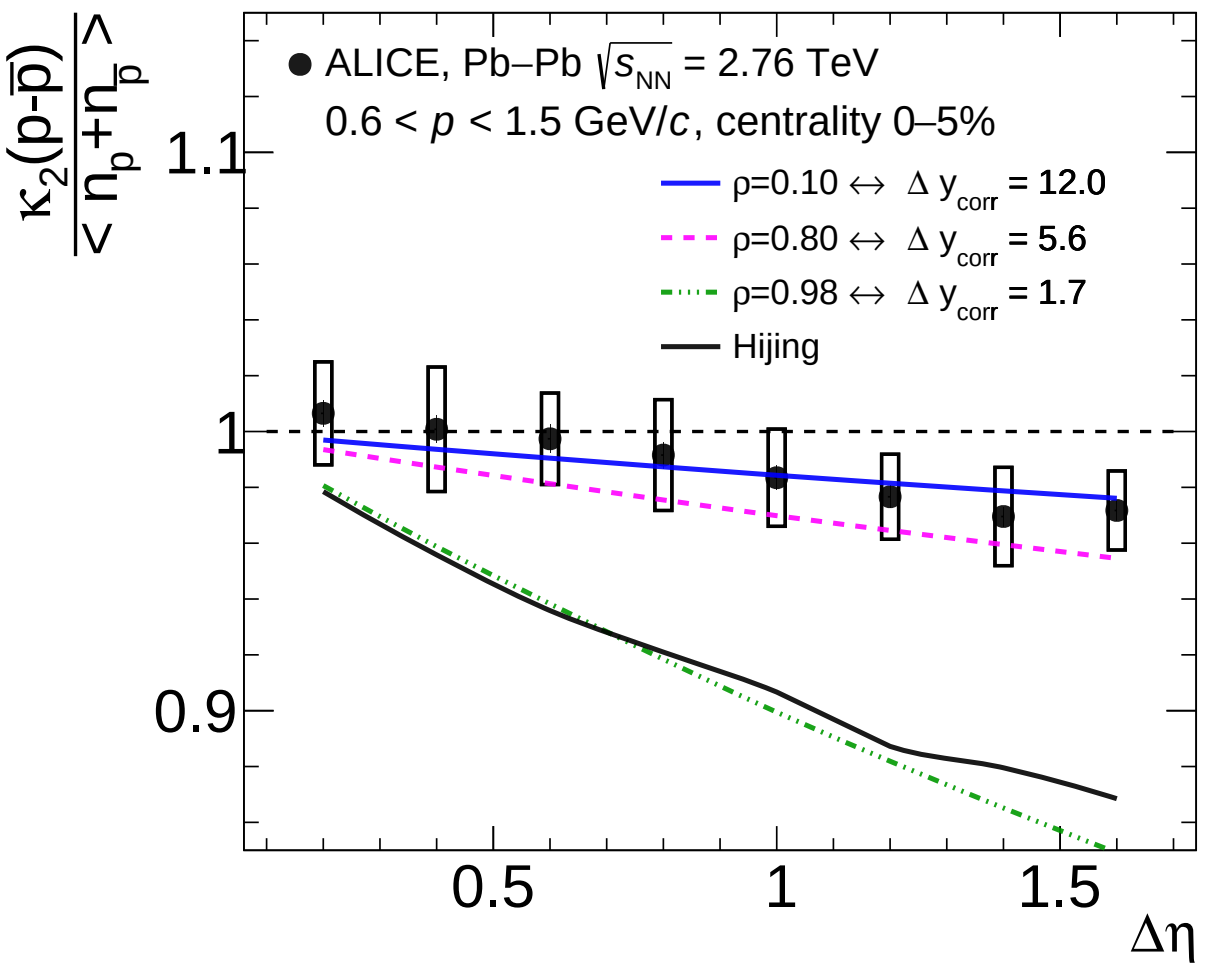

**Fig. 7.1.10** Scaled second order cumulants of the net-proton distribution as a function of the pseudo-rapidity acceptance measured by the ALICE experiment (black symbols) [2175]. The colored lines correspond to calculations accounting for baryon number conservation with different correlation length in rapidity space [2169]. The results of the HIJING event generator are presented with the black solid line.

All predicted signals are of generic nature and mostly based on searching for non-monotonic behavior in the excitation function of fourth order cumulants of, e.g., net-protons [2178]. A compilation of the respective measurements [2179, 2180] is presented in Fig. 7.1.11. The search for non-monotonic behaviour needs a starting point. In Fig. 7.1.11 two possibilities are presented, one corresponding to calculations in HRG within GCE (dashed line at unity) and the other the non-critical baseline introduced above where baryon number conservation is explicitly accounted for (red solid line or blue symbols). With respect to unity the data indeed exhibit an indication for non-monotonic behaviour with a significance corresponding to 3.1 standard deviations [2180]. However, a significant part of this deviation from unity is induced by non-critical effects, such as baryon number conservation. Therefore, one must search for non-monotonic behaviour with respect to the red solid line. Analysis of the data shows that there are no significant deviations from a statistical ensemble with event-by-event baryon number conservation, i.e, within the current precision of the data there is not yet evidence for the presence of a critical end point [2130, 2168]. The analysis of fourth order cumulants from a much higher statistics data set has just started and will be essential for a possible discovery of the critical point.

The current status on experimental verification of the nature of the chiral cross-over transition at vanishing or moderate $\mu_B$ is still rather open. Within QCD inspired model calculations [2181, 2182], based on O(4) scaling functions the predicted sixth order cumulants for net-baryon distributions exhibit negative values at $T_{pc}$ due to a singular term in the pressure. Similarly, the sixth order susceptibilities of baryon number resulting from lQCD calculations are also negative [2173, 2183] and this sign change (relative to the HRG baseline in GCE) has been linked to the critical component in the pressure present as a residue of the 2nd order chiral phase transition for vanishing (u,d) quark masses, due to the smallness of the physical masses. First experimental results on sixth order net-proton cumulants were reported by the STAR collaboration [2184] for Au–Au collisions, albeit with sizeable statistical uncertainties since the data analysis to determine high order cumulants is extremely statistics hungry. Qualitatively, the STAR results at $\sqrt{s_{\rm NN}}$ = 200 GeV are indeed consistent with the expectations for the crossover transition. At the same time, the experimentally measured energy dependence of $\kappa_6$ [2184] is at odds with both model and lQCD calculations. For a quantitative conclusion, in any case, the effects of baryon number conservation [2130] and transformation from net-protons (experiment) to net-baryons (theory) [2185] are still to be performed. So far, experimental insight into the nature of the chiral cross-over transition and the development towards low net-baryon densities remains inconclusive. It can be expected that ongoing and future high statistics measurement campaigns by the STAR and ALICE collaborations will elucidate the situation.

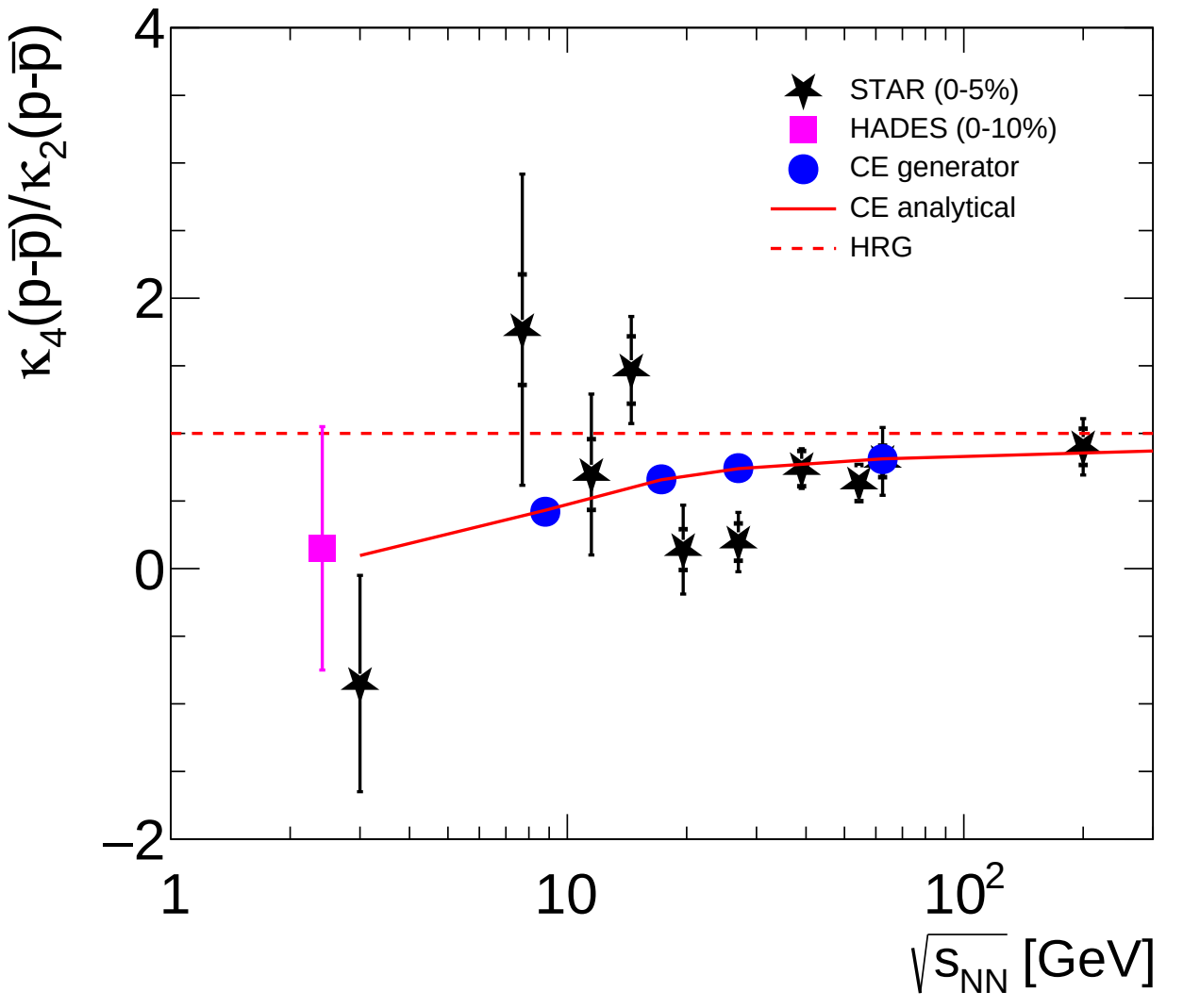

**Fig. 7.1.11** Collision energy dependence of the fourth to second order cumulants of net-proton distributions as measured by experiments. The STAR data are for $|y| < 0.5$ and $p_t$ = 0.4 - 2 GeV/c, the HADES data for $|y| < 0.4$ and $p_t$ = 0.4 - 1.6 GeV/c. The non-critical baseline induced by global baryon number conservation is indicated by the blue circles and the red line.

There is now significant experimental information, from relativistic nuclear collisions, not only on the production of hadrons composed of light (u,d,s) quarks, but also of open and hidden charm and beauty hadrons. In particular, there is good evidence, mainly from results obtained at the CERN Large Hadron Collider (LHC) [2186–2188], that charm quarks reach a large degree of thermal equilibrium, although charm quarks in the system are chemically far out of equilibrium. This is supported by heavy quark diffusion coefficients from lQCD [2189]. A strong indication for equilibration is the fact that J/$\psi$ mesons participate in the collective, anisotropic hydrodynamic expansion [2190, 2191].

To microscopically understand the production mechanism of charmed hadrons for systems ranging from pp to Pb–Pb, various forms of quark coalescence models have been developed [2192–2196]. This provides a natural way to study the dependence of production yields on hadron size and, hence, may help to settle the still open question whether the many exotic hadrons that have been observed recently are compact multi-quark states or hadronic molecules (see [2197, 2198] and refs. cited there). Conceptual difficulties with this approach are that energy is not conserved in the coalescence process and that color neutralization at hadronization requires additional assumptions about quark correlations in the QGP [2199].

Another approach, named SHMc, has been made possible by the extension of the SHM to also incorporate charm quarks. This was first proposed in [2200] and developed further in [2159, 2187, 2201–2204] to include all hadrons with hidden and open charm. The key idea is based on the recognition that, contrary to what happens in the (u,d,s) sector, the heavy (mass $\sim$ 1.2 GeV) charm quarks are not thermally produced. Rather, production takes place in initial hard collisions. The produced charm quarks then thermalize in the hot fireball, but the total number of charm quarks is conserved during the evolution of the fireball [2204] since charm quark annihilation is very small. In essence, this implies that charm quarks can be treated like impurities. Their thermal description then requires the introduction of a charm fugacity $g_c$ [2187, 2200]. The value of $g_c$ is not a free parameter but experimentally determined by measurement of the total charm cross section. For central Pb–Pb collisions at LHC energy, $g_c \approx 30$ [2187]. The charmed hadrons are, in the SHMc, all formed at the phase boundary, i.e. at hadronization, in the same way as all (u,d,s) hadrons.

In Fig. 7.1.8 it can be seen that, with that choice, the measured yield for J/$\psi$ mesons is very well reproduced,

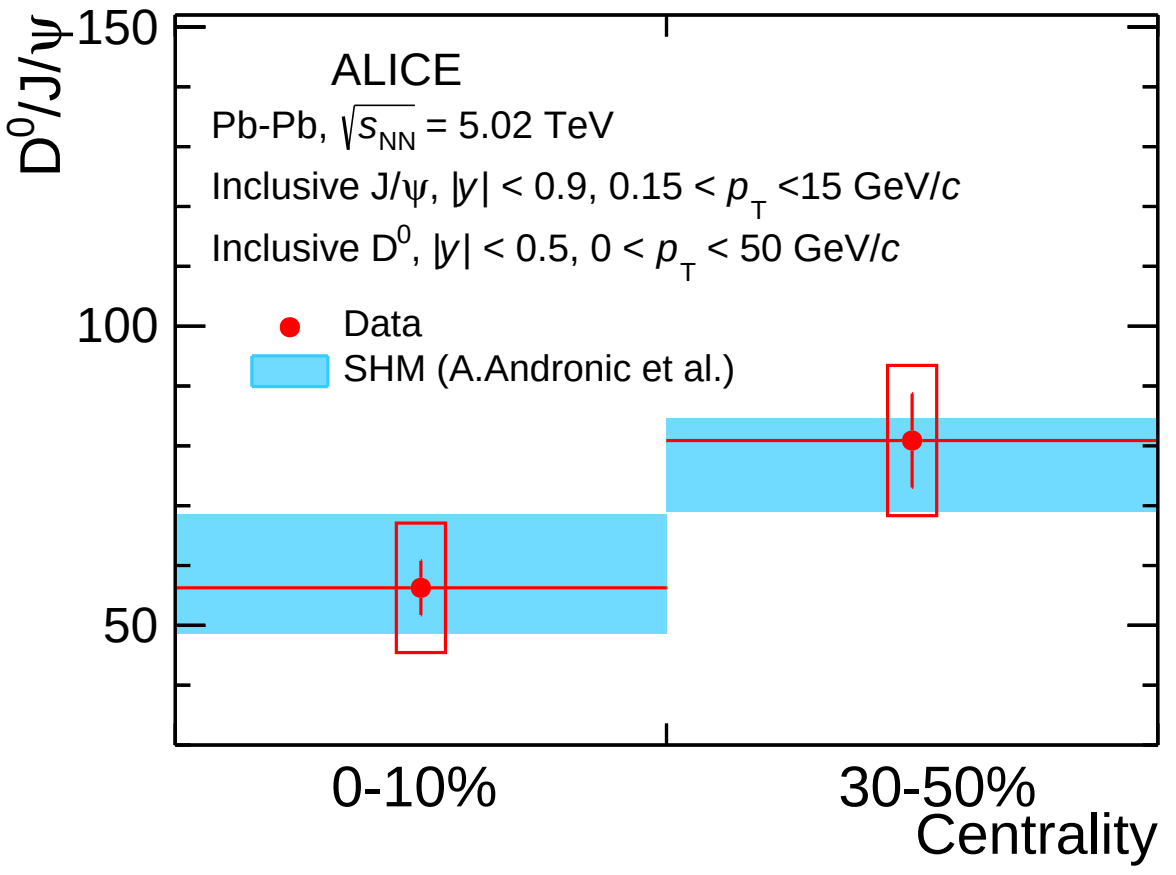

**Fig. 7.1.12** $D^0$ to J/$\psi$ yield ratio measured in Pb–Pb collisions at the LHC and predicted by the Statistical Hadronization Model with charm SHMc. Figure from [2205].

the uncertainty in the prediction is mainly caused by the uncertainty in the total charm cross section in Pb–Pb collisions. We note here that, because of the formation from deconfined charm quarks at the phase boundary, charmonia are unbound inside the QGP but their final yield exhibits enhancement compared to expectations using collision scaling from pp collisions, contrary to the original predictions based on [2112]. For a detailed discussion see [2159].

For the description of yields of charmonia, feeding from excited charmonia is very small because of their strong Boltzmann suppression. For open charm mesons and baryons, this is not the case and feeding from excited $D^*$ and $\Lambda_c^*$ is an essential ingredient for the description of open charm hadrons [2187]. Even though the experimental delineation of the mass spectrum of excited open charm mesons and baryons is currently far from complete, the prediction of yields for D-mesons and $\Lambda_c$ baryons compares very well with the measurements[75], both concerning transverse momentum and centrality dependence.

A particularly transparent way to look at the data for Pb–Pb collisions is obtained by analyzing the centrality dependence of the yield ratio $D^0/(J/\psi)$ and comparing the results to the predictions of the SHMc. Recently, both the $D^0$ and $J/\psi$ production cross sections have been well measured down to $p_t = 0$. The yield ratio $D^0/(J/\psi)$ is reproduced with very good precision for both measured centralities, as demonstrated in Fig. 7.1.12. This result lends strong support to the assumption that open and hidden charm states are both produced by statistical hadronization at the phase boundary. A more extensive comparison between SHMc and data for open charm hadrons is shown in [2187].

From the successful comparison of measured yields for the production of (u,d,s) as well as open and hidden charm hadrons obtained from the SHM or SHMc with essentially only the temperature as a free parameter at LHC energies, one may draw a number of important conclusions.

- First, we note that hadron production in relativistic nuclear collisions is described quantitatively by the chemical freeze-out parameters $(T_{chem}, \mu_B)$. Note that the fireball volume appearing in the partition function is determined by normalization to the measured number of charged particles. At least for energies $\sqrt{s_{NN}} \geq 10$ GeV these freeze-out parameters agree with good precision with the results from lQCD for the location of the chiral cross over transition. Under these conditions, hadronization is independent of particle species and only dependent on the values of $T$ and $\mu_B$ at the phase boundary. At LHC energy, the chemical potential vanishes, and only $T = T_{pc}$ is needed to describe hadronization.
- The mechanism implemented in the SHMc for the production of charmed hadrons implies that these particles are produced from uncorrelated, thermalized charm quarks as is expected for a strongly coupled, deconfined QGP (see also the discussion in [2187]). At LHC energy, where chemical freeze-out takes place for central Pb–Pb collisions in a volume per unit rapidity of $V \approx 4000$ fm$^3$, this implies that charm quarks can travel over linear distances of order 10 fm (see [2159, 2187] for more detail).

One may ask whether there is a possible contribution to the production of charmed hadrons (in particular of J/$\psi$) from the hadronic phase. At the phase boundary, assembly of J/$\psi$ from deconfined charm quarks or from all possible charmed hadrons is indistinguishable, as discussed in detail in [2159]. In fact, in [2165] it was demonstrated that multi-hadron collisions lead to very rapid thermal population, while within very few MeV below the phase boundary, the system falls out of equilibrium. Both is driven by the rapid drop of entropy and thereby particle density in the vicinity of $T_{pc}$. In the confined hadronic phase, i.e. for temperatures lower than $T_{pc}$, the hadron gas is off-equilibrium, and any calculation via reactions of the type $D\bar{D}^* \leftrightarrow n\pi$J/$\psi$ has to implement the back-reaction [2206]. Since predictions with the SHMc agree very well with the data for J/$\psi$ production at an accuracy of about 10%, and since any possible hadronic contribution has to be added to the SHMc value, we estimate any contribution to J/$\psi$ production from the confined phase to be less than 10%.

[75] For $\Lambda_c$ baryons on has to augment the currently measured charm baryon spectrum with additional states to achieve complete agreement with experimental data [2187].

Future measurement campaigns at the LHC will yield detailed information on the production cross sections of hadrons with multiple charm quarks as well as excited charmonia. The predictions from the SHMc for the relevant cross sections exhibit a rather dramatic hierarchy of enhancements [2187] for such processes. Experimental tests of these predictions would lead to a fundamental understanding of confinement/deconfinement and hadronization.

## 7.2 QCD at high density

**Kenji Fukushima**

### 7.2.1 QCD Phase Diagram

The QCD vacuum has rich contents, very different from an empty "vacuum" but rather close to a medium. The relevant physical degrees of freedom can change according to the probe resolution to the medium. As long as the typical momentum scale in physical processes is large compared to the QCD scale, i.e., $\Lambda_{\rm QCD} \sim 200\,\text{MeV}$, observed particles – all hadrons including mesons and baryons – are only color-singlet composites. The typical scale of hadronic masses and radii is characterized by $\Lambda_{\rm QCD}$ or $\Lambda_{\rm QCD}^{-1} \sim 1\,\text{fm}$. Therefore, if hadronic matter is compressed so that the interparticle distance becomes comparable to $\Lambda_{\rm QCD}^{-1}$, wavefunctions of hadrons overlap each other. Then, hadrons are no longer isolated and more elementary particles should take over the physical degrees of freedom.

High compression of QCD matter is achieved by increasing the particle number density. Actually, if matter is heated up, the density of massless thermal excitations increases as $\sim T^3$ which corresponds to the scaling of interparticle distance $\propto T^{-1}$. If the baryon density, $n_{\rm B}$, is increased in the same way, the average distance between baryons should scale as $\propto n_{\rm B}^{-1/3}$. It is hence natural to expect a phase boundary in the plane of $T$ and $n_{\rm B}$ from hadronic matter to a new state of matter composed of quarks and gluons, which portrays the QCD phase diagram.

The idea of the QCD phase diagram was first cast into a concrete shape by Cabibbo and Parisi [428] based on the conjecture of Hagedorn's limiting temperature. Let us briefly look over the theory foundations according to explanations in Ref. [2207]. The thermal partition function at finite $T$ but zero density reads:

$$\mathcal{Z}_{\rm M} = \int dm\, \rho_{\rm M}(m)\, e^{-m/T}\,, \tag{7.2.1}$$

where $dm\rho_{\rm M}(m)$ represents the number of mesonic states within the mass window, $m \sim m + dm$. The last exponential factor appears from the thermal Boltzmann weight. The density of states associated with degeneracy is increasing for larger eigen-energies, and so $\rho_{\rm M}(m)$ is an increasing function of $m$. It is empirically known that $\rho_{\rm M}(m) \sim e^{m/T_{\rm H}}$ with a phenomenological parameter $T_{\rm H}$ called the Hagedorn temperature. Because the logarithm of the combinatorial factor for a given energy is nothing but the entropy, this exponentially increasing $\rho_{\rm M}(m)$ means that the entropy grows linearly with $m$. As seen from Eq. (7.2.1), the $m$ integration in $\mathcal{Z}_{\rm M}$ blows up for $T > T_{\rm H}$ for which the entropy enhancement overwhelms the energy suppression and the free energy is bottomlessly pushed down with increasing $m$. Hagedorn proposed that $T_{\rm H}$ is interpreted as the upper bound of the physically possible temperature. Later on, a physically sensible interpretation was clarified that the singularity in $\mathcal{Z}_{\rm M}$ should be overridden by a phase transition, possibly the one to a state with more fundamental degrees of freedom. The critical temperature from mesonic matter to deconfined matter with quarks and gluons is thus $T_{\rm c}^{(\rm M)} = T_{\rm H}$.

The above mentioned argument can be generalized to the case at finite baryon density. Then, the partition function is

$$\mathcal{Z}_{\rm B} = \int dm\, \rho_{\rm B}(m)\, e^{-(m-\mu_{\rm B})/T}\,, \tag{7.2.2}$$

where the Boltzmann factor depends on the baryon chemical potential $\mu_{\rm B}$. The experimental data imply that the baryonic spectrum exhibits $\rho_{\rm B}(m) \propto e^{m/T'_{\rm H}}$ with the baryonic Hagedorn temperature, $T'_{\rm H}$, that is slightly different from $T_{\rm H}$. The critical temperature for baryons is deduced from the singularity as $T_{\rm c}^{(\rm B)} = T'_{\rm H} - (T'_{\rm H}/m_0)\mu_{\rm B}$, which is derived from an approximation that the Boltzmann factor is replaced by $e^{-m(1-\mu_{\rm B}/m_0)/T}$ with a phenomenological parameter, $m_0$ (see Ref. [2207] for detailed discussions).

Now, supposing that $T'_{\rm H} > T_{\rm H}$, the critical temperature for the deconfinement transition is dominantly characterized by mesonic $T_{\rm c}^{(\rm M)}$ in the low density region at $\mu_{\rm B} \ll T$. With increasing $\mu_{\rm B}$, the two lines of constant $T_{\rm c}^{(\rm M)}$ and decreasing $T_{\rm c}^{(\rm B)}$ cross eventually. This consideration leads us to a picture of the phase diagram on the plane of the baryon density (along the horizontal axis) and the temperature (along the vertical axis) as illustrated in Fig. 7.2.1. This QCD phase diagram handwritten by Gordon Baym (see Ref. [2208] for more historical backgrounds) has played a role of prototype of the contemporary QCD phase diagram.

So far, we addressed only the deconfinement phase transition associated with the liberation of quarks and gluons in hot and dense media. The theoretical description of deconfinement in the presence of dynamical quarks is subtle, however. One may think that each phase separated by phase boundaries should be distinctly defined by a different realization of some global symmetry but for the deconfinement phenomenon, the symmetry corresponding to quark confinement/deconfinement (known as *center symmetry* [439] that is a 1-form symmetry in finite-$T$ quantum field theory [1265]) is not exact but approximate. Still, as long as the approximate symmetry is barely broken, an approximate value of critical temperature called the *pseudo-critical temperature* can be prescribed. Therefore, the temperature at which the deconfinement takes place is not uniquely defined but the pseudo-critical temperature is inevitably scheme dependent. This is why some theoretical QCD phase diagrams show the phase boundary with an uncertainty band associated with non-unique pseudo-critical temperatures.

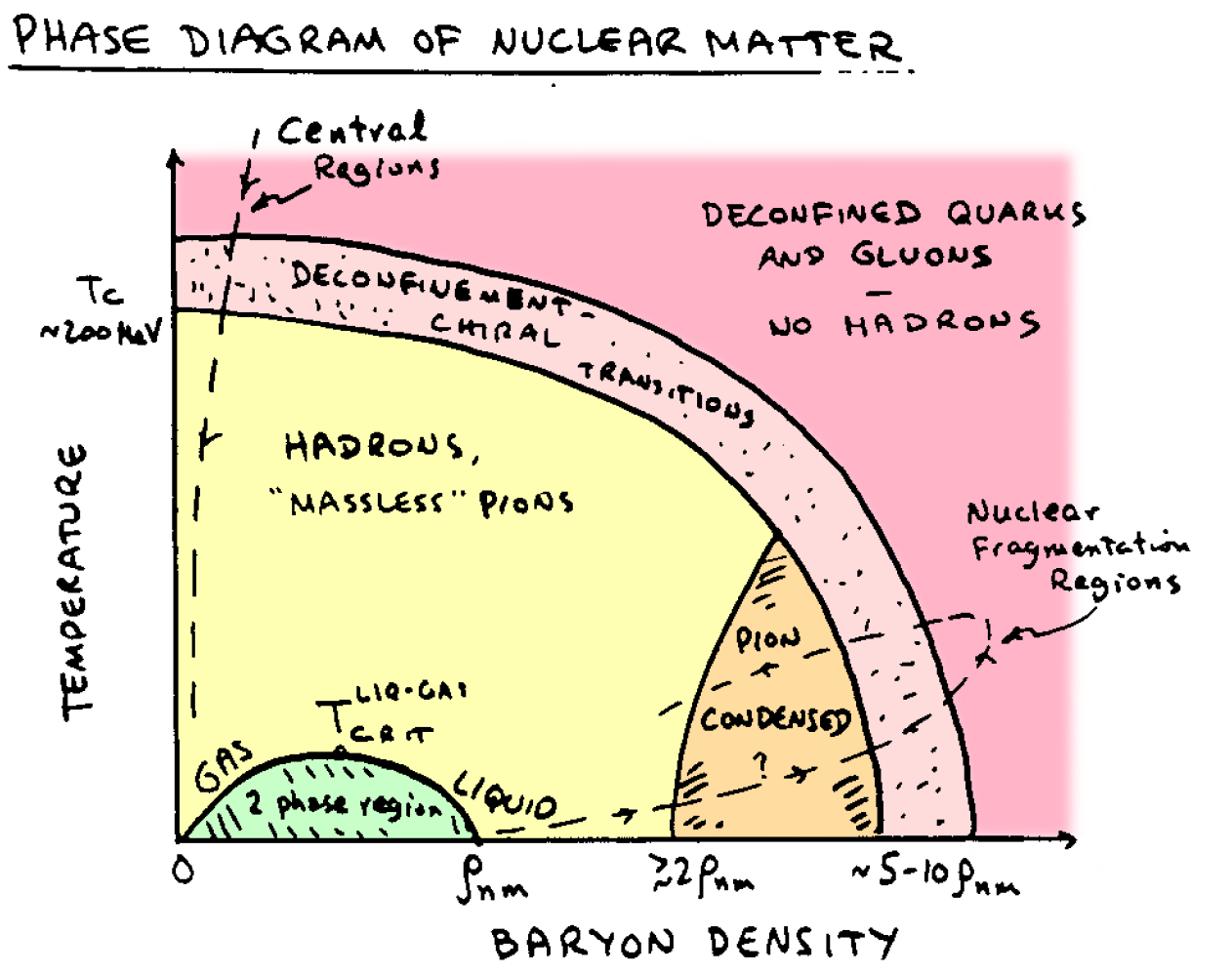

**Fig. 7.2.1** A prototype of the QCD phase diagram. The handwritten phase diagram recaptured in Ref. [2208] was adapted and colorized here.

The deconfined phase of gluons corresponds to the vacuum with spontaneous breaking of center symmetry, while quarks explicitly break this symmetry. Here, we shall avoid cumbersome mathematics and limit ourselves to pedagogical explanations about center symmetry. Let us consider a free energy gain, $f_{\mathrm{q}}(\boldsymbol{x})$, in response to a test quark placed at $\boldsymbol{x}$, and then construct a quantity called the Polyakov loop:

$$L(\boldsymbol{x}) = e^{-f_{\mathrm{q}}(\boldsymbol{x})/T}\,. \tag{7.2.3}$$

If the gluonic medium confines quarks, on the one hand, $f_{\mathrm{q}} \to \infty$ leads to $L \to 0$. Here we note that in a homogeneous system the $\boldsymbol{x}$ dependence of the expectation value can be safely dropped. On the other hand, $L$ remains nonvanishing in the deconfined phase with $f_{\mathrm{q}} < \infty$. Thus, $L$ serves as an order parameter for quark confinement, and in quantum field theory $L$ is expressed by an expectation value of a 1-form line operator (i.e., Wilson line), see Ref. [1265] for generalized higher-form symmetries.

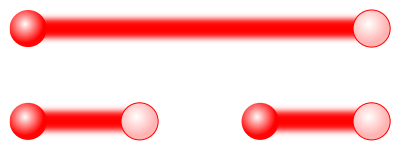

**Fig. 7.2.2** The QCD string extends between a pair of test quark and antiquark (upper figure). The string breaking occurs once the fluxtube energy exceeds the meson mass (lower figure).

In reality with dynamical quarks, $f_{\mathrm{q}}$ would never diverge. The reason is easy to understand. A single test quark is a source to which a color fluxtube is attached. The color field energy is proportional to the squeezed fluxtube length. Thus, in a purely gluonic medium, a test quark cannot be screened and the fluxtube goes to spatial infinity, yielding $f_{\mathrm{q}} \to \infty$ and thus $L \to 0$ if confined. Fluctuations of dynamical quarks allow for the creation of a quark and an antiquark pair once the fluxtube energy exceeds the mesonic mass threshold as illustrated in Fig. 7.2.2. Then, the color field energy stored between a test quark and a test antiquark cannot become greater than twice the mesonic mass $M_V$. That is, the clustering decomposition property indicates $L(0)L^\dagger(\boldsymbol{x}) \sim e^{-2M_V/T}$ for sufficiently large $|\boldsymbol{x}|$, and so $L \sim e^{-M_V/T} > 0$ even in the confined hadronic phase.

This argument implies that the QCD string is further breached by fluctuations of surrounding quarks and holes at finite density. In other words the explicit breaking of center symmetry is enlarged in high-density matter and the mathematical concept of quark confinement would be obscure. Not that we do not yet find a better order parameter. The absence of the deconfinement order parameter could be attributed to the profound nature of dense QCD matter; namely, duality from hadronic to quark matter.

Now, let us shift gear to another aspect of the QCD vacuum and the QCD phase transition. The QCD Lagrangian contains quark mass parameters $m_{\mathrm{q}}$. The bare values of up and down (i.e., $u$ and $d$) quark masses are only a few MeV, accounting for an only small fraction of the nucleon mass composed of $u$ and $d$ quarks. This huge discrepancy in the masses of quarks and baryons is explained by spontaneous breaking of *chiral symmetry*. Its order parameter is the chiral condensate $\langle\bar{q}q\rangle$ that gives rise to the dynamical mass, $M_{\mathrm{q}} \sim \Lambda_{\mathrm{QCD}} \gg m_{\mathrm{q}}$.

Almost all textbooks on quantum field theory affirm that the divergent zero-point oscillation energy is harmlessly discarded, but this common assertion is not valid for QCD because the mass is not a physical constant but is dynamically rooted in the QCD interactions. That is, the zero-point oscillation of quarks and antiquarks with

$N_c$ colors and $N_f$ flavors gives [54]

$$E_{\rm zero} = -2N_f N_c \int^{\Lambda} \frac{d^3p}{(2\pi)^3}\sqrt{p^2+M_q^2} \simeq -N_f N_c \frac{\Lambda^4}{8\pi^2}\left[2+\xi^2+\mathcal{O}(\xi^4)\right], \tag{7.2.4}$$

where $\Lambda$ is a ultraviolet (UV) cutoff and the dimensionless parameter, $\xi = M_q/\Lambda$, is assumed to be small. We see that the UV divergent term $\propto \Lambda^4$ is irrelevant to the dynamics, but we cannot drop another UV divergent term $\sim M_q^2\Lambda^2 \sim \Lambda^4\xi^2$. Because $M_q$ is related to the chiral condensate in the QCD vacuum, $\langle\bar{q}q\rangle$, the value of $M_q$ is dynamically determined to minimize the vacuum energy. The coefficient of the quadratic term, $\xi^2$, is negative in Eq. (7.2.4), so that $E_{\rm zero}$ energetically favors larger $M_q$. It is the condensation energy, $E_{\rm cond}$, that competes the zero-point oscillation energy. Let us postulate that gluon mediation induces a four-fermionic interaction term $\sim \lambda\bar{q}q\bar{q}q$ in the low-energy Lagrangian where the mass dimension of the coupling constant, $\lambda$, is $-2$. Thus, a dimensionless coupling, $\hat{\lambda} = \Lambda^2\lambda$, is useful, and the dimensional analysis hints at a relation $M_q = -2\lambda\langle\bar{q}q\rangle$. (In QCD $\langle\bar{q}q\rangle$ is known to take a negative value.) Then, the condensation energy from the interaction term is parametrically written as

$$E_{\rm cond} = N_f N_c \lambda \langle\bar{q}q\rangle^2 = N_f N_c \frac{M_q^2}{4\lambda} = N_f N_c \frac{\Lambda^4}{4\hat{\lambda}}\xi^2 . \tag{7.2.5}$$

Now, the balance between two energies gives a condition for the spontaneous generation of $M_q \neq 0$; that is, $\hat{\lambda} > 2\pi^2$, as first derived by Nambu and Jona-Lasinio [54, 2209]. For the four-fermionic interaction stronger than this threshold, the QCD vacuum accommodates a nonvanishing chiral condensate.

From the Dirac mass term $m_q\bar{q}q$ in the Lagrangian we see that the mass and the chiral condensate are conjugate to each other. It is thus evident that a nonzero $\langle\bar{q}q\rangle$ is a source to generate $M_q$ even from a massless theory with $m_q = 0$. The massless Dirac fermions are split into the right-handed and the left-handed components and they do not communicate. Therefore, for the theory with $N_f$ flavors of massless quarks, a unitary rotation in flavor space is a symmetry in each of the right-handed and the left-handed sectors, i.e., the system enjoys the symmetry of $U_R(N_f)\times U_L(N_f)$. Actually, the chiral condensate is decomposed as $\langle\bar{q}q\rangle = \langle q_R^\dagger q_L + q_L^\dagger q_R\rangle$ and it breaks the symmetry down to the vectorial one only, $U_V(N_f)$. Among these symmetries, conventionally, $SU_R(N_f)\times SU_L(N_f)$ is called *chiral symmetry* that is spontaneously broken so as to generate the dynamical mass, $M_q \sim \Lambda_{\rm QCD}$, out from the bare mass, $m_q \ll \Lambda_{\rm QCD}$.

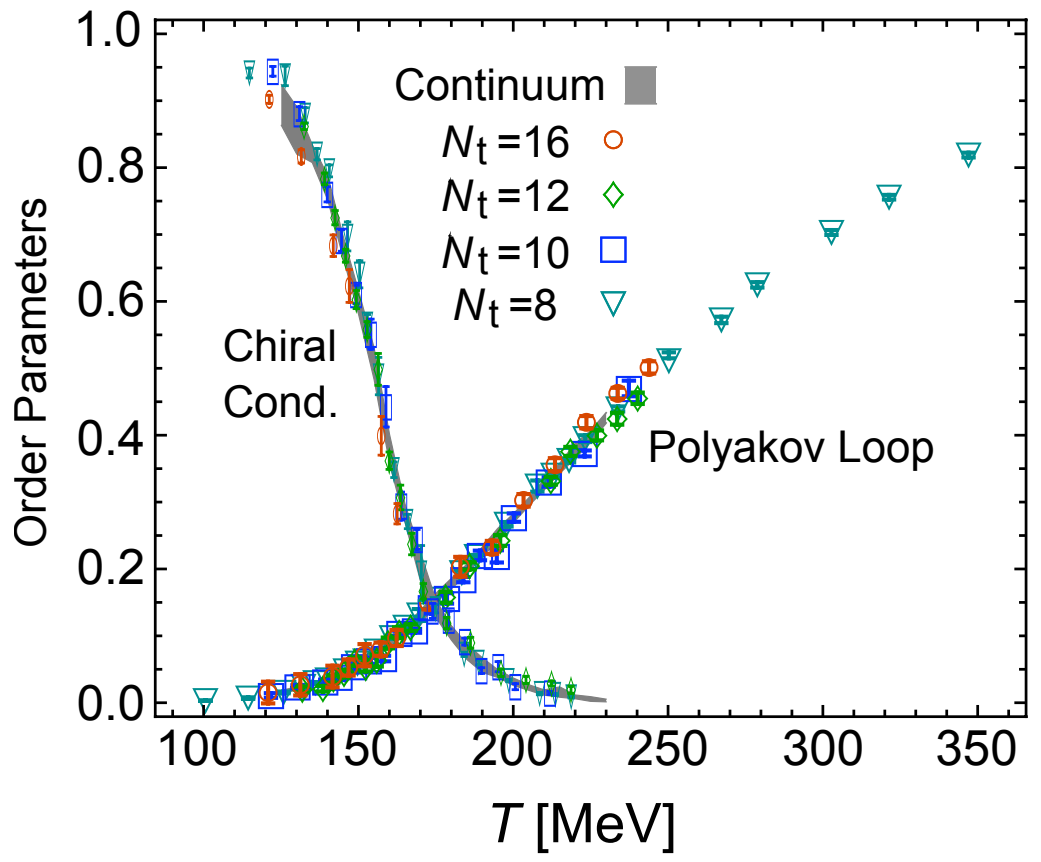

**Fig. 7.2.3** Two order parameters as functions of the temperature at zero density as measured in the lattice-QCD simulation. $N_t$ represents the site number along the temporal direction and the extrapolation to $N_t \to \infty$ defines the continuum limit. The figure and the lattice data are adapted from Ref. [2211].

We can expect, as elaborated below, that $\langle\bar{q}q\rangle$ should melt at high density and chiral symmetry should be restored then, which is commonly referred to as the *chiral phase transition*. It is the zero-point oscillation energy (7.2.4) that favors the symmetry breaking, and its expression involves the phase-space integration. At finite quark chemical potential $\mu_q$ which takes a larger value with increasing quark number density, the Fermi sphere is excluded from the phase-space integration due to the Pauli exclusion principle. Accordingly the symmetry breaking effect is diminished at finite $\mu_q$. Therefore, it is a reasonable educated guess that the chiral phase transition makes a boundary curve on the density-temperature plane just like the deconfinement phase transition, as already depicted in Fig. 7.2.1.

The exact relation between the deconfinement phase transition with an approximate order parameter $L$ and the chiral phase transition with another approximate order parameter $\langle\bar{q}q\rangle$ is a longstanding problem in QCD, and the satisfactory answer has not been found especially at finite density. As a function of $m_q$, actually, the deconfinement phase transition is exact only in the limit of $m_q \to \infty$, while the chiral phase transition is exact only in the opposite limit of $m_q \to 0$. The lattice-QCD data at finite $T$ suggest that these two conceptually distinct phase transitions at opposite limits be interpolated by a single line for arbitrary $m_q$ [2210].

Figure 7.2.3 shows the Polyakov loop and the chiral condensate as functions of $T$, normalized by the $T = 0$ values. We clearly notice that chiral symmetry is restored around $T_c \sim 150\,\text{MeV}$, and at the same time the Polyakov loop starts increasing from nearly zero, indicating a simultaneous deconfinement crossover. Thus, the lattice-QCD simulation at finite $T$ has led us to a conclusion that two phase transitions of chiral restoration and deconfinement are somehow locked together. Actually, the prototype phase diagram in Fig. 7.2.1 assumes such tight locking of two transitions on the entire plane. However, as mentioned earlier, the barrier for the QCD string breaking would be eased by the density effect and the deconfinement would be more and more blurred at higher density, which implies a modernized version of the phase diagram as shown in Fig. 7.2.4. Here, as compared to the prototype in Fig. 7.2.1, there are three new ingredients added to Fig. 7.2.4; namely, the color superconductivity, the quarkyonic regime, and the QCD Critical Point (QCP). Moreover, Fig. 7.2.4 shows a new label, "sQGP" at high $T$ and zero density, that refers to strongly correlated quark-gluon plasma. We will address only high-density aspects of QCD in this section, and for the physical interpretation of sQGP and the experimental characterization, see the previous section.

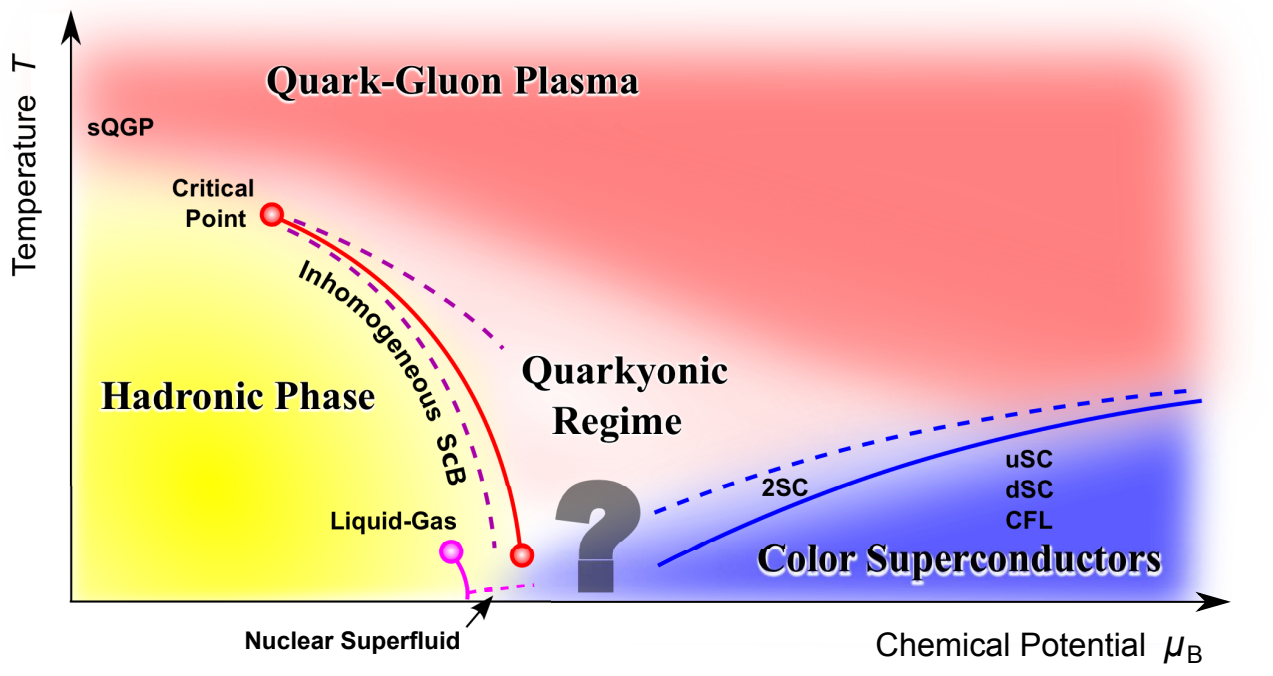

**Fig. 7.2.4** A modern phase diagram of QCD with blurred deconfinement at higher density represented by the color gradation. Unfortunately we are still unable to remove a big question mark.

### 7.2.2 Quark Matter

There is no clear definition that distinguishes nuclear and quark matter. In one working definition, quark matter is a state of matter whose properties are reasonably approximated by perturbative QCD (pQCD) calculations. The presence of quark matter in the neutron star (NS) has been proposed by Collins and Perry [427] based on the asymptotic freedom at high baryon density (see also Ref. [2212] for a preceding hypothesis on quark matter). If the momentum scale associated with the running strong coupling constant, $\alpha_s$, is characterized by the baryon chemical potential $\mu_B$ or the quark chemical potential $\mu_q = \mu_B/3$, asymptotically free quarks should be liberated from hadrons as the density goes above a certain threshold. In Ref. [427] the leading order (LO) contributions, i.e., thermodynamic quantities of free massless quarks, were considered:

$$P_{(0)} = \frac{N_c}{12\pi^2} \sum_{i=1}^{N_f} \mu_i^4 \,. \tag{7.2.6}$$

The next-to-leading order (NLO) diagrams add corrections of $\mathcal{O}(\alpha_s)$, i.e.,

$$P_{(2)} = -\left(\frac{\alpha_s}{\pi}\right) \frac{N_g}{16\pi^2} \sum_{i=1}^{N_f} \mu_i^4 \,, \tag{7.2.7}$$

where the adjoint color factor, $N_g = N_c^2 - 1$, was introduced. The next-to-next leading order ($N^2LO$) calculations produce a logarithmic term with $\mu_q$ dependence in the argument. All the terms are not listed up here (see Ref. [2213] for the formulation and Ref. [2214] for the QCD application); the logarithmic term looks like

$$P_{(4)} = +\left(\frac{\alpha_s}{\pi}\right)^2 \frac{N_g \beta_0}{64\pi^2} \sum_{i=1}^{N_f} \mu_i^4 \ln \frac{\mu_i^2}{\mu_0^2} + (\text{non-log terms}) \,, \tag{7.2.8}$$

where $\beta_0 = (11N_c - 2N_f)/3$. Non-logarithmic terms are omitted. Even if $\alpha_s$ is sufficiently small, $\alpha_s \ln(\mu_i^2/\mu_0^2)$ may become large, and then the perturbative expansion breaks down. A remedy for this problem of the singular logarithm is the resummation over the leading-log terms. For simplicity let us assume that all the quark chemical potentials are identical. (More generally one can introduce a flavor averaged value of the chemical potential.) Actually, it is easy to confirm that, if $\alpha_s$ is upgraded to the running one, i.e.,

$$\frac{\alpha_s(\mu_q)}{\pi} = \frac{\alpha_s}{\pi}\left[1 + \left(\frac{\alpha_s}{\pi}\right)\frac{\beta_0}{4}\ln\left(\mu_q^2/\mu_0^2\right)\right]^{-1} \,, \tag{7.2.9}$$

an expansion of Eq. (7.2.7) can reproduce Eq. (7.2.8). In other words, such dangerous logarithmic terms are absorbed into the density-dependent running coupling, $\alpha_s(\mu_q)$ (see Ref. [2214] for more details). In this way the perturbative calculation is justified at high enough density.

From this construction of the running coupling constant, one can easily imagine that the resummation is not free from an arbitrary choice of irrelevant constants. Instead of $\ln(\mu_q^2/\mu_0^2)$, one could try to make a resummation of $\ln(\mu_q^2/\mu_0^2) + C = \ln(\mu_q^2/\mu_1^2)$ with $\mu_1^2 = \mu_0^2 e^{-C}$. In principle, an optimal choice of $C$ could exist to reduce higher-order corrections. If $C$ is close to the optimal point, the results are expected to be flat against changes of $C$, and it is customary to check the stability of the results by changing $X$ of $\alpha_s(X\mu_q)$. Here, the logarithmic term in $\alpha_s(X\mu_q)$ takes the form of $\ln\left(X^2\mu_q^2/\Lambda_{\overline{\rm MS}}^2\right)$ in the $\overline{\rm MS}$ scheme [2215]. It is then found that such variation of $X = 1 \sim 4$ leads to huge uncertainty unless $\mu_q$ becomes unphysically large. This is sometimes referred to as the slow-convergence problem. The next correction, i.e., the $N^3LO$ contribution is expected to stabilize the results better, and indeed the soft $N^3LO$ part has been shown to cure the slow-convergence problem partially [2216, 2217].

### 7.2.3 Color-Superconducting Phases

The pQCD calculation is not capable of describing dynamical generation of $\langle \bar{q}q \rangle$, which is apparently consistent with melting chiral condensate at high density. However, even at high density, high enough to validate pQCD, the chiral condensate is not simply gone.

Quarks carry a fundamental charge in color SU(3), and so two charges of a pair of quarks (i.e., a diquark) connected by one-gluon exchange are coupled via

$$(t^a)_{ij}(t^a)_{kl} = -\frac{N_c+1}{4N_c}(\delta_{ij}\delta_{kl} - \delta_{il}\delta_{kj}) + \frac{N_c-1}{4N_c}(\delta_{ij}\delta kl + \delta_{il}\delta_{kj}) \tag{7.2.10}$$

corresponding to $\mathbf{3}\otimes\mathbf{3} = \bar{\mathbf{3}}\oplus\mathbf{6}$ in the group theoretical language. Interestingly, as implied from the sign of each term in the above decomposition, the inter-quark interaction in the $\bar{\mathbf{3}}$ channel is attractive, while the $\mathbf{6}$ channel interaction is repulsive. This attractive nature is intuitively understood as follows: Suppose that two quarks are infinitely separate (in the deconfined phase), then the total field energy is just twice of the field energy associated with a single quark. If two quarks approach and make a composite of $\bar{\mathbf{3}}$, the total field energy is the same as that of a single quark, that is, a half of the original total energy. So, the energy decreases as two quarks touch. Consequently two quarks in the $\bar{\mathbf{3}}$ channel should feel an attractive force to minimize the total energy.

The most favored diquark channel is color anti-triplet (anti-symmetric) and spin singlet (anti-symmetric) and thus the flavor must be anti-symmetric. The diquarks generally carry two color indices and two flavor indices, but the diquark matrix in the most favored color-flavor channel simplifies to

$$\Phi_{i\alpha} = \epsilon_{ijk}\epsilon_{\alpha\beta\gamma}\, q^T_{j\beta} C\gamma_5 q_{k\gamma}\,. \tag{7.2.11}$$

Here, $C = i\gamma^0\gamma^2$ is the charge conjugation matrix necessary to make the diquark a Lorentz scalar. The Latin and the Greek letters represent the indices in flavor and color space, respectively.

In the three-flavor symmetric limit with $m_u = m_d = m_s$, the flavor rotation as well as the color rotation is a symmetry of the system. Then, it is possible to choose the flavor and the color bases to diagonalize $\Phi_{i\alpha}$. Without loss of generality we can parametrize the diquark condensate as

$$\langle \Phi_{i\alpha} \rangle = \delta_{i\alpha}\, \Delta_i\,. \tag{7.2.12}$$

Under the identification of $i = 1$, $\alpha = 1$ for up ($u$) and red ($r$), $i = 2$, $\alpha = 2$ for down ($d$) and green ($g$), and $i = 3$, $\alpha = 3$ for strange ($s$) and blue ($b$), for example, $\Delta_1$ involves pairings of *gd*-*bs* and *gs*-*bd* quarks. A state of quark matter with $\Delta_1 \neq 0$, $\Delta_2 \neq 0$, and $\Delta_3 \neq 0$ is known as the color-flavor locking (CFL) phase. The CFL phase is considered to be the ground state as long as the strange quark mass is ignored. In the opposite limit of infinitely heavy strange quark mass, we can regard quark matter as composed from only light flavors. In this case only $\Delta_3$ (involving *ru*-*gd* and *rd*-*gu* quark pairings) can take a nonzero value, while $\Delta_1 = \Delta_2 = 0$ due to suppression of strange quarks. Such a state of $\Delta_1 = \Delta_2 = 0$ and $\Delta_3 \neq 0$ is called the two-flavor color-superconducting (2SC) phase.

Which symmetry should spontaneously be broken by the diquark condensate is a nontrivial question. Let us first consider the 2SC phase. We note that the local gauge symmetry is never broken. Then, the baryon $U_V(1)$ symmetry is not broken in the 2SC phase since its rotation on $\Delta_3$ can be always canceled by unbroken electromagnetic transformation. The same argument concludes that flavor (chiral) symmetry is not broken, either. Therefore, in the 2SC phase, all global symmetries are unbroken, only modified with a mixture of local symmetry. One might think that color-superconducting phases assume deconfined quark matter, but as shown in Ref. [2218], the low-energy physics in the 2SC phase is governed by ungapped gluons in the unbroken $SU(2)$ sector and color confinement persists. Theoretically speaking, there is no gauge-invariant order parameter to define the 2SC phase.

In reality, however, the 2SC phase is anyway taken over by the CFL phase at high density where the strange quark mass is negligible. The $N_f = 3$ world is drastically different from the 2SC phase. The $U_V(1)$ symmetry can no longer be restored by the electromagnetic symmetry because $\Delta_1$ and $\Delta_{2,3}$ are differently charged. Thus, the CFL phase has a superfluid, and a vortex configuration is topologically stabilized. Also, chiral symmetry is spontaneously broken. We note that the diquark condensate has both the left-handed and the right-handed components; that is, $\langle qq \rangle = \langle q_R q_R \rangle + \langle q_L q_L \rangle \neq 0$, and $\langle q_{R,L} q_{R,L} \rangle$ breaks $SU_{R,L}(3)$. The vectorial rotation in flavor space can still be canceled by unbroken color rotation, so the symmetry breaking pattern in the CFL phase turns out to be: $SU_{\rm R}(3) \times SU_{\rm L}(3) \to SU_{\rm V}(3)$. Interestingly, this is identical to the symmetry breaking in the hadronic phase. Actually the gauge-invariant order parameter of the CFL phase is, $\langle(\bar{q}\bar{q})(qq)\rangle \sim \langle(\bar{q}q)^2\rangle$ that induces $\langle\bar{q}q\rangle \neq 0$ unless the anomalous $U_{\rm A}(1)$ is restored. The observation of exactly the same symmetry properties has led to a conjecture of continuity between the hadronic phase (i.e., the confined phase) with superfluidity and the CFL phase (i.e., the Higgs phase) [2219].

We can develop an intuitive understanding of the continuity. In the case of electron superconductivity, there is no gauge-invariant order parameter, and one might think that the theoretical characterization is as problematic. In this case, however, the solution has already been known. Because the Cooper pairs have twice the elementary charge, they cannot completely screen a single elementary charge. This would lead to an emergent $Z_2$ symmetry and the superconducting state is unambiguously defined by the symmetry.

This argument makes it clear why the CFL phase is so special. As mentioned earlier, the most favored diquark is found in the color triplet (and the anti-triplet) channel made from $\mathbf{3} \otimes \mathbf{3} \to \bar{\mathbf{3}}$. So, the Cooper pairs (i.e, the diquarks) are charged just like the fundamental (anti-)charge. Thus, a fundamental charge can be screened by Cooper pairs and the definition of the CFL phase is obscured, which underlies the continuity scenario between hypernuclear matter and CFL quark matter.

The continuity scenario cannot be applied to the 2SC phase as it is, but it was pointed out in Ref. [2220] that the NS environment can realize continuity within the two-flavor sector only. The idea is that the electric neutrality requires twice more $d$-quarks than $u$-quarks, and free $d$-quarks (not paired with $u$-quarks) may form a condensate of $\langle dd\rangle$. In this exotic phase that may be called the 2SC+d phase, the electromagnetic rotation cannot cancel rotations in $\Delta_3$ and $\langle dd\rangle$ simultaneously, and so it is a superfluid with $U_{\rm V}(1)$ breaking, and also, it spontaneously breaks chiral symmetry. In this way, as illustrated in Fig. 7.2.5, the continuity can be formulated.

Recently, the quark-hadron continuity scenario is encountering a fatal challenge. As mentioned, the emergent $Z_2$ symmetry characterizes ordinary electron superconductivity, and to see it mathematically, a Wilson loop as a symmetry generator is acted on a magnetic vortex operator. The magnetic flux in a superconducting cylinder is quantized in units not of $2\pi\hbar c/e$ but of $\pi\hbar c/e$ due to doubly charged Cooper pairs. The symmetry operation with the Wilson loop, hence, results in a $Z_2$ phase. The same exercise in the CFL phase replaces the magnetic vortex with the non-Abelian CFL vortex [2221] that carries both the global phase as well as the chromo-magnetic flux. From the explicit expression of the non-Abelian CFL vortex, it has been shown in Ref. [2222] that a $Z_3$ symmetry emerges (see also Ref. [2223] for more mathematical discussions). The hadronic phase presumably confines any color degrees of freedom, and it is a natural anticipation (but not proven yet) that this $Z_3$ symmetry operation is merely trivial in the confined phase. If so, the spontaneous breaking of emergent $Z_3$ symmetry should result in a phase transition from nuclear to quark matter. It is not yet settled theoretically whether a phase transition really separates nuclear and quark matter. The symmetry arguments are convincing, but the calculations are feasible only at high enough density, not at intermediate density where a transitional change may occur. As we will argue later, astrophysical observations constrain the strength of first-order phase transition for the neutron-rich NS matter, and for the moment it disfavors the first-order phase transition.

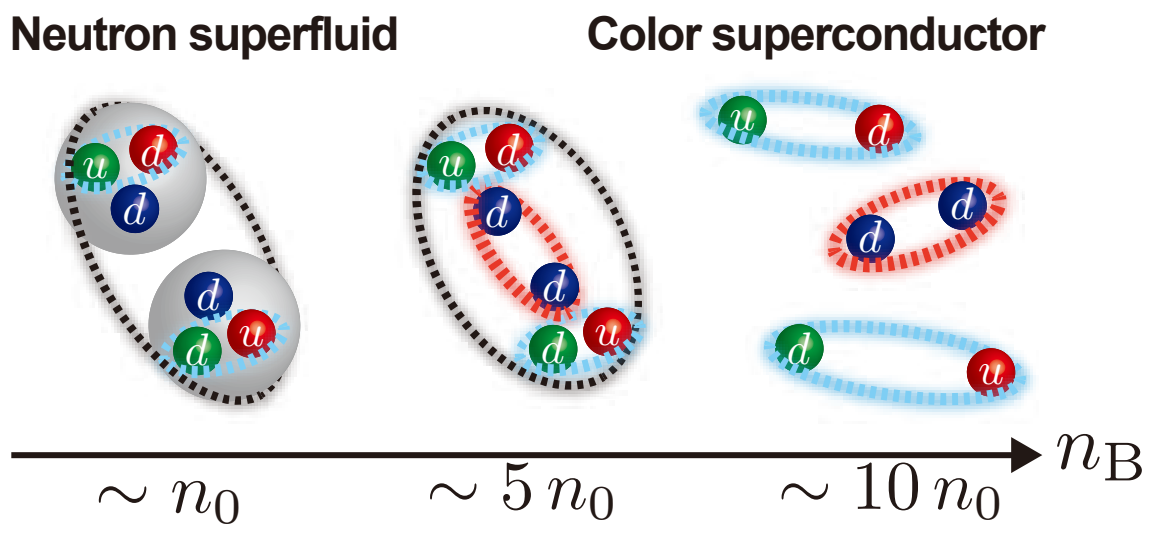

**Fig. 7.2.5** An illustration of the two-flavor continuity scenario between nuclear matter and 2SC+d quark matter in the NS environment in $\beta$ equilibrium. Figure taken from Ref. [2220].

### 7.2.4 Quarkyonic Regime

In the large-$N_{\rm c}$ limit the duality between nuclear and quark matter has been recognized by McLerran and Pisarski [2224] and they named the dual regime of matter Quarkyonic Matter after a combination of "quark" and "baryonic". It should be noted that Quarkyonic Matter is not a novel phase of matter but it refers to a regime in which the duality is manifested.

The conjectured duality is based on the large-$N_{\rm c}$ counting of the pressure. Along the temperature axis at zero baryon density, the pressure jumps from $\mathcal{O}(1)$ in the confined phase to $\mathcal{O}(N_{\rm c}^2)$ in the deconfined gluonic phase, which defines a first-order phase transition even with dynamical quarks. Then, along the axis of the baryon/quark chemical potential at zero temperature, one might also think of a phase transition from $\mathcal{O}(1)$ in confined nuclear matter to $\mathcal{O}(N_{\rm c})$ in deconfined quark matter. This naïve order counting implicitly neglects the contribution from interactions that could be dropped in the dilute/dense limits, but not in the intermediate density region. Actually, in the large-$N_c$ limit, the amplitude of meson scattering is suppressed so that mesons can be regarded as non-interacting particles, while baryons interact strongly. It is immediately understood why the strength of baryon interaction scales as $\mathcal{O}(N_c)$. The one pion exchange process for the two-body nucleon-nucleon ($NN$) interaction can be viewed microscopically as a quark hopping from one to the other baryon as shown in a schematic picture in Fig. 7.2.6. There are $N_c^2$ combinations of quark exchanges, among which color singlets are of $\mathcal{O}(N_c)$. In contrast, the $n$-point interaction vertices of mesons scale as $\mathcal{O}(N_c^{1-n/2})$ that goes to zero as $N_c \to \infty$ for $n \geq 3$. All the multi-body interactions of nucleons turn out to scale as $\mathcal{O}(N_c)$ which coincides with the scaling property peculiar to quark matter. In this way, in Quarkyonic Regime, the system is still in the confined phase and the relevant degrees of freedom are baryons, but the pressure is sensitive to quark degrees of freedom through inter-baryonic interactions.

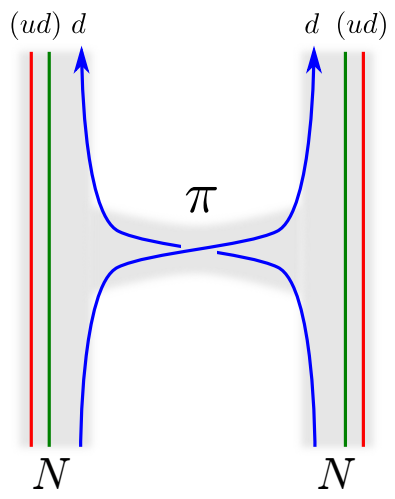

**Fig. 7.2.6** A quark description of two-body $NN$ interaction.

Now, we see that the deconfinement phenomenon induced by baryons at high density is far more nontrivial than the high temperature situation dominated by mesons. For weakly interacting mesons the onset of deconfinement can be approximated as an overlap of wavefunctions, that agrees with a picture of site percolation. For baryons, however, the onset of deconfinement would be rather located at the density where the $NN$, $NNN$, and arbitrary multi-body interactions become comparably strong, building a connected network of interacting bonds. In the language of percolation, hence, it would not be the site percolation but the bond percolation that is appropriate for high-density deconfinement. It has also been speculated that the deconfinement onset could be delayed toward higher density by quantum fluctuations as suggested in a quantum percolation picture [2225].

In Quarkyonc Regime the state of matter is not simply quark-like nor baryon-like, but something that shares both features. It is unlikely that there is any sharp deconfinement boundary in the phase diagram as drawn in the prototype in Fig. 7.2.1. This is why deconfinement is represented by smooth gradation over Quarkyonic Regime in Fig. 7.2.4. It is quite suggestive that both the CFL phase and Quarkyonic Regime favor smooth continuity between nuclear and quark matter in parallel, even though the diquark condensate is suppressed in the large-$N_c$ limit and color-superconducting matter and Quarkyonic Matter seem not to coexist.

### 7.2.5 Critical Point vs. Inhomogeneous States

So far, we have focused on deconfinement, and we shall now turn to the chiral phase transition at finite density. It has been established that the chiral phase transition at physical quark masses is a smooth crossover if the chiral restoration is induced by the temperature effect [2226]. Most chiral models predict that, as the baryon density increases, the behavior of the chiral condensate as a function of increasing $T$ becomes steeper. Eventually, in some chiral models, the chiral restoration occurs with a discontinuous jump in the chiral condensate, and the separation point between the smooth crossover and the first-order phase transition corresponds to the exact second-order critical point, which is commonly called the QCD Critical Point (QCP). It is sometimes referred to as the critical end point (CEP) of QCD matter as well. The presence of the QCP was first recognized in the Nambu–Jona-Lasinio model by Asakawa, Yazaki [2227], and in a QCD-like model by Barducci, Casalbuoni, De Crutis, Gatto, Pettini [2228], independently. For a comprehensive review on the order of chiral restoration at the early stage, see Ref. [2229].

In the language of the Ginzburg-Landau theory, the grand potential has an expansion,

$$\Omega = \frac{\alpha_2}{2} M^2 + \frac{\alpha_4}{4} M^4 + \frac{\alpha_6}{6} M^6 + \mathcal{O}(M^8)\,, \qquad (7.2.13)$$

with respect to an order parameter $M \sim \langle \bar{q} q \rangle$ (proportional to the constituent quark mass). For simplicity the bare quark mass effect that induces a symmetry-breaking term $\propto M$ is dropped. The coefficients, $\alpha_i$, are functions of $T$ and $\mu_B$. If $\alpha_2$ changes its sign while $\alpha_4 > 0$ is kept, a second-order phase transition is derived. If $\alpha_2 = 0$ and $\alpha_4$ changes its sign for $\alpha_6 > 0$, a tricritical point appears.

Interestingly, the QCP has nothing to do with the original chiral symmetry of QCD, and the universality class belongs to the same as the three-dimensional Ising model. Only when the bare quark mass is vanishing, as mentioned above, the QCP is located on the chiral phase transition, which exhibits tricriticality. At finite bare quark mass that explicitly breaks chiral symmetry, the QCP is identified as the $Z_2$ liquid-gas transition whose order parameter is the density, i.e., a conserved quantity coupled with the energy-momentum tensor, resulting in the dynamical universality class of the model H [2230].

The QCP can be an unambiguous landmark, if experimentally confirmed, on the QCD phase diagram. It is, however, quite nontrivial what plays a role of a signature. The most well-investigated quantity in the search for the QCP is the fluctuation observable. Because the correlation functions scale with the correlation length $\xi$, that ideally diverges near the critical point but does not in reality due to the critical slowing down, one can make a robust prediction for the critical behavior characterized by $\xi$ to the power of critical exponent. Although the time evolution away from the QCP may wash out the critical signature, the fluctuation of the conserved quantities such as the baryon number, the electric charge, and the strangeness (within the time scale of the strong interaction) could retain a trace of criticality if its value is frozen inside the critical region. This means that, to probe the QCP in a heavy-ion collision experiment, the created hot and dense matter must cool down along the trajectory hitting the critical region of the QCP, and the chemical freezeout (that fixes the ratio of the particle species) must be located sufficiently near the QCP. Such requirements may hold or may not.

There is no reliable QCD-based prediction for the location of the QCP due to the sign problem, but the virtue of the QCP search is that the critical theory provides us with unique theoretical prediction once its location on the phase diagram is experimentally constrained. We have already learnt a lot about fluctuations from nearly zero baryon density (at high collision energy $\sqrt{s_{NN}} \gtrsim 100\,\mathrm{GeV}$) to high density (at $\sqrt{s_{NN}} \sim 3\,\mathrm{GeV}$) from the heavy-ion collision experiments. See discussions in the previous section and the figure to show the data of $\kappa_4/\kappa_2$. For a review including related topics, especially the kurtosis (fourth order fluctuation of the proton number) and the skewness (third order fluctuation), see Ref. [2231] and references therein.

It is also mentioned that constructing an effective description of low-energy dynamics near the critical point is an intriguing theoretical challenge. Typically the time evolution of locally equilibrated matter is governed by undamped zero modes associated with conservation laws, which constitutes the hydrodynamic description based on the derivative expansion. In the vicinity of the critical point, the critical slowing down breaks the clear scale separation. Then, the correlation of the diffusive mode, that is the slowest one, should be coupled in the hydrodynamic equations, and such a generalized framework – called the "Hydro+" – has been proposed [2232].

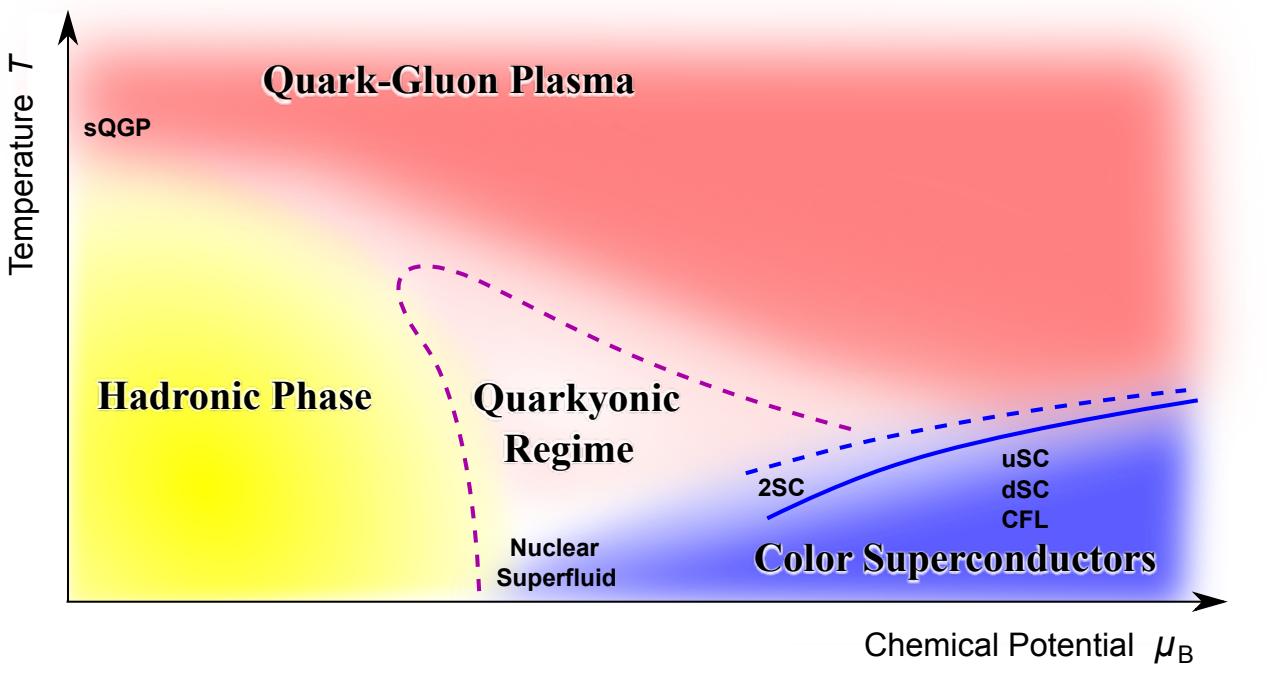

**Fig. 7.2.7** Another phase diagram of QCD matter without the QCD Critical Point, which may be the case in the neutron star matter in $\beta$ equilibrium for which the nuclear liquid-gas critical point is known to disappear.

One subtlety remains. The grand potential, $\Omega$, can be generally expressed as a functional of the order parameter, which is denoted by $M(\boldsymbol{x})$ here. If the spatial variation is smaller than other scales, the derivative expansion makes sense and the spatial profile of $M(\boldsymbol{x})$ is optimized to minimize the energy locally. It was Nickel [2233] who first recognized that in a typical chiral model in the mean-field approximation the local energy takes the following structured form:

$$\Omega = \frac{\alpha_2}{2} M(\boldsymbol{x})^2 + \frac{\alpha_4}{4}\left[M(\boldsymbol{x})^4 + (\boldsymbol{\nabla} M(\boldsymbol{x}))^2\right] + \frac{\alpha_6}{6}\left[M(\boldsymbol{x})^6 + 5(\boldsymbol{\nabla} M(\boldsymbol{x}))^2 M(\boldsymbol{x})^2 + \frac{1}{2}(\Delta M(\boldsymbol{x}))^2\right]. \tag{7.2.14}$$

This is a striking result. At the tricritical point (and near the QCP also) $\alpha_4$ should change its sign. For $\alpha_4 < 0$, the coefficient of the first derivative correction turns out to be negative too, which means that $\boldsymbol{\nabla} M(\boldsymbol{x}) \neq 0$ would lower the local energy. Therefore, the above form of the expanded energy indicates that the ground state should be spatially inhomogeneous.

The onset of inhomogeneity is called the Lifshitz point and Nickel's calculation was the first clarification for an explicit relation between the QCP and the Lifshitz point, though there were preceding works to hint at the possibility of inhomogeneous ground states [2234]. Whether the QCP and the Lifshitz point exactly coincide or not depends on the model choice and the approximation, and in more realistic model studies the QCP is overridden by the inhomogeneous states (see Ref. [2235] for a comprehensive review). Interestingly, such an inhomogeneous state is favored also in Quarkyonic Regime; the large-$N_c$ limit justifies an approximation of nuclear matter by a Skyrme crystal that inevitably gives rise to inhomogeneous chiral condensate.

Therefore, another view of the QCD phase diagram may look like Fig. 7.2.7 on which the QCP is taken over by an approximate triple point where the hadronic phase, the quark-gluon plasma, and Quarkyonic Regime (or the inhomogeneous state) meet [2207]. Once the large-$N_c$ approximation is relaxed, however, the thermal fluctuations of phonons and pions should be taken into account. It is known by now that inhomogeneous condensates are unstable and the quasi-long-range order (i.e., not exponential but algebraic decay of the order parameter correlation) could survive there [2236, 2237]. In contrast to the QCP on Fig. 7.2.4, it is a demanding question what can be an experimental signature to detect Quarkyonic Regime (or the quasi-long-range order) if the genuine phase diagram is like Fig. 7.2.7. Even without inhomogeneous condensates, for example for the theory proposal, the order parameter modes could be modified nontrivially to have a damped dispersion relation similar to the roton, which was discussed as a candidate for the observable signature [2238].

### 7.2.6 Astrophysical Constraints

Figure 7.2.7 looks like one variant of conjectured phase diagrams, but a special realization of dense matter in accord to Fig. 7.2.7 is known. That is, the state of dense matter in deep cores of a neutron star (NS) satisfies the $\beta$ equilibrium condition and contains more neutrons than protons due to the Coulomb interaction. This makes a sharp contrast to the heavy-ion collision whose time scale is shorter than the weak interaction, and flavor changing processes are negligible. It is important to note that the isospin contents would significantly affect the phase structure of QCD matter. A well-known example is that the first-order liquid-gas phase transition of symmetric nuclear matter in Fig. 7.2.4 does not exist any more in the NS matter; that is, pure neutron matter is not a self-bound fermionic system unlike symmetric nuclear matter. Then, it would be conceivable that the $\beta$ equilibrium condition simplifies the phase diagram from the conventional one as in Fig. 7.2.4 into a smoother shape without any solid phase boundary as in Fig. 7.2.7.

In fact, as we saw already before, the quark-hadron continuity scenario of the color-superconducting phase and the large-$N_c$ Quarkyonic Regime supports a picture of smooth crossover from nuclear to quark matter. Here, we discuss astrophysical constraints about the phase transition of QCD matter. The internal structure of the NS follows from the hydrostatic condition (called the Tolman-Oppenheimer-Volkoff equation) between the inward gravitational force and the outward pressure gradient. To this end, the calculation of the pressure gradient needs the relation of the pressure as a function of the baryon density, i.e., $p = p(\rho)$, or as a function of the energy density, $p = p(\varepsilon)$, which is referred to as the equation of state (EOS).

There is no first-principles derivation of the EOS except for the zero-density and the high-density limits and the EOS is the most crucial source of uncertainty in NS phenomenology. For a given $\varepsilon$, the EOS with larger $p$ (and smaller $p$) is called "stiff" (and "soft", respectively). Generally speaking, stiffer EOSs can support heavier NSs, and so the heaviest NS can give us the information about the EOS stiffness. If an assumed model cannot predict a stiff EOS enough to explain the experimentally confirmed largest NS mass, this model is falsified. In the presence of the first-order phase transition, $p = p(\varepsilon)$ should have a plateau, i.e., a window of $\varepsilon$ with a constant $p$, in the mixed-phase region, which generally makes the EOS softer.

In 2010 the mass measurement in a binary system (an NS and a white dwarf) by means of the Shapiro time delay established the existence of a two-solar-mass NS (PSR J1614-2230 [2239]). Later, similar massive NSs (PSR J0348+0432 [2240] and PSR J0740+6620 [2241]) have been discovered. These observations are extremely useful to make strict constraints and to exclude some of soft EOSs. In particular, the first-order phase transition is disfavored; it should be weak if the first-order phase transition takes place at moderate density reachable in the NS environment, or the first-order phase transition can occur only at large density beyond the NS region [2242]. In principle, a very rapid stiffening before/after the first-order phase transition could also yield an EOS that supports the massive NSs, but justification of the underlying mechanism needs further investigations. Actually, the *ab initio* estimates based on the chiral effective theory ($\chi$EFT) and the pQCD suggest that the nuclear EOS near the saturation density $n_0$ and quark EOS for high density $\gtrsim 5n_0$ are both softer than empirically adopted EOSs, and the stiffening should occur around 1.5-1.8 times $n_0$ [2243].

It is quite suggestive that such behavior of rapid stiffening from a low-density soft EOS is inferred from the experimental data, irrespective of any theoretical conjecture. The distribution of masses and radii of the observed NSs can be analyzed by probabilistic methods and the preferred EOS can be constructed from the observational data only. Figure 7.2.8 shows a specific combination of the EOS, i.e., $1/3 - p/\varepsilon$, as a function of dimensionless energy density $\varepsilon/\varepsilon_0$ with $\varepsilon_0 = 150\,\text{MeV/fm}^3$, that approaches zero in the conformal limit at high density. In Fig. 7.2.8 the orange, the green, and the red lines represent the results from the Bayesian analyses of the observational data in Ref. [2244], Ref. [2245], and Ref. [2246], respectively. The blue line represents the results from the neural network analysis in Ref. [2220].

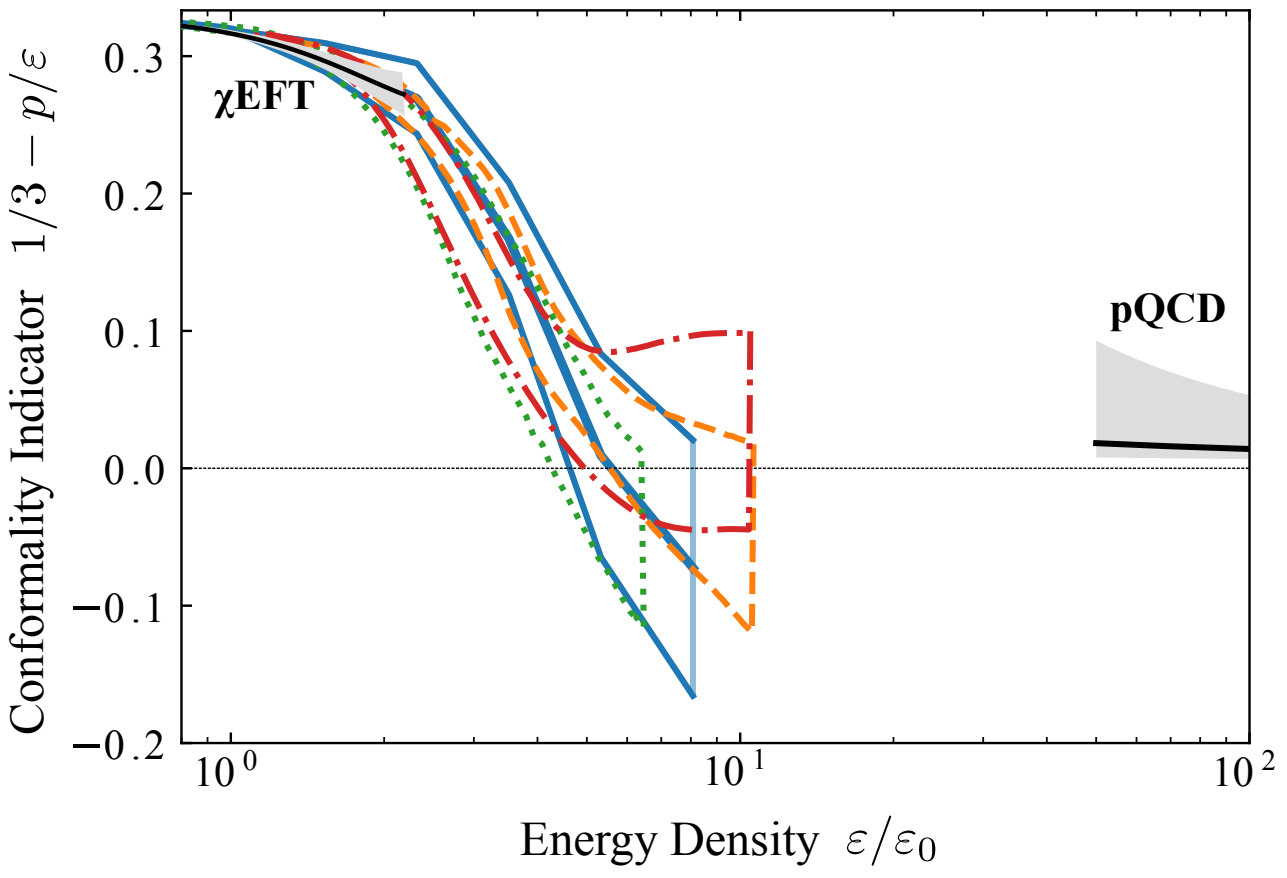

**Fig. 7.2.8** Conformality indicator deduced from the neutron star data as a function of the energy density normalized by $\varepsilon_0 = 150\,\mathrm{MeV/fm}^3$. Bands with different colors refer to the results from Refs. [2220, 2244–2246]. Figure adapted from Ref. [2247].

An intriguing finding is that the system seems to restore the conformality quite rapidly as first quantified in Ref. [2247]. The pQCD results at asymptotically high density as indicated in Fig. 7.2.8 are nearly conformal because the density scale is sufficiently larger than $\Lambda_{\mathrm{QCD}}$ and the system is weakly interacting. Thus, the NS experimental data imply the realization of strongly correlated conformal matter far earlier at not asymptotically high but just intermediate density. The microscopic origin of early conformality is to be identified by future studies.

Finally, let us briefly mention the impact of the gravitational wave signal from the NS merger. So far, the LIGO-Virgo collaboration reported two events of GW170817 and GW190425 as candidates of the NS-NS merger. In particular the former happened at a short distance of only 40 Mpc, and the electromagnetic counterpart (called the "kilonova") was also detected. For the moment only the signal from the inspiral stage before the merger has led to an EOS constraint in terms of the tidal deformability coefficient [2248, 2249], which turned out to be consistent with preceding constraints from the symmetry energy measurement [2250] as well as the NS mass-radius distributions. In the future the post-merger stage after the merger might be detected, and if so, an extremely dense state of matter, even denser than the largest density in the central core of the NS, could be probed, which will eventually clarify the nature of Quarkyonic Regime, quark matter, and hopefully color-superconducting states.

# 8 Mesons

*Conveners:*
**Eberhard Klempt and Curtis Meyer**

The Particle Data Group lists 78 light mesons with $u$ and $d$ quarks, 50 of them are "established", with 3* or 4* ratings. 25 mesons carry strangeness, 16 of them are established. Most mesons show a regular pattern, their masses are mostly compatible with a Regge behavior in $L$ and $N$. Curtis Meyer introduces the meson quantum numbers and their regularities. The scalar mesons of lowest mass have resisted for a long time an undisputed acceptance with proper poles in the complex energy plane. José Pelaez shows how unitarity, analyticity and dispersion relations are exploited to determine the scalar partial wave and to extract the poles with high precision. A driving force in meson spectroscopy is since long the quest for hybrids, in particular those with exotic quantum numbers, and for glueballs. Boris Grube and Eberhard Klempt present old and recent evidence for these states. 12 (7) established (candidate) charmed mesons are known at present, 7 (5) mesons with a bottom and a light quark, 6 (5) with a strange and 2 with a charm quark. Charmonium (and bottomonium) played a crucial role for the general acceptance of the quark model. Nowadays, 39 $c\bar{c}$ states are known, 25 of them established. The so-called XYZ states, unexpected states or states with unexpected properties, play an important role to understand the richness of QCD. Marco Pappagallo reports on the crime story of $X(3872)$ with its dual nature, on the unexpected $Y(4260)$ and the discovery of $Z_c^+(4430)$ and $T_{cc}(3875)$, both with minimal four-quark content. Nora Brambilla outlines the different approaches to identify the degrees of freedom driving the exotic states.

## 8.1 The meson mass spectrum, a survey

**Curtis Meyer**

### 8.1.1 Introduction

In the quark model, mesons are states containing quarks, antiquarks and gluons such that the net baryon number of the state is zero. Conventional mesons are described as bound states of a quarks and an antiquark ($q\bar{q}$) and can be viewed as similar to positronium ($e^+e^-$). Exotic mesons can include hybrids, which are $q\bar{q}g$ states with valence glue, four-quark states containing two quarks and two antiquarks, and glueballs containing only glue. These more exotic forms will be discussed in later sections, this section will deal with the ordinary mesons, referred to here as simply mesons. Mesons containing only $u$, $d$ and $s$ quarks are known as *light-quark mesons.* Given three quarks and three antiquarks, nine possible $q\bar{q}$ combinations can be made. These nine mesons form a so-called nonet where the members have the same well-defined quantum numbers: total spin $J$, parity $P$, and C-parity $C$, represented as $J^{PC}$.

### 8.1.2 Meson Quantum Numbers

The $J^{PC}$ quantum numbers of quark-antiquark systems are functions of the total spin, $S$, and the relative orbital angular momentum, $L$. The spin $S$ and angular momentum $L$ combine to yield the total spin $J$,

$$\vec{J} = \vec{L} \oplus \vec{S}\,, \tag{8.1.1}$$

where $L$ and $S$ add as two angular momenta.

Parity is the result of a mirror reflection of the wave function, taking $\vec{r}$ into $-\vec{r}$. It can be written as

$$P\left[\psi(\vec{r})\right] = \psi(-\vec{r}) = \eta_P \psi(\vec{r})\,, \tag{8.1.2}$$

where $\eta_P$ is the eigenvalue of parity. As application of parity twice must return the original state, $\eta_P = \pm 1$. In spherical coordinates, the parity operation reduces to the reflection of a $Y_{LM}$ function,

$$Y_{LM}(\pi - \theta, \pi + \phi) = (-1)^L Y_{LM}(\theta, \phi)\,. \tag{8.1.3}$$

From this, we conclude that $\eta_P = (-1)^L$. For a $q\bar{q}$ system, the intrinsic parity of the antiquark is opposite to that of the quark, which yields the total parity of a $q\bar{q}$ system as

$$P(q\bar{q}) = -(-1)^L\,. \tag{8.1.4}$$

Charge conjugation, $C$, is the result of a transformation that takes a particle into its antiparticle. For a $q\bar{q}$ system, only electrically-neutral states can be eigenstates of $C$. In order to determine the eigenvalues of $C$ ($\eta_C$), we need to consider a wave function that includes both spatial and spin information

$$\Psi(\vec{r}, \vec{s}) = R(r) Y_{lm}(\theta, \phi) \chi(\vec{s})\,. \tag{8.1.5}$$

If we consider a $u\bar{u}$ system, the $C$ operator reverses the meaning of $u$ and $\bar{u}$ which has the effect of mapping the vector $\vec{r}$ to the $u$ quark into $-\vec{r}$. Thus, following the arguments for parity, the spatial part of $C$ yields a factor of $(-1)^L$. The $C$ operator also reverses the two individual spins. For a symmetric $\chi$, we get a factor of 1, while for an antisymmetric $\chi$, we get a factor of $-1$. For two spin-$\frac{1}{2}$ particles, the $S = 0$ singlet is antisymmetric, while the $S = 1$ triplet is symmetric. Finally, there is an additional factor of $-1$ when we interchange two fermions. Combining all of this, we find that the C-parity of (a neutral) $q\bar{q}$ system is

$$C(q\bar{q}) = (-1)^{L+S}\,. \tag{8.1.6}$$

**Table 8.1.1** The allowed $J^{PC}$ quantum numbers for light-quark mesons with $L$ up to 4.

| $L$ | $S$ | $J^{PC}$ | $L$ | $S$ | $J^{PC}$ |
|---|---|---|---|---|---|
| 0 | 0 | $0^{-+}$ | 3 | 0 | $3^{+-}$ |
| 0 | 1 | $1^{--}$ | 3 | 1 | $2^{++}$ |
| 1 | 0 | $1^{+-}$ | 3 | 1 | $3^{++}$ |
| 1 | 1 | $0^{++}$ | 3 | 1 | $4^{++}$ |
| 1 | 1 | $1^{++}$ | 4 | 0 | $4^{+-}$ |
| 1 | 1 | $2^{++}$ | 4 | 1 | $3^{--}$ |
| 0 | 2 | $2^{-+}$ | 4 | 1 | $4^{--}$ |
| 1 | 2 | $1^{--}$ | 4 | 1 | $5^{--}$ |
| 1 | 2 | $2^{--}$ | | | |
| 1 | 2 | $3^{--}$ | | | |

In Table 8.1.1 are shown the $J^{PC}$s and the possible values of $L$ and $S$ up to $L$ of 3. From the list, the $J^{PC}$ values of $0^{--}$, $0^{+-}$, $1^{-+}$, $2^{+-}$ and $3^{+-}$ are missing. These missing $J^{PC}$ are referred to as *exotic*.

Because $C$-parity is only defined for neutral meson, we define $G$-parity to extend this to all non-strange $q\bar{q}$ states, independent of charge. For isovector states ($I = 1$), $C$ would transform a charged member into the oppositely charged state (*e.g.* $\pi^+ \to \pi^-$). In order to transform this back to the original charge, we would need to perform a rotation in isospin ($\pi^- \to \pi^+$). For a state of whose neutral partner has $C$-parity $C$, and whose total isospin is $I$, the $G$-parity is defined to be

$$G = C \cdot (-1)^I\,, \tag{8.1.7}$$

which can be generalized to

$$G(q\bar{q}) = (-1)^{L+S+I}\,. \tag{8.1.8}$$

The latter is valid for all of the $I = 0$ and $I = 1$ non-strange members of a nonet. In the limit of exact SU(3) symmetry, $G$ is conserved. Mesons with $G = +1$ decay into an even number of pions while mesons with $G = -1$ decay into an odd number of pions. From this, mesons have the following well defined quantum numbers: total angular momentum $J$, isospin $I$, parity $P$, C-parity $C$, and G-parity $G$. These are represented as $(I^G)J^{PC}$, or simply $J^{PC}$ for short. For the case of $L = 0$ and $S = 0$, we have $J^{PC} = 0^{-+}$, while for $L = 0$ and $S = 1$, $J^{PC} = 1^{--}$.

### 8.1.3 Light-quark meson names

Prior to 1986, there was no systematic naming scheme for mesons. Those who discovered new states often proposed what those states would be called. In 1986, the Particle Data Group [2251] proposed a naming scheme for mesons that is still in use today. This scheme is based on the total spin $J$, parity $P$ and charge conjugation $C$ of the nonet, and then the isospin of the nonet members. The base name is the same for all mesons of a given $I$ and $PC$, where there is a subscript denoting the total spin $J$. For the kaons, ($I = \frac{1}{2}$), those with $J^P = 0^-, 1^+, 2^-, \cdots$ are named $K_J$, while those with $J^P = 0^+, 1^-, 2^+, \cdots$ are named $K_J^*$. Table 8.1.2 lists the names of the light-quark mesons up to $L = 3$.

**Table 8.1.2** The naming scheme for light-quark mesons [939].

| $L$ | $S$ | $J^{PC}$ | $I=1$ | $G$ | $I=0$ | | $G$ | $K$ |
|---|---|---|---|---|---|---|---|---|
| 0 | 0 | $0^{-+}$ | $\pi$ | $-1$ | $\eta$ | $\eta'$ | $+1$ | $K$ |
| 0 | 1 | $1^{--}$ | $\rho$ | $+1$ | $\omega$ | $\phi$ | $-1$ | $K^*$ |
| 1 | 0 | $1^{+-}$ | $b_1$ | $+1$ | $h_1$ | $h_1'$ | $-1$ | $K_1$ |
| 1 | 1 | $0^{++}$ | $a_0$ | $-1$ | $f_0$ | $f_0'$ | $+1$ | $K_0^*$ |
| 1 | 1 | $1^{++}$ | $a_1$ | $-1$ | $f_1$ | $f_1'$ | $+1$ | $K_1$ |
| 1 | 1 | $2^{++}$ | $a_2$ | $-1$ | $f_2$ | $f_2'$ | $+1$ | $K_2^*$ |
| 2 | 0 | $2^{-+}$ | $\pi_2$ | $-1$ | $\eta_2$ | $\eta_2'$ | $+1$ | $K_2$ |
| 2 | 1 | $1^{--}$ | $\rho_1$ | $+1$ | $\omega_1$ | $\phi_1$ | $-1$ | $K_1^*$ |
| 2 | 1 | $2^{--}$ | $\rho_2$ | $+1$ | $\omega_2$ | $\phi_2$ | $-1$ | $K_2$ |
| 2 | 1 | $3^{--}$ | $\rho_3$ | $+1$ | $\omega_3$ | $\phi_3$ | $-1$ | $K_3^*$ |
| 3 | 0 | $3^{+-}$ | $b_3$ | $+1$ | $h_3$ | $h_3'$ | $-1$ | $K_3$ |
| 3 | 1 | $2^{++}$ | $a_2$ | $-1$ | $f_2$ | $f_2'$ | $+1$ | $K_2^*$ |
| 3 | 1 | $3^{++}$ | $a_3$ | $-1$ | $f_3$ | $f_3'$ | $+1$ | $K_3$ |
| 3 | 1 | $4^{++}$ | $a_4$ | $-1$ | $f_4$ | $f_4'$ | $+1$ | $K_4^*$ |
| 4 | 0 | $4^{-+}$ | $\pi_4$ | $-1$ | $\eta_4$ | $\eta_4'$ | $+1$ | $K_4$ |
| 4 | 1 | $3^{--}$ | $\rho_3$ | $+1$ | $\omega_3$ | $\phi_3$ | $-1$ | $K_3^*$ |
| 4 | 1 | $4^{--}$ | $\rho_4$ | $+1$ | $\omega_4$ | $\phi_4$ | $-1$ | $K_4$ |
| 4 | 1 | $5^{--}$ | $\rho_5$ | $+1$ | $\omega_5$ | $\phi_5$ | $-1$ | $K_5^*$ |

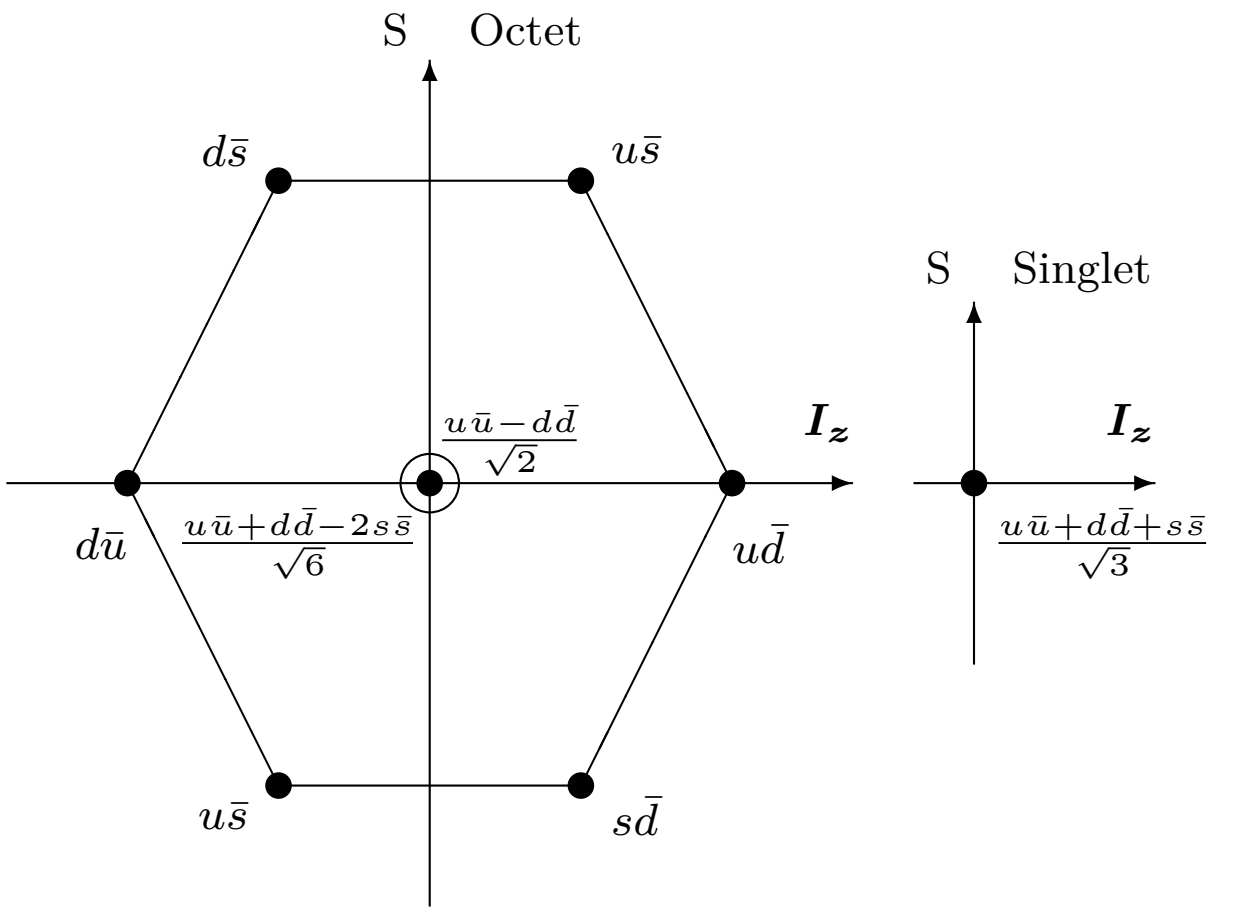

**Fig. 8.1.1** The SU(3) quark structure of the light-quark mesons. The mesons are plotted against strangeness $S$ on the vertical axis and the $z$ component of isospin, $I_z$ on the horizontal axis The left-hand plot shows the octet mesons, while the right-hand plot shows the singlet meson.

### 8.1.4 SU(3) flavor and light-quark mesons

Given three flavors of light quarks, there are nine possible $q\bar{q}$ combinations. SU(3) flavor groups these mesons into eight members of an SU(3) octet and one SU(3) singlet. Figure 8.1.1 shows these $q\bar{q}$ combinations plotted on a graph where the strangeness $S$ is plotted against the third component of isospin, $I_3$. There are four mesons with $S = 0$, three with isospin 1 and one with isospin 0. The SU(3) singlet state also has $I = 0$. The SU(3) singlet state with $I = 0$ is

$$\mid 1\rangle = \frac{1}{\sqrt{3}}\left(u\bar{u} + d\bar{d} + s\bar{s}\right) , \tag{8.1.9}$$

while the SU(3) octet state with $I = 0$ is

$$\mid 8\rangle = \frac{1}{\sqrt{6}}\left(u\bar{u} + d\bar{d} - 2\, s\bar{s}\right) . \tag{8.1.10}$$

**Table 8.1.3** The nonet mixing angles as reported in reference [939]. The linear formula is given by Eq. 8.1.12 while the quadratic angle is given by Eq. 8.1.13.

| $J^{PC}$ | $\theta_{lin}$ | $\theta_{quad}$ |
|---|---|---|
| $0^{-+}$ | $-24.5°$ | $-11.3°$ |
| $1^{--}$ | $36.5°$ | $39.2°$ |
| $2^{++}$ | $28.0°$ | $29.6°$ |
| $3^{--}$ | $30.8°$ | $31.8°$ |

The two physical $I = 0$ states are mixtures of the two SU(3) states. Following the Particle Data Group [939] notation with nonet mixing angle $\theta_n$, the physical isospin-zero states are

$$\begin{pmatrix} f \\ f' \end{pmatrix} = \begin{pmatrix} \cos\theta_n & \sin\theta_n \\ -\sin\theta_n & \cos\theta_n \end{pmatrix} \begin{pmatrix} \mid 1\rangle \\ \mid 8\rangle \end{pmatrix} . \tag{8.1.11}$$

Many of the known nonets have physical states that separate the light-quark states, $u\bar{u} + d\bar{d}$, and the states with hidden strangeness, $s\bar{s}$. This is known as ideal mixing and corresponds to $\tan\theta_n = \frac{1}{\sqrt{2}}$, or $\theta_n \approx 35.26°$. Contrary to this, the ground state mesons are almost pure SU(3) states. The $\eta'$ is nearly pure singlet and the $\eta$ is the octet state. The nonet mixing angles can be determined from masses of the member states. In the following, $m_a$ is the mass of the isospin 1 state, $m_K$ is the mass of the isospin-$\frac{1}{2}$, and $m_f$ and $m_{f'}$ are the masses of the two isospin-0 states. $\theta$ in equation 8.1.12 is known as the linear mixing angle,

$$\tan\theta = \frac{4m_K - m_a - 3m_{f'}}{2\sqrt{2}\,(m_a - m_K)} \tag{8.1.12}$$

while equation 8.1.13 is known to define the quadratic mixing angle.

$$\tan^2\theta = \frac{4m_K - m_a - 3m_{f'}}{-4m_K + m_a + 3m_f} \tag{8.1.13}$$

The Particle Data Group quotes mixing angles for four nonets which are listed in Table 8.1.3. With exception of the pseudoscalar $J^{PC} = 0^{-+}$ nonet, the others nonets are all fairly close to being ideally mixed.

**Table 8.1.4** The nonet mixing angles for mesons with orbital angular momentum less than 4. The lattice results are reported in reference [484], while the references for the decay rate determinations are given in the table.

| $J^{PC}$ | $\theta_n$ Lattice | $\theta_n$ Decays |
|---|---|---|
| $0^{-+}$ | $-11°$ | $-9.3°$ [2253] |
| $1^{--}$ | $33°$ | |
| $1^{+-}$ | $35°$ | |
| $1^{++}$ | $8°$ | |
| $2^{++}$ | $28°$ | $32.1°$ [2254] |
| $2^{-+}$ | $33°$ | $-6.7°$ [2255] |
| $1^{--}$ | $30°$ | |
| $2^{--}$ | $33°$ | |
| $3^{--}$ | $33°$ | $31.8°$ [2256] |
| $3^{-+}$ | $34°$ | |
| $2^{++}$ | $26°$ | |
| $3^{++}$ | $33°$ | |
| $4^{++}$ | $29°$ | |

**Table 8.1.5** The pseudoscalar mesons.

| Isospin | State(s) | Mass MeV | Width or Lifetime |
|---|---|---|---|
| 1 | $\pi^0$ | 134.9768 | $8.52 \times 10^{-17}$s |
| 1 | $\pi^{\pm}$ | 139.57039 | $2.6033 \times 10^{-8}$s |
| $\frac{1}{2}$ | $K^{\pm}$ | 493.677 | $1.238 \times 10^{-8}$s |
| $\frac{1}{2}$ | $K^0/\bar{K}^0$ | 497.611 | |
| 0 | $\eta$ | 547.862 | $0.00131\,MeV$ |
| 0 | $\eta'$ | 957.78 | $8.49\,MeV$ |

**Table 8.1.6** The radially excited pseudoscalar mesons according to reference [939].

| Isospin | State(s) | Mass MeV | Width MeV |
|---|---|---|---|
| 1 | $\pi(1300)$ | 1300 | 200 to 600 |
| 0 | $\eta(1295)$ | 1294 | 55 |
| 0 | $\eta(1475)$ | 1475 | 90 |
| $\frac{1}{2}$ | $K(1460)$ | 1482 | 335 |

**Table 8.1.7** A possible third nonet of pseudoscalar mesons.

| Isospin | State(s) | Mass MeV | Width MeV |
|---|---|---|---|
| 1 | $\pi(1800)$ | 1810 | 215 |
| 0 | $\eta(1760)$ | 1751 | 240 |
| 0 | | | |
| $\frac{1}{2}$ | $K(1830)$ | 1874 | 168 |

The most comprehensive predictions for nonet mixing angles comes from lattice QCD [484]. Those predictions are in good agreement with the known values. The mixing angles can also be determined using relative decay rates for the physical isospin 0 states to pairs of mesons in the same nonets, to two pseudoscalar mesons, or to a pseudoscalar and a vector $J^{PC} = 1^{--}$ meson. Determinations exploiting decay rates exist for several nonets [2252]. In Table 8.1.4 are listed the lattice QCD predictions as well as several determinations of mixing angles from decay measurements. The one discrepancy between lattice and decay rate predictions are in the $2^{-+}$ nonet; this may be due to incorrect assignments and is discussed later.

### 8.1.5 Light-quark mesons

*The pseudoscalar mesons*

The $J^{PC} = 0^{-+}$ mesons are spin singlets with 0 orbital angular momentum and are known as the pseudoscalar mesons. They are listed in Table 8.1.5. These are the lightest mesons, and with the exception of the $\eta'$, all their decays are either weak or electromagnetic. In addition, the mixing of this nonet is quite different from other nonets in that the mixing angle is small, and the $\eta$ and $\eta'$ are very close to being SU(3) octet and singlet states respectively.

In addition to the ground state pseudoscalar mesons, there can also be radially excited states. Both excited $\pi$s, the $\pi(1300)$ and $\pi(1800)$, and $\eta$s, the $\eta(1295)$, the $\eta(1405)$, the $\eta(1475)$, $\eta(1760)$ and the $\eta(2225)$ have been observed. The $K(1460)$ and $K(1830)$ are the observed $J^P = 0^-$ states. The lighter is consistent with the first radial excitation. The Particle Data Group [939] identifies the states listed in Table 8.1.6 as the nonet of radially excited pseudoscalar mesons. One could also associate the $\pi(1800)$, $\eta(1760)$ and $K(1830)$ together as a third nonet as listed in Table 8.1.7. However in additon to the second radial excitation, there is a predicted pseudoscalar glueball (see Section 8.4) as well as a nonet of $J^{PC} = 0^{-+}$ hybrid mesons (see Section 8.3).

With regard to the $\eta(1295)$ state, we believe that its status deserves some scrutiny and that the $\eta(1405)$ and $\eta(1475)$ should be the two $I = 0$ members of the radially excited pseudoscalar mesons. For the $\eta(1295)$, there is a single report in radiative $J/\psi$ decays [2257, 2258], but there is ambiguity about whether the signal is $\eta(1295)$ or $f_1(1285)$. It has not been reported in other $J/\psi$ measurements since then, while there has been extensive results of the $\eta'$, $\eta(1405)$ and $\eta(1475)$. The majority of the observations have been in pion production [2259–2263] where there are generally contributions from both the $\eta(1295)$ and the $f_1(1285)$. In $p\bar{p}$ annihilation, both the $\eta(1405)$ [2264] and the $\eta(1475)$ [2265] have been observed, but no observation of the $\eta(1295)$ has been reported.

*The vector mesons*

A spin triplet system with $L = 0$ forms the $J^{PC} = 1^{--}$ nonet, its members are known as vector mesons. These mesons are shown in Table 8.1.8. The dominant decay modes of the vector mesons are through the strong interaction to two or three pseudoscalar mesons and the states are nearly ideally mixed with the $\omega$ nearly all $u\bar{u}$ and $d\bar{d}$, while the $\phi$ is nearly all $s\bar{s}$.

**Table 8.1.8** The vector mesons.

| Isospin | State(s) | Mass MeV | Width MeV |
|---|---|---|---|
| 1 | $\rho(770)$ | 775.26 | 149.1 |
| 0 | $\omega(782)$ | 782.65 | 8.49 |
| 0 | $\phi(1020)$ | 1019.461 | 4.249 |
| $\frac{1}{2}$ | $K^{*\pm}(892)$ | 891.66 | 50.8 |
| $\frac{1}{2}$ | $K^{*0}(892)$ | 895.5 | 47.3 |

**Table 8.1.9** The radially excited vector mesons.

| Isospin | State(s) | Mass MeV | Width MeV |
|---|---|---|---|
| 1 | $\rho(1450)$ | 1465 | 400 |
| 0 | $\omega(1420)$ | 1410 | 290 |
| 0 | $\phi(1680)$ | 1680 | 150 |
| $\frac{1}{2}$ | $K^*(1410)$ | 1414 | 232 |

**Table 8.1.10** A possible fourth nonet of vector mesons.

| Isospin | State(s) | Mass MeV | Width MeV |
|---|---|---|---|
| 1 | $\rho(1900)$ | | |
| 0 | | | |
| 0 | $\phi(2170)$ | 2162 | 100 |
| $\frac{1}{2}$ | | | |

In addition to the expected radial excitations of the vector mesons, the $L = 2$, $S = 1$ $q\bar{q}$ system can also have $J^{PC} = 1^{--}$. Finally, there is a nonet of hybrid mesons expected with the same $J^{PC}$. Thus, we expect disentangling of the excited vector meson spectrum to be tricky. The reported states in $I = 1$ are $\rho(1450)$, $\rho(1570)$, $\rho(1700)$, $\rho(1900)$ and $\rho(2150)$. In the $I = 0$ system, $\omega(1420)$, $\omega(1680)$, $\phi(1680)$ and $\phi(2170)$ have been reported. Finally, for $I = \frac{1}{2}$, the $K^*(1410)$ and $K^*(1680)$ are known. The Particle Data Group identifies the radially excited states as in Table 8.1.9. The states identified with the D-wave nonet are listed in Table 8.1.16 and discussed later. Finally, the $\rho(1900)$ and $\phi(2170)$ could be part of another nonet; either the hybrid nonet or the second radial excitation of the ground-state vector mesons (Table 8.1.10).

*The pseudo vector mesons*

Spin singlet states with $L = 1$ form the $J^{PC} = 1^{+-}$ nonet, and are known as the pseudo vector mesons. These mesons are listed in Table 8.1.11. There is one known state beyond those listed in the table, the $h_1(1595)$ which has been reported in pion production [2266]. There is also an interesting complication with the kaonic states where $C$-parity is not defined. The states with open strangeness have $J^P = 1^+$ which is the same as those in the $J^{PC} = 1^{++}$ axial vector mesons. Because of this, the two states can mix, and it is believed that the physical states, $K_1(1270)$ and $K_1(1400)$, are mixtures of the SU(3) states, $K_{1A}$ and $K_{1B}$, with a mixing angle $\theta_{K_1} = -(33.6 \pm 4.3)^\circ$ [2267] conventionally defined by

$$\begin{pmatrix} \mid K_1(1270)\rangle \\ \mid K_1(1400)\rangle \end{pmatrix} = \begin{pmatrix} \sin\theta_{K_1} & \cos\theta_{K_1} \\ \cos\theta_{K_1} & -\sin\theta_{K_1} \end{pmatrix} \begin{pmatrix} \mid K_{1A}\rangle \\ \mid K_{1B}\rangle \end{pmatrix} .$$

**Table 8.1.11** The pseudo vector mesons.

| Isospin | State(s) | Mass MeV | Width MeV |
|---|---|---|---|
| 1 | $b_1(1235)$ | 1229.5 | 142 |
| 0 | $h_1(1170)$ | 1166 | 375 |
| 0 | $h_1(1415)$ | 1416 | 90 |
| $\frac{1}{2}$ | $K_{1A}$ | | |

**Table 8.1.12** The axial vector mesons.

| Isospin | State(s) | Mass MeV | Width MeV |
|---|---|---|---|
| 1 | $a_1(1260)$ | 1230 | 250 to 600 |
| 0 | $f_1(1285)$ | 1281.9 | 22, 7 |
| 0 | $f_1(1420)$ | 1426.3 | 54.5 |
| $\frac{1}{2}$ | $K_{1B}$ | | |

*The scalar mesons*

A spin triplet with $L = 1$ can form three possible $J^{PC}$s: $0^{++}$, $1^{++}$ and $2^{++}$. The $0^{++}$ states are known as scalar mesons and are discussed in Section 8.2 because there are added complications which make it difficult to discuss them with the other mesons. There is also significant discussion of the scalar states and their relation to the scalar glueball, see Section 8.4).

*The axial vector mesons*

The $L = 1$ $J^{PC} = 1^{++}$ mesons are known as the axial vector mesons and are listed in Table 8.1.12. As noted earlier in the discussion of the pseudo vector mesons, the SU(3) $K_{1B}$ state is a mixture of the physical $K_1(1270)$ and $K_1(1400)$ states. In addition to the states listed, two additional states have been reported. The $f_1(1510)$ has been seen in kaon production [2268, 2269] as well as pion production [2270] decaying to $K^*K$. These productions and decay would favor an $s\bar{s}$ interpretation of the $f_1(1510)$, but it is probably too light to be the radial excitation. A second state, the $a_1(1640)$ is identified as the radial excitation of the $a_1(1270)$. This has been observed in pion production with the most significant observation in reference [2271]. It has also been reported in $D$ decays [2272].

**Table 8.1.13** The tensor mesons.

| Isospin | State(s) | Mass MeV | Width MeV |
|---|---|---|---|
| 1 | $a_2(1320)$ | 1316.9 | 107 to 600 |
| 0 | $f_2(1270)$ | 1275.5 | 186.7 |
| 0 | $f_2'(1525)$ | 1517.4 | 86 |
| $\frac{1}{2}$ | $K_2^*(1430)$ | 1427 | 100 |

**Table 8.1.14** The radial excitations of the tensor mesons.

| Isospin | State(s) | Mass MeV | Width MeV |
|---|---|---|---|
| 1 | $a_2(1700)$ | 1698 | 265 to 600 |
| 0 | $f_2(1640)$ | 1639 | 99 |
| 0 | $f_2(1950)$ | 1936 | 464 |
| $\frac{1}{2}$ | $K_2^*(1980)$ | 1995 | 349 |

**Table 8.1.15** The pseudo tensor mesons.

| Isospin | State(s) | Mass MeV | Width MeV |
|---|---|---|---|
| 1 | $\pi_2(1670)$ | 1670.6 | 258 |
| 0 | $\eta_2(1645)$ | 1617 | 181 |
| 0 | $\eta_2(1870)$ | 1842 | 225 |
| $\frac{1}{2}$ | $K_2(1770)$ | 1773 | 186 |

**Table 8.1.16** The $L = 2$ $1^{--}$ vector mesons.

| Isospin | State(s) | Mass MeV | Width MeV |
|---|---|---|---|
| 1 | $\rho(1700)$ | 1720 | 250 |
| 0 | $\omega(1650)$ | 1670 | 315 |
| 0 | | | |
| $\frac{1}{2}$ | $K^*(1680)$ | 1718 | 322 |

*The tensor mesons*

The last $L = 1$ nonet contains the $J^{PC} = 2^{++}$ tensor mesons, where the states are listed in Table 8.1.13. This well-established nonet is close to ideally mixed as noted in Table 8.1.3. As with the vector mesons, there is a second $L$, $S$ combination that can exist for $J^{PC} = 2^{++}$, $L = 3$ and $S = 1$. In addition, one of the lightest glueballs is also expected to have these quantum numbers, and of course radial excitations should be present.

With regard to excited states, there is a second $a_2$ state, the $a_2(1700)$, which the Particle Data Group associates with the radial excitation of the tensor mesons. This assignment is based on mass, where we would expect the radial $a_2$ state to be close in mass to the $a_1(1640)$, the radial excitation of the $a_1$. The $L = 3$ state is expected to have similar mass to other $L = 3$ states, where here the $a_4(1970)$ anchors these nonet around 2 GeV. The $a_2(1700)$ has been observed in many production mechanisms including pion production [2271, 2273–2276], $p\bar{p}$ annihilation [2277–2280], two-photon production [2281, 2282] and $\psi'$ radiative decays [2283].

For the isospin 0 states, there is an overpopulation of $f_2$ states, with 10 additional states beyond the two ground state tensors reported. These include the $f_2(1430)$, $f_2(1565)$, $f_2(1640)$, $f_2(1810)$, $f_2(1910)$, $f_2(1950)$, $f_2(2010)$, $f_2(2150)$, $f_2(2300)$ and $f_2(2340)$. For $I = \frac{1}{2}$, there is a single state, the $K_2^*(1980)$. The Particle Data Group identifies the radially excited states as listed in Table 8.1.14. With the radial states accounting for 2 of the 10 extra states, a second pair in the $L = 4$ mesons, probably above 2 GeV in mass, and a glueball state, there are still 5 states. Presumably several of the reported states are all the same state, with low statistics and differences in production mechanisms accounting for the differences. Three of the isoscalar tensor states were observed in the OZI rule suppressed reaction $\pi^- p \to \phi\phi n$ [2284] and were discussed as one or three glueballs. This interpretation is supported by a recent analysis of BESIII data on radiative $J/\psi$ decays (see Section 8.4). In any case, a careful examination of the $I = 0$ $J^{PC} = 2^{++}$ data with high statistics experiments is merited.

*The pseudo tensor mesons*

Mesons formed with $S = 0$ and $L = 2$ have $J^{PC} = 2^{-+}$ and are known as the pseudo tensor mesons. The known states are listed in Table 8.1.15. In addition to the radial excitations of these states, there is also a nonet of hybrid mesons expected. The latter are likely slightly heavier than the mesons in Table 8.1.15. There are three known states beyond those in the table, the $\pi_2(1880)$, $\pi_2(2005)$ and the $\pi_2(2100)$. It is also interesting that the decay patterns of the $\eta_2(1645)$ and the $\eta_2(1870)$ both look like those for a $u\bar{u}/d\bar{d}$ state and not an $s\bar{s}$ state [2285]. This suggests that the $\eta_2(1870)$ might be paired with the $\pi_2(1880)$ in a third nonet. However, studies of the axial anomaly [2252] favor the assignment in Table 8.1.15, but with an unusual mixing angle that is inconsistent with lattice, as shown in Table 8.1.4.

*The D-state vector mesons*

The mesons formed from an $S = 1$, $L = 2$ $q\bar{q}$ system can have $J^{PC} = 1^{--}$, $2^{--}$ and $3^{--}$ and are referred to as vector mesons. The Particle Data Group identifies the states listed in Table 8.1.16 with the $1^{--}$ states, where there is no candidate for the $\phi$ state which is probably expected with a mass in the 1.8 to 1.9 GeV mass region. For the $J^{PC} = 2^{--}$ states, very little is known with the only assignment made by the Particle Data Group being the $K_2(1820)$. However, similar to the $K_{1A}$ and $K_{1B}$ of the $1^{+-}$ and $1^{++}$ nonets, the kaonic states from the $2^{-+}$ and $2^{--}$ nonets can also mix.

**Table 8.1.17** The $L = 2$ $3^{--}$ vector mesons.

| Isospin | State(s) | Mass MeV | Width MeV |
|---|---|---|---|
| 1 | $\rho_3(1690)$ | 1688.8 | 161 |
| 0 | $\omega_3(1670)$ | 1677 | 168 |
| 0 | $\phi_3(1850)$ | 1854 | 87 |
| $\frac{1}{2}$ | $K_3^*(1780)$ | 1776 | 159 |

**Table 8.1.18** The $L = 3$ $4^{++}$ mesons.

| Isospin | State(s) | Mass MeV | Width MeV |
|---|---|---|---|
| 1 | $a_4(1970)$ | 1967 | 324 |
| 0 | $f_4(2050)$ | 2018 | 237 |
| 0 | $f_4(2300)$ | 2320 | 260 |
| $\frac{1}{2}$ | $K_4^*(2045)$ | 2048 | 199 |

**Table 8.1.19** The $L = 4$ $5^{--}$ vector mesons.

| Isospin | State(s) | Mass MeV | Width MeV |
|---|---|---|---|
| 1 | $\rho_5(2350)$ | 2350 | 400 |
| 0 | | | |
| 0 | | | |
| $\frac{1}{2}$ | $K_5^*(2380)$ | 2382 | 178 |

The $J^{PC} = 3^{--}$ nonet is one of the well established nonets where a mixing angle is also reported. These states are listed in Table 8.1.17. In addition to the listed states, there are two additional $\rho_3$ states reported in literature. The $\rho_3(1990)$ is reported in $p\bar{p}$ annihilation [2286, 2287], and the $\rho_3(2250)$ reported in both $\bar{p}p$ annihilation and in $\psi'$ decays [2286, 2288]. The lighter state could be a radial excitation of the $L = 2$ $\rho_3(1690)$. The higher mass state is of similar mass to the $\rho_5(2350)$ and could be an $L = 4$ meson.

*Higher excitations*

Going beyond the $L = 2$ mesons, less is known, with the most information tending to be on the nonets with the largest $J$. For the case of $L = 3$, there are candidates for the $J^{PC} = 4^{++}$ mesons as shown in Table 8.1.18. There should also be a $2^{++}$ and $3^{++}$ nonet as well as a $J^{PC} = 3^{+-}$ nonet. While as noted in the tensor meson section, there are a large number of reported $f_2$ states, in particular the $f_2(2010)$, $f_2(2150)$, $f_2(2300)$ and $f_2(2340)$, assigning any of these to an $L = 3$ nonet is not clear. There is also a $J^P = 3^+$ kaonic state, the $K_3(2320)$ which could be a member of either of the spin 3 nonets.

For the $L = 4$ mesons, the highest spin is $J^{PC} = 5^{--}$, and a few states with these quantum numbers are known, as listed in Table 8.1.19. There should also be a $3^{--}$, $4^{--}$ and $4^{-+}$ nonet for which a few states are reported. For $I = \frac{1}{2}$ the $K_4(2500)$ which could be a member of either of the $J = 4$ nonets. There are also two $\rho_3$ states reported, the $\rho_3(1990)$ and the $\rho_3(2250)$. The latter state is of similar mass to the $\rho_5(2350)$ and could be an $L = 4$ meson. The lighter state could be a radial excitation of the $L = 2$ $\rho_3(1690)$.

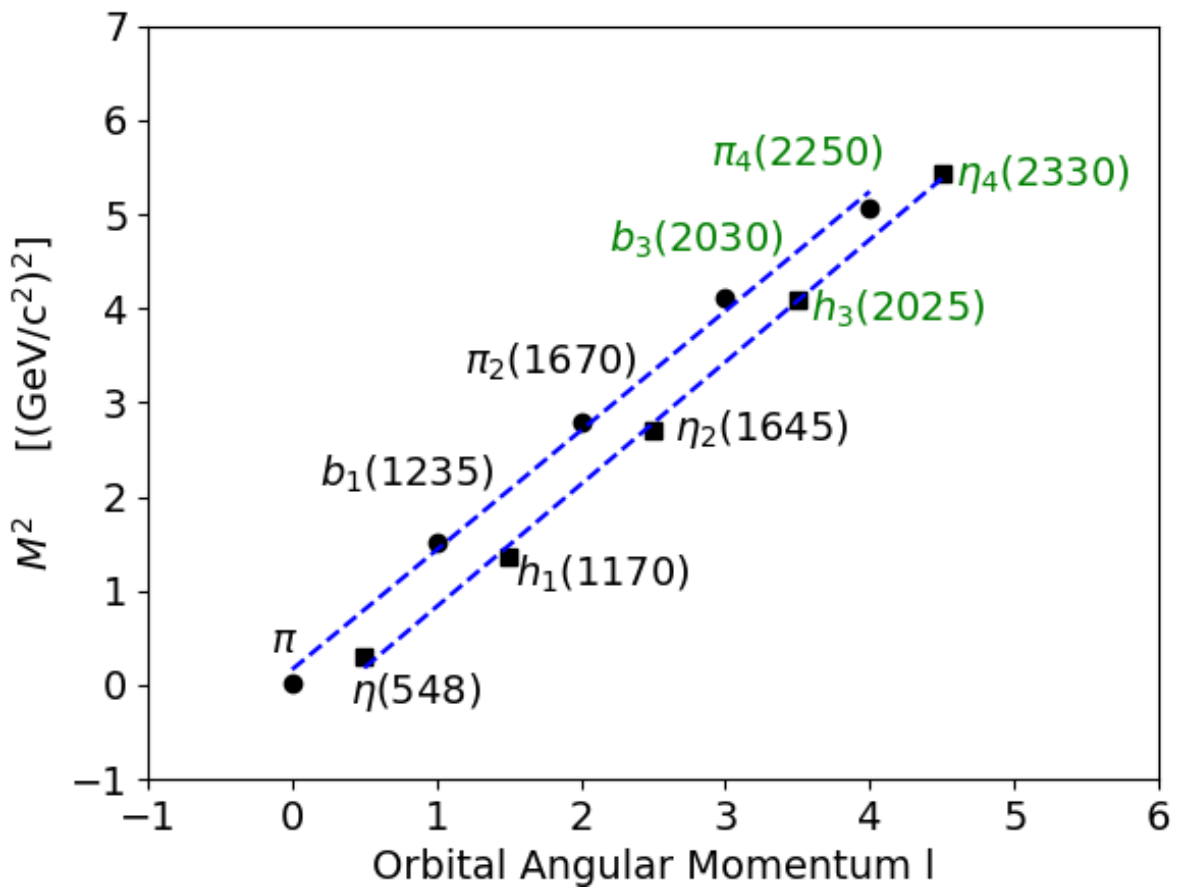

**Fig. 8.1.2** The pseudoscalar meson Regge trajectories as a function of the orbital angular momentum l. The isospin 0 trajectory has been shifted by $\frac{1}{2}$ unit in $l$ to the right. The established states, $\pi$, $b_1(1235)$, $\pi_2(1670)$, $\eta$, $h_1(1170)$ and $\eta_2(1645)$ are shown in blue. The states shown in green, $b_3(2030)$, $\pi_4(2250)$, $h_3(2025)$ and $\eta_4(2330)$, need confirmation. The fitted slopes are consistent with 1.20 GeV$^2$ for $I = 1$ and 1.37 GeV$^2$ for $I = 0$ as in reference [2289].

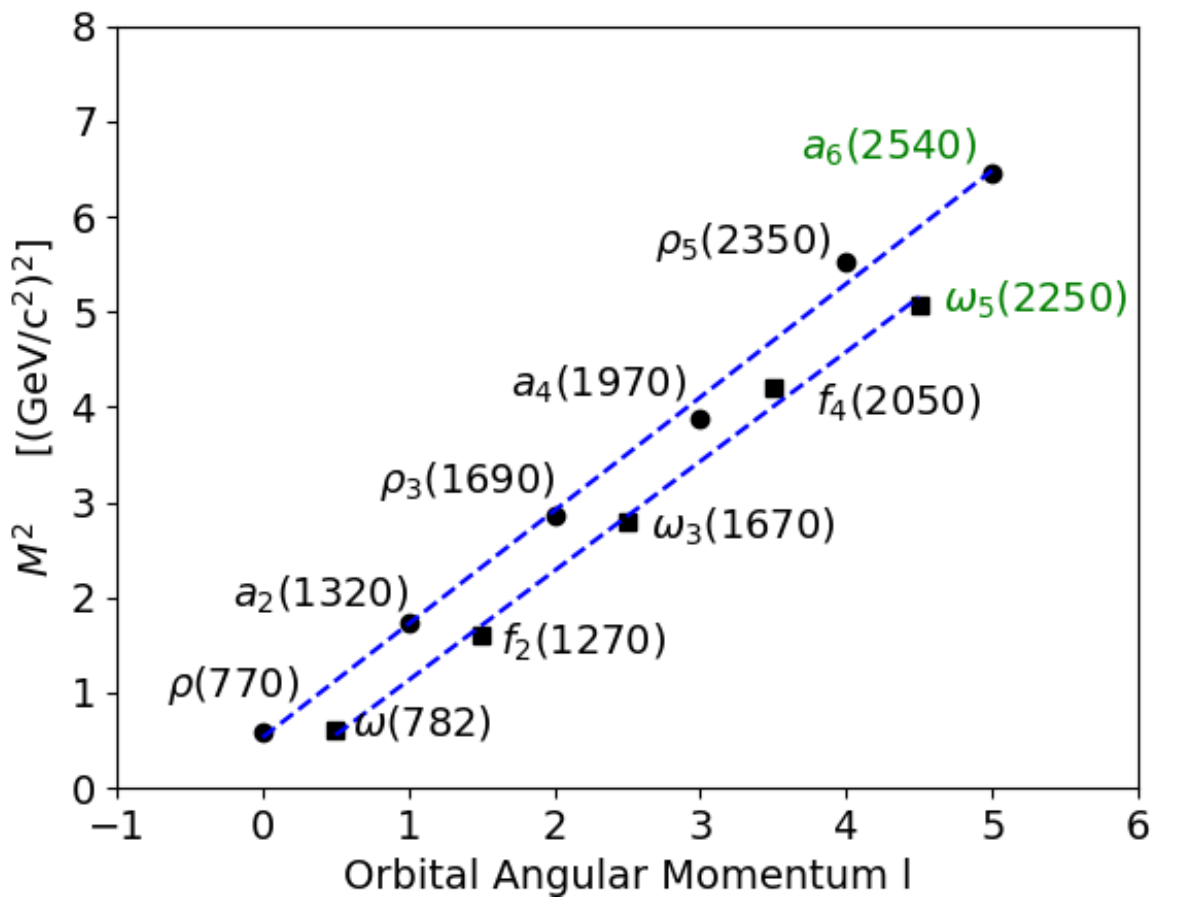

**Fig. 8.1.3** The vector meson Regge trajectories as a function of the orbital angular momentum l. The isospin 0 trajectory has been shifted by $\frac{1}{2}$ unit in $l$ to the right. The established states, $\rho(770)$, $a_2(1320)$, $\rho_3(1690)$, $a_4(1970)$, $\rho_5(2350)$, $\omega(781)$, $f_2(1270)$, $\omega_3(1670)$ and $f_4(2050)$ are shown in blue. The states shown in green, $a_6(2540)$ and $\omega_5(2250)$ need confirmation. The fitted slopes are consistent with 1.10 GeV$^2$ for $I = 1$ and 1.09 GeV$^2$ for $I = 0$ as in reference [2289].

### 8.1.6 The leading Regge trajectories

The original meson Regge trajectories[76] described a linear relation between the mass squared and the orbital angular momentum of mesons [1072, 2290], where the trajectories include $J^{PC} = 0^{-+}, 1^{+-}, 2^{-+}, 3^{+-}, \cdots$ and $1^{--}, 2^{++}, 3^{--}, 4^{++}, 5^{--}, \cdots$. In reality, the trajectories are often more complicated than the simple linear form. In a simplified picture when the quarks can be regarded as ultrarelativistic, a linear confining potential leads to linear Regge trajectories, while in the nonrelativistic regime, the trajectories would be nonlinear, and the intermediate regime would lead to a transition in the slope of the Regge trajectories. In the ultrarelativistic regime, the Regge slope depends on the string tension, while more generally it depends on both the quark masses and the tension. See reference [2289] for a more detailed discussion on this. In addition to the trajectories in orbital angular momentum $l$, there are also trajectories in the radial excitation quantum number, $n$. From a simple linear confining potential with string tension $\sigma$, the orbital trajectory is given in equation 8.1.14 and the radial as in equation 8.1.15, where one would expect universal slopes in both cases, with the slopes related by a factor of $\frac{\pi}{2}$.

$$M^2 = 8\sigma l + c_1 \,, \tag{8.1.14}$$

$$M^2 = 4\pi\sigma n + c_2 \,. \tag{8.1.15}$$

For light-quark mesons, the slopes are similar, but not universal. The orbital trajectories starting with the pseudoscalar mesons are shown in Fig. 8.1.2. The slopes for the two are found to be $1.20\,\mathrm{GeV}^2$ and $1.37\,\mathrm{GeV}^2$ respectively. The orbital trajectories starting with the vector mesons are shown in Fig. 8.1.3. In these cases, the slopes are found to be $1.10\,\mathrm{GeV}^2$ and $1.09\,\mathrm{GeV}^2$ respectively.

[76] See paragraph *The Regge approach and QCD* in Section 12.6 for an introduction to Regge phenomenolgy

## 8.2 The light scalars

**José R. Peláez**

### 8.2.1 Introduction

Light scalar mesons are treated in a separate subsection because, on the one hand, both their existence and nature have been the subject of a six-decade-long debate that predates QCD. On the other hand, they are particularly interesting because they play a very relevant role in several aspects, gathered below in seven items for concreteness, some of them already present before QCD, some others after.

First of all, in 1955, well before QCD was formulated, Johnson and Teller [2291] proposed the existence of a light scalar-isoscalar field to explain the attractive part of the nucleon-nucleon interaction. Two years later, Schwinger suggested that such a field, which he named $\sigma$, could be an isospin singlet, difficult to observe due to its huge width caused by its most likely very strong coupling to two pions.

Second, in the early sixties, Gell-Mann and Levy [2292] considered this field as the fourth member of a multiplet together with the three pions to build the popular "Linear sigma model" (L$\sigma$M). Such a state could also be generated dynamically in the Nambu and Jona-Lasinio (NJL) models [54, 2209]. These relatively simple models were able to explain the light masses of the pions, kaons and eta, and their mass gap with respect to the other hadrons, since they are the Nambu-Goldstone Bosons (NGB) of a spontaneous chiral symmetry breaking observed in the spectrum. Actually, they are pseudo-NGB, because they are not strictly massless. The masses of light-scalar mesons are closely related to the size of the non-zero vacuum expectation value, particularly those that share its same quantum numbers. Details of their interactions are also related to the specifics of the spontaneous breaking mechanism. The consequences of chiral symmetry were initially worked out with current-algebra methods as described in Section 1. Of particular interest for us will be the derivative interactions of NGB among themselves and the requirement of an Adler-zero below threshold in the NGB scattering amplitudes [20]. The leading order at low energies of those amplitudes was obtained by Weinberg in [22].

Third, since light scalars are the lightest states in the QCD spectrum that are not pseudo-NGB, we may expect them to fit as ordinary quark-antiquark mesons within the Quark Model that was proposed in the mid 60's [17, 18]. However, they do not, as we will see repeatedly below. Moreover, within the Quark Model, another scalar strange state, relatively similar to the $\sigma$ and called $\kappa$, was proposed by Dalitz in 1965 [2293], with a quark-antiquark assignment in a simple potential model, or more generally "simply on the basis of SU(3) symmetry". The existence of these two states, the $\sigma$ and $\kappa$, nowadays known as $f_0(500)$ and $K_0^*(700)$, has been very controversial until very recently, because they are extremely wide and difficult to observe. Actually, since they were first proposed, there were many experimental and phenomenological claims of such states, sometimes narrow, sometimes wide, sometimes lighter than 1 GeV, sometimes heavier and sometimes absent. The list of references is huge and we refer the reader to the Review of Particle Properties (RPP) [939] and the evolution of its "Note on Light Scalars" over the years, as well as the historical accounts in relatively recent reviews [2294, 2295]. An additional pair of scalar mesons, sitting very close to the $K\bar{K}$ threshold at 980 MeV, were soon identified, presently known as $f_0(980)$ and $a_0(980)$. These are narrower and their existence has not been controversial, although their mass and width values have changed slightly over the years. All in all they form the lightest scalar SU(3) nonet in Fig. 8.2.1. Note the largely broken flavor symmetry since the difference in the nominal masses is as large as 480 MeV. In addition, the mass hierarchy is inverted with respect to the naive expectations for an ordinary nonet of quark-antiquark states as in Fig. 8.1.1. For example, since in such a scheme the $a_0(980)$ would contain no strange valence quarks or antiquarks it should be about 200 MeV lighter than the $K_0^*(700)$, with one valence strange quark or antiquark. But this is precisely the opposite of what is found for the lightest scalars.

Fourth, light scalars, and particularly the $\sigma$ and $\kappa$, are difficult to include in the linear Regge trajectories that the other ordinary mesons follow [2296–2298] . These linear trajectories are related to the confinement mechanism. This difficulty became clear only around the time when the existence of the lightest scalars was being settled, and although QCD had already been formulated, it played no direct role in this discussion.

With the advent of QCD new interesting perspectives arose. In particular:

Fifth, one of the most attractive possibilities of a non-abelian gauge theory like QCD is the existence of glueballs, discussed in Section 8.4. The lightest one is expected to have scalar-isoscalar quantum numbers, and to appear as an "extra state" beyond the quark SU(3) multiplets. It is therefore important to identify all states within some light-scalar meson SU(3) nonets. For this, strange states are important, since they do not mix with glueballs and count how many quark nonets exist.

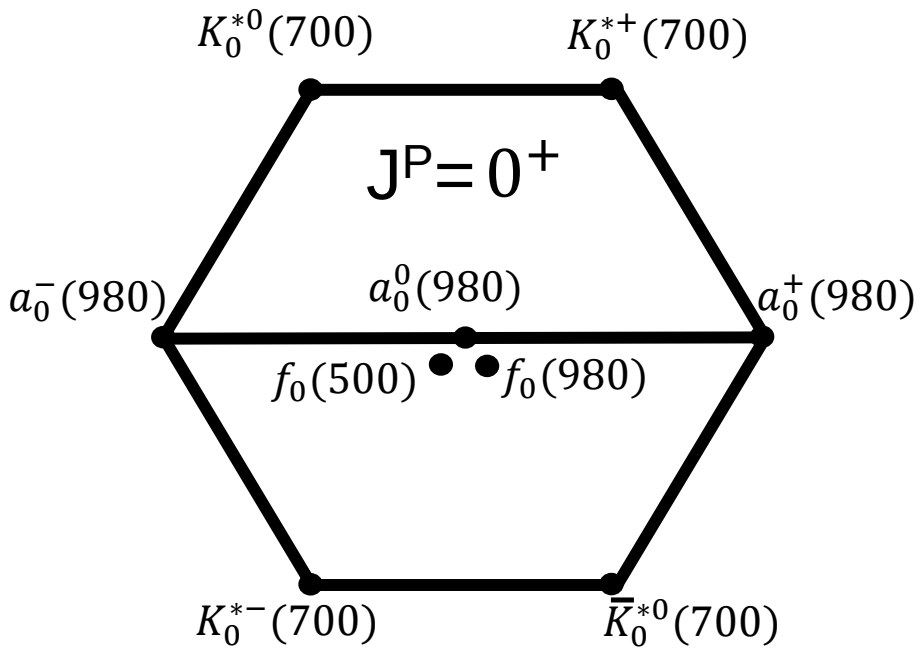

**Fig. 8.2.1** Light scalar nonet. Note the inverted hierarchy with respect to the naive $q\bar{q}$ assignment in Fig. 8.1.1, according to which the $a_0(980)$ should be $\sim$ 200 MeV lighter than the $K_0^*(700)$.

Sixth, given the quark constituent masses, tetraquarks would be naively expected to appear naturally around 1.4 GeV, if they appear at all. However, based on the dominance of the magnetic contribution of gluon interactions, Jaffe [2299] was able to build, within the "bag"model, tetraquarks well below 1 GeV. This suggests the existence of two $0^+$ nonets, one made of such tetraquarks, below 1 GeV, that on a first approximation could be identified with the nonet in Fig. 8.2.1 and another one made of ordinary $q\bar{q}$ above 1 GeV. This is how light scalars became the first non-ordinary-meson candidates, in the form of tetraquarks, or meson molecules. Still, they are not usually considered "exotics", but "crypto-exotics", since their quantum numbers can also be built with ordinary quark-antiquark configurations, with which they will necessarily mix, thus complicating this simple picture.

Seventh and final, despite QCD being non perturbative at low energies, its symmetries, and particularly the spontaneous symmetry breaking of chiral symmetry leading to a mass gap between NGB and other hadrons, allow for a systematic low-energy (and low-mass) expansion of amplitudes involving pions, kaons and the eta. The mathematical formulation in terms of an Effective Theory [1387] in the meson sector, has been presented in Section 6.2 and is called Chiral Perturbation Theory (ChPT) [64, 1570]. Being the next less massive states after the NGB, one would naively expect the lightest scalars to saturate the ChPT parameters at NLO discussed in Section 6.2. Once again, they do not, but the vector mesons do instead. This suggests once again that the dynamics that govern the formation of light scalars might be different from that of ordinary mesons like the vectors. With all those pieces of motivation in mind, the rest of the section is divided in two parts. First we will describe the light-scalars present status, paying attention to the dispersive and analytic methods used to settle the controversy about their existence, and other dispersive applications that are of relevance for the next part. Since the purpose of this work is to celebrate the 50th anniversary of QCD, we apologize for discussing in the second part only the most direct connections with it. Namely, their description in terms of (unitarized) ChPT and their dependence on the number of colors and quark masses. We will then discuss what can we conclude from these results.

| Isospin | State(s) | Mass (MeV) | Width (MeV) |
|---|---|---|---|
| 0 | $f_0(500)$ | 400-550 | 400-700 |
| 1/2 | $K_0^*(700)$ | 630-730 | 520-680 |
| 0 | $f_0(980)$ | 980-1010 | 40-70 |
| 1 | $a_0(980)$ | 960-1030 | 40-140 |
| 0 | $f_0(1370)$ | 1200-1500 | 200-500 |
| 1/2 | $K_0^*(1430)$ | $1425 \pm 50$ | $270 \pm 80$ |
| 1 | $a_0(1450)$ | $1474 \pm 19$ | $265 \pm 13$ |
| 0 | $f_0(1500)$ | $1506 \pm 6$ | $112 \pm 9$ |
| 0 | $f_0(1710)$ | $1704 \pm 12$ | $123 \pm 18$ |

**Table 8.2.1** Scalar light mesons below 1.9 GeV as listed in the RPP[278]. Note that for the first nonet we have taken the "$T$-matrix pole" parameters, not available for the rest. Also, there seems to be one $f_0$ state too many to form a second nonet.

*Present status*

As it is customary, and given the present precision, we consider the isospin limit. At present, 19 well-established scalar mesons are identified in the RPP [278] below 1.9 GeV, which we list in Table 8.2.1 with their present names. We have already classified 9 of them in the lightest scalar nonet in Fig. 8.2.1. The other 10 are 3 isoscalars, $f_0(1370)$, $f_0(1500)$, $f_0(1710)$, the $a_0(1450)$ isovectors with their 3 different charges, and the $K_0^*(1430)$ in 4 different combinations of strangeness amd charge. There are more scalar mesons, but they all lie nominally at or above 1.95 GeV. Hence, given the amount of strange scalars and isovectors, there must be two nonets below 1.75 GeV. Looking at their masses, one lies below 1 GeV and the other one around 1.4 GeV. Note, however, that above 1 GeV there seems to be one scalar state too many. This agrees with the expectation for the lightest QCD glueball, that will be discussed in Section 8.4.

Comparing with other mesons made of light quarks discussed in Section 8.1, we see that, for similar masses, they tend to have larger widths. The exception are the $f_0(980)$ and $a_0(980)$, which are narrower than the rest because their decay into $K\bar{K}$ is suppressed due to their proximity to the $K\bar{K}$ threshold. Given the $O(100)$ MeV width of most of these resonances, there must be some mixing between states with the same quantum numbers in different nonets. This mixing most likely distorts the mass hierarchies expected if they were narrow. Many mixing schemes have been proposed, but they only make sense for the flavor part of the wave function. We will see below one such treatment. Unfortunately, they are often used for the spatial or momentum part, which would only make sense in the narrow width approximation for almost stable mesons, which is not the case of any pair of light scalars with the same quantum numbers, and should be avoided (see Section 4.6.2 in [2294] and references therein).

In general, light-scalar-meson parameters have much bigger uncertainties than those of other mesons. This is because their large widths make them often overlap one another as well as with other analytic features like thresholds. As a consequence, in many analyses they do not show up as clean resonance peaks and their observed shapes can vary strongly, depending on specific features of their production, becoming dips or or being even completely masked. It is therefore essential to determine light-scalar-meson parameters from process-independent quantities. In particular, resonances are rigorously defined through their associated poles in the complex plane, that we briefly describe next.

### 8.2.2 Resonance poles and dispersive determinations

*Resonance poles*

These are poles appearing in the complex $s$-plane of any $T$-matrix element describing a process where a resonance $R$ is produced as an intermediate state. As a technical remark, these poles appear in conjugated pairs in the Riemann sheet that is reached when crossing continuously from above the square-root cuts associated to the center-of-mass (CM) momenta of the particles in the physically available intermediate states. This sheet is sometimes called "adjacent", "proximal" and in the elastic case "second" sheet. Out of the conjugate pair, it is the pole in the lower-half plane that most influences the behavior of the amplitude on the real axis. Then, its position $s_R$ is related to the resonance mass and width as $\sqrt{s_R} \equiv M - i\Gamma/2$. The familiar peak shape in the modulus squared of the amplitude is clearly observed for real-physical values of $s$ only when the resonance is narrow and well isolated from other singularities. Only in such cases the simple Breit-Wigner (BW) approximation, or models like K-matrices or isobar sums, etc... may be justified. However, this is not the case for most scalars and definitely not for the $f_0(500)$ and $K_0^*(700)$, which have been the most controversial and latest states to be accepted as well established in the RPP. This is the reason why in Table 8.2.1 we provide the "$T$-matrix pole" mass and widths, avoiding "Breit-Wigner" parameters.

Let us then briefly comment on how the poles of those states are determined by means of model independent dispersive and analytic techniques, although we first need to define partial waves.

*Partial waves*

Resonances and their quantum numbers are most easily identified using partial waves of definite isospin and angular momentum $\ell$. For rigorous determinations of the lightest scalar mesons, the most relevant process is meson-meson scattering, whose partial waves are defined as follows:

$$f_\ell^I(s) = \frac{1}{32\pi K}\int_{-1}^{1} dz_s P_\ell(z_s) F^I(s,t(z_s)), \qquad (8.2.1)$$

where $F^I(s,t)$ are the amplitudes, or elements of the $T$-matrix, of definite isospin $I$; $s,t$ are the usual Mandelstamm variables, $P_\ell$ the Legendre polynomials and $z_s$ the scattering angle in the $s$ channel. Note that $K=1,2$ for $K\pi$ and $\pi\pi$, respectively, because, for hadron interactions, pions are identical particles in the isospin limit that we will use.

It is convenient to recast partial waves in terms of the phase shift $\delta_\ell^I$ and elasticity $\eta_\ell^I$as follows:

$$f_\ell^I(s) = \frac{\eta_\ell^I(s)e^{i2\delta_\ell^I(s)}-1}{2i\sigma(s)}, \quad \sigma(s) = \frac{2q(s)}{\sqrt{s}}, \qquad (8.2.2)$$

where $q$ is the CM momentum of the scattering particles. In the elastic regime $\eta_\ell^I=1$ and we can write:

$$f_\ell^I(s) = \frac{e^{i\delta_\ell^I(s)}\sin\delta_\ell^I(s)}{\sigma(s)}. \qquad (8.2.3)$$

For later purposes it is important to recall that we are interested in poles in the second Riemann sheet. Let us illustrate the elastic case, where the analytic continuation to the second sheet through the physical cut is very simple. Moreover, it is the most relevant for the $\sigma/f_0(500)$ and $\kappa/K_0^*(700)$, since they appear in elastic $\pi\pi$ and $\pi K$ scattering, respectively, well below the next open threshold. For elastic partial waves, the following relation holds: $S_\ell = 1+2i\sigma(s)f_\ell$. Note that, in the partial-wave context, the $T$-matrix is actually called $f$. In addition, above threshold, the unitarity of the $S$-matrix implies

$$\mathrm{Im}\, f_\ell^I(s) = \sigma(s)|f_\ell^I(s)|^2, \quad \mathrm{Im}\, f_\ell^I(s)^{-1} = -\sigma(s), \qquad (8.2.4)$$

which in turn imposes the following unitarity bounds:

$$|f_\ell^I(s)| \leq 1/\sigma(s). \qquad (8.2.5)$$

Knowing the imaginary part of $f_\ell^I$ on the cut allows us to write a very simple relation between the $S$-matrix in the first (I) and second (II) sheet:

$$S_\ell^{(II)} = \frac{1}{S_\ell^{(I)}}, \quad f^{(II)}(s) = \frac{f^{(I)}}{1+2i\sigma(s)f^{(I)}}, \qquad (8.2.6)$$

where the isospin and angular-momentum indices have been momentarily suppressed for convenience. Note that in the second sheet $\sigma(s^*) = -\sigma(s)^*$.

Nevertheless, in the $\sigma$ and $\kappa$ case we still need to know the value of $f_\ell^I$ in the first Riemann sheet, but very deep in the complex plane. Unfortunately, the continuation to the complex plane is a hard and unstable mathematical problem. Different parameterizations or models, seemingly equivalent when describing data in a given region, may lead to different analytic continuations and different poles. The rigorous way of extending the amplitudes to the complex plane is through dispersion relations, if available, and analytic continuation techniques.

*Analyticity and dispersion relations*

Relativistic causality implies that the amplitude $F(s,t)$, for fixed $t$, must be analytic in the first Riemann sheet of the complex $s$-plane except for the real axis. In the absence of bound states in meson-meson scattering, only singularity cuts are present on the real axis. First of all, a right-hand-cut (RHC) appears from threshold to $+\infty$. Crossing this RHC continuously leads to the adjacent Riemann sheet, where resonance poles may exist. In turn, crossing symmetry implies that there is a left-hand-cut (LHC) from $-\infty$ to $s=-t$ due to cuts in the $u$ channels. In particular, the LHC extends up to $s=0$ for forward scattering ($t=0$) and for partial-wave amplitudes. Finally, for scattering of two particles with different masses, the $P_\ell(\cos\theta)$ integration in the partial wave definition yields a circular cut of radius $|m_1^2-m_2^2|$ centered at $s=0$. Then, Cauchy's integral formula relates the amplitude at any $s$ in the first Riemann sheet with integrals over the amplitude imaginary part along the cuts. These are called dispersion relations.

Since Cauchy's Integral formula applies to functions that depend on one variable, say $s$, the other variables have to be fixed or integrated over. Of particular interest are forward dispersion relations (FDRs), which correspond to the fixed-$t$ case with $t=0$. Also of interest for our discussion below are hyperbolic dispersion relations, obtained when $s,t,u$ are fixed to lie on an hyperbola $(s-a)(u-a)=b$. Any of these relations can also be integrated in $t$ as in Eq. (8.2.1) to obtain a partial-wave dispersion relation. In principle, forward dispersion relations are applicable at any $s$, but for different fixed-$t$ and hyperbolic cases the applicability is reduced. These applicability domains affect those of the partial waves, depending on how they have been obtained (see the appendix in [2295] for details).

Generically, the most complicated part of the calculation are the left and circular cuts. Within the context of light scalars, partial-wave dispersion relations are the most relevant and we can crudely group their most frequent uses into two categories precision dispersive approaches and unitarization techniques.

Before discussing these two uses in detail, let us just mention that dispersive approaches also constrain Regge trajectories and they hence can be used to calculate, not fit, the Regge parameters of resonances using their poles as input. While the resulting trajectories for ordinary mesons like the $\rho(770)$, $K^*(892)$, $f_2'(1525)$, $f_2'(1525)$ come out [2297, 2300] with a rather small imaginary part and a dominant real part, whose $s$ dependence is almost a straight line, as expected, those for the $f_0(500)$ and $\kappa$ come at odds with the ordinary behavior [2297, 2298]. This explains why the latter do not fit well in the usual phenomenological Regge plots.

*Precision dispersive approaches:*
We aim at mathematical rigor to establish the existence of the $\sigma$ and $\kappa$ poles and at precision to determine their parameters. Note that these are the poles closest to the left and circular cuts. Therefore, those cuts can be rewritten using the partial-wave expansion of the crossed channels. This complicates the integrands, and the new relations then couple different partial waves and channels. These relations are generically called Roy-like equations [2301]. There are variations like Roy-Steiner ([2302, 2303] for different masses and hyperbolic relations), GKPY ([2304] with minimal subtractions), etc. Their applicability is reduced in practice to energies around 1.1 GeV for $\pi\pi$ [2305, 2306] and $\pi K$ scattering [2307]. The inelastic, higher-energy, and higher-wave contributions are calculated from phenomenological fits. They have been used with two approaches:

– *Solving the equations* for the lowest partial waves $\ell = 0, 1$, in the region of interest, without using any data in that region. All other contributions come from phenomenological fits. Sometimes these are supplemented with ChPT constraints, which reduce considerably the uncertainties. Thus, poles and results in the resonance region could be considered as predictions from the equations and the other terms (and ChPT if used). The proof of the applicability of this approach to determine the existence of the $\sigma/f_0(500)$ and $\kappa/K_0^*$ and their resulting parameters were provided in [2308] and [2309], respectively.
– *Data driven approach.* Here Roy-like equations are used as constraints on fits to the $S$ and $P$ partial-wave data [2310]. Data sets that are largely inconsistent with these constraints are discarded. Additional contributions from higher energies and partial waves are constrained with forward dispersion relations and sum rules. Simple parameterizations are then fitted to the remaining data, but constrained to satisfy Roy-like equations in different versions and/or number of subtractions, as well as forward dispersion relations up to 1.42 GeV for $\pi\pi$ [2304] and up to 1.6 GeV for $\pi K$ [2311]. The latter was later coupled to $\pi\pi \to K\bar{K}$ and studied in [2295] with Roy-Steiner equations. With this approach the $\sigma/f_0(500)$ and $\kappa/K_0^*(700)$ poles were obtained in [2310] and [2295, 2312], respectively.

Recall that dispersion relations are written in the first Riemann sheet. However, in both approaches above, poles can be determined within a fully dispersive approach, because the second sheet can be easily obtained using Eq. (8.2.6). In contrast, accessing the “contiguous” sheet in the inelastic regime requires additional analytic continuation methods. Detailed reports on the dispersive determinations of the $\sigma/f_0(500)$ and $\kappa/K_0^*(700)$ poles can be found in [2294] and [2295], respectively. For convenience we have gathered their resulting poles in Tables 8.2.2 and 8.2.3. We also provide the modulus of the coupling to the dominant decay channel. Note that the uncertainty and spread of the dispersive results is much smaller than the RPP estimates in Table 8.2.1. This is because other non-dispersive and model-dependent determinations are included in the RPP estimate. However, the existence of two independent dispersive approaches was decisive to consider both resonances as well established in the RPP 2012 and 2020 editions, respectively, changing their nominal masses in their names to be closer to their pole values.

Note that the $f_0(980)$ pole was obtained simultaneously within the same framework [2304, 2308]. However, being a narrow resonance and further away from left cuts, its pole is more similar to those obtained with other methods. Finally, some of these analytic continuation methods — using dispersively constrained input — have been applied to determine the poles of further mesons in the inelastic regime, including the scalars $K_0(1430)$ [2315], $f_0(1370)$ and $f_0(1500)$ [2316]. In such case Eqs. (8.2.3), (8.2.4) and (8.2.6) do not hold and the use of analytic continuation methods is unavoidable to suppress any model dependence.

### 8.2.3 Light scalars and QCD

In the previous section, we have discussed how the rigorous dispersive approach was instrumental in settling the controversy about the existence and parameters of the $\sigma$ and $\kappa$. Once this is settled, we now concentrate on light scalars within the context of QCD, which is the subject of this volume.

*Unitarized Chiral Perturbation Theory (UChPT)*
Being so light, these resonances lie in the non perturbative region of QCD, and thus an effective treatment with ChPT seems appropriate. However, the ChPT series by itself cannot generate poles and also violates unitarity as the energy reaches the resonance region. The most successful approach is thus a combination of Chiral Perturbation Theory (ChPT) with dispersion relation. This is generically known as unitarized ChPT.

ChPT, which is the low-energy theory of QCD, and is formulated as an expansion in momenta or masses of the NGB, has been introduced for the meson sector in section 6.2.2. Meson-meson scattering partial waves are then expanded as $f(s) = f_2(s) + f_4(s) + ...$, where $f_{2n}(s) = O(p^2/F_\pi^2) \times O(p/\Lambda_\chi)^{2n-2}$, where $p$ are the meson CM momenta or masses, $\Lambda_\chi = 4\pi F_0$ and $F_0$ is the NGB decay constant at LO, common to all mesons at that order. Up to higher orders $F_0$ can be approximated by $F_\pi$, $F_K$.... Note that we have suppressed momentarily the isospin and angular momentum indices $I, \ell$. As an example, the $O(p^2)$ or LO $\pi\pi$ and $\pi K$ elastic partial waves in the scalar channel with lowest isospin are:

$$f_0^0(s) = \frac{2s - M_\pi^2}{32\pi F_\pi^2}, \quad (8.2.7)$$

$$f_0^{1/2}(s) = \frac{5s^2 - 2(M_K^2 + M_\pi^2)s - 3(M_K^2 - M_\pi^2)^2}{128\pi F_\pi^2 s}.$$

ChPT amplitudes are an expansion in powers of $p$ and cannot satisfy the unitarity condition in Eq. (8.2.4) exactly, but just perturbatively:

$$\mathrm{Im} f_2(s) = 0, \ \mathrm{Im} f_4(s) = \sigma(s) f_2(s)^2, ... \quad (8.2.8)$$

When $p/\Lambda_\chi$ is very small, this is not a problem, but the violation of unitarity grows with momenta or energy. This violation then becomes a severe caveat to describe resonances, since in the typical cases, resonant effects saturate the unitarity bound in Eq. 8.2.5. Even worse, the ChPT series cannot generate poles in $s$ and thus, in principle cannot generate resonances.

Therefore, if we want to describe resonances, we need to implement unitarity, but also analyticity if we want to study their associated poles. Let us now provide a simple, but formal, derivation of ChPT unitarization methods. The elastic unitarity condition in Eq. 8.2.4 fixes the imaginary part of the inverse partial wave. Hence, naively, we just have to use ChPT to calculate the real part of the inverse amplitude, and write: $\mathrm{Re}(1/f) = \mathrm{Re}\, 1/(f_2 + f_4 + ...) \simeq (1/f_2)(1 - \mathrm{Re} f_4/f_2 + ...)$, since $f_2$ is real from Eq. (8.2.7). Then we write a unitarized elastic partial wave at different orders as:

$$f^U_{LO}(s) = \frac{1}{1/f_2(s) - i\sigma(s)}, \quad (8.2.9)$$

$$f^U_{NLO}(s) = \frac{1}{1/f_2(s) - \mathrm{Re} f_4(s)/f_2(s)^2 - i\sigma(s)}, .. (8.2.10)$$

and similar expressions for NNLO, etc... Note that the ChPT series is recovered if re-expanding again. These expressions are unitary and can be recast in explicitly analytic forms. For instance, using Eqs. 8.2.8, the second one is $f^U_{NLO} = f_2^2/(f_2 - f_4)$, which is known as the NLO Inverse Amplitude Method (IAM). Similar analytic formulas for higher orders exist [2317–2319]. Thus these methods can be analytically continued to the complex plane and the second sheet using Eq. (8.2.6). This derivation is formal, because strictly speaking we could still not use the expansion of the real part beyond the applicability realm of ChPT into the resonance region. However, there are derivations [2320–2322] from partial-wave dispersion relations for the inverse partial wave and ChPT is only used in the subtraction constants at $s = 0$ or in the left and circular cuts. The use of several subtractions makes those cuts to be dominated by the low energies, where ChPT is applicable, thus justifying the use of the Inverse Amplitude Method.

**Table 8.2.2** $\sigma/f_0(500)$ pole determinations using Roy-Steiner equations and the conservative dispersive estimate [2294] which covers them. For the latter we have corrected a typo in the error of $\mathrm{Im}\sqrt{s_{pole}}$ which read $\pm 12$ MeV instead of $\pm 15$ MeV.

| $\sigma/f_0(500)$ | $\sqrt{s_{pole}}$ (MeV) | $\lvert g\rvert$ (GeV) |
|---|---|---|
| Refs. [2308, 2313] | $(441^{+16}_{-8}) - i(272^{+9}_{-12.5})$ | $3.31^{+0.35}_{-0.15}$ |
| Ref. [2314] | $(442^{+5}_{-8}) - i(274^{+6}_{-5})$ | - |
| Ref. [2310] | $(457^{+14}_{-13}) - i(279^{+11}_{-7})$ | $3.59^{+0.11}_{-0.13}$ |
| Conservative Dispersive Estimate | | |
| Ref. [2294] | $(449^{+22}_{-16}) - i(275 \pm 15)$ | $3.45^{+0.25}_{-0.29}$ |

**Table 8.2.3** $\kappa/K_0^*(700)$ dispersive pole determinations using Roy-Steiner equations.

| $\kappa/K_0^*(700)$ | $\sqrt{s_{pole}}$ (MeV) | $\lvert g\rvert$ (GeV) |
|---|---|---|
| Ref. [2309] | $(658 \pm 13) - i(279 \pm 12)$ | |
| Ref. [2295] | $(648 \pm 7) - i(280 \pm 16)$ | $3.81 \pm 0.09$ |

Interestingly, with the simplest possible calculation, i.e. using just the LO in Eqs. (8.2.9) and (8.2.7) in the chiral limit $M_\pi, M_K \to 0$, we find the following poles in the second Riemann sheets of the partial waves where the $\sigma/f_0(500)$ and $\kappa/K_0^*(700)$ are seen:

$$f_0^0 : \sqrt{s_\sigma} = (1-i)\sqrt{8\pi}F_0 \simeq (463 - i463)\,\mathrm{MeV}, \tag{8.2.11}$$
$$f_0^{1/2} : \sqrt{s_\kappa} = (1-i)8\sqrt{\pi/5}F_0 \simeq (638 - i638)\,\mathrm{MeV},$$

where for the numerical values of $F_0^2$ we have taken $F_\pi^2 \simeq 92.3\,\mathrm{MeV}$ for $\pi\pi$ and $F_\pi F_K$ for $\pi K$ scattering, with $F_K = 1.19 F_\pi$ [278]. Taking into account that this is the most naive LO unitarized calculation, with no free parameters, the lightest scalar masses come remarkably close to their actual values and their widths about a factor of 2 too wide. Note that the only dynamical information is the scale of the spontaneous chiral symmetry breaking, given by $F_0$. In contrast, if the same procedure is followed with the vector $\ell = 1$ channels, the resulting poles for the $\rho(770)$ and $K^*(892)$ come almost twice too heavy and their widths more than 16 times too wide. This is already an indication that the LO low-energy chiral dynamics plays a predominant role in the formation of light scalar resonances, and very little for other ordinary mesons.

The description of meson-meson scattering at NLO in UChPT is very succesful for both scalar and vector partial waves in all isospin combinations (tensor waves start at NNLO). In particular, now not only the pole-width of the scalars comes right, but also the vector meson poles and their parameters. Recall that, as explained in Section 6.2, the NLO ChPT calculations contain several Low Energy Constants $L_i(\mu)$, which multiply the terms in the NLO Lagrangian allowed by symmetry. They are scale dependent because they absorb, through renormalization, the loop divergences at previous orders. In addition, they contain the information about the underlying quark and gluon dynamics, namely, QCD. Only when these $L_i$ are taken into account it is possible to describe the "ordinary" quark-antiquark vector mesons with UChPT. However, the $L_i$ combinations that appear in the scalar channels are much less relevant numerically and that is why scalar poles come out fairly decent with just the LO UChPT and just information on the chiral breaking scale.

So far we have only discussed elastic unitarization. But exactly the same naive derivation can be followed in matrix form to obtain a coupled channel $T$-matrix formalism [2323–2325], only slightly more complicated. When this is done, besides the $f_0(500)$ and $K_0^*(700)$ poles, those associated to the $a_0(980)$ and $f_0(980)$ resonances also appear in the Inverse Amplitude Method [2325], completing the lightest scalar nonet, as well as those of the $\rho(770)$ and $K^*(892)$ vectors.

Many variations of ChPT unitarization techniques exist in the literature of which, together with the IAM, the simplest and most popular is the Chiral Unitary Approach [2326, 2327] (for other variations, see the reports [2294, 2328–2330]), which usually raises the caveat about some arbitrariness. However, all unitarization methods just correspond to finer or more crude approximations to $\mathrm{Re}(1/f)$ and its ChPT series or to different treatments of the left cut, or even including some additional heavier states. But as long as they contain the basic information about the chiral scale, or are equivalent to the ChPT LO, they all obtain a similar description of light scalars, whereas vectors or other resonances can be accommodated only when including enough NLO information.

Of course, since unitarization methods involve some truncation of ChPT and approximations, they are not competitive in precision and rigor with the precise dispersive approaches discussed before. They have, however, another advantage, which is that we can study the dependence of the resonances on QCD parameters, which we will describe next.

*Leading QCD $1/N_c$ behavior*

At leading order in the $1/N_c$ expansion [1163, 2331], ordinary $q\bar{q}$ mesons behave as $M \sim O(1)$ and $\Gamma \sim O(1/N_c)$. Genuine tetraquark states [2332, 2333] have at least that same $N_c$ behavior, which is even more suppressed for glueballs.

First of all, using meson-meson scattering and the light-resonance pole parameters it is possible to build observables whose sub-dominant $N_c$ corrections are highly suppressed [2334]. When evaluated for the $f_0(500)$ and $K_0^*(700)$ the resulting values are at odds with the ordinary meson or glueball behavior by several orders of magnitude.

Next, using the effective theory approach, the $1/N_c$ leading order of the NLO ChPT parameters is known from a model-independent analysis: $M_\pi, M_K \sim O(1)$, $F_0 \sim \sqrt{N_c}$ and the $L_i$ behavior is either $O(1)$ or $O(N_c)$ [64, 2335]. If in the UChPT amplitudes we then call $p$ a parameter whose behavior is $O(N_c^k)$ and change its value to $p \to p(N_c/3)^k$, we will obtain the leading $1/N_c$ behavior of resonances following their associated poles as $N_c$ is increased. Thus, already with the substitution $F_0 \to F_0\sqrt{N_c/3}$ in the LO UChPT results in Eqs. (8.2.11), we obtain a non-ordinary behavior for both the $\sigma$ and $\kappa$. Namely, their $M, \Gamma \sim O(\sqrt{N_c})$.

That was just the naive LO estimate, but the leading $1/N_c$ dependence within UChPT has been studied to NLO in [2325, 2336] and NNLO in [2337]. It is then possible to study the light vectors as well, and they come remarkably compatible with the expected ordinary behavior. This is shown for the $\rho(770)$ in the top panel of Fig. 8.2.2. However, the $\sigma$ and $\kappa$ poles, shown in the center and bottom panels of Fig. 8.2.2, respectively, display again a non-ordinary behavior, at least near the physical value of $N_c = 3$. This is a robust result also found in other approaches. Of course if $N_c$ is made very large, the dominance of meson loops governed just by $F_0$, which are suppressed by $1/N_c$, fades away. Then, even the tiniest mixture with an ordinary meson could dominate at sufficiently large $N_c$. We should remark that there is some uncertainty, that grows with $N_c$ due to the scale dependence of the $L_i$, illustrated for the $\sigma$ in Fig. 8.2.2. Indeed, Fig. 8.2.2 shows that the sigma pole could turn back [2338] to the real axis, well above 1 GeV. This could be a small mixture with an "ordinary" state around or above 1 GeV. This is also found in NNLO UChPT [2337]. Similarly, in other phenomenological approaches the $\sigma$ and $\kappa$ only appear when the unitarized meson-meson interaction is included, showing up as an additional pole due to unitarization, in addition to ordinary states above 1 GeV that are present even if meson-meson interactions are turned off (this was first proposed in [2339], for additional references see[2294]). Back to UChPT, the ordinary subdominant component restores the semi-local duality sum-rules [2338] that would be violated if the light scalars just disappeared from the spectrum by becoming too massive and wide. However, other analyses [2340, 2341], challenged in [2342], yield a $\sigma$ behavior closer to the one of the opposite side of the scale uncertainty in Fig. 8.2.2, reaching the third quadrant at very large $N_c$, which lacks a clear interpretation. One should nevertheless recall that the large-$N_c$ regime, although of mathematical interest, is not the one of relevance for the observed meson, but the leading $1/N_c$ behavior near $N_c = 3$.

*Quark-mass dependence and light-scalar multiplets*
The study of quark-mass dependence is of interest to understand the dynamics of their formation, to provide a guideline for lattice studies, and to check that the light scalar states that we have grouped in an octet are degenerate when the strange and non-strange quark masses are equal.

We have seen in Section 6.2 the relation between quark and meson masses. This allows us to study the quark mass dependence of the $\sigma$ at NLO [2343] and NNLO [2344] and $\kappa$ at NLO [2345]. A slight IAM modification is used to deal with subthreshold Adler zeros [2346]. Figure 8.2.3 thus shows the resulting $\sigma$ and $\kappa$ pion mass dependence. Note that beyond 300-350 MeV the results are at most qualitative. With increasing pion mass, the meson masses grow, although slower than the two-pion threshold, and their pole-widths decreases. When the pion mass is 2-3 times its physical value, the $2\pi$ threshold is above the pole-mass of these resonances. Then, their behavior differs dramatically from that of the $\rho(770)$ and $K^*(892)$ (the latter shown in Fig. 8.2.3). The width of these non-scalar mesons would tend to zero, and their conjugated pair of poles would meet at threshold [2343, 2349]. Right after that, one of their poles would jump to the first sheet, whereas the other would remain at a symmetric position in the second sheet, both below threshold. This is a bound state. In contrast, the $\sigma$ and $\kappa$ conjugated poles meet in the second sheet below threshold. The two branches observed in Fig. 8.2.3 correspond to these two poles in the second sheet, where at first one moves towards threshold and the other away form it. The closest one to threshold, influencing the most the physical region, is known as a "virtual" or quasi-bound state. Eventually, it reaches threshold and jumps to the first sheet, becoming a bound state. However, its second-sheet counterpart lies in a rather different position. The more asymmetric their positions, the more predominant is their "molecular" or "meson cloud" nature. Hence, UChPT suggests that, at high pion masses, both the $f_0(500)$ and $K_0^*(700)$ are closer to two-meson states than to ordinary mesons.

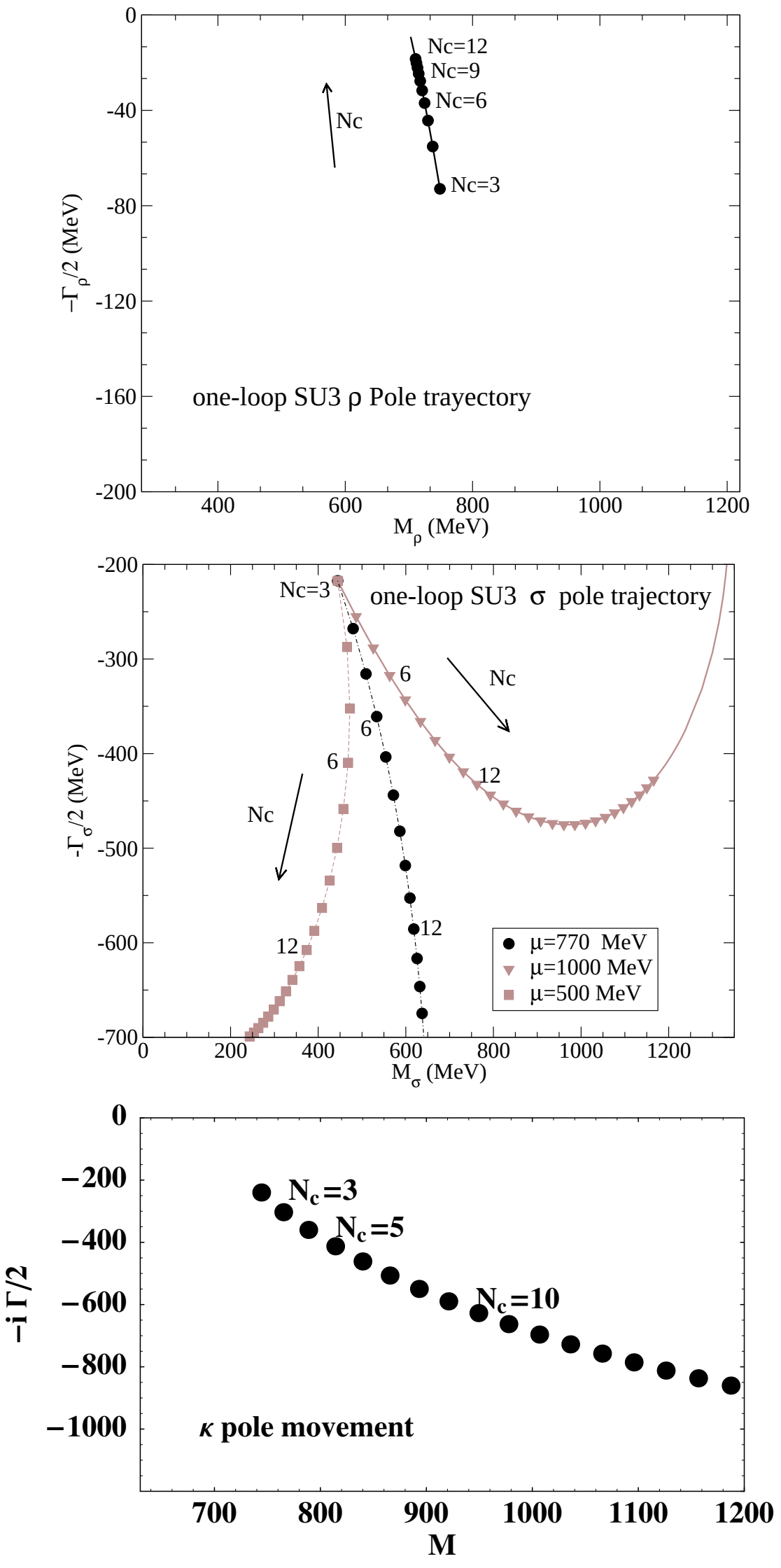

**Fig. 8.2.2** Trajectories of the $\rho(770)$ (top), $\sigma/f_0(500)$ (center) and $\kappa/K_0^*(700)$ (bottom) poles in the complex plane as $N_c$ is varied away from 3 within NLO ChPT unitarized with the IAM. The lighter curves in the center plot indicate the uncertainties when varying the regularization scale $\mu$ in the usual range, as recalculated in [2338]. In the case of the $\rho(770)$ the three lines almost overlap and are not plotted. Top and center figures taken from [2338] and bottom figure from [2325].

Quark masses can be changed on the lattice. Actually, calculations are not often done at physical masses, which are expensive numerically. Note also that analytic continuation to reach poles would be required, although models are often used to reach poles. There are lattice calculations for the $\sigma$ [2350, 2351], supporting its molecular picture at very large pion masses, where it is a bound state. The $\sigma$ is also found at moderately large pion masses [523, 2352] qualitatively consistent with UChPT. For $m_\pi = 236\,\text{MeV}$ [523, 524] lattice results are consistent with a pole now in the second Riemann sheet, also consistent with UChPT. A virtual state was found for $\kappa$ in $\pi K$ scattering on the lattice [516, 542], again in qualitative agreement with UChPT. However, as the pion mass becomes lighter, the $\sigma$ and $\kappa$ poles are plagued again with instabilities [517, 524, 525]. In [2353], Adler zeros, i.e. chiral symmetry, were found to be very relevant in the $\kappa$ determination. A dispersive "data driven" approach of the kind explained above may be relevant for a robust extraction of light scalar poles from lattice-QCD. We refer to Section 4 for further details.

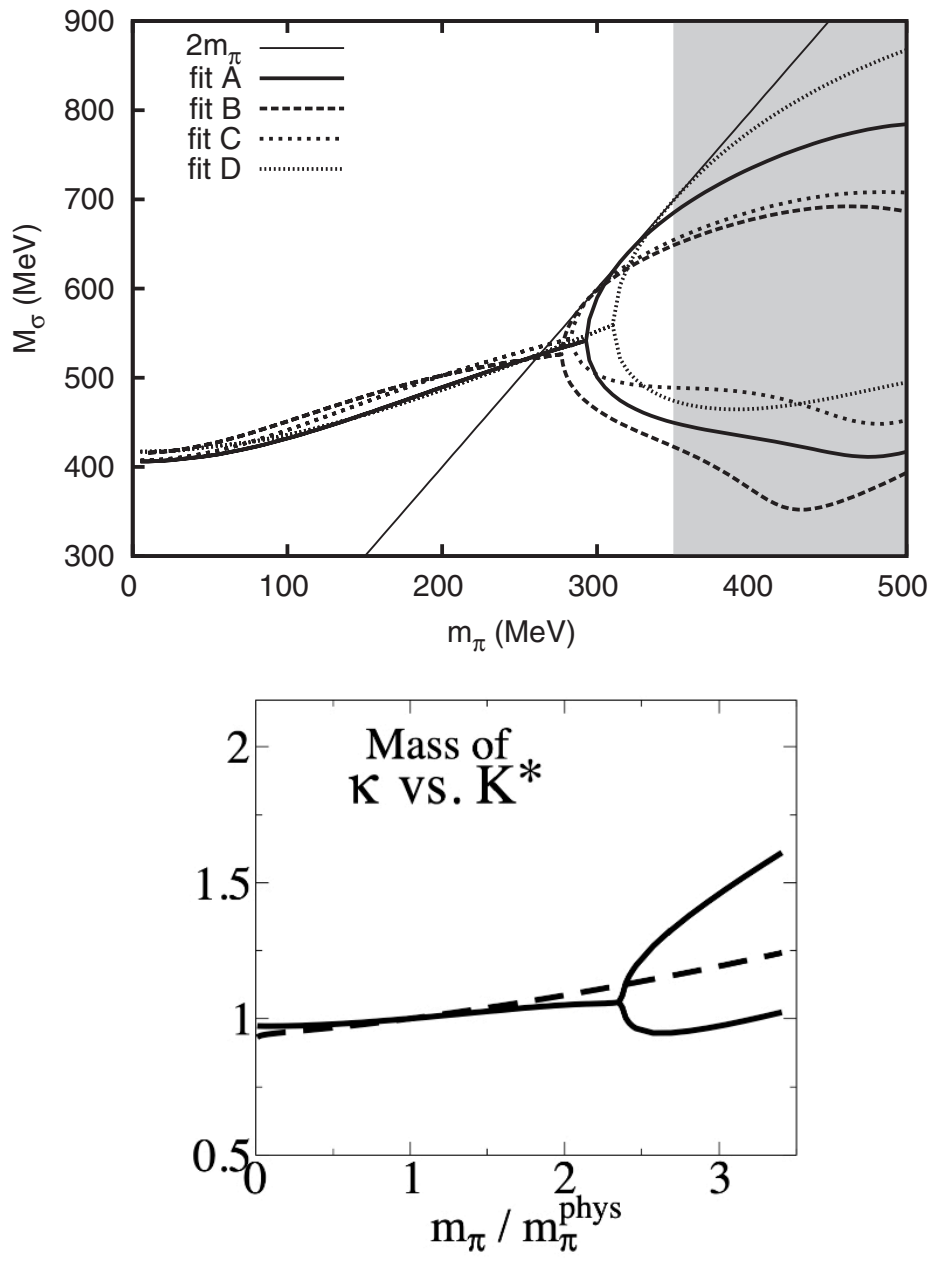

**Fig. 8.2.3** Top: Dependence of the sigma mass $M_\sigma$ on the pion mass, from the NNLO (two-loops) IAM [2337]. Different curves represent different fits on [2337]. The thin continuous line shows the $2m_\pi$ threshold. Bottom: $m_\pi$ dependence of the $\kappa$ (solid line) and $K^*(892)$ (dashed line) masses [2345]. All masses and widths are defined from the pole positions as obtained from NLO IAM fits. Figures taken from [2347] (top) and [2348] (bottom).

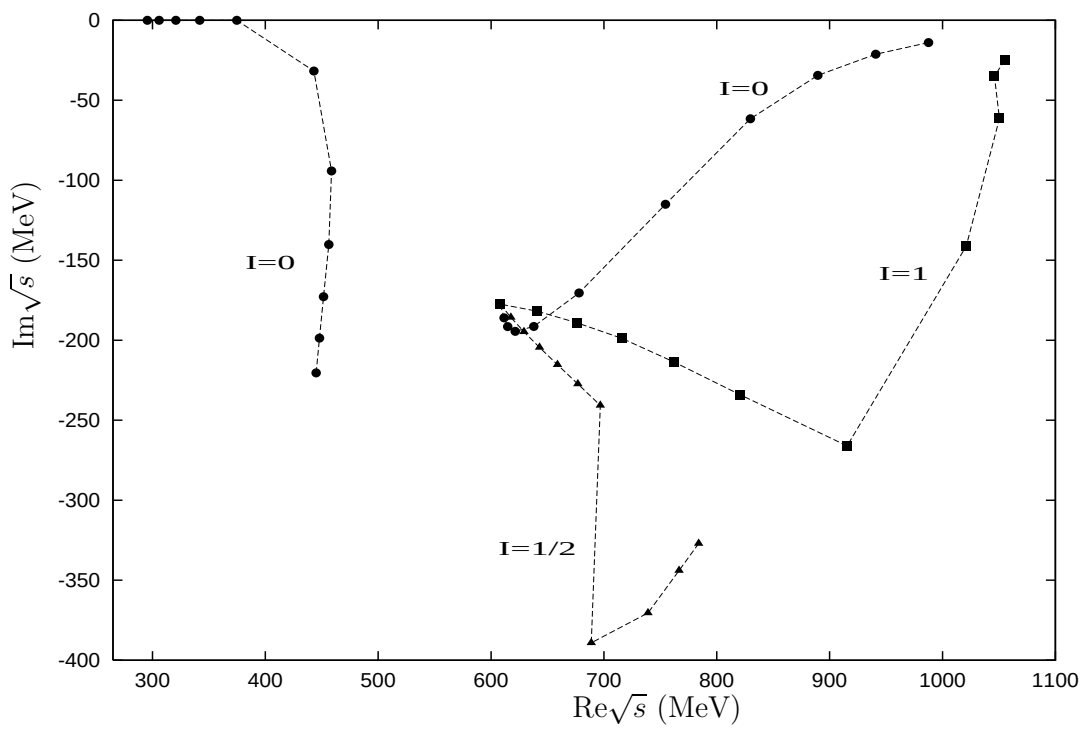

**Fig. 8.2.4** Trajectories of the poles that appear in coupled-channel unitarized amplitudes of different isospin as the pion, kaon and eta masses are varied from their physical values to a common value of 350 MeV [2354]. This shows that the lightest scalars actually belong to a nonet in the SU(3) limit. The two trajectories with $I = 0$ correspond to the singlet and octet states, not directly to the poles of the $\sigma$ or $f_0(980)$ resonances, which are a mixture of these two. Figure taken from [2354].

The strange-quark mass can also be varied [2345], but not much since it is already quite high, and thus the observed changes on scalars are very smooth. However, when changing both quark masses one can reach the degenerate pion-kaon mass limit. Fig. 8.2.4 shows that the trajectories of the $\kappa$ pole, and a combination of the $\sigma$ and $f_0(980)$ become degenerate with the $a_0(980)$ pole in that limit. This result has been obtained [2354] within the Unitary Chiral Approach, where the left cut is neglected and the effect of the $L_i$ are mimicked by a mass-independent cutoff. Still, this provides a strong support for the assignment of these states to the same lightest scalar octet.

### 8.2.4 Summary

Despite their relevant role in numerous aspects of hadron physics and QCD, the controversy about the existence and the parameters of the lightest scalar nonet, particularly for the $\sigma/f_0(500)$ and $\kappa/K_0^*(700)$, predated the establishment of QCD. The settling of this controversy was hindered by the conflicting available data sets and the by the use of models. We have provided here a brief account on how it has been settled recently by using rigorous dispersive techniques to constrain data analyses and to determine the poles associated to the light-scalar resonances. Many phenomenological approaches were able to describe to different degrees of accuracy these states. Here, we have focused on those most directly linked to QCD through the unitarization of Chiral Perturbation Theory, the $1/N_c$ behavior and the dependence on the quark masses. The general picture that arises is that there is one light scalar nonet below 1 GeV. Their non-ordinary $N_c$ behavior, quark mass dependence, Regge trajectories, and the fact that they do not saturate the ChPT constants strongly support that these mesons are not of the ordinary quark-antiquark type. Rather their predominant component would be of the meson-meson type (molecule, meson cloud, etc). Still they are most likely mixed with some companion bare or preexistent quark-antiquark state above 1 GeV. Indeed, a second scalar multiplet can be identified between 1.2 and 1.8 GeV. There is still ample room for refining this picture and a high expectation on further experiments and developments from lattice-QCD.

## 8.3 Exotic mesons

**Boris Grube**

### 8.3.1 Introduction

Already when Gell-Mann [17] and Zweig [18] formulated the constituent quark model they presumed that additional states beyond the baryonic $qqq$ and the mesonic $q\bar{q}$ combinations exist.[77] For a long time, the search for such states was unsuccessful and hence all hadronic states going beyond the constituent quark model were labelled *exotic*. However, rather recently experiments have found compelling evidence that exotic states indeed exist. Here, we will focus on exotic mesons, which can be divided into three categories: (i) *spin-exotic states*, which have $J^{PC}$ quantum number combinations that are not possible for ordinary $q\bar{q}$ states (cf. Tab. 8.1.1),[78] (ii) *flavor-exotic states*, which have flavor quantum numbers, such as isospin and/or strangeness, that are not possible for $q\bar{q}$ states, and (iii) *crypto-exotic states*, which have quantum numbers of ordinary $q\bar{q}$ states and are therefore able to mix with them.

Possible exotic mesonic configurations beyond $q\bar{q}$ are four-quark combinations such as tightly bound $qq\bar{q}\bar{q}$ *tetraquark states*, where the constituents are bound directly by the strong force, or more loosely bound $(q\bar{q})(q\bar{q})$ *molecular states*, which consist of a pair of mesons bound by nuclear forces. Also, the gluon fields are expected to manifest themselves in the meson spectrum either in the form of *hybrid states*, where, in addition to a $q\bar{q}$ pair, excited gluonic field configurations contribute to the quantum numbers of the meson, or in the form of *glueballs*, which are color-singlet bound states of gluons (see Sec. 8.4). However, in general, physical mesons are not pure realizations of single configurations but are instead mixtures of all possible configurations that are allowed for the given quantum numbers. Disentangling these different contributions is a highly difficult experimental and theoretical problem.

Crypto-exotic states will manifest themselves as supernumerary states compared to the spectrum expected from the quark model. This makes them rather difficult to establish. And even if experimental data unambiguously show an overpopulation of states in a certain mass range, the determination of the internal configuration of these states is an even harder problem. The prime example for such a situation is the sector of isoscalar scalar mesons discussed in Secs. 8.2 and 8.4. Therefore, the cleanest way to unambiguously establish the existence of exotic mesons is to search for spin-and/or flavor-exotic states. Presently, the clearest evidence for the existence of such states comes from the heavy-quark sector (see Secs. 8.5 and 8.6), where experiments have found several flavor-exotic states with a minimum quark content of four, for example, the charged charmonium and bottomonium states, $Z_c^{\pm}$ and $Z_b^{\pm}$ [1388, 2355], or the doubly-charmed state, $T_{cc}^{+}$ [1071].

Although mesons from the light-quark sector, i.e. mesons composed of up, down, or strange quarks, are usually easier to produce in experiments, the picture is less clear in this sector. This is mainly because light mesons have relatively large decay widths compared to their masses. As a consequence, these mesons usually do not appear as isolated and narrow peaks in the invariant mass spectra of their decay products. Instead, they often overlap and interfere with neighboring states, which makes their extraction from experimental data challenging. In addition, in most analyses models are required in order to extract resonances from the data and the results therefore depend on the employed model assumptions and approximations. In the following, we will confine the discussion to spin-exotic light mesons. More details on exotic light mesons can be found in the reviews in Refs. [385, 2356–2362].

[77] In Ref. [17], Gell-Mann writes: "Baryons can now be constructed from quarks by using the combinations $(qqq)$, $(qqqq\bar{q})$, etc., while mesons are made out of $(q\bar{q})$, $(qq\bar{q}\bar{q})$, etc." Similarly, Zweig writes in a footnote in Ref. [18]: "In general, we would expect that baryons are built not only from the product of three aces, $AAA$, but also from $\bar{A}AAAA$, $\bar{A}\bar{A}AAAAA$, etc., where $\bar{A}$ denotes an anti-ace. Similarly, mesons could be formed from $\bar{A}A$, $\bar{A}\bar{A}AA$ etc."

[78] More correctly, these states have forbidden $J^{PG}$ quantum numbers. However, here we use the common convention that the $C$-parity of a charged meson in an isospin triplet is given by the $C$-parity of its neutral partner state.

### 8.3.2 Predictions

*Model predictions*
Various models have been employed to study the light-meson spectrum. Some of these model approaches are discussed in more detail in Sec. 5. Further discussions can be found, e.g., in Refs. [385, 2362]. Most of the models that include exotic mesons predict the lightest spin-exotic state to be a hybrid meson with $J^{PC} = 1^{-+}$ quantum numbers.

The first detailed studies of hybrid light mesons were based on the bag model [750–752, 2363, 2364]. In this model, quarks and gluons are described by cavity modes in a confining vacuum bubble (see Sec. 5.1.3). Detailed predictions for the decays of hybrid light mesons were obtained using, for example, the fluxtube model [2365–2371]. This model extends the conventional quark model by explicitly modeling the gluonic fields in form of an oscillating flux tube described by single-phonon excitations. Decays of hybrid mesons were also studied in constituent-glue models [2372–2375], where one assumes that a massless gluon with $J^P = 1^-$ interacts with quarks via potentials that depend linearly on the distance of the constituents. Recently, also the Dyson-Schwinger/Bethe-Salpeter approach (see, e.g., Refs. [904, 2376, 2377] and also Sec. 5.3), basis light-front quantization (see, e.g., Ref. [958] and also Sec. 5.4), as well as the AdS/QCD correspondence (see, e.g., Ref. [1010] and also Sec. 5.5) were applied to study hybrid light mesons.

The models predict the mass of the lightest $1^{-+}$ state to be in the range from about 1.3 to 2.2 GeV and most model calculations find that $f_1(1285)\pi$ and $b_1(1235)\pi$ are the dominant decay modes for the lightest isovector $1^{-+}$ state. However, for the $\eta\pi$, $\eta'\pi$, and $\rho(770)\pi$ decay modes, discussed in Sec. 8.3.4 below, the model predictions diverge.

*Lattice QCD calculations*
In recent years, lattice QCD calculations of the hadron excitation spectrum have made tremendous progress (see Sec. 4, in particular Sec. 4.5). Currently, calculations that study the excitation spectrum of light mesons still have to be performed in an unphysical world, where the up and down quarks are much heavier in the simulation than in nature.[79] The main reason for this is that decays into multi-body hadronic final states, which for most excited states are the dominant decay modes, cannot yet be calculated on the lattice. By setting the quark masses to sufficiently high values and neglecting multi-hadron operators, the excited states become quasi stable and can be extracted from the simulation. Consequently, such calculations cannot predict widths and decay modes and also cannot take into account coupled-channel effects. Despite these limitations, lattice calculations have already provided important insights by making predictions for light-meson spectra and for two-body scattering processes [477, 498].

For example, the seminal calculation performed by the Hadron Spectrum collaboration [484] showed for the first time a nearly complete spectrum of isoscalar and isovector mesons covering a wide range of $J^{PC}$ quantum numbers up to $J = 4$ (see Fig. 4.5.2 on page 89). The lattice spectrum is qualitatively similar to the one obtained from quark-model calculations. However, the lattice calculation in addition revealed a whole supermultiplet of extra states [493] that lie about 1.3 GeV above the lightest $J^{PC} = 1^{--}$ state and that have quantum numbers of $0^{-+}$, $1^{--}$, $2^{-+}$, and $1^{-+}$, where the latter one is spin-exotic. Studying the overlap of these states with various operators used in the calculation allowed to probe their internal structure. All states in the supermultiplet have large overlaps with operators that correspond to a chromomagnetic gluonic excitation coupled to a color-octet $q\bar{q}$ pair in an $S$-wave and were therefore identified as hybrid states. Intriguingly, the spin-exotic $1^{-+}$ state was predicted to be the lightest hybrid state confirming many model calculations (see Sec. 8.3.2).

Recently, the Hadron Spectrum collaboration published results of the first lattice QCD calculation of the hadronic decays of the lightest $1^{-+}$ resonance using a two-body approximation for the decay [547]. They performed this calculation at the $SU(3)_{\text{flavor}}$ symmetric point, where up, down, and strange-quark masses are chosen to approximately match the physical strange-quark mass, corresponding to a large unphysical pion mass of about 700 MeV. Using a coupled-channel approach, the Hadron Spectrum collaboration studied the scattering amplitudes of eight meson-meson systems and extrapolated the extracted $1^{-+}$ resonance pole and its couplings to the physical light-quark masses. Doing so and assuming a $1^{-+}$ resonance mass of 1564 MeV (value taken from Ref [2276]), they found a broad $\pi_1$ resonance with a total width ranging between 139 and 590 MeV. The dominant decay mode of this resonance is $b_1(1235)\pi$ (partial width ranging from 139 to 529 MeV), in qualitative agreement with most model calculations (see Sec. 8.3.2). Compared to the $b_1(1235)\pi$ channel, the partial widths for the decays into $f_1(1285)\pi$, $\rho(770)\pi$, $\eta'\pi$, and $\eta\pi$ are much smaller. Although these results still have large uncertainties, they provide important guidance for experiments.

[79] This is often expressed in terms of an unphysically large pion mass.

The next great leap for lattice QCD is the calculation of three-body systems, which is already looming on the horizon (see Sec. 4.5.8 and Ref. [558]). First proof-of-principle calculations of three-body systems that do not contain any resonances (see, e.g., Fig. 4.5.9 on page 94) demonstrate the feasibility of the approach and are paving the way towards calculations of more interesting systems that contain two- and/or three-body resonances.

### 8.3.3 Experimental methods

Excited light mesons can be studied in many reactions. They are copiously produced in high-energy scattering reactions of meson beams on nucleon or nuclear targets, such as diffractive dissociation or charge exchange, as well as in central-production reactions in hadron-hadron scattering. Also, antiproton-nucleon annihilations are a source of light mesons. Complementary to these purely strong-interaction processes are photoproduction reactions, which are induced by photon or lepton beams, and $e^+e^-$ scattering reactions such as annihilation, initial-state radiation, or two-photon fusion. Finally, also multi-body decays of heavy particles, such as $\tau$, $J/\psi$, or $D$, are good laboratories to study light mesons. Conservation laws, couplings, and the available energy impose constraints that determine which excited states are allowed to be produced from the various initial states in these reactions. The study of the light-meson spectrum is a world-wide effort with experiments performed at all major particle-accelerator labs covering all the above reactions.

Excited light mesons decay via the strong interaction and are hence extremely short-lived. This is why these states are usually referred to as *resonances*, which are characterized by their nominal mass $m_0$, their total width $\Gamma_0$, and their quantum numbers. In the simplest case of an isolated resonance, its experimental signature is a *peak* at $m_0$ in the distribution of the invariant mass $m$ of the system of daughter particles that the resonance decays into. This peak is accompanied by a *phase motion*, i.e. an increase of the phase of the quantum mechanical amplitude of the studied process by 180° with increasing $m$, reaching 90° at $m_0$ (see Fig. 8.3.1). If the resonance is in addition narrow and $m_0$ is far away from kinematical thresholds, the resonance amplitude is well approximated by a Breit-Wigner amplitude. However, in general resonances are described by amplitudes that are analytical functions of $m^2$ and the resonance parameters are defined by the position of pole singularities of this amplitude in the complex $m^2$ plane (see, e.g., Ref. [2378] for more details).

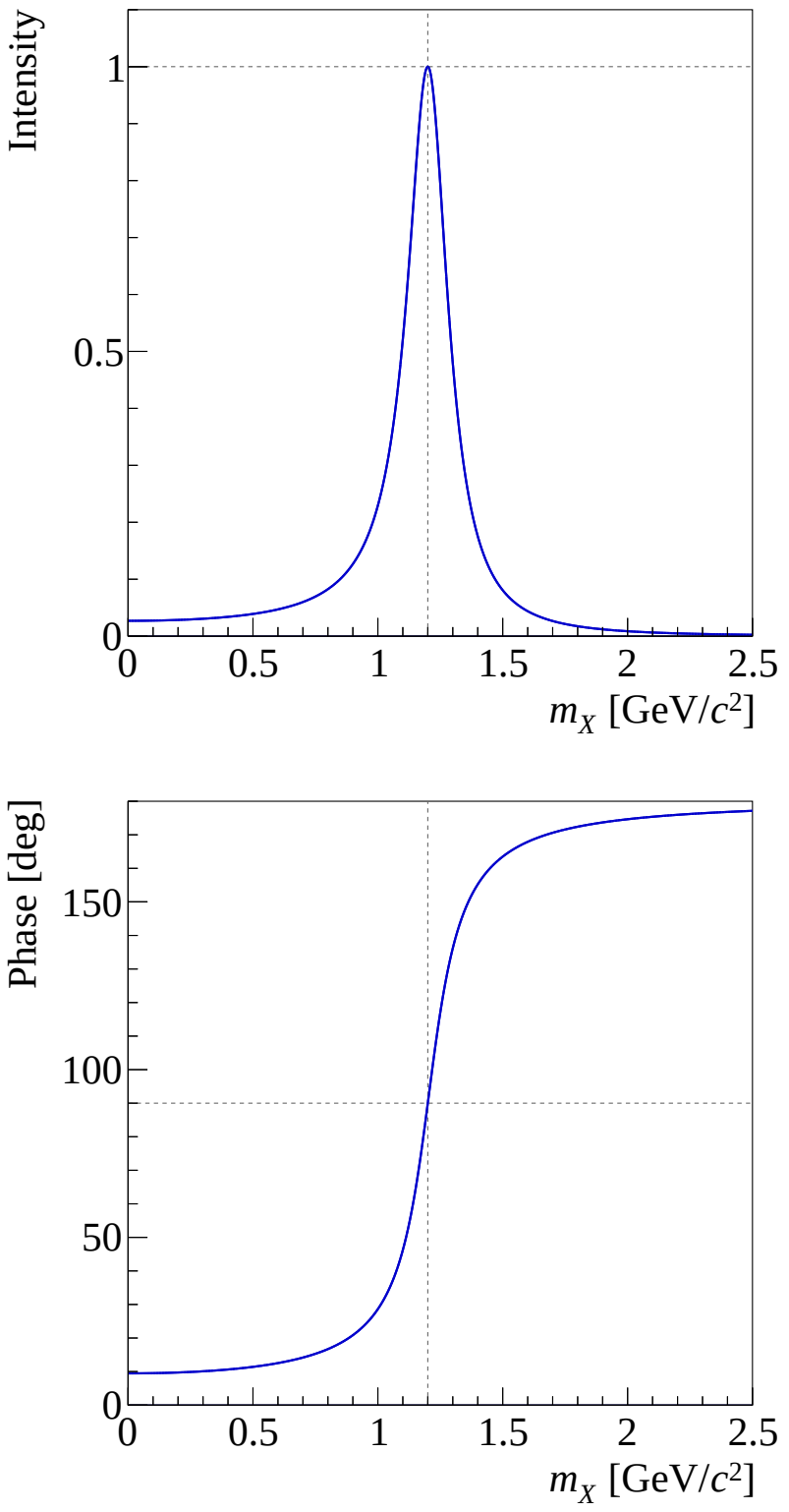

**Fig. 8.3.1** Example of a relativistic Breit-Wigner amplitude with constant width for a fictitious resonance with a mass of $m_0 = 1200\,\text{MeV}$ and a total width of $\Gamma_0 = 200\,\text{MeV}$. (Top) intensity, i.e., absolute value squared of the amplitude, (bottom) phase of the amplitude.

Depending on its mass and quantum numbers, a resonance may have several decay modes, which for highly excited states often lead to multi-body hadronic final states consisting mostly of $\pi$, $K$, $\eta$, and/or $\eta'$. Due to their short-lived nature, any information about resonances has to be inferred from the kinematic distribution of their decay products. To this end, *partial-wave analysis* (PWA) techniques are often employed, which take into account possible interferences of all the intermediate resonances produced in the reaction and exploit the full kinematic information contained in the data. For an $n$-body final state with given mass $m$, a set $\tau$ of $3n-4$ kinematic variables is needed to completely define the decay kinematics. In a simplified picture, a PWA model describes the measured *intensity distribution* $\mathcal{I}(m,\tau)$, i.e. the density distribution of the events in the $(3n-4)$-dimensional phase space of the final-state particles, in terms of *partial-wave amplitudes* $\mathcal{T}_i(m)$, which describe the strength and phase with which an intermediate state with given quantum numbers $i = \{J^{PC}\, M\}$ and mass $m$ is produced, and *decay amplitudes* $\Psi_i(m, \tau)$, which describe the decay of this intermediate state into the observed final state. Here, $M$ is the projection of the spin $J$ along the chosen quantization axis. High-energy scattering reactions, for which examples will be discussed below, are known to be dominated by natural-parity exchange.[80] When analyzing data from these reactions, it is hence advantageous to perform the PWA in the *reflectivity* basis [2379], where the spin state of a resonance is characterized by $M^\epsilon$ with $M \geq 0$ and $\epsilon = \pm 1$ such that the multiplicity of $2J+1$ of the spin state remains unchanged. Here, $\epsilon$ corresponds to the naturality of the exchange particle in the scattering reaction. By performing the PWA in this basis, it is therefore possible to separate the contributions from natural- and unnatural-parity exchange to the scattering reaction.

Since production and decay of a resonance are independent of each other, the total amplitude for an intermediate state $i$ is given by $\mathcal{T}_i(m)\, \Psi_i(m, \tau)$. In the simplest case, the amplitudes of the various allowed intermediate states $i$ are assumed to be fully coherent so that

$$\mathcal{I}(m, \tau) = \left| \sum_i \mathcal{T}_i(m)\, \Psi_i(m, \tau) \right|^2, \tag{8.3.1}$$

where the sum runs over all allowed states. It is important to note that in the above equation, the intensity is given by the sum of the contributing amplitudes, i.e. all intermediate states may interfere with each other. The decay amplitudes can be calculated using first principles and models. The analyses that will be discussed in Sec. 8.3.4 below use a two-stage procedure, where in the first stage the known decay amplitudes $\Psi_i$ are used to determine the partial-wave amplitudes $\mathcal{T}_i$ in narrow $m$ bins by fitting the PWA model in Eqn. (8.3.1) to the measured $\tau$ distributions. At this stage, no assumptions are made about the resonance content in the studied $n$-body system. In a second stage, a resonance model is fit to the $m$ dependence of selected partial-wave amplitudes in order to extract the resonances and their parameters. For high-energy scattering data, the resonance model also has to take into account contributions from *non-resonant processes*, i.e. processes where the measured $n$-body final state is produced without going through an intermediate $n$-body resonance. Unfortunately, in most cases no detailed theoretical models exist for these non-resonant contributions and one has to revert to empirical models. More details on the PWA procedure and the involved model assumptions can be found, e.g., in Ref. [2380].

### 8.3.4 Experimental evidence

More than three decades ago the GAMS experiment claimed the first observation of a spin-exotic resonance with $J^{PC} = 1^{-+}$ [2381]. Since then, many other experiments reported such signals. Currently, the Particle Data Group (PDG) lists three spin-exotic light-meson states: the $\pi_1(1400)$, the $\pi_1(1600)$, and the $\pi_1(2015)$ [476]. However, despite the seemingly large body of evidence, which includes data from pion diffraction, antiproton-nucleon annihilation, photoproduction, and charmonium decays covering several decay channels, the experimental situation is still puzzling and the interpretation of many of the observed signals is controversial.

The $\pi_1(1400)$ was observed nearly exclusively in the $\eta\pi$ decay channel produced in pion diffraction and antiproton-nucleon annihilation [2280, 2381–2387]. Only the OBELIX and Crystal Barrel experiments claimed to see the $\pi_1(1400)$ also in the $\rho(770)\pi$ decay channel in their antiproton-nucleon annihilation data [2388, 2389]. Surprisingly, the signal in the $\rho(770)\pi$ channel arises from antiproton-nucleon initial states with different quantum numbers than the signal in the $\eta\pi$ channel.[81] Since production and decay of a resonance are independent, the $\rho(770)\pi$ resonance claimed by OBELIX and Crystal Barrel cannot be the same $\pi_1(1400)$ state that is observed in $\eta\pi$—a puzzling result. The $\pi_1(1400)$ masses quoted by the various experiments are in fair agreement; the width values, however, scatter over a larger range. The PDG estimates for the $\pi_1(1400)$ mass and width are $m_0 = (1354 \pm 25)\,\text{MeV}$ and $\Gamma_0 = (330 \pm 35)\,\text{MeV}$ [476].

Compared to the $\pi_1(1400)$, the $\pi_1(1600)$ was seen in a much wider range of decay channels produced in pion diffraction, antiproton-nucleon annihilation, and $\chi_{c1}$ decays. Signals were reported in the $\rho(770)\pi$ [2271, 2390–2393], $\eta'\pi$ [2394–2398], $f_1(1285)\pi$ [2397, 2399], and $b_1(1235)\pi$ [2274, 2386, 2395–2397, 2400] decay channels. As for the $\pi_1(1400)$, the measured $\pi_1(1600)$ mass values are in better agreement with each other than the measured width values. The PDG estimates for the $\pi_1(1600)$ mass and width are $m_0 = (1661^{+15}_{-11})\,\text{MeV}$ and $\Gamma_0 = (240 \pm 50)\,\text{MeV}$ [476].

The $\pi_1(2015)$ was so far only observed by the BNL E852 experiment in the decay modes $f_1(1285)\pi$ [2399]

[80] The *naturality* is defined as $\epsilon = P\,(-1)^J$, i.e. $\epsilon = +1$ corresponds to the natural-parity series with $J^P = 0^+, 1^-, 2^+, \ldots$ and, correspondingly, $\epsilon = -1$ corresponds to the unnatural-parity series with $J^P = 0^-, 1^+, 2^-, \ldots$.

[81] In the $\rho(770)\pi$ channel, the $\pi_1(1400)$ is seen predominantly in $P$-wave antiproton-nucleon initial states, whereas in the $\eta\pi$ channel it is seen mainly in the $^3S_1$ initial state.

and $b_1(1235)\ \pi$ [2274]. It hence still needs to be confirmed by other experiments and is listed as a “further state” by the PDG.

Although on first sight there seems to be strong experimental evidence for the $\pi_1(1400)$ and the $\pi_1(1600)$, some analyses have issues and some experimental results are disputed. From a phenomenological standpoint, the properties of the $\pi_1(1400)$ are problematic. Compared to most of the predictions (see Sec. 8.3.2), it is too light. Also, the $\pi_1(1600)$ is too close in mass to the $\pi_1(1400)$ in order to be an excitation of the latter. Additionally, the fact that the $\pi_1(1400)$ seems to decay only to $\eta\pi$ is hard to explain.[82]

The analyses of some channels also face technical issues. For example, in order to extract the $\pi_1(1400)$ in the $\eta\pi$ channel and the $\pi_1(1600)$ in the $\eta'\pi$ channel, the phase motions of the $P$-wave amplitudes need to be measured. Often, this can be done only relative to the $D$-wave amplitudes. However, in the mass region of interest the $D$-waves contain contributions from the $a_2(1700)$, which is the first radial excitation of the $a_2(1320)$ ground state. Unfortunately, the $a_2(1700)$ is a rather broad state and its resonance parameters are not well known. For the widely used simple Breit-Wigner based resonance models, this may lead to systematic uncertainties that are hard to control.

The analysis of the data of the BNL E852 experiment yielded inconsistent results on the production properties of the $\pi_1(1600)$. Whereas in the $\eta'\pi$ [2394] and $f_1(1285)\pi$ [2399] channels the $\pi_1(1600)$ is observed to be produced only via natural-parity exchange, i.e. with $M^\epsilon = 1^+$, it appeared in the $\rho(770)\pi$ [2390, 2391] and $b_1(1235)\pi$ [2274] channels also in unnatural-parity exchange, i.e. in waves with $M^\epsilon = 0^-$ and $1^-$, with similar strength as in the $M^\epsilon = 1^+$ wave. This is hard to explain as production and decay of a resonance are independent processes.

One of the deepest puzzles, however, concerns the seemingly contradictory conclusions on the existence of the $\pi_1(1600)$ in the $\rho(770)\pi$ decay channel that were drawn from similar analyses. The BNL E852 experiment was the first to claim the observation of $\pi_1(1600) \to \rho(770)\pi$ based on a sample of about 250 000 $\pi^-p \to \pi^-\pi^-\pi^+p$ events and using a PWA model with 21 waves [2390, 2391]. The measured intensity distribution of the spin-exotic wave with $J^{PC} = 1^{-+}$ quantum numbers is shown in Fig. 8.3.2. It exhibits a pronounced peak at about 1.6 GeV that is accompanied by significant phase

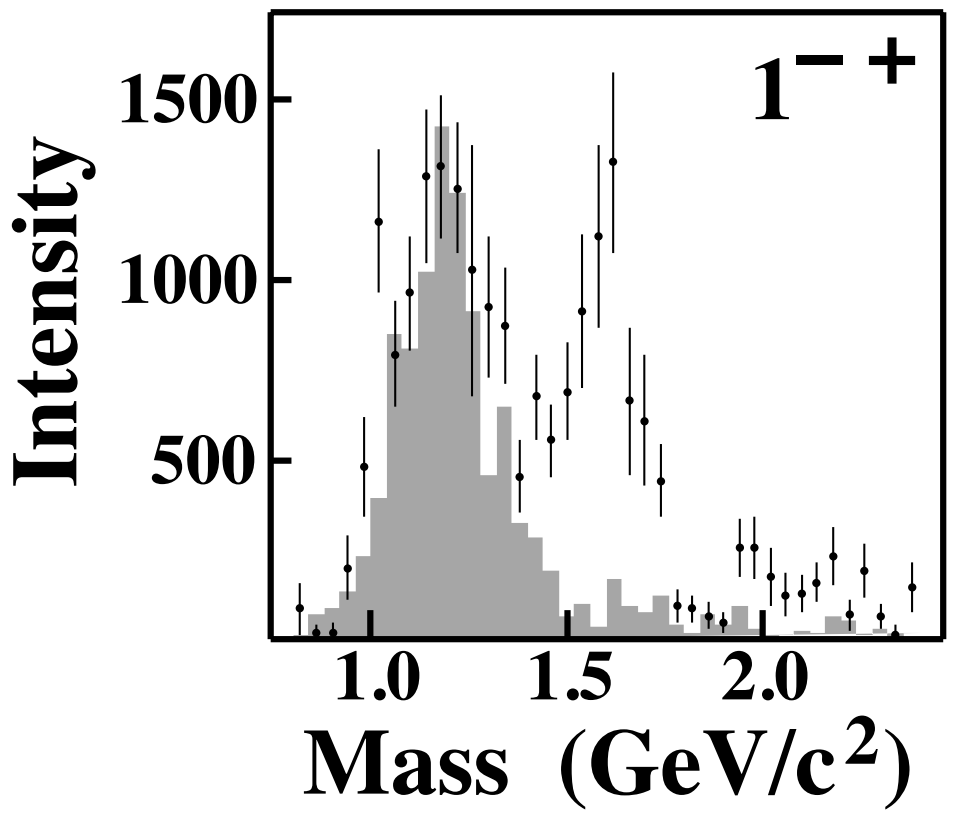

**Fig. 8.3.2** Intensity of the $\rho(770)\pi$ $P$-wave with spin-exotic $J^{PC} = 1^{-+}$ quantum numbers produced in natural-parity exchange (points with statistical uncertainties) as a function of the $\pi^-\pi^-\pi^+$ mass obtained by the BNL E852 collaboration. (Adapted from Fig. 3(b) in Ref. [2390])

motion with respect to other partial waves (see Fig. 19 in Ref. [2391]).[83] Based on a simultaneous resonance-model fit of the intensities of the $1^{-+}$ wave and of the $f_2(1270)\pi$ $S$-wave with $J^{PC} = 2^{-+}$ and their relative phase, the authors of Refs. [2390, 2391] claimed the observation of the $\pi_1(1600)$. However, they also observed a strong dependence of the shape and strength of the $\pi_1(1600)$ signal on the PWA model.

Surprisingly, an analysis of a more than 20 times larger data sample ($2.6 \times 10^6$ $\pi^-p \to \pi^-\pi^-\pi^+p$ events plus $3.0 \times 10^6$ $\pi^-p \to \pi^-\pi^0\pi^0p$ events) from the same experiment performed by Dzierba *et al.* came to a completely different conclusion [2401]. They performed the partial-wave analysis independently in 12 bins of the reduced four-momentum squared $t'$ that is transferred from the beam to the target recoil particle[84] in the range from 0.08 to 0.53 GeV$^2$ using a larger PWA model of 36 waves. The observed intensity distribution of the $1^{-+}$ wave exhibits a broad and structureless enhancement (see black points in Fig. 8.3.3; cf. Fig. 8.3.2). The shape of this enhancement was found to change strongly with $t'$ with intensity moving from the 1.2 GeV region towards higher masses with increasing $t'$. However, the peak at 1.6 GeV, which in Refs. [2390, 2391] was attributed to the $\pi_1(1600)$, had disappeared. By applying the 21-wave PWA model from Refs. [2390, 2391],

[82] If one would take the $\pi_1(1400) \to \rho(770)\pi$ claims of OBELIX and Crystal Barrel [2388, 2389] at face value, then even two mass-degenerate $\pi_1(1400)$ states would exist, one decaying to $\eta\pi$ the other to $\rho(770)\pi$—an even more puzzling scenario.

[83] The second peak at about 1.2 GeV was explained as an analysis artifact caused by intensity leaking from the dominant $1^{++}$ wave into the spin-exotic wave because of a non-uniform detector acceptance in combination with the finite experimental resolution. The gray-shaded histogram in Fig. 8.3.2 represents an estimate of this effect from Monte Carlo simulations.

[84] Here, $t' \equiv |t| - |t|_{\min}$ with $t = (p_{\text{beam}} - p_X)^2$ being the Mandelstam variable, $p_{\text{beam}}$ the four-momentum of the beam pion, and $p_X$ the total four-momentum of the produced $3\pi$ system.

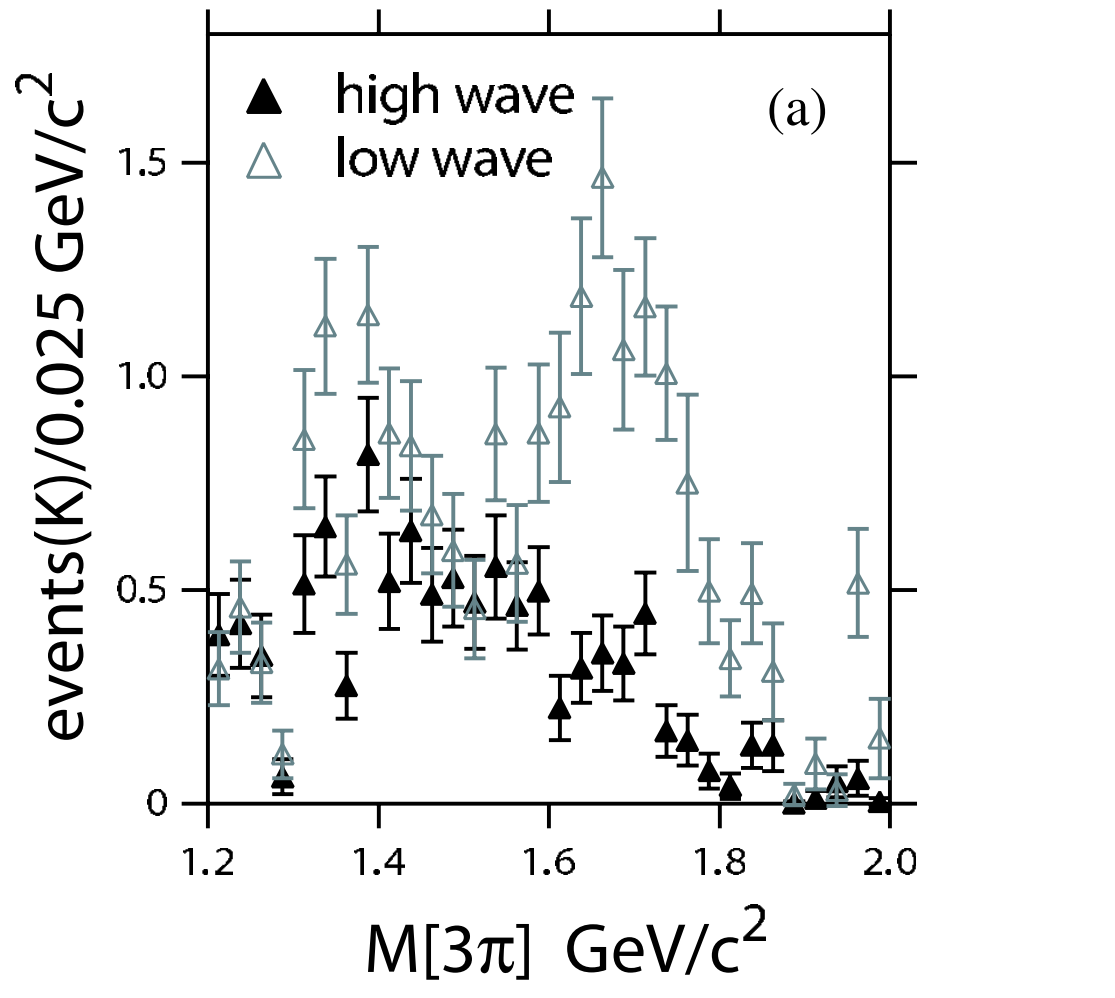

**Fig. 8.3.3** Intensity distribution of the $\rho(770)\pi$ $P$-wave with spin-exotic $J^{PC} = 1^{-+}$ quantum numbers produced in natural-parity exchange as obtained by Dzierba *et al.* using BNL E852 data on $\pi^- p \to \pi^-\pi^-\pi^+ p$ in the kinematic range $0.18 < t' < 0.23\,\text{GeV}^2$. The open gray points ("low wave") correspond to the 21-wave PWA model from Refs. [2390, 2391] (cf. Fig. 8.3.2), the solid black points ("high wave") correspond to the 36-wave PWA model from Ref. [2401]. (Taken from Fig. 25(a) in Ref. [2401])

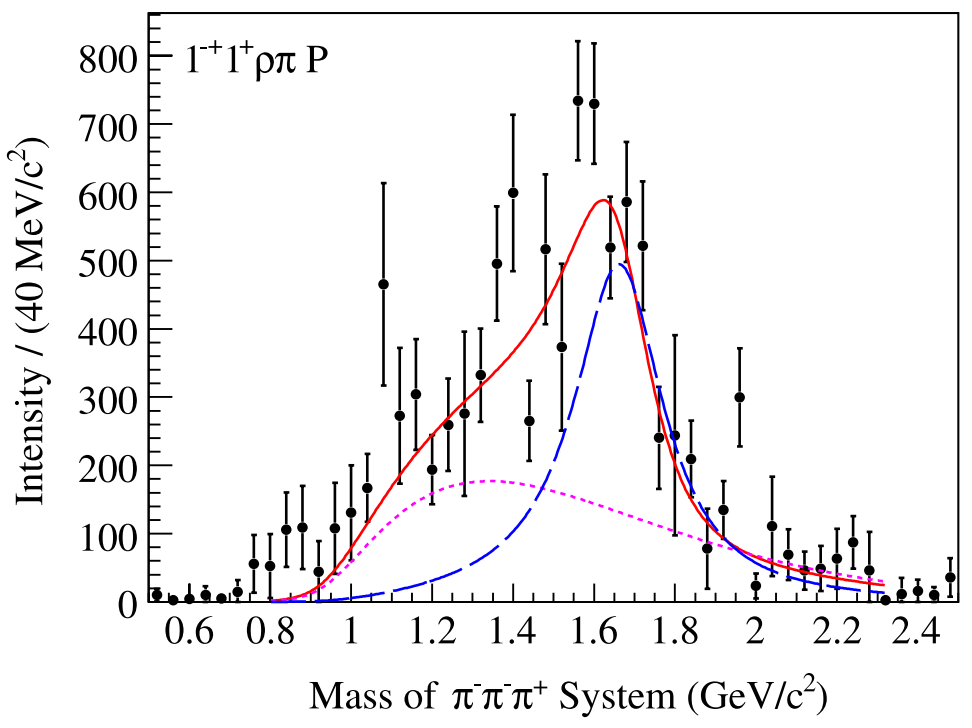

**Fig. 8.3.4** Intensity distribution of the $\rho(770)\pi$ $P$-wave with spin-exotic $J^{PC} = 1^{-+}$ quantum numbers produced in natural-parity exchange as obtained by the COMPASS experiment using data on $\pi^-\text{Pb} \to \pi^-\pi^-\pi^+\text{Pb}$ (points with statistical uncertainties). The red curve represents the result of a fit with a resonance model, which is the coherent sum of a Breit-Wigner amplitude for the $\pi_1(1600)$ (blue) and a non-resonant amplitude (magenta). (Taken from Fig. 2(d) in Ref. [2392])

Dzierba *et al.* were able to reproduce the results from Refs. [2390, 2391] (see gray points in Fig. 8.3.3; cf. Fig. 8.3.2). They also showed that the omission of important $2^{-+}$ waves in the 21-wave PWA model causes leakage from the $\pi_2(1670)$ producing an artificial peak at 1.6 GeV in the $1^{-+}$ wave. Based on these findings, Dzierba *et al.* concluded that the BNL E852 data provide no evidence for the existence of the $\pi_1(1600)$ in the $\rho(770)\pi$ decay channel and that the signal reported in Refs. [2390, 2391] was an artifact of a too restricted PWA model. However, this conclusion was not based on a resonance-model fit and did not take into account the phase motions of the $1^{-+}$ wave that were still present in the analysis of Dzierba *et al.*. In addition, Dzierba *et al.* only considered the kinematic region $t' < 0.53\,\text{GeV}^2$, which will become important in the discussion below.

The first results from the COMPASS experiment only added to the confusion. The authors of Ref. [2392] performed a partial-wave analysis of 420 000 events for the reaction $\pi^-\text{Pb} \to \pi^-\pi^-\pi^+\text{Pb}$ in the kinematic range $0.1 < t' < 1.0\,\text{GeV}^2$ using an even larger PWA model than Dzierba *et al.* consisting of 42 waves. This model is similar to the 36-wave PWA model used in Ref. [2401] and includes in particular the $2^{-+}$ waves that were found to cause leakage from the $\pi_2(1670)$ into the $1^{-+}$ wave. However, in contrast to Dzierba *et al.*, COMPASS observed an enhancement at 1.6 GeV in the intensity distribution of the $1^{-+}$ wave (see data points in Fig. 8.3.4; cf. black data points in Fig. 8.3.3). In the performed resonance-model fit, which describes the intensities and mutual interference terms of six waves simultaneously, the $1^{-+}$ amplitude is well described by a coherent sum of a non-resonant and a Breit-Wigner amplitude for the $\pi_1(1600)$ (see curves in Fig. 8.3.4) and the resulting resonance parameters are compatible with the previous measurements of the $\pi_1(1600)$. Hence, COMPASS claimed the observation $\pi_1(1600) \to \rho(770)\pi$.

These puzzling experimental findings were reconciled only recently by the results of a comprehensive partial-wave analysis performed on a highly precise sample of $46 \times 10^6$ $\pi^- p \to \pi^-\pi^-\pi^+ p$ events obtained by the COMPASS experiment [2271, 2393, 2402]. The PWA was performed independently in 11 $t'$ bins in the range $0.1 < t' < 1.0\,\text{GeV}^2$ using the so far largest PWA model with 88 waves. The intensity distribution of the $1^{-+}$ wave summed over the 11 $t'$ bins exhibits a broad enhancement from about 1.0 to 1.8 GeV but no peak at 1.6 GeV. This is consistent with the distribution observed by the VES experiment in a similar $t'$ range [2403]. The shape of the intensity distribution changes strongly with $t'$ confirming a similar observation made by Dzierba *et al.* in the BNL E852 data [2401]. At low $t'$, COMPASS observes a broad structure in the mass range from about 1.0 to 1.7 GeV (see Fig. 8.3.5(left)).[85] As $t'$ increases, this structure becomes narrower and its maximum moves to about 1.6 GeV so that it becomes similar to the distribution observed in the first COM-

[85] The distribution also exhibits a narrow peak at about 1.1 GeV, which, however, has no associated phase motion and depends on the PWA model. According to Refs. [2271, 2393] this peak is likely an artifact induced by imperfections of the analysis method.

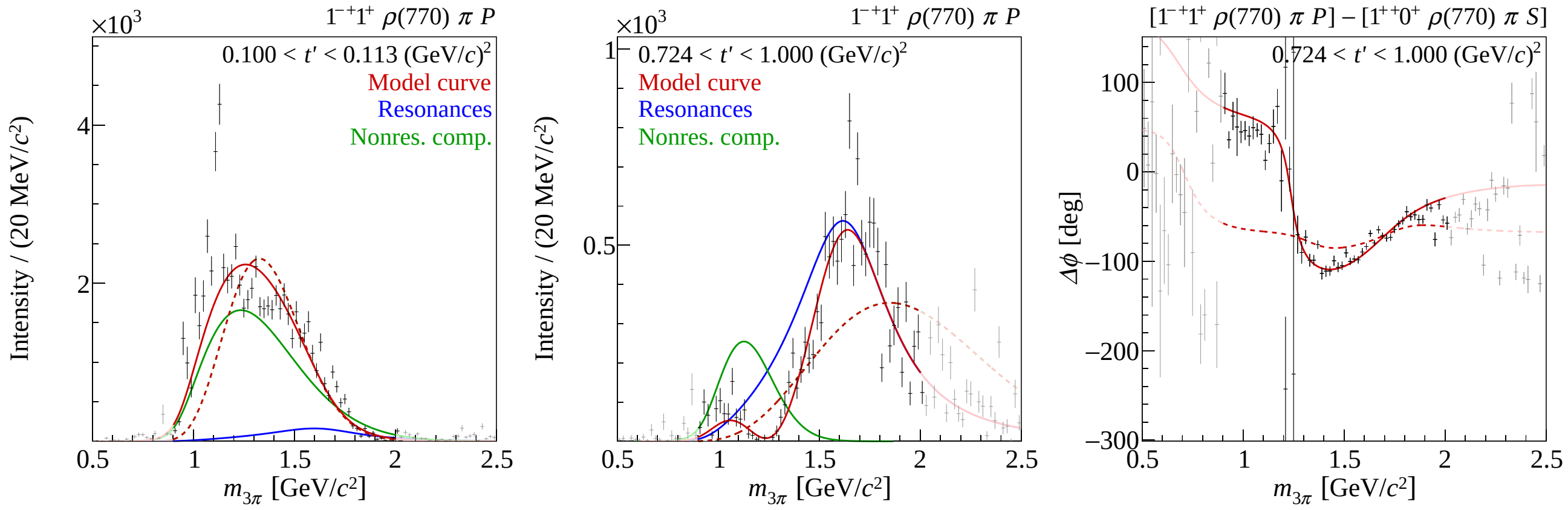

**Fig. 8.3.5** (Left) and (center) Intensity distribution of the $\rho(770)\pi$ $P$-wave with spin-exotic $J^{PC} = 1^{-+}$ quantum numbers produced in natural-parity exchange as obtained by the COMPASS experiment using data on $\pi^- p \to \pi^-\pi^-\pi^+ p$ at low and high $t'$. (Right) Phase of the $1^{-+}$ wave relative to the $\rho(770)\pi$ $S$-wave with $J^{PC} = 1^{++}$ at high $t'$. In the three diagrams, the points with statistical uncertainties represent the measured values. The red curves represent the results of fits with two resonance models. The continuous red curve corresponds to the coherent sum of a Breit-Wigner amplitude for the $\pi_1(1600)$ (blue) and a non-resonant amplitude (green). The dashed red curve corresponds to a model that contains only the non-resonant amplitude. (Taken from Figs. 48(b), (c), and (d) in Ref. [2271])

PASS data on the Pb target (see Fig. 8.3.5(center); cf. Fig. 8.3.4).

Since resonance parameters are independent of $t'$, the observed strong modulation of the intensity distribution with $t'$ hints at large contributions from non-resonant processes. This was confirmed by the resonance-model fit, which simultaneously describes the amplitudes of 14 selected partial waves. The large wave set provides tight constraints for the $1^{-+}$ amplitude via the mutual interference terms between the amplitudes. In addition, for the first time all 11 $t'$ bins were fit simultaneously, forcing the resonance parameters to be the same across the $t'$ bins. This $t'$-resolved approach leads to a much better disentanglement of the resonant and the non-resonant contributions, which have in general different dependences on $t'$. For $t' \lesssim 0.5\,\text{GeV}^2$, the fit finds that the $1^{-+}$ intensity is almost saturated by the non-resonant component (green curve in Fig. 8.3.5(left)) with only a small $\pi_1(1600)$ contribution (blue curve). With increasing $t'$ the strength of the non-resonant component decreases relative to that of the $\pi_1(1600)$, so that for $t' \gtrsim 0.5\,\text{GeV}^2$ the $\pi_1(1600)$ becomes the dominant component (see Fig. 8.3.5(center)).

Applying the 21- and 36-wave PWA models from the two analyses of BNL E852 data [2393] to the COMPASS data yields results consistent with those reported in Refs. [2390, 2391, 2401] confirming the observations by Dzierba *et al.* that the 21-wave model produces an artificial peak at 1.6 GeV in the $1^{-+}$ waves for natural- as well as unnatural-parity exchange due to leakage from the $\pi_2(1670)$. This explains the puzzling observation of a $\pi_1(1600) \to \rho(770)\pi$ signal in unnatural-parity exchange by the BNL E852 experiment [2390, 2391] as an artifact caused by leakage. In addition, the $t'$-resolved analysis of the COMPASS data shows that for $t' \lesssim 0.5\,\text{GeV}^2$ the $\pi_1(1600)$ signal is masked by the dominant non-resonant contribution. This explains why Dzierba *et al.*, who considered only the range $t' < 0.53\,\text{GeV}^2$, reported a non-observation of the $\pi_1(1600)$. However, in the kinematic range $t' \gtrsim 0.5\,\text{GeV}^2$ COMPASS observes a clear $\pi_1(1600) \to \rho(770)\pi$ signal and a $\pi_1(1600)$ resonance is indeed required to explain the COMPASS data. This is demonstrated by the dashed red curve in Fig. 8.3.5, which represents the result of a resonance-model fit, where the $1^{-+}$ amplitude was described using only the non-resonant component. At low $t'$, this model is able to describe the data fairly well (see Fig. 8.3.5(left)), but clearly fails at high $t'$ (see Fig. 8.3.5(center) and (right)). The $t'$-resolved COMPASS results in Refs. [2271, 2392, 2393] therefore establish unambiguously the $\rho(770)\pi$ decay mode of the $\pi_1(1600)$ and in addition resolve a long-standing controversy by showing that the data of previous experiments are indeed consistent and that the BNL E852 puzzle was caused by a too restricted PWA model on the one hand [2390, 2391] and a too restricted $t'$ range on the other hand [2401].

Another big step towards a better understanding of the $\pi_1$ states was the coupled-channel analysis of the $\eta\pi$ and $\eta'\pi$ $P$- and $D$-wave amplitudes measured by the COMPASS experiment [2405], which was performed by the JPAC collaboration [2276]. Using a unitary model based on $S$-matrix principles they find in the $D$-wave amplitudes two resonance poles, the $a_2(1320)$ and the $a_2(1700)$ and in the $P$-wave amplitudes a single resonance pole. The parameters of the $P$-wave resonance

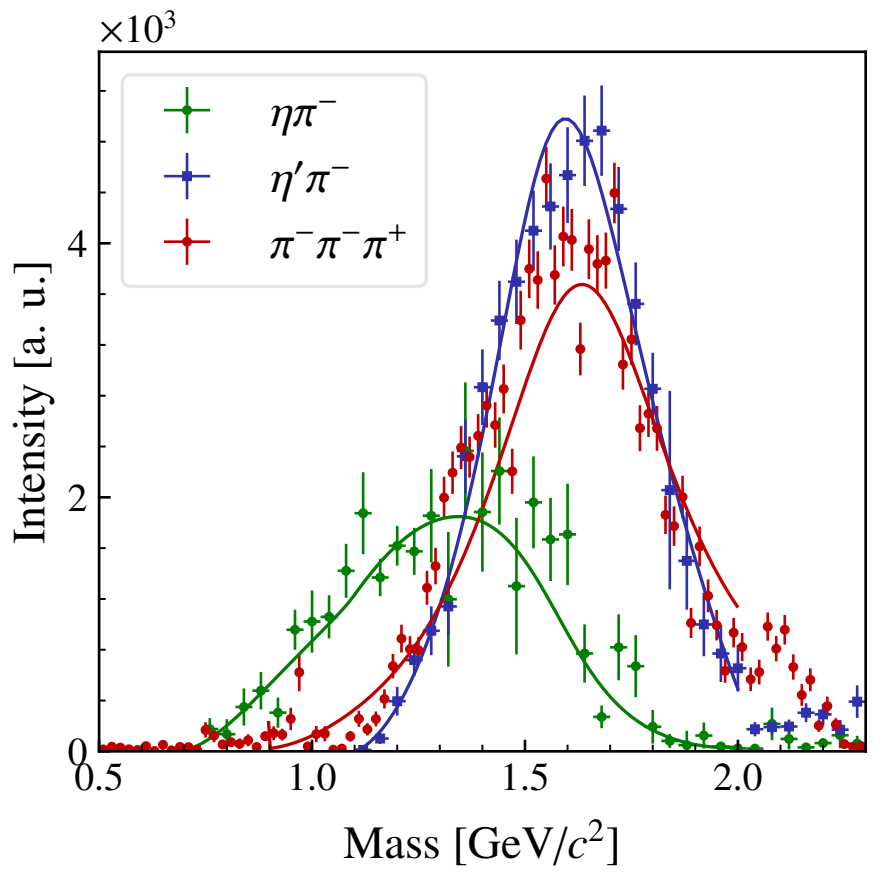

**Fig. 8.3.6** Intensity distributions of spin-exotic waves with $J^{PC} = 1^{-+}$ from COMPASS data. (Green points) $\eta\pi$ *P*-wave, (blue points) $\eta'\pi$ *P*-wave, both for $0.1 < t' < 1.0\,\text{GeV}^2$. (Red points) $\rho(770)\pi$ *P*-wave for $0.449 < t' < 0.724\,\text{GeV}^2$. The curves represent the results of the resonance-model fits from Refs. [2271, 2276]. (Taken from Fig. 2 in Ref. [2404])

pole are $m_0 = (1564^{+24}_{-86})$ MeV and $\Gamma_0 = (492^{+\ 54}_{-102})$ MeV, consistent with the $\pi_1(1600)$. Apart from determining the $\pi_1(1600)$ pole position for the first time using an analytic and unitary model, this result is in so far remarkable as only a single resonance pole is required to simultaneously describe the $\eta\pi$ and the $\eta'\pi$ *P*-wave amplitudes despite their rather different intensity distributions (see green and blue points and curves in Fig. 8.3.6). This is in contrast to most previous analyses, which considered two different resonance components in their models: a $\pi_1(1400)$ to describe the broad peak at 1.4 GeV in the $\eta\pi$ *P*-wave intensity and a $\pi_1(1600)$ to describe the narrower peak at 1.6 GeV in the $\eta'\pi$ *P*-wave intensity. It is interesting to note that in the COMPASS data the latter peak is nearly identical to the one observed in the $1^{-+}$ intensity in the high-$t'$ region of the $\pi^-\pi^-\pi^+$ data (cf. blue and red points and curves in Fig. 8.3.6). Since the COMPASS partial-wave data are consistent with previous experiments, the JPAC analysis raises serious doubts about the existence of the $\pi_1(1400)$ as a separate resonance. Recently, the JPAC results were confirmed by Kopf *et al.*, who performed a coupled-channel analysis that in addition to the COMPASS $\eta\pi$ and $\eta'\pi$ *P*- and *D*-wave amplitudes also includes Crystal Barrel data on $\bar{p}p \to \pi^0\pi^0\eta$, $\pi^0\eta\eta$, and $K^+K^-\pi^0$ as well as $\pi\pi$ scattering data [2406].

Both coupled-channel analyses favor a much simpler and more plausible picture with only one $\pi_1$ state below 2 GeV, the $\pi_1(1600)$, decaying into (at least) $\eta\pi$, $\eta'\pi$, $\rho(770)\pi$, $f_1(1285)\pi$, and $b_1(1235)\pi$. This scenario resolves the longstanding puzzle of two spin-exotic states having peculiar decay modes and lying unexpectedly close to each other. If interpreted in terms of hybrid states, this would also remove the discrepancy with lattice QCD and most model calculations, which predict the lightest hybrid state to have a mass substantially higher than that of the $\pi_1(1400)$ (see Sec. 8.3.2).

Up to now only isovector spin-exotic states were observed in the light-meson sector. However, models and lattice QCD predict that $\text{SU}(3)_\text{flavor}$ partner states of the $\pi_1$, i.e. $\eta_1$ and $\eta_1'$ as well as $K^*$ states,[86] should exist. In order to establish exotic resonances it is therefore important to find these states. A first sign that they indeed exist is the very recent first observation of a spin-exotic isocalar $\eta_1(1855)$ state in the $\eta\eta'$ decay channel produced in $J/\psi \to \gamma\eta\eta'$[87] reported by the BESIII experiment [2409, 2410]. The challenge is now to confirm this state in other experiments.

### 8.3.5 Summary and outlook

The dust of more than three decades of research on spin-exotic light mesons is starting to settle. For a long time, the experimental data were confusing leading to contradictory conclusions on the existence and properties of $\pi_1$ mesons. Recently, high-precision data and more advanced theory models helped to resolve many of these puzzles and a more coherent picture seems to be emerging, where instead of two low-lying states, $\pi_1(1400)$ and $\pi_1(1600)$, with hard to explain properties only the $\pi_1(1600)$ remains. However, there are at least two puzzles to be solved. The first is the unexpected production of the $\pi_1(1600)$ in unnatural-parity exchange claimed by the BNL E852 experiment in the $b_1(1235)\pi$ channel [2274]. This can be clarified by the COMPASS experiment using data on the same reaction at higher energy. The second remaining puzzle is the seeming non-observation of the $\pi_1(1600)$ in photon-induced reactions. Since the $\pi_1(1600)$ is observed to decay into $\rho(770)\pi$, it should couple to $\gamma\pi$ via vector-meson dominance. However, in the $\gamma + \pi^\pm \to \pi^\pm\pi^-\pi^+$ reaction studied by the CLAS and the COMPASS experiments[88]

[86] As kaons are neither eigenstates of *C* nor of *G* parity, there are no spin-exotic kaon states. Hence, the exotic $K^*$ states can be identified only as supernumerary states and via their couplings.

[87] This is an example for a radiative $J/\psi$ decay. Such decays are "gluon-rich" processes because in lowest order the $c\bar{c}$ pair in the $J/\psi$ annihilates into the measured photon and a pair of gluons that hadronize into the measured final state, here the $\eta\eta'$ system. The production of mesons with explicit gluonic degrees of freedom, i.e. hybrids and glueballs, is expected to be enhanced in these decays.

[88] CLAS measured the photoproduction reaction $\gamma + p \to \pi^+\pi^-\pi^+ + (n)_\text{miss}$, where a pion is exchanged between the target and the beam photon producing the $3\pi$ final state. In COMPASS data, the $\gamma\pi \to 3\pi$ reaction is embedded into the

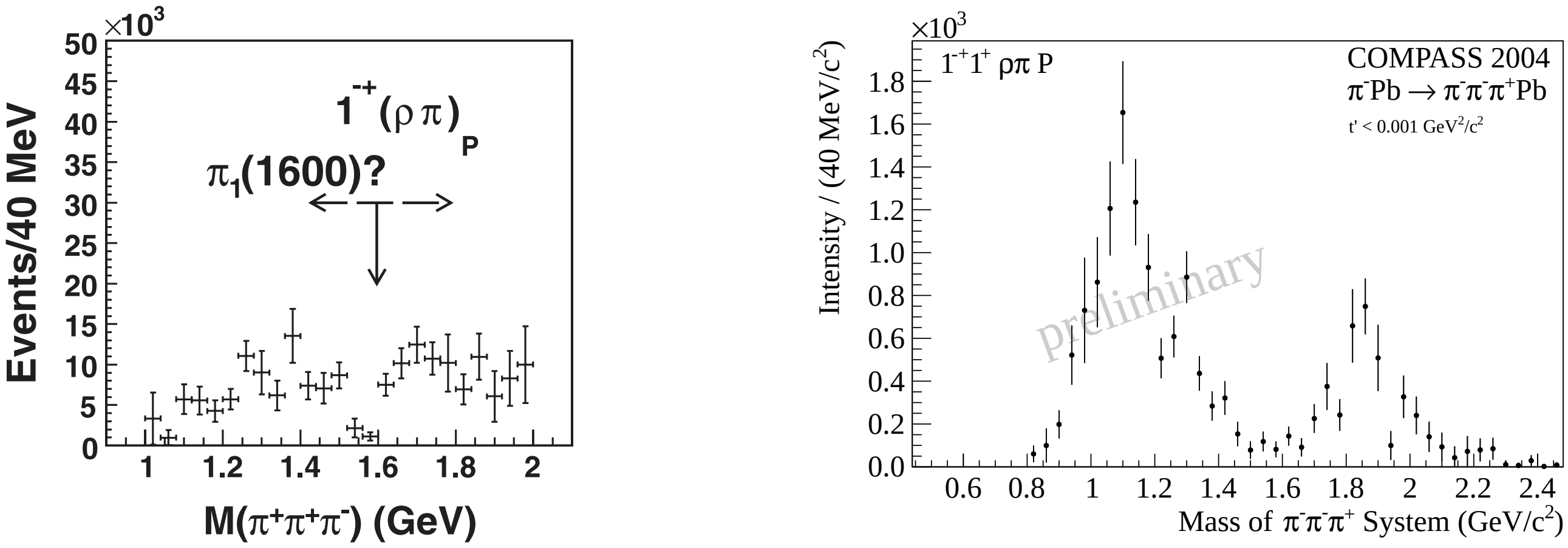

**Fig. 8.3.7** Intensity distributions of the $\rho(770)\pi$ $P$-wave with spin-exotic $J^{PC} = 1^{-+}$ quantum numbers produced in $\gamma + \pi^{\pm} \to \pi^{\pm}\pi^{-}\pi^{+}$ reactions. (Left) Result from the CLAS experiment [2407], where the process is embedded into $\gamma{+}p \to \pi^{+}\pi^{-}\pi^{+}{+}(n)_{\mathrm{miss}}$. (Right) Result from the COMPASS experiment [2361, 2408], where the process is embedded into $\pi^{-} + \mathrm{Pb} \to \pi^{-}\pi^{-}\pi^{+} + \mathrm{Pb}$. (Taken from Fig. 5(d) of Ref. [2407] and Fig. 7(a) of Ref. [2361])

nearly vanishing intensity was observed in the $J^{PC} = 1^{-+}$ wave in the mass range where a $\pi_1(1600)$ signal would be expected (see Fig. 8.3.7). The nearly vanishing intensity could be the result of a destructive interference of the $\pi_1(1600)$ amplitude with the one of non-resonant contributions. However, no resonance-model fits have been performed yet to test this hypothesis. In the future, much more precise photoproduction data from the GlueX experiment at JLab will help to clarify the situation.

Having established that spin-exotic $1^{-+}$ light-meson states do exist is, of course, only the starting point. The next goal is to study their properties in detail, in particular their couplings, by measuring them in various production and decay modes. Another goal is to find their excitations. A first step in this direction would be the confirmation of the $\pi_1(2015)$ signal in the $f_1(1285)\pi$ and $b_1(1235)\pi$ decay channels. In addition, it is important to search for the exotic $\mathrm{SU}(3)_{\mathrm{flavor}}$ partner states of the $\pi_1$. Here, the result by the BESIII experiment of a possible observation of an $\eta_1(1855)$ state could be a breakthrough. Last but not least, the search for states with other spin-exotic $J^{PC}$ quantum numbers such as $0^{+-}$ and $2^{+-}$ continues. These searches will also yield a more complete picture of the spectrum of states with ordinary quantum numbers, which not only helps to identify supernumerary states, but is also an important input to theory in order to improve our understanding of the non-perturbative regime of QCD.

In turn, the analysis of the extremely high-precision data from running and upcoming experiments requires more advanced theoretical models and in particular a more accurate understanding of the dynamics of hadrons. Close collaboration of theorists and experimentalists will help us to formulate, test, and apply detailed models for production reactions and for the interactions of final-state hadrons in order to overcome limitations of the currently available analysis approaches. Together with refined statistical tools and novel approaches such as Machine Learning, this will enable us to leverage the full potential of the data.

reaction $\pi^{-} + \mathrm{Pb} \to \pi^{-}\pi^{-}\pi^{+} + \mathrm{Pb}$, which was measured at very low squared four-momentum transfer, where the beam pion predominantly scatters off quasi-real photons from the Coulomb field of the Pb target nucleus.

## 8.4 Glueballs, a fulfilled promise of QCD?

**Eberhard Klempt**

### 8.4.1 Introduction

At the *Workshop on QCD: 20 Years Later* [2411] held in 1992 in Aachen, Heusch [2412] reported on searches for glueballs, gluonium, or glue states as Fritzsch and Gell-Mann [30, 50] had called this new form of matter. Glueballs are colorless bound states of gluons and should exist when their newly proposed quark-gluon field theory yields a correct description of the strong interaction. The title of Heusch's talk *Gluonium: An unfulfilled promise of QCD?* expressed the disappointment of a glueball hunter: At that time there was some - rather weak - evidence for glueball candidates but there was no convincing case. In 1973, the $e^{+}e^{-}$ storage ring SPEAR at the Stanford Linear Accelerator Center had come into operation and one year later, the $J/\psi$ resonance was discovered [75] - this was the very first SPEAR publication on physics. The $J/\psi$ resonance and its radiative decay became and still is the prime reaction for glueball searches.

One of the first glueball candidates was the $\iota(1440)$ [2413, 2414]. The name $\iota$ stood for the "number one" of all glueballs to be discovered. It was observed as very strong signal with pseudoscalar quantum numbers in the reaction $J/\psi \to \gamma K\bar{K}\pi$. Its mass was not too far from the bag-model prediction (1290 MeV) [747]. Now the $\iota(1440)$ is supposed to be split into two states, $\eta(1405)$ and $\eta(1475)$, where the lower-mass meson is still discussed as glueball candidate even though its mass is incompatible with lattice gauge calculations. They find the mass of the pseudoscalar glueball above 2 GeV.

A second candidate was a resonance called $\Theta(1640)$ [2415, 2416]. It was seen in the reaction $J/\psi \to \gamma\eta\eta$ and confirmed - as $G(1590)$ - by the GAMS collaboration in $\pi^-p \to \eta\eta n$ [2417]. Its quantum numbers shifted from $J^{PC} = 2^{++}$ to $0^{++}$, and its mass changed to 1710 MeV. This resonance still plays an important role in the glueball discussion.

A third candidate, or better three candidates, were observed in the OZI rule violating process $\pi^-p \to \phi\phi n$ [2284, 2418]. Three $\phi\phi$ resonances at 2050, 2300 and 2350 MeV were reported. I remember Armenteros saying: *When you have found one glueball, you have made a discovery. When you find three, you have a problem.* Now I believe that this was a very early manifestation of the tensor glueball.

The situation was not that easy at that time as described here. Nearly for each observation, there were contradicting facts, and Heusch concluded his talk at the QCD workshop with the statement: *there is no smoking-gun candidate for gluonium* $\cdots$. At this workshop, I had the honor to present the results of the Crystal Barrel experiment at LEAR and to report the discovery of two new scalar mesons, $f_0(1370)$ and $f_0(1500)$, and I was convinced, Heusch was wrong: $f_0(1500)$ was the glueball! And I turned down my internal critical voice which told me that in my understanding of $\bar{p}N$ annihilation, this process is not particularly suited to produce glueballs [2419, 2420]. Our glueball $f_0(1500)$ was not seen in radiative $J/\psi$ decays where a glueball should stick out like a tower in the landscape. The $f_0(1500)$ as scalar glueball? That could not be the full truth!

### 8.4.2 QCD predictions

*Glueball masses*

First estimates of the masses of glueballs were based on bag models. The color-carrying gluon fields were required to vanish on the surface of the bag. Transverse electric and transverse magnetic gluons were introduced populating the bag. The lowest excitation modes were predicted to have quantum numbers $J^{PC} = 0^{++}$ and $2^{++}$ and to be degenerate in mass with $M = 960$ MeV [747, 2421]. A very early review can be found in Ref. [2422].

The bag model is obsolete nowadays. Most reliable are presumably simulation of QCD an a lattice (see Section 4 and Ref. [2423] for an introduction). In lattice gauge theory, the spacetime is rotated into an Euclidean space by the transformation $t \to i\,t$ and then discretized into a lattice with sites separated by a distance in space and time. The gauge fields are defined as links between neighboring lattice points, closed loops of the link variables (Wilson loops) allow for the calculation of the action density. Technically, gluons on a space-time lattice struggle against large vacuum fluctuations of the correlation functions of their operators, the signal-to-noise ratio falls extremely rapidly as the separation between the source and sink is increased. These difficulties can be overcome by anisotropic space-times with coarser space and narrow time intervals [2424, 2425]. Fermion fields are defined at lattice sites. Different techniques have been developed to include fermions in lattice calculations [2426]. The effect of see quarks on glueball masses seems to be small [2427].

Recently, a number of different approaches were chosen to approximate QCD by a model that is solvable analytically. Szczepaniak and Swanson [2428] constructed a quasi-particle gluon basis for a QCD Hamiltonian in Coulomb gauge that was solved analytically. A full glueball spectrum was calculated with no free parameter. The authors of Ref. [2429] constructed relativistic two- and three-gluon glueball currents and applied them to perform QCD sum rule analyses of the glueball spectrum. The Gießen group calculated masses of ground and excited glueball states using a Yang-Mills theory and a functional approach based on a truncation of Dyson-Schwinger equations and a set of Bethe-Salpeter equations derived from a three-particle-irreducible effective action [2430, 2431].

AdS/QCD relies on a correspondence between a five dimensional classical theory with an AdS metric and a supersymmetric conformal quantum field theory in four dimensions. In the bottom-up approach, models with appropiate operators are constructed in the classical AdS theory with the aim of resembling QCD as much as possible. Confinement is generated by a hard wall cutting off AdS space in the infrared region, or spacetime is capped off smoothly by a soft wall to break the conformal invariance. Rinaldi and Vento [1107] calculated the glueball mass spectrum within AdS/QCD. The results on glueball masses are summarized in Table 8.4.1.

**Table 8.4.1** Masses of low-mass glueballs, in units of MeV. Lattice QCD results are taken from Refs. [2424, 2426] (quenched) and Ref. [2427] (unquenched). Szczepaniak and Swanson [2428] construct of a quasiparticle gluon basis for a QCD Hamiltonian. Results from QCD sum rule results are given in Ref. [2429], from using Dyson-Schwinger equations in [2430, 2431], and from a graviton-soft-wall model in Ref. [1107].

| Glueball | Ref. [2424] | Ref. [2426] | Ref. [2427] | Ref. [2428] | Ref. [2429] | Ref. [2430] | Ref. [1107] |
|---|---|---|---|---|---|---|---|
| $\|0^{++}\rangle$ | $1710 \pm 50 \pm 80$ | $1653 \pm 26$ | $1795 \pm 60$ | 1980 | $1780^{+140}_{-170}$ | $1850 \pm 130$ | 1920 |
| $\|2^{++}\rangle$ | $2390 \pm 30 \pm 120$ | $2376 \pm 32$ | $2620 \pm 50$ | 2420 | $1860^{+140}_{-170}$ | $2610 \pm 180$ | 2371 |
| $\|0^{-+}\rangle$ | $2560 \pm 40 \pm 120$ | $2561 \pm 40$ | – | 2220 | $2170 \pm 110$ | $2580 \pm 180$ | |

*The width of glueballs*
Glueballs are often assumed to be narrow. $\phi$ decays into $\rho\pi$ are suppressed since the primary $s\bar{s}$ pair needs to annihilate and a new $q\bar{q}$ pair needs to be created. In glueball decays, there is no pair to be annihilated but a $q\bar{q}$ pair needs to be created. If the OZI rule suppresses the decay by a factor 10 to 100, we might expect the width of glueballs to be suppressed by a factor 3 to 10. Assuming a "normal" width of 150 MeV, a glueball at 1600 MeV could have a width of 15 to 50 MeV. This argument is supported by arguments based on the $1/N_c$ expansion (see, e.g., Ref. [2338]).

Narison applied QCD sum rules [2432]. Assuming a mass of 1600 MeV, he calculated the $4\pi$ width of the scalar glueball to 60 to 138 MeV while the partial decay width of the tensor glueball at 2 GeV to pseudoscalar mesons should be less than 155 MeV. Calculations on the lattice gave a partial decay width for decays into pseudoscalar mesons of 108±29 MeV for a scalar glueball mass of 1700 MeV [2433]. In a semi-phenomenological model, Burakovsky and Page find that the width of the scalar glueball (at 1700 MeV) should exceed 250 to 390 MeV. A flux tube model predicted the mass of the glueball of lowest mass to 1680 MeV and its width to 300 MeV [2434]. In a field theoretical approach with an effective Coulomb gauge the glueball width was estimated to 100 MeV [2435].

*Radiative yields*
The study of radiative decays of the $J/\psi$ meson is the prime path to search for glueballs with masses of less than ~2500 MeV.

Gui *et al.* [2436] calculated the yield of a scalar glueball having a mass of 1710 MeV on lattice and found

$$BR_{J/\psi\to\gamma G_{0^{++}}}(TH) = (3.8 \pm 0.9)\cdot 10^{-3}. \quad (8.4.1)$$

For higher glueball masses the yield increases.

Narison gave a mass dependent formula derived from sum rules. For a mass of 1865 MeV, a yield of about $10^{-3}$ is predicted [2432].

The tensor glueball is expected [2437] to be observed with a branching ratio

$$BR_{J/\psi\to\gamma G_{2^{++}}}(TH) = (11 \pm 2)\cdot 10^{-3}. \quad (8.4.2)$$

Production of the pseudoscalar glueball seems to be considerably smaller. For a mass of 2395 (or 2560) MeV, the authors of Ref. [2438] find

$$BR_{J/\psi\to\gamma G_{0^{-+}}}(TH) = (0.231 \pm 0.080)\cdot 10^{-3}$$
$$\text{or} = (0.107 \pm 0.037)\cdot 10^{-3}. \quad (8.4.3)$$

These are very significant yields, and the glueballs must be found provided they can be identified convincingly as glueballs amidst their $q\bar{q}$ companions.

### 8.4.3 How to identify a glueball

Figure 8.4.1 shows the prime reactions in which glueballs have been searched for.

***$N\bar{N}$* annihilation**
A decisive step forward in the search for glueballs was the discovery of two scalar isoscalar states in $\bar{p}p$ annihilation at rest. With the large statistics available at the Low Energy Anitiproton Ring (LEAR) at CERN, $f_0(1370)$ and $f_0(1500)$ were identified in several final states. A large fraction of the data taken at LEAR is still used jointly with data on radiative $J/\psi$ decays in a coupled-channel analysis. Glueballs decay via $q\bar{q}$ pair creation. Hence they can be produced via $q\bar{q}$ annihilation. Meson production in $\bar{p}p$ annihilation was studied by the ASTERIX, OBELIX and Crystal Barrel experiments at LEAR and is a major objective of the PANDA collaboration at the GSI.

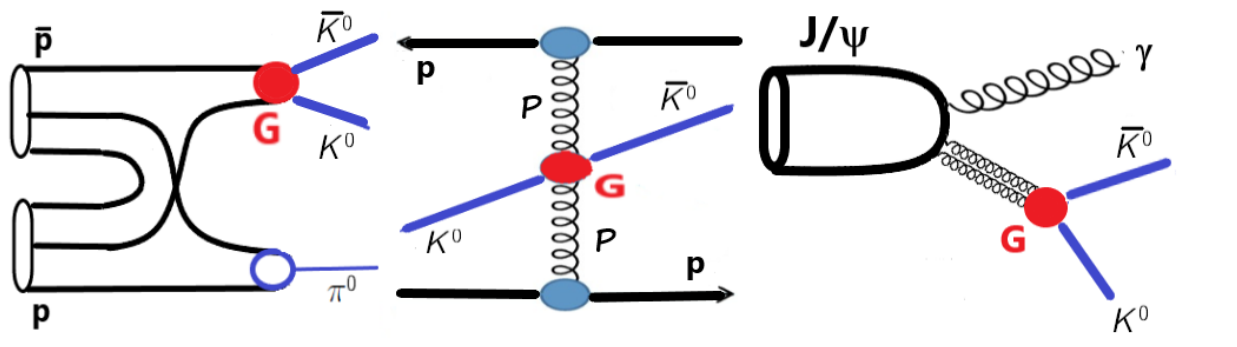

**Fig. 8.4.1** Reactions most relevant for glueball searches. Left: $\bar{p}p$ annihilation; middle: Pomeron-Pomeron fusion; right: radiative $J/\psi$ decays. The glueball is supposed to decay into $K^0\bar{K}^0$.

*Central production*
In central production, two hadrons (e.g. two protons) scatter in forward direction via the exchange of Pomerons. Pomerons are supposed to be glue-rich, hence glueballs can be formed in Pomeron-Pomeron fusion. This process was studied extensively at CERN by the WA76 and WA102 collaborations and is now investigated with the STAR detector at BNL. In the WA102 experiment, $f_0(1370)$ and $f_0(1500)$ were confirmed and $f_0(1710)$ was added to the number of scalar resonances.

*radiative $J/\psi$ decays*
In radiative $J/\psi$ decays, the primary $c\bar{c}$ pair converts into two gluons and a photon. The two gluons are mainly produced in $S$-wave, the two gluons can form scalar and tensor glueballs which should be produced abundantely. The large statistics available from BESIII at Beijing makes this reaction the most favorable one for glueball searches. Radiative decays of heavy mesons is the only process for which glueball yields have been calculated. The data will be discussed below in more detail.

*Decay analysis*
The decay of mesons into two pseudoscalar mesons is governed by $\mathrm{SU}(3)_F$. In a meson nonet, there are two isoscalar mesons, one lower in mass the other one higher, which both contain a $n\bar{n} = (u\bar{u} + d\bar{d})/\sqrt{2}$ and a $s\bar{s}$ component and are mixed with the mixing angle $\varphi$. Figure 8.4.2 shows the $\mathrm{SU}(3)_F$ squared matrix elements for meson decays into two pseudoscalar mesons as a function of the scalar mixing angle.

$$\begin{pmatrix} f^H \\ f^L \end{pmatrix} = \begin{pmatrix} \cos\varphi^s & -\sin\varphi^s \\ \sin\varphi^s & \cos\varphi^s \end{pmatrix} \begin{pmatrix} |n\bar{n}> \\ |s\bar{s}> \end{pmatrix} \quad (6)$$

*Supernumery*
The three scalar isoscalar mesons $f_0(1370)$, $f_0(1500)$ and $f_0(1710)$ played an important role in the glueball discussion. Amsler and Close [2440, 2441] suggested to interpret these three states as the result of mixing of the two expected isoscalar states with the scalar glueball.

$$\begin{pmatrix} f_0(1370) \\ f_0(1500) \\ f_0(1710) \end{pmatrix} = \begin{pmatrix} x_{11} & x_{12} & x_{13} \\ x_{21} & x_{22} & x_{23} \\ x_{31} & x_{32} & x_{33} \end{pmatrix} \begin{pmatrix} |n\bar{n}\rangle \\ |s\bar{s}\rangle \\ |gg\rangle \end{pmatrix} \quad (8.4.7)$$

These papers led to a large number of follow-up papers, references can be found in Ref. [2439]. In all these papers, these three mesons contain the full glueball, $\sum_j x_{ij}^2 = 1$ is imposed. Note that the squared mass difference between $f_0(1370)$ and $f_0(1710)$ is slightly above $1\,\mathrm{GeV}^2$, the $f_0(1710)$ could also be a radial excitation (and is interpreted as radial excitation below).

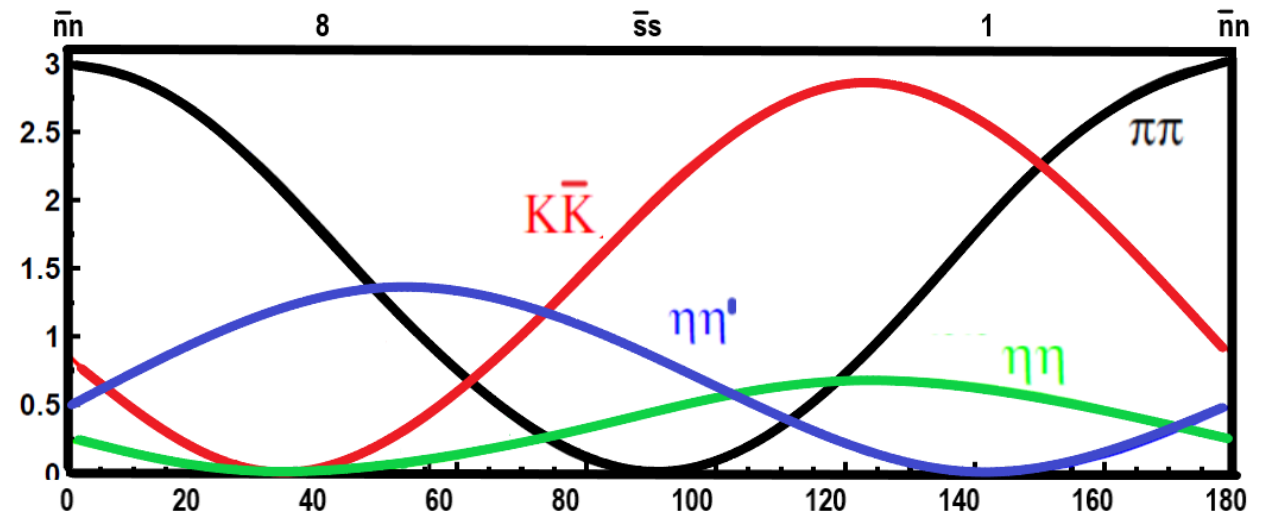

**Fig. 8.4.2** Decay probabilities of mesons for decays into two pseudoscalar mesons as a function of the scalar mixing angle [2439].

*Conclusions*
Identifying a glueball is a difficult task. The main argument in favor of a glueball interpretation is an anomalously large production rate in $J/\psi$ decays. It turns out that scalar mesons are organized like pseudoscalar mesons, into mainly singlet and mainly octet mesons. A large production rate of a mainly-octet scalar isoscalar meson in radiative $J/\psi$ decays directly points to a significant glueball content in its wave function. A second argument relies an meson decays into pseudoscalar mesons. In presence of a glueball, a pair of mesons assigned to the same multiplet should have a decay pattern that is incompatible with a $q\bar{q}$ interpretation for any mixing angle. Supernumery is a weak argument. It requires a full understanding of the regular excitation spectrum. Further studies are required to learn if central production is gluon-rich. The large production rates of $f_0(1500)$, $f_0(1710)$ and $f_0(2100)$ in $\bar{p}p$ annihilation at collision energies above 3 GeV encourages glueball searches at the FAIR facility with the PANDA detector (see Section 14.5).

### 8.4.4 Evidence for glueballs from coupled-channel analysis

We have performed a coupled-channel partial wave analysis of radiative $J/\psi$ decays into $\pi^0\pi^0$, $K_s^0K_s^0$, $\eta\eta$, and $\omega\phi$, constrained by the CERN-Munich data on $\pi N$ scattering, data from the GAMS collaboration at CERN, data from BNL on $\pi\pi \to K_s^0K_s^0$, and 15 Dalitz plots on $\bar{p}p$ annihilation at rest from LEAR. Data on $K_{e4}$ decays constrain the low-energy region. Fitting details and references to the data can be found in Ref. [2442] [89] Figure 8.4.3 shows the data on radiative $J/\psi$ decays

[89] The BESIII data were fitted by Rodas *et al.* [2443] with four scalar and three tensor resonances only. I have several objections against the fit. i) It uses an amplitude in which the $J/\psi$ converts into three gluons which hadronize. A final-state meson radiates off the photon. Since the photon is not an isospin

into $\pi^0\pi^0$, $K_s^0K_s^0$ and the fit. Ten scalar isoscalar resonances were included in the fit. Oller [2354] has shown that $f_0(500)$ is singlet-like, the $f_0(980)$ octet-like (see also [2444]). The $f_0(1500)$ is seen in Figure 8.4.3 as a dip. This pattern was reproduced in Ref. [2442] assuming that $f_0(1370)$ is a singlet state and $f_0(1500)$ an octet state. Hence we assumed that the ten mesons can be divided into two series of states, mainly-singlet states with lower masses and mainly-octet states with higher masses.

In a $(M^2, n)$ plot, the masses of singlet and octet states follow two linear trajectories (see Fig. 8.4.4). Remarkably, the slope ($1.1\,\mathrm{GeV}^{-2}$) is close to the slope of standard Regge trajectories. The separation between the two trajectories is given by the mass square difference between $\eta'$ and $\eta$-meson as suggested by instanton-induced interactions [2448]. The figure includes a meson reported by the BESIII collaboration studying $J/\psi \to \gamma\eta'\eta'$ [2447]. As $\eta'\eta'$ resonance, $f_0(2480)$ is very likely a SU(3) singlet state. Indeed, its mass is compatible with the "mainly-singlet" trajectory. The figure gives the pole positions of the eleven resonances as small inserts.

The total yields of scalar mesons in radiative $J/\psi$ decays - including decay modes not reported by the BESIII collaboration - was determined from the coupled-channel analysis [2442] that included also other data. The yield of mainly-octet and mainly-singlet mesons as a function of their mass is shown in Fig. 8.4.5. Mainly-octet mesons should not be produced (or at most weakly) in radiative $J/\psi$ decays. However, they are produced abundantly, in a limited mass range centered at about 1865 MeV. Mainly-singlet mesons are produced over the full mass range but show a peak structure at the same mass. This enhancement must be due to the scalar glueball mixing into the wave functions of scalar mainly-octet and mainly-singlet mesons. A Breit-Wigner fit to these distributions gives mass and width

$$M_G = (1865 \pm 25\,^{+10}_{-30})\,\mathrm{MeV} \qquad \Gamma_G = (370 \pm 50\,^{+30}_{-20})\,\mathrm{MeV}\,, \tag{8.4.8}$$

and the (observed) yield is determined to

$$Y_{J/\psi\to\gamma G} = (5.8 \pm 1.0)\,10^{-3}\,. \tag{8.4.9}$$

eigenstate, this amplitude can produce isovector mesons. This process is highly suppressed and experimentally absent. ii) the $f_0(1370) - f_0(1500)$ interference region is not well described, neither in the mass distribution nor in the $S-D$ phase difference. iii) The fit is neither constrained by the $\pi\pi$ $S$-wave from the CERN-Munich data nor by the data on $K_{e4}$ decays. A fit with the seven resonances used in Ref. [2443] without an isospin breaking amplitude leads to a $\pi\pi$ $S$-wave that is extremely incompatible with the known $\pi\pi$ $S$-wave (A.V. Sarantsev, private communication, October 2021.)

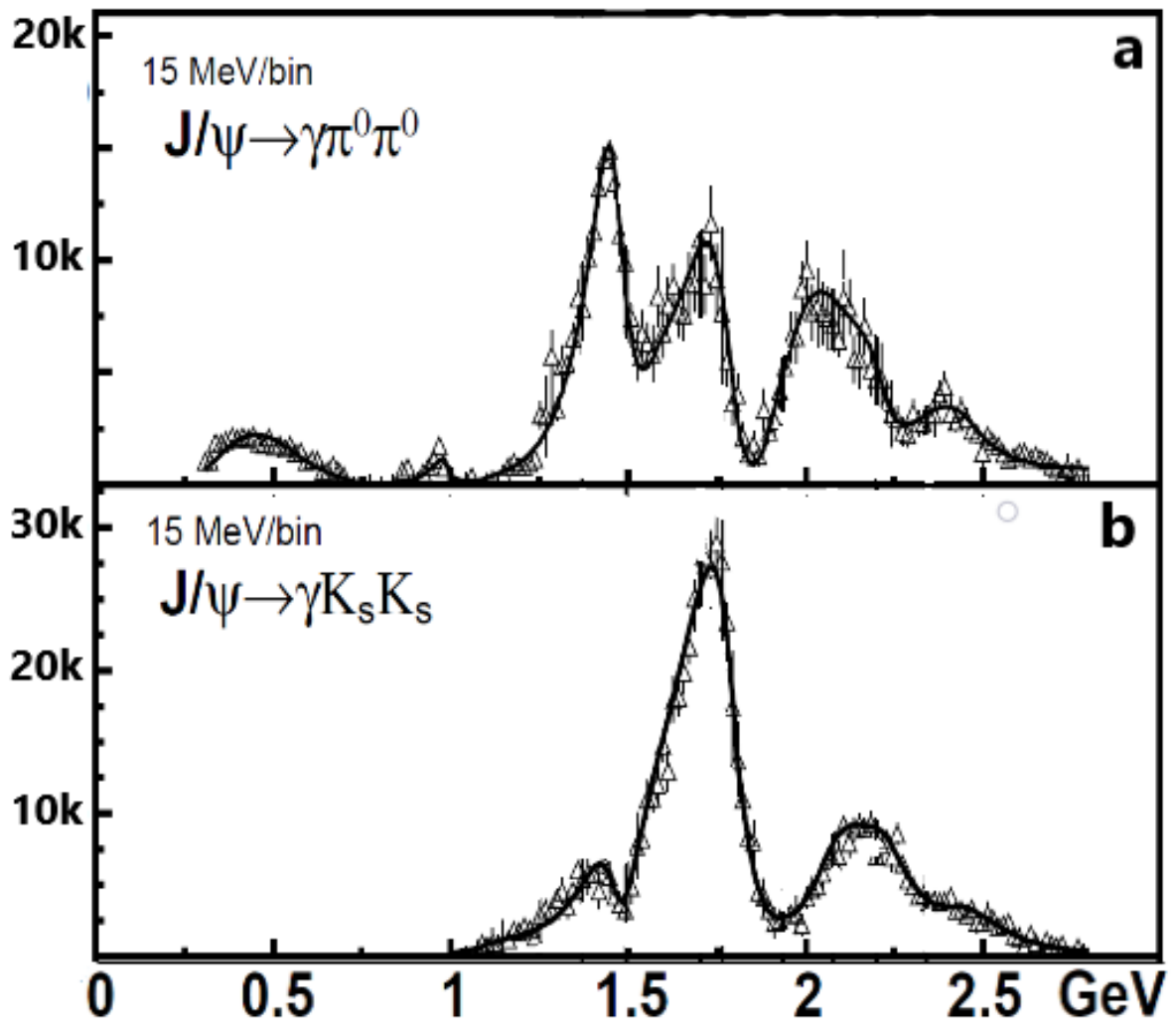

**Fig. 8.4.3** The squared $S$-wave tranisition amplitudes for $J/\psi \to \pi^0\pi^0$ (a) and $J/\psi \to K_s^0K_s^0$ (b). The data points are from an energy-independent partial-wave analysis [2445, 2446], the curve represents our fit [2442].

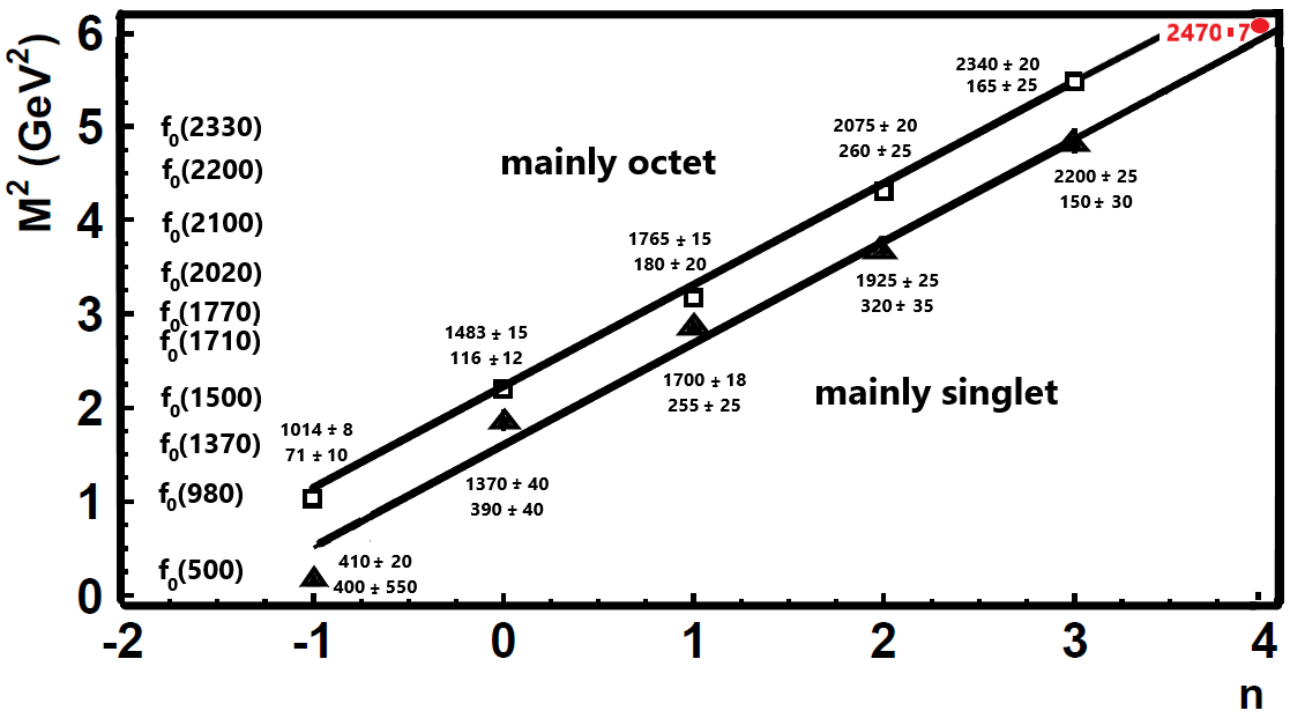

**Fig. 8.4.4** $M^2, n$ trajectories for mainly-singlet and mainly-octet scalar isoscalar resonances. The red dot at high masses represents a scalar state from $J/\psi \to \gamma\eta'\eta'$ [2447]. Adapted from Ref. [2442].

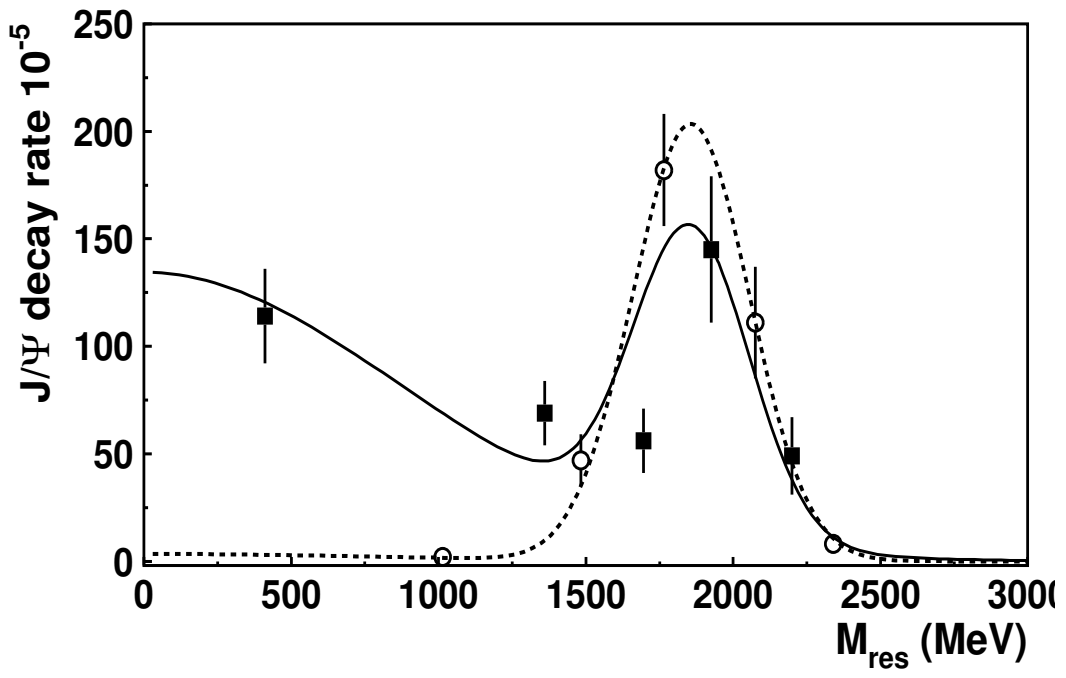

**Fig. 8.4.5** Yield of scalar isoscalar mesons in radiative $J/\psi$ decays into mainly-octet (open circles) and mainly-singlet mesons (full squares) as a function of their mass [2442].

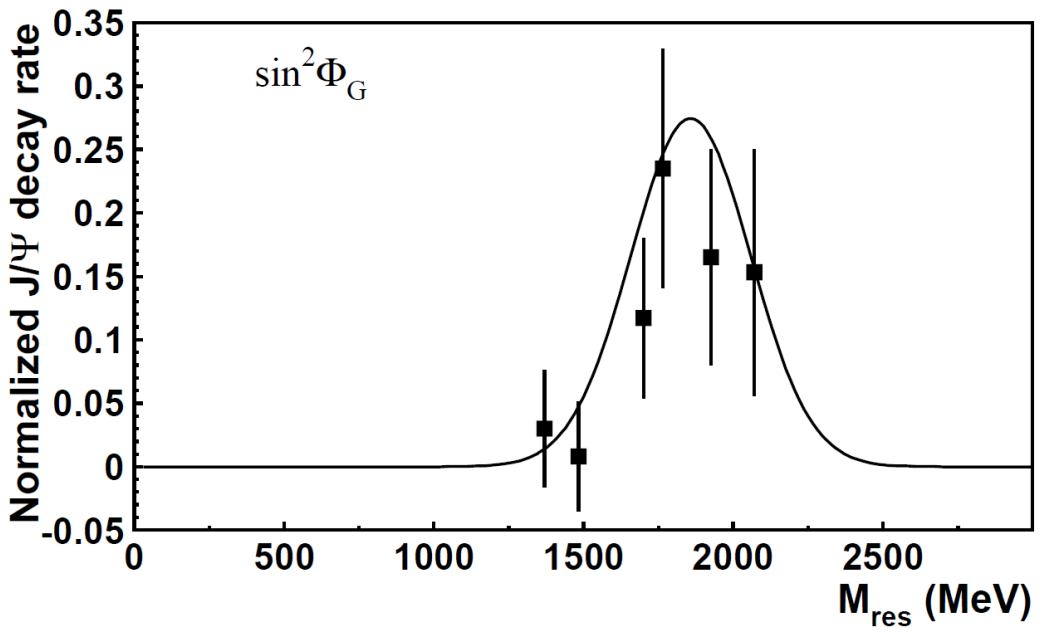

**Fig. 8.4.6** The glueball content of scalar mesons. Black squares: $\sin^2\varphi_{\rm n}^{\rm s}$, solid curve: Breit-Wigner resonance with area 1 [2439].

### 8.4.5 Meson-glueball mixing

Earlier attempts to identify the glueball have in common that the full glueball is distributed among the three states $f_0(1370)$, $f_0(1500)$ and $f_0(1710)$. Inspecting Fig. 8.4.3, this seems not to be obvious: Above 1 GeV, four peaks with three valleys are seen, and there is no reason why one particular region should be more gluish than the other ones. The yield of scalar mesons sees the glueball contribution distributed over several resonances.

We did not impose that the full glueball should be seen in these three states nor that we must see the full glueball at all. We fitted the decay modes of pairs of scalar mesons, one mainly-singlet one mainly-octet, and allowed for a glueball component [2439].

$$f_0^{\rm nH}(xxx) = (n\bar n\cos\varphi_{\rm n}^{\rm s} - s\bar s\sin\varphi_{\rm n}^{\rm s})\cos\phi_{\rm nH}^G + G\sin\phi_{\rm nH}^G$$
$$f_0^{\rm nL}(xxx) = (n\bar n\sin\varphi_{\rm n}^{\rm s} + s\bar s\cos\varphi_{\rm n}^{\rm s})\cos\phi_{\rm nL}^G + G\sin\phi_{\rm nL}^G \tag{8.4.10}$$

$\varphi_{\rm n}^{\rm s}$ is the scalar mixing angle, $\phi_{\rm nH}^G$ and $\phi_{\rm nL}^G$ are the meson-glueball mixing angles of the high-mass state H and of the low-mass state L in the nth nonet. The fractional glueball content of a meson is given by $\sin^2\phi_{\rm nH}^G$ or $\sin^2\phi_{\rm nL}^G$.

With this mixing scheme and the SU(3) coupling constant (see Fig. 8.4.2), we have fitted the meson decay modes and have thus determined the glueball content of the eight high-mass scalar mesons. Figure 8.4.6 shows the glueball fraction in the scalar mesons.

The glueball fractions derived from the decay analysis of pairs of scalar mesons add up to a sum that is compatible with 1. The distribution of the glueball fraction in Fig. 8.4.6 is identical to the distribution of yields in Fig. 8.4.5. This is a remarkable confirmation that the scalar glueball of lowest mass has been identified and has mass and width as given in Eqn. (8.4.8) and a yield as given in Eqn. (8.4.9).

### 8.4.6 Comparison with LHCb data

Most striking is the mountain landscape above 1500 MeV in the data on radiative $J/\psi$ decays. In these decays a $c\bar c$ pair converts into gluons which hadronize (see Fig. 8.4.7, left). The huge peak in the $K\bar K$ mass spectrum at 1750 MeV and the smaller one at 2100 MeV decay are produced with two gluons in the initial state. This is to be contrasted with data on $B_s^0$ and $\bar B_s^0$ decays into $J/\psi{+}\pi^+\pi^-$ [2449] and $K\bar K$ [2450]. In this reaction, a primary $s\bar s$ pair – recoiling against the $J/\psi$ – converts into the final state mesons (see Fig. 8.4.7, right). We have included the spherical harmonic moments in the coupled channel analysis that describes the radiative $J/\psi$ decays [2451]. High-mass scalar mesons are only weakly produced in $B_s^0$ decays with $s\bar s$ in the initial state. The strong peak in the $K\bar K$ invariant mass at 1750 MeV in Fig. 8.4.3 is nearly absent in $B_s^0 \to J/\psi\, K\bar K$!

Figure 8.4.8 shows the ratio of the decay frequencies of $J/\psi\to\gamma\, f_0$ and $B_s^0\to J/\psi\, f_0$ with $f_0$ decaying into $\pi\pi$ or $K\bar K$. The $f_0(980)$ is likely a mainly-octet state, little produced in radiative $J/\psi$ decays but strongly with $s\bar s$ in the initial state. On the contrary, $f_0(1770)$ is seen as strong peak in radiative $J/\psi$ but very weakly only in $B_s^0$ decays. The uncertainties are large, but the ratio of the decay frequencies is fully compatible with the shape of the glueball derived above.

This is highly remarkable: the two gluons in the initial state must be responsible for the production of resonances that decay strongly into $K\bar K$ but are nearly absent when $s\bar s$ pairs are in the initial state. Also, there is a rich structure in the $\pi\pi$ mass spectrum produced in radiative $J/\psi$ decays but little activity only when the initial state is an $s\bar s$ pair: The rich structure stems from gluon-gluon dynamics. Similar conclusions can be drawn [2444] from a comparison of the invariant mass distributions from radiative $J/\psi$ decays with the pion and kaon form factors [2452]. Their square is proportional to the cross sections. The $f_0(980)$ resonance dominates both formfactors but is nearly absent in radiative $J/\psi$ decays: The $f_0(980)$ has large $n\bar n$ and $s\bar s$ components mixed to a dominant SU(3) octet component. The large intensity above 1500 MeV in radiative $J/\psi$ decays is absent when not two gluons but an $s\bar{s}$ pair is in the initial state: the mountainous structure in radiative $J/\psi$ decays is produced by gluons and not by $q\bar{q}$ pairs: The structure is due to the scalar glueball.

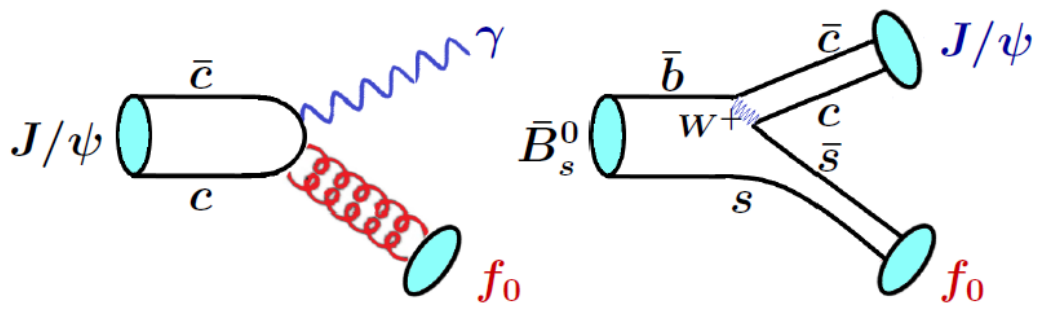

**Fig. 8.4.7** In radiative $J/\psi$ decays two gluons, in $\bar B_s^0\to J/\psi+$ $s\bar s$, a $s\bar s$ pair may convert into a scalar meson.

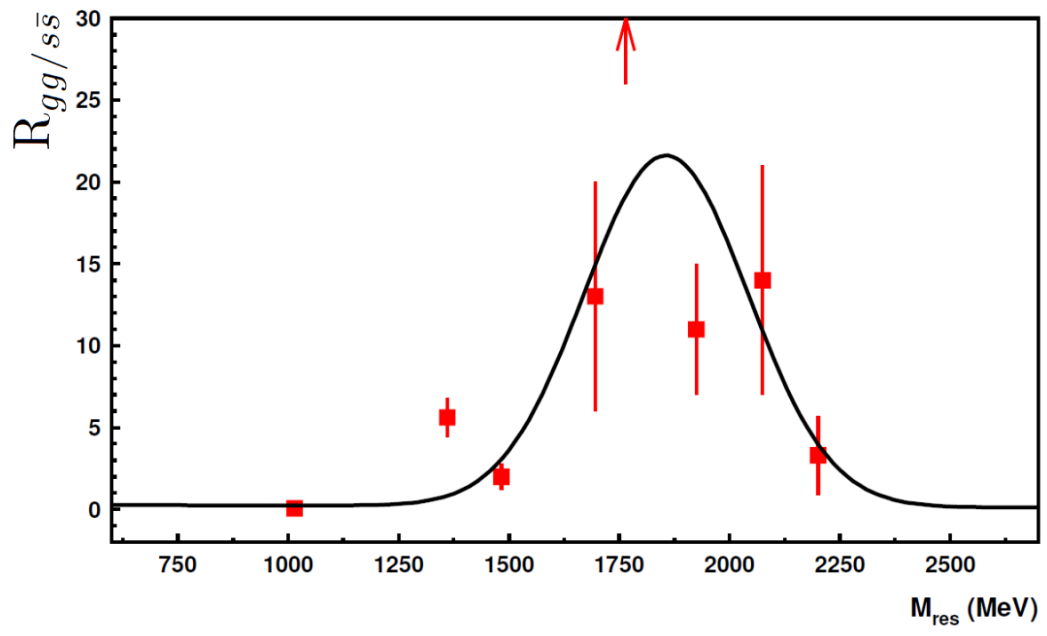

**Fig. 8.4.8** The ratio $R_{gg/s\bar{s}}$ of the frequencies for $J/\psi \to \gamma\, f_0$ and $B_s^0 \to J/\psi\, f_0$ with $f_0$ decaying into $\pi\pi$ or $K\bar{K}$.

### 8.4.7 A trace of the tensor glueball

The tensor glueball is predicted with an even higher yield [2437]:

$$\Gamma_{J/\psi\to\gamma/G_{2++}}/\Gamma_{\rm tot} = (11 \pm 2)10^{-3}\,. \qquad (8.4.11)$$

The yield of $f_2(1270)$ in radiative $J/\psi$ decays is $(1.64\pm 0.12)10^{-3}$, about six times weaker than the predicted rate for the tensor glueball! Bose symmetry implies that the $\pi^0\pi^0$ or $K_sK_s$ pairs are limited to even angular momenta, practically, only $S$ and $D$-waves contribute. The scalar intensity originates from the electric dipole transition $E0$. Three electromagnetic amplitudes $E1, M2$, and $E3$ excite tensor mesons. Figure 8.4.9 shows these three amplitudes and the relative phases.

Two fits were performed [2453]. One fit describes the mass distribution only. Apart from the well known $f_2(1270)$ and $f_2'(1525)$ the fit needs one high-mass resonance with

$$M = (2210\pm 60)\,\text{MeV};\quad \Gamma = (360\pm 120)\,\text{MeV}\,, \quad (8.4.12)$$

where the error includes systematic studies with or without additional low-yield resonances. The enhancement was called $X_2(2210)$. In this fit, the phases are not well described. Figure 8.4.9 shows a fit in which the 2200 MeV region is described by three tensor resonances with masses and widths of about $(M,\Gamma) = (2010, 200)$, $(2300, 150)$, and $(2340, 320)$ MeV. These resonances had been observed by Etkin *et al.* [2284] in the reaction $\pi^-p \to \phi\phi n$. The unusual production characteristics were interpreted in Ref. [2284] as evidence that *these states are produced by* $1-3$ *glueballs.*

The total observed yield of $X_2(2210)$ in $\pi\pi$ and $K\bar{K}$ is $(0.35 \pm 0.15)\,10^{-3}$, far below the expected glueball yield. We assume the glueball is – like the scalar glueball

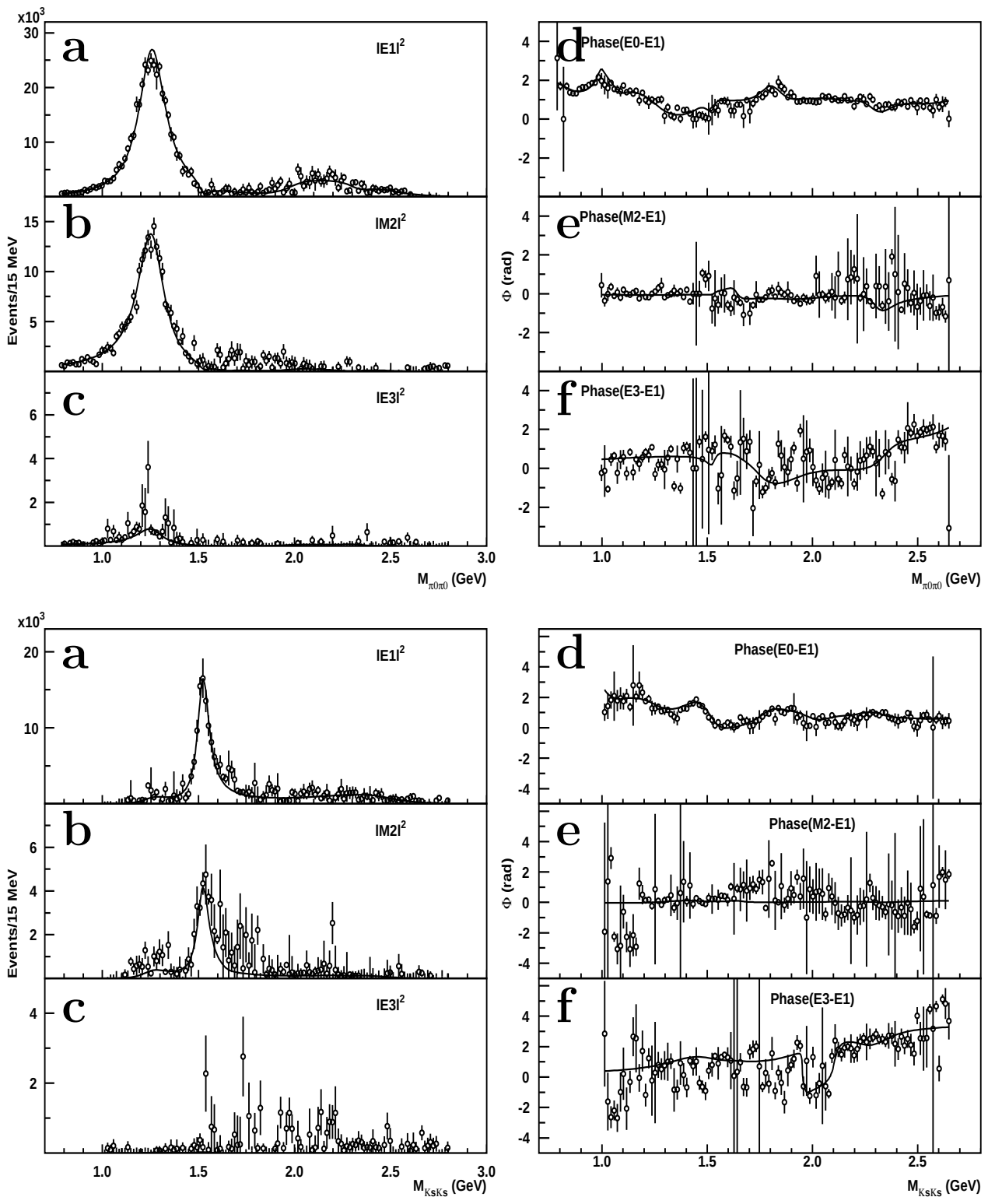

**Fig. 8.4.9** $D$-wave intensities and phases for radiative $J/\psi$ decays into $\pi^0\pi^0$ (top subfigures) and $K_s\,K_s$ (bottom subfigures) from Ref. [2445, 2446]. The subfigures show the $E1$ (a), $M2$ (b) and $E3$ (c) squared amplitudes and the phase differences between the $E0$ and $E1$ (d) amplitudes, the $M2$ and $E1$ (e) amplitudes, and the $E3$ and $E1$ (f) amplitudes as functions of the meson-meson invariant mass. The phase of the $E0$ amplitude is set to zero. The curve represents our best fit.

– distruted over several tensor mesons. Adding up all contributions from tensor states above 1900 MeV seen in radiative $J/\psi$ decays, one obtains

$$\sum_{M=1.9\,\text{GeV}}^{M=2.5\,\text{GeV}} Y_{J/\psi\to\gamma f_2} = (3.1 \pm 0.6)\,10^{-3}\,, \qquad (8.4.13)$$

which is a large yield even though still below the predicted yield.

### 8.4.8 How to find the pseudoscalar glueball

The BESIII collaboration has studied the reaction $J/\psi \to \pi^+\pi^-\eta'$ [2454]. The top panel of Fig. 8.4.10 shows the $\pi^+\pi^-\eta'$ invariant mass distributions with a series of peaks. Assuming that these are all pseudoscalar mesons, two trajectories can be drawn (bottom panel of Fig. 8.4.10). The figure suggests that the higher-mass structures could house two mesons, possibly singlet and octet states in SU(3). If this is true, a cut in the $\pi^+\pi^-$ invariant mass at about 1480 MeV would partly separate the two isobars, $X(2600) \to f_0(1370)\eta'$ and $X(2600) \to f_0(1500)\eta'$. We may expect a slight mass shift in the two $\pi^+\pi^-\eta'$ invariant mass distributions. The two mesons $f_0(1370)$ and $\eta'$ are both mainly singlet. The $f_0(1370)\eta'$ isobar as singlet meson in the $X(2600)$ complex should be slightly higher in mass than the $f_0(1500)\eta'$ mainly octet meson.

The total yields of the high-mass structures – including unseen decay modes – are not known. Nevertheless, their appearance above a comparatively low background is surprising. Personally, I suppose that the pseudoscalar glueball is rather wide, and that the structures are seen so clearly because of a small glueball content. More studies of theses data and of different channels are required to substantiate this conjecture.

### 8.4.9 Outlook

The data of the BESIII collaboration presented above are based on $1.3\cdot 10^9$ events taken at the $J/\psi$. Presently available are $10^{10}$ events. Based on this large statistics, rare radiative decays like $J/\psi \to \gamma\eta\eta'$ [2409, 2410] and $J/\psi \to \gamma\eta'\eta'$ [2447] have been analysed. Data on the different charge mode of $J/\psi \to \gamma 4\pi$ would be extremely important. In an ideal world, these data would be publicly available after publication and would be included in different coupled-channel partial-wave analyses.

Radiative decays of $\psi(2S)$ and of $\Upsilon(1S)$ open a wider range in invariant mass. The authors of Ref. [2455] used the data of the CLEO collaboration on radiative $\psi(2S)$ decays into $\pi^+\pi^-$ and $K^+K^-$. The data are shown in Fig. 8.4.11. The data are fit with known resonances, no partial-wave analysis was performed. The BaBar collaboration studied radiative $\Upsilon(1S)$ decays into $\pi^+\pi^-$ and $K^+K^-$ [2456]. The results are shown in Fig. 8.4.12. In all four distributions, there is not a single prominent peak in the $S$-wave contribution which would stick out as glueball candidate. The $S$-waves rather resembles the distributions observed in radiative $J/\psi$: three major enhancement in the 1500, 1750 and 2200 MeV region separated by dips. (With the larger statistics in $J/\psi$ decays, a fourth enhancement is seen at about 2350 MeV.) In Fig. 8.4.11, a peak is found at 1447 MeV and assigned to $f_0(1500)$. At 1500 MeV, there is the dip. The wrong mass is due to the neglect of interference: The phase between $f_0(1500)$ and the "background" (due to the wider $f_0(1370)$) is 180° [2442]. This phase difference and the significant $f_0(1500) \to \eta\eta'$ branching ratio identify $f_0(1500)$ as mainly $SU(3)_F$ octet state. The different masses for the high-mass state in the $\pi^+\pi^-$ and $K^+K^-$ invariant mass distributions point again to the neglect of interference between the prominent octet states and the singlet "background". Inspecting Figs. 8.4.11 and 8.4.12 shows: there is no striking isolated peak which could be interpreted as "the glueball". The glueball content must be distributed over a large number of states.

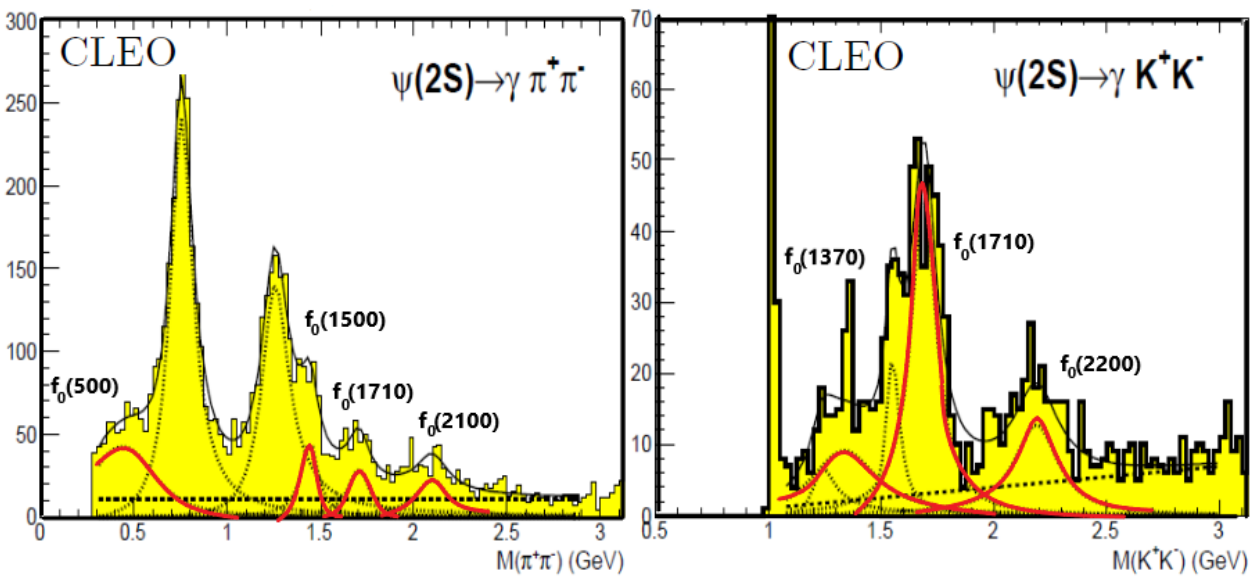

**Fig. 8.4.11** $\pi^+\pi^-$ (left) and $K^+K^-$ (right) invariant mass distributions from radiative deacys of $\psi(2S)$. The red curves represent the $S$-wave contributions. Adapted from [2455].

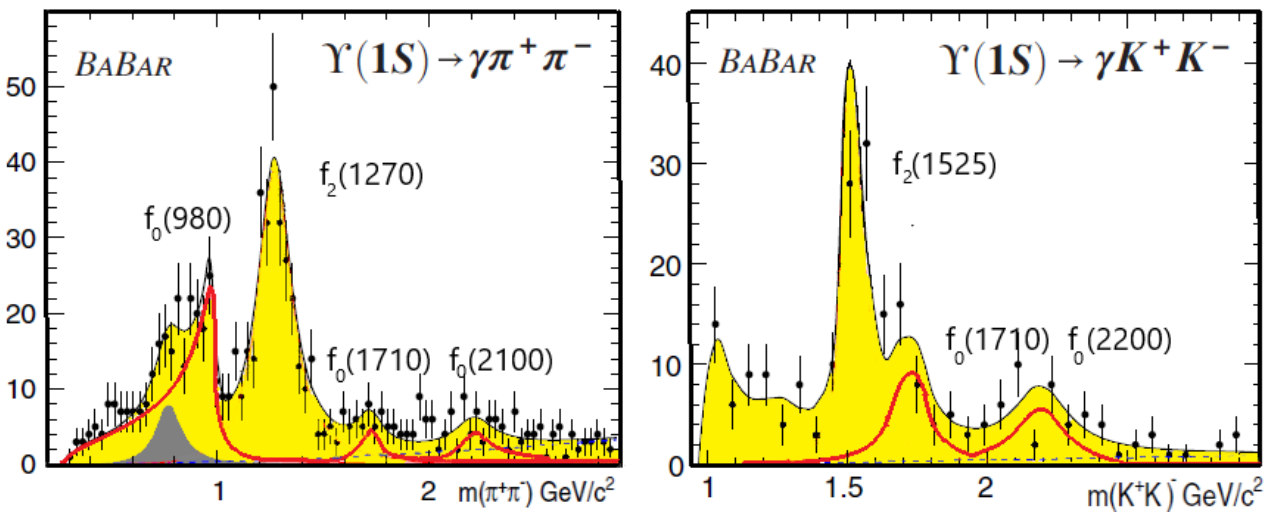

**Fig. 8.4.12** $\pi^+\pi^-$ (left) and $K^+K^-$ (right) invariant mass distributions from radiative deacys of $\Upsilon(1S)$. The $\Upsilon(1S)$ is observed in $\Upsilon(2S)/\Upsilon(3S) \to \pi^+\pi^-\Upsilon(1S)$. The red curves represent the $S$-wave contributions, the grey area the $\rho(770)$ contribution. Adapted from [2456].

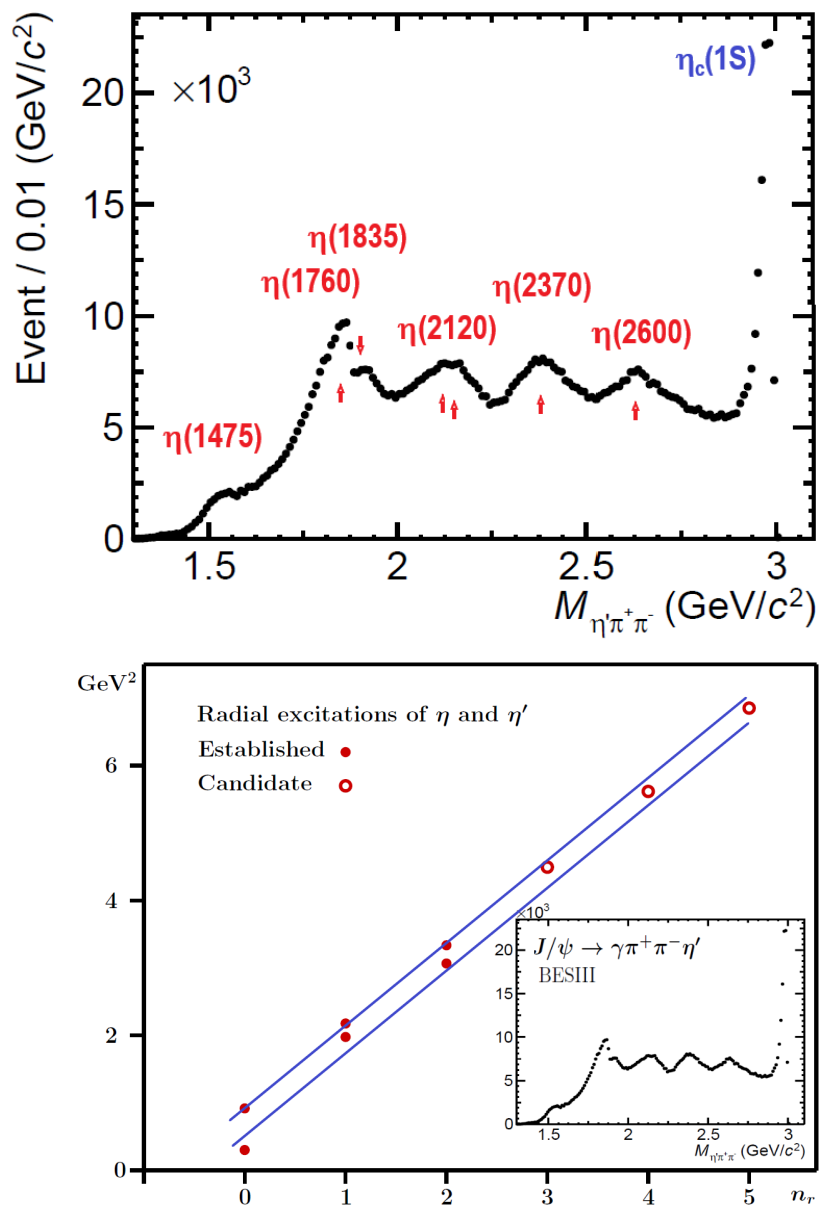

**Fig. 8.4.10** Top: The $\pi^+\pi^-\eta'$ mass distribution from radiative $J/\psi$ decays [2454]. The quantum numbers are not known. Bottom: $M^2$ versus n trajectories.

**Table 8.4.2** Radiative yields expected for $\psi(2S)$ and $\Upsilon(1S)$ radiative decays into the scalar glueball.

| | "Exp." | Theory | Ref. |
|---|---|---|---|
| $\psi(2S) \to \gamma G_0(1865)$ | $\sim 5 \cdot 10^{-4}$ | $(5.9^{+3.4}_{-1.4}) \cdot 10^{-4}$ | [2457] |
| $\Upsilon(1S) \to \gamma G_0(1865)$ | $\sim 3 \cdot 10^{-4}$ | $(1.3^{+0.7}_{-0.3}) \cdot 10^{-4}$ | [2457] |
| | | $(1-2) \cdot 10^{-3}$ | [2458] |

In $\psi(2S)$ radiative decays, the $f_0(1710) \to K\bar{K}$ is observed with a branching fraction of $(6.7\pm0.9)\cdot10^{-5}$, in $\Upsilon(1S)$ radiative decays, the $f_0(1710) \to K^+K^-$ is seen with a branching ratio of $(2.02\pm0.51\pm0.35)\cdot10^{-5}$. The comparison with the yield observed in Ref. [2442] allows us to calculated the branching ratio expected for $\psi(2S)$ and $\Upsilon(1S)$ decays when the full scalar glueball is covered, i.e. for $\Upsilon(1S) \to \gamma G_0(1865)$. The values are given in Table 8.4.2.

Clearly, a significant increase in statistics is required when these reactions should make in independent impact. The advantage of $\psi(2S)$ and $\Upsilon(1S)$ radiative decays is of course that phase space limitations play no role any more. This is particularly important for the search for the tensor and pseudoscalar glueball. The scalar glueball seems to be confirmed: there is not much intensity above 2500 MeV.

At the end I would like to give an answer to the question posed in the title: yes, I am convinced, the scalar glueball is identified, and the tensor glueball seems to have left first traces in the data.

## 8.5 Heavy quark-antiquark sector: experiment

**Marco Pappagallo**

### 8.5.1 Introduction

The term "quarkonium" is a collective name to denote heavy quark-antiquark bound states $Q\bar{Q}'$ ($Q, Q' = c, b$) where the masses of heavy (anti-)quarks are much larger than $\Lambda_{\mathrm{QCD}}$, the scale of non-perturbative physics. Therefore the velocities of the heavy (anti-)quark in quarkonium systems are small and a nonrelativistic potential between the heavy quark-antiquark can be employed to predict the properties of the quarkonium states. The spectra of the charmonium and bottomonium states, with quark content $c\bar{c}$ and $b\bar{b}$ respectively, have been extensively studied in the past years. All excited quarkonium states below the open-flavor $D\bar{D}^{(*)}$ or $B\bar{B}^{(*)}$ thresholds were predicted to be narrow. The observation of the $J/\psi$ meson in 1974 and the success, to predict the electromagnetic and hadronic transitions among the narrow quarkonium states, established the potential models as a tool to unravel the complicated QCD dynamics.

Starting from 2003, new states with masses above the $D\bar{D}^{(*)}$ and $B\bar{B}^{(*)}$ thresholds were observed. A common feature is the presence of a heavy quark $Q$ and anti-quark $\bar{Q}$ pair in the decay products. As a consequence, the constituent-quark content of the decaying meson has to include a heavy quark and a heavy antiquark. However, the properties of many of these states did not match to those of any conventional quarkonium state. So, what are they?

In addition to the conventional $q\bar{q}$ mesons and $qqq$ baryons, models based on QCD predict hadrons with different combinations of quarks $q$ and gluons $g$, such as: pentaquarks ($q\bar{q}qqq$), tetraquarks ($q\bar{q}q\bar{q}$), six-quark H-dibaryons ($q\bar{q}q\bar{q}q\bar{q}$), hydrids ($q\bar{q}g$) and glueballs ($ggg$), see Sections 8.3 and 8.4. The existence of such "exotic" hadrons has been debated for several years without reaching a general consensus. In the early 2000s new hadrons with unexpected features were observed, in particular the $D^*_{s0}(2317)^+$ [2459] and $\chi_{c1}(3872)$ [2460] mesons and the $\Theta^+$ baryon [2461]. While the first two candidates are still consistent with being conventional $c\bar{s}$ and $c\bar{c}$ states, the latter one is manifestly exotic with a minimal quark content $udd u\bar{s}$ since it was observed in the $nK^+$ and $pK^0_S$ final states. However, while the existence of the $D^*_{s0}(2317)^+$ and $\chi_{c1}(3872)$ mesons has been extensively confirmed by many experiments, the evidence of the $\Theta^+$ baryon has faded away with time [2462]. The discovery of the $\chi_{c1}(3872)$ drew a lot of attention due to the narrowness of the signal and the proximity of the mass to the $m(D^0)+m(\bar{D}^{*0})$ threshold. Soon after many other charmonium-like and bottomonium-like states were observed. While it is still not possible to rule out firmly a conventional nature for the majority of them, the observation of the $Z_c(4430)^+$ meson, an electrically charged charmonium-like state, and of the $T^+_{cc}$ state, a meson containing two charm quarks, established definitively the existence of QCD exotics. Many models have been proposed to explain the exotic nature of such a states: *hadronic molecules* [2463], whose constituents are color-singlet mesons bound by residual nuclear forces, *tetraquarks* [2464], bound states between a diquark and diantiquark, *hadro-quarkonium* [2465], a cloud of light quarks and gluons bound to a heavy $Q\bar{Q}$ core state via van-der-Waals forces, *threshold effects*, enhancements caused by threshold cusps [2466] or rescattering processes [2467].

The spectra of the conventional and exotic charmonium-like and bottomonium-like states are shown in Fig. 8.5.1. Many of them have been named $X$, $Y$ and $Z$ in the cor-responding discovery papers without a consistent criterion as a consequence of their uncertain nature. With the number of $X, Y, Z$ states growing, the need of an adequate naming scheme emerged. The current naming scheme in Particle Data Group extends the convention used for ordinary quarkonia by taking in account the isospin, spin and parity of the state [476]. The names are not related to the internal structure of the states given their nature is controversial. However, even the current scheme presents some limitation for the manifestly exotic states and a new scheme has been proposed recently [2468].

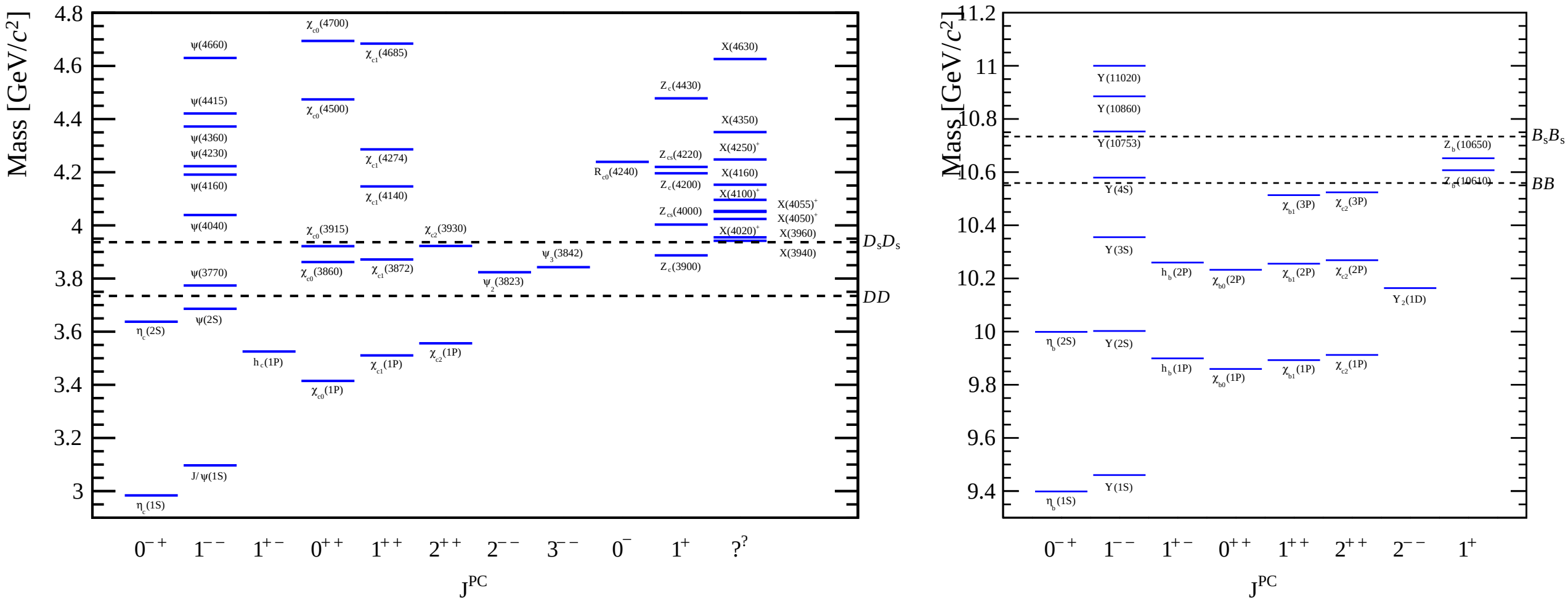

**Fig. 8.5.1** The spectrum of charmonium(-like) (left) and bottomonium(-like) (right) states. States are labeled according to the PDG naming scheme. Dashed horizontal lines show some relevant open-charm or open-bottom thresholds. The states shown in the right columns are manifestly exotic, i.e. the quark content can not be only $c\bar{c}$ or $b\bar{b}$ given their non-zero electrical charge.

### 8.5.2 $\chi_{c1}(3872)$: The renaissance of the exotic spectroscopy

In 2003, the Belle collaboration, while studying the $B^+ \to J/\psi\pi^-\pi^+K^+$ decays, observed two peaking structures in the $J/\psi\pi^-\pi^+$ mass projection (Fig. 8.5.2): the well known $\psi(2S)$ meson and a new state, originally dubbed $X(3872)$ [2460]. The new meson has been confirmed by many experiments [2469–2475] and observed in prompt production in $pp$, $p\bar{p}$, $p$Pb [2476] and PbPb [2477] collisions as well as in $B$ and $\Lambda_b^0$ hadron decays [2478, 2479]. The invariant mass distribution of the dipion system is consistent with originating from $\rho(770)^0 \to \pi^+\pi^-$ decays [2473, 2480]. Recently using a larger dataset the presence of a sizeable contribution of $\omega(782) \to \pi^+\pi^-$ decays has been established as well [2481]. For a pure charmonium state, the decays to $J/\psi\omega$ [2482, 2483] and $J/\psi\rho^0$ are isospin conserving and violating, respectively. Therefore the latter should be strongly suppressed, in contrast to the measured branching ratios. Later on, further decay modes have been reported: $D^0\bar{D}^0\pi^0$ [2484], $\bar{D}^0D^{*0}$ [2485], $\chi_{c1}\pi^0$ [2486], $J/\psi\gamma$ [2487] and $\psi(2S)\gamma$. The current branching-fraction measurements of the $\psi(2S)\gamma$ radiative decay [2487–2490] are, however, not fully consistent and further studies are needed to solve the emerging tension. Solving this puzzle will help to understand the nature of the $X(3872)$ meson, given that the predicted branching fractions span over a broad range of values depending if the $X(3872)$ state is a $D^{*0}\bar{D}^0$ molecule [2491, 2492] or a pure charmonium state [2493, 2494].

A study of the angular correlations among the final state particles from $X(3872) \to J/\psi\pi^+\pi^-$ decays constrained the possible $J^{PC}$ assignments for the $X(3872)$ to $J^{PC} = 1^{++}$ and $2^{-+}$ [2495]. The latter, disfavoured by the observation of the radiative decays, was definitively ruled out by the LHCb experiment [2496, 2497]. Once the quantum numbers $J^{PC} = 1^{++}$ have been firmly established, the name of $X(3872)$ turned into $\chi_{c1}(3872)$ according to the PDG naming scheme [476]. The identification of the $X(3872)$ with the the $2^3P_1$ $c\bar{c}$ state is disfavoured by the large branching fraction of $X(3872) \to J/\psi\rho^0$ and the large mass splitting with respect to the $2^3P_2$ state, identified with $\chi_{c2}(3930)$.

An intriguing feature of the $\chi_{c1}(3872)$ meson is the proximity of its mass to the $m(D^{*0}) + m(D^0)$ threshold. This characteristic has led to speculate that the $\chi_{c1}(3872)$ is a molecular state [2498] where the $D^{*0}$ and $\bar{D}^0$ mesons are bound by residual nuclear forces, similarly to a proton and a neutron in the nucleus of the deuterium. An important input for such an interpretation is the binding energy $E_b \equiv m_{D^0} + m_{D^{*0}} - m_{\chi_{c1}(3872)}$ which is still consistent with zero despite being mea-sured with a precision of $\mathcal{O}(100)$ keV [2499, 2500]. The analyses also reported a measurement of the natural width $\Gamma^{BW}_{\chi_{c1}(3872} = (1.39 \pm 0.24 \pm 0.10)$ MeV by using a Breit-Wigner lineshape for the $\chi_{c1}(3872)$ signal. However, since the $|E_b| < \Gamma^{BW}_{\chi_{c1}(3872)}$, coupled channel effects might distort the lineshape. Indeed a Flatté-inspired model returned a significantly smaller full width at half-maximum FWHM $= (0.22 \pm^{0.07}_{0.06} \pm^{0.11}_{0.13})$ MeV, highlighting the relevance of a physically well-motivated lineshape parameterization (see Section 14.5).

The smallness of the binding energy $E_b = (0.07 \pm 0.12)$ MeV [2499, 2500] implies a size of $\mathcal{O}(10)$ fm in a molecular scenario. The production of a large and weakly bound molecule is expected to be suppressed due to the interactions with comoving hadrons produced in the underlying event [2501]. The ratio of $\chi_{c1}(3872)$ to $\psi(2S)$ cross-sections for promptly produced particles has been measured at LHC [2502] and has been found to decrease with multiplicity. However the slope would seem not to agree with the expectations for a molecular state [2503]. In addition no enhancement of the $\chi_{c1}(3872)$ production has been observed in association to a pion [2504] as expected for a molecular state produced via the formation of a $D^*\bar{D}^*$ pair at short distance followed by the rescattering of the charmed mesons into $\chi_{c1}(3872)\pi$ [2505]. Finally, the relative production of $\chi_{c1}(3872)$ to $\psi(2S)$ mesons as a function of the transverse momentum and rapidity has shown a mild (or null) dependence for $\chi_{c1}(3872)$ (or $\psi(2S)$) mesons produced in prompt $pp$ collisions and from $b$-hadron decays, respectively [2473, 2506]. The CMS collaboration has measured a large production rate of the $\chi_{c1}(3872)$ mesons also at large transverse momenta while a suppression is expected for hadronic molecules [2507], as measured for the deuteron [2508].

In order to reconcile the molecular picture to the production measurements, it has been suggested that the physical $\chi_{c1}(3872)$ might be a quantum mechanical mixture of a $D^{*0}\bar{D}^0$ molecule and the $2^3P_1$ $c\bar{c}$ charmonium state [2509], where the production is dominated by the charmonium component. Alternatively, an interpretation has been proposed where the $\chi_{c1}(3872)$ meson is a tightly bound diquark-diantiquark system [2464] with a size of a few fermis. In this scenario, isospin partner states are expected to exist. A search for charged $X^-$ states has been carried out by studying the decays $B^0 \to X^- K^+$ and $B^- \to X^- K^0_S$, where $X^- \to J/\psi\pi^-\pi^0$ [2510, 2511]. No charged $X^-$ has been reported. Moreover no $X^- \to D^0 D^{*-}$ signal has been observed in the $D^0\bar{D}^0\pi^-$ mass spectrum [2512]. Another firm prediction of the compact tetraquark models is that hidden charm states must form complete flavor-SU(3) multiplets with mass differences determined by the quark mass difference $m_s - m_u$ [2513]. The $\chi_{c1}(3872)$ meson could belong to the same flavor multiplet of the X(4140) given both states have $J^{PC} = 1^{++}$, where the two mesons are interpreted as $[\bar{c}\bar{q}][cq]$ ($q = u$ or $d$) and $[\bar{c}\bar{s}][cs]$ bound states, respectively. As a consequence, a $[cs][\bar{c}\bar{q}]$ state with mass $(m_{X(4140)} + m_{\chi_{c1}(3872)})/2 =$ 4009 MeV should exist. Two exotic states, $Z_{cs}(3985)$ [2514] and $Z_{cs}(4000)$ [2515] have been observed in $D_s^+\bar{D}^{*0}$ and $J/\psi K^+$ mass spectra, respectively, with masses close to 4000 MeV, making them potential candidates to complete the $J^{PC} = 1^{++}$ tetraquark nonet, where $C = +1$ refers to the sign of charge conjugation of the neutral-non-strange members.

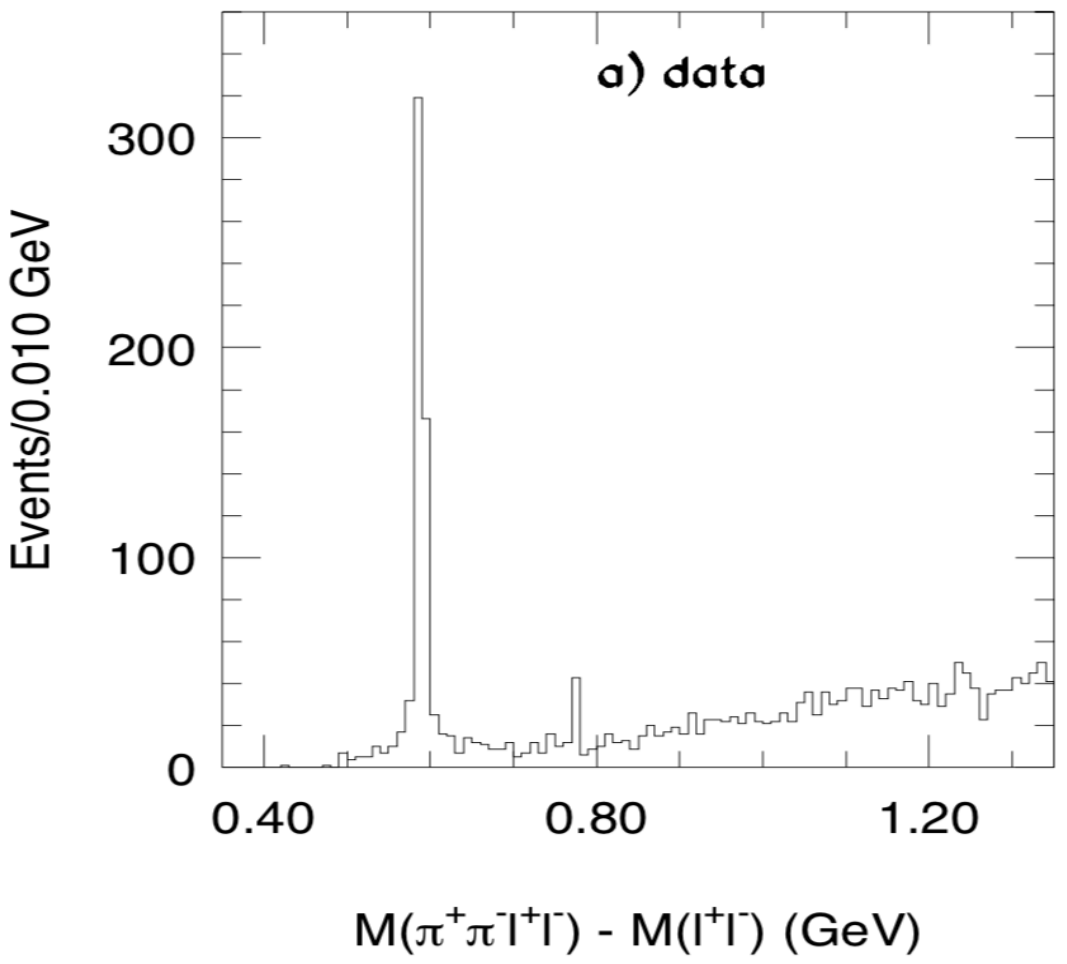

**Fig. 8.5.2** Distribution of the mass difference $M(\pi^+\pi^-\ell^+\ell^-) - M(\ell^+\ell^-)$, where $\ell = e, \mu$ and the invariant mass of the dilepton system is within a range around the $J/\psi$ mass. The two signals correspond to the $\psi(2S)$ and $\chi_{c1}(3872)$ mesons, respectively [2460].

A $C$-odd partner of the $\chi_{c1}(3872)$ state, dubbed $\tilde{X}(3872)$, is expected as well [2464, 2516]. Several experiments searched for a $\tilde{X}(3872)$ candidate in the $J/\psi\eta$ and $\chi_{c1}\gamma$ mass spectra in $B^+ \to J/\psi\eta/\chi_{c1}\gamma K^+$ decays but no signal was reported [2517–2520], even though many other charmonium states were observed. The COMPASS collaboration searched for muo-production of charmonia in the process $\mu^+ N \to \mu^+ X^0 \pi^\pm N'$ with $X^0 \to J/\psi\pi^+\pi^-$ where $N$ denotes the target nucleon, $N'$ the unobserved recoil system and $X^0$ an intermediate charmonium state [2521]. In addition to the observation of the $\psi(2S)$ meson, evidence of a narrow structure, peaking at about 3872 MeV in the $J/\psi\pi^+\pi^-$ spectrum, was reported. While the measured mass and width pointed to an interpretation of the signal as $\chi_{c1}(3872)$ meson, the $\pi^+\pi^-$ mass spectrum showed a rather flat distribution instead of the expected $\rho^0$-like shape thus disagreeing significantly with previous experimental re-sults. This surprising result led the authors to speculate that the observed state might be the $C$-odd partner $\tilde{X}(3872)$ decaying to the $J/\psi f_0(500)$ final state.

Assuming heavy flavor symmetry, a bottomonium counterpart $X_b$ of the $\chi_{c1}(3872)$ meson is expected. Searches for $X_b$, carried out by the CMS [2522] and ATLAS [2523] collaborations by studying the $\Upsilon\pi^+\pi^-$ final state, have not been successful. This result does not rule out the existence of an $X_b$ state since, contrary to the $\chi_{c1}(3872)$ case, the $\Upsilon\pi^+\pi^-$ decay mode is expected to be suppressed due to the smaller isospin breaking effect: the mass difference between the neutral and charged $B$ mesons is very small. Most likely, the $X_b$ state would decay into the $\Upsilon\omega$ and $\chi_b\pi^+\pi^-$ final states. The former decay has been recently studied by the Belle-II collaboration. No $X_b$ meson has been observed [2524].

### 8.5.3 $Z_c(4430)^+$ and the charmonium-like states

The observation of manifestly exotic candidates was a turning point in the discussion about the existence of non-conventional hadrons. Indeed, a peculiar characteristic of charmonium-like states is the possibility to observe states with non-zero electrical charge and quark content $c\bar{c}u\bar{d}$.

The first-ever candidate, the $Z_c(4430)^+$ meson, was observed by the Belle collaboration in the $\psi(2S)\pi^+$ projection of $\bar{B}^0 \to \psi(2S)K^-\pi^+$ and $B^+ \to \psi(2S)K^0_S\pi^+$ decays [2525]. The $m^2(K\pi^+)$ versus $m^2(\psi(2S)\pi^+)$ Dalitz-plot distributions show a continuous band (and a peak in the $m^2(\psi(2S)\pi^+)$ projection) together with two bands in the $m^2(K\pi^+)$ mass distributions corresponding to the $K^*(892)$ and $K^*_{0/2}(1430)$ resonances. After applying a veto on the $K^*$ regions, a one-dimensional fit to the $\psi(2S)\pi^+$ projection returned the mass and width (Table 8.5.1) of a signal that was interpreted as the first charmonium-like state with non-zero electrical charge (Fig. 8.5.3). Given that the decay modes have four degrees of freedom, the claim of a new exotic state based on the study of a one-dimensional projection received some criticism. In addition, excluding regions in the $K\pi^+$ invariant mass does not imply that interference effects are removed which could lead to peaking structures in other projections.

A model-independent approach was pursued by the BaBar collaboration which investigated the extent to which the reflections of mass and angular distribution of structures in the $K\pi^+$ system might describe the associated $\psi(2S)\pi^+$ mass distributions in $\bar{B}^0 \to \psi(2S)K^-\pi^+$ and $B^+ \to \psi(2S)K^0_S\pi^+$ decays [2526]. For this purpose, the $K\pi^+$ angular distribution was represented, at a given $m(K\pi^+)$, in terms of a Legendre polynomial expansion. The combinations of the first $N = 2J_{max}+1 = 7$ terms reproduces adequately the $\psi(2S)\pi^+$ mass distribution where $J_{max} = 3$ is the maximum spin of the excited $K^*$ resonances expected in the $K\pi^+$ spectrum. This result provided a hint that an exotic contribution may not be needed, but it cannot rule out the presence of the $Z_c(4430)^+$ meson either. Later on, the Belle collaboration performed a fit to the $m^2(K\pi^+)$ versus $m^2(\psi(2S)\pi^+)$ Dalitz-plot [2527] and finally a complete four-dimensional amplitude analysis [2528], both confirming the observation of an exotic state. The latter analysis quotes a natural width for the $Z_c(4430)^+$ much larger than the one reported in the discovery paper (Table 8.5.1) which highlights the relevance of performing full amplitude analyses to measure the physical parameters.

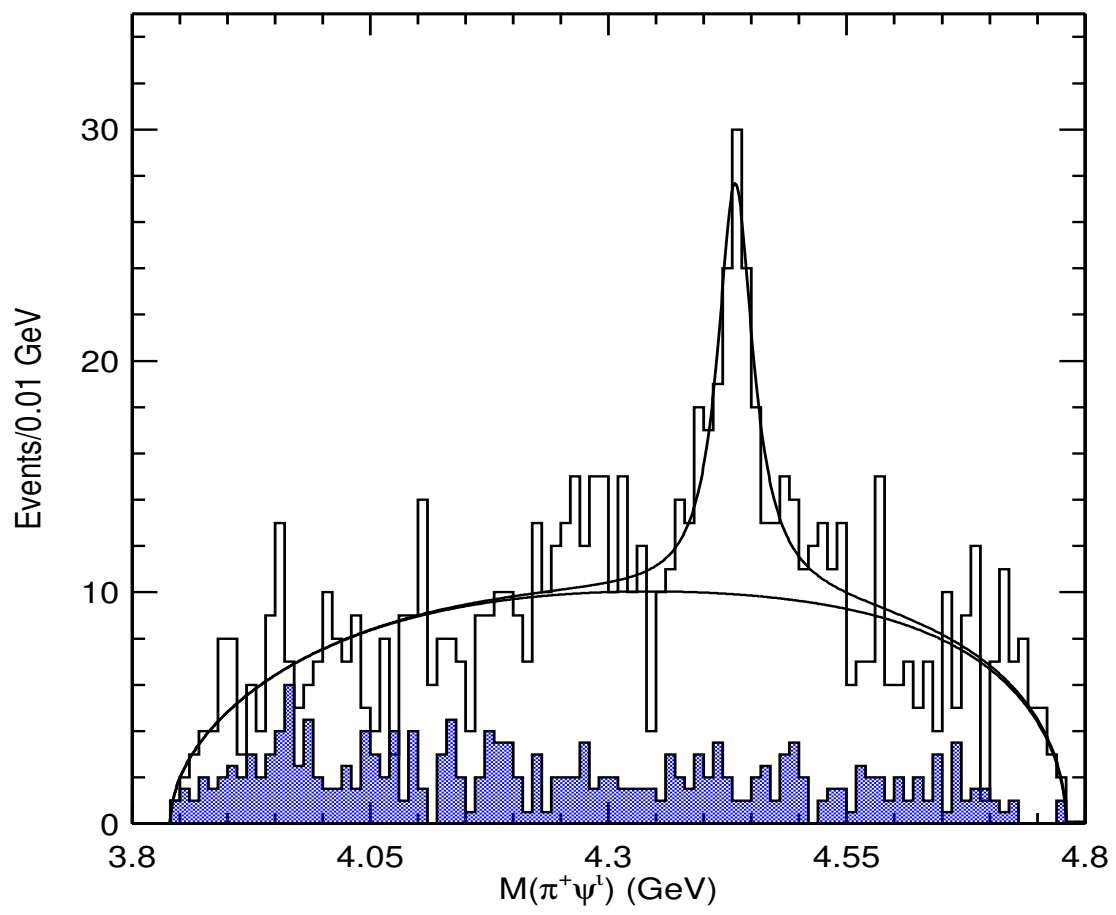

**Fig. 8.5.3** Observation of the $Z_c(4430)^+$ meson in the $m(\psi(2S)\pi^+)$ distribution of the $\bar{B}^0 \to \psi(2S)K^-\pi^+$ and $B^+ \to \psi(2S)K^0_S\pi^+$ decays after applying a veto on the $K^*(892)$ and $K^*_{0/2}(1430)$ states in the $K\pi^+$ systems [2525].

**Table 8.5.1** Measurements of mass and natural width of the $Z_c(4430)^+$ meson. The last column reports the number of dimensions considered in the corresponding amplitude analysis.

| | $Z_c(4430)^+$ | | |
|---|---|---|---|
| | Mass [MeV/c$^2$] | Width [MeV] | |
| Belle [2525] | $4433 \pm 4 \pm 2$ | $45^{+18+30}_{-13-13}$ | 1D |
| Belle [2527] | $4443^{+15+19}_{-12-13}$ | $107^{+86+74}_{-43-56}$ | 2D |
| Belle [2528] | $4485^{+22+28}_{-22-11}$ | $200^{+41+26}_{-46-35}$ | 4D |
| LHCb [2529] | $4475 \pm 7^{+15}_{-25}$ | $172 \pm 13^{+37}_{-34}$ | 4D |

The existence of the $Z_c(4430)^+$ exotic state was debated for many years until the LHCb collaboration also studied the $\bar{B}^0 \to \psi(2S)K^-\pi^+$ decays [2529]. The tenfold increase in signal yield over the previous measurements allowed the collaboration to confirm the $Z_c(4430)^+$ state firmly with an improved measurement of mass and width (Table 8.5.1) and to establish its spin and parity to $J^P = 1^+$.

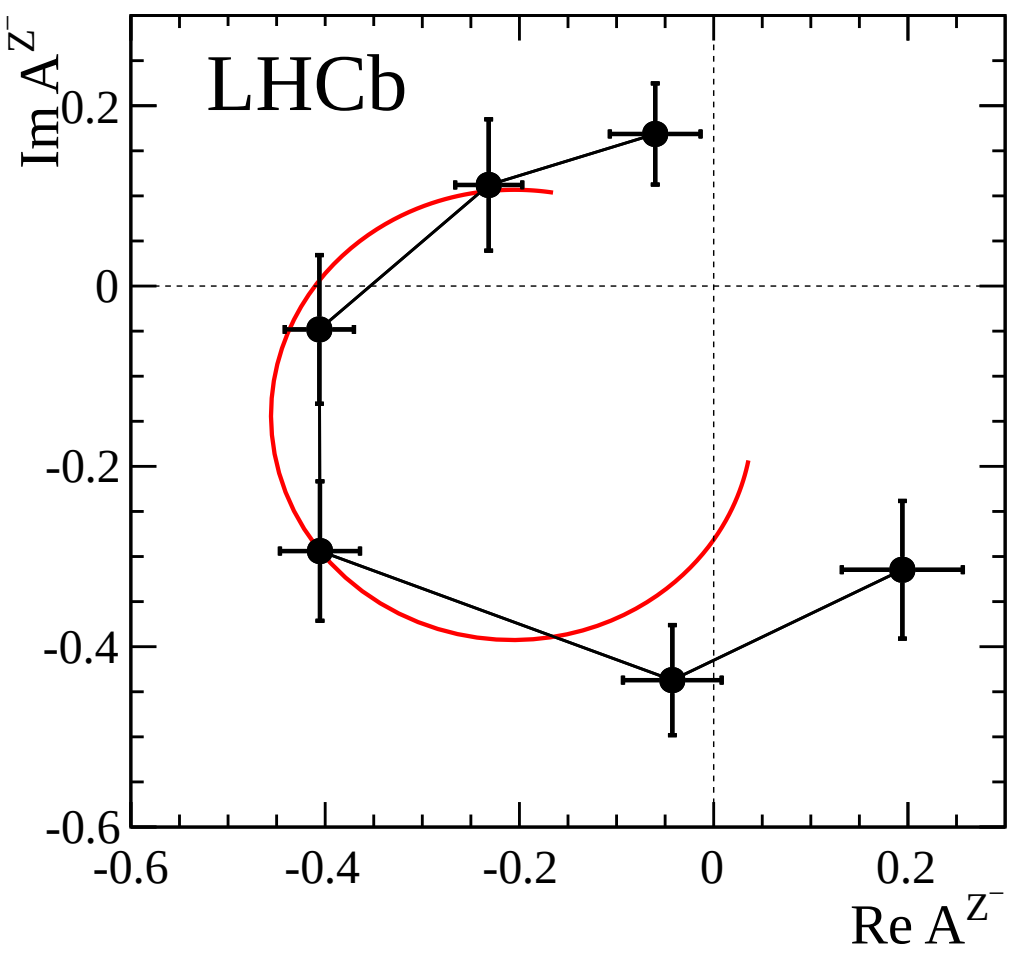

**Fig. 8.5.4** Argand diagram of the $Z_c(4430)^+$ meson by a Dalitz analysis of the $\bar{B}^0 \to \psi(2S)K^-\pi^+$ decays where the $Z_c(4430)^+$ amplitude is fitted in six independent $m^2(\psi(2S)\pi^+)$ bins. The red curve is the expected shape according to a Breit-Wigner function with a resonance mass (width) of 4475 (172) MeV. Units are arbitrary [2529].

In addition, the resonant character of a charged four-quark state is demonstrated for the first time by representing the $Z_c(4430)^+$ amplitude as the combination of independent complex amplitudes at six equidistant points in the $m^2(\psi(2S)\pi^+)$ spectrum. The resulting Argand diagram, shown in Fig. 8.5.4, is consistent with a rapid change of the $Z_c(4430)^+$ phase when its magnitude reaches the maximum, a behavior characteristic of a resonance. Finally, an analysis of the data, using the model-independent approach developed by the BaBar collaboration, shows significant inconsistencies in the $Z_c(4430)^+$ region between the data and a model introducing $K^*$ states with $J \leq 3$ [2530]. Evidence of the $Z_c(4430)^+ \to J/\psi\pi^+$ is also reported by an amplitude analysis of the $\bar{B}^0 \to J/\psi K^-\pi^+$ decays [2531]. After the discovery of the $Z_c(4430)^+$ meson, many further charged charmonium-like states have been reported (Table 8.5.2), including candidates with strangeness and isospin partners [2532, 2533].

### 8.5.4 The bottomonium-like $Z_b^+$ states

Few years after the discovery of the $Z_c(4430)^+$ meson, the Belle collaboration claimed the observation of two bottomonium-like states $Z_b(10610)^+$ and $Z_b(10650)^+$ in the $\Upsilon(nS)\pi^+$ (n=1, 2, 3) and $h_b(mP)\pi^+$ (m =1, 2) spectra by studying the exclusive processes $e^+e^- \to \Upsilon(nS)\pi^+\pi^-$ (n = 1, 2, 3) and $e^+e^- \to h_b(mP)\pi^+\pi^-$ (m = 1, 2) with data collected at the collision energy $\sqrt{s} = 10.865$ GeV [2544], the $\Upsilon(5S)$ mass. Amplitude analyses of the three-body $\Upsilon(nS)\pi^+\pi^-$ decays were performed by means of unbinned maximum likelihood fits to two-dimensional $m^2(\Upsilon(nS)\pi^+)$ versus $m^2(\Upsilon(nS)\pi^-)$ Dalitz distributions. Two narrow structures appear in the $m(\Upsilon(nS)\pi^\pm)$ spectrum (e.g. Fig. 8.5.5). The analyses of the $h_b(mP)\pi^\pm$ spectra returned compatible results. Weighted averages of mass and width measurements over all five channels yield for the $Z_b(10610)^+$

$$m = (10607.2 \pm 2.0)\,\text{MeV},\ \Gamma = (18.4 \pm 2.4)\,\text{MeV},$$

and for the $Z_b(10650)^+$

$$m = (10652.2 \pm 1.5)\,\text{MeV},\ \Gamma = (11.5 \pm 2.2)\,\text{MeV}.$$

Later on a six-dimensional amplitude analysis of the $\Upsilon(nS)\pi^+\pi^-$ (n = 1, 2, 3) three-body final states confirmed the existence of the two $Z_b^+$ states and strongly favored $I^G(J^P) = 1^+(1^+)$ quantum-number assignments for both of them [2545]. Finally the $Z_b(10610)^+$ and $Z_b(10650)^+$ mesons have been observed in the $B^+\bar{B}^{*0}$ and $B^{*+}\bar{B}^{*0}$ mass spectrum, respectively [2546]. Table 8.5.3 summarizes the branching fractions of $Z_b(10610)^+$ and $Z_b(10650)^+$ states by assuming that their sum is equal to one.

**Table 8.5.2** Decay modes and quantum numbers of manifestly exotic charmonium-like states.

| State | Decay modes | $I^G(J^{PC})$ |
|---|---|---|
| $Z_c(3900)^+$ | $J/\psi\pi^+$ [2534–2536] | $1^+(1^{+-})$ |
| | $\bar{D}^0D^{*+}, \bar{D}^{*0}D^+$ [2537, 2538] | |
| $X(4020)^+$ | $h_c\pi^+$ [2539], $D^{*+}\bar{D}^{*0}$ [2540] | $1^+(?^{?-})$ |
| $X(4050)^+$ | $\chi_{c1}(1P)\pi^+$ [2541] | $1^-(?^{?+})$ |
| $X(4055)^+$ | $\psi(2S)\pi^+$ [2542] | $1^+(?^{?-})$ |
| $X(4100)^+$ | $\eta_c(1S)\pi^+$ [2543] | $1^-(?^{??})$ |
| $Z_c(4200)^+$ | $J/\psi\pi^+$ [2531] | $1^+(1^{+-})$ |
| $R_{c0}(4240)^+$ | $\psi(2S)\pi^+$ [2529] | $1^+(0^{--})$ |
| $X(4250)^+$ | $\chi_{c1}(1P)\pi^+$ [2541] | $1^-(?^{?+})$ |
| $X(3985)^+$ | $D_s^+\bar{D}^{*0}, D_s^{*+}\bar{D}^0$ [2514] | $1/2(?^?)$ |
| $Z_{cs}(4000)^+$ | $J/\psi K^+$ [2515] | $1/2(1^+)$ |
| $Z_{cs}(4220)^+$ | $J/\psi K^+$ [2515] | $1/2(1^?)$ |

**Table 8.5.3** Branching fractions for the $Z_b(10610)^+$ and $Z_b(10650)^+$ decays. The first uncertainty is statistical while the second is systematic [2546].

| Channel | Fraction (%) | |
|---|---|---|
| | $Z_b(10610)^+$ | $Z_b(10650)^+$ |
| $\Upsilon(1\text{S})\pi^+$ | $0.60 \pm 0.17 \pm 0.07$ | $0.17 \pm 0.06 \pm 0.02$ |
| $\Upsilon(2\text{S})\pi^+$ | $4.05 \pm 0.81 \pm 0.58$ | $1.38 \pm 0.45 \pm 0.21$ |
| $\Upsilon(3\text{S})\pi^+$ | $2.40 \pm 0.58 \pm 0.36$ | $1.62 \pm 0.50 \pm 0.24$ |
| $h_b(1\text{P})\pi^+$ | $4.26 \pm 1.28 \pm 1.10$ | $9.23 \pm 2.88 \pm 2.28$ |
| $h_b(2\text{P})\pi^+$ | $6.08 \pm 2.15 \pm 1.63$ | $17.0 \pm 3.74 \pm 4.1$ |
| $B^+\bar{B}^{*0} + \bar{B}^0B^{*+}$ | $82.6 \pm 2.9 \pm 2.3$ | – |
| $B^{*+}\bar{B}^{*0}$ | – | $70.6 \pm 4.9 \pm 4.4$ |

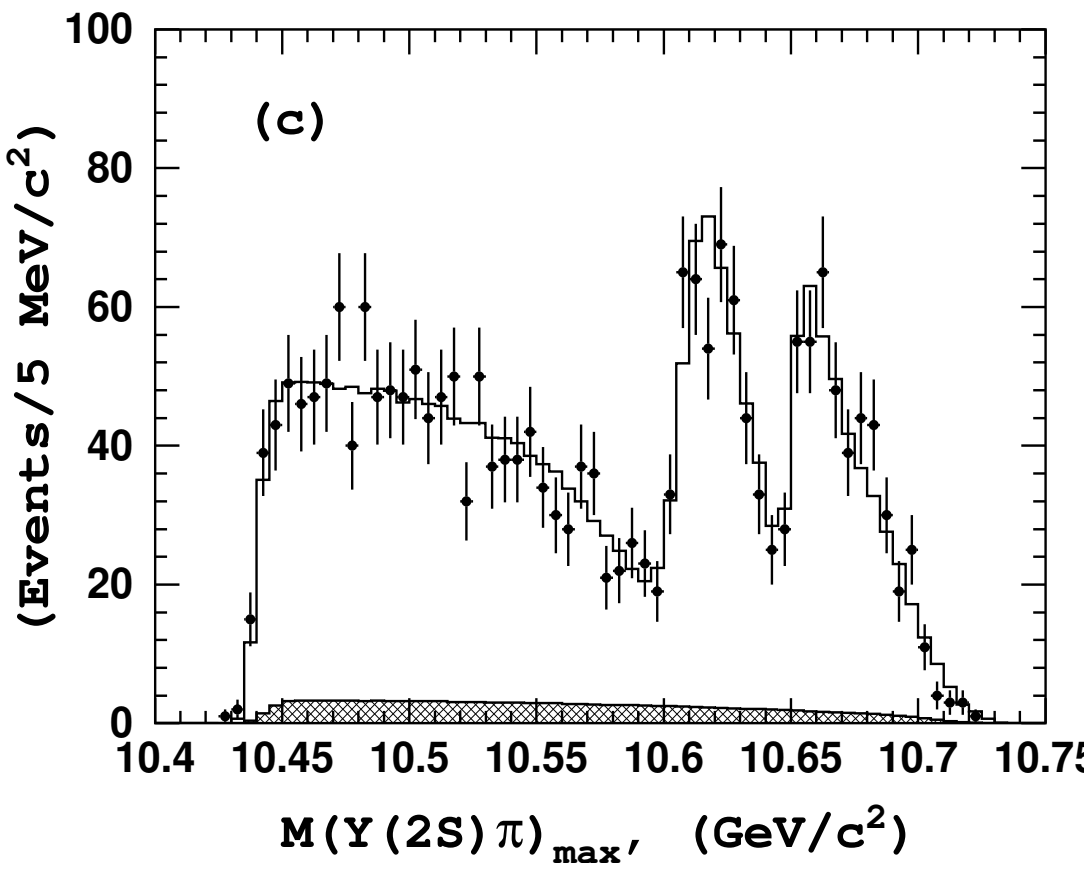

**Fig. 8.5.5** The maximum invariant mass of the two $\Upsilon(2S)\pi$ combinations of the $e^+e^- \to \Upsilon(2S)\pi^+\pi^-$ process at $\sqrt{s} = 10.865$ GeV. The two peaking structures are interpreted as the $Z_b(10610)^+$ and $Z_b(10650)^+$ bottomonium-like states. [2544].

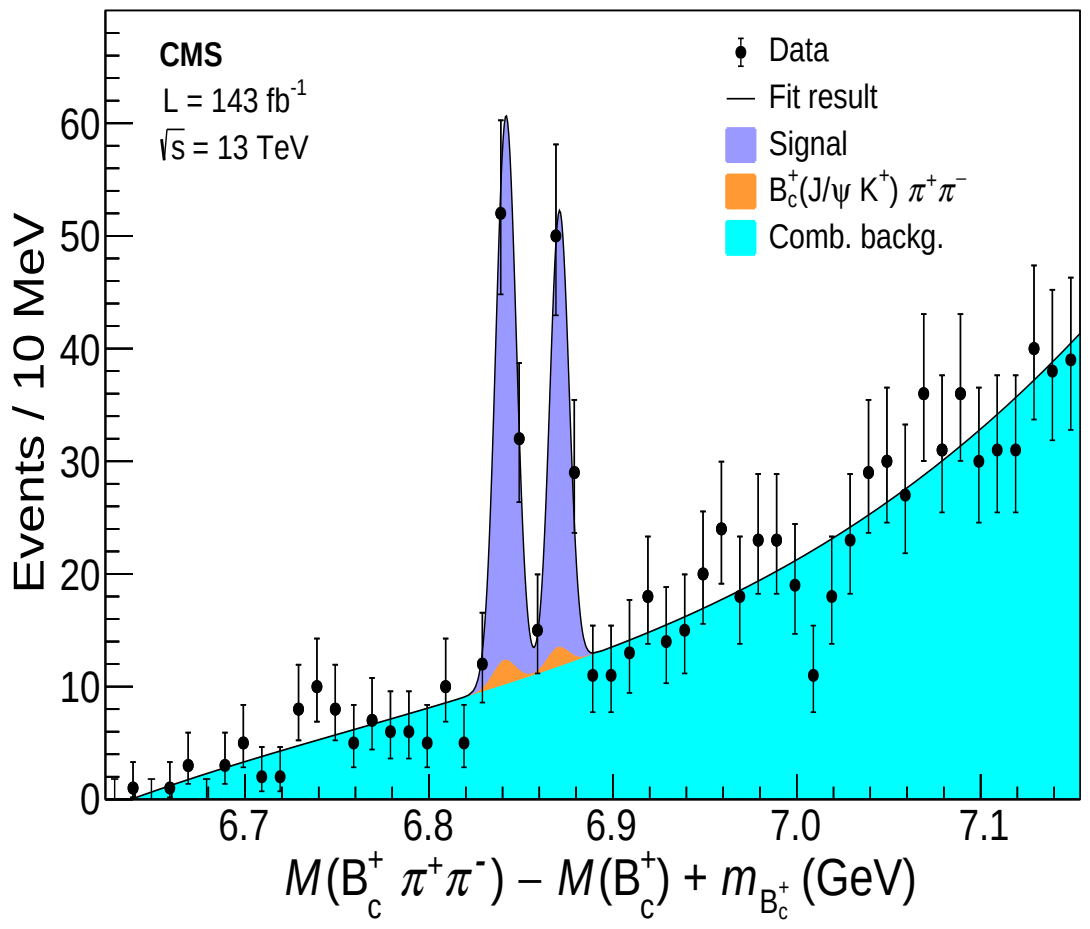

**Fig. 8.5.6** Observation of the $B_c^*(2S)^+$ (left-most peak) and $B_c(2S)^+$ (right-most peak) states in the $M(B_c^+\pi^+\pi^-) - M(B_c^+) + m_{B_c^+}$ mass distribution [2557].

The large branching fractions of the $B^{(*)}\bar{B}^*$ decay modes and the measured quantum numbers are consistent with the interpretation of the two states as $B\bar{B}^*$ and $B^*\bar{B}^*$ loosely bound molecular hadrons. However the measured mass for the $Z_b(10610)^+$ and $Z_b(10650)^+$ are both above the nearby open-flavor thresholds. This might be the result of using Breit-Wigner functions to parameterize the amplitudes of very near-threshold states. Indeed, when amplitudes consistent with unitarity and analyticity are used instead, lower masses are measured, typically below the thresholds [2547].

### 8.5.5 The $B_c^+$ mesons

Contrary to charmonium and bottomonium states, the $B_c^+$ mesons can not annihilate into gluons and thus these states are more stable. Indeed, apart from the ground state which decays weakly, all the excited states, with masses below the lowest strong decay $B^{(*)}D^{(*)}$ thresholds, are predicted to have narrow widths [2548, 2549].

Before the start of LHC, only the ground $B_c^+$ state was observed [2550] via few decays modes: $B_c^+ \to J/\psi\pi^+$ and $B_c^+ \to J/\psi\ell^+\nu$. The LHCb and CMS experiments have observed 15 new decays modes and have largely improved the precision of the $B_c^+$ mass [2551] and lifetime [2552–2554]. The production of the $B_c^+$ meson has been observed in $p\bar{p}$, $pp$ as well as in PbPb collisions [2555], where the measurement of the nuclear modification factor hints that effects of the hot and dense nuclear matter created in heavy ion collisions contribute to its production.

Despite the large number of expected excited states, only a few have been observed so far due to the small production cross sections of the $B_c^+$ mesons and the small branching ratios of the reconstructed decay chains. In 2014 the ATLAS collaboration reported the first observation of an excited $B_c^+$ state decaying to $B_c^+\pi^+\pi^-$ final state [2556]. Few years later the same mass spectrum was investigated by other LHC experiments [2557, 2558] and it turned out that the ATLAS structure was very likely the result of a superimposition of two narrower signals (Fig. 8.5.6), interpreted as the $B_c(2S)^+$ and $B_c^*(2S)^+$ states. The latter appears in the mass spectrum as a partially reconstructed decay $B_c^*(2S)^+ \to B_c^{*+}\pi^+\pi^-$, where the photon of the $B_c^{*+} \to B_c^+\gamma$ reaction is not reconstructed. Since the $B_c^{*+}$ meson has not been observed yet, the mass of the $B_c^*(2S)^+$ state can not be measured and it is not listed in the PDG. In the next years the upgraded LHC experiments will probe the largely unexplored spectrum of the excited $B_c^+$ mesons below and above the $B^{(*)}D^{(*)}$ thresholds with the intriguing possibility to observe exotic states as for the other quarkonium systems [2559].

### 8.5.6 The doubly charmed $T_{cc}(3875)^+$ state

All the exotic mesons described so far are featured by a heavy quark-antiquark pair $Q\bar{Q}$ and a light quark-antiquark pair $q\bar{q}$. The observation of several $Q\bar{Q}q\bar{q}$ state has revived the discussion on the existence of of $QQ\bar{q}\bar{q}$ states with two heavy quarks and two light antiquarks. In the limit of a large heavy-quark mass, the two heavy quarks $QQ$ form a heavy point-like color-antitriplet object, that behaves like an antiquark, and the corresponding four-quark state should be bound. The argument that such a state should exist, if the mass of the charm quark is enough, has been discussed ex-tensively, but a consensus was not reached. Even lattice QCD calculations had not provided a definite conclusion [2560]. The $cc\bar{u}\bar{d}$ ground state, hereafter denoted as $T^+_{cc}$, is predicted with spin-parity quantum numbers $J^P = 1^+$ and isospin $I = 0$. The only known hadron with a similar quark content is the $\Xi^{++}_{cc}$ baryon [2561–2563], a bound state of two $c$ quarks and one $u$ quark. Its measured mass [2564] implies that the mass of the $T^+_{cc}$ is close to the sum of masses of $D^0$ and $D^{*+}$ mesons [1070].

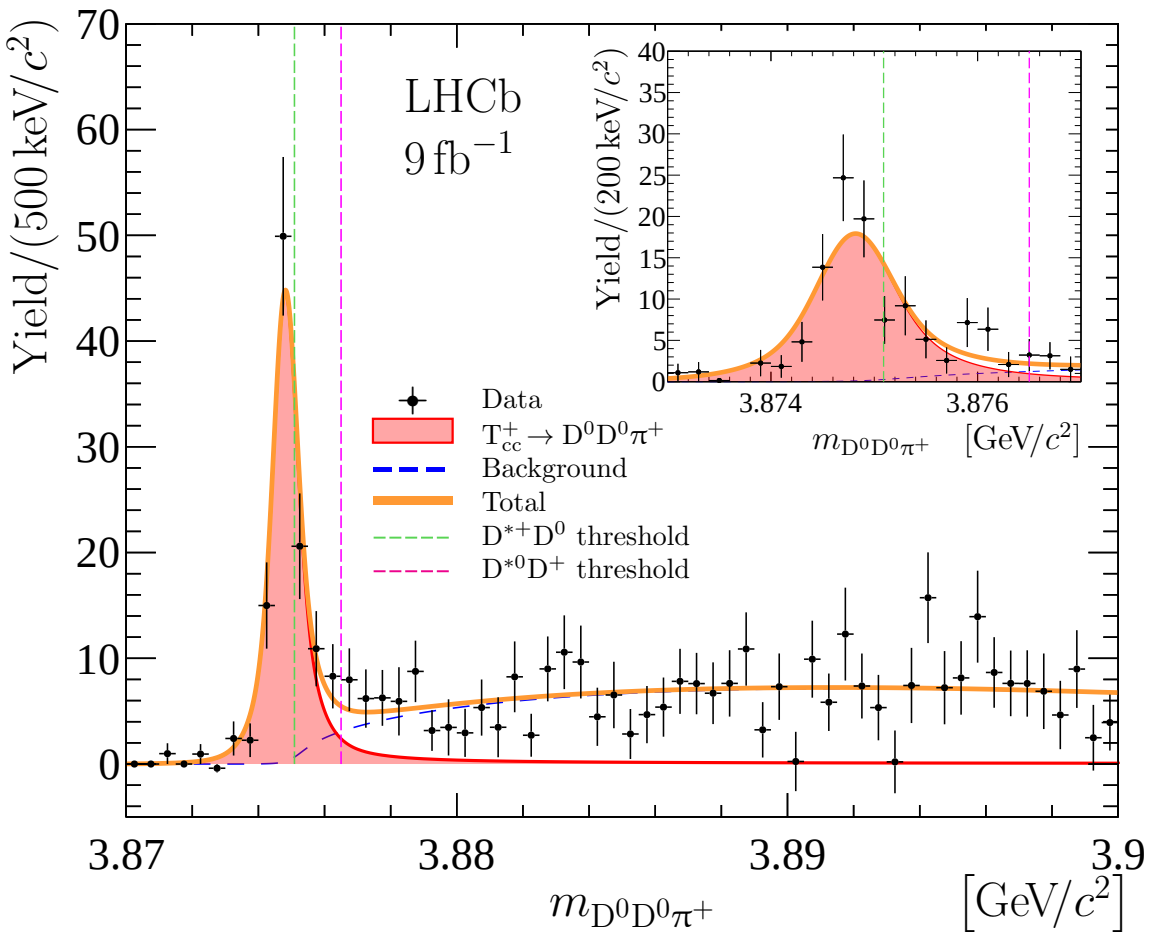

**Fig. 8.5.7** Distribution of $D^0D^0\pi^+$ mass where the contribution of the non-$D^0$ background has been statistically subtracted. The $D^{*+}D^0$ and $D^{*0}D^+$ thresholds are indicated with the vertical dashed lines. Inset shows a zoomed signal region with fine binning scheme [1071, 2512].

The LHCb experiment reported the observation of a narrow state in the $D^0D^0\pi^+$ mass spectrum near the $D^{*+}D^0$ mass threshold compatible with being a $T^+_{cc}$ tetraquark state [1071, 2512]. The $D^0D^0\pi^+$ final state is reconstructed by selecting events with two $D^0$ mesons and a positively charged pion, all produced at the same proton-proton interaction point. Both $D^0$ mesons are reconstructed in the $D^0 \to K^-\pi^+$ decay channel. The mass distribution of the selected $D^0D^0\pi^+$ candidates is shown in Fig. 8.5.7. A narrow peak near the $D^{*+}D^0$ mass threshold is clearly visible.

An extended unbinned maximum-likelihood fit to the $D^0D^0\pi^+$ mass spectrum is performed by modelling the signal with a Breit-Wigner function $\mathfrak{F}^{BW}$. The measured mass $\delta m$ and the full width at half maximum (FWHM) of the $T^+_{cc}$ state are reported in Table 8.5.4, where the uncertainties are statistical. The mass parameter $\delta m$ is defined relative to the $D^{*+}D^0$ mass threshold as $\delta m \equiv m - m_{D^{*+}} - m_{D^0}$, where $m_{D^{*+}}$ and $m_{D^0}$ denote the known masses of the $D^{*+}$ and $D^0$ mesons. The measured $\delta m$ value corresponds to a mass of approximately 3875 MeV. Though the use of a standard Breit-

**Table 8.5.4** Mass difference $\delta m \equiv m - m_{D^{*+}} - m_{D^0}$ and the full width at half maximum (FWHM) of the $T_{cc}(3875)^+$ state by fitting the $D^0D^0\pi^+$ mass spectrum with the $\mathfrak{F}^{BW}$ and $\mathfrak{F}^{U}$ models. The uncertainties are statistical only. See Refs. [1071, 2512] for a complete set of results.

| | $T_{cc}(3875)^+$ | |
|---|---|---|
| | $\delta m$ [keV/c$^2$] | FWHM [keV/c$^2$] |
| $\mathfrak{F}^{BW}$ | $-279 \pm 59$ | $409 \pm 163$ |
| $\mathfrak{F}^{U}$ | $-359 \pm 40$ | $47.8 \pm 1.9$ |

Wigner function is sufficient to reveal the existence of a state, it does not take in account the proximity to $D^*D$ thresholds. A more advanced parameterization is needed to probe the physical properties of the resonance. An unitarized Breit-Wigner profile $\mathfrak{F}^U$ is considered as an alternative model for the $T_{cc}(3875)^+$ signal, where the energy-dependent width accounts for the $T^+_{cc} \to D^0D^0\pi^+$, $T^+_{cc} \to D^0D^+\pi^0$ and $T^+_{cc} \to D^0D^+\gamma$ decays. The resulting mass, relative to $D^{*+}D^0$ threshold, and the FWHM of the signal are shown in Table 8.5.4 and compared to the results of the $\mathfrak{F}^{BW}$ model. The narrowness of the $T^+_{cc}$ state varies substantially highlighting the relevance of accounting for the $D^*D$ thresholds. Despite the difference in results, both models can describe the data adequately given the mass resolution of about 400 keV/c$^2$. The $T_{cc}(3875)^+$ state is the narrowest exotic state observed to date.

The $D^0D^0\pi^+$ events with a mass below the $D^{*+}D^0$ threshold (Fig. 8.5.7) are selected to study the $D^0\pi^+$ mass distribution which indicates that the $T^+_{cc} \to D^0D^0\pi^+$ decay proceeds via an intermediate off-shell $D^{*+}$ meson.

The peak in the $D^0D^0\pi^+$ could be interpreted as the $I_3 = 0$ component of an isotriplet $(\hat{T}^0_{cc}, \hat{T}^+_{cc}, \hat{T}^+_{cc}+)$ with $cc\bar{u}\bar{u}$, $cc\bar{u}\bar{d}$ and $cc\bar{d}\bar{d}$ quark content, respectively. A search for a $\hat{T}^{++}_{cc}$ state in the $D^+D^0\pi^+$ mass spectrum reports no signal. All the observed properties strongly support the interpretation of the new state as the isoscalar $J^P = 1^+$ $cc\bar{u}\bar{d}$ tetraquark ground state.

Using the $\mathfrak{F}^U$ model, the scattering length $a$, the effective range $r$ [2568], and the compositeness $Z$ [2569] are determined:

$$a = \Big[-(7.16 \pm 0.51) + i\,(1.85 \pm 0.28)\Big] \text{ fm}\,, \quad (8.5.1)$$

$$-r < 11.9\,(16.9) \text{ fm at } 90\,(95)\%\text{ CL}\,, \quad (8.5.2)$$

$$Z < 0.52\,(0.58) \text{ at } 90\,(95)\%\text{ CL}\,. \quad (8.5.3)$$

The real part of the scattering length $a$ can be interpreted as the characteristic size of the state $R_a \equiv -\mathcal{R}e[a] = 7.16 \pm 0.51$ fm which corresponds to a spatial extension as large as expected for molecular states. Within the $\mathfrak{F}^U$ model the resonance pole is found to be located on the second Riemann sheet at $\hat{s} = m_{\text{pole}} -$

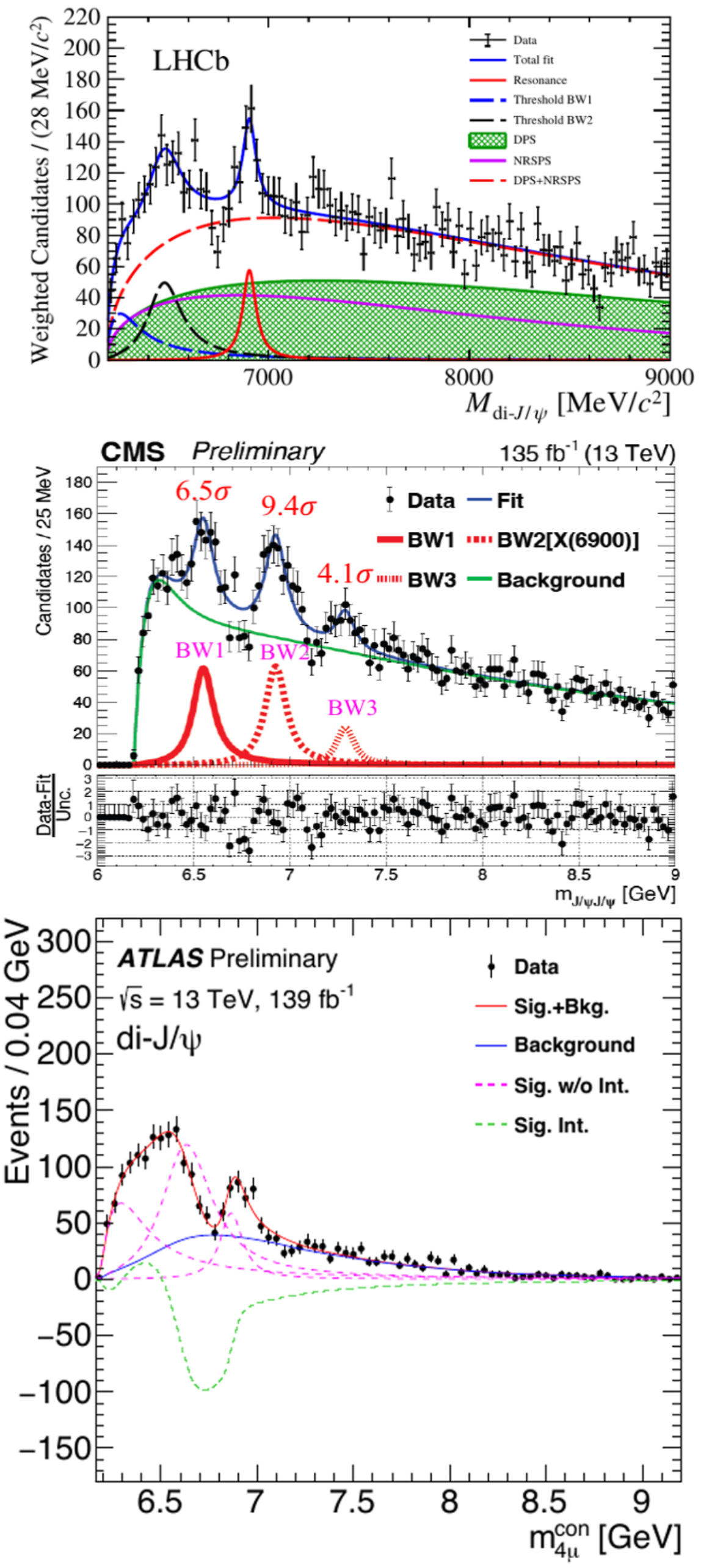

**Fig. 8.5.8** Invariant mass spectra of $J/\psi$ pair candidates by using data recorded bu the LHCb [2565] (top), CMS [2566] (middle) and ATLAS [2567] (bottom) experiments. The three spectra have been aligned for an easier comparison.

$\frac{i}{2}\Gamma_{\rm pole}$, where

$$\delta m_{\rm pole} = -360 \pm 40\,^{+4}_{-0}\ \ {\rm keV/c^2}\,, \tag{8.5.4}$$

$$\Gamma_{\rm pole} = \quad 48 \pm\ \ 2\,^{+0}_{-14}\ {\rm keV}\,. \tag{8.5.5}$$

All exotic hadrons observed so far predominantly decay via the strong interaction; their decay widths vary from a few to a few hundred MeV. The discovery of the $T_{cc}(3875)^+$ meson implies the existence of a $bb\bar{u}\bar{d}$ state that should be stable against strong and electromagnetic interactions: its mass is expected to fall below the $B^{*-}B^0$ and $B^-B^0$ mass thresholds. The observation of a long-lived exotic state will be an intriguing goal for future experiments.

### 8.5.7 The fully charmed tetraquark X(6900)

Many QCD-motivted phenomenological models [2570, 2571] have predicted the existence of states consisting of four heavy quarks $T_{QQ\bar{Q}\bar{Q}}$. In 2020 the LHCb collaboration reported the study of the invariant mass spectrum of the $J/\psi$ pairs where both $J/\psi$ mesons are reconstructed via the $\mu^+\mu^-$ decay [2565]. As a result, the reconstruction efficiency is large due to the presence of muons only in the final state. A pair of $J/\psi$ mesons can be produced in proton-proton collisions at LHC via single (SPS) or double (DPS) parton scattering processes, where the two $J/\psi$ are produced in a single or two separated interactions of gluons or quarks, respectively. The SPS process includes both resonant production via intermediate states, such as $T_{c\bar{c}c\bar{c}}$, and nonresonant production.

The $J/\psi J/\psi$ mass distribution (Fig. 8.5.8) shows a broad structure just above the kinematic threshold and a narrower peak at about 6.9 GeV, dubbed $X(6900)$. An unusual dip also appears between them. The broad structure can be modelled as a superimposition of two Breit-Wigner structures or as an interference between a Breit-Wigner function and the background. The latter model successes to describe also the dip adequately. The presence of the $X(6900)$ state is established in both models, though the natural width is twice larger in the latter.

Recently the CMS [2566] and ATLAS [2567] collaborations have presented preliminary studies of the $J/\psi J/\psi$ spectrum (Fig. 8.5.8). While the $X(6900)$ state is confirmed, there is no consensus on the fit model. Common features are the presence of dips in the spectrum and the need of interference terms to describe it properly. Interestingly the CMS collaboration also claimed the observation and the evidence of two new states $X(6600)$ and $X(7300)$, respectively. A hint of the latter structure was also pointed out by the LHCb collaboration.

Given no single light hadron can mediate the interaction between charmonia to generate a loosely bound molecule, the $X(6900)$ meson seems likely to be a compact tetraquark [2572]. The LHC experiments will profit of larger datasets in a near future which will help to investigate further the resonant nature of the peaks and eventually to measure their spins and parities [2573] in the $J/\psi J/\psi$ and $\psi(2S)J/\psi$ spectra.

Tetraquarks states containing only bottom quarks, $T_{bb\bar{b}\bar{b}}$, have been also searched for by the LHCb and CMS collaborations in the $\Upsilon\mu^+\mu^-$ decay [2574, 2575] but no signal have been observed.

### 8.5.8 Conclusions

The existence of exotic hadronic states with more than minimal quark content ($q\bar{q}$ or $qqq$) was proposed since the birth of the quark model [17, 18]. In the last decades samples of quarkonia larger and larger have been exploited to study their transition and production processes. New and fascinating exotic $X, Y, Z$ states have been observed at a large number of facilities and in different production processes: at tau-charm (BES experiment) and $B$ factories (BaBar and Belle experiments), in hadroproduction at Fermilab Tevatron and the Large Hadron Collider (LHC) at CERN, in photon-gluon fusion at DESY, photoproduction at JLab, and in heavy-ion production and suppression at RHIC, NA60, and LHC. In the upcoming years an unprecedented amount of data will be available from the upgraded experiments CMS, ATLAS, LHCb, ALICE, Belle II and BESIII [1424, 2576–2580] and more data will come in future from Panda at FAIR and the Electron Ion Collider (EIC) [2581, 2582]. The measurements of the quantum numbers and classification of the exotic hadrons in $SU(3)$ flavour multiplets will be an important step to understand the nature of the observed exotic states. In addition it will be important to identify observables which can discriminate between different models. For instance the measurement of the effective range has been suggested as a physical quantity able to determine if the $\chi_{c1}(3872)$ is a compact tetraquark or a loosely-bound molecular state [2583].

## 8.6 Heavy quark-antiquark sector: theory

Nora Brambilla

### 8.6.1 Introduction

Heavy quarks have been instrumental in accessing the strong interactions as they provide a mass scale $m_h$ which is bigger than $LQCD$: at such scale perturbation theory is valid and scale factorization is useful. Among the systems with heavy quarks, systems with two (or more) heavy quarks are very special, being endowed with a pattern of separated energy scales. Quarkonium in particular, a bound state of a heavy quark and a heavy anti-quark, provides a special tool to study strong interactions.

The 1974 discovery of the $J/\psi$ [75, 76], the charmonium ground state, drastically changed and shaped the Standard Model (SM) of particle physics: termed the November revolution, it represented the confirmation of the quark model, the discovery of the charm quark, the confirmation of the GIM mechanism [78] (the mechanism through which flavor-changing neutral currents are suppressed in loop diagrams), and the first discovery of a quark of large mass moving nonrelativistically. It was also the confirmation of QCD in its most peculiar properties of high-energy asymptotic freedom and low-energy confinement [80]. The small width can be explained by the fact that $J/\psi$ is the lowest $c\bar{c}$ energy level and can decay only via annihilation, which makes available in the process a large energy, of order of twice the mass of the charm quark (about 2 GeV). The annihilation width is then proportional to $\alpha_s^2(2m_c)$ which is small due to asymptotic freedom, since $m_c$ is bigger than $\Lambda_{QCD}$. Confinement becomes also manifest in the case of quarkonium, where the color-singlet static quark-antiquark interaction potential can be written in terms of a Wilson loop (see e.g. [2584, 2585]). Confinement emerges as an area law in the Wilson loop [80] , cf. Fig. 8.6.1, and correspondingly a linear potential grows with the distance between the quarks [2586] as $V_0 = \lim_{T\to\infty}(i/T)\ln W$, where $W = \mathrm{Tr}\,\mathrm{P}\exp\{ig_s \oint_{\Gamma_0} dz_\mu \mathcal{A}_\mu(z)\}$, see Fig. 8.6.2 and Section 6.1.

The energy scales involved in quarkonium span from the hard region, where an expansion in the coupling constant is possible and precision studies may be done, to the low-energy region, dominated by confinement and the many manifestations of nonperturbative dynamics. This property underlies its uniqueness and is the reason for which quarkonium plays a crucial role for a number of problems at the frontier of our research, from the investigation of the confinement dynamics in QCD to the study of deconfinement and the phase diagram of nuclear matter, from the precise determination of Standard Model parameters up to the emergence of exotics $X, Y, Z$ states of an unprecedented nature [1424, 1425, 2576–2578, 2585], as we will summarize in the next sections. It is also the reason for which quarkonium should be addressed with effective field-theory methods to take advantage of the scales separation.

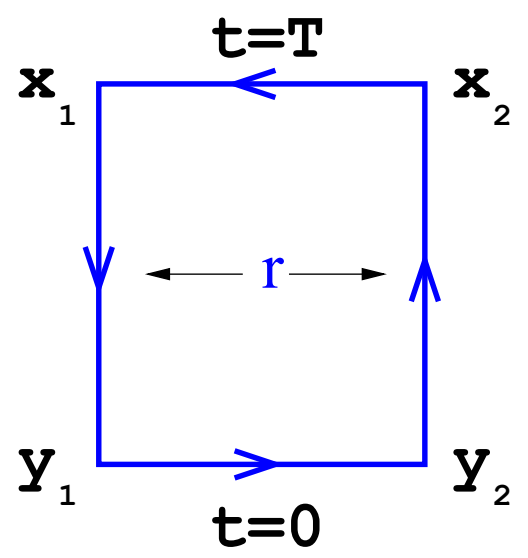

**Fig. 8.6.1** The static Wilson loop along the circuit $\Gamma_0$: it contains the interaction of a static quark-antiquark pair created at a time t = 0 (respectively at space points $\mathbf{y}_1$ and $\mathbf{y}_2$) annihilated at a subsequent large time T (at space points $\mathbf{x}_1$ and $\mathbf{x}_2$) Initial and final states are made gauge invariant by the presence of the Schwinger line. The Wilson area law says that the Wilson loop behavior at large distances is exponential in the area of the loop weighted by the string tension $\sigma$.

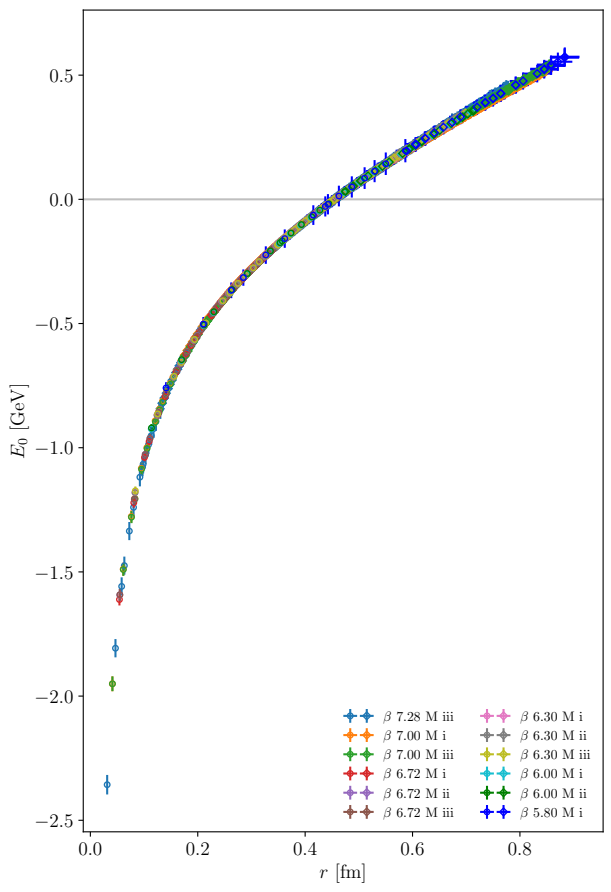

**Fig. 8.6.2** Results for the static potential in physical units for 2+1+1 dynamical quark flavors. The data are from twelve ensembles of varying lattice spacing (keyed by $\beta$) and three choices of light quark mass (denoted "M i", "M ii", "M iii"). Lattice units are eliminated via the $r_0/a$ scale setting, and an unphysical constant is eliminated by setting $V_0(r_0) = 0$. For details see [2587]. This is the first ever determination of the potential with 4 dynamical fermions.

### 8.6.2 Scales and Effective Field Theories

The multiscale nature of quarkonium has made a description within Quantum Field Theory particularly difficult until the advent of non-relativistic effective field theories (NREFTs), cf. Sec. 6.1. When in the eighties of last century, theorists set up to investigate the structure of the energy levels of charmonium and bottomonium, they noticed that it can be reproduced by using a Schrödinger equation with a static potential composed of an attractive Coulomb contribution (with the appropriate $SU(3)$ color factor for a singlet $Q\bar{Q}$) and a term linear in the distance: the famous Cornell potential (see Section 2.1 and Ref. [734, 2588]:

$$V_0(r) = -\frac{\kappa}{r} + \sigma r + \text{const.} \tag{8.6.1}$$

This was the quark model description with the potential inspired by QCD. The parameter $\kappa$ was identified with $\frac{4}{3}\alpha_s$, corresponding to a one-gluon exchange that should dominate at small distances due to asymptotic freedom. The string tension $\sigma$ corresponds to a constant energy density related to confinement and generates a potential growing with the interquark distance $r$ at large distances. A fit to the states gave $\kappa = 0.52$ and $\sigma = 0.182$ GeV$^2$. In order to describe the fine and hyperfine characteristics of the spectrum, relativistic corrections to the static potential have been introduced to account for effects of order $v^2$, i.e. 20% to 30% for the charmonium and up to 10% for the bottomonium spectrum. They appear at the order $1/m_h^2$ in the expansion, involving both spin dependent (spin-spin, tensor and spin-orbit couplings) and purely velocity dependent terms. They were derived in the eighties, either from the semirelativistic reduction of a Bethe–Salpeter equation [824] for the quark-antiquark Green functions (or, equivalently at this level, from the reduction of the quark-antiquark scattering amplitude with an effective exchange) or in some model description like the flux-tube model [2366], for a review see [2584, 2585, 2589]. The problem of these approaches is the lack of a precise connection to QCD. Taking advantage of NREFTs, quarkonium can be described directly in QCD, and in this way it becomes a probe of strong interactions.

The spectrum of quarkonium, see Fig. 8.1.1, clearly states that it is a nonrelativistic system: the difference in the orbital energy levels is much smaller than the quark mass. Defining $v$ as the heavy quark velocity in the rest frame of the meson in units of $c$ (with $v^2 \sim 0.1$ for the $b\bar{b}$, $v^2 \sim 0.3$ for $c\bar{c}$ systems) the energy levels scales like $m_h v^2$, while fine and hyperfine separations scale like $m_h v^4$. This is the same scaling as for the hydrogen atom (identifying $v$ with the fine structure constant $\alpha_{\text{em}}$). This scaling is the signature of a nonrelativistic system. Being nonrelativistic, quarkonia are characterized by a hierarchy of energy scales: the mass $m_h$ of the heavy quark (hard scale), the typical relative momentum $p \sim m_h v$ (in the meson rest frame) corresponding to the inverse Bohr radius $r \sim 1/(m_h v)$ (soft scale), and the typical binding energy $E \sim m_h v^2$ (ultrasoft scale). Of course, for quarkonium there is another scale that can never be switched off in QCD, i.e. $\Lambda_{\text{QCD}}$, the scale at which nonperturbative effects become dominant. A similar pattern of scales emerge in the case of baryons composed of two or three heavy quarks [1411, 1412] and for the just discovered state $X(6900)$ made by two charm and two anticharm quarks. The pattern of nonrelativistic scales makes all the difference between heavy quarkonia and heavy-light mesons, which are characterized by just two scales: $m_h$ and $\Lambda_{\text{QCD}}$.

The correct zero-order problem is thus the Schrödinger equation with potentials. These should, however, be defined and calculated directly in QCD, and nonpoten-tial corrections that should be accounted for. As explained in Sec. 6.1 using the EFT method to integrate out in QCD (in the sector with one heavy quark and one heavy antiquark) the hard scale $m_h$ and the soft scale $m_h v$, give origin to the NREFT called pNRQCD (potential Nonrelativistic QCD) [1422, 1450, 1452]. The pNRQCD description directly addresses the bound state dynamics, implements the Schrödinger equation as zero-order problem, properly defines the potentials as matching coefficients, and allows to systematically calculate relativistic and retardation corrections. Each correction has a size determined by the power counting in $v$ and in $\alpha_s$. The EFT allows us to make model-independent predictions and we can use the power counting to attach an error to the theoretical predictions.

When $mv \gg \Lambda_{\rm QCD}$, we speak about weakly-coupled pNRQCD because the soft scale is perturbative and the potentials can be calculated in perturbation theory. The lowest levels of quarkonium, like $J/\psi, \Upsilon(1S), \Upsilon(2S)\ldots$, may be described by weakly coupled pNRQCD, while the radii of the excited states are larger and presumably need to be described by strongly coupled pNRQCD. All this is valid for states away from the strong-decay threshold, i.e. the threshold for a decay into two heavy-light hadrons. In the first case the dynamical degrees of freedom are $Q\bar{Q}$ pairs in color singlet or color octet configuration and ultrasoft gluons, in the second case just $Q\bar{Q}$ pairs in color singlet. The details of the two theories have been presented in Section 6.1. The nonperturbative physics in pNRQCD is encoded in a few low-energy correlators that depend only on the gluons and are gauge invariant: these are objects in principle ideal for lattice calculations. Strongly coupled pNRQCD allows us to obtain a definition of the potentials that are given in terms of Wilson loops also in generalized form (i.e. with the insertion of chromoelectric and chromomagnetic field in the static loop). The static potential is given by the static Wilson loop described before that was calculated on the lattice since the inception of QCD [80, 326, 1528, 2586, 2590], up to the present state of the art that includes four dynamical quarks in the calculation, see Fig. 8.6.2. Some of these potentials have been obtained before the advent of the EFT in the so-called Wilson-loop approach [80, 768, 1461, 1462, 2591, 2592], but they were missing the contribution of the hard scale. Moreover in the EFT, new (spin-independent) contributions appear at the order $1/m_h$ and at the order $1/m_h^2$ [1516, 1517]. The results of strongly coupled pNRQCD – which are valid in the regime in which $m_h v$ is of order $\Lambda_{\rm QCD}$ and where strong decay thresholds are far away – justify the success of the quark model from the QCD perspective. In fact in this regime the only degree of freedom is the $Q\bar{Q}$ singlet, the dynamics is controlled by the Schrödinger equation and ultrasoft corrections are carried only by pions. The potentials, however, are calculated from QCD and they have a structure that is different from what one gets in models, especially for terms related to momentum dependent contributions. This EFT description allows for modifications that could be used to describe $X, Y, Z$ exotics and (combining with finite temperature QCD and open quantum system) the nonequilibrium evolution of quarkonium in medium, as it will summarized later.

### 8.6.3 The quarkonium potential and confinement

The lattice QCD evaluation of the static Wilson loop clearly displays an area law which is the sign for confinement. Still, it is relevant to investigate the nature of the confinement mechanism. Quarkonium is a golden tool for this aim. Stron/-gly-coupled pNRQCD realizes a scale factorization encoding the low-energy physics in the Wilson loop and its generalized versions, i.e. Wilson loops with insertions of chromoelectric and chromomagnetic fields. All the potentials, static ones and spin and velocity-dependent ones, are given in terms of these gauge invariant nonperturbative objects that no longer depend on the heavy quark degrees of freedom and on the quark flavor. This turns out to be a systematic method to study the QCD confinement properties and put them directly in relation to the quarkonium phenomenology.

The area law emerging in the static Wilson loop at large distance corresponds to the formation of a chromolectric flux tube between the quark and the antiquark that sweeps the area of the loop: this has been directly observed on the lattice, see Fig. 8.6.3. The effects originates from the nonperturbative QCD vacuum that can be imagined as a disordered medium with whirlpools of color on different scales, thus densely populated by fluctuating fields whose amplitude is so large that they cannot be described by perturbation theory [404]. A QCD vacuum model can be established by making an assumption on the behavior of the Wilson loop in the low energy. The relativistic corrections that involve insertions of gluonic fields in the Wilson loop follow via functional derivative with respect to the quark path see [2584, 2592]. One may notice that the part proportional to the square of the angular momentum in the velocity dependent potential at order $1/m_h^2$ takes into account the energy and the angular momentum of the flux tube, which is something that could not be obtained e.g. in any Bethe Salpeter approach with a confining interaction represented by a scalar convolution kernel. The action density or the energy density structure between the static quark and the static antiquark is currently studied both in the lowest energy configuration as well as in the hybrid configurations with excited glue [2593–2595]. The mechanism underlying confinement and flux tube formation has been investigated since long on the lattice [397] using Wilson loops and the 't Hooft abelian projection, to identify the roles of magnetic monopoles [2596, 2597] and center vortices [403], see e.g. the review [1425].

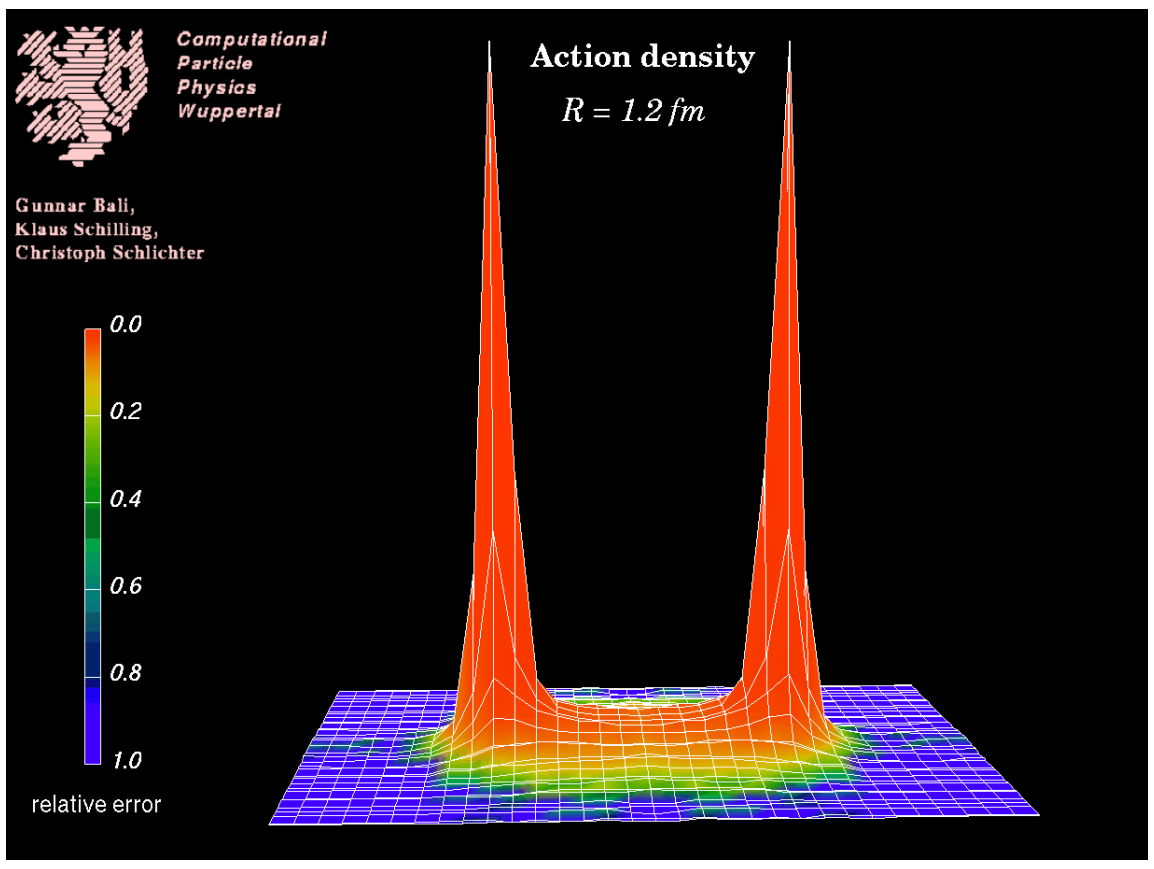

**Fig. 8.6.3** The origin of the linear potential between the static quark and antiquark may be traced back to a flux tube: a string of gluon energy between the quark pair. Here we present the hystorical picture of the action density distribution between a static quark antiquark couple in SU(2) at a physical distance of 1.2 fm, from [2606].

In the continuum, several models of low-energy QCD have been introduced to explain the flux-tube formation. The models vary from the dual Meissner effect and a dual-abelian Higgs-model picture, from dual QCD [2598] to the stochastic vacuum [2599], to the flux tube model [2366] and an effective low energy string description. Each of these models can be used to obtain analytic estimates of the behavior of the generalized Wilson loops for large distance, which in turn give the static potential and the relativistic corrections as function of $r$, see e.g. [1531, 2600–2603]. Similar nonperturbative configurations leading to confinement can be studied analyzing the Wilson loop in case of baryons with three or two heavy quarks [2604, 2605].

### 8.6.4 States below threshold:quarkonium

On the basis of EFTs and lattice calculations we have reached today a comprehensive understanding of the properties of quarkonium below the strong decay threshold.

*Spectra, transition and decays*
The power counting of the EFT allows us to attach an error to each prediction. For states with a small radius one can use weakly-coupled pNRQCD with potentials calculated at high order in perturbation theory (see Sec. 6.1) and retardation effects carried by local or non-local condensates. For states with larger radii, the potentials obtained in strongly-coupled pNRQCD (see Sec. 6.1) have been calculated on the lattice [769, 1526, 1528, 2607–2609], and the full quarkonium phenomenology may be obtained using such potentials in the Schrödinger equation. The imaginary parts of the potentials control the decays.

On the other hand, direct lattice calculations of quarkonium properties along the years have reached realistic values for the dynamical quark flavors, solid continuum limits and have been extended to the excited states, reaching a comprehensive and precise description [1434, 2610–2614]. Electromagnetic M1 and E1 transitions have been calculated in pNRQCD, see e.g. [1511–1514]. There are so many results that it is impossible to discuss all of them here and we refer to some reviews [1422, 1424, 1425, 1478, 2576].

Summarizing: today we understand in a precise way, on the basis of QCD, most of the properties of the quarkonium states lying below the strong decay threshold. This has a great impact for the physics of quarkonium, as we explain with examples in the next paragraph.

*Precise determination of SM parameters*
In the regime in which the soft scale is perturbative, pNRQCD enables precise and systematic higher order calculations of bound states allowing us to extract precisely standard-model parameters like the quark masses and $\alpha_s$. For example, it has been possible to extract a precise value of $\alpha_s$ at rather low energy by comparing a short-distance lattice calculation with 2+1 flavors of the static energy and a next-to-next-to-next leading order (NNNLO, $\alpha_s^4$) perturbative pNRQCD calculation of the same quantity, including also ultrasoft logs resummation. When $\alpha_s$ is extrapolated to the mass of the Z, $\alpha_s(M_Z) = 0.11660^{+0.00110}_{-0.00056}$ is obtained. This is a competitive extraction made at a pretty high order of the perturbative expansion [2615, 2616]. This method of $\alpha_s$ extraction is now used by several groups, see e.g. [2617, 2618]. The QCD static force, defined in terms of a single chromolectric insertion in the Wilson loop could be used as well to the same scope [2619, 2620].

In the same way precise values of the bottom and charm masses can be extracted from measurements of the masses of the lowest states and by comparing them to the formula for the energies in pNRQCD at NNNLO. The renormalon ambiguity between the mass and the static potential cancels and a pretty good determination is possible, see e.g. [1478, 1487, 1497] and references therein. Similar methods [1477, 1497] can be used to describe the top anti-top S-wave pair-production cross section near threshold in $e^+e^-$ annihilation and to study the possible achievable accuracy of top-quark mass measurement expected at a future linear collider. A precise determination of the top quark mass is very important for precision tests of the SM, and also due to its crucial role in the vacuum stability of the SM at a very high energy scale. Hence, progress in our understanding of heavy quarkonia leads to an access to key aspects of the SM.

### 8.6.5 Production

Production of heavy quarkonium has been extensively studied along the years at the Tevatron collider at Fermilab, at Hera at DESY, at B factories and in particular at the LHC where quarkonium production with high statistics at unprecedented values of $p_T$ is measured [1424, 1425, 1437, 1438, 2576–2578]. This is a complex problem encompassing many physical scales still not fully understood, and it can be used to test and extend our understanding of factorization theorems, which are the foundation for all the perturbative calculations in QCD. New theoretical concepts that have been developed here, e.g. arising from kinematic enhancements and from large endpoint logarithms, could have wider applicability in the calculation of high-energy cross sections. Quarkonium production is also relevant to BSM, as certain quarkonium production processes can be used to measure Higgs couplings.

The standard method for calculating quarkonium production rates is the NRQCD [1436] (see Sec. 6.1) factorization approach, where production rates are expressed as perturbatively calculable partonic cross sections multiplied by nonperturbative constants called NRQCD long-distance matrix elements (LDMEs) which are universal. The NRQCD factorization approach is a conjecture that has not been proven to all orders in $\alpha_s$. Another important theoretical development is the next-to-leading-power (NLP) fragmentation approach [1277, 2023], in which quarkonium production rates are expressed as perturbatively calculable partonic cross sections convolved with fragmentation functions, up to corrections suppressed by a factor $m_h^4/p_T^4$. The NLP fragmentation approach becomes more predictive if NRQCD factorization is used to express the fragmentation functions in terms of NRQCD matrix elements. This organizes the NRQCD factorization expression for the cross section according to powers of $m_h^2/p_T^2$, which simplifies the calculation of higher-order corrections and the resummation of large logarithms. NRQCD factorization predictions have now been computed at NLO in $\alpha_s$ for many production processes.

The NRQCD approach has brought a great progress into the field even though not all experimental data are understood coherently, and the extraction of the LDMEs remains a complex enterprise [2621–2627]. Recently, it has been possible to factorize the quarkonium production-cross sections at lower energy in pNRQCD [1533–1535], rewriting the octet NRQCD LDMEs, which are the nonperturbative unknowns, in terms of products of wave functions and gauge invariant low energy correlators depending only on the glue and not the on flavor quantum numbers. This allows to reduce by half the number of LDMEs, opens up the possibility of their lattice evaluation and may lead to further progress of the field.

### 8.6.6 Nonequilibrium evolution in medium

The properties of production and absorption of quarkonium in a nuclear and hot medium are crucial inputs for the study of QCD at high density and temperature (see Sec.7), reaching out to cosmology.

Heavy ions experiments at the LHC at CERN and at the RHIC at BNL aim at producing the Quark Gluon Plasma (QGP): heavy quarks are good probes of this hot QCD medium. They are produced at the beginning of the collision and remain up to the end. As we discussed, quarkonia are special hard probes as they are multi-scale systems. In the medium besides the energy scales of quarkonium, also the thermal scales of the QGP have to be considered (cf Sec. 6.5): the scale $\pi T$ related to the temperature, the Debye mass $m_D \sim gT$, with $\alpha_s = g^2/4\pi$, related to the (chromo) electric screening and the scale $g^2T$ related to the (chromo) magnetic screening. In a weakly-coupled plasma, the scales are separated and hierarchically ordered, in a strongly coupled plasma, $m_D \sim T$. To calculate QCD at finite $T$ in real time, Hard Thermal Loop EFT can be used to integrate out the temperature scale. Heavy quarkonium dissociation has been proposed a long time ago as a clear probe of QGP formation through the measurement of the dilepton decay-rate [2112]. The dissociation was related to the screening of the quark-antiquark interaction due to the Debye mass and it was suggested that dissociation would manifest itself in an exponential screening term $\exp(-m_D r)$ in the static potential. One of the key quantities measured in experiments is the nuclear modification factor $R_{AA}$, a measure for the difference of quarkonium production in pp and in nucleus-nucleus collisions. Since higher excited quarkonium states have larger radii, the expectation was that, as the temperature increases, quarkonium would dissociate first for the higher-mass and then for lower-mass states giving origin to sequential melting [2112].

In the last decades, using pNRQCD at finite T [1552, 2119], it has been possible to actually define and calculate the $Q\bar{Q}$ potential in medium. In perturbative calculations it was found that the thermal part of the static potential has a real part (roughly described by the free energy) and an imaginary part. The imaginary part comes from two effects: the Landau damping [1552, 1553, 2113], an effect existing also in QED, and the singlet to octet transition, existing only in QCD [1552]. Which one dominates depends on the ratio between $m_D$ and $E$. In the EFT one could show that the imaginary part of the potential related to the Landau damping comes from inelastic parton scattering [1556] and the singlet to octet transition from gluon dissociation [1555]. The existence of the imaginary part, first realized in Ref. [2113], changed our paradigm for quarkonium suppression: it has become clear that the state dissociates well before the conventional screening becomes active [1553, 2113]. A similar pattern emerges in lattice nonperturbative calculations of the potential [2628, 2629].

So far, we have discussed an equilibrium description. However, the evolution of quarkonium in the QGP is an out-of-equilibrium process in which many effects enter: the hydrodynamical evolution of the plasma and the production, dissociation and regeneration of quarkonium in the medium, to quote the most prominent ones. It is necessary therefore to introduce an appropriate framework to describe the real-time nonequilibrium evolution. Recently, using the formalism of open quantum system (see Sec. 6.6) and pNRQCD, it has been possible to describe the nonequilibrium evolution of bottomonium inside a strongly coupled QGP, in a way that incorporates the quantum effects, conserves the number of heavy quarks and considers both color singlet and color octet quarkonium degrees of freedom as well as their recombination [1559, 1560]. The results not only describe well the $R_{AA}$s measured at LHC [1562, 2110] (see also [2118, 2630]), but they allow also to establish a connection with QCD, since the quarkonium evolution depends only on two transport coefficients given in terms of QCD gluonic correlators characterizing the QGP [1559, 1560]. For a review of open quantum system approaches for quarkonium, see [1561, 2631].

### 8.6.7 States at and above threshold: $X, Y, Z$ Exotics: intro

As explained in Sec. 8.5, the spectroscopy of charmonium and bottomonium states at or above the open-heavy-flavor thresholds have reserved us several surprises. Experiments at $e^+e^-$ and hadron colliders have discovered many new, unexpected states in the last decades, cf. Sec 8.5 and Fig. 8.5.1. Many of these states are surprisingly narrow, and some have electric charge. The observations of these charged quarkonium states are the first definitive discoveries of manifestly exotic hadrons. These results challenge our understanding of the QCD spectrum. The $X, Y, Z$ offer us a unique opportunity to investigate the dynamical properties of strongly correlated systems in QCD.

As mentioned in Sec. 8.5, These states have been termed $X, Y, Z$ in the discovery publications, without any special criterion. Meanwhile, the Particle Data Group (PDG) has proposed a new naming scheme [2632], that extends the scheme used for ordinary quarkonia, in which the new names carry information on the $J^{PC}$ quantum numbers, see [1388] for more details. Since the situation is still in evolution we will stick to $X, Y, Z$ names. The field is in enormous and very fast development both experimentally and theoretically, with a continuous flux of new papers: we refer to reviews to account for this development [1388, 1424, 2576, 2633–2638]. The $X, Y, Z$ states offer us unique possibilities for the investigation of the dynamical properties of strongly correlated systems: we should develop the tools to gain a solid interpretation from the underlying field theory, QCD. This is a very significant problem with trade off to other fields featuring strong correlations and pretty interesting connections to heavy ion physics, as propagation of these states in medium may help us to scrutinize their structure and composition.

### 8.6.8 $X, Y, Z$ models and degrees of freedom

Since the $X(3872)$ discovery in 2003, a wealth of theoretical papers appeared to investigate the characteristics of the exotics. Most papers are based on models, which involve a choice of some dominant degrees of freedom and an assumption on their interaction hamiltonian. In the case of states particularly close to their heavy-light threshold, with a very small binding energy and a large scattering length, a more universal picture based on an effective-field-theory molecular description has been put forward [2638–2642] and along the years it has been refined arriving at detailed calculations of the line shapes and the production properties.

A priori the simplest system consisting of only two quarks and two antiquarks (generically called tetraquarks) is already a complicated object and it is unclear whether or not any kind of clustering occurs in it. To simplify the problem, models focus on certain substructures and investigate their implications: in hadroquarkonia the heavy quark and antiquark form a compact core surrounded by a light-quark cloud; in compact tetraquarks the relevant degrees of freedom are compact diquarks and antidiquarks; in the molecular picture two color singlet mesons are interacting at some typical distance: we have no chance here to illustrate all these models and we refer to some recent reviews [1388, 2634, 2636, 2638]. Discussions about exotics therefore often concentrate on the "pictures" of the states, for example the tetraquark interpretation against the molecular one (of which both several different realizations exist). However, as a matter of fact, all the light degrees of freedom (light quarks, glue, in different configurations) should be there in QCD close or above the strong decay threshold: it is a result of the strong dynamics which one sets in, and when, and which configuration dominates in a given regime.

Even in an ordinary quarkonium, which has a dominant $Q\bar{Q}$, subleading contributions of the Fock space may contribute, which have additional quark-antiquark pairs and active gluons. However, in the most interesting region, close or above the strong decay threshold, where the $X, Y, Z$ have been discovered, the situation is more complicated: there is no mass gap between quarkonium and the creation of a pair of heavy-light mesons, nor to gluon excitations, therefore many additional states appear and are dynamical degrees of freedom to be considered [2633]. Still, $m_h$ is a large scale, and a scale factorization is applicable so that nonrelativistic QCD is still valid. There is still another scale separation that can be used to introduce a description of the bound state similar to what is done in pNRQCD, in which the zero order problem is the Schrödinger equation. Let us consider bound states of two nonrelativistic particles and some light-quark degrees of freedom, e.g. molecules in QED or quarkonium hybrids ($Q\bar{Q}g$ states) or tetraquarks ($Q\bar{Q}q\bar{q}$ states) in QCD: electrons, gluon fields or light quarks fields change adiabatically in the presence of heavy quarks or nuclei. In this situation the interaction between the heavy quarks or the one between nuclei due to the electron cloud may be described at leading order of a nonrelativistic expansion by an effective static energy (or potential) $E_\kappa$ between the static sources where $\kappa$ labels different excitations of the light degrees of freedom. A plethora of states can be built on each of the static energies $E_\kappa$ by solving the corresponding Schrödinger equation, see Fig. 8.6.4 and 8.6.5. Based on this scale separation one may describe hybrids and tetraquarks using a Born-Oppenheimer (BO) description, similarly to what is done in molecular systems. On the basis of this, the QCD static energies in presence of a static quark and a static antiquark can be classified according to representations of the symmetry group $D_{\infty h}$, typical of diatomic molecules, and labeled by $\Lambda_\eta^\sigma$ (see Fig. 8.6.6): $\Lambda$ is the rotational quantum number $|\hat{\vec{r}} \cdot \vec{K}| = 0, 1, 2, \ldots$, with $\vec{K}$ the angular momentum of the gluons (or in general the nonperturbative collective degrees of freedom), that corresponds to $\Lambda = \Sigma, \Pi, \Delta, \ldots$; $\eta$ is the CP eigenvalue ($+1 \equiv g$ (gerade) and $-1 \equiv u$ (ungerade)); $\sigma$ is the eigenvalue of reflection with respect to a plane passing through the $Q\bar{Q}$ axis. The quantum number $\sigma$ is relevant only for $\Sigma$ states. In general there can be more than one state for each irreducible representation of $D_{\infty h}$: higher states are denoted by primes, e.g., $\Pi_u$, $\Pi_u'$, $\Pi_u''$, .... In presence of a light quark that takes part in the binding, isospin quantum numbers should be added. The QCD static energies, $E_\Gamma$ in Fig. 8.6.7, have been calculated on the lattice in NRQCD more than 20 years ago [2643], $\Gamma$ representing a choice for $\Lambda_\eta^\sigma$ or in short for the collective quantum number $\kappa$. These lattice calculations uses Wilson loops with initial and final states encoding the given quantum numbers, to select a given symmetry. The Born Oppenheimer approximation idea has been first exploited in phenomenological applications by [1546, 1548]. This picture may be made precise inside QCD using NREFTs and has the possibility to subdue many different models and pictures. In the next section the content of BOEFT and its implications will be presented.

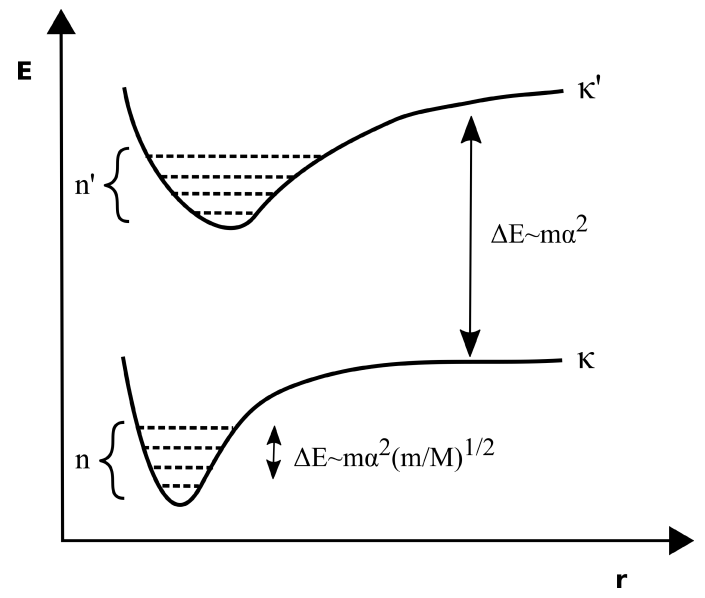

**Fig. 8.6.4** Pictorial view of electronic static energies in QED, labelled by a collective quantum number $\kappa$.

### 8.6.9 BO Effective Field Theory

Starting from pNRQED/pNRQCD the BO approximation can be made rigorous and cast into a suitable EFT called Born-Oppenheimer EFT (BOEFT) [1388, 1421, 1547, 1549, 2644–2646] which exploits the hierarchy of scales $\Lambda_{\rm QCD} \gg m_h v^2$.

In Ref. [1549] the BOEFT that describes hybrids has been obtained. In particular, the static potentials and a set of coupled Schrödinger equations were derived and solved to produce the hybrids multiplets for the two first static energies $\Sigma_u^-$ and $\Pi_u$. Such static energies are degenerated at short distance where the cylindrical symmetry gets subdue to a $O(3)$ symmetry and are then labelled by the quantum number of a gluonic oper-ator $1^{+-}$ called a gluelump. The hybrid static energies are described by a repulsive octet potential plus the gluelump mass in the short distance limit. The $O(3)$ symmetry is broken at order $r^2$ of the multipole expansion. In the long distance regime the static energies display a behavior linear in $r$ , cf. Fig. 8.6.7. The gluelump correlator can be calculated on the lattice to determine the gluelump mass. It depends on the scheme used (the scheme dependence cancels against the analogous dependence in the quark mass and in the octet static potential) but it is of the order of 800 MeV. The hybrid multiplets $H_i$ are constructed from the solution of the Schrödinger equations in correspondence to their $J^{PC}$ quantum numbers. The coupling between the different Schrödinger equations is induced by a non-adiabatic term, known in the Born-Oppenheimer description of diatomic molecules, induced by the non-commutation between the kinetic term and a projector of the cylindrical symmetry in the BOEFT lagrangian. The degeneracy of the static energies at small distance induces a phenomenon called $\Lambda$ doubling, removing the degeneration between multiplets of opposite parity. This phenomenon is known in molecular physics but with smaller size. This and the structure of the multiplets differ from what is obtained in models for the hybrids, cf. [1549]. The BOEFT hybrid multiplets can be compared to neutral exotic states measured in the bottomonium and charmonium sector [1388]: there are many experimental candidates and to make clear identifications one should study also the decay and production properties in the same framework.

The picture can be generalized to tetraquarks by considering static energies classified by the $D_\infty$ quantum numbers and isospin quantum numbers, extracted from lattice evaluation of the static energies of system of a heavy quark, a heavy antiquark and two light quarks [1421].

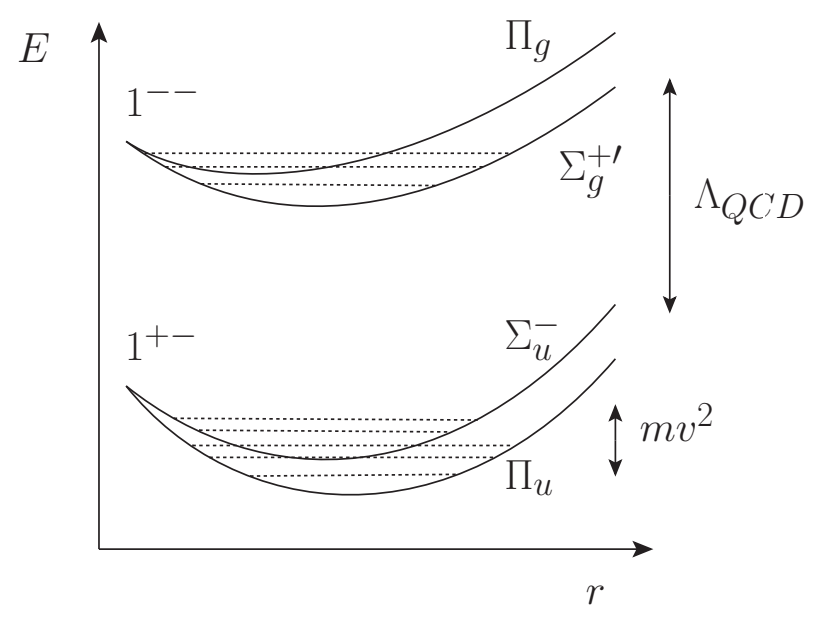

**Fig. 8.6.5** Pictorial view of the QCD static energies, $E_\Gamma$, in QCD. The collective quantum number $\kappa$ has been detailed in $\Lambda_\eta^\sigma$ as explained in the text. It corresponds to the actual lattice results in Fig. 8.6.7.

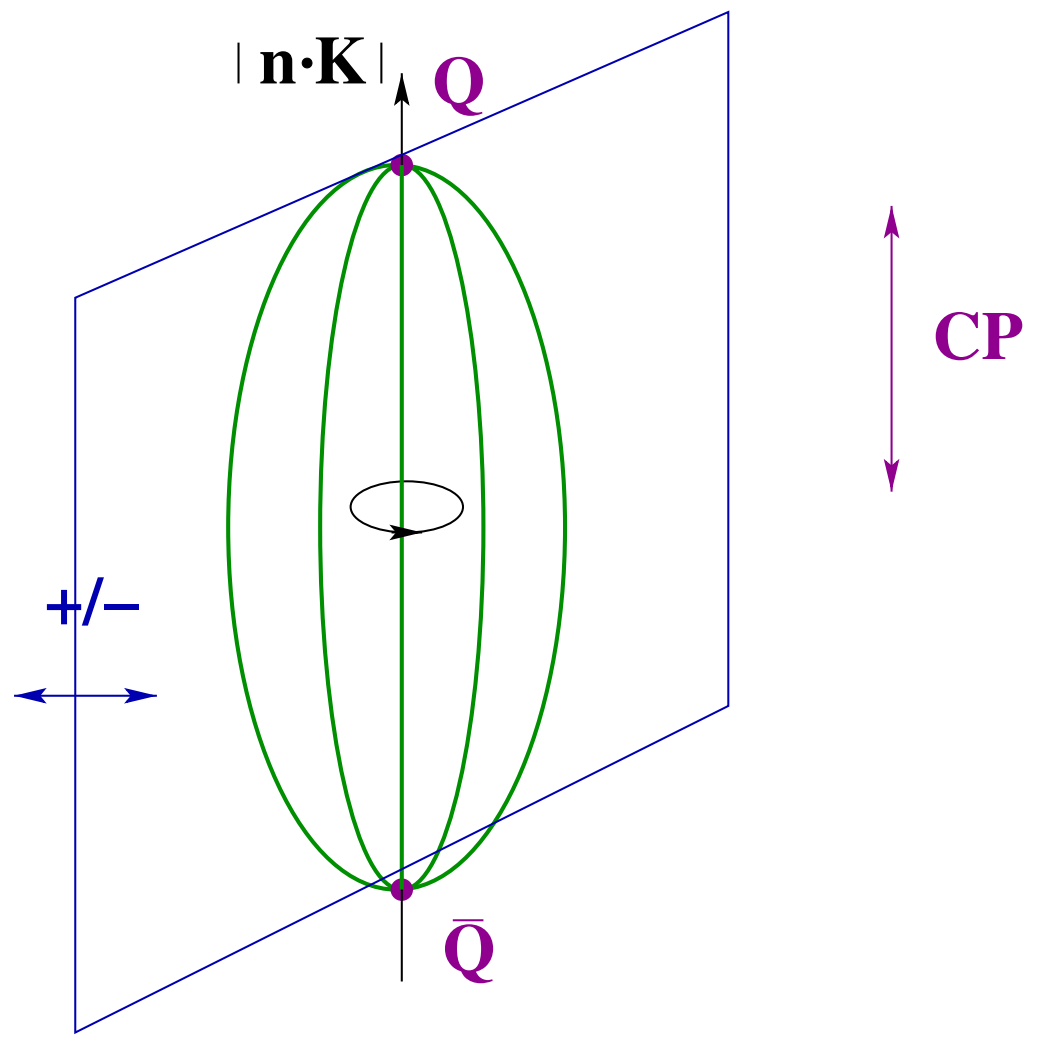

**Fig. 8.6.6** Symmetries of a system with a static quark and a static antiquark and a nonperturbative cloud (gluonic, light quarks) and respective quantum numbers.

*Exotic spin structures, decays*

In Refs. [2645, 2646] the spin-dependent potential of hybrids has been obtained at order $1/m_h$ and $1/m_h^2$ in the quark mass expansion. The potential turned out to be rather different from the spin potential known from standard quarkonium. In fact, a $1/m_h$ contribution appears due to the coupling of the angular momentum of the gluonic excitation (which is not suppressed in $m_h$) with the total spin of the heavy-quark-antiquark pair. Among the $1/m_h^2$ operators are the standard spin-orbit, total spin squared, and tensor spin operators respectively, which appear for standard quarkonia. But now three novel operators appear in addition. So interestingly and differently from the quarkonium case, the hybrid potential gets a first contribution already at order $\Lambda_{\rm QCD}^2/m$. Hence, spin splittings are remarkably less suppressed in heavy quarkonium hybrids than in heavy quarkonia: this will have a notable impact on the phenomenology of exotics. The nonperturbative low-energy correlators appearing in the factorization can be extracted by fixing them using lattice data on the masses of charmonium hybrids [2613], Then, all bottomonium-hybrid spin-multiplets (more difficult to evaluate on the lattice [1434]) can be predicted. The BOEFT is therefore able to predict all spin hybrid multiplets, including their decays and transitions [1388].

*Avoided Level Crossing*

In the BOEFT the information is carried by the QCD static energies and a few purely gluonic low-energy correlators. The information is relevant, however to describe the static energies in the region close to the threshold of two heavy-light mesons. A phenomenological de-scription has been put forward recently [2648, 2649] inspired by lattice calculation in which the avoided-level-crossing-effect is exploited to construct a set of coupled Schrödinger equations and a procedure for the calculation of open-flavor meson-meson scattering cross sections from diabatic potentials. In this framework, the $X(3872)$ naturally emerges [2648]. Such a description can be carried over to BOEFT.

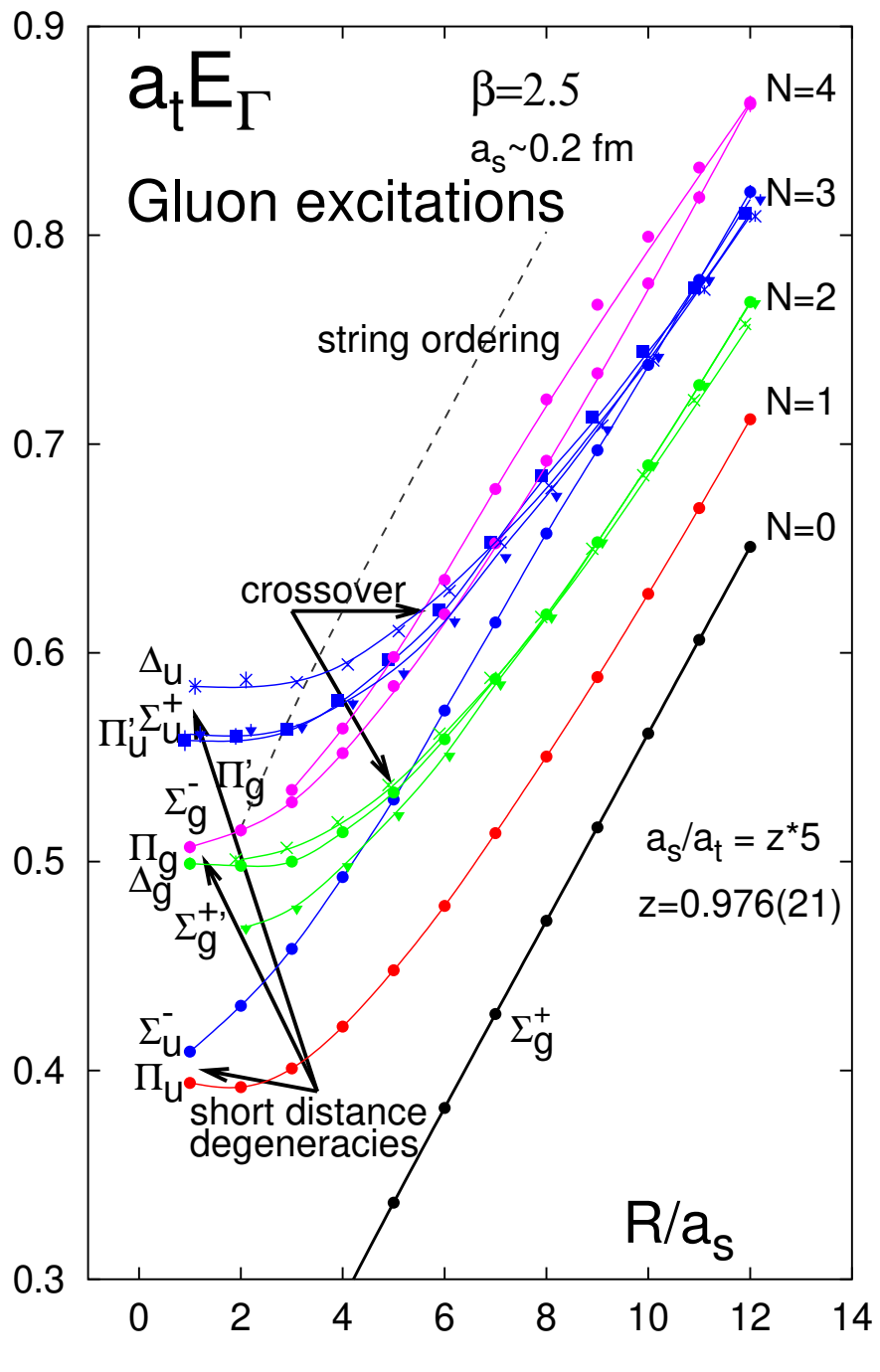

**Fig. 8.6.7** Hybrid static energies labeled by the $D_\infty$ group quantum number, $E_\Gamma$, in lattice units, from the historical Ref. [2643]. For updated calculations see [2647].

The BOEFT may be used to describe also tetraquarks, double heavy baryons and pentaquarks [1421, 1547]. In the case of tetraquarks, a necessary input of the theory is the calculation on the lattice of the generalized Wilson loops with appropriate symmetry and light quark operators. Note that besides the quantum number $\kappa$ also the isospin quantum numbers $I = 0, 1$ have to be considered. It is interesting to note that the BOEFT approach reconciles the different pictures of exotics based on tetraquarks, molecules, hadroquarkonium...In fact in the plot of the static energy as a function of $r$ for a state with $Q\bar{Q}q\bar{q}$ or $Q\bar{Q}g$ we will have different regions: for short distance a hadroquarkonium picture would emerge, then a tetraquark (or hybrid) one and, when passing the heavy-light mesons line, avoided-cross-level effects should be taken into account and a molecular picture would emerge. QCD would then dictate, through the lattice correlators and the BOEFT characteristics and power counting, which structure prevails and in which precise way. In addition production and suppression in medium may be described in the same approach [1534, 2109].

### 8.6.10 $X, Y, Z$ Lattice

Lattice QCD plays a key role for the description of exotics [2612, 2650]. For what concerns BOEFT, nonperturbative input from the lattice is needed in the form of static energies, gluelumps correlators, insertions of chromoelectric fields on hybrids states, for a full list see [2633]. Lattice groups have started to calculate some of these crucial quantities [2651–2654].

Direct lattice calculations of the spectrum and properties of exotic states at and above thresholds are extremely challenging. These states are resonances in the pertinent multi-hadron channels and to obtain their properties, scattering amplitudes in the relevant kinematic range should be calculated on the lattice. The approach is based on Lüscher's work. Later generalizations give access to scattering amplitudes of two-hadron elastic channels, of multiple coupled two-hadron inelastic channels, and of three-hadron channels. While the Lüscher method for a single two-hadron elastic channel provides a straightforward mapping between scattering amplitudes and finite-volume energies, this connection is lost for the multi-channel case, and a parameterization of the amplitude is needed. Abundant and precise energy eigenvalues in a given kinematic range should be obtained to constrain these multi-parameter forms, with solid systematic uncertainties. As the calculations move toward physical values of the light-quark masses, the multi-hadron thresholds move towards lower energies and the number of kinematically allowed hadronic channels that need to be included in the determination of scattering amplitudes increases, making everything more challenging. Still interesting information about some exotics mesons could be obtained in these direct lattice calculations, see e.g. [549, 2655].

### 8.6.11 Summary

Quarkonium has been at the origin of QCD. It has been a long way to arrive at describing quarkonium within QCD. It has payed off, making quarkonium a special probe of strong interactions at zero and at finite temperature. We are now in the process to attack the next frontier, i.e. to develop a coherent, field-theory-based description of exotic quarkonium states. Notice, that if new physics involves nonrelativistic bound states, then the techniques that have been developed for understanding quarkonia will be directly applicable. This holds for example for studies of pairs of heavy dark matter in the evolution in the early universe, that well match the studies of the nonequilibrium evolution of quarkonium in medium, or for the production and the spectroscopy of heavy particles of BSM.

# 9 Baryons

*Conveners:*

**Volker Burkert and Franz Gross**

As we are trying to make progress in the complex world of physical sciences, we should not lose sight of what physics is all about: understanding the origin and the history of our universe, and the laws underlying the observations. In this section we also address how excited states of the nucleon fit in to our understanding of the forces and the dynamics of matter in the history of the universe. On the internet we find beautiful representations of the phases through which the universe evolved from the Big Bang (BB) to our times as shown in Fig. 9.0.1.

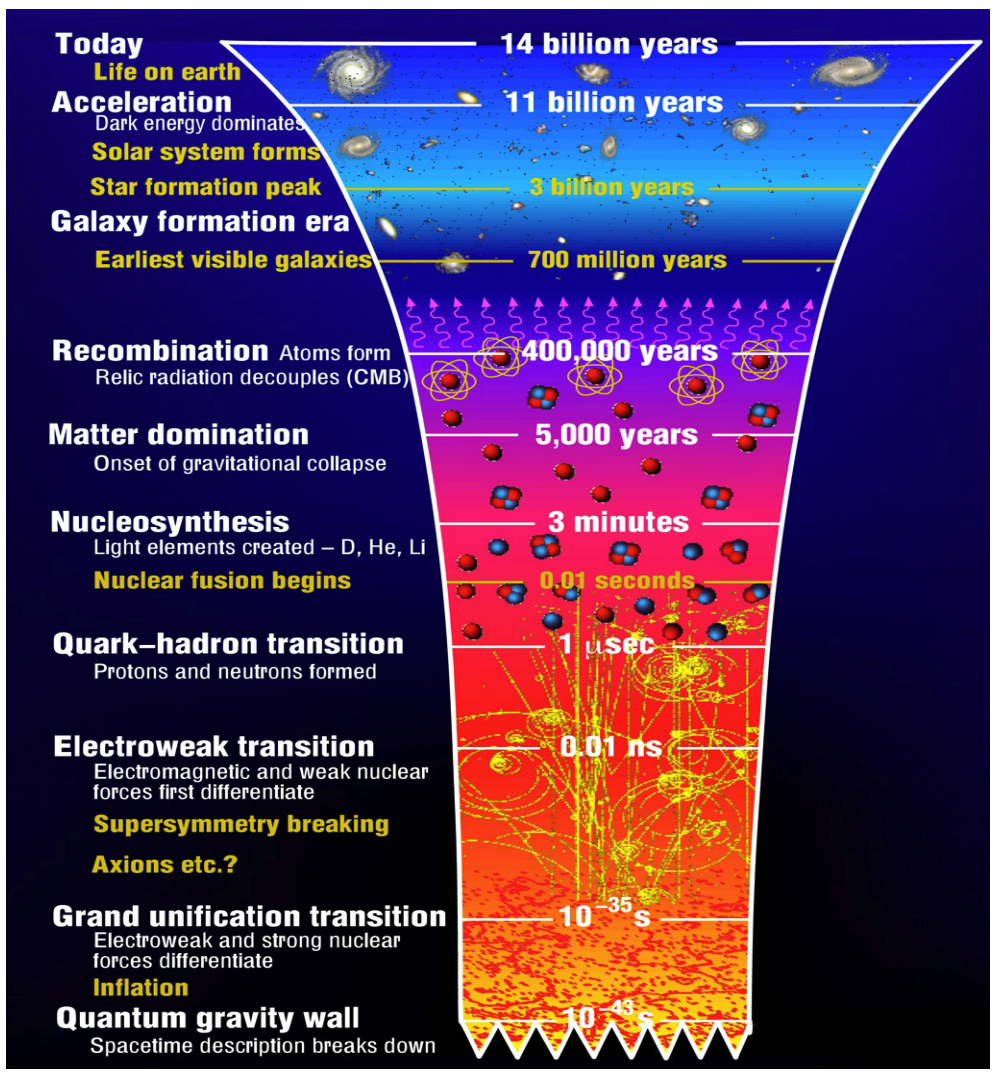

**Fig. 9.0.1** The evolution of the Universe. The line denoted as *Quark-hadron transition*, is where protons and neutrons are formed.

Existing electron accelerators as CEBAF, ELSA, and MAMI, and colliders as BES III have sufficient energy reach to access this region and study processes in isolation that occurred during this transition in the microsecond old universe and resulted in the freeze out of baryons. There are some marked events that have been of particular significance during the early phases of the its history, such as the quark-gluon plasma of non-interacting colored quarks and gluons, and the forming of protons and neutrons. During this transition dramatic events occur - chiral symmetry is broken, quarks acquire mass dynamically, baryon resonances occur abundantly, and colored quarks and gluons are confined. This crossover process is governed by the excited hadrons. During this period strong QCD (sQCD) emerges as the process describing the interaction of colored quarks and gluons.

These are the phenomena that we are exploring with electron and hadron accelerators - the full discovery of the baryon (and meson) spectrum, the role of chiral symmetry breaking and the generation of dynamical quark mass in confinement. While we can not recreate the exact condition in the laboratory, with existing accelerators we can explore these processes in isolation. With electron machines and high energy photon beams in the few GeV energy range we search for undiscovered excitation of nucleons and other baryons.

In this section, Capstick and Crede give an overview over the spectrum of light-quark baryons, followed by a review o the present experimental status by Burkert, Klempt and Thoma. The structure of baryon resonances is explored in electroproduction experiments (Burkert). The section ends with a review of baryons with heavy quarks.

## 9.1 Theoretical overview of the baryon spectrum

**Simon Capstick and Volker Crede**

### 9.1.1 Overview

This contribution examines the constraints on the baryon spectrum imposed by general considerations of flavor, rotational, parity, and particle-exchange symmetries, which lead to a classification scheme for excited baryons. Theoretical approaches to a description of the baryon spectrum based on constituent quark models with various methods for treating the short-range range interactions between quarks are described, and are contrasted to investigations of the spectrum based on lattice and Dyson-Schwinger equation approaches to QCD. Models predict more excited states than are present in the spectrum extracted from data; considerations of how these missing states decay point to alternative ways to produce them, and how to detect their presence once produced. Finally, hybrid baryons with explicit gluonic excitations and the prospect for their discovery are discussed. More detail is given in, for example, reviews of the theoretical approach to the baryon spectrum in Refs. [2656–2658], and reviews of recent experimental developments in Refs. [2659, 2660].

### 9.1.2 Symmetry, group theory, and constraints on the baryon spectrum

*Exchange symmetry, baryons, and the color degrees of freedom*

The development of $SU(3)_f$ and its isospin subgroup in order to understand the proliferation of what are now known as ground-state baryons led to an understanding that there are states with flavor wave functions that are totally symmetric under exchange of up and down quarks, which are identical in the isospin-symmetric limit. An example is the isospin-3/2 baryon, $\Delta$, with the four charge states $\Delta^{++}$, $\Delta^{+}$, $\Delta^{0}$, and $\Delta^{-}$, three of which were discovered in early $\pi N$ elastic scattering experiments with charged pion beams, and shown by examining their strong decays to have spin and parity $J^P = 3/2^+$. This led to a paradox: Ground states of few-body systems made up of identical particles usually have spatial wave functions with orbital angular momentum and parity $L^P = 0^+$, and are exchange symmetric. This implies a total quark spin $S = 3/2$, which is also totally symmetric under exchange of the spin-1/2 quarks. However, as fermions, the Pauli principle requires that the wave function of these baryons in the product flavor, spatial, and spin space is totally *antisymmetric.*

The solution is to assign to the quarks an additional degree of freedom, and a wave function in this degree of freedom which is totally antisymmetric under exchange of the quarks. The simplest way to do this is with a three-valued degree of freedom, now called color. QCD was developed when it was realized that this would result from an $SU(3)_c$ symmetry, where the strong interactions are independent of rotations of the quarks in the color space, with baryons as automatically totally antisymmetric color singlets. This naturally led to a gauge theory which could be the basis for the strong interactions between quarks, and by extension, between all hadrons.

*Flavor symmetry in baryons*

There is an approximate $SU(3)_f$ symmetry of the strong interactions under exchange of the quarks $u$, $d$, and $s$, which is broken by the higher mass of the strange quark. Gell-Mann [1564] and Okubo [15] were able to write a mass formula for ground-state decuplet ($J^P = 3/2^+$), shown in Fig. 9.1.1, and separately for octet ($J^P = 1/2^+$) baryons, shown in Fig. 9.1.2, in terms of the eigenvalues of the two generators of the Lie algebra of $SU(3)_f$ that can be simultaneously diagonalized. These generators are the third component of isospin $I_3$, and the hypercharge $Y = B + S$, where $B$ is baryon number and $S$ is strangeness ($S = -1$ for a strange quark, for historical reasons). Hypercharge is represented by the diagonal matrix $(1, 1, -2)$ in the $\{u, d, s\}$ flavor space. The Gell-Mann-Okubo mass formula ascribes the breaking of symmetry in hadrons to differences in the hypercharge, now understood to be due to the larger mass of the strange quark. It is realized in the ground-state octet baryons as

$$(M_N + M_\Xi)/2 = (3M_\Lambda + M_\Sigma)/4,$$

which holds to a fraction of a percent accuracy, and in ground-state decuplet baryons as the equal-spacing rule

$$M^*_\Sigma - M_\Delta = M^*_\Xi - M^*_\Sigma = M_\Omega - M^*_\Xi,$$

each approximately 147 MeV, which can be thought of as the difference in the strange and average light $(u, d)$ quark masses. The latter led to the prediction by Gell-Mann [2661] of the existence at around 1680 MeV of the decuplet $\Omega$ baryon, made up of three strange quarks. Although the formula is phenomenological, it is now understood in the context of chiral perturbation theory.

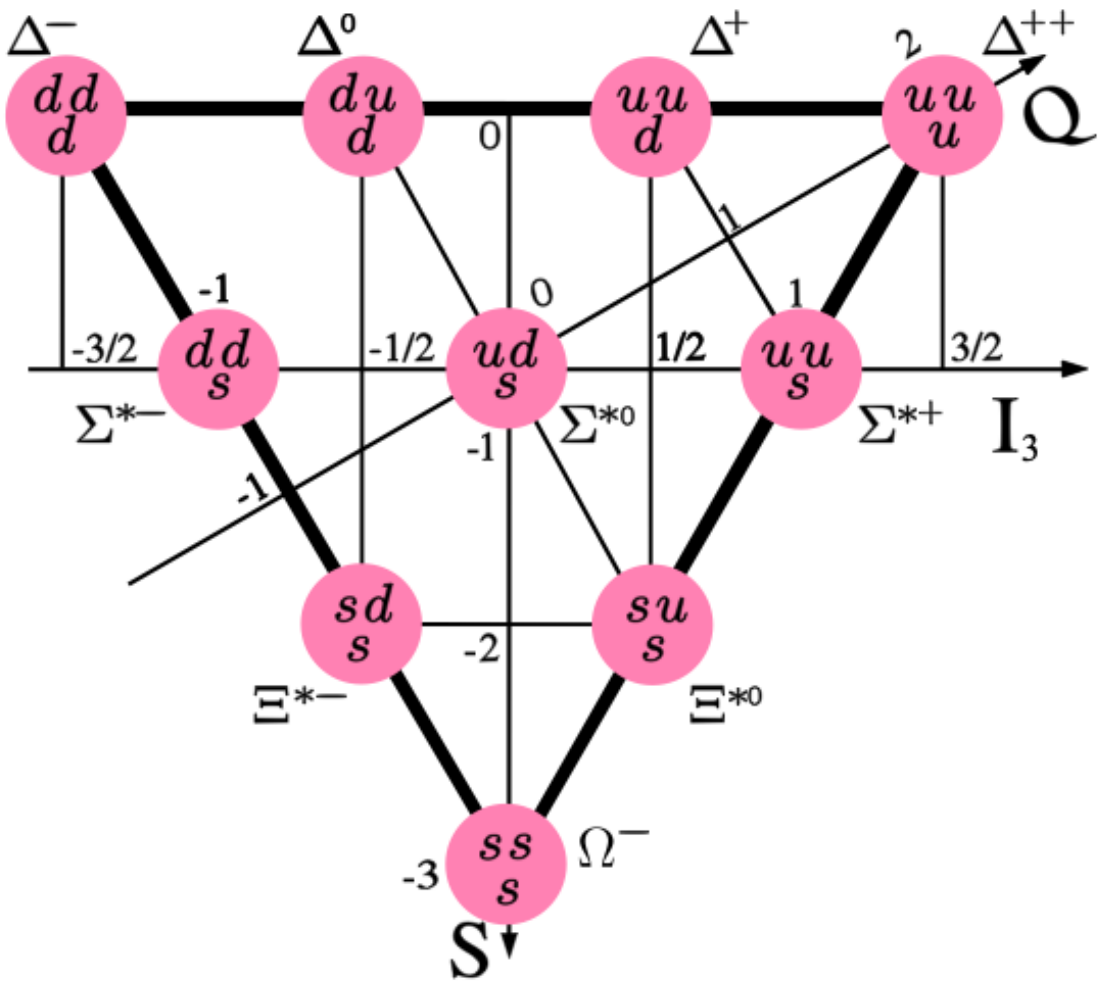

**Fig. 9.1.1** The ground-state baryon decuplet, with strangeness $(Y - B)$ plotted *vs.* the third component of isospin $I_3$.

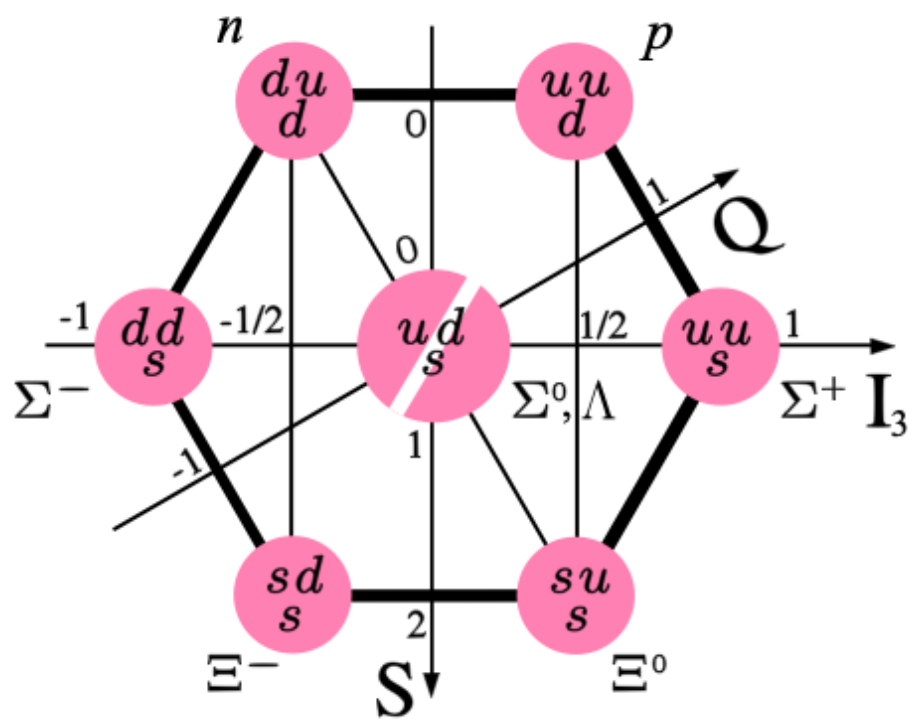

**Fig. 9.1.2** The ground-state baryon octet, with strangeness plotted *vs.* the third component of isospin.

De Rujula, Georgi and Glashow [728] were able to explain the above results in the context of a model of hadrons which confined the quarks with a flavor and spin-independent interaction, and used a short-distance interaction between the quarks that results from asymptotic freedom. This is the result of one-gluon exchange, and led to a short-distance potential between two quarks that was Coulomb in nature, and could be interpreted of as due to interactions between two colored spin-1/2 quarks. The mass dependence of the color-magnetic moments of the quarks led naturally to spin- and flavor-dependent interactions between the quarks, which
could also explain the mass differences between octet and decuplet baryons of the same flavor, and allow a qualitative understanding of the sign and size of the difference $\Sigma^0 - \Lambda^0$ between the masses of the $I = 1$ and $I = 0$ neutral strange baryons.

One consequence of this simple (additive) quark model description of baryons is an understanding of the magnetic moments of the nucleons $p, n$ and other octet and decuplet ground-state baryons. Using the total quark-spin $S = 1/2$ wave function and octet flavor wave functions for the three quarks in nucleons yields

$$\mu_p = \frac{4}{3}\mu_u - \frac{1}{3}\mu_d,$$

and, since the proton can be turned into the neutron by the transformation $u \leftrightarrow d$, we have

$$\mu_n = \frac{4}{3}\mu_d - \frac{1}{3}\mu_u.$$

Fitting this to the measured moments gives quark magnetic moments in the ratio of the quark charges to a good approximation, if we assume the quark masses are identical, and that this light quark mass is around one third of the mass of nucleons. This approach also leads to a qualitative understanding of the magnetic moments of other ground-state octet and decuplet baryons, and the transition moment that affects the rate of the decay $\Sigma^0 \to \Lambda^0\gamma$. Isgur and Karl [2662] added small contributions to baryon magnetic moments, due to configuration mixing, relativistic corrections, and violations of isospin symmetry, to refine these non-relativistic quark model estimates. The result was better agreement with the moments extracted from experimental data.

*Rotational and parity symmetries*

Ignoring for now interactions that couple the orbital and spin angular momenta of the quarks, rotational symmetry and the conservation of angular momentum also imply that ground and excited-state baryons should lie in multiplets with a given orbital angular momentum $L$ and total quark spin $S$, with the overall angular momentum of a baryon given by $\vec{J} = \vec{L} + \vec{S}$. The confining and spin-independent part of the short-range interaction will cause splittings between and within multiplets of states with different orbital angular momentum $L$, and the short distance interactions between the quarks, for example those in the work of De Rujula, Georgi and Glashow, will further split those multiplets into groups of states with the same total quark spin $S$.

It is always possible to describe the orbital angular momentum of a basis state used to describe the wave function of a baryon in terms of the angular momentum of the orbital wave functions in the two vectors required to describe the relative positions of the quarks. These can be conveniently chosen to be the Jacobi coordinates

$$\begin{aligned} \vec{\rho} &= \frac{1}{\sqrt{2}}\left(\vec{r}_1 - \vec{r}_2\right), \\ \vec{\lambda} &= \frac{1}{\sqrt{6}}\left(\vec{r}_1 + \vec{r}_2 - 2\vec{r}_3\right) \end{aligned} \tag{9.1.1}$$

shown in Fig. 9.1.3, where the $\vec{r}_i$ are the vector positions of the three quarks. The total orbital angular momentum is then $\vec{L} = \vec{l}_\rho + \vec{l}_\lambda$, and the parity of the resulting state is $P = (-1)^{l_\rho + l_\lambda}$. It is simple to show that all values of baryon spin and parity can be attained by various choices of the eigenvalues for quarks moving in a static potential; in contrast to the situation in mesons, there are no baryons with 'exotic' quantum numbers. Exotic quantum numbers in mesons require degrees of freedom like those of the glue binding the hadrons together to be in other than their $J^P = 0^+$ ground state.

This situation is complicated in the presence of spin-orbit (vector in spin and vector in space coupled to a scalar) and tensor ($S = 2$ and $L = 2$ coupled to a scalar) interactions between the quarks. These are present in models which have short-distance interactions between the quarks based on the exchange of a vector boson, such as those due to one-gluon exchange. The evidence for the presence of such interactions in the spectrum of baryons is weaker than that for the presence of interactions which are simultaneous spin and orbital angular momentum scalars; this is discussed in what follows.

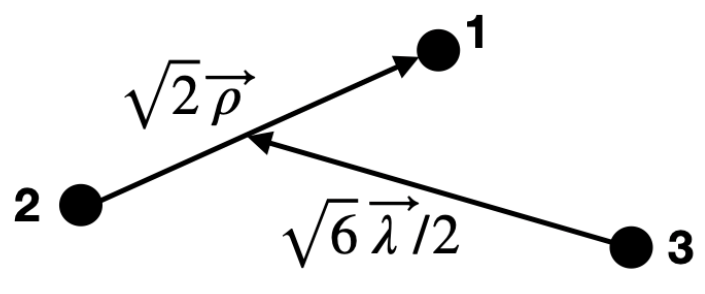

**Fig. 9.1.3** The three-body Jacobi coordinate vectors $\vec{\rho}$ and $\vec{\lambda}$.

The dominance in the baryon spectrum of simultaneous quark spin and orbital angular momentum scalar interactions, when combined with the observation that states assigned quark spin $S = 3/2$ are more massive than those with $S = 1/2$, allowed Klempt [2663] to fit the spectrum of baryons made of $\{u, d, s\}$ quarks with a mass formula. The squares of the masses of baryons are proportional to their orbital angular momenta $L$, as in Regge theory and, approximately, the spectrum of a linear confining potential. For a given flavor of baryon, more massive recurrences of the same $J^P$ quantum numbers were assigned the same gap in mass-squared as orbital excitations.

*Symmetry under particle exchange*

The requirement of Pauli symmetry implies, in the isospin symmetric limit, that the wave functions of baryons are totally symmetric under the exchange of light quarks, since the color wave function is totally antisymmetric in the absence of excitation of the gluon fields. For non-strange baryons made up of three light quarks, this means that each component of the wave function must be a basis function for a representation of the exchange group $S_3$. These basis functions are either totally symmetric ($S$) under particle exchange, totally antisymmetric ($A$), or one of a pair with mixed symmetry $\{M_\rho, M_\lambda\}$ that transform into each other under the elements of $S_3$ in a predictable way. The rules for this transformation can be found, for example, by examining the result of the various exchanges on the relative position vectors $\vec{\rho}$ and $\vec{\lambda}$.

The rules of combining the spin angular momentum of three $S = 1/2$ particles require that the overall spin wave function of the quarks for all values of the total quark spin projection $M_S$ be either totally symmetric, when $S = 3/2$, or be one of a pair of states of mixed symmetry when $S = 1/2$. The same rules apply for the isospin wave functions for baryons made up of three light quarks, $\Delta$-flavor baryons with $I = 3/2$ and $N$-flavor baryons with $I = 1/2$.

The lowest lying basis state for the spatial wave functions of ground-state baryons made up only of light quarks have $L^P = 0^+$, and are totally symmetric under quark exchange. Overall exchange symmetry then requires that the flavor and spin wave functions be combined by using the rules for combining two representations of $S_3$. For $\Delta$ baryons this is trivial, since both the spin and flavor wave functions are totally symmetric. For $N$ baryons, the mixed-symmetry spin ($\chi$) and flavor ($\phi$) wave functions are combined to the symmetric linear combination

$$\frac{1}{\sqrt{2}}\left(\chi^{M_\rho}\phi^{M_\rho} + \chi^{M_\lambda}\phi^{M_\lambda}\right).$$

The wave functions of baryons with a given flavor and spin-parity $J^P$ can be expanded in a basis of states that satisfy the requirements of antisymmetry under exchange of identical (or nearly identical, for $u \leftrightarrow d$) quarks. A convenient choice is to use a harmonic oscillator (HO) basis, which has the useful feature of being form invariant under Fourier transformation; another is the Sturmian basis, which has improved large momentum behavior useful for calculating decay form factors, but is harder to use in both coordinate and momentum space. Configuration mixing due to the confining potential and the short-range interactions between quarks can be implemented by diagonalization of the Hamiltonian matrix calculated in this basis.

The rules for combining representations of the exchange group $S_3$ are used to construct this basis from states with given values of radial $\{n_\rho, n_\lambda\}$, orbital $\{L, l_\rho, l_\lambda\}$, and spin $S$ quantum numbers (magnetic quantum numbers have been suppressed, and the sums and Clebsch-Gordan coefficients required to form states of definite $\{L, S, J\}$ are assumed). It is often convenient to expand the wave function up to a given energy, or equivalently polynomial order, which in the HO basis is

$$E = (2n_\rho + l_\rho + 3/2)\hbar\omega_\rho + (2n_\lambda + l_\lambda + 3/2)\hbar\omega_\lambda, \quad (9.1.2)$$

where $\omega_\rho$ and $\omega_\lambda$ are oscillator energies related by $\alpha^2_{\rho,\lambda} = m_{\rho,\lambda}\omega_{\rho,\lambda}$ to the scale $\alpha$ at which the radial wave functions fall with distance and the reduced masses, equal when all three quark masses are the same. Karl and Obryk [2664] give the general procedure up to fourth-order polynomials. Examples of how to construct these bases can be found in [729, 2665, 2666]; for a pedagogical overview see, for example, Ref. [2667].

It is not necessary to antisymmetrize the wave functions of baryons under the exchanges $u \leftrightarrow s$ or $d \leftrightarrow s$. It is convenient to use the '$uds$' basis [729] for baryons with $S = -1$ or $-2$, which uses basis states that have either symmetry or antisymmetry under these exchanges.

*Baryon resonance classification scheme*

The quantum numbers of the total orbital angular momentum $\vec{L} = \vec{l}_\rho + \vec{l}_\lambda$ and spin $\vec{S} = \sum_i \vec{s}_i$ are not good quantum numbers in a relativistic theory. The parity of such states is given by $P = (-1)^{l_\rho + l_\lambda}$. It is always possible to use a basis of states with specific values of $L$ and $S$ which can couple to the total angular momentum $J$ of the baryon being described. These are then mixed in the eigenstates of a Hamiltonian that includes interactions that are not simultaneous scalars in both spin and space, but are overall scalars of the form $\sum_q C_{kq} R_{k,q} S_{k,-q}$, where $R_k$ and $S_k$ are tensor operators of rank $k$ acting on the spatial and spin bases, and the $C_{kq}$ are the coefficients required to make the result an overall scalar. Examples are the tensor ($k = 2$) interactions which occur in models of the short-range interactions between quarks, and spin-orbit ($k = 1$) interactions. To the extent that these interactions are small, a classification scheme based on the $\{L, S\}$ values of the dominant component of a configuration-mixed eigenstate is useful.

It may also be useful to further break down sets of states with the same $\{L, S\}$ values into those with specific spatial symmetries. As an example, consider excitations of $N$ and $\Delta$ flavored baryons, which are made up of only $\{u, d\}$ quarks. It is useful to enumerate basis states in a harmonic-oscillator basis. Because of isospin symmetry, this basis has $\omega_\rho = \omega_\lambda$ in Eq. 9.1.2, so that

$$E = [2(n_\rho + n_\lambda) + l_\rho + l_\lambda + 3]\,\hbar\omega = [N + 3]\hbar\omega.$$

*Ground states*

Baryon states can be classified according to the flavor-spin $SU(6)$ multiplet in which they predominantly lie. This would be an exact symmetry of the Hamitonian if it was simultaneously invariant under both rotations of quark flavors in the $SU(3)_f$ space, and independent of the spin projections of the quarks. Although useful as a classification scheme, this is clearly only a very approximate symmetry: In addition to the flavor-symmetry breaking effect of the larger strange quark mass, the measured mass difference $M_\Delta - M_N \simeq 300$ MeV shows that interactions between quarks are not independent of their spin. In this scheme the ground-state nucleon lies in an $SU(3)_f$ octet of ground-state baryons with $J = S = 1/2$, which are, in order of increasing strangeness, $\{n, p, \Lambda^0, \Sigma^{+,0,-}, \Xi^{0,-}\}$, giving $(2S+1)\cdot 8 = 16$ states. The ground-state $\Delta$ is a member of an $SU(3)_f$ decuplet of baryons with $J = S = 3/2$, which are $\{\Delta, \Sigma^*, \Xi^*, \Omega\}$, giving $(2S + 1) \cdot 10 = 40$ states. Collectively, these dominantly $L^P = 0^+$ states make up the $SU(6)$ multiplet, labeled $[56, 0^+]$.

Following the notation of Isgur and Karl [2666], we can label harmonic-oscillator basis states by $|X^{2S+1}L_\pi J^P\rangle$, where $X$ is the flavor, $L$ is given in $\{S, P, D...\}$ notation, and $\pi$ is the spatial exchange symmetry, either totally symmetric ($S$), mixed symmetry ($M$), or totally anti-symmetric ($A$). In this notation, the dominant $N = 0$ components of the ground state non-strange baryons are

$$|N^2S_S 1/2^+\rangle, \quad |\Delta^4S_S 3/2^+\rangle.$$

*Negative-parity states*

The low lying (dominantly $N = 1$) negative-parity non-strange excitations are made up of a triplet of $S = 3/2$ $N$ states and a doublet of $S = 1/2$ $N$ states with mixed flavor symmetry

$$|N^4P_M(1/2^-, 3/2^-, 5/2^-)\rangle, \quad |N^2P_M(1/2^-, 3/2^-)\rangle.$$

Their spatial wave functions necessarily have mixed exchange symmetry, since they are proportional to $Y_{1m}(\Omega_\rho) \propto \rho_m$, where $\rho_m$ is a spherical component of the vector $\vec{\rho}$, or $Y_{1m}(\Omega_\lambda) \propto \lambda_m$. There is also a a doublet of $S = 1/2$ $\Delta$ states with $S$ flavor symmetry,

$$|\Delta^2P_M(1/2^-, 3/2^-)\rangle.$$

These negative-parity resonances are members of the $SU(6)$ multiplet $[70, 1^-]$, with two flavor octets of $S = \{1/2, 3/2\}$ states and a decuplet of $S = 1/2$ states [symmetric under $SU(3)_f$], plus a flavor singlet state $\Lambda$, with $S = 1/2$. References [729, 2664, 2667] show how the rules for combining representations of the exchange group $S_3$ can be applied to yield properly anti-symmetrized basis state wave functions. Basis states with the same flavor, spin, and parity can and will undergo configuration mixing when a model Hamiltonian is diagonalized.

*Positive-parity excited states*

This scheme can be extended to, for example, positive-parity non-strange excitations, dominantly described by $N = 2$ basis states. This is complicated by the presence of both radial and orbital excitations. There are three radial excitations that have $n_\rho = 1$ or $n_\lambda = 1$, and, in a three-body system, $l_\rho = l_\lambda = 1$ coupled to $L = 0$; linear combinations form $L = 0$ states with definite ($S$ or $M$) exchange symmetry. There are also orbital excitations with $l_\rho = 2$ or $l_\lambda = 2$, or $l_\rho = l_\lambda = 1$. Since they have positive parity, these baryons must be described by basis states with even $l_\rho + l_\lambda$.

After forming linear combinations of spatial basis states to form anti-symmetrized basis state wave functions, we have a total of 21 basis states that contribute at $N = 2$ to the wave functions of $N$ and $\Delta$ states. These are radial excitations

$$|N^2S_{S'} 1/2^+\rangle, \quad |\Delta^4S_{S'} 3/2^+\rangle$$

in the $SU(6)$ multiplet $[56', 0^+]$, and

$$|N^4S_M 3/2^+\rangle, \quad |\Delta^2S_M 1/2^+\}\rangle, \quad |N^2S_M 1/2^+\rangle$$

in the $SU(6)$ multiplet $[70, 0^+]$; $L = 2$ orbital excitations that are admixtures of $l_\rho = 2$ and $l_\lambda = 2$,

$$|\Delta^4D_S(1/2^+, 3/2^+, 5/2^+, 7/2^+)\rangle, \quad |N^2D_S(3/2^+, 5/2^+)\rangle$$

in the $SU(6)$ multiplet $[56, 2^+]$, and

$$|N^4D_M(1/2^+, 3/2^+, 5/2^+, 7/2^+)\rangle,\ |\Delta^2D_M(3/2^+, 5/2^+)\rangle,\ |N^2D_M(3/2^+, 5/2^+)\rangle,$$

in the $SU(6)$ multiplet $[70, 2^+]$, and $L = 1$ orbital excitations formed from $l_\rho = 1$ and $l_\lambda = 1$

$$|N^2P_A(1/2^+, 3/2^+)\rangle \tag{9.1.3}$$

in the $SU(6)$ multiplet $[20, 1^+]$.

The $J^P = 1/2^+$ nucleon and the $J^P$=3/2$^+$, isospin $I$=3/2 $\Delta$ ground states have dominant components with $N$=0, $L$=0 and $S$=1/2 and $S$=3/2, respectively. Spin-independent and spin-scalar (contact) interactions between the quarks (arising from their short-distance interactions and confinement) allow mixing between three basis states: the $N = 0$ ground state $|N^2S_S 1/2^+\rangle$, and the $L$=0, $S$=1/2 states $|N^2S_{S'} 1/2^+\rangle$ and $|N^2S_M 1/2^+\rangle$. Tensor (or spin-orbit) interactions cause mixings with the $L$=2, $S$=3/2 state $|N^4D_M 1/2^+\rangle$, and the $L$=1, $S$=1/2 state $|N^2P_A 1/2^+\rangle$. The situation is simpler for the $N$=0 ground state $|\Delta^4S_S 3/2^+\rangle$, which mixes with the $L$=0, $S$=3/2 radial excitation $|\Delta^4S_{S'} 3/2^+\rangle$, and the $L$=2, $S = 3/2$ orbital excitations $|\Delta^4D_S 3/2^+\rangle$ and $|\Delta^2D_M 3/2^+\rangle$. The resulting $D$-wave components in both the $N$ and $\Delta$ wave functions can lead to measurable consequences in the photo- and electro-production amplitudes for the transition $\gamma^{(*)}N \to \Delta$. For details, see the review on $N$ and $\Delta$ resonance electro-production in the 2022 RPP [476].

*Hyperons*

If we use a basis of states that imposes $SU(6)$ symmetry despite the larger strange quark mass, this classification scheme extends to the hyperons $\Lambda$ and $\Sigma$, $\Xi$, and $\Omega$. As an example, the notation of Isgur and Karl [2666] for the ground state $SU(3)_f$ singlet $\Lambda$ is $|\Lambda_1{}^2S_M 1/2^+\rangle$. The $SU(3)_f$ singlet wave function is totally antisymmetric under quark exchange, and so is included in the wave function of the radial excitation $|\Lambda_1{}^2P_A 1/2^+\rangle$, with its antisymmetric spatial wave function; other radial recurrences such as $|\Lambda_8{}^2S_S 1/2^+\rangle$ necessarily involve the $SU(3)_f$ octet flavor wave functions, so the notation is supplemented by the $SU(3)_f$ multiplet (singlet, octet, or decuplet) in which the state lies. The total number of basis states at each harmonic oscillator level for $\Xi$ baryons, containing two identical strange quarks, is the sum of the number of $N$ and $\Delta$ states at that same level. There is a one-to-one correspondence between basis states for $\Omega$ baryons and those of the $\Delta$ states.

Constructing the wave functions is made simpler by use of the *uds* basis [729], which requires overall antisymmetry only under exchange of equal mass quarks. In this basis, the exchange symmetry of the $\vec{\rho}$ and $\vec{\lambda}$ relative coordinate wave functions are specified separately. It is always possible to convert from the *uds* basis back to a basis with definite $SU(3)_f$ symmetry when that is convenient, for example when calculating strong decays [2668].

### 9.1.3 Constituent quark models

Constituent quark models treat a baryon as made up of three 'valence' quark degrees of freedom, with the gluon fields providing a static potential in which the quarks move. In flux-tube models [2366, 2669, 2670] this is treated as the lowest energy state of a system of three strings that meet at a junction, whose energy is proportional to their length. There are several approaches to the treatment of the short-range interactions between the quarks, which are responsible for splitting groups of states which would otherwise be degenerate or have their flavor dependence explained by violations of $SU(3)_f$ symmetry due to the additional mass of the strange quark. These approaches are briefly outlined here.

*One-gluon exchange models*

The earliest constituent quark models had short-distance interactions based on the exchange of a single gluon [728], which postulate that asymptotic freedom implies that high momentum transfer interactions between quarks are dominated by the exchange of a gluon. The result can be written as the interaction between two color-magnetic dipoles, with a $\vec{\lambda}_i \cdot \vec{\lambda}_j$ dependence on the colors of quarks $i$ and $j$, and spatial dependence given by the Fourier transform of the vector gluon propagator. Here the $\lambda_i$ are the generators of $SU(3)_c$ realized in the quark triplet basis. This naturally leads to a spin-independent Coulomb interaction at short range, and, with the assumption of point-like constituent quarks, a 'contact' interaction proportional to

$$\frac{2\alpha_s}{3m_im_j} \sum_{i<j} \vec{s}_i \cdot \vec{s}_j \delta^3(\vec{r}_{ij}), \tag{9.1.4}$$

where $\vec{r}_{ij} = \vec{r}_i - \vec{r}_j$ is the relative coordinate of quarks $i$ and $j$. This approach also results in tensor ($S = 2$ and $L = 2$ coupled to a scalar) and spin-orbit (vector in spin and vector in space coupled to a scalar) interactions between the quarks. There is some evidence for the former in the spectrum of $J^P = 1/2^-, 3/2^-$ nucleon resonances, and from patterns of strong decays of negative-parity excited baryons [2671], for example the $N\eta$ decays of the lightest non-strange $I = 1/2$ $(N^*)$ resonances with $J^P = 1/2^-$, nominally at 1535 and 1650 MeV. Isgur and Karl [729] noted a partial cancellation between spin-orbit interactions resulting from one-gluon exchange and from Thomas precession of the quarks in the confining potential, but the agreement with the spectrum of low-lying negative-parity baryons extracted from data and their one-gluon exchange model was best when they were left out altogether.

The Coulomb, contact and tensor interactions resulting from one gluon exchange were evaluated in low-lying negative parity excited baryons made up of $\{u, d, s\}$ quarks by Isgur and Karl [729]. This was extended to positive-parity excited baryons [2666], where the effects of the difference of the confining potential from that defining the harmonic oscillator basis were also evaluated using perturbation theory. The resulting parameters were fit to the spectrum extracted from data without needing to specify the form of the anharmonicities. In this work and a treatment of ground state baryons [2665], the effects of configuration mixing by the various potentials were taken into account by diagonalization of the Hamiltonian matrix, independently for each sector with $N = 2(n_\rho + n_\lambda) + l_\rho + l_\lambda = 0$ for the ground states, $N = 1$ for the low-lying negative-parity excited states, and $N = 2$ for the positive-parity excited states.

While diagonalization independently by sector has the advantage of simply describing the important physics, the parameters fit to each sector's spectrum may be inconsistent. Systems of light quarks are also relativistic, with $p/m \simeq 1$ when using constituent-quarks, which are effective degrees of freedom with masses that include the effects of sea quark and gluons. These theoretical problems can be solved by simultaneously diagonalizing the Hamiltonian in a large basis, using a relativistic kinetic energy and allowing for other relativistic effects, and using a consistent set of parameters for all baryon excitations [736].

*Pseudoscalar-meson exchange models*

Glozman and Riska [2672, 2673] emphasize the role of chiral symmetry in determining the baryon spectrum by using a short-range interaction between quarks similar to that of Eq. 9.1.4, but with the exchange of the 'chiral' octet of pseudoscalar mesons between quarks. This leads to a contact interaction between quarks $i$ and $j$ similar in form to that of Eq. 9.1.4, but proportional to the expectation of the product $\vec{\lambda}_i^f \cdot \vec{\lambda}_j^f$ of $SU(3)_f$ generators. A fit to the spectrum of low-lying negative and positive-parity baryons made up of $u$, $d$, and $s$ quarks with harmonic confinement allows first radial recurrence of the nucleon, corresponding to the Roper resonance $N(1440)$, to be lighter than the lightest $J^P = 1/2^-$ orbital excitation, corresponding to $N(1535)$, and the same behavior holds for the $\Lambda$ baryons, as seen in extractions from experimental data. There are no spin-orbit interactions that arise from pseudoscalar meson exchange; those from other sources are neglected, along with tensor forces that accompany the contact interaction. The calculation of the $N$ and $\Delta$ spectra was refined by performing three-body Faddeev calculations with a Goldstone-boson-exchange interaction plus linear confinement between the constituent quarks in Ref [2674].

Dziembowski, Fabre de la Ripelle, and Miller [2675] put these two approaches together by including the effects of pseudoscalar meson exchange and of one gluon exchange between quarks, neglecting the complexity introduced by tensor and spin-orbit interactions, in a hyper-spherical method calculation that goes beyond wave function perturbation theory. They showed that it is possible to describe the non-strange baryon spectrum using a quark-meson coupling constant that reproduces the measured pion-nucleon coupling constant, and a reasonably small value of the strong-coupling constant, which governs the strength of the one-gluon exchange terms.

*Instanton-induced interactions*

Instantons are topologically nontrivial gauge-field configurations in 4-dimensional Euclidean space, with field strengths that vanish at large spatial distances. These configurations are localized in both space and (Euclidean) time, and so are instantanous interactions, which gives rise to their name. They are crucial to understanding the formation of condensates in the QCD vacuum, and how the axial current anomaly gives mass to the $\eta'$ meson; their presence in QCD also yields short-range interactions between the quarks. Löring, Kretzschmar, Metsch and Petry [2676, 2677] investigated the spectrum of baryons in a relativistic model by solving the three-body Bethe-Salpeter equation. This model uses instantaneous pairwise linear confinement, with the Dirac structure required to make the confining potential spin-independent, and an instantaneous two-body interaction based on 't Hooft's residual interaction, which arises from QCD-instanton effects. This model was able to explain salient features of the non-strange baryon spectrum, such as the low mass of the Roper resonance $|N^2S_{S'}1/2^+\rangle$, and the presence of approximate parity doublets. This study was extended [2678] to the study of excited $\Lambda$ and $\Sigma$ hyperons, where the equivalent features of these spectra were also explained, and later to charmed baryons in Ref. [2679].

*The Dyson-Schwinger Bethe-Salpeter approach*

There has been significant recent progress in understanding the physics of baryons [830, 2680] by using the Dyson-Schwinger equations of QCD and Bethe-Salpeter equations [828, 2681]. In this approach baryons are relativistic bound states of three quarks, and the treatment of their interactions arising from QCD is non-perturbative, incorporating aspects of confinement and dynamical symmetry breaking. Two paths to solving the three-body problem are taken; direct solution of the three-body Faddeev equation, and decomposition of baryons into quark-diquark systems, with all quark pairs able to constitute the diquark. The latter path requires the calculation of diquark Bethe-Salpeter amplitudes, and diquark propagators. These depend on the quark and gluon propagators and quark-gluon vertex, which are consistent with those used for the Bethe-Salpeter equation for mesons, and with chiral symmetry. Due to the complexity of the three-body system, baryon calculations are performed using the rainbow-ladder approximation, where the $q-q$ kernel has the form of a single gluon exchange with a momentum-dependent vertex strength, summed by the Bethe-Salpeter equation into Feynman diagrams that take the form of a ladder, or rainbow. This construction preserves chiral symmetry.

Using this dynamical quark-diquark approach, the ground state nucleon, $\Delta(1232)\frac{3}{2}^+$, and Roper $N(1440)\frac{1}{2}^+$ resonances are described well [907], as their configurations are dominated by scalar and axial vector diquarks. However, other baryons are sensitive to other diquark channels, which are known to be too strongly bound in this approximation, as are the corresponding scalar and axial-vector mesons. The result is that the other excited baryon masses come out too low. Reducing the strength of the attraction in the pseudoscalar and vector di-quark kernels simulates effects beyond the rainbow-ladder approximation, and the result is good agreement between the calculated spectrum for excited $N$, $\Delta$, $\Lambda$, $\Sigma$, $\Xi$ and $\Omega$ baryons with $J^P = 1/2^{\pm}, 3/2^{\pm}$, with the exception of the $\Lambda(1405)1/2^-$, $\Lambda(1520)3/2^-$, and to a lesser extent the Roper resonance $N(1440)1/2^+$. The authors of Ref. [907] point out that this is likely due to the lack of a consistent treatment of baryon-meson coupled channel effects.

### 9.1.4 Missing states in the baryon spectrum

*Models of strong decays*

For ground-state and low-lying negative-parity excited state baryons made up of $\{u, d\}$ and a single $s$ quark, the spectrum of states extracted from experimental data can be matched to model predictions without ambiguity. (There is little experimental information about the spectrum of the excited $\Xi$, strangeness $S = -2$, and $\Omega$, strangeness $S = -3$ states.) However, for positive-parity excited states, more states are predicted by models that treat three quarks symmetrically than are present in analyses of the data. This is called the 'missing resonance' problem. One possible explanation is to postulate that they contain static, tightly-bound $ud$ di-quarks, which reduces the effective number of degrees of freedom and so the number of excitations in the spectrum [756]. However, lattice QCD calculations of nucleon stucture [2682], and of the entire excited baryon spectrum using a broad spread of operators [495, 2683] do not show the reduced number of states expected if only 'good' di-quarks prevail, and recent experimental evidence for the existence of states that are ruled out in such models is described below.

A solution to this problem is that, unlike those states seen in partial-wave analyses of elastic $\pi N$ and $\bar{K}N$ scattering data, these missing positive-parity excited states have weak couplings to the corresponding strong-interaction production channel [2684, 2685]. In any case, in order to make a detailed and exhaustive comparison between predictions of any model and the experimental spectrum of excited baryons, a model of the strong decay $B^* \to BM$ of baryons into a ground-state baryon and meson is required. For a detailed, comparative review of such models see Ref. [2656].

One approach is to couple point-like pseudoscalar mesons to the quarks in the decaying baryon, an elementary-meson emission model [2668, 2686, 2687]. As an example, Koniuk and Isgur [2668] modeled such decays, by coupling point-like pseudoscalar mesons to the quarks in the decaying baryon, and evaluating the transition amplitudes using the configuration-mixed wave functions resulting from the one-gluon exchange model of Isgur and Karl [2666]. They also examined baryon electromagnetic transition amplitudes that can be extracted from meson photo-production experiments. Many states were observed to have small $\pi N$ or $\bar{K}N$ amplitudes, which would lead to them decoupling from elastic-scattering partial-wave analyses, and the masses and decay amplitudes of those that did not corresponded to those of the observed states.

The internal structure of mesons can be taken into account if strong decays of excited baryons proceed *via* an operator that creates a $q\bar{q}$ pair with vacuum, $^{2S+1}L_J = {}^3P_0$, quantum numbers. The operator is assumed $SU(3)_f$ symmetric; the additional energy required to produce a strange quark pair is taken into account by the kinematics. Strong decay amplitudes are formed by evaluating the required spin and flavor overlaps, forming the expectation value of this operator between wave functions for the final state baryon and meson, and that of the initial excited baryon, and integrating over the relative momentum of the $q\bar{q}$ pair.

### 9.1.5 Decay-channel couplings

Models for the baryon spectrum that do not take into account decay-channel couplings effectively assume that baryons are infinitely long-lived bound states. In practice, excited baryons decay strongly, with decay widths that are significant fraction of their masses. Excited baryons can have large couplings to continuum states, which can and will affect the positions of the poles in scattering amplitudes that describe these resonances. This can be due to their proximity to decay-channel thresholds, or unusually large couplings to a decay channel, or both. Examples include the low-lying negative-parity resonances $\Lambda(1405)1/2^-$, which has a nominal mass below the $N\bar{K}$ threshold, and $N(1535)1/2^-$, which couples strongly to the $N\eta$ final state, for which it is just above threshold. The authors of Ref. [2688] were among the first to suggest that these states could be dynamically generated resonances. Hyodo and Meissner review the interesting physics of the $\Lambda(1405)$ state in the 2022 RPP [476]. There is also evidence of these effects from lattice QCD [2683] for states like the Roper resonance; its mass changes rapidly as pion mass in the calculation approaches the physical pion mass, due to strong $N\pi\pi$ channel coupling.

*Beyond elastic meson scattering*

The observation that baryon resonances could be missing due to weak couplings for both their strong-interaction production and decay in elastic meson-nucleon scattering led to the idea that $\gamma N$ photo-production experiments could excite missing resonances that had appreciable photo-couplings, which could be discovered via their strong decays to final states with more than one pseudoscalar meson. For example, missing $N$ and $\Delta$-flavored baryons could be searched for in proton-target photo-production experiments examining two or three pion final states resulting from the intermediate vector-mesons $\rho(770)$ and $\omega(782)$. In particular, certain missing, positive-parity resonances can be expected to decay to two-pion final states by simultaneous de-excitation of both $l_\rho = 1$ and $l_\lambda = 1$ excitations.

*Recent developments from photo-production experiments*

The Particle Data Group in its bi-annual updates of the Review of Particle Physics (RPP) lists the known baryon resonances, their properties, and the experimental evidence for their existence in terms of star assignments ranging from one star (poor evidence) to four stars (evidence is strong). Since the 2010 edition of the RPP [2689], much new information about $N$ and $\Delta$-flavored baryons has been added based on recent photo-production experiments. In particular, various polarization observables have played a crucial role in identifying new resonances, or consolidating the existence of those previously poorly known.

The experimental spectroscopy efforts to address the missing resonance problem have concentrated, for the most part, on the $N = 2$ positive-parity non-strange excitations, and on the $N = 1$ negative-parity singly- and doubly-strange states. While only one negative-parity $\Lambda$ state has yet to be identified to complete the first excitation band with many more positive-parity candidates known for the second excitation band, overall as many $\Sigma$ and $\Xi$ states are expected as $N^*$ and $\Delta$ states combined, and of these, many negative-parity states are still missing. The situation is worse for $\Xi$ and $\Omega$ resonances, since their spins and parities have been measured for very few states; speculative $J^P$ assignments based on quark model predictions are listed by the PDG for the majority of the observed states. The potential for new discoveries remains high in the hyperon sector.

Of particular interest in the non-strange sector are those multiplets of the second excitation band where both oscillators have a single orbital excitation, $l_\rho = l_\lambda = 1$, which combine to either $L = 0$, 1, or 2. Since both relative coordinate vectors must be excited in order to have the necessary exchange symmetry, their presence would rule out tightly-bound, static di-quarks. In fact, a quartet of $S = 3/2$ states

$$|N^4D_M(1/2^+, 3/2^+, 5/2^+, 7/2^+)\rangle\,,$$

has been proposed [2690] for the 70-plet $(70, 2^+)$ largely based on the photo-produced double-pion final state.

These states were expected to be seen in double-meson reactions since each oscillator can de-excite via the emission of a meson. There is a new $J^P = 1/2^+$ state, $N(1880)1/2^+$, and the state $N(1900)3/2^+$ has had its likely existence upgraded from 2 to 4 stars in the Review of Particle Properties (RPP) by the PDG [476]. Evidence for two other states in the quartet, $N(2000)5/2^+$ and $N(1990)7/2^+$, is strong in some partial wave analyses but requires additional confirmation, and so remains listed with weak evidence (two stars) in the RPP. Such double-meson reactions had been under-explored until recently, which would explain why these states escaped detection in the past. Another previously one-star resonance, $N(2100)1/2^+$, has been upgraded to three stars, and can be tentatively assigned to the doublet of states with $S = 1/2$ forming the SU(6) multiplet $[20, 1^+]$ of Eq. 9.1.3, where $l_\rho = l_\lambda = 1$ combine to $L = 1$. There remains at best weak evidence for the second state in this doublet. Although the assignment of experimental $N^*$ candidates to these multiplets is speculative, some optimism persists that the goal of completely mapping the second excitation band for non-strange baryons is within reach once all currently available (polarization) data have been analyzed.

There is also an interesting pattern of parity doublets of $N^*$ baryons with masses around 2 GeV, which might indicate the restoration of the chiral symmetry at higher energies [2691, 2692]. A similar pattern was observed for $\Delta$ resonances in the same mass region:

| | |
|---|---|
| $\Delta(1910)1/2^+$ | $\Delta(1900)1/2^-$ |
| $\Delta(1920)3/2^+$ | $\Delta(1940)3/2^-$ |
| $\Delta(1905)5/2^+$ | $\Delta(1930)5/2^-$ |
| $\Delta(1950)7/2^+$ | |

A detailed study has not yet revealed the missing $7/2^-$ state [2693]. Closest in mass is a previously poorly-known state, $\Delta(2200)7/2^-$, which has since been upgraded from one to three stars based on photo-production data. Interestingly, the corresponding mass difference is observed in the nucleon spectrum between $N(1990)7/2^+$ (two stars) and $N(2190)7/2^-$ (four stars).

The result is that, based on photo-production data, six completely new $N^*$ resonances have been proposed with
masses around 2 GeV, and three additional states have been upgraded. No new $\Delta$ state has been proposed, but four states have had their status upgraded by the PDG.

### 9.1.6 Baryons with excited glue

In a strongly-coupled system, *hybrid* baryons with excited gluon degrees of freedom must exist. Unlike in the spectrum of mesons, all $J^P$ quantum numbers are accessible via spatial excitation for a given flavor of baryon, as explained in Sec. 9.1.2. In the absence of exotic quantum numbers, another approach to the discovery of hybrid baryons might be to search for an over-population, relative to the expectations of constituent quark models, of states with a given flavor, spin, and parity quantum numbers. However, it is expected that the lowest-lying states with excited gluon degrees of freedom are positive-parity states that overlap in mass the region in the spectrum where there are already several missing conventional states.

Early approaches to the physics of hybrid baryons include those based on the MIT bag model [2694], large-$N_c$ QCD [2695], and QCD sum rules [2696]. As an example, the calculation of Ref. [2694] confined a constituent gluon and three quarks to an MIT bag, and used $\mathcal{O}(\alpha_s)$ interactions between the constituents. In these studies, the lightest hybrid baryons were found to have $N$ flavor and $J^P = \{1/2^+, 3/2^+\}$, with the lightest of these having $J^P = 1/2^+$ and a mass of approximately 1500 MeV, between those of the two lightest radial excitations of the nucleon, the Roper resonance at 1440 MeV, and the N(1710).

The flux-tube model developed to examine hybrid meson structure and decays by Isgur and Paton [2366] was applied to hybrid baryons in Refs. [2670, 2697]. An adiabatic approximation is employed, where a Y-shaped flux tube is allowed to move with the three quark positions fixed, except for center of mass corrections. This defines a potential in which the quarks move, for both conventional (glue in its ground state) and hybrid (glue in its lowest-lying excited state) baryons. The flux-tube dynamical problem can be reduced to the independent motion of the junction and the strings connecting the junction to the quarks. The seven low-lying hybrid baryons are found to be two doublets of $N^2 1/2^+$ and $N^2 3/2^+$ states with quark spin $S = 1/2$, and three states

$$^4\Delta(1/2^+, 3/2^+, 5/2^+)$$

with quark spin $S = 3/2$. Baryon masses are found by using a variational method to solve for the quark energies in these string potentials. Including the hyperfine contact spin-spin term in Eq. 9.1.4 lowers the mass of the quark-spin 1/2 hybrid states by 110 MeV to 1865 MeV, and raises the mass of the quark-spin 3/2 hybrid states, which coincide with the lightest $\Delta$ flavored hybrids, by a similar amount.

Lattice QCD approaches to describing the spectrum of conventional and hybrid baryons assuming isolated bound states [495, 2683] are able to determine the spectrum of baryon states up to $J^P = 7/2^\pm$. The results show the same number of states as non-relativistic models based on three-quark degrees of freedom [2683], with no signs of the reduced number of excitations predicted by di-quark models, or parity doubling. States in this spectrum can be grouped into $SU(6) \times O(3)$ multiplets, with weak mixing. Using many composite QCD interpolating fields, hybrid baryons of $N$ and $\Delta$ flavor were identified in Ref. [495] by searching for states with a substantial overlap with operators containing gluonic excitations. This led to doublets of $N 1/2^+$ and $N 3/2^+$ hybrids, and $N 5/2^+$, $\Delta 1/2^+$ and $\Delta 3/2^+$ states at energies above the center of the first band of conventional positive-parity excitations. This suggests that exciting the glue adds a color-octet effective degree of freedom, with roughly the same additional energy in mesons and baryons, that has $J^P = 1^+$, unlike the vector nature of this excitation in the flux-tube model. A $J^P = 1^+$ excitation is expected in the bag model of Ref. [2694], as these are the quantum numbers of the lowest energy, transverse electric mode of a gluon in a spherical bag.

This approach is extended to all baryons made from $u$, $d$, and $s$ quarks in Ref. [2683], using operators that lie in irreducible representations of $SU(3)_f$ symmetry, in addition to $SU(4)$ symmetry for the Dirac spins and $O(3)$ symmetry for the orbital state. The spectra that result for non-hybrid states are again consistent with quark model expectations based on weakly broken $SU(6)\otimes O(3)$ symmetry. States with strong hybrid content are usually at about 1 GeV above the corresponding conventional excited states, and the quantum numbers and multiplicity of the positive-parity hybrid states can be roughly predicted by combining a $J^P = 1^+$ gluonic excitation with non-relativistic quark spins, although some of the expected states are not found in the calculation performed at the lowest pion mass. The use of multi-hadron operators will allow the exploration of the energy dependence of and resonances in hadron scattering amplitudes.

A recent proposal prepared by the CLAS12 Collaboration and presented to the Jefferson Lab Physical Advisory Committee aims to experimentally search for hybrid baryon states in electro-produced $KY$ and $p\pi^+\pi^-$ final states by focusing on measurements for $Q^2 < 1.0$ GeV$^2$. Since the spin and parity of hybrid baryons are expected to be the same as those for conventional states, the experimental signature of hybrid baryons is the distinctively different low-$Q^2$ evolution of their electro-couplings that originate from the additional gluonic component of their wave function. More details are discussed in the contribution by V. Burkert.

## 9.2 Light-quark baryons

**Volker Burkert, Eberhard Klempt, Ulrike Thoma**

### 9.2.1 Why $N^*$'s ?

This was the question with which Nathan Isgur opened his talk at $N^*$2000 [2698] held at the Thomas Jefferson National Accelerator Facility in Newport News, VA, one year before he passed away, much too early. He gave three answers:

First, nucleons are the stuff of which our world is made. In the Introduction to this Section, two of us have outlined the importance of $N^*$'s and $\Delta^*$'s in the development of the Universe 9, when hadrons materialized from a soup of quarks and gluons at some 10 $\mu$s after the big bang. The full spectrum of excited baryon states including those carrying strangeness must be included in hadron gas models that simulate the freeze-out behavior observed in hot-QCD calculations. These simulations aim at finding the underlying processes, to pin-point the "critical point" of the phase transition that is expected to occur between the QGP phase and the hadron phase at a temperature near 155 MeV. Experiments are ongoing at CERN, RHIC and planned at FAIR to study the phase diagram of strongly interacting matter, e.g. by varying the collision energy.

Second, nucleons are the simplest system in which the non-abelian character of QCD is manifest. The proton consists of three (constituent) quarks since the number of colors is three.

Third, baryons are sufficiently complex to reveal physics to us hidden in the mesons. Gell-Mann and Zweig did not develop their quark model along mesons, their simple structure allowed for different interpretations. *Three* quarks resulted in a baryon structure that gave - within SU(3) symmetry - the octet and the decuplet containing the famous $\Omega^-$.

Isgur made many important contributions to the development of the quark model. With Karl he developed the idea that gluon-mediated interactions between quarks bind them into hadrons and constructed a quark model of baryons [2699]. This was a non-relativistic model, hardly justifiable. With Capstick he relativized the model [736], but surprisingly, the pattern of predicted resonances remained rather similar. Isgur always defended the basic principles: hadrons have to be understood in terms of constituent quarks bound in a confining potential and additionally interacting via the exchange of "effective" gluons.

Nearly 20 years later, Meißner ended his contribution [2700] to the $N^*$2019 conference held in Bonn, Germany, by stating: "Forget the quark model". We need to ask: What has happened in these two decades? What did we know before? What have we learned?

Mapping the excitation spectrum of the nucleon (protons and neutrons) and understanding the effective degrees of freedom are important and most challenging tasks of hadron physics. A quantitative description of the spectrum and properties of excited nucleons must eventually involve solving QCD for a complex strongly interacting multi-particle system. The experimental $N^*$ program currently focuses on the search for new excited states in the mass range just below and above 2 GeV using energy-tagged photon beams in the few GeV range, and on the study of resonances, their properties, and their internal structure, e.g. in cascade decays and in meson electro-production.

### 9.2.2 $N^*$'s: how?

In the previous contribution by Capstick and Crede 9.1 we have seen the complexity of the expected spectrum of nucleon and $\Delta$ excitations. Even in the lowest excitation mode with $l_\rho = 1$ or $l_\lambda = 1$, we expect five $N^*$ and two $\Delta^*$ states; they are all well established. But already in the second excitation mode, the quark model predicts 13 $N^*$ and 8 $\Delta^*$ states The resonances have quantum numbers $J^P = 1/2^+, \cdots, 7/2^+$ and isospin $I = 1/2$ or $3/2$, respectively. All these 21 resonances are expected to fall into a mass range of, let's say, 1600 - 2100 MeV. This complexity of the light-quark ($u$ & $d$ quarks) baryon excitation spectrum complicates the experimental search for individual states, especially since, as a result of the strong interaction, these states are broad, the typical width being 150-300 MeV. They overlap, interfere, and often several resonances show up in the same partial wave. Grube in his contribution 8.3 has convincingly demonstrated the difficulties of extracting the existence and properties of mesonic resonances from $\pi\pi$ scattering experiments. With nucleon resonances, additional complications due to the nucleon spin emerge: in $\pi N$ elastic scattering there are two complex amplitudes to be determined, for spin-flip and spin-non-flip scattering.

Pion scattering off nucleons was mostly performed in the pre-QCD era. Nearly all excited nucleon states listed in the Review of Particle Physics (RPP) prior to 2012 have been observed in elastic pion scattering $\pi N \to \pi N$. However there are important limitations in the sensitivity to the higher-mass nucleon states. These may have very small $\Gamma_{\pi N}$ decay widths, and their identification becomes exceedingly difficult in elastic scattering. Three groups extracted the real and imaginary parts of the $\pi N$ partial-wave amplitude from the data [2701–2703]. Their results are still used as constraints in all modern analyses of photo-induced reactions.

Figure 9.2.1a,b shows the real and imaginary part of the $S_{11}$ amplitude for $\pi N$ scattering. The imaginary part peaks at 1500 MeV and just below 1700 MeV indicating the presence of two resonances, $N(1535)1/2^-$ and $N(1650)1/2^-$. These are known since long and established. Above, there is no clearly visible sign for any additional resonance. Higher-mass resonances – if they exist – must have very small $\Gamma_{\pi N}$ decay widths.

Estimates for alternative decay channels have been made in quark model calculations [2709]. This has led to major experimental efforts at Jefferson Lab, ELSA and MAMI to determine differential cross sections and (double) polarization observables for a variety of meson photoproduction channels. Spring-8 at Sayo in Japan and the ESRF in Grenoble, France, made further contributions to the field.

Figure 9.2.1c,d shows an example. In Fig. 9.2.1c, the total cross section for $\eta$ photoproduction off protons and off neutrons is shown [2704, 2705]. They are dominated by $N(1535)1/2^- \to N\eta$ interfering with $N(1650)1/2^-$. The opening of important channels is indicated by vertical lines. At the $\eta'$ threshold, the intensity suddenly drops: significant intensity goes into the $N\eta'$ channel. This is a strong argument in favor of a resonance at or close to the $p\eta'$ threshold. It also clearly demonstrates the advantage of investigating different final states and production mechanisms. In contrast to the $\pi N$-$S_{11}$ scattering amplitude, here, already in the total $\eta$-photoproduction cross section, a structure relating to $N(1895)1/2^-$ becomes visible. Furthermore, in Fig. 9.2.1d, the result of a fit with Legendre moments to the so-called $\Sigma$ polarization observable for $\gamma p \to \eta p$ is compared to two energy-dependent solutions of the BnGa coupled-channel analysis. Plotted is the coefficient $(a_4)_4^{\Sigma}$ of the Legendre expansion which receives (among others) a contribution from the interference of the $S$-wave with the $G$-wave. Data from different experiments are given with their error bars. The curves represent BnGa fits with (solid curve) and without (dashed curve) inclusion of data on $\gamma p \to \eta' p$. The $N(2190)7/2^-$ ($G$-wave) was included in both fits. From 1750 MeV to the $p\eta'$-threshold the coefficient is approximately constant, then at the $p\eta'$-threshold, the fit result shows an almost linear rise towards positive values. This change of the coefficient at about 1.9 GeV indicates the presence of a cusp. The strong cusp is an effect of the $p\eta'$ threshold [$E_\gamma$ = 1447 MeV (W = 1896 MeV)], the $N\eta'$ amplitude must be strongly rising above threshold. Indeed, the inclusion of the full data set on $\gamma p \to p\eta'$ (cross sections, polarization observables) into the BnGa data base had already confirmed the existence of a new $N(1895)1/2^-$ resonance with a significant coupling to $p\eta$ and $p\eta'$ [2710, 2711], first observed in [2712].

This resonance was not seen in classical analyses of $\pi N$ elastic scattering data[90]. The example shows the importance of inelastic channels and of coupled-channel analyses. Thresholds can be identified by the missing intensity in other channels, cusp effects can show up, all these effects need to be considered and finally contribute to find the correct solution. High-precison and high-statistics data are required as well as a large body of different polarization data.

### 9.2.3 Photoproduction of exclusive final states

In the photoproduction of a single pseudoscalar meson like $\gamma p \to \eta p$, not only the proton has two spin states but also the photon has two possible spin orientations. In electroproduction, discussed by Burkert in the subsequent section 9.3, the virtual photon can also be polarized longitudinally. But even for experiments with real photons, there are four complex amplitudes to be determined. There is a large number of observables: the target nucleon can be polarized longitudinally, i.e. in beam direction, or transversely, the photon can carry

[90] Höhler and Manley has claimed a similar state that had been combined with Cutkovsky's result to $N(2090)$.

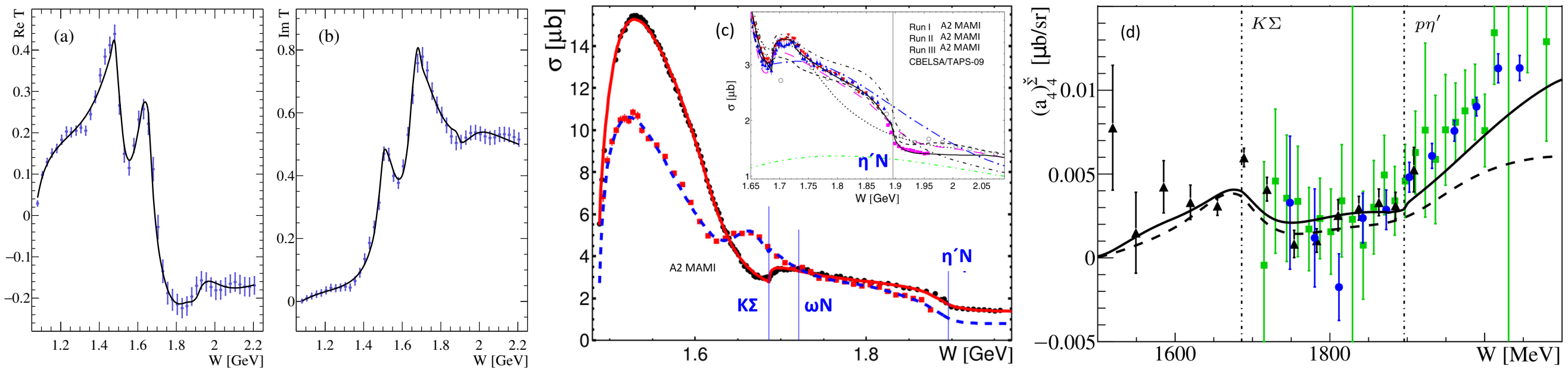

**Fig. 9.2.1** (a),(b): Real and imaginary part of the $S_{11}$ $\pi N$ scattering amplitude. Resonances in this partial wave have quantum numbers $J^P = 1/2^-$. Clearly seen are $N(1535)1/2^-$ and $N(1650)1/2^-$. There is no convincing evidence for any resonance above 1700 MeV. Data points are from [2701], errors are estimates, the curve represents a recent Bonn-Gatchina (BnGa) fit. (c): Total cross sections for $\gamma p \to \eta\, p$ and $\gamma n \to \eta\, n$. Important thresholds are marked by lines. The inset shows the $\eta'$ threshold region for $\eta$-photoproduction off the proton (picture adapted from [2704, 2705]). (d): The Legendre coefficient of the polarization observable $\Sigma$ $(a_4)_4^{\Sigma}$ exhibits a cusp at the $\eta'$ threshold [2706]. The data stems from GRAAL (black), CBELSA/TAPS (blue) and CLAS (green) Picture taken from [2706]. (c),(d): see publications[2704, 2705, 2707] for references to the data.

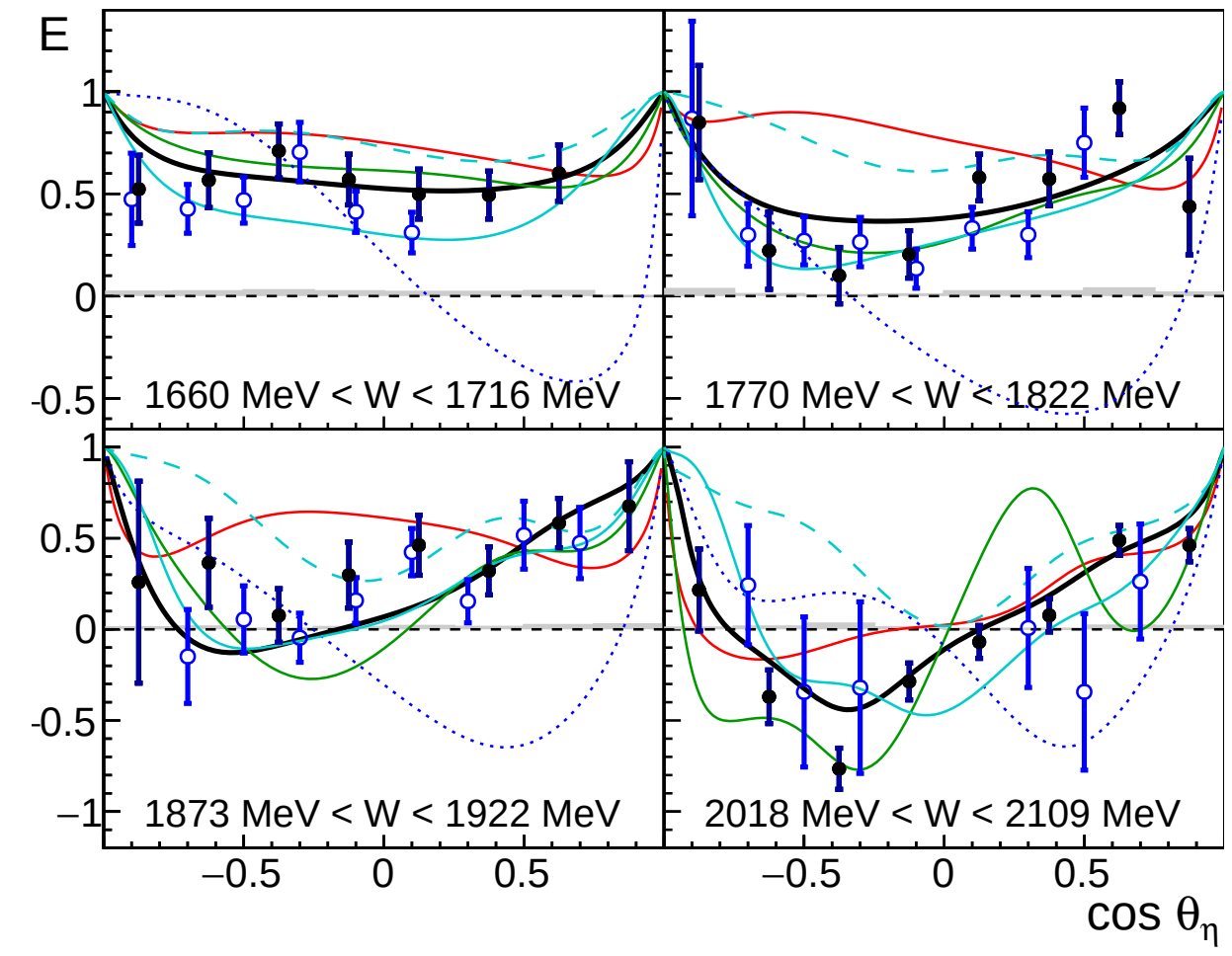

**Fig. 9.2.2** The double polarization observable E as a function of $\cos\theta_\eta$ in the cms for selected energy bins, black: CBELSA/TAPS [2707], blue: CLAS data [2708] (due to different binning, the energies differ by up to half of the bin size). Colored curves: Predictions from different PWAs (see publication for references), black: BnGa-fit including the data shown here and further new polarization data. Figure adapted from [2707].

linear or circular polarization. The final-state nucleon can carry polarization along its flight direction or perpendicular to the scattering plane. There is an intense discussion in the literature on how many independent measurements have to be performed to determine the four complex amplitudes, see Ref. [2660]. In practice, energy-independent analyses in bins of the invariant mass were only done for the very low energy region [2716, 2717] or with additional assumptions (see [2718–2720] and references therein).

In most cases, energy-dependent analyses have been performed to extract the information hidden in the photoproduction data. Here the polarization data, in particular those with polarized photon beam and polarized target nucleons, were decisive to reduce ambiguities of the solutions. The double polarisation observable $E$ is one of the beam-target-observables; it requires a circularly polarized photon beam and a longitudinaly polarized target. Examples of $E$ for selected W-bins are shown in Fig. 9.2.2 for $\gamma p \to p\eta$ [2707]. The data are compared to the predictions of different PWA solutions (colored curves). The curves scatter over a wide range indicating the high sensitivity of the polarisation observable on differences in the contributing amplitudes. A new BnGa fit returned masses and widths of $N^*$-resonances and their $N\eta$-branching fractions [2707], several of them unknown before. Interestingly a $N(1650)1/2^- \to N\eta$-branching fraction of $0.33 \pm 0.04$ was found while in the RPP'2010, a value of only $0.023 \pm 0.022$ was given. Recently, also within the Jülich-Bonn dynamical coupled channel approach, a $N\eta$-residue for $N(1650)1/2^-$ was found, larger by almost a factor of two compared to earlier analyses, after inclusion of the new polarisation data [2721]. Historically, the large $N(1535)1/2^- \to N\eta$ branching fraction and the small one for $N(1650)1/2^- \to N\eta$ has played a significant role in the development of the quark model [729], of theories based on coupled-channel chiral effective dynamics [2688] and led to several interesting interpretations of the low mass $1/2^-$-resonances (for references see [2707]). The old values from 2010 were obtained without the constraints provided by the new high quality (double) polarization data covering almost the complete solid angle. The impact of polarization observables on the convergence of different PWA-solutions was e.g. also very clearly demonstrated in a common study of pion-photoproduction [2722].

In hyperon decays, the polarization of the $\Lambda$ or $\Sigma^{\circ}$ can be determined by analyzing the parity violating

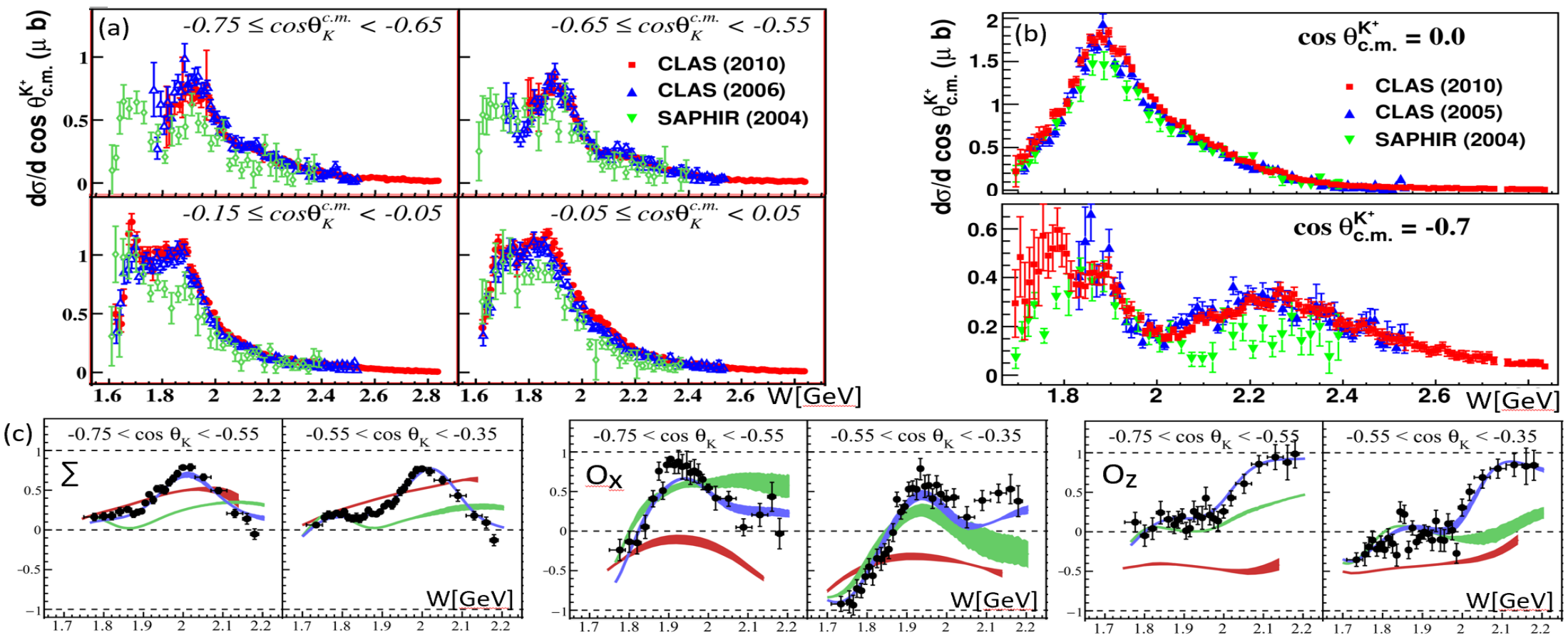

**Fig. 9.2.3** Invariant mass dependence of the $\gamma p \to K^+\Lambda$ [2713] (a) and $\gamma p \to K^+\Sigma$ [2714] (b) differential cross sections for selected bins in the polar angle. (c) Examples for polarization observables determined for $\gamma p \to K^+\Lambda$ (only selected bins shown) [2715]. Curves: PWA-predictions from ANL-Osaka (red) and BnGa 2014 (green). Blue: BnGa 2014-refit including the data shown. (a)-(c): For references to the data and the PWAs see [2713–2715], Picture adapted from [2713–2715]

decay $\Lambda \to p\pi^-$. Thus the spin orientation of the final state baryon (recoil polarization) can be determined. Kaon-hyperon production using a spin-polarized photon beam provides access to the beam-, recoil-, target-[91] and to beam-recoil polarization observables. The data had a significant impact on the determination of the resonance amplitudes in the mass range above 1.7 GeV. Precision cross section and polarization data, examples of which are shown in Figure 9.2.3, span the $K^+\Lambda$ and $K^+\Sigma$ invariant mass range from threshold to 2.9 GeV, hence covering the interesting domain where new states could be discovered. Clear resonance-like structures at 1.7 GeV and 1.9 GeV are seen in the $K^+\Lambda$-differential cross section that are particularly prominent and well-separated from other structures at backward angles. At more forward angles (not shown) t-channel processes become prominent and dominate the cross section. The broad enhancement at 2.2 GeV may also indicate resonant behavior although it is less visible at more central angles with larger background contributions. Similar resonance-like structures are observed in the $K\Sigma$ channel (Figure 9.2.3(b)). Examples for different polarisation observables determined for the reaction $\gamma p \to K^+\Lambda$ are shown in the lower row of Figure 9.2.3 for selected bins in the $K^+$-scattering angle in the $\gamma p$ center-of-mass frame. They are compared to predictions from ANL-Osaka, BnGa-2014 and to a refit from the BnGa-PWA. The large differences between the curves demonstrate the sensitivity of the data to the underlying dynamics. The $K\Lambda$ channel is somewhat easier to understand than the $K\Sigma$ channel, as the iso-scalar nature of the $\Lambda$ selects isospin-1/2 states to contribute to the $K\Lambda$ final state, while both isospin-1/2 and isospin-3/2 states can contribute to the $K\Sigma$ final state. Of course, here, as well as for other final states, only a full partial wave analysis can determine the underlying resonances, their masses and spin-parity. Polarization data are required to disentangle the different amplitudes.

Energy-dependent analyses have been performed e.g. at GWU [2723] as SAID, in Mainz as MAID [2705], at Kent [2724], at JLab [2725], by the BnGa [2712, 2726], the Jülich-Bonn (JüBo) [2721], the ANL-Osaka [2727] and by other groups. A short description of the different methods can be found in Ref. [2660]. Here we emphasize that the energy-dependence of a partial-wave amplitude for one particular channel is influenced by other reaction channels due to unitarity constraints. To fully describe the energy-dependence of a production amplitude, all (or at least the most significant) reaction channels must be included in a coupled-channel approach. Many different final states have been measured with high precision off protons and partly also off neutrons (bound in a deuteron with a quasi-free proton in the final state). Polarization data for meson photoproduction off neutrons are, however, still scarce. A fairly complete list of references can be found in [2660]. Most data are now included in single- and in multi-channel analyses [92].

The photoproduction data had a strong impact on the discovery of several new baryon states or provided new evidence for candidate states that had been ob-served previously but lacked confirmation (e.g. [2693, 2705, 2712]). Many new decay modes were discovered, in particular in the photoproduction of $2\pi^0$ and $\pi^0\eta$, [2726, 2728, 2729] and references therein. At the NSTAR'2000 worskhop, 12 $N^*$ and 8 $\Delta^*$ were considered to be established (4*,3*) by the Particle Data Group[93]. These numbers increased to 19 $N^*$ and 10 $\Delta^*$ two decades later. Table 9.2.1 lists the new resonances below 2300 MeV and those that had not a four-star status in 2010. Resonances which had four stars in 2010 are well established and kept their status. These are:

$$N(1440)1/2^+, N(1520)3/2^-, N(1535)1/2^-, N(1650)1/2^-,$$
$$N(1675)5/2^-, N(1680)5/2^+, N(1720)3/2^+, N(2190)7/2^-,$$
$$N(2220)9/2^+ N(2250)9/2^-, \Delta(1620)1/2^- \Delta(1700)3/2^-,$$
$$\Delta(1905)5/2^+, \Delta(1910)1/2^+, \Delta(1950)7/2^+.$$

[91] The target polarisation observable can also be accessed by performing a double polarization experiment using a linearly polarised photon beam and measuring the baryon polarisation in the final state.

[92] A list of data on photoproduction reactions including polarization and double-polarization observables can be found at the BnGa web page: https://pwa.hiskp.uni-bonn.de/

A few resonances were removed from the RPP tables. They often had wide-spread mass values, and the old results were redistributed according to their masses and the new findings. Even more impressive is the number of reported decay modes. Our knowledge on $N^*$ and $\Delta^*$ decays has at least been doubled.

### 9.2.4 Regge trajectories

Like mesons, baryons fall onto linear Regge trajectories when their squared masses are plotted as a function of their total spin $J$ or their intrinsic orbital angular momentum $L$. In the case of $\Delta^*$, the leading trajectory consists of $\Delta(1232)3/2^+$, $\Delta(1950)7/2^+$, $\Delta(2420)11/2^+$, $\Delta(2950)15/2^+$. In the quark model, these have intrinsic orbital angular momenta $L = 0, 2, 4, 6$. Figure 9.2.4 shows the squared $\Delta^*$-masses as a function of $L{+}N_{\rm radial}$, where $N_{\rm radial}$ indicates the intrinsic radial excitation. The resonances $\Delta(1910)1/2^+$, $\Delta(1920)3/2^+$, $\Delta(1905)5/2^+$ have intrinsic $L = 2$ like $\Delta(1950)7/2^+$, and fit onto the trajectory. Also, there are three positive-parity resonances that likely have $L = 4$ with the $5/2^+$ state missing. The two $L = 1$ resonances $\Delta(1620)1/2^-$ and $\Delta(1700)3/2^-$ also have masses close to the linear trajectory. Further, there are resonances in which the $\rho$ or $\lambda$ oscillator is excited radially to $n_\rho = 1$ or $n_\lambda = 1$ ($N_{\rm radial} = 1$). Quark models with a harmonic oscillator as confining potential predict that resonances belong to shells. Radial excitations are predicted in the shell $L + 2\,N_{\rm radial}$. This is not what we find experimentally: the masses are approximately proportional to $L{+}N_{\rm radial}$ if $N_{\rm radial} = 1$ is assigned to $\Delta(1600)3/2^+$, the first radial excitation of $\Delta(1232)3/2^+$, as well as to the $\Delta(1900)1/2^-$, $\Delta(1940)3/2^-$, $\Delta(1930)5/2^-$ triplet, to the two members of a partly unseen quartet $\Delta(2350)5/2^-$ and $\Delta(2400)9/2^-$, and to $\Delta(2750)13/2^-$ (with $L$=5, $S$=3/2 and $N_{\rm radial}$=1).

Clearly, this is a very simplified picture of the $\Delta^*$ spectrum. The picture is that of the non-relativistic quark model – nobody understands why it works[94]. Resonances – assumed to have the same mass if spin orbit-coupling is neglected– have indeed somewhat different masses. But the gross features of the spectrum of $\Delta^*$ resonances are well reproduced.

The nucleon spectrum is more complicated. First, there are more resonances, and second, there are two-quark configurations which are antisymmetric in spin and flavor[95]. Due to instanton induced interactions, the relativistic quark model [2677], expects a lowering of states with the respective symmetry. Indeed baryons with two-quark configurations which are antisymmetric in spin and flavor (*good diquarks*) seem to have lower masses than those having *bad diquarks* only. Attempts to include *good-diquark* effects were rather successful [2663, 2730]. The $\chi^2$ for the model-data comparison was twice better for the 2-parameter fit than for quark models [2731] when the same mass-uncertainties are assumed.

**Table 9.2.1** Baryon resonances above the $\Delta(1232)$ and below 2300 MeV given in the RPP'2022 in comparison to the resonances considered in the RPP'2010. Resonances with 4∗ in 2010 are not listed here. See text for further discussion.

| | RPP 2010 | RPP 2022 | | RPP 2010 | RPP 2022 |
|---|---|---|---|---|---|
| $N(1700)3/2^-$ | *** | *** | $\Delta(1600)3/2^+$ | *** | **** |
| $N(1710)1/2^+$ | *** | **** | $\Delta(1750)1/2^+$ | * | * |
| $N(1860)5/2^+$ | – | ** | $\Delta(1900)1/2^-$ | ** | *** |
| $N(1875)3/2^-$ | – | *** | $\Delta(1920)3/2^+$ | *** | *** |
| $N(1880)1/2^+$ | – | *** | $\Delta(1930)5/2^-$ | *** | *** |
| $N(1895)1/2^-$ | – | **** | $\Delta(1940)3/2^-$ | * | ** |
| $N(1900)3/2^+$ | ** | **** | $\Delta(2000)5/2^+$ | ** | ** |
| $N(1990)7/2^+$ | ** | ** | $\Delta(2150)1/2^-$ | * | * |
| $N(2000)5/2^+$ | ** | ** | $\Delta(2200)7/2^-$ | * | *** |
| $N(2040)3/2^+$ | – | * | | | |
| $N(2060)5/2^-$ | – | *** | N(2080)$3/2^-$ | ** | – |
| $N(2100)1/2^+$ | * | *** | N(2090)$1/2^-$ | * | – |
| $N(2120)3/2^-$ | – | *** | N(2200)$5/2^-$ | ** | – |

[93] In PDG notation: 4* Existence certain, 3* almost certain, 2* evidence fair, 1* poor

[94] In addition, we neglect the possible configuration mixing of states in our discussion.

[95] These two-quark configurations are often called *good diquarks*. They may carry orbital-angular momenta, these are not *frozen* diquarks.

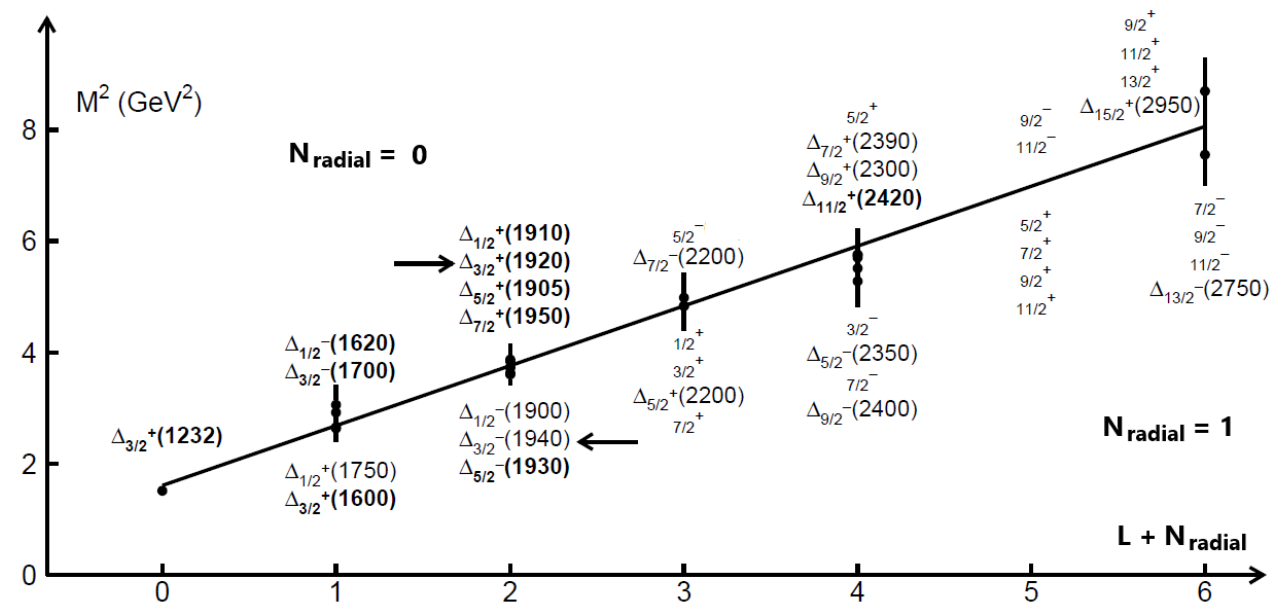

**Fig. 9.2.4** Regge-like trajectory of $\Delta^*$-resonances. Taken from [2732]

### 9.2.5 Hyperons

Nearly no new data on $\bar{K}N$ scattering have become available for several decades except some new data from BNL at very low energy (see Ref. [2733] and references therein). The reaction $\gamma p \to K^+\Sigma\pi$ was studied at JLab and helped to understand the low-energy region [2734]. However, four groups have re-analyzed $K^-p$ reactions using extensive collections of the old data. The new analysis progress was pioneered by the Kent group which performed a comprehensive partial wave analysis [2735, 2736]. Energy-independent amplitudes were constructed by starting from an energy-dependent fit and by freezing or releasing sets of amplitudes. The resulting amplitudes were then fit in a coupled-channel approach. The JPAC group performed coupled-channel fits to the partial waves of the Kent group. The fit described the Kent partial waves well while significant discrepancies showed up between data and the observables calculated from their partial-wave amplitudes [2737]. The ANL-Osaka group used the data set collected by the Kent group and derived energy-dependent amplitudes based on a phenomenological SU(3) Lagrangian. Two models were presented which agreed for the leading contributions but which showed strong deviations for weaker contributions [2738, 2739]. The BnGa group added further data and tested systematically the inclusion of additional states with any set of quantum numbers. Only small improvements in the fit were found [2740, 2741].

The new studies of old data did not change the situation significantly. Some new decay modes were reported, some new but faint signals were found, some were confirmed by one group and missed by others. Several *bumps* were removed from the RPP Tables (for details see [2742]). As a result, our picture of hyperons (with strangeness $S = -1$) remains unclear. Not even all states expected in the first $\Lambda$ and $\Sigma$ excitation shell have been seen. In Table 9.2.2 all candidates are included.

Very little is known about excited Cascade baryons. A few structures in invariant mass spectra were observed, nearly no spin-parities have been determined. The hope is that at FAIR, JLab and J-PARC (see Section 14) new $\Xi$'s and $\Omega$'s will be observed and their quantum numbers be determined.

### 9.2.6 QCD expectations

The spectrum of excited nucleons has been calculated in different approaches. We list a few here: QCD on a lattice has been used to calculate the spectrum of light-quark baryons including hybrid states (see Section 4.5 and [494]). In the Dyson-Schwinger/Bethe-Salpeter approach (see Section 5.3 and [2680]) the covariant three-body Fadeev-equation is solved in a rainbow-ladder approximation. The spectra of baryon resonances have been calculated for $J = 1/2^{\pm}$ and $J = 3/2^{\pm}$, reaching for the $N^*$- and $\Delta^*$-resonances to masses up to about 2000 MeV. AdS/QCD (see Section 5.5 and [1010]) predicts a spectrum of $N^*$ and $\Delta^*$ that is proportional to $L + N_{\text{radial}}$. Using chiral unitary approaches for the meson-baryon interactions, certain baryon resonances can be generated dynamically (see Section 6.2). Various quark models have been developed that treat baryons as bound states of three quarks with constituent masses, a confinement potential and residual quark-quark interactions. The models are discussed in Section 9.1. At present, they are still best suited to discuss what has been learned from recent results in the spectroscopy of light baryons.

### 9.2.7 What did we learn within the quark model?

*SU(6)⊗O(3) classification*

Table 9.2.2 lists the observed $N^*$-, $\Delta^*$-, $\Lambda^*$- and $\Sigma^*$-baryons in a SU(6)⊗O(3) classification. This classification assumes non-relativistic constituent quarks. It has been a miracle since the early times of the quark model that this scheme works so well. But baryon resonances often have a leading component in the wave function corresponding to the SU(6)⊗O(3) classification even in relativistic calculations.

The first excitation shell ($N$=1) is fairly complete. As expected, there are five $N^*$'s and two $\Delta^*$'s with negative parity. Of the $\Lambda$ and $\Sigma$ octet states with negative parity, only the $J^P = 3/2^-$ states are missing[96]. The two states $\Lambda(1800)1/2^-$ and $\Sigma(1750)1/2^-$ are interpreted as states with intrinsic spin 3/2: they seem to be spin partners of $\Lambda(1830)5/2^-$ and $\Sigma(1775)5/2^-$.

[96] The $N(1700)3/2^-$ is wider than its spin partners and more difficult to identify. This may also be the reason for the absence of the $J^P = 3/2^-$ $\Lambda$ and $\Sigma$ states.

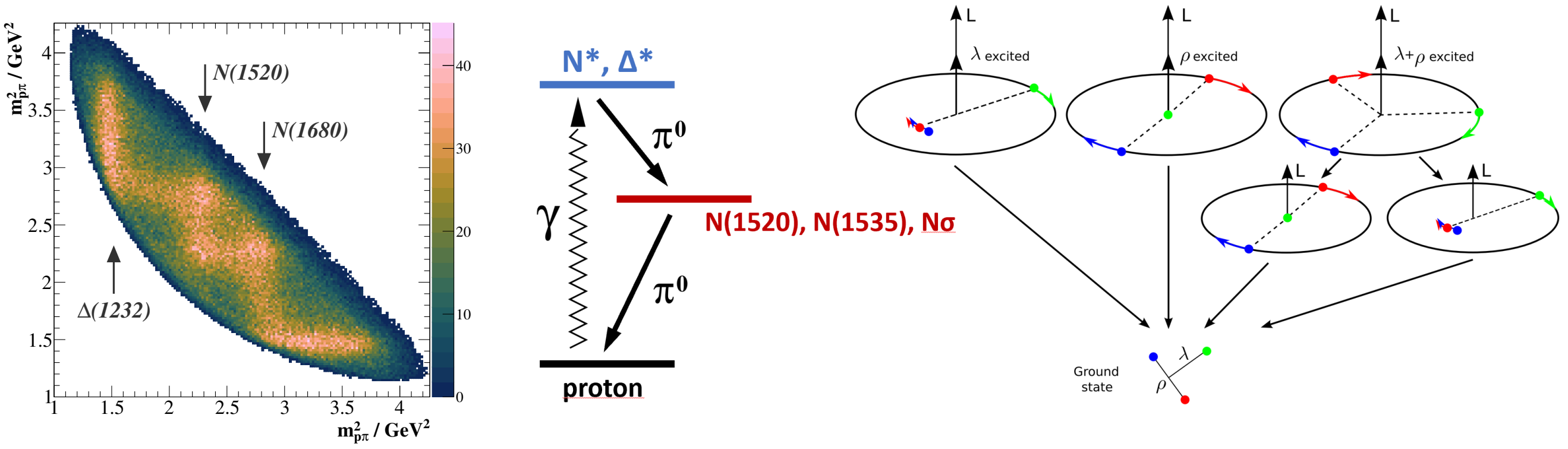

**Fig. 9.2.5** Left: $\gamma p \to p\pi^0\pi^0$-Dalitz plot for a selected $E_\gamma$-bin of 1900-2100 MeV (CBELSA/TAPS) [2743], Middle: Cascade decays of resonances via an immediate state. Right: Classical orbits of nucleon excitations with $L = 2$ (upper row) and $L = 1$ (middle row). Taken from [2729]. The first two pictures in the upper row show excitations of the $\rho$ and $\lambda$ oscillators, in the third picture both, $\rho$ and $\lambda$ are excited. When both oscillators are excited, de-excitation leads to an excited intermediate state (middle row).

The doublet of negative-parity decuplet $\Sigma$ states is not uniquely identified. Expected is this doublet at about 1750 MeV, and in the $(56,1_3^-)$-configuration a second doublet at about 2050 MeV and, finally, a triplet at about the same mass. The analysis found (poor) evidence for two doublets, marked $^a$ in Table 9.2.2. The singlet states $\Lambda(1405)1/2^-$ and $\Lambda(1520)3/2^-$ deserve a more detailed discussion.

At higher masses, some choices are a bit arbitrary: Because of its mass, $N(1900)3/2^+$ belongs to the second excitation shell. It may have intrinsic quark spin 1/2 or 3/2, both with $L = 2$. Further, there should be a $3/2^+$ radially excited state with $L = 0$. These three states can mix. Only one of the states is clearly identified. In any case, quark models predict three resonances with $J^P = 3/2^+$ in this mass range while only one is found. Also missing is a doublet of states with $L = 1$ belonging to the 20plet in SU(6)⊗O(3).[97] The production of this doublet is expected to be strongly suppressed for reasons to be discussed below.

Only few hyperons are known that can be assigned to the second excitation shell. The interpretation of some $\Lambda$ resonances as SU(3) singlet configuration is plausible but not at all compelling.

*Missing resonances*

In the spectrum of $N^*$ and $\Delta^*$, the first excitation shell is complete, in the second shell, 21 states are expected (two of them likely not observable in $\pi N$-elastic scattering or in single/double meson photoproduction), 16 are seen, three are missing. To a large extend, the *missing-resonance* problem is solved for $N^*$ and $\Delta^*$: there are no frozen diquarks. Admittedly, five of the resonances are not yet "established", i.e. have not (yet?) a 3* or 4* status.

[97] The RPP lists three more $N^*/\Delta^*$-resonances: $N(2040)3/2^+$, $\Delta(2150)1/2^-$, which need confirmation and $N(2100)1/2^+$ which we assign to the 4th shell.

In the third shell, only few resonances are known, but the number of expected resonances is quite large and the analysis challenging: 45 $N^*$ and $\Delta^*$, likely with widths often exceeding 300 MeV, are expected to populate an about 400 MeV wide mass range.

| $J^P$<br>Masses | 1. shell<br>1/2, 3/2, 5/2⁻<br>1500 - 1750 | 2. shell<br>1/2 3/2 5/2 7/2⁺<br>1700 - 2100 | 3. shell<br>1/2⁻ – 9/2⁻<br>1900 - 2300 |
|---|---|---|---|
| $N$ | 5: 2 2 1 | 13: 4 5 3 1 | 30: 7 9 8 5 1 |
| $\Delta$ | 2: 1 1 - | 8: 2 3 2 1 | 15: 3 5 4 2 1 |

*Three-quark dynamics in cascade decays*

The CBELSA/TAPS collaboration studied cascade decays of high mass resonances via an intermediate resonance down to the ground state nucleon. The analyses were based on a large data base of photoproduction data including final states such as $\gamma p \to p\pi^0\pi^0$ and $p\pi^0\eta$ (see [2726, 2728] and Refs. therein). The Dalitz plot of Fig. 9.2.5, shows very clearly band-like structures due to the occurrence of baryon resonances in the intermediate state. It was observed that the positive parity $N^*$- and $\Delta^*$-resonances at a mass of about 1900 MeV show a very different decay pattern. The four $N^*$-resonances:

$$N(1880)1/2^+, N(1900)3/2^+, N(2000)5/2^+, N(1990)7/2^+,$$

decay with an average branching fraction of $(34 \pm 6)\%$ into $N\pi$ and $\Delta\pi$ and with a branching fraction of $(21 \pm 5)\%$ into the orbitally excited states $N(1520)3/2^-\pi$, $N(1535)1/2^-\pi$, and $N\sigma$. The four $\Delta^*$-states:

$$\Delta(1910)1/2^+, \Delta(1920)3/2^+, \Delta(1905)5/2^+, \Delta(1950)7/2^+,$$

have an average decay branching fraction into $N\pi/\Delta\pi$ of $(44 \pm 7)\%$ while their branching fraction into the excited states mentioned above is almost negligible, only $(5\pm2)\%$ [2726]. At the first sight, this is very surprising.

**Table 9.2.2** The spectrum of $N$, $\Delta$, $\Lambda$ and $\Sigma$ excitations. The first row shows the quantum numbers of the SU(6)⊗O(3) symmetry group. $D$ is the dimensionality of the SU(6) group, $L$ the total internal quark orbital angular momentum, $P$ the parity, $N$ a shell index, $S$ the total quark spin, $J$ the total angular momentum. The assignment of particles to SU(6)⊗O(3) is an educated guess. In the first and second excitation band, all expected states are listed, missing resonances are indicated by a $-$ sign. The third band lists only bands for which at least one candidate exists. The states with an index are special: above 1700 MeV, one pair of $\Sigma$ states is expected at about 1750 to 1800 MeV, two pairs at about 2000 to 2050 MeV. Two pairs marked[a] are found only. The pairs are shown with the three possible assignments. Likewise, $N(2060)$ and $N(2190)$ marked[b] could form a spin-doublet or be members of a spin-quartet. Likely, the observed pairs of states are mixtures of these allowed configurations (Adapted from [2742]).

| $(D, L_N^P)$ | $S$ | $J^P$ | Singlet | Octet | | | Decuplet | |
|---|---|---|---|---|---|---|---|---|
| $(56, 0_0^+)$ | $\frac{1}{2}$ | $\frac{1}{2}^+$ | | $N(939)$ | $\Lambda(1116)$ | $\Sigma(1193)$ | | |
| | $\frac{3}{2}$ | $\frac{3}{2}^+$ | | | | | $\Delta(1232)$ | $\Sigma(1385)$ |
| $(70, 1_1^-)$ | $\frac{1}{2}$ | $\frac{1}{2}^-$ | $\Lambda(1405)$ | $N(1535)$ | $\Lambda(1670)$ | $\Sigma(1620)$ | $\Delta(1620)$ | $\Sigma(1900)^a$ |
| | | $\frac{3}{2}^-$ | $\Lambda(1520)$ | $N(1520)$ | $\Lambda(1690)$ | $\Sigma(1670)$ | $\Delta(1700)$ | $\Sigma(1910)^a$ |
| | $\frac{3}{2}$ | $\frac{1}{2}^-$ | | $N(1650)$ | $\Lambda(1800)$ | $\Sigma(1750)$ | | |
| | | $\frac{3}{2}^-$ | | $N(1700)$ | - | - | | |
| | | $\frac{5}{2}^-$ | | $N(1675)$ | $\Lambda(1830)$ | $\Sigma(1775)$ | | |
| $(56, 0_2^+)$ | $\frac{1}{2}$ | $\frac{1}{2}^+$ | | $N(1440)$ | $\Lambda(1600)$ | $\Sigma(1660)$ | | |
| | $\frac{3}{2}$ | $\frac{3}{2}^+$ | | | | | $\Delta(1600)$ | $\Sigma(1780)$ |
| $(70, 0_2^+)$ | $\frac{1}{2}$ | $\frac{1}{2}^+$ | $\Lambda(1710)$ | $N(1710)$ | $\Lambda(1810)$ | $\Sigma(1880)$ | $\Delta(1750)$ | - |
| | $\frac{3}{2}$ | $\frac{3}{2}^+$ | | - | - | - | | |
| $(56, 2_2^+)$ | $\frac{1}{2}$ | $\frac{3}{2}^+$ | | $N(1720)$ | $\Lambda(1890)$ | $\Sigma(1940)$ | | |
| | $\frac{1}{2}$ | $\frac{5}{2}^+$ | | $N(1680)$ | $\Lambda(1820)$ | $\Sigma(1915)$ | | |
| | $\frac{3}{2}$ | $\frac{1}{2}^+$ | | | | | $\Delta(1910)$ | - |
| | $\frac{3}{2}$ | $\frac{3}{2}^+$ | | | | | $\Delta(1920)$ | $\Sigma(2080)$ |
| | $\frac{3}{2}$ | $\frac{5}{2}^+$ | | | | | $\Delta(1905)$ | $\Sigma(2070)$ |
| | $\frac{3}{2}$ | $\frac{7}{2}^+$ | | | | | $\Delta(1950)$ | $\Sigma(2030)$ |
| $(70, 2_2^+)$ | $\frac{1}{2}$ | $\frac{3}{2}^+$ | $\Lambda(2070)$ | - | - | - | - | - |
| | | $\frac{5}{2}^+$ | $\Lambda(2110)$ | $N(1860)$ | - | - | $\Delta(2000)$ | - |
| | $\frac{3}{2}$ | $\frac{1}{2}^+$ | | $N(1880)$ | - | - | | |
| | | $\frac{3}{2}^+$ | | $N(1900)$ | - | - | | |
| | | $\frac{5}{2}^+$ | | $N(2000)$ | - | - | | |
| | | $\frac{7}{2}^+$ | | $N(1990)$ | $\Lambda(2085)$ | - | | |
| $(20, 1_2^+)$ | $\frac{1}{2}$ | $\frac{1}{2}^+$ | - | - | - | - | | |
| | | $\frac{3}{2}^+$ | - | - | - | - | | |
| | | $\frac{5}{2}^+$ | - | | | | | |
| $(56, 1_3^-)$ | $\frac{1}{2}$ | $\frac{1}{2}^-$ | | $N(1895)$ | $\Lambda(2000)$ | $\Sigma(1900)^a$ | | |
| | | $\frac{3}{2}^-$ | | $N(1875)$ | $\Lambda(2050)$ | $\Sigma(1910)^a$ | | |
| | $\frac{3}{2}$ | $\frac{1}{2}^-$ | | | | | $\Delta(1900)$ | $\Sigma(2110)^a$ |
| | | $\frac{3}{2}^-$ | | | | | $\Delta(1940)$ | $\Sigma(2010)^a$ |
| | | $\frac{5}{2}^-$ | | | | | $\Delta(1930)$ | - |
| $(70, 3_3^-)$ | $\frac{1}{2}$ | $\frac{5}{2}^-$ | $\Lambda(2080)$ | $N(2060)^b$ | - | - | - | - |
| | | $\frac{7}{2}^-$ | $\Lambda(2100)$ | $N(2190)^b$ | - | $\Sigma(2100)$ | $\Delta(2200)$ | - |
| $(70, 3_3^-)$ | $\frac{3}{2}$ | $\frac{3}{2}^-$ | | $N(2120)$ | - | - | - | - |
| | | $\frac{5}{2}^-$ | | $N(2060)^b$ | - | - | - | - |
| | | $\frac{7}{2}^-$ | | $N(2190)^b$ | - | - | - | - |
| | | $\frac{9}{2}^-$ | | $N(2290)$ | - | - | - | - |

The difference can be traced to the different wave functions. The spin and the flavor wave functions of the four $\Delta^*$-states are both symmetric with respect to the exchange of any two quarks, the spatial wave function needs to be symmetric as well. This means that - having a three-quark-picture in mind - that either the $\rho$- or the $\lambda$-oscillator is excited to $\ell = 2$, the other one is not excited. (There is a mixture of the two possibilities $\ell_\rho = 2$, $\ell_\lambda = 0$ or $\ell_\lambda = 2$, $\ell_\rho = 0$). If this state decays, the orbital angular momentum is carried away and the decay products are found preferentially in their ground state.

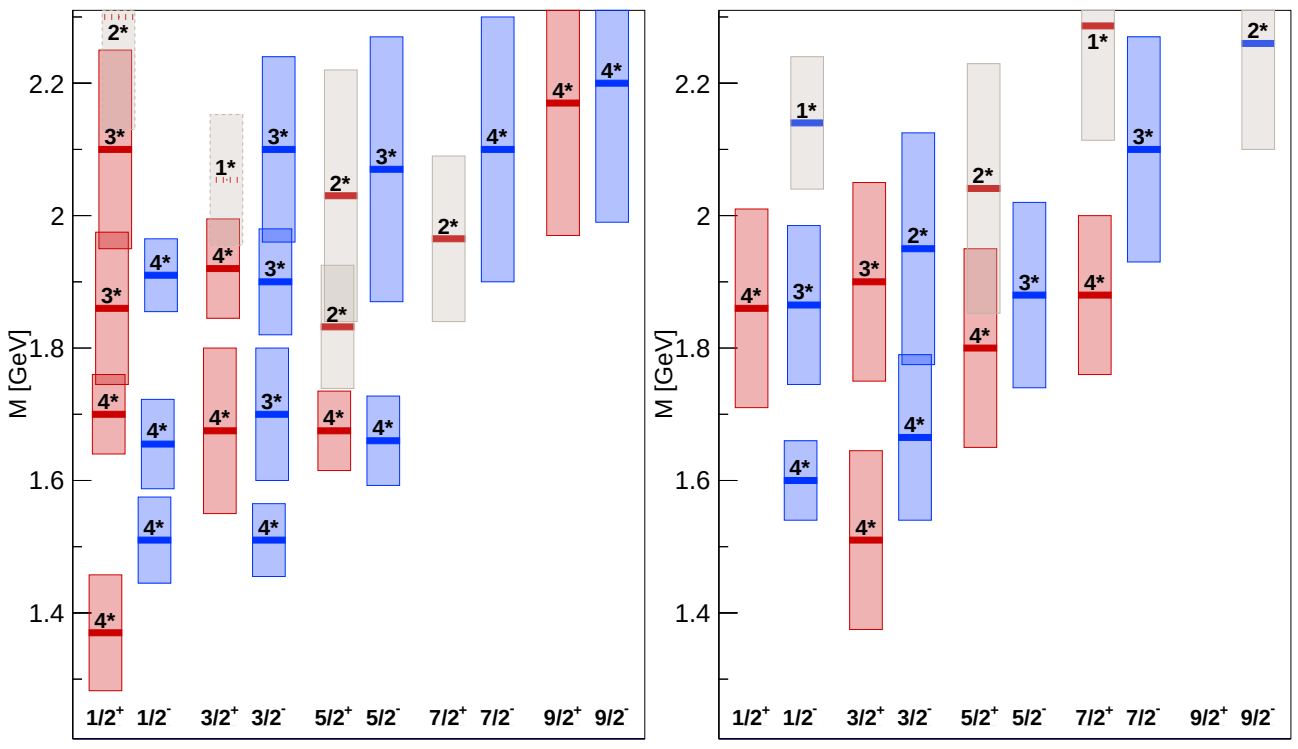

**Fig. 9.2.6** N*- (left) and $\Delta^*$-resonances (right) above $\Delta(1232)$ for different spin and parities $J^p$. For each resonance, the real part of the pole position $Re(M_R)$ is given together with a box of length $\pm Im(M_R)$, using the PDG estimates. $2 \cdot Im(M_R)$ corresponds to the total width of the resonance. RPP star ratings are also indicated. If no pole positions are given in the RPP (above the line), the RPP Breit-Wigner estimates for masses and widths are used instead. This is indicated by dashed resonance-mass lines and dashed lines surrounding the boxes. If no RPP-estimates are given, the values above the line have been averaged and the states are shown as gray boxes. This may indicate one measurement above the line only. $\Delta(1750)1/2^+$ is not included, as there is no RPP-value given above the line.

The four $N^*$-states have a spatial wave function with mixed symmetry. Thus the spatial wave function has one part which is mixed-symmetric and one part which is mixed anti-symmetric. In the latter one, both oscillators are excited simultaneously ($\ell_\rho = \ell_\lambda = 1$). If this state decays, one of the excitations remains in the decay product as illustrated in Fig. 9.2.5. A similar argument has been used by Hey and Kelly [2744] to explain why the 20'plet in the second excitation shell of Fig. 9.2.2 cannot be formed in a $\pi N$ scattering experiment. For the 20'plet the spacial wave function is entirely antisymmetric, both oscillators are excited simultaneously, and there is no other component in the wave function. A single-step excitation is suppressed.

*Parity doublets?*
The spontaneous breaking of the chiral symmetry leads to the large mass gap observed between chiral partners: the masses of the $\rho(770)$ meson with spin-parity $J^P = 1^-$ and its chiral partner $a_1(1260)$ with $J^P = 1^+$ differ by about 500 MeV, those of the $J^P = 1/2^+$ nucleon and $N(1535)1/2^-$ by about 600 MeV. In contrast to quark-models expectations and lattice QCD calculations [494] higher-mass baryons are often observed in parity doublets (see Fig. 9.2.6), in pairs of resonances having about the same mass, the same total spin $J$ and opposite parities.

This observation and similar observations in meson spectrum has led to the suggestion that chiral sym-metry might be effectively restored in highly excited hadrons [2691, 2745]. Then, all high-mass resonances should have a parity partner. This is a testable prediction.

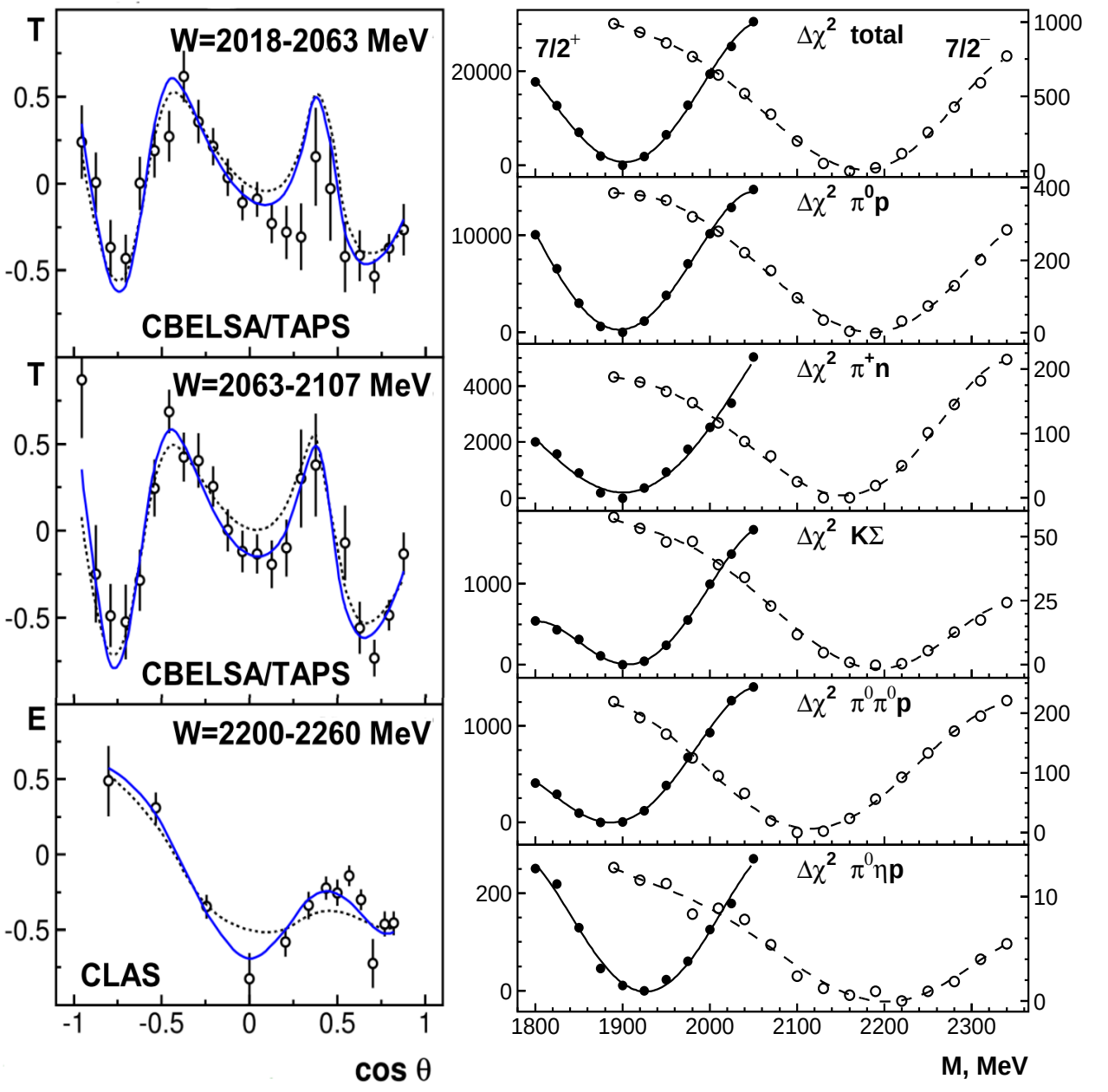

**Fig. 9.2.7** Left: The new polarization observables $T$ and $E$ shown for selected mass bins (see [2693] for refs. to the data). The fit curves represent the best fits with (solid) and without (dashed) inclusion of $\Delta(2200)7/2^-$. Right: The increase in pseudo-$\chi^2$ of the fit to a large body of pion- and photo-produced reactions when the mass of $\Delta(1950)7/2^+$ (solid points) or $\Delta(2200)7/2^-$ (open circles) is scanned. The scale on the left (right) abscissa refers to the $7/2^+$ ($7/2^-$) partial wave. The curves are to guide the eye. Adapted/taken from [2693].

In the mass region of 1900 MeV a quartet of well known positive parity $\Delta^*$ states exists, consisting of

$$\Delta(1910)1/2^+, \Delta(1920)3/2^+, \Delta(1905)5/2^+, \Delta(1950)7/2^+.$$

Figure 9.2.6 shows the parity partners of the first three states:

$$\Delta(1900)1/2^-, \Delta(1940)3/2^-, \Delta(1930)5/2^-.$$

However, the four-star $\Delta(1950)7/2^+$ has no close-by $\Delta(xxx)7/2^-$-state that could serve as parity partner. Where is the closest $\Delta^*$ with $J^P = 7/2^-$ ? Figure 9.2.7 shows a resonance scan over the mass region of interest [2693]. There is clear evidence for $\Delta(2200)7/2^-$ (which was upgraded from 1* to 3* based on this result). But its mass difference to $\Delta(1950)7/2^+$ is too large. These two states are no parity partners! Within the quark model and the SU(6)⊗O(3)-systematics, the four positive-parity $\Delta^*$'s have $L = 2, S = 3/2$ that couple to $J^P = 1/2^+, \cdots, 7/2^+$. The natural assignment for the three negative-parity $\Delta^*$'s is that they form a triplet with $L = 1$ and $S = 3/2$. Then, they must have one unit of radial excitation. The four positive-parity $\Delta$-states belong to the $2\hbar\omega$ shell and the negative-parity states to the $3\hbar\omega$ shell. With masses considered to be proportional to $L + N_{\rm radial}$, these seven states are expected to have about the same mass. $\Delta(2200)7/2^-$ has $L = 3, S = 1/2$ and its expected mass is higher. We note that $\Delta(2400)9/2^-$ has $L = 3, S = 3/2$, and we assume $N_{\rm radial} = 1$ for this state (as well as for $\Delta(2750)13/2^-$, see Fig. 9.2.4).

### 9.2.8 Dynamically generated resonances

*$N^*$'s and $\Delta^*$'s*

Apart from $\Lambda(1405)1/2^-$ that will be discussed below, the first dynamically generated resonance was the negative-parity $N(1535)1/2^-$ [2688]. At the 1995 International Conference on the Structure of Baryons, Santa Fe, New Mexico, there was a heated discussion between Weise, defending his new approach, and Isgur who argued that $N(1535)1/2^-$ is well understood within the quark model and no new approach is needed. For some time, there was even the idea that there could be two overlapping states but this is excluded by data. Later, in Refs. [2746, 2747], $N(1535)1/2^-$ and $N(1650)1/2^-$, were both shown to be generated dynamically. However, $\Delta(1620)1/2^-$ was not[98]. An important question remains: Are (qqq)-resonance poles and dynamically generated poles different descriptions of the same object or do they present different (orthogonal) states?

*The $\Lambda(1405)1/2^-$*

The $\Lambda(1405)1/2^-$ mass is very close to the $N\bar{K}$ threshold. Kaiser, Waas and Weise [2748] proved that the resonance can be generated dynamically from $N\bar{K} - \Sigma\pi$ coupled-channel dynamics. Oller and Meissner [2749] studied the $S$-wave $N\bar{K}$ interactions in a relativistic chiral unitary approach based on a chiral Lagrangian obtained from the interaction of the octet of pseudoscalar mesons and the ground state baryon octet and found two isoscalar resonances in the $\Lambda(1405)1/2^-$ mass region and one isovector state. In a subsequent paper [2750], Jido *et al.* studied the the effects of SU(3) breaking on the results in detail. These two papers had an immense impact on the further development. It is the only result in light-baryon spectroscopy that is in clear contradiction to the quark model. It introduces a new state $\Lambda(1380)1/2^-$, that has no role in a quark model, it enforces an interpretation of $\Lambda(1405)1/2^-$ as mainly SU(3) octet resonance, and it interprets $\Lambda(1670)1/2^-$ as high-mass partner of $\Lambda(1405)1/2^-$. The $\Lambda(1405)1/2^-$ and $\Lambda(1670)1/2^-$ would then be the strange partners of the $N(1535)1/2^-$ and the $N(1650)1/2^-$. In quark models, $\Lambda(1405)$ is a mainly SU(3) singlet resonance and the octet states $\Lambda(1670)1/2^-$ and $\Lambda(1800)1/2^-$ are the strange partners of $N(1535)1/2^-$ and $N(1650)1/2^-$ (see Table 9.2.2). In the quark-model interpretation, the hyperon states $\Lambda(1405)1/2^-$ and $\Lambda(1670)1/2^-$ have close-by $J^P = 3/2^-$ partners (the $J^P = 3/2^-$-partner of $\Lambda(1800)1/2^-$ is missing but there is $\Lambda(1830)5/2^-$). The masses of the mainly octet states are about 130 MeV above their non-strange partners.

[98] It should be mentioned that not only the SU(6)⊗O(3)-systematics in the spectrum seems to indicate a 3-quark-nature of $N(1535)1/2^-$ and $N(1650)1/2^-$ but also the electroproduction results discussed in the following section 9.3 indicate that $N(1535)1/2^-$ is a 3-quark state with little meson-baryon contribution only ($Q^2$ dependence of the transition form factor $A_{1/2}$).

This conflict initiated an attempt to fit (nearly) all existing data relevant for $\Lambda(1405)1/2^-$ in the BnGa approach [2751]. The data could be fit with one single resonance in the $\Lambda(1405)1/2^-$ region but were also compatible, with a slightly worsened $\chi^2$, with a description using two resonances with properties as obtained in the chiral unitary approach.

### 9.2.9 Outlook

There is not yet a unified picture of baryons. Regge-like trajectories ($M^2 \propto L + N_{\rm radial}$) are best described by AdS/QCD. Unitary effective field theories describe consistently meson-baryon interactions and some resonances are generated dynamically from their interaction. The quark model is useful to understand cascade decays of highly excited states and is indispensable to discuss the full spectrum including missing resonances.The symmetry of quark pairs, symmetric or anti-symmetric with respect to their exchange, has a significant impact on baryon masses. They could be due to effective gluon exchange. More likely seems an interpretation by quark and gluon condensates, e.g. by instanton-induced interactions. Based on the new high quality (polarized) photoproduction data, new baryon resonances were discovered and our knowledge of properties of existing resonances has increased considerably. Yet, our understanding is still unsatisfactory mirroring the complexity of QCD in the non-perturbative regime. New results from lattice QCD are eagerly awaited and new experiments are needed to understand the spectrum and the properties of baryon resonances in further detail. Those include further precise photoproduction experiments measuring polarisation observables not only off the proton but also off the neutron as well as multi-meson final states. Strange baryon resonances need to be addressed. Other production processes such as electroproduction, $\bar{p}p$-annihilation, experiments with $\pi$- or $K$-beams and baryon resonances produced in $J/\psi$ or $\psi'$-decays will also contribute to improve our understanding of the bound states of the strong interaction.

## 9.3 Nucleon Resonances and Transition Form Factors

**Volker D. Burkert**

Meson photoproduction has become an essential tool in the search for new excited light-quark baryon states. As discussed in the previous section, many new excited states have been discovered thanks to high precision photoproduction data in different final states [2718], and are now included in recent editions of the Review of Particle Physics (RPP) [476]. The exploration of the internal structure of excited states and the effective degrees of freedom contributing to s-channel resonance excitation requires the use of electron beams, which is the subject of this contribution, where the virtuality ($Q^2$) of the exchanged photon can be varied to pierce through the peripheral meson cloud and probe the quark core and its spatial structure. Electroproduction can thus say something about if a resonance is generated through short distance photon interaction with the small quark core, or through interaction with a more extended hadronic system.

The experimental exploration of resonance transition form factors reaches over 60 years with many review articles describing this history. Here we refer to a few recent ones [2752–2755]. A review of recent electroproduction experiments in hadron physics and their interpretation within modern approaches of strong interaction physics can be found in Ref. [2756].

Electroproduction of final states with pseudoscalar mesons (e.g. $N\pi$, $p\eta$, $K\Lambda$) have been employed at Jefferson Laboratory mostly with the CEBAF Large Acceptance Spectrometer (CLAS) operating at an instantaneous luminosity of $10^{34}$sec$^{-1}$cm$^{-2}$. In Hall A and Hall C, pairs of individual well-shielded focusing magnetic spectrometers are employed with more specialized aims and limited acceptance, but operating at much higher luminosity. This experimental program led to new insights into the scale dependence of effective degrees of freedom, e.g. meson-baryon, constituent quarks, and dressed quark contributions. Several excited states, shown in Fig. 9.3.1 assigned to their primary $SU(6)\otimes O(3)$ supermultiplets, have been studied this way, mostly with CLAS in Hall B. Most of the resonance couplings have been extracted from single pseudoscalar meson production. In electroproduction, there are 6 complex helicity amplitudes, requiring a minimum of 11 independent measurements for a complete [99] model-independent determination of the amplitudes. In addition, measurements of isospin amplitudes require additional measurements. Following this, the complex amplitudes would need to be subjected to analyses of their phase motions to determine resonance masses on the (real) energy axis, or poles in the (complex) energy plane. Fortunately, in the lower mass range a variety of constraints can be applied to limit the number of unknowns when fitting the cross section data. These include the masses of states quite well known from hadronic processes or from meson photoproduction. Also, the number of possible angular momenta is limited to $l_\pi \leq 2$ in the examples discussed in the following. Additional constraints come from the Watson theorem [2758] that relates the electromagnetic phases to the hadronic ones, and the use of dispersion relations, assuming the imaginary parts of the amplitude are given by the resonance contribution, and the real parts determined through dispersion integrals and additional pole terms. Other approaches use unitary isobar models that parameterize all known resonances and background terms, and unitarize the full amplitudes in a K-matrix procedure. In the following, we show results based on both approaches, where additional systematic uncertainties have been derived from the differences in the two procedures.

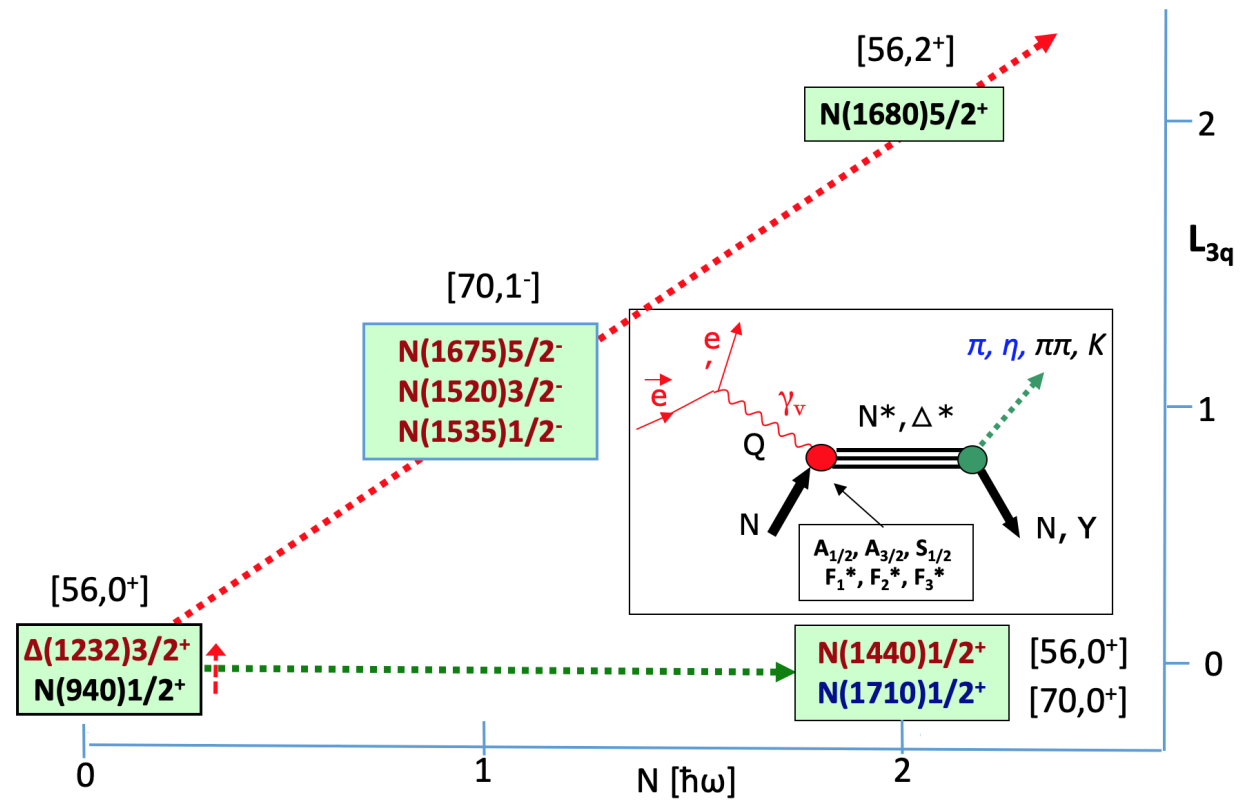

**Fig. 9.3.1** Excited states of the proton that have been studied in electroproduction to determine their resonance transition amplitudes or form factors. States highlighted in red are discussed in this subsection. Graphics from Ref. [2757].

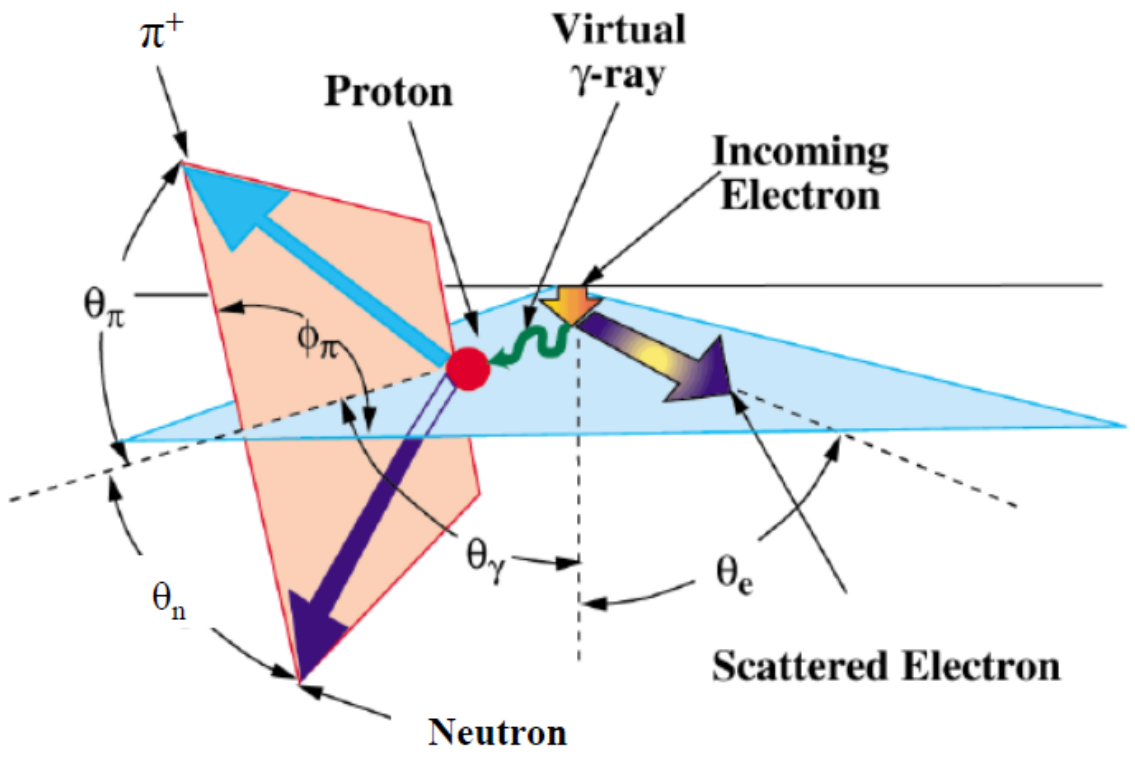

**Fig. 9.3.2** The kinematics of single $\pi^+$ electro-production off protons in the laboratory system.

The availability of electron accelerators with the possibility of generating high beam currents at CEBAF at Jefferson Lab in the US and MAMI at Mainz in Germany, has enabled precise studies of the internal structure of excited states in the $N^*$ and the $\Delta^*$ sectors employing s-channel resonance excitations in large ranges of photon virtuality $Q^2$. This has enabled probing the degrees of freedom relevant in the resonance excitation as a function of the distance scale probed. This is the topic we will elucidate in the following sections and the relevance to (approximations to) QCD. First we briefly discuss the formalism needed for a quantitative analysis of the single pseudoscalar meson electroproduction.

### 9.3.1 Formalism in the analysis of electroproduction of single pseudoscalar mesons

The simplest process used in the study of resonance transition amplitudes is single pion or kaon production, e.g. $ep \to e\pi^+ n$. Single $\pi^+$ and $\pi^0$ production are most suitable for the study of the lower-mass excited states as they couple dominantly to the excited states with masses up to $\approx 1.7$ GeV. It may then be useful to describe in more detail the analysis techniques and the formalism used. The unpolarized differential cross section for single pseudoscalar meson production can be written in the one-photon exchange approximation as:

$$\frac{d\sigma}{dE_f d\Omega_e d\Omega_\pi} = \Gamma \frac{d\sigma}{d\Omega_\pi}\,, \tag{9.3.1}$$

where $\Gamma$ is the virtual photon flux,

$$\Gamma = \frac{\alpha_{em}}{2\pi^2 Q^2}\frac{(W^2 - M^2)E_f}{2ME_i}\frac{1}{1-\epsilon}, \tag{9.3.2}$$

where $M$ is the proton mass, $W$ the mass of the hadronic final state, $\epsilon$ is the photon polarization parameter, $Q^2$ the photon virtuality, $E_i$ and $E_f$ represent the initial and the final electron energies, respectively. Moreover,

$$\epsilon = \left[1 + 2\left(1 + \frac{\nu^2}{Q^2}\right)\tan^2\frac{\theta_e}{2}\right]^{-1} \tag{9.3.3}$$

and

$$\frac{d\sigma}{d\Omega_\pi} = \sigma_T + \epsilon\sigma_L + \epsilon\sigma_{TT}\cos 2\phi_\pi + \sqrt{2\epsilon(1+\epsilon)}\sigma_{LT}\cos\phi_\pi\,.$$

[99] With the exception of an overall phase that cannot be determined

The kinematics for single $\pi^+$ production is shown in Fig. 9.3.2.
The observables of the process $\gamma_v p \to \pi N'$ can be expressed in terms of six parity-conserving helicity amplitudes [2754, 2759, 2760] :

$$H_i = \langle \lambda_\pi ; \lambda_N | T | \lambda_{\gamma_\nu} ; \lambda_p \rangle \,, \quad (9.3.4)$$

where $\lambda$ denotes the helicity of the respective particle, $\lambda_\pi = 0$, $\lambda_N = \pm\frac{1}{2}$, $\lambda_{\gamma_v} = \pm 1, 0$, and $\lambda_p = \pm\frac{1}{2}$, and $H_i$ are complex functions of $Q^2$, $W$, and $\theta^*_\pi$.

### 9.3.2 Multipoles and partial wave decompositions

The response functions in (1) are given by:

$$\sigma_T = \frac{\vec{p}_\pi W}{2KM}(|H_1|^2 + |H_2|^2 + |H_3|^2 + |H_4|^2), \quad (9.3.5)$$

$$\sigma_L = \frac{\vec{p}_\pi W}{2KM}(|H_5|^2 + |H_6|^2), \quad (9.3.6)$$

$$\sigma_{TT} = \frac{\vec{p}_\pi W}{2KM} Re(H_2 H_3^* - H_1 H_4^*), \quad (9.3.7)$$

$$\sigma_{LT} = \frac{\vec{p}_\pi W}{2KM} Re[H_5^*(H_1 - H_4) + H_6^*(H_2 + H_3)] \,, \quad (9.3.8)$$

where $\vec{p}_\pi$ is the pion 3-momentum in the hadronic center-of-mass system, and $K$ is the equivalent real photon lab energy needed to generate a state with mass $W$:

$$K = \frac{W^2 - M^2}{2M} \,. \quad (9.3.9)$$

The helicity amplitudes $H_i, i = 1$–$6$, can be expanded into Legendre polynomials:

$$H_1 = \frac{1}{\sqrt{2}} \sin\theta \cos\frac{\theta}{2} \sum_{l=1}^{\infty} (B_{l+} - B_{(l+1)-})(P_l'' - P_{l+1}'')$$

$$H_2 = \sqrt{2} \cos\frac{\theta}{2} \sum_{l=1}^{\infty} (A_{l+} - A_{(l+1)-})(P_l' - P_{l+1}')$$

$$H_3 = \frac{1}{\sqrt{2}} \sin\theta \sin\frac{\theta}{2} \sum_{l=1}^{\infty} (B_{l+} + B_{(l+1)-})(P_l'' + P_{l+1}'')$$

$$H_4 = \sqrt{2} \sin\frac{\theta}{2} \sum_{l=1}^{\infty} (A_{l+} + A_{(l+1)-})(P_l' + P_{l+1}')$$

$$H_5 = \sqrt{2} \cos\frac{\theta}{2} \sum_{l=1}^{\infty} (C_{l+} - C_{(l+1)-})(P_l' - P_{l+1}')$$

$$H_6 = \sqrt{2} \sin\frac{\theta}{2} \sum_{l=1}^{\infty} (C_{l+} + C_{(l+1)-})(P_l' + P_{l+1}') \,, \quad (9.3.10)$$

where the $A_{l+}$ and $B_{l+}$ etc., are the transverse partial wave helicity elements for $\lambda_{\gamma p} = \frac{1}{2}$ and $\lambda_{\gamma p} = \frac{3}{2}$, and $C_\pm$ the longitudinal partial wave helicity elements. In the subscript, $l+$ and $(l+1)-$ define the $\pi$ orbital angular momenta, and the sign $\pm$ is related to the total angular momentum $J = l_\pi \pm \frac{1}{2}$. In the analysis of data on the $N\Delta(1232)$ transition, linear combinations of partial wave helicity elements are expressed in terms of electromagnetic multipoles:

$$M_{l+} = \frac{1}{2(l+1)}[2A_{l+} - (l+2)B_{l+}] \quad (9.3.11)$$

$$E_{l+} = \frac{1}{2(l+1)}(2A_{l+} + lB_{l+}) \quad (9.3.12)$$

$$M_{l+1,-} = \frac{1}{2(l+1)}(2A_{l+1,-} - lB_{l+1,-}) \quad (9.3.13)$$

$$E_{l+1,-} = \frac{1}{2(l+1)}[-2A_{l+1,-} + (l+2)B_{l+1,-}] \quad (9.3.14)$$

$$S_{l+} = \frac{1}{l+1}\sqrt{\frac{\vec{Q}^{*2}}{Q^2}} C_{l+} \quad (9.3.15)$$

$$S_{l+1,-} = \frac{1}{l+1}\sqrt{\frac{\vec{Q}^{*2}}{Q^2}} C_{l+1,-} \,, \quad (9.3.16)$$

where $\vec{Q}^*$ is the photon 3-momentum in the hadronic rest frame. The electromagnetic multipoles are often used to describe the transition from the nucleon ground state to the $\Delta(1232)$, which is dominantly described as a magnetic dipole transition $M_{1+}$. The electromagnetic multipoles as well as the partial wave helicity elements are complex quantities and contain both non-resonant and resonant contributions. In order to compare the results to model predictions and LQCD, an additional analysis must be performed to separate the resonant parts $\hat{A}_\pm$, $\hat{B}_\pm$, etc., from the non-resonant parts of the amplitudes. In a final step, the known hadronic properties of a given resonance can be used to determine photocoupling helicity amplitudes that characterize the electromagnetic vertex:

$$\hat{A}_{l\pm} = \mp F C^I_{\pi N} A_{1/2}, \quad (9.3.17)$$

$$\hat{B}_{l\pm} = \pm F\sqrt{\frac{16}{(2j-1)(2j+3)}} C^I_{\pi N} A_{3/2}, \quad (9.3.18)$$

$$\hat{S}_{l\pm} = -F\frac{2\sqrt{2}}{2J+1} C^I_{\pi N} S_{1/2}, \quad (9.3.19)$$

$$F = \sqrt{\frac{1}{(2j+1)}\pi\frac{K}{p_\pi}\frac{\Gamma_\pi}{\Gamma^2}}$$

where the $C^I_{\pi N}$ are isospin coefficients. The total transverse absorption cross section for the transition into a specific resonance is given by:

$$\sigma_T = \frac{2M}{W_R \Gamma}(A^2_{1/2} + A^2_{3/2}). \quad (9.3.20)$$

Experiments in the region of the $\Delta(1232)\frac{3}{2}^+$ resonance often determine the electric quadrupole ratio $R_{EM}$

$$R_{EM} = \frac{Im(E_{1+})}{Im(M_{1+})} \quad (9.3.21)$$

and the scalar quadrupole ratio $R_{SM}$

$$R_{SM} = \frac{Im(S_{1+})}{Im(M_{1+})} \tag{9.3.22}$$

where $E_{1+}$, $S_{1+}$, and $M_{1+}$ are the electromagnetic transition multipoles at the mass of the $\Delta(1232)\frac{3}{2}^+$ resonance.

### 9.3.3 Resonance analysis tools

A model-independent determination of the amplitudes contributing to the electro-excitation of resonances in single pseudoscalar pion production $ep \to e'N\pi$ (see kinematics of single pion production in Fig. 9.3.2) requires a large number of independent measurements at each value of the electron kinematics $W$, $Q^2$, the hadronic cms angle $\cos\theta^\pi$, and the azimuthal angle $\phi^\pi$ describing the angle between the electron scattering plane and the hadronic decay plane. Such a measurement requires full exclusivity of the final state and employing both polarized electron beams and the measurements of the nucleon recoil polarization.

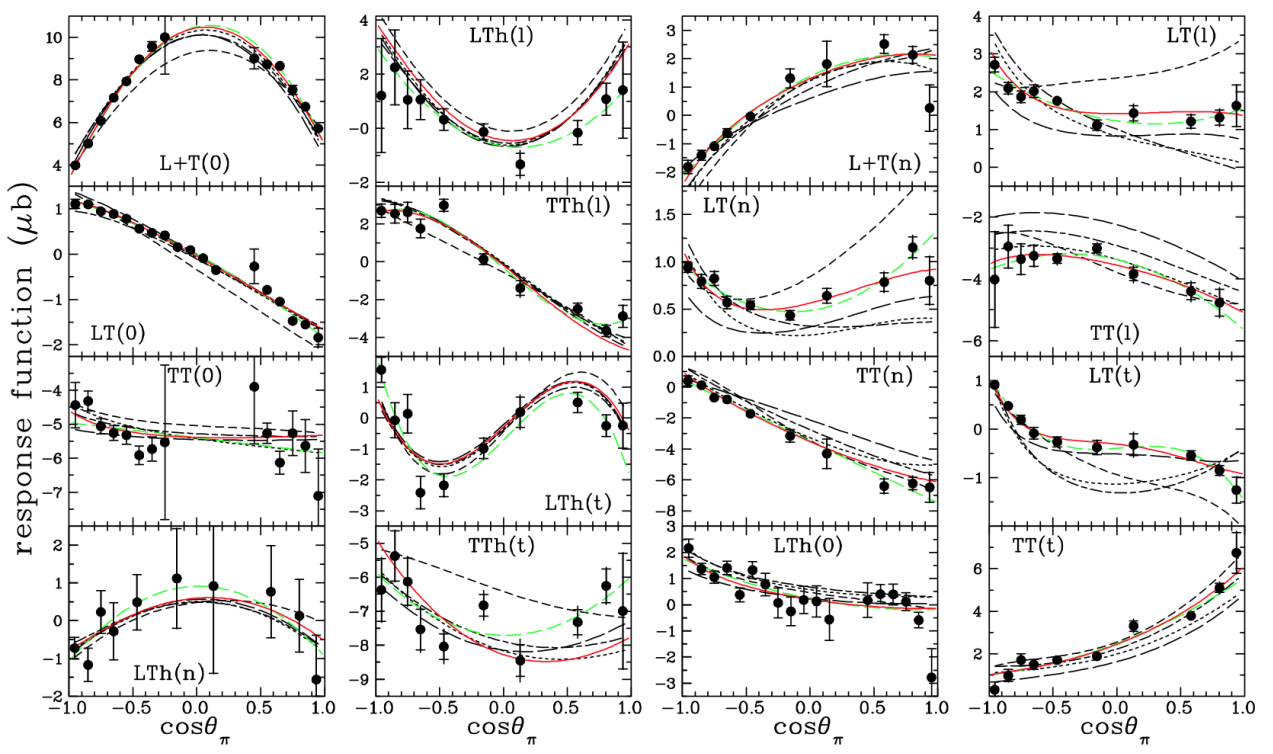

**Fig. 9.3.3** JLab/Hall A data for $\vec{e}p \to e\vec{p}\pi^0$ response functions at W=1.232 GeV and $Q^2 = 1.0$ [2761]. Notations refer to transverse (t), normal (n) and longitudinal (l) components of the proton recoil polarization. The curves correspond to results obtained using SAID (short dashed), MAID (dashed-dotted), and the dynamical models DMT [2762] (dotted), and SL [2763] (long-dashed/green). The other curves correspond to Legendre and multipole fits performed by the authors.

Such measurements would in general require full $4\pi$ coverage for the hadronic final state. The only measurement that could claim to be complete was carried out at Jefferson Lab in Hall A [2761] employing a limited kinematics centered at resonance for $\vec{e}p \to e'\vec{p}\pi^0$ at $W = 1.232$ GeV, and $Q^2 \approx 1$ GeV$^2$. Figure 9.3.3 shows the 16 response functions extracted from this measurement. The results of this measurement in terms of the magnetic $N\Delta$ transition form factor and the quadrupole ratios are included in Fig. 9.3.4 among other data. They coincide very well with results of other experiments [2764–2767] using different analysis techniques that may be also applied to broader kinematic conditions, especially higher mass resonances. Details of the latter are discussed in [2754, 2768]. We briefly summarize them here:

- *Dispersion Relations* have been employed in two ways: One is based on fixed-t dispersion relations for the invariant amplitudes and was successfully used throughout the nucleon resonance region. Another way is based on DR for the mulipole amplitudes of the $\Delta(1232)$ resonance, and allows getting functional forms of these amplitudes with one free parameter for each of them. It was employed for the analysis of the more recent data.
- *The Unitary Isobar Model (UIM)* was developed in [2774] from the effective Lagrangian approach for pion photoproduction [2775]. Background contributions from t-channel $\rho$ and $\omega$ exchanges are introduced and the overall amplitude is unitarized in a K-matrix approximation.
- *Dynamical Models* have been developed, as SAID from pion photoproduction data [2776], the Sato-Lee model was developed in [2777]. Its essential feature is the consistent description of $\pi N$ scattering and the pion electroproduction from nucleons. It was utilized in the study of $\Delta(1232)$ excitations in the $ep \to ep\pi^0$ channel [2763]. The Dubna-Mainz-Taipei model [2778] builds unitarity via direct inclusion of the $\pi N$ final state in the T-matrix of photo- and electroproduction.

### 9.3.4 Models for light-quark resonance electroproduction

In order to learn from the meson electroproduction data about the internal spin and spatial electromagnetic structure, it is essential to have advanced models available with links to the fundamentals of $QCD$.

While most of the analyses have focused on single pseudoscalar meson production, such as

$$\gamma_v p \to N\pi, p\eta, K\Lambda, K\Sigma,$$

more recent work included the $p\pi^+\pi^-$ final state both in real photoproduction [2780] as well as in electroproduction [2781]. The 2-pion final state has more sensitivity to excited $N^*$ and $\Delta^*$ states in the mass range above 1.6 GeV, with several states dominantly coupling to $N\pi\pi$ final states, enabling the study of their electromagnetic transition form factors in the future.

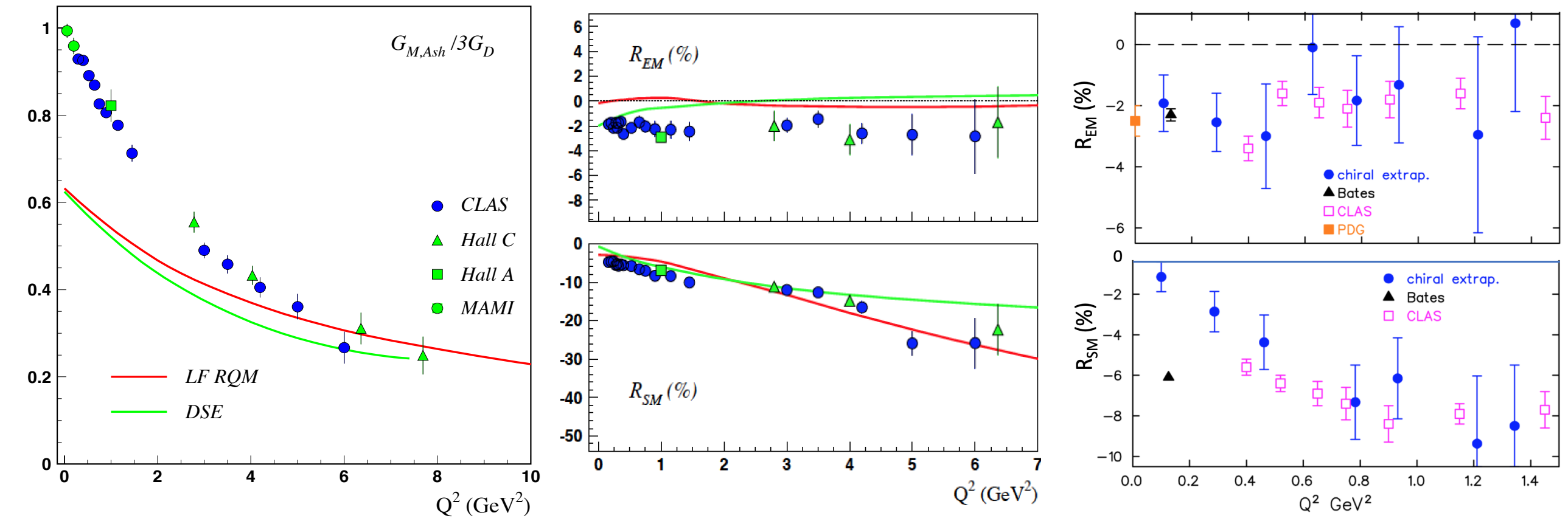

**Fig. 9.3.4** The $N\Delta(1232)$ transition amplitudes. Left: The magnetic N$\Delta$ transition form factor normalized to the dipole form factor and compared with the Light-Front Relativistic Quark Model (LFRQM)[2769, 2770] with running quark mass, and with results using the Dyson-Schwinger Equation [2771]. Both predictions are close to the data at high $Q^2$. At $Q^2 < 3\text{GeV}^2$ meson-baryon contributions are significant. Middle: The electric (top) and scalar (bottom) quadrupole/magnetic-dipole ratios $R_{EM}$ and $R_{SM}$. Right: $R_{EM}$ and $R_{SM}$ from Lattice QCD [2772, 2773] compared to data in the low $Q^2$ domain.

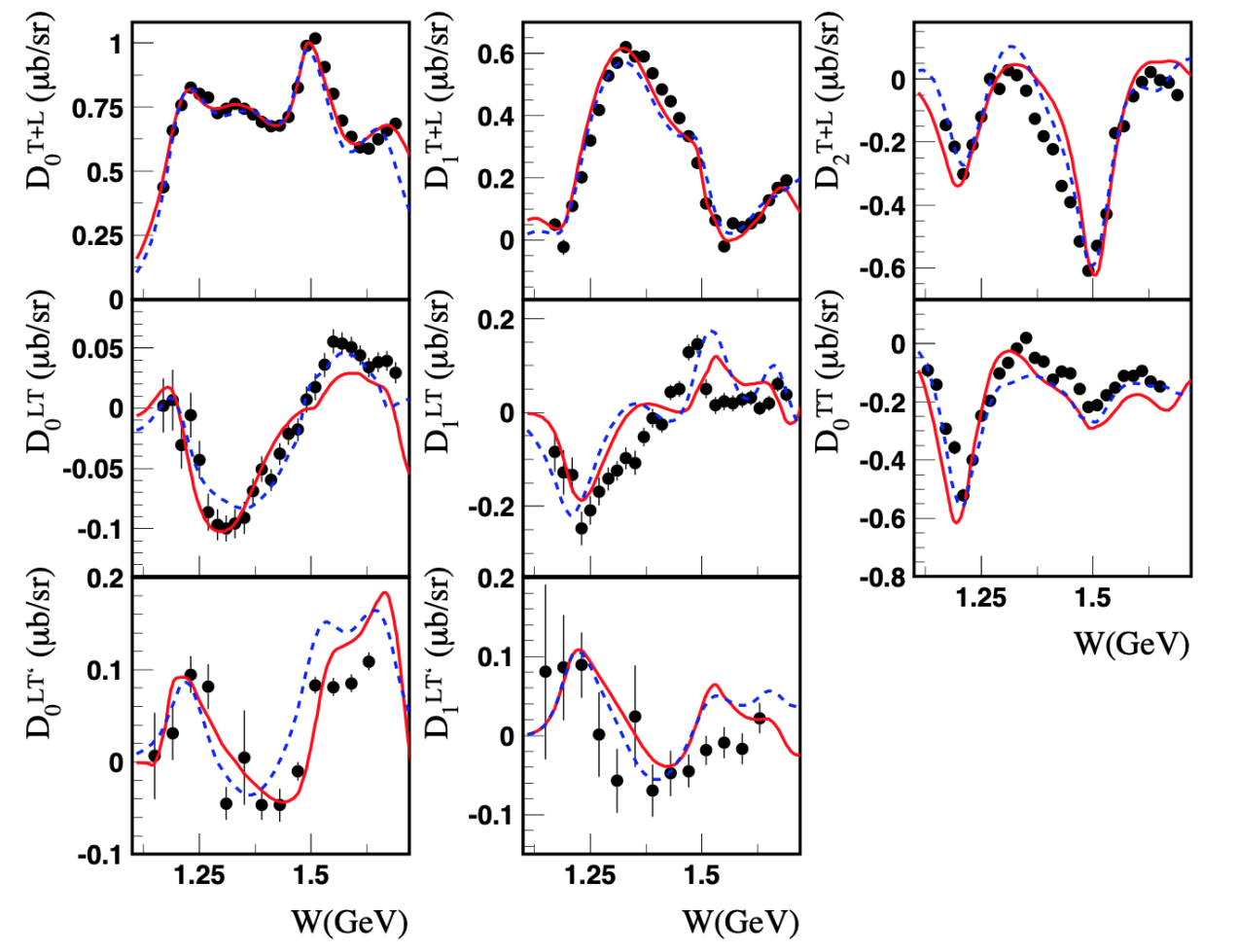

**Fig. 9.3.5** Sample of results of an analysis by the JLab group of the Legendre moments of $\vec{e}p \to e\pi^+n$ structure functions in comparison with experimental data [2779] at $Q^2 = 2.44$ GeV$^2$. The solid (dashed) curves correspond to results obtained using the DR (UIM) approach.

## 9.3.5 The $N\Delta(1232)\frac{3}{2}^+$ transition

The $\Delta^{++}$ isobar was first observed 70 years ago in Enrico Fermi's experiment that used a $\pi^+$ meson beam scattered off the protons in a hydrogen target [2782]. The cross section showed a sharp rise above threshold towards a mass near 1200 MeV. While the energy of the pion beam was not high enough to see the maximum and the fall-off following the peak, a strong indication of the first baryon resonance was observed. It took 12 more years and the development of the underlying symmetry in the quark model before a microscopic explanation of this observation could emerge. There was, however, a problem; while the existence of the $\Delta^{+,0,-}$ could be explained within the model, the existence of the $\Delta(1232)^{++}$, which within the quark model would correspond to a state $|u\uparrow u\uparrow u\uparrow\rangle$, was forbidden as it would have an overall symmetric wave function. It took the introduction of para Fermi statistics [26] what later became "color" (see Section 1.2), to make the overall wave function anti-symmetric. In this way the $\Delta^{++}(1232)$ resonance may be considered a harbinger of the development of QCD.

The nucleon to $\Delta(1232)\frac{3}{2}^+$ transition is now well measured in a large range of $Q^2$ [2765–2767]. At the real photon point, it is explained by a dominant magnetic transition from the nucleon ground state to the $\Delta(1232)$ excited state. Additional contributions are related to small D-wave components in both the nucleon and the $\Delta(1232)$ wave functions leading to electric quadrupole and scalar quadrupole transitions. These remain in the few % ranges at small $Q^2$. The magnetic transition is to $\approx 65\%$ given by a simple spin flip of one of the valence quarks as seen in Fig. 9.3.4. The remaining 35% of the magnetic dipole strength is attributed to meson-baryon contributions.

The electric quadrupole ratio $R_{EM}$ was found as:

$$R_{EM} \approx -0.02. \qquad (9.3.23)$$

There has been a longstanding prediction of asymptotic pQCD, that $R_{EM} \to +1$ at $Q^2 \to \infty$. Results on the magnetic transition form factor $G_{Mn,Ash}$, defined in the Ash convention [2783], and on the quadrupole transition ratios are shown in Fig. 9.3.4. $G_{Mn,Ash}$ is shown normalized to the dipole form factor, but shows a much faster $Q^2$ fall-off compared to that. In comparison to the advanced LF RQM with momentum-dependent constituent quark mass, and with the Dyson-Schwinger

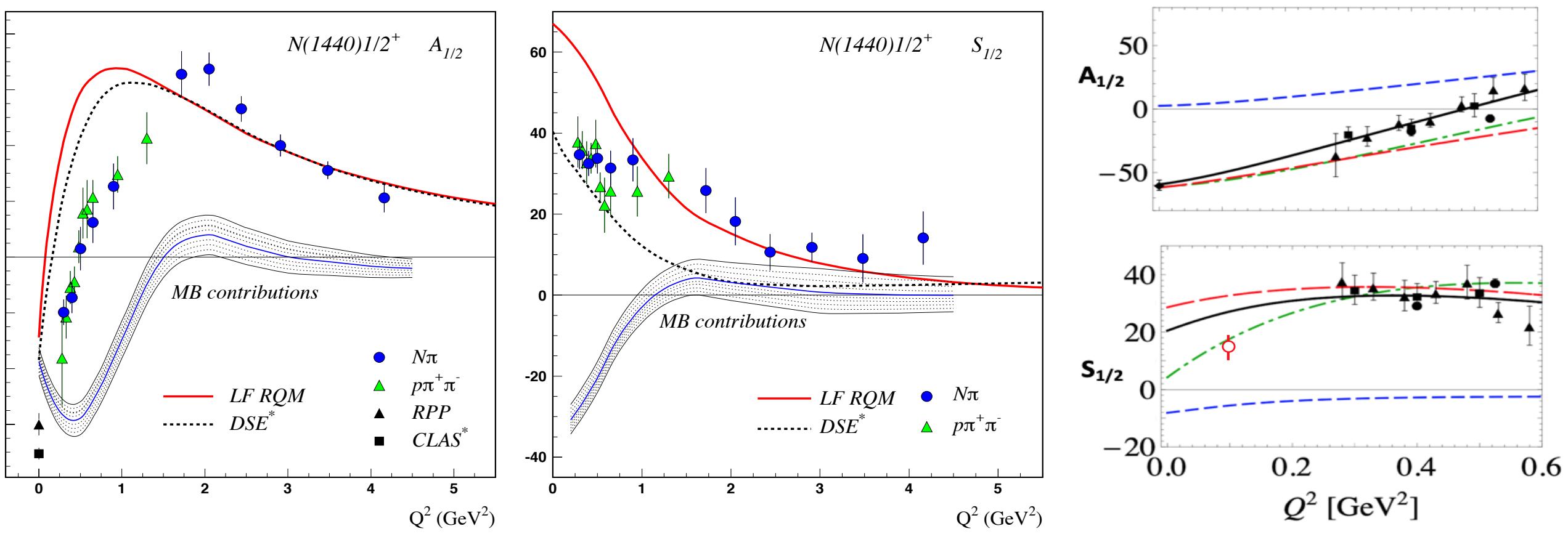

**Fig. 9.3.6** Helicity transition amplitudes for the proton to Roper $N(1440)\frac{1}{2}^+$ excitation compared to model calculations in LFRQM, DSE, and EFT; see text. Left: Transverse $A_{1/2}$ amplitude. Middle: Scalar $S_{1/2}$ amplitude. Right: Helicity amplitudes of the Roper resonance $N(1440)\frac{1}{2}^+$ at low $Q^2$. Data are compared to calculations within Effective Field Theory [2784], shown in solid black lines. The other broken lines are parts of the full calculations. The data are from [2764, 2785, 2786]. The open red circle at $Q^2 \approx 0.1\text{GeV}^2$ is the result of an analysis of $ep \to ep\pi^0$ data from MAMI [2787].

Equation (DSE-QCD) results, there is good agreement at the high-$Q^2$ end of the data. The discrepancy at small $Q^2 = 0$ is likely due to the meson-baryon contributions at small $Q^2$, which are not modeled in either of the calculations.

The quadrupole ratio $R_{EM}$ shows no sign of departing significantly from its value at $Q^2 = 0$, even at the highest $Q^2 \approx 6.5$ GeV$^2$. Both calculations barely depart from $R_{EM} = 0$, and remain near zero at all $Q^2 > 2$ GeV$^2$. This indicates that the negative constant value shown by the data is likely due to meson-baryon contributions that are not included in the theoretical models. For the scalar quadrupole ratio $R_{SM}$ the asymptotic prediction in holographic QCD (hQCD) [2788] is:

$$R_{SM} = \frac{ImS_{1+}}{ImM_{1+}} \to -1, \quad \text{at } \mathrm{Q}^2 \to \infty, \tag{9.3.24}$$

while $R_{EM}$ in hQCD is predicted to approach +1 asymptotically. The $R_{SM}$ data show indeed a strong trend towards increasing negative values at larger $Q^2$, semi-quantitatively described by both calculations at $Q^2 <$ 4 GeV$^2$. The Dyson-Schwinger equation approach predicts a flattening of $R_{SM}$ at $Q^2 > 4$ GeV$^2$, while the Light Front relativistic Quark Model predicts a near constant negative slope of $R_{SM}(Q^2)$ also at higher $Q^2$.

### 9.3.6 The Roper resonance $N(1440)\frac{1}{2}^+$

The Roper resonance, discovered in 1964 [2789] in a phase shift analysis of elastic $\pi N$ scattering data, has been differently interpreted for half a century. In the non-relativistic quark model (nrQM), the state is the first radial excitation of the nucleon ground state with a mass expected around 1750 MeV, much higher than the measured Breit-Wigner mass of $\approx$ 1440 MeV. This discrepancy is now understood as the consequence of a dynamical coupled channel effect that shifts the mass below the mass of the $N(1535)1/2^-$ state, the negative-parity partner of the nucleon [2790]. Another problem with the quark model was the sign of the transition form factor $A_{1/2}(Q^2 = 0)$, predicted in the nrQM as large and positive, while experimental analyses showed a negative value.

These discrepancies resulted in different interpretations of the state that could only be resolved with electroproduction data from CLAS at Jefferson Lab, the development of continuous QCD approximations in the Dyson-Schwinger equation approach [2791] and Light Front Relativistic QM with momentum-dependent quark masses [2769] shown in Fig. 9.3.6, and Lattice data [2792, 2793]. A recent review of the history and current status of the Roper resonance, is presented in a colloquium-style article published in Review of Modern Physics [2794].

Descriptions of the baryon resonance transitions form factors, including the Roper resonance $N(1440)\frac{1}{2}^+$, have also been carried out within holographic models [2795, 2796]. In the range $Q^2 < 0.6$ GeV$^2$, calculations based on meson-baryon degrees of freedom and effective field theory [2784] have been successfully performed, as may be seen in Fig. 9.3.6. Earlier model descriptions, such as the Isgur-Karl model that describe the nucleon as a system of 3 constituent quarks in a confining potential and a one-gluon exchange contribution leading to a magnetic hyperfine splitting of states [729, 2699], and the relativized version of Capstick [736] have popular-ized the model that became the basis for many further developments and variations, e.g. the light front relativistic quark model, and the hypercentral quark model [2797]. Other models were developed in parallel. The cloudy bag model [742] describes the nucleon as a bag of 3 constituent quarks surrounded by a cloud of pions. It has been mostly applied to nucleon resonance excitations in real photoproduction, $Q^2 = 0$ [742, 2798], with some success in the description of the $\Delta(1232)\frac{3}{2}^+$ and the Roper resonance transitions.

There is agreement with the data at $Q^2 > 1.5$ GeV$^2$ for these two states, while the meson-baryon contributions for the $\Delta(1232)$ are more extended, and agreement with the quark based calculations is reached at $Q^2 > 4$ GeV$^2$. The calculations deviate significantly from the data at lower $Q^2$, which indicates the presence of non-quark core effects. For the Roper resonance such contributions have been described successfully in dynamical meson-baryon models [2799] and in effective field theory [2784]. Calculations on the Lattice for the N-Roper transition form factors $F_1^{pR}$ and $F_2^{pR}$, which are combinations of the transition amplitudes $A_{1/2}$ and $S_{1/2}$, have been carried out with dynamical quarks [2793]. The results agree well with the data in the range $Q^2 < 1.0$ GeV$^2$, where data and calculations overlap Fig. 9.3.7.

New electroproduction data on the Roper [2787] and on several higher mass states have been obtained in the 2-pion channel, specifically in $ep \to e'p\pi^+\pi^-$ [2800].

The mass of the Roper state has been computed on the Lattice and extrapolated to the physical pion mass, showing good agreement with the physical value measured with a Breit-Wigner parametrization. It should be noted that the Roper mass measured at the pole in the complex plane is significantly different from the value obtained using the BW ansatz. Supported by an extensive amount of single pion electroproduction data, covering the full phase space in the pion polar and azimuthal center-of-mass angles, and accompanied by several theoretical modeling, we can summarize our current understanding of the $N(1440)\frac{1}{2}^+$ state as follows:

- The Roper resonance is, at heart, the first radial excitation of the nucleon.
- It consists of a well-defined dressed-quark core, which plays a role in determining the system's properties at all length scales, but exerts a dominant influence on probes with $Q^2 > m_N^2$, where $m_N$ is the nucleon mass.
- The core is augmented by a meson cloud, which both reduces the Roper's core mass by $\approx 20\%$, thereby solving the mass problem that was such a puzzle in constituent quark model treatments, and, at low $Q^2$, contributes an amount to the electroproduction transition form factors that is comparable in magnitude with that of the dressed quark core, but vanishes rapidly as $Q^2$ is increased beyond $m_N^2$.

As stated in the conclusions of [2794]: "The fifty years of experience with the Roper resonance have delivered lessons that cannot be emphasized too strongly. Namely, in attempting to predict and explain the QCD spectrum, one must fully consider the impact of meson-baryon final state interactions and the coupling between channels and states that they generate, and look beyond merely locating the poles in the S-matrix, which themselves reveal little structural information, to also consider the $Q^2$ dependencies of the residues, which serve as a penetrating scale-dependent probe of resonance composition."

### 9.3.7 Transition Form Factors of $N(1535)\frac{1}{2}^-$ - A state with a hard quark core.

This state is the parity partner state to the ground state nucleon, with the same spin 1/2 but with opposite parity, its quark content requires an orbital L=1 excitation in the transition from the proton. In the $SU(6) \otimes O(3)$ symmetry scheme, the state is a member of the $[70, 1^-]$ super multiplet. This state couples equally to $N\pi$ and to $N\eta$ final state. It has therefore be probed using both decay channels $ep \to ep\eta$ and $ep \to eN\pi^{+,0}$. Because of isospin $I = 1/2$ for nucleon states, the coupling to the charged $\pi^+n$ channel is preferred over $\pi^0 p$ owing to the Clebsch-Gordon coefficients.

The $A_{1/2}$ helicity amplitude for the $\gamma pN(1535)\frac{1}{2}^-$ resonance excitation shown in Fig. 9.3.7 represents the largest range in $Q^2$ of all nucleon states for which resonance transition form factors have been measured as part of the broad experimental program at JLab.

For this state, as well as for the $N(1440)\frac{1}{2}^+$ state, advanced relativistic quark model calculations [2803], DSE-QCD calculations [2791] and Light Cone sum rule results [2804] are available, employing QCD-based modeling of the excitation of the quark core for the first time.

The transverse transition amplitude $A_{1/2}$ of $N(1535)\frac{1}{2}^-$ is a prime example of the power of meson electroproduction to unravel the internal structure of the resonance transition. In the previous section 9.2, the nature of this state is discussed as a possible example of a dynamically generated resonance. The electroproduction data shown here reveal structural aspects of the state and its nature that require a different interpretation. The transition form factor $A_{1/2}$ of the state, shown in Fig. 9.3.7, is quantitatively reproduced over a large range in $Q^2$ by two alternative approaches, the LFRQM and the LCSR. Both calculations are based on the assumptions of the presence of a 3-quark core. Note that there is a deviation from the quark calculations at $Q^2 < 1-2$ GeV$^2$, highlighted as the shaded area in Fig. 9.3.7, which may be assigned to the presence of non-quark contributions. Attempts to compute the transition form factors within strictly dynamical models have not succeeded in explaining the data [2805]. The discrepancy could be resolved if the character of the probe, meson (pion) in the case of hadron interaction and short range photon interaction in the case of electroproduction, probe different parts of the resonance's spatial structure: peripheral in case of meson scattering and short distance behavior in electroproduction. The peripheral meson scattering and low $Q^2$ meson photoproduction reveal the dynamical features of the state, whereas high $Q^2$ electroproduction reveals the structure of the quark core.

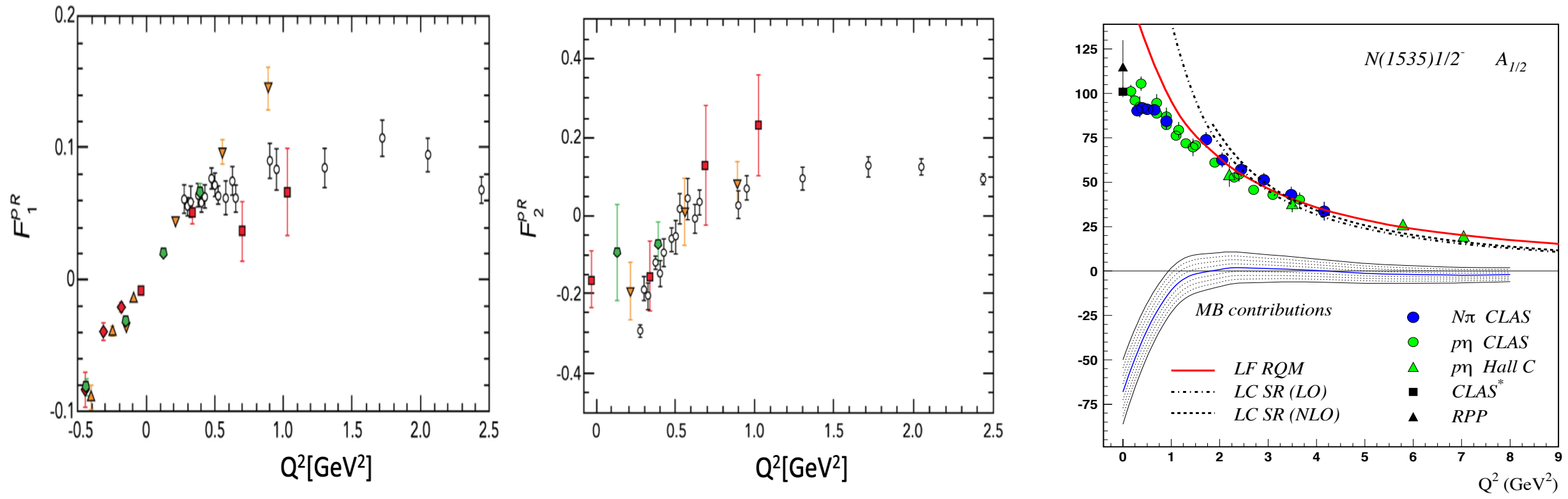

**Fig. 9.3.7** Left & middle: Dirac and Pauli transition form factors $F_1$ and $F_2$ for the proton to $N(1440)1/2^+$ transition compared to Lattice QCD calculations [2793] with pion masses (in GeV): 0.39 (red squares), 0.45 (orange triangles), and 0.875 (green circles) on the $N_f = 2+1$ anisotropic lattices, compared to CLAS results (black circles). The $F_1$ and $F_2$ form factors are linear combinations of the $A_{1/2}$ and $S_{1/2}$ amplitudes. Right: The transverse transition helicity amplitude $A_{1/2}$ versus $Q^2$. At $Q^2 > 2\text{GeV}^2$ the data are well described by the light-cone sum rules LCSR [2801]. The light front relativistic quark model (LFRQM) [2802] describes that data at $Q^2 > 1\text{GeV}^2$.

### 9.3.8 The helicity structure of the $N(1520)\frac{3}{2}^-$

The $N(1520)\frac{3}{2}^-$ state corresponds to the lowest excited nucleon resonance with $J^P = \frac{3}{2}^-$. Its helicity structure is defined by the $Q^2$ dependence of the two transverse transition amplitudes $A_{1/2}$ and $A_{3/2}$. They are both shown in Fig. 9.3.8. A particularly interesting feature of this state is that at the real photon point, $A_{3/2}$ is strongly dominant, while $A_{1/2}$ is very small. However, at high $Q^2$, $A_{1/2}$ is becoming dominant, while $A_{3/2}$ drops rapidly. This behavior is qualitatively consistent with the expectation of asymptotic QCD, which predicts the transition helicity amplitudes to behave like:

$$A_{1/2} \propto \frac{a}{Q^3}, A_{3/2} \propto \frac{b}{Q^5}\,. \qquad (9.3.25)$$

The helicity asymmetry

$$A_{hel} = \frac{A_{1/2}^2 - A_{3/2}^2}{A_{1/2}^2 + A_{3/2}^2}, \qquad (9.3.26)$$

shown in Fig. 9.3.8, illustrates this rapid change in the helicity structure of the $\gamma_v p N(1520)3/2^-$ transition. At $Q^2 > 2$ GeV$^2$, $A_{1/2}$ fully dominates the process.

### 9.3.9 The helicity transition amplitudes to the $N(1535)\frac{1}{2}^-$ resonance

The Roper $N(1440)\frac{1}{2}^+$ resonance, at the core, is a radial excitation. Its parity partner, the $N(1535)\frac{1}{2}^-$, in the quark model, is an orbitally excited quark state of the nucleon. It is then interesting to compare the transition amplitude to the $N(1535)\frac{1}{2}^-$ with the amplitude to the Roper resonance. The $N(1535)\frac{1}{2}^-$ is, together with the $\Delta(1232)\frac{3}{2}^+$, the best measured state, and both its transverse and longitudinal (scalar) amplitudes are well measured [2754]. Figure 9.3.9 shows the transverse amplitude $A_{1/2}$ versus $Q^2$. They reveal a very different behavior at low $Q^2$, where $N(1535)\frac{1}{2}^-$ indicates only small effects from meson-baryon contributions below $Q^2 \approx 1\text{GeV}^2$, while the $N(1440)\frac{1}{2}^+$ changes sign at small $Q^2$ and reveals a much more prominent impact of meson-baryon contributions. The $Q^2$ dependence of the $N(1535)\frac{1}{2}^-$ is well reproduced by LC SR in LO and NLO. There have been attempts to explain the transition form factor of the $N(1535)\frac{1}{2}^-$ as a dynamically generated resonance [2805] that does not achieve quantitative agreement with experiment and concludes that admixture with a genuine three-quark state is demanded that could help to better reproduce the magnitude or the $Q^2$ falloff of the $A_{1/2}$ helicity amplitude.

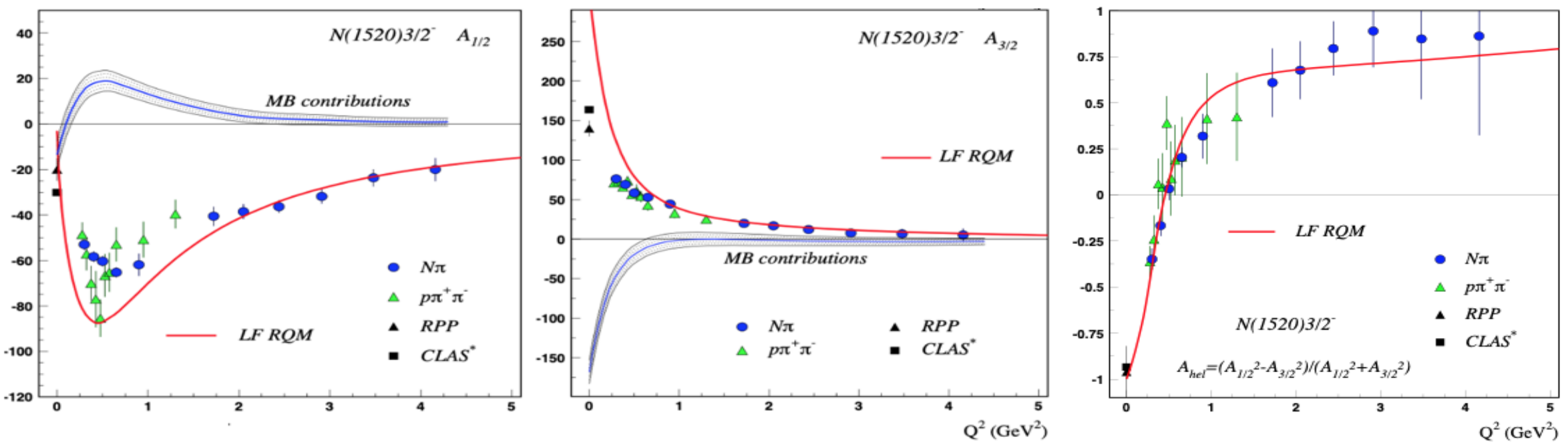

**Fig. 9.3.8** The transverse helicity transition amplitudes of $N(1520)\frac{3}{2}^-$ versus $Q^2$, compared to the LFRQM, $A_{1/2}$ (left), $A_{3/2}$ (middle). The shaded area indicates the contribution from non-quark contributions as estimated from the difference of the measured data points and the LF RQM contribution, likely due to hadronic contributions. Right: Helicity asymmetry $A_{hel}$, as defined in Eq. 9.3.26. Graphics from Ref. [2754]

### 9.3.10 The $N(1675)\frac{5}{2}^-$ state - revealing the meson-baryon contributions

In previous discussions we have concluded that meson-baryon degrees of freedom provide significant strength to the resonance excitation in the low-$Q^2$ domain where quark based approaches LF RQM, DSE/QCD, and LC SR calculations fail to reproduce the transition amplitudes quantitatively. Our conclusion rests, in part, with this assumption. But, how can we be certain of the validity of this conclusion?

The $N(1675)\frac{5}{2}^-$ resonance allows us to test this assumption, quantitatively. Figure 9.3.9 shows our current knowledge of the transverse helicity amplitudes $A_{1/2}(Q^2)$ and $A_{3/2}(Q^2)$, for proton target compared to RQM [2802] and hypercentral CQM [2807] calculations. The specific quark transition for a $J^P = \frac{5}{2}^-$ state belonging to the $[SU(6) \otimes O(3)] = [70, 1^-]$ supermultiplet configuration, in non-relativistic approximation prohibits the transition from the proton in a single quark transition. This suppression, known as the Moorhouse selection rule [725], is valid for the transverse transition amplitudes $A_{1/2}$ and $A_{3/2}$ at all $Q^2$. It should be noted that this selection rule does apply to the transition from a proton target, it does not apply to the transition from the neutron, which is consistent with the data. Modern quark models that go beyond single quark transitions, confirm quantitatively the suppression resulting in very small amplitudes from protons but large ones from neutrons. Furthermore, a direct computation of the hadronic contribution to the transition from protons confirms this (Fig. 9.3.9). The measured helicity amplitudes off the protons are almost exclusively due to meson-baryon contributions as the dynamical coupled channel (DCC) calculation indicates (dashed line). The close correlation of the DCC calculation and the measured data for the case when quark contributions are nearly absent, supports the phenomenological description of the helicity amplitudes in terms of a 3-quark core that dominate at high $Q^2$ and meson-baryon contributions that can make important contributions at lower $Q^2$.

### 9.3.11 Resonance lightfront transition charge densities.

Knowledge of the helicity amplitudes in a large $Q^2$ allows for the determination of the transition charge densities on the light cone in transverse impact parameter space $(b_x, b_y)$ [2809]. The relations between the helicity transition amplitudes and the Dirac and Pauli resonance transition form factors are given by:

$$A_{1/2} = e\frac{Q_-}{\sqrt{K}(4M_N M^*)^{1/2}}\{F_1^{NN^*} + F_2^{NN^*}\} \quad (9.3.27)$$

$$S_{1/2} = e\frac{Q_-}{\sqrt{K}(4M_N M^*)^{1/2}}\left(\frac{Q_+Q_-}{2M^*}\right)\frac{(M^*+M_N)}{Q^2} \times\{F_1^{NN^*} - \frac{Q^2}{(M^*+M_N)^2}F_2^{NN^*}\}, \quad (9.3.28)$$

where $M^*$ is the mass of the excited state $N^*$, $K = \frac{M^{*2}-M_N^2}{2M^*}$ is the equivalent photon energy, $Q_+$ and $Q_-$ are short hands for $Q_\pm \equiv \sqrt{M^* \pm M_N)^2 + Q^2}$. The charge and magnetic lightfront transition densities $\rho_0^{NN^*}$ and $\rho_T^{NN^*}$, respectively, are given as:

$$\rho_0^{NN^*}(\vec{b}) = \int_0^\infty \frac{dQ}{2\pi} J_0(bQ) F_1^{NN^*}(Q^2) \quad (9.3.29)$$

$$\rho_T^{NN^*}(\vec{b}) = \rho_0^{NN^*}(\vec{b}) + \sin(\phi_b - \phi_s) \times \int_0^\infty \frac{dQ}{2\pi}\frac{Q^2}{(M^*+M_N)} J_1(bQ) F_2^{NN^*}(Q^2)\,. \quad (9.3.30)$$

Similar transverse charge transition densities can be defined for $J^P = \frac{3}{2}^+$ states such as the $\Delta(1232)\frac{3}{2}^+$. This has been studied in [2810] and is shown in Fig. 9.3.11.

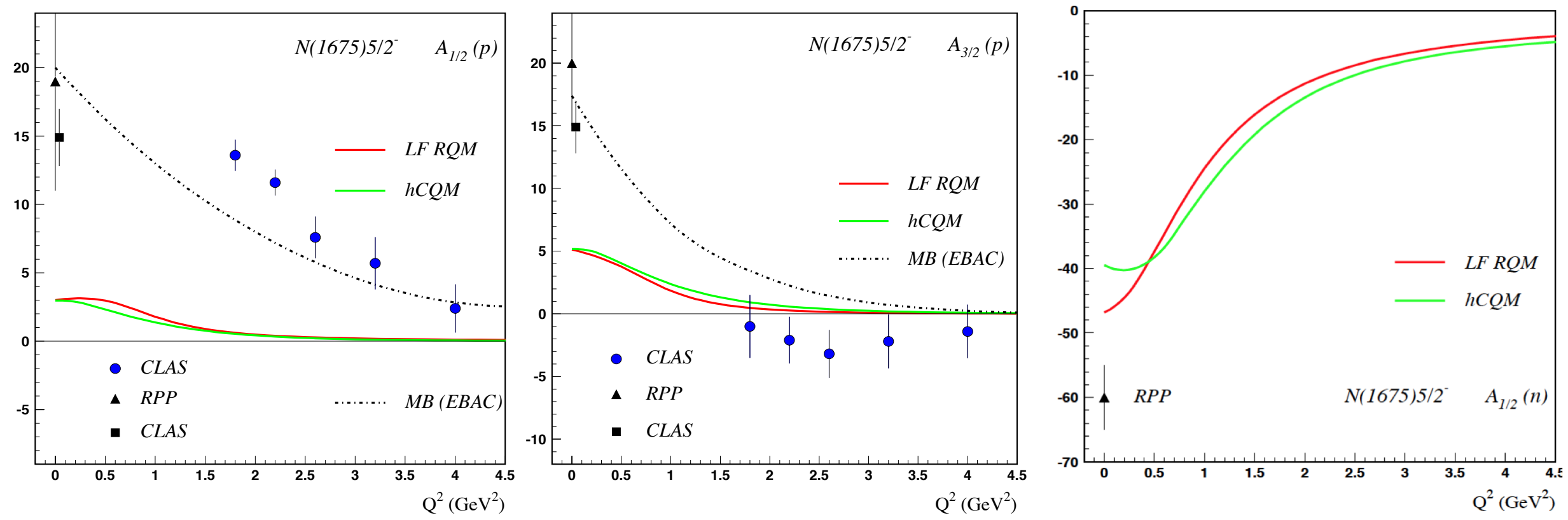

**Fig. 9.3.9** The transverse amplitudes of the proton to $N(1675)\frac{5}{2}^-$ transition compared to the LF RQM [2806], hypercentral QM [2807], and contributions from meson-baryon (MB) coupled channel dynamics [2808]. Left: $A_{1/2}$, Middle:$A_{3/2}$. Both quark models predict very small amplitudes for the proton, while the meson-baryon contributions estimate is large and is close to the data. Right: $A_{1/2}$ for neutron target (only photoproduction data available) compared to the LFRQM and hCQM. Both quark models predict large amplitudes for neutrons, more than factor 10 compared to protons at $Q^2 = 0$. Assuming similar meson-baryon contributions as in the proton case with opposite sign could quantitatively explain the single measured value at the photon point.

A comparison of $N(1440)\frac{1}{2}^+$ and $N(1535)\frac{1}{2}^-$ is shown in Figure 9.3.10. There are clear differences in the charge transition densities between the two states. The Roper state has a softer positive core and a wider negative outer cloud than $N(1535)\frac{1}{2}^-$ and develops a larger shift in $b_y$ when the proton is polarized along the $b_x$ axis.

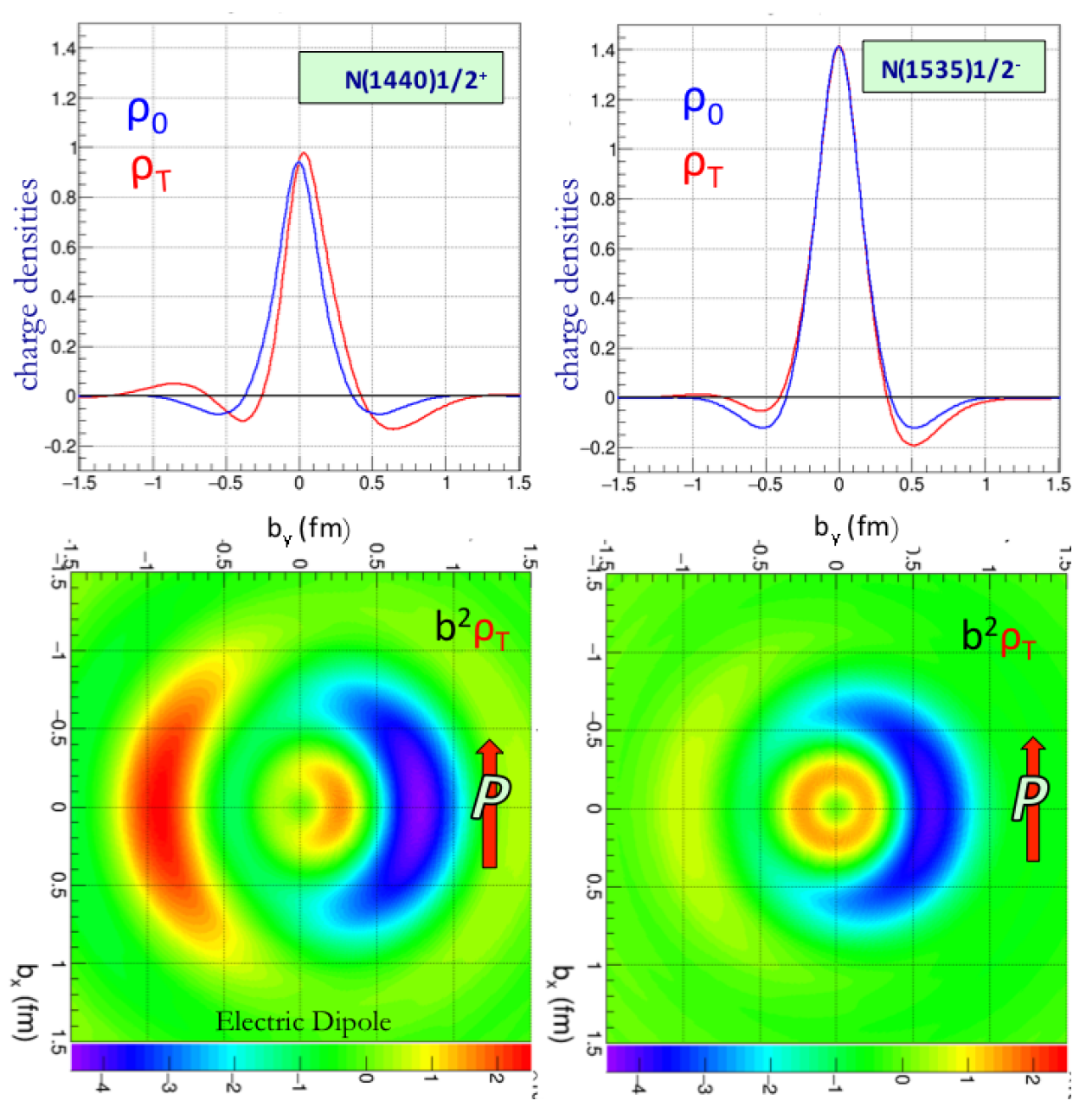

**Fig. 9.3.10** Left panels: $N(1440)$, top: projection of charge densities on $b_y$, bottom: transition charge densities when the proton is spin polarized along $b_x$. Right panels: same for $N(1535)$. Note that the densities are scaled with $b^2$ to emphasize the outer wings. Color code:negative charge is blue, positive charge is red. Note that all scales are the same for ease of comparison [2811]. Graphics credit: F.X. Girod.

### 9.3.12 Single Quark Transition Model

Many of the exited states for which there is information about the transition form factors available have been assigned as members of the $[SU(6), L^P] = [70, 1^-]$ super multiplet of the $[SU(6) \otimes O(3)]$ symmetry group. In a model, where only single quark transitions to the excited states are considered [2812–2814], only 3 of the amplitudes need to be known to determine the remaining 16 transverse helicity amplitudes for all states in $[70, 1^-]$ including on neutrons. However, the picture is now more complicated due to the strong admixture of meson-baryon components to the single quark transition especially in the lower $Q^2$ range. This requires a model to separate the single quark contributions from the hadronic part before projections for other states can be made [2815].

### 9.3.13 Higher mass baryons and hybrid baryons

The existence of baryons containing significant active gluonic components in the wave function has been predicted some decade ago [495] employing Lattice QCD simulations. The lowest such "hybrid" state is expected to be a $J^P = \frac{1}{2}^+$ nucleon state. LQCD projects a mass of 1.3 GeV above the nucleon mass, i.e. approximately 2.2-2.3 GeV, and several other states should appear close by in $J^P = \frac{1}{2}^+$ and $J^P = \frac{3}{2}^+$, as seen in Fig. 9.3.12.

How do we identify these states? Hybrid baryons have same spin-parity as other ordinary baryons. In contrast to hybrid mesons, there are no hybrid baryons with "exotic" quantum numbers. One possibility is to search for more states than the quark model predicts in some mass range. The other possibility is to study the transition form factors of excited states. Hybrid states may be identified as states with a different $Q^2$ behavior than what is expected from a 3-quark state. The sensitivity [2816] is demonstrated for the Roper resonance that projected a very rapid drop of the $A_{1/2}(Q^2)$ with $Q^2$, and $S_{1/2}(Q^2) \sim 0$ prediction. Both are incompatible with what we know today about the Roper resonance. Precision electroproduction data in the mass range above 2 GeV will be needed to test high mass states for their potential hybrid character, e.g. from experiments at CLAS12 [2817].

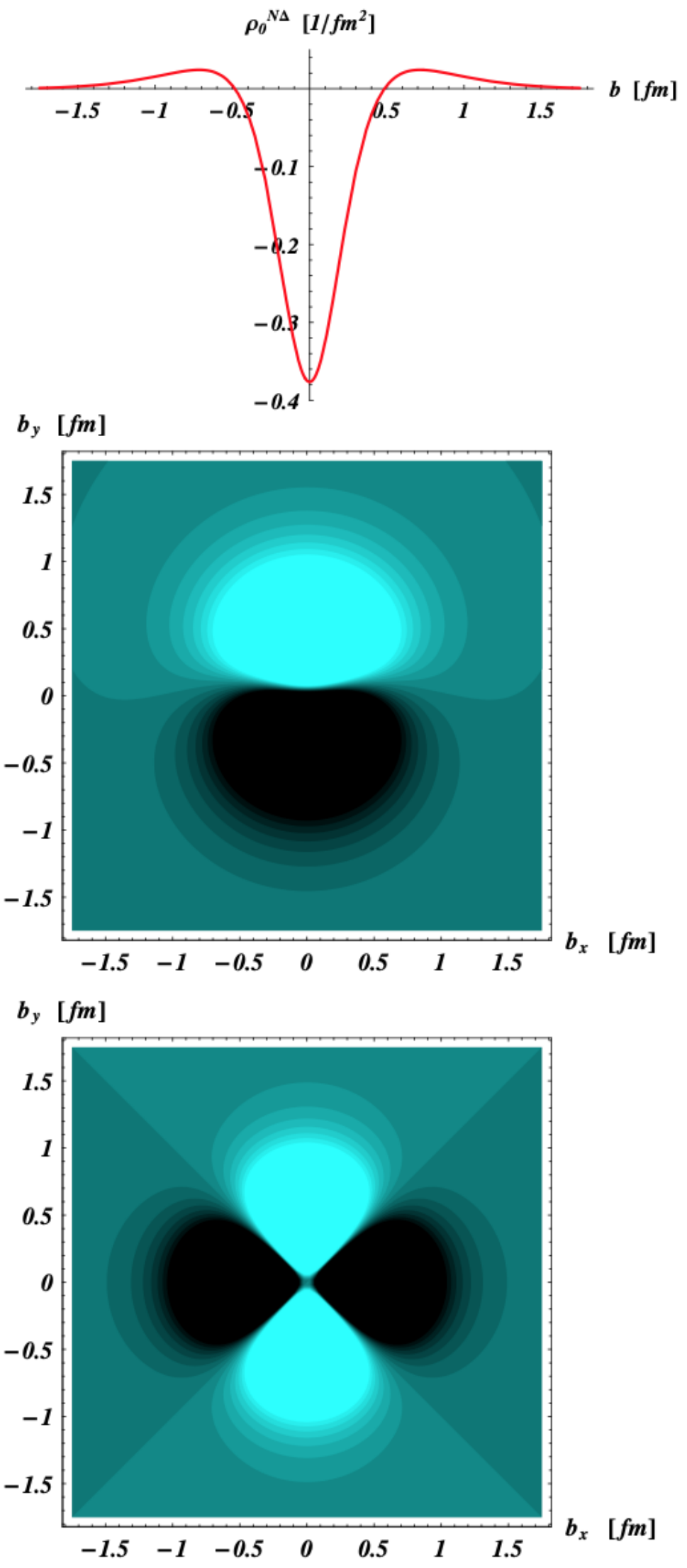

**Fig. 9.3.11** Quark transverse transition charge density corresponding to the $p \to \Delta^+$ transition. Light color indicates positive charge, dark color indicates negative charge. Top: p and $\Delta$ are unpolarized. Middle: p and $\Delta$ are polarized along $b_x$ axis generating an electric dipole along the $b_y$ axis. Bottom: Quadrupole contribution to transition density. Graphics adapted from [2810].

### 9.3.14 Conclusions and Outlook

In this contribution we have focused on more recent results of nucleon resonance transition amplitudes and their interpretation within LQCD and within most advanced approaches, e.g. in light front relativistic quark models and approaches with traceable links to first principle QCD such as Dyson-Schwinger Equations [2818] and light cone sum rules [2801]. These calculations describe the transition form factors at $Q^2 \geq 2$ GeV$^2$, while at lower $Q^2$ values hadronic degrees of freedom must be included and could even dominate contributions of the quark core.

For the lowest mass states, $\Delta(1232)\frac{3}{2}^+$ and the Roper $N(1440)\frac{1}{2}^+$, LQCD calculations have been carried out that are consistent with the data within large uncertainties. These calculations are about one decade old, and new data, with higher precision in more extended kinematic range have been added to the database that warrant new Lattice calculations at the physical pion mass to be carried out.

Over the past decade, eight baryon states in the mass range from 1.85 to 2.15 GeV have been either discovered or evidence for the existence of states has been significantly strengthened. Some of these states are in the mass range and have $J^{PC}$ quantum numbers that could have significant contributions of gluonic components. Such "hybrid" states are in fact predicted in LQCD [495]. These states appear with the same quantum numbers as ordinary quark excitations, and can only be isolated from ordinary states due to the $Q^2$ dependence of their helicity amplitudes [2816], which is expected to be quite different from ordinary 3-quark excitation. The study of hybrid baryon excitations then requires new electroproduction data especially at low $Q^2$ [2817] with different final states and with masses above 2 GeV. Despite the very significant progress made in recent years to further establish the light-quark baryon spectrum and explore the internal structure of excited states and the relationship to QCD [2756, 2819], much remains to be done. A vast amount of precision data already collected needs to be included in the multi-channel analysis frameworks, and polarization data are still to be analyzed. There are approved proposals to study resonance excitation at much higher $Q^2$ and with higher precision at Jefferson Lab with CLAS12 [2820, 2821], which may begin to reveal the transition to the bare quark core contributions at short distances.

A new avenue of experimental research has recently been opened up with the first data-based extraction of a gravitational property of the proton, its internal pressure distribution, which is represented by the gravita-tional form factor $D^q(t)$. It is one of the form factors of the QCD matrix element of the energy-momentum tensor, its internal pressure and shear stress distribution in space [2822, 2823]. These properties, as well as the distribution of mass and angular momentum, and are accessible directly in gravitational interaction, which is highly impractical. However the relevant gravitational form factor $D^q(t)$ for the ground state nucleon can be accessed indirectly through the process of deeply virtual Compton scattering and in time-like Compton scattering [1288, 2824]. Both processes, having a $J = 1$ photon in the initial state as well as in the final state, contain components of $J = 2$ that couple to the proton through a tensor interaction, as gravity does [2825].

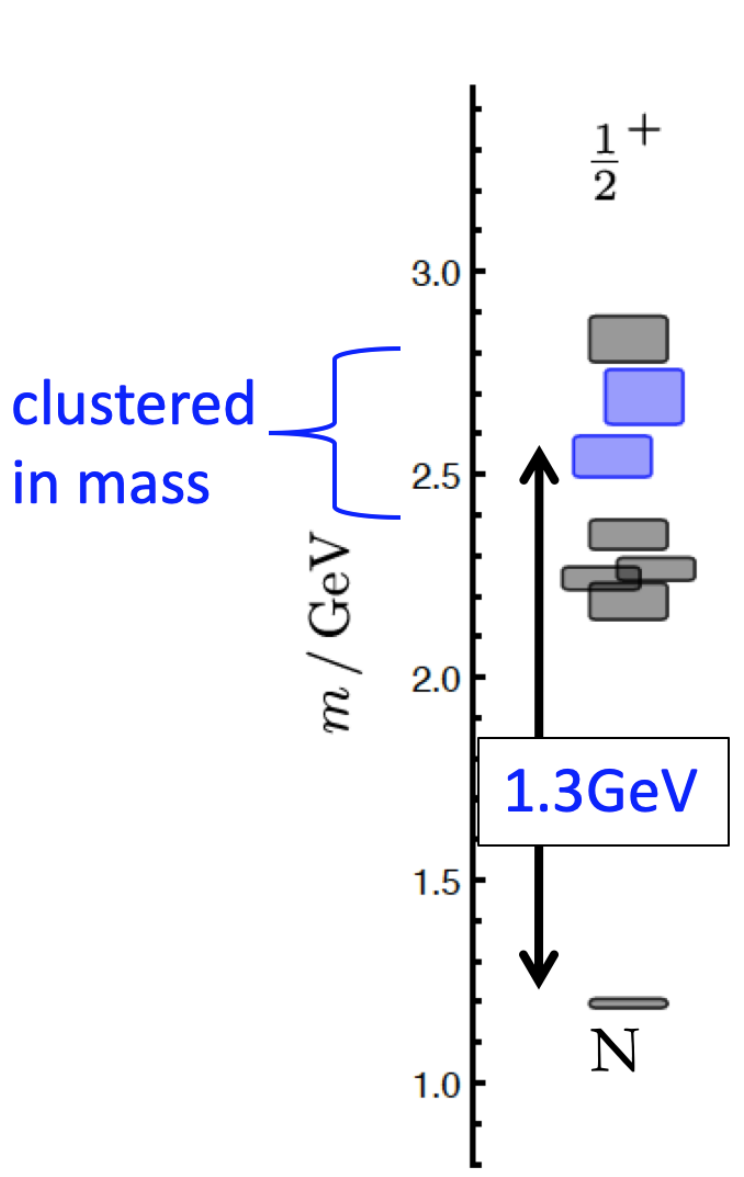

**Fig. 9.3.12** Projections of excited baryons with dominant gluonic components (marked in blue shades) in LQCD with 400MeV pions. The lowest hybrid baryon is projected with mass 1.3 GeV above the nucleon mass. The $1/2^+$ and $3/2^+$ states are clustered in a narrow mass range of about 200 MeV.

Mechanical properties of resonance transitions have recently been explored for the $N(1535)\frac{1}{2}^- \to N(938)$ gravitational transition form factors calculations in [2826] and in [2827]. In order to access these novel gravitational transition form factors experimentally, the nucleon to resonance transition generalized parton distributions must be studied. The framework for studying the $N \to N(1535)$ transition GPDs, which would enable experimental access to these mechanical properties, remains to be developed. The required effort is quite worthwhile as a new avenue of hadron physics has opened up that remains to be fully explored.

## 9.4 Heavy-flavor baryons

**Eberhard Klempt and Sebastian Neubert**

### 9.4.1 Introduction

Baryons with one heavy quark $Q$ and a light diquark $qq$ provide an ideal place to study diquark correlations and the dynamics of the light quarks in the environment of a heavy quark. The heavy quark is almost static and provides the color source to the light quarks. Here, we attempt to understand the dynamics leading to the spectrum of baryons with one heavy quark.

The Review of Particle Physics [939] lists 28 charmed baryons (16 with known spin-parity) and 22 bottom baryons (15 with known spin-parity). One doubly charmed state has been detected, the ground state $\Xi_{cc}^{++}$. (Its isospin partner $\Xi_{cc}^{+}$ is known as well, with poor evidence and one star in RPP, but we do not count isospin partners separately.) In the decays of the lightest bottom baryon, exotic $J/\psi p$ states, incompatible with a three-quark configuration, have been have been reported in studies of the reaction $\Lambda_b \to J/\psi p K^-$ [2828, 2829]. The search for further states and attempts to understand the underlying dynamics of heavy baryons are active fields in particle physics. New information can be expected from the upgrades of LHC, BELLE and J-PARC, and from the new FAIR facility at GSI (see Section 14).

### 9.4.2 Ground states of heavy baryons

*Masses and lifetimes*

Table 9.4.1 presents masses and life times of the ground states of heavy baryons containing a $b$-quark. Naively, one could expect all these life times to represent the life time of the $b$ quark, that they all agree with the life time of the $B^0$ meson. This life time is $\tau_{B^0}$=(1519±4) fs. Indeed, all life times agree within $\sim$ 10% percent.

This is not at all the case when the $b$-quark is replaced by a $c$-quark (see Table 9.4.2). The $D^0$ has a life time $\tau_{D^0} = (410.3 \pm 1.0)$ fs, the $D^+$ has $\tau_{D^+} = (1033 \pm 5)$ fs. The life times of charmed baryons are spread over a wide range and do not agree with the life times of $D$ mesons. In addition to the decay of the $c$-quark, the $c\bar{d}$ pair in a $D^0$ meson can annihilate into a $W^+$, a process forbidden for the $D^+$. In $B$ decays, the corresponding CKM matrix element is small, and this effect is suppressed. Further significant corrections are required to arrive at a consistent picture for the decays of charmed mesons and baryons. The authors of Ref. [1245] have performed an extensive study of the lifetimes within the heavy quark expansion, and have included all known corrections. The impact of the charmed-quark mass and of the wave-functions of charmed hadrons were carefully studied. Then, qualitative agreement between their calculations and the experimental data was achieved. For a more detailed discussion, see Section 5.8.

**Table 9.4.1** Masses and lifetimes of baryon ground states with one $b$-quark. The second line gives the mass in MeV, the third line the life time in fs.

| $\Lambda_b^0$ | $\Xi_b^-$ | $\Xi_b^0$ | $\Omega_b^-$ |
|---|---|---|---|
| 5619.60±0.17 | 5797.0±0.6 | 5791.9±0.5 | 6045.2±1.2 |
| 1464±11 | 1572±40 | 1480±0.030 | $1640^{+180}_{-170}$ |

**Table 9.4.2** Masses and lifetimes of baryon ground states with one $c$-quark. The second line gives the mass in MeV, the third line the life time in fs.

| $\Lambda_c^0$ | $\Xi_c^+$ | $\Xi_c^0$ | $\Omega_c^-$ |
|---|---|---|---|
| 2286.46±0.14 | 2467.71±0.23 | 2470.44±0.2 | 2695.2±1.7 |
| 201.5±2.7 | 453±5 | 151.9±2.4 | 268±26 |

The first state with two charmed quarks, the $\Xi_{cc}^+$ was reported by the SELEX collaboration in two decay modes at a mass of (3518.9±0.9) MeV and with 5-6$\sigma$ [2830, 2831]. In later searches, this state was never confirmed. The LHCb collaboration found its doubly charged partner $\Xi_{cc}^{++}$ [2564]. Its mass is (3621.6±0.4) MeV, its life time (25.6±2.7) fs. Later, the LHCb collaboration reported evidence for a $\Xi_{cc}^+$ baryon at (3623.0±1.4) MeV [2832]. It is seen with 3-4$\sigma$ only but its mass is better compatible with an interpretation of $\Xi_{cc}^+$ and $\Xi_{cc}^{++}$ as isospin partners. A search for the $\Xi_{bc}^+$ remained unsuccessful [2833].

*The flavor wave function: SU(4)*

In this contribution we discuss baryons with one heavy-quark flavor, with either a charm or a bottom quark. Overall, we consider five quarks, $u, d, s, c, b$, but we will not discuss baryons with one light ($q = u, d, s$) and two different heavy quarks like $\Xi_{cb}^+ = (ucb)$. Thus we can restrict ourselves to SU(4). The four quarks have very different masses, and the SU(4) symmetry is heavily broken, nevertheless it provides a guide to classify heavy-quark baryons. Three-quark baryons can classified according to

$$4 \otimes 4 \otimes 4 = 20_\mathrm{s} \oplus 20_\mathrm{m} \oplus 20_\mathrm{m} \oplus 4_\mathrm{a} \qquad (9.4.1)$$

into a fully symmetric 20-plet, two 20-plets of mixed symmetry and a fully antisymmetric 4-plet. In states with one heavy quark only, there is one light quark pair. The light diquark can be decomposed

$$3 \otimes 3 = \bar{3}_\mathrm{a} \oplus 6_\mathrm{s} \qquad (9.4.2)$$

The light diquark in the 6-plet is symmetric, in the $\bar{3}$-plet antisymmetric.

Figure 9.4.1a shows the symmetric 20-plet, which contains the well-known baryon decuplet and a sextet of charmed baryons. In addition to $\Xi_{cc}^+$ and $\Xi_{cc}^{++}$, a $\Omega_{cc}^+$ (with two charmed and one strange quarks) and a $\Omega_{ccc}^{++}$ are expected but not yet observed. All baryons in the symmetric 20-plet in the ground state have a total spin $J = 3/2$. The three quark pairs are symmetric with respect to (w.r.t.) their exchange, in particular the pair of light quarks is symmetric w.r.t. their exchange, they have $\mathrm{SU_F}(3)$ multiplicity 6. Baryons with three charmed quarks have not yet been discovered.

Figure 9.4.1b shows the mixed symmetry 20-plet of heavy baryons. In the ground state they have $J = 1/2$. Baryons with one heavy quark occupy the second layer. The 6-plet and the $\bar{3}$-plet are indicated. The sextet in the first floor has a a symmetric light-quark pair, the two light-heavy quark pairs are then antisymmetric in flavor. The 3-plet in the first floor has an antisymmetric light-quark pair, the light-heavy quark pairs are then symmetric in flavor.

Finally, there is a fully anti-symmetric 4-plet. It is shown in Fig. 9.4.1c. Ground-state baryons have a symmetric spatial wave function. A spin wave function of three fermions has mixed symmetry. A fully symmetry, a fully antisymmetric and a mixed-symmetry wave function cannot be coupled to a fully symmetric wave function. Hence ground-state baryons cannot be in the 4-plet. Only excited baryons can have a fully antisymmetric flavor wave function. Below, in Section 9.4.5, the wave functions and their symmetries are discussed in more detail.

### 9.4.3 Excited baryons: Selected experimental results

*BaBar, BELLE and LHCb:*

Most information on heavy baryons stems from three experiments, BaBar, BELLE and LHCb even though many discoveries had already been made before with the Split-Field-Magnet, by the SELEX, UA and LEP experiments at CERN, and by the CDF experiment at FERMILAB. BaBar at SLAC (US) and BELLE at KEK (Japan) study the decays of $B$ mesons produced in asymetric $e^+e^-$ storage rings with beam energies of 9 (KEK: 7) GeV for electrons and 3.1 (KEK: 4) GeV for positrons resulting in a center-of-mass energy equal to the $\Upsilon(4S)$ mass of 10.58 GeV. The LHCb experiment is placed at the Large Hadron Collider at CERN operating at $\sqrt{s} = 13.6$ GeV. The experiment is a single-arm forward spectrometer covering the pseudorapidity range $2 \leq \eta \leq 5$. It is designed for the study of particles containing $b$ or $c$ quarks. All three detectors have vertex reconstruction capabilities; BaBar and BELLE track charged particles in tracking chambers placed in the 1.5 T magnetic field of a superconducting solenoid. Particle identification is provided by a measurement of the specific ionization and by detection of the Cherenkov radiation in reflecting ring imaging Cherenkov detectors. CsI(Tl)-crystal electromagnetic calorimeters allow for energy measurements of electrons and photons. LHCb is equipped with silicon-strip detector located upstream and downstream of a dipole magnet with a bending power of about 4 Tm. Photons, electrons and hadrons are identified by a calorimeter system consisting of scintillating counters and pre-shower detectors, and an electromagnetic and a hadronic calorimeter. Muons are identified by a system composed of alternating layers of iron and multiwire proportional chambers.

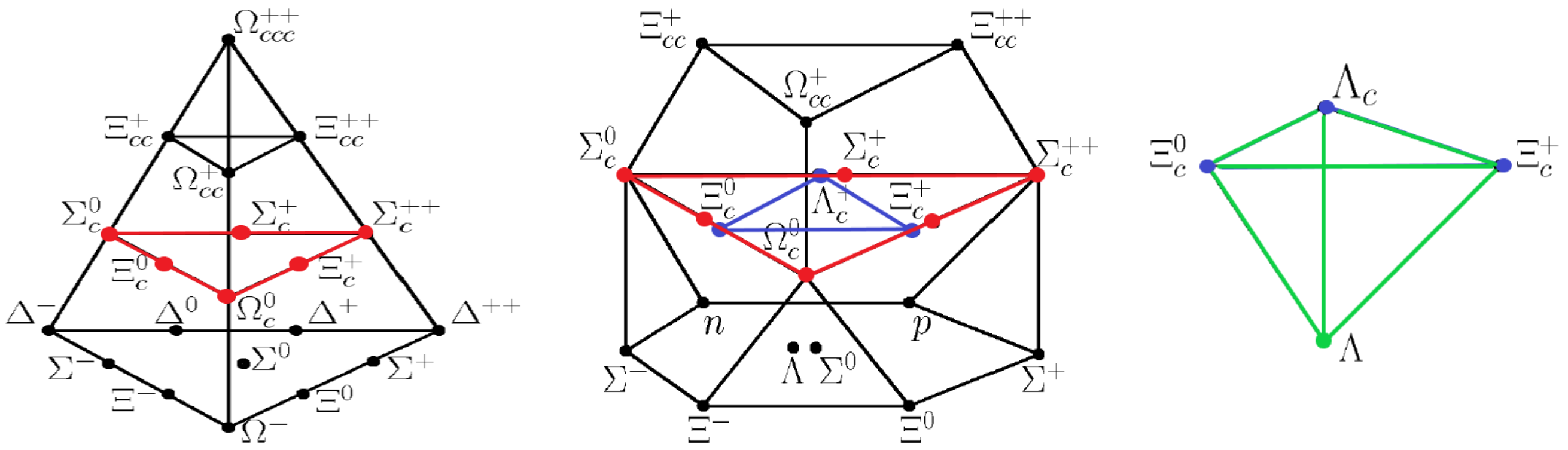

**Fig. 9.4.1** Ground-state heavy baryons in SU(4). Baryons with one charm quark are represented by colored dots. Left: The symmetric 20-plet. Center: Baryons in the mixed-symmetry 20-plet. The mixed-symmetry 20-plet contains a sextet with a symmetric light-quark pair ($SU_F(3)$ multiplicity 6) and a triplet with an anti-symmetric light-quark pair ($SU_F(3)$ multiplicity $\bar{3}$). Right: The fully antisymmetric 4-plet.

In the following we discuss three important results from these experiments that demonstrate the capabilities of the detectors.

*Observation of $\Omega_c^{*0}(2770)$ decaying to $\Omega_c^0\gamma$ by BaBar:* The Babar experiment studied the inclusive reaction $e^+e^- \to \Omega_c^{*0} X$ where $X$ denote the recoiling particles [2834]. $\Omega_c^0$ baryons are identified via different decay modes and reconstructed with a mass resolution $\sigma_{\rm RMS} = 13\,$MeV. The $\gamma$ is reconstructed in the $\Omega_c^0$ CsI(Tl) calorimeter. Figure 9.4.2 shows the reconstructed $\Omega_c^0$ and the $\Omega_c^{*0}$ in its $\Omega_c^{*0} \to \Omega_c^0\gamma$ decay. Obviously, the $\Omega_c^{*0}(2770)$ is equivalent to $\Delta^0(1232)$ with the $u, d, d$ quarks exchanged by $c, s, s$, and the transition corresponds to the $\Delta(1232) \to N\gamma$ decay.

*First determination of the spin and parity of the charmed-strange baryon $\Xi_c^+(2970)$ by BELLE.* The BELLE collaboration identified $\Xi_c^+(2970)$ in the decay chain $\Xi_c^+(2970) \to \Xi_c^0(2645)\pi^+ \to \Xi_c^+\pi^-\pi^+$; $\Xi_c^+$ is reconstructed from its decay into $\Xi^-\pi^+\pi^+$ [2835]. Due to its mass, $\Xi_c^0(2645)$ is likely the spin excitation with $J^P = 3/2^+$ of the $J^P = 1/2^+$ ground state $\Xi_c^0$. The helicity angle in the primary decay, i.e. the angle between the $\pi^+$ and the opposite of the boost direction in the c.m. frame both calculated in the $\Xi_c^+(2970)$ rest frame, proved to be insensitive to some likely $J^P$ combinations. However, the predictions for different $J^P$'s vary significantly for the angular distributions in the secondary decay (see Fig. 9.4.3).

The analysis shows that quantum numbers $J^P = 1/2^+$ are preferred for $\Xi_c^+(2970)$. These are the quantum numbers of the Roper resonance. The BELLE collaboration noted that its mass difference to the $\Xi_c$ ground state is about 500 MeV. The same excitation energy is required to excite the Roper resonance $N(1440)$, the $\Lambda(1600)$ and the $\Sigma(1660)$, all with $J^P = 1/2^+$.

*First observation of excited $\Omega_b$ states by LHCb.* The LHCb collaboration searched for narrow resonances in the $\Xi_b^0 K^-$ invariant mass distribution [2836]. The $\Xi_b^0$ has a lifetime of $(1.48\pm0.03)10^{-12}\,$s, $c\tau \approx 500\,\mu$m, which is sufficiently long to separate the interaction and the decay vertices. Four peaks can be seen (Fig. 9.4.4), which correspond to excited states of $\Omega_c$. With the given statistics, quantum numbers can not yet be determined.

### 9.4.4 The mass spectrum of excited heavy baryons

Figure 9.4.5 shows the mass spectrum of heavy baryons with a single charm or bottom quark. Established light baryons with strangeness are shown for comparison. The quantum numbers of low-mass heavy baryons are mostly known, for higher-mass states this information is often missing. The masses are given as excitation energies above the $\Lambda$ ($\Lambda_c$, $\Lambda_b$) mass.

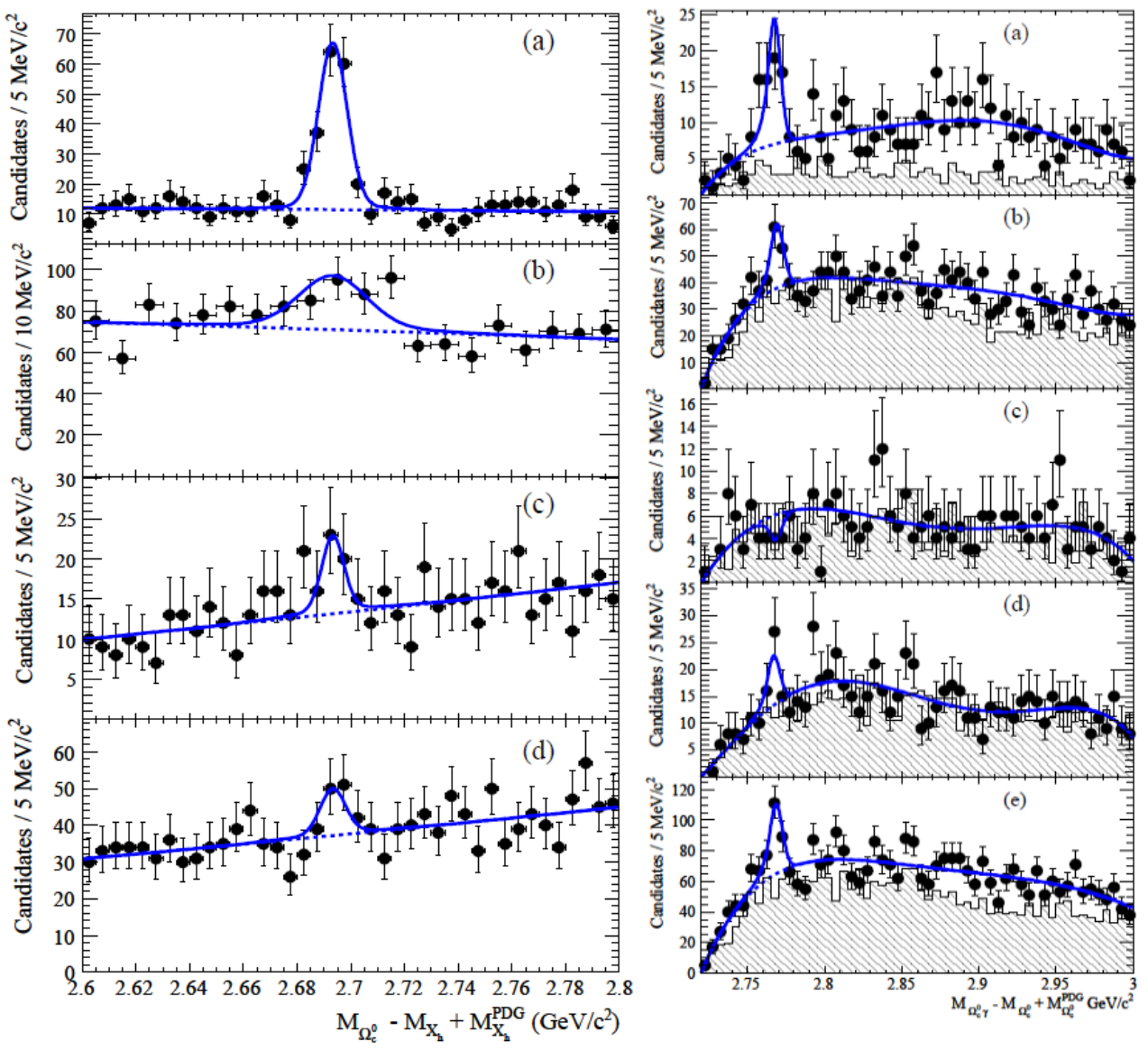

**Fig. 9.4.2** Left: The invariant mass distributions of $\Omega_c^0$ candidates in their decay to $\Omega^-\pi^+$ (a), $\Omega^-\pi^+\pi^0$ (b), $\Omega^-\pi^+\pi^-\pi^+$ (c), $\Xi^-K^-\pi^+\pi^+$ (d). $M_{\Omega_c^0}$ is the reconstructed mass of $\Omega_c^0$ candidates, $X_h$ denotes the daughter hyperon. Right: Invariant mass distribution of $\Omega_c^* \to \Omega_c\gamma$ for the indivual $\Omega_c^0$ decay modes (a-d) and for the sum (e). (Adapted from [2834].)

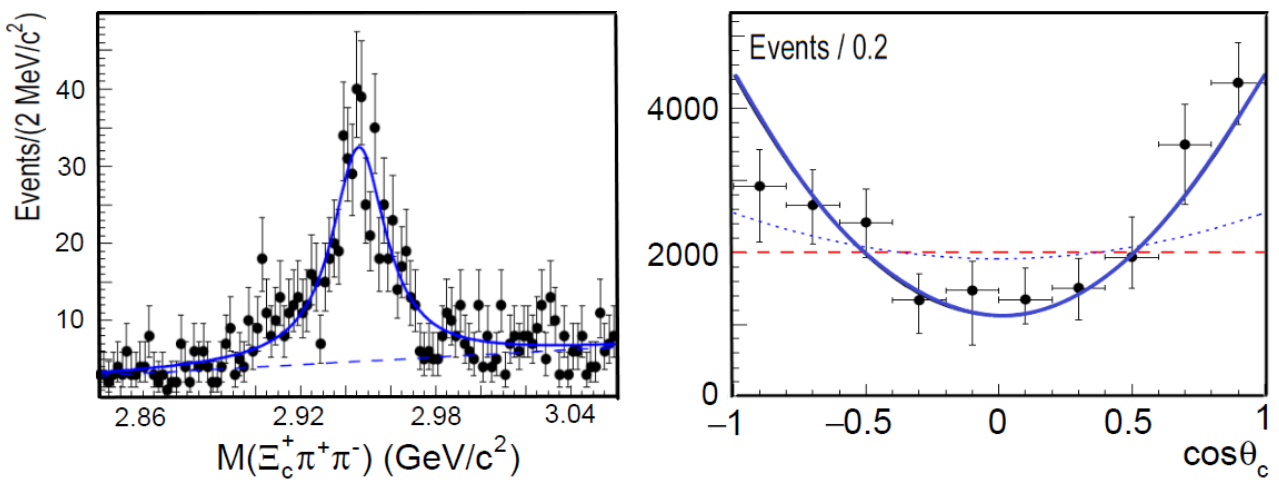

**Fig. 9.4.3** Left: The $\Xi_c^+\pi^-\pi^+$ invariant mass distribution for events in which the $\Xi_c^+\pi^-$ invariant mass is compatible with the $\Xi_c^0(2645)$ mass. Right: The helicity angle $\theta_c$ between the direction of the $\pi^-$ relative to the opposite direction of the $\Xi_c^+(2970)$ in the rest frame of the $\Xi_c^0(2645)$. (Adapted from Ref. [2835].)

At the first glance, the spectrum looks confusing. The $\Lambda$ spectrum is crowded, there is a low-mass negative-parity spin doublet, a second doublet – at about the same mass as a $\Sigma$ spin doublet – a pair with $J^P = 1/2^-$ and $5/2^-$ where a $3/2^-$ state seems to be missing, and then a positive-parity doublet with $J^P = 3/2^+, 5/2^+$. In the $\Lambda_c$ spectrum, the higher-mass negative-parity states and the positive-parity doublet are inverted in mass[100]. The $3/2^+ - 1/2^+$ hyperfine splitting decreases rapidly when going from $\Sigma$ and $\Xi$ to $\Sigma_c$ and $\Xi_c$ and from $\Sigma_b$ and $\Xi_b$. It is interesting to note that a similar pattern is observed in mesons: the hyperfine splitting decreases when going from $\rho - \pi$ to $D^* - D$ and to $B^* - B$. Also, there is one $\Xi$ $1/2^+$ ground state but two states for $\Xi_c$ and $\Xi_b$. The lowest-mass $\Omega$ has $J^P = 3/2^+$, in the charm sector, two low-mass $\Omega_c$ states are known with $J^P = 1/2^+$ and $3/2^+$, the $\Omega_b$ spectrum has just one low-mass state with $J^P = 1/2^+$.

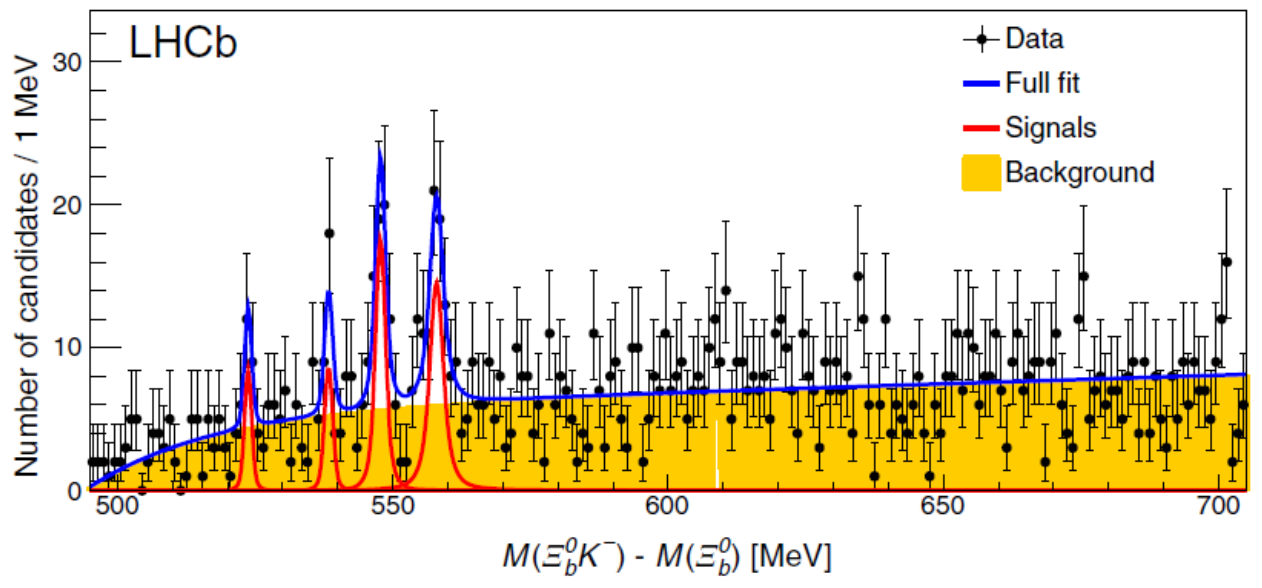

**Fig. 9.4.4** Distribution of the mass difference $M_{\Xi_b^0 K^-} - M_{\Xi_b^0}$ for $\Xi_b^0 K^-$ candidates. The background is given by the wrong-sign candidates $\Xi_b^0 K^+$. (From Ref. [2836].)

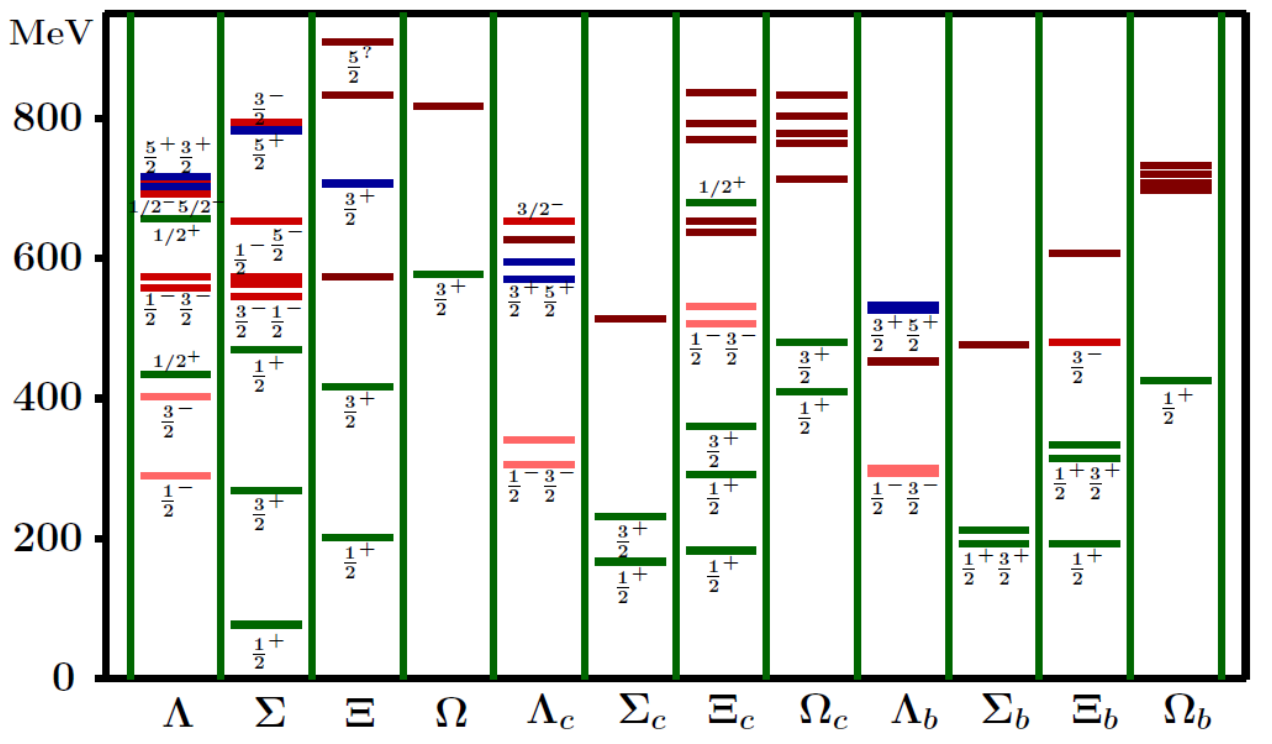

**Fig. 9.4.5** Heavy baryons with charm or bottomnes and a comparison with light baryons with strangeness. All heavy baryons are shown, light baryons are shown at the pole mass and are only included with 3* or 4* rating. When two quantum numbers are given, the first one refers to the lower-mass state, the second one to the state above. The states with $L = l_\rho = l_\lambda = 0$ are shown in green, states with $L = 1$ in red (orange for members of $\underline{4}_{\rm F}$), states $L = 2$ in blue, states with unknown spin-parity in brown.

### 9.4.5 Heavy baryons as three-quark systems

*The spatial wave function*

The orbital wave functions of excited states are classified into two kinds of orbital excitations, the $\lambda$-mode and the $\rho$-mode (see Eqn. (9.1.1)). In heavy baryons with one heavy quark, the $\lambda$-mode is the excitation of

100 This inversion was predicted by Capstick and Isgur long before the states were discovered [736]

the coordinate between the heavy quark and the light diquark, and the $\rho$-mode is the excitation of the diquark cluster. In light-baryon excitations, the $\lambda$ and $\rho$ oscillators are mostly both excited, e.g. to $l_\lambda = 1$, $l_\rho = 0$ and $l_\lambda = 0$, $l_\rho = 1$, the two components of the wave function having a relative $+$ or $-$ sign. In heavy baryons with one heavy quark, the mixing between these two configurations is small.

The two oscillators have different reduced masses, $m_\rho$ and $m_\lambda$:

$$m_\rho = \frac{m_q}{2} \quad , \quad m_\lambda = \frac{2m_q m_Q}{2m_q + m_Q}. \tag{9.4.3}$$

The ratio of harmonic oscillator frequencies is then given by

$$\frac{\omega_\lambda}{\omega_\rho} = \sqrt{\frac{1}{3}(1 + 2m_q/m_Q)} \leq 1. \tag{9.4.4}$$

In the heavy-quark limit ($m_Q \to \infty$), the excitation energies in the $\lambda$ oscillator are reduced by a factor $\sqrt{3}$.

*Diquarks*

We first consider the light diquark. The two light quarks can have either the symmetric flavor structure $\mathbf{6}_F$ or the antisymmetric flavor structure $\bar{\mathbf{3}}_F$. The spin of the light diquark can be $s_{qq} = s_l = 1$ or $s_l = 0$ leading to a symmetric or an antisymmetric spin wave function. The color part of the wave function is totally antisymmetric. Hence flavor and spin wave functions are linked. In an $S$-wave, scalar ("good" or g) and axial-vector ("bad" or b) diquarks can be formed. The intrinsic quark spins couple to the internal orbital angular momentum $l_\rho$, leading to excited diquarks with orbital excitations.

$$(l_\rho = 0,\ \mathbf{S}) \begin{cases} s_l = 0\ (\mathbf{A}),\ \bar{\mathbf{3}}_F\ (\mathbf{A}),\ j_{qq} = 0\,, & \text{(g)} \\ s_l = 1\ (\mathbf{S}),\ \mathbf{6}_F\ (\mathbf{S}),\ j_{qq} = 1\,, & \text{(b)} \end{cases}$$

$$(l_\rho = 1,\ \mathbf{A}) \begin{cases} s_l = 0\ (\mathbf{A}),\ \mathbf{6}_F\ (\mathbf{S}),\ j_{qq} = 1\,, & \text{(g)} \\ s_l = 1\ (\mathbf{S}),\ \bar{\mathbf{3}}_F\ (\mathbf{A}),\ j_{qq} = 0/1/2\,, & \text{(b)} \end{cases}$$

$$(l_\rho = 2,\ \mathbf{S}) \begin{cases} s_l = 0\ (\mathbf{A}),\ \bar{\mathbf{3}}_F\ (\mathbf{A}),\ j_{qq} = 2\,, & \text{(g)} \\ s_l = 1\ (\mathbf{S}),\ \mathbf{6}_F\ (\mathbf{S}),\ j_{qq} = 1/2/3\,, & \text{(b)} \end{cases}$$

$$\cdots$$

where we have denoted the total angular momentum of the light diquark as $j_{qq}$.

*Coupling of angular momenta*

Figure 9.4.6 shows how the orbital angular momentum and the diquark spin couple to the total diquark angular momentum $j_l$. This in turn couples to the heavy-quark spin $s_Q$ giving rise to spin doublets (or just spin-1/2 states for $j_l = 0$). Note the $\Lambda$ and $\Xi$ spin doublet with $s_l = 0$ and $\bar{3}_\mathrm{F}$. In this case the wave function is antisymmetric in spin and flavor, this is a "good" diquark.

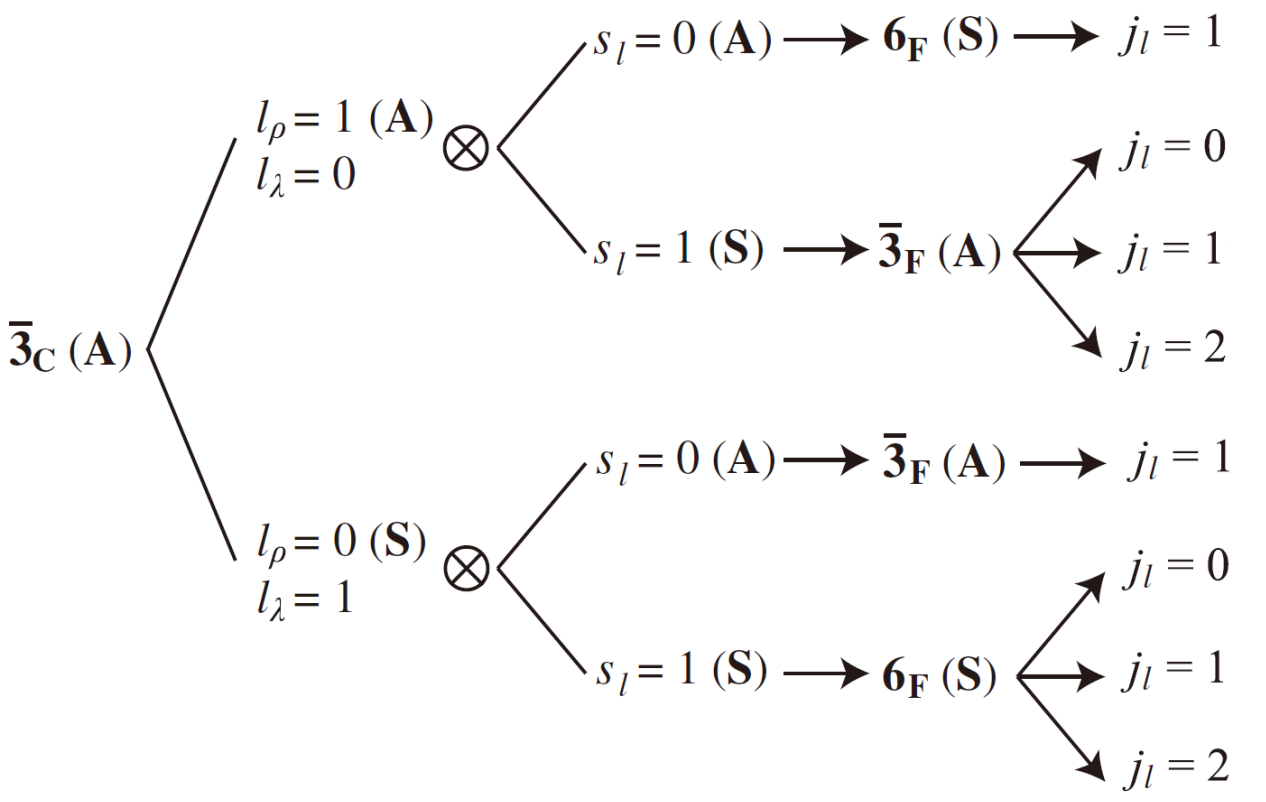

**Fig. 9.4.6** Heavy baryons in $P$-wave. The light diquark couples to the spin of the heavy quark. The light diquark of $\Lambda$ and $\Xi$ heavy baryons are in the antisymmetric flavor $\bar{3}_\mathrm{F}$ and in the symmetric $6_\mathrm{F}$ in the case of $\Sigma, \Xi'$ and $\Omega$ (Adapted from [2837]).

**Table 9.4.3** The $\lambda$- and $\rho$-mode assignments of the $P$ and $D$-wave excitations of singly-heavy baryons. $l_\rho, l_\lambda$ are orbital angular momenta of the two oscillators, $L$ the total orbital angular momentum, $s_q$ is the spin, $j_q$ the total angular momentum of the diquark, and $J$ the total spin.

| $l_\rho$ | $l_\lambda$ | $L$ | $s_q$ | $j_q$ | $\Lambda, \Xi$ | $\Sigma, \Xi', \Omega$ | $J^P$ |
|---|---|---|---|---|---|---|---|
| 0 | 1 | 1 | 0 | 1 | $\lambda$ | $\rho$ | $1/2^-, 3/2^-$ |
| 0 | 1 | 1 | 1 | 0 | $\rho$ | $\lambda$ | $1/2^-$ |
| 1 | 0 | 1 | 1 | 1 | $\rho$ | $\lambda$ | $1/2^-, 3/2^-$ |
| 1 | 0 | 1 | 1 | 2 | $\rho$ | $\lambda$ | $3/2^-, 5/2^-$ |
| 0 | 2 | 2 | 0 | 2 | $\lambda$ | - | $3/2^+, 5/2^+$ |
| 2 | 0 | 2 | 0 | 2 | $\rho$ | - | $3/2^+, 5/2^+$ |
| 0 | 2 | 2 | 1 | 1 | - | $\lambda$ | $1/2^+, 3/2^+$ |
| 0 | 2 | 2 | 1 | 2 | - | $\lambda$ | $3/2^+, 5/2^+$ |
| 0 | 2 | 2 | 1 | 3 | - | $\lambda$ | $5/2^+, 7/2^+$ |
| 2 | 2 | 2 | 1 | 1 | - | $\rho$ | $1/2^+, 3/2^+$ |
| 2 | 0 | 2 | 1 | 2 | - | $\rho$ | $3/2^+, 5/2^+$ |
| 2 | 0 | 2 | 1 | 3 | - | $\rho$ | $5/2^+, 7/2^+$ |

Only a few heavy baryons are known with $L = 2$: $\Lambda_c$ and $\Lambda_b$ with spin-parity $3/2^+$ and $5/2^+$. The other expected states seem to show up only in the higher-mass, less-explored region. The two observed doublets can be assigned to a configuration in which $l_\rho = 2$, $L_\lambda = 0$, and the diquark is in $\bar{3}_\mathrm{F}$ and $s_l = 0$.

$$\bar{\mathbf{3}}_\mathrm{F}\ (\mathbf{A}):\ L = 2 \otimes s_l = 0\ (\mathbf{A}) \;\text{——}\; j_l = 2$$

The diquark is a "good" diquark. Note that the states $\Lambda_c(2860)$, $\Lambda_c(2880)$ with spin-parity $3/2^+$ and $5/2^+$ ($L = 2$) are below $\Lambda_c(2940)$ with $3/2^-$. The latter state has a "bad" diquark and is excited to $L = 1$ in the $\rho$ oscillator. In this competition, the "good" diquark and $\lambda$ excitation with $L = 2$ wins over "bad" diquark and $\rho$ oscillator even though the orbital angular momentum of $\Lambda_c(2940)$ is $L = 1$!

Table 9.4.3 gives a survey of the coupling scheme of $Qqq$ baryons. The spin and orbital angular momentum of the two light quarks couple to $j_q$, and when combined with the heavy-quark spin $s_Q$, the final $J^P$ results. There are also states with mixed excitations like $l_\rho = 1$, $l_\lambda = 1$. These are unlikely to be produced (see Section 9.2) and are not included here. $\Lambda$ and $\Xi$ with $s_q = 0$ and $l_\rho = 0$ have a "good" light diquark. For the $\Lambda_c$ we denote the light diquark by $[u,d]$. Note that also one light and the heavy quark can be antisymmetric in their spin and flavor wave function. We write $\Sigma_b = [ub]s$.

*Heavy quark limit*
When $m_Q \to \infty$, the heavy quark spin $s_Q$ is conserved. Due to the conservation of the total angular momentum $J$, also the angular momentum carried by the light quarks is conserved. Hence all interactions which depend on the spin of the heavy quark disappear. Thus, the mass difference within a spin doublet with, e.g., $J^P = 3/2^+$ and $1/2^+$, will disappear in the heavy-quark limit. Indeed, the mass differences

$$\begin{aligned} M_{\Sigma(1520)3/2^+} - M_{\Sigma(1190)} &= 230\,\text{MeV} \\ M_{\Sigma_c(2520)3/2^+} - M_{\Sigma_c(2455)} &= \ 65\,\text{MeV} \\ M_{\Sigma_b(5830)3/2^-} - M_{\Sigma_b(5820)} &= \ 20\,\text{MeV} \end{aligned}$$

decrease as $m_Q$ becomes large.

### 9.4.6 A guide to the literature

The first prediction of the full spectrum of baryons including charmed and bottom baryons was presented by Capstick and Isgur [736], three years before the first baryon with bottomness was discovered. The publication remained a guideline for experimenters for now 36 years! Capstick and Isgur used a relativized quark model with a confining potential and effective one-gluon exchange. Based on the quark model, further studies of the mass spectra of heavy baryons were performed. They are numerous, and only a selction of papers can be mentioned here.

Ebert, Faustov and Galkin calculated the mass spectra for orbital and radial excitations and constructed Regge trajectories [2838]. Yu, Li, Wang, Lu, and Ya [2839] calculated the mass spectra and decays of heavy baryons excited in the $\lambda$-mode. Li, Yu, Wang, Lu, and Gu [2840] restricted the calculation - again based on the relativized quark model - to the $\Xi_c$ and $\Xi_b$ families. In their model, all excitations are in the $\lambda$-mode.

Migura, Merten, Metsch, and Petry [2841] calculated excitations of charmed baryons within a relativistically covariant quark model based on the Bethe-Salpeter-equation in instantaneous approximation. Interactions are given by a linearly rising three-body confinement potential and a flavor dependent two-body force derived from QCD instanton effects. Valcarce, Garcilazo and Vijande [2842] performed a comparative Faddeev study of heavy baryons with nonrelativistic and relativistic kinematics and different interacting potentials that differ in the description of the hyperfine splitting. The authors conclude that the mass difference between members of the same $\text{SU}_\text{F}(3)$ configuration, either $\bar{3}_F$ or $6_F$, is determined by the interaction in the light-heavy quark subsystem, and the mass difference between members of different representations is mainly determined by the dynamics of the light diquark.

Chen, Wei and Zhang [2843] derive a mass formula in a relativistic flux tube model to calculate mass spectra for $\Lambda$ and $\Xi$ heavy baryons and assign quantum numbers to states whose quantum numbers were not known. Faustov and Galkin [2844] assigned flavor- and symmetry dependent masses and form factors to diquarks and calculated the masses of heavy baryons within a relativistic quark-diquark picture. Quantum numbers are suggested for the $\Omega_c$ excitations [2845, 2846] and other states with unknown spin-parities. A further diquark model, again with adjusted diquark masses, is presented by Kim, Liu, Oka, and Suzuki [2847] exploiting a chiral effective theory of scalar and vector diquarks according to the linear sigma model.

QCD sum rules have been exploited to study P-wave heavy baryons and their decays within the heavy quark effective theory (see [2848] and refs. therein). The low-lying spectrum of charmed baryons has also been calculated in lattice QCD with a pion mass of 156 MeV [2849]. The results - comparing favorably with the data - are compared to earlier lattice studies that are not discussed here.

All calculations reproduce the observed spectrum with good success, with a large number of parameters. For the reader, it is often not easily seen what are the main driving forces that generate the mass spectrum. Clearly, a confinement potential is mandatory, spin dependent forces are necessary. In the following phenomenological part we try to identify the leading effects driving the resonance spectrum.

**Table 9.4.4** Increase of baryon masses with the number of strange quarks.

| | $n \to s$ | $2n \to 2s$ | $3n \to 3s$ |
|---|---|---|---|
| $\Delta^-(1232)3/2^+$ | $\Sigma^-(1385)3/2^+$ | $\Xi^-(1530)3/2^+$ | $\Omega^-3/2^+$ |
| | +155 MeV | +148 MeV | + 137 MeV |
| $\Sigma_c^0(2520)3/2^+$ | $\Xi_c^0(2645)3/2^+$ | $\Omega_c^0(2770)3/2^+$ | |
| | +128 MeV | + 120 MeV | |
| $\Sigma_c^0(2455)1/2^+$ | $\Xi_c'^0\ 1/2^+$ | $\Omega_c^0\ 1/2^+$ | |
| | +121 MeV | +116 MeV | |
| $\Sigma_b^-(5816)1/2^+$ | $\Xi_b'^0\ 1/2^+$ | $\Omega_b^-\,1/2^+$ | |
| | +120 MeV | +111 MeV | |

### 9.4.7 Phenomenology of heavy baryons

We start with a simple observation: masses of baryons increase when a $u$ or $d$ quark is replaced by an $s$ quark (see Table 9.4.4). For light baryons, this is known as $U$-spin rule. The constituent $s$-quark mass decreases in heavy baryons. Note that the difference of current quark masses is $m_s - m_n \approx 124\,\mathrm{MeV}$ (see Table 3.1.1 on page 46).

In Table 9.4.5 we show the mass difference of the lowest-mass $J^P = 3/2^-$ states with $(u,d,s,c)$ or $(u,d,s,b)$ quarks and the $J^P = 1/2^+$ ground states: The mass differences are surprisingly small. The $N(1520) - N$ mass difference is 580 MeV, much larger than the mass differences seen here. In the table, [ud] represents wave functions with a $u,d$ quark pair that is anti-symmetric in spin and flavor. These diquarks are often called *good diquarks*. The presence of good diquarks leads to a stronger binding. In the 4-plet, all three quark pairs have such a component w.r.t. their exchange. We denote this by [ud,us,ds]. Thus there are three good diquarks in the wave function. This fact leads to the low masses of the 4-plet members. The similarity of the mass splittings supports similar interpretations of the four resonances from $\Lambda(1520)$ to $\Xi^0_{b\ 3/2^-}$.

In most publications, both resonances, $\Lambda_c(2595)1/2^-$ and $\Lambda_c(2625)3/2^-$, are discussed as 3-quark baryons. However, Nieves and Pavao [2850] have studied these two resonances in an effective field theory that incorporates the interplay between $\Sigma_c^{(*)}\pi - ND^{(*)}$ baryon-meson dynamics and bare $P$-wave *cud* quark-model state and suggest that these two resonances are not heavy quark symmetry spin partners. Instead, they see

$$\Lambda_c(2625)3/2^-$$

as a dressed three-quark state while $\Lambda_c(2595)1/2^-$ is reported to have a predominant molecular structure. Nevertheless, the two states $\Lambda_c(2625)3/2^-$ and $\Lambda_c(2595)1/2^-$ obviously form a spin doublet.

The mass shift in H atoms between the two ground states with electron and proton spins parallel or antiparallel is called hyperfine splitting. We borrow this expression to discuss the difference between the ground states with all three quark spins adding to $J = 3/2$ (and belonging to the symmetric 20-plet) and with those having $J = 1/2$ (that belong to the mixed-symmetrx 20-plet). We thus compare masses of the fully symmetric $20_s$-plet with those from the $\bar{3}$-plet or 6-plet within the $20_m$-plet (see Table 9.4.6). The two configurations differ by the orientation of the heavy-quark spin relative to the spin of the light diquark. According to the heavy-quark-spin symmetry, this mass difference has to vanish with $m_Q \to \infty$. In the Table we assume constituent quark masses of 0.15 GeV $(u,d)$, 0.3 GeV $(s)$, 1.25 GeV $(c)$ and 4.1 GeV $(b)$.

The $J^P = 3/2^+$ states have a fully symmetric flavor wave function, the $J^P = 1/2^+$ states have an antisymmetric quark pair (a good diquark) that is indicated in the list. Their effect scales with $1/m_q$. The mass shift due to the presence of good diquarks is expected for instanton-induced interactions.

*Heavy baryons at higher mass:*
Next we discuss the higher-mass negative-parity states. In light-baryon spectroscopy, there are seven negative-parity $\Lambda$ states expected in the first excitation level: two singlet states with $J^P = 1/2^-, 3/2^-$, two octet states with intrinsic total quark spin $s = 1/2$ and $J^P = 1/2^-, 3/2^-$, and a $J^P = 1/2^-, 3/2^-, 5/2^-$ triplet with $s = 3/2$. In light baryons, both $\lambda$ and $\rho$ oscillator are coherently excited. In heavy-quark baryons, the two oscillators decouple, and the $\lambda$ and $\rho$ modes are well separated. The low-lying spin-doublet of $P$-wave $\Lambda_Q$ states is dominated by a $\lambda$-mode excitation, the other five expected states are excited in the $\rho$ mode.

Unfortunately, only one negative-parity state at a higher mass has been reported, the $\Lambda_c(2940)3/2^-$. Its mass is 653 MeV above the $\Lambda_c^+$. We interpret this state as $l_\rho$ excitation with a diquark spin $s = 1$. The $\Lambda(1690)3/2^-$

**Table 9.4.5** Mass splitting between baryon ground states belonging to the symmetric 20plet (with $J^P = 3/2^+$) and to the mixed-symmetry 20plet (with $J^P = 1/2^+$).

| | | | | |
|---|---|---|---|---|
| $\Xi^0_{b\ 3/2^-}$ | [us,ub,sb] | $\Xi^0_{b\ 1/2^+}$ | [us] | $\delta M = 310\,\mathrm{MeV}$ |
| $\Lambda^0_{b\ 3/2^-}$ | [ud,ub,db] | $\Lambda^0_{b\ 1/2^+}$ | [ud] | $\delta M = 300\,\mathrm{MeV}$ |
| $\Xi^+_{c\ 3/2^-}$ | [us,uc,sc] | $\Xi^+_{c\ 1/2^+}$ | [us] | $\delta M = 350\,\mathrm{MeV}$ |
| $\Lambda^+_{c\ 3/2^-}$ | [ud,uc,dc] | $\Lambda^+_{c\ 1/2^+}$ | [ud] | $\delta M = 400\,\mathrm{MeV}$ |
| $\Lambda(1520)$ | [ud,us,ds] | $\Lambda_{1/2^+}$ | [ud] | $\delta M = 400\,\mathrm{MeV}$ |

**Table 9.4.6** Mass splitting between baryons with fully symmetric wave functions and baryons with antisymmetric quark pairs. The [us] indicates an antismmetric quark pair.

| | | | $\delta M$ | $m_q$ | $\delta M \cdot m_q$ |
|---|---|---|---|---|---|
| $\Sigma_{b\ 3/2^+}$ | $\Lambda_b$ | $[ud]b$ | 0.211 MeV | ∼ 0.3 GeV | 0.063 |
| $\Sigma_{c\ 3/2^+}$ | $\Lambda_c$ | $[ud]c$ | 0.232 MeV | ∼ 0.3 GeV | 0.070 |
| $\Sigma_{\ 3/2^+}$ | $\Lambda$ | $[ud]s$ | 0.268 MeV | ∼ 0.3 GeV | 0.080 |
| $\Delta_{\ 3/2^+}$ | N | $[ud]u$ | 0.292 MeV | ∼ 0.3 GeV | 0.088 |
| $\Xi_{b\ 3/2^+}$ | $\Xi_b$ | $[us]b$ | 0.163 MeV | ∼ 0.45 GeV | 0.073 |
| $\Xi_{c\ 3/2^+}$ | $\Xi_c$ | $[us]c$ | 0.177 MeV | ∼ 0.45 GeV | 0.080 |
| $\Xi_{\ 3/2^+}$ | $\Xi$ | $[us]s$ | 0.217 MeV | ∼ 0.45 GeV | 0.098 |
| $\Sigma_{\ 3/2^+}$ | $\Sigma$ | $[us]u$ | 0.191 MeV | ∼ 0.45 GeV | 0.086 |
| $\Xi_{c\ 3/2^+}$ | $\Xi_c'$ | $[uc]s$ | 0.067 MeV | ∼ 1.4 GeV | 0.093 |
| $\Sigma_{c\ 3/2^+}$ | $\Sigma_c$ | $[uc]u$ | 0.065 MeV | ∼ 1.4 GeV | 0.090 |
| $\Xi_{b\ 3/2^+}$ | $\Xi_b'$ | $[ub]s$ | 0.020 MeV | ∼ 4.25 GeV | 0.085 |
| $\Sigma_{b\ 3/2^+}$ | $\Sigma_b$ | $[ub]u$ | 0.021 MeV | ∼ 4.25 GeV | 0.089 |

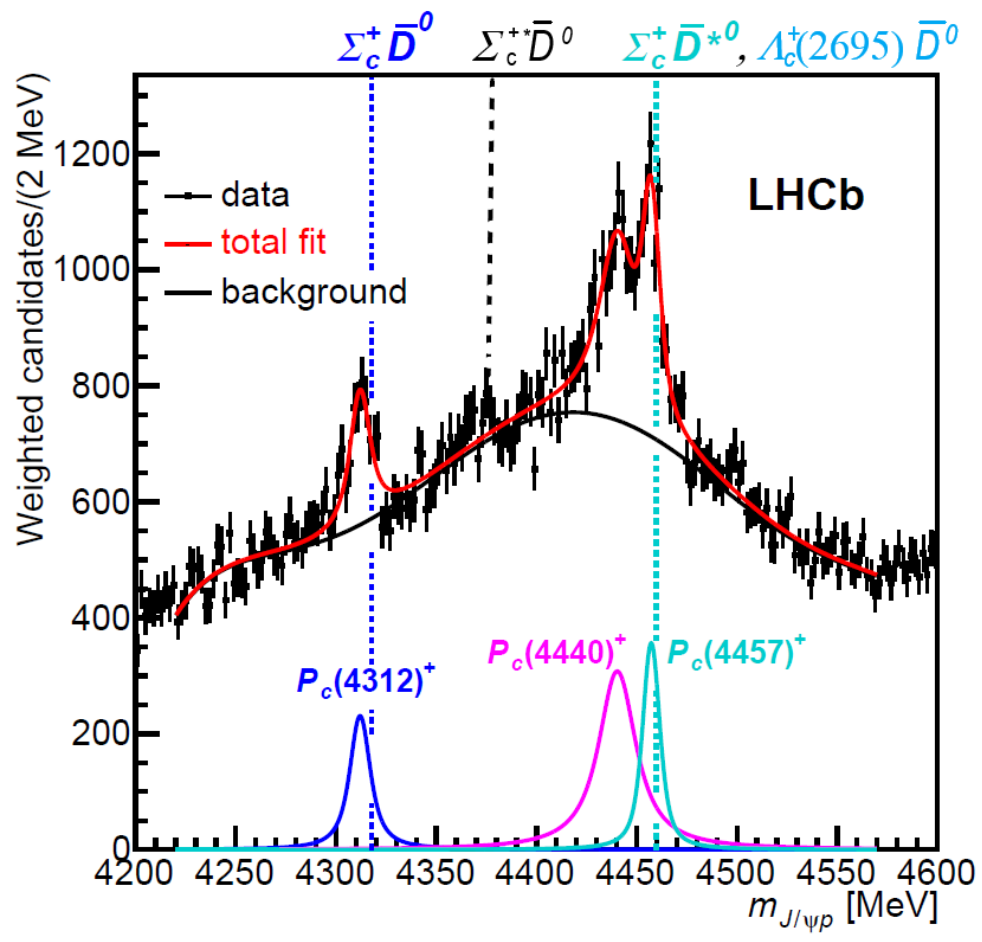

**Fig. 9.4.7** (Color online) The $J/\psi\, p$ mass distribution fitted with three BW amplitudes and a sixth-order polynomial background. The thresholds for the $\Sigma_c^+ \bar{D}^0$. (Adapted from [2829].)

is only 570 MeV above the $\Lambda$, it is excited in both the $\lambda$ and the $\rho$ mode.

The mass of $\Lambda_c(2940)3/2^-$ (with intrinsic orbital angular momentum $L=1$) is above the masses of the positive-parity states $\Lambda_c(2860)3/2^+$ and $\Lambda_c(2880)5/2^+$ (having $L=2$). Yet, the mass of $\Lambda(1690)3/2^-$ falls well below the masses of $\Lambda(1890)3/2^+$ and $\Lambda(1820)5/2^+$ for reasons discussed above.

### 9.4.8 Pentaquarks

In 2015, the LHCb collaboration reported the observation of two exotic structures in the $J/\psi p$ system, a broad resonant structure with a Breit-Wigner width of about 200 MeV called $P_c(4380)^+$ and a narrow state called $P_c(4450)^+$ [2828]. The exotic structures were observed in the reaction $\Lambda_b^0 \to J/\psi K^- p$. An excited three-quark nucleon cannot decay into $J/\psi\, p$, this would violate the OZI rule. Hence the minimal quark content is $(c\bar{c}uud)$. The findings met with great interest; the publication is quoted nearly 1500 times (2022, November). Indeed narrow baryonic resonances with hidden charm had been predicted several years before as dynamically generated states [2851–2853].

A multitude of different interpretations of the observed structures is offered in the literature, but none is accepted anonymously. There are numerous reviews on tetra- and pentaquarks and their possible interpretations [1388, 2638, 2854–2857].

With increased statistics, $P_c(4312)^+$ was confirmed and the higher-mass $P_c(4450)^+$ was shown to be split into two narrow overlapping structures, $P_c(4440)^+$ and $P_c(4457)^+$ [2829]. The existence of the broad resonance was not confirmed. The data and a fit are shown in

**Table 9.4.7** $J/\psi p$ and $J/\psi\Lambda$ pentaquarks found by the LHCb collaboration.

| | | | |
|---|---|---|---|
| $P_c(4312)^+$: | M = | $(4311.9 \pm 0.7\ ^{+6.8}_{-0.6})$ | MeV |
| [2829] | $\Gamma$ = | $(9.8 \pm 2.7\ ^{+3.7}_{-4.5})$ | MeV |
| $P_c(4380)^+$: | M = | $(4380 \pm 30)$ | MeV |
| [2828] | $\Gamma$ = | $(205 \pm 90)$ | MeV |
| $P_c(4440)^+$: | M = | $(4440.3 \pm 1.3\ ^{+4.1}_{-4.7})$ | MeV |
| [2829] | $\Gamma$ = | $(20.6 \pm 4.9\ ^{+8.7}_{-10.1})$ | MeV |
| $P_c(4457)^+$: | M = | $(4457.3 \pm 0.6\ ^{+4.1}_{-1.7})$ | MeV |
| [2829] | $\Gamma$ = | $(6.4 \pm 2.0\ ^{+5.7}_{-1.9})$ | MeV |
| $P_c(4337)^+$: | M = | $(4337\ ^{+7}_{-4}\ ^{+2}_{-2})$ | MeV |
| [2859] | $\Gamma$ = | $(29\ ^{+26}_{-12}\ ^{+14}_{-14})$ | MeV |
| $P_{cs}^0(4459)$: | M = | $(4458.8 \pm 2.9\ ^{+4.7}_{-1.1})$ | MeV |
| [2861] | $\Gamma$ = | $(17.3 \pm 6.5\ ^{+8.0}_{-5.7})$ | MeV |
| $P_{cs}^0(4338)$: | M = | $(4338.2 \pm 0.7 \pm 0.4)$ | MeV |
| [2862] | $\Gamma$ = | $(7.0 \pm 1.2 \pm 1.3)$ | MeV |

Fig. 9.4.7 which also displays some relevant thresholds. In addition, a further smaller structure can be seen at 4380 MeV, close to the $\Sigma_c^{+*}\bar{D}^0$ threshold. A narrow structure here is expected in molecular models (see e. g. [2858]), but due to limited statistics there was no attempt to describe it in the recent LHCb analysis [2829]. The resonant parameters – including the broad structure at 4380 MeV – are reproduced in Table 9.4.7.

Quantum numbers $J^P = 3/2^-$ and $5/2^+$ were preferred for $P_c(4380)^+$ and $P_c(4450)^+$. In the later publication [2829], no quantum numbers are determined.

In the reaction $B_s^0 \to J/\psi\,\bar{p}p$ a pentaquark-like structure, named $P_c(4337)^+$, was observed in the $J/\psi\,\bar{p}$ and $J/\psi\, p$ mass distributions [2859]. The significance, as determined from a 3-body amplitude analysis, is between 3.1 and 3.7 $\sigma$. Its Breit-Wigner parameters are incompatible with the structures observed in $\Lambda_b$ decays. The lighter state at 4312 MeV was not found in this reaction, highlighting the importance of the production mechanism for the formation of these resonances. However, it has been pointed out in [2860] that in a region with many close-by thresholds, the Breit-Wigner parameters measured in a particular channel may differ significantly from the pole location.

Strange counterparts to these pentaquark states, e.g. resonances in the $J/\psi\,\Lambda$ system, are denoted by $P_{cs}$ and have $(c\bar{c}uds)$ as minimal quark content. A peak have been reported by LHCb in the reaction $\Xi_b^- \to J/\psi\,\Lambda K^-$ [2861]. Close to the $\Xi_c^0\, D^{*0}$ threshold a further peak was found, with a mass and width given in Table 9.4.7, too.

The $J/\psi\Lambda$ system was also investigated in 2019 by CMS [2863], exploiting the small phase space available in the B-meson decay $B^- \to J/\psi\,\Lambda\bar{p}$. The analysis showed that the observed spectrum was incompatible with a pure phase space distribution. Very recently, the LHCb collaboration reported a new analysis of this process [2864]. Now, a signal in the $J/\psi\,\Lambda$ subsystem, with preferred quantum numbers $J^P = 1/2^-$, was established at high significance, named $P^0_{cs}(4338)$. Due to the presence of the second (anti)baryon, the phase space in the $B$-meson decay is too small to access the heavier pentaquark state found in the $\Xi_b$ decay.

These structures have stimulated an intense discussion of the nature of these structures. Do they originate from threshold singularities due to rescattering in the final state leading to a logarithmic branching point in the amplitude? Are they hadronic molecules like the deuteron? Are they compact or triple-quark–diquark systems or states where a $c\bar{c}$ center is surrounded by light quarks?

The peaks are mostly seen very close to important thresholds. Thus they could originate from threshold singularities. We refer to a few publications [2635, 2865–2867]. The LHCb collaboration studied this hypothesis and found it incompatible with the data, but the attempts continued [2868–2871].

Very popular are interpretations as bound states composed of charmed baryons and anti-charmed mesons or of charmonium states binding light-quark baryons. The pentaquark states are then seen to be of molecular nature and be bound by coupled-channel dynamics [2858, 2872–2882]. Diquark-triquark models were studied [2883–2886], and sum rules are exploited in Refs. [2887, 2888].

### 9.4.9 Concluding remarks:

The study of hadrons with heavy quarks has developed into a fascinating new field of particle physics. Particular excitement is due to the discovery of unconventional structures that are hotly debated. But also the "regular" heavy hadrons yield very useful information on the interactions of quarks in the confinement region.

# 10 Structure of the Nucleon

*Conveners:*
**Volker Burkert and Franz Gross**

After discussion of the baryon spectrum in the previous section, this section focuses on the nucleon, the most studied of all hadrons. Soon after the proton and neutron were established as the constituents of atomic nuclei, experiments measuring their magnetic moments $\mu_N$ found that these spin-1/2 particles are not point-like elementary fermions with expected $\mu_p = 1.0\mu_N$ for the proton, and $\mu_n = 0$ for the neutron. Instead $\mu_p \approx 2.5\mu_N$ and $\mu_n \approx -1.5\mu_N$, showing that the nucleons have significant structure. The discovery that the proton and the neutron are not point-like objects gave birth to the field of hadron structure explorations discussed in this section. Beginning with the Nobel prize winning measurement of the finite size of the proton in elastic electron-proton scattering experiments (Hofstadter, 1956) there have been generations of electron scattening measurements studying the proton and neutron form factors, reviewed by Andrew Puckett

In 1968 experiments employing high-energy electrons scattering from proton targets at SLAC found surprisingly large inelastic cross sections, or structure functions, which rather than falling rapidly with the exchanged four-momentum squared $Q^2$ (as would elastic cross sections) were observed to "scale" with $Q^2$. The observation of scaling suggested scattering from point-like quarks in the proton, which could most naturally be described in terms of parton distribution functions (PDFs). These PDF measurements have shed light on the momentum distributions of the different quark species (Wally Melnitchouk), and with the use of spin-polarized electrons and polarized nucleon targets the quark contributions to the nucleon spin have been precisely measured (Xiangdong Ji), putting significant challenges on the theory of QCD to reproduce or predict the results of these measurements.

As these studies continue, both in experiment with high precision measurements, and in theory, new challenges have arisen with the discovery of the generalized parton distributions that lead to the assembly of 3-dimensional tomographic images of the quark (and gluon) transverse spatial and longitudinal momentum distributions employing deeply virtual exclusive processes (Andreas Schafer and Feng Yuan). The challenges here will be on the experiments to access these generalized parton distributions (GPDs) and transverse momentum distributions (TMDs) from experiments like deeply virtual Compton scattering and deeply virtual meson production, and on phenomenology aiding the analysis. Some of the measurements are underway at Jefferson Lab in several experiment halls. The EIC will vastly extend the kinematic reach of the measurements into the gluon dominated regime.

## 10.1 Form factors

**Andrew Puckett**

### 10.1.1 Introduction

Elastic scattering of nucleons by point-like, leptonic probes is among the simplest observable processes sensitive to the nucleon's internal structure. The study of elastic electron-nucleon scattering started in the 1950's with the pioneering measurements by Robert Hofstadter and collaborators in the High Energy Physics Lab (HEPL) at Stanford [563] at incident electron energies of up to 550 MeV. Among the highlights of this work were the first conclusive demonstration of a deviation of the elastic electron-proton scattering cross section from point-like behavior, and the first direct measurement of the proton's finite size, leading to the awarding of the 1961 Nobel Prize in Physics to Hofstadter for *"for his pioneering studies of electron scattering in atomic nuclei and for his thereby achieved discoveries concerning the structure of the nucleons"*.

In the Standard Model, the lepton-nucleon interaction is purely electroweak. Due to the nucleon's finite size and complicated structure, the elastic scattering cross section falls much more rapidly as a function of the squared four-momentum transfer $Q^2$ than the point-like scattering cross section. Given the limitations of past, present, and planned lepton-hadron scattering facilities, *elastic* scattering of leptons by nucleons only occurs with sufficient probability to be practically measurable at energy scales where electomagnetic interactions are dominant; i.e., at four-momentum transfers $Q^2 \ll M_{W,Z}^2$, where $M_W(M_Z) \approx 80(91)$ GeV is the $W$ ($Z$) boson mass. As such, for most practical purposes this process can be interpreted in the framework of low-order perturbation theory in quantum electrodynamics (QED). However, the elastic form factors of the nucleon for charged- and neutral-current weak interactions are interesting in their own right and accessible even at relatively low energies in neutrino scattering [2889] and through parity-violating asymmetries in polarized electron scattering that are sensitive at leading order to the interference between photon and $Z$ exchange amplitudes [2890–2892].

The use of elastic lepton-nucleon scattering as a precision probe of nucleon structure and dynamics remains a highly active area of investigation at low and high energies. The dramatic improvements in energy reach and precision of these measurements over decades have led to many important discoveries and surprises that have dramatically reshaped our understanding of the nucleon. This section will present a brief summary of the status of the nucleon's elastic scattering form factors, their definition and physical interpretation, outstanding challenges and problems, and the near-future outlook for further advancements.

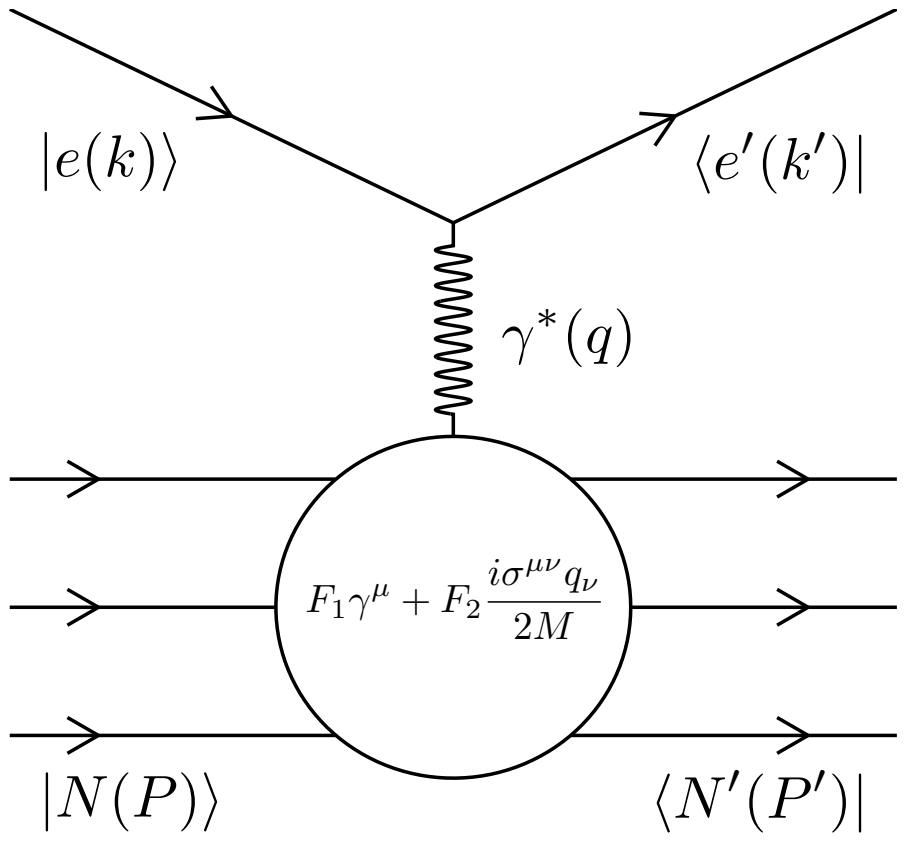

**Fig. 10.1.1** Feynman diagram for elastic $eN$ scattering in the one-photon-exchange approximation. Initial and final nucleons are represented by three lines each to indicate the nucleon's three-quark valence structure, while the circle represents the modification of the photon-nucleon vertex function by the nucleon's internal structure. See text for details.

### 10.1.2 Theoretical Formalism

The starting point for the interpretation of elastic electron-nucleon ($eN$) scattering is the one-photon-exchange (OPE) approximation, which is roughly analogous to the first Born approximation and/or the plane wave impulse approximation in non-relativistic quantum scattering theory. In the following discussion the terms OPE and Born approximation will be used interchangeably. The tree-level Feynman diagram for $eN \to eN$ is depicted schematically in Fig. 10.1.1. An incident electron of four-momentum $k \equiv (E_e, \mathbf{k})$ scatters from a nucleon of mass $M$, assumed to be initially at rest in the lab frame, with initial four-momentum $P \equiv (E_N, \mathbf{p}) = (M, \mathbf{0})$. The electron recoils with four-momentum $k' \equiv (E_e', \mathbf{k}')$ and the nucleon recoils with four-momentum $P' \equiv (E_N', \mathbf{p}')$ after absorbing the four-momentum transfer $q \equiv k - k'$. Energy and momentum conservation in this two-body scattering process dictate $P' = P + q = (M + E_e - E_e', \mathbf{k} - \mathbf{k}')$. Together with the requirement that the final-state particles be "on mass shell"; i.e., that they satisfy the relativistic energy-momentum relation for a free particle ($E^2 = p^2 + m^2$), the kinematics of the elastic $eN$ scattering process are entirely specified by just two independent variables, commonly chosen to be the incident electron energy $E_e$ and the electron scattering angle $\theta_e$ that are directly observed experimentally. Another main variable of interest is the squared four-momentum transfer $Q^2 \equiv -q^2 = -(k-k')^2 > 0$. In the nucleon rest frame (or in the center-of-momentum frame, or any other frame in which the initial momenta of the colliding particles are collinear), the scattering process is independent of the azimuthal scattering angle of the electron.

In most modern electron-nucleon scattering experiments, it is safe to use the ultrarelativistic approximation for the electron ($|\mathbf{k}| = E_e, |\mathbf{k}'| = E'_e, k^2 = k'^2 = 0$), as the incident beam energies required for sensitivity to the non-trivial details of nucleon structure are generally quite large compared to the electron mass. Moreover, the vast majority of elastic electron-nucleon scattering data come from fixed-target experiments, in which the target nucleus is at rest in the lab frame. Unless otherwise noted, all of the following expressions apply in the initial nucleon's rest frame.

To develop intuition for the physical interpretation of elastic $eN$ scattering, it is useful to consider the closely related process of ultrarelativistic electron scattering from a static charge distribution $\rho(\mathbf{r})$ with total charge $Ze$, given in the OPE approximation by:

$$\left(\frac{d\sigma}{d\Omega_e}\right) = \frac{Z\alpha^2}{4E_e^2}\frac{\cos^2\left(\frac{\theta_e}{2}\right)}{\sin^4\left(\frac{\theta_e}{2}\right)}|F(\mathbf{q})|^2 \tag{10.1.1}$$

$$\equiv \left(\frac{d\sigma}{d\Omega_e}\right)_{\text{Mott}}|F(\mathbf{q})|^2, \tag{10.1.2}$$

where $E_e$ is the incident electron energy, $\theta_e$ is the electron scattering angle, $\alpha$ is the fine structure constant, and $F(\mathbf{q})$ is the electron scattering *form factor* given by the Fourier transform of the charge distribution:

$$F(\mathbf{q}) \equiv \int \rho(\mathbf{r})e^{i\mathbf{q}\cdot\mathbf{r}}d^3\mathbf{r}, \tag{10.1.3}$$

with $\mathbf{q} \equiv \mathbf{k} - \mathbf{k}'$ the three-momentum transfer in the scattering process. The Mott cross section as defined in Eq. (10.1.2) describes the scattering of ultrarelativistic electrons from a point-like target of charge $Ze$ with zero spin and zero magnetic moment, in the limit where target recoil is negligible. In the electron-nucleon scattering case, this corresponds to the requirement $Q^2 \ll 2ME_e$. When target recoil is not negligible, the electron loses energy in the collision, and the Mott cross section is modified by the factor $E'_e/E_e$:

$$\left(\frac{d\sigma}{d\Omega_e}\right)_{\text{Mott}} = \frac{Z\alpha^2}{4E_e^2}\frac{\cos^2\left(\frac{\theta_e}{2}\right)}{\sin^4\left(\frac{\theta_e}{2}\right)}\frac{E'_e}{E_e} \tag{10.1.4}$$

In much of the modern literature, Eq. (10.1.4) is taken as the definition of the Mott cross section, whereas in Mott's original paper, the target recoil factor $E'_e/E_e$ is not included. Hereafter, we will use the definition (10.1.4) unless otherwise noted.

The most general form of the single-photon-exchange amplitude $\mathcal{M}$ for elastic $eN$ scattering consistent with Lorentz invariance, gauge invariance, and parity conservation as required by QED, and under the assumption that the nucleon is a spin-1/2 fermion obeying the Dirac equation, can be expressed using the Feynman rules of QED (see, e.g., [2100]) as follows (in "natural units" $\hbar = c = 1$):

$$\mathcal{M} = 4\pi\alpha\bar{u}(k')\gamma^\mu u(k)\left(\frac{g_{\mu\nu}}{q^2}\right)\bar{u}(P')\Gamma^\nu u(P) \tag{10.1.5}$$

Here $\mathcal{M}$ is the Lorentz-invariant single-photon-exchange amplitude, $\alpha$ is the fine-structure constant, $\bar{u}$ and $u$ represent free-particle Dirac spinors for the incoming and outgoing particles, evaluated at the relevant four-momenta, $\gamma^\mu$ is a Dirac $\gamma$ matrix, $g_{\mu\nu}$ is the Minkowski metric tensor, and $\Gamma^\nu$ represents the photon-nucleon vertex function, given by:

$$\Gamma^\mu = F_1(q^2)\gamma^\mu + \frac{i\sigma^{\mu\nu}q_\nu}{2M}F_2(q^2), \tag{10.1.6}$$

with $\sigma^{\mu\nu} \equiv \frac{i}{2}[\gamma^\mu, \gamma^\nu]$ the antisymmetric tensor formed from $\gamma^\mu, \gamma^\nu$. The form factors $F_1(q^2)$ (Dirac) and $F_2(q^2)$ (Pauli) can be thought of as matrix elements of the electromagnetic current operator between final and initial nucleon states. They are real-valued functions of $q^2$, which is the only independent Lorentz scalar variable on which the photon-nucleon vertex function $\Gamma^\mu$ can depend. The convention (10.1.6) for the $\gamma^* N$ vertex function is the most commonly used one in the literature, and is constructed such that the amplitude is real (assuming real-valued form factors)[101]. $F_1$ and $F_2$ represent the (electron) helicity-conserving and (electron) helicity-flip amplitudes, respectively. The nucleon's charge and Dirac ("non-anomalous") magnetic moment distributions determine the behavior of $F_1(q^2)$, while $F_2(q^2)$ measures the contribution of the "anomalous" magnetic moment distribution to the scattering.

Experimentally, the following linearly independent combinations of $F_1$ and $F_2$, known as the Sachs electric ($G_E$) and magnetic ($G_M$) form factors [2893], are more convenient:

$$G_E = F_1 - \tau F_2 \tag{10.1.7}$$

$$G_M = F_1 + F_2 \tag{10.1.8}$$

The differential cross section in OPE is given in terms of the Sachs form factors by [563, 2893–2895]

$$\frac{d\sigma}{d\Omega_e} = \left(\frac{d\sigma}{d\Omega_e}\right)_{\text{Mott}}\frac{\epsilon G_E^2 + \tau G_M^2}{\epsilon(1+\tau)}, \tag{10.1.9}$$

where $\tau$ and $\epsilon$ are kinematic parameters defined as

$$\tau \equiv \frac{Q^2}{4M^2} \tag{10.1.10}$$

$$\epsilon \equiv \left[1 + 2(1+\tau)\tan^2\left(\frac{\theta_e}{2}\right)\right]^{-1}. \tag{10.1.11}$$

[101] Since no other diagrams interfere with the OPE at the same order in $\alpha$, we are of course free to choose the phase of the OPE amplitude arbitrarily without affecting physical observables.

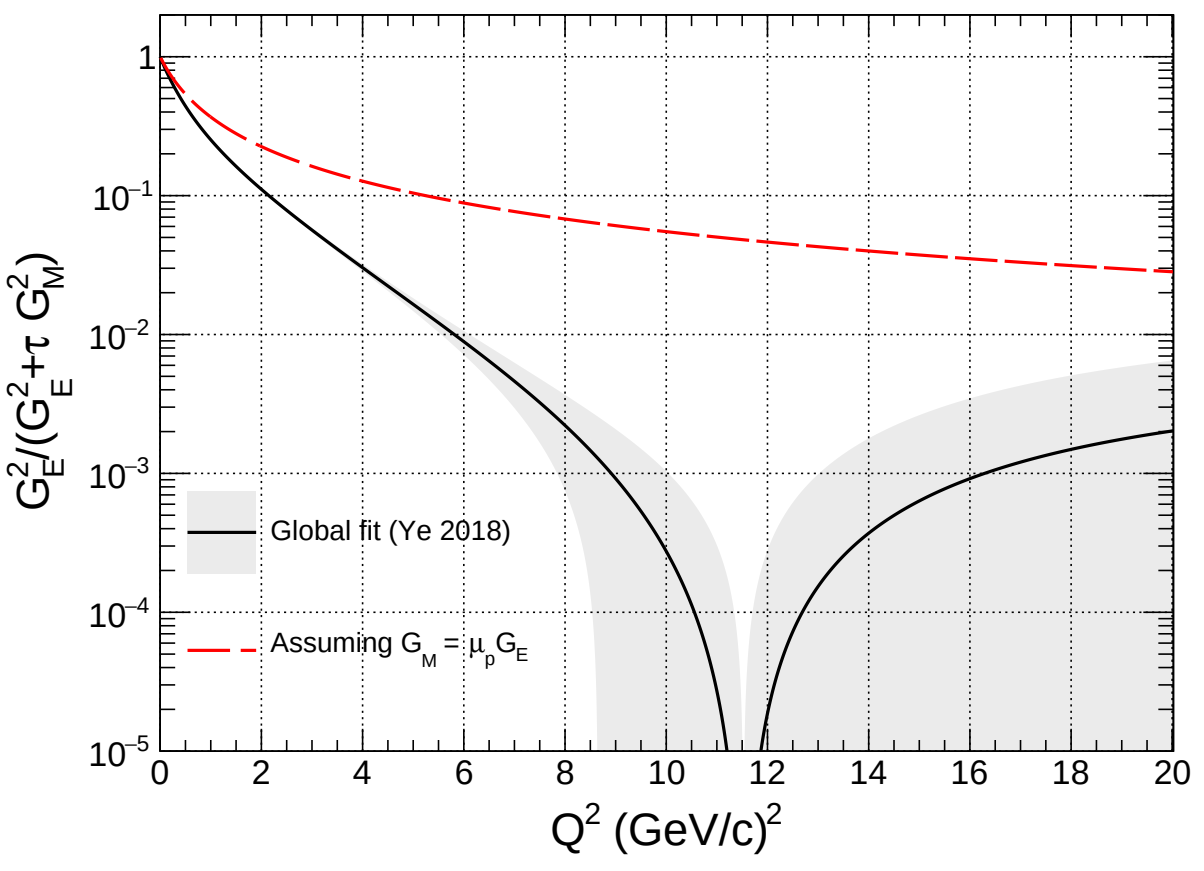

**Fig. 10.1.2** $Q^2$ dependence of the ratio $G_E^2/(G_E^2+\tau G_M^2)$ for the proton, representing the maximum fraction of the reduced cross section carried by the electric term (at $\epsilon = 1$). The central value and uncertainty band of the curve are calculated from the global fit of Ref. [1080]. The dashed line shows the ratio that would be obtained under the assumption of form factor scaling ($G_M^p = \mu_p G_E^p$).

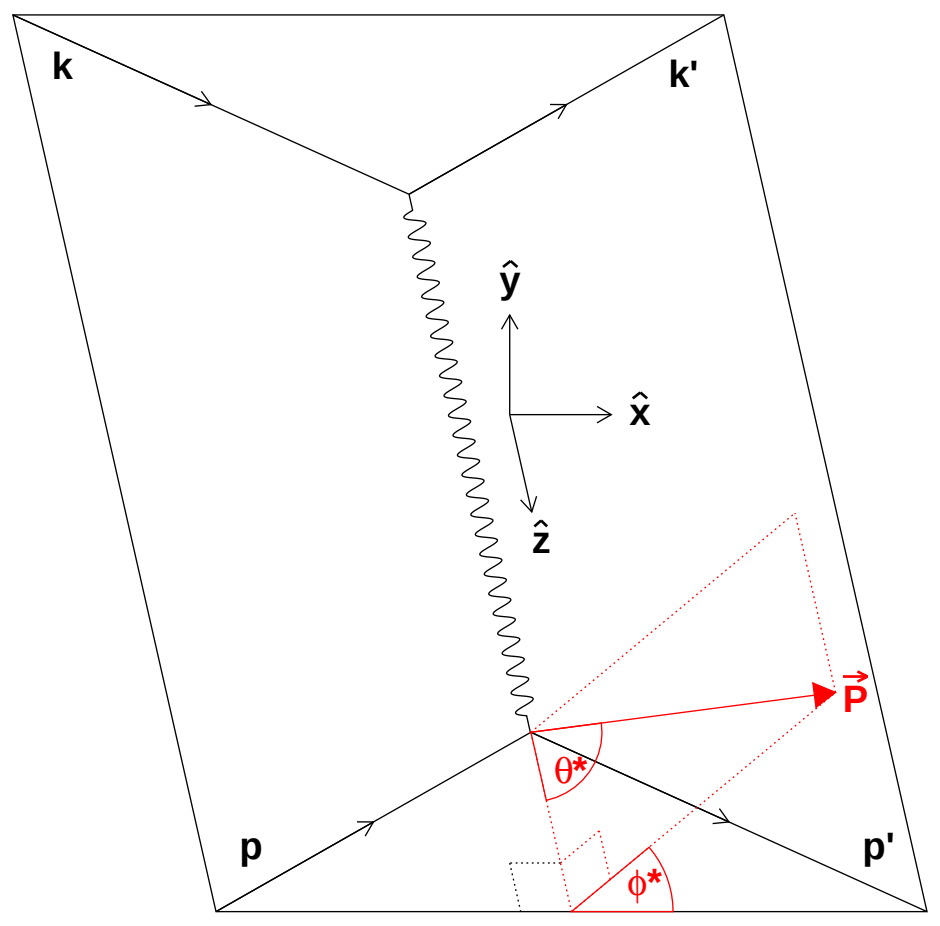

**Fig. 10.1.3** Standard coordinate system for nucleon polarization components in elastic $eN$ scattering. The arrow labeled $\vec{P}$ indicates the nucleon polarization direction and illustrates the definitions of the angles $\theta^*$ and $\phi^*$ between $\vec{P}$ and the momentum transfer $\mathbf{q}$. The $x$ or "$t$" (transverse) axis is parallel to the reaction plane but perpendicular to the momentum transfer. The $y$ or "$n$" (normal) axis is perpendicular to the reaction plane defined by $\hat{n} \equiv \hat{q} \times \hat{k}$. The $z$ or "$\ell$" (longitudinal) axis is along the momentum transfer direction, which coincides with the outgoing nucleon direction in the lab frame. The direction of the $x$ axis is chosen so that the Cartesian basis $(\hat{x}, \hat{y}, \hat{z})$ is right-handed.

In the OPE approximation, $\epsilon$ can be interpreted as the longitudinal polarization of the virtual photon [2895]. The electric and magnetic contributions to the scattering can be separated by measuring the cross section while varying the beam energy and the scattering angle in such a way as to hold $Q^2$ constant while varying $\epsilon$, a technique known as Longitudinal/Transverse (L/T) separation or Rosenbluth separation. The "reduced" cross section

$$\sigma_R \equiv \epsilon(1+\tau)\frac{(d\sigma/d\Omega_e)_{\text{Measured}}}{(d\sigma/d\Omega_e)_{\text{Mott}}},$$

is linear in $\epsilon$, with slope (intercept) equal to $G_E^2$ ($\tau G_M^2$).

In the limit of very small $Q^2$, corresponding to long-wavelength virtual photons, the cross section behaves as if the nucleon were a point particle of charge $ze$ ($z = +1(0)$ for proton (neutron)) and magnetic moment $\mu = (z+\kappa)$ (in units of the nuclear magneton), with $\kappa$ the anomalous magnetic moment. In this limit, the form factors thus become $G_E(0) = z$ and $G_M(0) = z+\kappa$. For small but finite $Q^2$ such that $\tau \ll \epsilon G_E^2/G_M^2$, the electric term dominates the cross section, and if target recoil is neglected, Eq. (10.1.9) takes the same form as Eq. (10.1.2), with $G_E \equiv F(\mathbf{q})$. Thus, in the low-energy limit, the electric form factor can be identified with the Fourier transform of the charge density. Similar reasoning leads to an interpretation of $G_M$ as a Fourier transform of the nucleon's magnetization density.

The Rosenbluth formula (10.1.9) describes *unpolarized* electron-nucleon scattering. At large values of $Q^2$, the magnetic term dominates the OPE cross section, and the sensitivity of the Rosenbluth method to $G_E$ vanishes (see Fig. 10.1.2). As the use of electron scattering to investigate nuclear structure expanded during the 1960s and 1970s, and as the technology to produce spin-polarized electron beams and nuclear targets was being developed and improved, several authors independently developed the theory of spin-polarized elastic $eN$ scattering in the OPE approximation and examined the implications for future measurements of polarization observables [2896–2899]. Nonzero asymmetries arise when the incident electron beam is longitudinally polarized and either the target nucleon is also polarized, or the recoil nucleon polarization is measured, or both. Asymmetries involving transverse electron beam polarization are generally suppressed by factors of $m_e/E_e$ relative to longitudinal asymmetries, and while such asymmetries have been measured and are interesting in their own right, they are not ideal observables for measuring electromagnetic form factors, and they will not be considered further in this section.

Figure 10.1.3 illustrates the "standard" coordinate system used in most of the literature on polarized elastic $eN$ scattering. In the case where the target nucleon is polarized, the asymmetry in the scattering cross section between positive and negative electron beam helicities

is given by

$$A_{eN} \equiv \frac{\sigma_+ - \sigma_-}{\sigma_+ + \sigma_-} \tag{10.1.12}$$

$$= P_{\rm beam} P_{\rm targ} \left[ A_t \sin\theta^* \cos\phi^* + A_\ell \cos\theta^* \right] , \tag{10.1.13}$$

where $P_{\rm beam}$ is the longitudinal electron beam polarization, $P_{\rm targ}$ is the magnitude of the target nucleon polarization, and the angles $\theta^*, \phi^*$ are defined in Fig. 10.1.3. The asymmetries $A_t$ and $A_\ell$ are given in terms of $\tau$, $\epsilon$, and the form factor ratio $r \equiv G_E/G_M$ by:

$$A_t = -\sqrt{\frac{2\epsilon(1-\epsilon)}{\tau}} \frac{r}{1 + \frac{\epsilon}{\tau} r^2}$$

$$A_\ell = -\frac{\sqrt{1-\epsilon^2}}{1 + \frac{\epsilon}{\tau} r^2} \tag{10.1.14}$$

Equations (10.1.14) show that the sensitivity of the double-spin asymmetry $A_{eN}$ to the form factor ratio is generally highest when the target is polarized perpendicular to the momentum transfer but parallel to the scattering plane; i.e., along the $x$ direction in Fig. 10.1.3. Note also that the asymmetries are sensitive to the *ratio* $G_E/G_M$, but not $G_E$ or $G_M$ separately. When the target is unpolarized, the longitudinally polarized electron transfers polarization to the outgoing nucleon. The non-vanishing components of the transferred polarization in OPE are

$$P_t = P_{\rm beam} A_t$$

$$P_\ell = -P_{\rm beam} A_\ell \tag{10.1.15}$$

Here $P_t$ and $P_\ell$ are the in-plane transverse and longitudinal components of the recoil nucleon's polarization, respectively. The sign change of $P_\ell$ relative to $A_\ell$ reflects the spin flip required to conserve angular momentum when the nucleon absorbs a transversely polarized virtual photon. The ratio $P_t/P_\ell$ is directly proportional to the form factor ratio $G_E/G_M$:

$$\frac{G_E}{G_M} = -\frac{P_t}{P_\ell}\sqrt{\frac{\tau(1+\epsilon)}{2\epsilon}} = -\frac{P_t}{P_\ell}\frac{E_e + E_e'}{2M}\tan\left(\frac{\theta_e}{2}\right) \tag{10.1.16}$$

Measurements of the differential cross sections, Eq. (10.1.9), and polarization observables, Eqs. (10.1.14) and (10.1.16), in elastic $eN$ scattering are the main source of knowledge of the nucleon's electromagnetic form factors, which are among the most important precision benchmarks for testing theoretical models of the nucleon. Moreover, precise knowledge of these form factors is required for the interpretation of many different experiments in nuclear and particle physics. In the next section, we summarize the existing data on nucleon form factors.

### 10.1.3 Experimental data

Figures 10.1.4, 10.1.5, 10.1.6, and 10.1.7 summarize the state of empirical knowledge of the proton electromagnetic form factors, as of this writing. The proton form factors $G_E^p$ and $G_M^p$ extracted from cross section measurements, as well as the neutron magnetic form factors $G_M^n$, can be described to within $\approx 10\%$ over most of the measured $Q^2$ range by $G_E^p \approx G_M^p/\mu_p \approx G_M^n/\mu_n \approx G_D$, where $G_D$ is the "dipole" form factor defined as

$$G_D = \left(1 + \frac{Q^2}{\Lambda^2}\right)^{-2} , \tag{10.1.17}$$

with the scale parameter $\Lambda^2 = 0.71\ ({\rm GeV/c})^2$ defining the so-called "standard dipole". The neutron electric form factor $G_E^n$ has a very different $Q^2$ dependence; since the neutron has zero *net* charge, $G_E^n(0) = 0$. Nevertheless, the neutron *rms* charge radius has been determined with good precision via neutron-electron scattering length measurements (see Ref. [278] and references therein). Existing measurements of $G_E^n$ in quasi-elastic electron scattering on bound neutrons in light nuclear targets, shown in Fig. 10.1.6, exhibit a rapid rise with $Q^2$ to an appreciable fraction of $G_D$ (nearly $\approx 50\%$ at the highest $Q^2$ for which we have reliable $G_E^n$ data). Precise high-$Q^2$ measurements of $G_E^p/G_M^p$ using the polarization transfer method revealed that $G_E^p$ starts falling much faster than $G_D$ above 1 $({\rm GeV/c})^2$, while $G_M^p/\mu_p$ falls to about 70% of $G_D$ at the highest measured $Q^2$ values. Reliable neutron form factor data only reach $Q^2 \approx 3.4(4.5)\ ({\rm GeV/c})^2$ for $G_E^n(G_M^n)$, but significant expansions in the $Q^2$ reach of the neutron data are anticipated in the near future.

The three-dimensional Fourier transform of $G_D$ gives an exponentially decreasing charge density as a function of the radial distance from the center of the nucleon, assuming a spherically symmetric density. The mean square radius of the nucleon charge density is related to the slope of the electric form factor in the limit $Q^2 \to 0$:

$$\langle r_E^2 \rangle = -6 \left.\frac{dG_E}{dQ^2}\right|_{Q^2=0} \tag{10.1.18}$$

For the standard dipole form factor, the implied charge radius is $\sqrt{\langle r_E^2 \rangle_D} = 0.81$ fm, which is in rough agreement with modern, precise determinations of the proton charge radius from electron scattering and the spectroscopy of electronic and muonic hydrogen. See Ref. [2905] for a very recent, in-depth review of the experimental and theoretical status of the proton charge radius.

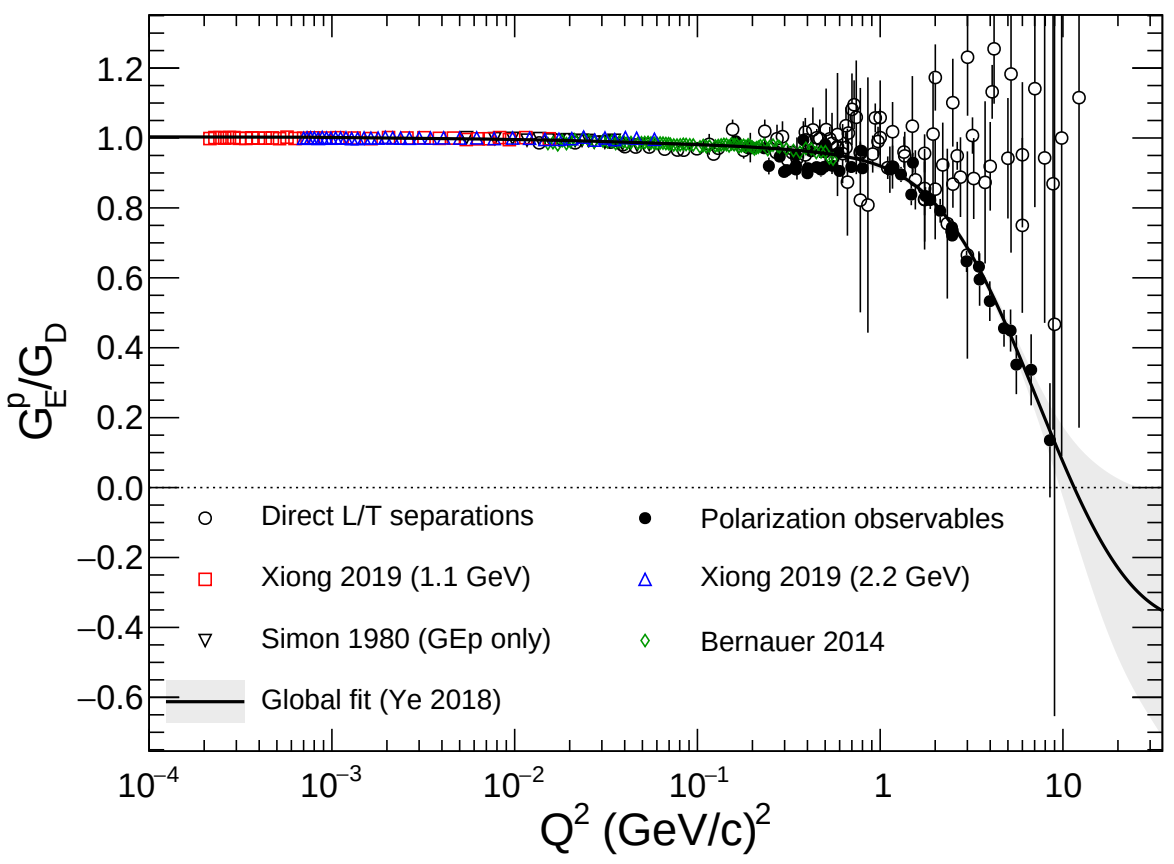

**Fig. 10.1.4** (approximate) World data for $G_E^p/G_D$. "Direct L/T separations" are published point extractions of $G_E^p$ from Rosenbluth plots. The points labeled "Bernauer 2014" are the direct Rosenbluth extractions from the Mainz A1 dataset [581, 2900]. The data labeled Xiong 2019 are from the PRad experiment [2901]. The global fit is from [1080]. See text for details.

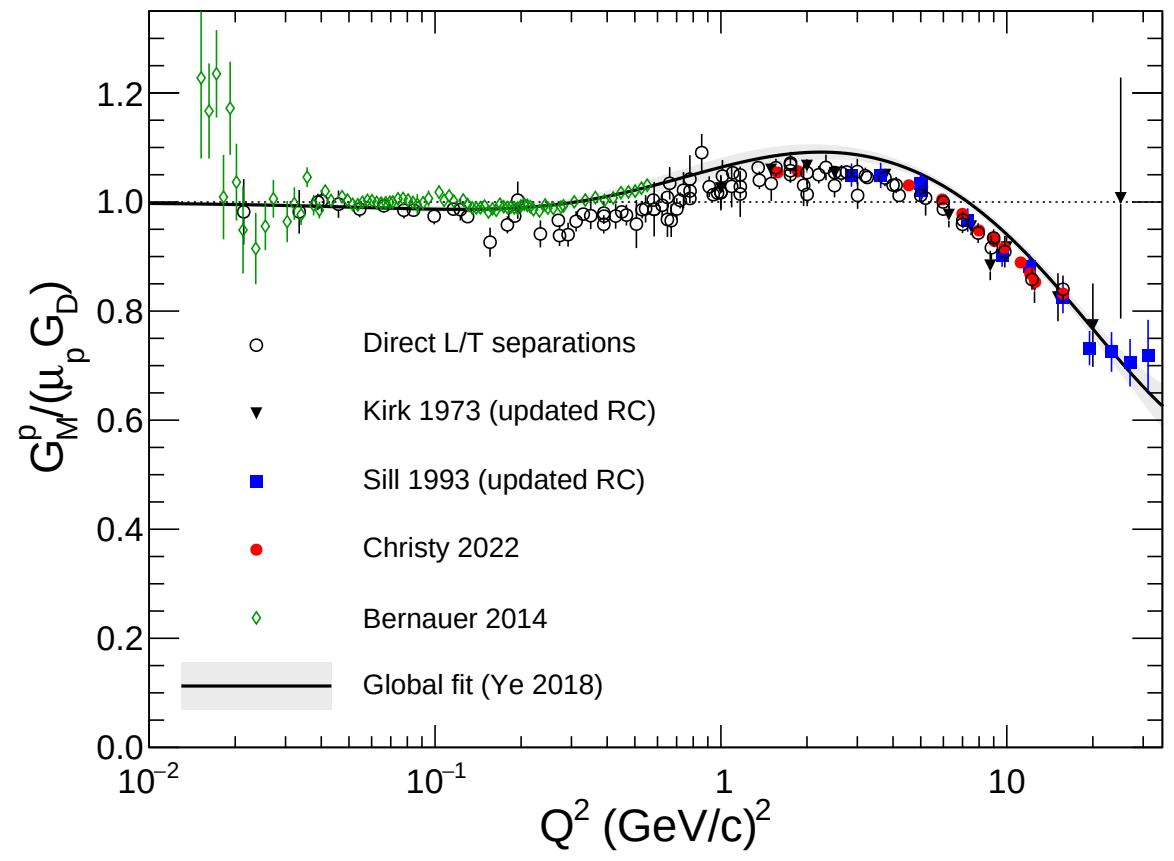

**Fig. 10.1.5** (approximate) World data for $G_M^p/(\mu_p G_D)$. The "Direct L/T separations" are published point extractions of $G_M^p$ from Rosenbluth plots. The Kirk 1973 data [2902] and the Sill 1993 data [2903] are point $G_M^p$ extractions from single cross section measurements, with updated radiative corrections as detailed in Ref. [2904]. The data labeled "Christy 2022" are the point $G_M^p$ extractions from the individual cross section measurements published in Ref. [2904]. The points labeled "Bernauer 2014" are the direct Rosenbluth extractions from the Mainz A1 dataset [581, 2900]. The global fit curve is that of Ref. [1080]. See text for details.

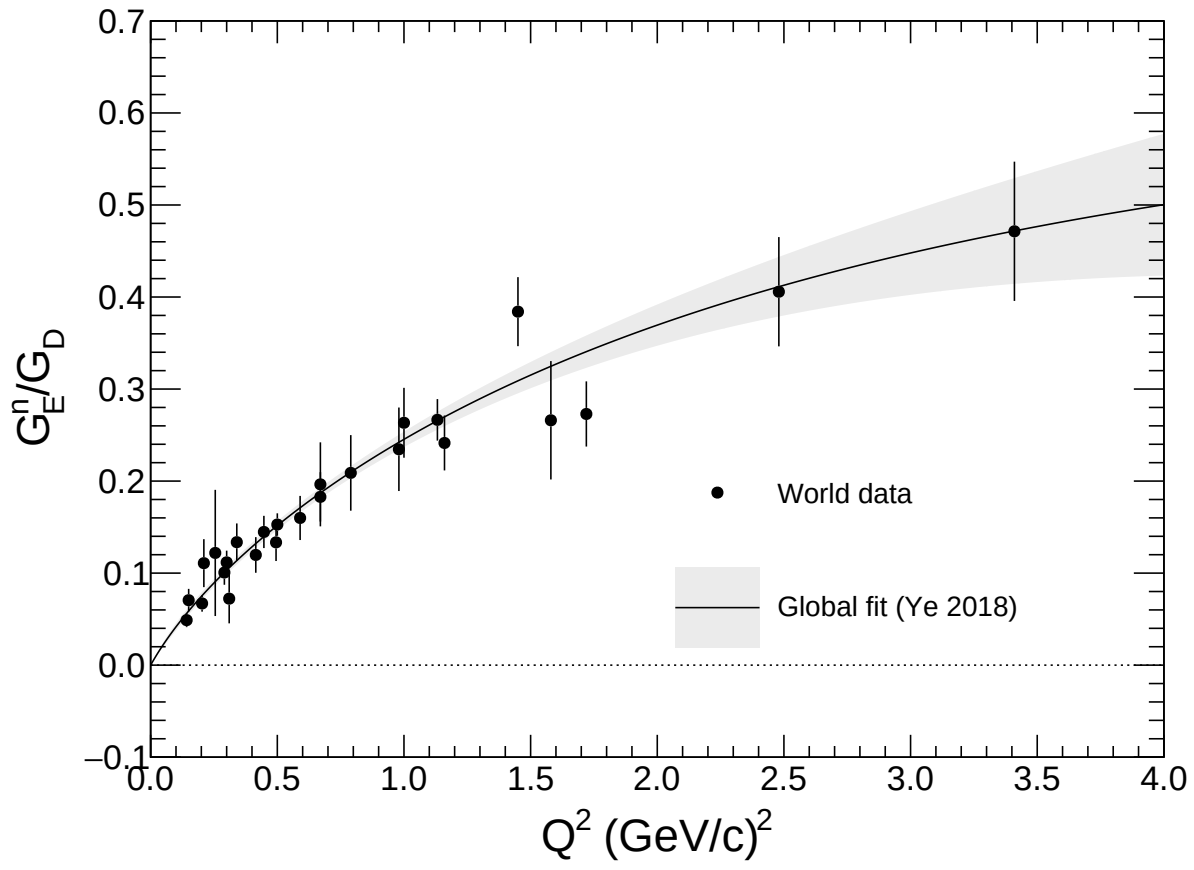

**Fig. 10.1.6** World data for neutron electric form factor $G_E^n/G_D$. See text for references, details.

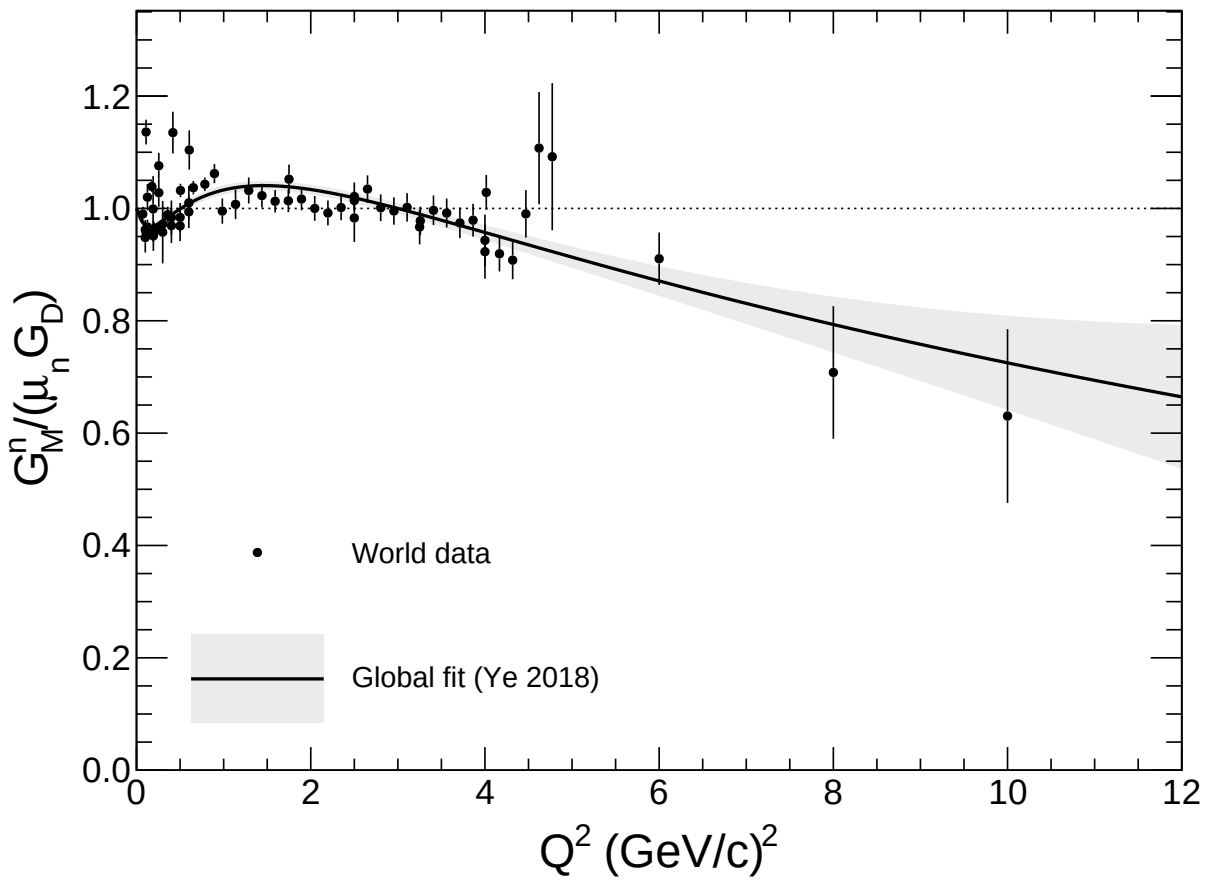

**Fig. 10.1.7** World data for neutron magnetic form factor $G_M^n/G_D$. See text for references, details.

*Proton data and discussion*

Figures 10.1.4 and 10.1.5 show most of the existing data for the proton electric and magnetic form factors $G_E^p$ and $G_M^p/\mu_p$, respectively, normalized to $G_D$, over the entire measured $Q^2$ range. While not comprehensive, the data shown are sufficiently representative of the $Q^2$ coverage and precision of the entire world data. The points shown as empty circles in Figs. 10.1.4 and 10.1.5 are published point extractions of $G_E^p$ and $G_M^p$ based on direct L/T separations from Rosenbluth plots, and are taken from Refs. [2904, 2906–2914]. These extractions are not entirely independent of each other in terms of cross section input, as several of the analyses combined data from multiple experiments at similar $Q^2$ values.

The points shown as filled circles in Fig. 10.1.4 are based on direct measurements of the ratio $G_E^p/G_M^p$ using polarization observables, converted to $G_E^p/G_D$ values using the global fit to $G_M^p$ from Ref. [1080]. The polarization data for $G_E^p$ include measurements based on the polarization transfer technique of Eq. (10.1.16) (Refs. [2915–2928]), and the beam-target double-spin asymmetry method, Eq. (10.1.14) [2929–2931]. The points labeled "Bernauer 2014" in Figures 10.1.4 and 10.1.5 are the direct Rosenbluth separations from the Mainz A1 dataset [581, 2900]. The points at very low $Q^2$ labeled "Simon 1980" and "Xiong 2019" in Fig. 10.1.4 are direct extractions of $G_E^p$ from individual cross section measurements based on the assumption of form factor scaling ($G_M^p = \mu_p G_E^p$) in the case of Ref. [2932], or using the Kelly fit to $G_M^p$( [574]) in the case of Ref. [2901]. In Fig. 10.1.5, the $G_M^p$ values extracted from the cross sections published in Refs. [2902, 2903] are based on the updated analysis in Ref. [2904], which used the "state-of-the-art" radiative corrections described in Ref. [2933]. It must also be noted that the global fits shown in Figs. 10.1.4,10.1.5 include phenomenological two-photon-exchange corrections that have *not* been applied to the published form factor extractions. These corrections tend to increase the value of $G_M^p$ by roughly 2-3% in the $Q^2$ range where the discrepancy between Rosenbluth and polarization results is largest.

The extraction of nucleon form factors from cross section measurements generally requires corrections to account for the effects of higher-order QED radiative processes in order to isolate the OPE term from which $G_E^2$ and/or $G_M^2$ can be determined. While each of these higher-order terms is at least $\mathcal{O}(\alpha)$ relative to the Born term, their combined effect on the observed cross sections can be significant; typically as much as 10-30% at modest-to-large $Q^2$ [2934]. As a general rule, the magnitude of the radiative correction (RC) to the elastic cross section tends to increase at large $Q^2$ values and/or large $\theta_e$/small $\epsilon$, and also depends on experiment-specific parameters including detector acceptance and resolution, electron beam properties, and target geometry, material, and density. Additionally, the calculation of the RC depends strongly on whether the experiment detects the scattered electron only (most common), the recoil proton only (see, e.g., Ref. [2913]), or both final-state particles. For many experiments, the RC calculation is an important source of uncertainty in the extraction of the Born cross section, which is not directly observable, and can dramatically change the slope of the Rosenbluth plot in converting measured cross sections to Born cross sections [2934].

At next-to-leading order in $\alpha$, the "standard", model-independent RC to $ep \to ep$ scattering include vacuum polarization, vertex, and self-energy terms that are purely virtual and depend only on $Q^2$, Bremsstrahlung (real photon emission), which depends strongly on both $Q^2$ and $\epsilon$ and modifies the reaction kinematics, and two-photon-exchange (TPE), in the limit where one of the two exchanged photons is "soft". The contribution of "hard" TPE, in which both exchanged photons carry a "large" momentum, cannot presently be calculated model-independently, and is neglected in the standard radiative correction procedures. It is thought to be largely responsible for the discrepancy between cross sections and polarization observables [2935] in high-$Q^2$ extractions of $G_E^p$, and is presently the subject of vigorous worldwide experimental and theoretical investigation. For a recent review of the subject, see Ref. [2936].

For conventional RC, most of the earlier published extractions of the proton form factors relied on the work of Tsai [2937] or Mo and Tsai [2938]. Following the initial discovery of the rapid fall-off of $G_E^p/G_M^p$ at large $Q^2$ using polarization transfer [2916], and the resulting large discrepancy between two different observables sensitive (in principle) to the same fundamental property of the proton, Maximon and Tjon [2939] refined the mathematical treatment of these corrections and removed many of the approximations made in the expressions of Mo and Tsai, including an exact calculation of the soft Bremsstrahlung contributions. Several authors [2904, 2933, 2940] have recently examined the quantitative differences between the calculations of Ref. [2938] and the more accurate approach of Ref. [2939], and studied the impact of these differences on previously published extractions of the form factors. Updating the published cross sections to use the more modern RC prescriptions is a non-trivial undertaking, especially for the older experiments, since the required modifications depend on details of the experiments and the associated data analyses that in some cases were not thoroughly documented in the final publications.

The most recent and comprehensive effort thus far to update published elastic *ep* cross sections to use "state-of-the-art" RC in the high-$Q^2$ region was described in Ref. [2904]. The reanalysis focused on a subset of high-$Q^2$ experiments from Jefferson Lab and SLAC for which the original publications provided sufficient details of the experimental parameters and the RC prescriptions and cutoffs used that they could be corrected in a self-consistent way [2904]. As noted by the authors of [2904] and earlier by [2933], the effect of updating the RC to the older SLAC data is to reduce, but not eliminate, the magnitude and significance of the discrepancy in the high-$Q^2$ region. The new, precise cross sections from Jefferson Lab's Hall A [2904] extend the $Q^2$ range for which a statistically significant discrepancy between cross sections and polarization observables is established.

In the polarization transfer method, the simultaneous measurement of the recoil nucleon polarization components $P_t$ and $P_\ell$ offers many advantages in the control of experimental uncertainties. In particular, the form factor ratio can be determined in a single measurement, eliminating uncertainties resulting from changes in experimental parameters such as the beam energy, detector angles, spectrometer magnetic field settings, target polarization and spin direction, and others. Moreover, the beam polarization and many other sources of systematic uncertainty associated with recoil nucleon polarimetry cancel in the ratio $P_t/P_\ell$, see, e.g., Ref. [2925], and reversal of the electron beam helicity reverses the direction of the recoil nucleon polarization while leaving all other experimental parameters unchanged, providing for robust cancellation of systematic effects associated with polarimeter acceptance and/or detection efficiency [2941]. The dramatically different behavior of $G_E^p$ implied by the polarization data has profound implications for theoretical modeling of nucleon structure, as discussed below.

While polarization measurements of $G_E/G_M$ are generally thought to have small systematic uncertainties, it must be noted that the published data exhibit significant internal tension in the region 0.1-1 GeV$^2$ where several high-precision experiments give somewhat conflicting results [2918, 2926–2928, 2930]. Despite this unresolved tension, polarization observables are generally regarded as giving the most reliable determination of $G_E^p$ at large $Q^2$ values, due to their superior sensitivity to $G_E$ as compared to the Rosenbluth method, and their relative insensitivity to radiative corrections [2942–2944] and higher-order QED corrections neglected by the standard RC procedures, such as hard Two-Photon-Exchange (TPE) [2924, 2925]. This property derives from the fact that polarization asymmetries are ratios of polarized and unpolarized cross sections, that tend to be affected similarly by radiative processes. The $P_t/P_\ell$ ratio in the polarization transfer method is a ratio of such ratios, and the model-independent RC to this ratio tend to be utterly negligible compared to the uncertainties in the presently measured range of $Q^2$ [2925]. Moreover, a precise search for evidence of hard TPE contributions in this observable found no significant effect [2924] at 2.5 GeV$^2$, with the ratio $\mu_p G_E^p/G_M^p$ showing no variation with $\epsilon$ in the range 0.15-0.8 with $\approx 1\%$ total uncertainties.

Assuming that polarization measurements give the "true" value of $G_E^p$, the fractional contribution of the $\epsilon G_E^2$ term to the OPE cross section falls rapidly with $Q^2$, as shown in Fig. 10.1.2. Based on the global fit of Ref. [1080], the electric term contributes at most 10% of $\sigma_R$ at 2 (GeV/c)$^2$, 2% at 5 (GeV/c)$^2$, and even less at higher $Q^2$, basically wiping out any meaningful sensitivity to $G_E$, since its contribution to $\sigma_R$ becomes comparable to the limits of experimental accuracy and to the expected magnitude of higher-order QED corrections that are theoretically and experimentally uncertain.

In addition to efforts to resolve the difficulties with $G_E^p$ at large $Q^2$, there has been a renewed effort to improve the precision of elastic $ep$ scattering data at very low $Q^2$, since the CREMA collaboration first published an extremely precise extraction of the proton radius from Lamb shift measurements in muonic hydrogen [2945], yielding a radius of about 0.84 fm, smaller by roughly seven standard deviations than the previous consensus value (at the time) of 0.88 fm from electron-proton scattering and spectroscopy of ordinary hydrogen. The Mainz A1 collaboration [581, 2900] carried out a systematic program of over 1,400 precision cross section measurements spanning the $Q^2$ range 0.003-1 GeV$^2$ using the "traditional" method based on magnetic spectrometers. They published several direct fits of $G_E^p$ and $G_M^p$ to their cross section data, testing various functional forms to accurately quantify the uncertainties. They also published direct L/T separations for $Q^2 \gtrsim 0.02$ GeV$^2$. While the Mainz $G_E^p$ extraction is in good agreement with the rest of the world data, their $G_M^p$ results, whether from global fits or direct L/T separations, are in significant tension with the other world data[102], as is evident from Fig. 10.1.5. The slower fall-off with $Q^2$ of the Mainz $G_M^p$ implies a smaller magnetic radius; indeed, the published Mainz extraction of the proton magnetic radius $r_M^p$ is about three standard deviations below the consensus of extractions based on other world data.

More recently, the PRad collaboration [2901] performed new $ep \to ep$ cross section measurements using a novel, magnetic-spectrometer-free method involving precision calorimetry, a windowless gas target, and a simultaneous measurement of the pure electroweak process of Möller scattering ($e^-e^- \to e^-e^-$) to constrain the absolute cross section normalization. Their measurements reached a minimum $Q^2$ of 0.0002 GeV$^2$ with small statistical and systematic uncertainties, achieving a proton radius measurement of $r_p \approx 0.83 \pm 0.01$ fm, consistent with the muonic hydrogen value.

Figure 10.1.8 shows the PRad and Mainz cross section data, normalized to the "standard dipole" cross section, calculated from Eq. (10.1.9) under the assumption $G_E^p = G_M^p/\mu_p = G_D$. The low end of the PRad $Q^2$ range is in a regime where the cross section is indistinguishable from point-like behavior within experimental precision; at the lowest $Q^2$ of the PRad dataset, $G_D^2 \approx 0.999$. This is unsurprising given that the de Broglie wavelength of the virtual photon $\lambda = \hbar c/\sqrt{Q^2} \approx 13$ fm is large compared to $r_p$ at this $Q^2$.

*Neutron data and discussion*

The neutron electromagnetic form factors are much more difficult to measure accurately than those of the proton, due primarily to the absence of free neutron targets of sufficient density for electron scattering experiments at large $Q^2$. The small cross sections for high-energy

[102] Note, however, that the Mainz dataset implies a $G_E^p/G_M^p$ ratio that is consistent with the high-precision polarization measurements by Zhan *et al.*, Ref. [2927]

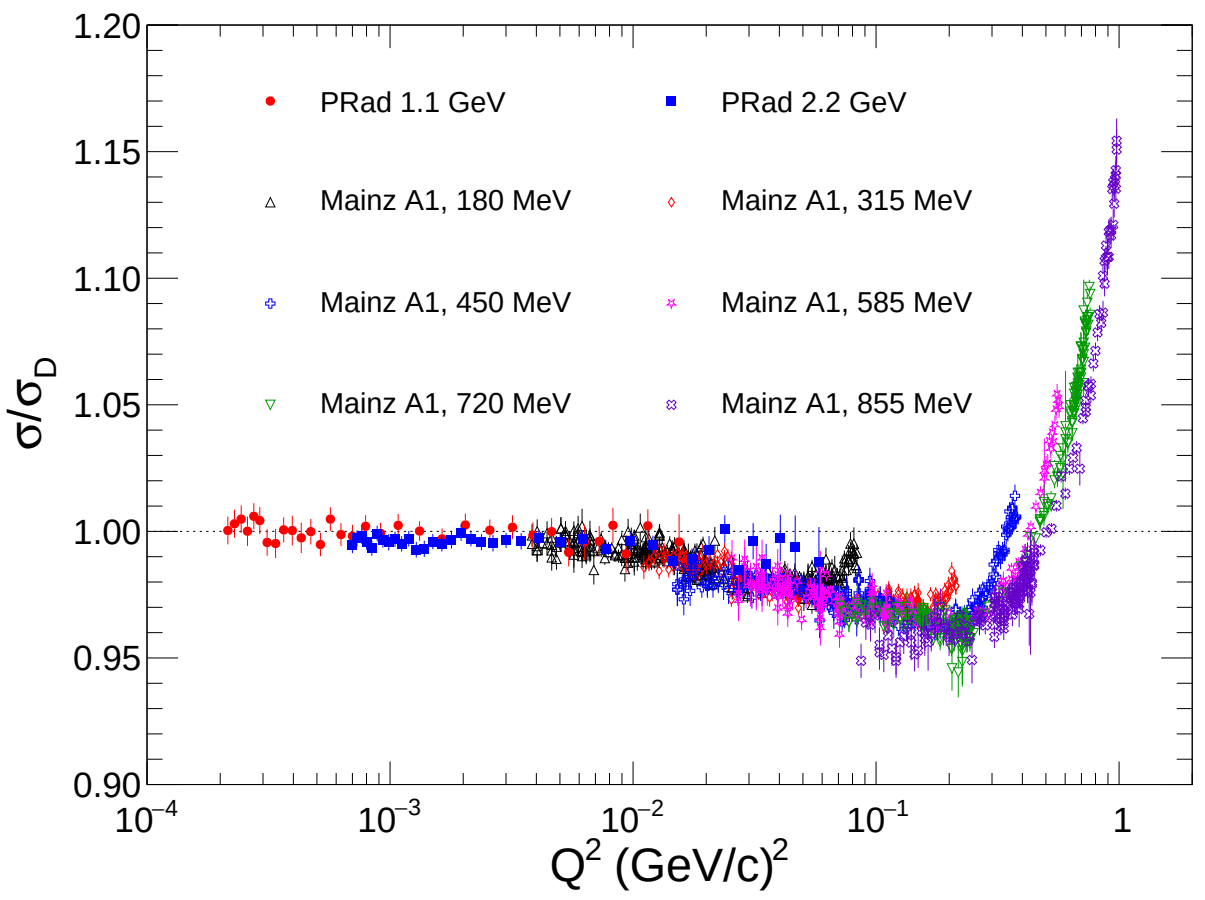

**Fig. 10.1.8** Comparison of PRad [2901] and Mainz A1 [581] elastic $ep \to ep$ cross sections, normalized to the "standard dipole" cross section $\sigma_D$, calculated from Eq. (10.1.9) assuming $G_E^p = G_M^p/\mu_p = G_D$; i.e., $\epsilon(1+\tau)\sigma_D/\sigma_{\rm Mott} = G_D^2\left(\epsilon + \mu_p^2\tau\right)$

electromagnetic interactions can generally only be measured accurately in high-luminosity experiments, and the neutron's instability and zero charge severely limit the number of free neutrons that can be collected in a suitably small volume for a suitable duration for such experiments. As such, essentially all knowledge of neutron electromagnetic form factors at meaningfully large $Q^2$ values comes from measurements of electron scattering on bound neutrons in light nuclear targets such as deuterium and $^3$He.

Since $G_E^n(0) = 0$, the cross section for elastic $en$ scattering is dominated by the magnetic term over essentially the entire measured $Q^2$ range, even at relatively low $Q^2$. The neutron form factors are accessible experimentally through a number of scattering observables on light nuclear targets, including cross sections and spin asymmetries. Model-dependent extractions of neutron elastic form factors from measurements of elastic electron-deuteron scattering have also been attempted at relatively low $Q^2$ values (see, e.g., [2946–2949]), but are subject to large theoretical and experimental systematic uncertainties, and are generally considered less reliable than extractions from measurements of quasi-elastic scattering on bound nucleons in deuterium and/or Helium-3, although they are qualitatively consistent.

Figures 10.1.6 and 10.1.7 show most of the existing data for $G_E^n$ and $G_M^n$, respectively, excluding extractions based on elastic $ed$ cross section measurements. For $G_E^n$, essentially all reliable data of reasonable precision come from measurements of polarization observables, since the (quasi)-elastic $(e, e'n)$ cross section has relatively low sensitivity to $G_E^n$ over the entire $Q^2$ range. The data shown in Fig. 10.1.6 include extractions from asymmetry measurements on polarized deuterium targets (Refs. [2950–2953]), polarized $^3$He targets (Refs. [2954–2958]), and via recoil neutron polarization on unpolarized deuterium (Refs. [2959–2961]).

The most reliable known method to determine the neutron magnetic form factor $G_M^n$ is the so-called "ratio" or "Durand" technique [2962], in which "neutron-tagged" and "proton-tagged" quasi-elastic electron scattering on a deuterium target are measured simultaneously, and the ratio of cross sections $^2{\rm H}(e,e'n)p/^2{\rm H}(e,e'p)n$ is measured. The simultaneous measurement of quasi-free scattering on bound protons and neutrons in deuterium, combined with the precise knowledge of the free proton cross section, allows a determination of the free neutron cross section with very small uncertainties. In particular, the electron acceptance and detection efficiency, the data acquisition deadtime, and the luminosity cancel exactly in the $n/p$ ratio, and nuclear effects such as Fermi motion and binding, final-state interactions, meson-exchange currents, and others, as well as QED radiative corrections, tend to affect the $d(e,e'n)p$ and $d(e,e'p)n$ cross sections nearly identically [2963], for sufficiently tight cuts on the photon-nucleon invariant mass $W^2$, and the angle $\theta_{pq}$ between the detected nucleon's momentum and the momentum transfer direction, determined from the scattered electron's kinematics, to ensure exclusivity of the reaction. The main source of experimental uncertainty with the ratio method is in the knowledge of the acceptance/detection efficiency for protons and neutrons. Of the data shown in Fig. 10.1.7, Refs. [2963–2967] used the ratio method, Refs. [2968–2971] extracted $G_M^n$ from the beam-target double-spin asymmetry in inclusive quasielastic electron scattering on polarized Helium-3, and Refs. [2972–2974] extracted $G_M^n$ from absolute cross section measurements in either inclusive scattering on deuterium or coincidence $d(e,e'n)p$ scattering. The low-$Q^2$ data for $G_M^n$ show some inconsistencies, suggesting underestimated theoretical or experimental systematic uncertainties in some of the older measurements. The Super BigBite Spectrometer (SBS) Collaboration in Jefferson Lab's Hall A recently collected data using the ratio method to extend the knowledge of $G_M^n$ to $Q^2 = 13.5$ GeV$^2$ with very small statistical and systematic uncertainties. The CEBAF Large Acceptance Spectrometer (CLAS) collaboration in Jefferson Lab's Hall B has also collected data for $G_M^n$ up to $Q^2 \approx 10$ GeV$^2$, with qualitatively different sources of systematic uncertainty. Both datasets are currently under analysis.

### 10.1.4 Theoretical interpretation of nucleon form factors

As the spacelike electromagnetic form factors are among the simplest, most clearly interpretable, and best-known measurable dynamical properties of the nucleon, they constitute important benchmarks for testing theoretical models. Figure 10.1.9 shows the world data for the nucleon's spacelike EMFFs together with selected theoretical models and the expected results from the ongoing high-$Q^2$ form factor program in Hall A at Jefferson Lab by the Super BigBite Spectrometer (SBS) collaboration. The SBS measurements of the neutron magnetic form factor were completed during the Oct. 2021-Feb. 2022 running period in Hall A, and the data are currently under analysis. The SBS measurement of $G_E^n/G_M^n$ using a polarized $^3$He target is underway as of October 2022 and will run through March of 2023, and the polarization transfer measurements of $G_E^n/G_M^n$ and $G_E^p/G_M^p$ are expected to take data in 2023-2024. The expansion of the $Q^2$ range and precision of the proton and neutron data will severely test theoretical models of nucleon structure.

The calculation of nucleon structure from first principles in QCD is presently only possible using the methods of lattice gauge theory. The accuracy of lattice QCD calculations is rapidly improving with increases in computing power and improvements in the control of systematic errors, and the range of measurable quantities lattice QCD can predict continues to expand. Nevertheless, the prediction of nucleon form factors and other observables of hard exclusive processes from lattice QCD (see Refs. [573, 575, 2979, 2980] and references therein for recent efforts at low and high $Q^2$) has not yet reached a level of precision and accuracy commensurate with that of the experimental data, particularly at high energies. As such, its predictions cannot yet be conclusively "tested" by the form factor data. Instead, the existing data serve to guide the improvement of the calculations. Meanwhile, the continued use of QCD-inspired phenomenological models, approximations, effective theories, and continuum methods provides valuable insight and improved understanding of the relevant degrees of freedom and dynamical effects at different energy scales when compared to the data.

For asymptotically large $Q^2$ values, perturbative QCD (pQCD) predicts the scaling behavior of the nucleon form factors based on simple constituent counting rules and helicity conservation [207]. The predictions for the nucleon, with its three-quark valence structure, are $F_1 \propto Q^{-4}$ and $F_2 \propto Q^{-6}$ (see the discussion in Sec. 5.10). While the proton data at the highest measured $Q^2$ values are in superficial qualitative agreement with the pQCD scaling predictions, it has been argued [2981, 2982] that the pQCD mechanism of multiple hard gluon exchange is not applicable to exclusive processes in the presently accessible range of $Q^2$. More recently, Belitsky *et al.* considered the effects of both leading and subleading twist contributions to the nucleon's light-cone wavefunctions in a pQCD analysis of the Pauli form factor $F_2$, deriving the modified logarithmic scaling expression $Q^2F_2/F_1 \propto \ln^2\left(Q^2/\Lambda^2\right)$, with a range of values of $\Lambda \approx 200-300$ MeV describing the proton data rather well [2983]. However, in an analysis of the quark flavor decomposition of the spacelike FFs [2984] shortly following the publication of data for $G_E^n/G_M^n$ up to 3.4 (GeV/c)$^2$ [2954] and $G_M^n$ up to 4.8 (GeV/c)$^2$ [2963], it was noted that the neutron $F_2^n/F_1^n$ data do not follow this logarithmic scaling, at least not for values of $\Lambda$ similar to those fitting the proton data.

Dispersion theoretical analysis, including models based on the assumption of Vector Meson Dominance (VMD) [2985], provide a coherent, self-consistent framework for the joint interpretation of spacelike and timelike nucleon form factors over the entire physical range of $Q^2$. VMD-based models were among the earliest to describe the global features of the nucleon form factors and predicted the high-$Q^2$ falloff of $G_E^p/G_M^p$ decades before the polarization transfer experiments. A key assumption of VMD and VMD-based models is that the virtual photon-nucleon interaction at low to intermediate $Q^2$ is dominated by vector meson pole terms, which contribute significantly to the dispersion integrals connecting the spacelike and timelike regions through the requirements of unitarity and analyticity of the form factors considered as functions of $q^2$ in the complex plane. For a recent review of the dispersion theoretical analysis of nucleon EMFFs, see Ref. [2986].

As mentioned above, in the very low-energy limit, when target recoil can be neglected, the form factors can be interpreted as three-dimensional Fourier transforms of the spatial distributions of charge ($G_E$) and current ($G_M$) in the nucleon. While this naive density interpretation is invalidated by relativity for finite momentum transfers, several authors have extracted three-dimensional rest-frame densities from the form factors using model-dependent relativistic prescriptions to relate the Sachs form factors measured at a four-momentum transfer $Q^2$ to the static rest frame densities. A common feature of such extractions is the identification of the Sachs form factors $G_E$ and $G_M$ with Fourier transforms of the Breit frame[103] charge and current densities. The Breit frame densities are then modified by a boost factor $k^2 = Q^2/(1+\tau)$ relating $Q^2$ to the wave number $k$ in the nucleon rest frame, and another model-dependent factor relating the Sachs FF to the so-called "intrinsic" form factors $\tilde{\rho}(k)$, defined as Fourier-Bessel transforms of the rest frame densities. The latter correction attempts to account for the $Q^2$-dependent boost of the nucleon wavefunction itself from the rest frame to the Breit frame. Kelly [2987] used expansions in a complete set of radial basis functions and a relativistic boost prescription consistent with the pQCD asymptotic behavior to minimize model-dependence and estimate the uncertainties in the radial densities due to the finite $Q^2$ range of the data. Among his key findings were a broader charge density for the proton compared to its magnetization density, consistent with the fall-off of the polarization data for $G_E^p/G_M^p$, and a neutron charge density described by a positive core surrounded by a negative exterior, consistent with pion-cloud models.

[103] The Breit or "brick-wall" frame in elastic $eN$ scattering is the frame in which there is no energy transfer in the collision. It is related to the nucleon rest frame by a boost along the momentum transfer direction, with a boost factor $\gamma = \sqrt{1+\tau}$.

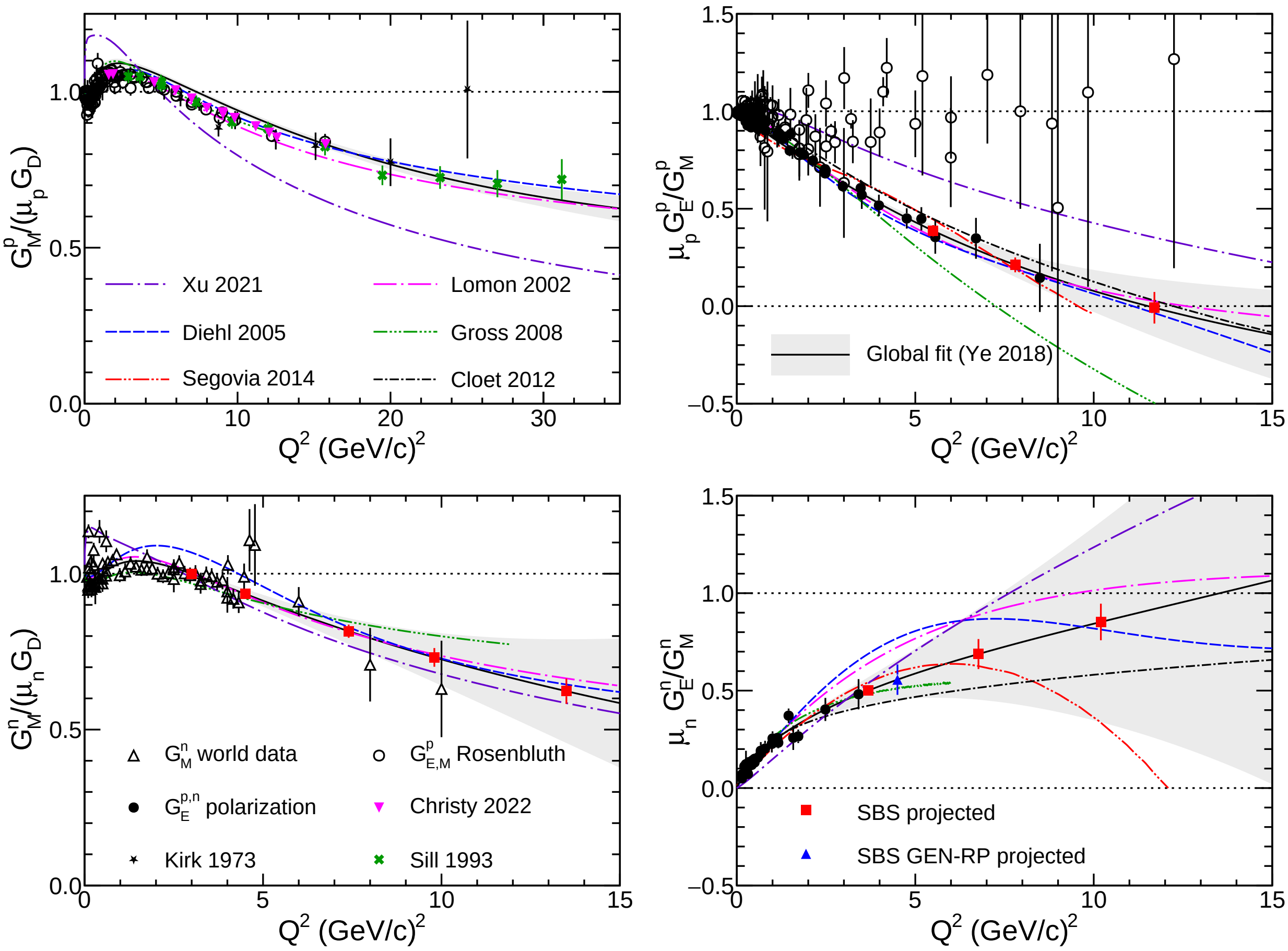

**Fig. 10.1.9** Data for all four nucleon electromagnetic form factors at spacelike $Q^2$ with selected theoretical models, the global fit from [1080], and the projected results from the ongoing SBS program at Jefferson Lab, plotted arbitrarily at the values from the global fit. Theoretical curves shown are the BLFQ calculations of Ref. [949] (Xu 2021), the VMD-based model of Ref. [2975] (Lomon 2002), the GPD-based model of Ref. [2976] (Diehl 2005), the covariant spectator model of Ref. [2977] (Gross 2008), the DSE-based calculation of Ref. [2771] (Segovia 2014), and the quark-diquark model of Ref. [2978] (Cloet 2012). Data references are the same as those given in the text and described in the captions of Figs. 10.1.4, 10.1.5, 10.1.6, and 10.1.7. See text for details.

While the three-dimensional radial densities extracted from the form factors are necessarily model-dependent, a model-independent density interpretation of the form factors exists through sum rules relating the form factors to moments of Generalized Parton Distributions (GPDs) [1085]. Miller [2988, 2989] showed that in the infinite momentum frame, the impact-parameter-space densities of charge and magnetization in the nucleon are two-dimensional Fourier-Bessel transforms of the Dirac ($F_1$) and Pauli ($F_2$) form factors, respectively. Examples of empirical extractions of the transverse densities from the form factor data can be found in Refs. [2810, 2990]. In apparent contrast to model-dependent extractions of 3D rest frame densities such as Kelly's, the neutron's transverse charge density exhibits a negative core surrounded by a positive exterior, contradicting the qualitative predictions of several models.

The form factors also play an important role in efforts to extract the GPDs from measurements of Deeply Virtual Compton Scattering (DVCS) and other hard exclusive processes. Through the aforementioned sum rules, the form factors $F_1$ and $F_2$ impose fairly powerful constraints on, respectively, the vector ($H$) and tensor ($E$) GPDs that enter the Ji sum rule for the nucleon spin decomposition [1085]. If good measurements and/or models of the GPDs exist, they can be used to predict the form factors [2991]. Alternatively, when combined with the forward parton distributions measured in deep inelastic scattering, the form factors can be used to constrain the GPDs [2976, 2992], particularly at high Bjorken $x$ and/or large $-t$. Apart from the direct constraints, precise knowledge of the form factors is also required for analysis of experiments attempting to measure GPDs, to separate the contributions of the DVCS and Bethe-Heitler processes, which interfere at the same order in $\alpha$ and are experimentally indistinguishable.

Constituent quark models (CQMs) have a long history in nuclear physics and predate the emergence of QCD as the accepted theory of strong interactions within the Standard Model. For a review and modern perspective on the role of the quark model in nuclear physics, see [476]. The early non-relativistic constituent quark model was successful in explaining the observed spectra of baryons and mesons as $qqq$ (fermionic) and $q\bar{q}$ (bosonic) bound states, and making qualitative predictions of meson and baryon masses and magnetic moments. Indeed, one of the original motivations for the introduction of the color quantum number prior to the development of QCD was to preserve the Pauli exclusion principle for low-lying baryon states, whose combined spin/flavor/orbital quantum numbers are symmetric under the exchange of any two quarks. This issue was particularly acute for the spin-3/2 baryon decuplet. To explain dynamical properties of hadrons in terms of constituent quarks, a model for the confining quark-quark interaction and the resulting quark wavefunctions is needed. The "bare" $u$ and $d$ valence quark constituents of nucleons appearing in the QCD Lagrangian are almost massless compared to the nucleon mass. As such, the nucleon, considered as the ground state of a bound system of three light quarks, is characterized by a large ratio of binding energy to constituent mass, making a fully relativistic treatment mandatory to obtain realistic phenomenology and accurate descriptions of the data. A common feature of CQM calculations of nucleon structure is the "dressing" of the bare, almost-massless valence quarks by gluons and quark-antiquark pairs, leading to massive constituent quarks and/or diquarks as effective degrees of freedom, often carrying their own internal structure. While a full review of relativistic constituent quark model calculations of nucleon form factors is beyond the scope of this section, a fairly comprehensive overview is given in Ref. [2993] (see also Sec. 5).

In recent years, Hamiltonian light-front field theory has emerged as a useful framework for the nonperturbative solution of invariant masses and correlated parton amplitudes of self-bound systems [929]. Xu *et al.* recently applied this framework to calculate the structure of the nucleon using the method of Basis Light Front Quantization (BLFQ) [949]; see also Sec. 5.4. Their calculation used an effective light-front Hamiltonian with quarks as the only effective degrees of freedom, a transverse confining potential from light-front holography supplemented by a longitudinal confinement, and a one-gluon-exchange interaction with a fixed coupling. The light-front wave functions resulting from the solution of this Hamiltonian were then used to calculate the nucleon form factors, parton distributions, and other dynamical properties. The first form factor results from BLFQ [949], solved in the valence space of three quarks, are compared to the data and a selection of other theoretical models in Fig. 10.1.9. Such comparisons indicate the need for improvements to the magnetic form factors within BLFQ, particularly in the low-$Q^2$ region. Augmenting the BLFQ basis with dynamical gluons may provide such improvements [961].

In recent years, significant progress has occurred in the explanation and prediction of a wide range of measurable dynamical properties of hadrons in continuum non-perturbative QCD [829], within the framework of Dyson-Schwinger Equations (DSE). In this framework, the high-$Q^2$ behavior of proton and neutron form factors is very sensitive to the behavior of the momentum-dependent dressed quark mass function that governs the transition from massive, constituent-quark-like behavior at low energies to light, parton-like behavior at high energies [2818]. Moreover, the flavor decomposition of the form factors enabled by combined proton and neutron measurements, soon to be extended to $Q^2$ values up to 10 GeV$^2$, has the potential to elucidate the importance of diquark correlations in nucleon structure [758, 2978]. Over the longer term, looking past the ongoing SBS program, major efforts are underway to establish intense polarized and unpolarized positron beams at Jefferson Lab, which will facilitate precise $e^+p/e^-p$ comparisons over a much larger range of $Q^2$ and $\epsilon$ than presently available, hopefully leading to a decisive resolution of the Rosenbluth/polarization discrepancy for the proton, as part of a larger physics pro-gram using positron beams [2994]. The planned Electron-Ion Collider at Brookhaven National Laboratory should be capable of measuring the elastic *ep* cross sections to a $Q^2$ of up to 40-50 $(\mathrm{GeV/c})^2$ [2995]. A proposed "low-cost" upgrade [2996] of Jefferson Lab's Continuous Electron Beam Accelerator Facility (CEBAF) to a maximum energy of 20+ GeV using fixed-field alternating gradient magnets to achieve 6-7 additional passes through the CEBAF linear accelerators would enable further expansions of the $Q^2$ reach for $G_E^p$, $G_E^n$, and $G_M^n$ to at least 20 GeV$^2$.

## 10.2 Parton distributions

**Wally Melnitchouk**

### 10.2.1 Theoretical foundations

Parton distribution functions are the prototypical examples of QCD quantum correlation functions, which allow high-energy lepton and/or hadron scattering processes to be described in terms of quarks and gluons (or partons) (for reviews see Refs. [2997–3001]). The PDF for a quark of flavor $i$ in a nucleon (moving with momentum $p$) is defined by the Fourier transform of a forward matrix element of quark bilinear operators, which in the $A^+ = 0$ gauge can be written as

$$f_{i/N}(x,\mu^2) = \frac{1}{4\pi}\int dz^-\, e^{-ixp^+z^-} \times \langle N(p)|\bar{\psi}_i(z^-)\,\gamma^+\,\psi_i(0)|N(p)\rangle, \quad (10.2.1)$$

where $\psi_i$ is the quark field operator, $x$ is the light-cone momentum fraction of the proton carried by the parton, and $\mu$ is the renormalization scale. Analagous expressions can be written for antiquark and gluon PDFs, the latter in terms of the gluon field strength tensor, $F^A_{\mu\nu}$.

The utility of PDFs is that they allow one to relate various high-energy scattering reactions, which would otherwise not be easily related to one another, and make predictions for new reactions in terms of the same set of PDFs obtained from previous experiments. The key to this is the ability to factorize the scattering process into a process-dependent, perturbatively calculable hard scattering cross section and the process-independent, nonperturbative function parametrized by the PDF. An important virtue of PDFs is that in the infinite momentum frame (or on the light-front) they can be simply interpreted as probability densities describing how the proton's momentum is shared amongst the different parton constituents, as a function of the fraction $x$ of the proton's momentum carried by the parton [1341].

Since quarks and gluons have nonzero spin, the fundamental distributions are the PDFs for a specific helicity (spin projection along the direction of motion), $f_i^\uparrow$ and $f_i^\downarrow$, corresponding to parton spins aligned and antialigned with the proton spin, respectively. Unpolarized scattering experiments are therefore only sensitive to sums of the helicity PDFs, $f_i = f_i^\uparrow + f_i^\downarrow$, while measurements involving polarized beams and/or targets are required to obtain information on differences, $\Delta f_i = f_i^\uparrow - f_i^\downarrow$.

Traditionally, PDFs have been determined in global QCD analyses by simultaneously fitting a wide variety of data for large momentum transfer processes. Typically, the PDFs are parametrized in terms of some functional form, the parameters of which are determined by fitting the calculated cross sections to data. Once the PDFs are determined at some initial momentum transfer scale, the DGLAP $Q^2$ evolution equations (see Sec. 2.3) are used to compute them at all other scales needed for the calculations. The standard data sets used in global analyses include deep-inelastic scattering (DIS) of charged leptons from proton or nuclear targets (or neutrinos from heavy nuclei), Drell-Yan (DY) inclusive lepton-pair production in hadron-hadron scattering, and the production of photons, $W^\pm$ or $Z$ bosons, or jets at large transverse momentum in hadronic collisions (see Sec. 10.2.2). We discuss the specific reactions and relevant data sets in more detail in the following.

### 10.2.2 Physical processes and experimental observables

Historically, the main source of information on proton PDFs has been the deep-inelastic scattering (DIS) of leptons from protons or nuclei, starting from the pioneering experiments at SLAC in the late 1960s. In the one-boson exchange approximation, the differential DIS cross section can be written as a product of leptonic and hadronic tensors,

$$\frac{d^2\sigma}{d\Omega dE'} \sim \alpha\, L^{\mu\nu}\, W_{\mu\nu},$$

where $\alpha$ is the fine structure constant, $\Omega = \Omega(\theta,\phi)$ is the laboratory solid angle of the scattered lepton, and $E'$ is the scattered lepton energy. Using constraints from Lorentz and gauge invariance, the hadronic tensor $W_{\mu\nu}$ can be decomposed into several independent terms,

$$W_{\mu\nu} = -\tilde{g}_{\mu\nu}\, F_1 \;+\; \frac{\tilde{p}_\mu \tilde{p}_\nu}{p\cdot q}\, F_2 \;+\; i\epsilon_{\mu\nu\alpha\beta}\, p^\alpha q^\beta\, F_3 + i\epsilon_{\mu\nu\alpha\beta}\frac{q^\alpha}{p\cdot q}\Big[s^\beta\, g_1 + \Big(s^\beta - \frac{s\cdot q}{p\cdot q}\,p^\beta\Big) g_2\Big], \quad (10.2.2)$$

where $p_\mu$ and $q_\mu$ are the nucleon and exchanged boson four-momenta, $\widetilde{g}_{\mu\nu} = g_{\mu\nu} - q_\mu q_\nu/q^2$, and $\widetilde{p}_\mu = p_\mu - (p \cdot q/q^2) q_\mu$. The nucleon polarization four-vector $s^\beta$ satisfies $s^2 = -1$ and $p \cdot s = 0$. The structure functions $F_{1,2,3}$ and $g_{1,2}$ contain the complete information about the structure of the nucleon in DIS, and are generally functions of two variables, conventionally chosen to be the Bjorken scaling variable $x = Q^2/2p \cdot q$ and the exchanged boson virtuality $Q^2$. In the Bjorken limit, in which both $Q^2$ and $p \cdot q \to \infty$ (or invariant final state hadron mass $W^2 = (p+q)^2 = M^2 + Q^2(1-x)/x \to \infty$), but $x$ is fixed, the structure functions become simple functions of $x$ only.

*Unpolarized scattering*

For spin-averaged scattering, the nucleon structure is parametrized in terms of the vector $F_1$ and $F_2$ structure functions, and the vector-axial vector interference $F_3$ structure function, which requires weak currents. According to QCD factorization theorems, the structure functions $F_j$ ($j = 1, 2, 3$), can be written in factorized form as convolutions of hard coefficient functions and PDFs, weighted by respective electroweak charges,

$$F_j(x, Q^2) = \sum_{i=q,\bar{q},g} e_i^2 \left[\mathcal{C}_i^j \otimes f_i\right](x, Q^2), \tag{10.2.3}$$

where the convolution symbol is defined by $[A \otimes B](x) = \int_x^1 (dy/y) A(x) B(x/y)$. The coefficient functions $\mathcal{C}_j^i$ can be computed perturbatively in a series in $\alpha_s$. At leading order (LO) in $\alpha_s$, $\mathcal{C}_j^i$ is a $\delta$ function, and the structure functions reduce to linear combinations of the PDFs,

$$F_1(x) = \frac{1}{2} \sum_q e_q^2\, q^+(x), \tag{10.2.4a}$$

$$F_2(x) = 2x F_1(x), \tag{10.2.4b}$$

$$F_3(x) = 2 \sum_q g_V^q\, g_A^q\, q^-(x), \tag{10.2.4c}$$

where $q^\pm = q \pm \bar{q}$ denote the $C$-even (odd) flavor combinations, and we use the short-hand notation $q(x) \equiv f_q(x)$ or $\bar{q}(x) \equiv f_{\bar{q}}(x)$ for a quark or antiquark PDF of flavor $q$ in the proton, and $g(x) \equiv f_g(x)$ for the gluon PDF. The $F_3$ structure function vanishes for photon exchange, but is nonzero for the exchange of weak bosons, with $g_{V(A)}^q$ representing the vector (axial vector) coupling of the boson to the quark $q$. Equations (10.2.4) correspond to the simple parton model of inclusive DIS, in which the structure functions are interpreted as parton densities. At finite energies, the logarithmic $Q^2$ dependence from the evolution equations described in Sec. 2.3, as well as residual $Q^2$ dependence associated with power corrections (see below), give corrections to the simple parton model expectations. (Note also that at LO the Bjorken $x$ variable coincides with the parton momentum fraction; however, at higher orders these are different.)

Many DIS experiments have been performed with charged lepton beams on proton targets, which for neutral currents in the one-photon exchange approximation constrain the flavor combination $4u^+ + d^+ + s^+$ ($Z$ boson exchange would involve a different linear combination of PDFs, involving the weak mixing angle, $\sin^2\theta_W$). For a neutron, the corresponding linear combination would be $4d^+ + u^+ + s^+$. In practice, free neutron targets do not exist, so deuterium is often used as a proxy, which then requires nuclear corrections be made to extract the free neutron structure information (see Sec. 10.2.3).

Charged current neutrino and antineutrino interactions constrain different combinations of $q^+$ or $q^-$ PDFs for the $F_{1,2}$ or $F_3$ structure functions, respectively, depending on the type of target used, so that by combining data on different targets and with different beams one can in principle isolate specific combinations of $q$ or $\bar{q}$. A special case is provided by charm production in $\nu$ and $\bar{\nu}$ DIS, which is sensitive to the $s$ and $\bar{s}$ PDFs, respectively (although in practice this involves heavy targets for which model-dependent nuclear corrections must be made).

The gluon PDF plays a lesser role in inclusive DIS, as it enters the cross sections at higher order, $\mathcal{O}(\alpha_s)$. In practice, it is mainly constrained through the $Q^2$ dependence of the structure functions, and the longitudinal structure function $F_L$, which depends on differences at higher order between the left- and right-hand sides of Eq. (10.2.4b). The strongest constraints on $g(x)$ in DIS have come from the HERA *ep* collider data at very small $x$ values [3002].

Since the PDFs are universal, the functions appearing in the DIS structure functions are the same as those that describe the structure of the incoming hadrons in hadronic collisions. In analogy with the QCD factorization for DIS, the cross section for the high energy scattering of hadron $A$ (momentum $p_A$) and hadron $B$ (momentum $p_B$) to an inclusive state in which a particle $C$ is identified (such as a vector boson, photon, or jet) can generally be written as

$$\sigma_{AB\to CX}(p_A, p_B) = \sum_{a,b} \int dx_a\, dx_b\, f_{a/A}(x_a)\, f_{b/B}(x_b) \times \hat{\sigma}_{ab\to CX}(x_a p_A, x_b p_B), \tag{10.2.5}$$

where $x_a$ and $x_b$ are the corresponding parton momentum fractions, and $\hat{\sigma}_{ab\to CX}$ is the partonic cross section. For vector boson production ($W^\pm$, $Z^0$, or lepton pairs produced from virtual photons), the process proceeds at LO through $q\bar{q}$ annihilation. In particular, the Drell-Yan lepton-pair production cross sections in *pp* and *pn* collisions depend on the combinations

$$\sigma^{pp} \sim 4u(x_a)\,\bar{u}(x_b) + d(x_a)\,\bar{d}(x_b) + (x_a \leftrightarrow x_b) + \cdots$$
$$\sigma^{pn} \sim 4d(x_a)\,\bar{d}(x_b) + u(x_a)\,\bar{u}(x_b) + (x_a \leftrightarrow x_b) + \cdots$$

where the ellipses indicate contributions from heavier quarks. (The $pn$ cross section is again obtained from deuterium data.) As we discuss below, ratios of these cross sections at kinematics such that $x_a \gg x_b$, where the hadron $A$ can be approximated by its valence structure, can be used to constrain the $\bar{d}/\bar{u}$ ratio. In contrast, the inclusive production of $W^{\pm}$ bosons constrains products of the form $q(x_a)\,\bar{q}'(x_b)$ with specific weights given by the appropriate CKM matrix elements.

For $p\bar{p}$ collisions at the Tevatron, for example, at large values of rapidity (very asymmetric values of $x_a$ and $x_b$) at LO one has

$$\sigma^{W^+} \sim u(x_a)\,d(x_b) + \bar{d}(x_a)\,\bar{u}(x_b) + \cdots$$
$$\sigma^{W^-} \sim d(x_a)\,u(x_b) + \bar{u}(x_a)\,\bar{d}(x_b) + \cdots$$

where the PDFs in the antiproton have been related to those in the proton. For large and positive rapidity, $x_a > x_b$ and the antiquark PDFs can be neglected, so that these cross sections depend only on the $u$ and $d$ PDFs. Because of the missing neutrino resulting from the $W$ decays, one cannot directly reconstruct the rapidity distributions, and typically the charged lepton rapidity asymmetry for $W^{\pm}$ production is presented. The decay process means that the constraints on the PDFs are less direct, but such measurements still provide useful constraints on the $d/u$ ratio at moderate values of $x$.

Recent data from the ATLAS Collaboration at the LHC [3003, 3004] on $W^{\pm}$ and $Z$ production and decay suggested a rather larger strange quark sea than traditionally obtained from neutrino scattering, with the ratio $R_s = (s+\bar{s})/(\bar{u}+\bar{d}) \approx 1.13$ at parton momentum fraction $x \approx 0.02$, compared with the traditionally accepted value of $R_s \approx 0.4$ from neutrino scattering. In contrast, a simultaneous analysis of PDFs and fragmentation functions including semi-inclusive $\pi^{\pm}$ and $K^{\pm}$ production data, along with single-inclusive $e^+e^-$ annihilation cross sections into hadrons [628, 3005], favored a strong suppression of the strange PDF at intermediate $x$ values, correlating with an enhancement of the $s \to K^-$ fragmentation function. The question of the magnitude and shape of the strange (and antistrange) PDF remains a topic of considerable phenomenological interest.

Other observables that can constrain PDFs are inclusive jet or photon, dijet, and photon + jet production cross sections. Generally, these have greater sensitivity to the gluon PDF at large $x$ than DIS reactions. Dijet production triple differential cross sections yield more information than single jet cross sections because the rapidity of the second jet is also constrained, thereby helping to constrain the momentum fractions of the PDFs. Direct or isolated photon production can also constrain the gluon PDF through the subprocess $qg \to \gamma q$ [3006]. Photon + jet production offers similar constraints, but now the subprocesses are weighted by the squared charge of the parton to which the photon couples. A summary of the kinematic coverage of the existing datasets used to constraint unpolarized PDFs is shown in Fig. 10.2.1.

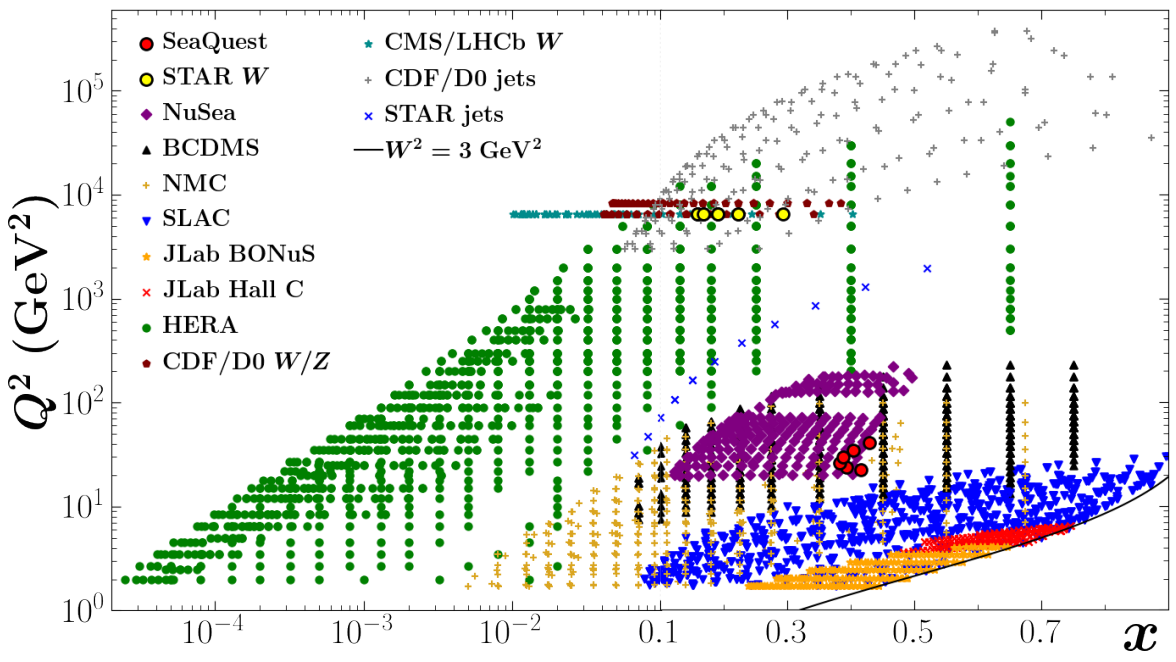

**Fig. 10.2.1** Kinematic coverage of datasets used in global QCD analyses. The variable $x$ represents Bjorken-$x$ for DIS and Feynman-$x$ for vector boson and jet production, while the scale $Q^2$ represents the four-momentum transfer squared for DIS, the mass squared of the intermediate boson for vector boson production, and the transverse momentum squared for jet production. A DIS cut of $W^2 = 3$ GeV$^2$ is indicated in the bottom right hand corner (solid back line).

*Polarized scattering*

For spin-dependent reactions, the structure functions $g_1$ and $g_2$ are extracted from DIS measurements with longitudinally polarized leptons scattered from a nucleon or nucleus that is polarized either longitudinally or transversely relative to the beam. For longitudinal beam and target polarization, the difference between the cross sections for spins aligned and antialigned is dominated by the $g_1$ structure function, while the $g_2$ structure function requires measurements with the target polarized transversely to the beam polarization. In practice one often measures the polarization asymmetry $A_1$, which is given as a ratio of spin-dependent and spin-averaged structure functions,

$$A_1 = \frac{1}{F_1(x)}\left[g_1(x) - \frac{4M^2x^2}{Q^2}g_2(x)\right], \qquad (10.2.6)$$

where $M$ is the nucleon mass. At small values of $x^2/Q^2$, the asymmetry simplifies to $A_1 \approx g_1/F_1$.

In analogy with the unpolarized $F_1$ structure function, the structure function $g_1$ can be expressed at LO

in terms of differences between quark distributions with spins aligned and antialigned with that of the nucleon,

$$g_1(x) = \frac{1}{2} \sum_q e_q^2 \, \Delta q^+(x). \tag{10.2.7}$$

The $g_2$ structure function, on the other hand, does not have a simple partonic interpretation. However, its measurement provides information on the subleading, higher-twist contributions which parametrize long-range multi-parton correlations in the nucleon. The dependence on both spin-dependent and spin-averaged structure functions in $A_1$ illustrates the need to consistently analyze both unpolarized and polarized PDFs simultaneously, as will be discussed below.

As with unpolarized measurements, historically most constraints on spin-dependent PDFs have come from polarized charged-lepton DIS experiments. For charged lepton scattering from polarized proton targets, the $g_1$ structure function depends on the combination $4\Delta u^+ + \Delta d^+ + \Delta s^+$, while for the neutron the combination would be $4\Delta d^+ + \Delta u^+ + \Delta s^+$. In practice, polarized $^3$He targets are usually used as effective sources of polarized neutron, since the neutron carries almost 90% of the $^3$He spin, while polarized deuterons, which have equal proton and neutron spin contributions, are sensitive to the isoscalar combination $5(\Delta u^+ + \Delta d^+) + 2\Delta s^+$.

At next-to-leading order (NLO), the polarized gluon distribution $\Delta g$ also enters in the $g_1$ structure function. The mixing with the quark flavor singlet contribution to $g_1$ under $Q^2$ evolution can then be used to provide constraints on $\Delta g$.

Semi-inclusive DIS (SIDIS) provides additional independent combinations of spin-dependent PDFs that can be used to separate individual quark and antiquark flavors. At high energies, production of hadrons $h$ in the current fragmentation region, primarily pions or kaons, is proportional to products of PDFs and quark → hadron fragmentation functions. Typically, such experiments measure the semi-inclusive polarization asymmetry, which at LO can be written as a ratio of spin-dependent to spin-averaged SIDIS cross sections,

$$A_1^h(x,z) = \frac{\sum_q e_q^2 \left(\Delta q(x)\, D_q^h(z) + \Delta \bar{q}(x)\, D_{\bar{q}}^h(z)\right)}{\sum_q e_q^2 \left(q(x)\, D_q^h(z) + \bar{q}(x)\, D_{\bar{q}}^h(z)\right)}, \tag{10.2.8}$$

where $D_q^h(z)$ is the fragmentation function for the scattered quark to produce a hadron $h$ with a fraction $z$ of the quark's energy. For large $z$, the produced hadron has a high probability of containing the scattered parton, providing a tag on the initial parton distribution.

The fragmentation functions $D_q^h$ can be determined from other reactions, such as inclusive single hadron production in $e^+e^-$ annihilation or $pp$ collisions. One can then weight particular quark or antiquark flavors by selecting favored (such as $D_u^{\pi^+}$ or $D_{\bar{d}}^{\pi^+}$) or unfavored ($D_d^{\pi^+}$ or $D_{\bar{u}}^{\pi^+}$) fragmentation functions for specific hadrons. The polarized strange quark PDF $\Delta s$, in particular, can be constrained from SIDIS $K$ production data. The polarized gluon distribution $\Delta g$ can also be constrained from SIDIS data on charmed or high-$p_T$ hadron production through the photon–gluon fusion process.

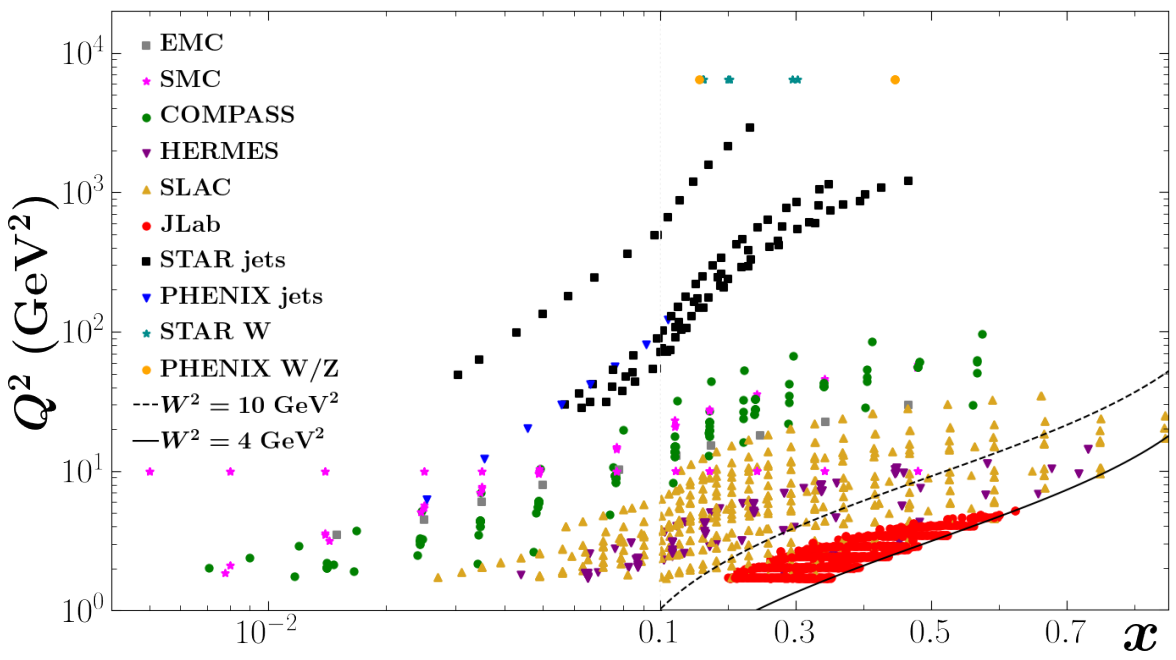

**Fig. 10.2.2** As in Fig. 10.2.1 but for spin-dependent observables.

Inclusive particle production in polarized proton–proton collisions provides an additional method of determining spin-dependent sea quark and gluon PDFs. The cross sections for the production of $W^\pm$ bosons in scattering longitudinally polarized protons from unpolarized protons, $\vec{p}p \to W^\pm X$, depend on products of spin-dependent and spin-averaged PDFs,

$$\Delta\sigma^{W^+} \sim \Delta\bar{d}(x_a)\, u(x_b) - \Delta u(x_a)\, \bar{d}(x_b),$$
$$\Delta\sigma^{W^-} \sim \Delta\bar{u}(x_a)\, d(x_b) - \Delta d(x_a)\, \bar{u}(x_b).$$

At large positive (negative) rapidities, $x_a \gg x_b$ ($x_a \ll x_b$), the cross sections are dominated by a single flavor, while at mid-rapidities both $u$ and $d$ flavors contribute.

Inclusive jet (or $\pi^0$) production in double-polarized proton–proton scattering, $\vec{p}\vec{p} \to$ jet (or $\pi^0$) $+ X$, is sensitive to the polarized gluon PDF. The first evidence for a small, but nonzero $\Delta g$ was observed by the STAR Collaboration at RHIC in jet data at $\sqrt{s} = 200$ GeV, although recent Monte Carlo analysis [984] suggests that the sign of $\Delta g$ is not unambiguously determined by these data. A summary of the kinematic coverage of the existing datasets used to constraint helicity PDFs is shown in Fig. 10.2.2.

### 10.2.3 Global QCD analysis

With the growing number of high energy scattering experiments in the 1970s and 1980s came the need to systematically and uniformly analyze the data with the tools that were being developed in perturbative QCD. The concept of fitting datasets from various experiments globally with a single set of quark, antiquark and gluon PDFs dates back to the early analyses of Duke and Owens [3007] and Morfin and Tung [3008]. Since then, a number of dedicated efforts have been made worldwide to fit both unpolarized and polarized scattering experiments in terms of spin-averaged and spin-dependent PDFs.

The standard paradigm has been to parametrize the PDFs at some input scale $Q_0$ and then evolve using the appropriate evolution equations to the scales needed for the calculation of each experimental observable. The parameters of the PDFs are estimated by comparing each calculated observable with the data using $\chi^2$ minimization techniques. All of the global PDF analysis groups use some variation of this approach, although the details of the implementation differ between different groups.

*PDF parametrizations and constraints*

A typical parametrization at the input scale $Q_0$ for a generic (unpolarized or polarized) PDF $f$ is

$$xf(x, Q_0^2) = a_0\, x^{a_1} (1-x)^{a_2}\, P(x), \qquad (10.2.9)$$

where $P(x)$ represents a smoothly varying function, such as a polynomial in $x$ or $\sqrt{x}$, or more elaborate forms based on neural networks [3009] or self-organizing maps [3010]. Some of the parameters in the input distributions can be determined from physical constraints. For example, in the unpolarized case the conservation of valence quark number gives for the first moments

$$\int_0^1 dx\, u^-(x, Q_0^2) = 2, \quad \int_0^1 dx\, d^-(x, Q_0^2) = 1, \quad (10.2.10)$$

and zero for all other flavors, while the momentum sum rule requires

$$\int_0^1 dx\, x \Big[ \sum_q^{n_f} q^+(x, Q_0^2) + g(x, Q_0^2) \Big] = 1, \qquad (10.2.11)$$

where the number of flavors at the input scale $Q_0^2$ is usually taken to be $n_f = 3$.

In the polarized case the first moments of the $C$-even distributions can be related to octet baryon weak decay constants. For the isovector combination, corresponding to the Bjorken sum rule,

$$\int_0^1 dx\, (\Delta u^+ - \Delta d^+)(x, Q_0^2) = g_A, \qquad (10.2.12)$$

where $g_A = 1.270(3)$ is the nucleon axial charge, while for the SU(3) octet one has

$$\int_0^1 dx\, (\Delta u^+ + \Delta d^+ - 2\Delta s^+)(x, Q_0^2) = a_8, \qquad (10.2.13)$$

where the octet axial charge $a_8 = 0.58(3)$ is extracted from hyperon $\beta$-decays assuming SU(3) flavor symmetry [3011]. Note that the sum rules (10.2.10)–(10.2.13) are preserved under $Q^2$ evolution.

*Power corrections*

We should note that the theoretical results summarized above have been obtained within the framework of perturbative QCD in the limit when both $Q^2$ and $W$ are much larger than all hadron mass scales, $Q^2, W^2 \gg M^2$, where the cross sections are dominated by their leading twist contributions. In actual experiments performed at finite beam energy $E$, the maximum values of $Q^2$ and $W$ are limited, which restricts the available coverage in Bjorken $x$. This is especially relevant at large $x$ in DIS, where for fixed $Q^2$, as $x \to 1$ the final state hadron mass $W$ decreases as one descends into the region dominated by nucleon resonances at $W \lesssim 2$ GeV. The resonance region may be treated using the concept of quark-hadron duality [3012], although this goes beyond the scope of the usual perturbative QCD analysis.

In the low-$Q^2$ region, power corrections to the Bjorken limit results that scale as powers of $\Lambda_{\rm QCD}^2/Q^2$ become increasingly important. In the operator product expansion, these are associated with matrix elements of higher twist operators, associated with multi-parton correlations which characterize the long-range nonperturbative interactions between quarks and gluons. While providing glimpses into the dynamics of quark confinement, the power corrections are viewed as unwelcome backgrounds in efforts aimed solely at extracting leading twist PDFs. Other subleading corrections are associated with target mass corrections (TMCs), which are of kinematical origin and arise from nonzero values of hadron masses [3013–3017].

Regardless of their origin, the various power suppressed corrections to the leading twist results can be absorbed into phenomenological functions, such as

$$F_i(x, Q^2) = F_i^{\rm LT}(x, Q^2) + \frac{h_i(x)}{Q^2} + \ldots, \qquad (10.2.14)$$

for an unpolarized structure function $F_i$, for example, where $F_i^{\rm LT}$ denotes the leading twist contribution. The higher twist corrections are sometimes assumed to be multiplicative, with the functions $h_i$ proportional to the leading twist contribution. Possible additional $Q^2$ dependence of the higher twist contributions, such as from radiative $\alpha_s(Q^2)$ corrections, is usually neglected.

*Nuclear corrections*

Since nucleons bound in a nucleus are not free, the parton distributions $f_{i/A}$ in a nucleus $A$ deviate from a simple sum of PDFs in the free proton and neutron, $f_{i/A} \neq Z f_{i/p} + (A - Z) f_{i/n}$, where $Z$ is the number of protons. This is especially relevant at small values of $x$, where nuclear shadowing effects suppress the nuclear to free isoscalar nucleon ($N$) ratio, $f_{i/A}/(A f_{i/N}) < 1$, and at large $x$, where the effects of Fermi motion, nuclear binding, and nucleon off-shellness give rise to the "nuclear EMC effect" [3018–3020]. For spin-dependent PDFs, the different polarizations of the bound nucleons and nuclei also need to be taken into account.

In the nuclear impulse approximation, where scattering takes place incoherently from partons inside individual nucleons, the PDF in a nucleus can be expressed as a convolution of the PDF in a bound nucleon and a momentum distribution function $f_{N/A}$ of nucleons in the nucleus [3021–3023]. The momentum distribution, or "smearing function", can be computed from nuclear wave functions, incorporating nuclear binding and Fermi motion effects. Coherent rescattering effects involving partons in two or more nucleons give rise to nuclear shadowing corrections to the impulse approximation, and such effects are typically important only in the small-$x$ region. In general, the relation between PDFs in a nucleus and in a nucleon can be written as

$$f_{i/A} = \sum_{N=p,n} \left[ f_{N/A} \otimes f_{i/N} \right] + \delta^{(\mathrm{off})} f_{i/A} + \delta^{(\mathrm{shad})} f_{i/A}, \tag{10.2.15}$$

where the term $\delta^{(\mathrm{off})} f_{i/A}$ represents nucleon off-shell or relativistic corrections that account for modification of the parton structure of the nucleon in the nuclear medium. A similar expression can be written for spin-dependent PDFs.

At large $Q^2$ the smearing function has a probabilistic interpretation in terms of the light-cone momentum fraction $y$ of the nucleus carried by the struck nucleon. Typically, the function $f_{N/A}$ is steeply peaked around $y \approx 1$, becoming broader with increasing mass number $A$ as the effects of binding and Fermi motion become more important. In the limit of zero binding, $f_{N/A}(y) \to \delta(1 - y)$, and one recovers the free nucleon case. This assumption has often been made in global PDF analyses. More recently, however, the important role of nuclear corrections has been more widely appreciated, especially in connection with extractions of the free neutron structure function data from measurements involving deuterons and other light nuclei [3023–3029].

For neutrino scattering, to increase the relatively low rates and obtain sufficient statistics for analyses such as strange PDF extraction [3030, 3031], experiments have usually resorted to using heavier nuclear targets, such as iron or lead. Extractions from such data are complicated by the presence of nuclear corrections in neutrino structure functions [3032, 3033], as well as effects of the nuclear medium on the charm quark propagation in the final state [3034].

For spin-dependent scattering, the scarcity of data and larger uncertainties at small $x$ and at high $x$, where nuclear corrections are most prominent, has meant that most global analyses have relied exclusively on the effective polarization *ansatz*, in which the polarized PDF in the nucleus $\Delta f_{i/A}$ is related to the polarized PDFs in the proton and neutron as $\Delta f_{i/A} \approx \langle\sigma\rangle^p \Delta f_{i/p} + \langle\sigma\rangle^n \Delta f_{i/n}$, where $\langle\sigma\rangle^{p(n)}$ is the average polarization of the proton (neutron) in the nucleus. In practice, along with polarized protons, only polarized deuterium and $^3$He nuclei have been used in DIS experiments. As experiments at high luminosity facilities such as Jefferson Lab at 12 GeV push to explore the higher-$x$ region, nuclear corrections from smearing and off-shell effects will become more relevant.

*Uncertainty quantification*

There are several sources of PDF uncertainties that enter in global QCD analyses. These include uncertainties on the experimental data, the approximations used in computing the partonic cross sections, and the parametrizations used to describe the PDFs. The experimental errors on the data can be directly propagated to the fitted PDFs. The most common method for implementing this is the Hessian method, described in Ref. [3035]. The elements of the Hessian matrix are given by partial derivatives of the $\chi^2$ function,

$$H_{ij} = \frac{1}{2} \frac{\partial^2 \chi^2}{\partial a_i \, \partial a_j}, \tag{10.2.16}$$

where $a_i$ denotes the $i^{\mathrm{th}}$ PDF parameter. The Hessian matrix is generated during the minimization procedure and its inverse gives the error matrix. The eigenvectors of the error matrix can then be used to define eigenvector parameter sets, from which the error bands for the PDFs or for specific processes are calculated. An important point to note is that the error bands generally depend on a $\chi^2$ tolerance. Mathematically, the expectation is that the $1\sigma$ parameter errors correspond to an increase of $\chi^2$ by one unit from the minimum value, $\Delta\chi^2 = 1$. However, it has been suggested [3036] that inconsistencies between different data sets should be handled by introducing a larger value to be used, $\Delta\chi^2 > 1$. This "$\chi^2$ tolerance" varies between groups ($\Delta\chi^2 \sim 10 - 100$), and allowance must be made for this when comparing the resulting error bands.

On the other hand, it has been argued [3037] that the tolerance criterion effectively changes the likelihood function, which is usually defined in terms of the $\chi^2$ function. In contrast, neural network based approaches suggest that the use of a tolerance criterion is not necessary [3009, 3038–3040]. In practice, the similar size of the uncertainties obtained in such different approaches may be coincidental and due to the likelihood deformation and resulting uncertainty inflation, as observed in a recent comparative study using toy data [3037]. Furthermore, concern has also been expressed [3041] that a meta-analysis, such as PDF4LHC [3042], that combines existing PDFs from different groups may obscure the fundamental connection between experimental data and theory and hide the true meaning of the uncertainties, if these ultimately originate from different choices of the likelihood function.

An alternative to the usual linear propagation of errors in the Hessian method which avoids ambiguities associated with tolerance criteria, and which is useful for minima that are not well behaved or defined, is the Monte Carlo method. To propagate the experimental errors a number of replica data sets are randomly generated within the original errors, and these replica sets are then fitted with the resulting replica PDF sets treated using standard statistics [3039]. The central values are computed as the averages over replicas, while the uncertainties are given by the envelope of predictions.

In practice, the data resampling method has been used by the NNPDF [983, 3043] and JAM [628, 984, 3029] collaborations, although these groups differ in their approach to PDF parametrizatios. While the JAM collaboration uses a traditional polynomial functional form for the function $P(x)$ in Eq. (10.2.9), the NNPDF group implements a similar basic parametric form that is supplemented by a series of trained neural network weights. The dependence on the functional form for the PDF can be minimized by choosing a flexible parametrization with parameters that are well-constrained by data. Outside of kinematic regions covered by data, the PDFs are not constrained, and care must be taken when using them in extrapolated regions at small or large $x$.

The approximations made in computing partonic cross sections naturally introduce uncertainties in PDFs, although these can be rather difficult to quantify reliably. One of these is the uncertainty arising from the truncation of the perturbative series. These can be estimated to some extent by comparing LO, NLO, NNLO, and recently even approximate $N^3$LO [3044] fits, although not all processes are known to the same accuracy. The topic of “missing higher order uncertainties” and how to estimate them has in fact attracted some attention recently in global PDF fitting efforts [3000].

Perturbative QCD calculations also depend to some degree on the choices made for the renormalization and factorization scales for each physical process. The choices will change the results for different processes, and the fitted PDFs must compensate these changes. A closely related issue is the choice of the strong running coupling $\alpha_s(M_Z)$, which is fitted together with the PDF parameters in some analyses, and fixed to the global average in others. Finally, the choice of data sets and kinematic cuts can of course affect the extracted PDFs, and these choices and the reasons for them need to be assessed when drawing conclusions from PDF comparisons.

### 10.2.4 Spin-averaged PDFs

Using the technology outlined in the previous sections, a number of global QCD analyses efforts have produced sets of unpolarized proton PDFs, with groups in Europe and the the US at the forefront of the data analyses. The European groups include the UK-based MSHT [1110] group and the ABM [3045] group, which use standard global fitting methodology; the HERAPDF [3002] analysis, which includes only data from the H1 and ZEUS experiments at HERA; and previously the Dortmund [3046] group, which pioneered the approach of dynamically generating PDFs through $Q^2$ evolution from a low input scale. More recently, the NNPDF [3043] collaboration introduced an approach based on neural networks.

US-based efforts have centered around the CTEQ collaboration, which involves two derivative analyses of nucleon PDFs, by the CT (CTEQ-Tung et al.) [626] and CJ (CTEQ-Jefferson Lab) [3027] groups. The former focuses more on LHC-related phenomenology, while the latter has developed methodologies needed for describing data over a broad energy range including the low-$Q^2$ and $W$ domain. The Jefferson Lab-based JAM [3029, 3047] collaboration uses a Monte Carlo approach with simultaneous determination of PDFs and other types of distributions, such as fragmentation functions and spin-dependent PDFs. In the following we illustrate the current state of knowledge of the spin-averaged proton PDFs, including the $u$ and $d$ valence quark distributions and the flavor structure of the proton sea.

*Valence quark distributions*

Valence quarks give the global properties of the nucleon, such as its baryon number and charge. Knowledge of their momentum distributions is important for many reasons, especially at high values of $x$, where a single quark carries most of the nucleon's momentum. The large-$x$ region is a unique testing ground, for example, for various nonperturbative models of the nucleon [1092, 1341, 3024, 3048, 3049]. Reliable determination of PDFs at large $x$ is also important for searches for new physics beyond the Standard Model in collider experiments at the LHC [3041, 3050].

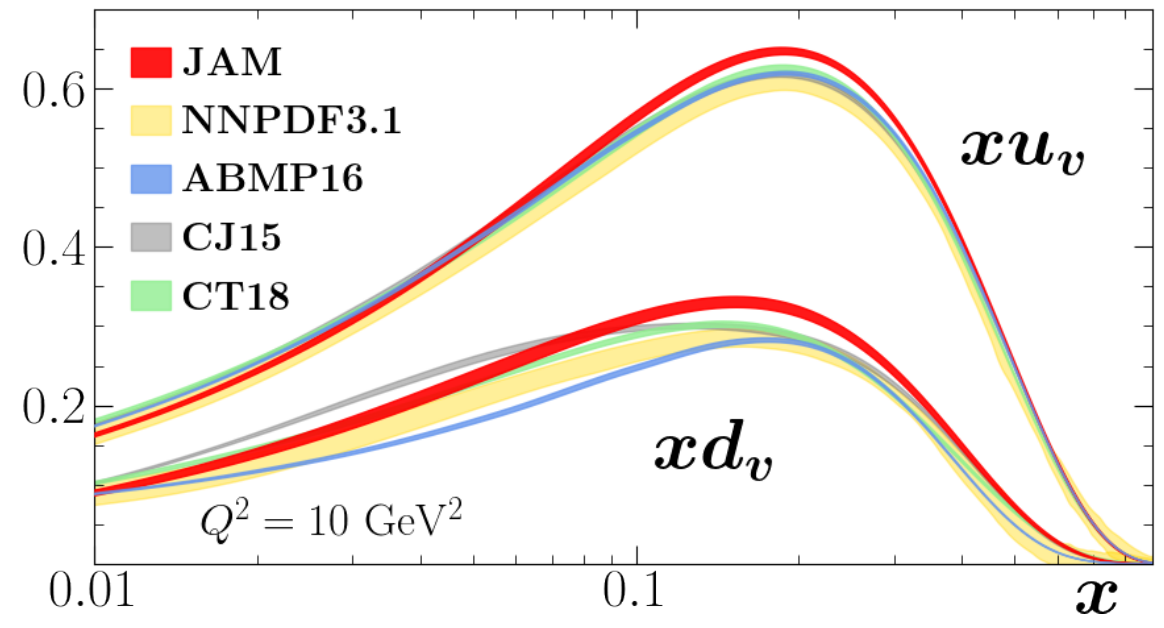

**Fig. 10.2.3** Valence $u$ and $d$ quark PDFs versus $x$ from several global QCD analyses: JAM21 [3029], NNPDF [627], ABMP [3045], CJ15 [3027], and CT18 [626] at a scale $Q^2 = 10$ GeV$^2$.

The valence $u$ and $d$ PDFs are illustrated in Fig. 10.2.3 from several PDF groups. The $u$ quark PDF is fairly well constrained (due to its larger charge) by the relatively abundant proton DIS data that have been collected over several decades at SLAC, CERN, DESY and Jefferson Lab. The $d$ quark distribution, on the other hand, relies in addition on neutron structure functions, whose determination requires both proton and deuteron DIS data. Studies of nuclear effects in the deuteron suggest that the uncertainties related to nucleon interactions increase significantly at large $x$ [3024], leading to large uncertainties in the $d/u$ PDF ratio for $x \gtrsim 0.6$, as Fig. 10.2.4 illustrates. Inclusion of tagged deuteron data from the BONuS experiment at Jefferson Lab [3051, 3052], and in particular the lepton and $W$ boson asymmetry data from $p\bar{p}$ collisions at the Tevatron [3053–3055], reduces the uncertainty considerably in the experimentally constrained region up to $x \sim 0.8$.

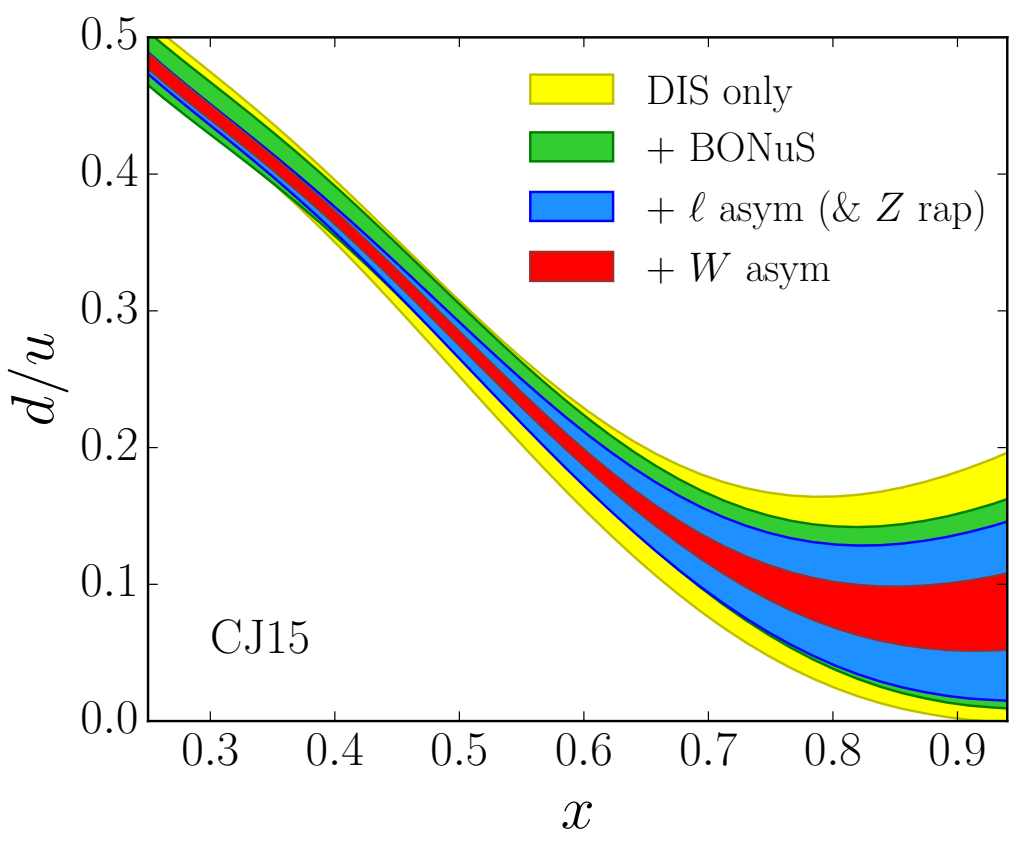

**Fig. 10.2.4** Impact of various data sets on the $d/u$ ratio at $Q^2 = 10$ GeV$^2$, using the CJ15 PDFs set [3027].

*Light quark sea*

Because inclusive DIS measures only C-even combinations of PDFs, $q^+$, to disentangle quark from antiquark contributions requires other types of observables, such as the DY cross sections, where the $q$ and $\bar{q}$ PDFs are weighted differently. As discussed in Sec. 10.2.2, ratios of $pd$ to $pp$ cross sections at $x_a \gg x_b$ are directly sensitive to the ratio $\bar{d}/\bar{u}$. The flavor asymmetry $\bar{d} - \bar{u}$ is illustrated in Fig. 10.2.5, which shows the impact of various data sets. Starting with inclusive DIS data only and excluding data from the NMC experiment, the asymmetry is consistent with zero within large uncertainties. Including the NMC data [3056, 3057], the asymmetry gives an indication of deviation from zero in the range $0.01 < x < 0.2$. When $W$-lepton, reconstructed $W$ and $Z$ boson, and jet production data from RHIC, Tevatron, and LHC are further included (but not the new STAR data [3058]), the asymmetry becomes significantly larger, and more distinguishable from zero below $x = 0.3$.

The new constraints come primarily from the high precision $W$ asymmetry measurements from the Tevatron and LHC, which are sensitive to $\bar{u}$ and $\bar{d}$. The further addition of the NuSea DY data [3059] greatly decreases the uncertainty, showing that these data provide a strong constraint on the asymmetry even when compared to the Tevatron and LHC $W$-lepton asymmetries. Finally, the inclusion of the new SeaQuest [3060] and STAR [3058] data reduces the uncertainty on the asymmetry even further, while increasing the magnitude at $x \gtrsim 0.2$. The behavior of the asymmetry seen in Fig. 10.2.5 is consistent with expectations from nonperturbative models of the nucleon in which the excess of $\bar{d}$ over $\bar{u}$ in the proton sea has been that associated with chiral symmetry breaking, and the consequent prevalence of the virtual $p \to n\pi^+$ dissociation [3061–3063].

*Strange quarks*

The strange quark distribution has generally been more difficult to determine experimentally than the nonstrange sea. While the size of the strange to nonstrange ratio $R_s$ has been controversial, with values ranging from $R_s \approx 0.4$ in neutrino DIS at $x \approx 0.02$ to $R_s \approx 1$ from ATLAS data on $W/Z$ production [3003, 3004], an independent and underutilized source of information at lower energies is semi-inclusive production of pions or kaons. Analysis of SIDIS data has often been complicated by the need to know both the PDFs of the initial state and the fragmentation functions describing hadronization to the final state, as assumptions about the latter can lead to significant differences in the extracted PDFs [3064, 3065]. For any definitive conclusion a combined analysis of PDFs and fragmentation

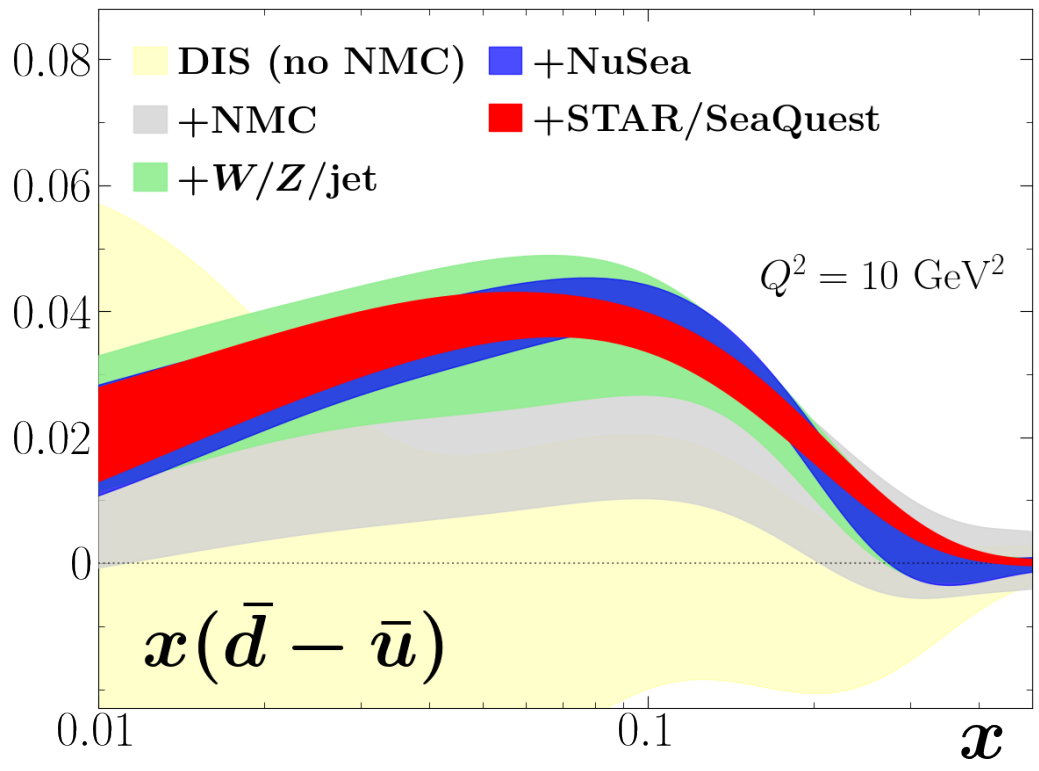

**Fig. 10.2.5** Comparison of $x(\bar{d} - \bar{u})$ with different combinations of datasets [3047]: DIS only, excluding NMC (gold band); with NMC (gray); with $W$, $Z$, and jet production from RHIC, Tevatron, and the LHC (green); with NuSea (blue); and finally with the SeaQuest DY and STAR $W$-lepton ratio (red).

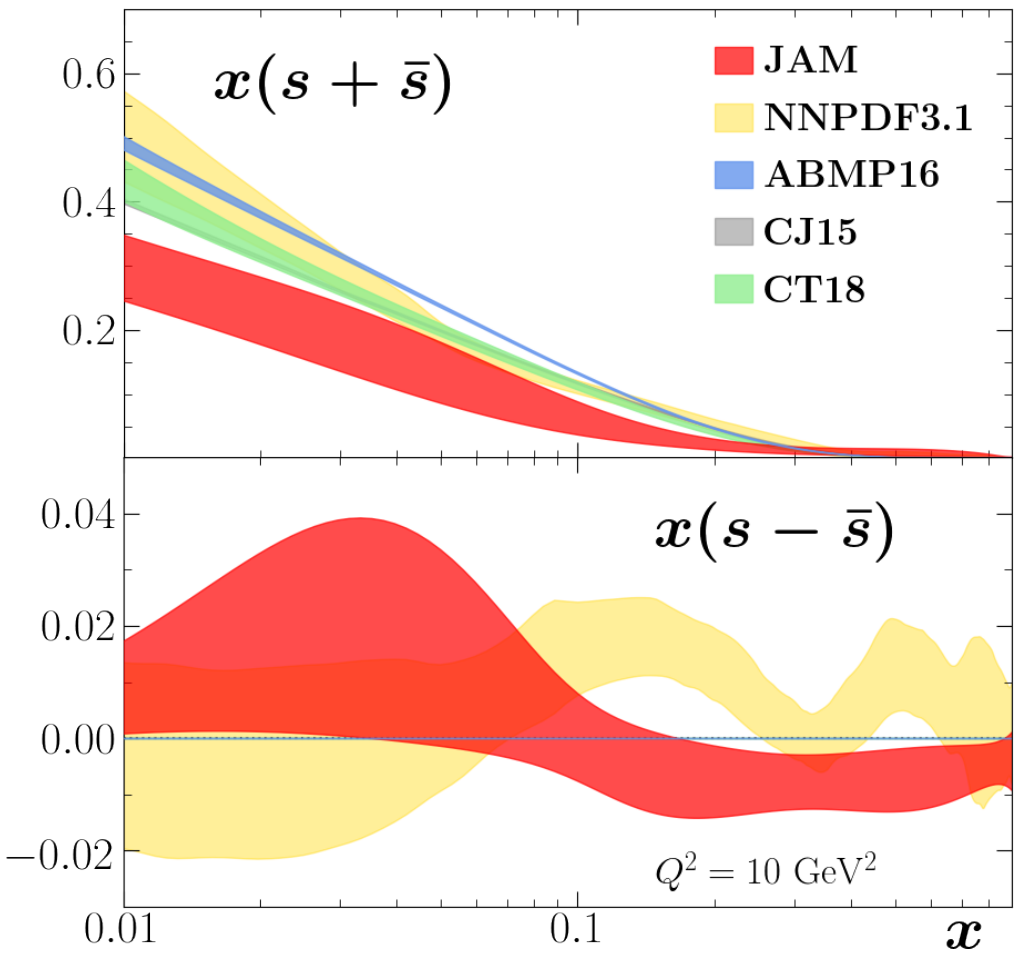

**Fig. 10.2.6** Sum and difference of the $s$ and $\bar{s}$ PDFs from several global QCD analyses, as in Fig. 10.2.3.

functions is necessary, which was first performed by the JAM group [628, 3005].

Including data from the standard datasets used for unpolarized PDFs, along with SIDIS multiplicities and $e^+e^-$ annihilation data to constrain the fragmentation functions [3066], the most striking result of the simultaneous JAM fit was a significantly reduced strange quark PDF compared with that reported by ATLAS, as Fig. 10.2.6 illustrates. The strange to nonstrange ratio was found to be $R_s \approx 0.2 - 0.3$ at $x \sim 0.02$, in contrast to values of $R_s \sim 1$ inferred from the ATLAS data, and closer to those extracted from neutrino experiments. The most significant source of the strange suppression is the SIDIS and SIA $K$ production data. Without these data, the $s^+$ PDF is poorly constrained, in contrast to the light flavor sea, which is not strongly affected by the SIDIS multiplicities. Consequently, while the ratio $R_s$ varies over a large range without SIDIS (and SIA) data, and at low $x$ is compatible with $R_s \approx 1$, once those data are included its spread becomes dramatically reduced.

The SIDIS $K^\pm$ production data could also in principle discriminate between the $s$ and $\bar{s}$ PDFs, which could have different $x$ dependence [3067–3072]. As shown in Fig. 10.2.6, however, the current data do not indicate any significant $s - \bar{s}$ asymmetry within uncertainties. Future high-precision SIDIS data from Jefferson Lab or the Electron-Ion Collider may allow more stringent determinations of the $s$ and $\bar{s}$ PDFs [3073], as would inclusion of $W$ + charm production data from the LHC [3074, 3075].

*Gluons and heavy quarks*

Gluons play an important role in the study of nucleon structure, contributing some 50% of the nucleon's overall (linear) momentum, and indirectly provide some con-straints on quark PDFs through the momentum sum rule, Eq. (10.2.11). Since photons do not couple directly to gluons, the constraints on the gluon PDF $g(x)$ from DIS come via the $Q^2$ evolution of the $F_2$ structure function at low values of $x$. In addition, measurements at HERA of the longitudinal structure function, which has a leading contribution at $\mathcal{O}(\alpha_s)$ through the $\gamma^* g \to q\bar{q}$ process, have allowed $g(x)$ to be relatively well determined at low $x$. More directly, inclusive jet and photon production cross sections at hadron colliders have constrained $g(x)$ at moderate $x$ values, although there is somewhat more uncertainty in the behavior at high $x$. A survey of various determinations of the gluon PDF at $Q^2 = 10$ GeV$^2$ is illustrated in Fig. 10.2.7 for the same set of PDF parametrizations as in Fig. 10.2.3.

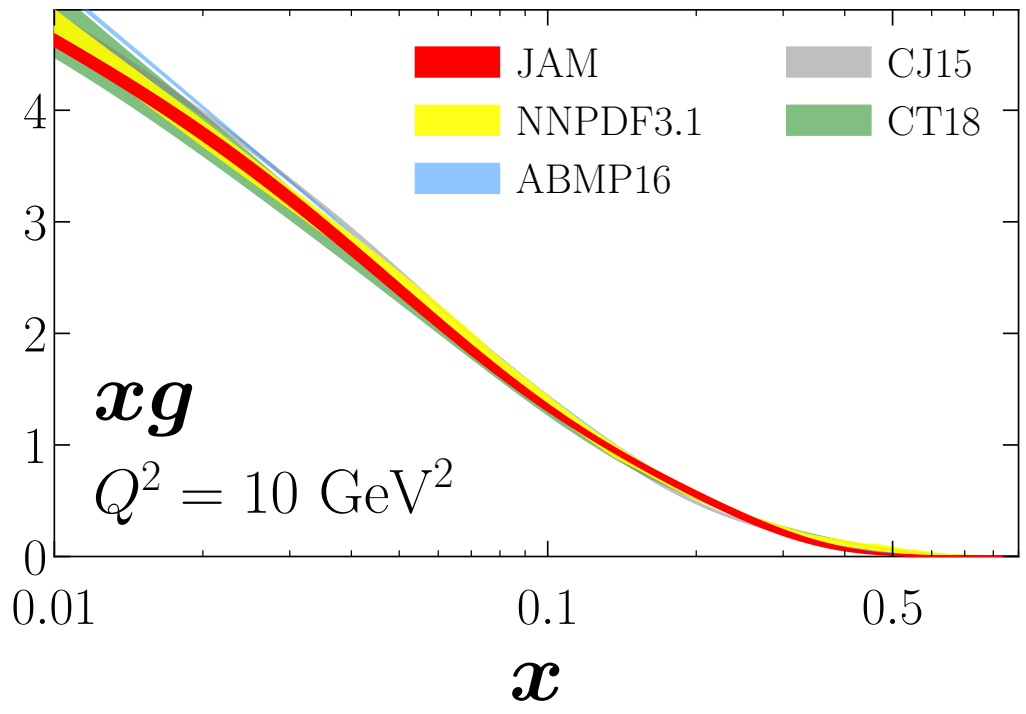

**Fig. 10.2.7** Unpolarized gluon PDF $xg$ from various QCD global analyses at a scale of $Q^2 = 10$ GeV$^2$ from several global QCD analyses, as in Fig. 10.2.3.

Since that the gluon PDF is accompanied by $\alpha_s$ in DIS structure functions, in practice there is a correlation between the value of $\alpha_s$ obtained in global PDF analyses and the shape of gluon distribution, with larger $\alpha_s$ leading to a smaller $g(x)$ at small $x$ and (via the momentum sum rule correlation) a larger $g(x)$ at large $x$. An interesting question is whether $\alpha_s$ should be fitted as a parameter in global analyses or, since it is a parameter of the QCD Lagrangian and should be the same for all processes, fixed to the world average value for $\alpha_s(M_Z)$. Comparisons of results with $\alpha_s(M_Z)$ fitted or fixed may indicate which processes are responsible for any differences [3046].

A related question is the shape of heavy quark PDFs, such as the charm distribution, which is known to contribute $\sim 30\%$ of the total $F_2$ measured at HERA at small $x$ values. Here the main production mechanism is photon-gluon fusion, so that data on inclusive charm production could also provide valuable constraints on the gluon PDF in the nucleon. The question of whether there is a sizable nonperturbative charm component at a low energy input scale [3076–3080] also remains controversial [3081–3083], with recent analyses claiming both positive [3084] and negative evidence [3085].

### 10.2.5 Spin-dependent PDFs

Considerable progress has been made in understanding the spin structure of the nucleon since the first precision polarized DIS experiments at CERN in the late 1980s indicated an anomalously small fraction of the proton spin carried by quarks. A rich program of spin-dependent inclusive and semi-inclusive DIS, as well as polarized proton-proton scattering experiments has followed, vastly improving our knowledge of spin-dependent PDFs of the nucleon over the last two decades. While the spin-dependent data have not been as abundant as those available for constraining spin-averaged PDFs, several dedicated global QCD analyses of spin-dependent PDFs to be performed. The main current global efforts include the DSSV group [1295, 3086, 3087], the NNPDF collaboration [983, 3088], and the JAM collaboration [982, 3089], extending earlier efforts by the LSS [3064], BB [3090], KATAO [3091] and AAC [3092] groups.

*Polarized valence quarks*

As for the unpolarized PDFs, the spin-dependent $\Delta u^+$ distribution is the most strongly constrained helicity PDF, largely by the proton $g_1$ structure function data. The corresponding $\Delta d^+$ distribution, which has a negative sign, is smaller in magnitude compared with $\Delta u^+$ and has larger relative uncertainties, especially at intermediate and large values of $x$. The size of the uncertainties depends somewhat on the theoretical assumptions made for the distributions. For example, if one assumes only the SU(2) symmetry constraint (10.2.12) for the difference $\Delta u^+ - \Delta d^+$, the uncertainties on the individual $\Delta u^+$ and $\Delta d^+$ PDFs are significantly larger than assuming in addition the SU(3) symmetry relation (10.2.13) involving also the strange polarization $\Delta s^+$.

This is illustrated in Fig. 10.2.8 for the JAM parametrization [984], which also shows the result of a fit that enforces in addition positivity constraints on the unpolarized PDFs. Whether spin-averaged PDFs need to be positive beyond LO in $\alpha_s$ has been debated recently in the literature [3093], and generally it is understood that the positivity constraint should hold only at LO [3094]. The general features of the $\Delta u^+$ and $\Delta d^+$ PDFs in Fig. 10.2.8 are similar to those found by other global QCD analysis groups [983, 3087], which reflects the common origin in the constraints on these PDFs from proton and neutron DIS data. In contrast, without the additional assumption of SU(3) symmetry [982, 3095], the strange helicity PDF remains largely unconstrained [984, 1296].

*Polarized sea quarks*

Since inclusive polarized DIS experiments measure C-even combinations of PDFs, $\Delta q^+$, additional constraints, either from theory or experiment, are needed to separate the individual quark and antiquark distributions. Additional experimental constraints come from the semi-inclusive production of hadrons, in which spin-dependent PDFs are weighted by fragmentation functions, as well as particle production in polarized hadron collisions, which involve products of spin-dependent (and spin-averaged) PDFs.

The strongest constraints on the polarization of the sea have come from recent $W$-lepton production data from polarized protons collisions at RHIC [3096–3098]. The effect of the polarized $W$ data is a clear nonzero antiquark asymmetry $\Delta\bar{u} - \Delta\bar{d}$ for $0.01 \lesssim x \lesssim 0.3$, as Fig. 10.2.9 illustrates for the recent JAM analysis [3089]. Qualitatively similar, although not as pronounced, behavior was also observed in the earlier DSSV [3087] and NNPDF [983] fits, although these made stronger theoretical assumptions about PDF positivity and SU(3) symmetry. The observed polarized sea asymmetry is also similar to expectations from some nonperturbative models of the nucleon [3099–3102].

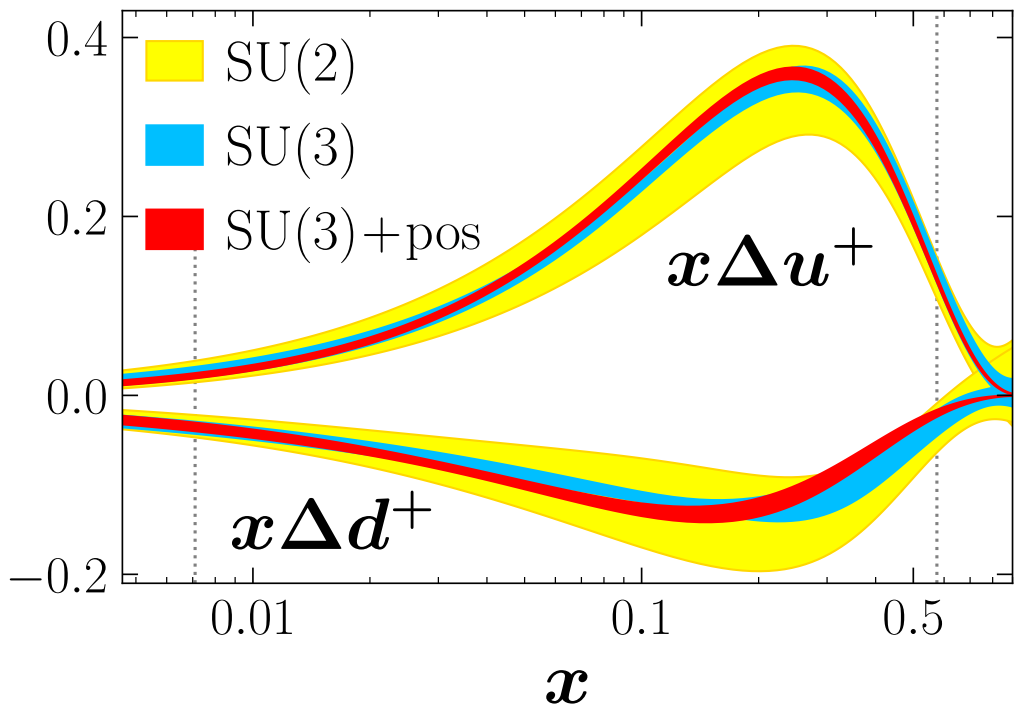

**Fig. 10.2.8** Polarized $x\Delta u^+$ and $x\Delta d^+$ PDFs from the JAM analysis [984] for various scenarios: assuming SU(2) symmetry (10.2.12) (yellow bands), SU(3) symmetry (10.2.13) (blue bands), and in addition the PDF positivity constraint (red bands).

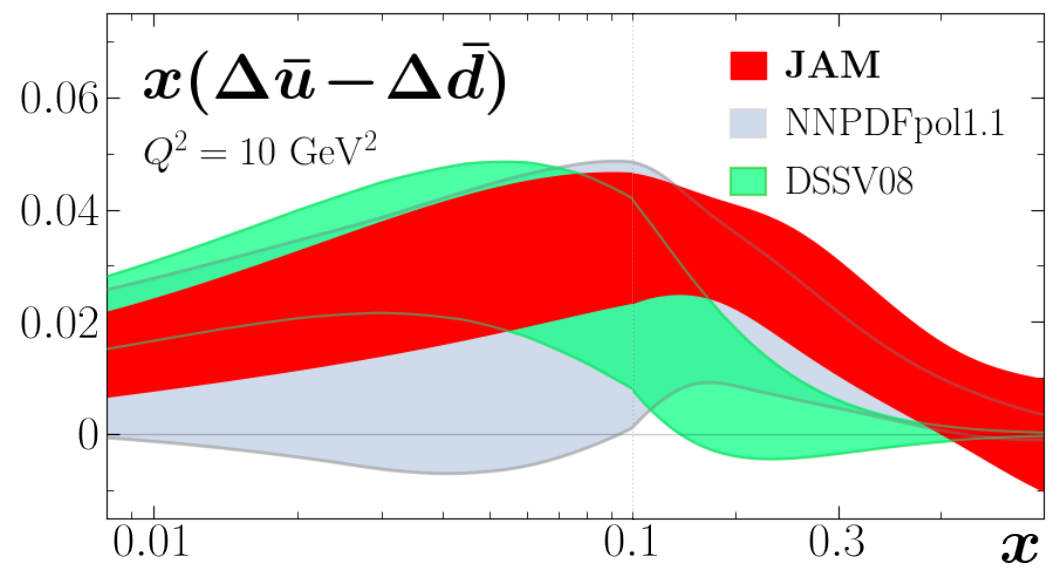

**Fig. 10.2.9** Polarized sea quark asymmetry $x(\Delta\bar{u}-\Delta\bar{d})$ from the JAM [3089], NNPDF [983] and DSSV [3087] analyses.

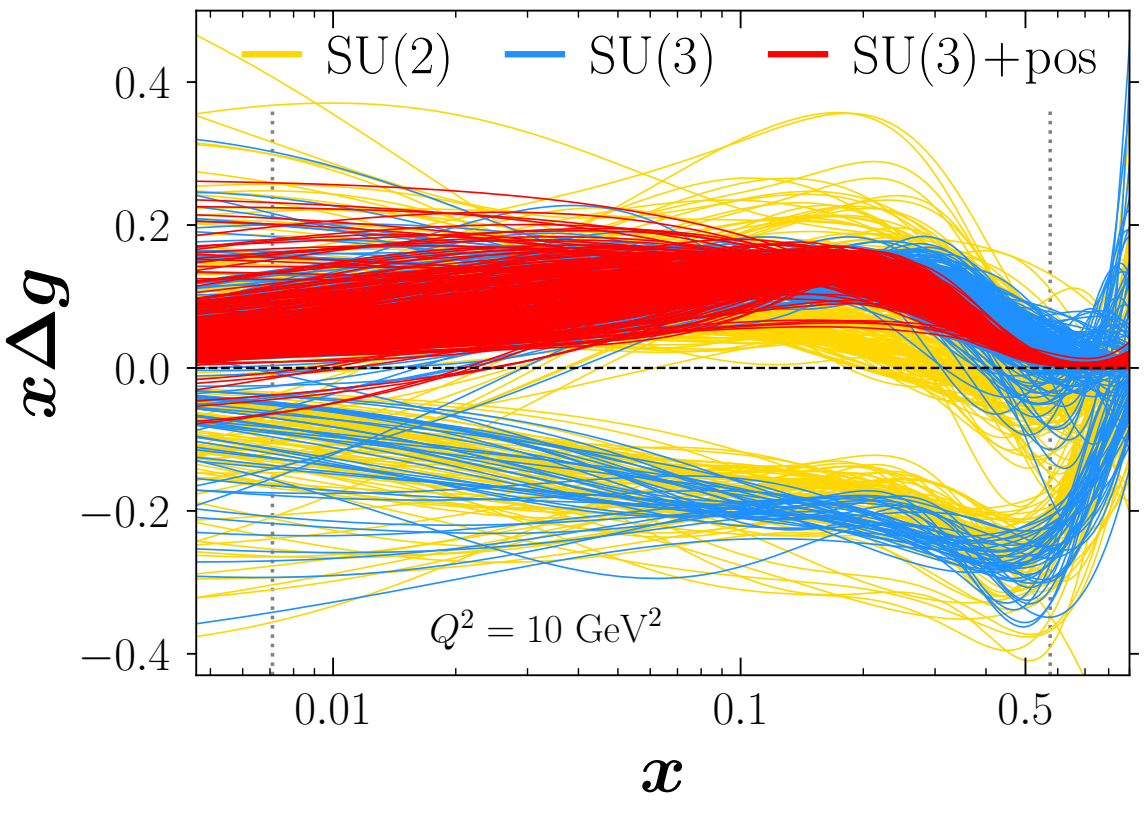

**Fig. 10.2.10** Monte Carlo replicas for the gluon helicity PDF $x\Delta g$ fitted under various theory assumptions according to the SU(2) (yellow lines), SU(3) (blue lines) and SU(3)+positivity (red lines) scenarios [984].

*Polarized gluons*
The sign and magnitude of the gluon polarization is a critical component to understanding the decomposition of the proton's spin amongst its quark and gluon constituents. The first clear indication of a positive $\Delta g$ came from analysis of RHIC jet production data in polarized proton-proton collisions, which were used by the DSSV group to extract a nonzero signal for gluon momentum fractions between $x \approx 0.05$ and $\approx 0.2$. More recently, the JAM collaboration [984] performed a simultaneous global fit to unpolarized and polarized data, testing in particular the sensitivity to theoretical assumptions about axial charges and PDF positivity.

The results of the simultaneous analysis, illustrated in Fig. 10.2.10, show that indeed the gluon helicity can depend strongly on the constraints imposed. Interestingly, without restricting PDFs to be positive and assuming SU(3) flavor symmetry for the axial vector charges, existing polarized data allow solutions containing negative gluon polarization, in addition to the standard positive gluon solutions found in earlier analyses, giving equally acceptable descriptions of the data. A negative gluon polarization would imply rather large quark or gluon orbital angular momentum contributions, in order to satisfy the proton spin sum rule. It will be important to verify the sign and magnitude of the gluon polarization in future experiments [3103], as well as explore possible insight gained from lattice QCD calculations [3104].

### 10.2.6 Outlook

Our knowledge of the detailed partonic structure of the nucleon has improved tremendously in recent years, thanks to high precision experiments and advances in computational and data analysis methods. With planned measurements at facilities such as Jefferson Lab, the LHC, and the future Electron-Ion Collider (EIC) [3105], we can look forward to further breakthroughs in addressing long-standing questions about the momentum and spin distributions of quarks and gluons in the nucleon.

The new experiments will probe hitherto unexplored corners of kinematics in which PDFs have been difficult to determine. An example is the behavior of PDFs and PDF ratios such as $\Delta q^+/q^+$ in the limit as $x \to 1$, which are particularly sensitive to the details of nonperturbative quark-gluon dynamics [3024, 3106]. The new data will allow one to test basic theoretical assumptions such as SU(2) and SU(3) symmetry, PDF positivity, and charge symmetry in PDFs. The latter, which is expected to be broken by light quark mass differences, $m_u \neq m_d$, and by electromagnetic corrections, will need to be taken into account if one hopes for PDF accuracy at the few-percent level. Further inroads into solving the proton spin puzzle, through the determination of the total spin contributions from quarks, antiquarks and gluons, will require measurements of spin structure functions down to smaller values of $x$ [3107, 3108], which will be one of the focuses of the EIC program [3105].

The aim of few-percent precision in PDFs will also require a more systematic treatment of radiative effects, which in the past have been treated using approximate prescriptions. Recently, a combined QED+QCD approach to factorization has been developed [3109], and while the differences with the traditional methods are not large for inclusive processes, for more exclu-sive reactions, such as semi-inclusive DIS [3110], the simultaneous paradigm of self-consistently incorporating QED and QCD effects and determining different types of distributions within the same analysis will be necessary.

Along with the new measurements, it is likely that complementary information will be needed from lattice QCD simulations, especially for quantities that will be difficult to access from experiment. Indeed, the first exploratory simultaneous analyses of experimental and lattice data have already been made recently [1094, 3111]. Future success in mapping out and understanding the quark and gluon structure of the proton will thus require a coordinated effort on the multiple fronts of experiment, theory, lattice simulation, and data analysis.

## 10.3 Spin structure

**Xiangdong Ji**

The nucleon (proton and neutron) is a spin-1/2 composite particle made from three valence quarks. Every model of the nucleon gives an explanation for its spin structure [3112–3116], from the Skyrme model [3112], to Gell-Mann and Zweig's quark model [3113, 3114], and to many other models popular in 70's and 80's [3115, 3116]. The simplest and most successful one is the quark model which, among others, inspired the discovery of QCD [50], predicted that the entire nucleon spin is carried out by the three valence quarks [26, 2666, 3117]. The non-relativistic quark model has indeed a simple explanation for the nucleon spin and the associate magnetic moments [26], also for their excited states [2666]: Three constituent quarks are all in the $s$-wave orbit in the nucleon, and their spins couple to 1/2 in a way consistent with the $\mathrm{SU}(2_{\mathrm{spin}} \times 3_{\mathrm{flavor}})$, a combined spin-flavor symmetry group [3117].

The quark-model picture for the spin was put under a test through polarized deep-inelastic scattering (DIS) on a polarized proton [3118]. The EMC collaboration made the first definitive measurement for the fraction of the proton spin carried by quarks in 1987 [3119, 3120], and the result

$$\Delta\Sigma(Q^2 = 10.7\mathrm{GeV}^2) = 0.060 \pm 0.047 \pm 0.069, \quad (10.3.1)$$

is consistent with zero. The discrepancy has inspired large amount of experimental and theoretical studies which have been summarized in a number of excellent reviews [3121–3125]. Perhaps the most important lesson we have learned is that the QCD quarks probed in polarized DIS are very different from those in the constituent quark models, and that QCD has a much more sophisticated way to build up the proton spin.

Understanding the nucleon spin in QCD remains an important challenge in hadron structure physics, particularly, in experiment. In the following, we will briefly review the current status and future perspective for this topic, focusing on the questions such as: does it make sense to talk about the different parts of the proton spin? What will be an interesting decomposition for the spin? To what extent do we believe that we can measure these parts experimentally? How can we calculate these contributions in fundamental theory and put them to experimental tests?

### 10.3.1 Spin sum rules in QCD

Angular momentum (AM) or spin structure of a composite system can be studied through various contributions to the total. In quantum field theories, the individual parts are renormalization scale and scheme dependent, although the total is not. The most popular convention in the literature is to use dimensional regularization and modified minimal subtraction, indicated by the dependence on the scale $\mu$. To understand the proton spin, we can start from QCD AM operator expressed in terms of individual sources,

$$\vec{J}_{\mathrm{QCD}} = \sum_\alpha \vec{J}_\alpha(\mu) \ . \quad (10.3.2)$$

Through the above, one can express the total spin 1/2 as contributions from different parts. This has been one of the main methods to explore the origins of the proton spin in the literature. Since the individual contributions are the expectation values of the AM sources in the entire wave function, they are neither integers nor half-integers: they are the quantum mechanical average of probability amplitudes.

There exists more than one way to split the AM and derive spin sum rules for the proton. However, a physically-interesting spin sum rule shall have the following properties:

*Experimental Measurability*
The overwhelming interest in the proton spin began with the EMC data. Much of the followup experiments, including polarized RHIC [3126], Jefferson Lab 12 GeV upgrade [3127] and Electron-Ion Collider (EIC) [820, 3128], have been partially motivated to search a full understanding of the proton spin.

*Frame Dependence*

Since spin is an intrinsic property of a particle, one naturally searches for a description of its structure independent of a reference frame. How the individual contributions depend on the proton momentum or reference frame requires understanding of the Lorentz transformation properties of $\vec{J}_\alpha$. Moreover, the longitudinal and transverse spins behave differently under frame transformation and therefore have very different experimental implications. Since the proton structure probed in high-energy scattering is best described in the infinite momentum frame (IMF), a partonic picture of the spin is phenomenologically interesting to explore.

In the rest frame, the proton state $|\vec{P}=0,\vec{s}\rangle$ can be defined with the angular momentum quantized along $\vec{s}$,

$$\vec{s}\cdot\vec{J}\left|\vec{P}=0,\vec{s}\right\rangle = 1/2\left|\vec{P}=0,\vec{s}\right\rangle \ , \tag{10.3.3}$$

where we have dropped the "QCD" subscript on $\vec{J}$. Boosting the above to an arbitrary Lorentz frame, one has ($\hbar=1$)

$$(-W^\mu S_\mu)|PS\rangle = 1/2|PS\rangle \ , \tag{10.3.4}$$

where $|PS\rangle$ have definite four-momentum $P^\mu$ and spin polarization four-vector $S^\mu$, $S^\mu=(\gamma\vec{s}\cdot\vec{\beta},\ \ \vec{s}+(\gamma-1)\vec{s}\cdot\hat{\beta}\hat{\beta})$ with $S^\mu S_\mu=-1$, $P^\mu S_\mu=0$, $\hat{\beta}$ the direction of $\vec{\beta}=\vec{v}/c$, $\gamma=(1-\beta^2)^{-1/2}$ the boost factor, and $W^\mu$ is the relativistic spin (or Pauli-Lubanski) four-vector ($\epsilon^{0123}=1$) [3129]

$$W^\mu = -\tfrac{1}{2}\epsilon^{\mu\alpha\lambda\sigma}J_{\alpha\lambda}P_\sigma/M, \tag{10.3.5}$$

$$= \gamma(\vec{J}\cdot\vec{\beta},\ \ \vec{J}+\vec{K}\times\vec{\beta}) \tag{10.3.6}$$

where $\vec{K}$ is the Lorentz-boost operator defined in terms of the $0\,i$ components of the Lorentz generator $J^{\alpha\beta}$. In the second line of the equation, we have replaced the four-momentum operator $P_\sigma$ by its eigenvalue specifying a Lorentz frame $\vec{\beta}$. One can use Eq. (10.3.4) to develop spin sum rules in any frame,

$$\langle PS|(-W^\mu S_\mu)|PS\rangle = 1/2 \ , \tag{10.3.7}$$

by expressing the left-hand side as the sums of expectation values. Thus the covariant spin is not only related to the AM operator but also to the boost $\vec{K}$. However, it is desirable to develop a spin picture in terms of the AM operator alone in a general Lorentz frame.

Without loss of generality, one can assume the proton momentum is along the $z$-direction $\vec{P}^z=(0,0,P^z)$. In the case of longitudinal polarization, one has $\vec{s}_z=(0,0,1)$, $-W^\mu S_\mu=J^z$, and Eq. (10.3.7) becomes the total helicity,

$$\langle PS_z|J^z|PS_z\rangle = 1/2 \ , \tag{10.3.8}$$

which is boost-invariant along the $z$-direction. This is a starting point to construct helicity sum rules. Since the helicity is independent of momentum, the individual contributions are generally sub-leading order in high-energy scattering.

For transverse polarization along the $x$-direction, $\vec{s}_x=(1,0,0)$ , and Eq. (10.3.7) becomes

$$\langle PS_x|\gamma(J^x-\beta K^y)|PS_x\rangle = 1/2 \ , \tag{10.3.9}$$

which contains the boost operator $K^y$ from the transformation of $J^x$ under the Lorentz boost along $z$. Since $\vec{K}$ and $\vec{J}$ transform under Lorentz tranformation as $(1,0)+(0,1)$, we can deduce separate relations:

$$\langle PS_x|J^x|PS_x\rangle = \gamma/2$$
$$\langle PS_x|K^y|PS_x\rangle = \gamma\beta/2\,,$$

true as expectation values. Therefore a transverse polarization sum rule from the AM operator starts from

$$\langle PS_x|J^x|PS_x\rangle = \gamma/2 \ . \tag{10.3.10}$$

Because the transverse angular momentum $J^x$ depends on the longitudinal momentum of the proton, its expectation value grows under boost, a fact less appreciated in the literature.

To obtain a spin sum rule, we need an expression for the QCD AM operator. It can be derived through Noether's theorem based on space-time symmetry of the QCD lagrangian density. Straightforward calculation yields the *canonical* AM expression [3130]

$$\vec{J}_{\rm QCD} = \int d^3\vec{x}\left[\psi_f^\dagger\tfrac{1}{2}\vec{\Sigma}\psi_f + \psi_f^\dagger\vec{x}\times(-i\vec{\partial})\psi_f + \vec{E}_a\times\vec{A}_a + E_a^i(\vec{x}\times\vec{\partial})A_a^i\right] \ , \tag{10.3.11}$$

where $\psi_f$ is a quark field of flavor $f$,$\vec{\Sigma}=\mathrm{diag}(\vec{\sigma},\vec{\sigma})$ with $\vec{\sigma}$ the Pauli matrices, $A_a^i$ vector potentials of gauge fields with color $a=1,...8$, $E_a^i$ color electric fields, and the contraction of flavor and color indices is implied. The above expression contains four different terms, each of which has clear physical meaning in free-field theory. The first term corresponds to the quark spin, the second to the quark orbital angular momentum (OAM), the third to the gluon spin, and the last one to the gluon OAM. Apart from the first term, the rest are not manifestly gauge-invariant under the general gauge transformation $A^\mu\to U(x)\,(A^\mu+(i/g)\partial^\mu)\,U^\dagger(x)$. However, the total is invariant under the gauge transformation up to a surface term at infinity which can be ignored in physical state matrix elements.

On the other hand, using the Belinfante improvement procedure (Belinfante, 1939) one can obtain a gauge-invariant form [1085],

$$\vec{J}_{\rm QCD} = \int d^3x \left[\psi_f^\dagger \frac{1}{2}\vec{\Sigma}\psi_f + \psi_f^\dagger \vec{x}\times(-i\vec{\nabla} - g\vec{A})\psi_f + \vec{x}\times(\vec{E}\times\vec{B})\right] , \quad (10.3.12)$$

All terms are manifestly gauge invariant, with the second term as mechanical or kinetic OAM, and the third term gluon AM.

*Helicity sum rule*

Using Eq. (10.3.8) and the gauge-invariant QCD AM in Eq.( 10.3.12), one can can write down a helicity sum rule [1085],

$$\frac{1}{2}\Delta\Sigma(\mu) + L_q^z(\mu) + J_g(\mu) = 1/2 \quad (10.3.13)$$

where $\Delta\Sigma/2$ is the quark helicity contribution, and $L_q^z$ is quark OAM contribution. Together, they give the total quark AM contribution $J_q$. The last term, $J_g$, is the gluon contribution. Both contributions can be obtained from the twist-two form factors of the energy-momentum tensor $T^{\mu\nu}$ [1085] (see below). One important feature of the above sum rule is that it is independent of the proton's momentum [3131]. This is an important feature because the sources of the proton spin does not depend on observer's reference frame so long as helicity is a good quantum number.

On the other hand, the canonical form of the AM operator in Eq.(10.3.11) allows deriving an infinite number of helicity sum rules with choices of gauges and/or frames of reference [3124, 3132]. The usefulness of such sum rules are questionable as they are not relevant to experiment. However, the gluon spin contribution in the IMF and light-cone gauge $A^+ = 0$ is measurable. Jaffe and Manohar proposed a canonical spin sum rule in a nucleon state with $P^z = \infty$ [3130],

$$\frac{1}{2}\Delta\Sigma + \Delta G + \ell_q + \ell_g = \frac{1}{2} \quad (10.3.14)$$

where $\Delta G$ is the gluon helicity and $\ell_{q,g}$ are quark and gluon OAM, respectively. Considerable attention has been given to the above sum rule because of its relevance to high-energy scattering. For example, the total quark helicity contribution can be written in terms of parton sum rule,

$$\Delta\Sigma = \int_{-1}^{1} dx(\Delta u(x) + \Delta d(x) + ....) , \quad (10.3.15)$$

where $\Delta q(x)$ is the quark helicity distribution function. Moreover, $\Delta G$ has been defined and measured experimentally as the first moment of the gauge-invariant polarized gluon distribution [3133]

$$\Delta G(Q^2) = \int_0^1 dx\ \Delta g(x, Q^2) ,$$
$$\Delta g(x) = \frac{i}{2x(P^+)^2}\int \frac{d\lambda}{2\pi} e^{i\lambda x} \times \langle PS|F^{+\alpha}(0)W(0,\lambda n)\tilde{F}_\alpha{}^+(\lambda n)|PS\rangle , \quad (10.3.16)$$

where $\tilde{F}^{\alpha\beta} = \frac{1}{2}\epsilon^{\alpha\beta\mu\nu}F_{\mu\nu}$, and the light-cone gauge link $W(\lambda n, 0)$ is defined in the adjoint representation of SU(3). In the light-cone gauge $A^+ = 0$, the nonlocal operator in Eq. (10.3.16) reduces to the free-field form in the Jaffe-Manohar sum rule. Additionally, one can write a parton sum rule for each of the OAM contributions

$$\ell_q = \int_{-1}^{1} dx \ell_q(x) , \quad (10.3.17)$$

$$\ell_g = \int_{-1}^{1} dx \ell_g(x) , \quad (10.3.18)$$

which give a more detailed picture of AM distributions in partons compared with the frame-independent sum rule above.

It appears that one can define a gauge-variant quantity which can be measured in experiment! This has inspired much debate about the gauge symmetry properties of the gluon spin operator and myriads of experimentally-unaccessible spin sum rules [3124]. It turns out, however, that the key is not about generalizing the concept of gauge invariance, it is about the proton state in the IMF [3134]. In particular, $A^+ = 0$ is a physical gauge as it leaves the transverse polarizations of the radiation field intact. This justifies the physical meaning of $\vec{E}\times\vec{A} = \vec{E}_\perp \times \vec{A}_\perp$ as the gluon spin (helicity) operator in the Jaffe-Manohar sum rule.

Comparing the two helicity sum rules Eqs. (10.3.13) and (10.3.14) above, they must be related in some way in the IMF. In fact, their relation is [3135, 3136]

$$J_g = \Delta G + \ell_g + \ell_{\rm int} \quad (10.3.19)$$

$$L_q = \ell_q - \ell_{\rm int} \quad (10.3.20)$$

where $\ell_{\rm int}$ represents the interaction AM and does not have a simple parton interpretation.

*Transverse spin sum rules*

For transverse polarization, a spin sum rule is less straightforward and much controversy exists in the literature [3124, 3137]. First of all, the transversely-polarized proton is not an eigenstate of the transverse AM operator. Second, the expectation value of the transverse AM has a intriguing frame dependence due to the center-of-mass contribution, which must be properly subtracted. Finally, there are two contributions to the transverse AM which transform differently under Lorentz boost and must combine properly to generate the total result. The delicate balance of two contributions entails two separate transverse spin sum rules.

The transverse spin has a simple frame-independent sum rule [3138],

$$J_q + J_g = 1/2 \ , \tag{10.3.21}$$

which is the same as the helicity sum rule due to Lorentz symmetry. One can separate the contributions to the quark into spin and orbit ones, however, such a separation is frame-dependent and therefore less interesting.

In the IMF, the above sum rule becomes partonic sum rules [3137, 3139],

$$J_q = \int_{-1}^{1} dx J_q(x) \ , \tag{10.3.22}$$

$$J_g = \int_{-1}^{1} dx J_g(x) \ , \tag{10.3.23}$$

where $J_q(x)$ and $J_g(x)$ are twist-2 transverse angular momentum densities of the quarks and gluons. They are related to quark and gluon unpolarized densities and generalized parton distributions through $J_q(x) = (1/2)x(q(x)+E_q(x))$ and $J_g(x) = (1/2)x(g(x)+E_g(x))$.

The second transverse spin sum rule can best be discussed in the IMF, where there is a sub-leading partonic sum rule for the transverse spin, corresponding to the twist-three part of the canonical angular momentum density $J^\perp$ in Eq. (10.3.11). In a simple form, one can write [3140]

$$\frac{1}{2}\Delta\Sigma_T + \Delta G_T + \ell_{qT} + \ell_{gT} = \frac{1}{2} \ . \tag{10.3.24}$$

The various terms have partonic interpretations in the IMF,

$$\Delta\Sigma_T = \int_{-1}^{1} dx g_T(x) \ , \tag{10.3.25}$$

$$\Delta G_T = \int_{-1}^{1} dx \Delta G_T(x) \ , \tag{10.3.26}$$

$$\ell_{qT} = \int_{-1}^{1} dx \ell_{qT}(x) \ , \tag{10.3.27}$$

$$\ell_{gT} = \int_{-1}^{1} dx \ell_{gT}(x) \ , \tag{10.3.28}$$

where $g_T(x) = g_1(x) + g_2(x)$ and $G_T(x)$ are transverse spin densities of quarks and gluons, respectively, and $\ell_{qT}$ and $\ell_{gT}$ are the corresponding twist-three transverse OAM densities. Because of Lorentz symmetry, the values of these integrated quantities with $T$ are exactly the same as the ones without $T$ in Jaffe-Manohar sum rule. However, the parton densities for the transversely polarized proton are different from those in the longitudinally polarized one. For instance, for the quark spin, the difference is the well-known $g_2(x)$ structure function.

### 10.3.2 Lattice Calculations

At present, the only systematic approach to solve the QCD proton structure is the lattice field theory [80], see, Sec. 4. There are less systematic approaches such as Schwinger-Dyson (Bethe-Salpeter) equations [765] and instanton liquid models [1371] in which a certain truncation is needed to find a solution, see, Sec. 5. Although much progress has been made in these other directions, we focus on the lattice QCD method.

A complete physical calculation on the lattice faces a number of obstacles. First the angular momentum is flavor-singlet quantity, and as such, one needs to compute the disconnected diagrams for the quarks. Since up and down quarks are light, computational demands at the physical pion mass are very high. Moreover, one also has to compute gluon observables to complete the picture, which is known to be very noisy. At the same time, one needs to keep the lattice space sufficiently small and the physical volume large enough. All of these add up to an extremely challenging task. However, a computation with all these issues considered has become possible recently, see for example Ref. [3146]. An additional challenge is present in computing light-cone correlations with a real time variable. The recent development of large-momentum effective theory (LaMET) has opened the door for such computations [600, 601, 609].

The matrix elements of local operators, $\Delta\Sigma$, $J_q$ and $J_g$ are relatively simple to calculate using the standard lattice QCD technology. Much progress has been made in understanding the content of manifestly gauge-invariant helicity sum rule in Eq.(10.3.13), and also the transverse spin sum rule in Eq.(10.3.21).

The first calculations have been about the $\Delta\Sigma$ from different quark flavors. A large amount of work has been summarized in a recent review [3147]. Three most recent calculations are in Refs. [3141–3143], with some at the physical quark mass. Table 10.3.1 is taken from Ref. [3145] and shows a summary of the recent lattice results on the quark helicity. The strange quark contribution was also calculated in Ref. [3148, 3149] through the anomalous Ward identity, and $\Delta s + \Delta\bar{s} =$

| | $\Delta u$ | $\Delta d$ | $\Delta s$ | $g_A^3 = \Delta u - \Delta d$ | $\Delta(u+d)$(CI) | $\Delta(u/d)$(DI) | $\Delta\Sigma$ |
|---|---|---|---|---|---|---|---|
| Cyprus | 0.830(26)(4) | -0.386(16)(6) | -0.042(10)(2) | 1.216(31)(7) | 0.598(24)(6) | -0.077(15)(5) | 0.402(34)(10) |
| $\chi$QCD | 0.846(18)(32) | -0.410(16)(18) | -0.035(8)(7) | 1.256(16)(30) | 0.580(16)(30) | -0.072(12)(15) | 0.401(25)(37) |
| PNDME | 0.777(25)(30) | -0.438(18)(30) | -0.053(8) | 1.218(25)(30) | | | 0.286(62)(72) |
| de Florian *et al.* ($Q^2$=10 GeV$^2$) | $0.793^{+0.011}_{-0.012}$ | $-0.416^{+0.011}_{-0.009}$ | $-0.012^{+0.020}_{-0.024}$ | | | | $0.366^{+0.015}_{-0.018}$ |
| NNPDFpol1.1 ($Q^2$=10 GeV$^2$) | 0.76(4) | -0.41(4) | -0.10(8) | | | | 0.25(10) |
| COMPASS ($Q^2$=3 GeV$^2$) | $[0.82, 0.85]$ | $[-0.45, -0.42]$ | $[-0.11, -0.08]$ | 1.22(5)(10) | | | $[0.26, 0.36]$ |

**Table 10.3.1** Results of quark spin for the $u, d$ and $s$ flavors from three recent lattice calculations by the Cyprus group [3141], $\chi$QCD[3142], PNDME [3143] in the $\overline{\rm MS}$ scheme at 2 GeV are listed. $\Delta(u+d)$(CI) and $\Delta(u+d)$(DI) are the spins of the $u$ and $d$ quarks in the connected insertion (CI) and disconnected insertion (DI). Three analyses of experiments from de Florian *et al.* [1295], NNPDF [983] and COMPASS [3144] are also listed for comparison. Source: Ref.[3145].

$-0.0403(44)(78)$. The total quark spin contribution to the proton helicity is about 40%.

To calculate the total quark orbital and gluon AM contributions, one can start with the AM density, $M^{\mu\nu\lambda}$, of QCD, from which the AM operator is defined. It is well-known that the AM density is related to the energy-momentum tensor (EMT) $T^{\mu\nu}$ through [3130],

$$M^{\mu\nu\lambda}(x) = x^\nu T^{\mu\lambda} - x^\lambda T^{\mu\nu} \,. \tag{10.3.29}$$

The individual contributions to the EMT, hence AM density, can be written as the sum of quark and gluon parts,

$$T^{\mu\nu} = T_q^{\mu\nu} + T_g^{\mu\nu} \,, \tag{10.3.30}$$

where

$$T_q^{\mu\nu} = \frac{1}{2}\left[\bar\psi\gamma^{(\mu} i\overrightarrow{D}^{\nu)}\psi + \bar\psi\gamma^{(\mu} i\overleftarrow{D}^{\nu)}\psi\right] \,, \tag{10.3.31}$$

$$T_g^{\mu\nu} = \frac{1}{4}F^2 g^{\mu\nu} - F^{\mu\alpha}F^{\nu}{}_{\alpha} \,, \tag{10.3.32}$$

where $T_q$ includes quarks of all flavor. The expectation values of the AM densities can be derived from the off-forward matrix elements of EMT [1085],

$$\begin{aligned}\langle P'S|T^{\mu\nu}_{q/g}(0)|PS\rangle = \bar U(P'S)\Big[&A_{q/g}(\Delta^2)\gamma^{(\mu}\bar P^{\nu)} \\ &+ B_{q/g}(\Delta^2)\frac{\bar P^{(\mu} i\sigma^{\nu)\alpha}\Delta_\alpha}{2M} + C_{q/g}(\Delta^2)\frac{\Delta^\mu\Delta^\nu - g^{\mu\nu}\Delta^2}{M} \\ &+ \bar C_{q/g}(\Delta^2) M g^{\mu\nu}\Big] U(PS) \,,\end{aligned} \tag{10.3.33}$$

where $\bar P = (P+P')/2$, $\Delta = P' - P$, and $A$, $B$, $C$ and $\bar C$ are four independent form factors. It has been shown that

$$J_q = 1/2(A_q(0) + B_q(0)) \tag{10.3.34}$$

and similarly for the gluon.

The calculation of the total quark and gluon angular momenta started from Ref. [3150] in which the quark part including the disconnected diagrams was calculated without dynamical quarks. The result is the total quark contribution is $J_q = 0.30 \pm 0.07$, i.e. 60%; therefore about 40% of the proton spin must be carried by the gluon. Following other quenched studies [3151, 3152], dynamical simulations took over [3153–3157]. A complete study of the angular momentum decomposition was made in Ref. [3148] in quenched formalism, and later in Ref. [3141]. It was found that the quark orbital angular momentum contributes about 47% and gluon angular momentum contributes 28%.

A complete dynamical simulation at the physical pion mass has been finished recently [3146]. It was found

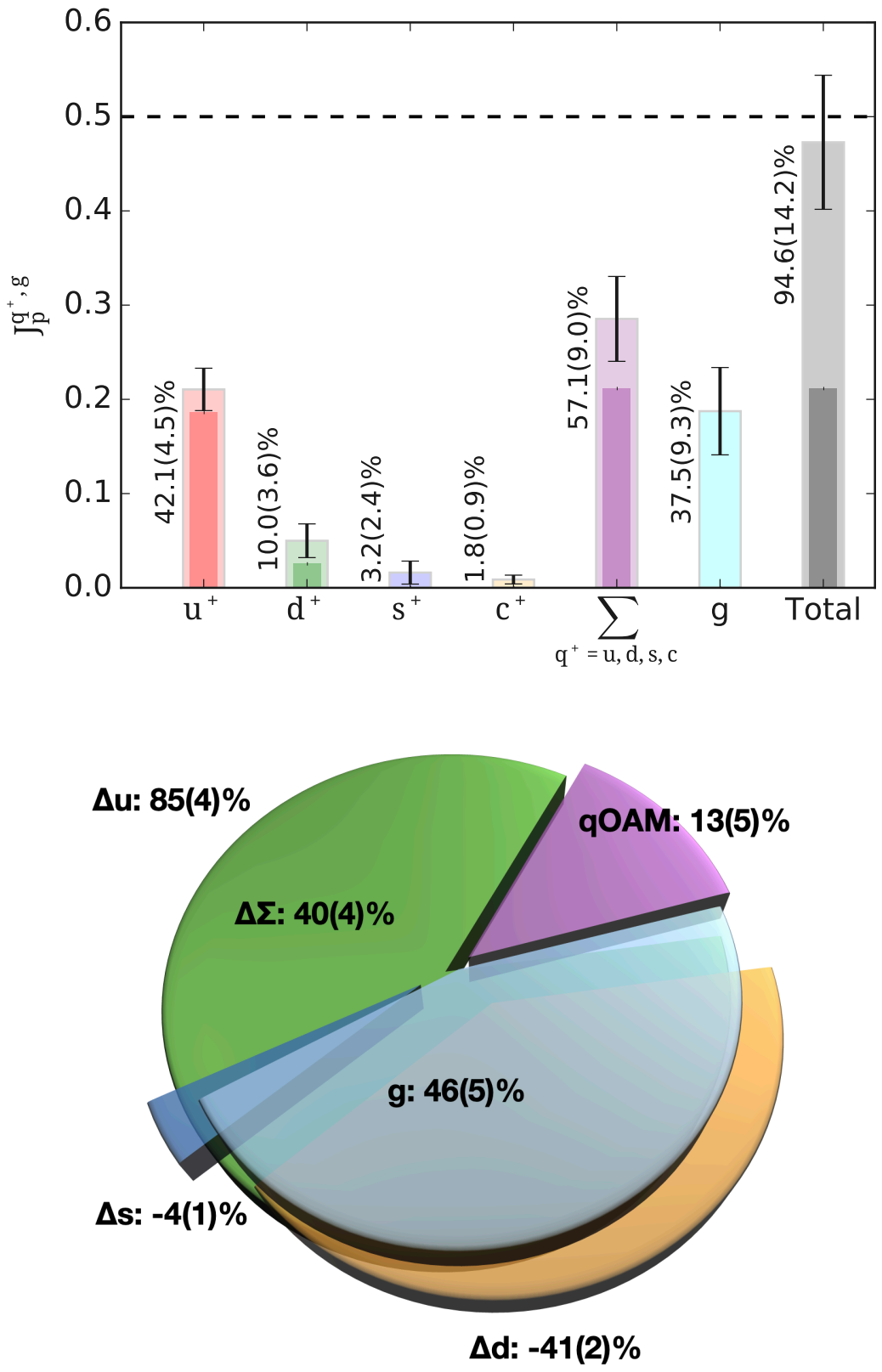

**Fig. 10.3.1** (upper) Proton spin decomposition in terms of different quark flavors and gluon from Ref. [3146]. (lower) Spin decomposition in terms of quark helicity, OAM and gluon contributions from Ref. [3158].

that the total quark spin contribution is about 38.2%, and the orbital angular momentum contribution of the quarks is about 18.8%, much reduced compared with quenched simulations. The total gluon contribution is 37.5%. The resulting picture is shown in Fig. 1. The total spin is 94.6% with an error bar of 14.2%. The spin decomposition in terms of the total quark helicity $\Delta\Sigma = \Delta u + \Delta d + \Delta s$, and quark OAM, and the gluon $J_g$ for $n_f = 2+1$ has been calculated in Ref. [3158].

Calculation of the gluon helicity has not been possible for many years because it is intrinsically a light-cone quantity. However, a progress in 2013 was made by studying the frame dependence of non-local matrix elements. One can match the large-momentum matrix element of a static "gluon spin" operator, which is calculable in lattice QCD, to $\Delta G$ in the IMF [3134]. This idea was a prototype of LaMET, which was soon put forward as a general approach to calculate all parton physics [600, 601]. Using LaMET, one can also calculate the polarized gluon helicity distribution $\Delta g(x)$ in a region of $x \sim 0.2-0.8$. However, the approach does not allow one to calculate the integrated $\Delta G$ starting from spatial correlation functions of gluon field strength.

The computation of parton OAM on lattice has been suggested in terms of lattice phase-space Wigner distribution, in which a quark bilinear non-local operator form factor is calculated [3159, 3160]. The non-local operator contains a Wilson line to make it gauge invariant. The canonical OAM can be constructed with Wilson lines along the main direction of the proton momentum going to infinity. One can in principle obtain the local gauge invariant OAM with a Wilson line connecting the two quark fields with a straight line. The result seems to be consistent with the calculation discussed above. The result in Ref. [3160] suggests that the isovector canonical OAM has a different sign from the mechanical one, and with a magnitude about 40% larger. One issue with this type of calculation is the renormalization, which can be done with LaMET matching.

One can also calculate the total parton OAM using local operators in a fixed gauge [3161] following the similar approach for the gluon helicity. Matching coefficients between IMF and finite momentum frame have been calculated. One particular feature of the calculation is fixed-gauge which is challenging both on lattice and QCD perturbation theory. On lattice, local gauge condition can lead to the Gribov copies; on the other hand, perturbation theory in a physical gauge requires better understanding at large orders.

Finally, the spin structure of the nucleon in the IMF requires calculations of various light-cone distributions, which include the quark and gluon helicity distributions $\Delta q(x)$, and $\Delta G(x)$, OAM distributions $J_q(x)$ and $J_g(x)$ through GPDs, and OAM distributions $\ell_q(x)$, $\ell_g(x)$, $g_2(x)$, $\Delta G_T(x)$, $\ell_{qT}(x)$, and $\ell_{gT}(x)$.

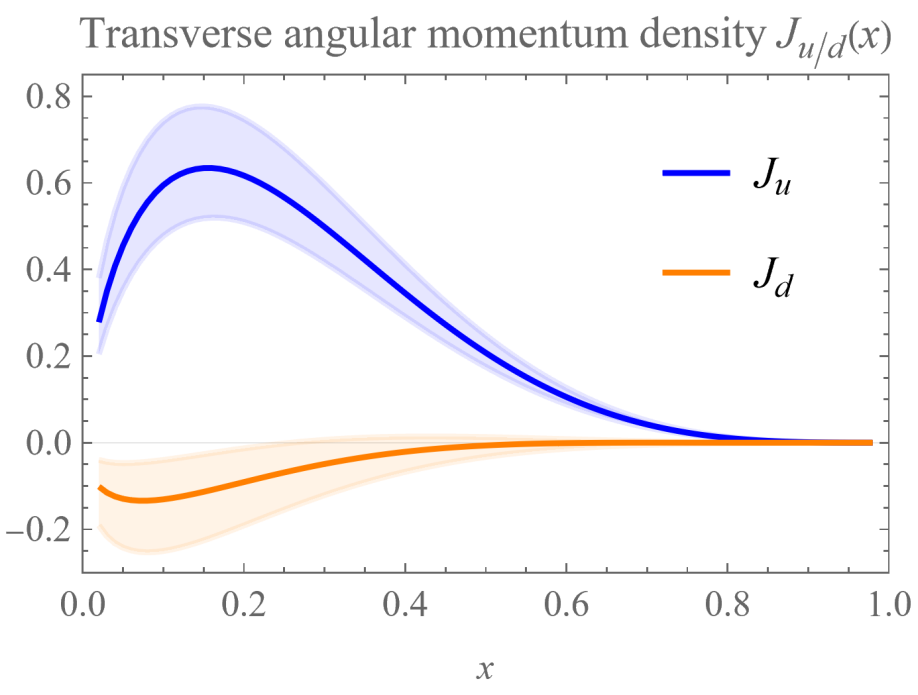

**Fig. 10.3.2** Angular momentum density distributions of up and down quarks in a transversely polarized proton, fitted to lattice form factors and GPDs [3162].

Shown in Fig. 10.3.2 are twist-2 angular momentum densities of up and down quarks in a transversely-polarized nucleon, obtained from phenomenological fit to lattice form factors and generelized parton distributions (GPDs) [3162]. They can be compared with direct lattice calculations and experimental data to be discussed below.

### 10.3.3 Experiments and phenomenology

Since the EMC experiments, there have been extensive experimental efforts around the globe to investigate the quark and gluon spin contributions to the proton spin, with two important improvements: higher precision and wider kinematic coverage. Majority of these efforts continued in line of the EMC experiment, measuring the polarized structure functions in inclusive DIS with polarized lepton on polarized target (proton, neutron, deuteron). Two important new initiatives have also emerged. First, the DIS experiment facilities extended their capabilities to measure the spin asymmetries in the semi-inclusive hadron production in DIS (SIDIS), which can help to identify the flavor structure in the polarized quark distributions. Second, the Relativistic Heavy Ion Collider (RHIC) at the Brookhaven National Laboratory (BNL) started the polarized proton-proton experiments. This facility opened new opportunities to explore the proton spin, in particular, for the helicity contributions from gluon and sea quarks (see the previous subsection for experimental data and analysis).

To take into account the constraints from all experiments, it is important to perform a global analysis of the polarized parton distributions from the world-wide data. In these analyses, one has to make some generic assumptions about the functional form (in terms of the

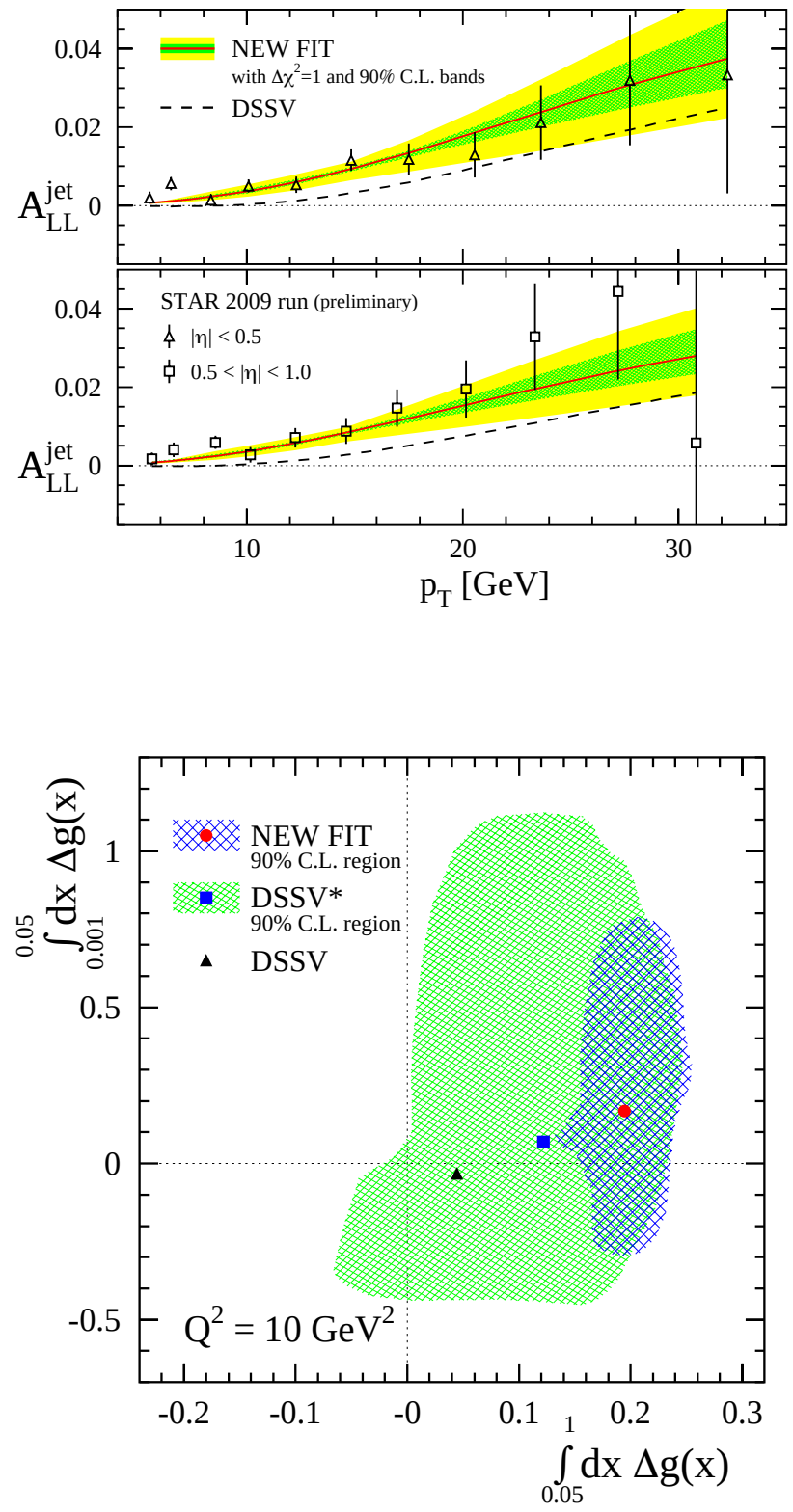

**Fig. 10.3.3** (upper) Double spin asymmetry in inclusive jet production at RHIC and (lower) constraints on the gluon helicity contribution to the proton spin. Source: Ref. [3086].

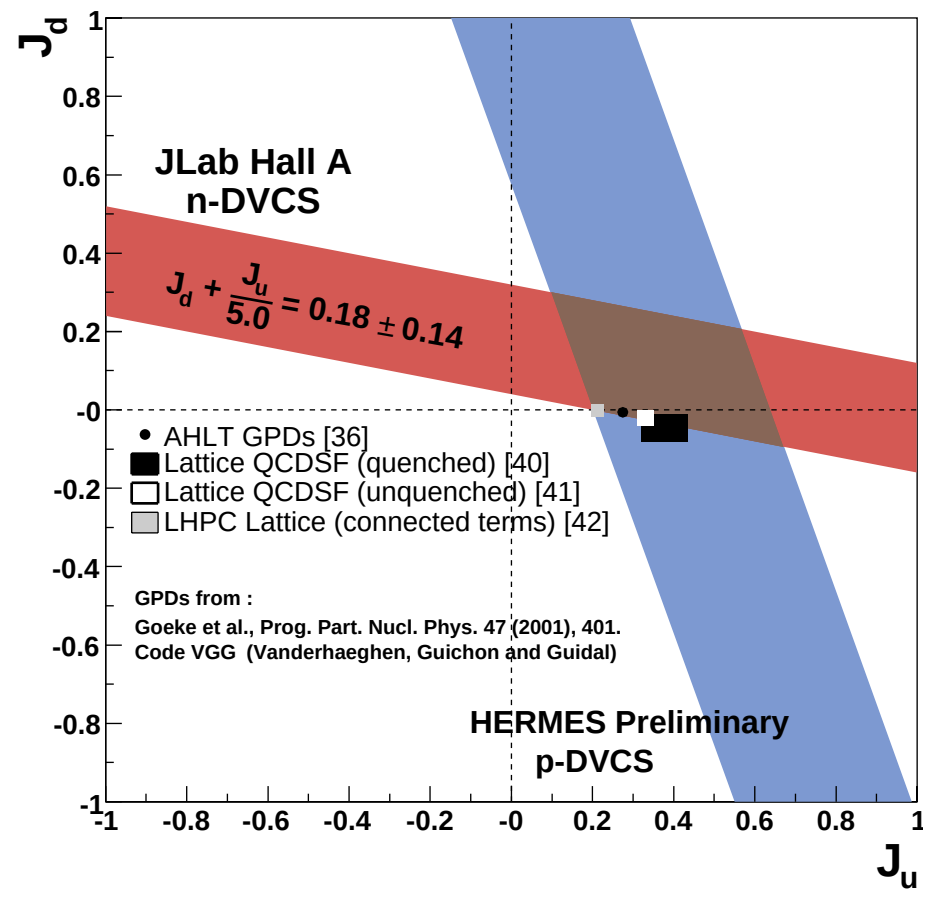

**Fig. 10.3.4** Model-dependent constraints on the up and down quark total angular momentum from DVCS measurement at JLab. Source: Ref. [3163].

unpolarized parton distributions) with a few parameters to fit to data, see, e.g., Refs. [983, 3086, 3144], where perturbative corrections have been included up to next-to-leading order. Very interesting results, in particular, for the double spin asymmetries in inclusive jet production from the RHIC experiments have provided more strong constraint on the gluon spin [3164], see Fig. 10.3.3. This promises great potential for future RHIC experiments to further reduce the uncertainties due to greater statistics [1278, 3165].

The total quark spin contribution to the proton spin $\Delta\Sigma$ has been well determined from the DIS measurements. For this quantity, all of the global fits agree well with each other, which essentially gives $\Sigma_q \approx 0.30$ with uncertainties around 0.05. However, for sea quark polarizations including $\bar{u}$, $\bar{d}$ and $s$ $(\bar{s})$, there exist great uncertainties, in particular, for the strange quark polarization [983, 1296, 3086], which mainly comes from SIDIS measurements from HERMES and COMPASS. Recently, it was also found that the $W$ boson spin asymmetries at $\sqrt{s} = 500$ GeV RHIC have also improved the constraints on $\bar{u}$ and $\bar{d}$ polarization [3166].

The OAM of the quarks may be extracted from measurement of GPD [1085],

$$J_q = \frac{1}{2}\Sigma_q + L_q = \lim_{t\to 0}\frac{1}{2}\int dx\,x\left[H^q(x,\xi,t) + E^q(x,\xi,t)\right] , \tag{10.3.35}$$

where $J_q$ is the total quark contribution to the proton spin, $H$ and $E$ are GPDs. After subtracting the helicity contribution $\Delta\Sigma$ from various experiments, the above equation will provide the quark OAM contribution to the proton spin. The GPDs can be measured in many different experiments, for example, deeply virtual compton scattering (DVCS) and hard exclusive meson production. Experimental efforts have been made at various facilities, including HERMES at DESY, Jefferson Lab, and COMPASS at CERN.

In real photon exclusive production in DIS process, the DVCS amplitude interferes with the Bethe-Heitler (BH) amplitude. This will, on one hand, complicate the analysis of the cross section, on the other hand, provide unique opportunities to direct access the DVCS amplitude through the interference. To obtain the constraints on the quark OAMs from these experiments, we need to find the observables which are sensitive to the GPD $E$s. Experiments on the DVCS from JLab 6 GeV Hall A [3163] and HERMES at DESY [3167] have shown strong sensitivity to the quark OAM in nucleon, see, e.g., Fig. 10.3.4. In these experiments, the single spin asymmetries associated with beam or target in DVCS processes are measured, including the beam (lepton)

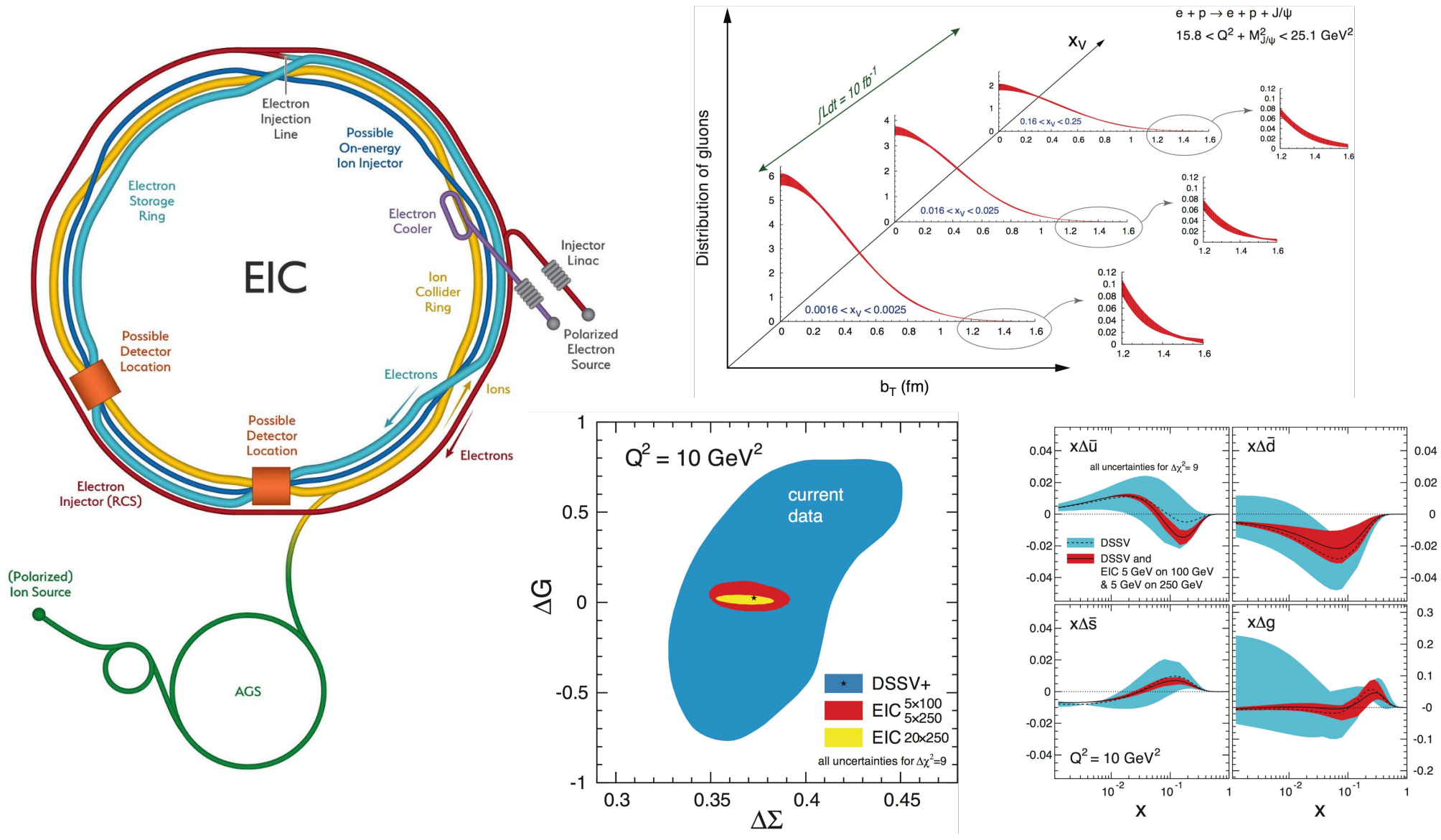

**Fig. 10.3.5** The planned electron-ion collider (EIC) at BNL, NY, USA. Highlights of the EIC impact on our understanding of nucleon spin: total quark/gluon helicity contributions to the proton spin; sea quark helicity distribution using semi-inclusive deep inelastic scattering; nucleon tomography of the 3D gluon density in the transverse plane for different momentum fractions. (This figure from Ref. [820]).

single spin asymmetry and (target) nucleon single spin (transverse or longitudinal) asymmetries.

A less model-dependent approach to extract the AM information from DVCS or similar experiments is to perform a global analysis. Several theory groups have been working on global analysis of the DVCS and DVEM processes [3168–3170]. Recently, a framework to make general analysis of GPDs similar to CTEQ program [1086], called GPDs through universal momentum parametrization (GUMP) [3162], has been proposed based on the previous work on conformal moments expansion [3171, 3172]. The framework, once including the $\xi$ dependence, can be used to fit experimental cross sections and asymmetries. In this way, the quark AM extracted will have less systematic error. In addition, this approach allows us to get the twist-2 quark AM densities, $J_q(x)$, with constraints from experimental data. A number of important AM densities in the spin sum rules depend on information from twist-3 GPDs, such as canonical OAM densities in both longitudinally and transversely polarized proton. Extracting the relevant GPDs from experimental data will be very challenging due to the kinematic suppression.

For the gluon GPDs and AM density $J_g(x)$, one of the most interesting processes is heavy quarkonium production in hard exclusive DIS. This is in particular important at the EIC machine. In early 2020, DOE announced that the next major facility for nuclear physics in US will be a high-energy high-luminosity polarized EIC to be built at BNL. The primary goal of the EIC is to precisely image gluon distributions in nucleons and nuclei, revealing the origin of the nucleon spin and exploring the new QCD frontier of cold nuclear matter [820, 3128].

The EIC will impact our understanding of nucleon spin in many different ways. In the following, we highlight some of these impacts. First, the quark and gluon helicity contributions to the proton spin is the major emphasis of the planned facility. With the unique coverage in both $x$ and $Q^2$, the EIC would provide the most powerful constraints on $\Delta\Sigma$ and $\Delta G$ [820]. Also shown in Fig. (10.3.5) are the projected uncertainty reductions with the proposed EIC machine. Clearly, the EIC will make a huge impact on our knowledge of these quantities, unmatched by any other existing or anticipated facility.

Second, the sea quark polarization will be very precisely determined through SIDIS. With much large $Q^2$ and $x$ coverage, SIDIS at EIC will provide unprecedented kinematic reach and improve the systematic uncertainties. In Fig. 10.3.5, we show the example of sea quark polarization constraints from the EIC pseudo-data simulations.

Third, there will be a comprehensive program on research of GPDs at the EIC. As discussed above, the GPDs provide first hand constraints on the total quark/gluon angular momentum contributions to the proton spin. Moreover, they also provide important information on the nucleon tomography, especially, the 3D imaging of partons inside the proton. With wide kinematic cov-erage at the EIC, a particular example was shown in Fig. 10.3.5 that the transverse imaging of the gluon can be precisely mapped out from the detailed measurement of hard exclusive $J/\psi$ production in DIS processes.

Finally, we would like to emphasize theoretical efforts are as important as the experiments to answer the nucleon spin puzzle. An important question concerns the asymptotic small-$x$ behavior for the spin sum rule. There have been some progresses to understand the proton spin structure at small-$x$ from the associated small-$x$ evolution equations [3173–3181]. More theoretical efforts are needed to resolve the controversial issues raised in these derivations. The final answer to these questions will provide important guidance for the future EIC, where proton spin rum rule is one of the major focuses.

For additional discussion of these issues, see Sec. 10.2.

## 10.4 Nucleon Tomography: GPDs, TMDs and Wigner Distributions

**Andreas Schafer and Feng Yuan**

Exploring the nucleon is of fundamental importance in science, starting from Rutherford's pioneering experiment one hundred years ago where he investigated the internal structure of atomic matter [3182]. Following this effort, the scientific developments in the last century have revealed the most fundamental structure of the matter in our universe: the nucleus is made of nucleons (protons and neutrons) and the nucleon is made of partons: quarks and gluons. In particular, inclusive DIS experiments probe the parton distribution functions which describe the momentum distributions of the partons inside the nucleon, see, Sec. 10.2.

On the other hand, the inclusive measurements of the above processes only probe one dimension of the parton distributions, where the PDF represents the probability distribution of a particular parton (quark or gluon) with a certain fraction $x$ of the nucleon momentum in the infinite momentum frame. In recent years, the hadron physics community is pursuing an extension of this picture to include the transverse direction. The goal is to obtain a three-dimensional tomography of parton densities inside the nucleon. In some sense, these efforts continue the original Rutherford experiment to map out the internal structure of a nucleon in three dimensions.

The nucleon is assumed to move in the $\hat{z}$-direction. Its structure in transverse direction can be either analysed in coordinate space using generalized parton distributions (GPDs) [1085, 1288, 3183–3188], or in momentum space using transverse momentum dependent parton distributions (TMDs) [1274, 1286, 3189, 3190]. In Refs. [972, 3191] introduce the impact parameter dependent parton distributions, which are Fourier transforms of GPDs in certain kinematics and which are the desired parton densities in coordinate space.

The information parametrized by GPDs and TMDs is contained in "mother distributions", the so-called Wigner distributions [3192, 3193]. Wigner distributions were introduced by Wigner in 1930s as phase space distributions in quantum mechanics,

$$W(r,p) = \int d\eta e^{ip\eta}\psi^*(r-\frac{\eta}{2})\psi(r+\frac{\eta}{2}) \ , \qquad (10.4.1)$$

where $r$ and $p$ represent the coordinate and momentum space variables, respectively, and $\psi$ is the wave function. When integrating over $r$ ($p$), one gets the momentum (probability) density from the wave function, which is positive definite. For arbitrary $r$ and $p$, the Wigner distribution is not positive definite and does not have a probability interpretation. This reflects the fact that the Wigner distribution contains all quantum mechanical information contained in $\psi$, which goes beyond probabilities.

Following this concept, we can define the Wigner distribution for a quark in a nucleon with momentum $P$ [3192, 3193],

$$W_\Gamma(x,k_\perp,\vec{r}) = \int \frac{d\eta^- d^2\eta_\perp}{(2\pi)^3} e^{ik\cdot\eta}\langle P|\overline{\Psi}(\vec{r}-\frac{\eta}{2})\Gamma\Psi(\vec{r}+\frac{\eta}{2})|P\rangle \ , \qquad (10.4.2)$$

where $x$ represents the longitudinal momentum fraction carried by the quark, $k_\perp$ is the transverse momentum, $\vec{r}$ the coordinate space variable, and $\Gamma$ the Dirac matrix to project out a particular quark distribution. The quark field $\Psi$ contains the relevant gauge link to guarantee gauge invariance of the above definition [3192]; see more discussions below. We can also define the Wigner distribution for gluons accordingly.

If we integrate the Wigner distribution over $r_z$, we obtain the transverse Wigner distribution,

$$\begin{aligned} & W_\Gamma^T(x,k_\perp,r_\perp) \\ &= \int \frac{dr_z d\eta^- d^2\eta_\perp}{(2\pi)^3} e^{ik\cdot\eta}\langle P|\overline{\Psi}(\vec{r}-\frac{\eta}{2})\Gamma\Psi(\vec{r}+\frac{\eta}{2})|P\rangle \ , \\ &= \int \frac{d^2q_\perp d\eta^- d^2\eta_\perp}{(2\pi)^5} e^{iq_\perp\cdot r_\perp} e^{ik\cdot\eta} \\ & \quad \times\langle P+\frac{q_\perp}{2}|\overline{\Psi}(-\frac{\eta}{2})\Gamma\Psi(\frac{\eta}{2})|P-\frac{q_\perp}{2}\rangle \ , \end{aligned}$$

where we have introduced a wave package for the nucleon state to derive the last equation. The Wigner distribution functions are also referred to as generalized

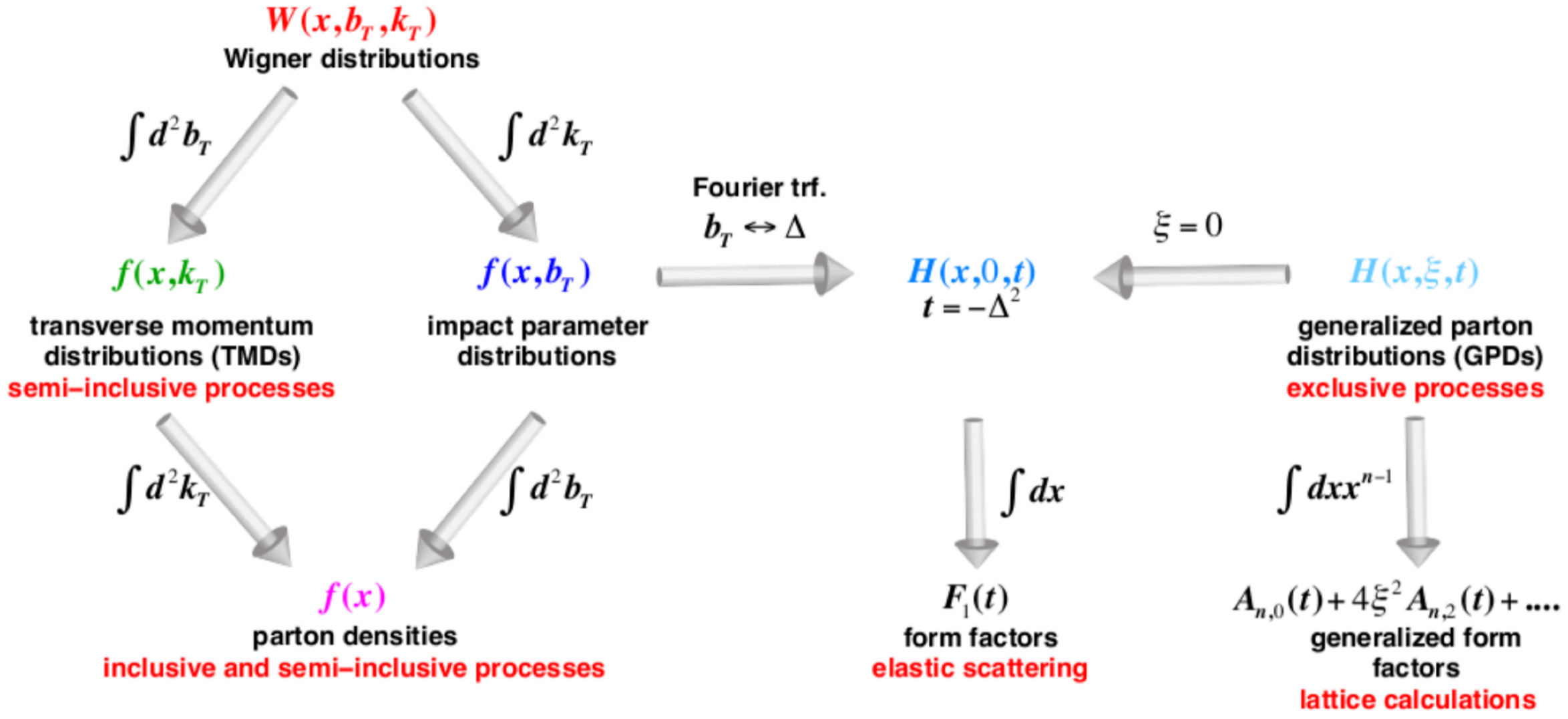

**Fig. 10.4.1** *Transverse momentum dependent parton distributions and the generalized parton distributions are unified in the Wigner distributions. This plot is adopted from Ref. [820].*

TMDs (GTMDs) [3194, 3195]. They can be interpreted as phase space ($r_\perp$,$k_\perp$) distributions of a parton in the transverse plane perpendicular to the nucleon momentum direction.

The Wigner distribution functions reduce to the TMDs and GPDs upon integration over certain kinematic variables. For example, when integrated over $r_\perp$, the above distribution leads to the transverse momentum dependent quark distributions,

$$f(x,k_\perp) = \int \frac{d\eta^- d^2\eta_\perp}{(2\pi)^3} e^{ik\cdot\eta} \langle P|\overline{\Psi}(-\frac{\eta}{2})\Gamma\Psi(\frac{\eta}{2})|P\rangle \ . \tag{10.4.3}$$

On the other hand, if we integrate out $k_\perp$, we obtain the impact parameter dependent quark distribution [972], which is the Fourier transform of the GPDs at $\xi = 0$,

$$\begin{aligned} f(x,b_\perp) &= \int \frac{d^2\Delta_\perp}{(2\pi)^2} e^{i\Delta_\perp\cdot b_\perp} \int \frac{d\eta^-}{2\pi} e^{ik\cdot\eta} \\ &\times \langle P + \frac{\Delta_\perp}{2}|\overline{\Psi}(-\frac{\eta^-}{2})\Gamma\Psi(\frac{\eta^-}{2})|P - \frac{\Delta_\perp}{2}\rangle \\ &= \int \frac{d^2\Delta_\perp}{(2\pi)^2} e^{i\Delta_\perp\cdot b_\perp} H(x,\xi,t)|_{\xi=0} \ . \end{aligned} \tag{10.4.4}$$

Here, $t = -\vec{\Delta}_\perp^2$ and $H(x,\xi,t)$ represents one of the GPDs (definitions will be given below).

The relations between these different functions is often illustrated by the cartoon in Fig. 10.4.1 which is, however, somewhat symbolic. Just like the Wigner distribution in quantum mechanics contains the full information of the wave function $\psi$, a Wigner function in quantum field theory (QFT) contains the full complexity of QFT, including its dependence on the chosen renormalization and factorization scheme. For example TMDs depend on the two scaling variables $\mu$ and $\zeta$, while PDFs depend only on $\mu$. Consequently equations like

$$f(x) \overset{?!}{=} \int d^2k_\perp f(x,k_\perp) \tag{10.4.5}$$

are only valid up to scheme dependent subtraction/renormalization factors or even matching functions. This has significant consequences. For example, usually, the lhs of Eq.(10.4.5) fulfills a different evolution equation than the rhs. Thus, when comparing the results of different phenomenological TMD fits or lattice calculations one has to convert them into the same scheme.

For other functions there is no such complication. For example, the $x$ integral of GPDs is equal to form factors, e.g., $F_1(Q^2) = \int dx H(x,\xi,t=-Q^2)$. This being said, such complications as well as the $\mu$ and $\zeta$ dependence are usually suppressed to simplify notation and we do the same in this review.

The status and perspective of both the collinear PDFs and nucleon form factors have been well covered in this review, see, Sec. 10.1 and Sec. 10.2.

The tomographical information inherent to Wigner distributions is best illustrated by the resulting intuitive and rigorous method to define the quark/gluon orbital angular momentum (OAM). This follows the concept of the Wigner distribution as a phase-space distribution, i.e., to compute the physical observable, one takes the average over the phase-space as if it were a classical distribution,

$$\langle\hat{O}(r,p)\rangle = \int drdpW(r,p)O(r,p) \ . \tag{10.4.6}$$

Since the orbital angular momentum represents the quantity $\vec{r} \times \vec{p}$, we obtain the quark/gluon OAM from the integral of $\vec{r} \times \vec{p}$ multiplied with the Wigner distribution.

For the parton Wigner distribution, one first realizes that a gauge invariant parton distribution must include a gauge link extending from the location of the parton to infinity. An optimal choice for high-energy collisions is a gauge link along the relevant light-cone direction $n^\mu$,

$$\Psi_{LC}(\xi) = P\left[\exp\left(-ig\int_0^{\pm\infty} d\lambda\, n\cdot A(\lambda n + \xi)\right)\right]\psi(\xi) \ . \tag{10.4.7}$$

where $P$ indicates path ordering. The above defined gauge link can go to $+\infty$ or $-\infty$; see more discussions below. In practical applications, we can also choose a straight-line gauge link along the direction of the space-time position $\xi^\mu$,

$$\Psi_{FS}(\xi) = P\left[\exp\left(-ig\int_0^{\infty} d\lambda\, \xi\cdot A(\lambda \xi)\right)\right]\psi(\xi) \ . \tag{10.4.8}$$

This link reduces to unity in Fock-Schwinger gauge, $\xi \cdot A(\xi) = 0$. With the above definitions, we can write down the quark Wigner distribution as,

$$W_{\mathcal{P}}(k^+ = xP^+, \vec{b}_\perp, \vec{k}_\perp) = \frac{1}{2}\int \frac{d^2\vec{q}_\perp}{(2\pi)^3}\int \frac{dk^-}{(2\pi)^3} e^{-i\vec{q}_\perp\cdot\vec{b}_\perp}\left\langle \frac{\vec{q}_\perp}{2}\right|\hat{\mathcal{W}}_{\mathcal{P}}(0,k)\left|-\frac{\vec{q}_\perp}{2}\right\rangle , \tag{10.4.9}$$

with the Wigner operator,

$$\hat{\mathcal{W}}_{\mathcal{P}}(\vec{r},k) = \int \overline{\Psi}_{\mathcal{P}}(\vec{r}-\xi/2)\gamma^+\Psi_{\mathcal{P}}(\vec{r}+\xi/2)e^{ik\cdot\xi}d^4\xi \ , \tag{10.4.10}$$

where $\mathcal{P}$ denotes the path and is either $LC$ or $FS$, $\vec{r}$ is the quark phase-space position, and $k$ the phase-space four-momentum.

It can be shown that the total OAM is given by the parton's Wigner distribution,

$$\begin{aligned} L_q &= \frac{\langle PS|\int d^3\vec{r}\,\overline{\psi}(\vec{r})\gamma^+(\vec{r}_\perp \times i\vec{D}_\perp)\psi(\vec{r})|PS\rangle}{\langle PS|PS\rangle} \\ &= \int (\vec{b}_\perp \times \vec{k}_\perp) W_{FS}(x,\vec{b}_\perp,\vec{k}_\perp)dxd^2\vec{b}_\perp d^2\vec{k}_\perp \end{aligned} \tag{10.4.11}$$

which provides a gauge-invariant expression for the parton's OAM [3138, 3197].

Similarly, the canonical OAM in light-cone gauge fulfills the simple but gauge-dependent parton sum rule in the quantum phase space [3195, 3196, 3198],

$$\begin{aligned} \ell_q &= \frac{\langle PS|\int d^3\vec{r}\,\overline{\psi}(\vec{r})\gamma^+(\vec{r}_\perp \times i\vec{\partial}_\perp)\psi(\vec{r})|PS\rangle}{\langle PS|PS\rangle} \\ &= \int (\vec{b}_\perp \times \vec{k}_\perp) W_{LC}(x,\vec{b}_\perp,\vec{k}_\perp)dxd^2\vec{b}_\perp d^2\vec{k}_\perp \ . \end{aligned} \tag{10.4.12}$$

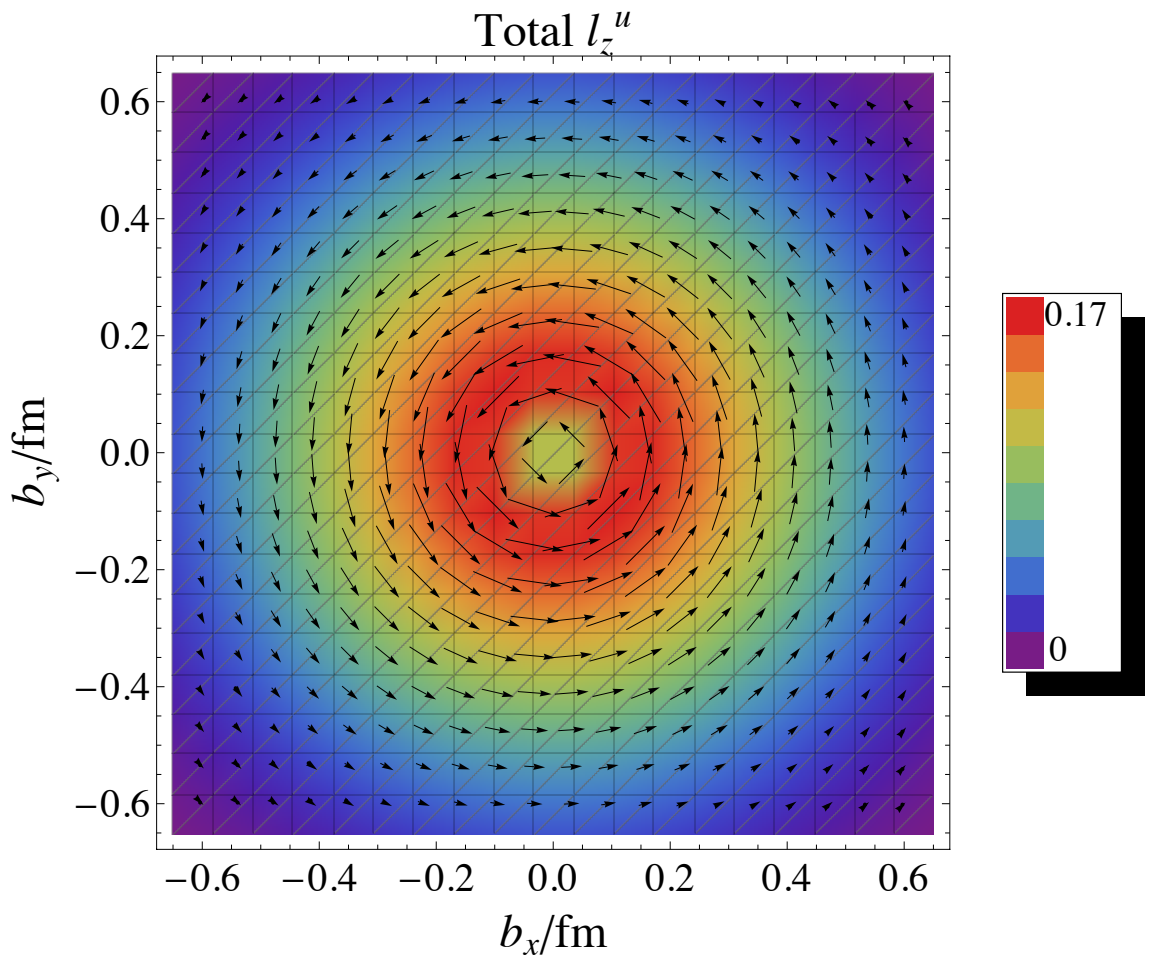

**Fig. 10.4.2** Distributions in impact parameter space of the mean transverse momentum of an unpolarized $u$-quark in a longitudinally polarized nucleon, taken from Ref. [3196]. The nucleon is polarized perpendicular to the plane, while the arrows show the size and direction of the mean transverse momentum of the quarks. This gives an intuitive picture of the quark orbital angular motion inside the nucleon.

The above two OAMs, $L_q$ and $\ell_q$, correspond to the quark OAMs in the Ji and Jaffe-Manohar spin sum rules, respectively, discussed in Sec. 10.3. Similar conclusions hold for the gluon OAMs as well.

Therefore, the Wigner distribution, to some extend, contains the parton OAMs in two different spin sum rules. This further illustrates that the difference between them comes from the gauge link direction. A recent lattice QCD calculation has shown that the quark OAMs can be obtained from the quark Wigner distributions and the difference between $L_q$ and $\ell_q$ has been demonstrated [3159, 3160].

In the last years a number of studies have directly probed the quark/gluon OAM contributions [3199–3203] applying the Wigner distribution for hard exclusive processes. For example, the single longitudinal target-spin asymmetries in hard exclusive dijet production in lepton-nucleon collisions [3199, 3200] and the double spin asymmetries in this process [3203] can provide crucial information on the gluon's canonical OAM contribution.

The determination of Wigner distributions is thus an important challenge for future studies; see discussions in the end of this subsection. The crucial point is that there exists a well-defined, standardized way to link nucleon tomography to Wigner distributions constructed from light-cone wave functions [3196]. As an example we show in Fig. 10.4.2 the average transverse momentum flow in impact parameter space for $u$-quarks inside the proton. While this result is model dependent, it has the great advantage of providing an intuitive im-age of the quark orbital motion distribution inside a hadron.

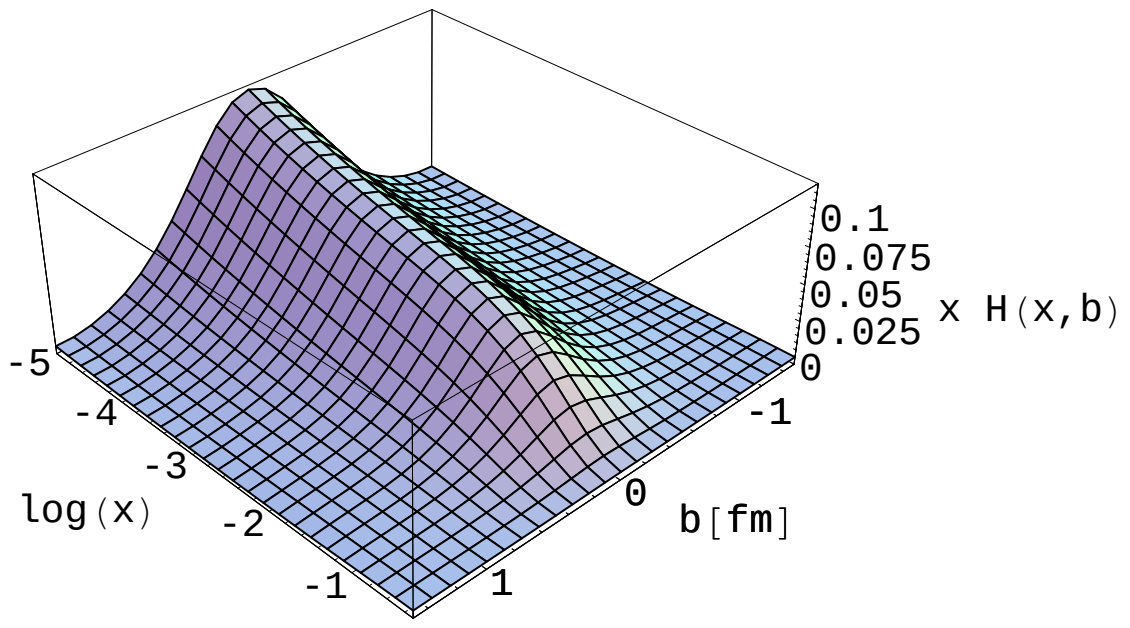

**Fig. 10.4.3** *Transverse profiles for the up quark distribution in transverse coordinate space as function of* $x$.

*Generalized Parton Distributions*

The GPDs are one of the projections from the Wigner distributions. They are extensions of the usual collinear parton distributions discussed in Sec. 10.2 and defined as off-forward matrix elements of the hadron. For example, for the quark GPDs, we have [1083, 1085, 1288, 3184–3187]

$$\int \frac{d\lambda}{2\pi} e^{i\lambda x} \langle P'S'|\overline{\Psi}_q\left(-\frac{\lambda}{2}n\right) \not{n} \Psi_q\left(\frac{\lambda}{2}n\right)|PS\rangle \quad (10.4.13)$$
$$= \overline{U}(P')\left[H_q(x,\xi,t)\not{n} + E_q(x,\xi,t)\frac{\sigma^{\alpha\beta}n_\alpha \Delta_\beta}{2M_p}\right]U(P) \ ,$$

where $\Delta = P' - P$ with $t = \Delta^2$, $x$ is the light-cone momentum fraction of the quark, and the skewness parameter $\xi$ is defined as $\xi = (P-P')\cdot n/(P+P')\cdot n$. In the forward limit, we have $\xi = 0$ and $t = 0$, and the GPDs reduce to the usual collinear PDFs. The $x$-moments of GPDs lead to not only the electromagnetic form factors but also the gravitational form factors [1085], one of which produces the spin sum rule as discussed in the previous subsection.

Depending on the polarization of the quark and the nucleon states, the leading-twist quark GPDs contain eight independent distributions. The GPDs can be measured in many different experiments, for example, DVCS and hard exclusive meson production. Experimental efforts have been made at various facilities, including HERMES at DESY, Jefferson Lab, and COMPASS at CERN. It will be a major focus of the future EIC as well.

Nucleon tomography in terms of the GPDs is best illustrated in the impact parameter dependent parton distirbution of Eq. (10.4.4). From that, we can define the transverse quark density profile [3191]:

$$\rho_q(x,\boldsymbol{b}) = \int \frac{\mathrm{d}^2\boldsymbol{\Delta}}{(2\pi)^2} e^{-i\boldsymbol{\Delta}\cdot\boldsymbol{b}} H_q(x,-\boldsymbol{\Delta}^2) \ . \quad (10.4.14)$$

An important feature of the above distribution is how it changes with longitudinal momentum fraction $x$. In Fig. 10.4.3, we show the transverse density profile for the up quark from the GPD parameterizations of [3168]. The plot shows that the transverse profile in coordinate space becomes wider at smaller $x$. At large $x$, however, it approaches a point-like structure, which means there is no $t$ dependence of the GPD quark distribution, a result consistent with large-$x$ power counting for GPDs [3204]. One of the primary goals of the GPD program at the JLab-12GeV and the EIC is to map out the $x$-dependence of the GPDs and the tomographic images for both quarks and gluons.

Most interestingly, when the nucleon is transversely polarized, the parton distribution in the transverse plane will be asymmetric due to the contribution from the GPD $E$ [3191],

$$\rho_q^X(x,\boldsymbol{b})$$
$$= \int \frac{\mathrm{d}^2\boldsymbol{\Delta}}{(2\pi)^2} e^{-i\boldsymbol{\Delta}\cdot\boldsymbol{b}} \left(H_q(x,-\boldsymbol{\Delta}^2) + \frac{i\Delta_Y}{2M} E_q(x,-\boldsymbol{\Delta}^2)\right)$$
$$= \mathcal{H}_q(x,\boldsymbol{b}) - \frac{1}{2M}\frac{\partial}{\partial b^Y}\mathcal{E}_q(x,\boldsymbol{b}) \ , \quad (10.4.15)$$

where $\mathcal{H}_q(x,\boldsymbol{b})$ and $\mathcal{E}_q(x,\boldsymbol{b})$ are the 2-dimensional Fourier transformations of $H_q(x,-\boldsymbol{\Delta}^2)$ and $E_q(x,-\boldsymbol{\Delta}^2)$, respectively, and the nucleon is polarized in the $X$ direction. This asymmetric distribution has attracted strong interest in the hadron physics community and it was argued that it might be related to the single spin asymmetry phenomena in hadronic processes [3191]. It has also been found in a lattice simulation [3205].

In order to factor out the transverse displacement from the nucleon's center of momentum and its contribution to the transverse polarization, one can introduce an intrinsic quark density [3140],

$$\rho_{q,\mathrm{In}}^X(x,\boldsymbol{b}) = \int \frac{\mathrm{d}^2\boldsymbol{\Delta}}{(2\pi)^2} e^{-i\boldsymbol{\Delta}\cdot\boldsymbol{b}} \Big[H_q(x,-\boldsymbol{\Delta}^2)$$
$$+ \frac{i\Delta_Y}{2M}\left(H_q(x,-\boldsymbol{\Delta}^2) + E_q(x,-\boldsymbol{\Delta}^2)\right)\Big]$$
$$= \mathcal{H}_q(x,\boldsymbol{b}) - \frac{1}{2M}\frac{\partial}{\partial b^Y}\left(\mathcal{H}_q(x,\boldsymbol{b}) + \mathcal{E}_q(x,\boldsymbol{b})\right) \quad (10.4.16)$$

from which one can reproduce the transverse polarization sum rule; see Sec. 10.3. In Fig. 10.4.4, we show the intrinsic transverse density for $u$ and $d$ quarks at $x = 0.3$ from the analysis of the GPD quark distribution of [3162]. Clearly, the quarks have non-zero transverse displacement, which contributes to the transverse angular momentum of the nucleon.

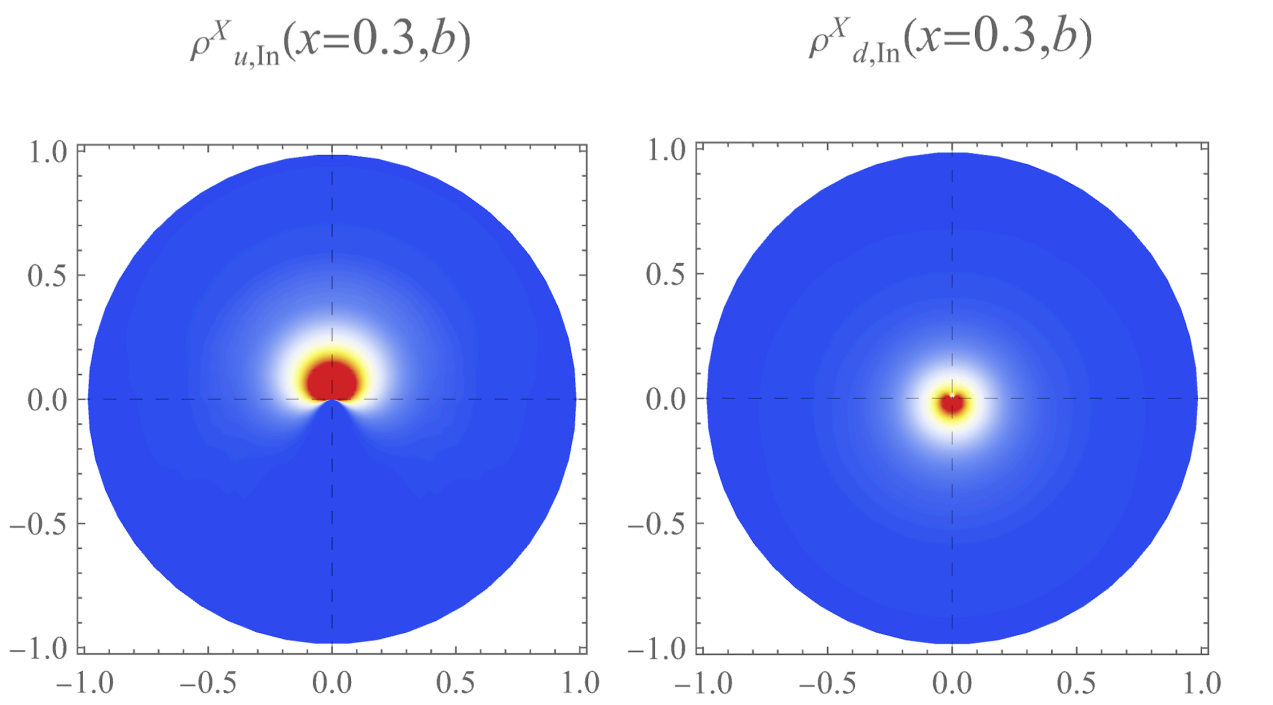

**Fig. 10.4.4** *Plots of the intrinsic quark densities $\rho^X_{q,\text{In}}(x, \boldsymbol{b})$ for both u and d quarks in a transversely polarized proton (in the X direction) at $x = 0.3$. Both the u and d quark densities are shifted in the Y direction and contribute to the angular momentum $J^X$. While the u contributions are positive (+Y direction) and the d contributions are negative (−Y direction). These plots are adopted from Ref. [3162].*

The theoretical framework has been well developed for the GPD studies with established QCD factorization for the associated exclusive processes [1290–1292]. Higher order perturbative QCD corrections have been calculated in a number of publications [1292, 3206–3214]. The first computation of next-to-next-leading order corrections for DVCS has also been reported recently [3215]. However, since GPDs depend on three variables $(x,\xi,t)$ in addition to the scale variable $\mu$, it is much more difficult to extract them from experiment than PDFs (which only depend on $x$).

Pioneering phenomenological work has been carried out in Refs. [1288, 3168, 3172, 3211, 3216, 3217]. In the last years, progress has also been made toward a global analysis of GPDs from a wide range of experiments [3162, 3218–3224]. Especially, the twist-2 and twist-3 results were re-derived with an optimal light cone coordinate and full kinematics adopted [3222–3224]. A dedicated program based on earlier developments of Ref. [3216] has been proposed in Ref. [3162]. All these theory advances are crucial for a successful campaign to determin GPDs from DVCS and other hard exclusive processes measured at JLab-12 GeV and the planned Electron-Ion Collider.

Lattice QCD can be used to study these GPDs as well. Employing the LaMET formalism, exciting results on the $x$-dependence of the GPD quark distributions have already been obtained [634, 3225]. We expect many more such simulations to emerge in the future, as well as combined fits to experimental and lattice data.

*Transverse Momentum Dependent Parton Distributions*
Theoretical studies of TMDs started long ago (see, for example, Ref.[1274]). In recent years great progress was made in the exploration of these distribution functions and the associated single spin asymmetry phenomena. In particular, TMDs provide not only an intuitive illustration of nucleon tomography, as we discussed above, but also the important opportunities to investigate the specific nontrivial QCD dynamics associated with their physics: QCD factorization, universality of the parton distributions and fragmentation functions, and their scale evolutions.

Different from the collinear PDFs discussed in Sec. 10.2, the TMD parton distributions can not be studied in inclusive processes. We have to go beyond that and explore semi-inclusive hard processes, where a hard momentum scale is involved in addition to the transverse momentum of the final-state particle produced. For example, we can study the TMD quark distributions in semi-inclusive DIS (SIDIS), where the virtual photon (with virtuality $Q$) scatters off the hadron and produces a final state hadron in the current fragmentation region. The hadron's transverse momentum $P_{h\perp}$ has to be much smaller than the hard momentum $Q$. Because of $P_{h\perp} \ll Q$, this process can be factorized into the TMD quark distribution convluted with the TMD fragmentation function. Similarly, the Drell-Yan lepton pair production (or $W/Z$-boson, Higgs boson production) in hadronic collisions can be described by the convolution of two TMD parton distributions with transverse momentum $q_\perp \ll Q$. A related process in $e^+e^-$ annihilation into two back-to-back hadrons can be factorized as a convolution of two TMD fragmentation functions.

The TMD quark distributions can be defined by the following matrix [1274, 1286, 1298, 1311, 3189, 3190],

$$\hat{\mathcal{M}}_{\alpha\beta}(x, k_\perp) = \int \frac{dy^- d^2 y_\perp}{(2\pi)^3} e^{-ixP^+\cdot y^- + i\vec{k}_\perp\cdot\vec{y}_\perp} \times \langle PS|\overline{\Psi}_\beta(y^-, y_\perp)\Psi_\alpha(0)|PS\rangle \;, \quad (10.4.17)$$

where $x$ is the longitudinal momentum fraction and $k_\perp$ the transverse momentum carried by the quark. The quark field $\Psi(y)$ contains a gauge link as defined in Eq. (10.4.7). This definition contains a light-cone singularity from higher order corrections. The regulation and subtraction procedure defines the scheme of the TMD distributions. Obviously, in the theoretical limit in which contributions from all orders and all twists are taken into account observable, physical quantities have to be scheme independent. (At the most simple level this was actually shown explicitly in Ref. [3226] but it has to be true also non-perturbatively.) Often, however (e.g. in event generators), rather specific models are used for which this is not the case. In these cases the fitted TMDs and thus the result of hadron tomography can be strongly scheme/model dependent (see e.g. Ref. [3227]). Calculating the model-specific matching factors or functions between such a scheme and the usual TMD factorization scheme is often not possible. This should flag a warning that one has to be careful when comparing fitted TMDs from different sources. The leading order expansion of the above matrix contains eight independent quark TMDs, depending on the polarization of the quark (varying horizontally) and the nucleon (varying vertically) in Fig. 10.4.5. The gauge link direction plays an essential role in the naive time-reversal-odd TMD quark distributions, including the quark Sivers function $f_{1T}^{\perp}(x, k_\perp)$ and Boer-Mulders function $h_1^{\perp}(x, k_\perp)$.

| | $U$ | $L$ | $T$ |
|---|---|---|---|
| $U$ | $f_1$ | | $h_1^{\perp}$ |
| $L$ | | $g_{1L}$ | $h_{1L}^{\perp}$ |
| $T$ | $f_{1T}^{\perp}$ | $g_{1T}$ | $h_1, h_{1T}^{\perp}$ |

**Fig. 10.4.5** The leading order transverse momentum dependent quark distributions depend on the polarization of the quark (rows) and the nucleon (columns).

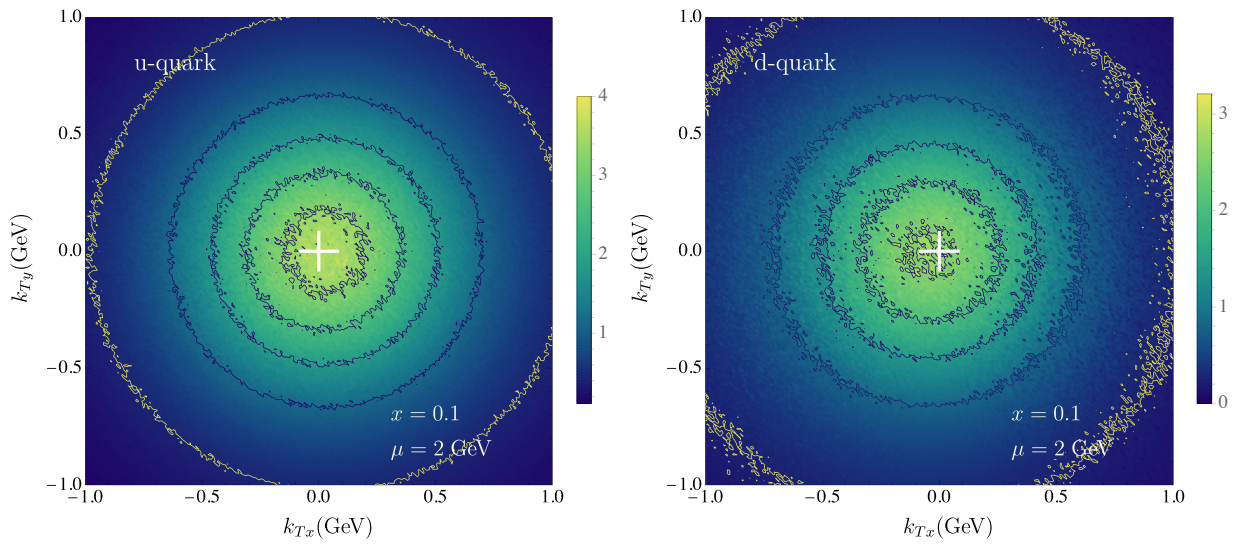

**Fig. 10.4.6** The quark Sivers function represents the asymmetric distribution of a quark in transverse momentum space when the nucleon is polarized along the $\hat{y}$-direction. This plot is adopted from a phenomenological study [3228].

The spin-average quark distributions are symmetric in the transverse plane. However, if the nucleon (or the quark) is transversely polarized, the quark distribution shows an azimuthal asymmetry. In particular, the TMD quark Sivers functions quantify these asymmetries in the transverse momentum space. In Fig. 10.4.6, we show one of the resulting distributions for the quark (averaged over $x$) in transverse momentum space in a transversely polarized nucleon. The TMD distribution comes from the fit to the associated single transverse spin asymmetries in semi-inclusive hard processes [3228].

For the quark Sivers function, because of the initial/final state interaction (represented by the gauge link pointing to $-\infty$ or $+\infty$ in the quark distribution definition) difference, they differ by signs for semi-inclusive hadron production in DIS (SIDIS) and Drell-Yan processes [1297, 1298, 1310, 1311, 3229, 3230]. This leads to a sign change between the SSAs in SIDIS and Drell-Yan processes,

$$\text{Sivers SSA}|_{\text{DY}} = -\text{Sivers SSA}|_{\text{DIS}} \, . \tag{10.4.18}$$

This nontrivial result still holds when gluon radiation contributions are taken into account [1321, 3231–3233]. It is very important to test this nontrivial QCD prediction by comparing the SSAs in these two processes. The Sivers single spin asymmetries in SIDIS processes have been observed by the HERMES [3234, 3235], COMPASS [3236–3239], and JLab [3240, 3241] collaborations. There have been significant efforts to measure the Sivers asymmetries in Drell-Yan process at COMPASS [3242] and that of $W^{\pm}$ production at RHIC [3243]. The analyses of these data provide an indication for a sign change [3244], but no proof. More precise measurements are needed to confirm this crucial property.

In TMD factorization for semi-inclusive hard processes [1269, 1274, 1282, 1285, 1871], collinear and soft gluon radiations are factorized into the TMD parton distributions or fragmenation functions and the associated soft factors. As for the integrated parton distribution functions, these gluon radiation contributions can be resummed to all orders by solving the relevant evolution equations. This resummation is referred to as TMD or Collins–Soper–Sterman resummation [1282]. As a result, the factorization simplifies the differential cross section to a convolution of soft factor–subtracted TMD distributions and/or fragmentation functions [1269], where the hard momentum scale is chosen as factorization scale $\mu_F = Q$. As an example, in Fig. 10.4.7, we show the TMD up quark distribution $f^{(sub)}(x = 0.1, k_T, \mu_F = Q)$ as a function of the transverse momentum at different scales. Clearly, the resummation/scale evolution leads to broadening effects for TMD distributions at higher scales. Based on these developments, recent global analyses have achieved high precision for the unpolarized TMD quark distribution and fragmentation functions fitted to data from various semi-inclusive hard processes [3245–3248]. Of course, further theoretical developments are still needed to answer crucial questions concerning TMD factorization at lower scale SIDIS and the question how non-perturbative effects affect the matching between the TMDs and collinear PDFs [3248–3251]. Upcoming data from the JLab-12 GeV program should significantly improve our understanding of these issues in the near future.

In addition, progress has been made in studying the scale evolution for the quark Sivers function and the associated quark-gluon-quark correlation functions [1318, 1319, 3252–3256], and the QCD resummation for the SSA observables [3256–3260]. These resummation effects have been taken into account in a recent phenomenological study of all single spin asymmetries associated with the quark Sivers function in a global analysis using (N$^3$LO) evolution for the TMDs [3228, 3244].

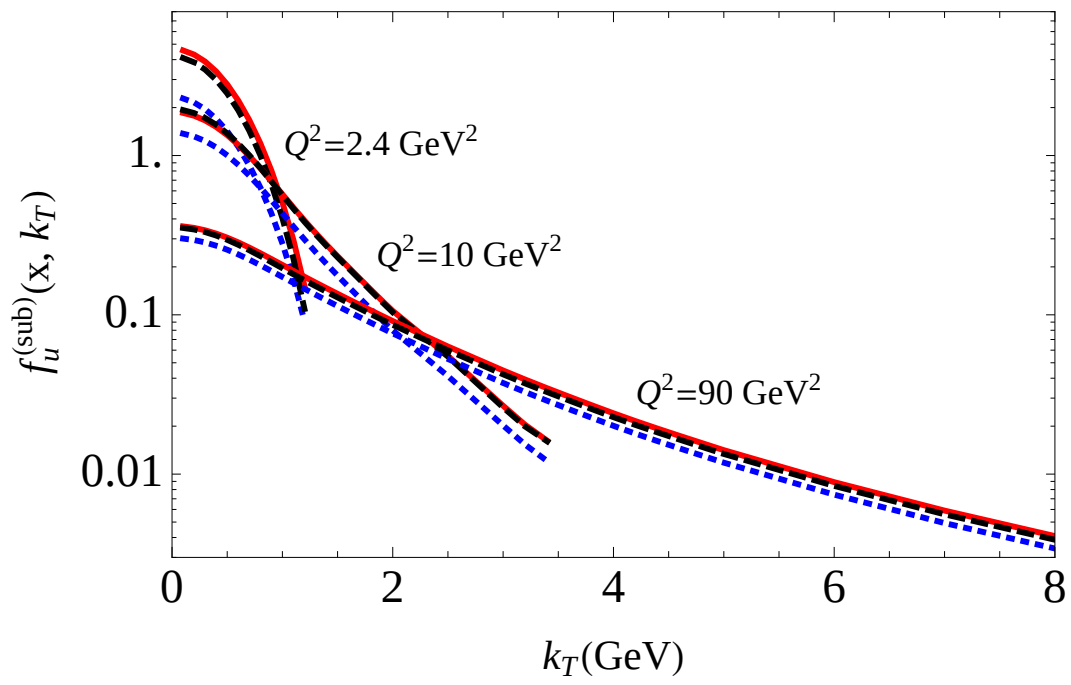

**Fig. 10.4.7** TMD up-quark distributions $f_u^{(sub.)}(x = 0.1, k_\perp)$ as functions of the transverse momentum $k_\perp$ (GeV) at three different scales $Q^2 = 2.4$, 10, 90 (GeV$^2$). This plot is adopted from Ref. [3226].

There has also been significant progresses toward lattice calculations of TMDs [635, 638, 639, 3261–3272] in the last few years, partially again based on the LaMET formalism. The TMD evolution kernel was calculated from lattice QCD [635, 637, 639, 3269, 3270] for perturbative and non-perturbative [3247, 3248, 3273] impact parameters $b$ and the result agreed with that of a fit to experimental data [3227]. This motivates great hopes for future combined TMD fits to experimental and lattice data. We also expect lattice simulation of the single spin asymmetries associated with the quark Sivers function, using the perturbative matching derived in Ref. [3265].

More recently, important developments have taken place addressing the connections between the TMD formalism and small-$x$ saturation physics. Small-$x$ gluon saturation is best described in the color-glass-condensate (CGC)/color-dipole formalism [3274–3278], for which the so-called unintegrated gluon distributions (UGDs) are essential elements. What has been shown in the recent papers [3279–3285] is that these UGDs are the same as the TMD gluon distribution functions at small-$x$. Meanwhile, considerable progress has also been made in computing Sudakov double logarithms in the small-$x$ formalism [3286–3291]. These computations provide a solid theoretical foundation for further rigorous investigations that probe the dynamics of the saturation regime with hard processes. We anticipate that in the foreseeable future a unified picture of nucleon structure will emerge that covers the whole kinematic domain, including small and large $x$.

*Direct access to the Wigner distributions*
It was generally believed that the parton Wigner distributions are not directly measurable in high energy scattering. However, it was realized recently that the Wigner distribution could be measured through hard exclusive processes [3292–3294]. In particular, it was shown in Ref. [3292] that the small-$x$ gluon Wigner distribution is connected to the color dipole S-matrix in the CGC formalism [3274–3278], that diffractive dijet production in $ep/eA$ collisions [3292, 3295–3299] may provide a direct probe of this gluon Wigner distribution. Additionally, semi-hard gluon radiation in this process or 'trijet' diffractive production has been shown to probe the color-dipole amplitude in the adjoint representation [3300, 3301]. This demonstrates that a new class of diffractive processes, including semi-inclusive diffractive DIS [3302] can provide crucial information on the gluon Wigner distributions at small-$x$. Extention to other processes, in particular, those at moderate and large $x$ will be interesting to follow as well. We expect more research along this direction in the future.

To summarize this subsection: There has been great progress in both experiment and theory for GPD and TMD physics. Of course, challenges are still there in both fields. We would like to emphasize that data from future experiments, including the 12 GeV upgrade of JLab, COMPASS and the planed EIC experiments, together with theory developments, will lead us to a complete 3D tomography of the nucleon.

# 11 QCD at high energy

*Conveners:*
**Gudrun Heinrich and Eberhard Klempt**

The core of high energy collisions consists in a hard scattering of two partons, where the momentum transfer is very large and therefore the process can be calculated perturbatively. The enormous progress in the calculation of QCD corrections beyond the leading order in perturbation theory is described by Gudrun Heinrich. The scattered partons can emit soft or nearly collinear gluons. In kinematic regions where the phase space for such emissions is restricted, large logarithms arise, which can spoil the perturbative convergence. Due to the universal structure of infrared divergent QCD radiation, such logarithms can be resummed analytically to all orders to restore the predictive power of the perturbative description in these kinematic regions, as described by Simone Marzani.

At the intermediate stage between the hard interaction and hadronization, the radiation of gluons from quarks and the splitting of gluons into secondary quarks and gluons, forming a cascade of emissions, can be described by parton showers. The development of these parton showers and our understanding of these processes are described by Frank Krauss.

Once these showers of partons have evolved to low energies, the process of hadron formation sets in. At these energies, the strong coupling is large, such that bound states are formed, which cannot be described perturbatively anymore. The description of hadronization needs to rely on parameters extracted from data. These parameters are tuned in Monte Carlo simulations. Torbjörn Sjöstrand gives a detailed view of different stages of the collision process and of their simulation.

The reconstruction of jets by reliable jet algorithms and the identification of the primary source, gluons or quarks of a certain flavor, is very important to extract information about the underlying particle dynamics from the data. Jet substructure variables can provide essential information about the decay of heavy particles leading to boosted jets, as described by Bogdan Malaescu, Dag Gillberg, Steven Schramm, and Chris Young.

## 11.1 Higher-order perturbative calculations

**Gudrun Heinrich**

### 11.1.1 Introduction

The property of asymptotic freedom of QCD, together with the fact that short- and long-distance effects in QCD can be factorized up to power corrections, allows us to describe processes with high momentum transfer as a perturbative series in the strong coupling $\alpha_s$, as illustrated in Eq. (11.1.1). For example, the cross section for a process such as the production of a Higgs boson through the collision of two protons with momenta $p_a$ and $p_b$, $p_a + p_b \to H + X$, has the form

$$\sigma_{pp\to H+X} = \sum_{i,j} \int_0^1 dx_1\, f_{i/p_a}(x_1, \alpha_s, \mu_F) \times \int_0^1 dx_2\, f_{j/p_b}(x_2, \alpha_s, \mu_F)\, \hat{\sigma}_{ij\to H+X}(\alpha_s(\mu_R), \mu_R, \mu_F) + \mathcal{O}\left(\frac{\Lambda}{Q}\right)^p , \tag{11.1.1}$$

where the partonic cross section $\hat{\sigma}_{ij\to H+X}$ can be expanded as

$$\hat{\sigma}_{ij\to H+X} = \alpha_s^2 \hat{\sigma}^{(0)} + \alpha_s^3 \hat{\sigma}^{\mathrm{NLO}} + \alpha_s^4 \hat{\sigma}^{\mathrm{NNLO}} + \ldots \tag{11.1.2}$$

The renormalization of ultraviolet singularites appearing in loop corrections leads to a dependence of both $\alpha_s$ and the partonic cross section on the renormalization scale $\mu_R$. Similarly, the absorption of collinear singularities into the "bare" parton distribution functions leads to a dependence on the factorization scale $\mu_F$. The functions $f_{i/p_a}(x, \alpha_s, \mu_F)$ are the (physical) parton distribution functions (PDFs), which can be interpreted as probabilities to find a parton of type $i$ with momentum fraction $x$ of the "parent" momentum $p_a$ in a proton (or, more generally, a hadron). This makes an assumption of collinearity of the parton's momentum with $p_a$, therefore the factorisation described by Eq. (11.1.1) is also called *collinear factorisation*. For more details about parton distribution functions we refer to Section 10.2. Factorization holds up to the so-called power corrections of order $(\Lambda/Q)^p$, where the power $p$ is process-dependent and usually larger than one, see however, Ref. [161].

In Eq. (11.1.2), the partonic cross section at leading order (LO) in an expansion in $\alpha_s$ is denoted by $\hat{\sigma}^{(0)}$, where for the sake of clarity the powers of the strong coupling have been extracted. The next-to-leading order (NLO) cross section comes with one more power of $\alpha_s$ relative to LO, the next-to-next-to-leading order (NNLO) cross section with two more $\alpha_s$ powers than LO, etc. Of course such an expansion also can be performed for the electroweak corrections, however, as $\alpha/\alpha_s(M_Z) \simeq 0.1$, the QCD corrections are usually larger, except in kinematic regions where logarithms of the form $\alpha \ln\left(M_W^2/\hat{s}\right)$ grow large. The dependence of the cross section $d\sigma_{pp\to H+X}$ on $\mu_R$ and $\mu_F$ is an artifact of the truncation of the perturbative series. Therefore, the dependence on these unphysical scales becomes weaker as more perturbative orders are calculated. The variation of the cross section as these scales are varied around a central scale – which should be chosen to be close to the energy at which the hard interaction takes place – therefore can be used as an estimate of the theoretical uncertainty due to missing higher orders.

Higgs boson production in gluon fusion is somewhat special, as the leading order amplitude is already loop-induced, and because the NLO QCD corrections are of the order of 100%, which makes the inclusion of QCD corrections beyond NLO a necessity for a satisfactory description of the data.

The perturbative expansion in powers of $\alpha_s$ is particularly reliable for inclusive observables. If the phase space for QCD radiation is restricted, large logarithms can appear, which spoil the convergence of the perturbative series in $\alpha_s$. This requires so-called resummation, as described in detail in Section 11.2. Here we will focus on calculations at a fixed order in the strong coupling.

### 11.1.2 Developments and status

*Next-to-leading order QCD corrections*
The development of systematic techniques for NLO QCD corrections started in the 1980ies with seminal work on $e^+e^-$–annihilation to jets [3303–3305] and hadron-hadron scattering [178], followed by pioneering developments of techniques for one-loop calculations based on Feynman diagrams and tensor reduction [3306–3309]. In parallel, subtraction methods for soft and collinear real radiation were established [180, 3310], leading to the first differential NLO calculations of 2-jet production in hadronic collisions [3311–3313], while the first NLO calculation of 3-jet production in hadronic collisions only appeared 10 years later [3314].

The calculation of NLO cross sections for $(n-2)$-jet production in hadronic collisions (or for $(n-1)$-jet production in $e^+e^-$–annihilation, as well as the calculation of amplitudes obtained by crossing) involves $n$-parton one-loop amplitudes and $(n+1)$-parton tree-level amplitudes with up to one unresolved (soft or collinear) parton, see Fig. 11.1.1. The efficient calculation of one-loop $n$-point amplitudes for $n \geq 5$ represented a major challenge in the 1990ies and led to the development of more efficient methods to calculate one-loop $n$-point amplitudes, based on the idea to exploit analytic properties of loop integrals if propagators are put on-shell (so-called "unitarity cuts") [181, 3315–3317]. The emergence of methods to perform these cuts numerically [3318–3320], together with the automation of subtraction methods for unresolved real radiation at NLO, led to a new level of efficiency, resulting in the availability of NLO QCD predictions for multi-particle scattering which were considered unfeasible some years before, such as 5-jet production at the LHC [3321], top-quark pair production with up to 3 jets [3322], $Wb\bar{b}$ production with up to 3 light jets [3323], or the NLO QCD and EW corrections to off-shell $t\bar{t}W$ production at the LHC, involving one-loop 10-point integrals [3324]. It also led to the development of automated tools providing one-loop amplitudes for fully differential NLO predictions [3319, 3325–3331]. This remarkable jump in efficiency is often called the "NLO revolution".

*Beyond NLO*
The next step, towards fully differential NNLO predictions, required not only major progress in the calculation of two-loop integrals, but also the development of subtraction schemes for infrared (IR) divergent real radiation where up to two particles can be unresolved.

*Multi-loop amplitudes*
First, the developments regarding loop integrals with two or more loops will be considered. A very important parameter characterising a Feynman integral besides the number of loops and legs is the number of kinematic/mass scales. As an example, it is instructive to consider the development of the calculation of 2-loop 4-point integrals (2 loops, 4 legs): the analytic calculation of the planar [3332] and non-planar [3333] two-loop 7-propagator diagrams with massless propagators and light-like legs has been performed in 1999, numerical checks of these results in the Euclidean region were performed in Ref. [3334], the calculation of such integrals with one off-shell leg was completed soon after [3335, 3336]. The first results for two-loop 4-point amplitudes with *massive* propagators have been achieved by a numerical method [3337, 3338], leading to NNLO predictions for top quark pair production in hadronic collisions [3339]. The analytic calculation of two-loop 4-point integrals with two massive legs and massless propagators, entering for example the production of $Z$-boson pairs or a $W^+W^-$ pair, was completed around the year 2015 for both the on-shell as well as for the off-shell case [3340–3345]. However, the step to include massive propagators leads out of the function class (so-called multiple polylogarithms) describing the above-mentioned objects analytically. Therefore, the calculation of two-loop 4-point integrals with both massive propagators as well as massive final state particles was performed numerically before analytic results appeared, examples are the two-loop QCD corrections to Higgs boson pair production [3346, 3347], Higgs+jet production [3348], $gg \to ZH$ [3349–3351] and $gg \to VV$ with massive loops [3352–3354], where $V$ denotes a massive vector boson. Thus, one can roughly say that it took almost twenty years to increase the number of independent mass scales entering these diagrams from two $(s_{12}, s_{23})$, to five
$(s_{12}, s_{23}, m_t, m_V, m_{V'})$, where $s_{ij} = (p_i + p_j)^2$ and $V'$ denotes a boson with invariant mass different from $V$. The timeline of available predictions for (differential) cross sections based on these integrals is shown in Table 11.1.1, illustrating how an additional mass scale increases the complexity. It is noteworthy that all integrals with massive propagators, i.e. with a non-zero entry in the third column, have been calculated with numerical methods. For processes with jets in the final state, the subtraction of IR divergent real radiation was the bottleneck, not the availability of the two-loop integrals.

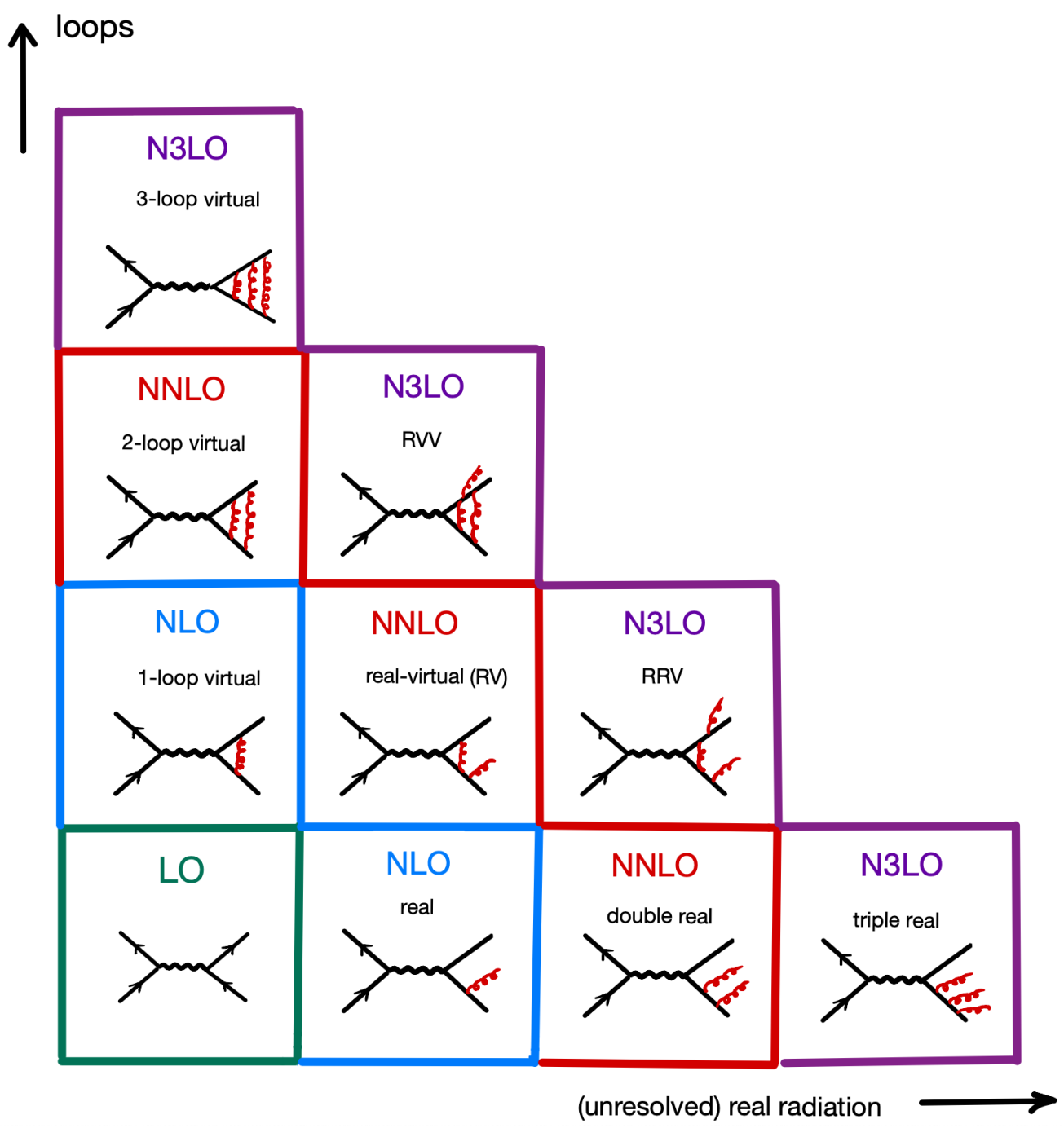

**Fig. 11.1.1** Building blocks of an N$^x$LO calculation for processes where the leading order is at tree level (in contrast to loop-induced). The higher order diagrams are only representatives of their class, the number of diagrams grows rapidly with the perturbative order.

| indep. kinem. scales | massive/ off-shell legs | internal masses | process | full $\sigma$ |
|---|---|---|---|---|
| | | $2 \to 2$ | | |
| 2 | 0 | 0 | $\gamma\gamma$ | 2011 |
| 2 | 0 | 0 | $j\,j$ | 2017 |
| 2 | 0 | 0 | $\gamma + j$ | 2017 |
| 3 | 2 | 1 | $t\bar{t}$ | 2013 |
| 3 | 2 | 0 | $VV$ | 2014 |
| 4 | 2 | 0 | $VV'$ | 2015 |
| 3 | 1 | 0 | $V + j$ | 2015 |
| 3 | 1 | 0 | $H + j$ (HTL) | 2015 |
| 4 | 2 | 1 | $HH$ | 2016 |
| 4 | 1 | 1 | $H + j$ | 2018 |
| 3 | 0 | 1 | $gg \to \gamma\gamma$ | 2019 |
| 4 | 2 | 1 | $gg \to ZZ$ | 2020 |
| 4 | 2 | 1 | $gg \to WW$ | 2020 |
| 5 | 2 | 1 | $gg \to ZH$ | 2021 |
| 4 | 2 | 1 | QCD-EW DY | 2022 |
| | | $2 \to 3$ | | |
| 4 | 0 | 0 | $3\gamma$ | 2019 |
| 4 | 0 | 0 | $\gamma\gamma j$ | 2021 |
| 4 | 0 | 0 | $3\,j$ | 2021 |
| 5 | 1 | 0 | $Wb\bar{b}$ | 2022 |

**Table 11.1.1** Timeline for the availability of full NNLO cross sections at hadron colliders (or NLO cross sections for loop induced processes) based on two-loop four-point or five-point integrals with an increasing number of mass scales. "HTL" denotes the heavy top limit, "QCD-EW DY" denotes mixed QCD-electroweak corrections to the Drell-Yan process.

More details on the methods employed for these calculations can be found e.g. in Refs. [196, 3355, 3356].

*Real radiation*

For many $1 \to 3$ or $2 \to 2$ processes, such as $e^+e^- \to$ 3 jets or di-jet production in hadronic collisions, the knowledge of the two-loop amplitudes was not the main bottleneck on the way to fully differential predictions at NNLO. Efficient schemes to treat the infrared (IR) divergent real radiation needed to be developed, and the emergence of several schemes led to an explosion in the availability of NNLO results for LHC processes with up to two particles or jets in the final state after 2015, a development which is sometimes referred to as the "NNLO revolution". The main methods to treat IR divergent real radiation beyond NLO can roughly be classified into two categories: (i) methods based on subtraction, and (ii) methods based on partitions of the phase space into IR-sensitive regions and hard regions, sometimes also called "slicing methods". The latter introduce a dependence on a resolution variable which cancels once the IR-sensitive and hard regions are combined. Subtraction methods aim at a local subtraction of the IR singular structures, i.e. a cancellation of singularities point-wise in the phase space, while for slicing methods the compensations are non-local. This non-locality can lead to large numerical cancellations, however, power corrections in the resolution variable can be included to mitigate their impact. Reviews about recent developments in IR subtraction schemes can be found e.g. in Refs. [195, 3356, 3357]. The main methods are summarized in Table 11.1.2.

The extension of methods to isolate IR divergent real radiation to N$^3$LO, i.e. the case of up to three unresolved partons, in particular in the presence of tagged coloured particles or jets, is one of the current challenges in the field of high precision perturbative QCD calculations.

While the complete automation of NNLO calculations is probably not feasible in view of efficiency optimisations that are process specific, libraries with a large collection of codes providing NNLO predictions are available, such as MATRIX [3395], NNLOJET [3362] or MCFM [3402].

*Current frontier and recent developments*

As shown in Fig. 11.1.1, the calculation of N$^x$LO corrections to processes with $(n-2)$ identified particles or jets in the final state in hadronic collisions (where the leading order is a tree amplitude, as contrasted to loop-induced amplitudes such as Higgs boson production in gluon fusion), requires the calculation of amplitudes with $x - j$ loops and $n + j$ legs, where $j = 0, \ldots, x$. The current frontier is, roughly speaking, $x + n \geq 6$, having in mind $2 \to 3$ processes at NNLO, $2 \to 1$ processes at N$^3$LO or 4-loop form factors. However, the type of the involved particles is very important for the complexity of the calculation: all available complete N$^3$LO results to date involve only color singlets in the final state, see e.g. Refs. [3403–3407] for the Drell-Yan process, Refs. [187, 3408–3411] for Higgs boson production in gluon fusion in the heavy top limit and Ref. [3412] for $VH$ production. Inclusive N$^3$LO results are also available for Higgs [3385] and Higgs pair [3413] production in vector boson fusion (VBF), Higgs pair production in gluon fusion in the heavy top limit [3414] and Higgs production in bottom quark fusion [3415, 3416]. The extension to colored final states requires advances in the treatment of IR divergent real radiation, for example $N$-jettiness soft and beam functions at this order, see e.g. Refs. [1987, 3417–3421] or triple collinear splitting functions [3422, 3423], see also Ref. [3357] for more details.

| method | NNLO examples |
|---|---|
| subtraction | |
| antenna subtraction [3360, 3361] | $e^+e^- \to 3$ jets [3358, 3359], $pp \to 2$ jets [3362, 3363], $pp \to WH\,j$ [3364] |
| sector-improved residue subtraction [3365–3367] | $pp \to t\bar{t}$ [3339], $pp \to W + c$-jet [3368], $pp \to 3$ jets [3369] |
| nested soft-collinear subtraction [3372–3374] | $pp \to VH$ [3370, 3371], VBF $H$ [3375], mixed QCD-EW to Drell-Yan [3376–3378] |
| ColorFul [3379, 3380] | $e^+e^- \to 3$ jets [3381], $H \to b\bar{b}$ [3382] |
| projection to Born [3383, 3385] | VBF $H$ [3383], VBF $HH$ [3384], single top [3386, 3387] |
| local analytic subtraction [3388–3390] | $e^+e^- \to 2$ jets [3388] |
| 4-dimensional schemes [195, 3392] | $\gamma^* \to t\bar{t}$ [3391] (inclusive) |
| slicing | |
| $q_T$ [3393, 3394] | $VV'$ [3395], $t\bar{t}$ [3396], mixed QCD-EW to Drell-Yan [3397, 3398] |
| N-jettiness [1776, 1980, 1981] | $V + j$ [1980, 3399], $H + j$ [3400, 3401], di-boson [3402] |

**Table 11.1.2** Methods for the isolation of IR divergent real radiation at NNLO and up to three examples of their application.

Another ingredient which is needed to be consistent at this order are N$^3$LO parton distribution functions, see Ref. [3044] for recent progress.

For processes such as Higgs boson decays into heavy quarks or the production of heavy quarks at $e^+e^-$ colliders at three loops, massive 3-loop form factors need to be calculated, and the presence of the additional mass scale substantially increases the complexity of the calculation. Analytical and semi-numerical methods have pushed these calculations quite far [3424–3430].

Only very few results for three-loop amplitudes with more than three legs exist. Remarkable recent results are the 3-loop amplitudes for $q\bar{q} \to \gamma\gamma$ [3431], $gg \to \gamma\gamma$ [3432], $q\bar{q} \to q'\bar{q}'$ [3433], $q\bar{q} \to gg$ [3434] and $gg \to gg$ [3435]. For the case of one massive external leg, results for planar master integrals exist [3436].

Another highlight on the 3-loop front is the calculation of the NNLO corrections to Higgs boson production in gluon fusion with full top quark mass dependence [3437], which involves the calculation of 3-loop integrals with two mass scales.

Considering $x = 2, n = 5$, i.e. processes involving 2-loop 5-point integrals, again the number of mass scales is the critical measure of complexity. Results for complete cross sections have been achieved for processes involving only massless particles: $pp \to 3\gamma$ [3438, 3439], $pp \to \gamma\gamma\,j$ [3440–3443] and $pp \to 3$ jets [3369], as well as for the process $pp \to Wb\bar{b}$ [3444, 3445].

At four loops, the computation of form factors has seen enormous progress in the past few years [3446], culminating in the calculation of the complete analytic expressions for the photon-quark and the Higgs-gluon form factors at 4-loop order [1919]. These form factors will serve as building blocks for a future complete N$^4$LO calculation of the Drell-Yan process and Higgs boson production in gluon fusion in the heavy top limit. N$^4$LO results for $gg \to H$ in the large-$N$ soft-virtual approximation already exist [3447]; results for soft corrections to deeply inelastic scattering (DIS) at 4-loop order are also available [3448].

Results at five and more loops mainly involve two-point functions, entering for example the calculation of $\beta$-functions, such as the 5-loop $\beta$-function in QCD [3449–3452] or in scalar theories [3453]. Five-loop contributions to the anomalous magnetic moment of the electron have been calculated in Refs. [3454–3457]. Results for anomalous dimensions at six [3458, 3459] or seven loops and beyond [3460, 3461] are available for scalar theories.

### 11.1.3 Phenomenology

The progress described above concerning precision calculations in QCD has led to a plethora of phenomenological results at unprecedented precision, such as determinations of the strong coupling described in Section 3.2, determinations of the $W$-boson mass, precision measurements in Higgs- and electroweak physics (see Sections 12.4, 12.3) and top quark physics (see Section 12.5). Advances in jet algorithms and jet substructure measurements (see Sections 12.2 and 11.5)) also play a major role in the LHC precision program. Cross sections for inclusive jet production can be measured at the LHC with an uncertainty of about 5% for central rapidities. This poses challenges on the theory side, in particular it requires a judicious choice of the central scale, as some choices can induce infrared-sensitive contributions [3462]. Furthermore, the transverse momenta of the jets can reach values around 4 TeV, making the combination of NNLO QCD corrections with NLO electroweak corrections indispensable to describe the high-$p_T$ region correctly. In order to make such precision calculations usable efficiently for PDF fits or $\alpha_s$ determinations, it is also important to have them available in a flexible format, for example in the form of fast interpolation grids, see e.g. Ref. [3463] for more details.

Together with ongoing progress in reducing PDF uncertainties, as well as in controlling non-perturbative effects and parton shower uncertainties (see e.g. Sections 11.4, 11.3), precision phenomenology at hadron colliders has reached a level which was unthinkable 50 years ago when QCD was "born".

### 11.1.4 Outlook

The calculation of perturbative higher order corrections in QCD at high energies is a success story. Inventive new methods have been developed to deal with the increasing level of complexity at higher perturbative orders. These technical advances were accompanied by a better understanding of important phenomenological concepts, such as infrared-safe observables and jet algorithms, and of the limitations of fixed-order perturbation theory. These developments went hand in hand with increasingly precise measurements of QCD processes at high energy colliders, and they are important pillars of the search for physics beyond the Standard Model.

While the uncertainties due to the truncation of the perturbative series were the dominant theory uncertainties for a long time in the 50-years history of QCD, for processes where the N$^3$LO level of QCD corrections is reached it became clear that other uncertainties, such as PDF uncertainties, parton shower uncertainties, quark mass effects, parametric uncertainties (e.g. in $\alpha_s, m_t$) or power-suppressed and non-perturbative contributions need to be considered with high priority as well. Being able to control them will play an important role in the next 50 years of QCD and in the search for physics beyond the Standard Model.

## 11.2 Analytic resummation

**Simone Marzani**

### 11.2.1 Large logarithms

QCD processes that involve high-momentum transfer, usually referred to as "hard processes", can be described in perturbation theory. In this framework, theoretical precision is achieved by computing the cross section $\sigma$ for an observable $\mathcal{V}$, which we assume having the dimension of an energy scale, including higher- and higher-order corrections in the strong coupling $\alpha_s$, i.e. the so-called fixed-order expansion:

$$\sigma\left(\mathcal{V}\right) = \sigma_0 + \alpha_s\, \sigma_1 + \alpha_s^2\, \sigma_2 + \alpha_s^3\, \sigma_3 + \mathcal{O}(\alpha_s^4), \qquad (11.2.1)$$

where the leading order (LO) contribution $\sigma_0$ is the Born-level cross section for the scattering process of interest. Subsequent contributions in the perturbative expansion $\sigma_x$ constitute the next-to$^x$-leading (N$^x$LO) corrections. In the language of Feynman diagrams, each power of $\alpha_s$ corresponds to the emission of an additional QCD parton, either a quark or a gluon, in the final state or to a virtual correction.

Calculations of Feynman diagrams are plagued by the appearance of divergences of different nature. Loop-diagrams can exhibit ultra-violet singularities. Because QCD is a renormalizable theory, such infinities can be absorbed into a redefinition of the parameters that enter the Lagrangian. Throughout this discussion, it is understood that such renormalization has already occurred. Real-emission diagrams exhibit singularities in particular corners of the phase-space. More specifically, these singular contributions have to do with collinear, i.e. small-angle, splittings of massless partons and emissions of soft gluons, either off massive or massless particles. Virtual diagrams also exhibit analogous infra-red and collinear (IRC) singularities, and rather general theorems [3464–3466] state that such infinities cancel when real and virtual corrections are added together, thus leading to observable transition probabilities that are free of IRC singularities. Moreover, in order to be able to use the perturbative expansion of Eqn. (11.2.1), one has to consider observables $\mathcal{V}$ that are "IRC safe", i.e. measurable quantities that do not spoil the above theorems.

The theoretical community has put a huge effort in computing higher-order corrections, as discussed in detail in Section 11.1. One of the main challenges in this enterprise is the treatment of the infra-red region and the cancellation of the singular contributions between real and virtual diagrams. Furthermore, the emissions of soft and/or collinear partons are also problematic because they can generate large logarithmic terms in the perturbative coefficients $\sigma_x$, thus invalidating the fixed-order approach. The expansion of Eqn. (11.2.1) works well if the measured value of the observable is $\mathcal{V} \simeq Q$, where $Q$ is the scale which characterizes the hard process. However, it loses its predictive power if the measurement of $\mathcal{V} \ll Q$ confines the real radiation into a small corner of phase-space, while clearly leaving virtual corrections unrestricted. For IRC safe observables, soft and collinear singularities cancel, but logarithmic corrections in the ratio $\mathcal{V}/Q$ are left behind, causing the coefficients $\sigma_x$ to become large, so that $\alpha_s^x \sigma_x \sim 1$. Because these logarithmic corrections are related to soft and/or collinear emissions, one can expect at most two powers of $L = \ln \frac{Q}{\mathcal{V}}$ for each power of the strong coupling:

$$\sigma(\mathcal{V}) = \sigma_0 + \alpha_s \left(\sigma_{12} L^2 + \sigma_{11} L + \dots\right) + \alpha_s^2 \left(\sigma_{24} L^4 + \sigma_{23} L^3 + \dots\right) + \mathcal{O}(\alpha_s^n L^{2n}). \tag{11.2.2}$$

All-order resummation is then a re-organization of the above perturbative series. For many observables of interest, the resummed expression exponentiates, leading to

$$\sigma(\mathcal{V}) = \sigma_0\, g_0(\alpha_s) \times \exp\left[L g_1(\alpha_s L) + g_2(\alpha_s L) + \alpha_s g_3(\alpha_s L) + \dots\right], \tag{11.2.3}$$

where $g_0$ is a constant contribution which admits an expansion in $\alpha_s$. In analogy to the fixed-order terminology, the inclusion of the contribution $g_{x+1}$, $i \geq 0$, leads to next-to$^x$-leading logarithmic (N$^x$LL) accuracy.

Fixed-order Eqn. (11.2.1) and resummed Eqn. (11.2.3) expansions are complementary. On the one hand, fixed-order calculations fail in particular limits of phase-space, indicating the need for an all-order approach. On the other hand, all-order calculations are only possible if particular assumptions on the emission kinematics are made. Thus, the most accurate theoretical description for the observable $\mathcal{V}$ is achieved by matching the two approaches

$$\sigma^{\text{matched}}(\mathcal{V}) = \sigma^{\text{f.o.}}(\mathcal{V}) + \sigma^{\text{res}}(\mathcal{V}) - \sigma^{\text{d.c.}}(\mathcal{V}), \tag{11.2.4}$$

where the third contribution corresponds to the expansion of the resummation to the order we are matching to and it is subtracted in order to avoid double counting. For instance, if we were to match the resummed expression Eqn. (11.2.3), computed to some logarithmic accuracy to a fixed-order calculation, see Eqn. (11.2.2), performed at NNLO, $\sigma^{\text{d.c.}}$ would correspond to the expansion of the resummed result up to second order in the strong coupling, relatively to Born term. Furthermore, we note that, if the resummation is computed at high-enough accuracy, the dangerous logarithmic corrections cancel between $\sigma^{\text{f.o.}}$ and $\sigma^{\text{d.c.}}$ and all the large contributions are resummed in $\sigma^{\text{res}}$.

All-order resummation is possible because (squared) matrix element and phase-space factorize in certain kinematic limits. Different methods to achieve such factorization have been developed in the literature. For instance, factorization can be obtained by studying directly QCD amplitudes and cross-sections in the soft and collinear limits. Then, resummation can be achieved by iteratively identifying factorization and exponentiation properties of QCD matrix elements and cross-sections [152, 3467, 3468]. Other approaches instead introduce non-local correlation operators, such as Wilson lines, and exploit their renormalization group evolution [3469]. Finally, one can construct soft-collinear effective field theories (SCET) to describe the soft and collinear degrees of freedom of QCD [1761–1764, 1791, 1861, 1863] (see Ref. [3470] for an extensive, review). There is a rich literature describing similarities and differences of the various resummation approaches, see e.g. [3471–3476]. In this presentation we will mostly follow the iterative point of view.

### 11.2.2 Transverse-momentum resummation

The transverse momentum ($Q_T = p_T^Z, p_T^W, p_T^H$) distribution of electroweak final states at hadron colliders is one of the most extensively investigated observables in QCD. Studies of $Q_T$ spectra and related angular correlations of lepton pairs produced via the Drell-Yan (DY) process provide us with a useful testing ground for an even more interesting Higgs and new physics program. These processes are characterised by the presence of two distinct scales: the measured $Q_T$ and the invariant mass of the final-state $Q$, which is close to the mass of the electroweak boson for on-shell production. Therefore, if we are interested in phase-space regions where $Q_T \ll Q$, large logarithmic corrections appear. They should be accounted for to all orders, in order to achieve an accurate theoretical description of these observable distributions.

Furthermore, one aspect of physics at hadron colliders that becomes important at small $Q_T$ is the role of non-perturbative effects commonly attributed to the intrinsic transverse motion of partons within the proton. One may therefore view any opportunity to compare precise perturbative predictions with accurate experimental data for DY lepton pairs as a chance to assess the size of non-perturbative physics; physics which also affects the Higgs $Q_T$ spectrum.

The literature on $Q_T$ resummation is vast and since the seminal papers, which date back to the late 1970s,

early 1980s, e.g. [1282, 3477], there has been a continuous effort in producing accurate theoretical predictions that can describe the experimental data. For example, high logarithmic accuracy [1907, 3478–3485] has been achieved and computer programs that allow one to compute NNLL predictions matched to next-to-leading order (NLO) for the $Q_T$ distribution in case of colorless final states in hadron collision have been available for a long time, e.g. [3479, 3480, 3486–3491]. Fixed-order predictions have reached NNLO accuracy and the resummation can be now performed to N$^3$LL accuracy [188, 1906, 3492–3495]. Results with partial N$^4$LL resummation have also been recently obtained [3407].

Moreover, observables such as $\phi^*$ [3496, 3497] that exploit angular correlations to probe similar physics as $Q_T$, while being measured with even better experimental resolution, have triggered theoretical studies to extend the formalism of $Q_T$ resummation to these new variables [3488, 3498–3501].

In this section we review the main ingredients of $Q_T$ resummation for an electroweak final state, i.e. Higgs or DY. For simplicity, we are going to consider distributions which are fully inclusive in the electroweak boson decay products, as well as integrated in the boson's rapidity. The extension to more differential distributions, including fiducial cuts on the final-state particles, is possible. For convenience, we work at NLL and, as further simplification, we explicitly consider only the flavor-diagonal contributions, while restoring full flavor-dependence in the end.

We compute the differential distribution for the transverse momentum of the boson (Higgs or $Z/\gamma^*$). At Born level, we have $gg \to h$ or $q\bar{q} \to Z/\gamma^*$, so the boson has no transverse momentum, i.e. the distribution is proportional to $\delta^{(2)}(\mathbf{Q}_T)$, where $\mathbf{Q}_T$ is the two-dimensional transverse-momentum. When computing higher perturbative orders, we must include contributions with additional partons $i$ in the final state. Thus, the boson can acquire a nonzero transverse momentum, such that $\mathbf{Q}_T = -\sum_i \mathbf{k}_{Ti}$. Resummation is relevant when the transverse momentum is much smaller than the mass (or virtuality) of the electroweak boson, $Q_T^2 = |\mathbf{Q}_T|^2 \ll Q^2$. This can happen in two situations: either all recoiling partons have small transverse momenta or their transverse momenta, although not individually small, mostly cancel in their vector sum. Both these mechanisms must be taken into account and, as we shall shortly see, this can be achieved if $Q_T$ resummation is performed in Fourier space. If we denote with $\mathbf{b}$ the conjugate variable to $\mathbf{Q}_T$, then the small-$Q_T$ region corresponds to large $b = |\mathbf{b}|$ and logarithms of $Q_T$ are mapped into logarithms of $1/b$.

Thus, we consider the emission of an arbitrary number of collinear gluons off the incoming hard legs. The partonic cross-section can be written as

$$\frac{d^2\sigma}{d\mathbf{Q}_T} = \sigma^{\text{born}}_{c\bar{c}\to F} \sum_{n=0}^{\infty} \frac{1}{n!} \prod_{i=1}^{n} \int [dk_i]\,(2C_c)\frac{\alpha_s(k_{Ti})}{2\pi} \left[ z_i^{N-1}\bar{P}^{\text{real}}(z_i)\,\delta^{(2)}\left(\mathbf{Q}_T + \sum_i \mathbf{k}_{Ti}\right) + \bar{P}^{\text{virtual}}(z_i)\,\delta^{(2)}(\mathbf{Q}_T)\right] \Theta(k_{Ti} - Q_0)\,\Theta\left(1 - z_i + \frac{k_{Ti}}{Q}\right), \tag{11.2.5}$$

where we have taken Mellin moments with respect to the longitudinal momentum fractions $z_i$. The emitted gluon phase space is $[dk_i] = dz_i \frac{dk_{Ti}^2}{k_{Ti}^2}\frac{d\phi_i}{2\pi}$ and $C_c = C_F, C_A$ is the appropriate color factor. The first $\Theta$ function expresses the fact that emissions below the cut-off $Q_0$ belong to the non-perturbative region of the proton wave-function, while the second one correctly accounts for the large-angle soft region of phase-space. In order to achieve NLL accuracy, the strong coupling $\alpha_s$ has to be evaluated at two loops, in the CMW scheme [152]. The emission probability is described by the real and virtual matrix elements (see e.g. App. E of Ref. [3502]):

$$\bar{P}^{\text{real}}(z) = \begin{cases} \frac{1+z^2}{1-z}, & \text{for a quark,} \\ \frac{2z}{1-z} + \frac{2(1-z)}{z} + 2z(1-z), & \text{for a gluon;} \end{cases} \tag{11.2.6}$$

$$\bar{P}^{\text{virtual}}(z) = (-1)\begin{cases} \frac{1+z^2}{1-z}, & \text{for a quark,} \\ \frac{2z}{1-z} + z(1-z) \\ +n_f T_R(z^2 + (1-z)^2), & \text{for a gluon.} \end{cases} \tag{11.2.7}$$

For later convenience, we also introduce the leading order regularized splitting functions

$$P_{qq}(z) = \frac{\alpha_s}{2\pi}C_F\left[\frac{1+z^2}{1-z}\right]_+,$$
$$P_{gg}(z) = \frac{\alpha_s}{2\pi}2C_A\left[\left(\frac{z}{1-z} + \frac{z(1-z)}{2}\right)_+ + \frac{1-z}{z} + \frac{z(1-z)}{2} - \frac{2}{3}n_f T_R\delta(1-z)\right], \tag{11.2.8}$$

and the corresponding anomalous dimensions

$$\gamma_{cc}(N,\alpha_s) = \int_0^1 z^{N-1}P_{cc}(z), \qquad c = q, g. \tag{11.2.9}$$

We note that virtual corrections in Eqn. (11.2.5) do not change the transverse momentum $Q_T$ and trivially exponentiate. The real-emission contribution is also factorized, with the exception of the two-dimensional delta-function constraint. This is where Fourier moments with respect to the two-dimensional vector $\mathbf{Q}_T$ become helpful. We can exploit the relation

$$\delta^{(2)}\left(\mathbf{Q}_T+\sum_i \mathbf{k}_{Ti}\right)=\frac{1}{4\pi^2}\int \mathrm{d}^2\mathbf{b}\, e^{i\mathbf{b}\cdot\mathbf{Q}_T}\prod_{i=1}^{n} e^{i\mathbf{b}\cdot\mathbf{k}_{Ti}}, \tag{11.2.10}$$

to fully factorize the real-contribution in Eqn. (11.2.5). We obtain

$$\begin{aligned}W^{\text{real}}(b,N)=&\sum_{n=0}^{\infty}\frac{1}{n!}\prod_i^n\int [dk_i]\, z_i^{N-1}(2C_c)\frac{\alpha_s(k_{Ti})}{2\pi}\\&\times \bar{P}^{\text{real}}(z_i)e^{i\mathbf{b}\cdot\mathbf{k}_{Ti}}\Theta\left(k_{Ti}-Q_0\right)\\&\times\Theta\left(1-z_i+\frac{k_{Ti}}{Q}\right).\end{aligned} \tag{11.2.11}$$

The series in Eqn. (11.2.11) sums to an exponential. Thus, the resummed exponent is obtained by putting together real, virtual and PDF ($k_{Ti}<Q_0$) contributions:

$$\begin{aligned}R(b,N)=&\,2C_c\int [dk]\frac{\alpha_s(k_T)}{2\pi}\Theta\left(k_T-Q_0\right)\Theta\left(1-z+\frac{k_T}{Q}\right)\\&\times\left(-z^{N-1}\bar{P}^{\text{real}}(z)e^{i\mathbf{b}\cdot\mathbf{k}_T}-\bar{P}^{\text{virtual}}(z)\right)\\&+2\int_{Q_0^2}^{Q^2}\frac{dk_T^2}{k_T^2}\gamma_{cc}(N,\alpha_s(k_T)).\end{aligned} \tag{11.2.12}$$

By rewriting $z^{N-1}=1+(z^{N-1}-1)$ and using the definitions in Eqs. (11.2.6), (11.2.7), and (11.2.8), we are able to reshuffle the contributions to the resummed exponent as follows

$$\begin{aligned}R(b,N)=&-\int_{Q_0^2}^{Q^2}\frac{dk_T^2}{k_T^2}\int_0^{2\pi}\frac{d\phi}{2\pi}\left(1-e^{i\mathbf{b}\cdot\mathbf{k}_T}\right)\\&\times\left[\int_0^{1-\frac{k_T}{Q}}dz\frac{\alpha_s(k_T)\,C_c}{\pi}\bar{P}^{\text{virtual}}(z)\right.\\&\left.-2\gamma_{cc}(N,\alpha_s(k_T))\right]+\mathcal{O}\left(\frac{k_T}{Q}\right).\end{aligned} \tag{11.2.13}$$

The factor $\left(1-e^{i\mathbf{b}\cdot\mathbf{k}_T}\right)$ essentially acts as a cut-off on the $k_T$ integral. At NLL we have [104]

$$\begin{aligned}R(b,N)=&-\int_{b_0^2/b^2}^{Q^2}\frac{dk_T^2}{k_T^2}\left[\int_0^{1-\frac{k_T}{Q}}dz\frac{\alpha_s(k_T)\,C_c}{\pi}\bar{P}^{\text{virtual}}(z)\right.\\&\left.-2\gamma_{cc}(N,\alpha_s(k_t))\right]\\=&-\ln S_c+2\int_{b_0^2/b^2}^{Q^2}\frac{dk_T^2}{k_t^2}\gamma_{cc}(N,\alpha_s(k_t)),\end{aligned} \tag{11.2.14}$$

$b_0=2e^{-\gamma_E}$, where $\gamma_E$ is the Euler constant. Thus, we have successfully separated two distinct contributions: the Sudakov form factor ($S_c$), computed here at NLL accuracy (and systematically improvable) and a DGLAP contribution, which evolves the PDFs from the hard scale $Q$ down to $b_0/b$. Note that here we have only considered flavor-diagonal splittings. Off-diagonal ones do not alter the Sudakov form factor and they are fully taken into account by the complete DGLAP evolution.

Taking into account all the above effects, the all-order transverse momentum distribution for the production of an electroweak final state $F$ from initial-state partons $c$ and $\bar{c}$ can be written

$$\begin{aligned}\frac{d\sigma}{dQ_T^2}=&\,\sigma^{\text{born}}_{c\bar{c}\to F}\int dx_1\int dx_1\int_0^\infty db\frac{b}{2}J_0(bQ_T)S_c(b,Q)\\&\times\int dz_1\int dz_2\,\delta\left(1-z_1z_2\frac{x_1x_2s}{Q^2}\right)\\&\times\left[H^F_{c\bar{c}}\left(\alpha_s(Q)\right)C_{ca_1}\left(z_1,\alpha_s\left(\tfrac{b_0}{b}\right)\right)C_{\bar{c}a_2}\left(z_2,\alpha_s\left(\tfrac{b_0}{b}\right)\right)\right.\\&\left.+\tilde{H}^F_{c\bar{c}}\left(\alpha_s(Q)\right)G_{ca_1}\left(z_1,\alpha_s\left(\tfrac{b_0}{b}\right)\right)G_{\bar{c}a_2}\left(z_2,\alpha_s\left(\tfrac{b_0}{b}\right)\right)\right]\\&\times f_{a_1}\left(x_1,\tfrac{b_0}{b}\right)f_{a_2}\left(x_2,\tfrac{b_0}{b}\right),\end{aligned} \tag{11.2.15}$$

where we have introduced the Bessel function $J_0$ and the sum over $a_1, a_2$ is understood. The functions $G_{ab}$, $C_{ab}$, $H^F_{ab}$, $\tilde{H}^F_{ab}$ can be computed in perturbation theory, while $f_a$ denotes the the parton distribution functions. For Standard Model Higgs production we have $F=h$, $c=\bar{c}=g$, and $H=\tilde{H}$, while for DY production we have $F=Z/\gamma^*$ and $c=q$, and $G_{q,a}=G_{\bar{q},a}=0$. As already mentioned, different resummation formalisms exist in the literature. They all agree at the perturbative accuracy they claim, but they may differ because of subleading effects. As an example, in Fig. 11.2.1 we show a comparison between the resummed and matched calculation of Ref. [3495] and LHC data, collected by the ATLAS collaboration [3504].

[104] See Ref. [3503] for a generalization of this approximation to higher-logarithmic accuracy.

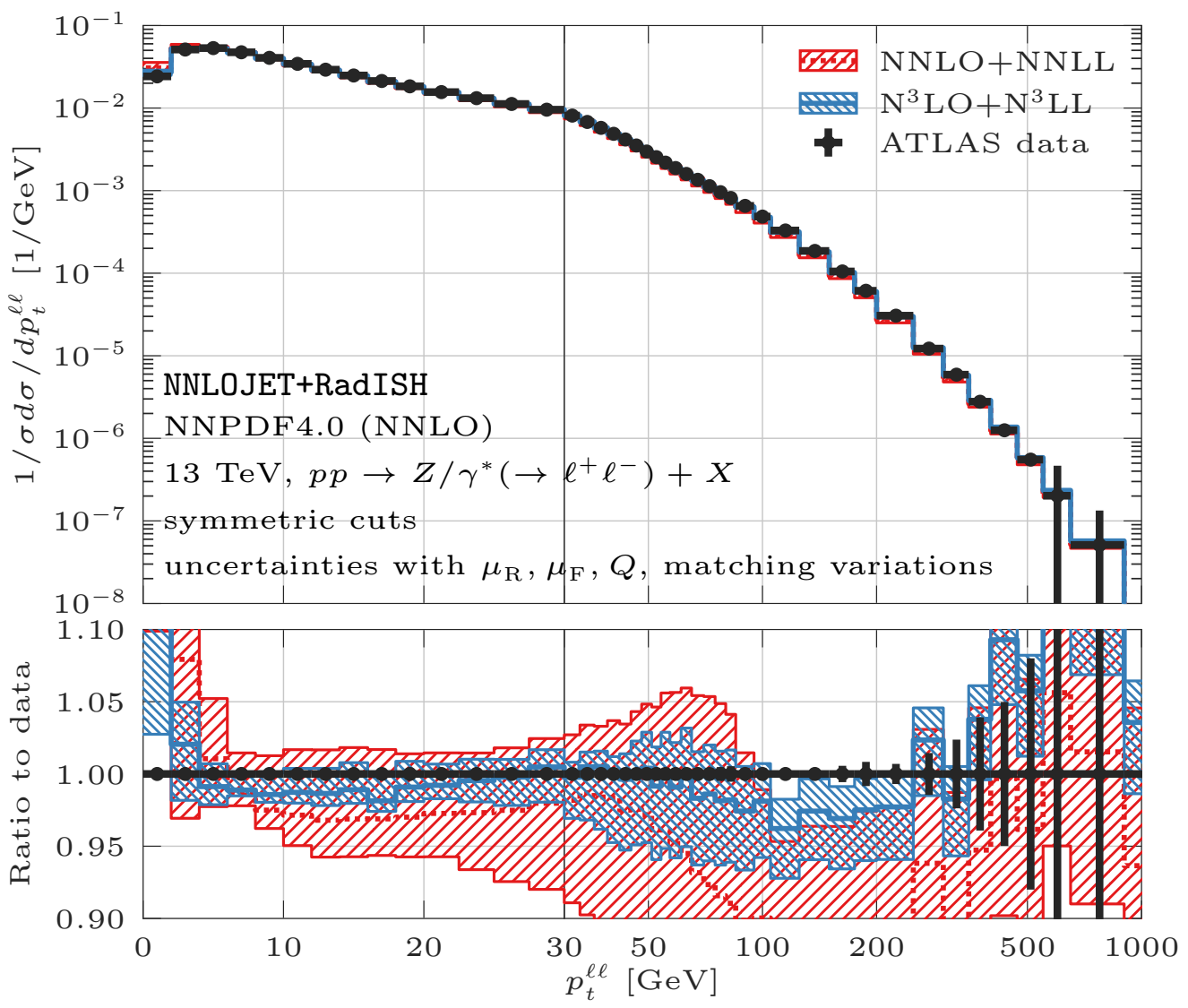

**Fig. 11.2.1** The lepton pair transverse momentum distribution measured by the ATLAS collaboration at the LHC [3504] is compared to a resummed and matched calculation. $Q_T$ resummation is performed at N$^3$LL logarithmic accuracy and it is matched to the N$^3$LO fixed-order result, with respect to the Born process, which corresponds to NNLO accuracy for the transverse momentum distribution. The plot is taken from Ref. [3495].

### 11.2.3 Jets and their substructure

All-order techniques not only allow us to probe the dynamics of electroweak bosons that recoil against QCD radiation, as discussed above, but can be employed to study the properties of the radiation itself in great detail. If we look at hadronic final states, we realise that QCD radiation is not uniformly distributed, but rather concentrated in collimated sprays of hadrons that are called jets. Jets really live at the boundary between experimental and theoretical particle physics and are abundantly used by both communities. They allow us to describe complex final states in terms of a few objects rather than hundreds of particles. Furthermore, from a theoretical point of view, jets are closely related to quarks and gluons, i.e. the degrees of freedom of perturbative QCD. Thus, jet definitions, commonly referred to as jet algorithms, must have good theoretical and experimental properties. For instance, jet algorithms should be IRC safe, so that they yield finite cross-sections when evaluated in perturbation theory [166].

Modern jet algorithms are based on the concept of sequential recombination. Pairwise distances between particles are evaluated in order to decide whether to recombine two particles. The metric used to evaluate these distances characterizes the jet algorithm. Nowadays, the most popular group of jet algorithm is the generalized $k_T$ family, for which the metric is defined by

$$d_{ij} = \min\left(p_{Ti}^{2p}, p_{Tj}^{2p}\right)\frac{\Delta R_{ij}^2}{R^2}, \quad d_{iB} = p_{Ti}^{2p}, \tag{11.2.16}$$

where $p_{Ti}, p_{Tj}$ are the particles' transverse momenta and $\Delta R_{ij}^2$ is their distance in the azimuth-rapidity plane. $R$ is an external parameter, which plays the role of the jet radius. Different choices for the parameter $p$ are possible. For instance, $p = 0$ corresponds to the so-called Cambridge-Aachen (C/A) algorithm [170, 171], with a purely geometrical distance. For $p = 1$ we have the $k_T$-algorithm [172, 3505], which by clustering particles at low $p_T$ first, is likely to faithfully reconstruct a QCD branching history. Finally, with the choice $p = -1$ we obtain the anti-$k_T$ algorithm [174], which clusters soft particles around a hard core, producing fairly round jets in the azimuth-rapidity plane. It is interesting to note that all algorithms of the generalized $k_T$ family act identically on a configuration with just two particles: they are recombined if $\Delta R_{ij} < R$. More details about jets can be found in Sec. 11.5. Although incredibly useful for phenomenology, jet algorithms introduce resolution parameters, such as the jet radius $R$, rendering the computation of jet properties a multi-scale problem.

In the past decade, many observables have been devised to study the internal properties of high-$p_T$ jets, see for instance [3506]. The simplest example of such observables is the jet invariant mass, which is defined as

$$m_{\text{jet}}^2 = \left(\sum_{i\in\text{jet}} p_i\right)^2, \quad \rho = \frac{m_{\text{jet}}^2}{R^2 p_T^2}, \tag{11.2.17}$$

where $p_i$ are the jet constituents' four-momenta and, in order to emphasise the multi-scale nature of the problem, we have also introduced the dimensionless ratio $\rho$. This ratio is small in the boosted regime $m_{\text{jet}} \ll Rp_t$, which is of particular interest at the LHC. As previously discussed, when scales become widely separated, logarithms (of $\rho$ in this case) become large and in order to obtain reliable predictions for this observable, we need to perform all-order calculations.

We do not report here the details of the resummed calculation for the jet mass distribution, which is closely related to the one of the thrust event shape [3468, 3507, 3508], but rather we stress similarities and differences with respect to $Q_T$ resummation, described above. Large logarithmic corrections always arise from the emission of soft and/or collinear partons. However, final-state four-momenta are combined differently in the two observables and therefore a different integral transform is needed to diagonalize the invariant mass $\delta$-function. Furthermore, because we are interested in the dynamics of a high-$p_T$ isolated jet, emissions collinear to the incoming legs do not significantly alter the jet properties, leading to a simplified treatment of the PDF contributions. However, there is a major complication that arises when performing calculations with jets. Only emissions that are recombined into the jet contribute to the invariant mass, making it an example of a non-global observable [3509]. As it turns out, the presence of phase-space boundaries noticeably complicates the structure of soft-emissions and essentially invalidates simple exponentiation. Furthermore, the actual shape of the boundary depends on the jet algorithm of choice. For instance, in the presence of many soft emissions together with a hard parton, the anti-$k_T$ algorithm will always cluster all soft gluons to the hard parton, behaving as a rigid cone algorithm, while the choice of different algorithms, such as C/A or $k_T$, can give rise to more complicated clustering sequences, see e.g. [3510] and references therein. The calculation of non-global logarithms constitutes the bottle-neck of jet calculations but thanks to an extraordinary effort of different groups, they can now be resummed at high accuracy [1784, 1997, 3511–3517].

The calculation techniques developed for the jet mass have been extended to other jet substructure observables. An interesting example is the family of jet angularities [3518]. These probe both the angular and the transverse momentum distribution of particles within a given jet. They are defined from the momenta of jet constituents as follows:

$$\lambda_\alpha^\kappa = \sum_{i\in\text{jet}} \left(\frac{p_{T,i}}{\sum_{j\in\text{jet}} p_{T,j}}\right)^\kappa \left(\frac{\Delta_i}{R}\right)^\alpha , \tag{11.2.18}$$

where

$$\Delta_i = \sqrt{(y_i - y_{\text{jet}})^2 + (\phi_i - \phi_{\text{jet}})^2} , \tag{11.2.19}$$

is the azimuth-rapidity distance of particle $i$ from the jet axis. Jet angularities are IRC safe for $\kappa = 1$ and $\alpha > 0$. In Fig. 11.2.2 we show a comparison between LHC data collected by the CMS collaboration [3519], for the so-called Les Houches Angularity (LHA), which corresponds to setting $\kappa = 1$ and $\alpha = 0.5$, and a resummed calculation performed at NLL accuracy [3520, 3521].

Despite the remarkable perturbative accuracy that can be achieved for jet observables, non-perturbative corrections due to the hadronization process or originating from multiple-parton interactions or pile-up, are rather large. Indeed, the resummed curve in Fig. 11.2.2 has been corrected for non-perturbative effects, which are important to ensure agreement with the data at small values of the angularity. The situation can be greatly improved if one considers "grooming" and "tagging" algorithms. Broadly speaking, a grooming procedure takes a jet as an input and tries to clean it up by removing constituents which, being at wide angle and relatively soft, are likely to come from contamination, such as the underlying event or pile-up. A tagging procedure instead focuses on some kinematic variable that is able to distinguish signal from background, such as, for instance, the energy sharing between two subjets within the jet, and cuts on it. Many of the algorithms on the market usually perform both grooming and tagging and a clear distinction between the two is difficult. Regardless of their nature, these algorithms try to resolve jets on smaller angular and energy scales, thereby introducing new parameters. This further challenges our ability of computing predictions in perturbative QCD. However, if these algorithms are properly designed, they can effectively reduce contamination from non-perturbative physics, while maintaining calculability. An example of this is SoftDrop [1875]. This procedure steps backward through the C/A clustering tree of a jet and iteratively checks whether the transverse momenta of the two branches satisfy the condition

$$\frac{\min(p_{T1}, p_{T2})}{p_{T1} + p_{T2}} > z_{\text{cut}} \left(\frac{\Delta_{12}}{R}\right)^\beta . \tag{11.2.20}$$

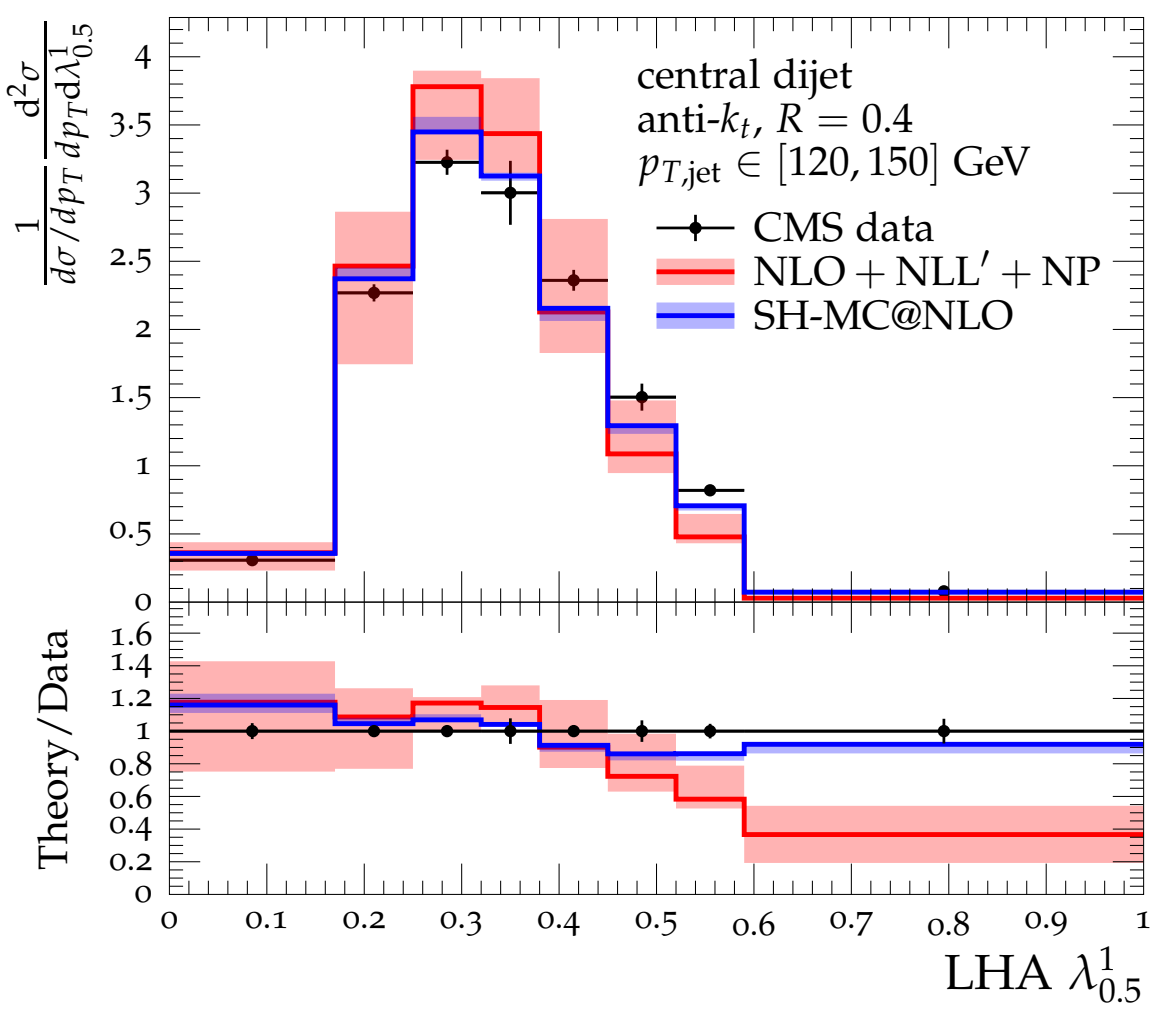

**Fig. 11.2.2** The Les Houches Angularity (LHA) distribution, which corresponds to $\kappa = 1$, $\alpha = 0.5$ in Eq. (11.2.18) measured by the CMS collaboration at the LHC [3519]. The data are compared to the resummed and matched calculation (NLL+NLO), supplemented by non-perturbative corrections [3520, 3521] and to the prediction obtain with a state-of-the-art Monte Carlo simulation using Sherpa at NLO QCD accuracy [3522]. The plot is taken from Ref. [3521].

The difficulty posed by substructure algorithms in general, and SoftDrop in particular, is the presence of new parameters, such as $z_{\rm cut}$ and $\beta$, that slice the phase-space for soft gluon emission in a non-trivial way, resulting in potentially complicated all-order behavior of the observable at hand. In the soft limit, the SoftDrop criterion reduces to

$$z > z_{\rm cut}\left(\frac{\theta}{R}\right)^{\beta} \quad \Rightarrow \quad \ln\frac{1}{z} < \ln\frac{1}{z_{\rm cut}} + \beta\ln\frac{R}{\theta}, \tag{11.2.21}$$

where $z$ is the momentum fraction and $\theta$ the opening angle. For $\beta > 0$, collinear splittings always satisfy the SoftDrop condition, so a SoftDrop jet still contains all of its collinear radiation. The amount of soft-collinear radiation that satisfies the SoftDrop condition depends on the relative scaling of the energy fraction $z$ to the angle $\theta$. As $\beta \to 0$, more of the soft-collinear radiation of the jet is removed, and in the $\beta = 0$ limit, all soft-collinear radiation is removed [1876, 3523]. Therefore, we expect the coefficient of the double logarithms in observables like the groomed jet mass, the origin of which is soft-collinear radiation, to be proportional to $\beta$. In the strict $\beta = 0$ limit, collinear radiation is only maintained if $z > z_{\rm cut}$. Because soft-collinear radiation is vetoed, the resulting jet mass distributions will only exhibit single logarithms, as emphasized in [1876, 3523]. Moreover, non-global logarithms are found to be power-suppressed for $\beta > 0$, and absent for $\beta = 0$. Finally, for $\beta < 0$, there are no logarithmic structures for observables like groomed jet mass at arbitrarily low values of the observable. For example, $\beta = -1$ roughly corresponds to a cut on the relative transverse momentum of the two subjets under scrutiny.

The above understanding can be formalized into actual calculations and the resummation of a variety of observables measured on SoftDrop jets has been performed to N$^3$LL [1786, 1899]. This outstanding theoretical accuracy, together with reduced sensitivity to non-perturbative corrections, make SoftDrop jets a particularly powerful way to probe QCD dynamics and jet formation.

### 11.2.4 Outlook

In this brief overview we have introduced resummation as a powerful tool that we can use to augment the ability of perturbative calculations to describe the data. We have given two examples of multi-scale processes, namely the transverse momentum of an electroweak boson and the Les Houches (jet) Angularity, for which the inclusion of all-order effects is mandatory in order to be able to describe the data.

Resummation provides us with the right tools to study emergent phenomena in QCD, such as jet formation and it allows us to scrutinise fundamental properties of the theory. The concept of factorization, i.e. the ability of separating physical effects happening at different energy scales, is the foundation of the whole resummation program that we have discussed. Even more generally, we can say any QCD calculation, being it done at fixed-order or at the resummed level, requires some notion of factorization. Of particular importance is the collinear factorization theorem [224] that allows us to separate the perturbative, i.e. calculable, part of a process from the non-perturbative one, which can be described in terms of parton distribution (or fragmentation) functions. Resummation techniques allows us to uncover limitations and possible breakdowns of factorization [1857, 3524], which typically happen at perturbative orders that are too high to be reached with fixed-order techniques. Thus, despite resummation being based on the soft/collinear approximation of the perturbative approximation of QCD, it opens up a window to fundamental aspects of the theory:

*Resummation just scratches the surface of QCD. But it makes a mark.* [105]

## 11.3 Parton showers

**Frank Krauss**

### 11.3.1 Motivation

Producing charged particles in a high-energy collision initiates the emission of secondary bremsstrahlung quanta. Due to the large strong coupling and because of the gluon self-coupling, the radiation of gluons is of particular relevance, and tens or even hundreds of secondary quarks and gluons can be produced in a cascade of emissions.

Apart from the wish to correctly describe particle production at collider experiments in all its facets, and preferably based on first principles, there is another, more practical reason why this process of multiple parton emission is of great phenomenological relevance. The confinement property of QCD prevents quarks and gluons to be directly observed and instead, they manifest themselves through hadrons, which constitute the observable final states. Unfortunately, to date, only phenomenological models for the dynamical transition from quarks and gluons to hadrons in a process aptly dubbed hadronization have been developed, which rely on a large number of parameters which have to be fitted –

[105] George Sterman, CTEQ school 2006.

"tuned" – to experimental data. Clearly, such a programme is sensible only, if the parameters are sufficiently independent from the hard process and rather depend on the properties of the parton ensembles at a common low scale. This is realized by casting the multiple emission of the secondary quanta, the parton cascade, into algorithms that systematically evolve the few original partons in the hard process at a scale of large momenta $Q$ into resulting many–parton ensembles resolved at a lower scale $Q_0$, at which hadronization sets in. The resulting algorithms are called parton showers, and one might think of them as numerical implementations of a renormalization–group equation that connects these two scales, $Q$ and $Q_0$. They form an integral part of modern event generators HERWIG [3525], PYTHIA [3526], and SHERPA [3522].

### 11.3.2 Parton Shower Realizations

Some first intuition about parton showers can be gained from the (quasi-classical) spectrum of gluons emitted by a fast moving color charge,

$$\mathrm{d}n_g = \frac{\alpha_S}{\pi}\frac{\mathrm{d}\omega}{\omega}\frac{\mathrm{d}^2p_\perp}{p_\perp^2}\,, \tag{11.3.1}$$

exhibiting its characteristic divergent structure in the limit where the emitted gluon becomes soft, with it energy $\omega \to 0$, or collinear with respect to the emitter, with its transverse momentum $p_\perp \to 0$. These well–known soft and collinear divergences, typical for quantum field theories with massless (vector) particles cancel for physically meaningful observables when both real and virtual emissions are taken into account [3527, 3528]. In parton showers, which aim to simulate the emission of real quanta, this is implicitly taken into account, by demanding that the emitted partons are resolvable with a minimal energy and transverse momentum; divergences in unresolvable emissions then cancel those from virtual corrections. Such a constraint is effectively realized for example by demanding a minimal transverse momentum, $k_\perp > Q_0$ in emissions. The integrated spectrum depends logarithmically on the cut-off, and small values of $Q_0$ overcoming the smallness of $\alpha_s$ necessitate the resummation of the infrared logarithms.

This physical picture is encoded in probabilistic language, by constructing a Sudakov form factor

$$\Delta_{N\to N+1}(Q,\,Q_0) = \exp\left\{-\int\limits_{Q_0^2}^{Q^2}\mathrm{d}\Phi_{N\to N+1}(t,\,z,\,\phi)\,\mathcal{K}_{N\to N+1}(\Phi_{N\to N+1})\right\}, \tag{11.3.2}$$

which yields the probability for an $N$-particle configuration with momenta $\{\tilde{p}\}$ *not* to emit another particle (and therefore *not* to turn it into an $(N+1)$-particle configuration with momenta $\{p\}$). The phase space element for the emission, $\Phi_{N\to N+1}(t,\,z,\,\phi)$, will depend on (1) the ordering parameter $t$ defined below; (2) the splitting parameter $z$ given by the light-cone momentum fraction or energy fraction of the emitted particle; and (3) the azimuth angle $\phi$, fixing the orientation of the emitted particle in the transverse plane of the mission. The emission kernel $\mathcal{K}_{N\to N+1}(\Phi_{N\to N+1})$ depends on the phase space of the emission and on the strong coupling $\alpha_S(p_\perp^2)$, with the transverse momentum as preferred scale choice. In the collinear limit, $t \to 0$ with finite $z$, the kernel for a specific emitter $(ij)$ splits into particles $i$ and $j$ and reduces to the well-known corresponding DGLAP splitting kernels. In the soft limit, $z \to 0$ which also forces $t \to 0$, the kernel should approach the eikonal form,

$$\lim_{z\to 0}\mathcal{K}_{N\to N+1}(\Phi_{N\to N+1}(t,\,z,\,\phi)) \propto \frac{(p_i\cdot p_k)}{(p_i\cdot p_j)(p_j\cdot p_k)}\,, \tag{11.3.3}$$

where $k$ denotes the color spectator. Owing to the current standard of using the leading-color approximation in the parton shower construction, $k$ can be uniquely chosen.

Individual simulated events are seeded by the hard process, evaluated at a fixed perturbative order, and dressed afterwards with the parton shower. In marked contrast to the forward evolution of the final-state parton shower, the parton shower in the initial state is described by a backward evolution, back to the initial beam particles and to a fixed, pre-defined state. To enforce that the backward evolution of the parton shower arrives at the correct initial state, while respecting the evolution of its internal structure, emissions are weighted by a ratio of parton distribution functions [3529],

$$\mathcal{K}_{N\to N+1}(\Phi_{N\to N+1}(t,\,z,\,\phi)) \propto \frac{f_i(x_{(ij)}/z,\,\mu_i^2)}{f_{(ij)}(x_{(ij)},\,\mu_{(ij)}^2)}\,. \tag{11.3.4}$$

In this way the particle $(ij)$, resolved at scale $\mu_{(ij)}^2$ and with momentum fraction $x_{(ij)}$, is replaced by the new initial-state particle $i$, resolved at a lower scale $\mu_i^2 < \mu_{(ij)}^2$ and with a larger momentum fraction $x_i = x_{(ij)}/z$.

The choice of a parton-shower realization has an impact on the accuracy with which the radiation pattern is simulated. In first-generation parton-shower implementations, the ordering parameter $t$ is either identified with the virtual mass of the parton before emission, $t = p_{(ij)}^2 = (p_i + p_j)^2$ [3530, 3531] or with the (scaled) opening angle of the emission, $t = E_{(ij)}^2(1 -$ $\cos\theta_{ij}$) [3532, 3533]. When the regular parts of the (massless) DGLAP splitting kernels at $\mathcal{O}(\alpha_S)$ are used, suitably limiting the allowed range for $z$ accounts for the effect of finite masses. Careful analysis of the radiation pattern indicated that angular ordering is an important ingredient to the correctness of the simulation. The ordering accounts for color coherence effects, and introduces an explicit veto on increasing opening angles of the virtuality-ordered parton showers. In contrast, the dipole shower formulation [3534] in ARIADNE [3535] explicitly fills the Lund plane [3536] in transverse momentum $p_\perp^2$ and rapidity $y$ of emissions. By setting the ordering parameter $t = p_\perp^2$ with the identification of $p_\perp^2$ as the inverse of the eikonal from Eq. (11.3.3), it fulfils the color coherence requirements that give rise to angular ordering [3537]. A similar approach has been chosen in VINCIA [3538], and extended to include initial state showering and other improvements. The same logic – using a form of transverse momentum as ordering parameter – was usually also chosen in the second-generation parton showers, for example in Refs. [3539–3542]. The explicit inclusion of mass effects in the splitting kernels forces to identify the splitting parameter $z$ with a light-cone momentum fraction. To systematically include universal higher-order effects $K$ from the two-loop cusp anomalous dimension, the customary CMW scheme [152] replaces

$$\alpha_S(p_\perp^2) \longrightarrow \alpha_S(p_\perp^2)\left[1 + K\frac{\alpha_S(p_\perp^2)}{2\pi}\right],$$
$$K = \left(\frac{67}{18} - \frac{\pi^2}{6}\right) C_A - \frac{10}{9} T_R n_f\,, \qquad (11.3.5)$$

where $n_f$ is the number of active flavors and $C_A$ and $T_R = 1/2$ are the usual QCD factors. Once an emission, parameterized by $t$, $z$, and $\phi$, has been found, the emission kinematics needs to be constructed, including the compensation of the transverse momentum of the emitted particle. Choices range from being local, *i.e.* contained to the splitter–spectator pair, to global, *i.e.* distributed over the full $N$-particle ensemble. They often reflect a preference for those schemes that lend themselves to a direct matching to infrared subtraction schemes for next-to leading order calculations such as the Catani-Seymour subtraction [180]. While these considerations sound like a minor technical detail, subtle differences in fact have an impact on the overall accuracy, as discussed below.

### 11.3.3 (N)NLO matching

Despite their success in describing the logarithmically-enhanced soft and collinear regimes of emission phase space, parton showers usually lack accuracy in the hard, wide-angle regions of phase space, the realm of fixed-order perturbative corrections, and they do not capture potentially large higher-order corrections to inclusive cross sections. Therefore the resummation implicit in the parton shower has to be matched to fixed-order calculations. Defining, respectively, $\mathcal{B}_N(\Phi_N)$, $\mathcal{V}_N(\Phi_N)$, and $\mathcal{R}_N(\Phi_{N+1})$ the Born-level, virtual and real corrections to a given process, and suppressing their phase space arguments in the following, a calculation – accurate in next-to leading order (NLO) – can schematically be written as

$$\mathrm{d}\sigma^{(NLO)} = \mathrm{d}\Phi_N\left[\mathcal{B}_N + \tilde{\mathcal{V}}_N\right] + \mathrm{d}\Phi_{N+1}\left[\mathcal{R}_N - \mathcal{D}_N\right], \qquad (11.3.6)$$

with the infrared subtracted virtual correction $\tilde{\mathcal{V}}_N(\Phi_N) = \mathcal{V}_N(\Phi_N) + \mathcal{B}_N(\Phi_N)\otimes\mathcal{I}(\Phi_N)$ and the real subtraction term $\mathcal{D}(\Phi_{N+1}) = \mathcal{B}(\Phi_N)\otimes\mathcal{S}_1(\Phi_{N\to N+1})$ both written in factorized form, and where $\mathcal{I}$ emerges from $\mathcal{S}_1$ by analytically integrating over the one-particle emission phase space.

This can be matched to a parton shower along two well–established algorithms. The MC@NLO method [3543] makes use of the fact that the parton shower correctly describes the soft and collinear divergent regions of phase space and the emission kernels $\mathcal{K}$ can thus be matched to the infrared subtraction terms $\mathcal{S}$ required in fixed-order calculations. Events that, at fixed-order, correspond to $N$-particle final states with Born-level kinematics, are denoted as "soft" events and the parton shower is attached to them in a way exactly like it would be attached to the Born-level leading-order events. Similarly, the $(N+1)$-particle events are dubbed "hard" events, and, again the parton shower starts like it would from any similar tree-level configuration. Simple expansion in $\alpha_S$ reveals that the MC@NLO scheme recovers the fixed-order results, and augments them with the resummation of higher-order terms from the parton shower. Despite its simplicity, the MC@NLO prescription has a practical downside, with the second term in Eq. (11.3.6) possibly leading to events with a negative weight, a typical feature of practically any higher-order calculation at fixed order.

This is alleviated in the POWHEG method [3544, 3545], which defines an NLO-accurate $N$-particle cross section, and dresses it, for its first emission, with a Sudakov form factor where the parton-shower splitting kernel is replaced with a ratio of real and Born contribution. However, the $N$-particle phase-space dependent $K$-factor implicit in the first square bracket is applied to the full $N+1$-particle spectrum, which may overestimate the hard region of emission phase space. To correct for this, in practical applications of the POWHEG method, the real-emission phase space is divided, with a suitable profile function, into a soft and a hard regime, $\mathcal{R}_N = \mathcal{R}_N^{(s)} + \mathcal{R}_N^{(h)}$. Schematically, then

$$\begin{aligned} \mathrm{d}\sigma^{(NLO)} = \mathrm{d}\Phi_N \Bigg[ & \mathcal{B}_N(\Phi_N) + \tilde{\mathcal{V}}_N(\Phi_N) \\ & + \int \mathrm{d}\Phi_1 \left( \mathcal{R}_N^{(s)}(\Phi_N \otimes \Phi_1) - \mathcal{D}_N(\Phi_N \otimes \Phi_1) \right) \Bigg] \\ & \times \exp\left[ - \int \mathrm{d}\Phi_1 \frac{\mathcal{R}_N^{(s)}(\Phi_N \otimes \Phi_1)}{\mathcal{B}_N(\Phi_N)} \right] \\ & + \mathrm{d}\Phi_{N+1}\, \mathcal{R}_N^{(h)}(\Phi_{N+1})\,. \end{aligned} \tag{11.3.7}$$

The regular parton shower is then applied to the $(N+1)$-particle configurations. Simple expansion shows, again, the overall cross section and the fixed-order emission spectrum at $\mathcal{O}(\alpha_S)$ are correctly reproduced.

NNLO calculations matched to parton shower so far have been solely available for the production of color singlets, **S**. The first realization was presented in Ref. [3546], based on the POWHEG method above. The underlying idea is to provide a POWHEG matching for $\mathbf{S}+p$ final states, with the additional parton $p$ filling the phase space down to the infrared cut-off of the parton shower and thereby providing NLO accuracy for the overall emission of the hardest particle. This sample is then reweighted to reproduce the inclusive NNLO cross section for the production of the singlet **S** - in the case of a single particle usually achieved by reproducing its rapidity spectrum at NNLO accuracy. Based on multijet merging introduced in the next section, the UNNLOPS method [3547] matches complementary phase spaces of color-singlet production for the emission 0, 1, and 2 additional particles, described by adequately subtracted matrix elements at the two-loop, one-loop, and tree-level respectively. There overall NNLO accuracy is obtained by defining a zero-emission bin and adjusting its cross section accordingly. An alternative approach has been presented in the GENEVA algorithm [1942] which matches the NNLO cross section for **S** production with NNLL resummation of 0-jettiness. Using this observable to define different regions of phase space allows to combine the resulting parton level configurations with a suitably vetoed parton shower.

### 11.3.4 Multijet Merging

Multijet merging provides another way to include exact fixed-order calculations into the parton shower, which is especially useful for the description of samples with large jet multiplicities. The underlying idea is to combine (merge) calculations with 0, 1, 2, etc. additional final state jets into one inclusive sample, by decomposing the parton emission phase space into two regimes, one of hard jet production and one of soft jet evolution. The algorithm achieving this at leading order [3548, 3549] proceeds in three steps. Once a parton-level event at fixed order has been produced, the jets are clustered back until a core process corresponding to the 0-additional jet configuration has been found. The differential cross section for this event is reweighted with ratios $\alpha_S(\mu^{(PS)})/\alpha_S(\mu^{(FO)})$ at each emission, with $\mu^{(PS)}$ the scale the parton shower would use and $\mu^{(FO))}$ the fixed-order scale used in the calculation. The cross section is corrected with Sudakov form factors for the internal and external lines, either with analytic expressions or by running the parton shower from the core process and vetoing those events with a emissions leading to additional jets. These steps transform the individual inclusive fixed-order calculations into exclusive calculations for exactly 0, 1, 2 etc. additional jets, and combine them with the resummation in the parton shower. The algorithm outlined above has been extended to a merging of towers of NLO calculations, effectively a merging of multiple MC@NLO simulations with increasing jet multiplicities in [3550, 3551].

### 11.3.5 Current Developments

Driven by the ever increasing requirements for improved theoretical accuracy, parton showers have come under increased scrutiny in the past few years, for example in Ref. [3552]. Recent studies revealed that currently used parton showers do not correctly fill the phase space in logarithmically enhanced regions of multiple emissions [3553], limiting their logarithmic accuracy. Criteria to systematically asses the logarithmic accuracy of parton showers and a solution to the problem above was presented in Ref. [183] and led to renewed activity in creating better, parton showers that are accurate at next-to leading logarithmic accuracy for critical observables. Including higher-order terms, *i.e.* $\mathcal{O}(\alpha_S^2)$ corrections, to the parton showers represents an important step to further increase the accuracy. The treatment of $\mathcal{O}(\alpha_S^2)$ splitting kernels has been discussed in Ref. [3554], and in Ref. [3555] the inclusion of differential two-loop soft corrections has been presented. But higher-order corrections in the strong coupling are not the only ordering parameter – the parton shower implicitly also resides on a leading-color approximation, and the impact of incorporating sub-leading color terms was studied for example in Ref. [3556, 3557]. This led to the development of a new paradigm, to describe parton splitting and ultimately to construct a parton shower at the level of amplitudes [3558].

While it is not certain where these activities will lead us in the future, they are testament to the im-portance and impact of the probabilistic description of the QCD radiation pattern in parton showers, which is nearly as old as QCD itself.

## 11.4 Monte Carlo event generators

**Torbjörn Sjöstrand**

A $pp$ collision at the LHC may lead to the production of hundreds of particles, via a multitude of processes that can range from the TeV scale down to below the confinement scale. While perturbative calculations can be used at high-momentum scales, currently there is no way to address lower ones directly from the QCD Lagrangian. Instead QCD-inspired models have been developed.

These models typically attempt to break down the full collision process into a combination of relevant mechanisms, that require separate descriptions. Each in its turn often can be formulated as an iterative procedure, where a set of rules are applied repeatedly. These rules are expected to represent quantum mechanical calculations that each gives a range of possible outcomes. The resulting complexity is such that analytical methods are of limited use. Instead the rules are coded up and combined within a bookkeeping framework, where the evolution from a primary perturbative collision to a final multiparticle state is traced. Such computer codes are called Monte Carlo Event Generators (MCEGs), where the "Monte Carlo" part refers to the frequent use of random numbers to pick outcomes according to the assumed quantum mechanical probabilities.

Such generators can be used in phenomenological studies, but the main application is within the experimental community, at all stages of the experiment. When an experiment is designed, it is important to check that the proposed detector has the capability to find key signals. When an experiment is run, triggers have to be optimized to catch the interesting event types. When data is analyzed, the impact of detector imperfections and background processes must be fully understood. In order to address these issues, the output of an MCEG is normally fed into a detector simulation program, that traces the fate of outgoing particles.

### 11.4.1 Event overview

Events come in many shapes, depending on the collider and the random nature of the collision process. As a starting point, consider a typical LHC $pp$ event, and what processes are involved for it. Below these are enumerated, starting from the shortest time/distance scales and progressing towards longer ones. This gives rise to a schematic picture with an onion-like structure in some approximate measure of invariant time, Fig. 11.4.1.

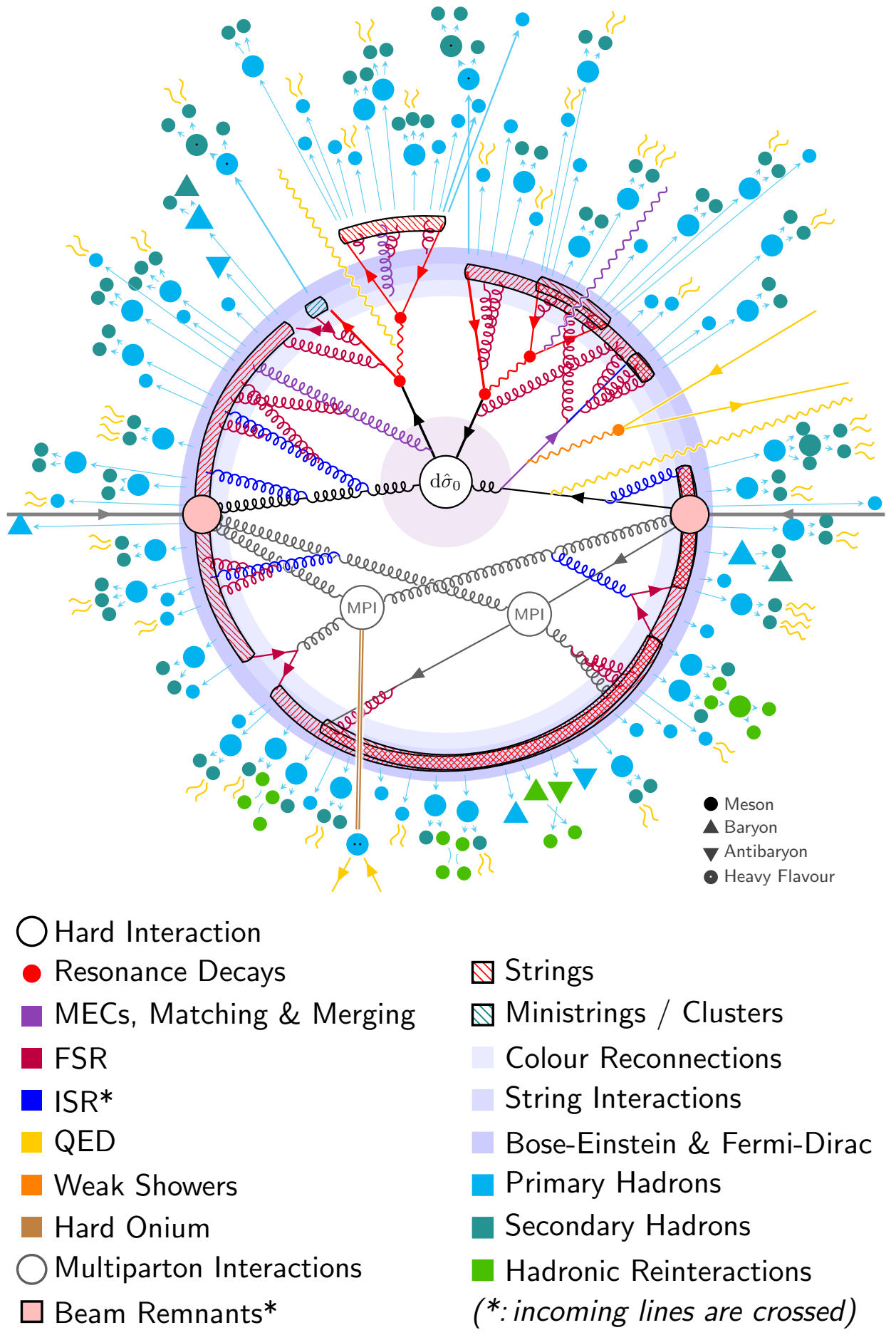

**Fig. 11.4.1** Schematic illustration of the structure of a $pp \to t\bar{t}$ event. Reproduced from [3526].

- At the center of a collision there is sometimes a hard interaction, *i.e.* one at a high-momentum scale, like in this case the production of a $t\bar{t}$ pair. Its cross section is obtained by a convolution of a matrix-element (ME) expression and parton distribution functions (PDFs). More common are events without any discernible hard interaction.
- The hard interaction may involve the decay of resonances like $t \to bW, W \to q\bar{q}'$ as shown in Fig. 11.4.1.
- The core hard interaction may be dressed up by higher-order corrections of matrix-elements. This partly overlaps with the subsequent showers, so a consistent transition (matching and merging) is required.
- Perturbative radiation from the scale of the (dressed) hard interaction down to a lower cutoff is usually subdivided into initial-state radiation (ISR) and final-state radiation (FSR). While partonic QCD branch-ings dominate, QED or even weak branchings may occur. Also some hadron production may be modelled as part of the perturbative stage, *e.g.* of charmonium and bottomonium.
- Since hadrons are composite objects, several of the incoming partons may undergo (more or less) separate perturbative subcollisions, so-called multiparton interactions (MPIs).
- Parts of the incoming hadrons pass unaffected through the hard-interaction region, and emerge as beam remnants.
- Typically colors are traced through the perturbative stage in the $N_c \to \infty$ limit. Apart from imperfections caused by this approximation, there may also be dynamical processes that lead to color reconnections relative to the naive assignments.
- The color assignments are used to combine partons into separate color singlet subsystems — strings or clusters — that each fragment into a set of primary hadrons.
- To first approximation each subsystem fragments independently, but there may be interactions between them.
- The primary hadrons may be unstable and decay further into secondary particles, in decay chains that span a wide range of time scales.
- Right after the fragmentation the hadrons may also be close-packed and rescatter against each other.

In most of the following subsections these aspects will be described in somewhat more detail. Examples of longer generator overviews are [3559, 3560].

### 11.4.2 A brief history

The first event generator of the QCD era probably is the 1974 one by Artru and Mennessier [3561]. It is based on the concept of linear confinement, originally introduced in pre-QCD string-theory models of hadrons, but later supported by the linear confinement found in quenched lattice QCD, see Section 4.3. It was not developed beyond a toy-model stage, however, and was largely forgotten. Instead it was the 1978 article by Field and Feynman [3562] that stimulated an interest in using Monte Carlo methods to simulate jet physics. Their iterative approach for the fragmentation of a single jet was extended to $e^+e^- \to q\bar{q}g$ three-jet events in the Hoyer *et al.* [3563] and Ali *et al.* [3564] codes, which played a key role in the discovery of gluon jets, see Section 2.2. The Lund string fragmentation model introduced the concept of a color flow in $q\bar{q}g$ events [3565], which was given experimental support by PETRA data [3566]. It helped establish the Jetset implementation as a main generator for subsequent $e^+e^-$ machines.

The first QCD-based generator for $pp/p\bar{p}$ physics was Isajet [3567], originally intended for the ISABELLE collider, but much used at the S$p\bar{p}$S and Tevatron colliders, and for SSC and LHC preparations. A few other generators were developed in the early eighties, but left little impact, except for Pythia, which was built on top of Jetset, with the same initial objective of modelling the color flow and its consequences. Later on the two programs were merged under the Pythia heading.

The earliest generators used leading-order matrix elements to describe the perturbative stage. This was insufficient to describe multijet topologies. The DGLAP equations and their extension to jet calculus [3568] suggested that parton showers could be used to generate multiparton topologies. Several early showers were constructed, but it was only with the Marchesini–Webber angularly-ordered shower [3532] that coherence phenomena were consistently handled. This was the starting point for the Herwig generator. An alternative was offered by transverse-momentum-ordered dipole showers, proposed by Gustafson [3537] and first implemented in Ariadne. Today various dipole formulations are the most common shower type.

The combination of hard interactions and parton showers gradually became more sophisticated as various matching and merging techniques were developed. The Sherpa program grew out of such efforts. It was also the first major generator written in C++ from scratch, whereas Herwig and Pythia had to be rewritten from Fortran to C++ to match LHC requirements.

Today Herwig [3525], Pythia [3526], and Sherpa [3522] are the three general-purpose generators used at LHC, or more generally for studies at $e^+e^-/ep/pp/p\bar{p}$ colliders. There also are important dedicated programs, *e.g.* for matrix element generation, such as MadGraph_aMC@NLO [3328] and the Powheg Box [3569].

Adjacent physics areas, such as heavy-ion collisions, cosmic ray cascades in the atmosphere, or neutrino interactions, started their generator development somewhat later, and partly under the influence of the general-purpose ones above, *e.g.* for the high-energy hadronization descriptions. Typically the hard-physics aspects become less relevant, and soft-physics ones more so. These issues will be briefly addressed towards the end.

### 11.4.3 The perturbative interface

A key task is to generate events of a predetermined type or types. This could be *e.g.* $W$ + jets, both as a signal and as a background to $t\bar{t}$ production. Typically there is a core hard interaction, that then is complemented by further perturbative QCD activity at varying scales. In such cases the core interaction provides the natural starting point for the description of the rest of the event. As already suggested above, one may discern three main stages:

1. the generation of partonic events purely based on matrix elements and parton distribution functions,
2. the matching and merging stage, where Sudakov form factors generated by parton showers are used to reject some of the events above, so as to avoid double counting, and
3. the subsequent pure parton shower evolution down to a lower cutoff somewhat above the $\Lambda$ scale.

The first of these is covered in Section 11.1, and in Section 10.2 for PDFs, while the second two are described in Section 11.3.

Of special interest for the continued story are the core $2 \to 2$ pure QCD interactions, $qq' \to qq'$, $q\bar{q} \to q'\bar{q}'$, $q\bar{q} \to gg$, $qg \to qg$, $gg \to gg$ and $gg \to q\bar{q}$. These are by far the dominant perturbative processes at hadron colliders. The main contribution is $t$-channel gluon exchange, which gives rise to a $\mathrm{d}p_T^2/p_T^4$ divergence in the $p_T \to 0$ limit.

### 11.4.4 Total cross sections and diffraction

Another key task, at the other extreme, is to generate the inclusive sample of all events at hadron colliders. In practice rare processes are generated separately, so the emphasis comes to lie on common QCD processes.

The total QCD cross section $\sigma_{\mathrm{tot}}$ is finite, related to a finite size of hadrons and a finite range of QCD interactions, owing to confinement. Currently there is no QCD-Lagrangian-based description of $\sigma_{\mathrm{tot}}$, but instead phenomenological models have been proposed based on Regge theory, with free parameters that have to be tuned to data. At a minimum one Pomeron and one Reggeon term is required to describe the energy dependence [1104], where the former can be associated with a trajectory of exchanged glueball states and the latter with one of mesonic states, see Section 8.1. More terms are needed in more realistic models. Notably, recent studies points towards the existence of an Odderon term, see Section 12.6.

The total cross section between two hadrons $A$ and $B$ can be subdivided into several partial ones, associated with different event topologies:

$$\begin{aligned}\sigma_{\mathrm{tot}}^{AB}(s) &= \sigma_{\mathrm{el}}^{AB}(s) + \sigma_{\mathrm{inel}}^{AB}(s) \\ &= \sigma_{\mathrm{el}}^{AB}(s) + \sigma_{\mathrm{sd}(XB)}^{AB}(s) + \sigma_{\mathrm{sd}(AX)}^{AB}(s) \\ &\quad + \sigma_{\mathrm{dd}}^{AB}(s) + \sigma_{\mathrm{cd}}^{AB}(s) + \sigma_{\mathrm{nd}}^{AB}(s) \ , \qquad (11.4.1)\end{aligned}$$

where $s$ is the squared collision energy in the rest frame. These topologies are illustrated schematically in Fig. 11.4.2.

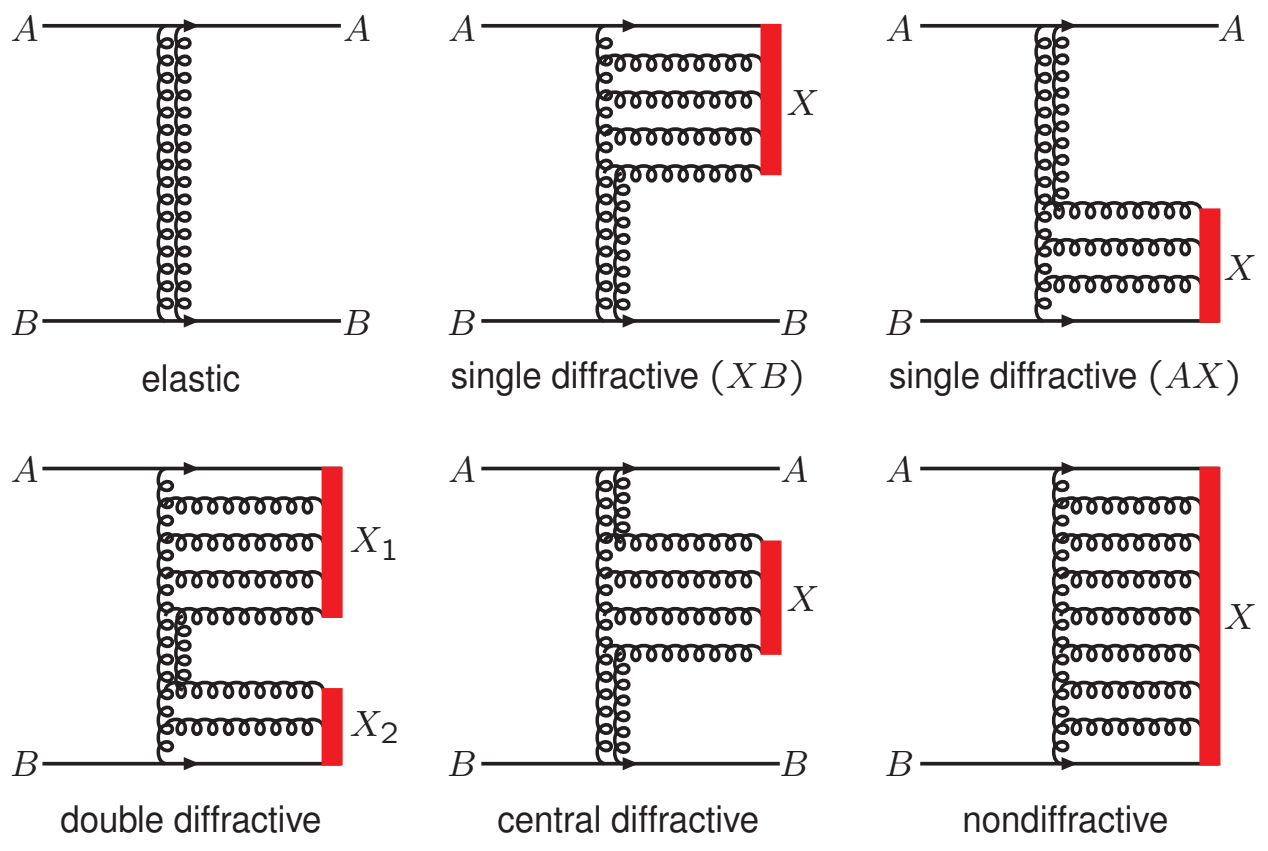

**Fig. 11.4.2** Main subclasses of the total cross section in $AB$ hadron collisions. The red bars represent the regions in rapidity between $A$ and $B$ where hadrons are produced. Reproduced from [3526].

In nondiffractive (nd) events the full rapidity range can be populated by particle production, whereas in single, double or central diffraction (sd, dd, or cd, respectively) only parts of this range are populated, and in elastic (el) events none of it is. The relative composition changes with energy, *e.g.* such that the elastic fraction is increasing. Roughly speaking, elastic is 25%, diffractive 20% and nondiffractive 55% at LHC energies.

Many approaches have been proposed to model these partial cross sections, both integrated and differential ones, notably again based on Regge theory. Common is that the mass $m_X$ of a diffractive system obeys an approximate $\mathrm{d}m_X^2/m_X^2 = \mathrm{d}y_X$ behaviour, where $y_X$ is the rapidity range of the $X$ system. An elastically scattered beam particle is also associated with a squared momentum transfer $t$ that obeys an approximate $\exp(Bt)$ shape at low $t$. The slope $B$ depends on the colliding hadron types, the event topology and the collision energy, but the order of magnitude is 10 GeV$^{-2}$, *i.e.* $\langle p_T \rangle \sim 0.3$ GeV.

At low energies also other collision types occur, such as resonant production and baryon annihilation.

The Ingelman–Schlein [3570] ansatz is commonly used for the description of diffractive systems. In it, a Pomeron is viewed as a hadronic state, with its own PDFs. Therefore the Pomeron–hadron subsystem can be described in the same way as we will introduce for nondiffractive events in the following, at least for reasonably large $m_X$, while a simpler description is called for at small $m_X$.

### 11.4.5 Multiparton interactions

All generators assume that a nondiffractive event can contain multiple parton–parton interactions, which can be viewed as the QCD reinterpretation of the cut Pomeron picture of olden days [3571]. MPIs are necessary to explain many aspects of hadronic collisions, such as the wide multiplicity distributions, where most of the multiplicity is related to low-$p_T$ processes. The case of two hard interactions, Double Parton Scattering, is well studied theoretically and experimentally [3572]. Different models have been developed starting from the same basic ideas. This section will begin with the PYTHIA approach, which is also used by SHERPA, and later the differences in HERWIG will be outlined.

It has already been noted that the perturbative QCD $2 \to 2$ cross section is divergent for $p_T \to 0$, on the one hand, and that the total $pp$ cross section is finite, on the other hand. The perturbative picture is based on the assumption of asymptotically free colored partons, however, while the reality is that of partons confined inside color singlet hadrons. Therefore a plausible regularization of the $p_T \to 0$ divergence is provided by color screening, *i.e.* that partons of opposite color gives destructive interference of scattering amplitudes. A parameter $p_{T0}$ is introduced in PYTHIA as the inverse of the spatial screening distance. This is used to dampen the conventional $2 \to 2$ QCD cross section by a factor

$$\left( \frac{\alpha_s(p_{T0}^2 + p_T^2)}{\alpha_s(p_T^2)} \frac{p_T^2}{p_{T0}^2 + p_T^2} \right)^2 , \tag{11.4.2}$$

which gives

$$\frac{d\sigma}{dp_T^2} \sim \frac{\alpha_s^2(p_T^2)}{p_T^4} \to \frac{\alpha_s^2(p_{T0}^2 + p_T^2)}{(p_{T0}^2 + p_T^2)^2} . \tag{11.4.3}$$

A tune to data gives a $p_{T0}$ of the order of 2 GeV, but slowly increasing with energy, consistent with an increasing screening, as lower-$x$ partons become accessible at higher energies.

The average number of MPIs in nondiffractive events is given by $\langle n_{\mathrm{MPI}} \rangle = \sigma_{\mathrm{pert}}(p_{T0})/\sigma_{\mathrm{nd}}$, neglecting a small correction from the part of $\sigma_{\mathrm{pert}}$ that should be associated with diffraction. Here $\sigma_{\mathrm{pert}}(p_{T0})$ is the integrated damped $2 \to 2$ QCD cross section. At first glance, the $n_{\mathrm{MPI}}$ should be distributed according to a Poissonian, with $n_{\mathrm{MPI}} = 0$ removed, since zero MPIs corresponds to the two hadrons passing through without any interactions.

This assumes that all collisions are equivalent, however. More plausibly, the impact parameter $b$ of the collision plays a role, where central collisions generate more activity than peripheral ones. To model this, an ansatz for the matter distribution inside a hadron is required. The simplest choice is a three-dimensional Gaussian, since then the convolution of two hadrons is easily integrated over the collision process to give a two-dimensional Gaussian $\mathcal{O}(b)$. Fits to data prefer a somewhat more uneven matter distribution, *e.g.* with "hot spots" of enhanced activity around the three valence quarks.

The actual generation of MPIs can conveniently be arranged in a falling sequence of transverse momenta with $\sqrt{s}/2 > p_{T1} > p_{T2} > \cdots > p_{Tn} > 0$. Neglecting the impact-parameter dependence for a moment, the probability for the $i$th MPI becomes

$$\frac{d\mathcal{P}}{dp_{Ti}} = \frac{1}{\sigma_{\mathrm{nd}}} \frac{d\sigma_{\mathrm{pert}}}{dp_{Ti}} \exp\left( - \int_{p_{Ti}}^{p_{T(i-1)}} \frac{1}{\sigma_{\mathrm{nd}}} \frac{d\sigma_{\mathrm{pert}}}{dp'_T} dp'_T \right) , \tag{11.4.4}$$

with a fictitious $p_{T0} = \sqrt{s}/2$. The exponential expresses the probability to have no MPIs between $p_{T(i-1)}$ and $p_{Ti}$, as comes out of Poissonian statistics and in exact analogy with the Sudakov form factor of parton showers. With impact parameter included, the $b$ is selected proportional to $\mathcal{O}(b)\, d^2b$, and the $p_T$ selection procedure acquires an enhancement/depletion factor of $\mathcal{O}(b)/\langle\mathcal{O}\rangle$. Sequences without any MPIs require a restart with a new $b$.

So far inclusive nondiffractive events have been considered. Alternatively one specific hard interaction is studied, and an underlying event should be added to it. Then again a $b$ is selected according to $\mathcal{O}(b)$, and an enhancement/depletion factor is defined as before. The upper $p_T$ limit for MPIs now depends on context. If the hard interaction is QCD $2 \to 2$ above some $p_{T\mathrm{min}}$ then its $p_T$ should be equated with a $p_{T1}$ of the MPI sequence, and subsequent ones be below that, or else high $p_T$ scales would be double counted. If the hard interaction is something else, then there is no such double counting, and MPIs can start from the highest possible scale.

The description of $n$ MPIs requires $n$-parton PDFs, $f(x_1, Q_1^2; x_2, Q_2^2; \ldots; x_n, Q_n^2)$, which are not known from first principles or from data. An approximate approach is to make use of the $p_T$-ordering of MPIs, such that the first interaction uses conventional PDFs, while subsequent MPIs use more and more modified ones. Thereby standard phenomenology is preserved in the hard region. Subsequently momentum conservation requires a gradually reduced $x_i$ range, within which PDFs are squeezed. Also flavor conservation must be respected. If a valence $u$ quark is taken out of a proton, say, then only one $u$ quark remains, and the valence $u$ distribution must be normalized to 1 rather than 2. If instead a sea $u$ quark is extracted, then the $\overline{u}$ sea must contain one parton more than $u$ sea, which can be implemented by having one valence-like $\overline{u}$ in addition to the normal $u$ and $\overline{u}$ sea distributions. Finally, when the valence-like distributions have been properly normalized, the gluon and sea distributions are uniformly rescaled so as to obey the momentum sum rule.

With the evolution of ISR and FSR parton showers usually formulated in terms of a decreasing sequence each of $p_T$ values, the MPIs now add a third sequence. In PYTHIA they are fully interleaved into one common sequence. Thus the key evolution equation is

$$\frac{\mathrm{d}\mathcal{P}}{\mathrm{d}p_T} = \left( \frac{\mathrm{d}\mathcal{P}_{\mathrm{MPI}}}{\mathrm{d}p_T} + \frac{\mathrm{d}\mathcal{P}_{\mathrm{ISR}}}{\mathrm{d}p_T} + \frac{\mathrm{d}\mathcal{P}_{\mathrm{FSR}}}{\mathrm{d}p_T} \right) \times S \qquad (11.4.5)$$

where $S$ represents the Sudakov factor, obtained by exponentiation of the real-emission rate, integrated from the previous $p_T$ scale to the current one, cf. Eqn. (11.4.4). In this way the harder part of the event sets the stage for what can occur at softer scales. Notably MPIs and ISR compete for the dwindling amount of momentum in the beams, as represented by the modified PDFs. The $p_T$ evolution should not be viewed as one in physical time; actually all MPIs occur at (almost) the same collision time $t = 0$, while lower $p_T$ scales means earlier times $t < 0$ for ISR and later times $t > 0$ for FSR.

The HERWIG description of MPIs [3573] splits them into a hard and a soft component, separated at a scale $p_T^{\mathrm{min}}(s)$. The perturbative cross section $\mathrm{d}\sigma_{\mathrm{QCD}}/\mathrm{d}p_T$ is recovered above $p_T^{\mathrm{min}}(s)$, while a simple tuneable shape $\mathrm{d}\sigma_{\mathrm{soft}}/\mathrm{d}p_T$ is used for $0 < p_T < p_T^{\mathrm{min}}(s)$, with the constraints that it must vanish at $p_T = 0$ and match $\mathrm{d}\sigma_{\mathrm{QCD}}/\mathrm{d}p_T$ at $p_T^{\mathrm{min}}(s)$. The electromagnetic form factor is used to represent the impact-parameter profile of protons. This gives an overlap function

$$A(b,\mu) = \frac{\mu^2}{96\,\pi}\,(\mu b)^3\,K_3(\mu b), \qquad (11.4.6)$$

where $\int \mathrm{d}^2b\,A(b) = 1$, and $\mu$ are used as free parameters, separately set for the hard and soft components, for more flexibility. Combining, an eikonal is defined as

$$\chi_{\mathrm{tot}}(b,s) = \frac{1}{2}\,A_{\mathrm{hard}}(b,\mu_{\mathrm{hard}})\,\sigma_{\mathrm{QCD}}(s,p_T^{\mathrm{min}}) \qquad (11.4.7)$$

$$+\,\frac{1}{2}\,A_{\mathrm{soft}}(b,\mu_{\mathrm{soft}})\sigma_{\mathrm{soft}}(s,p_T^{\mathrm{min}})\ , \qquad (11.4.8)$$

where $\sigma_{\mathrm{QCD}}$ and $\sigma_{\mathrm{soft}}$ are the respective $p_T$-integrated cross sections. The number of MPIs at a given $b$ is given by a Poissonian, as in PYTHIA, with $\langle n(b,s)\rangle = 2\chi(b,s)$. The eikonal formalism also predicts other quantities, such as total and elastic cross sections, and the elastic slope, that can be used to constrain the free parameters of the model.

When a hard interaction has been selected in HERWIG, and been associated with an impact parameter $b$, the number of hard and soft additional MPIs can be selected according to Poissonians. The hard interactions are generated first, and thereafter the soft ones. Unlike PYTHIA they are not ordered in $p_T$ within the hard and soft groups, and there is no rescaling of PDFs. Also the ISR and FSR associated with an interaction are reconstructed before the next is considered. For the hardest interaction the ISR is forced to reconstruct back to a valence quark, while for subsequent ones the ISR evolution is forced back to a gluon. This gluon can then be color-attached to the hardest interaction itself. The MPIs together may take more momentum out of the protons than is available, given the lack of PDF rescaling. When that happens, the latest MPI is regenerated, but if repeated attempts fail the MPI generation may be interrupted with a lower MPI number than intended.

### 11.4.6 Beam remnants and color reconnection

Since the MPI+ISR machinery in HERWIG reconstructs back the perturbative activity to one single valence quark, having been taken out of an incoming proton, the other two valence quarks together form a diquark remnant, with opposite color to the one quark taken out. Four-momentum conservation fixes the remnant momentum.

The situation is more complicated in PYTHIA, since the MPI+ISR can extract a variable number of "initiator" partons out of the incoming proton, leaving behind multiple quarks and antiquarks. Then ad hoc probability distributions are used to share the remnant longitudinal momentum between them. The initiator partons also carry a transverse momentum, a so-called primordial $k_T$, that is to be compensated by the remnant. When the remnant consists of the several partons, these may also have a relative $k_T$ component. The size of all these transverse kicks should be at or below the hadronic mass scale, though empirically they appear to be at the higher rather than at the lower end of the expected range.

The color lines of the initiator partons naively stretch from the remnants in through the hard interaction at the core of each MPI, *i.e.* usually fill the whole rapidity range. If so, the average charged multiplicity $n_{\mathrm{ch}}$ of an event increases linearly with the number of MPIs, up to corrections from momentum conservation and the details of the remnant handling. Since all MPIs will be equivalent, a constant average transverse momentum per hadron should result, *i.e.* a flat $\langle p_T\rangle(n_{\mathrm{ch}})$ curve. Instead an increasing $\langle p_T\rangle(n_{\mathrm{ch}})$ is observed at hadron colliders.

The natural explanation for this phenomenon is color reconnection (CR). Specifically, it is assumed that the color lines stretched between all final-state partons can be rearranged so as to reduce the overall length. The number of possible rearrangements increases with the number of MPIs, such that the $\langle n_{\mathrm{ch}} \rangle$ increase is smaller for each further MPI. The perturbative $p_T$ kick of each MPI remains, however, so when this $p_T$ is shared between fewer particle the result is an increasing $\langle p_T \rangle$.

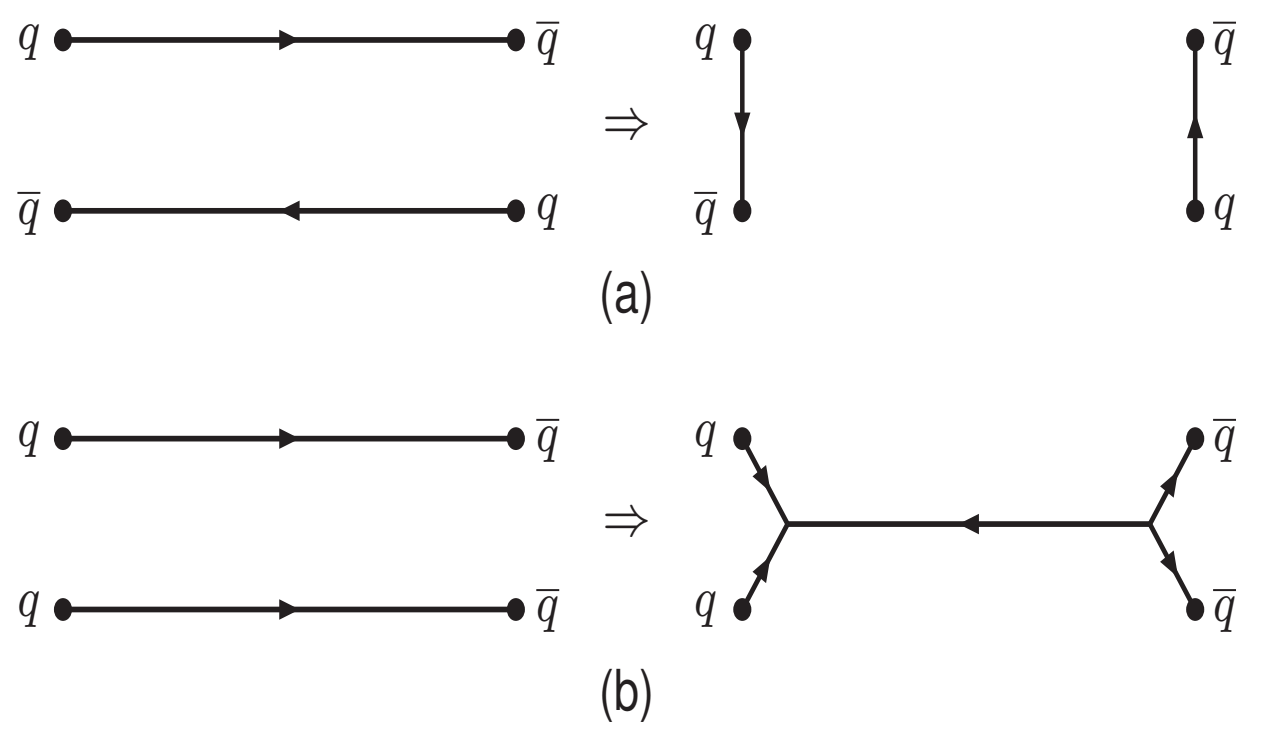

**Fig. 11.4.3** Schematic illustration of color reconnection. (a) Simple flip. The arrows indicate flow from color to anticolor. (b) Junction reconnection. Note changed direction of the long line, according to $3 \otimes 3 = \overline{3}\,(\oplus 6)$.

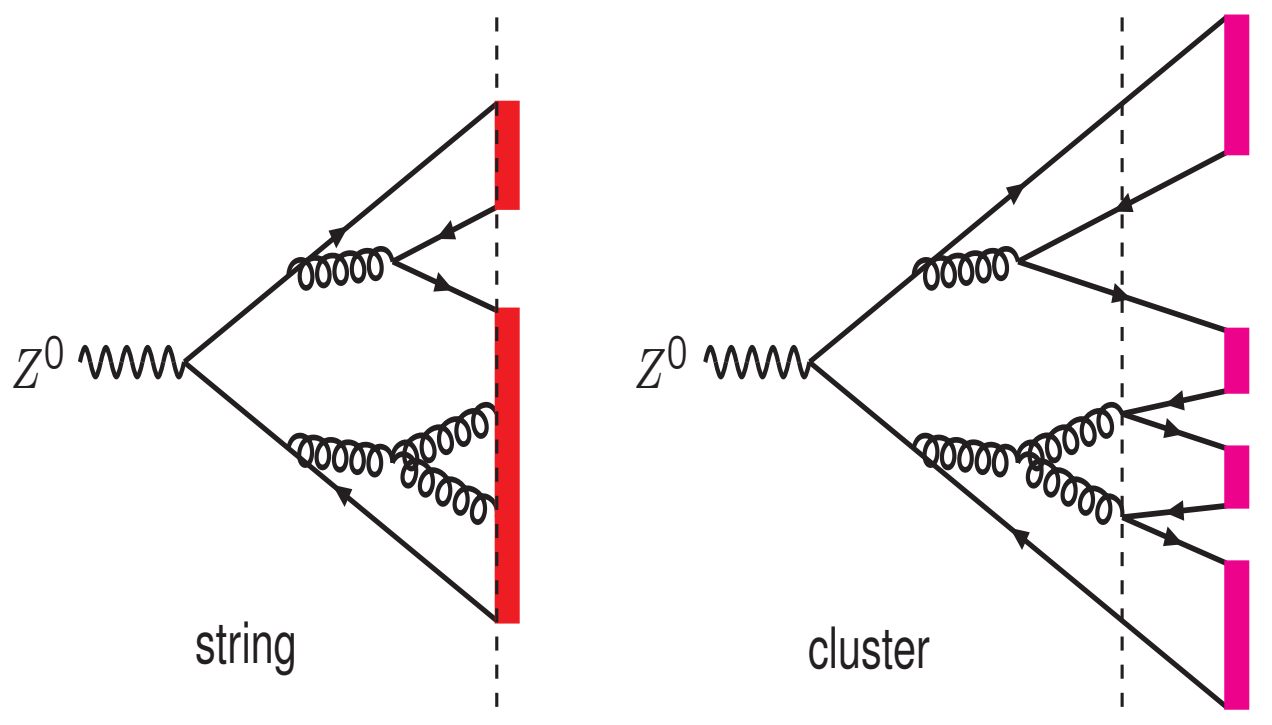

**Fig. 11.4.4** String versus cluster fragmentation. At the end of the perturbative evolution, the vertical dashed line, strings are directly attached, red regions. Alternatively, a nonperturbative $g \to q\overline{q}$ stage is inserted before clusters are formed, magenta regions.

Many CR models have been implemented over the years, in all three main generators, and it would carry too far to discuss each in detail. A frequent starting point is that standard parton showers operate in the $N_{\mathrm{c}} \to \infty$ limit, and thus miss corrections of order $1/N_{\mathrm{c}}^2$ at each shower branching. One possible approach is to do the evolution in color space more carefully, and thereby be able to formulate CR as a consequence of such subleading corrections. More common is to formulate CR on the nonperturbative level, but then color algebra should be used to restrict the rate at which it can occur.

Also common for many nonperturbative approaches is that a key role is assigned to the string length $\lambda$ between two color-connected partons $i$ and $j$

$$\lambda_{ij} \approx \ln\left(1 + \frac{m_{ij}^2}{m_0^2}\right) = \ln\left(1 + \frac{(p_i + p_j)^2}{m_0^2}\right) . \quad (11.4.9)$$

Here $m_0 \approx m_\rho$ is a typical hadron mass, and 1 has been added to ensure that $\lambda \geq 0$. With this definition $\lambda$ is a reasonable measure of how many hadrons typically will be produced by the string. A flip of two color-connected pairs $(i,j)$ and $(k,l)$ into $(i,l)$ and $(k,j)$, Fig. 11.4.3a, corresponds to a net change $\Delta\lambda = \lambda_{il} + \lambda_{kj} - \lambda_{ij} - \lambda_{kl}$. The assumption is that $\Delta\lambda < 0$ reconnections are favoured.

Further CR variants include ones that change the number of string pieces, say by taking a central gluon connected to both remnants and putting it on an already existing central string piece. Of recent special interest is junction reconnection [3574]. A junction is the center of a Y shape where three string pieces come together, and topologically is the carrier of the baryon number. Two strings can collapse to one in a central region with the production of a junction and antijunction near either end, Fig. 11.4.3b. This gives an enhanced baryon production. In the cluster model a similar effect can be obtained by letting three aligned $q\overline{q}$ clusters rearrange into one $qqq$ and one $\overline{qqq}$ cluster.

CR ought to be possible only when the strings concerned overlap in space–time. For normal $pp$ collisions this is almost automatic, since most strings run more-or-less parallel with the collision axis within a small transverse region. Space–time should be taken more seriously *e.g.* in $e^+e^- \to W^+W^- \to q_1\overline{q}_2 q_3 \overline{q_4}$, where the $W^\pm$ decay angles will influence the amount of overlap. Models have been developed to this end, and predictions agree well with the combined LEP data [3575]. The best description is obtained with an $\sim$ 50% CR rate, but unfortunately statistics is limited and a no-CR scenario is only disfavoured at the $2.2\sigma$ level.

### 11.4.7 Hadronization

There are two main fragmentation models in common use: strings and clusters. Both start out from the color flow topologies set up according to the previous sections, in the $N_{\mathrm{c}} \to \infty$ limit. Specifically, each $q \to qg$ and $g \to gg$ leads to a new uniquely defined color line between the two daughter partons. The string model retains all the gluons produced in the perturbative stage. A string can therefore be stretched *e.g.* like $q - g_1 - g_2 - g_3 - \overline{q}$, where each color line between two adjacent partons is unique. In the cluster model the perturbative shower is followed by a semi-perturbative step where all gluons branch by $g \to q\bar{q}$. The system therefore subdivides into smaller $q\bar{q}$ clusters.This key difference is illustrated in Fig. 11.4.4. The string is central in PYTHIA, while HERWIG and SHERPA implements clusters. In the latter program there is an interface to PYTHIA strings to allow comparisons. In the following key features will be presented in some more detail.

The string approach is based on the assumption of a linear confinement potential, as supported by quenched lattice QCD phenomenology. In a simple $q\bar{q}$ system studied in the rest frame, *e.g.* from a $Z^0$ decay, the potential can then be written as $V(r) = \kappa r$, where $r$ is the separation and $\kappa$ is the string tension. Empirically $\kappa \approx 1$ GeV/fm, determined mainly from hadron spectroscopy. The mathematical one-dimensional string stretched straight between the $q$ and the $\bar{q}$ can be viewed as defining the center of a physical chromoelectric flux tube, with transverse dimensions comparable to hadron sizes, *i.e.* with a radius of around 0.7 fm. It is not settled whether this tube should be viewed in analogy with a vortex line in a type II superconductor, or with an elongated bag in a type I one, or with an intermediate behaviour, but for the basic considerations this is also not important.

If the string does not break, it will undergo a yo-yo-like oscillatory motion, where initially the quarks carry the full energy of the system, but gradually lose it to the string being stretched out between them. Massless quarks will reach a maximal separation where all the energy is stored in the string, and then the string tension will pull them back, so that they again cross, carrying the full energy. The key relation is that the massless endpoint quarks, moving out along the $\pm z$ axis, obey

$$\left|\frac{\mathrm{d}E}{\mathrm{d}t}\right| = \left|\frac{\mathrm{d}E}{\mathrm{d}z}\right| = \left|\frac{\mathrm{d}p_z}{\mathrm{d}t}\right| = \left|\frac{\mathrm{d}p_z}{\mathrm{d}z}\right| = \kappa \tag{11.4.10}$$

(with $c = 1$). If such a system is boosted along the $z$ axis the $q$ and $\bar{q}$ start out with different energies, which means that their turning points occur at different times, which gives the expected net motion of the system as a whole. The string tension remains unchanged by the boost, and a string piece in the new frame still carries no three-momentum. This may seem counterintuitive, but note that the boost will take an equal-times string piece to one where the endpoints are at different times, and if viewed this way the boosted string piece will pick up the expected momentum.

Now introduce the possibility for a string to break by the production of a new $q_i\bar{q}_i$ pair somewhere along the string. In lattice QCD this corresponds to going from the quenched to the unquenched situation. Each break splits the original color singlet system into two separate smaller ones. A sequence of breaks thus gives an ordered singlet chain $q\bar{q}_1 - q_1\bar{q}_2 - q_2\bar{q}_3 - \ldots - q_{n-1}\bar{q}$, and these singlets can be associated with the primary (*i.e.* before any decays) hadrons. Such a sequence of breaks is illustrated in Fig. 11.4.5. Notice that, in this picture, each produced hadron undergoes a yo-yo motion of its own.

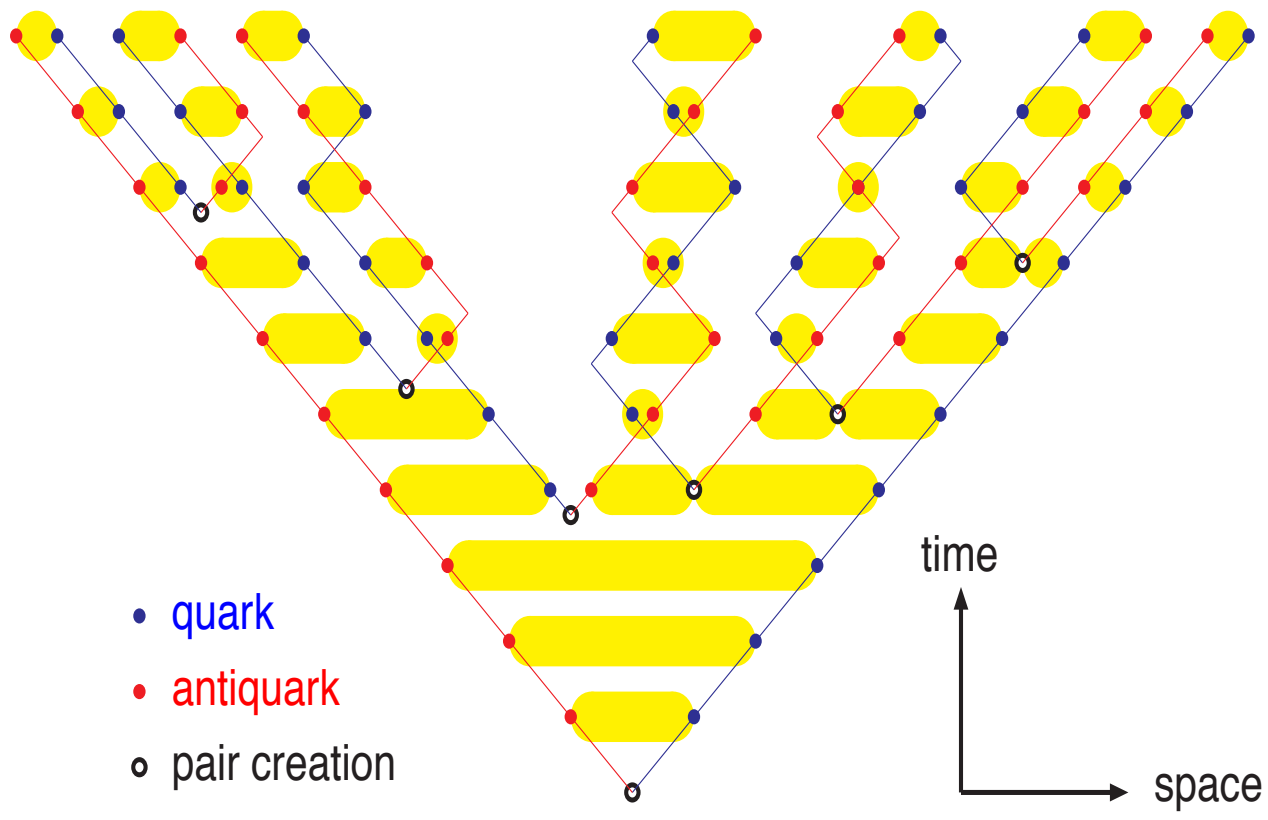

**Fig. 11.4.5** String fragmentation of a $q\bar{q}$ system, where yellow regions represents snapshots in time of the string pieces being stretched out. Reproduced from [3576].

If the $q_i$ have $m = p_T = 0$ then a $q_i\bar{q}_i$ pair can be produced on-shell in a single vertex, and afterwards be pulled apart. The partons are virtual initially when this is not the case, and each has to tunnel a distance $m_T/\kappa$ until it has absorbed enough energy from the string to come on-shell. This leads to a suppression factor

$$\exp\left(-\frac{\pi\, m_T^2}{\kappa}\right) = \exp\left(-\frac{\pi\, m^2}{\kappa}\right)\, \exp\left(-\frac{\pi\, p_T^2}{\kappa}\right)\,. \tag{11.4.11}$$

The transverse momentum kick can be compensated locally between the $q_i$ and $\bar{q}_i$, which defines a vector sum for each $q_i\bar{q}_{i+1}$ hadron. Empirically the observed $\langle p_T\rangle$ is somewhat higher than predicted this way, which could be related to the cutoff of soft gluons in the parton shower, so for tuning purposes the $p_T$ width of primary hadrons is considered a free parameter.

The tunneling also implies that nonperturbative production of heavier quarks is suppressed, for charm and bottom to a negligible level. For the strangeness suppression it is not clear what quark masses to use — the observed $s/u \approx 0.25$ production ratio is in between results for current algebra and constituent masses — so again it is considered a free parameter.

Neglecting orbitally and radially excited states, a produced meson belongs either to the pseudoscalar or to the vector multiplet. Naive spin counting would imply a $1:3$ production rate, but vectors are suppressed by the heavier mass, to an extent that is not easily calculated from first principles, so further parameters are introduced. The many flavor-related parameters is the biggest weakness of the string model.

For baryon production antidiquark–diquark string breaks are introduced, in analogy with quark–antiquark ones, with the diquark in a color antitriplet. Again tunneling, spin and mass factors are combined in production-rate parameters. The overall diquark break fraction needed to describe the observed baryon production rate is around 10%. A modified scenario is the popcorn one. In it, a $q\overline{q}$ pair can be produced with a color that does not screen the endpoint ones, such that it does not break the string. Inside that pair one or two further breaks may occur, where the latter would allow a meson to be produced between the baryon and the antibaryon.

String breaks on the average ought to be produced along a hyperbola of fixed invariant time, which translates into a flat rapidity plateau of produced hadrons. Then particle production would start in the middle of the event and spread outwards, Fig. 11.4.5. But actually all string breaks have a spacelike separation to each other, so there is no Lorentz-frame-independent definition of what comes first. It is then more convenient to begin at an endpoint quark and work inwards. The final result should be independent of the order used, which is satisfied for an almost unique fragmentation function shape

$$f(z) = \frac{1}{z}\,(1-z)^a \,\exp\left(-\frac{b\,m_T^2}{z}\right). \tag{11.4.12}$$

Here $a$ and $b$ are two free parameters and $m_T$ the transverse mass of the hadron considered. The $z$ variable parametrizes the fraction of remaining lightcone momentum that the hadron takes. That is, if the quark is moving in the $+z$ direction, then $z$ is the fraction of $E + p_z$ taken, with $1 - z$ the fraction remaining to be used in subsequent steps. Note that heavier hadrons on the average take a larger fraction $z$ than lighter ones. The $f(z)$ shape may be generalized slightly to take into account of the effect of different quark masses, notably for massive ($c$ or $b$) endpoint quarks.

The extension to more complicated string topologies involves no new principles or parameters. In a $q\overline{q}g$ event, Fig. 11.4.6, the $N_c \to \infty$ color algebra implies that one color is shared between $q$ and $g$, and another between $g$ and $\overline{q}$, with no direct connection between $q$ and $\overline{q}$. The strings pulled out now have a transverse motion and thus a higher energy per unit length, but less length is pulled out per unit time, and these two effect exactly compensate to give $|\mathrm{d}E/\mathrm{d}t| = \kappa$ for the quarks and twice that for the gluon. This is to be compared with the QCD color-charge ratio $N_c/C_F = 9/4$. Again

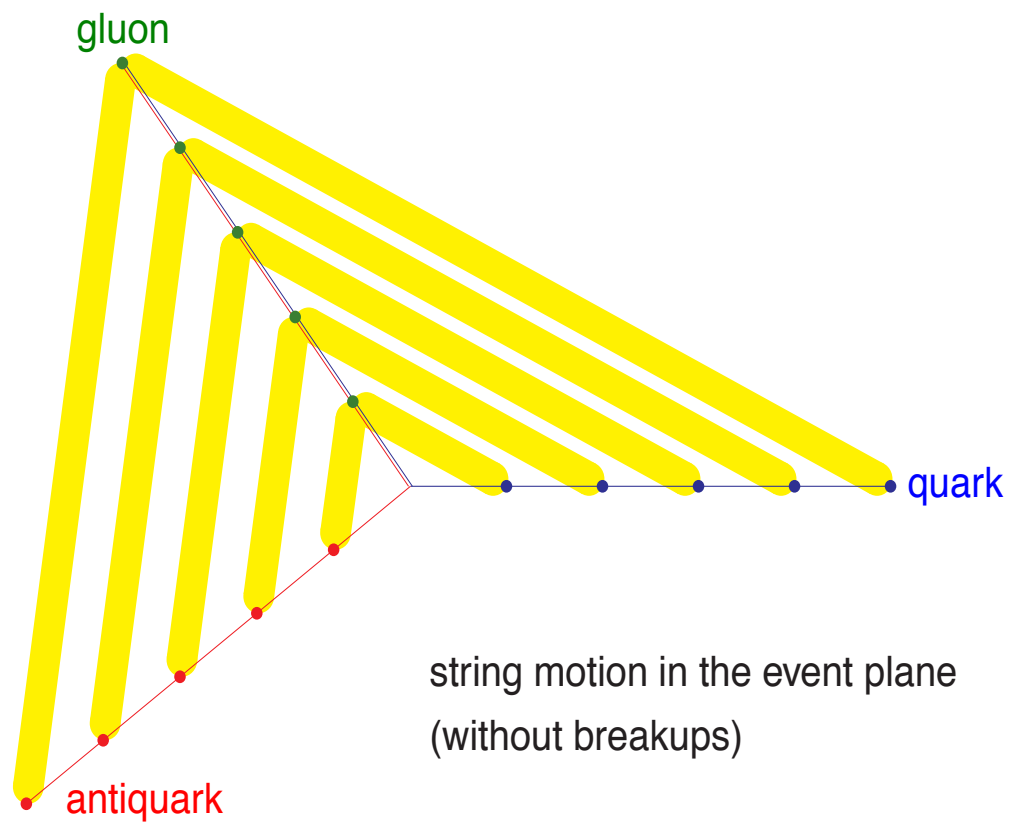

**Fig. 11.4.6** String motion in a $q\overline{q}g$ system. Yellow regions represent snapshots in time of the string pieces. The fragmentation of the strings is suppressed for clarity. Reproduced from [3576].

the two string pieces will fragment along their length, with one hadron formed around the gluon kink, obtaining four-momentum contributions from both pieces. The generalization to multigluon systems is obvious, and also closed gluon loops can be addressed in this way.

The string motion becomes more complicated when a gluon is soft or collinear, with new string regions arising. The punch line, however, is that the total string motion is only mildly affected by such gluon emissions, although technical complications arise.

The occasional low-mass string also needs some special care, and in the extreme case it may become necessary to force the endpoint flavor content to form a single hadron, with four-momentum conservation being ensured by exchange with another parton or hadron.

LHC $pp$ data has revealed several unexpected features, notably that fragmentation properties change when the multiplicity is increased, towards more strangeness and baryon production, and with signs of collective flow, both in the direction of the heavy-ion behaviour. Possibly a quark–gluon plasma is being formed, but in a string context it is also worthwhile to consider how the environment, *i.e.* the close-packing of strings at high multiplicities, could perturb the standard fragmentation picture. One such potential effect is color rope formation, *i.e.* that several parallel strings combine into an object in a higher color representation [3577]. Then baryon and strangeness production is enhanced. Baryons can in addition be enhanced by the aforementioned junction CR mechanism. There can also be a repulsive force between strings, so-called shove, that can give rise to collective flow [3578]. It remains to be seen whether these ideas can be combined into a new coherent framework in agreement with LHC observations.

The cluster model is based on the concept of preconfinement [3579] during the parton-shower evolution. That is, each color line (for $N_c \to \infty$) tends to correspond to a low-mass system, with only a small tail towards larger masses. The model becomes even more suggestive if it is assumed that all gluons branch into quarks, $g \to q\overline{q}$, at the end of the cascade, such that each color line is associated with a separate color singlet cluster. This would occur naturally if constituent masses obey $m_g \geq 2m_u = 2m_d$, as is supported by lattice QCD. Several cluster studies have been presented over the years. Here the generic features are outlined.

A gluon decays into any kinematically allowed $q\overline{q}$ pair according to its phase-space weight, which implies a dependence on the choice of gluon and quark constituent masses, notably whether $s\overline{s}$ can occur at this stage. Thereafter each $q_1\overline{q}_2$ cluster decays isotropically into a two-body state, hadrons $q_1\overline{q}_3$ and $q_3\overline{q}_2$, where $\overline{q}_3$ may also represent a diquark, resulting in baryon production. The hadrons are picked at random among all possibilities consistent with the flavors, according to relative weights. These weights are the product of the spin factor $2s+1$ for each final hadron and the phase-space factor $2p^*/m$, where $p^*$ is the common magnitude of the three-momentum of the hadrons in the rest frame of the cluster with mass $m$. In some cases, such as $\pi^0 - \eta - \eta'$, also the mixing of identical-flavor states needs to be included in the weight. It is also possible to allow an overall extra factor for a multiplet, notably to enhance baryon production.

A number of improvements have been introduced to this basic picture, as follows.

When the four-momenta of the cluster constituent $q_1$ and $\overline{q}_2$ are combined into the four-momentum of the cluster, the tail to large cluster masses is suppressed, but it is not completely absent. It is therefore assumed that such clusters can fission into two smaller ones, preferentially aligned along the $q_1\overline{q}_2$ axis. Flavor-dependent parameters are introduced to provide the mass above which a cluster must break, and others to describe the mass spectrum of the daughter clusters. The fission procedure can be repeated on the daughters, if necessary. In $e^+e^-$ events ~15% of the clusters need to be split, but these account for ~50% of the final hadrons.

If baryons only are produced as baryon–antibaryon pairs inside isotropically decaying clusters then that does not agree with observed anisotropies in $e^+e^-$ events. One solution is to allow $g \to qq + \overline{qq}$ branches in the final stages of the shower. This has been implemented both for Herwig and Sherpa, but has now been replaced by the next approach in Herwig. (The possibility to rearrange three mesonic clusters into two baryonic ones has already been mentioned, but is not relevant for $e^+e^-$.)

Isotropic cluster decays also give too soft charm and bottom hadron spectra in $e^+e^-$. Therefore such cluster decays are treated anisotropically, such that the heavy hadron is preferentially near the heavy-quark direction, when viewed in the cluster rest frame. Some further improvements can be obtained if also other cluster decays preferentially favor hadrons closer to the cluster end with the matching flavor.

There may be a small fraction where the cluster mass is not large enough to produce two hadrons with the required flavor content. In such cases the cluster can be allowed to collapse into a single hadron, with excess four-momentum shuffled to another nearby cluster. For heavy quarks one may also allow some such collapses above the two-body threshold, to further harden the heavy-hadron spectrum.

Further procedures exist both in Herwig and Sherpa to handle other special cases.

### 11.4.8 Decays, rescattering and Bose–Einstein

Many of the primary produced hadrons are unstable and decay further. Often the decay channels and their branching ratios are well-known, but for charm and especially bottom hadrons the picture is incomplete. Higher resonances are poorly known also in the light-quark sector. Furthermore, inclusive measurements of a given final state may need to be translated into a potential sequence of intermediate states, *e.g.* $K\pi\pi$ may receive contributions from $\rho$ and $K^*$ resonances. Once the decay sequence has been settled, angular correlations in the decays should be considered, where feasible. Especially for bottom the Evtgen package provides a large selection of relevant matrix elements, as does Tauola for $\tau$ lepton decay. The standard event generators also handle such nonisotropic decays to a varying degree.

The main $pp$ generators assume that particles are free-streaming once formed. This is not the case in heavy-ion collisions, where the particle density remains high a while after the hadronization stage, and hadrons therefore can rescatter against each other. Studies show that also $pp$ collisions are affected by rescattering, but not to a dramatic degree.

Another issue is Bose–Einstein (or Fermi–Dirac) correlations, present in the production of identical bosons (or fermions). Empirically this results in an enhancement (or depletion) of nearby pairs. Typical deduced emission source sizes range from somewhat below 1 fm in $e^+e^-$ to somewhat above that for $pp$. Such scales obviously overlap with the hadronization ones, but also with the decays of short-lived resonances such as $\rho$, and with hadronic rescattering. The modelling therefore is far from trivial, and no traditional generator includes Bose–Einstein effects by default.

### 11.4.9 Other collision types

While the emphasis of the description above has been on the three main *pp* generators at LHC, a few words on adjacent fields and other generators are in place [3560]. Many of these other programs do not address hard physics, but are intended to describe inclusive events dominated by low-$p_T$ QCD processes. Via an MPI machinery they may or may not contain a tail of harder QCD events.

Fields that can be covered by the $e^+e^-/pp$ generators include Deeply Inelastic Scattering and photoproduction in *ep* or $\mu p$. The latter is largely based on the concept of Vector Meson Dominance (VMD), *i.e.* that a real photon can fluctuate into a vector meson state. The transition between the two regions of photon virtualities remains less easily modelled. The VMD picture can also be used *e.g.* for ultraperipheral $\gamma\gamma$ collisions in heavy-ion beams. Work remains to extend the *ep* collision framework to $eA$, as required for the simulation EIC physics.

Generators for heavy-ion physics span a broad range. In one extreme models introduce nuclear geometry and multiple $pp/pn/nn$ collisions, but with each collision similar to a regular *pp* one, up to energy–momentum conservation effects and the like. The earliest such example is Fritiof, the Angantyr descendant of which is now included in Pythia. Others are Sibyll, Qgsjet and Dpmjet. Such models can be run reasonably fast, and the latter three therefore are commonly used for the hadronic part of cosmic-ray cascades in the atmosphere.

In the other extreme the formation and evolution of a quark–gluon plasma (QGP) is the key feature. This requires the combination of models for several stages of the evolution, notably the hydrodynamical evolution of the plasma, which can be quite time-consuming. Jetscape offers a common framework where models for the different stages can be combined at will.

A successful intermediate program is the core–corona EPOS one [3580]. In it, peripheral $pp/pn/nn$ collisions (corona) occur more-or-less separated from each other, while the central higher-density core region may form a QGP, which then decays to hadrons according to a statistical model. The core QGP component gains in relative importance when going from *pp* to $pA$ to $AA$, and from peripheral to central collisions. This gives a behaviour largely consistent with data.

Finally, generators for neutrino physics, like Genie, are largely separate from the ones above, in that an emphasis lies on interactions with nuclei at low energies. The separation into a primary physics process followed by a simulation of detector effects thereby is blurred.

### 11.4.10 Standardization

The main generators discussed here largely are separate codes. This allows for cross-checks where results should agree, and a spread of predictions where the physics is not well-specified. Comparisons are greatly simplified by common standards.

The oldest standard is the PDG particle numbering scheme, whereby observed and postulated particles are assigned unique integer numbers.

The transfer of information from matrix-element generators to general-purpose generators is defined in the Les Houches Accord (LHA), and the associated Les Houches Event File (LHEF) [3581]. It specifies in particular a listing of all incoming and outgoing partons of a hard interaction, with their four-momenta. Extensions include multiple event weights to represent scale and PDF variations.

The transfer of the much bigger complete events from generators to detector simulation, or straight to users, is handled by the HepMC standard [3582]. Again PDG particle codes and four-momenta provide the basic information. Also the step-by-step event history is documented, but cannot be made completely generator-independent since different physics is involved, *e.g.* strings versus clusters.

Parton distributions are widely used in generators, for hard interactions, MPIs and ISR. Today each new PDF set typically consists in the order of a hundred members, to provide a representation of the correlated uncertainties. Each member is stored as a file with the PDF value of all relevant partons in a grid in $(x, Q)$ space. LHAPDF [3583] specifies the file format, such that common interpolation routines can be used for the PDF evaluation for arbitrary $x$ and $Q$ values.

A major issue in the interpretation of data, not least for generator development and tuning, is the difficulty to reproduce all the methods and cuts used in the analysis, even after the data has been corrected for detector inefficiencies and smearing. Here the Rivet framework [3584] allows a standardized way for experiments to submit a code that takes generated (HepMC) events and analyzes them in such a way that the output can be directly compared with published data.

### 11.4.11 The future

Before the start of the LHC, we believed to have a fair understanding of the physics at high-energy $e^+e^-/pp/p\bar{p}$ colliders. The hadronization description developed in the light of PETRA worked surprisingly well also at LEP. By jet universality — the assumption that the same hadronization mechanisms are at play in different collision types — the same should hold also for hadron colliders, when extended by aspects such as multiparton interactions and color reconnection.

The shock of LHC then was that high-multiplicity $pp$ events were shown to behave surprisingly like heavy-ion collisions, with strangeness and baryon enhancement, both in the light-quark and in the charm/bottom sectors, and signs of collective flow such as ridge effects. From what we have been able to learn so far, it seems that high-$p_T$ physics remains unaffected, such that there perturbation theory still can be reliably combined with LEP-tuned hadronization models. This makes sense, in that partons in that region mainly evolve in a vacuum. But, at low $p_T$, we already knew that the multiparton interactions lead to a close-packing of fragmenting systems, whether strings or clusters. We just had not fully appreciated its consequences, in part lulled by the common belief in the heavy-ion community that time scales in $pp$ collisions are too short for a quark–gluon plasma to form. Now we are in the process of rethinking hadronization. One approach is the core–corona one, where a core part of the $pp$ event indeed behaves like a plasma, while the corona part does not. The alternative is to avoid the plama and introduce other possible mechanisms, such as junctions, ropes and shove. While some progress has been made, still no such coherent alternative exists. Anyway, the bottom line is that LHC has reinvigorated the study of the soft-physics aspects of event generators, in addition to obviously driving the hard-physics evolution, see Section 11.3.

While there is still much more to be learned from the LHC, attention is also turning to other future colliders. The one that may require most new generator development is the EIC, since it involves new physics scenarios not addressed before.

## 11.5 Jet reconstruction

**Bogdan Malaescu, Dag Gillberg**
**Steven Schramm, and Chris Young**

A QCD interaction at a very high energy, such as the hard process of an LHC collision, produces quarks and gluons that are asymptotically free at very short distances, but often result in a final state of hundreds of particles at the distance scales of detectors (>1 mm). It is highly desirable to reduce the complexity of the hadronic final state and map it onto a representation that mimics the kinematics of the short-distance hard process. This is the goal of *jet algorithms*. Jet algorithms are a set of rules used to group directionally nearby particles to form jets. A jet can hence be thought of as a collimated group of particles that might correspond to a high energy parton of the hard process. The particles used as input to form jets can be of several types: a set of partons, a consistent set of hadrons, or a set of detector objects such as reconstructed charged-particle tracks or localized calorimeter energy measurements.

### 11.5.1 Jet algorithms

There are a number of desirable features for a jet algorithm. It should be computationally robust and well specified, ideally with few parameters. It should be theoretically well behaved, and exhibit both *infrared* and *collinear safety*. The former refers to adding one or several particles with infinitesimal energy, and the latter to split any input particle into two. For both these kinds of alterations of the input particles, the resulting jet four-momenta will be identical if the jet algorithm is safe against said effects. The jet algorithm should further behave equivalently at different orders of the QCD evolution: at the parton, hadron and detector levels. Furthermore, it should not be tailored to a specific detector, but be useful and used both by theorists and by experimental collaborations.

One of the early jet algorithms was the Snowmass Cone Algorithm of 1990 [3585]. This algorithm, which used $E_T$ and operated in $(\eta, \phi)$-space[106], wrestled with several of the issues mentioned above. Complication arose due to choice of seeds and overlapping cones, which were dealt with by a merging and splitting stage of the jet algorithm, and which tried to find 'stable cones'. Similar cone algorithms with various improvements were employed by the CDF and DØ collaborations at Fermilab [3586, 3587]. The $k_t$ algorithm [172] was developed in 1993, inspired by QCD splittings scales (see Section 11.2). The advantages of the $k_t$ algorithm are that it has no split/merge stage, and jets are uniquely defined; disadvantages include the irregular jet shapes, and the difficulty to experimentally reconstruct and calibrate the jets.

[106] $E_T \equiv E\sin(\theta)$ and the pseudorapidity $\eta = -\ln(\tan(\theta/2))$, where $\theta$ is the angle to the beam pipe.

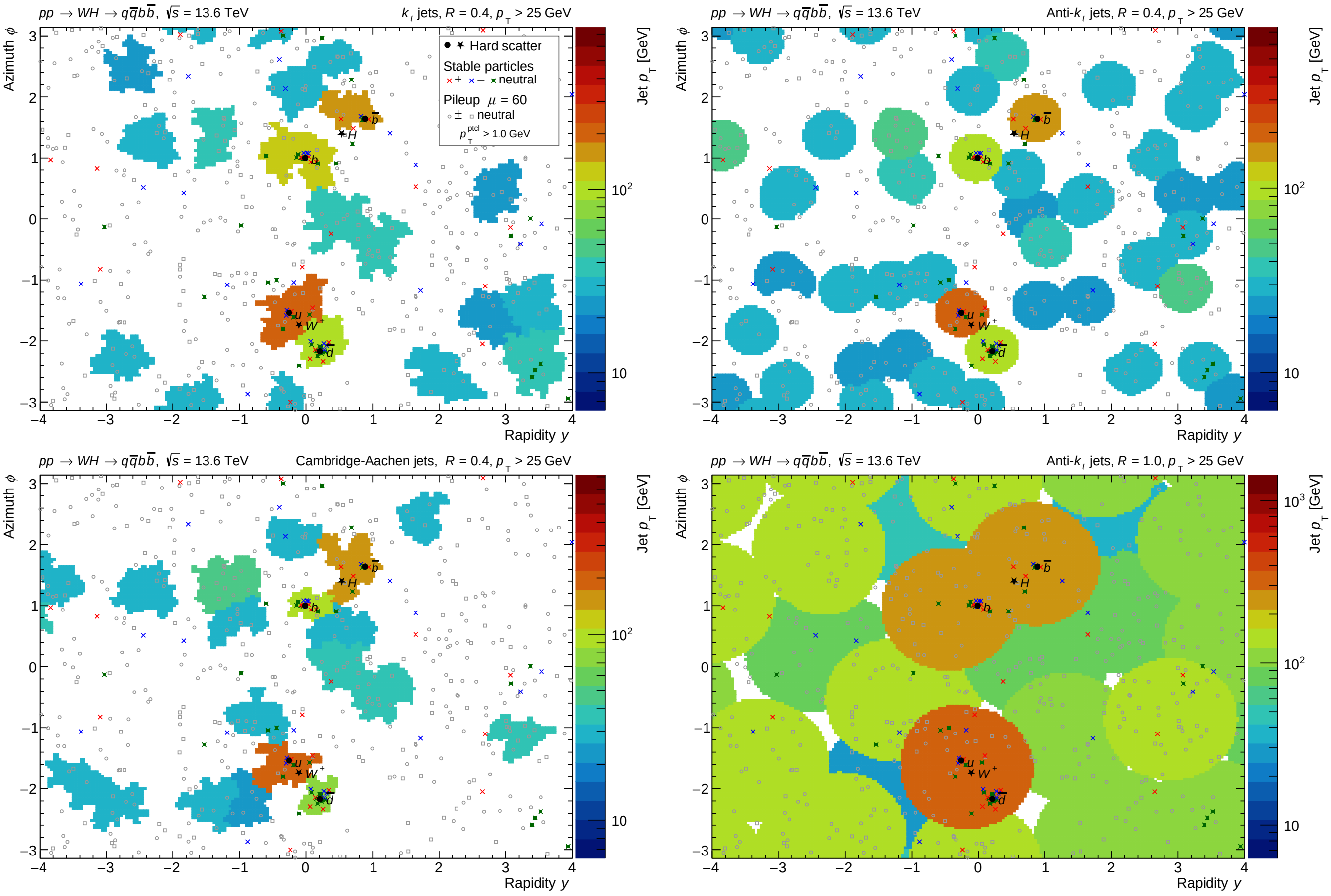

**Fig. 11.5.1** The same simulated $pp \to W^+H \to u\bar{d}\,b\bar{b}$ event, reconstructed with four different jet algorithms: $k_t$ (top left), anti-$k_t$ (top right) and Cambridge-Aachen (bottom left), all with radius parameter $R = 0.4$, and anti-$k_t$ with $R = 1.0$ (bottom right). The hard process particles are shown as black markers, while the final set of stable particles are displayed as crosses. Particles from pileup interactions, generated using a mean of $\mu = 60$ inelastic $pp$ collisions, are shown as grey open markers. Particles with $p_T < 1\,\text{GeV}$ are not displayed. The solid colored areas show the extension (catchment area) of each jet with $p_T > 25\,\text{GeV}$, and their colors indicate the jet $p_T$. The code needed to produce this plot is available as the example program `main95` in recent PYTHIA distributions.

Today, the most common method to build jets is the anti-$k_t$ algorithm [174], defined very similarly to the $k_t$ algorithm. Both algorithms start from a set of particles, each with associated four-momenta, and the following distance measures are calculated

$$d_{ij} = \min(p_{T,i}^{2p}, p_{T,j}^{2p})\frac{\Delta R_{ij}^2}{R^2}\,, \qquad d_{iB} = p_{T,i}^{2p}, \quad (11.5.1)$$

where $R$ is a radius parameter, $\Delta R_{ij}^2 = \Delta y_{ij}^2 + \Delta\phi_{ij}^2$ is the distance squared in $(y,\phi)$-space between particles $i$ and $j$, and the parameter $p$ is 1 for the $k_t$ algorithm, 0 for the Cambridge-Aachen [170] algorithm and $-1$ for the anti-$k_t$ algorithm. The distance $d_{ij}$ is calculated for all combinations of pairs of particles, and $d_{iB}$ once per particle. The smallest distance is found; if this is a $d_{iB}$ value, then particle $i$ will define a jet. If it is a $d_{ij}$ value, then particles $i$ and $j$ are merged, normally by four-momentum addition ($p_k = p_i + p_j$). In both cases, the list of particles and the associated distances are updated, and the algorithm proceeds with one less particle per iteration until all particles have been used. When finished, each input particle is uniquely part of a jet. An illustration of the produced jets for these three $k_t$-style jet algorithms is presented in Fig. 11.5.1, where the jets are built for stable particles produced by a simulated $pp \to W^+H \to u\bar{d}\,b\bar{b}$ event at the LHC with a pileup contribution corresponding to a mean number of inelastic $pp$ interactions of $\mu = 60$.

As is clear from Fig 11.5.1, jets do not provide a unique interpretation of any given event, rather they are a tool that can be optimized to best address the needs of a given task. Even if jet algorithms are intended to represent the underlying hard process of a given collision, the variety of possible hard processes necessitates the consideration of different jet algorithm configurations. In other words, a jet algorithm defines an event organization concept and it can be adapted for different physics processes.

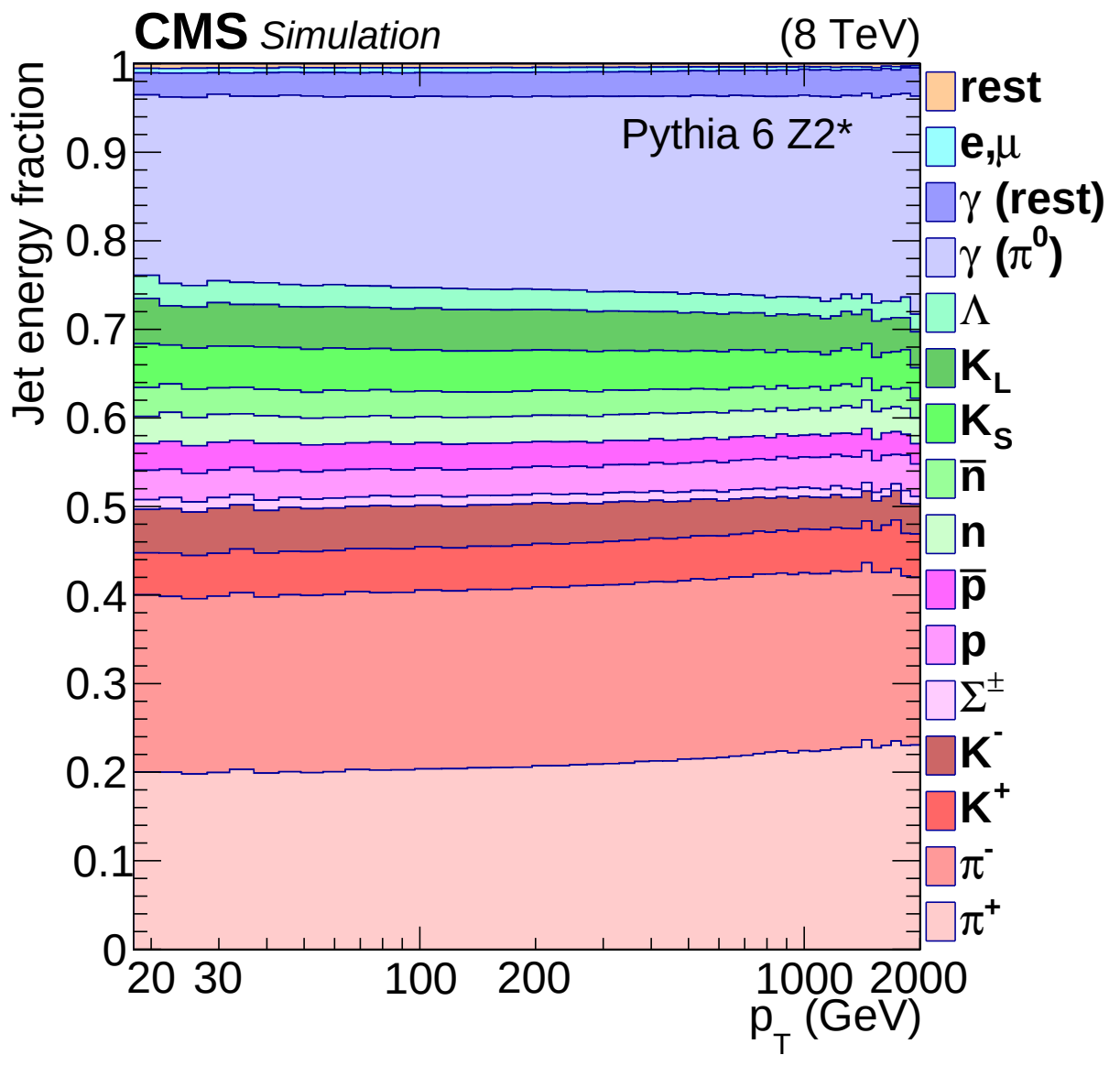

**Fig. 11.5.2** The fraction of the total jet energy carried by different types of particles of particle-level jets produced in simulated LHC dijet events. Particle-level jets are built from particles ($c\tau > 10\,\mathrm{mm}$). The ratio of charged-to-neutral pions is 2:1 due to isospin symmetry, while for baryons it is 1:1; the overall charged-to-neutral fraction of particles in a jet roughly averages between these two expectations. [3588]

The most common usage of jets in the collider context is to represent the collimated group of final state particles originating from individual quarks or gluons of the hard scatter. For this task, the preferred jet radius parameter has slightly changed during the last decades. Values of $R = 0.6$ or 0.7 have often been used for studies of events with well separated jets (e.g. dijet production), while smaller radii (0.4 or 0.5) are more appropriate to resolve more complex final states, such as $t\bar{t}$ or the example shown in Fig. 11.5.1. As is further discussed in Section 11.5.2, smaller radii makes jets less susceptible to pileup, which has become an important consideration at the LHC. Since the start of LHC Run 2, the anti-$k_t$ algorithm with a radius parameter of $R = 0.4$ has been the standard choice largely due to these reasons. The resulting jets are then interpreted as a set of quark- and gluon-initiated showers. Such jets are primarily composed of charged and neutral pions, but baryons and other types of mesons contribute a moderate fraction of the total jet energy, as shown in Fig. 11.5.2. Small energy fractions of electrons and muons can also be seen that originate from semi-leptonic heavy hadron decays.

A natural second-level question relating to such jets is to determine their underlying production mechanism. Is a given jet produced by a light-flavor quark ($u/d/s$), a gluon, a heavy-flavor quark ($c$ or $b$), or by some other process? Heavy-flavor jets are typically easier to define at all levels: they can be identified by whether or not the list of constituents the jet is composed of contains $b$ or $c$ quarks at parton-level; $B$ or $D$ hadrons (or their decay products) at particle-level; or, have associated charge-particle tracks originating from collision-point-displaced vertices at the experimental level. The difference between light-flavor-quark- and gluon-initiated showers is more subtle, and is not rigorously defined for particle- or experiment-level jets. Instead, the expected properties of quarks and gluons can be used to differentiate between such jets on average, noting that quarks have a single color charge and are thus expected to radiate less, resulting in more narrow showers containing fewer constituents than showers produced by gluons.

Another important concept, and which is of great relevance at the LHC, is to use jets to represent complex energy flow processes rather than individual showers. The high energy collisions at the LHC can result in the production of massive particles, such as $W$, $Z$ and $H$ bosons and top quarks, with high transverse momentum. Therefore they have a sizable Lorentz boost in the rest frame of the detector, and their decay products will be collimated. In the case of hadronic decay products, each daughter particle further produces showers of hadrons, which can overlap. Rather than reconstructing this complex structure of overlapping hadronic showers as separate jets, the entire decay of the massive particle can be treated as a single jet, and properties of that jet can be used to infer the nature of the originating particle [3589]. In such a scenario, it is useful to increase the distance parameter used to build jets to contain the entire hadronic decay, as shown for the anti-$k_t$ algorithm with $R = 1.0$ in the bottom right plot in Fig. 11.5.1 where the $W$ boson decay is within a single jet, while the $H$ boson decay is split between two jets. The collimation of the decay particles is related to the mass and momentum of the parent particle; for a two-body decay, this becomes:

$$\Delta R \gtrsim \frac{2\, m_{\mathrm{parent}}}{p_{\mathrm{T}}^{\mathrm{parent}}}\,, \tag{11.5.2}$$

where $\Delta R$ is the angular separation between the decay products in $(y, \phi)$-space. From this equation, it is clear that increased collision energies producing higher-momentum massive particles will result in increasingly collimated decays, and thus the importance of using a larger value of $R$ to represent a complex energy flow is related to the energy scale of the process under study. Jets built with this context in mind are typically referred to as large-radius or large-$R$ jets, where typical modern values are $R = 0.8$ for CMS or $R = 1.0$ for ATLAS; this is in contrast to the previously discussed $R = 0.4$ jets, which are referred to as small-radius or small-$R$ jets.

Using a larger distance parameter comes with several complications, both experimental and theoretical. From the purely algorithmic perspective, one challenge is that the catchment area [3590] of an individual jet grows dramatically, as clearly visible when comparing the top right and bottom right plots in Fig. 11.5.1. Among other effects, this increases the amount of energy from the underlying event included in the jet, which can hide the features of interest: for example, the mass of the jet should peak at the mass of the parent particle, but this is not the case due to the presence of the underlying event. This can be mitigated through the use of a variety of different grooming algorithms [3589, 3591–3594]. These algorithms are typically applied after building the initial jets. The objects clustered into the jet are then subject to a further selection, and those which appear to be inconsistent with originating from a hard-scattering process are removed, thus suppressing the underlying event and other undesired contributions while retaining the physics features of interest.

### 11.5.2 Jet reconstruction

*Inputs to jet reconstruction:*

Particle-level jets, often referred to as *truth jets*, are used as a theoretical reference for experimental measurements. These are jets built from *stable particles*, defined as those with lifetime $\tau$ such that $c\tau > 10\,\mathrm{mm}$ ($\tau > 33\,\mathrm{ps}$), which can be thought of as "what a perfect detector would see". It should be noted that neutral pions are not considered stable and hence their decay products (photons) will be used as input to truth jets (see Fig. 11.5.2). Only particles produced in the proton–proton interaction of interest are considered. These jets also form the reference for the calibration of reconstructed jets.

Experimental reconstruction of jets requires the definition of a given set of inputs, which will ideally represent the true particles of the jet or the energy flow. As jets consist of both charged and neutral hadrons, the simplest reconstruction makes use of the energy flow captured in a calorimeter, which measures the energy of both charged and neutral incident particles. However, as we will see in this section, the accuracy of jet reconstruction can be improved through the use of additional information from tracks reconstructed from charged particles.

At a hadron collider such as the LHC, a wide range of energies of jets need to be accurately reconstructed: from 20 GeV to above 4 TeV in $p_T$. This represents a significant challenge for the design of the detectors. Both ATLAS and CMS surround the interaction point with a tracking detector immersed in a magnetic field, such that the momentum of charged particles can be measured. Around this are the calorimeters. The innermost calorimeters are designed to reconstruct electromagnetically showering particles, such as electrons and photons, and will also capture some energy from charged and neutral hadrons. Radially outward of these detectors are hadronic calorimeters that measure the energy of showers from remaining charged and neutral hadrons.

An additional complication at the LHC is *pileup*. Each time two bunches of protons cross, multiple pairs of protons can collide. This is referred to as in-time pileup. The beam-spot, the region of interactions, is typically 30–50 mm in length along the beam direction. This means that such collisions are typically separated in this dimension and tracks originating from different collisions can be identified. A second effect is out-of-time pileup. The bunches of protons cross every 25 ns in the LHC, therefore there are still residual effects in many of the detectors from the previous (and subsequent for some systems with large integration times) bunch crossings. These residual signals are referred to as out-of-time pileup.

Throughout Run 1 of the LHC (2010–2012), ATLAS used solely calorimeter inputs to build their jets. The ATLAS calorimeters consist of over 100,000 cells. This fine cell granularity is used to suppress noise by constructing clusters of cells, which represent the energy flow. Cells with energy significantly greater than the expected background noise are used to seed such clusters, and adjacent cells are added iteratively, forming topologically connected clusters representing a shower [3595]. This process means that most cells in the calorimeter are not included in the event reconstruction, and hence their noise does not contribute to the jet resolution. As the calorimeters are non-compensating, showers caused by electromagnetically and hadronically interacting particles of identical initial energies have different energy responses. The jet resolution can therefore be improved by identifying which type of shower each cluster contains and calibrating it appropriately. In ATLAS the energy density of the cluster and its position in the calorimeter are used for classification and subsequent calibration [3595]. These calibrated clusters were the input signals to jet reconstruction for ATLAS in Run 1.

CMS has employed a particle-flow approach both in Run 1 and Run 2 [3597], and ATLAS also developed such an approach for Run 2 [3596]. The principle of particle flow is to supplement the information from the calorimeter with tracking information. Both collaborations match tracks reconstructed in the inner detector to calorimeter energy deposits from the same particle. The ability to do this depends on the granularity of the detector, the small transverse size of the showers in the calorimeter, and the separation of the particles. Figure 11.5.3 shows how this can be achieved by extrapolating tracks through the magnetic field to the calorimeter and matching them to calorimeter energy deposits. The CMS algorithm combines the measurements of tracks and matched calorimeter-energy deposits to create combined reconstructed charged hadrons with improved resolution. Calorimeter deposits without tracks are then identified as neutral hadrons. Situations where the showers of a charged hadron and a neutral hadron are overlapping are identified by the excess of energy in the calorimeter above what would be expected from the charged hadron. In ATLAS a choice is made between the calorimeter and track reconstruction. For low $p_T$ tracks, where the track resolution is significantly better than that of the calorimeter, the momentum measurement is taken from the reconstructed track and the corresponding shower created by that particle is removed from the calorimeter. The remaining calorimeter energy deposits then represent the energy flow from particles without tracks and those where the track is not selected.

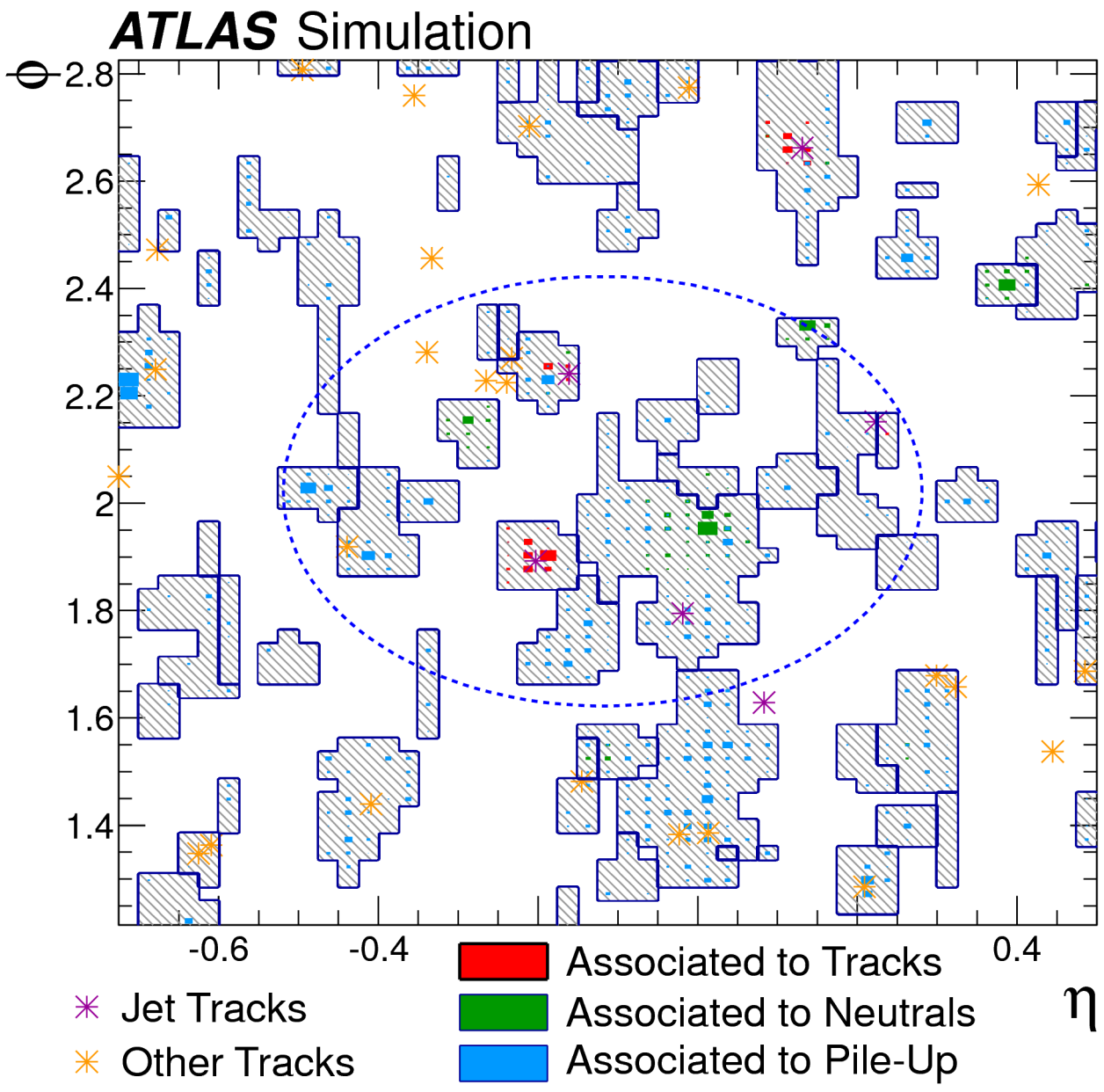

**Fig. 11.5.3** The simulated signals from a $p_T = 30\,\text{GeV}$ jet in the $(\eta, \phi)$ plane of the second layer of the ATLAS electromagnetic calorimeter. The shaded cells are those included in calorimeter topoclusters. Green deposits are from neutral hadrons within the jet, red deposits are from charged hadrons within the jet, and blue deposits are from pileup particles. The purple $*$ represents the tracks of charged hadrons within the jet after being extrapolated to the calorimeter, and the yellow $*$ represents tracks from pileup [3596].

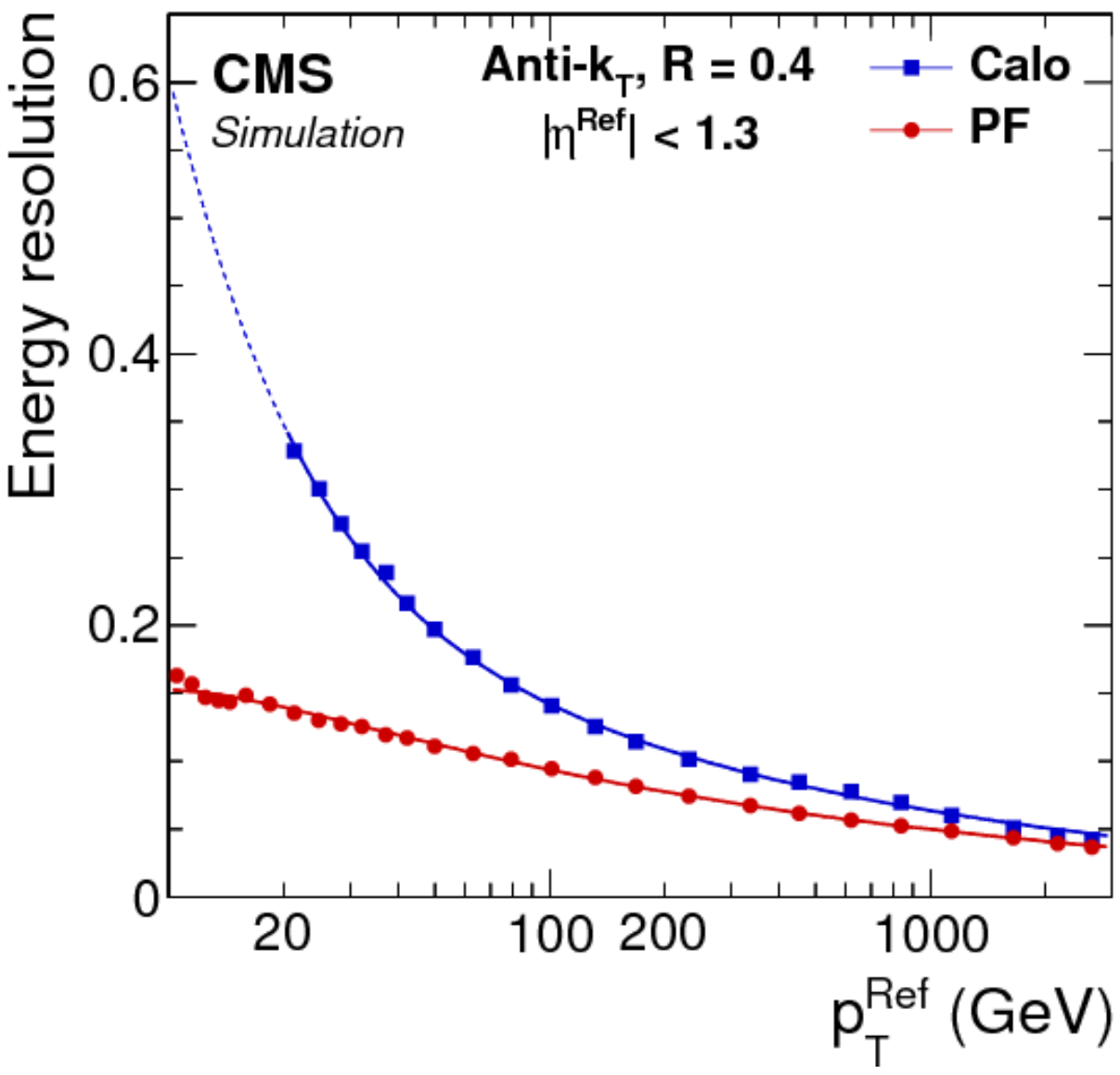

**Fig. 11.5.4** The jet resolution in the central region of the CMS detector when jets are reconstructed using calorimeter signals (Calo) or particle flow objects (PF). The simulated QCD events have $\sqrt{s} = 13\,\text{TeV}$ and there are no pileup effects present [3597].

Both collaborations see significant improvements in the $p_T$ and angular resolutions of jets reconstructed using particle flow. Fig. 11.5.4 shows the dramatic improvement in the energy resolution in CMS. In ATLAS the improvement is smaller, and primarily at lower $p_T$, due to the superior calorimeter resolution. However, the gains from the use of particle flow increase at higher pileup motivating its use in Run 2 and beyond.

*Jet algorithms in the experimental context:*

Having reconstructed either clusters or a set of particle flow objects, the jet algorithms featured in Sec. 11.5.1 can be used to build jets. A key advantage of using particle flow objects is that prior to building the jets, charged particles that are from in-time pileup interactions can be excluded. This is known as Charged Hadron Subtraction, and is performed by both experiments' particle flow algorithms [3596, 3597]. This removes the majority of the effects of charged pileup particles but the effects due to neutral pileup particles and out-of-time pileup remain. This explains why ATLAS observes increasing benefits of the particle flow approach at higher pileup. Additionally CMS employs PUPPI [3598] which uses the local information to try to identify neutral pileup energy deposits and weight these to lower significance prior to jet finding [3598, 3599].

Some small-$R$ jets reconstructed from either calorimeter or particle flow inputs will consist of only signals from pileup particles. These are referred to as pileup jets and can be the result of QCD jets from other in-time collisions, multiple particles from different in-time collisions, out-of-time pileup signals, or a combination of several of these effects. These jets will not have tracks pointing at them from the interaction vertex of interest, while they will in some cases have tracks from other vertices. These features are used by both ATLAS and CMS to reject such jets such that they are not used in analyses [3599, 3600].

Large-radius jets are much more susceptible to pileup, due to their larger catchment area. Most large-radius jets at the LHC will therefore contain a mixture of energy originating from multiple collisions (either in-time or out-of-time), and thus it is impractical to reject entire jets. Moreover, large-radius jet are typically used in situations where the internal structure of the large radius jet is of interest, and thus any constituents originating from other processes than the hard-scatter interaction must be suppressed to observe the jet's internal structure. Charged hadron subtraction, from particle flow algorithms, can help to remove charged contributions for separate collisions, but alternative strategies are needed to remove overlapping neutral contributions. Grooming algorithms, previously motivated in the context of suppressing underlying event contributions, are also useful in this context: the same criteria of suppressing soft and wide-angle radiation is also useful for mitigating pileup contributions. These grooming algorithms are applied after the jet is built, but the inputs to jet algorithms can also be corrected; various criteria can be used to suppress neutral jet inputs from vertices other than the one of interest, such as Constituent Subtraction (CS) [3601], Soft Killer (SK) [3602], PUPPI, or combinations thereof such as CS+SK. Currently, ATLAS makes use of CS+SK to modify the inputs to large-radius jet reconstruction [3603], while CMS makes use of PUPPI [3604].

### 11.5.3 Jet calibration

*Energy scale and resolution:*

Once jets are reconstructed, they need to be calibrated such that on average the reconstructed jet four-momenta match those at the particle level within the assigned uncertainties. At hadron colliders, the jet energy-scale (JES) calibration-correction is typically applied in a sequence of steps. Those account for (the mitigation of) contributions from additional proton–proton collisions, energy losses in the dead material of the detector, calorimeter non-compensation (where applicable), angular biases, etc. Several of these calibration steps rely on a detailed Monte Carlo simulation (MC) of detector effects. Modern techniques use jet and event properties (e.g. jet area, jet width, fraction of energy in the various layers of the calorimeters, average $p_T$ density) to improve resolution and to mitigate the dependence of the JES response on the jet flavor. The latter are sizable mainly at low jet transverse momentum ($p_T$) and yield one of the main modeling uncertainties impacting the JES calibration.

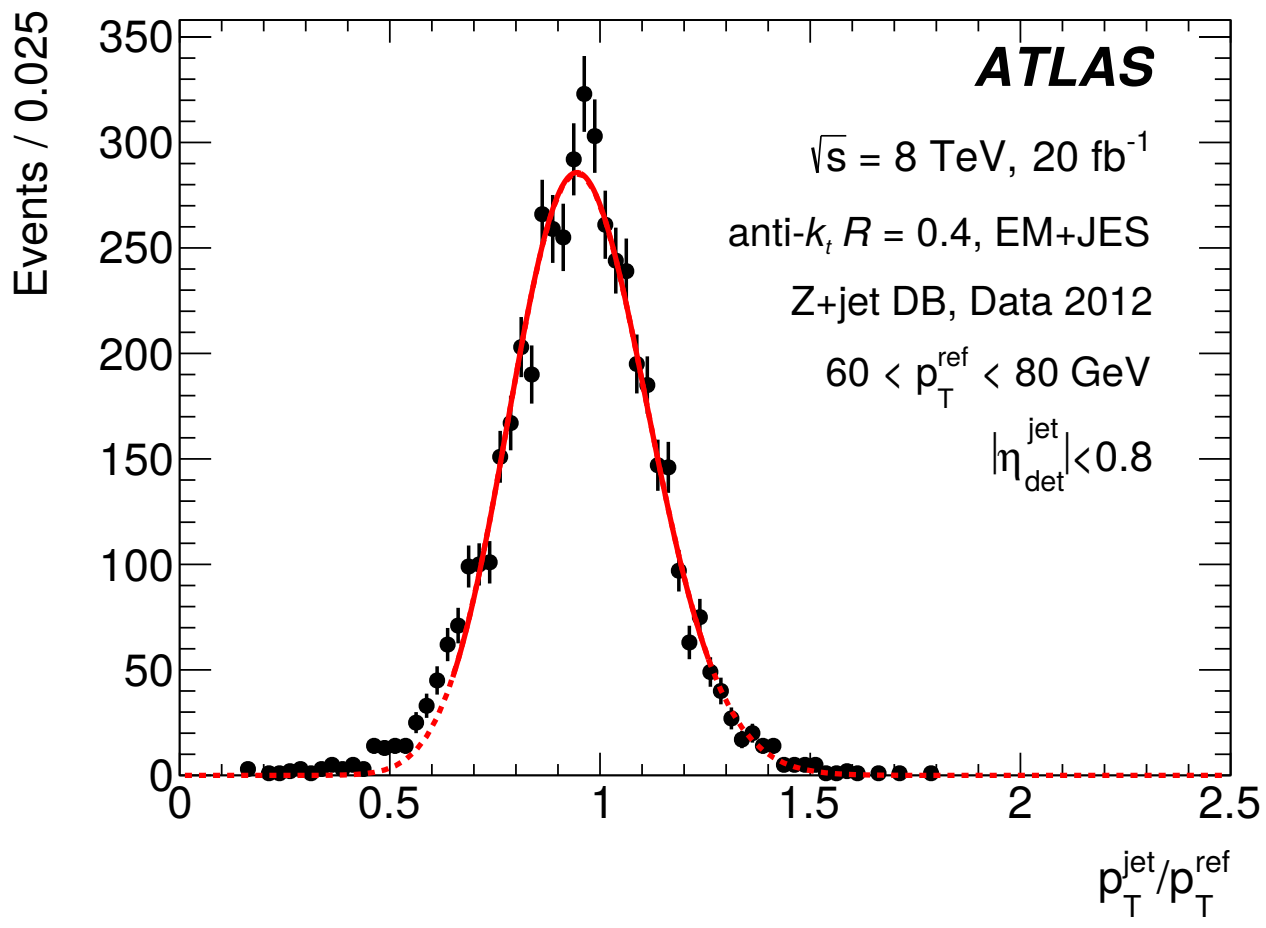

**Fig. 11.5.5** Distributions of $p_T^{\rm jet}/p_T^{\rm ref}$ in $Z$+jet events, where $p_T^{\rm ref}$ is defined by the reconstructed $Z$ boson $p_T$ and is required to be in the range $(60, 80)$ GeV. The dashed line shows the fitted distribution, from which the means are taken as the response measurements. The solid line indicates the fitting ranges. The markers are the data counts with error bars corresponding to the statistical uncertainties. Figure from Ref. [3605].

The calibration chain is completed by *in-situ* corrections that are most commonly derived by exploiting momentum balance between jets and well-measured reference objects. Selection criteria are applied to suppress extra radiation and obtain a sample of events where a probe jet is back-to-back with the reference object. A correction is then derived by comparing the measured balance in data relative to the expectations of MC simulation, and correcting for the residual difference:

$$\left(p_T^{\rm jet}/p_T^{\rm ref}\right)^{\rm data} \Big/ \left(p_T^{\rm jet}/p_T^{\rm ref}\right)^{\rm MC}. \tag{11.5.3}$$

This principle was developed for the calibration of small-radius ($R \in [0.4, 0.7]$) jets [3588, 3605, 3606] and has now also been used for large-radius jets [3607].

These *in-situ* methods employ, as reference objects, photons, $Z$ bosons decaying to charged leptons, and one or several pre-calibrated jets. Fig. 11.5.5 presents an example of $p_T^{\rm jet}/p_T^{\rm ref}$ distribution in data, the mean of which is used to derive the jet calibration. They also provide the main uncertainties impacting the JES calibration, reaching nowadays sub-percent precision across a broad phase-space, while being larger for relatively

low- and large-$p_T$ jets, as well as in the forward region of the detectors. While for large-$p_T$ jets these approaches are limited by the available statistics, for low-$p_T$ and forward jets they are limited by modeling effects, related to e.g. the emission of extra radiation impacting the $p_T$ balance. The use of *in-situ* techniques have allowed for significant improvements in precision compared to jet calibrations based on test-beam studies. The latter are still used in phase-space regions with little/no statistics coverage for the *in-situ* approaches.

Statistical combinations, with a full propagation of uncertainties and correlations, are generally employed and yield the necessary inputs for physics analyses. In these studies, uncertainties on the uncertainties and on the correlations have also been evaluated (see e.g. Ref. [3608]). This is an example where QCD studies trigger developments that set new standards on a topic of interest in other scientific areas too.

The width of the $p_T^{\rm jet}/p_T^{\rm ref}$ distributions, such as the one exemplified in Fig. 11.5.5, provides information about the jet energy resolution (JER). Indeed, the JER is determined in various $p_T^{\rm jet}$ ranges and detector regions, after subtracting statistically the smearing effect induced by the presence of extra radiation in the events. Afterwards, a statistical combination of several *in-situ* methods through a fit allows for the extraction of a parameterization of the JER in data, together with its uncertainties, readily usable in physics analyses accounting for detector smearing effects.

*Mass scale and resolution:*

To first order, calibrations derived to correct the energy of a jet are also important to use when correcting the mass of a jet, as these two quantities are related. However, the mass calculation includes both energy and angular components, and thus the jet mass must be further corrected after the energy has been addressed. Similarly to the energy, calibrating the mass of a jet begins with corrections based on simulated samples, to correct the average simulated jet mass to the particle-level scale. In the context of large-radius jets, it is very important to apply the same grooming algorithms to the truth jets and the reconstructed jets, as the grooming algorithm has a substantial impact on the mass of the jet built from particles, primarily due to the suppression of the underlying event contributions.

Following these simulation-based corrections, the resulting mass must be compared between data and simulation, but the strategies to evaluate differences between data and simulation necessarily differ. The jet energy corrections exploited the conservation of momentum in the transverse frame through the balance between probe and reference objects to obtain a precise calibration. There is no equivalent conservation law for jet mass, so a different approach is needed. Instead, the mass has a well-defined expectation value in specific cases, notably if a pure sample of $W$, $Z$ or $H$ bosons or top quarks can be obtained. $W$ bosons and top quarks are the easiest particles to identify in this context: semi-leptonic $t\bar{t}$ events provide an ideal means of identifying a high-purity selection of hadronically decaying top quarks, and the distinction between a full top decay and a $W$ boson decay can be made by requiring the $b$-quark from the top decay to be either inside or outside of the large-radius jet of interest. The resulting high-purity selection of $W$ bosons or top quarks can be compared between data and simulated events, where differences in the mass peak's central value (mass scale) and width (mass resolution) can be evaluated and corrected for; an example of the top quark selection is shown in Fig. 11.5.6, where it is clear that the selected events are very pure in the signal of interest.

This approach works well, but is limited to only a few possible jet mass values where we have a well-defined Standard Model expectation. Correcting the scale and resolution for other mass values is a much more complex task, and a robust, high-precision method to provide a general mass correction remains an open challenge.

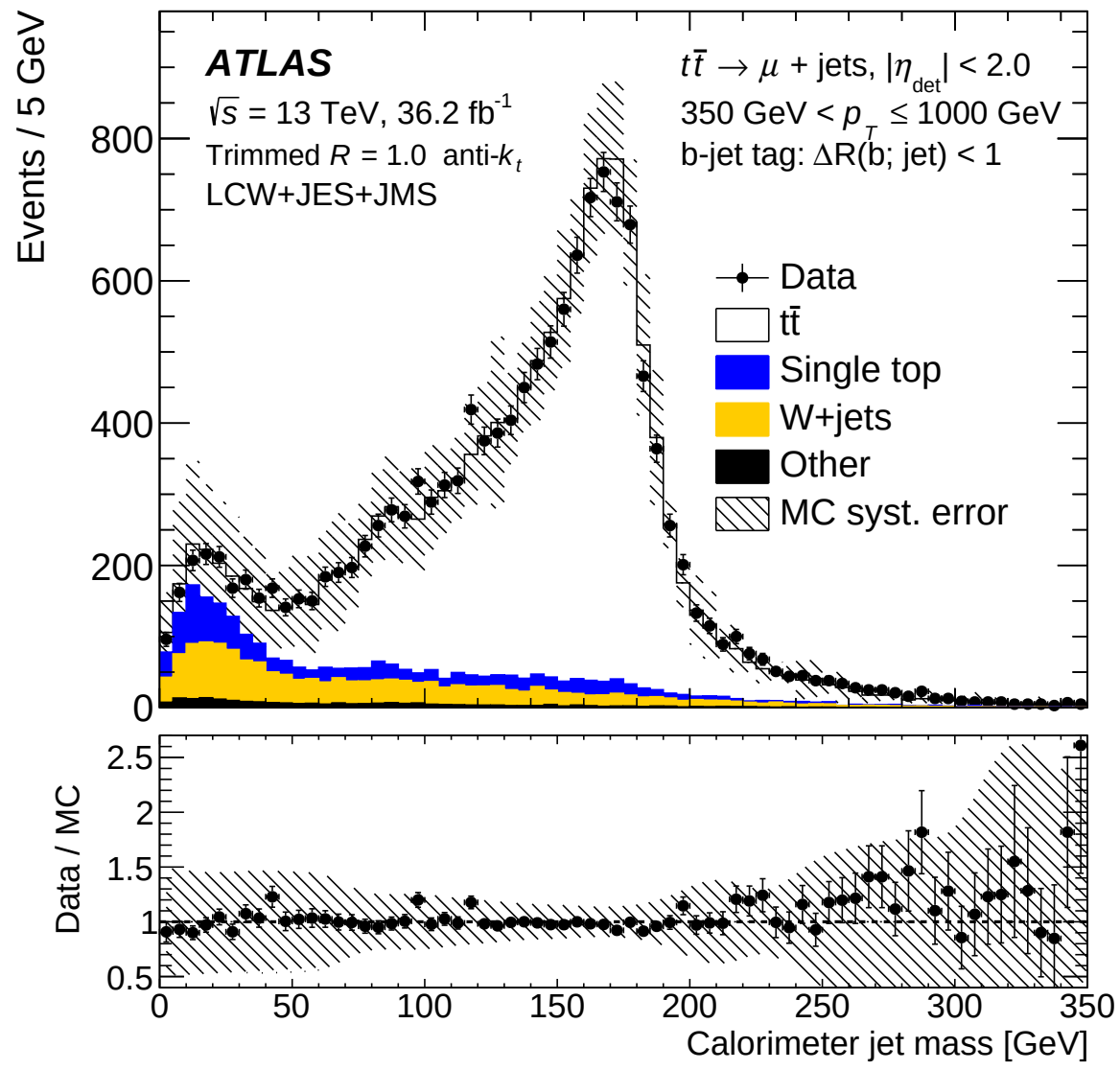

**Fig. 11.5.6** The mass of large-radius jets in a final state targeting semi-leptonic decays of $t\bar{t}$ events, where a $b$-tagged jet overlaps with the large-radius jet. This selection primarily identifies large-radius jets containing the decays of boosted top quarks, as is clear from the dominant peak structure consistent with the $t\bar{t}$ simulation expectation. Differences between data and simulation in both the jet mass scale and resolution can be extracted from such a plot. [3607]

### 11.5.4 Classifying hadronic decays of massive particles

The use of large-radius jets is overwhelmingly linked to the desire to represent the entire hadronic decay of a massive particle, such as (but not limited to) a $W/Z/H$ boson or a top quark. If the jet does contain all of the daughter particles and their corresponding showers, then the mass of the jet now has a well-defined prior, namely the mass of the parent particle. This prior holds so long as the large-radius jet represents only the process of interest: underlying event and pile-up contributions falling within the jet's catchment area can both obscure the internal structure of the jet, and must thus be mitigated, as previously discussed. The mass then becomes an excellent means of classifying jets based on the parent particle that they originate from.

While the jet mass provides a robust means of differentiating between different possible sources of large-radius jets, in many cases it is not sufficient, as light quarks and gluons from QCD multijet processes are produced in extreme abundance compared to the massive particle decays of interest. The mass distribution of light quarks and gluons is peaked at low values, well below the $W/Z/H$ boson or top quark masses, but the tail of the mass distribution extends to high masses, and these tails are still more probable than the production of the target massive particles.

Additional jet properties can be used to further classify the origin of a given large-radius jet. These properties are referred to as *jet substructure variables* and are designed to quantify the internal angular energy structure of a jet. Substructure variables are almost always correlated with the jet mass, and thus it is important to identify variables that are sufficiently distinct to provide further separation power. One commonly used example is the N-subjettiness ratios, $\tau_{xy} = \tau_x/\tau_y$, where $\tau_n$ is a projection of the constituents of a jet along $n$ axes, thereby evaluating the consistency of the jet containing $n$ or fewer decay particles. As an example, $\tau_{32}$ is commonly used to identify jets containing top quarks, as it differentiates 3-body decays from 2-or-fewer-body decays, as shown in Fig. 11.5.7. This is only one example out of the many different types of jet substructure variables that have been used to complement the jet mass in classifying the origin of large-radius jets.

While the jet mass and substructure variables provide a solid baseline, modern large-radius jet classifiers make use of machine learning techniques to maximally discriminate between different possible jet origin interpretations. There is a wide variety of machine-learning-based classifiers in use by both the ATLAS and CMS Collaborations, and they continue to become more powerful; a comparison of several such algorithms as used by CMS is provided in Fig. 11.5.8.

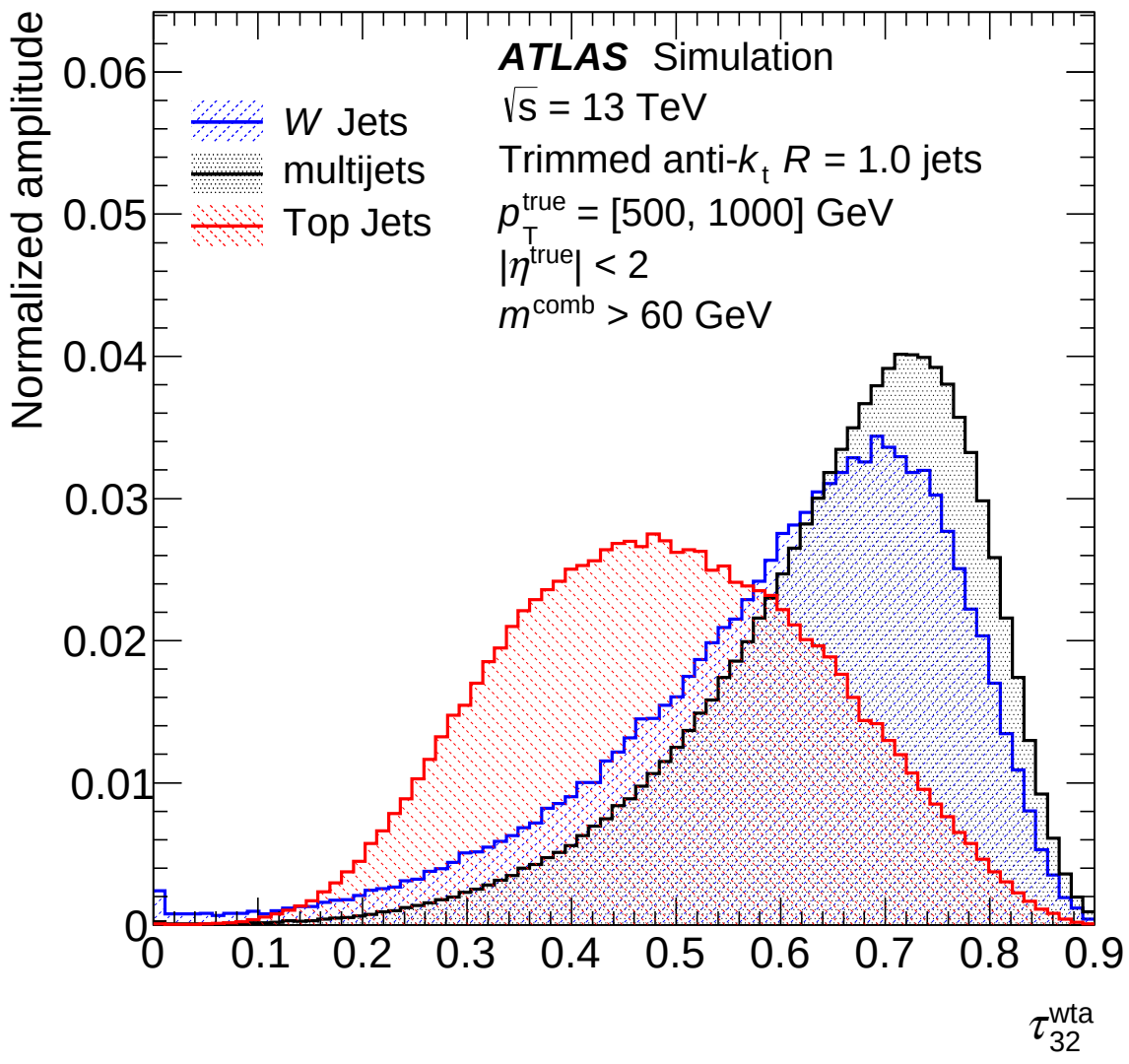

**Fig. 11.5.7** The N-subjettiness ratio $\tau_{32}$, with the *winner takes all* (wta) axis definition [**Bertolini:2013iqa**], showing the separation of jets containing three-body decays (top jets) against jets containing either two-body decays ($W$ jets) or individual quarks/gluons (multijets). This is after a selection criterion is applied on the jet mass, and thus the separation shown between the three jet types provides additional classification power. [3609].

Similar to the jet energy and mass calibrations, the difference between data and simulation must also be quantified when classifying the origins of large-radius jets. The algorithms used are usually optimized using simulated events, and there is no guarantee that the simulation properly describes the data, especially for the complex angular energy structure within a jet, which is what such classifiers rely upon to differentiate between different jet categories. Similar to the jet mass scale calibration, semi-leptonic $t\bar{t}$ events provide a useful signal-enriched region to evaluate the performance of both $W$ boson and top quark classifiers in simulation and data; other signal categories remain more challenging, as we do not yet have sufficiently signal-pure regions to perform similar comparisons. In contrast, comparing the differences between data and simulation for the misidentified background events is straightforward, as the QCD multijet and $\gamma$+jet processes have such large cross sections that they are highly background enriched by default. Any differences between data and simulation in the fraction of background events passing a given large-radius jet classifier can thus be evaluated using these two processes.

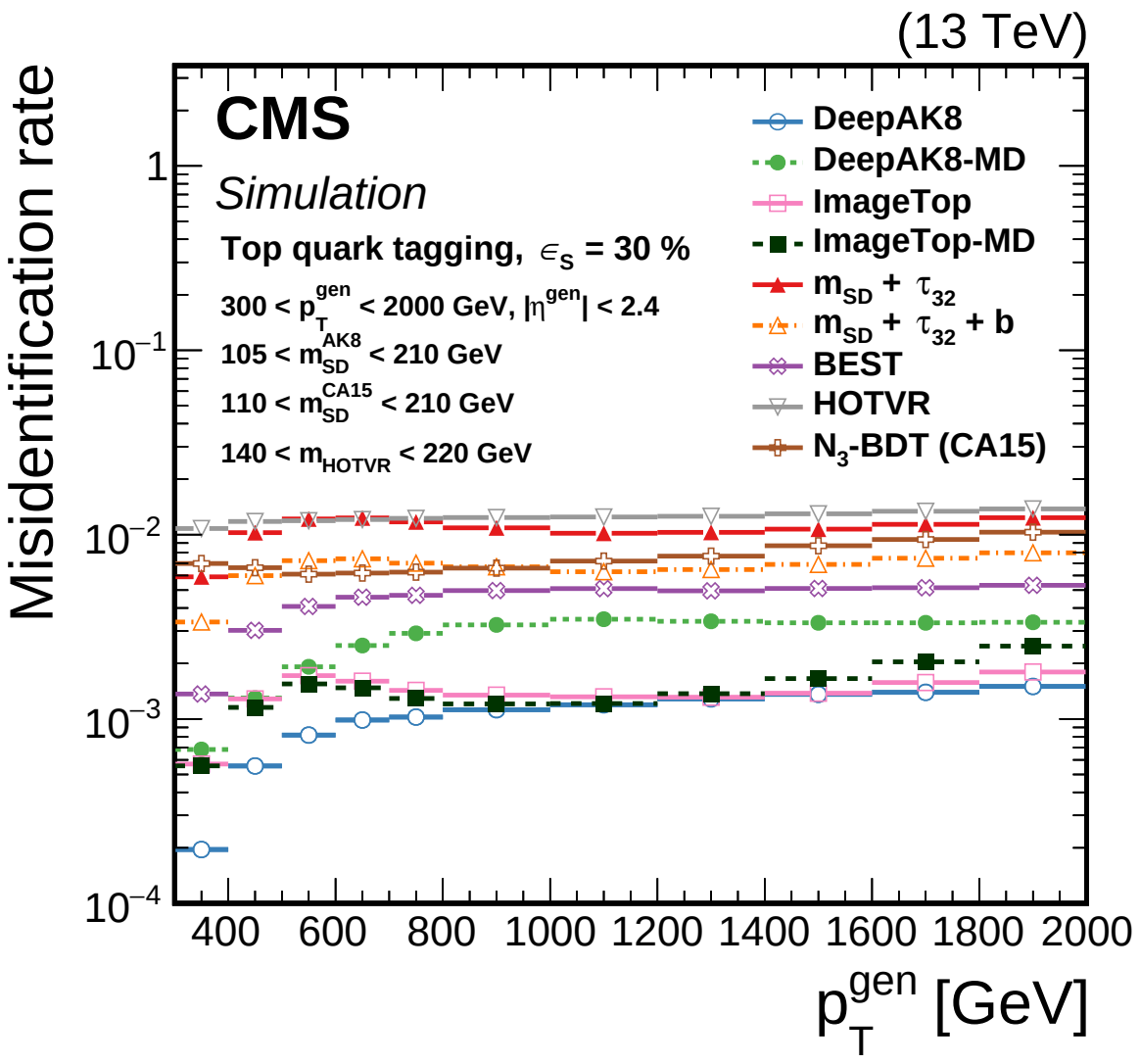

**Fig. 11.5.8** A comparison of many different algorithms designed to identify jets originating from hadronic decays of top quarks. A simple tagger based on the jet mass combined with the N-subjettiness ratio $\tau_{32}$ ($m_{\mathrm{SD}} + \tau_{32}$) is shown alongside many alternative classifiers which providing significantly better performance. This can be seen as they have much lower quark/gluon (background) misidentification rates for a fixed top quark (signal) efficiency. The large majority of the alternative classifiers make use of machine learning techniques. [3610]

### 11.5.5 Summary

Jets are crucial tools for numerous physics analyses performed at hadron colliders. During the last four decades, there has been significant development in this field and jet definitions that are robust for both experimental measurements and theoretical predictions have been identified. In addition, the improved particle detectors, with highly granular calorimeters and high resolution reconstruction of charged tracks are enabling reconstruction of the full jet four momentum, investigation of the jet internal structure and classification of jets via tagging. These developments are allowing us to expand the knowledge about QCD and to look for signatures of BSM physics, yielding greatly improved searches and measurements.

# 12 Measurements at colliders

*Conveners:*
**Karl Jakobs and Eberhard Klempt**

Since the successful operation of the $Sp\bar{p}S$ at CERN and the Tevatron at Fermilab, the high-energy frontier of particle physics is defined by colliders. The Large-Electron-Positron (LEP) collider, operating from 1989 to 2000, allowed the collaborations to confirm the gauge structure of QCD and to test QCD systematically (Stefan Kluth). The most significant results of HERA, the only electron-proton collider operating from 1992 to 2007, were studies of the structure of nucleons presented in Section 10. The physics of the Brookhaven Relativistic Heavy Ion Collider (RHIC) is discussed in Section 7.

The formation of jets, of streams of collimated hadrons, was first observed at SPEAR, later in all colliders. As outlined by Daniel Britzger, Klaus Rabbertz and Markus Wobisch, the production of jets developed to a QCD testing ground to searches for new phenomena up to the largest accelerator-based energies at the Large Hadron Collider (LHC). Jets initiated by gluons, quarks – including the heavy quarks $c$ and $b$ – can be produced jointly with the vector bosons $W^{\pm}$ and $Z^0$. The cross sections of all these processes are precisely reproduced by QCD calculations (Monica Dunford). The discovery of the Higgs in 2012 was a milestone for particle physics. Chiara Marotti describes with which surprising precision the properties of the Higgs boson follow the predictions of the SM. The top quark, discvovered in 1995 at Tevatron and discussed here by Marcel Vos, is identified in a large variety of reactions, from top-anti-top production to $t\bar{t}$ production associated with a vector boson or the production of two $t\bar{t}$ pairs. The cross sections for these processes span a wide range from nearly $10^3$ pb down to a few $10^{-2}$ pb.

## 12.1 The Legacy of LEP

**Stefan Kluth**

The large electron positron collider LEP was conceived and designed at CERN in the 1980's to study the then just discovered massive vector bosons of the Standard Model (SM), the neutral Z and the charged $W^{\pm}$ bosons [3611]. The four LEP experiments ALEPH, DELPHI, L3 and OPAL collected more than four million Z decays and about 10,000 W pairs each. The LEP 2 runs at centre-of-mass (cms) energies above the Z resonance up to 209 GeV provided samples of O(1000) hadronic final states from off-shell $(Z/\gamma)^*$ decays at each cms energy. These data, together with the extremely accurate LEP beam energy determination, established "electro-weak precision observables" (EWPO) and the confirmation of the SM at very high precision [3612].

Hadronic final states at LEP are also a great laboratory to study a large spectrum of QCD predictions. The missing initial- and final-state interference and the comparatively high energy lead to clearly interpretable hadronic final states and usually small corrections from non-perturbative effects. All LEP experiments have among their first few publications papers on properties of hadronic Z decays.

The detectors of the LEP experiments were significant improvements on their predecessors and offered an almost complete coverage of the solid angle with efficient and precise tracking and finely grained calorimeters with layers for electromagnetic and hadronic showers. All LEP experiments had silicon micro-vertex detectors and full coverage with muon detection systems outside of the calorimeters.

The $e^+e^-$ initial state with well known beam energies provides a strong constraint to improve energy measurements. For example the scaled jet energies in Z decays to 3-jets can be determined from jet angles only [3613]. Even without using the beam energy directly in a constraint the use of quantities scaled to the cms energy reduces dependence on the absolute energy scale of the detector. As explained below, jet definitions, event shape observables and particle spectra are normalised to the cms energy $Q = \sqrt{s}$. Note that in the measurements the normalisation to $Q$ is replaced by the measured total visible energy $E_{vis}$ which also partially removes the influence of statistical fluctuations.

Compared to previous experiments the LEP data have much larger event samples on the Z peak, low experimental systematic uncertainties and higher cms energies leading to smaller and well controlled hadronisation corrections.

The data taken on the Z peak (LEP 1) have favorable experimental conditions. The trigger efficiency for hadronic final states is essentially 100% and can be measured using redundant triggers. Backgrounds from hadronic decays of $\tau$ lepton pairs are suppressed by demanding more than four charged particles. Requirements on balance of observed momentum along the beam direction and total visible energy remove backgrounds from $e^+e^- \to 2\gamma \to$ hadrons interactions. There are corrections for initial state photon radiation effects, but on the Z peak these are small.

The data taken at $\sqrt{s} > m_Z$ but below the threshold for $W^+W^-$ pair production (LEP 1.5) at $\sqrt{s} = 130$ and 136 GeV already contain a substantial fraction of so-called "radiative return" interactions $e^+e^- \to \gamma_{\rm ISR} + Z \to$ hadrons[107]. Simply speaking, instead of a high-energy interaction near the nominal $\sqrt{s}$, a Z decay to hadrons recoiling against the ISR photon $\gamma_{ISR}$ is produced. The LEP collaborations developed algorithms to reconstruct the effective cms energy $\sqrt{s'}$ of the observed hadronic system by assuming a 2-body decay together with one or more high-energy ISR photons.

[107] ISR stands for initial state radiation.

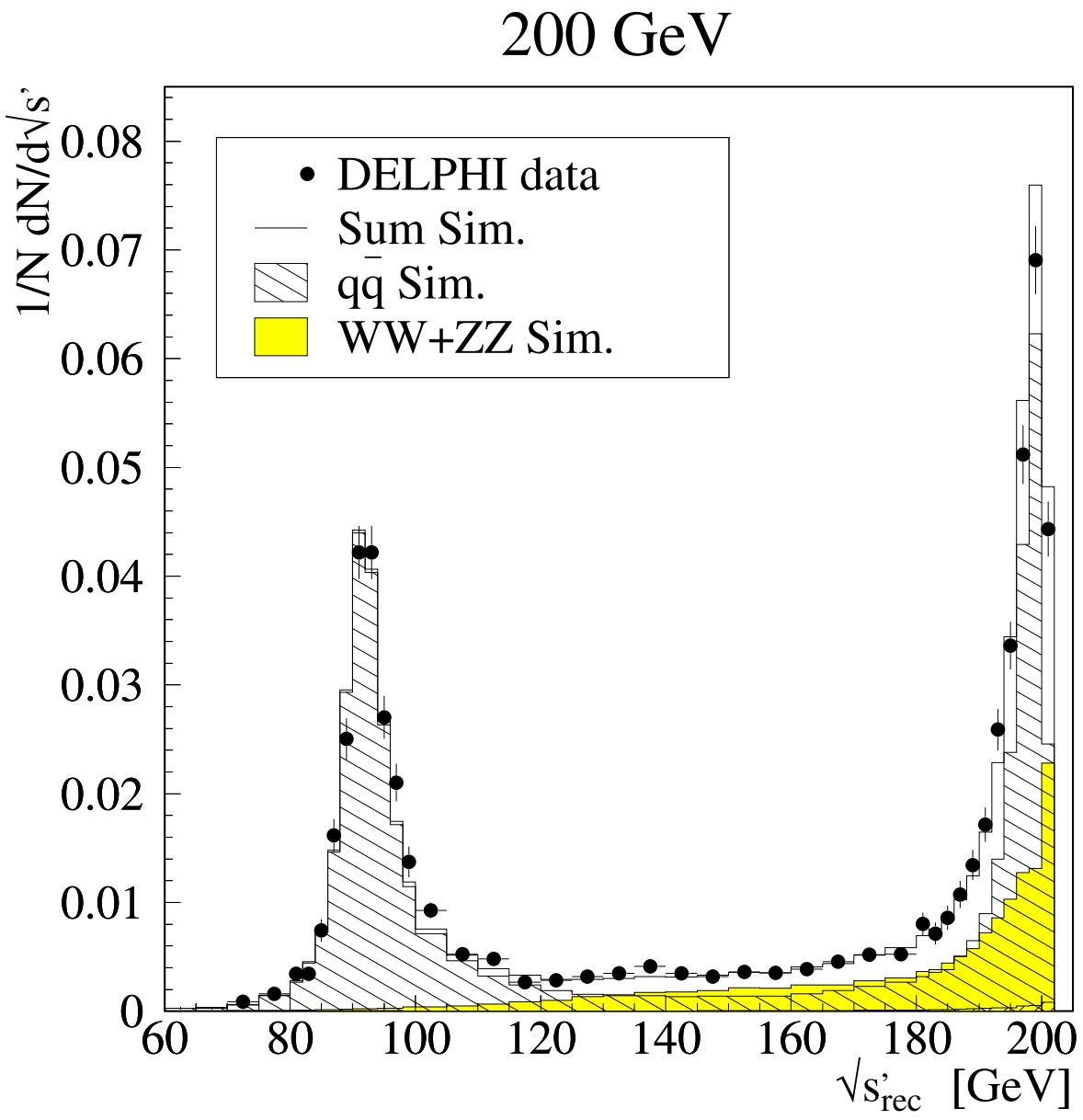

**Fig. 12.1.1** The figure shows the distribution of reconstructed effective cms energies $\sqrt{s'_{rec}}$ in hadronic final states in $e^+e^-$ collisions at $\sqrt{s} = 200$ GeV. The data are compared with simulations of hadronic final states mediated by a single $(Z/\gamma)^*$ ($q\bar{q}$ Sim.) and $W^+W^-$ or ZZ pair production (WW+ZZ Sim.) [3615].

The data taken at $\sqrt{s} \geq 2m_W$ (LEP 2) include an increasing fraction of so-called "4-fermion" final states including quarks. These 4-fermion final states are dominated by $W^+W^-$ pair production in the all-hadronic or lepton+jets channel depending on the decays of the W bosons. The di-lepton channel is after a hadronic preselection a rather small background. The LEP collaborations developed sophisticated selections for the $W^+W^-$ pairs for the precise measurements of W boson properties designed to reject "2-fermion" final states with quarks $e^+e^- \to (Z\gamma)^* \to q\bar{q} \to$ hadrons [3614]. These results are then basis of selections for hadronic final states produced via a $(Z/\gamma)^*$ at high energy. The remaining 4-fermion background in the data increases with $\sqrt{s}$ to about 10% at the highest LEP 2 energies but contributes mostly in regions of the distributions dominated by multi-jet topologies, see e.g. [3615]. Figure 12.1.1 shows the distribution of $\sqrt{s'_{rec}}$ observed for hadronic final states at $\sqrt{s} = 200$ GeV by DELPHI [3615]. The peak at $m_Z \simeq 91.2$ GeV is due to hadronic Z decays recoiling against photon ISR. The analysis imposes a cut on $\sqrt{s'_{rec}}$ to select the peak near the nominal $\sqrt{s} = 200$ GeV. The yellow shaded area shows the simulated background contribution of $W^+W^-$ and ZZ final states with hadrons.

### 12.1.1 Gluon properties

The gluon was established as one of the elementary particles of the SM by the PETRA experiments, see section 2.2. QCD requires for its gauge bosons that they have spin-1, and that they carry colour charge themselves manifesting in the three- and four-gluon vertices of the QCD Lagrangian.

The phenomenological analysis of [3616] based on measurements of jet axes in Υ decays to three gluons provided evidence for the spin-1 assignment. The method of calculating QCD predictions for spin-0 and spin-1 gluons of [3616] was the basis of an analysis by OPAL using the energy distribution of the 2nd jet after energy ordering in hadronic Z decays to three jets [3613]. The 2nd jet energy distribution after correction for experimental and hadronisation effects was in good agreement with a NLO QCD prediction while a MC based LO calculation with scalar gluons showed an estimated $\chi^2/dof = 44/14$. This is clearly well above requirements for a discovery. A similar study is discussed in [3617].

The search strategy for directly observable effects at LEP of the three gluon vertex of QCD was discussed in [3618], but convincing results could only be obtained after NLO calculations for the angular correlations between four jets in hadronic Z decays became available [3619]. The QCD predictions at NLO decompose into contributions proportional to (products of) the color factors $C_F$, $C_F C_A$, $C_F C_F$ and $C_F N_F T_F$, and two of them can be determined together with the strong coupling $\alpha_S(M_Z^2)$. The analyses by OPAL and ALEPH [3620, 3621] determine $C_A$ and $C_F$ corresponding to the contributions of three-gluon or quark-gluon vertices to the NLO predictions. The contribution of the three-gluon vertex proportional to $C_A$ is clearly observed. Since the result for the second color factor product can be recast as $C_F$ in these analyses, the color charge of quarks at the strength required by QCD is observed as well.

The analysis of event shape observables Thrust and C-parameter (see below for details) at several cms energy points from re-analysed JADE (at PETRA) data and OPAL data is based on the same decomposition of the NLO QCD prediction and also results in a clear observation of the three-gluon vertex contribution [3622]. A combination of these and other results for determinations of the color factors is discussed in [3623]. Figure 12.1.2 shows a summary of the results for $C_A$ and $C_F$ from 4-jet angular correlations, event shapes, and other analyses [3623].

The properties of quarks in the SM such as their spin-1/2 assignment and their electric charges have been studied at LEP and earlier collider experiments [105, 3624, 3625]. The other quark property directly connected with QCD is their mass, which will be discussed below in section 12.1.4.

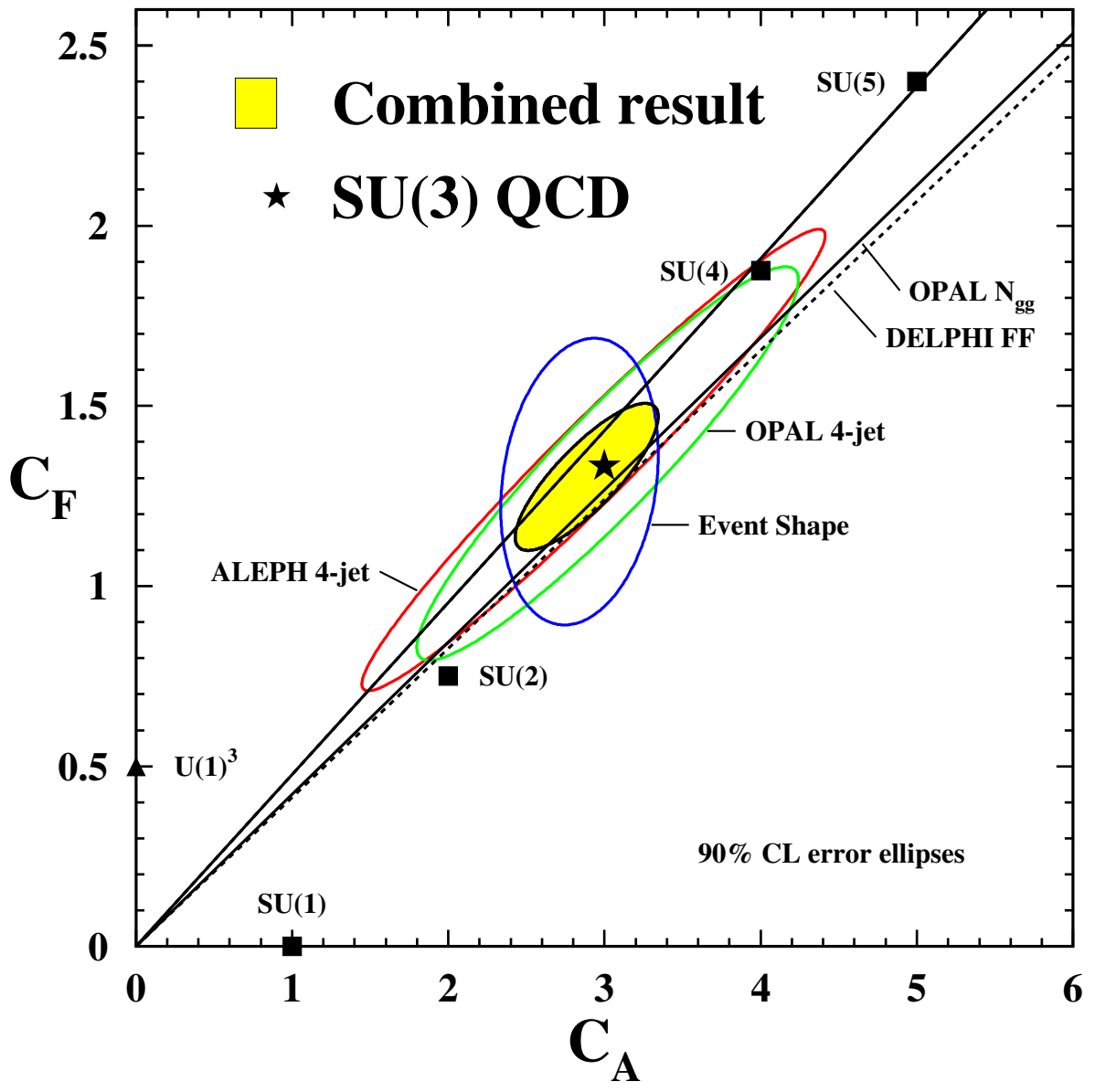

**Fig. 12.1.2** The figure shows results for the color factors $C_A$ and $C_F$ from various analyses as indicated [3623].

### 12.1.2 Jets and event shapes

Jet and event shape observables have been designed to study properties of hadronic final states at colliders. The aim generally is to classify hadronic final states according to their topology by introducing an additional energy scale. E.g. for clustering hadronic final states in $e^+e^-$ annihilation with the JADE algorithm [3626] $m_{ij}^2 = 2E_iE_j(1-\cos\theta_{ij})$ is the distance between two objects $i$ and $j$ with energies $E_i$ and $E_j$. At each iteration the pair $ij$ with the smallest distance $m_{ij}$ is merged by adding the pair's 4-vectors[108]. One can introduce the scaled quantity $y_{cut} = m_{cut}^2/s$ and count how many events have three jets when the clustering is stopped at $y_{cut}$. Alternatively, the value of $y_{23} = m_{23}^2/s$ where in each event the clustering goes from three to two jets can be used to classify events [3627]. In the first case jet rates are studied and in the second an event shape observable is used. The Thrust observable $T = \max_{\vec{n}} \sum_i |\vec{p}_i \cdot \vec{n}| / \sum_i |\vec{p}_i|$, where $i$ runs over all particles in the hadronic final state, $\vec{p}_i$ are the particle 3-momenta, and $\vec{n}$ is a unit vector, in a similar way defines after optimisation an energy scale given by the sum of projections of all 3-momenta on the thrust axis $\vec{n}_T$.

[108] This is the E-scheme, other merging schemes exist.

The value of an event shape observable is the classifier which can distinguish between e.g. collimated 2-jet like events and broader 3-jet (or multi-jet) like events. Their distributions reflect the proportion of 2-jet like vs. 3-jet or multi-jet like events in the data in a similar way as the fraction of 3-jet events at a fixed value of $y_{cut}$.

As discussed by Dokshitzer in section 2.3, it is the property of infrared-collinear safety which allows for stable prediction by perturbative QCD (pQCD) and thus a meaningful comparison between experimental observations and pQCD predictions. However, before a successful quantitative comparison of experiment and theory can be made the transition from the partons of pQCD calculations to the observed hadrons in the detector (hadronisation) must be accounted for. If there was a major redistribution of 4-momenta between partons and hadrons in a given final state due to hadronisation then a comparison of pQCD predictions with data would be highly problematic. Turning this argument around we must have a hadronisation process which is local in phase space. This is discussed as "local parton hadron duality" (LPHD) by Dokshitzer in section 2.3. Experimental evidence for the LPHD collected by the LEP experiments and previous studies is discussed below.

Figure 12.1.3 (left) shows as an example the measurements by OPAL of the event shape observable $y_{23}^D$ at cms energies $\sqrt{s}$ = 91.2, 133, 177 and 197 GeV. The cms energies are weighted averages of combined LEP runs with similar cms energies. The observable $y_{23}^D$ is the value of the jet distance in the Durham algorithm [168] $y = 2\min(E_i, E_J)^2(1 - \cos\theta_{ij})/s$ where the number of jets changes from three to two. Figure 12.1.3 (right) shows measurements by ALEPH [3628] of n-jet, $n = 1, \ldots, 6+$ production fractions using the Durham algorithm. These data show that at LEP hadronic final states with complex jet topologies can be measured well.

The reasonably successful comparisons of the data with simulations by the Monte Carlo event generators PYTHIA, HERWIG and ARIADNE validate the experimental corrections derived using these simulations after passing them through the simulations of the detectors. Furthermore, they pave the way for using these simulations to derive the hadronisation corrections needed to compare pQCD predictions with the data. The final LEP measurements and their comparison to the then relevant NLO+NLLA QCD predictions and determinations of $\alpha_S(m_Z)$ are discussed in [3623]. Improved determinations of $\alpha_S(m_Z)$ using NNLO QCD predictions combined with resummed NLLA calculation appeared soon after the NNLO predictions became available [3630–3633]

The QCD analyses of some jet rates and event shape distributions (starting at 3-jet final states) from LEP and previous $e^+e^-$ experiments today has reached percent level precision using pQCD predictions at NNLO combined with resummation up to N3LL. For example in [280] distributions of Thrust at $\sqrt{s}$ = 35 to 200 GeV are analysed in a global fit based on NNLO+N3LL QCD predictions[109]. The hadronisation corrections are applied using an analytic model integrated into the prediction. The final result is $\alpha_S(m_Z) = 0.1135 \pm 0.0011$ and has a relative uncertainty of 1%. A similar measurement using the C-parameter is [281], the energy-energy correlation EEC was analysed in NNLO+NNLL accuracy and the 2-jet rate with the Durham algorithm was studied with N3LO+NNLL predictions [3634].

Limitations for the ultimate accuracy of these studies are currently the uncertainties connected with hadronisation corrections, see e.g. [3635] for a recent study. An early study [3628] based on event shapes at all LEP energies and NLO+NLL pQCD found differences in $\alpha_S(m_Z)$ of about 10% between results using MC simulations or an analytic model to derive hadronisation corrections. These differences became smaller with more complete QCD predictions such as NNLO+NNLL or NNLO+N3LL. They also tend to reduce when MC simulations with NLO calculations matched to the parton shower are used. In both cases a larger fraction of the prediction is contributed by pQCD and thus only a smaller difference w.r.t. the data is left to be covered by hadronisation corrections. New studies show that the hadronisation corrections in an improved analytic model depend on the event shape value [3636], in contrast with the analytic models used so far.

The analyses of final states with four or more jets are based on the accurate measurements of multi-jet rates and corresponding event shape distributions at LEP. Similar to NLO QCD predictions for angular correlations in 4-jet final states also NLO predictions for 4-jet rates became possible [3637]. It is important to realise that for 4-jet final states the NLO QCD prediction is $O(\alpha_S^2) + O(\alpha_S^3)$ which implies a sensitivity to $\alpha_S$ larger by about a factor of 2 compared with a prediction for 3-jet final states. The higher sensitivity can compensate for the larger experimental uncertainties of the 4-jet measurements w.r.t. 3-jet measurements. The corresponding analyses with LEP data are [3615, 3638] while [3639] is a study based on re-analysed data from JADE at PETRA.

Automated NLO QCD calculations allowed predictions for 5-jet observables [3640] and the corresponding

[109] The exact power counting is explained in [280].

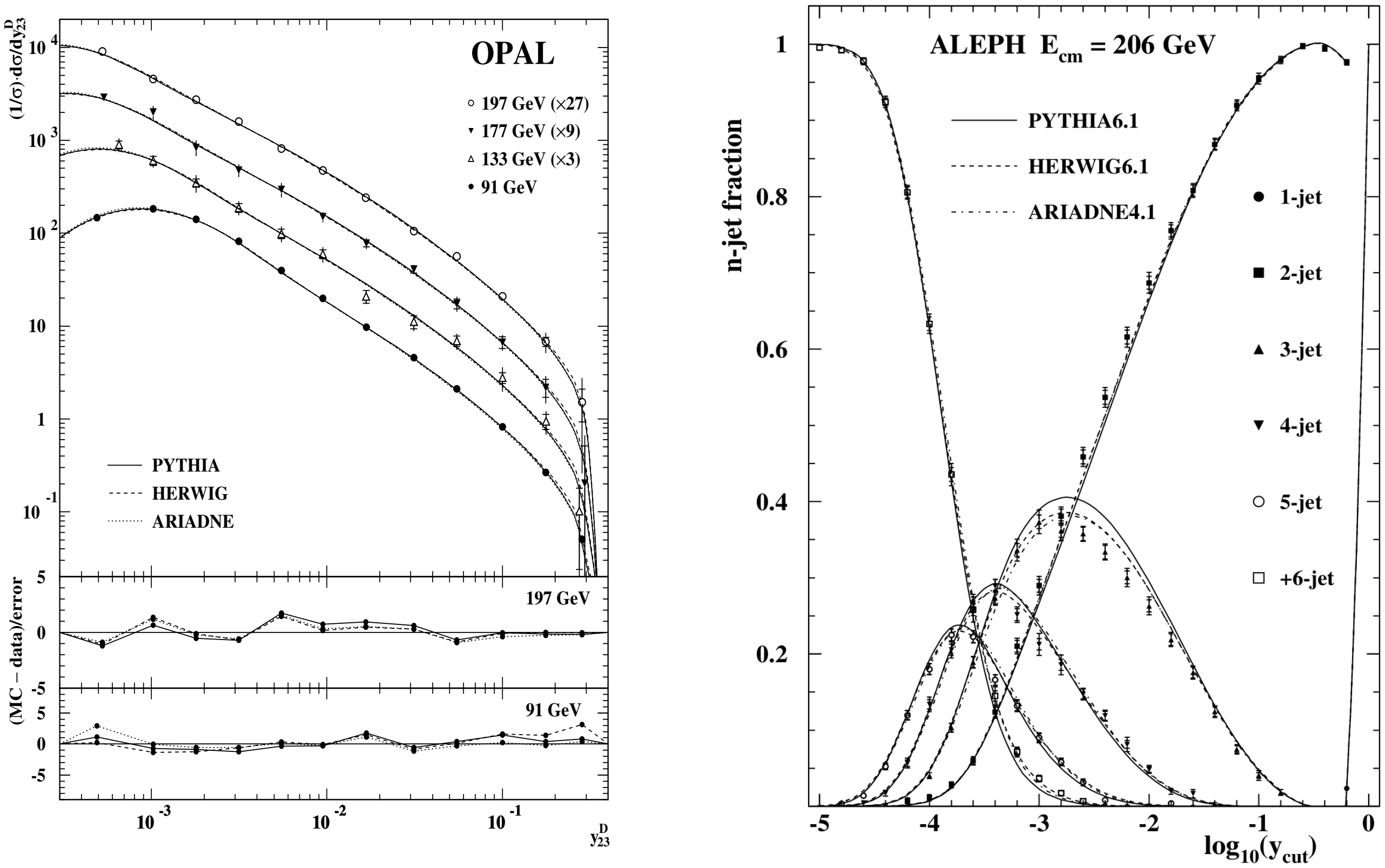

**Fig. 12.1.3** (left) The figure shows measurements of the event shape observable $y_{23}^D$ by OPAL at average cms energies as indicated. The measurements are corrected for experimental effects and are compared with simulations as indicated [3629]. (right) The figure displays measurements of n-jet production fractions as a function of $y_{cut}$ using the Durham algorithm by ALEPH at $\sqrt{s} = 206$ GeV. The measurements are compared with simulations [3628].

determination of $\alpha_S(m_Z)$. By the same argument as above the sensitivity to $\alpha_S(m_Z)$ is enhanced w.r.t. 3-jet observables which compensates for larger measurement uncertainties.

The review of measurements of $\alpha_S(Q)$ in section 3 shows clearly that the strong coupling strength decreases with increasing energy scale of the process, i.e. asymptotic freedom. Here we discuss direct experimental evidence without performing measurements of $\alpha_S$. Figure 12.1.3 (left) shows distributions of $y_{23}^D$ measured at cms energies from 91 to 197 GeV and a change in the distribution is clearly visible. A more direct way to observe a change of the strong coupling strength with the energy scale of the process $Q = \sqrt{s}$ is to use inclusive observables such as jet production rates at a fixed value of $y_{cut}$ or moments of event shape observables.

In QCD in LO the prediction for the mean value of e.g. the Thrust $1-T$ distribution is $\langle 1-T\rangle(Q) = \alpha_S(Q^2)A_{1-T}$ while the running coupling follows $\alpha_S(Q^2) = \alpha_S(\mu^2)/(1+\alpha_S(\mu^2)\beta_0\ln(x_\mu^2))$, $\beta_0 = (11C_A - 4T_F N_F)/(12\pi)$, $x_\mu = Q/\mu$, $\mu$ is the renormalisation scale. This implies $1/\langle 1-T\rangle \sim \ln Q$ at LO with $\mathrm{O}(\alpha_S^2)$ corrections. Figure 12.1.4 displays data from DELPHI and lower energy experiments for $1/\langle 1-T\rangle$ as a function of $Q$ on a logarithmic scale confirming the QCD prediction for the running coupling as measured by $\langle 1-T\rangle$. Hadronisation corrections to $\langle 1-T\rangle$ are predicted using simulations to only change the logarithmic slope, see e.g. [3641]. Earlier studies using JADE (at PETRA) data for 3-jet rates using the JADE algorithm as a function of cms energy had already proven the running strong coupling at the 4-$\sigma$ level [3642].

### 12.1.3 Fragmentation

The term fragmentation refers to measuring and predicting properties of the hadrons produced in hadronic final states. In contrast, hadronisation refers to modifications to observables derived from hadronic final states such as event shapes or jet rates. In studies of fragmentation of hadrons the energies or momentum components w.r.t. an event orientation or jet axis, or their multiplicity, are studied.

The scaled momentum fraction of a hadron with momentum $p$ is defined as $x = 2p/Q$. One expects in the quark-parton model, i.e. in the absence of strong interactions of the partons, that the $x$-spectra are independent of $\sqrt{s}$. This is analogous to the prediction of scaling for $x_{Bj}$ in lepton-hadron DIS, i.e. that distributions of $x_{Bj}$ are independent of the 4-momentum transfer $Q^2$ of the DIS process. Scaling violations are then due to scale-dependent strong interactions of the partons. Figure 12.1.5 shows as an example the ratio of measurements of $x$-spectra measured by ALEPH on the Z peak to corresponding measurements by TASSO (at PETRA) measured at $\sqrt{s}=22$ GeV [3643]. The scaling violations are clearly visible.

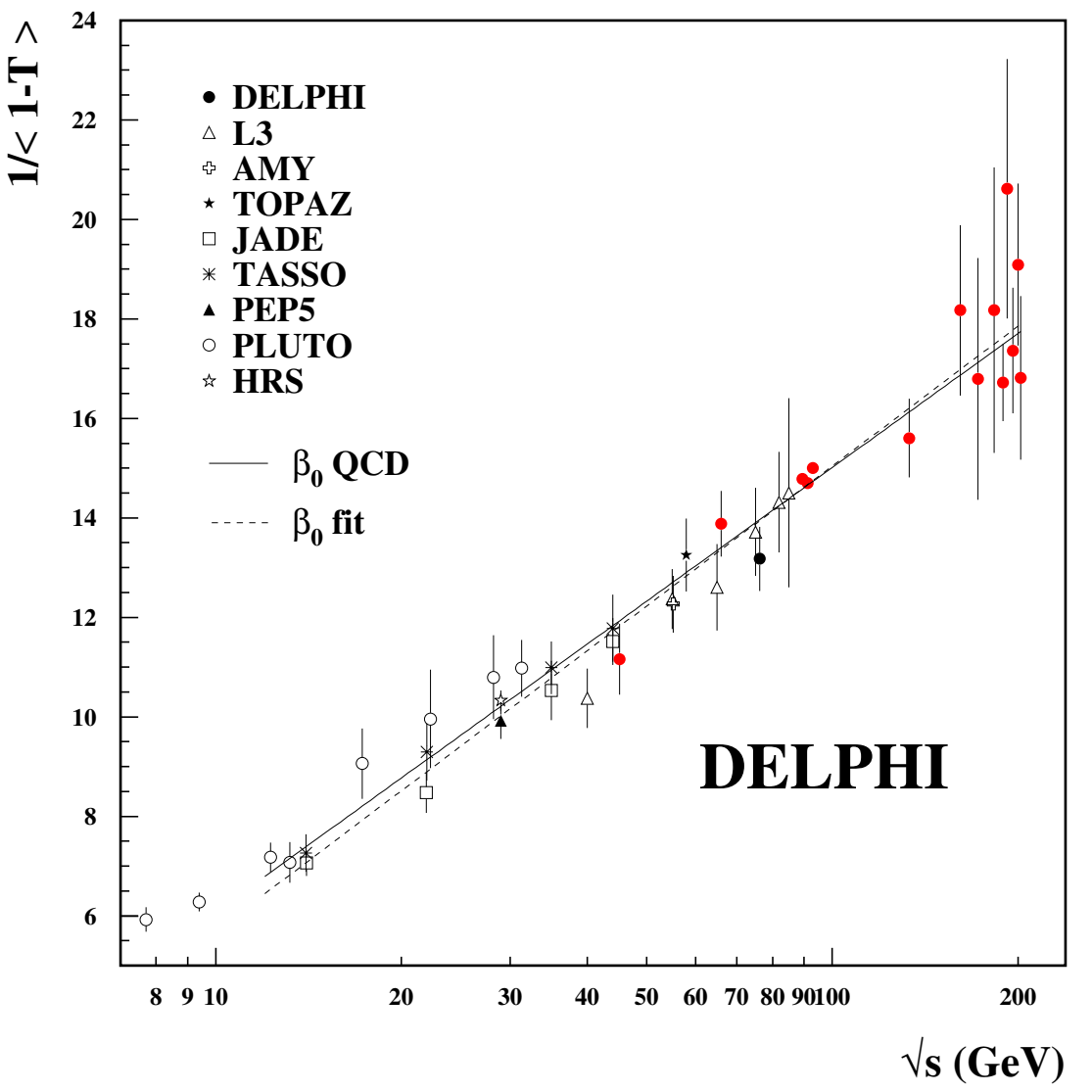

**Fig. 12.1.4** The figure shows measurements of $1/\langle 1-T\rangle$ as a function of $\sqrt{s}=Q$ on a logarithmic scale by DELPHI and lower energy experiments. The lines show a NLO QCD prediction and fit by DELPHI [3641].

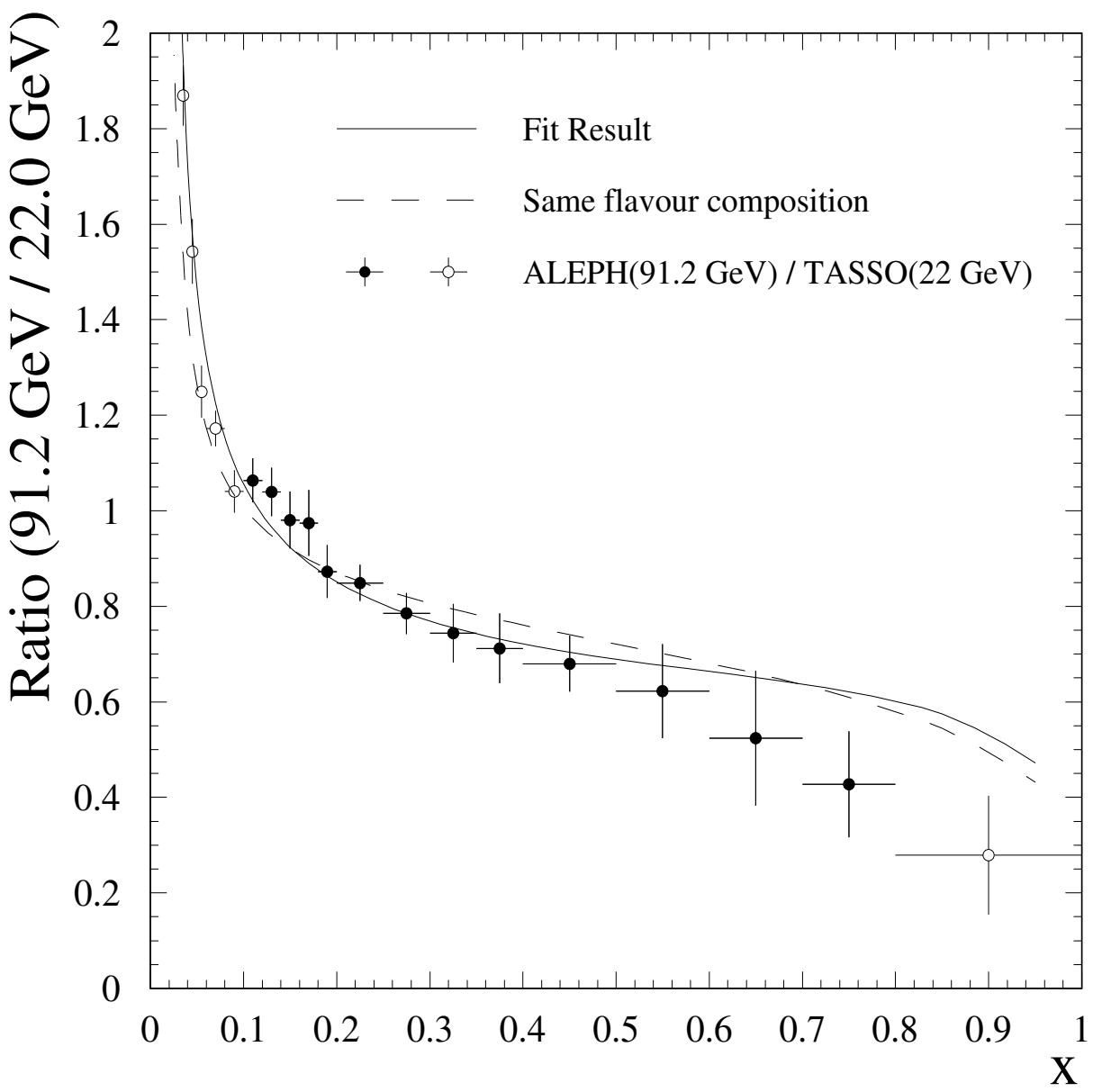

**Fig. 12.1.5** The figure shows the ratio of the scaled momentum spectra $1/\sigma_h d\sigma_h/dx$ of charged particles measured by ALEPH at $\sqrt{s}\simeq 91.2$ GeV to data from TASSO measured at $\sqrt{s}=$ 22 GeV [3643].

The QCD analysis of scaling violations of scaled momentum distributions measured at different cms energies is the $e^+e^-$ analog of the analysis of structure functions $F_2(Q^2,x_{Bj})$ in lepton-hadron DIS. The scaled momentum distribution is described by

$$\frac{1}{\sigma_h}\frac{d\sigma_h}{dx}=\int_0^1\sum_f C_f(z,\alpha_S(\mu),x_\mu)D_f(\frac{x}{z},\mu)\frac{dz}{z} \tag{12.1.1}$$

with the flavour index $f=u,d,s,c,b,g$. The $C_f$ are coefficient functions known in NNLO QCD, and the $D_f$ are non-perturbative fragmentation functions. The $D_f$ correspond to the probability to obtain a hadron with momentum fraction $x$ from a parton $f$ analogous to the parton density functions (PDF) of DIS. The rate of change with changing momentum scale $\mu$ of the $D_f$ is described by the DGLAP equations, see 2.3. A first NNLO framework for the analysis of scaled momentum distributions in $e^+e^-$ annihilation to hadrons is [3644].

It is interesting to focus on low momentum hadrons. To this end the variable $\xi=\ln(1/x)$ is introduced. The majority of hadrons is produced at low values of $x$ and by transforming to $\xi$ their properties can be studied in more detail. As an example figure 12.1.6 shows measurements of $\xi$ for charged hadrons at LEP by OPAL and also from previous experiments at lower energies [3645]. The distributions show a maximum and drop quickly towards small $\xi$ corresponding to large hadron momenta. At large $\xi$, i.e. for small momenta, the distributions fall off faster than expected from the kinematic limits from hadron masses.

This can be explained by destructive interference of multiple soft gluon radiation in the parton shower, often named soft gluon coherence. Under the assumption of LPHD the production of soft hadrons is driven by the production of soft gluons from the parton shower. The "QCD Chudakov effect" means that soft gluons cannot resolve the individual parton color charges and instead the smaller color charge before branchings is relevant. Based on these ideas detailed pQCD predictions for multiple soft gluon radiation are calculated. For figure 12.1.6 such predictions [3646] are shown by the solid and dashed lines, where the solid lines are fitted to the data and the dashed lines are extrapolations. The extrapolated QCD predictions at small $\xi$ (large $x$) are not expected to be a good approximation while at large $\xi$ (small $x$) the data are well described.

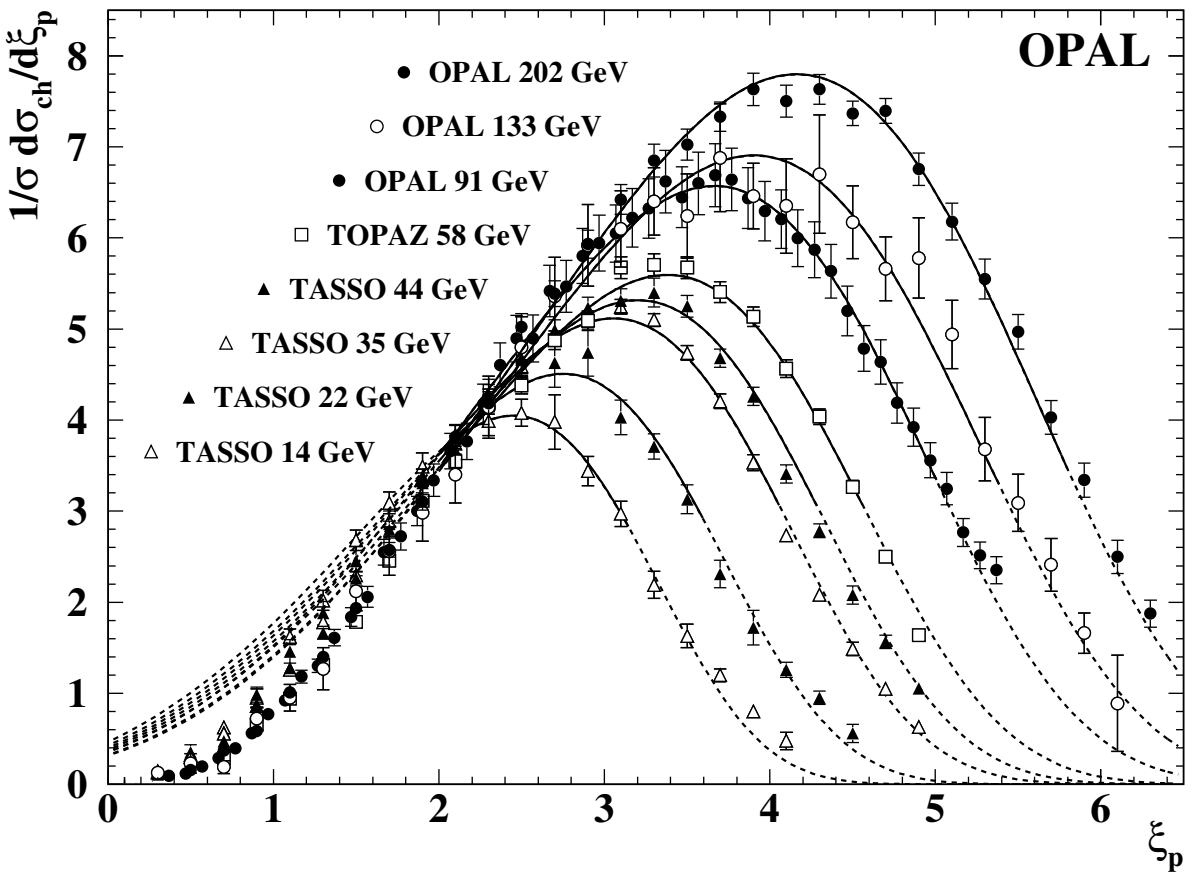

**Fig. 12.1.6** The figure shows the spectra of $\xi = \ln(1/x)$ measured by OPAL, TOPAZ and TASSO [3645]. The data are compared with fitted QCD predictions, see text for details.

The evolution of the peak position with cms energy extracted from the fits also follows the pQCD prediction, see e.g. [3645]. These measurements provide convincing experimental confirmation of the LPHD and the corresponding pQCD calculations. A recent analysis of the $\xi$ spectra measured in $e^+e^-$ annihilation and other processes including higher order corrections is [3647].

The interpretation of [3648], based on simulations with and without soft gluon interference (coherence) effects, that the data of scaled momentum spectra do not provide evidence for coherence has been discussed in [3649]. There it was pointed out that the hadronisation models of the simulation programs will compensate for the lack of coherence effects to still give a reasonable description of the data. The confirmation of the LPHD lies in the successful comparison of the corresponding QCD calculations with the data involving only two free normalisation and scale parameters.

The QCD parton shower picture, i.e. the idea of high-momentum partons radiating many times a gluon, and also gluons producing a $q\bar{q}$ pair, is the basis of the simulation programs, because it allows implementations as iterative probabilistic branchings. The implementations of the parton shower picture are approximations correctly summing leading logarithmic terms (LLA). In the LLA the soft gluon interference effects correspond to the angular ordering phenomenon: a subsequent parton branching must occur at branching angles smaller than the previous one. There are limitations to the angular ordering approximation for less inclusive observables [3650], in particular some which are used for tuning (optimisation of agreement with data) of the simulation programs.

The legacy of LEP in this area is the wealth of precise data on event shape observables, jet production, spectra of inclusive and identified hadrons, and multiplicities which can in many cases be interpreted with little ambiguities. These data are to a large part the basis for parameter settings of the popular simulation programs used in our field and in particular at the LHC [3559, 3576, 3651].

The topic of colour reconnection (CR) concerns possible changes to hadronisation effects if several colour singlet sources are produced in a collision. The question is: do the final partons after the parton showers of different colour singlet sources form hadrons together or not. At LEP 2 the production of $e^+e^- \to W^+W^- \to$ *hadrons* final states was an important contribution to the LEP 2 measurements of the mass and other properties of the W boson [3614]. The modeling uncertainties of CR effects gave rise to significant systematic errors on the W boson mass and width, after measurements of particle flow between the four jets of the two hadronic W decays were used to constrain different CR models. New models for CR were discussed in [3574] and compared with the LEP 2 measurements. CR also affects measurements of the top quark mass due to the intermediate colour singlet W boson in the top quark decay [3652] and due to interactions of proton remnants (multi parton interactions MPI) in pp collisions. Recent measurements from LHC take this into account [476, 3653]. The CR model with the biggest impact on the results of [3653] is also the only one in tension with LEP 2 data in [3574]. This shows that the LEP data can still help to constrain CR models.

### 12.1.4 Heavy quarks

In QCD with massless quarks the coupling constant is the only free parameter. Asymptotic freedom of the running strong coupling is one of the defining features of QCD and is well confirmed by experiments [3654] since LEP results contributed. Quark masses are also free parameters of the theory and subject to similar phenomena as asymptotic freedom for the strong coupling. The quark masses are predicted to depend on the energy scale of the process through so-called mass anomalous dimensions, the quark mass analogous of the beta-function.

The two main phenomenological predictions are firstly that effective quark mass values decrease with energy scale of the interaction, or, asymptotic freedom for quark masses, see e.g. [3655] for a review. The second prediction is the suppression of gluon radiation from massive quarks with angle $\Theta < \Theta_0 = m/E$, where $m$ is the heavy quark mass and $E$ the heavy quark energy. This is referred to as the "dead-cone" effect of QCD [3656].

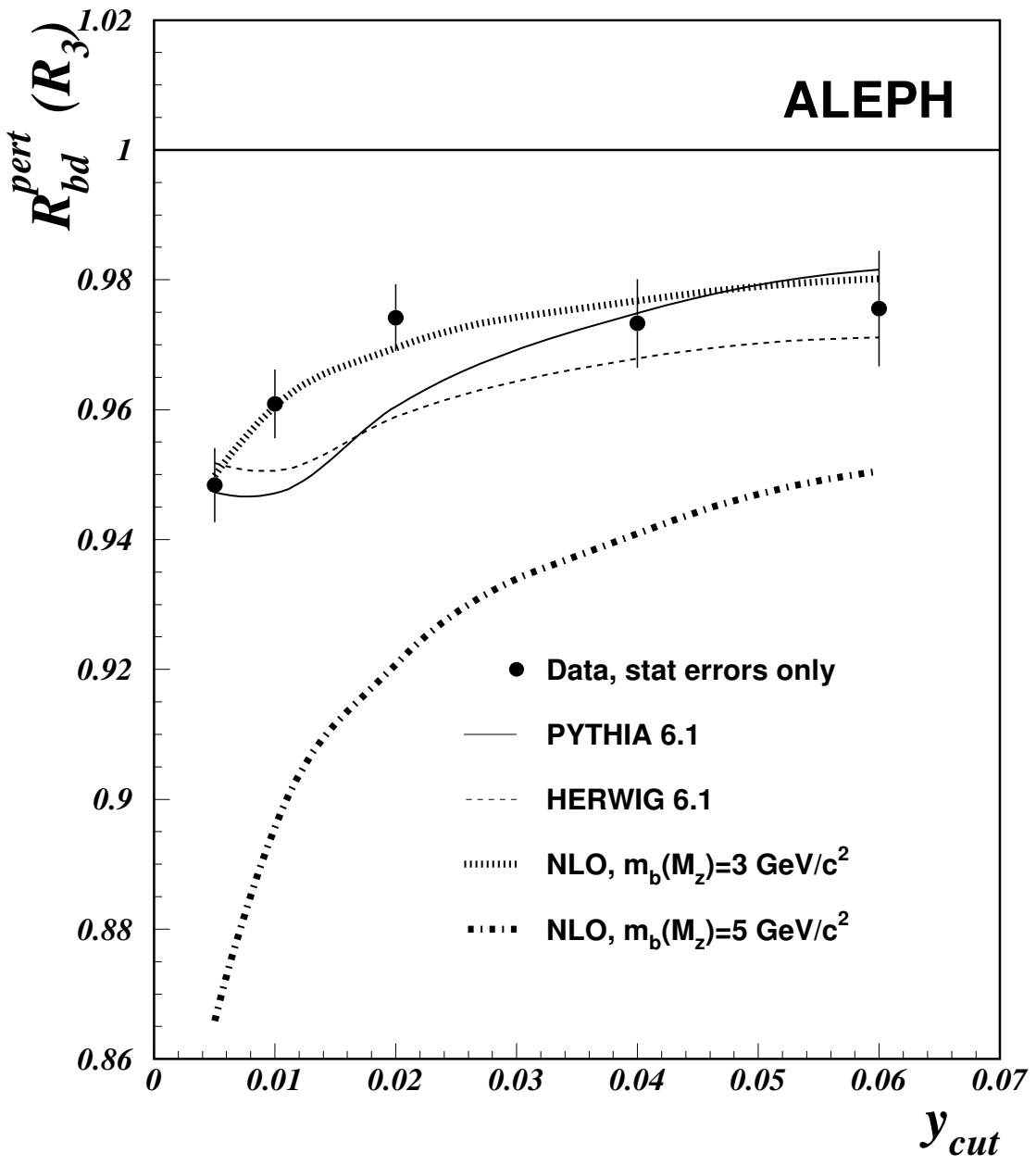

**Fig. 12.1.7** The diagram shows data for $B_3(y_{cut})$ by ALEPH corrected for experimental and hadronisation effects using the Durham algorithm. The lines show NLO QCD predictions for $m_b(M_Z)$ values as indicated, as well as predictions from simulations [3658].

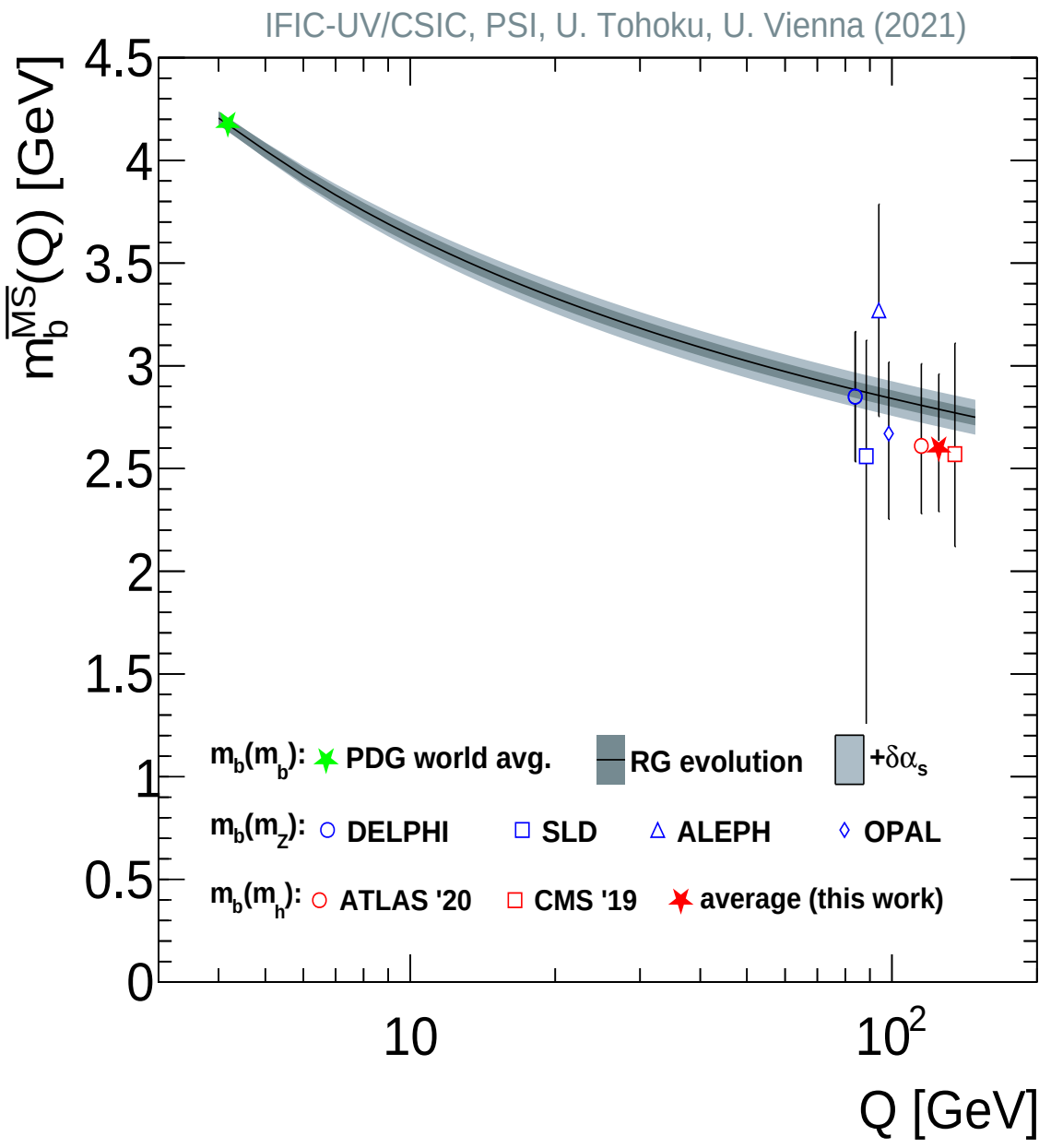

**Fig. 12.1.8** The figure shows determinations of $m_b(m_b)$, $m_b(M_Z)$ and $m_b(M_H)$ together with the QCD prediction for the running $m_b(Q)$ [3659].

The first prediction of running quark masses was studied at LEP using large samples of $O(10^6)$ hadronic Z decays with b-tags. Mass effects can be enhanced for observables like the 3-jet rate $R_3(y_{cut})$ due to their additional energy scale $y_{cut}$ [3657]. In order to reduce common experimental uncertainties a double ratio $B_3 = R_3^b/R_3^l$ is defined, with $R_3^{b(l)}$ the 3-jet rate in Z decays to $b$ (light) quarks. Figure 12.1.7 shows data for $B_3$ from ALEPH compared with NLO QCD predictions for values of the running b quark mass in the $\overline{\text{MS}}$ scheme $m_b(M_Z) = 3$ or 5 GeV [3658]. The data are consistent with the lower value of $m_b(M_Z)$.

The analyses by ALEPH, DELPHI, OPAL and SLD are summarised in [3623, 3659] with $m_b(M_Z) = 2.82 \pm 0.28$ GeV [3659]. With $m_b(m_b) = 4.18^{+0.03}_{-0.02}$ [476] a running b quark mass is observed with a significance of more than four standard deviations. The analysis [3659] adds a determination of $m_b(M_H)$ from measurements of the branching ratio of the Higgs boson to b quarks by the LHC experiments ATLAS and CMS assuming the Yukawa coupling of b quarks at its SM value. Figure 12.1.8 presents results for $m_b(m_b)$, $m_b(M_Z)$ and $m_b(M_H)$ together with the QCD prediction for the running $m_b(Q)$ [3659]. There is good agreement between the measurements and the QCD prediction.

The dead cone effect is not straightforward to study at LEP or other colliders. For example for b-jets from on-peak Z decays at LEP the dead cone angle is expected to be $\Theta_0 \simeq 2m_b/m_Z \simeq 0.1$ which is well inside typical jet energy profiles in hadronic Z decays [3660]. A recent analysis by ALICE has found evidence for reduced particle production inside angular regions consistent with the dead cone for charm tagged jets produced in pp collisions at the LHC [176]. The key to this observation was reversing a sequential jet clustering history using an angular distance definition[110] which enforces angular ordering by construction.

Predictions for phenomenology of the dead cone effect at LEP concentrate on multiple soft gluon production and thus on particle spectra or multiplicities [3656, 3661]. The so-called "leading particle effect" refers to large and mass dependent average scaled momenta of heavy hadrons (c or b). The leading particle effect is derived from pQCD as a direct consequence of the dead cone and shown to be consistent with data from LEP and previous $e^+e^-$ colliders [153].

The particle multiplicity in Z decays to b or light (u, d, s) quarks is sensitive to the dead cone effect due to its impact on soft gluon radiation, which is directly related to particle production via the LPHD. The pQCD prediction in the MLLA for the charged particle multiplicity difference in hadronic Z decays to b or light quarks is $\delta_{bl} = 4.4 \pm 0.4$ [3662]. A different model for $\delta_{bl}$ without dead cone contributions predicts a fast decrease with cms energy $\sqrt{s}$. The predictions for $\delta_{bl}$ and measurements by LEP experiments and previous experiments at different $\sqrt{s}$ are shown in figure 12.1.9 [3662]. The blue band corresponding to the QCD dead cone prediction is in agreement with the data within theoretical and experimental uncertainties. The alternative model is excluded by the high energy LEP 2 measurements at $\sqrt{s} \geq 183$ GeV with an estimated $\chi^2/dof \simeq 100/11$. The hypothesis that $\delta_{bl} \to 0$ for large $\sqrt{s}$ leads to an estimated $\chi^2/dof \simeq 43/11$ and is thus also clearly excluded.

[110] The Cambridge/Aachen (C/A) algorithm.

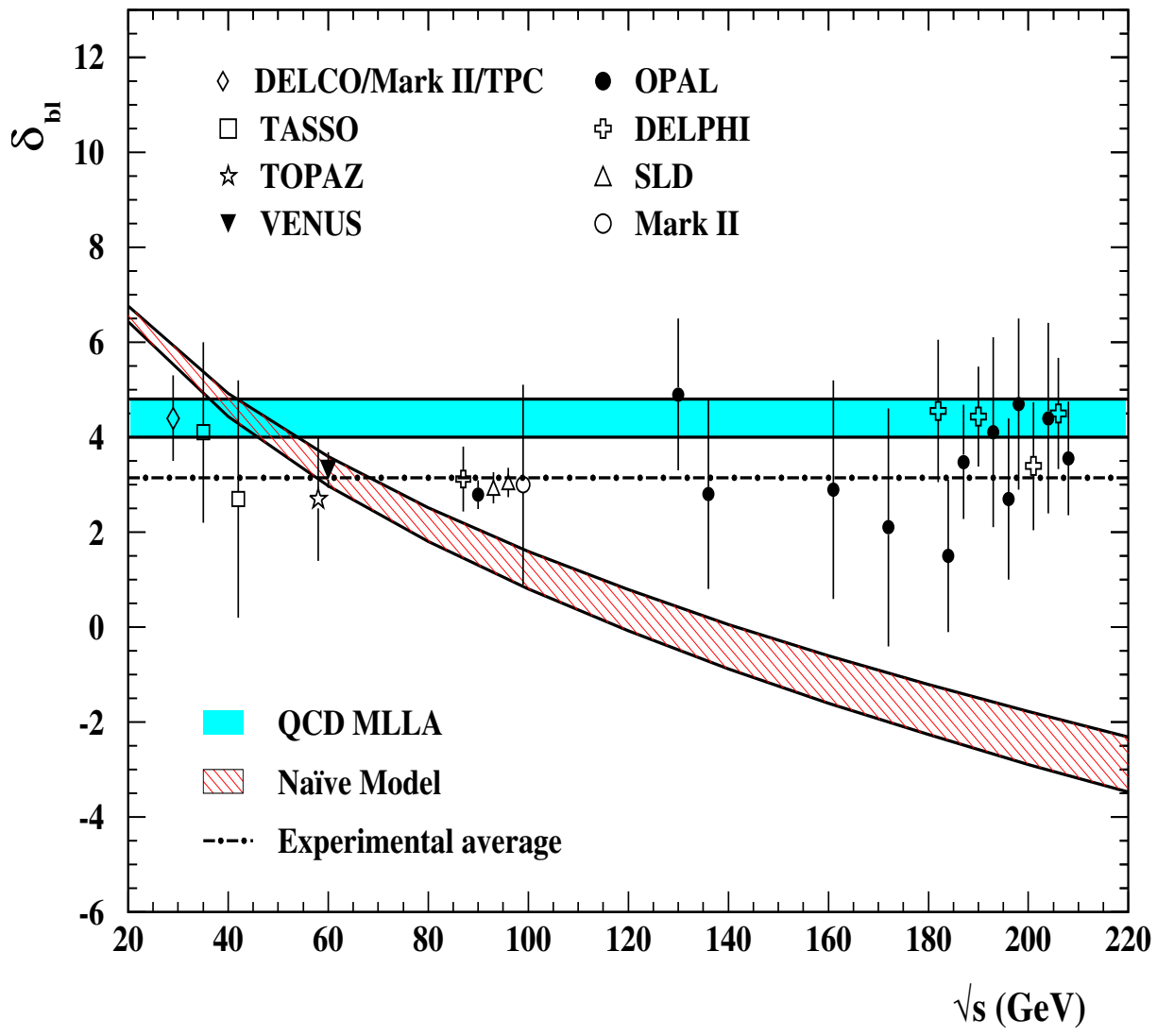

**Fig. 12.1.9** The figure presents measurements of $\delta_{bl}$ compared with QCD predictions and an alternative model as indicated [3662].

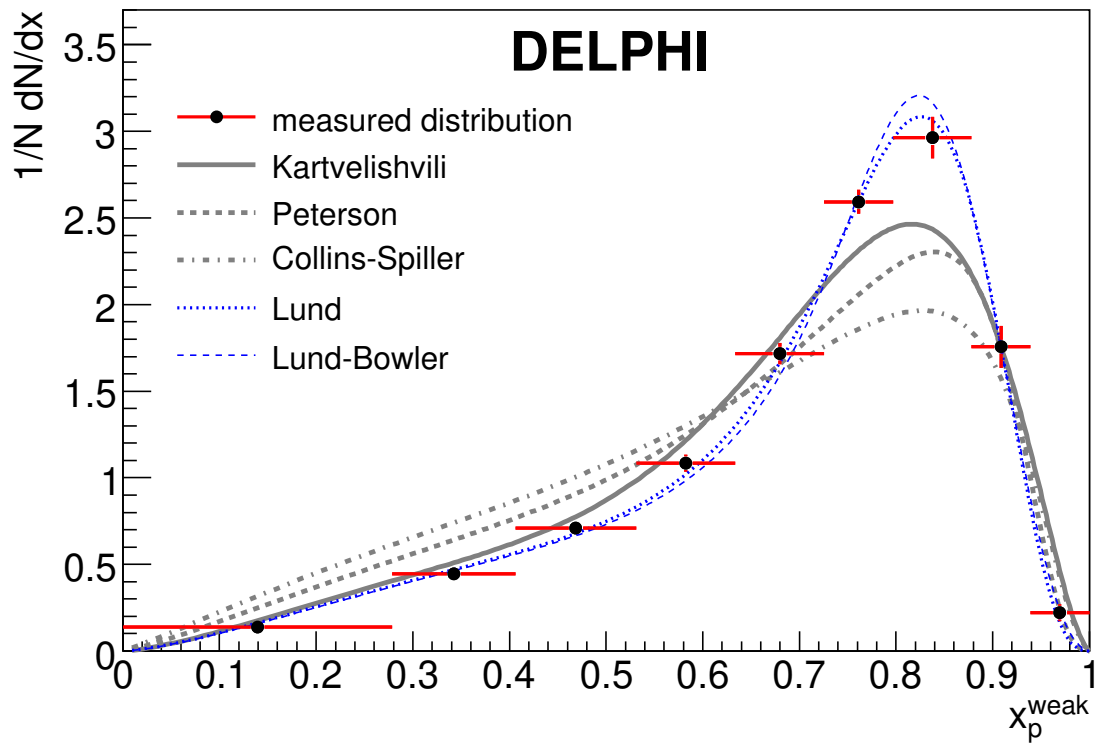

**Fig. 12.1.10** The figure shows the b quark to B hadron fragmentation function for weak B decays. The lines display predictions by simulations with a fixed perturbative component and different models for the fragmentation functions [3663].

Another example of precision measurements in the heavy flavour sector is the b quark to hadron fragmentation function. The measurement by DELPHI is shown in figure 12.1.10 [3663]. The quantity $x_p^{weak}$ refers to the scaled momentum of the B hadron reconstructed from its weak decay. In this way possible preceeding strong decays of excited B hadrons are accounted for. In the figure the data are compared with several models for the fragmentation functions folded with a fixed perturbative component. The data can clearly separate the different models. Recent parameter optimisations of e.g. the PYTHIA simulation take these results into account [3651].

### 12.1.5 Zedometry and hadronic $\tau$ decays

The EWPOs measured by the LEP experiments and SLD are the main legacy of the LEP program. The EWPOs are also a valuable legacy for the understanding and experimental verification of QCD. All EWPOs connected with quarks will have SM predictions with QCD corrections reflecting gluon radiation. Corrections to pure electroweak processes involving quarks scale typically like $1 + C\alpha_S(m_Z)/\pi$, where $C$ is a process specific constant, and are thus expected to modify electroweak EWPO predictions by a few %.

Figure 12.1.11 shows measurements of cross sections for the process $e^+e^- \to hadrons$ at cms energies around $E_{cm} = m_Z$ by the LEP experiments [3612]. The measurements map out the Z boson resonance in $e^+e^-$ annihilation in the hadronic channel. The lines show the result of a model-independent fit before and after QED corrections to these and other measurements to extract the Z boson resonance parameters such as the mass $m_Z$, the total width $\Gamma_Z$, the R-ratio $R_l^0 = \Gamma_{Z,had}/\Gamma_{Z,ll}$ and the hadronic pole cross section $\sigma_{had}^0$.

The extracted parameters are part of the set of EWPOs which can be compared with predictions by the SM including the QCD corrections. The QCD corrections for the EWPOs connected with the Z lineshape are known to N3LO, the corrections due to mixed and non-factorising electroweak and strong interaction diagrams are known up to $\alpha\alpha_S$ terms, and the QCD corrections for massive quarks are known up to $(m_q/Q)^4\alpha_S(Q)^3$, see [3664] for details.

Figure 12.1.12 shows the $\chi^2$ profile of a recent SM global fit as a function of the strong coupling $\alpha_S(m_Z)$ using the LEP data and other data for the masses of the top quark, the W boson and the Higgs boson [3664]. The blue band shows the $\chi^2$ of the global fit around the best value of $\alpha_S(m_Z)$. The grey lines show the contributions to this result of the most sensitive EWPOs. The width of the band reflects the theoretical uncertainties of the global SM fit. A comparison of the grey bands shows the consistency between the QCD corrections to the different EWPOs. The red data point is a direct measurement of $\alpha_S(m_Z)$ from the hadronic branching ratio of $\tau$ lepton decays measured mostly using LEP data.

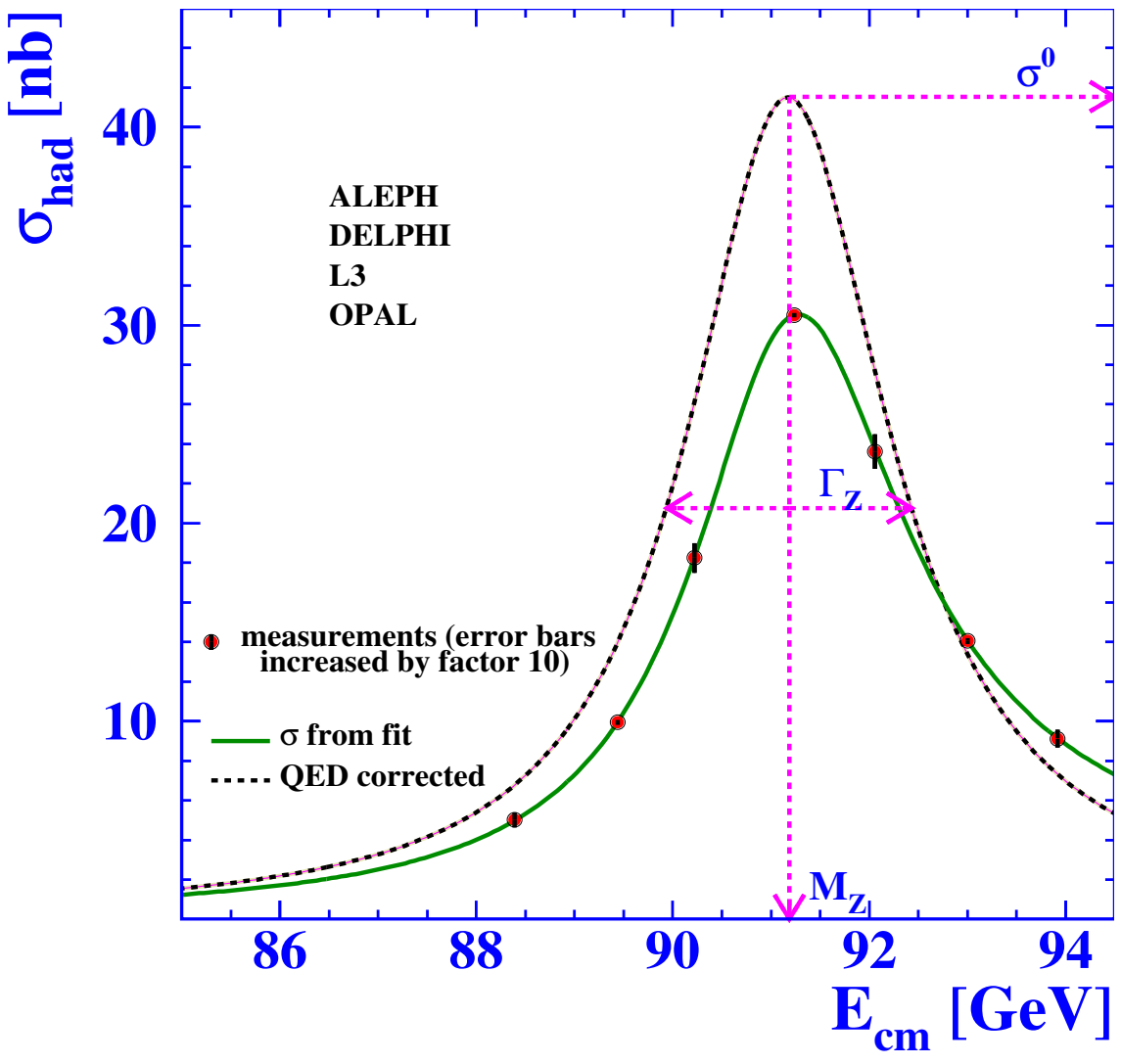

**Fig. 12.1.11** The figure displays measurements of the hadronic cross section in $e^+e^-$ annihilation at cms energies $E_{cm}$ around $m_Z$ measured by the LEP experiments. The lines show the model-indepedent fit to extract EWPOs before and after QED corrections [3612].

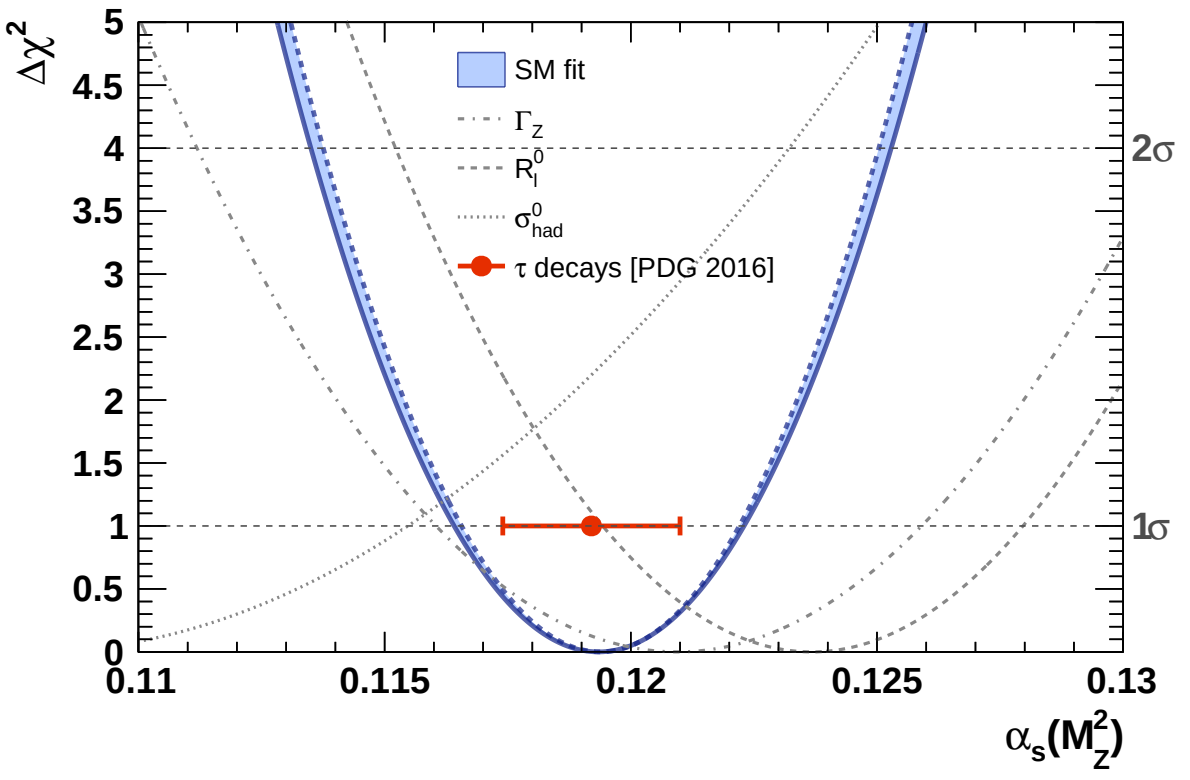

**Fig. 12.1.12** The figure shows with the blue band the $\chi^2$ profile of a global SM fit as a function of the value of the strong coupling $\alpha_S(m_Z)$. The grey lines are similar profiles for individual EWPOs as indicated. The red data point shows the value of $\alpha_S(m_Z)$ determined from hadronic $\tau$ lepton decays [3664].

The $\tau$ lepton weak decays to hadrons proceed in the SM description via a virtual W boson decaying to quarks. Similar to hadronic Z boson decays QCD corrections to the final state modify the predictions. At the scale of the $\tau$ lepton mass $m_\tau \simeq 1.78$ GeV the strong coupling $\alpha_S(m_\tau) \simeq 0.3$ such that large corrections are expected. The QCD corrections are also known to N3LO due to the similarity of the calculations. In addition non-perturbative effects are significant, while they are strongly suppressed for hadronic Z decays.

A recent analysis of the important theoretical issues for the extraction of $\alpha_S$ from hadronic $\tau$ lepton decays is [3665]. The data point shows the average of $\alpha_S(m_Z)$ determinations by the PDG from 2016 which has since been updated with only small changes [476]. The good consistency between these related determinations of $\alpha_S(m_Z)$ is a strong test of the consistent application of QCD corrections in the SM, as well as of the understanding of the evolution equations for the running of the strong coupling including the treatment of quark mass thresholds.

The large collection of measurements from the LEP experiments, SLD, and also the previous and partially re-analysed experiments at $e^+e^-$ colliders are a cornerstone of the experimental validation of the theory of strong interactions, QCD. Possible future $e^+e^-$ colliders are designed to deliver at least 1000 times the integrated luminosity w.r.t. LEP and with more advanced detectors. In addition large samples of Higgs and $W^+W^-$ bosons, and possibly of top-antitop quark pairs will open the door to many more tests of the SM including its QCD sector, and its proposed extensions.

## 12.2 High-$p_T$ jets

**Daniel Britzger, Klaus Rabbertz, and Markus Wobisch**

### 12.2.1 Introduction

One of the most fundamental testing grounds for the predictions of perturbative QCD (pQCD) are studies of the production rates of collimated sprays of hadrons, so-called *hadronic jets*. Although such jets are neither fundamental entities of the theory nor objects directly measurable in experiment, the notion of jets proved to be an extremely useful concept, because it allows to make the connection between the objects of pQCD, the quarks and gluons or, generically, *partons*, and the tracks and energy depositions in a detector. In a measured collision event, high-energetic jets can roughly be identified by eye for example when looking at an event display in the radial or the transverse plane. However, for an unambiguous attribution of each track and energy deposit to a jet, a mathematical prescription is required: a *jet algorithm*. Equally, to relate experimental measurements of such jets to production rates predictable in perturbative QCD, a precise definition of *partonic jets* is needed. To close the gap, for good jet algorithms it must also be demonstrated that corrections are under control that on the one hand unfold for detector effects to the level of stable hadrons as in Monte Carlo event generators, and on the other hand account for the non-perturbative transition from partons to the same stable-hadron level. History has shown that jet algorithms can be found that are suitable simultaneously for all three levels, measured tracks and energy clusters, the partons of perturbative calculations, and the hadrons of Monte Carlo event generators used in detector simulations. Alas, it took time approximately halfway through "the first 50 years of QCD" to evolve from first ideas to mature jet definitions used in today's precision phenomenology. In the following sections, the authors describe the essential steps of this evolution from their perspective of working at the LEP, HERA, Tevatron, and LHC colliders.

### 12.2.2 A hint of color: quark- and gluon-initiated jets

Establishing QCD as the theory of the strong interaction requires us not only to investigate the pattern of *colorless* hadronic particles and their properties, but to go beyond confinement and search for signs of the underlying dynamics of this asymptotically free quantum field theory. In other words, we need to find hints of color even though the confining property of QCD does not allow us to directly measure colored quarks — let alone gluons. Indirect evidence came in 1968 from the observation of Bjorken scaling in Deep-Inelastic Scattering (DIS) at SLAC [93, 148], where inelastic scattering of electrons on nucleons at large momentum-transfer squared, $Q^2$, is well described by the assumption of a virtual photon interacting with point-like constituents inside a nucleon. These constituents, named *partons* by Feynman, were later identified with the (valence) quarks of Gell-Mann and Zweig [18, 3113].

It is conjectured that the struck parton should manifest itself in the form of a collimated stream of hadrons moving along the direction of the primary parton with only a few hundred MeV of transverse momentum, like defined as *jet* in the introduction. This brings us to the second question implicit in this section's title "high-$p_\mathrm{T}$ jets": How high is "high"? The center-of-mass energies of a few GeV available at the time were insufficient to clearly observe well separated jets simply because the opening angles of the hadron streams were far too large and the "jets" interleaved with each other even though the back-to-back orientation of the primary $q\bar{q}$ pair should guarantee their maximal separation. A way out was found by focusing on the main interest to differentiate between a two-jet like structure favored by QCD and the expectations from other models. Instead of reconstructing jets or jet quantities explicitly, the strategy rather consists in searching for a principal event axis along which most of the momentum of each produced hadron is aligned. In 1975, the SLAC-LBL Mark I experiment at the $e^+e^-$ storage ring SPEAR used *sphericity* [3666, 3667], which defines such an event axis by minimizing the sum of squares of all momenta with respect to this axis. The *event shape* sphericity, $S$, is then defined as

$$S = 3\sum_i \left(p^2_{\mathrm{T},i}\right) \,/\, 2\sum_i |\vec{p}_i|^2 \;, \tag{12.2.1}$$

where the sum is over all particles $i$ in the event with 3-momenta $\vec{p}_i$ and transverse momenta $p_{\mathrm{T},i}$ with respect to the sphericity axis. Each event is characterized by one number $S$ ranging from zero, when all particles are fully aligned along the axis, up to unity for isotropic events. By means of defining such an event axis for their measurements at 3.0, 3.8, 4.8, 6.2, and 7.4 GeV center-of-mass energy, the Mark I experiment found first evidence for quark-initiated jet production emerging when going to the higher center-of-mass energies [105]. Moreover, profiting from polarized beams, by comparing the angular distribution of the sphericity axis of $q\bar{q}$ production to the one of $e^+e^- \to \mu^+\mu^-$ they concluded that the potential partons must have spin 1/2 rather than spin 0.

How about gluons then, the exchange quanta of QCD? Do they exist and, if yes, how do they manifest themselves? In 1976 Ellis, Gaillard, and Ross [104] argued gluon bremsstrahlung $e^+e^- \to q\bar{q}g$ to be the leading correction to $q\bar{q}$ dijet production. As a consequence, with increasing center-of-mass energy one of the two quark-initiated jets should exhibit signs of widening up with higher multiplicity until finally a third gluon-initiated jet emerges leading to planar 3-jet events. The center-of-mass energies available at SPEAR and also DORIS at DESY, however, were not sufficient to provide evidence for 3-jet production, although valuable results could be achieved by investigating the conjectured dominant decay of the upsilon resonance into three gluons, $\Upsilon \to ggg$, confirming predictions by QCD including the vector character of the gluons [3668]. Only the much higher center-of-mass energy of 27 GeV reached by the PETRA collider at DESY in spring 1979 could provide sufficiently high-energetic $e^+e^-$ collisions such that clearly identifiable 3-jet events could be produced. The first event display of the TASSO Collaboration was presented by Wiik at the "Neutrino 79" conference in Bergen [91] and, of course, is also reproduced in this commemorative work, see the section by S.L. Wu for a more personal recollection of events. Subsequently, all four experiments at PETRA published clear evidence for planar 3-jet events affirming the discovery of the gluon and gluon-induced jets [92, 109–111].

The increasing $e^+e^-$ center-of-mass energies at PETRA, TRISTAN, SLC, and LEP up to $\sqrt{s} = 209$ GeV

allowed a plethora of (multi-)jet measurements to be performed, all confirming the conjectures of QCD as theory of the strong interaction. Notably, the rate of events with three jets as compared to dijet production is to first order proportional to the strong coupling, which then can be extracted at each energy point to demonstrate its energy dependence or *running* as predicted by QCD.

Finally, angular correlations in 4-jet events are sensitive already at leading order (LO) to the color factors $C_A = 3$ and $C_F = 4/3$ of the non-Abelian special unitary group SU(3) of QCD and thus are probing its non-Abelian nature as described in the previous section. A compilation of constraints on these color factors is presented in Ref. [3623], where world average values are quoted that are in perfect agreement with the expectations from QCD.

### 12.2.3 Jets at hadron-hadron colliders

Despite great new insights obtained thanks to high-precision measurements at $e^+e^-$ colliders, the term of *discovery machines* generally is reserved for hadron-hadron colliders. Because of the much larger mass of protons as compared to electrons, the huge loss of energy per turn in circular storage rings due to synchrotron radiation can be avoided enabling much higher collision energies of e.g. $p\bar{p}$ accelerators than possible with circular $e^+e^-$ beams. The benchmark observable of jet physics at hadron-hadron colliders is the inclusive jet production cross section and in the early days the phase space was divided up into intervals of the jet transverse energy $E_\mathrm{T}$ and the jet pseudorapidity $\eta$ defined in terms of the polar angle $\theta$ as $\eta = -\ln\tan(\theta/2)$. Measured jet yields are transformed into a double-differential cross section via

$$\frac{\mathrm{d}^2\sigma}{\mathrm{d}E_\mathrm{T}\,\mathrm{d}\eta} = \frac{1}{\epsilon \cdot \mathcal{L}_\mathrm{int}} \cdot \frac{N_\mathrm{jets}}{\Delta E_\mathrm{T} \Delta\eta}\,, \qquad (12.2.2)$$

where $N_\mathrm{jets}$ is the number of jets counted within a bin, corrected for detector distortions, $\epsilon$ is the experimental efficiency, and $\Delta E_\mathrm{T}$ and $\Delta\eta$ are the respective bin widths.

The first such measurement of inclusive jet production was published in 1982 by the UA2 Collaboration with data recorded in the so-called *jet run* at the Sp$\bar{\text{p}}$S collider operating at 540 GeV center-of-mass energy [3669]. The observed steep decrease of the jet $E_\mathrm{T}$ spectrum proportional to $E_\mathrm{T}^{-n}$ with $n \approx 9$ was correctly predicted by QCD at LO [3670]. Firm conclusions on the absolute normalization, however, were not possible because of large experimental and theoretical uncertainties, and lack of a well-defined jet algorithm. The UA2 Collaboration employed a cell-based clustering of energy deposits in the calorimeters, where neighboring cells could be merged into one cluster. A "final" cluster could be split up again, if it contained multiple, well separated maxima. Instead of referring directly to cell geometry, the UA1 experiment used an algorithm based on cones of radius $R$ equal to unity in $(\eta, \phi)$ space in order to decide whether cells are merged or not [3671]. Here, $\phi$ is the azimuthal angle. To initiate a jet, cells exceeding a minimal transverse energy are taken in decreasing order of $E_\mathrm{T}$ as "seeds", around which cells within the defined cone are combined with this seed to form the jet. This algorithm corresponds already to a cone jet algorithm; alas, it suffers from a number of shortcomings like unclustered energy or sensitivity to collinear splittings further described in the next section. Nevertheless, at the level of the limited experimental precision and with only order-of-magnitude predictions at LO, jet measurements conducted at the Sp$\bar{\text{p}}$S and still at the Intersecting Storage Rings ISR [3672] were in agreement with expectations from QCD.

### 12.2.4 The evolution of jet algorithms

Until the end of the 1980s, a vast amount of jet data from hadron colliders were collected, reaching a level of precision of 10 %. Predictions at LO in pQCD, however, were very limited in precision by the uncompensated dependence on the *renormalization scale*, $\mu_r$, through the running strong coupling. The calculation of next-to-leading-order (NLO) corrections to jet production advanced the accuracy of perturbative predictions to a comparable level. This progress required a careful re-evaluation of the concept of jets and resulted into new classes of jet algorithms, since several shortcomings of previous jet definitions were identified, which limited their usability in higher-order pQCD predictions or in hadron-induced processes. Let us have a closer look into the evolution of jet algorithms over time.

The first jet algorithm was described in 1977 by Sterman and Weinberg for $e^+e^-$ collisions [166]. In their algorithm, particles with momenta pointing towards the same direction within some opening angle were clustered together. Most importantly, their jet definition made the result insensitive to the emission of either soft or collinear particles. This is called *infrared and collinear safety*, which is crucial to produce finite results at all orders in perturbation theory. Otherwise the cancellation of soft and collinear singularities associated with such partonic emissions in calculations of pQCD is spoiled leading to infinite results. To be useful in comparisons to pQCD, the outcome of a jet algo-rithm therefore must neither depend on the addition of arbitrarily soft clustering objects to the set of inputs, nor on the merging of two collinear input objects or the splitting of an input object into two collinear ones.

The following decade saw the proposal by Sterman and Weinberg to be generalized in order to analyze hadron-hadron collisions in terms of a number of cone-shaped jets of a chosen jet radius, $R$, pointing into the directions of the highest energy or momentum densities in an event. In the same period the JADE Collaboration at the PETRA collider introduced another type of jet algorithm based on iterative pairwise clusterings for the analysis of $e^+e^-$ events [3626]. Hence, two classes of jet algorithms emerged:

1. cone algorithms that assign objects to the leading energy-flow objects in an event based on geometrical criteria;
2. sequential-recombination algorithms that iteratively combine the closest pairs of objects.

A summary of jet algorithms discussed at the time is presented in the proceedings of the *Snowmass* "Summer Study on High Energy Physics" [3673].

Although introduced only in 2008 in its general form, one can determine the so-called *catchment area* of a jet, often just named *jet area*, for both classes provided the algorithm is infrared- and collinear-safe [3590]. For cone algorithms defined in $(\eta, \phi)$ space as used already by the UA1 Collaboration, this jet area formerly was identified with the circular area with jet radius $R$, which simplified considerably the task of jet energy calibration at hadron-hadron colliders.

In $e^+e^-$ collisions all final-state particles emerge from the hard subprocess. Therefore, in $e^+e^-$ measurements exclusive jet algorithms were applied, which assign each final-state particle to one of the high-$p_T$ jets. Hence, a collision event is classified as an exclusive jet final state, *e.g.* $e^+e^- \to n$ jets and nothing else.

Although being more costly in terms of computing time, it was affordable to use successive recombination algorithms because of the low multiplicity in $e^+e^-$ annihilations. Initially, the JADE algorithm was favored, where pairs of particles are clustered in the order of increasing invariant di-particle masses, assuming this would result in jets with small invariant masses. In the phenomenology of $e^+e^-$ physics, it was, however, discovered that the JADE algorithm frequently clusters soft particles at large angles, cf. also Fig. 12.2.1, which is very disadvantageous for precision calculations [3674]. This problem was addressed in the $k_t$ or "Durham algorithm" [168][111], in which the distance measure was changed from the invariant di-particle mass to the relative transverse momentum, $k_t$, of the particle pair. This version, also called the (exclusive) $k_t$ algorithm, was confirmed to have superior properties than the JADE algorithm in $e^+e^-$ annihilation.

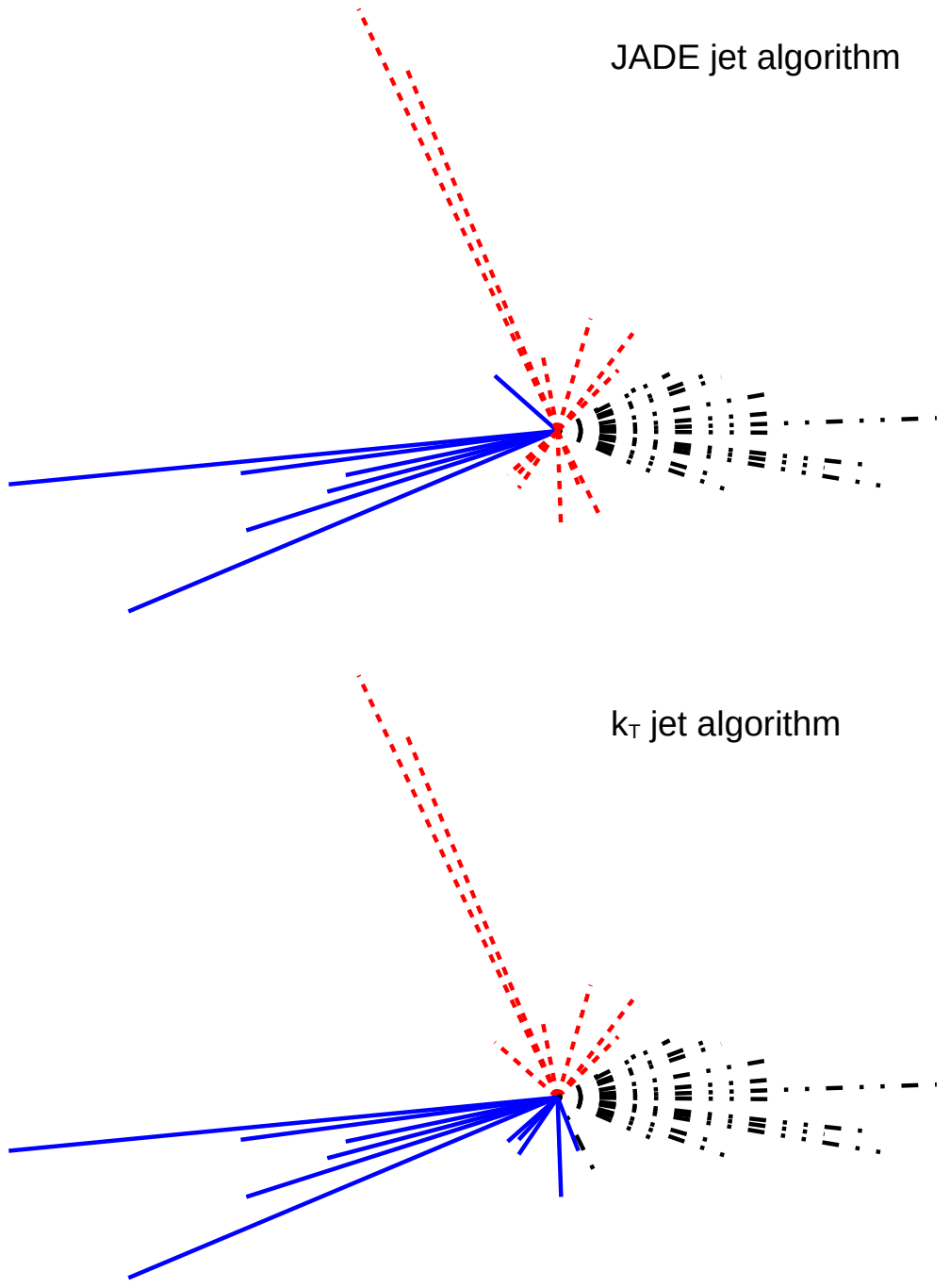

**Fig. 12.2.1** A 3-jet final state in $e^+e^-$ collisions as defined by the JADE (upper) and $k_t$ (lower) jet algorithms. The particle assignments to the three jets according to the algorithms are indicated by blue full, black dash-dotted, and red dashed lines. Figure redrawn from Ref. [3675].

When HERA, the first and only electron–proton ($ep$) collider, started in 1992, "standard" jet algorithms had been defined already for $e^+e^-$ annihilation as discussed. In hadron–hadron collisions cone-type algorithms were favored over sequential-recombination algorithms to avoid time-consuming repeated iterations over many final-state particles. Nothing yet had been developed for physics at an $ep$ collider such that many physicists coming from LEP experiments tried to adopt methods as they were used in $e^+e^-$ physics. So in the early HERA jet analyses, a modified version of the JADE algorithm was used (the "mJADE algorithm" [3676]), in which the proton beam remnant is treated by introducing a pseudo-particle (carrying the missing longitudinal momentum in the event), to which particles can be clustered. At the end, all particles are either assigned to the high-$p_T$ jets, or to the jet including the pseudo-particle. The former are considered as the $n$ high-$p_T$ jets, while the latter is considered to be the (one) beam

[111] Originally, $k_\perp$ was used as label instead of $k_t$. For simplicity we use $k_t$ throughout.

remnant. The final states are therefore classified as exclusive $(n+1)$-jet final states.

In reactions with initial state hadrons, *i.e. ep* and hadron-hadron collisions, collinear singularities in the matrix elements of the hard subprocess are factorized into process-independent parton distribution functions (PDFs), which depend on the *factorization scale*, $\mu_f$, that defines the limit between attribution to the perturbative hard process or the non-perturbative hadron structure in form of the PDFs. This factorization, however, only works, if it is not spoiled by the definition of the measured quantity that must not depend on the beam-remnant(s). For the mJADE algorithm, it was the inclusion of the kinematics of the beam-remnant that made the algorithm non-factorizable. This issue was fixed in the exclusive $k_t$ algorithm for *ep* and hadron-hadron collisions by treating the beam remnant(s) as particles of infinite momentum and thus independent of their actual kinematics. This exclusive $k_t$ algorithm was in use for some time within the HERA experiments and later was replaced by its inclusive counterpart.

Hadron-hadron and *ep* collisions share the common feature of having activity in their final states related to the remnant(s) of the beam hadrons. Therefore, the jet definitions used in hadron-hadron physics were based on the cone-type proposal by Sterman and Weinberg to define a jet by the transverse energies through a cone, which is moved so as to maximize the transverse energy flow through it. In this approach, only selected final-state particles are included in jets. Those, which are not assigned to jets are effectively interpreted to stem from the so-called *Underlying Event* that is related to soft processes involving interactions with the beam remnants. The jet final-states are thus classified as *inclusive* with respect to additional unclustered particles, *e.g.* $pp \to n$ jets plus additional activity, which could consist of additional jets and/or unclustered particles.

Another difference between $e^+e^-$ and hadron-hadron physics consists in the choice of variables. In hadron-hadron collisions, the center-of-mass frame of the hard subprocess is boosted longitudinally, *i.e.* along the beam direction with respect to the detector rest frame. Hence, instead of energies and angles as used in $e^+e^-$ collisions, transverse momenta and/or transverse energies are used, together with azimuthal angles and either the pseudorapidity $\eta$ as defined before, or the rapidity $y = 1/2 \cdot \ln[(E+p_z)/(E-p_z)]$, which coincides with $\eta$ for massless objects. As a consequence, cone-jet algorithms in hadron-hadron collisions are used with cone radii $R$ defined in the plane of azimuthal angle and (pseudo)rapidity.

Cone algorithms are, however, not as easy to implement as one would naïvely think. The basic idea of a cone-jet algorithm sounds rather simple: Decide on a cone radius, $R$, place it in the plane of azimuthal angle and (pseudo)rapidity, compute the transverse energy/momentum flow through the cone, and move the cone over the plane so as to maximize this flow. Before the end of the 1990s, experimental jet measurements used a large number of different implementations. These early cone algorithms suffered from a number of problems. Many were not infrared- or collinear safe, while others had undesired features. Some of the problems arise from the fact that a true, continuous maximization procedure of the energy flow through the cone required too much computing resources, and shortcuts were applied. Some versions simply defined the final jets by building cones around the particles/detector clusters of highest energy. Other versions used these clusters as starting points, or "seeds" for an iterative procedure. All of these algorithms were either not infrared-, or not collinear-safe, or even both. Other undesired features emerged through the treatment of overlapping cones. Sometimes, it happens that two resulting jet cones share a number of particles. To have a unique assignment of particles to jets, an *overlap treatment* is added to the algorithm, which assigns the particles in the overlap regions uniquely to one of the two jets. This overlap treatment depends on additional parameters (adding to the complexity of the algorithm) and in most cases it also introduced additional violations of infrared or collinear safety. These problems were ultimately addressed and solved with the Seedless Infrared-Safe Cone (SISCone) jet algorithm [3677]. By eliminating seeds, and using a refined overlap treatment, SISCone became the first (and so far, only) cone jet algorithm that is infrared- and collinear safe.

The SISCone algorithm was, however, never widely used since the rather late time it was introduced. Jet measurements had moved on to different jet algorithms: Soon after the introduction of the exclusive $k_t$ algorithm for $e^+e^-$ physics and the above-mentioned modifications for processes with initial-state hadrons, a similar inclusive algorithm was introduced the "Cambridge algorithm" [170]. This algorithm transferred the basic concepts of the exclusive $k_t$ algorithm consistently to hadron-hadron collider physics. In the same way that the Cambridge algorithm was a modification of the exclusive $k_t$ algorithm, a corresponding modification of the inclusive $k_t$ algorithm was introduced, called the "Aachen algorithm" or, later, the "Cambridge-Aachen algorithm" [171]. This algorithm recombines pairs of particles simply in the order of increasing distances in $(y, \phi)$ space. Both algorithms can be specified in a uni-fied way by defining the pairwise distance $d_{ij}$ between any two objects $i$ and $j$, and the *beam distance* $d_{iB}$ of each object $i$ as:

$$d_{ij} = \min\left(p_{\mathrm{T},i}^{2p}, p_{\mathrm{T},j}^{2p}\right) \frac{\Delta R_{ij}^2}{R^2}\,, \tag{12.2.3}$$

$$d_{iB} = p_{\mathrm{T},i}^{2p}\,. \tag{12.2.4}$$

Here, the power $p$ is the algorithm defining parameter, and $\Delta R_{ij}$ is the purely "angular" distance in $(y, \phi)$ space between $i$ and $j$:

$$(\Delta R_{ij})^2 = (y_i - y_j)^2 + (\phi_i - \phi_j)^2\,. \tag{12.2.5}$$

Then, each time the minimal distance of all pairwise and beam distances is a $d_{iB}$, object $i$ is declared a final jet and removed from the list of clustering objects. If the minimal distance is a $d_{ij}$ instead, the two objects are merged using four-vector addition into a new object that is added to the clustering list. This is repeated until no more input objects are left.

Setting $p$ equal to unity gives the $k_t$ algorithm, while $p = 0$ corresponds to the Cambridge-Aachen one that only considers $\Delta R_{ij}$ in the clustering and is frequently used for studies of jet substructure. Interestingly, as discovered in Ref. [174], the choice of $p = -1$ is also a valid option, where in contrast to the $k_t$ algorithm the clustering starts with the highest-$p_T$ objects and produces round-shaped jet areas as if from a cone jet algorithm! This third "family member" was dubbed the "anti-$k_t$ algorithm" and was quickly adopted as the main jet algorithm for jet physics at the LHC.

### 12.2.5 New physics with jets: excesses in jet cross sections

The next stage of establishing QCD as the theory of the strong interaction was triggered by two developments: the arrival of predictions at NLO in pQCD also for hadron-hadron collisions, and the start of the Tevatron collider at Fermilab with a $p\bar{p}$ center-of-mass energy ranging from 540 GeV up to 1.96 TeV. The by far dominating theoretical uncertainty caused by the large $\mu_r$ scale dependence of LO predictions was reduced from factors of roughly two to 10–30% [3678, 3679]. Additional uncertainties from non-perturbative effects and from the proton structure were estimated to lie between 5 and 20%, respectively. The latter uncertainty was derived from calculations using different extractions of the proton PDFs from data of deep-inelastic scattering of leptons on fixed targets [3680–3683]. First comparisons of these NLO predictions to $p\bar{p}$ collider data from UA2 and from the new CDF experiment at Tevatron exhibited a very nice agreement.

This picture changed suddenly in 1996 when the CDF Collaboration reported an excess in inclusive jet data at high $E_T$ beyond 200 GeV as shown in Fig. 12.2.2 [3684]. A possible explanation could be new phenomena at an energy scale $\Lambda$ far beyond reach to allow *e.g.* resonant production of new particles. Similarly to Fermi's low-energy four-fermion coupling to approximate weak interactions at scales well below the $W$ boson mass, such an excess can be described in terms of *contact interactions (CI)* [3680, 3685]. Speculations about such contact interactions as a possible explanation were, however, quickly abandoned and the results were scrutinized for effects not properly covered by uncertainties. With respect to the proton structure there was no other means than taking the spread in predictions using different proton PDFs, also shown in Fig. 12.2.2, as a proxy for the uncertainty, which now had become very relevant. As all the PDFs known at the time were potentially prone to the same biases, the association of the spread in the corresponding predictions with a PDF uncertainty could only be considered an educated guess or, in the words of Soper [3686]: "This is similar to estimating the size of a French mountain valley by taking the r.m.s. dispersion in the locations of individuals in a flock of sheep grazing in the valley."

The way forward was described in the seminal paper Ref. [3687], where a systematic approach was presented to derive parton distributions with reliable uncertainty estimations. Using the preliminary PDFs including experimental uncertainties derived in Ref. [3688] from DIS data, the authors demonstrated that the excess reported by the CDF Collaboration can be absorbed in updated parameter values for the strong coupling constant and the gluon distribution. While the quark parton distributions are directly determined in DIS, in particular with data from the new HERA collider as used in Ref. [3688], the DIS data are insufficient to also fix $\alpha_S(M_Z)$ and the gluon content in the proton. For both, jet cross sections measured at the Tevatron and at HERA, as described in the next section, are valuable input to the PDF fits.

### 12.2.6 The running coupling and the gluon content of the proton

HERA, approved in 1984, just became operational during 1992, the same year as the 20th anniversary of QCD was celebrated in Aachen [3689].[112] At that time, QCD was in a "transition from the stage of early exploratory studies to high precision analyses in QCD" as noted by Zerwas and Kastrup in the introduction to this workshop [2411]. A milestone for testing QCD was achieved by demonstrating experimentally the *running* of the strong coupling from the $\tau$ mass of around 2 GeV up to the $Z$ boson mass at 91 GeV using various observables and data from different experiments as reported in Ref. [3690]. A first summary of determinations of $\alpha_S(M_Z)$ was presented by Altarelli [3691] who concluded on $\alpha_S(M_Z) = 0.118 \pm 0.007$.

[112] This was the very first conference participation of KR triggering his profound interest in jets and QCD.

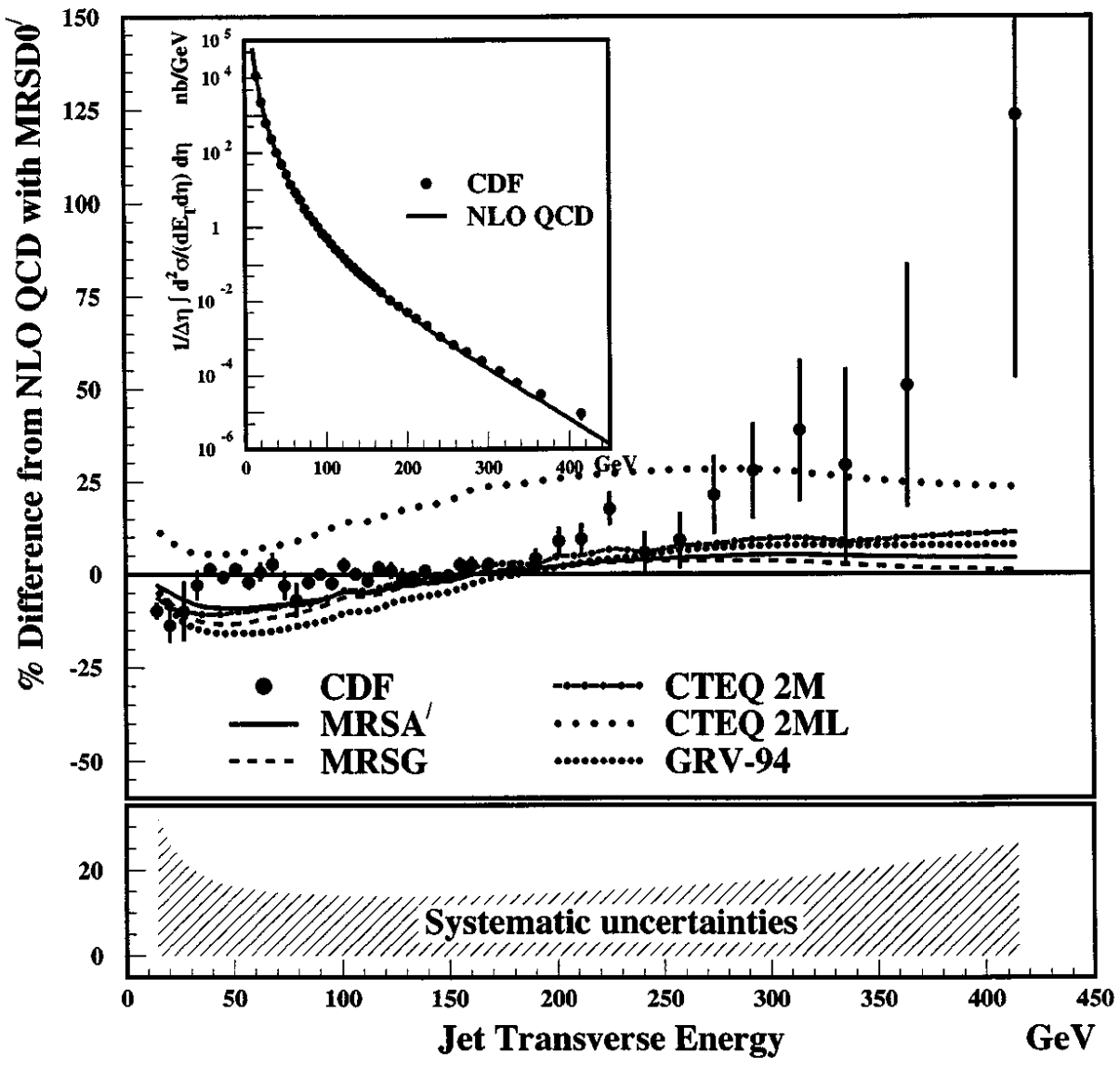

**Fig. 12.2.2** Percentual difference between the CDF inclusive jet cross section (points) and a prediction at NLO QCD using MRSD0' PDFs. Additional lines show predictions for a selection of alternative PDFs available at the time. The error bars represent uncorrelated uncertainties, while the quadratic sum of the correlated systematic uncertainties are shown in the bottom panel. The inset compares the absolute cross sections. Figure taken from Ref. [3684].

HERA was constructed as the paramount extension to the series of previous, very successful fixed-target lepton-nucleon scattering experiments, which have led the way to conceiving QCD as the theory of strong interactions. The increase by a factor of ten in the lepton–proton center-of-mass energy promised rich new data for testing many aspects of QCD. In particular, with $\sqrt{s} = 300$ GeV, HERA allowed to elucidate the structure of the proton and the running of $\alpha_S(\mu_r)$ by means of unique and detailed measurements of the hadronic final state in addition to the scattered lepton.

The HERA collider at DESY consisted of two independent accelerators designed to collide 30 GeV electron and 820 GeV proton beams. Two multi-purpose detectors, H1 and ZEUS, were conceived to precisely measure the hadronic final state with almost hermetic coverage. The main difference between the two experiments with respect to jets is given by their calorimeters. The H1 collaboration opted for a liquid argon calorimeter with electromagnetic and hadronic sections, both inside the solenoïd providing the magnetic field [3692, 3693]. The ZEUS collaboration optimized their calorimetric system for hadronic measurements and employed a compensating uranium plastic-scintillator sampling calorimeter [3694]. The overconstrained kinematics of neutral-current DIS events enabled precise in-situ calibrations for the electromagnetic and hadronic energy scales such that both collaborations could report a jet energy scale uncertainty of only 1 % for jets with transverse momenta exceeding 10 GeV in the laboratory rest frame [3695, 3696].

Already the first HERA data brought striking QCD results, like the confirmation of the logarithmic violation of Bjorken scaling shown by the $F_2$ structure function in dependence of the parton fractional momentum $x$ as predicted by QCD [3697, 3698], or support for the presence of a hadronic structure of quasi-real photons as a result of dijet events observed in photoproduction [3699, 3700]. Hence, jets were an integral part of the HERA physics program from the very beginning. The term *jet physics* quickly extended well beyond the simple picture of one "DIS jet", which is initiated by the struck quark in the Quark-Parton-Model (QPM) picture, or of dijet topologies in photoproduction. Studies of further properties like jet charge, substructure, fragmentation, or the heavy flavor content of jets led to many more interesting results, which, however, cannot be covered here. In the following we will limit ourselves to high-$p_{\mathrm{T}}$ jets in neutral-current DIS and will refer the interested reader to other sections in this book or to review articles [3701–3705].

At HERA, for the first time, it became possible to study large numbers of dijet events in neutral-current DIS, so-called (2+1) jet events. In pQCD the cross section for hard processes in DIS is given up to order $n$ in the perturbative expansion in $\alpha_S$ through the factorization theorem

$$\sigma = \sum_{n}^{\text{order}} \sum_{i}^{q,\bar{q},g} \left( \frac{\alpha_S(\mu_r)}{2\pi} \right)^{k+n} \int \mathrm{d}x f_{a/i}(x, \mu_f) \mathrm{d}\hat{\sigma}_i^{[n]}(x, \mu_r) , \tag{12.2.6}$$

where $i$ denotes the parton flavors in the proton PDF $f_a$, and $k$ corresponds to the power in $\alpha_S$ at leading order. The universal proton PDFs are convoluted in $x$ with the hard coefficients at a selected factorization scale $\mu_f$. At LO, pQCD predicts the (2+1) jet events to be produced proportional to $\alpha_S$ ($k$=1). At HERA, this process is mainly initiated by a gluon inside the proton and thus dijet data provide direct access to the gluon content of the proton down to $x \sim 10^{-3}$. A second LO contribution arises from gluon radiation off one of the quark-lines in the QPM diagram and becomes dominant at large $x$.

The first measurement of (2+1) jet rates by the H1 Collaboration [3706] employed the JADE jet algorithm [3626], while the ZEUS Collaboration [3707] opted for a cone jet algorithm following the *Snowmass convention* [3585]. The hadronization corrections were found to be reasonably small and the measured jet profiles could be directly related to the underlying hard process and the gluonic content of the proton. These early data strongly supported the QCD picture of jet-production in DIS and the data were found to be well described by first order QCD calculations supplemented with leading-logarithmic parton showers as an approximation of higher-order QCD corrections. Already at this stage a running coupling was significantly favored over a constant value of $\alpha_S$.

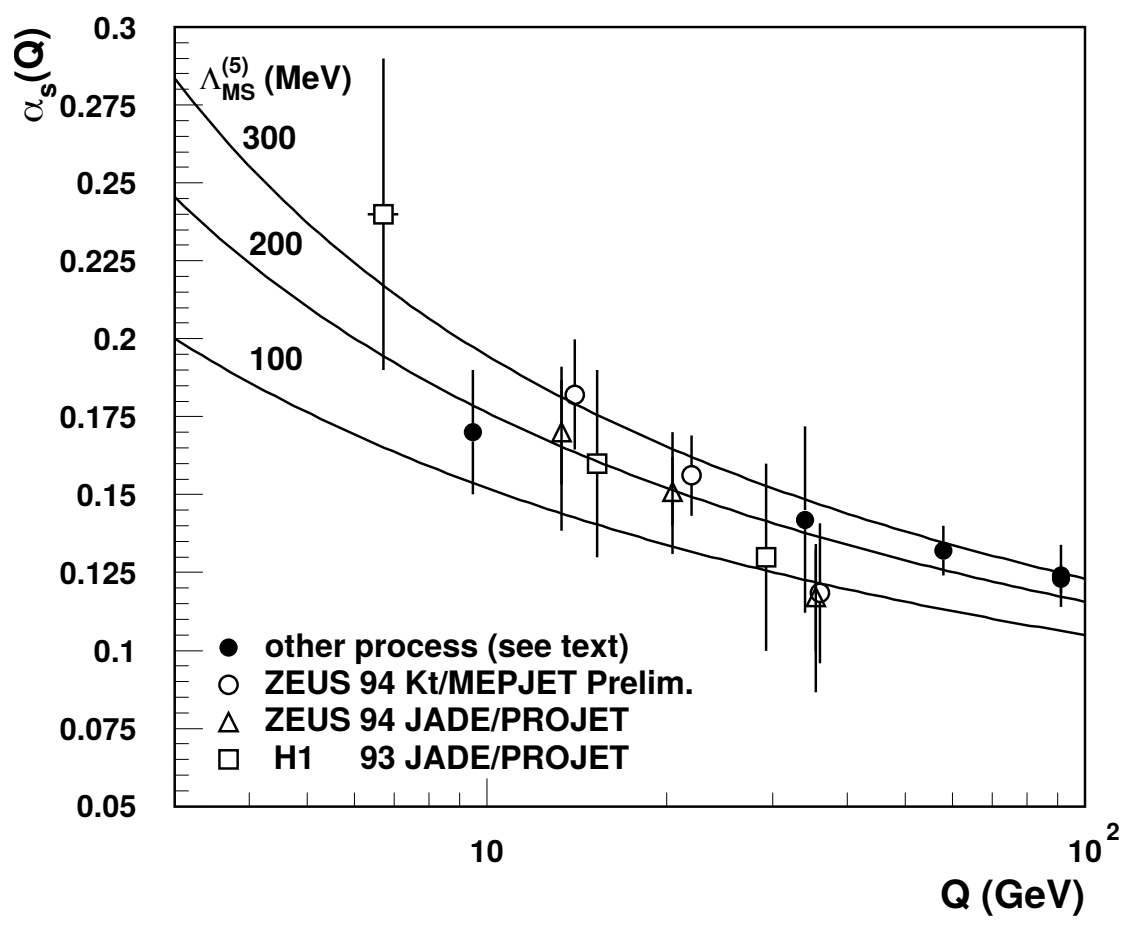

**Fig. 12.2.3** Dependence on the energy scale $Q$ of the strong coupling, $\alpha_S(\mu_r = Q)$, from early HERA data in comparison to other processes, see text for details. The predictions of QCD for three values of the $\alpha_S$-equivalent $\Lambda_{\overline{\rm MS}}$ parameter are superimposed as lines. Figure taken from Ref. [3708].

The inclusion of NLO QCD corrections in dijet calculations [3709] and an improved understanding of hadronization corrections of jet data together with refined and enlarged data sets, allowed for the first time the study of the running of the strong coupling constant in a single process using (2+1) jet rates based on the JADE algorithm [3710, 3711]. A summary of these results from H1 and ZEUS in comparison to measurements in $e^+e^-$ collisions is displayed in Fig. 12.2.3. The additional points are determined from $\Upsilon$ decays ($\Gamma_\Upsilon$), the ratio $R$ of hadronic over total cross section ($\sigma_{\rm had}/\sigma_{\rm tot}$), event shapes, and the ratio of hadronic over leptonic decay width of the $Z$ boson ($\Gamma_{\rm hadron}/\Gamma_{\rm lepton}$) as described in Ref. [3712]. An insight gained from these data and from subsequent studies with improved NLO calculations [3713] was that cone or $k_t$ jet algorithms seem to be preferred over the JADE algorithm for precision QCD analyses due to their improved perturbative stability in hadron-induced processes [172, 3714, 3715], as already outlined in the previous section. In addition, it became apparent that the choice of suitable renormalization and factorization scales is crucial to achieve reliable results for multi-scale processes such as jet production in DIS.

Despite these first successes it became rapidly clear that for jet measurements in the laboratory rest frame theoretical shortcomings prevent optimal comparisons to theory. Firstly, it is highly desirable for the jet observables to respect factorization, and secondly it is highly non-trivial to separate the hadronic final state from the beam remnant. A way forward is found by boosting every event to the Breit frame of reference [3715] using the reconstructed DIS kinematics. In the Breit frame the incoming parton collides head-on with the exchanged electroweak boson along the $z$ axis of this reference frame. Any significant transverse momentum is generated from QCD effects. High-$p_T$ jets primarily occur in dijet topologies, for which the LO QCD diagram is of $\mathcal{O}(\alpha_S^1)$, whereas LO DIS or the beam remnant do not contribute. First measurements of jet cross sections in the Breit frame using variants of the longitudinal invariant $k_t$ jet algorithm have been conducted by the H1 and ZEUS collaborations with a distance parameter of $R = 1.0$ [3716–3718]. This choice promises high accuracy of pQCD predictions and small non-perturbative corrections for hadronization effects. From data at high $Q^2 \gtrsim 150\,\mathrm{GeV}^2$, where scale choice ambiguities are reduced, since jet transverse momenta are of a similar size as the virtuality of the exchanged boson $\sqrt{Q^2}$, both collaborations determined $\alpha_S(M_Z)$ with NLO pQCD predictions at a precision of around 4 %. The uncertainty in $\alpha_S(M_Z)$ was comparable to the level of the LEP experiments [3719] and considerably outperformed the ongoing experiments CDF and D0 at the Tevatron. Moreover, the running of $\alpha_S$ could be successfully tested in the scale range from about 7 to 50 GeV. Together with inclusive neutral- and charged-current DIS data, even the first combined determination of the proton PDFs together with $\alpha_S(M_Z)$ was performed from data of a single experiment [3720].

In 1998, the beam energy of the HERA protons was raised to 920 GeV, corresponding to $\sqrt{s} \simeq 320$ GeV. The large amount of data recorded from 1998 to 2000, and during the HERA-II running period from 2003 to 2007, led to a multitude of measurements i.a. investigating the dependence of jet cross sections on the type of jet algorithm and the jet size $R$, or the benefits of normalizing to the DIS cross section. With respect to the strong coupling constant the development culminated in a determination of $\alpha_S(M_Z)$ with only 0.4 % of experimental uncertainty [3721]. Yet, in all these QCD analyses, the NLO scale uncertainties of roughly 5 % in the jet predictions remained the dominant uncertainty and, hence, the limiting factor preventing a higher precision for $\alpha_S(M_Z)$. The next decisive progress, then, should come from theory. After more than 15 years, the next-to-next-to-leading order (NNLO) corrections to jet production in DIS were finally calculated in Refs. [3722, 3723], which allowed to reduce the scale dependence of the predictions in the interpretation of the HERA jet data. The latest improved HERA-II measurements were then the first to be confronted with the new NNLO cross section predictions, which proved the corrections to be sizeable and could be as large as 40 % at low scales. Nevertheless, the NNLO predictions provided a very good description of the data over the entire accessible kinematic range [3721] and a significant improvement as compared to the long-standing NLO predictions. This NNLO revolution for single-jet inclusive predictions was the ultimate step to reduce the theoretical uncertainties to a level comparable to the experimental uncertainties. A full analysis of all inclusive jet data from H1 [3724], and an analysis of data from H1 and ZEUS [3725] demonstrated an excellent agreement between the data and the NNLO pQCD predictions. A comparison of selected inclusive jet cross section data with NNLO predictions is displayed in Fig. 12.2.5 below.

From inclusive jet data the value of $\alpha_S(M_Z)$ was finally determined at NNLO to be

$$\alpha_S(M_Z) = 0.1178 \pm 0.0015\,(\text{exp}) \pm 0.0021\,(\text{theo}) \tag{12.2.7}$$

with percent level experimental and theoretical uncertainties of similar size. Surprisingly, although jet data were believed to have a significant sensitivity to the gluon PDF, a complete analysis of jet data together with HERA inclusive DIS data at NLO [3002] or NNLO [3043, 3724, 3726] showed only little impact on the gluon density.

Finally, the inclusive jet data from HERA were able to unfold their full potential to test the running of the strong coupling from a single process using NNLO pQCD predictions [3724, 3725]. The results are found to be in excellent agreement with expectations from pQCD and are shown in Fig. 12.2.4, where the extracted values of $\alpha_S(\mu_r)$ from these data are compared additionally with the $\alpha_S(\mu_r)$ determinations from inclusive jet data of the CMS experiment [3727] and with analyses using jet-rate measurements in $e^+e^-$ collisions [3630, 3728, 3729].

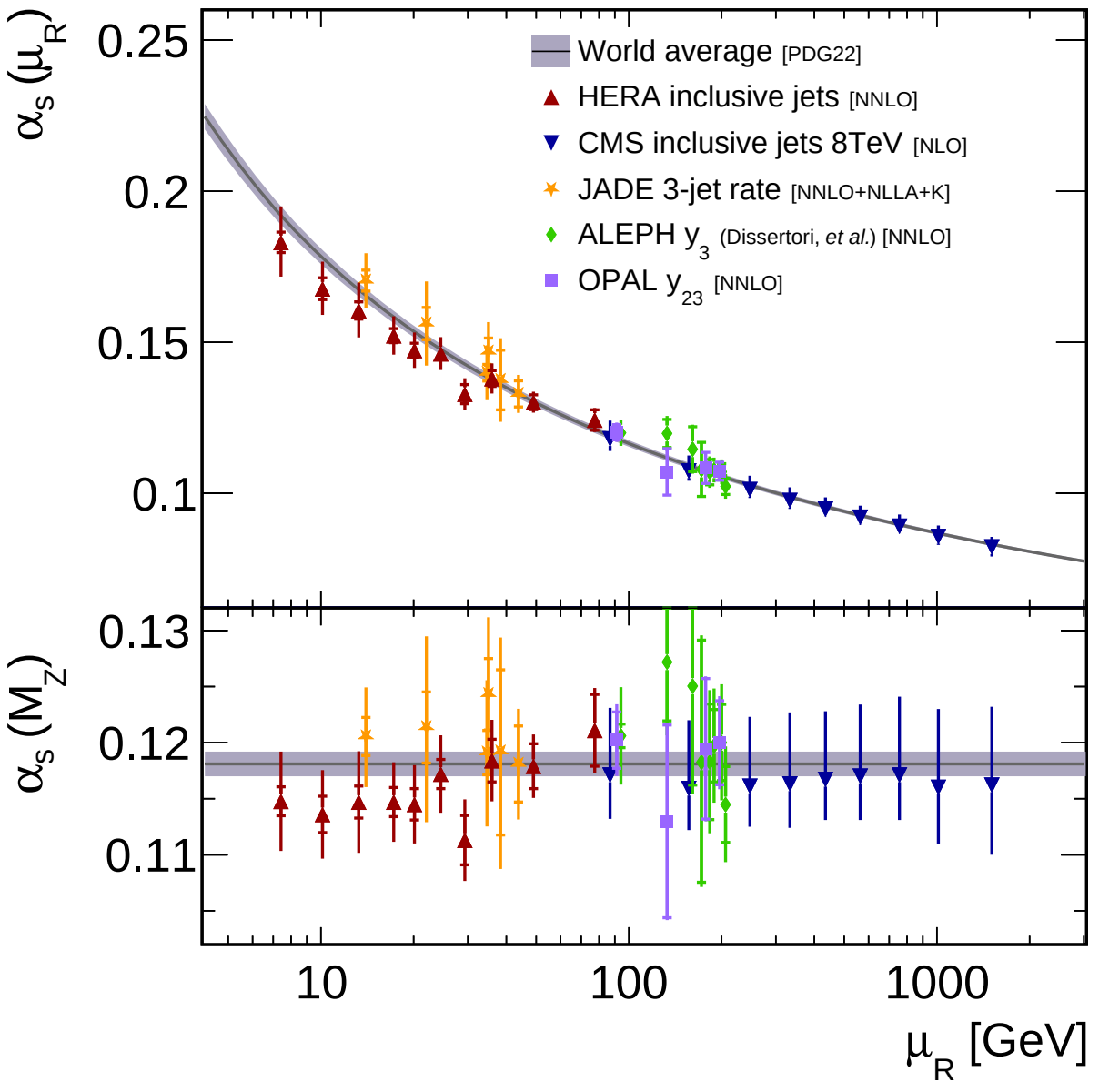

**Fig. 12.2.4** Tests of the running of the strong coupling from HERA and CMS inclusive jet cross section data using NNLO or NLO pQCD predictions, respectively. The data are compared to the expectation from QCD and measurements of jet-rates in $e^+e^-$. The lower panel displays the respective value of $\alpha_S(M_Z)$ for the representative value $\mu_r$ of the data.

Thus, the HERA inclusive jet data improve significantly over measurements from the JADE experiment in a similar region of $\mu_r$, and with its unique data bridges the gap between low-scale determinations of $\alpha_S$ from $\tau$-decays and the precision measurements at the $Z$-pole in $e^+e^-$ collisions.

### 12.2.7 Highest-$p_T$ jets at the LHC

From early exploratory up to the latest results, jet measurements have accumulated numerous successes: the gluon discovery at PETRA, the confirmation of the gauge structure of QCD at LEP, or the running of the strong coupling constant at HERA. So what is in store with the next-to-next hadron-hadron collider, the LHC? After 25 years from first concepts discussed in 1984, cf. Ref. [3730], up to first collisions at the LHC in 2009, and a similar timespan between the availability of NLO calculations for jet production in hadron-hadron collisions in 1989/1990 [3679, 3731] and the arrival of NNLO predictions in 2017 [3362] we are now in a much better position for precision comparisons. The dependence of the NNLO predictions on the choice of the renormalization scale is significantly reduced as compared to NLO. The required proton PDFs have much smaller uncertainties and were determined from a lot more and more accurate data in a more systematic way that considers and provides systematic uncertainties.

The modern experiments at the LHC deliver more precise data than at any other hadron-hadron collider before and include correlations as well as the full decomposition of systematic uncertainties. Figure 12.2.5 provides an overview of data-theory comparisons for the inclusive jet cross section versus jet $p_T$ as measured at the LHC and previous hadron colliders. Overall, the description of the data at various center-of-mass energies and covering many magnitudes in inclusive jet cross section and jet $p_T$ is excellent. Figure 12.2.6 summarises such measurements at the LHC in the form of a total inclusive jet cross section within a suitably defined fiducial phase space as a function of $\sqrt{s}$.

Despite the great success of pQCD for the description of jet data, a few concerns in particular on the theory side still persist. The scale dependence is just a proxy to estimate the effect of missing higher orders (MHO) and can be misleading if not combined with other insights into the process of interest like the relative sizes of the higher-order corrections or the absence of new process types at a given perturbative order. A newer approach [3755] makes use of Bayesian models assuming a specific behaviour of the coefficients of the perturbative series to estimate MHO uncertainties with the advantage that a proper description in statistical terms like credibility intervals becomes possible. Newer work in this direction can be found in Refs. [3756–3758], while Ref. [3759] follows a different technique to approximately complete the perturbative series. With respect to PDFs this uncertainty of purely theoretical nature only starts being considered in fits and the corresponding uncertainties [3044, 3760, 3761]. Another point of concern, which limits the precision of phenomenological analyses, is related to the uncertainties of non-perturbative effects, which are important specifically for small transverse momenta. Currently, they are "guesstimated" in a similar manner as PDF uncertainties 25 years ago, i.e. essentially the predictions by a number of MC event generators and their model parameter tunes are compared without systematic account of potential biases or correlations.

With the data from the LHC, it became possible for the first time to probe QCD and the running of the strong coupling from 100 GeV up to the TeV scale as shown in Fig. 12.2.4 using CMS inclusive jet data at $\sqrt{s} = 8$ TeV from Ref. [3727]. Notably, beyond 1 TeV of jet $p_T$, electroweak effects become significant and must be considered. Also, in a search for new phenomena with the so-called dijet angular distribution $\chi = \exp(|y_1 - y_2|)$ it was found that small deviations at low $\chi$ for dijet masses beyond 2 TeV could be accommodated by electroweak corrections [3762]. Otherwise such deviations from a mostly flat behaviour that is expected from Rutherford-like parton-parton scattering could again be an indication for contact interactions as an expression of new phenomena at a scale $\Lambda$. Similarly, excesses at large jet $p_T$ like the one by CDF discussed in Section 12.2.5 have to be considered carefully to avoid premature conclusions on new phenomena, or, much worse, fitting away first signs of new physics by absorbing them into PDFs! Again Ref. [3687] provides advice: "Note that once data is used in the PDF fit, it cannot be used for other purposes: specifically, setting limits on possible physics beyond the standard model. In that case, one should fit the PDFs and the new physics simultaneously." In the latest publication on inclusive jet production at $\sqrt{s} = 13$ TeV [3747] the CMS Collaboration performed such a fit in the framework of the effective field theory-improved standard model (SMEFT), where a perturbative coefficient $c_1$ representing potential contact interactions was used as a free fit parameter. It was found that the data are well described by the standard model alone and the $c_1$ coefficient was compatible with zero. A modification of the gluon PDF as before was not required as shown in Fig. 12.2.7. Once, it has been assured that new LHC jet data are consistent with the standard model, they can be used in combination with HERA data to simultaneously extract PDFs and the strong coupling constant at NNLO to

$$\alpha_S(M_Z) = 0.1170 \pm 0.0014\,(\text{exp}) \pm 0.0011\,(\text{theo})\,. \tag{12.2.8}$$

Also data from multiple reactions can be combined in PDF determinations as recently demonstrated by the ATLAS experiment [3763]. Yet, the best results of the LHC run 2 are still to come, since the data recorded from 2015–2018 are still in preparation by the collaborations for final calibration and publication.

### 12.2.8 Final words

The presented article tries to recount the story of jet measurements and their relevance for QCD. Specifically, we addressed what has been learned in the course of time from the interplay between theory and measurement at the highest jet $p_T$ available at each moment in time. We have selected a few key measurements for this purpose from a plethora of results achieved at the various colliders. For a more complete overview other sources may be consulted [3764, 3765].

For the future, of course, we expect to see more precise jet measurements at even higher jet $p_T$ with corresponding studies of their impact on searches for new phenomena, the running of the strong coupling, or the proton structure. Before concluding, we would like

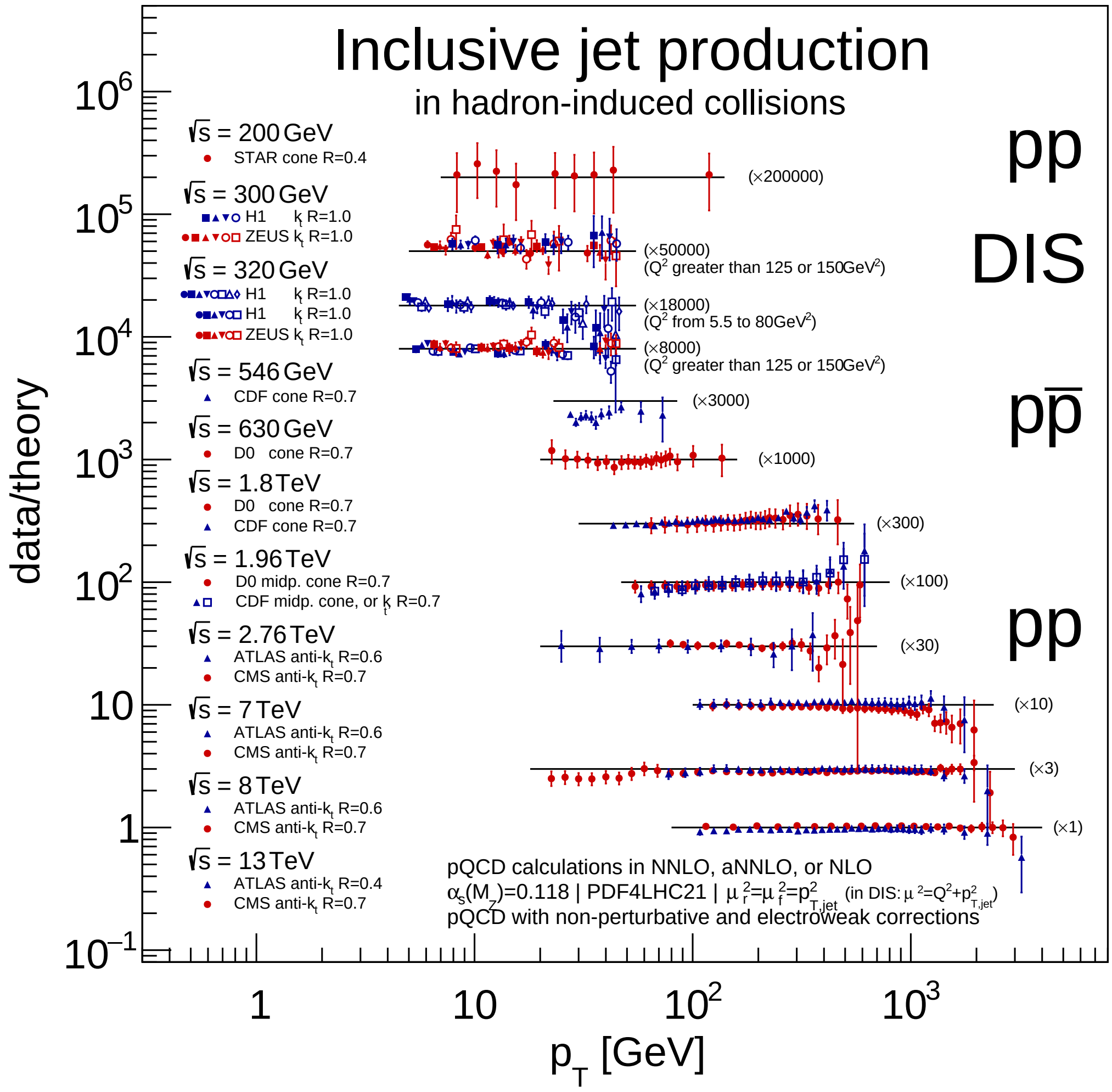

**Fig. 12.2.5** Ratios of cross-section measurements to predictions in perturbative QCD for inclusive jet production at central (pseudo-)rapidity as a function of the jet $p_{\mathrm{T}}$ or $E_{\mathrm{T}}$. The data were taken in $pp$, $p\bar{p}$, or $ep$ collisions by the ATLAS, CDF, CMS, D0, H1, STAR, and ZEUS experiments, at the RHIC, HERA, Tevatron, and LHC colliders [3696, 3716, 3718, 3721, 3727, 3732–3747]. From data available for multiple jet algorithms and/or distance parameters one particular choice has been made as indicated. The vertical error bars indicate the total experimental uncertainty of the data. The pQCD predictions are derived using the PDF4LHC21 PDF set [3748] for a value of $\alpha_S(M_Z) = 0.118$ at NNLO in QCD [3362, 3463, 3722, 3723, 3725, 3749–3751] unless indicated otherwise. The renormalization and factorization scales $\mu_r$ and $\mu_f$ are identified with $p_{\mathrm{T}}$ at hadron-hadron colliders, and $\sqrt{Q^2 + p_{\mathrm{T}}^2}$ in DIS. The predictions for $p\bar{p}$ are only in NLO QCD supplemented with 2-loop threshold corrections ($aNNLO$) [3314, 3752–3754], since most of the jet algorithms are IRC-unsafe. For STAR, the predictions are at NLO QCD only. The pQCD predictions are complemented with correction factors for non-perturbative and electroweak effects where applicable.

to point out explicitly three developments that might change how future analyses will be performed.

First, not only gluons can be radiated in large numbers by a (color) charge, but also photons by electric charges. So whenever comparing electrons in the final state to predictions including radiative corrections, one has to account for the effect that calorimeter cells add up the energies of *e.g.* an electron and all surrounding photons hitting the same cell. To avoid a potential mismatch between what experimentally is considered an electron and what is calculated in theory, one can define a cone around the electron and include all photon-like objects into the definition of the electron. This is then called a *dressed* electron or, more generally, a *dressed* lepton, since the same concept can be applied to muons, although the latter radiate less and are measured predominantly in tracking detectors. Essentially, this is again a kind of jet algorithm, but applied to leptons as primary particles [3766], raising the question "What is not a jet?".

Secondly, enormous technical progress not only allows us to produce jets at unprecedented transverse momenta of several TeV instead of GeV, we can also measure with much better precision such high-$p_T$ jets of order hundred or more tracks and clusters. This is especially important, since high-$p_T$ jets may not only be categorized into quark- or gluon-initiated jets, but also into *boosted jets* meaning that such jets may additionally contain the whole decay chain of massive boosted objects from either standard model $W$ and $Z$ bosons, and top quarks up to new hypothetical particles. A whole new field of QCD-focused analyses has been opened up here looking in detail into the substructure of jets asking the question “What is in a jet?”.

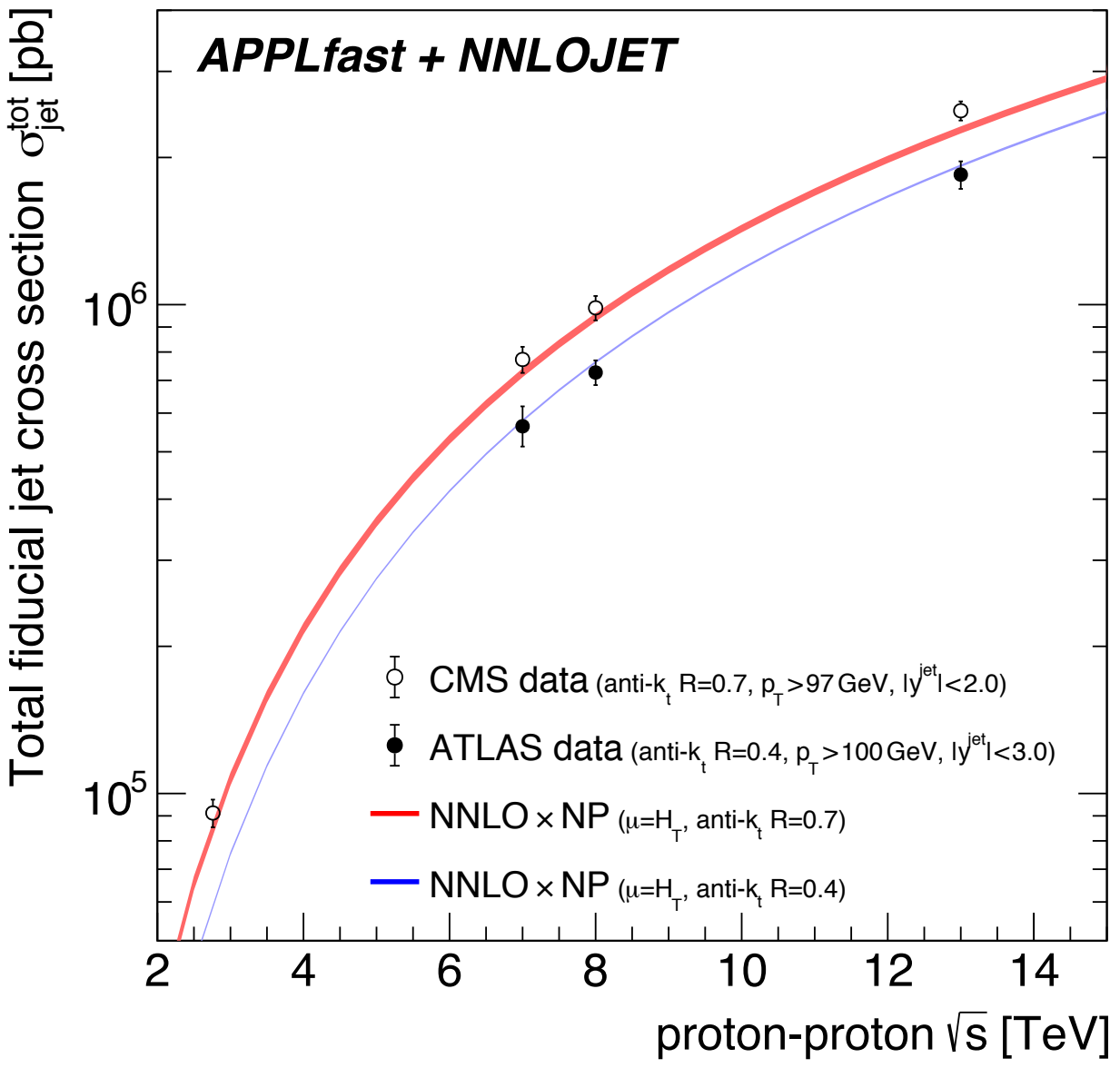

**Fig. 12.2.6** The total jet cross section as a function of the $pp$ center-of-mass energy for anti-$k_t$ jets with $R = 0.4$ and $0.7$. The predictions are compared to data from ATLAS ($R = 0.4$) and CMS ($R = 0.7$). The size of the shaded area indicates the scale uncertainty. Figure taken from Ref. [3463].

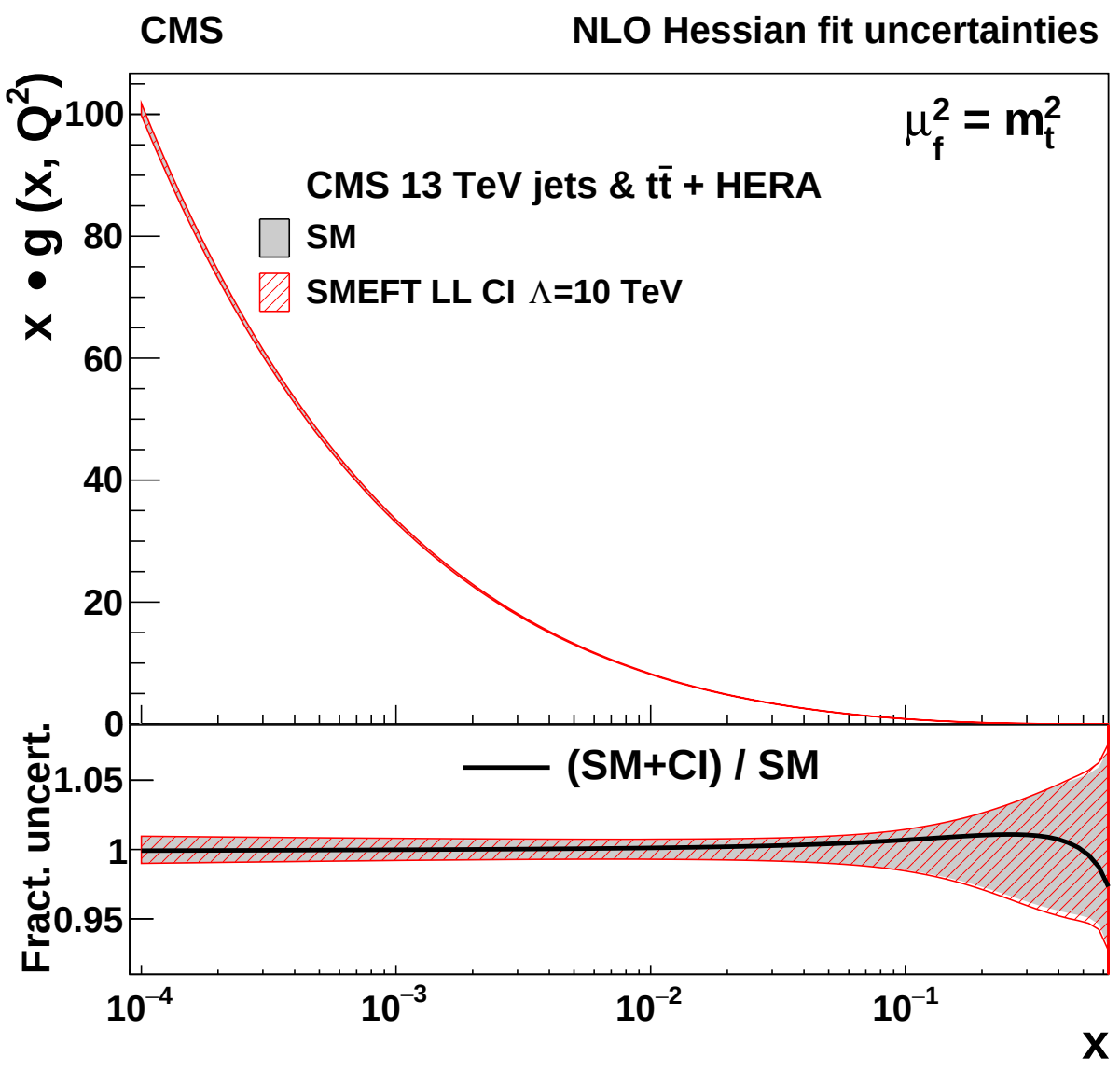

**Fig. 12.2.7** The gluon distribution as a function of the proton fractional momentum $x$ resulting from fits with and without contact interaction terms to HERA DIS data combined with data from $t\bar{t}$ and inclusive jet production as measured by CMS. The SMEFT fit here is performed with the left-handed CI model with $\Lambda = 10$ TeV. The gluon distribution is shown at a factorization scale $\mu_f$ chosen to be the top quark mass ($\mu_f^2 = {m_t}^2$). Figure taken from Ref. [3747].

Finally, progress in computing technology enabled large-scale application of neural network techniques and machine learning methods to jet physics and jet substructure. For order hundred and more jet components with kinematic properties and other chracteristics, deep learning techniques allow us to study all available information in its high dimensionality. This development has considerably increased the discrimination power among different jet types, and has the potential to genuinely improve our understanding of perturbative QCD, cf. for example the review in Ref. [1789].

This article certainly is not the final word on jets and QCD. It just starts the next decade.

## 12.3 Vector boson + jet production

### Monica Dunford

Measurements of single vector boson production in association with jets ($V$+jets production) play a central role in particle physics as they are sensitive probes to several different aspects of the Standard Model. With these measurements the predictions of perturbative QCD can be tested in new regions of phase space and with small statistical and systematic uncertainties. In many places, the experimental accuracy is better or comparable to that of the theoretical predictions. The studies of $W$ and $Z$ boson production with additional jets are sensitive tests of the dynamics of higher-order diagrams in the QCD predictions, of models of heavy-flavor production and of parton density functions (PDFs). These measurements are used to test the accuracy of the wide range of theoretical models available. This is especially important since $W$ and $Z$ boson productions are dominant backgrounds to measurements and to search for New Physics. Accurate simulations are necessary for everything from the calibration of the detector to modeling of the signal process of interest. Measurements of jets in $V$+jets production is one of the main processes used for simulation, defining event parameters (tuning), and the validation of the theoretical model. Excellent knowledge of QCD-related variables is also critical for precision measurements at hadron collider, such as measurements of the $W$ boson mass, which rely upon accurate modeling of the $W$ boson $p_T$ spectra.

### 12.3.1 Results from S$\bar{p}p$S and the Tevatron

The $W$ and $Z$ vector bosons were both discovered in 1983 by the UA1 and UA2 experiments at the Super Proton Synchrotron (S$\bar{p}p$S) at CERN. By today's standards, the number of vector boson events collected was miniscule; the UA1 detector for example collected 240 $W \to e\nu$ events and 57 $W \to \mu\nu$ events at a center-of-mass energy of 0.630 TeV [3767]. The data from these detectors permitted first tests of QCD in vector boson production. One of the immediate conclusions drawn from the data was that higher-order QCD corrections such as gluon radiation from an initial-state quark or anti-quark are needed to explain events where the vector boson has large a momentum in the transverse plane ($p_T$).

Since the dominant production of V+jets at the S$\bar{p}p$S collider is due to gluon radiation from an initial-state quark or anti-quark, these events are an ideal sample of gluon-initiated jets. Using $W$+1-jet events, measurements of the angular distribution of the jet are consistent with the expected bremsstrahlung-like radiation [3767]. In addition, the spin of the gluon was measured via the polarization of the $W$ boson. When a scalar gluon is radiated by an incoming quark or anti-quark, the helicity of the quark will be changed since the axial coupling is not conserved. In contrast, in the case of a vector gluon which conserves helicity, the quark's helicity will be preserved. The two cases lead to different polarizations of the $W$ boson. Although the gluon spin was determined at PETRA [3768] and using two-jet events at UA1 [3769], this test was an important confirmation that the gluon has a spin of one. Finally, the value of the strong coupling ($\alpha_s$) was determined by measuring the ratio of the number of $W$+1-jet events to $W$+0-jet events [3770]. Although the precision of these measurements could not compete with contemporary results from electron-positron colliders [3771], they verified that the value of $\alpha_s$ for events where a gluon is radiated in the initial state is consistent with other measurements.

The Tevatron collider, which ran at center-of-mass energies of 1.8 TeV and 1.96 TeV ushered in the era of large data samples of $W$ and $Z$ boson events and of increasing sophistication of the theoretical predictions used to describe that data. Since $V$+jets production is a dominant background to $t\bar{t}$ measurements and searches for the Higgs boson, the focus of the measurements shifted away from tests of the properties of QCD, such as $\alpha_s$ measurements, to tests of the dynamics of $V$+jets events. The large data samples collected by the CDF and D0 experiments allowed for measurements of $W$ and $Z$ boson production with up to four associated jets [3772]. Studies from the CDF and D0 experiments were expanded to include, for example, measurements of the differential cross sections as a function of the transverse momenta and rapidities of the jets, the angular separation of the two highest energy jets and the transverse momentum of the $Z$ boson.

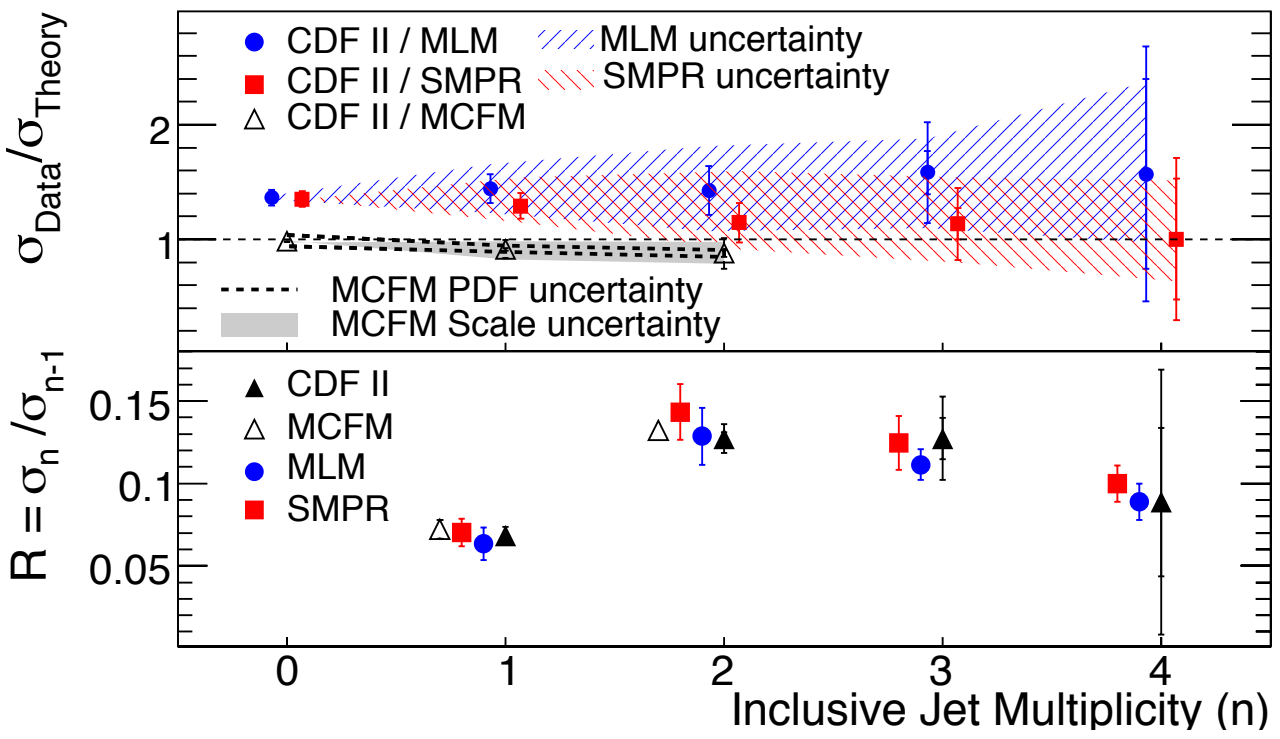

**Fig. 12.3.1** CDF [3773]: The top panel shows the ratio of data to the predictions for the cross section of $W$+jets production for different inclusive jet multiplicities. The bottom panel shows the ratio of the cross section for $n$ jets to $(n-1)$ jets. The NLO predictions (MCFM) are shown by the open triangles and the LO predictions (MLM, SMPR) are shown by the blue circles and red squares. The uncertainties on the data are indicted by the error bars, where the inner bars are the statistical uncertainty and the outer bars are the total uncertainties. The uncertainties on the predictions in the top panel are indicated by the hashed lines.

To describe these data, increasing sophisticated theoretical predictions were developed. The experimental and theoretical status at the time is nicely summarized in Figure 12.3.1, which compares a next-to-leading-order (NLO) calculation and two leading-order (LO) calculations to the data. The LO calculations, which included multiple partons in the matrix-element calculations, are able to describe the shape of many kinematic distributions up to an overall normalization factor for high numbers of associated jets but are plagued by large uncertainties. In contrast, the theoretical uncertainties for the NLO calculation are much improved but the predictions do not extend over the full kinematic range of the data. For many years this figure represented the state-of-the-art in theoretical predictions for $V$+jets production.

The large $W$ and $Z$ boson data samples produced at the Tevatron also allow detailed studies of vector boson production in association with heavy-flavor jets, where heavy-flavor jets refers to $c$- or $b$-quark initiated jets. These measurements are extremely important as these events provided the largest background contribution to measurements of $t\bar{t}$ production and searches of the Higgs boson via $WH(H \to bb)$ production. From the CDF and D0 collaborations, measurements of $W$ production in association with a charm quark and $W$ and $Z$ production in association with $b$ quarks were performed [3772]. One most notable result is the first measurement of $W + b$-jets production, which was done by the CDF collaboration, the measured cross section is 2.5–3.5 times larger than the various predictions with significance of 2.8 standard deviations. While the theoretical predictions used in this comparison did not fully account for $b$ quarks in the initial state, this is not expected to explain the difference. The data sample itself was too small to allow measurements of kinematic distributions to resolve the source of the discrepancy.

In summary, the experiments at the S$\bar{p}p$S and the Tevatron colliders provided important tests of QCD theory in $V$+jets production. However, the scope of these measurements, with the exception of Ref. [3774] focused largely on measurements of the cross section for different jet multiplicities and a handful of differential cross section measurements. These measurements are important in validating QCD theory for topologies with multiple low energy jets, where the highest jet energies are not much greater than the mass of the vector boson itself. Rare processes such as $W$+$b$-jets production were measured for the first time but the statistical precision of the data samples is not sufficient to probe the kinematic distributions of these events.

### 12.3.2 V+jets at the LHC

In $V$+jets production at the LHC, measurements of jets with a transverse momentum greater than 1 TeV, which is much beyond the mass of the vector boson, are now accessible. At these high energies, the calculations from perturbative QCD suffer from large logarithmic corrections and are themselves potentially unreliable [3775]. With the large data samples available from the LHC, we have entered an era of high precision differential measurements with which we can explore QCD at higher-orders and in extreme corners of the phase space. For the first time, we also have sufficient data samples to measure in detail heavy-flavor production in multiple differential distributions. All of these measurements also provide for better understanding of the PDFs.

In pace with the increase in data samples, a plethora of new, more precise theoretical predictions, all with slightly different focuses, exist today for $V$+jets production. A more detailed summary of the available predictions can be found in Refs. [3355, 3356]. In addition to LO matrix-element calculations, NLO calculations matched to parton shower models are now available; most notable for $V$+jets production are SHERPA, MADGRAPH5-AMC@NLO, MC@NLO and MEPS@NLO. NNLO calculations with next-to-next-to-leading logarithmic resummation and with parton showering are available using GENEVA. For fixed-order calculations, NLO predictions to five jets or more are available, such as BLACKHAT-SHERPA calculations, approximate NNLO predictions for jets with up to one jet, such as LOOPSIM calculations and NNLO predictions, such as $N_{jetti}$. Another calculation, HEJ, focuses on large rapidity separation and uses a resummation method to give an approximation to the hard-scattering matrix element for jet multiplicities of two or greater; in the limit of large rapidity separation between partons, this approximation becomes exact.

### 12.3.3 Tests of higher-orders

For our theoretical understanding of particle physics to keep pace with the improved accuracy of the measurements, theoretical predictions which include higher-order corrections are indispensable. Most of the measurements and searches performed today involve very high momenta jets, leptons or large amounts of missing transverse energy. In these regions, the high-order corrections play large and vital roles.

One important variable to test contributions from higher-order corrections is the observable of $H_T$, which is defined as the scalar sum of transverse momenta of the leptons, the missing transverse energy (for $W$+jets events) and the transverse momenta of all jets passing the selection criteria. At large values of $H_T$ the average number of associated jets in the event increases. LO matrix-element calculations which do not provide higher-order terms drastically underestimate the average jet multiplicity. Here NLO predictions are needed to fully model these distributions. These distributions, among others, have been measured for both $W$+jets and $Z$+jets production [3776–3779]. Compared to the previous colliders the increase in kinematic reach at the LHC is dramatic; Tevatron results reach up to $H_T$ values of 500 GeV, while the LHC result extends to 2 TeV.

The necessity of high-order corrections can readily be seen in measurements of the balance between the $Z$ boson and the jet transverse momenta. The so-called jet-$Z$ balance (JZB) is defined as the difference between the sum of the jet $p_T$s (with $p_T > 30$ GeV and rapidity within 2.4) and the $Z$ boson $p_T$. When all hadronic activity is contained within the selected jets, the JZB variable is zero. Figure 12.3.2 shows the measured data for events with $p_T(Z) < 50$ GeV compared to a LO and NLO MadGraph predictions and the GENEVA predictions [3780]. As seen in the figure, the distribution is not symmetric, with hadronic activity more dominantly pointing in the direction of the $Z$ boson (i.e. positive values in this definition). The low $p_T(Z)$ region is interesting: While larger $Z$ boson momenta can be described by fixed-order calculations, small values require resummation of multiple soft-gluon emissions to all orders in perturbation theory [1282, 3781] (see Sections 11.1 and 11.2). Different $p_T(Z)$ regions therefore test different theoretical treatments. The NLO predictions best describe the data and indicate that the NLO corrections are important to describe not only jet emission but also the hadronic activity outside of it.

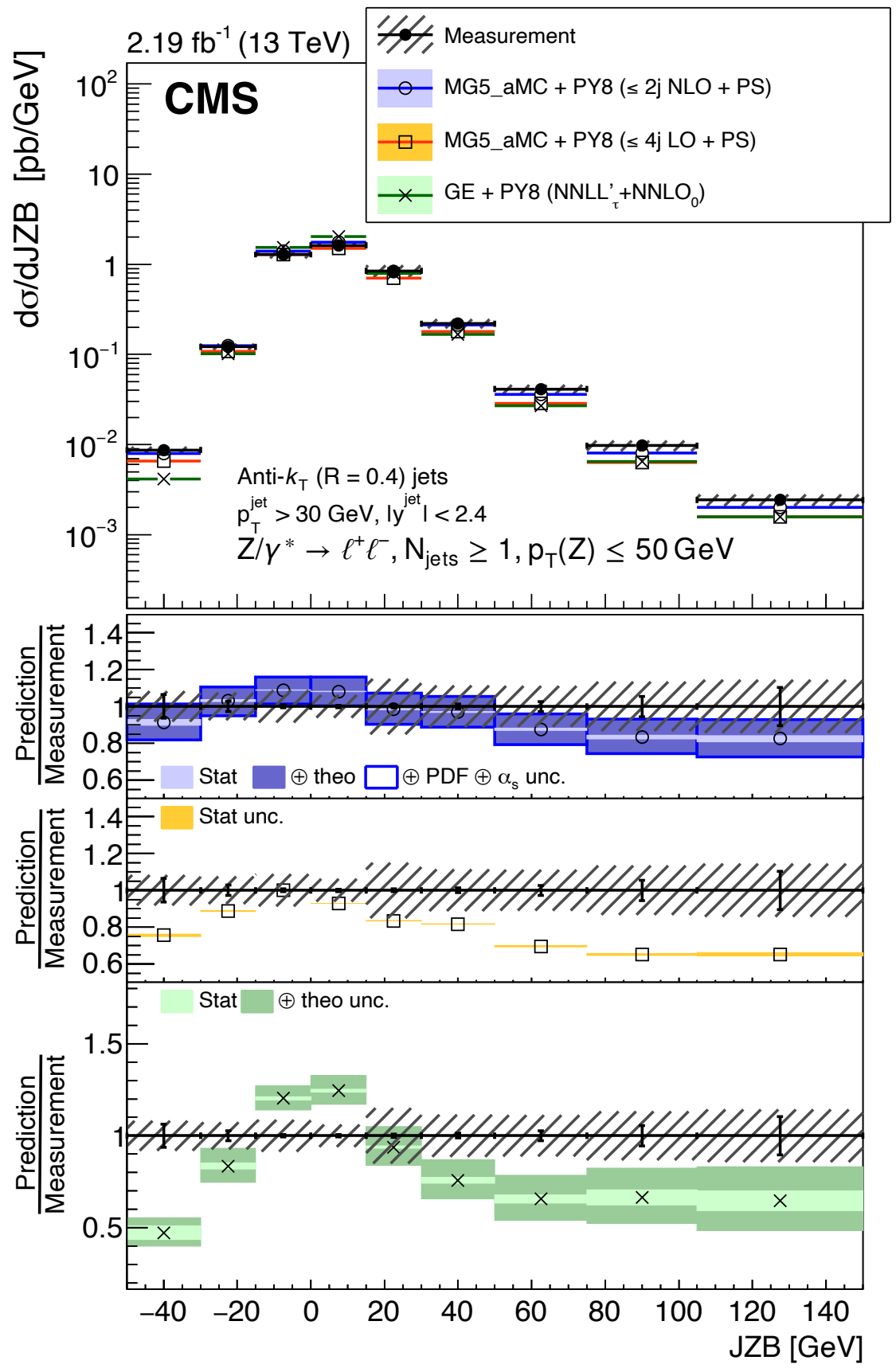

**Fig. 12.3.2** CMS [3780]: Measured $Z$+jets cross section as a function of the JZB variable for $P_T(Z) < 50$ GeV. The data are compared to LO, NLO and NNLO predictions. The lower panels show the ratio of the three predictions to the data. The error bars on the measurement represent the statistical uncertainty and the grey hatched bands represent the total uncertainty. The uncertainties included on the theory predictions are listed in the labels.

### 12.3.4 Modeling in extreme phase spaces

Extreme phase space regions, including events with high-$p_T$ jets or high boson momenta or events with small angular separation between objects in the final state, tend to be governed by a complex mixture of the number of jets contributing to the final state and contributions from QCD as well as EW processes. This makes measurements of this nature an ideal test bed for studying multiple theory approaches.

The study of $V$+jet production with small angles between the boson and jets is one such critical test [3782–3784]. At LO, $V$+1-jet production is simply described by a $V$ boson recoiling, in a back-to-back configuration with a jet. However, at NLO both real and virtual contributions to $V$+1 jet production appear via QCD and EW corrections. For real emissions of a $V$ boson, either from an initial- or final-state quark, these contributions lead to an enhancement in production that is proportional to $\alpha_s ln^2(p_{T,j}/m_V)$, where $\alpha_s$ is the strong coupling, $p_{T,j}$ is the transverse momentum of the jet, and $m_V$ is the mass of the $V$ boson. This effect becomes largest with high transverse-momentum jets and can be isolated by selecting events with small angular separation between a jet and the $V$ boson. In this region, the cancellation between real and virtual correction is incomplete, making the region ideal to probe. However, other processes such as $V$+2-jet production also play a role in this region and must be considered.

To study these effects, the ratio (so-called $r_{Z,j}$) of the $Z$ boson $p_T$ to the closet-jet $p_T$ is defined. For collinear radiation of a $Z$ boson, where the $Z$ boson is expected to have a lower transverse momentum, this ratio results in smaller values. Figure 12.3.3 shows the $r_{Z,j}$ distribution for events where the angular separation, $\Delta R$, between the jet and the $Z$ boson is less than 1.4, corresponding to the region where the $Z$ boson has a small angular separation from the jet (the collinear region) [3782]. While regions with back-to-back topologies (not shown here) are better modeled by predictions, the figure shows that most predictions fail to model the collinear region.

There are many other examples of tests in extreme regions including specific tests to isolate matrix element and parton shower effects [3785], measurements of probability of producing an additional jet in a rapidity gap of two jets [3786–3791], measurements in the forward region [3792, 3793] and tests of QCD with sensitivity to physics beyond the Standard Model [3794, 3795]. All of these measurements are critical for understanding QCD in these difficult-to-model regions.

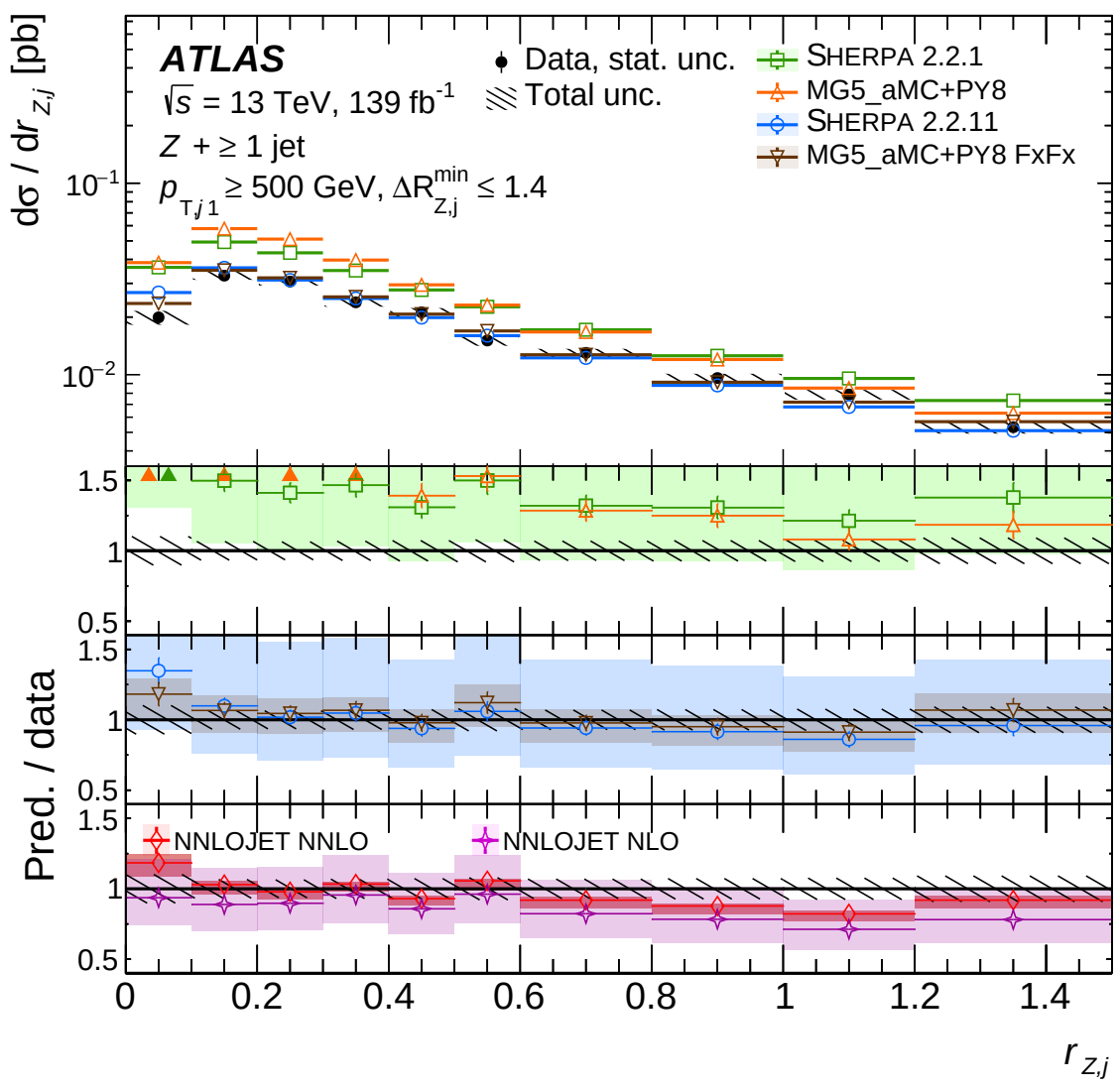

**Fig. 12.3.3** ATLAS [3782]: $Z$+jets cross section in the collinear region as a function of the $r_{Z,j}$ variable. The data are compared to LO and NLO predictions. The lower panels show the ratio of the predictions to the data. The uncertainties shown on the data include the statistical uncertainty, as indicated by the error bar and the total uncertainty, as indicated by the hatched region. The uncertainties on the predictions are indicated by the colored bands.

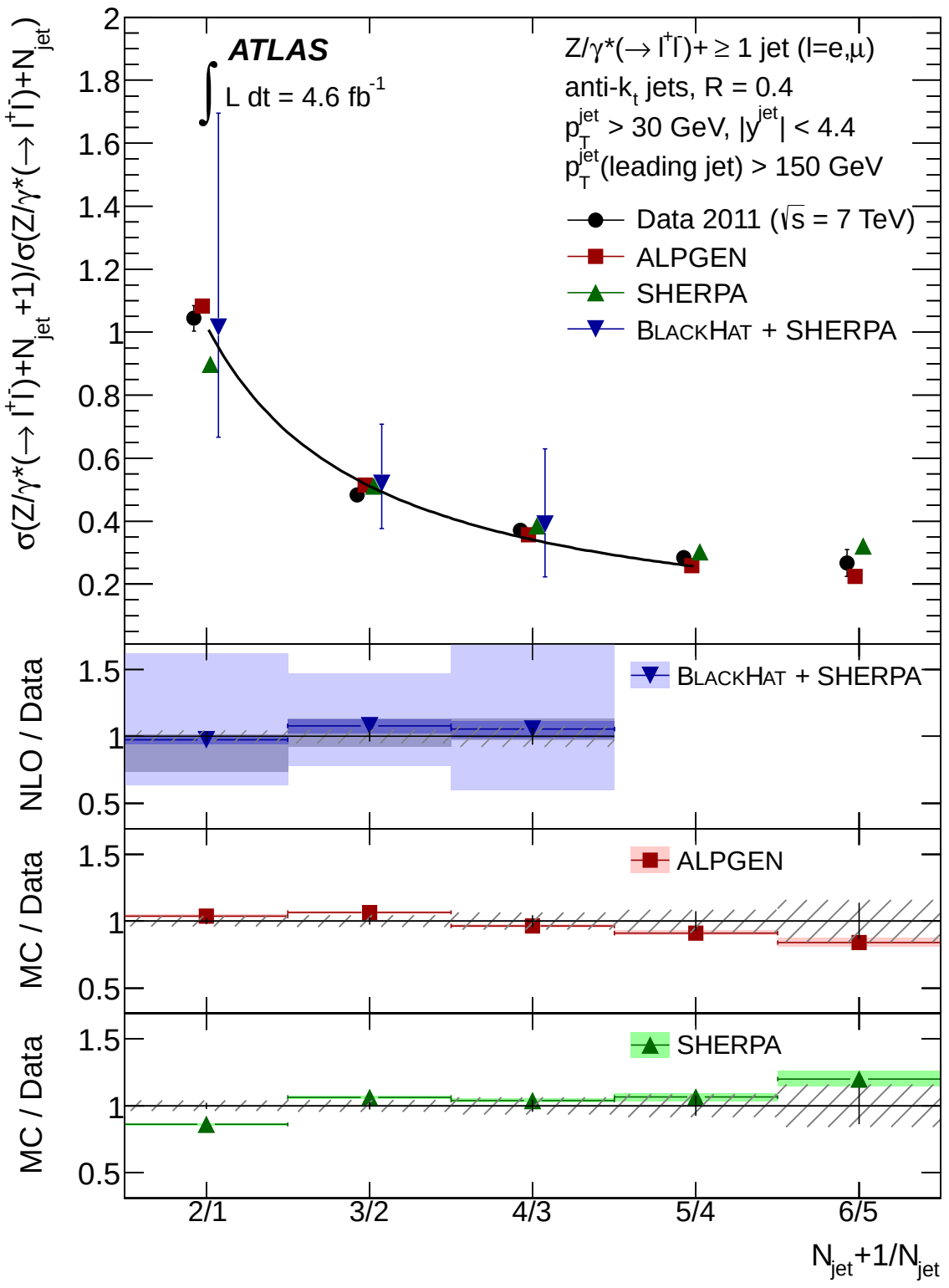

**Fig. 12.3.4** ATLAS [3797]: Measurement of the ratio of the exclusive $n$-jet and $(n+1)$-jet cross sections for events where the highest $p_T$ jet must have an energy above 150 GeV. The data are compared to the predictions from BLACKHAT-SHERPA, ALPGEN and SHERPA. The lower panels show the ratio of the three predictions to the data. The error bars indicate the statistical uncertainty on the data, and the hatched bands the statistical and systematic uncertainties on data. Uncertainties on the theory predictions are statistical only except for those of BLACKHAT-SHERPA.

### 12.3.5 Tests of QCD emissions

As demonstrated by the results from the UA1 and UA2 experiments, radiation of additional quarks and gluons is necessary in order to describe the events where the vector boson has a large transverse momentum. These higher-order QCD corrections consist of two classes; terms with a virtual loop which do not directly affect the boson $p_T$ and terms with a real emission which do so but result in a jet which could be recorded by the detector. Measurements of the V+jet cross section for each jet multiplicity is therefore a direct test of QCD radiation. Measurements of the jet multiplicity ratios at the Tevatron and then at the LHC showed a striking feature: the ratio of the $n$-jet cross section to the $(n+1)$-jet cross section is a constant. This implies that the $V$+jet cross section follows the form

$$\sigma^{LO}_{V+n-jet} \sim \sigma_0 e^{-an} \tag{12.3.1}$$

with $a$ is an experimentally-determined constant values and depends on the exact definition of the jets and $\sigma_0$ is representing the zero-jet exclusive cross section.

Although the constant scaling is a well established experimental observation, this behavior is not a priori expected [3796]. Assume, for example, that the probability of radiating a gluon from a quark follows the theory of Quantum Electrodynamics (QED), such that the gluon cannot radiate another gluon. In this case, the probability of radiating a gluon is dictated by a Poisson distribution, which implies that the cross section for an $n$-jet exclusive final state is

$$\sigma^{LO}_{V+n-jet} \sim \frac{\bar{n}e^{-\bar{n}}}{n!}\sigma_{tot}\,, \tag{12.3.2}$$

where $\sigma_{tot}$ is the total cross section and $\bar{n}$ is the expectation value of the Poisson distribution, which also depends on the exact definition and selection of the jets. However, the gluon follows the non-abelian QCD theory and can radiate an additional gluon from itself. Therefore, at higher jet multiplicities the scaling would become constant.

The observation of a constant scaling for all jet multiplicities is instead a subtle cancellation of two different and opposite-sign effects. At low jet multiplicities, the Poisson scaling is present but cancelled by effects from the PDFs. To understand this effect, consider the case of high jet multiplicities with a cross section ratio of $n$-jet events to $(n+1)$-jet events, where $n$ is a large number of jets. Here, the parton momentum fraction, $x$, for the involved partons is similar between the two jet multiplicities and therefore any effects on the cross section due to the PDFs essentially cancel in the ratio. In contrast, at low jet multiplicities, the relative difference in $x$ for the involved partons between 2-jet events and 1-jet events is larger. Due to the steeply falling $x$-distribution of the gluon PDF, this implies that the production of 2-jet events compared to 1-jet events is suppressed by the PDFs. Depending on the exact selection criteria, this suppression cancels the increase in the production cross sections, which arises from the Poisson scaling.

Based on the work of Ref. [3796], the Poisson nature can be seen directly by selecting events with one very energetic jet. In these events, the effect on the cross sections from the PDFs is reduced; as seen in Figure 12.3.4, the jet multiplicity cross section follows the expected Poisson distribution. This measurement is a nice validation of the nature of QCD emissions from first principles.

### 12.3.6 Differential heavy flavor results

The associated production of vector bosons with heavy flavor jets is an important precision test of perturbative QCD in the presence of two mass scales - the vector boson mass and the $c$- or $b$-quark masses. Measurements of this nature also provide critical input to charm and strange distributions inside the proton, as discussed more in the next section. At LO, heavy-flavor production stems from either a gluon in the final state splitting into a heavy-flavor quark-antiquark pair or a heavy-flavor quark produced in the initial state. At the Tevatron, corrections to the cross section from the latter contributions are small, but at the LHC, these processes can lead to corrections of up to 50% [3798–3801]. Theory predictions for heavy-flavor production consist of 5-flavor-scheme models, where the $b$-quark is included in the PDF itself or 4-flavor-scheme models, where it is not. The two schemes, however, are equivalent if the calculations included all orders of $\alpha_s$.

With the large data samples available from the LHC, these processes can be studied for a variety of differential observables [3772, 3802–3807] and also in the forward region and in phase spaces with very energetic, boosted jets [3808–3811]. In general, 5-flavor-scheme predictions are better at describing the data compared to 4-flavor ones. However, there are sizable differences even between predictions of a similar nature. Figure 12.3.5 shows the separation between the two $b$-quarks, which is a sensitive variable to gluon splitting [3802]. The NLO Sherpa simulation estimates this observable well but fails to get the overall cross section correct (not seen in this figure). In contrast, the LO and NLO MadGraph predictions are less able to model the shapes of the kinematic observables but estimate well the overall cross section. In regimes where the vector boson has a large $p_T$, the predictions generally perform worse; for example they underestimate events with high m(bb) by about half [3804]. Work is on-going to combine massless NNLO with a massive NLO computation, with promising comparisons to data [3812].

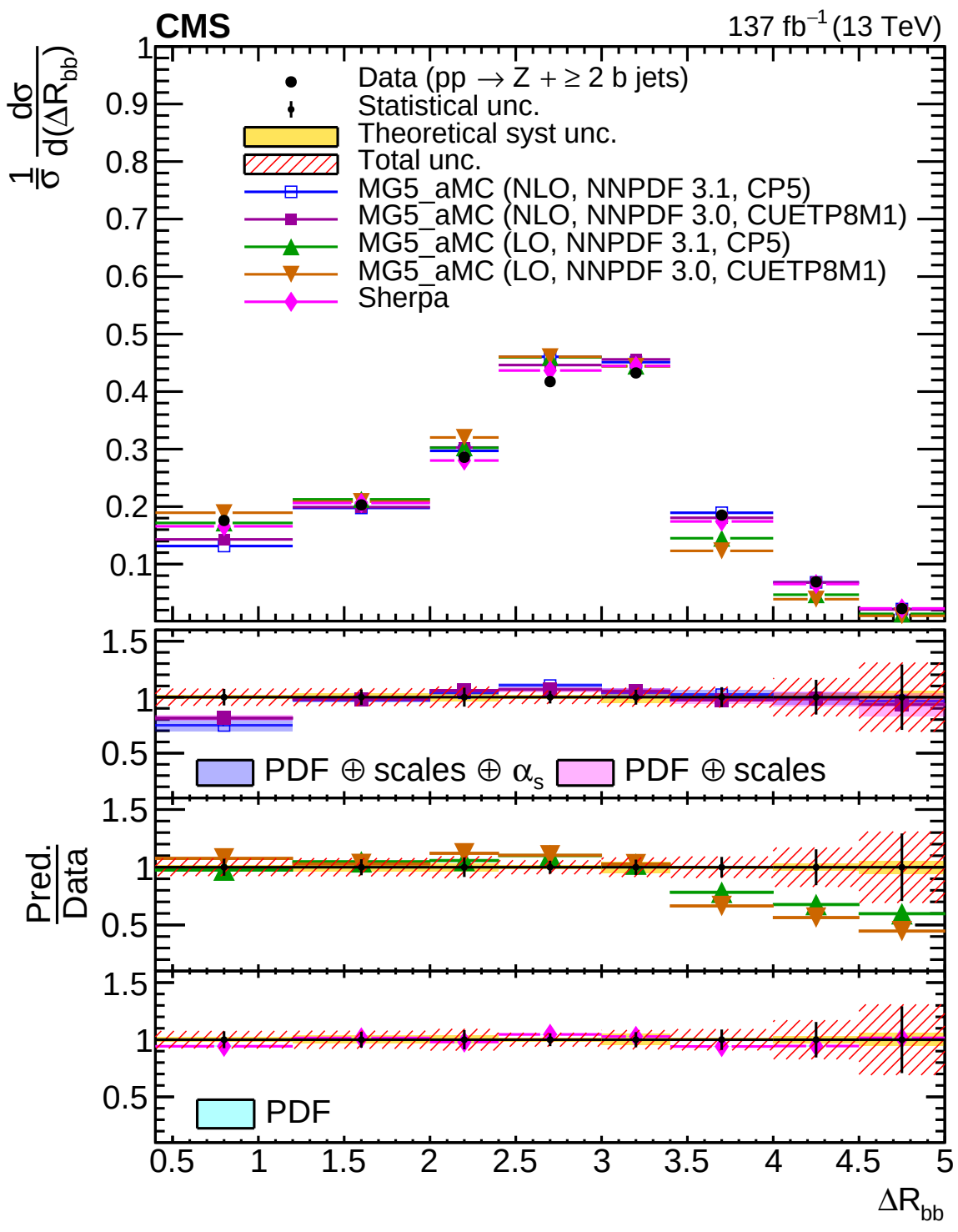

**Fig. 12.3.5** CMS [3802]: Normalized differential cross section as function of the angular separation between two $b$ jets, $\Delta R_{bb}$ for $Z+\geq 2$ $b$-jets events. The uncertainties in the predictions are shown as colored bands in the bottom panel. The statistical, theoretical, and total uncertainties in data are indicated by the vertical bars and the hatched bands.

### 12.3.7 Probes of parton density functions

A major source of uncertainty in all hadron collider measurements stems from knowledge of the PDFs. As our knowledge of QCD deepens, better knowledge of the PDFs are needed to continue to be sensitive to deviations from Standard Model predictions [3000, 3001]. Deep inelastic scattering data from the HERA experiments provided some of the best data for PDF determination over a wide range of $Q^2$ and $x$. In addition to these data, data from various experiments, such as those from neutrino and hadron collider experiments. The LHC offers a unique opportunity in that it provides a diverse set of processes, such as jet, photon, vector boson or top production, which can be used to constrain different regions with the PDFs. Today, PDFs can be determined at up to NNLO precision in perturbative QCD. The input data span the range of $10^{-5} \lesssim x \lesssim 1$ and $1 \lesssim Q^2 \lesssim 10^6$ GeV$^2$.

Measurements of $V$+jet production are particularly important since these processes can probe $u$ and $d$ quarks and also contributions from $s$, $c$ and $b$ quarks. By considering vector boson processes with additional jets, the measurements are sensitive to higher values of $x$, accessing $x \approx 0.1 - 0.3$ [3813], compared to inclusive $W$ and $Z$ measurements. Measurements of this nature constrain the light-quark sea at higher $x$ as well as the strangeness contributions and help to better understand the gluon distribution at high $x$ [3763]. The LHCb experiment, with its precision tracking coverage in the forward region, offers new possibilities here in that its $V$+jets measurements are sensitive to PDFs at different $x$ ranges compared to the ATLAS and CMS experiments [3792]. These measurements probe PDFs at $x$ as low as $10^{-4}$ and at high $x > 0.5$ [3814].

The contributions to the proton from strange quarks can be probed through measurements of vector boson production in association with $c$-quark, as was done at the Tevatron and the LHC [3074, 3772, 3803, 3810]. In past years, whether or not strangeness contributions are suppressed in the proton is a topic of debate, with mainly the ATLAS data preferring less suppression as compared to neutrino scattering and CMS data. However, with more data, in particular measurements of $W$+jets and $W+c$ production have provide powerful input on the strangeness contribution. Today, there is general agreement by modern PDFs that strangeness is not strongly suppressed at low $x$ but has substantial suppression at high $x$.

It has been a decade long debate if the proton may contain an 'intrinsic' charm component in addition to that from gluon splitting, which decreases sharply at large values of $x$ [3815]. Such models, like the BHPS model, predict that protons would have a valence-like charm content. Global PDF analyses are generally inconclusive and therefore more direct probes are needed [3083, 3816]. Since intrinsic charm contributions are enhanced in $Z$+jet production where the $Z$ boson has large rapidity, the LHCb experiment is perfectly suited for these measurements [3809]. As seen in Figure 12.3.6, the data at forward $Z$ rapidities from a recent $Z$+$c$ measurement, are consistent with an intrinsic charm contribution as predicted by BHPS models. Future analysis of the effect of these data within PDF fits themselves, however, is still needed.

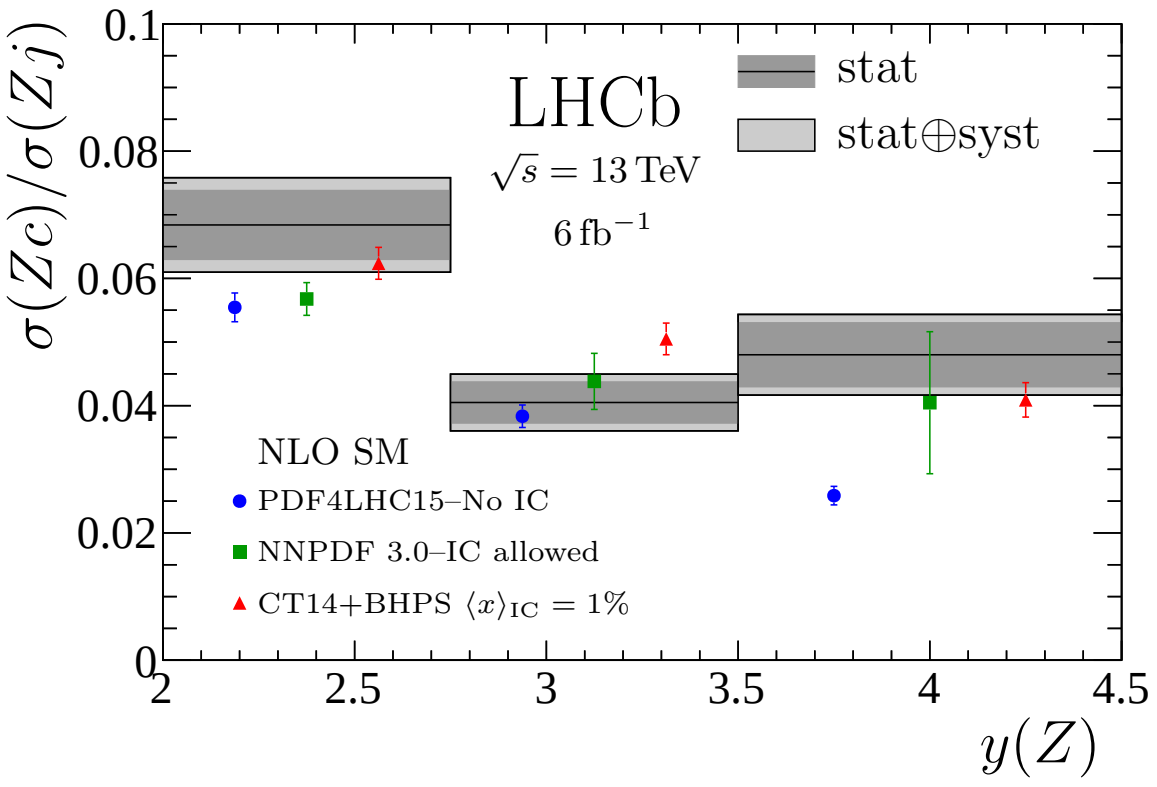

**Fig. 12.3.6** LHCb [3809]: Measured cross section ratio of $Z+c/Z$+jet production for three intervals of forward $Z$ rapidity, compared to NLO predictions with and without IC, and with IC as predicted by BHPS with a mean momentum fraction of 1%.

Since their discovery in the early 1980s, the $W$ and $Z$ bosons are important probes to understanding QCD. The early measurements at the S$\bar{p}p$S and the Tevatron were critical in establishing the dynamics of these processes, while at the LHC, $V$ + jets production is now explored at the highest available energies. To step up with experimental precision, a suite of versatile and precise theory predictions have been developed to compare to the data. Future measurements of $V$+jets production are needed to better understand QCD theory in very energetic regions of phase space, to measure electroweak corrections, to improve PDFs and for a better understanding of heavy flavor production.

## 12.4 Higgs production

**Chiara Mariotti**

In July 2012, the ATLAS and CMS Collaborations at the CERN Large Hadron Collider (LHC) announced the discovery of the last missing piece of the Standard Model (SM) of elementary particles: the Higgs boson [120, 121, 3818]. The discovery arrived about 50 years after theorists had postulated its existence to explain the mechanism by which the elementary particles acquire mass.

The Higgs boson is the excitation of a field, called Brout-Englert-Higgs (BEH) field. The field name comes from the theoreticians who first introduced the mechanism [38, 39, 3819]. The BEH field filled the entire universe less than a picoseconds after the Big-Bang. The elementary particles interacting with it acquire mass.

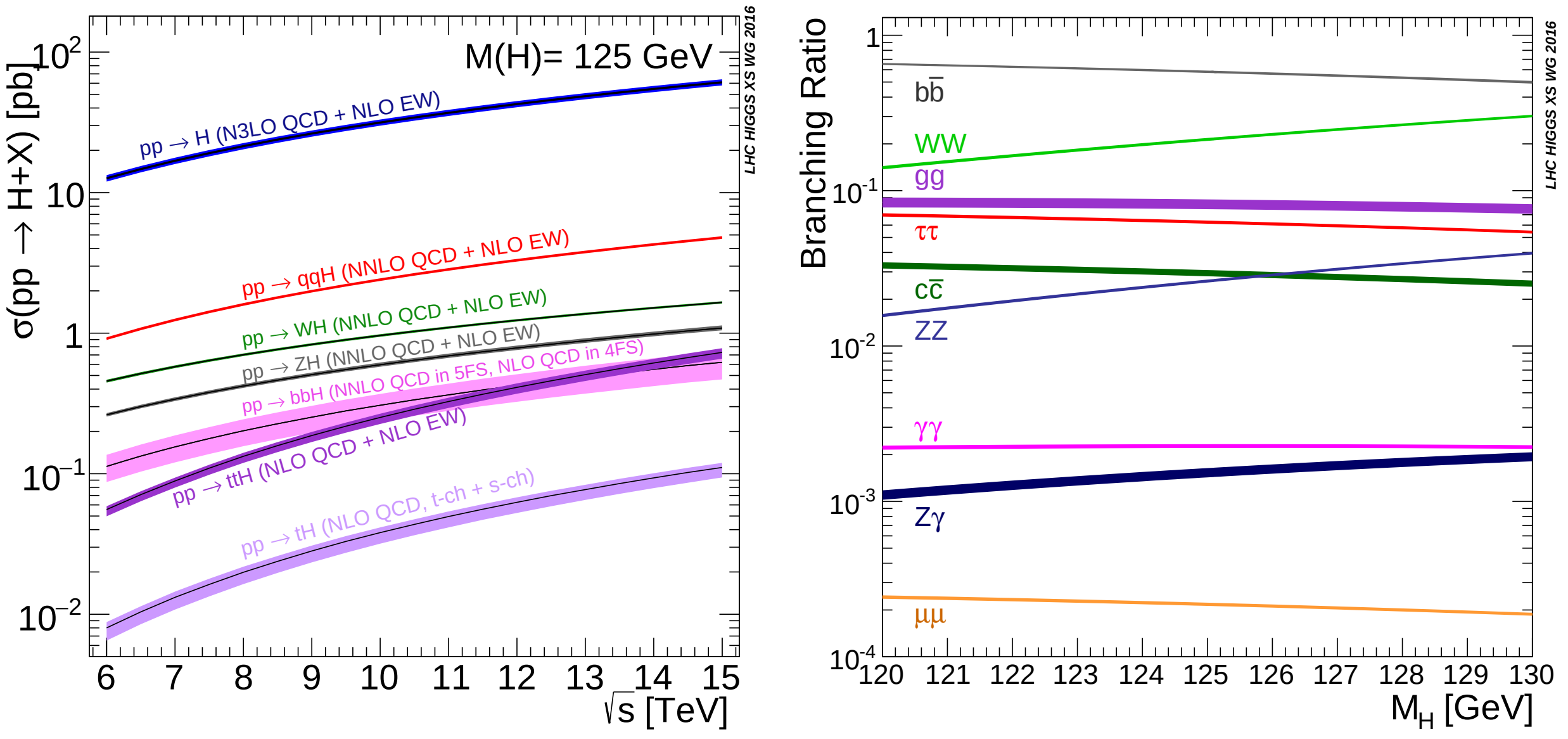

**Fig. 12.4.1** (Left) Cross section for a Higgs boson of 125 GeV of mass as a function of the center-of-mass energy at the LHC [3817]. (Right) Branching ratios as a function of the Higgs boson mass [3817].

Without this field the world would not be the same, as an example the electron would be massless and atoms could not be formed.

The Higgs boson has unique quantum numbers: $J^{PC} = 0^{++}$, since the field must be the same everywhere in the space and should not depend on the reference frame.

Since the time of the discovery, the ATLAS and CMS experiments have accumulated data during the Run 1 (2009–2012) at 7 and 8 TeV proton-proton center-of-mass energy and Run 2 (2015–2018) at 13 TeV. The two collaborations observed the Higgs boson in numerous bosonic ($ZZ$, $WW$, $\gamma\gamma$), and fermionic decay channels ($\tau^+\tau^-$, $b\bar{b}$ quark), measured its mass and width, determined its spin-parity quantum numbers, and measured its production cross sections in various modes (gluon-gluon fusion, vector boson fusion, associated production with a W or a Z, associated production with 2 top quarks). Within the uncertainties, all these observations are compatible with the predictions of the SM.

Finding the Higgs boson has been very demanding. Its production cross section is 12 orders of magnitude smaller than the proton-proton inelastic cross section at LHC energies. Few hundreds of particles are produced at each collision, and there can be several simultaneous proton-proton collisions at each proton bunch crossing (pileup). It is thus fundamental to have a very good understanding of the resonant and non-resonant hadronic background: production of background processes via QCD interactions has to be well understood and modeled.

Because of its large mass, the Higgs boson could not be discovered at LEP [3820] at CERN, and because of its very low production cross section it was very challenging to observe it at the Tevatron [3821] at Fermilab. Only at LHC, thanks to the energy available in the center-of-mass, and to the exceptionally high luminosity, it was possible to produce it with a rate sufficient to discover it.

Precise theoretical calculations for the Higgs boson production modes and decay channels have been performed; the results are shown in Fig. 12.4.1. The dominant production mode at the proton-proton LHC collider is the gluon-gluon fusion (ggF, or $pp \to H$) as shown in Fig. 12.4.1(left), followed by the vector boson fusion (VBF, or $pp \to qqH$), the associated $H$ production with vector bosons ($pp \to ZH, WH$), and the associated production with two $b$ quarks or two top quarks or just one top quark. Many groups contributed to the computation of these production cross sections over many years [3817, 3822–3824]. The perturbative order of the calculations in QCD and EW is indicated in the figure. The thickness of the line represents the uncertainty of the calculation.

The cross section of the ggF process is known at N3LO with very good precision (5% in total, of which 3% are due to missing higher order effects). The calculation of the higher perturbative orders in QCD, as well as the resummation (see sections 11.1 and 11.2), contribute substantially to the precision as shown in Fig.12.4.2 [3817]. The parton distribution functions (PDFs) have been determined with very good accuracy by sev-eral groups at NNLO in QCD and reached a precision of $\sim 2\%$ for the gluon-gluon luminosity over a wide range of energy [3817].

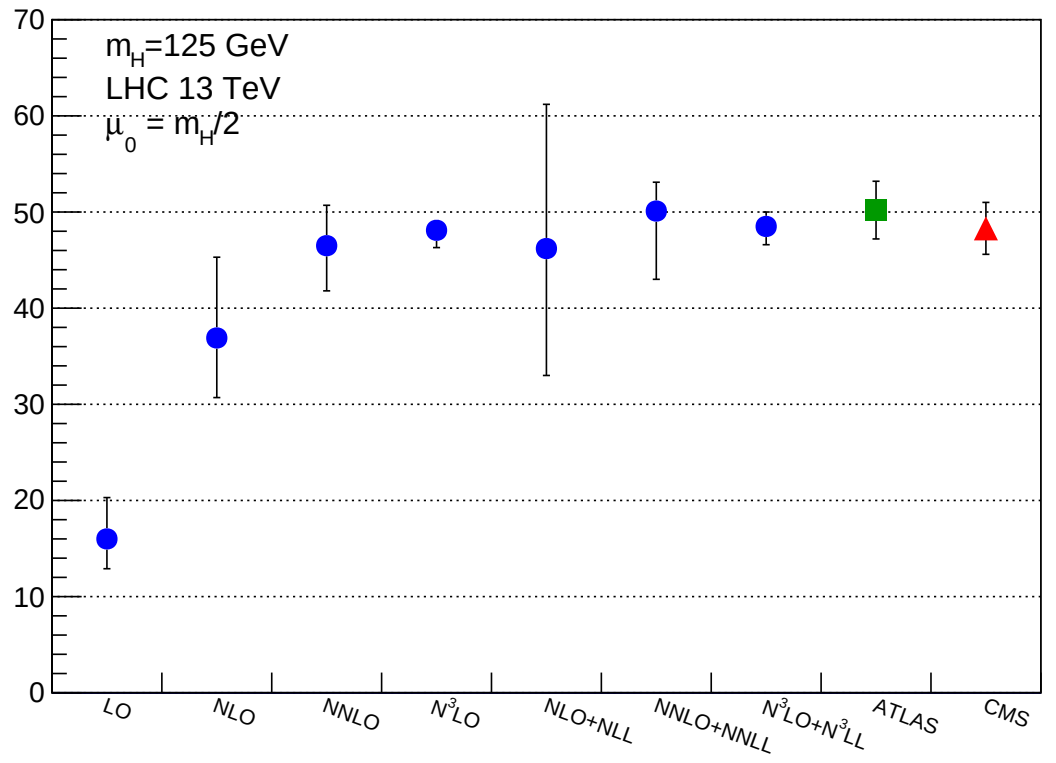

**Fig. 12.4.2** Calculated theoretical ggF cross-section values (blue circles) at various perturbation orders [3817]. The latest ATLAS (green square) [3825] and CMS (red triangle) [3826] results from Run2 are also shown.

The strength of the Standard Model Higgs boson coupling is proportional to the mass of the fermions, and to the mass squared of the vector bosons. Thus it will decay predominantly to the available elementary particle with larger mass: for a Higgs boson of $m_H = 125$ GeV, the largest branching ratio (BR) is to $b\bar{b}$, followed by $W^*W$. The various BRs have been computed at least at NLO precision for both QCD and EW corrections, and are shown in Fig. 12.4.1(right) [3817].

Calculation of the background processes for the various Higgs boson decay channels have been and are being computed with increasing precision at higher order in perturbation theory. In parallel, experiments have developed methods to estimate the various sources of background in a data-driven way, not to depend on the availability of Monte Carlo (MC) simulations, or on precise theoretical calculations and modeling.

### 12.4.1 Higgs boson properties

The ATLAS and CMS experiments, with the data collected during the Run 1 and Run 2, measured with very good precision the properties of the Higgs boson: the mass is measured with a precision better than 0.2% in the $H \to \gamma\gamma$ and $H \to ZZ \to 4\ell$ final states:
ATLAS ($H \to ZZ \to 4\ell$ final state only) [3827]:

$$m_H = 124.94 \pm 0.17 \text{ GeV}$$

CMS [3828]:

$$m_H = 125.38 \pm 0.14 \text{ GeV}\,.$$

As an example, Fig. 12.4.3 shows the diphoton invariant mass distribution targeting the study of the decay channel $H \to \gamma\gamma$ in ATLAS [3829], and the invariant mass distribution of four charged leptons targeting the study of the decay channel $H \to ZZ^* \to 4\ell$ in CMS [3828].

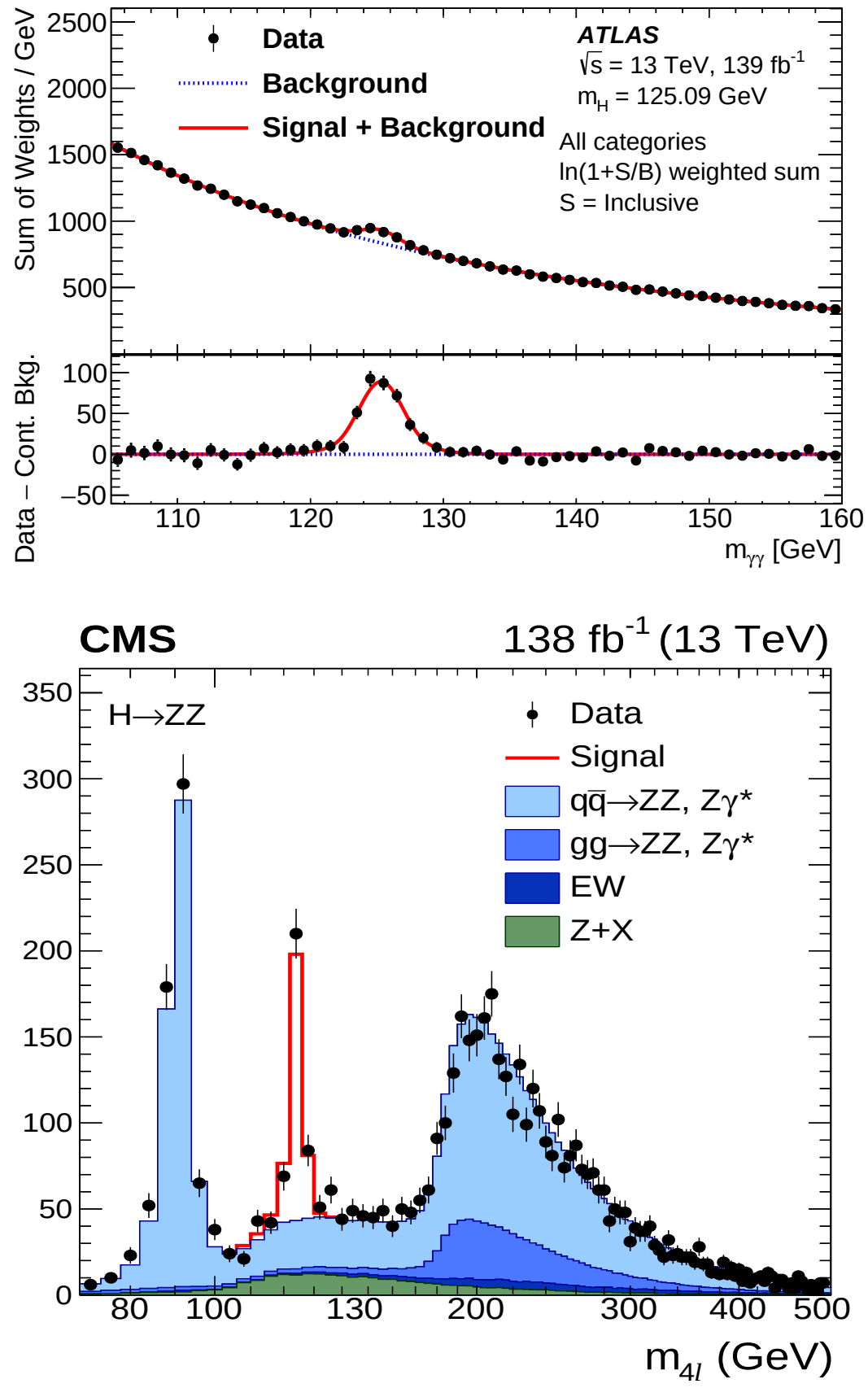

**Fig. 12.4.3** (Upper) The diphoton invariant mass distribution in ATLAS [3829]. The data events (dots) are weighted by $\ln(1 + |S|/|B|)$, where S and B are the expected signal and background. (Lower) The invariant mass distribution of four charged leptons targeting the study of the decay channel $H \to ZZ^* \to 4\ell$ in CMS [3828].

The spin and parity have been measured and found to be compatible with the SM prediction, $J^P = 0^+$, at $> 99.9\%$ confidence level (CL) [3830, 3831]. The width of the Higgs boson has been measured to be $\Gamma_H = 3.2^{+2.4}_{-1.7}$ MeV by using off-mass-shell and on-mass-shell Higgs boson production, in final states with four charged leptons or $2\ell 2\nu$ [3832], with the assumption that on-shell and off-shell effective couplings are the same.

All the production modes (except $tH$ and $bbH$) have been observed with a significance larger than $5\sigma$, as well as several decay channels: $WW$, $ZZ$, $\gamma\gamma$, $\tau\tau$, $b\bar{b}$. A $3\sigma$ evidence for the $\mu\mu$ final state was found by the CMS experiment [3826]. ATLAS and CMS have recently presented results on the search for the $\ell\bar{\ell}\gamma$ final state, reaching about $3\sigma$ significance [3833, 3834].

The experiments test the compatibility of their measurements with the SM, and the results are generally presented in two ways: by means of *signal-strength modifiers* $\mu$ (defined as $\mu = \sigma \times BR/(\sigma \times BR)_{SM}$, or *coupling-strength modifiers* $\kappa$ (defined as $\kappa^2 = \sigma/\sigma_{SM}$, or $\kappa^2 = \Gamma/\Gamma_{SM}$) [3824]. Fitting the data from all production modes and decay channels with a common signal strength $\mu$, the experiment found the following results:
ATLAS [3825]:

$\mu = 1.05 \pm 0.04(th) \pm 0.03(exp) \pm 0.03(stat)$,

CMS [3826]:

$\mu = 1.002 \pm 0.036(th) \pm 0.033(exp) \pm 0.029(stat)$,

showing a very good agreement with the SM, within the uncertainty. The theoretical (th) uncertainty has decreased by about a factor of 2 with respect to Run1, thanks to the huge effort of the theoretical community; the huge increase in statistics (i.e. 30 times more Higgs boson events), a better understanding of the detector, and more sophisticated methods (like Boosted Decision Trees, Deep Neural Network and Advanced Machine Learning) have helped to decrease the experimental (exp) and statistical (stat) uncertainty by a factor of more than two.

For a given production and decay, $i \to H \to f$, two parameters $\mu_i$ and $\mu_f$ are defined as $\mu_i = \sigma_i/(\sigma_i)_{SM}$ and $\mu^f = BR^f/(BR^f)_{SM}$. Many initial states $i$ and final states $f$ share the same coupling, e.g. VBF $H$ production and $H \to VV$ decay both involve the $HVV$ coupling ($V = W, Z$). Another example is the $H \to \gamma\gamma$, that proceeds via a loop of $W$ bosons or top quarks, thus involving the $HWW$ and $Htt$ couplings. Each $iiH$ and $Hff$ coupling is multiplied by a scaling factor $\kappa$, thus defined as $\kappa_j^2 = \sigma^j/\sigma^j_{SM}$, or $\kappa_j^2 = \Gamma^j/\Gamma^J_{SM}$. The experiments have presented results on the $\kappa_j$ with the full Run2 statistic [3825, 3826].

In the presence of new physics, new particles could contribute to the loops, affecting the various couplings and modifying the SM relations. Thus an alternative fit could be performed assuming non resolved loop for the coupling of the Higgs boson with photons or gluons, and thus assuming effective couplings $\kappa_\gamma$ and $\kappa_g$. The results are shown in Fig. 12.4.4 [3825, 3826]. Moreover, in the fit the possibility of the Higgs boson decaying to invisible particles (i.e. neutrinos or dark matter candidates), $B_{inv}$, or to undetected particles, $B_u$ or $B_{Undet.}$ (i.e. particles that may or may not leave a trace in the detector, and the experiments do not have dedicated searches looking for these) is allowed. The presence of invisible or undetected decays can be inferred indirectly from a reduction in the branching fraction for SM decays or by an increase in the total Higgs boson width. In this interpretation, the total width becomes $\Gamma_H = \sum \Gamma_f(\kappa)/(1 - B_{inv.} - B_u)$.

Figure 12.4.5(left) shows that indeed the Higgs boson couples with the fermion and boson masses as predicted by the SM. The very good agreement spans over many orders of magnitude. The results are shown for CMS [3826], and ATLAS has presented similar results [3825]. Figure 12.4.5(right) shows the observed and projected values resulting from the fit in the $\kappa$-framework in different data sets: at the time of the Higgs boson discovery, using the full data from LHC Run 1, in the Run2 data set (*"This paper"*), and the expected 1 standard deviation uncertainty at the high-luminosity run (HL-LHC) for an integrated luminosity of 3000 fb$^{-1}$ [3826].

### 12.4.2 Cross section measurements

With the data collected during Run 1 and Run 2, the ATLAS and CMS experiments measured the Higgs boson ggF production cross section with about 6% precision. The total cross section measurement from ATLAS [3825] at $\sqrt{s} = 13$ TeV is $50.2 \pm 3.0$ pb, and CMS measures $48.3 \pm 2.7$ pb [3826], both in agreement with the SM prediction of $48.5^{+1.5}_{-1.9}$ pb, as shown in Fig. 12.4.2.

Figure 12.4.6 shows the cross sections for different production processes and the branching fractions for different decay modes, as measured by the ATLAS experiment [3825].

### 12.4.3 The Simplified Template Cross Section

The simplified template cross section (STXS) method has been developed at the Les Houches 2015 workshop, and within the LHC Higgs Cross Section Working Group [3817] with the aims to separate more cleanly measurement and interpretation steps in order to reduce the theory dependencies that are folded into the measurements (including the dependence on theoretical uncertainties and on the underlying physics model). Its primary goals are to maximize the sensitivity of the measurements and to minimize their theory dependence. The method is designed to measure cross sections separated into production modes (instead of signal strengths), in mutually exclusive regions of phase space, and to be inclusive over Higgs boson decays, allowing to perform a global combination of all decay channels and to ease interpretation and search for BSM phenomena. Figure 12.4.7 shows the results of ATLAS for the LHC Run2 data [3825].

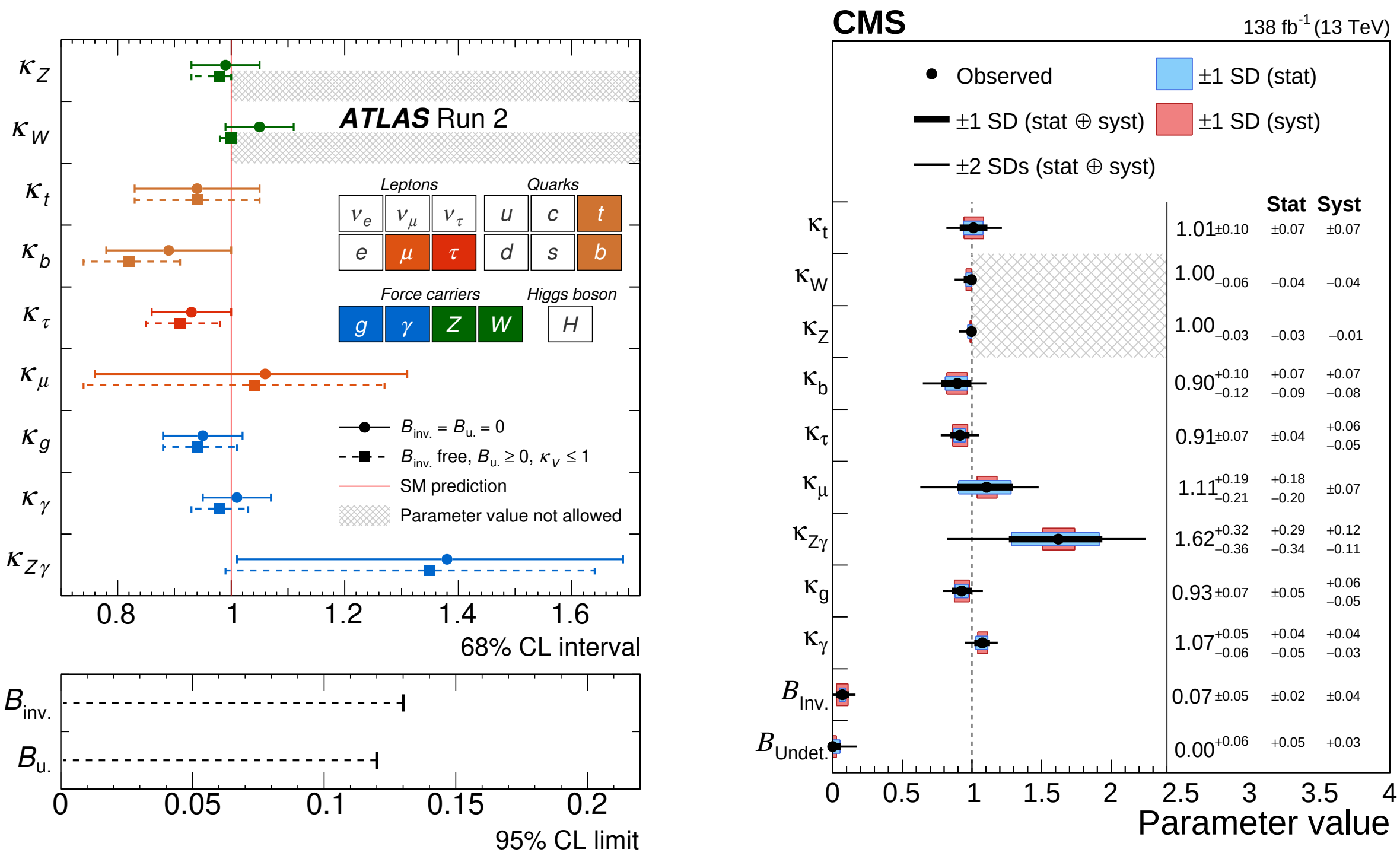

**Fig. 12.4.4** (Left) Coupling-strength modifiers and their uncertainties per particle type with effective photon, $Z\gamma$ and gluon couplings in the ATLAS experiment [3825]. The horizontal bars on each point denote the 68% confidence interval. The scenario where $B_{inv.} = B_{u.} = 0$ is assumed is shown as solid lines with circle markers. The p-value for compatibility with the SM prediction is 61% in this case. The scenario where $B_{inv.}$ and $B_{u.}$ are allowed to contribute to the total Higgs boson decay width while assuming that $\kappa_V \leq 1$ and $B_{u.} \geq 0$ is shown as dashed lines with square markers. The lower panel shows the 95% CL upper limits on $B_{inv.}$ and $B_{u.}$. (Right) Results of a fit to the coupling-strength modifiers $\kappa$ allowing both invisible and the undetected decay modes, with the SM value used as an upper bound on both $\kappa_W$ and $\kappa_Z$ in the CMS experiment [3826]. The thick (thin) black lines indicate the 1 (2) standard deviation confidence intervals, with the systematic and statistical components of the 1 standard deviation interval indicated by the red and blue bands, respectively. The p-value with respect to the SM prediction is 33%.

### 12.4.4 Differential distributions

The large data set accumulated during the LHC Run 2 allowed the experiments to do the first studies of differential distributions. A convenient set of kinematic variables to describe the Higgs boson production in hadronic collisions, and to test QCD consists of the transverse momentum $p_T$, the rapidity $y$, and the azimuthal angle $\phi$. The first two variables allow to understand many important QCD effects. The $p_T$ distribution is sensitive to perturbative QCD, and at low value it is strictly connected with the resummation of the leading logarithms, while at large values new physics could manifest. The $y$ distribution is sensitive to the parton distribution functions. At LHC the processes should not depend on $\phi$. Two important additional variables, that probe the theoretical modeling of high-$p_T$ QCD radiations in Higgs boson production, are the number of jets in the event, $N_{jet}$, and the transverse momentum of the leading jet, $p_T^{lead.jet}$.

Differential distributions are usually measured unfolding the detector resolution and efficiency effects and calculating "fiducial" cross sections. Cross sections are measured in a fiducial phase space, which is defined to closely match the experimental acceptance in terms of the physics object kinematics and topological event selection. This approach is chosen in order to reduce the systematic uncertainty associated with the underlying model and with the extrapolation to non-measured regions. As an example, the fiducial phase space for $H \to 4\ell$ constitutes approximately 50% of the total phase space. The fiducial differential cross sections are then compared with the various MC simulations and analytical calculations.

Figure 12.4.8 shows the differential cross section for the processes $pp \to H \to 4\ell$, $pp \to H \to \gamma\gamma$, and their combination as a function of the Higgs boson transverse momentum, its rapidity, the number of jets in the event, and the leading jet $p_T$ as measured by the ATLAS experiment [3835]. The data are compared with various theoretical predictions, all normalized to the total cross section, where the dominant ggF contribution is calculated at fixed order N3LO.

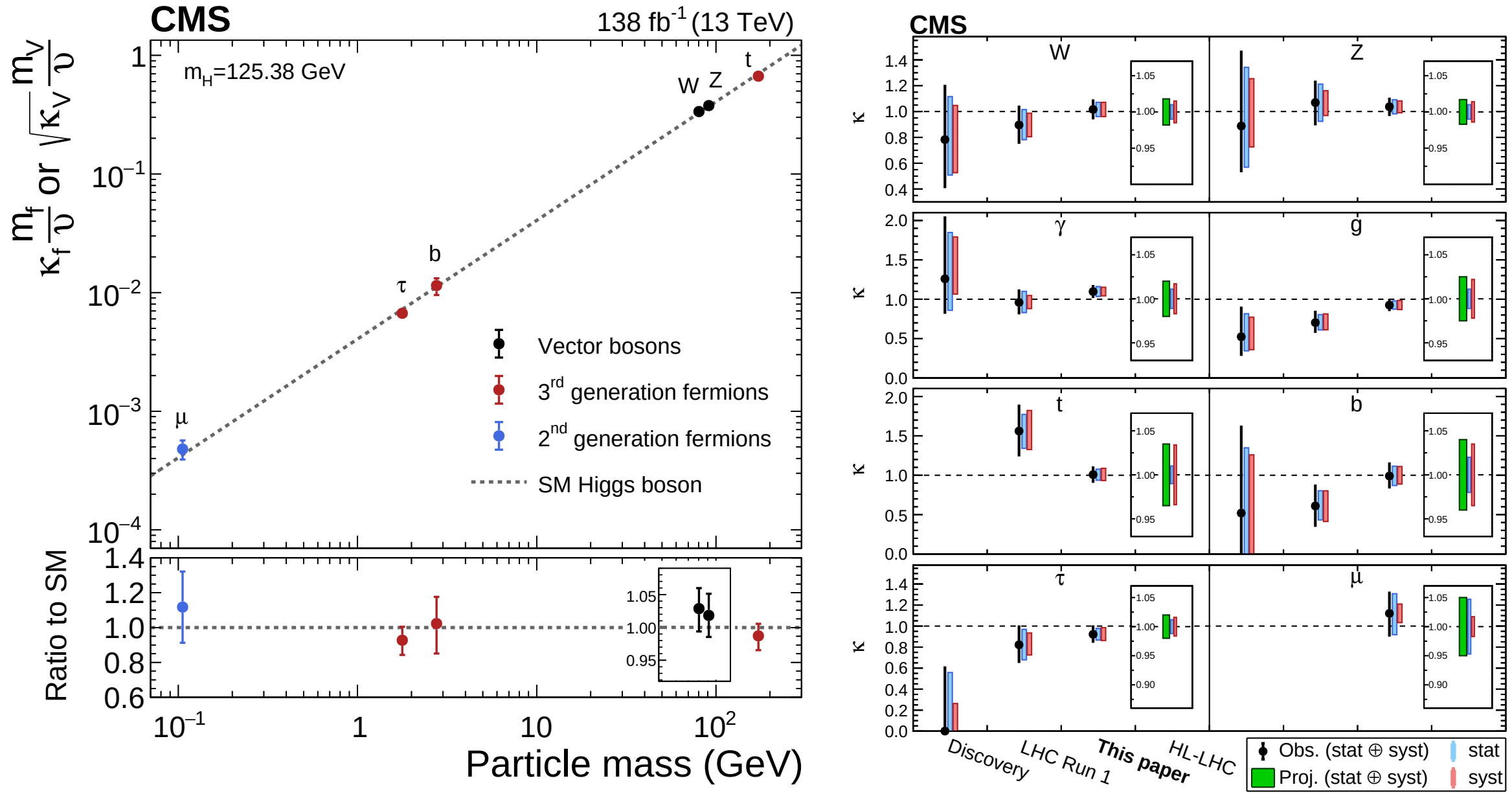

**Fig. 12.4.5** (Left) Measured Higgs boson couplings to fermions and gauge bosons as a function of the fermion or gauge boson mass, where $v$ is the vacuum expectation value of the BEH field, and $\kappa_i$ are the coupling modifiers as described in the text [3826]. (Right) Observed and projected values resulting from the fit in the $\kappa$-framework in different data sets: at the time of the Higgs boson discovery, using the full data from LHC Run 1, in the Run 2 data set (*this paper*), and the expected 1 standard deviation uncertainty at the HL-LHC for an integrated luminosity of 3000 fb$^{-1}$ [3826]. These results assume that no contributions from BSM is present in loops.

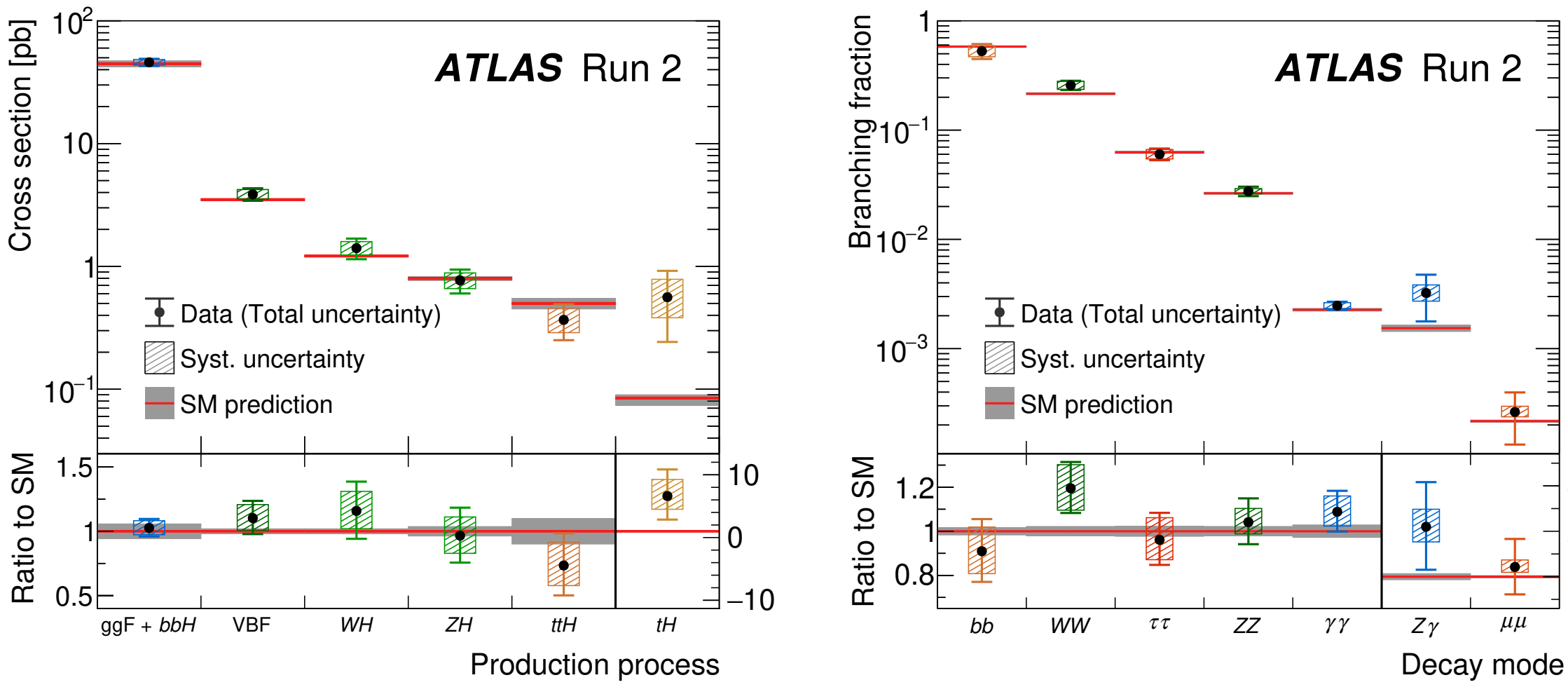

**Fig. 12.4.6** (Left) Observed and predicted cross section for different Higgs boson production modes, measured assuming SM values for the decay branching fractions in ATLAS [3825]. (Right) Observed and predicted branching fractions for different Higgs boson decay channels. The lower panels show the ratio of the measured values to their SM predictions [3825].

Figure 12.4.9 shows the double differential fiducial cross section measured in bins of $p_T^{\gamma\gamma}$ and $n_{jets}$ for $H \to \gamma\gamma$ events in the CMS experiment [3836]. The data are compared to the predictions from different setups of the event generator MadGraph5_aMC@NLO (version 2.6.5) [3328].

### 12.4.5 The Higgs boson and heavy quarks

The dominant decay of the SM Higgs boson is into pairs of $b$ quarks, with an expected branching fraction of approximately 58% for a mass of 125 GeV, but the large background from multi-jet (QCD) production makes the search in ggF very challenging. The decay of the

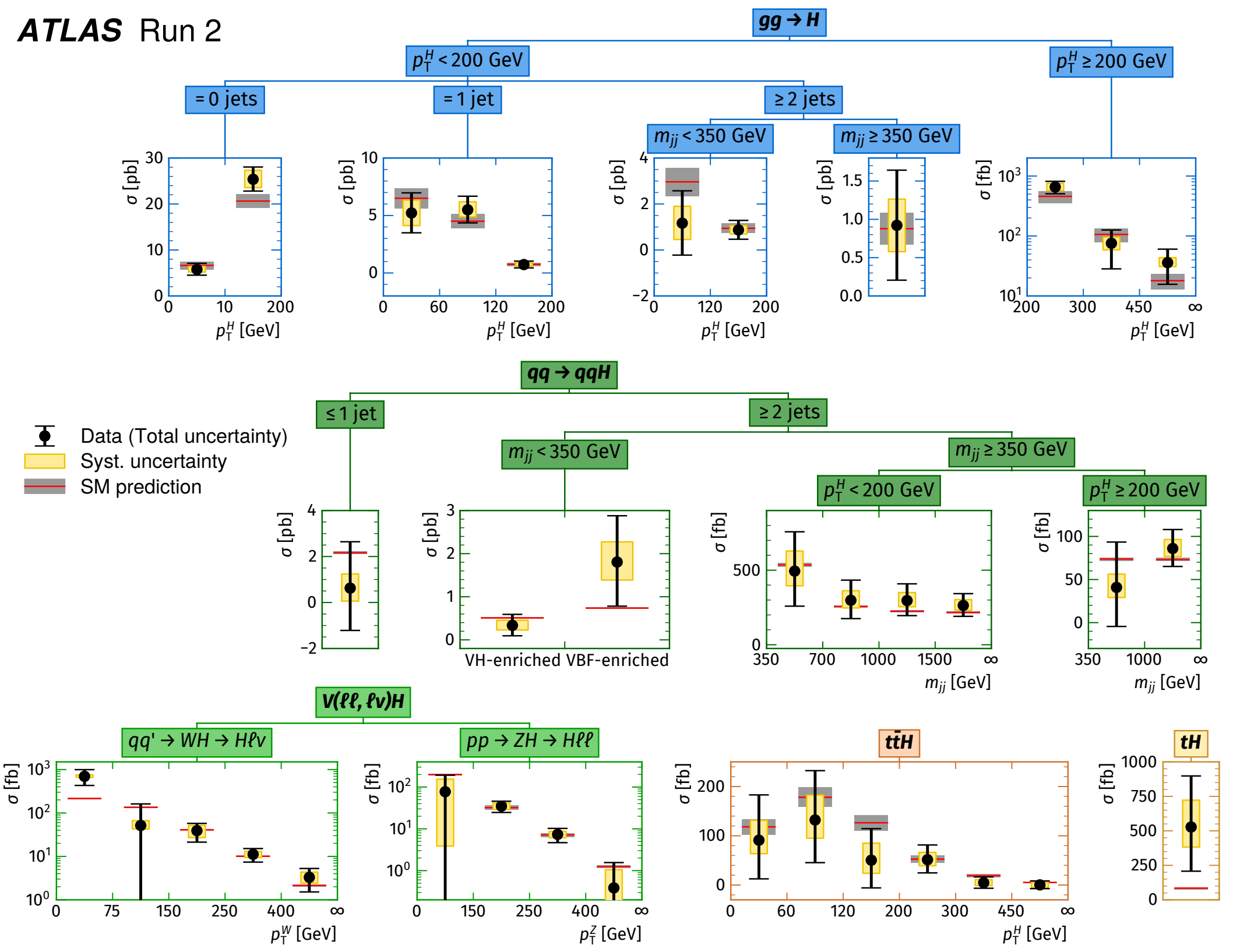

**Fig. 12.4.7** Observed and predicted Higgs boson production cross sections in different kinematic regions [3825]. The vertical bar on each point denotes the 68% confidence interval. The p-value for compatibility of the combined measurement and the SM prediction is 94%. Kinematic regions are defined separately for each production process, based on the jet multiplicity, the transverse momentum of the Higgs boson $p_T(H)$ and vector bosons $p_T(W)$ and $p_T(Z)$ and the two-jet invariant mass $(m_{jj})$. The $VH$-enriched and VBF-enriched regions with the respective requirements of $60 < m_{jj} < 120$ GeV and $m_{jj} < 60$, $m_{jj} > 120$ GeV are enhanced in signal events from $VH$ and VBF productions, respectively.

Higgs boson to $b\bar{b}$ was observed during Run 2 by ATLAS and CMS, in events where the $H$ is produced in association with a vector boson, i.e. in the $WH$ and $ZH$ production modes [3837, 3839]. In these events, the leptonic decay of the vector boson allows for efficient triggering and a significant reduction of the multi-jet background. In addition, two identified jets coming from the hadronization of $b$ quarks from the Higgs boson decay are required. The dominant background processes after the event selection are $V$+jets, $t\bar{t}$, single-top, diboson process and multi-jets.

Benefiting from multivariate techniques (MVA) and new machine learning algorithms, the experiments are now developing analyses to search for $H \to b\bar{b}$ inclusively in the production mode. Highly Lorentz-boosted Higgs bosons decaying to $b\bar{b}$, recoiling against a hadronic system, are reconstructed as single large-radius jets, which are identified using jet substructure algorithms and a dedicated $b$ tagging technique based on a deep neural network (see Sect. 11.5). The jet mass is required to be consistent with that of the observed Higgs boson, and the jet transverse momentum is required to be $p_T > 400 - 450$ GeV. The analyses are validated using $Z \to b\bar{b}$ events. The measured cross section is compatible with the SM one, but for the moment the uncertainty is still very large, i.e. around 10% [3840, 3841]. Figure 12.4.10(left) shows the reconstructed $b\bar{b}$ invariant mass for the selected $VH$ events in the ATLAS experiment [3837].

The decay branching fraction of the SM Higgs boson into a pair of $c$ quarks is slightly less than 3%. The difficulties to measure this channel are even larger than for the $b$ quark final state, because the main background to $c$ jet identification is indeed from $b$ jets. Higgs boson candidates, produced in association with a $W$ or a $Z$ boson, are constructed from the two jets with the highest $p_T$, with at least one jet identified as originating from a $c$ quark. [3838, 3842]. In CMS the search is extended to events in which the $H$ boson decays to a single large-radius jet. Additionally, a $b$–jet identification algorithm is used to veto $b$ jets. Novel charm jet identification and analysis methods using machine learning techniques are employed. In Fig. 12.4.10(center) the $c\bar{c}$ tagging efficiency is shown versus the efficiency of misindentifying quarks and gluons from $V$+jet and $H \to b\bar{b}$ in CMS. The analysis is validated by searching for $Z \to c\bar{c}$ decays in the $VZ$ process, leading to the first observation of this process at a hadron collider with a significance of 5.7 standard deviations, as shown in Fig.12.4.10(right) [3838]. The observed upper limit on $\sigma(VH)BR(H \to c\bar{c})$ is ranging from 14 to 26 times the SM prediction, for an expected limit that ranges from 7 to 31 for CMS and ATLAS, respectively.

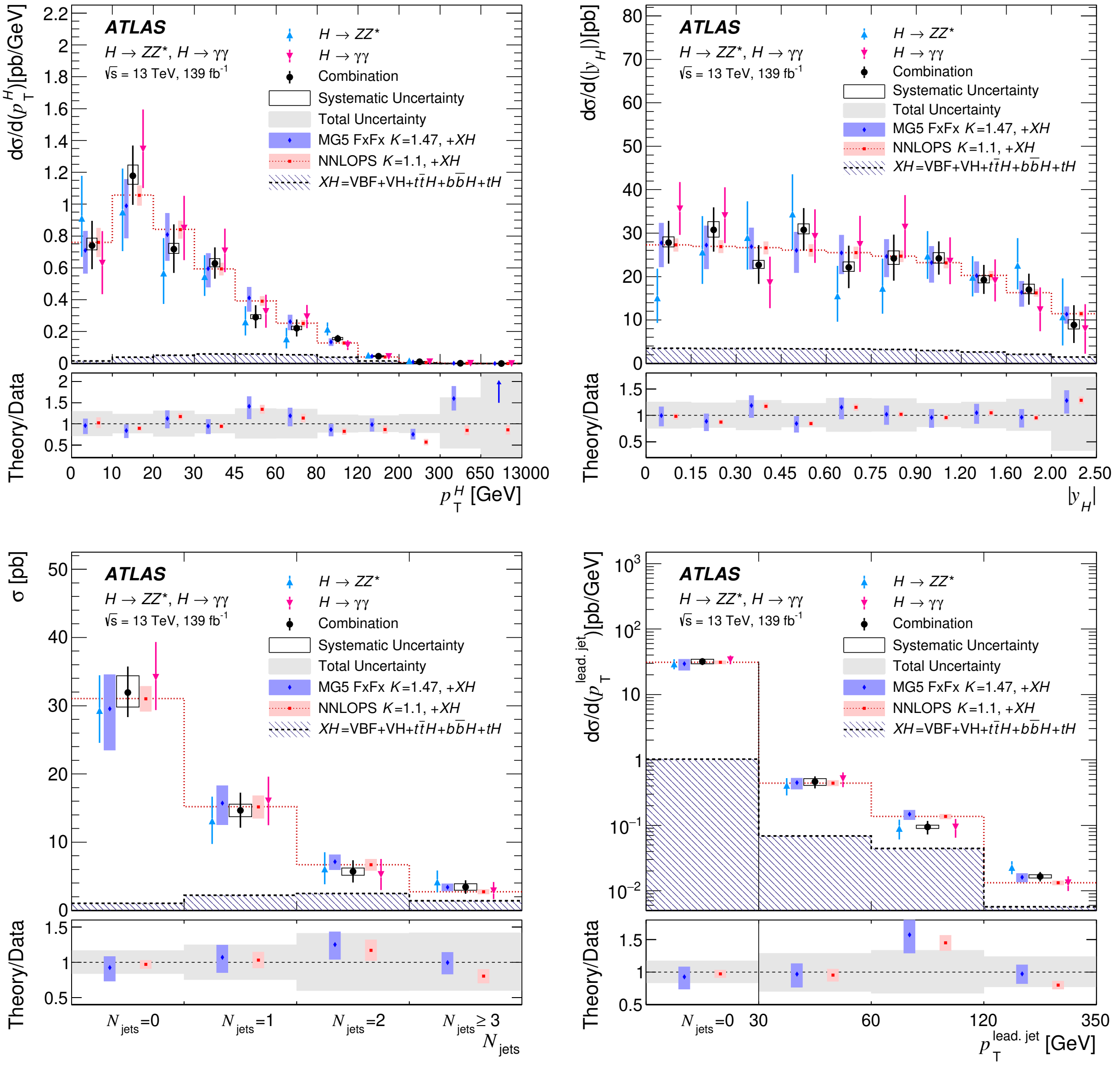

**Fig. 12.4.8** Differential $pp \to H + X$ cross-sections, in the full phase space, as a function of variables characterising the Higgs boson kinematics in ATLAS [3835]: (a) Higgs boson transverse momentum $p_T^H$, (b) Higgs boson rapidity $y^H$, (c) number of jets and (d) $p_T$ of the leading jet, compared with the Standard Model prediction. The $H \to ZZ^* \to 4\ell$ (blue triangles), $H \to \gamma\gamma$ (magenta inverted triangles), and combined (black squares) measurements are shown. The error bars on the data points show the total uncertainties, while the systematic uncertainties are indicated by the boxes. The measurements are compared with two predictions, obtained by summing the ggF predictions of NNLOPS or MG5 FxFx, normalised to the fixed order N3LO total cross-section, and MC predictions for the other production processes $XH$. The shaded bands indicate the relative impact of the PDF and scale systematic uncertainties in the prediction. These include the uncertainties related to the $XH$ production modes. The dotted red histogram corresponds to the central value of the prediction that uses NNLOPS for the modelling of the ggF component. The bottom panels show the ratios between the predictions and the combined measurement. The grey area represents the total uncertainty of the measurement. For better visibility, all bins are shown as having the same size, independent of their numerical width.

The $t\bar{t}H$ and $tH$ production channels probe the coupling of the Higgs boson to the top quarks. The large mass of the top quark may indicate that it plays a special role in the mechanism of electroweak symmetry breaking. Deviations from the SM prediction would indicate the presence of physics beyond the SM. The measurement of the Higgs boson production rate in association with a top quark pair ($t\bar{t}H$) provides a model-independent determination of the magnitude of the top quark Yukawa coupling $y_t$. The sign of $y_t$ is determined from the associated production of a Higgs boson with a single top quark ($tH$). The $t\bar{t}H$ and $tH$ production channels are studied in the case where the Higgs boson and the top quarks subsequently decay into final states with several leptons (including taus, also when they decay hadronically), complementing dedicated studies of the $H \to \gamma\gamma$, $H \to ZZ \to 4\ell$, and $H \to b\bar{b}$ decay modes. Several MVA techniques are employed to better separate the $t\bar{t}H$ and $tH$ production modes. The $ttH$ production modes has been observed in Run 2 [3843, 3844]. The precision on the top Yukawa coupling and on $t\bar{t}H$ cross section measurement is presented in figures 12.4.4 and 12.4.6.

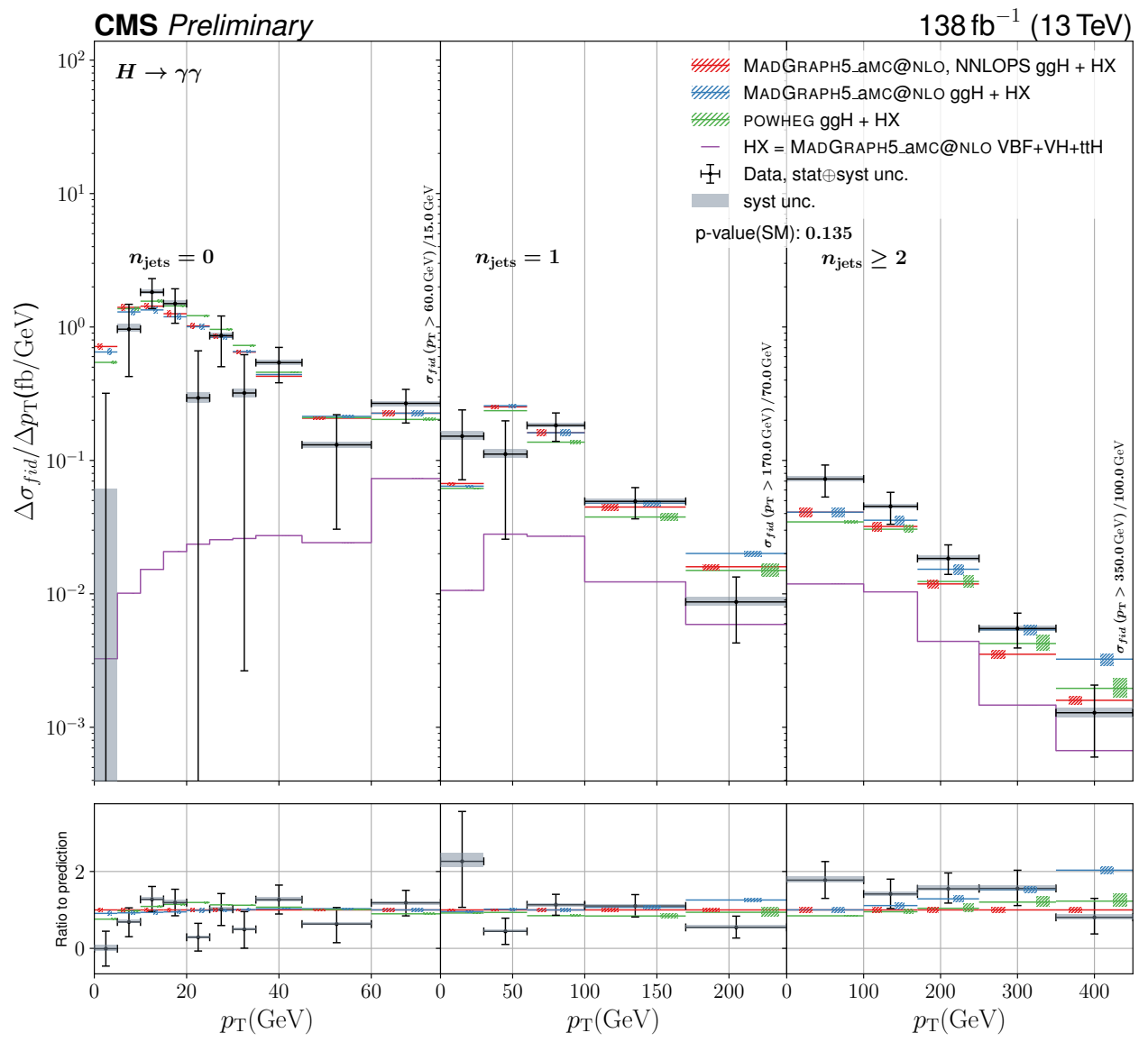

**Fig. 12.4.9** Double differential fiducial cross section measured in bins of $p_T^{\gamma\gamma}$ and $n_{jets}$ [3836]. The observed differential fiducial cross section values are shown as black points with the vertical error bars showing the full uncertainty, the horizontal error bars show the width of the respective bin. The grey shaded areas visualize the systematic component of the uncertainty. The coloured lines denote the predictions from different setups of the event generator. All of them have the $HX = VBF+VH+ttH$ component from MadGraph5_aMC@NLO in common. The green lines show the sum of $HX$ and the ggF component from MadGraph5_aMC@NLO reweighted to match the nnlops prediction. For the orange lines no nnlops reweighting is done and the purple lines take the prediction for the ggF production mode from POWHEG. The hatched areas show the uncertainties on theoretical predictions. Only effects coming from varying the set of PDF replicas, the $\alpha_S$ value and the QCD renormalization and factorization scales that impact the shape are taken into account here, the total cross section is kept constant.

### 12.4.6 Precision Higgs boson physics

The Higgs boson was discovered by the ATLAS and CMS experiments in 2012 at the LHC. With the data taken during the Run1 and Run2 the two experiments successfully tested the SM Higgs boson. The precision on the couplings is ranging from 3% for the coupling to the $Z$, to 10% for the coupling to $b\bar{b}$ and $t\bar{t}$, to 20-30% for the couplings to muons and $Z\gamma$. Run3 and the high-luminosity LHC (HL-LHC) will deliver approximately 3000 fb$^{-1}$ of luminosity to each experiment. By the end of HL-LHC, rare decays channels such as $H \to \mu\mu$ and $H \to Z\gamma$ will be observed and studied, the SM Higgs boson pair production is estimated to be observed with a significance of 4 to 5 standard deviations, when combining the results of the two experiments, as well as the Higgs boson coupling to charm quarks. As of today, the experiments have analysed only 3% of the Higgs boson events that they will have at the end of LHC. By then, most of the couplings measurements will reach the 2 to 3% precision, sufficient to start exploring contributions from physics beyond the SM in the Higgs boson area. A detailed discussion on the physics reach at HL-LHC is given in Sect. 14.9.

## 12.5 Top quark physics

Marcel Vos

### 12.5.1 A brief history of the top quark

The late 1960s and early 1970s established the quark model, as described in Section 1. After the discovery of the charm quark [75, 76] in 1974, and the bottom quark [3845] in 1977, the hunt for the sixth quark was open. The LEP and SLC experiments could feel the effect of radiative corrections involving the top quark in the $Z$-pole data collected in the 1980s and 1990s, and could infer limits on the top quark mass from a fit of electroweak precision observables [3612, 3846], but could not produce top quarks. Finally, in 1995, the CDF and D0 experiments at the Tevatron, Fermilab's 1.9 TeV $p\bar{p}$ collider, observed the top quark directly [3847, 3848]. The two experiments could also demonstrate the existence of the electro-weak single-top-quark production processes in the $t-$channel [3849, 3850] and $s-$channel [3851]. Precise measurements confirmed key SM predictions, such as the forward-backward asymmetry in $t\bar{t}$ production [3852] and the $W$-boson helicity fractions

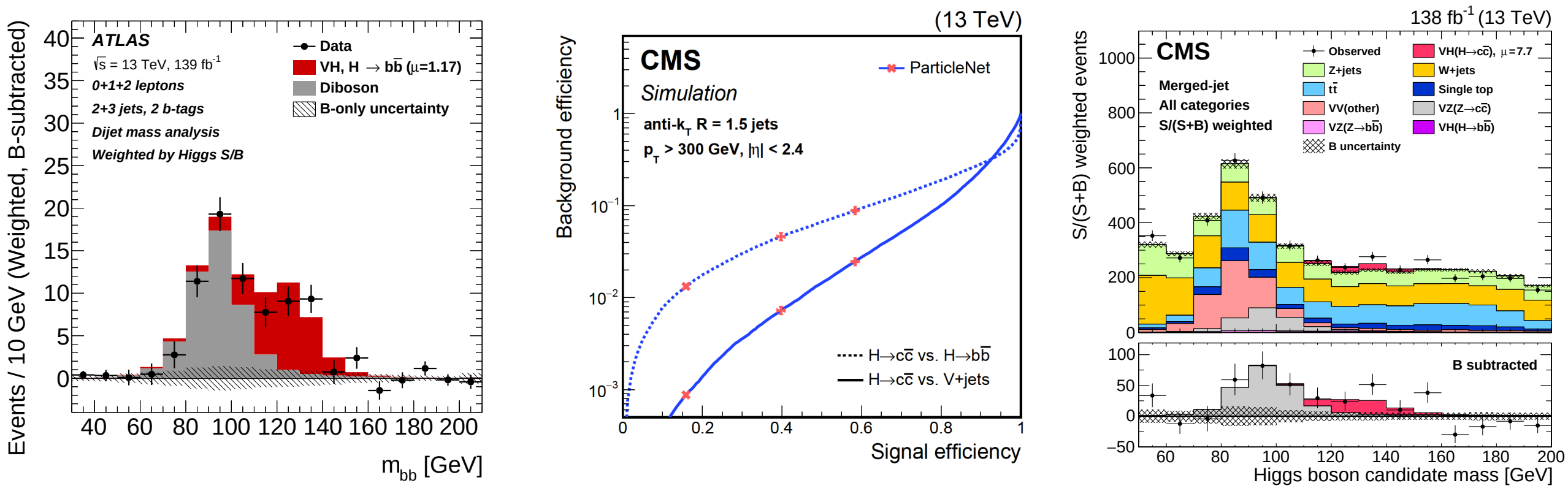

**Fig. 12.4.10** (left) The distribution of $m_{bb}$ in data after subtraction of all backgrounds except for the $WZ$ and $ZZ$ diboson processes. All the contributions are summed and weighted by their respective S/B ratios, with S being the total fitted signal and B the total fitted background. The expected contribution of the associated $WH$ and $ZH$ production of a SM Higgs boson with $m_H = 125$ GeV is shown scaled by the measured signal strength ($\mu = 1.17$). The size of the combined statistical and systematic uncertainty for the fitted background is indicated by the hatched band [3837]. (center) Performance of CMS algorithm ParticleNet for identifying a $c\bar{c}$ pair for large-radius jets with $p_T > 300$ GeV. The solid (dashed) line shows the efficiency to correctly identify $H \to c\bar{c}$ vs the efficiency of misidentifying quarks or gluons from the $V$+jets process ($H \to b\bar{b}$). The red crosses represent the three working points used in the large-radius jet analysis [3838]. (right) Invariant mass distribution of the selected $c\bar{c}$ events [3838]. The lower panel shows the data (points) and the fitted $VH(H \to c\bar{c})$ (red) and $VZ(Z \to c\bar{c})$ (grey) distributions after subtracting all other processes. Error bars represent pre-subtraction statistical uncertainties in data, while the gray hatching indicates the total uncertainty in the signal and all background processes.

in top quark decay [3853]. Last but not least, the Tevatron legacy includes a top quark mass combination with a sub-GeV precision [3854].

The Large Hadron Collider [3855] at CERN entered operation in 2010 with $pp$ runs at 7 TeV and 8 TeV. Data taking resumed at 13 TeV in 2015, and the ATLAS and CMS experiments had harvested 140 $fb^{-1}$ by 2018. At the time of writing, in summer 2022, run 3 has just started with $pp$ collisions at 13.6 TeV. The large center-of-mass energy strongly enhances the top quark production cross section. In combination with the large instantaneous luminosity, the LHC is a genuine ”top factory”. More than 100 million $t\bar{t}$ pairs have been produced in run 1 and 2 and more than 1 billion are expected in future runs. The LHC therefore marks a new era in experimental top-quark physics and dominates the summary in this chapter.

Its properties make the top quark an ideal laboratory for studies of the electro-weak and strong interactions. As the top quark mass of approximately 172 GeV exceeds that of the $W$-boson, the decay $t \to Wb$ is kinematically allowed and makes up nearly 100% of the branching ratio (with sub-% fractions of top quarks decaying to $Wd$ and $Ws$ in the Standard Model). The subsequent $W^+ \to l^+\nu_l$ and $W^- \to l^-\bar{\nu}_l$ decays of the $W$-boson yield an isolated charged lepton $l^\pm = e^\pm, \mu^\pm, \tau^\pm$. These are a key signature to trigger and select events with top quarks at hadron colliders. The charge of the lepton furthermore reveals whether the decay corresponds to that of a top-quark or anti-quark, providing an efficient ”tag” for asymmetry measurements [3856]. Finally, the charged lepton is an efficient polarimeter that enables studies of top quark polarization [3857], spin correlations [3858, 3859] and quantum entanglement [3860]. All these features lead to a rich and varied experimental top quark physics programme.

### 12.5.2 Precise predictions for top quark physics

The calculability of top quark production is one of the keys to the top quark physics programme at hadron colliders. The large top quark mass regulates perturbative calculations, enabling precise predictions of QCD processes with colored objects in the final state.

The fully differential top quark pair production cross section at hadron colliders is known to NNLO accuracy in the strong coupling [3339, 3396, 3861]. Electro-weak corrections are available at NLO [3862] and NNLL resummations are available. Predictions of the inclusive $pp \to t\bar{t}$ production rate reach an uncertainty of 4-5%. The uncertainty is dominated by the scale uncertainties, that estimate the impact of higher-order QCD corrections, followed closely by the PDF uncertainties.

While the NNLO calculation of top quark pair production is a major milestone, it remains a considerable challenge to meet the experimental precision that can be achieved at the LHC. The most precise measurements reach an uncertainty a bit over 2%, half that of the predictions. The NNLO QCD corrections have a sizable impact on the shape of differential measurements, in particular on the top $p_t$ and related distributions [3396, 3861–3863]. Fully differential NNLO fixed-order calculations and Monte Carlo generators are required to provide an adequate description of the data collected by ATLAS [3864–3866] and CMS [3867–3870].

Associated top quark production processes with electro-weak bosons become accessible at the LHC and provide a direct probe of the top quark interactions with the Higgs boson and the neutral electro-weak gauge bosons (see for instance Ref. [3871] and references therein). The $t\bar{t}X$ processes at the LHC are known to NLO accuracy, and uncertainties on the inclusive production rates are well below 10%. The experimental results for these rare processes are improving rapidly and already challenge the precision of the best SM predictions. Resummation to NNLL and NLO electroweak corrections are available [1953, 3872] and elements of the NNLO calculations for $t\bar{t}H$ production are known [3873]. A complete NNLO description is required for all $t\bar{t}X$ processes to take full advantage of the HL-LHC programme [3874].

NLO calculations are available for $2 \to N$ processes that include top quark decays and off-shell effects [3875]. These provide sizable corrections for the top quark pair production process and associated production processes.

Predictions at the particle- and detector-level play an important role in measurements of top quark cross sections and properties and in searches for rare processes. State-of-the-art Monte Carlo generators match NLO matrix elements to the parton shower and hadronization models implemented in Pythia8 [3876] or Herwig [3877]. The work horse implementations for the LHC programme during run 1 and run 2 are provided by the Powheg-box [3544, 3545, 3569], where resonance-aware matching is an important recent addition for top physics [3878], and the MG5_aMCNLO [3328] package, that can include also NLO electroweak corrections [3330]. SHERPA [3522] offers multi-leg generation for top quark pair production and other high-jet-multiplicity processes involving top quarks. The MINNLOps package [3879, 3880] provides a Monte Carlo event generator at NNLO accuracy for top quark pair production that can be interfaced to Parton Shower and hadronization programmes.

### 12.5.3 Precision measurements at hadron colliders

The measurements of top quark production cross sections in the ATLAS experiment are summarized in Fig. 12.5.1. The measurements cover four different center-of-mass energies (5, 7, 8 and 13 TeV) and span over five decades in production rate: from $\mathcal{O}(1\text{ nb})$ for top quark pair production to $\mathcal{O}(10\text{ fb})$ for $t\bar{t}t\bar{t}$ production. The experimental results indicated by the colored markers are compared to the best available Standard Model predictions in grey.

The measurements of the production cross section for the classical top quark production processes have become precision measurements, with the measurement of the inclusive cross section reaching 2.4% precision [3865]. The result is limited by the knowledge of the integrated luminosity delivered by the LHC. Progress in the understanding of the luminosity calibration is expected to reduce this uncertainty to about 1%, but this is likely to remain the limiting factor for the most precise inclusive measurements.

Also electro-weak single top production is characterized precisely in the $t$-channel and $tW$ associated production channel. With a precision of less than 7% for the $t$-channel [3881], the Cabibbo-Kobayashi-Maskawa matrix element $V_{tb}$ is determined as: $|f_{LV}V_tb| = 1.02 \pm 0.04$, where the uncertainty includes contributions from experiment and predictions and $f_{LV}$ is a form factor, identical to 1 in the SM, that parameterizes the possible presence of anomalous left-handed vector couplings. This result is in good agreement with the determinations from $b-$physics [278].

The LHC programme has eclipsed the Tevatron measurements in nearly all processes and measurements. However, the Tevatron legacy remains very relevant, as the different initial states ($p\bar{p}$ instead of $pp$) and center-of-mass energy lead to important complementarities and Tevatron data continue to provide important inputs for global analyses. Several highlights of the Tevatron programme remain unrivalled to this day, as the dominance of $q\bar{q}$-initiated production provides an ideal laboratory for certain measurements. Good examples are the study of $s-$channel single top quark production [3851] and the measurement of the forward-backward asymmetry in $t\bar{t}$ production, that reached a high significance for the SM effect at the Tevatron [3852].

### 12.5.4 Boosting sensitivity

The enormous sample of top quark pairs collected at the LHC enables precise differential cross section measurements. Many measurements extend well into the *boosted regime*, where the top quark transverse momentum significantly exceeds the top quark mass and the collimated hadronic top quark decays are reconstructed as a single large-radius jet.

From the first observation of boosted top quark candidates at the start of the LHC, the study of their production has come a long way. An avalanche of new techniques has been developed [3882], from pile-up mit-igation to top tagging algorithms, and the experiments have carefully characterized jet substructure [3883, 3884] and the experimental response [3607]. With the large samples of boosted top quarks, these developments have enabled precise measurements of top quark interactions in the most energetic collisions at the LHC. The most recent measurements of top quark pair production in the boosted regime [3885, 3886] yield precise bounds on the Wilson coefficients of the $q\bar{q}t\bar{t}$ operators in the Standard Model Effective Field Theory, as the energy-growth of their impact boosts the sensitivity of these measurements.

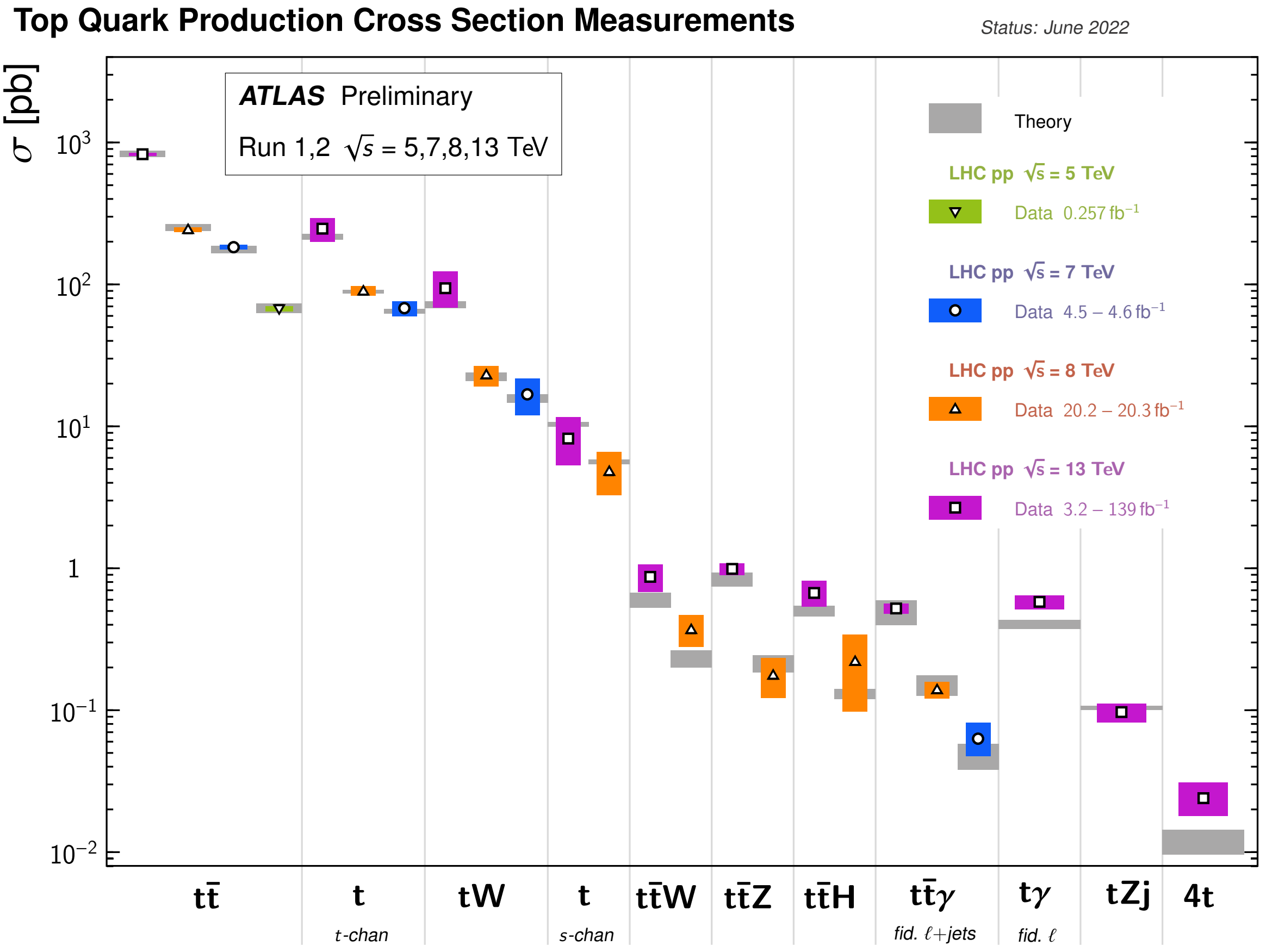

**Fig. 12.5.1** Measurements of production cross sections for processes with top quarks in the final state by the ATLAS and CMS experiments. The experimental results indicated by the colored markers are compared to the best available Standard Model predictions in grey. Figure courtesy of the ATLAS experiment.

### 12.5.5 New rare top quark production processes

The right half of Fig. 12.5.1 is devoted to the new, rare associated top quark production processes that were observed by the LHC experiments in run 2. Many of these measurements scrutinize aspects of the Standard Model description of the top quark interactions that were not, or not directly, tested at previous facilities. The associated production processes of a top quark pair with a photon [3887, 3888] or $Z$-boson [3889, 3890] offer a new, direct probe to the neutral-current interactions of the top quark [3871]. These processes are well-established and differential measurements are available. More recently, the single top production process in association with a $Z$-boson [3891, 3892] and a photon [3893] were observed as well. The observation of the $pp \to t\bar{t}H$ production process [3843, 3894] confirms unambiguously that the heaviest SM particle indeed couples to the Higgs boson. The combination of rate measurements in different production and decay channels yields a robust estimate of the top-quark Yukawa coupling [3825, 3826]. In run 2 the first evidence was found for four-top-quark-production [3895]. With more data and improved experimental techniques this process can be observed before the end of the HL-LHC programme and will provide a probe for the four-heavy-quark vertex. These rare top quark production processes provide qualitatively new information on previously unprobed interactions and form a valuable input to fits of the Standard-Model-Effective-Field-Theory parameters to collider data.

### 12.5.6 New physics searches with top quarks

Beyond Standard Model (BSM) searches in final states with top quarks have pioneered the development of tools for boosted object tagging. Thus prepared, the experiments have been able to take advantage of the LHC centre-of-mass energy to push bounds on new massive states beyond 1 TeV and in many cases into the multi-TeV range. The $t\bar{t}$ resonance searches by ATLAS and CMS indicate that new narrow massive states that decay to top quark pairs or a top and bottom quark cannot have a production cross section times branching ratio greater than 0.1 pb in the mass region from 1.5 to 4-5 TeV. Concrete scenarios such as the bulk RS KK gluon [3896] and $W'$ bosons are excluded for resonance masses around 4 TeV[3897]. Searches for vector-like quarks decaying to a top quark and Higgs or gauge boson yield lower limits greater than 1 TeV for the mass of the vector-like quarks.

The integrated luminosity is a key for the search for flavor-changing-neutral-current (FCNC) interactions of the top quark. The branching fractions $t \to qX$ decays (with $X = \gamma, Z, g, H$) are suppressed well beyond the experimental sensitivity in the SM, but can be enhanced to $\mathcal{O}(10^{-5})$ in several extensions [3898]. An even larger branching ratio $BR(t \to cH) \sim 10^{-3}$ can be present in certain two-Higgs-doublet models [3899, 3900]. The observation of these FCNC interactions would be an unambiguous sign of physics beyond the Standard Model.

Searches have advanced rapidly in sensitivity in run 2 and the exclusion bounds in Figure 12.5.2 are reaching $\mathcal{O}(10^{-4} - 10^{-5})$, scratching the surface of the branching ratios predicted in viable models. The inclusion of single top production in association with a Higgs or gauge boson have been important to improve the bounds, in particular for the FCNC vertex with an up-quark.

### 12.5.7 The top quark mass

The top quark mass is a fundamental parameter of the SM Lagrangian that must be determined experimentally. As any other quark mass, its definition generally depends on the renormalization scheme and the value of the renormalization scale at which it is evaluated. The pole mass scheme is used in Monte Carlo generators and many fixed-order calculations. The $\overline{\text{MS}}$ mass is extracted from the top quark pair production cross section [3901].

Three main classes of measurements of the top quark mass at hadron colliders are discussed below. A selection of results obtained with each approach is presented in the rightmost panel of Fig. 12.5.2.

The first class of measurements extracts the top quark mass from the comparison of top quark pair cross section measurements (corrected to the parton level) to SM predictions at NNLO+NNLL accuracy. The uncertainty of the mass determined from the total cross section a 13 TeV is around 2 GeV. This includes a theoretical uncertainty, estimated by varying the renormalization and factorization scales and propagating uncertainties from the parton distribution functions of the proton. Importantly, recent cross section measurements have a much reduced dependence on the mass assumption in the correction of detector acceptance and efficiency, such that in practice the dependence on the MC mass parameter can be ignored to good precision. There is a broad consensus that this methods yields a solid measurement with a rigorous interpretation. Future progress is expected from improving fixed-order calculations and PDFs, and from a reduction of the luminosity uncertainty on the experimental side.

A more precise determination is possible based on measurements of differential cross sections [3902]. These enhance the mass sensitivity in e.g. the threshold region. In the shape analysis of the differential cross section important uncertainties in the absolute cross section and integrated luminosity cancel, leading to a precision of about 1 GeV for the most precise measurements [3903]. The theory uncertainty is accounted for in the same way as in the inclusive measurements and the method retains some flexibility in the choice of the mass scheme. More work is required, however, to account for bound-state effects in the threshold region [3904].

The third, and experimentally most precise, approach determines the mass parameter of the Monte Carlo generator that yields the best fit to the observed distribution of top quark decay products. The 2014 combination of Tevatron and LHC run 1 results yields $m_t = 173.3 \pm 0.8$ [3905] and is used as a reference in Fig. 12.5.2, indicated by the magenta area. The most precise single measurements by CDF, D0, ATLAS and CMS have since reached an uncertainty of approximately 600 MeV.

The results of direct mass measurements are interpreted as the top quark pole mass. An additional uncertainty of 500 MeV is assigned to cover the ambiguity in this interpretation [3906]. Further theory work is needed to improve Monte Carlo templates [3907], refine the understanding of the Monte Carlo mass parameter [3908] and derive a quantitative relation with a field-theoretical mass scheme [3909, 3910].

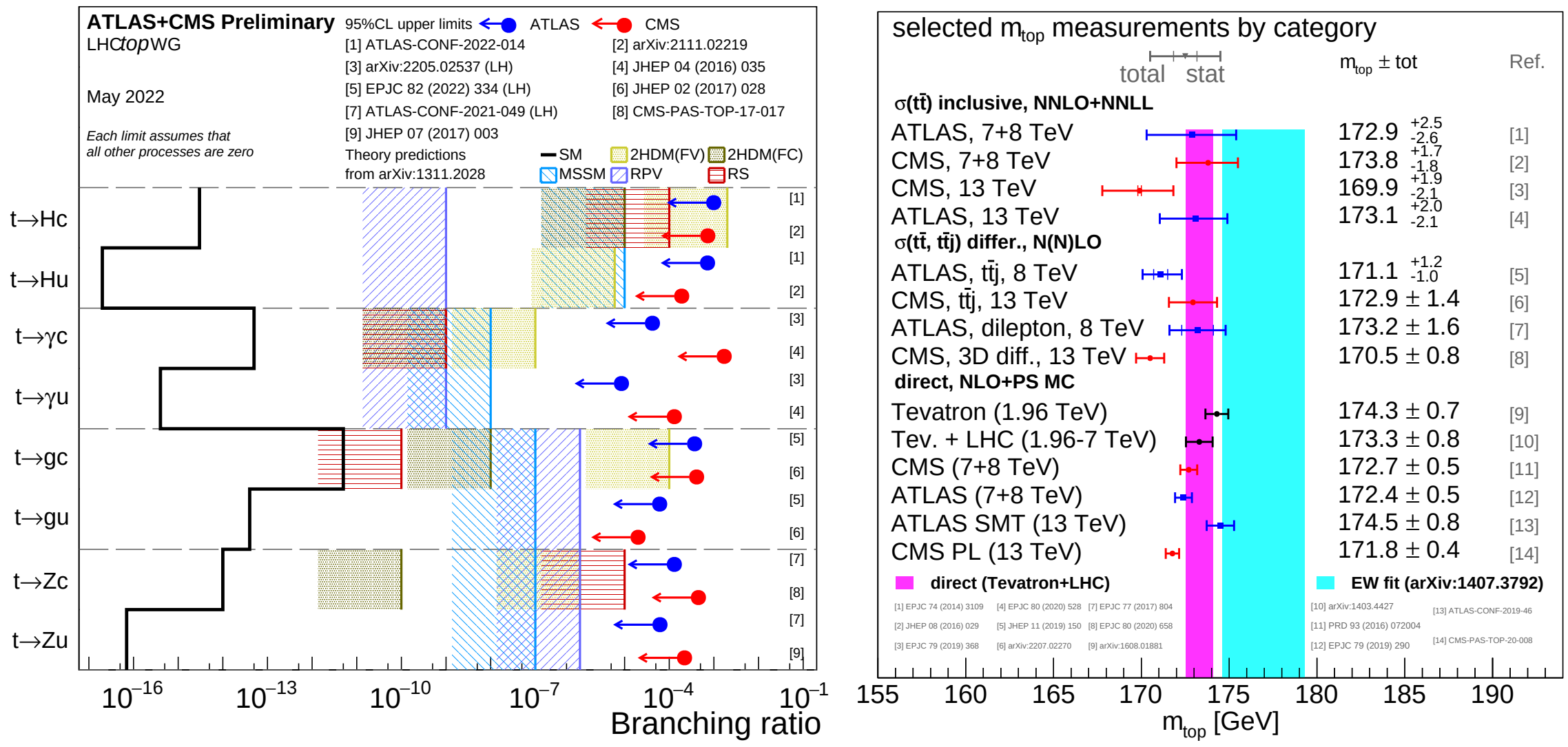

**Fig. 12.5.2** Leftmost panel: summary of the searches for FCNC interactions of the top quark with the Higgs boson, photon, gluon and $Z$-boson. 95% confidence limits are derived on the equivalent branching ratio $t \to Xu$ and $t \to Xc$, and in some cases for left-handed and right-handed couplings (left-handed couplings are assumed for the limits collected in the summary plot in those cases). Figure courtesy of the LHC top Working Group. Rightmost panel: selection of top quark mass measurements at the Tevatron and LHC, by category. ATLAS measurements are indicated with blue markers, CMS measurements in red and the Tevatron or combined Tevatron-LHC results in black. The 2014 world average is given by the pink bar, and the indirect determination of the top quark mass from the electroweak fit with the cyan band. Figure prepared by the author based on data collected by the LHC top Working Group.

The analysis of run 2 results is in full swing. The last two points in the rightmost panel of Fig. 12.5.2 correspond to two innovative analyses on partial run 2 data. ATLAS published an analysis based on a purely leptonic observable, the invariant mass of the system formed by the prompt lepton from W-decay and the soft muon found in the $b$-jet, shifting the systematic uncertainties from jet response to fragmentation and B-decay modelling. A recent preliminary result by CMS [3911] based on a profile-likelihood fit reaches an uncertainty below 400 MeV. This result demonstrates the power of the profile-likelihood-fit approach in top quark mass measurements, but also emphasizes the importance of a robust uncertainty model for MC-related uncertainties. A combination of all measurements collected can reach an experimental precision of 300 MeV.

Projections of future improvements are notoriously hard in this area, where a detailed understanding of the limitations of Monte Carlo generators is key. Direct measurements can potentially be improved to an experimental precision below 200 MeV in the remainder of the LHC programme [3874], while cross-section-based mass extractions can reach a total uncertainty below 500 MeV [3912].

### 12.5.8 Top quark data in global analyses

The top quark and the results in the top quark sector presented in previous sections are inevitably part of the "global" interpretations of collider data. In this section, three examples are briefly discussed.

Most recent analyses of the parton density functions [3748] consider also data on top quark production [113], that provide an important constraint on the high-$x$ gluon content of the proton. The ATLAS PDF fit [3763] includes differential measurements of $t\bar{t}$ production, using NNLO predictions with a fixed top quark mass. The CMS experiment has performed a global analysis [3913] with partial run 2 data, where the top quark pole mass is floated, as well as the PDFs and the strong coupling. The analysis is based on NLO predictions for the top quark pair production process and threshold corrections remain to be included.

Radiative effects connect electroweak precision observables at the $Z$-pole to precise measurements of $\alpha_s$ and the $W$-boson, Higgs boson and top quark masses. The electroweak fit (see e.g. Ref. [3914]) tests the relations among these parameters predicted by the SM and forms a stringent check of the internal consistency of the theory. There is a very mild tension between direct top quark mass measurements and the mass inferred from precision electroweak data (e.g. the magenta and cyan bands in Fig. 12.5.2). The main avenue towards a tighter test should focus on better measurements of the $W$-boson mass, but eventually also the precision of measurements of other parameters, among which the top quark mass, must be improved.

[113] To avoid absorbing potential BSM constributions to top quark production in the PDFs, care is taken to select differential measurements that are less likely affected. PDF results without top data are available in at least one PDF set and allow for important cross-checks.

The legacy data from collider experiments are collected in the framework of the Standard Model Effective Field Theory. Global fits of the top quark sector have been performed by several groups [3915, 3916]. State-of-the-art fits include a combined analysis of Higgs, electroweak and top data [3917, 3918], showing an interesting interplay between the top and Higgs sectors through the effect of operators involving top quarks in loops, for instance in $gg \to H$ and $H \to \gamma\gamma$, through the box diagram contribution to di-Higgs production and through the associated production of top quarks and a Higgs boson.

### 12.5.9 Outlook

A vibrant top quark physics programme was started at the Tevatron and has culminated in a broad and rich programme at the LHC. Direct searches in final states with top quarks explore the multi-TeV regime looking for signs of new resonances and exotic phenomena. Precise measurements of the classical top quark production processes and many new rare processes involving top quarks and the Higgs and gauge bosons form a powerful set of constraints on top quark couplings. The top quark mass is known to a precision of less than 0.5%.

The upcoming runs of the LHC and its high-luminosity upgrade are expected to improve considerably on current run 2 results [3874], increasing the precision, pushing differential measurements further into the high-energy regime, and probing ever more rare processes. Projections are particularly encouraging for rare associated production processes, where statistical limitations remain important and theory predictions are currently only available at NLO accuracy.

Top quark physics is an important consideration also for a new facility in high energy physics beyond the HL-LHC. A new electron-positron collider is identified as the highest-priority installation in the European, American and Asian road maps for particle physics. All projects for such a Higgs/ EW/top factory envisage operation at and above the $t\bar{t}$ production threshold. This enables scrutiny of the top quark in $e^+e^-$ collisions and provides precision measurements of the top quark mass, with $\mathcal{O}(50\text{MeV})$ precision [3919], and electroweak couplings, that improve by an order of magnitude compared to the HL-LHC projections [3920].

A new $pp$ collider at the energy frontier can potentiall push the discovery reach for massive particles by a further order of magnitude. It can also unlock ultra-rare SM processes, such as six-top-production and $t\bar{t}HH$. Quantitative projections remain to be made, as well as more detailed studies of top quark reconstruction in this challenging environment. Also the top quark physics of a multi-TeV lepton collider, be it the CLIC high-energy stage, a muon collider or a novel installation based on plasma-wakefield acceleration, remains to be explored in detail. High-energy lepton collisions, well above 1 TeV, offer the possibility to constrain four-fermion operators with two light particles and two top quark to unprecedented precision [3921], and provide an exquisite precision probe for new physics [3922].

## 12.6 Soft QCD and elastic scattering

**Per Grafstrom**

### 12.6.1 Introduction

Soft QCD has become a term covering many different topics. Elastic scattering and diffraction are central topics associated with soft QCD but in addition there is a long list of different areas associated with the term Soft QCD e.g. particle correlations, multiple parton interactions, particle densities, the underlying event. The list is not inclusive and could be made longer. It covers an enormous amount of different processes and concepts and just the elastic and diffractive part represents by itself more than 30-40% of the total cross section ($\sigma_{tot}$) at high energy hadron colliders. What basically unifies all those different processes is a large distance scale or equivalently a relative small momentum transferred in the reaction.

Another way of expressing the same criteria is to say that "Soft QCD" deals with processes for which the perturbative approach of QCD is not applicable due to the size of the strong coupling at small momentum transfer. This is a direct consequence of the running of the strong coupling $\alpha_s$. In this low momentum transfer regime more phenomenological approaches have to be applied. However, while using phenomenological methods, the aim is always to try to provide a smooth transition to harder and thus perturbative QCD processes.

Soft QCD processes have an interest in their own right representing a particular challenging part of QCD. However, there are a number of other reasons that motivate trying to achieve a better understanding of Soft QCD processes. The Soft QCD processes represent often the most significant background in searches for new physics. The so called underlying event stands for everything which is produced in a *pp* collision except for the hard scatterer. The better one understands the underlying event the easier it is to extract signals for new physics. There is also the phenomena of pile-up at modern colliders. In order to push the instantaneous luminosity to such high values that very rare processes can be detected, the colliders have to be operated with such high bunch population that several hundreds of separate interactions occur during one and the same bunch crossing. Most of those interactions are soft and produce what is called "pile-up" background in the different detectors and this background has to be separated from the signal.

Understanding of Soft QCD processes are also important in the context of cosmic rays. Monte-Carlo event generators used for simulation of the forward cascades in air showers have to be tuned in order to extract the essential physics parameters in cosmic ray studies.

Here we will start with a discussion of elastic scattering and the total cross section in proton-proton collisions, and in a second part some other typical Soft QCD topics will be addressed. It will be impossible to cover all the topics nowadays associated with Soft QCD in this short review and we have to make a biased selection. A very good and more extensive summary of Soft QCD is given in the article "High Energy Soft QCD and Diffraction" written by V.A. Khoze, M. Ryskin and M.Taševský published in PDG [939].

### 12.6.2 Elastic proton-proton scattering and the total cross section

*First principles*

Elastic scattering is the simplest process possible at a hadron-hadron collider. Two incoming protons scattering at the Interaction Point (IP) giving two outgoing protons and nothing more. It is the most simple process possible involving strongly interacting particles but still it can not be described directly by QCD. However there are first principles or fundamental concepts which are relevant for elastic scattering and the total cross section. Those principles have to be fulfilled by any theory of strong interactions and must obviously also be fulfilled by QCD. Principles like unitarity, crossing symmetry and analyticity of the elastic scattering amplitude are of importance. Those principles connect elastic scattering with the total cross section in different manners.

The most straight forward is the optical theorem that connects the total cross section with the imaginary part of the scattering amplitude in the forward direction. The high energy form of the optical theorem can be written:

$$\sigma_{tot} = \frac{ImF_{el}(t=0)}{s}, \tag{12.6.1}$$

where $t$ is the four momentum squared which at high energies can be written as -$t$ = $(p\theta)^2$ with $p$ being the momentum and $\theta$ the scattering angle. The Mandelstam variable $s$ represents the centre of mass energy squared. The optical theorem is based upon probability conservation in the scattering process and is easily derived using Quantum Mechanics.

The optical theorem has been used to experimentally determine the total cross section via measurement of the differential elastic cross section from ISR times to LHC today. From the optical theorem one derives the formula

$$\sigma^2_{tot} = \frac{16\pi}{1+\rho^2}\frac{d\sigma_{el}}{dt}(t=0), \tag{12.6.2}$$

where $\frac{d\sigma_{el}}{dt}(t=0)$ is the elastic differential cross section extrapolated to $t=0$ and $\rho$ is the ratio of the real to imaginary part of the elastic scattering amplitude in the forward direction i.e.

$$\rho = \frac{ReF_{el}(t=0)}{ImF_{el}(t=0)}. \tag{12.6.3}$$

However the optical theorem is not the only connection between $\sigma_{tot}$ and elastic scattering. Using the concepts of analyticity and crossing symmetry, dispersion relations for elastic scattering can be derived. Dispersion relations connect the $\rho$-parameter at a certain energy to the energy evolution of $\sigma_{tot}$ both below and above this energy and are a very powerful tool which play a crucial role in the understanding of elastic scattering. Dispersion relations thus imply that the $\rho$-value at a certain energy is sensitive to the energy evolution of $\sigma_{tot}$ beyond the energy at which $\rho$ is measured. The $\rho$-value is accessible experimentally and can be measured by measuring elastic scattering as such small angles where the Coulomb amplitude starts to be significant. The Coulomb amplitude is proportional to $1/t$ and dominates in the very forward direction. The interference between the Coulomb amplitude and the strong amplitude permits a measurement of $\rho$. Using the measurement of $\rho$ and dispersion relations one can make predictions of $\sigma_{tot}$ to an energy of the order 10 times higher than the energy at which $\rho$ has been measured. This has been done several times in the past [3923, 3924].

The Froissart-Martin bound [3925, 3926] is another example of an important consequence derived from first principles. Based upon axiomatic quantum field theory it was shown that $\sigma_{tot}$ can not grew faster than

$$\sigma_{tot} < \frac{\pi}{m_\pi^2} ln^2 s. \tag{12.6.4}$$

As will be discussed in the paragraph "The total cross section" below, this bound is not very constraining given the energy scales available today.

*The Regge approach and QCD*

The principles discussed above generate bounds and relations between important entities but do not lead to a concrete proposal for the scattering amplitude. For this, one still has to rely upon phenomenological approaches. The phenomenology of Regge theory dominated the description of high energy scattering process in pre-QCD times (see for instance [3927],[3928] and references therein). With the advent of QCD as the theory for strong interaction in the 70th, Regge theory started to loose its role. The obvious wish was of course to try to derive Regge theory from QCD. Due to the non-perturbative character of low $p_T$ reactions this turned out to be extremely difficult and still today Regge concepts are the basis of the phenomenology used to describe soft processes. However, whenever possible one tries to connect to QCD in a smooth way.

The key concept in Regge theory is singularities of the amplitude in the complex angular moment plane, the so called $j$-plane. The most straightforward singularity is a simple pole. Using this concept for a given scattering process has as consequence that the scattering amplitude in the $t$-channel can be be calculated using an exchange of so called Regge-trajectories which replaces a single particle exchange. A Regge trajectory composes of many particles with the same quantum numbers except for the spin. The particles are organised in increasing spin with increasing mass on the trajectories. A trajectory is represented by the function $\alpha(t)$ where $\alpha(t)$ is the pole position in the $j$-plane and is usually parameterized as a linear function of $t$ :

$$\alpha(t) = \alpha(0) + \alpha^{'} t. \tag{12.6.5}$$

The exchange of a Regge trajectory or a Reggeon leads to a power-like growth of the amplitude with $s$ and an exponent $\alpha(t)$ i.e.

$$A(s,t) \propto s^{\alpha(t)}. \tag{12.6.6}$$

Using the optical theorem one then gets for the corresponding cross section

$$\sigma \propto s^{\alpha(0)-1}. \tag{12.6.7}$$

The contribution of a given Regge trajectory factorizes in general, i.e. the amplitude is a product of two factors depending only on the coupling of the exchanged object to the scattered particles at each vertex.

At energies around 20-50 GeV, corresponding to the ISR and at energies below, several different leading Regge trajectories contribute to the amplitude. Experimentally it turns out that the leading trajectories in $pp$ scattering have an intercept $\alpha(0) \approx 0.5$ (see section 51 in [3929]). Thus the corresponding contributions all vanish with increasing energy in an inverse power law according to Eqn. 12.6.7. At higher energies only the so called Pomeron trajectory survives. The Pomeron carries the quantum numbers of vacuum with $CP = ++$ and was proposed in the 1960s to explain the asymptotic behaviour of the total cross section as will be discussed in the following paragraph. The Pomeron is a good example how Regge theory connects to QCD. The Pomeron has now been identified as a two gluon state in QCD (see e.g. references in [3930]) and some of the properties of the Pomeron has been derived in QCD.This will be discussed in Sect. 12.6.4. QCD also predicts the possible existence of a three gluon state with $CP = --$. Such a state corresponds to a trajectory proposed in Regge theory in the 1970s [3931], the so called Odderon. The Pomeron and the Odderon will be discussed more in detail later.

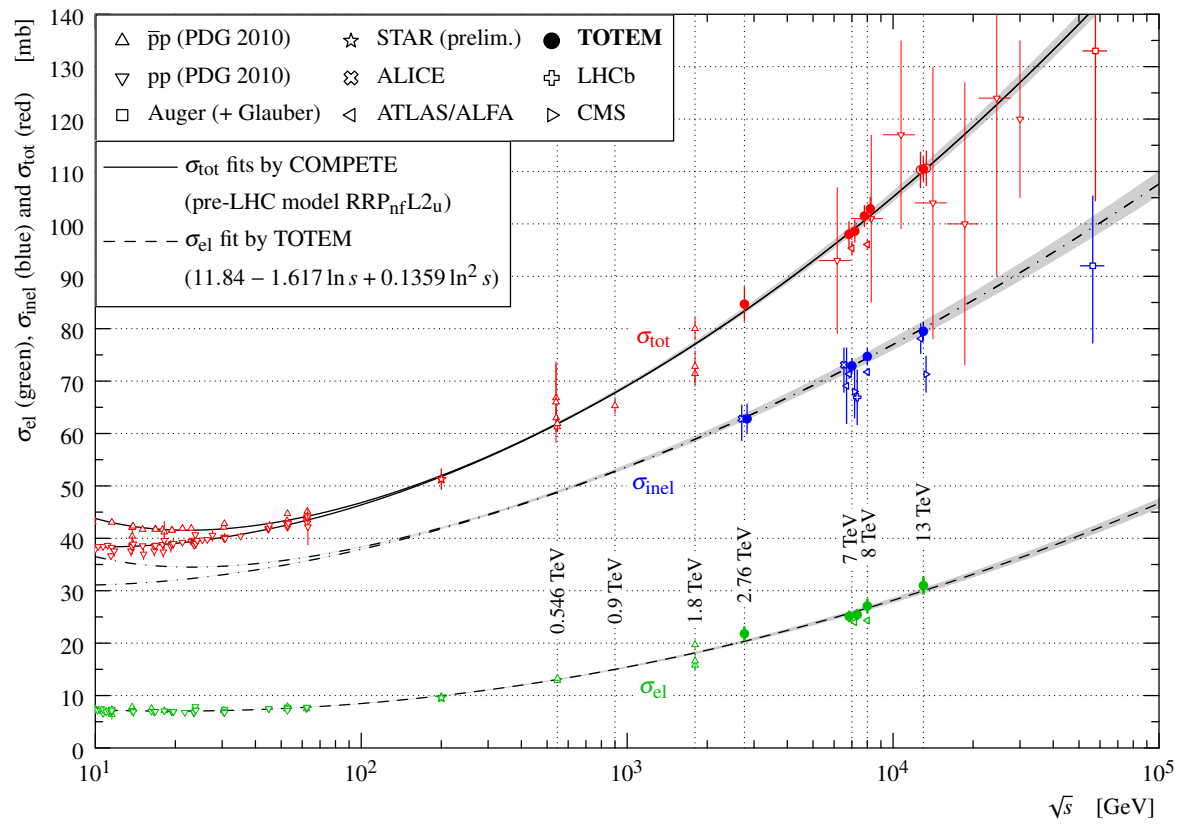

**Fig. 12.6.1** The total cross section for $pp$ and $\bar{p}p$ as a function of the center-of-mass energy. In the figure is also shown the energy dependence of the elastic and inelastic cross sections. Figure is taken from Ref. [3932]. More details about the figure can be found in [3932].

*The total cross section*

In Fig. 12.6.1 is shown all the world data of the total $pp$ and $\bar{p}p$-cross section from the ISR to LHC. Low energy proton data from fixed target experiments are also shown in the figure. The total cross section starts to rise at ISR. The rise of $\sigma_{tot}$ was not at all expected. Still today this rise can not be derived from first principles and is not satisfactory solved within QCD.

At the time of the discovery of the surprising rise of the total cross section with increasing centter-of-mass energy, Regge theory was the relevant theory for strong interaction. The Pomeron was thought to dominate the high energy behaviour of $\sigma_{tot}$ and the s-dependence was thus given by ( see Eqn. 12.6.7)

$$\sigma_{tot} \propto s^{\alpha_P(0)-1} \,, \tag{12.6.8}$$

where $\alpha_P(0)$ is the intercept of the Pomeron trajectory. The intercept was believed to be exactly 1 thus giving a constant total cross section asymptotically. Later when it was discovered that $\sigma_{tot}$ was starting to rise an intercept $\alpha_P(0)$ just above 1 was introduced. Taking into account the data from the SPS collider and the Tevatron, $\sigma_{tot}$ was parameterized in terms of an "effective" Pomeron trajectory $\alpha(t) = 1+\Delta+\alpha^{'} t \approx 1+0.08+0.25t$ where $t$ is given in $GeV^2$ [1104]. However it was always clear that at some very high energy such a power growth in $s$ will violate unitarity and the Froissart bound (see Eqn. 12.6.4). This problem is addressed by also considering cuts in the $j$-plane in addition to the simple poles. The cuts describe multi-pomeron exchanges and it turns out that those multi-pomeron exchanges tame the growth of the total cross section and thus restore unitarity.

Actually the possibility of a rising total cross section had been outlined already by Heisenberg in 1952 [3933]. He used a very simple argument based upon the range of the strong interaction and the pion mass to indicate a possibility of a $ln^2(s)$ rise of $\sigma_{tot}$. This argument had fallen into oblivion in the mid's of the seventies. Now it has turned out at each new collider energy that $\sigma_{tot}$ essentaily rise as $ln^2(s)$. The full line drawn in Fig. 12.6.1 represents one of many $ln^2(s)$ fit to the data. In this case it is one of the COMPETE parametrisations [3934]. In Regge theory such a $ln^2(s)$behaviour can only appear if the the singularity in the $j$-plane is a pole of order 3 i.e. a triple pole.

Does the $ln^2(s)$ rise mean that the Froissart-Martin bound mentioned in the previous paragraph is saturated? Actually we are far away from a saturation today. The coefficient in front of $ln^2(s)$ term in the Froissart-Martin bound is 60 mb and typically $ln^2(s)$ fits to the data give coefficients O(0.1 mb). Thus the bound is far away from being saturated at today's energies.

The rise of $\sigma_{tot}$ as $ln^2(s)$ cannot be derived from QCD today. However it is interesting to note that there have been quite successful attempts in lattice QCD to reproduce an $ln^2(s)$ behaviour with a coefficient not so far away from the one experimentally found [3935]. There has also been attempts to generate a $ln^2(s)$ behaviour using gluon saturation in Coulor Glass Saturation models [3936]. This is a good example of how perturbative and non-perturbative physics meet giving an interesting result.

*Elastic scattering*

Figure 12.6.2 shows a couple of examples of the differential elastic cross section and its $t$-dependence at different energies at the LHC. The measurements have been done by the TOTEM collaboration [3937]. As mentioned in the previous paragraph the Pomeron trajectory dominates at energies of the LHC. In terms of QCD this means a dominance of two gluon exchange. The gross features of the $t$-dependence of differential elastic cross section at high energies can be described in terms of the Pomeron or a two gluon exchange. The cross section falls close to exponential in the forward direction. This means that the Pomeron–proton coupling has an exponential fall off. There are small deviations from the exponential that are not completely

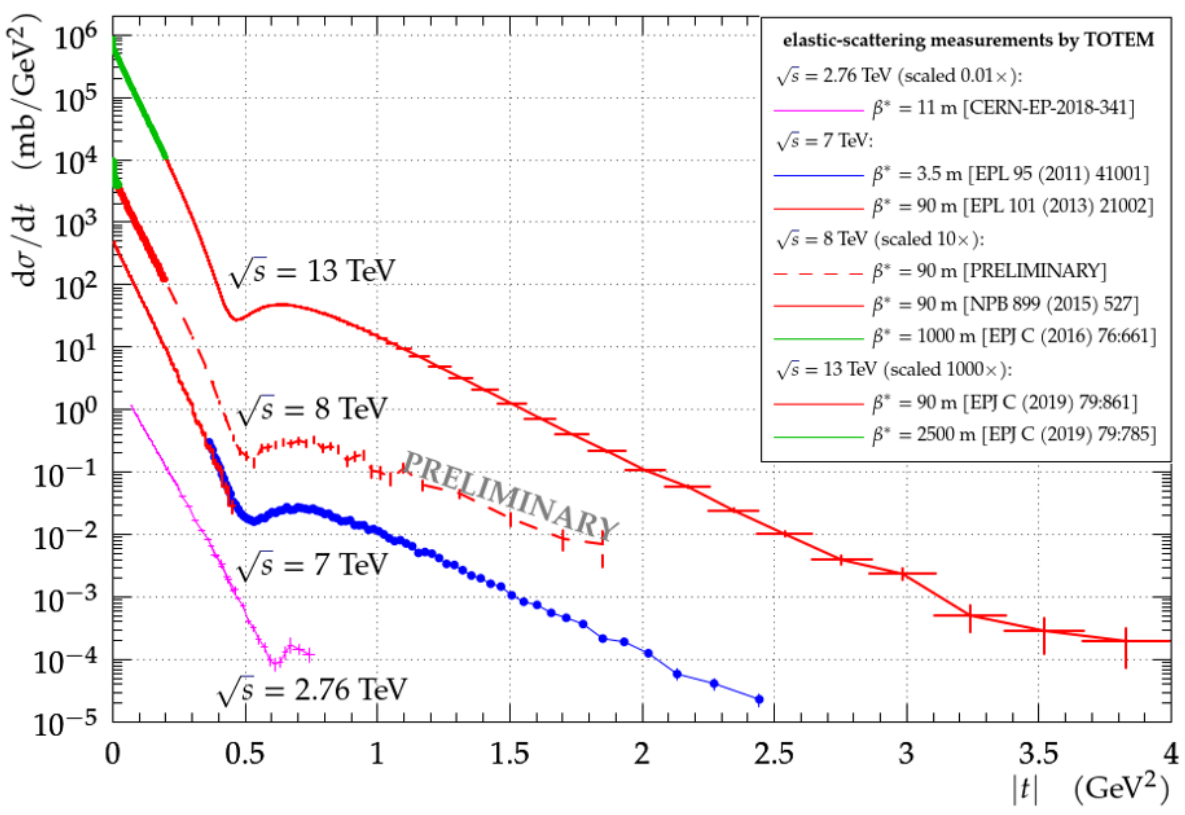

**Fig. 12.6.2** The differential elastic cross section as a function of the four momentum transfer $t$ for different energies at the LHC as measured by the TOTEM collaboration. Figure is taken from Ref. [3938].

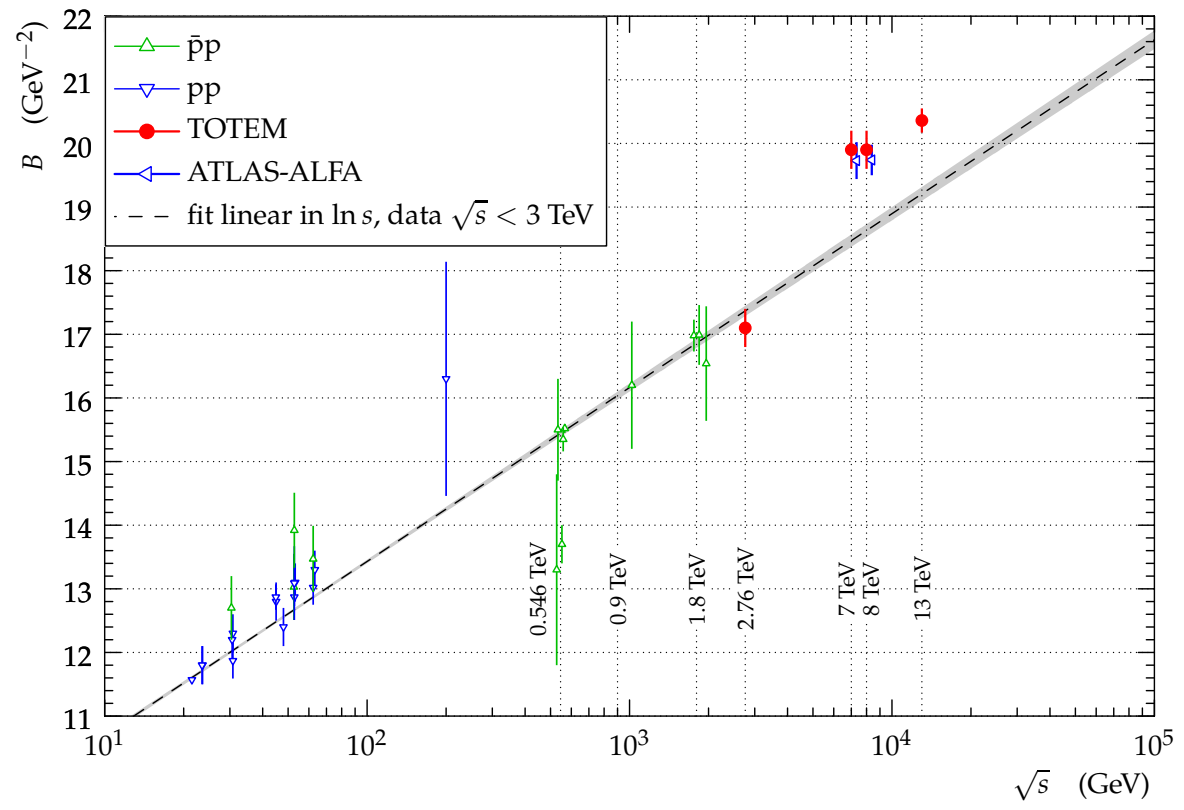

**Fig. 12.6.3** Measurements of the slope-parameter $B$ as a function of the center-of-mass energy $\sqrt{s}$ for $pp$ and $\bar{p}p$ scattering . The straight line represents a linear fit in $\ln(s)$ of the data below $\sqrt{s}$=3 TeV. Figure taken from Ref. [3939].

understood but might at least partly be due to multi-pomeron exchanges.

The exponential fall-off parameter, often called the $t$-slope $B$, also has an energy dependence. The energy dependence of $B$ is plotted in Fig. 12.6.3. The straight line corresponds to a linear dependence in $ln(s)$

$$B = B_0 + 2\alpha_P^{'} \ln(s). \tag{12.6.9}$$

This linear dependence in $ln(s)$ is a direct consequence of the exchange of a Pomeron in the $t$-channel. The $\alpha_P^{'}$ parameter represents the slope of the Pomeron trajectory $\alpha_P(t) = \alpha_P(0) + \alpha_P^{'} t$ .

As can be seen in the figure this linear relation works well for energies below LHC. However at the LHC the increase with $s$ starts to accelerate. Also this can be explained in terms of multi-pomeron exchanges mentioned above [3940].

In Fig. 12.6.2 is also seen that after the exponential decrease, the differential cross section exhibits a dip that moves towards smaller $t$-values when the energy increase. In the Pomeron language this is interpreted as an interference between one pomeron exchange amplitude and multi-pomeron exchange amplitudes making essential the imaginary part of the total amplitude disappear at the dip. This mechanism generates a dip which correctly moves towards smaller $t$-values with energy.

At high values of $t$, beyond the dip, the cross section decreases further in a smooth way. Here one moves away from the non-perturbative regime and instead one might see signs of perturbative QCD. The triple gluon exchange proposed in Ref. [3941] could be a manifestation of this.

*The Odderon*

As seen in the previous paragraphs the Pomeron plays an essential role in the description of elastic scattering and the total cross section. The situation is very different concerning the Odderon. The Odderon is the $CP = --$ counter-partner of the Pomeron and contributes with a different sign to the amplitude for $pp$-scattering relative to $\bar{p}p$-scattering. The Odderon is both controversial and non-controversial. It is non-controversial in the sense that no one really doubts its existence. It is a firm prediction of QCD and represents a three gluon state in contrast to the two gluon state of the Pomeron. What is somewhat controversial is the size of its coupling and its importance in the elastic amplitude. To what extent the Odderon really has manifested itself in the available experimental data is debatable (see e.g Ref. [3942]) though the authors of Ref. [3943] claim a discovery.

Experimentally there are two different signals that have been evoked as a sign of an Odderon. The most convincing is probably the difference between $\bar{p}p$-scattering and $pp$-scattering observed in the dip region of elastic scattering. The $\bar{p}p$ data from the D0 experiment at the Tevatron at 1.96 TeV has been compared to the $pp$ data at 2.76 TeV from the TOTEM experiment at the LHC [3943]. The dip is supposed to be filled partly by the real part of the Odderon amplitude having a different sign for $pp$ and $\bar{p}p$-scattering. The two distributions are shown in Fig. 12.6.4. Ideally the comparison $pp$ and $\bar{p}p$ should be done at the same energy. However, the authors have taken great care to compare the D0 measurement with TOTEM data extrapolated to the 1.96 TeV

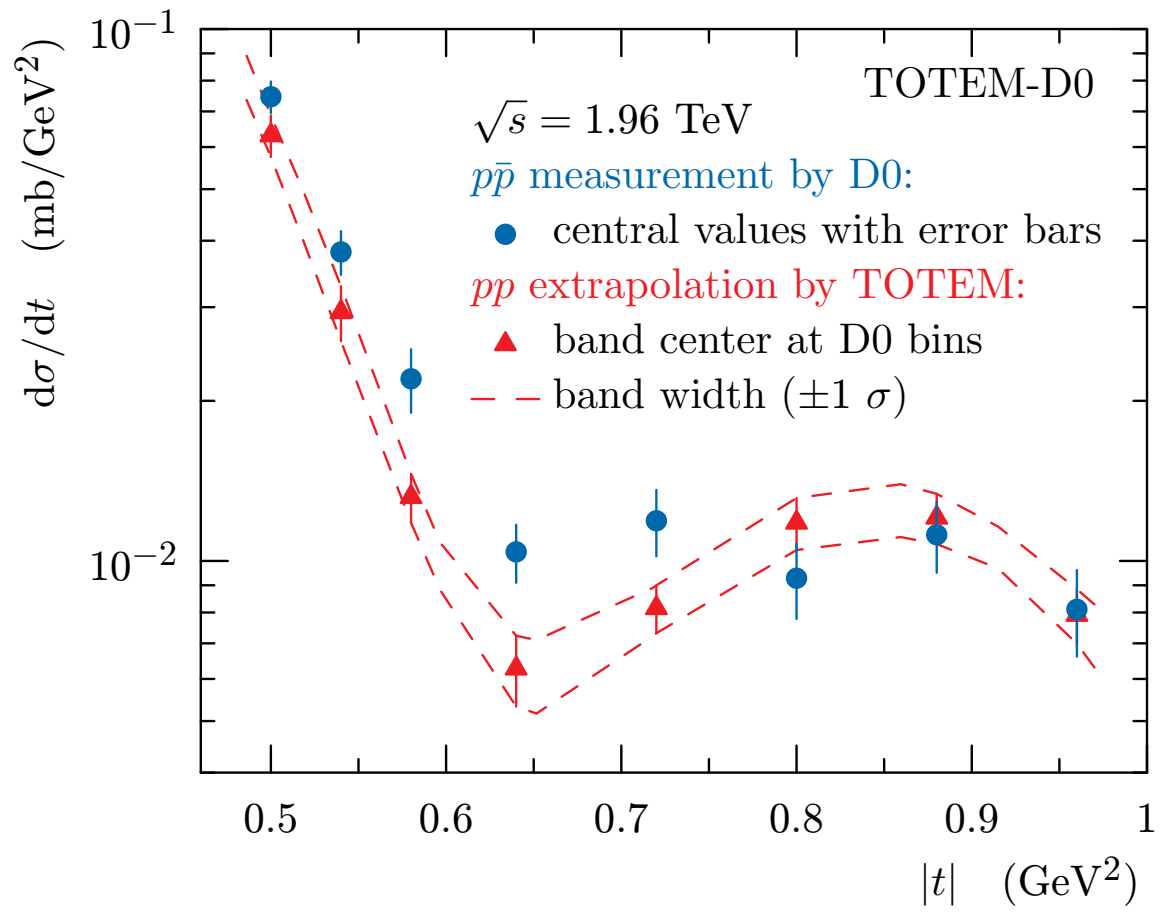

**Fig. 12.6.4** Comparison between the D0 $\bar{p}p$ measurement at $\sqrt{s}$=1.96 TeV and the extrapolated TOTEM $pp$ cross section rescaled to match the optical point of the D0 measurement. The dashed lines show the 1$\sigma$ uncertainty band. Figure is taken from Ref. [3943] where more details are given.

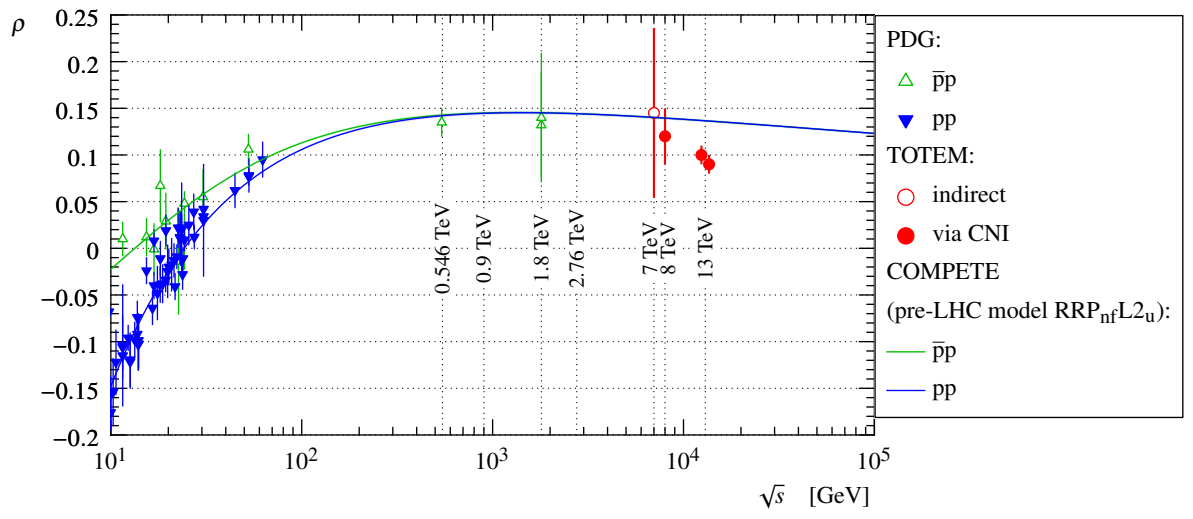

**Fig. 12.6.5** Dependence of the $\rho$-parameter on center-of-mass energy. The $pp$(blue) and $\bar{p}p$(green) curves are taken from Ref. [3934]. Figure is taken from Ref. [3932].

of the Tevatron. They find a 3.4$\sigma$ difference between the two distributions in Fig. 12.6.4.

The second possible experimental manifestation of the Odderon is a measurement of the TOTEM experiment which has measured the $\rho$ parameter at 13 TeV to be $\rho = 0.09$ [3932]. This result is in contradiction to dispersion relation calculations assuming that the standard $ln^2(s)$ behaviour of $\sigma_{tot}$ continues beyond LHC and assuming that the elastic amplitude only contains the Pomeron contribution. Those calculations give $\rho = 0.13 - 0.14$ ( see Fig. 12.6.5 and Ref. [3934] ) thus signifcantly higher than the TOTEM result.

The TOTEM result could therefore be an indication that $\sigma_{tot}$ starts to grow somewhat slower beyond the LHC energies. However an alternative explanation might well be that the low $\rho$ value is produced by an Odderon effect. An Odderon contribution to the amplitude can modify the dispersion relation calculation in a way to give a better agreement with the data. The effect depends on the size of the Odderon contribution at a certain energy. The so called maximal Odderon [3944] is one example that actually produces an effect agreeing with the TOTEM data.

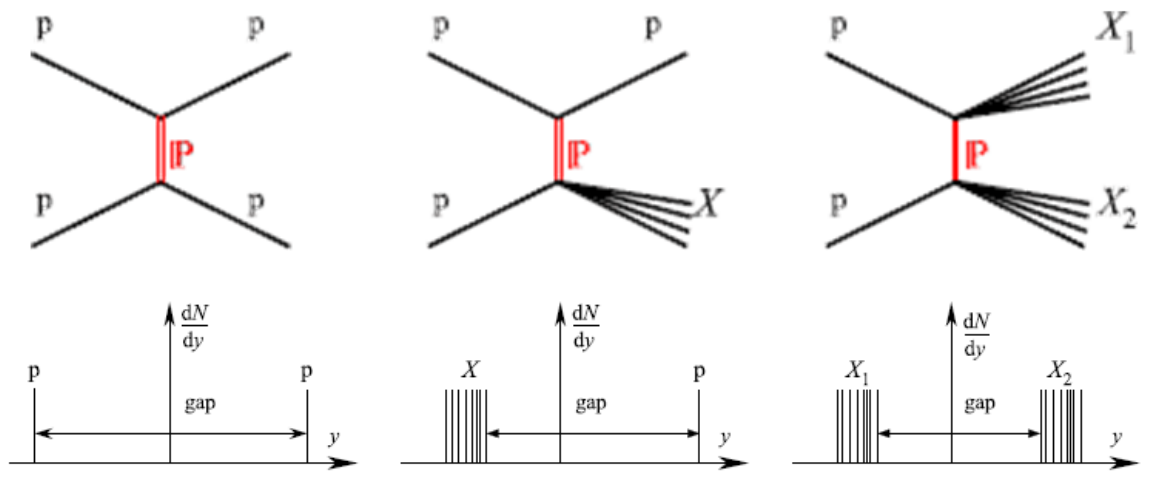

**Fig. 12.6.6** Feynman diagrams for different diffractive topologies. $I\!P$ stands for Pomeron and $p$ for proton while $X$ represents the diffractive systems. Below each diagram is also shown the corresponding rapidity distribution of the outgoing particles. Figure taken from Ref. [3945].

To summarise: the measurement of $\rho$ at 13 TeV may be an indication of the Odderon but the fact that an alternative explanation exists means that this signal can not be taken as a hard proof of the Odderon.

### 12.6.3 Diffraction

In this article we have separated the discussion of elastic scattering and diffraction but actually elastic scattering is the dominant diffractive process. There is no unique definition of diffraction, neither theoretically nor experimentally. A key concept when talking about diffraction is rapidity gaps. For elastic scattering the size of the rapidity gap (a rapidity[114] region void of particles) between the two outgoing protons is at its kinematical limit. In general a diffractive event is characterized by a rapidity gap which is significant larger than possible fluctuations in the hadronisation process.Typical this means rapidity gaps bigger than 4-5 units of rapidity at LHC energies. Depending on the topology of the rapidity gaps one talks about different types of diffractive events. This is illustrated in Fig. 12.6.6 where the topologies, elastic, single dissociation and double dissociation are shown. Below the Feynman diagrams are also illustrated schematically the corresponding rapidity distributions of the outgoing particles.

All these topologies are characterized by the exchange of the Pomeron in the $t$-channel or in other words an exchange of a a color singlet state of two gluons.

[114] When dealing with a particle whose mass is negligible compared with its energy, the pseudorapidity = -ln (tan($\theta$/2)) is a good approximation to the rapidity. Here $\theta$ is the polar angle of the particle. In this article we do not make the distinction between rapidity and pseudorapidity.

It is not always possible to map the experimental data directly to the different topologies seen in Fig. 12.6.6. In general it is difficult to measure diffraction at high energy colliders, especially diffractive system with a low mass. For low mass systems a large fraction of the diffractively produced particles are emitted in the very forward direction and lost in the beam pipe.

Experimentally there are two ways to select diffractive events. Either to look for rapidity gaps or use so called proton tagging. Proton tagging implies that one or both of the two intact protons actually are detected and measured. This requires small and sophisticated detectors situated as close as possible to the beam line. In practice, the detectors have to be placed at distances of a mm or smaller from the beam and thus the vessels containing the detectors have to be integrated in the beampipe. This technique, using so called Roman Pots, was introduced by the CERN–Rome group at the ISR half a century ago and is still used as the main technique to approach the beam [3923].

Here we will limit ourselves to discuss the simplest topology in Fig. 12.6.6, i.e. single diffraction. As an example we show in Fig. 12.6.7 the distribution of the experimental gap size $\Delta\eta_f$ measured by ATLAS at 7 TeV for particles with $p_T > 200\,\text{MeV}$[3946]. The true gap size is $\Delta\eta{=}\Delta\eta_f{+}4$ in this example. The size of the rapidity gap is directly related to the mass $M_x$ of the diffracted system.

$$\Delta\eta \backsimeq -ln(\xi_x) \tag{12.6.10}$$

with

$$\xi_x = \frac{M_x^2}{s} \tag{12.6.11}$$

The larger the rapidity gap is the smaller is the produced mass.

In Fig. 12.6.7 one can clearly see the difference between non diffractive and diffractive events. At small gap sizes $\Delta\eta_f < 2$, non diffractive events dominate and the expected exponential decrease of the cross section with increasing gap size which characterize the fluctuations of the hadronisation is the dominant feature. On the other hand for gap sizes $\Delta\eta_f > 3$ there is a rather flat plateau, which corresponds mainly to single diffractive processes. The largest rapidity gap size bin at the end of the plateau in Fig. 12.6.7 corresponds to diffractive masses larger than about 15 GeV and thus the plateau corresponds to masses above and equal 15 GeV. The cross section on the plateau is roughly 1 mb per unit of rapidity gap size. In the Regge theory, such type of high mass diffraction is characterized by a triple Pomeron coupling which actually predicts such a plateau. In perturbative QCD a triple Pomeron coupling of the same order of magnituded is found [3947].

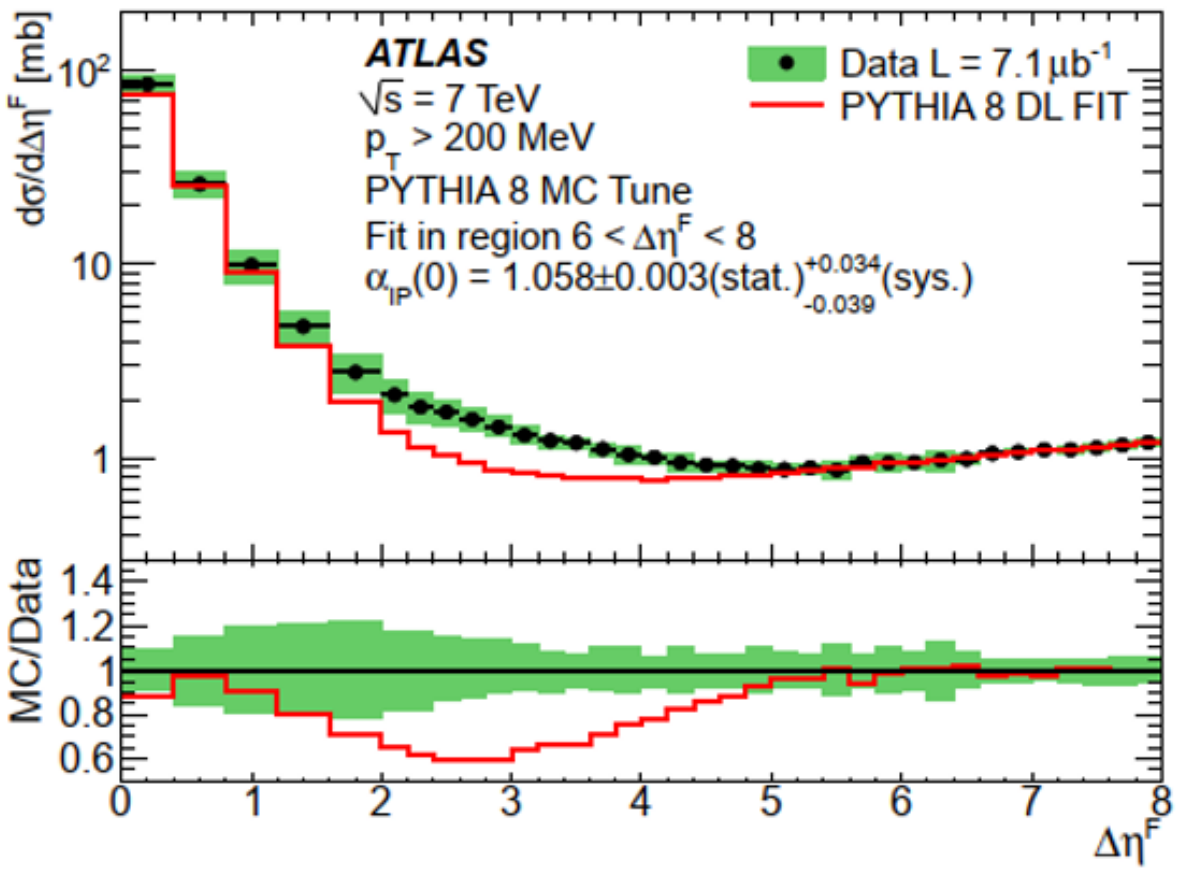

**Fig. 12.6.7** Inelastic cross section differential in the experimental gap size $\Delta\eta_f$ measured by ATLAS at 7 TeV for particles with $p_T > 200$ MeV. Figure taken from Ref. [3946]. See text in Ref. [3946] for further details.

Measurements at masses much lower than this are very difficult at high center of mass energies. There exist measurements down to masses of 3-4 GeV but they are scarce and often contradictory. Moreover also theoretically the estimates of low mass diffraction are notorious difficult and uncertain. Actually the uncertainties related to low mass diffraction constitute the largest uncertainty of the total cross section measurements by the TOTEM experiment using the so called luminosity independent method which requires an estimate of the total inelastic cross section including low mass diffraction (see e.g Ref. [3939]).

### 12.6.4 ”Soft” and ”hard” diffraction

This article deals with ”soft” diffraction but this is of course a somewhat arbitrary classification and in this section the concept of ”hard” diffraction will also briefly be touched upon. A part of diffractive events has a hard scale present.The hard scale is often given by a diffractively produced heavy system such as for example dijets, $W$ or $Z$ bosons, or heavy quarks. With such a hard scale present, perturbative QCD is applicable. There is actually no sharp distinction between what is called ”soft” diffraction and ”hard” diffraction but rather a smooth transition between the two. Often the perturbative approach is extended into the soft domain in a gradual manner using a unified framework. This has lead to the concept of a ”soft” Pomeron and a separate and different ”hard” QCD Pomeron. This distinction between different Pomerons is very likely an oversimplification of a more complex situation.

The data seem to indicate a ”hard” Pomeron with the intercept $\alpha(0) \backsim 1.3 - 1.5$ with a small slope $\alpha'$ in contrast to the "effective" Pomeron which is relevant for elastic scattering and the total cross section with an intercept of $\alpha(0) = 1.08$ and a slope $\alpha' = 0.25$ as mentioned in Section 12.6.2. This means that when a hard scale is involved the energy dependence is steeper relative to soft diffraction (see Eqn. 12.6.7). Taking diffractive vector meson production at HERA as an example: the energy dependence of the $\gamma^*$p cross section for $J/\psi$ production corresponds to an intercept $\alpha(0) \backsim 1.4$ (see Refs. [3948],[3949]). Such an intercept of the hard Pomeron, represented by a two gluon state, agrees with what has been calculated in perturbative QCD by resummation of the leading logarithms. The small slope $\alpha'$ of the "hard" Pomeron is also reproduced in perturbative QCD calculations [3950].

Hard diffraction has extensively been studied at HERA in $\gamma^* p$ processes and the results have been interpreted in terms of diffractive Parton Distribution Functions of the Pomeron and a Pomeron flux factor based upon Regge theory [3570]. It was shown within QCD that factorization is valid for diffractive hard scattering in $\gamma^* p$ processes [3951]. However, using the same formalism and using the DPDF's determined at HERA from $\gamma^* p$ processes applied to $\bar{p}p$ processes at the Tevatron gives about an order of magnitude too high cross section for QCD jet production (see e.g. [3952]). At the Tevatron the process is completely hadronic, and the reduction of the cross section is thought to be due to the fact that sometimes the rapidity gap is filled or partially filled by hadron remnants which are not present in $\gamma^* p$ processes. In this case, the factorisation using diffractive DPDF's suggested by the HERA data breaks down.

### 12.6.5 The Underlying Event

The underlying event is not to be confused with Minimum Bias Events. As the name indicates, minimum bias events are events collected with as little bias as possible. The concept of Underlying Event (UE) is different. Here one refers to events that contain a hard parton-parton interaction and the term underlying event refers to all the activity that accompanies the hard scatter but is not a part of it. The Underlying Event has several different components. There are contributions from initial and final-state interaction but also particles from the proton break-up so called beam–beam remnants contribute. An important part of the Underlying Events consists of Multiple Parton Interactions (MPI) i.e. two or more soft or semi hard interactions within the same $pp$ interaction. In Fig. 12.6.8 a typical UE is shown in schematized way.

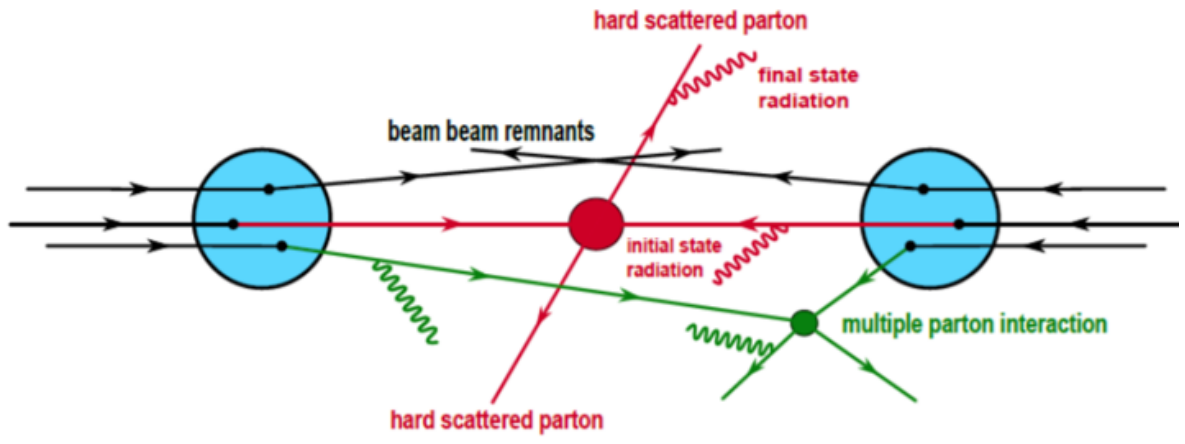

**Fig. 12.6.8** Schematic drawing illustrating a typical Underlying Event. Figure taken from ATLAS slides ATL-PHYS-SLIDE-2013-330.

A good description of the UE is needed to extract the relevant signals from the hard scatter and rely upon MC generators which are based upon different phenomenological approaches. Many input parameters in the MC generators parameters have to be tuned with data. To get information relevant for the UE the event is often divided into different regions of the phase space as indicated in Fig. 12.6.9. Normally a "transverse" region is defined relative to the azimuthal angle of the leading $p_T$ particle. This region is then taken as the reference region for the underlying event. In Fig. 12.6.10 is shown an example of the mean charged-particle multiplicity as a function of the leading $p_T$ for the different regions around the leading particle [3953]. All regions exhibit a fast rise at low $p_T$ up to a $p_T$ of about 5 GeV. Here there are no real "hard" processes and we see the rise of activity due to the increase of MPI. At higher $p_T$, hard processes start to dominate and the transverse region which is decoupled from the hard scatter reach a plateau.

### 12.6.6 Charged particle density

The charged particle density as a function of rapidity is an important observable in $pp$ collisions. The measurement covering the largest rapidity interval has been done by a combination of CMS and TOTEM at the LHC [3954, 3955]. The result of their measurement is shown in Fig. 12.6.11.

To describe the entire rapidity interval models must be able to combine and connect perturbative QCD with non-perturbative approaches. The experimental points are compared to a number of different models which are available. The approaches are different but there are also several common elements in the models. As can be seen the gross features of the distribution are reasonably well described by the models.

The density at $\eta = 0$ as a function of the centre of mass energy has been plotted in Fig. 12.6.12 using data from the Sp$\bar{p}$S collider and the Tevatron in addition to the LHC data [3955]. The data points have been fitted

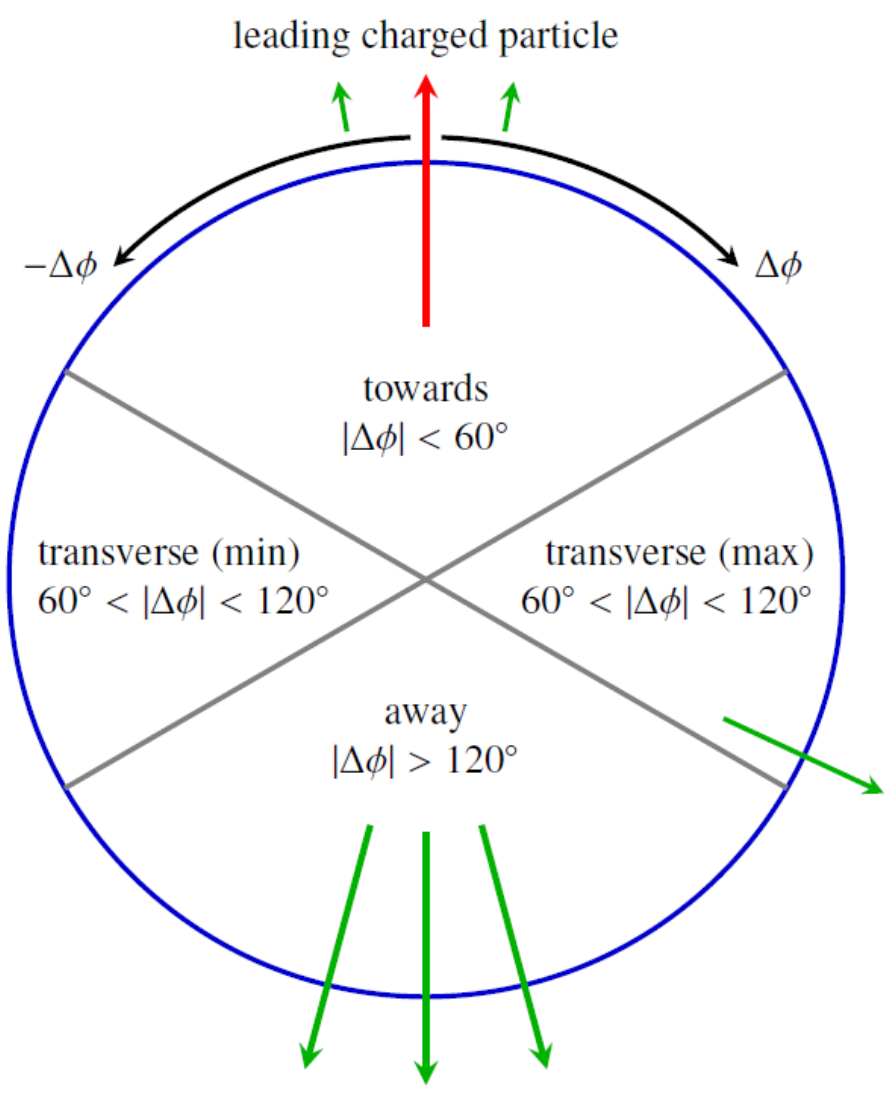

**Fig. 12.6.9** Definition of regions in the azimuthal angle with respect to the leading (highest-$p_T$) charged particle, with arrows representing particles associated with the hard scattering process and the leading charged particle highlighted in red. Conceptually, the presence of a hard-scatter particle on the right-hand side of the transverse region, increasing its $\Sigma p_T$, typically leads to that side being identified as the "trans-max" and hence the left-hand side as the "trans-min", with maximum sensitivity to the UE. Figure taken from Ref. [3953].

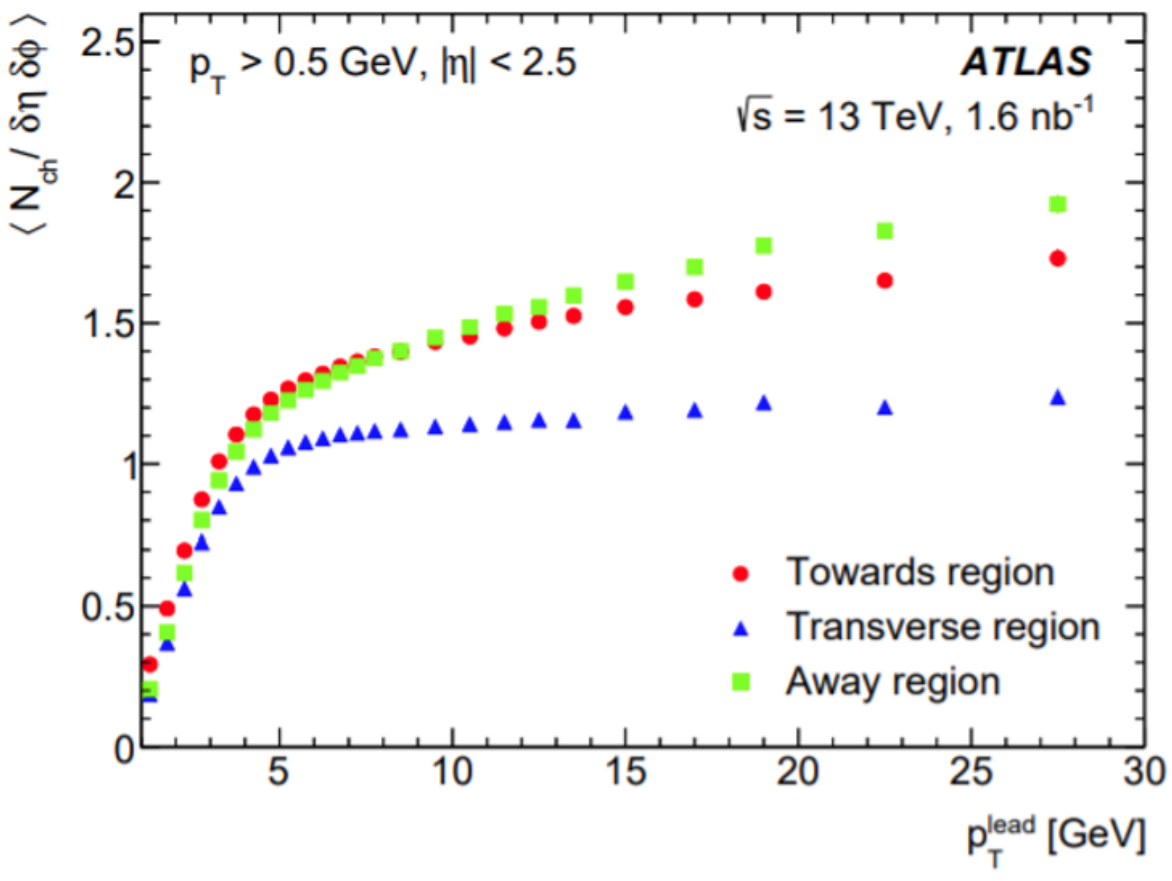

**Fig. 12.6.10** : Mean $\eta - \phi$ densities of charged-particle multiplicities as a function of the transverse momentum of the leading charged particle in the transverse, towards, and away azimuthal regions. The error bars, which are mostly hidden by the data markers, represent combined statistical and systematic uncertainty. Figure taken from Ref. [3953].

with a power law. It is interesting to note that the increase of the density at $\eta = 0$ is faster than the increase of the total cross section with energy. This can be understood in terms of Pomeron interaction. To calculate an inclusive cross section like the density at $\eta = 0$ it is enough to use a one-pomeron exchange diagram. For the total cross section on the other hand one has to take

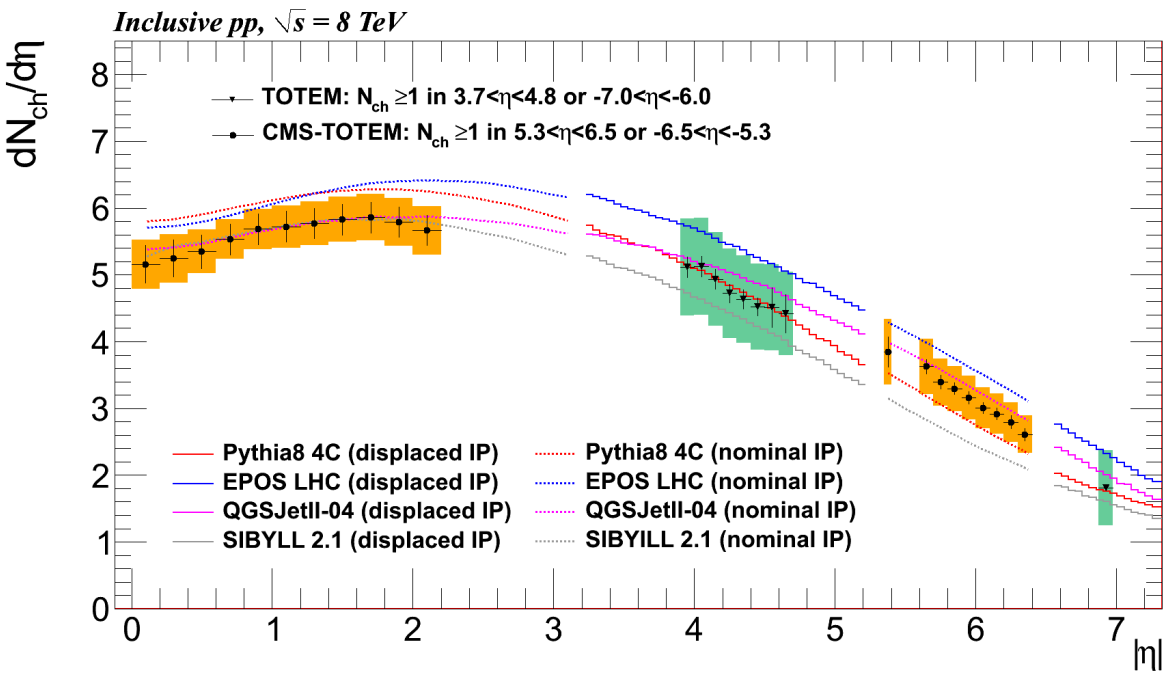

**Fig. 12.6.11** Charged particle pseudorapidity distributions obtained in $pp$ collisions at $\sqrt{s} = 8$ TeV for inelastic events as measured by the CMS and TOTEM experiments. The coloured bands show the combined systematic and statistical uncertainties and the error bars represent the $\eta$ uncorrelated uncertainties. The colored lines represents different model predictions. Figure taken from Ref. [3954] .

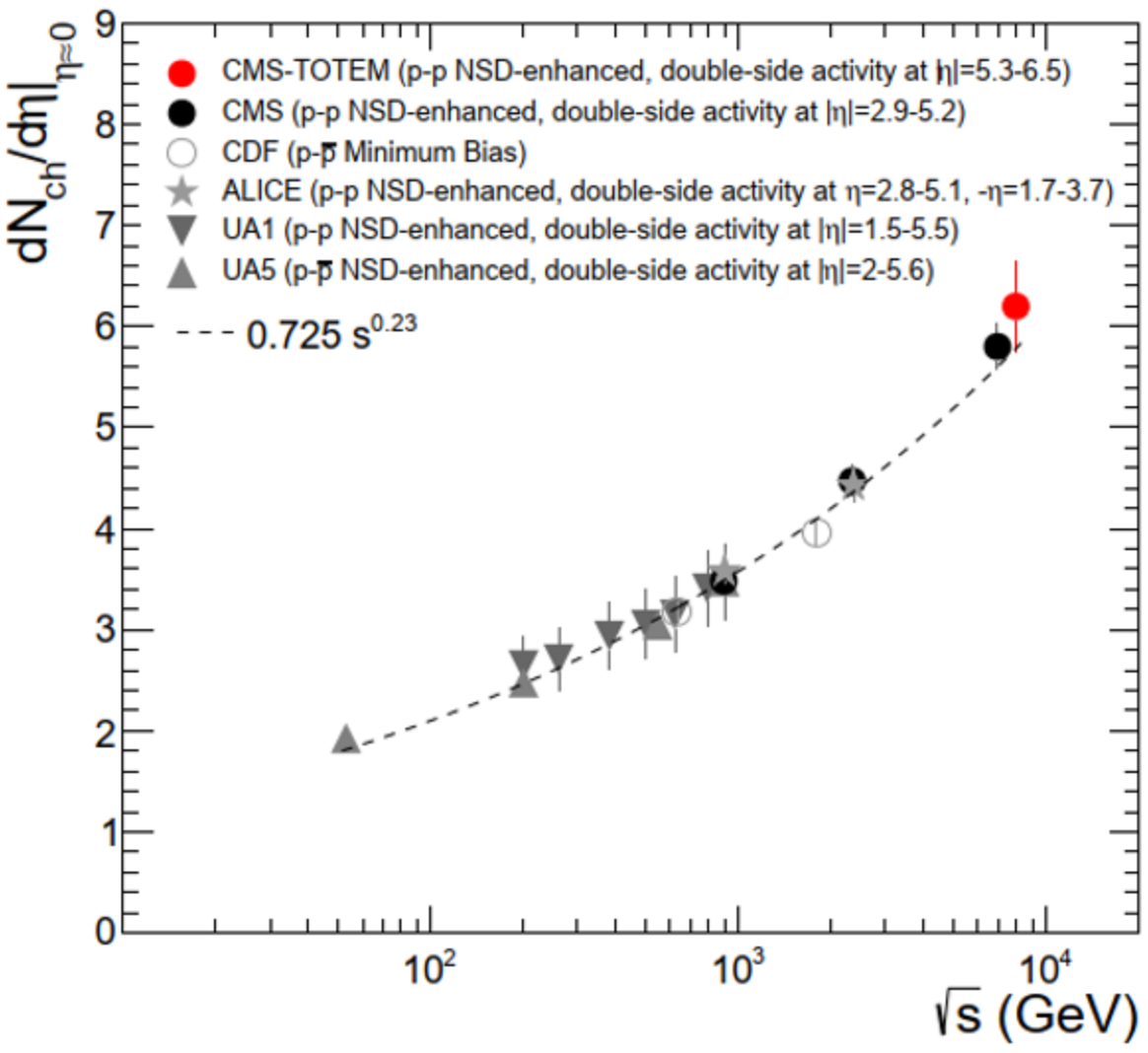

**Fig. 12.6.12** Value of $\mathrm{dN_{ch}}/\mathrm{d}\eta$ at $\eta \approx 0$ as a function of the center-of-mass energy for $pp$ and $\bar{p}p$ collisions. Shown are measurements performed with different event selections from a number of experiments listed in the figure. The dashed line is a power-law fit to the data. Figure taken from Ref. [3955].

into account multi-pomeron exchanges which tames the rise of the total cross section.

### 12.6.7 Conclusion

As mentioned in the introduction "soft" processes cover a large part of the total cross section. Collider experiments, at HERA and the Tevatron and now also at LHC, have produced a large amount of measurements related to low $p_T$ reactions. The large rapidity coverage of the LHC detectors, and dedicated small angle experiments such as TOTEM, have offered new possibilities and there is still more to come. Moreover, the high center-of-mass energy of the LHC means that kinematically a larger rapidity range is available which opens up a window of studies where a separation between diffractive and non-diffractive events is somewhat easier, at least for what concerns high mass diffraction.

The richness of the data at the LHC also implies that there are a number of aspects that we have not been able to treat in this short overview. For instance, the interesting topic of particle correlations has not been discussed and neither has multiple parton interactions been considered (For those topics and also for other topics that have not been discussed here, see e.g PDG [939]).

In the 1970s and 1980s the interest moved from Regge theory and low $p_T$ physics to high $p_T$ reactions and perturbative QCD. The “old“ physics lost considerable interest. However it turns out that the tools of the “old” physics work remarkably well also today. The current theoretical efforts try to bridge this gap between ”old” physics and ”new” and produce convincing descriptions of soft processes in terms of QCD. A lot of theoretical efforts have occurred over the years trying to make the transition from Regge poles and Regge Field Theory to QCD. Some attempts in this direction have been mentioned in this overview, but far from all.

With the abundant data from LHC available today the study of soft interactions has become a more vigorous field again. The hope is that “old” and “new” physics will meet and that a proper calculational framework based upon QCD will be developed in the close future leading to a better understanding of soft processes. A lot of progress have been made until today but the challenge is still there to incorporate a full understanding of soft processes in QCD.

# 13 Weak decays and quark mixing

*Conveners:*
**Andrzej J. Buras and Eberhard Klempt**

One of the main frontiers in the elementary particle physics is the search for new particles and new forces beyond those present in the Standard Model (SM) of particle physics. As the direct searches at Large Hadron Collider (LHC) at CERN, even ten years after the Higgs discovery, did not provide any clue what these new particles and forces could be, the indirect searches for new physics (NP) through very rare processes caused by virtual exchanges of heavy particles gained in importance. They allow in fact to see footprints of new particles and forces acting at much shorter distance scales than it is possible to explore at the LHC and presently planned high energy colliders. While the LHC can explore distance scales as short as $10^{-19}$m, the indirect search with the help of suitably chosen processes can offer us the information about scales as short as $10^{-21}$m which cannot be probed even by the planned 100 TeV collider at CERN. Also shorter scales can be explored in this manner.

In fact rare processes like $K_L \to \mu^+\mu^-$ known since the early 1970s implied the existence of the charm quark prior to its discovery in 1974 as only then its branching ratio could be suppressed in the SM with the help of the Glashow-Iliopoulos-Maiani (GIM) mechanism [78], to agree with experiment. Moreover, it was possible to predict successfully its mass with the help of the $K_L - K_S$ mass difference $\Delta M_K$ in the $K^0 - \bar{K}^0$ mixing prior to its discovery [3956]. Similary the size of the $B_d^0 - \bar{B}_d^0$ mixing[115], discovered in the late 1980s, implied a heavy top quark that has been confirmed only in 1995. It is then natural to expect that this indirect search for NP will also be successful at much shorter distance scales.

In this context, rare weak decays of mesons play a prominent role besides the transitions between particles and antiparticles in which flavors of quarks are changed. In particular $K^+ \to \pi^+\nu\bar{\nu}$, $K_L \to \pi^0\nu\bar{\nu}$, $K_S \to \mu^+\mu^-$, $B_s^0 \to \mu^+\mu^-$, $B_d^0 \to \mu^+\mu^-$ and $B_d^0 \to K(K^*)\nu\bar{\nu}$ but also $B_s^0 - \bar{B}_s^0$, $B_d^0 - \bar{B}_d^0$, $K^0 - \bar{K}^0$ mixings and CP-violation in $K \to \pi\pi$, $B_d \to \pi K$ decays among others provide important constraints on NP. Most of these transitions are very strongly loop-suppressed within the SM due to the GIM mechanism and also due to small elements $V_{cb}$, $V_{ub}$, $V_{td}$ and $V_{ts}$ of the CKM matrix [3957, 3958]. The predicted branching ratios for some of them are as low as $10^{-11}$. But as the GIM mechanism is generally violated by NP contributions these branching ratios could in fact be much larger.

The first step in this indirect strategy is to search for the departures of the measurements of the branching ratios of the decays in question from SM predictions and similar for mass differences like $\Delta M_K$, and analogous mass differences $\Delta M_s$ and $\Delta M_d$ in $B_s^0 - \bar{B}_s^0$ and $B_d^0 - \bar{B}_d^0$ mixings, respectively. But while these processes are governed by quark interactions at the fundamental level, the decaying objects are mesons, the bound states of quarks and antiquarks. In particular in the case of non-leptonic transitions like $B_s^0 - \bar{B}_s^0$, $B_d^0 - \bar{B}_d^0$, $K^0 - \bar{K}^0$ mixings and CP-violation in $K \to \pi\pi$ and $B \to \pi K$ decays, QCD plays an important role. It enters at short distance scales, where due to the asymptotic freedom in QCD perturbative calculations can be performed, and at long distance scales where non-perturbative methods are re-quired. QCD has also an impact on semi-leptonic decays like $K^+ \to \pi^+\nu\bar{\nu}$, $K_L \to \pi^0\nu\bar{\nu}$, $B \to K(K^*)\nu\bar{\nu}$ and even on leptonic ones like $K_S \to \mu^+\mu^-$, $B_s^0 \to \mu^+\mu^-$, $B_d^0 \to \mu^+\mu^-$ and $B \to K(K^*)\nu\bar{\nu}$. In order to be able to identify the departures of various experimental results from the SM predictions that would signal NP at work, the latter predictions must be accurate, and this means the effects of QCD have to be brought under control. But this is not the whole story. To make predictions for rare processes in the SM one has to determine the four parameters of the unitary CKM matrix

$$V_{us}, \quad V_{cb}, \quad V_{ub}, \quad \gamma \tag{13.0.1}$$

with $\gamma$ being the sole phase in this matrix.

[115] The $B_d^0 = (d\bar{b})$ is listed as $B^0$ in the Review of Particle Physics.

This section is divided into five parts. We present first the effective weak Hamiltonians both in the SM and beyond. We summarize briefly the history of the efforts to construct them and present their status. Here, renormalization-group (RG) methods - used to calculate QCD impact on the Wilson coefficients (WC) of local operators - are essential but also the non-perturbative evaluation of their hadronic matrix elements. This will be followed by the discussion of the present status of the CKM matrix (see Section 13.2) which will demonstrate the role of QCD in the determination of its elements. Subsequently, in Section 13.3, we will first summarize briefly the impact of QCD effects on rare leptonic and semileptonic decays. Here, these effects are mostly moderate, with the exception of radiative $B$ decays like the one into final states with open strangeness, $B \to X_s\gamma$, and $B \to K^*\gamma$. The efforts to calculate QCD corrections to $B \to X_s\gamma$ will be briefly described. Subsequently, two examples will be discussed where the control over non-perturbative contributions is mandatory to find out whether the SM is able to describe the experimental data or not: the $\Delta I = 1/2$ rule in $K \to \pi\pi$ decays and the ratio $\varepsilon'/\varepsilon$ related to the direct CP violation in $K_L \to \pi\pi$ decays. The last two presentations deal with the role of QCD in the context of the presently most pronounced anomalies in flavor physics: the violation of lepton flavor universality in tree-level $B$-meson decays (Section 13.4) and the departure of data from the SM predictions for $(g-2)_{e,\mu}$ (Section 13.5).

## 13.1 Effective Hamiltonians in the Standard Model and Beyond

**Andrzej J. Buras**

The basis for any serious phenomenology of weak decays of hadrons is the *Operator Product Expansion* (OPE) [25, 3959], which allows us to write down the effective weak Hamiltonian in full generality simply as follows

$$\mathcal{H}_{\text{eff}} = \sum_i C_i \mathcal{O}_i^{\text{SM}} + \sum_j C_j^{\text{NP}} \mathcal{O}_j^{\text{NP}}, \qquad C_i = C_i^{\text{SM}} + \Delta_i^{\text{NP}}. \tag{13.1.1}$$

Here

- $\mathcal{O}_i^{\text{SM}}$ are local operators present in the SM and $\mathcal{O}_j^{\text{NP}}$ are new local operators having typically new Dirac structures, in particular scalar-scalar and tensor-tensor ones.
- $C_i$ and $C_j^{\text{NP}}$ are the Wilson coefficients (WCs) of these operators. NP effects modify not only the WCs of the SM operators but also generate new operators with non-vanishing $C_j^{\text{NP}}$.

Examples of operators contributing to $K^0-\bar{K}^0$ mixing observables in the SM and in any of its extensions are given as follows

$$Q_1^{\text{VLL}} = (\bar{s}\gamma_\mu P_L d)(\bar{s}\gamma^\mu P_L d), \tag{13.1.2a}$$
$$Q_1^{\text{VRR}} = (\bar{s}\gamma_\mu P_R d)(\bar{s}\gamma^\mu P_R d), \tag{13.1.2b}$$
$$Q_1^{\text{LR}} = (\bar{s}\gamma_\mu P_L d)(\bar{s}\gamma^\mu P_R d), \tag{13.1.2c}$$
$$Q_2^{\text{LR}} = (\bar{s}P_L d)(\bar{s}P_R d), \tag{13.1.2d}$$

$$Q_1^{\text{SLL}} = (\bar{s}P_L d)(\bar{s}P_L d), \tag{13.1.3a}$$
$$Q_1^{\text{SRR}} = (\bar{s}P_R d)(\bar{s}P_R d), \tag{13.1.3b}$$
$$Q_2^{\text{SLL}} = (\bar{s}\sigma_{\mu\nu} P_L d)(\bar{s}\sigma^{\mu\nu} P_L d), \tag{13.1.3c}$$
$$Q_2^{\text{SRR}} = (\bar{s}\sigma_{\mu\nu} P_R d)(\bar{s}\sigma^{\mu\nu} P_R d), \tag{13.1.3d}$$

where

$$P_{R,L} = \frac{1}{2}(1 \pm \gamma_5), \qquad \sigma_{\mu\nu} = i\frac{1}{2}[\gamma_\mu, \gamma_\nu], \tag{13.1.4}$$

and we suppressed color indices as they are summed up in each factor. For instance $\bar{s}\gamma_\mu P_L d$ stands for $\bar{s}_\alpha \gamma_\mu P_L d_\alpha$ and similarly for other factors. Only $Q_1^{\text{VLL}}$ is present in the SM. For meson decays the number of operators in the SM is larger. This is also the case for the number of NP operators. We will encounter some of them in Section 13.3.

The amplitude for a decay of a given meson $M = K, B, ..$ into a final state $F = \mu^+\mu^-$, $\pi\nu\bar{\nu}$, $\pi\pi$, $DK$ is then simply given by

$$A(M \to F) = \langle F|\mathcal{H}_{\text{eff}}|M\rangle = \textstyle\sum_i C_i(\mu)\langle F|\mathcal{O}_i^{\text{SM}}(\mu)|M\rangle + \sum_j C_j^{\text{NP}}(\mu)\langle F|\mathcal{O}_j^{\text{NP}}(\mu)|M\rangle \tag{13.1.5}$$

where $\langle F|\mathcal{O}_i(\mu)|M\rangle$ are the matrix elements of $\mathcal{O}_i$ between $M$ and $F$, evaluated at the renormalization scale $\mu$. The WCs $C_i(\mu)$ describe the strength with which a given operator enters the Hamiltonian. They can be considered as scale dependent "couplings" related to "vertices" $\mathcal{O}_i$ and can be calculated using perturbative methods as long as the scale $\mu$ is not too small. In the case of $K^0-\bar{K}^0$ mixing, matrix elements $\langle \bar{K}^0|\mathcal{O}_i(\mu)|K^0\rangle$ are present. Other particle-antiparticle mixings have similar matrix elements.

The essential virtue of the OPE is this one. It allows us to separate the problem of calculating the amplitude $A(M \to F)$ into two distinct parts: the *short distance* (perturbative) calculation of the coefficients $C_i(\mu)$ and the *long-distance* (generally non-perturbative) calculation of the matrix elements $\langle \mathcal{O}_i(\mu)\rangle$. The scale $\mu$ separates, roughly speaking, the physics contributions into short distance contributions contained in $C_i(\mu)$ and the long distance contributions contained in $\langle \mathcal{O}_i(\mu)\rangle$.

It should be stressed that this separation of short and long distance contribution is only useful due to the asymptotic freedom in QCD [48, 49] that allows us to calculate the WCs by means of ordinary or RG-improved perturbation theory. On the other hand, the matrix elements $\langle \mathcal{O}_i(\mu)\rangle$ can only be calculated by non-perturbative methods like numerical Lattice QCD computations and analytic methods like Dual QCD (DQCD) [3960, 3961] and Chiral Perturbation Theory (ChPT) [64, 1570].

Experimentally, the $\pi\pi$ system in $K \to \pi\pi$ decays was often found to have isospin $I = 0$ and rarely $I = 2$, an effect which is called $\Delta I = 1/2$ rule; $\Delta I = 1/2$ decays are enhanced over the $\Delta I = 3/2$ ones by a factor of 22.4. Altarelli and Maiani [1212] and Gaillard and Lee [1211] made a first unsuccessful attempt to explain this huge enhancement through short distance QCD effects. The precision of the calculation of the WCs increased considerably in the last fifty years since this first pioneering calculation. The basic QCD dynamics behind this rule - contained in the hadronic matrix elements of current-current operators - has been identified analytically first in 1986 in the framework of the Dual QCD in [3960] with some improvements in 2014 [3961]. This has been confirmed more than 30 years later by the RBC-UKQCD collaboration [3962] although the modest accuracy of both approaches still allows for some NP contributions. See [3963] for the most recent summary and Section 13.3.

Now, the coefficients $C_i$ include, in addition to tree-level contributions from the $W$-exchange, virtual top quark contributions and contributions from other heavy particles such as W, Z bosons, charged Higgs particles, supersymmetric particles and other heavy objects in numerous extensions of this model. Consequently, $C_i(\mu)$ generally depend on $m_t$ and also on the masses of new particles if extensions of the SM are considered. This dependence can be found by evaluating one-loop diagrams, so-called *box* and *penguin* diagrams with full W,

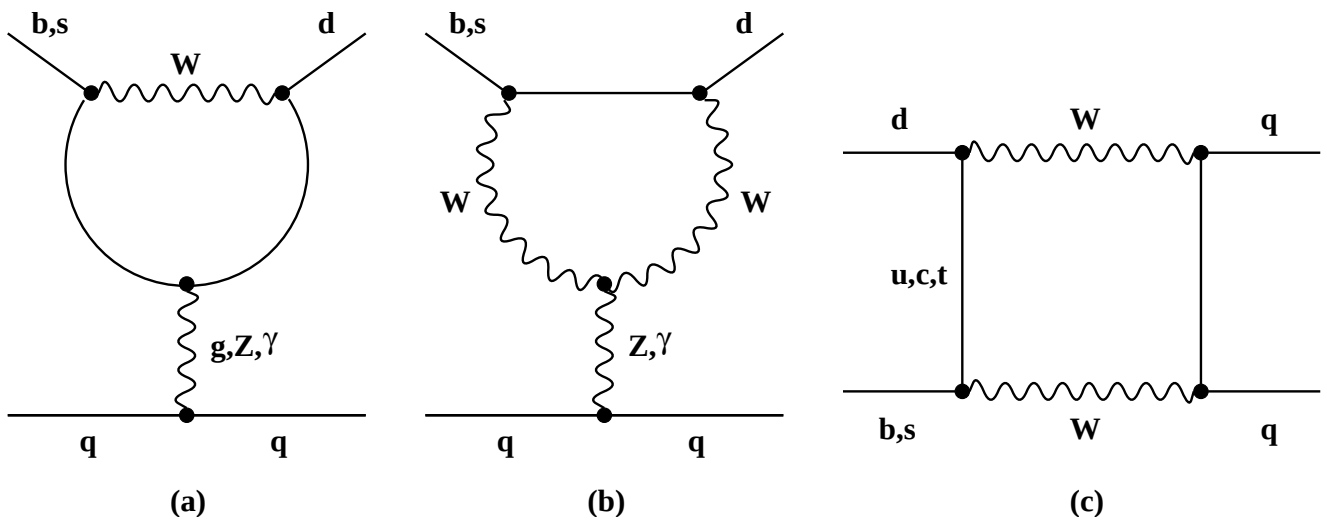

**Fig. 13.1.1** *Penguin and Box Diagrams. From [3964]*

Z, top quark and new particles exchanges and *properly* including short distance QCD effects. The latter govern the $\mu$-dependence of $C_i(\mu)$. In models in which the GIM mechanism [78] is absent, also *tree* diagrams can contribute to flavor changing neutral current (FCNC) processes. The point is that a given $C_i$ generally receives contributions from all these three classes of diagrams.

The value of $\mu$ can be chosen arbitrarily but the final result must be $\mu$-independent. Therefore the $\mu$-dependence of $C_i(\mu)$ has to cancel the $\mu$-dependence of $\langle Q_i(\mu)\rangle$. In other words as far as heavy-mass-independent terms are concerned, it is a matter of choice what exactly belongs to $C_i(\mu)$ and what to $\langle Q_i(\mu)\rangle$. This cancellation of the $\mu$-dependence involves generally several terms in the expansion in Eqn. (13.1.5). $C_i(\mu)$ depend also on the renormalization scheme used in the calculation of QCD effects. This scheme-dependence must also be canceled by the one of $\langle Q_i(\mu)\rangle$ so that the physical amplitudes are renormalization-scheme independent. Again, as in the case of the $\mu$-dependence, the cancellation of the renormalization-scheme-dependence involves generally several terms in the expansion in Eqn. (13.1.5). One of the types of scheme-dependence is the manner in which $\gamma_5$ is defined in $D = 4 - 2\varepsilon$ dimensions implying various renormalization schemes as analyzed first in the context of weak decays in [3965]. A pedagogical presentation of these issues can be found in [3966].

### 13.1.1 Renormalization Group Improved Perturbation Theory

Generally in weak decays several vastly different scales are involved. These are the hadronic scales of a few GeV, scales like $M_W$ or $m_t$ and - in extensions of the SM - not only of a few TeV but even 100 TeV. Already within the SM, but in particular in its NP extensions, the ordinary perturbation theory in $\alpha_s$ is spoiled by the appearance of large logarithms of the ratios of two very different scales that multiply $\alpha_s$. Such logarithms have to be summed to all orders of perturbation theory which can be efficiently done by means of renormalization-group methods. Denoting the lower scale simply by $\mu$ and the high scale by $\Lambda$ the general expression for $C_i(\mu)$ is given by:

$$\vec{C}(\mu) = \hat{U}(\mu, \Lambda)\vec{C}(\Lambda) \ , \tag{13.1.6}$$

where $\vec{C}$ is a column vector built out of $C_i$. $\vec{C}(\Lambda)$ are the initial conditions for the RG evolution down to low energy scale $\mu$. They depend on the short distance physics at high energy scales. In particular they depend on $m_t$ and the masses and couplings of new heavy particles.

The evolution matrix $\hat{U}(\mu, \Lambda)$ sums large logarithms $\log \Lambda/\mu$ which appear for $\mu \ll \Lambda$. In the so-called leading logarithmic approximation (LO) terms $(g_s^2 \log \Lambda/\mu)^n$ are summed. The next-to-leading logarithmic correction (NLO) to this result involves summation of terms $(g_s^2)^n (\log \Lambda/\mu)^{n-1}$ and so on. This hierarchical structure gives the RG-improved perturbation theory.

As an example let us consider only a single operator so that Eqn. (13.1.6) reduces to

$$C(\mu) = U(\mu, \Lambda)C(\Lambda) \tag{13.1.7}$$

with $C(\mu)$ denoting the coefficient of the operator in question.

Keeping the first two terms in the expansions of the anomalous dimension of this operator $\gamma(g_s)$ and in $\beta(g_s)$, that governs the evolution of $\alpha_s$, in powers of $\alpha_s$ and $g_s$,

$$\gamma(g_s) = \gamma^{(0)} \frac{\alpha_s}{4\pi} + \gamma^{(1)} \left(\frac{\alpha_s}{4\pi}\right)^2 , \tag{13.1.8}$$

$$\beta(g_s) = -\beta_0 \frac{g_s^3}{16\pi^2} - \beta_1 \frac{g_s^5}{(16\pi^2)^2} \tag{13.1.9}$$

gives:

$$U(\mu, \Lambda) = \left[1 + \frac{\alpha_s(\mu)}{4\pi} J_1\right] \left[\frac{\alpha_s(\Lambda)}{\alpha_s(\mu)}\right]^P \left[1 - \frac{\alpha_s(\Lambda)}{4\pi} J_1\right] \tag{13.1.10}$$

where

$$P = \frac{\gamma^{(0)}}{2\beta_0}, \qquad J_1 = \frac{P}{\beta_0}\beta_1 - \frac{\gamma^{(1)}}{2\beta_0}. \tag{13.1.11}$$

General formulae for the evolution matrix $\hat{U}(\mu, \Lambda)$ in the case of operator mixing and valid also for electroweak effects at the NLO level can be found in [3967]. The corresponding NNLO formulae are rather complicated and were given for the first time in [3968].

While by now NLO and NNLO QCD contributions to almost all weak decays are known within the SM, the pioneering LO calculations for current-current operators [1211, 1212], penguin operators [3969, 3970], $\Delta S = 2$ operators [3971] and rare $K$ decays [3972] should not be forgotten. The first review of NLO QCD calculations can be found in [3967] and more recently including NNLO corrections in [3966, 3973].

It should be stressed that at the NLO level not only two-loop anomalous dimensions of operators have to be known but also QCD corrections to the WCs at $\mu = \Lambda$. Only then renormalization-scheme independent results can be obtained. They are known for most processes of interest and this technology is explained in details in [3964, 3966]

On the whole, the status of present short distance (SD) contributions within the SM is satisfactory. Let us then see what is the status of these calculations beyond the SM.

### 13.1.2 QCD Effects Beyond the SM

As already stated at the beginning, NP contributions can affect the WCs of the SM operators. This modification takes place at the NP scale $\Lambda$ so that after the RG evolution, the $C_i(\mu)$ in Eqn. (13.1.5) are modified. But in addition new operators with different Dirac structure, with examples given in Eqns. (13.1.2) and (13.1.3), can contribute if their coefficients $C_j^{\rm NP}(\Lambda)$ are non-vanishing or if they are generated by mixing of different operators in the process of the RG evolution. The inclusion of these contributions in the RG analysis requires at the NLO level the calculations of their one-loop and two-loop anomalous dimensions. While the one-loop anomalous dimensions of such operators have been calculated in [680], the first two-loop calculations have been presented in [3974, 3975]. Recently, these NLO calculations have been generalized for both $\Delta F = 1$ and $\Delta F = 2$ transitions in the so-called Weak Effective Theory (WET) [3976, 3977] and also for the Standard Model Effective Field Theory (SMEFT) [3978]. It turns out that the anomalous dimensions of operators involving both left-handed and right-handed currents, the so-called left-right operators, are much larger than those of most operators within the SM except for QCD-penguin operators. Thus even if their WCs could be small at the scale $\Lambda$ they can be enhanced at scales of the order of a few GeV. The same applies also to scalar operators.

### 13.1.3 Hadronic Matrix Elements

The WCs, that include in the SM the CKM factors, are not the whole story. To obtain the results for the decay amplitudes and the quark mixing observables, also hadronic matrix elements of local operators, like the ones in Eqns. (13.1.2) and (13.1.3), have to be calculated. The present status can be sumarized as follows.

– For leptonic decays like $B_{s,d} \to \mu^+\mu^-$ and $K_{L,S} \to \mu^+\mu^-$ only the weak decay constants $f_{B_s}$, $f_{B_d}$ and $f_K$ are required. They are defined e.g. by

$$\langle 0|(\bar{s}\gamma^\mu(1-\gamma_5)u)|K^+\rangle = i f_K p_K^\mu, \tag{13.1.12}$$

where $p_K^\mu$ is the four-momentum of the decaying $K^+$ mesons. Similar for $f_{B_s}$ and $f_{B_d}$.
They are known from LQCD calculations already with an impressive precision [63, 685, 3979]

$$f_{B_s} = 230.3(1.3)\mathrm{MeV},\ f_{B_d} = 190.0(1.3)\mathrm{MeV},$$
$$f_K = 155.7(3)\mathrm{MeV}, \tag{13.1.13}$$

although in the case of $K_{L,S} \to \mu^+\mu^-$ also genuine long distance QCD contributions enter. They cannot be described by matrix elements of local operators and one has to develop some strategies to isolate the contribution described by the effective Hamiltonian discussed by us. In $B_{s,d}$ and $B^\pm$ decays such effects are much smaller. However, they are significant in charm meson decays.

– In semileptonic decays like $K^+ \to \pi^+\nu\bar{\nu}$, $K_L \to \pi^0\nu\bar{\nu}$, $K_L \to \pi^0\ell^+\ell^-$, $B \to K(K^*)\ell^+\ell^-$, $B \to D(D^*)$ $\ell^+\ell^-$ and $B \to K(K^*)\nu\bar{\nu}$ the formfactors for the transitions $K \to \pi$, $B \to K(K^*)$, $B \to D(D^*)$ enter. For $K$ decays these formfactors can even be extracted from data on leading decays with the help of ChPT and isospin symmetry [3980–3982]. Those that enter $B$ decays they are usually calculated using lightcone sum rules for low momentum transfer squared $q^2$ [3983] and LQCD for large $q^2$ [3984, 3985]. Significant progress has been made here by now with most recent analyses in [703, 3986–3988] where more information can be found.

– Moreover Heavy Quark Effective Theory (HQET) and Heavy Quark Expansions (HQE) play an important roles here. HQET represents a static approximation for the heavy quark, covariantly formulated in the language of an effective field theory. It allows us to extract the dependence of hadronic matrix elements on the heavy quark mass and to exploit the simplifications that arise in QCD in the static limit. The most important application of HQET has been to the analysis of exclusive semileptonic transitions involving heavy quarks, where this formalism allows us to exploit the consequences of heavy quark symmetry to relate formfactors and provides a basis for systematic corrections to the $m \to \infty$ limit. There are several excellent reviews on this subject [674, 1390, 3989, 3990].

– For the calculation of the width differences in $B^0_{s,d}-\bar{B}^0_{s,d}$ mixing $\Delta\Gamma_{s,d}$, lifetimes and totally inclusive decay rates of heavy hadrons, the so-called heavy quark expansion (HQE) has been developed by several authors. It relies on the smallness of the parameter $\Lambda_{\mathrm{QCD}}/m_b$, where $\Lambda_{\mathrm{QCD}}$ is a hadronic scale. The coefficients in this expansion can be calculated by LQCD. Nice reviews with some details are the ones in [674, 1225, 1239, 3991] and a nice summary of the present situation including historical development can be found in [3992].

– For $\Delta M_{s,d}$ significant progress has been made by LQCD in the recent years. Here the relevant hadronic matrix elements are parametrized by $f_{B_s}\sqrt{\hat{B}_s}$ and $f_{B_d}\sqrt{\hat{B}_d}$ with $\hat{B}_s$ and $\hat{B}_d$ close to unity. Presently the most accurate results are those from HPQCD collaboration [685]

$$f_{B_s}\sqrt{\hat{B}_s} = 256.1(5.7)\mathrm{MeV},$$
$$f_{B_d}\sqrt{\hat{B}_d} = 210.6(5.5)\mathrm{MeV} \tag{13.1.14}$$

that in addition to light quarks includes charm quarks. Also corresponding matrix elements for BSM operators are already known but their precision should be still improved. Similarly, the relevant hadronic matrix elements for the parameter $\varepsilon_K$ describing the indirect CP-violation in $K_L \to \pi\pi$ decay are already known with respectable precision from LQCD both in the SM and beyond [684, 3993, 3994]. Some physics insight into the numerical LQCD results has also been gained with the help of the DQCD approach [3995].

– The calculations of hadronic matrix elements for non-leptonic decays like $K \to \pi\pi$, $B \to \pi K$ etc. are much more involved. For $K \to \pi\pi$ the only approaches providing matrix elements that can be consistently combined (matched) with the WCs are LQCD, lead by the RBC-UKQCD collaboration and the DQCD approach. But while from LQCD only the matrix elements of SM operators are known, all matrix elements of BSM operators have been calculated using the DQCD approach [3996]. Yet, the accuracy of the latter calculations have to be improved, and one should hope that also LQCD collaborations will calculate these matrix elements one day. However, based on the time required to compute the matrix elements of SM operators using LQCD, it could take even a decade to obtain satisfactory results on these matrix elements from LQCD. This is important in view of the present status of the direct CP violation in $K_L \to \pi\pi$ decay represented by the ratio $\varepsilon'/\varepsilon$. We will return to this issue in Section 13.3.

– For non-leptonic exclusive $B$ decays LQCD cannot provide the hadronic matrix elements directly but can help in calculating non-perturbative parameters in the context of the so-called *QCD factorization* (QCDF) [3997, 3998]. This approach can be applied to $B \to \pi\pi$, but also to rare and radiative decays, such as $B \to K^*\gamma$ or $B \to K^* l^+ l^-$. In the heavy-quark limit, that is up to relative corrections of order $\Lambda_{\rm QCD}/m_b$, the problem of computing exclusive hadronic decay amplitudes simplifies considerably. A nice review by Buchalla can be found in Section 7.4 of [3966], and also the one by Beneke [3999] can be strongly recommended. There, also the so-called soft-collinear effective theory (SCET) [1762, 1764] is briefly discussed.

## 13.2 The quark mixing matrix

**Paolo Gambino**

The rich flavor structure of the Standard Model (SM) and its CP violation both follow from the matrices of Yukawa couplings between the fermions (down and up quarks and charged leptons) and the Higgs boson. The diagonalisation of these matrices determines the fermion masses and brings us to the flavor basis, where the charged weak current is no longer diagonal: as first understood in the hadronic sector by Cabibbo [3957] and extended to three generations by Kobayashi and Maskawa [3958], charged currents mix the quarks of different generations in a way parameterized by the Cabibbo-Kobayashi-Maskawa (CKM) quark mixing matrix. Interestingly, its elements display a remarkable hierarchy, possibly indicative of the unknown mechanism of flavor breaking [4000]:

$$\hat{V}_{\rm CKM} = \begin{pmatrix} V_{ud} & V_{us} & V_{ub} \\ V_{cd} & V_{cs} & V_{cb} \\ V_{td} & V_{ts} & V_{tb} \end{pmatrix} \tag{13.2.1}$$

$$= \begin{pmatrix} 1-\lambda^2/2 & \lambda & A\lambda^3(\rho - i\eta) \\ -\lambda & 1-\lambda^2/2 & A\lambda^2 \\ A\lambda^3(1-\rho-i\eta) & -A\lambda^2 & 1 \end{pmatrix} + O(\lambda^4)$$

where $\lambda = \sin\theta_c \simeq 0.22$ is a small expansion parameter and $A \simeq 0.8$, $\rho \simeq 0.16$, $\eta \simeq 0.36$. As a unitary matrix, $\hat{V}_{\rm CKM}$ has in principle nine free parameters but some of them can be absorbed by phase redefinitions. In the end, $\hat{V}_{\rm CKM}$ has only four independent real parameters: three Euler angles and a phase, or equivalently $\lambda$, $A$, $\rho$ and $\eta$. The presence of a nonvanishing phase, *i.e.* of an imaginary part, implies CP violation. Since unitarity is specific to the three generations of the SM and to the absence of additional flavor violation, testing $\hat{V}^\dagger_{\rm CKM}\hat{V}_{\rm CKM} = 1$ is an important step in the verification of the SM and represents the modern equivalent of the tests of the universality of the charged currents. Any of the off-diagonal relations can be represented by a triangle in the complex plane whose area is a measure of CP violation. In particular, the triangle

$$1 + \frac{V_{ud}V^*_{ub}}{V_{cd}V^*_{cb}} + \frac{V_{td}V^*_{tb}}{V_{cd}V^*_{cb}} = 0 \tag{13.2.2}$$

is frequently considered because it has sides of comparable length, see Fig.13.2.1, and its parameters can all be well determined in $B$ decays. The angles $\beta$ and $\gamma$ at the basis of this triangle correspond to the phases of the elements $V_{ub}$ and $V_{td}$: $V_{ub} = |V_{ub}|e^{-i\gamma}$, $V_{td} = |V_{td}|e^{-i\beta}$. Various observables constrain the apex of this triangle. The results of a global fit are shown in Fig. 13.2.1, where one can see that different constraints agree well, verifying unitarity and determining the apex of the triangle with high accuracy. As we will see below, there are tests of the unitarity of $\hat{V}_{\rm CKM}$ that cannot be represented in this plot.

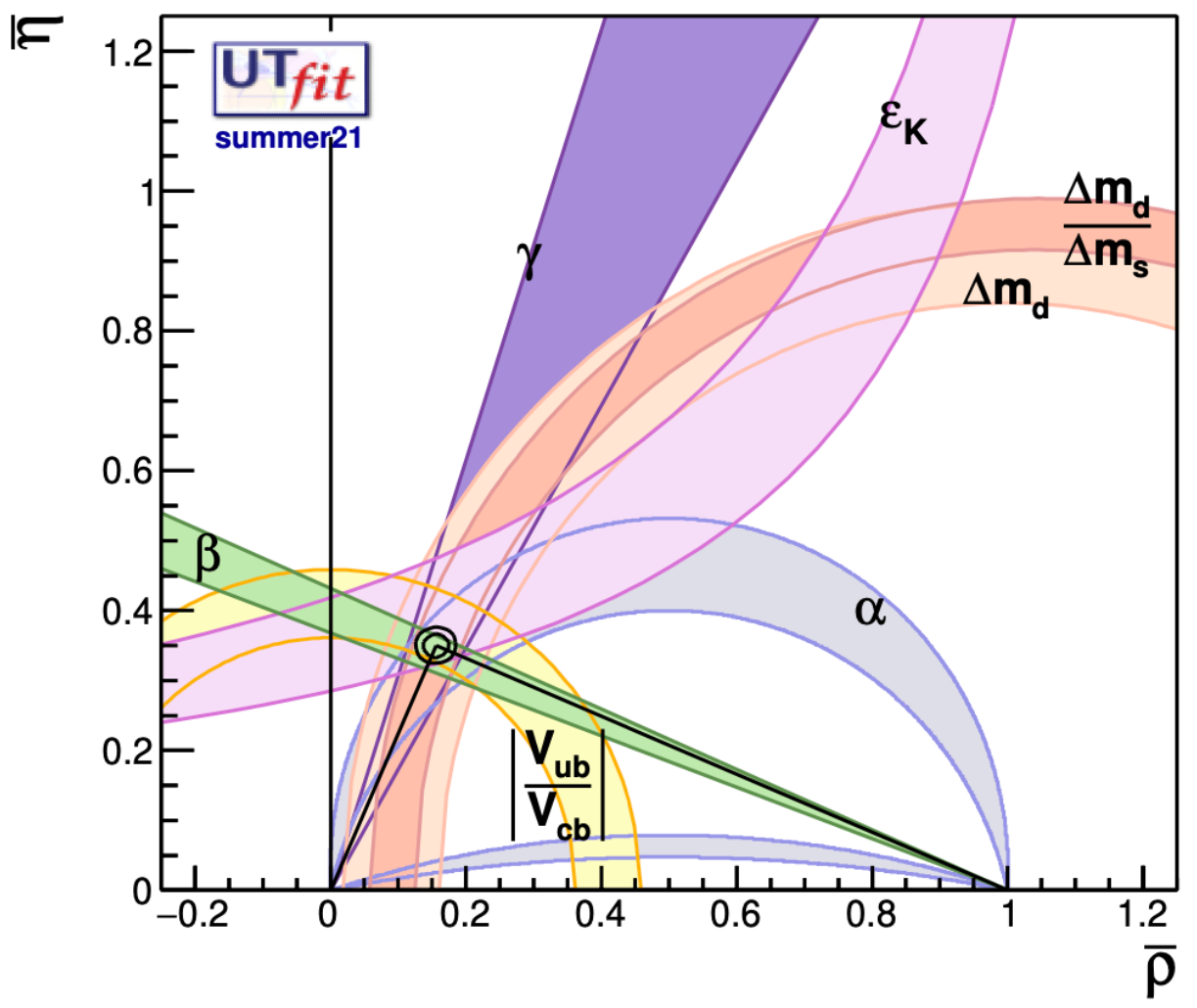

**Fig. 13.2.1** Constraints on the apex of the Unitarity Triangle of (13.2.2) and their combination according to the UTFit collaboration. Figure taken from Ref. [4001]. $\bar{\rho} = \rho(1-\lambda^2/2), \bar{\eta} = \eta(1-\lambda^2/2)$

The role of QCD in the determination of the CKM elements and in testing the CKM mechanism is crucial, with important perturbative and nonperturbative aspects depending on the observable; some of the nonperturbative methods have already been discussed in Secs. 4.7 and 5.8.

The experimental and theoretical progress made in the last 30 years is enormous and was mostly driven by lattice QCD; it allows for very precise tests of the CKM mechanism, as is apparent from Fig. 13.2.1. Further improvements will be possible with LCHb and Belle II data, but will generally require an effective synergy of theory and experiment. In this section I will focus on measurements where QCD effects are most relevant and where tensions have appeared with the SM.

### 13.2.1 The Cabibbo angle and the first row unitarity

The parameter $\lambda$ in Eq. (13.2.1) corresponds to the sine of the Cabibbo angle and is determined, up to very small higher orders in $\lambda$, by $|V_{us}|$ or $|V_{cd}|$. The high precision with which $|V_{ud}|$ is known also allows for a competitive $\lambda$ determination. The unitarity of the CKM matrix implies for the first row the relation

$$\Sigma_1 = |V_{ud}|^2 + |V_{us}|^2 + |V_{ub}|^2 = 1, \tag{13.2.3}$$

but since $|V_{ub}| \approx 0.004$ only the first two terms are relevant. Precise measurements of $|V_{us}|$ and $|V_{ud}|$ therefore lead to a first important check of the CKM mechanism.

The most precise determination of $|V_{ud}|$ comes from superallowed Fermi transitions (SFT), i.e. $0^+ \to 0^+$ nuclear $\beta$ decays. At the tree level, these decays are mediated by the vector current, whose conservation allows for a particularly clean theoretical description. Among recent refinements, hadronic effects in the radiative corrections, in particular in the $\gamma W$ box, have been studied with dispersive methods [4002, 4003], and the effect of nuclear polarizability, which depends on nuclear structure (NS), has been exposed [4004]. Considering 15 different superallowed transitions gives a consistent result and the error of the final value [4005],

$$|V_{ud}| = 0.97367(32) \qquad (0^+ \to 0^+) \tag{13.2.4}$$

is dominated by the NS effects. Neutron $\beta$ decay depends on the nucleon isovector axial charge $g_A/g_V$ and has recently become competitive, $|V_{ud}| = 0.97413(43)$, if one includes only the current best experiments [4006]. Theoretically the cleanest channel is $\pi^+ \to \pi^0 e\nu$, which is however penalized by a $10^{-8}$ BR. The present uncertainty based on PIBETA results [4007], $\delta V_{ud} \sim 0.003$, is still far from being competitive, but there are plans to improve drastically on that [4008].

$|V_{us}|$ can be directly accessed from kaon, hyperon, and tau semileptonic decays. The kaon decays, $K \to \pi\ell\nu$ or $K_{\ell 3}$ are measured in five channels ($K_{L,S}, K^+$ with electron and muons) affected by different systematics, with $K \to \pi$ form factors computed on the lattice, as discussed in Section 4.7. Combining experimental data and the average of several $N_f = 2{+}1{+}1$ lattice results one obtains [476]

$$|V_{us}| = 0.2231(4)_{exp}(4)_{lat} \qquad (K_{\ell 3}), \tag{13.2.5}$$

see also [4006]. At this level of precision, however, a consistent treatment of QED effects in the lattice calculation becomes mandatory [63]. Hyperon decays give a consistent $|V_{us}|$ but are presently not competitive with the above result. The ratio of inclusive tau decays into strange and non-strange hadrons can also be used to extract $|V_{us}|/|V_{ud}|$, employing experimental data and Finite Energy Sum Rules, without lattice input. Recent results tend to be over $2\sigma$ lower than Eq. (13.2.5) and subject to debate [4009, 4010], but a combination of experimental and lattice data on the hadronic vacuum polarization functions gives $|V_{us}| = 0.2245(11)_{exp}(13)_{th}$ [4011], in agreement with Eq. (13.2.5). Exclusive tau decay channels or ratio such as $\mathcal{B}(\tau \to K\nu)/\mathcal{B}(\tau \to \pi\nu)$ can also be used together with $f_{K,\pi}$ computed on the lattice, see Section 4.7, to obtain $|V_{us}| = 0.2229(19)$ [4012], again consistent with Eq. (13.2.5).

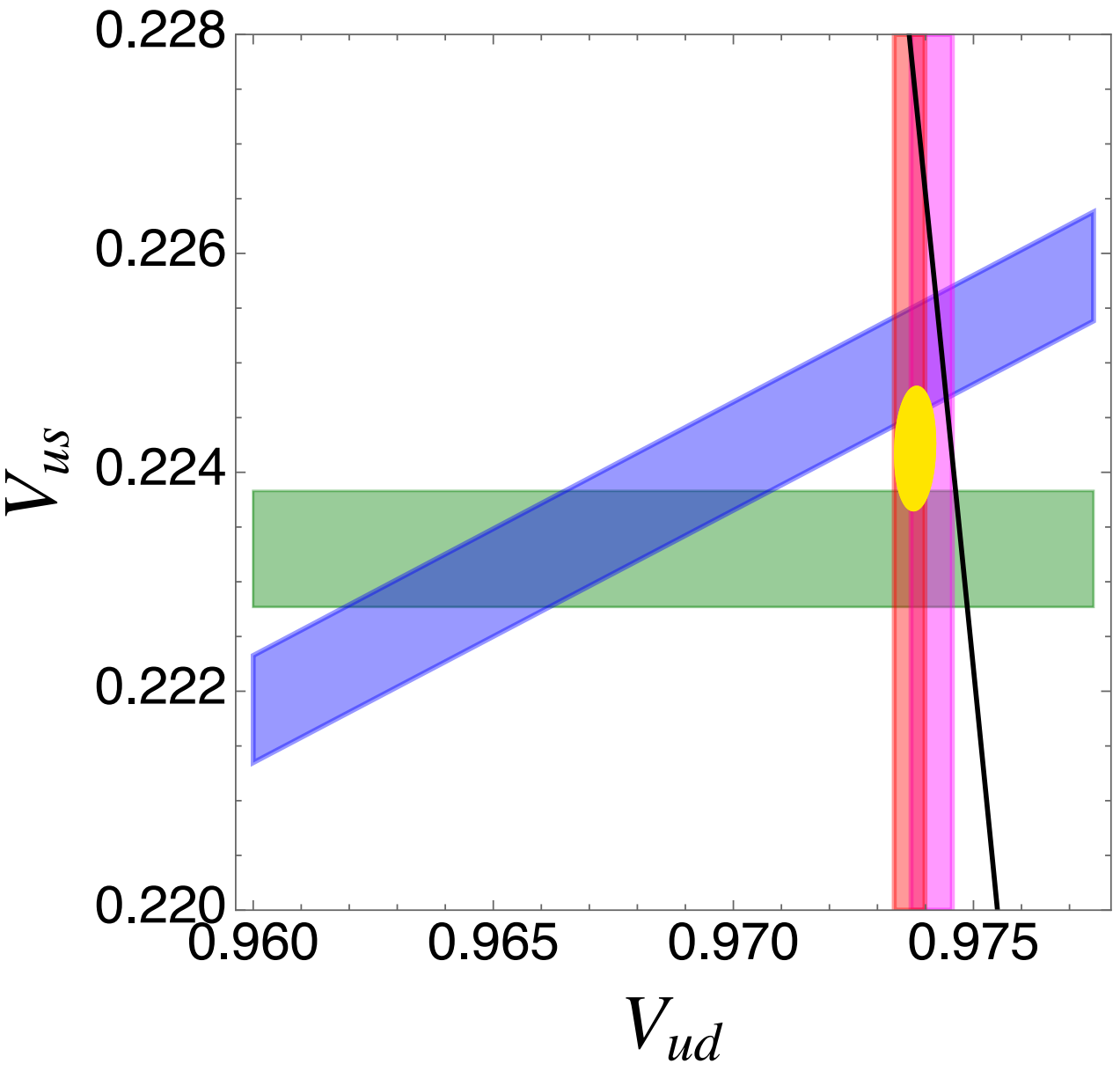

**Fig. 13.2.2** $1\sigma$ constraints in the $(|V_{ud}|, |V_{us}|)$ plane from superallowed Fermi transitions (red), from neutron decay (violet), $K_{\ell 3}$ (green), $K_{\mu 2}$ (blue) and the 68% CL contour of the combined fit (yellow). The black line marks the unitarity relation between $|V_{ud}|$ and $|V_{us}|$. Figure taken from [4006].

A very precise determination of the ratio $|V_{us}|/|V_{ud}|$ can be obtained from the ratio of $K \to \mu\nu(\gamma)$ to $\pi \to \mu\nu(\gamma)$ (decays [656]. Here nonperturbative QCD sits almost completely in the ratio of $f_K$ and $f_\pi$, which is known with a 0.2% uncertainty in 2+1+1 lattice QCD [63]. It then follows [4006]

$$\left|\frac{V_{us}}{V_{ud}}\right| = 0.2311(5) \qquad (K_{\mu 2}) \tag{13.2.6}$$

with the uncertainty dominated by lattice QCD. Using unitarity this is equivalent to $|V_{us}| = 0.2245(5)$ and in some tension with Eq. (13.2.5).

The most precise constraints can be combined in the $(|V_{ud}|, |V_{us}|)$ plane, see Fig. 13.2.2. We observe a clear tension between the best fit and unitarity, mostly driven by the kaon determinations, which cross far from the unitarity line, and by the superallowed Fermi transitions, which under unitarity imply a very high $|V_{us}|$. On the other hand, $|V_{us}|$ from $K_{\ell 2}$ and the neutron $|V_{ud}|$ are compatible with unitarity. Taking the average of the determinations from Fermi and $n$ decay, $|V_{ud}| = 0.97384(26)$, the actual deviation of $\Sigma_1$ from 1 varies between about $1.5\sigma$ using Eq. (13.2.6) and $\sim 3\sigma$ using Eq. (13.2.5) and it is sometimes referred to as the *Cabibbo anomaly.* It could be due to underestimated uncertainties in the NS correction, in the lattice calculations, in the experimental results, or due to New Physics [4013, 4014], and a renewed campaign of $K_{\mu 3}$ and $K_{\mu 2}$ measurements will be crucial to clarify the situation [4006].

As mentioned above, $\lambda$ can also be determined from $D_{(s)} \to \ell\nu$ and $D \to \pi(K)\ell\nu$. Concerning the former, as lattice calculations for $f_D$ have become very precise, the uncertainties in $|V_{cs}| = 0.982(10)_{exp}(2)_{lat}$ and $|V_{cd}| = 0.2181(49)_{exp}(7)_{lat}$ [4012] are dominated by experiment. These results are consistent with Eqs. (13.2.4, 13.2.5). FLAG has performed a combined fit to lattice and experimental data for the two $D$ semileptonic decays that yields $|V_{cs}| = 0.971(7)$ and $|V_{cd}| = 0.234(7)$ [63], but $|V_{cd}|$ is about $2\sigma$ above its $D \to \mu\nu$ value. Averaging all these results, one can check the unitarity of the second row of the CKM matrix [63],

$$\Sigma_2 = |V_{cd}|^2 + |V_{cs}|^2 + |V_{cb}|^2 = 1 + 0.001(11), \quad (13.2.7)$$

where again the last term in the sum is negligible at the present accuracy. Neutrino Deep Inelastic Scattering is also used to extract a consistent but less precise value of $|V_{cd}|$. The second row of $\hat{V}_{CKM}$ appears to be consistent with unitarity, but the accuracy is much lower than for the first row.

### 13.2.2 Determination of $V_{cb}$ and $V_{ub}$

The magnitudes of two of the elements of the CKM matrix, $|V_{ub}|$ and $|V_{cb}|$, can be directly extracted from semileptonic $b$-hadron (mostly $B$ meson) decays. In exclusive decays one looks at specific hadronic final states, while inclusive decays sum over all decays channels to a certain flavor (*i.e.* $b \to c$). Inclusive and exclusive semileptonic decays are subject to very different theoretical and experimental systematics, see Refs. [4015, 4016] for recent reviews.

The results of the $B$ factories, analysed in the light of the most recent theoretical calculations, are puzzling, because – especially for $|V_{cb}|$ – the determinations from exclusive and inclusive decays are in strong tension, and despite recent new experimental and theoretical results the situation remains unclear. While in principle New Physics may explain the tensions, it is significantly constrained by the measured differential distributions in $B \to D^{(*)}\ell\nu$ [4017] and, in the context of the SM Effective Theory or SMEFT, by LEP data [4018]. This tension is all the more relevant as measurements in the semitauonic channels at Belle, Babar, and LHCb show discrepancies with the SM predictions, pointing to a possible violation of lepton-flavor universality. This $V_{cb}$ puzzle casts a shadow on our understanding of semitauonic decay as well. The inability to determine precisely $V_{cb}$ also hampers significantly NP searches in Flavor Changing Neutral Currents processes: the uncertainty on the value of $V_{cb}$ dominates the theoretical uncertainty in the SM predictions for several observables, from $\varepsilon_K$ to the branching fraction of $B_s \to \mu^+\mu^-$.

Our understanding of inclusive semileptonic $B$ decays, see also Section 5.8, is based on a simple idea: since inclusive decays sum over all possible hadronic final states, the quark in the final state hadronizes with unit probability and the transition amplitude is sensitive only to the long-distance dynamics of the initial $B$ meson. Thanks to the large hierarchy between the typical energy release, of $O(m_b)$, and the hadronic scale $\Lambda_{\rm QCD}$, and to asymptotic freedom, any residual sensitivity to non-perturbative effects is suppressed by powers of $\Lambda_{\rm QCD}/m_b$. From a phenomenological point of view, it is remarkable that the linear preasymptotic correction is actually absent and that the leading nonperturbative corrections are $O(\Lambda_{\rm QCD}^2/m_b^2)$. This is due to the Operator Product Expansion (OPE) that allows us to express the nonperturbative physics in terms of $B$ meson matrix elements of local operators of dimension $d \geq 5$, while the Wilson coefficients can be expressed as a perturbative series in $\alpha_s$ [1255–1257, 4019, 4020]. The OPE disentangles the physics associated with *soft* scales of order $\Lambda_{\rm QCD}$ (parameterized by the matrix elements of the local operators) from that associated with *hard* scales $\sim m_b$, which determine the Wilson coefficients. Inclusive observables such as the total semileptonic width and the moments of the kinematic distributions are therefore double expansions in $\alpha_s$ and $\Lambda_{\rm QCD}/m_b$, with a leading term that is given by the free $b$ quark decay. As already noted, the power corrections start at $O(\Lambda_{\rm QCD}^2/m_b^2)$ and are comparatively suppressed. At higher orders in the OPE, terms suppressed by powers of $m_c$ also appear, starting with $O(\Lambda_{\rm QCD}^3/m_b^3 \times \Lambda_{\rm QCD}^2/m_c^2)$ [4021]. The expansion for the total semileptonic width is

$$\begin{aligned}\Gamma_{sl} = \Gamma_0 \Big[ & 1 + a^{(1)} \frac{\alpha_s(m_b)}{\pi} + a^{(2)} \Big(\frac{\alpha_s}{\pi}\Big)^2 + + a^{(3)} \Big(\frac{\alpha_s}{\pi}\Big)^3 \\ & + \Big(-\frac{1}{2} + p^{(1)} \frac{\alpha_s}{\pi}\Big) \frac{\mu_\pi^2}{m_b^2} + \Big(g^{(0)} + g^{(1)} \frac{\alpha_s}{\pi}\Big) \frac{\mu_G^2(m_b)}{m_b^2} \\ & + d^{(0)} \frac{\rho_D^3}{m_b^3} - g^{(0)} \frac{\rho_{LS}^3}{m_b^3} + \text{higher orders} \Big] \end{aligned} \quad (13.2.8)$$

where $\Gamma_0$ is the tree-level free-quark decay width, and $\mu_\pi^2$, $\mu_G^2$, $\rho_D^3$ and $\rho_{LS}^3$ are hadronic parameters that have to be determined from experimental data, i.e. from the moments of differential distributions, which can be expanded in the same way as the total width. The perturbative corrections are known up to $O(\alpha_s^3)$ and $O(\alpha_s/m_b^3)$ for the total width [1241, 4022] and up to $O(\alpha_s^2)$ and $O(\alpha_s/m_b^2)$ for the moments [4023–4026]. In line with the discussion of Section 5.8, it is important that $m_b$ and the other Heavy-Quark Expansion (HQE) parameters are free from renormalon ambiguities. The kinetic scheme [4027, 4028], for instance, employs a Wilsonian cutoff $\mu \sim 1\,\text{GeV}$. Higher power corrections have been considered in [4029–4031] and appear to have a negligible impact on $|V_{cb}|$. Although the moments are rather sensitive to the difference $m_b - m_c$, a more precise determination of $|V_{cb}|$ can be obtained taking advantage of the precise lattice determinations of the charm and bottom masses, see [476] for a review. The most recent global analysis in the kinetic scheme [4032] gives

$$|V_{cb}| = 42.16(51) \times 10^{-3}, \qquad (B \to X_c \ell \nu) \qquad (13.2.9)$$

where the uncertainty follows from the combination of theoretical and experimental uncertainties. A consistent but less precise result has been recently obtained from an analysis of the new Belle and Belle II measurements of the $q^2$ moments [4033]. While the estimate in Eq.(13.2.9) appears solid, new measurements at Belle II will provide welcome checks and may reduce the experimental uncertainty. There are also a few more higher order effects worth computing, and QED effects should be understood better. Most importantly, however, lattice calculations of inclusive quantities are now possible and may soon complement the OPE approach [713, 4034].

The inclusive determination of $|V_{ub}|$ from $B \to X_u \ell \nu$ decays differs from that of $|V_{cb}|$ mostly because of the experimental cuts necessary to suppress the large $b \to c \ell \nu$ background: the local OPE does not converge well in the restricted phase space. The modern description of these inclusive decays is therefore based on a non-local OPE [1259, 1260], where nonperturbative shape functions (SFs) play the role of parton distribution functions of the b quark inside the $B$ meson. While the first few moments of the SFs are expressed in terms of the same HQE parameters extracted in $B \to X_c \ell \nu$, direct experimental information on the SFs is limited to the $B \to X_s \gamma$ photon spectrum, to which they are only related in the $m_b \to \infty$ limit. There are a few frameworks that incorporate the above picture with a range of additional assumptions: BLNP [4035] and GGOU [4036] use a large set of models for the SFs, while DGE [4037] computes the leading SF in resummed perturbative QCD. Another potential source of theoretical uncertainty in all approaches is represented by the so called Weak Annihilation contributions, namely nonperturbative contributions at high $q^2$ arising from $b\bar{q}$ weak annihilations (WA) in the $B$ meson, where the $\bar{q}$ is not necessarily the light valence quark [4038]. Charm decays, and particularly moments of the inclusive leptonic spectrum, constrain them effectively, and one can conclude that the WA correction to the total rate of $B \to X_u \ell \nu$ must be smaller than about 2% [4039, 4040]. Its localisation at high $q^2$ and the sensitivity of the $q^2$ tail to higher power corrections suggest that an upper cut on $q^2$ would be useful in future analyses.

A few experimental analyses extend the measurement into the phase space region dominated by $b \to c$ transitions, which are then modelled, trading part of the theory uncertainty for a larger systematic experimental uncertainty (in particular, $D^{**}$ and multihadron final states are not known very well): agreement among the various analyses should then increase our confidence in the result, but one should be aware that the reconstruction efficiencies depend on the modelling of the signal, *i.e.* again on the SFs. The latest Heavy Flavour Averaging Group (HFLAV) $|V_{ub}|$ world averages in the three above frameworks [4012] are based on a number of different experimental results with different kinematic cuts and read

$$\begin{aligned} |V_{ub}|^{\text{BLNP}} &= 4.28(13)^{+20}_{-21} \times 10^{-3}, \\ |V_{ub}|^{\text{GGOU}} &= 4.19(12)^{+11}_{-12} \times 10^{-3}, \\ |V_{ub}|^{\text{DGE}} &= 3.93(10)^{+9}_{-10} \times 10^{-3}, \end{aligned} \qquad (13.2.10)$$

where the first uncertainty is experimental and the second comes from theory. Unfortunately, they do not agree well with each other. Moreover, the values obtained from different experimental analyses are not always compatible within their stated theoretical and experimental uncertainties. The latest electron endpoint analysis by BaBar [4041], in particular, shows a dependence on the model used to simulate the signal and leads to sharply different results in BLNP and GGOU. This is the most precise analysis to date; in GGOU it favours a lower $|V_{ub}| = 3.96(10)(17) \times 10^{-3}$ while in BLNP the result is $|V_{ub}| = 4.41(12)(27) \times 10^{-3}$. While it is possible that modelling the signal has biased previous endpoint results, we stress that analyses involving a larger fraction of the phase space are generally less sensitive to SFs and other theoretical systematics, which are inherently difficult to estimate. In this respect, applying a cut on the hadronic invariant mass $M_X < 1.7\,\mathrm{GeV}$ seems to be the safest approach, as it depends little on the reconstruction of the $b \to c$ background, captures almost 60% of the phase space, and strikes a balance between experimental and theoretical uncertainties. In the recent Belle analysis [4042], where machine learning techniques and hadronic tagging were used to reduce backgrounds, the result in GGOU (very much consistent with BLNP and DGE) is

$$|V_{ub}| = 3.97(18)(17)\times 10^{-3}, \qquad (B \to X_u \ell\nu) \quad (13.2.11)$$

which in my opinion represents the current state of the art. Improvements will certainly come from the higher statistics available at Belle II and from the implementation of higher order calculations such as [4043]. For instance, the complete $O(\alpha_s^2)$ perturbative contributions to the triple differential rate is still missing, despite numerical results for the moments [4044]. A precise study of the differential spectra, recently measured at Belle for the first time [4045], will validate the theoretical frameworks and help constrain the SFs. The SIMBA [1800] and NNVub [4046] methods are well posed to analyse the Belle II data in a model independent and efficient way. In the longer run, lattice studies like those mentioned for inclusive $b \to c$ transitions should also become possible.

The exclusive $B \to D\ell\nu$ and $B \to D^*\ell\nu$ channels are also used to extract $|V_{cb}|$. These decays are described by nonperturbative form factors which are computed in lattice QCD (as discussed in Section 4.7) as well as with approximate methods like Light Cone Sum Rules (LCSR), see Section 5.8. Typically, the lattice calculations are better under control at large or maximal $q^2$, corresponding to small or vanishing recoil, while LCSR prefer the small $q^2$ range and are less precise. Moreover, heavy quark symmetry guarantees that the form factors at zero recoil are absolutely normalised in the heavy quark limit. As the rates vanish at zero recoil in both cases, see Eq. (4.7.13), the experimental data are much less precise at low recoil and one needs to parameterize the form factors in a model independent way in order to describe the form factors in the whole kinematic range and to interpolate between the small and large recoil regions. Model independent parametrizations based on a dispersive approach have been developed in the 1990s and the two most relevant ones are known as BGL and BCL [4047, 4048]; the form factors are expressed, up to known prefactors, as series

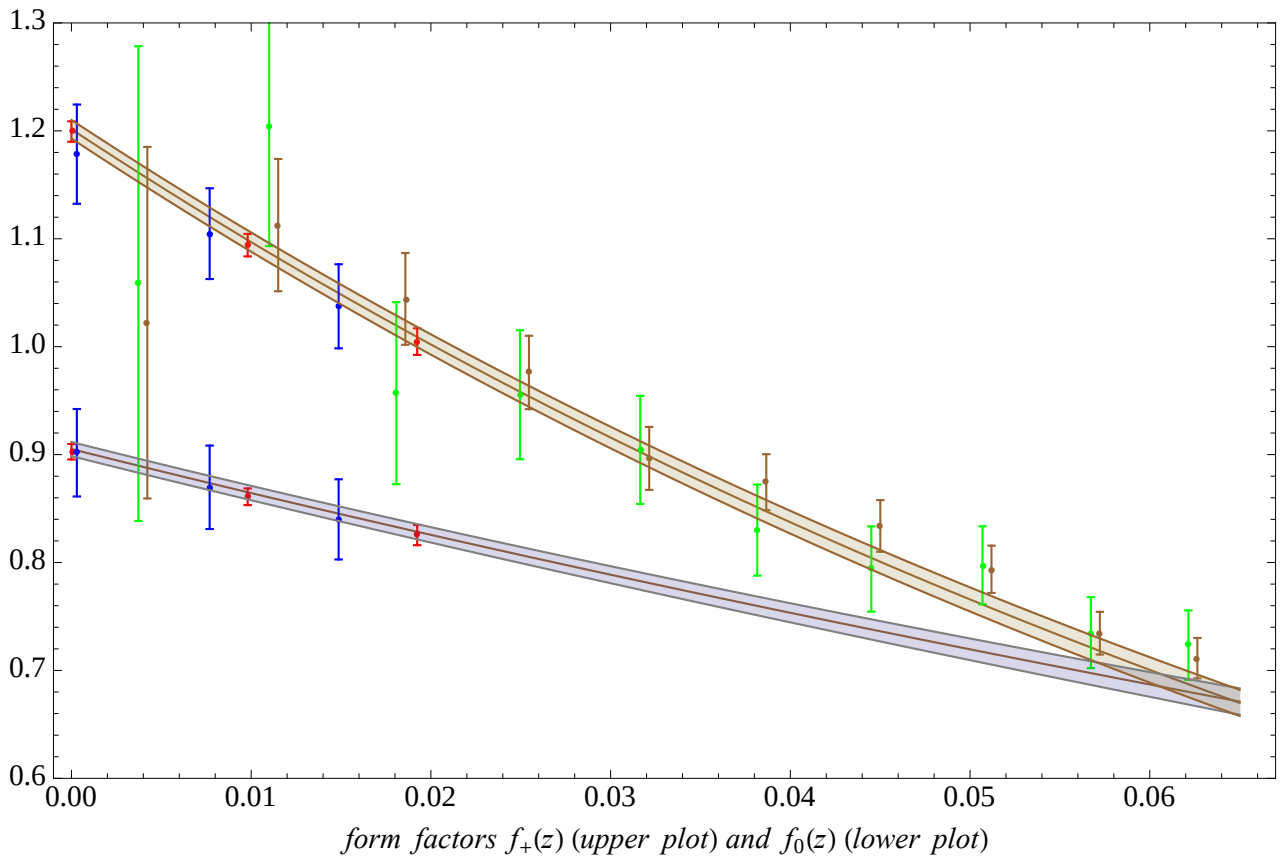

**Fig. 13.2.3** Form factors $f_{+,0}(z)$ for the $B \to D$ transitions computed by FNAL/MILC [4057] (red) and HPQCD [4058] (blue) and experimental data from Belle (brown) and BaBar (green) normalised by the fitted value of $|V_{cb}|$. The bands show the results of the global fit. Figure from Ref. [4059].

in the variable

$$z = \frac{\sqrt{w+1}-\sqrt{2}}{\sqrt{w+1}+\sqrt{2}}, \qquad (13.2.12)$$

where $w = (m_B^2 + M_{D^{(*)}} - q^2)/(2m_B m_{D^{(*)}})$. In the physical range $z$ is small, $< 0.07$, and unitarity puts constraints on the size of the series coefficients. A variant, proposed in [4049] and known as CLN, additionally employs Next to Leading Order Heavy Quark Effective Theory relations and QCD sum rules to reduce the number of relevant parameters to two. These additional inputs imply an uncertainty that can no longer be neglected, see [4050–4053] for updates and improvements on the CLN approach. It is then unfortunate that prior to 2016 the experimental results were generally given in terms of fits to the CLN parametrization, without accounting for this uncertainty. More recent measurements [4054–4056] provide the differential $q^2$ and angular (for $B \to D^*\ell\nu$) unfolded distributions or the necessary ingredients (efficiencies and response functions) to *fold* theoretical predictions and get the yields in each bin.

In the $B \to D\ell\nu$ case precise lattice calculations at small but non-zero recoil are available since several years [4058, 4060] and have been combined with the experimental results of Refs. [4054, 4061] to get [4059]

$$|V_{cb}| = 40.5(1.0)10^{-3} \qquad (B \to D\ell\nu). \qquad (13.2.13)$$

A similar value is found in [63]. Indeed, the lattice and experimental form factor shapes are in good agreement, satisfy the unitarity constraints, and the overall fit is good and stable, see Fig. 13.2.3. The BGL and BCL parametrizations give identical results and the fit also provides a SM prediction for the Lepton Flavor Universality ratio $R(D) = \Gamma(B \to D\tau\nu)/\Gamma(B \to D\mu\nu) = 0.299(3)$ [4059], in reasonable agreement with the experimental world average $R(D)_{exp} = 0.339(30)$ [4012].

In the $B \to D^*\ell\nu$ channel the situation is more complicated. From the experimental point of view this channel allows for a more precise determination of $|V_{cb}|$ than the $B \to D$ channel and angular distributions can be studied in addition to the $q^2$ distribution. On the other hand, the $D^*$ meson decays strongly to $D\pi$ (it cannot be considered stable) and three (four) different form factors contribute for a massless (massive) lepton. The only lattice calculation of these form factors away from the zero-recoil point has been published so far by the Fermilab-MILC Collaboration [708], although JLQCD and HPQCD calculation are in their final stage [710, 4062]. Restricting to experimental analyses that provide data in a model independent way, Belle has presented a tagged [4055] and an untagged analysis [4056]. The dataset of [4055] showed for the first time that the extraction of $|V_{cb}|$ could strongly depend on the parametrization employed: BGL and CLN both gave reasonable fits with $|V_{cb}|$ values differing by about 6% [4063, 4064]. It has subsequently been disowned by the collaboration, but the point remains valid: parametrizations matter and the related uncertainties have to be carefully considered. The more precise dataset of the untagged analysis [4056], despite a few problems [4065], did not show any parametrization dependence. A global fit based on [4066] that includes the Fermilab calculation [708], unitarity constraints, and the Belle untagged data only, while adjusting for the D'Agostini bias [4067], leads to

$$|V_{cb}| = 39.3(9)10^{-3} \qquad (B \to D^*\ell\nu), \tag{13.2.14}$$

but the agreement between the Fermilab form factor shape and the experimental distributions is not good and the total $\chi^2$ is large.[116] An additional uncertainty of $\sim 0.5\%$ for missing QED corrections should be added to Eq. (13.2.14), as well as to Eqs. (13.2.9) and (13.2.13). There is also a troubling tension between the Fermilab results and the ratio of form factors computed in NLO HQET. Preliminary results for the $B \to D^*$ form factors have also been disclosed by the JLQCD collaboration [4068] and in this case the agreement with Belle data is much better, with a final $|V_{cb}| = 40.7(^{+1.0}_{-0.9})10^{-3}$. One can also add LCSR constraints on the form factors [3987], with minimal change in $|V_{cb}|$. Despite these latest developments, HFLAV also quotes an average of experimental results in the CLN parametrization based on the form factor at zero recoil only, $|V_{cb}| = 38.46(68)\,10^{-3}$, but this result is subject to uncontrolled uncertainties related to the way the CLN parametrization has been used. The two Belle datasets have also been analysed in the Dispersive Matrix approach [4069], where the form factors are constrained by the Fermilab lattice data and unitarity only; tensions with the experimental data are observed here as well. The fit that originates Eq. (13.2.14) gives also $R(D^*) = \Gamma(B \to D^*\tau\nu)/\Gamma(B \to D^*\mu\nu) = 0.249(1)$, confirming the tension with the experimental world average $R(D^*)_{exp} = 0.295(14)$ [4012].

LHCb has recently performed the first determination of $|V_{cb}|$ using $B_s^0$ decays [4070]. Using both $B_s^0 \to D_s^{(*)-}\mu^+\nu$ and the lattice results from Refs. [709, 4071], they obtain $|V_{cb}| = 41.7(0.8)(0.9)(1.1)10^{-3}$. On the other hand, Babar using a simplified BGL parametrization finds $|V_{cb}| = 38.4(9)10^{-3}$ [4072]. In summary, the situation for the exclusive determination of $|V_{cb}|$ is still unsettled, but a tension with the inclusive determination of Eq. (13.2.9) is undisputable. New lattice calculations performed with relativistic heavy quarks such as [710] will extend their $q^2$ range, making it possible to extract $|V_{cb}|$ at large recoil, where experimental data are more accurate. New experimental analyses of Belle and Belle II data are also expected soon. As this is paralleled by a renewed experimental and theoretical activity on the inclusive front, we can hope that the $V_{cb}$ puzzle will find its resolution.

Moving to the exclusive determination of $|V_{ub}|$, it proceeds through the $B \to \pi$ channel. In analogy to the $B \to D$ case, only one form factor is relevant for massless leptons and it is standard practice to perform a BCL fit to lattice [704, 711, 4073] and LCSR calculations and to experimental data from several experiments, see [4012]. HFLAV employs the Fermilab and RBC/UKQCD form factors and the LCSR calculation of [4074] to find $|V_{ub}| = 3.67(15)10^{-3}$. An updated LCSR result is presented in [4075] and leads to

$$|V_{ub}| = 3.77(15)10^{-3} \qquad (B \to \pi\ell\nu). \tag{13.2.15}$$

The recent JLQCD form factor $f_+(q^2)$ [711] is slightly lower than the Fermilab and RBC/UKQCD and also implies a higher $|V_{ub}|$. The fits in [4012, 4075] are both consistent, but there are two outliers which drive the value of $|V_{ub}|$ down. Removing the outliers the result increases $|V_{ub}|$ by about one sigma [4076]. We can conclude that the agreement between inclusive and exclusive determinations of $|V_{ub}|$ has become acceptable, but more stringent tests will be possible in the next few years. With the large statistics that will be available at Belle II the channel $B \to \tau\nu$ will become competitive with $B \to \pi\ell\nu$ for the extraction of $|V_{ub}|$. To this end, neglecting QED effects, the only QCD input is the decay constant $f_B$, which is already known to better than 1%, see Section 4.7.

[116] The result in (13.2.14) differs from that reported in [708] and adopted in [476], $|V_{cb}| = 38.4(7)10^{-3}$, mostly because of the D'Agostini bias (not considered in [708]), of the way unitarity constraints are implemented, and of the QED Coulomb factor that is included in [708], neglecting however other QED corrections.

Finally, two recent semileptonic measurements at LHCb place constraints on $|V_{ub}/V_{cb}|$. The first concerns the ratio of $\Lambda_b \to p\mu\nu$ to $\Lambda_b \to \Lambda_c\mu\nu$ decays [715] and makes use of a pioneering lattice calculation of baryonic form factors [714]; the result is [4012]

$$\frac{|V_{ub}|}{|V_{cb}|} = 0.079(4)(4) \qquad (\Lambda_b \to p\mu\nu) \tag{13.2.16}$$

where the uncertainties are experimental and from the form factors. The second is the first measurement of $B_s \to K\mu\nu$; the decay is normalised to $B_s \to D_s\mu\nu$ in two bins of $q^2$ [4077]. Using lattice results from the FNAL/MILC Collaboration [4078] for the high $q^2$ bin and LCSR [4079] for the low $q^2$ bin, one obtains values of $|V_{ub}/V_{cb}|$ in sharp disagreement with each other, which requires further scrutiny. Averaging Ref. [4078] with older results in the high $q^2$ bin of Ref. [4077], FLAG finds $|V_{ub}/V_{cb}| = 0.086(5)$ [63]. We can compare this and Eq. (13.2.16) with Eqs. (13.2.9,): from inclusive decays we get $|V_{ub}/V_{cb}| = 0.094(6)$, from exclusive decays $|V_{ub}/V_{cb}| = 0.094(4)$, and in both cases the tension with Eq. (13.2.16) is over $2\sigma$. The agreement improves for lower $|V_{ub}|$ or higher $|V_{cb}|$. This is another puzzling issue: hopefully, future measurements and lattice calculations of baryonic and mesonic form factor will clarify the situation.

As mentioned above, semileptonic $b$ decays are not the only observables sensitive to $|V_{cb}|$ and $|V_{ub}|$. Assuming the validity of the SM and therefore the unitarity of the CKM matrix, one can also extract $V_{cb}$ from loop induced observables like $\varepsilon_K$ and $B_{(s)} - \bar{B}_{(s)}$ mixing, as well as from rare kaon and B decays [4001, 4080–4084], and the precision starts to be competitive. For instance, the $B_{(s)}$ meson mass differences are proportional to $|V_{cb}|^2$: $\Delta M_{(d,s)} \propto |V_{td,ts}|^2$ and $|V_{ts}|^2 \approx |V_{cb}|^2$, $|V_{td}|^2 = \lambda^2 \sin^2\gamma |V_{cb}|^2$. $\varepsilon_K$ is even more sensitive, $\varepsilon_K \propto |V_{cb}|^{3.4}$, and the branching fraction for $K_L \to \pi^0\nu\bar{\nu}$ is proportional to $|V_{cb}|^4$. Deviations from the direct (semileptonic) determinations would signal New Physics. The present situation is illustrated in Fig. 13.2.4, where the constraints from some of these observables in the $(\gamma, |V_{cb}|)$ plane are shown, with a clear preference for a high $|V_{cb}|$. As far as $|V_{ub}|$ is concerned, global fits performed without its direct determination tend to return values close to Eq. (13.2.15).

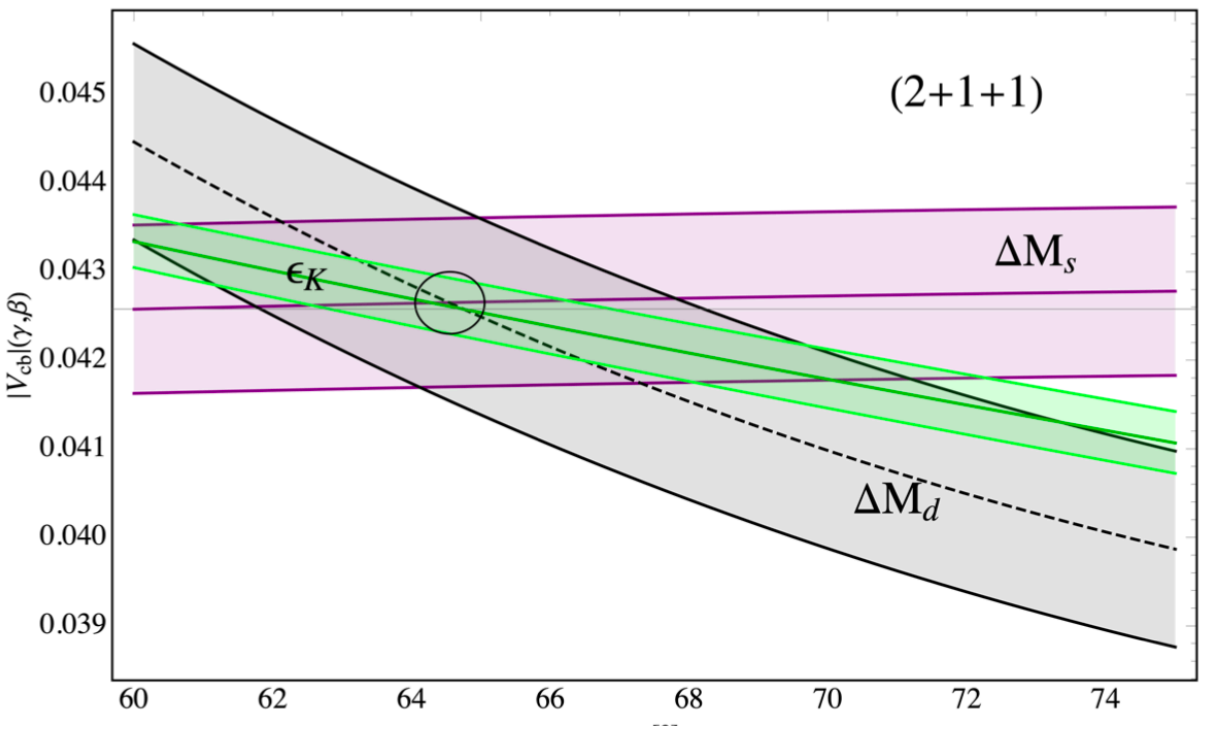

**Fig. 13.2.4** Present constraints from $\varepsilon_K$, $\Delta M_d$, and $\Delta M_s$ in the $(\gamma, |V_{cb}|)$ plane, see Ref. [4082] for details.

### 13.2.3 Meson mixing and CP asymmetries

So far we have discussed the elements of the first two rows of $\hat{V}_{CKM}$: their magnitudes determine precisely $\lambda$ and $A$ in Eq. (13.2.1), and the ratio $|V_{ub}/V_{cb}|$ constrains the apex of the unitarity triangle, as shown in Fig. 13.2.1. In order to determine completely the remaining parameters $\rho$ and $\eta$, however, one needs additional information. As the elements of the third row cannot be precisely measured yet, we now turn to loop mediated $B_{(s)}$ mixing and rare decays, and CP asymmetries, focussing only on the most constraining observables.

In the SM the mass difference $\Delta M_{d,s}$ between the two mass eigenstates of the $B^0$ and $B_s^0$ systems is proportional to $|V_{td}|^2$ and $|V_{ts}|^2$, respectively, and the relevant nonperturbative QCD physics is all contained in the product $f_{B_q}^2 \hat{B}_{B_q}$ of decay constants and bag parameters, see Eq. (4.7.11). The ratio $\Delta M_s/\Delta M_d$ is particularly interesting because some uncertainty cancels out: the latest $N_f = 2+1+1$ value [685] for $\xi = f_{B_s}/f_{B_d}\sqrt{B_{B_s}/B_{B_d}}$ is $\xi = 1.216(16)$, which together with accurate measurements [4012] allows for the very strong constraint shown in red in Fig. 13.2.1. Individually, $\Delta M_{d,s}$ are slightly less precise but have a very different sensitivity to $|V_{cb}|$, see Fig. 13.2.4. In the kaon sector one looks at CP-violation in mixing, quantified by $\varepsilon_K$, see Subsec. 13.3.3, which is sensitive to a combination of CKM elements. The bulk of $\varepsilon_K$ is due to its *short-distance* component, whose uncertainty is dominated by the bag parameter $\hat{B}_K$, see *e.g.* [3966]. The recent average of lattice calculations reported in Section 4.7, $\hat{B}_K = 0.7625(97)$, leads to the constraints shown in Fig. 13.2.1 and 13.2.4. Finally, different CP asymmetries allow for a direct extraction of the phase of some CKM element, with minimal or no QCD input. In particular, the measurement of the time-dependent CP asymmetry in $B \to J/\psi K_S$ gives $\sin 2\beta = 0.699(17)$ (green band in Fig. 13.2.1); the study of the interference between the tree-level decays $B^- \to D^0 K-$ and $B^- \to \bar{D}^0 K-$ gives $\gamma = 66.1(3.5)^\circ$ [4012] (blue band in Fig. 13.2.1); the time-dependent asymmetries in $B \to \pi\pi, \rho\rho$ have been used to extract $\alpha = 85.4(4.6)^\circ$ (gray bands in Fig. 13.2.1).

The global picture that emerges from all these and additional less important inputs is summarised by the global fit that gives the apex of the unitarity triangle in Fig. 13.2.1: $\bar{\rho} = 0.156(12)$ and $\bar{\eta} = 0.350(10)$ [4001]. The consistency between the various constraints is impressive and in the last 18 years the overall precision has improved by a factor 4(3) for $\bar{\rho}(\bar{\eta})$. One can compare some of the above inputs with the values obtained from a global fit performed without them: the results are $\sin 2\beta = 0.750(27)$, $\gamma = 66.1(2.1)^\circ$, $\alpha = 90.5(2.1)^\circ$ [4001]. Very similar results are also obtained by the CKMFitter Collaboration [4084], which reports $\bar{\rho} = 0.157(^{+8}_{-5})$ and $\bar{\eta} = 0.348(^{+12}_{-5})$.

In summary, the CKM mechanism describes successfully a host of data, in many cases with crucial QCD input. As discussed in Subsections 13.2.1 and 13.2.2, there are potential problems that require further scrutiny, and more serious anomalies will be discussed in Section 13.4, but it is premature to attribute them to New Physics. On the contrary, present data place very strong constraints on a variety of New Physics scenarios, in particular on those that modify the CKM mechanism more radically, see *e.g.* [3966, 4085]. From an effective field theory point of view, the measurements we have considered in this Section imply that the scale $\Lambda$ of New Physics with a generic flavor structure must be well beyond the TeV range.

## 13.3 The Important Role of QCD in flavor Physics

**Andrzej J. Buras**

The importance of QCD effects depends on processes considered. While their inclusion in processes like $K^+ \to \pi^+ \nu\bar{\nu}$, $K_L \to \pi^0 \nu\bar{\nu}$, $B^0_{s,d} \to \mu^+\mu^-$ is important in order to increase the precision of SM predictions, neglecting them would result in uncertainties in the ballpark of at most 30%, significant but not crucial if one wants to get a rough idea what are the SM predictions for such decays. There are extensive reviews on them and most of these decays are discussed in [3966]. Here we want to confine our presentation to cases in which QCD plays an essential role and neglecting QCD effects one would fail the description of the data not by 30%, but by factors of at least two and sometimes even by an order of magnitude.

### 13.3.1 $B \to X_s\gamma$ Decay

The calculations of NLO and NNLO QCD corrections to $B \to X_s\gamma$ decay are probably the best known to the physics community among all QCD calculations in the field of weak decays. One of the reasons is the fact that the $b \to s\gamma$ transition was the first penguin-mediated transition in $B$ physics to be discovered in 1993 in the exclusive decay channel $B \to K^*\gamma$ measured in the CLEO experiment [4086]. The inclusive branching ratio $B \to X_s\gamma$ has been measured in 1994 by the same group [4087]. The other reason is the particular structure of the QCD corrections to this decay that requires a two-loop calculation in order to obtain the anomalous dimension matrix in the LO approximation. Because of this it took six years after the first QCD calculations in ordinary perturbation theory to obtain the correct result for the QCD corrections to $B \to X_s\gamma$ in the RG-improved perturbation theory at LO. It involved 5 groups and 16 physicists. It is not then surprising that the corresponding NLO calculations took nine years. In 2022 this decay is known including NNLO corrections. A detailed historical account of NLO calculations can be found in [3973] and an introduction to technical details in [3966]. Most extensive NNLO calculations have been reported first in [4088], and after a number of updates the last one has been presented in [4089]

$$\mathcal{B}(B \to X_s\gamma)_{\rm SM} = (3.36 \pm 0.23) \times 10^{-4}\,, \qquad (13.3.1)$$

for $E_\gamma \geq 1.6\,\text{GeV}$. It agrees very well with experiment which reached the accuracy of 4.5% [4090]

$$\mathcal{B}(B \to X_s\gamma)_{\rm exp} = (3.32 \pm 0.15) \times 10^{-4}\,, \qquad (13.3.2)$$

where again $E_\gamma \geq 1.6\,\text{GeV}$ has been imposed. One expects that in this decade the Belle II experiment will reach the accuracy of 3% so that very precise tests of the SM will be possible. Already now this decay provides an important constraint on new physics.

In order to appreciate these results let us briefly describe why these very difficult calculations were crucial. Indeed in 1987 two groups [4091, 4092] calculated $\mathcal{O}(\alpha_s)$ QCD corrections to the $B \to X_s\gamma$ rate finding a huge enhancement of this rate relative to the partonic result without QCD corrections. In 1987, when $m_t \leq M_W$ was still considered, this enhancement was almost by an order of magnitude. With the increased value of $m_t$ in the 1990's also the partonic rate increased, and in 2022 the dominant additive QCD corrections, although still very important, amount roughly to a factor of 2.5.

The additive QCD corrections in question originate in the mixing of the leading current-current operator $Q_2$ like the one in Eqn. (13.3.5) with the magnetic-photon penguin operator $Q_{7\gamma}$ that is directly responsible for the decay $b \to s\gamma$. The calculation of the relevant anomalous dimensions at LO is a two-loop affair and consequently it took some time before the correct result had been obtained. An important role in resolving these inconsistencies present in the literature played Mikolaj Misiak [4093, 4094]. But the final LO result has been provided by the Rome group [4095, 4096].

Once this issue had been solved it was possible to outline an NLO calculation in [4097]. Such a calculation was motivated by the finding in [4098] that the LO rate for $B \to X_s\gamma$ exhibited a very large renormalization-scale dependence. Changing the scale $\mu_b$ in the Wilson coefficient from $m_b/2$ to $2m_b$ changed the rate of $B \to X_s\gamma$ by roughly 60% making a detailed comparison of theory with experiment impossible.

A large number of authors contributed to the calculation of NLO corrections, with their names and references listed in Table 5 of the review in [3973]. See also the 2002 summary of NLO calculation in [4099].

Yet already one year before a motivation for a NNLO calculation was born. While the NLO calculations decreased the $\mu_b$-dependence present in the LO expressions significantly, a new uncertainty had been pointed out by Paolo Gambino and Mikolaj Misiak in 2001 [4100]. It turns out that the $B \to X_s\gamma$ rate suffers at the NLO from a significant, $\pm 6\%$, uncertainty due to the choice of the charm quark mass in the two-loop matrix elements of the four quark operators, in particular in $\langle s\gamma|Q_2|B\rangle$. In the following years, considerable progress in the NNLO program of $B \to X_s\gamma$ was made. It was an effort of 17 theorists [4088] and lead eventually to the result in Eqn. (13.3.1) summarized in [4089].

### 13.3.2 QCD dynamics and the $\Delta I = 1/2$ rule

One of the puzzles of the 1950s was a large disparity between the measured values of the real parts of the isospin amplitudes $A_0$ and $A_2$ in $K \to \pi\pi$ decays, which on the basis of usual isospin considerations were expected to be of the same order. In 2022 we know the experimental values of the real parts of these amplitudes very precisely [4101]

$$\mathrm{Re}A_0 = 27.04(1) \times 10^{-8}\ \ \mathrm{GeV},$$
$$\mathrm{Re}A_2 = 1.210(2) \times 10^{-8}\ \ \mathrm{GeV}. \qquad (13.3.3)$$

As $\mathrm{Re}A_2$ is dominated by $\Delta I = 3/2$ transitions but $\mathrm{Re}A_0$ receives contributions also from $\Delta I = 1/2$ transitions, the latter transitions dominate $\mathrm{Re}A_0$ which expresses the so-called $\Delta I = 1/2$ rule [4102, 4103]

$$R = \frac{\mathrm{Re}A_0}{\mathrm{Re}A_2} = 22.35. \qquad (13.3.4)$$

In the 1950s QCD and the Operator Product Expansion did not exist and clearly one did not know that $W^\pm$ bosons exist in nature, but using the ideas of Fermi [4104], Feynman and Gell-Mann [4105] and Marshak and Sudarshan [4106] one could still evaluate the amplitudes $\mathrm{Re}A_0$ and $\mathrm{Re}A_2$ to find out that such a high value of $R$ is a real puzzle.

In modern times we can reconstruct this puzzle by evaluating the simple $W^\pm$ boson exchange between the relevant quarks which after integrating out $W^\pm$ generates the current-current operator $Q_2$:

$$Q_2 = (\bar{s}\gamma_\mu(1-\gamma_5)u)\ (\bar{u}\gamma^\mu(1-\gamma_5)d)\ . \qquad (13.3.5)$$

With only $Q_2$ contributing we have

$$\mathrm{Re}A_{0,2} = \frac{G_F}{\sqrt{2}} V_{ud}V^*_{us}\langle Q_2\rangle_{0,2}\,. \qquad (13.3.6)$$

Calculating the matrix elements $\langle Q_2\rangle_{0,2}$ in the strict large $N$ limit, which corresponds to factorization of matrix elements of $Q_2$ into the product of matrix elements of currents, we find

$$\langle Q_2\rangle_0 = \sqrt{2}\langle Q_2\rangle_2 = \frac{2}{3}\,f_\pi(m_K^2 - m_\pi^2), \qquad (13.3.7)$$

and consequently

$$\mathrm{Re}A_0 = 3.59 \times 10^{-8}\,\mathrm{GeV},$$
$$\mathrm{Re}A_2 = 2.54 \times 10^{-8}\,\mathrm{GeV}\ ,\ R = \sqrt{2}, \qquad (13.3.8)$$

in plain disagreement with the data in Eqns. (13.3.3) and (13.3.4). It should be emphasized that the explanation of the missing enhancement factor of 15.8 in $R$ through some dynamics must simultaneously give the correct values for $\mathrm{Re}A_0$ and $\mathrm{Re}A_2$. This means that this dynamics should suppress $\mathrm{Re}A_2$ by a factor of 2.1, not more, and enhance $\mathrm{Re}A_0$ by a factor of 7.5. This tells us that while the suppression of $\mathrm{Re}A_2$ is an important ingredient in the $\Delta I = 1/2$ rule, it is not the main origin of this rule. It is the enhancement of $\mathrm{Re}A_0$ as already emphasized in [1209] even if, in contrast to this paper, as pointed out first in 1986 [3960] and demonstrated in the context of the Dual QCD approach, the current-current operators, like $Q_2$, are responsible dominantly for this rule and not QCD penguins. An update and improvements over the 1986 analysis appeared in 2014 [3961] with the result

$$R \approx 16.0 \pm 1.5\ , \qquad \mathrm{DQCD}\ \ (1986, 2014), \qquad (13.3.9)$$

that is one order of magnitude enhancement over the result in Eqn. (13.3.8) without QCD up to confinement of quarks in mesons. The missing piece could come from final state interactions as pointed out first by nuclear physicists [4107] and stressed much later by ChPT experts [4108]. Also $1/N^2$ corrections could also change this result but are unknown.

Meanwhile the RBC-UKQCD LQCD collaboration confirmed in 2012 the 1986 DQCD finding that current-current operators dominate the $\Delta I = 1/2$ rule. But the results from the series of their three papers show how difficult these calculations on the lattice are: $R = 12 \pm 1.7$ [4109], $R = 31.0 \pm 11.1$ [691] and finally [3962]

$$\frac{\mathrm{ReA}_0}{\mathrm{ReA}_2} = 19.9(2.3)(4.4), \qquad \mathrm{RBC-UKQCD} \quad (2020) \tag{13.3.10}$$

that is consistent with the DQCD value and in agreement with the experimental value 22.4.

While the RBC-UKQCD result is closer to the data than the DQCD one, the dynamics behind this rule, except for the statement that it is QCD, has not been provided by these authors. To this end it is necessary to switch off QCD interactions which can be done in the large $N$ limit in DQCD but it seems to be impossible or very difficult on the lattice.

The anatomy of QCD dynamics as seen within the DQCD approach has been presented in [3960, 3961] and in particular in Section 7.2.3 of [3966]. Here we just present an express view of this dynamics.

Starting with the values in Eqn. (13.3.8), the first step is to include the short-distance RG-evolution of WCs from scales $\mathcal{O}(M_W)$ down to scales in the ballpark of 1 GeV. This is the step made already in the pioneering 1974 calculations in [1211, 1212] except that they were done at LO in the RG-improved perturbation theory and now can be done at the NLO level. These 1974 papers have shown that the short distance QCD effects enhance $\mathrm{Re}A_0$ and suppress $\mathrm{Re}A_2$. However, the inclusion of NLO QCD corrections to WCs of $Q_2$ and $Q_1$ operators [3965, 4110] made it clear, as stressed in particular in [3965], that the $K \to \pi\pi$ amplitudes without the proper calculation of hadronic matrix elements of $Q_i$ are both scale and renormalization-scheme dependent. Moreover, further enhancement of $\mathrm{Re}A_0$ and further suppression of $\mathrm{Re}A_2$ are needed in order to be able to understand the $\Delta I = 1/2$ rule.

This brings us to the second step first performed in 1986 in [3960] within the DQCD approach. Namely, the RG-evolution down to the scales $\mathcal{O}(1\,\mathrm{GeV})$ is continued as a short but fast *meson evolution* down to zero momentum scales at which the factorization of hadronic matrix elements is at work and one can in no time calculate the hadronic matrix elements in terms of meson mases and weak decay constants as seen in (13.3.7). Equivalently, starting with factorizable hadronic matrix elements of current-current operators at $\mu \approx 0$ and evolving them to $\mu = \mathcal{O}(1\,\mathrm{GeV})$ at which the WCs are evaluated one is able to calculate the matrix elements of these operators at $\mu = \mathcal{O}(1\,\mathrm{GeV})$ and properly combine them with their WCs evaluated at this scale. The final step is the inclusion of QCD penguin operators that provide an additional enhancement of $A_0$ by roughly 10% without changing $A_2$.

In [3960] only the pseudoscalar meson contributions to meson evolution have been included and the *quark evolution*, RG evolution above $\mu = \mathcal{O}(1\,\mathrm{GeV})$, has been performed at LO. The improvements in 2014 [3961] were the inclusion of vector meson contributions to the meson evolution and the NLO corrections to quark evolution. These improvements practically removed scale and renormalization-scheme dependences and brought the theory closer to data.

Based on DQCD and RBC-UKQCD results we conclude that the QCD dynamics is dominantly responsible for the $\Delta I = 1/2$ rule. However, in view of large uncertainties in both DQCD and RBC-UKQCD results, NP contributions at the level of 15% could still be present. See [4111] to find out what this NP could be.

Finally other authors suggested different explanations of the $\Delta I = 1/2$ rule within QCD that were published dominantly in the 1990s and their list can be found in [3966]. But in my view the DQCD picture of what is going on is more beautiful and transparent as asymptotic freedom and related non-factorizable QCD interactions are primarly responsible for this rule. It is simply the *quark evolution* from $M_W$ down to scale $\mathcal{O}(1\,\mathrm{GeV})$ as analysed first by Altarelli and Maiani [1212] and Gaillard and Lee [1211], followed by the *meson evolution* [3960, 3961] down to very low scales at which QCD becomes a theory of weakly interacting mesons and a free theory of mesons in the strict large $N$ limit, a point made by 'tHooft and Witten in 1970s.

### 13.3.3 QCD Dynamics and the Ratio $\varepsilon'/\varepsilon$

While, the parameter $\varepsilon \equiv \varepsilon_K$ measures the indirect CP-violation in $K_L \to \pi\pi$ decays, that is originating in $K^0 - \bar{K}^0$ mixing, the parameter $\varepsilon'$ describes the direct CP violation, that is in the decay itself.

Experimentally $\varepsilon$ and $\varepsilon'$ can be found by measuring the ratios

$$\eta_{00} = \frac{A(K_\mathrm{L} \to \pi^0\pi^0)}{A(K_\mathrm{S} \to \pi^0\pi^0)}, \qquad \eta_{+-} = \frac{A(K_\mathrm{L} \to \pi^+\pi^-)}{A(K_\mathrm{S} \to \pi^+\pi^-)}.$$

(13.3.11)

Indeed, assuming $\varepsilon$ and $\varepsilon'$ to be small numbers one finds

$$\eta_{00} = \varepsilon - \frac{2\varepsilon'}{1-\sqrt{2\omega}}, \qquad \eta_{+-} = \varepsilon + \frac{\varepsilon'}{1+\omega/\sqrt{2}}, \tag{13.3.12}$$

where $\omega = \mathrm{ReA}_2/\mathrm{ReA}_0 = 0.045$. In the absence of direct CP violation $\eta_{00} = \eta_{+-}$. The ratio $\varepsilon'/\varepsilon$ can then be measured through

$$\mathrm{Re}(\varepsilon'/\varepsilon) = \frac{1}{6(1+\omega/\sqrt{2})}\left(1 - \left|\frac{\eta_{00}}{\eta_{+-}}\right|^2\right) . \tag{13.3.13}$$

The story of $\varepsilon'/\varepsilon$ both in the theory and experiment has been described in detail in [4112]. On the experimental side the chapter on $\varepsilon'/\varepsilon$ seems to be closed for the near future. After heroic efforts, lasting 15 years, the experimental world average of $\varepsilon'/\varepsilon$ from NA48 [4113] and KTeV [4114, 4115] collaborations reads

$$(\varepsilon'/\varepsilon)_{\mathrm{exp}} = (16.6 \pm 2.3) \times 10^{-4} \, . \tag{13.3.14}$$

On the theoretical side the first calculation of $\varepsilon'/\varepsilon$ that included RG QCD effects to QCD penguin (QCDP) contributions is due to Gilman and Wise [4116] who - following Shifman, Vainshtein and Zakharov [1209] - assumed that the $\Delta I = 1/2$ rule is explained by QCDP. Using the required values of the QCDP matrix elements for the explanation of this rule, they predicted $\varepsilon'/\varepsilon$ to be in the ballpark of $5 \times 10^{-2}$. During the 1980s this value decreased by roughly a factor of 50 dominantly due to three effects:

- The first calculation of hadronic matrix elements of QCDP operators in QCD - carried out in the framework of the DQCD [3960, 4117, 4118] in the strict large $N$ limit of colors - proved that QCDPs are not responsible for the $\Delta I = 1/2$ rule and their hadronic matrix elements are much smaller.
- The QCDP contribution to $\varepsilon'/\varepsilon$ through isospin breaking in the quark masses [4119, 4120] is suppressed.
- Enhancement of the suppression of $\varepsilon'/\varepsilon$ by electroweak penguin (EWP) contributions by the large top quark mass [4121, 4122].

In the 1990s these calculations have been refined through NLO QCD calculations to both QCDP and EWP contributions by the Munich and Rome teams [4123–4126] and [4127, 4128], respectively. In [4129] the NNLO QCD effects on EWP contributions have been calculated. The NNLO QCD effects on QCDP contributions are expected to be known in 2022.

These NLO and NNLO QCD contributions decreased various scale and renormalization-scheme uncertainties and suppressed $\varepsilon'/\varepsilon$ within the SM further so that already in 2000 we knew that this ratio should be of the order of $1.0 \times 10^{-3}$. Unfortunately even today the theorists do not agree on whether the SM agrees with the experimental value in (13.3.14) or not. The reason are different estimates of non-perturbative hadronic QCD effects. This has been summarized recently in [3963]. We recall only the main points below.

$\varepsilon'$ is governed by the real and imaginary parts of the isospin amplitudes $A_0$ and $A_2$ so that $\varepsilon'/\varepsilon$ is given by [4130]

$$\frac{\varepsilon'}{\varepsilon} = -\frac{\omega_+}{\sqrt{2}\,|\varepsilon|}\left[\frac{\mathrm{Im}A_0}{\mathrm{Re}A_0}\,(1-\hat{\Omega}_{\mathrm{eff}}) - \frac{1}{a}\frac{\mathrm{Im}A_2}{\mathrm{Re}A_2}\right], \tag{13.3.15}$$

with $(\omega_+, a)$ and $\hat{\Omega}_{\mathrm{eff}}$ given in 2022 as follows

$$\omega_+ = a\,\frac{\mathrm{ReA}_2}{\mathrm{ReA}_0} = (4.53 \pm 0.02) \times 10^{-2} \tag{13.3.16}$$

with $a = 1.017$ and

$$\hat{\Omega}_{\mathrm{eff}} = (29 \pm 7) \times 10^{-2} \, . \tag{13.3.17}$$

Here $a$ and $\hat{\Omega}_{\mathrm{eff}}$ summarize isospin breaking corrections and include strong isospin violation ($m_u \neq m_d$), the correction to the isospin limit coming from $\Delta I = 5/2$ transitions and electromagnetic corrections [4131–4133]. The most recent value for $\hat{\Omega}_{\mathrm{eff}}$ given above includes the nonet of pseudoscalar mesons and $\eta - \eta'$ mixing [4134]. If only the octet of pseudoscalar mesons is included so that $\eta - \eta'$ mixing does not enter, as presently done in ChPT, one finds $\hat{\Omega}_{\mathrm{eff}} = (17 \pm 9)\, 10^{-2}$ [4135], a value called $\hat{\Omega}^{(8)}_{\mathrm{eff}}$ here. The inclusion of $\eta - \eta'$ mixing yields $\hat{\Omega}^{(9)}_{\mathrm{eff}}$ in (13.3.17). This contribution is important, a fact known already for 35 years [4119, 4120].

$\mathrm{Im}A_0$ receives dominantly contributions from QCDP but also from EWP. $\mathrm{Im}A_2$ receives contributions exclusively from EWP. Keeping this in mind it is useful to write [4136]

$$\left(\frac{\varepsilon'}{\varepsilon}\right)_{\mathrm{SM}} = \left(\frac{\varepsilon'}{\varepsilon}\right)_{\mathrm{QCDP}} - \left(\frac{\varepsilon'}{\varepsilon}\right)_{\mathrm{EWP}} \tag{13.3.18}$$

with

$$\left(\frac{\varepsilon'}{\varepsilon}\right)_{\mathrm{QCDP}} = \mathrm{Im}\lambda_{\mathrm{t}} \cdot \big(1-\hat{\Omega}_{\mathrm{eff}}\big)\big[15.4\,\mathrm{B}_6^{(1/2)}(\mu^*) - 2.9\big], \tag{13.3.19}$$

$$\left(\frac{\varepsilon'}{\varepsilon}\right)_{\mathrm{EWP}} = \mathrm{Im}\lambda_{\mathrm{t}} \cdot \big[8.0\,\mathrm{B}_8^{(3/2)}(\mu^*) - 2.0\big]. \tag{13.3.20}$$

This formula includes NLO QCD corrections to the QCDP contributions and NNLO contributions to EWP

ones mentioned previously. The coefficients in this formula and the parameters $B_6^{(1/2)}$ and $B_8^{(3/2)}$, conventionally normalized to unity at the factorization scale, are scale dependent. Here we will set $\mu^* = 1\,\text{GeV}$ because at this scale it is most convenient to compare the values for $B_6^{(1/2)}$ and $B_8^{(3/2)}$ obtained in the three non-perturbative approaches LQCD, ChPT and DQCD that we already encountered in the context of the $\Delta I = 1/2$ rule.

The $B_6^{(1/2)}$ and $B_8^{(3/2)}$ represent the relevant hadronic matrix elements of the dominant QCDP and EWP operators, respectively:

$$Q_6 = (\bar{s}_\alpha d_\beta)_{V-A} \sum_{q=u,d,s,c,b} (\bar{q}_\beta q_\alpha)_{V+A}, \tag{13.3.21}$$

$$Q_8 = \frac{3}{2}\,(\bar{s}_\alpha d_\beta)_{V-A} \sum_{q=u,d,s,c,b} e_q\,(\bar{q}_\beta q_\alpha)_{V+A}, \tag{13.3.22}$$

with $V-A = \gamma_\mu(1-\gamma_5)$ and $V+A = \gamma_\mu(1+\gamma_5)$. They are then left-right operators with large hadronic matrix elements which assures their dominance over left-left operators. The remaining QCDP and EWP operators, represented here by $-2.9$ and $-2.0$, respectively, play subleading roles. Current-current operators $Q_{1,2}$ that played crucial role in the case of the $\Delta I = 1/2$ rule do not contribute to $\varepsilon'/\varepsilon$ because their WCs are real. In obtaining the formulae in Eqns. (13.3.19) and (13.3.20) it is common to use the experimental values for the real parts of $A_{0,2}$ in Eqn. (13.3.3). Finally, $\text{Im}\lambda_\text{t} = \text{Im}(\text{V}_\text{ts}^*\text{V}_\text{td}) \approx 1.4 \times 10^{-4}$.

There are two main reasons why $Q_8$ can compete with $Q_6$ here despite the smallness of the electroweak couplings in the WC of $Q_8$ relative to the QCD one in the WC of $Q_6$. In the basic formula (13.3.15) for $\varepsilon'/\varepsilon$ its contribution is enhanced relative to the $Q_6$'s one by the factor $\text{ReA}_0/\text{ReA}_2 = 22.4$. In addition its WC is enhanced for the large top-quark mass which is not the case for $Q_6$ [4121, 4122].

In the three non-perturbative approaches the values of $B_6^{(1/2)}$ and $B_8^{(3/2)}$ were found at $\mu = 1\,\text{GeV}$ to be:

$$\begin{aligned} &B_6^{(1/2)}(1\,\text{GeV}) = 1.49 \pm 0.25, \quad (\text{RBC-UKQCD}-2020)\\ &B_8^{(3/2)}(1\,\text{GeV}) = 0.85 \pm 0.05\,.\\ &B_6^{(1/2)}(1\,\text{GeV}) = 1.35 \pm 0.20,, \qquad (\text{ChPT}-2019)\\ &B_8^{(3/2)}(1\,\text{GeV}) = 0.55 \pm 0.20\,.\\ &B_6^{(1/2)}(1\,\text{GeV}) \le 0.6, \qquad (\text{DQCD}-2015)\\ &B_8^{(3/2)}(1\,\text{GeV}) = 0.80 \pm 0.10\,. \end{aligned} \tag{13.3.23}$$

While the large $B_6^{(1/2)}$ and $B_8^{(3/2)} < 1.0$ from LQCD has until now no physical interpretation, the pattern found in ChPT results apparently from final state interactions (FSI) that enhance $B_6^{(1/2)}$ above unity and suppress $B_8^{(3/2)}$ below it [4137–4140]. The suppression of $B_6^{(1/2)}$ and $B_8^{(3/2)}$ below unity in the DQCD approach comes from the meson evolution [4141] which is required to have a proper matching with the WCs of QCDP and EWP operators. The meson evolution is absent in present ChPT calculations and it is argued in [4142] that including it in ChPT calculations will lower $B_6^{(1/2)}$ below unity. On the other hand adding non-leading FSI in the DQCD approach would raise $B_6^{(1/2)}$ above 0.6. Nevertheless $B_6^{(1/2)} \le 1.0$ is expected to be satisfied even after the inclusion of FSI in DQCD.

Moreover, while ChPT and DQCD use $\hat{\Omega}_\text{eff}^{(8)} = (17 \pm 9)\,10^{-2}$ and $\hat{\Omega}_\text{eff}^{(9)} = (29 \pm 7)\,10^{-2}$, respectively, as already stated above, RBC-UKQCD still uses $\hat{\Omega}_\text{eff} = 0$.

These differences in the values of $B_6^{(1/2)}$, $B_8^{(3/2)}$ and $\hat{\Omega}_\text{eff}$ imply significant differences in $\varepsilon'/\varepsilon$ presented by these three groups:

$$(\varepsilon'/\varepsilon)_\text{SM} = (21.7 \pm 8.4) \times 10^{-4} \tag{13.3.24}$$

from the RBC-UKQCD collaboration [3962] which uses $\hat{\Omega}_\text{eff} = 0$. Here statistical, parametric and systematic uncertainties have been added in quadrature. Next

$$(\varepsilon'/\varepsilon)_\text{SM} = (14 \pm 5) \times 10^{-4} \tag{13.3.25}$$

from ChPT [4135]. The large error is related to the problematic matching of LD and SD contributions in this approach which can be traced back to the absence of meson evolution in this approach. Finally

$$(\varepsilon'/\varepsilon)_\text{SM} = (5 \pm 2) \cdot 10^{-4}, \tag{13.3.26}$$

from DQCD [4112, 4141, 4142], where $B_6^{(1/2)} \le 1.0$ has been used.

While the results in Eqns. (13.3.24) and (13.3.25) are fully consistent with the data shown in Eqn. (13.3.14), the DQCD result in Eqn. (13.3.26) implies a significant anomaly and NP at work. Clearly, the confirmation of the DQCD result is highly important.

Let us end this presentation with good news. There is a very good agreement between LQCD and DQCD as far as EWP contribution to $\varepsilon'/\varepsilon$ is concerned. This implies that this contribution to $\varepsilon'/\varepsilon$, that is unaffected by leading isospin breaking corrections, is already known within the SM with acceptable accuracy:

$$(\varepsilon'/\varepsilon)_\text{SM}^\text{EWP} = -(7 \pm 1) \times 10^{-4}\,, \quad (\text{LQCD and DQCD}). \tag{13.3.27}$$

Because both LQCD and DQCD can perform much better in the case of EWP than in the case of QCDP I expect that this result will remain with us for the coming years. On the other hand, the value from ChPT of $B_8^{(3/2)} \approx 0.55$ [4135] implies using Eqn. (13.3.20) that the EWP contribution is roughly by a factor of 2 below the result in Eqn. (13.3.27).

Let us hope that at the 60th birthday of QCD we will know which prediction is right. Further summaries can be found in [3963, 3966, 4112] and details in original references.

## 13.4 The role of QCD in $B$ physics anomalies

**Danny van Dyk and Javier Virto**

The so-called $b \to s\ell^+\ell^-$ anomalies present one of the few current tensions between theory predictions within the SM and experimental measurements. They represent long-standing tensions that first presented themselves in a 2013 publication by the LHCb collaboration [4143]. Here, we discuss how QCD plays a central role at every stage of the interpretation of these anomalies.

QCD and hadronic physics enter the theory predictions, both in the SM and beyond, in one of three ways:

- First, they enter the Weak Effective field Theory (WET) description of neutral-current processes, such as $b \to s\ell^+\ell^-$. The effective Hamiltonian at the leading-mass dimension six reads

$$\mathcal{H}_{\text{WET}} = \frac{4G_F}{\sqrt{2}} V_{tb}V_{ts}^* \sum_i \mathcal{C}_i \mathcal{Q}_i \,, \tag{13.4.1}$$

  with local operators $\mathcal{Q}_i$ and Wilson coefficients $\mathcal{C}_i$. It includes semileptonic operators,

$$\mathcal{Q}_{9(10)} = \frac{e^2}{16\pi^2} [\bar{s}\gamma^\mu P_L b] [\bar{\mu}\gamma_\mu(\gamma_5)\mu] \,, \tag{13.4.2}$$

  electromagnetic dipole operators,

$$\mathcal{Q}_7 = \frac{e}{16\pi^2} [\bar{s}\sigma_{\mu\nu} P_R b] F^{\mu\nu} \,, \tag{13.4.3}$$

  and four-quark operators

$$\mathcal{Q}_{1q(2q)} = [\bar{q}\gamma^\mu P_L b] [\bar{s}\gamma_\mu P_L q] \,. \tag{13.4.4}$$

  QCD has a substantial effect on the matching of the WET to the SM [4144–4146]. For instance, at the low scale $\mu_b \simeq 5$ GeV, about half of the value of $\mathcal{C}_9$ is generated by QCD effects due to operator running and mixing of the four-quark operators into $\mathcal{Q}_9$ [4144].
  Here we discuss only the numerically leading operators needed for a description within the SM. BSM effects are encoded in the values of the Wilson coefficients or through additional operators with a different spin structure.
- Second, they enter the hadronic matrix elements of local $\bar{s}b$ operators, c.f. Eq. (13.4.8). These matrix elements are then expressed in terms of scalar-valued form factors, which are functions of the momentum transfer (typically: $q^2$). The $\bar{s}b$ form factors are very similar to the form factors arising in the description of exclusive charged-current semileptonic processes such as $b \to c\mu^-\bar{\nu}$.
- Third, they enter the hadronic form factors of non-local $\bar{s}b$ operators, c.f. Eq. (13.4.10). These operators arise in the time-ordered product of the four-quark operators and the electromagnetic current. They have no correspondence in charged-current semileptonic decays and currently present the biggest obstacle to accurate and precise theoretical predictions of exclusive $b \to s\mu^+\mu^-$ decays.

In the following, we do not further discuss the effective field theory description, which is well established. The matching coefficients to NNLO in QCD can be found in Refs. [4144–4146]. Instead, we focus on the second and third type of QCD effects in exclusive $b \to s\ell^+\ell^-$ processes.

### 13.4.1 Anatomy of exclusive $b \to s\ell^+\ell^-$ processes

**$\bar{B}_s \to \mu^+\mu^-$**

Amongst the exclusive $b \to s\ell^+\ell^-$ decays, the cleanest ones from a theory perspective are the purely leptonic decays $\bar{B}_s \to \ell^+\ell^-$. Up to QED corrections [4147], all QCD effects are contained in a single local hadronic matrix element. This matrix element is commonly parametrised in terms of the $B_q$-meson decay constant $f_{B_q}$ [279]

$$\langle 0|\, \bar{q}\gamma^\mu\gamma_5 b \,|\bar{B}_q(p)\rangle = i f_{B_q} p^\mu \,. \tag{13.4.5}$$

It has been calculated ab-initio from lattice QCD simulations. Several analyses with $N_f = 2+1+1$ light quark flavours have become available [655, 672, 673, 1432, 4148]. Their world average [279]

$$f_{B_s} = 230.3 \pm 1.3\,\text{MeV} \,, \tag{13.4.6}$$

is dominated by a single analysis published by the Fermilab/MILC collaboration [655].

This constant has been computed using a variety of lattice QCD techniques, which have presently reached a precision of 0.5%. The current theoretical uncertainty on the muonic branching ratio is no longer governed by hadronic physics. Instead, it is dominated by CKM matrix elements. The theory predictions have reached the level of 5% [4147], which is much smaller than the uncertainty of the average of the results by the LHC experiments of $\sim 13\%$ [4149]. While $\bar{B}_s \to \mu^+\mu^-$ is not sensitive to the Wilson coefficient $\mathcal{C}_9$ (to leading order in QED [4147]), it does constrain very strongly the scalar and pseudoscalar operators, and indirectly also $\mathcal{C}_{10}$, which has an impact on the global interpretations of the $b \to s\mu^+\mu^-$ anomalies.

### $\bar{B} \to M\mu^+\mu^-$

Amongst the exclusive semileptonic $b \to s\ell^+\ell^-$ decays, $B$-meson decays to either a pseudoscalar ($P$) or a vector ($V$) meson are presently the best understood. Compared to the purely leptonic decay $\bar{B}_s \to \mu^+\mu^-$, the additional meson in the final state provides the opportunity to test the SM through a larger number of observables that arise in the differential decay rates. The downside for is – generally – an increased sensitivity to QCD effects in their theoretical description, which leads to larger theoretical uncertainties.

To leading order in QED, the matrix elements of the semileptonic and radiative operators $\mathcal{Q}_{7,9,10}$ factorise. A useful schematic decomposition of the amplitude is given by [4150]

$$A(\bar{B} \to M\ell^+\ell^-) \sim G_F\, V_{tb}V_{ts}^* \Big[(\mathcal{C}_9\, L_V^\mu + \mathcal{C}_{10}\, L_A^\mu)\, \mathcal{F}^\mu - \frac{L_V^\mu}{q^2} 2im_b\mathcal{C}_7\, \mathcal{F}^{T,\mu} + 16\pi^2\mathcal{H}^\mu\Big] . \quad (13.4.7)$$

Here $L^\mu_{V(A)} = [\bar{\ell}\gamma^\mu(\gamma_5)\ell]$ are leptonic currents, and a generalization to operators beyond the SM can be found in Ref. [4151]. In the above, we use the hadronic matrix elements

$$\mathcal{F}^\mu_{B\to M}(k,q) \equiv \langle M(k)|\, \bar{s}\gamma_\mu P_L\, b\, |\bar{B}(p)\rangle \ , \quad (13.4.8)$$

$$\mathcal{F}^{T,\mu}_{B\to M}(k,q) \equiv \langle M(k)|\, \bar{s}\sigma_{\mu\nu}q^\nu P_R\, b\, |\bar{B}(p)\rangle \ , \quad (13.4.9)$$

$$\mathcal{H}^\mu_{B\to M}(k,q) \equiv i\int d^4x\, e^{iq\cdot x} \langle M(k)|\, T\{j^{\text{em}}_\mu(x), \sum_i \mathcal{C}_i Q_i(0)\}\, |\bar{B}(p)\rangle \quad (13.4.10)$$

with $i = 1q, 2q, \dots$, which arise from the semileptonic, radiative, and four-quark operators in that order

The first two matrix elements are classified as local matrix elements, and the last one as a non-local matrix element. Both types of matrix elements are needed for reliable and accurate predictions of the amplitudes and therefore of the observables in semileptonic decays. For phenomenological discussions, one commonly encounters projections of the hadronic amplitudes onto some basis of scalar form factors, either the helicity basis [4152] or more commonly the transversity basis [4153–4155]. The number of independent amplitudes depends on the angular momentum of the initial and final state hadrons. The form factors are functions of the momentum transfer from the hadronic system to the leptons. This functional dependence is commonly expressed in terms of $q^2$, the squared mass of the lepton pair.

The process $B \to K\ell^+\ell^-$ is the most reliably understood one amongst the exclusive semileptonic $b \to s\ell^+\ell^-$ decays. Both the $B$ and $K$ meson are stable in the absence of weak interactions, which facilitates the determination of their hadronic form factors. Conservation of angular momentum limits this process to two amplitudes: the dominant longitudinally polarized amplitude and the lepton-mass suppressed time-like amplitude [4156]. As a consequence, the process provides only a few independent observables.

The processes $B \to K^*\ell^+\ell^-$ and $B_s \to \phi\ell^+\ell^-$ both feature a vector meson in the final state. Compared to $B \to K\ell^+\ell^-$, two further transversely-polarized amplitudes can contribute. This more complex structure leads to numerous independent observables arising from the differential decay rate [4153–4155, 4157]. However, this enriched phenomenological reach comes at the expense of somewhat larger uncertainties in the individual hadronic form factors. Since both the $K^*$ and $\phi$ are not stable in the absence of weak interactions, their description as a "quasi stable" state incurs additional theoretical uncertainty [4158]. Here, the $K^*$ is substantially more affected than the $\phi$, due to the hierarchy of their hadronic decay widths.

## 13.4.2 Hadronic Matrix Elements

**Local form factors** Local form factors for $B \to K$, $B \to K^*$ and $B_s \to \phi$ transitions are accessible at low values of $q^2 \lesssim 10\,\text{GeV}^2$ [1231] with two different continuum QCD methods.

First, QCD factorisation (QCDF) provides a means to relate the various form factors to each other. This relation emerged from a symmetry amongst currents involving one collinear and one heavy quark field [4159]. The breaking of this symmetry occurs due to two effects: (a) contributions beyond leading order in the strong coupling constant, which involves interactions between the quarks inherent to the transition with the specator quark [4160]; and (b) contributions beyond leading power in the double expansion in the $b$-quark mass and the energy $E$ of the final-state hadron within the $B$-meson rest frame. Early predictions for exclusive $b \to s\ell^+\ell^-$ decays relied heavily on the QCDF relations, to construct so-called "clean" observables; i.e., observables in which local hadronic form factors cancel approximately [4161–4163]. Most famously, the $P_i'$ basis of observables in the $\bar{B} \to K^*\ell^+\ell^-$ angular distribution [4163] makes use of this cancellation. The $P_5'$ ob-servable [4164] is commonly used to illustrate the tensions between SM predictions and measurements.

Second, light-cone QCD sum rules (LCSR) are used to predict the full set of local form factors in $B \to K$, $B \to K^*$ and $B_s \to \phi$ transitions. Two different versions of LCSRs can be employed [1230, 4165], which differ in the choice of the interpolating current. The LCSRs with $B$-meson interpolation involve hadronic matrix elements for the final-state hadron, i.e., the $K$, $K^*$ and $\phi$. These sum rules are presently better understood than their competitors, leading to overall smaller *parametric* uncertainties. However, the sum rules with vector-meson final states suffer from hard-to-quantify systematic uncertainties due to the unstable nature of these state. The competing LCSRs with interpolation of the final-state hadrons $K$, $K^*$, and $\phi$ have not yet reached the same level of sophistication [3987].

It remains to be emphasized that both types of LCSRs suffer from systematic uncertainties that are difficult to assess. It is commonly understood that the LCSR results serve as a stop gap, to be replaced by results from more systematic approaches to QCD.

Lattice QCD provides such a systematic approach to the local form factors. However, it is presently limited to large values of $q^2 \gtrsim 12\,\text{GeV}^2$ [3984, 4166, 4167], which is not an inherent limitation of the method. Lattice QCD results for the decays $B \to K^*\ell^+\ell^-$ and $B_s \to \phi\ell^+\ell^-$, which are of great phenomenological interest are restricted to this range. However, a very recent study of the $B \to K$ form factors [3988] for the first time accesses the full $q^2$ range available to the semileptonic decay. Their results are in good agreement with previous LCSR estimates, with smaller uncertainties.

Having constraints on the form factors at opposite ends of the semileptonic phase space it is natural to ask if these constraints are mutually compatible. This poses an interpolation problem. For $B$-meson decays, this problem is usually addressed using the so-called $z$-expansion [4169]. Using

$$q^2 \mapsto z(q^2; t_0, t_+) \equiv \frac{\sqrt{t_+ - q^2} - \sqrt{t_+ - t_0}}{\sqrt{t_+ - q^2} + \sqrt{t_+ - t_0}} \qquad (13.4.11)$$

the first Riemann sheet of the complex $q^2$ plane is mapped onto the unit disk in $z$. A Taylor expansion of the form-factors in $z$, after removal of any physical poles, converges quickly and provides some control of the interpolation error. Studies of the $B \to V$ form factors find reasonable to good agreement between the available LCSR and lattice QCD results [3987, 4165, 4168], which is not surprising given the large uncertainties attached to the former. An example of such a fit from Ref. [4168] is displayed in Figure 13.4.1, showcasing the agreement between lattice QCD and LCSR results.

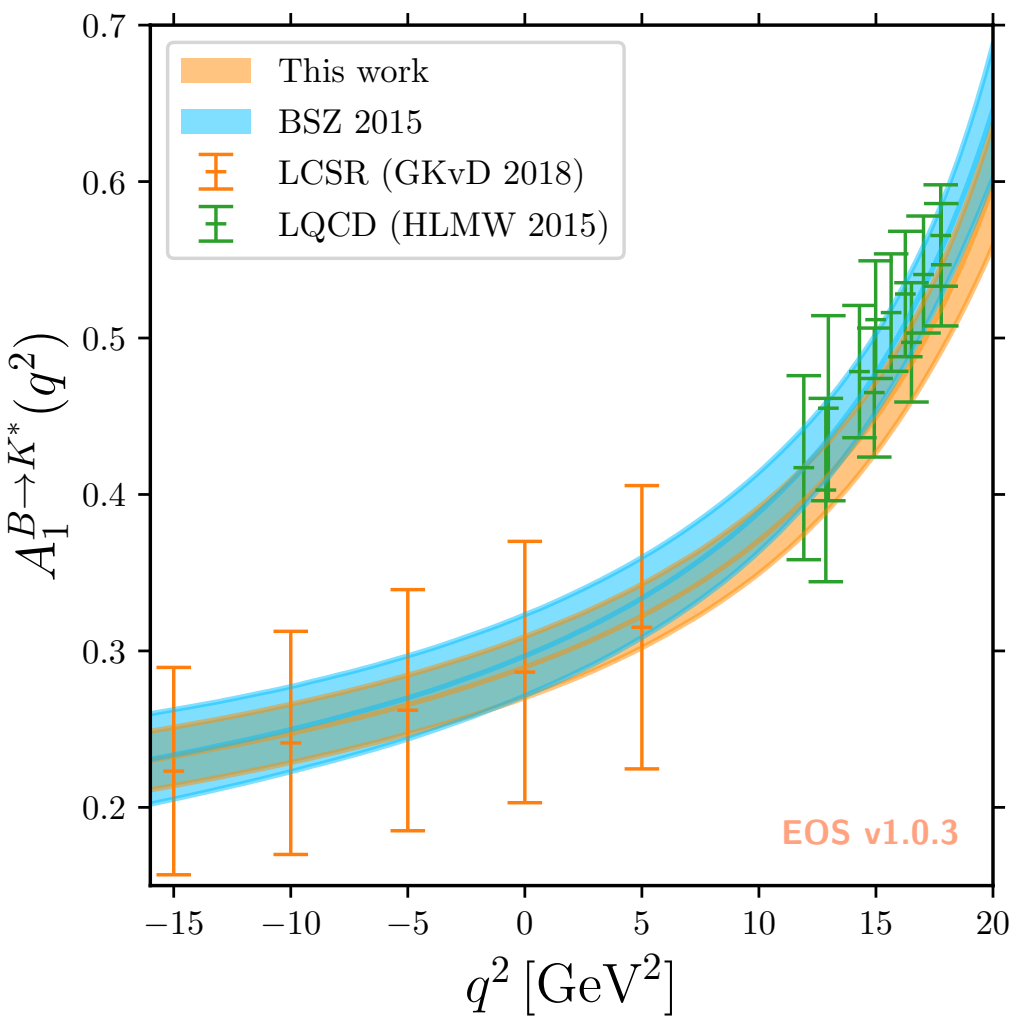

**Fig. 13.4.1** Simultaneous fit to lattice QCD and LCSR results for the local $B \to K^*$ form factor $A_1 \propto \mathcal{F}_\parallel$, taken from Ref. [4168].

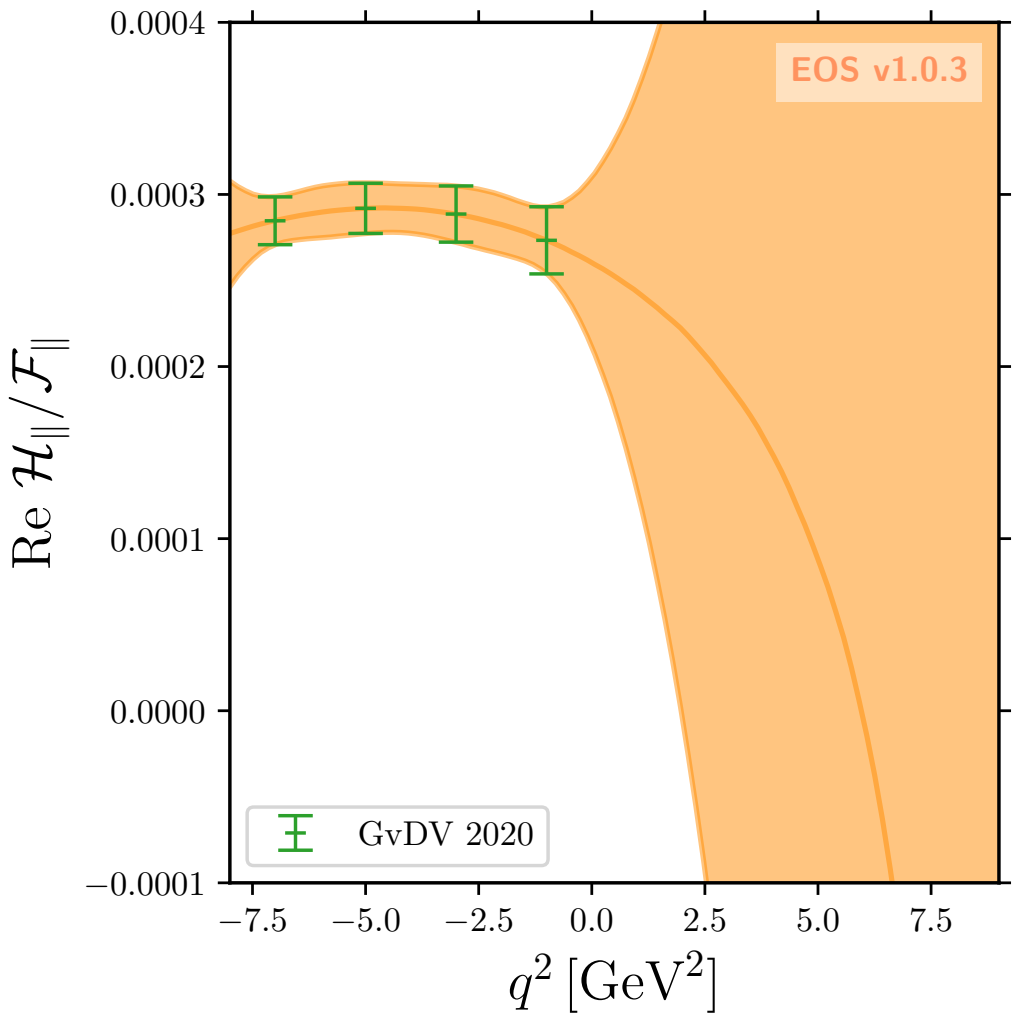

**Fig. 13.4.2** Fit to the non-local $B \to K^*$ form factor $\mathcal{H}_\parallel$, produced from Ref. [4168].

Future prospects on the theoretical precision for local form factors rely dominantly on the expected improvements from the Lattice QCD side. These include enlarging the accessible $q^2$ range (as recently achieved for the $B \to K$ form factors) and accounting for the non-zero width of the vector final states [509]. The effect due to a non-zero $\rho$ and $K^*$ width on the $B \to \pi\pi$ and $B \to K\pi$ form factors was recently critically discussed within the setup of LCSRs with final-state interpolation, estimating corrections to the zero-width limit of up to 10% in the case of the $K^*$ [4158, 4170, 4171].

*Non-local Hadronic Matrix Elements*

Non-local form factors are significantly more difficult to approach theoretically [4172–4175]. The reason is the large number of virtual and on-shell intermediate states that contribute to the time-ordered product in Eq. (13.4.10)s. This non-local operator is commonly separated by the electric charge of the quark flavor to which the electro-magnetic current couples:

$$T\left\{ j_\mu^{\rm em}(x), \sum_{i=1q,2q,\ldots} \mathcal{C}_i Q_i(0) \right\} \equiv \mathcal{K}(x)$$
$$\equiv Q_c\,\mathcal{K}_c(x) + Q_{bs}\,\mathcal{K}_{bs}(x) + \ldots . \qquad (13.4.12)$$

In the above, the dots indicate contributions due to up and down quarks, which are suppressed by CKM matrix element or the small Wilson coefficients of QCD-penguin four-quark operators. The terms proportional to bottom and strange-quark charges are only gauge invariant when considered in sum, leading to the joint description with label $bs$. Our labelling of the non-local form factors follows from the above, i.e., $\mathcal{H}_{\lambda,c}$ arises from the hadronic matrix element of the operator $\mathcal{K}_c$.

The first systematic approach to the non-local form factors has been provided in Ref. [4172, 4176], which is expected to work for small values of $q^2$ sufficiently far below the open charm threshold. This approach was subsequently developed into a light-cone Operator Product Expansion (OPE) of the non-local operator Eq. (13.4.12) [4172, 4173]. This expansion is shown to break down as $q^2$ approaches the partonic open charm threshold from below. The hadronic matrix elements of the next-to-leading operator in this light-cone OPE have been calculated within a LCSR approach [4173, 4177]. The most recent calculation indicated that the term at next-to-leading power is negligible in comparison to the leading-power term.

At $q^2 = \mathcal{O}(m_b^2) \gtrsim 4m_c^2$, an OPE in term of local operators applies [4174, 4175]. The simple structure of the OPE leads to phenomenologically powerful theory predictions [4162, 4178, 4179]. However, the fact that this region of phase space lays on the open-charm branch cut leads to considerable complications in the interpretation of experimental measurements. Chiefly, one cannot expect that the OPE result agrees with nature *locally*, i.e., in every $q^2$ point [4175]. Instead of such local duality, *semi-local* quark-hadron duality is assumed, i.e., the OPE prediction integrated over a sufficiently large $q^2$ range is expected to correspond to the $q^2$ integrated observables [4175]. Nevertheless, this approach gives rise to large unquantifiable systematic uncertainties in the theory predictions [4180, 4181]. Due to these limitations, commonly a single bin covering the whole low-$q^2$ region is used in the BSM analyses. However, the $q^2$ spectrum can be used to test the level of "duality violation", i.e., the disagreement between the perturbative partonic prediction and the hadronic spectrum. In this way, reliable estimates of these intrinsically non-perturbative effects are obtained. Ref. [4181] uses all currently available data on $B \to K^*\mu\mu$ at low recoil and finds agreement between data and the OPE prediction within $\sim 20\%$ in all the bins.

The first parametrizations of the $q^2$ dependence of the non-local form factors $\mathcal{H}_{\lambda,c}$ are based on a dispersion relation [4173] or an expansion in powers of $q^2$ [4152]. A subsequent publication proposes to apply a conformal mapping similar to Eq. (13.4.11) [4182], very similar to what is done for the local form factors. The dispersive and $z$-expansion approaches are consistent with analyticity and therefore permit using additional data, such as measurements of the branching ratios and angular distributions of $B \to \psi M$ processes, were $\psi = \{J/\psi, \psi(2S)\}$. In Ref. [4182] it is shown quantitatively how this information can be used *a priori* to produce data-assisted theory predictions for the non-local effect independent of NP, or *a posteriori* to fit all the $B \to \psi K^*$ and $B \to K^*\mu^+\mu^-$ spectra up to $q^2 = m^2_{\psi(2S)}$ simultaneously to the hadronic parameters and NP. In this last approach, short- and long-distance effects are disentangled by the experimental input from $B \to \psi K^*$, the fixed $q^2$ dependence of the NP contribution, and by the theory constraints at negative $q^2$. A notable byproduct is the fact that experimental data *between* the two narrow charmonia can be used in the analyses. An application of the $z$-expansion, including newly derived dispersive bounds on the expansion coefficients [4177], has been used in Ref. [4168] to challenge the experimental measurements of various exclusive semileptonic $b \to s\ell^+\ell^-$ decays. This parametrization yields results that are compatible with analyses based on a perturbative treatment, albeit with somewhat larger uncertainties. A representative example of the non-local form factors obtained in this way is shown in Figure 13.4.2. The impact of these improved determinations of non-local form factors on the global fits to separate exclusive $b \to s\mu\mu$ modes has been studied in Ref. [4168] and it is shown in Figure 13.4.3. The overall picture of significant tensions between data and the SM expectation seen in the literature [4183–4187] are confirmed.

The prospects for this data-driven approach with the future data from LHCb, including the prospects of doing without theory input altogether, have been studied in [4188]. The conclusion is that unbinned analyses can infer knowledge about both QCD and potential BSM effects in these decays *simultaneously*. The high statistics studies of $b \to s\mu\mu$ exclusive transitions at the LHC, either with fine $q^2$ binning or unbinned, will therefore not only probe for BSM effects but also further our understanding of the non-local form factors. While current global fits to different $q^2$ bins show consistency with the current treatment of non-local effects [4189], future LHC data will require, and provide, a higher level of control over them.

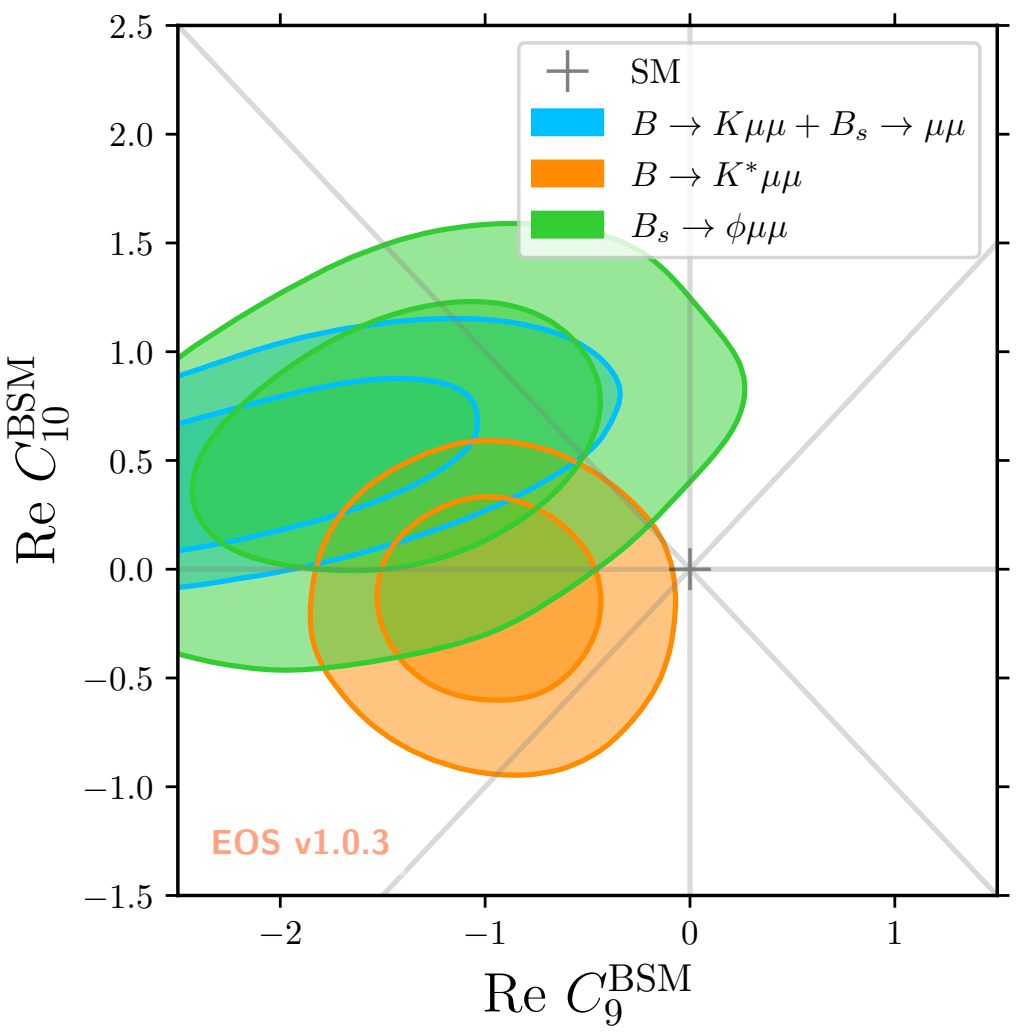

**Fig. 13.4.3** Overview of the tensions between NP parameters and the SM expectation for three representative processes. Taken from Ref. [4168], which takes into account a parametrization of the non-local effects in the fits.

Data-driven and joint theoretical and data-driven methods have been proposed in an effort to control the uncertainties [4177, 4182, 4190–4192]. Some of these methods will be possible and improve significantly with the high statistics collected at LHCb after the upgrade. They are all based on precise measurements of the $q^2$ spectra, together with a theoretically motivated parametrization of the $q^2$ dependence of the amplitudes and a theory benchmark that allows to separate short- from long-distance contributions.

Finally, various hadronic models have been proposed to analyse parts or the entire $q^2$ phase space. Some of these analyses are carried out within the "Krüger-Sehgal" (naive factorization) approach [4193], which allows to use data on the $R(s)$ ratio in $e^+e^-$ annihilation [4175, 4180, 4181]. These models have recently been refined to account also for light-meson intermediate states [4194]. Notably, future precision data from the LHC with the expected fine binning will be essential in refining these data-driven methods and disentangle potential BSM contributions, with the prospects of confirming or refuting a BSM origin to the $b \to s\mu\mu$ anomalies.

## 13.5 QCD and $(g-2)$ of the muon

**Achim Denig and Harvey Meyer**

The anomalous magnetic moment of the muon, as one of the most precisely measured quantities in fundamental physics, has been at the forefront of testing the Standard Model (SM) of particle physics for decades [4195]. The proportionality factor $g \cdot e/(2m)$ between the spin and the magnetic moment of an elementary particle is predicted in Dirac's theory of the electron to satisfy $g = 2$. Already the deviation of the electron's $g$ factor from this prediction played a central role in testing Quantum Electrodynamics at one loop [4196]. It was understood early on [4197, 4198] that the contribution of virtual particles much heavier than the lepton $l$ would be suppressed as $(m_l/m_{\text{heavy}})^2$. Hence the strong interest in the analogous property of the muon, denoted $a_\mu = (g-2)_\mu/2$, given that the 207 times larger mass of the muon strongly enhances the virtual contributions from particles upward from the mass scale of a few MeV$/c^2$, and thus provides access to potential new-physics contributions. Since the very first measurement of 1960 [4199], experiments have refined their sensitivity to $a_\mu$, thereby successively testing contributions from all sectors of the SM, and making this observable the paradigmatic example of searching for new physics at the precision frontier.

The experimental measurements of $a_\mu$ [4200] rely on the muon spin precessing relative to the direction of the muon momentum under the influence of a static magnetic field: the precession frequency is directly proportional to $a_\mu$. The observation that the (undesirable) impact of an electric field on the muon spin precession is suppressed at a special muon momentum of 3.1 GeV/c [4201] eventually led to the third muon storage ring experiment at CERN [4202], which for the first time probed hadronic effects, among which the hadronic vacuum polarization (HVP) provides the leading contribution. Progress in the experimental techniques culminated in the Brookhaven E821 experiment [4203], which achieved a precision of 0.54 ppm on $a_\mu$.

Meanwhile, the SM prediction for $a_\mu$ had been worked out to a very similar degree of precision, as described in the 2009 review [4204]. The QED contribution, by far dominant, and the weak contribution having been calculated to sufficiently high order, the uncertainty of the SM prediction has been entirely dominated by the hadronic contributions, specifically by the HVP and by the hadronic light-by-light (HLbL) contributions, which are both illustrated in Fig. 13.5.1. A tension at the level of 3.2 standard deviations was found between the experimental and the theoretical value of $a_\mu$ [4204].

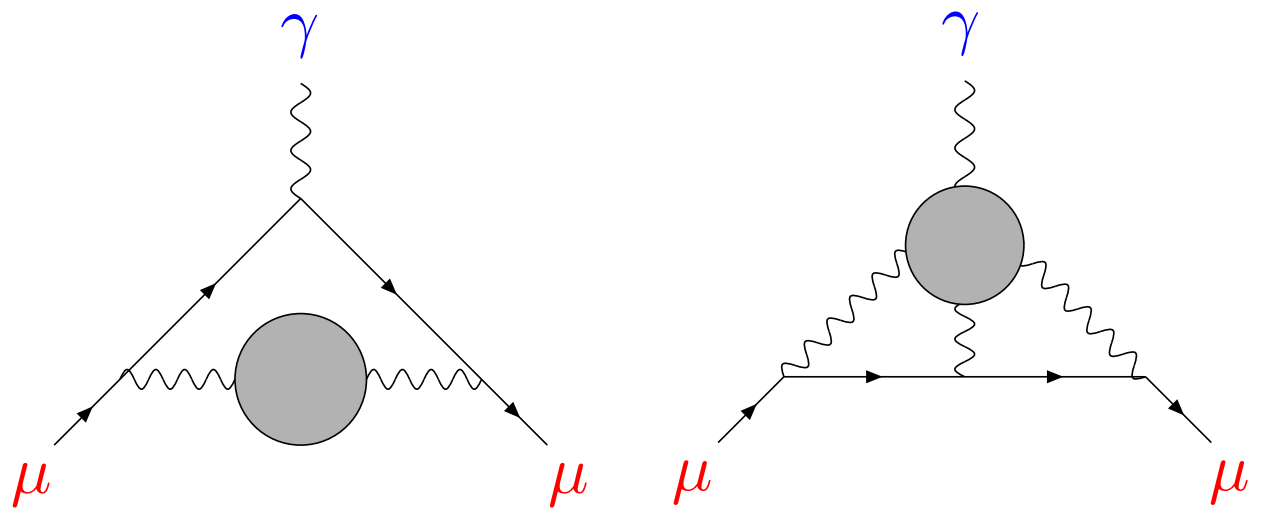

**Fig. 13.5.1** Feynman diagrams representing the two contributions that currently saturate the uncertainty of the SM prediction for the muon $(g-2)$: the hadronic vacuum polarization (left), $a_\mu^{\mathrm{HVP,LO}}$, and the hadronic light-by-light contributions (right), $a_\mu^{\mathrm{HLbL}}$. Solid lines represent muon propagators and wavy lines photon propagators. The external photon line represents the magnetic field of the experiment, which probes the magnetic moment of the muon.

In the past decade, a new experimental effort was undertaken in an attempt to clarify the situation. The Fermilab experiment E989 [4205] was designed with the goal of reaching a precision of 0.14 ppm on $a_\mu$. In order to arrive at an up-to-date prediction before the announcement of the first results by the Fermilab experiment, the $(g-2)$ Theory Initiative was launched in 2017, which led to the 2020 Theory White Paper [4206]. The theory precision had by then improved to the level of 0.37 ppm, and the tension with the world experimental average (dominated by the Brookhaven measurement) was found to be at the $3.7\,\sigma$ level.

The Fermilab $(g-2)$ experiment announced its first result on April 7, 2021. Its measurement of $a_\mu$ [4207] at the 0.46 ppm level slightly surpassed the precision of the Brookhaven measurement [4203] and led to the situation illustrated in Fig. 13.5.2. The new measurement agrees well with the older Brookhaven one, and the tension with the SM prediction (from the 2020 White Paper [4206]) has increased to the level of $4.2\,\sigma$, or

$$a_\mu(\mathrm{Exp}) - a_\mu(\mathrm{WP\ 2020}) = (25.1 \pm 5.9) \times 10^{-10} \quad (13.5.1)$$

in absolute size. From here, it might seem like the next experimental update by the Fermilab experiment could finally raise the tension above the conventional 'discovery' level of five standard deviations.

However, on the same day as the announcement of the experimental result from Fermilab, a lattice QCD calculation of the HVP contribution with a competitive precision was published [4208], which, taken at face value, would increase the SM prediction for $a_\mu$ and bring it into better agreement (at the $1.5\,\sigma$ level) with the experimental world average. Instead, the tension between this lattice QCD calculation and the dispersive, data-driven evaluation underlying the White Paper prediction of $a_\mu$ amounts to $2.1\,\sigma$ (see Eq. (13.5.7) below).

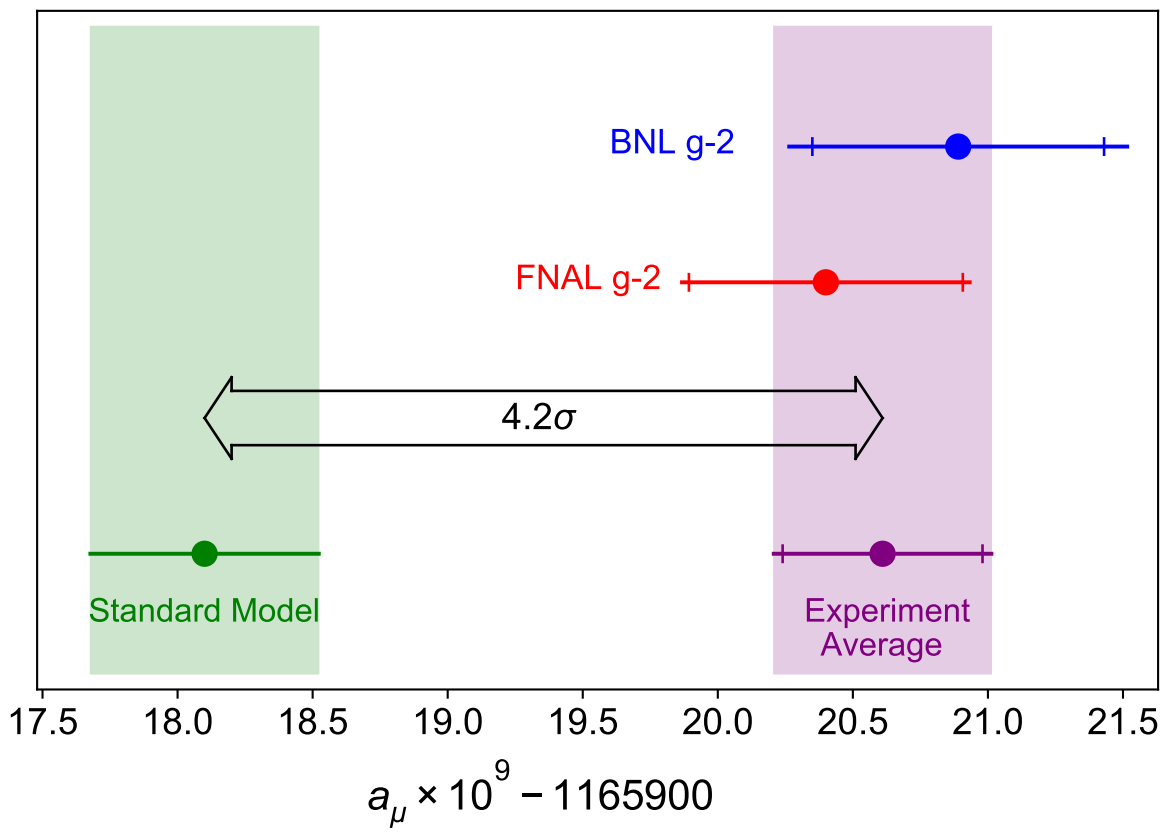

**Fig. 13.5.2** Status of $a_\mu$ after the 2021 FNAL measurement. The tension between the experimental average of the FNAL and the 2001 BNL measurements with the Standard Model prediction provided by the Theory White Paper amounts to 4.2 standard deviations. Figure from [4207].

Thus it is the intricacies of hadron-photon interactions that are currently limiting the resolving power of the muon $(g-2)$ to probe new physics. In section 13.5.1, we describe how the evidence for a genuine difference between lattice calculations of the HVP and its dispersive evaluation has strengthened significantly in the past eighteen months. Obviously, finding the origin of this difference is of utmost importance in the ongoing saga of the muon $(g-2)$.

We begin by reviewing the status of the HVP contribution to $a_\mu$ in section 13.5.1, whereafter we describe the progress made in the HLbL contribution in section 13.5.2. We close with some concluding remarks and an outlook on the near future of the subject.

### 13.5.1 The hadronic vacuum polarization contribution

The leading contribution to $a_\mu$ is given by Schwinger's result $\alpha/(2\pi) \simeq 0.00116$ [4196]. In contrast, the HVP contribution to $a_\mu$ only amounts to about $700 \times 10^{-10}$, but given the precision expected from the ongoing Fermilab experiment and the upcoming J-PARC [4209] experiment, the target for the HVP contribution $a_\mu^{\mathrm{HVP,LO}}$ is a precision of $1.5 \times 10^{-10}$, or 0.2%. This represents a major challenge for a strong-interaction effect, which has been addressed by the long-established data-driven dispersive method and by ab initio lattice QCD methods.

*Dispersive determination*

The dispersive approach to computing $a_\mu^{\rm HVP,LO}$ is based on the expression

$$a_\mu^{\rm HVP,LO} = \left(\frac{\alpha m_\mu}{3\pi}\right)^2 \int_{m_{\pi_0}^2}^{\infty} \frac{ds}{s^2}\, \widehat{K}(s/m_\mu^2)\, R(s), \qquad (13.5.2)$$

$$R(s) = \frac{\sigma(e^+e^- \to \text{hadrons})}{4\pi\alpha(s)^2/(3s)}. \qquad (13.5.3)$$

The dimensionless function $\widehat{K}$ is a smooth function that increases monotonically from the value 0.63 at the $4m_\pi^2$ threshold to unity in the limit $s \to \infty$. The determination of $R(s)$ requires measurements of the hadronic cross section in $e^+e^-$ collisions, $\sigma(e^+e^- \to \text{hadrons})$. Given the $1/s^2$ dependence in the dispersion integrand, low-energy contributions of the hadronic cross section have a very strong weight and therefore have to be known to high accuracy. The most relevant channels are the exclusive reactions $e^+e^- \to \pi^+\pi^-, 3\pi, 4\pi$, and $K\bar{K}$, for all of which the cross section is peaked at $\sqrt{s}$ <2 GeV.

The channel $e^+e^- \to \pi^+\pi^-$ is dominated by the $\rho(770)$ intermediate state and contributes to more than 70% to the dispersion integral. Fig. 13.5.3 shows various recent measurements of the two-pion cross section in the $\rho$ peak region between 600 and 900 MeV. Two classes of measurements are shown in Fig. 13.5.3. These are energy scan measurements (CMD-2 [4210–4213], SND [4214]), in which the center-of-mass energy of the collider (in this case the VEPP-2M collider in Novosibirsk) is systematically varied to cover the energy range under study. A second class of measurements (KLOE [4215], BABAR [4216, 4217], BESIII [4218]) is carried out with the colliders running at a fixed center-of-mass energy and by exploiting events in which the initial beam electrons or positrons have radiated a highly energetic photon, lowering in such a way the available hadronic mass in the final state. This method is called initial-state radiation (ISR) or radiative return and has been applied most successfully at modern particle factories [4219]. In the past, also spectral functions from hadronic $\tau$ decays have been used [4220] in the phenomenological determination of HVP, since these can be related to $R(s)$ via the *Conserved Vector Current* theorem. However, since the phenomenological estimates of the isospin corrections are not well understood, the recent determinations of HVP were obtained without the use of hadronic $\tau$ data.

Fig. 13.5.3 demonstrates the very high precision of the data. However, sizeable discrepancies have been observed for the cross-section integral contributing to Eq. (13.5.2). This is demonstrated in Fig. 13.5.4, where the two-pion contribution to HVP, $a_\mu^{\pi\pi,\rm LO}$, in the $\rho$ peak region between 600 and 900 MeV is shown for the individual experiments as well as for two combinations of the data sets (KNT 19 [4223] and DHMZ 19 [4224]). Especially the two most precise determinations of the two-pion cross section from the KLOE [4215] and BABAR [4216, 4217] collaborations happen to exhibit a significant deviation, which currently limits the overall precision of the dispersive determination of HVP. Furthermore, given the tensions in the experimental data sets, systematic effects have to be considered in the averaging procedures. In Ref. [4206] a conservative merging procedure was applied to reflect the differences between the evaluations in Refs. [4223] and [4224]. The Theory White Paper [4206] estimate for the LO HVP contribution is solely based on the dispersive approach [4223–4228] and reads $a_\mu^{\rm HVP,LO} = (693.1 \pm 4.0) \times 10^{-10}$.

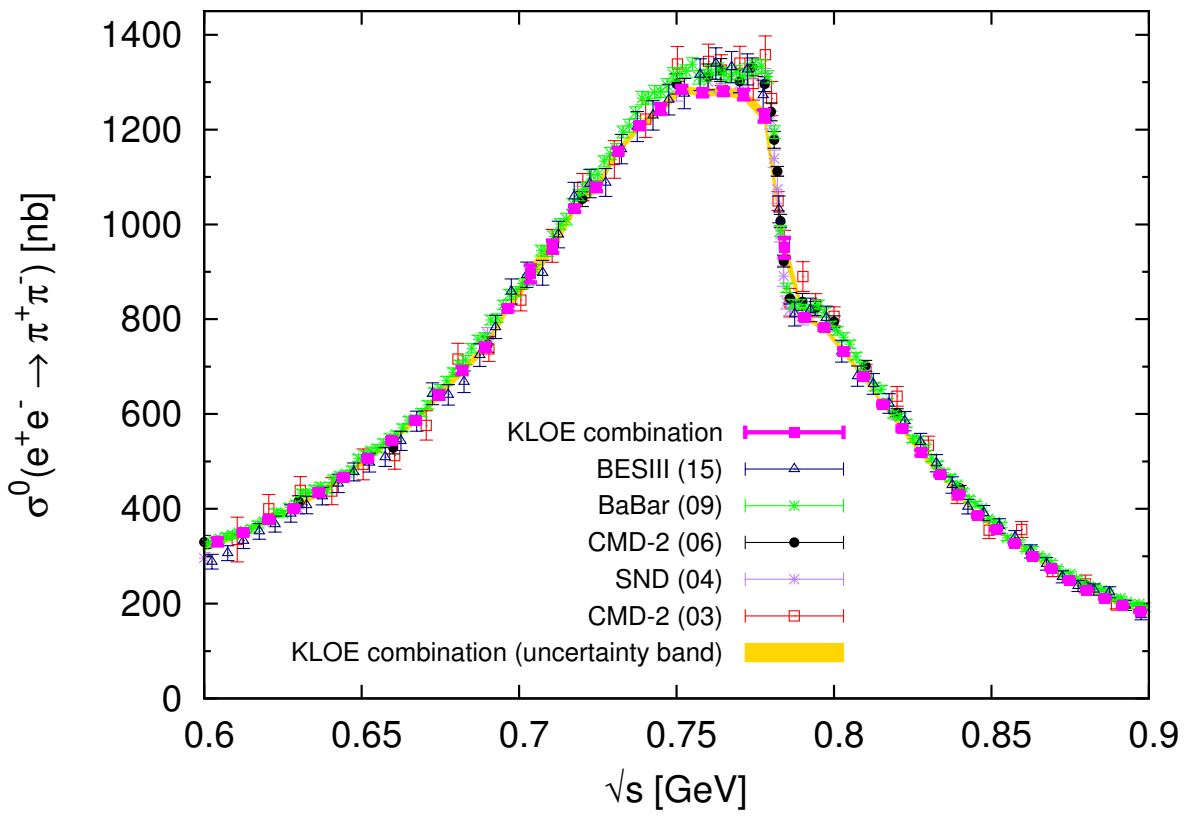

**Fig. 13.5.3** Recent experimental data on the cross section $\sigma(e^+e^- \to \pi^+\pi^-)$ in the energy range between 600 and 900 MeV. The interference of the $\rho$ decay with the two-pion decay of the $\omega(780)$ is well visible as a structure around the $\omega$ mass. Figure taken from [4215]; a new SND analysis [4221] from the VEPP-2000 collider and an ISR analysis from CLEO [4222] are not yet shown.

Fortunately, new experimental measurements of the two-pion channel are expected in the near future by CMD-3, SND, BABAR, BESIII, and BELLE-II. It remains to be seen whether the currently existing discrepancy between BABAR and KLOE can be resolved. Provided the upcoming data sets reach the precision level of 0.5% and agree with each other, the total uncertainty of the HVP contribution obtained via the dispersive approach would decrease from currently 0.6% to 0.3% or better.

*Lattice QCD calculation*

Since the HVP contribution to the muon $(g-2)$ involves only spacelike photons, it is a natural quantity to be calculated in lattice QCD [4232], which is formulated in Euclidean space. Although initially expressed in momentum space, the master formula now used almost exclusively is in the 'time-momentum representa-

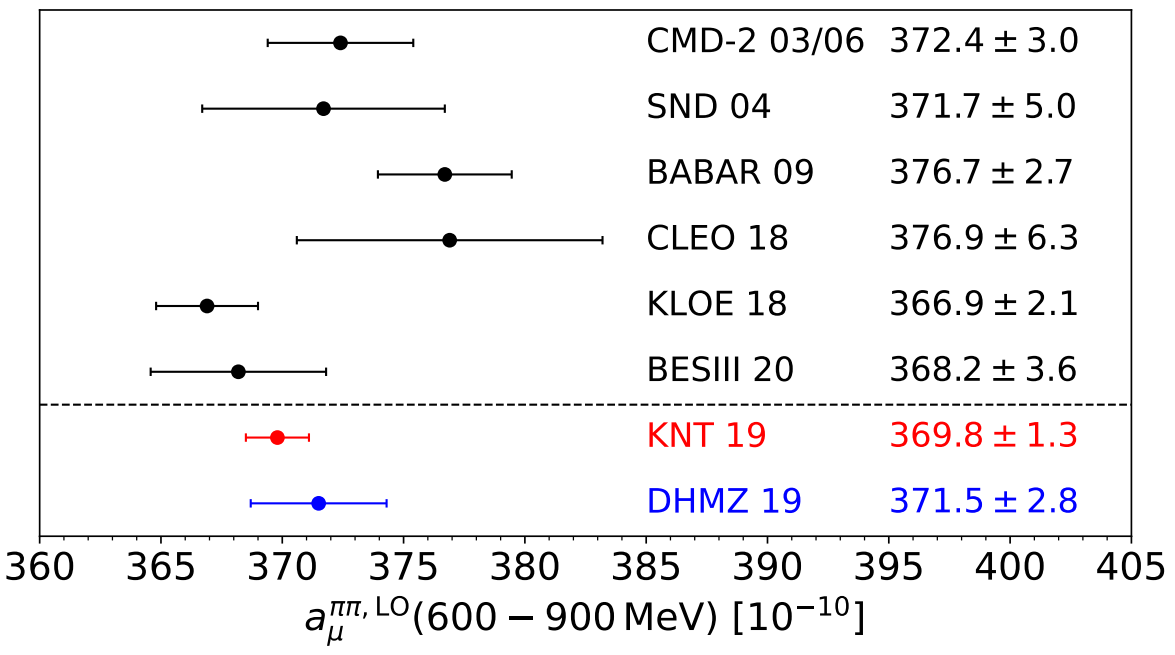

**Fig. 13.5.4** Comparison of $a_\mu^{\pi\pi,\mathrm{LO}}$ in the energy range between 600 and 900 MeV. The upper part of the plot shows the values of recent experimental measurements in this energy range [4210, 4213–4218, 4222], while the lower two values in red and blue are the estimates of the KNT [4206, 4223] and DHMZ [4206, 4224] groups, which carry out a merging procedure of the available data. In the case of DHMZ an additional systematic uncertainty has been included to account for the KLOE/BABAR tension. Please note that the KLOE value is the combination of the three analyses published in Ref.s [4229–4231]

tion' [4233],

$$a_\mu^{\mathrm{HVP,LO}} = \left(\frac{\alpha}{\pi m_\mu}\right)^2 \int_0^\infty dt\; G(t)\;\mathcal{K}(m_\mu t), \qquad (13.5.4)$$

$$G(t) = \frac{1}{3}\sum_{i=k}^{3}\int d^3x\;\langle j_k^{\mathrm{em}}(t,\vec{x})\; j_k^{\mathrm{em}\dagger}(0)\rangle, \qquad (13.5.5)$$

where $j_k^{\mathrm{em}} = \frac{2}{3}\bar{u}\gamma_k u - \frac{1}{3}\bar{d}\gamma_k d - \frac{1}{3}\bar{s}\gamma_k s + \ldots$ is a spatial component of the electromagnetic current carried by the quarks, and the dimensionless weight function $\mathcal{K}(\hat{t})$ is known analytically in terms of Meijer's function [4234]. It is proportional to $\hat{t}^4$ for arguments well below unity, and to $\hat{t}^2$ for arguments well above unity, thus strongly enhancing the long-distance contribution. The spectral representation [4233]

$$G(t) = \int_0^\infty ds\;\frac{s\,R(s)}{12\pi^2}\;\frac{e^{-\sqrt{s}t}}{2\sqrt{s}} \qquad (13.5.6)$$

between the Euclidean correlator and the $R$ ratio allows for detailed comparisons between the dispersive and the lattice approach.

The recipe for computing $a_\mu^{\mathrm{HVP,LO}}$ on the lattice thus appears remarkably simple. However, many effects must be controlled to reach the subpercent level of precision, including discretization and finite-size effects, as well as the leading effects of the unequal up and down quark masses and of the electromagnetic interactions among quarks. The state-of-the-art lattice calculations available at the time of the 2020 White Paper had uncertainties of two percent and larger [4235–4243]. While they had a tendency to lie above the dispersive estimates, they were broadly consistent with them. The BMW collaboration achieved a reduction of the uncertainty of its lattice calculation down to the 0.8% level and published its result in 2021 [4208]. The difference with the White Paper result amounts to

$$a_\mu^{\mathrm{HVP,LO}}(\mathrm{BMW'21}) - a_\mu^{\mathrm{HVP,LO}}(\mathrm{WP'20}) = (14.4 \pm 6.8)\times 10^{-10}. \qquad (13.5.7)$$

At this point, an independent lattice calculation at the same level of precision would be extremely desirable to help clarify the situation.

Both the very short and the very long distances pose distinct challenges to a lattice calculation [4233]. Given the difficulties associated with controlling the statistical and systematic errors of the tail of the correlator $G(t)$, the lattice community has adopted the strategy of partitioning the Euclidean-time axis into intervals, whose contributions to $a_\mu^{\mathrm{HVP,LO}}$ are individually more tractable. This strategy was first applied in Ref. [4237]. In particular, an intermediate interval from 0.4 to 1.0 fm (with smooth edges of width 0.15 fm) was chosen, thus defining the 'window observable', which represents about one third of the total $a_\mu^{\mathrm{HVP,LO}}$. This quantity has received a lot of attention, especially since the BMW collaboration found a discrepancy of 3.7 standard deviations with the dispersive estimate [4208]. Since then, the Mainz/CLS [4244] and the ETM collaboration have computed the window observable on the lattice. The results are summarized in Fig. 13.5.5. The RBC/UKQCD collaboration has recently presented an update [4245] based on a blinded analysis, indicating an upward shift in the (dominant) light-quark connected contribution from $(202.9\pm1.4)\times10^{-10}$ to $(206.5\pm0.7)\times 10^{-10}$ (where we have added their errors in quadrature) and bringing their result into good agreement with the other lattice calculations displayed in Fig. 13.5.5.

*Discussion HVP*

Beyond the 2.1 $\sigma$ tension of Eq. (13.5.7)) between the data-driven evaluation of $a_\mu^{\mathrm{HVP,LO}}$ [4206] and the lattice QCD based BMW calculation [4208], a statistically more significant tension between lattice QCD and dispersion theory has arisen in the partial contribution known as the 'window quantity'. The latter has been computed independently by several lattice collaborations, whose results are in good mutual agreement but disagree with the $R$-ratio based evaluation of [4247], at the level of 3.1, 3.7 and 3.8 $\sigma$ respectively for Refs. [4246], [4208], [4244].

If one assumes that the tension is due to an erroneous cross section measurement in a certain interval of $\sqrt{s}$, it is important to clarify which interval and which hadronic channel it might be. In this regard, we note that the window observable receives a contribution of about 55% from the $\sqrt{s}$ interval between 0.6 and 0.9 GeV, while about 40% come from higher center-of-mass energies [4244]. Its relative sensitivity to the $(\rho,\omega)$-meson region is thus similar to the full $a_\mu^{\mathrm{HVP,LO}}$. If one therefore assumes the $2\pi$ channel to be responsible for the tension, this would require shifts of the $2\pi$ cross section which exceed by far the claimed systematic errors of the experiments as well as the observed discrepancies between the various experiments.

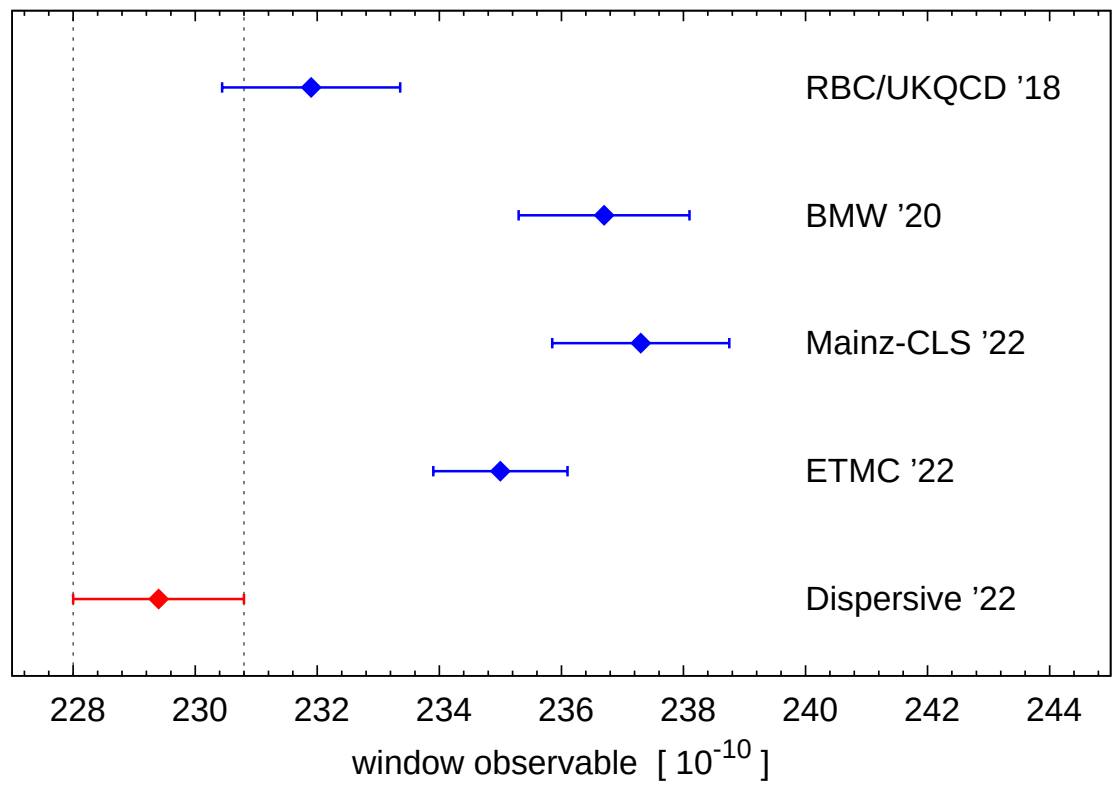

**Fig. 13.5.5** The partial contribution to $a_\mu^{\mathrm{HVP,LO}}$ called 'window observable', as computed by four lattice collaborations [4208, 4237, 4244, 4246], compared to its dispersive determination [4247]. Further recent lattice results, particularly for the (dominant) 'light-quark connected contribution', can be found in [4248–4250] as well as in the update [4245] of the RBC/UKQCD '18 result.

On the other hand, one might ask what could go wrong in the lattice calculations of the window quantity. Perhaps the most critical common source of systematic error among lattice calculations is the one associated with taking the continuum limit. After all, the ranges of lattice spacing used by the different collaborations as well as their fit ansätze in the lattice spacing are fairly similar. Thus, new cross-section measurements as well as additional lattice calculations of the full $a_\mu^{\mathrm{HVP,LO}}$ will give important indications as to the origin of the current tension.

In case of an eventual consolidation of the isospin breaking corrections, e.g. by means of auxiliary lattice QCD calculations [4251], the use of hadronic $\tau$ decays in the HVP dispersion integral might be reconsidered for the future. New and high–statistics measurements of spectral functions of hadronic $\tau$ decays are indeed expected from BELLE-II in the upcoming years. It is going to be exciting to see whether such a $\tau$-based dispersive analysis of HVP will be in agreement with the current $e^+e^-$-based methodology.

### 13.5.2 Hadronic light-by-light scattering in the muon ($g$-2)

The HLbL contribution $a_\mu^{\mathrm{HLbL}}$ is of order $\alpha^3$, and thus of one order higher than $a_\mu^{\mathrm{HVP,LO}}$ in the expansion of $a_\mu$ in the fine-structure constant. The absolute precision target is to reach a level under $1\times10^{-10}$, which given the contribution's approximate size, $a_\mu^{\mathrm{HLbL}} \simeq 10\times10^{-10}$, amounts to a result with a precision under 10%. While this requirement is much less stringent than for $a_\mu^{\mathrm{HVP,LO}}$, the physics and kinematics involved in $a_\mu^{\mathrm{HLbL}}$ are also much more complex. We first review the model and dispersive calculations before describing the status of the lattice QCD approach.

*Data-driven determination*

The hadronic blob on the right-hand side diagram of Fig. 13.5.1 can be decomposed into subgraphs with intermediate pseudoscalar meson exchanges ($\pi^0$,$\eta$, $\eta'$) as well as exchanges of heavier scalar, axial-vector, or tensor mesons. Furthermore, intermediate pion, kaon, and even quark loop exchanges need to be considered. In the past, many of these individual contributions were estimated using hadronic models [4204, 4252–4255], for which an estimate of the model uncertainty is notoriously difficult and for which possible double counting issues have been discussed as an additional source of uncertainty. A consensus exists among all the various estimates that the exchange of pseudoscalar mesons, particularly the $\pi^0$, is the dominant contribution to HLbL. For years, the so called Glasgow consensus value [4256] of $a_\mu^{\mathrm{HLbL}} = (10.5\pm2.6)\cdot10^{-10}$ was considered as a benchmark estimate and was found to be in good agreement with other estimates (see e.g. [4257]), although the individual subgraphs were partly in conflict with each other.

Developing a predictive dispersive representation for the LbL scattering amplitude with three spacelike photons represents a much more complex theoretical task than in the case of the HVP (see Eq. 13.5.2). The recent developments of dispersion relations for the pseudoscalar and the pion-loop subgraphs within the Refs. [4258, 4259] can therefore be considered as a major breakthrough in the analytical treatment of HLbL (see also Ref. [4260] for an alternative representation). Indeed, for the first time an unambiguous definition of individual contributions became possible together with an exact relation to experimental data to be used as input, namely a relation to meson transition form factors (TFFs), which encode the coupling of two virtual photons to mesons. Besides the TFFs, which depend on the two photon virtualities, also meson decays, certain $e^+e^-$ annihilation reactions and Primakoff mea-surements have been found to be highly relevant. As pointed out in Ref. [4261], the most relevant photon virtualities for $a_\mu^{\mathrm{HLbL}}$ are on the GeV scale and below, an observation that calls for a dedicated campaign of experimental measurements in this energy range. The BESIII collaboration has recently presented a new high-quality measurement [4262] of the singly-virtual TFF of the $\pi^0$, which is shown in Fig. 13.5.6, where it is compared with older data [4263, 4264] as well as a calculation of this form factor in lattice QCD [4265], a phenomenological estimate based on Canterbury approximants [4266], and with a dispersive treatment of the TFF [4267]. The agreement between data and theory is very good. Unfortunately, at low energies experiments have not been able yet to provide data with two photon virtualities, as needed for the new dispersive treatment of the pseudoscalar and pion loop contributions. Dispersive evaluations of the TFFs [4268] and lattice QCD calculations [4265] have been used instead. The good agreement shown in Fig. 13.5.6 and the overall consistency found elsewhere indicate the robustness of the theoretical descriptions of the TFFs. For the future, the first double-virtual TFF measurements are expected from BELLE-II and BESIII.

Currently, in the Theory White Paper, the new dispersive treatments have led to a major reduction of the uncertainties of the pseudoscalar exchanges and pion and kaon loop subgraphs. For the remaining scalar, axial vector, and tensor exchange graphs as well as the short-distance contributions, a conservative error estimate has been applied and future research in experiment and theory will eventually lead to a further reduction of the uncertainty of those contributions. The dispersive result arrived at in Ref. [4206] amounts to $a_\mu^{\mathrm{HLbL}} = (9.2 \pm 1.9) \cdot 10^{-10}$ [4195, 4265–4267, 4269–4277] and is found to be in good agreement with the Glasgow consensus value with a slightly reduced uncertainty, but with a significant reduction of the model dependence compared to this older value.

*Lattice QCD calculation*

The first proposal for computing the hadronic light-by-light contribution in lattice QCD dates back to 2005 [4278]. The subject lay dormant for some years until 2013 [4279], the new effort leading to first results for the quark-connected contribution at a pion mass of 330 MeV/$c^2$ [4280]. Important technical improvements to the original methods were made in [4281]. The leading disconnected contribution was calculated for the first time in [4282], along with the connected part, at the physical pion mass. Finally, this multi-year effort culminated into a full calculation [4283] in the $(u, d, s)$ quark sector. This result, displayed in Fig. 13.5.7 as RBC/UKQCD '18, contributed to the White Paper 2020 theory average, together with the dispersive estimate quoted above.

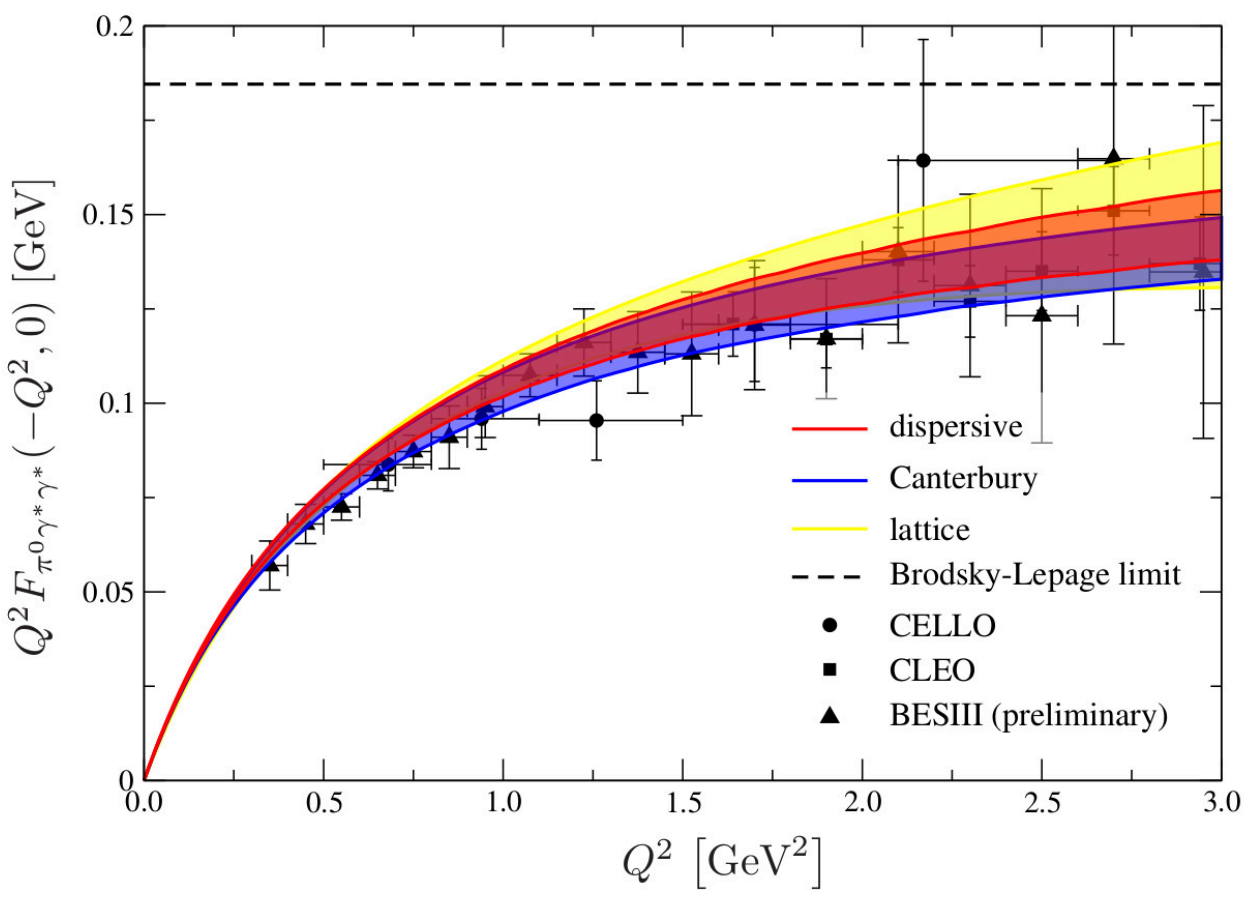

**Fig. 13.5.6** The single-virtual pion form factor $F_{\pi^0\gamma^*\gamma^*(-Q^2,0)}$ as a function of $Q^2$ measured by the CELLO [4264], CLEO [4263], and BESIII [4262] experiments as well as phenomenological predictions using a dispersive analysis [4267] and Canterbury approximants [4266]; shown is furthermore an ab-initio calculation within Lattice QCD [4265].

The treatment of massless internal photons is an important technical issue in lattice QCD. In the publications cited in the previous paragraph, the photons were treated on the same lattice as the QCD degrees of freedom. In [4284–4286], a position-space method allowing for the photons to be treated in infinite volume was proposed and worked out. Meanwhile, similar methods were also developed by members of the RBC/UKQCD collaboration [4287]. Altogether, the development of optimized position-space methods led to the calculations of [4288–4290] by the Mainz-CLS group. The result, displayed in Fig. 13.5.7, has an uncertainty very similar to the dispersive result.

*Discussion HLbL*

Fig. 13.5.7 illustrates the good consistency among the data-driven, lattice and earlier hadronic model determinations. This is a good sign, since the dominant sources of uncertainty are very different in the different determinations: for instance, the RBC/UKQCD calculation involves a fairly long extrapolation to infinite volume, while the Mainz-CLS determination results from an extrapolation over a sizeable interval of pion masses. Updates of the lattice calculations are planned in the near future.

In the dispersive data-driven approach, further progress can be achieved by improved TFF measurements and calculations for the $\eta$ and $\eta'$ mesons. Most important, however, is a future experimental program of mea-surements of the two-photon couplings of mesons in the (1-2) GeV/$c^2$ range, where especially axial vector mesons play an important role and for which the current data base is limited. New results are expected in the future by the BESIII collaboration in a range of momentum transfer similar to the one shown in Fig. 13.5.6. Moreover, also BABAR and BELLE-II will be able to provide new measurements at a higher momentum transfer. New TFF data will also be crucial for a matching of individual hadronic channels to the short-distance behaviour of HLbL.

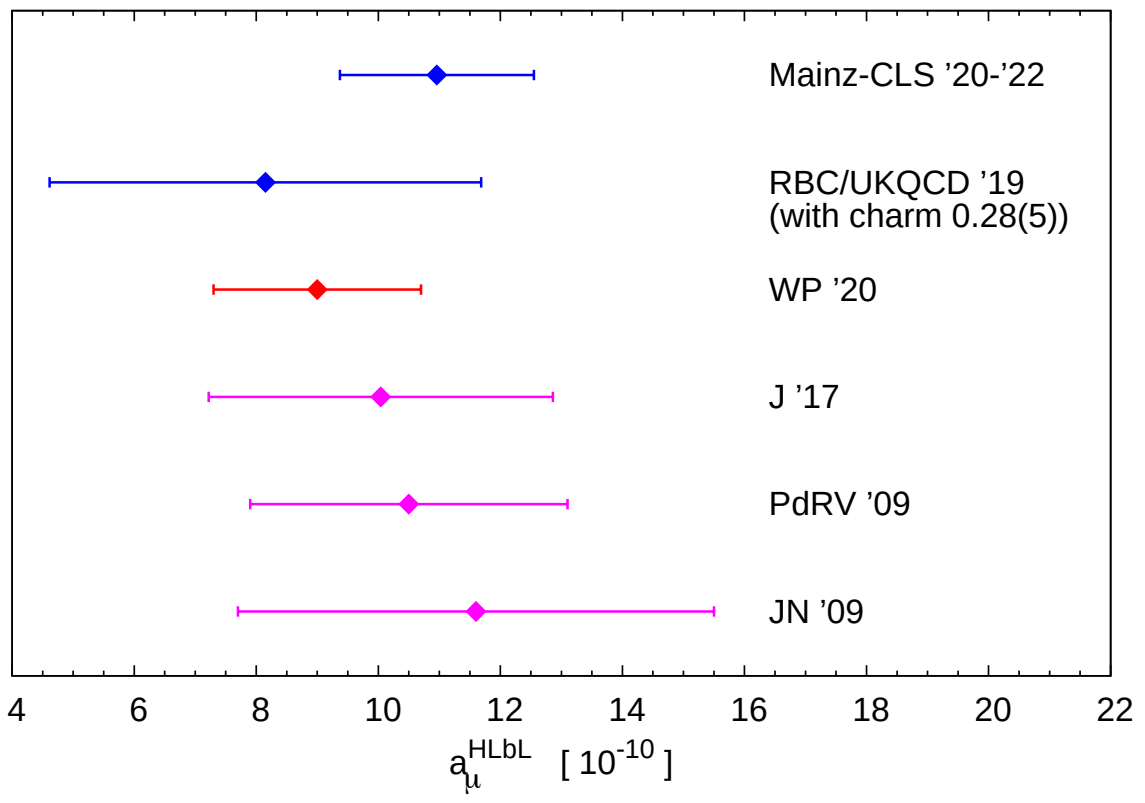

**Fig. 13.5.7** Overview of results obtained for the hadronic light-by-light contribution to the muon $(g-2)$: the Mainz-CLS [4289, 4290] and RBC/UKQCD lattice results [4283], the Theory White Paper 2020 average [4206], and previous model estimates by Jegerlehner [4195], Prades–de Rafael–Vainshtein [4256] (the 'Glasgow consensus') and Jegerlehner–Nyffeler [4204, 4291]. We have supplemented the RBC/UKQCD result with the charm contribution computed in [4290]. The WP average is based on the dispersive [4195, 4265–4267, 4269–4277] and the RBC/UKQCD [4283] lattice result.

Given the ongoing program of various groups in experiment, hadron phenomenology and lattice QCD, we expect an improvement of the HLbL error from currently 20% to 10% or lower. An agreement between an ab-initio lattice QCD calculation with a data-driven estimate on such a level will represent a non-trivial cross-check between two completely independent methods.

### 13.5.3 Conclusions and Outlook

Many theoretical and experimental developments have taken place in the past five years on the anomalous magnetic moment of the muon $a_\mu$. The direct measurement of $a_\mu$ [4203] has been confirmed and improved [4207], while the $(g-2)$ Theory Initiative has helped coordinate many activities to improve the Standard Model prediction for $a_\mu$ [4206]. Hadronic effects limit the precision of this prediction, especially the hadronic vacuum polarization (HVP) and the hadronic light-by-light (HLbL) contributions reviewed above.

In the immediate future, the top priority is to clarify the tensions that have emerged in partial and full HVP determinations. Additional lattice QCD calculations of the full $a_\mu^{\mathrm{HVP,LO}}$ contribution are eagerly awaited, in conjunction with a strategy to identify the origin of the existing strong tension with the dispersive approach for the 'intermediate window' subcontribution. On the data-driven side, the accuracy of the dispersive approach for obtaining $a_\mu^{\mathrm{HVP,LO}}$ is currently hampered by inconsistencies in the experimental data bases. The most problematic issues arise from the tension in the determination of the $e^+e^- \to \pi^+\pi^-$ cross section (KLOE/BABAR puzzle), but also in other exclusive channels, e.g. in the process $e^+e^- \to K^+K^-$, inconsistencies have been observed. The clarification of these issues is one of the most important challenges for an improved determination of the SM prediction of $(g-2)_\mu$ and will be addressed by several existing and upcoming $e^+e^-$ experiments in future. In that respect, since the cross section measurements heavily rely on high-precision Monte-Carlo generators [4292], it is of utmost importance to maintain and to refine the PHOKHARA [4293–4310] generator as well as other Monte Carlo programs [4311–4316] for future applications.

As an alternative to the program of hadronic cross section measurements at $e^+e^-$ colliders, it has been proposed [4317] to carry out a spacelike measurement of the effective electromagnetic coupling via a scattering experiment providing thereby input to a dispersion integral for HVP. The MUonE collaboration is currently preparing the design of a detector [4318] at the muon beam of SPS/CERN towards the final approval of the project. Provided that the differential cross section of the $\mu e$ scattering process can be measured to the desired accuracy, this will allow for an entirely new determination of HVP.

In summary, controlling hadronic effects in the muon $(g-2)$ to match the absolute experimental precision represents a major challenge. Overcoming this challenge will demonstrate that strong-interaction contributions to precision observables can be controlled with the required level of accuracy and consistency between data-driven and lattice QCD approaches. This ability will be crucial to maximize the science output of a future lepton collider.

# 14 The future

*Conveners:*
**Franz Gross and Eberhard Klempt**

Higher energy, higher intensity, higher precision. These are the frontiers at which New Physics beyond the Standard Model is expected. This last section of this volume describes the status and the prospects at new facilities which recently came into operation or which are presently under construction.

The 12 GeV project at JLab, presented by Patrizia Rossi, is dedicated to a study of the structure of nucleons and nuclei, to an intense search for gluonic degrees of freedom in meson and baryon spectroscopy, to a search for new physics in parity violating processes, and to a search for dark matter. The electron-ion collider (EIC) will provide electron-proton and electron-nuclei collisions at CM energies $\sqrt{s} = 20$–$100$ GeV, later possibly up to 140 GeV. Global properties and the partonic structure of hadrons and nuclei will be studied (Christian Weiss). The study of in-medium properties of hadrons and the nuclear matter Equation of State (EoS) and a search for possible signals of a deconfinement and a chiral-symmetry-restoration phase transitions are at the heart of the NICA (Nuclotron-based Ion Collider fAcility) program at the Joint Institute for Nuclear Research in Dubna and of the J-PARC hadron facility at Tokai. NICA provides beams of nuclei with 4.5 GeV per nucleon and protons up to 12.6 GeV. Using polarized beams, the internal structure of the proton and deuteron will also be studied (Alexey Guskov). Strange nuclear matter, hypernuclei and the study of hyperons are further research areas (Shinzo Kumano).

The $e^+e^-$ colliders in Beijing and Tsukuba had delivered a large number of unexpected results. BES III will increase further the statistics of $J\psi$ from now $10^{10}$ and $\psi(2S)$ $(2.7 \cdot 10^9)$ decays and extend its program to cover the full range up to 5.6 GeV in mass. Meson and baryon spectroscopy form the core of the program with extensions to mesonic and baryonic form factors and to $\tau$ decays (Hai-Bo Li, Ryan Edward Mitchell and Xiaorong Zhou). The BELLE II program, presented by Toru Iijima, has a strong part in spectroscopy as well. The experiment operates at an asymmetric $e^+e^-$ collider mostly at the $\Upsilon(4S)$ mass. In addition to the spectroscopy program, BELLE III will search for non-SM contributions in hadronic, semileptonic and leptonic $b$-quark decays, determine quark mixing parameters, determine parameters in $\tau$ physics to precisions and perform searches for dark-sector particles. The new international Facility for Antiproton and Ion Research (FAIR), presently under construction at Darmstadt, is presented by Johan Messchendorp, Frank Nerling and Joachim Stroth. Its program encompasses hadron physics using anti-proton annihilation, heavy-ion reactions at relativistic energies, and nuclear structure physics at the limit of stability using rare isotope beams. The High-Luminosity Large Hadron Collider (HL-LHC) will have a five time larger luminosity than LHC. Major goals are improved tests of the Standard Model, searches for beyond the Standard Model (BSM) physics, studies of the properties of the Higgs boson, flavor physics of heavy quarks and leptons, and studies of QCD matter at high density and temperature. Project and prospects of HL-LHC are summarized by Massimiliana Grazzini and Tim Gershon.

## 14.1 JLab: the 12 GeV project and beyond

**Patrizia Rossi**

### 14.1.1 Jefferson Lab and CEBAF

Jefferson Lab (JLab), is a US National Lab located in Newport News - Virginia. It is a world-leading research laboratory for exploring the nature of matter in depth, providing unprecedented insight into the details of the particles and forces that build our visible universe inside the nucleus of the atom. Its scientific program spans the study of hadronic physics, the physics of complex nuclei, the hadronization of colored constituents, and precision tests of the Standard Model of particle physics. Fig. 14.1.1 shows an areal view of the laboratory with the accelerator complex in the foreground. The core of Jefferson Lab is the Continuous Electron Beam Accelerator Facility (CEBAF). It operates as a pair of superconducting radio frequency linear accelerators (linacs) in a "racetrack" configuration and is designed to circulate a near continuous-wave electron beam through one to five passes recirculating arcs (see Fig. 14.1.2).

Jefferson Lab started physics operations in 1995, providing up to 6-GeV electron beams to three experimental halls, Halls A, B and C, simultaneously. In May 2012, the 6-GeV beam operations were stopped, with Jefferson Lab upgrading its facility to expand opportunity for discovery. In addition to the accelerator scope of doubling the energy, from 6 GeV to 12 GeV, the upgrade included the addition of a new fourth experimental hall, Hall D, and the construction of upgraded/new detectors hardware in the other halls. In two of the existing halls new spectrometers were added, the large acceptance-device CLAS12 in Hall B [4319] and the precision magnetic spectrometer Super High Momentum Spectrometer, or SHMS, in Hall C. The new experimental Hall D makes use of a tagged bremsstrahlung photon beam and solenoidal detector to house the GlueX experiment.The initial energy upgraded program in Hall A made use of both the existing High Resolution Spectrometers.

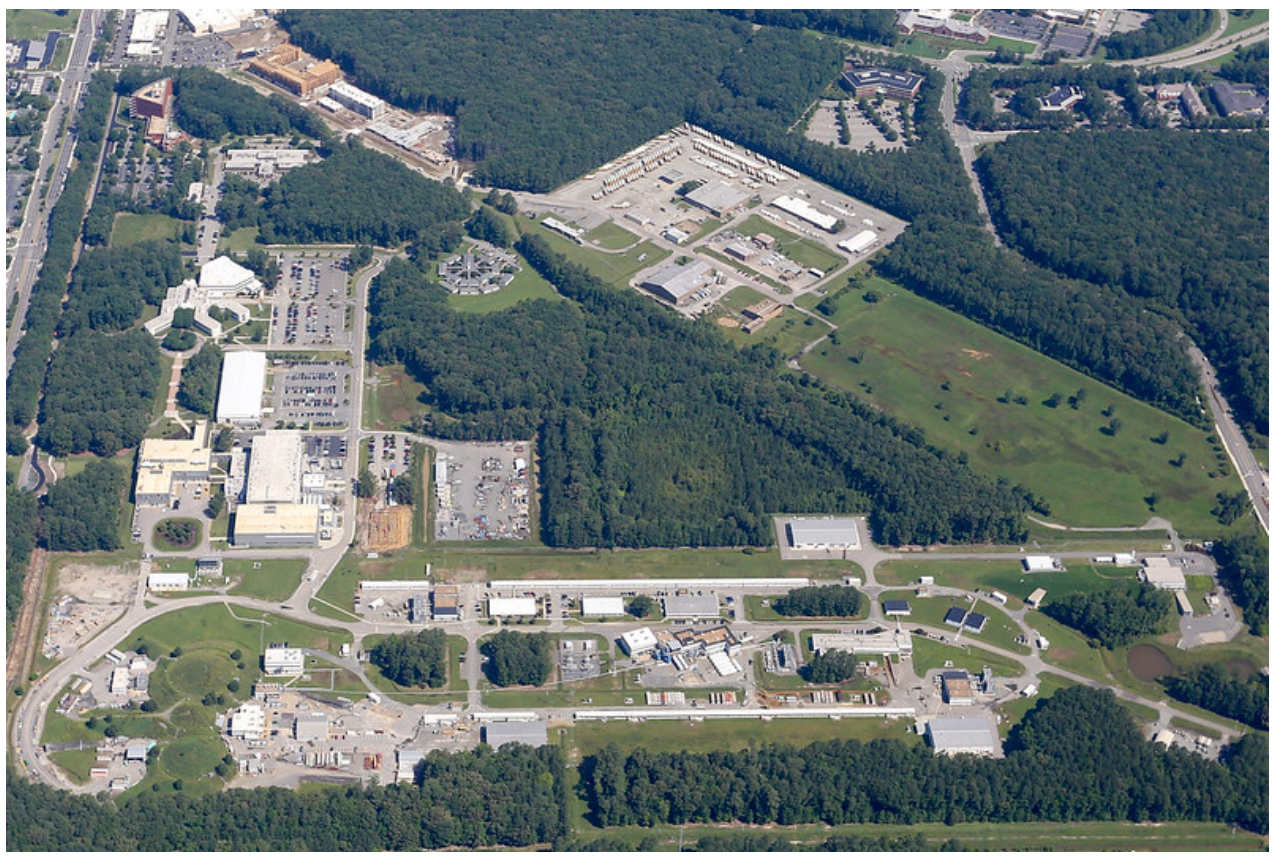

**Fig. 14.1.1** Areal view of Jefferson Lab with the accelerator complex in the foreground.

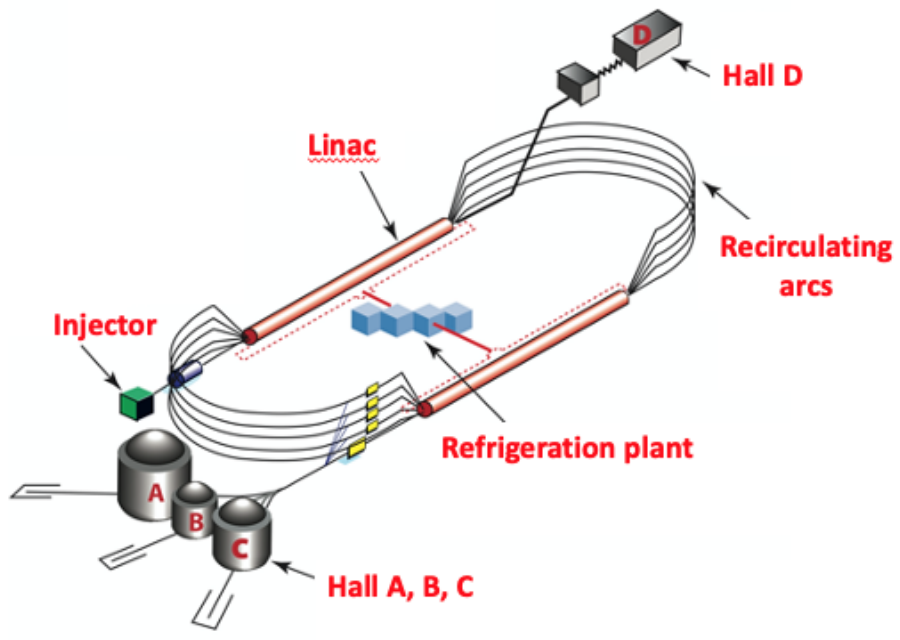

**Fig. 14.1.2** CEBAF accelerator concept.

The equipment in the four halls is well matched to the demands of the broad 12 GeV scientific program [3127] with complementary capabilities of acceptance, precision and required luminosity: high luminosity in Halls A and C and large acceptance detectors in Halls B and D. The upgraded CEBAF accelerator, which can deliver a maximum energy of 12 GeV to Hall D and 11 GeV to Halls A, B, C, delivered the first beam to Halls A and D in the spring of 2014. The full project was completed in spring 2017 with the commissioning of the two remainig halls.

In the meantime, Jefferson Lab has been continuing actively to invest in facilities that make optimum use of CEBAF's capabilities and the existing equipment, to produce science with high impact in Nuclear Physics as well as High Energy Physics and Astrophisics. In Hall A the Super Big Bite spectrometer (SBS) was installed in 2021, while the Measurement Of Lepton-Lepton Elastic Reaction (MOLLER) equipment is under construction with completion date foreseen for late 2026. On a longer term, Hall A plans to host the SOLenoidal Large Intensity Device (SoLID). Future additions include also: new large angle tagging detectors (TDIS in Hall A and ALERT in Hall B); the neutral particle spectrometer (NPS) and the compact photon source (CPS) in Hall C; and an intense $K_L$ beamline that would serve new experiments in the GlueX spectrometer in Hall D.

### 14.1.2 The 12 GeV Physics Program

CEBAF has been delivering the world's highest intensity and highest precision CW multi-GeV electron beams for more than 25 years. The capabilities of the upgraded CEBAF represent a significant leap over previous technology, with an unmatched combination of beam energy, quality and intensity. At Jefferson Lab experiments can run at luminosity up to $10^{38}$ cm$^{-2}$s$^{-1}$ using a highly polarized electron beam (up to 90%), high power cryogenic targets, and several polarized targets using $NH_3$, $ND_3$, and $^3He$ to support a broad range of polarization measurements. This combination of beam, targets and large acceptance and high precision detectors, offers a powerful set of experimental tools that enables unprecedented studies of the inner structure of nucleons and nuclei and allows to push the limits of our understanding of the Standard Model.

The facility serves an international scientific user community of $\sim$ 1700 scientists which, in collaboration with the laboratory and with the guidance of the Jefferson Lab Program Advisory Committee (PAC), develops the scientific program. Following the last PAC meeting in 2022, there are a total of 90 approved experiments in the 12 GeV program [117], of which more than 1/3 have received the highest scientific rating of A. There are 61 approved experiments still waiting to run, representing at least a decade of running in the future. Furthermore, PAC meetings are expected to continue each summer, with a call for new proposals for beam time. Clearly, CEBAF is a facility in high demand.

The JLab physics program falls into four main categories:

- the study of the transverse, longitudinal and 3-dimensional structure of the nucleon through the measurements of the elastic and transition form factors (FFs), the (un)polarized parton distribution functions (PDFs), and the Transverse Momentum Dependent (TMDs) and Generalized Parton Distributions functions (GPDs), respectively.

[117] A list of approved experiments is available on the JLab website.

- The study of hadron spectroscopy and the search for exotic mesons to explore the nature of confinement.
- The study of the QCD structure in nuclei; its connection with the nucleon-nucleon interactions, including the modification of the valence quark PDFs in a dense nuclear medium, and the investigation of the quark hadronization properties. The neutron distribution radius in medium heavy nuclei, is also part of the program.
- The search of physics beyond the Standard Model in high-precision parity-violating processes and in the search for signals of dark matter.

Due to the limited space, only few selected highlights of the scientific agenda and present results of the JLab 12 GeV rich program are presented in this review. Some key results of the earlier JLab 6 GeV program are also reported for completeness when needed. The part related to the search of physics beyond the Standard Model,instead, are not discussed since it is somewhat beyond the scope of this volume. A more complete summary of the ongoing scientific program of the 12 GeV CEBAF and an outlook into future opportunities can be found in Ref. [4320].

#### 14.1.3 *The structure of the nucleon.*

For the theoretical formalism and a general overview of the structure of the nucleon, the reader should refer to Sec. 10 of this volume.

*Elastic Form Factors at high and ultra low $Q^2$*
Since Hofstadter's pioneering experiment in the '50s, the measurements of the electromagnetic space-like nucleon FFs have been a crucial source of information for our understanding of the internal structure of the nucleons. In 2000 Jefferson Lab rewrote the textbook of the proton and neutron form factors when precise data for the proton's electric to magnetic form factor ratio, $G_E^p/G_M^p$ from double polarization experiments at $Q^2$ up to 5.6 $GeV^2$[2916], didn't show the scaling behavior observed using the Rosenbluth separation method and subsequently confirmed by experiments with improved precision [2914],[4321]. According to the pQCD predictions the ratio $Q^2\frac{F_{2p}}{F_{1p}}$, where $F_{1p}$ and $F_{2p}$ are the Dirac and Pauli form factors, respectively, would reach a constant value at high enough $Q^2$. The data clearly indicate that this asympvtotic regime has not been reached yet.[2917]. These observations suggest the presence of orbital angular momentum in the leading 3-quark component of the nucleon wave function in QCD Ref. [2983]. Another explanation of this discrepancy has been attributed to "two–photon" exchange (TPE) or higher order corrections to the cross sections. Jefferson Lab is tackling these questions and in the coming years will offer unprecedented opportunities to extend the current proton and neutron FF's measurements to higher momentum transfer $Q^2$ and to improve statistical and uncertainties at very low $Q^2$, where the nucleon size can be accurately investigated. The measurements at high $Q^2$ will also contribute to constraint two of the nucleon Generalized Parton Distributions, and in general will test the validity of quite a few fundamental nucleon models in a region of transition between perturbative and non-perturbative regimes.

One of the first completed experiments in Hall A with the upgraded CEBAF accelerator was a precision measurement of the proton magnetic form factor up to $Q^2$ = 16 $GeV^2$[2904]. This experiment nearly doubled the $Q^2$ range over which direct Rosenbluth separations of $G_E$ and $G_M$ can be performed. It confirmed the discrepancy with polarization measurements to larger $Q^2$ values and attributed it to hard TPE. These new, high-precision cross section measurement provides also an important baseline for the nucleon form factors program.

A series of experiments [4322–4327] for the measurements of the proton and neutron magnetic and electric form factors, has started at the end of 2021 using the Super Bigbite Spectrometer (SBS) and the upgraded BigBite Spectrometer in Hall A. This facility provides large acceptance at high luminosity so that small cross sections can be measured with high precision allowing a determination of the flavor separated form factors to $Q^2$ = 10-12 $GeV^2$. A complementary measurement of the neutron magnetic form factor will be performed with CLAS12 in Hall B [4328]. The SBS form factor experiments will push into a $Q^2$ regions in which theory expects new degrees of freedom to emerge in our understanding of QCD non-perturbative phenomena in nucleon structure as predicted in Ref. [2983].

From the perspective of QCD in exclusive processes, another important measurement is accessing the structure of the pion and kaon. The E12-06-101 experiment [4329] in Hall C will extract the pion form factor through $p(e, e'\pi^+)n$ and $d(e, e'\pi^-)pp$ with $Q^2$ extending to 6 GeV$^2$ from 2 GeV$^2$ and $-t_{\rm min} \sim 0.005 \sim 0.2$ GeV$^2$. The proposed separation of longitudinal and transverse structure functions is a critical check of the reaction dynamics. The charged pion electric form factor is a topic of fundamental importance to our understanding of hadronic structure. There is a robust pQCD prediction in the asymptotic limit where $Q^2 \to \infty$: $Q^2F_\pi(Q^2) \to 16\pi\alpha_s(Q^2)f_\pi^2$. Therefore it is an interesting question at what $Q^2$ this pQCD result will become dominant. The available data indicate that the form factor at $Q^2$ = 2 GeV is at least a factor of 3-4 larger. The new data will provide improved understanding of the non-perturbative contribution to this important property of the pion as well as mapping out the transition to the perturbative regime.

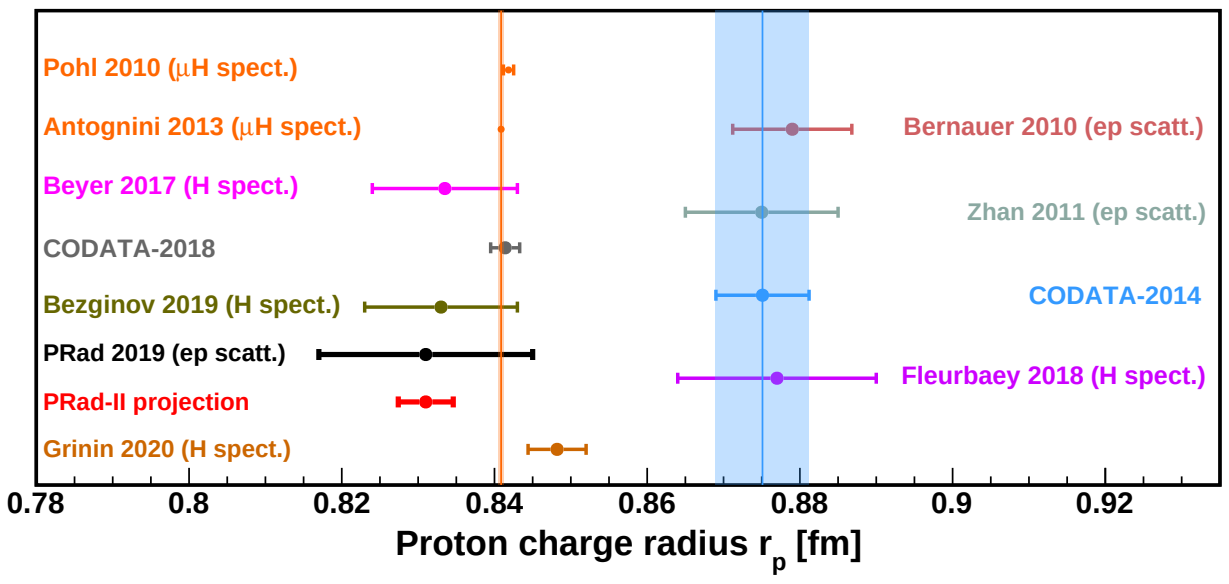

**Fig. 14.1.3** The projected $r_p$ result from PRad-II, shown along with the result from PRad and other measurements (see text).

A high precision measurement of the elastic cross section on the proton at ultra low $Q^2$, the *PRad* experiment, was performed in 2016 with the aim to solve the proton charge radius puzzle triggered by the muonic hydrogen spectroscopic measurements. To improve the precision of the measurement, the experiment utilized a new type of windowless target system flowing the hydrogen gas directly into the stream of CEBAF's 1.1 and 2.2 GeV electrons, and a calorimeter to detect the scattered electrons, rather than the traditionally used magnetic spectrometer. Moreover, the experiment was able to measure the scattered electron at very low ($Q^2$), facilitating a highly accurate extrapolation to $Q^2 = 0$ and extraction of the proton charge radius. The new value obtained for the proton radius is 0.831 fm [2901], which is smaller than the previous electron-scattering values and is, within its experimental uncertainty, in agreement with recent muonic atomic spectroscopy results.

To reach the ultimate precision offered by this new method, an enhanced version of *PRad*, the $PRad-II$ experiment [4330] has been approved. It will deliver the most precise measurement of $G_E^p$ reaching the lowest ever $Q^2$ value ($10^{-5}GeV^2$) in lepton scattering experiments, critical for the model independent extraction of $r_p$. The projected $r_p$ from $PRad-II$ is shown Fig. 14.1.3 along with the *PRad* result, recent electron scattering extractions, atomic physics measurements on ordinary hydrogen and muonic hydrogen, and the CODATA values (see [2901] for references of these measurements).

*Quark parton distributions at high $x$*

The quark and gluon structure of the proton has been under intense experimental and theoretical investigation for more than five decades. Nevertheless, even for the distributions of the well-studied valence quarks, challenges such as the value of the down quark to up quark ratio at high fractional momenta $x$ ($x \geq 0.5$), where a single parton carries most of the nucleon's momentum, remain. Recently, three JLab unpolarized DIS experiments, MARATHON [4331] in Hall A, BoNUS12 [4332] in Hall B, and $F_{2d}/F_{2p}$ [4333] in Hall-C completed data taking. These experiments aim to provide data to constrain PDFs in the high-$x$ region, especially the $d/u$ PDF ratio.

The experiments in Hall A and Hall B used two different approaches to minimizing nuclear effects in extracting the neutron information: MARATHON measured the ratio of $^3H$ to $^3He$ structure functions, while BONUS12 tagged slow recoiling protons in the deuteron. The Hall-C experiment measured H(e, e') and D(e, e') inclusive cross sections in the resonance region and beyond. While there will be nuclear effects in the deuterium data, the experiment provides a significant large $x$ range and reduced uncertainty to be combined with the large global data set of inclusive cross sections for PDF extraction. Fig. 14.1.4 shows the MARATHON $F_2^d/F_2^p$ results [4331], along with data from the JLab BoNUS experiment [3052] for $W \geq 1.84GeV/c^2$, evolved to the $Q^2$ of MARATHON, and results from early SLAC measurements with $W \geq 1.84GeV/c^2$ [4334] presented as a band. The results, which cover the Bjorken scaling variable range $0.19 < x < 0.83$, represent a significant improvement compared to previous measurements for the ratio. The results are expected to improve our knowledge of the nucleon PDFs, and to be used in algorithms which fit hadronic data to properly determine the essentially unknown $(u+\bar{u})/(d+\bar{d})$ ratio at large x. A planned experiment using Parity Violation in Deep Inelastic Scattering (PVDIS) [4335] on the proton, with the proposed SoLID [4336] spectrometer, will provide input on the $d/u$ ratio at high $x$ without contamination from nuclear corrections by measuring the ratio of $\gamma Z$ interference to total structure functions.

An extensive experimental program on spin physics at low and moderate $Q^2$, has been pursued by JLab during the 6 GeV era. The main focus of the DIS experiments has been the $x-$dependence of virtal photon asymmetry $A_1 = g_1/F_1$, to determine the contributions of quark spins to the spin of nucleon. In addition, the high statistical precision data and kinematic coverage allowed an accurate study of sum rules in the parton to hadron transition region as well as higher twist contributions (see Ref. [4337] for a review). A spin physics program has been approved to run with the upgrade CEBAF which extends the kinematical coverage to higher $x$ and can, among other things, answer the key question on what happens when a single quark carries nearly all (more than 80%) of the momentum of the nucleon. This region is well suited to test various theoretical predictions including those from the relativistic constituent quark model and perturbative QCD. The $A_1^n$ high-impact experiment in Hall C [4338] completed data taking in 2020. The experiment ran at a luminosity of 2x$10^{36} cm^{-2}s^{-1}$ thanks to the upgraded polarized $^3He$ target [4339]. The new precision measurement will expand knowledge of the extracted $g_1^n$ structure function to $x = 0.75$. Combined with the currently running experiments to measure the proton and deuteron asymmetries $A_1^p$ and $A_1^d$ with CLAS12 [4329], new global analyses will be able to extract the $\Delta u$ and $\Delta d$ quark helicity distributions in the high-$x$ region with much improved precision.

*Nuclear Femtography:TMDs and GPDs*
Pioneering measurements to access Generalized Parton Distributions (GPDs) and Transverse Momentum Distributions (TMDs) were provided by the HERMES, COMPASS, and the JLab 6 GeV program, among others. For recent reviews see Refs. [4340, 4341]. The upgraded detectors and CEBAF beam energy and intensity, promise to provide a more detailed three-dimensional (3D) mapping of the nucleon over wider ranges of the relevant kinematic variables. Indeed, this is a major thrust of the 12 GeV program accounting, so far, for almost $\sim 1/3$ of the whole approved experimental program.

Experimentally GPDs are accessible through deep exclusive processes, the most prominent ones being Deeply Virtual Compton Scattering (DVCS), and Deeply Virtual Meson Production (DVMP). TMDs, at JLab, are accessed through Semi-Inclusive Deep Inelastic Scattering (SIDIS), in which the nucleon is no longer intact and one or two of the outgoing hadrons are detected in coincidence with the scattered lepton. GPDs and TMDs are not measured directly. They are extracted through global fits to experimental data of Compton Form Factors (CFFs) for GPDs and Structure Functions for TMDs, and model dependent techniques with various assumptions involved. Therefore, accessing them demands not only a structured connection between theory, experiment and phenomenology, but availability of high precision data in a wide kinematical range and from different targets and several target/beam polarization combinations. A 3D description of the nucleon internal structure comes at the price of an unprecedented complexity. Therefore, for a correct interpretation of the data and a detailed comparison between results and theoretical models, a full differential analysis, using multi-dimensional information is crucial. The high-intensity, high-polarization electron beam provided by CEBAF with the complementary equipment of halls, A, B, C, makes JLab an ideal place for these studies.

SIDIS experiments provide access to the nucleon spin-orbit correlations. Observables are spin azimuthal asymmetries, and in particular single spin azimuthal asymmetries (SSAs), of the detected hadron. SSAs are due to the correlation between the quark transverse momentum and the spin of the quark/nucleon and early measurements indicated that they become larger with increasing $x$, i.e in the region where valence quarks have visible presence. Measurements of SSAs at JLab with the 6 GeV beam, performed with longitudinally polarised $NH_3$ [4342], and transversely polarised $^3He$ [3240] [4343] [3241] [4344] indicate that spin orbit correlations may be significant for certain combinations of spins of quarks and nucleons and transverse momentum of scattered quarks.

Large spin-azimuthal asymmetries have been observed at JLab also for a longitudinally polarised beam [4345] and a transversely polarised $^3He$ target [4346], which have been interpreted in terms of higher-twist contributions related to quark-gluon correlations and novel aspects of emergent hadron mass. At JLab with upgraded energy, three experimental halls, A, B, and C are involved in TMDs studies. The measurements aim to access leading and higher twists TMDs and their flavour and spin dependence, in multi-dimentinal binning of $x, Q^2, z, P_T$. The joint efforts of the three halls, where the high-precision, high-statistics measurements in Hall A and C will be combined with the wide kinematics ones performed in Hall B, by using different targets and several target/beam polarization combinations, will allow a thorough exploration of the 3D structure of the nucleon in momentum space. The program inludes the BigBite spectrometer and SBS[4347], as well as, the SoLID detector at Hall A [4348–4350], CLAS12 at Hall B [4351–4355], and High Momentum Spectrometer (HMS) and Super HMS at Hall C [4356–4358].

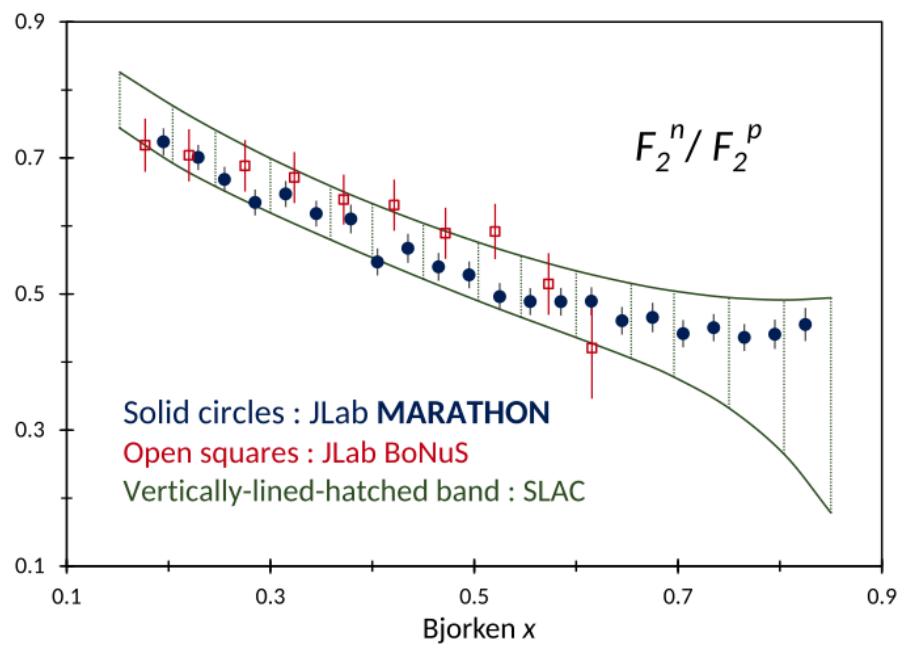

**Fig. 14.1.4** The $F_2^d/F_2^p$ ratio versus Bjorken $x$ from the JLab MARATHON experiment [4331], together with data from BoNUS [3052] and a band based on the fit of the SLAC data as provided in Ref. [4334], for the MARATHON kinematics $Q^2 = 14x$ (GeV/c)$^2$ . All three experimental data-sets include statistical, point to point systematic, and normalization uncertainties.

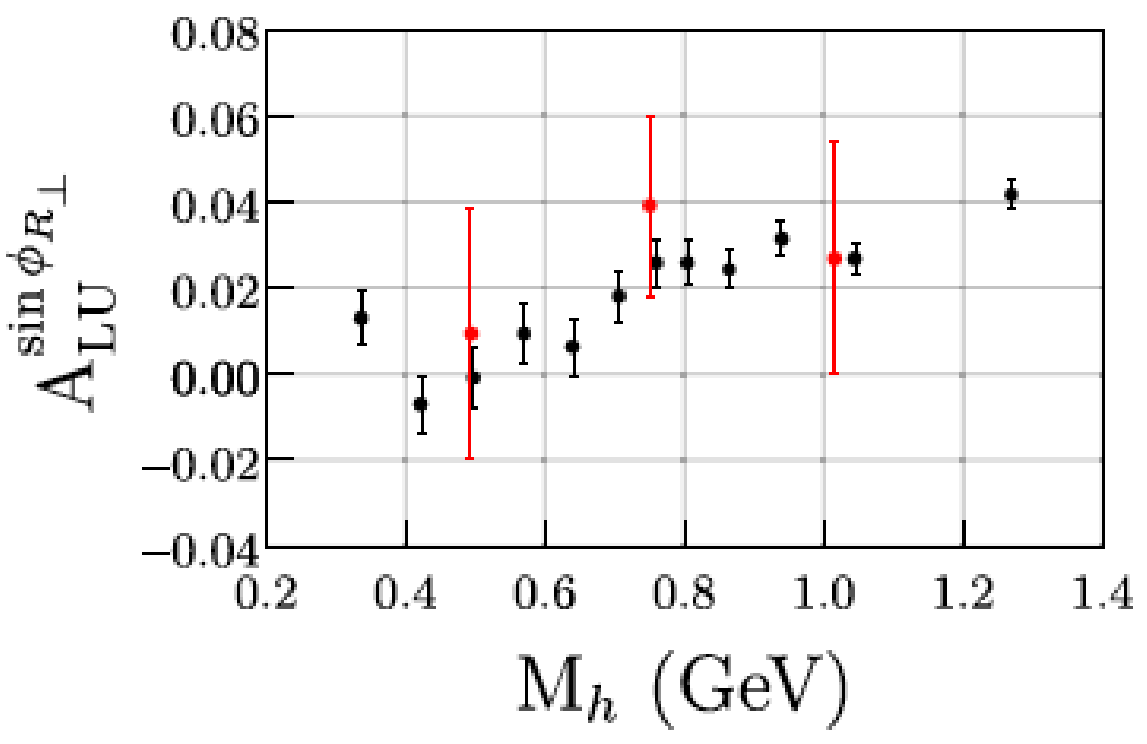

**Fig. 14.1.5** The new CLAS12 results on beam helicity asymmetry in two-pion semi-inclusive deep inelastic electroproduction [4360] as a function of the invariant mass of pion pairs. The red points are from CLAS6 measurements [4362].

The first SIDIS publications of the 12 GeV era were reported by the CLAS12 collaboration on measurements of beam SSA for sigle pion [4359], two-pion [4360] and back-to-back dihadron [4361] productions off an unpolarized proton target using 10.6 and 10.2 GeV longitudinally spin-polarized electron beams. The sigle $\pi^+$ production was measured over a wide range of kinematics in a fully multidimensional study. The comparison with calculations shows the promise of high-precision data to enable differentiation between competing reaction models and effects.

The first significant beam spin asymmetries observed in two-pion production provide the first opportunity to extract the higher-twist parton distribution function $e(x)$, interpreted in terms of the average transverse forces acting on a quark after it absorbs the virtual photon. Moreover, this measurement constitutes the first ever signal sensitive to the helicity-dependent two-pion fragmentation function $G_1^\perp$. The comparison of the 6 GeV and 12 GeV measurements shown in Fig. 14.1.5) demonstrates the impact of the beam energy on the phase space for production of multiple hadrons in the final state and the huge reduction in the corresponding error bars. Finally, the measured beam-spin asymmetries in back-to-back dihadron electroproduction, $ep \to' p\pi^+X$, with the first hadron produced in the current-fragmentation region and the second in the target-fragmentation region, provide a first access in dihadron production to a previously unobserved leading-twist spin- and transverse-momentum-dependent fracture functions [4363].

A comprehensive program is carried out at JLab in deeply virtual exclusive scattering processes (DVCS and DVMP) with the goal to create the transverse spacial images of quarks and gluons as a function of their longitudinal momentum fraction in the proton, neutron and nuclei through the study of the GPDs. The physical content of the GPDs is quite rich. Among other features, they give access to the contribution of the orbital momentum of the quarks and gluons to the nucleon, and the D-term, a poorly known element of GPD parametrizations, which gives valuable insights to the mechanical properties of the nucleon [2822, 4364–4366]. The study of the deeply exclusive processes and the GPDs extraction started, at JLab, in the 6 GeV era. After the first publication by CLAS in 2001 [4367], a series of high-statistics DVCS-dedicated experiments in Hall A and B followed at moderate $Q^2$ $(1-3GeV^2)$ and in a $x_B$ range centered around $x_B \sim 0.3$ (for a recent review see[4368]).

The polarized and unpolarized cross sections measured at in Hall A at 6 GeV [4369, 4370] indicate, via a $Q^2$-scaling test, that the factorization and the hypothesis of leading-twist dominance are valid already at relatively low $Q^2$ $(\sim 1-2GeV^2)$ and thus the applicability of the GPD-based description. Covering a range in $x_B$ from 0.1 to 0.7 and in $Q^2$ from 1 to 10 $GeV^2$, the upgraded JLab is very well matched to study GPDs in the valence regime. The program is executed in the three experimental halls, A, B, C, and aims to measure accurately fully differential beam-polarized cross section differences and unpolarized cross sections, longitudinally polarized target-spin asymmetries along with double polarization observables.

The first result of the 12 GeV era was reported by Hall A on the DVCS cross section measurement at high Bjorken $x_B$ off an unpolarized proton target [4371]. The work presents the first experimental extraction of the four helicity-conserving nucleon Compton Form Factors (CFFs) as a function of $x_B$. A similar experiment, which will complement the kinematic coverage of the Hall A, is planned to run in Hall C with the HMS and NPS in 2024 [4372]. In Hall B two experiments measuring DVCS off an unpolarized proton target at 11 GeV [4373] and 6.6 and 8.8 GeV [4374] will allow a larger kinematical coverage, while the measurement of the beam-spin asymmetry off a deuteron target, with detected neutron, will allow to constrain the poorly known GPD E, related to the quark orbital angular momentum through the Ji's sum rule, and to perform the GPDs quark-flavor separation. These experiments will release their results soon. Finally, an experiment using longitudinally polarized $NH_3$ and $ND_3$ target [4329] is currently running in Hall B and one has been proposed to use a transversely polarized proton [4375]. The precision and kinematical coverage of these asymmetries obtained with different combination of targets and polarization will bring stringent constraints to GPD parametrizations.

Meson production at JLab at 6 GeV has not yet shown parton dominance of scattering. Experimental data from 11 GeV beam will provide important test of the deep-exclusive meson production mechanism. Hall A recently published deep exclusive electroproduction of $\pi^0$ at high $Q^2$ [4376] using the 11 GeV beam off an unpolarized proton target. The results suggest the amplitude for transversely polarized virtual photons continues to dominate the cross section throughout this kinematic range. Experiments have also been approved in Hall B for $\pi^0$, $\eta$ [4377] and $\phi$ production [4378], the latter with the hope to determine the t-slope of the gluon GPDs. In Hall C, it is important to mention the precise measurement of the L/T separation on kaon and pion electroproduction [4379, 4380] and the neutral pion cross-section measurements [4372].

Finally, DVCS and DVMP will be measured on the $^4He$ nucleus (with emphasis on $\phi$ production) [4381], with the aim of comparing a) the quark and gluon radii of the helium nucleus, b) GPDs of the bound proton and neutron with the free proton and quasi-free neutron.

While the most attention so far is on studies of GPD using spin (beam/target) observables and cross-sections in DVCS, also the Time-like Compton Scattering (TCS), the time-reversal symmetric process of DVCS where the incoming photon is real and the outgoing photon has large time-like virtuality, has much to offer. The first ever measurement of TCS on the proton $\gamma p \to p'\gamma^*(\gamma^* \to e^+e^-)$ has been obtained with CLAS12 [2824]. Both the photon circular polarization and forward/backward asymmetries were measured. The comparison of the measured polarization asymmetries with model predictions points toward the interpretation of GPDs as universal functions. Fig. 14.1.6 shows the photon polarization asymmetry $A_{\odot U}$ as a function of $-t$ at the averaged kinematic point $E\gamma = 7.29 \pm 1.55 GeV; M = 1.80 \pm 0.26 GeV$, compared with GPDs based models.

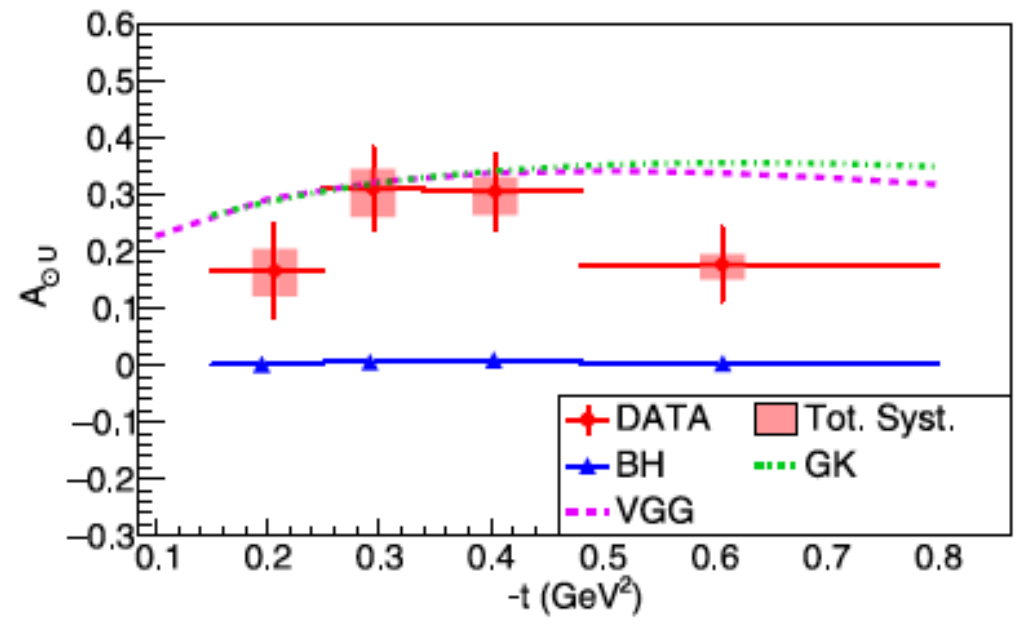

**Fig. 14.1.6** Photon polarization asymmetry as a function of -t. The dashed and dashed-dotted lines are the predictions of GPDs based models, respectively, the VGG [4382] and the GK [4383] models, evaluated at the average kinematics. For detailed explanation see [2824].

### 14.1.4 *Hadron Spectroscopy*

For the theoretical formalism and a general overview of hadron spectroscopy, the reader should refer to Sec. 8 of this volume.

This is an exciting period in hadron spectroscopy. The last two decades witnessed the discovery of many states that challenged the basic model of hadron physics according to which particles are made of $3q$ (baryons) or a $q\bar{q}$ (mesons), and pointed to states with multi-quark content, or with explicit gluonic components (glueballs and hybrids). Mapping states with explicit gluonic degrees of freedom in the light sector is a challenge.

One example is the $\pi_1$ state which has led to controversies. Experiments have reported two different hybrid candidates with spin-exotic signature, $\pi_1(1400)$ and $\pi_1(1600)$, which couple separately to $\eta\pi$ and $\eta'\pi$ (for a review see Ref. [2362]). This picture is not compatible with recent Lattice QCD estimates for hybrid states, nor with most phenomenological models. A recent work by the JPAC [4384] provides a robust extraction of a single exotic $\pi_1$ resonant pole, but no evidence for a second exotic state. The main goal of the GlueX experiment [4385, 4386] in Hall D is to search for exotic mesons, and together with CLAS12 MesonEx experiment [4387] in Hall B, to provide a unique contribution to the landscape of experimental meson spectroscopy through the novel photoproduction mechanism previously relatively unexplored. Utilizing a real, linearly-polarized photon beam in GlueX and quasi-real, low-$Q^2$ photons in CLAS12, this program covers a wide range of beam energies from $E_\gamma$ = 3-12 GeV.

GlueX has already collected high-statistics, high-quality photoproduction data and published various results on photoproduction cross sections for several single pseudoscalar mesons including the $\pi^0$, $\pi^-$, $K^+$, $\eta$, $\eta'$ over a broad range of momentum transfer [4388–4391], focused on a quantitative understanding of the meson photoproduction mechanism. Polarization observables, such as spin-density matrix elements, provide also valuable input for the theoretical description of the production mechanism, which is essential for the interpretation of possible exotic meson signals. Moreover, these studies require a complete understanding of the detector acceptance and efficiencies in fits to multi-dimensional data and therefore are essential for assessing the Partial Wave Analysis (PWA) machinery.

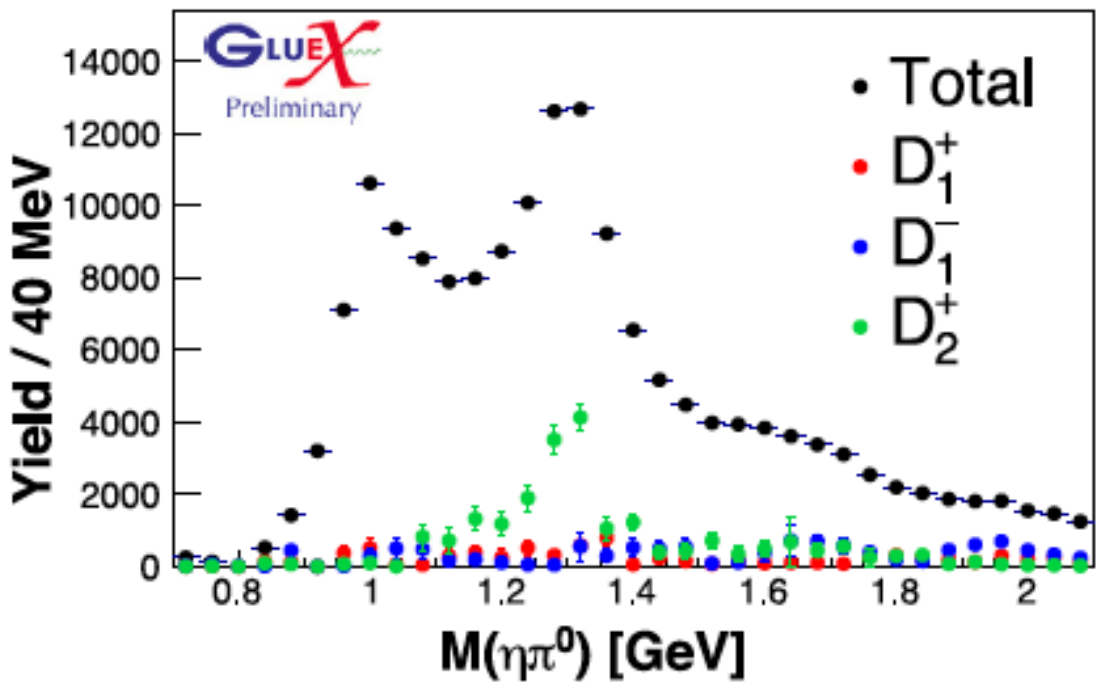

**Fig. 14.1.7** Preliminary mass spectra and amplitude analysis results from GlueX for the reactions $\gamma p \to \eta^{(\prime)}\pi^0 p$, with $0.1 < -t < 0.3 GeV^2$ and $8.2 < E_\gamma < 8.8 GeV$.

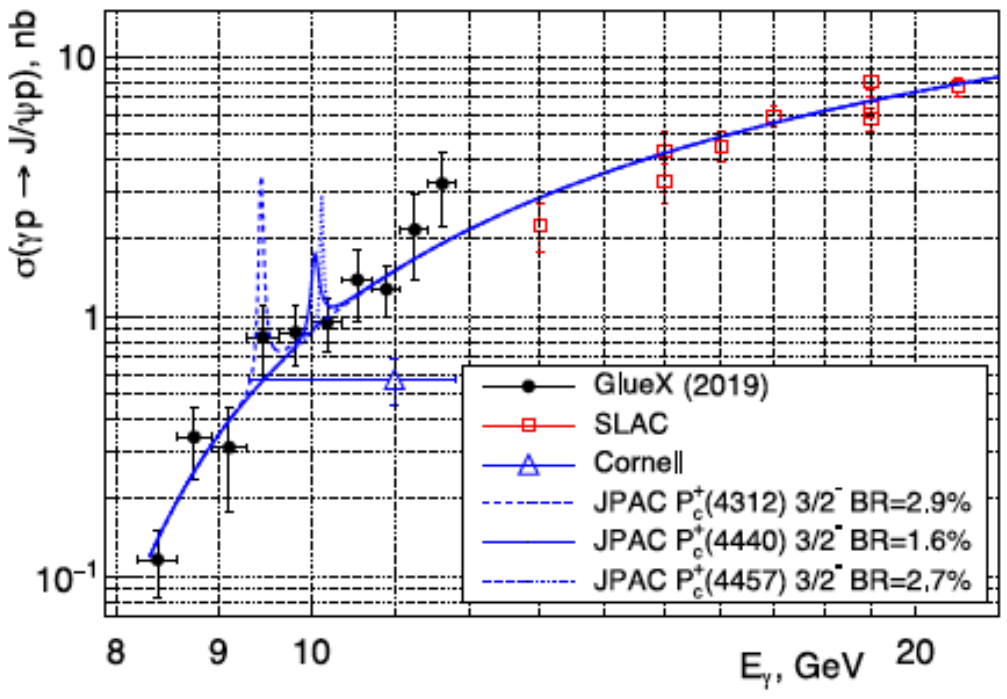

**Fig. 14.1.8** GlueX results for the $J/\psi$ total cross section vs beam energy, compared to the JPAC model with hypothetical branching ratios provided in the legend for $P_c^+$ with $J^P = 3/2^-$ as described in Ref. [4404].

GlueX published the first measurement of spin density matrix elements of the $\Lambda(1520)$ in the energy range $E_\gamma = 8.2$-$8.8$ GeV [4392] and released preliminary results on spin-density matrix elements of the vector mesons $\rho(770)$, $\phi(1020)$ and $\omega(782)$ [4393]. The final analysis with the full data set will surpass previous measurements by orders of magnitude. The search for hybrid mesons has started in GlueX by studying $\eta^{(\prime)}\pi$ final-states to eventually confirm the $\pi_1$ pole position extracted by JPAC. With a large acceptance to both charged and neutral particles, GlueX has access to both neutral $\gamma p \to \eta^{(\prime)}\pi^0 p$ and charged $\gamma p \to \eta^{(\prime)}\pi^- \Delta^{++} p$ exchanges. Fig. 14.1.7) shows preliminary results for the measured intensity of the dominant waves in the $\gamma p \to \eta^{(\prime)}\pi^0 p$ channel.

JLab at 12 GeV will continue the program to study the spectrum and structure of excited nucleon states, which in the last 15 years have provided critical input to global analyses to elucidate the $N^*$ spectrum (see Refs. [2819, 4394] for recent reviews). Detailed electrocouplings measurements through exclusive electroproduction study of both strange and non-strange final states, will be extended with the new CLAS12 detector and the upgraded energy beam which will significantly extend the kinematic range to $Q^2 > 5GeV^2$ [4395, 4396]. The program comprises also the search of hybrid baryons with constituent gluonic excitations, for which a rich spectrum is predicted by Lattice QCD. Finally, many hyperon spectroscopy measurements are expected from the GlueX and CLAS12 measurements, including the $\Xi$ and $\Omega$ [4397, 4398]. This program will be expanded by proposal to perform hyperon spectroscopy with the $K_L$ neutral kaon beam in Hall D, which was recently approved by the PAC[4399].

Over the past several years there has been a renewed interest in studying near-threshold $J/\psi$ photoproduction as a tool to experimentally probe important properties of the nucleon target related to its mass and gluon content. Moreover, in the beam energy region of $E_\gamma = 9.4$-$10.1$ GeV, the $\gamma p \to J/\psi p$ process can be used to search, directly in a simple $2 \to 2$ body kinematics [4400–4403] for the pentaquark candidates, $P_c^+(4312)$, $P_c^+(4440)$, and $P_c^+(4457)$, reported by the LHCb experiment but still under debate [2828, 2829]. JLab has an active $J/\psi$ physics program. There are either published, ongoing, or planned future $J/\psi$ experiments in each experimental hall. The first measurement was performed by GlueX [4404] and is shown in Fig. 14.1.8 , with curves depicting the strength of hypothetical $P_c$ signals. No structures are observed in the measured cross section, however model-dependent upper limits are set on the branching ratio of the possible $P_C \to J/\psi p$ decays. Preliminary results from the $J\psi - 007$ experiment in Hall C also observe no $P_c$ signal and will set more restrictive limits on the branching ratio [4405]. In Hall B analysis of data are ongoing [4406] and in Hall A an experiment has been approved to run with SoLID[4407].

### 14.1.5 *QCD and Nuclei*

Nuclear interactions are described using effective models that are well constrained at typical internucleon distances in nuclei but not at shorter distances. The strong component of the nucleon-nucleon potential associated with hard, intermediate short-distance interactions between pairs of nucleons, called Short-Range Correlated (SRC) pairs, is a poorly understood parts of nuclear structure and generates a high-momentum tail to the nucleon momentum distribution. The existence and characteristics of SRC pairs are related to outstanding issues in particle, nuclear, and astrophysics, among which are the modification of the internal structure of nucleons bound in atomic nuclei (the EMC ef-fect) [4408] and the nuclear symmetry energy governing neutron star properties [4409].

The studies of SRCs are a sizeable part of the JLab program that started already in the 6 GeV era. After the initial observation of identical structure in the high-momentum components of nuclei at SLAC[4410], electron-scattering measurements at JLab have identified the kinematic region where SRCs dominate[4411, 4412] and mapped out the contribution of SRCs in various light and heavy nuclei relative to the deuteron[795, 4413]. Data demonstrated also that the contribution is sensitive to details of the nuclear structure [4414, 4415] rather than the previously assumed average nuclear density [4416]. In addition, they showed a clear correlation between the contribution of SRCs [795] and the size of the EMC effect [4414]. To study the isospin dependence of the SRCs, measurements of two-nucleon knock-out were carried out. These experiments showed dominance of *np*-SRC pairs over *pp* and *nn*-SRC pairs by a factor of about 20 [796, 4417, 4418]. The result was confirmed in measurements of quasi-elastic knock-out of protons and neutrons from medium and heavy nuclei [4419], and later through inclusive measurements of the $^{48}Ca/^{40}Ca$ cross section ratio [4420] taking advantage of the target isospin structure.

The first measurement using a novel technique to extract the $np/pp$ ratio of SRCs taking advantage of the isospin structure of the mirror nuclei $^3H$ and $^3He$ was carried out in the 12 GeV era[4421]. The $np/pp$ SRC ratio obtained is an order of magnitude more precise than previous experiments, and shows a dramatic deviation from the near-total *np* dominance observed in heavy nuclei (see Fig. 14.1.9). This result implies an unexpected structure in the high-momentum wave-function for $^3He$ and $^3H$. Finally, measurements at $x > 2$ carried out with the 6 GeV beam, tried to establish the presence of three-nucleon SRCs [795, 4422], but didn't come to a definitive conclusion. Experiment [4423] with the 11 GeV beam will provide the first significant test by taking high-statistics $A/^3He$ ratio data at $x > 2$ and $Q^2$ =3 GeV$^2$.

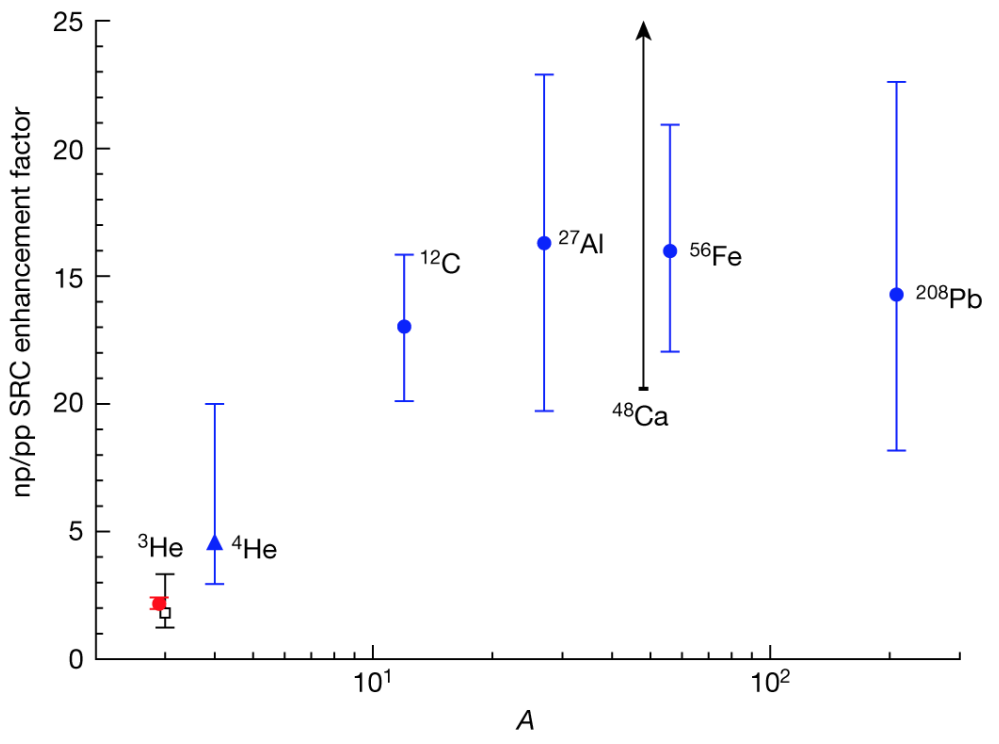

**Fig. 14.1.9** Ratio of np-SRCs to pp-SRCs relative to the total number of np and pp pairs, for the new inclusive data (red circle), compared with previous measurements [4421].

Determining the origin of the EMC effect, i.e. the modification of nuclear PDFs relative to the sum of the individual nucleon PDFs, is one of the major unsolved problems in the field of nuclear physics and is still a puzzle after forty years. Measurement at Jlab at 6 GeV in light nuclei demonstrated the correlation between the size of the EMC effect and the contribution of SRCs [795]. The JLab12 program addresses the three open questions of the EMC effect: *i)* the isospin dependence; *ii)* the spin dependence; *iii)* the configuration/distance dependence. The isospin dependence has been investigated with the already mentioned experiment using mirror nuclei [4421]. Polarization measurements can also help to understand the origin of the EMC effect [4424, 4425]. An 11 GeV experiment will measure the EMC effect in polarized $^7Li$ [4426] with the goal to distinguish between mean-field models with explanations based on SRCs. Tagging of recoil nuclei in deep inelastic reactions will be used in [4427] to address point *iii)*. This is a powerful technique to provide unique information about the nature of medium modifications, through the measurement of the EMC ratio and its dependence on the nucleon off-shellness.

There are several ways to study QCD in nuclei. One is through the hadronization process, a mechanism by which quarks struck in hard processes form the hadrons observed in the final state. This is a poorly known mechanism and more insight can be obtained by systematically studying production of different baryon and meson types using large and small nuclear systems, and observing the multi-variable dependence of observables, such as multiplicity ratios and transverse momentum broadening. These studies started with CLAS at 6 GeV [4428] and will continue with CLAS12[4426].

Hadron propagation in the medium can also be studied by searching for color transparency, where the final (and/or initial) state interactions of hadrons with the nuclear medium must vanish for exclusive processes at high momentum transfers. Color transparency for pions [4429] and $\rho$ mesons [4430] was observed at 6 GeV while the 11 GeV experiment [1331] ruled out color transparency in quasielastic $^{12}C(e, e'p)$ up to $Q^2$ of 14.2 GeV$^2$. These results impose strict constraints on models of color transparency for protons.

Measurements on nuclei which are directly relevant for understanding aspects of astrophysics and neutrino physics are also part of the JLab program. One of the early experiments of the 12 GeV era was the measurements of inclusive quasi-elastic scattering and single proton knockout on $^{40}Ar$ [4431, 4432].These data will allow for tests of $\nu - ^{40}Ar$ scattering simulations needed for the DUNE experiment. Another experiment [4433] measured electron scattering from a variety of targets and different beam energies in CLAS12 in order to test neutrino event selection and energy reconstruction techniques and to benchmark neutrino event generators.

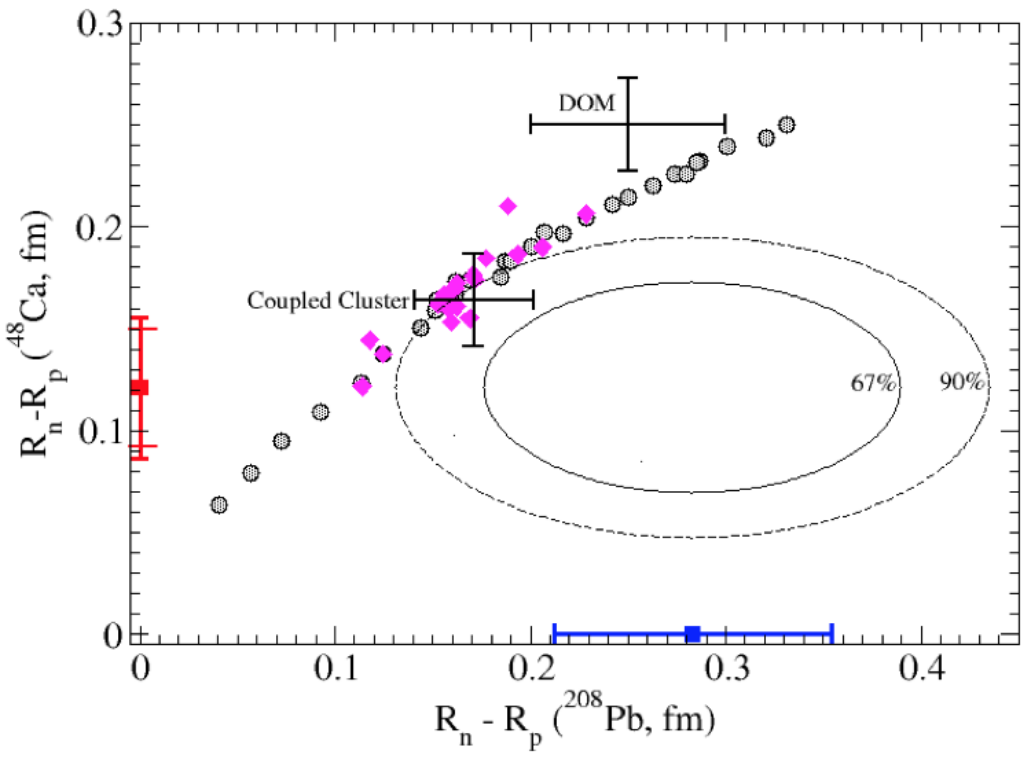

**Fig. 14.1.10** $^{48}Ca$ neutron minus proton radius (red square) versus that for $^{208}Pb$ (blue square). The ellipses are joint PREX-II and CREX 67% and 90% probability contours.The gray circles (magenta diamonds) show a variety of relativistic (non-relativistic) density functionals (see Ref. [4435]).

Thanks to the intense and highly polarized CEBAF electron beams, measurements of the parity-violating electron scattering asymmetry from $^{208}Pb$ and $^{48}Ca$ have demonstrated a new opportunity to measure the weak charge distribution and hence pin down the neutron radius in nuclei in a relatively clean and model-independent way. A precise measurement of the neutron radius, and hence of the neutron skin thickness, helps to constrain the density dependence of the symmetry energy of neutron rich nuclear matter, which has implications on neutron stars and supernova. The PREX-II experiment [4434] measured the "neutron skin thickness" of $^{208}Pb$ while CREX[4435] measured that of $^{48}Ca$. For CREX, the extracted neutron skin can be directly compared to microscopic calculations [4436] providing a bridge between medium nuclei ab initio calculations and heavy nuclei Density Functional Theory calculations. The extremely precise CREX measurement indicates a thin neutron skin around its nucleus, in contrast with the PREX measurement which revealed a thicker skin (see Fig. 14.1.10). This discrepancy is exciting and presents the opportunity for further exploration to determine why there's such a big difference between the medium-density calcium nucleus and the high-density lead nucleus.

### 14.1.6 *Future Opportunities*

With a fixed target program at the "luminosity frontier," up to $10^{39}$ $cm^{-2}s^{-1}$, and large acceptance detection systems, CEBAF will continue to offer unique opportunities to illuminate the nature of QCD and the origin of confinement for decades to come. In fact, CEBAF operates with several orders of magnitude higher in luminosity than the Electron-Ion Collider (EIC) and exciting scientific opportunities using CEBAF beyond the currently planned decade of experiments can provide very complementary capabilities, even in the era of EIC operations. A discovery science program utilizing CEBAF in the EIC era has been developing jointly between JLab and its user community towards exploring both the science and technical case for moving beyond 12 GeV. A series of upgrades to increase luminosity, enable positron beams, and double the energy of CEBAF is envisioned [4320].

- An increase in luminosity with modest detector upgrades will facilitate double DVCS (DDVCS) studies in experimental Halls A and B. DDVCS can bring significant additional information to the three dimensional imaging of the quark structure. This is a process with interaction rates a factor of 100 lower than DVCS. Therefore it is not viable at EIC and must be studied using CEBAF.
- Positron beams, both polarized and unpolarized, are identified as an essential ingredient for the hadronic physics program at JLab, and they are important tools for a precise understanding of the electromagnetic structure of the nucleon, in both the elastic and the deep-inelastic regimes. Proof of principle of a new concept for creating polarized positron beams at CEBAF has been demonstated and a scinetific prgram has been developed [4437].
- Encouraged by recent success of CBETA at Cornell, a proposal was formulated to increase the CEBAF energy from the present 12 GeV to 20-24 GeV by replacing the highest-energy arcs with Fixed Field Alternating Gradient (FFA) arcs but using the existing CEBAF SRF cavity system. This exciting new technology would be a cost-effective method to double the energy of CEBAF, enabling new scientific opportunities in meson spectroscopy and extending the kinematic range of nucleon imaging studies. Technical studies of the implementation of FFA technology at CEBAF are in progress.

### 14.1.7 Conclusions

Jefferson Lab is a world-leading research laboratory for exploring the nature of matter in depth. Its powerful experimental program at 12 GeV will advance our understanding of the quark/gluon structure of hadronic matter, the nature of Quantum Chromodynamics, and the properties of a new extended standard model of particle interactions. CEBAF at Jefferson Lab is a facility in high demand due to its unique capability to operate with a fixed target program at the "luminosity frontier" up to $10^{39} cm^{-2} s^{-1}$, with exciting scientific opportunities beyond the currently planned decade of experiments. Potential upgrades of CEBAF and their impact on scientific reach are being discussed, such as higher luminosity, the addition of polarized and upolarized postron beams, and doubling the beam energy. They will keep CEBAF uniquely capable of a large number of important measurements in nuclear and hadronic physics.

## 14.2 The EIC program

**Christian Weiss**

The Electron-Ion Collider (EIC) at Brookhaven National Lab (BNL) is planned as a next-generation facility for high-energy $ep/eA$ scattering experiments supporting basic research in hadronic/nuclear physics and QCD. The design combines the RHIC superconducting proton/ion accelerator ring with an electron storage ring in the same tunnel and an injector for on-energy injection of polarized bunches and enables collisions at one (possibly two) interaction points (see Fig. 14.2.1) [4438]. It provides $ep$ collisions at CM energies $\sqrt{s}$ = 20–100 GeV, upgradable to 140 GeV, using various combinations of beam energies; for $eA$ collisions with the same setup the CM energy per nucleon is lower by a factor $\sqrt{Z/A}$. It is projected to achieve peak luminosities in the range $\sim 10^{33}$–$10^{34}$ cm$^{-2}$ s$^{-1}$ and deliver an integrated lifetime luminosity $\sim$10–100 fb$^{-1}$. It accelerates ion species including the proton ($p$), light ions (D, $^3$He, others), and heavy ions (Au, U, others). Polarization is available for the electron and the light ion beams ($p$ and $^3$He) with an average ion polarization $\sim$70%. The EIC will be the first colliding beam facility enabling electron collisions with ion beams ($A > 1$), and with polarized proton/ion beams. Its luminosity will exceed that of the HERA $ep$ collider by 100-1000. As such it will provide qualitatively new capabilities for physics research [3105].

The concept of a polarized electron-ion collider was inspired by the results of the fixed-target spin physics experiments (CERN, SLAC, DESY), the DESY HERA $ep$ collider, and the BNL RHIC polarized $pp$ and $AA$ collider, and motivated by advances in theoretical concepts for hadron structure and high-energy QCD. The developments began with planning exercises in the 1990s and advanced through extensive community efforts (science studies, program development) [820, 3128] and technical design work (accelerator, facility) at BNL, JLab, and other laboratories in the 2000s and 2010s. Important milestones were the recommendation in 2015 Nuclear Science Advisory Committee Long-Range Plan [4439] and the endorsement by a study of the U.S. National Academy of Sciences 2018 [4440]. The EIC was granted Critical Decision Zero (CD-0) by the U.S. Department of Energy in December 2019 and is now an official project of the U.S. Government. It is executed according to project management principles and passed CD-1 in 2021. Completion of construction and begin of operations are expected around 2034.

The EIC will enable a comprehensive science program aimed at understanding hadrons and nuclei as emergent phenomena of QCD. Scattering experiments will be performed at momentum transfers $Q^2 \sim 10^1$-$10^2$ GeV$^2$, corresponding resolution scales where the quark and gluon degrees of freedom are manifest and methods of QCD factorization can be applied (see Fig. 14.2.2). The partonic content will be sampled at momentum fractions down to $x \sim 10^{-3}$–$10^{-4}$, where gluons and sea quarks are abundant and dominate hadron structure. The wide kinematic coverage will enable study of scale dependence and radiation processes building up the parton densities, which provide essential insight into the dynamics. The luminosity and detection systems will permit measurements of the final states of deep-inelastic processes in unprecedented detail (exclusive processes, semi-inclusive production, jets, nuclear breakup, diffraction, etc.) and enable analysis using modern theoretical concepts (GPDs, TMDs, jets).

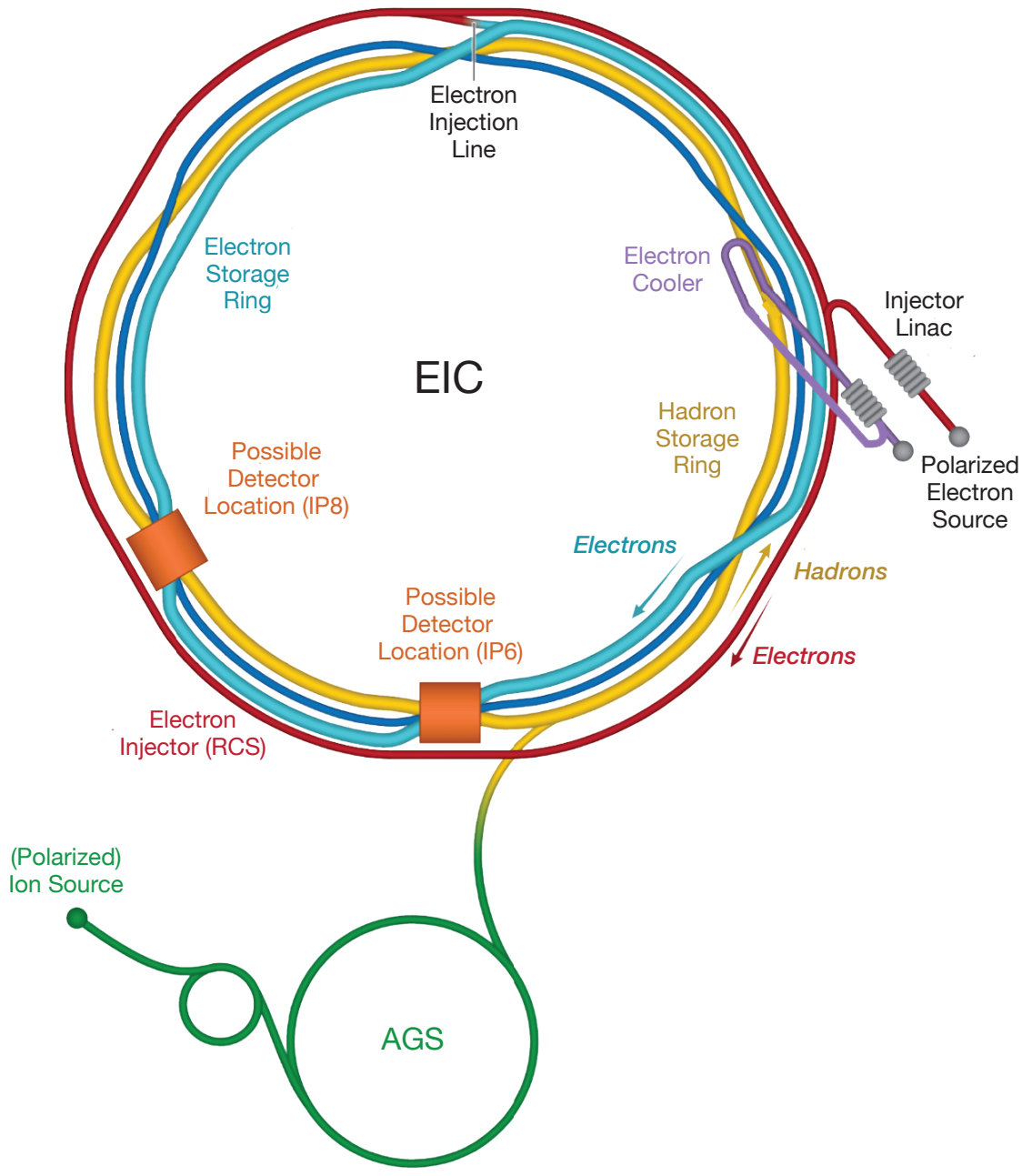

**Fig. 14.2.1** Schematic of the EIC accelerator complex [3105, 4438].

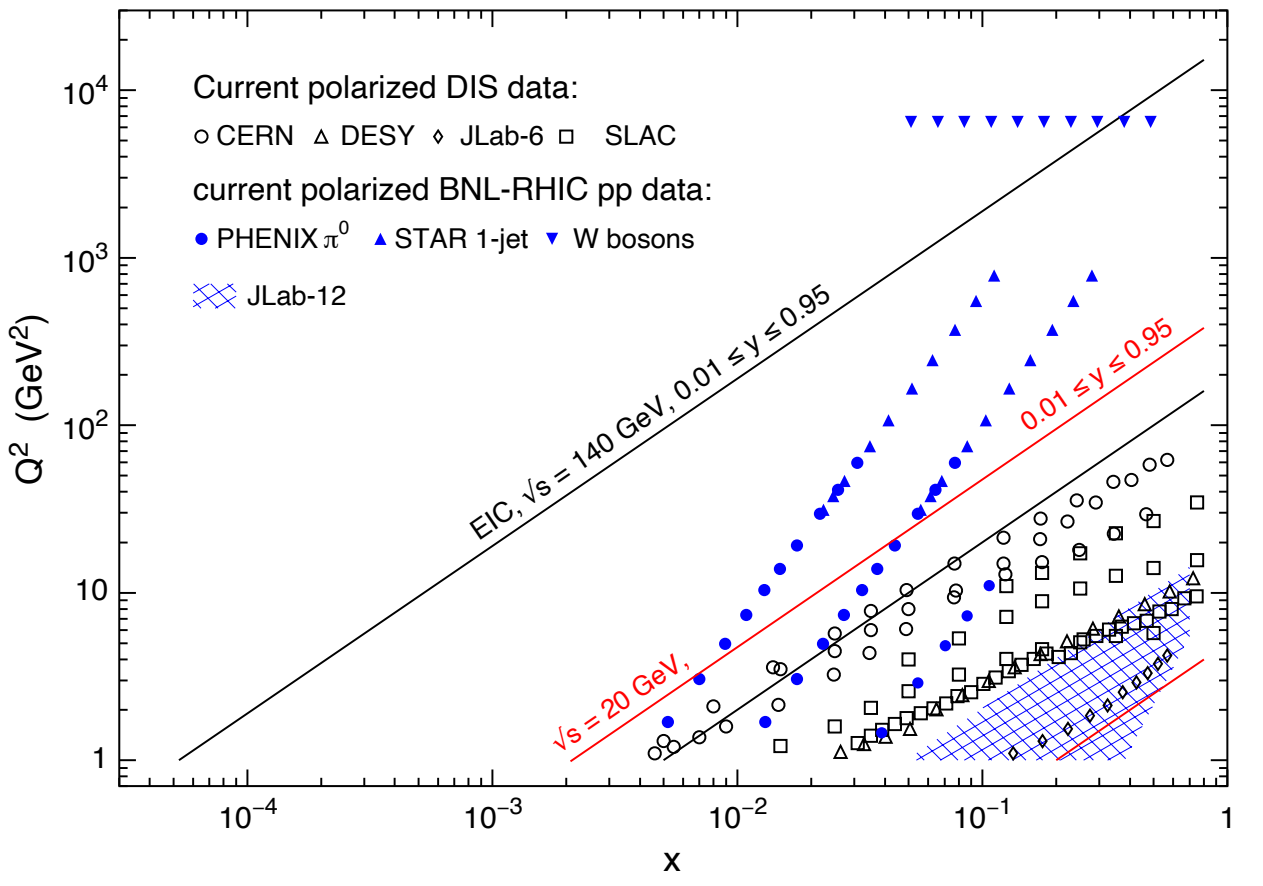

**Fig. 14.2.2** Kinematic coverage in $x$ and $Q^2$ in DIS experiments with the EIC at CM energies of 20 GeV and 140 GeV [3105].

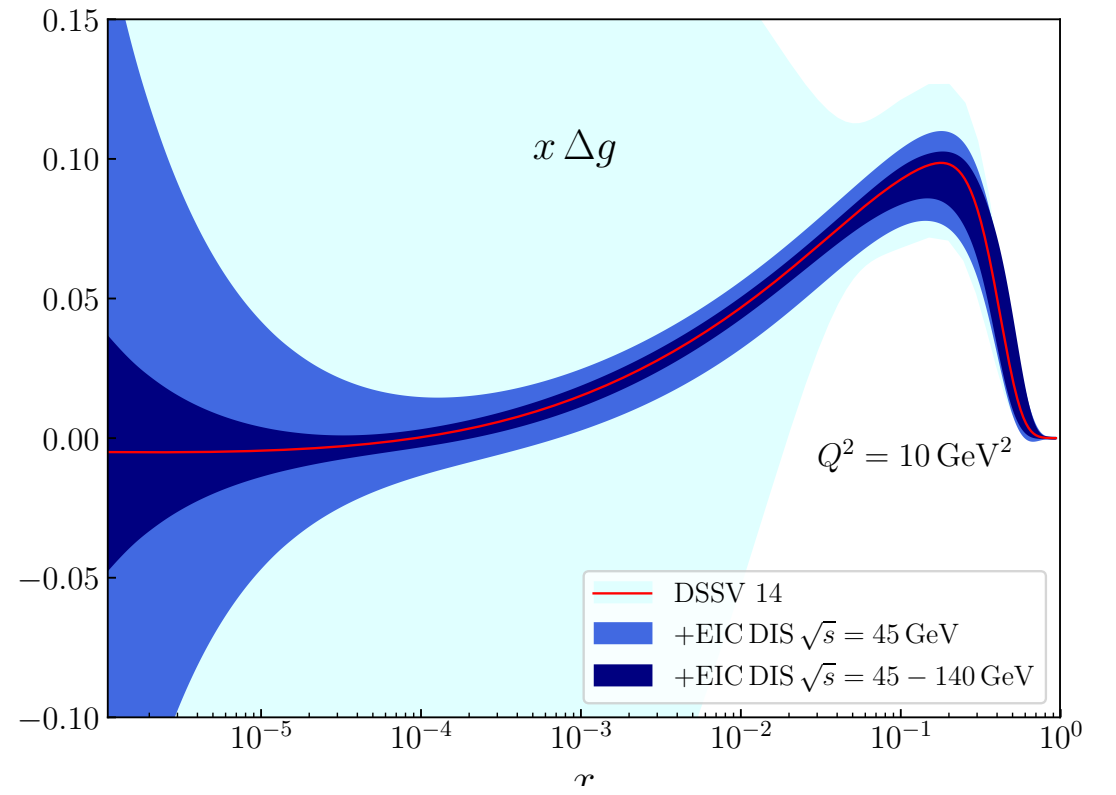

**Fig. 14.2.3** Gluon spin PDF extracted from polarized inclusive DIS pseudodata at EIC [3073, 3105]. Similar results are obtained in studies using other PDF parametrizations [3105].

The EIC science program is organized in four broad themes, defined by basic physics questions and concepts that are explored using various measurements:

- Global properties and partonic structure of hadrons
- Multi-dimensional imaging of hadrons and nuclei
- Nuclear high-energy scattering in QCD
- Emergence of hadrons from QCD

The boundaries between them are not strict, as some measurements serve to answer questions in more than one area. In the following we briefly summarize the objectives and main measurements in each of the themes; further information can be found in Refs. [820, 3105, 3128].[118] The program and its organization are still evolving; new topics are being discussed and proposed in response to developments in theory and detector design.

### 14.2.1 Global properties and partonic structure

One basic objective is to understand how the global properties of hadrons such as spin, mass, charges, and other characteristics emerge from the quark/gluon fields of QCD and their interactions (see Sec. 10.3). The quantities are expressed as matrix elements of QCD composite operators between hadronic states, $\langle h|\mathcal{O}_{\rm QCD}|h\rangle$, some of which can be measured in deep-inelastic processes. For some quantities the operators have a partonic interpretation, and the matrix elements and can be expressed as integrals of the PDFs/GPDs (sum rules). For other quantities the operators involve interactions (higher twist), and the interpretation is more indirect. The EIC will advance this program through several measurements:

*Gluon polarization and nucleon spin*

The quark and gluon contributions to the nucleon spin are expressed as the integrals of the quark and gluon spin PDFs, which are measured in various polarized scattering experiments (see Sec. 10.3). Despite much effort, the contributions to the spin sum rule are still poorly known. While fixed-target DIS measurements have determined the quark spin densities, and the RHIC spin program has provided evidence of nonzero gluon spin, the distributions are known with good precision only at $x \gtrsim 0.01$, so that the integrals suffer from large uncertainties (see Sec. 10.2). At EIC, measurements of inclusive polarized *ep* DIS will accurately determine the quark and gluon spin densities down to $x \gtrsim 10^{-4}$. The wide kinematic coverage will make it possible to determine the gluon spin density indirectly through DGLAP evolution (see Fig. 14.2.3) [3073, 3105, 3107]. Complementary information will come from direct measurements of the gluon spin density using dijets or heavy flavor production [4441]. The gluon and quark spin PDFs extracted in this way will permit accurate evaluation of quark and gluon spin contributions to the spin sum rule. The results will also constrain the possible contribution of quark/gluon orbital angular momentum to the nucleon spin (see Fig. 14.2.4).

[118] The literature supporting the concepts and measurements of the EIC physics program is very extensive. In this summary we refer to the other sections of the review article for concepts and previous results whenever possible; we refer directly to the literature for simulation and impact studies for the EIC, and for topics not covered elsewhere in the review.

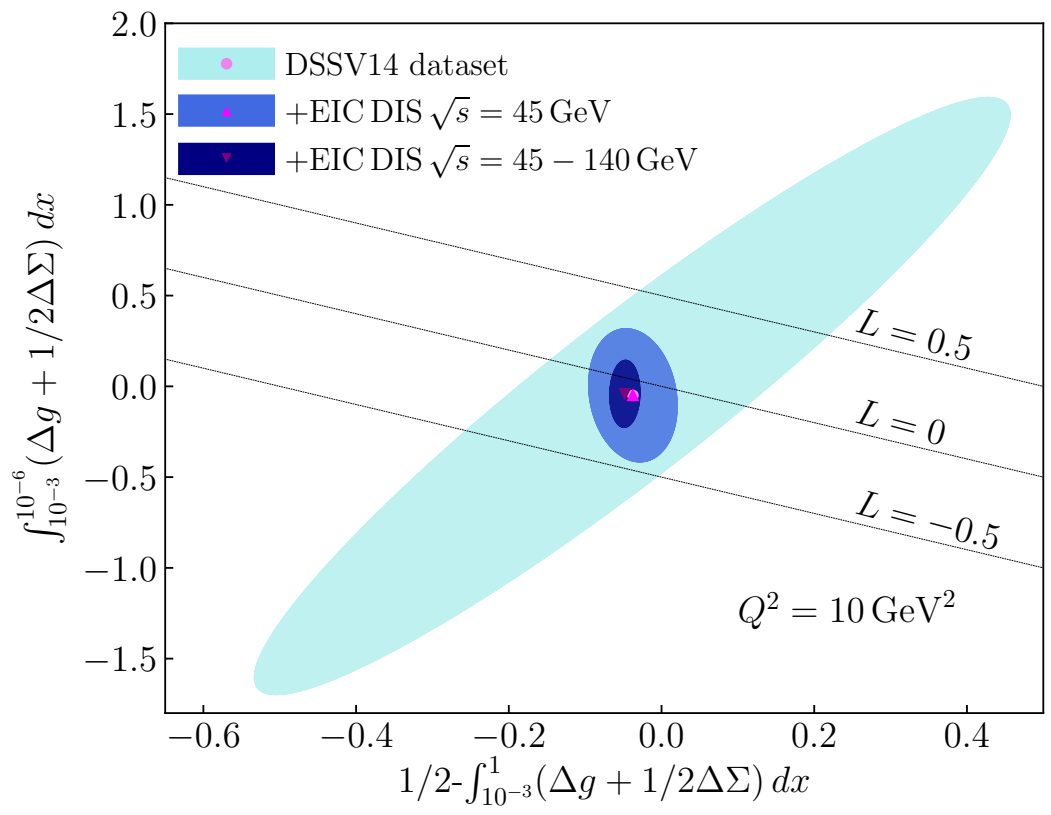

**Fig. 14.2.4** Room left for potential orbital angular momentum contributions to the proton spin after determining the quark and gluon spin contributions at EIC [3073, 3105].

*Sea quark spin and flavor distributions*
Equally important are the spin distributions of the sea quarks in the nucleon, which exhibit flavor dependence ($\Delta\bar{u} \neq \Delta\bar{d} \neq \Delta\bar{s}, \Delta s \neq \Delta\bar{s}$) and attest to flavor-dependent non-perturbative interactions with the valence quarks in the nucleon. Present results on the flavor dependence from fixed-target semi-inclusive DIS and the RHIC $W^{\pm}$ production data show large uncertainties (see Sec. 10.2). EIC will determine the polarized sea quark distributions and their flavor dependence through polarized $ep$ semi-inclusive DIS, taking advantage of large phase space for fragmentation (see Fig. 14.2.5) [3073, 3105]. Complementary information will come from DIS on the neutron measured with polarized $^3$He beams. The determination of the flavor structure of the polarized sea will also indirectly improve the extraction of the gluon spin distribution and the spin sum rule (separation of flavor singlet and non-singlet distributions). EIC will also enable novel studies of the flavor structure of the unpolarized sea using charged-current DIS.

*Orbital angular momentum*
The total angular momentum of quarks and gluons in the nucleon can be expressed through integrals of the GPDs (see Sec. 10.3). This representation provides alternative insight into the role of orbital angular momentum in the nucleon spin decomposition. The GPDs appear in the amplitudes of hard exclusive processes (deeply virtual Compton scattering or DVCS, meson production) and can be accessed experimentally in this way; see Refs. [3185–3187, 4442] for a review. While the hard exclusive processes sample the GPDs in a restricted domain of variables that is not sufficient for evaluating the angular momentum sum rule, it is possible to establish a connection in the context of dynamical models of the GPDs, or a global analysis recruiting other data. EIC will advance this program through measurements of DVCS and meson production over a wide kinematic range; the same data will be used for the 3D spatial imaging (see below).

*Energy-momentum tensor*
Other global properties follow from the nucleon matrix elements of the QCD energy-momentum tensor and can be studied by using the connection with scattering processes. The D-term of the energy-momentum tensor, which expresses certain mechanical properties of the nucleon, appears as a subtraction constant in the dispersion relations for the DVCS amplitude and can be extracted from fits to DVCS data with minimal model dependence; see Refs. [2825, 4443] for a review. EIC measurements will allow one to precisely determine the D-term, taking advantage of the wide energy coverage of the data in evaluating the dispersion integral.

The trace of the QCD energy-momentum tensor contains important information on the emergence of the nucleon mass from QCD; see Refs. [4444–4446] for recent discussion and review. The breaking of scale invariance through the UV divergences of QCD implies that the trace is proportional to the twist-4 gluonic operator $G^2_{\mu\nu}$ (trace anomaly). An interesting question is how much this effect contributes to nucleon mass. It has been suggested that the twist-4 gluonic operator could be accessed in exclusive photo/electroproduction of heavy quarkonia at near-threshold energies [4447–4449]; however, this connection relies on the questionable assumption of vector meson dominance [4450], and the mechanism of heavy quarkonium production near threshold is a matter of current research and discussion; see e.g. Refs. [4451–4454]. EIC will contribute to this program by measuring exclusive $\Upsilon$ production near threshold (measuring $J/\psi$ production near threshold is very challenging with the high-energy collider) [3105, 4455]. With a future theoretical framework, these data will constrain the gluonic structure of the nucleon at the higher-twist level and contribute to the understanding of the origin of its mass.

*Pion and kaon structure*
The spontaneous breaking of chiral symmetry in QCD generates most of the light hadron masses and governs the effective dynamics of strong interactions at low energies (see Secs. 6.2 and 6.3). The pion and kaon are the Goldstone bosons of chiral symmetry, and their quark/gluon structure provides insight into the microscopic mechanism of symmetry breaking. The EIC will pursue a program of pion and kaon structure studies using exclusive scattering to measure the pion/kaon form factor, and peripheral deep-inelastic $ep$ scattering to probe the pion/kaon partonic structure [3105, 4456]. The extraction of pion/kaon structure from $ep/eA$ scattering data requires theoretical methods that can be tested with the EIC data.

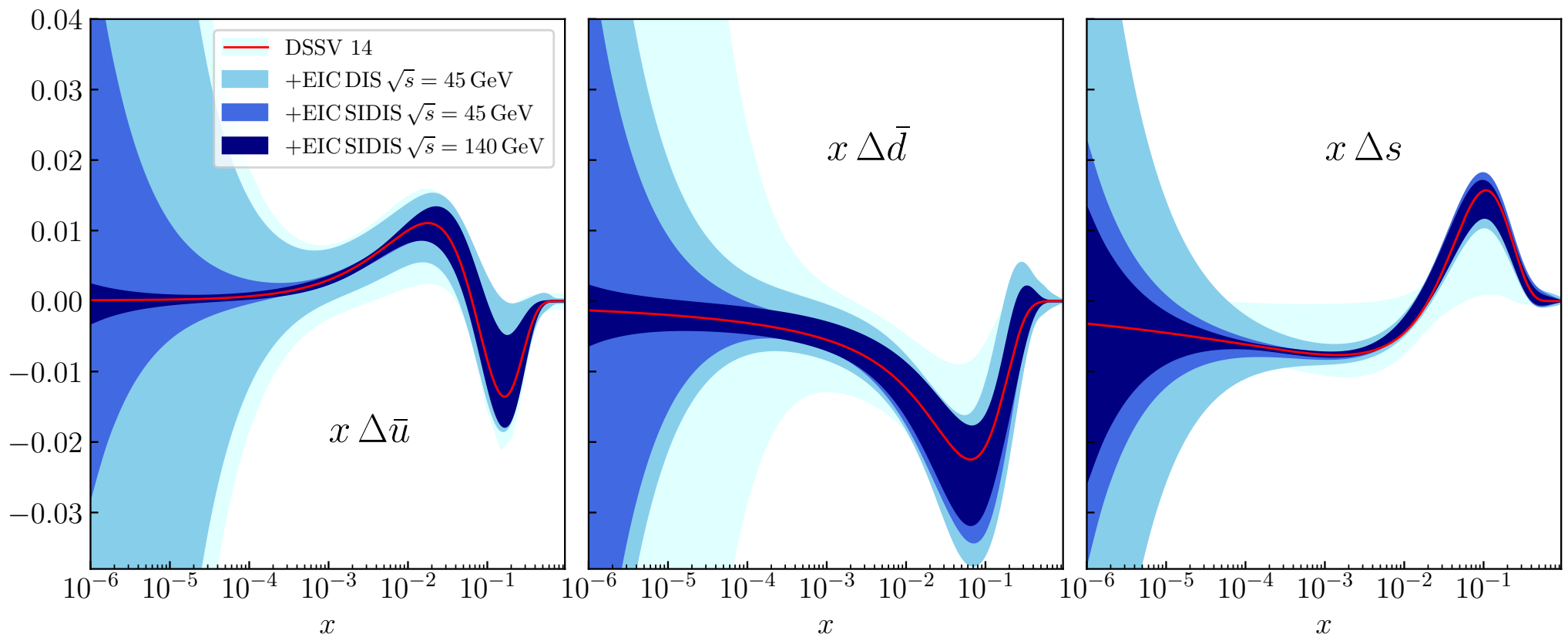

**Fig. 14.2.5** Flavor decomposition of the polarized sea quark distributions in the proton with projected EIC SIDIS data [3073, 3105]. Similar results are obtained in studies using other PDF parametrizations [3105].

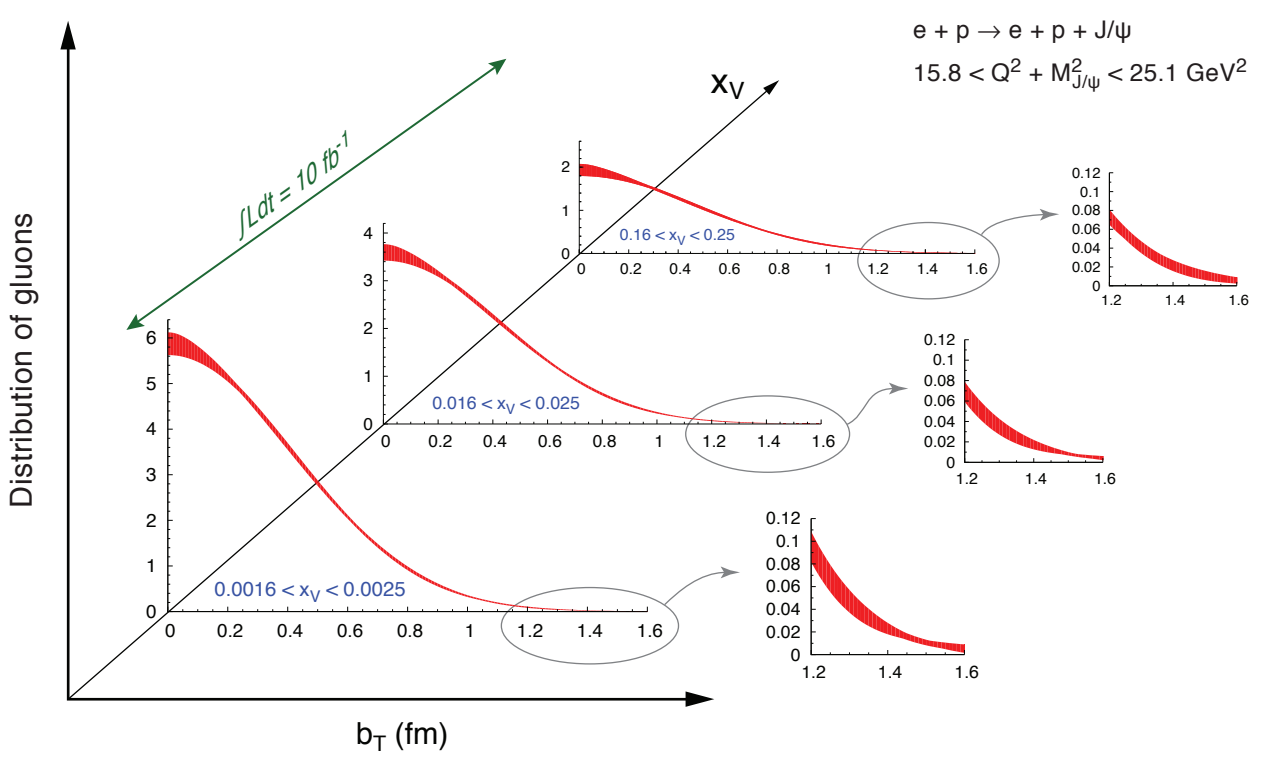

**Fig. 14.2.6** Transverse spatial distribution of gluons in the nucleon determined from projected EIC exclusive $J/\psi$ electroproduction data [820, 3105].

### 14.2.2 Multidimensional imaging of hadrons and nuclei

Another basic objective is to understand and visualize hadrons as extended systems in space. This can be accomplished using the concepts of GPDs (transverse coordinate space imaging) and TMDs (momentum space imaging), which provide a spatial representation consistent with the relativistic and quantum nature of the dynamics (see Sec. 10.4). Measurements at EIC will allow one to employ these concepts in regions where they are practically applicable and realize their full potential.

*Transverse quark/gluon imaging of the nucleon*
The transverse spatial distributions of quarks/gluons and their dependence on $x$ represent the size and shape of the nucleon in QCD (see Sec. 10.4 and Refs. [3186, 3187] for a review) and contain rich information about dynamics (parton diffusion, chiral dynamics). Exclusive $J/\psi$ electro- and photoproduction at EIC provides a clean probe of the gluon GPD and will determine transverse spatial distribution of gluons from the $t$-slope of the differential cross section (see Fig. 14.2.6) [820, 3105, 3128]. DVCS offers direct access to the quark GPDs and their spin dependence, and provides indirect information on the gluon GPD through NLO effects and $Q^2$ evolution [3105, 4457]. The combination of both will allow for an accurate determination of the quark and gluon GPDs, including validation of the factorized approximation and tests of the universality of the extracted structures. Essential capabilities for this program are the kinematic coverage (probing quarks/gluons down to $x \sim 10^{-3}$, $Q^2$ dependence in electroproduction), luminosity (differential measurements, e.g. $t$-dependence at fixed $x$ and $Q^2$), far-forward proton detection (recoil, exclusivity), and beam polarization (polarization observables). The results can be synthesized in comprehensive transverse images of nucleon structure (see Sec. 10.4).

*Transverse quark/gluon imaging of nuclei*
The same concepts and measurements can be used to create images of nuclei ($A > 1$) in terms of quark/gluon degrees of freedom. Such studies provide new insight into nuclear structure (comparison of $q - \bar{q}$, $q + \bar{q}$, and $g$ spatial distributions in the nucleus) and a new avenue for studying nuclear modifications of partonic structure (comparison of nucleus with non-interacting ensemble of nucleons) [4458–4464]. EIC measurements of coherent $J/\psi$ [4465] and $\gamma$ production on nuclei probe the nuclear GPDs, $\langle A'|\mathcal{O}_{\text{partonic}}|A\rangle$, and can be analyzed in the same way as measurements on the proton. The identification of coherent nuclear scattering events places strong demands on the far-forward detection system and is a matter of on-going development (active detection of recoiling nucleus for light nuclei; veto detection of breakup for heavy nuclei) [4466]. A new aspect of light nuclei is that they cover a variety of spins (Spin-1 D, Spin-1/2 $^3$He, Spin-0 4He) and express it in the GPD structure and the transverse images.

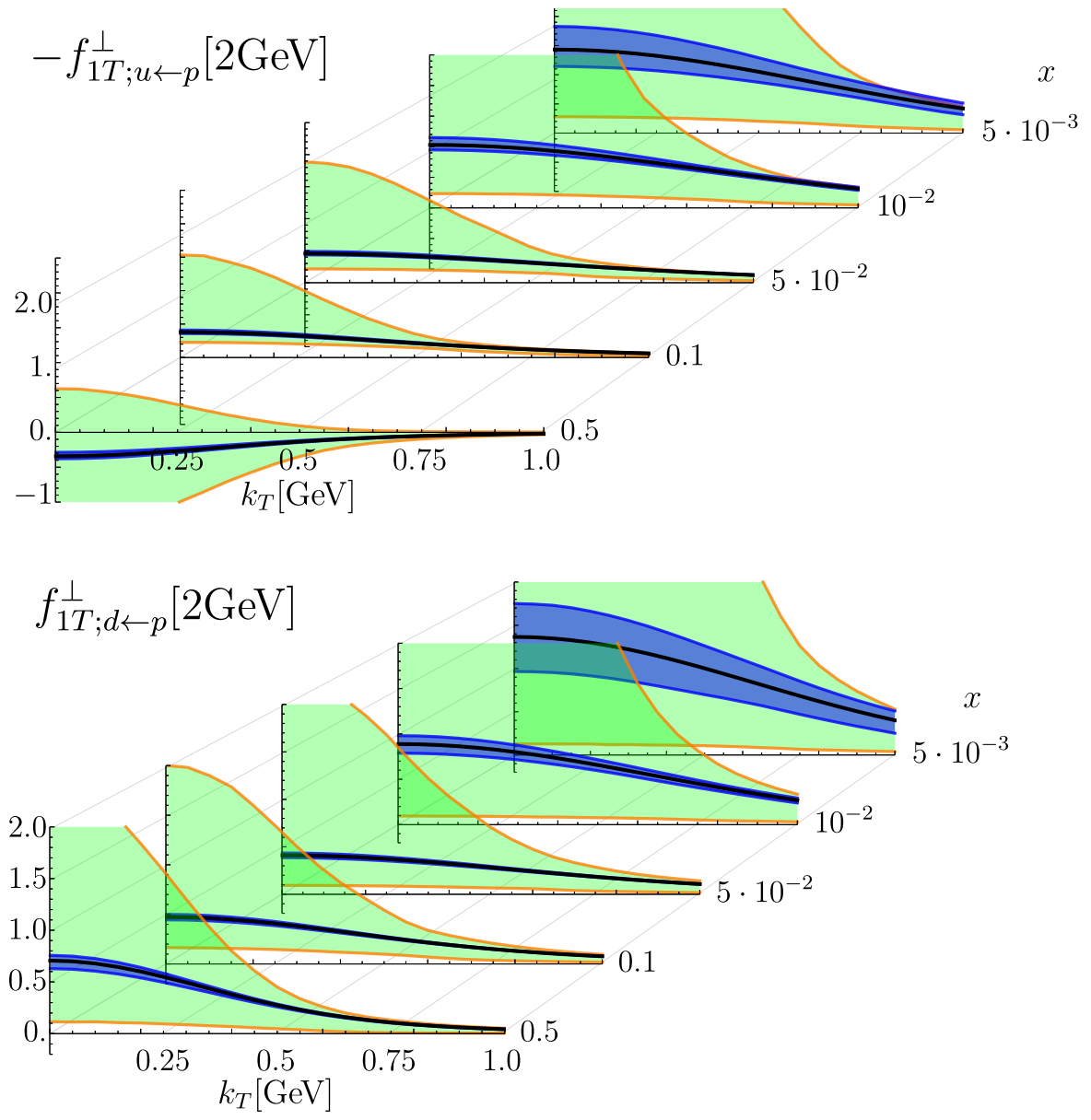

**Fig. 14.2.7** Expected impact of EIC pseudodata on the determination of the $u$ and $d$ quark Sivers distribution [3105]. Green bands: Present uncertainties [3244]. Blue: Uncertainties when including EIC pseudodata [3105].

*Evolution of TMD distributions*
The theoretical formulation of the transverse momentum dependence of partons has made substantial progress in the last decade (see Sec. 10.4). Factorization and renormalization predict a distinctive scale and rapidity dependence of the TMD distributions, generated by gluon radiation with Sudakov suppression, and described by the CSS evolution equations. The EIC will allow one to test these predictions in measurements of semi-inclusive hadron production $\gamma^* + N \to h + X, h = \pi, K, \ldots$ The wide kinematic range accessible with EIC is essential for observing the logarithmic dependencies implied by the evolution equation and separating perturbative and nonperturbative dynamics (see Fig. 14.2.2). The results will provide crucial insight into the theory of CSS-type radiation and its applicability to DIS-type processes.

*Spin-orbit correlations in TMD distributions*
An interesting feature of the transverse momentum dependence of partons is that it is correlated with the nucleon and parton spin, giving rise to observable spin-orbit effects that provide insights into nucleon structure and color field dynamics (see Sec. 10.4). At EIC these effects can be studied in measurements of hadron production (semi-inclusive DIS, jets) with polarized electron and proton beams. Measurement of the Sivers and Collins asymmetries are possible with the transverse proton beam polarization readily available at collider (see Fig. 14.2.7) [3105]. The results will provide extensive information on orbital angular momentum, final state interactions, and the quark transversity distributions in nucleon.

### 14.2.3 Nuclear high-energy scattering in QCD

High-energy scattering on nuclei ($A > 1$) provides a wealth of information on the effective dynamics emerging from QCD at various energy and distance scales. Depending on the kinematic regime, such processes reveal the QCD substructure of individual nucleon interactions (intermediate/large $x$) or coherent QCD phenomena involving the entire nucleus (small $x$). The EIC will realize the first electron-nucleus collisions in colliding beam experiments, combining the kinematic reach of colliding beams with the precision and control of electromagnetic scattering, and thus transform this field of study.

*Nuclear quark/gluon densities*
The nuclear PDFs describe the basic particle content of the nucleus in QCD degrees of freedom [4467–4470]. Comparison with the PDFs of an ensemble of non-interacting nucleons provides insight into nucleon interactions and coherent phenomena. Many aspects of the nuclear PDFs are still poorly known, esp. the nuclear gluons and the charge and flavor dependence of the nuclear quarks at $x \lesssim 0.1$. The EIC will determine the nuclear PDFs using inclusive DIS on a broad range of nuclei [3105, 4471]. The nuclear gluon PDF will be determined indirectly through the $Q^2$ dependence of the nuclear DIS cross section (DGLAP evolution), using the wide kinematic coverage available with the collider. It will also be determined directly through measurements of heavy flavor production in nuclear DIS, taking advantage of the high production rates and next-generation reconstruction capabilities provided by the EIC. The results will establish whether the nuclear gluons are suppressed at $x > 0.3$ like the valence quarks (EMC effect), and whether they are enhanced at $x \sim 0.1$ (anti-shadowing) as suggested by theoretical arguments; both phenomena reveal aspects of the QCD substructure of nucleon interactions.

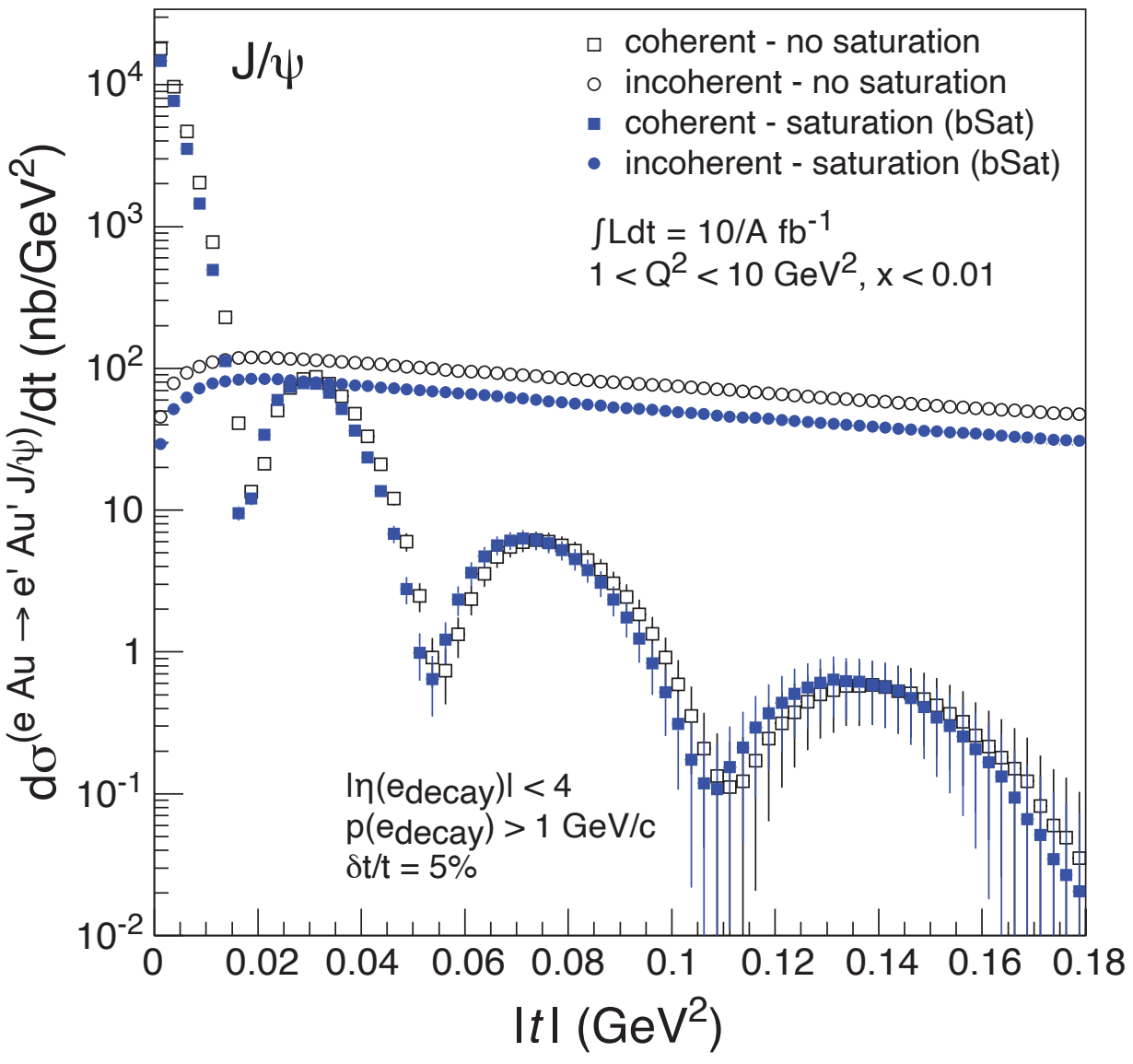

**Fig. 14.2.8** Differential cross section of coherent and incoherent $J/\psi$ production on a Au nucleus, as a function of the momentum transfer $t$ [3105, 4472, 4473]. The diffraction pattern in coherent scattering is sensitive to the impact parameter dependence of shadowing and saturation effects in the nuclear gluon density.

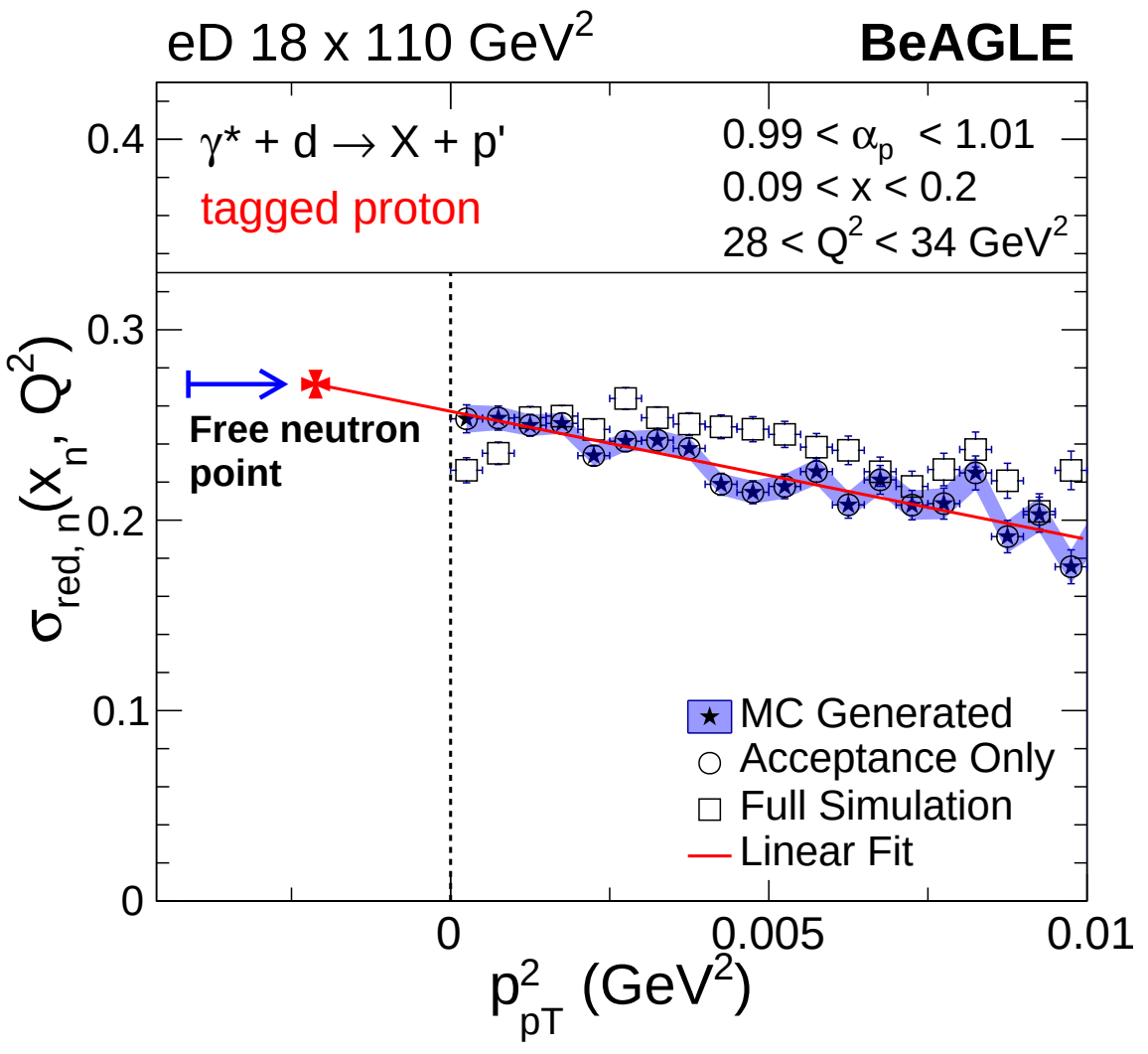

**Fig. 14.2.9** Simulation of free neutron structure extraction through DIS on the deuteron with proton spectator tagging at EIC [4487]. The neutron reduced cross section is measured as a function of the spectator proton transverse momentum $p_{pT}^2$ and extrapolated to the "free neutron point" at $p_{pT}^2 < 0$, corresponding to $pn$ configurations of infinite size.

*Shadowing and saturation*

In high-energy scattering at $x \ll 0.1$ the coherence length of the process becomes larger than the size of the nucleus, and the high-energy probe interacts with all nucleons along its path. In this regime the gluons "seen" by the probe can no longer be attached to individual nucleons but represent a property of the whole nucleus, giving rise to striking new phenomena. Shadowing is the reduction of the leading-twist nuclear gluon density resulting from destructive interference of amplitudes with gluons attached to different nucleons; see Ref. [4474] for a review. Saturation is the appearance of a new dynamical scale in the form of the transverse density of gluons per area. It emerges from nonlinear QCD evolution equations including gluon recombination [4475–4480] and can be used as the basis of an effective field theory description of strong interactions at small $x$ – the Color Glass Condensate [3276], leading to many interesting predictions; see Refs. [4481–4483] for reviews. Both phenomena are connected, as shadowing reduces the gluon density and modifies the expected $Q_{\rm sat}^2 \sim A^{1/3}$ scaling of the saturation scale. Exploring these phenomena will be a prime task of the EIC.

Basic information will come from the behavior of the nuclear gluon PDF at $x \ll 0.1$ [3105]. More detailed tests of the small-$x$ gluon dynamics will be possible with dijet and dihadron production [3281, 4484, 4485]. Further insight can be gained from studies of diffractive scattering on nuclei. Measurements of coherent heavy vector meson production on nuclei probe the impact parameter dependence of the shadowing and/or saturation effects through the diffraction pattern in the momentum transfer $|t|$ (see Fig. 14.2.8) [3105, 4472, 4473]. Similar studies can be performed in measurements of coherent inclusive diffraction on nuclei [4486]. The EIC provides the necessary energy for diffractive scattering, and the ability to identify coherent processes through forward detection.

*Nuclear breakup and spectator tagging*

In high-energy scattering on light ions, detection of the nuclear breakup state provides information on the nuclear configuration present during the high-energy process [4488]. In the case of the deuteron, detection of the "spectator" proton identifies events with scattering on the neutron and fixes the relative momentum of the proton-neutron configuration. This can be used to select scattering in large-size nuclear configurations, where interactions are absent and the neutron is free [4489, 4490], or small-size configurations, where the $pn$ system strongly interacts and the partonic structure is modified (short-range nucleon-nucleon correlations) [4491]. The EIC will enable a program of high-energy scattering on the deuteron with proton or neutron spectator tagging. In the collider kinematics the spectator nucleon appears in the forward ion direction and is detected with far-forward detectors (magnetic spectrometer for protons, zero-degree calorimeter for neutrons) [3105]. The setup can be used to extract free neutron structure functions (see Fig. 14.2.9) [4487], study the configuration dependence of EMC effect, or explore short-range nucleon-nucleon correlations in deuteron breakup in diffractive scattering [4492].

### 14.2.4 Emergence of hadrons from QCD

Understanding hadronization – the emergence of hadrons from the energetic quarks/gluons produced in deep-inelastic processes – remains a major challenge of strong interaction physics. The hadronization process is "reciprocal" to the partonic structure of hadrons but much less understood theoretically, because it involves time-like momentum transfers and propagation over large distances, and methods based on imaginary-time (Euclidean) quantum field theory such as Lattice QCD are generally not applicable (see Sec. 4). Basic open questions are the time/distance scales of parton fragmentation and hadron formation; the role of non-perturbative dynamics (chiral symmetry breaking, vacuum fields; see Sec. 5.11), and the effects of the nuclear medium on the hadronization process. In addition to the scientific interest, these topics are of eminent practical importance for the development of event generators describing strong interaction dynamics in high-energy collisions (see Sec. 11.4).

*Fragmentation functions*
Basic information on the hadronization process is summarized in the quark/gluon fragmentation functions, describing the probability for single-inclusive hadron production by an energetic color charge; see Ref. [4493] for a review. While much information on the fragmentation functions has been extracted from $e^+e^-$ annihilation, $pp$ collisions, and fixed-target semi-inclusive DIS experiments, several features remain poorly known, such as the quark charge dependence (so-called unfavored vs. favored fragmentation), strangeness fragmentation and kaon production, and gluon fragmentation [3066, 4494–4496]. The EIC will determine the fragmentation functions from semi-inclusive DIS in $ep$ and $en$ scattering over a broad kinematic range [3105]. These measurements will be able to separate the quark charges in the initial state, extract the gluon through NLO effects, and study the $Q^2$ evolution of the fragmentation functions. The spin dependence of quark fragmentation will be investigated through measurements of $\Lambda$ fragmentation [4497]. Precise knowledge of the fragmentation functions will in turn improve the extraction of the flavor dependence of the quark/antiquark spin PDFs from polarized semi-inclusive DIS data.

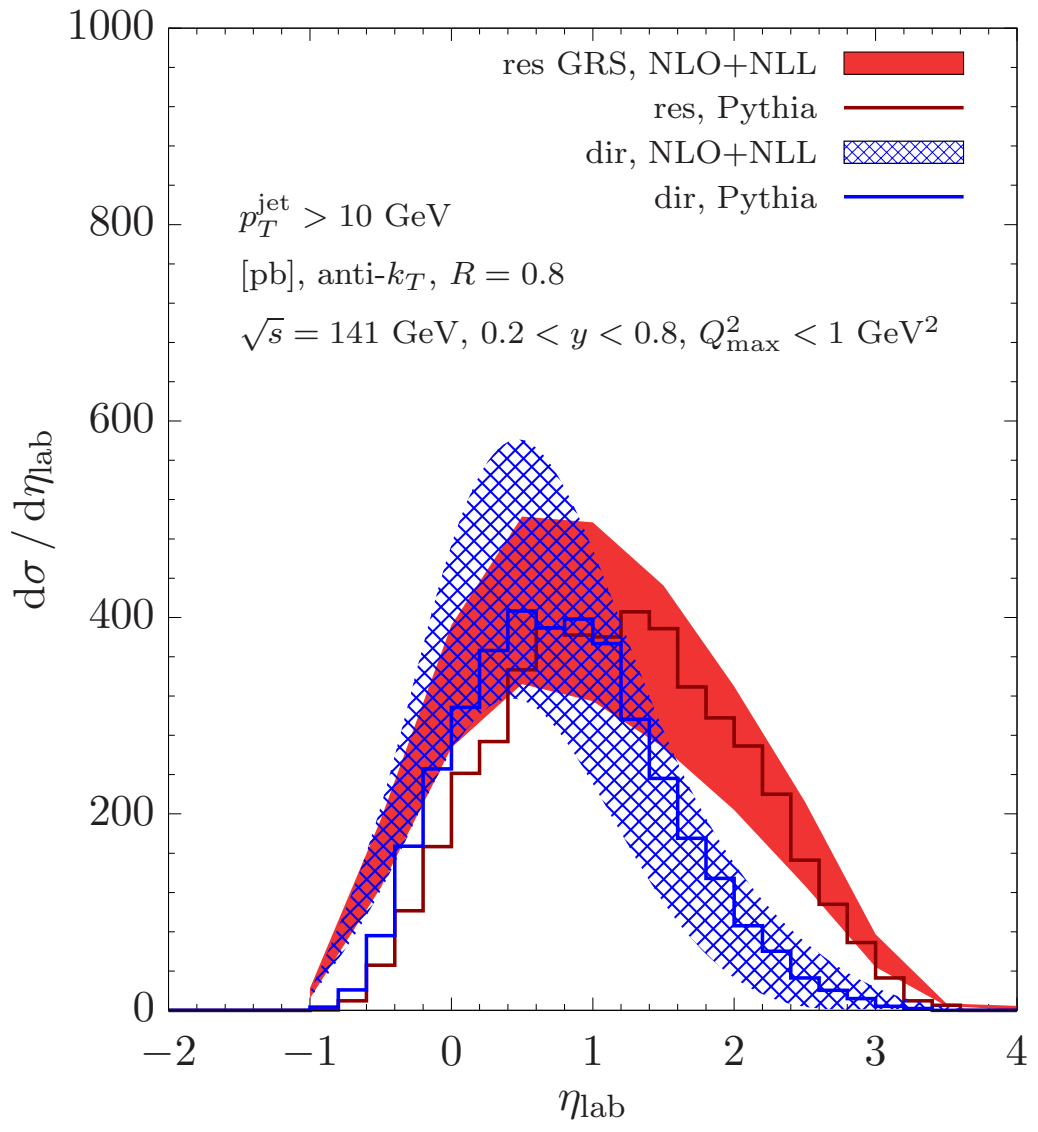

**Fig. 14.2.10** Inclusive production cross section of jets in photoproduction at EIC, as a function of the pseudorapidity $\eta$ in the laboratory frame (see Fig. 14.2.12) [3105, 4501].

*Dihadron correlations*
More detailed information on the fragmentation process comes from measurements of hadron correlations, described by the theoretical framework of dihadron fragmentation functions [4498–4500]. The EIC will measure dihadron fragmentation functions in DIS and allow for the new theoretical concepts to be applied and tested. The kinematic coverage provided by the EIC will ensure that the picture of independent fragmentation remains applicable even in multi-hadron measurements.

*Jets and heavy flavors*
An alternative view of the hadronization process is obtained by applying the concepts of jet physics, where one defines a system of collinear partons according to quantitative observable criteria without reference to non-perturbative fragmentation functions (see Secs. 6.4, 11.5 and Sec. 12). These concepts and methods have been developed for $pp/p\bar{p}$ scattering at hadron colliders (LHC, Tevatron) but can be extended to $ep$ scattering at EIC at lower energies. This extension opens up several new directions for studying the internal properties of jets and using them as a probe of partonic structure. In $ep$ collisions where the scattered electron is detected, it defines the jet energy and scale, and the concepts for leading jets can be applied to the DIS current jet with known initial conditions, providing new possibilities to test the dynamics [4502–4504]. In addition, jet substructure can be investigated [4501]. Jets can also be studied in $ep$ collisions where the scattered electron is not detected, or in $\gamma p$ collisions, where the jet transverse momentum serves as the hard scale (see Fig. 14.2.10 as an example). Particularly interesting are jets induced by heavy quarks, which remain stable under strong interactions and create distinct signals in the detector ($D, B$ meson decays). The EIC will support this program through a comprehensive set of measurements of leading jets, jet substructure, heavy flavor jets, and studies of partonic structure and TMD distributions using jets [3105]. This is a rapidly evolving field, where new theoretical methods will become available until the EIC experiments are performed.

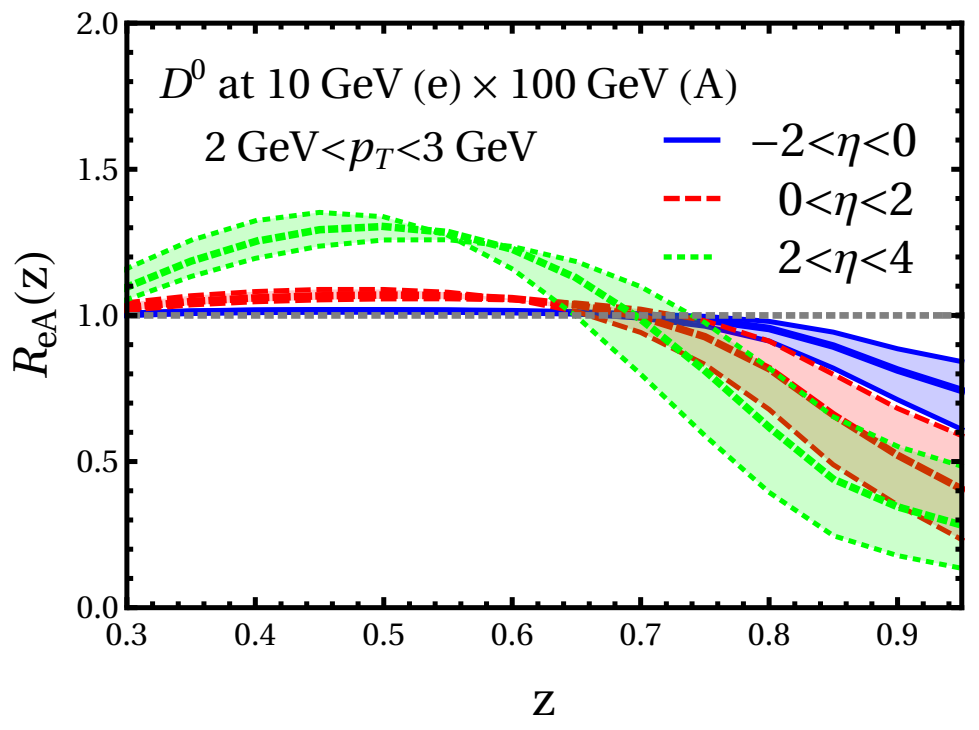

**Fig. 14.2.11** Medium modification of the $D^0$ production cross section expected at EIC, as a function of $z$, in different regions of pseudoradpidity $\eta$ [3105, 4508].

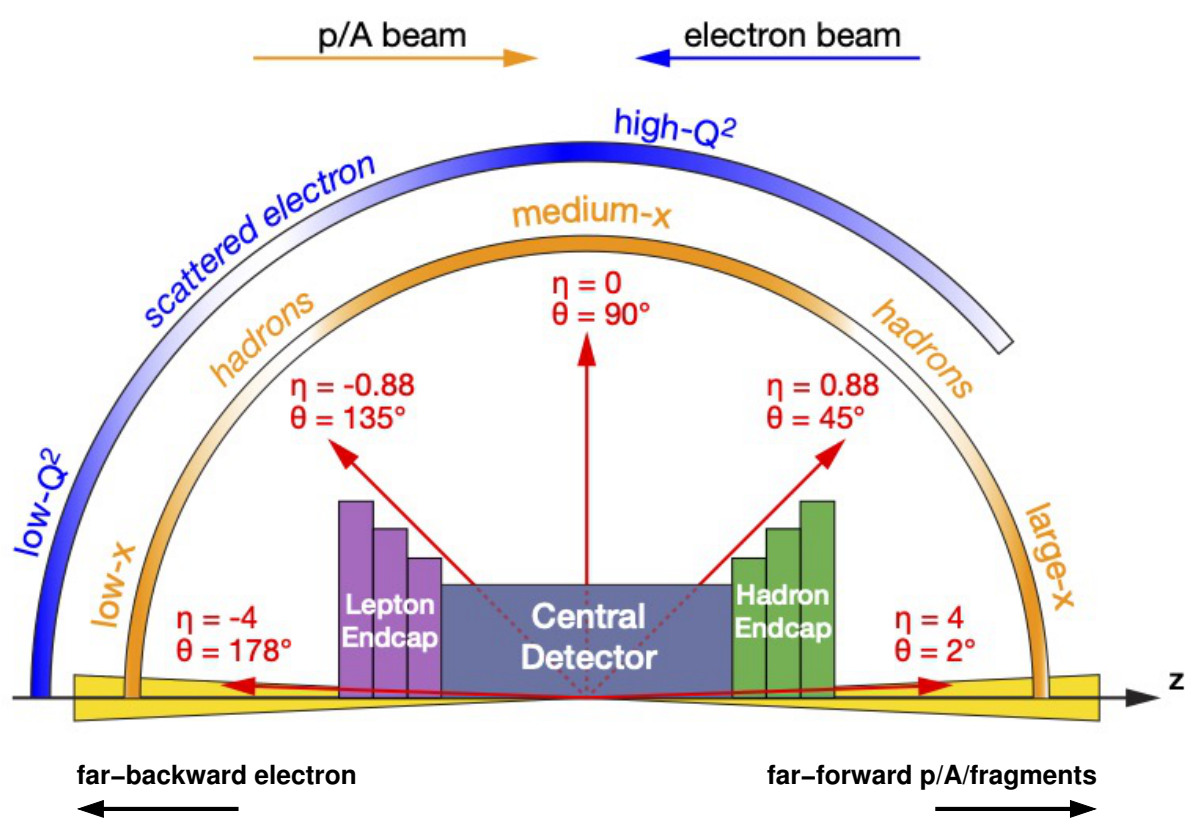

**Fig. 14.2.12** Schematic of the EIC detector concept.

*Target fragmentation*
Equally interesting is the hadronization of the target remnant in DIS processes (target fragmentation). It can be regarded as the materialization of a nucleon with a "hole" in its color wave function (created by the removed parton) and provides information on baryon number transport, multiparton correlations [4505], haronization dynamics, and spin-orbit effects. A framework for QCD analysis of target fragmentation is provided by the generalized factorization theorems [3951, 4506]. The EIC will enable a comprehensive program of nucleon target fragmentation studies, using the detectors in forward pseudorapidity region [3105]. Spin effects in target fragmentation can be studied using polarized proton beams and/or fragmentation into $\Lambda$ baryons [4507]. Important advantages of the collider compared to fixed-target experiments are that there is no material surrounding the target, and that the fragments move forward with a finite fraction of the proton beam momentum.

*Hadronization in medium*
The hadronization studies described above can be extended from $ep$ to $eA$ scattering, to investigate the influence of the nuclear medium on the hadronization process. The medium effects depend essentially on the energy of the fragmenting parton in the nuclear rest frame $E_h = z\nu$. The wide range of scattering energies available at EIC will allow one to move the fragmentation process "in" and "out" of the nucleus, enabling controlled and detailed studies of the medium effects. This will make it possible to test various hadronization models and determine the time/distance scale parameters. The study of nuclear final-state interactions will also improve the modeling of nuclear breakup in DIS processes, which in turn will help with the analysis of coherent nuclear scattering and spectator tagging. Particularly useful for the study of medium effects are heavy-quark probes (see Fig. 14.2.11 for an example) [4508, 4509].

*Hadron spectroscopy*
Hadron production in high-energy $ep/eA$ scattering at EIC can also be used for spectroscopy, complementing experiments using $pp$ and $e^+e^-$ scattering. Exotic heavy quarkonium states (XYZ states, see Secs. 8.5 and 8.6) can be produced in exclusive photo/electroproduction processes $\gamma^* + p \to M + N$. The production rates and reconstruction efficiency with the EIC detector are presently under study [3105, 4510, 4511]. At EIC, new possibilities arise from measurements of the spin density matrix elements of heavy vector states, target polarization observables, and the $Q^2$ dependence in electroproduction. These unique capabilities of the EIC could be used as the focus shifts from spectroscopy to investigations of the structure of exotic states.

### 14.2.5 Detectors and collaboration

The EIC science program requires a general-purpose detector with large acceptance and high resolution to reconstruct the scattered electron and the multiple different hadronic final states over a wide range of rapidities and energies/momenta. The physics requirements and detector concept are described in detail in the EIC Yellow Report [3105]. A schematic is shown in Fig. 14.2.12. The pseudorapidity region $-1 \lesssim \eta \lesssim 1$ is covered by the central "barrel" detector with a solenoidal magnetic field; the regions $-4 \lesssim \eta \lesssim -1$ and $1 \lesssim \eta \lesssim 4$ are covered by the "lepton endcap" and "hadron endcap" detectors; the detectors provide capabilities for tracking and vertex detection, electromagnetic and hadronic calorimetry, and particle identification. These systems capture the scattered electron and the final state produced by the struck parton in typical DIS events. The far-backward region (outgoing electron beam direction) is instrumented with a low-$Q^2$ electron tagger for photoproduction. The far-forward region (outgoing proton/ion beam direction) is equipped with an elaborate detection system for charged and neutral beam fragments, integrated in the interaction region, involving a magnetic dipole spectrometer with tracking detector for charged particles and a zero-degree calorimeter for neutral particles. This system provides essential capabilities for detecting far-forward protons and neutrons in exclusive/diffractive processes on the proton, spectator nucleons or nuclear fragments in scattering on nuclei, and coherent nuclear recoil. It presents a major challenge for design, integration, and engineering, and is critical for large part of the physics program. Further information on the EIC detector requirements and conceptual design can be found in Ref. [3105]. The technical design and formation of a detector collaboration are in progress. The addition of a second detector with complementary capabilities is planned as a future upgrade.

The EIC User Group is an international affiliation of scientists promoting scientific, technological, and educational efforts in the development of the EIC facility and science program. It presently has more than 1200 members from more than 250 institutions (laboratories, universities) worldwide. Resources and information about activities and events can be found on the webpages [4512].

## 14.3 J-PARC hadron physics

**Shunzo Kumano**

Hadron physics is the field to understand our visible universe, namely hadronic many-body systems from low to high densities, from low to high temperatures, and from low to high energies, in terms of fundamental particles of quarks and gluons and their interactions. With the significant developments of perturbative QCD during 50 years of QCD, asymptotic freedom and scaling violation are now basically understood. On the other hand, the nonperturbative region is still under investigations by phenomenological models and lattice QCD. One may note that at present lattice QCD cannot be applied to finite density systems, which makes it difficult to predict precisely hadronic and nuclear phenomena at low energies.

Although QCD is known as the correct theory of strong interactions, there are unexpected experimental discoveries of new hadronic and nuclear forms which were not predicted by theorists. Therefore, experimental projects are essential for a deeper understanding and for further developments of QCD beyond the 50-years history. The Japan Proton Accelerator Research Complex (J-PARC) as one of the flagship facilities in hadron physics should play a key role in hadron physics from the low to the medium-energy region, by supplying precise experimental information on new forms of matters, as illustrated in Fig. 14.3.1.

The J-PARC is located at Tokai in Japan. It is operated by both the High Energy Accelerator Research Organization (KEK) and the Japan Atomic Energy Agency (JAEA). J-PARC is responsible to coordinate the efforts of KEK and JAEA. KEK is in charge of nuclear and particle-physics projects by using the 30-GeV proton accelerator. J-PARC is a multi-purpose facility to investigate a wide range of scientific topics from life sciences to condensed-matter, nuclear, and particle physics [4513].

The J-PARC accelerator consists of a 400-MeV linac as an injector, a 3-GeV rapid-cycling synchrotron (RCS), and the 30 GeV main-ring synchrotron. The RCS accelerates the protons up to 3 GeV as shown in Fig. 14.3.2. Its beam pulses are delivered mostly to the materials and life-science experimental facility, and a small portion is injected to the main ring. The protons are accelerated to 30 GeV in the main ring, and they are de-livered to the neutrino experimental facility and the hadron experimental facility. The beam reached an energy of 30 GeV in 2008, its power was increased towards the design intensity of 0.75 MW. In the near future, we expect to have about 1 MW for the neutrino facility and about 100 kW for the hadron one [4513].

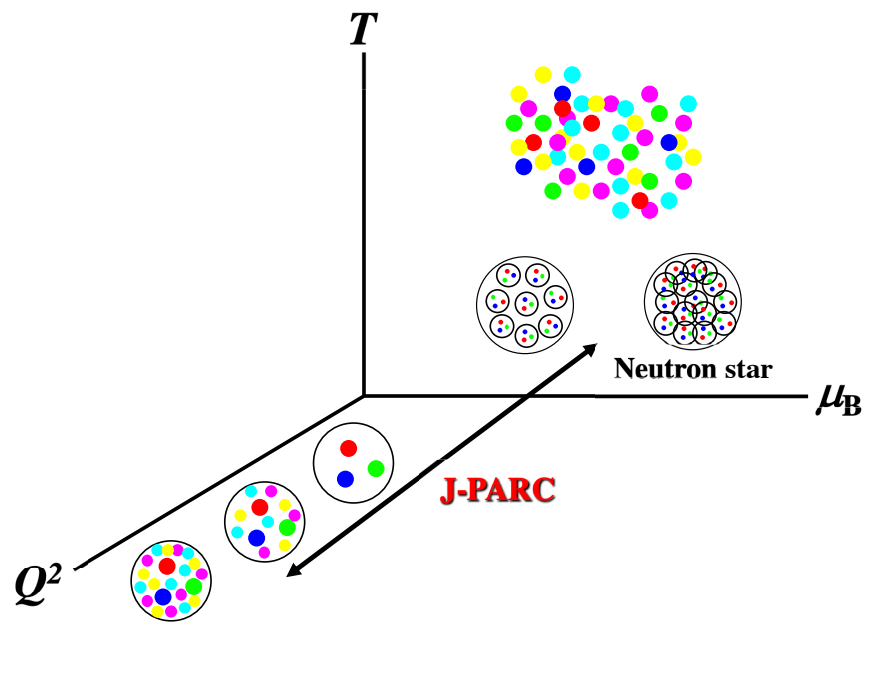

**Fig. 14.3.1** QCD phase diagram and J-PARC hadron projects.

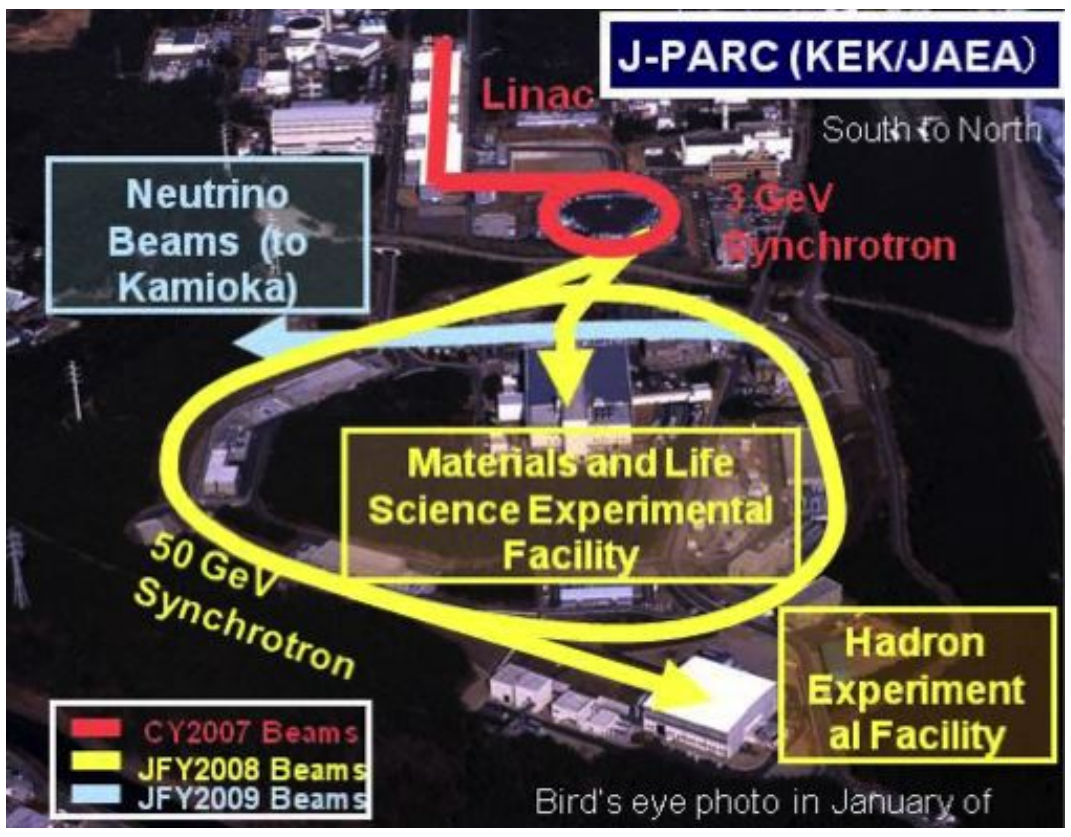

**Fig. 14.3.2** Aerial view of J-PARC [4513].

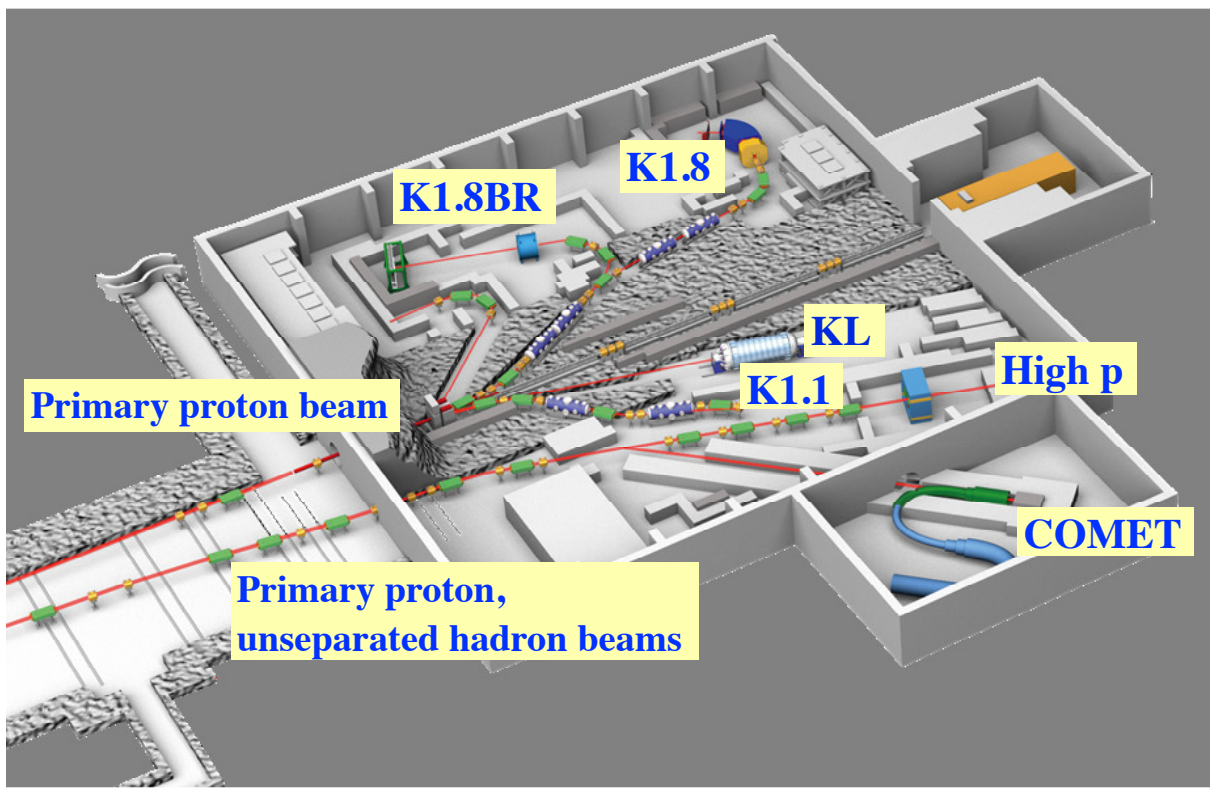

**Fig. 14.3.4** J-PARC hadron hall.

The J-PARC is the most intense accelerator above the multi-GeV energy region. Its aim is to investigate a wide range of nuclear and particle physics by using secondary beams of kaons, pions, antiprotons, neutrinos, and muons as well as the primary proton beam as shown in Fig. 14.3.3. There are particle physics experiments on neutrino oscillations, lepton-flavor violation, $g-2$, rare kaon decays, and the neutron electric-dipole moment to search for physics beyond the Standard Model. Since the purpose of this report is to discuss QCD-related topics, only the hadron-physics projects are explained.

### 14.3.1 J-PARC hadron facility

The layout of the J-PARC hadron facility is shown in Fig. 14.3.4 with the hall size of 60 m width and 56 m length. Nuclear and particle physics experiments are done by using the primary proton beam and secondary beams of pions, kaons, antiprotons, and muons. Unique points of this proton accelerator facility are (1) high intensity and (2) intermediate energy. The first point indicates the decisive advantage when secondary beams or the primary proton beam are used for precision experiments. Intermediate energies are important since low-energy hadron projects can bridge the transition region from hadrons to quarks and gluons by variation of the momentum transfer in the QCD phase diagram, as illustrated in Fig. 14.3.1. The facility should be able to contribute to the development of QCD from the non-perturbative region to the transition region, then to the perturbative one.

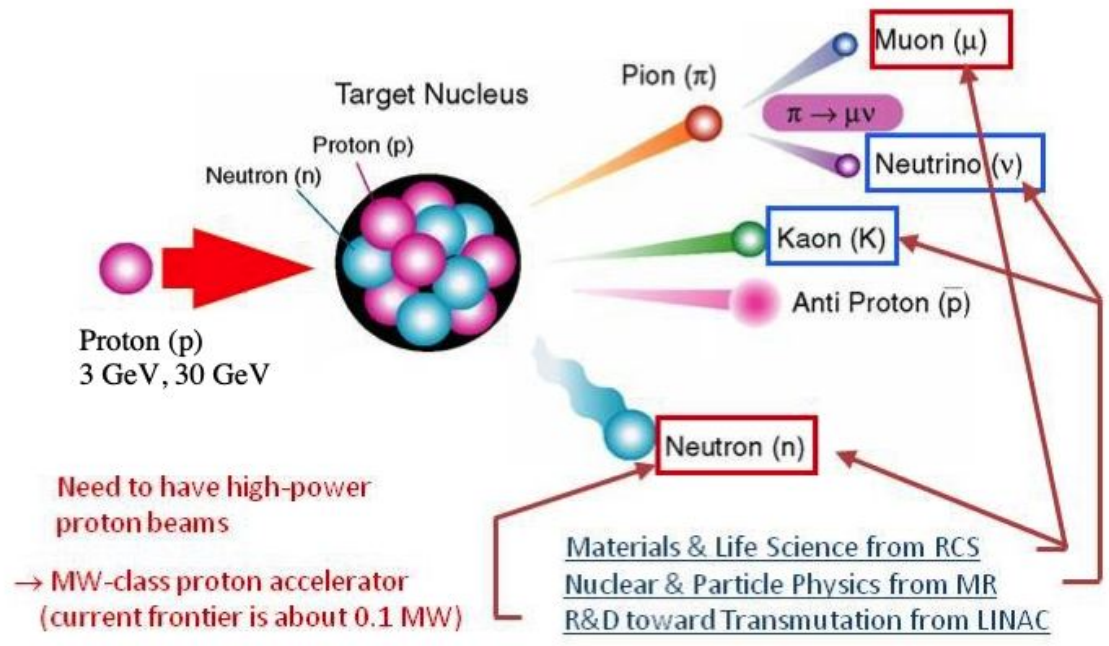

**Fig. 14.3.3** Secondary beams at J-PARC [4513].

Particle-physics experiments in the hadron hall are lepton-flavor violation (COMET) and rare kaon decays (KL). The COMET experiment uses muons from the decays of pions produced by 8 GeV proton collisions on a production target. COMET will search for the lepton-flavor violation process, the conversion of muons into electrons in the field of a nucleus, $\mu^- + A \to e^- + A$. The KOTO experiment uses the neutral-kaon beamline KL for measuring the frequency of the CP-violating decay $K_L^0 \to \pi^0 \nu\bar{\nu}$. These projects are intended to find a signature beyond the Standard Model in particle physics.

Hadron-physics experiments are done at the beamlines K1.8, K1.8BR, K1.1, and High p, see Fig. 14.3.4 [4514]. The K1.1 beamline is yet to be constructed. The K1.8 beamline supplies kaons with the momentum of about 1.8 GeV and is used to study hypernuclei, e.g. $\Xi$ hypernuclei, by $(K^-, K^+)$ reactions. One may note that the cross section of $p(K^-, K^+)\Xi$ reaches a maximum at a momentum of 1.8 GeV. The K1.8BR is a branch line of K1.8 to supply kaons with low momenta of 0.7-1.1 GeV. The cross section of the quasi-elastic reaction $K^-N \to \bar{K}N$ maximizes at 1 GeV momentum, so that this beamline is intended to study $\bar{K}N$ interactions and kaonic nuclei by $(K^-, N)$ reactions with light nuclei.

The K1.1 beamline supplies kaons with momentum around 1.1 GeV for measurements of $\Lambda$ hypernuclei. Because of the space interference between the K1.1 and high-p beamlines, K1.1 experiments will be done after the first stage of the high-p experiment. These strange nuclear physics projects are explained in Sec. 14.3.3.

The high-momentum beamline provides 30 GeV protons and unseparated hadrons up to 20 GeV. The beam of unseparated hadrons, to be prepared in the near future, consists mainly of pions. The first experiment in this beamline will measure hadron mass modifications in a nuclear medium to study chiral-symmetry breaking and hadron-mass generation (see Sec. 14.3.4).

Then, charmed baryon spectroscopy will be investigated by $(\pi^-, D^{*-})$ reactions. This experiment intends to find di-quark degrees of freedom, which are not easily found in hadrons consisting of light quarks only, as explained below in Sec. 14.3.5. The hadron tomography project will be performed together with this spectroscopy experiment by studying generalized parton distributions (GPDs) as discussed in Sec. 14.3.6. This experiment is set up to find the origin of hadron masses and spins by the tomography technique. In future, separated hadron beams could become possible; an extension plan of this hadron hall is discussed in the next subsection 14.3.2.

More details of each hadron project are explained in the following sections. The first major experiment will study the role of strangeness in nuclear physics. The next experiment is devoted to hadron mass modifications in the nuclear medium, and then the charmed-baryon project will start. The GPD tomography experiment is expected to join this baryon-spectroscopy project. The scope of the hadron physics projects at J-PARC is thus expanding in the near future.

Furthermore, there is a significant interest to build a new heavy-ion facility at J-PARC to investigate the phase diagram in the low-temparature and high-density region in contrast to the kinematical region of RHIC and LHC. There are interesting topics in cold and dense matters, such as the end point of the phase transition and color superconductor, as explained in Sec. 14.3.7.

When the hadron program will be completed, the heavy-ion facility will be built. This is expected in the 2030's. J-PARC will then become a leading hadron accelerator facility. It will investigate QCD in a wide kinematical region and for a wide range of topics, from strangeness in nuclear physics, charmed-baryon spectroscopy, nucleon structure at intermediate energies, and quark-hadron matter.

### 14.3.2 Hadron-hall extension

The current hadron hall cannot accommodate enough projects in nuclear and particle physics. The experimental hall size and beamlines are much smaller than, for example, the BNL-AGS facility. The efficient way for utilizing the full ability of the J-PARC is to expand its space and to build additional beamlines.

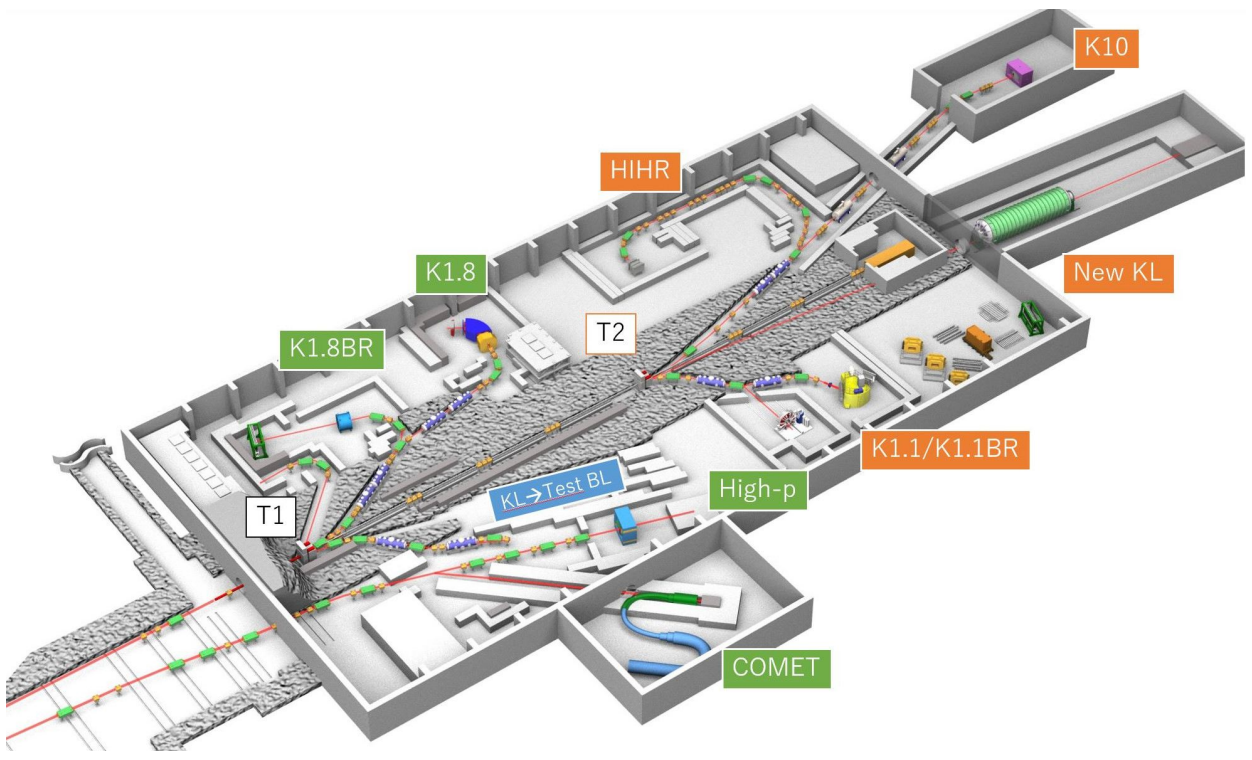

**Fig. 14.3.5** Extension plan of the J-PARC hadron hall [4515].

This extension project, as shown in Fig. 14.3.5, was proposed together with the current hall [4515]. The area of the hall becomes twice larger to accommodate new experiments. A new production target T2 will be prepared. The beamlines with orange color are new ones in the extended hall. They are designed for the following topics.

1. HIHR
   This HIHR (High Intensity High Resolution) beamline is intended for precision spectroscopy of $\Lambda$ hypernuclei through $(\pi^\pm, K^+)$ reactions by using high-intensity and high-resolution charged pions up to 2 GeV momentum with an excellent momentum resolution of $10^{-4}$ and a missing-mass resolution of a few hundred keV.
2. K10
   This beamline will be used to investigate $S = -3$ strangeness physics and charm physics by using separated secondary hadron beams of high-momentum $(2 - 10\,\text{GeV})$ charged kaons and anti-protons.
3. K1.1
   This beamline will be prepared for physics with strangeness $S = -1$ using charged kaons with momenta of less than 1.2 GeV. The branch beamline K1.1BR is for the stopped kaon experiments.
4. KL2
   The frequency of the kaon rare decay $K_L^0 \to \pi^0 \nu \bar{\nu}$ will be measured. It may provide a hint for New

Physics beyond the Standard Model by using this high-intensity neutral kaon beamline.

This extension project was selected as one of top priority projects of KEK in 2022. After the financial approval, it will take 6 years for its construction. When it is realized, it will provide excellent opportunities for nuclear and particle physicists to create innovative fields with unprecedented precision. The following major physics purposes are presently considered for this extension project: (1) precise spectroscopy of hypernuclei to understand neutron stars, (2) novel aspects of charmed baryons, and (3) New Physics beyond the Standard Model. The details of the topics (1) and (2) are discussed in Sec. 14.3.3 and Sec. 14.3.5, respectively, along with past J-PARC experiments on hypernuclei.

Because the J-PARC is an intermediate-energy facility, the current scope of physics could be extended in future, for example, by including projects of high-energy QCD such as on nucleon structure, exotic hadrons by the constituent counting rule, and color transparency [4516]. Furthermore, if the heavy-ion accelerator will be built [4517], the unexplored cold and dense region of the QCD phase diagram will be investigated.

Here, we briefly summarize the major purposes related to the hadron-hall extension including possible future topics.

*Establishing the role of strangeness in nuclear physics*
The nuclear physics without strangeness has been established by precise information on the fundamental $NN$ potentials from abundant experimental measurements on $NN$ scatterings and deuteron properties, whereas the $YN$ scattering information is in a poor situation. The J-PARC will supply precise data on the fundamental $YN$ interactions and also properties of hypernuclei. We expect that spectroscopy of hypernuclei could become a precision field by the J-PARC experiments.

*Applications to neutron stars*
The existence of strangeness inside neutron stars would make their equations of state much softer. This is in conflict with astrophysical observations of neutron-star masses. By establishing strangeness nuclear physics, we expect that this issue will be solved.

*Creation of a di-fermion field in hadron physics*
The di-fermion physics has been investigated in quantum many-body systems, especially condensed-matter physics. In hadron physics, the color superconductor, for example, is investigated in such a context. The J-PARC intends to create a new di-fermion field by the spectroscopy of the charmed baryons.

*Emergence of hadron masses and spins*
Hadron masses and spins are fundamental physics quantities to constitute our visible universe. However, their origins are not understood easily from quark and gluon degrees of freedom. They should originate as emergent phenomena of nontrivial quark-gluon dynamics within hadrons. These should be clarified by the J-PARC projects on hadron-mass modifications in nuclear medium and by hadron tomography via GPDs.

*Understanding cold and dense QCD matters*
From the RHIC and LHC, the high-temparature region of the QCD phase diagram has been investigated and evidence for quark-gluon-plasma formation was found. J-PARC will clarify the cold and dense region, where interesting phase properties, such as the end point of the phase transition and color superconductor, are theoretically expected.

### 14.3.3 Strangeness nuclear physics

Major properties of stable nuclei are now relatively well understood, whereas unstable nuclei are still under investigations especially in connection with the nucleosynthesis in astrophysics. One of the major purposes of the J-PARC hadron program is to investigate nuclei by including new flavor degrees of freedom, strangeness and charm [4514, 4515].

Under the flavor SU(3) symmetry, nucleons and a part of hyperons constitute a flavor octet. Two-baryon interactions are decomposed into symmetric (under the exchange of baryons) states $\mathbf{27} \oplus \mathbf{8} \oplus \mathbf{1}$ and antisymmetric ones $\mathbf{10} \oplus \mathbf{10}^* \oplus \mathbf{8}$ as

$$\mathbf{8} \otimes \mathbf{8} = \mathbf{27_S} \oplus \mathbf{10_A} \oplus \mathbf{10^*_A} \oplus \mathbf{8_S} \oplus \mathbf{8_A} \oplus \mathbf{1_S}. \quad (14.3.1)$$

Nucleon-nucleon ($NN$) interactions provide information only on the $\mathbf{27_S}$ and $\mathbf{10^*_A}$ states. Therefore, hyperon interactions need to be investigated to understand the other terms and to find possible new hadronic many-body systems. These new interactions are relevant in neutron stars. This nuclear-physics project with strangeness has the following advantages [4518].

1. SU(3) flavor symmetry and new interactions
   The new interaction terms $\mathbf{10_A}$, $\mathbf{8_S}$, $\mathbf{8_A}$, and $\mathbf{1_S}$ can be investigated by the hyperon ($Y$) interactions. In general, $YN$ interactions are expected to be weaker than the $NN$ ones, so that new forms of baryonic many-body systems should be created.
2. Probe of short-range interactions
   Since the pion isospin is 1 and the $\Lambda$ isospin is 0, the $\pi\Lambda\Lambda$ coupling constant vanishes. Because of its low mass, the pion contributes to the long-range part of the baryon interactions. Without the pion contribution, medium- and short-range baryon interactions should become more apparent when compared to the $NN$ case.

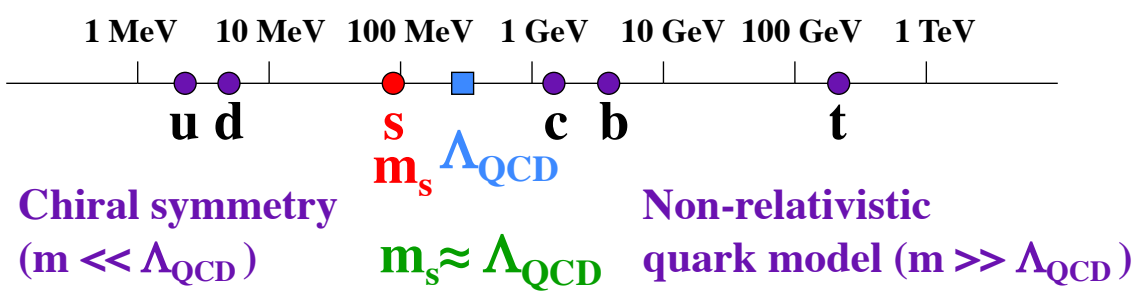

**Fig. 14.3.6** Strangeness as a probe of QCD dynamics.

3. Probe of QCD dynamics
The quark masses and the QCD scale parameter $\Lambda$ are shown in Fig. 14.3.6. We notice that the strange-quark mass is of the order of the scale parameter. This fact suggests an advantage that the strange quark could be a good probe of QCD dynamics. However, it may also indicate difficulties for describing hadrons with strangeness.
4. New forms of hadronic matters
Ordinary nuclei consist mainly of up and down quarks. The interactions of hyperons or cascade particles with nucleons are still unexplored. With strangeness, new forms of nuclei should be created such as $\bar{K}NN$, and so on. Another important topic is the possible existence of a H dibaryon with isospin 0, spin 0, and strangeness $-2$. It corresponds to the term $\mathbf{1_S}$ in Eq. (14.3.1).
5. Probe of deep regions in nuclei
The Pauli exclusion plays an important role in nuclear physics. Although nuclei are strongly-interacting systems with nucleons close together, they are often described by a non-interacting Fermi gas model or an independent particle model. It is justified by solving the Bethe-Goldstone equation. Hyperons do not suffer from such an exclusion effect, which indicates the advantage of probing deep regions of nuclei, as the shell structure should become obvious as visualized in Fig. 14.3.7 [4515, 4519].
6. Equation of state for neutron stars
Neutron-star physics has significantly developed recently due to new astrophysical experiments and observations of gravitational waves. In the inner high-density region of the neutron stars, the reactions $p+e^- \to \Lambda+\nu_e$ and $n+e^- \to \Sigma^- +\nu_e$ could occur because the changes of the Fermi energies of neutrons, protons, and electrons exceed the mass gap of the reactions. The equation of state of neutron stars should be significantly softened by the possible existence of hyperons, which contradicts the neutron-star observations. The appearance of hyperons in the neutron stars is affected by the details of hyperon interactions, which are investigated at J-PARC.

We introduce some of the major experimental results on strangeness in nuclear physics from J-PARC.

*Charge symmetry breaking*
Charge symmetry is taken as granted as a good symmetry for ordinary nuclei as typically shown in mirror nuclei with exchange of a proton and a neutron. For example, the binding energy difference between $^3$He and $^3$H is merely 0.07 MeV after removing QED effects. However, a significant breaking was found by the E13 experiment at J-PARC. The $1^+$ excited state of $^4_\Lambda$He was produced in the $^4\text{He}\,(K^-,\pi^-)^4_\Lambda\text{He}$ reaction with a 1.5 GeV $K^-$ beam. Then, by a measurement of the $\gamma$ rays for the $1^+ \to 0^+$ transition, a $1.406 \pm 0.002 \pm 0.002$ MeV energy spacing was found. With other measurements, the spectra of $^4_\Lambda$He and $^4_\Lambda$H are compared in Fig. 14.3.8 [4520]. The binding energy difference between $^4_\Lambda$He and $^4_\Lambda$H was $0.35\pm0.05$ MeV, which indicates a significant charge-symmetry breaking in hypernuclei. It provided a valuable information on the nature of $\Lambda$N interactions which are different from the NN ones. Theoretically, the breaking is considered to come from $\Lambda - \Sigma^0$ mixing.

*Double $\Lambda$ hypernuclei*
One of the major purposes of J-PARC program on hypernuclei is to investigate strangeness $-2$ systems. The J-PARC-E07 experiment was done at the K1.8 beamline with the $K^-$ beam of 1.8 GeV. By using nuclear emulsions tagged by the $(K^-, K^+)$ reaction, the double-$\Lambda$ hypernucleus $_{\Lambda\Lambda}$Be was found [4521]. It is produced as $^{10}_{\Lambda\Lambda}$Be by $\Xi^-+^{16}$O $\to$ $^{10}_{\Lambda\Lambda}$Be $+^4$He$+t$,
as $^{11}_{\Lambda\Lambda}$Be by $\Xi^-+^{16}$O $\to$ $^{11}_{\Lambda\Lambda}$Be$+^4$He$+d$, or
as $^{12}_{\Lambda\Lambda}$Be$^*$ by $\Xi^-+^{16}$O $\to$ $^{12}_{\Lambda\Lambda}$Be$^*+^4$He$+p$,

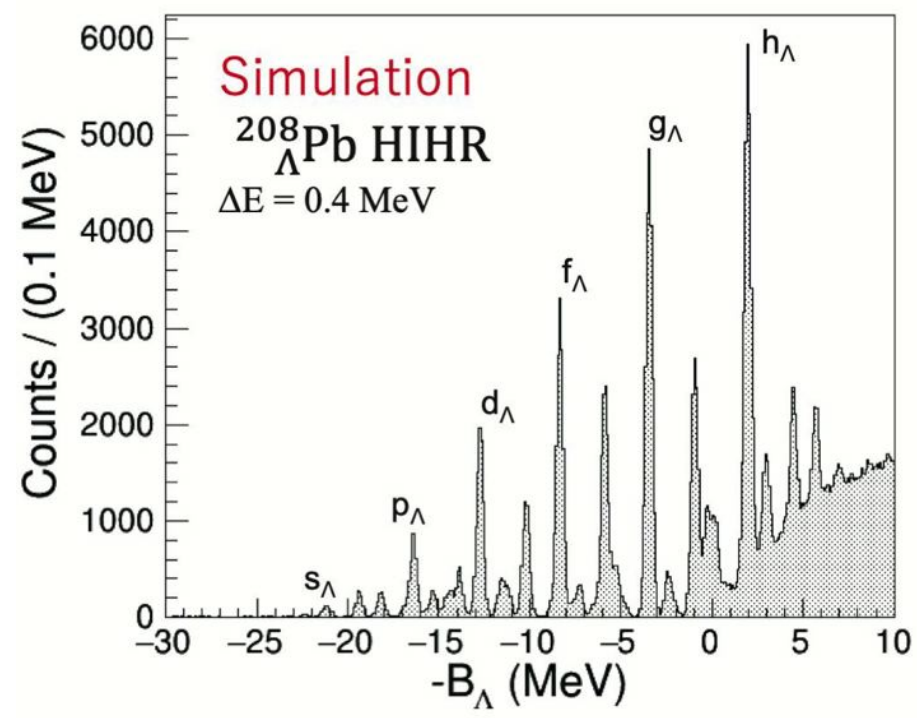

**Fig. 14.3.7** Simulation for the $\Lambda$ binding energy spectra of $^{208}_{\Lambda}$Pb for the hadron-extension program [4519]

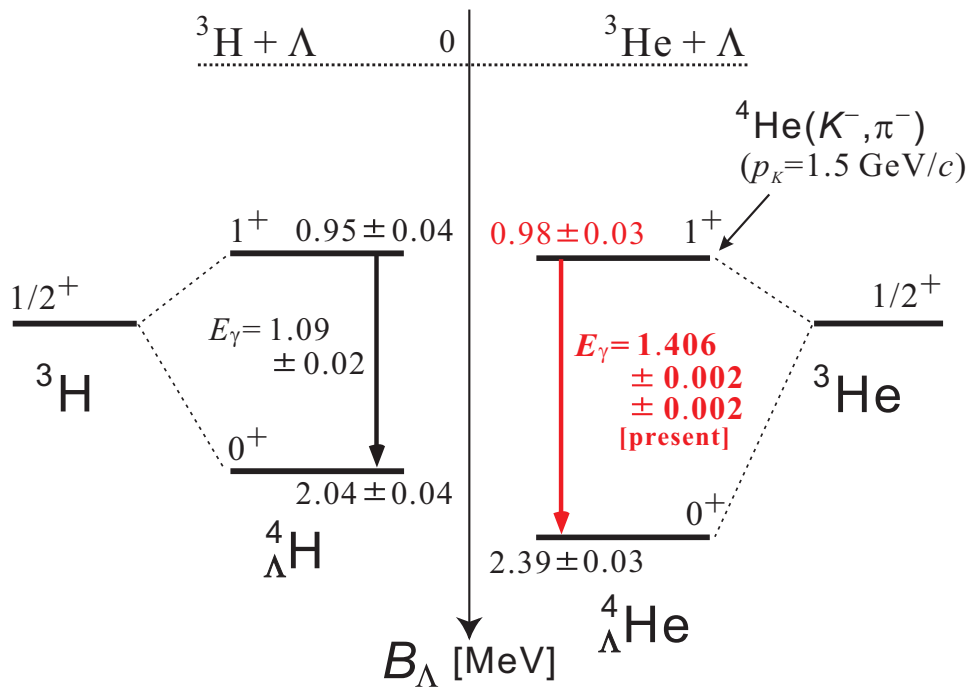

**Fig. 14.3.8** ${}^4_\Lambda$He and ${}^4_\Lambda$H spectra [4520]. (Used with the copyright permission of American Physical Society)

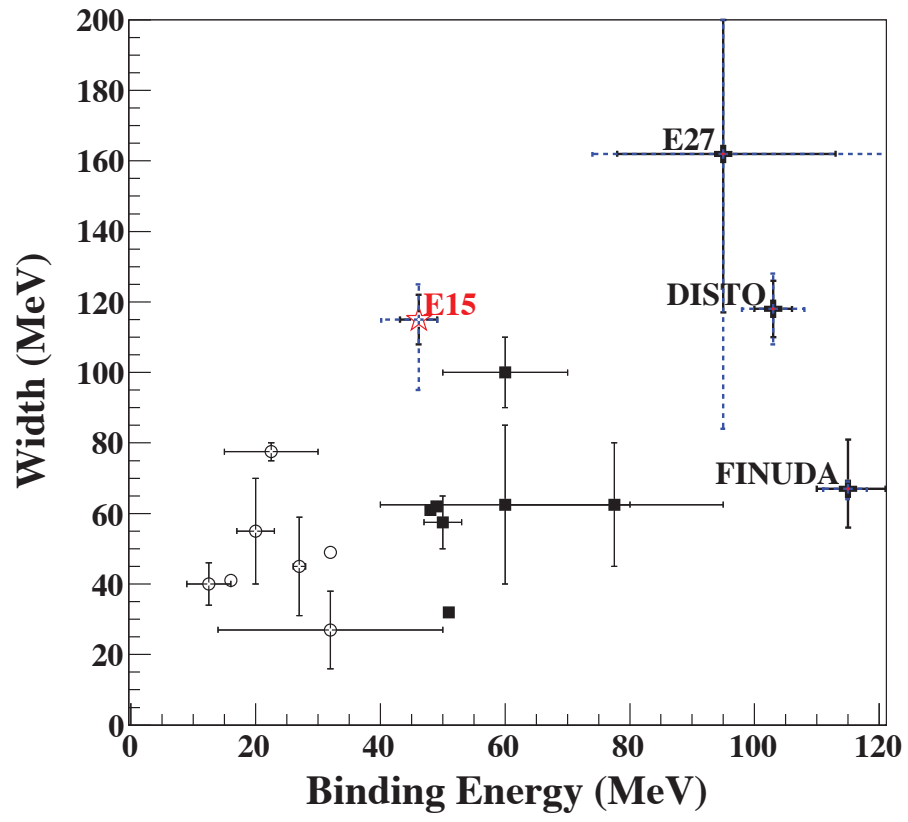

**Fig. 14.3.9** Situation for the $K^-pp$-bound state. [4514]. (Used with the permission of the Elsevier Science.)

and the binding energy of two $\Lambda$ hyperons is 15.05±0.11 MeV, 19.07±0.11 MeV, or 13.68±0.11 MeV, respectively. This result improves our understanding of the $\Lambda\Lambda$ interaction and double-strange hypernulcei.

*Ξ hypernuclei*

The J-PARC-E07 collaboration used the 1.81 GeV $K^-$ beam for observing the reaction $\Xi^- + {}^{14}\text{N} \rightarrow {}^{10}_{\Lambda}\text{Be} + {}^{5}_{\Lambda}\text{He}$. From the measurements, the $\Xi^-$ binding energy in the $\Xi^-$-${}^{14}$N system was determined to 1.27±0.21 MeV [4522]. From the experimental data and theoretical calculations, the energy level of the $\Xi^-$ is interpreted as $1p$ state; the $\Xi N$-$\Lambda\Lambda$ coupling must be weak.

Next, $\Xi^-$ capture was studied in the $\Xi^-$-${}^{14}$N system. Two events were found by analyzing KEK-E373 and J-PARC-E07 data signaling deep $\Xi^-$ bound states [4523]. One event from the reaction

$$\Xi^- + {}^{14}\text{N} \rightarrow {}^{5}_{\Lambda}\text{He} + {}^{5}_{\Lambda}\text{He} + {}^{4}\text{He} + \text{n}$$

yields a binding energy in the ${}^{14}$N nucleus of $B_{\Xi^-} = 6.27 \pm 0.27\,\text{MeV}$. The other event in

$$\Xi^- + {}^{14}\text{N} \rightarrow {}^{9}_{\Lambda}\text{Be} + {}^{5}_{\Lambda}\text{He} + \text{n}$$

yields $B_{\Xi^-}$ given by either $8.00 \pm 0.77\,\text{MeV}$ or $4.96 \pm 0.77\,\text{MeV}$, depending on the final-state ${}^{9}_{\Lambda}$Be nucleus which can be in the ground or an excited state. These binding energies are larger than the preceding value 1.27 MeV; likely, these events come from the $1s$ state of the $\Xi$ hypernucleus ${}^{15}_{\Xi}$C.

*Kaonic nuclei*

Kaonic nuclei are new forms of hadronic many-body systems with strangeness. Since $\Lambda(1405)$ can be considered as a $\bar{K}N$ molecule state, a few nucleon systems with a kaon should exist as bound states. The J-PARC-E15 collaboration used the K1.8BR beamline for measuring the reaction $K^- + {}^3\text{He} \rightarrow \Lambda + p + n$ with a kaon momentum of 1 GeV. In the $\Lambda p$ invariant mass spectrum, a clear peak was observed. It indicates a kaonic $\bar{K}NN$ nucleus with a binding energy $B_K = 42 \pm 3(\text{stat.})^{+3}_{-4}(\text{syst.})\,\text{MeV}$ and the decay width $\Gamma_K = 100 \pm 7(\text{stat.})^{+10}_{-9}(\text{syst.})\,\text{MeV}$ [4524]. The current situation is shown in Fig. 14.3.9 for energies and widths of possible $K^-pp$ bound states. The experimental data are shown with the collaboration names, and the other points are theoretical calculations. As it is obvious, the world data do not agree with each other and they are also different from the theoretical results, so that further J-PARC experiments are needed for clarifying the situation.

The J-PARC-E62 collaboration used the $K^-$ beam with 900 MeV momentum at the K1.8BR beamline. The negative kaons were stopped in a liquid-helium target [4525]. They obtained the energies and widths of the $3d \rightarrow 2p$ transition X-rays of kaonic ${}^3$He and ${}^4$He atoms with 10 times higher accuracy than previous data. On the other hand, using the $K^-$ beam with the momentum 1.8 GeV at the K1.8 beamline, the J-PARC-E05 collaboration measured the missing-mass spectrum of ${}^{12}\text{C}(K^-, p)$ and observed a quasi-elastic peak from $K^-p \rightarrow K^-p$ [4526]. Then, they extracted differential cross sections of the $K^-p$ elastic scattering. These experimental measurements impose a constraint on theoretical models of kaonic nuclei.

$\Sigma^\pm p$ scattering cross sections

Good data were not available for hyperon-nucleon and hyperon-hyperon scattering. So far, these interactions had been investigated mainly within hypernuclear models. This approach makes it difficult to establish hypernuclear physics as a precision field on the same level as the $NN$-interaction and ordinary nuclear physics. Furthermore, hyperon interactions are also essential for applications to neutron stars. Now, the situation is changing due to new results on $\Sigma p$ scattering data from J-PARC.

First, $\Sigma^- p$ elastic scattering data were reported for a $\Sigma^-$ momentum range from 470 to 850 MeV by the J-PARC-E40 collaboration [4527]. A $\pi^-$ beam in the K.18 beamline with a momentum of 1.33 GeV impinged on liquid hydrogen target, where $\Sigma^-$ particles were produced in the reaction $\pi^- p \to K^+ \Sigma^-$. 4500 events were identified and differential cross sections for $\Sigma^- p$ elastic scattering were determined. Second, this collaboration reported differential cross sections of $\Sigma^- p \to \Lambda n$ in the $\Sigma^-$ momentum range from 470 to 650 MeV [4528]. About 100 events were identified and angular distributions were obtained for the first time. Third, differential cross sections were measured for the $\Sigma^+ p$ elastic scattering in the momentum range from 0.44 to 0.80 GeV [4529]. The $\pi^+$ beam with the momentum 1.41 GeV was used to produce $\Sigma^+$ in the reaction $\pi^+ p \to K^+ \Sigma^+$ . About 2400 $\Sigma^+ p$ elastic scattering events were identified, and the $^3S_1$ and $^1P_1$ phase shifts were obtained from the precise data for the first time.

These data are valuable for building the full baryon-baryon interactions of the SU(3) multiplets, see Eq. (14.3.1). With such experimental information, the Nijmegen-type baryon models should become much accurate and lead to a better understanding of hadronic and nuclear many-body systems and neutron stars.

### 14.3.4 Hadrons in nuclear medium

Hadron masses in nuclear medium will be measured by using the primary protons of 30 GeV at the high-momentum beamline as the J-PARC-E16 experiment [4530]. This project is intended to investigate the role of chiral symmetry in hadron properties. The study is thus related to a clarification of the origin of hadron masses. The discovery of the Higgs particle clarified the origin of the masses of quarks and leptons. However, this does not imply that masses of our nature, for example, the nucleon mass, are understood. The “god” particle cannot create the hadron masses.

Since the nucleon mass is defined by the matrix element of $\int d^3x T^{00}(x)$, where $T^{\mu\nu}$ is the energy-momentum tensor, it is decomposed into four terms [4531]:

$$M = \text{quark energy} + \text{gluon energy} + \text{quark mass} + \text{trace anomaly}. \qquad (14.3.2)$$

Current masses of up- and down-quarks are very small, so their simple summation is much smaller than the nucleon mass. To understand the origin of hadron masses, it is necessary to clarify the complicated emergence of mass from confined quarks and gluons. The clarification of this mass emergence is one of top priority projects for building electron-ion colliders for physics in 2030’s [4532, 4533]. In the mass decomposition of Eq. (14.3.2), the trace anomaly term and the gluon condensate could play an important role in hadron masses. These will be investigated by the $J/\psi$ production process at charged-lepton accelerator facilities, such as the JLab, CERN-AMBER, and EICs. On the other hand, this topic has already been investigated by spacelike GPDs at JLab and CERN-COMPASS and also by timelike GPDs at KEKB. In fact, gravitational form factors of a hadron were already extracted from actual experimental data [4534]. This E16 experiment is intimately related to these world projects.

The original idea for generating the hadron masses is to use chiral-symmetry breaking. It gives rise to a non-vanishing $\langle \bar{q}q \rangle$ condensate [4535, 4536], which is called scalar quark condensate. It plays a role of an order parameter for the chiral phase transition. It cannot be directly measured in experiments, so that we have to rely on actual observables. One of such quantities are vector-meson masses in a nuclear medium, they will be measured by the E16 experiment. There are theoretical estimates on their mass modifications from the partial restoration of chiral symmetry inside the nuclear medium [4535, 4536].

As for the experimental side, there were already measurements on the masses of vector mesons. For example, the KES-PS with the primary 12-GeV proton beam provided data on the processes $p + A \to V + X$ ($V = \rho,\ \omega,\ \phi\ \to e^+ e^-$) [4537, 4538]. They indicated 9% mass shifts for $\omega$ ($\rho$) and 3% for $\phi$-mesons, respectively. From a comparison of theoretical models with the mass-modification data, one can find that the quark condensate provides an important clue for mass generation.

Precise measurements are expected for these mass modifications from the E16 experiment at J-PARC. The first physics run will be taken with C and Cu targets with limited detector acceptance, and then more measurements will be done with the H and Pb targets and full detector acceptance. The expected outcome for the $\phi$ meson spectrum from the reaction $p + A \to \phi + X$ for the first run with a copper target and 30 GeV protons was simulated using GEANT4, see Fig. 14.3.10 [4539]. The momentum distribution of the $\phi$ meson was evaluated by using the code JAM (Jet AA Microscopic transport model) [4540], and the mass-modification parameter deduced by KEK-E325 [4538] was used. The figure is shown for slowly moving $\phi$ mesons ($\beta\gamma < 1.25$), the mass resolution is expected as 5.8 MeV. In this slow-$\phi$ case, nuclear medium effects are large and the spec-trum is modified significantly as shown in Fig. 14.3.10. The difference between the simulated data and the red spectrum should come from nuclear medium effects. As the $\phi$ velocity becomes larger, the spectrum modification becomes smaller. From these simulated data, the mass of $\phi$-meson at rest in a nuclear medium can be deduced. In Fig. 14.3.11, the mass is extracted by using a theoretical dispersion relation. The KEK-E325 data is shown for comparison. The KEK data was taken at only one point and the errors are large. We notice that the J-PARC data are much more accurate even at the first stage and that four data points will enable us to extrapolate the momentum dependence for determining the $\phi$ mass at zero momentum.

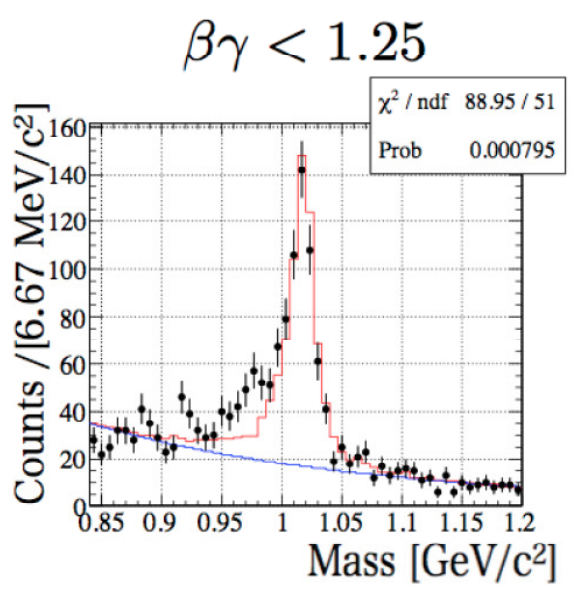

**Fig. 14.3.10** Expected $\phi$ meson spectrum with the copper target by the J-PARC-E16 experiment [4539].

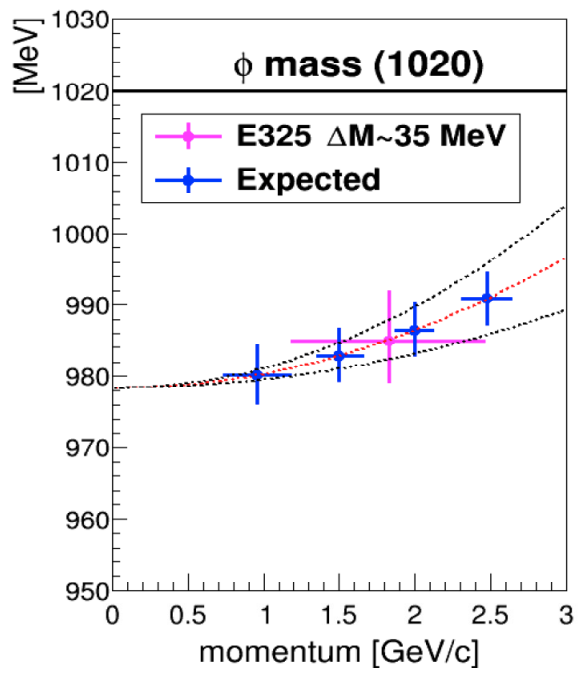

**Fig. 14.3.11** Expected $\phi$-mass data by the J-PARC-E16 and the KEK-E325 one [4539].

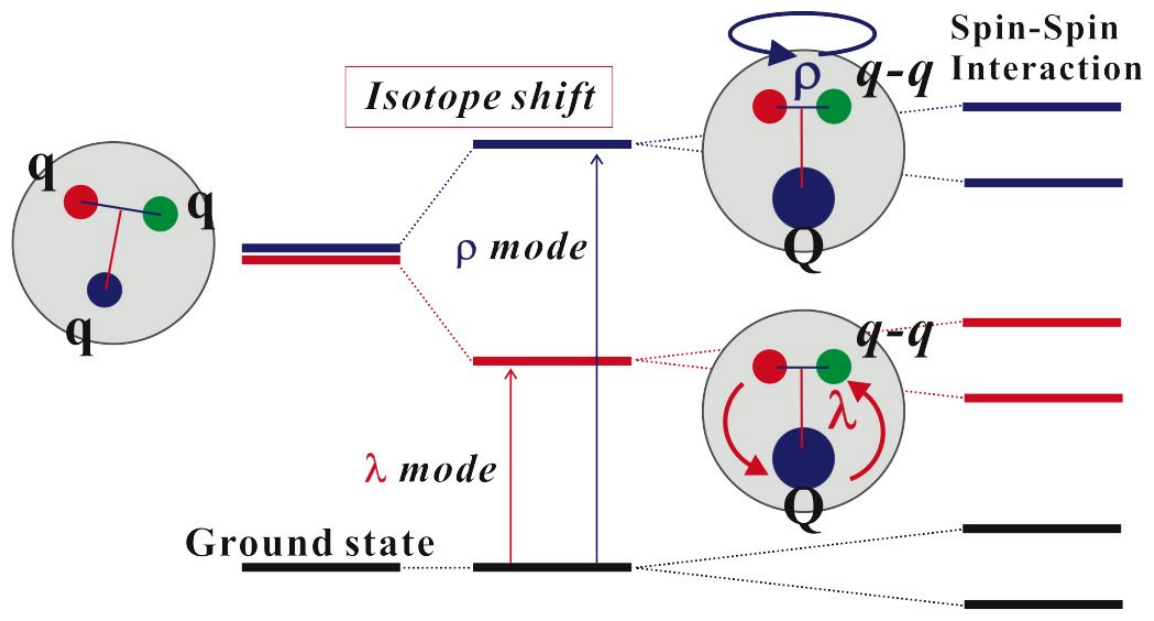

**Fig. 14.3.12** Expected excitations of $N^*(qqq)$ and $Y_c^*(qqQ)$ [4542].

To relate the actual experimental data of E16 to the quark condensate, it is important to understand hadron interactions in nuclear medium because the $\phi$ meson is produced with the momentum 1–2.5 GeV/$c$ and it decays into $e^+e^-$ outside or inside of the nucleus. Such an effort to describe the momentum dependence is in progress by transport simulations by using the Hadron-String Dynamics model [4541], where $\phi$-meson spectral function and their density dependence can be specified. Therefore, new J-PARC data should provide a clue in understanding the role of chiral symmetry breaking for the hadron masses.

### 14.3.5 Hadron spectroscopy

Hadron spectroscopy entered into the new era in the last decade in the sense that there have been many reports on exotic hadron candidates. Exotic hadrons were expected already when the quark model was proposed in 1964. The status of exotic mesons with quantum numbers not accessible within the quark model is reported in Section 8.3. In heavy-quark spectroscopy, a large number of states, both mesons (see Sections 8.5, 8.6) and baryons (see Section 9.4) have been found with unusual properties. However, it is often not easy to distinguish so-called cryptoexotic hadrons, i.e. hadrons with quantum numbers compatible with regular hadrons, from ordinary ones because they may have similar masses. Examples are $f_0(980)$, $a_0(980)$ and $\Lambda(1405)$ in the 1 GeV mass region. It took rather a long time to accumulate signatures from various observables for their tetra- or penta-quark-like (or hadron molecular) nature .

In these days, exotic-hadron studies tend to focus on the heavy-quark sector due to KEKB and LHCb discoveries on exotic hadron candidates with charm and bottom quarks (see Section 9.4). Since charmed baryons will be copiously produced at J-PARC, it is a good opportunity to investigate details of charmed baryon spectroscopy including exotic candidates. At J-PARC, charmed baryons consist of two light quarks and one heavy quark. These will be investigated by the E50 collaboration. Due to the existence of a heavy quark within a baryon, there are specific interactions and internal configurations, which do not exist in baryons with only light quarks. In the extended hadron hall, $\Xi$ and $\Omega$ excitation spectra will be also investigated. Physics motivations of this project include the following.

1. Di-quark correlations in hadrons
   The color magnetic interaction between quarks with indices $i$ and $j$ is given by $V_{\rm mag} \sim \alpha_s(\lambda_i \cdot \lambda_j)(\vec{\sigma}_i \cdot \vec{\sigma}_j)/(m_i m_j)$ where $\lambda$ is the color SU(3) Gell-Mann matrix, $\vec{\sigma}$ is the Pauli spin matrix, and $m$ is the quark mass. Because it is proportional to $1/(m_i m_j)$, the interaction becomes weak for a heavy quark. Let us denote $q$ and $Q$ for light and heavy quarks, respectively. For a $qqQ$-type baryon, the $qq$ interaction should be much stronger than the $qQ$ one. It means that a strong $qq$ diquark correlation could appear in such a baryon. Its expected spectrum for $qqQ$-type baryon in comparison with the $qqq$-type baryon is shown in Fig. 14.3.12 [4542], where $\rho$ and $\lambda$ are the Jacobi coordinates. The $\rho$ is defined as coordinate between the two quarks $qq$, and the $\lambda$ is between $qq$ and $Q$. The spectrum splits into $\rho$- and $\lambda$-mode excitations, called isotope shift. The $\rho$ mode corresponds to a rotation of the diquark $qq$, and the $\lambda$ mode to an orbital excitation between $qq$ and $Q$. These levels are further split by spin-spin interactions. These studies will lead to new dynamical aspects in hadron physics and, more in general, to di-fermion physics in quantum-many-body systems.

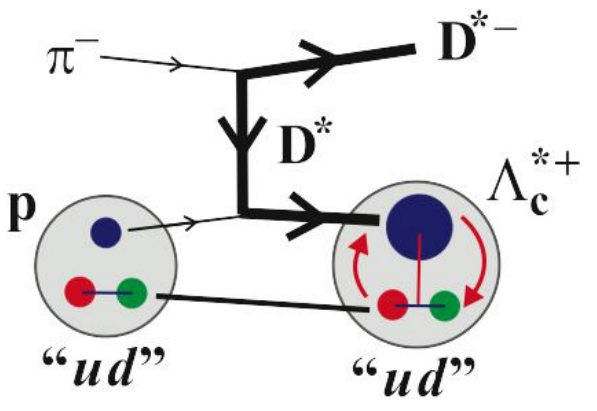

**Fig. 14.3.13** Schematic picture of $\pi^- p \to D^{*-} \Lambda_c^{*+}$ [4542].

2. $\Xi$ and $\Omega$ baryon spectra and their properties
The details of the $\Xi$ and $\Omega$ spectroscopy will be investigated. In addition, the $\Omega$ electric quadrupole moment is highly interesting. Observations of quadrupole moments provide us information on the nature of interactions among constituents and on system deformations. A finite quadrupole moment suggests that a non-central force should exist. Indeed, the tensor force in the one-gluon-exchange potential leads to the expectation that hadrons should be deformed. The $\Omega$ quadruple moment could be measured at J-PARC due to its "stable" nature. The quadrupole moment has never been measured for any hadrons including $\Delta$ [4543], it is an ambitious project.

The charmed-baryon-spectroscopy experiment will start in the hadron hall at the high-momentum beamline by using a beam of unseparated hadrons, essentially pions, with momenta up to 20 GeV . The reaction $\pi^- + p \to D^{*-} + \Lambda_c^{*+}$ is used, as illustrated in Fig. 14.3.13, for measuring the $\Lambda_c^{*+}$ spectrum by the $p(\pi^-, D^{*-})$ missing mass. The simulation is shown for the $\Lambda_c^{*+}$ spectrum in Fig. 14.3.14 by considering the pion momentum of 20 GeV and 100-day beam time. A new field of di-quark physics should be developed by this project.

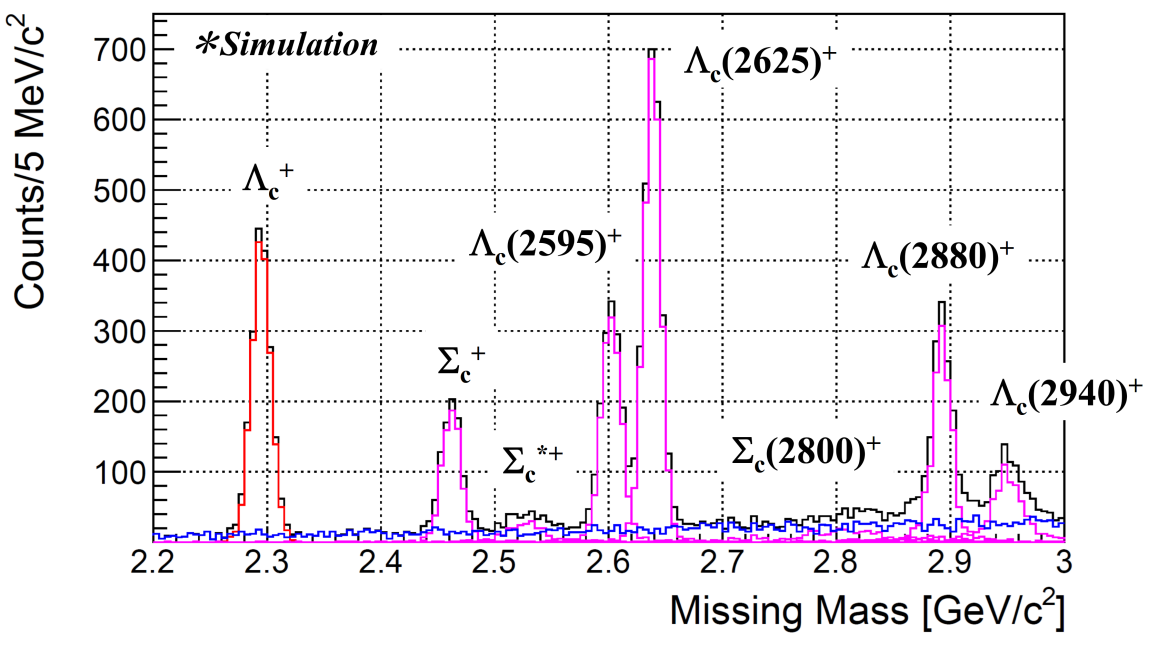

**Fig. 14.3.14** Simulation for the $\Lambda_c^{*+}$ spectrum by the K10 beamline experiment at the extended hadron hall [4515].

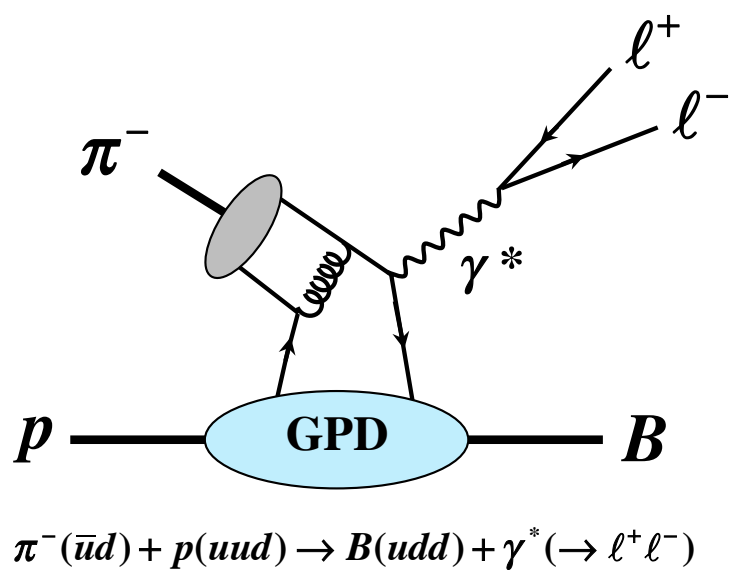

**Fig. 14.3.15** Exclusive Drell-Yan process for measuring GPDs.

### 14.3.6 Hadron structure functions

The J-PARC proton-beam energy of 30 GeV covers the intermediate region from hadron degrees of freedom (d.o.f.) to quark d.o.f. described by perturbative QCD. In addition to hypernuclear and charmed-baryon physics at low energies, the higher-energy region should therefore also be investigated, as illustrated in Fig. 14.3.1. The situation is similar to JLab projects, and J-PARC is complimentary to JLab in the sense that different observables are available in hadron reactions.

The first experiment on hadron structure functions will be on the GPDs for the proton [4544]. A proposal is being prepared [4516] to study exclusive Drell-Yan processes. The GPDs are observables to probe the three-dimensional structure, namely the transverse structure, in addition to the longitudinal parton distribution functions, and the nucleon spin and mass compositions. This project should be able to contribute to the clarification of the hadron spin and mass in terms of quarks and gluons.

At the J-PARC high-momentum beamline, the exclusive Drell-Yan process $\pi^- p \to \mu^+ \mu^- B$ is considered as shown in Fig. 14.3.15. The "pion" beam momentum is up to about 20 GeV. If the baryon $B$ is a neutron, the nucleonic GPDs will be measured, and transition GPDs will be investigated if $B$ is different from the neutron. This process is complementary to the pion-production experiment $\gamma^* + p \to \pi + N$ at JLab with spacelike virtual photon, whereas the J-PARC process is with the timelike one.

At the high-momentum beamline, there is an approved experiment E50 for investigating charmed baryons [4545]. The GPD experiment will be proposed as a col-laboration project with this E50 experiment by supplying a dimuon detector. The dimuons could come from various sources; however, the exclusive Drell-Yan process should be identified by the missing-mass ($M_X$) spectra as shown in Fig. 14.3.16. Here, the Monte-Carlo simulation is given for the pion momentum $p_\pi = 15$ GeV. The exclusive peak is obvious just below 1 GeV, and it should be separated from other processes like inclusive Drell-Yan, $J/\psi$ production, or random backgrounds. In this experiment, the GPDs will be measured for $0.1 < x < 0.3$ and timelike photons in contrast to the JLab experiment on the pion production for larger $x$ and spacelike photons.

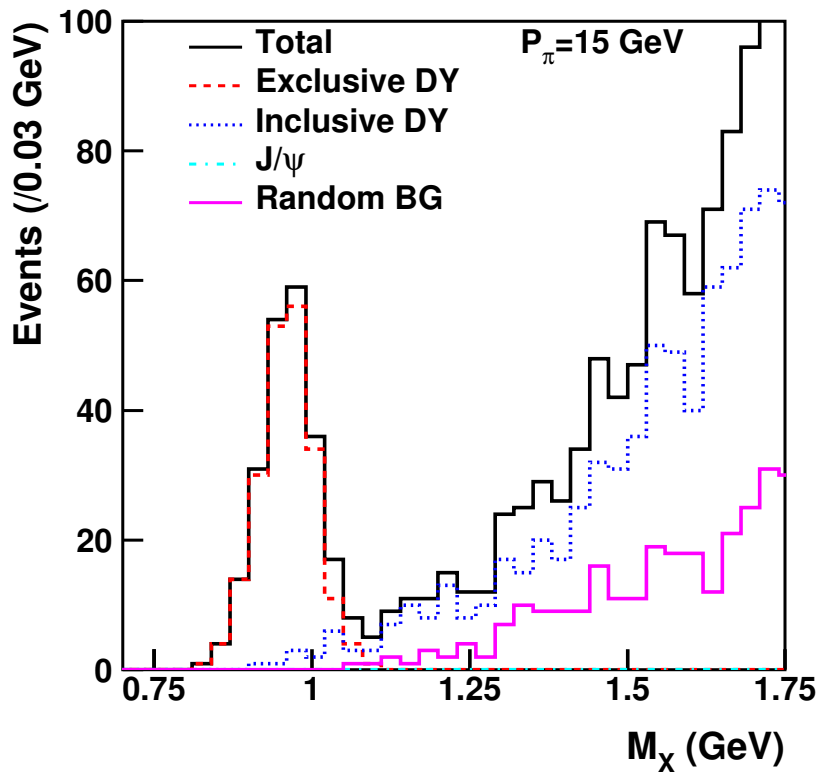

**Fig. 14.3.16** Simulation for the missing-mass spectra [4544]. (Used with the copyright permission of American Physical Society)

In future, there are further possibilities to extend this project on GPD-related studies and, more generally, on high-energy hadron physics [4546–4548]. We explain some examples.

1. Pion-nucleon transition distribution amplitudes
   By backward charmonium production in pion-nucleon collisions, pion-to-nucleon transition-distribution amplitudes can be investigated.
2. GPDs in the ERBL region
   The primary proton beam can be used to measure GPDs by using the $2 \to 3$ process $p+p \to p+\pi+B$. If the final pion and proton have nearly opposite and large transverse momenta with a large invariant energy, the cross section is sensitive to the GPDs in the special kinematical region of ERBL (Efremov-Radyushkin-Brodsky-Lepage).
3. Exotic hadrons by constituent counting rule
   The determination of exotic hadrons is not easy in low-energy global observables, and a much clearer determination could be done by using the constituent counting rule in perturbative QCD. Actually, the structure of the exotic-hadron candidate $\Lambda(1405)$ could be determined by the exclusive process $\pi^- + p \to K^0 + \Lambda(1405)$ at J-PARC.
4. Color transparency
   The color transparency indicates that a hadron passes freely through the nuclear medium at large momentum transfer. It is a unique feature of QCD. There was a mysterious BNL-EVA measurement that the transparency drops at a proton momentum $p > 10$ GeV. The J-PARC should be able to clarify this issue.

In future, we expect that a separated high-momentum kaon beam will become available as the hadron-hall extension program in addition to the protons and pions, so that a variety of these type experiments should become possible.

### 14.3.7 Heavy-ion physics

The purpose of the J-PARC hadron physics is to contribute to our understanding of quantum many-body systems in a wide kinematical range of the phase diagram by precision measurements of new observables as explained in the beginning of this Section. Presently, the physics of dense QCD matters is an important missing program in the current J-PARC experiments.

Dense hadronic systems have been investigated by heavy-ion collisions at RHIC and LHC in the high-temperature and low-density region as shown in Fig. 14.3.17 [4517, 4549]. The creation of a quark-gluon plasma (QGP) was established in the RHIC project by observables such as the collective flow of hadrons and medium modifications of jets. It was surprising to find a small viscosity for the QGP, which initiated interdisciplinary studies with the string theory through the AdS/CFT correspondence (see Section 5.5). Higher-energy collisions are now under investigations at LHC. In addition, the signature of the color-glass condensate has been investigated at these facilities.

At zero baryon density, lattice QCD suggests that the phase transition is a crossover, whereas theoretical models indicate that at high densities the phase transition should be a first-order transition [4550]. This implies that an endpoint of the first-order transition should exist as shown in Fig. 14.3.17. There are also interesting topics on color superconductivity in the cold and dense matter region. After the QGP discovery and studies of its properties, the frontier of heavy-ion physics should be this unexplored region. In fact, there are projects at FAIR and NICA to investigate this region in the near future.

In order to realize such experiments at J-PARC, an additional facility is needed to accelerate heavy ions.

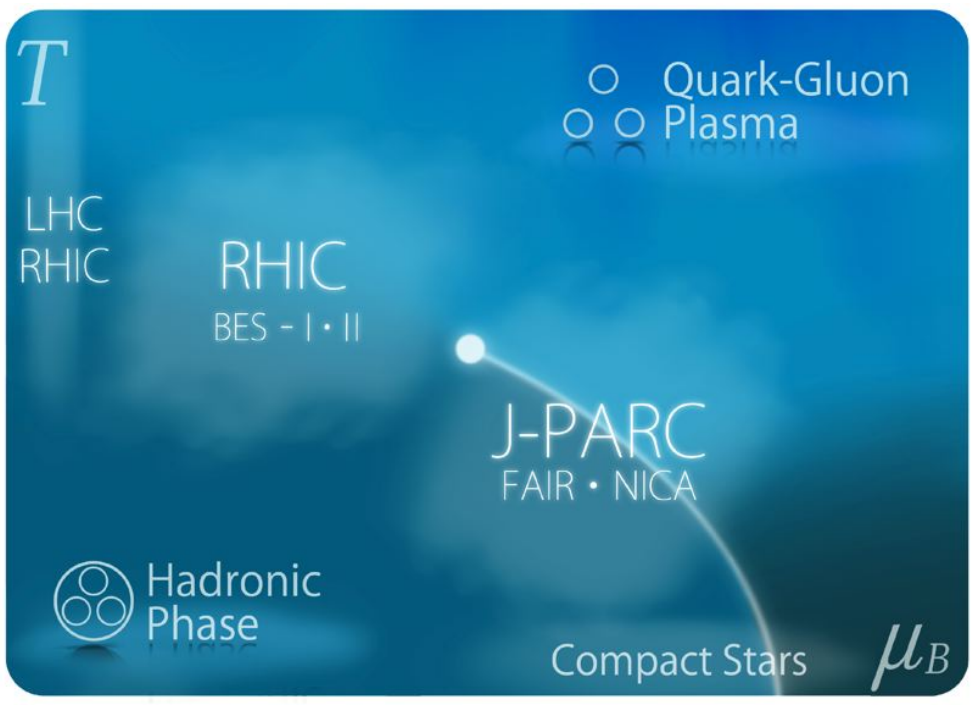

**Fig. 14.3.17** QCD phase diagram with heavy-ion facilities [4517]

The possibility of the heavy-ion experiment was studied in a letter of intent in 2016 [4551]; the proposal was submitted to the J-PARC PAC in 2021 [4517, 4552]. For this project, it is necessary to construct a new linac and a new booster synchrotron. With this injector consists of the linac and the synchrotron together with the rapid-cycling and the main-ring synchrotrons (see Fig. 14.3.2), high-intensity heavy-ion beams with 2-12 A GeV will be obtained. The energies of the heavy-ion facilities for the cold and dense experiments are shown in Fig. 14.3.18 The J-PARC-HI (heavy ion) project is a unique position as the highest-intensity facility in the several GeV region.

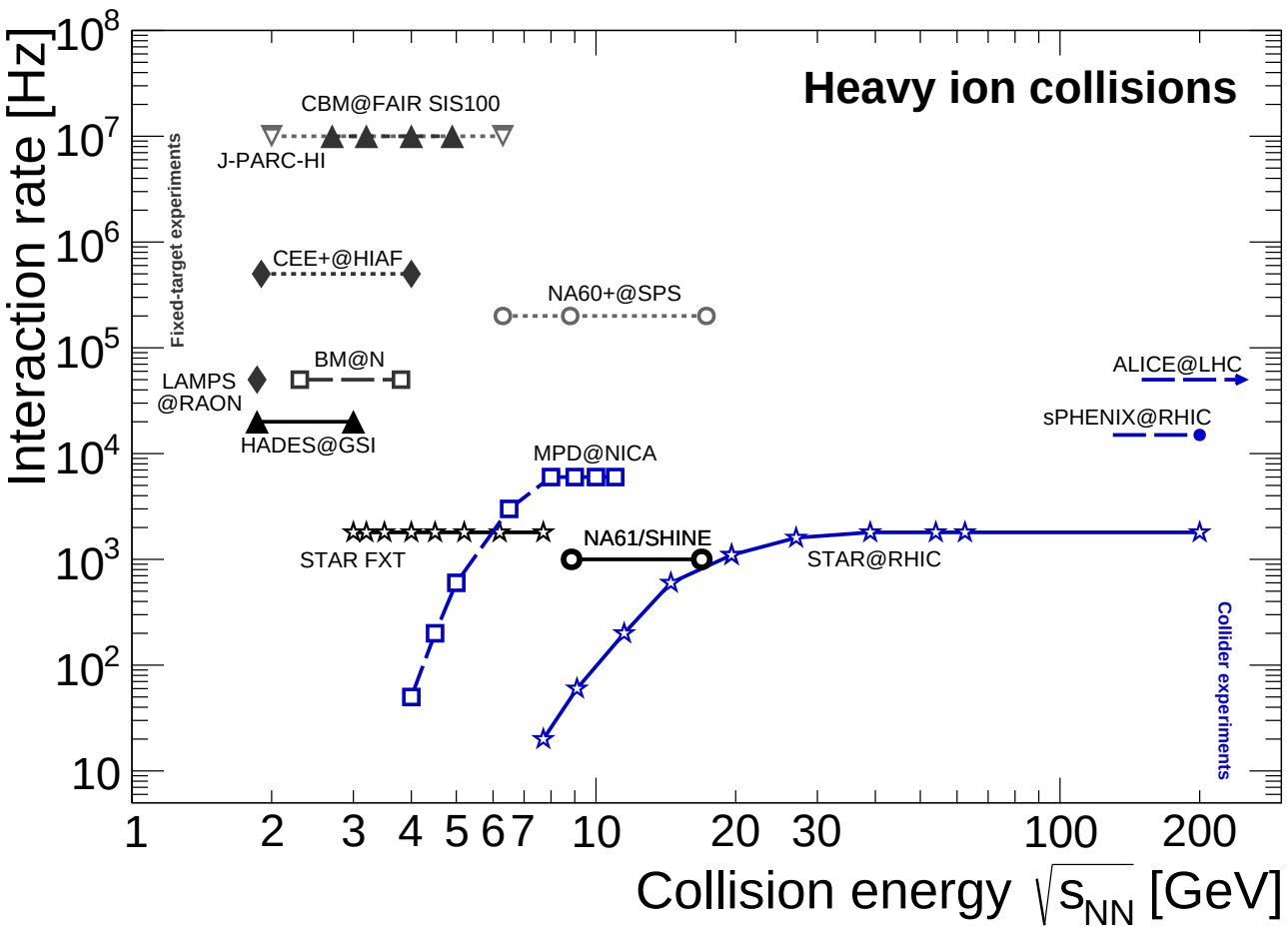

**Fig. 14.3.18** Maximum instantaneous interaction rates recorded by various existing (full lines), under construction (dashed) and proposed fixed-target (black) and collider (blue) experiments addressing the high-$\mu_B$ region of the QCD phase diagram (from [4553], consistently updated based on [4554])

The first purpose of this new facility is to find the phase transition to deconfined quarks and gluons at high densities, by measuring di-electrons, which originate from the virtual photon emission in the hot medium. The advantage of the di-electron measurement is that the virtual photon does not suffer from strong final-state interactions in the medium, so that it directly reflects the information on the QCD matter.

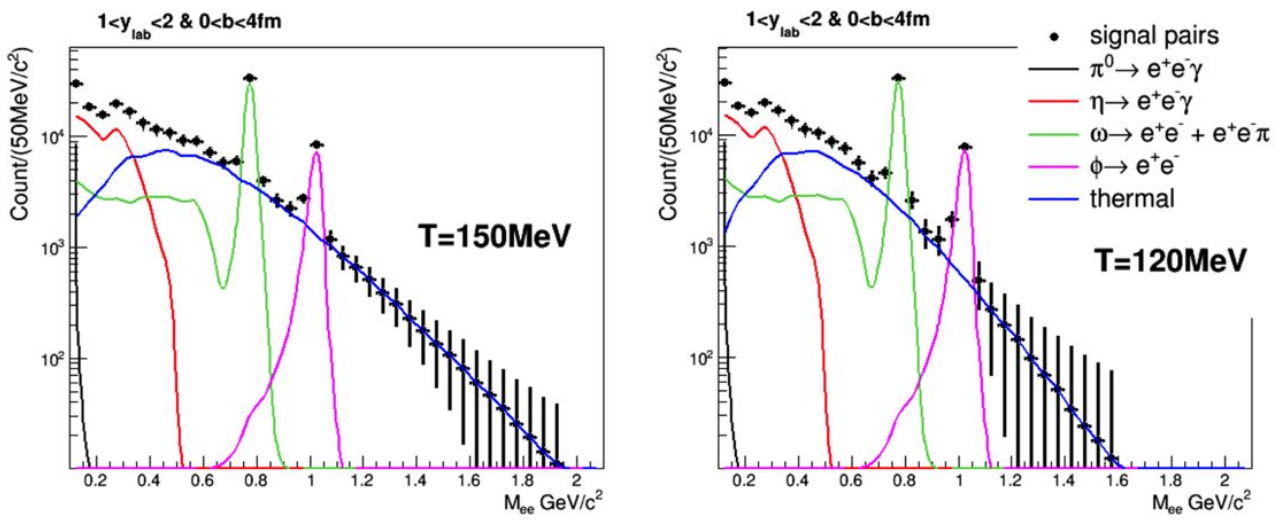

**Fig. 14.3.19** Simulations for the di-electron mass spectra [4517].

Two simulation studies are shown in Fig. 14.3.19 for the di-electron spectrum [4517]. The left-hand side presents the case of no phase transition at $T = 150\,\text{MeV}$, and the right-hand side the case for a first-order phase transition at $T = 120\,\text{MeV}$. The invariant mass spectrum of the di-electrons was taken as $(M_{ee}T)^{3/2}\exp(-M_{ee}/T)$. These results were obtained for the mid-rapity region ($1 \leq y_{\text{lab}} \leq 2$) with 100-day beam time. From such measurements, a determination of the temperature should be possible with the 10% accuracy by the spectrum slope at the mass range $M_{ee} > 1.1\,\text{GeV}$ for the left-side case of Fig. 14.3.19. In the right-hand side, 10% accuracy is possible if $M_{ee} > 0.7\,\text{GeV}$ data are selected. This ambitious J-PARC project makes it possible to find new phenomena of cold and dense matter.

## 14.4 The NICA program

### Alexey Guskov

The Nuclotron-based Ion Collider fAcility (NICA) is a new research complex for studying the fundamental properties of the strong interaction under development as a flagship project at the Joint Institute for Nuclear Research [4555–4557]. The heart of NICA is the Nuclotron – a superconducting ion synchrotron put in operation in 1993. It will be equipped with two injection chains: for heavy (including a booster – a small superconducting synchrotron) and light ions, and a storage ring where particle collisions are planned. The storage ring of racetrack shape has a maximum magnetic rigidity of 45 T×m and a circumference of 503 m. The maximum field of superconducting dipole magnets is 1.8 T.

NICA will provide a variety of heavy-ion beams up to $Au^{79+}$ with a kinetic energy up to 4.5 GeV/u. Collisions of high-intensity proton beams with a high degree of longitudinal or transverse polarization and with total energy up to 13.5 GeV will also be available [4558]. Major accelerator challenges include strong intra-beam scattering and space-charge effects which will be partially compensated by extensive use of electron and stochastic cooling systems.

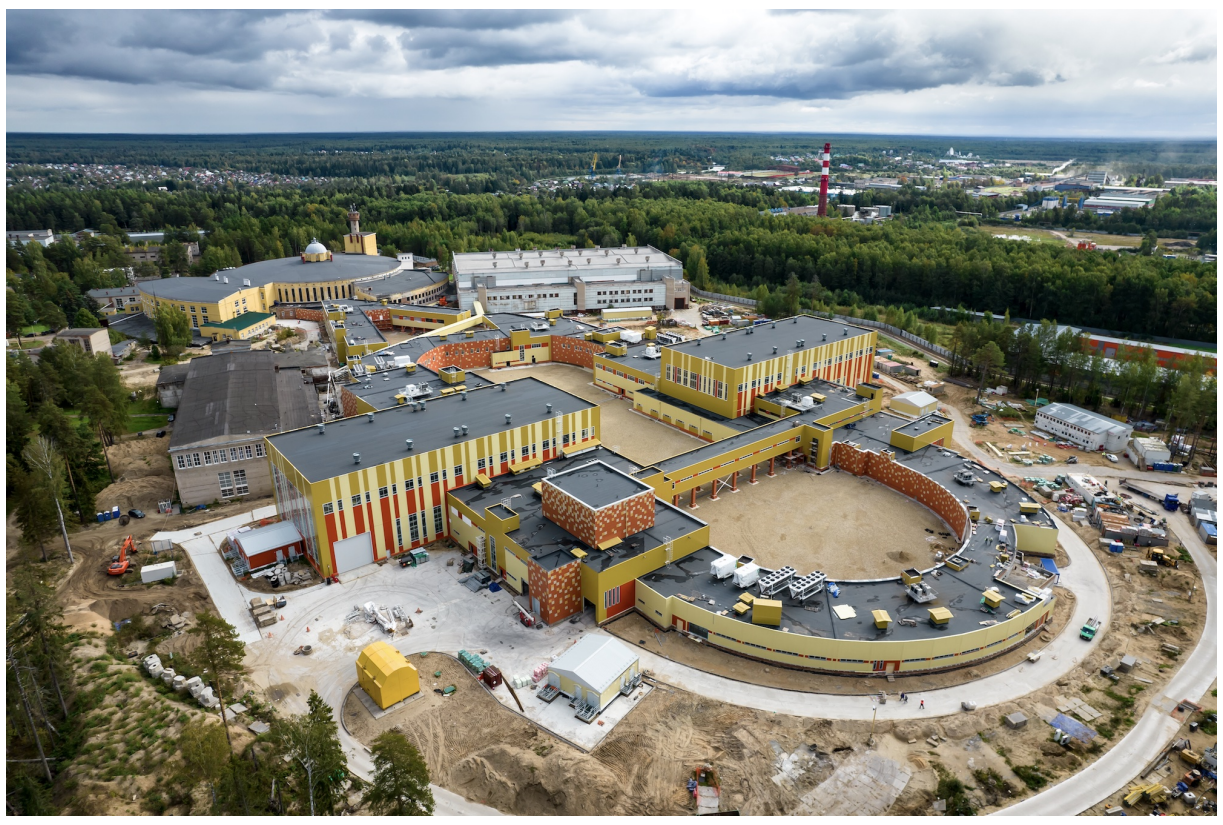

**Fig. 14.4.1** View of the NICA site.

Two experimental setups with different physics programs will run at two interaction points located in the opposite straight sections of the racetrack ring. The MultiPurpose Detector (MPD) placed at the first interaction point will study hot and dense baryonic matter in heavy-ion collisions with luminosity up to $10^{27}$ cm$^{-2}$ s$^{-1}$. The Spin Physics Detector (SPD) in the second interaction point is dedicated to the study of the spin structure of the proton and deuteron and other spin-related phenomena in *p*-*p* and *d*-*d* collisions with luminosity up to $10^{32}$ cm$^{-2}$ s$^{-1}$. In addition, the heavy-ion beams can be extracted to the fixed-target experimental setup BM@N (Baryonic Matter at Nuclotron) whose main goals are investigations of strange/multi-strange hyperons, hypernuclei production, and short-range correlations. Extracted beams will also be used for applied research. A view of the NICA site is shown in Fig. 14.4.1 while Fig. 14.4.2 represents the schematic layout of the accelerator complex.

The implementation of the physic program of the NICA complex is envisioned in three main stages: i) heavy-ion physics with a fixed target (BM@N), ii) heavy-ion physics in the colliding mode (MPD), and iii) spin physics (SPD). The possibility of using NICA in the electron-ion collider mode in the future is under discussion.

### 14.4.1 The study of dense and hot strongly interacting matter at NICA

Asymptotic freedom has a very deep importance for hadronic matter under extreme conditions. At sufficiently high nuclear density or temperature, average inter-parton distances become small and their interaction strength weakens. Above a critical energy density of about 0.3 GeV/fm$^3$, a gas of hadrons passes through a deconfinement transition and becomes a system of unbounded quarks and gluons called quark-gluon plasma (QGP). An evidence of this transition has been obtained from lattice simulations of QCD, in the form of a rapid increase of the entropy density around the critical energy density. The deconfinement of quarks and gluons is accompanied by a restoration of chiral symmetry, spontaneously broken in the QCD vacuum.

The phase diagram (see Fig. 7.1.9 translates the properties of strong interactions and their underlying QCD theory into a visible pattern. Recent lattice calculations have shown that for vanishing baryon chemical potential, $\mu_B$, and at a pseudocritical temperature 156.5 ± 1.5 MeV, a crossover transition happens from the phase with a broken chiral symmetry to the restored chiral symmetry phase [448, 4559]. Different effective models conclude that at higher $\mu_B$, the transition from the ordinary hadron-matter phase to a phase, where chiral symmetry is restored, is of first order. The corresponding critical endpoint is an object of desire of experimenters and theorists, however, its existence is not established yet.

The major goal of MPD and BM@N experiments at NICA is to explore the QCD phase diagram by the study of in-medium properties of hadrons and the nuclear matter Equation of State (EoS), including a search for possible signals of deconfinement and/or chiral symmetry restoration phase transitions, and the QCD critical endpoint. The range of energies and interaction rates covered in different heavy-ion collision experiments including MPD and BM@N experiments at NICA is presented in Fig. 14.3.18.

*The BM@N experiment*

BM@N is a fixed-target experimental setup operating with extracted ion beams from the upgraded Nuclotron. The main final goal of the BM@N experiment is the comprehensive study of the early phase of nuclear interaction at high densities of nuclear matter (3-4$n_0$) via registration of strange and multi-strange particles (kaons, $\Lambda$, $\Xi$ and $\Omega$ hyperons, double hypernuclei, etc.) production with enormous statistical precision. Investigation of the reaction dynamics and nuclear equation of state, as well as the study of the in-medium properties of hadrons, are also planned. In order to provide normalization for the measured A+A spectra, a study of elementary reactions (p+p, p+n(d)) will be performed.

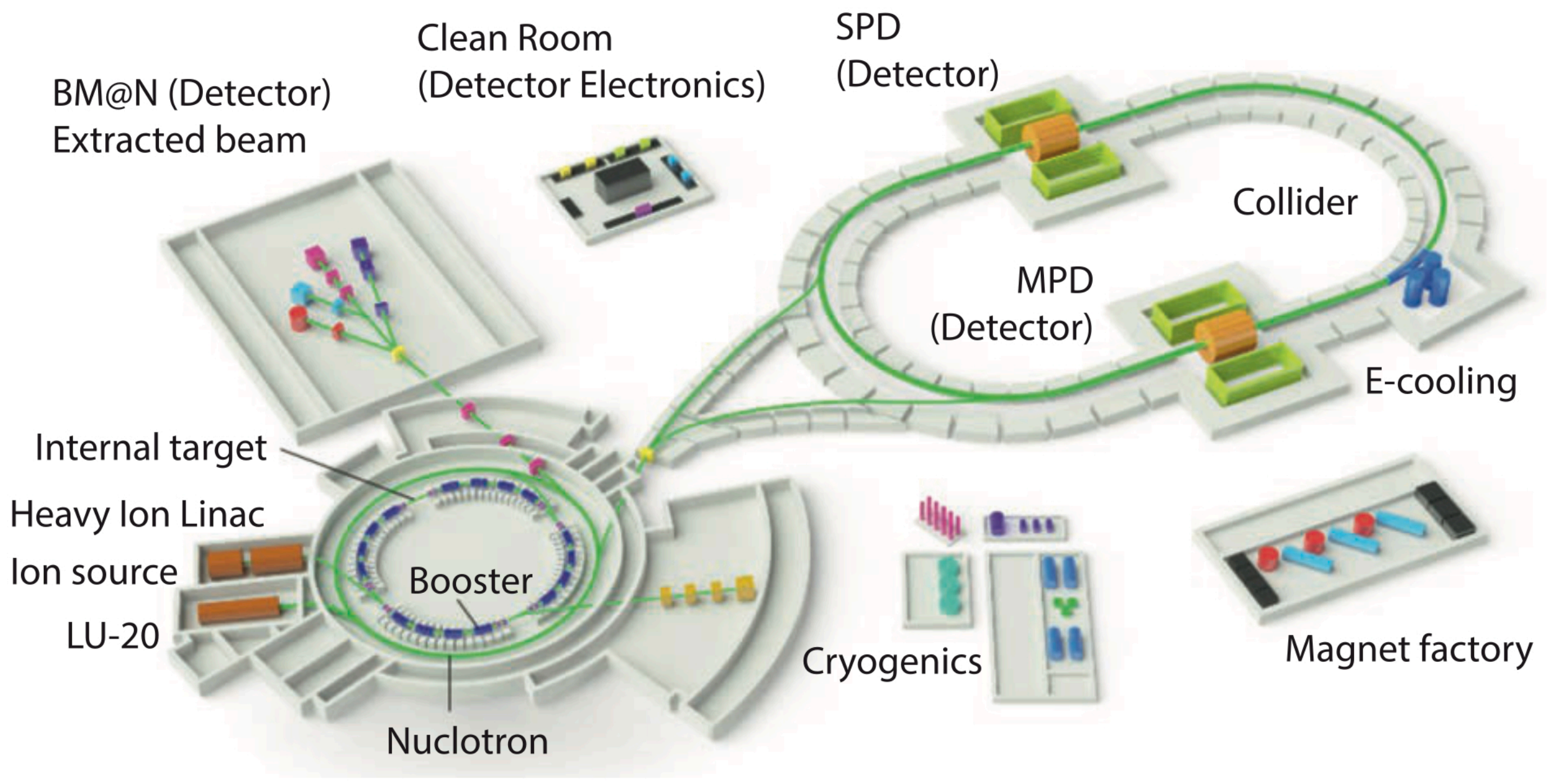

**Fig. 14.4.2** The NICA accelerator complex at JINR.

The layout of the expected full configuration of the BM@N setup is shown in Fig. 14.4.3. The tracking system consists of the silicon strip sensors, and gaseous detectors and is partially placed inside the analyzing magnet with a field up to 1.2 T. Particle identification is provided by the multi-gap Resistive Plate Chamber-based Time-of-Flight system. A Zero Degree Calorimeter is foreseen for the extraction of the collision impact parameter and centrality determination. The BM@N setup currently operates in test mode.

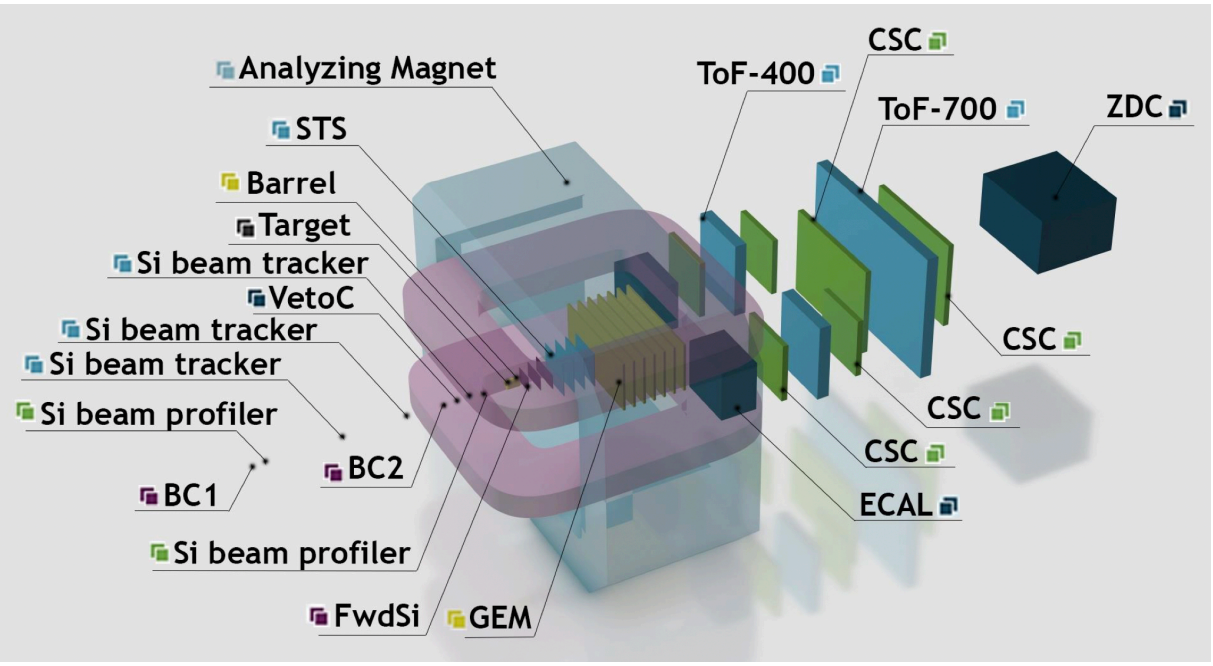

**Fig. 14.4.3** Layout of the BM@N detector [4560].

The relevant degrees of freedom at the Nuclotron energies are first of all nucleons and their excited states followed by light and strange mesons [4561]. The focus of experimental studies at BM@N will be on hadrons with strangeness, which are early produced in the collision and not present in the initial state of two colliding nuclei. The measured production yields of light and strange mesons, as well as of hyperons and anti-hyperons are shown in Fig. 14.4.4 as a function of the nucleon-nucleon collision energy. The Nuclotron heavy-ion beam-energy range corresponds to $\sqrt{s_{NN}} = 2.3$–$3.5$ GeV. It is well suited for studies of strange mesons and multi-strange hyperons which are produced in nucleus-nucleus collisions close to the kinematic threshold. Heavy-ion collisions are a rich source of strangeness, and capturing $\Lambda$-hyperons by nucleons can produce a variety of light hyper-nuclei [4562, 4563]. In heavy-ion collisions, light hypernuclei are expected to be abundantly produced at low energies due to the high baryon density. However, the production mechanisms of hypernuclei in heavy-ion collisions are not well understood, due to the scarcity of data. The study of hyper-nuclei production is expected to provide new insights into the properties of the hyperon-nucleon and hyperon-hyperon interactions. Figure 14.4.5 presents the yields of hyper-nuclei as a function of the nucleon-nucleon collision energy in the center-of-mass system in Au+Au collisions, predicted by a thermal model [4564]. The maximum in the hyper-nuclei production rate is predicted at $\sqrt{s_{NN}}$ = 4-5 GeV, which is close to the Nuclotron energy range.

Short-range correlations in nuclei (SRC) are an additional topic for study at BM@N. In an attempt to simplify the description of the nuclei as complex strongly interacting systems, we tend to separate their short- and long-range structure. Effective field theories describe the long-range structure using a mean-field approximation. The short-range structure of nuclei can be described in terms of nucleon-nucleon short-range correlations. SRC are brief fluctuations of two nucleons with high and opposite momenta, where each of them is higher than the Fermi momentum for the given nucleus.

Hard knock-out reactions where the beam probe interacts with a single nucleon are the standard way to study the properties of SRC pairs. In the pilot studies at BM@N the new approach with the inverted kinematics was used [4565]: a carbon beam with the momentum of 4 GeV/c per nucleon scatter off a liquid hydrogen target. A proton with momentum from the SRC pair is scattered off a target proton. Two protons from the (p,2p) reaction were detected by a two-arm spectrometer while a $A-2$ nuclear fragment was identified via $p/Z$ ratio. The events with $^{10}$B and $^{10}$Be fragments corresponded to *p*-*n* and *p*-*p* SRC pairs, respectively. The direct experimental evidence for the separation of the pair wavefunction from that of the residual many-body nuclear system was obtained. All measured reactions are well described by theoretical calculations that include no distortions from the initial- and final-state interactions (Fig. 14.4.6). The obtained results illustrate the ability to study the short-distance structure of short-lived radioactive nuclei at the forthcoming FAIR and FRIB facilities.

*The MPD experiment*

MPD is a collider experiment designed to perform a comprehensive scan of the QCD phase diagram with beam species from protons to gold by varying the center-of-mass collision energy from 4 to 11 GeV per nucleon which is complementary to the RHIC beam energy scan towards lower energies. The unique feature of MPD as a collider experiment is the invariant acceptance at different beam energies as compared to fixed-target experiments [4567].

To reach this goal, the experimental program includes the simultaneous measurement of the observables which are sensitive to high density effects and phase transitions. The observables measured on event-by-event basis are particle yields and ratios, correlations and fluctuations. Different species probe different stages of the nucleus-nucleus interaction due to their differences in mass, energy and interaction cross-sections.

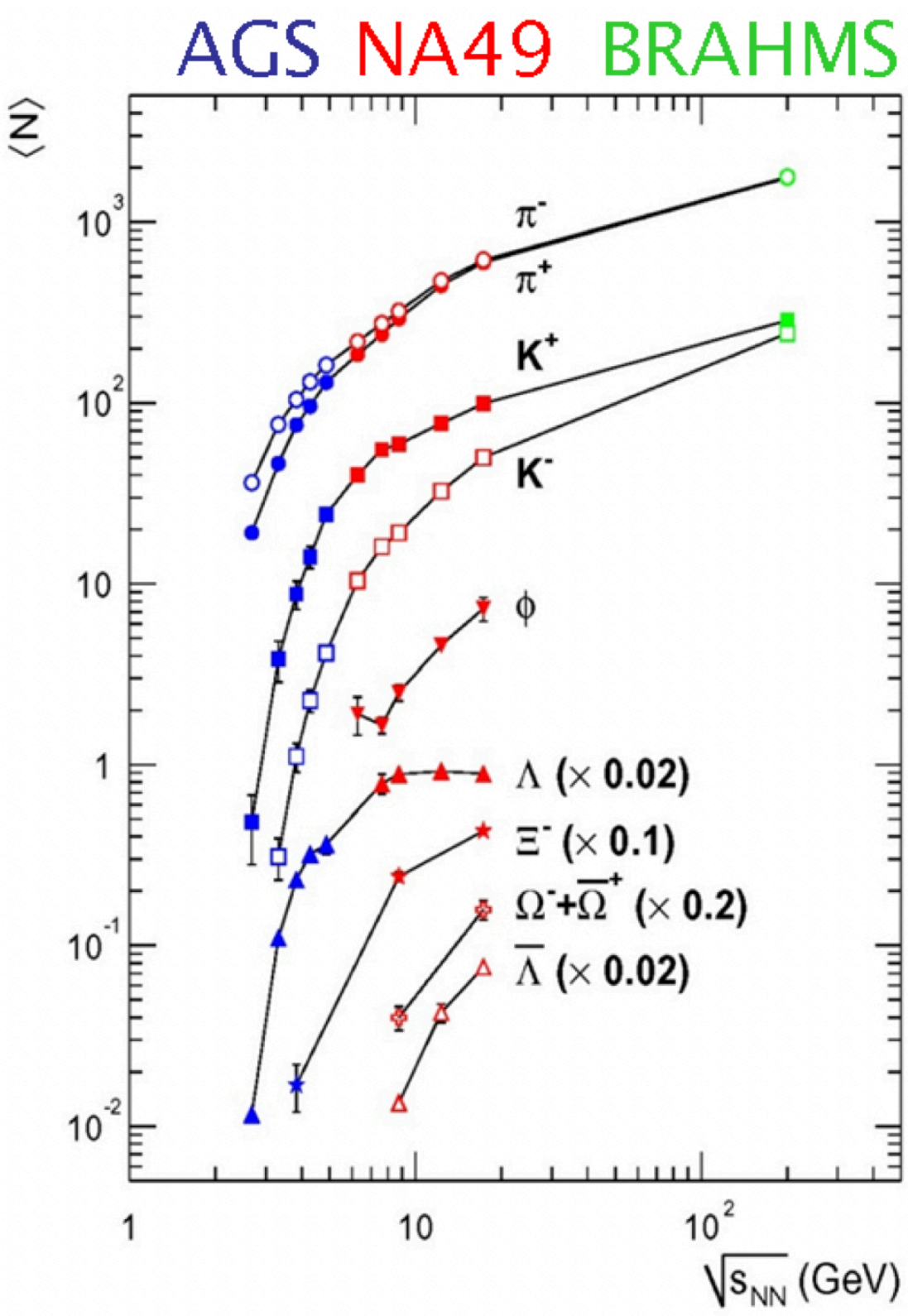

**Fig. 14.4.4** Yields of mesons and (anti-)hyperons measured in different experiments as a function of the collision energy $\sqrt{s_{NN}}$ for Au+Au and Pb+Pb collisions [4566].

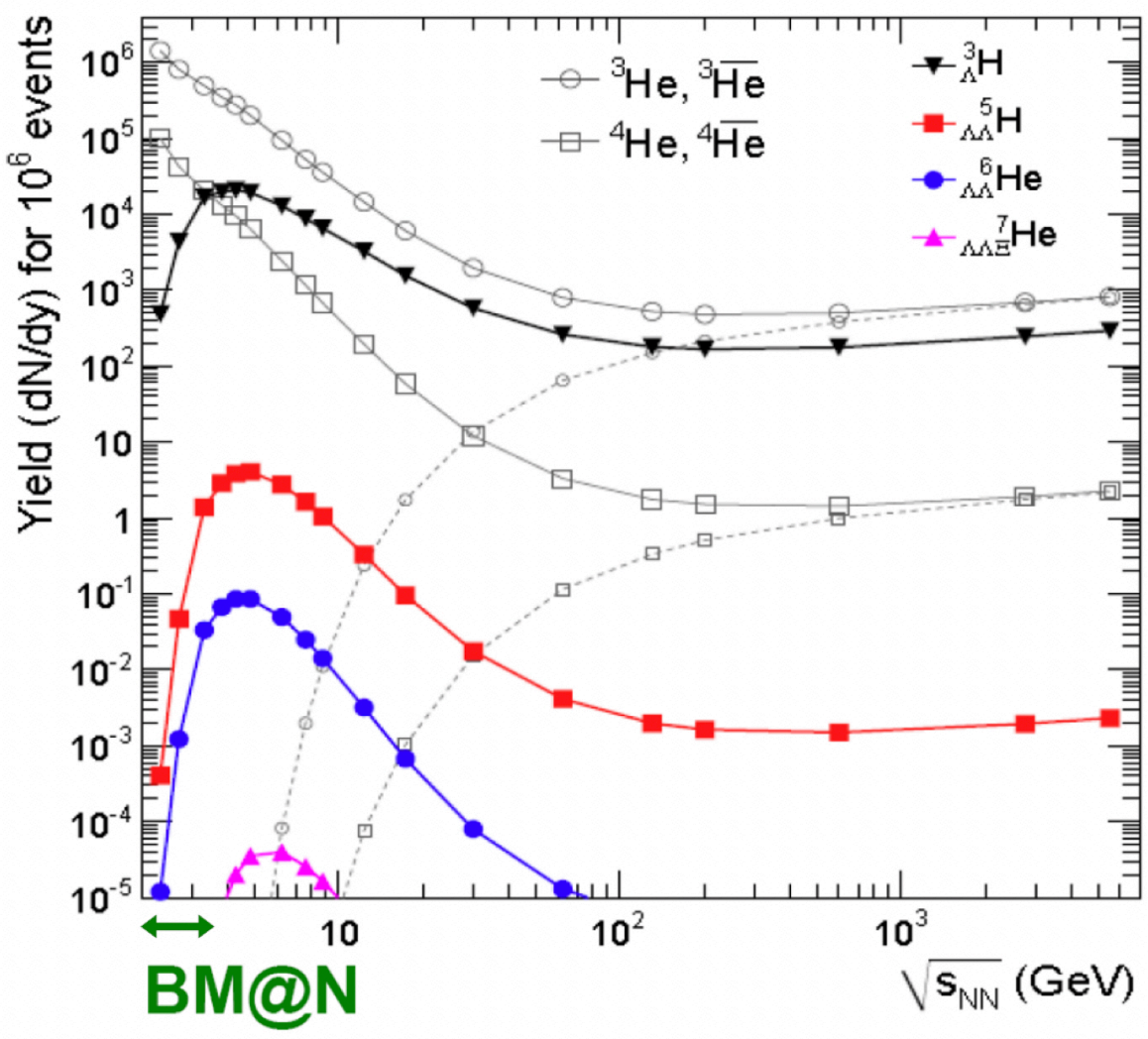

**Fig. 14.4.5** Yields of hyper-nuclei predicted by the thermal model in Ref. [4564] as a function of the $\sqrt{s_{NN}}$ for Au+Au collisions. Predictions for the yields of $^3$He and $^4$He nuclei are presented for comparison.

The hadrons containing heavy strange quarks are especially interesting. These strange heavy hadrons are created in the early high-temperature and high-density stage but may quickly decouple due to their low interaction cross section with the surrounding matter. Among various characteristics, the elliptic flow deserves special attention because this collective motion is formed mainly in the early stage of the collision. The spatio-temporal information on the particle freeze-out source, which depends on the preceding evolution of the system, is provided by the measurement of identical particles interference. The direct information on hot and dense transient matter is provided by penetrating probes, photons and leptons. In this respect, vector mesons which contain information on chiral symmetry restoration are very attractive. Measurement of the positive/negative pion asymmetry with respect to the reaction plane as a function of centrality of heavy-ion collisions opens a possibility to touch such fundamental problem as spontaneous violation of CP parity in strong interactions.

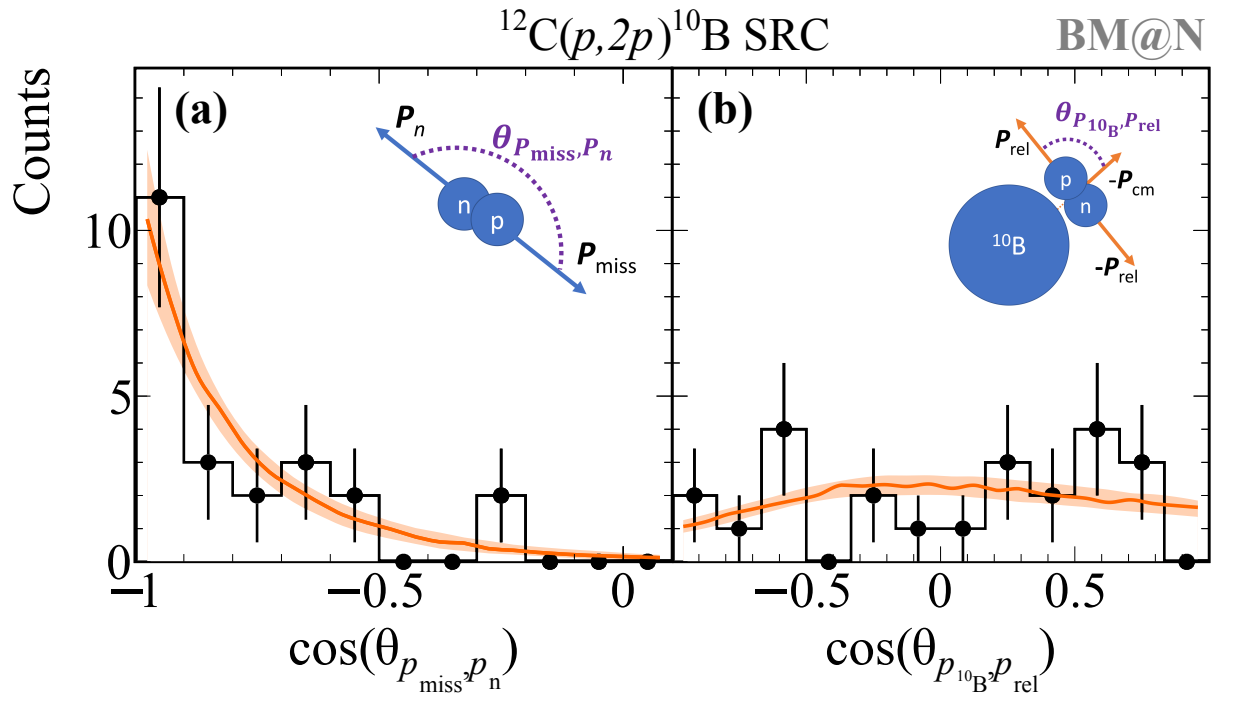

**Fig. 14.4.6** Opening angle in SRC p-n pair (left) and the angle between the $^{10}B$ fragment and pair relative momentum (right). The model calculations are shown in orange [4565].

The physics program of the first stage of the MPD experiment includes the following items [4568]:

– multiplicity and spectral characteristics of the identified hadrons including strange particles, multi-strange baryons and antibaryons characterizing entropy production and system temperature at freeze-out;
– event-by-event fluctuations in multiplicity, charges, transverse momenta and $K/\pi$ ratios as a generic property of critical phenomena;
– collective flow effects (directed, elliptic and higher ones) for hadrons including strange particles;
– femtoscopy with identified particles and particle correlations.

In the second stage, the physics with electromagnetic probes (photons and dileptons) will be accessed.

The behaviour of hadron abundances along the hydrodynamic trajectories of heavy-ion experiments is closely related with the properties of the strongly interacting matter near the phase transition. For example, a promising observable to study the onset of deconfinement is the pion-to-kaon ratio. The $K^+$ yield is proportional to the overall strangeness production and pions can be associated with the total entropy produced in the reaction. Thus, the $K^+/\pi^+$ production ratio can be a good measure of strangeness-to-entropy ratio, which is different in the confined phase and the QGP. The experimental results for $K^+/\pi^+$, $K^-/\pi^-$ and $\Lambda/\pi^+$ ratios as a function of collision energies in the wide energy range are shown in Fig. 14.4.7. The experimental points in the most interesting region around $\sqrt{s_{NN}} = 10$ GeV have large uncertainties that could be significantly reduced by the measurements at MPD.

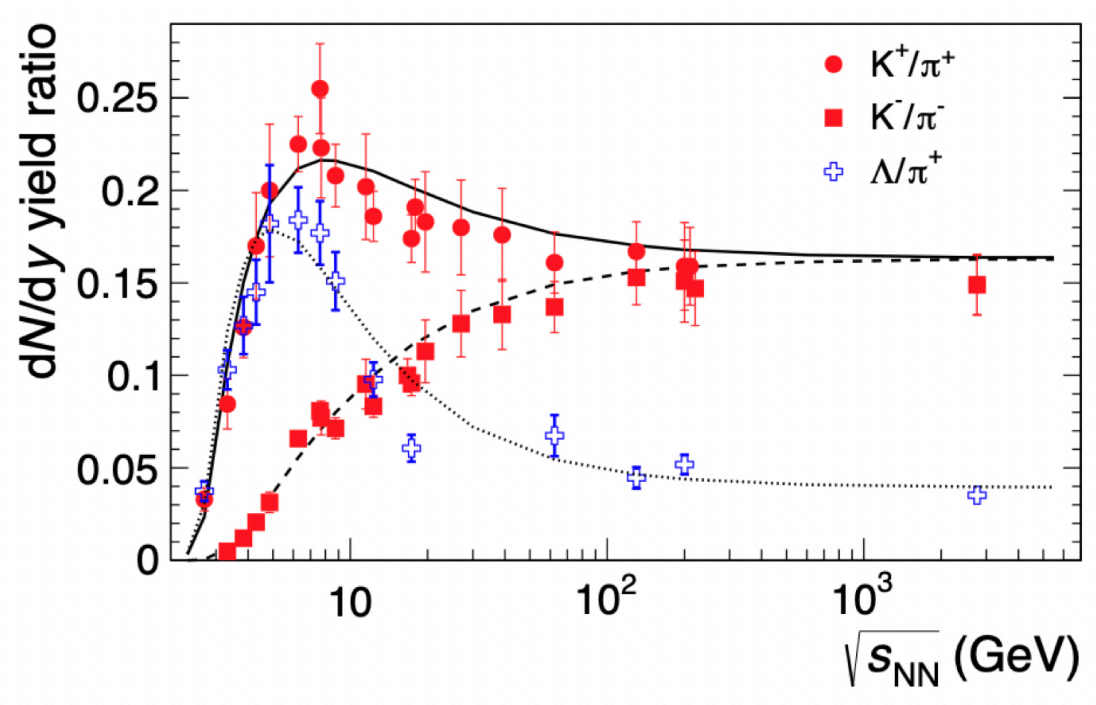

**Fig. 14.4.7** $K^+/\pi^+$, $K^-/\pi^-$ and $\Lambda/\pi^+$ ratios as a function of $\sqrt{s_{NN}}$ [2159].

Measurements of event-by-event fluctuations have been performed by the numerous fixed-target and collider experiments. Recent STAR measurements from the RHIC-BES program [2180] indicate a non-monotonic behaviour of the excitation function for the net-proton moments in central Au+Au collisions in the region below $\sqrt{s_{NN}} = 20$ GeV, which can be a hint for the critical point in the range of finite baryon number density. At MPD the region below 11 GeV will be scanned with much higher precision.

The main task of femtoscopy, the technique of two-particle correlations in momentum space, is to measure the space-time evolution of the system created in particle collisions. The two-pion correlation functions are excellent candidates for first-day physics measurements at MPD. Femtoscopy measurements for pions have been performed in several previous experiments. Fig. 14.4.8 presents the energy dependence of the freeze-out volume, obtained from two-pion interferometry. A non-monotonic behavior of this volume in the NICA energy range raises interest in such measurements at MPD.

The anisotropic collective flow is also one of the promising observables sensitivite to the transport properties of the strongly interacting matter, in particular, the speed of sound, and the specific shear and viscosities. It can be quantified by the Fourier coefficients $v_n$ in the expansion of the particles azimuthal distribution. Relativistic viscous hydrodynamic models have been successful in describing the observed anisotropy $v_n$ for the produced particles in the collisions of heavy ions at RHIC and the LHC [2150, 4570, 4571]. The directed flow $v_1$ can probe the very early stages of the collision as it is generated during the passage time of the two colliding nuclei. The results of a model-to-data comparison for the elliptic flow $v_2$ at $\sqrt{s_{NN}} = 7.7$ GeV and 4.5 GeV may indicate that at NICA energies a transition occurs from partonic to hadronic matter. The high-statistics differential measurements of $v_n$, that are anticipated from the MPD experiment at NICA, are expected to provide valuable information about this parton-hadron transient energy domain [4572, 4573].

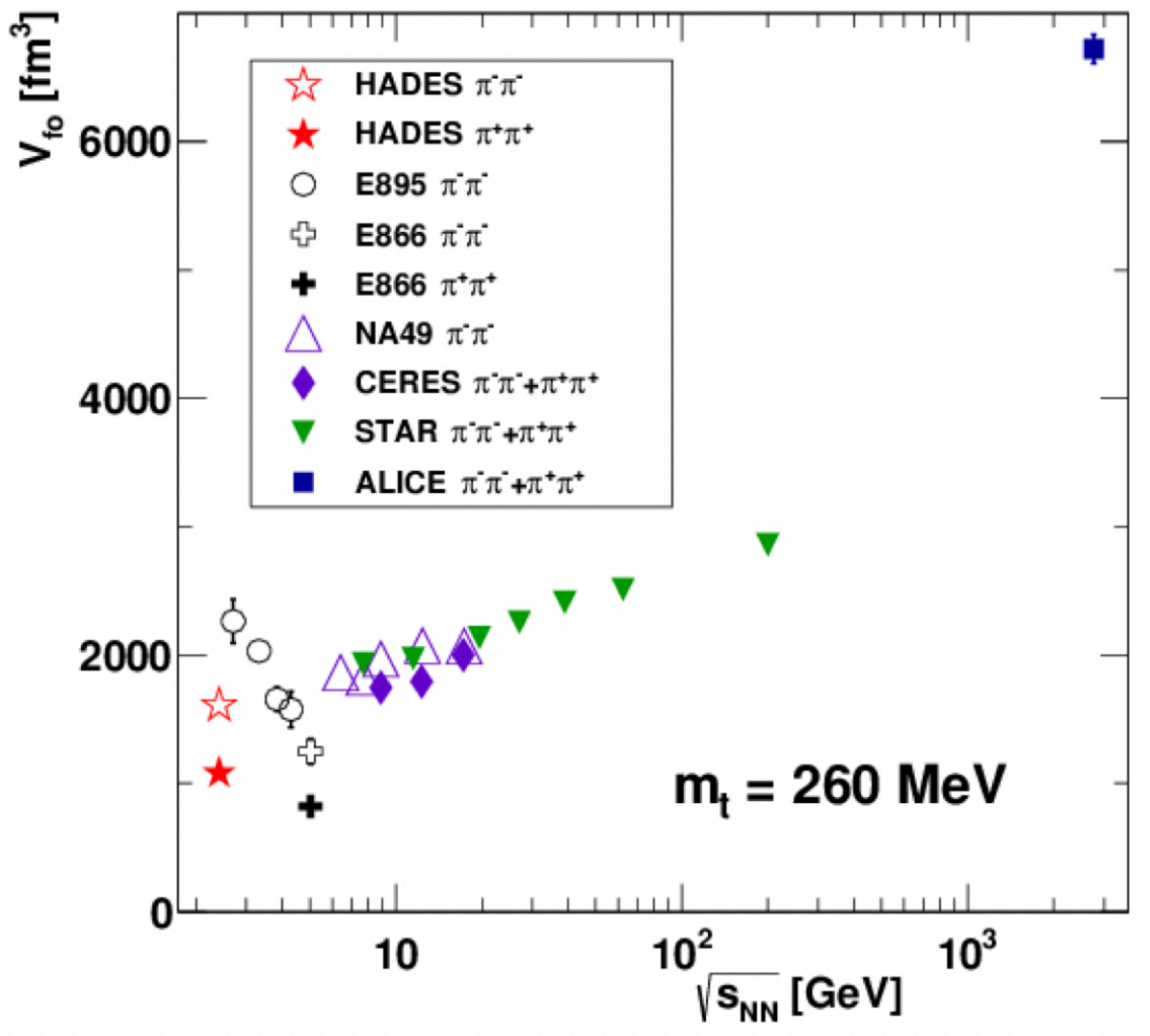

**Fig. 14.4.8** Freeze-out volume for pions as a function of the collision energy [4569].

The layout of the MPD setup is shown in Fig. 14.4.9 [4573]. The components of the MPD barrel part have an approximate cylindrical symmetry. The beam line is surrounded by the large gaseous Time Projection Chamber (TPC) which is enclosed by the TOF barrel. The TPC is the main tracker, and in conjunction with the TOF they will provide precise momentum measurements and particle identification. It is placed in a highly homogeneous magnetic field of up to 0.57 T. The Electromagnetic Calorimeter (ECal) is placed in between the TOF and the MPD Magnet. It will be used for detection of electromagnetic showers, and will play the central role in photon and electron measurements. In the forward direction, the Fast Forward Detector (FFD) is located still within the TPC barrel. It will play the role of a wake-up trigger. The Forward Hadronic Calorimeter (FHCal) for determination of the collision centrality and the orientation of the reaction plane is located near the Magnet end-caps. At the moment, this detector configuration is at the assembling stage.

Additional detectors like the silicon-based Inner Tracker System for precision secondary vertex reconstruction, the miniBeBe detector for triggering and start time determination, and the cosmic ray detector on the outside of the magnet yoke are proposed for the later stages.

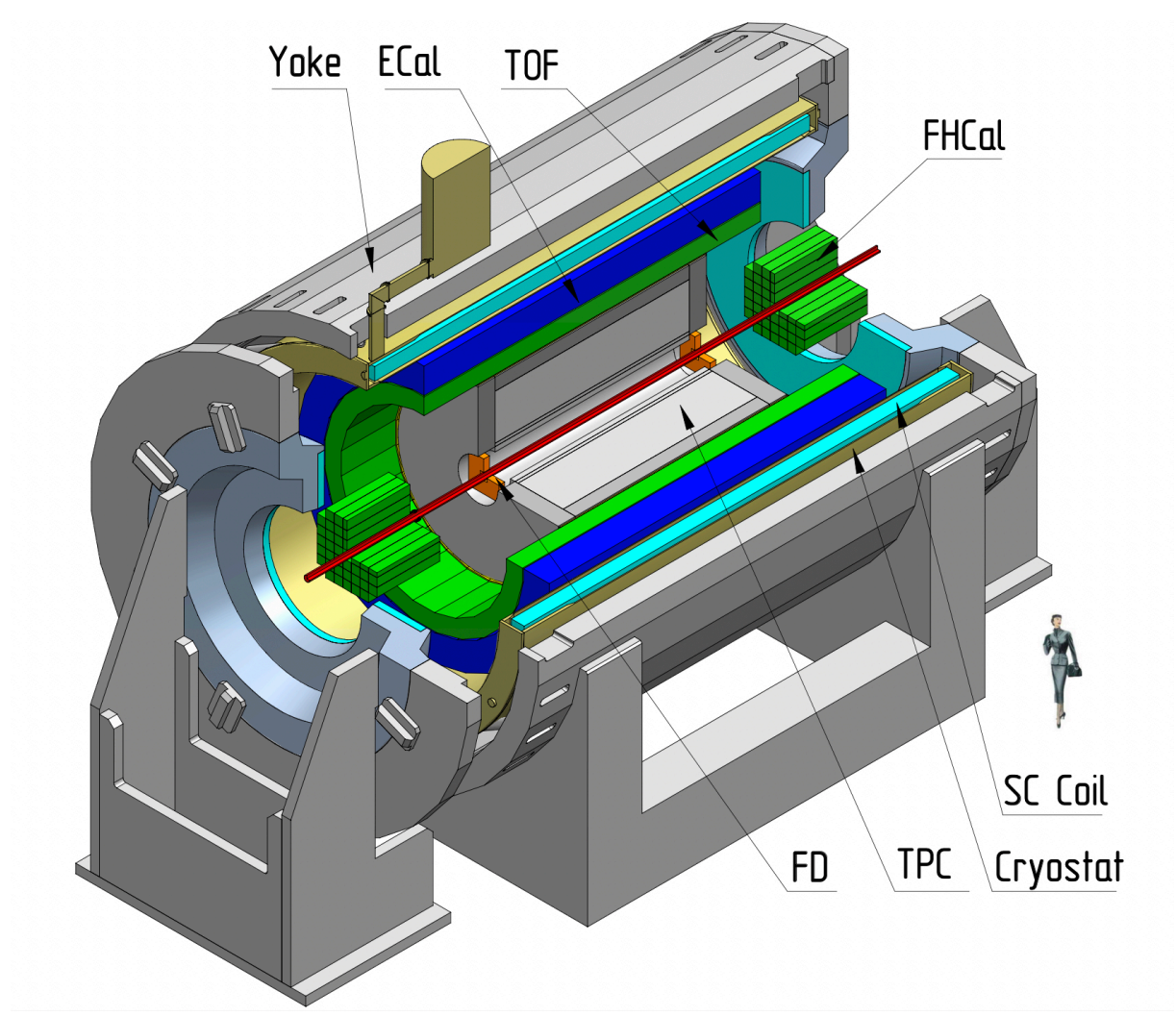

**Fig. 14.4.9** Layout of the MPD experimental setup [4573].

## 14.4.2 The spin structure of proton and deuteron in the SPD experiment

While the main goal of the BM@N and MPD experiments is to study deconfinement, the third experiment, SPD, aims to study the internal structure of the proton and deuteron using polarized beams. In the polarized proton-proton collisions, the SPD experiment [4574] will cover the kinematic gap between the low-energy measurements at ANKE-COSY and SATURNE and the high-energy measurements at the Relativistic Heavy Ion Collider, as well as the planned fixed-target experiments at the LHC (see Fig. 14.4.10). The possibility for NICA to operate with polarized deuteron beams at such energies is unique. SPD is planned to be operated as a universal facility for comprehensive tests of the basics of the QCD. The main efforts, however, will be devoted to the study of the unpolarized and polarized gluon content of the proton at large Bjorken-$x$, using different complementary probes [4575].

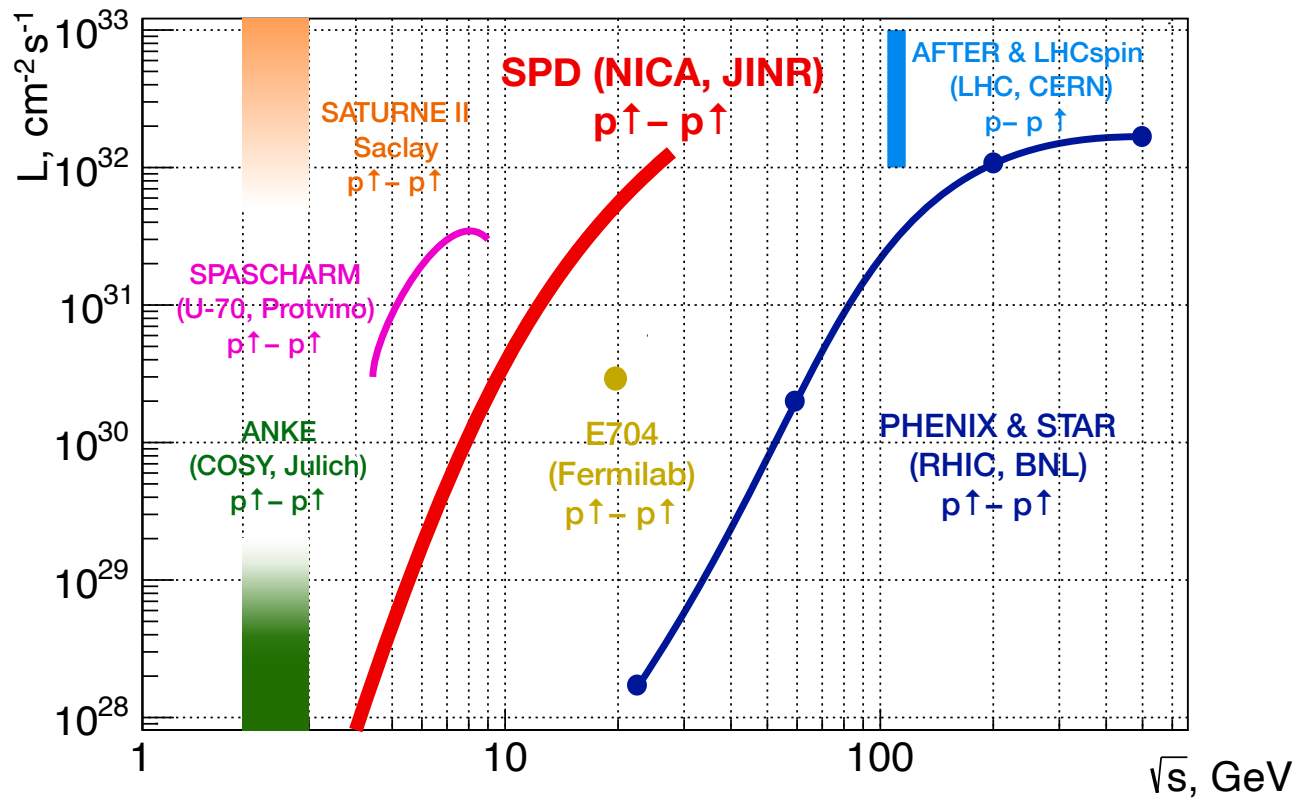

**Fig. 14.4.10** NICA SPD and the other past, present, and future experiments with polarized protons.

Quantum chromodynamics has remarkable success in describing the high-energy and large-momentum transfer processes, where quarks and gluons that are the fundamental constituents of hadrons, behave, to some extent, as free particles and, therefore, the perturbative QCD approach can be used. The cross-section of a process in QCD is factorized into two parts: the process-dependent perturbatively-calculable short-distance partonic cross-section (the hard part) and universal long-distance functions, PDFs, and FFs (the soft part), see Section 11. The parton distributions could be applied also to describe the spin structure of the nucleon that is built up from the intrinsic spin of the valence and sea quarks (spin-1/2), gluons (spin-1), and their orbital angular momenta.

In recent years, the three-dimensional partonic structure of the nucleon became a subject of careful studies. Precise mapping of the three-dimensional structure of the nucleon is crucial for our understanding of QCD. One of the ways to go beyond the usual collinear approximation is to describe the nucleon content in the momentum space by employing the so-called Transverse-Momentum-Dependent Parton Distribution Functions (TMD PDFs) [1286, 3189, 3190, 4576–4578].

Considerable progress has been achieved during the last decades in the understanding of the quark contribution to the nucleon spin, yet the gluon sector is much less developed. One of the difficulties is the lack of direct probes to access the gluon content in high-energy processes.

The final goal of the SPD experiment is to provide access to the gluon TMD PDFs (see Table 14.4.1) in the proton and deuteron via the measurement of specific single and double spin asymmetries in the production of charmonia, open charm, and high-$p_T$ prompt photons. The kinematic region to be covered by SPD for these processes (Fig. 14.4.11) is unique and has never been accessed purposefully in polarized hadronic collisions. Quark TMD PDFs, as well as spin-dependent fragmentation functions, could also be studied. The results expected to be obtained by SPD will play an important role in the general understanding of the nucleon gluon content and will serve as a complementary input to the ongoing and planned studies at RHIC, and future measurements at the EIC (BNL) and fixed-target facilities at the LHC (CERN). Simultaneous measurement of the same quantities using different processes at the same experimental setup is of key importance for the minimization of possible systematic effects.

**Table 14.4.1** Gluon TMD PDFs at twist-2. The columns represent gluon polarization, while the rows represent hadron polarization.

| | Unpolarized | Circular | Linear |
|---|---|---|---|
| Unpolarized | g(x) density | | $h_1^{\perp g}(x, k_T)$ Boer-Mulders function |
| Longitudinal | | $\Delta g(x)$ helicity | Kotzinian-Mulders function |
| Transverse | $\Delta_N^g(x, k_T)$ Sivers function | Worm-gear function | $\Delta_T g(x)$ transversity, pretzelosity |

The naive model describes the deuteron as a weakly-bound state of a proton and a neutron mainly in S-state with a small admixture of the D-state. However, such a simplified picture failed to describe the HERMES experimental results on the $b_1$ tensor structure function [819]. A unique possibility to operate with polarized deuteron beams brings us to the world of the tensor structure of the deuteron (tensor PDFs). A possible non-baryonic content in the deuteron could be accessed via the measurement of the gluon transversity distribution and the comparison of the unpolarized gluon PDFs in the nucleon and deuteron at high values of $x$.

Nevertheless, the largest fraction of hadronic interactions involves low-momentum transfer processes in which the effective strong coupling constant is large and the description within a perturbative approach is not adequate. A number of (semi-)phenomenological approaches have been developed through the years to describe strong interaction in the non-perturbative domain starting from the very basic principles. They successfully describe such crucial phenomena as the nuclear properties and interactions, hadronic spectra, deconfinement, various polarized and unpolarized effects in hadronic interaction, etc. The transition between the perturbative and non-perturbative QCD is also a subject of special attention. In spite of a large set of experimental data and huge experience in a few-GeV region with fixed-target experiments worldwide, this energy range still attracts both experimentalists and theoreticians.

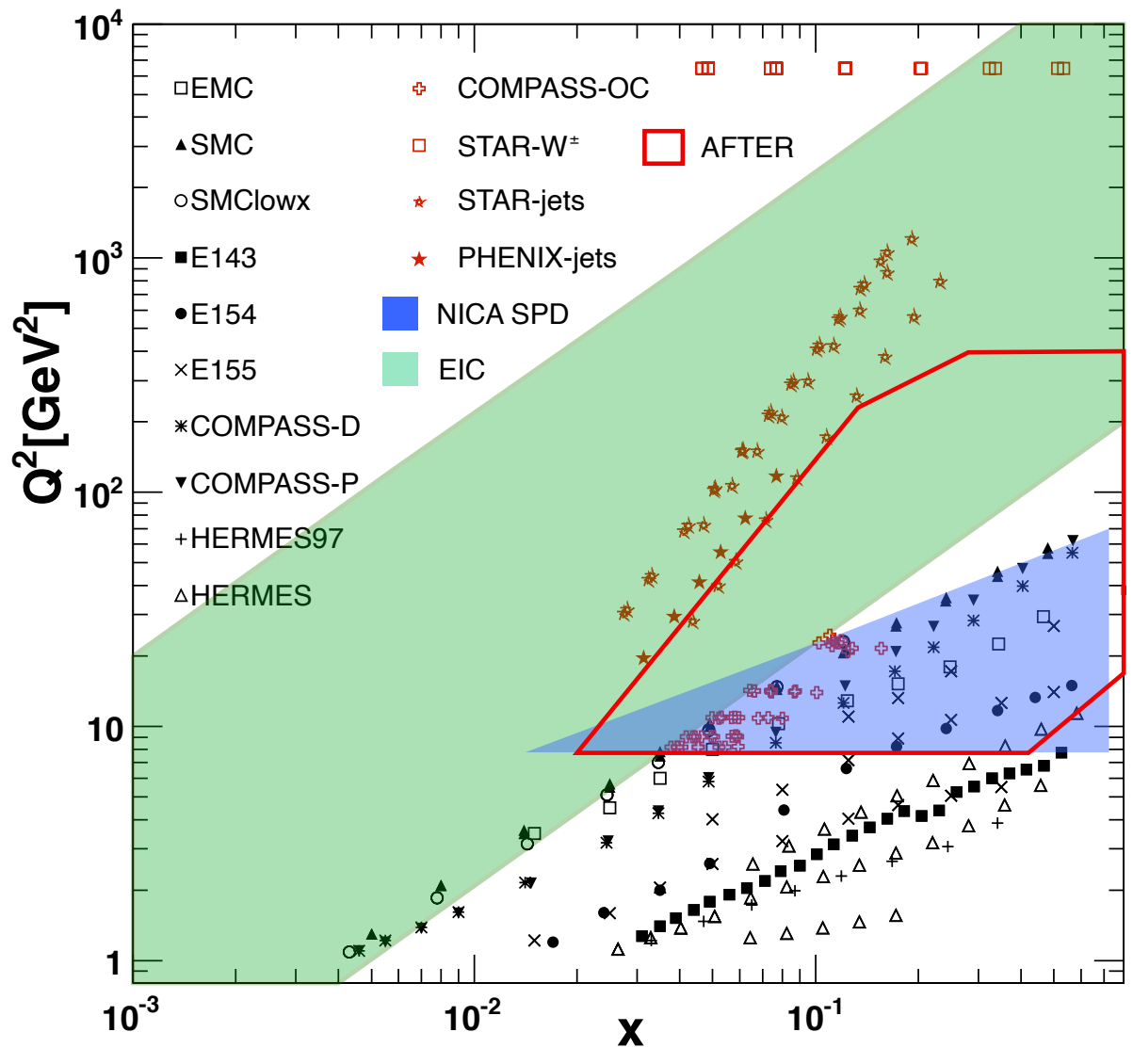

**Fig. 14.4.11** Kinematic coverage of SPD in the charmonia, open charm, and prompt photon production processes.

SPD has an extensive physics program for the first stage of the NICA collider operation with reduced luminosity and collision energy of the proton and ion beams, devoted to comprehensive tests of the various phenomenological models in the non-perturbative and transitional kinematic domain. It includes such topics as the spin effects in elastic scattering, in exclusive reactions as well as in hyperons production, multiquark correlations and dibaryon resonances, charmonia and open charm production, physics of light and intermediate nuclei collision, hypernuclei, etc. [4579]. The proposed program covers up to 5 years of the NICA collider running.

The SPD experimental setup, shown in Fig. 14.4.12, is designed as a universal $4\pi$ detector with advanced tracking and particle identification capabilities based on modern technologies, consisting of the barrel part and two end-caps. The silicon vertex detector will provide a reconstruction of secondary vertices of $D$-meson decays. The straw-tube-based tracking system placed within a solenoidal magnetic field of up to 1 T should provide tracking capability. The time-of-flight system will provide $\pi/K$ and $K/p$ separation together with an aerogel-based Cherenkov detector in the end-caps. Detection of photons will be provided by the sampling electromagnetic calorimeter. To minimize multiple scattering and photon conversion effects for photons, the detector material will be kept to a minimum throughout the internal part of the detector. The muon (range) system is planned for muon identification. It can also act as a rough hadron calorimeter. The pair of beam-beam counters and zero-degree calorimeters will be responsible for the local polarimetry and luminosity control. To minimize possible systematic effects, SPD will be equipped with a free-running (triggerless) DAQ system. The SPD experimental setup is currently in the phase of the technical project preparation.

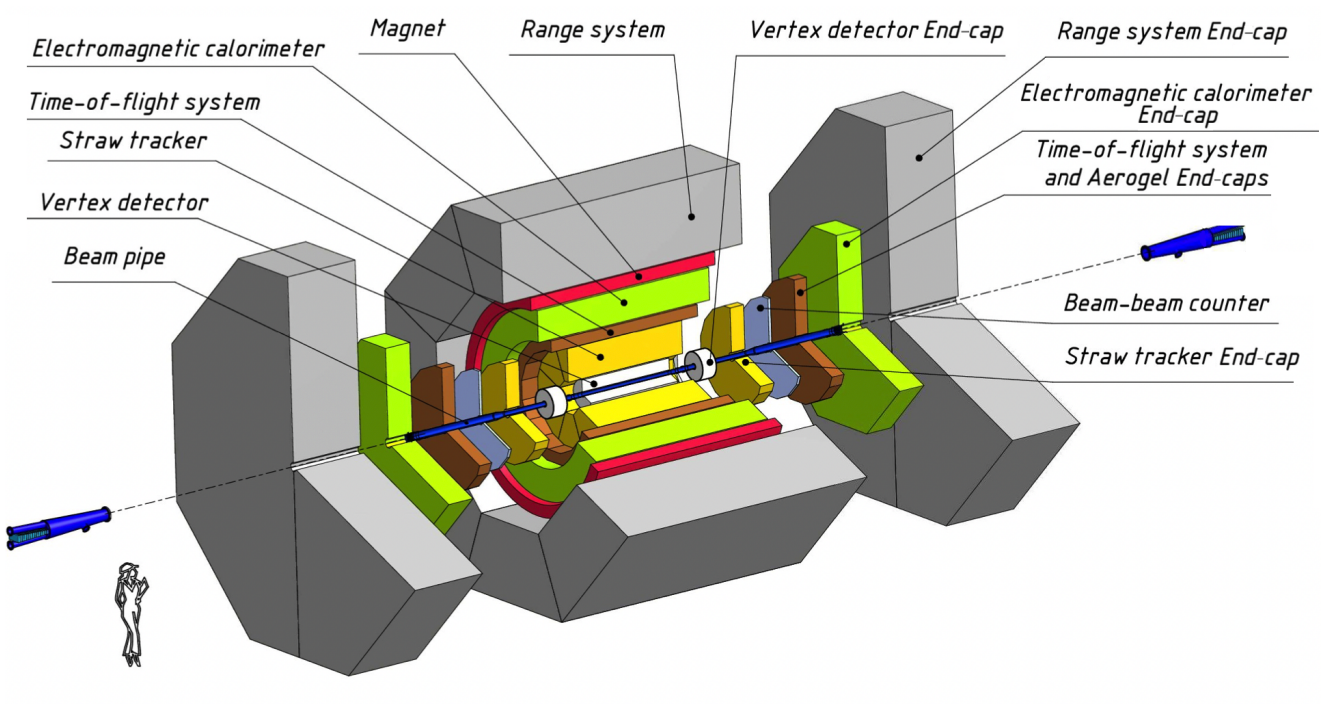

**Fig. 14.4.12** Layout of the SPD experimental setup.

## 14.5 QCD at FAIR

**Johan Messchendorp, Frank Nerling, and Joachim Stroth**

### 14.5.1 The FAIR facility

The international Facility for Antiproton and Ion Research FAIR (14.5.1) is an accelerator complex currently constructed at the site of the national GSI Helmholtz Center for Heavy-ion Research, Germany. It is composed of a rapid cycling synchrotron with maximum rigidity 100 Tm providing beams directly to experimental halls and to production targets for secondary ion and anti-proton beams [4580]. A high-energy storage ring (HESR) enables experiments with antiproton and rare radioactive isotope beams. The latter are selected out of either nuclear fragments or fission products, emerging from reactions of e.g. relativistic uranium beams by the Super Fragment Separator (S-FRS), providing high transmission for reaction products and high selectivity and purity for selected rare isotopes [4581].

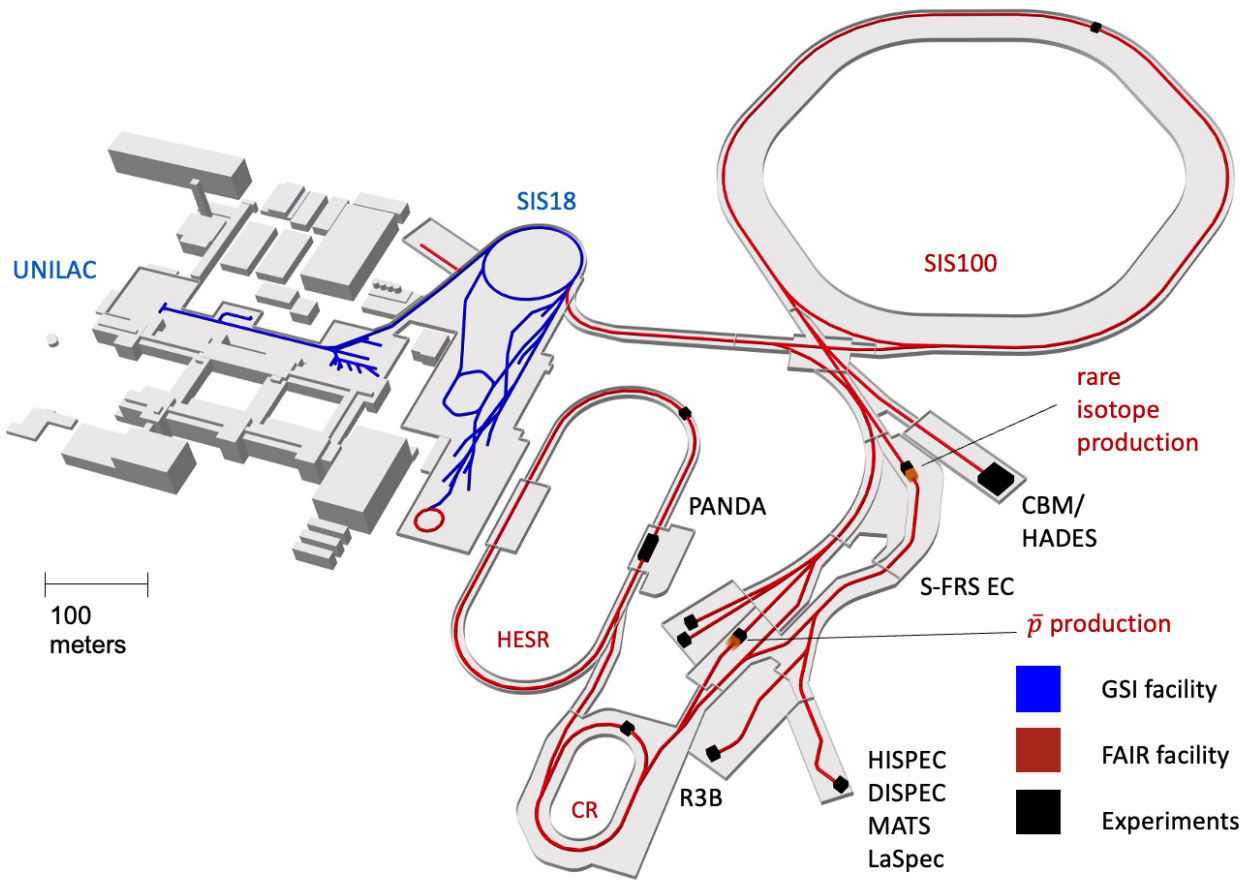

**Fig. 14.5.1** Layout of the FAIR accelerator complex. See text for the meaning of the various acronyms.

The scientific goals encompass many open questions connected with the formation of matter and the role of the strong force herein. The respective activities are organized in three pillars, hadron physics using antiproton annihilation (PANDA), heavy-ion reactions at relativistic energies (CBM), and nuclear structure physics at the limit of stability using relativistic, stored or decelerated rare isotope beams (NUSTAR). For the latter, not discussed in the remainder of this section, FAIR will pursue a unique approach enabling nuclear structure studies of e.g. the r-process isotopes relevant for the third r-process abundance peak. Acceleration of 28+ uranium ions in the SIS100 will push the space charge limit and yet provide beam energies around 1 $A$ GeV [4582]. SIS100 is particularly designed to accelerate medium charge state ions with a fast cycling rate of 1 Hz. This is achieved ramping the superconducting dipole magnets with 4 T/s to a maximum field of 1.9 T [4583]. Combined with the large acceptance and transmission of the Super-FRS, separated fission fragments will provide fully stripped isotope beams up to the neutron drip line. Such beams can be transferred to a storage ring for precision mass measurements (ILIMA), directed to a secondary target in the high-energy experiment hall for reaction experiments (R3B), or to experiments utilizing $\gamma$-spectroscopy in flight (HISPEC) or with stopped beams (DISPEC). Complementary experiments can also be performed at the Super-FRS operating the second half of the separator as high-resolution forward spectrometer and using a secondary target in the middle section of the separator (Super-FRS EC). Last not least isotope beams can also be decelerated and trapped (MATS) or investigated using laser spectroscopy (LaSpec). FAIR will also give home to many other experimental collaborations working in fields of atomic physics, radio biology, plasma physics and material science (APPA).

Civil construction of the accelerator complex has been started in 2017 focusing on the north area of the complex. As of 2022, the shell construction of the ring tunnel, the transfer buildings, the reaction experiment cave and the Super-FRS is mostly finished and the technical building installation has been started. The facility will be completed in a staged approach aligned with the funding profile and first beam from SIS100 to the CBM cave is anticipated for 2028. A FAIR early science program will be started as soon as the Super-FRS is installed providing uranium beam from SIS18 directly to the separator. Already now, a rich research program is ongoing at GSI and various other international facilities employing instrumentation developed for FAIR (FAIR Phase-0).

### 14.5.2 CBM - QCD studies under extreme baryon conditions

The research pillar *Compressed Baryonic Matter* (CBM) is addressing the physics of QCD matter under extreme conditions of baryon density and temperature. In a dedicated experiment hall, ion beams extracted from SIS100 will be directed onto stationary targets to form transients states of QCD matter in central collisions. The formation process is expected to reach maximum baryon densities of around five times the nuclear ground-state density at temperatures of up to 100 MeV. Model calculations suggest that e.g. in a Au+Au collision at a few $A$ GeV, the incoming nucleons are stopped to a large extent in the collision zone and that the nuclear matter is compressed to densities of $\rho_{\max} \simeq 1\ \mathrm{fm}^{-3}$ [4584]. It is expected that the formed hadronic system is approaching local equilibrium before it freezes out chemically at densities around $\rho_{\mathrm{ch}} \simeq 0.1\ \mathrm{fm}^{-3}$ (see Section 7.1). At such initial densities, the system can no longer be understood as resonance gas, but rather as an entangled meson cloud surrounding the baryonic cores (see Section 7.2).

Figure 14.5.3 demonstrates the world-wide efforts that explore the high-$\mu_B$-region (high net-baryon density) obtained at lower beam energy (c.f. 7.1) of the QCD phase diagram by means of heavy-ion collisions. Please note that by today no experiment has crossed the 50 kHz line.

The CBM collaboration has designed an experiment to investigate heavy-ion collisions with emphasis on the detection of rare and penetrating probes. Figure 14.5.2 shows the configuration of the Compressed Baryonic

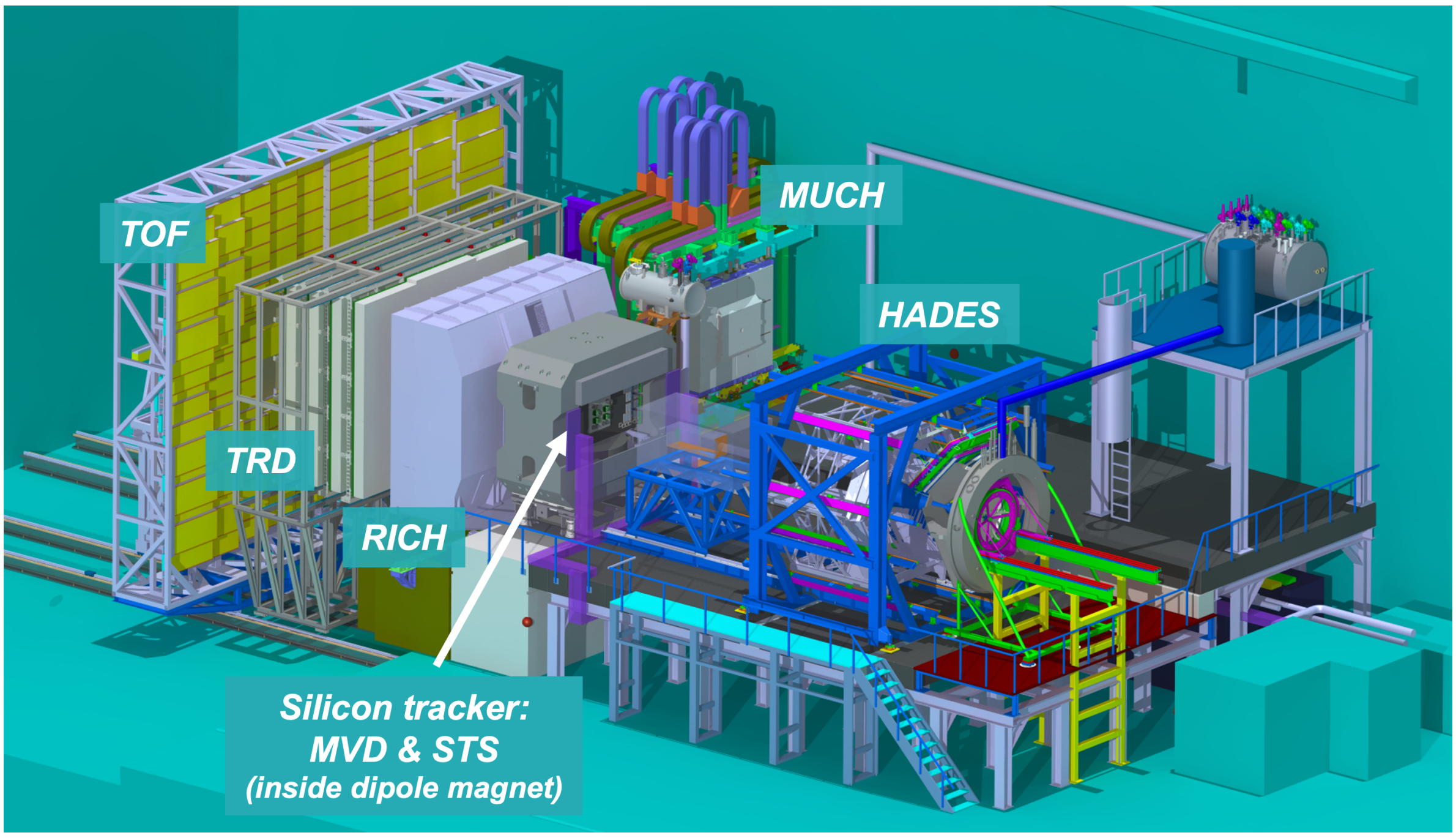

**Fig. 14.5.2** Computer rendering of the two experiments CBM and HADES installed in the FAIR fixed-target experimental hall. In case CBM is operated, the beam pipe is continuing through the center of the HADES experiment up to the CBM dipole (target vacuum chamber and beam pipe are not drawn). In case HADES is taking beam, a beam stop is placed between the two experiments (half transparent cube shown on a stand). The HADES setup is shown with blue support structure.

Matter experiment, together with the already existing HADES experiment placed at the same beam line delivering slow-extracted beam from the heavy-ion synchrotron SIS-100 The unique features of this fixed-target experiment are the rate capability reaching 10 MHz of inspected reactions and a modular composition of detectors for particle identification. The high-rate capability is achieved by performing tracking of charged particles in a compact configuration of 12 planes of silicon detectors placed in a 1 Tm dipole field. The planes are arranged over 1 m downstream the target. The first four planes are composed of monolithic pixel sensors, manufactured in a 180 nm CMOS process, and provide a total of 140 M-Pixels right behind the target and placed inside the beam vacuum (MVD). Behind, and outside the vacuum region, eight planes of silicon strip sensors constitute the core tracking system (STS). This tracking system is contained in a magnetic dipole field providing a maximum bending power of 1 Tm. Behind the tracking station different detector systems can be placed, depending on the observables to be addressed. In the standard configuration, a ring-imaging Cherenkov detector (RICH) provides superb electron/positron identification up to momenta of around 4 GeV. Behind, four stations of transition radiation detector enable intermediate tracking, energy loss measurement and additional electron/positron identification for high momentum tracks (TRD). The last detector is a wall of multichannel resistive-plate counters (TOF) covering about 20 $m^2$ in the transverse plane. It provides a high-precision time signal to enable particle identification by velocity vs momentum of charged particles. The CBM detector uses a trigger-less data acquisition system where every individual detector cell is digitized and where signals passing their thresholds receive a timestamp. Data streams of up to a TeraByte per second are transferred to the online compute cluster where real-time event building and feature extraction is performed. By selecting events with signatures of interest, the data stream is reduced to a level that allows storage on disks. Up to 40.000 compute nodes will be needed to accomplish this task in the case of operating at the highest interaction rate. The compute cluster will be installed in the FAIR Green Cube. The online event selection and rejection requires a high level of understanding and monitoring of the detector performance at the time of the data taking. To gain experiences and to prepare all software and firmware for fast calibration and event reconstruction, the CBM collaboration has installed a small version of the CBM detector at SIS18 beam line of GSI. This mini-CBM setup is composed out of prototypes or first-of-a-series modules of each detector system of CBM. The detectors are arranged as a single arm telescope and are operated without mag-netic field. The performance of the online event selection is benchmarked by investigating the production of hyperons. Their particular decay topology is used as identification.

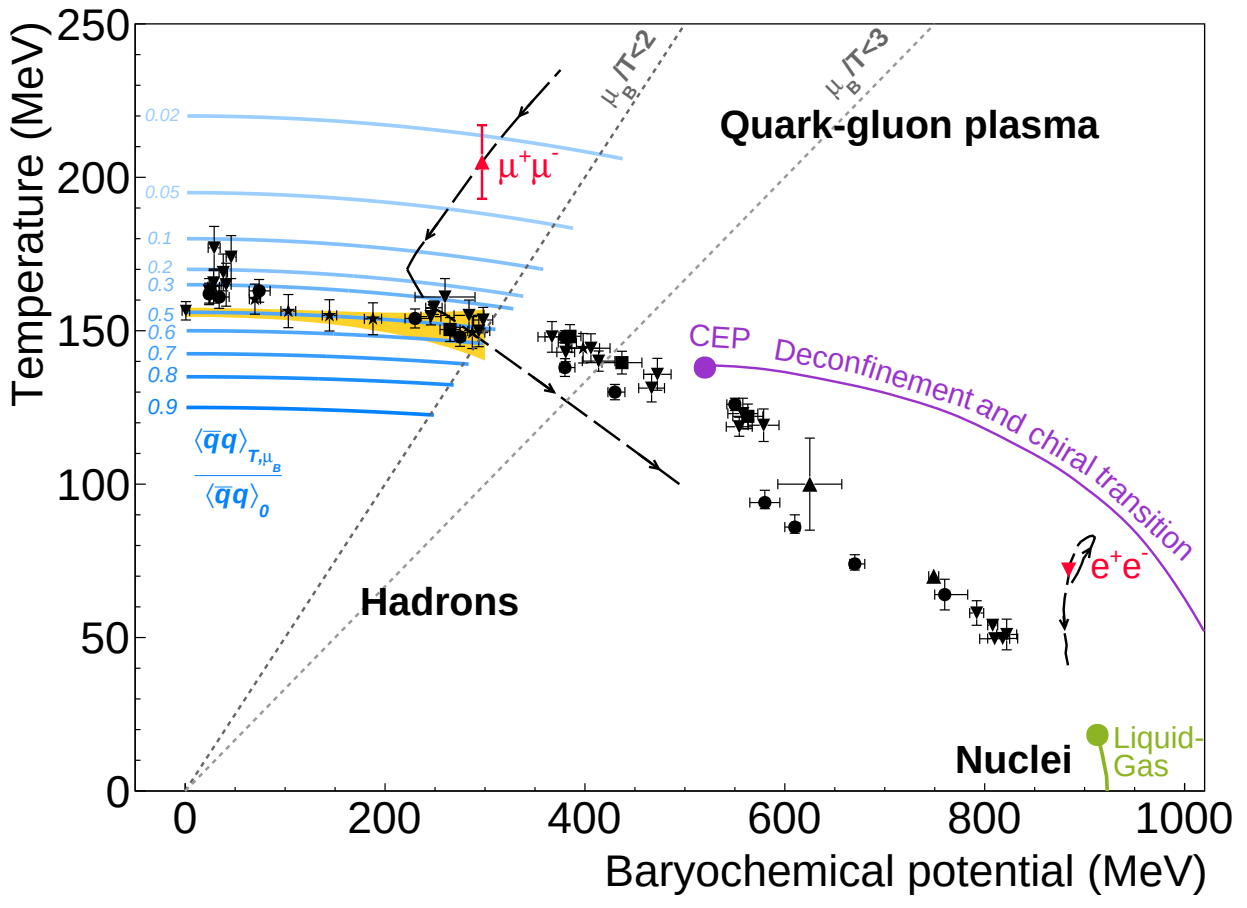

**Fig. 14.5.3** The QCD phase diagram as function of temperature and baryo-chemical potential. Freeze-out configurations extracted from particle yields assuming sudden freeze-out of a hadron resonance gas are shown as green circles (*cf* 7.1.9). Expectation values of the chiral condensate deduced from lattice calculations as sky-blue lines. Measurements of the mean fireball temperature based on the dilepton continuum radiation are shown as red squares together with the expected trajectory of the expanding and radiating system.

The prime goal of the CBM program at FAIR is to search for signatures of a first-order phase transition, separating the hadron resonance gas region from a likely novel state of matter (*cf.* 7.2). The established strategy for this is to search for non-monotonic behavior of the excitation function of various observables, or more general for trends signaling a change in the number of degrees of freedom of the transient system like the (dis)appearance of a certain scaling behavior. An example is the excitation function of the multiplicity of multi-strange hyperons. The high-rate capability of CBM will enable such measurements well below the proton-proton production threshold[119].

Indeed, the region in the QCD phase diagram at high baryo-chemical potential is by large *terra incognita*. Figure 14.5.3 depicts the QCD phase diagram with experimental landmarks and predictions by lattice QCD. The landmarks include, first, chemical-freezeout points that characterize the temperature and the baryochemical potential below which the system can be understood as an expanding hadron gas in which inelastic collisions no longer occur. Two additional points are shown which depict an average temperature of the dense and hot system prior to freeze out. This very promising observable, so far not addressed in excitation functions with the needed precision, is the spectral distribution and yield of dileptons emitted from the dense and hot stage of the collision. Such dileptons couple via virtual intermediary photons directly to the in-medium hadronic current-current correlator and thus probe the microscopic structure of the medium they are expelled from [4585, 4586]. In the so-called low-mass region (LMR), i.e. for dilepton invariant masses around the vector-meson pole masses $\rho, \omega$ and $\phi$ and below, the spectral distribution encodes the "melting" of the vector mesons embedded in a hot and dense hadronic environment, while the dilepton spectrum from a purely partonic medium would not feature any particular structure. Moreover, the integral yield of continuum dileptons in the LMR dominantly depends on the size, the lifetime and the temperatures of the emitting source. It has been demonstrated using a hydro model that the fireball ball evolution can significantly change if during the evolution the system experiences a phase transition from a QGP-like to a hadronic equation-of-state. The study observed an increase of the yield by roughly a factor of two in the case of a first-order phase transition [4587]. Dilepton continuum radiation also provides a model independent measurement of the average temperature of the emitting source. This is possible if the imaginary part of the in-medium current-current correlator is sufficiently featureless and approaching a dependence $\propto T^2/M^2$. In that case, the spectral distribution is defined essentially by the thermal Bose factor and the invariant-mass distribution takes the form of black-body radiation, i.e. $\propto (MT)^{3/2} \exp(-M/T)$ [4588]. A fit of a Planck distribution function to the spectral distribution in the respective invariant mass reveal an invariant measurement of that average temperature, unaffected by any blue shift due to rapid expansion of the emitting source. The two measurements of the average temperature shown in Fig.14.5.3 were obtained by the NA60 collaboration in the dimuon channel [4589] and by the HADES collaboration in the dielectron channel [4590]. The "trajectories" indicated as dashed-dotted lines depict the evolution of the fireball used to integrate the emissivity over the four-volume characterizing the evolution of the collision zone. For details see [4590].

In order to obtain the continuum radiation, contributions to the dilepton invariant-mass distribution from the early pre-equilibrium stage and from late decays of long-lived hadronic states have to be determined and subtracted [4591]. An important part of the CBM program are therefore reference measurements of elementary collision systems or the production of dileptons in collisions of protons on nuclei. For this, the HADES detector will be moved to the SIS100 experimental hall where it will be installed in front of the CBM detector. HADES, with its large polar acceptance, is well suited to study in particular the production and propagation of vector mesons in cold nuclear matter. The feasibility of reconstructing the dilepton continuum radiation in heavy-ion collisions at energies SIS18 energies has been demonstrated for the system Au+Au at $\sqrt{s_{\rm NN}} = 2.42$ GeV. Figure 14.5.4 depicts the respective invariant-mass distribution together with various model calculations. It is important to note that at this collision energy, the $\rho$ meson is substantially broadened due to the high baryon density, thus satisfying the criteria for temperature measurement outlined above also in the LMR.

[119] The threshold is here defined as the energy needed to produce a given hyperon in an elementary proton-proton collision and the beam energy is referred to as $\sqrt{s_{\rm NN}}$.

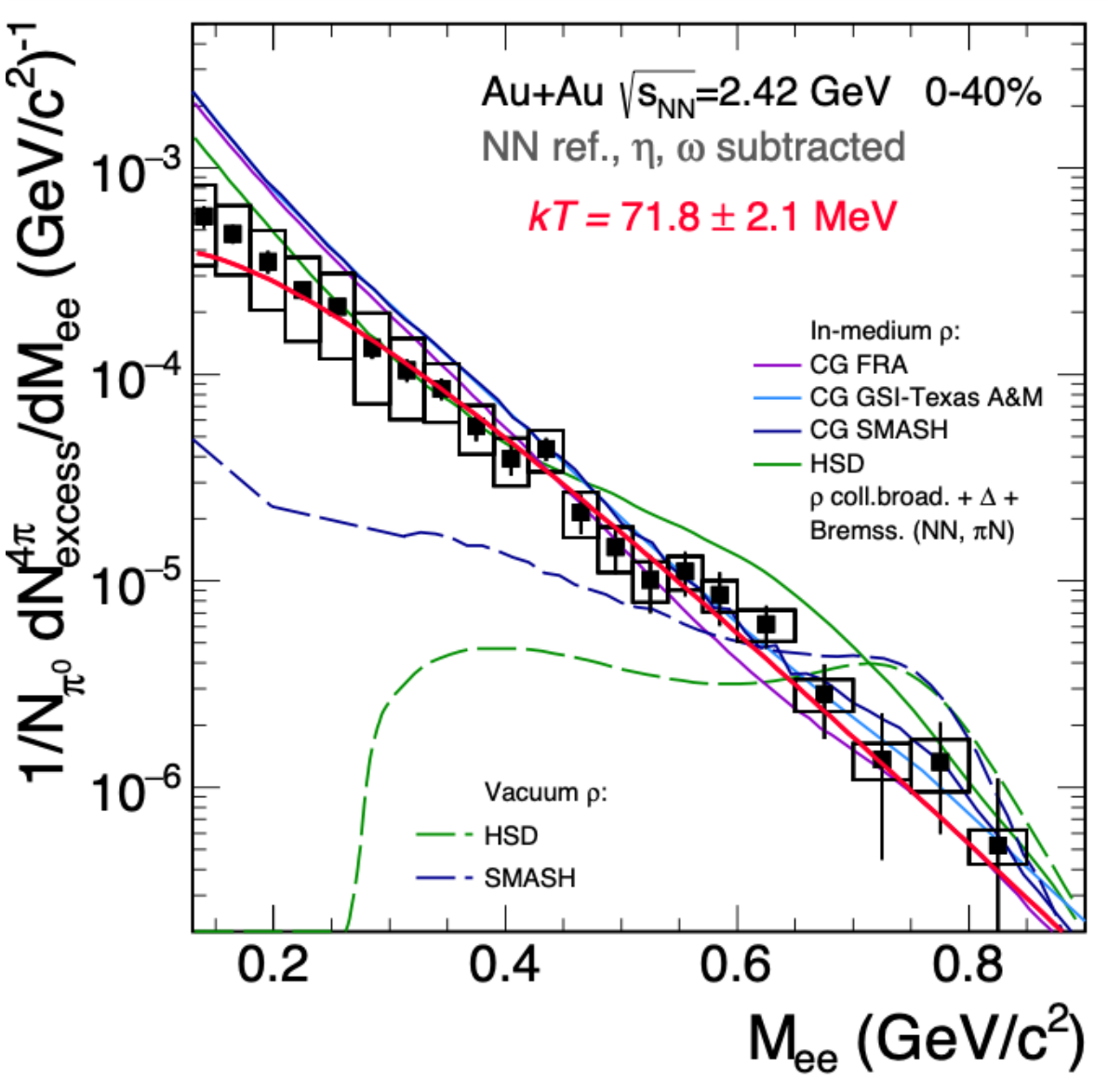

**Fig. 14.5.4** Di-electron excess radiation measured by HADES for the collision system Au+Au at $\sqrt{s_{\rm NN}} = 2.42$ GeV (black squares). Systematic uncertainties are depicted as open boxes while the statistical errors are shown as vertical lines. Various model calculations are shown as colored lines (see inserts for explanation). Lines labeled CG refer to calculations using coarse grained microscopic transport calculations for the fireball evolution folded with thermal emissivities derived from many-body theory. The line labeled HSD is the result of a full microscopic transport simulation treating the dilepton emission perturbatively, i.e. after the full hadron cascade has been processed. Also shown as dashed lines are the descriptions of dilepton emission from $\rho$-meson decay used in the full microscopic (shining) approach. The spectrum has been obtained by subtracting from the total yield in the centrality class 0-40% the contributions from late hadron neutron meson decays (cocktail) and from first-chance collisions.

### 14.5.3 PANDA - Hadron structure & spectroscopy studies using antiprotons

*Physics with antiprotons and PANDA*

The ambition of PANDA is to exploit the annihilation of antiprotons with protons and nuclei to study the properties of hadrons and their interactions with unprecedented precision and coverage in parity, spin, and gluon and quark flavor contents. Partly as the successor of the successful LEAR facility at CERN, PANDA will combine a high-resolution and intense antiproton beam with a state-of-art detector system. The experiment is designed to produce hadrons with masses of up to about 5.5 GeV/$c^2$ and to unambiguously detect a large variety of final-state particles with excellent momentum resolution, particle identification capabilities, and exclusivity.

PANDA will be an internal-target experiment installed at the High Energy Storage Ring (HESR). The antiproton beam from HESR has several key advantages, namely i) the production cross sections of hadrons are generally large, resulting in large data samples; ii) meson-like states of any quark-antiquark spin-parity combination can be produced in formation with a superb mass resolution; iii) baryon-antibaryon pairs, including multi-strange and charm, can be produced in two-body reactions, which provide clean conditions for baryon studies; iv) proton-antiproton annihilations constitute a gluon-rich environment.

In the initial phase, HESR will be able to store $10^{10}$ antiprotons with momenta $p$ from 1.5 GeV/$c$ up to 15 GeV/$c$. By making use of the stochastic cooling technique, the relative beam-momentum spread ($\Delta p/p$) will be $< 5\times10^{-5}$. The antiprotons will interact with a cluster jet target or pellet target, which results in a luminosity during the first phase (Phase One) of data taking of about $10^{31}$ s$^{-1}$cm$^{-2}$. The final goal is a luminosity of up to $2\times10^{32}$ s$^{-1}$cm$^{-2}$, referred to as Phase Three.

The PANDA detector is designed to measure momenta of charged and neutral final-state particles with 1-2% resolution and with excellent particle identification, vertex reconstruction, and count-rate capabilities. The nearly $4\pi$ acceptance allows to study exclusive reactions covering a large part of their phase spaces, thereby enabling a conclusive partial-wave analysis. The detector consists of a Target Spectrometer (TS) and a Forward Spectrometer (FS). The TS provides precise vertex tracking by the micro vertex detector, surrounded by straw tube trackers and gas electron multiplier detectors in the forward direction. The trajectories of charged particles in the TS are bent by the field of a solenoid magnet providing a field of 2 T, with muon detectors within the segmented yoke. For particle identification, the TS will consist of time-of-flight and Cherenkov detectors and an electromagnetic calorimeter composed of $PbWO_2$ crystals. With the electromagnetic calorimeter, nearly covering the full phase space using a barrel and two endcaps, the measurement of energies and scattering angles of photons, electrons, and positrons will become possible.

The FS consists of straw tube stations for tracking, a dipole magnet, a ring imaging Cherenkov detector, a forward time-of-flight system and a Shashlyk electromagnetic calorimeter, followed by a muon range system. The luminosity at PANDA will be determined by using elastic antiproton-proton scattering as the reference channel registered by a dedicated luminosity detector.

The combination of the intense, high resolution antiproton beam with the nearly $4\pi$ PANDA detector, opens up unprecedented possibilities with a very rich physics program, particularly suited to provide a deeper understanding of QCD in the non-perturbative regime. In the following, we discuss some of the QCD-driven highlights from the various pillars of the physics program of PANDA. We note that PANDA has a more extensive physics program that includes various nuclear physics aspects as well, such as the foreseen hypernuclei and hyperatom topics. We limit ourselves here to those topics in which the quarks, gluon, and their interactions are expected to be the most important degrees of freedom. For a more detailed description of the complete physics program at the first phase of the experiment, we refer to [4592].

*Hidden charm and exotics*

PANDA will be devoted to provide precision data for hadron spectroscopy with light to charm constituent quarks, and gluons. Given the anti-proton beam momentum range of up to 15 GeV/$c$, the accessible invariant-mass range in direct formation is about 2 - 5.5 GeV/$c^2$, and the PANDA experiment is thus designed and optimized to cover the charmonium mass region. In addition, the light quark sector can be explored via the production with recoil particles.

The cross sections associated with antiproton-proton annihilations are generally several orders of magnitude larger than those of experiments using electromagnetic probes, allowing for excellent statistical precision already at moderate luminosities available in the initial Phase One ($\sim 10^{31}$ cm$^{-2}s^{-1}$).

In the charmonium mass region, different unexpected charmonium-like states have been discovered since the beginning of the millenium. Some of these so-called XYZ states are electrically charged and in combination with the mass those are manifestly exotic states. They have unambiguously a minimum quark content of four quarks (e.g. $c\bar{c}d\bar{u}$) and are, among others, discussed to be tetraquark or molecular states in form of a loosely bound di-meson system. PANDA will contribute to solve the puzzle of the nature of these unexpected charmonium-like XYZ states. Moreover, there is a number of pentaquark states and other exotic candidates reported by LHCb recently that will be accessible with PANDA.

In order to understand the nature of the XYZ states, e.g. which of the different four-quark configurations are realized by nature, and to confirm further candidates reported, PANDA will play an unique role. The different multiplets need to be completed, especially the corresponding high-spin states. Those can uniquely be addressed by PANDA, since there is no restriction in produced $J^{PC}$ quantum numbers in $\bar{p}p$ annihilation and thanks to the mostly $4\pi$ acceptance of the detector. Given the excellent electromagnetic calorimetry in the barrel as well as in the forward part of the detector, PANDA will have full acceptance not only for charged but also for neutral final-state particles.

Another crucial and unique tool are precision line-shape measurements. The energy-dependent resonance cross sections of these states are strongly connected with the inner structure of such states – theoretical interpretations come along with predictions for absolute decay widths and line shapes. The narrow and famous $X(3872)$, meanwhile renamed by the PDG to $\chi_{c1}(3872)$, was the first of these XYZ states discovered in 2003 [4596]. Its nature is still not understood.

As shown by a comprehensive Monte Carlo based feasibility study [4593], the line shape of narrow states, particularly the $X(3872)$, can be measured precisely and directly by PANDA with sub-MeV resolution, Fig. 14.5.5, allowing for sorting out models, Fig. 14.5.5, right. Thanks to the unprecedented beam momentum and energy resolution of the HESR of up to $\Delta p/p = 2 \times 10^{-5}$ and $\Delta E_{\rm cms}/E_{\rm cms} = 34$ keV, even very similar line-shape models can be discriminated by employing the technique of a resonance energy scan [4593].

At LHCb, it was not possible to distinguish between a Breit-Wigner and a Flatté-like line-shape for the $X(3872)$ even though huge statistics has been accumulated [4594]. This state cannot be produced in direct formation at LHCb, and the energy-scan technique cannot be employed. Consequently, the resolution of the measurement is dominated by the detector resolution (order of a few MeV) and the LHCb data are equally well described using both line-shape models (Fig. 14.5.6).

As an addendum to the published sensitivity study [4593], the expected PANDA performance in distin-guishing these two different line-shape models has been investigated and quantified [4595]. The achievable performance has been evaluated in terms of the mis-identification probability $P_{\rm mis}$ to assign the wrong line-shape model, namely the Breit-Wigner line shape for Monte Carlo data generated using a Flatté line shape, and vice versa. The outcome is summarized in Fig. 14.5.7, where the resultant sensitivities in assigning the correct line shape (shown here for the Flatté-like line shape) are better than 90% and 98%, depending on the given accelerator operation mode (Fig. 14.5.7, left). For this figure of merit, a mis-identification probability of $P_{\rm mis} = 50\%$ corresponds to "indistinguishable". To answer the question, how much better the expected PANDA performance is as compared to "indistinguishable", one may consider the so-called "odds" defined as the number of correct assignments per wrong one: $odds := (1 - P_{\rm mis})/P_{\rm mis}$. The corresponding results are shown in Fig. 14.5.7 (right). Using this measure, PANDA is expect to be at least a factor of 10 better than "indistinguishable", a feature that is only possible due to PANDA's excellent beam-momentum resolution and the direct formation of the $X(3872)$ state in antiproton-proton annihilations.

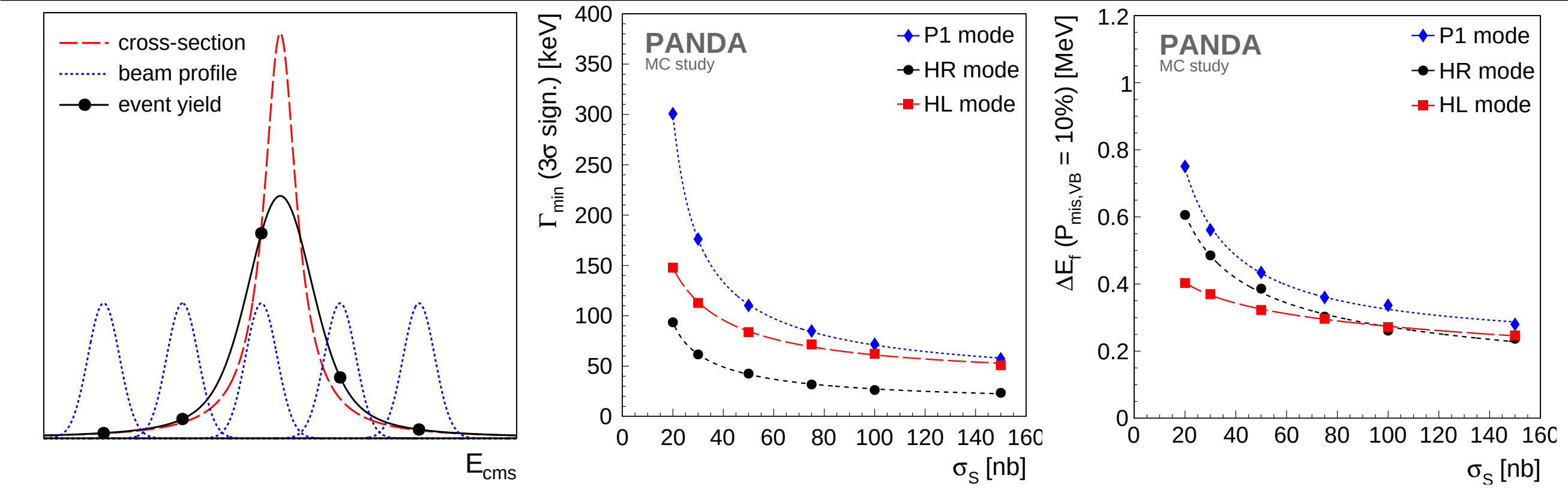

**Fig. 14.5.5** Illustration and summary of a comprehensive Monte Carlo simulated scan experiment study for PANDA [4593]. Schematic of the resonance energy scan principle *(left)*. Summary of the sensitivity study for an absolute (Breit-Wigner) decay width measurement in terms of the minimum decay width $\Gamma_{\min}$ that can be measured with an relative precision of 33% as a function of the assumed input $\sigma_S$ *(center)*. Summary of the sensitivity study for line-shape measurements via the $E_f$ parameter (Molecule case) to distinguish between a bound and a virtual state scenario in terms of the probability to mis-identify a virtual as a bound state *(right)* .

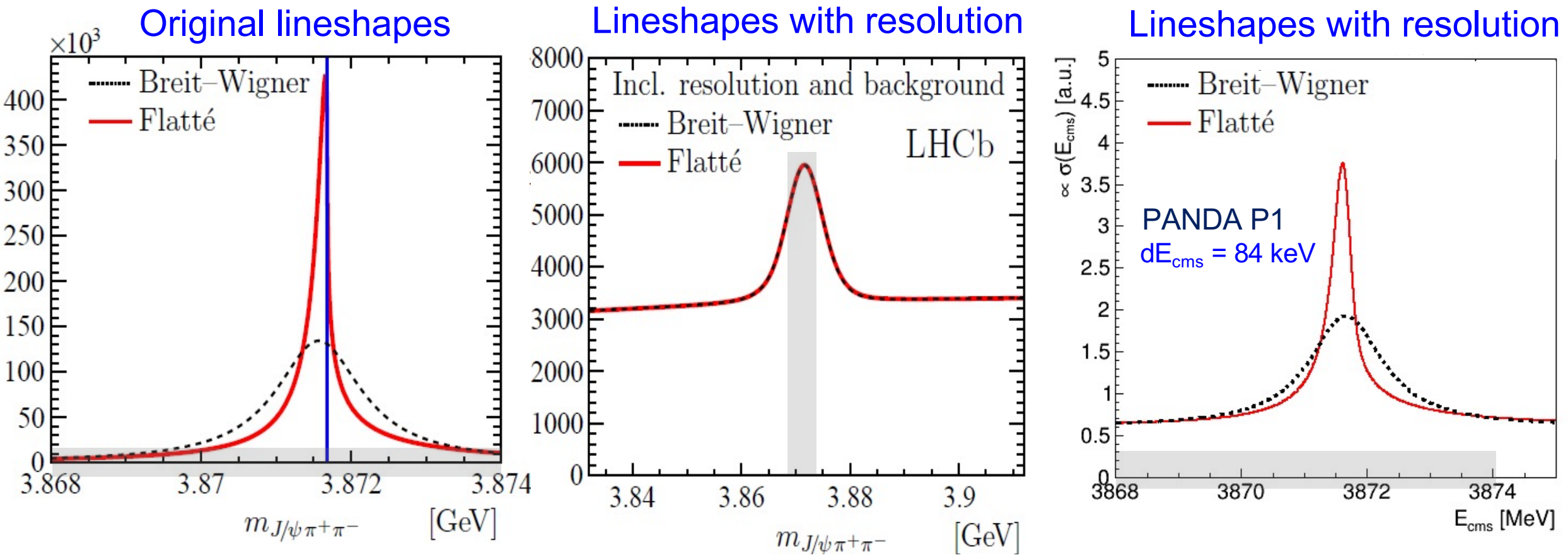

**Fig. 14.5.6** Comparison of the Breit-Wigner and Flatté-like line shapes without and with the LHCb and PANDA resolutions convolved. *Left:* The two line shapes (Breit-Wigner vs. Flatté-like) obtained from the fit to the LHCb data [4594]. *Center:* The same two line shapes when including backgrounds and resolution, *i.e.* convolved with the detector resolution. Due to the resolution, the two line shapes are just indistinguishable based on the LHCb data [4594]. *Right:* The same two line shapes (Breit-Wigner vs. Flatté-like) convolved with the foreseeen beam-energy resolution expected for the initial phase of the experiment. Thanks to the excellent beam energy resolution, they are well distinguishable with PANDA at HESR [4595].

Concerning the light-quark and gluon sector, PANDA will search for exotic forms of matter such as hybrid mesons and glueballs. In the mass range accessible at FAIR, a large number of glueballs is expected and some of them might be narrow. Their SU(3) structure can be determined from an analysis of their decay modes.

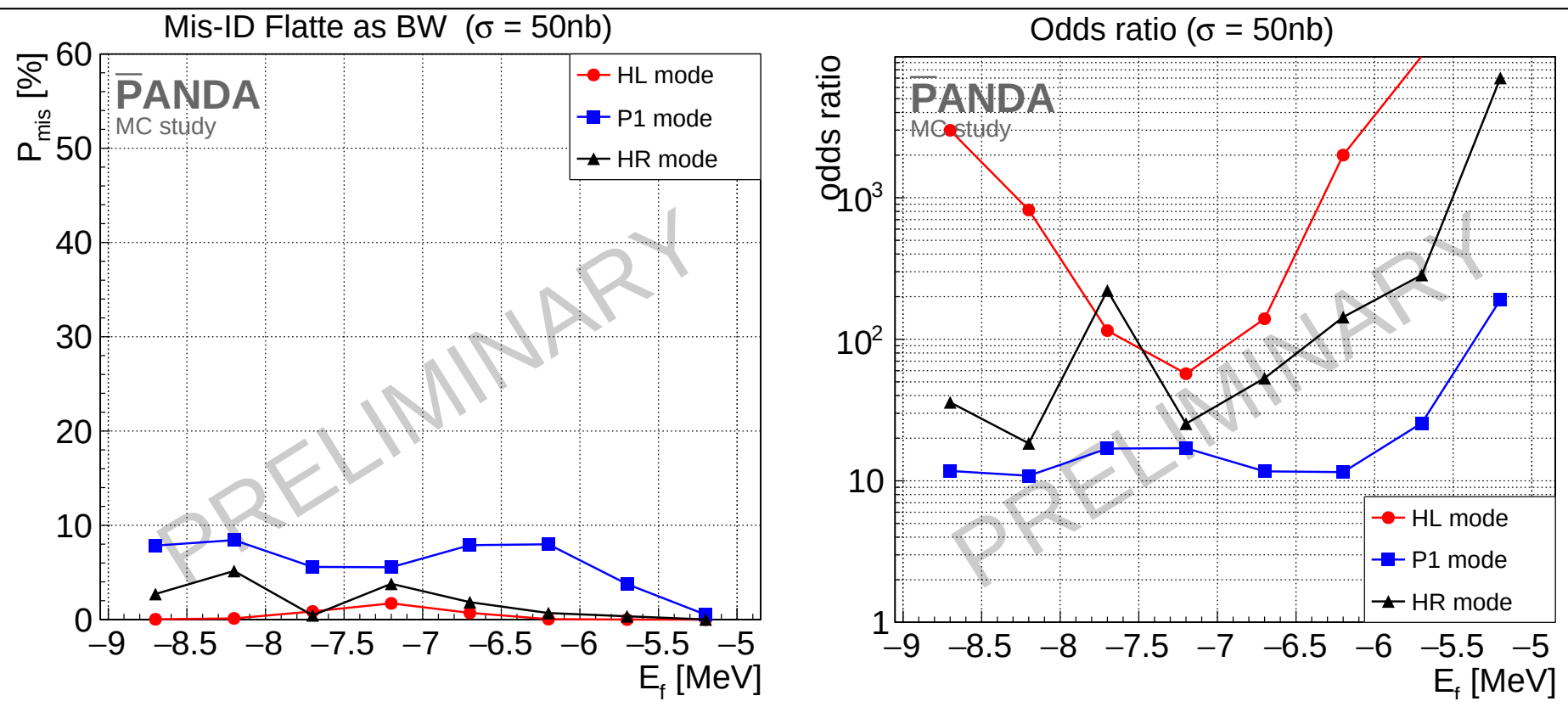

**Fig. 14.5.7** Performances to distinguish between a Breit-Wigner and a Flatté-like line shape with PANDA/HESR at FAIR. *Left:* Sensitivity in terms of the mis-identification probability $P_{\rm mis}$ to wrongly assign the Breit-Wigner line shape instead of the correct Flatté-like line shape as a function of the Flatté energy parameter $E_{\rm f}$, whereas $P_{\rm mis} = 50\,\%$ corresponds to "indistinguishable". *Right:* The correspondingly computed so-called "odds", *i.e.* the number of correct assignments per wrong one, defined as *odds* $:= (1 - P_{\rm mis})/P_{\rm mis}$. Using this measure, the expected performance is at least ten times better than "indistinguishable", i.e. as it is achieved based on the LHCb data [4594], see also [4595].

For light hybrid mesons, such as the $\pi_1(1400)$ and $\pi_1(1600)$, the most conclusive results so far have been provided by the COMPASS experiment at CERN/SPS, employing a 190 GeV/$c$ pion beam, see e.g. [4597–4599]. The GlueX photoproduction experiment has been constructed and is dedicated to map the full spectrum of hybrid mesons with masses of up to about 2.5 GeV/$c^2$. The findings by both of these experiments and others on hybrids as well as on non-exotic new light meson states, such as the [4600], will complementary be addressed in $\bar{p}p$ annihilation processes at PANDA. These kind of investigations will moreover be extended to the charmonium region, for which several glueball and hybrid states are predicted, e.g. a spin-exotic state at about 4.2 GeV/$c^2$ [4601].

Presently, there is no experiment dedicated to glueballs. In comparison to glueball searches in $J/\psi$ decays e.g. at BESIII, they are expected to be produced with orders of magnitude higher production rate in $\bar{p}p$ annihilation [4602]. In particular in the charm region, glueball candidates with masses above 4 GeV/$c^2$ are predicted, some of which might be narrow and could thus be found. An analysis of their decay fraction could be used to decide if the state has a large glueball component.

*Strangeness physics*

With antiproton-proton annihilations and baryon number conservation, the final state has zero total baryon number. This feature has the advantage that relatively clean two-body final-state topologies may emerge involving exclusively a baryon together with its antibaryon. The maximum center-of-mass foreseen with PANDA amounts to 5.5 GeV/$c^2$ which provides access to produce pairs of various hadrons including strange and charm quarks such as $\bar{p}p \to \Lambda\bar{\Lambda}$, $\Sigma\bar{\Sigma}$, $\Xi\bar{\Xi}$, $\Omega\bar{\Omega}$, $\Lambda_c\bar{\Lambda}_c$, $\Sigma_c\bar{\Sigma}_c$, $\Xi_c\bar{\Xi}_c$, $\Omega_c\bar{\Omega}_c$, together with various excited states of these hadrons. The production of these pairs has various benefits, namely i) close to the appropriate production threshold, the identification and analysis of these reactions are fairly simple, since one may apply tagging methods, deal with limited number of partial waves, and with a good signal-to-background level; ii) combined with the excellent momentum resolution of the initial antiproton beam, a near-threshold scan allows to determine basic properties, such as mass and width, of these states, and their excitations very accurately [4603]; iii) the self-analyzing feature of the weak decays of these (anti)baryons can be exploited to study spin degrees-of-freedom of their production process. The latter feature is a powerful tool that can be used for various physics aspects ranging from particle physics (test CP conservation in the hyperon sector), spectroscopy studies (baryon resonances with strangeness), and spin physics (detailed study of hyperon production and interactions). In the following, we highlight two aspects that will be foreseen with PANDA, namely the spin-physics and hyperon-spectroscopy programs.

The spin-physics program of PANDA aims to measure accurately differential cross sections and spin observables such as polarization and spin correlations. These observables provide a deeper understanding of the spin production mechanisms or, more generally, of the dynamics that lead to the production of hyperons in antiproton proton collisions. Which effective degrees of freedom are adequate to describe the hadronic reaction dynamics: quarks and gluons or mesons and baryons? And how does this picture change with center-of-mass energy? The high production rates of hyperon and antihyperon pairs in combination with the excellent signal to background yield give perfect conditions to perform these measurements. Already with moderate initial luminosities, a spectacular production rate of hyperon and antihyperon pairs are to be expected. The reaction $\bar{p}p \to \Lambda\bar{\Lambda}$, with $\Lambda \to p\pi^-$ and $\bar{\Lambda} \to \bar{p}\pi^+$, was studied in detailed Monte Carlo simulations. At a luminosity of $10^{31}$ cm$^{-2}$s$^{-1}$ and at a antiproton beam momentum of 1.64 GeV/$c$ we expect $3.8\times10^6$ of fully reconstructed $\Lambda\bar{\Lambda}$ pairs per day. For strangeness $|S| = 2$ baryon pairs via $\bar{p}p \to \bar{\Xi}^+\Xi^-$ at a beam momentum of 4.6 GeV/$c$, the expected rate is about $2.6\times10^4$/day exclusively reconstructed pairs in the $\Xi^- \to \Lambda\pi^-$ and $\bar{\Xi}^+ \to \bar{\Lambda}\pi^+$ decay modes. Moreover, the signal-to-background ratio is estimated to be better than 100 (250) for the $\bar{\Lambda}\Lambda$ ($\bar{\Xi}^+\Xi^-$) channel. With the perspectives of PANDA to reach the high luminosity conditions at HESR at Phase Three, precision studies of hyperons with charm contents will become feasible and CP violation tests will become competitive [4604].

PANDA's environment to produce abundantly pairs of hyperons and antihyperon is also the ideal setting to carry out detailed spectroscopy studies of these baryons. The underlying physics motivation is to understand the internal structure of baryons. For this purposes, baryon spectroscopy has demonstrated to be a very powerful tool. In the case of PANDA, the conceptual idea is to replace light valence quarks of the (anti)proton with heavier strange and charm ones via the processes sketches above, measure the excitation spectrum of excited hyperon states, determine their properties such as mass, width, spin, parity, and decay modes, and compare such observations between the various baryonic systems including those of the light-quark sector, i.e. $N^*$ and $\Delta$ resonance levels. With these measurements some of the open questions will be addressed, such as i) Which baryonic excitations are efficiently and well described in a three-quark picture and which are generated by coupled-channel effects of hadronic interactions? ii) To which extent do the excitation spectra of baryons consisting of *u, d, s* obey SU(3) flavor symmetry? iii) Are there exotic baryon states, *e.g.* pentaquarks or dibaryons? iv) What is the role of diquark correlations inside baryons? v) Can we understand the missing resonance phenomena and the observed level ordering in the light-quark baryon sector? PANDA has the potential to be the key player in providing conclusive data for the strangeness $|S| = 2, 3$ (anti)baryons thereby complementary to the future activities planned at J-PARC [4605] and the wealth of baryon spectroscopy data that have been obtained with photo- and pion-induced reactions at JLab, ELSA, MAMI, GRAAL, Spring-8, HADES, etc.. As an illustration of the capabilities of PANDA to determine spin-parity assignment of excited $\Xi^*$ states, we refer to the results of a preliminary feasible study described in [4606].

*Nucleon structure*

In the past 60 years, the structure of the proton has been extensively studied with great success exploiting lepton-hadron scattering (see Section 10). With the annihilation of antiproton with protons, it will be possible to extract electromagnetic form factors (EMFF) and structure functions of the (anti)proton in a region of phase space not accessible using electromagnetic probes.

EMFFs quantify the hadron structure as a function of the four-momentum transfer squared $q^2$ and are defined on the complex $q^2$ plane. Space-like EMFFs ($q^2 < 0$) are real functions of $q^2$ and have been studied extensively using elastic electron-hadron scattering. Time-like EMFFs are complex and will be studied at PANDA using different processes in various $q^2$ regions. Figure 14.5.8 sketches the various processes that can be exploited to study EMFFs for various $q^2$ regions. Here, $B$, $B_1$ and $B_2$ denote various baryons. With antiproton-proton annihilations, EMFFs of the (anti)proton will be probed for the $q^2$ range starting from the unphysical region, using the reaction $\bar{p}p \to e^+e^-\pi^0$, to high-$q^2$ via $\bar{p}p \to \ell^+\ell^-$ whereby $\ell$ refers to both electrons and muons. Detailed Monte Carlo simulations demonstrated that both $G_E$ and $G_M$ can be measured with a precision of about 3% in the $e^+e^-$ final state at $q^2$ around 5 (GeV/$c$)$^2$ and with a total integrated luminosity of 0.1 fb$^{-1}$, which is well suitable for the first years of data taking. Figure 14.5.9 depicts the present state-of-the-art of the $R = |G_E|/|G_M|$ measurements as a function of $q^2$ together with the precision perspectives of PANDA for the early phases of the experiment (green band) and for the high luminosity mode (purple band). PANDA will be able to harvest more precise form factor data compared to today's measurement and extend the measurements towards higher values of $q^2$ including, for the first time, both the di-electron and di-muon as probes. Being analytic functions of $q^2$, space-like and time-like form factors are related by dispersion theory. With the future data taken at PANDA and the various other complementary facilities, it will become feasibly to rigorously test the analyticity and universality of the measured EMFFs. Besides measuring the EMFFs of the (anti)proton, also transition form factors ($B_1 \neq B_2$) are accessible. With the copious production of hyperons and antihyperons in antiproton-proton collisions, PANDA will provide unique data to extract transition form factors of various hyperons and their corresponding antihyperons.

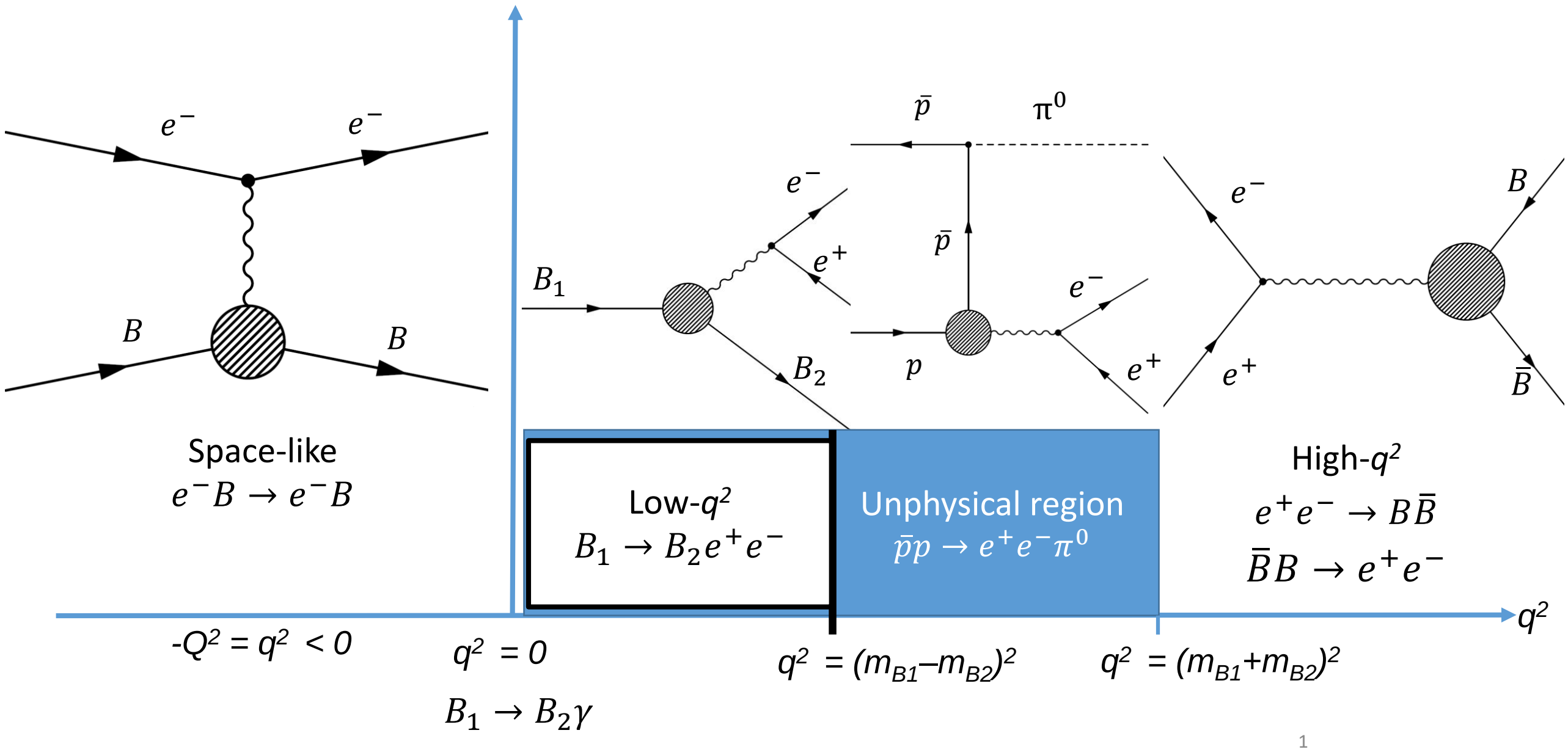

**Fig. 14.5.8** The various processes that are used to extract information about the EMFF in the space-like ($q^2 < 0$) and time-like ($q^2 > 0$) regions. The time-like region $0 < q^2 < (M_{B1} - M_{B2})^2$) is studied by Dalitz decays. The so-called unphysical region ($4m_e^2 < q^2 < (M_{B1} + M_{B2})^2$) by $\bar{p}p \to \ell^+\ell^-\pi^0$ and the high-$q^2$ region ($q^2 > (M_{B1} + M_{B2})^2$) by $B\bar{B} \leftrightarrow e^+e^-$. Figure is taken from [4592].

With PANDA operating at the highest beam energies, the partonic degrees of freedom at distances much smaller than the size of the proton can be studied via measurements of various structure functions. A key in such studies is the factorization theorem stating that the interaction can be factorized into a hard, reaction-specific but perturbative and hence calculable part and a soft, reaction-universal and measurable part. In the space-like region, probed by deep inelastic lepton-hadron scattering, the structure is described by parton distribution functions (PDFs), generalized parton distributions (GPDs), transverse-momentum-dependent parton distribution functions (TMDs), and transition distribution amplitudes (TDAs). These observables extend the information provided by EMFFs and give further insight in the spatial and momentum distributions of the constituent partons and the spin structure. With PANDA, the time-like counterpart becomes experimentally accessible via hard proton-antiproton annihilations. Detailed studies to access $\pi N$ TDAs at PANDA in the reactions $\bar{p}p \to \gamma\pi^0 \to e^+e^-\pi^0$ and $\bar{p}p \to J/\Psi\pi^0 \to e^+e^-\pi^0$ can be found in [4607, 4608]. For these measurements, as well as for the TMD studies, the designed high luminosity of PANDA is needed to accumulate reasonable statistics. The counterparts of the GPDs in the annihilation processes are the generalized distribution amplitudes (GDAs). They can be measured in the hard exclusive processes $\bar{p}p \to \gamma\gamma$ [4609] and $\bar{p}p \to \gamma M$ [4610, 4611], where $M$ could be a pseudo-scalar or vector meson (e.g. $\pi^0$, $\eta$, $\rho^0$, $\phi$). Differential cross section measurements become already feasible to study with the Phase One luminosity of PANDA during the first years of data taking.

## 14.6 BESIII

**Hai-Bo Li, Ryan Edward Mitchell, and Xiaorong Zhou**

### 14.6.1 Introduction to the BESIII Experiment

The BESIII collaboration, which operates the BESIII spectrometer (Fig. 14.6.1) at the Beijing Electron Positron Collider (BEPCII), uses $e^+e^-$ collisions with center-of-mass (CM) energies ranging from 2.0 to 5.0 GeV to study the broad spectrum of physics accessible in the tau-charm energy region. Since the start of operations in 2009, BESIII has collected more than 40 fb$^{-1}$ of data, comprising several world-leading data samples, including:

– 10 billion $J/\psi$ decays, giving unprecedented access to the light hadron spectrum;
– 2.7 billion $\psi(2S)$ decays, allowing precision studies of charmonium and its transitions;

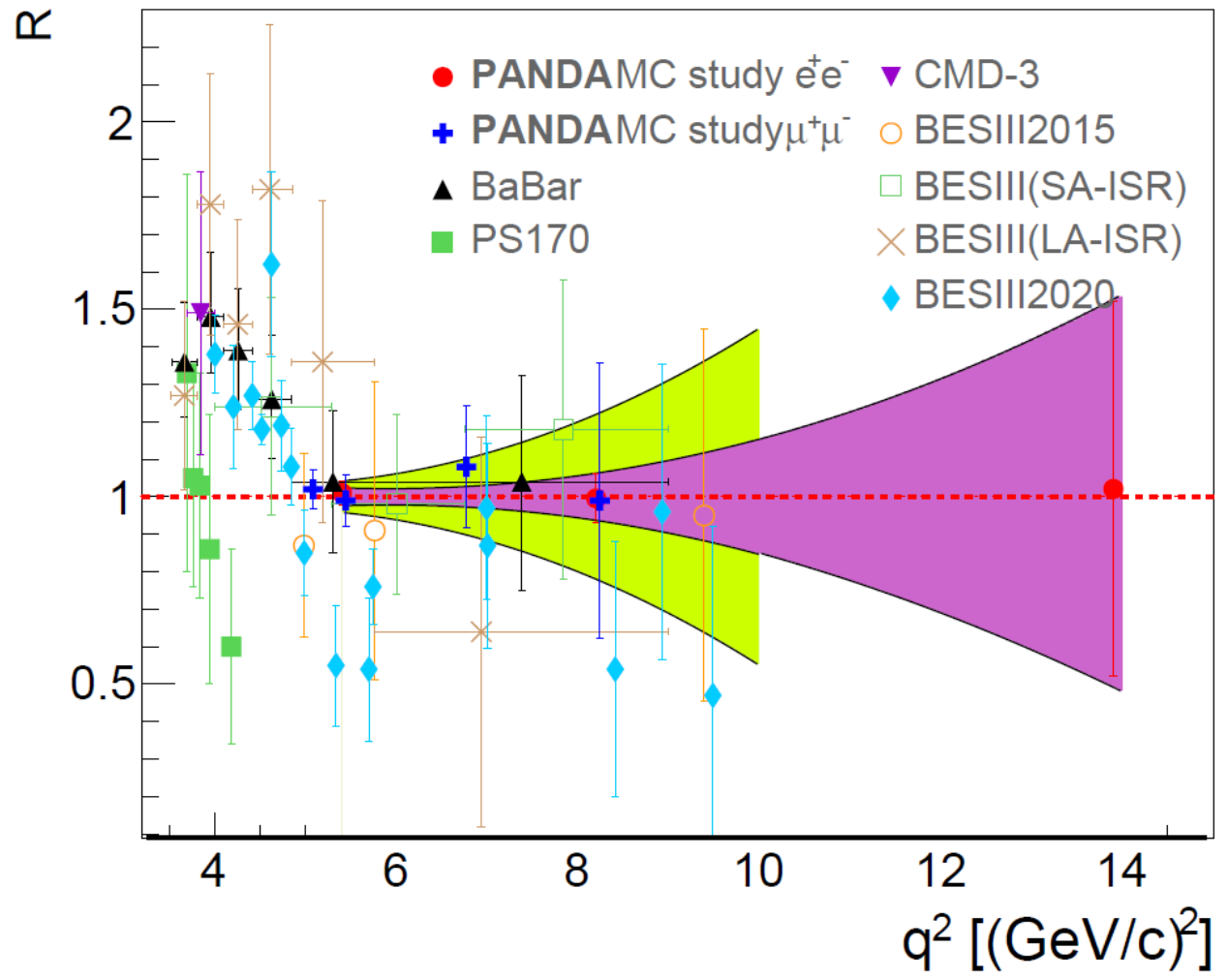

**Fig. 14.5.9** The form factor ratio $R = |G_E|/|G_M|$ of the proton as function of the square of the four momentum, $q^2$. The data are from PS170 [4612], BaBar [4613, 4614], BESIII [4615–4618], CMD-3 [4619]. The expected precisions of PANDA for the $e^+e^-$ final state are indicated as shaded areas for Phase One corresponding to an integrated luminosity of 0.1 fb$^{-1}$ (green band) and high luminosity phase with an integrated luminosity of 2 fb$^{-1}$ (purple band and red filled circles). Also shown are the expected performances for the di-muon channel for the high luminosity phase (dark blue crosses).

- targeted data samples above 4 GeV, providing unique access to exotic $XYZ$ hadrons;
- 8.6 fb$^{-1}$ of data at the $\psi(3770)$ mass, providing a large sample of $D$ decays and quantum-correlated $D^0\bar{D}^0$ pairs, crucial for global flavor physics efforts;
- 3 fb$^{-1}$ at 4.18 GeV, near the peak of the $D_s^{\pm}D_s^{*\mp}$ cross section, for $D_s$ studies;
- more than 3 fb$^{-1}$ above $\Lambda_c\bar{\Lambda}_c$ threshold for precision $\Lambda_c$ studies; and
- fine-scan samples for measurements of $R$, the mass of the $\tau$, and electromagnetic form factors.

The program will continue for at least the next 5-10 years, building on the data sets already collected, and ensuring the BESIII collaboration will remain a key player in future global efforts in hadron spectroscopy, flavor physics, and searches for new physics. The maximum energy of BEPCII will soon be upgraded to 5.6 GeV, and there are plans to more than double the BEPCII luminosity at high CM energies by increasing the maximum achievable beam currents. Below we briefly outline a few highlights from BESIII, how these achievements have contributed to global physics efforts, and how the next era at BESIII will build on this momentum. More details and references can be found in a recent white paper describing the future physics program at BESIII [4620] and in a recent contribution to the 2021 Snowmass process [**snowmass**].

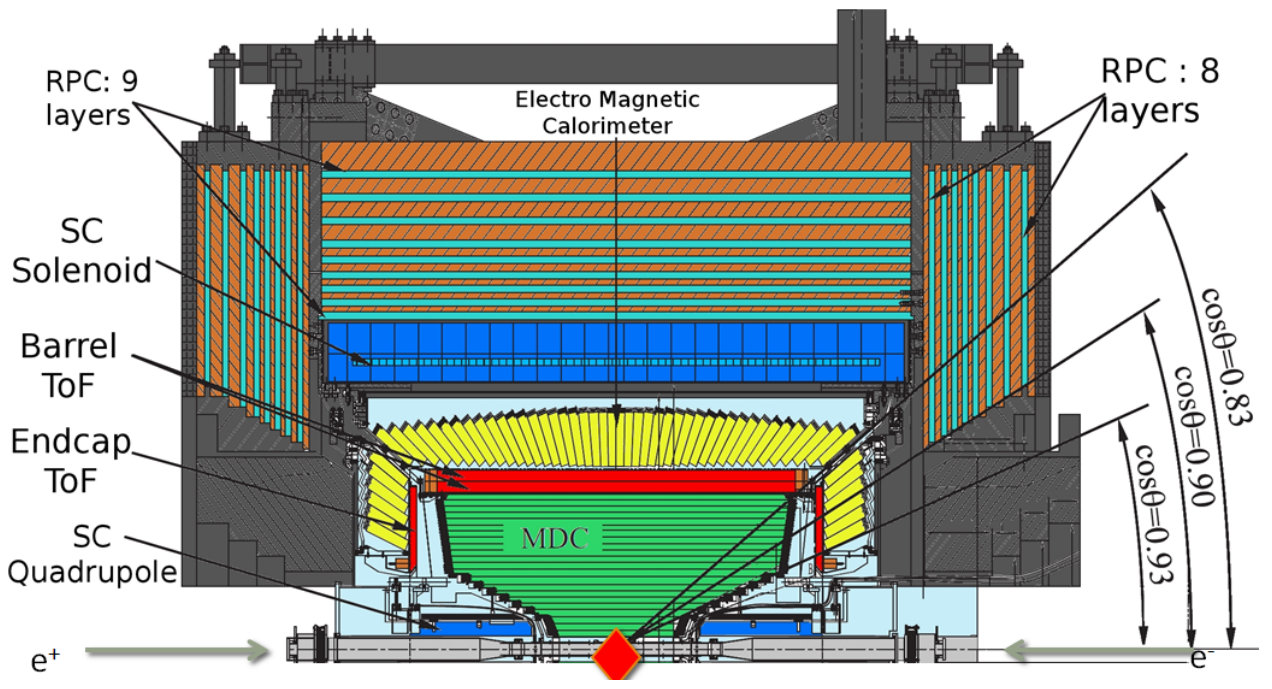

**Fig. 14.6.1** Schematic view of the BESIII dector, covering 93% of the 4$\pi$ solid angle. It consists of a Helium-gas based drift chamber, a Time-of-Flight system, a CsI(Tl) crystal calorimeter and a 9-layer RPC-based muon chamber. Figure taken from the official BESIII website.

### 14.6.2 The BEPCII-U Upgrade

BEPCII delivered its first physics data in 2009 on the $\psi(2S)$ resonance. Since then, BESIII has collected more than 40 fb$^{-1}$ of integrated luminosity at different CM energies from 2.0 to 4.95 GeV. In order to extend the physics potential of BESIII, two upgrade plans for BEPCII were proposed and approved in 2020. The first upgrade will increase the maximum beam energy to 2.8 GeV (corresponding to a CM energy of 5.6 GeV), which will expand the energy reach of the collider into new territory. The second upgrade will increase the peak luminosity by a factor of 3 for beam energies from 2.0 to 2.8 GeV (CM energies from 4.0 to 5.6 GeV).

To perform these upgrades, BEPCII will increase the beam current and suppress bunch lengthening, which will require higher RF voltage. The RF, cryogenic, and feedback systems will be upgraded accordingly. Nearly all of the photon absorbers along the ring and some vacuum chambers will also be replaced in order to protect the machine from the heat of synchrotron radiation. The budget is estimated to be about 200 million CNY and it will take about 3 years to prepare the upgraded components and half a year for installation and commissioning, which will start in June 2024 and finish in December 2024. With these upgrades, BESIII will enhance its capabilities to explore $XYZ$ physics and will have the unique ability to perform precision measurements of the production and decays of charmed mesons and baryons at threshold.

### 14.6.3 Hadronic Production: via direct $e^+e^-$ annihilation

Precision measurements of hadron production help make QCD-related models more reliable and help test SM parameters with an unprecedented sensitivity. BESIII has advanced our knowledge of hadron production using both inclusive and exclusive approaches, mainly via direct production in $e^+e^-$ collisions.

*R value measurement*

The $R$ ratio, defined as the lowest-order cross section for inclusive hadron production, $e^+e^- \to hadrons$, normalized by the lowest-order cross section for the QED process $e^+e^- \to \mu^+\mu^-$, is a central quantity in particle physics. Precision measurements of the $R$ ratio below 5 GeV contribute to the SM prediction of the muon anomalous magnetic moment. The $R$ ratio also contributes in the determination of the QED running coupling constant evaluated at the $Z$ pole. In a first measurement at BESIII [4621], 14 data points with CM energies from 2.2324 to 3.6710 GeV are used for the inclusive $R$ value measurement. An accuracy of better than 2.6% below 3.1 GeV and 3.0% above is achieved in the $R$ ratios, as shown in Fig. 14.6.2. Previous results had uncertainties at the level of 3-6%. The average $R$ value in the CM range from 3.4 to 3.6 GeV obtained by BESIII is larger than the corresponding KEDR result and the theoretical expectation by 1.9 and 2.7 standard deviations, respectively.

The complete data set for the $R$ value measurement at BESIII consists in a total of 130 energy points with an integrated luminosity of about 1300 pb$^{-1}$, corresponding to more than $10^5$ hadronic events at each of the points between 2 and 4.6 GeV. Thus, the final result is expected to be dominated by a systematic uncertainty of less than 3%.

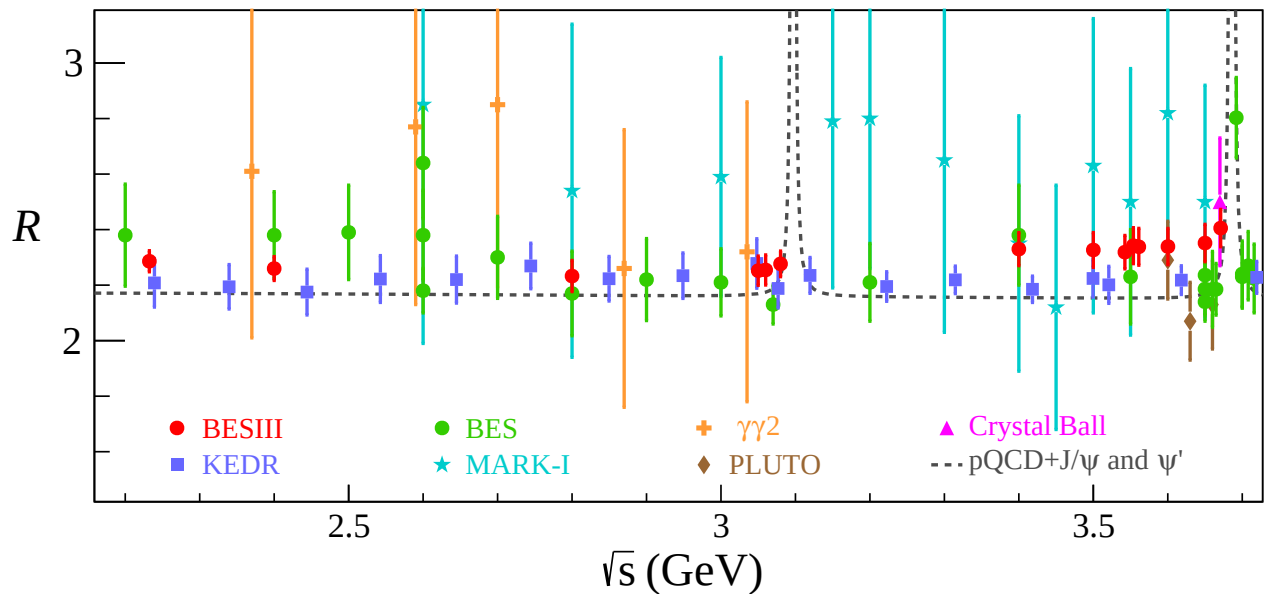

**Fig. 14.6.2** Comparison of $R$ values in the CM energy from 2.2 to 3.7 GeV. Figure taken from Ref. [4621].

*Fragmentation functions*

Fragmentation functions describe the probability of finding a given hadron within the fragmentation of a quark, and carrying a given fraction of the quark momentum. Precise knowledge of fragmentation functions are essential ingredients for studies of the internal structure of the nucleon as carried out by semi-inclusive deep inelastic scattering (SIDIS) experiments (*e.g.* at a future Electron-Ion Collider). At BESIII, using data collected in the continuum energy region, unpolarized fragmentation functions are extracted from inclusive hadron production processes $e^+e^- \to h + X$, where $h$ denotes $\pi^0$, $\eta$, $K_S$, or charged hadrons. Polarized fragmentation functions, *i.e.* the Collins effects, have been obtained by BESIII using pairs of pions produced at $\sqrt{s} = 3.65$ GeV [4622]. In the future, the Collins effect for strange quarks could be studied in $e^+e^- \to \pi K + X$ and $e^+e^- \to KK + X$. It is also interesting to study the Collins effect in neutral hadrons like $e^+e^- \to PP' + X$ with $P/P' = \pi^0/\eta$.

*Exclusive cross section measurements using initial state radiation*

The dispersive integral formalism used to determine the HVP contribution to $a_\mu$ relies heavily on the hadronic $e^+e^-$ cross sections at CM energies $\sqrt{s} \leq 2$ GeV. At BESIII, these energies are only accessible by exploiting the initial state radiation (ISR) method. With an initial data set of 2.83 fb$^{-1}$ at $\sqrt{s} = 3.773$ GeV, this technique already produces results competitive with the B-factories for hadronic masses above approximately 1.3 GeV.

In a first measurement by BESIII, the largest hadronic cross section, for $e^+e^- \to \pi^+\pi^-$, was measured in the mass region from 600 to 900 MeV by reconstructing the ISR photon at large angles only [4218]. With 20 fb$^{-1}$ of data at $\sqrt{s} = 3.773$ GeV expected soon, a new measurement of the $\pi^+\pi^-$ cross section will use the improved statistical accuracy to implement an alternative normalization scheme relative to the muon yield. With this approach, the largest uncertainties will cancel, bringing the expected final uncertainty down to 0.5%, as illustrated in Fig. 14.6.3. Additionally, the multi-meson cross sections for $e^+e^- \to \pi^+\pi^-\pi^0$ as well as $e^+e^- \to \pi^+\pi^-\pi^0\pi^0$ have been measured using the same analysis strategy. Uncertainties of approximately 3% were achieved. These cross sections can be used to study resonances in the final state as well as in the intermediate states. Further improvements are expected with additional data at $\sqrt{s} = 3.773$ GeV.

*Meson transition form factors*

Transition form factors (TFF) of mesons $M$ describe the effects of the strong interaction on the $\gamma^*\gamma^*M$ vertex. At BESIII, TFFs are studied in the region of timelike virtualities through meson Dalitz decays and radiative meson production in $e^+e^-$ annihilations. Spacelike virtualities are studied in two-photon fusion reactions, which in principle give access to TFFs over a wide range of virtualities by measuring the momentum transfer of the scattered electrons. Due to the rapid drop of the cross section with $Q_i^2 = -q_i^2$, BESIII currently uses single-tagged measurements, where the TFF is only studied depending on one of the virtualities.

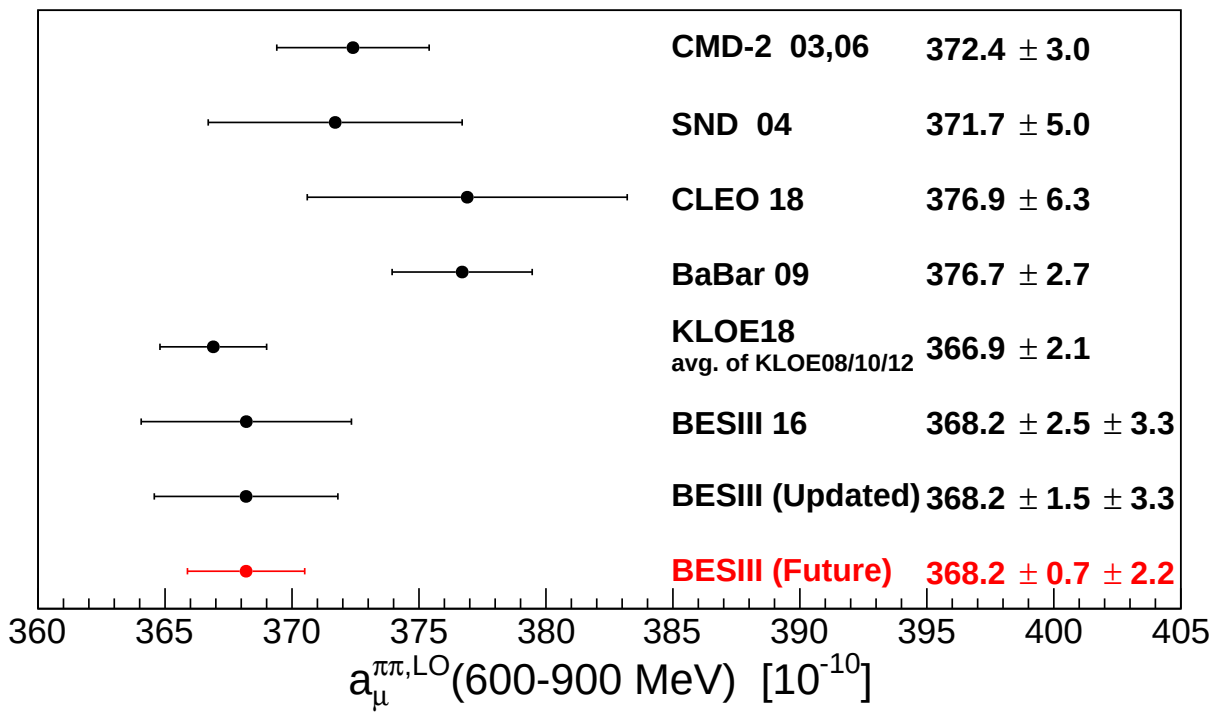

**Fig. 14.6.3** Comparison of the leading-order hadronic vacuum polarization contribution to $(g-2)_\mu$ due to $\pi^+\pi^-$ in the energy range 600-900 MeV from vaious experiments and the prospect result with 20 fb$^{-1}$ of data at $\sqrt{s} = 3.773$ GeV at BESIII. Figure modified according to Ref. [4218].

A first measurement of the $\pi^0$ TFF based on 2.83 fb$^{-1}$ of data at $\sqrt{s} = 3.773$ GeV covers virtualities from 0.3 GeV$^2$ to 3.1 Gev$^2$. The results confirm the recent calculations in disperion theory and on the lattice. Analogous studies are performed for $\eta$ and $\eta'$ mesons, and also for multi-meson systems. The production of charged and neutral two-pion systems in two-photon fusion gives access to pion masses from threshold to 2 GeV and virtualities from 0.2 Gev$^2$ to 3 Gev$^2$ at a full coverage of the pion helicity angle. The results will be complementary to all previous measurements, which have mostly been performed with quasi-real photons. The production of higher meson multiplicities in two-photon fusion allows access to scalar, tensor and axial resonances. The single-tagged strategy allows for the production of axial mesons due to the presence of a highly virtual photon. A first measurement of the $f_1(1285)$ will be performed using the $\pi^+\pi^-\eta$ final state for reconstruction. With the upcoming data set of 20 fb$^{-1}$ at $\sqrt{s} = 3.773$ GeV all two-photon fusion analysis will benefit from higher statistics, which will be sensitive to higher virtualities.

*Time-like baryon form factors*

The simplest observables for nucleon structure are the electromagnetic form factors (EMFFs) that arise from their charge and magnetization distributions, and provide a crucial testing ground for QCD-related models. At BESIII, the $|G_E/G_M|$ of the proton in the time-like region is determined over a large $q^2$ from threshold to 9.5 GeV$^2$ with the best precision reaching 3.7% [4623]. The cross section of $e^+e^- \to n\bar{n}$ [4624] is found to be smaller than that of $e^+e^- \to p\bar{p}$. The effective FFs of the neutron show a periodic behavior, similar to earlier observations of proton FFs reported by BaBar. The energy region of BESIII covers the production threshold of all SU(3) octet hyperons and several charmed baryons. At BESIII, the Born cross sections of electron-positron annihilation to various baryon pairs are measured from threshold [4625], including $\Lambda\bar{\Lambda}$, $\Sigma\bar{\Sigma}$, $\Xi\bar{\Xi}$ and $\Lambda_c\bar{\Lambda}_c^+$. Obvious threshold effects are observed. The $|G_E/G_M|$ of the $\Lambda$, $\Sigma^+$, and $\Lambda_c$ are obtained from angular analyses while effective FFs are extracted for other baryons. More precise data or finer scans are necessary for deeper insight into these results. The hyperon EMFFs and the cross section line shapes can also be studied with improved precision via ISR approaches with a 20 fb$^{-1}$ data set collected at $\sqrt{s} = 3.773$ GeV.

The EMFFs in the time-like region are complex and the relative phase between $G_E$ and $G_M$ will lead to the transverse polarization of the final baryons. At BESIII, the relative phase of the $\Lambda$ is determined at $\sqrt{s} = 2.396$ GeV with a joint angular distribution analysis, to be $\Delta\Phi = 37° \pm 12° \pm 6°$ [4626]. Combining with the obtained $|G_E/G_M|$ at the same CM energy, the complete EMFFs are determined for the first time. Similarly, the relative phase of the $\Lambda_c$ is determined at $\sqrt{s} = 4.60$ GeV [4627]. The currently available data set from $\sqrt{s} = 4.6$ to 4.95 GeV will help complete determinations of $\Lambda_c$ EMFFs in a wide $q^2$ range. As the energy dependence of the relative phase is essential for distinguishing various theoretical predictions, a complete determination of EMFFs for SU(3) octet hyperons are necessary in the future.

*Precision measurement of the $\tau$ mass*

The $\tau$ lepton is one of three charged elementary leptons in nature, and its mass is an important parameter of the Standard Model. The $\tau$ mass can and should be provided by experiment precisely. Precision $\tau$ mass measurements probe lepton universality, which is a basic ingredient in the Standard Model.

To aid in the $\tau$ mass measurement, a high-accuracy beam energy measurement system (BEMS), located at the north crossing point of BEPCII, was designed, constructed, and finally commissioned at the end of 2010. By comparing a $\psi(2S)$ scan result with the PDG value of the $\psi(2S)$ mass, the relative accuracy of the BEMS was determined to be at the level of $2 \times 10^{-5}$ [4628]. The BESIII collaboration performed a fine mass scan experiment in the spring of 2018. The $\tau$ mass scan data were collected at five scan points near the $\tau$ pair production threshold with total luminosity of 137 pb$^{-1}$. The analysis is in progress. The uncertainty, including statistical and systematic error, will be less than 0.1 MeV.

### 14.6.4 Hadron Spectroscopy: from light to heavy

*Light Hadron Physics*

QCD allows for a richer meson spectrum than the conventional quark model predicts, including tetraquark states, mesonic molecules, hybrid mesons and glueballs.

Lattice QCD predicts the lightest glueballs to be scalar, tensor and pseudo-scalar, allowing mixing with the conventional mesons of the same quantum numbers. Generally, glueballs are expected to be produced in gluon-rich processes such as radiative $J/\psi$ decays, so that the high-statistics $J/\psi$ sample puts BESIII in a unique position to study glueball candidates. Partial wave analyses (PWA) of the radiative decays $J/\psi \to \gamma\pi^0\pi^0$, $\gamma K^0_S K^0_S$ and $\gamma\eta\eta$ reveal a strong production of the $f_0(1710)$ and $f_0(2100)$ [4629]. One might speculate that these resonances have a large gluonic component. Similarly, the tensor meson $f_2(2340)$ is strongly produced in the radiative decays $J/\psi \to \gamma\eta\eta$ and $\gamma\phi\phi$ [4629], rendering it a good candidate for a tensor glueball. Two recent coupled channel analyses [4630, 4631] of BESIII data on radiative $J/\psi$ decays came to different conclusions concerning the number of contributing resonances and the identification of a glueball candidate, so that additional studies using the full 10 billion $J/\psi$ data sample will be of high importance in the future.

Based on 10 billion $J/\psi$ events, the decay $J/\psi \to \gamma f_0(1500) \to \gamma\eta\eta'$ has been observed with a significance over $30\sigma$ while $J/\psi \to \gamma f_0(1710) \to \gamma\eta\eta'$ is found to be insignificant [2409, 2410]. The suppressed decay rate of the $f_0(1710)$ into $\eta\eta'$ lends further support to the hypothesis that $f_0(1710)$ has a large overlap with the ground state scalar glueball [4632].

In the search for the pseudo-scalar glueball, the decay $J/\psi \to \gamma\eta'\pi^+\pi^-$ has proven to be particularly interesting [4629]. Here, the $X(1835)$ can be observed with a lineshape that appears to be distorted at the proton anti-proton threshold, indicating a potential $p\bar{p}$ bound-state or resonance. In addition, the higher mass structures $X(2120)$, $X(2370)$ and $X(2600)$ are observed, as shown in Fig. 14.6.4, although their spin-parity remains to be determined, a task that will be possible using the new, high precision $J/\psi$ data.

Motivated by multiple studies of the hybrid meson candidate $\pi_1(1600)$, a recent search for the isoscalar partner states $\eta_1$ and $\eta_1'$ in the radiative decays $J/\psi \to \gamma\eta\eta'$ revealed a significant contribution from a new structure $\eta_1(1855)$ with exotic quantum numbers $J^{PC} = 1^{-+}$ [2409, 2410]. While it is too early to say whether the $\eta_1(1855)$ is indeed an isoscalar hybrid meson, future studies of alternative decay modes will help reveal its nature.

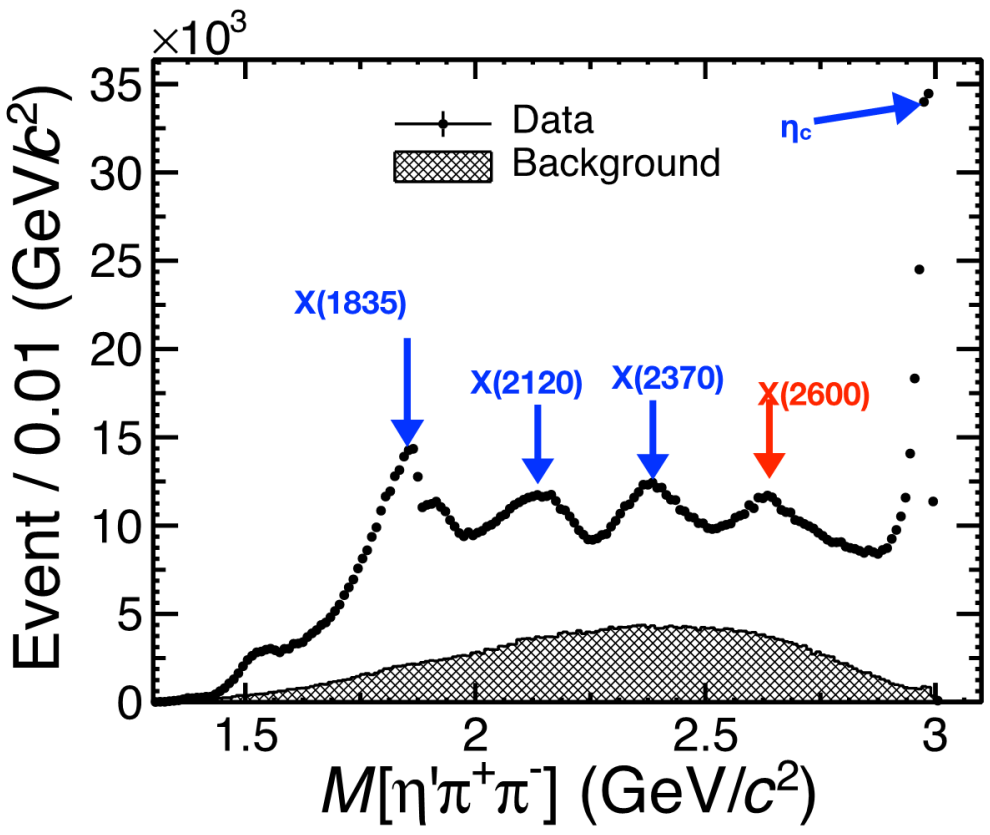

**Fig. 14.6.4** The invariant mass spectrum of the final state $\pi^+\pi^-\eta'$ for $J/\psi \to \gamma\pi^+\pi^-\eta'$ candidates. A series of new particles are observed including $X(1835)$, $X(2100)$, $X(2370)$ and $X(2600)$. Figure taken from Ref. [4633].

The light scalar mesons $f_0(980)$ and $a_0(980)$ are frequently discussed as potential multiquark candidates, either as $K\bar{K}$ molecules or as compact tetraquark states. One possible way to probe their structure is the study of $f_0(980)$ - $a_0(980)$ mixing first observed by BESIII in the isospin-violating processes $J/\psi \to \phi a^0_0(980)$ and $\chi_{c1} \to \pi^0 f_0(980)$ [4629]. These results provide constraints in the development of theoretical models concerning the $f_0(980)$ and $a_0(980)$.

With 10 billion $J/\psi$ decays and the newly acquired 2.7 billion $\psi(2S)$, precision studies of conventional and exotic mesons, including multiquark states, glueballs and hybrid mesons, in radiative and hadronic $J/\psi$, $\psi(2S)$ and $\chi_{cJ}$ decays will be key tasks in the coming years.

*Light baryon spectroscopy*

The high production rate of baryons in charmonium decays, combined with the large data samples of $J/\psi$ and $\psi(2S)$ decays produced from $e^+e^-$ annihilations, provides excellent opportunities for studying excited baryons. Therefore, the BES experiment launched a program to study the excited baryon spectrum. At present, the search for hyperon resonances remains an important challenge. Some of the lowest excitation resonances have not yet been experimentally resolved, which are necessary to establish the spectral pattern of hyperon resonances. The large data samples of $J/\psi$ and $\psi(2S)$ decays accumulated by the BESIII experiment enable us to complete the hyperon (e.g., $\Lambda^*$, $\Sigma^*$ and $\Xi^*$) spectrum and examine various pictures for their internal structures. Such pictures include a simple $3q$ quark structure or a more complicated structure with pentaquark components dominating. In particular, $\psi(2S)$ decays, because of the larger mass of the $\psi(2S)$, have great potential to uncover new higher excitations of hyperons.

At BESIII, $10^{10}$ $J/\psi$ and $2.7 \times 10^9$ $\psi(2S)$ decays are now available, which offer great additional opportunities for investigating baryon spectroscopy. Together with other high-precision experiments, such as GlueX and JPARC, these very abundant and clean event samples will bring the study of baryon spectroscopy into a new era, and will make significant contributions to our understanding of hadron physics in the non-perturbative regime.

*Charmonium physics*

Below the open-charm threshold, the spin-triplet charmonium states are produced copiously in $e^+e^-$ annihilation and in $B$ decays so they are understood much better than the spin-singlet charmonium states, including the lowest lying S-wave state $\eta_c$, its radial excited partner $\eta_c(2S)$, and the P-wave spin-singlet state $h_c$. The 2.7 billion $\psi(2S)$ decays at BESIII make it possible to study the properties of these states with improved precision. In addition, the unexpectedly large production cross section for $e^+e^- \to \pi^+\pi^-h_c$ in the BESIII high-energy region provides a new mechanism for studying the $h_c$ and $\eta_c$ (from $h_c \to \gamma\eta_c$).

The coupling of vector charmonium states to the open-charm meson pairs will provide crucial information in identifying the states in this region. The hadronic and radiative transitions between the (excited) charmonium states can be investigated to study the transition rates and decay dynamics. The cross section of $e^+e^- \to \eta J/\psi$ [4634] shows an enhancement around the $\psi(4040)$ mass, while the cross sections of $e^+e^- \to \pi^+\pi^-\psi(3770)$ [4635] and $e^+e^- \to \pi^+\pi^-\psi_2(3823)$ [4636] show an enhancement around the $\psi(4415)$ mass. The process $e^+e^- \to \gamma\chi_{cJ}$ is studied to search for radiative transitions between the excited vector charmonium states and the $\chi_{cJ}$ [4637]. Whether they are produced via hadronic transitions from the excited vector charmonium states or via vector charmonium-like states is not yet clear and can be addressed using improved luminosity and more decay channels.

Using the $e^+e^- \to \pi^+\pi^-\psi_2(3823)$ process, the most precise mass of the $\psi_2(3823)$ has been determined [4636] and new decay modes of the $\psi_2(3823)$ have been searched for [4638]. These recent measurements at BESIII are examples that the transitions between charmonium states can also serve as production sources of non-vector charmonium states, and can be used to study the properties (mass, width and decay modes) of non-vector charmonium states. They will also be important study topics in the future at BESIII.

With a dedicated data sample taken in the $\chi_{c1}$ mass region, the direct production of the $C$-even resonance, $\chi_{c1}$, in $e^+e^-$ annihilation is observed for the first time with a statistical significance larger than $5\sigma$ [4639]. A typical interference pattern around the $\chi_{c1}$ mass is observed as shown in Fig. 14.6.5. The electronic width of the $\chi_{c1}$ has been determined for the first time from a common fit to the four scan samples to be $\Gamma_{ee} = (0.12^{+0.13}_{-0.08})$ eV, in contrast of a few keV for vector states, which is 4 orders of magnitude smaller. This observation proves that the direct production of $C$-even resonances through two virtual photons is accessible and measurable at the current generation of electron-positron colliders.

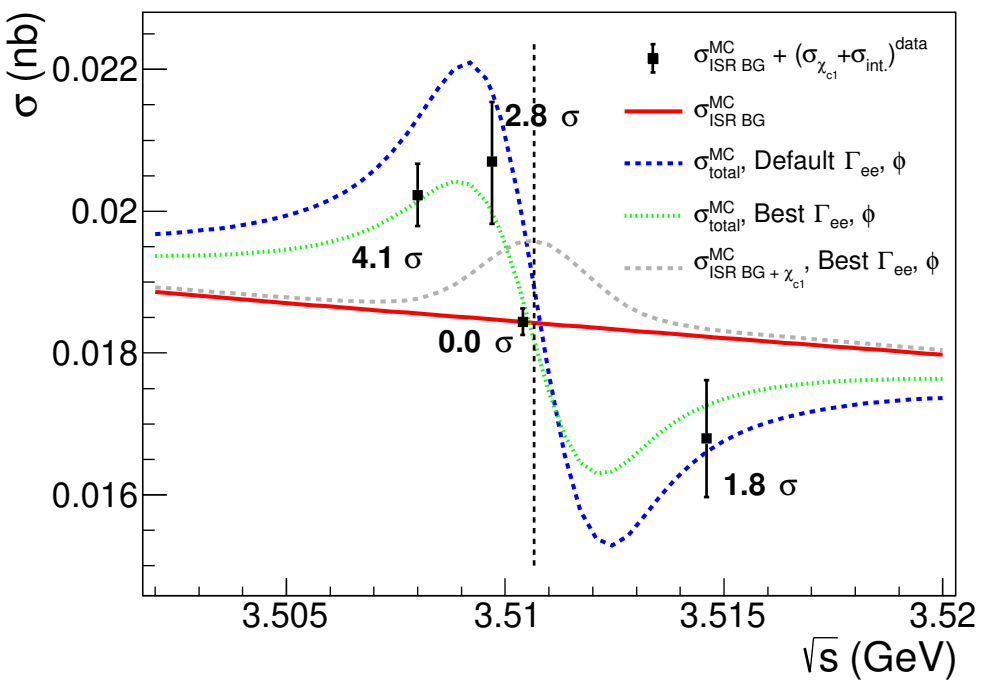

**Fig. 14.6.5** The energy-dependent cross sections of $e^+e^- \to \gamma J/\psi \to \gamma\mu^+\mu^-$ including (blue and green curves) and not including (red curve) the signal process $e^+e^- \to \chi_{c1}(1P)$. The gray curve denotes the signal strength in the hypothetical case of no interference. The black dots with error bars are measured results from data. Figure taken from Ref. [4639].

*XYZ physics*

The discovery of the XYZ states has revolutionized traditional studies of the charmonium spectrum [4640]. These exotic states cannot be embedded in the conventional charm-anticharm potential model framework, but instead point towards novel quark configurations, such as tetraquarks, hybrids, or hadronic molecules. Studying them opens a new window into nonperturbative QCD, which underlies the formation of hadrons via the strong interaction. The existence of the XYZ states poses several problems, which are addressed as the "Y problem", "Z problem", and "X problem" below.

*The Y problem*

BESIII has systematically measured the cross sections of various exclusive $e^+e^-$ annihilations with hidden charm, open charm, and light hadronic final states [4640], and has shown that the lineshapes are complicated as a function of CM energy. The masses and widths of various structures appearing in these cross sections are shown in Fig. 14.6.6. However, the extracted parameters of these $Y$ states are not consistent with each other in different channels. Furthermore, they deviate from the resonances observed in inclusive channels, such as the $\psi(4040)$, $\psi(4160)$, and $\psi(4415)$, that are believed to be conventional charmonia. This leads to the $Y$ problem. What are the exact lineshapes of these cross sections? Are these observed structures new resonances or just results of some subtle kinematic effects? To address these issues, a detailed scan between 4.0 and 4.6 GeV is proposed [4620], with 500 pb$^{-1}$ per point, for points spaced at 10 MeV intervals. This target has been partially achieved with about 22 fb$^{-1}$ integrated luminosity, and will be updated with larger maximum energy (5.6 GeV) after the upgrade of the BEPCII.

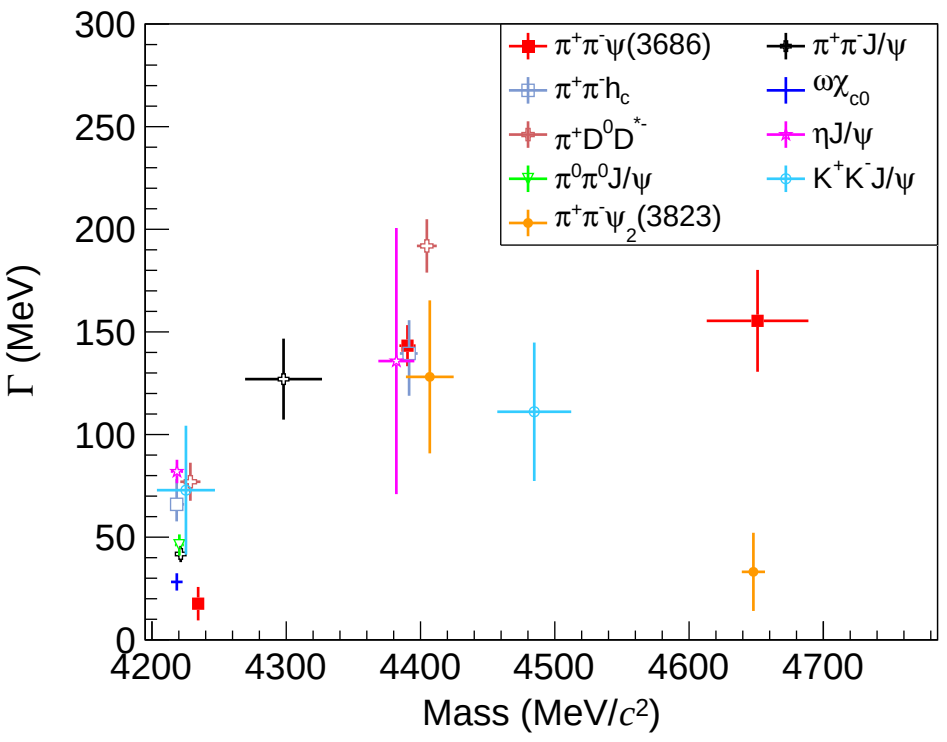

**Fig. 14.6.6** Masses versus widths of the Y states obtained from different processes at BESIII. Figure modified according to Ref. [4641].

*The Z problem*

The $Z_c(3900)$ [4640] was discovered at BESIII in the process $e^+e^- \to \pi^\mp Z_c^\pm$ with $Z_c^\pm \to \pi^\pm J/\psi$, and the $Z_c(4020)$ was discovered in the process $e^+e^- \to \pi^\mp Z_c^\pm$ with $Z_c^\pm \to \pi^\pm h_c$. The $Z_c(3900)$ has also been observed in the open-charm channel $(D\bar{D}^* + c.c.)^\pm$, similarly the $Z_c(4020)$ was seen via the open-charm channel $(D^*\bar{D}^*)^\pm$. Furthermore, neutral partners of these charged $Z_c$ states have been observed at BESIII via processes $e^+e^- \to \pi^0\pi^0 J/\psi$ and $e^+e^- \to \pi^0\pi^0 h_c$. BESIII has also determined the quantum numbers of the $Z_c(3900)$ to be $J^P = 1^+$. Recently, BESIII has observed a new near-threshold structure in the $K^+$ recoil-mass spectra in $e^+e^- \to K^+(D_s^- D^{*0} + D_s^{*-} D^0)$ [2514]. This structure, named $Z_{cs}(3985)$, is a good candidate for a charged hidden-charm tetraquark with strangeness. Besides, the evidence for its neutral partner, $Z_{cs}(3985)^0$ is observed via $e^+e^- \to K_S(D_s^+ D^{*-} + D_s^{*+} D^-)$ [4642].

However, at the energy region higher than 4.3 GeV the data have revealed more complex structure in the Daliz plots of $e^+e^- \to \pi^+\pi^- J/\psi$. A similar situation is found in the $e^+e^- \to \pi^+\pi^-\psi(2S)$ [4643]. This is the Z problem. Are the properties of these $Z_c$ states constant (corresponding to real resonant states) or energy dependent (corresponding to kinematic effects such as cusps or singularities)? What are the exact lineshapes of them? Can we find more decay patterns for them, especially for the newly discovered $Z_{cs}$ states? Are there spin multiplets of these $Z_c$ states? To answer these questions, BESIII may take advantage of the fine scan data mentioned before, but at a few points, a set of samples with very high statistics will be very helpful. BESIII currently has 1 fb$^{-1}$ of data for $e^+e^-$ cms energy at 4.23 and 4.42 GeV. Additional data including three or four points with an order of 5 fb$^{-1}$ or more per point is proposed to guarantee adequate statistics for amplitude analyses [4620]. After the upgrade of BEPCII with triple the luminosity, this goal will be achieved more easily.

*The X problem*

For the $X(3872)$, BESIII has discovered the process $e^+e^- \to \gamma X(3872)$, studied the open-charm decay and radiative transitions of the $X(3872)$, and has observed the hadronic transitions $X(3872) \to \pi^0\chi_{c1}(1P)$ and $X(3872) \to \omega J/\psi$ [4640]. The $X(3872)$, with its quantum numbers $J^{PC} = 1^{++}$, has a mass very close to the predicted $\chi_{c1}(2P)$ state with a very narrow width. Then the X problem is finding a way to separate the $X(3872)$ from the $\chi_{c1}(2P)$. Is the $X(3872)$ really exotic or conventional, or even a mixture state? Can we measure the line shape of the $X(3872)$? Are there other $X$ states (for example close to the $D^*\bar{D}^*$ threshold) that have not been observed yet? The related studies will benefit from the large scan and other data samples mentioned before. Furthermore, at $E_{cm} > 4.7$ GeV with highly excited $\psi$ or $Y$ states produced, the hadronic transitions, that take larger production rates than the radiative transitions, are accessible. After the upgrade of BEPCII to its maximum CM energy, BESIII will have the ability to search for the $J^{++}$ states via hadronic transitions such as the processes $e^+e^- \to \omega X$ and $e^+e^- \to \phi X$.

*Relationships*

There are two kinds of relationships that deserve discussion. One is the relationship between XYZ states and conventional charmonia. For example, the $\chi_{c1}(2P)$ has a similar mass and the same $J^{PC}$ as the $X(3872)$. So a detailed understanding of the spectrum of the conventional $2P$ charmonium states, that include the spin triplet $\chi_{cJ}(2P)$ and singlet $h_c(2P)$, is crucial for understanding the nature of the $X(3872)$. This is also true for the other conventional charmonia and XYZ states under similar conditions. The studies of the conventional charmonia and exotic XYZ are complementary to each other. Understanding the relations between the two kinds of states, even the possible mixing between them, will be helpful for understanding the properties of the XYZ states. The other relationship is among the XYZ states. The analyses of processes $e^+e^- \to \gamma X(3872)$ and $e^+e^- \to \pi^0\pi^0 J/\psi$ have already shown that there is evidence for the radiative transition $Y(4230) \to \gamma X(3872)$ and the hadronic transition [4640]

$$Y(4230) \to \pi^0 Z_c^0(3900).$$

Searching for new transition modes and confirming these relations may be a unique chance for BESIII to reveal the nature of the internal structure of the XYZ states [4644].

*Pentaquark states*
The LHCb experiment reported the observation of three pentaquark states with a $c\bar{c}$ component in the $J/\psi p$ system via $\Lambda_b^0 \to J/\psi K^- p$. To confirm these states, further experimental research should be pursued with the current available and the forthcoming experimental data [4645]. BESIII may search for such and similar states with data to be collected at CM energies above 5 GeV in the processes $e^+e^- \to J/\psi p + X$, $\chi_{cJ} p + X$, $J/\psi \Lambda + X$, $\bar{D}^{(*)} p + X$, $D^{(*)} p + X$, and so on. It is clear that a systematic search for baryon-meson resonances should be pursed in various processes, where the baryon could be $p$, $\Lambda$, $\Sigma$, $\Sigma_c$, ..., and the meson could be $\eta_c$, $J/\psi$, $\chi_{cJ}$, $D^{(*)}$, etc. It is worth pointing out that the tetraquark and pentaquark candidates mentioned above have a pair of charm-anticharm quarks which may annihilate. Observations of states like $T_{cc}^+$ ($cc\bar{u}\bar{d}$) or $\Theta_c^0$ ($uudd\bar{c}$) or $P_{cc}^0$ ($ccdd\bar{u}$) or similar serve as more direct evidence for multiquark states. The BES experiment pioneered a search for the pentaquark candidate $\Theta(1540)$ in $\psi(2S)$ and $J/\psi$ decays to $K_S p K^- \bar{n}$ and $K_S p K^+ n$ [4646]. More attempts will be performed with 10 billion $J/\psi$ and 3 billion $\psi(2S)$ at BESIII.

### 14.6.5 Hadron Decay: from light to heavy

*Light meson decays*
The $\eta$ and $\eta'$ mesons, the neutral members of the ground state pseudoscalar nonet, are important for understanding low energy quantum QCD [4647]. The 10 billion $J/\psi$ events collected at BESIII offer an unique opportunity to investigate all these aspects, as well as the search for rare $\eta$ and $\eta'$ decays needed to test fundamental QCD symmetries and probe physics beyond the SM. The decays $J/\psi \to \gamma\eta(\eta')$ and $J/\psi \to \phi\eta(\eta')$ provide clean and efficient sources of $\eta/\eta'$ mesons for the decay studies.

The observation of new $\eta'$ decay modes [4648], including $\eta' \to \rho^\mp\pi^\pm$, $\eta' \to \gamma e^+e^-$, and $\eta' \to 4\pi$ have been reported for the first time using about $10^9$ $J/\psi$ decays. Using the same data set, the branching fractions of the five dominant decay channels of the $\eta'$ were measured for the first time using events in which the radiative photon converts to $e^+e^-$.

The double Dalitz decay $\eta' \to e^+e^+e^-e^-$ is of great interest for understanding the pseudoscalar transition form factor and the interaction between pseudoscalar and virtual photons. This process has not been observed to date, while the predicted branching fraction is of the order of $2\times10^{-6}$ [4649, 4650]. Another interesting study is the hadronic decay $\eta' \to \pi^0\pi^0\eta$ which is sensitive to the elastic $\pi\pi$ S-wave scattering lengths, and causes a prominent cusp effect in the $\pi^0\pi^0$ invariant mass spectrum at the $\pi^+\pi^-$ mass threshold [4651]. The full $J/\psi$ data set collected by BESIII offers unique opportunities to investigate the cusp effect in this decay for which no evidence has yet been found.

The absolute branching fraction of the decay $J/\psi \to \gamma\eta$ has been measured with high precision using radiative photon conversions [4648], and the four dominant $\eta$ decays have been measured for the first time. The $\eta/\eta' \to \gamma\pi^+\pi^-$ decay results are related to details of chiral dynamics; $\eta/\eta' \to 3\pi$ decays provide information on the up and down quark masses; and the decay widths of $\eta/\eta' \to \gamma\gamma$ are related to the quark content of the two mesons. Despite the impressive progress in the last years, many $\eta$ and $\eta'$ decays are still to be observed and explored. The full $J/\psi$ data set now available at BESIII makes possible more detailed studies with unprecedented precision. It allows, in addition, an intensive investigation of the properties of the pseudoscalar states $\eta(1405)/\eta(1475)$ [4648]; a thorough study of all states observed in the $1.4 - 1.5\,\mathrm{GeV/c^2}$ mass region; a deep investigation of the $\omega \to \pi^+\pi^-\pi^0$ Dalitz plot; and searches for rare $\omega$ decays.

*Hyperon decays*
Observation of a significant polarization of the $\Lambda$ and $\bar{\Lambda}$ from $J/\psi \to \Lambda\bar{\Lambda}$ led to the revision of the decay asymmetry parameter $\alpha_\Lambda$ [4652, 4653], and has shown BESIII has the potential to study properties of the ground-state (anti)hyperons. Moreover, the cascade decays of $J/\psi \to \Xi^-\Xi^+$ made it possible to measure the strong and weak phases of the $\Xi^-$ decay [4604]. The branching fractions for $J/\psi$ decays into a hyperon–antihyperon pair are relatively large, $\mathcal{O}(10^{-3})$, and thus the collected 10 billion $J/\psi$ decays can be used for precision studies of hyperon decays and tests of $CP$ symmetry. The hyperon–antihyperon pair is produced in a well-defined spin-entangled state based on the two possible partial waves (parity symmetry in this strong decay allows for an $S$- and a $D$-wave). The charge-conjugated decay modes of the hyperon and antihyperon can be measured simultaneously and their properties compared directly. In the first round of analyses both the hyperon and antihyperon decay via the common pionic modes. The full data set will be used to improve the precision of the $CP$-violation searches within these decays. The next stage will be to use a common decay of one of the (anti)hyperons to study rare decays of the produced partner. For example, the kinematical constraints make it possible to perform complete reconstruction of the semileptonic decays and radiative decays of polarized hyperons.

*Leptonic decays of charm mesons*

In the SM, the partial widths of the leptonic decay $D^+_{(s)} \to \ell^+\nu_\ell$ can be expressed in terms of the $D^+_{(s)}$ decay constant $f_{D^+_{(s)}}$ and the CKM matrix element $|V_{cd(s)}|$. Using the measured branching fractions of the leptonic $D^+_{(s)}$ decays, the product $f_{D^+_{(s)}}|V_{cs(d)}|$ can be determined. By taking the $f_{D^+_{(s)}}$ calculated by LQCD with a precision of 0.2% [4654, 4655], one can precisely determine the CKM matrix elements $|V_{cs}|$ and $|V_{cd}|$.Conversely, taking the $|V_{cs}|$ and $|V_{cd}|$ from the standard model global fit, one can precisely measure the $D^+_{(s)}$ decay constants, which are crucial to calibrate LQCD for heavy-quark studies. Comparing the obtained branching fractions of $D^+_{(s)} \to \tau^+\nu_\tau$ and $D^+_{(s)} \to \mu^+\nu_\mu$ gives an important comprehensive test of $\tau - \mu$ lepton-flavor universality.

In recent years, BESIII reported the most precise experimental studies of $D^+_{(s)} \to \ell^+\nu_\ell$ by using 2.93, 0.48, and 6.32 fb$^{-1}$ of data taken at $\sqrt{s} = 3.773$, 4.009, and 4.178-4.226 GeV [4656]. However, the statistical uncertainty still dominates studies of $D^+ \to \ell^+\nu_\ell$ decays, whereas the statistical and systematic uncertainties are comparable in measurements of $D^+_s \to \ell^+\nu_\ell$ decays. The full BESIII data samples to be collected in the coming years allow improvements in the precision of these important constants. The current results of $f_{D^+}$ and $|V_{cd}|$ and their expected precision are shown in Fig. 14.6.7. Furthermore, the accuracy of the lepton-flavor universality tests in $D^+ \to \ell^+\nu_\ell$ and $D^+_s \to \ell^+\nu_\ell$ decays are expected to be reduced from 24.0% and 4.0% to about 10.0% and 3.0%, respectively.

*Semileptonic decays of charm mesons*

Over the years, BESIII reported experimental studies of the semi-leptonic $D^{0(+)}_{(s)}$ decays into $P$, $V$, $S$, and $A$ [4656], where $P$ denotes pseudoscalar mesons of $K$, $\pi$, $\eta$, $\eta'$; $V$ denotes vector mesons of $K^*$, $\rho$, $\omega$, and $\phi$; $S$ denotes scalar mesons of $f_0$ and $a_0$; and $A$ denotes axial vector mesons of $K_1$ and $b_1$. These measurements were carried out by using 2.93, 0.48, and 6.32 fb$^{-1}$ of data taken at $\sqrt{s} = 3.773$, 4.009, and 4.178-4.226 GeV, respectively.

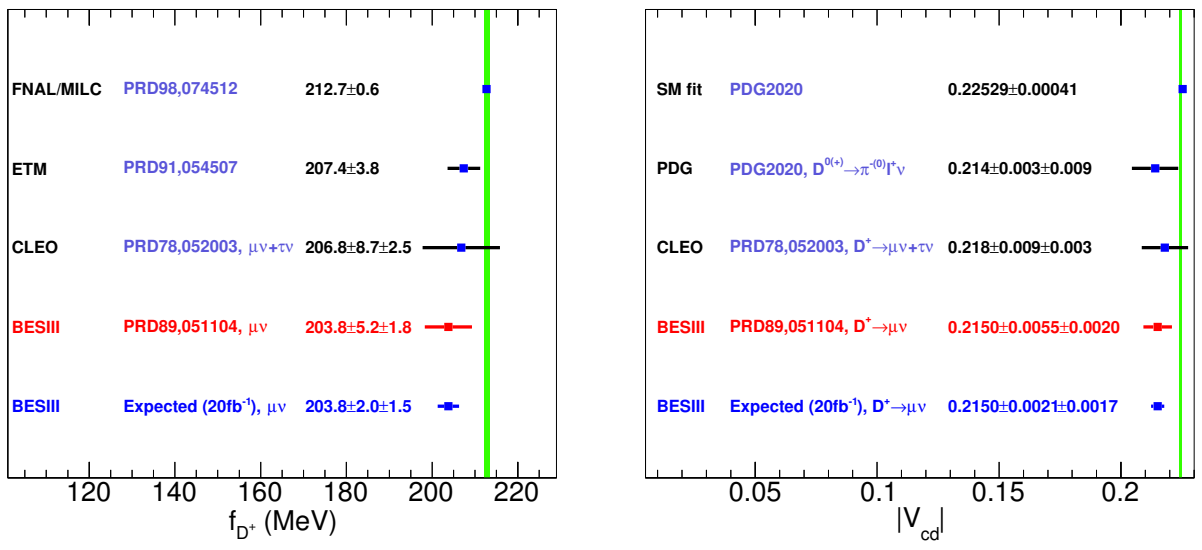

**Fig. 14.6.7** Comparison of extracted $D^+$ decay constant and $|V_{cd}|$ from various experiments and the expected precision with 20 fb$^{-1}$ $\psi(3770)$ data at BESIII.

Except for the $D^{0(+)} \to K$ and $D^{0(+)} \to K^*$ form factors, the precision of all other measurements of the $D^{0(+)}_{(s)} \to P$ and $D^{0(+)}_{(s)} \to V$ form factors are restricted due to the limited size of the data sets. Therefore, with the full BESIII data samples, all the form-factor measurement uncertainties that are limited by the size of the data sample will improve by factors of up to 2.6 for semi-leptonic $D^{0(+)}$ and 1.4 for semi-leptonic $D^+_s$ decays. Complementary studies of the semi-muonic charmed meson decays further improve the form factor knowledge. In addition, we plan to extract the $D \to S$ and $D \to A$ form factors for the first time.

The best precision in the $c \to s$ and $c \to d$ semi-leptonic $D^{0(+)}$ decay form factors will be from the studies of $D^{0(+)} \to \bar{K}\ell^+\nu_\ell$ and $D^{0(+)} \to \pi\ell^+\nu_\ell$. Combining analysis of semi-electronic and semi-muonic $D^0$, as well as $D^+$ decays will give more precise results. The experimental uncertainties are expected to be reduced from 0.6% to 0.4% on $c \to s$ decays and from 1.5% to 0.7% on $c \to d$ decays, as indicated in Fig. 14.6.8.

For semi-leptonic $D^{0(+)}_{(s)}$ decays, the best test of $\mu - e$ lepton-flavor universality is expected to be from $D \to \bar{K}\ell^+\nu_\ell$ decays, where the test precision can be reduced from 1.3% to the level of 0.8% in the near future. At present, it is not conclusive whether the $\mu - e$ lepton-flavor universality always holds in semi-leptonic $D^{0(+)}_{(s)}$ decays, because there are still many unobserved semi-muonic decays such as

$$D^+ \to \eta'\mu^+\nu_\mu,\ D^{0(+)} \to a_0(980)\mu^+\nu_\mu,$$

$$D^{0(+)} \to K_1(1270)\mu^+\nu_\mu,\ D^+ \to f_0(500)\mu^+\nu_\mu,$$

$$D^+_s \to K^0\mu^+\nu_\mu,\ D^+_s \to K^{*0}\mu^+\nu_\mu,$$

$$D_s^+ \to f_0(980)\mu^+\nu_\mu,\ D_s^+ \to \eta'\mu^+\nu_\mu.$$

Larger data samples provide improved opportunities to search for these decays, whose observation will help clarify if there is violation of $\mu - e$ lepton-flavor universality in the charm sector.

Moreover, the studies on the intermediate resonances in hadronic final states, e.g., $K_1(1270)$ and $a_0(980)$, in the semi-leptonic $D_{(s)}^{0(+)}$ decays provide a clean environment to explore meson spectroscopy, as no other particles interfere. This corresponds to a much simpler treatment than those studies in charmonium decays or hadronic $D_{(s)}^{0(+)}$ decays.

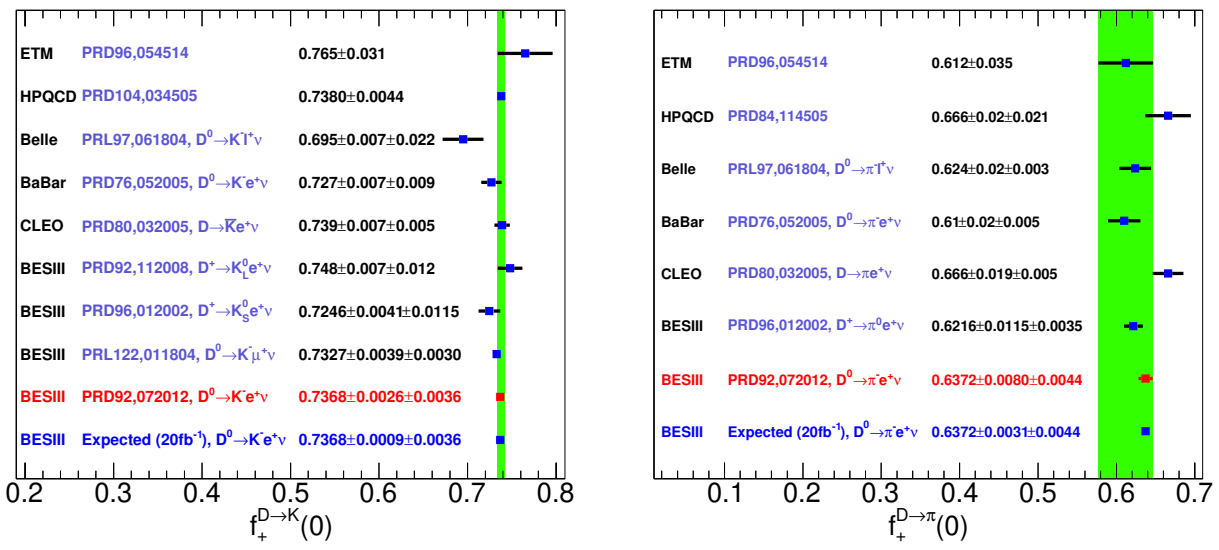

**Fig. 14.6.8** Comparison of $f_+^{\pi}(0)$ and $f_+^{K}(0)$ from various experiments and the expected precision with 20 fb$^{-1}$ $\psi(3770)$ data at BESIII.

*Hadronic decays of charm mesons*

Some experiments, for example LHCb, have the ability to measure a large number of charm and beauty hadron relative branching-fraction ratios due to the high yields given by the large charm and beauty production cross section. The conversion from the branching-fraction ratio to the absolute branching fraction incurs the uncertainty of the branching fraction of the reference mode, such as, $D^0 \to K^-\pi^+$, $D^0 \to K^-\pi^+\pi^+\pi^-$, $D^+ \to K^-\pi^+\pi^+$, $D_s^+ \to K^-K^+\pi^+$, and $\Lambda_c^+ \to pK^-\pi^+$. Improved measurements of these absolute branching fractions at BESIII will be highly beneficial to some key measurements at LHCb. With 20 fb$^{-1}$ data taken around $\sqrt{s}$ = 3.773 and 4.18 GeV at BESIII, these decays are expected to be measured with an uncertainty of about 1%.

At present, the sum of the branching fractions for the known exclusive decays of $D^0$, $D^+$ and $D_s^+$ are more than 80%. However, there is still significant room to explore more hadronic decays to increase the known branching fractions for $D^0$, $D^+$ and $D_s^+$. A 20 fb$^{-1}$ dataset will allow the determination of the absolute branching fractions of those missing decays $K\pi\pi\pi$, $KK\pi\pi$, and $KK\pi\pi\pi$ and exploring the sub-structures in these decays using amplitude analyses is also interesting. In addition, precise measurements of the branching fractions for $D^0$, $D_s^+$ and $D^+$ inclusive decays to three charged pions and other neutral particles, and exclusive decays to final states with neutral kaons and pions (e.g. $D_s^+ \to \eta'\pi^+\pi^0$, $D^+ \to \bar{K}^0\pi^+\pi^+\pi^-\pi^0$ and decay modes contributing to $D^{0(+)} \to \eta X$) are also highly desirable to better understand backgrounds in several measurements, particularly $B \to D^*\tau^+\nu_\tau$.

Studies of such multi-body decays benefit from amplitude analyses to understand the intermediate resonances. Even though it is possible to accumulate large samples of singly tagged $D$ mesons, they have very high backgrounds making them unsuitable to perform amplitude analyses. In contrast to this, the doubly tagged $D\bar{D}$ mesons can provide clean $D$ samples with low backgrounds. However, the sample size limits the precision with the current data. Therefore, such measurements will be significantly improved with the full BESIII data sets.

*Decays of charmed baryons*

The lightest charmed baryon, $\Lambda_c^+$, which was observed in 1979, is the cornerstone of the charmed baryon spectra. The improved knowledge of $\Lambda_c^+$ decays, especially for the normalization mode $\Lambda_c^+ \to pK^-\pi^+$, is key for the studies of the charmed baryon family. Moreover, the $\Lambda_c^+$ decays can also open a window upon a deeper understanding of strong and weak interactions in the charm sector. In addition, these will provide important inputs for the studies of beauty baryons that decay into final states involving $\Lambda_c^+$.

Compared to the significant progress in the study of charmed mesons, the advancements in the knowledge of the charmed baryons are relatively slow in the past 40 years. Before 2014, almost all the decays of $\Lambda_c^+$ were measured relative to the normalization mode $\Lambda_c^+ \to pK^-\pi^+$, whose branching fraction suffered a large uncertainty of 25%. Moreover, no data sample taken around the $\Lambda_c^+\bar{\Lambda}_c^-$ pair production threshold had been used to study the $\Lambda_c^+$ decays.

BESIII have already collected 4.4 fb$^{-1}$ of data above $\Lambda_c\bar{\Lambda}_c$ threshold, which will provide the most precise values of many absolute branching fractions and polarization parameters [4620]. Future running with the upgraded BEPC-II will allow large samples of $\Sigma_c$ and $\Xi_c$ pairs to be collected, which will lead to many absolute branching fractions of charm baryon decays to be determined for the first time [4620].

*The "post-BEPCII era"*

The super $\tau$-Charm facility (STCF) [4657] is one of the major options for future accelerator-based high energy projects in China. The proposed STCF is a symmetric double ring electron-positron collider that would operate in the CM region $\sqrt{s} = 2 \sim 7$ GeV with a peaking luminosity of $0.5 \times 10^{35}$ cm$^{-2}$s$^{-1}$ or higher. It is expected to deliver more than 1 ab$^{-1}$ of integrated luminosity per year. Huge samples of exotic charmonium-like states ($XYZ$), $J/\psi$, $D$, $D_s$ and $\Lambda_c$ decays could be used to make precision measurements of the properties of $XYZ$ particles, and map out the spectroscopies of QCD hybrids and glueballs. High statistics data samples could also be used to search for new sources of $CP$ violation in the hyperon and $\tau$-lepton sectors with unprecedented sensitivity and search for anomalous decays of various hadrons with sensitivities extending down to the level of SM-model expectations.

Since 2012, when the STCF was proposed, the Chinese STCF working group, together with international teams, have carried out a series of feasibility studies, completed the preliminary Conceptual Design Report (CDR) and made significant progress. Compared to the BEPCII/BESIII experiments, the substantial improvement in the performance of the STCF will lay the foundation for breakthroughs in the relevant frontiers of research. Meanwhile, it will pose major technical challenges in accelerator and detector development. At present, the STCF project for the research and development of key technologies is actively performed with the support of Anhui Province of China. More efforts are being made to promote the implementation and construction of the STCF project.

## 14.7 BELLE II

**Toru Iijima**

The Belle II experiment is a particle-physics experiment operating at the SuperKEKB collider built in the KEK laboratory in Japan (Figure 14.7.1). It is a successor of the Belle experiment at the KEKB collider, which experimentally established the Kobayashi-Maskawa theory of the $CP$ violation, together with the BaBar experiment at the SLAC PEP II collider. Over the next decades, Belle II will record the decay of billions of bottom mesons, charm hadrons, and $\tau$ leptons produced in electron-positron collisions at and near the $\Upsilon(4S)$ energy. The ultimate goal is to accumulate 50 ab$^{-1}$ data of $e^+e^-$ collisions, which is about 50 times larger than the data set of the Belle experiment. These data, collected in the low background and kinematically known conditions, will provide a complementary approach to experiments at hadron machines. It will allow us to critically test the standard model (SM) and search for new particles through processes sensitive to virtual heavy particles at mass scale orders of magnitudes higher than direct searches at the energy frontier experiment.

The Belle II physics program includes variety of subjects in the areas of;

- Precision CKM measurements to critically test SM and find or constrain non-SM physics contribution in a model-independent way.
- Search for non-SM $CP$ violation in rare $B$ processes, such as $b \to q\bar{q}s$.
- Search for non-SM physics in semileptonic, radiative and other rare $B$ decays, including precision tests of the lepton-universality in $b \to c\ell\nu$ and $b \to s\ell^+\ell^-$, where $\ell$ stands for either of $e$, $\mu$ and $\tau$.
- Measurements of many parameters in decays of charm hadrons and the $\tau$ leptons with world-leading precisions, including their masses, lifetimes, $CP$ violation parameters, and branching fractions for charged-lepton-flavor-violating decays.
- Unique searches for dark-sector particles with masses in the MeV-GeV range, where some of them are possible dark matter candidates.
- Broad spectroscopy program for both conventional and multi-quark $c\bar{c}$ and $b\bar{b}$ states using different production processes; through $B$ decays, through initial state radiation processes, two-photon collisions and double charmonia productions.
- Provide essential inputs to sharpen the interpretation of results for the anomalous magnetic moment of the muon $(g-2)_\mu$, which indicates $4.2\sigma$ deviation from the SM.

In these physics studies at Belle II, the importance of QCD is two-fold. First, better understandings of non-perturbative QCD properties associated with particle decays are essential ingredients for sharpening the SM predictions as references for non-SM physics searches. Second, a variety of low-energy QCD phenomena, such as the $c\bar{c}$ and $b\bar{b}$ spectroscopy as mentioned above, are the subjects that could be uniquely studied at the Belle II experiment. Also, the $e^+e^-$ collisions to hadron final states offer unique opportunities to study hadronization processes like the Collins effect. The variety of physics studies that can be carried out at Belle II is discussed in detail in Ref. [4658]. In the subsections following 14.7.2, we describe only a brief summary for subjects that are of primary relevance to QCD, where Belle II will be unique and will be world-leading.

### 14.7.1 SuperKEKB/Belle II experiment

The SuperKEKB accelerator is an asymmetric energy collider of 4.0 GeV $e^+$ and 7.0 GeV $e^-$. The target instantaneous luminosity is $\sim 6 \times 10^{35}$ cm$^{-2}$ s$^{-1}$, enabling accumulation of 50 ab $^{-1}$ over the next decade. It is the world's leading luminosity machine with an innovative "nano-beam scheme", where the two beams collide with a large horizontal crossing angle and the vertical beam size is squeezed down to a level of 50-60 nm at the interaction point (IP).

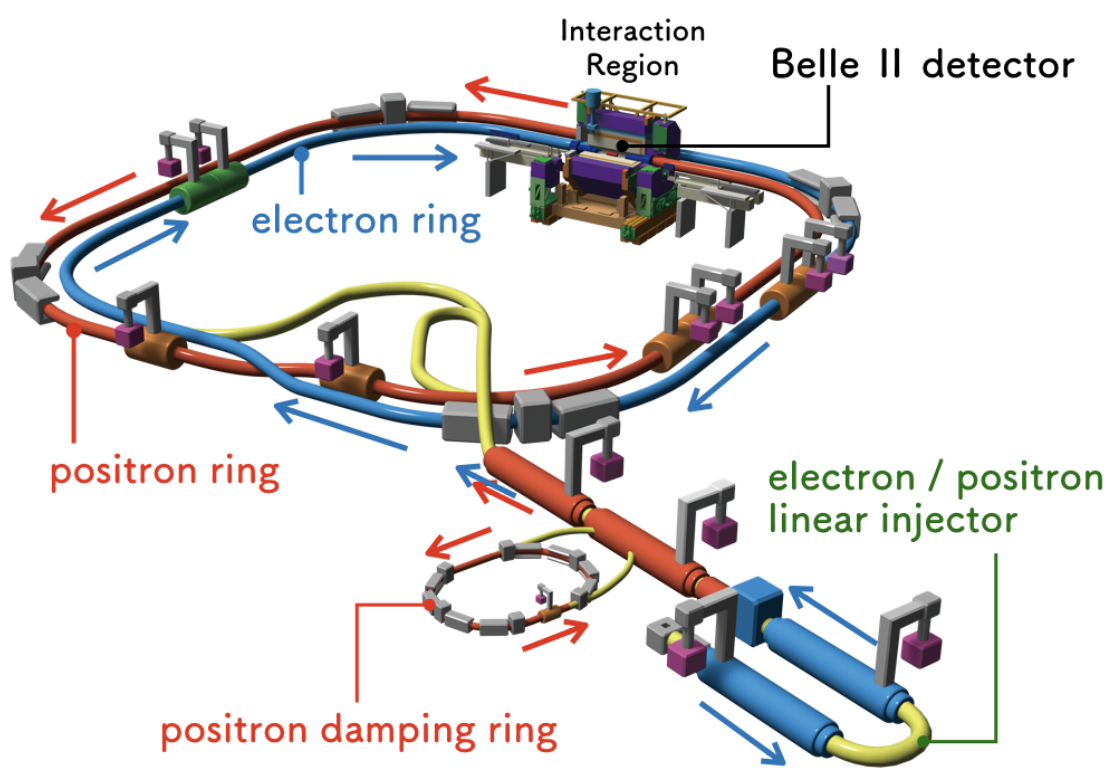

**Fig. 14.7.1** Layout of the SuperKEKB accelerator.

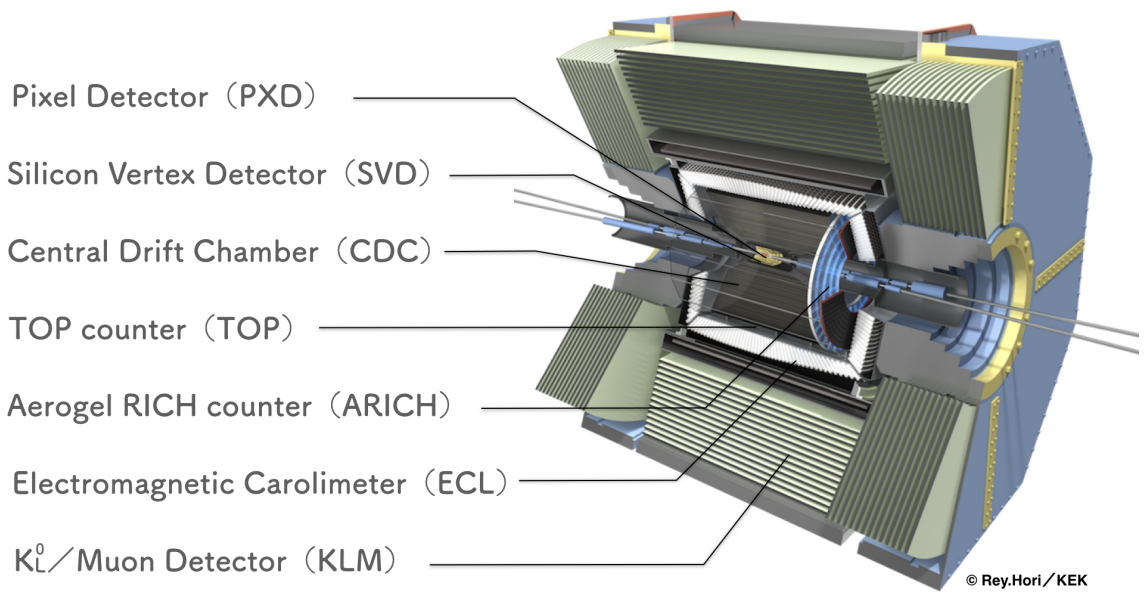

**Fig. 14.7.2** The Belle II detector which consists of seven subsystems.

The Belle II detector, as shown in Figure 14.7.2, is located at the single collision point (IP) of the SuperKEKB. It is nearly a $4\pi$ magnetic spectrometer surrounded by a calorimeter and muon detectors and comprises several subdetectors arranged cylindrically around IP and with a polar structure reflective of the asymmetric distribution of final-state particles resulting from the asymmetric energy collision. From the innermost out, these subdetectors are the vertex detector (VXD), central drift chamber (CDC), electromagnetic calorimeter (ECL), and K-long and muon detector (KLM). In between CDC and ECL, are charged-particle-identification subdetectors: a time-of-propagation Cherenkov counter (TOP) in the barrel, and an aerogel ring-imaging Cherenkov detector (ARICH) in the forward region. Between ECL and KLM, is a solenoid coil that provides a 1.5 T axial magnetic field for measurements of the momenta and electric charge of charged particles. The vertex detector consists of two layers of pixel sensors (PXD) surrounded by four layers of microstrip sensors (SVD) to determine the positions of decaying particles with the typical impact-parameter resolution of $10-15\mu m$, resulting in $20-30\mu m$ typical vertex resolution [120] . The small-cell helium-ethane central drift chamber measures the positions of charged particles at large radii and their energy losses due to ionization. The relative charged-particle transverse momentum resolution is typically 0.4%/pT [GeV/c]. The observed hadron identification efficiencies are typically 90% at 10% contamination. Typical uncertainties in hadron-identification performance are 1%. The CsI(Tl)-crystal electromagnetic calorimeter measures the energies of electrons and photons with energy-dependent resolutions in the 1.6-4% range. Layers of plastic scintillators and resistive-plate chambers interspersed between the magnetic flux-return yoke's iron plates allow us to identify KL0 and muons. Our observed lepton-identification performance shows 0.5% pion contamination at 90% electron efficiency, and 7% kaon contamination at 90% muon efficiency. Typical uncertainties in lepton-identification performance are $1\% - 2\%$.

[120] The second pixel layer is currently incomplete, covering approximately 15% of the azimuthal acceptance. Installation of the pixel detector will be completed in 2023.

The Belle II experiment has unique advantages over hadron-collider experiments, such as the LHCb experiment. Despite having comparatively less data and fewer accessible initial states;

- It produces heavy flavor particles in a less background environment, which enables efficient detection of neutral particles, such as $\gamma$, $\pi^0$, $K^0_S$, $K^0_L$.
- It produces quantum correlated $B^0$-$\bar{B}^0$ pairs, by which we can tag the $B$ meson flavor with high effective efficiency. We can also measure precisely $B$ decay modes with neutrinos in the final state, by fully reconstructing one of the $B$ mesons, referred to as "full reconstruction tagging".
- It provides a large sample of $\tau$ leptons obtained, which allows us to study in detail the property of the $\tau$ lepton, including Lepton-Flavor-Violating (LFV) decays.

As for the full reconstruction tagging, a new "Full Event Interpretation (FEI)" tool has been developed [4659]. The basic idea of FEI is to reconstruct, in a hierarchical manner, individual particle decay channels that occur in the decay chain of the $B$ meson. For each unique decay channel of a particle, a multivariate classifier (MVC) is trained using simulated events. Both hadronic and semileptonic $B$ decays are used. The typical tag-side efficiency, defined as the number of correctly reconstructed tag-side B mesons divided by the total number of $\Upsilon(4S)$ events, is 0.61% (0.34%) for hadronic $B^+$ ($B^0$) decays and 1.45%(1.25%) for semileptonic $B^+$ ($B^0$) decays. The full reconstruction tagging provides unique methods to measure $B$ decays with neutrinos in the final states, such as $B \to \pi\ell\nu$, $B \to D^{(*)}\tau\nu$ and $B \to K\nu\bar{\nu}$.

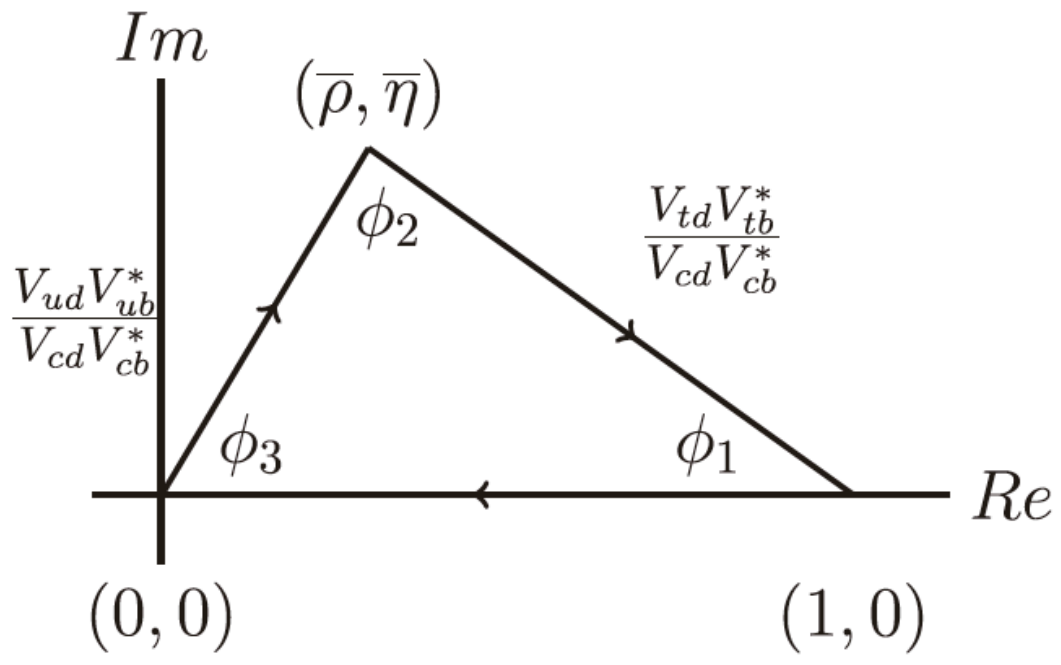

**Fig. 14.7.3** The unitarity triangle.

### 14.7.2 Precision CKM measurements

In the Standard Model (SM), $CP$ violation in the $K/B$ meson decays can occur as the complex phase in the Cabibbo-Kobayashi-Maskawa (CKM) quark mixing matrix [3957, 3958]. The high luminosity data at Belle II enable precision measurements of the three internal angles, $(\phi_1, \phi_2, \phi_3) \equiv (\alpha, \beta, \gamma)$, and the three sides of the unitarity triangle, which represents the unitarity condition of the CKM matrix elements, $V_{ud}^*V_{ub} + V_{cd}^*V_{cb} + V_{td}^*V_{tb} = 0$, in the complex plane with the three terms divided by $V_{cd}V_{cb}^*$, as shown in Figure 14.7.3.

*Measurement of $\phi_1$*

The internal angle $\phi_1 \equiv \arg(-V_{cd}V_{cb}^*/V_{td}V_{tb}^*)$ is determined from measurements of time-dependent $CP$ asymmetries, which occures via interference between $B_d - \bar{B}_d$ oscillation and $b \to c\bar{c}s$ decay amplitudes. Most of the hadronic uncertainties cancel out in the $CP$ asymmetry, therefore, these measurements provide very clean and precise determinations of $\phi_1$. In the experiment, after the $B^0 - \bar{B}^0$ system is coherently produced from an $\Upsilon(4S)$ decay, one of the B mesons, $B_{CP}$, decays to a $CP$ eigenstate $f_{CP}$ at $t = t_{CP}$ whereas the other, $B_{\rm tag}$, may decay to favor specific final state at $t = t_{\rm tag}$. The distribution of the proper-time difference $\Delta t \equiv t_{CP} - t_{\rm tag}$ is expressed by

$$\mathcal{P}_{f_{CP}}(\Delta t, q) = \frac{e^{-|\Delta t|/\tau_{B^0}}}{4\tau_{B^0}}\{1 + q[\mathcal{A}_{f_{CP}}\cos(\Delta m_d \Delta t) + \mathcal{S}_{f_{CP}}\sin(\Delta m_d \Delta t)]\} \quad , \tag{14.7.1}$$

where $\tau_{B^0}$ and $\Delta m_d$ are the average lifetime and mass difference between neutral B physical states, respectively, and $\mathcal{A}_{f_{CP}}$ and $\mathcal{S}_{f_{CP}}$ are the direct and mixing-induced CP-violating asymmetries, respectively. The B meson flavor q takes values $+1(-1)$ when $B_{\rm tag}$ is $B^0(\bar{B}^0)$ and it is statistically determined from the favor tagging algorithm based on final-state information [4660]. The time-difference $\Delta t$ is approximated by the distance between the two B-meson decay vertices divided by the speed of the $\Upsilon(4S)$ projected onto the boost axis.

The previous experiments Belle, BaBar, and LHCb achieved determination of $\phi_1$ at 2.4% precision [4661], using tree dominated $(c\bar{c})K^0$ decays, such as $J/\psi K_S^0$, $\psi(2S)K_S^0, \chi_{c1}K_S^0$ and $J/\psi K_L^0$. The error is still dominated by systematic uncertainties, associated with imperfections in vertex reconstruction and flavor tagging. The precision is expected to further improve to below 1% in the next decade, and it will provide a firm basis to search for non-SM contributions.

*Measurement of $\phi_2$*

Studies of $b \to u$ charmless $B$ decays give access to $\phi_2 \equiv \arg[-V_{tb}^*V_{td}/V_{ub}^*V_{ud}]$, the least known angle of the CKM unitarity triangle, and probe non-SM contributions in processes mediated by loop decay-amplitudes. However, clean extraction of $\phi2$ is not trivial due to hadronic uncertainties, which are hardly tractable in perturbative calculations. Appropriate combinations of measurements from decays related by flavor (isospin) symmetries reduce the impact of such uncertainties [4662]. The most promising determination of $\phi_2$ relies on the combined analysis of the decays $B^+ \to \rho^+\rho^0$, $B^0 \to \rho^+\rho^-$, $B^0 \to \rho^0\rho^0$, and corresponding decay into pions. The current global precision of 4 degrees is dominated by $B \to \rho\rho$ data [4661]. Leveraging efficient reconstruction of low-energy $\pi^0$, improved measurements in $B^+ \to \rho^+\rho^0$ and $B^0 \to \rho^+\rho^-$ decays will be unique to Belle II. The expected experimental accuracy for the $\phi_2$ determination is less than 1° at 50ab$^{-1}$.

*Measurement of $\phi_3$*

The third internal angle $\phi_3 \equiv \arg[-V_{ud}V_{ub}^*/V_{cd}V_{cb}^*]$ is accessible via tree-level decays, such as $B \to DK$, where $D$ represents a generic superposition of $D^0$ and $\overline{D}^0$. Assuming that non-SM amplitudes do not affect appreciably tree-level processes, precise measurements of $\phi_3$ and $|V_{\rm ub}/V_{\rm cb}|$ set strong constraints on the SM description of $CP$ violation, to be compared with measurements from higher-order processes potentially sensitive to non-SM amplitudes, such as mixing-induced $CP$ violation through $\sin 2\phi_1$. Extraction of $\phi_3$ involves measurement of $B^- \to \overline{D}^0K^-$ and $B^- \to D^0K^-$ ampli-tudes, which are expressed as

$$\frac{\mathcal{A}(B^- \to \overline{D}^0 K^-)}{\mathcal{A}(B^- \to D^0 K^-)} = r_B e^{i(\delta_B - \phi_3)}, \tag{14.7.2}$$

where $r_B \approx 0.1$ is the ratio of amplitude magnitudes and $\delta_B$ is the strong-phase difference. Since the hadronic parameters, $r_B$ and $\delta_B$ can be determined from data together with $\phi_3$, these measurements are essentially free of theoretical uncertainties [4663]. The precision of $\phi_3$ is mostly limited by the small branching fractions of the decays involved (around $10^{-7}$). The current world average is $\phi_3 = (66.2^{+3.4}_{-3.6})^\circ$ [4661], whereas the indirect determination is $(63.4 \pm 0.9)^\circ$ [4080]. Various methods with different choices of final states accessible to both $D^0$ and $\overline{D}^0$ have been proposed to extract $\phi_3$. They include $CP$-eigenstates (GLW method) [4664, 4665], Cabibbo-favoured (CF) and doubly-Cabibbo-suppressed (DCS) decays (ADS method) [4666], self-conjugate modes (BPGGSZ method) [4667–4669], and singly Cabibbo-suppressed (SCS) decays (GLS method) [4670].

Currently, precision is dominated by measurements based on $B^- \to D(K^0_S\pi^+\pi^-)K^-$ as well as $B^- \to D(K^0_L\pi^+\pi^-)K^-$ decays [4667–4669]. Belle II will be competitive in this mode and others involving final-state $K^0_S$, $\pi^0$, and $\gamma$ such as $K^0_S\pi^0$, $K^0_S\pi^+\pi^-\pi^0$ or $B^- \to D^*(D(\gamma,\pi^0))h^-$. Precision will further improve following the expected three-fold improvements on the external charm-strong-phase inputs from BESIII [4658]. In addition, $B^- \to D(K^0_S\pi^+\pi^-\pi^0)K^-$ is promising at Belle II due to its sizable branching fraction and rich resonance substructures, as shown by Belle [4671]. Improved charm-strong-phase inputs, availability of a suitable amplitude model of $D \to K^0_S\pi^+\pi^-\pi^0$ and a larger $B$ decay sample will render $B^- \to D(K^0_S\pi^+\pi^-\pi^0)K^-$ a strong contributor for determination of $\phi_3$. The precision of $\phi_3$ is expected to be $\mathcal{O}(1^\circ)$ with the full 50 $^{-1}$ data set.

*Determination of* $|V_{cb}|$ *and* $|V_{ub}|$
The magnitudes of the CKM matrix elements $|V_{cb}|$ and $|V_{ub}|$ can be deduced from tree $b \to c$ and $b \to u$ processes and provide reliable SM references to test non-SM contributions. The most precise determinations of $|V_{cb}|$ and $|V_{ub}|$ come from measurements of semileptonic transitions $b \to cl\nu$ and $b \to ul\nu$, either in inclusive or exclusive final states, combined with theoretical inputs to characterize the QCD effects associated with $B$ decays. There has been significant disagreement in the results obtained from exclusive and inclusive measurements [4661]. The reason for this discrepancy is unknown and has been a long-standing issue. It can be possibly inconsistent experimental or theory inputs, but also interpretations in terms of non-SM physics cannot be excluded [4015]. The large data set at Belle II will offer more precise and richer experimental information to test theoretical investigations and to clarify the issue.

*Exclusive* $|V_{ub}|$*:*
Belle-II will provide a variety of ways for exclusive $|V_{\rm ub}|$ determinations. While $\overline{B}^0 \to \pi^+\ell^-\bar{\nu}_\ell$ is currently the most effective in terms of availability of experimental data and theoretical calculations of the form factor, Belle II will also measure other exclusive $b \to u\ell\nu_\ell$ modes with good precision, in particular those involving neutral final-state particles such as

$$B^- \to (\pi^0, \rho^0, \omega, \eta, \eta')\ell^-\nu_\ell$$

and $\overline{B}^0 \to \rho^+\ell^-\nu_\ell$. The excellent resolution in $q^2 \equiv (p_\ell + p_\nu)^2$ also gives access to the decay form factors equally important for determining $|V_{ub}|$. Typically, experimental uncertainties are smallest for low $q^2$ whereas uncertainties in the form factors from lattice QCD are smallest at high $q^2$. Improvements in the experimental constraints will be driven mainly by data set sizes. Belle II can also measure the variety of exclusive decays with high purities in analyses, where the (non-signal) partner $B$-meson is reconstructed [4672]. Belle II will double the global precision in exclusive $|V_{ub}|$ results below 3%. Expected progress in lattice QCD [4658] will offer further significant improvement.

*Inclusive* $|V_{ub}|$*:*
Belle II will provide a unique opportunity to measure inclusive $B \to X_u\ell\nu$ decays, where $X_u$ is a charmless hadronic system. Taking advantage of the $B\overline{B}$ threshold experiment, after reconstructing a signal lepton and the partner $B$ meson, all remaining tracks and energy clusters can be associated with the $X_u$ candidate. Measurements require accurate modeling of the $b \to u$ signal and the $b \to c$ background as demonstrated in the latest Belle measurement of $B \to X_u\ell\nu$, which indeed reports results closer to exclusive [4042]. With larger sample sizes and continuing developments in reconstruction algorithms (*e.g.*, improved partner $B$ reconstruction), Belle II will accomplish measurements of inclusive $|V_{ub}|$ to $\mathcal{O}(1)$% precision. Belle II can also explore novel ideas of measurements, such as the measurement of differential branching fractions of $B \to X_u\ell\nu$ which enables shape-function model-independent determinations of $|V_{\rm ub}|$ as demonstrated by Ref. [4045, 4046, 4673].

*Determination of* $|V_{cb}|$*:*
Belle II will be able to improve also determinations of $|V_{cb}|$ from exclusive $B \to D^{(*)}\ell\nu$ decays and inclusive

$B \to X_c \ell \nu$ decays. For exclusive analyses, the key experimental challenges will be to understand better the composition and form factors of $B \to D^{**} \ell \nu$ decays and reduce relevant systematic uncertainties as those associated with lepton identification and low-momentum-pion reconstruction for $B \to D^* \ell \nu$ decays. Belle II will tackle this with a detailed program based on dedicated auxiliary studies of $B \to D^{**} \ell \nu$ decays. The precision of inclusive determinations, which is limited by theory, will benefit from measurements of the kinematic moments of $B \to X_c \ell \nu$ decays that will constrain hadronic matrix elements in the operator-product-expansion based theory. Ultimately Belle II will accomplish measurements of $|V_{cb}|$ to $\mathcal{O}(1)\%$ precision.

*Summary of CKM measurements*

Figure 14.7.4 presents the improvements of the CKM measurements, currently achieved and expected at Belle II. The `CKMFitter` group has performed analyses of non-SM physics in mixing, assuming that tree decays are not affected by non-SM effects. Within this framework, non-SM contributions to the $B_d$ mixing amplitudes can be parametrized as

$$M_{12}^d = (M_{12}^d)_{SM} \times (1 + h_d e^{2i\sigma_d}) \tag{14.7.3}$$

Here $h_d$ and $\sigma_d$ stand for the amplitude and phase of the non-SM physics, which are related to the mass-scale parameter $\Lambda$ via

$$h \simeq \frac{|C_{ij}|^2}{|\lambda_{ij}^t|^2} \left( \frac{4.5\,\text{TeV}}{\Lambda} \right) \tag{14.7.4}$$

$$\sigma = \arg(C_{ij} \lambda_{ij}^{t*}), \tag{14.7.5}$$

where $\lambda_{ij}^t = V_{ti}^* V_{tj}$ and $V$ is the CKM matrix. The scales $\Lambda$ probed in $B_d$ mixing by the end of the Belle II data-taking will be 17 TeV and 1.4 TeV for CKMI-like couplings in a tree and one-loop-level non-SM interactions respectively. For a scenario with no hierarchy, i.e. $|C_{ij}| = 1$, the corresponding scale of operators probed will be $2 \times 10^3$ TeV and $2 \times 10^2$ TeV in a tree- and one-loop-level non-SM interactions respectively.

### 14.7.3 Search for non-SM $CP$ violation in rare $B$ processes

In order to search for the non-SM contribution, the most promising channel is $B^0 \to \eta' K_S^0$; it has a sizable decay rate dominated by the $b \to s$ loop amplitude, where non-SM physics can contribute, and its associated hadronic uncertainties is relatively small. The quantity of interest is $\Delta\mathcal{S}_{\eta' K_S^0} \equiv \mathcal{S}_{\eta' K_S^0} - \sin\phi_1$. The SM predictions that include a systematic treatment of low-energy QCD amplitudes assuming factorization yield $0.00 < \Delta\mathcal{S}_{\eta' K_S^0} < 0.03$ [4674]. The current world average of $\Delta\mathcal{S}_{\eta' K_S^0}$ is $-0.07 \pm 0.06$ [4661]. Low backgrounds and a high-resolution electromagnetic calorimeter offer Belle II unique access to this measurement. Similarly promising is the channel $B^0 \to \phi K_S^0$, whose final state makes Belle II strongly competitive despite challenges associated with model-related systematic uncertainties from the Dalitz plot analysis. The expected experimental accuracy at 50ab$^{-1}$ is $\sim 0.01 (\sim 0.02)\%$ for $\mathcal{S}_{\eta' K_S^0} (\mathcal{S}_{\phi K_S^0})$. Figure 14.7.5 demonstrates the time-dependent $CP$ asymmetry for the final state $\eta' K_S^0$ compared to $J/\psi K_S^0$, using $\mathcal{S}_{\eta' K S^0} = 0.55$ and $\mathcal{S}_{J/\psi K_S^0} = 0.70$ in a Monte Carlo simulation with the integrated luminosity of 50 ab$^{-1}$, where the two values would be unambiguously distinguishable, signifying the existence of new physics. In addition, the processes $B^0 \to K_S^0 \pi^0 \gamma$, $B^0 \to K_S^0 \pi^+ \pi^- \gamma$, and $B^0 \to \rho^0 \gamma$ are greatly sensitive to non-SM physics through $b \to s$ and $b \to d$ loops and offer Belle II further exclusive opportunities.

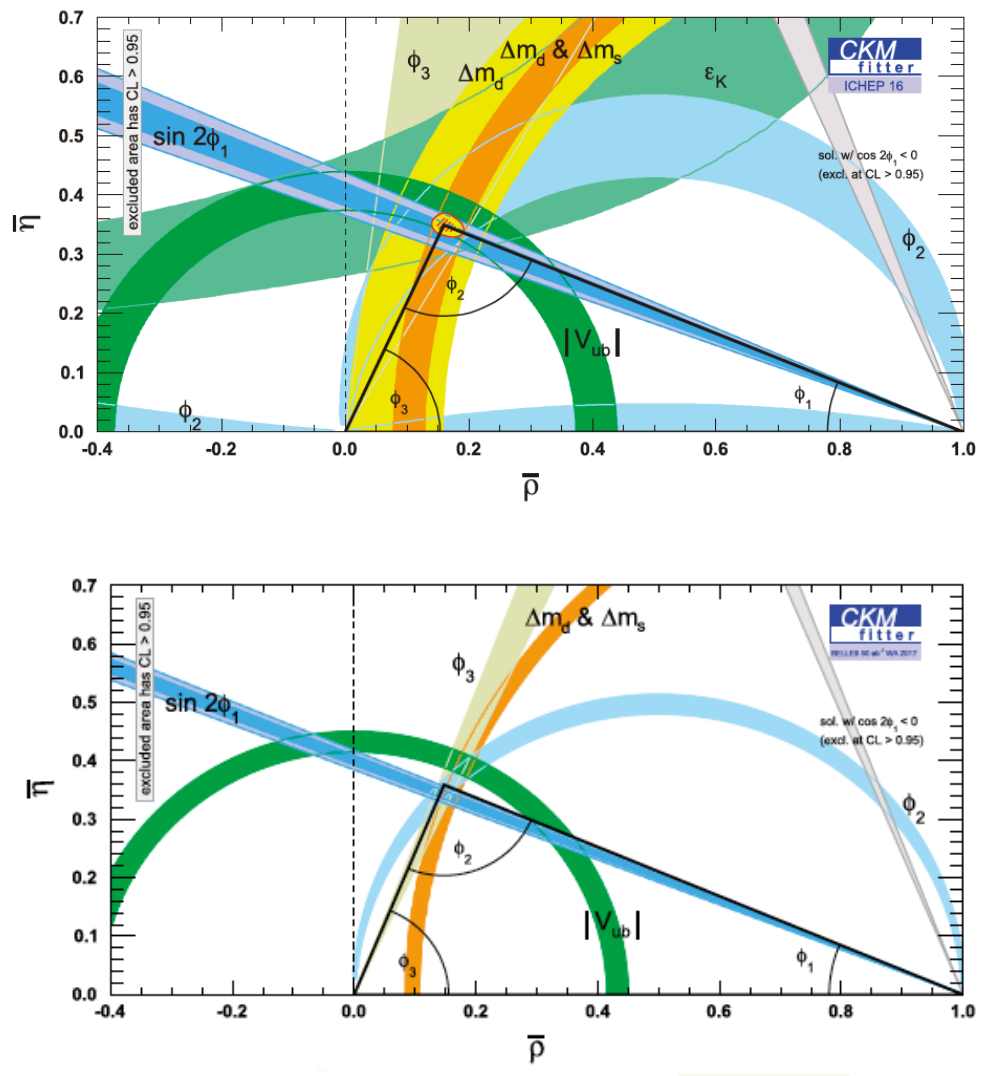

**Fig. 14.7.4** Current unitarity triangle fit (top) and extrapolated to 50 ab$^{-1}$ (bottom) [4658].

### 14.7.4 Search for non-SM physics in semileptonic and radiative $B$ decays

A number of persistent anomalies have been observed in semileptonic $B$ meson decays; deviation from lepton-flavor universality in the decays $B \to D^{(*)} \tau \nu_\tau$ consistently stayed at the $3\sigma$ level since these decays were first measured [4661]. Another case of lepton-flavor universality violation has been seen in $B \to K^{(*)} \ell^+ \ell^-$. The unique capability of Belle II to reconstruct final states with missing energy and identify efficiently all species of leptons will considerably improve the understanding of these anomalies.

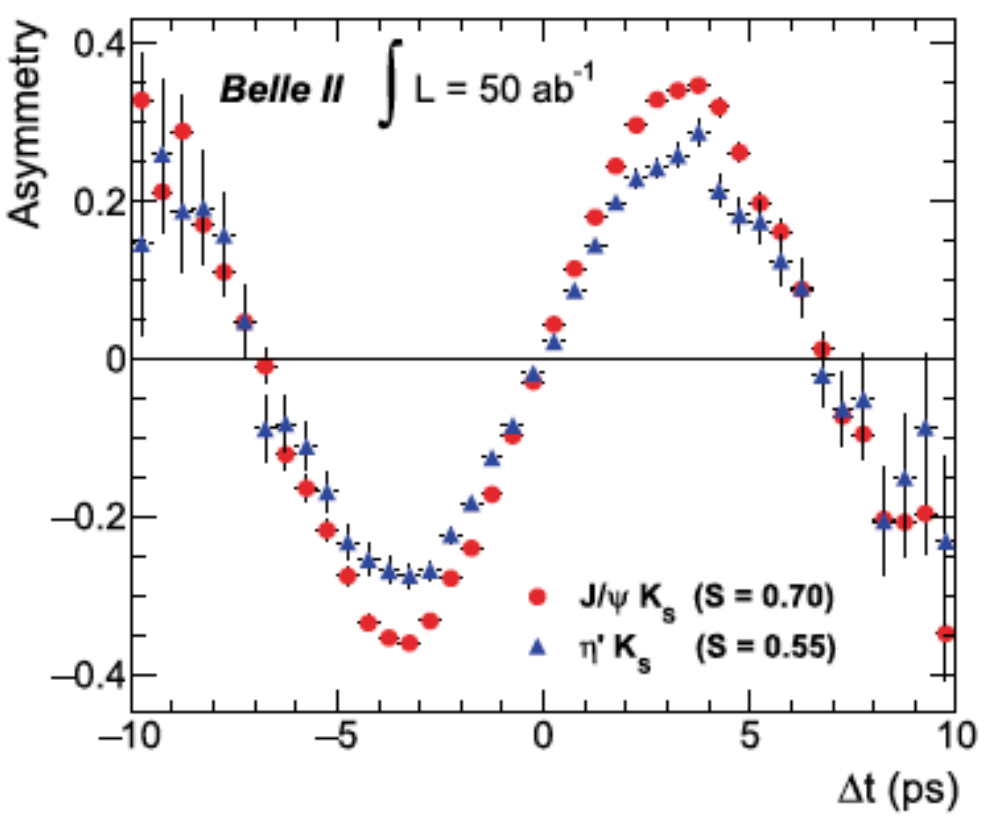

**Fig. 14.7.5** Time-dependent $CP$ asymmetry for the final state $\eta' K_S^0$ compared to $J/\psi K_S^0$, using $\mathcal{S}_{\eta' K_S^0} = 0.55$ and $\mathcal{S}_{J/\psi K_S^0} = 0.70$ in a Monte Carlo simulation with the integrated luminosity of 50 ab$^{-1}$ [4658].

*Semitauonic B decays*

Decays $B \to D^{(*)}\tau\nu_\tau$ offer precious opportunities for testing lepton-flavor universality at high precision opening a window onto lower-mass (TeV range) non-SM particles. Sensitive observables are the ratio $R(D)$ and $R(D^*)$ of branching fractions of $B \to D^{(*)}\tau\nu_\tau$ to those of $B \to D^{(*)}\ell\nu_\ell$ decays, where $\ell = e$ or $\mu$. There have been numerous SM calculations of $R(D^{(*)})$ and experimentally, the ratio allows for numerous systematic uncertainties to cancel. The SM predictions for the ratios $R(D)$ and $R(D^*)$ are:

$$R(D) = 0.299 \pm 0.011 \tag{14.7.6}$$

$$R(D^*) = 0.252 \pm 0.003 \tag{14.7.7}$$

Current best results on $R(D^{(*)})$ are reported by the Belle experiment [4675] and are consistent with previous measurements [4676–4680] in showing a (combined) $3.1\sigma$ excess with respect to the SM expectation [4661].

$$R(D) = 0.349 \pm 0.027_{(stat)} \pm 0.015_{(syst)} \tag{14.7.8}$$

$$R(D^*) = 0.298 \pm 0.011_{(stat)} \pm 0.007_{(syst)} \tag{14.7.9}$$

This deviation has attracted significant interest in the community as it could be a potential indication of non-SM dynamics.

The main experimental challenge is achieving a detailed understanding of poorly known $B \to D^{**}\ell\nu$ backgrounds, whose feed-down may bias the results. The anticipated data set size will allow for accurate tagged measurements of $B \to D^{**}\ell\nu$ decays for several $D^{**}$ states using samples reconstructing on the signal-side a lepton, a $D^{(*)}$ meson and $n$ pions. If a non-SM source of the anomaly would be established, angle-dependent asymmetries and differences between forward-backward asymmetries observed in muons and electrons, which are ideally suited for Belle II, may offer insight into the properties of the non-SM couplings involved.

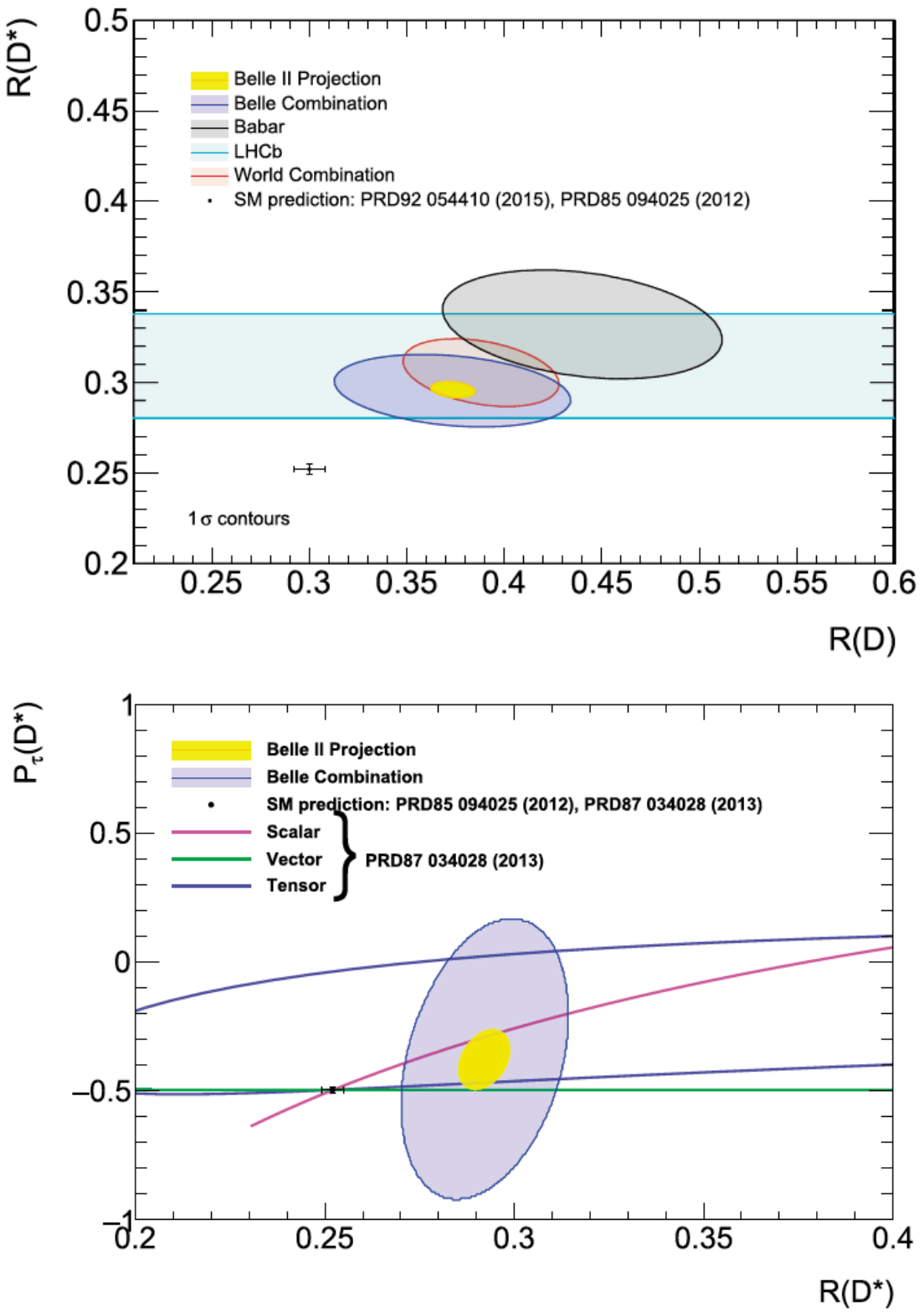

**Fig. 14.7.6** Expected Belle II constraints on the $R(D) - R(D^*)$ plane (top) and the $R(D^*) - P_\tau(D^*)$ plane (bottom) compared to existing experimental constraints from Belle. The SM predictions are indicated by the black points with theoretical error bars [4658].

Measurements of polarization of the $\tau$ lepton $((\Gamma^+ - \Gamma^-)/(\Gamma^+ + \Gamma^-))$ and $D^*$ mesons $(\Gamma_L/(\Gamma_T + \Gamma_L))$ provide supplementary sensitivity to non-SM physics. Here, $\Gamma^+(\Gamma^-)$ is the semitauonic decay rate where the $\tau$ has $+\frac{1}{2}$ $(-\frac{1}{2})$ helicity and $\Gamma_L(\Gamma_T)$ is the rate where the $D^*$ has longitudinal (transverse) polarization. Figure 14.7.6 shows the expected Belle II constraints on the $R(D) - R(D^*)$ plane (top) and the $R(D^*) - P_\tau(D^*)$ plane (bottom). Furthermore, differential angular distributions in $B \to D^{(*)}\tau\nu$, usually studied as functions of $q^2$, may also be important to decipher the dynamics and are distinctive to Belle II.

*$B \to K^*\ell^+\ell^-$ decays*

The transitions $b \to s\mu\mu$ and $b \to see$ are under extensive experimental investigation due to several observed anomalies [4681–4685] that prompted interpretations in terms of $\mathcal{O}(10)$ TeV non-SM particles. The unique feature of Belle II is its high efficiency and similar performance for muons and electrons, along with access to absolute branching fractions. Based on a recent Belle

II analysis [4686], we expect to provide distinctive information to assess independently the existence of the anomalies (at current central values) with samples of 5 ab$^{-1}$ to 10 ab$^{-1}$ of data. Belle II can provide also results based on inclusive $B \to X_S \ell^+ \ell^-$ decays, which do not specify the final strange hadronic states $X_S$ and has fewer theoretical ambiguities.

Belle II can reach also $b \to s\tau\tau$ transitions. These can be enhanced, by up to three orders of magnitude, in several SM extensions that allow for lepton-flavor universality violation in the third generation [4687, 4688]. The SM branching fraction for the $B \to K^* \tau\tau$ decay is around $10^{-7}$ [4689], much smaller than current experimental upper limits, which are at around $2.0 \times 10^{-3}$ at 90% CL [4690, 4691]. The presence of two $\tau$ leptons in the final state makes access to these decays ideally suited to Belle II.

*Radiative B decays*

Radiative $b \to s\gamma$ transitions are dominated by a one-loop amplitude involving a $t$ quark and $W$ boson. Extensions of the SM predict particles that can contribute to the loop, potentially altering various observables from their SM predictions [4692, 4693]. Belle II has a unique capability to study these transitions both inclusively and using specific channels.

The availability of precise and reliable SM predictions of inclusive $B \to X_S \gamma$ rates, where $X_s$ identifies a particle with strangeness, make these rates sensitive probes for non-SM physics. In addition, these analyses enable the determination of observables like the $b$-quark mass and can provide input to inclusive determinations of $|V_{ub}|$ [4658]. Ability to measure precisely the decay properties of the partner $B$ recoiling against the signal $B$ is key for inclusive analyses[4659]. Current best results show 10% fractional precision mostly limited by systematic uncertainties associated with understanding the large backgrounds. The expected relative uncertainties on the branching fractions are $\sim 6\%$ at 5 ab$^{-1}$ and $\sim 2\%$ at 50 ab$^{-1}$ slightly depending on the lower $E_\gamma$ thereshold. The construction of relative quantities like asymmetries will offer a further reduction of systematic uncertainties and enhanced reach. Inclusive analyses of radiative $B$ decays will offer unique windows over non-SM physics throughout the next decade.

### 14.7.5 Hadron Spectroscopy

While many hadron states are categorized into mesons and baryons containing constituent quark-antiquark ($q\bar{q}$) and three quarks (qqq), respectively, there is no proof in QCD to exclude the hadrons having other structures than the ordinary mesons and baryons. The situation has largely changed by the series of discoveries of charmonium-like states, $X(3872)$ [2460], $Y_c(4260)$ [4694], $Z_c^\pm(3900)$ [2534], and several others that do not fit the well-established quark model. Analogous discoveries containing bottom quarks (*e.g.*, $\Upsilon(5S)$ decays to $Z_b^\pm(10610/50)$ [2544]) indicate a similar unexplored family of particles in the bottomonium sector. The Belle II experiment offers several unique opportunities in this domain. It will exploit 40 times more data than the previous generation $B$-factories and, compared with hadron-collisions experiments, leverages a greater variety of quarkonium production mechanisms including $B$ meson decays, initial state radiation (ISR), double $c\bar{c}$ processes, two-photon processes, and direct production by changing collider center-of-mass energy [4658]. Belle II is the only experiment with the ability to operate at tuneable center-of-mass energy near the $\Upsilon(4S)$ resonance, providing direct access to multi-quark states containing bottom quarks. In addition, Belle II's good efficiency for reconstructing neutral final-state particles opens the pathway for first observations of the predicted neutral partners of charged tetraquark states.

Belle II has the unique opportunity to explore bottomonium(-like) states by operating at center-of-mass energies around 10 GeV, where only small samples exist worldwide: $\mathcal{O}(10)$ fb$^{-1}$ at $\Upsilon(1S, 2S, 3S, 6S)$, $\mathcal{O}(100)$ fb$^{-1}$ at $\Upsilon(5S)$, and typically less than 1 fb$^{-1}$ at intermediate points. This opens a fruitful program, as demonstrated by previous discoveries at $e^+e^-$ colliders that yielded first observations of predicted bottomonia ($\eta_b(1S, 2S)$, $h_b(1P, 2P)$, and $\Upsilon(1D_2)$) and unexpected four-quark states ($Z_b^\pm(10610, 10650)$, $Y_b(10753)$) [4695, 4696]. Collisions at energies below the $\Upsilon(4S)$ allow for testing non-SM predictions in $\Upsilon$ decays to invisible or lepton-flavor-violating final states [4697, 4698].

### 14.7.6 Constraining hadronic vacuum-polarization in muon g-2

The anomalous magnetic moment of the muon often parametrized as $a_\mu = (g-2)_\mu/2$, is one of the observables which indicate significant deviation from the SM and has attracted much attention from the community. The current experimental value (combining the BNL E821 result with the first result from the Fermilab $g-2$ experiment) differs from SM predictions based on dispersion relations by $4.2\sigma$, $a_\mu(\text{exp}) - a_\mu(\text{theory}) = (26.0 \pm 7.9) \times 10^{-10}$ [4206, 4207]. In order to clarify the deviation, it is important to improve the precision of both experiments and the SM predictions. On the experimental side, the experiment at Fermilab will provide results by further accumulated data and also an experiment with different methods and thus have dif-ferent systematic errors has been proposed and is being prepared at J-PARC [4699]. The uncertainty in the SM prediction is dominated by the leading-order hadronic contribution (HVP), which can be calculated from the cross-section $\sigma(e^+e^- \to \text{hadrons})$ measured in $e^+e^-$ experiments. The result, HVP=$(693.1 \pm 4.0) \times 10^{-10}$, is dominated by BaBar and KLOE measurements of $\sigma(e^+e^- \to \pi^+\pi^-)$. However, the BaBar and KLOE measurements notably differ. This difference introduces a systematic uncertainty of $2.8 \times 10^{-10}$ [4224].

Belle II will perform these measurements with larger data sets, and at least comparable systematic uncertainty, to resolve this discrepancy. Furthermore, large statistics data at Belle II will allow us to use new approaches to suppress systematic uncertainties, particularly from particle identification. Although the specific systematic studies still need to be refined, the goal for the final accuracy including both statistical and systematic uncertainties is to be 0.5% or lower [4658]. This will match the expected experimental precision on $g-2$[4206, 4658]. Belle II's operation at the highest luminosity $\boldsymbol{e^+e^-}$ collider, as well as its excellent particle-identification capabilities, places it in a unique position to further the studies of the HVP contribution to $(g-2)_\mu$ in the next decade. HVP can be estimated also by $\tau$ hadronic spectral functions and CVC, together with isospin-breaking corrections.

### 14.7.7 Status and outlook

The physics data taking with all the Belle II subdetector components started in March 2019, following the SuperKEKB main ring commissioning run in 2016, and the collision test runs in 2018. At the time when this article is written, the SuperKEKB accelerator has achieved the peak luminosity of $4.7 \times 10^{34}$ cm$^{-2}$s$^{-1}$, more than two times higher than the record of the previous KEKB accelerator. The Belle II experiment has accumulated 428 fb$^{-1}$, almost similar to the BaBar and about half of the Belle experiments. Some results are already world-leading thanks to the efficiency and resolution improved significantly compared to the previous experiments. The operation is suspended since June 2022 for the upgrade work on the SuperKEKB and Belle II instrumentations. The operation is planned to resume in autumn 2023. Many world-leading results in heavy flavor decays will be obtained with $\mathcal{O}(1)$ ab$^{-1}$ data in the near future, and then with $O(10)$ ab$^{-1}$ toward the next decade.

## 14.8 Heavy flavors at the HL-LHC

**Tim Gershon**

Proton-proton collisions at energies of the LHC collider result in production of vast quantities of beauty and charm quarks. The production cross-sections at centre-of-mass collision energies of 7–14 TeV are around 100 $\mu$b for beauty hadrons and an order of magnitude larger for charm hadrons [4700, 4701]. Thus, for each fb$^{-1}$ of integrated luminosity, there are around $10^{11}$ beauty hadrons and around $10^{12}$ charm hadrons produced. As there are no constraints on the quantum numbers of the particles that emerge from the primary interaction followed by hadronization, essentially all physically possible hadrons are produced in LHC collisions. Since effects of double parton scattering, where multiple heavy quark-antiquark pairs are produced in the same proton-proton interaction, are significant, this includes states with more than one heavy-flavor quark.

The LHC and its high luminosity upgrade therefore provide a unique and unprecedented opportunity to learn about QCD from the production and decays of these hadrons. However, in order for this experimental program to be realized, it is necessary to have dedicated and state-of-the-art detection capability. In particular, focusing on charged particle detection, one needs:

- acceptance, with good reconstruction efficiency, in the kinematic region that the majority of the decay products will travel through (production of beauty and charm hadrons at the LHC predominantly occurs at small angles to the beam axis);
- good momentum resolution, so that narrow signal peaks in invariant mass distributions originating from states close to each other in mass to be resolved;
- capability to discriminate between different final-state charged particles, in particular electrons, pions, muons, kaons and protons;
- ability to reject background from random combinations of particles, which must be achieved in real-time (online) in order to avoid the data rate overwhelming the available computing resources.

As regards the last point, the presence of one or more well-identified muons in the decay, above a $p_T$ threshold of typically a few GeV/$c$, is a signature which has traditionally been used in triggers for heavy-flavor physics in hadron collider experiments. This signature continues to be exploited at the LHC, and will be throughout the HL-LHC era. However, the fact that the ground-state hadrons with heavy-flavor quantum numbers can only decay by the weak interaction provides an extremely valuable handle, as their non-negligible lifetimes cause a significant — and potentially measurable – displace-ment between the production and decay vertices. Consider for example a state of mass 5 GeV/$c^2$ and lifetime $\tau = 1$ ps. If produced with 50 GeV/$c$ momentum, corresponding to a Lorentz boost factor of $\beta\gamma = 10$, it will travel a mean distance of $\beta\gamma c\tau \approx 3$ mm before decaying. Therefore if the vertex position can be reconstructed with resolution significantly better than this, the potentially huge background from combinations of the large numbers of tracks produced at the primary proton-proton interaction point can be removed. Indeed, while proton-proton collisions are generally considered a difficult (or "dirty") environment due to the large numbers of particles produced, if one only needs to consider particles originating from displaced secondary vertices the signatures can be extremely clean.

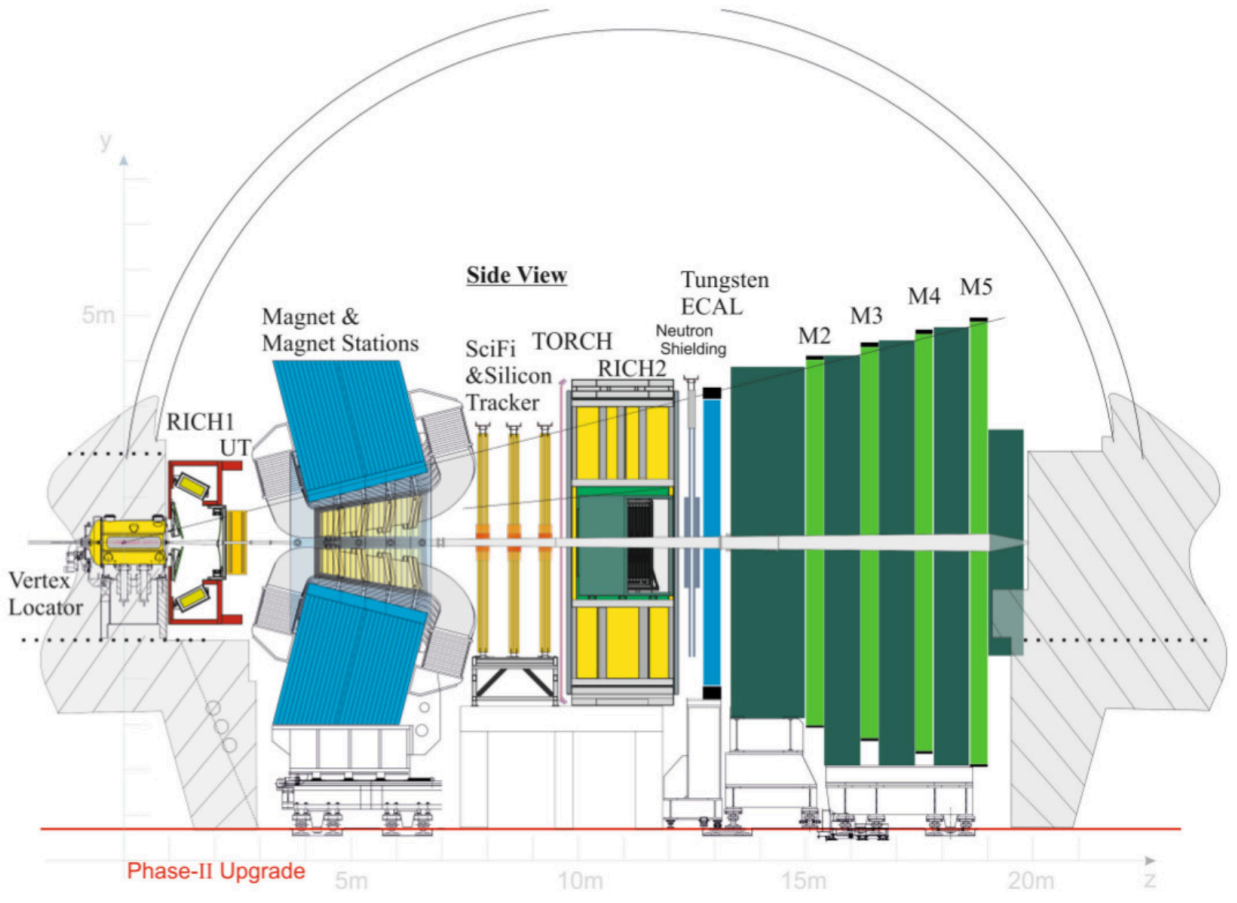

**Fig. 14.8.1** The proposed LHCb Upgrade II detector [4702].

The LHCb detector is designed in order to provide this detection capability. It is the only dedicated heavy-flavor experiment at the LHC, although ALICE, ATLAS and CMS all have some ability to reconstruct heavy-flavor hadrons. The original LHCb detector operated during Runs 1 and 2 of the LHC, 2011–12 and 2015–18 respectively, enabling the collection of a data sample corresponding to 9 fb$^{-1}$ of proton-proton collisions. This has led to a wealth of publications on a diverse range of topics. An upgraded detector has been installed during the LHC long shutdown 2 (2019 - 21) and is designed for the collection of a sample of 50 fb$^{-1}$ during Runs 3 and 4, with significantly improved efficiencies for many channels of interest. In order to exploit fully the flavor-physics potential of the HL-LHC, a second major upgrade of the LHCb detector is now being planned [4702]; this will allow 300 fb$^{-1}$ to be collected in the final operational periods of the HL-LHC. Together with the 3 ab$^{-1}$ anticipated to be collected by ATLAS and CMS, this provides exciting potential in heavy-flavor physics.

The above discussion focussed on charged particles. For neutral particles it is much harder both to obtain good momentum resolution and to associate them correctly to the vertex they originated from, particularly bearing in mind that they will be reconstructed in the forward kinematic region. Nonetheless, information from calorimeters can be used to broaden the flavor-physics program to include decays with photons in the final state, including those from neutral pion decays and from bremstrahlung emission from electrons. Moreover, timing information can be used to provide some capability to associate calorimeter clusters with reconstructed vertices; indeed the addition of timing capability is central to the plans for LHCb Upgrade II, not only for the calorimeter but also for the vertex and charged hadron identification detectors [4702].

The opportunities in flavor physics at the HL-LHC are discussed in Ref. [4703], while the LHCb Upgrade II physics program is described in Ref. [2579]. Here only a brief summary of some aspects that are most interesting with regard to QCD are discussed. The focus is primarily on LHCb, but areas where other LHC experiments can contribute are also mentioned.

*CP violation*

Violation of symmetry under the combined charge conjugation and parity ($CP$) operation can occur in the Standard Model as the complex phase in the Cabibbo–Kobayashi–Maskawa (CKM) quark mixing matrix [3957, 3958] results in the charged-current weak-interaction coupling constants being different for quarks and antiquarks. The uniqueness of the origin of all $CP$ violating effects in the SM – and the knowledge that additional sources must be present in nature in order to explain the baryon asymmetry of the Universe – make experimental probes of $CP$-violating phenomena a well-motivated way to search for physics beyond the SM.

There are a number of theoretically clean probes of $CP$ violation, where QCD effects that may otherwise render the interpretation of results difficult are either minimal or can be determined directly from data. In particular, the determination of the phase

$$\gamma \equiv \arg\left(-\frac{V_{ud}V_{ub}^*}{V_{cd}V_{cb}^*}\right)$$

from $B \to DK$ and similar processes is essentially unaffected by theoretical uncertainties in the SM [4663]. However, there are many more measurements where uncertainties related to QCD need to be reduced in order to obtain the best sensitivity to physics beyond the SM. An interesting class of such measurements are those where decays can be related by flavor symmetries, as the breaking of this symmetry by QCD can often be calculated theoretically. The fact that both $B^0$ and $B_s^0$ mesons can be studied at the LHC opens a number of possibilities involving U-spin symmetry, related to interchange of $d$ and $s$ quarks. For example, the determination of the phase

$$2\beta \equiv 2 \arg \left( -\frac{V_{cd}V_{cb}^*}{V_{td}V_{tb}^*} \right)$$

from $B^0 \to J/\psi K_S^0$ decays has a small but hard-to-quantify uncertainty due to subleading amplitudes; the size of this effect can be constrained using the U-spin partner $B_s^0 \to J/\psi K_S^0$ decays [4704, 4705]. In a similar way, the $B_s^0 \to K^{*0}\overline{K}^{*0}$ decay is considered a golden channel to probe for $CP$-violation effects beyond the SM, as theoretical uncertainties can be constrained from the U-spin partner $B^0 \to K^{*0}\overline{K}^{*0}$ decay [4706–4708].

The above examples are special cases where the final state is left unchanged by U-spin. Similar ideas can be also exploited for U-spin pairs where this is not the case, such as $B^0 \to D^+D^- \leftrightarrow B_s^0 \to D_s^+D_s^-$, $B^0 \to \pi^+\pi^- \leftrightarrow B_s^0 \to K^+K^-$ and $B^0 \to K^+\pi^- \leftrightarrow B_s^0 \to K^-\pi^+$ [4709–4714]. In these cases however the U-spin breaking effects can be larger, making it harder to use them for precise tests of the SM. However, with the data samples available at the HL-LHC it will be possible to reverse the argument: assuming the SM, the extent of U-spin breaking in these decays can be precisely measured and compared to theoretical calculations. Moreover, the samples will be large enough that similar exercises can also be done for suppressed partner decays (*e.g.* $B^0 \to D_s^+D_s^- \leftrightarrow B_s^0 \to D^+D^-$ and $B^0 \to K^+K^- \leftrightarrow B_s^0 \to \pi^+\pi^-$) where effects of subleading amplitudes are enhanced. Studies of U-spin breaking and its influence on $CP$ violation in the charm meson decays $D^0 \to K^+K^-$, $\pi^+\pi^-$, $K^-\pi^+$ and $K^+\pi^-$ provide a complementary probe [4715–4718]. These measurements will provide a unique handle on our understanding of flavor symmetry breaking effects in QCD.

A number of null tests of the SM can be made by testing the prediction of small or vanishing $CP$-violating effects in specific processes. In such cases it is necessary to ensure that theoretical uncertainties in the prediction are well under control. One example is the determination of the phase $\phi_s$ through $B_s^0 \to J/\psi\phi$ and similar processes, where LHCb, ATLAS and CMS all have potential to reach sufficient precision to observe a non-zero effect at the SM rate [4719–4721]. Another example is the corresponding phase in the neutral charm system, $\phi_D$, where recent progress measuring the mixing parameters has set the stage for precise determinations when more data are available [4722, 4723]. It remains an open question to what extent QCD effects can enhance SM $CP$ violation in the charm sector [4724], and further progress on this front will be essential.

Data on two-body decays are in general easier to interpret than those in three- or multi-body decays (including quasi-two-body resonant contributions). Nevertheless, the latter remain of great interest as interference effects can provide sensitivity to additional $CP$-violating observables: the range of effects observed in three-body $B$ meson decays illustrate this clearly [4725–4729]. Overcoming hadronic uncertainties is challenging, but with HL-LHC data ambitious coupled-channel analyses will allow additional constraints. In particular, effects related to $\pi\pi \leftrightarrow KK$ scattering can be fitted for directly in coupled-channel analyses of $B^0$ and (separately or simultaneously) $B_s^0$ decays to the $J/\psi\pi^+\pi^-$ and $J/\psi K^+K^-$ final states [4730]. Similar analyses can also be carried out in $B_{(s)}^0 \to \overline{D}^0\pi^+\pi^-$ and $\overline{D}^0K^+K^-$ decays, and in $B^+ \to K^+\pi^+\pi^-$ and $K^+K^+K^-$ decays. The latter, and also the more suppressed $B^+ \to \pi^+\pi^+\pi^-$ and $\pi^+K^+K^-$ decays, are known to feature regions of phase space with large $CP$ violation, which could be used to test the SM if theoretical uncertainties can be controlled sufficiently.

As mentioned above, the CKM angle $\gamma$ can be determined with negligible uncertainty using $B \to DK$ and related decays. The reason for this is that by combining results with multiple different $D$ decay modes, all hadronic parameters can be determined from data. Recent examples of such combinations can be found in Refs. [4731, 4732]. From the point of view of understanding QCD, this provides an opportunity to compare the values of the hadronic parameters obtained from the combinations to those from theoretical calculations. In the case of multibody decays such as $B \to DK\pi$, the parameters that can be obtained include those related to variation of hadronic phases across the phase-space of the decay [4733, 4734]. These can be determined model-independently as a by-product of the measurement of $\gamma$, thus providing insight into a poorly understand aspect of QCD.

*Semileptonic decays and form factors*

As discussed in Sec. 13.2.2, the rates of semileptonic $b$-hadron decays, $X_b \to X_c\ell^-\overline{\nu}_\ell$ depend on the square of the magnitude of the CKM matrix element $V_{cb}$. Here, $X_b$ represents a hadron containing a $b$ quark, $X_c$ the corresponding hadron with $b$ replaced by $c$, $\ell^-$ a negatively charged lepton and $\overline{\nu}_\ell$ the corresponding antineutrino. Thus, measurements of the rates can allow $|V_{cb}|^2$ to be determined if the form factors, which encode the probability for the $X_c$ hadron to be produced in the final state as a function of the $\ell^-\overline{\nu}_\ell$ invariant mass squared ($q^2$), are known from theoretical calculations.

Likewise, studies of $X_b \to X_u \ell^- \overline{\nu}_\ell$ transitions, with obvious definition of $X_u$, provide sensitivity to $|V_{ub}|^2$.

The reconstruction of decays involving neutrinos in the final state is challenging in the environment of a hadron collider, as one cannot exploit the kinematic constraints that are available in the $e^+e^- \to \Upsilon(4S) \to B\overline{B}$ system. Nonetheless, exploiting LHCb's capability in reconstruction of vertices and charged hadron identification, it has been possible to study semileptonic $\Lambda_b^0$ (to $p\mu^-\overline{\nu}_\mu$ and $\Lambda_c^+\mu^-\overline{\nu}_\mu$) and $\overline{B}{}_s^0$ (to $K^+\mu^-\overline{\nu}_\mu$ and $D_s^+\mu^-\overline{\nu}_\mu$) decays [715, 4077]. In each case measuring the ratio allows the cancellation of some potential sources of systematic uncertainty, leading to competitive measurements of $|V_{ub}/V_{cb}|^2$.

With the full HL-LHC statistics it will be possible to extend this program to the full range of $b$ hadrons. This will provide complementary information to the determinations using $B$ mesons alone, and will test QCD by comparison of the form factors in heavy-to-light transitions (such as $B \to \pi$) with those in heavy-to-heavy transitions. A particularly interesting example occurs in $B_c^-$ decays, where study of $B_c^- \to D^0\mu^-\overline{\nu}_\mu$ could potentially allow a theoretically clean determination of $|V_{ub}|^2$. In fact, the large samples of $B_c^-$ mesons that will be available at HL-LHC present a further opportunity, since these particles preferentially decay through transitions of the charm quark. Thus, $B_c^- \to \overline{B}{}_s^0\mu^-\overline{\nu}_\mu$ and $\overline{B}{}^0\mu^-\overline{\nu}_\mu$ decays could be used to make novel measurements of the squared magnitudes of $V_{cs}$ and $V_{cd}$, respectively, thereby allowing a quantitative comparison of the form factors observed in data with those calculated from first principles QCD.

Understanding QCD effects encoded in form factors and, more generally, the effects of hadronization in semileptonic $b$-hadron decays, will also be crucial for tests of lepton universality at HL-LHC. Within the Standard Model the $W$ and $Z$ couplings to all lepton flavors are identical; any deviation from this prediction would provide a clear signature of non-SM physics contributing to the decay amplitude. Due to the heavier $\tau$ mass, compared to the electron and muon, contributions from different form factors have to be understood in order to predict the SM value of the ratio of branching fractions [4051–4053]. Given the indications of potential violation on lepton universality in previous measurements of these processes at the BaBar, Belle and LHCb experiments [4675–4680] there is intense interest in the significantly more precise results that the HL-LHC can potentially provide. The challenge will be to control experimental systematic uncertainties to the required level; this is even harder for ATLAS and CMS than for LHCb, but if the background composition can be understood then all three experiments may be able to test the SM in this sector.

*Rare decays*

Decays which proceed by flavor-changing neutral currents are highly suppressed in the Standard Model as they involve loop diagrams, typically with additional CKM suppression factors. As physics beyond the SM does not have to have the same structure, the rates and phase space distributions of these channels allow detailed tests for new contributions to the amplitudes.

In order to obtain the best sensitivity from these measurements, it is necessary to have QCD uncertainties, related to the hadrons in initial, intermediate and final states of the decay, under excellent control. Thus, typically the theoretically cleanest probes are decays involving leptons or photons. However, even in these cases there can be residual QCD effects that must be well understood. Recent progress is therefore focussed mainly on theoretically clean channels and data-driven approaches to constrain hadronic parameters.

The purely leptonic $B_{(s)}^0$ meson decays are a good example of channels where theoretically clean predictions are possible. Moreover, the helicity-suppression of these processes that occurs in the SM — resulting in small branching fractions for the dimuon and, especially, dielectron, processes — need not be replicated in beyond SM contributions to the amplitudes, so that large deviations from the SM predictions are possible in principle. The decay rates for these processes depend on the $B_{(s)}^0$ decay constants, which can be (and have been) calculated in lattice QCD to good precision [279]. The experimentally most amenable channel is the dimuon final state; the $B_s^0 \to \mu^+\mu^-$ decay has been observed by LHCb, CMS and ATLAS, and the sensitivity to the $B^0$ decay branching fraction approaches the level required to observe it at the SM expectation [4735–4737]. The limits on decays to dielectron and ditau final states remain considerably above the SM expectations [4738, 4739].

Further improvement in the knowledge of the $B_{(s)}^0 \to \mu^+\mu^-$ branching fractions and their ratio is well motivated, as the experimental uncertainties remain larger than those for theory. These measurements can be expected as a key component of the HL-LHC era heavy-flavor physics programs of all of the LHCb, CMS and ATLAS experiments: it is anticipated that relative uncertainties on $\mathcal{B}\left(B_s^0 \to \mu^+\mu^-\right)$ of 4%, 7% and 12–15% can be achieved by each of the three experiments, respectively [4702, 4740, 4741]. In addition, the increasingly large sample sizes will make additional probes possible. In particular, the $B_s^0 \to \mu^+\mu^-$ effective lifetime can be used as an independent probe for physics be-yond the SM [4742], with first measurements already available, albeit with large uncertainties. With the full HL-LHC statistics it will also be possible to measure $CP$ violation parameters in this decay, providing one more independent probe, also with negligible theoretical uncertainty.

The $b \to s\ell^+\ell^-$ and $b \to d\ell^+\ell^-$ processes can also be studied through decays in which the $s$ or $d$ quark is found in the final state. These do not have the helicity suppression of the purely leptonic decays, but as a corollary have sensitivity to additional effective field theory operators. A large range of final states and a large number of observables can be studied. Those related to angular distributions in $B \to V\ell^+\ell^-$ processes are particularly interesting (where $V$ is a vector meson, *i.e.* decays such as $B^0 \to K^{*0}\ell^+\ell^-$). In these measurements, all relevant operators can be constrained from data. Indeed, as discussed in Sec. 13.4, existing measurements of the rates and of angular observables in $B^0 \to K^{*0}\mu^+\mu^-$ and $B^0_s \to \phi\mu^+\mu^-$ decays constrain possible contributions from physics beyond the SM and, excitingly, hint at these contributions being non-zero [4685, 4743–4746]. However, the possibility of these effects being caused by larger than expected non-perturbative QCD corrections is not yet ruled out [4190, 4192].

Progress in this area, with the larger data samples available at the HL-LHC, can be expected in two complementary approaches. Firstly, model-dependent fits to the data can be used to attempt to constrain the non-perturbative QCD effects within specific parameterizations [4182, 4188, 4194, 4747]. Secondly, the SM property of lepton universality in these processes can be tested – comparison of equivalent parameters for decays involving $\mu^+\mu^-$ and $e^+e^-$ pairs provide theoretically clean tests of the SM. While the second case can provide an unambiguous signal of physics beyond the SM, this is only possible if the new physics violates lepton universality. Progress on both fronts is therefore essential in order to be able to constrain the full range of potential operators. Early measurements from LHCb of the ratios of decay rates for $B^+ \to K^+\ell^+\ell^-$ and $B^0 \to K^{*0}\ell^+\ell^-$ (with $\ell = e, \mu$) give tantalizing hints of disagreement with SM predictions, but do not reach a level of significance for which strong claims would be justified [4682, 4748]. In addition to larger data samples, improved electron reconstruction can help to reduce the uncertainties in future measurements. The range of lepton universality tests can also be expected to be increased in future beyond the rates alone to include also angular observables.

A further way to test the SM is through its prediction that the photon emitted in $b \to s\gamma$ flavor-changing neutral-current transitions should be predominantly left-handed, as a consequence of the V−A structure of the SM weak interaction. This can be tested in a number of ways, including through studies of the decay-time dependence of $B^0 \to K^{*0}\gamma$ and $B^0_s \to \phi\gamma$ decays, and of the angular distributions in $\Lambda^0_b \to \Lambda\gamma$ decays [4749–4752]. The angular distribution of $B^0 \to K^{*0}e^+e^-$ decays at very low $e^+e^-$ invariant mass also probes the same physics [4753]. However, the statistically most powerful approach involves analysis of the phase-space distribution of $B^+ \to K^+\pi^+\pi^-\gamma$ decays, complemented by measurement of the decay-time dependence of the $B^0 \to K^0_S\pi^+\pi^-\gamma$ process [4754–4758]. To realise the full potential of this method will require improved understanding of hadronic effects in the $K\pi\pi$ system. The large data samples available at the HL-LHC will provide a number of ways to acquire such knowledge, including measurement of the corresponding processes where the final-state photon is replaced by a $J/\psi$ meson.

*Hadron spectroscopy*

As mentioned previously, the copious production of beauty and charm quarks in LHC collisions provides opportunities for detailed studies of hadron spectroscopy, including discoveries of previously unmeasured states. Various production mechanisms are available, including central exclusive production. However, the two mechanisms for which studies have proved most productive to date are so-called prompt production, where a hadron is produced directly in a proton-proton collision (including via strongly decaying resonances), and production in weak decays of a heavier hadron. Prompt decays tend to have large backgrounds, and are limited to cases with a distinctive signature – but they provide the only possible approach for hadrons too heavy to be produced in weak decays. Weak decays of heavy hadrons can provide an extremely clean environment; moreover this approach makes possible determination of the quantum numbers of intermediate resonances produced in multibody final states.

At the time of writing, 67 hadrons have been observed for the first time at the LHC as illustrated in Fig. 14.8.2. As discussed in Secs. 8.5 and 9.4, these include a number of states that do not fit into the conventional scheme of $q\overline{q}$ mesons and $qq'q''$ baryons. One of the most exciting topics, related to furthering knowledge of QCD, is what new hadrons may be discovered at the HL-LHC. This is, of course, impossible to predict with confidence; nonetheless there are certain areas where progress appears likely. In what follows states with four and five quarks are referred to as tetraquarks and pentaquarks respectively, with no prejudice as to their internal binding mechanisms – indeed, addressing the question of how such states are bound is one of the main goals for the HL-LHC in this area – and the naming convention of Ref. [2468] is used.

Perhaps the most striking discovery of exotic hadrons to date is that of the $P_\psi$ states, observed as resonances decaying to $J/\psi p$, and hence with minimal quark content $c\bar{c}uud$ in $\Lambda_b^0 \to J/\psi p K^-$ decays [2828, 2829]. The proximity of the $P_\psi$ masses to $\Sigma_c D$ thresholds has led to much speculation on their nature. Further progress requires the determination of the $P_\psi$ spin-parity quantum numbers. Discoveries of other production modes and decays to other final states will also provide insight. The data samples of the HL-LHC should allow LHCb to perform such studies, and also to make detailed studies of lineshapes.

The $P_\psi$ pentaquarks contain a $c\bar{c}$ pair, as do all tetraquarks that had been observed prior to 2020. This fueled theoretical speculation that a $c\bar{c}$ component, or at least the presence of two heavy quarks or antiquarks, was necessary for the formation of exotic hadrons. Such models were, however, ruled out by the observation of $T_{cs}$ tetraquarks decaying to $D^+K^-$, produced in $B^- \to D^-D^+K^-$ decays [4760, 4761]. This observation implies the existence of many more tetraquarks, containing different sets of quark flavors, which may be discoverable with the HL-LHC. As such states are observed and can be arranged in families, it will allow for a new understanding of strong interactions in much the same way as occurred for the "particle zoo" in the 1960s and 70s.

Even if a $c\bar{c}$ component is not required for the formation of exotic hadrons, a $J/\psi$ meson in the final state facilitates the observation of new particles due to the clean signature provided by the $J/\psi$ dimuon decay. This has been exploited in the observations of $T_{\psi\psi}$ states decaying to $J/\psi J/\psi$ [2565, 4762, 4763]. The discovery of states with minimal quark content of $cc\bar{c}\bar{c}$ motivates searches for partner states, including decays to final states such as $J/\psi\chi_{c1}$, which may cause feed-down into the $J/\psi J/\psi$ spectrum, as well as for tetraquarks with other fully heavy-quark content (*e.g.* $b\bar{b}c\bar{c}$). Knowledge of bottomonia decays to double charmonia final states will also be necessary for a full understanding in this area.

The first doubly charmed hadron, the $\Xi_{cc}^{++}$ state, was observed by LHCb in 2017 [2561], and precise measurements of its mass and lifetime have followed [2563, 2564]. Its flavor partners, the $\Xi_{cc}^{+}$ and $\Omega_{cc}^{+}$ baryons have also been searched for, but not yet discovered [2832, 4764, 4765]. The reason for this may be the shorter lifetimes that are expected for these states, since a short lifetime makes it harder to separate signal from background. The improved vertex resolution of the upgraded LHCb detector, together with larger data samples, will hence provide excellent prospects for discovery. Doubly heavy states containing beauty and charm quarks also appear within reach, while double beauty states appear more challenging.

The discovery of the $T_{cc}^{+}$ tetraquark, seen in prompt production as a narrow structure decaying to $D^0D^0\pi^+$ [1071, 2512], complements both the previous observations of the $\Xi_{cc}^{++}$ baryon and of tetraquarks with $c\bar{c}$ content. Its mass is only just above threshold for $D^0D^{*+}$ decays, supporting the hypothesis that ground-state tetraquarks containing beauty and charm or double beauty ($T_{bc}$ or $T_{bb}$), which are expected to be more tightly bound, may be stable to strong decays. If so, they would decay only via the weak interaction and hence have lifetimes comparable to those of ground state beauty and charm hadrons. As such, they may have displaced vertex signatures that could be exploited in the LHCb experiment to enhance their observability [4766]. It is also possible that $P_{cc}$, $P_{bc}$ and $P_{bb}$ pentaquarks could be detected, with the appropriate analysis strategy depending on whether or not they are stable against strong decay. Furthermore, it is plausible (albeit speculative) that six quark, dibaryon states containing at least two beauty or charm quarks may be measurable. Studies of hadron spectroscopy with the HL-LHC data sample may therefore provide dramatic breakthroughs in the knowledge of the possible range of states that can be bound together within QCD.

## 14.9 High-$p_\mathrm{T}$ physics at HL-LHC

**Massimiliano Grazzini and Gudrun Heinrich**

### 14.9.1 Introduction

The High-Luminosity LHC (HL-LHC) is scheduled to start operation in 2029. By colliding protons with an instantaneous luminosity that is five times higher than what is achieved at the LHC, the HL-LHC is expected to deliver data corresponding to an integrated luminosity of 3000 fb$^{-1}$ by the end of the 2030s, which is a factor of 20 more than what has been collected so far. Despite the highly challenging experimental environment, such an increased dataset – collected with upgraded detectors – has an immense physics potential: it will give access to the rarest phenomena, and will be critical to reduce systematic uncertainties or bypass their limitations with new analyses, leading to measurements of unprecedented precision. It will allow us to achieve a sensitivity to sectors of Beyond-the-Standard-Model (BSM) phenomena that are beyond the reach of current analyses, and will ultimately help us to get closer to answering fundamental questions of particle physics.

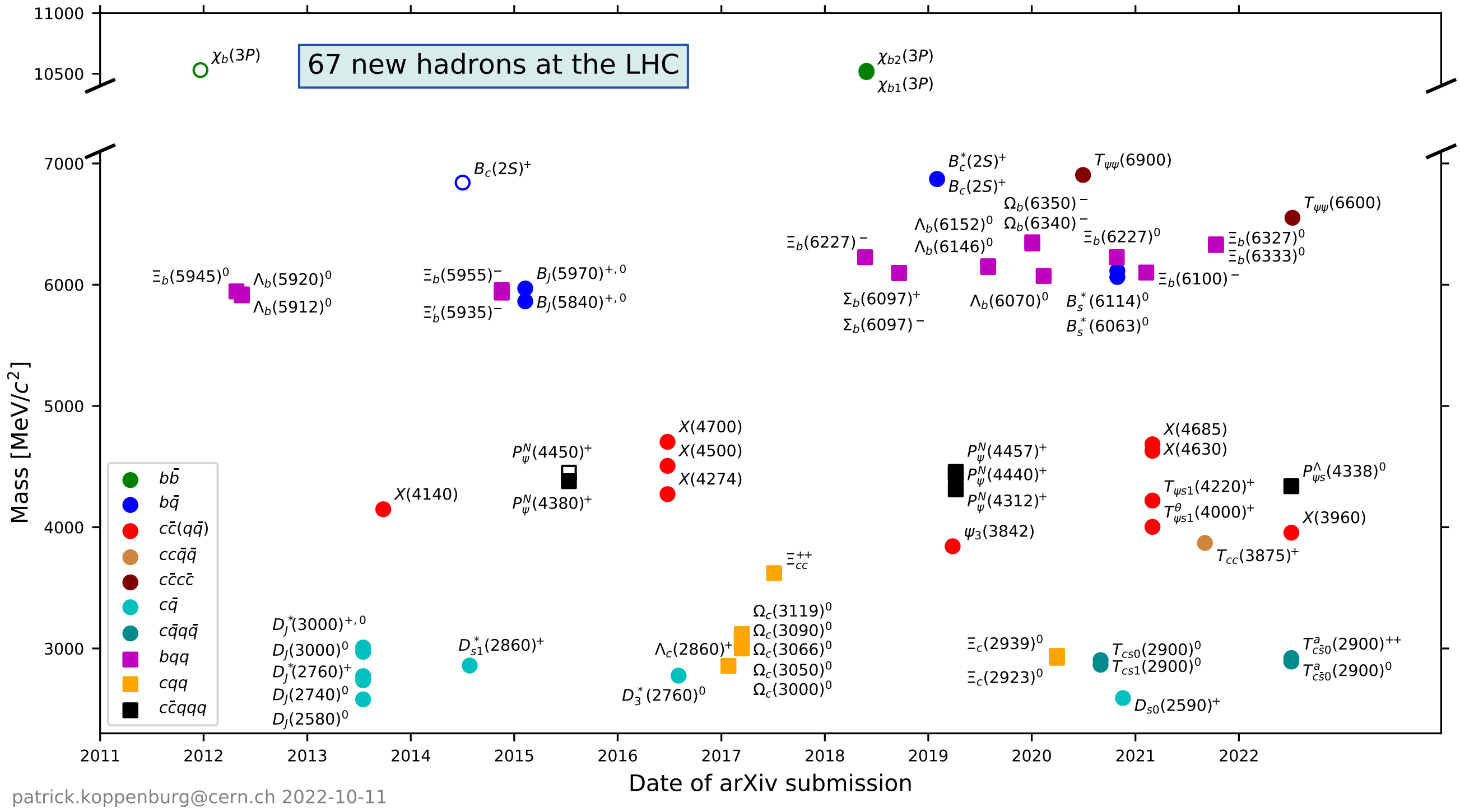

**Fig. 14.8.2** Discoveries of hadrons at the LHC, by year of arXiv submission [4759]. Only states observed with significance larger than $5\sigma$ are included.

### 14.9.2 Higgs properties

The study of Higgs boson ($H$) properties is central in the HL-LHC physics programme. Since its discovery in 2012, analyses related to the Higgs boson have significantly expanded, and have now turned into a vast campaign of precision measurements, with fundamental opportunities to indirectly constrain the Higgs boson width and to access its trilinear coupling. Small deviations from the SM can be described in a consistent framework by using effective field theory (EFT).

The main measurements of Higgs boson properties are based on five production modes (gluon fusion $ggF$, vector boson fusion VBF, associated production with a $W$ or $Z$ vector boson or with a top-quark pair) and five decay modes: $H \to \gamma\gamma$, $ZZ$, $WW$, $\tau\tau$, $b\bar{b}$. The $H \to \mu\mu$ and $Z\gamma$ channels should become visible in the future. The rate measurements in the production and decay channels mentioned above yield measurements of the Higgs boson couplings in the so-called "$\kappa$-framework" [4767]. The latter introduces a set of scaling factors $\kappa_i$ that linearly modify the couplings of the Higgs boson to the corresponding SM elementary particles, including the effective couplings to gluons and photons. The projected uncertainties, combining ATLAS and CMS, are summarized in Figure 14.9.1. Note that theory uncertainties are assumed to be halved with respect to their current values. Except for rare decays, the overall uncertainties will be dominated by the theoretical systematics, with a precision close to the percent level. These coupling measurements assume the absence of sizable additional contributions to $\Gamma_H$. As observed in Ref. [4769], the signal-background interference in the production of $Z$-boson pairs is sensitive to $\Gamma_H$. Measuring the off-shell four-lepton final states and assuming that the Higgs boson couplings can be extrapolated in the off-shell region from their SM values, the HL-LHC will extract $\Gamma_H$ using this indirect measurement with a 20% precision at 68% CL [4768].

The production of Higgs boson pairs is a central process to access the Higgs trilinear coupling. The Run 2 experience in searches for Higgs boson pair production led to a reassessment of the HL-LHC sensitivity, including additional channels that were not considered in previous projections. ATLAS and CMS anticipate a sensitivity to the $HH$ signal of approximately $3\sigma$ per experiment, leading to a combined observation sensitivity of $4\sigma$. These analyses lead to the combined likelihood profile as a function of $\kappa_\lambda$ shown in Figure 14.9.2.

It should be noted that the upper limit on the signal strength for $HH$ production can reach the SM expecta-tion already for Run 3 by combining ATLAS and CMS results if the improvements in the reconstruction and analysis techniques continue at the same pace (see e.g. Elisabeth Brost, talk at Higgs10 meeting, CERN, July 2022).

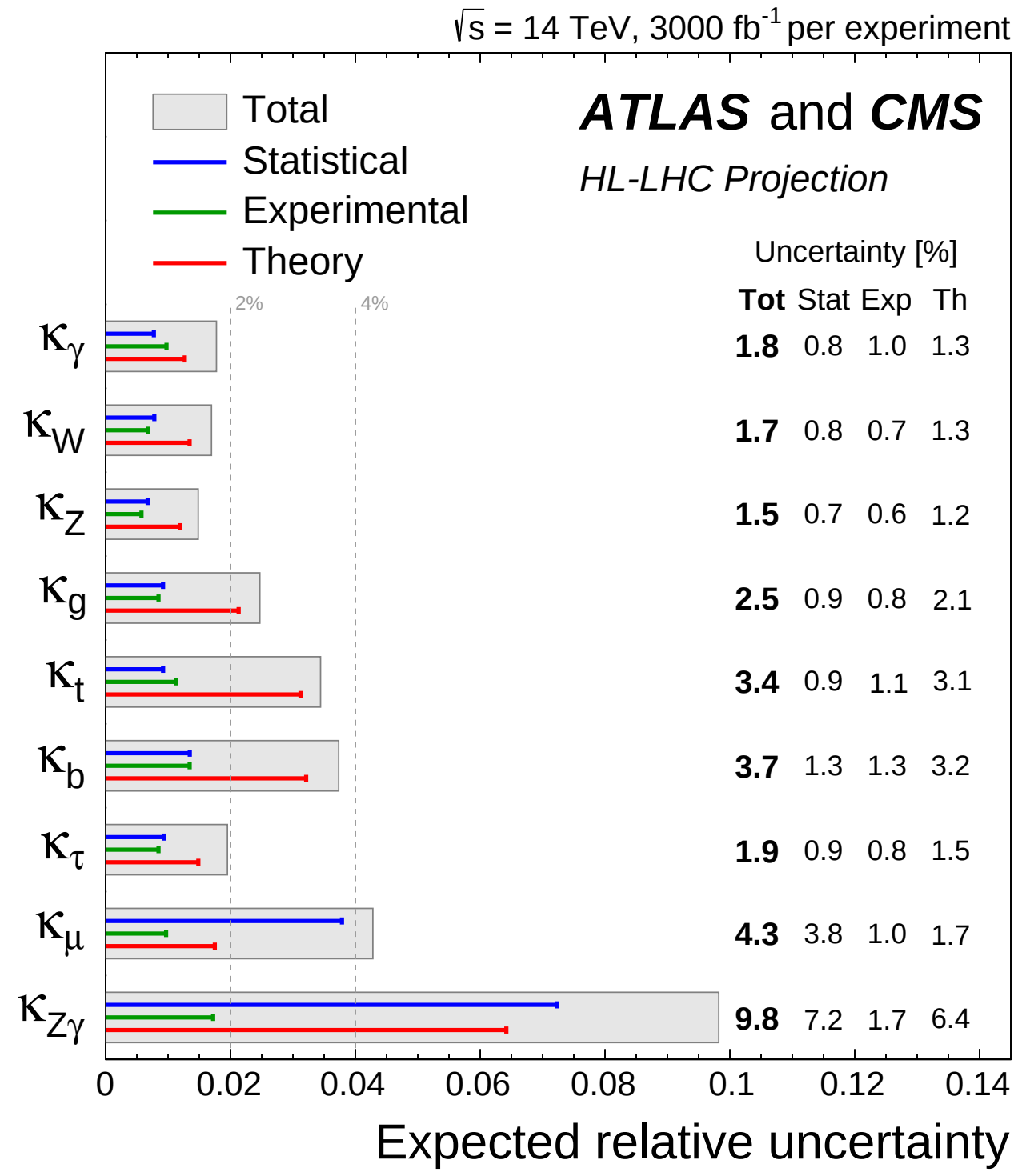

**Fig. 14.9.1** Projected uncertainties for the scaling parameters $\kappa_i$, combining ATLAS and CMS: total (grey box), statistical (blue), experimental (green) and theory (red) uncertainties. From Ref. [4768].

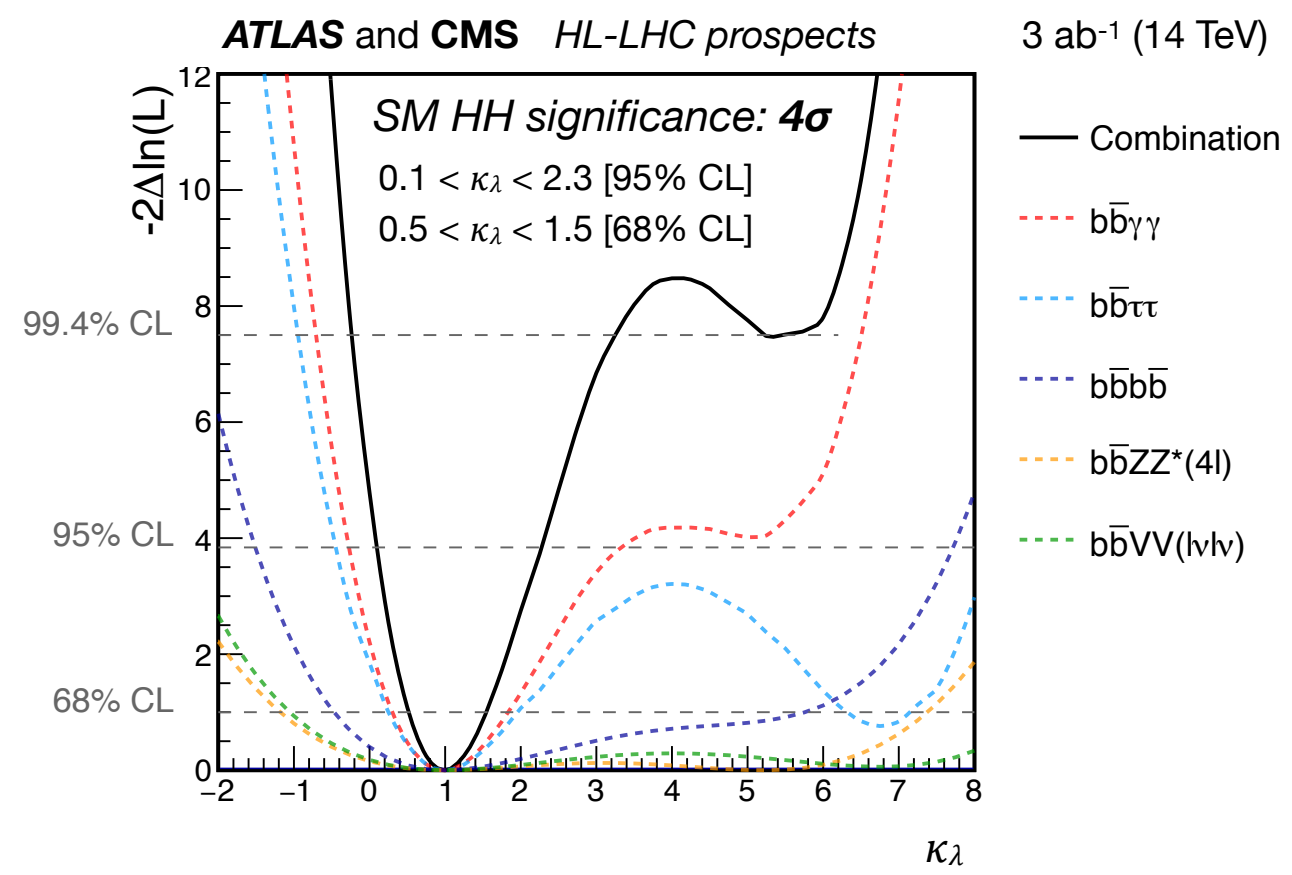

**Fig. 14.9.2** Projected combined HL-LHC sensitivity to the Higgs boson trilinear coupling expressed in terms of $\kappa_\lambda$, from direct search channels. From Ref. [4768].

### 14.9.3 Multiple gauge bosons

The study of multiple gauge boson production is of crucial importance to test the EW gauge symmetry, since it can signal the presence of anomalous gauge couplings [4770]. At HL-LHC, evidence for the production of three gauge bosons can be obtained at the $3\sigma$ level in the $WWZ$ and $WZZ$ channels and at the $5\sigma$ level in the $WWW$ channel considering the fully leptonic decay modes [3874]. Following the first observation of vector-boson scattering (VBS) at the LHC, the HL-LHC is expected to provide a more complete picture of these processes, including the option to measure polarized components, thanks to the higher statistics and improved detectors.

### 14.9.4 New-physics searches

The HL-LHC will allow us to test BSM phenomena that are beyond the reach of current analyses [1281]. Many BSM models predict the existence of new particles, which can be searched for at HL-LHC, exploiting the much larger statistics and detector upgrades.

In the case of supersymmetry (SUSY), the extension of the kinematic reach is reflected in improved sensitivity to sleptons, gluinos and squarks. In the strong SUSY sector, HL-LHC will probe gluino masses up to 3.2 TeV, in R-parity conserving scenarios and under several possible assumptions on the gluino prompt decay mode. This significantly extends the reach of LHC Run 2. In the context of R-parity conserving models, scenarios in which the mass difference between the produced superpartners and the lightest superpartner (LSP) they decay into is small (usually called *compressed* SUSY) are the most difficult to study experimentally, and have been barely covered at the LHC till now. At the HL-LHC, these scenarios will be studied by using mono-jet and mono-photon signatures as well as VBF production.

An interesting scenario in the search for dark matter is the one containing a dark photon that couples very weakly to charged particles. Prospects for an inclusive search for dark photons decaying into muon or electron pairs indicate that the HL-LHC could cover a large fraction of the theoretically favored parameter space.

The flavor anomalies in B-decays suggest the possible presence of new states, such as $Z'$ or leptoquarks (LQ), coupling to second and/or third generation SM fermions. The HL-LHC will be able to cover a significant portion of the parameter space of these models, with an exclusion reach up to 4 TeV for the $Z'$. Pair produced scalar LQs coupling to $\mu$ ($\tau$) and $b$-quarks, on the other hand, can be excluded up to masses of 2.5 (1.5) TeV, depending on the assumptions on the couplings.

### 14.9.5 QCD challenges

Already now the LHC experiments have reached a very high level of sophistication in the reconstruction of collision events, thereby making precise measurements possible despite the complex environment and substantial pileup.

Even though significant progress has been made in QCD and electro-weak (EW) calculations for hard processes in the last few years (see Section 11.1), further progress will be needed to avoid theory uncertainties to become the limiting factor in interpreting a wide range of HL-LHC data. For example, in the case of Higgs boson couplings, the projections of Figure 14.9.1 show that theory uncertainties will be a limiting factor even if reduced by a factor of two with respect to their current values. Progress on the theory side is therefore needed and it is indeed expected in the following areas:

1. **Parton Distribution functions:** All hard scattering reactions at the LHC are eventually initiated by a partonic collision. The parton scattering rate, which is computed perturbatively, is weighted by the PDFs, whose knowledge is therefore required to extract fundamental couplings from cross section measurements or from kinematic distributions. PDFs are also a fundamental input to predict the tails of the distributions of SM processes at high $Q^2$ or high $p_T$, which in turn allow us to probe possible new physics effects. The current knowledge of PDFs will be improved at HL-LHC by accurate measurements of SM processes with jets, vector bosons and top quarks. LHCb data also have the potential to further constrain the PDFs. At scales $Q > 100$ GeV the HL-LHC data can reduce PDF uncertainties by a factor between 2 and 4, depending on the process and on the assumptions on systematic uncertainties [3874].
2. **Benchmark processes at high accuracy:** The experimental precision for many benchmark $2 \to 1$ and $2 \to 2$ processes (the most significant example being Drell-Yan lepton pair production) is likely to approach the 1% level, over a substantial range of phase space. Perturbative QCD predictions at next-to-next-leading order (NNLO) normally do not reach 1% precision, and N$^3$LO accuracy might be needed for a range of $2 \to 1$ and $2 \to 2$ processes. For example, N$^3$LO predictions for Higgs and vector boson production are already available [1909, 3404, 3405, 3411, 3412, 3495, 4771] and are crucial to control perturbative uncertainties. The improved theoretical control of simple processes will in turn improve our knowledge of PDFs, allowing N$^3$LO PDF fits, with impact on the whole range of LHC processes, and will also increase the sensitivity to BSM effects manifesting themselves as small deviations from SM predictions. A first approximate N$^3$LO PDF fit has been recently presented in Ref. [3044].
3. **$2 \to 3$ processes at few-percent accuracy:** There are a number of crucially important signal and background processes that involve a $2 \to 3$ scattering structure at parton level; these are at the current frontier of NNLO calculations. While calculations of 3-jet production rates became recently available [3369], processes like $t\bar{t}H$, $t\bar{t}V$, $H + 2\,$jets are only known up to NLO and would benefit from the extension to NNLO[121]. The $t\bar{t}H$ cross section, e.g., is now measured with roughly 15% statistical precision and is expected to be known with a statistical precision of $\sim 2\%$ at the end of the HL-LHC. Without NNLO QCD and NLO EW accurate calculations for signal and backgrounds, this experimental precision cannot be matched on the theory side, thereby limiting the exploitation of the results for physics studies.
   A significant amount of work is currently being devoted to break the $2 \to 3$ barrier for two-loop amplitudes involving massive particles. At the same time, an effort is ongoing to improve available methods to isolate and cancel infrared singularities (see Section 11.1 for more details). In the HL-LHC era the complete availability of combined NNLO precision in the strong coupling and NLO precision in the EW coupling would be desirable.
4. **Accuracy at high $p_T$:** Current measurements have only explored a limited range of the available phase space. NNLO accurate differential cross sections pave the way to more detailed data/theory comparisons in less populated phase-space regions where new physics effects could be hidden. An important example is provided by high-$p_T$ Higgs production. The ATLAS and CMS collaborations anticipate an $\mathcal{O}(10\%)$ precision in the Higgs boson production rate for $p_T \geq 350$ GeV at the end of the HL phase of the LHC [4768].
   The recent computations of $2 \to 2$ amplitudes mediated by massive quarks [3348, 4773], combined with NNLO calculations in the heavy-top limit [3400, 3401, 4774–4776] offer a comparable precision in the SM prediction, and will therefore allow us to disentangle possible new physics effects in this region.

[121] First NNLO results for inclusive $t\bar{t}H$ production have been recently presented in Ref. [4772].

5. **Bottlenecks in NLO multi-particle simulations:** The full deployment of NLO precision through automated MC frameworks in the huge range of HL-LHC analyses raises important technical challenges. Establishing the predictivity of MC tools at precision levels of order 10% – as well as their correct usage within the experiments – will require quantitatively and qualitatively unprecedented validation work. Already now, the accuracy at which event samples for $2 \to 4$ processes can be calculated at NLO is limited by dramatic efficiency bottlenecks related to the poor convergence of the phase-space integration and by various other technical aspects. The HL-LHC era will require efficiency improvements by an order of magnitude. This can only be achieved through a significant step forward in the optimization of event generators and new techniques in the calculation of amplitudes.
6. **Theory systematics:** The appropriate estimate of theory uncertainties in the presence of experimental cuts or in the context of sophisticated multi-variate analyses is a challanging problem. A typical example is provided by $t\bar{t}H$ analyses in the $H \to b\bar{b}$ decay mode. The sensitivity is presently limited by theory uncertainties in the $t\bar{t}b\bar{b}$ QCD background. In this kind of analyses, MC predictions for the large QCD background are constrained by data through a profile likelihood fit of several kinematic distributions in different event categories. In this context, theoretical predictions for the correlations across different categories and kinematic regions play a key role. All related uncertainties, e.g. at the level of NLO matrix elements, parton showers and NLO matching, need to be properly identified and modelled. This task is further complicated by the presence of multiple scales, which may require resummations. This type of problem is characteristic for a broad range of LHC analyses; its solution will require a joint effort between theorists and experimentalists.
7. **Non-perturbative effects:** While the perturbative computations follow a systematic approach based on perturbation theory and factorization, our understanding of non-perturbative effects is still quite rough. With the increasing accuracy of perturbative calculations, which in some cases now reach the N$^3$LO level, non-perturbative effects might become relevant, also in inclusive observables. Moreover, in the case of measurements dealing with hadronic final states, the poor control of the hadronization stage limits the precision that can be attained, thereby potentially affecting the extraction of important parameters, such as the top quark mass.
8. **Resummation and parton showers:** For key observables depending on disparate scales, advances in the all-order resummation of large logarithmic corrections will be crucial. Such advances require to increase the logarithmic accuracy of the resummed calculations, but also the extension to multiply-differential resummations, the inclusion of power suppressed effects, as well as the understanding of sub-leading and super-leading structures (see Section 11.2). In parallel, work towards the extension of the logarithmic accuracy of parton showers will be essential (see Section 11.3).
9. **BSM effects:**
The great success of the SM in describing all phenomena observed at the LHC suggests that the key to a potential discovery of new physics is precision. Precision measurements indeed provide an important tool to search for BSM physics associated to mass scales beyond direct reach of the LHC. EFT frameworks, where the SM Lagrangian is supplemented with additional operators built from SM fields, consistent with gauge symmetries and based on a well-defined counting scheme, allow us to systematically parameterize BSM effects and their modifications to SM processes. These operators can either modify existing SM couplings, or generate new couplings. In the case of BSM operators that mix with the SM ones, if $r$ is the relative precision on a given physical observable, the new physics mass scale $\Lambda$ that can be probed with this observable will scale as $1/\sqrt{r}$ in the generic case.

### 14.9.6 Outlook

While the HL-LHC offers great opportunities due to the enormous reduction of statistical uncertainties compared to previous LHC runs, some measurements remain difficult and will leave questions that could be addressed more straightforwardly with the great precision that future lepton colliders, such as the ILC [4777], CEPC [4778], FCC-ee [4779] or CLIC [4780] could achieve, or with the impressive energy reach and statistics a future hadron collider (FCC-hh [4781]) could provide. For example, the trilinear Higgs boson self-coupling – a parameter which is crucial to probe the mechanism of EW symmetry breaking – is expected to be constrained with an uncertainty of 50% after the HL-LHC runs, as shown in Figure 14.9.2, while a combination of FCC-ee and HL-LHC results could reach a precision of about 30%, and a future hadron collider operating at a center-of-mass energy of 100 TeV could achieve a clear measurement with a precision of about 5% [4782]. Similar arguments hold for other quantities that are important to probe the SM at an unprecedented level of precision, such as the $W$-boson mass, the couplings of the Higgs boson to light fermions, or the line-shape and therefore the total width of the Higgs boson [4783].

Apart from the potential of future lepton colliders to find hints for new phenomena through a scrutiny of the Higgs sector and other SM particles and interactions, they offer new possibilities to search for physics beyond the SM, including the production of dark matter particles at colliders, taking advantage from the fact that the final state can be fully reconstructed. Direct searches for additional gauge bosons, such as $Z'$, or for heavy neutral leptons, could also shed light on the flavor anomalies, thereby providing complementary information to experiments at lower energies, to give just some examples. Finally, FCC-hh energies would give access to a huge, so far uncharted energy range and parton kinematic region, offering the possibility of a direct production of so far unknown particles.

This review shows how multi-faceted QCD is, as well as its embedding in the SM. The quest to answer fundamental questions about matter, its interactions and, on a large scale, the origin and evolution of the Universe, needs to be addressed by a diverse experimental programme, and high-energy colliders are just one part of it. However, they offer the unique possibility to produce particles that are simply inaccessible by other means in a controlled way. Therefore, high energy colliders form an important building block in a coordinated global effort towards a more complete theory of fundamental interactions, where the Standard Model might be embedded as a sub-part, as much as QCD today is embedded in the Standard Model.

# Postscript

This volume tries to give a comprehensive and balanced view of the progress in the development of QCD since its inception. To do so presented many challenges: are all important topics adequately covered, are all opposing views represented, and is all important work included? As the volume was being developed, we often added new material that our conveners suggested (see the title page for the names of the conveners). This process was greatly aided by the use of Overleaf, which allowed all of the contributors to follow developments. In a real sense, this volume is the work of many people who often worked together to shape the final result even though they were under the intense pressures of their very busy professional schedules. We thank all of them; this volume is truly a collective effort. Still, we leave it to you to judge if we succeeded.

Another goal was to produce a coherent discussion useful for new Ph.D's and postdocs. Here we know our efforts were only partially successful. There was never enough time to fully coordinate all of the contributions, and we are sure you will find many places where more cross references would have been helpful. Again, it is up to our intended audience to judge the extent to which we were successful.

Finally, as we reflect back on this effort, we realize that the timing of this volume was more urgent that we realized at the start. Fifty years is a long time, and many who have made important contributions to the subject are no longer alive. This was never more apparent than when we learned of Harald Fritzsch's untimely death on August 16. We were delighted when he accepted our invitation to write his contribution, guided by his early and helpful suggestions, and surprised at how quickly he completed his work. His contribution was among those that were completed very early and could serve as examples for other contributors.

Both of us learned a lot about QCD as we edited the contributions and participated in the discussions. This was a great pleasure, for which we thank all of the contributors.

# Acknowledgements

The help of many people is acknowledged: Chiara Mariotti thanks A.C. Marini and A. Mecca; Per Grafstrom thanks Peter Jenni, Valery Khoze and Mikhail Ryskin who were kind enough to read his text and given him many useful comments and suggestions to improve his contribution; Andreas Schafer thanks the University of the Basque Country, Bilbao for hospitality; Mikhail Shifman is grateful to Alexander Khodjamirian, Alexander Lenz and Blaženka Melić for very useful discussions and comments; S. Kumano thanks A. Dote, M. Kitazawa, K. Ozawa, S. Sawada, T. Takahashi, M. Takizawa, and S. Yokkaichi for suggestions on the J-PARC experiments; Stanley J. Brodsky, Guy F. de Téramond and Hans Günter Dosch are grateful to Tianbo Liu, Raza Sabbir Sufian and Alexandre Deur, who have greatly contributed to the new applications of the holographic ideas reviewed in this volume. Eberhard Klempt and Ulrike Thoma thank Andrey V. Sarantsev for many years of collaboration on meson and baryon spectroscopy. Peter Braun-Munzinger, Anar Rustamov, Johanna Stachel acknowledge continued and long-term collaboration with Anton Andronic and Krzysztof Redlich on many of the topics described in their contribution; Daniel Britzger, Klaus Rabbertz and Markus Wobish would like to thank Monica Dunford, Karl Jakobs, and Jürgen Scheins for their careful reading of the manuscript; Kostas Orginos thanks Carl Carlson for many discussions on the charge radius and Anatoly Radyushkin for many discussions on aspects of hadron structure. Volker Burkert expresses his gratitude to Inna Aznauryan for many years of collaboration on the subject of electroproduction of nucleon resonances, which has lead to many of the results discussed in his contribution. He also wishes to acknowledge Victor Mokeev for numerous discussions and collaboration on many aspects of resonance electroproduction. Finally, Volker Burkert thanks Francois-Xavier Girod for contributing Fig. 9.3.10 to his section.

The editors wish to thank Brad Sawatzky for setting up the original Overleaf site and for many hours of invaluable technical help, essential to the production of this volume, and Nora Brambilla, Karl Jakobs, and J. Peter LePage for several long discussions at the early stages in the preparation of this volume that influenced its structure and content.

The editors express their gratitude to Dieter Haidt, the Review Editor of the European Physical Journal C, for continuous encouragement and valuable suggestions. The authors received funding from

*Australia:*
National Computational Infrastructure (NCI) and the Australian Research Council through Grants No. DP190102215 and DP210103706 (*D. Leinweber*).

*PR of China:*
National Natural Science Foundation of China (NSFC) under Contracts Nos. 11935018, 11875054 (*H-B. Li*).

*France*
CNRS and ANR (*B. Malaescu*).

*Germany:*
The Bundesministerium für Bildung und Forschung (BMBF) (*M. Dunford, K. Jakobs, S. Neubert, K. Rabbertz*);
Gesellschaft für Schwerionenforschung Gmb (GSI), Darmstadt, Germany (*J. Messchendorp*);
The Deutsche Forschungsgemeinschaft (DFG, German Research Foundation) through the funds provided to the Sino-German Collaborative Research Center TRR110 “Symmetries and the Emergence of Structure in QCD” (DFG Project-ID 196253076 - TRR 110) (*N. Brambilla, U. Thoma, A. Vairo*);
The DFG Collaborative Research Centre “SFB 1225 (ISOQUANT)” (*P. Braun-Munzinger, A. Rustamov, J. Stachel*);
The DFG Collaborative Research Centre 396021762 - TRR 257, ”Particle Physics Phenomenology after the Higgs Discovery” (*G. Heinrich*);
The DFG Collaborative Research Centre 315477589-TRR 211, ”Strong interaction matter under extreme conditions” (*F. Karsch*);
The DFG Emmy-Nöther project NE2185/1-1: ”Spektroskopie exotischer Baryonen mit LHCb” (*S. Neubert*)
DFG individual grant, Project Number 455635585 (*A. Denig*);
The Excellence Cluster ORIGINS (www.origins-cluster.de), funded by the DFG, German Research Foundation, Excellence Strategy, EXC-2094, 390783311 (*N. Brambilla, A. J. Buras, A. Vairo*);
The Helmholtz Forschungsakademie Hessen für FAIR (HFHF) (*F. Nerling, J. Stroth*);
The Ministry of Culture and Science of the State of North Rhine-Westphalia (MKW NRW, Germany), project ”NRW-FAIR” (*S. Neubert, U. Thoma*);
European Research Council (ERC) under the European Union’s Horizon 2020 research and innovation program through grant agreements 771971-SIMDAMA (*H. Meyer*);
European Union’s Horizon 2020 research and innovation program under grant agreement STRONG-2020 No. 824093 (*U. Thoma*).

*Italy:*
Italian Ministry of Research (MIUR) under grant PRIN 20172LNEEZ (*P. Gambino and S. Marzani*)

*Japan:*
Japan Society for the Promotion of Science (JSPS) Grants-in-Aid for Scientific Research (KAKENHI):
Grant Number 19K03830 (*S. Kumano*);
Grant Number 18H05226(*T. Iijima*).

*Spain*
MINECO through the “Ramón y Cajal” program RYC-2017-21870, the “Unit of Excellence María de Maeztu 2020-2023” award to the Institute of Cosmos Sciences (CEX2019-000918-M) and from the grants PID2019-105614GB-C21 and 2017-SGR-929 (*J. Varto*);
The Spanish Ministerio de Ciencia e Innovación grant PID2019- 106080GB-C21 and the European Union’s Horizon 2020 research and innovation program under grant agreement No 824093 (STRONG2020) (*J. R. Peláez*);
European Research Council project ERC-2018-ADG-835105 YoctoLHC, by Maria de Maetzu excellence program under project CEX2020-001035-M, by Spanish Research State Agency under project PID2020-119632GB-I00, and by Xunta de Galicia (Centro singular de investigación de Galicia accreditation 2019-2022), by European Union ERDF (*M. Escobedo*);
The Spanish Ministerio de Ciencia e Innovación grant PID2021-122134NB-C21 and by the Generalitat Valenciana under grant CIPROM/2021/073 and by CSIC under grant LINKB20065 (*M. Vos*).

*Sweden:*
The Swedish Research Council, contract number 2016-05996 (*T. Sjöstrand*).

*Switzerland:*
The European Research Council (ERC) under the European Union's Horizon 2020 research and innovation programme (Grant agreement No. 948254) and from the Swiss National Science Foundation (SNSF) under Eccellenza grant number PCEFP2-194658 (*S. Schramm*).

*UK*
Science and Technology Facilities Council (STCF): Ernest Rutherford Fellowship grants
ST/T000945/1 and ST/P000746/1 (*C. Davies*);
ST/P000630/1 (*L. Del Debbio*);
ST/V003941/1 (*D. Van Dyk*);
Royal Society Wolfson fellowship and STFC (*F. Krauss*).

*USA:*
The US Department of Energy, Office of Science, Office of Nuclear Physics (contract DE-AC05-06 OR23177- under which Jefferson Science Associates operates, LLC, operates Jefferson Lab) (*V. Burkert, J. Dudek, F. Gross, W. Melnitchouk, P. Rossi*);
DE-AC02-76SF00515 (*S. J. Brodsky*);
Early Career Award under Grant No. DE-SC0020405 (*M. Constantinou*);
DE-FG02-92ER40735 *V. Crede*);
DE-SC0018416 (*J. Dudek*);
DESC0018223 (*P. Maris and J. Vary*)
DE-SC0023495 (*P. Maris and J. Vary*)
DE-FG02-87ER40315 (*C. Meyer*);
DE-FG02-04ER41302 (*K. Orginos*);
DE-SC0021027 (*S. Pastore*);
DE-SC0021200 (*A. Puckett*);
DE-SC0019647 (*M. Schindler*);
DE-AC02-05CH11231 (*A. Schafer, F. Yuan*);
DE-SC0011842 (*M. Shifman*);
DE-SC0013470 (*M. Strickland*)
DE-SC0011090 and Simons Foundation Investigator grant 327942 (*I. Stewart*);
DE-FG02-87ER40371 (*J. Vary*)
DE-FG02-95ER40896, (*S. L. Wu*);
University of Wisconsin through the Wisconsin Alumni Research Foundation and the Vilas Foundation (*S. L. Wu*);
The National Science Foundation under grants PHY-1915093 and PHY-2210533 (*G. Sterman*);
NSF grant PHY20-13064 (*C. DeTar*).

# References

[1] H. Leutwyler. "On the history of the strong interaction". In: *Mod. Phys. Lett. A* 29 (2014). Ed. by Antonino Zichichi, p. 1430023.
[2] J. Chadwick. "The Existence of a Neutron". In: *Proc. Roy. Soc. Lond. A* 136.830 (1932), pp. 692–708.
[3] W. Heisenberg. "On the structure of atomic nuclei". In: *Z. Phys.* 77 (1932), pp. 1–11.
[4] Hideki Yukawa. "On the Interaction of Elementary Particles I". In: *Proc. Phys. Math. Soc. Jap.* 17 (1935), pp. 48–57.
[5] E. C. G. Stueckelberg. "Interaction energy in electrodynamics and in the field theory of nuclear forces". In: *Helv. Phys. Acta* 11 (1938), pp. 225–244, 299–328.
[6] Herbert Pietschmann. "On the Early History of Current Algebra". In: *Eur. Phys. J. H* 36 (2011), pp. 75–84.
[7] G. Veneziano. "Construction of a crossing - symmetric, Regge behaved amplitude for linearly rising trajectories". In: *Nuovo Cim. A* 57 (1968), pp. 190–197.
[8] M. Gell-Mann. "Isotopic Spin and New Unstable Particles". In: *Phys. Rev.* 92 (1953), pp. 833–834.
[9] T. Nakano and K. Nishijima. "Charge Independence for V-particles". In: *Prog. Theor. Phys.* 10 (1953), pp. 581–582.
[10] M. L. Goldberger and S. B. Treiman. "Decay of the $\pi$ meson". In: *Phys. Rev.* 110 (1958), pp. 1178–1184.
[11] Yoichiro Nambu. "Axial vector current conservation in weak interactions". In: *Phys. Rev. Lett.* 4 (1960). Ed. by T. Eguchi, pp. 380–382.
[12] Jeffrey Goldstone, Abdus Salam, and Steven Weinberg. "Broken Symmetries". In: *Phys. Rev.* 127 (1962), pp. 965–970.
[13] Yuval Ne'eman. "Derivation of strong interactions from a gauge invariance". In: *Nucl. Phys.* 26 (1961). Ed. by R. Ruffini and Y. Verbin, pp. 222–229.
[14] Murray Gell-Mann. "Symmetries of baryons and mesons". In: *Phys. Rev.* 125 (1962), pp. 1067–1084.
[15] Susumu Okubo. "Note on unitary symmetry in strong interactions". In: *Prog. Theor. Phys.* 27 (1962), pp. 949–966.
[16] V. E. Barnes et al. "Observation of a Hyperon with Strangeness Minus Three". In: *Phys. Rev. Lett.* 12 (1964), pp. 204–206.

[17] Murray Gell-Mann. "A Schematic Model of Baryons and Mesons". In: *Phys. Lett.* 8 (1964), pp. 214–215.

[18] G. Zweig. "An SU(3) model for strong interaction symmetry and its breaking. Version 1". In: ed. by D. B. Lichtenberg and Simon Peter Rosen. Developments in the Quark Theory of Hadron. Vol. 1. 1964 - 1978, pp 22–101, January 1964, CERN-TH-401, see also: Version 2, CERN-TH-412 and NP-14146, PRINT-64-170.

[19] Murray Gell-Mann. "The Symmetry group of vector and axial vector currents". In: *Physics Physique Fizika* 1 (1964), pp. 63–75.

[20] Stephen L. Adler. "Consistency conditions on the strong interactions implied by a partially conserved axial vector current". In: *Phys. Rev.* 137 (1965), B1022.

[21] William I. Weisberger. "Renormalization of the Weak Axial Vector Coupling Constant". In: *Phys. Rev. Lett.* 14 (1965), pp. 1047–1051.

[22] Steven Weinberg. "Pion scattering lengths". In: *Phys. Rev. Lett.* 17 (1966), pp. 616–621.

[23] J. D. Bjorken. "Asymptotic Sum Rules at Infinite Momentum". In: *Phys. Rev.* 179 (1969), pp. 1547–1553.

[24] Jerome Isaac Friedman, H. W. Kendall, and R. E. Taylor. In: (). For an account of the experimental developments, see their *Nobel lectures in physics 1990*, SLAC-REPRINT-1991-019.

[25] Kenneth G. Wilson. "Nonlagrangian models of current algebra". In: *Phys. Rev.* 179 (1969), 1499
–1512.

[26] O. W. Greenberg. "Spin and Unitary Spin Independence in a Paraquark Model of Baryons and Mesons". In: *Phys. Rev. Lett.* 13 (1964), pp. 598–602.

[27] N. N. Bogolyubov, B. V. Struminsky, and Tavkhelidze. "On composite models in the theory of elementary particles. " In: (1965).

[28] M. Y. Han and Yoichiro Nambu. "Three Triplet Model with Double SU(3) Symmetry". In: *Phys. Rev.* 139 (1965), B1006–B1010.

[29] Y. Miyamoto. "Three Kinds of Triplet Mode". In: *Prog. Theor. Phys. Suppl., Extra Number* (1965), pp. 187.

[30] Harald Fritzsch and Murray Gell-Mann. "Current algebra: Quarks and what else?" In: *Proceedings of the XIV International Conference on High Energy Physics, Chicago 1972, Volume 2.* 1972, pp. 135–165.

[31] M. Gell-Mann. "Quarks". In: *Acta Phys. Austriaca Suppl.* 9 (1972). Ed. by Harald Fritzsch and Murray Gell-Mann, pp. 733–761.

[32] V. Fock. "On the invariant form of the wave equation and the equations of motion for a charged point mass. (In German and English)". In: *Z. Phys.* 39 (1926). Ed. by J. C. Taylor. [Surveys H. E. Phys. **5** (1986) 245]. For a discussion of the significance of this paper, see, L. B. Okun, "V A Fock and gauge symmetry," Phys. Usp. **53** (2010) 835 [Usp. Fiz. Nauk **180** (2010) 871]., pp. 226–232.

[33] Th. Kaluza. "Zum Unitätsproblem der Physik". In: *Sitzungsber. Preuss. Akad. Wiss. Berlin (Math. Phys. )* 1921 (1921), pp. 966–972.

[34] Oskar Klein. "Quantum Theory and Five-Dimensional Theory of Relativity. (In German and English)". In: *Z. Phys.* 37 (1926). Ed. by J. C. Taylor, pp. 895–906.

[35] Norbert Straumann. "Wolfgang Pauli and Modern Physics". In: *Space Sci. Rev.* 148 (2009), pp. 25–36.

[36] Chen-Ning Yang and Robert L. Mills. "Conservation of Isotopic Spin and Isotopic Gauge Invariance". In: *Phys. Rev.* 96 (1954). Ed. by Jong-Ping Hsu and D. Fine, pp. 191–195.

[37] R. Shaw. "The problem of particle types and other contributions to the theory of elementary particles". In: *Cambridge PhD thesis (unpublished)* (1955).

[38] Peter W. Higgs. "Broken symmetries, massless particles and gauge fields". In: *Phys. Lett.* 12 (1964), pp. 132.

[39] F. Englert and R. Brout. "Broken Symmetry and the Mass of Gauge Vector Mesons". In: *Phys. Rev. Lett.* 13 (1964). Ed. by J. C. Taylor, pp. 321–323.

[40] G. S. Guralnik, C. R. Hagen, and T. W. B. Kibble. "Global Conservation Laws and Massless Particles". In: *Phys. Rev. Lett.* 13 (1964). Ed. by J. C. Taylor, pp. 585–587.

[41] S. L. Glashow. "Partial Symmetries of Weak Interactions". In: *Nucl. Phys.* 22 (1961), pp. 579–588.

[42] Steven Weinberg. "A Model of Leptons". In: *Phys. Rev. Lett.* 19 (1967), pp. 1264–1266.

[43] Abdus Salam. "Weak and Electromagnetic Interactions". In: *Conf. Proc. C* 680519 (1968), pp. 367–377.

[44] V. S. Vanyashin and M. V. Terentev. "The Vacuum Polarization of a Charged Vector Field". In: *Zh. Eksp. Teor. Fiz.* 48.2 (1965), pp. 565–573.

[45] I. B. Khriplovich. "Green's functions in theories with non-abelian gauge group." In: *Sov. J. Nucl. Phys.* 10 (1969), pp. 235–242.

[46] Gerard 't Hooft. "Renormalizable Lagrangians for Massive Yang-Mills Fields". In: *Nucl. Phys. B* 35 (1971). Ed. by J. C. Taylor, pp. 167–188.

[47] Gerard 't Hooft. "Renormalization of Massless Yang-Mills Fields". In: *Nucl. Phys. B* 33 (1971), pp. 173–199.

[48] David J. Gross and Frank Wilczek. "Ultraviolet Behavior of Nonabelian Gauge Theories". In: *Phys. Rev. Lett.* 30 (1973). Ed. by J. C. Taylor, pp. 1343–1346.

[49] H. David Politzer. "Reliable Perturbative Results for Strong Interactions?" In: *Phys. Rev. Lett.* 30 (1973). Ed. by J. C. Taylor, pp. 1346–1349.

[50] Harald Fritzsch, Murray Gell-Mann, and Heinrich Leutwyler. "Advantages of the Color Octet Gluon Picture". In: *Phys. Lett. B* 47 (1973), pp. 365–368.

[51] Jogesh C. Pati and Abdus Salam. "Unified Lepton-Hadron Symmetry and a Gauge Theory of the Basic Interactions". In: *Phys. Rev. D* 8 (1973), pp. 1240–1251.

[52] Steven Weinberg. "Nonabelian Gauge Theories of the Strong Interactions". In: *Phys. Rev. Lett.* 31 (1973), pp. 494.

[53] H. Leutwyler. "Is the quark mass as small as 5 MeV?" In: *Phys. Lett. B* 48 (1974), pp. 431–434.

[54] Yoichiro Nambu and G. Jona-Lasinio. "Dynamical Model of Elementary Particles Based on an Analogy with Superconductivity. 1." In: *Phys. Rev.* 122 (1961). Ed. by T. Eguchi, pp. 345–358.

[55] S. Okubo. "Asymptotic SU(6)$_w$ spectral sum rules. ii. Applications and bare quark masses". In: *Phys. Rev.* 188 (1969), pp. 2300–2307.

[56] W. N. Cottingham. "The neutron proton mass difference and electron scattering experiments". In: *Annals Phys.* 25 (1963), pp. 424–432.

[57] J. Gasser and H. Leutwyler. "Implications of Scaling for the Proton - Neutron Mass - Difference". In: *Nucl. Phys. B* 94 (1975), pp. 269–310.

[58] Roger F. Dashen. "Chiral SU(3) x SU(3) as a symmetry of the strong interactions". In: *Phys. Rev.* 183 (1969), pp. 1245–1260.

[59] Steven Weinberg. "The Problem of Mass". In: *Trans. New York Acad. Sci.* 38 (1977), pp. 185.

[60] Gerard 't Hooft. "Dimensional regularization and the renormalization group". In: *Nucl. Phys. B* 61 (1973), pp. 455–468.

[61] Steven Weinberg. "New approach to the renormalization group". In: *Phys. Rev. D* 8 (1973), pp. 3497–3509.

[62] William A. Bardeen et al. "Deep Inelastic Scattering Beyond the Leading Order in Asymptotically Free Gauge Theories". In: *Phys. Rev. D* 18 (1978), p. 3998.

[63] Y. Aoki et al. "FLAG Review 2021". In: (Nov. 2021).

[64] J. Gasser and H. Leutwyler. "Chiral Perturbation Theory: Expansions in the Mass of the Strange Quark". In: *Nucl. Phys. B* 250 (1985), pp. 465–516.

[65] Gilberto Colangelo et al. "Dispersive analysis of $\eta \to 3\pi$". In: *Eur. Phys. J. C* 78.11 (2018), pp. 947.

[66] B. Ananthanarayan et al. "Analytic representations of $m_K$, $F_K$, $m_\eta$ and $F_\eta$ in two loop $SU(3)$ chiral perturbation theory". In: *Phys. Rev. D* 97 (2018), pp. 114004.

[67] Stephen L. Adler. "Axial vector vertex in spinor electrodynamics". In: *Phys. Rev.* 177 (1969), pp. 2426–2438.

[68] J. S. Bell and R. Jackiw. "A PCAC puzzle: $\pi^0 \to \gamma\gamma$ in the $\sigma$ model". In: *Nuovo Cim. A* 60 (1969), pp. 47–61.

[69] William A. Bardeen. "Anomalous Ward identities in spinor field theories". In: *Phys. Rev.* 184 (1969), pp. 1848–1857.

[70] Harald Fritzsch and Murray Gell-Mann. "Light cone current algebra". In: *Proceedings of the International Conference on Duality and Symmetry in Hadron Physics, Tel Aviv 1971, Weizmann Science Press.* 1971, pp. 317–374.

[71] W. Bardeen, H. Fritzsch, and M. Gell-Mann. *Light-Cone current Algebra, $\pi^0$ Decay and $e^+e^-$ Annihilation.* in: Scale and Conformal Symmetry ind Hadron Physics, ed. R. Gatto, John Wiley - Sons, Inc. 1973), pp.139 - 151. 1973.

[72] John B. Kogut and Davison E. Soper. "Quantum Electrodynamics in the Infinite Momentum Frame". In: *Phys. Rev. D* 1 (1970), pp. 2901–2913.

[73] S. M. Berman, J. D. Bjorken, and John B. Kogut. "Inclusive Processes at High Transverse Momentum". In: *Phys. Rev. D* 4 (1971), p. 3388.

[74] Thomas Appelquist and H. David Politzer. "Heavy Quarks and $e^+e^-$ Annihilation". In: *Phys. Rev. Lett.* 34 (1 Jan. 1975), pp. 43–45.

[75] J. E. Augustin et al. "Discovery of a Narrow Resonance in $e^+e^-$ Annihilation". In: *Phys. Rev. Lett.* 33 (1974), pp. 1406–1408.

[76] J. J. Aubert et al. "Experimental Observation of a Heavy Particle $J$". In: *Phys. Rev. Lett.* 33 (1974), p. 1404.

[77] G. S. Abrams et al. "The Discovery of a Second Narrow Resonance in $e^+e^-$ Annihilation". In: *Phys. Rev. Lett.* 33 (1974), pp. 1453–1455.

[78] S. L. Glashow, J. Iliopoulos, and L. Maiani. "Weak Interactions with Lepton-Hadron Symmetry". In: *Phys. Rev. D* 2 (1970), pp. 1285–1292.

[79] A. Casher, John B. Kogut, and Leonard Susskind. "Vacuum polarization and the quark parton puzzle". In: *Phys. Rev. Lett.* 31 (1973), pp. 792–795.

[80] Kenneth G. Wilson. "Confinement of Quarks". In: *Phys. Rev. D* 10 (1974). Ed. by J. C. Taylor, pp. 2445–2459.

[81] John B. Kogut and Leonard Susskind. "Hamiltonian Formulation of Wilson's Lattice Gauge Theories". In: *Phys. Rev. D* 11 (1975), pp. 395–408.

[82] E. Eichten et al. "The Spectrum of Charmonium". In: *Phys. Rev. Lett.* 34 (1975). [Erratum: Phys.Rev.Lett. 36, 1276 (1976)], pp. 369–372.

[83] Thomas Appelquist et al. "Charmonium Spectroscopy". In: *Phys. Rev. Lett.* 34 (1975), p. 365.

[84] Shawna Williams. "The Ring on the parking lot". In: *CERN Cour.* 43N5 (2003), pp. 16–18.

[85] D. Hitlin. "SPEAR MARK-II Magnetic Detector". In: *SLAC Beam Line* 7.6 (1976), S1–s4.

[86] M. Oreglia et al. "A Study of the Reaction $\psi' \to \gamma\gamma J/\psi$". In: *Phys. Rev. D* 25 (1982), p. 2259.

[87] G. Goldhaber et al. "Observation in $e^+e^-$ Annihilation of a Narrow State at 1865-MeV/c$^2$ Decaying to $K\pi$ and $K\pi\pi\pi$". In: *Phys. Rev. Lett.* 37 (1976), p. 255.

[88] John B. Kogut. "An Introduction to Lattice Gauge Theory and Spin Systems". In: *Rev. Mod. Phys.* 51 (1979), p. 659.

[89] Christian W. Bauer. "Quantum Simulations for High Energy Physics". In: *submitted to the Proceedings of the US Community Study on the Future of Particle Physics (Snowmass 2021).* 2022.

[90] *Quantum Simulation of Strong Interactions (QuaSI), Workshop 2 (online): Implementation Strategies for Gauge Theories.* 2022.

[91] B. H. Wiik. "First Results from PETRA". In: *Conf. Proc. C* 7906181 (1979), pp. 113–154.

[92] R. Brandelik et al. "Evidence for Planar Events in $e^+e^-$ Annihilation at High-Energies". In: *Phys. Lett. B* 86 (1979), pp. 243–249.

[93] Elliott D. Bloom et al. "High-Energy Inelastic e p Scattering at 6-Degrees and 10-Degrees". In: *Phys. Rev. Lett.* 23 (1969), pp. 930–934.

[94] Henry W. Kendall. "DEEP INELASTIC ELECTRON SCATTERING IN THE CONTINUUM REGION". In: vol. 710823. 1971, pp. 248–261.

[95] Jerome I. Friedman and Henry W. Kendall. "Deep inelastic electron scattering". In: *Ann. Rev. Nucl. Part. Sci.* 22 (1972), pp. 203–254.

[96] Guthrie Miller et al. "Inelastic electron-Proton Scattering at Large Momentum Transfers". In: *Phys. Rev. D* 5 (1972), p. 528.

[97] E. Eichten, F. Feinberg, and J. F. Willemsen. "Current and constituent quarks in the light cone quantization". In: *Phys. Rev. D* 8 (1973), pp. 1204–1219.

[98] Y. Watanabe et al. "Test of Scale Invariance in Ratios of Muon Scattering Cross-Sections at 150-GeV and 56-GeV". In: *Phys. Rev. Lett.* 35 (1975), p. 898.

[99] C. Chang et al. "Observed Deviations from Scale Invariance in High-Energy Muon Scattering". In: *Phys. Rev. Lett.* 35 (1975), p. 901.

[100] W. B. Atwood et al. "Inelastic electron Scattering from Hydrogen at 50-Degrees and 60-Degrees". In: *Phys. Lett. B* 64 (1976), pp. 479–482.

[101] P. C. Bosetti et al. "Analysis of Nucleon Structure Functions in CERN Bubble Chamber Neutrino Experiments". In: *Nucl. Phys. B* 142 (1978), pp. 1–28.

[102] J. G. H. de Groot et al. "QCD Analysis of Charged Current Structure Functions". In: *Phys. Lett. B* 82 (1979), pp. 456–460.

[103] J. G. H. de Groot et al. "Inclusive Interactions of High-Energy Neutrinos and anti-neutrinos in Iron". In: *Z. Phys. C* 1 (1979), p. 143.

[104] John R. Ellis, Mary K. Gaillard, and Graham G. Ross. "Search for Gluons in e+ e- Annihilation". In: *Nucl. Phys. B* 111 (1976). [Erratum: Nucl.Phys.B 130, 516 (1977)], p. 253.

[105] G. Hanson et al. "Evidence for Jet Structure in Hadron Production by $e^+e^-$ Annihilation". In: *Phys. Rev. Lett.* 35 (1975). Ed. by Martha C. Zipf, pp. 1609–1612.

[106] Sau Lan Wu. "Discovery of the first Yang–Mills gauge particle — The gluon". In: *Int. J. Mod. Phys. A* 30.34 (2015), p. 1530066.

[107] Sau Lan Wu and Georg Zobernig. “A Method of Three Jet Analysis in $e^+e^-$ Annihilation”. In: *Z. Phys. C* 2 (1979), p. 107.

[108] P. Soding. “Jet analysis”. In: *1979 EPS High-Energy Physics Conference*. Vol. 1. Geneva, Switzerland: CERN, 1979, pp. 271–281.

[109] D. P. Barber et al. “Discovery of Three Jet Events and a Test of Quantum Chromodynamics at PETRA Energies”. In: *Phys. Rev. Lett.* 43 (1979), p. 830.

[110] Christoph Berger et al. “Evidence for Gluon Bremsstrahlung in $e^+e^-$ Annihilations at High-Energies”. In: *Phys. Lett. B* 86 (1979), pp. 418–425.

[111] W. Bartel et al. “Observation of Planar Three Jet Events in $e^+e^-$ Annihilation and Evidence for Gluon Bremsstrahlung”. In: *Phys. Lett. B* 91 (1980), pp. 142–147.

[112] G. Arnison et al. “Experimental Observation of Isolated Large Transverse Energy Electrons with Associated Missing Energy at $\sqrt{s} = 540$ GeV”. In: *Phys. Lett. B* 122 (1983), pp. 103–116.

[113] G. Arnison et al. “Experimental Observation of Lepton Pairs of Invariant Mass Around 95-GeV/$c^2$ at the CERN SPS Collider”. In: *Phys. Lett. B* 126 (1983), pp. 398–410.

[114] M. Banner et al. “Observation of Single Isolated Electrons of High Transverse Momentum in Events with Missing Transverse Energy at the CERN anti-p p Collider”. In: *Phys. Lett. B* 122 (1983), pp. 476–485.

[115] P. Bagnaia et al. “Evidence for $Z^0 \to e^+e^-$ at the CERN $\bar{p}p$ Collider”. In: *Phys. Lett. B* 129 (1983), pp. 130–140.

[116] C. N. Yang. “Selected Papers (1945-1980) With Commentary”. In: *World Scientific* (2005).

[117] F. Abe et al. “Observation of Top Quark Production in p anti-p Collisions with the Collider Detector at Fermilab”. In: *Phys. Rev. Lett.* 74 (1995), p. 2626.

[118] S. Abachi et al. “Observation of the Top Quark”. In: *Phys. Rev. Lett.* 74 (1995), p. 2632.

[119] S. W. Herb, D. Hom, L. Lederman et al. “Observation of a Dimuon Resonance at 9.5 GeV in 400-GeV Proton-Nucleus Collisions”. In: *Phys. Rev. Lett.* 39 (1977), p. 252.

[120] Georges Aad et al. “Observation of a new particle in the search for the Standard Model Higgs boson with the ATLAS detector at the LHC”. In: *Phys. Lett. B* 716 (2012), pp. 1–29.

[121] Serguei Chatrchyan et al. “Observation of a New Boson at a Mass of 125 GeV with the CMS Experiment at the LHC”. In: *Phys. Lett. B* 716 (2012), pp. 30–61.

[122] T. T. Wu and S. L. Wu. In: *F. C. Chen et al.(Ed), A Festschrift for Yang Centenary: Scientific Papers, World Scientific Publisher.* 2022.

[123] T. T. Wu and S. L. Wu. “Concept of the basic standard model and a relation between the three gauge coupling constants”. In: (2022). to be published.

[124] J. D. Bjorken and Emmanuel A. Paschos. “Inelastic Electron Proton and gamma Proton Scattering, and the Structure of the Nucleon”. In: *Phys. Rev.* 185 (1969), pp. 1975–1982.

[125] Elliott D. Bloom and Frederick J. Gilman. “Scaling, Duality, and the Behavior of Resonances in Inelastic electron-Proton Scattering”. In: *Phys. Rev. Lett.* 25 (1970), p. 1140.

[126] Mikhail A. Shifman, A. I. Vainshtein, and Valentin I. Zakharov. “QCD and Resonance Physics. Theoretical Foundations”. In: *Nucl. Phys. B* 147 (1979), pp. 385–447.

[127] E. Braaten, Stephan Narison, and A. Pich. “QCD analysis of the tau hadronic width”. In: *Nucl. Phys. B* 373 (1992), pp. 581–612.

[128] Patricia Ball, M. Beneke, and Vladimir M. Braun. “Resummation of (beta0 alpha-s)**n corrections in QCD: Techniques and applications to the tau hadronic width and the heavy quark pole mass”. In: *Nucl. Phys. B* 452 (1995), pp. 563–625.

[129] Michel Davier, Andreas Hocker, and Zhiqing Zhang. “The Physics of Hadronic Tau Decays”. In: *Rev. Mod. Phys.* 78 (2006), pp. 1043–1109.

[130] Siegfried Bethke. “Experimental tests of asymptotic freedom”. In: *Prog. Part. Nucl. Phys.* 58 (2007), pp. 351–386.

[131] E. M. Levin and L. L. Frankfurt. “The Quark hypothesis and relations between cross-sections at high-energies”. In: *JETP Lett.* 2 (1965), pp. 65–70.

[132] M. Gupta. “Baryon magnetic moments and the naive quark model”. In: *J. Phys. G* 16 (1990), pp. L213–L217.

[133] V. A. Matveev, R. M. Muradyan, and A. N. Tavkhelidze. “Automodelity in strong interactions”. In: *Lett. Nuovo Cim.* 5S2 (1972), pp. 907–912.

[134] Stanley J. Brodsky and Glennys R. Farrar. “Scaling Laws at Large Transverse Momentum”. In: *Phys. Rev. Lett.* 31 (1973), pp. 1153–1156.

[135] C. Bochna et al. “Measurements of deuteron photodisintegration up to 4.0-GeV”. In: *Phys. Rev. Lett.* 81 (1998), pp. 4576–4579.

[136] Yuri L. Dokshitzer. “QCD phenomenology”. In: *2002 European School of High-Energy Physics*. June 2003, pp. 1–33.

[137] F. E. Low. “A Model of the Bare Pomeron”. In: *Phys. Rev. D* 12 (1975), pp. 163–173.

[138] S. Nussinov. “Colored Quark Version of Some Hadronic Puzzles”. In: *Phys. Rev. Lett.* 34 (1975), pp. 1286–1289.

[139] J. F. Gunion and Davison E. Soper. “Quark Counting and Hadron Size Effects for Total Cross-Sections”. In: *Phys. Rev. D* 15 (1977), pp. 2617–2621.

[140] Victor S. Fadin, E. A. Kuraev, and L. N. Lipatov. “On the Pomeranchuk Singularity in Asymptotically Free Theories”. In: *Phys. Lett. B* 60 (1975), pp. 50–52.

[141] V. N. Gribov. “Interaction of gamma quanta and electrons with nuclei at high-energies”. In: *Zh. Eksp. Teor. Fiz.* 57 (1969), pp. 1306–1323.

[142] Richard P. Feynman. “Very high-energy collisions of hadrons”. In: *Phys. Rev. Lett.* 23 (1969). Ed. by L. M. Brown, pp. 1415–1417.

[143] Gerard ’t Hooft and M. J. G. Veltman. “Regularization and Renormalization of Gauge Fields”. In: *Nucl. Phys. B* 44 (1972), pp. 189–213.

[144] C. G. Bollini and J. J. Giambiagi. “Dimensional Renormalization: The Number of Dimensions as a Regularizing Parameter”. In: *Nuovo Cim. B* 12 (1972), pp. 20–26.

[145] Warren Siegel. “Supersymmetric Dimensional Regularization via Dimensional Reduction”. In: *Phys. Lett. B* 84 (1979), pp. 193–196.

[146] G. Grunberg. “Renormalization Scheme Independent QCD and QED: The Method of Effective Charges”. In: *Phys. Rev. D* 29 (1984), pp. 2315–2338.

[147] A. C. Mattingly and Paul M. Stevenson. “Optimization of R(e+ e-) and ’freezing’ of the QCD couplant at low-energies”. In: *Phys. Rev. D* 49 (1994), pp. 437–450.

[148] J. D. Bjorken. “Applications of the Chiral U(6) x (6) Algebra of Current Densities”. In: *Phys. Rev.* 148 (1966), pp. 1467–1478.

[149] Thomas Appelquist, Michael Dine, and I. J. Muzinich. “The Static Potential in Quantum Chromodynamics”. In: *Phys. Lett. B* 69 (1977), pp. 231–236.

[150] A. Czarnecki, K. Melnikov, and N. Uraltsev. “NonAbelian dipole radiation and the heavy quark expansion”. In: *Phys. Rev. Lett.* 80 (1998), pp. 3189–3192.

[151] P. A. Baikov, K. G. Chetyrkin, and J. H. Kuhn. “Adler Function, Bjorken Sum Rule, and the Crewther Relation to Order $\alpha_s^4$ in a General Gauge Theory”. In: *Phys. Rev. Lett.* 104 (2010), p. 132004.

[152] S. Catani, B. R. Webber, and G. Marchesini. “QCD coherent branching and semiinclusive processes at large x”. In: *Nucl. Phys. B* 349 (1991), pp. 635–654.

[153] Yuri L. Dokshitzer, Valery A. Khoze, and S. I. Troian. “Specific features of heavy quark production. LPHD approach to heavy particle spectra”. In: *Phys. Rev. D* 53 (1996), pp. 89–119.

[154] G. P. Korchemsky. “Asymptotics of the Altarelli-Parisi-Lipatov Evolution Kernels of Parton Distributions”. In: *Mod. Phys. Lett. A* 4 (1989), pp. 1257–1276.

[155] G. P. Korchemsky and G. Marchesini. “Structure function for large x and renormalization of Wilson loop”. In: *Nucl. Phys. B* 406 (1993), pp. 225–258.

[156] F. E. Low. “Bremsstrahlung of very low-energy quanta in elementary particle collisions”. In: *Phys. Rev.* 110 (1958), pp. 974–977.

[157] D. M. Howe and C. J. Maxwell. “All orders infrared freezing of observables in perturbative QCD”. In: *Phys. Rev. D* 70 (2004), p. 014002.

[158] Stanley J. Brodsky, Guy F. de Teramond, and Alexandre Deur. “Nonperturbative QCD Coupling and its $\beta$-function from Light-Front Holography”. In: *Phys. Rev. D* 81 (2010), p. 096010.

[159] G. Parisi and R. Petronzio. “Small Transverse Momentum Distributions in Hard Processes”. In: *Nucl. Phys. B* 154 (1979), pp. 427–440.

[160] Yuri L. Dokshitzer, G. Marchesini, and B. R. Webber. “Dispersive approach to power behaved contributions in QCD hard processes”. In: *Nucl. Phys. B* 469 (1996), pp. 93–142.

[161] Fabrizio Caola et al. “On linear power corrections in certain collider observables”. In: *JHEP* 01 (2022), p. 093.

[162] S. D. Drell, Donald J. Levy, and Tung-Mow Yan. “A Theory of Deep Inelastic Lepton-Nucleon Scattering and Lepton Pair Annihilation Processes. 3. Deep Inelastic electron-Positron Annihilation”. In: *Phys. Rev. D* 1 (1970), pp. 1617–1639.

[163] N. Cabibbo, G. Parisi, and M. Testa. “Hadron Production in e+ e- Collisions”. In: *Lett. Nuovo Cim.* 4S1 (1970), pp. 35–39.

[164] B. L. Ioffe. “Space-time picture of photon and neutrino scattering and electroproduction cross-section asymptotics”. In: *Phys. Lett. B* 30 (1969), pp. 123–125.

[165] Bo Andersson et al. “Parton Fragmentation and String Dynamics”. In: *Phys. Rept.* 97 (1983), pp. 31–145.

[166] George F. Sterman and Steven Weinberg. “Jets from Quantum Chromodynamics”. In: *Phys. Rev. Lett.* 39 (1977), p. 1436.

[167] W. James Stirling. “Hard QCD working group: Theory summary”. In: *J. Phys. G* 17 (1991), pp. 1567–1574.

[168] S. Catani et al. “New clustering algorithm for multi - jet cross-sections in e+ e- annihilation”. In: *Phys. Lett. B* 269 (1991), pp. 432–438.

[169] S. Bethke et al. “New jet cluster algorithms: Next-to-leading order QCD and hadronization corrections”. In: *Nucl. Phys. B* 370 (1992). [Erratum: Nucl.Phys.B 523, 681–681 (1998)], pp. 310–334.

[170] Yuri L. Dokshitzer et al. “Better jet clustering algorithms”. In: *JHEP* 08 (1997), p. 001.

[171] M. Wobisch and T. Wengler. “Hadronization corrections to jet cross-sections in deep inelastic scattering”. In: *Workshop on Monte Carlo Generators for HERA Physics (Plenary Starting Meeting).* Apr. 1998, pp. 270–279.

[172] S. Catani et al. “Longitudinally invariant $K_t$ clustering algorithms for hadron hadron collisions”. In: *Nucl. Phys. B* 406 (1993), pp. 187–224.

[173] Matteo Cacciari and Gavin P. Salam. “Dispelling the $N^3$ myth for the $k_t$ jet-finder”. In: *Phys. Lett. B* 641 (2006), pp. 57–61.

[174] Matteo Cacciari, Gavin P. Salam, and Gregory Soyez. “The anti-$k_t$ jet clustering algorithm”. In: *JHEP* 04 (2008), p. 063.

[175] Gavin P. Salam. “Towards Jetography”. In: *Eur. Phys. J. C* 67 (2010), pp. 637–686.

[176] Shreyasi Acharya et al. “Direct observation of the dead-cone effect in quantum chromodynamics”. In: *Nature* 605.7910 (2022), pp. 440–446.

[177] Guido Altarelli, R. Keith Ellis, and G. Martinelli. “Large Perturbative Corrections to the Drell-Yan Process in QCD”. In: *Nucl. Phys. B* 157 (1979), pp. 461–497.

[178] R. Keith Ellis et al. “Large Corrections to High p(T) Hadron-Hadron Scattering in QCD”. In: *Nucl. Phys. B* 173 (1980), pp. 397–421.

[179] W. T. Giele and E. W. Nigel Glover. “Higher order corrections to jet cross-sections in e+ e- annihilation”. In: *Phys. Rev. D* 46 (1992), pp. 1980–2010.

[180] S. Catani and M. H. Seymour. “A General algorithm for calculating jet cross-sections in NLO QCD”. In: *Nucl. Phys. B* 485 (1997). [Erratum: Nucl.Phys.B 510, 503–504 (1998)], pp. 291–419.

[181] R. Keith Ellis et al. “One-loop calculations in quantum field theory: from Feynman diagrams to unitarity cuts”. In: *Phys. Rept.* 518 (2012), pp. 141–250.

[182] R. Keith Ellis, W. T. Giele, and Giulia Zanderighi. “The One-loop amplitude for six-gluon scattering”. In: *JHEP* 05 (2006), p. 027.

[183] Mrinal Dasgupta et al. “Parton showers beyond leading logarithmic accuracy”. In: *Phys. Rev. Lett.* 125.5 (2020), p. 052002.

[184] S. Amoroso et al. “Les Houches 2019: Physics at TeV Colliders: Standard Model Working Group Report”. In: *11th Les Houches Workshop on Physics at TeV Colliders: PhysTeV Les Houches.* Mar. 2020.

[185] Charalampos Anastasiou et al. “Higgs Boson GluonFfusion Production Beyond Threshold in $N^3LO$ QCD”. In: *JHEP* 03 (2015), p. 091.

[186] Andrea Banfi et al. “Jet-vetoed Higgs cross section in gluon fusion at N$^3$LO+NNLL with small-$R$ resummation”. In: *JHEP* 04 (2016), p. 049.

[187] Bernhard Mistlberger. “Higgs boson production at hadron colliders at N$^3$LO in QCD”. In: *JHEP* 05 (2018), p. 028.

[188] Stefano Camarda, Leandro Cieri, and Giancarlo Ferrera. “Drell–Yan lepton-pair production: qT resummation at N3LL accuracy and fiducial cross sections at N3LO”. In: *Phys. Rev. D* 104.11 (2021), p. L111503.

[189] J. Currie et al. “N$^3$LO corrections to jet production in deep inelastic scattering using the Projection-to-Born method”. In: *JHEP* 05 (2018), p. 209.

[190] Maurizio Abele, Daniel de Florian, and Werner Vogelsang. “Threshold resummation at NLL3 accuracy and approximate N3LO corrections to semi-inclusive DIS”. In: *Phys. Rev. D* 106.1 (2022), p. 014015.

[191] C. Gnendiger et al. “To $d$, or not to $d$: recent developments and comparisons of regularization schemes”. In: *Eur. Phys. J. C* 77.7 (2017), p. 471.

[192] Zvi Bern, Lance J. Dixon, and David A. Kosower. “Progress in one loop QCD computations”. In: *Ann. Rev. Nucl. Part. Sci.* 46 (1996), pp. 109–148.

[193] Isabella Bierenbaum et al. “A Tree-Loop Duality Relation at Two Loops and Beyond”. In: *JHEP* 10 (2010), p. 073.

[194] German F. R. Sborlini et al. "Four-dimensional unsubtraction from the loop-tree duality". In: *JHEP* 08 (2016), p. 160.

[195] W. J. Torres Bobadilla et al. "May the four be with you: Novel IR-subtraction methods to tackle NNLO calculations". In: *Eur. Phys. J. C* 81.3 (2021), p. 250.

[196] Gabriele Travaglini et al. "The SAGEX Review on Scattering Amplitudes". In: (Mar. 2022).

[197] Stefano Catani et al. "Diphoton production at hadron colliders: a fully-differential QCD calculation at NNLO". In: *Phys. Rev. Lett.* 108 (2012). [Erratum: Phys.Rev.Lett. 117, 089901 (2016)], p. 072001.

[198] M. Beneke. "Renormalons". In: *Phys. Rept.* 317 (1999), pp. 1–142.

[199] Yuri L. Dokshitzer, Dmitri Diakonov, and S. I. Troian. "Hard Semiinclusive Processes in QCD". In: *Phys. Lett. B* 78 (1978), pp. 290–294.

[200] Yuri L. Dokshitzer, Dmitri Diakonov, and S. I. Troian. "Hard Processes in Quantum Chromodynamics". In: *Phys. Rept.* 58 (1980), pp. 269–395.

[201] V. N. Gribov. "Bremsstrahlung of hadrons at high energies". In: *Sov. J. Nucl. Phys.* 5 (1967), p. 280.

[202] R. Kirschner and L. n. Lipatov. "Double Logarithmic Asymptotics and Regge Singularities of Quark Amplitudes with Flavor Exchange". In: *Nucl. Phys. B* 213 (1983), pp. 122–148.

[203] Nikolaos Kidonakis, Gianluca Oderda, and George F. Sterman. "Evolution of color exchange in QCD hard scattering". In: *Nucl. Phys. B* 531 (1998), pp. 365–402.

[204] Andrey Grozin et al. "The three-loop cusp anomalous dimension in QCD and its supersymmetric extensions". In: *JHEP* 01 (2016), p. 140.

[205] Yu. L. Dokshitzer and G. Marchesini. "Soft gluons at large angles in hadron collisions". In: *JHEP* 01 (2006), p. 007.

[206] A. V. Efremov and A. V. Radyushkin. "Factorization and Asymptotical Behavior of Pion Form-Factor in QCD". In: *Phys. Lett. B* 94 (1980), pp. 245–250.

[207] G. Peter Lepage and Stanley J. Brodsky. "Exclusive Processes in Perturbative Quantum Chromodynamics". In: *Phys. Rev. D* 22 (1980), p. 2157.

[208] B. Z. Kopeliovich et al. "Decisive test of color transparency in exclusive electroproduction of vector mesons". In: *Phys. Lett. B* 324 (1994), pp. 469–476.

[209] Stanley J. Brodsky et al. "Diffractive leptoproduction of vector mesons in QCD". In: *Phys. Rev. D* 50 (1994), pp. 3134–3144.

[210] E. M. Aitala et al. "Observation of color transparency in diffractive dissociation of pions". In: *Phys. Rev. Lett.* 86 (2001), pp. 4773–4777.

[211] V. N. Gribov and L. N. Lipatov. "Deep inelastic e p scattering in perturbation theory". In: *Sov. J. Nucl. Phys.* 15 (1972), pp. 438–450.

[212] V. N. Gribov and L. N. Lipatov. "e+ e- pair annihilation and deep inelastic e p scattering in perturbation theory". In: *Sov. J. Nucl. Phys.* 15 (1972), pp. 675–684.

[213] L. N. Lipatov. "The parton model and perturbation theory". In: *Yad. Fiz.* 20 (1974), pp. 181–198.

[214] Guido Altarelli and G. Parisi. "Asymptotic Freedom in Parton Language". In: *Nucl. Phys. B* 126 (1977), pp. 298–318.

[215] Yuri L. Dokshitzer. "Calculation of the Structure Functions for Deep Inelastic Scattering and e+ e- Annihilation by Perturbation Theory in Quantum Chromodynamics." In: *Sov. Phys. JETP* 46 (1977), pp. 641–653.

[216] S. Moch, J. A. M. Vermaseren, and A. Vogt. "The Three loop splitting functions in QCD: The Nonsinglet case". In: *Nucl. Phys. B* 688 (2004), pp. 101–134.

[217] A. Vogt, S. Moch, and J. A. M. Vermaseren. "The Three-loop splitting functions in QCD: The Singlet case". In: *Nucl. Phys. B* 691 (2004), pp. 129–181.

[218] I. I. Balitsky and L. N. Lipatov. "The Pomeranchuk Singularity in Quantum Chromodynamics". In: *Sov. J. Nucl. Phys.* 28 (1978), pp. 822–829.

[219] Alfred H. Mueller. "A Simple derivation of the JIMWLK equation". In: *Phys. Lett. B* 523 (2001), pp. 243–248.

[220] G. Camici and M. Ciafaloni. "Model (in)dependent features of the hard pomeron". In: *Phys. Lett. B* 395 (1997), pp. 118–122.

[221] Alfred H. Mueller. "Limitations on using the operator product expansion at small values of x". In: *Phys. Lett. B* 396 (1997), pp. 251–256.

[222] Alfred H. Mueller and H. Navelet. "An Inclusive Minijet Cross-Section and the Bare Pomeron in QCD". In: *Nucl. Phys. B* 282 (1987), pp. 727–744.

[223] Grigory Safronov. "Beyond-DGLAP searches with Mueller-Navelet jets, and measurements of low-$p_T$ and forward jets at CMS". In: *PoS* DIS2014 (2014), p. 094.

[224] John C. Collins, Davison E. Soper, and George F. Sterman. "Factorization of Hard Processes in QCD". In: *Adv. Ser. Direct. High Energy Phys.* 5 (1989), pp. 1–91.

[225] A. V. Kotikov and L. N. Lipatov. "DGLAP and BFKL equations in the $N = 4$ supersymmetric gauge theory". In: *Nucl. Phys. B* 661 (2003). [Erratum: Nucl.Phys.B 685, 405–407 (2004)], pp. 19–61.

[226] M. Z. Akrawy et al. "A Study of coherence of soft gluons in hadron jets". In: *Phys. Lett. B* 247 (1990), pp. 617–628.

[227] K. Charchula. "Coherence effects in the current fragmentation region at HERA". In: *J. Phys. G* 19 (1993), pp. 1587–1593.

[228] Yakov I. Azimov et al. "Similarity of Parton and Hadron Spectra in QCD Jets". In: *Z. Phys. C* 27 (1985), pp. 65–72.

[229] A. Korytov. "Inclusive momentum distributions of charged particles in jets at CDF". In: *Nucl. Phys. B Proc. Suppl.* 54 (1997). Ed. by Z. Ajduk and A. K. Wroblewski, pp. 67–70.

[230] D. Acosta et al. "Momentum Distribution of Charged Particles in Jets in Dijet Events in $p\bar{p}$ Collisions at $\sqrt{s} = 1.8$ TeV and Comparisons to Perturbative QCD Predictions". In: *Phys. Rev. D* 68 (2003), p. 012003.

[231] Yuri L. Dokshitzer et al. *Basics of perturbative QCD.* [available on the web]. 1991.

[232] Valery A. Khoze, Sergio Lupia, and Wolfgang Ochs. "Perturbative universality in soft particle production". In: *Eur. Phys. J. C* 5 (1998), pp. 77–90.

[233] J. Abdallah et al. "Coherent soft particle production in Z decays into three jets". In: *Phys. Lett. B* 605 (2005). Ed. by D. Bruncko, J. Ferencei, and P. Strizenec, pp. 37–48.

[234] Yuri L. Dokshitzer, Valery A. Khoze, and S. I. Troian. "On the concept of local parton hadron duality". In: *J. Phys. G* 17 (1991), pp. 1585–1587.

[235] Yuri L. Dokshitzer and Dmitri Diakonov. "Angular distribution of energy in jets". In: *Phys. Lett. B* 84 (1979), pp. 234–236.

[236] G. Curci, W. Furmanski, and R. Petronzio. "Evolution of Parton Densities Beyond Leading Order: The Nonsinglet Case". In: *Nucl. Phys. B* 175 (1980), pp. 27–92.

[237] Luigi Del Debbio and Alberto Ramos. "Lattice determinations of the strong coupling". In: *Physics Reports* 920 (Jan. 2021), pp. 1–71.

[238] Martin Luscher, Peter Weisz, and Ulli Wolff. "A Numerical method to compute the running coupling in asymptotically free theories". In: *Nucl. Phys. B* 359 (1991), pp. 221–243.

[239] Mattia Dalla Brida et al. "Non-perturbative renormalization by decoupling". In: *Phys. Lett. B* 807 (2020), p. 135571.

[240] Mattia Dalla Brida. "Past, present, and future of precision determinations of the QCD parameters from lattice QCD". In: *Eur. Phys. J. A* 57.2 (2021), p. 66.

[241] Andreas Athenodorou et al. "How perturbative are heavy sea quarks?" In: *Nucl. Phys. B* 943 (2019), p. 114612.

[242] Mattia Dalla Brida et al. "Determination of $\alpha_s(m_Z)$ by the non-perturbative decoupling method". In: (Sept. 2022).

[243] P. Petreczky and J. H. Weber. "Strong coupling constant and heavy quark masses in ( 2+1 )-flavor QCD". In: *Phys. Rev. D* 100.3 (2019), p. 034519.

[244] E. Shintani et al. "Strong coupling constant from vacuum polarization functions in three-flavor lattice QCD with dynamical overlap fermions". In: *Phys. Rev.* D82.7 (2010). [Erratum: Phys. Rev. D89, no.9, 099903 (2014)], p. 074505.

[245] Renwick J. Hudspith et al. "$\alpha_s$ from the Lattice Hadronic Vacuum Polarisation". In: (2018).

[246] Katsumasa Nakayama, Hidenori Fukaya, and Shoji Hashimoto. "Lattice computation of the Dirac eigenvalue density in the perturbative regime of QCD". In: *Phys. Rev.* D98.1 (2018), p. 014501.

[247] Z. Fodor et al. "Up and down quark masses and corrections to Dashen's theorem from lattice QCD and quenched QED". In: *Phys. Rev. Lett.* 117.8 (2016), p. 082001.

[248] D. Giusti et al. "Leading isospin-breaking corrections to pion, kaon and charmed-meson masses with Twisted-Mass fermions". In: *Phys. Rev. D* 95.11 (2017), p. 114504.

[249] M. Bruno et al. "Light and strange quark masses from $N_f = 2{+}1$ simulations with Wilson fermions". In: *PoS* LATTICE2018 (2019), p. 220.

[250] S. Durr et al. "Lattice QCD at the physical point: light quark masses". In: *Phys. Lett. B* 701 (2011), pp. 265–268.

[251] S. Durr et al. "Lattice QCD at the physical point: Simulation and analysis details". In: *JHEP* 08 (2011), p. 148.

[252] C. McNeile et al. "High-Precision c and b Masses, and QCD Coupling from Current-Current Correlators in Lattice and Continuum QCD". In: *Phys. Rev.* D82 (2010), p. 034512.

[253] A. T. Lytle et al. "Determination of quark masses from $\mathbf{n_f = 4}$ lattice QCD and the RI-SMOM

intermediate scheme". In: *Phys. Rev. D* 98.1 (2018), p. 014513.
[254] Bipasha Chakraborty et al. "High-precision quark masses and QCD coupling from $n_f = 4$ lattice QCD". In: *Phys. Rev.* D91.5 (2015), p. 054508.
[255] Yi-Bo Yang et al. "Charm and strange quark masses and $f_{D_s}$ from overlap fermions". In: *Phys. Rev. D* 92.3 (2015), p. 034517.
[256] Katsumasa Nakayama, Brendan Fahy, and Shoji Hashimoto. "Short-distance charmonium correlator on the lattice with M00F6bius domain-wall fermion and a determination of charm quark mass". In: *Phys. Rev.* D94.5 (2016), p. 054507.
[257] C. Alexandrou et al. "Baryon spectrum with $N_f = 2 + 1 + 1$ twisted mass fermions". In: *Phys. Rev. D* 90.7 (2014), p. 074501.
[258] D. Hatton et al. "Charmonium properties from lattice $QCD$+QED : Hyperfine splitting, $J/\psi$ leptonic width, charm quark mass, and $a_\mu^c$". In: *Phys. Rev. D* 102.5 (2020), p. 054511.
[259] D. Hatton et al. "Determination of $\overline{m}_b/\overline{m}_c$ and $\overline{m}_b$ from $n_f = 4$ lattice QCD+QED". In: *Phys. Rev. D* 103.11 (2021), p. 114508.
[260] B. Colquhoun et al. "$\Upsilon$ and $\Upsilon'$ Leptonic Widths, $a_\mu^b$ and $m_b$ from full lattice QCD". In: *Phys. Rev. D* 91.7 (2015), p. 074514.
[261] N. Carrasco et al. "Up, down, strange and charm quark masses with $\mathrm{N}_f$ = 2+1+1 twisted mass lattice QCD". In: *Nucl. Phys. B* 887 (2014), pp. 19–68.
[262] A. Bussone et al. "Mass of the b quark and B -meson decay constants from $N_f = 2 + 1 + 1$ twisted-mass lattice QCD". In: *Phys. Rev. D* 93.11 (2016), p. 114505.
[263] P. Gambino, A. Melis, and S. Simula. "Extraction of heavy-quark-expansion parameters from unquenched lattice data on pseudoscalar and vector heavy-light meson masses". In: *Phys. Rev. D* 96.1 (2017), p. 014511.
[264] T. Blum et al. "Domain wall QCD with physical quark masses". In: *Phys. Rev. D* 93.7 (2016), p. 074505.
[265] A. Bazavov et al. "MILC results for light pseudoscalars". In: *PoS* CD09 (2009), p. 007.
[266] A. Bazavov et al. "Up-, down-, strange-, charm-, and bottom-quark masses from four-flavor lattice QCD". In: *Phys. Rev. D* 98.5 (2018), p. 054517.
[267] Rainer Sommer. "Introduction to Non-perturbative Heavy Quark Effective Theory". In: *Les Houches Summer School: Session 93: Modern perspectives in lattice QCD: Quantum field theory and high performance computing.* Aug. 2010, pp. 517–590.
[268] G. Peter Lepage et al. "Improved nonrelativistic QCD for heavy quark physics". In: *Phys. Rev. D* 46 (1992), pp. 4052–4067.
[269] A. Bazavov et al. "Up-, down-, strange-, charm-, and bottom-quark masses from four-flavor lattice QCD". In: *Phys. Rev.* D98 (2018), p. 054517.
[270] Agostino Patella. "QED Corrections to Hadronic Observables". In: *PoS* LATTICE2016 (2017), p. 020.
[271] Martin Luscher et al. "A Precise determination of the running coupling in the SU(3) Yang-Mills theory". In: *Nucl. Phys. B* 413 (1994), pp. 481–502.
[272] Michele Della Morte et al. "Computation of the strong coupling in QCD with two dynamical flavors". In: *Nucl. Phys. B* 713 (2005), pp. 378–406.
[273] S. Aoki et al. "Precise determination of the strong coupling constant in N(f) = 2+1 lattice QCD with the Schrodinger functional scheme". In: *JHEP* 10 (2009), p. 053.
[274] Mattia Dalla Brida et al. "Slow running of the Gradient Flow coupling from 200 MeV to 4 GeV in $N_\mathrm{f} = 3$ QCD". In: *Phys. Rev. D* 95.1 (2017), p. 014507.
[275] Mattia Bruno et al. "QCD Coupling from a Nonperturbative Determination of the Three-Flavor $\Lambda$ Parameter". In: *Phys. Rev. Lett.* 119.10 (2017), p. 102001.
[276] Isabel Campos et al. "Non-perturbative quark mass renormalisation and running in $N_f = 3$ QCD". In: *Eur. Phys. J. C* 78.5 (2018), p. 387.
[277] Jochen Heitger, Fabian Joswig, and Simon Kuberski. "Determination of the charm quark mass in lattice QCD with $2 + 1$ flavours on fine lattices". In: *JHEP* 05 (2021), p. 288.
[278] P. A. Zyla et al. "Review of Particle Physics". In: *PTEP* 2020.8 (2020), p. 083C01.
[279] S. Aoki et al. "FLAG Review 2019: Flavour Lattice Averaging Group (FLAG)". In: *Eur. Phys. J. C* 80.2 (2020), p. 113.
[280] Riccardo Abbate et al. "Thrust at N$^3$LL with Power Corrections and a Precision Global Fit for $\alpha_s(mZ)$". In: *Phys. Rev.* D83 (2011), p. 074021.
[281] Andr00E9 H. Hoang et al. "Precise determination of $\alpha_s$ from the $C$-parameter distribution". In: *Phys. Rev.* D91.9 (2015), p. 094018.
[282] D. d'Enterria et al. "The strong coupling constant: State of the art and the decade ahead". In: (Mar. 2022).
[283] Gionata Luisoni, Pier Francesco Monni, and Gavin P. Salam. "C-parameter hadronisation

in the symmetric 3-jet limit and impact on $\alpha_s$ fits". In: *Eur. Phys. J. C* 81.2 (2021), p. 158.

[284] Kenneth G. Wilson. "Quarks and Strings on a Lattice". In: *13th International School of Subnuclear Physics: New Phenomena in Subnuclear Physics.* Nov. 1975.

[285] B. Sheikholeslami and R. Wohlert. "Improved Continuum Limit Lattice Action for QCD with Wilson Fermions". In: *Nucl. Phys. B* 259 (1985), p. 572.

[286] M. Luscher and P. Weisz. "O(a) improvement of the axial current in lattice QCD to one loop order of perturbation theory". In: *Nucl. Phys. B* 479 (1996), pp. 429–458.

[287] R. Wohlert. "Improved limit lattice action for quarks". In: (July 1987).

[288] Martin Luscher et al. "Chiral symmetry and O(a) improvement in lattice QCD". In: *Nucl. Phys. B* 478 (1996), pp. 365–400.

[289] Martin Luscher et al. "Nonperturbative O(a) improvement of lattice QCD". In: *Nucl. Phys. B* 491 (1997), pp. 323–343.

[290] K. Symanzik. "Cutoff dependence in lattice $\phi^4$ in four-dimensions theory". In: *NATO Sci. Ser. B* 59 (1980). Ed. by Gerard 't Hooft et al., pp. 313–330.

[291] K. Symanzik. "Continuum Limit and Improved Action in Lattice Theories. 1. Principles and phi**4 Theory". In: *Nucl. Phys. B* 226 (1983), pp. 187–204.

[292] William A. Bardeen et al. "Light quarks, zero modes, and exceptional configurations". In: *Phys. Rev. D* 57 (1998), pp. 1633–1641.

[293] Roberto Frezzotti et al. "Lattice QCD with a chirally twisted mass term". In: *JHEP* 08 (2001), p. 058.

[294] Tom Banks, Leonard Susskind, and John B. Kogut. "Strong Coupling Calculations of Lattice Gauge Theories: (1+1)-Dimensional Exercises". In: *Phys. Rev. D* 13 (1976), p. 1043.

[295] Tom Banks et al. "Strong Coupling Calculations of the Hadron Spectrum of Quantum Chromodynamics". In: *Phys. Rev. D* 15 (1977), p. 1111.

[296] Leonard Susskind. "Lattice Fermions". In: *Phys. Rev. D* 16 (1977), pp. 3031–3039.

[297] Thomas DeGrand and Carleton E. DeTar. *Lattice methods for quantum chromodynamics.* Singapore: World Scientific, 2006.

[298] J. Smit. *Introduction to quantum fields on a lattice: A robust mate.* Vol. 15. Cambridge University Press, Jan. 2011.

[299] A. Bazavov et al. "Nonperturbative QCD Simulations with 2+1 Flavors of Improved Staggered Quarks". In: *Rev. Mod. Phys.* 82 (2010), pp. 1349–1417.

[300] Satchidananda Naik. "On-shell Improved Lattice Action for QCD With Susskind Fermions and Asymptotic Freedom Scale". In: *Nucl. Phys. B* 316 (1989), pp. 238–268.

[301] M. Luscher and P. Weisz. "On-Shell Improved Lattice Gauge Theories". In: *Commun. Math. Phys.* 97 (1985), p. 59.

[302] M. Luscher and P. Weisz. "Computation of the Action for On-Shell Improved Lattice Gauge Theories at Weak Coupling". In: *Phys. Lett. B* 158 (1985), pp. 250–254.

[303] Tom Blum et al. "Improving flavor symmetry in the Kogut-Susskind hadron spectrum". In: *Phys. Rev. D* 55 (1997), R1133–R1137.

[304] J. F. Lagae and D. K. Sinclair. "Improved staggered quark actions with reduced flavor symmetry violations for lattice QCD". In: *Phys. Rev. D* 59 (1999), p. 014511.

[305] Peter Lepage. "Perturbative improvement for lattice QCD: An Update". In: *Nucl. Phys. B Proc. Suppl.* 60 (1998). Ed. by Y. Iwasaki and A. Ukawa, pp. 267–278.

[306] Kostas Orginos and Doug Toussaint. "Testing improved actions for dynamical Kogut-Susskind quarks". In: *Phys. Rev. D* 59 (1999), p. 014501.

[307] Kostas Orginos, Doug Toussaint, and R. L. Sugar. "Variants of fattening and flavor symmetry restoration". In: *Phys. Rev. D* 60 (1999), p. 054503.

[308] E. Follana et al. "Highly improved staggered quarks on the lattice, with applications to charm physics". In: *Phys. Rev. D* 75 (2007), p. 054502.

[309] A. Bazavov et al. "Lattice QCD Ensembles with Four Flavors of Highly Improved Staggered Quarks". In: *Phys. Rev. D* 87.5 (2013), p. 054505.

[310] K. C. Bowler et al. "Quenched QCD with O(a) improvement. 1. The Spectrum of light hadrons". In: *Phys. Rev. D* 62 (2000), p. 054506.

[311] Sara Collins et al. "Comparing Wilson and Clover quenched SU(3) spectroscopy with an improved gauge action". In: *Nucl. Phys. B Proc. Suppl.* 53 (1997). Ed. by C. Bernard et al., pp. 877–879.

[312] Paul H. Ginsparg and Kenneth G. Wilson. "A Remnant of Chiral Symmetry on the Lattice". In: *Phys. Rev. D* 25 (1982), p. 2649.

[313] Rajamani Narayanan and Herbert Neuberger. "Infinitely many regulator fields for chiral fermions". In: *Phys. Lett. B* 302 (1993), pp. 62–69.

[314] Rajamani Narayanan and Herbert Neuberger. "Chiral fermions on the lattice". In: *Phys. Rev. Lett.* 71.20 (1993), p. 3251.

[315] Rajamani Narayanan and Herbert Neuberger. "Chiral determinant as an overlap of two vacua". In: *Nucl. Phys. B* 412 (1994), pp. 574–606.

[316] Rajamani Narayanan and Herbert Neuberger. "A Construction of lattice chiral gauge theories". In: *Nucl. Phys. B* 443 (1995), pp. 305–385.

[317] Herbert Neuberger. "Exactly massless quarks on the lattice". In: *Phys. Lett. B* 417 (1998), pp. 141–144.

[318] David B. Kaplan. "A Method for simulating chiral fermions on the lattice". In: *Phys. Lett. B* 288 (1992), pp. 342–347.

[319] Yigal Shamir. "Chiral fermions from lattice boundaries". In: *Nucl. Phys. B* 406 (1993), pp. 90–106.

[320] Yigal Shamir. "Constraints on the existence of chiral fermions in interacting lattice theories". In: *Phys. Rev. Lett.* 71 (1993), pp. 2691–2694.

[321] Vadim Furman and Yigal Shamir. "Axial symmetries in lattice QCD with Kaplan fermions". In: *Nucl. Phys. B* 439 (1995), pp. 54–78.

[322] Christof Gattringer and Christian B. Lang. *Quantum chromodynamics on the lattice*. Vol. 788. Berlin: Springer, 2010.

[323] Michael Creutz. *Quarks, gluons and lattices*. Cambridge Monographs on Mathematical Physics. Cambridge, UK: Cambridge Univ. Press, June 1985.

[324] Michael Creutz, ed. *Quantum Fields on the Computer*. Singapore: WSP, 1992.

[325] Thomas A. DeGrand and D. Toussaint, eds. *From actions to answers. Proceedings, Theoretical Advanced Study Institute in Elementary Particle Physics, Boulder, USA, June 5-30, 1989*. 1990.

[326] Heinz J. Rothe. *Lattice Gauge Theories : An Introduction (Fourth Edition)*. Vol. 43. World Scientific Publishing Company, 2012.

[327] I. Montvay and G. Munster. *Quantum fields on a lattice*. Cambridge Monographs on Mathematical Physics. Cambridge University Press, Mar. 1997.

[328] M. Creutz. "Monte Carlo Study of Quantized SU(2) Gauge Theory". In: *Phys. Rev. D* 21 (1980), pp. 2308–2315.

[329] Michael Creutz. "Asymptotic Freedom Scales". In: *Phys. Rev. Lett.* 45 (1980). Ed. by J. Julve and M. Ramón-Medrano, p. 313.

[330] Richard Phillips Feynman and Albert Roach Hibbs. *Quantum mechanics and path integrals*. International series in pure and applied physics. New York, NY: McGraw-Hill, 1965.

[331] Mark Byrd. "The Geometry of SU(3)". In: (Aug. 1997).

[332] W. K. Hastings. "Monte Carlo Sampling Methods Using Markov Chains and Their Applications". In: *Biometrika* 57 (1970), pp. 97–109.

[333] S. Duane et al. "Hybrid Monte Carlo". In: *Phys. Lett. B* 195 (1987), pp. 216–222.

[334] Steven A. Gottlieb et al. "Hybrid Molecular Dynamics Algorithms for the Numerical Simulation of Quantum Chromodynamics". In: *Phys. Rev. D* 35 (1987), pp. 2531–2542.

[335] Tatsuhiro Misumi and Jun Yumoto. "Varieties and properties of central-branch Wilson fermions". In: *Phys. Rev. D* 102.3 (2020), p. 034516.

[336] Richard C. Brower et al. "Multigrid for chiral lattice fermions: Domain wall". In: *Phys. Rev. D* 102.9 (2020), p. 094517.

[337] Christof Gattringer and Stefan Solbrig. "Remnant index theorem and low-lying eigenmodes for twisted mass fermions". In: *Phys. Lett. B* 621 (2005), pp. 195–200.

[338] M. A. Clark and A. D. Kennedy. "The RHMC algorithm for two flavors of dynamical staggered fermions". In: *Nucl. Phys. B Proc. Suppl.* 129 (2004). Ed. by S. Aoki et al., pp. 850–852.

[339] E. I. Zolotarev. "Application of elliptic functions to the questions of functions deviating least and most from zero". In: *Zap. Imp. Akad. Nauk. St. Petersburg* 30 (1877). reprinted in his Collected works, Vol. 2, Izdat, Akad. Nauk SSSR, Moscow, 1932, p. 1-59.

[340] Martin Hasenbusch. "Speeding up the hybrid Monte Carlo algorithm for dynamical fermions". In: *Phys. Lett. B* 519 (2001), pp. 177–182.

[341] J. Brannick et al. "Adaptive Multigrid Algorithm for Lattice QCD". In: *Phys. Rev. Lett.* 100 (2008), p. 041601.

[342] R. Babich et al. "Adaptive multigrid algorithm for the lattice Wilson-Dirac operator". In: *Phys. Rev. Lett.* 105 (2010), p. 201602.

[343] Bálint Jóo. personal communication. 2019.

[344] Richard C. Brower et al. "Multigrid algorithm for staggered lattice fermions". In: *Phys. Rev. D* 97.11 (2018), p. 114513.

[345] Peter Boyle and Azusa Yamaguchi. "Comparison of Domain Wall Fermion Multigrid Methods". In: (Mar. 2021).

[346] A. D. Kennedy and Brian Pendleton. "Cost of the generalized hybrid Monte Carlo algorithm for free field theory". In: *Nucl. Phys. B* 607 (2001), pp. 456–510.

[347] Tuan Nguyen et al. "Riemannian Manifold Hybrid Monte Carlo in Lattice QCD". In: *PoS* LATTICE2021 (2022), p. 582.

[348] Hermann Nicolai. "On a New Characterization of Scalar Supersymmetric Theories". In: *Phys. Lett. B* 89 (1980), p. 341.

[349] Antonio Aurilia, H. Nicolai, and P. K. Townsend. "Hidden Constants: The Theta Parameter of QCD and the Cosmological Constant of N=8 Supergravity". In: *Nucl. Phys. B* 176 (1980), pp. 509–522.

[350] Martin Luscher. "Trivializing maps, the Wilson flow and the HMC algorithm". In: *Commun. Math. Phys.* 293 (2010), pp. 899–919.

[351] Ryan Abbott et al. "Sampling QCD field configurations with gauge-equivariant flow models". In: *39th International Symposium on Lattice Field Theory*. Aug. 2022.

[352] Derek B. Leinweber and Ethan Puckridge. *Structure of the QCD Vacuum - CSSM Visualisations - YouTube*. https://youtu.be/WZgZI5vymiM. 2019.

[353] M. C. Chu et al. "Evidence for the role of instantons in Hadron structure from lattice QCD". In: *Nucl. Phys. B Proc. Suppl.* 34 (1994), pp. 170–175.

[354] Margarita Garcia Perez et al. "Instantons from over - improved cooling". In: *Nucl. Phys. B* 413 (1994), pp. 535–552.

[355] Sundance O. Bilson-Thompson, Derek B. Leinweber, and Anthony G. Williams. "Highly improved lattice field strength tensor". In: *Annals Phys.* 304 (2003), pp. 1–21.

[356] Sundance O. Bilson-Thompson et al. "Comparison of $|Q| = 1$ and $|Q| = 2$ gauge-field configurations on the lattice four-torus". In: *Annals Phys.* 311 (2004), pp. 267–287.

[357] Peter J. Moran and Derek B. Leinweber. "Over-improved stout-link smearing". In: *Phys. Rev. D* 77 (2008), p. 094501.

[358] Derek B. Leinweber. *Visual QCD Archive*. http://www.physics.adelaide.edu.au/theory/staff/leinweber/VisualQCD/QCDvacuum/. 2002.

[359] Derek Leinweber. *QCD Lava Lamp*. http://www.physics.adelaide.edu.au/theory/staff/leinweber/VisualQCD/QCDvacuum/su3b600s24t36cool30action.gif. 2004.

[360] Frank Wilczek. *2004 Nobel Prize Lecture*. https://www.nobelprize.org/prizes/physics/2004/wilczek/lecture/. 2004.

[361] Derek B. Leinweber. *Visualizations of Quantum Chromodynamics*. http://www.physics.adelaide.edu.au/theory/staff/leinweber/VisualQCD/Nobel/. 2004.

[362] Frederic D. R. Bonnet, Derek B. Leinweber, and Anthony G. Williams. "General algorithm for improved lattice actions on parallel computing architectures". In: *J. Comput. Phys.* 170 (2001), pp. 1–17.

[363] Derek Muller. *Empty Space is NOT Empty*. https://www.youtube.com/watch?v=S1tFT4smd6E. 2013.

[364] Derek Muller. *Your Mass is NOT From the Higgs Boson*. https://www.youtube.com/watch?v=Ztc6QP-NUqls. 2013.

[365] James Biddle et al. "Publicizing Lattice Field Theory through Visualization". In: *PoS* LATTICE2018 (2019), p. 325.

[366] Dallas DeMartini and Edward Shuryak. "Deconfinement phase transition in the SU(3) instanton-dyon ensemble". In: *Phys. Rev. D* 104.5 (2021), p. 054010.

[367] Frederic D. R. Bonnet et al. "Discretization errors in Landau gauge on the lattice". In: *Austral. J. Phys.* 52 (1999), pp. 939–948.

[368] Peter Hasenfratz, Victor Laliena, and Ferenc Niedermayer. "The Index theorem in QCD with a finite cutoff". In: *Phys. Lett. B* 427 (1998), pp. 125–131.

[369] E. -M. Ilgenfritz et al. "Vacuum structure revealed by over-improved stout-link smearing compared with the overlap analysis for quenched QCD". In: *Phys. Rev. D* 77 (2008). [Erratum: Phys.Rev.D 77, 099902 (2008)], p. 074502.

[370] I. Horvath et al. "The Negativity of the overlap-based topological charge density correlator in pure-glue QCD and the non-integrable nature of its contact part". In: *Phys. Lett. B* 617 (2005), pp. 49–59.

[371] Peter J. Moran and Derek B. Leinweber. "Impact of Dynamical Fermions on QCD Vacuum Structure". In: *Phys. Rev. D* 78 (2008), p. 054506.

[372] I. Horvath et al. "Low dimensional long range topological charge structure in the QCD vacuum". In: *Phys. Rev. D* 68 (2003), p. 114505.

[373] R. Horsley et al. "Isospin splittings of meson and baryon masses from three-flavor lattice QCD + QED". In: *J. Phys.* G43.10 (2016), 10LT02.

[374] R. Horsley et al. "QED effects in the pseudoscalar meson sector". In: *JHEP* 04 (2016), p. 093.

[375] Christof Gattringer and Alexander Schmidt. "Center clusters in the Yang-Mills vacuum". In: *JHEP* 01 (2011), p. 051.

[376] Finn M. Stokes, Waseem Kamleh, and Derek B. Leinweber. "Visualizations of coherent center domains in local Polyakov loops". In: *Annals Phys.* 348 (2014), pp. 341–361.

[377] Colin Morningstar and Mike J. Peardon. "Analytic smearing of SU(3) link variables in lattice QCD". In: *Phys. Rev. D* 69 (2004), p. 054501.

[378] F. Bissey et al. "Gluon flux-tube distribution and linear confinement in baryons". In: *Phys. Rev. D* 76 (2007), p. 114512.

[379] Finn M. Stokes, Waseem Kamleh, and Derek B. Leinweber. *Centre Domains in the QCD Vacuum - Smeared Phase*. https://youtu.be/KkiOQOOb69k. 2014.

[380] Ph. de Forcrand and Oliver Jahn. "The Baryon static potential from lattice QCD". In: *Nucl. Phys. A* 755 (2005). Ed. by M. Guidal et al., pp. 475–480.

[381] Jonathan M. M. Hall et al. "Lattice QCD Evidence that the $\Lambda(1405)$ Resonance is an Antikaon-Nucleon Molecule". In: *Phys. Rev. Lett.* 114.13 (2015), p. 132002.

[382] Jonathan M. M. Hall et al. "Light-quark contributions to the magnetic form factor of the Lambda(1405)". In: *Phys. Rev. D* 95.5 (2017), p. 054510.

[383] Gunnar S. Bali et al. "Observation of string breaking in QCD". In: *Phys. Rev. D* 71 (2005), p. 114513.

[384] John Bulava et al. "String breaking by light and strange quarks in QCD". In: *Phys. Lett. B* 793 (2019), pp. 493–498.

[385] Eberhard Klempt and Alexander Zaitsev. "Glueballs, Hybrids, Multiquarks. Experimental facts versus QCD inspired concepts". In: *Phys. Rept.* 454 (2007), pp. 1–202.

[386] Claude W. Bernard et al. "The QCD spectrum with three quark flavors". In: *Phys. Rev. D* 64 (2001), p. 054506.

[387] C. Aubin et al. "Light hadrons with improved staggered quarks: Approaching the continuum limit". In: *Phys. Rev. D* 70 (2004), p. 094505.

[388] Gerard 't Hooft. "On the Phase Transition Towards Permanent Quark Confinement". In: *Nucl. Phys. B* 138 (1978), pp. 1–25.

[389] Gerard 't Hooft. "A Property of Electric and Magnetic Flux in Nonabelian Gauge Theories". In: *Nucl. Phys. B* 153 (1979), pp. 141–160.

[390] Holger Bech Nielsen and P. Olesen. "A Quantum Liquid Model for the QCD Vacuum: Gauge and Rotational Invariance of Domained and Quantized Homogeneous Color Fields". In: *Nucl. Phys. B* 160 (1979), pp. 380–396.

[391] L. Del Debbio et al. "Center dominance and Z(2) vortices in SU(2) lattice gauge theory". In: *Phys. Rev. D* 55 (1997), pp. 2298–2306.

[392] Manfried Faber, J. Greensite, and S. Olejnik. "Casimir scaling from center vortices: Towards an understanding of the adjoint string tension". In: *Phys. Rev. D* 57 (1998), pp. 2603–2609.

[393] L. Del Debbio et al. "Detection of center vortices in the lattice Yang-Mills vacuum". In: *Phys. Rev. D* 58 (1998), p. 094501.

[394] R. Bertle et al. "The Structure of projected center vortices in lattice gauge theory". In: *JHEP* 03 (1999), p. 019.

[395] Manfried Faber et al. "The Vortex finding property of maximal center (and other) gauges". In: *JHEP* 12 (1999), p. 012.

[396] M. Engelhardt and H. Reinhardt. "Center projection vortices in continuum Yang-Mills theory". In: *Nucl. Phys. B* 567 (2000), p. 249.

[397] J. Greensite. "The Confinement problem in lattice gauge theory". In: *Prog. Part. Nucl. Phys.* 51 (2003), p. 1.

[398] Amalie Trewartha, Waseem Kamleh, and Derek Leinweber. "Evidence that centre vortices underpin dynamical chiral symmetry breaking in SU(3) gauge theory". In: *Phys. Lett.* B747 (2015), pp. 373–377.

[399] Amalie Trewartha, Waseem Kamleh, and Derek Leinweber. "Centre vortex removal restores chiral symmetry". In: *J. Phys.* G44.12 (2017), p. 125002.

[400] Adam Virgili, Waseem Kamleh, and Derek Leinweber. "Impact of centre vortex removal on the Landau-gauge quark propagator in dynamical QCD". In: *PoS* LATTICE2021 (2021), p. 082.

[401] Kurt Langfeld. "Vortex structures in pure SU(3) lattice gauge theory". In: *Phys. Rev.* D69 (2004), p. 014503.

[402] Patrick O. Bowman et al. "Role of center vortices in chiral symmetry breaking in SU(3) gauge theory". In: *Phys. Rev. D* 84 (2011), p. 034501.

[403] James C. Biddle, Waseem Kamleh, and Derek B. Leinweber. "Static quark potential from center vortices in the presence of dynamical fermions". In: *Phys. Rev. D* 106.5 (2022), p. 054505.

[404] Derek Leinweber, James Biddle, and Waseem Kamleh. "Centre vortex structure of QCD-vacuum fields and confinement". In: *SciPost Phys. Proc.* 6 (2022), p. 004.

[405] K. Langfeld, H. Reinhardt, and J. Gattnar. "Gluon propagators and quark confinement". In: *Nucl. Phys. B* 621 (2002), pp. 131–156.

[406] James C. Biddle, Waseem Kamleh, and Derek B. Leinweber. "Gluon propagator on a center-vortex background". In: *Phys. Rev.* D98.9 (2018), p. 094504.

[407] James C. Biddle, Waseem Kamleh, and Derek B. Leinweber. "Impact of dynamical fermions on the center vortex gluon propagator". In: *Phys. Rev. D* 106.1 (2022), p. 014506.
[408] Alan O'Cais et al. "Preconditioning Maximal Center Gauge with Stout Link Smearing in SU(3)". In: *Phys. Rev. D* 82 (2010), p. 114512.
[409] Amalie Trewartha, Waseem Kamleh, and Derek Leinweber. "Connection between center vortices and instantons through gauge-field smoothing". In: *Phys. Rev.* D92.7 (2015), p. 074507.
[410] James C. Biddle, Waseem Kamleh, and Derek B. Leinweber. "Visualization of center vortex structure". In: *Phys. Rev. D* 102.3 (2020), p. 034504.
[411] Elyse-Ann O'Malley et al. "SU(3) centre vortices underpin confinement and dynamical chiral symmetry breaking". In: *Phys. Rev. D* 86 (2012), p. 054503.
[412] James C. Biddle, Waseem Kamleh, and Derek B. Leinweber. "Emergent Structure in QCD". In: *EPJ Web Conf.* 245 (2020). Ed. by C. Doglioni et al., p. 06009.
[413] Derek Leinweber, James Biddle, and Waseem Kamleh. "Impact of Dynamical Fermions on Centre Vortex Structure". In: *PoS* LATTICE2021 (2021), p. 197.
[414] L. Del Debbio et al. "Center dominance and Z(2) vortices in SU(2) lattice gauge theory". In: *Phys. Rev.* D55 (1997), pp. 2298–2306.
[415] Kurt Langfeld, Hugo Reinhardt, and Oliver Tennert. "Confinement and scaling of the vortex vacuum of SU(2) lattice gauge theory". In: *Phys. Lett.* B419 (1998), pp. 317–321.
[416] Falk Bruckmann and Michael Engelhardt. "Writhe of center vortices and topological charge: An Explicit example". In: *Phys. Rev.* D68 (2003), p. 105011.
[417] Michael Engelhardt. "Center vortex model for the infrared sector of SU(3) Yang-Mills theory: Topological susceptibility". In: *Phys. Rev.* D83 (2011), p. 025015.
[418] Michael Engelhardt. "Center vortex model for the infrared sector of Yang-Mills theory: Topological susceptibility". In: *Nucl. Phys.* B585 (2000), p. 614.
[419] S. Aoki et al. "2+1 Flavor Lattice QCD toward the Physical Point". In: *Phys. Rev. D* 79 (2009), p. 034503.
[420] Massimo D'Elia and Maria-Paola Lombardo. "Finite density QCD via imaginary chemical potential". In: *Phys. Rev. D* 67 (2003), p. 014505.
[421] Philippe de Forcrand and Owe Philipsen. "The QCD phase diagram for small densities from imaginary chemical potential". In: *Nucl. Phys. B* 642 (2002), pp. 290–306.
[422] Rajiv V. Gavai and Sourendu Gupta. "Quark number susceptibilities, strangeness and dynamical confinement". In: *Phys. Rev. D* 64 (2001), p. 074506.
[423] C. R. Allton et al. "The QCD thermal phase transition in the presence of a small chemical potential". In: *Phys. Rev. D* 66 (2002), p. 074507.
[424] Larry D. McLerran and Benjamin Svetitsky. "A Monte Carlo Study of SU(2) Yang-Mills Theory at Finite Temperature". In: *Phys. Lett. B* 98 (July 1981), pp. 195–198.
[425] J. Kuti, J. Polonyi, and K. Szlachanyi. "Monte Carlo Study of SU(2) Gauge Theory at Finite Temperature". In: *Phys. Lett. B* 98 (Sept. 1981), pp. 199–204.
[426] John B. Kogut et al. "Deconfinement and Chiral Symmetry Restoration at Finite Temperatures in SU(2) and SU(3) Gauge Theories". In: *Phys. Rev. Lett.* 50 (1983), p. 393.
[427] John C. Collins and M. J. Perry. "Superdense Matter: Neutrons Or Asymptotically Free Quarks?" In: *Phys. Rev. Lett.* 34 (1975), p. 1353.
[428] N. Cabibbo and G. Parisi. "Exponential Hadronic Spectrum and Quark Liberation". In: *Phys. Lett. B* 59 (1975), pp. 67–69.
[429] Gordon Baym. "Confinement of quarks in nuclear matter". In: *Physica A* 96.1-2 (1979). Ed. by S. Deser, pp. 131–135.
[430] Robert D. Pisarski and Frank Wilczek. "Remarks on the Chiral Phase Transition in Chromodynamics". In: *Phys. Rev. D* 29 (1984), pp. 338–341.
[431] Edward V. Shuryak. "Quark-Gluon Plasma and Hadronic Production of Leptons, Photons and Psions". In: *Phys. Lett. B* 78 (1978), p. 150.
[432] K. Kajantie, C. Montonen, and E. Pietarinen. "Phase Transition of SU(3) Gauge Theory at Finite Temperature". In: *Z. Phys. C* 9 (1981), p. 253.
[433] L.G. Yaffe and B. Svetitsky. "First Order Phase Transition in the SU(3) Gauge Theory at Finite Temperature". In: *Phys. Rev. D* 26 (1982), p. 963.
[434] J. Engels et al. "High Temperature SU(2) Gluon Matter on the Lattice". In: *Phys. Lett. B* 101 (1981). Ed. by J. Julve and M. Ramón-Medrano, p. 89.
[435] J. Engels et al. "Gauge Field Thermodynamics for the SU(2) Yang-Mills System". In: *Nucl. Phys. B* 205 (1982), pp. 545–577.

[436] Olaf Kaczmarek and Felix Zantow. "Static quark anti-quark interactions in zero and finite temperature QCD. I. Heavy quark free energies, running coupling and quarkonium binding". In: *Phys. Rev. D* 71 (2005), p. 114510.

[437] Krishna Rajagopal and Frank Wilczek. "Static and dynamic critical phenomena at a second order QCD phase transition". In: *Nucl. Phys. B* 399 (1993), pp. 395–425.

[438] Krishna Rajagopal and Frank Wilczek. "The Condensed matter physics of QCD". In: *At the frontier of particle physics. Handbook of QCD. Vol. 1-3.* Ed. by M. Shifman and Boris Ioffe. Nov. 2000, pp. 2061–2151.

[439] Benjamin Svetitsky and Laurence G. Yaffe. "Critical Behavior at Finite Temperature Confinement Transitions". In: *Nucl. Phys. B* 210 (1982), pp. 423–447.

[440] F. R. Brown et al. "Nature of the Deconfining Phase Transition in SU(3) Lattice Gauge Theory". In: *Phys. Rev. Lett.* 61 (1988), p. 2058.

[441] J. Engels, J. Fingberg, and M. Weber. "Finite Size Scaling Analysis of SU(2) Lattice Gauge Theory in (3+1)-dimensions". In: *Nucl. Phys. B* 332 (1990), pp. 737–759.

[442] H. T. Ding et al. "Chiral Phase Transition Temperature in ( 2+1 )-Flavor QCD". In: *Phys. Rev. Lett.* 123.6 (2019), p. 062002.

[443] Olaf Kaczmarek et al. "Universal scaling properties of QCD close to the chiral limit". In: *Acta Phys. Polon. Supp.* 14 (2021), p. 291.

[444] Francesca Cuteri, Owe Philipsen, and Alessandro Sciarra. "On the order of the QCD chiral phase transition for different numbers of quark flavours". In: *JHEP* 11 (2021), p. 141.

[445] Frank R. Brown et al. "On the existence of a phase transition for QCD with three light quarks". In: *Phys. Rev. Lett.* 65 (1990), pp. 2491–2494.

[446] Lorenzo Dini et al. "Chiral phase transition in three-flavor QCD from lattice QCD". In: *Phys. Rev. D* 105.3 (2022), p. 034510.

[447] Andrey Yu. Kotov, Maria Paola Lombardo, and Anton Trunin. "QCD transition at the physical point, and its scaling window from twisted mass Wilson fermions". In: *Phys. Lett. B* 823 (2021), p. 136749.

[448] A. Bazavov et al. "Chiral crossover in QCD at zero and non-zero chemical potentials". In: *Phys. Lett. B* 795 (2019), pp. 15–21.

[449] Claudio Bonati et al. "Curvature of the chiral pseudocritical line in QCD: Continuum extrapolated results". In: *Phys. Rev. D* 92.5 (2015), p. 054503.

[450] Claudio Bonati et al. "Curvature of the pseudocritical line in QCD: Taylor expansion matches analytic continuation". In: *Phys. Rev. D* 98.5 (2018), p. 054510.

[451] Szabolcs Borsanyi et al. "QCD Crossover at Finite Chemical Potential from Lattice Simulations". In: *Phys. Rev. Lett.* 125.5 (2020), p. 052001.

[452] Riccardo Guida and Jean Zinn-Justin. "Critical exponents of the N vector model". In: *J. Phys. A* 31 (1998), pp. 8103–8121.

[453] Andrea Pelissetto and Ettore Vicari. "Relevance of the axial anomaly at the finite-temperature chiral transition in QCD". In: *Phys. Rev. D* 88.10 (2013), p. 105018.

[454] Anirban Lahiri. "Aspects of finite temperature QCD towards the chiral limit". In: *PoS* LATTICE2021 (2022), p. 003.

[455] Gert Aarts et al. "Hyperons in thermal QCD: A lattice view". In: *Phys. Rev. D* 99.7 (2019), p. 074503.

[456] Alexei Bazavov et al. "Meson screening masses in (2+1)-flavor QCD". In: *Phys. Rev. D* 100.9 (2019), p. 094510.

[457] Simon Dentinger, Olaf Kaczmarek, and Anirban Lahiri. "Screening masses towards chiral limit". In: *Acta Phys. Polon. Supp.* 14 (2021), p. 321.

[458] Carleton E. Detar and John B. Kogut. "The Hadronic Spectrum of the Quark Plasma". In: *Phys. Rev. Lett.* 59 (1987), p. 399.

[459] Carleton E. Detar and John B. Kogut. "Measuring the Hadronic Spectrum of the Quark Plasma". In: *Phys. Rev. D* 36 (1987), p. 2828.

[460] Guido Cossu et al. "Finite temperature study of the axial U(1) symmetry on the lattice with overlap fermion formulation". In: *Phys. Rev. D* 87.11 (2013). [Erratum: Phys.Rev.D 88, 019901 (2013)], p. 114514.

[461] A. Tomiya et al. "Evidence of effective axial U(1) symmetry restoration at high temperature QCD". In: *Phys. Rev. D* 96.3 (2017). [Addendum: Phys.Rev.D 96, 079902 (2017)], p. 034509.

[462] Michael I. Buchoff et al. "QCD chiral transition, U(1)A symmetry and the dirac spectrum using domain wall fermions". In: *Phys. Rev. D* 89.5 (2014), p. 054514.

[463] Viktor Dick et al. "Microscopic origin of $U_A(1)$ symmetry violation in the high temperature phase of QCD". In: *Phys. Rev. D* 91.9 (2015), p. 094504.

[464] Adam Miklos Halasz et al. "On the phase diagram of QCD". In: *Phys. Rev. D* 58 (1998), p. 096007.

[465] Michael Buballa and Stefano Carignano. "Inhomogeneous chiral phases away from the chiral limit". In: *Phys. Lett. B* 791 (2019), pp. 361–366.

[466] A. Bazavov et al. "Equation of state in ( 2+1 )-flavor QCD". In: *Phys. Rev. D* 90 (2014), p. 094503.

[467] Szabocls Borsanyi et al. "Full result for the QCD equation of state with 2+1 flavors". In: *Phys. Lett. B* 730 (2014), pp. 99–104.

[468] D. Bollweg et al. "Taylor expansions and Padé approximants for cumulants of conserved charge fluctuations at nonvanishing chemical potentials". In: *Phys. Rev. D* 105.7 (2022), p. 074511.

[469] C. M. Hung and Edward V. Shuryak. "Hydrodynamics near the QCD phase transition: Looking for the longest lived fireball". In: *Phys. Rev. Lett.* 75 (1995), pp. 4003–4006.

[470] Sz. Borsanyi et al. "QCD equation of state at nonzero chemical potential: continuum results with physical quark masses at order $mu^2$". In: *JHEP* 08 (2012), p. 053.

[471] Swagato Mukherjee and Vladimir Skokov. "Universality driven analytic structure of the QCD crossover: radius of convergence in the baryon chemical potential". In: *Phys. Rev. D* 103.7 (2021), p. L071501.

[472] Sourav Mondal, Swagato Mukherjee, and Prasad Hegde. "Lattice QCD Equation of State for Nonvanishing Chemical Potential by Resumming Taylor Expansions". In: *Phys. Rev. Lett.* 128.2 (2022), p. 022001.

[473] P. Dimopoulos et al. "Contribution to understanding the phase structure of strong interaction matter: Lee-Yang edge singularities from lattice QCD". In: *Phys. Rev. D* 105.3 (2022), p. 034513.

[474] Szabolcs Borsanyi et al. "Resummed lattice QCD equation of state at finite baryon density: Strangeness neutrality and beyond". In: *Phys. Rev. D* 105.11 (2022), p. 114504.

[475] Saumen Datta, Rajiv V. Gavai, and Sourendu Gupta. "Quark number susceptibilities and equation of state at finite chemical potential in staggered QCD with Nt=8". In: *Phys. Rev. D* 95.5 (2017), p. 054512.

[476] R. L. Workman. "Review of Particle Physics". In: *PTEP* 2022 (2022), p. 083C01.

[477] Matthew R. Shepherd, Jozef J. Dudek, and Ryan E. Mitchell. "Searching for the rules that govern hadron construction". In: *Nature* 534.7608 (2016), pp. 487–493.

[478] Andreas S. Kronfeld. "Twenty-first Century Lattice Gauge Theory: Results from the QCD Lagrangian". In: *Ann. Rev. Nucl. Part. Sci.* 62 (2012), pp. 265–284.

[479] S. Aoki et al. "1+1+1 flavor QCD + QED simulation at the physical point". In: *Phys. Rev. D* 86 (2012), p. 034507.

[480] Sz. Borsanyi et al. "Isospin splittings in the light baryon octet from lattice QCD and QED". In: *Phys. Rev. Lett.* 111.25 (2013), p. 252001.

[481] Sz. Borsanyi et al. "Ab initio calculation of the neutron-proton mass difference". In: *Science* 347 (2015), pp. 1452–1455.

[482] R. Horsley et al. "Isospin splittings in the decuplet baryon spectrum from dynamical QCD+QED". In: *J. Phys. G* 46 (2019), p. 115004.

[483] Michael Peardon et al. "A Novel quark-field creation operator construction for hadronic physics in lattice QCD". In: *Phys. Rev. D* 80 (2009), p. 054506.

[484] Jozef J. Dudek et al. "Toward the excited isoscalar meson spectrum from lattice QCD". In: *Phys. Rev. D* 88.9 (2013), p. 094505.

[485] Colin Morningstar et al. "Improved stochastic estimation of quark propagation with Laplacian Heaviside smearing in lattice QCD". In: *Phys. Rev. D* 83 (2011), p. 114505.

[486] Christopher Michael. "Adjoint Sources in Lattice Gauge Theory". In: *Nucl. Phys. B* 259 (1985), pp. 58–76.

[487] Martin Luscher and Ulli Wolff. "How to Calculate the Elastic Scattering Matrix in Two-dimensional Quantum Field Theories by Numerical Simulation". In: *Nucl. Phys. B* 339 (1990), pp. 222–252.

[488] Benoit Blossier et al. "On the generalized eigenvalue method for energies and matrix elements in lattice field theory". In: *JHEP* 04 (2009), p. 094.

[489] Jozef J. Dudek et al. "Highly excited and exotic meson spectrum from dynamical lattice QCD". In: *Phys. Rev. Lett.* 103 (2009), p. 262001.

[490] Jozef J. Dudek et al. "Toward the excited meson spectrum of dynamical QCD". In: *Phys. Rev. D* 82 (2010), p. 034508.

[491] Jozef J. Dudek et al. "Isoscalar meson spectroscopy from lattice QCD". In: *Phys. Rev. D* 83 (2011), p. 111502.

[492] Christopher E. Thomas, Robert G. Edwards, and Jozef J. Dudek. "Helicity operators for mesons

in flight on the lattice". In: *Phys. Rev. D* 85 (2012), p. 014507.

[493] Jozef J. Dudek. "The lightest hybrid meson supermultiplet in QCD". In: *Phys. Rev. D* 84 (2011), p. 074023.

[494] Robert G. Edwards et al. "Excited state baryon spectroscopy from lattice QCD". In: *Phys. Rev. D* 84 (2011), p. 074508.

[495] Jozef J. Dudek and Robert G. Edwards. "Hybrid Baryons in QCD". In: *Phys. Rev. D* 85 (2012), p. 054016.

[496] M. Luscher. "Volume Dependence of the Energy Spectrum in Massive Quantum Field Theories. 2. Scattering States". In: *Commun. Math. Phys.* 105 (1986), pp. 153–188.

[497] Martin Luscher. "Two particle states on a torus and their relation to the scattering matrix". In: *Nucl. Phys. B* 354 (1991), pp. 531–578.

[498] Raul A. Briceno, Jozef J. Dudek, and Ross D. Young. "Scattering processes and resonances from lattice QCD". In: *Rev. Mod. Phys.* 90.2 (2018), p. 025001.

[499] Jozef J. Dudek, Robert G. Edwards, and Christopher E. Thomas. "Energy dependence of the $\rho$ resonance in $\pi\pi$ elastic scattering from lattice QCD". In: *Phys. Rev. D* 87.3 (2013). [Erratum: Phys.Rev.D 90, 099902 (2014)], p. 034505.

[500] Jozef J. Dudek, Robert G. Edwards, and Christopher E. Thomas. "S and D-wave phase shifts in isospin-2 pi pi scattering from lattice QCD". In: *Phys. Rev. D* 86 (2012), p. 034031.

[501] David J. Wilson et al. "Coupled $\pi\pi, K\bar{K}$ scattering in $P$-wave and the $\rho$ resonance from lattice QCD". In: *Phys. Rev. D* 92.9 (2015), p. 094502.

[502] S. Aoki et al. "Lattice QCD Calculation of the rho Meson Decay Width". In: *Phys. Rev. D* 76 (2007), p. 094506.

[503] Xu Feng, Karl Jansen, and Dru B. Renner. "Resonance Parameters of the rho-Meson from Lattice QCD". In: *Phys. Rev. D* 83 (2011), p. 094505.

[504] C. B. Lang et al. "Coupled channel analysis of the rho meson decay in lattice QCD". In: *Phys. Rev. D* 84.5 (2011). [Erratum: Phys.Rev.D 89, 059903 (2014)], p. 054503.

[505] S. Aoki et al. "$\rho$ Meson Decay in 2+1 Flavor Lattice QCD". In: *Phys. Rev. D* 84 (2011), p. 094505.

[506] Craig Pelissier and Andrei Alexandru. "Resonance parameters of the rho-meson from asymmetrical lattices". In: *Phys. Rev. D* 87.1 (2013), p. 014503.

[507] Gunnar S. Bali et al. "$\rho$ and $K^*$ resonances on the lattice at nearly physical quark masses and $N_f = 2$". In: *Phys. Rev. D* 93.5 (2016), p. 054509.

[508] John Bulava et al. "$I = 1$ and $I = 2$ $\pi - \pi$ scattering phase shifts from $N_{\mathrm{f}} = 2 + 1$ lattice QCD". In: *Nucl. Phys. B* 910 (2016), pp. 842–867.

[509] Constantia Alexandrou et al. "$P$-wave $\pi\pi$ scattering and the $\rho$ resonance from lattice QCD". In: *Phys. Rev. D* 96.3 (2017), p. 034525.

[510] Christian Andersen et al. "The $I = 1$ pion-pion scattering amplitude and timelike pion form factor from $N_{\mathrm{f}} = 2 + 1$ lattice QCD". In: *Nucl. Phys. B* 939 (2019), pp. 145–173.

[511] Markus Werner et al. "Hadron-Hadron Interactions from $N_f = 2 + 1 + 1$ Lattice QCD: The $\rho$-resonance". In: *Eur. Phys. J. A* 56.2 (2020), p. 61.

[512] Matthias Fischer et al. "The $\rho$-resonance from $N_f = 2$ lattice QCD including the physical pion mass". In: *Phys. Lett. B* 819 (2021), p. 136449.

[513] C. B. Lang et al. "$K\pi$ scattering for isospin 1/2 and 3/2 in lattice QCD". In: *Phys. Rev. D* 86 (2012), p. 054508.

[514] Ziwen Fu and Kan Fu. "Lattice QCD study on $K^*(892)$ meson decay width". In: *Phys. Rev. D* 86 (2012), p. 094507.

[515] Sasa Prelovsek et al. "K $\pi$ Scattering and the K* Decay width from Lattice QCD". In: *Phys. Rev. D* 88.5 (2013), p. 054508.

[516] Ruairí Brett et al. "Determination of $s$- and $p$-wave $I = 1/2$ $K\pi$ scattering amplitudes in $N_{\mathrm{f}} = 2+1$ lattice QCD". In: *Nucl. Phys. B* 932 (2018), pp. 29–51.

[517] David J. Wilson et al. "The quark-mass dependence of elastic $\pi K$ scattering from QCD". In: *Phys. Rev. Lett.* 123.4 (2019), p. 042002.

[518] Jozef J. Dudek et al. "Phase shift of isospin-2 $\pi\pi$ scattering from lattice QCD". In: *Phys. Rev. D* 83 (2011), p. 071504.

[519] S. R. Beane et al. "The I=2 pipi S-wave Scattering Phase Shift from Lattice QCD". In: *Phys. Rev. D* 85 (2012), p. 034505.

[520] C. Culver et al. "Pion scattering in the isospin $I = 2$ channel from elongated lattices". In: *Phys. Rev. D* 100.3 (2019), p. 034509.

[521] Matthias Fischer et al. "Scattering of two and three physical pions at maximal isospin from lattice QCD". In: *Eur. Phys. J. C* 81.5 (2021), p. 436.

[522] T. Blum et al. "Lattice determination of I=0 and 2 $\pi\pi$ scattering phase shifts with a physical pion mass". In: *Phys. Rev. D* 104.11 (2021), p. 114506.

[523] Raul A. Briceno et al. “Isoscalar $\pi\pi$ scattering and the $\sigma$ meson resonance from QCD”. In: *Phys. Rev. Lett.* 118.2 (2017), p. 022002.

[524] Dehua Guo et al. “Extraction of isoscalar $\pi\pi$ phase-shifts from lattice QCD”. In: *Phys. Rev. D* 98.1 (2018), p. 014507.

[525] Raul A. Briceno et al. “Isoscalar $\pi\pi, K\overline{K}, \eta\eta$ scattering and the $\sigma, f_0, f_2$ mesons from QCD”. In: *Phys. Rev. D* 97.5 (2018), p. 054513.

[526] Daniel Mohler, Sasa Prelovsek, and R. M. Woloshyn. “$D\pi$ scattering and $D$ meson resonances from lattice QCD”. In: *Phys. Rev. D* 87.3 (2013), p. 034501.

[527] Sasa Prelovsek and Luka Leskovec. “Evidence for X(3872) from DD* scattering on the lattice”. In: *Phys. Rev. Lett.* 111 (2013), p. 192001.

[528] C. B. Lang et al. “Ds mesons with DK and D*K scattering near threshold”. In: *Phys. Rev. D* 90.3 (2014), p. 034510.

[529] C. B. Lang et al. “Vector and scalar charmonium resonances with lattice QCD”. In: *JHEP* 09 (2015), p. 089.

[530] C. B. Lang, Daniel Mohler, and S. Prelovsek. “$B_s\pi^+$ scattering and search for X(5568) with lattice QCD”. In: *Phys. Rev. D* 94 (2016), p. 074509.

[531] Graham Moir et al. “Coupled-Channel $D\pi$, $D\eta$ and $D_s\bar{K}$ Scattering from Lattice QCD”. In: *JHEP* 10 (2016), p. 011.

[532] Luke Gayer et al. “Isospin-1/2 D$\pi$ scattering and the lightest $D_0^*$ resonance from lattice QCD”. In: *JHEP* 07 (2021), p. 123.

[533] Liuming Liu et al. “Interactions of charmed mesons with light pseudoscalar mesons from lattice QCD and implications on the nature of the $D_{s0}^*(2317)$”. In: *Phys. Rev. D* 87.1 (2013), p. 014508.

[534] Daniel Mohler et al. “$D_{s0}^*(2317)$ Meson and $D$-Meson-Kaon Scattering from Lattice QCD”. In: *Phys. Rev. Lett.* 111.22 (2013), p. 222001.

[535] Gunnar S. Bali et al. “Masses and decay constants of the $D_{s0}^*(2317)$ and $D_{s1}(2460)$ from $N_f = 2$ lattice QCD close to the physical point”. In: *Phys. Rev. D* 96.7 (2017), p. 074501.

[536] Constantia Alexandrou et al. “Tetraquark interpolating fields in a lattice QCD investigation of the $D_{s0}^*(2317)$ meson”. In: *Phys. Rev. D* 101.3 (2020), p. 034502.

[537] Gavin K. C. Cheung et al. “DK I $= 0, D\overline{K}$I $=$ 0, 1 scattering and the $D_{s0}^*(2317)$ from lattice QCD”. In: *JHEP* 02 (2021), p. 100.

[538] Nicolas Lang and David J. Wilson. “Axial-vector $D_1$ hadrons in $D^*\pi$ scattering from QCD”. In: (May 2022).

[539] C. B. Lang et al. “Pion-nucleon scattering in the Roper channel from lattice QCD”. In: *Phys. Rev. D* 95.1 (2017), p. 014510.

[540] Christian Walther Andersen et al. “Elastic $I = 3/2$ $p$-wave nucleon-pion scattering amplitude and the $\Delta(1232)$ resonance from N$_f$=2+1 lattice QCD”. In: *Phys. Rev. D* 97.1 (2018), p. 014506.

[541] Giorgio Silvi et al. “$P$-wave nucleon-pion scattering amplitude in the $\Delta(1232)$ channel from lattice QCD”. In: *Phys. Rev. D* 103.9 (2021), p. 094508.

[542] Jozef J. Dudek et al. “Resonances in coupled $\pi K - \eta K$ scattering from quantum chromodynamics”. In: *Phys. Rev. Lett.* 113.18 (2014), p. 182001.

[543] David J. Wilson et al. “Resonances in coupled $\pi K, \eta K$ scattering from lattice QCD”. In: *Phys. Rev. D* 91.5 (2015), p. 054008.

[544] Jozef J. Dudek, Robert G. Edwards, and David J. Wilson. “An $a_0$ resonance in strongly coupled $\pi\eta$, $K\overline{K}$ scattering from lattice QCD”. In: *Phys. Rev. D* 93.9 (2016), p. 094506.

[545] Antoni Woss et al. “Dynamically-coupled partial-waves in $\rho\pi$ isospin-2 scattering from lattice QCD”. In: *JHEP* 07 (2018), p. 043.

[546] Antoni J. Woss et al. “$b_1$ resonance in coupled $\pi\omega$, $\pi\phi$ scattering from lattice QCD”. In: *Phys. Rev. D* 100.5 (2019), p. 054506.

[547] Antoni J. Woss et al. “Decays of an exotic 1−+ hybrid meson resonance in QCD”. In: *Phys. Rev. D* 103.5 (2021), p. 054502.

[548] Christopher T. Johnson and Jozef J. Dudek. “Excited $J^{--}$ meson resonances at the SU(3) flavor point from lattice QCD”. In: *Phys. Rev. D* 103.7 (2021), p. 074502.

[549] Sasa Prelovsek et al. “Charmonium-like resonances with J$^{PC}$ = 0$^{++}$, 2$^{++}$ in coupled D$\overline{\text{D}}$, D$_\text{s}\overline{\text{D}}_\text{s}$ scattering on the lattice”. In: *JHEP* 06 (2021), p. 035.

[550] Raul A. Briceno et al. “The resonant $\pi^+\gamma \to \pi^+\pi^0$ amplitude from Quantum Chromodynamics”. In: *Phys. Rev. Lett.* 115 (2015), p. 242001.

[551] Raúl A. Briceño et al. “The $\pi\pi \to \pi\gamma^\star$ amplitude and the resonant $\rho \to \pi\gamma^\star$ transition from lattice QCD”. In: *Phys. Rev. D* 93.11 (2016), p. 114508.

[552] Constantia Alexandrou et al. “$\pi\gamma \to \pi\pi$ transition and the $\rho$ radiative decay width from lattice QCD”. In: *Phys. Rev. D* 98.7 (2018), p. 074502.

[553] Laurent Lellouch and Martin Luscher. “Weak transition matrix elements from finite volume

correlation functions”. In: *Commun. Math. Phys.* 219 (2001), pp. 31–44.

[554] Raúl A. Briceño, Maxwell T. Hansen, and André Walker-Loud. “Multichannel $1 \to 2$ transition amplitudes in a finite volume”. In: *Phys. Rev. D* 91.3 (2015), p. 034501.

[555] Raúl A. Briceño and Maxwell T. Hansen. “Multichannel $0 \to 2$ and $1 \to 2$ transition amplitudes for arbitrary spin particles in a finite volume”. In: *Phys. Rev. D* 92.7 (2015), p. 074509.

[556] Maxwell T. Hansen et al. “Energy-Dependent $\pi^+\pi^+\pi^+$ Scattering Amplitude from QCD”. In: *Phys. Rev. Lett.* 126 (2021), p. 012001.

[557] Alessandro Baroni et al. “Form factors of two-hadron states from a covariant finite-volume formalism”. In: *Phys. Rev. D* 100.3 (2019), p. 034511.

[558] Maxwell T. Hansen and Stephen R. Sharpe. “Lattice QCD and Three-particle Decays of Resonances”. In: *Ann. Rev. Nucl. Part. Sci.* 69 (2019), pp. 65–107.

[559] Tyler D. Blanton, Fernando Romero-López, and Stephen R. Sharpe. “$I = 3$ Three-Pion Scattering Amplitude from Lattice QCD”. In: *Phys. Rev. Lett.* 124.3 (2020), p. 032001.

[560] Maxim Mai et al. “Three-Body Dynamics of the a1(1260) Resonance from Lattice QCD”. In: *Phys. Rev. Lett.* 127.22 (2021), p. 222001.

[561] Ruairí Brett et al. “Three-body interactions from the finite-volume QCD spectrum”. In: *Phys. Rev. D* 104.1 (2021), p. 014501.

[562] Tyler D. Blanton et al. “Interactions of two and three mesons including higher partial waves from lattice QCD”. In: *JHEP* 10 (2021), p. 023.

[563] Robert Hofstadter. “Electron scattering and nuclear structure”. In: *Rev. Mod. Phys.* 28 (1956), pp. 214–254.

[564] Subhasish Basak et al. “Lattice QCD determination of patterns of excited baryon states”. In: *Phys. Rev. D* 76 (2007), p. 074504.

[565] Stefan Sint. “Nonperturbative renormalization in lattice field theory”. In: *Nucl. Phys. B Proc. Suppl.* 94 (2001). Ed. by T. Bhattacharya, R. Gupta, and A. Patel, pp. 79–94.

[566] Martha Constantinou et al. “Parton distributions and lattice QCD calculations: toward 3D structure”. In: (June 2020).

[567] R. Acciarri et al. “Long-Baseline Neutrino Facility (LBNF) and Deep Underground Neutrino Experiment (DUNE): Conceptual Design Report, Volume 2: The Physics Program for DUNE at LBNF”. In: (Dec. 2015).

[568] K. Abe et al. “Hyper-Kamiokande Design Report”. In: (May 2018).

[569] Aaron S. Meyer, André Walker-Loud, and Callum Wilkinson. “Status of Lattice QCD Determination of Nucleon Form Factors and their Relevance for the Few-GeV Neutrino Program”. In: (Jan. 2022).

[570] Sungwoo Park et al. “Precision nucleon charges and form factors using (2+1)-flavor lattice QCD”. In: *Phys. Rev. D* 105.5 (2022), p. 054505.

[571] Yong-Chull Jang et al. “Nucleon electromagnetic form factors in the continuum limit from ( 2+1+1 )-flavor lattice QCD”. In: *Phys. Rev. D* 101.1 (2020), p. 014507.

[572] Richard J. Hill and Gil Paz. “Model independent extraction of the proton charge radius from electron scattering”. In: *Phys. Rev. D* 82 (2010), p. 113005.

[573] Dalibor Djukanovic. “Recent progress on nucleon form factors”. In: *PoS* LATTICE2021 (2022), p. 009.

[574] J. J. Kelly. “Simple parametrization of nucleon form factors”. In: *Phys. Rev. C* 70 (2004), p. 068202.

[575] D. Djukanovic et al. “Isovector electromagnetic form factors of the nucleon from lattice QCD and the proton radius puzzle”. In: *Phys. Rev. D* 103.9 (2021), p. 094522.

[576] C. Alexandrou et al. “Proton and neutron electromagnetic form factors from lattice QCD”. In: *Phys. Rev. D* 100.1 (2019), p. 014509.

[577] Eigo Shintani et al. “Nucleon form factors and root-mean-square radii on a $(10.8\ \mathrm{fm})^4$ lattice at the physical point”. In: *Phys. Rev. D* 99.1 (2019). [Erratum: Phys.Rev.D 102, 019902 (2020)], p. 014510.

[578] Constantia Alexandrou et al. “Model-independent determination of the nucleon charge radius from lattice QCD”. In: *Phys. Rev. D* 101.11 (2020), p. 114504.

[579] Ken-Ichi Ishikawa et al. “Calculation of the derivative of nucleon form factors in Nf=2+1 lattice QCD at Mπ=138 MeV on a (5.5 fm)3 volume”. In: *Phys. Rev. D* 104.7 (2021), p. 074514.

[580] H. Atac et al. “Measurement of the neutron charge radius and the role of its constituents”. In: *Nature Commun.* 12.1 (2021), p. 1759.

[581] J. C. Bernauer et al. “Electric and magnetic form factors of the proton”. In: *Phys. Rev. C* 90.1 (2014), p. 015206.

[582] Aaron S. Meyer et al. “Deuterium target data for precision neutrino-nucleus cross sections”. In: *Phys. Rev. D* 93.11 (2016), p. 113015.

[583] Gunnar S. Bali et al. “Nucleon axial structure from lattice QCD”. In: *JHEP* 05 (2020), p. 126.

[584] Ken-Ichi Ishikawa et al. "Nucleon form factors on a large volume lattice near the physical point in 2+1 flavor QCD". In: *Phys. Rev. D* 98.7 (2018), p. 074510.

[585] Aaron S. Meyer et al. "Nucleon Axial Form Factor from Domain Wall on HISQ". In: *PoS* LATTICE2021 (2022), p. 081.

[586] Nesreen Hasan et al. "Computing the nucleon charge and axial radii directly at $Q^2 = 0$ in lattice QCD". In: *Phys. Rev. D* 97.3 (2018), p. 034504.

[587] C. Alexandrou et al. "Nucleon axial and pseudoscalar form factors from lattice QCD at the physical point". In: *Phys. Rev. D* 103.3 (2021), p. 034509.

[588] Stefano Capitani and Giancarlo Rossi. "Deep inelastic scattering in improved lattice QCD. 1. The First moment of structure functions". In: *Nucl. Phys. B* 433 (1995), pp. 351–389.

[589] Giuseppe Beccarini et al. "Deep inelastic scattering in improved lattice QCD. 2. The second moment of structure functions". In: *Nucl. Phys. B* 456 (1995), pp. 271–295.

[590] M. Gockeler et al. "Lattice operators for moments of the structure functions and their transformation under the hypercubic group". In: *Phys. Rev. D* 54 (1996), pp. 5705–5714.

[591] Stefano Capitani. "Perturbative renormalization of the first two moments of nonsinglet quark distributions with overlap fermions". In: *Nucl. Phys. B* 592 (2001), pp. 183–202.

[592] Stefano Capitani. "Perturbative renormalization of moments of quark momentum, helicity and transversity distributions with overlap and Wilson fermions". In: *Nucl. Phys. B* 597 (2001), pp. 313–336.

[593] Keh-Fei Liu and Shao-Jing Dong. "Origin of difference between anti-d and anti-u partons in the nucleon". In: *Phys. Rev. Lett.* 72 (1994), pp. 1790–1793.

[594] K. F. Liu et al. "Valence QCD: Connecting QCD to the quark model". In: *Phys. Rev. D* 59 (1999), p. 112001.

[595] Keh-Fei Liu. "Parton degrees of freedom from the path integral formalism". In: *Phys. Rev. D* 62 (2000), p. 074501.

[596] William Detmold and C. J. David Lin. "Deep-inelastic scattering and the operator product expansion in lattice QCD". In: *Phys. Rev. D* 73 (2006), p. 014501.

[597] William Detmold et al. "Parton physics from a heavy-quark operator product expansion: Formalism and Wilson coefficients". In: *Phys. Rev. D* 104.7 (2021), p. 074511.

[598] V. Braun and Dieter Müller. "Exclusive processes in position space and the pion distribution amplitude". In: *Eur. Phys. J. C* 55 (2008), pp. 349–361.

[599] Zohreh Davoudi and Martin J. Savage. "Restoration of Rotational Symmetry in the Continuum Limit of Lattice Field Theories". In: *Phys. Rev. D* 86 (2012), p. 054505.

[600] Xiangdong Ji. "Parton Physics on a Euclidean Lattice". In: *Phys. Rev. Lett.* 110 (2013), p. 262002.

[601] Xiangdong Ji. "Parton Physics from Large-Momentum Effective Field Theory". In: *Sci. China Phys. Mech. Astron.* 57 (2014), pp. 1407–1412.

[602] Anatoly Radyushkin. "Nonperturbative Evolution of Parton Quasi-Distributions". In: *Phys. Lett. B* 767 (2017), pp. 314–320.

[603] Yan-Qing Ma and Jian-Wei Qiu. "Extracting Parton Distribution Functions from Lattice QCD Calculations". In: *Phys. Rev. D* 98.7 (2018), p. 074021.

[604] Yan-Qing Ma and Jian-Wei Qiu. "QCD Factorization and PDFs from Lattice QCD Calculation". In: *Int. J. Mod. Phys. Conf. Ser.* 37 (2015). Ed. by Alexei Prokudin, Anatoly Radyushkin, and Leonard Gamberg, p. 1560041.

[605] Yan-Qing Ma and Jian-Wei Qiu. "Exploring Partonic Structure of Hadrons Using ab initio Lattice QCD Calculations". In: *Phys. Rev. Lett.* 120.2 (2018), p. 022003.

[606] A. J. Chambers et al. "Nucleon Structure Functions from Operator Product Expansion on the Lattice". In: *Phys. Rev. Lett.* 118.24 (2017), p. 242001.

[607] Christopher Monahan. "Recent Developments in $x$-dependent Structure Calculations". In: *PoS* LATTICE2018 (2018), p. 018.

[608] Krzysztof Cichy and Martha Constantinou. "A guide to light-cone PDFs from Lattice QCD: an overview of approaches, techniques and results". In: *Adv. High Energy Phys.* 2019 (2019), p. 3036904.

[609] Xiangdong Ji et al. "Large-momentum effective theory". In: *Rev. Mod. Phys.* 93.3 (2021), p. 035005.

[610] Martha Constantinou. "The x-dependence of hadronic parton distributions: A review on the progress of lattice QCD". In: *Eur. Phys. J. A* 57.2 (2021), p. 77.

[611] Krzysztof Cichy. "Progress in $x$-dependent partonic distributions from lattice QCD". In: *PoS* LATTICE2021 (2022), p. 017.

[612] Constantia Alexandrou et al. "Light-Cone Parton Distribution Functions from Lattice QCD". In: *Phys. Rev. Lett.* 121.11 (2018), p. 112001.

[613] Constantia Alexandrou et al. "Systematic uncertainties in parton distribution functions from lattice QCD simulations at the physical point". In: *Phys. Rev. D* 99.11 (2019), p. 114504.

[614] Bálint Joó et al. "Parton Distribution Functions from Ioffe Time Pseudodistributions from Lattice Calculations: Approaching the Physical Point". In: *Phys. Rev. Lett.* 125.23 (2020), p. 232003.

[615] Manjunath Bhat et al. "Parton distribution functions from lattice QCD at physical quark masses via the pseudo-distribution approach". In: (May 2020).

[616] Huey-Wen Lin et al. "Proton Isovector Helicity Distribution on the Lattice at Physical Pion Mass". In: *Phys. Rev. Lett.* 121.24 (2018), p. 242003.

[617] Constantia Alexandrou et al. "Transversity parton distribution functions from lattice QCD". In: *Phys. Rev. D* 98.9 (2018), p. 091503.

[618] Colin Egerer et al. "Transversity parton distribution function of the nucleon using the pseudodistribution approach". In: *Phys. Rev. D* 105.3 (2022), p. 034507.

[619] V. Braun, P. Gornicki, and L. Mankiewicz. "Ioffe - time distributions instead of parton momentum distributions in description of deep inelastic scattering". In: *Phys. Rev. D* 51 (1995), pp. 6036–6051.

[620] V. N. Gribov, B. L. Ioffe, and I. Ya. Pomeranchuk. "On the total annihilation cross-section of electron - positron pairs into hadrons at high-energies". In: *Yad. Fiz.* 6 (1967), p. 587.

[621] V. N. Gribov, B. L. Ioffe, and I. Ya. Pomeranchuk. "What is the range of interactions at high-energies". In: *Yad. Fiz.* 2 (1965), pp. 768–776.

[622] Tanjib Khan et al. "Unpolarized gluon distribution in the nucleon from lattice quantum chromodynamics". In: *Phys. Rev. D* 104.9 (2021), p. 094516.

[623] Zhouyou Fan and Huey-Wen Lin. "Gluon parton distribution of the pion from lattice QCD". In: *Phys. Lett. B* 823 (2021), p. 136778.

[624] Colin Egerer et al. "Distillation at High-Momentum". In: *Phys. Rev. D* 103.3 (2021), p. 034502.

[625] Joseph Karpie et al. "The continuum and leading twist limits of parton distribution functions in lattice QCD". In: *JHEP* 11 (2021), p. 024.

[626] Tie-Jiun Hou et al. "New CTEQ global analysis of quantum chromodynamics with high-precision data from the LHC". In: *Phys. Rev. D* 103.1 (2021), p. 014013.

[627] Richard D. Ball et al. "Parton distributions from high-precision collider data". In: *Eur. Phys. J. C* 77.10 (2017), p. 663.

[628] Eric Moffat et al. "Simultaneous Monte Carlo analysis of parton densities and fragmentation functions". In: *Phys. Rev. D* 104.1 (2021), p. 016015.

[629] Constantia Alexandrou et al. "Flavor decomposition for the proton helicity parton distribution functions". In: *Phys. Rev. Lett.* 126.10 (2021), p. 102003.

[630] Constantia Alexandrou et al. "Flavor decomposition of the nucleon unpolarized, helicity, and transversity parton distribution functions from lattice QCD simulations". In: *Phys. Rev. D* 104.5 (2021), p. 054503.

[631] Rui Zhang, Huey-Wen Lin, and Boram Yoon. "Probing nucleon strange and charm distributions with lattice QCD". In: *Phys. Rev. D* 104.9 (2021), p. 094511.

[632] Shohini Bhattacharya et al. "Generalized Parton Distributions from Lattice QCD with Asymmetric Momentum Transfer: Unpolarized Quarks". In: (Sept. 2022).

[633] Constantia Alexandrou et al. "Transversity GPDs of the proton from lattice QCD". In: *Phys. Rev. D* 105.3 (2022), p. 034501.

[634] Huey-Wen Lin. "Nucleon Tomography and Generalized Parton Distribution at Physical Pion Mass from Lattice QCD". In: *Phys. Rev. Lett.* 127.18 (2021), p. 182001.

[635] Phiala Shanahan, Michael Wagman, and Yong Zhao. "Collins-Soper kernel for TMD evolution from lattice QCD". In: *Phys. Rev. D* 102.1 (2020), p. 014511.

[636] Qi-An Zhang et al. "Lattice-QCD Calculations of TMD Soft Function Through Large-Momentum Effective Theory". In: *Phys. Rev. Lett.* 125.19 (2020), p. 192001.

[637] Maximilian Schlemmer et al. "Determination of the Collins-Soper Kernel from Lattice QCD". In: *JHEP* 08 (2021), p. 004.

[638] Yuan Li et al. "Lattice QCD Study of Transverse-Momentum Dependent Soft Function". In: *Phys. Rev. Lett.* 128.6 (2022), p. 062002.

[639] Phiala Shanahan, Michael Wagman, and Yong Zhao. "Lattice QCD calculation of the Collins-Soper kernel from quasi-TMDPDFs". In: *Phys. Rev. D* 104.11 (2021), p. 114502.

[640] Shohini Bhattacharya et al. "Insights on proton structure from lattice QCD: The twist-3 parton

distribution function $g_T(x)$". In: *Phys. Rev. D* 102.11 (2020), p. 111501.
[641] Markus A. Ebert, Iain W. Stewart, and Yong Zhao. "Determining the Nonperturbative Collins-Soper Kernel From Lattice QCD". In: *Phys. Rev. D* 99.3 (2019), p. 034505.
[642] I. I. Balitsky and Vladimir M. Braun. "Evolution Equations for QCD String Operators". In: *Nucl. Phys. B* 311 (1989), pp. 541–584.
[643] Matthias Burkardt. "Transverse force on quarks in deep-inelastic scattering". In: *Phys. Rev. D* 88 (2013), p. 114502.
[644] Chien-Yeah Seng. "Relating hadronic CP-violation to higher-twist distributions". In: *Phys. Rev. Lett.* 122.7 (2019), p. 072001.
[645] Shohini Bhattacharya et al. "One-loop matching for the twist-3 parton distribution $g_T(x)$". In: *Phys. Rev. D* 102.3 (2020), p. 034005.
[646] Shohini Bhattacharya et al. "The role of zero-mode contributions in the matching for the twist-3 PDFs $e(x)$ and $h_L(x)$". In: *Phys. Rev. D* 102 (2020), p. 114025.
[647] Shohini Bhattacharya et al. "Parton distribution functions beyond leading twist from lattice QCD: The hL(x) case". In: *Phys. Rev. D* 104.11 (2021), p. 114510.
[648] S. Wandzura and Frank Wilczek. "Sum Rules for Spin Dependent Electroproduction: Test of Relativistic Constituent Quarks". In: *Phys. Lett.* 72B (1977), pp. 195–198.
[649] Vladimir M. Braun, Yao Ji, and Alexey Vladimirov. "QCD factorization for twist-three axial-vector parton quasidistributions". In: *JHEP* 05 (2021), p. 086.
[650] Vladimir M. Braun, Yao Ji, and Alexey Vladimirov. "QCD factorization for chiral-odd parton quasi- and pseudo-distributions". In: *JHEP* 10 (2021), p. 087.
[651] Matthew Wingate. "Quark flavor physics and lattice QCD". In: *Eur. Phys. J. A* 57.7 (2021), p. 239.
[652] C. Aubin et al. "Light pseudoscalar decay constants, quark masses, and low energy constants from three-flavor lattice QCD". In: *Phys. Rev. D* 70 (2004), p. 114501.
[653] Gilberto Colangelo, Stephan Durr, and Christoph Haefeli. "Finite volume effects for meson masses and decay constants". In: *Nucl. Phys. B* 721 (2005), pp. 136–174.
[654] C. McNeile et al. "High-Precision $f_{B_s}$ and HQET from Relativistic Lattice QCD". In: *Phys. Rev. D* 85 (2012), p. 031503.
[655] A. Bazavov et al. "$B$ and $D$ meson leptonic decay constants from four-flavor lattice QCD". In: *Phys. Rev. D* 98.7 (2018), p. 074512.
[656] William J. Marciano. "Precise determination of $|V(us)|$ from lattice calculations of pseudoscalar decay constants". In: *Phys. Rev. Lett.* 93 (2004), p. 231803.
[657] R. J. Dowdall et al. "$V_{us}$ from pi and K decay constants in full lattice QCD with physical u, d, s and c quarks". In: *Phys. Rev. D* 88 (2013), p. 074504.
[658] N. Carrasco et al. "Leptonic decay constants $f_K, f_D$, and $f_{D_s}$ with $N_f = 2 + 1 + 1$ twisted-mass lattice QCD". In: *Phys. Rev. D* 91.5 (2015), p. 054507.
[659] Nolan Miller et al. "$F_K/F_\pi$ from Möbius Domain-Wall fermions solved on gradient-flowed HISQ ensembles". In: *Phys. Rev. D* 102.3 (2020), p. 034507.
[660] M. Di Carlo et al. "Light-meson leptonic decay rates in lattice QCD+QED". In: *Phys. Rev. D* 100.3 (2019), p. 034514.
[661] C. McNeile et al. "Heavy meson masses and decay constants from relativistic heavy quarks in full lattice QCD". In: *Phys. Rev. D* 86 (2012), p. 074503.
[662] V. Lubicz, A. Melis, and S. Simula. "Masses and decay constants of $\mathrm{D^*}_{(s)}$ and $\mathrm{B^*}_{(s)}$ mesons with $\mathrm{N}_f = 2+1+1$ twisted mass fermions". In: *Phys. Rev. D* 96.3 (2017), p. 034524.
[663] B. Colquhoun et al. "B-meson decay constants: a more complete picture from full lattice QCD". In: *Phys. Rev. D* 91.11 (2015), p. 114509.
[664] Bipasha Chakraborty et al. "Nonperturbative comparison of clover and highly improved staggered quarks in lattice QCD and the properties of the $\phi$ meson". In: *Phys. Rev. D* 96.7 (2017), p. 074502.
[665] D. Hatton et al. "Bottomonium precision tests from full lattice QCD: Hyperfine splitting, $\Upsilon$ leptonic width, and b quark contribution to $e^+e^- \to$ hadrons". In: *Phys. Rev. D* 103.5 (2021), p. 054512.
[666] C. T. H. Davies et al. "Update: Precision $D_s$ decay constant from full lattice QCD using very fine lattices". In: *Phys. Rev. D* 82 (2010), p. 114504.
[667] Estia Eichten and Brian Russell Hill. "An Effective Field Theory for the Calculation of Matrix Elements Involving Heavy Quarks". In: *Phys. Lett. B* 234 (1990), pp. 511–516.
[668] Aida X. El-Khadra, Andreas S. Kronfeld, and Paul B. Mackenzie. "Massive fermions in lattice gauge theory". In: *Phys. Rev. D* 55 (1997), pp. 3933–3957.

[669] Estia Eichten and Brian Russell Hill. "Renormalization of Heavy - Light Bilinears and F($B$) for Wilson Fermions". In: *Phys. Lett. B* 240 (1990), pp. 193–199.

[670] Colin J. Morningstar and J. Shigemitsu. "One loop matching of lattice and continuum heavy light axial vector currents using NRQCD". In: *Phys. Rev. D* 57 (1998), pp. 6741–6751.

[671] Junpei Harada et al. "Application of heavy quark effective theory to lattice QCD. 2. Radiative corrections to heavy light currents". In: *Phys. Rev. D* 65 (2002). [Erratum: Phys.Rev.D 71, 019903 (2005)], p. 094513.

[672] A. Bussone et al. "Mass of the b quark and B - meson decay constants from $\mathrm{N}_f$=2+1+1 twisted-mass lattice QCD". In: *Phys. Rev. D* 93.11 (2016), p. 114505.

[673] R. J. Dowdall et al. "B-Meson Decay Constants from Improved Lattice Nonrelativistic QCD with Physical u, d, s, and c Quarks". In: *Phys. Rev. Lett.* 110.22 (2013), p. 222003.

[674] Aneesh V. Manohar and Mark B. Wise. *Heavy quark physics.* Vol. 10. Cambridge University Press, 2000.

[675] G. C. Donald et al. "Prediction of the $D_s^*$ width from a calculation of its radiative decay in full lattice QCD". In: *Phys. Rev. Lett.* 112 (2014), p. 212002.

[676] Colin J. Morningstar and J. Shigemitsu. "Perturbative matching of lattice and continuum heavy light currents with NRQCD heavy quarks". In: *Phys. Rev. D* 59 (1999), p. 094504.

[677] G. Martinelli et al. "A General method for nonperturbative renormalization of lattice operators". In: *Nucl. Phys. B* 445 (1995), pp. 81–108.

[678] C. Sturm et al. "Renormalization of quark bilinear operators in a momentum-subtraction scheme with a nonexceptional subtraction point". In: *Phys. Rev. D* 80 (2009), p. 014501.

[679] D. Hatton et al. "Renormalizing vector currents in lattice QCD using momentum-subtraction schemes". In: *Phys. Rev. D* 100.11 (2019), p. 114513.

[680] F. Gabbiani et al. "A Complete analysis of FCNC and CP constraints in general SUSY extensions of the standard model". In: *Nucl. Phys. B* 477 (1996), pp. 321–352.

[681] Jack Laiho and Ruth S. Van de Water. "Pseudoscalar decay constants, light-quark masses, and $B_K$ from mixed-action lattice QCD". In: *PoS* LATTICE2011 (2011). Ed. by Pavlos Vranas, p. 293.

[682] S. Durr et al. "Precision computation of the kaon bag parameter". In: *Phys. Lett. B* 705 (2011), pp. 477–481.

[683] Benjamin J. Choi et al. "Kaon BSM B-parameters using improved staggered fermions from $N_f = 2+1$ unquenched QCD". In: *Phys. Rev. D* 93.1 (2016), p. 014511.

[684] N. Carrasco et al. "$\Delta$S=2 and $\Delta$C=2 bag parameters in the standard model and beyond from $\mathrm{N}_f$=2+1+1 twisted-mass lattice QCD". In: *Phys. Rev. D* 92.3 (2015), p. 034516.

[685] R. J. Dowdall et al. "Neutral B-meson mixing from full lattice QCD at the physical point". In: *Phys. Rev. D* 100.9 (2019), p. 094508.

[686] Christopher Monahan et al. "Matching lattice and continuum four-fermion operators with nonrelativistic QCD and highly improved staggered quarks". In: *Phys. Rev. D* 90.5 (2014), p. 054015.

[687] Elvira Gamiz et al. "Neutral $B$ Meson Mixing in Unquenched Lattice QCD". In: *Phys. Rev. D* 80 (2009), p. 014503.

[688] Yasumichi Aoki et al. "Neutral $B$ meson mixings and $B$ meson decay constants with static heavy and domain-wall light quarks". In: *Phys. Rev. D* 91.11 (2015), p. 114505.

[689] A. Bazavov et al. "$B^0_{(s)}$-mixing matrix elements from lattice QCD for the Standard Model and beyond". In: *Phys. Rev. D* 93.11 (2016), p. 113016.

[690] A. Bazavov et al. "Short-distance matrix elements for $D^0$-meson mixing for $N_f = 2+1$ lattice QCD". In: *Phys. Rev. D* 97.3 (2018), p. 034513.

[691] T. Blum et al. "$K \to \pi\pi$ $\Delta I = 3/2$ decay amplitude in the continuum limit". In: *Phys. Rev. D* 91.7 (2015), p. 074502.

[692] R. Abbott et al. "Direct CP violation and the $\Delta I = 1/2$ rule in $K \to \pi\pi$ decay from the standard model". In: *Phys. Rev. D* 102.5 (2020), p. 054509.

[693] P. A. Boyle et al. "Emerging understanding of the $\Delta I = 1/2$ Rule from Lattice QCD". In: *Phys. Rev. Lett.* 110.15 (2013), p. 152001.

[694] Z. Bai et al. "$K_L - K_S$ Mass Difference from Lattice QCD". In: *Phys. Rev. Lett.* 113 (2014), p. 112003.

[695] Asobu Suzuki et al. "Four quark operators for kaon bag parameter with gradient flow". In: *Phys. Rev. D* 102.3 (2020), p. 034508.

[696] Heechang Na et al. "The $D \to K, l\nu$ Semileptonic Decay Scalar Form Factor and $|V_{cs}|$ from Lattice QCD". In: *Phys. Rev. D* 82 (2010), p. 114506.

[697] Bipasha Chakraborty et al. "Improved Vcs determination using precise lattice QCD form fac-

tors for D→K$\ell\nu$". In: *Phys. Rev. D* 104.3 (2021), p. 034505.
[698] Chien-Yeah Seng et al. "Update on |Vus| and |Vus/Vud| from semileptonic kaon and pion decays". In: *Phys. Rev. D* 105.1 (2022), p. 013005.
[699] N. Carrasco et al. "$K \to \pi$ semileptonic form factors with $N_f = 2{+}1{+}1$ twisted mass fermions". In: *Phys. Rev. D* 93.11 (2016), p. 114512.
[700] A. Bazavov et al. "$|V_{us}|$ from $K_{\ell 3}$ decay and four-flavor lattice QCD". In: *Phys. Rev. D* 99.11 (2019), p. 114509.
[701] Paulo F. Bedaque. "Aharonov-Bohm effect and nucleon nucleon phase shifts on the lattice". In: *Phys. Lett. B* 593 (2004), pp. 82–88.
[702] L. Riggio, G. Salerno, and S. Simula. "Extraction of $|V_{cd}|$ and $|V_{cs}|$ from experimental decay rates using lattice QCD $D \to \pi(K)\ell\nu$ form factors". In: *Eur. Phys. J. C* 78.6 (2018), p. 501.
[703] W. G. Parrott, C. Bouchard, and C. T. H. Davies. "$B \to K$ and $D \to K$ form factors from fully relativistic lattice QCD". In: (July 2022).
[704] J. M. Flynn et al. "$B \to \pi\ell\nu$ and $B_s \to K\ell\nu$ form factors and $|V_{ub}|$ from 2+1-flavor lattice QCD with domain-wall light quarks and relativistic heavy quarks". In: *Phys. Rev. D* 91.7 (2015), p. 074510.
[705] Jon A. Bailey et al. "$|V_{ub}|$ from $B \to \pi\ell\nu$ decays and (2+1)-flavor lattice QCD". In: *Phys. Rev. D* 92.1 (2015), p. 014024.
[706] Jon A. Bailey et al. "Update of $|V_{cb}|$ from the $\bar{B} \to D^*\ell\bar{\nu}$ form factor at zero recoil with three-flavor lattice QCD". In: *Phys. Rev. D* 89.11 (2014), p. 114504.
[707] Judd Harrison, Christine Davies, and Matthew Wingate. "Lattice QCD calculation of the $B_{(s)} \to D^*_{(s)}$ form factors at zero recoil and implications for $|V_{cb}|$". In: *Phys. Rev. D* 97.5 (2018), p. 054502.
[708] A. Bazavov et al. "Semileptonic form factors for $B \to D^*\ell\nu$ at nonzero recoil from 2 + 1-flavor lattice QCD". In: (May 2021).
[709] E. McLean et al. "$B_s \to D_s\ell\nu$ Form Factors for the full $q^2$ range from Lattice QCD with non-perturbatively normalized currents". In: *Phys. Rev. D* 101.7 (2020), p. 074513.
[710] Judd Harrison and Christine T. H. Davies. "$B_s \to D_s^*$form factors for the full $q^2$ range from lattice QCD". In: *Phys. Rev. D* 105.9 (2022), p. 094506.
[711] Brian Colquhoun et al. "Form factors of $B \to \pi\ell\nu$ and a determination of $|V_{ub}|$ with Möbius domain-wall-fermions". In: (Mar. 2022).
[712] Andreas S. Kronfeld et al. "Lattice QCD and Particle Physics". In: (July 2022).
[713] Paolo Gambino et al. "Lattice QCD study of inclusive semileptonic decays of heavy mesons". In: *JHEP* 07 (2022), p. 083.
[714] William Detmold, Christoph Lehner, and Stefan Meinel. "$\Lambda_b \to p\ell^-\bar{\nu}_\ell$ and $\Lambda_b \to \Lambda_c\ell^-\bar{\nu}_\ell$ form factors from lattice QCD with relativistic heavy quarks". In: *Phys. Rev. D* 92.3 (2015), p. 034503.
[715] Roel Aaij et al. "Determination of the quark coupling strength $|V_{ub}|$ using baryonic decays". In: *Nature Phys.* 11 (2015), pp. 743–747.
[716] M. Gell-Mann. "A Schematic Model of Baryons and Mesons". In: *Phys. Lett.* 8 (1964), pp. 214–215.
[717] G. Zwieg. "an SU3 model for strong interactions symmetry and its breaking". In: (1964).
[718] G. Zweig. "Origins of the Quark Model". In: (1980).
[719] R.H. Dalitz. *in Proceedings of the Thirteenth International Conference on High Energy Physics, Berkeley.* Berkeley, CA: University of California Press, 1967, p. 215.
[720] C. Becchi and G. Morpurgo. "Vanishing of the E2 part of the $N_{33} \to N + \gamma$ amplitude in the non-relativistic quark model of "elementary" particles". In: *Phys. Lett.* 17 (1965), pp. 352–354.
[721] H. R. Rubinstein, F. Scheck, and R. H. Socolow. "Electromagnetic Properties of Hadrons in the Quark Model". In: *Phys. Rev.* 154 (1967), pp. 1608–1616.
[722] H. J. Lipkin and F. Scheck. "Quark model for forward scattering amplitudes". In: *Phys. Rev. Lett.* 16 (1966), pp. 71–75.
[723] J.J.J. Kokkedee. *The Quark Model.* New York: W.A. Benjamin, 1969.
[724] R.H. Dalitz. *High Energy Physics.* page 251. New York, 1965.
[725] R. G. Moorhouse. "Photoproduction of N* resonances in the quark model". In: *Phys. Rev. Lett.* 16 (1966), pp. 772–774.
[726] R. Van Royen and V. F. Weisskopf. "Hadron Decay Processes and the Quark Model". In: *Nuovo Cim. A* 50 (1967). [Erratum: Nuovo Cim.A 51, 583 (1967)], pp. 617–645.
[727] L. A. Copley, G. Karl, and E. Obryk. "Single pion photoproduction in the quark model". In: *Nucl. Phys. B* 13 (1969), pp. 303–319.
[728] A. De Rujula, Howard Georgi, and S. L. Glashow. "Hadron Masses in a Gauge Theory". In: *Phys. Rev. D* 12 (1975), pp. 147–162.

[729] Nathan Isgur and Gabriel Karl. “P Wave Baryons in the Quark Model”. In: *Phys. Rev. D* 18 (1978), p. 4187.

[730] Howard J. Schnitzer. “Inverted Charmed Meson Multiplets as a Test for Scalar Confinement”. In: *Phys. Lett. B* 76 (1978), pp. 461–465.

[731] W. Buchmuller. “Fine and Hyperfine Structure of Quarkonia”. In: *Phys. Lett. B* 112 (1982), pp. 479–483.

[732] L. Micu. “Decay rates of meson resonances in a quark model”. In: *Nucl. Phys. B* 10 (1969), pp. 521–526.

[733] H. R. Rubinstein and I. Talmi. “Quark model decay rates for spin $2^+$ mesons”. In: *Phys. Lett.* 23 (1966), pp. 693–695.

[734] E. Eichten et al. “Charmonium: The Model”. In: *Phys. Rev. D* 17 (1978). [Erratum: Phys.Rev.D 21, 313 (1980)], p. 3090.

[735] S. Godfrey and Nathan Isgur. “Mesons in a Relativized Quark Model with Chromodynamics”. In: *Phys. Rev. D* 32 (1985), pp. 189–231.

[736] Simon Capstick and Nathan Isgur. “Baryons in a relativized quark model with chromodynamics”. In: *Phys. Rev. D* 34.9 (1986), pp. 2809–2835.

[737] D. P. Stanley and D. Robson. “Nonperturbative Potential Model for Light and Heavy Quark anti-Quark Systems”. In: *Phys. Rev. D* 21 (1980), pp. 3180–3196.

[738] A. Chodos et al. “A New Extended Model of Hadrons”. In: *Phys. Rev. D* 9 (1974), pp. 3471–3495.

[739] William A. Bardeen et al. “Heavy Quarks and Strong Binding: A Field Theory of Hadron Structure”. In: *Phys. Rev. D* 11 (1975), p. 1094.

[740] Carleton E. DeTar and John F. Donoghue. “BAG MODELS OF HADRONS”. In: *Ann. Rev. Nucl. Part. Sci.* 33 (1983), pp. 235–264.

[741] R. Friedberg and T. D. Lee. “QCD and the Soliton Model of Hadrons”. In: *Phys. Rev. D* 18 (1978), p. 2623.

[742] Anthony William Thomas, S. Theberge, and Gerald A. Miller. “The Cloudy Bag Model of the Nucleon”. In: *Phys. Rev. D* 24 (1981), p. 216.

[743] Anthony William Thomas. “Chiral Symmetry and the Bag Model: A New Starting Point for Nuclear Physics”. In: *Adv. Nucl. Phys.* 13 (1984), pp. 1–137.

[744] Alan Chodos and Charles B. Thorn. “Chiral Hedgehogs in the Bag Theory”. In: *Phys. Rev. D* 12 (1975), p. 2733.

[745] Thomas A. DeGrand and R. L. Jaffe. “Excited States of Confined Quarks”. In: *Annals Phys.* 100 (1976), p. 425.

[746] P. Hasenfratz et al. “The Effects of Colored Glue in the QCD Motivated Bag of Heavy Quark - anti-Quark Systems”. In: *Phys. Lett. B* 95 (1980), pp. 299–305.

[747] R. L. Jaffe and K. Johnson. “Unconventional States of Confined Quarks and Gluons”. In: *Phys. Lett. B* 60 (1976), pp. 201–204.

[748] Ted Barnes, F. E. Close, and S. Monaghan. “Hyperfine Splittings of Bag Model Gluonia”. In: *Nucl. Phys. B* 198 (1982), pp. 380–406.

[749] Ted F. E. Barnes. “Quarks, gluons, bags, and hadrons”. PhD thesis. Caltech, 1977.

[750] Michael S. Chanowitz and Stephen R. Sharpe. “Hybrids: Mixed States of Quarks and Gluons”. In: *Nucl. Phys. B* 222 (1983). [Erratum: Nucl.Phys.B 228, 588–588 (1983)], pp. 211–244.

[751] Ted Barnes, F. E. Close, and F. de Viron. “Q anti-Q G Hermaphrodite Mesons in the MIT Bag Model”. In: *Nucl. Phys. B* 224 (1983), p. 241.

[752] Ted Barnes and F. E. Close. “A LIGHT EXOTIC q anti-q g HERMAPHRODITE MESON?” In: *Phys. Lett. B* 116 (1982), pp. 365–368.

[753] R. L. Jaffe. “Exotica”. In: *Phys. Rept.* 409 (2005). Ed. by Teiji Kunihiro et al., pp. 1–45.

[754] Masakuni Ida and Reido Kobayashi. “Baryon resonances in a quark model”. In: *Prog. Theor. Phys.* 36 (1966), p. 846.

[755] D. B. Lichtenberg and L. J. Tassie. “Baryon Mass Splitting in a Boson-Fermion Model”. In: *Phys. Rev.* 155 (1967), pp. 1601–1606.

[756] Mauro Anselmino et al. “Diquarks”. In: *Rev. Mod. Phys.* 65 (1993), pp. 1199–1234.

[757] D. B. Lichtenberg. In: *Phys. Rev.* 178 (1969), p. 2197.

[758] M. Yu. Barabanov et al. “Diquark correlations in hadron physics: Origin, impact and evidence”. In: *Prog. Part. Nucl. Phys.* 116 (2021), p. 103835.

[759] Anthony Francis et al. “Good and bad diquark properties and spatial correlations in lattice QCD”. In: *Rev. Mex. Fis. Suppl.* 3.3 (2022), p. 0308082.

[760] L. Maiani et al. “Diquark-antidiquark states with hidden or open charm and the nature of $X(3872)$”. In: *Phys. Rev. D* 71 (1 Jan. 2005), p. 014028.

[761] Martin J. Savage and Mark B. Wise. “Spectrum of baryons with two heavy quarks”. In: *Phys. Lett. B* 248 (1990), pp. 177–180.

[762] Anthony Francis et al. “Lattice Prediction for Deeply Bound Doubly Heavy Tetraquarks”. In: *Phys. Rev. Lett.* 118.14 (2017), p. 142001.

[763] N. Isgur. “Nuclear Physics from the Quark Model with Chromodynamics”. In: *Acta Phys. Austriaca Suppl.* 27 (1985). Ed. by H. Mitter and Willibald Plessas, pp. 177–266.

[764] Adam P. Szczepaniak and Eric S. Swanson. “Chiral extrapolation, renormalization, and the viability of the quark model”. In: *Phys. Rev. Lett.* 87 (2001), p. 072001.

[765] Pieter Maris and Craig D. Roberts. “Dyson-Schwinger equations: A Tool for hadron physics”. In: *Int. J. Mod. Phys. E* 12 (2003), pp. 297–365.

[766] Adam P. Szczepaniak and Eric S. Swanson. “From current to constituent quarks: A Renormalization group improved Hamiltonian based description of hadrons”. In: *Phys. Rev. D* 55 (1997), pp. 1578–1591.

[767] Adam P. Szczepaniak and Eric S. Swanson. “On the Dirac structure of confinement”. In: *Phys. Rev. D* 55 (1997), pp. 3987–3993.

[768] E. Eichten and F. Feinberg. “Spin Dependent Forces in QCD”. In: *Phys. Rev. D* 23 (1981), p. 2724.

[769] Yoshiaki Koma and Miho Koma. “Spin-dependent potentials from lattice QCD”. In: *Nucl. Phys. B* 769 (2007), pp. 79–107.

[770] Olga Lakhina and Eric S. Swanson. “Dynamic properties of charmonium”. In: *Phys. Rev. D* 74 (2006), p. 014012.

[771] K. J. Juge, J. Kuti, and C. J. Morningstar. “Gluon excitations of the static quark potential and the hybrid quarkonium spectrum”. In: *Nucl. Phys. B Proc. Suppl.* 63 (1998). Ed. by C. T. H. Davies et al., pp. 326–331.

[772] K. J. Juge, J. Kuti, and C. J. Morningstar. “Ab initio study of hybrid anti-b g b mesons”. In: *Phys. Rev. Lett.* 82 (1999), pp. 4400–4403.

[773] N. A. Campbell, I. H. Jorysz, and Christopher Michael. “The Adjoint Source Potential in SU(3) Lattice Gauge Theory”. In: *Phys. Lett. B* 167 (1986), pp. 91–93.

[774] M. Foster and Christopher Michael. “Hadrons with a heavy color adjoint particle”. In: *Phys. Rev. D* 59 (1999), p. 094509.

[775] Gunnar S. Bali and Antonio Pineda. “QCD phenomenology of static sources and gluonic excitations at short distances”. In: *Phys. Rev. D* 69 (2004), p. 094001.

[776] Liuming Liu et al. “Excited and exotic charmonium spectroscopy from lattice QCD”. In: *JHEP* 07 (2012), p. 126.

[777] Gavin K. C. Cheung et al. “Excited and exotic charmonium, $D_s$ and $D$ meson spectra for two light quark masses from lattice QCD”. In: *JHEP* 12 (2016), p. 089.

[778] Francesco Knechtli. “Charmonium and Exotics from Lattice QCD”. In: *EPJ Web Conf.* 202 (2019). Ed. by A. Bondar and S. Eidelman, p. 01006.

[779] Stanley J. Brodsky and B. T. Chertok. “The Deuteron Form-Factor and the Short Distance Behavior of the Nuclear Force”. In: *Phys. Rev. Lett.* 37 (1976), p. 269.

[780] Stanley J. Brodsky and B. T. Chertok. “The Asymptotic Form-Factors of Hadrons and Nuclei and the Continuity of Particle and Nuclear Dynamics”. In: *Phys. Rev. D* 14 (1976), pp. 3003–3020.

[781] Victor A. Matveev and Paul Sorba. “Is Deuteron a Six Quark System?” In: *Lett. Nuovo Cim.* 20 (1977), p. 435.

[782] M. Harvey. “On the Fractional Parentage Expansions of Color Singlet Six Quark States in a Cluster Model”. In: *Nucl. Phys. A* 352 (1981). [Erratum: Nucl.Phys.A 481, 834 (1988)], p. 301.

[783] M. Harvey. “Effective nuclear forces in the quark model with Delta and hidden color channel coupling”. In: *Nucl. Phys. A* 352 (1981), pp. 326–342.

[784] M. Harvey, J. Letourneux, and B. Lorazo. “Nucleon nucleon scattering in the quark cluster model”. In: *Nucl. Phys. A* 424 (1984), pp. 428–446.

[785] Stanley Brodsky, Guy de Teramond, and Marek Karliner. “Puzzles in Hadronic Physics and Novel Quantum Chromodynamics Phenomenology”. In: *Ann. Rev. Nucl. Part. Sci.* 62 (2011), p. 2082.

[786] Stanley J. Brodsky and Chueng-Ryong Ji. “Applications of Quantum Chromodynamics to Hadronic and Nuclear Interactions”. In: *Lect. Notes Phys.* 248 (1986). Ed. by C. A. Engelbrecht, pp. 153–245.

[787] Stanley J. Brodsky, Chueng-Ryong Ji, and G. Peter Lepage. “Quantum Chromodynamic Predictions for the Deuteron Form-Factor”. In: *Phys. Rev. Lett.* 51 (1983), p. 83.

[788] Stanley J. Brodsky and Chueng-Ryong Ji. “Evolution of Relativistic Multi - Quark Systems”. In: *Phys. Rev. D* 33 (1986), p. 1406.

[789] Chueng-Ryong Ji and Stanley J. Brodsky. “Quantum Chromodynamic Evolution of Six Quark States”. In: *Phys. Rev. D* 34 (1986), p. 1460.

[790] Bernard L. G. Bakker and Chueng-Ryong Ji. “Nuclear chromodynamics: Novel nuclear phenomena predicted by QCD”. In: *Prog. Part. Nucl. Phys.* 74 (2014), pp. 1–34.

[791] Stanley J. Brodsky and John R. Hiller. “Reduced Nuclear Amplitudes in Quantum Chromodynamics”. In: *Phys. Rev. C* 28 (1983). [Erratum: Phys.Rev.C 30, 412–412 (1984)], p. 475.
[792] Stanley J. Brodsky, Hans-Christian Pauli, and Stephen S. Pinsky. “Quantum chromodynamics and other field theories on the light cone”. In: *Phys. Rept.* 301 (1998), pp. 299–486.
[793] Stanley J. Brodsky, Alexandre Deur, and Craig D. Roberts. “Artificial dynamical effects in quantum field theory”. In: *Nature Rev. Phys.* 4.7 (2022), pp. 489–495.
[794] Gerald A. Miller. “Pionic and Hidden-Color, Six-Quark Contributions to the Deuteron b1 Structure Function”. In: *Phys. Rev. C* 89.4 (2014), p. 045203.
[795] N. Fomin et al. “New measurements of high-momentum nucleons and short-range structures in nuclei”. In: *Phys. Rev. Lett.* 108 (2012), p. 092502.
[796] R. Subedi et al. “Probing Cold Dense Nuclear Matter”. In: *Science* 320 (2008), pp. 1476–1478.
[797] M. Bashkanov, Stanley J. Brodsky, and H. Clement. “Novel Six-Quark Hidden-Color Dibaryon States in QCD”. In: *Phys. Lett. B* 727 (2013), pp. 438–442.
[798] I. Vidaña et al. “The $d^*(2380)$ in neutron stars - a new degree of freedom?” In: *Phys. Lett. B* 781 (2018), pp. 112–116.
[799] M. Bashkanov et al. “Double-Pionic Fusion of Nuclear Systems and the ABC Effect: Aproaching a Puzzle by Exclusive and Kinematically Complete Measurements”. In: *Phys. Rev. Lett.* 102 (2009), p. 052301.
[800] P. Adlarson et al. “ABC Effect in Basic Double-Pionic Fusion — Observation of a new resonance?” In: *Phys. Rev. Lett.* 106 (2011), p. 242302.
[801] P. Adlarson et al. “Isospin Decomposition of the Basic Double-Pionic Fusion in the Region of the ABC Effect”. In: *Phys. Lett. B* 721 (2013), pp. 229–236.
[802] P. Adlarson et al. “Measurement of the $pn \to pp\pi^0\pi^-$ reaction in search for the recently observed resonance structure in $d\pi^0\pi^0$ and $d\pi^+\pi^-$ systems”. In: *Phys. Rev. C* 88.5 (2013), p. 055208.
[803] P. Adlarson et al. “Evidence for a New Resonance from Polarized Neutron-Proton Scattering”. In: *Phys. Rev. Lett.* 112.20 (2014), p. 202301.
[804] P. Adlarson et al. “Neutron-proton scattering in the context of the d* (2380) resonance”. In: *Phys. Rev. C* 90.3 (2014), p. 035204.
[805] P. Adlarson et al. “Measurement of the $np \to np\pi^0\pi^0$ Reaction in Search for the Recently Observed $d^*(2380)$ Resonance”. In: *Phys. Lett. B* 743 (2015), pp. 325–332.
[806] P. Adlarson et al. “Measurement of the $\vec{n}p \to d\pi^0\pi^0$ reaction with polarized beam in the region of the d*(2380) resonance”. In: *Eur. Phys. J. A* 52.5 (2016), p. 147.
[807] M. Bashkanov et al. “Signatures of the $d^*(2380)$ Hexaquark in d($\gamma$,$p\vec{n}$)”. In: *Phys. Rev. Lett.* 124.13 (2020), p. 132001.
[808] X. Q. Yuan et al. “Deltaron dibaryon structure in chiral SU(3) quark model”. In: *Phys. Rev. C* 60 (1999), p. 045203.
[809] Q. B. Li and P. N. Shen. “Possible Delta-Delta dibaryons in the quark cluster model”. In: *J. Phys. G* 26 (2000), pp. 1207–1216.
[810] Q. B. Li et al. “Dibaryon systems in chiral SU(3) quark model”. In: *Nucl. Phys. A* 683 (2001), pp. 487–509.
[811] Fei Huang et al. “Is d* a candidate for a hexaquark-dominated exotic state?” In: *Chin. Phys. C* 39.7 (2015), p. 071001.
[812] Yubing Dong et al. “Theoretical study of the $d^*(2380) \to d\pi\pi$ decay width”. In: *Phys. Rev. C* 91.6 (2015), p. 064002.
[813] Yubing Dong et al. “Decay width of $d^*(2380) \to NN\pi\pi$ processes”. In: *Phys. Rev. C* 94.1 (2016), p. 014003.
[814] M. Bashkanov, D. P. Watts, and A. Pastore. “Electromagnetic properties of the $d^*(2380)$ hexaquark”. In: *Phys. Rev. C* 100.1 (2019), p. 012201.
[815] H. Clement and T. Skorodko. “Dibaryons: Molecular versus Compact Hexaquarks”. In: *Chin. Phys. C* 45.2 (2021), p. 022001.
[816] A. J. Krasznahorkay et al. “New anomaly observed in He4 supports the existence of the hypothetical X17 particle”. In: *Phys. Rev. C* 104.4 (2021), p. 044003.
[817] Jennifer Rittenhouse West et al. “QCD hidden-color hexadiquark in the core of nuclei”. In: *Nucl. Phys. A* 1007 (2021), p. 122134.
[818] Valery Kubarovsky, Jennifer Rittenhouse West, and Stanley J. Brodsky. “Quantum Chromodynamics Resolution of the ATOMKI Anomaly in $^4$He Nuclear Transitions”. In: (June 2022).
[819] A. Airapetian et al. “First measurement of the tensor structure function b(1) of the deuteron”. In: *Phys. Rev. Lett.* 95 (2005), p. 242001.
[820] A. Accardi et al. “Electron Ion Collider: The Next QCD Frontier: Understanding the glue that binds us all”. In: *Eur. Phys. J. A* 52.9 (2016). Ed. by A. Deshpande, Z. E. Meziani, and J. W. Qiu, p. 268.

[821] F. J. Dyson. "The S matrix in quantum electrodynamics". In: *Phys. Rev.* 75 (1949), pp. 1736–1755.

[822] Julian S. Schwinger. "On the Green's functions of quantized fields. 1." In: *Proc. Nat. Acad. Sci.* 37 (1951), pp. 452–455.

[823] Julian S. Schwinger. "On the Green's functions of quantized fields. 2." In: *Proc. Nat. Acad. Sci.* 37 (1951), pp. 455–459.

[824] E. E. Salpeter and H. A. Bethe. "A Relativistic equation for bound state problems". In: *Phys. Rev.* 84 (1951), pp. 1232–1242.

[825] Craig D. Roberts and Sebastian M. Schmidt. "Dyson-Schwinger equations: Density, temperature and continuum strong QCD". In: *Prog. Part. Nucl. Phys.* 45 (2000), S1–S103.

[826] Reinhard Alkofer and Lorenz von Smekal. "The Infrared behavior of QCD Green's functions: Confinement dynamical symmetry breaking, and hadrons as relativistic bound states". In: *Phys. Rept.* 353 (2001), p. 281.

[827] Reinhard Alkofer et al. "The Quark-gluon vertex in Landau gauge QCD: Its role in dynamical chiral symmetry breaking and quark confinement". In: *Annals Phys.* 324 (2009), pp. 106–172.

[828] Adnan Bashir et al. "Collective perspective on advances in Dyson-Schwinger Equation QCD". In: *Commun. Theor. Phys.* 58 (2012), pp. 79–134.

[829] Ian C. Cloet and Craig D. Roberts. "Explanation and Prediction of Observables using Continuum Strong QCD". In: *Prog. Part. Nucl. Phys.* 77 (2014), pp. 1–69.

[830] Gernot Eichmann et al. "Baryons as relativistic three-quark bound states". In: *Prog. Part. Nucl. Phys.* 91 (2016), pp. 1–100.

[831] C. Itzykson and J. B. Zuber. *Quantum Field Theory.* International Series In Pure and Applied Physics. New York: McGraw-Hill, 1980.

[832] Franz Gross. *Relativistic quantum mechanics and field theory.* New York, 1993.

[833] G. C. Wick. "Properties of Bethe-Salpeter Wave Functions". In: *Phys. Rev.* 96 (1954), pp. 1124–1134.

[834] R. E. Cutkosky. "Solutions of a Bethe-Salpeter equations". In: *Phys. Rev.* 96 (1954), pp. 1135–1141.

[835] Noboru Nakanishi. "A General survey of the theory of the Bethe-Salpeter equation". In: *Prog. Theor. Phys. Suppl.* 43 (1969), pp. 1–81.

[836] Kensuke Kusaka, Ken M. Simpson, and Anthony G. Williams. "Solving the Bethe-Salpeter equation for bound states of scalar theories in Minkowski space". In: *Phys. Rev. D* 56 (1997), pp. 5071–5085.

[837] V. Sauli. "Solving the Bethe-Salpeter equation for a pseudoscalar meson in Minkowski space". In: *J. Phys. G* 35 (2008), p. 035005.

[838] J. Carbonell and V. A. Karmanov. "Solving Bethe-Salpeter equation for two fermions in Minkowski space". In: *Eur. Phys. J. A* 46 (2010), pp. 387–397.

[839] W. de Paula et al. "Advances in solving the two-fermion homogeneous Bethe-Salpeter equation in Minkowski space". In: *Phys. Rev. D* 94.7 (2016), p. 071901.

[840] A. Castro et al. "The Bethe-Salpeter approach to bound states: from Euclidean to Minkowski space". In: *J. Phys. Conf. Ser.* 1291.1 (2019). Ed. by Valdir Guimaraes et al., p. 012006.

[841] Gernot Eichmann, Eduardo Ferreira, and Alfred Stadler. "Going to the light front with contour deformations". In: *Phys. Rev. D* 105.3 (2022), p. 034009.

[842] Steven Weinberg. "Dynamics at Infinite Momentum". In: *Phys. Rev.* 150 (4 Oct. 1966), pp. 1313–1318.

[843] L. A. Kondratyuk and M. I. Strikman. "Relativistic correction to the deuteron magnetic moment and angular condition". In: *Nucl. Phys. A* 426 (1984), pp. 575–598.

[844] Lei Chang et al. "Imaging dynamical chiral symmetry breaking: pion wave function on the light front". In: *Phys. Rev. Lett.* 110.13 (2013), p. 132001.

[845] Chen Chen et al. "Valence-quark distribution functions in the kaon and pion". In: *Phys. Rev. D* 93.7 (2016), p. 074021.

[846] Chao Shi et al. "Spatial and Momentum Imaging of the Pion and Kaon". In: *Phys. Rev. D* 101.7 (2020), p. 074014.

[847] Franz Gross. "Three-dimensional covariant integral equations for low-energy systems". In: *Phys. Rev.* 186 (1969), pp. 1448–1462.

[848] Yu. A. Simonov and J. A. Tjon. "The Feynman-Schwinger representation for the relativistic two particle amplitude in field theory". In: *Annals Phys.* 228 (1993), pp. 1–18.

[849] Taco Nieuwenhuis and J. A. Tjon. "Nonperturbative study of generalized ladder graphs in a $\phi^2\chi$ theory". In: *Phys. Rev. Lett.* 77 (1996), pp. 814–817.

[850] Mariane Mangin-Brinet and Jaume Carbonell. "Solutions of the Wick-Cutkosky model in the light front dynamics". In: *Phys. Lett. B* 474 (2000), pp. 237–244.

[851] V. A. Karmanov and P. Maris. "Manifestation of three-body forces in three-body Bethe-Salpeter and light-front equations". In: *Few Body Syst.* 46 (2009), pp. 95–113.

[852] P. C. Tiemeijer and J. A. Tjon. "Meson mass spectrum from relativistic equations in configuration space". In: *Phys. Rev. C* 49 (1994), pp. 494–512.

[853] Cetin Savkli, Franz Gross, and John Tjon. "The Role of interaction vertices in bound state calculations". In: *Phys. Lett. B* 531 (2002), pp. 161–166.

[854] Franz Gross and Alfred Stadler. "Covariant spectator theory of np scattering: Phase shifts obtained from precision fits to data below 350-MeV". In: *Phys. Rev. C* 78 (2008), p. 014005.

[855] Franz Gross. "Removal of singularities from the covariant spectator theory". In: *Phys. Rev. D* 104.5 (2021), p. 054020.

[856] Franz Gross. "Covariant Spectator Theory of np scattering: Deuteron Quadrupole Moment". In: *Phys. Rev. C* 91.1 (2015), p. 014005.

[857] E. O. Alt, P. Grassberger, and W. Sandhas. "Reduction of the three - particle collision problem to multichannel two - particle Lippmann-Schwinger equations". In: *Nucl. Phys. B* 2 (1967), pp. 167–180.

[858] R. A. Malfliet and J. A. Tjon. "Three-nucleon calculations with realistic forces". In: *Annals Phys.* 61 (1970), pp. 425–450.

[859] G. Rupp and J. A. Tjon. "Bethe-Salpeter Calculation of Three Nucleon Observables With Rank One Separable Potentials". In: *Phys. Rev. C* 37 (1988), p. 1729.

[860] L. E. Marcucci et al. "Electromagnetic Structure of Few-Nucleon Ground States". In: *J. Phys. G* 43 (2016), p. 023002.

[861] Alfred Stadler, Franz Gross, and Michael Frank. "Covariant equations for the three-body bound state". In: *Phys. Rev. C* 56 (1997), p. 2396.

[862] Alfred Stadler and Franz Gross. "Relativistic calculation of the triton binding energy and its implications". In: *Phys. Rev. Lett.* 78 (1997), pp. 26–29.

[863] Sergio Alexandre Pinto, Alfred Stadler, and Franz Gross. "Covariant spectator theory for the electromagnetic three-nucleon form factors: Complete impulse approximation". In: *Phys. Rev. C* 79 (2009), p. 054006.

[864] Sergio Alexandre Pinto, Alfred Stadler, and Franz Gross. "First results for electromagnetic three-nucleon form factors from high-precision two-nucleon interactions". In: *Phys. Rev. C* 81 (2010), p. 014007.

[865] Franz Gross and Joseph Milana. "Covariant, chirally symmetric, confining model of mesons". In: *Phys. Rev. D* 43 (7 Apr. 1991), pp. 2401–2417.

[866] Cetin Savkli and Franz Gross. "Quark - antiquark bound states in the relativistic spectator formalism". In: *Phys. Rev. C* 63 (2001), p. 035208.

[867] Elmar P. Biernat et al. "Confinement, quark mass functions, and spontaneous chiral symmetry breaking in Minkowski space". In: *Phys. Rev. D* 89.1 (2014), p. 016005.

[868] Sofia Leitão et al. "Linear confinement in momentum space: singularity-free bound-state equations". In: *Phys. Rev. D* 90.9 (2014), p. 096003.

[869] Sofia Leitão et al. "Covariant spectator theory of quark-antiquark bound states: Mass spectra and vertex functions of heavy and heavy-light mesons". In: *Phys. Rev. D* 96.7 (2017), p. 074007.

[870] Sofia Leitão et al. "Comparison of two Minkowski-space approaches to heavy quarkonia". In: *Eur. Phys. J. C* 77.10 (2017), p. 696.

[871] Franz Gross. "The CST: Its Achievements and Its Connection to the Light Cone". In: *Few Body Syst.* 58.2 (2017), p. 39.

[872] R. Blankenbecler et al. "Singularities of Scattering Amplitudes on Unphysical Sheets and Their Interpretation". In: *Phys. Rev.* 123 (1961), pp. 692–699.

[873] Franz Gross and D. O. Riska. "Current Conservation and Interaction Currents in Relativistic Meson Theories". In: *Phys. Rev. C* 36 (1987), p. 1928.

[874] A. Bender, Craig D. Roberts, and L. Von Smekal. "Goldstone theorem and diquark confinement beyond rainbow ladder approximation". In: *Phys. Lett. B* 380 (1996), pp. 7–12.

[875] Axel Bender et al. "Bethe-Salpeter equation and a nonperturbative quark gluon vertex". In: *Phys. Rev. C* 65 (2002), p. 065203.

[876] Pieter Maris and Peter C. Tandy. "Bethe-Salpeter study of vector meson masses and decay constants". In: *Phys. Rev. C* 60 (1999), p. 055214.

[877] P. Maris and P. C. Tandy. "QCD modeling of hadron physics". In: *Nucl. Phys. B Proc. Suppl.* 161 (2006). Ed. by D. B. Leinweber, L. von Smekal, and A. G. Williams, pp. 136–152.

[878] Christian S. Fischer and Reinhard Alkofer. "Non-perturbative propagators, running coupling and

dynamical quark mass of Landau gauge QCD". In: *Phys. Rev. D* 67 (2003), p. 094020.
[879] Frederic D. R. Bonnet et al. "Overlap quark propagator in Landau gauge". In: *Phys. Rev. D* 65 (2002), p. 114503.
[880] Patrick O. Bowman, Urs M. Heller, and Anthony G. Williams. "Lattice quark propagator with staggered quarks in Landau and Laplacian gauges". In: *Phys. Rev. D* 66 (2002), p. 014505.
[881] Reinhard Alkofer et al. "Analytic properties of the Landau gauge gluon and quark propagators". In: *Phys. Rev. D* 70 (2004), p. 014014.
[882] Pieter Maris and Craig D. Roberts. "Pi- and K meson Bethe-Salpeter amplitudes". In: *Phys. Rev. C* 56 (1997), pp. 3369–3383.
[883] H. David Politzer. "Effective Quark Masses in the Chiral Limit". In: *Nucl. Phys. B* 117 (1976), pp. 397–406.
[884] D. C. Curtis and M. R. Pennington. "Truncating the Schwinger-Dyson equations: How multiplicative renormalizability and the Ward identity restrict the three point vertex in QED". In: *Phys. Rev. D* 42 (1990), pp. 4165–4169.
[885] Reijiro Fukuda and Taichiro Kugo. "Schwinger-Dyson Equation for Massless Vector Theory and Absence of Fermion Pole". In: *Nucl. Phys. B* 117 (1976), pp. 250–264.
[886] D. Atkinson and D. W. E. Blatt. "Determination of the Singularities of the Electron Propagator". In: *Nucl. Phys. B* 151 (1979), pp. 342–352.
[887] P. Maris. "Analytic structure of the full fermion propagator in quenched and unquenched QED". In: *Phys. Rev. D* 50 (1994), pp. 4189–4193.
[888] S. J. Stainsby and R. T. Cahill. "Is space-time Euclidean 'inside' hadrons?" In: *Phys. Lett. A* 146 (1990), pp. 467–470.
[889] G. Krein, Craig D. Roberts, and Anthony G. Williams. "On the implications of confinement". In: *Int. J. Mod. Phys. A* 7 (1992), pp. 5607–5624.
[890] Shaoyang Jia et al. "Minkowski-space solutions of the Schwinger-Dyson equation for the fermion propagator with the rainbow-ladder truncation". In: *18th International Conference on Hadron Spectroscopy and Structure*. 2020, pp. 560–564.
[891] James S. Ball and Ting-Wai Chiu. "Analytic Properties of the Vertex Function in Gauge Theories. 1." In: *Phys. Rev. D* 22 (1980), p. 2542.
[892] Pieter Maris, Craig D. Roberts, and Peter C. Tandy. "Pion mass and decay constant". In: *Phys. Lett. B* 420 (1998), pp. 267–273.
[893] J. J. Sakurai. "Theory of strong interactions". In: *Annals Phys.* 11 (1960), pp. 1–48.
[894] M. S. Bhagwat and P. Maris. "Vector meson form factors and their quark-mass dependence". In: *Phys. Rev. C* 77 (2008), p. 025203.
[895] Pieter Maris and Peter C. Tandy. "The pi, K+, and K0 electromagnetic form-factors". In: *Phys. Rev. C* 62 (2000), p. 055204.
[896] Jiangshan Lan et al. "Light mesons with one dynamical gluon on the light front". In: *Phys. Lett. B* 825 (2022), p. 136890.
[897] S. R. Amendolia et al. "A Measurement of the Space - Like Pion Electromagnetic Form-Factor". In: *Nucl. Phys. B* 277 (1986). Ed. by S. C. Loken, p. 168.
[898] G. M. Huber et al. "Charged pion form-factor between Q**2 = 0.60-GeV**2 and 2.45-GeV**2. II. Determination of, and results for, the pion form-factor". In: *Phys. Rev. C* 78 (2008), p. 045203.
[899] Pieter Maris and Peter C. Tandy. "Electromagnetic transition form-factors of light mesons". In: *Phys. Rev. C* 65 (2002), p. 045211.
[900] Stephen R. Cotanch and Pieter Maris. "Ladder Dyson-Schwinger calculation of the anomalous gamma-3pi form-factor". In: *Phys. Rev. D* 68 (2003), p. 036006.
[901] Pedro Bicudo et al. "Chirally symmetric quark description of low-energy pi pi scattering". In: *Phys. Rev. D* 65 (2002), p. 076008.
[902] Lei Chang and Craig D. Roberts. "Sketching the Bethe-Salpeter kernel". In: *Phys. Rev. Lett.* 103 (2009), p. 081601.
[903] Daniele Binosi et al. "Symmetry preserving truncations of the gap and Bethe-Salpeter equations". In: *Phys. Rev. D* 93.9 (2016), p. 096010.
[904] Richard Williams, Christian S. Fischer, and Walter Heupel. "Light mesons in QCD and unquenching effects from the 3PI effective action". In: *Phys. Rev. D* 93.3 (2016), p. 034026.
[905] Ángel S. Miramontes, Hèlios Sanchis Alepuz, and Reinhard Alkofer. "Elucidating the effect of intermediate resonances in the quark interaction kernel on the timelike electromagnetic pion form factor". In: *Phys. Rev. D* 103.11 (2021), p. 116006.
[906] Nico Santowsky and Christian S. Fischer. "Light scalars: Four-quark versus two-quark states in the complex energy plane from Bethe-Salpeter equations". In: *Phys. Rev. D* 105.3 (2022), p. 034025.
[907] Gernot Eichmann, Christian S. Fischer, and Helios Sanchis-Alepuz. "Light baryons and their excitations". In: *Phys. Rev. D* 94.9 (2016), p. 094033.

[908] J. R. Hiller. “Nonperturbative light-front Hamiltonian methods”. In: *Prog. Part. Nucl. Phys.* 90 (2016), pp. 75–124.

[909] Paul A. M. Dirac. “Forms of Relativistic Dynamics”. In: *Rev. Mod. Phys.* 21 (1949), pp. 392–399.

[910] D. E. Soper. “Infinite-momentum helicity states”. In: *Phys. Rev. D* 5 (1972), pp. 1956–1962.

[911] Yang Li et al. “Introduction to Basis Light-Front Quantization Approach to QCD Bound State Problems”. In: *International Conference on Nuclear Theory in the Supercomputing Era.* 2013, p. 136.

[912] E. Tomboulis. “Quantization of the yang-mills field in the null-plane frame”. In: *Phys. Rev. D* 8 (1973), pp. 2736–2740.

[913] Aharon Casher. “Gauge Fields on the Null Plane”. In: *Phys. Rev. D* 14 (1976), p. 452.

[914] V. A. Karmanov, J. -F. Mathiot, and A. V. Smirnov. “Systematic renormalization scheme in light-front dynamics with Fock space truncation”. In: *Phys. Rev. D* 77 (2008), p. 085028.

[915] Stanislaw D. Glazek and Kenneth G. Wilson. “Renormalization of Hamiltonians”. In: *Phys. Rev. D* 48 (1993), pp. 5863–5872.

[916] Kenneth G. Wilson et al. “Nonperturbative QCD: A Weak coupling treatment on the light front”. In: *Phys. Rev. D* 49 (1994), pp. 6720–6766.

[917] Robert J. Perry. “A Renormalization group approach to Hamiltonian light front field theory”. In: *Annals Phys.* 232 (1994), pp. 116–222.

[918] Stanislaw D. Glazek. “Perturbative formulae for relativistic interactions of effective particles”. In: *Acta Phys. Polon. B* 43 (2012), pp. 1843–1862.

[919] Stanisław D. Głazek et al. “Renormalized quark–antiquark Hamiltonian induced by a gluon mass ansatz in heavy-flavor QCD”. In: *Phys. Lett. B* 773 (2017), pp. 172–178.

[920] María Gómez-Rocha and Stanisław D. Głazek. “Asymptotic freedom in the front-form Hamiltonian for quantum chromodynamics of gluons”. In: *Phys. Rev. D* 92.6 (2015), p. 065005.

[921] Robert J. Perry, Avaroth Harindranath, and Kenneth G. Wilson. “Light front Tamm-Dancoff field theory”. In: *Phys. Rev. Lett.* 65 (1990), pp. 2959–2962.

[922] Yang Li et al. “Ab Initio Approach to the Non-Perturbative Scalar Yukawa Model”. In: *Phys. Lett. B* 748 (2015), pp. 278–283.

[923] S. S. Chabysheva and J. R. Hiller. “A Light-Front Coupled-Cluster Method for the Nonperturbative Solution of Quantum Field Theories”. In: *Phys. Lett. B* 711 (2012), pp. 417–422.

[924] M. Krautgartner, H. C. Pauli, and F. Wolz. “Positronium and heavy quarkonia as testing case for discretized light cone quantization. 1.” In: *Phys. Rev. D* 45 (1992), pp. 3755–3774.

[925] Kent Hornbostel, Stanley J. Brodsky, and Hans Christian Pauli. “Light Cone Quantized QCD in (1+1)-Dimensions”. In: *Phys. Rev. D* 41 (1990), p. 3814.

[926] Matthias Burkardt and Simon Dalley. “The Relativistic bound state problem in QCD: Transverse lattice methods”. In: *Prog. Part. Nucl. Phys.* 48 (2002), pp. 317–362.

[927] Dipankar Chakrabarti, A. Harindranath, and James P. Vary. “A Study of q anti-q states in transverse lattice QCD using alternative fermion formulations”. In: *Phys. Rev. D* 69 (2004), p. 034502.

[928] A. Harindranath, R. J. Perry, and J. Shigemitsu. “Bound state problem in the light front Tamm-Dancoff approximation: Numerical study in (1+1)-dimensions”. In: *Phys. Rev. D* 46 (1992), pp. 4580–4602.

[929] J. P. Vary et al. “Hamiltonian light-front field theory in a basis function approach”. In: *Phys. Rev. C* 81 (2010), p. 035205.

[930] H. Honkanen et al. “Electron in a transverse harmonic cavity”. In: *Phys. Rev. Lett.* 106 (2011), p. 061603.

[931] Xingbo Zhao et al. “Electron Anomalous Magnetic Moment in Basis Light-Front Quantization Approach”. In: *Few Body Syst.* 52 (2012). Ed. by Simon Dalley, pp. 339–344.

[932] Xingbo Zhao et al. “Electron g-2 in Light-Front Quantization”. In: *Phys. Lett. B* 737 (2014), pp. 65–69.

[933] D. Chakrabarti et al. “Generalized parton distributions in a light-front nonperturbative approach”. In: *Phys. Rev. D* 89.11 (2014), p. 116004.

[934] Zhi Hu et al. “Transverse structure of electron in momentum space in basis light-front quantization”. In: *Phys. Rev. D* 103.3 (2021), p. 036005.

[935] Paul Wiecki et al. “Non-perturbative Calculation of the Positronium Mass Spectrum in Basis Light-Front Quantization”. In: *Few Body Syst.* 56.6-9 (2015). Ed. by Chueng-Ryong Ji, pp. 489–494.

[936] Hans A. Bethe and Edwin E. Salpeter. *Quantum Mechanics of One- and Two-Electron Atoms.* 1957.

[937] Yang Li, Pieter Maris, and James P. Vary. “Quarkonium as a relativistic bound state on the light front”. In: *Phys. Rev. D* 96 (2017), p. 016022.

[938] Christian S. Fischer, Stanislav Kubrak, and Richard Williams. "Spectra of heavy mesons in the Bethe-Salpeter approach". In: *Eur. Phys. J. A* 51 (2015), p. 10.

[939] R. L. Workman et al. "Review of Particle Physics". In: *PTEP* 2022 (2022), p. 083C01.

[940] Paul Wiecki et al. "Basis Light-Front Quantization Approach to Positronium". In: *Phys. Rev. D* 91.10 (2015), p. 105009.

[941] Dipankar Chakrabarti and A. Harindranath. "Mesons in light front QCD(2+1): Investigation of a Bloch effective Hamiltonian". In: *Phys. Rev. D* 64 (2001), p. 105002.

[942] Lekha Adhikari et al. "Form Factors and Generalized Parton Distributions in Basis Light-Front Quantization". In: *Phys. Rev. C* 93.5 (2016), p. 055202.

[943] Sreeraj Nair et al. "Basis light-front quantization approach to photon". In: *Phys. Lett. B* 827 (2022), p. 137005.

[944] Yang Li, Meijian Li, and James P. Vary. "Two-photon transitions of charmonia on the light front". In: *Phys. Rev. D* 105.7 (2022), p. L071901.

[945] J. P. Lees et al. "Measurement of the $\gamma\gamma * -- > \eta_c$ transition form factor". In: *Phys. Rev. D* 81 (2010), p. 052010.

[946] Meijian Li et al. "Radiative transitions between $0^{-+}$ and $1^{--}$ heavy quarkonia on the light front". In: *Phys. Rev. D* 98.3 (2018), p. 034024.

[947] Jiangshan Lan et al. "Parton Distribution Functions of Heavy Mesons on the Light Front". In: *Phys. Rev. D* 102.1 (2020), p. 014020.

[948] Jiangshan Lan et al. "Parton Distribution Functions from a Light Front Hamiltonian and QCD Evolution for Light Mesons". In: *Phys. Rev. Lett.* 122.17 (2019), p. 172001.

[949] Siqi Xu et al. "Nucleon structure from basis light-front quantization". In: *Phys. Rev. D* 104.9 (2021), p. 094036.

[950] Yang Li et al. "Heavy Quarkonium in a Holographic Basis". In: *Phys. Lett. B* 758 (2016), pp. 118–124.

[951] Lekha Adhikari et al. "Form factors and generalized parton distributions of heavy quarkonia in basis light front quantization". In: *Phys. Rev. C* 99.3 (2019), p. 035208.

[952] Meijian Li et al. "Frame dependence of transition form factors in light-front dynamics". In: *Phys. Rev. D* 100.3 (2019), p. 036006.

[953] Shuo Tang et al. "Semileptonic decay of Bc to $\eta$c and J/$\psi$ on the light front". In: *Phys. Rev. D* 104.1 (2021), p. 016002.

[954] Shuo Tang et al. "$B_c$ mesons and their properties on the light front". In: *Phys. Rev. D* 98.11 (2018), p. 114038.

[955] Shuo Tang et al. "Heavy-light mesons on the light front". In: *Eur. Phys. J. C* 80.6 (2020), p. 522.

[956] Pieter Maris et al. "On the light-front wave functions of quarkonia". In: *PoS* LC2019 (2020), p. 007.

[957] Shaoyang Jia and James P. Vary. "Basis light front quantization for the charged light mesons with color singlet Nambu–Jona-Lasinio interactions". In: *Phys. Rev. C* 99.3 (2019), p. 035206.

[958] Wenyang Qian et al. "Light mesons within the basis light-front quantization framework". In: *Phys. Rev. C* 102.5 (2020), p. 055207.

[959] Jiangshan Lan et al. "Pion and kaon parton distribution functions from basis light front quantization and QCD evolution". In: *Phys. Rev. D* 101.3 (2020), p. 034024.

[960] Jiangshan Lan et al. "Light meson parton distribution functions from basis light-front quantization and QCD evolution". In: *18th International Conference on Hadron Spectroscopy and Structure*. 2020, pp. 581–585.

[961] Siqi Xu et al. "Nucleon spin decomposition with one dynamical gluon". In: (Sept. 2022).

[962] Chandan Mondal et al. "Proton structure from a light-front Hamiltonian". In: *Phys. Rev. D* 102.1 (2020), p. 016008.

[963] Zhi Hu et al. "Transverse momentum structure of proton within the basis light-front quantization framework". In: *Phys. Lett. B* 833 (2022), p. 137360.

[964] Tiancai Peng et al. "Basis light-front quantization approach to $\Lambda$ and $\Lambda_c$ and their isospin triplet baryons". In: (July 2022).

[965] Richard J. Hill et al. "Nucleon Axial Radius and Muonic Hydrogen — A New Analysis and Review". In: *Rept. Prog. Phys.* 81.9 (2018), p. 096301.

[966] S. D. Drell and Tung-Mow Yan. "Connection of Elastic Electromagnetic Nucleon Form-Factors at Large Q**2 and Deep Inelastic Structure Functions Near Threshold". In: *Phys. Rev. Lett.* 24 (1970), pp. 181–185.

[967] Geoffrey B. West. "Phenomenological model for the electromagnetic structure of the proton". In: *Phys. Rev. Lett.* 24 (1970), pp. 1206–1209.

[968] Stanley J. Brodsky, Matthias Burkardt, and Ivan Schmidt. "Perturbative QCD constraints on the shape of polarized quark and gluon distributions". In: *Nucl. Phys. B* 441 (1995), pp. 197–214.

[969] M. G. Alekseev et al. "Quark helicity distributions from longitudinal spin asymmetries in muon-proton and muon-deuteron scattering". In: *Phys. Lett. B* 693 (2010), pp. 227–235.

[970] Raphael Dupre, Michel Guidal, and Marc Vanderhaeghen. "Tomographic image of the proton". In: *Phys. Rev. D* 95.1 (2017), p. 011501.

[971] Yiping Liu et al. "Angular momentum and generalized parton distributions for the proton with basis light-front quantization". In: *Phys. Rev. D* 105.9 (2022), p. 094018.

[972] Matthias Burkardt. "Impact parameter dependent parton distributions and off forward parton distributions for zeta —> 0". In: *Phys. Rev. D* 62 (2000). [Erratum: Phys.Rev.D 66, 119903 (2002)], p. 071503.

[973] Zhongkui Kuang et al. "All-charm tetraquark in front form dynamics". In: *Phys. Rev. D* 105.9 (2022), p. 094028.

[974] M. Kaluza and H. C. Pauli. "Discretized light cone quantization: e+ e- (gamma) model for positronium". In: *Phys. Rev. D* 45 (1992), pp. 2968–2981.

[975] Kaiyu Fu et al. "Positronium on the light front". In: *18th International Conference on Hadron Spectroscopy and Structure*. 2020, pp. 550–554.

[976] Xingbo Zhao et al. "Positronium: an illustration of nonperturbative renormalization in a basis light-front approach". In: *PoS* LC2019 (2020), p. 090.

[977] Xingbo Zhao, Kaiyu Fu, and James P. Vary. "Bound States in QED from a light-front approach". In: *Rev. Mex. Fis. Suppl.* 3.3 (2022), p. 0308105.

[978] P. C. Barry et al. "First Monte Carlo Global QCD Analysis of Pion Parton Distributions". In: *Phys. Rev. Lett.* 121.15 (2018), p. 152001.

[979] Ivan Novikov et al. "Parton Distribution Functions of the Charged Pion Within The xFitter Framework". In: *Phys. Rev. D* 102.1 (2020), p. 014040.

[980] D. Abrams et al. "Measurement of the Nucleon $F_2^n/F_2^p$ Structure Function Ratio by the Jefferson Lab MARATHON Tritium/Helium-3 Deep Inelastic Scattering Experiment". In: *Phys. Rev. Lett.* 128.13 (2022), p. 132003.

[981] L. A. Harland-Lang et al. "Parton distributions in the LHC era: MMHT 2014 PDFs". In: *Eur. Phys. J. C* 75.5 (2015), p. 204.

[982] Nobuo Sato et al. "Iterative Monte Carlo analysis of spin-dependent parton distributions". In: *Phys. Rev. D* 93.7 (2016), p. 074005.

[983] Emanuele R. Nocera et al. "A first unbiased global determination of polarized PDFs and their uncertainties". In: *Nucl. Phys. B* 887 (2014), pp. 276–308.

[984] Y. Zhou, N. Sato, and W. Melnitchouk. "How well do we know the gluon polarization in the proton?" In: *Phys. Rev. D* 105.7 (2022), p. 074022.

[985] James P. Vary et al. "Critical coupling for two-dimensional $\phi 4$ theory in discretized light-cone quantization". In: *Phys. Rev. D* 105.1 (2022), p. 016020.

[986] Michael Kreshchuk et al. "Quantum simulation of quantum field theory in the light-front formulation". In: *Phys. Rev. A* 105.3 (2022), p. 032418.

[987] Yang Li, Pieter Maris, and James P. Vary. "Chiral sum rule on the light front". In: (Mar. 2022).

[988] Silas R. Beane. "Broken Chiral Symmetry on a Null Plane". In: *Annals Phys.* 337 (2013), pp. 111–142.

[989] Yang Li and James P. Vary. "Light-front holography with chiral symmetry breaking". In: *Phys. Lett. B* 825 (2022), p. 136860.

[990] Stanley J. Brodsky et al. "Essence of the vacuum quark condensate". In: *Phys. Rev. C* 82 (2010), p. 022201.

[991] Xingbo Zhao et al. "Scattering in Time-Dependent Basis Light-Front Quantization". In: *Phys. Rev. D* 88 (2013), p. 065014.

[992] Weijie Du et al. "Coulomb excitation of the deuteron in peripheral collisions with a heavy ion". In: *Phys. Rev. C* 97.6 (2018), p. 064620.

[993] Meijian Li et al. "Ultrarelativistic quark-nucleus scattering in a light-front Hamiltonian approach". In: *Phys. Rev. D* 101.7 (2020), p. 076016.

[994] Stanley J. Brodsky, Tao Huang, and G. Peter Lepage. "The Hadronic Wave Function in Quantum Chromodynamics". In: June 1980.

[995] Meijian Li, Tuomas Lappi, and Xingbo Zhao. "Scattering and gluon emission in a color field: A light-front Hamiltonian approach". In: *Phys. Rev. D* 104.5 (2021), p. 056014.

[996] Xingbo Zhao et al. "Non-perturbative quantum time evolution on the light-front". In: *Phys. Lett. B* 726 (2013), pp. 856–860.

[997] Bolun Hu, Anton Ilderton, and Xingbo Zhao. "Scattering in strong electromagnetic fields: Transverse size effects in time-dependent basis light-front quantization". In: *Phys. Rev. D* 102.1 (2020), p. 016017.

[998] Zhiyu Lei, Bolun Hu, and Xingbo Zhao. "Pair production in strong electric fields". In: (Jan. 2022).

[999] Guangyao Chen et al. "Particle distribution in intense fields in a light-front Hamiltonian approach". In: *Phys. Rev. D* 95.9 (2017), p. 096012.

[1000] Juan Martin Maldacena. "The Large N limit of superconformal field theories and supergravity". In: *Adv. Theor. Math. Phys.* 2 (1998), pp. 231–252.

[1001] S. S. Gubser, Igor R. Klebanov, and Alexander M. Polyakov. "Gauge theory correlators from noncritical string theory". In: *Phys. Lett. B* 428 (1998), pp. 105–114.

[1002] Edward Witten. "Anti-de Sitter space and holography". In: *Adv. Theor. Math. Phys.* 2 (1998), pp. 253–291.

[1003] Ofer Aharony et al. "Large N field theories, string theory and gravity". In: *Phys. Rept.* 323 (2000), pp. 183–386.

[1004] G. Mack and Abdus Salam. "Finite component field representations of the conformal group". In: *Annals Phys.* 53 (1969), pp. 174–202.

[1005] Joseph Polchinski and Matthew J. Strassler. "Hard scattering and gauge/string duality". In: *Phys. Rev. Lett.* 88 (2002), p. 031601.

[1006] Anton Rebhan. "The Witten-Sakai-Sugimoto model: A brief review and some recent results". In: *EPJ Web Conf.* 95 (2015). Ed. by L. Bravina, Y. Foka, and S. Kabana, p. 02005.

[1007] Joshua Erlich et al. "QCD and a holographic model of hadrons". In: *Phys. Rev. Lett.* 95 (2005), p. 261602.

[1008] Leandro Da Rold and Alex Pomarol. "Chiral symmetry breaking from five dimensional spaces". In: *Nucl. Phys. B* 721 (2005), pp. 79–97.

[1009] Andreas Karch et al. "Linear confinement and AdS/QCD". In: *Phys. Rev. D* 74 (2006), p. 015005.

[1010] Stanley J. Brodsky et al. "Light-Front Holographic QCD and Emerging Confinement". In: *Phys. Rept.* 584 (2015), pp. 1–105.

[1011] Guy F. de Téramond and Stanley J. Brodsky. "Light-Front Holography: A First Approximation to QCD". In: *Phys. Rev. Lett.* 102 (2009), p. 081601.

[1012] V. A. Matveev, R. M. Muradian, and A. N. Tavkhelidze. "Automodellism in the large - angle elastic scattering and structure of hadrons". In: *Lett. Nuovo Cim.* 7 (1973), pp. 719–723.

[1013] Joseph Polchinski and Matthew J. Strassler. "Deep inelastic scattering and gauge/string duality". In: *JHEP* 05 (2003), p. 012.

[1014] Stanley J. Brodsky and Guy F. de Téramond. "Hadronic spectra and light-front wave functions in holographic QCD". In: *Phys. Rev. Lett.* 96 (2006), p. 201601.

[1015] Stanley J. Brodsky and Guy F. de Téramond. "Light-Front Dynamics and AdS/QCD Correspondence: The Pion Form Factor in the Space- and Time-Like Regions". In: *Phys. Rev. D* 77 (2008), p. 056007.

[1016] Stanley J. Brodsky and Guy F. de Téramond. "Light-Front Dynamics and AdS/QCD Correspondence: Gravitational Form Factors of Composite Hadrons". In: *Phys. Rev. D* 78 (2008), p. 025032.

[1017] Vittorio de Alfaro, S. Fubini, and G. Furlan. "Conformal Invariance in Quantum Mechanics". In: *Nuovo Cim. A* 34 (1976), p. 569.

[1018] S. Fubini and E. Rabinovici. "Super Conformal Quantum Mechanics". In: *Nucl. Phys. B* 245 (1984), p. 17.

[1019] V. P. Akulov and A. I. Pashnev. "Quantum Superconformsl Model in (1,2) Space". In: *Theor. Math. Phys.* 56 (1983), pp. 862–866.

[1020] Stanley J. Brodsky, Guy F. de Téramond, and Hans Günter Dosch. "Threefold Complementary Approach to Holographic QCD". In: *Phys. Lett. B* 729 (2014), pp. 3–8.

[1021] Guy F. de Téramond, Hans Gunter Dosch, and Stanley J. Brodsky. "Baryon Spectrum from Superconformal Quantum Mechanics and its Light-Front Holographic Embedding". In: *Phys. Rev. D* 91.4 (2015), p. 045040.

[1022] Hans Gunter Dosch, Guy F. de Téramond, and Stanley J. Brodsky. "Superconformal Baryon-Meson Symmetry and Light-Front Holographic QCD". In: *Phys. Rev. D* 91.8 (2015), p. 085016.

[1023] H. Miyazawa. "Baryon Number Changing Currents". In: *Prog. Theor. Phys.* 36.6 (1966), pp. 1266–1276.

[1024] Sultan Catto and Feza Gursey. "Algebraic Treatment of Effective Supersymmetry". In: *Nuovo Cim. A* 86 (1985), p. 201.

[1025] D. B. Lichtenberg. "Whither hadron supersymmetry?" In: *International Conference on Orbis Scientiae 1999: Quantum Gravity, Generalized Theory of Gravitation and Superstring Theory Based Unification (28th Conference on High-Energy Physics and Cosmology Since 1964)*. Dec. 1999, pp. 203–208.

[1026] Guy F. de Teramond et al. "Universality of Generalized Parton Distributions in Light-Front Holographic QCD". In: *Phys. Rev. Lett.* 120.18 (2018), p. 182001.

[1027] Tianbo Liu et al. "Unified Description of Polarized and Unpolarized Quark Distributions in the Proton". In: *Phys. Rev. Lett.* 124.8 (2020), p. 082003.

[1028] Guy F. de Téramond et al. "Gluon matter distribution in the proton and pion from extended holographic light-front QCD". In: *Phys. Rev. D* 104.11 (2021), p. 114005.

[1029] Liping Zou and H. G. Dosch. "A very Practical Guide to Light Front Holographic QCD". In: (Jan. 2018).

[1030] Stanley J. Brodsky. "Supersymmetric and Conformal Features of Hadron Physics". In: *Universe* 4.11 (2018), p. 120.

[1031] Stanley J. Brodsky, Guy F. de Téramond, and Hans Gunter Dosch. "Light-Front Holography and Supersymmetric Conformal Algebra: A Novel Approach to Hadron Spectroscopy, Structure, and Dynamics". In: Apr. 2020.

[1032] U. Gursoy, E. Kiritsis, and F. Nitti. "Exploring improved holographic theories for QCD: Part II". In: *JHEP* 02 (2008), p. 019.

[1033] Youngman Kim, Ik Jae Shin, and Takuya Tsukioka. "Holographic QCD: Past, Present, and Future". In: *Prog. Part. Nucl. Phys.* 68 (2013), pp. 55–112.

[1034] Joshua Erlich. "An Introduction to Holographic QCD for Nonspecialists". In: *Contemp. Phys.* 56.2 (2015), pp. 159–171.

[1035] Martin Ammon and Johanna Erdmenger. *Gauge/gravity duality: Foundations and applications*. Cambridge: Cambridge University Press, Apr. 2015.

[1036] Guy F. de Téramond, Hans Günter Dosch, and Stanley J. Brodsky. "Kinematical and Dynamical Aspects of Higher-Spin Bound-State Equations in Holographic QCD". In: *Phys. Rev. D* 87.7 (2013), p. 075005.

[1037] Peter Breitenlohner and Daniel Z. Freedman. "Stability in Gauged Extended Supergravity". In: *Annals Phys.* 144 (1982), p. 249.

[1038] Ingo Kirsch. "Spectroscopy of fermionic operators in AdS/CFT". In: *JHEP* 09 (2006), p. 052.

[1039] Zainul Abidin and Carl E. Carlson. "Nucleon electromagnetic and gravitational form factors from holography". In: *Phys. Rev. D* 79 (2009), p. 115003.

[1040] Thomas Gutsche et al. "Dilaton in a soft-wall holographic approach to mesons and baryons". In: *Phys. Rev. D* 85 (2012), p. 076003.

[1041] Guy F. de Téramond and Stanley J. Brodsky. "Hadronic Form Factor Models and Spectroscopy Within the Gauge/Gravity Correspondence". In: *Ferrara International School Niccolò Cabeo 2011: Hadronic Physics*. 2011, pp. 54–109.

[1042] Edward Witten. "Dynamical Breaking of Supersymmetry". In: *Nucl. Phys. B* 188 (1981), p. 513.

[1043] Hans Gunter Dosch, Guy F. de Teramond, and Stanley J. Brodsky. "Supersymmetry Across the Light and Heavy-Light Hadronic Spectrum II". In: *Phys. Rev. D* 95.3 (2017), p. 034016.

[1044] Stanley J. Brodsky et al. "Universal Effective Hadron Dynamics from Superconformal Algebra". In: *Phys. Lett. B* 759 (2016), pp. 171–177.

[1045] S. W. MacDowell. "Analytic Properties of Partial Amplitudes in Meson-Nucleon Scattering". In: *Phys. Rev.* 116 (1959), pp. 774–778.

[1046] E. Klempt and B. C. Metsch. "Multiplet classification of light-quark baryons". In: *Eur. Phys. J. A* 48 (2012), p. 127.

[1047] Stanley J. Brodsky and Guy F. de Téramond. "AdS/CFT and Light-Front QCD". In: *Subnucl. Ser.* 45 (2009). Ed. by Antonino Zichichi, pp. 139–183.

[1048] Hans Gunter Dosch, Guy F. de Téramond, and Stanley J. Brodsky. "Supersymmetry Across the Light and Heavy-Light Hadronic Spectrum". In: *Phys. Rev. D* 92.7 (2015), p. 074010.

[1049] J. R. Forshaw and R. Sandapen. "An AdS/QCD holographic wavefunction for the rho meson and diffractive rho meson electroproduction". In: *Phys. Rev. Lett.* 109 (2012), p. 081601.

[1050] Edward V. Shuryak. "Hadrons Containing a Heavy Quark and QCD Sum Rules". In: *Nucl. Phys. B* 198 (1982), pp. 83–101.

[1051] Nathan Isgur and Mark B. Wise. "Spectroscopy with heavy quark symmetry". In: *Phys. Rev. Lett.* 66 (1991), pp. 1130–1133.

[1052] Thomas Gutsche et al. "Chiral Symmetry Breaking and Meson Wave Functions in Soft-Wall AdS/QCD". In: *Phys. Rev. D* 87.5 (2013), p. 056001.

[1053] Marina Nielsen et al. "Supersymmetry in the Double-Heavy Hadronic Spectrum". In: *Phys. Rev. D* 98.3 (2018), p. 034002.

[1054] H. G. Dosch et al. "Exotic states in a holographic theory". In: *Nucl. Part. Phys. Proc.* 312-317 (2021), pp. 135–139.

[1055] Arkadiusz P. Trawiński et al. "Effective confining potentials for QCD". In: *Phys. Rev. D* 90.7 (2014), p. 074017.

[1056] Sergey Afonin and Timofey Solomko. "Cornell potential in generalized soft wall holographic model". In: *J. Phys. G* 49.10 (2022), p. 105003.

[1057] Gerard 't Hooft. "A Two-Dimensional Model for Mesons". In: *Nucl. Phys. B* 75 (1974), pp. 461–470.

[1058] S. S. Chabysheva and John R. Hiller. "Dynamical model for longitudinal wave functions in light-front holographic QCD". In: *Annals Phys.* 337 (2013), pp. 143–152.

[1059] Guy F. de Teramond and Stanley J. Brodsky. "Longitudinal dynamics and chiral symmetry breaking in holographic light-front QCD". In: *Phys. Rev. D* 104.11 (2021), p. 116009.

[1060] Mohammad Ahmady et al. "Extending light-front holographic QCD using the 't Hooft Equation". In: *Phys. Lett. B* 823 (2021), p. 136754.

[1061] Mohammad Ahmady et al. "Hadron spectroscopy using the light-front holographic Schrödinger equation and the 't Hooft equation". In: *Phys. Rev. D* 104.7 (2021), p. 074013.

[1062] Edward Shuryak and Ismail Zahed. "Hadronic structure on the light-front II: QCD strings, Wilson lines and potentials". In: (Nov. 2021).

[1063] Edward Shuryak and Ismail Zahed. "Meson structure on the light-front III : The Hamiltonian, heavy quarkonia, spin and orbit mixing". In: (Dec. 2021).

[1064] Colin M. Weller and Gerald A. Miller. "Confinement in two-dimensional QCD and the infinitely long pion". In: *Phys. Rev. D* 105.3 (2022), p. 036009.

[1065] Yang Li and James P. Vary. "Longitudinal dynamics for mesons on the light cone". In: *Phys. Rev. D* 105.11 (2022), p. 114006.

[1066] Valery E. Lyubovitskij and Ivan Schmidt. "Meson masses and decay constants in holographic QCD consistent with ChPT and HQET". In: *Phys. Rev. D* 105.7 (2022), p. 074009.

[1067] Matteo Rinaldi, Federico Alberto Ceccopieri, and Vicente Vento. "The pion in the graviton soft-wall model: phenomenological applications". In: *Eur. Phys. J. C* 82.7 (2022), p. 626.

[1068] Mohammad Ahmady et al. "Pion spectroscopy and dynamics using the holographic light-front Schrödinger equation and the 't Hooft equation". In: (Aug. 2022).

[1069] Marina Nielsen and Stanley J. Brodsky. "Hadronic superpartners from a superconformal and supersymmetric algebra". In: *Phys. Rev. D* 97.11 (2018), p. 114001.

[1070] Marek Karliner and Jonathan L. Rosner. "Discovery of doubly-charmed $\Xi_{cc}$ baryon implies a stable $(bb\bar{u}\bar{d})$ tetraquark". In: *Phys. Rev. Lett.* 119.20 (2017), p. 202001.

[1071] Roel Aaij et al. "Observation of an exotic narrow doubly charmed tetraquark". In: (Sept. 2021).

[1072] G. F. Chew and Steven C. Frautschi. "Regge Trajectories and the Principle of Maximum Strength for Strong Interactions". In: *Phys. Rev. Lett.* 8 (1962), pp. 41–44.

[1073] T. Regge. "Introduction to complex orbital momenta". In: *Nuovo Cim.* 14 (1959), p. 951.

[1074] R. Dolen, D. Horn, and C. Schmid. "Finite-energy sum rules and their application to $\pi N$ charge exchange". In: *Phys. Rev.* 166 (1968), pp. 1768–1781.

[1075] Massimo Bianchi et al. "Partonic behavior of string scattering amplitudes from holographic QCD models". In: *JHEP* 05 (2022), p. 058.

[1076] H. R. Grigoryan and A. V. Radyushkin. "Structure of vector mesons in holographic model with linear confinement". In: *Phys. Rev. D* 76 (2007), p. 095007.

[1077] Raza Sabbir Sufian et al. "Analysis of nucleon electromagnetic form factors from light-front holographic QCD : The spacelike region". In: *Phys. Rev. D* 95.1 (2017), p. 014011.

[1078] M. Ademollo and E. Del Giudice. "Nonstrong amplitudes in a Veneziano-type model". In: *Nuovo Cim. A* 63 (1969), pp. 639–656.

[1079] P. V. Landshoff and J. C. Polkinghorne. "The scaling law for deep inelastic scattering in a new veneziano-like amplitude". In: *Nucl. Phys. B* 19 (1970), pp. 432–444.

[1080] Zhihong Ye et al. "Proton and Neutron Electromagnetic Form Factors and Uncertainties". In: *Phys. Lett. B* 777 (2018), pp. 8–15.

[1081] Raza Sabbir Sufian et al. "Nonperturbative strange-quark sea from lattice QCD, light-front holography, and meson-baryon fluctuation models". In: *Phys. Rev. D* 98.11 (2018), p. 114004.

[1082] Stanley J. Brodsky and G. Peter Lepage. "Exclusive Processes and the Exclusive Inclusive Connection in Quantum Chromodynamics". In: *Workshop on the Baryon Number of the Universe and Unified Theories.* Mar. 1979.

[1083] Dieter Müller et al. "Wave functions, evolution equations and evolution kernels from light ray operators of QCD". In: *Fortsch. Phys.* 42 (1994), pp. 101–141.

[1084] A. V. Radyushkin. "Scaling limit of deeply virtual Compton scattering". In: *Phys. Lett. B* 380 (1996), pp. 417–425.

[1085] Xiang-Dong Ji. "Gauge-Invariant Decomposition of Nucleon Spin". In: *Phys. Rev. Lett.* 78 (1997), pp. 610–613.

[1086] Sayipjamal Dulat et al. "New parton distribution functions from a global analysis of quantum chromodynamics". In: *Phys. Rev.* D93.3 (2016), p. 033006.

[1087] Richard D. Ball et al. "Parton distributions for the LHC Run II". In: *JHEP* 04 (2015), p. 040.

[1088] K. Wijesooriya, P. E. Reimer, and R. J. Holt. "The pion parton distribution function in the

valence region". In: *Phys. Rev. C* 72 (2005), p. 065203.

[1089] Matthias Aicher, Andreas Schafer, and Werner Vogelsang. "Soft-gluon resummation and the valence parton distribution function of the pion". In: *Phys. Rev. Lett.* 105 (2010), p. 252003.

[1090] J. S. Conway et al. "Experimental Study of Muon Pairs Produced by 252-GeV Pions on Tungsten". In: *Phys. Rev. D* 39 (1989), pp. 92–122.

[1091] Alexandre Deur, Stanley J. Brodsky, and Guy F. de Teramond. "On the Interface between Perturbative and Nonperturbative QCD". In: *Phys. Lett. B* 757 (2016), pp. 275–281.

[1092] Glennys R. Farrar and Darrell R. Jackson. "Pion and Nucleon Structure Functions Near $x = 1$". In: *Phys. Rev. Lett.* 35 (1975), p. 1416.

[1093] P. C. Barry et al. "Global QCD Analysis of Pion Parton Distributions with Threshold Resummation". In: *Phys. Rev. Lett.* 127.23 (2021), p. 232001.

[1094] P. C. Barry et al. "Complementarity of experimental and lattice QCD data on pion parton distributions". In: *Phys. Rev. D* 105.11 (2022), p. 114051.

[1095] Raza Sabbir Sufian et al. "Constraints on charm-anticharm asymmetry in the nucleon from lattice QCD". In: *Phys. Lett. B* 808 (2020), p. 135633.

[1096] Stanley J. Brodsky and Guy F. de Téramond. "Onset of Color Transparency in Holographic Light-Front QCD". In: *MDPI Physics* 4.2 (2022), pp. 633–646.

[1097] Dmitriy N. Kim and Gerald A. Miller. "Light-Front Holography Model of the EMC Effect". In: (Sept. 2022).

[1098] Richard C. Brower et al. "The Pomeron and gauge/string duality". In: *JHEP* 12 (2007), p. 005.

[1099] Lorenzo Cornalba and Miguel S. Costa. "Saturation in Deep Inelastic Scattering from AdS/CFT". In: *Phys. Rev. D* 78 (2008), p. 096010.

[1100] Sophia K. Domokos, Jeffrey A. Harvey, and Nelia Mann. "The Pomeron contribution to p p and p anti-p scattering in AdS/QCD". In: *Phys. Rev. D* 80 (2009), p. 126015.

[1101] Richard C. Brower et al. "String-Gauge Dual Description of Deep Inelastic Scattering at Small-$x$" In: *JHEP* 11 (2010), p. 051.

[1102] Miguel S. Costa and Marko Djuric. "Deeply Virtual Compton Scattering from Gauge/Gravity Duality". In: *Phys. Rev. D* 86 (2012), p. 016009.

[1103] David Jorrin and Martin Schvellinger. "Scope and limitations of a string theory dual description of the proton structure". In: *Phys. Rev. D* 106.6 (2022), p. 066024.

[1104] A. Donnachie and P. V. Landshoff. "Total cross-sections". In: *Phys. Lett. B* 296 (1992), pp. 227–232.

[1105] Zainul Abidin and Carl E. Carlson. "Gravitational form factors of vector mesons in an AdS/QCD model". In: *Phys. Rev. D* 77 (2008), p. 095007.

[1106] Csaba Csaki et al. "Glueball mass spectrum from supergravity". In: *JHEP* 01 (1999), p. 017.

[1107] Matteo Rinaldi and Vicente Vento. "Meson and glueball spectroscopy within the graviton soft wall model". In: *Phys. Rev. D* 104.3 (2021), p. 034016.

[1108] P. E. Shanahan and W. Detmold. "Gluon gravitational form factors of the nucleon and the pion from lattice QCD". In: *Phys. Rev. D* 99.1 (2019), p. 014511.

[1109] Dimitra A. Pefkou, Daniel C. Hackett, and Phiala E. Shanahan. "Gluon gravitational structure of hadrons of different spin". In: *Phys. Rev. D* 105.5 (2022), p. 054509.

[1110] S. Bailey et al. "Parton distributions from LHC, HERA, Tevatron and fixed target data: MSHT20 PDFs". In: *Eur. Phys. J. C* 81.4 (2021), p. 341.

[1111] N. Y. Cao et al. "Towards the three-dimensional parton structure of the pion: Integrating transverse momentum data into global QCD analysis". In: *Phys. Rev. D* 103.11 (2021), p. 114014.

[1112] E. A. Kuraev, L. N. Lipatov, and Victor S. Fadin. "The Pomeranchuk Singularity in Non-abelian Gauge Theories". In: *Sov. Phys. JETP* 45 (1977). [Zh. Eksp. Teor. Fiz.72,377(1977)], pp. 199–204.

[1113] Hans Günter Dosch et al. "Towards a single scale-dependent Pomeron in holographic light-front QCD". In: *Phys. Rev. D* 105.3 (2022), p. 034029.

[1114] H. G. Dosch and E. Ferreira. "Diffractive Electromagnetic Processes from a Regge Point of View". In: *Phys. Rev. D* 92 (2015), p. 034002.

[1115] Roel Aaij et al. "Updated measurements of exclusive $J/\psi$ and $\psi(2S)$ production cross-sections in pp collisions at $\sqrt{s} = 7$ TeV". In: *J. Phys. G* 41 (2014), p. 055002.

[1116] Betty Bezverkhny Abelev et al. "Exclusive J/$\psi$ photoproduction off protons in ultra-peripheral p-Pb collisions at $\sqrt{s_{\rm NN}} = 5.02$ TeV". In: *Phys. Rev. Lett.* 113.23 (2014), p. 232504.

[1117] A. Deur et al. "Experimental determination of the QCD effective charge $\alpha_{g_1}(Q)$". In: *Particles* 5 (2022), p. 171.

[1118] Jacqueline Stern and Gerard Clement. “Quarks, gluons, pions and the Gottfried sum rule”. In: *Phys. Lett. B* 264 (1991), pp. 426–431.

[1119] S. J. Brodsky et al. “Angular distributions of massive quarks and leptons close to threshold”. In: *Phys. Lett. B* 359 (1995), pp. 355–361.

[1120] Stefano Catani et al. “The Resummation of soft gluons in hadronic collisions”. In: *Nucl. Phys. B* 478 (1996), pp. 273–310.

[1121] Daniele Binosi et al. “Bridging a gap between continuum-QCD and ab initio predictions of hadron observables”. In: *Phys. Lett. B* 742 (2015), pp. 183–188.

[1122] Craig D. Roberts et al. “Insights into the emergence of mass from studies of pion and kaon structure”. In: *Prog. Part. Nucl. Phys.* 120 (2021), p. 103883.

[1123] J. C. Taylor. “Ward Identities and Charge Renormalization of the Yang-Mills Field”. In: *Nucl. Phys. B* 33 (1971), pp. 436–444.

[1124] A. A. Slavnov. “Ward Identities in Gauge Theories”. In: *Theor. Math. Phys.* 10 (1972), pp. 99–107.

[1125] John Clive Ward. “An Identity in Quantum Electrodynamics”. In: *Phys. Rev.* 78 (1950), p. 182.

[1126] Y. Takahashi. “On the generalized Ward identity”. In: *Nuovo Cim.* 6 (1957), p. 371.

[1127] Murray Gell-Mann and F. E. Low. “Quantum electrodynamics at small distances”. In: *Phys. Rev.* 95 (1954), pp. 1300–1312.

[1128] Alexandre Deur, Stanley J. Brodsky, and Guy F. de Teramond. “The QCD Running Coupling”. In: *Nucl. Phys.* 90 (2016), p. 1.

[1129] William E. Caswell. “Asymptotic Behavior of Nonabelian Gauge Theories to Two Loop Order”. In: *Phys. Rev. Lett.* 33 (1974), p. 244.

[1130] A. I. Sanda. “A Nonperturbative Determination of $\alpha(q^2)$ and Its Experimental Implications”. In: *Phys. Rev. Lett.* 42 (1979), p. 1658.

[1131] Tom Banks and A. Zaks. “On the Phase Structure of Vector-Like Gauge Theories with Massless Fermions”. In: *Nucl. Phys. B* 196 (1982), pp. 189–204.

[1132] John M. Cornwall. “Dynamical Mass Generation in Continuum QCD”. In: *Phys. Rev. D* 26 (1982), p. 1453.

[1133] John L. Richardson. “The Heavy Quark Potential and the Upsilon, J/psi Systems”. In: *Phys. Lett. B* 82 (1979), pp. 272–274.

[1134] D. V. Shirkov and I. L. Solovtsov. “Analytic model for the QCD running coupling with universal alpha-s (0) value”. In: *Phys. Rev. Lett.* 79 (1997), pp. 1209–1212.

[1135] Philippe Boucaud et al. “The Strong coupling constant at small momentum as an instanton detector”. In: *JHEP* 04 (2003), p. 005.

[1136] Stanley J. Brodsky, G. Peter Lepage, and Paul B. Mackenzie. “On the Elimination of Scale Ambiguities in Perturbative Quantum Chromodynamics”. In: *Phys. Rev.* D28 (1983), p. 228.

[1137] Stanley J. Brodsky and Xing-Gang Wu. “Scale Setting Using the Extended Renormalization Group and the Principle of Maximum Conformality: the QCD Coupling Constant at Four Loops”. In: *Phys. Rev. D* 85 (2012). [Erratum: Phys.Rev.D 86, 079903 (2012)], p. 034038.

[1138] Christoph Lerche and Lorenz von Smekal. “On the infrared exponent for gluon and ghost propagation in Landau gauge QCD”. In: *Phys. Rev. D* 65 (2002), p. 125006.

[1139] William Celmaster and Frank S. Henyey. “The Quark - Anti-quark Interaction at All Momentum Transfers”. In: *Phys. Rev. D* 18 (1978), p. 1688.

[1140] Robert Levine and Yukio Tomozawa. “An Effective Potential for Heavy Quark - Anti-quark Bound Systems”. In: *Phys. Rev. D* 19 (1979), p. 1572.

[1141] W. Buchmuller, G. Grunberg, and S. H. H. Tye. “The Regge Slope and the Lambda Parameter in QCD: An Empirical Approach via Quarkonia”. In: *Phys. Rev. Lett.* 45 (1980). [Erratum: Phys.Rev.Lett. 45, 587 (1980)], p. 103.

[1142] W. Buchmuller and S. H. H. Tye. “Quarkonia and Quantum Chromodynamics”. In: *Phys. Rev. D* 24 (1981), p. 132.

[1143] Daniele Binosi et al. “Process-independent strong running coupling”. In: *Phys. Rev. D* 96.5 (2017), p. 054026.

[1144] Zhu-Fang Cui et al. “Effective charge from lattice QCD”. In: *Chin. Phys. C* 44.8 (2020), p. 083102.

[1145] A. Deur et al. “Experimental determination of the effective strong coupling constant”. In: *Phys. Lett. B* 650 (2007), pp. 244–248.

[1146] A. Deur et al. “Determination of the effective strong coupling constant alpha(s,g(1))(Q**2) from CLAS spin structure function data”. In: *Phys. Lett. B* 665 (2008), pp. 349–351.

[1147] S. B. Gerasimov. “A Sum rule for magnetic moments and the damping of the nucleon magnetic moment in nuclei”. In: *Yad. Fiz.* 2 (1965), pp. 598–602.

[1148] S. D. Drell and Anthony C. Hearn. “Exact Sum Rule for Nucleon Magnetic Moments”. In: *Phys. Rev. Lett.* 16 (1966), pp. 908–911.

[1149] Amanda W. Peet and Joseph Polchinski. "UV / IR relations in AdS dynamics". In: *Phys. Rev. D* 59 (1999), p. 065011.
[1150] Alexandre Deur, Stanley J. Brodsky, and Guy F. de Teramond. "Connecting the Hadron Mass Scale to the Fundamental Mass Scale of Quantum Chromodynamics". In: *Phys. Lett. B* 750 (2015), pp. 528–532.
[1151] A. Deur, S. J. Brodsky, and G. F. de Teramond. "Determination of $\Lambda_{\bar{MS}}$ at five loops from holographic QCD". In: *J. Phys. G* 44.10 (2017), p. 105005.
[1152] K. Raya et al. "Structure of the neutral pion and its electromagnetic transition form factor". In: *Phys. Rev. D* 93.7 (2016), p. 074017.
[1153] Khepani Raya et al. "Partonic structure of neutral pseudoscalars via two photon transition form factors". In: *Phys. Rev. D* 95.7 (2017), p. 074014.
[1154] Jose Rodríguez-Quintero et al. "Process-independent effective coupling. From QCD Green's functions to phenomenology". In: *Few Body Syst.* 59.6 (2018). Ed. by R. Gothe et al., p. 121.
[1155] Chao Shi et al. "Kaon and pion parton distribution amplitudes to twist-three". In: *Phys. Rev. D* 92 (2015), p. 014035.
[1156] Minghui Ding et al. "Leading-twist parton distribution amplitudes of S-wave heavy-quarkonia". In: *Phys. Lett. B* 753 (2016), pp. 330–335.
[1157] Minghui Ding et al. "Drawing insights from pion parton distributions". In: *Chin. Phys. C* 44.3 (2020), p. 031002.
[1158] Minghui Ding et al. "Symmetry, symmetry breaking, and pion parton distributions". In: *Phys. Rev. D* 101.5 (2020), p. 054014.
[1159] Lei Chang and Craig D. Roberts. "Tracing masses of ground-state light-quark mesons". In: *Phys. Rev. C* 85 (2012), p. 052201.
[1160] Stanley J. Brodsky and Robert Shrock. "Maximum Wavelength of Confined Quarks and Gluons and Properties of Quantum Chromodynamics". In: *Phys. Lett. B* 666 (2008), pp. 95–99.
[1161] D. J. Gross and Frank Wilczek. "Asymptotically Free Gauge Theories - I". In: *Phys. Rev. D* 8 (1973), pp. 3633–3652.
[1162] D. J. Gross and Frank Wilczek. "Asymptotically Free Gauge Theories - II". In: *Phys. Rev. D* 9 (1974), pp. 980–993.
[1163] Gerard 't Hooft. "A Planar Diagram Theory for Strong Interactions". In: *Nucl. Phys. B* 72 (1974). Ed. by J. C. Taylor, p. 461.
[1164] Elizabeth Ellen Jenkins and Richard F. Lebed. "Baryon mass splittings in the 1/N(c) expansion". In: *Phys. Rev. D* 52 (1995), pp. 282–294.
[1165] Roland Kaiser and H. Leutwyler. "Large N(c) in chiral perturbation theory". In: *Eur. Phys. J. C* 17 (2000), pp. 623–649.
[1166] Sidney Coleman. *Aspects of Symmetry: Selected Erice Lectures.* Cambridge, U.K.: Cambridge University Press, 1985.
[1167] Biagio Lucini and Marco Panero. "SU(N) gauge theories at large N". In: *Phys. Rept.* 526 (2013), pp. 93–163.
[1168] N. Matagne and Fl. Stancu. "Baryon resonances in large $N_c$ QCD". In: *Rev. Mod. Phys.* 87 (2015), pp. 211–245.
[1169] Edward Witten. "Baryons in the 1/n Expansion". In: *Nucl. Phys. B* 160 (1979), pp. 57–115.
[1170] Thomas D. Cohen. "Quantum number exotic hybrid mesons and large N(c) QCD". In: *Phys. Lett. B* 427 (1998), pp. 348–352.
[1171] S. Okubo. "Phi meson and unitary symmetry model". In: *Phys. Lett.* 5 (1963), pp. 165–168.
[1172] Jugoro Iizuka. "Systematics and phenomenology of meson family". In: *Prog. Theor. Phys. Suppl.* 37 (1966), pp. 21–34.
[1173] Thomas D. Cohen and Richard F. Lebed. "Are There Tetraquarks at Large $N_c$ in QCD(F)?" In: *Phys. Rev. D* 90.1 (2014), p. 016001.
[1174] Gerard 't Hooft. "How Instantons Solve the U(1) Problem". In: *Phys. Rept.* 142 (1986), pp. 357–387.
[1175] Edward Witten. "Current Algebra Theorems for the U(1) Goldstone Boson". In: *Nucl. Phys. B* 156 (1979), pp. 269–283.
[1176] A. R. Zhitnitsky. "On Chiral Symmetry Breaking in QCD in Two-dimensions ($N_c \to$ Infinity)". In: *Phys. Lett. B* 165 (1985), pp. 405–409.
[1177] Sidney R. Coleman. "There are no Goldstone bosons in two-dimensions". In: *Commun. Math. Phys.* 31 (1973), pp. 259–264.
[1178] Edward Witten. "Chiral Symmetry, the 1/n Expansion, and the SU(N) Thirring Model". In: *Nucl. Phys. B* 145 (1978), pp. 110–118.
[1179] V. L. Berezinski. In: *Phys. JETP* 32 (1970), p. 493.
[1180] J. M. Kosterlitz and D. J. Thouless. "Ordering, metastability and phase transitions in two-dimensional systems". In: *J. Phys. C* 6 (1973), pp. 1181–1203.
[1181] Gregory S. Adkins and Chiara R. Nappi. "The Skyrme Model with Pion Masses". In: *Nucl. Phys. B* 233 (1984), pp. 109–115.
[1182] Gregory S. Adkins, Chiara R. Nappi, and Edward Witten. "Static Properties of Nucleons

in the Skyrme Model". In: *Nucl. Phys. B* 228 (1983), p. 552.

[1183] I. Zahed and G. E. Brown. "The Skyrme Model". In: *Phys. Rept.* 142 (1986), pp. 1–102.

[1184] G. E. Brown, ed. *Selected papers, with commentary, of Tony Hilton Royle Skyrme.* 1995.

[1185] Gregory S. Adkins and Chiara R. Nappi. "Model Independent Relations for Baryons as Solitons in Mesonic Theories". In: *Nucl. Phys. B* 249 (1985), pp. 507–518.

[1186] Jean-Loup Gervais and B. Sakita. "Large N QCD Baryon Dynamics: Exact Results from Its Relation to the Static Strong Coupling Theory". In: *Phys. Rev. Lett.* 52 (1984), p. 87.

[1187] Roger F. Dashen and Aneesh V. Manohar. "Baryon - pion couplings from large N(c) QCD". In: *Phys. Lett. B* 315 (1993), pp. 425–430.

[1188] Roger F. Dashen, Elizabeth Ellen Jenkins, and Aneesh V. Manohar. "The 1/N(c) expansion for baryons". In: *Phys. Rev. D* 49 (1994). [Erratum: Phys.Rev.D 51, 2489 (1995)], p. 4713.

[1189] Roger F. Dashen and Aneesh V. Manohar. "1/N(c) corrections to the baryon axial currents in QCD". In: *Phys. Lett. B* 315 (1993), pp. 438–440.

[1190] Thomas D. Cohen and Boris A. Gelman. "Nucleon-nucleon scattering observables in large N(c) QCD". In: *Phys. Lett. B* 540 (2002), pp. 227–232.

[1191] Thomas D. Cohen and Boris A. Gelman. "Total nucleon-nucleon cross sections in large N(c) QCD". In: *Phys. Rev. C* 85 (2012), p. 024001.

[1192] David B. Kaplan and Martin J. Savage. "The Spin flavor dependence of nuclear forces from large n QCD". In: *Phys. Lett. B* 365 (1996), pp. 244–251.

[1193] David B. Kaplan and Aneesh V. Manohar. "The Nucleon-nucleon potential in the 1/N(c) expansion". In: *Phys. Rev. C* 56 (1997), pp. 76–83.

[1194] Manoj K. Banerjee, Thomas D. Cohen, and Boris A. Gelman. "The Nucleon nucleon interaction and large N(c) QCD". In: *Phys. Rev. C* 65 (2002), p. 034011.

[1195] G. Veneziano. "Some Aspects of a Unified Approach to Gauge, Dual and Gribov Theories". In: *Nucl. Phys. B* 117 (1976), pp. 519–545.

[1196] A. Armoni, M. Shifman, and G. Veneziano. "SUSY relics in one flavor QCD from a new 1/N expansion". In: *Phys. Rev. Lett.* 91 (2003), p. 191601.

[1197] A. Armoni, M. Shifman, and G. Veneziano. "Exact results in non-supersymmetric large N orientifold field theories". In: *Nucl. Phys. B* 667 (2003), pp. 170–182.

[1198] A. Armoni, M. Shifman, and G. Veneziano. "From superYang-Mills theory to QCD: Planar equivalence and its implications". In: *From fields to strings: Circumnavigating theoretical physics. Ian Kogan memorial collection (3 volume set).* Ed. by M. Shifman, A. Vainshtein, and J. Wheater. Mar. 2004, pp. 353–444.

[1199] Thomas D. Cohen and Richard F. Lebed. "Tetraquarks with exotic flavor quantum numbers at large $N_c$ in QCD(AS)". In: *Phys. Rev. D* 89.5 (2014), p. 054018.

[1200] Thomas D. Cohen, Daniel L. Shafer, and Richard F. Lebed. "Baryons in QCD(AS) at Large N(c): A Roundabout Approach". In: *Phys. Rev. D* 81 (2010), p. 036006.

[1201] Aleksey Cherman, Thomas D. Cohen, and Richard F. Lebed. "All you need is N: Baryon spectroscopy in two large N limits". In: *Phys. Rev. D* 80 (2009), p. 036002.

[1202] "Proceedings of the 16th International Conference On High-Energy Physics". In: ed. by A. Roberts J.D. Jackson. Batavia, Illinois, Sept. 1972.

[1203] L. D. Faddeev and V. N. Popov. "Feynman Diagrams for the Yang-Mills Field". In: *Phys. Lett. B* 25 (1967), pp. 29–30.

[1204] David J. Gross. "Twenty five years of asymptotic freedom". In: *Nucl. Phys. B Proc. Suppl.* 74 (1999), pp. 426–446.

[1205] M. Shifman. "Historical curiosity: How asymptotic freedom of the Yang-Mills theory could have been discovered three times before Gross, Wilczek, and Politzer, but was not". In: *At the Frontier of Particle Physics. Handbook of QCD,* (World Scientific) (2001). Ed. by M. Shifman, pp. 126–130.

[1206] Steven Weinberg. "The U(1) Problem". In: *Phys. Rev. D* 11 (1975), pp. 3583–3593.

[1207] E. B. Bogomolnyi, V. A. Novikov, and Mikhail A. Shifman. "Behaviour of the physical charge at small distances in nonabelian gauge theories". In: *Sov. J. Nucl. Phys.* 20 (1974), p. 110.

[1208] A. I. Vainshtein, Valentin I. Zakharov, and Mikhail A. Shifman. "A Possible mechanism for the Delta T = 1/2 rule in nonleptonic decays of strange particles". In: *JETP Lett.* 22 (1975), pp. 55–56.

[1209] Mikhail A. Shifman, A.I. Vainshtein, and Valentin I. Zakharov. "Light Quarks and the Origin of the $\Delta I = 1/2$ Rule in the Nonleptonic Decays of Strange Particles". In: *Nucl. Phys.* B120 (1977), p. 316.

[1210] Arkady I. Vainshtein. "How penguins started to fly". In: *Int. J. Mod. Phys. A* 14 (1999), pp. 4705–4719.

[1211] M.K. Gaillard and Benjamin W. Lee. "$\Delta I = 1/2$ Rule for Nonleptonic Decays in Asymptot-

ically Free Field Theories". In: *Phys. Rev. Lett.* 33 (1974), p. 108.

[1212] Guido Altarelli and L. Maiani. "Octet Enhancement of Nonleptonic Weak Interactions in Asymptotically Free Gauge Theories". In: *Phys. Lett.* B52 (1974), pp. 351–354.

[1213] A. I. Vainshtein, Valentin I. Zakharov, and Mikhail A. Shifman. "Gluon condensate and lepton decays of vector mesons. (In Russian)". In: *JETP Lett.* 27 (1978), pp. 55–58.

[1214] V. A. Novikov et al. "Wilson's Operator Expansion: Can It Fail?" In: *Nucl. Phys. B* 249 (1985), pp. 445–471.

[1215] Mikhail A. Shifman. "Snapshots of hadrons or the story of how the vacuum medium determines the properties of the classical mesons which are produced, live and die in the QCD vacuum". In: *Prog. Theor. Phys. Suppl.* 131 (1998), pp. 1–71.

[1216] M. Shifman. "Resurgence, operator product expansion, and remarks on renormalons in supersymmetric Yang-Mills theory". In: *J. Exp. Theor. Phys.* 120.3 (2015), pp. 386–398.

[1217] N. Seiberg and Edward Witten. "Electric - magnetic duality, monopole condensation, and confinement in N=2 supersymmetric Yang-Mills theory". In: *Nucl. Phys. B* 426 (1994). [Erratum: Nucl.Phys.B 430, 485–486 (1994)], pp. 19–52.

[1218] N. Seiberg and Edward Witten. "Monopoles, duality and chiral symmetry breaking in N=2 supersymmetric QCD". In: *Nucl. Phys. B* 431 (1994), pp. 484–550.

[1219] Mikhail A. Shifman, A. I. Vainshtein, and Valentin I. Zakharov. "QCD and Resonance Physics: Applications". In: *Nucl. Phys. B* 147 (1979), pp. 448–518.

[1220] Mikhail A. Shifman, A. I. Vainshtein, and Valentin I. Zakharov. "QCD and Resonance Physics. The rho-omega Mixing". In: *Nucl. Phys. B* 147 (1979), pp. 519–534.

[1221] V. A. Novikov et al. "Are All Hadrons Alike? " In: *Nucl. Phys. B* 191 (1981), pp. 301–369.

[1222] M. Shifman. "New and Old about Renormalons: in Memoriam of Kolya Uraltsev". In: *Int. J. Mod. Phys. A* 30.10 (2015), p. 1543001.

[1223] Ikaros I. Y. Bigi et al. "The Pole mass of the heavy quark. Perturbation theory and beyond". In: *Phys. Rev. D* 50 (1994), pp. 2234–2246.

[1224] M. Beneke and Vladimir M. Braun. "Heavy quark effective theory beyond perturbation theory: Renormalons, the pole mass and the residual mass term". In: *Nucl. Phys. B* 426 (1994), pp. 301–343.

[1225] Ikaros I. Y. Bigi, Mikhail A. Shifman, and N. Uraltsev. "Aspects of heavy quark theory". In: *Ann. Rev. Nucl. Part. Sci.* 47 (1997), pp. 591–661.

[1226] Daniel Schubring, Chao-Hsiang Sheu, and Mikhail Shifman. "Treating divergent perturbation theory: Lessons from exactly solvable 2D models at large N". In: *Phys. Rev. D* 104.8 (2021), p. 085016.

[1227] Mikhail A. Shifman, ed. *Vacuum structure and QCD sum rules*. North-Holland, Elsevier Science Publishers, 1992.

[1228] M. Shifman. "Vacuum structure and QCD sum rules: Introduction". In: *Int. J. Mod. Phys. A* 25 (2010), pp. 226–235.

[1229] Vladimir M. Braun. "Light cone sum rules". In: 1997, pp. 105–118.

[1230] Alexander Khodjamirian, Thomas Mannel, and Nils Offen. "Form-factors from light-cone sum rules with B-meson distribution amplitudes". In: *Phys. Rev. D* 75 (2007), p. 054013.

[1231] Pietro Colangelo and Alexander Khodjamirian. "QCD sum rules, a modern perspective". In: (2000), pp. 1495–1576.

[1232] W. Braunschweig et al. "Radiative Decays of the J/psi and Evidence for a New Heavy Resonance". In: *Phys. Lett. B* 67 (1977), pp. 243–248.

[1233] Mikhail A. Shifman et al. "eta(c) Puzzle in Quantum Chromodynamics". In: *Phys. Lett. B* 77 (1978), pp. 80–83.

[1234] Richard Partridge et al. "Observation of an eta(c) Candidate State with Mass 2978-MeV +- 9-MeV". In: *Phys. Rev. Lett.* 45 (1980), pp. 1150–1153.

[1235] Mikhail A. Shifman and M. B. Voloshin. "Preasymptotic Effects in Inclusive Weak Decays of Charmed Particles". In: *Sov. J. Nucl. Phys.* 41 (1985), p. 120.

[1236] Mikhail A. Shifman and M. B. Voloshin. "Hierarchy of Lifetimes of Charmed and Beautiful Hadrons". In: *Sov. Phys. JETP* 64 (1986), p. 698.

[1237] Nikolai Uraltsev. "Topics in the heavy quark expansion". In: (2000), pp. 1577–1670.

[1238] Mikhail A. Shifman. "Recent progress in the heavy quark theory". In: *PASCOS / HOPKINS 1995 (Joint Meeting of the International Symposium on Particles, Strings and Cosmology and the 19th Johns Hopkins Workshop on Cur-*

*rent Problems in Particle Theory).* Mar. 1995, pp. 0069–94.

[1239] Alexander Lenz. "Lifetimes and heavy quark expansion". In: *Int. J. Mod. Phys. A* 30.10 (2015), p. 1543005.

[1240] M. Kirk, A. Lenz, and T. Rauh. "Dimension-six matrix elements for meson mixing and lifetimes from sum rules". In: *JHEP* 12 (2017). [Erratum: JHEP 06, 162 (2020)], p. 068.

[1241] Matteo Fael, Kay Schönwald, and Matthias Steinhauser. "Third order corrections to the semileptonic b→c and the muon decays". In: *Phys. Rev. D* 104.1 (2021), p. 016003.

[1242] Alexander Lenz, Maria Laura Piscopo, and Aleksey V. Rusov. "Contribution of the Darwin operator to non-leptonic decays of heavy quarks". In: *JHEP* 12 (2020), p. 199.

[1243] Thomas Mannel, Daniel Moreno, and Alexei Pivovarov. "Heavy quark expansion for heavy hadron lifetimes: completing the $1/m_b^3$ corrections". In: *JHEP* 08 (2020), p. 089.

[1244] Daniel King et al. "Revisiting Inclusive Decay Widths of Charmed Mesons". In: (2021).

[1245] James Gratrex, Blaženka Melić, and Ivan Nišandžić "Lifetimes of singly charmed hadrons". In: (2022).

[1246] Roel Aaij et al. "Measurement of the $Omega_c^0$ baryon lifetime". In: *Phys. Rev. Lett.* 121.9 (2018), p. 092003.

[1247] Roel Aaij et al. "Precision measurement of the $Lambda_c^+$, $Xi_c^+$ and $Xi_c^0$ baryon lifetimes". In: *Phys. Rev. D* 100.3 (2019), p. 032001.

[1248] Roel Aaij et al. "Measurement of the lifetimes of promptly produced $Omega_c^0$ and $Xi_c^0$ baryons". In: *Sci. Bull.* 67.5 (2022), pp. 479–487.

[1249] Boris Blok and Mikhail A. Shifman. "Lifetimes of charmed hadrons revisited. Facts and fancy". In: *3rd Workshop on the Tau-Charm Factory.* 1993.

[1250] Shmuel Nussinov and Werner Wetzel. "Comparison of Exclusive Decay Rates for b —> u and b —> c Transitions". In: *Phys. Rev. D* 36 (1987), p. 130.

[1251] Mikhail A. Shifman and M. B. Voloshin. "On Production of d and D* Mesons in B Meson Decays". In: *Sov. J. Nucl. Phys.* 47 (1988), p. 511.

[1252] Nathan Isgur and Mark B. Wise. "Weak Decays of Heavy Mesons in the Static Quark Approximation". In: *Phys. Lett. B* 232 (1989), pp. 113–117.

[1253] Nathan Isgur and Mark B. Wise. "Weak transition form-factors between heavy mesons". In: *Phys. Lett. B* 237 (1990), pp. 527–530.

[1254] Howard Georgi. "An Effective Field Theory for Heavy Quarks at Low-energies". In: *Phys. Lett. B* 240 (1990), pp. 447–450.

[1255] Junegone Chay, Howard Georgi, and Benjamin Grinstein. "Lepton energy distributions in heavy meson decays from QCD". In: *Phys. Lett. B* 247 (1990), pp. 399–405.

[1256] Ikaros I. Y. Bigi, N. G. Uraltsev, and A. I. Vainshtein. "Nonperturbative corrections to inclusive beauty and charm decays: QCD versus phenomenological models". In: *Phys. Lett. B* 293 (1992). [Erratum: Phys.Lett.B 297, 477–477 (1992)], pp. 430–436.

[1257] Ikaros I. Y. Bigi et al. "QCD predictions for lepton spectra in inclusive heavy flavor decays". In: *Phys. Rev. Lett.* 71 (1993), pp. 496–499.

[1258] R. L. Jaffe and Lisa Randall. "Heavy quark fragmentation into heavy mesons". In: *Nucl. Phys. B* 412 (1994), pp. 79–105.

[1259] Matthias Neubert. "QCD based interpretation of the lepton spectrum in inclusive anti-B —> X(u) lepton anti-neutrino decays". In: *Phys. Rev. D* 49 (1994), pp. 3392–3398.

[1260] Ikaros I. Y. Bigi et al. "On the motion of heavy quarks inside hadrons: Universal distributions and inclusive decays". In: *Int. J. Mod. Phys. A* 9 (1994), pp. 2467–2504.

[1261] M. B. Voloshin and Mikhail A. Shifman. "On the annihilation constants of mesons consisting of a heavy and a light quark, and $B^0 \leftrightarrow \overline{B}^0$ oscillations". In: *Sov. J. Nucl. Phys.* 45 (1987), p. 292.

[1262] H. David Politzer and Mark B. Wise. "Leading Logarithms of Heavy Quark Masses in Processes with Light and Heavy Quarks". In: *Phys. Lett. B* 206 (1988), pp. 681–684.

[1263] Mikhail A. Shifman. "Quark hadron duality". In: *8th International Symposium on Heavy Flavor Physics.* Vol. 3. Singapore: World Scientific, July 2000, pp. 1447–1494.

[1264] Mikhail A. Shifman. "QCD sum rules: The Second decade". In: *Workshop on QCD: 20 Years Later.* Apr. 1993, pp. 775–794.

[1265] Davide Gaiotto et al. "Generalized Global Symmetries". In: *JHEP* 02 (2015), p. 172.

[1266] Davide Gaiotto et al. "Theta, Time Reversal, and Temperature". In: *JHEP* 05 (2017), p. 091.

[1267] Stephen B. Libby and George F. Sterman. "Jet and Lepton Pair Production in High-Energy Lepton-Hadron and Hadron-Hadron Scattering". In: *Phys. Rev. D* 18 (1978), p. 3252.

[1268] Stephen B. Libby and George F. Sterman. "Mass Divergences in Two Particle Inelastic Scattering". In: *Phys. Rev. D* 18 (1978), p. 4737.

[1269] John Collins. *Foundations of perturbative QCD*. Vol. 32. Cambridge University Press, Nov. 2013.

[1270] Jian-Wei Qiu. "Twist Four Contributions to the Parton Structure Functions". In: *Phys. Rev. D* 42 (1990), pp. 30–44.

[1271] E. Reya. "Perturbative Quantum Chromodynamics". In: *Phys. Rept.* 69 (1981), p. 195.

[1272] Alfred H. Mueller. "Perturbative QCD at High-Energies". In: *Phys. Rept.* 73 (1981), p. 237.

[1273] Guido Altarelli. "Partons in Quantum Chromodynamics". In: *Phys. Rept.* 81 (1982), p. 1.

[1274] John C. Collins and Davison E. Soper. "Parton Distribution and Decay Functions". In: *Nucl. Phys. B* 194 (1982), pp. 445–492.

[1275] Raymond Brock et al. "Handbook of perturbative QCD: Version 1.0". In: *Rev. Mod. Phys.* 67 (1995), pp. 157–248.

[1276] Jian-Wei Qiu et al. "Factorization of jet cross sections in heavy-ion collisions". In: *Phys. Rev. Lett.* 122.25 (2019), p. 252301.

[1277] Gouranga C. Nayak, Jian-Wei Qiu, and George F. Sterman. "Fragmentation, NRQCD and NNLO factorization analysis in heavy quarkonium production". In: *Phys. Rev.* D72 (2005), p. 114012.

[1278] Elke-Caroline Aschenauer et al. "The RHIC SPIN Program: Achievements and Future Opportunities". In: (Jan. 2015).

[1279] P. Aurenche et al. "Large p(T) inclusive pi0 cross-sections and next-to-leading-order QCD predictions". In: *Eur. Phys. J. C* 13 (2000), pp. 347–355.

[1280] Daniel de Florian and Werner Vogelsang. "Threshold resummation for the inclusive-hadron cross-section in pp collisions". In: *Phys. Rev. D* 71 (2005), p. 114004.

[1281] Xabier Cid Vidal et al. "Report from Working Group 3: Beyond the Standard Model physics at the HL-LHC and HE-LHC". In: *CERN Yellow Rep. Monogr.* 7 (2019). Ed. by Andrea Dainese et al., pp. 585–865.

[1282] John C. Collins, Davison E. Soper, and George F. Sterman. "Transverse Momentum Distribution in Drell-Yan Pair and W and Z Boson Production". In: *Nucl. Phys.* B250 (1985), pp. 199–224.

[1283] Dennis W. Sivers. "Single Spin Production Asymmetries from the Hard Scattering of Point-Like Constituents". In: *Phys. Rev. D* 41 (1990), p. 83.

[1284] John C. Collins. "Fragmentation of transversely polarized quarks probed in transverse momentum distributions". In: *Nucl. Phys. B* 396 (1993), pp. 161–182.

[1285] Xiang-dong Ji, Jian-ping Ma, and Feng Yuan. "QCD factorization for semi-inclusive deep-inelastic scattering at low transverse momentum". In: *Phys. Rev. D* 71 (2005), p. 034005.

[1286] Alessandro Bacchetta et al. "Semi-inclusive deep inelastic scattering at small transverse momentum". In: *JHEP* 02 (2007), p. 093.

[1287] Markus Diehl. "Introduction to GPDs and TMDs". In: *Eur. Phys. J. A* 52.6 (2016), p. 149.

[1288] Xiang-Dong Ji. "Deeply virtual Compton scattering". In: *Phys. Rev. D* 55 (1997), pp. 7114–7125.

[1289] John C. Collins, L. Frankfurt, and M. Strikman. "Proof of factorization for exclusive deep inelastic processes". In: *Madrid Workshop on Low x Physics*. June 1997, pp. 296–303.

[1290] John C. Collins, Leonid Frankfurt, and Mark Strikman. "Factorization for hard exclusive electroproduction of mesons in QCD". In: *Phys. Rev. D* 56 (1997), pp. 2982–3006.

[1291] John C. Collins and Andreas Freund. "Proof of factorization for deeply virtual Compton scattering in QCD". In: *Phys. Rev. D* 59 (1999), p. 074009.

[1292] Xiang-Dong Ji and Jonathan Osborne. "One loop corrections and all order factorization in deeply virtual Compton scattering". In: *Phys. Rev. D* 58 (1998), p. 094018.

[1293] Jian-Wei Qiu and Zhite Yu. "Exclusive production of a pair of high transverse momentum photons in pion-nucleon collisions for extracting generalized parton distributions". In: *JHEP* 08 (2022), p. 103.

[1294] Jian-Wei Qiu and Zhite Yu. "Single diffractive hard exclusive processes for the study of generalized parton distributions". In: (Oct. 2022).

[1295] Daniel de Florian et al. "Extraction of Spin-Dependent Parton Densities and Their Uncertainties". In: *Phys. Rev. D* 80 (2009), p. 034030.

[1296] J. J. Ethier, N. Sato, and W. Melnitchouk. "First simultaneous extraction of spin-dependent parton distributions and fragmentation functions from a global QCD analysis". In: *Phys. Rev. Lett.* 119.13 (2017), p. 132001.

[1297] Stanley J. Brodsky, Dae Sung Hwang, and Ivan Schmidt. "Final state interactions and single spin asymmetries in semiinclusive deep inelastic scattering". In: *Phys. Lett. B* 530 (2002), pp. 99–107.

[1298] Xiang-dong Ji and Feng Yuan. "Parton distributions in light cone gauge: Where are the fi-

nal state interactions?" In: *Phys. Lett. B* 543 (2002), pp. 66–72.

[1299] John C. Collins and Andreas Metz. "Universality of soft and collinear factors in hard-scattering factorization". In: *Phys. Rev. Lett.* 93 (2004), p. 252001.

[1300] A. Bacchetta et al. "Single spin asymmetries in hadron-hadron collisions". In: *Phys. Rev. D* 72 (2005), p. 034030.

[1301] A. V. Efremov and O. V. Teryaev. "QCD Asymmetry and Polarized Hadron Structure Functions". In: *Phys. Lett. B* 150 (1985), p. 383.

[1302] Jian-wei Qiu and George F. Sterman. "Single transverse spin asymmetries". In: *Phys. Rev. Lett.* 67 (1991), pp. 2264–2267.

[1303] Jian-wei Qiu and George F. Sterman. "Single transverse spin asymmetries in direct photon production". In: *Nucl. Phys. B* 378 (1992), pp. 52–78.

[1304] Jian-wei Qiu and George F. Sterman. "Single transverse spin asymmetries in hadronic pion production". In: *Phys. Rev. D* 59 (1999), p. 014004.

[1305] Chris Kouvaris et al. "Single transverse-spin asymmetry in high transverse momentum pion production in pp collisions". In: *Phys. Rev. D* 74 (2006), p. 114013.

[1306] Yuji Koike and Kazuhiro Tanaka. "Master Formula for Twist-3 Soft-Gluon-Pole Mechanism to Single Transverse-Spin Asymmetry". In: *Phys. Lett. B* 646 (2007). [Erratum: Phys.Lett.B 668, 458–459 (2008)], pp. 232–241.

[1307] Jian-Wei Qiu, Werner Vogelsang, and Feng Yuan. "Asymmetric di-jet production in polarized hadronic collisions". In: *Phys. Lett. B* 650 (2007), pp. 373–378.

[1308] Zhong-Bo Kang and Jian-Wei Qiu. "Testing the Time-Reversal Modified Universality of the Sivers Function". In: *Phys. Rev. Lett.* 103 (2009), p. 172001.

[1309] Zhong-Bo Kang, Feng Yuan, and Jian Zhou. "Twist-three fragmentation function contribution to the single spin asymmetry in p p collisions". In: *Phys. Lett. B* 691 (2010), pp. 243–248.

[1310] Andrei V. Belitsky, X. Ji, and F. Yuan. "Final state interactions and gauge invariant parton distributions". In: *Nucl. Phys. B* 656 (2003), pp. 165–198.

[1311] John C. Collins. "Leading twist single transverse-spin asymmetries: Drell-Yan and deep inelastic scattering". In: *Phys. Lett. B* 536 (2002), pp. 43–48.

[1312] Jian-Wei Qiu and George F. Sterman. "Power corrections to hadronic scattering. 2. Factorization". In: *Nucl. Phys. B* 353 (1991), pp. 137–164.

[1313] Jian-Wei Qiu and George F. Sterman. "Power corrections in hadronic scattering. 1. Leading 1/Q**2 corrections to the Drell-Yan cross-section". In: *Nucl. Phys. B* 353 (1991), pp. 105–136.

[1314] Zhong-Bo Kang et al. "Heavy Quarkonium Production at Collider Energies: Factorization and Evolution". In: *Phys. Rev. D* 90.3 (2014), p. 034006.

[1315] Xiang-Dong Ji. "Gluon correlations in the transversely polarized nucleon". In: *Phys. Lett. B* 289 (1992), pp. 137–142.

[1316] Yuji Koike and Kazuhiro Tanaka. "Universal structure of twist-3 soft-gluon-pole cross-sections for single transverse-spin asymmetry". In: *Phys. Rev. D* 76 (2007), p. 011502.

[1317] A. Metz and D. Pitonyak. "Fragmentation contribution to the transverse single-spin asymmetry in proton-proton collisions". In: *Phys. Lett. B* 723 (2013). [Erratum: Phys.Lett.B 762, 549–549 (2016)], pp. 365–370.

[1318] Zhong-Bo Kang and Jian-Wei Qiu. "Evolution of twist-3 multi-parton correlation functions relevant to single transverse-spin asymmetry". In: *Phys. Rev. D* 79 (2009), p. 016003.

[1319] V. M. Braun, A. N. Manashov, and B. Pirnay. "Scale dependence of twist-three contributions to single spin asymmetries". In: *Phys. Rev. D* 80 (2009). [Erratum: Phys.Rev.D 86, 119902 (2012)], p. 114002.

[1320] Zhong-Bo Kang. "QCD evolution of naive-time-reversal-odd fragmentation functions". In: *Phys. Rev. D* 83 (2011), p. 036006.

[1321] Xiangdong Ji et al. "A Unified picture for single transverse-spin asymmetries in hard processes". In: *Phys. Rev. Lett.* 97 (2006), p. 082002.

[1322] Alessandro Bacchetta et al. "Matches and mismatches in the descriptions of semi-inclusive processes at low and high transverse momentum". In: *JHEP* 08 (2008), p. 023.

[1323] M.L. Perl. "High energy hadron physics". In: (1974).

[1324] T. T. Chou and Chen-Ning Yang. "Model of Elastic High-Energy Scattering". In: *Phys. Rev.* 170 (1968), pp. 1591–1596.

[1325] G. Antchev et al. "Proton-proton elastic scattering at the LHC energy of s** (1/2) = 7-TeV". In: *EPL* 95.4 (2011), p. 41001.

[1326] Stanley J. Brodsky and Glennys R. Farrar. "Scaling Laws for Large Momentum Transfer Processes". In: *Phys. Rev. D* 11 (1975), p. 1309.

[1327] George Sterman. "Fixed Angle Scattering and the Transverse Structure of Hadrons". In: *4th*

*Workshop on Exclusive Reactions at High Momentum Transfer*. 2011, pp. 16–25.
[1328] Stanley J. Brodsky. "Exclusive Processes and the Fundamental Structure of Hadrons". In: *Int. J. Mod. Phys. A* 30.02 (2015), p. 1530014.
[1329] C. White et al. "Comparison of 20 exclusive reactions at large t". In: *Phys. Rev. D* 49 (1994), pp. 58–78.
[1330] Stanley J. Brodsky and Alfred H. Mueller. "Using Nuclei to Probe Hadronization in QCD". In: *Phys. Lett. B* 206 (1988), pp. 685–690.
[1331] D. Bhetuwal et al. "Ruling out Color Transparency in Quasielastic $^{12}$C(e,e'p) up to $Q^2$ of 14.2 (GeV/c)$^2$". In: *Phys. Rev. Lett.* 126.8 (2021), p. 082301.
[1332] L. Frankfurt, G. A. Miller, and M. Strikman. "Coherent nuclear diffractive production of mini - jets: Illuminating color transparency". In: *Phys. Lett. B* 304 (1993), pp. 1–7.
[1333] Pankaj Jain, Bernard Pire, and John P. Ralston. "Quantum color transparency and nuclear filtering". In: *Phys. Rept.* 271 (1996), pp. 67–179.
[1334] P. V. Landshoff. "Model for elastic scattering at wide angle". In: *Phys. Rev. D* 10 (1974), pp. 1024–1030.
[1335] E. Nagy et al. "Measurements of Elastic Proton Proton Scattering at Large Momentum Transfer at the CERN Intersecting Storage Rings". In: *Nucl. Phys. B* 150 (1979), pp. 221–267.
[1336] W. Faissler et al. "Large Angle Proton Proton Elastic Scattering at 201-GeV/$c$ and 400-GeV/$c$". In: *Phys. Rev. D* 23 (1981), p. 33.
[1337] Ashoke Sen. "Asymptotic Behavior of the Wide Angle On-Shell Quark Scattering Amplitudes in Nonabelian Gauge Theories". In: *Phys. Rev. D* 28 (1983), p. 860.
[1338] Yao Ma. "A Forest Formula to Subtract Infrared Singularities in Amplitudes for Wide-angle Scattering". In: *JHEP* 05 (2020), p. 012.
[1339] Neelima Agarwal et al. "The Infrared Structure of Perturbative Gauge Theories". In: (Dec. 2021).
[1340] Lance J. Dixon, Lorenzo Magnea, and George F. Sterman. "Universal structure of subleading infrared poles in gauge theory amplitudes". In: *JHEP* 08 (2008), p. 022.
[1341] R. P. Feynman. "Photon-hadron interactions". In: (1973).
[1342] V. A. Nesterenko and A. V. Radyushkin. "Sum Rules and Pion Form-Factor in QCD". In: *Phys. Lett. B* 115 (1982), p. 410.
[1343] A. Duncan and Alfred H. Mueller. "Asymptotic Behavior of Composite Particle Form-Factors and the Renormalization Group". In: *Phys. Rev. D* 21 (1980), p. 1636.
[1344] Bijoy Kundu et al. "The Perturbative proton form-factor reexamined". In: *Eur. Phys. J. C* 8 (1999), pp. 637–642.
[1345] Sumeet K. Dagaonkar, Pankaj Jain, and John P. Ralston. "Uncovering the Scaling Laws of Hard Exclusive Hadronic Processes in a Comprehensive Endpoint Model". In: *Eur. Phys. J. C* 74.8 (2014), p. 3000.
[1346] James Botts and George F. Sterman. "Hard Elastic Scattering in QCD: Leading Behavior". In: *Nucl. Phys. B* 325 (1989), pp. 62–100.
[1347] John C. Collins and Davison E. Soper. "Back-To-Back Jets in QCD". In: *Nucl. Phys.* B193 (1981). [Erratum: Nucl. Phys. B213, 545 (1983)], p. 381.
[1348] Gerard 't Hooft. "Magnetic Monopoles in Unified Gauge Theories". In: *Nucl. Phys. B* 79 (1974). Ed. by J. C. Taylor, pp. 276–284.
[1349] Alexander M. Polyakov. "Particle Spectrum in Quantum Field Theory". In: *JETP Lett.* 20 (1974). Ed. by J. C. Taylor, pp. 194–195.
[1350] A. A. Belavin et al. "Pseudoparticle Solutions of the Yang-Mills Equations". In: *Phys. Lett. B* 59 (1975). Ed. by J. C. Taylor, pp. 85–87.
[1351] Yoichiro Nambu. "Strings, Monopoles and Gauge Fields". In: *Phys. Rev. D* 10 (1974). Ed. by T. Eguchi, p. 4262.
[1352] Thomas C. Kraan and Pierre van Baal. "Monopole constituents inside SU(n) calorons". In: *Phys. Lett. B* 435 (1998), pp. 389–395.
[1353] Edward V. Shuryak. "Theory of Hadronic Plasma". In: *Sov. Phys. JETP* 47 (1978), pp. 212–219.
[1354] S. Mandelstam. "Vortices and Quark Confinement in Nonabelian Gauge Theories". In: *Phys. Rept.* 23 (1976), pp. 245–249.
[1355] Gerard 't Hooft. "On the Phase Transition Towards Permanent Quark Confinement". In: *Nucl. Phys. B* 138 (1978), pp. 1–25.
[1356] Paul Adrien Maurice Dirac. "Quantised singularities in the electromagnetic field". In: *Proc. Roy. Soc. Lond. A* 133.821 (1931), pp. 60–72.
[1357] Yakov M. Shnir. *Magnetic Monopoles*. Text and Monographs in Physics. Berlin/Heidelberg: Springer, 2005.
[1358] Gunnar S. Bali. "The Mechanism of quark confinement". In: *3rd International Conference in Quark Confinement and Hadron Spectrum (Confinement III)*. June 1998, pp. 17–36.

[1359] Alessio D'Alessandro, Massimo D'Elia, and Edward V. Shuryak. "Thermal Monopole Condensation and Confinement in finite temperature Yang-Mills Theories". In: *Phys. Rev. D* 81 (2010), p. 094501.

[1360] Alessio D'Alessandro and Massimo D'Elia. "Magnetic monopoles in the high temperature phase of Yang-Mills theories". In: *Nucl. Phys. B* 799 (2008), pp. 241–254.

[1361] Jinfeng Liao and Edward Shuryak. "Magnetic Component of Quark-Gluon Plasma is also a Liquid!" In: *Phys. Rev. Lett.* 101 (2008), p. 162302.

[1362] Jinfeng Liao and Edward Shuryak. "Angular Dependence of Jet Quenching Indicates Its Strong Enhancement Near the QCD Phase Transition". In: *Phys. Rev. Lett.* 102 (2009), p. 202302.

[1363] Shiin-Shen Chern and James Simons. "Characteristic forms and geometric invariants". In: *Annals Math.* 99 (1974), pp. 48–69.

[1364] D. M. Ostrovsky, G. W. Carter, and E. V. Shuryak. "Forced tunneling and turning state explosion in pure Yang-Mills theory". In: *Phys. Rev. D* 66 (2002), p. 036004.

[1365] Roger F. Dashen, Brosl Hasslacher, and Andre Neveu. "Nonperturbative Methods and Extended Hadron Models in Field Theory. 3. Four-Dimensional Nonabelian Models". In: *Phys. Rev. D* 10 (1974), p. 4138.

[1366] Frans R. Klinkhamer and N. S. Manton. "A Saddle Point Solution in the Weinberg-Salam Theory". In: *Phys. Rev. D* 30 (1984), p. 2212.

[1367] Edward Shuryak and Ismail Zahed. "How to observe the QCD instanton/sphaleron processes at hadron colliders?" In: (Jan. 2021).

[1368] Edward V. Shuryak. "The Role of Instantons in Quantum Chromodynamics. 3. Quark - Gluon Plasma". In: *Nucl. Phys. B* 203 (1982), pp. 140–156.

[1369] Derek B. Leinweber. "Visualizations of the QCD vacuum". In: *Workshop on Light-Cone QCD and Nonperturbative Hadron Physics.* Dec. 1999, pp. 138–143.

[1370] Gerard 't Hooft. "Computation of the Quantum Effects Due to a Four-Dimensional Pseudoparticle". In: *Phys. Rev. D* 14 (1976). Ed. by Mikhail A. Shifman. [Erratum: Phys.Rev.D 18, 2199 (1978)], pp. 3432–3450.

[1371] Thomas Sch¥"afer and Edward V. Shuryak. "Instantons in QCD". In: *Rev. Mod. Phys.* 70 (1998), pp. 323–426.

[1372] Edward V. Shuryak. "Correlation functions in the QCD vacuum". In: *Rev. Mod. Phys.* 65 (1993), pp. 1–46.

[1373] Edward V. Shuryak and J. J. M. Verbaarschot. "Quark propagation in the random instanton vacuum". In: *Nucl. Phys. B* 410 (1993), pp. 37–54.

[1374] Edward V. Shuryak and J. J. M. Verbaarschot. "Mesonic correlation functions in the random instanton vacuum". In: *Nucl. Phys. B* 410 (1993), pp. 55–89.

[1375] Thomas Sch¥"afer, Edward V. Shuryak, and J. J. M. Verbaarschot. "Baryonic correlators in the random instanton vacuum". In: *Nucl. Phys. B* 412 (1994), pp. 143–168.

[1376] Vesteinn Thorsson and Ismail Zahed. "Diquarks in the Nambu-Jona-Lasinio Model". In: *Phys. Rev. D* 41 (1990), p. 3442.

[1377] Thomas Sch¥"afer and Edward V. Shuryak. "Phases of QCD at high baryon density". In: *Lect. Notes Phys.* 578 (2001). Ed. by D. Blaschke, N. K. Glendenning, and A. Sedrakian, pp. 203–217.

[1378] Ki-Myeong Lee and Chang-hai Lu. "SU(2) calorons and magnetic monopoles". In: *Phys. Rev. D* 58 (1998), p. 025011.

[1379] Kurt Langfeld and Ernst-Michael Ilgenfritz. "Confinement from semiclassical gluon fields in SU(2) gauge theory". In: *Nucl. Phys. B* 848 (2011), pp. 33–61.

[1380] Rasmus N. Larsen, Sayantan Sharma, and Edward Shuryak. "The topological objects near the chiral crossover transition in QCD". In: *Phys. Lett. B* 794 (2019), pp. 14–18.

[1381] Rasmus N. Larsen, Sayantan Sharma, and Edward Shuryak. "Towards a semiclassical description of QCD vacuum around $T_c$". In: *Phys. Rev. D* 102.3 (2020), p. 034501.

[1382] Rasmus Larsen and Edward Shuryak. "Interacting ensemble of the instanton-dyons and the deconfinement phase transition in the SU(2) gauge theory". In: *Phys. Rev. D* 92.9 (2015), p. 094022.

[1383] Rasmus Larsen and Edward Shuryak. "Instanton-dyon Ensemble with two Dynamical Quarks: the Chiral Symmetry Breaking". In: *Phys. Rev. D* 93.5 (2016), p. 054029.

[1384] Dallas DeMartini and Edward Shuryak. "Chiral symmetry breaking and confinement from an interacting ensemble of instanton dyons in two-flavor massless QCD". In: *Phys. Rev. D* 104.9 (2021), p. 094031.

[1385] Nick Dorey and Andrei Parnachev. "Instantons, compactification and S duality in N=4 SUSY Yang-Mills theory. 2." In: *JHEP* 08 (2001), p. 059.

[1386] Edward Shuryak. *Nonperturbative Topological Phenomena in QCD and Related Theories*. Vol. 977. Lecture Notes in Physics. Mar. 2021.

[1387] Steven Weinberg. "Phenomenological Lagrangians". In: *Physica A* 96.1-2 (1979). Ed. by S. Deser, pp. 327–340.

[1388] Nora Brambilla et al. "The $XYZ$ states: experimental and theoretical status and perspectives". In: *Phys. Rept.* 873 (2020), pp. 1–154.

[1389] Nathan Isgur and Mark B. Wise. "Weak Transition form-factors between heavy mesons". In: *Phys. Lett.* B237 (1990), p. 527.

[1390] Matthias Neubert. "Heavy quark symmetry". In: *Phys. Rept.* 245 (1994), p. 259.

[1391] W. E. Caswell and G. P. Lepage. "Effective Lagrangians for Bound State Problems in QED, QCD, and Other Field Theories". In: *Phys. Lett.* 167B (1986), p. 437.

[1392] Ayesh Gunawardana and Gil Paz. "On HQET and NRQCD Operators of Dimension 8 and Above". In: *JHEP* 07 (2017), p. 137.

[1393] Andrew Kobach and Sridip Pal. "Hilbert series and operator basis for NRQED and NRQCD/HQET". In: *Phys. Lett.* B772 (2017), p. 225.

[1394] Aneesh V. Manohar. "Heavy quark effective theory and nonrelativistic QCD Lagrangian to order $\alpha_s/m^3$". In: *Phys. Rev.* D56 (1997), p. 230.

[1395] A. G. Grozin et al. "Three-loop chromomagnetic interaction in HQET". In: *Nucl. Phys.* B789 (2008), p. 277.

[1396] Christopher Balzereit. "Spectator effects in heavy quark effective theory at $O(1/m_Q^3)$". In: *Phys. Rev.* D59 (1999), p. 094015.

[1397] Daniel Moreno and Antonio Pineda. "Chromopolarizabilities of a heavy quark at weak coupling". In: *Phys. Rev.* D97 (2018). [Erratum: Phys. Rev. D98 (2018) 059902], p. 016012.

[1398] Michael E. Luke and Aneesh V. Manohar. "Reparametrization invariance constraints on heavy particle effective field theories". In: *Phys. Lett. B* 286 (1992), pp. 348–354.

[1399] Nora Brambilla, Dieter Gromes, and Antonio Vairo. "Poincaré invariance constraints on NRQCD and potential NRQCD". In: *Phys. Lett.* B576 (2003), p. 314.

[1400] Johannes Heinonen, Richard J. Hill, and Mikhail P. Solon. "Lorentz invariance in heavy particle effective theories". In: *Phys. Rev. D* 86 (2012), p. 094020.

[1401] Adam F. Falk and Matthias Neubert. "Second-order power corrections in the heavy-quark effective theory. 1. Formalism and meson form factors". In: *Phys. Rev.* D47 (1993), p. 2965.

[1402] Peter Marquard et al. "Quark Mass Relations to Four-Loop Order in Perturbative QCD". In: *Phys. Rev. Lett.* 114.14 (2015), p. 142002.

[1403] Peter Marquard et al. "$\overline{\text{MS}}$-on-shell quark mass relation up to four loops in QCD and a general SU($N$) gauge group". In: *Phys. Rev.* D94 (2016), p. 074025.

[1404] Nikolai Uraltsev. "BLM resummation and OPE in heavy flavor transitions". In: *Nucl. Phys.* B491 (1997), p. 303.

[1405] M. Beneke. "A quark mass definition adequate for threshold problems". In: *Phys. Lett.* B434 (1998), p. 115.

[1406] A. H. Hoang. "$1S$ and $\bar{M}S$ bottom quark masses from $\Upsilon$ sum rules". In: *Phys. Rev.* D61 (2000), p. 034005.

[1407] Antonio Pineda. "Determination of the bottom quark mass from the $\Upsilon(1S)$ system". In: *JHEP* 06 (2001), p. 022.

[1408] Andre H. Hoang et al. "Infrared Renormalization Group Flow for Heavy Quark Masses". In: *Phys. Rev. Lett.* 101 (2008), p. 151602.

[1409] N. Brambilla et al. "Relations between heavy-light meson and quark masses". In: *Phys. Rev.* D97 (2018), p. 034503.

[1410] William A. Bardeen, Estia J. Eichten, and Christopher T. Hill. "Chiral multiplets of heavy-light mesons". In: *Phys. Rev.* D68 (2003), p. 054024.

[1411] Nora Brambilla, Antonio Vairo, and Thomas Rosch. "Effective field theory Lagrangians for baryons with two and three heavy quarks". In: *Phys. Rev.* D72 (2005), p. 034021.

[1412] Sean Fleming and Thomas Mehen. "Doubly heavy baryons, heavy quark-diquark symmetry and NRQCD". In: *Phys. Rev.* D73 (2006), p. 034502.

[1413] Thomas Mehen and Brian C. Tiburzi. "Doubly heavy baryons and quark-diquark symmetry in quenched and partially quenched chiral perturbation theory". In: *Phys. Rev.* D74 (2006), p. 054505.

[1414] Yong-Liang Ma and Masayasu Harada. "Degeneracy of doubly heavy baryons from heavy quark symmetry". In: *Phys. Lett.* B754 (2016), p. 125.

[1415] Thomas Mehen. "Implications of heavy quark-diquark symmetry for excited doubly heavy baryons and tetraquarks". In: *Phys. Rev.* D96 (2017), p. 094028.

[1416] Yong-Liang Ma and Masayasu Harada. "Chiral partner structure of doubly heavy baryons with heavy quark spin-flavor symmetry". In: *J. Phys.* G45 (2018), p. 075006.

[1417] Hai-Yang Cheng and Yan-Liang Shi. "Lifetimes of doubly charmed baryons". In: *Phys. Rev.* D98 (2018), p. 113005.

[1418] Thomas C. Mehen and Abhishek Mohapatra. "Perturbative Corrections to Heavy Quark-Diquark Symmetry Predictions for Doubly Heavy Baryon Hyperfine Splittings". In: *Phys. Rev. D* 100.7 (2019), p. 076014.

[1419] Joan Soto and Jaume Tarrús Castellà. "Effective field theory for double heavy baryons at strong coupling". In: *Phys. Rev. D* 102.1 (2020). [Erratum: Phys.Rev.D 104, 059901 (2021)], p. 014013.

[1420] Estia J. Eichten and Chris Quigg. "Heavy-quark symmetry implies stable heavy tetraquark mesons $Q_iQ_j\bar{q}_k\bar{q}_l$". In: *Phys. Rev. Lett.* 119 (2017), p. 202002.

[1421] Joan Soto and Jaume Tarrús Castellà. "Nonrelativistic effective field theory for heavy exotic hadrons". In: *Phys. Rev. D* 102.1 (2020), p. 014012.

[1422] Nora Brambilla et al. "Effective field theories for heavy quarkonium". In: *Rev. Mod. Phys.* 77 (2005), p. 1423.

[1423] N. Brambilla et al. "Heavy quarkonium physics". In: (2004).

[1424] N. Brambilla et al. "Heavy quarkonium: progress, puzzles, and opportunities". In: *Eur. Phys. J.* C71 (2011), p. 1534.

[1425] N. Brambilla et al. "QCD and Strongly Coupled Gauge Theories: Challenges and Perspectives". In: *Eur. Phys. J.* C74 (2014), p. 2981.

[1426] Antonio Vairo. "Non-relativistic bound states: the long way back from the Bethe-Salpeter to the Schröedinger equation". In: (2009).

[1427] T. Kinoshita and M. Nio. "Radiative corrections to the muonium hyperfine structure. 1. The alpha**2 (Z-alpha) correction". In: *Phys. Rev. D* 53 (1996), pp. 4909–4929.

[1428] Patrick Labelle. "Effective field theories for QED bound states: Extending nonrelativistic QED to study retardation effects". In: *Phys. Rev. D* 58 (1998), p. 093013.

[1429] B. A. Thacker and G. Peter Lepage. "Heavy quark bound states in lattice QCD". In: *Phys. Rev. D* 43 (1991), pp. 196–208.

[1430] Ciaran Hughes et al. "Hindered M1 Radiative Decay of $\Upsilon(2S)$ from Lattice NRQCD". In: *Phys. Rev.* D92 (2015), p. 094501.

[1431] Brian Colquhoun et al. "Phenomenology with Lattice NRQCD b Quarks". In: *PoS* LATTICE2015 (2016), p. 334.

[1432] C. Hughes, C. T. H. Davies, and C. J. Monahan. "New methods for $B$ meson decay constants and form factors from lattice NRQCD". In: *Phys. Rev.* D97 (2018), p. 054509.

[1433] Andrew Lytle et al. "$B_c$ spectroscopy using highly improved staggered quarks". In: *36th International Symposium on Lattice Field Theory (Lattice 2018) East Lansing, MI, United States, July 22-28, 2018.*

[1434] Sinéad M. Ryan and David J. Wilson. "Excited and exotic bottomonium spectroscopy from lattice QCD". In: *JHEP* 02 (2021), p. 214.

[1435] Geoffrey T. Bodwin, Eric Braaten, and G. Peter Lepage. "Rigorous QCD predictions for decays of $P$-wave quarkonia". In: *Phys. Rev.* D46 (1992), R1914.

[1436] Geoffrey T. Bodwin, Eric Braaten, and G. Peter Lepage. "Rigorous QCD analysis of inclusive annihilation and production of heavy quarkonium". In: *Phys. Rev.* D51 (1995). [Erratum: Phys. Rev. D55 (1997) 5853], p. 1125.

[1437] Geoffrey T. Bodwin et al. "Quarkonium at the Frontiers of High Energy Physics: A Snowmass White Paper". In: *Community Summer Study 2013: Snowmass on the Mississippi.* July 2013.

[1438] Hee Sok Chung. "Review of quarkonium production: status and prospects". In: *PoS* Confinement2018 (2018), p. 007.

[1439] Jean-Philippe Lansberg. "New Observables in Inclusive Production of Quarkonia". In: *Phys. Rept.* 889 (2020), pp. 1–106.

[1440] Gouranga C. Nayak, Jian-Wei Qiu, and George F. Sterman. "Fragmentation, factorization and infrared poles in heavy quarkonium production". In: *Phys. Lett.* B613 (2005), p. 45.

[1441] Yan-Qing Ma et al. "Factorized power expansion for high-$p_T$ heavy quarkonium production". In: *Phys. Rev. Lett.* 113 (2014), p. 142002.

[1442] Zhong-Bo Kang et al. "Heavy Quarkonium Production at Collider Energies: Partonic Cross Section and Polarization". In: *Phys. Rev.* D91 (2015), p. 014030.

[1443] A. Pineda and J. Soto. "Matching at one loop for the four quark operators in NRQCD". In: *Phys. Rev.* D58 (1998), p. 114011.

[1444] Antonio Vairo. "A theoretical review of heavy quarkonium inclusive decays". In: *Mod. Phys. Lett.* A19 (2004), p. 253.

[1445] Nora Brambilla, Emanuele Mereghetti, and Antonio Vairo. "Electromagnetic quarkonium decays at order $v^7$". In: *JHEP* 08 (2006). [Erratum: JHEP 04 (2011) 058], p. 039.

[1446] Nora Brambilla, Emanuele Mereghetti, and Antonio Vairo. "Hadronic quarkonium decays at

order $v^7$". In: *Phys. Rev.* D79 (2009). [Erratum: Phys. Rev. D83 (2011) 079904], p. 074002.

[1447] Matthias Berwein et al. "Poincaré invariance in NRQCD and pNRQCD revisited". In: *Phys. Rev.* D99 (2019), p. 094008.

[1448] Nora Brambilla et al. "Inclusive decays of heavy quarkonium to light particles". In: *Phys. Rev.* D67 (2003), p. 034018.

[1449] A. Pineda and J. Soto. "Effective field theory for ultrasoft momenta in NRQCD and NRQED". In: *Nucl. Phys. Proc. Suppl.* 64 (1998), p. 428.

[1450] Nora Brambilla et al. "Potential NRQCD: An Effective theory for heavy quarkonium". In: *Nucl. Phys.* B566 (2000), p. 275.

[1451] Nora Brambilla et al. "QCD static energy at next-to-next-to-next-to leading-logarithmic accuracy". In: *Phys. Rev.* D80 (2009), p. 034016.

[1452] Nora Brambilla et al. "Infrared behavior of the static potential in perturbative QCD". In: *Phys. Rev.* D60 (1999), p. 091502.

[1453] C. Anzai, Y. Kiyo, and Y. Sumino. "Static QCD potential at three-loop order". In: *Phys. Rev. Lett.* 104 (2010), p. 112003.

[1454] Alexander V. Smirnov, Vladimir A. Smirnov, and Matthias Steinhauser. "Three-loop static potential". In: *Phys. Rev. Lett.* 104 (2010), p. 112002.

[1455] Nora Brambilla et al. "The Logarithmic contribution to the QCD static energy at $N^4LO$". In: *Phys. Lett.* B647 (2007), p. 185.

[1456] Bernd A. Kniehl et al. "Non-Abelian $\alpha_s^3/(m_q r^2)$ heavy quark anti-quark potential". In: *Phys. Rev.* D65 (2002), p. 091503.

[1457] Bernd A. Kniehl et al. "Potential NRQCD and heavy quarkonium spectrum at next-to-next-to-next-to-leading order". In: *Nucl. Phys.* B635 (2002), p. 357.

[1458] Nora Brambilla and Antonio Vairo. "The $B_c$ mass up to order $\alpha_s^4$". In: *Phys. Rev.* D62 (2000), p. 094019.

[1459] Nora Brambilla, Dieter Gromes, and Antonio Vairo. "Poincaré invariance and the heavy quark potential". In: *Phys. Rev.* D64 (2001), p. 076010.

[1460] Clara Peset, Antonio Pineda, and Maximilian Stahlhofen. "Potential NRQCD for unequal masses and the $B_c$ spectrum at $N^3LO$". In: *JHEP* 05 (2016), p. 017.

[1461] Dieter Gromes. "Spin Dependent Potentials in QCD and the Correct Long Range Spin Orbit Term". In: *Z. Phys.* C26 (1984), p. 401.

[1462] A. Barchielli, N. Brambilla, and G. M. Prosperi. "Relativistic Corrections to the Quark - anti-Quark Potential and the Quarkonium Spectrum". In: *Nuovo Cim.* A103 (1990), p. 59.

[1463] Bernd A. Kniehl and Alexander A. Penin. "Ultrasoft effects in heavy quarkonium physics". In: *Nucl. Phys.* B563 (1999), p. 200.

[1464] Nora Brambilla et al. "The heavy quarkonium spectrum at order $m\alpha_s^5 ln\alpha_s$". In: *Phys. Lett.* B470 (1999), p. 215.

[1465] Antonio Pineda and Joan Soto. "The Renormalization group improvement of the QCD static potentials". In: *Phys. Lett.* B495 (2000), p. 323.

[1466] Antonio Pineda. "Renormalization group improvement of the NRQCD Lagrangian and heavy quarkonium spectrum". In: *Phys. Rev.* D65 (2002), p. 074007.

[1467] Antonio Pineda. "Next-to-leading ultrasoft running of the heavy quarkonium potentials and spectrum: Spin-independent case". In: *Phys. Rev.* D84 (2011), p. 014012.

[1468] Clara Peset, Antonio Pineda, and Jorge Segovia. "$P$-wave heavy quarkonium spectrum with next-to-next-to-next-to-leading logarithmic accuracy". In: *Phys. Rev.* D98 (2018), p. 094003.

[1469] C. Anzai, D. Moreno, and A. Pineda. "$S$-wave heavy quarkonium spectrum with next-to-next-to-next-to-leading logarithmic accuracy". In: *Phys. Rev.* D98 (2018), p. 114034.

[1470] Thomas Appelquist, Michael Dine, and I. J. Muzinich. "The Static Limit of Quantum Chromodynamics". In: *Phys. Rev. D* 17 (1978), p. 2074.

[1471] M. Beneke, A. Signer, and Vladimir A. Smirnov. "Top quark production near threshold and the top quark mass". In: *Phys. Lett. B* 454 (1999), pp. 137–146.

[1472] A. H. Hoang et al. "Top - anti-top pair production close to threshold: Synopsis of recent NNLO results". In: *Eur. Phys. J. direct* 2.1 (2000), p. 3.

[1473] Antonio Pineda and Adrian Signer. "Heavy Quark Pair Production near Threshold with Potential Non-Relativistic QCD". In: *Nucl. Phys.* B762 (2007), p. 67.

[1474] M. Beneke et al. "Hadronic top-quark pair production with NNLL threshold resummation". In: *Nucl. Phys. B* 855 (2012), pp. 695–741.

[1475] André H. Hoang and Maximilian Stahlhofen. "The Top-Antitop Threshold at the ILC: NNLL QCD Uncertainties". In: *JHEP* 05 (2014), p. 121.

[1476] M. Beneke, J. Piclum, and T. Rauh. "P-wave contribution to third-order top-quark pair production near threshold". In: *Nucl. Phys. B* 880 (2014), pp. 414–434.

[1477] Martin Beneke et al. "Next-to-Next-to-Next-to-Leading Order QCD Prediction for the Top Antitop $S$-Wave Pair Production Cross Sec-

tion Near Threshold in $e^+e^-$ Annihilation". In: *Phys. Rev. Lett.* 115.19 (2015), p. 192001.

[1478] Antonio Pineda. "Review of heavy quarkonium at weak coupling". In: *Prog. Part. Nucl. Phys.* 67 (2012), p. 735.

[1479] N. Brambilla, Y. Sumino, and A. Vairo. "Quarkonium spectroscopy and perturbative QCD: A new perspective". In: *Phys. Lett.* B513 (2001), p. 381.

[1480] N. Brambilla, Y. Sumino, and A. Vairo. "Quarkonium spectroscopy and perturbative QCD: Massive quark loop effects". In: *Phys. Rev.* D65 (2002), p. 034001.

[1481] S. Recksiegel and Y. Sumino. "Improved perturbative QCD prediction of the bottomonium spectrum". In: *Phys. Rev.* D67 (2003), p. 014004.

[1482] Cesar Ayala, Gorazd Cvetic, and Antonio Pineda. "The bottom quark mass from the $\boldsymbol{\Upsilon}(1S)$ system at NNNLO". In: *JHEP* 09 (2014), p. 045.

[1483] M. Beneke et al. "The bottom-quark mass from non-relativistic sum rules at NNNLO". In: *Nucl. Phys.* B891 (2015), p. 42.

[1484] Y. Kiyo, G. Mishima, and Y. Sumino. "Determination of $m_c$ and $m_b$ from quarkonium $1S$ energy levels in perturbative QCD". In: *Phys. Lett.* B752 (2016). [Erratum: Phys. Lett. B772 (2017) 878], p. 122.

[1485] Cesar Ayala, Gorazd Cvetic, and Antonio Pineda. "Mass of the bottom quark from $\Upsilon(1S)$ at NNNLO: an update". In: *J. Phys. Conf. Ser.* 762 (2016), p. 012063.

[1486] Vicent Mateu and Pablo G. Ortega. "Bottom and Charm Mass determinations from global fits to $Q\bar{Q}$ bound states at N$^3$LO". In: *JHEP* 01 (2018), p. 122.

[1487] Clara Peset, Antonio Pineda, and Jorge Segovia. "The charm/bottom quark mass from heavy quarkonium at N$^3$LO". In: *JHEP* 09 (2018), p. 167.

[1488] S. Recksiegel and Y. Sumino. "Fine and hyperfine splittings of charmonium and bottomonium: An improved perturbative QCD approach". In: *Phys. Lett.* B578 (2004), p. 369.

[1489] Nora Brambilla and Antonio Vairo. "The $1P$ quarkonium fine splittings at NLO". In: *Phys. Rev.* D71 (2005), p. 034020.

[1490] Bernd A. Kniehl et al. "$M(\eta_b)$ and $\alpha_s$ from nonrelativistic renormalization group". In: *Phys. Rev. Lett.* 92 (2004). [Erratum: Phys. Rev. Lett. 104 (2010) 199901], p. 242001.

[1491] A. A. Penin et al. "$M(B_c^*) - M(B_c)$ splitting from nonrelativistic renormalization group". In: *Phys. Lett.* B593 (2004). [Erratum: Phys. Lett. B677 (2009) 343], p. 124.

[1492] Alexander A. Penin and Matthias Steinhauser. "Heavy quarkonium spectrum at $O(\alpha_s^5 m_q)$ and bottom/top quark mass determination". In: *Phys. Lett.* B538 (2002), p. 335.

[1493] A. A. Penin, Vladimir A. Smirnov, and M. Steinhauser. "Heavy quarkonium spectrum and production/annihilation rates to order $\beta_0^3\alpha_s^3$". In: *Nucl. Phys.* B716 (2005), p. 303.

[1494] M. Beneke, Y. Kiyo, and K. Schuller. "Third-order coulomb corrections to the $S$-wave Green function, energy levels and wave functions at the origin". In: *Nucl. Phys.* B714 (2005), p. 67.

[1495] M. Beneke, Y. Kiyo, and K. Schuller. "Third-order non-Coulomb correction to the $S$-wave quarkonium wave functions at the origin". In: *Phys. Lett.* B658 (2008), p. 222.

[1496] Y. Kiyo and Y. Sumino. "Perturbative heavy quarkonium spectrum at next-to-next-to-next-to-leading order". In: *Phys. Lett.* B730 (2014), p. 76.

[1497] Y. Kiyo and Y. Sumino. "Full Formula for Heavy Quarkonium Energy Levels at Next-to-next-to-next-to-leading Order". In: *Nucl. Phys.* B889 (2014), p. 156.

[1498] A. A. Penin et al. "Spin dependence of heavy quarkonium production and annihilation rates: Complete next-to-next-to-leading logarithmic result". In: *Nucl. Phys.* B699 (2004). [Erratum: Nucl. Phys. B829 (2010) 398], p. 183.

[1499] Yuichiro Kiyo, Antonio Pineda, and Adrian Signer. "New determination of inclusive electromagnetic decay ratios of heavy quarkonium from QCD". In: *Nucl. Phys.* B841 (2010), p. 231.

[1500] M. Beneke, Y. Kiyo, and A. A. Penin. "Ultrasoft contribution to quarkonium production and annihilation". In: *Phys. Lett.* B653 (2007), p. 53.

[1501] Martin Beneke et al. "Leptonic decay of the $\Upsilon(1S)$ meson at third order in QCD". In: *Phys. Rev. Lett.* 112 (2014), p. 151801.

[1502] A. Pineda. "Next-to-leading nonperturbative calculation in heavy quarkonium". In: *Nucl. Phys.* B494 (1997), p. 213.

[1503] T. Rauh. "Higher-order condensate corrections to $\Upsilon$ masses, leptonic decay rates and sum rules". In: *JHEP* 05 (2018), p. 201.

[1504] Christian W. Bauer et al. "Resumming the color octet contribution to radiative $\Upsilon$ decay". In: *Phys. Rev.* D64 (2001), p. 114014.

[1505] Sean Fleming and Adam K. Leibovich. "Resummed Photon Spectrum in Radiative $\Upsilon$ Decays". In: *Phys. Rev. Lett.* 90 (2003), p. 032001.

[1506] Sean Fleming and Adam K Leibovich. "The photon spectrum in $\Upsilon$ decays". In: *Phys. Rev.* D67 (2003), p. 074035.

[1507] Xavier Garcia i Tormo and Joan Soto. "Soft, collinear and nonrelativistic modes in radiative decays of very heavy quarkonium". In: *Phys. Rev.* D69 (2004), p. 114006.

[1508] Xavier Garcia i Tormo and Joan Soto. "Semi-inclusive radiative decays of $\Upsilon(1S)$". In: *Phys. Rev.* D72 (2005), p. 054014.

[1509] Xavier Garcia i Tormo and Joan Soto. "Radiative decays and the nature of heavy quarkonia". In: *Phys. Rev. Lett.* 96 (2006), p. 111801.

[1510] Nora Brambilla et al. "Extraction of alpha(s) from radiative Upsilon(1S) decays". In: *Phys. Rev.* D75 (2007), p. 074014.

[1511] Nora Brambilla, Yu Jia, and Antonio Vairo. "Model-independent study of magnetic dipole transitions in quarkonium". In: *Phys. Rev. D* 73 (2006), p. 054005.

[1512] Nora Brambilla, Piotr Pietrulewicz, and Antonio Vairo. "Model-independent study of electric dipole transitions in quarkonium". In: *Phys. Rev.* D85 (2012), p. 094005.

[1513] Antonio Pineda and J. Segovia. "Improved determination of heavy quarkonium magnetic dipole transitions in potential nonrelativistic QCD". In: *Phys. Rev.* D87 (2013), p. 074024.

[1514] Jorge Segovia, Sebastian Steinbeißer, and Antonio Vairo. "Electric dipole transitions of $1P$ bottomonia". In: *Phys. Rev.* D99 (2019), p. 074011.

[1515] Nora Brambilla, Pablo Roig, and Antonio Vairo. "Precise determination of the $\eta_c$ mass and width in the radiative $J/\psi \to \eta_c\gamma$ decay". In: *AIP Conf. Proc.* 1343 (2011), p. 418.

[1516] Nora Brambilla et al. "QCD potential at $O(1/m)$". In: *Phys. Rev.* D63 (2001), p. 014023.

[1517] Antonio Pineda and Antonio Vairo. "The QCD potential at $O(1/m^{2)}$ : Complete spin dependent and spin independent result". In: *Phys. Rev.* D63 (2001). [Erratum: Phys. Rev. D64 (2001) 039902], p. 054007.

[1518] Nora Brambilla et al. "New predictions for inclusive heavy quarkonium $P$-wave decays". In: *Phys. Rev. Lett.* 88 (2002), p. 012003.

[1519] Nora Brambilla et al. "The $\sqrt{m\Lambda_{QCD}}$ scale in heavy quarkonium". In: *Phys. Lett.* B580 (2004), p. 60.

[1520] Gunnar S. Bali et al. "Static potentials and glueball masses from QCD simulations with Wilson sea quarks". In: *Phys. Rev.* D62 (2000), p. 054503.

[1521] Nora Brambilla et al. "Long-range properties of $1S$ bottomonium states". In: *Phys. Rev.* D93 (2016), p. 054002.

[1522] Leonard Susskind. "Coarse Grained Quantum Chromodynamics". In: *Ecole d'Ete de Physique Theorique - Weak and Electromagnetic Interactions at High Energy Les Houches, France, July 5-August 14, 1976*, p. 207.

[1523] W. Fischler. "Quark-antiquark potential in QCD". In: *Nucl. Phys.* B129 (1977), p. 157.

[1524] Lowell S. Brown and William I. Weisberger. "Remarks on the Static Potential in Quantum Chromodynamics". In: *Phys. Rev.* D20 (1979), p. 3239.

[1525] Clara Peset, Antonio Pineda, and Maximilian Stahlhofen. "Relativistic corrections to the static energy in terms of Wilson loops at weak coupling". In: *Eur. Phys. J.* C77 (2017), p. 681.

[1526] Yoshiaki Koma, Miho Koma, and Hartmut Wittig. "Nonperturbative determination of the QCD potential at $O(1/m)$". In: *Phys. Rev. Lett.* 97 (2006), p. 122003.

[1527] Yoshiaki Koma, Miho Koma, and Hartmut Wittig. "Relativistic corrections to the static potential at $O(1/m)$ and $O(1/m^2)$". In: *PoS* LATTICE2007 (2007), p. 111.

[1528] Gunnar S. Bali. "QCD forces and heavy quark bound states". In: *Phys. Rept.* 343 (2001), pp. 1–136.

[1529] John B. Kogut and G. Parisi. "Long Range Spin Spin Forces in Gauge Theories". In: *Phys. Rev. Lett.* 47 (1981), p. 1089.

[1530] Guillem Perez-Nadal and Joan Soto. "Effective string theory constraints on the long distance behavior of the subleading potentials". In: *Phys. Rev.* D79 (2009), p. 114002.

[1531] Nora Brambilla et al. "Effective string theory and the long-range relativistic corrections to the quark-antiquark potential". In: *Phys. Rev.* D90 (2014), p. 114032.

[1532] Nora Brambilla et al. "Decay and electromagnetic production of strongly coupled quarkonia in pNRQCD". In: *JHEP* 04 (2020), p. 095.

[1533] Nora Brambilla, Hee Sok Chung, and Antonio Vairo. "Inclusive Hadroproduction of $P$-Wave Heavy Quarkonia in Potential Nonrelativistic QCD". In: *Phys. Rev. Lett.* 126.8 (2021), p. 082003.

[1534] Nora Brambilla, Hee Sok Chung, and Antonio Vairo. "Inclusive production of heavy quarkonia in pNRQCD". In: *JHEP* 09 (2021), p. 032.

[1535] Nora Brambilla et al. “Production and polarization of S-wave quarkonia in potential nonrelativistic QCD”. In: *Phys. Rev. D* 105.11 (2022), p. L111503.

[1536] Nora Brambilla et al. “Inclusive production of $J/\psi$, $\psi(2S)$, and $\Upsilon$ states in pNRQCD”. In: (Oct. 2022).

[1537] Antonio Pineda and Jaume Tarrús Castellà. “Novel implementation of the multipole expansion to quarkonium hadronic transitions”. In: *Phys. Rev. D* 100.5 (2019), p. 054021.

[1538] Nora Brambilla, Jacopo Ghiglieri, and Antonio Vairo. “Three-quark static potential in perturbation theory”. In: *Phys. Rev.* D81 (2010), p. 054031.

[1539] Nora Brambilla, Felix Karbstein, and Antonio Vairo. “Symmetries of the three-heavy-quark system and the color-singlet static energy at next-to-next-to-leading logarithmic order”. In: *Phys. Rev.* D87 (2013), p. 074014.

[1540] Toru T. Takahashi and Hideo Suganuma. “Gluonic excitation of the three-quark system in SU(3) lattice QCD”. In: *Phys. Rev. Lett.* 90 (2003), p. 182001.

[1541] Toru T. Takahashi and Hideo Suganuma. “Detailed analysis of the gluonic excitation in the three-quark system in lattice QCD”. In: *Phys. Rev.* D70 (2004), p. 074506.

[1542] Yoshiaki Koma and Miho Koma. “Precise determination of the three-quark potential in SU(3) lattice gauge theory”. In: *Phys. Rev.* D95 (2017), p. 094513.

[1543] Maxim V. Polyakov and Peter Schweitzer. “Determination of $J/\psi$ chromoelectric polarizability from lattice data”. In: *Phys. Rev. D* 98.3 (2018), p. 034030.

[1544] Nora Brambilla et al. “Effective field theories for van der Waals interactions”. In: *Phys. Rev.* D95 (2017), p. 116004.

[1545] Jaume Tarrús Castellà and Gastão Krein. “Effective field theory for the nucleon-quarkonium interaction”. In: *Phys. Rev.* D98 (2018), p. 014029.

[1546] Eric Braaten, Christian Langmack, and D. Hudson Smith. “Born-Oppenheimer Approximation for the XYZ Mesons”. In: *Phys. Rev. D* 90.1 (2014), p. 014044.

[1547] Nora Brambilla et al. “Born-Oppenheimer approximation in an effective field theory language”. In: *Phys. Rev. D* 97.1 (2018), p. 016016.

[1548] Eric Braaten, Christian Langmack, and D. Hudson Smith. “Selection Rules for Hadronic Transitions of XYZ Mesons”. In: *Phys. Rev. Lett.* 112 (2014), p. 222001.

[1549] Matthias Berwein et al. “Quarkonium Hybrids with Nonrelativistic Effective Field Theories”. In: *Phys. Rev.* D92 (2015), p. 114019.

[1550] Jaume Tarrús Castellà and Emilie Passemar. “Exotic to standard bottomonium transitions”. In: *Phys. Rev. D* 104.3 (2021), p. 034019.

[1551] Jaume Tarrús Castellà. “Heavy meson thresholds in Born-Oppenheimer Effective field theory”. In: (July 2022).

[1552] Nora Brambilla et al. “Static quark-antiquark pairs at finite temperature”. In: *Phys. Rev. D* 78 (2008), p. 014017.

[1553] Miguel Angel Escobedo and Joan Soto. “Non-relativistic bound states at finite temperature (I): The Hydrogen atom”. In: *Phys. Rev. A* 78 (2008), p. 032520.

[1554] Miguel Angel Escobedo and Joan Soto. “Non-relativistic bound states at finite temperature (II): the muonic hydrogen”. In: *Phys. Rev. A* 82 (2010), p. 042506.

[1555] Nora Brambilla et al. “Thermal width and gluo-dissociation of quarkonium in pNRQCD”. In: *JHEP* 12 (2011), p. 116.

[1556] Nora Brambilla et al. “Thermal width and quarkonium dissociation by inelastic parton scattering”. In: *JHEP* 05 (2013), p. 130.

[1557] Simone Biondini et al. “Momentum anisotropy effects for quarkonium in a weakly-coupled quark-gluon plasma below the melting temperature”. In: *Phys. Rev. D* 95.7 (2017), p. 074016.

[1558] Yukinao Akamatsu. “Heavy quark master equations in the Lindblad form at high temperatures”. In: *Phys. Rev. D* 91.5 (2015), p. 056002.

[1559] Nora Brambilla et al. “Quarkonium suppression in heavy-ion collisions: an open quantum system approach”. In: *Phys. Rev. D* 96.3 (2017), p. 034021.

[1560] Nora Brambilla et al. “Heavy quarkonium suppression in a fireball”. In: *Phys. Rev. D* 97.7 (2018), p. 074009.

[1561] Alexander Rothkopf. “Heavy Quarkonium in Extreme Conditions”. In: *Phys. Rept.* 858 (2020), pp. 1–117.

[1562] Nora Brambilla et al. “Heavy quarkonium dynamics at next-to-leading order in the binding energy over temperature”. In: *JHEP* 08 (2022), p. 303.

[1563] Stefan Scherer and Matthias R. Schindler. *A Primer for Chiral Perturbation Theory*. Vol. 830. 2012.

[1564] Murray Gell-Mann. “The Eightfold Way: A Theory of strong interaction symmetry”. In: (Mar. 1961).

[1565] C. Abel et al. “Measurement of the Permanent Electric Dipole Moment of the Neutron”. In: *Phys. Rev. Lett.* 124.8 (2020), p. 081803.

[1566] Aneesh Manohar and Howard Georgi. “Chiral Quarks and the Nonrelativistic Quark Model”. In: *Nucl. Phys. B* 234 (1984), pp. 189–212.

[1567] H. Leutwyler. “Theoretical chiral dynamics”. In: *3rd Workshop on Chiral Dynamics - Chiral Dynamics 2000: Theory and Experiment.* July 2000, pp. 3–17.

[1568] Murray Gell-Mann and Yuval Ne’eman. *The Eightfold Way.* Benjamin, Sept. 1964.

[1569] Stephen L. Adler and Roger F. Dashen. *Current Algebras and Applications to Particle Physics.* Benjamin, 1968.

[1570] J. Gasser and H. Leutwyler. “Chiral Perturbation Theory to One Loop”. In: *Annals Phys.* 158 (1984), p. 142.

[1571] H. Leutwyler. “On the foundations of chiral perturbation theory”. In: *Annals Phys.* 235 (1994), pp. 165–203.

[1572] Sidney R. Coleman, J. Wess, and Bruno Zumino. “Structure of phenomenological Lagrangians. 1.” In: *Phys. Rev.* 177 (1969), pp. 2239–2247.

[1573] Y. Aoki et al. *FLAG Review 2021.* Nov. 2021.

[1574] Murray Gell-Mann, R. J. Oakes, and B. Renner. “Behavior of current divergences under SU(3) x SU(3)”. In: *Phys. Rev.* 175 (1968), pp. 2195–2199.

[1575] Steven Weinberg. “Dynamical approach to current algebra”. In: *Phys. Rev. Lett.* 18 (1967), pp. 188–191.

[1576] Julian S. Schwinger. “Chiral dynamics”. In: *Phys. Lett. B* 24 (1967), pp. 473–476.

[1577] Steven Weinberg. *The Quantum theory of fields. Vol. 1: Foundations.* Cambridge University Press, June 2005.

[1578] Johan Bijnens and Gerhard Ecker. “Mesonic low-energy constants”. In: *Ann. Rev. Nucl. Part. Sci.* 64 (2014), pp. 149–174.

[1579] Ling-Fong Li and Heinz Pagels. “Perturbation theory about a Goldstone symmetry”. In: *Phys. Rev. Lett.* 26 (1971), pp. 1204–1206.

[1580] J. Wess and B. Zumino. “Consequences of anomalous Ward identities”. In: *Phys. Lett. B* 37 (1971), pp. 95–97.

[1581] Edward Witten. “Global Aspects of Current Algebra”. In: *Nucl. Phys. B* 223 (1983), pp. 422–432.

[1582] J. L. Manes. “Differential Geometric Construction of the Gauged Wess-Zumino Action”. In: *Nucl. Phys. B* 250 (1985), pp. 369–384.

[1583] J. Bijnens. “Chiral perturbation theory and anomalous processes”. In: *Int. J. Mod. Phys. A* 8 (1993), pp. 3045–3105.

[1584] T. Ebertshauser, H. W. Fearing, and S. Scherer. “The Anomalous chiral perturbation theory meson Lagrangian to order p**6 revisited”. In: *Phys. Rev. D* 65 (2002), p. 054033.

[1585] J. Bijnens, L. Girlanda, and P. Talavera. “The Anomalous chiral Lagrangian of order p**6”. In: *Eur. Phys. J. C* 23 (2002), pp. 539–544.

[1586] J. Gasser, M. E. Sainio, and A. Svarc. “Nucleons with Chiral Loops”. In: *Nucl. Phys. B* 307 (1988), pp. 779–853.

[1587] V. Bernard, Norbert Kaiser, and Ulf-G. Meissner. “Chiral dynamics in nucleons and nuclei”. In: *Int. J. Mod. Phys. E* 4 (1995), pp. 193–346.

[1588] Lisheng Geng. “Recent developments in SU(3) covariant baryon chiral perturbation theory”. In: *Front. Phys. (Beijing)* 8 (2013), pp. 328–348.

[1589] G. Ecker. “Chiral perturbation theory”. In: *Prog. Part. Nucl. Phys.* 35 (1995), pp. 1–80.

[1590] Elizabeth Ellen Jenkins and Aneesh V. Manohar. “Baryon chiral perturbation theory using a heavy fermion Lagrangian”. In: *Phys. Lett. B* 255 (1991), pp. 558–562.

[1591] Paul J. Ellis and Hua-Bin Tang. “Pion nucleon scattering in a new approach to chiral perturbation theory”. In: *Phys. Rev. C* 57 (1998), pp. 3356–3375.

[1592] Thomas Becher and H. Leutwyler. “Baryon chiral perturbation theory in manifestly Lorentz invariant form”. In: *Eur. Phys. J. C* 9 (1999), pp. 643–671.

[1593] J. Gegelia and G. Japaridze. “Matching heavy particle approach to relativistic theory”. In: *Phys. Rev. D* 60 (1999), p. 114038.

[1594] J. Gegelia, G. Japaridze, and X. Q. Wang. “Is Heavy baryon approach necessary?” In: *J. Phys. G* 29 (2003), pp. 2303–2309.

[1595] T. Fuchs et al. “Renormalization of relativistic baryon chiral perturbation theory and power counting”. In: *Phys. Rev. D* 68 (2003), p. 056005.

[1596] Nadia Fettes et al. “The Chiral effective pion nucleon Lagrangian of order p**4”. In: *Annals Phys.* 283 (2000). [Erratum: Annals Phys. 288, 249–250 (2001)], pp. 273–302.

[1597] Stefan Scherer. “Introduction to chiral perturbation theory”. In: *Adv. Nucl. Phys.* 27 (2003). Ed. by John W. Negele and E. W. Vogt, p. 277.

[1598] Veronique Bernard. “Chiral Perturbation Theory and Baryon Properties”. In: *Prog. Part. Nucl. Phys.* 60 (2008), pp. 82–160.

[1599] Matthias R. Schindler, Jambul Gegelia, and Stefan Scherer. "Infrared and extended on mass shell renormalization of two loop diagrams". In: *Nucl. Phys. B* 682 (2004), pp. 367–376.

[1600] Thomas Fuchs et al. "Power counting in baryon chiral perturbation theory including vector mesons". In: *Phys. Lett. B* 575 (2003), pp. 11–17.

[1601] Matthias R. Schindler, Jambul Gegelia, and Stefan Scherer. "Infrared regularization of baryon chiral perturbation theory reformulated". In: *Phys. Lett. B* 586 (2004), pp. 258–266.

[1602] Peter C. Bruns and Ulf-G. Meissner. "Infrared regularization for spin-1 fields". In: *Eur. Phys. J. C* 40 (2005), pp. 97–119.

[1603] Sven Steininger, Ulf-G. Meissner, and Nadia Fettes. "On wave function renormalization and related aspects in heavy fermion effective field theories". In: *JHEP* 09 (1998), p. 008.

[1604] Thomas Fuchs, Jambul Gegelia, and Stefan Scherer. "Structure of the nucleon in chiral perturbation theory". In: *Eur. Phys. J. A* 19 (2004), pp. 35–42.

[1605] Judith A. McGovern and Michael C. Birse. "On the absence of fifth order contributions to the nucleon mass in heavy baryon chiral perturbation theory". In: *Phys. Lett. B* 446 (1999), pp. 300–305.

[1606] M. R. Schindler et al. "Chiral expansion of the nucleon mass to order(q**6)". In: *Phys. Lett. B* 649 (2007), pp. 390–393.

[1607] Matthias R. Schindler et al. "Infrared renormalization of two-loop integrals and the chiral expansion of the nucleon mass". In: *Nucl. Phys. A* 803 (2008). [Erratum: Nucl.Phys.A 1010, 122175 (2021)], pp. 68–114.

[1608] Maarten Golterman. "Applications of chiral perturbation theory to lattice QCD". In: *Les Houches Summer School: Session 93: Modern perspectives in lattice QCD: Quantum field theory and high performance computing.* Dec. 2009, pp. 423–515.

[1609] Ulf-G. Meissner. "Quark mass dependence of baryon properties". In: *PoS* LAT2005 (2006). Ed. by Christopher Michael, p. 009.

[1610] D. Djukanovic, J. Gegelia, and S. Scherer. "Probing the convergence of perturbative series in baryon chiral perturbation theory". In: *Eur. Phys. J. A* 29 (2006), pp. 337–342.

[1611] Veronique Bernard, Thomas R. Hemmert, and Ulf-G. Meissner. "Infrared regularization with spin 3/2 fields". In: *Phys. Lett. B* 565 (2003), pp. 137–145.

[1612] C. Hacker et al. "Including the Delta(1232) resonance in baryon chiral perturbation theory". In: *Phys. Rev. C* 72 (2005), p. 055203.

[1613] William Rarita and Julian Schwinger. "On a theory of particles with half integral spin". In: *Phys. Rev.* 60 (1941), p. 61.

[1614] P. A. Moldauer and K. M. Case. "Properties of Half-Integral Spin Dirac-Fierz-Pauli Particles". In: *Phys. Rev.* 102 (1956), pp. 279–285.

[1615] L. M. Nath, B. Etemadi, and J. D. Kimel. "Uniqueness of the interaction involving spin 3/2 particles". In: *Phys. Rev. D* 3 (1971), pp. 2153–2161.

[1616] Hua-Bin Tang and Paul J. Ellis. "Redundance of Delta isobar parameters in effective field theories". In: *Phys. Lett. B* 387 (1996), pp. 9–13.

[1617] Thomas R. Hemmert, Barry R. Holstein, and Joachim Kambor. "Chiral Lagrangians and delta(1232) interactions: Formalism". In: *J. Phys. G* 24 (1998), pp. 1831–1859.

[1618] V. Pascalutsa. "Quantization of an interacting spin - 3 / 2 field and the Delta isobar". In: *Phys. Rev. D* 58 (1998), p. 096002.

[1619] N. Wies, J. Gegelia, and S. Scherer. "Consistency of the pi Delta interaction in chiral perturbation theory". In: *Phys. Rev. D* 73 (2006), p. 094012.

[1620] H. Krebs, E. Epelbaum, and U. -G. Meissner. "Redundancy of the off-shell parameters in chiral effective field theory with explicit spin-3/2 degrees of freedom". In: *Phys. Lett. B* 683 (2010), pp. 222–228.

[1621] Vladimir Pascalutsa and Daniel R. Phillips. "Effective theory of the delta(1232) in Compton scattering off the nucleon". In: *Phys. Rev. C* 67 (2003), p. 055202.

[1622] T. Papenbrock. "Effective theory for deformed nuclei". In: *Nucl. Phys. A* 852 (2011), p. 36.

[1623] H. -W. Hammer, C. Ji, and D. R. Phillips. "Effective field theory description of halo nuclei". In: *J. Phys. G* 44.10 (2017), p. 103002.

[1624] Eric Braaten and H. -W. Hammer. "Universality in few-body systems with large scattering length". In: *Phys. Rept.* 428 (2006), pp. 259–390.

[1625] H. -W. Hammer, S. König, and U. van Kolck. "Nuclear effective field theory: status and perspectives". In: *Rev. Mod. Phys.* 92.2 (2020), p. 025004.

[1626] Steven Weinberg. "Nuclear forces from chiral Lagrangians". In: *Phys. Lett. B* 251 (1990), pp. 288–292.

[1627] Steven Weinberg. “Effective chiral Lagrangians for nucleon - pion interactions and nuclear forces”. In: *Nucl. Phys. B* 363 (1991), pp. 3–18.

[1628] Evgeny Epelbaum, Jambul Gegelia, and Ulf-G Meißner. “Wilsonian renormalization group versus subtractive renormalization in effective field theories for nucleon–nucleon scattering”. In: *Nucl. Phys. B* 925 (2017), pp. 161–185.

[1629] E. Epelbaum et al. “Effective Field Theory for Shallow P-Wave States”. In: *Few Body Syst.* 62.3 (2021), p. 51.

[1630] Ingo Tews et al. “Nuclear Forces for Precision Nuclear Physics – a collection of perspectives”. In: (Feb. 2022).

[1631] Michael C. Birse, Judith A. McGovern, and Keith G. Richardson. “A Renormalization group treatment of two-body scattering”. In: *Phys. Lett. B* 464 (1999), pp. 169–176.

[1632] David B. Kaplan, Martin J. Savage, and Mark B. Wise. “A New expansion for nucleon-nucleon interactions”. In: *Phys. Lett. B* 424 (1998), pp. 390–396.

[1633] David B. Kaplan, Martin J. Savage, and Mark B. Wise. “Two nucleon systems from effective field theory”. In: *Nucl. Phys. B* 534 (1998), pp. 329–355.

[1634] Thomas D. Cohen and James M. Hansen. “Low-energy theorems for nucleon-nucleon scattering”. In: *Phys. Rev. C* 59 (1999), pp. 13–20.

[1635] Sean Fleming, Thomas Mehen, and Iain W. Stewart. “NNLO corrections to nucleon-nucleon scattering and perturbative pions”. In: *Nucl. Phys. A* 677 (2000), pp. 313–366.

[1636] Silas R. Beane, David B. Kaplan, and Aleksi Vuorinen. “Perturbative nuclear physics”. In: *Phys. Rev. C* 80 (2009), p. 011001.

[1637] E. Epelbaum et al. “$^1S_0$ nucleon-nucleon scattering in the modified Weinberg approach”. In: *Eur. Phys. J. A* 51.6 (2015), p. 71.

[1638] David B. Kaplan. “Convergence of nuclear effective field theory with perturbative pions”. In: *Phys. Rev. C* 102.3 (2020), p. 034004.

[1639] G. P. Lepage. “How to renormalize the Schrodinger equation”. In: *8th Jorge Andre Swieca Summer School on Nuclear Physics.* Feb. 1997, pp. 135–180.

[1640] Evgeny Epelbaum, Hermann Krebs, and Patrick Reinert. “High-precision nuclear forces from chiral EFT: State-of-the-art, challenges and outlook”. In: *Front. in Phys.* 8 (2020), p. 98.

[1641] A. M. Gasparyan and E. Epelbaum. “Nucleon-nucleon interaction in chiral effective field theory with a finite cutoff: Explicit perturbative renormalization at next-to-leading order”. In: *Phys. Rev. C* 105.2 (2022), p. 024001.

[1642] C. Ordonez and U. van Kolck. “Chiral lagrangians and nuclear forces”. In: *Phys. Lett. B* 291 (1992), pp. 459–464.

[1643] C. Ordonez, L. Ray, and U. van Kolck. “The Two nucleon potential from chiral Lagrangians”. In: *Phys. Rev. C* 53 (1996), pp. 2086–2105.

[1644] Norbert Kaiser, R. Brockmann, and W. Weise. “Peripheral nucleon-nucleon phase shifts and chiral symmetry”. In: *Nucl. Phys. A* 625 (1997), pp. 758–788.

[1645] S. Pastore, R. Schiavilla, and J. L. Goity. “Electromagnetic two-body currents of one- and two-pion range”. In: *Phys. Rev. C* 78 (2008), p. 064002.

[1646] S. Pastore et al. “Electromagnetic Currents and Magnetic Moments in (chi)EFT”. In: *Phys. Rev. C* 80 (2009), p. 034004.

[1647] S. Pastore et al. “The two-nucleon electromagnetic charge operator in chiral effective field theory ($\chi$EFT) up to one loop”. In: *Phys. Rev. C* 84 (2011), p. 024001.

[1648] A. Baroni et al. “Nuclear Axial Currents in Chiral Effective Field Theory”. In: *Phys. Rev. C* 93.1 (2016). [Erratum: Phys.Rev.C 93, 049902 (2016), Erratum: Phys.Rev.C 95, 059901 (2017)], p. 015501.

[1649] E. Epelbaum, Walter Gloeckle, and Ulf-G. Meissner. “Nuclear forces from chiral Lagrangians using the method of unitary transformation. 1. Formalism”. In: *Nucl. Phys. A* 637 (1998), pp. 107–134.

[1650] E. Epelbaum. “Four-nucleon force using the method of unitary transformation”. In: *Eur. Phys. J. A* 34 (2007), pp. 197–214.

[1651] V. Bernard et al. “Subleading contributions to the chiral three-nucleon force. I. Long-range terms”. In: *Phys. Rev. C* 77 (2008), p. 064004.

[1652] V. Bernard et al. “Subleading contributions to the chiral three-nucleon force II: Short-range terms and relativistic corrections”. In: *Phys. Rev. C* 84 (2011), p. 054001.

[1653] Hermann Krebs, A. Gasparyan, and Evgeny Epelbaum. “Chiral three-nucleon force at $\mathrm{N}^4$LO I: Longest-range contributions”. In: *Phys. Rev. C* 85 (2012), p. 054006.

[1654] Hermann Krebs, A. Gasparyan, and Evgeny Epelbaum. “Chiral three-nucleon force at $N^4LO$ II: Intermediate-range contributions”. In: *Phys. Rev. C* 87.5 (2013), p. 054007.

[1655] S. Kolling et al. “Two-nucleon electromagnetic current in chiral effective field theory: One-pion

exchange and short-range contributions”. In: *Phys. Rev. C* 84 (2011), p. 054008.

[1656] S. Kolling et al. “Two-pion exchange electromagnetic current in chiral effective field theory using the method of unitary transformation”. In: *Phys. Rev. C* 80 (2009), p. 045502.

[1657] H. Krebs, E. Epelbaum, and U. -G. Meißner. “Nuclear axial current operators to fourth order in chiral effective field theory”. In: *Annals Phys.* 378 (2017), pp. 317–395.

[1658] H. Krebs, E. Epelbaum, and U. -G. Meißner. “Nuclear Electromagnetic Currents to Fourth Order in Chiral Effective Field Theory”. In: *Few Body Syst.* 60.2 (2019), p. 31.

[1659] Hermann Krebs, Evgeny Epelbaum, and Ulf-G. Meißner. “Subleading contributions to the nuclear scalar isoscalar current”. In: *Eur. Phys. J. A* 56.9 (2020), p. 240.

[1660] Evgeny Epelbaum. “Nuclear Forces from Chiral Effective Field Theory: A Primer”. In: Jan. 2010.

[1661] E. Epelbaum. “Four-nucleon force in chiral effective field theory”. In: *Phys. Lett. B* 639 (2006), pp. 456–461.

[1662] James Lewis Friar and S. A. Coon. “Non-adiabatic contributions to static two-pion-exchange nuclear potentials”. In: *Phys. Rev. C* 49 (1994), pp. 1272–1280.

[1663] V. Baru et al. “The Multiple-scattering series in pion-deuteron scattering and the nucleon-nucleon potential: Perspectives from effective field theory”. In: *Eur. Phys. J. A* 48 (2012), p. 69.

[1664] C. Ditsche et al. “Roy-Steiner equations for pion-nucleon scattering”. In: *JHEP* 06 (2012), p. 043.

[1665] Martin Hoferichter et al. “Matching pion-nucleon Roy-Steiner equations to chiral perturbation theory”. In: *Phys. Rev. Lett.* 115.19 (2015), p. 192301.

[1666] D. Siemens et al. “Reconciling threshold and subthreshold expansions for pion–nucleon scattering”. In: *Phys. Lett. B* 770 (2017), pp. 27–34.

[1667] V. Bernard, Norbert Kaiser, and Ulf-G. Meissner. “Aspects of chiral pion - nucleon physics”. In: *Nucl. Phys. A* 615 (1997), pp. 483–500.

[1668] Norbert Kaiser. “Chiral 2 pi exchange N N potentials: Two loop contributions”. In: *Phys. Rev. C* 64 (2001), p. 057001.

[1669] Norbert Kaiser. “Chiral 2 pi exchange NN potentials: Relativistic 1 /M**2 corrections”. In: *Phys. Rev. C* 65 (2002), p. 017001.

[1670] D. R. Entem et al. “Peripheral nucleon-nucleon scattering at fifth order of chiral perturbation theory”. In: *Phys. Rev. C* 91.1 (2015), p. 014002.

[1671] D. R. Entem et al. “Dominant contributions to the nucleon-nucleon interaction at sixth order of chiral perturbation theory”. In: *Phys. Rev. C* 92.6 (2015), p. 064001.

[1672] P. Reinert, H. Krebs, and E. Epelbaum. “Semilocal momentum-space regularized chiral two-nucleon potentials up to fifth order”. In: *Eur. Phys. J. A* 54.5 (2018), p. 86.

[1673] Evgeny Epelbaum, Hans-Werner Hammer, and Ulf-G. Meissner. “Modern Theory of Nuclear Forces”. In: *Rev. Mod. Phys.* 81 (2009), pp. 1773–1825.

[1674] R. Machleidt and D. R. Entem. “Chiral effective field theory and nuclear forces”. In: *Phys. Rept.* 503 (2011), pp. 1–75.

[1675] U. van Kolck. “Few nucleon forces from chiral Lagrangians”. In: *Phys. Rev. C* 49 (1994), pp. 2932–2941.

[1676] E. Epelbaum et al. “Three nucleon forces from chiral effective field theory”. In: *Phys. Rev. C* 66 (2002), p. 064001.

[1677] S. Ishikawa and M. R. Robilotta. “Two-pion exchange three-nucleon potential: O(q**4) chiral expansion”. In: *Phys. Rev. C* 76 (2007), p. 014006.

[1678] L. Girlanda, A. Kievsky, and M. Viviani. “Subleading contributions to the three-nucleon contact interaction”. In: *Phys. Rev. C* 84.1 (2011). [Erratum: Phys.Rev.C 102, 019903 (2020)], p. 014001.

[1679] E. Epelbaum et al. “Three-nucleon force at large distances: Insights from chiral effective field theory and the large-$N_c$ expansion”. In: *Eur. Phys. J. A* 51.3 (2015), p. 26.

[1680] Jordy de Vries et al. “Parity- and Time-Reversal-Violating Nuclear Forces”. In: *Front. in Phys.* 8 (2020), p. 218.

[1681] Tae-Sun Park, Dong-Pil Min, and Mannque Rho. “Chiral dynamics and heavy fermion formalism in nuclei. 1. Exchange axial currents”. In: *Phys. Rept.* 233 (1993), pp. 341–395.

[1682] Tae-Sun Park, Dong-Pil Min, and Mannque Rho. “Chiral Lagrangian approach to exchange vector currents in nuclei”. In: *Nucl. Phys. A* 596 (1996), pp. 515–552.

[1683] Martin Hoferichter, Philipp Klos, and Achim Schwenk. “Chiral power counting of one- and two-body currents in direct detection of dark matter”. In: *Phys. Lett. B* 746 (2015), pp. 410–416.

[1684] Hermann Krebs. “Nuclear Currents in Chiral Effective Field Theory”. In: *Eur. Phys. J. A* 56.9 (2020), p. 234.

[1685] Norbert Kaiser, S. Gerstendorfer, and W. Weise. “Peripheral NN scattering: Role of delta excitation, correlated two pion and vector meson exchange”. In: *Nucl. Phys. A* 637 (1998), pp. 395–420.

[1686] Hermann Krebs, Evgeny Epelbaum, and Ulf-G. Meissner. “Nuclear forces with Delta-excitations up to next-to-next-to-leading order. I. Peripheral nucleon-nucleon waves”. In: *Eur. Phys. J. A* 32 (2007), pp. 127–137.

[1687] E. Epelbaum, H. Krebs, and Ulf-G. Meissner. “Delta-excitations and the three-nucleon force”. In: *Nucl. Phys. A* 806 (2008), pp. 65–78.

[1688] H. Krebs, A. M. Gasparyan, and E. Epelbaum. “Three-nucleon force in chiral EFT with explicit $\Delta(1232)$ degrees of freedom: Longest-range contributions at fourth order”. In: *Phys. Rev. C* 98.1 (2018), p. 014003.

[1689] Henk Polinder, Johann Haidenbauer, and Ulf-G. Meissner. “Hyperon-nucleon interactions: A Chiral effective field theory approach”. In: *Nucl. Phys. A* 779 (2006), pp. 244–266.

[1690] J. Haidenbauer et al. “Hyperon-nucleon interaction at next-to-leading order in chiral effective field theory”. In: *Nucl. Phys. A* 915 (2013), pp. 24–58.

[1691] Stefan Petschauer et al. “Leading three-baryon forces from SU(3) chiral effective field theory”. In: *Phys. Rev. C* 93.1 (2016), p. 014001.

[1692] J. Haidenbauer and U. -G. Meißner. “Status of the hyperon-nucleon interaction in chiral effective field theory”. In: *14th International Conference on Hypernuclear and Strange Particle Physics.* Aug. 2022.

[1693] E. Epelbaum, Walter Gloeckle, and Ulf-G. Meissner. “Nuclear forces from chiral Lagrangians using the method of unitary transformation. 2. The two nucleon system”. In: *Nucl. Phys. A* 671 (2000), pp. 295–331.

[1694] Evgeny Epelbaum, Walter Gloeckle, and Ulf-G. Meissner. “Improving the convergence of the chiral expansion for nuclear forces. 1. Peripheral phases”. In: *Eur. Phys. J. A* 19 (2004), pp. 125–137.

[1695] A. Gezerlis et al. “Quantum Monte Carlo Calculations with Chiral Effective Field Theory Interactions”. In: *Phys. Rev. Lett.* 111.3 (2013), p. 032501.

[1696] E. Epelbaum, H. Krebs, and U. G. Meißner. “Improved chiral nucleon-nucleon potential up to next-to-next-to-next-to-leading order”. In: *Eur. Phys. J. A* 51.5 (2015), p. 53.

[1697] M. Piarulli et al. “Minimally nonlocal nucleon-nucleon potentials with chiral two-pion exchange including $\Delta$ resonances”. In: *Phys. Rev. C* 91.2 (2015), p. 024003.

[1698] A. Dyhdalo et al. “Regulator Artifacts in Uniform Matter for Chiral Interactions”. In: *Phys. Rev. C* 94.3 (2016), p. 034001.

[1699] D. R. Entem, R. Machleidt, and Y. Nosyk. “High-quality two-nucleon potentials up to fifth order of the chiral expansion”. In: *Phys. Rev. C* 96.2 (2017), p. 024004.

[1700] Ingo Tews et al. “New Ideas in Constraining Nuclear Forces”. In: *J. Phys. G* 47.10 (2020), p. 103001.

[1701] P. Reinert, H. Krebs, and E. Epelbaum. “Precision determination of pion-nucleon coupling constants using effective field theory”. In: *Phys. Rev. Lett.* 126.9 (2021), p. 092501.

[1702] E. Epelbaum, H. Krebs, and P. Reinert. “Semi-local nuclear forces from chiral EFT: State-of-the-art & challenges”. In: (June 2022).

[1703] R. J. Furnstahl et al. “Quantifying truncation errors in effective field theory”. In: *Phys. Rev. C* 92.2 (2015), p. 024005.

[1704] Evgeny Epelbaum. “High-precision nuclear forces : Where do we stand?” In: *PoS* CD2018 (2019), p. 006.

[1705] E. Epelbaum et al. “Towards high-order calculations of three-nucleon scattering in chiral effective field theory”. In: *Eur. Phys. J. A* 56.3 (2020), p. 92.

[1706] A. A. Filin et al. “Extraction of the neutron charge radius from a precision calculation of the deuteron structure radius”. In: *Phys. Rev. Lett.* 124.8 (2020), p. 082501.

[1707] A. A. Filin et al. “High-accuracy calculation of the deuteron charge and quadrupole form factors in chiral effective field theory”. In: *Phys. Rev. C* 103.2 (2021), p. 024313.

[1708] Krzysztof Pachucki, Vojtěch Patkóš, and Vladimir A. Yerokhin. “Three-photon exchange nuclear structure correction in hydrogenic systems”. In: *Phys. Rev. A* 97.6 (2018), p. 062511.

[1709] Mariusz Puchalski, Jacek Komasa, and Krzysztof Pachucki. “Hyperfine Structure of the First Rotational Level in $H_2$, $D_2$ and HD Molecules and the Deuteron Quadrupole Moment”. In: *Phys. Rev. Lett.* 125.25 (2020), p. 253001.

[1710] M. C. M. Rentmeester et al. “Chiral two pion exchange and proton proton partial wave anal-

ysis". In: *Phys. Rev. Lett.* 82 (1999), pp. 4992–4995.

[1711] Michael C. Birse and Judith A. McGovern. "On the effectiveness of effective field theory in peripheral nucleon nucleon scattering". In: *Phys. Rev. C* 70 (2004), p. 054002.

[1712] E. Epelbaum, H. Krebs, and U. G. Meißner. "Precision nucleon-nucleon potential at fifth order in the chiral expansion". In: *Phys. Rev. Lett.* 115.12 (2015), p. 122301.

[1713] Maria Piarulli and Ingo Tews. "Local Nucleon-Nucleon and Three-Nucleon Interactions Within Chiral Effective Field Theory". In: *Front. in Phys.* 7 (2020), p. 245.

[1714] W. G. Jiang et al. "Accurate bulk properties of nuclei from $A = 2$ to $\infty$ from potentials with $\Delta$ isobars". In: *Phys. Rev. C* 102.5 (2020), p. 054301.

[1715] P. Maris et al. "Light nuclei with semilocal momentum-space regularized chiral interactions up to third order". In: *Phys. Rev. C* 103.5 (2021), p. 054001.

[1716] E. Epelbaum et al. "Few- and many-nucleon systems with semilocal coordinate-space regularized chiral two- and three-body forces". In: *Phys. Rev. C* 99.2 (2019), p. 024313.

[1717] P. Maris et al. "Nuclear properties with semilocal momentum-space regularized chiral interactions beyond N2LO". In: (June 2022).

[1718] Bruce R. Barrett, Petr Navratil, and James P. Vary. "Ab initio no core shell model". In: *Prog. Part. Nucl. Phys.* 69 (2013), pp. 131–181.

[1719] G. Hagen, G. R. Jansen, and T. Papenbrock. "Structure of $^{78}$Ni from first principles computations". In: *Phys. Rev. Lett.* 117.17 (2016), p. 172501.

[1720] Eskendr Gebrerufael et al. "Ab Initio Description of Open-Shell Nuclei: Merging No-Core Shell Model and In-Medium Similarity Renormalization Group". In: *Phys. Rev. Lett.* 118.15 (2017), p. 152503.

[1721] D. Lonardoni et al. "Properties of nuclei up to $A = 16$ using local chiral interactions". In: *Phys. Rev. Lett.* 120.12 (2018), p. 122502.

[1722] M. Piarulli et al. "Light-nuclei spectra from chiral dynamics". In: *Phys. Rev. Lett.* 120.5 (2018), p. 052503.

[1723] H. Hergert. "A Guided Tour of *ab initio* Nuclear Many-Body Theory". In: *Front. in Phys.* 8 (2020), p. 379.

[1724] Ingo Tews. "Quantum Monte Carlo Methods for Astrophysical Applications". In: *Front. in Phys.* 8 (2020), p. 153.

[1725] N. Kalantar-Nayestanaki et al. "Signatures of three-nucleon interactions in few-nucleon systems". In: *Rept. Prog. Phys.* 75 (2012), p. 016301.

[1726] Hans-Werner Hammer, Andreas Nogga, and Achim Schwenk. "Three-body forces: From cold atoms to nuclei". In: *Rev. Mod. Phys.* 85 (2013), p. 197.

[1727] S. Pastore et al. "Quantum Monte Carlo calculations of electromagnetic transitions in $^{8}$Be with meson-exchange currents derived from chiral effective field theory". In: *Phys. Rev. C* 90.2 (2014), p. 024321.

[1728] Sonia Bacca and Saori Pastore. "Electromagnetic reactions on light nuclei". In: *J. Phys. G* 41.12 (2014), p. 123002.

[1729] R. Schiavilla et al. "Local chiral interactions and magnetic structure of few-nucleon systems". In: *Phys. Rev. C* 99.3 (2019), p. 034005.

[1730] N. Nevo Dinur et al. "Zemach moments and radii of $^{2,3}$H and $^{3,4}$He". In: *Phys. Rev. C* 99.3 (2019), p. 034004.

[1731] Garrett B. King et al. "Weak Transitions in Light Nuclei". In: *Front. in Phys.* 8 (2020), p. 363.

[1732] S. Pastore et al. "Quantum Monte Carlo calculations of weak transitions in $A = 610$ nuclei". In: *Phys. Rev. C* 97.2 (2018), p. 022501.

[1733] L. E. Marcucci et al. "Implication of the proton-deuteron radiative capture for Big Bang Nucleosynthesis". In: *Phys. Rev. Lett.* 116.10 (2016). [Erratum: Phys.Rev.Lett. 117, 049901 (2016)], p. 102501.

[1734] L. Ceccarelli et al. "Muon capture on deuteron using local chiral potentials". In: (Sept. 2022).

[1735] S. Pastore et al. "Quantum Monte Carlo calculations of electromagnetic moments and transitions in A $\leq$ 9 nuclei with meson-exchange currents derived from chiral effective field theory". In: *Phys. Rev. C* 87.3 (2013), p. 035503.

[1736] Yutaka Utsuno. "Anomalous magnetic moment of C-9 and shell quenching in exotic nuclei". In: *Phys. Rev. C* 70 (2004), p. 011303.

[1737] G. B. King et al. "Chiral Effective Field Theory Calculations of Weak Transitions in Light Nuclei". In: *Phys. Rev. C* 102.2 (2020), p. 025501.

[1738] W. -T. Chou, E. K. Warburton, and B. Alex Brown. "Gamow-Teller beta-decay rates for A <= 18 nuclei". In: *Phys. Rev. C* 47 (1993), pp. 163–177.

[1739] P. Gysbers et al. "Discrepancy between experimental and theoretical $\beta$-decay rates resolved from first principles". In: *Nature Phys.* 15.5 (2019), pp. 428–431.

[1740] Zohreh Davoudi et al. "Nuclear matrix elements from lattice QCD for electroweak and beyond-

Standard-Model processes". In: *Phys. Rept.* 900 (2021), pp. 1–74.

[1741] Evgeny Epelbaum, Ulf-G. Meissner, and Walter Gloeckle. "Nuclear forces in the chiral limit". In: *Nucl. Phys. A* 714 (2003), pp. 535–574.

[1742] Silas R. Beane and Martin J. Savage. "The Quark mass dependence of two nucleon systems". In: *Nucl. Phys. A* 717 (2003), pp. 91–103.

[1743] Jiunn-Wei Chen et al. "On the Quark Mass Dependence of Two Nucleon Observables". In: *Phys. Rev. C* 86 (2012), p. 054001.

[1744] J. C. Berengut et al. "Varying the light quark mass: impact on the nuclear force and Big Bang nucleosynthesis". In: *Phys. Rev. D* 87.8 (2013), p. 085018.

[1745] X. L. Ren, E. Epelbaum, and J. Gegelia. "$\Lambda$ -nucleon scattering in baryon chiral perturbation theory". In: *Phys. Rev. C* 101.3 (2020), p. 034001.

[1746] Qian-Qian Bai et al. "Pion-mass dependence of the nucleon-nucleon interaction". In: *Phys. Lett.* B (2020), p. 135745.

[1747] V. Baru et al. "Low-energy theorems for nucleon-nucleon scattering at unphysical pion masses". In: *Phys. Rev. C* 92.1 (2015), p. 014001.

[1748] V. Baru, E. Epelbaum, and A. A. Filin. "Low-energy theorems for nucleon-nucleon scattering at $M_\pi = 450$ MeV". In: *Phys. Rev. C* 94.1 (2016), p. 014001.

[1749] Moti Eliyahu, Betzalel Bazak, and Nir Barnea. "Extrapolating Lattice QCD Results using Effective Field Theory". In: *Phys. Rev. C* 102.4 (2020), p. 044003.

[1750] W. Detmold and P. E. Shanahan. "Few-nucleon matrix elements in pionless effective field theory in a finite volume". In: *Phys. Rev. D* 103.7 (2021), p. 074503.

[1751] Xiangkai Sun et al. "Finite-volume pionless effective field theory for few-nucleon systems with differentiable programming". In: *Phys. Rev. D* 105.7 (2022), p. 074508.

[1752] Lu Meng and E. Epelbaum. "Two-particle scattering from finite-volume quantization conditions using the plane wave basis". In: *JHEP* 10 (2021), p. 051.

[1753] N. Barnea et al. "Effective Field Theory for Lattice Nuclei". In: *Phys. Rev. Lett.* 114.5 (2015), p. 052501.

[1754] Timo A. Lähde, Ulf-G. Meißner, and Evgeny Epelbaum. "An update on fine-tunings in the triple-alpha process". In: *Eur. Phys. J. A* 56.3 (2020), p. 89.

[1755] Evgeny Epelbaum et al. "Ab initio calculation of the Hoyle state". In: *Phys. Rev. Lett.* 106 (2011), p. 192501.

[1756] Serdar Elhatisari et al. "Ab initio alpha-alpha scattering". In: *Nature* 528 (2015), p. 111.

[1757] Timo A. Lähde and Ulf-G. Meißner. *Nuclear Lattice Effective Field Theory: An introduction.* Vol. 957. Springer, 2019.

[1758] Dillon Frame et al. "Eigenvector continuation with subspace learning". In: *Phys. Rev. Lett.* 121.3 (2018), p. 032501.

[1759] S. König et al. "Eigenvector Continuation as an Efficient and Accurate Emulator for Uncertainty Quantification". In: *Phys. Lett. B* 810 (2020), p. 135814.

[1760] R. J. Furnstahl et al. "Efficient emulators for scattering using eigenvector continuation". In: *Phys. Lett. B* 809 (2020), p. 135719.

[1761] Christian W. Bauer, Sean Fleming, and Michael E. Luke. "Summing Sudakov logarithms in $B \to X_s\gamma$ in effective field theory". In: *Phys. Rev.* D63 (2000), p. 014006.

[1762] Christian W. Bauer et al. "An Effective field theory for collinear and soft gluons: Heavy to light decays". In: *Phys. Rev.* D63 (2001), p. 114020.

[1763] Christian W. Bauer and Iain W. Stewart. "Invariant operators in collinear effective theory". In: *Phys. Lett.* B516 (2001), pp. 134–142.

[1764] Christian W. Bauer, Dan Pirjol, and Iain W. Stewart. "Soft collinear factorization in effective field theory". In: *Phys. Rev.* D65 (2002), p. 054022.

[1765] G. Peter Lepage and Stanley J. Brodsky. "Exclusive Processes in Quantum Chromodynamics: Evolution Equations for Hadronic Wave Functions and the Form-Factors of Mesons". In: *Phys. Lett. B* 87 (1979), pp. 359–365.

[1766] John C. Collins, Davison E. Soper, and George Sterman. "Soft gluons and factorization". In: *Nucl. Phys.* B308 (1988), p. 833.

[1767] Christian W. Bauer et al. "Hard scattering factorization from effective field theory". In: *Phys. Rev.* D66 (2002), p. 014017.

[1768] Markus A. Ebert, Anjie Gao, and Iain W. Stewart. "Factorization for azimuthal asymmetries in SIDIS at next-to-leading power". In: *JHEP* 06 (2022), p. 007.

[1769] Sean Fleming et al. "Jets from massive unstable particles: Top-mass determination". In: *Phys. Rev.* D77 (2008), p. 074010.

[1770] Matthew D. Schwartz. "Resummation and NLO Matching of Event Shapes with Effective Field Theory". In: *Phys. Rev.* D77 (2008), p. 014026.

[1771] Christian W. Bauer et al. “Factorization of $e^+e^-$ Event Shape Distributions with Hadronic Final States in Soft Collinear Effective Theory”. In: *Phys. Rev.* D78 (2008), p. 034027.

[1772] Thomas Becher and Matthew D. Schwartz. “A precise determination of $\alpha_s$ from LEP thrust data using effective field theory”. In: *JHEP* 07 (2008), p. 034.

[1773] Xiaohui Liu and Frank Petriello. “Resummation of jet-veto logarithms in hadronic processes containing jets”. In: *Phys.Rev.* D87 (2013), p. 014018.

[1774] Teppo T. Jouttenus et al. “Jet mass spectra in Higgs boson plus one jet at next-to-next-to-leading logarithmic order”. In: *Phys. Rev.* D88.5 (2013), p. 054031.

[1775] Randall Kelley and Matthew D. Schwartz. “1-loop matching and NNLL resummation for all partonic 2 to 2 processes in QCD”. In: *Phys.Rev.* D83 (2011), p. 045022.

[1776] Iain W. Stewart, Frank J. Tackmann, and Wouter J. Waalewijn. “N-Jettiness: An Inclusive Event Shape to Veto Jets”. In: *Phys. Rev. Lett.* 105 (2010), p. 092002.

[1777] Stephen D. Ellis et al. “Jet Shapes and Jet Algorithms in SCET”. In: *JHEP* 11 (2010), p. 101.

[1778] Christian W. Bauer et al. “Factorization and Resummation for Dijet Invariant Mass Spectra”. In: *Phys.Rev.* D85 (2012), p. 074006.

[1779] Ilya Feige et al. “Precision Jet Substructure from Boosted Event Shapes”. In: *Phys. Rev. Lett.* 109 (2012), p. 092001.

[1780] Wouter J. Waalewijn. “Calculating the Charge of a Jet”. In: *Phys. Rev. D* 86 (2012), p. 094030.

[1781] Andrew J. Larkoski, Ian Moult, and Duff Neill. “Power Counting to Better Jet Observables”. In: *JHEP* 12 (2014), p. 009.

[1782] Andrew J. Larkoski, Ian Moult, and Duff Neill. “Analytic Boosted Boson Discrimination”. In: *JHEP* 05 (2016), p. 117.

[1783] Yang-Ting Chien, Andrew Hornig, and Christopher Lee. “Soft-collinear mode for jet cross sections in soft collinear effective theory”. In: *Phys. Rev. D* 93.1 (2016), p. 014033.

[1784] Thomas Becher et al. “Effective Field Theory for Jet Processes”. In: *Phys. Rev. Lett.* 116.19 (2016), p. 192001.

[1785] Andrew Hornig, Yiannis Makris, and Thomas Mehen. “Jet Shapes in Dijet Events at the LHC in SCET”. In: *JHEP* 04 (2016), p. 097.

[1786] Christopher Frye et al. “Factorization for groomed jet substructure beyond the next-to-leading logarithm”. In: *JHEP* 07 (2016), p. 064.

[1787] Piotr Pietrulewicz, Frank J. Tackmann, and Wouter J. Waalewijn. “Factorization and Resummation for Generic Hierarchies between Jets”. In: *JHEP* 08 (2016), p. 002.

[1788] Andrew J. Larkoski, Ian Moult, and Duff Neill. “Analytic Boosted Boson Discrimination at the Large Hadron Collider”. In: (Aug. 2017).

[1789] Andrew J. Larkoski, Ian Moult, and Benjamin Nachman. “Jet Substructure at the Large Hadron Collider: A Review of Recent Advances in Theory and Machine Learning”. In: *Phys. Rept.* 841 (2020), pp. 1–63.

[1790] André H. Hoang et al. “Nonperturbative Corrections to Soft Drop Jet Mass”. In: *JHEP* 12 (2019), p. 002.

[1791] M. Beneke et al. “Soft-collinear effective theory and heavy-to-light currents beyond leading power”. In: *Nucl. Phys.* B643 (2002), pp. 431–476.

[1792] Christian W. Bauer, Dan Pirjol, and Iain W. Stewart. “Factorization and endpoint singularities in heavy to light decays”. In: *Phys. Rev.* D67 (2003), p. 071502.

[1793] M. Beneke and T. Feldmann. “Factorization of heavy to light form-factors in soft collinear effective theory”. In: *Nucl. Phys. B* 685 (2004), pp. 249–296.

[1794] Christian W. Bauer, Dan Pirjol, and Iain W. Stewart. “A Proof of factorization for $B \to D\pi$”. In: *Phys. Rev. Lett.* 87 (2001), p. 201806.

[1795] Sonny Mantry, Dan Pirjol, and Iain W. Stewart. “Strong phases and factorization for color suppressed decays”. In: *Phys. Rev.* D68 (2003), p. 114009.

[1796] Christian W. Bauer et al. “$B \to M_1 M_2$: Factorization, charming penguins, strong phases, and polarization”. In: *Phys. Rev.* D70 (2004), p. 054015.

[1797] Keith S. M. Lee and Iain W. Stewart. “Factorization for power corrections to $B \to X_s\gamma$ and $B \to X_u\ell\bar{\nu}$”. In: *Nucl. Phys.* B721 (2005), pp. 325–406.

[1798] Stefan W. Bosch, Matthias Neubert, and Gil Paz. “Subleading shape functions in inclusive B decays”. In: *JHEP* 11 (2004), p. 073.

[1799] M. Beneke et al. “Power corrections to $\bar{B} \to X_u\ell\bar{\nu}(X_s\gamma)$ decay spectra in the 'shape-function' region”. In: *JHEP* 06 (2005), p. 071.

[1800] Zoltan Ligeti, Iain W. Stewart, and Frank J. Tackmann. “Treating the b quark distribution function with reliable uncertainties”. In: *Phys. Rev.* D78 (2008), p. 114014.

[1801] Michael Benzke et al. “Factorization at Subleading Power and Irreducible Uncertainties in $\bar{B} \to X_s\gamma$ Decay”. In: *JHEP* 08 (2010), p. 099.

[1802] Sean Fleming, Adam K. Leibovich, and Thomas Mehen. “Resumming the color-octet contribution to $e^+e^- \to J/\psi + X$”. In: *Phys. Rev.* D68 (2003), p. 094011.

[1803] Sean Fleming, Adam K. Leibovich, and Thomas Mehen. “$J/\psi$ photo-production at large Z in soft collinear effective theory”. In: (2005), pp. 239–252.

[1804] Sean Fleming, Adam K. Leibovich, and Thomas Mehen. “Resummation of Large Endpoint Corrections to Color-Octet $J/\psi$ Photoproduction”. In: *Phys.Rev.* D74 (2006), p. 114004.

[1805] Adam K. Leibovich and Xiaohui Liu. “The Color-singlet contribution to $e^+e^- \to J/\psi + X$ at the endpoint”. In: *Phys. Rev. D* 76 (2007), p. 034005.

[1806] Sean Fleming, Christopher Lee, and Adam K. Leibovich. “Exclusive radiative decays of Upsilon in SCET”. In: *Phys.Rev.* D71 (2005), p. 074002.

[1807] Sean Fleming and Adam K. Leibovich. “Flavor-singlet light-cone amplitudes and radiative Upsilon decays in SCET”. In: *Phys. Rev. D* 70 (2004), p. 094016.

[1808] Xavier Garcia i Tormo and Joan Soto. “Soft, collinear and nonrelativistic modes in radiative decays of very heavy quarkonium”. In: *Phys. Rev. D* 69 (2004), p. 114006.

[1809] Xavier Garcia i Tormo and Joan Soto. “Semi-inclusive radiative decays of Upsilon(1S)”. In: *Phys. Rev. D* 72 (2005), p. 054014.

[1810] Ira Z. Rothstein and Iain W. Stewart. “An Effective Field Theory for Forward Scattering and Factorization Violation”. In: *JHEP* 08 (2016), p. 025.

[1811] Ian Moult et al. “Fermionic Glauber Operators and Quark Reggeization”. In: *JHEP* 02 (2018), p. 134.

[1812] Arindam Bhattacharya, Aneesh V. Manohar, and Matthew D. Schwartz. “Quark-gluon backscattering in the Regge limit at one-loop”. In: *JHEP* 02 (2022), p. 091.

[1813] Ian Moult et al. “Anomalous Dimensions from Soft Regge Constants”. In: (July 2022).

[1814] Francesco D’Eramo, Hong Liu, and Krishna Rajagopal. “Transverse Momentum Broadening and the Jet Quenching Parameter, Redux”. In: *Phys. Rev. D* 84 (2011), p. 065015.

[1815] Grigory Ovanesyan and Ivan Vitev. “An effective theory for jet propagation in dense QCD matter: jet broadening and medium-induced bremsstrahlung”. In: *JHEP* 1106 (2011), p. 080.

[1816] Grigory Ovanesyan and Ivan Vitev. “Medium-induced parton splitting kernels from Soft Collinear Effective Theory with Glauber gluons”. In: *Phys.Lett.* B706 (2012), pp. 371–378.

[1817] Michael Benzke et al. “Gauge invariant definition of the jet quenching parameter”. In: *JHEP* 02 (2013), p. 129.

[1818] Varun Vaidya and Xiaojun Yao. “Transverse momentum broadening of a jet in quark-gluon plasma: an open quantum system EFT”. In: *JHEP* 10 (2020), p. 024.

[1819] Varun Vaidya. “Effective Field Theory for jet substructure in heavy ion collisions”. In: *JHEP* 11 (2021), p. 064.

[1820] Martin Beneke and Grisha Kirilin. “Soft-collinear gravity”. In: *JHEP* 09 (2012), p. 066.

[1821] Timothy Cohen, Gilly Elor, and Andrew J. Larkoski. “Soft-Collinear Supersymmetry”. In: *JHEP* 03 (2017), p. 017.

[1822] Takemichi Okui and Arash Yunesi. “Soft collinear effective theory for gravity”. In: *Phys. Rev. D* 97.6 (2018), p. 066011.

[1823] Timothy Cohen et al. “Navigating Collinear Superspace”. In: *JHEP* 02 (2020), p. 146.

[1824] Sabyasachi Chakraborty, Takemichi Okui, and Arash Yunesi. “Topics in soft collinear effective theory for gravity: The diffeomorphism invariant Wilson lines and reparametrization invariance”. In: *Phys. Rev. D* 101.6 (2020), p. 066019.

[1825] Martin Beneke, Patrick Hager, and Robert Szafron. “Soft-collinear gravity beyond the leading power”. In: *JHEP* 03 (2022), p. 080.

[1826] Jui-yu Chiu et al. “Electroweak Sudakov corrections using effective field theory”. In: *Phys. Rev. Lett.* 100 (2008), p. 021802.

[1827] Jui-yu Chiu et al. “Electroweak Corrections in High Energy Processes using Effective Field Theory”. In: *Phys. Rev. D* 77 (2008), p. 053004.

[1828] Jui-yu Chiu, Randall Kelley, and Aneesh V. Manohar. “Electroweak Corrections using Effective Field Theory: Applications to the LHC”. In: *Phys. Rev. D* 78 (2008), p. 073006.

[1829] Andreas Fuhrer et al. “Radiative Corrections to Longitudinal and Transverse Gauge Boson and Higgs Production”. In: *Phys. Rev. D* 81 (2010), p. 093005.

[1830] Thomas Becher and Xavier Garcia i Tormo. “Electroweak Sudakov effects in $W, Z$ and $\gamma$ production at large transverse momentum”. In: *Phys. Rev. D* 88.1 (2013), p. 013009.

[1831] Aneesh V. Manohar and Wouter J. Waalewijn. “Electroweak Logarithms in Inclusive Cross Sections”. In: *JHEP* 08 (2018), p. 137.

[1832] Bartosz Fornal, Aneesh V. Manohar, and Wouter J. Waalewijn. "Electroweak Gauge Boson Parton Distribution Functions". In: *JHEP* 05 (2018), p. 106.

[1833] Matthew Baumgart, Ira Z. Rothstein, and Varun Vaidya. "Calculating the Annihilation Rate of Weakly Interacting Massive Particles". In: *Phys. Rev. Lett.* 114 (2015), p. 211301.

[1834] Martin Bauer et al. "Soft Collinear Effective Theory for Heavy WIMP Annihilation". In: *JHEP* 01 (2015). Ed. by Monica Tecchio and Daniel Levin, p. 099.

[1835] Grigory Ovanesyan, Tracy R. Slatyer, and Iain W. Stewart. "Heavy Dark Matter Annihilation from Effective Field Theory". In: *Phys. Rev. Lett.* 114.21 (2015), p. 211302.

[1836] Matthew Baumgart et al. "Resummed Photon Spectra for WIMP Annihilation". In: *JHEP* 03 (2018), p. 117.

[1837] Martin Beneke, Stefan Lederer, and Kai Urban. "Sommerfeld enhancement of resonant dark matter annihilation". In: (Sept. 2022).

[1838] Oleksandr Tomalak et al. "Theory of QED radiative corrections to neutrino scattering at accelerator energies". In: (Apr. 2022).

[1839] Oleksandr Tomalak et al. "QED radiative corrections for accelerator neutrinos". In: *Nature Commun.* 13.1 (2022), p. 5286.

[1840] Aneesh V. Manohar et al. "Reparameterization invariance for collinear operators". In: *Phys. Lett.* B539 (2002), pp. 59–66.

[1841] Junegone Chay and Chul Kim. "Collinear effective theory at subleading order and its application to heavy-light currents". In: *Phys. Rev.* D65 (2002), p. 114016.

[1842] Christian W. Bauer, Dan Pirjol, and Iain W. Stewart. "Power counting in the soft collinear effective theory". In: *Phys. Rev.* D66 (2002), p. 054005.

[1843] Prem P. Srivastava and Stanley J. Brodsky. "Light front quantized QCD in covariant gauge". In: *Phys. Rev. D* 61 (2000), p. 025013.

[1844] Claudio Marcantonini and Iain W. Stewart. "Reparameterization Invariant Collinear Operators". In: *Phys.Rev.* D79 (2009), p. 065028.

[1845] Ian Moult et al. "Employing Helicity Amplitudes for Resummation". In: *Phys. Rev.* D93.9 (2016), p. 094003.

[1846] Daniel W. Kolodrubetz, Ian Moult, and Iain W. Stewart. "Building Blocks for Subleading Helicity Operators". In: *JHEP* 05 (2016), p. 139.

[1847] Ian Moult, Iain W. Stewart, and Gherardo Vita. "A subleading operator basis and matching for $gg \to H$". In: *JHEP* 07 (2017), p. 067.

[1848] Ilya Feige et al. "A Complete Basis of Helicity Operators for Subleading Factorization". In: *JHEP* 11 (2017), p. 142.

[1849] Arindam Bhattacharya et al. "Helicity Methods for High Multiplicity Subleading Soft and Collinear Limits". In: *JHEP* 05 (2019), p. 192.

[1850] Aneesh V. Manohar and Iain W. Stewart. "The zero-bin and mode factorization in quantum field theory". In: *Phys. Rev.* D76 (2007), p. 074002.

[1851] Christian W. Bauer, Bjorn O. Lange, and Grigory Ovanesyan. "On Glauber modes in Soft-Collinear Effective Theory". In: *JHEP* 07 (2011), p. 077.

[1852] Matthew D. Schwartz, Kai Yan, and Hua Xing Zhu. "Collinear factorization violation and effective field theory". In: *Phys. Rev. D* 96.5 (2017), p. 056005.

[1853] John C. Collins, Davison E. Soper, and George F. Sterman. "Factorization for One Loop Corrections in the Drell-Yan Process". In: *Nucl. Phys. B* 223 (1983), pp. 381–421.

[1854] Geoffrey T. Bodwin. "Factorization of the Drell-Yan Cross-Section in Perturbation Theory". In: *Phys. Rev. D* 31 (1985). [Erratum: Phys.Rev.D 34, 3932 (1986)], p. 2616.

[1855] John Collins and Jian-Wei Qiu. "$k_T$ factorization is violated in production of high-transverse-momentum particles in hadron-hadron collisions". In: *Phys. Rev. D* 75 (2007), p. 114014.

[1856] Ted C. Rogers and Piet J. Mulders. "No Generalized TMD-Factorization in Hadro-Production of High Transverse Momentum Hadrons". In: *Phys. Rev. D* 81 (2010), p. 094006.

[1857] Stefano Catani, Daniel de Florian, and German Rodrigo. "Space-like (versus time-like) collinear limits in QCD: Is factorization violated?" In: *JHEP* 07 (2012), p. 026.

[1858] Matthew D. Schwartz, Kai Yan, and Hua Xing Zhu. "Factorization Violation and Scale Invariance". In: *Phys. Rev. D* 97.9 (2018), p. 096017.

[1859] Matthew Baumgart et al. "Breakdown of the naive parton model in super-weak scale collisions". In: *Phys. Rev. D* 100.9 (2019), p. 096008.

[1860] Thomas Becher, Matthias Neubert, and Ding Yu Shao. "Resummation of Super-Leading Logarithms". In: *Phys. Rev. Lett.* 127.21 (2021), p. 212002.

[1861] M. Beneke and T. Feldmann. "Multipole-expanded soft-collinear effective theory with non-abelian

gauge symmetry". In: *Phys. Lett.* B553 (2003), pp. 267–276.

[1862] Christian W. Bauer, Dan Pirjol, and Iain W. Stewart. "On Power suppressed operators and gauge invariance in SCET". In: *Phys. Rev.* D68 (2003), p. 034021.

[1863] Richard J. Hill and Matthias Neubert. "Spectator interactions in soft-collinear effective theory." In: *Nucl. Phys.* B657 (2003), pp. 229–256.

[1864] Dan Pirjol and Iain W. Stewart. "A Complete basis for power suppressed collinear ultrasoft operators". In: *Phys. Rev.* D67 (2003). [Erratum: Phys. Rev.D69,019903(2004)], p. 094005.

[1865] S. W. Bosch et al. "Factorization and Sudakov resummation in leptonic radiative B decay". In: *Phys. Rev. D* 67 (2003), p. 094014.

[1866] M. Beneke, Y. Kiyo, and D. s. Yang. "Loop corrections to subleading heavy quark currents in SCET". In: *Nucl. Phys. B* 692 (2004), pp. 232–248.

[1867] R. J. Hill et al. "Sudakov resummation for subleading SCET currents and heavy-to-light form-factors". In: *JHEP* 07 (2004), p. 081.

[1868] Andre H. Hoang and Iain W. Stewart. "Designing gapped soft functions for jet production". In: *Phys. Lett.* B660 (2008), pp. 483–493.

[1869] Riccardo Abbate et al. "Precision Thrust Cumulant Moments at $N^3$LL". In: *Phys. Rev.* D86 (2012), p. 094002.

[1870] Thomas Becher, Matthias Neubert, and Daniel Wilhelm. "Electroweak Gauge-Boson Production at Small $q_T$: Infrared Safety from the Collinear Anomaly". In: *JHEP* 02 (2012), p. 124.

[1871] Miguel G. Echevarria, Ahmad Idilbi, and Ignazio Scimemi. "Factorization Theorem For Drell-Yan At Low $q_T$ And Transverse Momentum Distributions On-The-Light-Cone". In: *JHEP* 07 (2012), p. 002.

[1872] Jui-Yu Chiu et al. "A Formalism for the Systematic Treatment of Rapidity Logarithms in Quantum Field Theory". In: *JHEP* 05 (2012), p. 084.

[1873] Ye Li, Duff Neill, and Hua Xing Zhu. "An exponential regulator for rapidity divergences". In: *Nucl. Phys. B* 960 (2020), p. 115193.

[1874] Jui-yu Chiu et al. "The Rapidity Renormalization Group". In: *Phys. Rev. Lett.* 108 (2012), p. 151601.

[1875] Andrew J. Larkoski et al. "Soft Drop". In: *JHEP* 05 (2014), p. 146.

[1876] Mrinal Dasgupta et al. "Towards an understanding of jet substructure". In: *JHEP* 09 (2013), p. 029.

[1877] Zhong-Bo Kang, Felix Ringer, and Ivan Vitev. "The semi-inclusive jet function in SCET and small radius resummation for inclusive jet production". In: *JHEP* 10 (2016), p. 125.

[1878] Zhong-Bo Kang et al. "The groomed and ungroomed jet mass distribution for inclusive jet production at the LHC". In: *JHEP* 10 (2018), p. 137.

[1879] Andre H. Hoang et al. "Extracting a Short Distance Top Mass with Light Grooming". In: *Phys. Rev.* D100.7 (2019), p. 074021.

[1880] Andrew J. Larkoski, Ian Moult, and Duff Neill. "Factorization and Resummation for Groomed Multi-Prong Jet Shapes". In: *JHEP* 02 (2018), p. 144.

[1881] Yiannis Makris, Duff Neill, and Varun Vaidya. "Probing Transverse-Momentum Dependent Evolution With Groomed Jets". In: *JHEP* 07 (2018), p. 167.

[1882] Jeremy Baron, Simone Marzani, and Vincent Theeuwes. "Soft-Drop Thrust". In: *JHEP* 08 (2018). [Erratum: JHEP 05, 056 (2019)], p. 105.

[1883] Yiannis Makris and Varun Vaidya. "Transverse Momentum Spectra at Threshold for Groomed Heavy Quark Jets". In: *JHEP* 10 (2018), p. 019.

[1884] Zhong-Bo Kang et al. "Soft drop groomed jet angularities at the LHC". In: *Phys. Lett. B* 793 (2019), pp. 41–47.

[1885] Christopher Lee, Prashant Shrivastava, and Varun Vaidya. "Predictions for energy correlators probing substructure of groomed heavy quark jets". In: *JHEP* 09 (2019), p. 045.

[1886] Daniel Gutierrez-Reyes et al. "Probing Transverse-Momentum Distributions With Groomed Jets". In: *JHEP* 08 (2019), p. 161.

[1887] Yang-Ting Chien and Iain W. Stewart. "Collinear Drop". In: *JHEP* 06 (2020), p. 064.

[1888] Zhong-Bo Kang et al. "The soft drop groomed jet radius at NLL". In: *JHEP* 02 (2020), p. 054.

[1889] Pedro Cal et al. "Calculating the angle between jet axes". In: *JHEP* 04 (2020), p. 211.

[1890] Pedro Cal et al. "Jet energy drop". In: *JHEP* 11 (2020), p. 012.

[1891] Aditya Pathak et al. "EFT for Soft Drop Double Differential Cross Section". In: *JHEP* 04 (2021), p. 032.

[1892] Yiannis Makris. "Revisiting the role of grooming in DIS". In: *Phys. Rev. D* 103.5 (2021), p. 054005.

[1893] Pedro Cal et al. "The soft drop momentum sharing fraction zg beyond leading-logarithmic accuracy". In: *Phys. Lett. B* 833 (2022), p. 137390.

[1894] Piotr Pietrulewicz et al. "Variable Flavor Number Scheme for Final State Jets in Thrust". In: *Phys. Rev. D* 90.11 (2014), p. 114001.

[1895] Yang-Ting Chien and Matthew D. Schwartz. "Resummation of heavy jet mass and comparison to LEP data". In: *JHEP* 08 (2010), p. 058.

[1896] Andr00E9 H. Hoang et al. "$C$-parameter distribution at $\mathrm{N}^3$LL including power corrections". In: *Phys. Rev.* D91.9 (2015), p. 094017.

[1897] Markus A. Ebert, Bernhard Mistlberger, and Gherardo Vita. "The Energy-Energy Correlation in the back-to-back limit at $\mathrm{N}^3$LO and $\mathrm{N}^3$LL"'. In: *JHEP* 08 (2021), p. 022.

[1898] Vicent Mateu and Germán Rodrigo. "Oriented Event Shapes at $\mathrm{N}^3$LL $+O(\alpha_S^2)$". In: *JHEP* 11 (2013), p. 030.

[1899] Adam Kardos, Andrew J. Larkoski, and Zoltán Trócsányi. "Groomed jet mass at high precision". In: *Phys. Lett. B* 809 (2020), p. 135704.

[1900] Brad Bachu et al. "Boosted top quarks in the peak region with NL3L resummation". In: *Phys. Rev. D* 104.1 (2021), p. 014026.

[1901] Zhong-Bo Kang, Sonny Mantry, and Jian-Wei Qiu. "N-Jettiness as a Probe of Nuclear Dynamics". In: *Phys.Rev.* D86 (2012), p. 114011.

[1902] Daekyoung Kang, Christopher Lee, and Iain W. Stewart. "1-Jettiness in DIS: Measuring 2 Jets in 3 Ways". In: *PoS* DIS2013 (2013), p. 158.

[1903] Zhong-Bo Kang, Xiaohui Liu, and Sonny Mantry. "1-jettiness DIS event shape: NNLL+NLO results". In: *Phys. Rev. D* 90.1 (2014), p. 014041.

[1904] Daekyoung Kang, Christopher Lee, and Iain W. Stewart. "DIS Event Shape at N3LL". In: *PoS* DIS2015 (2015), p. 142.

[1905] Thomas Becher and Tobias Neumann. "Fiducial $q_T$ resummation of color-singlet processes at $\mathrm{N}^3$LL+NNLO". In: *JHEP* 03 (2021), p. 199.

[1906] Markus A. Ebert et al. "Drell-Yan $q_T$ resummation of fiducial power corrections at $\mathrm{N}^3$LL". In: *JHEP* 04 (2021), p. 102.

[1907] Duff Neill, Ira Z. Rothstein, and Varun Vaidya. "The Higgs Transverse Momentum Distribution at NNLL and its Theoretical Errors". In: *JHEP* 12 (2015), p. 097.

[1908] Xuan Chen et al. "Precise QCD Description of the Higgs Boson Transverse Momentum Spectrum". In: *Phys. Lett. B* 788 (2019), pp. 425–430.

[1909] Georgios Billis et al. "Higgs pT Spectrum and Total Cross Section with Fiducial Cuts at Third Resummed and Fixed Order in QCD". In: *Phys. Rev. Lett.* 127.7 (2021), p. 072001.

[1910] Markus A. Ebert, Johannes K. L. Michel, and Frank J. Tackmann. "Resummation Improved Rapidity Spectrum for Gluon Fusion Higgs Production". In: *JHEP* 05 (2017), p. 088.

[1911] Carola F. Berger et al. "Higgs Production with a Central Jet Veto at NNLL+NNLO". In: *JHEP* 1104 (2011), p. 092.

[1912] Thomas Becher and Matthias Neubert. "Factorization and NNLL Resummation for Higgs Production with a Jet Veto". In: *JHEP* 07 (2012), p. 108.

[1913] Frank J. Tackmann, Jonathan R. Walsh, and Saba Zuberi. "Resummation Properties of Jet Vetoes at the LHC". In: *Phys. Rev. D* 86 (2012), p. 053011.

[1914] Thomas Becher, Matthias Neubert, and Lorena Rothen. "Factorization and $N^3LL_p$+NNLO predictions for the Higgs cross section with a jet veto". In: *JHEP* 10 (2013), p. 125.

[1915] Iain W. Stewart et al. "Jet $p_T$ resummation in Higgs production at $NNLL' + NNLO$". In: *Phys. Rev.* D89.5 (2014), p. 054001.

[1916] Johannes K. L. Michel, Piotr Pietrulewicz, and Frank J. Tackmann. "Jet Veto Resummation with Jet Rapidity Cuts". In: *JHEP* 04 (2019), p. 142.

[1917] Claude Duhr, Bernhard Mistlberger, and Gherardo Vita. "The Four-Loop Rapidity Anomalous Dimension and Event Shapes to Fourth Logarithmic Order". In: (May 2022).

[1918] Bakul Agarwal et al. "Four-loop collinear anomalous dimensions in QCD and N=4 super Yang-Mills". In: *Phys. Lett. B* 820 (2021), p. 136503.

[1919] Roman N. Lee et al. "Quark and Gluon Form Factors in Four-Loop QCD". In: *Phys. Rev. Lett.* 128.21 (2022), p. 212002.

[1920] Ian Moult, Hua Xing Zhu, and Yu Jiao Zhu. "The four loop QCD rapidity anomalous dimension". In: *JHEP* 08 (2022), p. 280.

[1921] Johannes M. Henn, Gregory P. Korchemsky, and Bernhard Mistlberger. "The full four-loop cusp anomalous dimension in $\mathcal{N} = 4$ super Yang-Mills and QCD". In: *JHEP* 04 (2020), p. 018.

[1922] F. Herzog et al. "Five-loop contributions to low-N non-singlet anomalous dimensions in QCD". In: *Phys. Lett. B* 790 (2019), pp. 436–443.

[1923] Ming-xing Luo et al. "Quark Transverse Parton Distribution at the Next-to-Next-to-Next-to-Leading Order". In: *Phys. Rev. Lett.* 124.9 (2020), p. 092001.

[1924] Markus A. Ebert, Bernhard Mistlberger, and Gherardo Vita. “Transverse momentum dependent PDFs at $N^3LO$”. In: *JHEP* 09 (2020), p. 146.

[1925] Ming-xing Luo et al. “Unpolarized quark and gluon TMD PDFs and FFs at $N^3LO$”. In: *JHEP* 06 (2021), p. 115.

[1926] Markus A. Ebert, Bernhard Mistlberger, and Gherardo Vita. “TMD fragmentation functions at $N^3LO$”. In: *JHEP* 07 (2021), p. 121.

[1927] Aneesh V. Manohar. “Deep inelastic scattering as $x \to 1$ using soft collinear effective theory”. In: *Phys. Rev. D* 68 (2003), p. 114019.

[1928] Ahmad Idilbi et al. “Threshold resummation for Higgs production in effective field theory”. In: *Phys. Rev. D* 73 (2006), p. 077501.

[1929] Ahmad Idilbi, Xiang-dong Ji, and Feng Yuan. “Resummation of threshold logarithms in effective field theory for DIS, Drell-Yan and Higgs production”. In: *Nucl. Phys. B* 753 (2006), pp. 42–68.

[1930] Thomas Becher, Matthias Neubert, and Ben D. Pecjak. “Factorization and Momentum-Space Resummation in Deep-Inelastic Scattering”. In: *JHEP* 01 (2007), p. 076.

[1931] Thomas Becher, Christian Lorentzen, and Matthew D. Schwartz. “Resummation for W and Z production at large pT”. In: *Phys. Rev. Lett.* 108 (2012), p. 012001.

[1932] Thomas Becher, Guido Bell, and Matthias Neubert. “Factorization and Resummation for Jet Broadening”. In: *Phys. Lett. B* 704 (2011), pp. 276–283.

[1933] Thomas Becher, Christian Lorentzen, and Matthew D. Schwartz. “Precision Direct Photon and $W$-Boson Spectra at High $p_T$ and Comparison to LHC Data”. In: *Phys. Rev. D* 86 (2012), p. 054026.

[1934] S. Dawson, Ian M. Lewis, and Mao Zeng. “Threshold resummed and approximate next-to-next-to-leading order results for $W^+W^-$ pair production at the LHC”. In: *Phys. Rev. D* 88.5 (2013), p. 054028.

[1935] Valentin Ahrens et al. “Renormalization-Group Improved Predictions for Top-Quark Pair Production at Hadron Colliders”. In: *JHEP* 09 (2010), p. 097.

[1936] Hua Xing Zhu et al. “Transverse-momentum resummation for top-quark pairs at hadron colliders”. In: *Phys. Rev. Lett.* 110.8 (2013), p. 082001.

[1937] Yang-Ting Chien et al. “Resummation of Jet Mass at Hadron Colliders”. In: *Phys. Rev.* D87.1 (2013), p. 014010.

[1938] S. Dawson et al. “Resummation Effects in Vector-Boson and Higgs Associated Production”. In: *Phys. Rev. D* 86 (2012), p. 074007.

[1939] Sean Fleming and Ou Z. Labun. “Rapidity Divergences and Deep Inelastic Scattering in the Endpoint Region”. In: *Phys. Rev. D* 91.9 (2015), p. 094011.

[1940] Thomas Becher and Guido Bell. “NNLL Resummation for Jet Broadening”. In: *JHEP* 11 (2012), p. 126.

[1941] Yang-Ting Chien and Ivan Vitev. “Jet Shape Resummation Using Soft-Collinear Effective Theory”. In: *JHEP* 12 (2014), p. 061.

[1942] Simone Alioli et al. “Drell-Yan production at NNLL +NNLO matched to parton showers”. In: *Phys. Rev.* D92.9 (2015), p. 094020.

[1943] Sean Fleming and Ou Z. Labun. “Rapidity regulators in the semi-inclusive deep inelastic scattering and Drell-Yan processes”. In: *Phys. Rev. D* 95.11 (2017), p. 114020.

[1944] Zhong-Bo Kang, Felix Ringer, and Wouter J. Waalewijn. “The Energy Distribution of Subjets and the Jet Shape”. In: *JHEP* 07 (2017), p. 064.

[1945] Daniel Gutierrez-Reyes et al. “Transverse momentum dependent distributions with jets”. In: *Phys. Rev. Lett.* 121.16 (2018), p. 162001.

[1946] Andrew Hornig et al. “Transverse Vetoes with Rapidity Cutoff in SCET”. In: *JHEP* 12 (2017), p. 043.

[1947] Edmond L. Berger, Jun Gao, and Hua Xing Zhu. “Differential Distributions for t-channel Single Top-Quark Production and Decay at Next-to-Next-to-Leading Order in QCD”. In: *JHEP* 11 (2017), p. 158.

[1948] Guido Bell et al. “$e^+e^-$ angularity distributions at NNLL′ accuracy”. In: *JHEP* 01 (2019), p. 147.

[1949] Christian W. Bauer and Pier Francesco Monni. “A numerical formulation of resummation in effective field theory”. In: *JHEP* 02 (2019), p. 185.

[1950] Gillian Lustermans et al. “Joint two-dimensional resummation in $q_T$ and 0-jettiness at NNLL”. In: *JHEP* 03 (2019), p. 124.

[1951] Daniel Gutierrez-Reyes et al. “Transverse momentum dependent distributions in $e^+e^-$ and semi-inclusive deep-inelastic scattering using jets”. In: *JHEP* 10 (2019), p. 031.

[1952] Christian W. Bauer and Pier Francesco Monni. “A formalism for the resummation of non-factorizable observables in SCET”. In: *JHEP* 05 (2020), p. 005.

[1953] Alessandro Broggio et al. “Top-quark pair hadroproduction in association with a heavy boson at

NLO+NNLL including EW corrections". In: *JHEP* 08 (2019), p. 039.

[1954] Alejandro Bris, Vicent Mateu, and Moritz Preisser. "Massive event-shape distributions at $N^2LL$". In: *JHEP* 09 (2020), p. 132.

[1955] Shireen Gangal et al. "Higgs Production at NNLL′+NNLO using Rapidity Dependent Jet Vetoes". In: *JHEP* 05 (2020), p. 054.

[1956] Yiannis Makris, Felix Ringer, and Wouter J. Waalewijn. "Joint thrust and TMD resummation in electron-positron and electron-proton collisions". In: *JHEP* 02 (2021), p. 070.

[1957] Lin Dai, Chul Kim, and Adam K. Leibovich. "Heavy quark jet production near threshold". In: *JHEP* 09 (2021), p. 148.

[1958] Kees Benkendorfer and Andrew J. Larkoski. "Grooming at the cusp: all-orders predictions for the transition region of jet groomers". In: *JHEP* 11 (2021), p. 188.

[1959] Yang-Ting Chien et al. "Precision boson-jet azimuthal decorrelation at hadron colliders". In: (May 2022).

[1960] Martin Beneke et al. "Anomalous dimension of subleading-power N-jet operators". In: *JHEP* 03 (2018), p. 001.

[1961] Cyuan-Han Chang, Iain W. Stewart, and Gherardo Vita. "A Subleading Power Operator Basis for the Scalar Quark Current". In: *JHEP* 04 (2018), p. 041.

[1962] Simon M. Freedman and Raymond Goerke. "Renormalization of Subleading Dijet Operators in Soft-Collinear Effective Theory". In: *Phys. Rev. D* 90.11 (2014), p. 114010.

[1963] Raymond Goerke and Matthew Inglis-Whalen. "Renormalization of dijet operators at order $1/Q^2$ in soft-collinear effective theory". In: *JHEP* 05 (2018), p. 023.

[1964] Martin Beneke et al. "Anomalous dimension of subleading-power $N$-jet operators. Part II". In: *JHEP* 11 (2018), p. 112.

[1965] Ian Moult et al. "First Subleading Power Resummation for Event Shapes". In: *JHEP* 08 (2018), p. 013.

[1966] Markus A. Ebert et al. "Subleading power rapidity divergences and power corrections for $q_T$". In: *JHEP* 04 (2019), p. 123.

[1967] Ian Moult, Iain W. Stewart, and Gherardo Vita. "Subleading Power Factorization with Radiative Functions". In: *JHEP* 11 (2019), p. 153.

[1968] Ze Long Liu et al. "Renormalization and Scale Evolution of the Soft-Quark Soft Function". In: *JHEP* 07 (2020), p. 104.

[1969] Andrew J. Larkoski, Duff Neill, and Iain W. Stewart. "Soft Theorems from Effective Field Theory". In: *JHEP* 06 (2015), p. 077.

[1970] Ze Long Liu and Matthias Neubert. "Factorization at subleading power and endpoint-divergent convolutions in $h \to \gamma\gamma$ decay". In: *JHEP* 04 (2020), p. 033.

[1971] Ze Long Liu et al. "Factorization at subleading power, Sudakov resummation, and endpoint divergences in soft-collinear effective theory". In: *Phys. Rev. D* 104.1 (2021), p. 014004.

[1972] Matthew Inglis-Whalen et al. "Factorization of power corrections in the Drell-Yan process in EFT". In: *Phys. Rev. D* 104.7 (2021), p. 076018.

[1973] Michael Luke, Jyotirmoy Roy, and Aris Spourdalakis. "Factorization at subleading power in Deep Inelastic Scattering in the $x \to 1$ limit". In: (Oct. 2022).

[1974] Ian Moult et al. "The Soft Quark Sudakov". In: *JHEP* 05 (2020), p. 089.

[1975] M. Beneke et al. "Next-to-leading power endpoint factorization and resummation for off-diagonal "gluon" thrust". In: *JHEP* 07 (2022), p. 144.

[1976] Martin Beneke et al. "Leading-logarithmic threshold resummation of the Drell-Yan process at next-to-leading power". In: *JHEP* 03 (2019), p. 043.

[1977] Martin Beneke et al. "Threshold factorization of the Drell-Yan process at next-to-leading power". In: *JHEP* 07 (2020), p. 078.

[1978] Martin Beneke et al. "Leading-logarithmic threshold resummation of Higgs production in gluon fusion at next-to-leading power". In: *JHEP* 01 (2020), p. 094.

[1979] Ian Moult, Gherardo Vita, and Kai Yan. "Subleading power resummation of rapidity logarithms: the energy-energy correlator in $\mathcal{N} = 4$ SYM". In: *JHEP* 07 (2020), p. 005.

[1980] Radja Boughezal et al. "$W$-boson production in association with a jet at next-to-next-to-leading order in perturbative QCD". In: *Phys. Rev. Lett.* 115.6 (2015), p. 062002.

[1981] Jonathan Gaunt et al. "N-jettiness Subtractions for NNLO QCD Calculations". In: *JHEP* 09 (2015), p. 058.

[1982] Ian Moult et al. "Subleading Power Corrections for N-Jettiness Subtractions". In: (2016).

[1983] Radja Boughezal, Xiaohui Liu, and Frank Petriello. "Power Corrections in the N-jettiness Subtraction Scheme". In: *JHEP* 03 (2017), p. 160.

[1984] Ian Moult et al. "N -jettiness subtractions for $gg \to H$ at subleading power". In: *Phys. Rev.* D97.1 (2018), p. 014013.

[1985] Radja Boughezal, Andrea Isgrò, and Frank Petriello. "Next-to-leading-logarithmic power corrections for $N$-jettiness subtraction in color-singlet production". In: *Phys. Rev. D* 97.7 (2018), p. 076006.

[1986] Markus A. Ebert et al. "Power Corrections for N-Jettiness Subtractions at $\mathcal{O}(\alpha_s)$". In: *JHEP* 12 (2018), p. 084.

[1987] Georgios Billis et al. "A toolbox for $q_T$ and 0-jettiness subtractions at N$^3$LO". In: *Eur. Phys. J. Plus* 136.2 (2021), p. 214.

[1988] Markus A. Ebert and Frank J. Tackmann. "Impact of isolation and fiducial cuts on q$_T$ and N-jettiness subtractions". In: *JHEP* 03 (2020), p. 158.

[1989] Radja Boughezal, Andrea Isgrò, and Frank Petriello. "Next-to-leading power corrections to $V + 1$ jet production in $N$-jettiness subtraction". In: *Phys. Rev. D* 101.1 (2020), p. 016005.

[1990] Gillian Lustermans, Johannes K. L. Michel, and Frank J. Tackmann. "Generalized Threshold Factorization with Full Collinear Dynamics". In: (Aug. 2019).

[1991] Markus A. Ebert, Bernhard Mistlberger, and Gherardo Vita. "Collinear expansion for color singlet cross sections". In: *JHEP* 09 (2020), p. 181.

[1992] Randall Kelley et al. "The two-loop hemisphere soft function". In: *Phys. Rev. D* 84 (2011), p. 045022.

[1993] Andrew Hornig et al. "Non-global Structure of the $\mathcal{O}(\alpha_s^2)$ Dijet Soft Function". In: *JHEP* 08 (2011). [Erratum: JHEP 10, 101 (2017)], p. 054.

[1994] Andrew Hornig et al. "Double Non-Global Logarithms In-N-Out of Jets". In: *JHEP* 01 (2012), p. 149.

[1995] Randall Kelley et al. "Jet Mass with a Jet Veto at Two Loops and the Universality of Non-Global Structure". In: *Phys. Rev. D* 86 (2012), p. 054017.

[1996] Matthew D. Schwartz and Hua Xing Zhu. "Non-global logarithms at three loops, four loops, five loops, and beyond". In: *Phys. Rev. D* 90.6 (2014), p. 065004.

[1997] Andrew J. Larkoski, Ian Moult, and Duff Neill. "Non-Global Logarithms, Factorization, and the Soft Substructure of Jets". In: *JHEP* 09 (2015), p. 143.

[1998] Duff Neill. "The Edge of Jets and Subleading Non-Global Logs". In: (Aug. 2015).

[1999] Andrew J. Larkoski and Ian Moult. "Nonglobal correlations in collider physics". In: *Phys. Rev. D* 93.1 (2016), p. 014012.

[2000] Thomas Becher et al. "Factorization and Resummation for Jet Processes". In: *JHEP* 11 (2016). [Erratum: JHEP 05, 154 (2017)], p. 019.

[2001] Andrew J. Larkoski, Ian Moult, and Duff Neill. "The Analytic Structure of Non-Global Logarithms: Convergence of the Dressed Gluon Expansion". In: *JHEP* 11 (2016), p. 089.

[2002] Duff Neill. "The Asymptotic Form of Non-Global Logarithms, Black Disc Saturation, and Gluonic Deserts". In: *JHEP* 01 (2017), p. 109.

[2003] Duff Neill. "Non-Global and Clustering Effects for Groomed Multi-Prong Jet Shapes". In: *JHEP* 02 (2019), p. 114.

[2004] Duff Neill and Felix Ringer. "Soft Fragmentation on the Celestial Sphere". In: *JHEP* 06 (2020), p. 086.

[2005] Duff Neill, Felix Ringer, and Nobuo Sato. "Leading jets and energy loss". In: *JHEP* 07 (2021), p. 041.

[2006] Christopher Lee and George F. Sterman. "Momentum Flow Correlations from Event Shapes: Factorized Soft Gluons and Soft-Collinear Effective Theory". In: *Phys. Rev. D* 75 (2007), p. 014022.

[2007] Vicent Mateu, Iain W. Stewart, and Jesse Thaler. "Power Corrections to Event Shapes with Mass-Dependent Operators". In: *Phys. Rev. D* 87.1 (2013), p. 014025.

[2008] Iain W. Stewart, Frank J. Tackmann, and Wouter J. Waalewijn. "Dissecting Soft Radiation with Factorization". In: *Phys. Rev. Lett.* 114.9 (2015), p. 092001.

[2009] Massimiliano Procura and Iain W. Stewart. "Quark Fragmentation within an Identified Jet". In: *Phys. Rev. D* 81 (2010). [Erratum: Phys.Rev.D 83, 039902 (2011)], p. 074009.

[2010] Ambar Jain, Massimiliano Procura, and Wouter J. Waalewijn. "Parton Fragmentation within an Identified Jet at NNLL". In: *JHEP* 05 (2011), p. 035.

[2011] Massimiliano Procura and Wouter J. Waalewijn. "Fragmentation in Jets: Cone and Threshold Effects". In: *Phys. Rev. D* 85 (2012), p. 114041.

[2012] Ambar Jain et al. "Fragmentation with a Cut on Thrust: Predictions for B-factories". In: *Phys. Rev. D* 87.7 (2013), p. 074013.

[2013] Christian W. Bauer and Emanuele Mereghetti. "Heavy Quark Fragmenting Jet Functions". In: *JHEP* 04 (2014), p. 051.

[2014] Matthew Baumgart et al. "Probing Quarkonium Production Mechanisms with Jet Substructure". In: *JHEP* 11 (2014), p. 003.

[2015] Mathias Ritzmann and Wouter J. Waalewijn. "Fragmentation in Jets at NNLO". In: *Phys. Rev. D* 90.5 (2014), p. 054029.

[2016] Reggie Bain et al. "Analytic and Monte Carlo Studies of Jets with Heavy Mesons and Quarkonia". In: *JHEP* 06 (2016), p. 121.

[2017] Reggie Bain, Yiannis Makris, and Thomas Mehen. "Transverse Momentum Dependent Fragmenting Jet Functions with Applications to Quarkonium Production". In: *JHEP* 11 (2016), p. 144.

[2018] Lin Dai, Chul Kim, and Adam K. Leibovich. "Fragmentation of a Jet with Small Radius". In: *Phys. Rev. D* 94.11 (2016), p. 114023.

[2019] Lin Dai, Chul Kim, and Adam K. Leibovich. "Fragmentation to a jet in the large $z$ limit". In: *Phys. Rev. D* 95.7 (2017), p. 074003.

[2020] Lin Dai, Chul Kim, and Adam K. Leibovich. "Heavy Quark Jet Fragmentation". In: *JHEP* 09 (2018), p. 109.

[2021] Sean Fleming, Yiannis Makris, and Thomas Mehen. "An effective field theory approach to quarkonium at small transverse momentum". In: *JHEP* 04 (2020), p. 122.

[2022] Aneesh V. Manohar and Wouter J. Waalewijn. "A QCD Analysis of Double Parton Scattering: Color Correlations, Interference Effects and Evolution". In: *Phys. Rev. D* 85 (2012), p. 114009.

[2023] Sean Fleming et al. "The Systematics of Quarkonium Production at the LHC and Double Parton Fragmentation". In: *Phys. Rev. D* 86 (2012), p. 094012.

[2024] Sean Fleming et al. "Anomalous dimensions of the double parton fragmentation functions". In: *Phys. Rev. D* 87 (2013), p. 074022.

[2025] Massimiliano Procura, Wouter J. Waalewijn, and Lisa Zeune. "Resummation of Double-Differential Cross Sections and Fully-Unintegrated Parton Distribution Functions". In: *JHEP* 02 (2015), p. 117.

[2026] Ian Moult and Hua Xing Zhu. "Simplicity from Recoil: The Three-Loop Soft Function and Factorization for the Energy-Energy Correlation". In: *JHEP* 08 (2018), p. 160.

[2027] Lance J. Dixon, Ian Moult, and Hua Xing Zhu. "Collinear limit of the energy-energy correlator". In: *Phys. Rev. D* 100.1 (2019), p. 014009.

[2028] Hao Chen et al. "Three point energy correlators in the collinear limit: symmetries, dualities and analytic results". In: *JHEP* 08.08 (2020), p. 028.

[2029] Hao Chen et al. "Rethinking jets with energy correlators: Tracks, resummation, and analytic continuation". In: *Phys. Rev. D* 102.5 (2020), p. 054012.

[2030] AnJie Gao et al. "Precision QCD Event Shapes at Hadron Colliders: The Transverse Energy-Energy Correlator in the Back-to-Back Limit". In: *Phys. Rev. Lett.* 123.6 (2019), p. 062001.

[2031] Kyle Lee, Bianka Meçaj, and Ian Moult. "Conformal Colliders Meet the LHC". In: (May 2022).

[2032] Iain W. Stewart, Frank J. Tackmann, and Wouter J. Waalewijn. "Factorization at the LHC: From PDFs to Initial State Jets". In: *Phys. Rev.* D81 (2010), p. 094035.

[2033] Iain W. Stewart, Frank J. Tackmann, and Wouter J. Waalewijn. "The Beam Thrust Cross Section for Drell-Yan at NNLL Order". In: *Phys. Rev. Lett.* 106 (2011), p. 032001.

[2034] Jesse Thaler and Ken Van Tilburg. "Identifying Boosted Objects with N-subjettiness". In: *JHEP* 03 (2011), p. 015.

[2035] Reggie Bain et al. "NRQCD Confronts LHCb Data on Quarkonium Production within Jets". In: *Phys. Rev. Lett.* 119.3 (2017), p. 032002.

[2036] Andrew J. Larkoski, Ian Moult, and Duff Neill. "Building a Better Boosted Top Tagger". In: *Phys. Rev. D* 91.3 (2015), p. 034035.

[2037] Ian Moult, Lina Necib, and Jesse Thaler. "New Angles on Energy Correlation Functions". In: *JHEP* 12 (2016), p. 153.

[2038] Pedro Cal, Jesse Thaler, and Wouter J. Waalewijn. "Power counting energy flow polynomials". In: *JHEP* 09 (2022), p. 021.

[2039] Andrew J. Larkoski, Duff Neill, and Jesse Thaler. "Jet Shapes with the Broadening Axis". In: *JHEP* 04 (2014), p. 017.

[2040] Duff Neill, Ignazio Scimemi, and Wouter J. Waalewijn. "Jet axes and universal transverse-momentum-dependent fragmentation". In: *JHEP* 04 (2017), p. 020.

[2041] Duff Neill et al. "Phenomenology with a recoil-free jet axis: TMD fragmentation and the jet shape". In: *JHEP* 01 (2019), p. 067.

[2042] Hsi-Ming Chang et al. "Calculating Track-Based Observables for the LHC". In: *Phys. Rev. Lett.* 111 (2013), p. 102002.

[2043] Hsi-Ming Chang et al. "Calculating Track Thrust with Track Functions". In: *Phys. Rev. D* 88 (2013), p. 034030.

[2044] Yibei Li et al. "Extending Precision Perturbative QCD with Track Functions". In: *Phys. Rev. Lett.* 128.18 (2022), p. 182001.

[2045] Max Jaarsma et al. "Renormalization group flows for track function moments". In: *JHEP* 06 (2022), p. 139.

[2046] Iain W. Stewart et al. "XCone: N-jettiness as an Exclusive Cone Jet Algorithm". In: *JHEP* 11 (2015), p. 072.

[2047] Iain W. Stewart and Xiaojun Yao. "Pure quark and gluon observables in collinear drop". In: *JHEP* 09 (2022), p. 120.

[2048] Jack Holguin et al. "A New Paradigm for Precision Top Physics: Weighing the Top with Energy Correlators". In: (Jan. 2022). arXiv:2201.08393

[2049] Xiaohui Liu and Hua Xing Zhu. "The Nucleon Energy Correlators". In: (Sept. 2022).

[2050] Joseph I. Kapusta and Charles Gale. *Finite-temperature field theory: Prinicples and Applications. Cambridge Monographs on Mathematical Physics.* 2nd ed. Cambridge monographs on mathematical physics. Previous edition: 1989. Cambridge: Cambridge University Press, 2006.

[2051] Jacopo Ghiglieri et al. "Perturbative Thermal QCD: Formalism and Applications". In: *Phys. Rept.* 880 (2020), pp. 1–73.

[2052] Michael Strickland. *Relativistic Quantum Field Theory, Volume 3*. 2053-2571. Institute of Physics (Morgan & Claypool), 2019.

[2053] Andrei D. Linde. "Infrared Problem in Thermodynamics of the Yang-Mills Gas". In: *Phys. Lett. B* 96 (1980), pp. 289–292.

[2054] David J. Gross, Robert D. Pisarski, and Laurence G. Yaffe. "QCD and Instantons at Finite Temperature". In: *Rev. Mod. Phys.* 53 (1981), p. 43.

[2055] Mark Abraao York and Guy D. Moore. "Second order hydrodynamic coefficients from kinetic theory". In: *Phys. Rev. D* 79 (2009), p. 054011.

[2056] K. Farakos et al. "3-d physics and the electroweak phase transition: A Framework for lattice Monte Carlo analysis". In: *Nucl. Phys. B* 442 (1995), pp. 317–363.

[2057] Pablo Navarrete and York Schröder. "Tackling the infamous $g^6$ term of the QCD pressure". In: *16th DESY Workshop on Elementary Particle Physics: Loops and Legs in Quantum Field Theory 2022*. July 2022.

[2058] Eric Braaten and Robert D. Pisarski. "Soft Amplitudes in Hot Gauge Theories: A General Analysis". In: *Nucl. Phys. B* 337 (1990), pp. 569–634.

[2059] Eric Braaten and Robert D. Pisarski. "Resummation and Gauge Invariance of the Gluon Damping Rate in Hot QCD". In: *Phys. Rev. Lett.* 64 (1990), p. 1338.

[2060] Eric Braaten and Robert D. Pisarski. "Calculation of the gluon damping rate in hot QCD". In: *Phys. Rev. D* 42 (1990), pp. 2156–2160.

[2061] Peter Brockway Arnold and Cheng-Xing Zhai. "The Three loop free energy for pure gauge QCD". In: *Phys. Rev. D* 50 (1994), pp. 7603–7623.

[2062] Peter Brockway Arnold and Cheng-xing Zhai. "The Three loop free energy for high temperature QED and QCD with fermions". In: *Phys. Rev. D* 51 (1995), pp. 1906–1918.

[2063] Cheng-xing Zhai and Boris M. Kastening. "The Free energy of hot gauge theories with fermions through g**5". In: *Phys. Rev. D* 52 (1995), pp. 7232–7246.

[2064] Eric Braaten and Agustin Nieto. "Effective field theory approach to high temperature thermodynamics". In: *Phys. Rev. D* 51 (1995), pp. 6990–7006.

[2065] Eric Braaten and Agustin Nieto. "Free energy of QCD at high temperature". In: *Phys. Rev. D* 53 (1996), pp. 3421–3437.

[2066] K. Kajantie et al. "3-D SU(N) + adjoint Higgs theory and finite temperature QCD". In: *Nucl. Phys. B* 503 (1997), pp. 357–384.

[2067] Jens O. Andersen, Eric Braaten, and Michael Strickland. "Hard thermal loop resummation of the free energy of a hot gluon plasma". In: *Phys. Rev. Lett.* 83 (1999), pp. 2139–2142.

[2068] Jens O. Andersen, Eric Braaten, and Michael Strickland. "Hard thermal loop resummation of the thermodynamics of a hot gluon plasma". In: *Phys. Rev. D* 61 (2000), p. 014017.

[2069] Jens O. Andersen, Eric Braaten, and Michael Strickland. "Hard thermal loop resummation of the free energy of a hot quark - gluon plasma". In: *Phys. Rev. D* 61 (2000), p. 074016.

[2070] J. P. Blaizot, Edmond Iancu, and A. Rebhan. "The Entropy of the QCD plasma". In: *Phys. Rev. Lett.* 83 (1999), pp. 2906–2909.

[2071] J. P. Blaizot, Edmond Iancu, and A. Rebhan. "Selfconsistent hard thermal loop thermodynamics for the quark gluon plasma". In: *Phys. Lett. B* 470 (1999), pp. 181–188.

[2072] J. P. Blaizot, Edmond Iancu, and A. Rebhan. "Approximately selfconsistent resummations for the thermodynamics of the quark gluon plasma. 1. Entropy and density". In: *Phys. Rev. D* 63 (2001), p. 065003.

[2073] J. P. Blaizot, Edmond Iancu, and A. Rebhan. "Quark number susceptibilities from HTL resummed thermodynamics". In: *Phys. Lett. B* 523 (2001), pp. 143–150.

[2074] Jens O. Andersen et al. "HTL perturbation theory to two loops". In: *Phys. Rev. D* 66 (2002), p. 085016.

[2075] Jens O. Andersen, Emmanuel Petitgirard, and Michael Strickland. “Two loop HTL thermodynamics with quarks”. In: *Phys. Rev. D* 70 (2004), p. 045001.

[2076] Jens O. Andersen, Michael Strickland, and Nan Su. “Three-loop HTL gluon thermodynamics at intermediate coupling”. In: *JHEP* 08 (2010), p. 113.

[2077] Jens O. Andersen et al. “Three-loop HTL QCD thermodynamics”. In: *JHEP* 08 (2011), p. 053.

[2078] Najmul Haque et al. “Three-loop pressure and susceptibility at finite temperature and density from hard-thermal-loop perturbation theory”. In: *Phys. Rev. D* 89.6 (2014), p. 061701.

[2079] Najmul Haque et al. “Three-loop HTLpt thermodynamics at finite temperature and chemical potential”. In: *JHEP* 05 (2014), p. 027.

[2080] Eric Braaten and Robert D. Pisarski. “Simple effective Lagrangian for hard thermal loops”. In: *Phys. Rev. D* 45.6 (1992), R1827.

[2081] Paul H. Ginsparg. “First Order and Second Order Phase Transitions in Gauge Theories at Finite Temperature”. In: *Nucl. Phys. B* 170 (1980), pp. 388–408.

[2082] Thomas Appelquist and Robert D. Pisarski. “High-Temperature Yang-Mills Theories and Three-Dimensional Quantum Chromodynamics”. In: *Phys. Rev. D* 23 (1981), p. 2305.

[2083] K. Kajantie et al. “Generic rules for high temperature dimensional reduction and their application to the standard model”. In: *Nucl. Phys. B* 458 (1996), pp. 90–136.

[2084] Jens O. Andersen et al. “N=4 supersymmetric Yang-Mills thermodynamics from effective field theory”. In: *Phys. Rev. D* 105.1 (2022), p. 015006.

[2085] Szabolcs Borsanyi et al. “The QCD equation of state with dynamical quarks”. In: *JHEP* 11 (2010), p. 077.

[2086] Szabolcs Borsanyi. “Thermodynamics of the QCD transition from lattice”. In: *Nucl. Phys. A* 904-905 (2013). Ed. by Thomas Ullrich, Bolek Wyslouch, and John W. Harris, pp. 270c–277c.

[2087] S. Borsanyi et al. “Freeze-out parameters: lattice meets experiment”. In: *Phys. Rev. Lett.* 111 (2013), p. 062005.

[2088] A. Bazavov et al. “Quark number susceptibilities at high temperatures”. In: *Phys. Rev. D* 88.9 (2013), p. 094021.

[2089] Paolo Cea, Leonardo Cosmai, and Alessandro Papa. “Critical line of 2+1 flavor QCD: Toward the continuum limit”. In: *Phys. Rev. D* 93.1 (2016), p. 014507.

[2090] A. Bazavov et al. “Chiral crossover in QCD at zero and non-zero chemical potentials”. In: *Phys. Lett. B* 795 (2019), pp. 15–21.

[2091] Najmul Haque and Michael Strickland. “Next-to-next-to leading-order hard-thermal-loop perturbation-theory predictions for the curvature of the QCD phase transition line”. In: *Phys. Rev. C* 103.3 (2021), p. 031901.

[2092] Jürgen Berges et al. “QCD thermalization: Ab initio approaches and interdisciplinary connections”. In: *Rev. Mod. Phys.* 93.3 (2021), p. 035003.

[2093] Michael Strickland. “Pseudothermalization of the quark-gluon plasma”. In: *J. Phys. Conf. Ser.* 1602.1 (2020). Ed. by Rene Bellwied et al., p. 012018.

[2094] Jean-Yves Ollitrault and Fernando G. Gardim. “Hydro overview”. In: *Nucl. Phys. A* 904-905 (2013). Ed. by Thomas Ullrich, Bolek Wyslouch, and John W. Harris, pp. 75c–82c.

[2095] Paul Romatschke and Ulrike Romatschke. *Relativistic Fluid Dynamics In and Out of Equilibrium.* Cambridge Monographs on Mathematical Physics. Cambridge University Press, May 2019.

[2096] Mubarak Alqahtani, Mohammad Nopoush, and Michael Strickland. “Relativistic anisotropic hydrodynamics”. In: *Prog. Part. Nucl. Phys.* 101 (2018), pp. 204–248.

[2097] Jorge Casalderrey-Solana and Carlos A. Salgado. “Introductory lectures on jet quenching in heavy ion collisions”. In: *Acta Phys. Polon. B* 38 (2007). Ed. by Michal Praszalowicz, Marek Kutschera, and Edward Malec, pp. 3731–3794.

[2098] Jorge Casalderrey-Solana and Derek Teaney. “Heavy quark diffusion in strongly coupled N=4 Yang-Mills”. In: *Phys. Rev. D* 74 (2006), p. 085012.

[2099] Michel Le Bellac. *Thermal Field Theory.* Cambridge Monographs on Mathematical Physics. Cambridge University Press, Mar. 2011.

[2100] Michael E. Peskin and Daniel V. Schroeder. *An Introduction to quantum field theory.* Reading, USA: Addison-Wesley, 1995.

[2101] Avinash Baidya et al. “Renormalization in open quantum field theory. Part I. Scalar field theory”. In: *JHEP* 11 (2017), p. 204.

[2102] Felix M. Haehl, R. Loganayagam, and Mukund Rangamani. “Schwinger-Keldysh formalism. Part I: BRST symmetries and superspace”. In: *JHEP* 06 (2017), p. 069.

[2103] Michael Crossley, Paolo Glorioso, and Hong Liu. “Effective field theory of dissipative fluids”. In: *JHEP* 09 (2017), p. 095.

[2104] Kristan Jensen, Natalia Pinzani-Fokeeva, and Amos Yarom. "Dissipative hydrodynamics in superspace". In: *JHEP* 09 (2018), p. 127.

[2105] H. P. Breuer and F. Petruccione. *The theory of open quantum systems.* 2002.

[2106] Vittorio Gorini, Andrzej Kossakowski, and E. C. G. Sudarshan. "Completely Positive Dynamical Semigroups of N Level Systems". In: *J. Math. Phys.* 17 (1976), p. 821.

[2107] Goran Lindblad. "On the Generators of Quantum Dynamical Semigroups". In: *Commun. Math. Phys.* 48 (1976), p. 119.

[2108] Takahiro Miura et al. "Simulation of Lindblad equations for quarkonium in the quark-gluon plasma". In: (May 2022).

[2109] Nora Brambilla et al. "Bottomonium suppression in an open quantum system using the quantum trajectories method". In: *JHEP* 05 (2021), p. 136.

[2110] Nora Brambilla et al. "Bottomonium production in heavy-ion collisions using quantum trajectories: Differential observables and momentum anisotropy". In: *Phys. Rev. D* 104.9 (2021), p. 094049.

[2111] Hisham Ba Omar et al. "QTRAJ 1.0: A Lindblad equation solver for heavy-quarkonium dynamics". In: *Comput. Phys. Commun.* 273 (2022), p. 108266.

[2112] T. Matsui and H. Satz. "$J/\psi$ Suppression by Quark-Gluon Plasma Formation". In: *Phys. Lett. B* 178 (1986), pp. 416–422.

[2113] M. Laine et al. "Real-time static potential in hot QCD". In: *JHEP* 03 (2007), p. 054.

[2114] Miguel Angel Escobedo, Joan Soto, and Massimo Mannarelli. "Non-relativistic bound states in a moving thermal bath". In: *Phys. Rev. D* 84 (2011), p. 016008.

[2115] Miguel Angel Escobedo et al. "Heavy Quarkonium moving in a Quark-Gluon Plasma". In: *Phys. Rev. D* 87.11 (2013), p. 114005.

[2116] Nora Brambilla et al. "The spin-orbit potential and Poincaré invariance in finite temperature pNRQCD". In: *JHEP* 07 (2011), p. 096.

[2117] Xiaojun Yao and Berndt Müller. "Approach to equilibrium of quarkonium in quark-gluon plasma". In: *Phys. Rev. C* 97.1 (2018). [Erratum: Phys.Rev.C 97, 049903 (2018)], p. 014908.

[2118] Xiaojun Yao and Thomas Mehen. "Quarkonium in-medium transport equation derived from first principles". In: *Phys. Rev. D* 99.9 (2019), p. 096028.

[2119] Nora Brambilla et al. "Heavy Quarkonium in a weakly-coupled quark-gluon plasma below the melting temperature". In: *JHEP* 09 (2010), p. 038.

[2120] Gyan Bhanot and Michael E. Peskin. "Short Distance Analysis for Heavy Quark Systems. 2. Applications". In: *Nucl. Phys. B* 156 (1979), pp. 391–416.

[2121] Ahmad Idilbi and Abhijit Majumder. "Extending Soft-Collinear-Effective-Theory to describe hard jets in dense QCD media". In: *Phys. Rev. D* 80 (2009), p. 054022.

[2122] Ivan Vitev. "Hard probes in heavy ion collisions: current status and prospects for application of QCD evolution techniques". In: *Int. J. Mod. Phys. Conf. Ser.* 37 (2015). Ed. by Alexei Prokudin, Anatoly Radyushkin, and Leonard Gamberg, p. 1560059.

[2123] Varun Vaidya. "Radiative corrections for factorized jet observables in heavy ion collisions". In: (June 2021).

[2124] Gerard 't Hooft and M. J. G. Veltman. "DIAGRAMMAR". In: *NATO Sci. Ser. B* 4 (1974), pp. 177–322.

[2125] Christof Gattringer and Kurt Langfeld. "Approaches to the sign problem in lattice field theory". In: *Int. J. Mod. Phys. A* 31.22 (2016), p. 1643007.

[2126] Barbara V. Jacak and Berndt Muller. "The exploration of hot nuclear matter". In: *Science* 337 (2012), pp. 310–314.

[2127] Berndt Muller, Jurgen Schukraft, and Boleslaw Wyslouch. "First Results from Pb+Pb collisions at the LHC". In: *Ann. Rev. Nucl. Part. Sci.* 62 (2012), pp. 361–386.

[2128] Peter Braun-Munzinger et al. "Properties of hot and dense matter from relativistic heavy ion collisions". In: *Phys. Rept.* 621 (2016), pp. 76–126.

[2129] Wit Busza, Krishna Rajagopal, and Wilke van der Schee. "Heavy Ion Collisions: The Big Picture, and the Big Questions". In: *Ann. Rev. Nucl. Part. Sci.* 68 (2018), pp. 339–376.

[2130] P. Braun-Munzinger et al. "Relativistic nuclear collisions: Establishing a non-critical baseline for fluctuation measurements". In: *Nucl. Phys. A* 1008 (2021), p. 122141.

[2131] K. Abe et al. "Leading Particle Distributions in 200-GeV/c $P$ + a Interactions". In: *Phys. Lett. B* 200 (1988), pp. 266–271.

[2132] J. Benecke et al. "Hypothesis of Limiting Fragmentation in High-Energy Collisions". In: *Phys. Rev.* 188 (1969), pp. 2159–2169.

[2133] H. Appelshauser et al. "Baryon stopping and charged particle distributions in central Pb + Pb collisions at 158-GeV per nucleon". In: *Phys. Rev. Lett.* 82 (1999), pp. 2471–2475.

[2134] I. C. Arsene et al. “Nuclear stopping and rapidity loss in Au+Au collisions at s(NN)**(1/2) = 62.4-GeV”. In: *Phys. Lett. B* 677 (2009), pp. 267–271.

[2135] J. D. Bjorken. “Highly Relativistic Nucleus-Nucleus Collisions: The Central Rapidity Region”. In: *Phys. Rev. D* 27 (1983), pp. 140–151.

[2136] Serguei Chatrchyan et al. “Measurement of the pseudorapidity and centrality dependence of the transverse energy density in PbPb collisions at $\sqrt{s_{NN}} = 2.76$ TeV”. In: *Phys. Rev. Lett.* 109 (2012), p. 152303.

[2137] J. Barrette et al. “Measurement of transverse energy production with Si and Au beams at relativistic energy: Towards hot and dense hadronic matter”. In: *Phys. Rev. Lett.* 70 (1993), pp. 2996–2999.

[2138] M. M. Aggarwal et al. “Scaling of particle and transverse energy production in Pb-208 + Pb-208 collisions at 158-A-GeV”. In: *Eur. Phys. J. C* 18 (2001), pp. 651–663.

[2139] A. Adare et al. “Transverse energy production and charged-particle multiplicity at midrapidity in various systems from $\sqrt{s_{NN}} = 7.7$ to 200 GeV”. In: *Phys. Rev. C* 93.2 (2016), p. 024901.

[2140] Andreas Bauswein et al. “Identifying a first-order phase transition in neutron star mergers through gravitational waves”. In: *Phys. Rev. Lett.* 122.6 (2019), p. 061102.

[2141] Gordon Baym et al. “New Neutron Star Equation of State with Quark-Hadron Crossover”. In: *Astrophys. J.* 885 (2019), p. 42.

[2142] Chun Shen et al. “Radial and elliptic flow in Pb+Pb collisions at the Large Hadron Collider from viscous hydrodynamic”. In: *Phys. Rev. C* 84 (2011), p. 044903.

[2143] J. E. Parkkila et al. “New constraints for QCD matter from improved Bayesian parameter estimation in heavy-ion collisions at LHC”. In: (Nov. 2021).

[2144] S. S. Adler et al. “Elliptic flow of identified hadrons in Au+Au collisions at s(NN)**(1/2) = 200-GeV”. In: *Phys. Rev. Lett.* 91 (2003), p. 182301.

[2145] J. Adams et al. “Azimuthal anisotropy in Au+Au collisions at s(NN)**(1/2) = 200-GeV”. In: *Phys. Rev. C* 72 (2005), p. 014904.

[2146] K Aamodt et al. “Elliptic flow of charged particles in Pb-Pb collisions at 2.76 TeV”. In: *Phys. Rev. Lett.* 105 (2010), p. 252302.

[2147] P. Danielewicz and M. Gyulassy. “Dissipative Phenomena in Quark Gluon Plasmas”. In: *Phys. Rev. D* 31 (1985), pp. 53–62.

[2148] P. Kovtun, Dan T. Son, and Andrei O. Starinets. “Viscosity in strongly interacting quantum field theories from black hole physics”. In: *Phys. Rev. Lett.* 94 (2005), p. 111601.

[2149] Rudolf Baier et al. “Relativistic viscous hydrodynamics, conformal invariance, and holography”. In: *JHEP* 04 (2008), p. 100.

[2150] Jonah E. Bernhard, J. Scott Moreland, and Steffen A. Bass. “Bayesian estimation of the specific shear and bulk viscosity of quark–gluon plasma”. In: *Nature Phys.* 15.11 (2019), pp. 1113–1117.

[2151] R. Baier et al. “Radiative energy loss of high-energy quarks and gluons in a finite volume quark - gluon plasma”. In: *Nucl. Phys. B* 483 (1997), pp. 291–320.

[2152] Serguei Chatrchyan et al. “Study of high-pT charged particle suppression in PbPb compared to *pp* collisions at $\sqrt{s_{NN}} = 2.76$ TeV”. In: *Eur. Phys. J. C* 72 (2012), p. 1945.

[2153] Morad Aaboud et al. “Measurement of the nuclear modification factor for inclusive jets in Pb+Pb collisions at $\sqrt{s_{\mathrm{NN}}} = 5.02$ TeV with the ATLAS detector”. In: *Phys. Lett. B* 790 (2019), pp. 108–128.

[2154] Korinna Zapp et al. “A Monte Carlo Model for 'Jet Quenching”'. In: *Eur. Phys. J. C* 60 (2009), pp. 617–632.

[2155] Nestor Armesto et al. “Comparison of Jet Quenching Formalisms for a Quark-Gluon Plasma 'Brick”'. In: *Phys. Rev. C* 86 (2012), p. 064904.

[2156] S. Cao et al. “Determining the jet transport coefficient $\hat{q}$ from inclusive hadron suppression measurements using Bayesian parameter estimation”. In: *Phys. Rev. C* 104.2 (2021), p. 024905.

[2157] Wei-tian Deng and Xin-Nian Wang. “Multiple Parton Scattering in Nuclei: Modified DGLAP Evolution for Fragmentation Functions”. In: *Phys. Rev. C* 81 (2010), p. 024902.

[2158] Peng Ru et al. “Global extraction of the jet transport coefficient in cold nuclear matter”. In: *Phys. Rev. D* 103.3 (2021), p. L031901.

[2159] Anton Andronic et al. “Decoding the phase structure of QCD via particle production at high energy”. In: *Nature* 561.7723 (2018), pp. 321–330.

[2160] Roger Dashen, Shang-Keng Ma, and Herbert J. Bernstein. “S Matrix formulation of statistical mechanics”. In: *Phys. Rev.* 187 (1969), pp. 345–370.

[2161] Pok Man Lo et al. “S-matrix analysis of the baryon electric charge correlation”. In: *Phys. Lett. B* 778 (2018), pp. 454–458.

[2162] Anton Andronic et al. "The thermal proton yield anomaly in Pb-Pb collisions at the LHC and its resolution". In: *Phys. Lett. B* 792 (2019), pp. 304–309.

[2163] L. Adamczyk et al. "Bulk Properties of the Medium Produced in Relativistic Heavy-Ion Collisions from the Beam Energy Scan Program". In: *Phys. Rev. C* 96.4 (2017), p. 044904.

[2164] Joachim Stroth (HADES Collaboration). Private Communications.

[2165] P. Braun-Munzinger, J. Stachel, and Christof Wetterich. "Chemical freezeout and the QCD phase transition temperature". In: *Phys. Lett. B* 596 (2004), pp. 61–69.

[2166] Adam Bzdak, Volker Koch, and Nils Strodthoff. "Cumulants and correlation functions versus the QCD phase diagram". In: *Phys. Rev. C* 95.5 (2017), p. 054906.

[2167] Adam Bzdak et al. "Mapping the Phases of Quantum Chromodynamics with Beam Energy Scan". In: *Phys. Rept.* 853 (2020), pp. 1–87.

[2168] Peter Braun-Munzinger, Anar Rustamov, and Johanna Stachel. "Experimental results on fluctuations of conserved charges confronted with predictions from canonical thermodynamics". In: *Nucl. Phys. A* 982 (2019). Ed. by Federico Antinori et al., pp. 307–310.

[2169] Peter Braun-Munzinger, Anar Rustamov, and Johanna Stachel. "The role of the local conservation laws in fluctuations of conserved charges". In: (July 2019).

[2170] Volodymyr Vovchenko, Roman V. Poberezhnyuk, and Volker Koch. "Cumulants of multiple conserved charges and global conservation laws". In: *JHEP* 10 (2020), p. 089.

[2171] P. Braun-Munzinger, A. Rustamov, and J. Stachel. "Bridging the gap between event-by-event fluctuation measurements and theory predictions in relativistic nuclear collisions". In: *Nucl. Phys. A* 960 (2017), pp. 114–130.

[2172] V. Skokov, B. Friman, and K. Redlich. "Volume Fluctuations and Higher Order Cumulants of the Net Baryon Number". In: *Phys. Rev. C* 88 (2013), p. 034911.

[2173] A. Bazavov et al. "Skewness, kurtosis, and the fifth and sixth order cumulants of net baryon-number distributions from lattice QCD confront high-statistics STAR data". In: *Phys. Rev. D* 101.7 (2020), p. 074502.

[2174] Anar Rustamov. "Net-baryon fluctuations measured with ALICE at the CERN LHC". In: *Nucl. Phys. A* 967 (2017). Ed. by Ulrich Heinz, Olga Evdokimov, and Peter Jacobs, pp. 453–456.

[2175] Shreyasi Acharya et al. "Global baryon number conservation encoded in net-proton fluctuations measured in Pb-Pb collisions at $\sqrt{s_{\rm NN}}$ = 2.76 TeV". In: *Phys. Lett. B* 807 (2020), p. 135564.

[2176] Anar Rustamov. "Overview of fluctuation and correlation measurements". In: *Nucl. Phys. A* 1005 (2021). Ed. by Feng Liu et al., p. 121858.

[2177] Anar Rustamov. "Deciphering the phases of QCD matter with fluctuations and correlations of conserved charges". In: (Oct. 2022).

[2178] M. A. Stephanov. "On the sign of kurtosis near the QCD critical point". In: *Phys. Rev. Lett.* 107 (2011), p. 052301.

[2179] J. Adamczewski-Musch et al. "Proton-number fluctuations in $\sqrt{s_{NN}}$ =2.4 GeV Au + Au collisions studied with the High-Acceptance DiElectron Spectrometer (HADES)". In: *Phys. Rev. C* 102.2 (2020), p. 024914.

[2180] J. Adam et al. "Nonmonotonic Energy Dependence of Net-Proton Number Fluctuations". In: *Phys. Rev. Lett.* 126.9 (2021), p. 092301.

[2181] B. Friman et al. "Fluctuations as probe of the QCD phase transition and freeze-out in heavy ion collisions at LHC and RHIC". In: *Eur. Phys. J. C* 71 (2011), p. 1694.

[2182] Gabor Andras Almasi, Bengt Friman, and Krzysztof Redlich. "Baryon number fluctuations in chiral effective models and their phenomenological implications". In: *Phys. Rev. D* 96.1 (2017), p. 014027.

[2183] Szabolcs Borsanyi et al. "Higher order fluctuations and correlations of conserved charges from lattice QCD". In: *JHEP* 10 (2018), p. 205.

[2184] Mohamed Abdallah et al. "Measurement of the sixth-order cumulant of net-proton multiplicity distributions in Au+Au collisions at $\sqrt{s_{NN}}$ = 27, 54.4, and 200 GeV at RHIC". In: *Phys. Rev. Lett.* 127.26 (2021), p. 262301.

[2185] Masakiyo Kitazawa and Masayuki Asakawa. "Relation between baryon number fluctuations and experimentally observed proton number fluctuations in relativistic heavy ion collisions". In: *Phys. Rev. C* 86 (2012). [Erratum: Phys.Rev.C 86, 069902 (2012)], p. 024904.

[2186] Shreyasi Acharya et al. "Prompt $D^0$, $D^+$, and $D^{*+}$ production in Pb–Pb collisions at $\sqrt{s_{\rm NN}}$ = 5.02 TeV". In: *JHEP* 01 (2022), p. 174.

[2187] Anton Andronic et al. "The multiple-charm hierarchy in the statistical hadronization model". In: *JHEP* 07 (2021), p. 035.

[2188] Anton Andronic et al. "Transverse momentum distributions of charmonium states with the sta-

tistical hadronization model". In: *Phys. Lett. B* 797 (2019), p. 134836.
[2189] Luis Altenkort et al. "Heavy quark momentum diffusion from the lattice using gradient flow". In: *Phys. Rev. D* 103.1 (2021), p. 014511.
[2190] Ehab Abbas et al. "J/Psi Elliptic Flow in Pb-Pb Collisions at $\sqrt{s_{\mathrm{NN}}} = 2.76$ TeV". In: *Phys. Rev. Lett.* 111 (2013), p. 162301.
[2191] Min He, Biaogang Wu, and Ralf Rapp. "Collectivity of J/$\psi$ Mesons in Heavy-Ion Collisions". In: *Phys. Rev. Lett.* 128.16 (2022), p. 162301.
[2192] Sungtae Cho et al. "Charmed hadron production in an improved quark coalescence model". In: *Phys. Rev. C* 101.2 (2020), p. 024909.
[2193] Jiaxing Zhao et al. "Sequential Coalescence with Charm Conservation in High Energy Nuclear Collisions". In: (May 2018).
[2194] Sungtae Cho et al. "Exotic hadrons from heavy ion collisions". In: *Prog. Part. Nucl. Phys.* 95 (2017), pp. 279–322.
[2195] Kai Zhou et al. "Medium effects on charmonium production at ultrarelativistic energies available at the CERN Large Hadron Collider". In: *Phys. Rev. C* 89.5 (2014), p. 054911.
[2196] V. Greco, C. M. Ko, and R. Rapp. "Quark coalescence for charmed mesons in ultrarelativistic heavy ion collisions". In: *Phys. Lett. B* 595 (2004), pp. 202–208.
[2197] G. Aarts et al. "Heavy-flavor production and medium properties in high-energy nuclear collisions - What next?" In: *Eur. Phys. J. A* 53.5 (2017), p. 93.
[2198] Luciano Maiani and Alessandro Pilloni. "GGI Lectures on Exotic Hadrons". In: July 2022.
[2199] Taesoo Song and Gabriele Coci. "Prerequisites for heavy quark coalescence in heavy-ion collisions". In: (Apr. 2021).
[2200] P. Braun-Munzinger and J. Stachel. "(Non)thermal aspects of charmonium production and a new look at J / psi suppression". In: *Phys. Lett. B* 490 (2000), pp. 196–202.
[2201] A. Andronic et al. "Statistical hadronization of charm at SPS, RHIC and LHC". In: *Nucl. Phys. A* 715 (2003). Ed. by H. Gutbrod, J. Aichelin, and K. Werner, pp. 529–532.
[2202] Loic Grandchamp, Ralf Rapp, and Gerald E. Brown. "In medium effects on charmonium production in heavy ion collisions". In: *Phys. Rev. Lett.* 92 (2004), p. 212301.
[2203] Francesco Becattini. "Production of multiply heavy flavored baryons from quark gluon plasma in relativistic heavy ion collisions". In: *Phys. Rev. Lett.* 95 (2005), p. 022301.
[2204] A. Andronic et al. "Statistical hadronization of heavy quarks in ultra-relativistic nucleus-nucleus collisions". In: *Nucl. Phys. A* 789 (2007), pp. 334–356.
[2205] Shreyasi Acharya et al. "Measurement of inclusive $J/\psi$ production at midrapidity and forward rapidity in Pb-Pb collisions at $\sqrt{s_{NN}} = = 5.02$ TeV". In: (2022).
[2206] Ralf Rapp and Edward V. Shuryak. "Resolving the anti-baryon production puzzle in high-energy heavy ion collisions". In: *Phys. Rev. Lett.* 86 (2001), pp. 2980–2983.
[2207] A. Andronic et al. "Hadron Production in Ultra-relativistic Nuclear Collisions: Quarkyonic Matter and a Triple Point in the Phase Diagram of QCD". In: *Nucl. Phys. A* 837 (2010), pp. 65–86.
[2208] Gordon Baym. "RHIC: From dreams to beams in two decades". In: *Nucl. Phys. A* 698 (2002). Ed. by T. J. Hallman et al., pp. XXIII–XXXII.
[2209] Yoichiro Nambu and G. Jona-Lasinio. "Dynamical Model of Elementary Particles Based on an Analogy with Superconductivity. II". In: *Phys. Rev.* 124 (1961). Ed. by T. Eguchi, pp. 246–254.
[2210] Yoshitaka Hatta and Kenji Fukushima. "Linking the chiral and deconfinement phase transitions". In: *Phys. Rev. D* 69 (2004), p. 097502.
[2211] Szabolcs Borsanyi et al. "Is there still any $T_c$ mystery in lattice QCD? Results with physical masses in the continuum limit III". In: *JHEP* 09 (2010), p. 073.
[2212] N. Itoh. "Hydrostatic Equilibrium of Hypothetical Quark Stars". In: *Prog. Theor. Phys.* 44 (1970), p. 291.
[2213] Barry A. Freedman and Larry D. McLerran. "Fermions and Gauge Vector Mesons at Finite Temperature and Density. 1. Formal Techniques". In: *Phys. Rev. D* 16 (1977), p. 1130.
[2214] Barry A. Freedman and Larry D. McLerran. "Fermions and Gauge Vector Mesons at Finite Temperature and Density. 3. The Ground State Energy of a Relativistic Quark Gas". In: *Phys. Rev. D* 16 (1977), p. 1169.
[2215] Aleksi Kurkela, Paul Romatschke, and Aleksi Vuorinen. "Cold Quark Matter". In: *Phys. Rev. D* 81 (2010), p. 105021.
[2216] Tyler Gorda et al. "Soft Interactions in Cold Quark Matter". In: *Phys. Rev. Lett.* 127.16 (2021), p. 162003.
[2217] Tyler Gorda et al. "Cold quark matter at N3LO: Soft contributions". In: *Phys. Rev. D* 104.7 (2021), p. 074015.

[2218] D. H. Rischke, D. T. Son, and Misha A. Stephanov. "Asymptotic deconfinement in high density QCD". In: *Phys. Rev. Lett.* 87 (2001), p. 062001.

[2219] Thomas Schäfer and Frank Wilczek. "Continuity of quark and hadron matter". In: *Phys. Rev. Lett.* 82 (1999), pp. 3956–3959.

[2220] Yuki Fujimoto, Kenji Fukushima, and Wolfram Weise. "Continuity from neutron matter to two-flavor quark matter with $^1S_0$ and $^3P_2$ superfluidity". In: *Phys. Rev. D* 101.9 (2020), p. 094009.

[2221] A. P. Balachandran, S. Digal, and T. Matsuura. "Semi-superfluid strings in high density QCD". In: *Phys. Rev. D* 73 (2006), p. 074009.

[2222] Aleksey Cherman, Srimoyee Sen, and Laurence G. Yaffe. "Anyonic particle-vortex statistics and the nature of dense quark matter". In: *Phys. Rev. D* 100.3 (2019), p. 034015.

[2223] Yuji Hirono and Yuya Tanizaki. "Quark-Hadron Continuity beyond the Ginzburg-Landau Paradigm". In: *Phys. Rev. Lett.* 122.21 (2019), p. 212001.

[2224] Larry McLerran and Robert D. Pisarski. "Phases of cold, dense quarks at large N(c)". In: *Nucl. Phys. A* 796 (2007), pp. 83–100.

[2225] Kenji Fukushima, Toru Kojo, and Wolfram Weise. "Hard-core deconfinement and soft-surface delocalization from nuclear to quark matter". In: *Phys. Rev. D* 102.9 (2020), p. 096017.

[2226] Y. Aoki et al. "The QCD transition temperature: Results with physical masses in the continuum limit". In: *Phys. Lett. B* 643 (2006), pp. 46–54.

[2227] M. Asakawa and K. Yazaki. "Chiral Restoration at Finite Density and Temperature". In: *Nucl. Phys. A* 504 (1989), pp. 668–684.

[2228] A. Barducci et al. "Chiral Symmetry Breaking in QCD at Finite Temperature and Density". In: *Phys. Lett. B* 231 (1989), pp. 463–470.

[2229] Frank Wilczek. "Remarks on the phase transition in QCD". In: *1st IFT Workshop: Dark Matter.* Mar. 1992.

[2230] D. T. Son and M. A. Stephanov. "Dynamic universality class of the QCD critical point". In: *Phys. Rev. D* 70 (2004), p. 056001.

[2231] Kenji Fukushima, Bedangadas Mohanty, and Nu Xu. "Little-Bang and Femto-Nova in Nucleus-Nucleus Collisions". In: *AAPPS Bull.* 31 (2021), p. 1.

[2232] M. Stephanov and Y. Yin. "Hydrodynamics with parametric slowing down and fluctuations near the critical point". In: *Phys. Rev. D* 98.3 (2018), p. 036006.

[2233] Dominik Nickel. "How many phases meet at the chiral critical point?" In: *Phys. Rev. Lett.* 103 (2009), p. 072301.

[2234] E. Nakano and T. Tatsumi. "Chiral symmetry and density wave in quark matter". In: *Phys. Rev. D* 71 (2005), p. 114006.

[2235] Michael Buballa and Stefano Carignano. "Inhomogeneous chiral condensates". In: *Prog. Part. Nucl. Phys.* 81 (2015), pp. 39–96.

[2236] Yoshimasa Hidaka et al. "Phonons, pions and quasi-long-range order in spatially modulated chiral condensates". In: *Phys. Rev. D* 92.3 (2015), p. 034003.

[2237] Tong-Gyu Lee et al. "Landau-Peierls instability in a Fulde-Ferrell type inhomogeneous chiral condensed phase". In: *Phys. Rev. D* 92.3 (2015), p. 034024.

[2238] Robert D. Pisarski and Fabian Rennecke. "Signatures of Moat Regimes in Heavy-Ion Collisions". In: *Phys. Rev. Lett.* 127.15 (2021), p. 152302.

[2239] Paul Demorest et al. "Shapiro Delay Measurement of A Two Solar Mass Neutron Star". In: *Nature* 467 (2010), pp. 1081–1083.

[2240] John Antoniadis et al. "A Massive Pulsar in a Compact Relativistic Binary". In: *Science* 340 (2013), p. 6131.

[2241] H. T. Cromartie et al. "Relativistic Shapiro delay measurements of an extremely massive millisecond pulsar". In: *Nature Astron.* 4.1 (2019), pp. 72–76.

[2242] Mark G. Alford et al. "Constraining and applying a generic high-density equation of state". In: *Phys. Rev. D* 92.8 (2015), p. 083002.

[2243] Christian Drischler et al. "Limiting masses and radii of neutron stars and their implications". In: *Phys. Rev. C* 103.4 (2021), p. 045808.

[2244] Mohammad Al-Mamun et al. "Combining Electromagnetic and Gravitational-Wave Constraints on Neutron-Star Masses and Radii". In: *Phys. Rev. Lett.* 126.6 (2021), p. 061101.

[2245] Ernesto Arganda and Maria J. Herrero. "Testing supersymmetry with lepton flavor violating tau and mu decays". In: *Phys. Rev.* D73 (2006), p. 055003.

[2246] Tyler Gorda, Oleg Komoltsev, and Aleksi Kurkela. "Ab-initio QCD calculations impact the inference of the neutron-star-matter equation of state". In: (Apr. 2022).

[2247] Yuki Fujimoto et al. "Trace anomaly as signature of conformality in neutron stars". In: (July 2022).

[2248] Eemeli Annala et al. "Gravitational-wave constraints on the neutron-star-matter Equation

of State". In: *Phys. Rev. Lett.* 120.17 (2018), p. 172703.

[2249] B. P. Abbott et al. "GW170817: Measurements of neutron star radii and equation of state". In: *Phys. Rev. Lett.* 121.16 (2018), p. 161101.

[2250] S. Gandolfi, J. Carlson, and Sanjay Reddy. "The maximum mass and radius of neutron stars and the nuclear symmetry energy". In: *Phys. Rev. C* 85 (2012), p. 032801.

[2251] M. Aguilar-Benitez et al. "Review of Particle Properties. Particle Data Group". In: *Phys. Lett. B* 170 (1986), pp. 1–350.

[2252] Francesco Giacosa, Adrian Koenigstein, and Robert D. Pisarski. "How the axial anomaly controls flavor mixing among mesons". In: *Phys. Rev. D* 97.9 (2018), p. 091901.

[2253] Denis Parganlija et al. "Meson vacuum phenomenology in a three-flavor linear sigma model with (axial)-vector mesons". In: *Phys. Rev. D* 87.1 (2013), p. 014011.

[2254] Shahriyar Jafarzade et al. "From well-known tensor mesons to yet unknown axial-tensor mesons". In: *Phys. Rev. D* 106.3 (2022), p. 036008.

[2255] Adrian Koenigstein and Francesco Giacosa. "Phenomenology of pseudotensor mesons and the pseudotensor glueball". In: *Eur. Phys. J. A* 52.12 (2016), p. 356.

[2256] Shahriyar Jafarzade, Adrian Koenigstein, and Francesco Giacosa. "Phenomenology of $J^{PC} = 3^{--}$ tensor mesons". In: *Phys. Rev. D* 103.9 (2021), p. 096027.

[2257] J. E. Augustin et al. "Radiative Decay of $J/\psi$ Into $\eta(1430)$ and Nearby States". In: *Phys. Rev. D* 42 (1990), pp. 10–19.

[2258] J. E. Augustin et al. "Partial wave analysis of DM2 data in the $\eta(1430)$ energy range". In: *Phys. Rev. D* 46 (1992), pp. 1951–1958.

[2259] N. R. Stanton et al. "Evidence for Axial Vector and Pseudoscalar Resonances Near 1.275-GeV in $\eta\pi^+\pi^-$". In: *Phys. Rev. Lett.* 42 (1979), pp. 346–349.

[2260] S. Fukui et al. "Study on the $\eta\pi^+\pi^-$ system in the $\pi^-p$ charge exchange reaction at 8.95-GeV/c". In: *Phys. Lett. B* 267 (1991). Ed. by K. Nakai and T. Ohshima, pp. 293–298.

[2261] D. Alde et al. "Partial-wave analysis of the $\eta\pi^0\pi^0$ system produced in $\pi^-p$ charge exchange collisions at 100-GeV/c". In: *Phys. Atom. Nucl.* 60 (1997), pp. 386–390.

[2262] J. J. Manak et al. "Partial-wave analysis of the $\eta\pi^+\pi^-$ system produced in the reaction $\pi^-p \to \eta\pi^+\pi^-n$ at 18-GeV/c". In: *Phys. Rev. D* 62 (2000), p. 012003.

[2263] G. S. Adams et al. "Observation of pseudoscalar and axial vector resonances in $\pi^-p \to \eta K^+K^-n$ at 18-GeV". In: *Phys. Lett. B* 516 (2001), pp. 264–272.

[2264] C. Amsler et al. "E decay to $\eta\pi\pi$ in $\bar{p}p$ annihilation at rest". In: *Phys. Lett. B* 358 (1995), pp. 389–398.

[2265] A. Bertin et al. "A search for axial vectors in $\bar{p}p \to K^\pm K^0_{\text{miss}}\pi^\mp\pi^+\pi^-$ annihilations at rest in gaseous hydrogen at NTP". In: *Phys. Lett. B* 400 (1997), pp. 226–238.

[2266] P. Eugenio et al. "Observation of a new $J^(PC) = 1^{+-}$ isoscalar state in the reaction $\pi^-$ proton $\to \omega\eta$ neutron at 18-GeV/c". In: *Phys. Lett. B* 497 (2001), pp. 190–198.

[2267] Florian Divotgey, Lisa Olbrich, and Francesco Giacosa. "Phenomenology of axial-vector and pseudovector mesons: decays and mixing in the kaonic sector". In: *Eur. Phys. J. A* 49 (2013), p. 135.

[2268] P. Gavillet et al. "Evidence for a New $K^*\bar{K}$ State at a Mass of 1530-MeV With $J^{PC} = 1^{++}$ Observed in $K^-p$ Interactions at 4.2-GeV/$c$". In: *Z. Phys. C* 16 (1982), p. 119.

[2269] D. Aston et al. "Evidence for Two Strangeonium Resonances With $J^{PC} = 1^{++}$ and $1^{+-}$ in $K^-p$ Interactions at 11-GeV/$c$". In: *Phys. Lett. B* 201 (1988), pp. 573–578.

[2270] A. Birman et al. "Partial Wave Analysis of the $K^+\bar{K}^0\pi^-$ System". In: *Phys. Rev. Lett.* 61 (1988). [Erratum: Phys.Rev.Lett. 62, 1577 (1989)], p. 1557.

[2271] M. Aghasyan et al. "Light isovector resonances in $\pi^-p \to \pi^-\pi^-\pi^+p$ at 190 GeV/$c$". In: *Phys. Rev. D* 98.9 (2018), p. 092003.

[2272] Philippe d'Argent et al. "Amplitude Analyses of $D^0 \to \pi^+\pi^-\pi^+\pi^-$ and $D^0 \to K^+K^-\pi^+\pi^-$ Decays". In: *JHEP* 05 (2017), p. 143.

[2273] V. K. Grigorev et al. "Investigation of a resonance structure in the system of two $K_s$ mesons in the mass region around 1775-MeV". In: *Phys. Atom. Nucl.* 62 (1999), pp. 470–478.

[2274] M. Lu et al. "Exotic meson decay to $\omega\pi^0\pi^-$". In: *Phys. Rev. Lett.* 94 (2005), p. 032002.

[2275] J. Nys et al. "Features of $\pi\Delta$ Photoproduction at High Energies". In: *Phys. Lett. B* 779 (2018), pp. 77–81.

[2276] A. Rodas et al. "Determination of the pole position of the lightest hybrid meson candidate". In: *Phys. Rev. Lett.* 122.4 (2019), p. 042002.

[2277] A. Abele et al. "Observation of resonances in the reaction $\bar{p}p \to \pi^0\eta\eta$ at 1.94-GeV/c". In: *Eur. Phys. J. C* 8 (1999), pp. 67–79.

[2278] Claude Amsler et al. "Proton anti-proton annihilation at 900-MeV/c into $\pi^0\pi^0\pi^0$, $\pi^0\pi^0\eta$ and $\pi^0\eta\eta$". In: *Eur. Phys. J. C* 23 (2002), pp. 29–41.

[2279] V. V. Anisovich and A. V. Sarantsev. "The combined analysis of $\pi N \to$ two mesons + $N$ reactions within Reggeon exchanges and data for $\bar{p}p$ (at rest) $\to$ three mesons". In: *Int. J. Mod. Phys. A* 24 (2009), pp. 2481–2549.

[2280] M. Albrecht et al. "Coupled channel analysis of $\bar{p}p \to \pi^0\pi^0\eta$, $\pi^0\eta\eta$ and $K^+K^-\pi^0$ at 900 MeV/c and of $\pi\pi$-scattering data". In: *Eur. Phys. J. C* 80.5 (2020), p. 453.

[2281] M. Acciarri et al. "Resonance formation in the $\pi^+\pi^-\pi^0$ final state in two photon collisions". In: *Phys. Lett. B* 413 (1997), pp. 147–158.

[2282] M. Acciarri et al. "$K^0_S K^0_S$ final state in two photon collisions and implications for glueballs". In: *Phys. Lett. B* 501 (2001), pp. 173–182.

[2283] Medina Ablikim et al. "Amplitude analysis of the $\chi_{c1} \to \eta\pi^+\pi^-$ decays". In: *Phys. Rev. D* 95.3 (2017), p. 032002.

[2284] A. Etkin et al. "Increased Statistics and Observation of the $g(T)$, $g'_T$, and $g''_T$ $2^{++}$ Resonances in the Glueball Enhanced Channel $\pi^-p \to \phi\phi n$". In: *Phys. Lett. B* 201 (1988), pp. 568–572.

[2285] J. Adomeit et al. "Evidence for two isospin zero $J^{PC} = 2^{-+}$ mesons at 1645-MeV and 1875-MeV". In: *Z. Phys. C* 71 (1996), pp. 227–238.

[2286] A. Hasan and D. V. Bugg. "Amplitudes for $\bar{p}p \to \pi\pi$ from 0.36-GeV/c to 2.5-GeV/c". In: *Phys. Lett. B* 334 (1994), pp. 215–219.

[2287] A. V. Anisovich et al. "Combined analysis of meson channels with I = 1, C = -1 from 1940 to 2410 MeV". In: *Phys. Lett. B* 542 (2002), pp. 8–18.

[2288] M. Ablikim et al. "Partial wave analysis of $\psi(3686) \to K^+K^-\eta$". In: *Phys. Rev. D* 101.3 (2020), p. 032008.

[2289] Jiao-Kai Chen. "Structure of the meson Regge trajectories". In: *Eur. Phys. J. A* 57.7 (2021), p. 238.

[2290] G. F. Chew and Steven C. Frautschi. "Principle of Equivalence for All Strongly Interacting Particles Within the S Matrix Framework". In: *Phys. Rev. Lett.* 7 (1961), pp. 394–397.

[2291] M.H. Johnson and E. Teller. "Classical Field Theory of Nuclear Forces". In: *Phys. Rev.* 98 (1955), pp. 783–787.

[2292] Murray Gell-Mann and Maurice Levy. "The axial vector current in beta decay". In: *Nuovo Cim.* 16 (1960), p. 705.

[2293] R. H. Dalitz. "Resonant states and strong interactions". In: *Oxford International Conference on Elementary Particles.* 1966, pp. 157–181.

[2294] J. R. Pelaez. "From controversy to precision on the sigma meson: a review on the status of the non-ordinary $f_0(500)$ resonance". In: *Phys. Rept.* 658 (2016), p. 1.

[2295] José R. Peláez and Arkaitz Rodas. "Dispersive $\pi K \to \pi K$ and $\pi\pi \to K\bar{K}$ amplitudes from scattering data, threshold parameters, and the lightest strange resonance $\kappa$ or $K^*_0(700)$". In: *Phys. Rept.* 969 (2022), pp. 1–126.

[2296] A. V. Anisovich, V. V. Anisovich, and A. V. Sarantsev. "Systematics of $q\bar{q}$ states in the $(n, M^2)$ and $(J, M^2)$ planes". In: *Phys. Rev.* D62 (2000), p. 051502.

[2297] J. T. Londergan et al. "Identification of non-ordinary mesons from the dispersive connection between their poles and their Regge trajectories: The $f_0(500)$ resonance". In: *Phys. Lett. B* 729 (2014), pp. 9–14.

[2298] J. R. Pelaez and A. Rodas. "The non-ordinary Regge behavior of the $K^*_0(800)$ or $\kappa$ -meson versus the ordinary $K^*_0(1430)$". In: *Eur. Phys. J. C* 77.6 (2017), p. 431.

[2299] Robert L. Jaffe. "Multi-Quark Hadrons. 1. The Phenomenology of $Q^2\bar{Q}^2$ mesons". In: *Phys.Rev.* D15 (1977), p. 267.

[2300] J. A. Carrasco et al. "Dispersive calculation of complex Regge trajectories for the lightest $f_2$ resonances and the K$^?$(892)". In: *Phys. Lett. B* 749 (2015), pp. 399–406.

[2301] S. M. Roy. "Exact integral equation for pion pion scattering involving only physical region partial waves". In: *Phys.Lett.* 36B (1971), pp. 353–356.

[2302] J. Baacke and F. Steiner. "$\pi$N partial wave relations from fixed-$t$ dispersion relations". In: *Fortsch. Phys.* 18 (1970), pp. 67–87.

[2303] F. Steiner. "Partial wave crossing relations for meson-baryon scattering". In: *Fortsch. Phys.* 19 (1971), pp. 115–159.

[2304] R. García-Martín et al. "The Pion-pion scattering amplitude. IV: Improved analysis with once subtracted Roy-like equations up to 1100 MeV". In: *Phys.Rev.* D83 (2011), p. 074004.

[2305] B. Ananthanarayan et al. "Roy equation analysis of $\pi\pi$ scattering". In: *Phys.Rept.* 353 (2001), pp. 207–279.

[2306] G. Colangelo, J. Gasser, and H. Leutwyler. "$\pi\pi$ scattering". In: *Nucl. Phys.* B603 (2001), pp. 125–179.

[2307] Paul Buettiker, S. Descotes-Genon, and B. Moussallam. "A new analysis of $\pi K$ scattering from Roy and Steiner type equations". In: *Eur. Phys. J. C* 33 (2004), pp. 409–432.

[2308] Irinel Caprini, Gilberto Colangelo, and Heinrich Leutwyler. "Mass and width of the lowest resonance in QCD". In: *Phys.Rev.Lett.* 96 (2006), p. 132001.

[2309] S. Descotes-Genon and B. Moussallam. "The $K_0^*(800)$ scalar resonance from Roy-Steiner representations of $\pi K$ scattering". In: *Eur. Phys. J.* C48 (2006), p. 553.

[2310] R. García-Martín et al. "Precise determination of the $f_0(600)$ and $f_0(980)$ pole parameters from a dispersive data analysis". In: *Phys.Rev.Lett.* 107 (2011), p. 072001.

[2311] J. R. Pelaez and A. Rodas. "$\pi\pi \to K\bar{K}$ scattering up to 1.47 GeV with hyperbolic dispersion relations". In: *Eur. Phys. J. C* 78.11 (2018), p. 897.

[2312] J. R. Peláez and A. Rodas. "Determination of the lightest strange resonance $K_0^*(700)$ or $\kappa$, from a dispersive data analysis". In: *Phys. Rev. Lett.* 124.17 (2020), p. 172001.

[2313] H. Leutwyler. "Model independent determination of the sigma pole". In: *AIP Conf. Proc.* 1030.1 (2008). Ed. by George Rupp et al., pp. 46–55.

[2314] B. Moussallam. "Couplings of light I=0 scalar mesons to simple operators in the complex plane". In: *Eur. Phys. J.* C71 (2011), p. 1814.

[2315] J. R. Peláez, A. Rodas, and J. Ruiz de Elvira. "Strange resonance poles from $K\pi$ scattering below 1.8 GeV". In: *Eur. Phys. J. C* 77.2 (2017), p. 91.

[2316] Jose Ramon Pelaez, Arkaitz Rodas, and Jacobo Ruiz de Elvira. "The $f_0(1370)$ controversy from dispersive meson-meson scattering data analyses". In: (June 2022).

[2317] A. Dobado and J. R. Pelaez. "The Inverse amplitude method in chiral perturbation theory". In: *Phys. Rev.* D56 (1997), pp. 3057–3073.

[2318] J. Nieves, M. Pavon Valderrama, and E. Ruiz Arriola. "The Inverse amplitude method in $\pi\pi$ scattering in chiral perturbation theory to two loops". In: *Phys. Rev. D* 65 (2002), p. 036002.

[2319] A. Dobado and J.R. Pelaez. "Chiral perturbation theory and the $f_2(1270)$ resonance". In: *Phys. Rev. D* 65 (2002), p. 077502.

[2320] Tran N. Truong. "Chiral Perturbation Theory and Final State Theorem". In: *Phys. Rev. Lett.* 61 (1988), p. 2526.

[2321] A. Dobado, Maria J. Herrero, and Tran N. Truong. "Unitarized Chiral Perturbation Theory for Elastic Pion-Pion Scattering". In: *Phys. Lett.* B235 (1990), pp. 134–140.

[2322] A. Dobado and J.R. Pelaez. "A Global fit of $\pi\pi$ and $\pi K$ elastic scattering in ChPT with dispersion relations". In: *Phys. Rev. D* 47 (1993), pp. 4883–4888.

[2323] Francisco Guerrero and Jose Antonio Oller. "$K\bar{K}$ scattering amplitude to one loop in chiral perturbation theory, its unitarization and pion formfactors". In: *Nucl. Phys. B* 537 (1999). [Erratum: Nucl.Phys.B 602, 641–643 (2001)], pp. 459–476.

[2324] A. Gomez Nicola and J. R. Pelaez. "Meson meson scattering within one loop chiral perturbation theory and its unitarization". In: *Phys. Rev. D* 65 (2002), p. 054009.

[2325] J. R. Pelaez. "Light scalars as tetraquarks or two-meson states from large $N_c$ and unitarized chiral perturbation theory". In: *Mod. Phys. Lett.* A19 (2004), p. 2879.

[2326] J. A. Oller and E. Oset. "Chiral symmetry amplitudes in the S wave isoscalar and isovector channels and the $\sigma$, $\mathrm{f}_0(980)$, $\mathrm{a}_0(980)$ scalar mesons". In: *Nucl. Phys.* A620 (1997). [Erratum: Nucl. Phys.A652,407(1999)], pp. 438–456.

[2327] J. A. Oller, E. Oset, and J. R. Pelaez. "Nonperturbative approach to effective chiral Lagrangians and meson interactions". In: *Phys. Rev. Lett.* 80 (1998), pp. 3452–3455.

[2328] J. A. Oller. "Coupled-channel approach in hadron-hadron scattering". In: *Prog. Part. Nucl. Phys.* 110 (2020), p. 103728.

[2329] Jose R. Peláez, Arkaitz Rodas, and Jacobo Ruiz de Elvira. "Precision dispersive approaches versus unitarized chiral perturbation theory for the lightest scalar resonances $\sigma/f_0(500)$ and $\kappa/K_0^*(700)$". In: *Eur. Phys. J. ST* 230.6 (2021), pp. 1539–1574.

[2330] J. A. Oller. "Unitarization Technics in Hadron Physics with Historical Remarks". In: *Symmetry* 12.7 (2020), p. 1114.

[2331] Edward Witten. "Large $N$ Chiral Dynamics". In: *Annals Phys.* 128 (1980), p. 363.

[2332] Steven Weinberg. "Tetraquark Mesons in Large $N$ Quantum Chromodynamics". In: *Phys. Rev. Lett.* 110 (2013), p. 261601.

[2333] Marc Knecht and Santiago Peris. "Narrow Tetraquarks at Large N". In: *Phys. Rev. D* 88 (2013), p. 036016.

[2334] J. Nebreda, J. R. Pelaez, and G. Rios. "Enhanced non-quark-antiquark and non-glueball

$N_c$ behavior of light scalar mesons". In: *Phys. Rev. D* 84 (2011), p. 074003.

[2335] Santiago Peris and Eduardo de Rafael. "On the large $N_c$ behavior of the $L(7)$ coupling in $\chi(PT)$". In: *Phys. Lett. B* 348 (1995), pp. 539–542.

[2336] J. R. Pelaez. "On the Nature of light scalar mesons from their large $N_c$ behavior". In: *Phys. Rev. Lett.* 92 (2004), p. 102001.

[2337] J. R. Pelaez and G. Rios. "Nature of the $f_0(600)$ from its $N_c$ dependence at two loops in unitarized Chiral Perturbation Theory". In: *Phys. Rev. Lett.* 97 (2006), p. 242002.

[2338] J. Ruiz de Elvira et al. "Chiral Perturbation Theory, the $1/N_c$ expansion and Regge behaviour determine the structure of the lightest scalar meson". In: *Phys. Rev. D* 84 (2011), p. 096006.

[2339] E. van Beveren et al. "A Low Lying Scalar Meson Nonet in a Unitarized Meson Model". In: *Z. Phys. C* 30 (1986), pp. 615–620.

[2340] Zhi-Hui Guo and J. A. Oller. "Resonances from meson-meson scattering in U(3) CHPT". In: *Phys. Rev. D* 84 (2011), p. 034005.

[2341] Zhi-Hui Guo, J. A. Oller, and J. Ruiz de Elvira. "Chiral dynamics in form factors, spectral-function sum rules, meson-meson scattering and semi-local duality". In: *Phys. Rev. D* 86 (2012), p. 054006.

[2342] J. Nieves, A. Pich, and E. Ruiz Arriola. "Large-$N_c$ Properties of the $\rho$ and $f_0(600)$ Mesons from Unitary Resonance Chiral Dynamics". In: *Phys. Rev. D* 84 (2011), p. 096002.

[2343] C. Hanhart, J. R. Pelaez, and G. Rios. "Quark mass dependence of the $\rho$ and $\sigma$ from dispersion relations and Chiral Perturbation Theory". In: *Phys. Rev. Lett.* 100 (2008), p. 152001.

[2344] J. Nebreda, J. R. Pelaez, and G. Rios. "Chiral extrapolation of pion-pion scattering phase shifts within standard and unitarized Chiral Perturbation Theory". In: *Phys. Rev. D* 83 (2011), p. 094011.

[2345] J. Nebreda and J. R. Pelaez. "Strange and non-strange quark mass dependence of elastic light resonances from SU(3) Unitarized Chiral Perturbation Theory to one loop". In: *Phys. Rev.* D81 (2010), p. 054035.

[2346] A. Gomez Nicola, J. R. Pelaez, and G. Rios. "The Inverse Amplitude Method and Adler Zeros". In: *Phys. Rev. D* 77 (2008), p. 056006.

[2347] J.R. Pelaez and G. Rios. "Chiral extrapolation of light resonances from one and two-loop unitarized Chiral Perturbation Theory versus lattice results". In: *Phys. Rev. D* 82 (2010), p. 114002.

[2348] Jose R. Pelaez et al. "Unitarized Chiral Perturbation Theory and the Meson Spectrum". In: *AIP Conf. Proc.* 1257.1 (2010). Ed. by Volker Crede, Paul Eugenio, and Alexander Ostrovidov, pp. 141–148.

[2349] C. Hanhart, J. R. Pelaez, and G. Rios. "Remarks on pole trajectories for resonances". In: *Phys. Lett. B* 739 (2014), pp. 375–382.

[2350] Teiji Kunihiro et al. "Scalar mesons in lattice QCD". In: *Phys. Rev. D* 70 (2004), p. 034504.

[2351] Masayuki Wakayama et al. "Lattice QCD study of four-quark components of the isosinglet scalar mesons: Significance of disconnected diagrams". In: *Phys. Rev. D* 91.9 (2015), p. 094508.

[2352] Sasa Prelovsek et al. "Lattice study of light scalar tetraquarks with I=0,2,1/2,3/2: Are $\sigma$ and $\kappa$ tetraquarks?" In: *Phys. Rev. D* 82 (2010), p. 094507.

[2353] Gumaro Rendon et al. "$I = 1/2$ $S$-wave and $P$-wave $K\pi$ scattering and the $\kappa$ and $K^*$ resonances from lattice QCD". In: *Phys. Rev. D* 102.11 (2020), p. 114520.

[2354] Jose A. Oller. "The Mixing angle of the lightest scalar nonet". In: *Nucl. Phys. A* 727 (2003), pp. 353–369.

[2355] *"Heavy Non-$q\bar{q}$ Mesons" in R. L. Workman, et al. (Particle Data Group).* Prog. Theor. Exp. Phys. 2022 (2022), 083C01.

[2356] *"Spectroscopy of Light Meson Resonances" in R. L. Workman, et al. (Particle Data Group).* Prog. Theor. Exp. Phys. 2022 (2022), 083C01.

[2357] Frank E. Close and Nils A. Tornqvist. "Scalar mesons above and below 1 GeV". In: *J. Phys. G* 28 (2002), R249–R267.

[2358] Claude Amsler and N. A. Tornqvist. "Mesons beyond the naive quark model". In: *Phys. Rept.* 389 (2004), pp. 61–117.

[2359] D. V. Bugg. "Four sorts of meson". In: *Phys. Rept.* 397 (2004), pp. 257–358.

[2360] C. A. Meyer and Y. Van Haarlem. "The status of exotic-quantum-number mesons". In: *Phys. Rev. C* 82 (2010), p. 025208.

[2361] Bernhard Ketzer. "Hybrid Mesons". In: *PoS* QNP2012 (2012), p. 025.

[2362] C. A. Meyer and E. S. Swanson. "Hybrid mesons". In: *Prog. Part. Nucl. Phys.* 82 (2015), pp. 21–58.

[2363] Ted Barnes. "Coloured quark and gluon constituents in the MIT bag model: A model of mesons". In: *Nucl. Phys. B* 158 (1979), pp. 171–188.

[2364] M. Flensburg, C. Peterson, and L. Skold. “Applications of an improved bag model”. In: *Z. Phys. C* 22 (1984), p. 293.

[2365] Nathan Isgur and Jack E. Paton. “A flux tube model for hadrons”. In: *Phys. Lett. B* 124 (1983), pp. 247–251.

[2366] Nathan Isgur and Jack E. Paton. “A flux-tube model for hadrons in QCD”. In: *Phys. Rev. D* 31 (1985), p. 2910.

[2367] Nathan Isgur, Richard Kokoski, and Jack Paton. “Gluonic excitations of mesons: Why they are missing and where to find them”. In: *Phys. Rev. Lett.* 54 (1985), p. 869.

[2368] Richard Kokoski and Nathan Isgur. “Meson decays by flux-tube breaking”. In: *Phys. Rev. D* 35 (1987), p. 907.

[2369] Frank E. Close and Philip R. Page. “The production and decay of hybrid mesons by flux-tube breaking”. In: *Nucl. Phys. B* 443 (1995), pp. 233–254.

[2370] Eric S. Swanson and Adam P. Szczepaniak. “Decays of hybrid mesons”. In: *Phys. Rev. D* 56 (1997), pp. 5692–5695.

[2371] Philip R. Page, Eric S. Swanson, and Adam P. Szczepaniak. “Hybrid meson decay phenomenology”. In: *Phys. Rev. D* 59 (1999), p. 034016.

[2372] D. Horn and J. Mandula. “Model of mesons with constituent gluons”. In: *Phys. Rev. D* 17 (1978), p. 898.

[2373] Morimitsu Tanimoto. “Decay patterns of $q\bar{q}g$ hybrid mesons”. In: *Phys. Lett. B* 116 (1982), pp. 198–202.

[2374] A. Le Yaouanc et al. “$q\bar{q}g$ Hybrid mesons in $\psi \to \gamma +$ hadrons”. In: *Z. Phys. C* 28 (1985), pp. 309–315.

[2375] F. Iddir et al. “$q\bar{q}g$ hybrid and $qq\bar{q}\bar{q}$ diquonium interpretation of the GAMS $1^{-+}$ resonance”. In: *Phys. Lett. B* 205 (1988), pp. 564–568.

[2376] Christian S. Fischer, Stanislav Kubrak, and Richard Williams. “Mass spectra and Regge trajectories of light mesons in the Bethe-Salpeter approach”. In: *Eur. Phys. J. A* 50 (2014), p. 126.

[2377] Zhen-Ni Xu et al. “Bethe-Salpeter kernel and properties of strange-quark mesons”. In: (2022).

[2378] *“Resonances” in R. L. Workman, et al. (Particle Data Group).* Prog. Theor. Exp. Phys. 2022 (2022), 083C01.

[2379] S. U. Chung and T. L. Trueman. “Positivity conditions on the spin density matrix: A simple parametrization”. In: *Phys. Rev. D* 11 (1975), p. 633.

[2380] Bernhard Ketzer, Boris Grube, and Dmitry Ryabchikov. “Light-meson spectroscopy with COMPASS”. In: *Prog. Part. Nucl. Phys.* 113 (2020), p. 103755.

[2381] D. Alde et al. “Evidence for a $1^{-+}$ exotic meson”. In: *Phys. Lett. B* 205 (1988), p. 397.

[2382] H. Aoyagi et al. “Study of the $\eta\pi^-$ system in the $\pi^- p$ reaction at 6.3 GeV/$c$”. In: *Phys. Lett. B* 314 (1993), pp. 246–254.

[2383] D. R. Thompson et al. “Evidence for Exotic Meson Production in the Reaction $\pi^- p \to \eta\pi^- p$ at 18 GeV/$c$”. In: *Phys. Rev. Lett.* 79 (1997), pp. 1630–1633.

[2384] A. Abele et al. “Exotic $\eta\pi$ state in $\bar{p}d$ annihilation at rest into $\pi^-\pi^0\eta p_{\text{spectator}}$”. In: *Phys. Lett. B* 423 (1998), pp. 175–184.

[2385] A. Abele et al. “Evidence for a $\pi\eta$-$P$-wave in $\bar{p}p$-annihilations at rest into $\pi^0\pi^0\eta$”. In: *Phys. Lett. B* 446 (1999), pp. 349–355.

[2386] Valery Dorofeev et al. “The $J^{PC} = 1^{-+}$ hunting season at VES”. In: *AIP Conf. Proc.* 619 (2002), pp. 143–154.

[2387] G. S. Adams et al. “Confirmation of the $1^{-+}$ meson exotics in the $\eta\pi^0$ system”. In: *Phys. Lett. B* 657 (2007), pp. 27–31.

[2388] P. Salvini et al. “$\bar{p}p$ annihilation into four charged pions at rest and in flight”. In: *Eur. Phys. J. C* 35 (2004), pp. 21–33.

[2389] W. Dünnweber and F. Meyer-Wildhagen. “Exotic States in Crystal Barrel Analyses of Annihilation Channels”. In: *AIP Conf. Proc.* 717 (2004), pp. 388–393.

[2390] G. S. Adams et al. “Observation of a New $J^{PC} = 1^{-+}$ Exotic State in the Reaction $\pi^- p \to \pi^+\pi^-\pi^- p$ at 18 GeV/$c$”. In: *Phys. Rev. Lett.* 81 (1998), pp. 5760–5763.

[2391] S. U. Chung et al. “Exotic and $q\bar{q}$ resonances in the $\pi^+\pi^-\pi^-$ system produced in $\pi^- p$ collisions at 18 GeV/$c$”. In: *Phys. Rev. D* 65 (2002), p. 072001.

[2392] M. Alekseev et al. “Observation of a $J^{PC} = 1^{-+}$ Exotic Resonance in Diffractive Dissociation of 190 GeV/$c$ $\pi^-$ into $\pi^-\pi^-\pi^+$”. In: *Phys. Rev. Lett.* 104 (2010), p. 241803.

[2393] M. G. Alexeev et al. “Exotic meson $\pi_1(1600)$ with $J^{PC} = 1^{-+}$ and its decay into $\rho(770)\pi$”. In: *Phys. Rev. D* 105 (2022), p. 012005.

[2394] E. I. Ivanov et al. “Observation of Exotic Meson Production in the Reaction $\pi^- p \to \eta'\pi^- p$ at 18 GeV/$c$”. In: *Phys. Rev. Lett.* 86 (2001), pp. 3977–3980.

[2395] Valeri Dorofeev. “New results from VES”. In: *Frascati Phys. Ser.* 15 (1999), pp. 3–12.

[2396] Yu. A. Khokhlov. "Study of $X(1600)$ $1^{-+}$ hybrid". In: *Nucl. Phys. A* 663 (2000), pp. 596–599.

[2397] D. V. Amelin et al. "Investigation of hybrid states in the VES experiment at the Institute for High Energy Physics (Protvino)". In: *Phys. Atom. Nucl.* 68 (2005). Ed. by Yu. G. Abov, pp. 359–371.

[2398] G. S. Adams et al. "Amplitude analyses of the decays $\chi_{c1} \to \eta\pi^+\pi^-$ and $\chi_{c1} \to \eta'\pi^+\pi^-$". In: *Phys. Rev. D* 84 (2011), p. 112009.

[2399] Joachim Kuhn et al. "Exotic meson production in the $f_1(1285)\pi^-$ system observed in the reaction $\pi^- p \to \eta\pi^+\pi^-\pi^- p$ at 18 GeV/$c$". In: *Phys. Lett. B* 595 (2004), pp. 109–117.

[2400] C. A. Baker et al. "Confirmation of $a_0(1450)$ and $\pi_1(1600)$ in $\bar{p}p \to \omega\pi^+\pi^-\pi^0$ at rest". In: *Phys. Lett. B* 563 (2003), pp. 140–149.

[2401] A. R. Dzierba et al. "A Partial wave analysis of the $\pi^-\pi^-\pi^+$ and $\pi^-\pi^0\pi^0$ systems and the search for a $J^{PC} = 1^{-+}$ meson". In: *Phys. Rev. D* 73 (2006), p. 072001.

[2402] C Adolph et al. "Resonance Production and $\pi\pi$ $S$-wave in $\pi^- + p \to \pi^-\pi^-\pi^+ + p_{\text{recoil}}$ at 190 GeV/$c$". In: *Phys. Rev. D* 95 (2017), p. 032004.

[2403] A. Zaitsev et al. "Study of exotic resonances in diffractive reactions". In: *Nucl. Phys. A* 675 (2000), pp. 155C–160C.

[2404] Boris Grube. "Light-Meson Spectroscopy at Lepto- and Hadroproduction Experiments". In: *18th International Conference on Hadron Spectroscopy and Structure (HADRON 2019)*. 2020, pp. 43–49.

[2405] C. Adolph et al. "Odd and even partial waves of $\eta\pi^-$ and $\eta'\pi^-$ in $\pi^- p \to \eta^{(\prime)}\pi^- p$ at 191 GeV/$c$". In: *Phys. Lett. B* 740 (2015). [Erratum: *Phys. Lett. B* 811 (2020), p. 135913], pp. 303–311.

[2406] B. Kopf et al. "Investigation of the lightest hybrid meson candidate with a coupled-channel analysis of $\bar{p}p$, $\pi^- p$ and $\pi\pi$ data". In: *Eur. Phys. J. C* 81 (2021), p. 1056.

[2407] M. Nozar et al. "Search for the Photoexcitation of Exotic Mesons in the $\pi^+\pi^+\pi^-$ System". In: *Phys. Rev. Lett.* 102 (2009), p. 102002.

[2408] Stefanie Grabmüller. "Cryogenic Silicon Detectors and Analysis of Primakoff Contributions to the Reaction $\pi^-$Pb $\to \pi^-\pi^-\pi^+$Pb at COMPASS". CERN-THESIS-2012-170. PhD thesis. Technische Universität München, 2012.

[2409] M. Ablikim et al. "Observation of an Isoscalar Resonance with Exotic $J^{PC} = 1^{-+}$ Quantum Numbers in $J/\psi \to \gamma\eta\eta'$". In: *Phys. Rev. Lett.* 129.19 (2022), p. 192002.

[2410] M. Ablikim et al. "Partial wave analysis of $J/\psi \to \gamma\eta\eta'$". In: *Phys. Rev. D* 106.7 (2022), p. 072012.

[2411] P. M. Zerwas and H. A. Kastrup, eds. *Proceedings, QCD : 20 Years Later : Aachen, June 9-13, 1992*. World Scientific, June 1993.

[2412] Clemens A. Heusch. "Gluonium: An Unfulfilled promise of QCD?" In: *Workshop on QCD: 20 Years Later*. Nov. 1992, pp. 555–574.

[2413] D. L. Scharre et al. "Observation of the Radiative Transition $\psi \to \gamma E(1420)$". In: *Phys. Lett. B* 97 (1980), pp. 329–332.

[2414] C. Edwards et al. "Observation of a Pseudoscalar State at 1440-MeV in $J/\psi$ Radiative Decays". In: *Phys. Rev. Lett.* 49 (1982). [Erratum: Phys.Rev.Lett. 50, 219 (1983)], p. 259.

[2415] C. Edwards et al. "Observation of an $\eta\eta$ Resonance in $J/\psi$ Radiative Decays". In: *Phys. Rev. Lett.* 48 (1982), p. 458.

[2416] W. Dunwoodie. "$J/\psi$ radiative decay to two pseudoscalar mesons from MARK III". In: *AIP Conf. Proc.* 432.1 (1998). Ed. by S. U. Chung and H. J. Willutzki, pp. 753–757.

[2417] F. G. Binon et al. "$G(1590)$: A Scalar Meson Decaying Into Two $\eta$ Mesons". In: *Nuovo Cim. A* 78 (1983), p. 313.

[2418] A. Etkin et al. "The Reaction $\pi^- p \to \phi\phi n$ and Evidence for Glueballs". In: *Phys. Rev. Lett.* 49 (1982), p. 1620.

[2419] E. Klempt. In: *Phys. Lett. B* 308 (1993), pp. 179–185.

[2420] Eberhard Klempt, Chris Batty, and Jean-Marc Richard. "The Antinucleon-nucleon interaction at low energy : Annihilation dynamics". In: *Phys. Rept.* 413 (2005), pp. 197–317.

[2421] K. Johnson. "The M.I.T. Bag Model". In: *Acta Phys. Polon. B* 6 (1975), p. 865.

[2422] D. Robson. "A Basic Guide for the Glueball Spotter". In: *Nucl. Phys. B* 130 (1977), pp. 328–348.

[2423] Heinz J. Rothe. Vol. 82 (2012). World Scientific Publishing Company, 2012, pp. 1–606.

[2424] Y. Chen et al. "Glueball spectrum and matrix elements on anisotropic lattices". In: *Phys. Rev. D* 73 (2006), p. 014516.

[2425] Colin J. Morningstar and Mike J. Peardon. "The Glueball spectrum from an anisotropic lattice study". In: *Phys. Rev. D* 60 (1999), p. 034509.

[2426] Andreas Athenodorou and Michael Teper. "The glueball spectrum of SU(3) gauge theory in 3 + 1 dimensions". In: *JHEP* 11 (2020), p. 172.

[2427] E. Gregory et al. "Towards the glueball spectrum from unquenched lattice QCD". In: *JHEP* 10 (2012), p. 170.

[2428] Adam P. Szczepaniak and Eric S. Swanson. "The Low lying glueball spectrum". In: *Phys. Lett. B* 577 (2003), pp. 61–66.

[2429] Hua-Xing Chen, Wei Chen, and Shi-Lin Zhu. "Two- and three-gluon glueballs of $C = +$". In: *Phys. Rev. D* 104.9 (2021), p. 094050.

[2430] Markus Q. Huber, Christian S. Fischer, and Helios Sanchis-Alepuz. "Higher spin glueballs from functional methods". In: *Eur. Phys. J. C* 81.12 (2021). [Erratum: Eur.Phys.J.C 82, 38 (2022)], p. 1083.

[2431] Markus Q. Huber, Christian S. Fischer, and Hèlios Sanchis-Alepuz. "Spectrum of scalar and pseudoscalar glueballs from functional methods". In: *Eur. Phys. J. C* 80.11 (2020), p. 1077.

[2432] Stephan Narison. "Masses, decays and mixings of gluonia in QCD". In: *Nucl. Phys. B* 509 (1998), pp. 312–356.

[2433] J. Sexton, A. Vaccarino, and D. Weingarten. "Coupling constants for scalar glueball decay". In: *Nucl. Phys. B Proc. Suppl.* 47 (1996). Ed. by T. D. Kieu, B. H. J. McKellar, and A. J. Guttmann, pp. 128–135.

[2434] Masaharu Iwasaki et al. "A Flux tube model for glueballs". In: *Phys. Rev. D* 68 (2003), p. 074007.

[2435] Pedro Bicudo et al. "The BES $f_0(1810)$: A New Glueball Candidate". In: *Eur. Phys. J. C* 52 (2007), pp. 363–374.

[2436] Long-Cheng Gui et al. "Scalar Glueball in Radiative $J/\psi$ Decay on the Lattice". In: *Phys. Rev. Lett.* 110.2 (2013), p. 021601.

[2437] Ying Chen et al. "Glueballs in charmonia radiative decays". In: *PoS* LATTICE2013 (2014), p. 435.

[2438] Long-Cheng Gui et al. "Study of the pseudoscalar glueball in $J/\psi$ radiative decays". In: *Phys. Rev. D* 100.5 (2019), p. 054511.

[2439] Eberhard Klempt and Andrey V. Sarantsev. "Singlet-octet-glueball mixing of scalar mesons". In: *Phys. Lett. B* 826 (2022), p. 136906.

[2440] Claude Amsler and Frank E. Close. "Evidence for a scalar glueball". In: *Phys. Lett. B* 353 (1995), pp. 385–390.

[2441] Claude Amsler and Frank E. Close. "Is $f_0(1500)$ a scalar glueball?" In: *Phys. Rev. D* 53 (1996), pp. 295–311.

[2442] A. V. Sarantsev et al. "Scalar isoscalar mesons and the scalar glueball from radiative $J/\psi$ decays". In: *Phys. Lett. B* 816 (2021), p. 136227.

[2443] A. Rodas et al. "Scalar and tensor resonances in $J/\psi$ radiative decays". In: *Eur. Phys. J. C* 82.1 (2022), p. 80.

[2444] Eberhard Klempt. "Scalar mesons and the fragmented glueball". In: *Phys. Lett. B* 820 (2021), p. 136512.

[2445] M. Ablikim et al. "Amplitude analysis of the $\pi^0\pi^0$ system produced in radiative $J/\psi$ decays". In: *Phys. Rev. D* 92.5 (2015). [Erratum: Phys.Rev.D 93, 039906 (2016)], p. 052003.

[2446] M. Ablikim et al. "Amplitude analysis of the $K_S K_S$ system produced in radiative $J/\psi$ decays". In: *Phys. Rev. D* 98.7 (2018), p. 072003.

[2447] M. Ablikim et al. "Partial wave analysis of $J/\psi \to \gamma\eta'\eta'$". In: *Phys. Rev. D* 105.7 (2022), p. 072002.

[2448] E. Klempt et al. "Scalar mesons in a relativistic quark model with instanton induced forces". In: *Phys. Lett. B* 361 (1995), pp. 160–166.

[2449] Roel Aaij et al. "Measurement of resonant and CP components in $\bar{B}^0_s \to J/\psi\pi^+\pi^-$ decays". In: *Phys. Rev. D* 89.9 (2014), p. 092006.

[2450] Roel Aaij et al. "Resonances and $CP$ violation in $B^0_s$ and $\overline{B}^0_s \to J/\psi K^+K^-$ decays in the mass region above the $\phi(1020)$". In: *JHEP* 08 (2017), p. 037.

[2451] A. V. Sarantsev and E. Klempt. "Scalar and tensor mesons in $d\bar{d}$, $s\bar{s}$ and $gg \to f_0, f_2$". In: (Nov. 2022).

[2452] Stefan Ropertz, Christoph Hanhart, and Bastian Kubis. "A new parametrization for the scalar pion form factors". In: *Eur. Phys. J. C* 78.12 (2018), p. 1000.

[2453] E. Klempt et al. "Search for the tensor glueball". In: *Phys. Lett. B* 830 (2022), p. 137171.

[2454] M. Ablikim et al. "Observation of a State $X(2600)$ in the $\pi^+\pi^-\eta'$ System in the Process $J/\psi \to \gamma\pi^+\pi^-\eta'$". In: *Phys. Rev. Lett.* 129.4 (2022), p. 042001.

[2455] S. Dobbs et al. "Comprehensive Study of the Radiative Decays of $J/\psi$ and $\psi(2S)$ to Pseudoscalar Meson Pairs, and Search for Glueballs". In: *Phys. Rev. D* 91.5 (2015), p. 052006.

[2456] J. P. Lees et al. "Study of $\Upsilon(1S)$ radiative decays to $\gamma\pi^+\pi^-$ and $\gamma K^+K^-$". In: *Phys. Rev. D* 97.11 (2018), p. 112006.

[2457] Ruilin Zhu. "Factorization for radiative heavy quarkonium decays into scalar Glueball". In: *JHEP* 09 (2015), p. 166.

[2458] X. G. He, H. Y. Jin, and J. P. Ma. "Radiative decay of $\Upsilon$ into a scalar glueball". In: *Phys. Rev. D* 66 (2002), p. 074015.

[2459] B. Aubert and others [BaBar collaboration]. "Observation of a narrow meson decaying to $D_s^+\pi^0$ at a mass of 2.32 GeV/c$^2$". In: *Phys. Rev. Lett.* 90 (2003), p. 242001.

[2460] S. K. Choi and others [Belle collaboration]. "Observation of a narrow charmonium-like state in exclusive $B^{\pm} \to K^{\pm}\pi^{+}\pi^{-}J/\psi$ decays". In: *Phys. Rev. Lett.* 91 (2003), p. 262001.

[2461] T. Nakano and others [LEPS collaboration]. "Evidence for a narrow $S = +1$ baryon resonance in photoproduction from the neutron". In: *Phys. Rev. Lett.* 91 (2003), p. 012002.

[2462] K. H. Hicks. "On the conundrum of the pentaquark". In: *Eur. Phys. J. H* 37 (2012), pp. 1–31.

[2463] F. E. Close and P. R. Page. "The $D^{*0}\bar{D}^{0}$ threshold resonance". In: *Phys. Lett. B* 578 (2004), pp. 119–123.

[2464] L. Maiani et al. "Diquark-antidiquarks with hidden or open charm and the nature of $X(3872)$". In: *Phys. Rev. D* 71 (2005), p. 014028.

[2465] S. Dubynskiy and M. B. Voloshin. "Hadro-Charmonium". In: *Phys. Lett. B* 666 (2008), pp. 344–346.

[2466] D. V. Bugg. "An explanation of Belle states $Z_b(10610)$ and $Z_b(10650)$". In: *EPL* 96.1 (2011), p. 11002.

[2467] P. Pakhlov and T. Uglov. "Charged charmonium-like $\mathrm{Z}^{+}(4430)$ from rescattering in conventional $B$ decays". In: *Phys. Lett. B* 748 (2015), pp. 183–186.

[2468] [LHCb collaboration]. "Exotic hadron naming convention". In: (June 2022).

[2469] D. Acosta and others [CDF collaboration]. "Observation of the narrow state $X(3872) \to J/\psi\pi^{+}\pi^{-}$ in $\bar{p}p$ collisions at $\sqrt{s} = 1.96$ TeV". In: *Phys. Rev. Lett.* 93 (2004), p. 072001.

[2470] V. M. Abazov and others [D0 collaboration]. "Observation and properties of the $X(3872)$ decaying to $J/\psi\pi^{+}\pi^{-}$ in $p\bar{p}$ collisions at $\sqrt{s} = 1.96$ TeV". In: *Phys. Rev. Lett.* 93 (2004), p. 162002.

[2471] B. Aubert and others [BaBar collaboration]. "Study of the $B^{-} \to J/\psi K^{-}\pi^{+}\pi^{-}$ decay and measurement of the $B^{-} \to X(3872)K^{-}$ branching fraction". In: *Phys. Rev. D* 71 (2005), p. 071103.

[2472] R. Aaij and others [LHCb collaboration]. "Observation of $X(3872)$ production in $pp$ collisions at $\sqrt{s} = 7$ TeV". In: *Eur. Phys. J. C* 72 (2012), p. 1972.

[2473] S. Chatrchyan and others [CMS collaboration]. "Measurement of the $X(3872)$ production cross section via decays to $J/\psi\pi^{+}\pi^{-}$ in $pp$ collisions at $\sqrt{s} = 7$ TeV". In: *JHEP* 04 (2013), p. 154.

[2474] M. Aaboud and others [ATLAS collaboration]. "Measurements of $\psi(2S)$ and $X(3872) \to J/\psi\pi^{+}\pi^{-}$ production in $pp$ collisions at $\sqrt{s} = 8$ TeV with the ATLAS detector". In: *JHEP* 01 (2017), p. 117.

[2475] M. Ablikim and others [BESIII collaboration]. "Observation of $e^{+}e^{-} \to \gamma X(3872)$ at BESIII". In: *Phys. Rev. Lett.* 112.9 (2014), p. 092001.

[2476] LHCb collaboration. "Modification of $\chi_{c1}(3872)$ and $\psi(2S)$ production in $p$Pb collisions at $\sqrt{s_{NN}} = 8.16$ TeV". In: (2022).

[2477] A. M. Sirunyan and others [CMS collaboration]. "Evidence for $X(3872)$ in Pb-Pb collisions and studies of its prompt production at $\sqrt{s_{NN}}$=5.02 TeV". In: *Phys. Rev. Lett.* 128.3 (2022), p. 032001.

[2478] A. M Sirunyan and others [CMS collaboration]. "Observation of the $B_{\mathrm{s}}^{0} \to$X(3872)$\phi$ decay". In: *Phys. Rev. Lett.* 125.15 (2020), p. 152001.

[2479] R. Aaij and others [LHCb collaboration]. "Observation of the $\Lambda_b^0 \to \chi_{c1}(3872)pK^{-}$ decay". In: *JHEP* 09 (2019), p. 028.

[2480] A. Abulencia and others [CDF collaboration]. "Measurement of the dipion mass spectrum in $X(3872) \to J/\psi\pi^{+}\pi^{-}$ decays." In: *Phys. Rev. Lett.* 96 (2006), p. 102002.

[2481] LHCb collaboration. "Observation of sizeable $\omega$ contribution to $\chi_{c1}(3872) \to \pi^{+}\pi^{-}J/\psi$ decays". In: (Apr. 2022).

[2482] P. del Amo Sanchez and others [BaBar collaboration]. "Evidence for the decay $X(3872) \to J/\psi\omega$". In: *Phys. Rev. D* 82 (2010), p. 011101.

[2483] M. Ablikim and others [BESIII collaboration]. "Study of $e^{+}e^{-} \to \gamma\omega J/\psi$ and observation of $X(3872) \to \omega J/\psi$". In: *Phys. Rev. Lett.* 122.23 (2019), p. 232002.

[2484] G. Gokhroo and others [Belle collaboration]. "Observation of a near-threshold $D^{0}\bar{D}^{0}\pi^{0}$ enhancement in $B \to D^{0}\bar{D}^{0}\pi^{0}K$ decay". In: *Phys. Rev. Lett.* 97 (2006), p. 162002.

[2485] T. Aushev and others [Belle collaboration]. "Study of the $B \to X(3872)(D^{*0}\bar{D}^{0})K$ decay". In: *Phys. Rev. D* 81 (2010), p. 031103.

[2486] M. Ablikim and others [BESIII collaboration]. "Observation of the decay $X(3872) \to \pi^{0}\chi_{c1}(1P)$". In: *Phys. Rev. Lett.* 122.20 (2019), p. 202001.

[2487] V. Bhardwaj and others [Belle collaboration]. "Observation of $X(3872) \to J/\psi\gamma$ and search for $X(3872) \to \psi'\gamma$ in B decays". In: *Phys. Rev. Lett.* 107 (2011), p. 091803.

[2488] R. Aaij and others [LHCb collaboration]. "Evidence for the decay $X(3872) \to \psi(2S)\gamma$". In: *Nucl. Phys. B* 886 (2014), pp. 665–680.

[2489] B. Aubert and others [BaBar collaboration]. "Evidence for $X(3872) \to \psi(2S)\gamma$ in $B^{\pm} \to X(3872)K^{\pm}$ decays, and a study of $B \to c\bar{c}\gamma K$". In: *Phys. Rev. Lett.* 102 (2009), p. 132001.

[2490] M. Ablikim and others [BESIII collaboration]. "Study of open-charm decays and radiative transitions of the $X(3872)$". In: *Phys. Rev. Lett.* 124.24 (2020), p. 242001.

[2491] E. S. Swanson. "Diagnostic decays of the $X(3872)$". In: *Phys. Lett. B* 598 (2004), pp. 197–202.

[2492] J. Ferretti, G. Galatà, and E. Santopinto. "Quark structure of the $X(3872)$ and $\chi_b(3P)$ resonances". In: *Phys. Rev. D* 90.5 (2014), p. 054010.

[2493] T. Barnes, S. Godfrey, and E. S. Swanson. "Higher charmonia". In: *Phys. Rev. D* 72 (2005), p. 054026.

[2494] B.-Q. Li and K.-T. Chao. "Higher charmonia and X, Y, Z states with screened potential". In: *Phys. Rev. D* 79 (2009), p. 094004.

[2495] A. Abulencia and others [CDF collaboration]. "Analysis of the quantum numbers $J^{PC}$ of the $X(3872)$". In: *Phys. Rev. Lett.* 98 (2007), p. 132002.

[2496] R Aaij and others [LHCb collaboration]. "Determination of the $X(3872)$ meson quantum numbers". In: *Phys. Rev. Lett.* 110 (2013), p. 222001.

[2497] R. Aaij and others [LHCb collaboration]. "Quantum numbers of the $X(3872)$ state and orbital angular momentum in its $\rho^0 J\psi$ decay". In: *Phys. Rev. D* 92.1 (2015), p. 011102.

[2498] N. A. Tornqvist. "Isospin breaking of the narrow charmonium state of Belle at 3872 MeV as a deuson". In: *Phys. Lett. B* 590 (2004), pp. 209–215.

[2499] R. Aaij and others [LHCb collaboration]. "Study of the $\psi_2(3823)$ and $\chi_{c1}(3872)$ states in $B^+ \to (J\psi\pi^+\pi^-)K^+$ decays". In: *JHEP* 08 (2020), p. 123.

[2500] R. Aaij and others [LHCb collaboration]. "Study of the lineshape of the $\chi_{c1}(3872)$ state". In: *Phys. Rev. D* 102.9 (2020), p. 092005.

[2501] E. Braaten, H.-W. Hammer, and T. Mehen. "Scattering of an ultrasoft pion and the $X(3872)$". In: *Phys. Rev. D* 82 (2010), p. 034018.

[2502] R. Aaij and others [LHCb collaboration]. "Observation of multiplicity dependent prompt $\chi_{c1}(3872)$ and $\psi(2S)$ production in *pp* collisions". In: *Phys. Rev. Lett.* 126.9 (2021), p. 092001.

[2503] A. Esposito et al. "The nature of $X(3872)$ from high-multiplicity pp collisions". In: *Eur. Phys. J. C* 81.7 (2021), p. 669.

[2504] V. M. Abazov and others [D0 collaboration]. "Studies of $X(3872)$ and $\psi(2S)$ production in $p\bar{p}$ collisions at 1.96 TeV". In: *Phys. Rev. D* 102.7 (2020), p. 072005.

[2505] E. Braaten, L. P. He, and K. Ingles. "Production of $X(3872)$ accompanied by a soft pion at hadron colliders". In: *Phys. Rev. D* 100.9 (2019), p. 094006.

[2506] R. Aaij and others [LHCb collaboration]. "Measurement of $\chi_{c1}(3872)$ production in proton-proton collisions at $\sqrt{s} = 8$ and 13 TeV". In: *JHEP* 01 (2022), p. 131.

[2507] A. Esposito et al. "Observation of light nuclei at ALICE and the $X(3872)$ conundrum". In: *Phys. Rev. D* 92.3 (2015), p. 034028.

[2508] J. Adam and others [ALICE collaboration]. "Production of light nuclei and anti-nuclei in pp and Pb-Pb collisions at energies available at the CERN Large Hadron Collider". In: *Phys. Rev. C* 93.2 (2016), p. 024917.

[2509] C. Hanhart, Yu. S. Kalashnikova, and A. V. Nefediev. "Interplay of quark and meson degrees of freedom in a near-threshold resonance: multi-channel case". In: *Eur. Phys. J. A* 47 (2011), pp. 101–110.

[2510] B. Aubert and others [BaBar collaboration]. "Search for a charged partner of the $X(3872)$ in the $B$ meson decay $B \to X^- K$, $X^- \to J/\psi\pi^-\pi^0$". In: *Phys. Rev. D* 71 (2005), p. 031501.

[2511] S. -K. Choi and others [Belle collaboration]. "Bounds on the width, mass difference and other properties of $X(3872) \to \pi^+\pi^- J/\psi$ decays". In: *Phys. Rev. D* 84 (2011), p. 052004.

[2512] R. Aaij and others [LHCb collaboration]. "Study of the doubly charmed tetraquark $T_{cc}^+$". In: *Nature Commun.* 13.1 (2022), p. 3351.

[2513] L. Maiani, A. D. Polosa, and V. Riquer. "The new resonances $Z_{cs}(3985)$ and $Z_{cs}(4003)$ (almost) fill two tetraquark nonets of broken $\mathrm{SU}(3)_f$". In: *Sci. Bull.* 66 (2021), pp. 1616–1619.

[2514] M. Ablikim and others [BESIII collaboration]. "Observation of a near-threshold structure in the $K^+$ recoil-mass spectra in $e^+e^- \to K^+(D_s^- D^{*0} + D_s^{*-} D^0)$". In: *Phys. Rev. Lett.* 126.10 (2021), p. 102001.

[2515] R. Aaij and others [LHCb collaboration]. "Observation of new resonances decaying to $J/\psi K^+$ and $J/\psi\phi$". In: *Phys. Rev. Lett.* 127.8 (2021), p. 082001.

[2516] K. Terasaki. "$X(3872)$ and its iso-triplet partners". In: *Prog. Theor. Phys.* 127 (2012), pp. 577–582.

[2517] B. Aubert and others [BaBar collaboration]. "Observation of the decay $B \to J/\psi\eta K$ and search for $X(3872) \to J/\psi\eta$". In: *Phys. Rev. Lett.* 93 (2004), p. 041801.

[2518] T. Iwashita and others [Belle collaboration]. "Measurement of branching fractions for $B \to J/\psi\eta K$ decays and search for a narrow resonance in the $J/\psi\eta$ final state". In: *PTEP* 2014.4 (2014), p. 043C01.

[2519] R. Aaij and others [LHCb collaboration]. "Study of charmonium and charmonium-like contributions in $B^+ \to J/\psi\eta K^+$ decays". In: *JHEP* 22 (2020), p. 046.

[2520] V. Bhardwaj and others [Belle collaboration]. "Evidence of a new narrow resonance decaying to $\chi_{c1}\gamma$ in $B \to \chi_{c1}\gamma K$". In: *Phys. Rev. Lett.* 111.3 (2013), p. 032001.

[2521] M. Aghasyan and others [COMPASS collaboration]. "Search for muoproduction of $X(3872)$ at COMPASS and indication of a new state $\widetilde{X}(3872)$". In: *Phys. Lett. B* 783 (2018), pp. 334–340.

[2522] S. Chatrchyan and others [CMS collaboration]. "Search for a new bottomonium state decaying to $\Upsilon(1S)\pi^+\pi^-$ in pp collisions at $\sqrt{s}$ = 8 TeV". In: *Phys. Lett. B* 727 (2013), pp. 57–76.

[2523] G. Aad and others [ATLAS collaboration]. "Search for the $X_b$ and other hidden-beauty states in the $\pi^+\pi^-\Upsilon$(1S) channel at ATLAS". In: *Phys. Lett. B* 740 (2015), pp. 199–217.

[2524] I. Adachi and others ['Belle-II collaboration]. "Observation of $e^+e^- \to \omega\chi_{bJ}(1P)$ and search for $X_b \to \omega\Upsilon(1S)$ at $\sqrt{s}$ near 10.75 GeV". In: (Aug. 2022).

[2525] S. K. Choi and others [Belle collaboration]. "Observation of a resonance-like structure in the $\pi^\pm\psi'$ mass distribution in exclusive $B \to K\pi^\pm\psi'$ decays". In: *Phys. Rev. Lett.* 100 (2008), p. 142001.

[2526] B. Aubert and others [BaBar collaboration]. "Search for the $Z(4430)^-$ at BABAR". In: *Phys. Rev. D* 79 (2009), p. 112001.

[2527] R. Mizuk and others [Belle collaboration]. "Dalitz analysis of $B \to K\pi^+\psi'$ decays and the $Z(4430)^+$". In: *Phys. Rev. D* 80 (2009), p. 031104.

[2528] K. Chilikin and others [Belle collaboration]. "Experimental constraints on the spin and parity of the $Z(4430)^+$". In: *Phys. Rev. D* 88.7 (2013), p. 074026.

[2529] R. Aaij and others [LHCb collaboration]. "Observation of the resonant character of the $Z(4430)^-$ state". In: *Phys. Rev. Lett.* 112.22 (2014), p. 222002.

[2530] R. Aaij and others [LHCb collaboration]. "Model-independent confirmation of the $Z(4430)^-$ state". In: *Phys. Rev. D* 92.11 (2015), p. 112009.

[2531] K. Chilikin and others [Belle collaboration]. "Observation of a new charged charmoniumlike state in $\bar{B}^0 \to J/\psi K^-\pi^+$ decays". In: *Phys. Rev. D* 90.11 (2014), p. 112009.

[2532] M. Ablikim and others [BESIII collaboration]. "Observation of $e^+e^- \to \pi^0\pi^0 h_c$ and a neutral charmoniumlike structure $Z_c(4020)^0$". In: *Phys. Rev. Lett.* 113.21 (2014), p. 212002.

[2533] M. Ablikim and others [BESIII collaboration]. "Observation of a neutral charmoniumlike state $Z_c(4025)^0$ in $e^+e^- \to (D^*\bar{D}^*)^0\pi^0$". In: *Phys. Rev. Lett.* 115.18 (2015), p. 182002.

[2534] M. Ablikim and others [BESIII collaboration]. "Observation of a charged charmoniumlike structure in $e^+e^- \to \pi^+\pi^- J/\psi$ at $\sqrt{s}$ =4.26 GeV". In: *Phys. Rev. Lett.* 110 (2013), p. 252001.

[2535] Z. Q. Liu and others [Belle collaboration]. "Study of $e^+e^- \to \pi^+\pi^- J/\psi$ and observation of a charged charmoniumlike state at Belle". In: *Phys. Rev. Lett.* 110 (2013). [Erratum: Phys.Rev.Lett. 111, 019901 (2013)], p. 252002.

[2536] M. Ablikim and other [BESIII collaboration]. "Determination of the spin and parity of the $Z_c(3900)$". In: *Phys. Rev. Lett.* 119.7 (2017), p. 072001.

[2537] M. Ablikim and others [BESIII collaboration]. "Observation of a charged $(D\bar{D}^*)^\pm$ mass peak in $e^+e^- \to \pi D\bar{D}^*$ at $\sqrt{s}$ = 4.26 GeV". In: *Phys. Rev. Lett.* 112.2 (2014), p. 022001.

[2538] M. Ablikim and others [BESIII collaboration]. "Confirmation of a charged charmoniumlike state $Z_c(3885)^\mp$ in $e^+e^- \to \pi^\pm(D\bar{D}^*)^\mp$ with double $D$ tag". In: *Phys. Rev. D* 92.9 (2015), p. 092006.

[2539] M. Ablikim and others [BESIII collaboration]. "Observation of a charged charmoniumlike structure $Z_c(4020)$ and search for the $Z_c(3900)$ in $e^+e^- \to \pi^+\pi^- h_c$". In: *Phys. Rev. Lett.* 111.24 (2013), p. 242001.

[2540] M. Ablikim and others [BESIII collaboration]. "Observation of a charged charmoniumlike structure in $e^+e^- \to (D^*\bar{D}^*)^\pm\pi^\mp$ at $\sqrt{s}$ = 4.26 GeV". In: *Phys. Rev. Lett.* 112.13 (2014), p. 132001.

[2541] R. Mizuk and others [Belle collaboration]. "Observation of two resonance-like structures in the $\pi^+\chi_{c1}$ mass distribution in exclusive $\bar{B}^0 \to K^-\pi^+\chi_{c1}$ decays". In: *Phys. Rev. D* 78 (2008), p. 072004.

[2542] X. L. Wang and others [Belle collaboration]. "Measurement of $e^+e^- \to \pi^+\pi^-\psi(2S)$ via initial state radiation at Belle". In: *Phys. Rev. D* 91 (2015), p. 112007.

[2543] R. Aaij and others [LHCb collaboration]. "Evidence for an $\eta_c(1S)\pi^-$ resonance in $B^0 \to \eta_c(1S)K^+\pi^-$ decays". In: *Eur. Phys. J. C* 78.12 (2018), p. 1019.

[2544] A. Bondar and others [Belle collaboration]. "Observation of two charged bottomonium-like resonances in $\Upsilon(5S)$ decays". In: *Phys. Rev. Lett.* 108 (2012), p. 122001.

[2545] A. Garmash and others [Belle collaboration]. "Amplitude analysis of $e^+e^- \to \Upsilon(nS)\pi^+\pi^-$ at $\sqrt{s}$ = 10.865 GeV". In: *Phys. Rev. D* 91.7 (2015), p. 072003.

[2546] A. Garmash and others [Belle collaboration]. "Observation of $Z_b(10610)$ and $Z_b(10650)$ decaying to $B$ mesons". In: *Phys. Rev. Lett.* 116.21 (2016), p. 212001.

[2547] F. -K. Guo et al. "Interplay of quark and meson degrees of freedom in near-threshold states: A practical parametrization for line shapes". In: *Phys. Rev. D* 93.7 (2016), p. 074031.

[2548] E. J. Eichten and C. Quigg. "Mesons with beauty and charm: Spectroscopy". In: *Phys. Rev. D* 49 (1994), pp. 5845–5856.

[2549] S. S. Gershtein et al. "$B_c$ spectroscopy". In: *Phys. Rev. D* 51 (1995), pp. 3613–3627.

[2550] F. Abe and others [CDF collaboration]. "Observation of the $B_c$ meson in $p\bar{p}$ collisions at $\sqrt{s} = 1.8$ TeV". In: *Phys. Rev. Lett.* 81 (1998), pp. 2432–2437.

[2551] R. Aaij and others [LHCb collaboration]. "Precision measurement of the $B_c^+$ meson mass". In: *JHEP* 07 (2020), p. 123.

[2552] R. Aaij and others [LHCb collaboration]. "Measurement of the $B_c^+$ meson lifetime using $B_c^+ \to J/\psi\mu^+\nu_\mu X$ decays". In: *Eur. Phys. J. C* 74.5 (2014), p. 2839.

[2553] R. Aaij and others [LHCb collaboration]. "Measurement of the lifetime of the $B_c^+$ meson using the $B_c^+ \to J/\psi\pi^+$ decay mode". In: *Phys. Lett. B* 742 (2015), pp. 29–37.

[2554] A. M. Sirunyan and others [CMS collaboration]. "Measurement of $b$ hadron lifetimes in $pp$ collisions at $\sqrt{s} = 8$ TeV". In: *Eur. Phys. J. C* 78.6 (2018). [Erratum: Eur.Phys.J.C 78, 561 (2018)], p. 457.

[2555] A. Tumasyan and others [CMS collaboration]. "Observation of the $B_c^+$ meson in PbPb and $pp$ collisions at $\sqrt{s_{NN}}$=5.02 TeV and measurement of its nuclear modification factor". In: *Phys. Rev. Lett.* 128.25 (2022), p. 252301.

[2556] G. Aad and others [ATLAS collaboration]. "Observation of an excited $B_c^\pm$ meson state with the ATLAS detector". In: *Phys. Rev. Lett.* 113.21 (2014), p. 212004.

[2557] A. Sirunyan and others [CMS collaboration]. "Observation of two excited $B_c^+$ States and measurement of the $\mathrm{B}_c^+(2\mathrm{S})$ mass in $pp$ collisions at $\sqrt{s} = 13$ TeV". In: *Phys. Rev. Lett.* 122.13 (2019), p. 132001.

[2558] R. Aaij and others [LHCb collaboration]. "Observation of an excited $B_c^+$ state". In: *Phys. Rev. Lett.* 122.23 (2019), p. 232001.

[2559] P. G. Ortega et al. "Spectroscopy of $B_c$ mesons and the possibility of finding exotic $B_c$-like structures". In: *Eur. Phys. J. C* 80.3 (2020), p. 223.

[2560] Y. Ikeda et al. "Charmed tetraquarks $T_{cc}$ and $T_{cs}$ from dynamical lattice QCD simulations". In: *Phys. Lett. B* 729 (2014), pp. 85–90.

[2561] R. Aaij and others [LHCb collaboration]. "Observation of the doubly charmed baryon $\Xi_{cc}^{++}$". In: *Phys. Rev. Lett.* 119.11 (2017), p. 112001.

[2562] R. Aaij and others [LHCb collaboration]. "Observation of the doubly charmed baryon decay $\Xi_{cc}^{++} \to \Xi_c'^{+}\pi^+$". In: *JHEP* 05 (2022), p. 038.

[2563] R. Aaij and others [LHCb collaboration]. "Measurement of the lifetime of the doubly charmed baryon $\Xi_{cc}^{++}$". In: *Phys. Rev. Lett.* 121.5 (2018), p. 052002.

[2564] R. Aaij and others [LHCb collaboration]. "Precision measurement of the $\Xi_{cc}^{++}$ mass". In: *JHEP* 02 (2020), p. 049.

[2565] R. Aaij and others [LHCb collaboration]. "Observation of structure in the $J/\psi$ -pair mass spectrum". In: *Sci. Bull.* 65.23 (2020), pp. 1983–1993.

[2566] CMS collaboration. "Observation of new structures in the J/$\psi$J/$\psi$ mass spectrum in pp collisions at $\sqrt{s} = 13$ TeV". In: *CMS-PAS-BPH-21-003* (2022).

[2567] ATLAS collaboration. "Observation of an excess of di-charmonium events in the four-muon final state with the ATLAS detector". In: *ATLAS-CONF-2022-040* (2022).

[2568] H. A. Bethe. "Theory of the effective range in nuclear scattering". In: *Phys. Rev.* 76 (1949), pp. 38–50.

[2569] S. Weinberg. "Evidence that the deuteron is not an elementary particle". In: *Phys. Rev.* 137 (1965), B672–B678.

[2570] J. P. Ader, J. M. Richard, and P. Taxil. "Do narrow heavy multiquark states exist?" In: *Phys. Rev. D* 25 (1982), p. 2370.

[2571] M. Karliner, S. Nussinov, and J. L. Rosner. "$QQ\bar{Q}\bar{Q}$ states: Masses, production, and decays". In: *Phys. Rev. D* 95.3 (2017), p. 034011.

[2572] A. Esposito et al. "Hunting for tetraquarks in ultraperipheral heavy ion collisions". In: *Phys. Rev. D* 104.11 (2021), p. 114029.

[2573] M. Mikhasenko, L. An, and R. McNulty. "The determination of the spin and parity of a vector-vector system". In: ().

[2574] R. Aaij and others [LHCb collaboration]. "Search for beautiful tetraquarks in the $\Upsilon(1S)\mu^+\mu^-$ invariant-mass spectrum". In: *JHEP* 10 (2018), p. 086.

[2575] A. M. Sirunyan and others [CMS collaboration]. "Measurement of the $\Upsilon$(1S) pair production cross section and search for resonances decaying to $\Upsilon(1\mathrm{S})\mu^+\mu^-$ in proton-proton colli-

sions at $\sqrt{s} = 13$ TeV". In: *Phys. Lett. B* 808 (2020), p. 135578.
[2576] N. Brambilla et al. "Heavy quarkonium physics". In: (Dec. 2004).
[2577] A. Andronic et al. "Heavy-flavour and quarkonium production in the LHC era: from proton–proton to heavy-ion collisions". In: *Eur. Phys. J. C* 76.3 (2016), p. 107.
[2578] Emilien Chapon et al. "Prospects for quarkonium studies at the high-luminosity LHC". In: *Prog. Part. Nucl. Phys.* 122 (2022), p. 103906.
[2579] R. Aaij and others [LHCb collaboration]. "Physics case for an LHCb Upgrade II - Opportunities in flavour physics, and beyond, in the HL-LHC era". In: (Aug. 2018).
[2580] M. Ablikim and others [BESIII collaboration]. "Future physics programme of BESIII". In: *Chin. Phys. C* 44.4 (2020), p. 040001.
[2581] G. Barucca et al. "PANDA Phase One". In: *Eur. Phys. J. A* 57.6 (2021), p. 184.
[2582] Oliver Brüning, Andrei Seryi, and Silvia Verdú-Andrés. "Electron-Hadron Colliders: EIC, LHeC and FCC-eh". In: *Front. in Phys.* 10 (2022), p. 886473.
[2583] A. Esposito et al. "From the line shape of the $X(3872)$ to its structure". In: *Phys. Rev. D* 105.3 (2022), p. L031503.
[2584] Nora Brambilla and Antonio Vairo. "Quark confinement and the hadron spectrum". In: *13th Annual HUGS AT CEBAF (HUGS 98).* May 1999, pp. 151–220.
[2585] Nora Brambilla. "Quark Nuclear Physics with Heavy Quarks". In: Apr. 2022.
[2586] Michael Creutz. "Gauge Fixing, the Transfer Matrix, and Confinement on a Lattice". In: *Phys. Rev. D* 15 (1977). Ed. by J. Julve and M. Ramón-Medrano, p. 1128.
[2587] Nora Brambilla et al. "Static Energy in (2 + 1 + 1)-Flavor Lattice QCD: Scale Setting and Charm Effects". In: (June 2022).
[2588] E. Eichten et al. "Charmonium: Comparison with experiment". In: *Phys. Rev.* D21 (1980), p. 203.
[2589] W. Lucha, F. F. Schoberl, and D. Gromes. "Bound states of quarks". In: *Phys. Rept.* 200 (1991), pp. 127–240.
[2590] M. Campostrini et al. "Dynamical Quark Effects on the Hadronic Spectrum and $Q\bar{Q}$ Potential in Lattice QCD". In: *Phys. Lett. B* 193 (1987), pp. 78–84.
[2591] A. Barchielli, E. Montaldi, and G. M. Prosperi. "On a Systematic Derivation of the Quark - Anti-quark Potential". In: *Nucl. Phys. B* 296 (1988). [Erratum: Nucl.Phys.B 303, 752 (1988)], p. 625.
[2592] N. Brambilla, P. Consoli, and G. M. Prosperi. "Consistent derivation of the quark-antiquark and three-quark potentials in a Wilson loop context". In: *Phys. Rev.* D50 (1994), p. 5878.
[2593] P. Bicudo, N. Cardoso, and M. Cardoso. "Color field densities of the quark-antiquark excited flux tubes in SU(3) lattice QCD". In: *Phys. Rev. D* 98.11 (2018), p. 114507.
[2594] Ryosuke Yanagihara and Masakiyo Kitazawa. "A study of stress-tensor distribution around the flux tube in the Abelian–Higgs model". In: *PTEP* 2019.9 (2019). [Erratum: PTEP 2020, 079201 (2020)], 093B02.
[2595] Marshall Baker et al. "The flux tube profile in full QCD". In: *PoS* LATTICE2021 (2022), p. 355.
[2596] Kazuhisa Amemiya and Hideo Suganuma. "Off diagonal gluon mass generation and infrared Abelian dominance in the maximally Abelian gauge in lattice QCD". In: *Phys. Rev. D* 60 (1999), p. 114509.
[2597] Shoichi Sasaki, Hideo Suganuma, and Hiroshi Toki. "Dual Ginzburg-Landau theory with QCD monopoles for dynamical chiral symmetry breaking". In: *Prog. Theor. Phys.* 94 (1995), pp. 373–384.
[2598] M. Baker, James S. Ball, and F. Zachariasen. "Dual QCD: A Review". In: *Phys. Rept.* 209 (1991), pp. 73–127.
[2599] Hans Gunter Dosch and Yu. A. Simonov. "The Area Law of the Wilson Loop and Vacuum Field Correlators". In: *Phys. Lett.* B205 (1988), pp. 339–344.
[2600] M. Baker et al. "Confinement: Understanding the relation between the Wilson loop and dual theories of long distance Yang-Mills theory". In: *Phys. Rev. D* 54 (1996). [Erratum: Phys.Rev.D 56, 2475 (1997)], pp. 2829–2844.
[2601] N. Brambilla and A. Vairo. "Heavy quarkonia: Wilson area law, stochastic vacuum model and dual QCD". In: *Phys. Rev. D* 55 (1997), pp. 3974–3986.
[2602] M. Baker et al. "Field strength correlators and dual effective dynamics in QCD". In: *Phys. Rev. D* 58 (1998), p. 034010.
[2603] Guillem Perez-Nadal and Joan Soto. "Effective string theory constraints on the long distance behavior of the subleading potentials". In: *Phys. Rev. D* 79 (2009), p. 114002.

[2604] Kanabu Nawa, Hideo Suganuma, and Toru Kojo. "Baryons in holographic QCD". In: *Phys. Rev. D* 75 (2007), p. 086003.

[2605] Joan Soto and Jaume Tarrús Castellà. "Effective QCD string and doubly heavy baryons". In: *Phys. Rev. D* 104 (2021), p. 074027.

[2606] G. S. Bali, K. Schilling, and C. Schlichter. "Observing long color flux tubes in SU(2) lattice gauge theory". In: *Phys. Rev. D* 51 (1995), pp. 5165–5198.

[2607] K. D. Born et al. "Spin dependence of the heavy quark potential: A QCD lattice analysis". In: *Phys. Lett. B* 329 (1994), pp. 332–337.

[2608] Gunnar S. Bali, Klaus Schilling, and Armin Wachter. "Complete O (v**2) corrections to the static interquark potential from SU(3) gauge theory". In: *Phys. Rev. D* 56 (1997), pp. 2566–2589.

[2609] "A combination of preliminary electroweak measurements and constraints on the standard model". In: (2003).

[2610] Jonna Koponen et al. "Properties of low-lying charmonia and bottomonia from lattice QCD + QED". In: *Rev. Mex. Fis. Suppl.* 3.3 (2022), p. 0308018.

[2611] A. Gray et al. "The Upsilon spectrum and m(b) from full lattice QCD". In: *Phys. Rev. D* 72 (2005), p. 094507.

[2612] John Bulava et al. "Hadron Spectroscopy with Lattice QCD". In: *2022 Snowmass Summer Study*. Mar. 2022.

[2613] David Tims et al. "Charmonium and charmed meson spectroscopy from lattice QCD". In: *PoS* LATTICE2016 (2017), p. 137.

[2614] Cian O'Hara et al. "Towards Radiative Transitions in Charmonium". In: *PoS* Lattice2016 (2016), p. 120.

[2615] Alexei Bazavov et al. "Determination of the QCD coupling from the static energy and the free energy". In: *Phys. Rev. D* 100.11 (2019), p. 114511.

[2616] Alexei Bazavov et al. "Determination of $\alpha_s$ from the QCD static energy". In: *Phys. Rev.* D86 (2012), p. 114031.

[2617] Cesar Ayala, Xabier Lobregat, and Antonio Pineda. "Determination of $\alpha(M_z)$ from an hyperasymptotic approximation to the energy of a static quark-antiquark pair". In: *JHEP* 09 (2020), p. 016.

[2618] Hiromasa Takaura et al. "Determination of $\alpha_s$ from static QCD potential: OPE with renormalon subtraction and lattice QCD". In: *JHEP* 04 (2019), p. 155.

[2619] Nora Brambilla et al. "Lattice gauge theory computation of the static force". In: *Phys. Rev. D* 105.5 (2022), p. 054514.

[2620] Antonio Vairo. "Strong coupling from the QCD static energy". In: *Mod. Phys. Lett. A* 31.34 (2016), p. 1630039.

[2621] Yan-Qing Ma, Kai Wang, and Kuang-Ta Chao. "A complete NLO calculation of the $J/\psi$ and $\psi'$ production at hadron colliders". In: *Phys. Rev. D* 84 (2011), p. 114001.

[2622] Hao Han et al. "$\Upsilon(nS)$ and $\chi_b(nP)$ production at hadron colliders in nonrelativistic QCD". In: *Phys. Rev. D* 94.1 (2016), p. 014028.

[2623] Mathias Butenschoen and Bernd A. Kniehl. "World data of J/psi production consolidate NRQCD factorization at NLO". In: *Phys. Rev. D* 84 (2011), p. 051501.

[2624] Geoffrey T. Bodwin et al. "Fragmentation contributions to $J/\psi$ production at the Tevatron and the LHC". In: *Phys. Rev. Lett.* 113.2 (2014), p. 022001.

[2625] Bin Gong et al. "Complete next-to-leading-order study on the yield and polarization of $\Upsilon(1S, 2S, 3S)$ at the Tevatron and LHC". In: *Phys. Rev. Lett.* 112.3 (2014), p. 032001.

[2626] Mathias Butenschoen and Bernd A. Kniehl. "Fits of $\psi(2S)$ NRQCD LDMEs to global hadroproduction data at NLO". In: (July 2022).

[2627] Yan-Qing Ma, Kai Wang, and Kuang-Ta Chao. "$J/\psi(\psi')$ production at the Tevatron and LHC at $\mathcal{O}(\alpha_s^4 v^4)$ in nonrelativistic QCD". In: *Phys. Rev. Lett.* 106 (2011), p. 042002.

[2628] Alexander Rothkopf, Tetsuo Hatsuda, and Shoichi Sasaki. "Complex Heavy-Quark Potential at Finite Temperature from Lattice QCD". In: *Phys. Rev. Lett.* 108 (2012), p. 162001.

[2629] Dibyendu Bala et al. "Static quark-antiquark interactions at nonzero temperature from lattice QCD". In: *Phys. Rev. D* 105.5 (2022), p. 054513.

[2630] Xiaojun Yao. "Open quantum systems for quarkonia". In: *Int. J. Mod. Phys. A* 36.20 (2021), p. 2130010.

[2631] Yukinao Akamatsu. "Quarkonium in quark–gluon plasma: Open quantum system approaches re-examined". In: *Prog. Part. Nucl. Phys.* 123 (2022), p. 103932.

[2632] M. Tanabashi et al. "Review of Particle Physics". In: *Phys. Rev. D* 98.3 (2018), p. 030001.

[2633] Nora Brambilla. "Effective Field Theories and Lattice QCD for the X Y Z frontier". In: *PoS* LATTICE2021 (2022), p. 020.

[2634] Nora Brambilla et al. “Substructure of Multiquark Hadrons (Snowmass 2021 White Paper)”. In: (Mar. 2022).

[2635] Ahmed Ali, Jens Sören Lange, and Sheldon Stone. “Exotics: Heavy Pentaquarks and Tetraquarks”. In: *Prog. Part. Nucl. Phys.* 97 (2017), pp. 123–198.

[2636] Ahmed Ali, Luciano Maiani, and Antonio D. Polosa. *Multiquark Hadrons.* Cambridge University Press, June 2019.

[2637] Richard F. Lebed, Ryan E. Mitchell, and Eric S. Swanson. “Heavy-Quark QCD Exotica”. In: *Prog. Part. Nucl. Phys.* 93 (2017), pp. 143–194.

[2638] Feng-Kun Guo et al. “Hadronic molecules”. In: *Rev. Mod. Phys.* 90.1 (2018). [Erratum: Rev.Mod.Phys. 94, 029901 (2022)], p. 015004.

[2639] Mohammad T. AlFiky, Fabrizio Gabbiani, and Alexey A. Petrov. “X(3872): Hadronic molecules in effective field theory”. In: *Phys. Lett. B* 640 (2006), pp. 238–245.

[2640] Eric Braaten and Meng Lu. “Line shapes of the X(3872)”. In: *Phys. Rev. D* 76 (2007), p. 094028.

[2641] Eric Braaten and Masaoki Kusunoki. “Low-energy universality and the new charmonium resonance at 3870-MeV”. In: *Phys. Rev. D* 69 (2004), p. 074005.

[2642] Sean Fleming and Thomas Mehen. “The decay of the $X(3872)$ into $\chi_{cJ}$ and the Operator Product Expansion in XEFT”. In: *Phys. Rev. D* 85 (2012), p. 014016.

[2643] K. Jimmy Juge, Julius Kuti, and Colin Morningstar. “Fine structure of the QCD string spectrum”. In: *Phys. Rev. Lett.* 90 (2003), p. 161601.

[2644] Rubén Oncala and Joan Soto. “Heavy Quarkonium Hybrids: Spectrum, Decay and Mixing”. In: *Phys. Rev. D* 96.1 (2017), p. 014004.

[2645] Nora Brambilla et al. “QCD spin effects in the heavy hybrid potentials and spectra”. In: *Phys. Rev. D* 101.5 (2020), p. 054040.

[2646] Nora Brambilla et al. “Spin structure of heavy-quark hybrids”. In: *Phys. Rev. D* 99.1 (2019). [Erratum: Phys.Rev.D 101, 099902 (2020)], p. 014017.

[2647] Carolin Schlosser and Marc Wagner. “Hybrid static potentials in SU(3) lattice gauge theory at small quark-antiquark separations”. In: *Phys. Rev. D* 105.5 (2022), p. 054503.

[2648] R. Bruschini and P. González. “Is $\chi$c1(3872) generated from string breaking?” In: *Phys. Rev. D* 105.5 (2022), p. 054028.

[2649] R. Bruschini and P. González. “Coupled-channel meson-meson scattering in the diabatic framework”. In: *Phys. Rev. D* 104 (2021), p. 074025.

[2650] Zohreh Davoudi et al. “Report of the Snowmass 2021 Topical Group on Lattice Gauge Theory”. In: *2022 Snowmass Summer Study.* Sept. 2022.

[2651] Mitja Sadl and Sasa Prelovsek. “Tetraquark systems $\bar{b}\bar{b}d\bar{u}$ in the static limit and lattice QCD”. In: *Phys. Rev. D* 104.11 (2021), p. 114503.

[2652] S. Prelovsek, H. Bahtiyar, and J. Petkovic. “Zb tetraquark channel from lattice QCD and Born-Oppenheimer approximation”. In: *Phys. Lett. B* 805 (2020), p. 135467.

[2653] Pedro Bicudo et al. “Bottomonium resonances with $I = 0$ from lattice QCD correlation functions with static and light quarks”. In: *Phys. Rev. D* 101.3 (2020), p. 034503.

[2654] Pedro Bicudo et al. “Doubly heavy tetraquark resonances in lattice QCD”. In: *J. Phys. Conf. Ser.* 1137.1 (2019). Ed. by Fernando Bar~ao et al., p. 012039.

[2655] M. Padmanath and S. Prelovsek. “Signature of a Doubly Charm Tetraquark Pole in DD* Scattering on the Lattice”. In: *Phys. Rev. Lett.* 129.3 (2022), p. 032002.

[2656] Simon Capstick and W. Roberts. In: *Prog. Part. Nucl. Phys.* 45 (2000), S241–S331.

[2657] Eberhard Klempt and Jean-Marc Richard. In: *Rev. Mod. Phys.* 82 (2010), pp. 1095–1153.

[2658] V. Crede and W. Roberts. In: *Rept. Prog. Phys.* 76 (2013), p. 076301.

[2659] David G. Ireland, Eugene Pasyuk, and Igor Strakovsky. In: *Prog. Part. Nucl. Phys.* 111 (2020), p. 103752.

[2660] Annika Thiel, Farah Afzal, and Yannick Wunderlich. In: *Prog. Part. Nucl. Phys.* 125 (2022), p. 103949.

[2661] M. Gell-Mann. In: *11th International Conference on High-Energy Physics.* 1962, pp. 533–542.

[2662] Nathan Isgur and Gabriel Karl. In: *Phys. Rev. D* 21 (1980), p. 3175.

[2663] Eberhard Klempt. In: *Phys. Rev. C* 66 (2002), p. 058201.

[2664] G. Karl and E. Obryk. In: *Nucl. Phys. B* 8 (1968), pp. 609–621.

[2665] Nathan Isgur and G. Karl. In: *Phys. Rev. D* 20 (1979), pp. 1191–1194.

[2666] Nathan Isgur and Gabriel Karl. “Positive Parity Excited Baryons in a Quark Model with Hyperfine Interactions”. In: *Phys. Rev. D* 19 (1979). [Erratum: Phys.Rev.D 23, 817 (1981)], p. 2653.

[2667] S. Capstick. In: *Nato Advanced Study Institute: Hadron Spectroscopy and the Confinement Problem.* June 1995, pp. 329–344.

[2668] Roman Koniuk and Nathan Isgur. In: *Phys. Rev. D* 21 (1980). [Erratum: Phys.Rev.D 23, 818 (1981)], p. 1868.
[2669] J. Carlson, J. B. Kogut, and V. R. Pandharipande. In: *Phys. Rev. D* 28 (1983), p. 2807.
[2670] Simon Capstick and Philip R. Page. In: *Phys. Rev. C* 66 (2002), p. 065204.
[2671] Nathan Isgur, Gabriel Karl, and Roman Koniuk. In: *Phys. Rev. Lett.* 41 (1978). [Erratum: Phys.Rev.Lett. 45, 1738 (1980)], p. 1269.
[2672] L. Ya. Glozman and D. O. Riska. In: *PiN Newslett.* 10 (1995), pp. 115–120.
[2673] L. Ya. Glozman and D. O. Riska. In: (Dec. 1994).
[2674] L. Ya. Glozman, Z. Papp, and Willibald Plessas. In: *Phys. Lett. B* 381 (1996), pp. 311–316.
[2675] Z. Dziembowski, M. Fabre de la Ripelle, and Gerald A. Miller. In: *Phys. Rev. C* 53 (1996), R2038–R2042.
[2676] Ulrich Loring et al. In: *Eur. Phys. J. A* 10 (2001), pp. 309–346.
[2677] Ulrich Loring, Bernard C. Metsch, and Herbert R. Petry. "The Light baryon spectrum in a relativistic quark model with instanton induced quark forces: The Nonstrange baryon spectrum and ground states". In: *Eur. Phys. J. A* 10 (2001), pp. 395–446.
[2678] Ulrich Loring, Bernard C. Metsch, and Herbert R. Petry. In: *Eur. Phys. J. A* 10 (2001), pp. 447–486.
[2679] Sascha Migura et al. In: *Eur. Phys. J. A* 28 (2006), p. 41.
[2680] Gernot Eichmann and Christian S. Fischer. In: *Few Body Syst.* 60.1 (2019). Ed. by R. Gothe et al., p. 2.
[2681] Craig D. Roberts. In: *IRMA Lect. Math. Theor. Phys.* 21 (2015), pp. 355–458.
[2682] Derek B. Leinweber. In: *Phys. Rev. D* 47 (1993), pp. 5096–5103.
[2683] Robert G. Edwards et al. In: *Phys. Rev. D* 87.5 (2013), p. 054506.
[2684] David Faiman and Archibald W. Hendry. In: *Phys. Rev.* 173 (1968), pp. 1720–1729.
[2685] D. Faiman and A. W. Hendry. In: *Phys. Rev.* 180 (1969), pp. 1609–1610.
[2686] R. Bijker, F. Iachello, and A. Leviatan. In: *Phys. Rev. D* 55 (1997), pp. 2862–2873.
[2687] R. Sartor and F. Stancu. In: *Phys. Rev. D* 34 (1986), pp. 3405–3413.
[2688] Norbert Kaiser, P. B. Siegel, and W. Weise. In: *Phys. Lett. B* 362 (1995), pp. 23–28.
[2689] K. Nakamura et al. In: *J. Phys. G* 37 (2010), p. 075021.
[2690] V. Sokhoyan et al. In: *Eur. Phys. J. A* 51.8 (2015). [Erratum: Eur.Phys.J.A 51, 187 (2015)], p. 95.
[2691] L. Ya. Glozman. In: *Phys. Lett. B* 475 (2000), pp. 329–334.
[2692] Thomas D. Cohen and Leonid Ya. Glozman. In: *Phys. Rev. D* 65 (2001), p. 016006.
[2693] A. V. Anisovich et al. In: *Phys. Lett. B* 766 (2017), pp. 357–361.
[2694] Ted Barnes and F. E. Close. In: *Phys. Lett. B* 123 (1983), pp. 89–92.
[2695] Chi-Keung Chow, Dan Pirjol, and Tung-Mow Yan. In: *Phys. Rev. D* 59 (1999), p. 056002.
[2696] L. S. Kisslinger and Z. P. Li. In: *Phys. Rev. D* 51 (1995), R5986–R5989.
[2697] Simon Capstick and Philip R. Page. In: *Phys. Rev. D* 60 (1999), p. 111501.
[2698] Nathan Isgur. "Why $N^*$'s are important". In: July 2000.
[2699] Nathan Isgur and Gabriel Karl. "Hyperfine Interactions in Negative Parity Baryons". In: *Phys. Lett. B* 72 (1977), p. 109.
[2700] Ulf-G Meißner. "Towards a theory of baryon resonances". In: *EPJ Web Conf.* 241 (2020). Ed. by R. Beck et al., p. 02003.
[2701] G. Hohler et al. *Handbook of pion nucleon scattering.* Vol. 12N1. 1979.
[2702] R. E. Cutkosky et al. "Pion - Nucleon Partial Wave Amplitudes". In: *Phys. Rev. D* 20 (1979), p. 2839.
[2703] R. A. Arndt et al. "Extended partial-wave analysis of $\pi N$ scattering data". In: *Phys. Rev. C* 74 (2006), p. 045205.
[2704] L. Tiator et al. "Eta and Etaprime Photoproduction on the Nucleon with the Isobar Model EtaMAID2018". In: *Eur. Phys. J. A* 54.12 (2018), p. 210.
[2705] V. L. Kashevarov et al. "Study of $\eta$ and $\eta'$ Photoproduction at MAMI". In: *Phys. Rev. Lett.* 118.21 (2017), p. 212001.
[2706] F. Afzal et al. "Observation of the $p\eta'$ Cusp in the New Precise Beam Asymmetry $\Sigma$ Data for $\gamma p \to p\eta$". In: *Phys. Rev. Lett.* 125.15 (2020), p. 152002.
[2707] J. Müller et al. "New data on $\vec{\gamma}\vec{p} \to \eta p$ with polarized photons and protons and their implications for $N^* \to N\eta$ decays". In: *Phys. Lett. B* 803 (2020), p. 135323.
[2708] I. Senderovich et al. "First measurement of the helicity asymmetry $E$ in $\eta$ photoproduction on the proton". In: *Phys. Lett. B* 755 (2016), pp. 64–69.

[2709] Simon Capstick and Winston Roberts. “Quasi two-body decays of nonstrange baryons”. In: *Phys. Rev. D* 49 (1994), pp. 4570–4586.

[2710] A. V. Anisovich et al. “$N^* \to N\eta'$ decays from photoproduction of $\eta'$-mesons off protons”. In: *Phys. Lett. B* 772 (2017), pp. 247–252.

[2711] A. V. Anisovich et al. “Proton-$\eta'$ interactions at threshold”. In: *Phys. Lett. B* 785 (2018), pp. 626–630.

[2712] A. V. Anisovich et al. “Properties of baryon resonances from a multichannel partial wave analysis”. In: *Eur. Phys. J. A* 48 (2012), p. 15.

[2713] M. E. McCracken et al. “Differential cross section and recoil polarization measurements for the $\gamma p \to K^+\Lambda$ reaction using CLAS at Jefferson Lab”. In: *Phys. Rev. C* 81 (2010), p. 025201.

[2714] B. Dey et al. “Differential cross sections and recoil polarizations for the reaction $\gamma p \to K^+\Sigma^0$”. In: *Phys. Rev. C* 82 (2010), p. 025202.

[2715] C. A. Paterson et al. “Photoproduction of $\Lambda$ and $\Sigma^0$ hyperons using linearly polarized photons”. In: *Phys. Rev. C* 93.6 (2016), p. 065201.

[2716] H. Osmanović et al. “Single-energy partial wave analysis for $\pi^0$ photoproduction on the proton with fixed-$t$ analyticity imposed”. In: *Phys. Rev. C* 100.5 (2019), p. 055203.

[2717] H. Osmanović et al. “Single-energy partial-wave analysis for pion photoproduction with fixed-t analyticity”. In: *Phys. Rev. C* 104.3 (2021), p. 034605.

[2718] A. V. Anisovich et al. “Strong evidence for nucleon resonances near 1900\,MeV”. In: *Phys. Rev. Lett.* 119.6 (2017), p. 062004.

[2719] J. Hartmann et al. “The $N(1520)3/2^-$ helicity amplitudes from an energy-independent multipole analysis based on new polarization data on photoproduction of neutral pions”. In: *Phys. Rev. Lett.* 113 (2014), p. 062001.

[2720] A. Švarc, Y. Wunderlich, and L. Tiator. “Application of the single-channel, single-energy amplitude and partial-wave analysis method to $K^+\Lambda$ photoproduction”. In: *Phys. Rev. C* 105.2 (2022), p. 024614.

[2721] Deborah Rönchen et al. “Light baryon resonances from a coupled-channel study including $K\Sigma$ photoproduction”. In: (July 2022).

[2722] A. V. Anisovich et al. “The impact of new polarization data from Bonn, Mainz and Jefferson Laboratory on $\gamma p \to \pi N$ multipoles”. In: *Eur. Phys. J. A* 52.9 (2016), p. 284.

[2723] Ron L. Workman et al. “Unified Chew-Mandelstam SAID analysis of pion photoproduction data”. In: *Phys. Rev. C* 86 (2012), p. 015202.

[2724] B. C. Hunt and D. M. Manley. “Updated determination of $N^*$ resonance parameters using a unitary, multichannel formalism”. In: *Phys. Rev. C* 99.5 (2019), p. 055205.

[2725] B. Julia-Diaz et al. “Dynamical coupled-channel model of $\pi N$ scattering in the $W \leq 2$-GeV nucleon resonance region”. In: *Phys. Rev. C* 76 (2007), p. 065201.

[2726] T. Seifen et al. “Polarization observables in double neutral pion photoproduction”. In: (July 2022).

[2727] H. Kamano et al. “The ANL-Osaka Partial-Wave Amplitudes of $\pi N$ and $\gamma N$ Reactions”. In: (Sept. 2019).

[2728] E. Gutz et al. “High statistics study of the reaction $\gamma p \to p\pi^0\eta$”. In: *Eur. Phys. J. A* 50 (2014), p. 74.

[2729] A. Thiel et al. “Three-body nature of $N^*$ and $\Delta^*$ resonances from sequential decay chains”. In: *Phys. Rev. Lett.* 114.9 (2015), p. 091803.

[2730] Hilmar Forkel and Eberhard Klempt. “Diquark correlations in baryon spectroscopy and holographic QCD”. In: *Phys. Lett. B* 679 (2009), pp. 77–80.

[2731] Eberhard Klempt. “Nucleon Excitations”. In: *Chin. Phys. C* 34.9 (2010), pp. 1241–1246.

[2732] E. Klempt. “Delta resonances, Quark models, chiral symmetry and AdS/QCD”. In: *Eur. Phys. J. A* 38 (2008). Ed. by Luigi Benussi et al., pp. 187–194.

[2733] S. Prakhov et al. “Measurement of $\pi^0\Lambda, \bar{K}^0 n$, and $\pi^0\Sigma^0$ production in $K^-p$ interactions for $p_{K^-}$ between 514 and 750-MeV/c”. In: *Phys. Rev. C* 80 (2009), p. 025204.

[2734] K. Moriya et al. “Measurement of the $\Sigma\pi$ photoproduction line shapes near the $\Lambda(1405)$”. In: *Phys. Rev. C* 87.3 (2013), p. 035206.

[2735] H. Zhang et al. “Partial-wave analysis of $\bar{K}N$ scattering reactions”. In: *Phys. Rev. C* 88.3 (2013), p. 035204.

[2736] H. Zhang et al. “Multichannel parametrization of $\bar{K}N$ scattering amplitudes and extraction of resonance parameters”. In: *Phys. Rev. C* 88.3 (2013), p. 035205.

[2737] C. Fernandez-Ramirez et al. “Coupled-channel model for $\bar{K}N$ scattering in the resonant region”. In: *Phys. Rev. D* 93.3 (2016), p. 034029.

[2738] H. Kamano et al. “Dynamical coupled-channels model of $K^-p$ reactions: Determination of partial-wave amplitudes”. In: *Phys. Rev. C* 90.6 (2014), p. 065204.

[2739] H. Kamano et al. “Dynamical coupled-channels model of $K^-p$ reactions. II. Extraction of $\Lambda^*$ and $\Sigma^*$ hyperon resonances”. In: *Phys. Rev. C*

92.2 (2015). [Erratum: Phys.Rev.C 95, 049903 (2017)], p. 025205.

[2740] M. Matveev et al. "Hyperon I: Partial-wave amplitudes for $K^-p$ scattering". In: *Eur. Phys. J. A* 55.10 (2019), p. 179.

[2741] A. V. Sarantsev et al. "Hyperon II: Properties of excited hyperons". In: *Eur. Phys. J. A* 55.10 (2019), p. 180.

[2742] E. Klempt et al. "$\Lambda$ and $\Sigma$ Excitations and the Quark Model". In: *Eur. Phys. J. A* 56.10 (2020), p. 261.

[2743] Philipp Mahlberg. In: *PhD-thesis Bonn, in preparation* 114.9 (2023), p. 091803.

[2744] Anthony J. G. Hey and Robert L. Kelly. "Baryon spectroscopy". In: *Phys. Rept.* 96 (1983), p. 71.

[2745] L. Ya. Glozman. "Chiral multiplets of excited mesons". In: *Phys. Lett. B* 587 (2004), pp. 69–77.

[2746] Peter C. Bruns, Maxim Mai, and Ulf G. Meissner. "Chiral dynamics of the S11(1535) and S11(1650) resonances revisited". In: *Phys. Lett. B* 697 (2011), pp. 254–259.

[2747] Maxim Mai, Peter C. Bruns, and Ulf-G. Meissner. "Pion photoproduction off the proton in a gauge-invariant chiral unitary framework". In: *Phys. Rev. D* 86 (2012), p. 094033.

[2748] Norbert Kaiser, T. Waas, and W. Weise. "SU(3) chiral dynamics with coupled channels: Eta and kaon photoproduction". In: *Nucl. Phys. A* 612 (1997), pp. 297–320.

[2749] J. A. Oller and Ulf G. Meissner. "Chiral dynamics in the presence of bound states: Kaon nucleon interactions revisited". In: *Phys. Lett. B* 500 (2001), pp. 263–272.

[2750] D. Jido et al. "Chiral dynamics of the two $\Lambda(1405)$ states". In: *Nucl. Phys. A* 725 (2003), pp. 181–200.

[2751] A. V. Anisovich et al. "Hyperon III: $K^-p-\pi\Sigma$ coupled-channel dynamics in the $\Lambda(1405)$ mass region". In: *Eur. Phys. J. A* 56.5 (2020), p. 139.

[2752] P. Stoler. In: *Phys. Rept.* 226 (1993), pp. 103–171.

[2753] V. D. Burkert and T. S. H. Lee. In: *Int. J. Mod. Phys. E* 13 (2004), pp. 1035–1112.

[2754] I. G. Aznauryan and V. D. Burkert. In: *Prog. Part. Nucl. Phys.* 67 (2012), pp. 1–54.

[2755] I. G. Aznauryan et al. In: *Int. J. Mod. Phys. E* 22 (2013), p. 1330015.

[2756] S. J. Brodsky et al. "Strong QCD from Hadron Structure Experiments: Newport News, VA, USA, November 4-8, 2019". In: *Int. J. Mod. Phys. E* 29.08 (2020), p. 2030006.

[2757] V.D. Burkert. "$N^*$ Experiments and what they tell us about Strong QCD Physics". In: *EPJ Web Conf.* 241 (2020). Ed. by R. Beck et al., p. 01004.

[2758] Kenneth M. Watson. "Some general relations between the photoproduction and scattering of pi mesons". In: *Phys. Rev.* 95 (1954), pp. 228–236.

[2759] R. L. Walker. In: *Phys. Rev.* 182 (1969), pp. 1729–1748.

[2760] Frits A. Berends, A. Donnachie, and D. L. Weaver. In: *Nucl. Phys. B* 4 (1967), pp. 1–53.

[2761] J. J. Kelly et al. In: *Phys. Rev. C* 75 (2007), p. 025201.

[2762] Sabit S. Kamalov et al. "Gamma* N —> Delta transition form-factors: A New analysis of the JLab data on p (e, e-prime p) pi0 at Q**2=(2.8-(GeV/c)**2 and 4.0-(GeV/c)**2)". In: *Phys. Rev. C* 64 (2001), p. 032201.

[2763] T. Sato and T. S. H. Lee. In: *Phys. Rev. C* 63 (2001), p. 055201.

[2764] I. G. Aznauryan et al. In: *Phys. Rev. C* 80 (2009), p. 055203.

[2765] M. Ungaro et al. In: *Phys. Rev. Lett.* 97 (2006), p. 112003.

[2766] K. Joo et al. In: *Phys. Rev. Lett.* 88 (2002), p. 122001.

[2767] V. V. Frolov et al. In: *Phys. Rev. Lett.* 82 (1999), pp. 45–48.

[2768] L. Tiator et al. "Electromagnetic Excitation of Nucleon Resonances". In: *Eur. Phys. J. ST* 198 (2011), pp. 141–170.

[2769] I. G. Aznauryan and V. D. Burkert. In: *Phys. Rev. C* 85 (2012), p. 055202.

[2770] I. G. Aznauryan and V. D. Burkert. "Configuration mixings and light-front relativistic quark model predictions for the electroexcitation of the $\Delta(1232)3/2^+$, $N(1440)1/2^+$, and $\Delta(1600)3/2^+$". In: (Mar. 2016).

[2771] Jorge Segovia et al. "Nucleon and $\Delta$ elastic and transition form factors". In: *Few Body Syst.* 55 (2014), pp. 1185–1222.

[2772] C. Alexandrou et al. In: *Phys. Rev. D* 77 (2008), p. 085012.

[2773] Klaus Behrndt and Mirjam Cvetic. In: *Phys. Rev. Lett.* 95 (2005), p. 021601.

[2774] D. Drechsel et al. In: *Nucl. Phys. A* 645 (1999), pp. 145–174.

[2775] R. D. Peccei. In: *Phys. Rev.* 181 (1969), pp. 1902–1904.

[2776] R. A. Arndt et al. In: *Phys. Rev. C* 66 (2002), p. 055213.

[2777] T. Sato and T-. S. H. Lee. In: *Phys. Rev. C* 54 (1996), pp. 2660–2684.

[2778] S. S. Kamalov and Shin Nan Yang. In: *Phys. Rev. Lett.* 83 (1999), pp. 4494–4497.

[2779] K. Park et al. "Cross sections and beam asymmetries for $\vec{e}p \to n\pi^+$ in the nucleon resonance region for $1.7 < Q^2 \leq 4.5\,(\mathrm{GeV})^2$". In: *Phys. Rev. C* 77 (2008), p. 015208.

[2780] E. Golovatch et al. "First results on nucleon resonance photocouplings from the $\gamma p \to \pi^+\pi^- p$ reaction". In: *Phys. Lett. B* 788 (2019), pp. 371–379.

[2781] V. I. Mokeev et al. "Evidence for the $N'(1720)3/2^+$ Nucleon Resonance from Combined Studies of CLAS $\pi^+\pi^- p$ Photo- and Electroproduction Data". In: *Phys. Lett. B* 805 (2020), p. 135457.

[2782] H. L. Anderson et al. In: *Phys. Rev.* 85 (1952), p. 936.

[2783] W. W. Ash et al. In: *Phys. Lett. B* 24 (1967), pp. 165–168.

[2784] T. Bauer, S. Scherer, and L. Tiator. "Electromagnetic transition form factors of the Roper resonance in effective field theory". In: *Phys. Rev. C* 90.1 (2014), p. 015201.

[2785] V. I. Mokeev et al. "Experimental Study of the $P_{11}(1440)$ and $D_{13}(1520)$ resonances from CLAS data on $ep \to e'\pi^+\pi^- p'$". In: *Phys. Rev. C* 86 (2012), p. 035203.

[2786] D. Drechsel, S. S. Kamalov, and L. Tiator. "Unitary Isobar Model - MAID2007". In: *Eur. Phys. J. A* 34 (2007), pp. 69–97.

[2787] S. Štajner et al. "Beam-Recoil Polarization Measurement of $\pi^0$ Electroproduction on the Proton in the Region of the Roper Resonance". In: *Phys. Rev. Lett.* 119.2 (2017), p. 022001.

[2788] Hovhannes R. Grigoryan, T. -S. H. Lee, and Ho-Ung Yee. In: *Phys. Rev. D* 80 (2009), p. 055006.

[2789] L. David Roper. In: *Phys. Rev. Lett.* 12 (1964), pp. 340–342.

[2790] N. Suzuki et al. In: *Phys. Rev. Lett.* 104 (2010), p. 042302.

[2791] Jorge Segovia et al. In: *Phys. Rev. Lett.* 115.17 (2015), p. 171801.

[2792] N. Mathur et al. In: *Phys. Lett. B* 605 (2005), pp. 137–143.

[2793] Huey-Wen Lin and Saul D. Cohen. In: *AIP Conf. Proc.* 1432.1 (2012). Ed. by Volker Burkert et al., pp. 305–308.

[2794] Volker D. Burkert and Craig D. Roberts. "Colloquium : Roper resonance: Toward a solution to the fifty year puzzle". In: *Rev. Mod. Phys.* 91.1 (2019), p. 011003.

[2795] Guy F. de Teramond and Stanley J. Brodsky. "Excited Baryons in Holographic QCD". In: *AIP Conf. Proc.* 1432.1 (2012). Ed. by Volker Burkert et al., pp. 168–175.

[2796] G. Ramalho and D. Melnikov. "Valence quark contributions for the $\gamma^* N \to N(1440)$ form factors from light-front holography". In: *Phys. Rev. D* 97.3 (2018), p. 034037.

[2797] M. M. Giannini, E. Santopinto, and A. Vassallo. "An Overview of the hypercentral constituent quark model". In: *Prog. Part. Nucl. Phys.* 50 (2003). Ed. by A. Faessler, pp. 263–272.

[2798] K. Bermuth et al. "Photoproduction of $\Delta$ and Roper Resonances in the Cloudy Bag Model". In: *Phys. Rev. D* 37 (1988), pp. 89–100.

[2799] I. T. Obukhovsky et al. In: *Phys. Rev. D* 84 (2011), p. 014004.

[2800] V. I. Mokeev et al. In: *Phys. Rev. C* 93.2 (2016), p. 025206.

[2801] V. M. Braun et al. "Electroproduction of the N*(1535) resonance at large momentum transfer". In: *Phys. Rev. Lett.* 103 (2009), p. 072001.

[2802] Inna G. Aznauryan and Volker Burkert. In: *Phys. Rev. C* 95.6 (2017), p. 065207.

[2803] I. G. Aznauryan and V. D. Burkert. In: *Phys. Rev. C* 92.3 (2015), p. 035211.

[2804] I. V. Anikin, V. M. Braun, and N. Offen. In: *Phys. Rev. D* 92.1 (2015), p. 014018.

[2805] D. Jido, M. Doering, and E. Oset. "Transition form factors of the N*(1535) as a dynamically generated resonance". In: *Phys. Rev. C* 77 (2008), p. 065207.

[2806] I. G. Aznauryan and V. D. Burkert. "Extracting meson-baryon contributions to the electroexcitation of the $N(1675)\frac{5}{2}^-$ nucleon resonance". In: *Phys. Rev. C* 92.1 (2015), p. 015203.

[2807] E. Santopinto and M. M. Giannini. In: *Phys. Rev. C* 86 (2012), p. 065202.

[2808] B. Julia-Diaz et al. "Dynamical coupled-channels effects on pion photoproduction". In: *Phys. Rev. C* 77 (2008), p. 045205.

[2809] Lothar Tiator and Marc Vanderhaeghen. In: *Phys. Lett. B* 672 (2009), pp. 344–348.

[2810] Carl E. Carlson and Marc Vanderhaeghen. "Empirical transverse charge densities in the nucleon and the nucleon-to-Delta transition". In: *Phys. Rev. Lett.* 100 (2008), p. 032004.

[2811] Volker D. Burkert. "N* Experiments and Their Impact on Strong QCD Physics". In: *Few Body Syst.* 59.4 (2018). Ed. by R. Gothe et al., p. 57.

[2812] A. J. G. Hey and J. Weyers. “Quarks and the helicity structure of photoproduction amplitudes”. In: *Phys. Lett. B* 48 (1974), pp. 69–72.
[2813] W. N. Cottingham and I. H. Dunbar. “Baryon Multipole Moments in the Single Quark Transition Model”. In: *Z. Phys. C* 2 (1979), p. 41.
[2814] V. D. Burkert et al. “Single quark transition model analysis of electromagnetic nucleon resonance transitions in the [70,1-] supermultiplet”. In: *Phys. Rev. C* 67 (2003), p. 035204.
[2815] G. Ramalho. “Using the Single Quark Transition Model to predict nucleon resonance amplitudes”. In: *Phys. Rev. D* 90.3 (2014), p. 033010.
[2816] Zhen-ping Li, Volker Burkert, and Zhu-jun Li. In: *Phys. Rev. D* 46 (1992), pp. 70–74.
[2817] L. Lanza and A. D’Angelo. In: *Nuovo Cim. C* 44.2-3 (2021), p. 51.
[2818] C. D. Roberts. “Hadron Properties and Dyson-Schwinger Equations”. In: *Prog. Part. Nucl. Phys.* 61 (2008). Ed. by Amand Faessler, pp. 50–65.
[2819] D. S. Carman, K. Joo, and V. I. Mokeev. “Strong QCD Insights from Excited Nucleon Structure Studies with CLAS and CLAS12”. In: *Few Body Syst.* 61.3 (2020), p. 29.
[2820] Y. Tian et al. “Exclusive $\pi^-$ Electroproduction off the Neutron in Deuterium in the Resonance Region”. In: (Mar. 2022).
[2821] Volker D. Burkert. In: *Ann. Rev. Nucl. Part. Sci.* 68 (2018), pp. 405–428.
[2822] V. D. Burkert, L. Elouadrhiri, and F. X. Girod. In: *Nature* 557.7705 (2018), pp. 396–399.
[2823] V. D. Burkert, L. Elouadrhiri, and F. X. Girod. “Determination of shear forces inside the proton”. In: (Apr. 2021).
[2824] P. Chatagnon et al. In: *Phys. Rev. Lett.* 127.26 (2021), p. 262501.
[2825] Maxim V. Polyakov and Peter Schweitzer. In: *Int. J. Mod. Phys. A* 33.26 (2018), p. 1830025.
[2826] U. Özdem and K. Azizi. In: *Phys. Rev. D* 101.5 (2020), p. 054031.
[2827] Maxim V. Polyakov and Asli Tandogan. “Comment on “Gravitational transition form factors of $N(1535) \to N$””. In: *Phys. Rev. D* 101.11 (2020), p. 118501.
[2828] Roel Aaij et al. “Observation of $J/\psi p$ Resonances Consistent with Pentaquark States in $\Lambda_b^0 \to J/\psi K^- p$ Decays”. In: *Phys. Rev. Lett.* 115 (2015), p. 072001.
[2829] Roel Aaij et al. “Observation of a narrow pentaquark state, $P_c(4312)^+$, and of two-peak structure of the $P_c(4450)^+$”. In: *Phys. Rev. Lett.* 122.22 (2019), p. 222001.
[2830] M. Mattson et al. “First Observation of the Doubly Charmed Baryon $\Xi_{cc}^+$”. In: *Phys. Rev. Lett.* 89 (2002), p. 112001.
[2831] A. Ocherashvili et al. “Confirmation of the double charm baryon $\Xi^+(cc)(3520)$ via its decay to $pD^+K^-$”. In: *Phys. Lett. B* 628 (2005), pp. 18–24.
[2832] Roel Aaij et al. “Search for the doubly charmed baryon $\Xi_{cc}^+$ in the $\Xi_c^+\pi^-\pi^+$ final state”. In: *JHEP* 12 (2021), p. 107.
[2833] Roel Aaij et al. “Search for the doubly heavy baryon $\Xi_{bc}^+$ decaying to $J/\psi\Xi_c^+$”. In: (Apr. 2022).
[2834] Bernard Aubert et al. “Observation of an excited charm baryon $\Omega_c^*$ decaying to $\Omega_c^0\gamma$”. In: *Phys. Rev. Lett.* 97 (2006). Ed. by Alexey Sissakian, Gennady Kozlov, and Elena Kolganova, p. 232001.
[2835] T. J. Moon et al. “First determination of the spin and parity of the charmed-strange baryon $\Xi_c(2970)^+$”. In: *Phys. Rev. D* 103.11 (2021), p. L111101.
[2836] Roel Aaij et al. “First observation of excited $\Omega_b^-$ states”. In: *Phys. Rev. Lett.* 124.8 (2020), p. 082002.
[2837] Hua-Xing Chen et al. “A review of the open charm and open bottom systems”. In: *Rept. Prog. Phys.* 80.7 (2017), p. 076201.
[2838] D. Ebert, R. N. Faustov, and V. O. Galkin. “Spectroscopy and Regge trajectories of heavy baryons in the relativistic quark-diquark picture”. In: *Phys. Rev. D* 84 (2011), p. 014025.
[2839] Guo-Liang Yu et al. “Systematic analysis of single heavy baryons $\Lambda_Q$, $\Sigma_Q$ and $\Omega_Q$”. In: (June 2022).
[2840] Zhen-Yu Li et al. “Systematic analysis of strange single heavy baryons”. In: (July 2022).
[2841] Sascha Migura et al. “Semileptonic decays of baryons in a relativistic quark model”. In: *Eur. Phys. J. A* 28 (2006), p. 55.
[2842] A. Valcarce, H. Garcilazo, and J. Vijande. “Heavy baryon spectroscopy with relativistic kinematics”. In: *Phys. Lett. B* 733 (2014), pp. 288–295.
[2843] Bing Chen, Ke-Wei Wei, and Ailin Zhang. “Assignments of $\Lambda_Q$ and $\Xi_Q$ baryons in the heavy quark-light diquark picture”. In: *Eur. Phys. J. A* 51 (2015), p. 82.
[2844] Rudolf N. Faustov and Vladimir O. Galkin. “Heavy Baryon Spectroscopy in the Relativistic Quark Model”. In: *Particles* 3.1 (2020), pp. 234–244.
[2845] Roel Aaij et al. “Observation of five new narrow $\Omega_c^0$ states decaying to $\Xi_c^+K^-$”. In: *Phys. Rev. Lett.* 118.18 (2017), p. 182001.

[2846] J. Yelton et al. "Observation of Excited $\Omega_c$ Charmed Baryons in $e^+e^-$ Collisions". In: *Phys. Rev. D* 97.5 (2018), p. 051102.

[2847] Yonghee Kim et al. "Heavy baryon spectrum with chiral multiplets of scalar and vector diquarks". In: *Phys. Rev. D* 104.5 (2021), p. 054012.

[2848] Hui-Min Yang et al. "Decay properties of $P$-wave bottom baryons within light-cone sum rules". In: *Eur. Phys. J. C* 80.2 (2020), p. 80.

[2849] Huseyin Bahtiyar et al. "Charmed baryon spectrum from lattice QCD near the physical point". In: *Phys. Rev. D* 102.5 (2020), p. 054513.

[2850] Juan Nieves and Rafael Pavao. "Nature of the lowest-lying odd parity charmed baryon $\Lambda_c(2595)$ and $\Lambda_c(2625)$ resonances". In: *Phys. Rev. D* 101.1 (2020), p. 014018.

[2851] J. Hofmann and M. F. M. Lutz. "D-wave baryon resonances with charm from coupled-channel dynamics". In: *Nucl. Phys. A* 776 (2006), pp. 17–51.

[2852] Jia-Jun Wu et al. "Dynamically generated $N^*$ and $\Lambda^*$ resonances in the hidden charm sector around 4.3 GeV". In: *Phys. Rev. C* 84 (2011), p. 015202.

[2853] Jia-Jun Wu, T. -S. H. Lee, and B. S. Zou. "Nucleon Resonances with Hidden Charm in Coupled-Channel Models". In: *Phys. Rev. C* 85 (2012), p. 044002.

[2854] Hua-Xing Chen et al. "The hidden-charm pentaquark and tetraquark states". In: *Phys. Rept.* 639 (2016), pp. 1–121.

[2855] Stephen Lars Olsen, Tomasz Skwarnicki, and Daria Zieminska. "Nonstandard heavy mesons and baryons: Experimental evidence". In: *Rev. Mod. Phys.* 90.1 (2018), p. 015003.

[2856] Yan-Rui Liu et al. "Pentaquark and Tetraquark states". In: *Prog. Part. Nucl. Phys.* 107 (2019), pp. 237–320.

[2857] T. J. Burns and E. S. Swanson. "Production of $P_c$ states in $\Lambda_b$ decays". In: *Phys. Rev. D* 106.5 (2022), p. 054029.

[2858] Meng-Lin Du et al. "Revisiting the nature of the $\mathrm{P_c}$ pentaquarks". In: *JHEP* 08 (2021), p. 157.

[2859] Roel Aaij et al. "Evidence for a new structure in the $J/\psi p$ and $J/\psi\bar{p}$ systems in $B_s^0 \to J/\psi p\bar{p}$ decays". In: *Phys. Rev. Lett.* 128.6 (2022), p. 062001.

[2860] Jun-Zhang Wang, Xiang Liu, and Takayuki Matsuki. "Evidence supporting the existence of $P_c(4380)^+$ from the recent measurements of $B_s \to J/\psi p\bar{p}$". In: *Phys. Rev. D* 104.11 (2021), p. 114020.

[2861] Roel Aaij et al. "Evidence of a $J/\psi\Lambda$ structure and observation of excited $\Xi^-$ states in the $\Xi_b^- \to J/\psi\Lambda K^-$ decay". In: *Sci. Bull.* 66 (2021), pp. 1278–1287.

[2862] Chen Chen and Elisabetta Spadaro Norella. "Particle Zoo 2.0: New Tetra- and Pentaquarks at LHCb". In: *CERN Seminar* (2022), July, 5.

[2863] Albert M Sirunyan et al. "Study of the $\mathrm{B}^+ \to \mathrm{J}/\psi\bar{\Lambda}\mathrm{p}$ decay in proton-proton collisions at $\sqrt{s}$ = 8 TeV". In: *JHEP* 12 (2019), p. 100.

[2864] "Observation of a $J/\psi\Lambda$ resonance consistent with a strange pentaquark candidate in $B^- \to J/\psi\Lambda\bar{p}$ decays". In: (Oct. 2022).

[2865] Feng-Kun Guo et al. "How to reveal the exotic nature of the $\mathrm{P_c}(4450)$". In: *Phys. Rev. D* 92.7 (2015), p. 071502.

[2866] Xiao-Hai Liu, Qian Wang, and Qiang Zhao. "Understanding the newly observed heavy pentaquark candidates". In: *Phys. Lett. B* 757 (2016), pp. 231–236.

[2867] Melahat Bayar et al. "A Discussion on Triangle Singularities in the $\Lambda_b \to J/\psi K^- p$ Reaction". In: *Phys. Rev. D* 94.7 (2016), p. 074039.

[2868] Feng-Kun Guo, Xiao-Hai Liu, and Shuntaro Sakai. "Threshold cusps and triangle singularities in hadronic reactions". In: *Prog. Part. Nucl. Phys.* 112 (2020), p. 103757.

[2869] Xiang-Kun Dong, Feng-Kun Guo, and Bing-Song Zou. "Explaining the Many Threshold Structures in the Heavy-Quark Hadron Spectrum". In: *Phys. Rev. Lett.* 126.15 (2021), p. 152001.

[2870] Chao-Wei Shen et al. "Exploring Possible Triangle Singularities in the $\Xi_b^- \to K^- J/\psi\Lambda$ Decay". In: *Symmetry* 12.10 (2020), p. 1611.

[2871] Satoshi X. Nakamura. "$P_c(4312)^+$, $P_c(4380)^+$, and $P_c(4457)^+$ as double triangle cusps". In: *Phys. Rev. D* 103 (2021), p. 111503.

[2872] Michael I. Eides, Victor Yu. Petrov, and Maxim V. Polyakov. "Narrow Nucleon-$\psi(2S)$ Bound State and LHCb Pentaquarks". In: *Phys. Rev. D* 93.5 (2016), p. 054039.

[2873] Feng-Kun Guo et al. "Isospin breaking decays as a diagnosis of the hadronic molecular structure of the $P_c(4457)$". In: *Phys. Rev. D* 99.9 (2019), p. 091501.

[2874] Meng-Lin Du et al. "Interpretation of the LHCb $P_c$ States as Hadronic Molecules and Hints of a Narrow $P_c(4380)$". In: *Phys. Rev. Lett.* 124.7 (2020), p. 072001.

[2875] Hao Xu et al. "Recently observed $P_c$ as molecular states and possible mixture of $P_c(4457)$". In: *Phys. Rev. D* 101.5 (2020), p. 054037.

[2876] Rui Chen. "Can the newly reported $P_{cs}(4459)$ be a strange hidden-charm $\Xi_c\bar{D}^*$ molecular pentaquark?" In: *Phys. Rev. D* 103.5 (2021), p. 054007.

[2877] Hua-Xing Chen et al. "Establishing the first hidden-charm pentaquark with strangeness". In: *Eur. Phys. J. C* 81.5 (2021), p. 409.

[2878] Qi Wu, Dian-Yong Chen, and Ran Ji. "Production of $P_{cs}(4459)$ from $\Xi_b$ Decay". In: *Chin. Phys. Lett.* 38.7 (2021), p. 071301.

[2879] Jun-Xu Lu et al. "Understanding Pcs(4459) as a hadronic molecule in the $\Xi_b^- \to J/\psi \Lambda K^-$ decay". In: *Phys. Rev. D* 104.3 (2021), p. 034022.

[2880] Jun-Tao Zhu, Lin-Qing Song, and Jun He. "$P_{cs}(4459)$ and other possible molecular states from $\Xi_c^{(*)}\bar{D}^{(*)}$ and $\Xi'_c\bar{D}^{(*)}$ interactions". In: *Phys. Rev. D* 103.7 (2021), p. 074007.

[2881] Brenda B. Malabarba, K. P. Khemchandani, and A. Martinez Torres. "$N^*$ states with hidden charm and a three-body nature". In: *Eur. Phys. J. A* 58.2 (2022), p. 33.

[2882] Nijiati Yalikun et al. "Coupled-channel effects of the $\Sigma_c^* D^{*-}\Lambda_(2595)D^-$ system and molecular nature of the $P_c$ pentaquark states from one-boson exchange model". In: *Phys. Rev. D* 104.9 (2021), p. 094039.

[2883] Ruilin Zhu and Cong-Feng Qiao. "Pentaquark states in a diquark–triquark model". In: *Phys. Lett. B* 756 (2016), pp. 259–264.

[2884] Ahmed Ali and Alexander Ya. Parkhomenko. "Interpretation of the narrow $J/\psi p$ Peaks in $\Lambda_b \to J/\psi p K^-$ decay in the compact diquark model". In: *Phys. Lett. B* 793 (2019), pp. 365–371.

[2885] Pan-Pan Shi, Fei Huang, and Wen-Ling Wang. "Hidden charm pentaquark states in a diquark model". In: *Eur. Phys. J. A* 57.7 (2021), p. 237.

[2886] K. Azizi, Y. Sarac, and H. Sundu. "Investigation of $P_{cs}(4459)^0$ pentaquark via its strong decay to $\Lambda J/\Psi$". In: *Phys. Rev. D* 103.9 (2021), p. 094033.

[2887] Zhi-Gang Wang. "Analysis of the $P_c(4312)$, $P_c(4440)$, $P_c(4457)$ and related hidden-charm pentaquark states with QCD sum rules". In: *Int. J. Mod. Phys. A* 35.01 (2020), p. 2050003.

[2888] Zhi-Gang Wang. "Analysis of the $P_{cs}(4459)$ as the hidden-charm pentaquark state with QCD sum rules". In: *Int. J. Mod. Phys. A* 36.10 (2021), p. 2150071.

[2889] Ulrich Mosel. "Neutrino Interactions with Nucleons and Nuclei: Importance for Long-Baseline Experiments". In: *Ann. Rev. Nucl. Part. Sci.* 66 (2016), pp. 171–195.

[2890] D. S. Armstrong and R. D. McKeown. "Parity-Violating Electron Scattering and the Electric and Magnetic Strange Form Factors of the Nucleon". In: *Ann. Rev. Nucl. Part. Sci.* 62 (2012), pp. 337–359.

[2891] Jens Erler et al. "Weak Polarized Electron Scattering". In: *Ann. Rev. Nucl. Part. Sci.* 64 (2014), pp. 269–298.

[2892] Roger D. Carlini et al. "Determination of the Proton's Weak Charge and Its Constraints on the Standard Model". In: *Ann. Rev. Nucl. Part. Sci.* 69 (2019), pp. 191–217.

[2893] F. J. Ernst, R. G. Sachs, and K. C. Wali. "Electromagnetic form factors of the nucleon". In: *Phys. Rev.* 119 (1960), pp. 1105–1114.

[2894] M. N. Rosenbluth. "High Energy Elastic Scattering of Electrons on Protons". In: *Phys. Rev.* 79 (1950), pp. 615–619.

[2895] L. N. Hand, D. G. Miller, and Richard Wilson. "Electric and Magnetic Formfactor of the Nucleon". In: *Rev. Mod. Phys.* 35 (1963), p. 335.

[2896] A. I. Akhiezer and Mikhail. P. Rekalo. "Polarization phenomena in electron scattering by protons in the high energy region". In: *Sov. Phys. Dokl.* 13 (1968), p. 572.

[2897] Norman Dombey. "Scattering of polarized leptons at high energy". In: *Rev. Mod. Phys.* 41 (1969), pp. 236–246.

[2898] R. G. Arnold, Carl E. Carlson, and Franz Gross. "Polarization Transfer in Elastic electron Scattering from Nucleons and Deuterons". In: *Phys. Rev. C* 23 (1981), p. 363.

[2899] T. W. Donnelly and A. S. Raskin. "Considerations of Polarization in Inclusive electron Scattering from Nuclei". In: *Annals Phys.* 169 (1986), pp. 247–351.

[2900] J. C. Bernauer et al. "High-precision determination of the electric and magnetic form factors of the proton". In: *Phys. Rev. Lett.* 105 (2010), p. 242001.

[2901] W. Xiong et al. "A small proton charge radius from an electron–proton scattering experiment". In: *Nature* 575.7781 (2019), pp. 147–150.

[2902] Paul N. Kirk et al. "Elastic electron - Proton Scattering at Large Four Momentum Transfer". In: *Phys. Rev. D* 8 (1973), pp. 63–91.

[2903] A. F. Sill et al. "Measurements of elastic electron - proton scattering at large momentum transfer". In: *Phys. Rev. D* 48 (1993), pp. 29–55.

[2904] M. E. Christy et al. "Form Factors and Two-Photon Exchange in High-Energy Elastic Electron-Proton Scattering". In: *Phys. Rev. Lett.* 128.10 (2022), p. 102002.

[2905] Haiyan Gao and Marc Vanderhaeghen. "The proton charge radius". In: *Rev. Mod. Phys.* 94.1 (2022), p. 015002.

[2906] T. Janssens et al. "Proton form factors from elastic electron-proton scattering". In: *Phys. Rev.* 142 (1966), pp. 922–931.

[2907] W. Bartel et al. "Measurement of proton and neutron electromagnetic form-factors at squared four momentum transfers up to 3-GeV/c$^2$". In: *Nucl. Phys. B* 58 (1973), pp. 429–475.

[2908] C Berger et al. "Electromagnetic form-factors of the proton at squared four momentum transfers between 10-fm$^{-2}$ and 50-fm$^{-2}$". In: *Phys. Lett. B* 35 (1971), pp. 87–89.

[2909] L. E. Price et al. "Backward-angle electron-proton elastic scattering and proton electromagnetic form-factors". In: *Phys. Rev. D* 4 (1971), pp. 45–53.

[2910] F. Borkowski et al. "Electromagnetic Form-Factors of the Proton at Low Four-Momentum Transfer". In: *Nucl. Phys. B* 93 (1975), pp. 461–478.

[2911] R. C. Walker et al. "Measurements of the proton elastic form-factors for 1-GeV/c$^2$ $\leq Q^2 \leq$ 3-GeV/c$^2$ at SLAC". In: *Phys. Rev. D* 49 (1994), pp. 5671–5689.

[2912] L. Andivahis et al. "Measurements of the electric and magnetic form-factors of the proton from $Q^2$ = 1.75-GeV/c$^2$ to 8.83-GeV/$^2$". In: *Phys. Rev. D* 50 (1994), pp. 5491–5517.

[2913] I. A. Qattan et al. "Precision Rosenbluth measurement of the proton elastic form-factors". In: *Phys. Rev. Lett.* 94 (2005), p. 142301.

[2914] M. E. Christy et al. "Measurements of electron proton elastic cross-sections for $0.4 < Q^2 < 5.5$ (GeV/c)$^2$". In: *Phys. Rev. C* 70 (2004), p. 015206.

[2915] B. D. Milbrath et al. "A Comparison of polarization observables in electron scattering from the proton and deuteron". In: *Phys. Rev. Lett.* 80 (1998). [Erratum: Phys.Rev.Lett. 82, 2221 (1999)], pp. 452–455.

[2916] M. K. Jones et al. "$G_{Ep}/G_{Mp}$ ratio by polarization transfer in $\vec{e}p \to e\vec{p}$". In: *Phys. Rev. Lett.* 84 (2000), pp. 1398–1402.

[2917] O. Gayou et al. "Measurement of $G_{Ep}/G_{Mp}$ in $\vec{e}p \to e\vec{p}$ to $Q^2$ = 5.6-GeV$^2$". In: *Phys. Rev. Lett.* 88 (2002), p. 092301.

[2918] V. Punjabi et al. "Proton elastic form-factor ratios to $Q^2$ = 3.5-GeV$^2$ by polarization transfer". In: *Phys. Rev. C* 71 (2005). [Erratum: Phys.Rev.C 71, 069902 (2005)], p. 055202.

[2919] O. Gayou et al. "Measurements of the elastic electromagnetic form-factor ratio $\mu_p G_{Ep}/G_{Mp}$ via polarization transfer". In: *Phys. Rev. C* 64 (2001), p. 038202.

[2920] Th. Pospischil et al. "Measurement of $G_{Ep}/G_{Mp}$ via polarization transfer at $Q^2$ = 0.4-GeV/c$^2$". In: *Eur. Phys. J. A* 12 (2001), pp. 125–127.

[2921] G. MacLachlan et al. "The ratio of proton electromagnetic form factors via recoil polarimetry at $Q^2$ = 1.13 (GeV/c)$^2$". In: *Nucl. Phys. A* 764 (2006), pp. 261–273.

[2922] A. J. R. Puckett et al. "Recoil Polarization Measurements of the Proton Electromagnetic Form Factor Ratio to $Q^2$ = 8.5 GeV$^2$". In: *Phys. Rev. Lett.* 104 (2010), p. 242301.

[2923] A. J. R. Puckett et al. "Final Analysis of Proton Form Factor Ratio Data at $Q^2$ = 4.0, 4.8 and 5.6 GeV$^2$". In: *Phys. Rev. C* 85 (2012), p. 045203.

[2924] M. Meziane et al. "Search for effects beyond the Born approximation in polarization transfer observables in $\vec{e}p$ elastic scattering". In: *Phys. Rev. Lett.* 106 (2011), p. 132501.

[2925] A. J. R. Puckett et al. "Polarization Transfer Observables in Elastic Electron Proton Scattering at $Q^2$ =2.5, 5.2, 6.8, and 8.5 GeV$^2$". In: *Phys. Rev. C* 96.5 (2017). [Erratum: Phys.Rev.C 98, 019907 (2018)], p. 055203.

[2926] G. Ron et al. "Low $Q^2$ measurements of the proton form factor ratio $mu_p G_E/G_M$". In: *Phys. Rev. C* 84 (2011), p. 055204.

[2927] X. Zhan et al. "High-Precision Measurement of the Proton Elastic Form Factor Ratio $\mu_p G_E/G_M$ at low $Q^2$". In: *Phys. Lett. B* 705 (2011), pp. 59–64.

[2928] M. Paolone et al. "Polarization Transfer in the $^4He(\vec{e}, e'\vec{p})^3H$ Reaction at $Q^2$ = 0.8 and 1.3 (GeV/c)$^2$". In: *Phys. Rev. Lett.* 105 (2010), p. 072001.

[2929] M. K. Jones et al. "Proton G(E)/G(M) from beam-target asymmetry". In: *Phys. Rev. C* 74 (2006), p. 035201.

[2930] Christopher B. Crawford et al. "Measurement of the proton electric to magnetic form factor ratio from vector H-1(vector e, e' p)". In: *Phys. Rev. Lett.* 98 (2007), p. 052301.

[2931] A. Liyanage et al. "Proton form factor ratio $\mu_p G_E^p/G_M^p$ from double spin asymmetry". In: *Phys. Rev. C* 101.3 (2020), p. 035206.

[2932] G. G. Simon et al. "Absolute electron Proton Cross-Sections at Low Momentum Transfer Measured with a High Pressure Gas Target System". In: *Nucl. Phys. A* 333 (1980), pp. 381–391.

[2933] A. V. Gramolin and D. M. Nikolenko. "Reanalysis of Rosenbluth measurements of the proton

form factors". In: *Phys. Rev. C* 93.5 (2016), p. 055201.

[2934] C. F. Perdrisat, V. Punjabi, and M. Vanderhaeghen. "Nucleon Electromagnetic Form Factors". In: *Prog. Part. Nucl. Phys.* 59 (2007), pp. 694–764.

[2935] Pierre A. M. Guichon and M. Vanderhaeghen. "How to reconcile the Rosenbluth and the polarization transfer method in the measurement of the proton form-factors". In: *Phys. Rev. Lett.* 91 (2003), p. 142303.

[2936] A. Afanasev et al. "Two-photon exchange in elastic electron–proton scattering". In: *Prog. Part. Nucl. Phys.* 95 (2017), pp. 245–278.

[2937] Yung-Su Tsai. "Radiative Corrections to Electron-Proton Scattering". In: *Phys. Rev.* 122 (1961), pp. 1898–1907.

[2938] Luke W. Mo and Yung-Su Tsai. "Radiative Corrections to Elastic and Inelastic e p and mu p Scattering". In: *Rev. Mod. Phys.* 41 (1969), pp. 205–235.

[2939] L. C. Maximon and J. A. Tjon. "Radiative corrections to electron proton scattering". In: *Phys. Rev. C* 62 (2000), p. 054320.

[2940] R. E. Gerasimov and V. S. Fadin. "Analysis of approximations used in calculations of radiative corrections to electron-proton scattering cross section". In: *Phys. Atom. Nucl.* 78.1 (2015), pp. 69–91.

[2941] D. Besset et al. "A set of efficient estimators for polarization measurements". In: *Nucl. Instrum. Meth.* 166 (1979), pp. 515–520.

[2942] A. Afanasev, I. Akushevich, and N. Merenkov. "Model independent radiative corrections in processes of polarized electron nucleon elastic scattering". In: *Phys. Rev. D* 64 (2001), p. 113009.

[2943] A. V. Afanasev et al. "QED radiative corrections to asymmetries of elastic e p scattering in hadronic variables". In: *Phys. Lett. B* 514 (2001), pp. 269–278.

[2944] I. Akushevich et al. "Monte Carlo Generator ELRADGEN 2.0 for Simulation of Radiative events in Elastic ep-Scattering of Polarized Particles". In: *Comput. Phys. Commun.* 183 (2012), pp. 1448–1467.

[2945] Randolf Pohl et al. "The size of the proton". In: *Nature* 466 (2010), pp. 213–216.

[2946] S. Platchkov et al. "Deutron $A(Q^2)$ Stucture Function and the Neutron Electric Form-Factor". In: *Nucl. Phys. A* 510 (1990), pp. 740–758.

[2947] R. Schiavilla and I. Sick. "Neutron charge form-factor at large $q^2$". In: *Phys. Rev. C* 64 (2001), p. 041002.

[2948] S. Galster et al. "Elastic electron-deuteron scattering and the electric neutron form factor at four-momentum transfers $5\text{fm}^{-2} < q^2 < 14\text{fm}^{-2}$". In: *Nucl. Phys. B* 32 (1971), pp. 221–237.

[2949] G. G. Simon, C. Schmitt, and V. H. Walther. "Elastic Electric and Magnetic $eD$ Scattering at Low Momentum Transfer". In: *Nucl. Phys. A* 364 (1981), pp. 285–296.

[2950] E. Geis et al. "The Charge Form Factor of the Neutron at Low Momentum Transfer from the $^2$H-$(\vec{e}, e'n)p$ Reaction". In: *Phys. Rev. Lett.* 101 (2008), p. 042501.

[2951] G. Warren et al. "Measurement of the electric form-factor of the neutron at $Q^2 = 0.5$ and 1.0 $GeV^2/c^2$". In: *Phys. Rev. Lett.* 92 (2004), p. 042301.

[2952] H. Zhu et al. "A Measurement of the electric form-factor of the neutron through $\vec{d}(\vec{e}, e'n)p$ at $Q^2 = 0.5\,(\text{GeV/c})^2$". In: *Phys. Rev. Lett.* 87 (2001), p. 081801.

[2953] I. Passchier et al. "The Charge form-factor of the neutron from the reaction polarized $^2\text{H}\vec{e}, e'n)p$". In: *Phys. Rev. Lett.* 82 (1999), pp. 4988–4991.

[2954] S. Riordan et al. "Measurements of the Electric Form Factor of the Neutron up to $Q^2 = 3.4GeV^2$ using the Reaction $^3\vec{\text{He}}(\vec{e}, e'n)pp$". In: *Phys. Rev. Lett.* 105 (2010), p. 262302.

[2955] V. Sulkosky et al. "Extraction of the Neutron Electric Form Factor from Measurements of Inclusive Double Spin Asymmetries". In: *Phys. Rev. C* 96.6 (2017), p. 065206.

[2956] B. S. Schlimme et al. "Measurement of the neutron electric to magnetic form factor ratio at $Q^2 = 1.58GeV^2$ using the reaction $^3\vec{He}(\vec{e}, e'n)pp$". In: *Phys. Rev. Lett.* 111.13 (2013), p. 132504.

[2957] J. Bermuth et al. "The Neutron charge form-factor and target analyzing powers from polarized-He-3 (polarized-e,e-prime n) scattering". In: *Phys. Lett. B* 564 (2003), pp. 199–204.

[2958] J. Becker et al. "Determination of the neutron electric form-factor from the reaction $^3$He $(e, e'n)$ at medium momentum transfer". In: *Eur. Phys. J. A* 6 (1999), pp. 329–344.

[2959] D. I. Glazier et al. "Measurement of the electric form-factor of the neutron at $Q^2 = 0.3\,(\text{GeV/c})^2$ to $0.8\,(\text{GeV/c})^2$". In: *Eur. Phys. J. A* 24 (2005), pp. 101–109.

[2960] C. Herberg et al. "Determination of the neutron electric form-factor in the $D(e, e'n)p$ reaction and the influence of nuclear binding". In: *Eur. Phys. J. A* 5 (1999), pp. 131–135.

[2961] B. Plaster et al. "Measurements of the neutron electric to magnetic form-factor ratio $G_{En}/G_{Mn}$

via the $^2$H($\vec{e}, e'\vec{n}$)$^1$H reaction to $Q^2 = 1.45$ (GeV/c)$^2$". In: *Phys. Rev. C* 73 (2006), p. 025205.
[2962] Loyal Durand. "Inelastic Electron-Deuteron Scattering Cross Sections at High Energies". In: *Phys. Rev.* 115 (1959), pp. 1020–1038.
[2963] J. Lachniet et al. "A Precise Measurement of the Neutron Magnetic Form Factor $G_M^n$in the Few-GeV$^2$ Region". In: *Phys. Rev. Lett.* 102 (2009), p. 192001.
[2964] H. Anklin et al. "Precision measurement of the neutron magnetic form-factor". In: *Phys. Lett. B* 336 (1994), pp. 313–318.
[2965] H. Anklin et al. "Precise measurements of the neutron magnetic form-factor". In: *Phys. Lett. B* 428 (1998), pp. 248–253.
[2966] E. E. W. Bruins et al. "Measurement of the neutron magnetic form-factor". In: *Phys. Rev. Lett.* 75 (1995), pp. 21–24.
[2967] G. Kubon et al. "Precise neutron magnetic form-factors". In: *Phys. Lett. B* 524 (2002), pp. 26–32.
[2968] W. Xu et al. "The Transverse asymmetry $A_{T'}$ from quasielastic polarized $^3$He ($\vec{e}, e'$) process and the neutron magnetic form-factor". In: *Phys. Rev. Lett.* 85 (2000), pp. 2900–2904.
[2969] W. Xu et al. "PWIA extraction of the neutron magnetic form-factor from quasielastic $^3\vec{\mathrm{He}}(\vec{e}, e')$ at $Q^2 = 0.3$-(GeV/c)$^2$–0.6-(GeV/c)$^2$". In: *Phys. Rev. C* 67 (2003), p. 012201.
[2970] B. Anderson et al. "Extraction of the Neutron Magnetic Form Factor from Quasi-elastic $^3\vec{He}(\vec{e}, e')$ at Q$^2$ = 0.1 - 0.6 (GeV/c)$^2$". In: *Phys. Rev. C* 75 (2007), p. 034003.
[2971] H. Gao et al. "Measurement of the neutron magnetic form-factor from inclusive quasielastic scattering of polarized electrons from polarized $^3$He". In: *Phys. Rev. C* 50 (1994), R546–R549.
[2972] A. Lung et al. "Measurements of the electric and magnetic form-factors of the neutron from $Q^2 = 1.75$-GeV/c$^2$ to 4-GeV/c$^2$". In: *Phys. Rev. Lett.* 70 (1993), pp. 718–721.
[2973] Stephen Rock et al. "Measurement of elastic electron - neutron scattering and inelastic electron - deuteron scattering cross-sections at high momentum transfer". In: *Phys. Rev. D* 46 (1992), pp. 24–44.
[2974] P. Markowitz et al. "Measurement of the magnetic form factor of the neutron". In: *Phys. Rev. C* 48.1 (1993), R5–R9.
[2975] Earle L. Lomon. "Effect of recent R(p) and R(n) measurements on extended Gari-Krumpelmann model fits to nucleon electromagnetic form-factors". In: *Phys. Rev. C* 66 (2002), p. 045501.
[2976] M. Diehl et al. "Generalized parton distributions from nucleon form-factor data". In: *Eur. Phys. J. C* 39 (2005), pp. 1–39.
[2977] Franz Gross, G. Ramalho, and M. T. Pena. "A Pure S-wave covariant model for the nucleon". In: *Phys. Rev. C* 77 (2008), p. 015202.
[2978] Ian C. Cloet and Gerald A. Miller. "Nucleon form factors and spin content in a quark-diquark model with a pion cloud". In: *Phys. Rev. C* 86 (2012), p. 015208.
[2979] A. J. Chambers et al. "Electromagnetic form factors at large momenta from lattice QCD". In: *Phys. Rev. D* 96.11 (2017), p. 114509.
[2980] Mischa Batelaan et al. "Nucleon Form Factors from the Feynman-Hellmann Method in Lattice QCD". In: *PoS* LATTICE2021 (2022), p. 426.
[2981] Nathan Isgur and C. H. Llewellyn Smith. "The Applicability of Perturbative QCD to Exclusive Processes". In: *Nucl. Phys. B* 317 (1989), pp. 526–572.
[2982] Nathan Isgur and C. H. Llewellyn Smith. "Perturbative QCD in exclusive processes". In: *Phys. Lett. B* 217 (1989), pp. 535–538.
[2983] Andrei V. Belitsky, Xiang-dong Ji, and Feng Yuan. "A Perturbative QCD analysis of the nucleon's Pauli form-factor $F_2(Q^2)$". In: *Phys. Rev. Lett.* 91 (2003), p. 092003.
[2984] G. D. Cates et al. "Flavor decomposition of the elastic nucleon electromagnetic form factors". In: *Phys. Rev. Lett.* 106 (2011), p. 252003.
[2985] Earle L. Lomon and Simone Pacetti. "Time-like and space-like electromagnetic form factors of nucleons, a unified description". In: *Phys. Rev. D* 85 (2012). [Erratum: Phys.Rev.D 86, 039901 (2012)], p. 113004.
[2986] Yong-Hui Lin, Hans-Werner Hammer, and Ulf-G. Meißner. "Dispersion-theoretical analysis of the electromagnetic form factors of the nucleon: Past, present and future". In: *Eur. Phys. J. A* 57.8 (2021), p. 255.
[2987] James J. Kelly. "Nucleon charge and magnetization densities from Sachs form-factors". In: *Phys. Rev. C* 66 (2002), p. 065203.
[2988] Gerald A. Miller. "Charge Density of the Neutron". In: *Phys. Rev. Lett.* 99 (2007), p. 112001.
[2989] Gerald A. Miller. "Transverse Charge Densities". In: *Ann. Rev. Nucl. Part. Sci.* 60 (2010), pp. 1–25.
[2990] Siddharth Venkat et al. "Realistic Transverse Images of the Proton Charge and Magnetic Densities". In: *Phys. Rev. C* 83 (2011), p. 015203.

[2991] M. Guidal et al. “Nucleon form-factors from generalized parton distributions”. In: *Phys. Rev. D* 72 (2005), p. 054013.

[2992] Markus Diehl and Peter Kroll. “Nucleon form factors, generalized parton distributions and quark angular momentum”. In: *Eur. Phys. J. C* 73.4 (2013), p. 2397.

[2993] V. Punjabi et al. “The Structure of the Nucleon: Elastic Electromagnetic Form Factors”. In: *Eur. Phys. J. A* 51 (2015), p. 79.

[2994] A. Accardi et al. “An experimental program with high duty-cycle polarized and unpolarized positron beams at Jefferson Lab”. In: *Eur. Phys. J. A* 57.8 (2021), p. 261.

[2995] B. Schmookler et al. “High $Q^2$ electron-proton elastic scattering at the future Electron-Ion Collider”. In: (July 2022).

[2996] Alex Bogacz et al. “20-24 GeV FFA CEBAF Energy Upgrade”. In: *JACoW* IPAC2021 (2021), MOPAB216.

[2997] Johannes Blumlein. “The Theory of Deeply Inelastic Scattering”. In: *Prog. Part. Nucl. Phys.* 69 (2013), pp. 28–84.

[2998] P. Jimenez-Delgado, W. Melnitchouk, and J. F. Owens. “Parton momentum and helicity distributions in the nucleon”. In: *J. Phys. G* 40 (2013), p. 093102.

[2999] Stefano Forte and Graeme Watt. “Progress in the Determination of the Partonic Structure of the Proton”. In: *Ann. Rev. Nucl. Part. Sci.* 63 (2013), pp. 291–328.

[3000] Jun Gao, Lucian Harland-Lang, and Juan Rojo. “The Structure of the Proton in the LHC Precision Era”. In: *Phys. Rept.* 742 (2018), pp. 1–121.

[3001] Jacob J. Ethier and Emanuele R. Nocera. “Parton Distributions in Nucleons and Nuclei”. In: *Ann. Rev. Nucl. Part. Sci.* 70 (2020), pp. 43–76.

[3002] H. Abramowicz et al. “Combination of measurements of inclusive deep inelastic $e^{\pm}p$ scattering cross sections and QCD analysis of HERA data”. In: *Eur. Phys. J. C* 75.12 (2015), p. 580.

[3003] Georges Aad et al. “Determination of the strange quark density of the proton from ATLAS measurements of the $W \to \ell\nu$ and $Z \to \ell\ell$ cross sections”. In: *Phys. Rev. Lett.* 109 (2012), p. 012001.

[3004] Morad Aaboud et al. “Precision measurement and interpretation of inclusive $W^+$ , $W^-$ and $Z/\gamma^*$ production cross sections with the ATLAS detector”. In: *Eur. Phys. J. C* 77.6 (2017), p. 367.

[3005] N. Sato et al. “Strange quark suppression from a simultaneous Monte Carlo analysis of parton distributions and fragmentation functions”. In: *Phys. Rev. D* 101.7 (2020), p. 074020.

[3006] David d’Enterria and Juan Rojo. “Quantitative constraints on the gluon distribution function in the proton from collider isolated-photon data”. In: *Nucl. Phys.* B860 (2012), pp. 311–338.

[3007] D. W. Duke and J. F. Owens. “$Q^2$ Dependent Parametrizations of Parton Distribution Functions”. In: *Phys. Rev. D* 30 (1984), pp. 49–54.

[3008] Jorge G. Morfin and W.-K. Tung. “Parton distributions from a global QCD analysis of deep inelastic scattering and lepton pair production”. In: *Z. Phys. C* 52 (1991), pp. 13–30.

[3009] Stefano Forte et al. “Neural network parametrization of deep inelastic structure functions”. In: *JHEP* 05 (2002), p. 062.

[3010] H. Honkanen et al. “New avenue to the Parton Distribution Functions: Self-Organizing Maps”. In: *Phys. Rev. D* 79 (2009), p. 034022.

[3011] F. E. Close and R. G. Roberts. “Consistent analysis of the spin content of the nucleon”. In: *Phys. Lett. B* 316 (1993), pp. 165–171.

[3012] W. Melnitchouk, R. Ent, and C. Keppel. “Quark-hadron duality in electron scattering”. In: *Phys. Rept.* 406 (2005), pp. 127–301.

[3013] Howard Georgi and H. David Politzer. “Freedom at Moderate Energies: Masses in Color Dynamics”. In: *Phys. Rev. D* 14 (1976), p. 1829.

[3014] R. Keith Ellis, W. Furmanski, and R. Petronzio. “Unraveling Higher Twists”. In: *Nucl. Phys.* B212 (1983), p. 29.

[3015] M. A. G. Aivazis, Frederick I. Olness, and Wu-Ki Tung. “Leptoproduction of heavy quarks. 1. General formalism and kinematics of charged current and neutral current production processes”. In: *Phys. Rev. D* 50 (1994), pp. 3085–3101.

[3016] Ingo Schienbein et al. “A Review of Target Mass Corrections”. In: *J. Phys. G* 35 (2008), p. 053101.

[3017] E. Moffat et al. “What does kinematical target mass sensitivity in DIS reveal about hadron structure?” In: *Phys. Rev. D* 99.9 (2019), p. 096008.

[3018] J. J. Aubert et al. “Measurement of the deuteron structure function $F2$ and a comparison of proton and neutron structure”. In: *Phys. Lett. B* 123 (1983), pp. 123–126.

[3019] Donald F. Geesaman, K. Saito, and Anthony William Thomas. “The nuclear EMC effect”. In: *Ann. Rev. Nucl. Part. Sci.* 45 (1995), pp. 337–390.

[3020] P. R. Norton. "The EMC effect". In: *Rept. Prog. Phys.* 66 (2003), pp. 1253–1297.

[3021] W. Melnitchouk, Andreas W. Schreiber, and Anthony William Thomas. "Deep inelastic scattering from off-shell nucleons". In: *Phys. Rev. D* 49 (1994), pp. 1183–1198.

[3022] Sergey A. Kulagin, G. Piller, and W. Weise. "Shadowing, binding and off-shell effects in nuclear deep inelastic scattering". In: *Phys. Rev. C* 50 (1994), pp. 1154–1169.

[3023] Sergey A. Kulagin and R. Petti. "Global study of nuclear structure functions". In: *Nucl. Phys.* A765 (2006), pp. 126–187.

[3024] W. Melnitchouk and Anthony William Thomas. "Neutron / proton structure function ratio at large x". In: *Phys. Lett. B* 377 (1996), pp. 11–17.

[3025] J. F. Owens, A. Accardi, and W. Melnitchouk. "Global parton distributions with nuclear and finite-$Q^2$ corrections". In: *Phys. Rev. D* 87.9 (2013), p. 094012.

[3026] A. D. Martin et al. "Extended Parameterisations for MSTW PDFs and their effect on Lepton Charge Asymmetry from W Decays". In: *Eur. Phys. J. C* 73.2 (2013), p. 2318.

[3027] A. Accardi et al. "Constraints on large-$x$ parton distributions from new weak boson production and deep-inelastic scattering data". In: *Phys. Rev. D* 93.11 (2016), p. 114017.

[3028] S. I. Alekhin, S. A. Kulagin, and R. Petti. "Nuclear Effects in the Deuteron and Constraints on the d/u Ratio". In: *Phys. Rev. D* 96.5 (2017), p. 054005.

[3029] C. Cocuzza et al. "Isovector EMC Effect from Global QCD Analysis with MARATHON Data". In: *Phys. Rev. Lett.* 127.24 (2021), p. 242001.

[3030] A. O. Bazarko et al. "Determination of the strange quark content of the nucleon from a next-to-leading order QCD analysis of neutrino charm production". In: *Z. Phys. C* 65 (1995), pp. 189–198.

[3031] D. Mason et al. "Measurement of the Nucleon Strange-Antistrange Asymmetry at Next-to-Leading Order in QCD from NuTeV Dimuon Data". In: *Phys. Rev. Lett.* 99 (2007), p. 192001.

[3032] Sergey A. Kulagin and R. Petti. "Neutrino inelastic scattering off nuclei". In: *Phys. Rev. D* 76 (2007), p. 094023.

[3033] Narbe Kalantarians, Cynthia Keppel, and M. Eric Christy. "Comparison of the Structure Function F2 as Measured by Charged Lepton and Neutrino Scattering from Iron Targets". In: *Phys. Rev. C* 96.3 (2017), p. 032201.

[3034] A. Accardi et al. "Parton Propagation and Fragmentation in QCD Matter". In: *Riv. Nuovo Cim.* 32.9-10 (2009), pp. 439–554.

[3035] J. Pumplin et al. "Uncertainties of predictions from parton distribution functions. 2. The Hessian method". In: *Phys. Rev. D* 65 (2001), p. 014013.

[3036] J. Pumplin et al. "New generation of parton distributions with uncertainties from global QCD analysis". In: *JHEP* 07 (2002), p. 012.

[3037] N. T. Hunt-Smith et al. "On the determination of uncertainties in parton densities". In: (June 2022).

[3038] Luigi Del Debbio et al. "Unbiased determination of the proton structure function $F_2^p$ with faithful uncertainty estimation". In: *JHEP* 03 (2005), p. 080.

[3039] Luigi Del Debbio et al. "Neural network determination of parton distributions: The Nonsinglet case". In: *JHEP* 03 (2007), p. 039.

[3040] Richard D. Ball et al. "A Determination of parton distributions with faithful uncertainty estimation". In: *Nucl. Phys.* B809 (2009). [Erratum: Nucl.Phys.B 816, 293 (2009)], pp. 1–63.

[3041] A. Accardi et al. "A Critical Appraisal and Evaluation of Modern PDFs". In: *Eur. Phys. J. C* 76.8 (2016), p. 471.

[3042] Jon Butterworth et al. "PDF4LHC recommendations for LHC Run II". In: *J. Phys.* G43 (2016), p. 023001.

[3043] Richard D. Ball et al. "The path to proton structure at 1% accuracy". In: *Eur. Phys. J. C* 82.5 (2022), p. 428.

[3044] J. McGowan et al. "Approximate $N^3LO$ Parton Distribution Functions with Theoretical Uncertainties: MSHT20a$N^3LO$ PDFs". In: (July 2022).

[3045] S. Alekhin et al. "Parton distribution functions, $\alpha_s$, and heavy-quark masses for LHC Run II". In: *Phys. Rev.* D96.1 (2017), p. 014011.

[3046] Pedro Jimenez-Delgado and Ewald Reya. "Delineating parton distributions and the strong coupling". In: *Phys. Rev.* D89.7 (2014), p. 074049.

[3047] C. Cocuzza et al. "Bayesian Monte Carlo extraction of the sea asymmetry with SeaQuest and STAR data". In: *Phys. Rev. D* 104.7 (2021), p. 074031.

[3048] F. E. Close. "Nu w(2) at small omega' and resonance form-factors in a quark model with broken su(6)". In: *Phys. Lett. B* 43 (1973), pp. 422–426.

[3049] Roy J. Holt and Craig D. Roberts. "Distribution Functions of the Nucleon and Pion in the Valence Region". In: *Rev. Mod. Phys.* 82 (2010), pp. 2991–3044.

[3050] L. T. Brady et al. "Impact of PDF uncertainties at large x on heavy boson production". In: *JHEP* 06 (2012), p. 019.

[3051] N. Baillie et al. "Measurement of the neutron F2 structure function via spectator tagging with CLAS". In: *Phys. Rev. Lett.* 108 (2012). [Erratum: Phys.Rev.Lett. 108, 199902 (2012)], p. 142001.

[3052] S. Tkachenko et al. "Measurement of the structure function of the nearly free neutron using spectator tagging in inelastic $^2$H(e, e'p)X scattering with CLAS". In: *Phys. Rev. C* 89 (2014). [Addendum: Phys.Rev.C 90, 059901 (2014)], p. 045206.

[3053] T. Aaltonen et al. "Direct Measurement of the $W$ Production Charge Asymmetry in $p\bar{p}$ Collisions at $\sqrt{s} = 1.96$ TeV". In: *Phys. Rev. Lett.* 102 (2009), p. 181801.

[3054] Victor Mukhamedovich Abazov et al. "Measurement of the W Boson Production Charge Asymmetry in $p\bar{p} \to W + X \to e\nu + X$ Events at $\sqrt{s} = 1.96$ TeV". In: *Phys. Rev. Lett.* 112.15 (2014). [Erratum: Phys.Rev.Lett. 114, 049901 (2015)], p. 151803.

[3055] Timo Antero Aaltonen et al. "Measurement of $d\sigma/dy$ of Drell-Yan $e^+e^-$ pairs in the $Z$ Mass Region from $p\bar{p}$ Collisions at $\sqrt{s} = 1.96$ TeV". In: *Phys. Lett. B* 692 (2010), pp. 232–239.

[3056] M. Arneodo et al. "Measurement of the proton and the deuteron structure functions, $F_2^p$ and $F_2^d$)". In: *Phys. Lett. B* 364 (1995), pp. 107–115.

[3057] M. Arneodo et al. "Measurement of the proton and deuteron structure functions, $F_2^p$ and $F_2^d$, and of the ratio $\sigma_L/\sigma_T$". In: *Nucl. Phys.* B483 (1997), pp. 3–43.

[3058] Jaroslav Adam et al. "Measurements of $W$ and $Z/\gamma^*$ cross sections and their ratios in p+p collisions at RHIC". In: *Phys. Rev. D* 103.1 (2021), p. 012001.

[3059] R. S. Towell et al. "Improved measurement of the $\bar{d}/\bar{u}$ asymmetry in the nucleon sea". In: *Phys. Rev. D* 64 (2001), p. 052002.

[3060] J. Dove et al. "The asymmetry of antimatter in the proton". In: *Nature* 590.7847 (2021). [Erratum: Nature 604, E26 (2022)], pp. 561–565.

[3061] Anthony William Thomas. "A Limit on the Pionic Component of the Nucleon Through SU(3) Flavor Breaking in the Sea". In: *Phys. Lett. B* 126 (1983), pp. 97–100.

[3062] J. Speth and Anthony William Thomas. "Mesonic contributions to the spin and flavor structure of the nucleon". In: *Adv. Nucl. Phys.* 24 (1997). Ed. by John W. Negele and E. Vogt, pp. 83–149.

[3063] Yusupujiang Salamu et al. "$\bar{d} - \bar{u}$ asymmetry in the proton in chiral effective theory". In: *Phys. Rev. Lett.* 114 (2015), p. 122001.

[3064] Elliot Leader, Aleksander V. Sidorov, and Dimiter B. Stamenov. "Determination of Polarized PDFs from a QCD Analysis of Inclusive and Semi-inclusive Deep Inelastic Scattering Data". In: *Phys. Rev. D* 82 (2010), p. 114018.

[3065] Elliot Leader, Alexander V. Sidorov, and Dimiter B. Stamenov. "A Possible Resolution of the Strange Quark Polarization Puzzle ?" In: *Phys. Rev. D* 84 (2011), p. 014002.

[3066] Nobuo Sato et al. "First Monte Carlo analysis of fragmentation functions from single-inclusive $e^+e^-$ annihilation". In: *Phys. Rev. D* 94.11 (2016), p. 114004.

[3067] A. I. Signal and Anthony William Thomas. "Possible Strength of the Nonperturbative Strange Sea of the Nucleon". In: *Phys. Lett. B* 191 (1987), p. 205.

[3068] Stefano Catani et al. "Perturbative generation of a strange-quark asymmetry in the nucleon". In: *Phys. Rev. Lett.* 93 (2004), p. 152003.

[3069] X. G. Wang et al. "Strange quark asymmetry in the proton in chiral effective theory". In: *Phys. Rev. D* 94.9 (2016), p. 094035.

[3070] Y. Salamu et al. "Parton distributions from nonlocal chiral SU(3) effective theory: Flavor asymmetries". In: *Phys. Rev. D* 100.9 (2019), p. 094026.

[3071] X. G. Wang et al. "Strange quark helicity in the proton from chiral effective theory". In: *Phys. Rev. D* 102.11 (2020), p. 116020.

[3072] Fangcheng He et al. "Helicity-dependent distribution of strange quarks in the proton from nonlocal chiral effective theory". In: *Phys. Rev. D* 105.9 (2022), p. 094007.

[3073] Elke C. Aschenauer et al. "Semi-inclusive Deep-Inelastic Scattering, Parton Distributions and Fragmentation Functions at a Future Electron-Ion Collider". In: *Phys. Rev. D* 99.9 (2019), p. 094004.

[3074] Georges Aad et al. "Measurement of the production of a $W$ boson in association with a charm quark in $pp$ collisions at $\sqrt{s} = 7$ TeV with the ATLAS detector". In: *JHEP* 05 (2014), p. 068.

[3075] Serguei Chatrchyan et al. "Measurement of Associated W + Charm Production in pp Collisions at $\sqrt{s} = 7$ TeV". In: *JHEP* 02 (2014), p. 013.

[3076] S. J. Brodsky et al. "The Intrinsic Charm of the Proton". In: *Phys. Lett. B* 93 (1980), pp. 451–455.

[3077] F. S. Navarra et al. "On the intrinsic charm component of the nucleon". In: *Phys. Rev. D* 54 (1996), pp. 842–846.

[3078] W. Melnitchouk and Anthony William Thomas. "HERA anomaly and hard charm in the nucleon". In: *Phys. Lett. B* 414 (1997), pp. 134–139.

[3079] J. Pumplin, H. L. Lai, and W. K. Tung. "The Charm Parton Content of the Nucleon". In: *Phys. Rev. D* 75 (2007), p. 054029.

[3080] T. J. Hobbs, J. T. Londergan, and W. Melnitchouk. "Phenomenology of nonperturbative charm in the nucleon". In: *Phys. Rev. D* 89.7 (2014), p. 074008.

[3081] P. Jimenez-Delgado et al. "New limits on intrinsic charm in the nucleon from global analysis of parton distributions". In: *Phys. Rev. Lett.* 114.8 (2015), p. 082002.

[3082] P. Jimenez-Delgado et al. "Reply to Comment on "New limits on intrinsic charm in the nucleon from global analysis of parton distributions"". In: *Phys. Rev. Lett.* 116.1 (2016), p. 019102.

[3083] Richard D. Ball et al. "A Determination of the Charm Content of the Proton". In: *Eur. Phys. J. C* 76.11 (2016), p. 647.

[3084] Richard D. Ball et al. "Evidence for intrinsic charm quarks in the proton". In: *Nature* 608.7923 (2022), pp. 483–487.

[3085] Marco Guzzi et al. "The persistent nonperturbative charm enigma". In: (Nov. 2022).

[3086] Daniel de Florian et al. "Evidence for polarization of gluons in the proton". In: *Phys. Rev. Lett.* 113.1 (2014), p. 012001.

[3087] Daniel De Florian et al. "Monte Carlo sampling variant of the DSSV14 set of helicity parton densities". In: *Phys. Rev. D* 100.11 (2019), p. 114027.

[3088] Richard D. Ball et al. "Unbiased determination of polarized parton distributions and their uncertainties". In: *Nucl. Phys.* B874 (2013), pp. 36–84.

[3089] C. Cocuzza et al. "Polarized Antimatter in the Proton from Global QCD Analysis". In: (Feb. 2022).

[3090] Johannes Blumlein and Helmut Bottcher. "QCD Analysis of Polarized Deep Inelastic Scattering Data". In: *Nucl. Phys. B* 841 (2010), pp. 205–230.

[3091] Ali N. Khorramian et al. "Polarized Deeply Inelastic Scattering (DIS) Structure Functions for Nucleons and Nuclei". In: *Phys. Rev. D* 83 (2011), p. 054017.

[3092] M. Hirai, S. Kumano, and N. Saito. "Determination of polarized parton distribution functions with recent data on polarization asymmetries". In: *Phys. Rev. D* 74 (2006), p. 014015.

[3093] Alessandro Candido, Stefano Forte, and Felix Hekhorn. "Can $\overline{\text{MS}}$ parton distributions be negative?" In: *JHEP* 11 (2020), p. 129.

[3094] John Collins, Ted C. Rogers, and Nobuo Sato. "Positivity and renormalization of parton densities". In: *Phys. Rev. D* 105.7 (2022), p. 076010.

[3095] P. Jimenez-Delgado, A. Accardi, and W. Melnitchouk. "Impact of hadronic and nuclear corrections on global analysis of spin-dependent parton distributions". In: *Phys. Rev. D* 89.3 (2014), p. 034025.

[3096] Jaroslav Adam et al. "Measurement of the longitudinal spin asymmetries for weak boson production in proton-proton collisions at $\sqrt{s} = 510$ GeV". In: *Phys. Rev. D* 99.5 (2019), p. 051102.

[3097] A. Adare et al. "Measurement of parity-violating spin asymmetries in $W^{\pm}$ production at midrapidity in longitudinally polarized $p + p$ collisions". In: *Phys. Rev. D* 93.5 (2016), p. 051103.

[3098] A. Adare et al. "Cross section and longitudinal single-spin asymmetry $A_L$ for forward $W^{\pm} \to \mu^{\pm}\nu$ production in polarized $p + p$ collisions at $\sqrt{s} = 510$ GeV". In: *Phys. Rev. D* 98.3 (2018), p. 032007.

[3099] Andreas W. Schreiber, A. I. Signal, and Anthony William Thomas. "Structure functions in the bag model". In: *Phys. Rev. D* 44 (1991), pp. 2653–2662.

[3100] Dmitri Diakonov et al. "Unpolarized and polarized quark distributions in the large $N_c$ limit". In: *Phys. Rev. D* 56 (1997), pp. 4069–4083.

[3101] M. Wakamatsu and T. Watabe. "Do we expect light flavor sea quark asymmetry also for the spin dependent distribution functions of the nucleon?" In: *Phys. Rev. D* 62 (2000), p. 017506.

[3102] Claude Bourrely and Jacques Soffer. "New developments in the statistical approach of parton distributions: tests and predictions up to LHC energies". In: *Nucl. Phys.* A941 (2015), pp. 307–334.

[3103] R. M. Whitehill et al. "Accessing gluon polarization with high-$P_T$ hadrons in SIDIS". In: (Oct. 2022).

[3104] Colin Egerer et al. "Towards the determination of the gluon helicity distribution in the nucleon from lattice quantum chromodynamics". In: (July 2022).

[3105] R. Abdul Khalek et al. "Science Requirements and Detector Concepts for the Electron-Ion Collider: EIC Yellow Report". In: (Mar. 2021).

[3106] P. Jimenez-Delgado, H. Avakian, and W. Melnitchouk. "Constraints on spin-dependent parton distributions at large x from global QCD analysis". In: *Phys. Lett. B* 738 (2014), pp. 263–267.

[3107] Y. Zhou et al. "Revisiting quark and gluon polarization in the proton at the EIC". In: *Phys. Rev. D* 104.3 (2021), p. 034028.

[3108] Daniel Adamiak et al. "First analysis of world polarized DIS data with small-x helicity evolution". In: *Phys. Rev. D* 104.3 (2021), p. L031501.

[3109] Tianbo Liu et al. "Factorized approach to radiative corrections for inelastic lepton-hadron collisions". In: *Phys. Rev. D* 104.9 (2021), p. 094033.

[3110] Tianbo Liu et al. "A new approach to semi-inclusive deep-inelastic scattering with QED and QCD factorization". In: *JHEP* 11 (2021), p. 157.

[3111] J. Bringewatt et al. "Confronting lattice parton distributions with global QCD analysis". In: *Phys. Rev. D* 103.1 (2021), p. 016003.

[3112] T. H. R. Skyrme. "A Unified Field Theory of Mesons and Baryons". In: *Nucl. Phys.* 31 (1962), pp. 556–569.

[3113] Murray Gell-Mann. "A Schematic Model of Baryons and Mesons". In: *Phys. Lett.* 8 (1964), pp. 214–215.

[3114] G. Zweig. "An SU(3) model for strong interaction symmetry and its breaking. Version 1". In: (Jan. 1964).

[3115] R. K. Bhaduri. *Models of the nucleon: from quarks to soliton*. 1988.

[3116] Anthony William Thomas and Wolfram Weise. *The Structure of the Nucleon*. Germany: Wiley, 2001.

[3117] F. E. Close. *An Introduction to Quarks and Partons*. 1979.

[3118] V. W. Hughes and J. Kuti. "Internal Spin Structure of the Nucleon". In: *Ann. Rev. Nucl. Part. Sci.* 33 (1983). Ed. by V. W. Hughes and C. Cavata, pp. 611–644.

[3119] J. Ashman et al. "A Measurement of the Spin Asymmetry and Determination of the Structure Function $g_1$ in Deep Inelastic Muon-Proton Scattering". In: *Phys. Lett. B* 206 (1988). Ed. by V. W. Hughes and C. Cavata, p. 364.

[3120] J. Ashman et al. "An Investigation of the Spin Structure of the Proton in Deep Inelastic Scattering of Polarized Muons on Polarized Protons". In: *Nucl. Phys. B* 328 (1989). Ed. by V. W. Hughes and C. Cavata, p. 1.

[3121] B. W. Filippone and Xiang-Dong Ji. "The Spin structure of the nucleon". In: *Adv. Nucl. Phys.* 26 (2001), p. 1.

[3122] Steven D. Bass. "The Spin structure of the proton". In: *Rev. Mod. Phys.* 77 (2005), pp. 1257–1302.

[3123] Christine A. Aidala et al. "The Spin Structure of the Nucleon". In: *Rev. Mod. Phys.* 85 (2013), pp. 655–691.

[3124] E. Leader and C. Lorc¥'e. "The angular momentum controversy: What's it all about and does it matter?" In: *Phys. Rept.* 541.3 (2014), pp. 163–248.

[3125] Alexandre Deur, Stanley J. Brodsky, and Guy F. De T¥'eramond. "The Spin Structure of the Nucleon". In: (July 2018).

[3126] Gerry Bunce et al. "Prospects for spin physics at RHIC". In: *Ann. Rev. Nucl. Part. Sci.* 50 (2000), pp. 525–575.

[3127] Jozef Dudek et al. "Physics Opportunities with the 12 GeV Upgrade at Jefferson Lab". In: *Eur. Phys. J. A* 48 (2012), p. 187.

[3128] Daniel Boer et al. "Gluons and the quark sea at high energies: Distributions, polarization, tomography". In: (Aug. 2011).

[3129] W. K. Tung. *Group theory in physics*. 1985.

[3130] R. L. Jaffe and Aneesh Manohar. "The $G_1$ Problem: Fact and Fantasy on the Spin of the Proton". In: *Nucl. Phys. B* 337 (1990), pp. 509–546.

[3131] Xiang-Dong Ji. "Lorentz symmetry and the internal structure of the nucleon". In: *Phys. Rev. D* 58 (1998), p. 056003.

[3132] Xiangdong Ji, Yang Xu, and Yong Zhao. "Gluon Spin, Canonical Momentum, and Gauge Symmetry". In: *JHEP* 08 (2012), p. 082.

[3133] Aneesh V. Manohar. "Polarized parton distribution functions". In: *Phys. Rev. Lett.* 66 (1991), pp. 289–292.

[3134] Xiangdong Ji, Jian-Hui Zhang, and Yong Zhao. "Physics of the Gluon-Helicity Contribution to Proton Spin". In: *Phys. Rev. Lett.* 111 (2013), p. 112002.

[3135] M. Wakamatsu. "On Gauge-Invariant Decomposition of Nucleon Spin". In: *Phys. Rev. D* 81 (2010), p. 114010.

[3136] Matthias Burkardt. "Parton Orbital Angular Momentum and Final State Interactions". In: *Phys. Rev. D* 88.1 (2013), p. 014014.

[3137] Xiangdong Ji and Feng Yuan. "Transverse spin sum rule of the proton". In: *Phys. Lett. B* 810 (2020), p. 135786.

[3138] Xiangdong Ji, Xiaonu Xiong, and Feng Yuan. "Proton Spin Structure from Measurable Parton Distributions". In: *Phys. Rev. Lett.* 109 (2012), p. 152005.

[3139] Xiangdong Ji, Xiaonu Xiong, and Feng Yuan. "Transverse Polarization of the Nucleon in Parton Picture". In: *Phys. Lett. B* 717 (2012), pp. 214–218.

[3140] Yuxun Guo, Xiangdong Ji, and Kyle Shiells. "Novel twist-three transverse-spin sum rule for the proton and related generalized parton distributions". In: *Nucl. Phys. B* 969 (2021), p. 115440.

[3141] C. Alexandrou et al. "Nucleon Spin and Momentum Decomposition Using Lattice QCD Simulations". In: *Phys. Rev. Lett.* 119.14 (2017), p. 142002.

[3142] Jian Liang et al. "Quark spins and Anomalous Ward Identity". In: *Phys. Rev. D* 98.7 (2018), p. 074505.

[3143] Huey-Wen Lin et al. "Quark contribution to the proton spin from 2+1+1-flavor lattice QCD". In: *Phys. Rev. D* 98.9 (2018), p. 094512.

[3144] C. Adolph et al. "The spin structure function $g_1^{rmp}$ of the proton and a test of the Bjorken sum rule". In: *Phys. Lett. B* 753 (2016), pp. 18–28.

[3145] Keh-Fei Liu. "Status on lattice calculations of the proton spin decomposition". In: *AAPPS Bull.* 32.1 (2022), p. 8.

[3146] C. Alexandrou et al. "Complete flavor decomposition of the spin and momentum fraction of the proton using lattice QCD simulations at physical pion mass". In: *Phys. Rev. D* 101.9 (2020), p. 094513.

[3147] Huey-Wen Lin et al. "Parton distributions and lattice QCD calculations: a community white paper". In: *Prog. Part. Nucl. Phys.* 100 (2018), pp. 107–160.

[3148] M. Deka et al. "Lattice study of quark and glue momenta and angular momenta in the nucleon". In: *Phys. Rev. D* 91.1 (2015), p. 014505.

[3149] Ming Gong et al. "Strange and charm quark spins from the anomalous Ward identity". In: *Phys. Rev. D* 95.11 (2017), p. 114509.

[3150] N. Mathur et al. "Quark orbital angular momentum from lattice QCD". In: *Phys. Rev. D* 62 (2000), p. 114504.

[3151] Philipp Hagler et al. "Moments of nucleon generalized parton distributions in lattice QCD". In: *Phys. Rev. D* 68 (2003), p. 034505.

[3152] M. Gockeler et al. "Generalized parton distributions from lattice QCD". In: *Phys. Rev. Lett.* 92 (2004), p. 042002.

[3153] Dirk Brommel et al. "Moments of generalized parton distributions and quark angular momentum of the nucleon". In: *PoS* LATTICE2007 (2007). Ed. by Gunnar Bali et al., p. 158.

[3154] J. D. Bratt et al. "Nucleon structure from mixed action calculations using 2+1 flavors of asqtad sea and domain wall valence fermions". In: *Phys. Rev. D* 82 (2010), p. 094502.

[3155] S. N. Syritsyn et al. "Quark Contributions to Nucleon Momentum and Spin from Domain Wall fermion calculations". In: *PoS* LATTICE2011 (2011). Ed. by Pavlos Vranas, p. 178.

[3156] C. Alexandrou et al. "Moments of nucleon generalized parton distributions from lattice QCD". In: *Phys. Rev. D* 83 (2011), p. 114513.

[3157] C. Alexandrou et al. "Nucleon form factors and moments of generalized parton distributions using $N_f = 2 + 1 + 1$ twisted mass fermions". In: *Phys. Rev. D* 88.1 (2013), p. 014509.

[3158] Gen Wang et al. "Proton momentum and angular momentum decompositions with overlap fermions". In: *Phys. Rev. D* 106.1 (2022), p. 014512.

[3159] M. Engelhardt. "Quark orbital dynamics in the proton from Lattice QCD – from Ji to Jaffe-Manohar orbital angular momentum". In: *Phys. Rev. D* 95.9 (2017), p. 094505.

[3160] M. Engelhardt et al. "From Ji to Jaffe-Manohar orbital angular momentum in lattice QCD using a direct derivative method". In: *Phys. Rev. D* 102.7 (2020), p. 074505.

[3161] Xiangdong Ji, Jian-Hui Zhang, and Yong Zhao. "Justifying the Naive Partonic Sum Rule for Proton Spin". In: *Phys. Lett. B* 743 (2015), pp. 180–183.

[3162] Yuxun Guo, Xiangdong Ji, and Kyle Shiells. "Generalized parton distributions through universal moment parameterization: zero skewness case". In: (July 2022).

[3163] M. Mazouz et al. "Deeply virtual compton scattering off the neutron". In: *Phys. Rev. Lett.* 99 (2007), p. 242501.

[3164] L. Adamczyk et al. "Precision Measurement of the Longitudinal Double-spin Asymmetry for Inclusive Jet Production in Polarized Proton Collisions at $sqrts = 200$ GeV". In: *Phys. Rev. Lett.* 115.9 (2015), p. 092002.

[3165] J. Adam et al. "Longitudinal double-spin asymmetry for inclusive jet and dijet production in pp collisions at $sqrts = 510$ GeV". In: *Phys. Rev. D* 100.5 (2019), p. 052005.

[3166] Jaroslav Adam et al. "Measurement of the longitudinal spin asymmetries for weak boson pro-

duction in proton-proton collisions at $sqrts =$ 510 GeV". In: *Phys. Rev. D* 99.5 (2019), p. 051102.

[3167] A. Airapetian et al. "Measurement of Azimuthal Asymmetries With Respect To Both Beam Charge and Transverse Target Polarization in Exclusive Electroproduction of Real Photons". In: *JHEP* 06 (2008), p. 066.

[3168] Kresimir Kumericki, Dieter Mueller, and Andreas Schafer. "Neural network generated parametrizations of deeply virtual Compton form factors". In: *JHEP* 07 (2011), p. 073.

[3169] S. V. Goloskokov and P. Kroll. "The Target asymmetry in hard vector-meson electroproduction and parton angular momenta". In: *Eur. Phys. J. C* 59 (2009), pp. 809–819.

[3170] Gary R. Goldstein, J. OsvaldoGonzalez Hernandez, and Simonetta Liuti. "Flexible Parametrization of Generalized Parton Distributions from Deeply Virtual Compton Scattering Observables". In: *Phys. Rev. D* 84 (2011), p. 034007.

[3171] Dieter Mueller and A. Schafer. "Complex conformal spin partial wave expansion of generalized parton distributions and distribution amplitudes". In: *Nucl. Phys. B* 739 (2006), pp. 1–59.

[3172] Kresimir Kumeri¥vcki and Dieter Mueller. "Deeply virtual Compton scattering at small $x_B$ and the access to the GPD H". In: *Nucl. Phys. B* 841 (2010), pp. 1–58.

[3173] Yuri V. Kovchegov, Daniel Pitonyak, and Matthew D. Sievert. "Helicity Evolution at Small-x". In: *JHEP* 01 (2016). [Erratum: JHEP 10, 148 (2016)], p. 072.

[3174] Yuri V. Kovchegov, Daniel Pitonyak, and Matthew D. Sievert. "Small-$x$ asymptotics of the quark helicity distribution". In: *Phys. Rev. Lett.* 118.5 (2017), p. 052001.

[3175] Yuri V. Kovchegov and Matthew D. Sievert. "Small-$x$ Helicity Evolution: an Operator Treatment". In: *Phys. Rev. D* 99.5 (2019), p. 054032.

[3176] Renaud Boussarie, Yoshitaka Hatta, and Feng Yuan. "Proton Spin Structure at Small-$x$". In: *Phys. Lett. B* 797 (2019), p. 134817.

[3177] Yuri V. Kovchegov, Andrey Tarasov, and Yossathorn Tawabutr. "Helicity evolution at small x: the single-logarithmic contribution". In: *JHEP* 03 (2022), p. 184.

[3178] Andrey Tarasov and Raju Venugopalan. "Role of the chiral anomaly in polarized deeply inelastic scattering. II. Topological screening and transitions from emergent axionlike dynamics". In: *Phys. Rev. D* 105.1 (2022), p. 014020.

[3179] Andrey Tarasov and Raju Venugopalan. "Role of the chiral anomaly in polarized deeply inelastic scattering: Finding the triangle graph inside the box diagram in Bjorken and Regge asymptotics". In: *Phys. Rev. D* 102.11 (2020), p. 114022.

[3180] Florian Cougoulic et al. "Quark and gluon helicity evolution at small x: revised and updated". In: *JHEP* 07 (2022), p. 095.

[3181] Yoshitaka Hatta and Jian Zhou. "Small-$x$ evolution of the gluon GPD $E_g$". In: (July 2022).

[3182] E. Rutherford. "The scattering of alpha and beta particles by matter and the structure of the atom". In: *Phil. Mag. Ser. 6* 21 (1911), pp. 669–688.

[3183] Dieter Müller et al. "Wave functions, evolution equations and evolution kernels from light ray operators of QCD". In: *Fortsch. Phys.* 42 (1994), pp. 101–141.

[3184] A. V. Radyushkin. "Nonforward parton distributions". In: *Phys. Rev. D* 56 (1997), pp. 5524–5557.

[3185] K. Goeke, Maxim V. Polyakov, and M. Vanderhaeghen. "Hard exclusive reactions and the structure of hadrons". In: *Prog. Part. Nucl. Phys.* 47 (2001), pp. 401–515.

[3186] M. Diehl. "Generalized parton distributions". In: *Phys. Rept.* 388 (2003), pp. 41–277.

[3187] A. V. Belitsky and A. V. Radyushkin. "Unraveling hadron structure with generalized parton distributions". In: *Phys. Rept.* 418 (2005), pp. 1–387.

[3188] Sigfrido Boffi and Barbara Pasquini. "Generalized parton distributions and the structure of the nucleon". In: *Riv. Nuovo Cim.* 30.9 (2007), pp. 387–448.

[3189] P. J. Mulders and R. D. Tangerman. "The Complete tree level result up to order 1/Q for polarized deep inelastic leptoproduction". In: *Nucl. Phys. B* 461 (1996). [Erratum: Nucl.Phys.B 484, 538–540 (1997)], pp. 197–237.

[3190] Daniel Boer and P. J. Mulders. "Time reversal odd distribution functions in leptoproduction". In: *Phys. Rev. D* 57 (1998), pp. 5780–5786.

[3191] Matthias Burkardt. "Impact parameter space interpretation for generalized parton distributions". In: *Int. J. Mod. Phys. A* 18 (2003), pp. 173–208.

[3192] Xiang-dong Ji. "Viewing the proton through 'color' filters". In: *Phys. Rev. Lett.* 91 (2003), p. 062001.

[3193] Andrei V. Belitsky, Xiang-dong Ji, and Feng Yuan. "Quark imaging in the proton via quan-

tum phase space distributions”. In: *Phys. Rev. D* 69 (2004), p. 074014.

[3194] S. Meissner, A. Metz, and K. Goeke. “Relations between generalized and transverse momentum dependent parton distributions”. In: *Phys. Rev. D* 76 (2007), p. 034002.

[3195] C. Lorce and B. Pasquini. “Quark Wigner Distributions and Orbital Angular Momentum”. In: *Phys. Rev. D* 84 (2011), p. 014015.

[3196] Cedric Lorce et al. “The quark orbital angular momentum from Wigner distributions and light-cone wave functions”. In: *Phys. Rev. D* 85 (2012), p. 114006.

[3197] Xiangdong Ji, Xiaonu Xiong, and Feng Yuan. “Probing Parton Orbital Angular Momentum in Longitudinally Polarized Nucleon”. In: *Phys. Rev. D* 88.1 (2013), p. 014041.

[3198] Yoshitaka Hatta. “Notes on the orbital angular momentum of quarks in the nucleon”. In: *Phys. Lett. B* 708 (2012), pp. 186–190.

[3199] Xiangdong Ji, Feng Yuan, and Yong Zhao. “Hunting the Gluon Orbital Angular Momentum at the Electron-Ion Collider”. In: *Phys. Rev. Lett.* 118.19 (2017), p. 192004.

[3200] Yoshitaka Hatta et al. “Gluon orbital angular momentum at small-$x$”. In: *Phys. Rev. D* 95.11 (2017), p. 114032.

[3201] Aurore Courtoy et al. “On the Observability of the Quark Orbital Angular Momentum Distribution”. In: *Phys. Lett. B* 731 (2014), pp. 141–147.

[3202] Aurore Courtoy et al. “Identification of Observables for Quark and Gluon Orbital Angular Momentum”. In: (Dec. 2014).

[3203] Shohini Bhattacharya, Renaud Boussarie, and Yoshitaka Hatta. “Signature of the Gluon Orbital Angular Momentum”. In: *Phys. Rev. Lett.* 128.18 (2022), p. 182002.

[3204] Feng Yuan. “Generalized parton distributions at x —> 1”. In: *Phys. Rev. D* 69 (2004), p. 051501.

[3205] M. Göckeler et al. “Transverse spin structure of the nucleon from lattice QCD simulations”. In: *Phys. Rev. Lett.* 98 (2007), p. 222001.

[3206] Andrei V. Belitsky and Dieter Mueller. “Predictions from conformal algebra for the deeply virtual Compton scattering”. In: *Phys. Lett. B* 417 (1998), pp. 129–140.

[3207] Xiang-Dong Ji and Jonathan Osborne. “One loop QCD corrections to deeply virtual Compton scattering: The Parton helicity independent case”. In: *Phys. Rev. D* 57 (1998), pp. 1337–1340.

[3208] L. Mankiewicz et al. “NLO corrections to deeply virtual Compton scattering”. In: *Phys. Lett. B* 425 (1998). [Erratum: Phys.Lett.B 461, 423–423 (1999)], pp. 186–192.

[3209] Dieter Mueller. “Next-to-next-to leading order corrections to deeply virtual Compton scattering: The Non-singlet case”. In: *Phys. Lett. B* 634 (2006), pp. 227–234.

[3210] K. Kumericki et al. “Deeply virtual Compton scattering beyond next-to-leading order: the flavor singlet case”. In: *Phys. Lett. B* 648 (2007), pp. 186–194.

[3211] K. Kumericki, Dieter Mueller, and K. Passek-Kumericki. “Towards a fitting procedure for deeply virtual Compton scattering at next-to-leading order and beyond”. In: *Nucl. Phys. B* 794 (2008), pp. 244–323.

[3212] B. Pire, L. Szymanowski, and J. Wagner. “NLO corrections to timelike, spacelike and double deeply virtual Compton scattering”. In: *Phys. Rev. D* 83 (2011), p. 034009.

[3213] V. M. Braun et al. “Three-loop evolution equation for flavor-nonsinglet operators in off-forward kinematics”. In: *JHEP* 06 (2017), p. 037.

[3214] V. M. Braun et al. “Two-loop coefficient function for DVCS: vector contributions”. In: *JHEP* 09 (2020), p. 117.

[3215] V. M. Braun, Yao Ji, and Jakob Schoenleber. “Deeply-virtual Compton scattering at the next-to-next-to-leading order”. In: (July 2022).

[3216] Andrei V. Belitsky, Dieter Mueller, and A. Kirchner. “Theory of deeply virtual Compton scattering on the nucleon”. In: *Nucl. Phys. B* 629 (2002), pp. 323–392.

[3217] A. V. Belitsky and Dieter Mueller. “Exclusive electroproduction revisited: treating kinematical effects”. In: *Phys. Rev. D* 82 (2010), p. 074010.

[3218] Brandon Kriesten et al. “Extraction of generalized parton distribution observables from deeply virtual electron proton scattering experiments”. In: *Phys. Rev. D* 101.5 (2020), p. 054021.

[3219] Brandon Kriesten and Simonetta Liuti. “Theory of deeply virtual Compton scattering off the unpolarized proton”. In: *Phys. Rev. D* 105.1 (2022), p. 016015.

[3220] Jake Grigsby et al. “Deep learning analysis of deeply virtual exclusive photoproduction”. In: *Phys. Rev. D* 104.1 (2021), p. 016001.

[3221] Brandon Kriesten et al. “Parametrization of quark and gluon generalized parton distributions in a dynamical framework”. In: *Phys. Rev. D* 105.5 (2022), p. 056022.

[3222] Yuxun Guo, Xiangdong Ji, and Kyle Shiells. "Higher-order kinematical effects in deeply virtual Compton scattering". In: *JHEP* 12 (2021), p. 103.

[3223] Kyle Shiells, Yuxun Guo, and Xiangdong Ji. "On extraction of twist-two Compton form factors from DVCS observables through harmonic analysis". In: *JHEP* 08 (2022), p. 048.

[3224] Yuxun Guo et al. "Twist-three cross-sections in deeply virtual Compton scattering". In: *JHEP* 06 (2022), p. 096.

[3225] Constantia Alexandrou et al. "Unpolarized and helicity generalized parton distributions of the proton within lattice QCD". In: *Phys. Rev. Lett.* 125.26 (2020), p. 262001.

[3226] Alexei Prokudin, Peng Sun, and Feng Yuan. "Scheme dependence and transverse momentum distribution interpretation of Collins–Soper–Sterman resummation". In: *Phys. Lett. B* 750 (2015), pp. 533–538.

[3227] Armando Bermudez Martinez and Alexey Vladimirov. "Determination of Collins-Soper kernel from cross-sections ratios". In: (June 2022).

[3228] Marcin Bury, Alexei Prokudin, and Alexey Vladimirov. "Extraction of the Sivers function from SIDIS, Drell-Yan, and $W^{\pm}/Z$ boson production data with TMD evolution". In: *JHEP* 05 (2021), p. 151.

[3229] Stanley J. Brodsky, Dae Sung Hwang, and Ivan Schmidt. "Initial state interactions and single spin asymmetries in Drell-Yan processes". In: *Nucl. Phys. B* 642 (2002), pp. 344–356.

[3230] Daniel Boer, P. J. Mulders, and F. Pijlman. "Universality of T odd effects in single spin and azimuthal asymmetries". In: *Nucl. Phys. B* 667 (2003), pp. 201–241.

[3231] Xiangdong Ji et al. "Single Transverse-Spin Asymmetry in Drell-Yan Production at Large and Moderate Transverse Momentum". In: *Phys. Rev. D* 73 (2006), p. 094017.

[3232] Xiangdong Ji et al. "Single-transverse spin asymmetry in semi-inclusive deep inelastic scattering". In: *Phys. Lett. B* 638 (2006), pp. 178–186.

[3233] Yuji Koike, Werner Vogelsang, and Feng Yuan. "On the Relation Between Mechanisms for Single-Transverse-Spin Asymmetries". In: *Phys. Lett. B* 659 (2008), pp. 878–884.

[3234] A. Airapetian et al. "Observation of the Naive-T-odd Sivers Effect in Deep-Inelastic Scattering". In: *Phys. Rev. Lett.* 103 (2009), p. 152002.

[3235] A. Airapetian et al. "Azimuthal single- and double-spin asymmetries in semi-inclusive deep-inelastic lepton scattering by transversely polarized protons". In: *JHEP* 12 (2020), p. 010.

[3236] M. Alekseev et al. "Collins and Sivers asymmetries for pions and kaons in muon-deuteron DIS". In: *Phys. Lett. B* 673 (2009), pp. 127–135.

[3237] C. Adolph et al. "Collins and Sivers asymmetries in muonproduction of pions and kaons off transversely polarised protons". In: *Phys. Lett. B* 744 (2015), pp. 250–259.

[3238] C. Adolph et al. "II – Experimental investigation of transverse spin asymmetries in $\mu$ -p SIDIS processes: Sivers asymmetries". In: *Phys. Lett. B* 717 (2012), pp. 383–389.

[3239] C Adolph et al. "Sivers asymmetry extracted in SIDIS at the hard scales of the Drell–Yan process at COMPASS". In: *Phys. Lett. B* 770 (2017), pp. 138–145.

[3240] X. Qian et al. "Single Spin Asymmetries in Charged Pion Production from Semi-Inclusive Deep Inelastic Scattering on a Transversely Polarized $^3$He Target". In: *Phys. Rev. Lett.* 107 (2011), p. 072003.

[3241] Y. X. Zhao et al. "Single spin asymmetries in charged kaon production from semi-inclusive deep inelastic scattering on a transversely polarized $^3He$ target". In: *Phys. Rev. C* 90.5 (2014), p. 055201.

[3242] M. Aghasyan et al. "First measurement of transverse-spin-dependent azimuthal asymmetries in the Drell-Yan process". In: *Phys. Rev. Lett.* 119.11 (2017), p. 112002.

[3243] L. Adamczyk et al. "Measurement of the transverse single-spin asymmetry in $p^{\uparrow}+p \to W^{\pm}/Z^0$ at RHIC". In: *Phys. Rev. Lett.* 116.13 (2016), p. 132301.

[3244] Marcin Bury, Alexei Prokudin, and Alexey Vladimirov. "Extraction of the Sivers Function from SIDIS, Drell-Yan, and $W^{\pm}/Z$ Data at Next-to-Next-to-Next-to Leading Order". In: *Phys. Rev. Lett.* 126.11 (2021), p. 112002.

[3245] Alessandro Bacchetta et al. "Extraction of partonic transverse momentum distributions from semi-inclusive deep-inelastic scattering, Drell-Yan and Z-boson production". In: *JHEP* 06 (2017). [Erratum: JHEP 06, 051 (2019)], p. 081.

[3246] Ignazio Scimemi and Alexey Vladimirov. "Analysis of vector boson production within TMD factorization". In: *Eur. Phys. J. C* 78.2 (2018), p. 89.

[3247] Ignazio Scimemi and Alexey Vladimirov. "Non-perturbative structure of semi-inclusive deep-inelastic and Drell-Yan scattering at small transverse momentum". In: *JHEP* 06 (2020), p. 137.

[3248] Alessandro Bacchetta et al. "Unpolarized Transverse Momentum Distributions from a global fit

of Drell-Yan and Semi-Inclusive Deep-Inelastic Scattering data". In: (June 2022).
[3249] Peng Sun et al. "Nonperturbative functions for SIDIS and Drell–Yan processes". In: *Int. J. Mod. Phys. A* 33.11 (2018), p. 1841006.
[3250] Alessandro Bacchetta et al. "Difficulties in the description of Drell-Yan processes at moderate invariant mass and high transverse momentum". In: *Phys. Rev. D* 100.1 (2019), p. 014018.
[3251] J. O. Gonzalez-Hernandez et al. "Challenges with Large Transverse Momentum in Semi-Inclusive Deeply Inelastic Scattering". In: *Phys. Rev. D* 98.11 (2018), p. 114005.
[3252] Werner Vogelsang and Feng Yuan. "Next-to-leading Order Calculation of the Single Transverse Spin Asymmetry in the Drell-Yan Process". In: *Phys. Rev. D* 79 (2009), p. 094010.
[3253] Jian Zhou, Feng Yuan, and Zuo-Tang Liang. "QCD Evolution of the Transverse Momentum Dependent Correlations". In: *Phys. Rev. D* 79 (2009), p. 114022.
[3254] Andreas Schafer and Jian Zhou. "A Note on the scale evolution of the ETQS function $T_F(x,x)$". In: *Phys. Rev. D* 85 (2012), p. 117501.
[3255] Zhong-Bo Kang and Jian-Wei Qiu. "QCD evolution of naive-time-reversal-odd parton distribution functions". In: *Phys. Lett. B* 713 (2012), pp. 273–276.
[3256] Ignazio Scimemi, Andrey Tarasov, and Alexey Vladimirov. "Collinear matching for Sivers function at next-to-leading order". In: *JHEP* 05 (2019), p. 125.
[3257] Ahmad Idilbi et al. "Collins-Soper equation for the energy evolution of transverse-momentum and spin dependent parton distributions". In: *Phys. Rev. D* 70 (2004), p. 074021.
[3258] Zhong-Bo Kang, Bo-Wen Xiao, and Feng Yuan. "QCD Resummation for Single Spin Asymmetries". In: *Phys. Rev. Lett.* 107 (2011), p. 152002.
[3259] S. Mert Aybat et al. "The QCD Evolution of the Sivers Function". In: *Phys. Rev. D* 85 (2012), p. 034043.
[3260] Peng Sun and Feng Yuan. "Transverse momentum dependent evolution: Matching semi-inclusive deep inelastic scattering processes to Drell-Yan and W/Z boson production". In: *Phys. Rev. D* 88.11 (2013), p. 114012.
[3261] Xiangdong Ji et al. "Soft factor subtraction and transverse momentum dependent parton distributions on the lattice". In: *Phys. Rev. D* 91 (2015), p. 074009.
[3262] Xiangdong Ji et al. "Transverse momentum dependent parton quasidistributions". In: *Phys. Rev. D* 99.11 (2019), p. 114006.
[3263] Xiangdong Ji, Yizhuang Liu, and Yu-Sheng Liu. "TMD soft function from large-momentum effective theory". In: *Nucl. Phys. B* 955 (2020), p. 115054.
[3264] Xiangdong Ji, Yizhuang Liu, and Yu-Sheng Liu. "Transverse-momentum-dependent parton distribution functions from large-momentum effective theory". In: *Phys. Lett. B* 811 (2020), p. 135946.
[3265] Xiangdong Ji et al. "Single Transverse-Spin Asymmetry and Sivers Function in Large Momentum Effective Theory". In: *Phys. Rev. D* 103.7 (2021), p. 074005.
[3266] Markus A. Ebert, Iain W. Stewart, and Yong Zhao. "Towards Quasi-Transverse Momentum Dependent PDFs Computable on the Lattice". In: *JHEP* 09 (2019), p. 037.
[3267] Markus A. Ebert, Iain W. Stewart, and Yong Zhao. "Renormalization and Matching for the Collins-Soper Kernel from Lattice QCD". In: *JHEP* 03 (2020), p. 099.
[3268] Markus A. Ebert et al. "One-loop Matching for Spin-Dependent Quasi-TMDs". In: *JHEP* 09 (2020), p. 099.
[3269] Phiala Shanahan, Michael L. Wagman, and Yong Zhao. "Nonperturbative renormalization of staple-shaped Wilson line operators in lattice QCD". In: *Phys. Rev. D* 101.7 (2020), p. 074505.
[3270] Min-Huan Chu et al. "Nonperturbative Determination of Collins-Soper Kernel from Quasi Transverse-Momentum Dependent Wave Functions". In: (Apr. 2022).
[3271] Markus A. Ebert et al. "Factorization connecting continuum & lattice TMDs". In: *JHEP* 04 (2022), p. 178.
[3272] Stella T. Schindler, Iain W. Stewart, and Yong Zhao. "One-loop matching for gluon lattice TMDs". In: *JHEP* 08 (2022), p. 084.
[3273] Alexey A. Vladimirov and Andreas Schäfer. "Transverse momentum dependent factorization for lattice observables". In: *Phys. Rev. D* 101.7 (2020), p. 074517.
[3274] Alfred H. Mueller. "Soft gluons in the infinite momentum wave function and the BFKL pomeron". In: *Nucl. Phys. B* 415 (1994), pp. 373–385.
[3275] Alfred H. Mueller. "Parton saturation at small x and in large nuclei". In: *Nucl. Phys. B* 558 (1999), pp. 285–303.
[3276] Larry D. McLerran and Raju Venugopalan. "Computing quark and gluon distribution functions

for very large nuclei". In: *Phys. Rev. D* 49 (1994), pp. 2233–2241.

[3277] Larry D. McLerran and Raju Venugopalan. "Gluon distribution functions for very large nuclei at small transverse momentum". In: *Phys. Rev. D* 49 (1994), pp. 3352–3355.

[3278] Larry D. McLerran and Raju Venugopalan. "Green's functions in the color field of a large nucleus". In: *Phys. Rev. D* 50 (1994), pp. 2225–2233.

[3279] Cyrille Marquet, Bo-Wen Xiao, and Feng Yuan. "Semi-inclusive Deep Inelastic Scattering at small x". In: *Phys. Lett. B* 682 (2009), pp. 207–211.

[3280] Fabio Dominguez, Bo-Wen Xiao, and Feng Yuan. "$k_t$-factorization for Hard Processes in Nuclei". In: *Phys. Rev. Lett.* 106 (2011), p. 022301.

[3281] Fabio Dominguez et al. "Universality of Unintegrated Gluon Distributions at small x". In: *Phys. Rev. D* 83 (2011), p. 105005.

[3282] Andreas Metz and Jian Zhou. "Distribution of linearly polarized gluons inside a large nucleus". In: *Phys. Rev. D* 84 (2011), p. 051503.

[3283] I. Balitsky and A. Tarasov. "Gluon TMD in particle production from low to moderate x". In: *JHEP* 06 (2016), p. 164.

[3284] Tolga Altinoluk, Renaud Boussarie, and Piotr Kotko. "Interplay of the CGC and TMD frameworks to all orders in kinematic twist". In: *JHEP* 05 (2019), p. 156.

[3285] Tolga Altinoluk and Renaud Boussarie. "Low $x$ physics as an infinite twist (G)TMD framework: unravelling the origins of saturation". In: *JHEP* 10 (2019), p. 208.

[3286] A. H. Mueller, Bo-Wen Xiao, and Feng Yuan. "Sudakov Resummation in Small-$x$ Saturation Formalism". In: *Phys. Rev. Lett.* 110.8 (2013), p. 082301.

[3287] A. H. Mueller, Bo-Wen Xiao, and Feng Yuan. "Sudakov double logarithms resummation in hard processes in the small-x saturation formalism". In: *Phys. Rev. D* 88.11 (2013), p. 114010.

[3288] I. Balitsky and A. Tarasov. "Rapidity evolution of gluon TMD from low to moderate x". In: *JHEP* 10 (2015), p. 017.

[3289] Bo-Wen Xiao, Feng Yuan, and Jian Zhou. "Transverse Momentum Dependent Parton Distributions at Small-x". In: *Nucl. Phys. B* 921 (2017), pp. 104–126.

[3290] Ian Balitsky. "Gauge-invariant TMD factorization for Drell-Yan hadronic tensor at small x". In: *JHEP* 05 (2021), p. 046.

[3291] Pieter Taels et al. "Dijet photoproduction at low $x$ at next-to-leading order and its back-to-back limit". In: (Apr. 2022).

[3292] Yoshitaka Hatta, Bo-Wen Xiao, and Feng Yuan. "Probing the Small- x Gluon Tomography in Correlated Hard Diffractive Dijet Production in Deep Inelastic Scattering". In: *Phys. Rev. Lett.* 116.20 (2016), p. 202301.

[3293] Shohini Bhattacharya, Andreas Metz, and Jian Zhou. "Generalized TMDs and the exclusive double Drell–Yan process". In: *Phys. Lett. B* 771 (2017), pp. 396–400.

[3294] Shohini Bhattacharya et al. "Exclusive double quarkonium production and generalized TMDs of gluons". In: (Feb. 2018).

[3295] Tolga Altinoluk et al. "Diffractive Dijet Production in Deep Inelastic Scattering and Photon-Hadron Collisions in the Color Glass Condensate". In: *Phys. Lett. B* 758 (2016), pp. 373–383.

[3296] Jian Zhou. "Elliptic gluon generalized transverse-momentum-dependent distribution inside a large nucleus". In: *Phys. Rev. D* 94.11 (2016), p. 114017.

[3297] Yoshikazu Hagiwara et al. "Accessing the gluon Wigner distribution in ultraperipheral $pA$ collisions". In: *Phys. Rev. D* 96.3 (2017), p. 034009.

[3298] Heikki Mäntysaari, Niklas Mueller, and Björn Schenke. "Diffractive Dijet Production and Wigner Distributions from the Color Glass Condensate". In: *Phys. Rev. D* 99.7 (2019), p. 074004.

[3299] Heikki Mäntysaari et al. "Multigluon Correlations and Evidence of Saturation from Dijet Measurements at an Electron-Ion Collider". In: *Phys. Rev. Lett.* 124.11 (2020), p. 112301.

[3300] E. Iancu, A. H. Mueller, and D. N. Triantafyllopoulos. "Probing Parton Saturation and the Gluon Dipole via Diffractive Jet Production at the Electron-Ion Collider". In: *Phys. Rev. Lett.* 128.20 (2022), p. 202001.

[3301] E. Iancu et al. "Gluon dipole factorisation for diffractive dijets". In: (July 2022).

[3302] Yoshitaka Hatta, Bo-Wen Xiao, and Feng Yuan. "Semi-inclusive Diffractive Deep Inelastic Scattering at Small-$x$". In: (May 2022).

[3303] R. Keith Ellis, D. A. Ross, and A. E. Terrano. "The Perturbative Calculation of Jet Structure in $e^+e^-$ Annihilation". In: *Nucl. Phys. B* 178 (1981), pp. 421–456.

[3304] Zoltan Kunszt. "Comment on the $O(\alpha_s^2)$ Corrections to Jet Production in $e^+e^-$ Annihilation". In: *Phys. Lett. B* 99 (1981), pp. 429–432.

[3305] K. Fabricius et al. "Higher Order Perturbative QCD Calculation of Jet Cross-Sections in $e^+e^-$ Annihilation". In: *Z. Phys. C* 11 (1981), p. 315.

[3306] G. Passarino and M. J. G. Veltman. "One Loop Corrections for $e^+e^-$ Annihilation Into $\mu^+\mu^-$

in the Weinberg Model". In: *Nucl. Phys. B* 160 (1979), pp. 151–207.

[3307] Gerard 't Hooft and M. J. G. Veltman. "Scalar One Loop Integrals". In: *Nucl. Phys. B* 153 (1979), pp. 365–401.

[3308] G. J. van Oldenborgh and J. A. M. Vermaseren. "New Algorithms for One Loop Integrals". In: *Z. Phys. C* 46 (1990), pp. 425–438.

[3309] Zvi Bern, Lance J. Dixon, and David A. Kosower. "Dimensionally regulated pentagon integrals". In: *Nucl. Phys. B* 412 (1994), pp. 751–816.

[3310] S. Frixione, Z. Kunszt, and A. Signer. "Three jet cross-sections to next-to-leading order". In: *Nucl. Phys.* B467 (1996), pp. 399–442.

[3311] F. Aversa et al. "Jet Production in Hadronic Collisions to O $(\alpha_s^3)$". In: *Z. Phys. C* 46 (1990), p. 253.

[3312] Stephen D. Ellis, Zoltan Kunszt, and Davison E. Soper. "Two jet production in hadron collisions at order $\alpha_s^3$ in QCD". In: *Phys. Rev. Lett.* 69 (1992), pp. 1496–1499.

[3313] W. T. Giele, E. W. Nigel Glover, and David A. Kosower. "Higher order corrections to jet cross-sections in hadron colliders". In: *Nucl. Phys. B* 403 (1993), pp. 633–670.

[3314] Zoltan Nagy. "Next-to-leading order calculation of three jet observables in hadron hadron collision". In: *Phys. Rev.* D68 (2003), p. 094002.

[3315] Zvi Bern et al. "One loop n point gauge theory amplitudes, unitarity and collinear limits". In: *Nucl. Phys.* B425 (1994), pp. 217–260.

[3316] Ruth Britto, Freddy Cachazo, and Bo Feng. "Generalized unitarity and one-loop amplitudes in N=4 super-Yang-Mills". In: *Nucl. Phys.* B725 (2005), pp. 275–305.

[3317] Charalampos Anastasiou et al. "D-dimensional unitarity cut method". In: *Phys. Lett. B* 645 (2007), pp. 213–216.

[3318] Giovanni Ossola, Costas G. Papadopoulos, and Roberto Pittau. "Reducing full one-loop amplitudes to scalar integrals at the integrand level". In: *Nucl. Phys.* B763 (2007), pp. 147–169.

[3319] C. F. Berger et al. "An Automated Implementation of On-Shell Methods for One-Loop Amplitudes". In: *Phys. Rev. D* 78 (2008), p. 036003.

[3320] R. Keith Ellis et al. "Masses, fermions and generalized $D$-dimensional unitarity". In: *Nucl. Phys.* B822 (2009), pp. 270–282.

[3321] Simon Badger et al. "Next-to-leading order QCD corrections to five jet production at the LHC". In: *Phys. Rev.* D89.3 (2014), p. 034019.

[3322] Stefan Höche et al. "Next-to-leading order QCD predictions for top-quark pair production with up to three jets". In: *Eur. Phys. J. C* 77.3 (2017), p. 145.

[3323] F. R. Anger et al. "NLO QCD predictions for $Wb\bar{b}$ production in association with up to three light jets at the LHC". In: *Phys. Rev. D* 97.3 (2018), p. 036018.

[3324] Ansgar Denner and Giovanni Pelliccioli. "Combined NLO EW and QCD corrections to off-shell $t\bar{t}W$ production at the LHC". In: *Eur. Phys. J. C* 81.4 (2021), p. 354.

[3325] G. Bevilacqua et al. "HELAC-NLO". In: *Comput. Phys. Commun.* 184 (2013), pp. 986–997.

[3326] Gavin Cullen et al. "Automated One-Loop Calculations with GoSam". In: *Eur. Phys. J. C* 72 (2012), p. 1889.

[3327] Fabio Cascioli, Philipp Maierhofer, and Stefano Pozzorini. "Scattering Amplitudes with Open Loops". In: *Phys. Rev. Lett.* 108 (2012), p. 111601.

[3328] J. Alwall et al. "The automated computation of tree-level and next-to-leading order differential cross sections, and their matching to parton shower simulations". In: *JHEP* 07 (2014), p. 079.

[3329] Stefano Actis et al. "RECOLA: REcursive Computation of One-Loop Amplitudes". In: *Comput. Phys. Commun.* 214 (2017), pp. 140–173.

[3330] R. Frederix et al. "The automation of next-to-leading order electroweak calculations". In: *JHEP* 07 (2018), p. 185.

[3331] Steve Honeywell et al. "NLOX, a one-loop provider for Standard Model processes". In: *Comput. Phys. Commun.* 257 (2020), p. 107284.

[3332] Vladimir A. Smirnov. "Analytical result for dimensionally regularized massless on shell double box". In: *Phys. Lett. B* 460 (1999), pp. 397–404.

[3333] J. B. Tausk. "Nonplanar massless two loop Feynman diagrams with four on-shell legs". In: *Phys. Lett. B* 469 (1999), pp. 225–234.

[3334] T. Binoth and G. Heinrich. "An automatized algorithm to compute infrared divergent multiloop integrals". In: *Nucl. Phys.* B585 (2000), pp. 741–759.

[3335] L. W. Garland et al. "The Two loop QCD matrix element for $e^+e^- \to 3$ jets". In: *Nucl. Phys. B* 627 (2002), pp. 107–188.

[3336] Sven Moch, Peter Uwer, and Stefan Weinzierl. "Two loop amplitudes with nested sums: Fermionic contributions to $e^+e^- \to q\bar{q}g$". In: *Phys. Rev. D* 66 (2002), p. 114001.

[3337] M. Czakon, A. Mitov, and S. Moch. "Heavy-quark production in massless quark scattering

at two loops in QCD". In: *Phys. Lett. B* 651 (2007), pp. 147–159.
[3338] M. Czakon, A. Mitov, and S. Moch. "Heavy-quark production in gluon fusion at two loops in QCD". In: *Nucl. Phys. B* 798 (2008), pp. 210–250.
[3339] Micha0142 Czakon, Paul Fiedler, and Alexander Mitov. "Total Top-Quark Pair-Production Cross Section at Hadron Colliders Through $O(\frac{4}{S})$". In: *Phys. Rev. Lett.* 110 (2013), p. 252004.
[3340] T. Gehrmann et al. "$W^+W^-$ Production at Hadron Colliders in Next to Next to Leading Order QCD". In: *Phys. Rev. Lett.* 113.21 (2014), p. 212001.
[3341] F. Cascioli et al. "ZZ production at hadron colliders in NNLO QCD". In: *Phys. Lett.* B735 (2014), pp. 311–313.
[3342] Fabrizio Caola et al. "QCD corrections to $W^+W^-$ production through gluon fusion". In: *Phys. Lett. B* 754 (2016), pp. 275–280.
[3343] Fabrizio Caola et al. "QCD corrections to ZZ production in gluon fusion at the LHC". In: *Phys. Rev. D* 92.9 (2015), p. 094028.
[3344] Thomas Gehrmann, Andreas von Manteuffel, and Lorenzo Tancredi. "The two-loop helicity amplitudes for $q\bar{q}' \to V_1V_2 \to 4$ leptons". In: *JHEP* 09 (2015), p. 128.
[3345] Andreas von Manteuffel and Lorenzo Tancredi. "The two-loop helicity amplitudes for $gg \to V_1V_2 \to 4$ leptons". In: *JHEP* 06 (2015), p. 197.
[3346] S. Borowka et al. "Higgs Boson Pair Production in Gluon Fusion at Next-to-Leading Order with Full Top-Quark Mass Dependence". In: *Phys. Rev. Lett.* 117.1 (2016). [Erratum: Phys. Rev. Lett. 117, no.7, 079901 (2016)], p. 012001.
[3347] Julien Baglio et al. "Gluon fusion into Higgs pairs at NLO QCD and the top mass scheme". In: *Eur. Phys. J. C* 79.6 (2019), p. 459.
[3348] S. P. Jones, M. Kerner, and G. Luisoni. "Next-to-Leading-Order QCD Corrections to Higgs Boson Plus Jet Production with Full Top-Quark Mass Dependence". In: *Phys. Rev. Lett.* 120.16 (2018), p. 162001.
[3349] Joshua Davies, Go Mishima, and Matthias Steinhauser. "Virtual corrections to $gg \to ZH$ in the high-energy and large-$m_t$ limits". In: *JHEP* 03 (2021), p. 034.
[3350] Long Chen et al. "$ZH$ production in gluon fusion: two-loop amplitudes with full top quark mass dependence". In: *JHEP* 03 (2021), p. 125.
[3351] Lina Alasfar et al. "Virtual corrections to $gg \to ZH$ via a transverse momentum expansion". In: *JHEP* 05 (2021), p. 168.
[3352] C. Brønnum-Hansen and C.-Y. Wang. "Contribution of third generation quarks to two-loop helicity amplitudes for W boson pair production in gluon fusion". In: *JHEP* 01 (2021), p. 170.
[3353] B. Agarwal, S.P. Jones, and A. von Manteuffel. "Two-loop helicity amplitudes for $gg \to ZZ$ with full top-quark mass effects". In: *JHEP* 05 (2021), p. 256.
[3354] C. Brønnum-Hansen and C.-Y. Wang. "Top quark contribution to two-loop helicity amplitudes for $Z$ boson pair production in gluon fusion". In: *JHEP* 05 (2021), p. 244.
[3355] Gudrun Heinrich. "Collider Physics at the Precision Frontier". In: *Phys. Rept.* 922 (2021), pp. 1–69.
[3356] Alexander Huss et al. "Les Houches 2021: Physics at TeV Colliders: Report on the Standard Model Precision Wishlist". In: (July 2022).
[3357] Fabrizio Caola et al. "The Path forward to $N^3LO$". In: *2022 Snowmass Summer Study.* Mar. 2022.
[3358] A. Gehrmann-De Ridder et al. "Infrared structure of $e^+e^- \to 3$ jets at NNLO". In: *JHEP* 11 (2007), p. 058.
[3359] Stefan Weinzierl. "NNLO corrections to 3-jet observables in electron-positron annihilation". In: *Phys. Rev. Lett.* 101 (2008), p. 162001.
[3360] A. Gehrmann-De Ridder, T. Gehrmann, and E. W. Nigel Glover. "Antenna subtraction at NNLO". In: *JHEP* 09 (2005), p. 056.
[3361] James Currie, E. W. N. Glover, and Steven Wells. "Infrared Structure at NNLO Using Antenna Subtraction". In: *JHEP* 04 (2013), p. 066.
[3362] J Currie, E. W. N. Glover, and J Pires. "Next-to-Next-to Leading Order QCD Predictions for Single Jet Inclusive Production at the LHC". In: *Phys. Rev. Lett.* 118.7 (2017), p. 072002.
[3363] X. Chen et al. "NNLO QCD corrections in full colour for jet production observables at the LHC". In: *JHEP* 09 (2022), p. 025.
[3364] R. Gauld et al. "Precise predictions for WH+jet production at the LHC". In: *Phys. Lett. B* 817 (2021), p. 136335.
[3365] M. Czakon. "A novel subtraction scheme for double-real radiation at NNLO". In: *Phys. Lett.* B693 (2010), pp. 259–268.
[3366] Radja Boughezal, Kirill Melnikov, and Frank Petriello. "A subtraction scheme for NNLO computations". In: *Phys. Rev. D* 85 (2012), p. 034025.
[3367] M. Czakon and D. Heymes. "Four-dimensional formulation of the sector-improved residue subtraction scheme". In: *Nucl. Phys. B* 890 (2014), pp. 152–227.

[3368] Michał Czakon et al. "NNLO QCD predictions for W+c-jet production at the LHC". In: (Nov. 2020).

[3369] Michal Czakon, Alexander Mitov, and Rene Poncelet. "Tour de force in Quantum Chromodynamics: A first next-to-next-to-leading order study of three-jet production at the LHC". In: (June 2021).

[3370] Fabrizio Caola et al. "NNLO QCD corrections to associated $WH$ production and $H \to b\bar{b}$ decay". In: *Phys. Rev. D* 97.7 (2018), p. 074022.

[3371] Wojciech Bizon et al. "Anomalous couplings in associated VH production with Higgs boson decay to massive b quarks at NNLO in QCD". In: *Phys. Rev. D* 105.1 (2022), p. 014023.

[3372] Fabrizio Caola, Kirill Melnikov, and Raoul Röntsch. "Nested soft-collinear subtractions in NNLO QCD computations". In: *Eur. Phys. J. C* 77.4 (2017), p. 248.

[3373] Maximilian Delto and Kirill Melnikov. "Integrated triple-collinear counter-terms for the nested soft-collinear subtraction scheme". In: *JHEP* 05 (2019), p. 148.

[3374] Wojciech Bizoń and Maximilian Delto. "Analytic double-soft integrated subtraction terms for two massive emitters in a back-to-back kinematics". In: *JHEP* 07 (2020), p. 011.

[3375] Konstantin Asteriadis et al. "NNLO QCD corrections to weak boson fusion Higgs boson production in the $H \to b\bar{b}$ and $H \to WW^* \to 4l$ decay channels". In: *JHEP* 02 (2022), p. 046.

[3376] Maximilian Delto et al. "Mixed QCD⊗QED corrections to on-shell $Z$ boson production at the LHC". In: *JHEP* 01 (2020), p. 043.

[3377] Arnd Behring et al. "Mixed QCD-electroweak corrections to $W$-boson production in hadron collisions". In: *Phys. Rev. D* 103.1 (2021), p. 013008.

[3378] Federico Buccioni et al. "Mixed QCD-electroweak corrections to dilepton production at the LHC in the high invariant mass region". In: *JHEP* 06 (2022), p. 022.

[3379] Gabor Somogyi, Zoltan Trocsanyi, and Vittorio Del Duca. "Matching of singly- and doubly-unresolved limits of tree-level QCD squared matrix elements". In: *JHEP* 06 (2005), p. 024.

[3380] Gabor Somogyi. "A subtraction scheme for computing QCD jet cross sections at NNLO: integrating the doubly unresolved subtraction terms". In: *JHEP* 04 (2013), p. 010.

[3381] Vittorio Del Duca et al. "Three-Jet Production in Electron-Positron Collisions at Next-to-Next-to-Leading Order Accuracy". In: *Phys. Rev. Lett.* 117.15 (2016), p. 152004.

[3382] Vittorio Del Duca et al. "Higgs boson decay into b-quarks at NNLO accuracy". In: *JHEP* 04 (2015), p. 036.

[3383] Matteo Cacciari et al. "Fully Differential Vector-Boson-Fusion Higgs Production at Next-to-Next-to-Leading Order". In: *Phys. Rev. Lett.* 115.8 (2015). [Erratum: Phys. Rev. Lett. 120, no. 13, 139901 (2018)], p. 082002.

[3384] Frédéric A. Dreyer and Alexander Karlberg. "Fully differential Vector-Boson Fusion Higgs Pair Production at Next-to-Next-to-Leading Order". In: *Phys. Rev. D* 99.7 (2019), p. 074028.

[3385] Fr00E9d00E9ric A. Dreyer and Alexander Karlberg. "Vector-Boson Fusion Higgs Production at Three Loops in QCD". In: *Phys. Rev. Lett.* 117.7 (2016), p. 072001.

[3386] Edmond L. Berger et al. "NNLO QCD Corrections to t-channel Single Top-Quark Production and Decay". In: *Phys. Rev.* D94.7 (2016), p. 071501.

[3387] John Campbell, Tobias Neumann, and Zack Sullivan. "Single-top-quark production in the $t$-channel at NNLO". In: *JHEP* 02 (2021), p. 040.

[3388] L. Magnea et al. "Local analytic sector subtraction at NNLO". In: *JHEP* 12 (2018). [Erratum: JHEP 06, 013 (2019)], p. 107.

[3389] Lorenzo Magnea et al. "Factorisation and Subtraction beyond NLO". In: *JHEP* 12 (2018), p. 062.

[3390] Lorenzo Magnea et al. "Analytic integration of soft and collinear radiation in factorised QCD cross sections at NNLO". In: *JHEP* 02 (2021), p. 037.

[3391] Zeno Capatti, Valentin Hirschi, and Ben Ruijl. "Local Unitarity: cutting raised propagators and localising renormalisation". In: (Mar. 2022).

[3392] Zeno Capatti et al. "Local Unitarity: a representation of differential cross-sections that is locally free of infrared singularities at any order". In: *JHEP* 04 (2021), p. 104.

[3393] Stefano Catani and Massimiliano Grazzini. "An NNLO subtraction formalism in hadron collisions and its application to Higgs boson production at the LHC". In: *Phys. Rev. Lett.* 98 (2007), p. 222002.

[3394] Stefano Catani et al. "Vector boson production at hadron colliders: a fully exclusive QCD calculation at NNLO". In: *Phys. Rev. Lett.* 103 (2009), p. 082001.

[3395] Massimiliano Grazzini, Stefan Kallweit, and Marius Wiesemann. "Fully differential NNLO computations with MATRIX". In: *Eur. Phys. J.* C78.7 (2018), p. 537.

[3396] Stefano Catani et al. "Top-quark pair production at the LHC: Fully differential QCD predictions at NNLO". In: *JHEP* 07 (2019), p. 100.

[3397] Roberto Bonciani et al. "Mixed strong−electroweak corrections to the Drell−Yan process". In: (June 2021).

[3398] Tommaso Armadillo et al. "Two-loop mixed QCD-EW corrections to neutral current Drell-Yan". In: *JHEP* 05 (2022), p. 072.

[3399] John M. Campbell, R. Keith Ellis, and Ciaran Williams. "Direct Photon Production at Next-to2013Next-to-Leading Order". In: *Phys. Rev. Lett.* 118.22 (2017), p. 222001.

[3400] Radja Boughezal et al. "Higgs boson production in association with a jet at NNLO using jettiness subtraction". In: *Phys. Lett.* B748 (2015), pp. 5–8.

[3401] John M. Campbell, R. Keith Ellis, and Satyajit Seth. "H + 1 jet production revisited". In: *JHEP* 10 (2019), p. 136.

[3402] John Campbell and Tobias Neumann. "Precision Phenomenology with MCFM". In: *JHEP* 12 (2019), p. 034.

[3403] Claude Duhr, Falko Dulat, and Bernhard Mistlberger. "Drell-Yan Cross Section to Third Order in the Strong Coupling Constant". In: *Phys. Rev. Lett.* 125.17 (2020), p. 172001.

[3404] Claude Duhr, Falko Dulat, and Bernhard Mistlberger. "Charged current Drell-Yan production at N$^3$LO". In: *JHEP* 11 (2020), p. 143.

[3405] Claude Duhr and Bernhard Mistlberger. "Lepton-pair production at hadron colliders at N$^3$LO in QCD". In: *JHEP* 03 (2022), p. 116.

[3406] Xuan Chen et al. "Transverse Mass Distribution and Charge Asymmetry in W Boson Production to Third Order in QCD". In: (May 2022).

[3407] Tobias Neumann and John Campbell. "Fiducial Drell-Yan production at the LHC improved by transverse-momentum resummation at N$^4$LL + N$^3$LO". In: (July 2022).

[3408] Charalampos Anastasiou et al. "Higgs Boson Gluon-Fusion Production in QCD at Three Loops". In: *Phys. Rev. Lett.* 114 (2015), p. 212001.

[3409] Falko Dulat, Bernhard Mistlberger, and Andrea Pelloni. "Precision predictions at N$^3$LO for the Higgs boson rapidity distribution at the LHC". In: *Phys. Rev.* D99.3 (2019), p. 034004.

[3410] Leandro Cieri et al. "Higgs boson production at the LHC using the $q_T$ subtraction formalism at N$^3$LO QCD". In: *JHEP* 02 (2019), p. 096.

[3411] X. Chen et al. "Fully Differential Higgs Boson Production to Third Order in QCD". In: (Feb. 2021).

[3412] Julien Baglio et al. "Inclusive Production Cross Sections at N3LO". In: (Sept. 2022).

[3413] Frédéric A. Dreyer and Alexander Karlberg. "Vector-Boson Fusion Higgs Pair Production at N$^3$LO". In: *Phys. Rev. D* 98.11 (2018), p. 114016.

[3414] Long-Bin Chen et al. "Higgs boson pair production via gluon fusion at N$^3$LO in QCD". In: *Phys. Lett. B* 803 (2020), p. 135292.

[3415] Claude Duhr, Falko Dulat, and Bernhard Mistlberger. "Higgs Boson Production in Bottom-Quark Fusion to Third Order in the Strong Coupling". In: *Phys. Rev. Lett.* 125.5 (2020), p. 051804.

[3416] A. H. Ajjath et al. "Resummed prediction for Higgs boson production through $b\bar{b}$ annihilation at N$^3$LL". In: *JHEP* 11 (2019), p. 006.

[3417] Markus A. Ebert, Bernhard Mistlberger, and Gherardo Vita. "$N$-jettiness beam functions at N$^3$LO". In: *JHEP* 09 (2020), p. 143.

[3418] Claude Duhr, Bernhard Mistlberger, and Gherardo Vita. "Soft Integrals and Soft Anomalous Dimensions at N$^3$LO and Beyond". In: (May 2022).

[3419] Arnd Behring. "Zero-jettiness beam functions at N$^3$LO". In: *16th DESY Workshop on Elementary Particle Physics: Loops and Legs in Quantum Field Theory 2022*. July 2022.

[3420] Daniel Baranowski et al. "Same-hemisphere three-gluon-emission contribution to the zero-jettiness soft function at N3LO QCD". In: *Phys. Rev. D* 106.1 (2022), p. 014004.

[3421] Wen Chen et al. "Double-Real-Virtual and Double-Virtual-Real Corrections to the Three-Loop Thrust Soft Function". In: (June 2022).

[3422] Oscar Braun-White and Nigel Glover. "Decomposition of Triple Collinear Splitting Functions". In: (Apr. 2022).

[3423] Vittorio Del Duca et al. "Tree-level soft emission of a quark pair in association with a gluon". In: (June 2022).

[3424] Johannes Henn et al. "Massive three-loop form factor in the planar limit". In: *JHEP* 01 (2017), p. 074.

[3425] AndreyG. Grozin. "Heavy-quark form factors in the large $\beta_0$ limit". In: *Eur. Phys. J. C* 77.7 (2017), p. 453.

[3426] Roman N. Lee et al. "Three-loop massive form factors: complete light-fermion and large-N$_c$ corrections for vector, axial-vector, scalar and pseudo-scalar currents". In: *JHEP* 05 (2018), p. 187.

[3427] J. Ablinger et al. "Heavy quark form factors at three loops in the planar limit". In: *Phys. Lett. B* 782 (2018), pp. 528–532.

[3428] J. Blümlein et al. “The Heavy Fermion Contributions to the Massive Three Loop Form Factors”. In: *Nucl. Phys. B* 949 (2019), p. 114751.

[3429] Matteo Fael et al. “Massive Vector Form Factors to Three Loops”. In: *Phys. Rev. Lett.* 128.17 (2022), p. 172003.

[3430] Matteo Fael et al. “Singlet and non-singlet three-loop massive form factors”. In: (June 2022).

[3431] Fabrizio Caola, Andreas Von Manteuffel, and Lorenzo Tancredi. “Diphoton Amplitudes in Three-Loop Quantum Chromodynamics”. In: *Phys. Rev. Lett.* 126.11 (2021), p. 112004.

[3432] Piotr Bargiela et al. “Three-loop helicity amplitudes for diphoton production in gluon fusion”. In: *JHEP* 02 (2022), p. 153.

[3433] Fabrizio Caola et al. “Three-loop helicity amplitudes for four-quark scattering in massless QCD”. In: *JHEP* 10 (2021), p. 206.

[3434] Fabrizio Caola et al. “Three-loop helicity amplitudes for quark-gluon scattering in QCD”. In: (July 2022).

[3435] Fabrizio Caola et al. “Three-Loop Gluon Scattering in QCD and the Gluon Regge Trajectory”. In: *Phys. Rev. Lett.* 128.21 (2022), p. 212001.

[3436] Dhimiter D. Canko and Nikolaos Syrrakos. “Planar three-loop master integrals for $2 \to 2$ processes with one external massive particle”. In: *JHEP* 04 (2022), p. 134.

[3437] M. Czakon et al. “Exact top-quark mass dependence in hadronic Higgs production”. In: (May 2021).

[3438] Herschel A. Chawdhry et al. “NNLO QCD corrections to three-photon production at the LHC”. In: *JHEP* 02 (2020), p. 057.

[3439] Stefan Kallweit, Vasily Sotnikov, and Marius Wiesemann. “Triphoton production at hadron colliders in NNLO QCD”. In: *Phys. Lett. B* 812 (2021), p. 136013.

[3440] Simon Badger et al. “Next-to-leading order QCD corrections to diphoton-plus-jet production through gluon fusion at the LHC”. In: *Phys. Lett. B* 824 (2022), p. 136802.

[3441] Simon Badger et al. “Virtual QCD corrections to gluon-initiated diphoton plus jet production at hadron colliders”. In: *JHEP* 11 (2021), p. 083.

[3442] Bakul Agarwal et al. “Two-Loop Helicity Amplitudes for Diphoton Plus Jet Production in Full Color”. In: *Phys. Rev. Lett.* 127.26 (2021), p. 262001.

[3443] Herschel A. Chawdhry et al. “NNLO QCD corrections to diphoton production with an additional jet at the LHC”. In: (May 2021).

[3444] Simon Badger, Heribertus Bayu Hartanto, and Simone Zoia. “Two-Loop QCD Corrections to $Wb\bar{b}$ Production at Hadron Colliders”. In: *Phys. Rev. Lett.* 127.1 (2021), p. 012001.

[3445] Heribertus Bayu Hartanto et al. “NNLO QCD corrections to $Wb\bar{b}$ production at the LHC”. In: (May 2022).

[3446] Andreas von Manteuffel, Erik Panzer, and Robert M. Schabinger. “Cusp and collinear anomalous dimensions in four-loop QCD from form factors”. In: *Phys. Rev. Lett.* 124.16 (2020), p. 162001.

[3447] G. Das, S. Moch, and A. Vogt. “Approximate four-loop QCD corrections to the Higgs-boson production cross section”. In: *Phys. Lett. B* 807 (2020), p. 135546.

[3448] Goutam Das, Sven-Olaf Moch, and Andreas Vogt. “Soft corrections to inclusive deep-inelastic scattering at four loops and beyond”. In: *JHEP* 03 (2020), p. 116.

[3449] P. A. Baikov, K. G. Chetyrkin, and J. H. Kühn. “Five-Loop Running of the QCD coupling constant”. In: *Phys. Rev. Lett.* 118.8 (2017), p. 082002.

[3450] F. Herzog et al. “The five-loop beta function of Yang-Mills theory with fermions”. In: *JHEP* 02 (2017), p. 090.

[3451] Thomas Luthe et al. “The five-loop Beta function for a general gauge group and anomalous dimensions beyond Feynman gauge”. In: *JHEP* 10 (2017), p. 166.

[3452] K. G. Chetyrkin et al. “Five-loop renormalisation of QCD in covariant gauges”. In: *JHEP* 10 (2017). [Addendum: JHEP12,006(2017)], p. 179.

[3453] M. Borinsky et al. “Five-loop renormalization of $\phi^3$ theory with applications to the Lee-Yang edge singularity and percolation theory”. In: *Phys. Rev. D* 103.11 (2021), p. 116024.

[3454] Tatsumi Aoyama et al. “Tenth-Order QED Contribution to the Electron g-2 and an Improved Value of the Fine Structure Constant”. In: *Phys. Rev. Lett.* 109 (2012), p. 111807.

[3455] Tatsumi Aoyama et al. “Tenth-Order Electron Anomalous Magnetic Moment — Contribution of Diagrams without Closed Lepton Loops”. In: *Phys. Rev. D* 91.3 (2015). [Erratum: Phys. Rev. D 96, 019901 (2017)], p. 033006.

[3456] Tatsumi Aoyama, Toichiro Kinoshita, and Makiko Nio. “Revised and Improved Value of the QED Tenth-Order Electron Anomalous Magnetic Moment”. In: *Phys. Rev. D* 97.3 (2018), p. 036001.

[3457] Sergey Volkov. “Calculating the five-loop QED contribution to the electron anomalous magnetic moment: Graphs without lepton loops”. In: *Phys. Rev. D* 100.9 (2019), p. 096004.

[3458] Mikhail V. Kompaniets and Erik Panzer. “Minimally subtracted six loop renormalization of $O(n)$-symmetric $\phi^4$ theory and critical exponents”. In: *Phys. Rev. D* 96.3 (2017), p. 036016.

[3459] Alexander Bednyakov and Andrey Pikelner. “Six-loop anomalous dimension of the $\phi^Q$ operator in the $O(N)$ symmetric model”. In: (Aug. 2022).

[3460] David J. Broadhurst and D. Kreimer. “Knots and numbers in $\phi^4$ theory to 7 loops and beyond”. In: *Int. J. Mod. Phys. C* 6 (1995). Ed. by Bruce H. Denby and D. Perret-Gallix, pp. 519–524.

[3461] Oliver Schnetz. “Numbers and Functions in Quantum Field Theory”. In: *Phys. Rev. D* 97.8 (2018), p. 085018.

[3462] James Currie et al. “Infrared sensitivity of single jet inclusive production at hadron colliders”. In: *JHEP* 10 (2018), p. 155.

[3463] D. Britzger et al. “NNLO interpolation grids for jet production at the LHC”. In: (July 2022).

[3464] F. Bloch and A Nordsieck. “Note on the radiation field of the electron”. In: *Phys. Rev.* 52 (1937), p. 54.

[3465] T. Kinoshita. “Mass singularities of Feynman amplitudes”. In: *J. Math. Phys.* 3 (1962), p. 650.

[3466] T.D. Lee and M Nauenberg. “Degenerate systems and mass singularities”. In: *Phys. Rev.* B133 (1964), p. 1549.

[3467] S. Catani and L. Trentadue. “Resummation of the QCD Perturbative Series for Hard Processes”. In: *Nucl. Phys. B* 327 (1989), pp. 323–352.

[3468] S. Catani et al. “Resummation of large logarithms in e+ e- event shape distributions”. In: *Nucl. Phys.* B407 (1993), pp. 3–42.

[3469] Nikolaos Kidonakis and George F. Sterman. “Resummation for QCD hard scattering”. In: *Nucl. Phys. B* 505 (1997), pp. 321–348.

[3470] Thomas Becher, Alessandro Broggio, and Andrea Ferroglia. “Introduction to Soft-Collinear Effective Theory”. In: *Lect. Notes Phys.* 896 (2015), pp.1–206.

[3471] George Sterman and Mao Zeng. “Quantifying Comparisons of Threshold Resummations”. In: *JHEP* 05 (2014), p. 132.

[3472] Leandro G. Almeida et al. “Comparing and counting logs in direct and effective methods of QCD resummation”. In: *JHEP* 04 (2014), p. 174.

[3473] Marco Bonvini et al. “Resummation prescriptions and ambiguities in SCET vs. direct QCD: Higgs production as a case study”. In: *JHEP* 01 (2015), p. 046.

[3474] Marco Bonvini et al. “The scale of soft resummation in SCET vs perturbative QCD”. In: *Nucl. Phys. B Proc. Suppl.* 241-242 (2013). Ed. by G. Ricciardi et al., pp. 121–126.

[3475] Marco Bonvini et al. “Threshold Resummation in SCET vs. Perturbative QCD: An Analytic Comparison”. In: *Nucl. Phys. B* 861 (2012), pp. 337–360.

[3476] Marco Bonvini and Luca Rottoli. “Three loop soft function for $\mathrm{N}^3\mathrm{LL}'$ gluon fusion Higgs production in soft-collinear effective theory”. In: *Phys. Rev. D* 91.5 (2015), p. 051301.

[3477] Guido Altarelli, G. Parisi, and R. Petronzio. “Transverse Momentum in Drell-Yan Processes”. In: *Phys. Lett. B* 76 (1978), pp. 351–355.

[3478] G. Bozzi et al. “The q(T) spectrum of the Higgs boson at the LHC in QCD perturbation theory”. In: *Phys. Lett. B* 564 (2003), pp. 65–72.

[3479] Giuseppe Bozzi et al. “Transverse-momentum resummation and the spectrum of the Higgs boson at the LHC”. In: *Nucl. Phys.* B737 (2006), pp. 73–120.

[3480] Giuseppe Bozzi et al. “Production of Drell-Yan lepton pairs in hadron collisions: Transverse-momentum resummation at next-to-next-to-leading logarithmic accuracy”. In: *Phys. Lett.* B696 (2011), pp. 207–213.

[3481] Stefano Catani and Massimiliano Grazzini. “QCD transverse-momentum resummation in gluon fusion processes”. In: *Nucl. Phys. B* 845 (2011), pp. 297–323.

[3482] Thomas Becher and Matthias Neubert. “Drell-Yan Production at Small $q_T$, Transverse Parton Distributions and the Collinear Anomaly”. In: *Eur. Phys. J.* C71 (2011), p. 1665.

[3483] Thomas Becher, Matthias Neubert, and Daniel Wilhelm. “Higgs-Boson Production at Small Transverse Momentum”. In: *JHEP* 05 (2013), p. 110.

[3484] Stefano Catani et al. “Universality of transverse-momentum resummation and hard factors at the NNLO”. In: *Nucl. Phys. B* 881 (2014), pp. 414–443.

[3485] Thomas Gehrmann, Thomas Luebbert, and Li Lin Yang. “Calculation of the transverse parton distribution functions at next-to-next-to-leading order”. In: *JHEP* 06 (2014), p. 155.

[3486] G. A. Ladinsky and C. P. Yuan. “The Nonperturbative regime in QCD resummation for gauge boson production at hadron colliders”. In: *Phys. Rev. D* 50 (1994), R4239.

[3487] F. Landry et al. “Tevatron Run-1 $Z$ boson data and Collins-Soper-Sterman resummation formalism”. In: *Phys. Rev. D* 67 (2003), p. 073016.

[3488] Andrea Banfi et al. “Predictions for Drell-Yan $\phi^*$ and $Q_T$ observables at the LHC”. In: *Phys. Lett.* B715 (2012), pp. 152–156.

[3489] Daniel de Florian et al. “Transverse-momentum resummation: Higgs boson production at the Tevatron and the LHC”. In: *JHEP* 11 (2011), p. 064.

[3490] D. de Florian et al. “Higgs boson production at the LHC: transverse momentum resummation effects in the H->2gamma, H->WW->lnu lnu and H->ZZ->4l decay modes”. In: *JHEP* 06 (2012), p. 132.

[3491] Stefano Camarda et al. “DYTurbo: Fast predictions for Drell-Yan processes”. In: *Eur. Phys. J. C* 80.3 (2020). [Erratum: Eur. Phys. J. C 80, 440 (2020)], p. 251.

[3492] Wojciech Bizon et al. “Momentum-space resummation for transverse observables and the Higgs $p_\perp$ at $N^3$LL+NNLO”. In: *JHEP* 02 (2018), p. 108.

[3493] Wojciech Bizoń et al. “Fiducial distributions in Higgs and Drell-Yan production at $N^3$LL+NNLO”. In: *JHEP* 12 (2018), p. 132.

[3494] Wojciech Bizon et al. “The transverse momentum spectrum of weak gauge bosons at N $^3$ LL + NNLO”. In: *Eur. Phys. J. C* 79.10 (2019), p. 868.

[3495] Xuan Chen et al. “Third-Order Fiducial Predictions for Drell-Yan Production at the LHC”. In: *Phys. Rev. Lett.* 128.25 (2022), p. 252001.

[3496] M. Vesterinen and T. R. Wyatt. “A Novel Technique for Studying the Z Boson Transverse Momentum Distribution at Hadron Colliders”. In: *Nucl. Instrum. Meth. A* 602 (2009), pp. 432–437.

[3497] A. Banfi et al. “Optimisation of variables for studying dilepton transverse momentum distributions at hadron colliders”. In: *Eur. Phys. J. C* 71 (2011), p. 1600.

[3498] Andrea Banfi, Mrinal Dasgupta, and Rosa Maria Duran Delgado. “The a(T) distribution of the Z boson at hadron colliders”. In: *JHEP* 12 (2009), p. 022.

[3499] Andrea Banfi et al. “Probing the low transverse momentum domain of Z production with novel variables”. In: *JHEP* 01 (2012), p. 044.

[3500] Andrea Banfi, Mrinal Dasgupta, and Simone Marzani. “QCD predictions for new variables to study dilepton transverse momenta at hadron colliders”. In: *Phys. Lett. B* 701 (2011), pp. 75–81.

[3501] Marco Guzzi, Pavel M. Nadolsky, and Bowen Wang. “Nonperturbative contributions to a resummed leptonic angular distribution in inclusive neutral vector boson production”. In: *Phys. Rev. D* 90.1 (2014), p. 014030.

[3502] Andrea Banfi, Gavin P. Salam, and Giulia Zanderighi. “Principles of general final-state resummation and automated implementation”. In: *JHEP* 03 (2005), p. 073.

[3503] Stefano Catani et al. “Soft gluon resummation for Higgs boson production at hadron colliders”. In: *JHEP* 07 (2003), p. 028.

[3504] Georges Aad et al. “Measurement of the transverse momentum distribution of Drell–Yan lepton pairs in proton–proton collisions at $\sqrt{s}$ = 13 TeV with the ATLAS detector”. In: *Eur. Phys. J. C* 80.7 (2020), p. 616.

[3505] Stephen D. Ellis and Davison E. Soper. “Successive combination jet algorithm for hadron collisions”. In: *Phys. Rev.* D48 (1993), pp. 3160–3166.

[3506] Simone Marzani, Gregory Soyez, and Michael Spannowsky. “Looking inside jets: an introduction to jet substructure and boosted-object phenomenology”. In: (2019). [Lect. Notes Phys.958,pp.(2019)].

[3507] Edward Farhi. “A QCD Test for Jets”. In: *Phys. Rev. Lett.* 39 (1977), pp. 1587–1588.

[3508] S. Catani et al. “Thrust distribution in e+ e- annihilation”. In: *Phys. Lett. B* 263 (1991), pp. 491–497.

[3509] M. Dasgupta and G. P. Salam. “Resummation of nonglobal QCD observables”. In: *Phys. Lett. B* 512 (2001), pp. 323–330.

[3510] Andrea Banfi et al. “Non-global logarithms and jet algorithms in high-pT jet shapes”. In: *JHEP* 08 (2010), p. 064.

[3511] A. Banfi, G. Marchesini, and G. Smye. “Away from jet energy flow”. In: *JHEP* 08 (2002), p. 006.

[3512] Jeffrey R. Forshaw, A. Kyrieleis, and M. H. Seymour. “Super-leading logarithms in non-global observables in QCD”. In: *JHEP* 08 (2006), p. 059.

[3513] J. R. Forshaw, A. Kyrieleis, and M. H. Seymour. “Super-leading logarithms in non-global observables in QCD: Colour basis independent calculation”. In: *JHEP* 09 (2008), p. 128.

[3514] Heribert Weigert. “Nonglobal jet evolution at finite N(c)”. In: *Nucl. Phys. B* 685 (2004), pp. 321–350.

[3515] Yoshitaka Hatta and Takahiro Ueda. “Resummation of non-global logarithms at finite $N_c$”. In: *Nucl. Phys. B* 874 (2013), pp. 808–820.

[3516] Simon Caron-Huot. “Resummation of non-global logarithms and the BFKL equation”. In: *JHEP* 03 (2018), p. 036.

[3517] Andrea Banfi, Frédéric A. Dreyer, and Pier Francesco Monni. "Next-to-leading non-global logarithms in QCD". In: *JHEP* 10 (2021), p. 006.

[3518] Andrew J. Larkoski, Jesse Thaler, and Wouter J. Waalewijn. "Gaining (Mutual) Information about Quark/Gluon Discrimination". In: *JHEP* 11 (2014), p. 129.

[3519] Armen Tumasyan et al. "Study of quark and gluon jet substructure in Z+jet and dijet events from pp collisions". In: *JHEP* 01 (2022), p. 188.

[3520] Simone Caletti et al. "Jet angularities in Z+jet production at the LHC". In: *JHEP* 07 (2021), p. 076.

[3521] Daniel Reichelt et al. "Phenomenology of jet angularities at the LHC". In: *JHEP* 03 (2022), p. 131.

[3522] Enrico Bothmann et al. "Event Generation with Sherpa 2.2". In: *SciPost Phys.* 7.3 (2019), p. 034.

[3523] Mrinal Dasgupta et al. "Jet substructure with analytical methods". In: *Eur. Phys. J. C* 73.11 (2013), p. 2623.

[3524] Jeffrey R. Forshaw, Michael H. Seymour, and Andrzej Siodmok. "On the Breaking of Collinear Factorization in QCD". In: *JHEP* 11 (2012), p. 066.

[3525] Johannes Bellm et al. "Herwig 7.2 release note". In: *Eur. Phys. J. C* 80.5 (2020), p. 452.

[3526] Christian Bierlich et al. "A comprehensive guide to the physics and usage of PYTHIA 8.3". In: (Mar. 2022).

[3527] T. Kinoshita. "Mass singularities of Feynman amplitudes". In: *J. Math. Phys.* 3 (1962), pp. 650–677.

[3528] T. D. Lee and M. Nauenberg. "Degenerate Systems and Mass Singularities". In: *Phys. Rev.* 133 (1964). Ed. by G. Feinberg, B1549–B1562.

[3529] Mats Bengtsson, Torbjorn Sjostrand, and Maria van Zijl. "Initial State Radiation Effects on $W$ and Jet Production". In: *Z. Phys. C* 32 (1986). Ed. by S. C. Loken, p. 67.

[3530] Geoffrey C. Fox and Stephen Wolfram. "A Model for Parton Showers in QCD". In: *Nucl. Phys. B* 168 (1980), pp. 285–295.

[3531] Mats Bengtsson and Torbjorn Sjostrand. "Coherent Parton Showers Versus Matrix Elements: Implications of PETRA - PEP Data". In: *Phys. Lett. B* 185 (1987), p. 435.

[3532] G. Marchesini and B. R. Webber. "Simulation of QCD Jets Including Soft Gluon Interference". In: *Nucl. Phys. B* 238 (1984), pp. 1–29.

[3533] B. R. Webber. "A QCD Model for Jet Fragmentation Including Soft Gluon Interference". In: *Nucl. Phys.* B238 (1984), pp. 492–528.

[3534] Gosta Gustafson and Ulf Pettersson. "Dipole Formulation of QCD Cascades". In: *Nucl. Phys. B* 306 (1988), pp. 746–758.

[3535] Leif Lonnblad. "ARIADNE version 4: A Program for simulation of QCD cascades implementing the color dipole model". In: *Comput. Phys. Commun.* 71 (1992), pp. 15–31.

[3536] Bo Andersson. *The Lund model.* Vol. 7. Cambridge University Press, July 2005.

[3537] G. Gustafson. "Dual Description of a Confined Color Field". In: *Phys. Lett. B* 175 (1986). Ed. by J. Tran Thanh Van, p. 453.

[3538] Nadine Fischer et al. "Vincia for Hadron Colliders". In: *Eur. Phys. J. C* 76.11 (2016), p. 589.

[3539] Zoltan Nagy and Davison E. Soper. "Matching parton showers to NLO computations". In: *JHEP* 10 (2005), p. 024.

[3540] Steffen Schumann and Frank Krauss. "A Parton shower algorithm based on Catani-Seymour dipole factorisation". In: *JHEP* 03 (2008), p. 038.

[3541] Simon Platzer and Stefan Gieseke. "Coherent Parton Showers with Local Recoils". In: *JHEP* 01 (2011), p. 024.

[3542] Stefan Höche and Stefan Prestel. "The midpoint between dipole and parton showers". In: *Eur. Phys. J. C* 75.9 (2015), p. 461.

[3543] Stefano Frixione and Bryan R. Webber. "Matching NLO QCD computations and parton shower simulations". In: *JHEP* 06 (2002), p. 029.

[3544] Paolo Nason. "A New method for combining NLO QCD with shower Monte Carlo algorithms". In: *JHEP* 11 (2004), p. 040.

[3545] Stefano Frixione, Paolo Nason, and Carlo Oleari. "Matching NLO QCD computations with Parton Shower simulations: the POWHEG method". In: *JHEP* 11 (2007), p. 070.

[3546] Keith Hamilton et al. "Merging H/W/Z + 0 and 1 jet at NLO with no merging scale: a path to parton shower + NNLO matching". In: *JHEP* 05 (2013), p. 082.

[3547] Stefan Höche, Ye Li, and Stefan Prestel. "Drell-Yan lepton pair production at NNLO QCD with parton showers". In: *Phys. Rev. D* 91.7 (2015), p. 074015.

[3548] S. Catani et al. "QCD matrix elements + parton showers". In: *JHEP* 11 (2001), p. 063.

[3549] Leif Lonnblad. "Correcting the color dipole cascade model with fixed order matrix elements". In: *JHEP* 05 (2002), p. 046.

[3550] Stefan Hoeche et al. "QCD matrix elements + parton showers: The NLO case". In: *JHEP* 04 (2013), p. 027.

[3551] Rikkert Frederix and Stefano Frixione. "Merging meets matching in MC@NLO". In: *JHEP* 12 (2012), p. 061.

[3552] Stefan Höche, Daniel Reichelt, and Frank Siegert. "Momentum conservation and unitarity in parton showers and NLL resummation". In: *JHEP* 01 (2018), p. 118.

[3553] Mrinal Dasgupta et al. "Logarithmic accuracy of parton showers: a fixed-order study". In: *JHEP* 09 (2018). [Erratum: JHEP 03, 083 (2020)], p. 033.

[3554] Stefan Höche, Frank Krauss, and Stefan Prestel. "Implementing NLO DGLAP evolution in Parton Showers". In: *JHEP* 10 (2017), p. 093.

[3555] Falko Dulat, Stefan Höche, and Stefan Prestel. "Leading-Color Fully Differential Two-Loop Soft Corrections to QCD Dipole Showers". In: *Phys. Rev. D* 98.7 (2018), p. 074013.

[3556] Simon Platzer and Malin Sjodahl. "Subleading $N_c$ improved Parton Showers". In: *JHEP* 07 (2012), p. 042.

[3557] Zoltan Nagy and Davison E. Soper. "Effects of subleading color in a parton shower". In: *JHEP* 07 (2015), p. 119.

[3558] René Ángeles Martínez et al. "Soft gluon evolution and non-global logarithms". In: *JHEP* 05 (2018), p. 044.

[3559] Andy Buckley et al. "General-purpose event generators for LHC physics". In: *Phys. Rept.* 504 (2011), pp. 145–233.

[3560] J. M. Campbell et al. "Event Generators for High-Energy Physics Experiments". In: *2022 Snowmass Summer Study.* Mar. 2022.

[3561] X. Artru and G. Mennessier. "String model and multiproduction". In: *Nucl. Phys. B* 70 (1974), pp. 93–115.

[3562] R. D. Field and R. P. Feynman. "A Parametrization of the Properties of Quark Jets". In: *Nucl. Phys. B* 136 (1978). Ed. by L. M. Brown, p. 1.

[3563] P. Hoyer et al. "Quantum Chromodynamics and Jets in e+ e-". In: *Nucl. Phys. B* 161 (1979), pp. 349–372.

[3564] Ahmed Ali et al. "QCD Predictions for Four Jet Final States in e+ e- Annihilation". In: *Nucl. Phys. B* 167 (1980), pp. 454–478.

[3565] Bo Andersson, Gosta Gustafson, and Torbjörn Sjöstrand. "How to Find the Gluon Jets in e+ e- Annihilation". In: *Phys. Lett. B* 94 (1980), pp. 211–215.

[3566] W. Bartel et al. "Experimental Study of Jets in electron - Positron Annihilation". In: *Phys. Lett. B* 101 (1981), pp. 129–134.

[3567] Frank E. Paige and Serban D. Protopopescu. "Isajet: A Monte Carlo Event Generator for Isabelle, Version 2". In: (Oct. 1981).

[3568] K. Konishi, A. Ukawa, and G. Veneziano. "Jet Calculus: A Simple Algorithm for Resolving QCD Jets". In: *Nucl. Phys. B* 157 (1979), pp. 45–107.

[3569] Simone Alioli et al. "A general framework for implementing NLO calculations in shower Monte Carlo programs: the POWHEG BOX". In: *JHEP* 06 (2010), p. 043.

[3570] G. Ingelman and P. E. Schlein. "Jet Structure in High Mass Diffractive Scattering". In: *Phys. Lett. B* 152 (1985), pp. 256–260.

[3571] V. A. Abramovsky, V. N. Gribov, and O. V. Kancheli. "Character of Inclusive Spectra and Fluctuations Produced in Inelastic Processes by Multi - Pomeron Exchange". In: *Yad. Fiz.* 18 (1973). [Sov. J. Nucl. Phys.18,308(1974)], pp. 595–616.

[3572] Paolo Bartalini and Jonathan Richard Gaunt, eds. *Multiple Parton Interactions at the LHC.* Vol. 29. WSP, 2019.

[3573] Johannes Bellm, Stefan Gieseke, and Patrick Kirchgaesser. "Improving the description of multiple interactions in Herwig". In: *Eur. Phys. J. C* 80.5 (2020), p. 469.

[3574] Jesper R. Christiansen and Peter Z. Skands. "String Formation Beyond Leading Colour". In: *JHEP* 08 (2015), p. 003.

[3575] J. Alcaraz et al. "A Combination of preliminary electroweak measurements and constraints on the standard model". In: (Dec. 2006).

[3576] Torbjörn Sjöstrand. "Status and developments of event generators". In: *PoS* LHCP2016 (2016). Ed. by Johan Bijnens, Andreas Hoecker, and Jim Olsen, p. 007.

[3577] Christian Bierlich et al. "Effects of Overlapping Strings in pp Collisions". In: *JHEP* 03 (2015), p. 148.

[3578] Christian Bierlich, Gösta Gustafson, and Leif Lönnblad. "A shoving model for collectivity in hadronic collisions". In: (Dec. 2016).

[3579] D. Amati and G. Veneziano. "Preconfinement as a Property of Perturbative QCD". In: *Phys. Lett. B* 83 (1979), pp. 87–92.

[3580] T. Pierog et al. "EPOS LHC: Test of collective hadronization with data measured at the CERN Large Hadron Collider". In: *Phys. Rev. C* 92.3 (2015), p. 034906.

[3581] Johan Alwall et al. "A Standard format for Les Houches event files". In: *Comput. Phys. Commun.* 176 (2007), pp. 300–304.

[3582] Andy Buckley et al. “The HepMC3 event record library for Monte Carlo event generators”. In: *Comput. Phys. Commun.* 260 (2021), p. 107310.

[3583] Andy Buckley et al. “LHAPDF6: parton density access in the LHC precision era”. In: *Eur. Phys. J. C* 75 (2015), p. 132.

[3584] Christian Bierlich et al. “Robust Independent Validation of Experiment and Theory: Rivet version 3”. In: *SciPost Phys.* 8 (2020), p. 026.

[3585] John E. Huth et al. “Toward a standardization of jet definitions”. In: *1990 DPF Summer Study on High-energy Physics: Research Directions for the Decade (Snowmass 90)*. Dec. 1990, pp. 0134–136.

[3586] Gerald C. Blazey et al. “Run II jet physics”. In: *Physics at Run II: QCD and Weak Boson Physics Workshop: Final General Meeting*. May 2000, pp. 47–77.

[3587] S. D. Ellis, J. Huston, and M. Tonnesmann. “On building better cone jet algorithms”. In: *eConf* C010630 (2001). Ed. by Norman Graf, p. 513.

[3588] Vardan Khachatryan et al. “Jet energy scale and resolution in the CMS experiment in pp collisions at 8 TeV”. In: *JINST* 12.02 (2017), P02014.

[3589] Jonathan M. Butterworth et al. “Jet substructure as a new Higgs search channel at the LHC”. In: *Phys. Rev. Lett.* 100 (2008), p. 242001.

[3590] Matteo Cacciari, Gavin P. Salam, and Gregory Soyez. “The Catchment Area of Jets”. In: *JHEP* 04 (2008), p. 005.

[3591] Stephen D. Ellis, Christopher K. Vermilion, and Jonathan R. Walsh. “Techniques for improved heavy particle searches with jet substructure”. In: *Phys. Rev. D* 80 (2009), p. 051501.

[3592] David Krohn, Jesse Thaler, and Lian-Tao Wang. “Jet Trimming”. In: *JHEP* 02 (2010), p. 084.

[3593] Mrinal Dasgupta et al. “Towards an understanding of jet substructure”. In: *JHEP* 09 (2013), p. 029.

[3594] Andrew J. Larkoski et al. “Soft Drop”. In: *JHEP* 05 (2014), p. 146.

[3595] Georges Aad et al. “Topological cell clustering in the ATLAS calorimeters and its performance in LHC Run 1”. In: *Eur. Phys. J. C* 77 (2017), p. 490.

[3596] Morad Aaboud et al. “Jet reconstruction and performance using particle flow with the ATLAS Detector”. In: *Eur. Phys. J. C* 77.7 (2017), p. 466.

[3597] A. M. Sirunyan et al. “Particle-flow reconstruction and global event description with the CMS detector”. In: *JINST* 12.10 (2017), P10003.

[3598] Daniele Bertolini et al. “Pileup Per Particle Identification”. In: *JHEP* 10 (2014), p. 059.

[3599] Albert M Sirunyan et al. “Pileup mitigation at CMS in 13 TeV data”. In: *JINST* 15.09 (2020), P09018.

[3600] Georges Aad et al. “Performance of pile-up mitigation techniques for jets in $pp$ collisions at $\sqrt{s} = 8$ TeV using the ATLAS detector”. In: *Eur. Phys. J. C* 76.11 (2016), p. 581.

[3601] Peter Berta et al. “Particle-level pileup subtraction for jets and jet shapes”. In: *JHEP* 06 (2014), p. 092.

[3602] Matteo Cacciari, Gavin P. Salam, and Gregory Soyez. “SoftKiller, a particle-level pileup removal method”. In: *Eur. Phys. J. C* 75.2 (2015), p. 59.

[3603] ATLAS Collaboration. “Optimisation of large-radius jet reconstruction for the ATLAS detector in 13 TeV proton–proton collisions”. In: *Eur. Phys. J. C* 81.4 (2021), p. 334.

[3604] CMS Collaboration. “Pileup mitigation at CMS in 13 TeV data”. In: *JINST* 15.09 (2020), P09018.

[3605] Morad Aaboud et al. “Determination of jet calibration and energy resolution in proton-proton collisions at $\sqrt{s} = 8$ TeV using the ATLAS detector”. In: *Eur. Phys. J. C* 80.12 (2020), p. 1104.

[3606] Georges Aad et al. “Jet energy scale and resolution measured in proton–proton collisions at $\sqrt{s} = 13$ TeV with the ATLAS detector”. In: *Eur. Phys. J. C* 81.8 (2021), p. 689.

[3607] Morad Aaboud et al. “In situ calibration of large-radius jet energy and mass in 13 TeV proton–proton collisions with the ATLAS detector”. In: *Eur. Phys. J. C* 79.2 (2019), p. 135.

[3608] Georges Aad et al. “Jet energy measurement and its systematic uncertainty in proton-proton collisions at $\sqrt{s} = 7$ TeV with the ATLAS detector”. In: *Eur. Phys. J. C* 75 (2015), p. 17.

[3609] Morad Aaboud et al. “Performance of top-quark and $W$-boson tagging with ATLAS in Run 2 of the LHC”. In: *Eur. Phys. J. C* 79.5 (2019), p. 375.

[3610] CMS Collaboration. “Identification of heavy, energetic, hadronically decaying particles using machine-learning techniques”. In: *JINST* 15.06 (2020), P06005.

[3611] R. Assmann, M. Lamont, and S. Myers. “A brief history of the LEP collider”. In: *Nucl. Phys. B Proc. Suppl.* 109 (2002). Ed. by F. L. Navarria, M. Paganoni, and P. G. Pelfer, pp. 17–31.

[3612] S. Schael et al. “Precision electroweak measurements on the $Z$ resonance”. In: *Phys. Rept.* 427 (2006), pp. 257–454.

[3613] G. Alexander et al. “Measurement of three jet distributions sensitive to the gluon spin in $e^+e^-$ annihilations at $s^{1/2} = 91$-GeV”. In: *Z. Phys. C* 52 (1991), pp. 543–550.

[3614] S. Schael et al. “Electroweak Measurements in Electron-Positron Collisions at W-Boson-Pair Energies at LEP”. In: *Phys. Rept.* 532 (2013), pp. 119–244.

[3615] J. Abdallah et al. “Measurement of the energy dependence of hadronic jet rates and the strong coupling $\alpha_s$ from the four-jet rate with the DELPHI detector at LEP”. In: *Eur. Phys. J. C* 38 (2005), pp. 413–426.

[3616] Karl Koller and Hartmut Krasemann. “Excluding Scalar Gluons”. In: *Phys. Lett. B* 88 (1979), pp. 119–122.

[3617] R. Keith Ellis, W. James Stirling, and B. R. Webber. *QCD and collider physics.* Vol. 8. Cambridge University Press, Feb. 2011.

[3618] S. Bethke, A. Ricker, and P. M. Zerwas. “Four jet decays of the Z0: Prospects of testing the triple gluon coupling”. In: *Z. Phys. C* 49 (1991), pp. 59–72.

[3619] Zoltan Nagy and Zoltan Trocsanyi. “Four jet angular distributions and color charge measurements: Leading order versus next-to-leading order”. In: *Phys. Rev. D* 57 (1998), pp. 5793–5802.

[3620] G. Abbiendi et al. “A Simultaneous measurement of the QCD color factors and the strong coupling”. In: *Eur. Phys. J. C* 20 (2001), pp. 601–615.

[3621] A. Heister et al. “Measurements of the strong coupling constant and the QCD color factors using four jet observables from hadronic Z decays”. In: *Eur. Phys. J. C* 27 (2003), pp. 1–17.

[3622] S. Kluth et al. “A Measurement of the QCD color factors using event shape distributions at s**(1/2) = 14-GeV to 189-GeV”. In: *Eur. Phys. J. C* 21 (2001), pp. 199–210.

[3623] Stefan Kluth. “Tests of Quantum Chromo Dynamics at e+ e- Colliders”. In: *Rept. Prog. Phys.* 69 (2006), pp. 1771–1846.

[3624] G. Abbiendi et al. “Measurement of the longitudinal cross-section using the direction of the thrust axis in hadronic events at LEP”. In: *Phys. Lett. B* 440 (1998), pp. 393–402.

[3625] P. D. Acton et al. “A Study of the electric charge distributions of quark and gluon jets in hadronic Z0 decays”. In: *Phys. Lett. B* 302 (1993), pp. 523–532.

[3626] W. Bartel et al. “Experimental Studies on Multi-Jet Production in $e^+e^-$ Annihilation at PETRA Energies”. In: *Z. Phys. C* 33 (1986). Ed. by J. Tran Thanh Van, p. 23.

[3627] Sachio Komamiya et al. “Determination of $\alpha^-s$ From a Differential Jet Multiplicity Distribution at SLC and PEP”. In: *Phys. Rev. Lett.* 64 (1990), p. 987.

[3628] A. Heister et al. “Studies of QCD at $e^+e^-$ centre-of-mass energies between 91-GeV and 209-GeV”. In: *Eur. Phys. J. C* 35 (2004), pp. 457–486.

[3629] G. Abbiendi et al. “Measurement of event shape distributions and moments in $e^+e^- \to$ hadrons at 91-GeV - 209-GeV and a determination of $\alpha_s$”. In: *Eur. Phys. J. C* 40 (2005), pp. 287–316.

[3630] G. Abbiendi et al. “Determination of $alpha_s$ using OPAL hadronic event shapes at $\sqrt{s} = 91$ - 209 GeV and resummed NNLO calculations”. In: *Eur. Phys. J.* C71 (2011), p. 1733.

[3631] G. Dissertori et al. “Determination of the strong coupling constant using matched NNLO+NLLA predictions for hadronic event shapes in e+e- annihilations”. In: *JHEP* 08 (2009), p. 036.

[3632] G. Dissertori et al. “Precise determination of the strong coupling constant at NNLO in QCD from the three-jet rate in electron–positron annihilation at LEP”. In: *Phys. Rev. Lett.* 104 (2010), p. 072002.

[3633] S. Bethke et al. “Determination of the Strong Coupling $\alpha_s$ from hadronic Event Shapes with $O(\alpha_s^3)$ and resummed QCD predictions using JADE Data”. In: *Eur. Phys. J. C* 64 (2009), pp. 351–360.

[3634] Andrii Verbytskyi et al. “High precision determination of $\alpha_s$ from a global fit of jet rates”. In: *JHEP* 08 (2019), p. 129.

[3635] Adam Kardos, Gábor Somogyi, and Andrii Verbytskyi. “Determination of $\alpha_s$ beyond NNLO using event shape averages”. In: *Eur. Phys. J. C* 81.4 (2021), p. 292.

[3636] Fabrizio Caola et al. “Linear power corrections to $e^+e^-$ shape variables in the three-jet region”. In: (Apr. 2022).

[3637] Zoltan Nagy and Zoltan Trocsanyi. “Next-to-leading order calculation of four jet shape variables”. In: *Phys. Rev. Lett.* 79 (1997), pp. 3604–3607.

[3638] G. Abbiendi et al. “Measurement of the Strong Coupling $\alpha_s$ from four-jet observables in $e^+e^-$ annihilation”. In: *Eur. Phys. J. C* 47 (2006), pp. 295–307.

[3639] J. Schieck et al. "Measurement of the strong coupling $\alpha_s$ from the four-jet rate in $e^+e^-$ annihilation using JADE data". In: *Eur. Phys. J. C* 48 (2006). [Erratum: Eur.Phys.J.C 50, 769 (2007)], pp. 3–13.

[3640] Rikkert Frederix et al. "NLO QCD corrections to five-jet production at LEP and the extraction of $\alpha_s(M_Z)$". In: *JHEP* 11 (2010), p. 050.

[3641] J. Abdallah et al. "A Study of the energy evolution of event shape distributions and their means with the DELPHI detector at LEP". In: *Eur. Phys. J. C* 29 (2003), pp. 285–312.

[3642] S. Bethke et al. "Experimental Investigation of the Energy Dependence of the Strong Coupling Strength". In: *Phys. Lett. B* 213 (1988), pp. 235–241.

[3643] D. Buskulic et al. "Measurement of alpha-s from scaling violations in fragmentation functions in $e^+e^-$ annihilation". In: *Phys. Lett. B* 357 (1995). [Erratum: Phys.Lett.B 364, 247–248 (1995)], pp. 487–499.

[3644] Daniele P. Anderle, Felix Ringer, and Marco Stratmann. "Fragmentation Functions at Next-to-Next-to-Leading Order Accuracy". In: *Phys. Rev. D* 92.11 (2015), p. 114017.

[3645] G. Abbiendi et al. "Charged particle momentum spectra in $e^+e^-$- annihilation at $s^{1/2}$ = 192-GeV to 209-GeV". In: *Eur. Phys. J. C* 27 (2003), pp. 467–481.

[3646] C. P. Fong and B. R. Webber. "One and two particle distributions at small x in QCD jets". In: *Nucl. Phys. B* 355 (1991), pp. 54–81.

[3647] Redamy Perez-Ramos and David d'Enterria. "Energy evolution of the moments of the hadron distribution in QCD jets including NNLL resummation and NLO running-coupling corrections". In: *JHEP* 08 (2014), p. 068.

[3648] Edouard R. Boudinov, P. V. Chliapnikov, and V. A. Uvarov. "Is there experimental evidence for coherence of soft gluons from the momentum spectra of hadrons in $e^+e^-$ data?" In: *Phys. Lett. B* 309 (1993), pp. 210–221.

[3649] Valery A. Khoze and Wolfgang Ochs. "Perturbative QCD approach to multiparticle production". In: *Int. J. Mod. Phys. A* 12 (1997), pp. 2949–3120.

[3650] Andrea Banfi, Gennaro Corcella, and Mrinal Dasgupta. "Angular ordering and parton showers for non-global QCD observables". In: *JHEP* 03 (2007), p. 050.

[3651] Peter Skands, Stefano Carrazza, and Juan Rojo. "Tuning PYTHIA 8.1: the Monash 2013 Tune". In: *Eur. Phys. J. C* 74.8 (2014), p. 3024.

[3652] Spyros Argyropoulos and Torbjörn Sjöstrand. "Effects of color reconnection on $t\bar{t}$ final states at the LHC". In: *JHEP* 11 (2014), p. 043.

[3653] Albert M Sirunyan et al. "Measurement of the top quark mass with lepton+jets final states using p p collisions at $\sqrt{s} = 13\,\text{TeV}$". In: *Eur. Phys. J. C* 78.11 (2018). [Erratum: Eur.Phys.J.C 82, 323 (2022)], p. 891.

[3654] Frank Wilczek. "Asymptotic freedom: From paradox to paradigm". In: *Proc. Nat. Acad. Sci.* 102 (2005), pp. 8403–8413.

[3655] K. G. Chetyrkin, Johann H. Kuhn, and A. Kwiatkowski. "QCD corrections to the $e^+e^-$ cross-section and the $Z$ boson decay rate". In: *Phys. Rept.* 277 (1996), pp. 189–281.

[3656] Yuri L. Dokshitzer, Valery A. Khoze, and S. I. Troian. "On specific QCD properties of heavy quark fragmentation ('dead cone')". In: *J. Phys. G* 17 (1991), pp. 1602–1604.

[3657] Mikhail S. Bilenky, German Rodrigo, and Arcadi Santamaria. "Three jet production at LEP and the bottom quark mass". In: *Nucl. Phys. B* 439 (1995), pp. 505–535.

[3658] R. Barate et al. "A Measurement of the b quark mass from hadronic Z decays". In: *Eur. Phys. J. C* 18 (2000), pp. 1–13.

[3659] Javier Aparisi et al. "mb at mH: The Running Bottom Quark Mass and the Higgs Boson". In: *Phys. Rev. Lett.* 128.12 (2022), p. 122001.

[3660] R. Akers et al. "QCD studies using a cone based jet finding algorithm for $e^+e^-$ collisions at LEP". In: *Z. Phys. C* 63 (1994), pp. 197–212.

[3661] Yuri L. Dokshitzer, Valery A. Khoze, and S. I. Troian. "Particle spectra in light and heavy quark jets". In: *J. Phys. G* 17 (1991), pp. 1481–1492.

[3662] Yuri L. Dokshitzer et al. "Multiplicity difference between heavy and light quark jets revisited". In: *Eur. Phys. J. C* 45 (2006), pp. 387–400.

[3663] J. Abdallah et al. "A study of the b-quark fragmentation function with the DELPHI detector at LEP I and an averaged distribution obtained at the Z Pole". In: *Eur. Phys. J. C* 71 (2011), p. 1557.

[3664] Johannes Haller et al. "Update of the global electroweak fit and constraints on two-Higgs-doublet models". In: *Eur. Phys. J.* C78.8 (2018), p. 675.

[3665] Miguel A. Benitez-Rathgeb et al. "Reconciling the contour-improved and fixed-order approaches for $\tau$ hadronic spectral moments. Part

II. Renormalon norm and application in $\alpha_s$ determinations". In: *JHEP* 09 (2022), p. 223.
[3666] S. Brandt et al. "The Principal axis of jets. An Attempt to analyze high-energy collisions as two-body processes". In: *Phys. Lett.* 12 (1964), pp. 57–61.
[3667] J. D. Bjorken and Stanley J. Brodsky. "Statistical Model for electron-Positron Annihilation Into Hadrons". In: *Phys. Rev. D* 1 (1970), p. 1416.
[3668] Christoph Berger et al. "Topology of the $\Upsilon$ Decay". In: *Z. Phys. C* 8 (1981), p. 101.
[3669] M. Banner et al. "Observation of Very Large Transverse Momentum Jets at the CERN anti-p p Collider". In: *Phys. Lett. B* 118 (1982), pp. 203–210.
[3670] R. Horgan and M. Jacob. "Jet Production at Collider Energy". In: *Nucl. Phys. B* 179 (1981), p. 441.
[3671] G. Arnison et al. "Observation of Jets in High Transverse Energy Events at the CERN Proton - anti-Proton Collider". In: *Phys. Lett. B* 123 (1983), pp. 115–122.
[3672] T. Akesson et al. "Direct Evidence for the Emergence of Jets in Events Triggered on Large Transverse Energy in $pp$ Collisions at $\sqrt{s} = 63$ GeV". In: *Phys. Lett. B* 118 (1982), pp. 185–192.
[3673] Brenna Flaugher and Karlheinz Meier. "A Compilation of jet finding algorithms". In: *Proceedings, 5th DPF Summer Study on High-energy Physics: Research Directions for the Decade (Snowmass 90).* Snowmass, CO, USA, Jun 25-Jul 13, 1990, p. 128.
[3674] N. Brown and W. James Stirling. "Finding jets and summing soft gluons: A New algorithm". In: *Z. Phys. C* 53 (1992), p. 629.
[3675] T. Hebbeker. "Tests of quantum chromodynamics in hadronic decays of $Z^0$ bosons produced in $e^+e^-$ annihilation". In: *Phys. Rept.* 217 (1992), p. 69.
[3676] D. Graudenz and N. Magnussen. "Jet cross-sections in deeply inelastic scattering at HERA". In: *Workshop on Physics at HERA*. 1991.
[3677] Gavin P. Salam and Gregory Soyez. "A Practical Seedless Infrared-Safe Cone jet algorithm". In: *JHEP* 05 (2007), p. 086.
[3678] F. Aversa et al. "Jet inclusive production to O($alpha-s^3$): Comparison with data". In: *Phys. Rev. Lett.* 65 (1990), pp. 401–403.
[3679] Stephen D. Ellis, Zoltan Kunszt, and Davison E. Soper. "One-Jet Inclusive Cross Section at Order $\alpha_s^3$: Quarks and Gluons". In: *Phys. Rev. Lett.* 64 (1990), p. 2121.
[3680] E. Eichten et al. "Super Collider Physics". In: *Rev. Mod. Phys.* 56 (1984). [Erratum: Rev. Mod. Phys. 56, 579 (1984)], p. 579.
[3681] M. Diemoz et al. "Parton Densities from Deep Inelastic Scattering to Hadronic Processes at Super Collider Energies". In: *Z. Phys. C* 39 (1988), p. 21.
[3682] Alan D. Martin, R. G. Roberts, and W. James Stirling. "Structure Function Analysis and psi, Jet, W, Z Production: Pinning Down the Gluon". In: *Phys. Rev. D* D37 (1988), p. 1161.
[3683] Alan D. Martin, W. James Stirling, and R. G. Roberts. "Benchmark cross sections for $p\bar{p}$ collisions at 1.8 TeV". In: *Z. Phys. C* 42 (1989), p. 277.
[3684] F. Abe et al. "Inclusive jet cross section in $\bar{p}p$ collisions at $\sqrt{s} = 1.8$ TeV". In: *Phys. Rev. Lett.* 77 (1996), p. 438.
[3685] E. Eichten, Kenneth D. Lane, and Michael E. Peskin. "New Tests for Quark and Lepton Substructure". In: *Phys. Rev. Lett.* 50 (1983), pp. 811–814.
[3686] Davison E. Soper. "Summary of the XXX Rencontres de Moriond QCD session". In: *30th Rencontres de Moriond: QCD and High-energy Hadronic Interactions.* 1995, pp. 615–628.
[3687] Walter T. Giele and Stephane Keller. "Implications of hadron collider observables on parton distribution function uncertainties". In: *Phys. Rev. D* 58 (1998), p. 094023.
[3688] Sergei Alekhin. "Extraction of parton distributions and alpha-s from DIS data within the Bayesian treatment of systematic errors". In: *Eur. Phys. J. C* 10 (1999), pp. 395–403.
[3689] G. E. Wolf. "First results from HERA". In: *Proceedings, QCD: 20 Years Later : Aachen, June 9-13, 1992.* Ed. by P. M. Zerwas and H. A. Kastrup. World Scientific, 1993, pp. 335–384.
[3690] S. Bethke. "Jets in $Z^0$ decays". In: *Proceedings, QCD : 20 Years Later : Aachen, June 9-13, 1992.* Ed. by P. M. Zerwas and H. A. Kastrup. World Scientific, 1993, pp. 43–72.
[3691] G. Altarelli. "QCD and Experiment: Status of $\alpha_S$". In: *Proceedings, QCD : 20 Years Later : Aachen, June 9-13, 1992.* Ed. by P. M. Zerwas and H. A. Kastrup. World Scientific, 1993, pp. 172–202.
[3692] I. Abt et al. "The Tracking, calorimeter and muon detectors of the H1 experiment at HERA". In: *Nucl. Instrum. Meth. A* 386 (1997), pp. 348–396.

[3693] I. Abt et al. “The H1 detector at HERA”. In: *Nucl. Instrum. Meth. A* 386 (1997), pp. 310–347.

[3694] U. Holm et al. *The ZEUS detector: Status report 1993.* Tech. rep. ZEUS-STATUS-REPT-1993. 1993, p. 597.

[3695] H. Abramowicz et al. “Inclusive dijet cross sections in neutral current deep inelastic scattering at HERA”. In: *Eur. Phys. J. C* 70 (2010), pp. 965–982.

[3696] V. Andreev et al. “Measurement of multijet production in *ep* collisions at high $Q^2$ and determination of the strong coupling $\alpha_s$”. In: *Eur. Phys. J.* C75.2 (2015), p. 65.

[3697] M. Derrick et al. “Measurement of the proton structure function F2 in e p scattering at HERA”. In: *Phys. Lett. B* 316 (1993), pp. 412–426.

[3698] I. Abt et al. “Measurement of the proton structure function $F_2(x, Q^2)$ in the low x region at HERA”. In: *Nucl. Phys. B* 407 (1993), pp. 515–538.

[3699] T. Ahmed et al. “Hard scattering in gamma p interactions”. In: *Phys. Lett. B* 297 (1992), pp. 205–213.

[3700] M. Derrick et al. “Observation of hard scattering in photoproduction at HERA”. In: *Phys. Lett. B* 297 (1992), pp. 404–416.

[3701] M. Kuhlen. “QCD at HERA: The hadronic final state in deep inelastic scattering”. In: *Springer Tracts Mod. Phys.* 150 (1999), pp. 1–172.

[3702] H. Abramowicz and A. Caldwell. “HERA collider physics”. In: *Rev. Mod. Phys.* 71 (1999), pp. 1275–1410.

[3703] M. Klein and R. Yoshida. “Collider Physics at HERA”. In: *Prog. Part. Nucl. Phys.* 61 (2008), pp. 343–393.

[3704] Paul Newman and Matthew Wing. “The Hadronic Final State at HERA”. In: *Rev. Mod. Phys.* 86.3 (2014), p. 1037.

[3705] O. Behnke, A. Geiser, and M. Lisovyi. “Charm, Beauty and Top at HERA”. In: *Prog. Part. Nucl. Phys.* 84 (2015), pp. 1–72.

[3706] I. Abt et al. “A Measurement of multi - jet rates in deep inelastic scattering at HERA”. In: *Z. Phys. C* 61 (1994), pp. 59–66.

[3707] M. Derrick et al. “Observation of two jet production in deep inelastic scattering at HERA”. In: *Phys. Lett. B* 306 (1993), pp. 158–172.

[3708] Tancredi Carli. “Hadronic final state in deep inelastic scattering at HERA”. In: *16th International Conference on Physics in Collision (PIC 96).* June 1996, pp. 415–438.

[3709] Dirk Graudenz. “Three jet production in deep inelastic electron - proton scattering to order $\alpha_s^2$”. In: *Phys. Lett. B* 256 (1991), pp. 518–522.

[3710] T. Ahmed et al. “Determination of the strong coupling constant from jet rates in deep inelastic scattering”. In: *Phys. Lett. B* 346 (1995), pp. 415–425.

[3711] M. Derrick et al. “Measurement of alpha-s from jet rates in deep inelastic scattering at HERA”. In: *Phys. Lett. B* 363 (1995), pp. 201–216.

[3712] R. Michael Barnett et al. “Review of particle physics. Particle Data Group”. In: *Phys. Rev. D* 54 (1996), pp. 1–720.

[3713] Erwin Mirkes and Dieter Zeppenfeld. “Dijet production at HERA in next-to-leading order”. In: *Phys. Lett. B* 380 (1996), pp. 205–212.

[3714] S. Catani, Yuri L. Dokshitzer, and B. R. Webber. “The $K^-$ perpendicular clustering algorithm for jets in deep inelastic scattering and hadron collisions”. In: *Phys. Lett. B* 285 (1992), pp. 291–299.

[3715] B. R. Webber. “Factorization and jet clustering algorithms for deep inelastic scattering”. In: *J. Phys. G* 19 (1993), pp. 1567–1575.

[3716] C. Adloff et al. “Measurement and QCD analysis of jet cross-sections in deep inelastic positron - proton collisions at $s^{1/2}$ of 300-GeV”. In: *Eur. Phys. J. C* 19 (2001), pp. 289–311.

[3717] C. Adloff et al. “Measurement of inclusive jet cross-sections in deep inelastic ep scattering at HERA”. In: *Phys. Lett. B* 542 (2002), pp. 193–206.

[3718] S. Chekanov et al. “Inclusive jet cross-sections in the Breit frame in neutral current deep inelastic scattering at HERA and determination of alpha(s)”. In: *Phys. Lett. B* 547 (2002), pp. 164–180.

[3719] S. Bethke. “Determination of the QCD coupling $\alpha_s$”. In: *J. Phys.* G26 (2000), R27.

[3720] S. Chekanov et al. “An NLO QCD analysis of inclusive cross-section and jet-production data from the zeus experiment”. In: *Eur. Phys. J. C* 42 (2005), pp. 1–16.

[3721] V. Andreev et al. “Measurement of Jet Production Cross Sections in Deep-inelastic ep Scattering at HERA”. In: *Eur. Phys. J. C* 77.4 (2017). [Erratum: Eur.Phys.J.C 81, 739 (2021)], p. 215.

[3722] James Currie, Thomas Gehrmann, and Jan Niehues. “Precise QCD predictions for the production of dijet final states in deep inelastic scattering”. In: *Phys. Rev. Lett.* 117.4 (2016), p. 042001.

[3723] Jan Niehues et al. "Precise QCD predictions for the production of dijet final states in deep inelastic scattering". In: *52nd Rencontres de Moriond on QCD and High Energy Interactions*. SISSA, 2017, pp. 199–202.

[3724] V. Andreev et al. "Determination of the strong coupling constant $\alpha_s(m_Z)$ in next-to-next-to-leading order QCD using H1 jet cross section measurements". In: *Eur. Phys. J. C* 77.11 (2017). [Erratum: Eur.Phys.J.C 81, 738 (2021)], p. 791.

[3725] D. Britzger et al. "Calculations for deep inelastic scattering using fast interpolation grid techniques at NNLO in QCD and the extraction of $\alpha_s$ from HERA data". In: *Eur. Phys. J. C* 79.10 (2019). [Erratum: arXiv: 1906.05303v3], p. 845.

[3726] I. Abt et al. "Impact of jet-production data on the next-to-next-to-leading-order determination of HERAPDF2.0 parton distributions". In: *Eur. Phys. J. C* 82.3 (2022), p. 243.

[3727] Vardan Khachatryan et al. "Measurement and QCD analysis of double-differential inclusive jet cross-sections in pp collisions at $\sqrt{s} = 8$ TeV and ratios to 2.76 and 7 TeV". In: *JHEP* 03 (2017), p. 156.

[3728] G. Dissertori et al. "First determination of the strong coupling constant using NNLO predictions for hadronic event shapes in $e^+e^-$ annihilations". In: *JHEP* 02 (2008), p. 040.

[3729] Jochen Schieck et al. "Measurement of the strong coupling alpha$_S fromthethree - jetrateine + e - -annihilationusingJADEdata$". In: *Eur. Phys. J.* C73.3 (2013), p. 2332.

[3730] M. Jacob, ed. *Proceedings of the ECFA-CERN Workshop: Large Hadron Collider in the LEP tunnel: Lausanne and Geneva, Switzerland 21 - 27 Mar 1984*. CERN. Geneva: CERN, Sept. 1984.

[3731] F. Aversa et al. "QCD Corrections to Parton-Parton Scattering Processes". In: *Nucl. Phys. B* 327 (1989), p. 105.

[3732] F. Abe et al. "Comparison of jet production in $\bar{p}p$ collisions at $\sqrt{s} = 546$ GeV and 1800 GeV". In: *Phys. Rev. Lett.* 70 (1993), pp. 1376–1380.

[3733] B. Abbott et al. "High-$p_T$ jets in $\bar{p}p$ collisions at $\sqrt{s} = 630$ GeV and 1800 GeV". In: *Phys. Rev. D* 64 (2001), p. 032003.

[3734] B. Abbott et al. "Inclusive jet production in $p\bar{p}$ collisions". In: *Phys. Rev. Lett.* 86 (2001), pp. 1707–1712.

[3735] T. Affolder et al. "Measurement of the Inclusive Jet Cross Section in $\bar{p}p$ Collisions at $\sqrt{s} = 1.8$ TeV". In: *Phys. Rev. D* 64 (2001). [Erratum: Phys.Rev.D 65, 039903 (2002)], p. 032001.

[3736] B. I. Abelev et al. "Longitudinal double-spin asymmetry and cross section for inclusive jet production in polarized proton collisions at $s^{1/2}$ = 200-GeV". In: *Phys. Rev. Lett.* 97 (2006), p. 252001.

[3737] S. Chekanov et al. "Inclusive-jet and dijet cross-sections in deep inelastic scattering at HERA". In: *Nucl. Phys.* B765 (2007), p. 1.

[3738] A. Abulencia et al. "Measurement of the Inclusive Jet Cross Section using the $k_{\mathrm{T}}$ algorithmin $p\bar{p}$ Collisions at $\sqrt{s} = 1.96$ TeV with the CDF II Detector". In: *Phys. Rev. D* 75 (2007). [Erratum: Phys.Rev.D 75, 119901 (2007)], p. 092006.

[3739] T. Aaltonen et al. "Measurement of the Inclusive Jet Cross Section at the Fermilab Tevatron p anti-p Collider Using a Cone-Based Jet Algorithm". In: *Phys. Rev. D* 78 (2008). [Erratum: Phys.Rev.D 79, 119902 (2009)], p. 052006.

[3740] V. M. Abazov et al. "Measurement of the inclusive jet cross-section in $p\bar{p}$ collisions at $s^{(1/2)}$ =1.96-TeV". In: *Phys. Rev. Lett.* 101 (2008), p. 062001.

[3741] Serguei Chatrchyan et al. "Measurements of Differential Jet Cross Sections in Proton-Proton Collisions at $\sqrt{s} = 7$ TeV with the CMS Detector". In: *Phys. Rev. D* 87.11 (2013). [Erratum: Phys.Rev.D 87, 119902 (2013)], p. 112002.

[3742] Georges Aad et al. "Measurement of the inclusive jet cross section in pp collisions at $\sqrt{s} =$ 2.76 TeV and comparison to the inclusive jet cross section at $\sqrt{s} = 7$ TeV using the ATLAS detector". In: *Eur. Phys. J. C* 73.8 (2013), p. 2509.

[3743] Georges Aad et al. "Measurement of the inclusive jet cross-section in proton-proton collisions at $\sqrt{s} = 7$ TeV using 4.5 fb$^{-1}$ of data with the ATLAS detector". In: *JHEP* 02 (2015). [Erratum: JHEP 09, 141 (2015)], p. 153.

[3744] Vardan Khachatryan et al. "Measurement of the inclusive jet cross section in pp collisions at $\sqrt{s} = 2.76$ TeV". In: *Eur. Phys. J. C* 76.5 (2016), p. 265.

[3745] M. Aaboud et al. "Measurement of inclusive jet and dijet cross-sections in proton-proton collisions at $\sqrt{s} = 13$ TeV with the ATLAS detector". In: *JHEP* 05 (2018), p. 195.

[3746] Morad Aaboud et al. "Measurement of the inclusive jet cross-sections in proton-proton collisions at $\sqrt{s} = 8$ TeV with the ATLAS detector". In: *JHEP* 09 (2017), p. 020.

[3747] Armen Tumasyan et al. "Measurement and QCD analysis of double-differential inclusive jet cross

sections in proton-proton collisions at $\sqrt{s}$ = 13 TeV". In: *JHEP* 02 (Nov. 2022), p. 142.

[3748] Richard D. Ball et al. "The PDF4LHC21 combination of global PDF fits for the LHC Run III". In: *J. Phys. G* 49.8 (2022), p. 080501.

[3749] Tancredi Carli et al. "A posteriori inclusion of parton density functions in NLO QCD final-state calculations at hadron colliders: The APPLGRID Project". In: *Eur. Phys. J. C* 66 (2010), pp. 503–524.

[3750] Daniel Britzger et al. "New features in version 2 of the fastNLO project". In: *Proceedings, XX. International Workshop on Deep-Inelastic Scattering and Related Subjects (DIS 2012).* Bonn, Germany, March 26-30, 2012, p. 217.

[3751] Thomas Gehrmann et al. "Jet cross sections and transverse momentum distributions with NNLOJET". In: *PoS* RADCOR2017 (2018). Ed. by Andre Hoang and Carsten Schneider, p. 074.

[3752] Nikolaos Kidonakis and J. F. Owens. "Effects of higher order threshold corrections in high $E_T$ jet production". In: *Phys. Rev. D* 63 (2001), p. 054019.

[3753] Zoltan Nagy. "Three jet cross-sections in hadron hadron collisions at next-to-leading order". In: *Phys. Rev. Lett.* 88 (2002), p. 122003.

[3754] M. Wobisch et al. "Theory-Data Comparisons for Jet Measurements in Hadron-Induced Processes". In: (2011).

[3755] Matteo Cacciari and Nicolas Houdeau. "Meaningful characterisation of perturbative theoretical uncertainties". In: *JHEP* 09 (2011), p. 039.

[3756] Emanuele Bagnaschi et al. "An extensive survey of the estimation of uncertainties from missing higher orders in perturbative calculations". In: *JHEP* 02 (2015), p. 133.

[3757] Marco Bonvini. "Probabilistic definition of the perturbative theoretical uncertainty from missing higher orders". In: *Eur. Phys. J. C* 80.10 (2020), p. 989.

[3758] Claude Duhr et al. "An analysis of Bayesian estimates for missing higher orders in perturbative calculations". In: (June 2021).

[3759] Andr00E9 David and Giampiero Passarino. "How well can we guess theoretical uncertainties?" In: *Phys. Lett.* B726 (2013), pp. 266–272.

[3760] Fredrick I. Olness and Davison E. Soper. "Correlated theoretical uncertainties for the one-jet inclusive cross section". In: *Phys. Rev. D* 81 (2010), p. 035018.

[3761] Rabah Abdul Khalek et al. "A first determination of parton distributions with theoretical uncertainties". In: *Eur. Phys. J.* C (2019), 79:838.

[3762] Vardan Khachatryan et al. "Search for quark contact interactions and extra spatial dimensions using dijet angular distributions in proton–proton collisions at $\sqrt{s} = 8$ TeV". In: *Phys. Lett. B* 746 (2015), p. 79.

[3763] ATLAS Collaboration. "Determination of the parton distribution functions of the proton using diverse ATLAS data from $pp$ collisions at $\sqrt{s}$ = 7, 8 and 13 TeV". In: *Eur. Phys. J. C* 82.5 (2022), p. 438.

[3764] John Campbell, Joey Huston, and Frank Krauss. *The Black Book of Quantum Chromodynamics: A Primer for the LHC Era.* Oxford University Press, Dec. 2017.

[3765] Richard Keith Ellis, William James Stirling, and Bryan R. Webber. *QCD and Collider Physics.* Cambridge Monographs on Particle Physics, Nuclear Physics and Cosmology. Cambridge: Cambridge University Press, 1996.

[3766] Ansgar Denner and Stefan Dittmaier. "Electroweak Radiative Corrections for Collider Physics". In: *Phys. Rept.* 864 (2020), pp. 1–163.

[3767] UA1 Collaboration. "Studies of intermediate vector boson production and decay in UA1 at the CERN proton-antiproton collider". In: *Zeitschrift for Physik C Particles and Fields* 44.1 (1989).

[3768] TASSO Collaboration. "Evidence for a Spin One Gluon in Three Jet Events". In: *Phys. Lett. B* 97 (1980), p. 453.

[3769] UA1 Collaboration. "Angular Distributions and Structure Functions from Two Jet Events at the CERN SPS p anti-p Collider". In: *Phys. Lett. B* 136 (1984), p. 294.

[3770] UA2 Collaboration. "A Determination of the strong coupling constant alpha-s from W production at the CERN p anti-p collider". In: *Phys. Lett.* B263 (1991), pp. 563–572.

[3771] P.D. Acton et al. "A Global determination of $\alpha^{-}s$ (M(z0) ) at LEP". In: *Z. Phys. C* 55 (1992), pp. 1–24.

[3772] Matthias Schott and Monica Dunford. "Review of single vector boson production in pp collisions at $\sqrt{s}$ = 7 TeV". In: *Eur. Phys. J. C* 74 (2014), p. 2916.

[3773] CDF Collaboration. "Measurement of the cross section for $W^-$ boson production in association with jets in $p\bar{p}$ collisions at $\sqrt{s} = 1.96$-TeV". In: *Phys. Rev. D* 77 (2008), p. 011108.

[3774] DO Collaboration. "Studies of W boson plus jets production in $p\bar{p}$ collisions at $\sqrt{s}$ = 1.96 TeV". In: *Phys. Rev.* D88.9 (2013), p. 092001.

[3775] Z.a Bern et al. "Missing Energy and Jets for Supersymmetry Searches". In: *Phys. Rev.* D87.3 (2013), p. 034026.

[3776] CMS Collaboration. "Measurement of differential cross sections for the production of a Z boson in association with jets in proton-proton collisions at $\sqrt{s} = 13$ TeV". In: (May 2022).

[3777] CMS Collaboration. "Measurement of the differential cross sections for the associated production of a $W$ boson and jets in proton-proton collisions at $\sqrt{s} = 13$ TeV". In: *Phys. Rev. D* 96.7 (2017), p. 072005.

[3778] ATLAS Collaboration. "Measurements of the production cross section of a $Z$ boson in association with jets in pp collisions at $\sqrt{s} = 13$ TeV with the ATLAS detector". In: *Eur. Phys. J. C* 77.6 (2017), p. 361.

[3779] ATLAS Collaboration. "Measurement of differential cross sections and $W^+/W^-$ cross-section ratios for $W$ boson production in association with jets at $\sqrt{s} = 8$ TeV with the ATLAS detector". In: *JHEP* 05 (2018). [Erratum: JHEP 10, 048 (2020)], p. 077.

[3780] CMS Collaboration. "Measurement of differential cross sections for Z boson production in association with jets in proton-proton collisions at $\sqrt{s} = 13$ TeV". In: *Eur. Phys. J. C* 78.11 (2018), p. 965.

[3781] Yuri L. Dokshitzer, Dmitri Diakonov, and S. I. Troian. "On the Transverse Momentum Distribution of Massive Lepton Pairs". In: *Phys. Lett. B* 79 (1978), pp. 269–272.

[3782] ATLAS Collaboration. "Cross-section measurements for the production of a $Z$ boson in association with high-transverse-momentum jets in $pp$ collisions at $\sqrt{s} = 13$ TeV with the ATLAS detector". In: (May 2022).

[3783] CMS Collaboration. "Measurements of the differential cross sections of the production of Z + jets and $\gamma$ + jets and of Z boson emission collinear with a jet in pp collisions at $\sqrt{s} = 13$ TeV". In: *JHEP* 05 (2021), p. 285.

[3784] ATLAS Collaboration. "Measurement of $W$ boson angular distributions in events with high transverse momentum jets at $\sqrt{s} = 8$ TeV using the ATLAS detector". In: *Phys. Lett. B* 765 (2017), pp. 132–153.

[3785] CMS Collaboration. "Measurements of angular distance and momentum ratio distributions in three-jet and Z + two-jet final states in pp collisions". In: *Eur. Phys. J. C* 81.9 (2021), p. 852.

[3786] ATLAS Collaboration. "Measurement of the electroweak production of dijets in association with a Z-boson and distributions sensitive to vector boson fusion in proton-proton collisions at $\sqrt{s} = 8$ TeV using the ATLAS detector". In: *JHEP* 04 (2014), p. 031.

[3787] ATLAS Collaboration. "Measurements of electroweak $Wjj$ production and constraints on anomalous gauge couplings with the ATLAS detector". In: *Eur. Phys. J. C* 77.7 (2017), p. 474.

[3788] CMS Collaboration. "Measurement of the Hadronic Activity in Events with a Z and Two Jets and Extraction of the Cross Section for the Electroweak Production of a Z with Two Jets in $pp$ Collisions at $\sqrt{s} = 7$ TeV". In: *JHEP* 10 (2013), p. 062.

[3789] CMS Collaboration. "Measurement of electroweak production of a W boson and two forward jets in proton-proton collisions at $\sqrt{s} = 8$ TeV". In: *JHEP* 11 (2016), p. 147.

[3790] CMS Collaboration. "Electroweak production of two jets in association with a Z boson in proton–proton collisions at $\sqrt{s} = 13$ TeV". In: *Eur. Phys. J. C* 78.7 (2018), p. 589.

[3791] CMS Collaboration. "Measurement of electroweak production of a W boson in association with two jets in proton–proton collisions at $\sqrt{s} = 13$ TeV". In: *Eur. Phys. J. C* 80.1 (2020), p. 43.

[3792] LHCb Collaboration. "Measurement of forward $W$ and $Z$ boson production in association with jets in proton-proton collisions at $\sqrt{s} = 8$ TeV". In: *JHEP* 05 (2016), p. 131.

[3793] LHCb Collaboration. "Study of forward Z + jet production in pp collisions at $\sqrt{s} = 7$ TeV". In: *JHEP* 01 (2014), p. 033.

[3794] ATLAS Collaboration. "Measurement of detector-corrected observables sensitive to the anomalous production of events with jets and large missing transverse momentum in $pp$ collisions at $\sqrt{\mathbf{s}} = \mathbf{13}$ TeV using the ATLAS detector". In: *Eur. Phys. J. C* 77.11 (2017), p. 765.

[3795] ATLAS Collaboration. "A measurement of the ratio of the production cross sections for $W$ and $Z$ bosons in association with jets with the ATLAS detector". In: *Eur. Phys. J. C* 74.12 (2014), p. 3168.

[3796] Erik Gerwick et al. "Scaling Patterns for QCD Jets". In: *JHEP* 1210 (2012), p. 162.

[3797] Georges Aad et al. "Measurement of the production cross section of jets in association with a Z boson in pp collisions at $\sqrt{s} = 7$ TeV with the ATLAS detector". In: *JHEP* 07 (2013), p. 032.

[3798] John M. Campbell et al. "Associated Production of a $W$ Boson and One $b$ Jet". In: *Phys. Rev. D* 79 (2009), p. 034023.

[3799] Simon Badger, John M. Campbell, and R. K. Ellis. "QCD Corrections to the Hadronic Production of a Heavy Quark Pair and a W-Boson Including Decay Correlations". In: *JHEP* 03 (2011), p. 027.

[3800] Fernando Febres Cordero, Laura Reina, and Doreen Wackeroth. "Associated Production of a W or Z Boson with Bottom Quarks at the Tevatron and the LHC". In: *PoS* RADCOR2009 (2010), p. 055.

[3801] J. M. Campbell et al. "NLO QCD predictions for $W + 1$ jet and $W + 2$ jet production with at least one $b$ jet at the 7 TeV LHC". In: *Phys. Rev. D* 86 (2012), p. 034021.

[3802] CMS Collaboration. "Measurement of the production cross section for Z+b jets in proton-proton collisions at $\sqrt{s} = 13$ TeV". In: *Phys. Rev. D* 105.9 (2022), p. 092014.

[3803] CMS Collaboration. "Measurements of the associated production of a W boson and a charm quark in proton-proton collisions at $\sqrt{s} = 8$ TeV". In: (Dec. 2021).

[3804] ATLAS Collaboration. "Measurements of the production cross-section for a $Z$ boson in association with $b$-jets in proton-proton collisions at $\sqrt{s} = 13$ TeV with the ATLAS detector". In: *JHEP* 07 (2020), p. 044.

[3805] CMS Collaboration. "Measurement of differential cross sections for Z bosons produced in association with charm jets in pp collisions at $\sqrt{s} = 13$ TeV". In: *JHEP* 04 (2021), p. 109.

[3806] CMS Collaboration. "Measurements of the associated production of a Z boson and b jets in pp collisions at $\sqrt{s} = 8$ TeV". In: *Eur. Phys. J. C* 77.11 (2017), p. 751.

[3807] ATLAS Collaboration. "Measurement of differential production cross-sections for a $Z$ boson in association with $b$-jets in 7 TeV proton-proton collisions with the ATLAS detector". In: *JHEP* 10 (2014), p. 141.

[3808] ATLAS Collaboration. "Measurement of cross-sections for production of a $Z$ boson in association with a flavor-inclusive or doubly $b$-tagged large-radius jet in proton-proton collisions at $\sqrt{s} = 13$ TeV with the ATLAS experiment". In: (Apr. 2022).

[3809] LHCb Collaboration. "Study of Z Bosons Produced in Association with Charm in the Forward Region". In: *Phys. Rev. Lett.* 128.8 (2022), p. 082001.

[3810] LHCb Collaboration. "Study of $W$ boson production in association with beauty and charm". In: *Phys. Rev. D* 92.5 (2015), p. 052001.

[3811] LHCb Collaboration. "Measurement of the Z+b-jet cross-section in pp collisions at $\sqrt{s} = 7$ TeV in the forward region". In: *JHEP* 01 (2015), p. 064.

[3812] R. Gauld et al. "Predictions for $Z$ -Boson Production in Association with a $b$-Jet at $\mathcal{O}(\alpha_s^3)$". In: *Phys. Rev. Lett.* 125.22 (2020), p. 222002.

[3813] Sarah Alam Malik and Graeme Watt. "Ratios of W and Z Cross Sections at Large Boson $p_T$ as a Constraint on PDFs and Background to New Physics". In: *JHEP* 02 (2014), p. 025.

[3814] Stephen Farry and Rhorry Gauld. "Leptonic W± boson asymmetry in association with jets at LHCb and parton distribution function constraints at large x". In: *Phys. Rev. D* 93.1 (2016), p. 014008.

[3815] S. J. Brodsky et al. "A review of the intrinsic heavy quark content of the nucleon". In: *Adv. High Energy Phys.* 2015 (2015), p. 231547.

[3816] Tie-Jiun Hou et al. "CT14 Intrinsic Charm Parton Distribution Functions from CTEQ-TEA Global Analysis". In: *JHEP* 02 (2018), p. 059.

[3817] LHC Higgs Cross Section Working Group. "Handbook of LHC Higgs Cross Sections: 4. Deciphering the nature of the Higgs sector". In: *CERN Yellow Rep. Monogr.* 2 (2017).

[3818] Serguei Chatrchyan et al. "Observation of a new boson with mass near 125 GeVin $pp$ collisions at $\sqrt{s} = 7$ and 8 TeV". In: *JHEP* 06 (2013), p. 081.

[3819] Peter W. Higgs. "Broken symmetries and the masses of gauge bosons". In: *Phys. Rev. Lett.* 13 (1964). Ed. by J. C. Taylor, p. 508.

[3820] R. Barate et al. "Search for the standard model Higgs boson at LEP". In: *Phys. Lett. B* 565 (2003), p. 61.

[3821] T. Aaltonen et al. "Evidence for a particle produced in association with weak bosons and decaying to a bottom-antibottom quark pair in Higgs boson searches at the Tevatron". In: *Phys. Rev. Lett.* 109 (2012), p. 071804.

[3822] LHC Higgs Cross Section Working Group. "Handbook of LHC Higgs cross sections: 1. Inclusive observables". In: *CERN Yellow Rep. Monogr.* 2 (2011).

[3823] LHC Higgs Cross Section Working Group. "Handbook of LHC Higgs cross sections: 2. Differential distributions". In: *CERN Yellow Rep. Monogr.* 2 (2012).

[3824] LHC Higgs Cross Section Working Group. "Handbook of LHC Higgs cross sections: 3. Higgs properties". In: *CERN Yellow Rep. Monogr.* 4 (2013). Ed. by S Heinemeyer et al.

[3825] ATLAS collaboration. "A detailed map of Higgs boson interactions by the ATLAS experiment ten years after the discovery". In: *Nature* 607.7917 (2022), pp. 52–59.

[3826] CMS collaboration. "A portrait of the Higgs boson by the CMS experiment ten years after the discovery". In: *Nature* 607.7917 (2022), pp. 60–68.

[3827] "Measurement of the Higgs boson mass in the $H \to ZZ^* \to 4\ell$ decay channel using 139 fb$^{-1}$ of $\sqrt{s} = 13$ TeV $pp$ collisions recorded by the ATLAS detector at the LHC". In: (July 2022).

[3828] Albert M Sirunyan et al. "A measurement of the Higgs boson mass in the diphoton decay channel". In: *Phys. Lett. B* 805 (2020), p. 135425.

[3829] "Measurement of the properties of Higgs boson production at $\sqrt{s} = 13$ TeV in the $H \to \gamma\gamma$ channel using 139 fb$^{-1}$ of $pp$ collision data with the ATLAS experiment". In: (July 2022).

[3830] Vardan Khachatryan et al. "Constraints on the spin-parity and anomalous $HVV$ couplings of the Higgs boson in proton collisions at 7 and 8 TeV". In: *Phys. Rev. D* 92 (2015), p. 012004.

[3831] Georges Aad et al. "Study of the spin and parity of the Higgs boson in diboson decays with the ATLAS detector". In: *Eur. Phys. J. C* 75 (2015). [Erratum: DOI: 10.1140/epjc/s10052-016-3934-y], p. 476.

[3832] Armen Tumasyan et al. "First evidence for off-shell production of the Higgs boson and measurement of its width". Submitted to *Nature Phys.* 2022.

[3833] Georges Aad et al. "A search for the $Z\gamma$ decay mode of the Higgs boson in $pp$ collisions at $\sqrt{s}$ = 13 TeV with the ATLAS detector". In: *Phys. Lett. B* 809 (2020), p. 135754.

[3834] "Search for Higgs boson decays to a Z boson and a photon in proton-proton collisions at $\sqrt{s}$ = 13 TeV". In: (Apr. 2022).

[3835] "Measurement of the total and differential Higgs boson production cross-sections at $\sqrt{s} = 13$ TeV with the ATLAS detector by combining the $H \to ZZ^* \to 4\ell$ and $H \to \gamma\gamma$ decay channels". In: (July 2022).

[3836] "Measurement of the Higgs boson inclusive and differential fiducial production cross sections in the diphoton decay channel with pp collisions at $\sqrt{s} = 13$ TeV with the CMS detector". In: (2022).

[3837] Georges Aad et al. "Measurements of $WH$ and $ZH$ production in the $H \to b\bar{b}$ decay channel in $pp$ collisions at 13 TeV with the ATLAS detector". In: *Eur. Phys. J. C* 81.2 (2021), p. 178.

[3838] "Search for Higgs boson decay to a charm quark-antiquark pair in proton-proton collisions at $\sqrt{s}$ = 13 TeV". In: (May 2022).

[3839] A. M. Sirunyan et al. "Observation of Higgs boson decay to bottom quarks". In: *Phys. Rev. Lett.* 121.12 (2018), p. 121801.

[3840] Georges Aad et al. "Constraints on Higgs boson production with large transverse momentum using $H \to b\bar{b}$ decays in the ATLAS detector". In: *Phys. Rev. D* 105 (2022), p. 092003.

[3841] Albert M Sirunyan et al. "Inclusive search for highly boosted Higgs bosons decaying to bottom quark-antiquark pairs in proton-proton collisions at $\sqrt{s} = 13$ TeV". In: *JHEP* 12 (2020), p. 085.

[3842] Georges Aad et al. "Direct constraint on the Higgs-charm coupling from a search for Higgs boson decays into charm quarks with the ATLAS detector". In: (Jan. 2022).

[3843] ATLAS collaboration. "Observation of Higgs boson production in association with a top quark pair at the LHC with the ATLAS detector". In: *Phys. Lett. B* 784 (2018), pp. 173–191.

[3844] Albert M Sirunyan et al. "Measurement of the Higgs boson production rate in association with top quarks in final states with electrons, muons, and hadronically decaying tau leptons at $\sqrt{s}$ = 13 TeV". In: *Eur. Phys. J. C* 81.4 (2021), p. 378.

[3845] S. W. Herb et al. "Observation of a Dimuon Resonance at 9.5-GeV in 400-GeV Proton-Nucleus Collisions". In: *Phys. Rev. Lett.* 39 (1977), pp. 252–255.

[3846] "Precision Electroweak Measurements and Constraints on the Standard Model". In: (Dec. 2010).

[3847] F. Abe et al. "Observation of top quark production in $\bar{p}p$ collisions". In: *Phys. Rev. Lett.* 74 (1995), pp. 2626–2631.

[3848] S. Abachi et al. "Observation of the top quark". In: *Phys. Rev. Lett.* 74 (1995), pp. 2632–2637.

[3849] T. Aaltonen et al. "First Observation of Electroweak Single Top Quark Production". In: *Phys. Rev. Lett.* 103 (2009), p. 092002.

[3850] V. M. Abazov et al. "Observation of Single Top Quark Production". In: *Phys. Rev. Lett.* 103 (2009), p. 092001.

[3851] Timo Antero Aaltonen et al. "Observation of s-channel production of single top quarks at the Tevatron". In: *Phys. Rev. Lett.* 112 (2014), p. 231803.

[3852] J. A. Aguilar-Saavedra et al. "Asymmetries in top quark pair production at hadron colliders". In: *Rev. Mod. Phys.* 87 (2015), pp. 421–455.

[3853] T. Aaltonen et al. “Combination of CDF and D0 measurements of the $W$ boson helicity in top quark decays”. In: *Phys. Rev. D* 85 (2012), p. 071106.

[3854] “Combination of CDF and D0 Results on the Mass of the Top Quark using up to 9.7 $\text{fb}^{-1}$ at the Tevatron”. In: (July 2014).

[3855] “LHC Machine”. In: *JINST* 3 (2008). Ed. by Lyndon Evans and Philip Bryant, S08001.

[3856] Johann H. Kuhn and German Rodrigo. “Charge asymmetry in hadroproduction of heavy quarks”. In: *Phys. Rev. Lett.* 81 (1998), pp. 49–52.

[3857] Gordon L. Kane, G. A. Ladinsky, and C. P. Yuan. “Using the Top Quark for Testing Standard Model Polarization and CP Predictions”. In: *Phys. Rev. D* 45 (1992), pp. 124–141.

[3858] Vernon D. Barger, J. Ohnemus, and R. J. N. Phillips. “Spin Correlation Effects in the Hadroproduction and Decay of Very Heavy Top Quark Pairs”. In: *Int. J. Mod. Phys. A* 4 (1989), p. 617.

[3859] Gregory Mahlon and Stephen J. Parke. “Spin Correlation Effects in Top Quark Pair Production at the LHC”. In: *Phys. Rev. D* 81 (2010), p. 074024.

[3860] Yoav Afik and Juan Ramón Muñoz de Nova. “Entanglement and quantum tomography with top quarks at the LHC”. In: *Eur. Phys. J. Plus* 136.9 (2021), p. 907.

[3861] Michal Czakon, David Heymes, and Alexander Mitov. “High-precision differential predictions for top-quark pairs at the LHC”. In: *Phys. Rev. Lett.* 116.8 (2016), p. 082003.

[3862] Michal Czakon et al. “Top-pair production at the LHC through NNLO QCD and NLO EW”. In: *JHEP* 10 (2017), p. 186.

[3863] Michal Czakon, David Heymes, and Alexander Mitov. “Dynamical scales for multi-TeV top-pair production at the LHC”. In: *JHEP* 04 (2017), p. 071.

[3864] ATLAS Collaboration. “Measurements of top-quark pair differential and double-differential cross-sections in the $\ell$+jets channel with $pp$ collisions at $\sqrt{s} = 13\,\text{TeV}$ using the ATLAS detector”. In: *Eur. Phys. J. C* 79 (2019), p. 1028.

[3865] ATLAS Collaboration. “Measurement of the $t\bar{t}$ production cross-section and lepton differential distributions in $e\mu$ dilepton events from $pp$ collisions at $\sqrt{s} = 13\,\text{TeV}$ with the ATLAS detector”. In: *Eur. Phys. J. C* 80 (2020), p. 528.

[3866] ATLAS Collaboration. “Measurements of top-quark pair differential cross-sections in the lepton+jets channel in $pp$ collisions at $\sqrt{s} = 8\,\text{TeV}$ using the ATLAS detector”. In: *Eur. Phys. J. C* 76 (2016), p. 538.

[3867] CMS Collaboration. “Measurement of differential $t\bar{t}$ production cross sections in the full kinematic range using lepton+jets events from proton–proton collisions at $\sqrt{s} = 13\,\text{TeV}$”. In: *Phys. Rev. D* 104 (2021), p. 092013.

[3868] CMS Collaboration. “Measurement of differential $t\bar{t}$ production cross sections using top quarks at large transverse momenta in $pp$ collisions at $\sqrt{s} = 13\,\text{TeV}$”. In: *Phys. Rev. D* 103 (2021), p. 052008.

[3869] CMS Collaboration. “Measurements of $t\bar{t}$ differential cross sections in proton–proton collisions at $\sqrt{s} = 13\,\text{TeV}$ using events containing two leptons”. In: *JHEP* 02 (2019), p. 149.

[3870] CMS Collaboration. “Measurement of the differential cross section for top quark pair production in $pp$ collisions at $\sqrt{s} = 8\,\text{TeV}$”. In: *Eur. Phys. J. C* 75 (2015), p. 542.

[3871] Victor Miralles et al. “The top quark electro-weak couplings after LHC Run 2”. In: *JHEP* 02 (2022), p. 032.

[3872] Anna Kulesza et al. “Associated top quark pair production with a heavy boson: differential cross sections at NLO+NNLL accuracy”. In: *Eur. Phys. J. C* 80.5 (2020), p. 428.

[3873] Stefano Catani et al. “$t\bar{t}H$ production at NNLO: the flavour off-diagonal channels”. In: *Eur. Phys. J. C* 81.6 (2021), p. 491.

[3874] P. Azzi et al. “Report from Working Group 1: Standard Model Physics at the HL-LHC and HE-LHC”. In: *CERN Yellow Rep. Monogr.* 7 (2019). Ed. by Andrea Dainese et al., pp. 1–220.

[3875] Giuseppe Bevilacqua et al. “Complete off-shell effects in top quark pair hadroproduction with leptonic decay at next-to-leading order”. In: *JHEP* 02 (2011), p. 083.

[3876] Torbj00F6rn Sjostrand et al. “An Introduction to PYTHIA 8.2”. In: *Comput. Phys. Commun.* 191 (2015), pp. 159–177.

[3877] Johannes Bellm et al. “Herwig 7.0/Herwig++ 3.0 release note”. In: *Eur. Phys. J. C* 76.4 (2016), p. 196.

[3878] Tomáš Ježo and Paolo Nason. “On the Treatment of Resonances in Next-to-Leading Order Calculations Matched to a Parton Shower”. In: *JHEP* 12 (2015), p. 065.

[3879] Javier Mazzitelli et al. “Top-pair production at the LHC with MINNLO$_{PS}$”. In: *JHEP* 04 (2022), p. 079.

[3880] Javier Mazzitelli et al. "Next-to-next-to-leading order event generation for top-quark pair production". In: (Dec. 2020).

[3881] ATLAS and CMS collaborations. "Combinations of single-top-quark production cross-section measurements and $|f_{LV}V_{tb}|$ determinations at $\sqrt{s} = 7$ and 8 TeV with the ATLAS and CMS experiments". In: *JHEP* 05 (2019), p. 088.

[3882] Roman Kogler et al. "Jet Substructure at the Large Hadron Collider: Experimental Review". In: *Rev. Mod. Phys.* 91.4 (2019), p. 045003.

[3883] ATLAS collaboration. "Jet mass and substructure of inclusive jets in $\sqrt{s} = 7$ TeV *pp* collisions with the ATLAS experiment". In: *JHEP* 05 (2012), p. 128.

[3884] CMS collaboration. "Measurement of jet substructure observables in $\mathrm{t\bar{t}}$ events from proton-proton collisions at $\sqrt{s}$ = 13TeV". In: *Phys. Rev. D* 98.9 (2018), p. 092014.

[3885] CMS collaboration. "Measurement of differential $t\bar{t}$ production cross sections in the full kinematic range using lepton+jets events from proton-proton collisions at $\sqrt{s} = 13$ TeV". In: *Phys. Rev. D* 104.9 (2021), p. 092013.

[3886] ATLAS collaboration. "Measurements of differential cross-sections in top-quark pair events with a high transverse momentum top quark and limits on beyond the Standard Model contributions to top-quark pair production with the ATLAS detector at $\sqrt{s} = 13$ TeV". In: (Feb. 2022).

[3887] ATLAS collaboration. "Measurements of inclusive and differential cross-sections of combined $t\bar{t}\gamma$ and $tW\gamma$ production in the $e\mu$ channel at 13 TeV with the ATLAS detector". In: *JHEP* 09 (2020), p. 049.

[3888] Armen Tumasyan et al. "Measurement of the inclusive and differential $\mathrm{t\bar{t}}\gamma$ cross sections in the dilepton channel and effective field theory interpretation in proton-proton collisions at $\sqrt{s}$ =13 TeV". In: *JHEP* 05 (2022), p. 091.

[3889] CMS collaboration. "Measurement of top quark pair production in association with a Z boson in proton-proton collisions at $\sqrt{s} = 13$ TeV". In: *JHEP* 03 (2020), p. 056.

[3890] ATLAS collaboration. "Measurements of the inclusive and differential production cross sections of a top-quark–antiquark pair in association with a Z boson at $\sqrt{s} = 13$ TeV with the ATLAS detector". In: *Eur. Phys. J. C* 81.8 (2021), p. 737.

[3891] ATLAS collaboration. "Observation of the associated production of a top quark and a $Z$ boson in *pp* collisions at $\sqrt{s} = 13$ TeV with the ATLAS detector". In: *JHEP* 07 (2020), p. 124.

[3892] CMS collaboration. "Inclusive and differential cross section measurements of single top quark production in association with a Z boson in proton-proton collisions at $\sqrt{s} = 13$ TeV". In: *JHEP* 02 (2022), p. 107.

[3893] *Observation of single-top-quark production in association with a photon at the ATLAS detector.* Tech. rep. All figures including auxiliary figures are available at https://atlas.web.cern.ch/Atlas/GROUPS/PHYSICS/CONFNOTES/ATLAS-CONF-2022-013. Geneva: CERN, Mar. 2022.

[3894] CMS collaboration. "Observation of $\mathrm{t\bar{t}H}$ production". In: *Phys. Rev. Lett.* 120.23 (2018), p. 231801.

[3895] ATLAS collaboration. "Measurement of the $\mathrm{t}\bar{t}\mathrm{t}\bar{t}$ production cross section in *pp* collisions at $\sqrt{s}$ = 13 TeV with the ATLAS detector". In: *JHEP* 11 (2021), p. 118.

[3896] Ben Lillie, Lisa Randall, and Lian-Tao Wang. "The Bulk RS KK-gluon at the LHC". In: *JHEP* 09 (2007), p. 074.

[3897] CMS collaboration. "Search for resonant $\mathrm{t\bar{t}}$ production in proton-proton collisions at $\sqrt{s} = 13$ TeV". In: *JHEP* 04 (2019), p. 031.

[3898] K. Agashe et al. "Working Group Report: Top Quark". In: *Community Summer Study 2013: Snowmass on the Mississippi.* Nov. 2013.

[3899] G. C. Branco et al. "Theory and phenomenology of two-Higgs-doublet models". In: *Phys. Rept.* 516 (2012), pp. 1–102.

[3900] F. J. Botella et al. "Flavour Changing Higgs Couplings in a Class of Two Higgs Doublet Models". In: *Eur. Phys. J. C* 76.3 (2016), p. 161.

[3901] U. Langenfeld, S. Moch, and P. Uwer. "Measuring the running top-quark mass". In: *Phys. Rev. D* 80 (2009), p. 054009.

[3902] Simone Alioli et al. "A new observable to measure the top-quark mass at hadron colliders". In: *Eur. Phys. J. C* 73 (2013), p. 2438.

[3903] ATLAS collaboration. "Measurement of the top-quark mass in $t\bar{t}$ + 1-jet events collected with the ATLAS detector in *pp* collisions at $\sqrt{s} = 8$ TeV". In: *JHEP* 11 (2019), p. 150.

[3904] Y. Kiyo et al. "Top-quark pair production near threshold at LHC". In: *Eur. Phys. J. C* 60 (2009), pp. 375–386.

[3905] "First combination of Tevatron and LHC measurements of the top-quark mass". In: (Mar. 2014).

[3906] André H. Hoang. “What is the Top Quark Mass?” In: *Ann. Rev. Nucl. Part. Sci.* 70 (2020), pp. 225–255.

[3907] Tomáš Ježo et al. “An NLO+PS generator for $t\bar{t}$ and $Wt$ production and decay including non-resonant and interference effects”. In: *Eur. Phys. J. C* 76.12 (2016), p. 691.

[3908] André H. Hoang, Simon Plätzer, and Daniel Samitz. “On the Cutoff Dependence of the Quark Mass Parameter in Angular Ordered Parton Showers”. In: *JHEP* 10 (2018), p. 200.

[3909] Mathias Butenschoen et al. “Top Quark Mass Calibration for Monte Carlo Event Generators”. In: *Phys. Rev. Lett.* 117.23 (2016), p. 232001.

[3910] *A precise interpretation for the top quark mass parameter in ATLAS Monte Carlo simulation.* Tech. rep. All figures including auxiliary figures are available at https://atlas.web.cern.ch/Atlas/GROUPS/PHYSICS/PUBNOTES/ATL-PHYS-PUB-2021-034. Geneva: CERN, July 2021.

[3911] *A profile likelihood approach to measure the top quark mass in the lepton+jets channel at* $\sqrt{s} =$ 13 TeV. Tech. rep. Geneva: CERN, 2022.

[3912] R. Schwienhorst, D. Wackeroth, et al. “Top Quark Physics and Heavy Flavor Production”. In: (2022). Summary of the EF03 topical group at Snowmass 2021.

[3913] Albert M Sirunyan et al. “Measurement of $\mathrm{t\bar{t}}$ normalised multi-differential cross sections in pp collisions at $\sqrt{s} = 13$ TeV, and simultaneous determination of the strong coupling strength, top quark pole mass, and parton distribution functions”. In: *Eur. Phys. J. C* 80.7 (2020), p. 658.

[3914] M. Baak et al. “The global electroweak fit at NNLO and prospects for the LHC and ILC”. In: *Eur. Phys. J.* C74 (2014), p. 3046.

[3915] Nathan P. Hartland et al. “A Monte Carlo global analysis of the Standard Model Effective Field Theory: the top quark sector”. In: *JHEP* 04 (2019), p. 100.

[3916] Ilaria Brivio et al. “O new physics, where art thou? A global search in the top sector”. In: *JHEP* 02 (2020), p. 131.

[3917] Jacob J. Ethier et al. “Combined SMEFT interpretation of Higgs, diboson, and top quark data from the LHC”. In: *JHEP* 11 (2021), p. 089.

[3918] John Ellis et al. “Top, Higgs, Diboson and Electroweak Fit to the Standard Model Effective Field Theory”. In: *JHEP* 04 (2021), p. 279.

[3919] H. Abramowicz et al. “Top-Quark Physics at the CLIC Electron-Positron Linear Collider”. In: *JHEP* 11 (2019), p. 003.

[3920] Gauthier Durieux et al. “The electro-weak couplings of the top and bottom quarks — Global fit and future prospects”. In: *JHEP* 12 (2019). [Erratum: JHEP 01, 195 (2021)], p. 98.

[3921] Gauthier Durieux et al. “Snowmass White Paper: prospects for the measurement of top-quark couplings”. In: *2022 Snowmass Summer Study.* May 2022.

[3922] J. de Blas et al. “The CLIC Potential for New Physics”. In: 3/2018 (Dec. 2018).

[3923] U. Amaldi et al. “The Real Part of the Forward Proton Proton Scattering Amplitude Measured at the CERN Intersecting Storage Rings”. In: *Phys. Lett. B* 66 (1977), pp. 390–394.

[3924] C. Augier et al. “Predictions on the total cross-section and real part at LHC and SSC”. In: *Phys. Lett. B* 315 (1993), pp. 503–506.

[3925] Marcel Froissart. “Asymptotic Behavior and Subtractions in the Mandelstam Representation”. In: *Phys. Rev.* 123 (3 1961), pp. 1053–1057.

[3926] Andre Martin. “Extension of the axiomatic analyticity domain of scattering amplitudes by unitarity-I”. In: *Il Nuovo Cimento A (1965-1970)* 42.4 (1966), pp. 930–953.

[3927] P. D. B. Collins. *An Introduction to Regge Theory and High-Energy Physics.* Cambridge Monographs on Mathematical Physics. Cambridge, UK: Cambridge Univ. Press, May 2009.

[3928] S. Donnachie et al. *Pomeron physics and QCD.* Vol. 19. Cambridge University Press, Dec. 2004.

[3929] C. Patrignani et al. “Review of Particle Physics”. In: *Chin. Phys. C* 40.10 (2016), p. 100001.

[3930] Otto Nachtmann. “Pomeron physics and QCD”. In: *Ringberg Workshop on New Trends in HERA Physics 2003.* 2004, pp. 253–267.

[3931] L. Lukaszuk and B. Nicolescu. “A Possible interpretation of p p rising total cross-sections”. In: *Lett. Nuovo Cim.* 8 (1973), pp. 405–413.

[3932] G. Antchev et al. “First determination of the $\rho$ parameter at $\sqrt{s} = 13$ TeV: probing the existence of a colourless C-odd three-gluon compound state”. In: *Eur. Phys. J. C* 79.9 (2019), p. 785.

[3933] W. Heisenberg. “Mesonenerzeugung als Stosswellenproblem”. In: *Z. Phys.* 133 (1952), p. 65.

[3934] J. R. Cudell et al. “Benchmarks for the forward observables at RHIC, the Tevatron Run II and the LHC”. In: *Phys. Rev. Lett.* 89 (2002), p. 201801.

[3935] Matteo Giordano, Enrico Meggiolaro, and Niccolo Moretti. “Asymptotic Energy Dependence of Hadronic Total Cross Sections from Lattice QCD”. In: *JHEP* 09 (2012), p. 031.

[3936] Elena Ferreiro et al. “Froissart bound from gluon saturation”. In: *Nucl. Phys. A* 710 (2002), pp. 373–414.

[3937] G. Anelli et al. “The TOTEM experiment at the CERN Large Hadron Collider”. In: *JINST* 3 (2008), S08007.

[3938] Jan Kaspar. “Slides presented at the LHC Working Group of Forward Physics and Diffraction, 16-17 December 2019”. In: ().

[3939] G. Antchev et al. “First measurement of elastic, inelastic and total cross-section at $\sqrt{s} = 13$ TeV by TOTEM and overview of cross-section data at LHC energies”. In: *Eur. Phys. J. C* 79.2 (2019), p. 103.

[3940] V. A. Schegelsky and M. G. Ryskin. “The diffraction cone shrinkage speed up with the collision energy”. In: *Phys. Rev. D* 85 (2012), p. 094024.

[3941] A. Donnachie and P. V. Landshoff. “The Interest of large - t elastic scattering”. In: *Phys. Lett. B* 387 (1996), pp. 637–641.

[3942] A. Donnachie and P. V. Landshoff. “Small $t$ elastic scattering and the $\rho$ parameter”. In: *Phys. Lett. B* 798 (2019), p. 135008.

[3943] V. M. Abazov et al. “Odderon Exchange from Elastic Scattering Differences between $pp$ and $p\bar{p}$ Data at 1.96 TeV and from pp Forward Scattering Measurements”. In: *Phys. Rev. Lett.* 127.6 (2021), p. 062003.

[3944] Evgenij Martynov and Basarab Nicolescu. “Did TOTEM experiment discover the Odderon?” In: *Phys. Lett. B* 778 (2018), pp. 414–418.

[3945] Betty Abelev et al. “Measurement of inelastic, single- and double-diffraction cross sections in proton–proton collisions at the LHC with ALICE”. In: *Eur. Phys. J. C* 73.6 (2013), p. 2456.

[3946] Georges Aad et al. “Rapidity gap cross sections measured with the ATLAS detector in $pp$ collisions at $\sqrt{s} = 7$ TeV”. In: *Eur. Phys. J. C* 72 (2012), p. 1926.

[3947] J. Bartels, M. G. Ryskin, and G. P. Vacca. “On the triple pomeron vertex in perturbative QCD”. In: *Eur. Phys. J. C* 27 (2003), pp. 101–113.

[3948] S. Chekanov et al. “Exclusive electroproduction of J/psi mesons at HERA”. In: *Nucl. Phys. B* 695 (2004), pp. 3–37.

[3949] A. Aktas et al. “Elastic J/psi production at HERA”. In: *Eur. Phys. J. C* 46 (2006), pp. 585–603.

[3950] L. Frankfurt, M. McDermott, and M. Strikman. “A Fresh look at diffractive $J/\psi$ photoproduction at HERA, with predictions for THERA”. In: *JHEP* 03 (2001), p. 045.

[3951] John C. Collins. “Proof of factorization for diffractive hard scattering”. In: *Phys. Rev. D* 57 (5 Mar. 1998), pp. 3051–3056.

[3952] T. Affolder et al. “Diffractive dijets with a leading antiproton in $\bar{p}p$ collisions at $\sqrt{s} = 1800$ GeV”. In: *Phys. Rev. Lett.* 84 (2000), pp. 5043–5048.

[3953] Morad Aaboud et al. “Measurement of charged-particle distributions sensitive to the underlying event in $\sqrt{s} = 13$ TeV proton-proton collisions with the ATLAS detector at the LHC”. In: *JHEP* 03 (2017), p. 157.

[3954] G. Antchev et al. “Measurement of the forward charged particle pseudorapidity density in pp collisions at $\sqrt{s} = 8$ TeV using a displaced interaction point”. In: *Eur. Phys. J. C* 75.3 (2015), p. 126.

[3955] Serguei Chatrchyan et al. “Measurement of pseudorapidity distributions of charged particles in proton-proton collisions at $\sqrt{s} = 8$ TeV by the CMS and TOTEM experiments”. In: *Eur. Phys. J. C* 74.10 (2014), p. 3053.

[3956] M.K. Gaillard and Benjamin W. Lee. “Rare Decay Modes of the K-Mesons in Gauge Theories”. In: *Phys. Rev.* D10 (1974), p. 897.

[3957] N. Cabibbo. “Unitary Symmetry and Leptonic Decays”. In: *Phys. Rev. Lett.* 10 (1963), pp. 531–533.

[3958] M. Kobayashi and T. Maskawa. “$CP$-violation in the renormalizable theory of weak interaction”. In: *Prog. Theor. Phys.* 49 (1973), pp. 652–657.

[3959] Wolfhart Zimmermann. “Normal products and the short distance expansion in the perturbation theory of renormalizable interactions”. In: *Annals Phys.* 77 (1973). [Lect. Notes Phys.558,278(2000)], pp. 570–601.

[3960] William A. Bardeen, Andrzej J. Buras, and Jean-Marc Gérard. “A Consistent Analysis of the $\Delta I = 1/2$ Rule for K Decays”. In: *Phys. Lett.* B192 (1987), p. 138.

[3961] Andrzej J. Buras, Jean-Marc Gérard, and William A. Bardeen. “Large $N$ Approach to Kaon Decays and Mixing 28 Years Later: $\Delta I = 1/2$ Rule, $\hat{B}_K$ and $\Delta M_K$”. In: *Eur. Phys. J.* C74.5 (2014), p. 2871.

[3962] Ryan Abbott et al. “Direct CP violation and the $\Delta I = 1/2$ rule in $K \to \pi\pi$ decay from the Standard Model”. In: (Apr. 2020).

[3963] Andrzej J. Buras. “$\varepsilon'/\varepsilon$ in the Standard Model and Beyond: 2021”. In: *11th International Workshop on the CKM Unitarity Triangle.* Mar. 2022.

[3964] Andrzej J. Buras. “Weak Hamiltonian, CP violation and rare decays”. In: *Probing the standard model of particle interactions. Proceedings, Summer School in Theoretical Physics, NATO Advanced Study Institute, 68th session, Les Houches, France, July 28-September 5, 1997. Pt. 1, 2.* 1998, pp. 281–539.

[3965] Andrzej J. Buras and Peter H. Weisz. “QCD Nonleading Corrections to Weak Decays in Dimensional Regularization and ’t Hooft-Veltman Schemes”. In: *Nucl. Phys.* B333 (1990), pp. 66–99.

[3966] Andrzej Buras. *Gauge Theory of Weak Decays.* Cambridge University Press, June 2020.

[3967] Gerhard Buchalla, Andrzej J. Buras, and Markus E. Lautenbacher. “Weak decays beyond leading logarithms”. In: *Rev. Mod. Phys.* 68 (1996), pp. 1125–1144.

[3968] Andrzej J. Buras et al. “Charm quark contribution to $K^+ \to \pi^+ \nu\bar{\nu}$ at next-to-next-to-leading order”. In: *JHEP* 11 (2006), p. 002.

[3969] Mikhail A. Shifman, A. I. Vainshtein, and Valentin I. Zakharov. “Nonleptonic Decays of K Mesons and Hyperons”. In: *Sov. Phys. JETP* 45 (1977). [Zh. Eksp. Teor. Fiz.72,1275(1977)], p. 670.

[3970] Frederick J. Gilman and Mark B. Wise. “Effective Hamiltonian for $\Delta s = 1$ Weak Nonleptonic Decays in the Six Quark Model”. In: *Phys. Rev.* D20 (1979), p. 2392.

[3971] Frederick J. Gilman and Mark B. Wise. “$K^0 - \bar{K}^0$ Mixing in the Six Quark Model”. In: *Phys. Rev. D* 27 (1983), p. 1128.

[3972] Claudio Dib, Isard Dunietz, and Frederick J. Gilman. “Strong Interaction Corrections to the Decay $K \to \pi$ Neutrino Anti-neutrino for Large M(t)”. In: *Mod. Phys. Lett.* A6 (1991), pp. 3573–3582.

[3973] Andrzej J. Buras. “Climbing NLO and NNLO Summits of Weak Decays”. In: (2011).

[3974] Marco Ciuchini et al. “Next-to-leading order QCD corrections to $\Delta F = 2$ effective Hamiltonians”. In: *Nucl. Phys.* B523 (1998), pp. 501–525.

[3975] Andrzej J. Buras, Mikolaj Misiak, and Jorg Urban. “Two loop QCD anomalous dimensions of flavor changing four quark operators within and beyond the standard model”. In: *Nucl. Phys.* B586 (2000), pp. 397–426.

[3976] Jason Aebischer et al. “General non-leptonic $\Delta F = 1$ WET at the NLO in QCD”. In: *JHEP* 11 (2021), p. 227.

[3977] Jason Aebischer et al. “SMEFT ATLAS of $\Delta$F = 2 transitions”. In: *JHEP* 12 (2020), p. 187.

[3978] Jason Aebischer, Andrzej J. Buras, and Jacky Kumar. “NLO QCD Renormalization Group Evolution for Non-Leptonic $\Delta F = 2$ Transitions in the SMEFT”. In: (Mar. 2022).

[3979] A. Bazavov et al. “$B^0_{(s)}$-mixing matrix elements from lattice QCD for the Standard Model and beyond”. In: *Phys. Rev.* D93.11 (2016), p. 113016.

[3980] Federico Mescia and Christopher Smith. “Improved estimates of rare K decay matrix-elements from $K_{\ell 3}$ decays”. In: *Phys. Rev.* D76 (2007), p. 034017.

[3981] Rui-Xiang Shi et al. “Revisiting the new-physics interpretation of the $b \to c\tau\nu$ data”. In: *JHEP* 12 (2019), p. 065.

[3982] P. Colangelo, F. De Fazio, and F. Loparco. “Probing New Physics with $\bar{B} \to \rho(770)\,\ell^-\bar{\nu}_\ell$ and $\bar{B} \to a_1(1260)\,\ell^-\bar{\nu}_\ell$”. In: *Phys. Rev. D* 100.7 (2019), p. 075037.

[3983] Patricia Ball and Roman Zwicky. “New results on $B \to \pi, K, \eta$ decay formfactors from light-cone sum rules”. In: *Phys. Rev.* D71 (2005), p. 014015.

[3984] Chris Bouchard et al. “Rare decay $B \to K\ell^+\ell^-$ form factors from lattice QCD”. In: *Phys. Rev. D* 88.5 (2013). [Erratum: Phys.Rev.D 88, 079901 (2013)], p. 054509.

[3985] R. R. Horgan et al. “Rare $B$ decays using lattice QCD form factors”. In: *PoS* LATTICE2014 (2015), p. 372.

[3986] Aoife Bharucha, David M. Straub, and Roman Zwicky. “$B \to V\ell^+\ell^-$ in the Standard Model from light-cone sum rules”. In: *JHEP* 08 (2016), p. 098.

[3987] Nico Gubernari, Ahmet Kokulu, and Danny van Dyk. “$B \to P$ and $B \to V$ Form Factors from $B$-Meson Light-Cone Sum Rules beyond Leading Twist”. In: *JHEP* 01 (2019), p. 150.

[3988] W. G. Parrott, C. Bouchard, and C. T. H. Davies. “Standard Model predictions for $B \to K\ell^+\ell^-$, $B \to K\ell_1^-\ell_2^+$ and $B \to K\nu\bar{\nu}$ using form factors from $N_f = 2 + 1 + 1$ lattice QCD”. In: (July 2022).

[3989] Nathan Isgur and Mark B. Wise. “Heavy quark symmetry”. In: *Adv. Ser. Direct. High Energy Phys.* 10 (1992), pp. 234–285.

[3990] Thomas Mannel, Winston Roberts, and Zbigniew Ryzak. “A Derivation of the heavy quark effective Lagrangian from QCD”. In: *Nucl. Phys.* B368 (1992), pp. 204–217.

[3991] Ikaros I. Y. Bigi et al. “The Question of CP Noninvariance - as Seen Through the Eyes of Neutral Beauty”. In: *Adv. Ser. Direct. High Energy Phys.* 3 (1989), pp. 175–248.

[3992] Marina Artuso, Guennadi Borissov, and Alexander Lenz. "CP violation in the $B_s^0$ system". In: *Rev. Mod. Phys.* 88.4 (2016), p. 045002.

[3993] Benjamin J. Choi et al. "Kaon BSM B-parameters using improved staggered fermions from $N_f = 2+1$ unquenched QCD". In: *Phys. Rev.* D93.1 (2016), p. 014511.

[3994] Peter A. Boyle et al. "Neutral kaon mixing beyond the Standard Model with $\mathrm{n}_f = 2+1$ chiral fermions. Part 2: non perturbative renormalisation of the $\Delta F = 2$ four-quark operators". In: *JHEP* 10 (2017), p. 054.

[3995] Andrzej J. Buras and Jean-Marc Gerard. "Dual QCD Insight into BSM Hadronic Matrix Elements for $K^0 - \bar{K}^0$ Mixing from Lattice QCD". In: (2018).

[3996] Jason Aebischer, Andrzej J. Buras, and Jean-Marc Gérard. "BSM hadronic matrix elements for $\epsilon'/\epsilon$ and $K \to \pi\pi$ decays in the Dual QCD approach". In: *JHEP* 02 (2019), p. 021.

[3997] M. Beneke et al. "QCD factorization for $B \to K\pi, \pi\pi$ decays: Strong phases and CP violation in the heavy quark limit". In: *Phys. Rev. Lett.* 83 (1999), pp. 1914–1917.

[3998] M. Beneke et al. "QCD factorization for exclusive, nonleptonic B meson decays: General arguments and the case of heavy light final states". In: *Nucl. Phys.* B591 (2000), pp. 313–418.

[3999] M. Beneke. "Soft-collinear factorization in B decays". In: *Nucl. Part. Phys. Proc.* 261-262 (2015), pp. 311–337.

[4000] Lincoln Wolfenstein. "Parametrization of the Kobayashi-Maskawa Matrix". In: *Phys. Rev. Lett.* 51 (1983), p. 1945.

[4001] Marcella Bona et al. "Unitarity Triangle global fits testing the Standard Model: UT$fit$ 2021 SM update". In: *PoS* EPS-HEP2021 (2022), p. 512.

[4002] Chien-Yeah Seng et al. "Reduced Hadronic Uncertainty in the Determination of $V_{ud}$". In: *Phys. Rev. Lett.* 121.24 (2018), p. 241804.

[4003] Andrzej Czarnecki, William J. Marciano, and Alberto Sirlin. "Radiative Corrections to Neutron and Nuclear Beta Decays Revisited". In: *Phys. Rev. D* 100.7 (2019), p. 073008.

[4004] Mikhail Gorchtein. "$\gamma$W Box Inside Out: Nuclear Polarizabilities Distort the Beta Decay Spectrum". In: *Phys. Rev. Lett.* 123.4 (2019), p. 042503.

[4005] J. C. Hardy and I. S. Towner. "Superallowed $0^+ \to 0^+$ nuclear $\beta$ decays: 2020 critical survey, with implications for $\mathrm{V}_{ud}$ and CKM unitarity". In: *Phys. Rev. C* 102.4 (2020), p. 045501.

[4006] Vincenzo Cirigliano et al. "Scrutinizing CKM unitarity with a new measurement of the $K_{\mu 3}/K_{\mu 2}$ branching fraction". In: (Aug. 2022).

[4007] D. Pocanic et al. "Precise measurement of the $\pi^+ \to \pi^0 e^+ \nu$ branching ratio". In: *Phys. Rev. Lett.* 93 (2004), p. 181803.

[4008] W. Altmannshofer et al. "PIONEER: Studies of Rare Pion Decays". In: (Mar. 2022).

[4009] Elvira Gamiz et al. "$V_{us}$ and $m_s$ from hadronic $\tau$ decays". In: *Phys. Rev. Lett.* 94 (2005), p. 011803.

[4010] Renwick J. Hudspith et al. "A resolution of the inclusive flavor-breaking $\tau$ $|V_{us}|$ puzzle". In: *Phys. Lett. B* 781 (2018), pp. 206–212.

[4011] Peter Boyle et al. "Novel $|V_{us}|$ Determination Using Inclusive Strange $\tau$ Decay and Lattice Hadronic Vacuum Polarization Functions". In: *Phys. Rev. Lett.* 121.20 (2018), p. 202003.

[4012] Y. Amhis et al. "Averages of $b$-hadron, $c$-hadron, and $\tau$-lepton properties as of 2021". In: (June 2022).

[4013] Claudio Andrea Manzari, Antonio M. Coutinho, and Andreas Crivellin. "Modified lepton couplings and the Cabibbo-angle anomaly". In: *PoS* LHCP2020 (2021). Ed. by Bruno Mansoulie et al., p. 242.

[4014] Bernat Capdevila et al. "Explaining $b \to s\ell^+\ell^-$ and the Cabibbo angle anomaly with a vector triplet". In: *Phys. Rev. D* 103.1 (2021), p. 015032.

[4015] P. Gambino et al. "Challenges in semileptonic $B$ decays". In: *Eur. Phys. J. C* 80.10 (2020), p. 966.

[4016] Giulia Ricciardi and Marcello Rotondo. "Determination of the Cabibbo-Kobayashi-Maskawa matrix element $|V_{cb}|$". In: *J. Phys. G* 47 (2020), p. 113001.

[4017] Martin Jung and David M. Straub. "Constraining new physics in $b \to c\ell\nu$ transitions". In: *JHEP* 01 (2019), p. 009.

[4018] Andreas Crivellin and Stefan Pokorski. "Can the differences in the determinations of $V_{ub}$ and $V_{cb}$ be explained by New Physics?" In: *Phys. Rev. Lett.* 114.1 (2015), p. 011802.

[4019] B. Blok et al. "Differential distributions in semileptonic decays of the heavy flavors in QCD". In: *Phys.Rev.* D49 (1994), p. 3356.

[4020] Aneesh V. Manohar and Mark B. Wise. "Inclusive semileptonic $B$ and polarized $\Lambda_b$ decays from QCD". In: *Phys.Rev.* D49 (1994), pp. 1310–1329.

[4021] Ikaros Bigi et al. "The Two Roads to 'Intrinsic Charm' in $B$ Decays". In: *JHEP* 1004 (2010), p. 073.

[4022] Thomas Mannel and Alexei A. Pivovarov. "QCD corrections to inclusive heavy hadron weak decays at $\Lambda^3_{\mathrm{QCD}}/m^3_Q$". In: *Phys. Rev. D* 100.9 (2019), p. 093001.

[4023] Kirill Melnikov. "$O({\alpha_s}^2)$ corrections to semileptonic decay $b \to c\ell\bar{\nu}$". In: *Phys.Lett.* B666 (2008), pp. 336–339.

[4024] Alexey Pak and Andrzej Czarnecki. "Mass effects in muon and semileptonic $b \to c$ decays". In: *Phys.Rev.Lett.* 100 (2008), p. 241807.

[4025] Paolo Gambino. "$B$ semileptonic moments at NNLO". In: *JHEP* 1109 (2011), p. 055.

[4026] Andrea Alberti, Paolo Gambino, and Soumitra Nandi. "Perturbative corrections to power suppressed effects in semileptonic $B$ decays". In: *JHEP* 1401 (2014), pp. 1–16.

[4027] Ikaros I.Y. Bigi et al. "High power n of $m_b$ in beauty widths". In: *Phys.Rev.* D56 (1997), pp. 4017–4030.

[4028] Matteo Fael, Kay Schönwald, and Matthias Steinhauser. "Kinetic Heavy Quark Mass to Three Loops". In: *Phys. Rev. Lett.* 125.5 (2020), p. 052003.

[4029] Thomas Mannel, Sascha Turczyk, and Nikolai Uraltsev. "Higher Order Power Corrections in Inclusive $B$ Decays". In: *JHEP* 1011 (2010), p. 109.

[4030] Paolo Gambino, Kristopher J. Healey, and Sascha Turczyk. "Taming the higher power corrections in semileptonic $B$ decays". In: *Phys. Lett.* B763 (2016), pp. 60–65.

[4031] Matteo Fael, Thomas Mannel, and K. Keri Vos. "$V_{cb}$ determination from inclusive $b \to c$ decays: an alternative method". In: *JHEP* 02 (2019), p. 177.

[4032] Marzia Bordone, Bernat Capdevila, and Paolo Gambino. "Three loop calculations and inclusive Vcb". In: *Phys. Lett. B* 822 (2021), p. 136679.

[4033] Florian Bernlochner et al. "First extraction of inclusive $V_{cb}$ from $q^2$ moments". In: (May 2022).

[4034] Paolo Gambino and Shoji Hashimoto. "Inclusive Semileptonic Decays from Lattice QCD". In: *Phys. Rev. Lett.* 125.3 (2020), p. 032001.

[4035] Bjorn O. Lange, Matthias Neubert, and Gil Paz. "Theory of charmless inclusive $B$ decays and the extraction of $V_{ub}$". In: *Phys. Rev.* D72 (2005), p. 073006.

[4036] P. Gambino et al. "Inclusive semileptonic $B$ decays and the determination of $|V_{ub}|$". In: *JHEP* 0710 (2007), p. 058.

[4037] Jeppe R. Andersen and Einan Gardi. "Inclusive spectra in charmless semileptonic $B$ decays by dressed gluon exponentiation". In: *JHEP* 01 (2006), p. 097.

[4038] Ikaros I. Y. Bigi and N. G. Uraltsev. "Weak annihilation and the endpoint spectrum in semileptonic $B$ decays". In: *Nucl. Phys. B* 423 (1994), pp. 33–55.

[4039] Zoltan Ligeti, Michael Luke, and Aneesh V. Manohar. "Constraining weak annihilation using semileptonic D decays". In: *Phys. Rev. D* 82 (2010), p. 033003.

[4040] Paolo Gambino and Jernej F. Kamenik. "Lepton energy moments in semileptonic charm decays". In: *Nucl. Phys. B* 840 (2010), pp. 424–437.

[4041] J. P. Lees et al. "Measurement of the inclusive electron spectrum from $B$ meson decays and determination of $|V_{ub}|$". In: *Phys. Rev. D* 95.7 (2017), p. 072001.

[4042] L. Cao et al. "Measurements of Partial Branching Fractions of Inclusive $B \to X_u\, \ell^+\, \nu_\ell$ Decays with Hadronic Tagging". In: *Phys. Rev. D* 104.1 (2021), p. 012008.

[4043] Bernat Capdevila, Paolo Gambino, and Soumitra Nandi. "Perturbative corrections to power suppressed effects in $\bar{B} \to X_u \ell\nu$". In: *JHEP* 04 (2021), p. 137.

[4044] Mathias Brucherseifer, Fabrizio Caola, and Kirill Melnikov. "On the $O(\alpha_s^2)$ corrections to $b \to X_u e\bar{\nu}$ inclusive decays". In: *Phys.Lett.* B721 (2013), pp. 107–110.

[4045] L. Cao et al. "Measurement of Differential Branching Fractions of Inclusive $B \to X_u\, \ell^+\, \nu_\ell$ Decays". In: *Phys. Rev. Lett.* 127.26 (2021), p. 261801.

[4046] Paolo Gambino, Kristopher J. Healey, and Cristina Mondino. "Neural network approach to $B \to X_u\ell\nu$". In: *Phys. Rev. D* 94.1 (2016), p. 014031.

[4047] C. Glenn Boyd, Benjamin Grinstein, and Richard F. Lebed. "Precision corrections to dispersive bounds on form-factors". In: *Phys. Rev.* D56 (1997), pp. 6895–6911.

[4048] Claude Bourrely, Irinel Caprini, and Laurent Lellouch. "Model-independent description of $B \to \pi\ell\nu$ decays and a determination of $|V_{ub}|$". In: *Phys. Rev. D* 79 (2009). [Erratum: Phys.Rev.D 82, 099902 (2010)], p. 013008.

[4049] Irinel Caprini, Laurent Lellouch, and Matthias Neubert. "Dispersive bounds on the shape of $\bar{B} \to D^*\ell\bar{\nu}$ form-factors". In: *Nucl. Phys.* B530 (1998), pp. 153–181.

[4050] Florian U. Bernlochner et al. "Combined analysis of semileptonic $B$ decays to $D$ and $D^*$: $R(D^{(*)})$, $|V_{cb}|$, and new physics". In: *Phys. Rev.* D95.11 (2017), p. 115008.

[4051] Dante Bigi, Paolo Gambino, and Stefan Schacht. "$R(D^*)$, $|V_{cb}|$, and the Heavy Quark Symmetry

relations between form factors". In: *JHEP* 11 (2017), p. 061.

[4052] Sneha Jaiswal, Soumitra Nandi, and Sunando Kumar Patra. "Extraction of $|V_{cb}|$ from $B \to D^{(*)}\ell\nu_\ell$ and the Standard Model predictions of $R(D^{(*)})$". In: *JHEP* 12 (2017), p. 060.

[4053] Marzia Bordone, Martin Jung, and Danny van Dyk. "Theory determination of $\bar{B} \to D^{(*)}\ell^-\bar{\nu}$ form factors at $\mathcal{O}(1/m_c^2)$". In: *Eur. Phys. J. C* 80.2 (2020), p. 74.

[4054] R. Glattauer et al. "Measurement of the decay $B \to D\ell\nu_\ell$ in fully reconstructed events and determination of the Cabibbo-Kobayashi-Maskawa matrix element $|V_{cb}|$". In: *Phys. Rev.* D93.3 (2016), p. 032006.

[4055] A. Abdesselam et al. "Precise determination of the CKM matrix element $|V_{cb}|$ with $\bar{B}^0 \to D^{*\,+}\,\ell^-\,\bar{\nu}_\ell$ decays with hadronic tagging at Belle". In: (2017).

[4056] E. Waheed et al. "Measurement of the CKM matrix element $|V_{cb}|$ from $B^0 \to D^{*-}\ell^+\nu_\ell$ at Belle". In: *Phys. Rev. D* 100.5 (2019). [Erratum: Phys.Rev.D 103, 079901 (2021)], p. 052007.

[4057] Jon A. Bailey et al. "$B \to D\ell\nu$ form factors at nonzero recoil and $|V_{cb}|$ from 2+1-flavor lattice QCD". In: *Phys. Rev. D* 92.3 (2015), p. 034506.

[4058] Heechang Na et al. "$B \to Dl\nu$ form factors at nonzero recoil and extraction of $|V_{cb}|$". In: *Phys. Rev.* D92.5 (2015). [Erratum: Phys. Rev.D93,no.11,119906(2016)], p. 054510.

[4059] Dante Bigi and Paolo Gambino. "Revisiting $B \to D\ell\nu$". In: *Phys. Rev.* D94.9 (2016), p. 094008.

[4060] Jon A. Bailey et al. "B→D form factors at nonzero recoil and $|V_{cb}|$ from 2+1-flavor lattice QCD". In: *Phys. Rev.* D92.3 (2015), p. 034506.

[4061] Bernard Aubert et al. "Measurement of $|V_{cb}|$ and the Form-Factor Slope in $\bar{B} \to Dl^-\bar{\nu}$ Decays in Events Tagged by a Fully Reconstructed B Meson". In: *Phys. Rev. Lett.* 104 (2010), p. 011802.

[4062] Takashi et al. Kaneko. "$B \to D^{(*)}\ell\nu$ semileptonic decays in lattice QCD with domain-wall heavy quarks". In: *PoS* LATTICE2021 (2022), p. 561.

[4063] Dante Bigi, Paolo Gambino, and Stefan Schacht. "A fresh look at the determination of $|V_{cb}|$ from $B \to D^*\ell\nu$". In: *Phys. Lett. B* 769 (2017), pp. 441–445.

[4064] Benjamin Grinstein and Andrew Kobach. "Model-Independent Extraction of $|V_{cb}|$ from $\bar{B} \to D^*\ell\bar{\nu}$". In: *Phys. Lett.* B771 (2017), pp. 359–364.

[4065] Christoph Bobeth et al. "Lepton-flavour non-universality of $\bar{B} \to D^*\ell\bar{\nu}$ angular distributions in and beyond the Standard Model". In: *Eur. Phys. J. C* 81.11 (2021), p. 984.

[4066] Paolo Gambino, Martin Jung, and Stefan Schacht. "The $V_{cb}$ puzzle: An update". In: *Phys. Lett. B* 795 (2019), pp. 386–390.

[4067] G. D'Agostini. "On the use of the covariance matrix to fit correlated data". In: *Nucl. Instrum. Meth.* A346 (1994), pp. 306–311.

[4068] Takashi Kaneko. "private communication". In: (2022).

[4069] G. Martinelli, S. Simula, and L. Vittorio. "Exclusive determinations of $|V_{cb}|$ and $R(D^*)$ through unitarity". In: (Sept. 2021).

[4070] Roel Aaij et al. "Measurement of $|V_{cb}|$ with $B_s^0 \to D_s^{(*)-}\mu^+\nu_\mu$ decays". In: *Phys. Rev. D* 101.7 (2020), p. 072004.

[4071] E. McLean et al. "Lattice QCD form factor for $B_s \to D_s^* l\nu$ at zero recoil with non-perturbative current renormalisation". In: *Phys. Rev. D* 99.11 (2019), p. 114512.

[4072] J. P. Lees et al. "Extraction of form Factors from a Four-Dimensional Angular Analysis of $\overline{B} \to D^*\ell^-\overline{\nu}_\ell$". In: *Phys. Rev. Lett.* 123.9 (2019), p. 091801.

[4073] Jon A. Bailey et al. "$|V_{ub}|$ from $B \to \pi\ell\nu$ decays and (2+1)-flavor lattice QCD". In: *Phys. Rev.* D92.1 (2015), p. 014024.

[4074] Aoife Bharucha. "Two-loop Corrections to the $B \to \pi$ Form Factor from QCD Sum Rules on the Light-Cone and $|V_{ub}|$". In: *JHEP* 05 (2012), p. 092.

[4075] Domagoj Leljak, Blaženka Melić, and Danny van Dyk. "The $\overline{B} \to \pi$ form factors from QCD and their impact on $|V_{ub}|$". In: *JHEP* 07 (2021), p. 036.

[4076] Aritra Biswas et al. "A closer look at the extraction of $|V_{ub}|$ from $B \to$ $pi\ell\nu$". In: *JHEP* 07 (2021), p. 082.

[4077] Roel Aaij et al. "First observation of the decay $B_s^0 \to K^-\mu^+\nu_\mu$ and Measurement of $|V_{ub}|/|V_{cb}|$". In: *Phys. Rev. Lett.* 126.8 (2021), p. 081804.

[4078] Alexei Bazavov et al. "$B_s \to K\ell\nu$ decay from lattice QCD". In: *Phys. Rev. D* 100.3 (2019), p. 034501.

[4079] Alexander Khodjamirian and Aleksey V. Rusov. "$B_s \to K\ell\nu_\ell$ and $B_{(s)} \to \pi(K)\ell^+\ell^-$ decays at large recoil and CKM matrix elements". In: *JHEP* 08 (2017), p. 112.

[4080] Daniel King et al. "$|V_{cb}|$ and $\gamma$ from $B$-mixing - Addendum to "$B_s$ mixing observables and $|V_{td}/V_{ts}|$ from sum rules"". In: (Nov. 2019). [Addendum: JHEP 03, 112 (2020)].

[4081] Wolfgang Altmannshofer and Nathan Lewis. “Loop-induced determinations of $V_{ub}$ and $V_{cb}$”. In: *Phys. Rev. D* 105.3 (2022), p. 033004.

[4082] Andrzej J. Buras and Elena Venturini. “The exclusive vision of rare $K$ and $B$ decays and of the quark mixing in the standard model”. In: *Eur. Phys. J. C* 82.7 (2022), p. 615.

[4083] Andrzej J. Buras. “On the superiority of the $|V_{cb}| - \gamma$ plots over the unitarity triangle plots in the 2020s”. In: *Eur. Phys. J. C* 82.7 (2022), p. 612.

[4084] J. Charles and *et al.* “CP violation and the CKM matrix: Assessing the impact of the asymmetric $B$ factories”. In: *Eur. Phys. J. C* 41.1 (2005), pp. 1–131.

[4085] Gino Isidori. “Flavour Physics and Implication for New Phenomena”. In: *Adv. Ser. Direct. High Energy Phys.* 26 (2016), pp. 339–355.

[4086] R. Ammar et al. “Evidence for penguins: First observation of B —> K* (892) gamma”. In: *Phys. Rev. Lett.* 71 (1993), pp. 674–678.

[4087] M. S. Alam et al. “First measurement of the rate for the inclusive radiative penguin decay $b \to s\gamma$”. In: *Phys. Rev. Lett.* 74 (1995), pp. 2885–2889.

[4088] M. Misiak et al. “Estimate of $\mathcal{B}(\bar{B} \to X(s)\gamma)$ at $\mathcal{O}(\alpha_s^2)$”. In: *Phys. Rev. Lett.* 98 (2007), p. 022002.

[4089] M. Misiak et al. “Updated NNLO QCD predictions for the weak radiative B-meson decays”. In: *Phys. Rev. Lett.* 114.22 (2015), p. 221801.

[4090] Y. Amhis et al. “Averages of $b$-hadron, $c$-hadron, and $\tau$-lepton properties as of summer 2016”. In: (2016).

[4091] S. Bertolini, Francesca Borzumati, and A. Masiero. “QCD Enhancement of Radiative b Decays”. In: *Phys. Rev. Lett.* 59 (1987), p. 180.

[4092] N. G. Deshpande et al. “$B \to K^*\gamma$ and the Top Quark Mass”. In: *Phys. Rev. Lett.* 59 (1987), pp. 183–185.

[4093] M. Misiak. “QCD corrected effective Hamiltonian for the $b \to s\gamma$ decay”. In: *Phys. Lett. B* 269 (1991), pp. 161–168.

[4094] Mikolaj Misiak. “The $b \to se^+e^-$ and $b \to s\gamma$ decays with next-to-leading logarithmic QCD corrections”. In: *Nucl. Phys.* B393 (1993). [Erratum: Nucl. Phys.B439,461(1995)], pp. 23–45.

[4095] Marco Ciuchini et al. “Scheme independence of the effective Hamiltonian for $b \to s\gamma$ and $b \to sg$ decays”. In: *Phys. Lett.* B316 (1993), pp. 127–136.

[4096] Marco Ciuchini et al. “Leading order QCD corrections to $b \to s\gamma$ and $b \to sg$ decays in three regularization schemes”. In: *Nucl. Phys.* B421 (1994), pp. 41–64.

[4097] A. J. Buras et al. “Theoretical uncertainties and phenomenological aspects of $B \to X_s\gamma$ decay”. In: *Nucl. Phys.* B424 (1994), pp. 374–398.

[4098] Ahmed Ali, C. Greub, and T. Mannel. “Rare B decays in the Standard Model”. In: (1993).

[4099] Andrzej J. Buras and Mikolaj Misiak. “$\bar{B} \to X_s\gamma$ after completion of the NLO QCD calculations”. In: *Acta Phys. Polon.* B33 (2002), pp. 2597–2612.

[4100] Paolo Gambino and Mikolaj Misiak. “Quark mass effects in $\bar{B} \to X_s\gamma$”. In: *Nucl. Phys.* B611 (2001), pp. 338–366.

[4101] M. Tanabashi et al. “Review of Particle Physics”. In: *Phys. Rev.* D98.3 (2018), p. 030001.

[4102] Murray Gell-Mann and A. Pais. “Behavior of neutral particles under charge conjugation”. In: *Phys. Rev.* 97 (1955), pp. 1387–1389.

[4103] M. Gell-Mann and A.H. Rosenfeld. “Hyperons and heavy mesons (systematics and decay)”. In: *Ann.Rev.Nucl.Part.Sci.* 7 (1957), pp. 407–478.

[4104] E. Fermi. “An attempt of a theory of beta radiation. 1.” In: *Z. Phys.* 88 (1934), pp. 161–177.

[4105] R. P. Feynman and Murray Gell-Mann. “Theory of Fermi interaction”. In: *Phys. Rev.* 109 (1958), pp. 193–198.

[4106] E. C. G. Sudarshan and R. e. Marshak. “Chirality invariance and the universal Fermi interaction”. In: *Phys. Rev.* 109 (1958), pp. 1860–1860.

[4107] G.E. Brown et al. “Final state interactions in K meson decays”. In: *Phys. Lett. B* 238 (1990), pp. 20–24.

[4108] Elisabetta Pallante and Antonio Pich. “Final state interactions in kaon decays”. In: *Nucl. Phys.* B592 (2001), pp. 294–320.

[4109] P.A. Boyle et al. “Emerging understanding of the $\Delta I = 1/2$ Rule from Lattice QCD”. In: (2012).

[4110] Guido Altarelli et al. “QCD Nonleading Corrections to Weak Decays as an Application of Regularization by Dimensional Reduction”. In: *Nucl. Phys.* B187 (1981), pp. 461–513.

[4111] Andrzej J. Buras, Fulvia De Fazio, and Jennifer Girrbach. “$\Delta I = 1/2$ rule, $\varepsilon'/\varepsilon$ and $K \to \pi\nu\bar{\nu}$ in $Z'(Z)$ and $G'$ models with FCNC quark couplings”. In: *Eur. Phys. J.* C74 (2014), p. 2950.

[4112] Andrzej J. Buras. “The $\epsilon'/\epsilon$-Story: 1976-2021”. In: *Acta Phys. Polon. B* 52.1 (2021), pp. 7–41.

[4113] J.R. Batley et al. “A Precision measurement of direct CP violation in the decay of neutral

kaons into two pions". In: *Phys. Lett.* B544 (2002), pp. 97–112.
[4114] A. Alavi-Harati et al. "Measurements of direct CP violation, CPT symmetry, and other parameters in the neutral kaon system". In: *Phys. Rev.* D67 (2003), p. 012005.
[4115] E. Abouzaid et al. "Precise Measurements of Direct CP Violation, CPT Symmetry, and Other Parameters in the Neutral Kaon System". In: *Phys. Rev.* D83 (2011), p. 092001.
[4116] Frederick J. Gilman and Mark B. Wise. "The $\Delta I = 1/2$ Rule and Violation of CP in the Six Quark Model". In: *Phys. Lett.* B83 (1979), pp. 83–86.
[4117] William A. Bardeen, Andrzej J. Buras, and Jean-Marc Gérard. "The $\Delta I = 1/2$ Rule in the Large $N$ Limit". In: *Phys. Lett.* B180 (1986), p. 133.
[4118] William A. Bardeen, Andrzej J. Buras, and Jean-Marc Gérard. "The $K \to \pi\pi$ Decays in the Large N Limit: Quark Evolution". In: *Nucl. Phys.* B293 (1987), p. 787.
[4119] John F. Donoghue et al. "Electromagnetic and Isospin Breaking Effects Decrease $\epsilon'/\epsilon$". In: *Phys. Lett.* B179 (1986). [Erratum: Phys. Lett.B188,511(1987)], p. 361.
[4120] A. J. Buras and J. M. Gérard. "Isospin Breaking Contributions to $\varepsilon'/\varepsilon$". In: *Phys. Lett.* B192 (1987), p. 156.
[4121] Joanathan M. Flynn and Lisa Randall. "The Electromagnetic Penguin Contribution to $\varepsilon'/\varepsilon$ for Large Top Quark Mass". In: *Phys. Lett.* B224 (1989), p. 221.
[4122] Gerhard Buchalla, Andrzej J. Buras, and Michaela K. Harlander. "The Anatomy of $\varepsilon'/\varepsilon$ in the Standard Model". In: *Nucl. Phys.* B337 (1990), pp. 313–362.
[4123] Andrzej J. Buras et al. "Effective Hamiltonians for $\Delta S = 1$ and $\Delta B = 1$ nonleptonic decays beyond the leading logarithmic approximation". In: *Nucl. Phys.* B370 (1992). [Addendum: Nucl. Phys.B375,501(1992)], pp. 69–104.
[4124] Andrzej J. Buras et al. "Two loop anomalous dimension matrix for $\Delta S = 1$ weak nonleptonic decays. 1. $\mathcal{O}(\alpha_s^2)$". In: *Nucl. Phys.* B400 (1993), pp. 37–74.
[4125] Andrzej J. Buras, Matthias Jamin, and Markus E. Lautenbacher. "Two loop anomalous dimension matrix for $\Delta S = 1$ weak nonleptonic decays. 2. $\mathcal{O}(\alpha\alpha_s)$". In: *Nucl. Phys.* B400 (1993), pp. 75–102.
[4126] Andrzej J. Buras, Matthias Jamin, and Markus E. Lautenbacher. "The Anatomy of $\varepsilon'/\varepsilon$ beyond leading logarithms with improved hadronic matrix elements". In: *Nucl. Phys.* B408 (1993), pp. 209–285.
[4127] Marco Ciuchini et al. "$\varepsilon'/\varepsilon$ at the Next-to-leading order in QCD and QED". In: *Phys. Lett.* B301 (1993), pp. 263–271.
[4128] Marco Ciuchini et al. "The $\Delta S = 1$ effective Hamiltonian including next-to-leading order QCD and QED corrections". In: *Nucl. Phys.* B415 (1994), pp. 403–462.
[4129] Andrzej J. Buras, Paolo Gambino, and Ulrich A. Haisch. "Electroweak penguin contributions to non-leptonic $\Delta F = 1$ decays at NNLO". In: *Nucl. Phys.* B570 (2000), pp. 117–154.
[4130] Andrzej J. Buras et al. "Improved anatomy of $\varepsilon'/\varepsilon$ in the Standard Model". In: *JHEP* 11 (2015), p. 202.
[4131] V. Cirigliano et al. "Isospin violation in $\epsilon'$". In: *Phys. Rev. Lett.* 91 (2003), p. 162001.
[4132] V. Cirigliano et al. "Isospin breaking in $K \to \pi\pi$ decays". In: *Eur. Phys. J.* C33 (2004), pp. 369–396.
[4133] Johan Bijnens and Fredrik Borg. "Isospin breaking in $K \to 3\pi$ decays III: Bremsstrahlung and fit to experiment". In: *Eur. Phys. J.* C40 (2005), pp. 383–394.
[4134] Andrzej J. Buras and Jean-Marc Gerard. "Isospin-breaking in $\varepsilon'/\varepsilon$: Impact of $\eta_0$ at the Dawn of the 2020s". In: (May 2020).
[4135] V. Cirigliano et al. "Theoretical status of $\varepsilon'/\varepsilon$". In: *J. Phys. Conf. Ser.* 1526 (2020). Ed. by Patrizia Cenci and Mauro Piccini, p. 012011.
[4136] Jason Aebischer, Christoph Bobeth, and Andrzej J. Buras. "$\varepsilon'/\varepsilon$ in the Standard Model at the Dawn of the 2020s". In: *Eur. Phys. J. C* 80.8 (2020), p. 705.
[4137] V. Antonelli et al. "The $\Delta I = 1/2$ selection rule". In: *Nucl. Phys.* B469 (1996), pp. 181–201.
[4138] S. Bertolini, J. O. Eeg, and M. Fabbrichesi. "A New estimate of $\varepsilon'/\varepsilon$". In: *Nucl. Phys.* B476 (1996), pp. 225–254.
[4139] Elisabetta Pallante and Antonio Pich. "Strong enhancement of $\varepsilon'/\varepsilon$ through final state interactions". In: *Phys. Rev. Lett.* 84 (2000), pp. 2568–2571.
[4140] E. Pallante, A. Pich, and I. Scimemi. "The Standard model prediction for $\varepsilon'/\varepsilon$". In: *Nucl. Phys.* B617 (2001), pp. 441–474.
[4141] Andrzej J. Buras and Jean-Marc Gérard. "Upper Bounds on $\varepsilon'/\varepsilon$ Parameters $B_6^{(1/2)}$ and $B_8^{(3/2)}$ from Large N QCD and other News". In: *JHEP* 12 (2015), p. 008.

[4142] Andrzej J. Buras and Jean-Marc Gérard. "Final state interactions in $K \to \pi\pi$ decays: $\Delta I = 1/2$ rule vs. $\varepsilon'/\varepsilon$". In: *Eur. Phys. J.* C77.1 (2017), p. 10.

[4143] R Aaij et al. "Measurement of Form-Factor-Independent Observables in the Decay $B^0 \to K^{*0}\mu^+\mu^-$". In: *Phys. Rev. Lett.* 111 (2013), p. 191801.

[4144] Christoph Bobeth, Mikolaj Misiak, and Jorg Urban. "Photonic penguins at two loops and $m_t$-dependence of $BR(B \to X_s\ell^+\ell^-)$". In: *Nucl. Phys.* B574 (2000), pp. 291–330.

[4145] Martin Gorbahn and Ulrich Haisch. "Effective Hamiltonian for non-leptonic $|\Delta F| = 1$ decays at NNLO in QCD". In: *Nucl. Phys.* B713 (2005), pp. 291–332.

[4146] Tobias Huber et al. "Electromagnetic logarithms in $\bar{B} \to X(s)l^+l^-$". In: *Nucl. Phys.* B740 (2006), pp. 105–137.

[4147] Martin Beneke, Christoph Bobeth, and Robert Szafron. "Enhanced electromagnetic correction to the rare $B$-meson decay $B_{s,d} \to \mu^+\mu^-$". In: *Phys. Rev. Lett.* 120.1 (2018), p. 011801.

[4148] N. Carrasco et al. "A $N_f = 2 + 1 + 1$ "twisted" determination of the $b$-quark mass, $f_B$ and $f_{B_s}$". In: *PoS* LATTICE2013 (2014), p. 313.

[4149] *Combination of the ATLAS, CMS and LHCb results on the $B^0_{(s)} \to \mu^+\mu^-$ decays.* LHCb-CONF-2020-002, CERN-LHCb-CONF-2020-002. Aug. 2020.

[4150] Johannes Albrecht, Danny van Dyk, and Christoph Langenbruch. "Flavour anomalies in heavy quark decays". In: *Prog. Part. Nucl. Phys.* 120 (2021), p. 103885.

[4151] Christoph Bobeth, Gudrun Hiller, and Danny van Dyk. "General analysis of $\bar{B} \to \bar{K}^{(*)}\ell^+\ell^-$ decays at low recoil". In: *Phys. Rev. D* 87.3 (2013), p. 034016.

[4152] S. Jäger and J. Martin Camalich. "On $B \to V\ell\ell$ at small dilepton invariant mass, power corrections, and new physics". In: *JHEP* 05 (2013), p. 043.

[4153] Frank Kruger and Joaquim Matias. "Probing new physics via the transverse amplitudes of $B^0 \to K^{*0}(\to K^-\pi^+)l^+l^-$ at large recoil". In: *Phys. Rev. D* 71 (2005), p. 094009.

[4154] Christoph Bobeth, Gudrun Hiller, and Giorgi Piranishvili. "CP Asymmetries in bar $B \to \bar{K}^*(\to \bar{K}\pi)\bar{\ell}\ell$ and Untagged $\bar{B}_s, B_s \to \phi(\to K^+K^-)\bar{\ell}\ell$ Decays at NLO". In: *JHEP* 07 (2008), p. 106.

[4155] Wolfgang Altmannshofer et al. "Symmetries and Asymmetries of $B \to K^*\mu^+\mu^-$ Decays in the Standard Model and Beyond". In: *JHEP* 01 (2009), p. 019.

[4156] Christoph Bobeth, Gudrun Hiller, and Giorgi Piranishvili. "Angular distributions of $\bar{B} \to \bar{K}\ell^+\ell^-$ decays". In: *JHEP* 12 (2007), p. 040.

[4157] Joaquim Matias. "On the S-wave pollution of B-> K* l+l- observables". In: *Phys. Rev. D* 86 (2012), p. 094024.

[4158] Sébastien Descotes-Genon, Alexander Khodjamirian, and Javier Virto. "Light-cone sum rules for $B \to K\pi$ form factors and applications to rare decays". In: *JHEP* 12 (2019), p. 083.

[4159] J. Charles et al. "Heavy to light form-factors in the heavy mass to large energy limit of QCD". In: *Phys. Rev. D* 60 (1999), p. 014001.

[4160] M. Beneke and T. Feldmann. "Symmetry breaking corrections to heavy to light B meson form-factors at large recoil". In: *Nucl. Phys. B* 592 (2001), pp. 3–34.

[4161] U. Egede et al. "New observables in the decay mode $\bar{B}_d \to \bar{K}^{*0}l^+l^-$". In: *JHEP* 11 (2008), p. 032.

[4162] Christoph Bobeth, Gudrun Hiller, and Danny van Dyk. "The Benefits of $\bar{B}- > \bar{K}^*l^+l^-$ Decays at Low Recoil". In: *JHEP* 07 (2010), p. 098.

[4163] Sebastien Descotes-Genon et al. "Optimizing the basis of $B \to K^*ll$ observables in the full kinematic range". In: *JHEP* 05 (2013), p. 137.

[4164] Sebastien Descotes-Genon et al. "Implications from clean observables for the binned analysis of $B- > K * \mu^+\mu^-$ at large recoil". In: *JHEP* 01 (2013), p. 048.

[4165] Aoife Bharucha, David M. Straub, and Roman Zwicky. "$B \to V\ell^+\ell^-$ in the Standard Model from light-cone sum rules". In: *JHEP* 08 (2016), p. 098.

[4166] Ronald R. Horgan et al. "Lattice QCD calculation of form factors describing the rare decays $B \to K^*\ell^+\ell^-$ and $B_s \to \phi\ell^+\ell^-$". In: *Phys. Rev. D* 89.9 (2014), p. 094501.

[4167] Jon A. Bailey et al. "$B \to Kl^+l^-$ Decay Form Factors from Three-Flavor Lattice QCD". In: *Phys. Rev. D* 93.2 (2016), p. 025026.

[4168] Nico Gubernari et al. "Improved Theory Predictions and Global Analysis of Exclusive $\boldsymbol{b \to s\mu^+\mu^-}$ Processes". In: (June 2022).

[4169] Irinel Caprini. *Functional Analysis and Optimization Methods in Hadron Physics.* SpringerBriefs in Physics. Springer, 2019.

[4170] Shan Cheng, Alexander Khodjamirian, and Javier Virto. "$B \to \pi\pi$ Form Factors from Light-Cone Sum Rules with $B$-meson Distribution Amplitudes". In: *JHEP* 05 (2017), p. 157.

[4171] Shan Cheng, Alexander Khodjamirian, and Javier Virto. "Timelike-helicity $B \to \pi\pi$ form factor

from light-cone sum rules with dipion distribution amplitudes". In: *Phys. Rev. D* 96.5 (2017), p. 051901.

[4172] M. Beneke, T. Feldmann, and D. Seidel. "Systematic approach to exclusive $B \to Vl^+l^-$, $V\gamma$ decays". In: *Nucl. Phys. B* 612 (2001), pp. 25–58.

[4173] A. Khodjamirian et al. "Charm-loop effect in $B \to K^{(*)}\ell^+\ell^-$ and $B \to K^*\gamma$". In: *JHEP* 09 (2010), p. 089.

[4174] Benjamin Grinstein and Dan Pirjol. "Exclusive rare $B \to K^*\ell^+\ell^-$ decays at low recoil: Controlling the long-distance effects". In: *Phys. Rev. D* 70 (2004), p. 114005.

[4175] M. Beylich, G. Buchalla, and T. Feldmann. "Theory of $B \to K^{(*)}\ell^+\ell^-$ decays at high $q^2$: OPE and quark-hadron duality". In: *Eur. Phys. J. C* 71 (2011), p. 1635.

[4176] M. Beneke, Th. Feldmann, and D. Seidel. "Exclusive radiative and electroweak $b \to d$ and $b \to s$ penguin decays at NLO". In: *Eur. Phys. J. C* 41 (2005), pp. 173–188.

[4177] Nico Gubernari, Danny van Dyk, and Javier Virto. "Non-local matrix elements in $B_{(s)} \to \{K^{(*)}, \phi\}\ell^+\ell^-$". In: *JHEP* 02 (2021), p. 088.

[4178] Christoph Bobeth, Gudrun Hiller, and Danny van Dyk. "More Benefits of Semileptonic Rare B Decays at Low Recoil: CP Violation". In: *JHEP* 07 (2011), p. 067.

[4179] Christoph Bobeth et al. "The Decay $B \to K\ell^+\ell^-$ at Low Hadronic Recoil and Model-Independent $\Delta B = 1$ Constraints". In: *JHEP* 01 (2012), p. 107.

[4180] James Lyon and Roman Zwicky. "Resonances gone topsy turvy - the charm of QCD or new physics in $b \to s\ell^+\ell^-$?" In: (June 2014).

[4181] Simon Braß, Gudrun Hiller, and Ivan Nisandzic. "Zooming in on $B \to K^*\ell\ell$ decays at low recoil". In: *Eur. Phys. J. C* 77.1 (2017), p. 16.

[4182] Christoph Bobeth et al. "Long-distance effects in $B \to K^*\ell\ell$ from analyticity". In: *Eur. Phys. J. C* 78.6 (2018), p. 451.

[4183] Marcel Algueró et al. "$b \to s\ell^+\ell^-$ global fits after $R_{K_S}$ and $R_{K^{*+}}$". In: *Eur. Phys. J. C* 82.4 (2022), p. 326.

[4184] Wolfgang Altmannshofer and Peter Stangl. "New physics in rare B decays after Moriond 2021". In: *Eur. Phys. J. C* 81.10 (2021), p. 952.

[4185] Marco Ciuchini et al. "Lessons from the $B^{0,+} \to K^{*0,+}\mu^+\mu^-$ angular analyses". In: *Phys. Rev. D* 103.1 (2021), p. 015030.

[4186] T. Hurth et al. "More indications for lepton nonuniversality in $b \to s\ell^+\ell^-$". In: *Phys. Lett. B* 824 (2022), p. 136838.

[4187] Daping Du et al. "Phenomenology of Semileptonic B-Meson Decays with Form Factors from Lattice QCD". In: (2015).

[4188] Marcin Chrzaszcz et al. "Prospects for disentangling long- and short-distance effects in the decays $B \to K^*\mu^+\mu^-$". In: *JHEP* 10 (2019), p. 236.

[4189] Sébastien Descotes-Genon et al. "Global analysis of $b \to s\ell\ell$ anomalies". In: *JHEP* 06 (2016), p. 092.

[4190] Marco Ciuchini et al. "$B \to K^*\ell^+\ell^-$ decays at large recoil in the Standard Model: a theoretical reappraisal". In: *JHEP* 06 (2016), p. 116.

[4191] Ulrik Egede, Mitesh Patel, and Konstantinos A. Petridis. "Method for an unbinned measurement of the q$^2$ dependent decay amplitudes of $\overline{B}^0 \to K^{*0}\mu^+\mu^-$ decays". In: *JHEP* 06 (2015), p. 084.

[4192] T. Hurth, F. Mahmoudi, and S. Neshatpour. "On the anomalies in the latest LHCb data". In: *Nucl. Phys. B* 909 (2016), pp. 737–777.

[4193] F. Kruger and L. M. Sehgal. "Lepton polarization in the decays $b \to X_s\mu^+\mu^-$ and $B \to X_s\tau^+\tau^-$". In: *Phys. Lett. B* 380 (1996), pp. 199–204.

[4194] Thomas Blake et al. "An empirical model to determine the hadronic resonance contributions to $\overline{B}^0 \to \overline{K}^{*0}\mu^+\mu^-$ transitions". In: *Eur. Phys. J. C* 78.6 (2018), p. 453.

[4195] Friedrich Jegerlehner. "The Anomalous Magnetic Moment of the Muon". In: *Springer Tracts Mod. Phys.* 274 (2017), pp.1–693.

[4196] Julian S. Schwinger. "On Quantum electrodynamics and the magnetic moment of the electron". In: *Phys. Rev.* 73 (1948), pp. 416–417.

[4197] V. B. Berestetskii, O. N. Krokhin, and A. K. Khlebnikov. "Concerning the Radiative Correction to the $\mu$-Meson Magnetic Moment". In: *JETP* 3.5 (1956). [Zh. Eksp. Teor. Fiz. **30**, 788 (1956)], p. 761.

[4198] W. S. Cowland. "On Schwinger's theory of the muon". In: *Nucl. Phys.* 8 (1958), pp. 397–401.

[4199] R. L. Garwin et al. "Accurate Determination of the mu+ Magnetic Moment". In: *Phys. Rev.* 118 (1960), pp. 271–283.

[4200] B. Lee Roberts. "The History of the Muon ($g-2$) Experiments". In: *SciPost Phys. Proc.* 1 (2019), p. 032.

[4201] V. Bargmann, Louis Michel, and V. L. Telegdi. "Precession of the polarization of particles mov-

ing in a homogeneous electromagnetic field". In: *Phys. Rev. Lett.* 2 (1959). [,92(1959)], pp. 435–436.

[4202] J. Bailey et al. "Final Report on the CERN Muon Storage Ring Including the Anomalous Magnetic Moment and the Electric Dipole Moment of the Muon, and a Direct Test of Relativistic Time Dilation". In: *Nucl. Phys. B* 150 (1979), pp. 1–75.

[4203] G. W. Bennett et al. "Final Report of the Muon E821 Anomalous Magnetic Moment Measurement at BNL". In: *Phys. Rev.* D73 (2006), p. 072003.

[4204] Fred Jegerlehner and Andreas Nyffeler. "The Muon $g-2$". In: *Phys. Rept.* 477 (2009), pp. 1–110.

[4205] Graziano Venanzoni. "The New Muon $g-2$ experiment at Fermilab". In: (2014).

[4206] T. Aoyama et al. "The anomalous magnetic moment of the muon in the Standard Model". In: *Phys. Rept.* 887 (2020), pp. 1–166.

[4207] B. Abi et al. "Measurement of the Positive Muon Anomalous Magnetic Moment to 0.46 ppm". In: *Phys. Rev. Lett.* 126.14 (2021), p. 141801.

[4208] Sz. Borsanyi et al. "Leading hadronic contribution to the muon magnetic moment from lattice QCD". In: *Nature* 593.7857 (2021), pp. 51–55.

[4209] Tsutomu Mibe. "Measurement of muon $g-2$ and EDM with an ultra-cold muon beam at J-PARC". In: *Nucl. Phys. Proc. Suppl.* 218 (2011), pp. 242–246.

[4210] R. R. Akhmetshin et al. "Reanalysis of hadronic cross-section measurements at CMD-2". In: *Phys. Lett. B* 578 (2004), pp. 285–289.

[4211] V. M. Aul'chenko et al. "Measurement of the pion form-factor in the range 1.04-GeV to 1.38-GeV with the CMD-2 detector". In: *JETP Lett.* 82 (2005), pp. 743–747.

[4212] V. M. Aul'chenko et al. "Measurement of the $e^+e^- \to \pi^+\pi^-$ cross section with the CMD-2 detector in the 370 - 520-MeV c.m. energy range". In: *JETP Lett.* 84 (2006), pp. 413–417.

[4213] R. R. Akhmetshin et al. "High-statistics measurement of the pion form factor in the rho-meson energy range with the CMD-2 detector". In: *Phys. Lett. B* 648 (2007), pp. 28–38.

[4214] M. N. Achasov et al. "Study of the process e+ e- —> pi+ pi- in the energy region 400 < s**(1/2) < 1000-MeV". In: *J. Exp. Theor. Phys.* 101.6 (2005), pp. 1053–1070.

[4215] A. Anastasi et al. "Combination of KLOE $\sigma\big(e^+e^- \to \pi^+\pi^-\gamma(\gamma)\big)$ measurements and determination of $a_\mu^{\pi^+\pi^-}$ in the energy range $0.10 < s < 0.95$ GeV$^2$". In: *JHEP* 03 (2018), p. 173.

[4216] Bernard Aubert et al. "Precise measurement of the $e^+e^- \to \pi^+\pi^-\gamma$ cross section with the Initial State Radiation method at BABAR". In: *Phys. Rev. Lett.* 103 (2009), p. 231801.

[4217] J. P. Lees et al. "Precise Measurement of the $e^+e^- \to \pi^+\pi^-(\gamma)$ Cross Section with the Initial-State Radiation Method at BABAR". In: *Phys. Rev. D* 86 (2012), p. 032013.

[4218] M. Ablikim et al. "Measurement of the $e^+e^? \to \pi^+\pi^?$ cross section between 600 and 900 MeV using initial state radiation". In: *Phys. Lett. B* 753 (2016). [Erratum: Phys.Lett.B 812, 135982 (2021)], pp. 629–638.

[4219] V. P. Druzhinin et al. "Hadron Production via e+e- Collisions with Initial State Radiation". In: *Rev. Mod. Phys.* 83 (2011), p. 1545.

[4220] Ricard Alemany, Michel Davier, and Andreas Höcker. "Improved determination of the hadronic contribution to the muon (g-2) and to alpha (M(z)) using new data from hadronic tau decays". In: *Eur. Phys. J.* C2 (1998), pp. 123–135.

[4221] M. N. Achasov et al. "Measurement of the $e^+e^- \to \pi^+\pi^-$ process cross section with the SND detector at the VEPP-2000 collider in the energy region $0.525 < \sqrt{s} < 0.883$ GeV". In: *JHEP* 01 (2021), p. 113.

[4222] T. Xiao et al. "Precision Measurement of the Hadronic Contribution to the Muon Anomalous Magnetic Moment". In: *Phys. Rev. D* 97.3 (2018), p. 032012.

[4223] Alexander Keshavarzi, Daisuke Nomura, and Thomas Teubner. "The muon $g-2$ and $\alpha(M_Z^2)$: a new data-based analysis". In: *Phys. Rev.* D97.11 (2018), p. 114025.

[4224] M. Davier et al. "A new evaluation of the hadronic vacuum polarisation contributions to the muon anomalous magnetic moment and to $\alpha(m_Z^2)$". In: *Eur. Phys. J.* C80.3 (2020). [Erratum: Eur. Phys. J. **C80**, 410 (2020)], p. 241.

[4225] Michel Davier et al. "Reevaluation of the hadronic vacuum polarisation contributions to the Standard Model predictions of the muon $g-2$ and $\alpha(m_Z^2)$ using newest hadronic cross-section data". In: *Eur. Phys. J.* C77.12 (2017), p. 827.

[4226] Gilberto Colangelo, Martin Hoferichter, and Peter Stoffer. "Two-pion contribution to hadronic vacuum polarization". In: *JHEP* 02 (2019), p. 006.

[4227] Martin Hoferichter, Bai-Long Hoid, and Bastian Kubis. "Three-pion contribution to hadronic vacuum polarization". In: *JHEP* 08 (2019), p. 137.

[4228] Alexander Keshavarzi, Daisuke Nomura, and Thomas Teubner. "The $g-2$ of charged leptons,

$\alpha(M_Z^2)$ and the hyperfine splitting of muonium". In: *Phys. Rev.* D101 (2020), p. 014029.

[4229] F. Ambrosino et al. "Measurement of $\sigma(e^+e^- \to \pi^+\pi^-\gamma(\gamma)$ and the dipion contribution to the muon anomaly with the KLOE detector". In: *Phys. Lett. B* 670 (2009), pp. 285–291.

[4230] F. Ambrosino et al. "Measurement of $\sigma(e^+e^- \to \pi^+\pi^-)$ from threshold to 0.85 GeV$^2$ using Initial State Radiation with the KLOE detector". In: *Phys. Lett. B* 700 (2011), pp. 102–110.

[4231] D. Babusci et al. "Precision measurement of $\sigma(e^+e^- \to \pi^+\pi^-\gamma)/\sigma(e^+e^- \to \mu^+\mu^-\gamma)$ and determination of the $\pi^+\pi^-$ contribution to the muon anomaly with the KLOE detector". In: *Phys. Lett. B* 720 (2013), pp. 336–343.

[4232] T. Blum. "Lattice calculation of the lowest order hadronic contribution to the muon anomalous magnetic moment". In: *Phys. Rev. Lett.* 91 (2003), p. 052001.

[4233] David Bernecker and Harvey B. Meyer. "Vector Correlators in Lattice QCD: Methods and applications". In: *Eur. Phys. J.* A47 (2011), p. 148.

[4234] M. Della Morte et al. "The hadronic vacuum polarization contribution to the muon $g-2$ from lattice QCD". In: *JHEP* 10 (2017), p. 020.

[4235] Bipasha Chakraborty et al. "Strong-isospin-breaking correction to the muon anomalous magnetic moment from lattice QCD at the physical point". In: *Phys. Rev. Lett.* 120.15 (2018), p. 152001.

[4236] Sz. Borsanyi et al. "Hadronic vacuum polarization contribution to the anomalous magnetic moments of leptons from first principles". In: *Phys. Rev. Lett.* 121 (2018), p. 022002.

[4237] T. Blum et al. "Calculation of the hadronic vacuum polarization contribution to the muon anomalous magnetic moment". In: *Phys. Rev. Lett.* 121 (2018), p. 022003.

[4238] D. Giusti et al. "Electromagnetic and strong isospin-breaking corrections to the muon $g-2$ from Lattice QCD+QED". In: *Phys. Rev.* D99.11 (2019), p. 114502.

[4239] Eigo Shintani and Yoshinobu Kuramashi. "Hadronic vacuum polarization contribution to the muon $g-2$ with 2+1 flavor lattice QCD on a larger than (10 fm)$^4$ lattice at the physical point". In: *Phys. Rev.* D100.3 (2019), p. 034517.

[4240] C. T. H. Davies et al. "Hadronic-vacuum-polarization contribution to the muon's anomalous magnetic moment from four-flavor lattice QCD". In: *Phys. Rev. D* 101.3 (2020), p. 034512.

[4241] Antoine Gérardin et al. "The leading hadronic contribution to $(g-2)_\mu$ from lattice QCD with $N_\mathrm{f} = 2+1$ flavours of O($a$) improved Wilson quarks". In: *Phys. Rev. D* 100.1 (2019), p. 014510.

[4242] Christopher Aubin et al. "Light quark vacuum polarization at the physical point and contribution to the muon $g-2$". In: *Phys. Rev. D* 101.1 (2020), p. 014503.

[4243] D. Giusti and S. Simula. "Lepton anomalous magnetic moments in Lattice QCD+QED". In: *PoS* LATTICE2019 (2019), p. 104.

[4244] Marco Cè et al. "Window observable for the hadronic vacuum polarization contribution to the muon $g-2$ from lattice QCD". In: (June 2022).

[4245] Christoph Lehner. "The hadronic vacuum polarization (RBC/UKQCD). Talk at the Fifth Plenary Workshop of the Muon g–2 Theory Initiative, Edinburgh". In: (September 5, 2022).

[4246] C. Alexandrou et al. "Lattice calculation of the short and intermediate time-distance hadronic vacuum polarization contributions to the muon magnetic moment using twisted-mass fermions". In: (June 2022).

[4247] G. Colangelo et al. "Data-driven evaluations of Euclidean windows to scrutinize hadronic vacuum polarization". In: *Phys. Lett. B* 833 (2022), p. 137313.

[4248] Gen Wang et al. "Muon g-2 with overlap valence fermion". In: (Apr. 2022).

[4249] Christopher Aubin et al. "Muon anomalous magnetic moment with staggered fermions: Is the lattice spacing small enough?" In: *Phys. Rev. D* 106.5 (2022), p. 054503.

[4250] C. T. H. Davies et al. "Windows on the hadronic vacuum polarisation contribution to the muon anomalous magnetic moment". In: (July 2022).

[4251] Mattia Bruno et al. "On isospin breaking in $\tau$ decays for $(g-2)_\mu$ from Lattice QCD". In: *PoS* LATTICE2018 (2018), p. 135.

[4252] M. Hayakawa, T. Kinoshita, and A. I. Sanda. "Hadronic light by light scattering contribution to muon g-2". In: *Phys. Rev.* D54 (1996), pp. 3137–3153.

[4253] Masashi Hayakawa and Toichiro Kinoshita. "Comment on the sign of the pseudoscalar pole contribution to the muon g-2". In: (Dec. 2001).

[4254] Johan Bijnens, Elisabetta Pallante, and Joaquim Prades. "Analysis of the hadronic light by light contributions to the muon g-2". In: *Nucl. Phys.* B474 (1996), pp. 379–420.

[4255] Johan Bijnens, Elisabetta Pallante, and Joaquim Prades. "Comment on the pion pole part of the

light by light contribution to the muon g-2". In: *Nucl. Phys.* B626 (2002), pp. 410–411.

[4256] Joaquim Prades, Eduardo de Rafael, and Arkady Vainshtein. "The Hadronic Light-by-Light Scattering Contribution to the Muon and Electron Anomalous Magnetic Moments". In: *Adv. Ser. Direct. High Energy Phys.* 20 (2009), pp. 303–317.

[4257] Fred Jegerlehner. "The role of mesons in muon $g-2$". In: *EPJ Web Conf.* 199 (2019), p. 01010.

[4258] Gilberto Colangelo et al. "Dispersion relation for hadronic light-by-light scattering: theoretical foundations". In: *JHEP* 09 (2015), p. 074.

[4259] Gilberto Colangelo et al. "Dispersive approach to hadronic light-by-light scattering". In: *JHEP* 09 (2014), p. 091.

[4260] Vladyslav Pauk and Marc Vanderhaeghen. "Anomalous magnetic moment of the muon in a dispersive approach". In: *Phys. Rev.* D90.11 (2014), p. 113012.

[4261] Andreas Nyffeler. "Precision of a data-driven estimate of hadronic light-by-light scattering in the muon $g-2$: Pseudoscalar-pole contribution". In: *Phys. Rev.* D94.5 (2016), p. 053006.

[4262] Igor Danilkin, Christoph Florian Redmer, and Marc Vanderhaeghen. "The hadronic light-by-light contribution to the muon's anomalous magnetic moment". In: *Prog. Part. Nucl. Phys.* 107 (2019), pp. 20–68.

[4263] J. Gronberg et al. "Measurements of the meson - photon transition form-factors of light pseudoscalar mesons at large momentum transfer". In: *Phys. Rev. D* 57 (1998), pp. 33–54.

[4264] H. J. Behrend et al. "A Measurement of the pi0, eta and eta-prime electromagnetic form-factors". In: *Z. Phys.* C49 (1991), pp. 401–410.

[4265] Antoine Gérardin, Harvey B. Meyer, and Andreas Nyffeler. "Lattice calculation of the pion transition form factor with $N_f = 2+1$ Wilson quarks". In: *Phys. Rev.* D100.3 (2019), p. 034520.

[4266] Pere Masjuan and Pablo Sanchez-Puertas. "Pseudoscalar-pole contribution to the $(g_\mu - 2)$: a rational approach". In: *Phys. Rev. D* 95.5 (2017), p. 054026.

[4267] Martin Hoferichter et al. "Dispersion relation for hadronic light-by-light scattering: pion pole". In: *JHEP* 10 (2018), p. 141.

[4268] Martin Hoferichter et al. "Pion-pole contribution to hadronic light-by-light scattering in the anomalous magnetic moment of the muon". In: *Phys. Rev. Lett.* 121.11 (2018), p. 112002.

[4269] Kirill Melnikov and Arkady Vainshtein. "Hadronic light-by-light scattering contribution to the muon anomalous magnetic moment revisited". In: *Phys. Rev.* D70 (2004), p. 113006.

[4270] Gilberto Colangelo et al. "Dispersion relation for hadronic light-by-light scattering: two-pion contributions". In: *JHEP* 04 (2017), p. 161.

[4271] Johan Bijnens, Nils Hermansson-Truedsson, and Antonio Rodríguez-Sánchez. "Short-distance constraints for the HLbL contribution to the muon anomalous magnetic moment". In: *Phys. Lett. B* 798 (2019), p. 134994.

[4272] Gilberto Colangelo et al. "Longitudinal short-distance constraints for the hadronic light-by-light contribution to $(g-2)_\mu$ with large-$N_c$ Regge models". In: *JHEP* 03 (2020), p. 101.

[4273] Vladyslav Pauk and Marc Vanderhaeghen. "Single meson contributions to the muon's anomalous magnetic moment". In: *Eur. Phys. J.* C74.8 (2014), p. 3008.

[4274] Igor Danilkin and Marc Vanderhaeghen. "Light-by-light scattering sum rules in light of new data". In: *Phys. Rev. D* 95.1 (2017), p. 014019.

[4275] M. Knecht et al. "Scalar meson contributions to a $\mu$ from hadronic light-by-light scattering". In: *Phys. Lett. B* 787 (2018), pp. 111–123.

[4276] Gernot Eichmann, Christian S. Fischer, and Richard Williams. "Kaon-box contribution to the anomalous magnetic moment of the muon". In: *Phys. Rev. D* 101.5 (2020), p. 054015.

[4277] Pablo Roig and Pablo Sanchez-Puertas. "Axial-vector exchange contribution to the hadronic light-by-light piece of the muon anomalous magnetic moment". In: *Phys. Rev. D* 101.7 (2020), p. 074019.

[4278] Masashi Hayakawa et al. "Hadronic light-by-light scattering contribution to the muon $g-2$ from lattice QCD: Methodology". In: *PoS* LAT2005 (2006), p. 353.

[4279] T. Blum, M. Hayakawa, and T. Izubuchi. "Hadronic corrections to the muon anomalous magnetic moment from lattice QCD". In: *PoS* LATTICE2012 (2012). Ed. by Derek Leinweber et al., p. 022.

[4280] Thomas Blum et al. "Hadronic light-by-light scattering contribution to the muon anomalous magnetic moment from lattice QCD". In: *Phys. Rev. Lett.* 114.1 (2015), p. 012001.

[4281] Thomas Blum et al. "Lattice Calculation of Hadronic Light-by-Light Contribution to the Muon Anomalous Magnetic Moment". In: *Phys. Rev.* D93.1 (2016), p. 014503.

[4282] Thomas Blum et al. "Connected and Leading Disconnected Hadronic Light-by-Light Contribution to the Muon Anomalous Magnetic Mo-

ment with a Physical Pion Mass". In: *Phys. Rev. Lett.* 118.2 (2017), p. 022005.

[4283] Thomas Blum et al. "The hadronic light-by-light scattering contribution to the muon anomalous magnetic moment from lattice QCD". In: *Phys. Rev. Lett.* 124.13 (2020), p. 132002.

[4284] Jeremy Green et al. "Direct calculation of hadronic light-by-light scattering". In: *PoS* LATTICE2015 (2016), p. 109.

[4285] Nils Asmussen et al. "Position-space approach to hadronic light-by-light scattering in the muon $g-2$ on the lattice". In: *PoS* LATTICE2016 (2016), p. 164.

[4286] Nils Asmussen et al. "Exploratory studies for the position-space approach to hadronic light-by-light scattering in the muon $g-2$". In: *EPJ Web Conf.* 175 (2018), p. 06023.

[4287] Thomas Blum et al. "Using infinite volume, continuum QED and lattice QCD for the hadronic light-by-light contribution to the muon anomalous magnetic moment". In: *Phys. Rev.* D96.3 (2017), p. 034515.

[4288] En-Hung Chao et al. "Hadronic light-by-light contribution to $(g-2)_\mu$ from lattice QCD with SU(3) flavor symmetry". In: *Eur. Phys. J. C* 80.9 (2020), p. 869.

[4289] En-Hung Chao et al. "Hadronic light-by-light contribution to $(g-2)_\mu$ from lattice QCD: a complete calculation". In: *Eur. Phys. J. C* 81.7 (2021), p. 651.

[4290] En-Hung Chao et al. "The charm-quark contribution to light-by-light scattering in the muon $(g-2)$ from lattice QCD". In: *Eur. Phys. J. C* 82.8 (2022), p. 664.

[4291] Andreas Nyffeler. "Hadronic light-by-light scattering in the muon g-2: A New short-distance constraint on pion-exchange". In: *Phys. Rev.* D79 (2009), p. 073012.

[4292] S. Actis et al. "Quest for precision in hadronic cross sections at low energy: Monte Carlo tools vs. experimental data". In: *Eur. Phys. J. C* 66 (2010), pp. 585–686.

[4293] S. Binner, Johann H. Kuhn, and K. Melnikov. "Measuring sigma(e+ e- —> hadrons) using tagged photon". In: *Phys. Lett. B* 459 (1999), pp. 279–287.

[4294] German Rodrigo et al. "Radiative return at NLO and the measurement of the hadronic cross-section in electron positron annihilation". In: *Eur. Phys. J. C* 24 (2002), pp. 71–82.

[4295] Johann H. Kuhn and German Rodrigo. "The Radiative return at small angles: Virtual corrections". In: *Eur. Phys. J. C* 25 (2002), pp. 215–222.

[4296] Henryk Czyz et al. "The Radiative return at phi and B factories: Small angle photon emission at next-to-leading order". In: *Eur. Phys. J. C* 27 (2003), pp. 563–575.

[4297] Henryk Czyz et al. "The Radiative return at Phi and B factories: FSR at next-to-leading order". In: *Eur. Phys. J. C* 33 (2004), pp. 333–347.

[4298] Henryk Czyz et al. "Nucleon form-factors, B meson factories and the radiative return". In: *Eur. Phys. J. C* 35 (2004), pp. 527–536.

[4299] Henryk Czyz et al. "The Radiative return at phi and B factories: FSR for muon pair production at next-to-leading order". In: *Eur. Phys. J. C* 39 (2005), pp. 411–420.

[4300] Henryk Czyz, Agnieszka Grzelinska, and Johann H. Kuhn. "Charge asymmetry and radiative phi decays". In: *Phys. Lett. B* 611 (2005), pp. 116–122.

[4301] Henryk Czyz et al. "Electron-positron annihilation into three pions and the radiative return". In: *Eur. Phys. J. C* 47 (2006), pp. 617–624.

[4302] Henryk Czyz, Agnieszka Grzelinska, and Johann H. Kuhn. "Spin asymmetries and correlations in lambda-pair production through the radiative return method". In: *Phys. Rev. D* 75 (2007), p. 074026.

[4303] Henryk Czyz and Johann H. Kuhn. "Strong and Electromagnetic J/psi and psi(2S) Decays into Pion and Kaon Pairs". In: *Phys. Rev. D* 80 (2009), p. 034035.

[4304] Henryk Czyz, Agnieszka Grzelinska, and Johann H. Kuhn. "Narrow resonances studies with the radiative return method". In: *Phys. Rev. D* 81 (2010), p. 094014.

[4305] Henryk Czyz, Johann H. Kuhn, and Agnieszka Wapienik. "Four-pion production in tau decays and e+e- annihilation: An Update". In: *Phys. Rev. D* 77 (2008), p. 114005.

[4306] H. Czyż, M. Gunia, and J. H. Kühn. "Simulation of electron-positron annihilation into hadrons with the event generator PHOKHARA". In: *JHEP* 08 (2013), p. 110.

[4307] F. Campanario et al. "Complete QED NLO contributions to the reaction $e^+e^- \to \mu^+\mu^-\gamma$ and their implementation in the event generator PHOKHARA". In: *JHEP* 02 (2014), p. 114.

[4308] Henryk Czyż, Johann H. Kühn, and Szymon Tracz. "Nucleon form factors and final state radiative corrections to $e^+e^- \to p\bar{p}$?". In: *Phys. Rev. D* 90.11 (2014), p. 114021.

[4309] Henryk Czyż, Patrycja Kisza, and Szymon Tracz. "Modeling interactions of photons with pseudoscalar and vector mesons". In: *Phys. Rev. D* 97.1 (2018), p. 016006.

[4310] Francisco Campanario et al. "Standard model radiative corrections in the pion form factor measurements do not explain the $a_\mu$ anomaly". In: *Phys. Rev. D* 100.7 (2019), p. 076004.

[4311] Giovanni Balossini et al. "Matching perturbative and parton shower corrections to Bhabha process at flavour factories". In: *Nucl. Phys. B* 758 (2006), pp. 227–253.

[4312] Henryk Czyz and Patrycja Kisza. "EKHARA 3.0: an update of the EKHARA Monte Carlo event generator". In: *Comput. Phys. Commun.* 234 (2019), pp. 245–255.

[4313] A. B. Arbuzov et al. "Monte-Carlo generator for e+e- annihilation into lepton and hadron pairs with precise radiative corrections". In: *Eur. Phys. J. C* 46 (2006), pp. 689–703.

[4314] V. P. Druzhinin, L. V. Kardapoltsev, and V. A. Tayursky. "GGRESRC: A Monte Carlo generator for the two-photon process $e^+e^- \to e^+e^- R(J^{PC} = 0^{-+})$ in the single-tag mode". In: *Comput. Phys. Commun.* 185 (2014), pp. 236–243.

[4315] Elisabetta Barberio and Zbigniew Was. "PHOTOS: A Universal Monte Carlo for QED radiative corrections. Version 2.0". In: *Comput. Phys. Commun.* 79 (1994), pp. 291–308.

[4316] S. Jadach, B. F. L. Ward, and Z. Was. "The Precision Monte Carlo event generator K K for two fermion final states in e+ e- collisions". In: *Comput. Phys. Commun.* 130 (2000), pp. 260–325.

[4317] C. M. Carloni Calame et al. "A new approach to evaluate the leading hadronic corrections to the muon $g$-2". In: *Phys. Lett. B* 746 (2015), pp. 325–329.

[4318] G. Abbiendi et al. "Measuring the leading hadronic contribution to the muon g-2 via $\mu e$ scattering". In: *Eur. Phys. J. C* 77.3 (2017), p. 139.

[4319] V. D. Burkert et al. "The CLAS12 Spectrometer at Jefferson Laboratory". In: *Nucl. Instrum. Meth. A* 959 (2020), p. 163419.

[4320] J. Arrington et al. "Physics with CEBAF at 12 GeV and Future Opportunities". In: (Nov. 2021).

[4321] Issam A. Qattan. "Precision Rosenbluth Measurement of the Proton Elastic Electromagnetic Form Factors and Their Ratio at $Q^2 = 2.64$-GeV$^2$, 3.20-GeV$^2$ and 4.10-GeV$^2$". Other thesis. Dec. 2005.

[4322] E. Cisbani et al. *Jefferson Lab Experiment E12-07-109*. 2007.

[4323] T. Averett et al. *Jefferson Lab Experiment E12-09-016*. 2009.

[4324] D. Hamilton, B. Quinn, and B. Wojtsekhowski. *Jefferson Lab Experiment E12-09-019*. 2009.

[4325] B.D. Anderson et al. *Jefferson Lab Experiment E12-11-009*. 2011.

[4326] V. Bellini et al. *Jefferson Lab Experiment E12-17-004*. 2017.

[4327] S. Alsalmi, E. Fuchey, and B. Wojtsekhowski. *Jefferson Lab Experiment E12-20-012*. 2020.

[4328] G. Gilfoyle et al. *Jefferson Lab Experiment E12-07-104*. 2007.

[4329] S. Kuhn et al. *The Longitudinal Spin Structure of the Nucleon*. 2006.

[4330] A. Gasparian et al. "PRad-II: A New Upgraded High Precision Measurement of the Proton Charge Radius". In: (Sept. 2020).

[4331] D. Abrams et al. "Measurement of the Nucleon $F_2^n/F_2^p$ Structure Function Ratio by the Jefferson Lab MARATHON Tritium/Helium-3 Deep Inelastic Scattering Experiment". In: *Phys. Rev. Lett.* 128.13 (2022), p. 132003.

[4332] S. Bultmann et al. *The Structure of the Free Neutron at Large x-Bjorken*. 2006.

[4333] I.M. Niculescu, S.P. Malace, and C. Keppel. *Jefferson Lab Experiment E12-10-002*. 2010.

[4334] A. Bodek et al. "Experimental Studies of the Neutron and Proton Electromagnetic Structure Functions". In: *Phys. Rev. D* 20 (1979), pp. 1471–1552.

[4335] P. Souder et al. *Precision Measurement of Parity-violation in Deep Inelastic Scattering Over a Broad Kinematic Range*. 2010.

[4336] *SoLID Updated Preliminary Conceptual Design Report*. 2019.

[4337] Alexandre Deur. "Nucleon spin structure measurements at Jefferson Lab". In: *13th Conference on the Intersections of Particle and Nuclear Physics*. Oct. 2018.

[4338] X. Zheng et al. *Measurement of Neutron Spin Asymmetry $A_1^n$ in the Valence Quark Region Using an 11 GeV Beam and a Polarized $^3$He Target in Hall C*. 2006.

[4339] J. H. Chen et al. *Upgraded Polarized Helium-3 Target and Its Performance in Experiments at Jefferson Lab*. 2021.

[4340] Rossi P. Anselmino M. Guidal M. "Topical Issue on the 3-D Structure of the Nucleon". In: *Eur. Phys. J. A* 52 (2016), p. 150.

[4341] Harut Avakian, Bakur Parsamyan, and Alexey Prokudin. "Spin orbit correlations and the struc-

ture of the nucleon". In: *Riv. Nuovo Cim.* 42.1 (2019), pp. 1–48.

[4342] H. Avakian et al. "Measurement of Single and Double Spin Asymmetries in Deep Inelastic Pion Electroproduction with a Longitudinally Polarized Target". In: *Phys. Rev. Lett.* 105 (2010), p. 262002.

[4343] J. Huang et al. "Beam-Target Double Spin Asymmetry $A_{LT}$ in Charged Pion Production from Deep Inelastic Scattering on a Transversely Polarized $^3$He Target at $1.4 < Q^2 < 2.7$ GeV$^2$". In: *Phys. Rev. Lett.* 108 (2012), p. 052001.

[4344] Y. Zhang et al. "Measurement of pretzelosity asymmetry of charged pion production in Semi-Inclusive Deep Inelastic Scattering on a polarized $^3$He target". In: *Phys. Rev. C* 90.5 (2014), p. 055209.

[4345] H. Avakian et al. "Measurement of beam-spin asymmetries for $\pi^+$ electroproduction above the baryon resonance region". In: *Phys. Rev. D* 69 (2004), p. 112004.

[4346] Y. X. Zhao et al. "Double Spin Asymmetries of Inclusive Hadron Electroproductions from a Transversely Polarized $^3$He Target". In: *Phys. Rev. C* 92.1 (2015), p. 015207.

[4347] G. Gates et al. *E12-09-018: Measurement of the Semi-Inclusive $\pi$ and K electro-production in DIS regime from transversely polarized 3He target with the SBS & BB spectrometers in Hall A.* 2009.

[4348] H. Gao et al. *Target Single Spin Asymmetry in Semi-Inclusive Deep-Inelastic (e, e$\pi^\pm$) Reaction on a Transversely Polarized Proton Target.* 2011.

[4349] H. Gao et al. *Target Single Spin Asymmetry in Semi-Inclusive Deep-Inelastic Electro Pion Production on a Trasversely Polarized $^3$He Target at 8.8 and 11 GeV.* 2010.

[4350] J. P. Chen et al. *Asymmetries in Semi-Inclusive Deep-Inelastic (e,e$'\pi^\pm$) Reactions on a Longitudinally Polarized $^3$He Target at 8.8 and 11 GeV.* 2011.

[4351] H. Avakian et al. *E12-06-112: Probing the Proton's Quark Dynamics in Semi-Inclusive Pion Production at 11 GeV.* 2006.

[4352] H. Avakian et al. *E12-07-107: Studies of Spin-Orbit Correlations with Longitudinally Polarized Target.* 2007.

[4353] H. Avakian et al. *E12-09-008: Boer-Mulders asymmetry in K SIDIS w/ H and D targets.* 2009.

[4354] H. Avakian et al. *E12-09-009: Spin-Orbit correlations in K production w/ pol. targets.* 2009.

[4355] H Avakian et al. *C12-20-002:A program of spin-dependent electron scattering from a polarized $^3$He target in CLAS12.* 2020.

[4356] R Ent and H Mkrtchyan. *E12-06-104: Measurement of the Ratio $R = \sigma_L/\sigma_T$ in Semi-Inclusive Deep-Inelastic Scattering.* 2006.

[4357] R Ent, P. Bosted, and H Mkrtchyan. *E12-09-017:Transverse Momentum Dependence of Semi-Inclusive Pion Production.* 2009.

[4358] R. Ent et al. *E12-13-007:Measurement of Semi-Inclusive $\pi^0$ Production as Validation of Factorization.* 2014.

[4359] S. Diehl et al. "Multidimensional, High Precision Measurements of Beam Single Spin Asymmetries in Semi-inclusive $\pi^+$ Electroproduction off Protons in the Valence Region". In: *Phys. Rev. Lett.* 128.6 (2022), p. 062005.

[4360] T. B. Hayward et al. "Observation of Beam Spin Asymmetries in the Process $ep \to e^{'}\pi^+\pi^-X$ with CLAS12". In: *Phys. Rev. Lett.* 126 (2021), p. 152501.

[4361] H. Avakian et al. "First observation of correlations between spin and transverse momenta in back-to-back dihadron production at CLAS12". In: (Aug. 2022).

[4362] M. Mirazita et al. "Beam Spin Asymmetry in Semi-Inclusive Electroproduction of Hadron Pairs". In: *Phys. Rev. Lett.* 126.6 (2021), p. 062002.

[4363] M. Anselmino, V. Barone, and A. Kotzinian. "Double hadron lepto-production in the current and target fragmentation regions". In: *Physics Letters B* 706.1 (2011), pp. 46–52.

[4364] M. V. Polyakov. "Generalized parton distributions and strong forces inside nucleons and nuclei". In: *Phys. Lett. B* 555 (2003), pp. 57–62.

[4365] Krešimir Kumerički. "Measurability of pressure inside the proton". In: *Nature* 570.7759 (2019), E1–E2.

[4366] H. Dutrieux et al. "Phenomenological assessment of proton mechanical properties from deeply virtual Compton scattering". In: *Eur. Phys. J. C* 81.4 (2021), p. 300.

[4367] S. Stepanyan et al. "Observation of exclusive deeply virtual Compton scattering in polarized electron beam asymmetry measurements". In: *Phys. Rev. Lett.* 87 (2001), p. 182002.

[4368] Nicole d'Hose, Silvia Niccolai, and Armine Rostomyan. "Experimental overview of Deeply Virtual Compton Scattering". In: *Eur. Phys. J. A* 52.6 (2016), p. 151.

[4369] C. Muñoz Camacho et al. "Scaling tests of the cross-section for deeply virtual compton scattering". In: *Phys. Rev. Lett.* 97 (2006), p. 262002.

[4370] M. Defurne et al. “E00-110 experiment at Jefferson Lab Hall A: Deeply virtual Compton scattering off the proton at 6 GeV”. In: *Phys. Rev. C* 92.5 (2015), p. 055202.

[4371] F. Georges et al. “Deeply Virtual Compton Scattering Cross Section at High Bjorken xB”. In: *Phys. Rev. Lett.* 128.25 (2022), p. 252002.

[4372] C. Munoz Camacho et al. *PR12-13-010: Exclusive Deeply Virtual Compton and Neutral Pion Cross-Section Measurements in Hall C.* 2013.

[4373] L. Elouadrhiri et al. *E12-06-119:Deeply Virtual Compton Scattering with CLAS12 at 11 GeV.* 2006.

[4374] L. Elouadrhiri and F.-X. Girod. *E12-16-010B:Deeply Virtual Compton Scattering with CLAS12 at 6.6 GeV and 8.8 GeV.* 2016.

[4375] L. Elouadrhiri et al. *E12-12-010:Deeply Virtual Compton Scattering at 11 GeV with transversely polarized target using the CLAS12 Detector.* 2012.

[4376] M. Dlamini et al. “Deep Exclusive Electroproduction of $\pi 0$ at High Q2 in the Quark Valence Regime”. In: *Phys. Rev. Lett.* 127.15 (2021), p. 152301.

[4377] P. Stoler et al. *E12-06-108:Hard Exclusive Electroproduction of $\pi^0$ and $\eta$ with CLAS12.* 2006.

[4378] P. Stoler et al. *E12-12-007:Exclusive Phi Meson Electroproduction with CLAS12.* 2012.

[4379] T. Horn, G.M. Huber, and P. Markowitz. *E12-09-011: Studies of the L-T Separated Kaon Electroproduction Cross Section from 5-11 GeV.* 2009.

[4380] T. Horn et al. *E12-07-105: Scaling Study of the L-T Separated Pion Electroproduction Cross Section at 11 GeV.* 2007.

[4381] Z. Meziani et al. *E12-17-012: Partonic Structure of Light Nuclei.* 2017.

[4382] M. Vanderhaeghen, Pierre A. M. Guichon, and M. Guidal. “Hard electroproduction of photons and mesons on the nucleon”. In: *Phys. Rev. Lett.* 80 (1998), pp. 5064–5067.

[4383] S. V. Goloskokov and P. Kroll. “Vector meson electroproduction at small Bjorken-x and generalized parton distributions”. In: *Eur. Phys. J. C* 42 (2005), pp. 281–301.

[4384] A. Rodas et al. “Determination of the pole position of the lightest hybrid meson candidate”. In: *Phys. Rev. Lett.* 122.4 (2019), p. 042002.

[4385] C. Meyer et al. *E12-06-102:Mapping the Spectrum of Light Quark Mesons and Gluonic Excitations with Linearly Polarized Photons.* 2006.

[4386] S. Adhikari et al. “The GLUEX beamline and detector”. In: *Nucl. Instrum. Meth. A* 987 (2021), p. 164807.

[4387] M. Battaglieri et al. *E12-11-005:Meson Spectroscopy with low $Q^2$ electron scattering in CLAS12.* 2011.

[4388] S. Adhikari et al. “Measurement of beam asymmetry for $\pi^- \Delta^{++}$ photoproduction on the proton at $E_\gamma$=8.5 GeV”. In: *Phys. Rev. C* 103.2 (2021), p. L022201.

[4389] S. Adhikari et al. “Beam Asymmetry $\Sigma$ for the Photoproduction of $\eta$ and $\eta'$ Mesons at $E_\gamma = 8.8$ GeV”. In: *Phys. Rev. C* 100.5 (2019), p. 052201.

[4390] H. Al Ghoul et al. “Measurement of the beam asymmetry $\Sigma$ for $\pi^0$ and $\eta$ photoproduction on the proton at $E_\gamma = 9$ GeV”. In: *Phys. Rev. C* 95.4 (2017), p. 042201.

[4391] S. Adhikari et al. “Measurement of the photon beam asymmetry in $\vec{\gamma} p \to K^+ \Sigma^0$ at $E_\gamma = 8.5$ GeV”. In: *Phys. Rev. C* 101.6 (2020), p. 065206.

[4392] S. Adhikari et al. “Measurement of spin density matrix elements in $\Lambda(1520)$ photoproduction at 8.2–8.8 GeV”. In: *Phys. Rev. C* 105.3 (2022), p. 035201.

[4393] Alexander Austregesilo. “Spin-density matrix elements for vector meson photoproduction at GlueX”. In: *AIP Conf. Proc.* 2249.1 (2020). Ed. by Curtis Meye and Reinhard A. Schumacher, p. 030005.

[4394] Daniel S. Carman. “Excited nucleon spectrum and structure studies with CLAS and CLAS12”. In: *AIP Conf. Proc.* 2249.1 (2020). Ed. by Curtis Meye and Reinhard A. Schumacher, p. 030004.

[4395] R.W. Gothe et al. *E12-09-003:Nucleon Resonance Studies with CLAS12.* 2009.

[4396] D.S. Carman, R.W. Gothe, and V.I. Mokeev. *E12-06-108A:Exclusive $N^* \to KY$ Studies with CLAS12.* 2006.

[4397] M. Dugger et al. “A study of decays to strange final states with GlueX in Hall D using components of the BaBar DIRC”. In: *arXiv:1408.0215* (2014).

[4398] L. Guo et al. *E12-12-008:Photoproduction of the Very Strangest Baryons on a Proton Target in CLAS12.* 2012.

[4399] Moskov Amaryan et al. “Strange Hadron Spectroscopy with Secondary $K_L$ Beam in Hall D”. In: *arXiv:2008.08215* (2020).

[4400] Qian Wang, Xiao-Hai Liu, and Qiang Zhao. “Photoproduction of hidden charm pentaquark states $P_c^+(4380)$ and $P_c^+(4450)$”. In: *Phys. Rev. D* 92 (2015), p. 034022.

[4401] V. Kubarovsky and M. B. Voloshin. “Formation of hidden-charm pentaquarks in photon-

nucleon collisions". In: *Phys. Rev. D* 92.3 (2015), p. 031502.

[4402] Marek Karliner and Jonathan L. Rosner. "Photoproduction of Exotic Baryon Resonances". In: *Phys. Lett. B* 752 (2016), pp. 329–332.

[4403] A. N. Hiller Blin et al. "Studying the $P_c(4450)$ resonance in $J/\psi$ photoproduction off protons". In: *Phys. Rev. D* 94.3 (2016), p. 034002.

[4404] A. Ali et al. "First Measurement of Near-Threshold $J/\psi$ Exclusive Photoproduction off the Proton". In: *Phys. Rev. Lett.* 123.7 (2019), p. 072001.

[4405] Sylvester Joosten. "Quarkonium production near threshold at JLab and EIC". In: *9th Workshop of the APS Topical Group on Hadronic Physics* (2021).

[4406] Mattaglieri et al. *E12-12-001A:Near threshold $J/\psi$ photoproduction and study of LHCb pentaquarks with CLAS12.* 2017.

[4407] Z-E. Meziani et al. *E12-12-006:Near Threshold Electroproduction of $J/\psi$ at 11 GeV.* 2012.

[4408] O. Hen et al. "Nucleon-Nucleon Correlations, Short-lived Excitations, and the Quarks Within". In: *Rev. Mod. Phys.* 89.4 (2017), p. 045002.

[4409] Leonid Frankfurt, Misak Sargsian, and Mark Strikman. "Recent observation of short range nucleon correlations in nuclei and their implications for the structure of nuclei and neutron stars". In: *Int. J. Mod. Phys. A* 23 (2008), pp. 2991–3055.

[4410] L. L. Frankfurt et al. "Evidence for short range correlations from high $Q^2(e,e')$ reactions". In: *Phys. Rev. C* 48 (1993), pp. 2451–2461.

[4411] K. S. Egiyan et al. "Observation of nuclear scaling in the $A(e,e')$ reaction at $x_B > 1$". In: *Phys. Rev. C* 68 (2003), p. 014313.

[4412] K. S. Egiyan et al. "Measurement of 2- and 3-nucleon short range correlation probabilities in nuclei". In: *Phys. Rev. Lett.* 96 (2006), p. 082501.

[4413] J. Arrington et al. "$x$- and $\xi$-scaling of the nuclear structure function at large $x$". In: *Phys. Rev. C* 64 (2001), p. 014602.

[4414] J. Seely et al. "New measurements of the EMC effect in very light nuclei". In: *Phys. Rev. Lett.* 103 (2009), p. 202301.

[4415] J. Arrington et al. "Measurement of the EMC effect in light and heavy nuclei". In: *Phys. Rev. C* 104.6 (2021), p. 065203.

[4416] J. Gomez et al. "Measurement of the A-dependence of deep inelastic electron scattering". In: *Phys. Rev. D* 49 (1994), pp. 4348–4372.

[4417] I. Korover et al. "Probing the Repulsive Core of the Nucleon-Nucleon Interaction via the $^4$He(e,e′pN) Triple-Coincidence Reaction". In: *Phys. Rev. Lett.* 113.2 (2014), p. 022501.

[4418] M. Duer et al. "Direct Observation of Proton-Neutron Short-Range Correlation Dominance in Heavy Nuclei". In: *Phys. Rev. Lett.* 122.17 (2019), p. 172502.

[4419] M. Duer et al. "Probing high-momentum protons and neutrons in neutron-rich nuclei". In: *Nature* 560.7720 (2018), pp. 617–621.

[4420] D. Nguyen et al. "Novel observation of isospin structure of short-range correlations in calcium isotopes". In: *Phys. Rev. C* 102.6 (2020), p. 064004.

[4421] S. Li et al. "Revealing the short-range structure of the mirror nuclei $^3$H and $^3$He". In: *Nature* 609.7925 (2022), pp. 41–45.

[4422] Z. Ye et al. "Search for three-nucleon short-range correlations in light nuclei". In: *Phys. Rev. C* 97.6 (2018), p. 065204.

[4423] J Arrington et al. *E12-06-105:Inclusive Scattering from Nuclei at $x > 1$ in the quasielastic and deeply inelastic regimes.* 2006.

[4424] I. C. Cloët, Wolfgang Bentz, and Anthony William Thomas. "EMC and polarized EMC effects in nuclei". In: *Phys. Lett.* B642 (2006), pp. 210–217.

[4425] Stephen Tronchin, Hrayr H. Matevosyan, and Anthony W. Thomas. "Polarized EMC Effect in the QMC Model". In: *Phys. Lett. B* 783 (2018), pp. 247–252.

[4426] W. Brooks and S. Kuhn. *The EMC Effect in Spin Structure Functions.* 2012.

[4427] R. Dupre et al. *PR12-16-011: Nuclear Exclusive and Semi-inclusive Measurements with a New CLAS12 Low Energy Recoil Tracker.* 2016.

[4428] S. Moran et al. "Measurement of charged-pion production in deep-inelastic scattering off nuclei with the CLAS detector". In: *Phys. Rev. C* 105.1 (2022), p. 015201.

[4429] X. Qian et al. "Experimental study of the $A(e,e'\pi^+)$ Reaction on $^1$H, $^2$H, $^{12}$C, $^{27}$Al, $^{63}$Cu and $^{197}$Au". In: *Phys. Rev. C* 81 (2010), p. 055209.

[4430] L. El Fassi et al. "Evidence for the onset of color transparency in $\rho^0$ electroproduction off nuclei". In: *Phys. Lett. B* 712 (2012), pp. 326–330.

[4431] L. Gu et al. "Measurement of the Ar(e,e′p) and Ti(e,e′p) cross sections in Jefferson Lab Hall A". In: *Phys. Rev. C* 103.3 (2021), p. 034604.

[4432] M. Khachatryan et al. "Electron-beam energy reconstruction for neutrino oscillation measurements". In: *Nature* 599 (2021), pp. 565–570.

[4433] O. Hen et al. *E12-17-006: Electrons for Neutrinos: Addressing Critical Neutrino-Nucleus Issues.* 2017.

[4434] D. Adhikari et al. "Accurate Determination of the Neutron Skin Thickness of $^{208}$Pb through Parity-Violation in Electron Scattering". In: *Phys. Rev. Lett.* 126.17 (2021), p. 172502.

[4435] D. Adhikari et al. "Precision Determination of the Neutral Weak Form Factor of Ca48". In: *Phys. Rev. Lett.* 129.4 (2022), p. 042501.

[4436] G. Hagen et al. "Neutron and weak-charge distributions of the $^{48}$Ca nucleus". In: *Nature Phys.* 12.2 (2015), pp. 186–190.

[4437] N. Alemanos et al. "Topical Issue on an experimental program with positron beams at Jefferson Lab". In: *Eur. Phys. J. A* 58 (2022).

[4438] Ferdinand Willeke and J. Beebe-Wang. *Electron Ion Collider Conceptual Design Report 2021.*

[4439] A. Aprahamian et al. *Reaching for the horizon: The 2015 long range plan for nuclear science.* DOE/NSF Nuclear Science Advisory Panel Report. 2015.

[4440] National Academies of Sciences, Engineering, and Medicine. *An Assessment of U.S.-Based Electron-Ion Collider Science.* Washington, DC: The National Academies Press, 2018.

[4441] Felix Hekhorn and Marco Stratmann. "Next-to-Leading Order QCD Corrections to Inclusive Heavy-Flavor Production in Polarized Deep-Inelastic Scattering". In: *Phys. Rev. D* 98.1 (2018), p. 014018.

[4442] Kresimir Kumericki, Simonetta Liuti, and Herve Moutarde. "GPD phenomenology and DVCS fitting: Entering the high-precision era". In: *Eur. Phys. J. A* 52.6 (2016), p. 157.

[4443] Cédric Lorcé, Hervé Moutarde, and Arkadiusz P. Trawiński. "Revisiting the mechanical properties of the nucleon". In: *Eur. Phys. J. C* 79.1 (2019), p. 89.

[4444] Cédric Lorcé. "On the hadron mass decomposition". In: *Eur. Phys. J. C* 78.2 (2018), p. 120.

[4445] Yoshitaka Hatta, Abha Rajan, and Kazuhiro Tanaka. "Quark and gluon contributions to the QCD trace anomaly". In: *JHEP* 12 (2018), p. 008.

[4446] Andreas Metz, Barbara Pasquini, and Simone Rodini. "Revisiting the proton mass decomposition". In: *Phys. Rev. D* 102.11 (2021), p. 114042.

[4447] D. Kharzeev. "Quarkonium interactions in QCD". In: *Proc. Int. Sch. Phys. Fermi* 130 (1996). Ed. by A. Di Giacomo and Dmitri Diakonov, pp. 105–131.

[4448] Yoshitaka Hatta and Di-Lun Yang. "Holographic $J/\psi$ production near threshold and the proton mass problem". In: *Phys. Rev. D* 98.7 (2018), p. 074003.

[4449] Renaud Boussarie and Yoshitaka Hatta. "QCD analysis of near-threshold quarkonium leptoproduction at large photon virtualities". In: *Phys. Rev. D* 101.11 (2020), p. 114004.

[4450] D. Kharzeev et al. "*J/psi* photoproduction and the gluon structure of the nucleon". In: *Eur. Phys. J. C* 9 (1999), pp. 459–462.

[4451] Oleksii Gryniuk and Marc Vanderhaeghen. "Accessing the real part of the forward $J/\psi$-p scattering amplitude from $J/\psi$ photoproduction on protons around threshold". In: *Phys. Rev. D* 94.7 (2016), p. 074001.

[4452] Meng-Lin Du et al. "Deciphering the mechanism of near-threshold $J/\psi$ photoproduction". In: *Eur. Phys. J. C* 80.11 (2020), p. 1053.

[4453] Kiminad A. Mamo and Ismail Zahed. "Diffractive photoproduction of $J/\psi$ and $\Upsilon$ using holographic QCD: gravitational form factors and GPD of gluons in the proton". In: *Phys. Rev. D* 101.8 (2020), p. 086003.

[4454] Oleksii Gryniuk et al. "$\Upsilon$ photoproduction on the proton at the Electron-Ion Collider". In: *Phys. Rev. D* 102.1 (2020), p. 014016.

[4455] S. Joosten and Z. E. Meziani. "Heavy Quarkonium Production at Threshold: from JLab to EIC". In: *PoS* QCDEV2017 (2018), p. 017.

[4456] J. Arrington et al. "Revealing the structure of light pseudoscalar mesons at the electron–ion collider". In: *J. Phys. G* 48.7 (2021), p. 075106.

[4457] Elke-Caroline Aschenauer et al. "Deeply Virtual Compton Scattering at a Proposed High-Luminosity Electron-Ion Collider". In: ().

[4458] Edgar R. Berger et al. "Generalized parton distributions in the deuteron". In: *Phys. Rev. Lett.* 87 (2001), p. 142302.

[4459] V. Guzey and M. Strikman. "DVCS on spinless nuclear targets in impulse approximation". In: *Phys. Rev. C* 68 (2003), p. 015204.

[4460] A. Kirchner and Dieter Mueller. "Deeply virtual Compton scattering off nuclei". In: *Eur. Phys. J. C* 32 (2003), pp. 347–375.

[4461] S. Liuti and S. K. Taneja. "Microscopic description of deeply virtual Compton scattering off spin-0 nuclei". In: *Phys. Rev. C* 72 (2005), p. 032201.

[4462] M. Rinaldi and S. Scopetta. "Neutron orbital structure from generalized parton distributions of 3He". In: *Phys. Rev. C* 85 (2012), p. 062201.

[4463] Sara Fucini, Matteo Rinaldi, and Sergio Scopetta. "Generalized parton distributions of light nuclei". In: *Few Body Syst.* 62.1 (2021), p. 3.

[4464] Sara Fucini, Sergio Scopetta, and Michele Viviani. “Coherent deeply virtual Compton scattering off $^4$He”. In: *Phys. Rev. C* 98.1 (2018), p. 015203.

[4465] V. Guzey et al. “Coherent $J/\psi$ electroproduction on $^4$He and $^3$He at the Electron-Ion Collider: probing nuclear shadowing one nucleon at a time”. In: (Feb. 2022).

[4466] Bamunuvita Gamage et al. “Design Concept for the Second Interaction Region for Electron-Ion Collider”. In: *JACoW* IPAC2021 (2021), TUPAB040.

[4467] Kari J. Eskola et al. “EPPS16: Nuclear parton distributions with LHC data”. In: *Eur. Phys. J. C* 77.3 (2017), p. 163.

[4468] Daniel de Florian et al. “Global Analysis of Nuclear Parton Distributions”. In: *Phys. Rev. D* 85 (2012), p. 074028.

[4469] K. Kovarik et al. “nCTEQ15 - Global analysis of nuclear parton distributions with uncertainties in the CTEQ framework”. In: *Phys. Rev. D* 93.8 (2016), p. 085037.

[4470] Hamzeh Khanpour and S. Atashbar Tehrani. “Global Analysis of Nuclear Parton Distribution Functions and Their Uncertainties at Next-to-Next-to-Leading Order”. In: *Phys. Rev. D* 93.1 (2016), p. 014026.

[4471] E. C. Aschenauer et al. “Nuclear Structure Functions at a Future Electron-Ion Collider”. In: *Phys. Rev. D* 96.11 (2017), p. 114005.

[4472] Tobias Toll and Thomas Ullrich. “Exclusive diffractive processes in electron-ion collisions”. In: *Phys. Rev. C* 87.2 (2013), p. 024913.

[4473] Wan Chang et al. “Investigation of the background in coherent J/$\psi$ production at the EIC”. In: *Phys. Rev. D* 104.11 (2021), p. 114030.

[4474] L. Frankfurt, V. Guzey, and M. Strikman. “Leading Twist Nuclear Shadowing Phenomena in Hard Processes with Nuclei”. In: *Phys. Rept.* 512 (2012), pp. 255–393.

[4475] I. Balitsky. “Operator expansion for high-energy scattering”. In: *Nucl. Phys.* B463 (1996), pp. 99–160.

[4476] Yuri V. Kovchegov. “Small x F(2) structure function of a nucleus including multiple pomeron exchanges”. In: *Phys. Rev.* D60 (1999), p. 034008.

[4477] Jamal Jalilian-Marian et al. “The BFKL equation from the Wilson renormalization group”. In: *Nucl. Phys. B* 504 (1997), pp. 415–431.

[4478] Jamal Jalilian-Marian, Alex Kovner, and Heribert Weigert. “The Wilson renormalization group for low x physics: Gluon evolution at finite parton density”. In: *Phys. Rev. D* 59 (1998), p. 014015.

[4479] Edmond Iancu, Andrei Leonidov, and Larry D. McLerran. “Nonlinear gluon evolution in the color glass condensate. 1.” In: *Nucl. Phys. A* 692 (2001), pp. 583–645.

[4480] Elena Ferreiro et al. “Nonlinear gluon evolution in the color glass condensate. 2.” In: *Nucl. Phys. A* 703 (2002), pp. 489–538.

[4481] Jamal Jalilian-Marian and Yuri V. Kovchegov. “Saturation physics and deuteron-Gold collisions at RHIC”. In: *Prog. Part. Nucl. Phys.* 56 (2006), pp. 104–231.

[4482] Francois Gelis et al. “The Color Glass Condensate”. In: *Ann. Rev. Nucl. Part. Sci.* 60 (2010), pp. 463–489.

[4483] Javier L. Albacete and Cyrille Marquet. “Gluon saturation and initial conditions for relativistic heavy ion collisions”. In: *Prog. Part. Nucl. Phys.* 76 (2014), pp. 1–42.

[4484] Dmitri Kharzeev, Eugene Levin, and Larry McLerran. “Jet azimuthal correlations and parton saturation in the color glass condensate”. In: *Nucl. Phys. A* 748 (2005), pp. 627–640.

[4485] L. Zheng et al. “Probing Gluon Saturation through Dihadron Correlations at an Electron-Ion Collider”. In: *Phys. Rev. D* 89.7 (2014), p. 074037.

[4486] Néstor Armesto et al. “Inclusive diffraction in future electron-proton and electron-ion colliders”. In: *Phys. Rev. D* 100.7 (2019), p. 074022.

[4487] Alexander Jentsch, Zhoudunming Tu, and Christian Weiss. “Deep-inelastic electron-deuteron scattering with spectator nucleon tagging at the future Electron Ion Collider: Extracting free nucleon structure”. In: *Phys. Rev. C* 104.6 (2021), p. 065205.

[4488] L. L. Frankfurt and M. I. Strikman. “High-Energy Phenomena, Short Range Nuclear Structure and QCD”. In: *Phys. Rept.* 76 (1981), pp. 215–347.

[4489] Misak Sargsian and Mark Strikman. “Model independent method for determination of the DIS structure of free neutron”. In: *Phys. Lett.* B639 (2006), pp. 223–231.

[4490] W. Cosyn and C. Weiss. “Polarized electron-deuteron deep-inelastic scattering with spectator nucleon tagging”. In: *Phys. Rev. C* 102 (2020), p. 065204.

[4491] W. Melnitchouk, M. Sargsian, and M. I. Strikman. “Probing the origin of the EMC effect via tagged structure functions of the deuteron”. In: *Z. Phys.* A359 (1997), pp. 99–109.

[4492] Zhoudunming Tu et al. “Probing short-range correlations in the deuteron via incoherent diffrac-

tive J/$\psi$ production with spectator tagging at the EIC". In: *Phys. Lett. B* 811 (2020), p. 135877.
[4493] Andreas Metz and Anselm Vossen. "Parton Fragmentation Functions". In: *Prog. Part. Nucl. Phys.* 91 (2016), pp. 136–202.
[4494] Valerio Bertone et al. "A determination of the fragmentation functions of pions, kaons, and protons with faithful uncertainties". In: *Eur. Phys. J. C* 77.8 (2017), p. 516.
[4495] R. J. Hernández-Pinto et al. "Global extraction of the parton-to-pion fragmentation functions at NLO accuracy in QCD". In: *J. Phys. Conf. Ser.* 761.1 (2016). Ed. by Eduard de la Cruz Burelo, Arturo Fernandez Tellez, and Pablo Roig, p. 012037.
[4496] R. J. Hernández-Pinto et al. "Global extraction of the parton-to-kaon fragmentation functions at NLO in QCD". In: *J. Phys. Conf. Ser.* 912.1 (2017). Ed. by I. Bautista et al., p. 012043.
[4497] Zhong-Bo Kang et al. "Transverse Lambda production at the future Electron-Ion Collider". In: *Phys. Rev. D* 105.9 (2022), p. 094033.
[4498] Alessandro Bacchetta and Marco Radici. "Partial wave analysis of two hadron fragmentation functions". In: *Phys. Rev. D* 67 (2003), p. 094002.
[4499] Stephen Gliske, Alessandro Bacchetta, and Marco Radici. "Production of two hadrons in semi-inclusive deep inelastic scattering". In: *Phys. Rev. D* 90.11 (2014). [Erratum: Phys.Rev.D 91, 019902 (2015)], p. 114027.
[4500] Hrayr H. Matevosyan, Aram Kotzinian, and Anthony W. Thomas. "Accessing Quark Helicity through Dihadron Studies". In: *Phys. Rev. Lett.* 120.25 (2018), p. 252001.
[4501] Elke-Caroline Aschenauer et al. "Jet angularities in photoproduction at the Electron-Ion Collider". In: *Phys. Rev. D* 101.5 (2020), p. 054028.
[4502] Mrinal Dasgupta et al. "Small-radius jets to all orders in QCD". In: *JHEP* 04 (2015), p. 039.
[4503] Darren J. Scott and Wouter J. Waalewijn. "The leading jet transverse momentum in inclusive jet production and with a loose jet veto". In: *JHEP* 03 (2020), p. 159.
[4504] Duff Neill, Felix Ringer, and Nobuo Sato. *Calculating the energy loss of leading jets.* 2020.
[4505] P. Schweitzer, M. Strikman, and C. Weiss. "Intrinsic transverse momentum and parton correlations from dynamical chiral symmetry breaking". In: *JHEP* 01 (2013), p. 163.
[4506] L. Trentadue and G. Veneziano. "Fracture functions: An Improved description of inclusive hard processes in QCD". In: *Phys. Lett. B* 323 (1994), pp. 201–211.
[4507] Federico Alberto Ceccopieri and Davide Mancusi. "QCD analysis of Lambda hyperon production in DIS target-fragmentation region". In: *Eur. Phys. J. C* 73 (2013), p. 2435.
[4508] Hai Tao Li, Ze Long Liu, and Ivan Vitev. "Heavy meson tomography of cold nuclear matter at the electron-ion collider". In: *Phys. Lett. B* 816 (2021), p. 136261.
[4509] Xuan Li. "Heavy flavor and jet studies for the future Electron-Ion Collider". In: *PoS* HardProbes2020 (2021), p. 175.
[4510] Spencer R. Klein and Ya-Ping Xie. "Photoproduction of charged final states in ultraperipheral collisions and electroproduction at an electron-ion collider". In: *Phys. Rev. C* 100.2 (2019), p. 024620.
[4511] M. Albaladejo et al. "XYZ spectroscopy at electron-hadron facilities: Exclusive processes". In: *Phys. Rev. D* 102 (2020), p. 114010.
[4512] *EIC User Group Webpages.*
[4513] J-PARC Center, J-PARC web page: https://j-parc.jp
/researcher/index-e.html; Program Advisory Committee for Nuclear and Particle Physics Experiments: https://j-parc.jp/researcher/Hadron/en/PAC_for_N clPart_e.html.
[4514] H. Ohnishi, F. Sakuma, and T. Takahashi. "Hadron Physics at J-PARC". In: *Progress in Particle and Nuclear Physics* 113 (2020), p. 103773.
[4515] Kazuya Aoki et al. "Extension of the J-PARC Hadron Experimental Facility: Third White Paper". In: (Oct. 2021).
[4516] J. K. Ahn *et al. Studying Generalized Parton Distributions with Exclusive Drell-Yan process at J-PARC.* Letter of Intent (2018), 7th J-PARC PAC meeting, January 16-18, 2019.
[4517] K. Aoki *et al.* (J-PARC-HI collaboration). *Proposal for dielectron measurements in heavy-ion collisions at J-PARC with E16 upgrades, June 14, 2021.* J-PARC PAC: https://j-parc.jp/researcher
/Hadron/en/Proposal_e.html#2107.
[4518] S. Kumano. *Nuclear Physics (in Japanese).* KEK Physics Series, Volume 2, Kyoritsu Shuppan Co., Ltd., June 10, 2015.
[4519] S. N. Nakamura. *Future prospects of spectroscopy of Lambda hypernuclei at JLab and J-PARC HIHR.* 14th International Conference on Hypernuclear and Strange Particle Physics (HYP2022), June 27 - July 1, 2022, Prague, Czech Republic.

[4520] T. O. Yamamoto et al. "Observation of Spin-Dependent Charge Symmetry Breaking in $\Lambda N$ Interaction: Gamma-Ray Spectroscopy of ${}^4_\Lambda$He". In: *Phys. Rev. Lett.* 115 (22 Nov. 2015), p. 222501.

[4521] H. Ekawa et al. "Observation of a Be double-Lambda hypernucleus in the J-PARC E07 experiment". In: *Progress of Theoretical and Experimental Physics* 2019.2 (Feb. 2019). 021D02.

[4522] S. H. Hayakawa et al. "Observation of Coulomb-Assisted Nuclear Bound State of $\Xi^- - -^{14}$N System". In: *Phys. Rev. Lett.* 126 (6 Feb. 2021), p. 062501.

[4523] M. Yoshimoto et al. "First observation of a nuclear s-state of a $\Xi$ hypernucleus, ${}^{15}_{\Xi}$C". In: *Progress of Theoretical and Experimental Physics* 2021.7 (July 2021). 073D02.

[4524] T. Yamaga et al. "Observation of a $\overline{K}NN$ bound state in the ${}^3$He$(K^-, \Lambda p)n$ reaction". In: *Phys. Rev. C* 102 (4 Oct. 2020), p. 044002.

[4525] T. Hashimoto et al. "Measurements of Strong-Interaction Effects in Kaonic-Helium Isotopes at Sub-eV Precision with X-Ray Microcalorimeters". In: *Phys. Rev. Lett.* 128 (11 Mar. 2022), p. 112503.

[4526] Yudai Ichikawa et al. "An event excess observed in the deeply bound region of the 12C ($K^-$, p) missing-mass spectrum". In: *Progress of Theoretical and Experimental Physics* 2020.12 (Dec. 2020). 123D01.

[4527] K. Miwa et al. "Measurement of the differential cross sections of the $\Sigma^- p$ elastic scattering in momentum range 470 to 850 MeV/$c$". In: *Phys. Rev. C* 104 (4 Oct. 2021), p. 045204.

[4528] K. Miwa et al. "Precise Measurement of Differential Cross Sections of the $\Sigma^- p \to \Lambda n$ Reaction in Momentum Range $470--650$ MeV/$c$". In: *Phys. Rev. Lett.* 128 (7 Feb. 2022), p. 072501.

[4529] PARC E40 Collaboration et al. *Measurement of differential cross sections for $\Sigma^+ p$ elastic scattering in the momentum range 0.44-0.80 GeV/c.* 2022.

[4530] S. Yokkaichi *et al. Proposal, Electron pair spectrometer at the J-PARC 50-GeV PS to explore the chiral symmetry in QCD.* http://j-parc.jp /researcher/Hadron/en/pac_0606/pdf /p16-Yokkaichi_2.pdf, http://j-parc.jp /researcher/Hadron/en/pac_1707/pdf /E16_2017-10.pdf, see E16 home page: https://ribf.riken.jp/~yokkaich/E16 /E16-index.html.

[4531] Xiangdong Ji. "QCD Analysis of the Mass Structure of the Nucleon". In: *Phys. Rev. Lett.* 74 (7 Feb. 1995), pp. 1071–1074.

[4532] R. Abdul Khalek *et al. Science Requirements and Detector Concepts for the Electron-Ion Collider: EIC Yellow Report.* 2021.

[4533] Daniele P. Anderle *et al.* "Electron-ion collider in China". In: *Frontiers of Physics* 16.6, 64701 (2021), p. 64701.

[4534] S. Kumano, Qin-Tao Song, and O. V. Teryaev. "Hadron tomography by generalized distribution amplitudes in pion-pair production process $\gamma^*\gamma \to \pi^0\pi^0$ and gravitational form factors for pion". In: *Phys. Rev. D* 97.1 (2018), p. 014020.

[4535] Ryugo S. Hayano and Tetsuo Hatsuda. "Hadron properties in the nuclear medium". In: *Rev. Mod. Phys.* 82 (2010), p. 2949.

[4536] Philipp Gubler and Daisuke Satow. "Recent Progress in QCD Condensate Evaluations and Sum Rules". In: *Prog. Part. Nucl. Phys.* 106 (2019), pp. 1–67.

[4537] M. Naruki et al. "Experimental Signature of Medium Modifications for $\rho$ and $\omega$ Mesons in the 12 GeV $p + A$ Reactions". In: *Phys. Rev. Lett.* 96 (9 Mar. 2006), p. 092301.

[4538] R. Muto et al. "Evidence for In-Medium Modification of the $\phi$ Meson at Normal Nuclear Density". In: *Phys. Rev. Lett.* 98 (4 Jan. 2007), p. 042501.

[4539] Sakiko Ashikaga et al. "Measurement of Vector Meson Mass in Nuclear Matter at J-PARC". In: *Proceedings of the 8th International Conference on Quarks and Nuclear Physics (QNP2018).*

[4540] Y. Nara et al. "Relativistic nuclear collisions at 10$A$ GeV energies from $p+$Be to Au+Au with the hadronic cascade model". In: *Phys. Rev. C* 61 (2 Dec. 1999), p. 024901.

[4541] Philipp Gubler. "The phi meson in nuclear matter in a transport approach". In: *PoS* PANIC2021 (2022), p. 215.

[4542] Kotaro Shirotori et al. "Charmed Baryon Spectroscopy Experiment at J-PARC". In: *Proceedings of the 2nd International Symposium on Science at J-PARC – Unlocking the Mysteries of Life, Matter and the Universe –.*

[4543] L. Heller et al. "Pion - Nucleon Bremsstrahlung and $\Delta$ Electromagnetic Moments". In: *Phys. Rev. C* 35 (1987), p. 718.

[4544] Takahiro Sawada et al. "Accessing proton generalized parton distributions and pion distribution amplitudes with the exclusive pion-induced Drell-Yan process at J-PARC". In: *Phys. Rev. D* 93.11 (2016), p. 114034.

[4545] Y. Morino *et al. Charmed Baryon Spectroscopy via the ($\pi$,$D^{*-}$) reaction.* P50 proposal (2013), 16th J-PARC PAC meeting, January 9-11, 2013,

http://www.j-parc.jp/researcher/Hadron/en/Proposal_e.html, see also https://www.rcnp.osaka-u.ac.jp/∼noumi/puki/E50/.
[4546] B. Pire, K. Semenov-Tian-Shansky, and L. Szymanowski. “Backward charmonium production in $\pi N$ collisions”. In: *Phys. Rev. D* 95.3 (2017), p. 034021.
[4547] S. Kumano, M. Strikman, and K. Sudoh. “Novel two-to-three hard hadronic processes and possible studies of generalized parton distributions at hadron facilities”. In: *Phys. Rev. D* 80 (2009), p. 074003.
[4548] Hiroyuki Kawamura, Shunzo Kumano, and Takayasu Sekihara. “Determination of exotic hadron structure by constituent-counting rule for hard exclusive processes”. In: *Phys. Rev. D* 88 (2013), p. 034010.
[4549] Kenji Fukushima and Tetsuo Hatsuda. “The phase diagram of dense QCD”. In: *Rept. Prog. Phys.* 74 (2011), p. 014001.
[4550] Asakawa Masayuki and Yazaki Koichi. “Chiral restoration at finite density and temperature”. In: *Nuclear Physics A* 504.4 (1989), pp. 668–684.
[4551] H. Sako *et al.* (J-PARC-HI collaboration). *Letter of Intent for J-PARC Heavy-Ion Program (J-PARC-HI), July 25, 2016.* J-PARC PAC: https://j-parc.jp/researcher/Hadron/en /Proposal_e.html#1607.
[4552] J-PARC-HI collaboration. *J-PARC Heavy Ion Project.* URL: https://asrc.jaea.go.jp/soshiki /gr/hadron/jparc-hi/.
[4553] T. Galatyuk. “https://github.com/tgalatyuk/interaction_rate_facilities/blob/main/hist_rates_detectors_2022_jun.pdf”. In: ().
[4554] Tetyana Galatyuk. “Future facilities for high $\mu_B$ physics”. In: *Nucl. Phys. A* 982 (2019). Ed. by Federico Antinori et al., pp. 163–169.
[4555] V. D. Kekelidze et al. “Three stages of the NICA accelerator complex”. In: *Eur. Phys. J. A* 52.8 (2016), p. 211.
[4556] V. D. Kekelidze. “NICA project at JINR: status and prospects”. In: *JINST* 12.06 (2017). Ed. by Lev Shekhtman, p. C06012.
[4557] Evgeny Syresin et al. “NICA Ion Coolider at JINR”. In: *27th Russian Particle Accelerator Conference.* Oct. 2021.
[4558] E. M. Syresin et al. “Formation of Polarized Proton Beams in the NICA Collider-Accelerator Complex”. In: *Phys. Part. Nucl.* 52.5 (2021), pp. 997–1017.
[4559] Y. Aoki et al. “The Order of the quantum chromodynamics transition predicted by the standard model of particle physics”. In: *Nature* 443 (2006), pp. 675–678.
[4560] Peter Senger et al. “Upgrading the Baryonic Matter at the Nuclotron Experiment at NICA for Studies of Dense Nuclear Matter”. In: *Particles* 2.4 (2019), pp. 481–490.
[4561] Mikhail Kapishin. “The fixed target experiment for studies of baryonic matter at the Nuclotron (BM@N)”. In: *EPJ Web Conf.* 182 (2018). Ed. by Y. Aharonov, L. Bravina, and S. Kabana, p. 02061.
[4562] Jan Steinheimer et al. “Strangeness at the international Facility for Antiproton and Ion Research”. In: *Prog. Part. Nucl. Phys.* 62 (2009). Ed. by Amand Faessler, pp. 313–317.
[4563] J. Steinheimer et al. “Hypernuclei, dibaryon and antinuclei production in high energy heavy ion collisions: Thermal production versus Coalescence”. In: *Phys. Lett. B* 714 (2012), pp. 85–91.
[4564] A. Andronic et al. “Production of light nuclei, hypernuclei and their antiparticles in relativistic nuclear collisions”. In: *Phys. Lett. B* 697 (2011), pp. 203–207.
[4565] M. Patsyuk et al. “Unperturbed inverse kinematics nucleon knockout measurements with a 48 GeV/c carbon beam”. In: *Nature Phys.* 17 (2021), p. 693.
[4566] C. Blume. “Energy dependence of hadronic observables”. In: *J. Phys. G* 31 (2005). Ed. by F. Antinori et al., S57–S68.
[4567] V. Golovatyuk et al. “The Multi-Purpose Detector (MPD) of the collider experiment”. In: *Eur. Phys. J. A* 52.8 (2016), p. 212.
[4568] Kh. Abraamyan et al. “The MultiPurpose Detector – MPD to study Heavy Ion Collisions at NICA (Conceptual Design Report)”. In: (Feb. 2014).
[4569] J. Adamczewski-Musch et al. “Identical pion intensity interferometry in central Au + Au collisions at 1.23 A GeV”. In: *Phys. Lett. B* 795 (2019), pp. 446–451.
[4570] Charles Gale et al. “Event-by-event anisotropic flow in heavy-ion collisions from combined Yang-Mills and viscous fluid dynamics”. In: *Phys. Rev. Lett.* 110.1 (2013), p. 012302.
[4571] Ulrich Heinz and Raimond Snellings. “Collective flow and viscosity in relativistic heavy-ion collisions”. In: *Ann. Rev. Nucl. Part. Sci.* 63 (2013), pp. 123–151.

[4572] P. Parfenov et al. “The comparison of methods for anisotropic flow measurements with the MPD Experiment at NICA”. In: Dec. 2020.

[4573] V. Abgaryan et al. “Status and initial physics performance studies of the MPD experiment at NICA”. In: (Feb. 2022).

[4574] V. M. Abazov et al. “Conceptual design of the Spin Physics Detector”. In: (Jan. 2021).

[4575] A. Arbuzov et al. “On the physics potential to study the gluon content of proton and deuteron at NICA SPD”. In: *Prog. Part. Nucl. Phys.* 119 (2021), p. 103858.

[4576] Aram Kotzinian. “New quark distributions and semiinclusive electroproduction on the polarized nucleons”. In: *Nucl. Phys. B* 441 (1995), pp. 234–248.

[4577] K. Goeke, A. Metz, and M. Schlegel. “Parameterization of the quark-quark correlator of a spin-1/2 hadron”. In: *Phys. Lett. B* 618 (2005), pp. 90–96.

[4578] R. Angeles-Martinez et al. “Transverse Momentum Dependent (TMD) parton distribution functions: status and prospects”. In: *Acta Phys. Polon. B* 46.12 (2015), pp. 2501–2534.

[4579] V. V. Abramov et al. “Possible Studies at the First Stage of the NICA Collider Operation with Polarized and Unpolarized Proton and Deuteron Beams”. In: *Phys. Part. Nucl.* 52.6 (2021), pp. 1044–1119.

[4580] P. Spiller and G. Franchetti. “The FAIR accelerator project at GSI”. In: *Nucl. Instrum. Meth. A* 561 (2006), pp. 305–309.

[4581] H. Geissel et al. “The Super-FRS project at GSI”. In: *Nucl. Instrum. Meth. B* 204 (2003), pp. 71–85.

[4582] Peter Spiller et al. “Acceleration of Intermediate Charge State Heavy Ions in SIS18”. In: *Conf. Proc. C* 100523 (2010). Ed. by Akira Noda et al., MOPD002.

[4583] Florian Kaether et al. “Superconducting Dipole Magnets for the SIS100 Synchrotron”. In: *12th International Particle Accelerator Conference.* Aug. 2021.

[4584] I. C. Arsene et al. “Dynamical phase trajectories for relativistic nuclear collisions”. In: *Physical Review C* 75.3 (Mar. 2007).

[4585] R. Rapp and J. Wambach. “Chiral symmetry restoration and dileptons in relativistic heavy ion collisions”. In: *Adv.Nucl. Phys.* 25 (2000), p. 1.

[4586] Larry D. McLerran and T. Toimela. “Photon and Dilepton Emission from the Quark - Gluon Plasma: Some General Considerations”. In: *Phys. Rev.* D31 (1985), p. 545.

[4587] Florian Seck et al. “Dilepton Signature of a First-Order Phase Transition”. In: (Oct. 2020).

[4588] Ralf Rapp and Hendrik van Hees. “Thermal Dileptons as Fireball Thermometer and Chronometer”. In: *Phys. Lett. B* 753 (2016), pp. 586–590.

[4589] R Arnaldi et al. “Evidence for the production of thermal-like muon pairs with masses above 1-GeV/c**2 in 158-A-GeV Indium-Indium Collisions”. In: *Eur. Phys. J. C* 59 (2009), pp. 607–623.

[4590] J. Adamczewski-Musch et al. “Probing dense baryon-rich matter with virtual photons”. In: *Nature Phys.* 15.10 (2019), pp. 1040–1045.

[4591] Piotr Salabura and Joachim Stroth. “Dilepton radiation from strongly interacting systems”. In: *Prog. Part. Nucl. Phys.* 120 (2021), p. 103869.

[4592] PANDA Collaboration. “PANDA Phase One: PANDA collaboration”. English. In: *European Physical Journal A* 57.6 (June 2021).

[4593] PANDA Collaboration. “Precision resonance energy scans with the PANDA experiment at FAIR”. In: *The European Physical Journal A* 55 (2019), p. 42.

[4594] LHCb Collaboration. “Study of the lineshape of the $\chi_{c1}(3872)$ state”. In: *Physical Review D* 102 (2020), p. 092005.

[4595] F. Nerling for the PANDA Collaboration. “Charm (-onium) physics at PANDA”. In: *Proceedings of Science* 385 (2021).

[4596] Belle Collaboration. “Observation of a Narrow Charmoniumlike State in Exclusive $B^{\pm}\to K^{\pm}\pi^{+}\pi^{-}J/\psi$ Decays”. In: 91.26, 262001 (Dec. 2003), p. 262001.

[4597] COMPASS Collaboration. “Odd and even partial waves of $\eta\pi^-$ and $\eta'\pi^-$ in $\pi^- p \to \eta^{(\prime)}\pi^- p$ at 191 GeV/$c$”. In: *Physics Letters B* 740 (2015), pp. 303–311.

[4598] JPAC and COMPASS Collaborations. “New analysis of $\eta\pi$ tensor resonances measured at the COMPASS experiment”. In: *Physics Letters B* 779 (2018), p. 464.

[4599] JPAC Collaboration. “Determination of the Pole Position of the Lightest Hybrid Meson Candidate”. In: *Physical Review Letters* 122 (2019), p. 042002.

[4600] COMPASS Collaboration. “Observation of a New Narrow Axial-Vector Meson $a_1(1420)$”. In: *Physical Review Letters* 115 (2015), p. 082001.

[4601] Nora Brambilla et al. “Spin structure of heavy-quark hybrids”. In: *Physical Review D* 99 (2019), p. 014017.

[4602] Claude Amsler. "Nucleon-antinucleon annihilation at LEAR". In: (2019).
[4603] Marius Christian Mertens for the PANDA Collaboration. "Determination of the $D_{s0}(2317)$ width with the PANDA detector". In: *Hyperfine Interact* 209 (2012), pp. 111–115.
[4604] BESIII Collaboration. "Probing CP symmetry and weak phases with entangled double-strange baryons". In: *Nature* 606 (2022), pp. 64–69.
[4605] Kazuya Aoki et al. "Extension of the J-PARC Hadron Experimental Facility: Third White Paper". In: (2021).
[4606] PANDA Collaboration. "Prospects for Spin-Parity Determination of Excited Baryons via the $^{+}$ $\mathbf{K}^{-}$ Final State with PANDA". In: (2022).
[4607] PANDA Collaboration. "Experimental access to Transition Distribution Amplitudes with the PANDA experiment at FAIR". In: *European Physical Journal A* 51 (2015), p. 107.
[4608] PANDA Collaboration. "Feasibility study for the measurement of $\pi N$ transition distribution amplitudes at PANDA in $\bar{p}p \to J/\Psi\pi^0$". In: *Physical Review D* 95 (2017), p. 032003.
[4609] A. Freund et al. "Exclusive Annihilation $p\bar{p} \to \gamma\gamma$ in a Generalized Parton Picture". In: *Physical Review Letters* 90 (2003), p. 092001.
[4610] P. Kroll and A. Schäfer. "The process $p\bar{p} \to \gamma\pi^0$ within the handbag approach". In: *European Physical Journal A* 26 (2005), p. 89.
[4611] P. Kroll and A. Schäfer. "Probing moments of baryon-antibaryon generalized parton distributions at BELLE and FAIR". In: *European Physical Journal A* 50 (2014), p. 1.
[4612] G. Bardin et al. "Determination of the electric and magnetic form factors of the proton in the time-like region". In: *Nuclear Physics B* 411.1 (1994), pp. 3–32.
[4613] BABAR Collaboration. "Study of $e^+e^- \to p\bar{p}$ via initial-state radiation at BABAR". In: *Phys. Rev. D* 87 (9 May 2013), p. 092005.
[4614] BABAR Collaboration. "Measurement of the $e^+e^- \to p\bar{p}$ cross section in the energy range from 3.0 to 6.5 GeV". In: *Phys. Rev. D* 88 (7 Oct. 2013), p. 072009.
[4615] BESIII Collaboration. "Observation and study of the decay $J/\psi \to \phi\eta\eta^{'}$". In: *Phys. Rev. D* 99 (11 June 2019), p. 112008.
[4616] BESIII Collaboration. "First Observation of $D^+ \to \eta\mu^+\nu_\mu$ and Measurement of Its Decay Dynamics". In: *Phys. Rev. Lett.* 124 (23 June 2020), p. 231801.
[4617] BESIII Collaboration. "Study of the process $e^+e^- \to p\bar{p}$ via initial state radiation at BESIII". In: *Phys. Rev. D* 99 (9 May 2019), p. 092002.
[4618] BESIII Collaboration. "Measurement of proton electromagnetic form factors in the time-like region using initial state radiation at BESIII". In: *Physics Letters B* 817 (2021), p. 136328.
[4619] R.R. Akhmetshin et al. "Study of the process $e^+e^- \to p\bar{p}$ in the c.m. energy range from threshold to 2 GeV with the CMD-3 detector". In: *Physics Letters B* 759 (2016), pp. 634–640.
[4620] M. Ablikim et al. "Future Physics Programme of BESIII". In: *Chin. Phys. C* 44.4 (2020), p. 040001.
[4621] M. Ablikim et al. "Measurement of the Cross Section for $e^+e^- \to$Hadrons at Energies from 2.2324 to 3.6710 GeV". In: *Phys. Rev. Lett.* 128.6 (2022), p. 062004.
[4622] M. Ablikim et al. "Measurement of azimuthal asymmetries in inclusive charged dipion production in $e^+e^-$ annihilations at $\sqrt{s} = 3.65$ GeV". In: *Phys. Rev. Lett.* 116.4 (2016), p. 042001.
[4623] Medina Ablikim et al. "Measurement of proton electromagnetic form factors in $e^+e^- \to p\bar{p}$ in the energy region 2.00 - 3.08 GeV". In: *Phys. Rev. Lett.* 124.4 (2020), p. 042001.
[4624] M. Ablikim et al. "Oscillating features in the electromagnetic structure of the neutron". In: *Nature Phys.* 17.11 (2021), pp. 1200–1204.
[4625] Guangshun Huang and Rinaldo Baldini Ferroli. "Probing the internal structure of baryons". In: *Natl. Sci. Rev.* 8.11 (2021), nwab187.
[4626] M. Ablikim et al. "Complete Measurement of the $\Lambda$ Electromagnetic Form Factors". In: *Phys. Rev. Lett.* 123.12 (2019), p. 122003.
[4627] Medina Ablikim et al. "Measurements of Weak Decay Asymmetries of $\Lambda_c^+ \to pK_S^0$, $\Lambda\pi^+$, $\Sigma^+\pi^0$, and $\Sigma^0\pi^+$". In: *Phys. Rev. D* 100.7 (2019), p. 072004.
[4628] E. V. Abakumova et al. "The Beam Energy Measurement System for the Beijing Electron-Positron Collider". In: *Nucl. Instrum. Meth. A* 659 (2011), pp. 21–29.
[4629] Shan Jin and Xiaoyan Shen. "Highlights of light meson spectroscopy at the BESIII experiment". In: *Natl. Sci. Rev.* 8.11 (2021), nwab198.
[4630] A. V. Sarantsev et al. "Scalar isoscalar mesons and the scalar glueball from radiative $J/\psi$ decays". In: *Phys. Lett. B* 816 (2021), p. 136227.
[4631] A. Rodas et al. "Scalar and tensor resonances in $J/\psi$ radiative decays". In: *Eur. Phys. J. C* 82.1 (2022), p. 80.
[4632] Frederic Brünner and Anton Rebhan. "Constraints on the $\eta\eta'$ decay rate of a scalar glue-

ball from gauge/gravity duality”. In: *Phys. Rev. D* 92.12 (2015), p. 121902.

[4633] M. Ablikim et al. “Observation of a State $X(2600)$ in the $\pi^+\pi^-\eta'$ System in the Process $J/\psi \to \gamma\pi^+\pi^-\eta'$”. In: *Phys. Rev. Lett.* 129.4 (2022), p. 042001.

[4634] Medina Ablikim et al. “Observation of the $Y(4220)$ and $Y(4360)$ in the process $e^+e^- \to \eta J/\psi$”. In: *Phys. Rev. D* 102.3 (2020), p. 031101.

[4635] M. Ablikim et al. “Observation of $e^+e^- \to \pi^+\pi^-\psi(3770)$ and $D_1(2420)^0\bar{D}^0+c.c.$”. In: *Phys. Rev. D* 100.3 (2019), p. 032005.

[4636] M. Ablikim et al. “Observation of resonance structures in $e^+e^- \to \pi^+\pi^-\psi_2(3823)$ and mass measurement of $\psi_2(3823)$”. In: *Phys. Rev. Lett.* 129.10 (2022), p. 102003.

[4637] M. Ablikim et al. “Measurement of $e^+e^- \to \gamma\chi_{c0,c1,c2}$ cross sections at center-of-mass energies between 3.77 and 4.60 GeV”. In: *Phys. Rev. D* 104.9 (2021), p. 092001.

[4638] Medina Ablikim et al. “Search for new decay modes of the $\psi_2(3823)$ and the process $e^+e^- \to \pi^0\pi^0\psi_2(3823)$”. In: *Phys. Rev. D* 103.9 (2021), p. L091102.

[4639] “First observation of the direct production of the $\chi_{c1}$ in $e^+e^-$ annihilation”. In: (Mar. 2022).

[4640] Chang-Zheng Yuan. “Charmonium and charmoniumlike states at the BESIII experiment”. In: *Natl. Sci. Rev.* 8.11 (2021), nwab182.

[4641] M. Ablikim et al. “Cross section measurement of $e^+e^- \to \pi^+\pi^-(3686)$ from $\sqrt{S} = 4.0076$ to 4.6984 GeV”. In: *Phys. Rev. D* 104.5 (2021), p. 052012.

[4642] M. Ablikim et al. “Evidence for a Neutral Near-Threshold Structure in the $K^0_S$ recoil-mass spectra in $e^+e^- \to K^0_S D^+_s D^{*-}$ and $e^+e^- \to K^0_S D^{*+}_s D^-$”. In: *Phys. Rev. Lett.* 129.11 (2022), p. 112003.

[4643] M. Ablikim et al. “Measurement of $e^+e^- \to \pi^+\pi^-\psi(3686)$ from 4.008 to 4.600~GeV and observation of a charged structure in the $\pi^\pm\psi(3686)$ mass spectrum”. In: *Phys. Rev. D* 96.3 (2017). [Erratum: Phys.Rev.D 99, 019903 (2019)], p. 032004.

[4644] Kai Zhu. “Triangle relations for XY Z states”. In: *Int. J. Mod. Phys. A* 36.14 (2021), p. 2150126.

[4645] Cheng-Ping Shen and Chang-Zheng Yuan. “Observation of pentaquark states and perspectives of further studies”. In: *Science Bulletin* 60.17 (2015), pp. 1549–1550.

[4646] J. Z. Bai et al. “Search for the pentaquark state in psi(2S) and J / psi decays to K0(S)pK- anti-n and K0(S) anti-p K+ n”. In: *Phys. Rev. D* 70 (2004), p. 012004.

[4647] Steven D. Bass and Pawel Moskal. “$\eta'$ and $\eta$ mesons with connection to anomalous glue”. In: *Rev. Mod. Phys.* 91 (1 Feb. 2019), p. 015003.

[4648] Shuang-Shi Fang. “Light meson physics at BESIII”. In: *Natl. Sci. Rev.* 8.11 (2021), nwab052.

[4649] Rafel Escribano and Sergi Gonzàlez-Solís. “A data-driven approach to $\pi^0\,\eta$ and $\eta'$ single and double Dalitz decays”. In: *Chinese Physics C* 42.2 (Jan. 2018), p. 023109.

[4650] Thimo Petri. “Anomalous decays of pseudoscalar mesons”. In: (2010).

[4651] Kubis, Bastian and Schneider, Sebastian P. “The cusp effect in $\eta' \to \eta\pi\pi$ decays”. In: *Eur. Phys. J. C* 62.3 (2009), pp. 511–523.

[4652] M. Ablikim et al. “Polarization and Entanglement in Baryon-Antibaryon Pair Production in Electron-Positron Annihilation”. In: *Nature Phys.* 15 (2019), pp. 631–634.

[4653] M. Ablikim et al. “Precise Measurements of Decay Parameters and $CP$ Asymmetry with Entangled $\Lambda - \bar{\Lambda}$ Pairs Pairs”. In: *Phys. Rev. Lett.* 129.13 (2022), p. 131801.

[4654] A. Bazavov et al. “$B$- and $D$-meson leptonic decay constants from four-flavor lattice QCD”. In: *Phys. Rev. D* 98.7 (2018), p. 074512.

[4655] N. Carrasco et al. “Leptonic decay constants $f_K, f_D$, and $f_{D_s}$ with $N_f = 2+1+1$ twisted-mass lattice QCD”. In: *Phys. Rev. D* 91.5 (2015), p. 054507.

[4656] Hai-Bo Li and Xiao-Rui Lyu. “Study of the standard model with weak decays of charmed hadrons at BESIII”. In: *Natl. Sci. Rev.* 8.11 (2021), nwab181.

[4657] Hai Ping Peng, Yang Heng Zheng, and Xiao Rong Zhou. “Super Tau-Charm Facility of China”. In: *Physics* 49.8 (2020), pp. 513–524.

[4658] W. Altmannshofer et al. “The Belle II Physics Book”. In: *PTEP* 2019.12 (2019). Ed. by E. Kou and P. Urquijo. [Erratum: PTEP 2020, 029201 (2020)], p. 123C01.

[4659] T. Keck et al. “The Full Event Interpretation: An Exclusive Tagging Algorithm for the Belle II Experiment”. In: *Comput. Softw. Big Sci.* 3.1 (2019), p. 6.

[4660] F. Abudinén et al. “$B$-flavor tagging at Belle II”. In: *Eur. Phys. J. C* 82.4 (2022), p. 283.

[4661] Y. Amhis *et al.* “Averages of b-hadron, c-hadron, and $\tau$-lepton properties as of 2018”. In: *Eur. Phys. J. C* 81.3 (2021), p. 226.

[4662] J. Charles et al. “Isospin analysis of charmless B-meson decays”. In: *Eur. Phys. J. C* 77.8 (2017), p. 574.

[4663] J. Brod and J. Zupan. "The ultimate theoretical error on $\gamma$ from $B \to DK$ decays". In: *J. High Energy Phys.* 01 (2014), p. 051.

[4664] M. Gronau and D. London. "How to determine all the angles of the unitarity triangle from B(d)0 —> D K(s) and B(s)0 —> D0". In: *Phys. Lett. B* 253 (1991), pp. 483–488.

[4665] M. Gronau and D. Wyler. "On determining a weak phase from CP asymmetries in charged B decays". In: *Phys. Lett. B* 265 (1991), pp. 172–176.

[4666] I. Dunietz D. Atwood and A. Soni. "Improved methods for observing CP violation in B+- —> K D and measuring the CKM phase gamma". In: *Phys. Rev. D* 63 (2001), p. 036005.

[4667] A. Giri et al. "Determining gamma using B+- —> DK+- with multibody D decays". In: *Phys. Rev. D* 68 (2003), p. 054018.

[4668] A. Bondar. September 2002.

[4669] A. Poluektov *et al.* "Measurement of $\phi_3$ with Dalitz plot analysis of $B^{+-} \to D^{(*)}K^{+-}$ decay". In: *Phys. Rev. D* 70 (2004), p. 072003.

[4670] Z. Ligeti Y. Grossman and A. Soffer. "Measuring $\gamma$ in $B^{+-} \to K^{+-}(KK^*)_D$ decays". In: *Phys. Rev. D* 67 (2003), p. 071301.

[4671] P. K. Resmi *et al.* "First measurement of the CKM angle $\phi_3$ with $B^\pm \to D(K^0_S\pi^+\pi^-\pi^0)K^\pm$ decays". In: *J. High Energy Phys.* 10 (2019), p. 178.

[4672] A. Sibidanov *et al.* "Study of Exclusive $B \to X_u\ell\nu$ Decays and Extraction of $\|V_{ub}\|$ using Full Reconstruction Tagging at the Belle Experiment". In: *Phys. Rev. D* 88.3 (2013), p. 032005.

[4673] F. U. Bernlochner et al. "Precision Global Determination of the B→Xs$\gamma$ Decay Rate". In: *Phys. Rev. Lett.* 127.10 (2021), p. 102001.

[4674] M. Beneke. "Corrections to sin(2$\beta$) from CP asymmetries in $B^0 \to (\pi^0, \rho^0, \eta, \eta', \omega, \phi)K_S$ decays". In: *Phys. Lett. B* 620 (2005), pp. 143–150.

[4675] G. Caria *et al.* "Measurement of $\mathcal{R}(D)$ and $\mathcal{R}(D^*)$ with a semileptonic tagging method". In: *Phys. Rev. Lett.* 124.16 (2020), p. 161803.

[4676] J. P. Lees *et al.* "Measurement of an Excess of $\bar{B} \to D^{(*)}\tau^-\bar{\nu}_\tau$ Decays and Implications for Charged Higgs Bosons". In: *Phys. Rev. D* 88.7 (2013), p. 072012.

[4677] R. Aaij *et al.* "Measurement of the ratio of branching fractions $\mathcal{B}(\bar{B}^0 \to D^{*+}\tau^-\bar{\nu}_\tau)/\mathcal{B}(\bar{B}^0 \to D^{*+}\mu^-\bar{\nu}_\mu)$". In: *Phys. Rev. Lett.* 115.11 (2015). [Erratum: Phys. Rev. Lett. **115**, 159901 (2015)], p. 111803.

[4678] R. Aaij *et al.* "Test of Lepton Flavor Universality by the measurement of the $B^0 \to D^{*-}\tau^+\nu_\tau$ branching fraction using three-prong $\tau$ decays". In: *Phys. Rev. D* 97.7 (2018), p. 072013.

[4679] M. Huschle *et al.* "Measurement of the branching ratio of $\bar{B} \to D^{(*)}\tau^-\bar{\nu}_\tau$ relative to $\bar{B} \to D^{(*)}\ell^-\bar{\nu}_\ell$ decays with hadronic tagging at Belle". In: *Phys. Rev. D* 92.7 (2015), p. 072014.

[4680] S. Hirose *et al.* "Measurement of the $\tau$ lepton polarization and $R(D^*)$ in the decay $\bar{B} \to D^*\tau^-\bar{\nu}_\tau$ with one-prong hadronic $\tau$ decays at Belle". In: *Phys. Rev. D* 97.1 (2018), p. 012004.

[4681] Roel Aaij et al. "Search for lepton-universality violation in $B^+ \to K^+\ell^+\ell^-$ decays". In: *Phys. Rev. Lett.* 122.19 (2019), p. 191801.

[4682] R. Aaij et al. "Test of lepton universality with $B^0 \to K^{*0}\ell^+\ell^-$ decays". In: *JHEP* 08 (2017), p. 055.

[4683] R. Aaij et al. "Differential branching fractions and isospin asymmetries of $B \to K^{(*)}\mu^+\mu^-$ decays". In: *JHEP* 06 (2014), p. 133.

[4684] Roel Aaij et al. "Angular analysis of the $B^0 \to K^{*0}\mu^+\mu^-$ decay using 3 fb$^{-1}$ of integrated luminosity". In: *JHEP* 02 (2016), p. 104.

[4685] Roel Aaij et al. "Measurement of $CP$-Averaged Observables in the $B^0 \to K^{*0}\mu^+\mu^-$ Decay". In: *Phys. Rev. Lett.* 125.1 (2020), p. 011802.

[4686] E. Manoni (Belle II Collaboration). *Electroweak penguins and radiative decays at Belle II.* 12–19 March 2022, https://moriond.in2p3.fr/2022/EW/slides/5/2/3_EManoni-v2.pdf.

[4687] J. P. Lees et al. "Search for $B^+ \to K^+\tau^+\tau^-$ at the BaBar experiment". In: *Phys. Rev. Lett.* 118.3 (2017), p. 031802.

[4688] Rodrigo Alonso, Benjamín Grinstein, and Jorge Martin Camalich. "Lepton universality violation and lepton flavor conservation in $B$-meson decays". In: *JHEP* 10 (2015), p. 184.

[4689] JoAnne L. Hewett. "Tau polarization asymmetry in $B \to X_s\tau^+\tau^-$". In: *Phys. Rev. D* 53 (1996), pp. 4964–4969.

[4690] T. V. Dong et al. "Search for the decay $B^0 \to K^{*0}\tau^+\tau^-$ at the Belle experiment". In: (Oct. 2021).

[4691] J. P. Lees *et al.* "Search for $B^+ \to K^+\tau^+\tau^-$ at the Experiment". In: *Phys. Rev. Lett.* 118 (3 Jan. 2017), p. 031802.

[4692] Michael Benzke et al. "Long-Distance Dominance of the CP Asymmetry in $B \to X_{s,d} + \gamma$ Decays". In: *Phys. Rev. Lett.* 106 (2011), p. 141801.

[4693] James Lyon and Roman Zwicky. "Isospin asymmetries in $B \to (K^*, \rho)\gamma/l^+l^-$ and $B \to Kl^+l^-$

in and beyond the standard model". In: *Phys. Rev. D* 88.9 (2013), p. 094004.

[4694] Bernard Aubert et al. "Observation of a broad structure in the $\pi^+\pi^- J/\psi$ mass spectrum around 4.26-GeV/c$^2$". In: *Phys. Rev. Lett.* 95 (2005), p. 142001.

[4695] A. Abdesselam et al. "Energy scan of the $e^+e^- \to h_b(nP)\pi^+\pi^-$ $(n = 1, 2)$ cross sections and evidence for $\Upsilon(11020)$ decays into charged bottomonium-like states". In: *Phys. Rev. Lett.* 117.14 (2016), p. 142001.

[4696] R. Mizuk et al. "Observation of a new structure near 10.75 GeV in the energy dependence of the $e^+e\Upsilon(nS)\pi^+\pi$ (n = 1, 2, 3) cross sections". In: *JHEP* 10 (2019), p. 220.

[4697] Bob McElrath. "Invisible quarkonium decays as a sensitive probe of dark matter". In: *Phys. Rev. D* 72 (2005), p. 103508.

[4698] Derek E. Hazard and Alexey A. Petrov. "Lepton flavor violating quarkonium decays". In: *Phys. Rev. D* 94.7 (2016), p. 074023.

[4699] M. Abe *et al.* "A new approach for measuring the muon anomalous magnetic moment and electric dipole moment". In: *PTEP* 2019.5 (2019), p. 052C02.

[4700] Roel Aaij et al. "Measurement of the $b$-quark production cross-section in 7 and 13 TeV $pp$ collisions". In: *Phys. Rev. Lett.* 118.5 (2017). [Erratum: Phys.Rev.Lett. 119, 169901 (2017)], p. 052002.

[4701] Roel Aaij et al. "Measurements of prompt charm production cross-sections in $pp$ collisions at $\sqrt{s} =$ 13 TeV". In: *JHEP* 03 (2016). [Erratum: JHEP 09, 013 (2016), Erratum: JHEP 05, 074 (2017)], p. 159.

[4702] LHCb Collaboration. "Framework TDR for the LHCb Upgrade II — Opportunities in flavour physics, and beyond, in the HL-LHC era". In: (2021).

[4703] A. Cerri et al. "Report from Working Group 4: Opportunities in Flavour Physics at the HL-LHC and HE-LHC". In: *CERN Yellow Rep. Monogr.* 7 (2019). Ed. by Andrea Dainese et al., pp. 867–1158.

[4704] Kristof De Bruyn, Robert Fleischer, and Patrick Koppenburg. "Extracting $\gamma$ and Penguin Topologies through CP Violation in $B_s^0 \to J/\psi K_S$". In: *Eur. Phys. J. C* 70 (2010), pp. 1025–1035.

[4705] Kristof De Bruyn and Robert Fleischer. "A Roadmap to Control Penguin Effects in $B_d^0 \to J/\psi K_S^0$ and $B_s^0 \to J/\psi\phi$". In: *JHEP* 03 (2015), p. 145.

[4706] M. Ciuchini, M. Pierini, and L. Silvestrini. "B(s) to K(*)0 anti-K(*)0 decays: The Golden channels for new physics searches". In: *Phys. Rev. Lett.* 100 (2008), p. 031802.

[4707] Sebastien Descotes-Genon, Joaquim Matias, and Javier Virto. "An analysis of $B_{d,s}$ mixing angles in presence of New Physics and an update of $B_s \to \bar{K}^{0*} anti - K^{0*}$". In: *Phys. Rev. D* 85 (2012), p. 034010.

[4708] Bhubanjyoti Bhattacharya et al. "Measuring $\beta_s$ with $B_s \to K^{0(*)}\bar{K}^{0(*)}$ – a Reappraisal". In: *Phys. Lett. B* 717 (2012), pp. 403–408.

[4709] Robert Fleischer. "Exploring CP violation and penguin effects through $B_d^0 \to D^+D^-$ and $B_s^0 \to D_s^+D_s^-$". In: *Eur. Phys. J. C* 51 (2007), pp. 849–858.

[4710] Martin Jung and Stefan Schacht. "Standard model predictions and new physics sensitivity in $B \to DD$ decays". In: *Phys. Rev. D* 91.3 (2015), p. 034027.

[4711] Lennaert Bel et al. "Anatomy of $B \to D\overline{D}$ decays". In: *JHEP* 07 (2015), p. 108.

[4712] Xiao-Gang He. "SU(3) analysis of annihilation contributions and CP violating relations in B —> P P decays". In: *Eur. Phys. J. C* 9 (1999), pp. 443–448.

[4713] Michael Gronau and Jonathan L. Rosner. "The Role of $B_s \to K\pi$ in determining the weak phase $\gamma$". In: *Phys. Lett. B* 482 (2000), pp. 71–76.

[4714] Harry J. Lipkin. "Is observed direct CP violation in B(d) —> K+ pi- due to new physics? Check standard model prediction of equal violation in B(s) —> K- pi+". In: *Phys. Lett. B* 621 (2005), pp. 126–132.

[4715] Hai-Yang Cheng and Cheng-Wei Chiang. "SU(3) symmetry breaking and CP violation in D -> PP decays". In: *Phys. Rev. D* 86 (2012), p. 014014.

[4716] Michael Gronau. "High order U-spin breaking: A precise amplitude relation in $D^0$ decays"". In: *Phys. Lett. B* 730 (2014). [Addendum: Phys.Lett.B 735, 282 (2014)], pp. 221–225.

[4717] Sarah Müller, Ulrich Nierste, and Stefan Schacht. "Sum Rules of Charm CP Asymmetries beyond the SU(3)$_F$ Limit". In: *Phys. Rev. Lett.* 115.25 (2015), p. 251802.

[4718] Margarita Gavrilova, Yuval Grossman, and Stefan Schacht. "The mathematical structure of U-spin amplitude sum rules". In: *JHEP* 08 (2022), p. 278.

[4719] Roel Aaij et al. "Updated measurement of time-dependent CP-violating observables in $B_s^0 \to J/\psi K^+K^-$ decays". In: *Eur. Phys. J. C* 79.8 (2019). [Erratum: Eur.Phys.J.C 80, 601 (2020)], p. 706.

[4720] Georges Aad et al. "Measurement of the $CP$-violating phase $\phi_s$ in $B_s^0 \to J/\psi\phi$ decays in ATLAS at 13 TeV". In: *Eur. Phys. J. C* 81.4 (2021), p. 342.

[4721] Albert M Sirunyan et al. "Measurement of the $CP$-violating phase $\phi_s$ in the $\mathrm{B_s^0} \to \mathrm{J}/\psi\,\phi(1020) \to \mu^+\mu^-\mathrm{K^+K^-}$ channel in proton-proton collisions at $\sqrt{s} = 13$ TeV". In: *Phys. Lett. B* 816 (2021), p. 136188.

[4722] Roel Aaij et al. "Observation of the Mass Difference Between Neutral Charm-Meson Eigenstates". In: *Phys. Rev. Lett.* 127.11 (2021), p. 111801.

[4723] Roel Aaij et al. "Measurement of the charm mixing parameter $y_{CP} - y_{CP}^{K\pi}$ using two-body $D^0$ meson decays". In: *Phys. Rev. D* 105.9 (2022), p. 092013.

[4724] Alexander Lenz and Guy Wilkinson. "Mixing and CP Violation in the Charm System". In: *Ann. Rev. Nucl. Part. Sci.* 71 (2021), pp. 59–85.

[4725] Roel Aaij et al. "Measurements of $CP$ violation in the three-body phase space of charmless $B^\pm$ decays". In: *Phys. Rev. D* 90.11 (2014), p. 112004.

[4726] Roel Aaij et al. "Amplitude analysis of $B^\pm \to \pi^\pm K^+K^-$ decays". In: *Phys. Rev. Lett.* 123.23 (2019), p. 231802.

[4727] Roel Aaij et al. "Observation of Several Sources of $CP$ Violation in $B^+ \to \pi^+\pi^+\pi^-$ Decays". In: *Phys. Rev. Lett.* 124.3 (2020), p. 031801.

[4728] Roel Aaij et al. "Amplitude analysis of the $B^+ \to \pi^+\pi^+\pi^-$ decay". In: *Phys. Rev. D* 101.1 (2020), p. 012006.

[4729] "Search for direct $CP$ violation in charged charmless $B \to PV$ decays". In: (June 2022).

[4730] J. T. Daub, C. Hanhart, and B. Kubis. "A model-independent analysis of final-state interactions in $\overline{B}{}^0_{d/s} \to J/\psi\pi\pi$". In: *JHEP* 02 (2016), p. 009.

[4731] Roel Aaij et al. "Simultaneous determination of CKM angle $\gamma$ and charm mixing parameters". In: *JHEP* 12 (2021), p. 141.

[4732] "Simultaneous determination of the CKM angle $\gamma$ and parameters related to mixing and CP violation in the charm sector". In: (2022).

[4733] Tim Gershon and Anton Poluektov. "Double Dalitz Plot Analysis of the Decay $B^0 \to DK^+\pi^-$, $D \to K_S^0\pi^+\pi^-$". In: *Phys. Rev. D* 81 (2010), p. 014025.

[4734] D. Craik, T. Gershon, and A. Poluektov. "Optimising sensitivity to $\gamma$ with $B^0 \to DK^+\pi^-$, $D \to K_S^0\pi^+\pi^-$ double Dalitz plot analysis". In: *Phys. Rev. D* 97 (2018), p. 056002.

[4735] Albert M Sirunyan et al. "Measurement of properties of $\mathrm{B_s^0} \to \mu^+\mu^-$ decays and search for $\mathrm{B^0} \to \mu^+\mu^-$ with the CMS experiment". In: *JHEP* 04 (2020), p. 188.

[4736] Morad Aaboud et al. "Study of the rare decays of $B_s^0$ and $B^0$ mesons into muon pairs using data collected during 2015 and 2016 with the ATLAS detector". In: *JHEP* 04 (2019), p. 098.

[4737] R. Aaij et al. "Analysis of Neutral B-Meson Decays into Two Muons". In: *Phys. Rev. Lett.* 128.4 (2022), p. 041801.

[4738] Roel Aaij et al. "Search for the Rare Decays $B_s^0 \to e^+e^-$ and $B^0 \to e^+e^-$". In: *Phys. Rev. Lett.* 124.21 (2020), p. 211802.

[4739] Roel Aaij et al. "Search for the decays $B_s^0 \to \tau^+\tau^-$ and $B^0 \to \tau^+\tau^-$". In: *Phys. Rev. Lett.* 118.25 (2017), p. 251802.

[4740] "Measurement of rare $B \to \mu^+\mu^-$ decays with the Phase-2 upgraded CMS detector at the HL-LHC". In: (2018).

[4741] "Prospects for the $\mathcal{B}(B^0_{(s)} \to \mu^+\mu^-)$ measurements with the ATLAS detector in the Run 2 and HL-LHC data campaigns". In: (2018). All figures including auxiliary figures are available at https://atlas.web.cern.ch/Atlas/GROUPS/PHYSICS/PUBNOTES/ATL-PHYS-PUB-2018-005.

[4742] Kristof De Bruyn et al. "Probing New Physics via the $B_s^0 \to \mu^+\mu^-$ Effective Lifetime". In: *Phys. Rev. Lett.* 109 (2012), p. 041801.

[4743] Roel Aaij et al. "Angular analysis of the rare decay $B_s^0 \to \phi\mu^+\mu$". In: *JHEP* 11 (2021), p. 043.

[4744] Roel Aaij et al. "Branching Fraction Measurements of the Rare $B_s^0 \to \phi\mu^+\mu^-$ and $B_s^0 \to f_2'(1525)\mu^+\mu^-$- Decays". In: *Phys. Rev. Lett.* 127.15 (2021), p. 151801.

[4745] Morad Aaboud et al. "Angular analysis of $B_d^0 \to K^*\mu^+\mu^-$ decays in *pp* collisions at $\sqrt{s} = 8$ TeV with the ATLAS detector". In: *JHEP* 10 (2018), p. 047.

[4746] Albert M Sirunyan et al. "Measurement of angular parameters from the decay $\mathrm{B^0} \to \mathrm{K^{*0}}\mu^+\mu^-$ in proton-proton collisions at $\sqrt{s} = 8$ TeV". In: *Phys. Lett. B* 781 (2018), pp. 517–541.

[4747] T. Hurth, C. Langenbruch, and F. Mahmoudi. "Direct determination of Wilson coefficients using $B^0 \to K^{*0}\mu^+\mu^-$ decays". In: *JHEP* 11 (2017), p. 176.

[4748] Roel Aaij et al. "Test of lepton universality in beauty-quark decays". In: *Nature Phys.* 18.3 (2022), pp. 277–282.

[4749] Bernard Aubert et al. "Measurement of Time-Dependent CP Asymmetry in $B^0 \to K_S^0\pi^0\gamma$ Decays". In: *Phys. Rev. D* 78 (2008), p. 071102.

[4750] Y. Ushiroda et al. "Time-Dependent CP Asymmetries in $B^0 \to K^0_S \pi^0 \gamma$ transitions". In: *Phys. Rev. D* 74 (2006). Ed. by Alexey Sissakian, Gennady Kozlov, and Elena Kolganova, p. 111104.
[4751] Roel Aaij et al. "Measurement of $CP$-violating and mixing-induced observables in $B^0_s \to \phi\gamma$ decays". In: *Phys. Rev. Lett.* 123.8 (2019), p. 081802.
[4752] Roel Aaij et al. "Measurement of the photon polarization in $\Lambda^0_b \to \Lambda\ \gamma$ decays". In: *Phys. Rev. D* 105.5 (2022), p. L051104.
[4753] Roel Aaij et al. "Strong constraints on the $b \to s\gamma$ photon polarisation from $B^0 \to K^{*0}e^+e^-$ decays". In: *JHEP* 12 (2020), p. 081.
[4754] Roel Aaij et al. "Observation of Photon Polarization in the b→sγ Transition". In: *Phys. Rev. Lett.* 112.16 (2014), p. 161801.
[4755] Michael Gronau et al. "Measuring the photon polarization in B —> K pi pi gamma". In: *Phys. Rev. Lett.* 88 (2002), p. 051802.
[4756] Emi Kou, Alain Le Yaouanc, and Andrey Tayduganov. "Determining the photon polarization of the b -> s gamma using the B -> K1(1270) gamma -> (K pi pi) gamma decay". In: *Phys. Rev. D* 83 (2011), p. 094007.
[4757] Michael Gronau and Dan Pirjol. "Reexamining the photon polarization in B→Kππγ". In: *Phys. Rev. D* 96.1 (2017), p. 013002.
[4758] Simon Akar et al. "The time-dependent $CP$ asymmetry in $B^0 \to K_{\rm res}\gamma \to \pi^+\pi^- K^0_S \gamma$ decays". In: *JHEP* 09 (2019), p. 034.
[4759] Patrick Koppenburg. *List of hadrons observed at the LHC*. LHCb-FIGURE-2021-001. See 2022 update online. 2021.
[4760] Roel Aaij et al. "Amplitude analysis of the $B^+ \to D^+D^-K^+$ decay". In: *Phys. Rev. D* 102 (2020), p. 112003.
[4761] Roel Aaij et al. "A model-independent study of resonant structure in $B^+ \to D^+D^-K^+$ decays". In: *Phys. Rev. Lett.* 125 (2020), p. 242001.
[4762] "Observation of an excess of di-charmonium events in the four-muon final state with the ATLAS detector". In: (2022). All figures including auxiliary figures are available at https://atlas.web.cern.ch/Atlas/GROUPS/PHYSICS/CONFNOTES/ATLAS-CONF-2022-040.
[4763] "Observation of new structures in the J/ψJ/ψ mass spectrum in pp collisions at $\sqrt{s} = 13$ TeV". In: (2022).
[4764] Roel Aaij et al. "Search for the doubly charmed baryon $\Xi^+_{cc}$". In: *Sci. China Phys. Mech. Astron.* 63.2 (2020), p. 221062.
[4765] Roel Aaij et al. "Search for the doubly charmed baryon $\Omega^+_{cc}$". In: *Sci. China Phys. Mech. Astron.* 64.10 (2021), p. 101062.
[4766] T. Gershon and A. Poluektov. "Displaced $B_c$ mesons as an inclusive signature of weakly decaying double beauty hadrons". In: *JHEP* 01 (2019), p. 019.
[4767] A. David et al. "LHC HXSWG interim recommendations to explore the coupling structure of a Higgs-like particle". In: (Sept. 2012).
[4768] M. Cepeda et al. "Report from Working Group 2: Higgs Physics at the HL-LHC and HE-LHC". In: *CERN Yellow Rep. Monogr.* 7 (2019). Ed. by Andrea Dainese et al., pp. 221–584.
[4769] Fabrizio Caola and Kirill Melnikov. "Constraining the Higgs boson width with ZZ production at the LHC". In: *Phys. Rev. D* 88 (2013), p. 054024.
[4770] Roberto Covarelli, Mathieu Pellen, and Marco Zaro. "Vector-Boson scattering at the LHC: Unraveling the electroweak sector". In: *Int. J. Mod. Phys. A* 36.16 (2021), p. 2130009.
[4771] Charalampos Anastasiou et al. "High precision determination of the gluon fusion Higgs boson cross-section at the LHC". In: *JHEP* 05 (2016), p. 058.
[4772] Stefano Catani et al. "$t\bar{t}H$ production in NNLO QCD". In: (Oct. 2022).
[4773] R. Bonciani et al. "Next-to-leading-order QCD Corrections to Higgs Production in association with a Jet". In: (June 2022).
[4774] Radja Boughezal et al. "Higgs boson production in association with a jet at next-to-next-to-leading order". In: *Phys. Rev. Lett.* 115.8 (2015), p. 082003.
[4775] X. Chen et al. "NNLO QCD corrections to Higgs boson production at large transverse momentum". In: *JHEP* 10 (2016), p. 066.
[4776] X. Chen et al. "Top-quark mass effects in H+jet and H+2 jets production". In: *JHEP* 03 (2022), p. 096.
[4777] Philip Bambade et al. "The International Linear Collider: A Global Project". In: (Mar. 2019).
[4778] J. Gao. "CEPC and SppC Status — From the completion of CDR towards TDR". In: *Int. J. Mod. Phys. A* 36.22 (2021), p. 2142005.
[4779] A. Abada et al. "FCC-ee: The Lepton Collider: Future Circular Collider Conceptual Design Report Volume 2". In: *Eur. Phys. J. ST* 228.2 (2019), pp. 261–623.
[4780] O. Brunner et al. "The CLIC project". In: (Mar. 2022).

[4781] A. Abada et al. “FCC-hh: The Hadron Collider: Future Circular Collider Conceptual Design Report Volume 3”. In: *Eur. Phys. J. ST* 228.4 (2019), pp. 755–1107.

[4782] Michelangelo L. Mangano, Giacomo Ortona, and Michele Selvaggi. “Measuring the Higgs self-coupling via Higgs-pair production at a 100 TeV p-p collider”. In: *Eur. Phys. J. C* 80.11 (2020), p. 1030.

[4783] J. de Blas et al. “Higgs Boson Studies at Future Particle Colliders”. In: *JHEP* 01 (2020), p. 139.